\pdfoutput=1
\documentclass[12pt,twoside]{article}
\usepackage[utf8]{inputenc}
\usepackage[T1]{fontenc}
\usepackage{amsmath}
\usepackage{amssymb}
\usepackage[nottoc,notlot,notlof]{tocbibind}
\usepackage{calc}
\usepackage[pdftex]{graphicx} % PN ADDED pdftex option, as dvips is the default for this package
\usepackage[figuresright]{rotating}
\usepackage{authblk}
\usepackage{caption}
\usepackage{array}
\usepackage{longtable}
\usepackage{etoolbox}
\usepackage{float}

\usepackage{adjustbox}
\usepackage{tabularx}
\usepackage{multirow}
\usepackage{booktabs}
\usepackage{color, colortbl}
\usepackage{footmisc}
\usepackage{afterpage}
\usepackage{relsize}                                      %SCB for Babar
\usepackage{wasysym}

\usepackage{lineno}  % for line numbering during review

\definecolor{Gray}{gray}{0.9}
\definecolor{LightGray}{gray}{0.97}

\usepackage{hyperref} %PN REMOVED [dvips] option as we have moved to pdflatex              %SCB should come late in the packages http://www.tug.org/applications/hyperref/manual.html
\usepackage[all]{hypcap} % Internal hyperlinks to floats.
\usepackage{cite}    % G. Cibinetto
\usepackage{mciteplus}

\newcommand{\mysection}[1]{\section{\boldmath #1}}
\newcommand{\mysubsection}[1]{\subsection[#1]{\boldmath #1}}
\newcommand{\mysubsubsection}[1]{\subsubsection[#1]{\boldmath #1}}
\newcommand{\mysubsubsubsection}[1]{\subsubsubsection{\boldmath #1}}

\def\dof{{\rm dof}}

\newcommand\VCKM{{V}}
\newcommand\etacpf{{\eta_f}}
\newcommand\etacp{{\eta}}

\renewcommand\Im{{\rm Im}} 
\renewcommand\Re{{\rm Re}}

\newcommand\Abar{\kern 0.18em\overline{\kern -0.18em A}{}}
\newcommand\Af{A_f}
\newcommand\Abarf{\Abar_f}
\newcommand\Afbar{A_{\bar f}}
\newcommand\Abarfbar{\Abar_{\bar f}}
\newcommand\Acp{{\cal A}}
\newcommand\Adirnoncp{\ensuremath{\langle{\cal A}_{f\bar f}\rangle}\xspace}
\newcommand\mc{\multicolumn}

\newcommand {\cbf}{\ensuremath{{\cal B}}}

\newcommand {\vcb}{\ensuremath{|V_{cb}|}}
\newcommand {\vub}{\ensuremath{|V_{ub}|}}

\def\Bp      {\ensuremath{B^{+}}}
\def\Bm      {\ensuremath{B^{-}}}
\def\Bz      {\ensuremath{B^{0}}}
\def\Bs      {\ensuremath{B_{s}}}

\newcommand{\BzbDplnu}    {\ensuremath{\Bzb \to D^{+}\ell^{-}\nub_\ell}}
\newcommand{\BzbDstarlnu} {\ensuremath{\Bzb \to D^{*+}\ell^{-}\nub_\ell}}

\newcommand {\rhoz} {\ensuremath{\rho^0}\hbox{ }}

\usepackage{relsize}
\def\beq{\begin{equation}}
\def\eeq#1{\label{#1}\end{equation}}
\def\eeqn{\end{equation}}

\def\beqa{\begin{eqnarray}}
\def\eeqa#1{\label{#1}\end{eqnarray}}
\def\eeqan{\end{eqnarray}}

\let\bar=\overbar

\def\ie{{\it i.e.}}
\def\eg{{\it e.g.}}
\def\etc{{\it etc.}}
\def\cf{{\it cf.}}

\def\Dslash{\ensuremath{\not{\hbox{\kern-4pt $D$}}}\xspace}
\def\dslash{\not{\hbox{\kern-2pt $\del$}}}

\def\BR{\mbox{\rm BR}}
\def\ee{e^+e^-}

\def\alphas{\alpha_s}
\def\msb{{\bar{\ssstyle M \kern -1pt S}}}

\RequirePackage{xspace}

\usepackage{relsize}
\def\babar{\mbox{\slshape B\kern-0.1em{\smaller A}\kern-0.1em
    B\kern-0.1em{\smaller A\kern-0.2em R}}\xspace}
\def\belle{\mbox{\normalfont Belle}\xspace}

\def\dzero{\mbox{\normalfont D0}\xspace} % TJG 30/4/2012: D0 themselves prefer use of standard ``0'' rather than \zero
\def\lhcb{\mbox{\normalfont LHCb}\xspace}

\def\epem       {\ensuremath{e^+e^-}\xspace}
\def\ee         {\ensuremath{e^-e^-}\xspace}

\def\mup        {\ensuremath{\mu^+}\xspace}
\def\mun        {\ensuremath{\mu^-}\xspace}    % muon negative (\mum is taken)
\def\mumu       {\ensuremath{\mu^+\mu^-}\xspace}
\def\mtau       {\ensuremath{\tau}\xspace}

\def\nub        {\ensuremath{\overline{\nu}}\xspace}

\def\nub        {\ensuremath{\overline{\nu}}\xspace}

\def\nut        {\ensuremath{\nu_\tau}\xspace}

\def\nul        {\ensuremath{\nu_\ell}\xspace}

\def\g     {\ensuremath{\gamma}\xspace}
\def\Z      {\ensuremath{Z^0}\xspace}

\def\ubar  {\ensuremath{\overline u}\xspace}

\def\dbar  {\ensuremath{\overline d}\xspace}
\def\ddbar {\ensuremath{d\overline d}\xspace}

\def\sbar  {\ensuremath{\overline s}\xspace}

\def\b  {\ensuremath{b}\xspace}
\def\bbar  {\ensuremath{\overline b}\xspace}

\def\piz   {\ensuremath{\pi^0}\xspace}

\def\pip   {\ensuremath{\pi^+}\xspace}
\def\pim   {\ensuremath{\pi^-}\xspace}

\def\pipm  {\ensuremath{\pi^\pm}\xspace}
\def\pimp  {\ensuremath{\pi^\mp}\xspace}

\def\etapr {\ensuremath{\eta^{\prime}}\xspace}

\def\Kbar  {\kern 0.2em\overline{\kern -0.2em K}{}\xspace}

\def\Kpm   {\ensuremath{K^\pm}\xspace}
\def\Kmp   {\ensuremath{K^\mp}\xspace}
\def\Kp    {\ensuremath{K^+}\xspace}
\def\Km    {\ensuremath{K^-}\xspace}
\def\KS    {\ensuremath{K^0_{\scriptscriptstyle S}}\xspace} 
\def\KL    {\ensuremath{K^0_{\scriptscriptstyle L}}\xspace} 
\def\Kstarz  {\ensuremath{K^{*0}}\xspace}
\def\Kstarzb  {\ensuremath{\Kbar^{*0}}\xspace}
\def\Kstar   {\ensuremath{K^*}\xspace}

\def\Kstarp   {\ensuremath{K^{*+}}\xspace}

\def\Kstarpm   {\ensuremath{K^{*\pm}}\xspace}

\def\Kz   {\ensuremath{K^0}\xspace}
\def\Kzb   {\ensuremath{\Kbar^0}\xspace}
\def\KzKzb {\ensuremath{K^0 \kern -0.16em \Kzb}\xspace}

\def\KorKstar   {\ensuremath{K^{(*)}}\xspace}

\def\KorKstarp  {\ensuremath{K^{(*)+}}\xspace}

\def\Dz    {\ensuremath{D^0}\xspace}
\def\Dbar  {\kern 0.2em\overline{\kern -0.2em D}{}\xspace}

\def\Dzb   {\ensuremath{\Dbar^0}\xspace}
\def\DzDzb {\ensuremath{D^0 {\kern -0.16em \Dzb}}\xspace}
\def\Dp    {\ensuremath{D^+}\xspace}
\def\Dm    {\ensuremath{D^-}\xspace}

\def\Dmp   {\ensuremath{D^\mp}\xspace}

\def\Dstar   {\ensuremath{D^*}\xspace}

\def\Dstarp  {\ensuremath{D^{*+}}}
\def\Dstarm  {\ensuremath{D^{*-}}}
\def\DorDstar   {\ensuremath{D^{(*)}}\xspace}
\def\DorDstarz  {\ensuremath{D^{(*)0}}\xspace}
\def\DorDstarzb {\ensuremath{\Dbar^{(*)0}}\xspace}

\def\Ds    {\ensuremath{D^+_s}\xspace}
\def\Dsp   {\ensuremath{D^+_s}\xspace}
\def\Dsm   {\ensuremath{D^-_s}\xspace}

\def\Bz    {\ensuremath{B^0}\xspace}
\def\B     {\ensuremath{B}\xspace}
\def\Bbar  {\kern 0.18em\overline{\kern -0.18em B}{}\xspace}
\def\Bb    {\ensuremath{\Bbar}\xspace}
\def\Bzb   {\ensuremath{\Bbar^0}\xspace}
\def\Bu    {\ensuremath{B^+}\xspace}

\def\Bpm   {\ensuremath{B^\pm}\xspace}
\def\Bmp   {\ensuremath{B^\mp}\xspace}
\def\Bs    {\ensuremath{B_s}\xspace}
\def\Bsb   {\ensuremath{\Bbar_s^0}\xspace}
\def\BB    {\ensuremath{B\Bbar}\xspace} 
\def\BzBzb {\ensuremath{B^0 {\kern -0.16em \Bzb}}\xspace}

\def\jpsi  {\ensuremath{{J\mskip -3mu/\mskip -2mu\psi\mskip 2mu}}\xspace}

\mathchardef\Upsilon="7107
\def\Y#1S{\ensuremath{\Upsilon{(#1S)}}\xspace}% no space before {...}!

\mathchardef\Deltares="7101
\mathchardef\Xi="7104
\mathchardef\Lambda="7103
\mathchardef\Sigma="7106
\mathchardef\Omega="710A
\def\Deltabar   {\kern 0.25em\overline{\kern -0.25em \Deltares}{}\xspace}
\def\Lbar {\kern 0.2em\overline{\kern -0.2em\Lambda\kern 0.05em}\kern-0.05em{}\xspace}
\def\Sigbar{\kern 0.2em\overline{\kern -0.2em \Sigma}{}\xspace}
\def\Xibar{\kern 0.2em\overline{\kern -0.2em \Xi}{}\xspace}
\def\Obar{\kern 0.2em\overline{\kern -0.2em \Omega}{}\xspace}
\def\Nbar{\kern 0.2em\overline{\kern -0.2em N}{}\xspace}
\def\Xb{\kern 0.2em\overline{\kern -0.2em X}{}}

\newcommand{\particle}[1]{\ensuremath{#1}\xspace}
\renewcommand{\ee}{\particle{e^+e^-}}
\newcommand{\Ups}{\particle{\Upsilon(4S)}}
\newcommand{\Upsfive}{\particle{\Upsilon(5S)}}
\renewcommand{\b}{\particle{b}}
\renewcommand{\B}{\particle{B}}
\newcommand{\Bd}{\particle{B^0}}
\renewcommand{\Bs}{\particle{B^0_s}}
\renewcommand{\Bu}{\particle{B^+}}
\newcommand{\Bc}{\particle{B^+_c}}
\newcommand{\Bdbar}{\particle{\bar{B}^0}}
\newcommand{\Bsbar}{\particle{\bar{B}^0_s}}
\newcommand{\Lb}{\particle{\Lambda_b^0}}
\newcommand{\Xib}{\particle{\Xi_b}}
\newcommand{\Xibd}{\particle{\Xi_b^-}}
\newcommand{\Xibu}{\particle{\Xi_b^0}}
\newcommand{\Omegab}{\particle{\Omega_b^-}}
\newcommand{\Lc}{\particle{\Lambda_c^+}}

\def\BR{{\ensuremath{\cal B}}}

\def\Btopilnu   {\ensuremath{B \to \pi l\nu}}

\newcommand{\tev}{\ensuremath{\mathrm{Te\kern -0.1em V}}\xspace}
\newcommand{\gev}{\ensuremath{\mathrm{Ge\kern -0.1em V}}\xspace}
\newcommand{\mev}{\ensuremath{\mathrm{Me\kern -0.1em V}}\xspace}
\newcommand{\kev}{\ensuremath{\mathrm{ke\kern -0.1em V}}\xspace}
\newcommand{\ev}{\ensuremath{\mathrm{e\kern -0.1em V}}\xspace}
\newcommand{\gevc}{\ensuremath{{\mathrm{Ge\kern -0.1em V\!/}c}}\xspace}
\newcommand{\mevc}{\ensuremath{{\mathrm{Me\kern -0.1em V\!/}c}}\xspace}
\newcommand{\gevcc}{\ensuremath{{\mathrm{Ge\kern -0.1em V\!/}c^2}}\xspace}
\newcommand{\gevgevcccc}{\ensuremath{{\mathrm{Ge\kern -0.1em V^2\!/}c^4}}\xspace}
\newcommand{\mevcc}{\ensuremath{{\mathrm{Me\kern -0.1em V\!/}c^2}}\xspace}
\def\pb {\ensuremath{\rm \,pb}\xspace}

\def\fb   {\ensuremath{\mbox{\,fb}}\xspace}
\def\invfb   {\ensuremath{\mbox{\,fb}^{-1}}\xspace}
\def\mus  {\ensuremath{\rm \,\mus}\xspace}

\def\ps   {\ensuremath{\rm \,ps}\xspace}

\def\mus        {\ensuremath{\,\mu{\rm s}}\xspace}    %% microsecond
\def\ps         {\ensuremath{{\rm \,ps}}\xspace}  %% picosecond
\def\degrees{\ensuremath{^{\circ}}\xspace}

\def\gsim{{~\raise.15em\hbox{$>$}\kern-.85em
          \lower.35em\hbox{$\sim$}~}\xspace}
\def\lsim{{~\raise.15em\hbox{$<$}\kern-.85em
          \lower.35em\hbox{$\sim$}~}\xspace}

\def\CP                 {\ensuremath{C\!P}\xspace}
\def\CPT                {\ensuremath{C\!PT}\xspace}
\def\ra                 {\ensuremath{\to}\xspace}
\def\pep2{PEP-II}

\newcommand{\chisq}{\ensuremath{\chi^2}\xspace}

\def\rhobar {\ensuremath{\overline{\rho}}\xspace}
\def\etabar {\ensuremath{\overline{\eta}}\xspace}

\def\Vud  {\ensuremath{|V_{ud}|}\xspace}

\def\Vus  {\ensuremath{|V_{us}|}\xspace}

\def\Vub  {\ensuremath{|V_{ub}|}\xspace}
\def\Vcb  {\ensuremath{|V_{cb}|}\xspace}

\def\stwob{\ensuremath{\sin\! 2 \beta   }\xspace}

\def\deltamd{\ensuremath{{\rm \Delta}m_d}\xspace}
\xspace
\newcommand{\fds}{\ensuremath{f_{D_s}}\xspace}

\def\jetset74   {\mbox{\tt Jetset \hspace{-0.5em}7.\hspace{-0.2em}4}}
\newcommand{\aerr}[4]   {\mbox{${{#1}^{+ #2}_{- #3}\pm #4}$}}
\newcommand{\berr}[4]   {\mbox{${{#1}\pm #2^{+ #3}_{- #4}}$}}
\newcommand{\cerr}[3]   {\mbox{${{#1}^{+ #2}_{- #3}}$}}
\newcommand{\aerrsy}[5] {\mbox{${{#1}^{+ #2 + #4}_{- #3 - #5}}$}}

\newcommand{\cerrsyt}[5] {\mbox{${{#1}^{+ #2}_{- #3}\pm{#4}\pm{#5}}$}}
\newcommand{\derrsyt}[5] {\mbox{${{#1}\pm{#2}\pm{#3}^{+ #4}_{- #5}}$}}
\newcommand{\ferrsyt}[5] {\mbox{${{#1}\pm{#2}\pm{#3}\pm{#4}\pm{#5}}$}}
\newcommand{\gerrsyt}[4] {\mbox{${{#1}\pm{#2}\pm{#3}\pm{#4}}$}}
\newcommand{\err}[3]   {\mbox{${{#1}\pm{#2}\pm{#3}}$}}

\newcommand{\nodata}{$$}
\newcommand{\vs}{\mbox{$vs.$}}

\def\sgline{\noalign{\vskip 0.10truecm\hrule\vskip 0.10truecm}}
\def\sglinespt{\noalign{\vskip 0.05truecm\hrule}}
\def\sglinespb{\noalign{\hrule\vskip 0.05truecm}}
\newcommand{\ks}    {\KS}
\newcommand{\kz}    {\Kz}
\newcommand{\kzb}   {\Kzb}

\renewcommand{\mysection}[1]{\section[#1]{#1}} % SCB override the mysection definition for hyperref compatibility

\newcommand\red[1]{{\color{red}#1}}
\newcommand\blue[1]{{\color{blue}#1}}

\newif\ifref
\newif\ifhtml
\reftrue 

\begin{document}

\setcounter{page}{1}
\thispagestyle{empty}
\renewcommand\Affilfont{\itshape\small}

\title{
  Averages of $b$-hadron, $c$-hadron, and $\tau$-lepton properties
  as of summer 2016
\vskip0.20in
\large{\it Heavy Flavor Averaging Group (HFLAV):}
%%\author{set one author to remove the spurious ``immediate''}
\vspace*{-0.20in}}
\author[1]{Y.~Amhis}\affil[1]{LAL, Universit\'{e} Paris-Sud, CNRS/IN2P3, Orsay, France}
\author[2]{Sw.~Banerjee}\affil[2]{University of Louisville, Louisville, Kentucky, USA}
\author[3]{E.~Ben-Haim}\affil[3]{LPNHE, Universit\'{e} Pierre et Marie Curie, Universit\'{e} Paris Diderot, CNRS/IN2P3, Paris, France}
\author[4]{F.~Bernlochner}\affil[4]{University of Bonn, Bonn, Germany}
\author[5]{A.~Bozek}\affil[5]{H. Niewodniczanski Institute of Nuclear Physics, Krak\'{o}w, Poland}
\author[6]{C.~Bozzi}\affil[6]{Universita e INFN, Ferrara, Ferrara, Italy}
\author[5,7]{M.~Chrz\k{a}szcz}\affil[7]{Physik-Institut, Universit\"at Z\"urich, Z\"urich, Switzerland}
\author[4]{J.~Dingfelder}
\author[4]{S.~Duell}
\author[8]{M.~Gersabeck}\affil[8]{School of Physics and Astronomy, University of Manchester, Manchester, UK}
\author[9]{T.~Gershon}\affil[9]{Department of Physics, University of Warwick, Coventry, UK}
\author[10]{D.~Gerstel}\affil[10]{Aix Marseille Univ, CNRS/IN2P3, CPPM, Marseille, France}
\author[11]{P.~Goldenzweig}\affil[11]{Institut f\"{u}r Experimentelle Kernphysik, Karlsruher Institut f\"{u}r Technologie, Karlsruhe, Germany}
\author[12]{R.~Harr}\affil[12]{Wayne State University, Detroit, Michigan, USA}
\author[13]{K.~Hayasaka}\affil[13]{Niigata University, Niigata, Japan}
\author[14]{H.~Hayashii}\affil[14]{Nara Women's University, Nara, Japan}
\author[15]{M.~Kenzie}\affil[15]{Cavendish Laboratory, University of Cambridge, Cambridge, UK}
\author[16]{T.~Kuhr}\affil[16]{Ludwig-Maximilians-University, Munich, Germany}
\author[10]{O.~Leroy}
\author[17,18]{A.~Lusiani}\affil[17]{Scuola Normale Superiore, Pisa, Italy}\affil[18]{Sezione INFN di Pisa, Pisa, Italy}
\author[19]{X.R.~Lyu}\affil[19]{University of Chinese Academy of Sciences, Beijing, China}
\author[13]{K.~Miyabayashi}
\author[20]{P.~Naik}\affil[20]{H.H.~Wills Physics Laboratory, University of Bristol, Bristol, UK}
\author[21]{T.~Nanut}\affil[21]{J. Stefan Institute, Ljubljana, Slovenia}
\author[22]{A.~Oyanguren Campos}\affil[22]{Instituto de Fisica Corpuscular, Centro Mixto Universidad de Valencia - CSIC, Valencia, Spain}
\author[23]{M.~Patel}\affil[23]{Imperial College London, London, UK}
\author[24]{D.~Pedrini} \affil[24]{INFN Sezione di Milano-Bicocca, Milano, Italy}
\author[25]{M.~Petri\v{c}}\affil[25]{European Organization for Nuclear Research (CERN), Switzerland}
\author[18]{M.~Rama}
\author[26]{M.~Roney}\affil[26]{University of Victoria, Victoria, British Columbia, Canada}
\author[27]{M.~Rotondo}\affil[27]{Laboratori Nazionali dell'INFN di Frascati, Frascati, Italy}
\author[28]{O.~Schneider}\affil[28]{Institute of Physics, Ecole Polytechnique F\'{e}d\'{e}rale de Lausanne (EPFL), Lausanne, Switzerland}
\author[29]{C.~Schwanda}\affil[29]{Institute of High Energy Physics, Vienna, Austria}
\author[30]{A.~J.~Schwartz}\affil[30]{University of Cincinnati, Cincinnati, Ohio, USA}
\author[10]{J.~Serrano}
\author[31,32]{B.~Shwartz}\affil[31]{Budker Institute of Nuclear Physics (SB RAS), Novosibirsk, Russia}\affil[32]{Novosibirsk State University, Novosibirsk, Russia}
\author[33]{R.~Tesarek}\affil[33]{Fermi National Accelerator Laboratory, Batavia, Illinois, USA}
\author[18]{D.~Tonelli}
\author[34,35]{K.~Trabelsi}\affil[34]{High Energy Accelerator Research Organization (KEK), Tsukuba, Japan}\affil[35]{SOKENDAI (The Graduate University for Advanced Studies), Hayama, Japan}
\author[36]{P.~Urquijo}\affil[36]{School of Physics, University of Melbourne, Melbourne, Victoria, Australia}
\author[37]{R.~Van Kooten}\affil[37]{Indiana University, Bloomington, Indiana, USA}
\author[38]{J.~Yelton}\affil[38]{University of Florida, Gainesville, Florida, USA}
\author[21,39]{A.~Zupanc}\affil[39]{Faculty of Mathematics and Physics, University of Ljubljana, Ljubljana, Slovenia}

\date{December 21, 2017} % replace \date{\today} at arXiv submission time
\maketitle

\begin{abstract}
\noindent
This article reports world averages of measurements of $b$-hadron, $c$-hadron,
and $\tau$-lepton properties obtained by the Heavy Flavor Averaging Group using results available through summer 2016. 
For the averaging, common input parameters used in the various analyses are adjusted (rescaled) to common values, and known correlations are taken into account.
The averages include branching fractions, lifetimes, neutral meson mixing
parameters, \CP~violation parameters, parameters of semileptonic decays, and 
Cabbibo-Kobayashi-Maskawa matrix elements.
\end{abstract}

\newpage
\tableofcontents
\newpage

%% Uncomment during review phase. 
%% Comment before a final submission.
%% \linenumbers

% Introduction
%\documentclass[12pt]{article}
%\begin{document}

\mysection{Introduction}
\label{sec:intro}

Flavor dynamics plays an important role in elementary particle interactions. 
%is an important element in understanding the nature of particle physics.  
The accurate knowledge of properties of heavy flavor
hadrons, especially $b$ hadrons, plays an essential role for
determining the elements of the Cabibbo-Kobayashi-Maskawa (CKM)
quark-mixing matrix~\cite{Cabibbo:1963yz,Kobayashi:1973fv}. 
The operation of the \belle\ and \babar\ $e^+e^-$ $B$ factory 
experiments led to a large increase in the size of available 
$B$-meson, $D$-hadron and $\tau$-lepton samples, 
enabling dramatic improvement in the accuracies of related measurements.
The CDF and \dzero\ experiments at the Fermilab Tevatron 
have also provided important results in heavy flavour physics,
most notably in the $B^0_s$ sector.
In the $D$-meson sector, the dedicated $e^+e^-$ charm factory experiments
CLEO-c and BESIII have made significant contributions.
Run~I of the CERN Large Hadron Collider delivered high luminosity, 
enabling the collection of even larger samples of $b$ and $c$ hadrons, and
thus a further leap in precision in many areas, at the ATLAS, CMS, and
(especially) LHCb experiments.  
With the LHC Run~II ongoing, further improvements are keenly anticipated.

The Heavy Flavor Averaging Group (HFLAV)\footnote{
  The group was originally known by the acronym ``HFAG.''  
  Following feedback from the community, this was changed to HFLAV in 2017.
} 
was formed in 2002 to 
continue the activities of the LEP Heavy Flavor Steering 
Group~\cite{Abbaneo:2000ej_mod,*Abbaneo:2001bv_mod_cont}. 
This group was responsible for calculating averages of measurements of $b$-flavor related quantities. 
HFLAV has evolved since its inception and currently consists of seven subgroups:
\begin{itemize}
\item the ``$B$ Lifetime and Oscillations'' subgroup provides 
averages for $b$-hadron lifetimes, $b$-hadron fractions in 
$\Upsilon(4S)$ decay and $pp$ or $p\bar{p}$ collisions, and various 
parameters governing $\Bz$--$\Bzb$ and $\Bs$--$\Bsb$ mixing;

\item the ``Unitarity Triangle Parameters'' subgroup provides
averages for parameters associated with time-dependent $\CP$ 
asymmetries and $B \to DK$ decays, and resulting determinations 
of the angles of the CKM unitarity triangle;

\item the ``Semileptonic $B$ Decays'' subgroup provides averages
for inclusive and exclusive $B$-decay branching fractions, and
subsequent determinations of the CKM matrix element magnitudes
$|V_{cb}|$ and $|V_{ub}|$;

\item the ``$B$ to Charm Decays'' subgroup provides averages of 
branching fractions for $B$ decays to final states involving open 
charm or charmonium mesons;

\item the ``Rare Decays'' subgroup provides averages of branching 
fractions and $\CP$ asymmetries for charmless, radiative, 
leptonic, and baryonic $B$-meson and \b-baryon decays;

\item the ``Charm Physics'' subgroup provides averages of numerous 
quantities in the charm sector, including branching fractions; 
properties of excited $D^{**}$ and $D^{}_{sJ}$ mesons; 
properties of charm baryons;
$\Dz$--$\Dzb$ mixing, $\CP$, and $T$ violation parameters;
and $D^+$ and $D^+_s$ decay constants $f^{}_{D}$ and~$f^{}_{D_s}$.

\item the ``Tau Physics'' subgroup provides averages for \mtau
  branching fractions using a global fit and elaborates the results
  to test lepton universality and to determine the CKM matrix element
  magnitude $|V_{us}|$; furthermore, it lists the \mtau lepton-flavor-violating
  upper limits and computes the combined upper limits.

\end{itemize}
%The ``Lifetime and Oscillations'' and ``Semileptonic'' subgroups were 
%formed from the merger of four LEP working groups.
%% with some reorganization, \ie\ merging four groups into two. 
%The ``Unitary Triangle,'' ``$B$ to Charm Decays,'' and ``Rare Decays''
%subgroups were formed to provide averages for new results obtained
%from the $B$ factory experiments (and now also from the Fermilab 
%Tevatron and CERN LHC experiments).
%The ``Charm'' and ``Tau''  subgroups were formed more recently in 
%response to the wealth of new data concerning $D$ and $\tau$ physics. 
Subgroups consist of representatives from experiments producing 
relevant results in that area, \ie, representatives from
\babar, \belle, BESIII, CDF, CLEO(c), \dzero, and LHCb.

This article is an update of the last HFLAV preprint, which used results available by summer 2014~\cite{Amhis:2014hma}. 
% Previous HFLAV reports were available with results up to summer
% 2012~\cite{Amhis:2012bh}, ...
Here we report world averages using results available by summer 2016.
In some cases, important new results made available in the latter part of 2016 have been included, or there have been minor revisions in the averages since summer 2016.
All plots carry a timestamp indicating when they were produced.
In general, we use all publicly available results that are supported by
written documentation, including preliminary results presented at conferences
or workshops.
However, we do not use preliminary results that remain unpublished 
for an extended period of time, or for which no publication is planned. 
Close contacts have been established between representatives from the
experiments and members of subgroups that perform averaging to ensure that the
data are prepared in a form suitable for combinations.  

Section~\ref{sec:method} describes the methodology used for calculating
averages. In the averaging procedure, common input parameters used in 
the various analyses are adjusted (rescaled) to common values, and, 
where possible, known correlations are taken into account. 
Sections~\ref{sec:life_mix}--\ref{sec:tau} present world 
average values from each of the subgroups listed above. 
A brief 
summary of the averages presented is given in Section~\ref{sec:summary}.   
A complete listing of the averages and plots,
including updates since this document was prepared,
are also available on the HFLAV web site:
\vskip0.15in\hskip0.75in
\vbox{
  \href{http://www.slac.stanford.edu/xorg/hflav}{\tt http://www.slac.stanford.edu/xorg/hflav} 
}

\clearpage
% Methodology
\section{Averaging methodology} 
\label{sec:method} 

The main task of HFLAV is to combine independent but possibly
correlated measurements of a parameter to obtain the world's 
best estimate of that parameter's value and uncertainty. These
measurements are typically made by different experiments, or by the
same experiment using different data sets, or sometimes by the same
experiment using the same data but using different analysis methods.
In this section, the general approach adopted by HFLAV is outlined.
For some cases, somewhat simplified or more complex algorithms are 
used; these are noted in the corresponding sections. 
% Some examples for extensions of the standard method for extracting
% averages are given here. These include the case where measurement errors
% depend on the measured value, \ie\ are relative errors, unknown
% correlation coefficients and the breakdown of error sources.
%The methodology described below in Sec.~\ref{sec:method:corrSysts} focuses on
%the problems of combining measurements performed with different systematic
%assumptions and with potentially correlated systematic uncertainties. 
%Our methodology relies on the close involvement, in the averaging process, 
%of the people performing the measurements.

Our methodology focuses on the problem of combining measurements 
obtained with different assumptions about external (or ``nuisance'') 
parameters and with potentially correlated systematic uncertainties.
%It is easily extended to the case where multiple
%parameters are simultaneously averaged, taking account of correlations.
%Such simultaneous averages are performed for quantities where correlations are
%known to be important.
It is important for any averaging procedure that the quantities
measured by experiments be statistically well-behaved, which in this 
context means having a (one- or multi-dimensional) Gaussian likelihood
function that is described by the central value(s) $\boldsymbol{x}_i$ 
and covariance matrix $\boldsymbol{V}_{\!i}$.
In what follows we assume $\boldsymbol{x}$ does not 
contain redundant information, \ie, if it contains $n$ 
elements then $n$ is the number of parameters being determined.
%also the number of degrees of freedom (\dof).
A $\chi^2$ statistic is constructed as
%obtained by minimising  
\begin{equation}
  \chi^2(\boldsymbol{x}) = \sum_i^N 
  \left( \boldsymbol{x}_i - \boldsymbol{x} \right)^{\rm T} 
  \boldsymbol{V}_{\!i}^{-1}  
  \left( \boldsymbol{x}_i - \boldsymbol{x} \right) \, ,
\end{equation}
where the sum is over the $N$ independent determinations of the quantities
$\boldsymbol{x}$. These are typically from different experiments; possible
correlations of the systematic uncertainties are discussed below.
%\eg, $i$ runs over the different measurements.
The results of the average are the central values $\boldsymbol{\hat{x}}$, 
which are the values of $\boldsymbol{x}$ at the minimum of
$\chi^2(\boldsymbol{x})$, and their covariance matrix
\begin{equation}
  \boldsymbol{\hat{V}}^{-1} = \sum_i^N \boldsymbol{V}_{\!i}^{-1}  \, .
\end{equation}
We report the covariance matrices or the correlation matrices derived from
the averages whenever possible. 
In some cases where the matrices are large, it is inconvenient to report them
in this document; however, all results can be found on the HFLAV web pages. 
%Some discussion of cases where the likelihood function is non-Gaussian
%can be found in Sec.~\ref{sec:method:nonGaussian}.

The value of $\chi^2(\boldsymbol{\hat{x}})$ provides a measure of the
consistency of the independent measurements of $\boldsymbol{x}$ after
accounting for the number of degrees of freedom ($\dof$), which is the 
difference between the number of measurements and the number of
fitted parameters: $N\cdot n - n$.
% \ie\ $(N-1)\times n$.
The values of $\chi^2(\boldsymbol{\hat{x}})$ and $\dof$ are typically 
converted to a confidence level (C.L.) and reported together with the 
averages. In cases where $\chi^2/\dof > 1$, 
we do not usually scale the resulting uncertainty, in contrast
to what is done by the Particle Data Group~\cite{PDG_2016}.
Rather, we examine the systematic uncertainties of each measurement 
to better understand them. Unless we find systematic discrepancies 
among the measurements, we do not apply any additional correction 
to the calculated error. 
% We provide the confidence level of the fit as an indicator for the 
% consistency of the measurements included in the average. 
If special treatment is necessary to calculate an average, or 
if an approximation used in the calculation might not be sufficiently
accurate (\eg, assuming Gaussian errors when the likelihood function 
exhibits non-Gaussian behavior), we include a warning message. 
Further modifications to the averaging procedures for non-Gaussian
situations are discussed in Sec.~\ref{sec:method:nonGaussian}.

For observables such as branching fractions, experiments typically 
report upper limits when the signal is not significant.  
Sometimes there is insufficient information available to combine 
upper limits on a parameter obtained by different experiments;
in this case we usually report 
%which may be obtained using different statistical approaches. 
%Therefore, we usually follow the convention of reporting 
only the most restrictive upper limit. 
For branching fractions of lepton-flavor-violating decays of 
tau leptons, we calculate combined upper limits as discussed
in Sec.~\ref{sec:tau:lfv-combs}.
%The reported upper limits are at 90\% C.L. unless stated otherwise.

\subsection{Treatment of correlated systematic uncertainties}
\label{sec:method:corrSysts} 

Consider two hypothetical measurements of a parameter $x$, which can
be summarized as
\begin{align*}
 & x_1 \pm \delta x_1 \pm \Delta x_{1,1} \pm \Delta x_{1,2} \ldots \\
 & x_2 \pm \delta x_2 \pm \Delta x_{2,1} \pm \Delta x_{2,2} \ldots \, ,
\end{align*}
where the $\delta x_k$ are statistical uncertainties and
the $\Delta x_{k,i}$ are contributions to the systematic
uncertainty. The simplest approach is to combine statistical 
and systematic uncertainties in quadrature:
\begin{align*}
 & x_1 \pm \left(\delta x_1 \oplus \Delta x_{1,1} \oplus \Delta x_{1,2} \oplus \ldots\right) \\
 & x_2 \pm \left(\delta x_2 \oplus \Delta x_{2,1} \oplus \Delta x_{2,2} \oplus \ldots\right) \, ,
\end{align*}
and then perform a weighted average of $x_1$ and $x_2$ using their
combined uncertainties, treating the measurements as independent. This 
approach suffers from two potential problems that we try to address. 
First, the values $x_k$ may have been obtained using different
assumptions for nuisance parameters; \eg, different values of the \Bz
lifetime may have been used for different measurements of the
oscillation frequency $\deltamd$. The second potential problem 
is that some systematic uncertainties may be correlated
between measurements. For example, different measurements of 
$\deltamd$ may depend on the same branching fraction 
used to model a common background.

The above two problems are related, as any quantity $y_i$
upon which $x_k$ depends gives a contribution $\Delta x_{k,i}$ to the
systematic error that reflects the uncertainty $\Delta y_i$ on $y_i$. 
We thus use the values of $y_i$ and
$\Delta y_i$ assumed by each measurement in our averaging (we refer 
to these values as $y_{k,i}$ and $\Delta y_{k,i}$). 
To properly treat correlated systematic uncertainties among measurements
requires decomposing the overall systematic uncertainties into correlated
and uncorrelated components.
%Furthermore, since we do not lump all the systematics together,
%we require that each measurement used in an average have a consistent
%definition of the various contributions to the systematic uncertainty.
As different measurements often quote different types of systematic
uncertainties, achieving consistent definitions in order to properly 
treat correlations
%for any potentially correlated contributions 
requires close coordination between HFLAV and the experiments. 
In some cases, a group of
systematic uncertainties must be combined into a coarser
description in order to obtain an average that is consistent 
among measurements. Systematic uncertainties
that are uncorrelated with any other source of uncertainty are 
combined together with the statistical error, so that the only
systematic uncertainties treated explicitly are those that are
correlated with at least one other measurement via a consistently-defined
external parameter $y_i$. When asymmetric statistical or systematic
uncertainties are quoted by experiments, we symmetrize them since our 
combination method implicitly assumes Gaussian likelihoods 
(or parabolic log likelihoods) for each measurement.

\begin{figure}[!tb]
\centering
\includegraphics[width=5.0in]{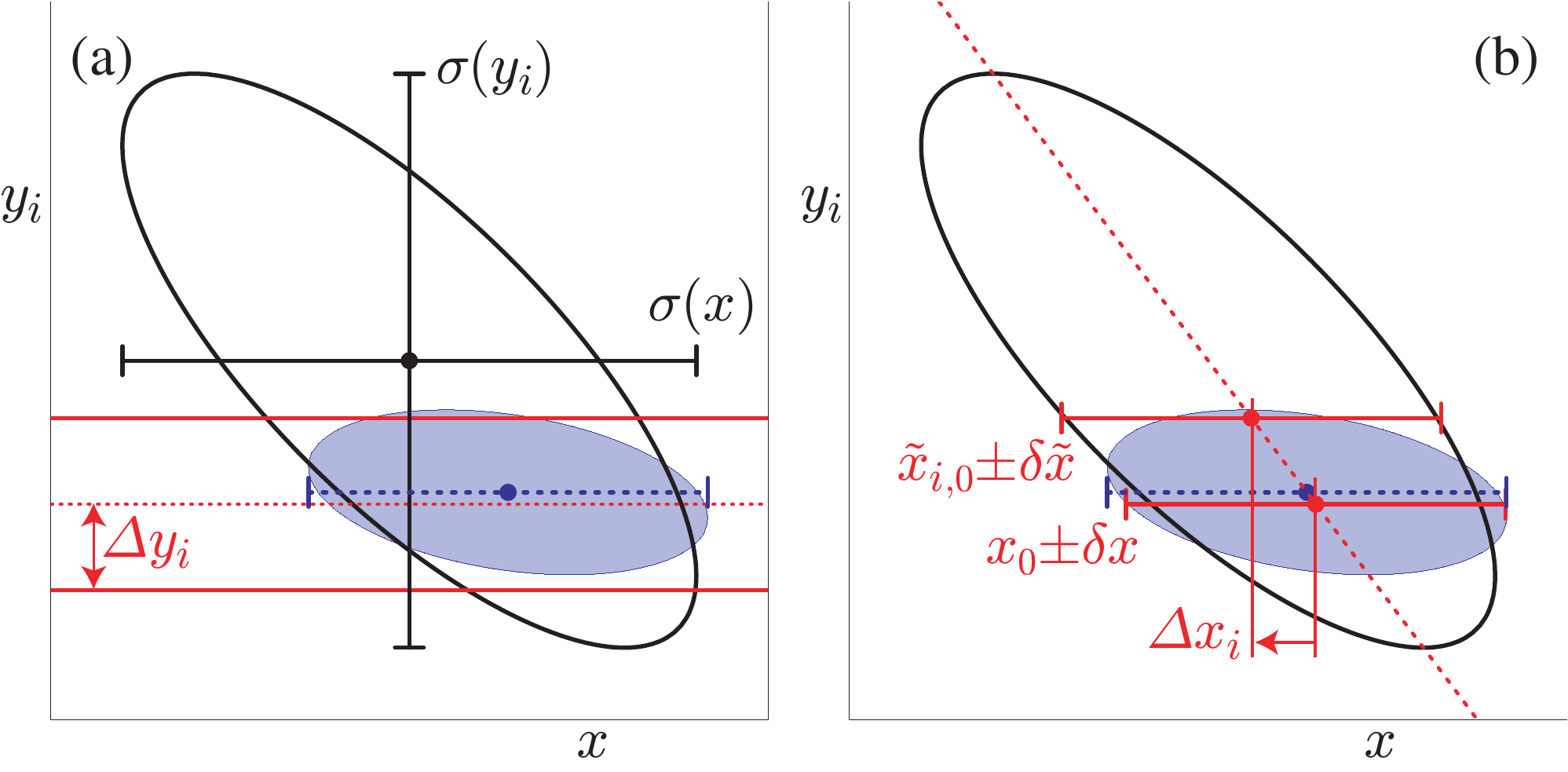}
\caption{
  Illustration of the possible dependence of a measured quantity $x$ on a
  nuisance parameter $y_i$.
  The left-hand plot (a) compares the 68\% confidence level contours of a
  hypothetical measurement's unconstrained (large ellipse) and
  constrained (filled ellipse) likelihoods, using the Gaussian
  constraint on $y_i$ represented by the horizontal band. 
  The solid error bars represent the statistical uncertainties $\sigma(x)$ and
  $\sigma(y_i)$ of the unconstrained likelihood. 
  The dashed error bar shows the statistical error on $x$ from a
  constrained simultaneous fit to $x$ and $y_i$. 
  The right-hand plot (b) illustrates the method described in the text of
  performing fits to $x$ with $y_i$ fixed at different values. 
  The dashed diagonal line between these fit results has the slope
  $\rho(x,y_i)\sigma(y_i)/\sigma(x)$ in the limit of an unconstrained
  parabolic log likelihood. 
  The result of the constrained simultaneous fit from (a) is shown as a dashed
  error bar on $x$.
}
\label{fig:singlefit}
\end{figure}

The fact that a measurement of $x$ is sensitive to $y_i$
indicates that, in principle, the data used to measure $x$ could
also be used for a simultaneous measurement of $x$ and $y_i$. This
is illustrated by the large contour in Fig.~\ref{fig:singlefit}(a).
% for a hypothetical measurement. 
However, there often exists an external constraint $\Delta y_i$ 
on $y_i$ (represented by the horizontal band in
Fig.~\ref{fig:singlefit}(a)) that is more precise than the constraint
$\sigma(y_i)$ from the $x$ data alone. In this case one can perform 
a simultaneous fit to $x$ and $y_i$, including the external 
constraint, and obtain the filled $(x,y)$ contour and dashed 
one-dimensional estimate of $x$ shown in Fig.~\ref{fig:singlefit}(a). 
For this procedure one usually takes the external constraint 
$\Delta y_i$ to be Gaussian.

When the external constraints $\Delta y_i$ are significantly more
precise than the sensitivity $\sigma(y_i)$ of the data alone, 
the additional complexity of a constrained fit with extra free 
parameters may not be justified by the resulting increase in
sensitivity. In this case the usual procedure 
%adopted by the experiments
is to perform a baseline fit with all $y_i$ fixed
to nominal values $y_{i,0}$, obtaining $x = x_0 \pm \delta x$. 
This baseline fit neglects the uncertainty due to $\Delta y_i$, but
this error is subsequently recovered by repeating the fit separately 
for each external parameter $y_i$, with its value fixed to 
$y_i = y_{i,0}\pm \Delta y_i$. This gives the result
$x = \tilde{x}_{0,i} \pm \delta\tilde{x}$ as
illustrated in Fig.~\ref{fig:singlefit}(b). The shift
%$|\tilde{x}_{i,0} - x_0|$ in the central value is 
in the central value $\Delta x_i = \tilde{x}_{0,i} - x_0$ is 
%what the experiments 
usually quoted as the systematic uncertainty due 
to the unknown value of $y_i$. 
%This procedure requires that one know not only the magnitude 
%of this shift but also its sign.
If the unconstrained data can be represented by a Gaussian
likelihood function, the shift will equal
\begin{equation}
\Delta x_i = \rho(x,y_i)\frac{\sigma(x)}{\sigma(y_i)}\,\Delta y_i \,,
\end{equation}
where $\sigma(x)$ and $\rho(x,y_i)$ are the statistical uncertainty on
$x$ and the correlation between $x$ and $y_i$ in the unconstrained data,
respectively. 
This procedure gives very similar results to that of the 
constrained fit with extra parameters: 
%it yields (in the limit of an unconstrained  parabolic log likelihood) 
the central values $x_0$ agree to ${\cal O}(\Delta y_i/\sigma(y_i))^2$, 
and the uncertainties $\delta x \oplus \Delta x_i$ agree to 
${\cal O}(\Delta y_i/\sigma(y_i))^4$.

% commented out by AJS 6/1/17 - this seems to be 
% the same procedure as method #1 above

%Another approach, frequently adopted by experiments, is to allow $y_i$ to vary
%within its uncertainty $\Delta y_i$ in the fit, by means of a Gaussian
%constraint added to the likelihood function.  
%In this approach, the resulting uncertainty on the measured parameter 
%$\Delta x_i$ is absorbed into the statistical uncertainty $\delta x_i$.
%In case this source of systematic uncertainty is shared between experiments,
%its proper treatment again requires that experiments provide information about
%the magnitude and sign of $\Delta x_i$.

To combine two or more measurements that share systematic
uncertainty due to the same external parameter(s) $y_i$, we try 
to perform a constrained simultaneous fit of all measurements 
to obtain values of $x$ and $y_i$.
%, being careful to only apply the constraint on each $y_i$ once. 
%This is usually not practical since we generally do not have
When this is not practical, \eg\ if we do not have sufficient 
information to reconstruct the %unconstrained 
likelihoods corresponding to each measurement, we perform 
the two-step approximate procedure described below.

\begin{figure}[!tb]
\centering
\includegraphics[width=5.0in]{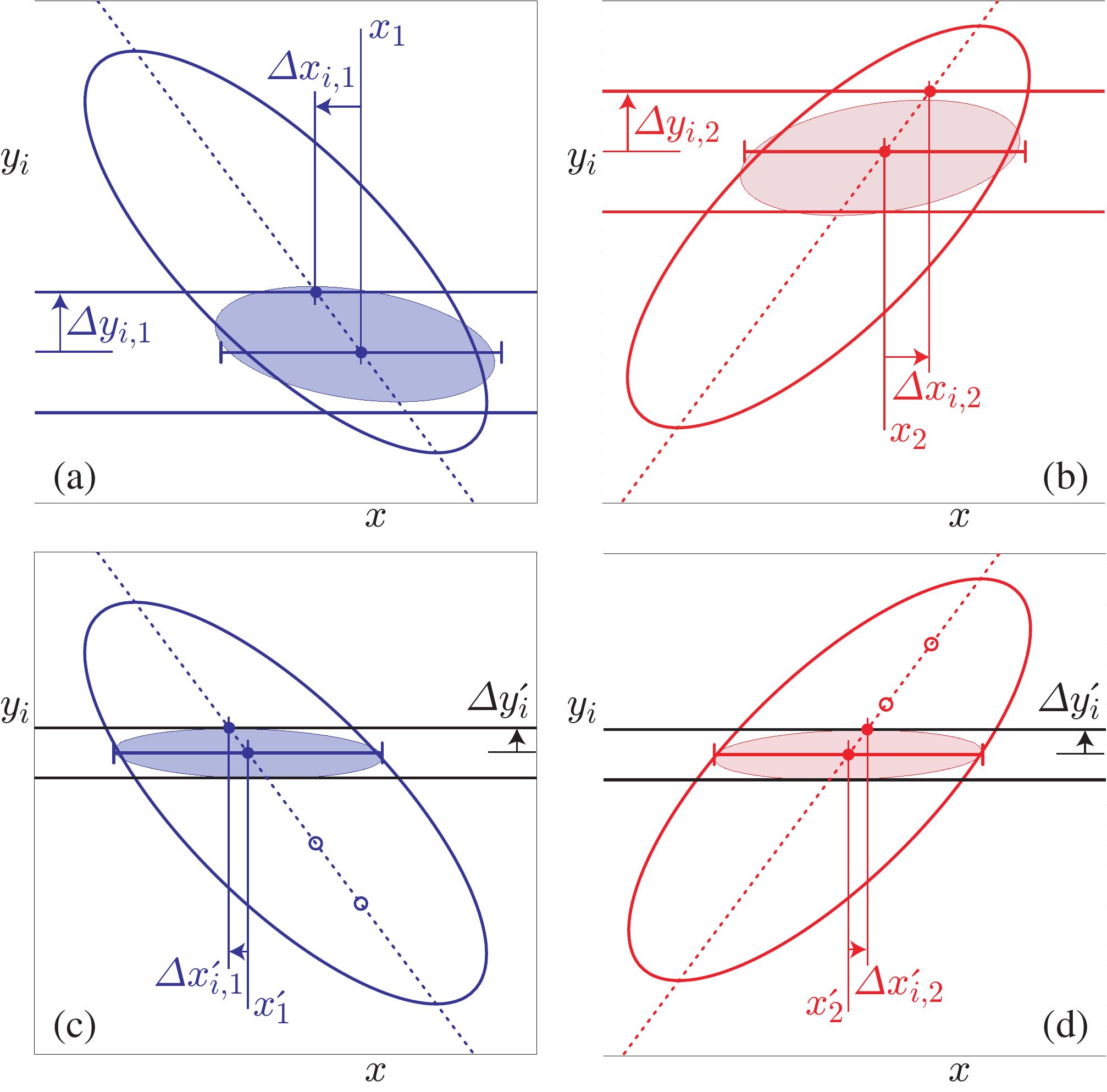}
\caption{
  Illustration of the HFLAV combination procedure for correlated systematic uncertainties.
  Upper plots (a) and (b) show examples of two individual measurements to be
  combined. 
  The large (filled) ellipses represent their unconstrained (constrained)
  likelihoods, while horizontal bands indicate the different assumptions about
  the value and uncertainty of $y_i$ used by each measurement. 
  The error bars show the results of the method described in the text for
  obtaining $x$ by performing fits with $y_i$ fixed to different values. 
  Lower plots (c) and (d) illustrate the adjustments to accommodate updated
  and consistent knowledge of $y_i$. 
  Open circles mark the central values of the unadjusted fits to $x$ with $y$
  fixed; these determine the dashed line used to obtain the adjusted values. 
}
\label{fig:multifit}
\end{figure}

Consider two statistically-independent measurements, 
$x_1 \pm (\delta x_1 \oplus \Delta x_{1,i})$ and 
$x_2\pm(\delta x_2\oplus \Delta x_{2,i})$, of 
the quantity $x$ as shown in Figs.~\ref{fig:multifit}(a,b).
For simplicity we consider only one correlated systematic 
uncertainty for each external parameter $y_i$.
As our knowledge of the $y_i$ improves, 
the measurements of $x$ will shift to different central
values and uncertainties. The first step of our procedure is 
to adjust the values of each measurement to reflect the current 
best knowledge of the external parameters $y_i'$ and their
ranges $\Delta y_i'$, as illustrated in Figs.~\ref{fig:multifit}(c,d). 
We adjust the central values $x_k$ and correlated systematic uncertainties
$\Delta x_{k,i}$ linearly for each measurement (indexed by $k$) and each
external parameter (indexed by $i$):
\begin{align}
x_k' &= x_k + \sum_i\,\frac{\Delta x_{k,i}}{\Delta y_{k,i}}\left(y_i'-y_{k,i}\right)\\
\Delta x_{k,i}'&= \Delta x_{k,i} \frac{\Delta y_i'}{\Delta y_{k,i}} \, .
\end{align}
This procedure is exact in the limit that the unconstrained
likelihood of each measurement is Gaussian.

\begin{figure}[!tb]
\centering
\includegraphics[width=3.0in]{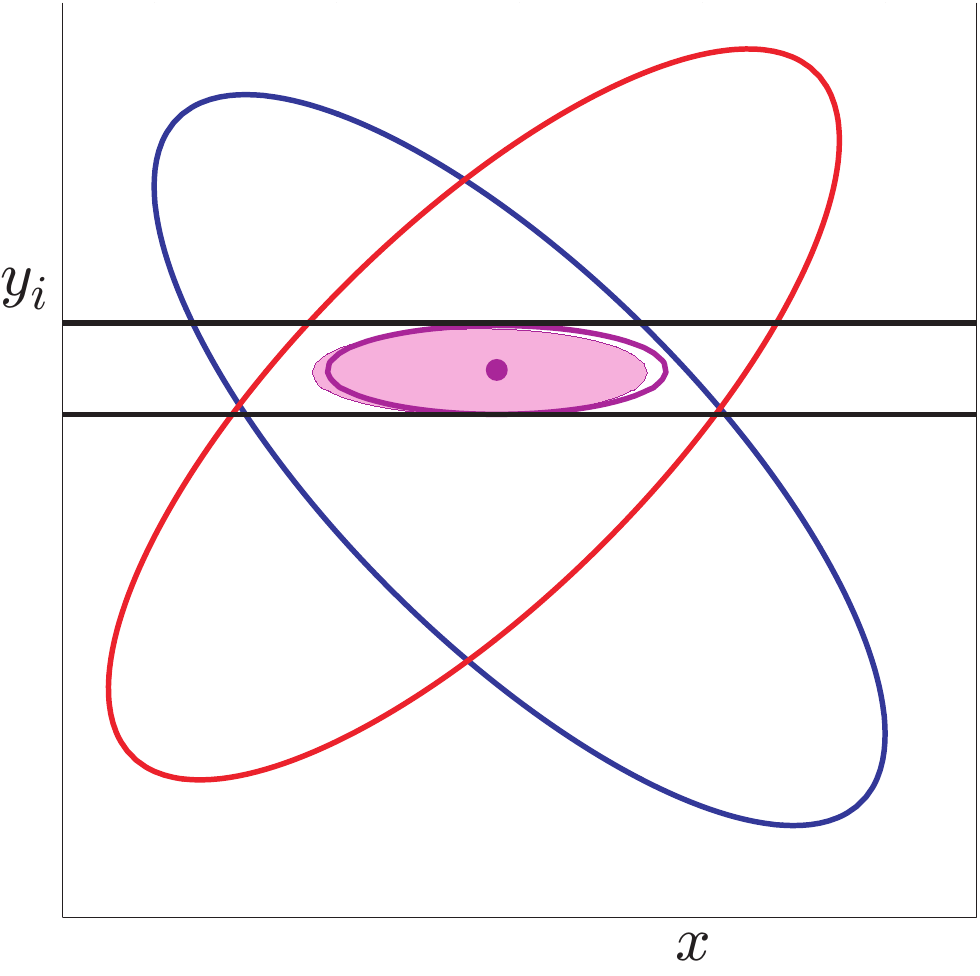}
\caption{
 Illustration of the combination of two hypothetical measurements of $x$
 using the method described in the text. 
 The ellipses represent the unconstrained likelihoods of each measurement,
 and the horizontal band represents the latest knowledge about $y_i$ that is
 used to adjust the individual measurements.
 The filled small ellipse shows the result of the exact method using 
 ${\cal L}_{\text{comb}}$, and the hollow small ellipse and dot show the
 result of the approximate method using $\chi^2_{\text{comb}}$. 
}
\label{fig:fit12}
\end{figure}

The second step is to combine the adjusted
measurements, $x_k'\pm (\delta x_k\oplus \Delta x_{k,1}'\oplus \Delta
x_{k,2}'\oplus\ldots)$ by constructing the goodness-of-fit statistic
\begin{equation}
\chi^2_{\text{comb}}(x,y_1,y_2,\ldots) \equiv \sum_k\,
\frac{1}{\delta x_k^2}\left[
x_k' - \left(x + \sum_i\,(y_i-y_i')\frac{\Delta x_{k,i}'}{\Delta y_i'}\right)
\right]^2 + \sum_i\,
\left(\frac{y_i - y_i'}{\Delta y_i'}\right)^2 \; .
\end{equation}
We minimize this $\chi^2$ to obtain the best values of $x$ and
$y_i$ and their uncertainties, as shown in Fig.~\ref{fig:fit12}. 
Although this method determines new values for
the $y_i$, we typically do not report them.
% since the $\Delta x_{i,k}$ reported
%by each experiment are generally not intended for this purpose (for
%example, they may represent a conservative upper limit rather than a
%true reflection of a 68\% confidence level).

For comparison, the exact method we perform if the 
unconstrained likelihoods ${\cal L}_k(x,y_1,y_2,\ldots)$ are
available is to minimize the simultaneous likelihood
\begin{equation}
{\cal L}_{\text{comb}}(x,y_1,y_2,\ldots) \equiv \prod_k\,{\cal
  L}_k(x,y_1,y_2,\ldots)\,\prod_{i}\,{\cal 
  L}_i(y_i) \; ,
\end{equation}
with an independent Gaussian constraint for each $y_i$:
\begin{equation}
{\cal L}_i(y_i) = \exp\left[-\frac{1}{2}\,\left(\frac{y_i-y_i'}{\Delta
 y_i'}\right)^2\right] \; .
\end{equation}
The results of this exact method 
%are illustrated by the filled ellipses in Figs.~\ref{fig:fit12}(a,b) and 
agree with those of the approximate method when the ${\cal L}_k$ are 
Gaussian and $\Delta y_i' \ll \sigma(y_i)$. 
If the likelihoods are non-Gaussian,,
%In the case of an unconstrained  non-parabolic log likelihood, 
experiments need to provide ${\cal L}_k$ in order to perform
a combination.
%itself to allow an improved combination. 
If $\sigma(y_i)\approx \Delta y_i'$, experiments are encouraged 
to perform a simultaneous measurement of $x$ and $y_i$ so that their 
data will improve the world knowledge of~$y_i$. 

For averages where common sources of systematic uncertainty are important,
central values and uncertainties are rescaled to a common set of input 
parameters following the prescription above.
We use the most up-to-date values for common inputs, consistently across
subgroups, taking values from within HFLAV or from the Particle Data Group
when possible.
%from external groups such as the PDG otherwise.
The parameters and values used are listed in each subgroup section.

\subsection{Treatment of non-Gaussian likelihood functions}
\label{sec:method:nonGaussian} 

For measurements with Gaussian errors, the usual estimator for the
average of a set of measurements is obtained by minimizing
\begin{equation}
  \chi^2(x) = \sum_k^N \frac{\left(x_k-x\right)^2}{\sigma^2_k} \, ,
\label{eq:chi2t}
\end{equation}
where $x_k$ is the $k$-th measured value of $x$ and $\sigma_k^2$ is the
variance of the distribution from which $x_k$ was drawn.  
%Here, a one-dimensional problem is considered for simplicity.
The value $\hat{x}$ at minimum $\chi^2$ is the estimate for the parameter $x$.
The true $\sigma_k$ are unknown but typically the error as assigned by the
experiment $\sigma_k^{\rm raw}$ is used as an estimator for it.
However, caution is advised when $\sigma_k^{\rm raw}$
depends on the measured value $x_k$. 
Examples of this are multiplicative systematic uncertainties such as those
due to acceptance, or the $\sqrt{N}$
%include an uncertainty in any multiplicative factor (like
%an acceptance) that enters the determination of $x_i$, \ie\ the $\sqrt{N}$
dependence of Poisson statistics for which $x_k \propto N$
and $\sigma_k \propto \sqrt{N}$.
Failing to account for this type of dependence when averaging leads to a
biased average. Such biases can be avoided by minimizing
%Biases in the average can be avoided (or at least reduced) by minimizing
\begin{equation}
  \chi^2(x) = \sum_k^N \frac{\left(x_k-x\right)^2}{\sigma^2_k(\hat{x})} \,,
\label{eq:chi2that}
\end{equation}
where $\sigma_k(\hat{x})$ is the uncertainty on $x_k$ that includes 
the dependence of the uncertainty on the value measured.  As an example, 
consider the error due to acceptance for which
$\sigma_k(\hat{x}) = (\hat{x} / x_k)\times\sigma_k^{\rm raw}$.
Inserting this into Eq.~(\ref{eq:chi2that}) leads to
$$ 
\hat{x} = \frac{\sum_k^N x_k^3/(\sigma_k^{\rm raw})^2}
{\sum_k^N x_k^2/(\sigma_k^{\rm raw})^2} \, ,
$$
which is the correct behavior, \ie, weighting by the inverse 
square of the fractional uncertainty $\sigma_k^{\rm raw}/x_k$.
It is sometimes difficult to assess the dependence of $\sigma_k^{\rm raw}$ 
on $\hat{x}$ from the errors quoted by the experiments.  

%An example is the
%uncertainty on a branching fraction, $\mathcal{B}=(N-B)/\eff$, due to
%a change in the background modeling.   
%As a result, the sensitivity
%to different assumptions on these dependences has been
%studied for the averages given in this section.

Another issue that needs careful treatment is that of correlations
among measurements, \eg, due to using the same decay model for 
intermediate states to calculate acceptances.
A common practice is to set the correlation
coefficient to unity to indicate full correlation. However, this is
not necessarily conservative and can result in 
%'' thing to do, and can in fact lead to a significantly
underestimated uncertainty on the average.  
%In the absence of better information, 
The most conservative choice of correlation coefficient
between two measurements $i$ and $j$
is that which maximizes the uncertainty on $\hat{x}$
due to the pair of measurements,
\begin{equation}
\sigma_{\hat{x}(i,j)}^2 = \frac{\sigma_i^2\,\sigma_j^2\,(1-\rho_{ij}^2)}
   {\sigma_i^2 + \sigma_j^2 - 2\,\rho_{ij}\,\sigma_i\,\sigma_j} \, ,
\label{eq:correlij}
\end{equation}
namely
\begin{equation}
\rho_{ij} =
\mathrm{min}\left(\frac{\sigma_i}{\sigma_j},\frac{\sigma_j}{\sigma_i}\right)
\, .
\label{eq:correlrho}
\end{equation}
This corresponds to setting 
$\sigma_{\hat{x}(i,j)}^2=\mathrm{min}(\sigma_i^2,\sigma_j^2)$.
Setting $\rho_{ij}=1$ when $\sigma_i\ne\sigma_j$ can lead to a significant
underestimate of the uncertainty on $\hat{x}$, as can be seen
from Eq.~(\ref{eq:correlij}).

Finally, we carefully consider the various errors
contributing to the overall uncertainty of an average. The 
covariance matrix describing the uncertainties of different 
measurements and their correlations is constructed, \ie,
$\boldsymbol{V} = 
\boldsymbol{V}_{\rm stat} + \boldsymbol{V}_{\rm sys} + 
\boldsymbol{V}_{\rm theory}$.
If the measurements are from independent data samples, then
$\boldsymbol{V}_{\rm stat}$ is diagonal, but
$\boldsymbol{V}_{\rm sys}$ and $\boldsymbol{V}_{\rm theory}$ 
may contain correlations.
The variance on the average $\hat{x}$ can be written
\begin{eqnarray}
\sigma^2_{\hat{x}} 
 &=& 
\frac{ \sum_{i,j}
  \left(\boldsymbol{V}^{-1}\, 
    \left[ \boldsymbol{V}_{\rm stat}+ \boldsymbol{V}_{\rm sys}+
      \boldsymbol{V}_{\rm theory} \right] \, \boldsymbol{V}^{-1}\right)_{ij} }
{\left(\sum_{i,j} \boldsymbol{V}^{-1}_{ij}\right)^2}
\ =\ \sigma^2_{\text{stat}} + \sigma^2_{\text{sys}} + \sigma^2_{\text{th}} \, .
\end{eqnarray}
%This expression can be used to determine the contribution of each 
%source of uncertainty to the total uncertainty of an average.  
This breakdown of uncertainties is used in certain cases,
% in the following sections, 
but usually only a single, total uncertainty is quoted for an average.

\clearpage
% %% % b-hadron production fractions, lifetimes and mixing parameters
%%%%%%%%%%%%%%%%%%%%%%%%%%%%%%%%%%%%%%
%
% This is file life_mix.tex
% containing the chapter on b-hadron fractions,
% lifetimes and mixing parameters
%
% Olivier Schneider, EPFL
%   - first version Jun 23, 2004
%   - updated Dec 16, 2004
%   - updated Apr 17, 2005
%   - updated Nov 15, 2005
%   - updated Nov 26, 2006
%   - updated Mar 28, 2007
%   - updated Mar 30, 2007 (RvK)
%   - updated Dec 21, 2007 (RvK)
%   - updated Mar 26, 2010 (RJT) (fractions)
%   - updated Jun 12, 2010 (RvK) 
%   - updated Jun 21, 2010 (Remi Louvot, EPFL) (fs with fit)
%
%%%%%%%%%%%%%%%%%%%%%%%%%%%%%%%%%%%%%%%%%%%%%%%
%

% -----------------------------
% LaTeX macros for this chapter
% -----------------------------

%%%%%%%%%%%%%%%%%%%%%%%%%%%%%%%%%%%%%%%%%%%%%%%
%
% This is file life_mix_defintions.tex
%
% Olivier Schneider, EPFL, updated Feb 19, 2017
%
%%%%%%%%%%%%%%%%%%%%%%%%%%%%%%%%%%%%%%%%%%%%%%%
%
% Define "history" and "unpublished" modes
%

\newboolean{history}\setboolean{history}{false}

\newboolean{unpublished}\setboolean{unpublished}{false}

\newcommand{\unpublished}[2]{\ifthenelse{\boolean{unpublished}}{#2}{#1}}
\newcommand{\history}[2]{\ifthenelse{\boolean{history}}{#2}{#1}}
\newcommand{\citehistory}[2]{\ifthenelse{\boolean{history}}{\cite{#2}}{\cite{#1}}}

%%%%%%%%%%%%%%%%%%%%%%%%%%%%%%%%%%%%%%%%%%%%%%%
%
% Load all numerical averages
%

% File generated in directory /afs/cern.ch/phys/lep/lepbosc/combined_results/publication_2017/WRITEUP on dim jun 4 22:47:44 CEST 2017

\newcommand{\definemath}[2]{\newcommand{#1}{\ensuremath{#2}\xspace}}

% HFLAV results (fragments)
\definemath{\hflavCHIBARLEPval}{0.1259}% from ../EXT_INPUT/log
\definemath{\hflavCHIBARLEPerr}{\pm0.0042}% from ../EXT_INPUT/log
\definemath{\hflavNSIGMATAULBEXCLSEMI}{3.1}% from ../LIFETIMES/log
\definemath{\hflavTAUBDval}{1.520}% from ../LIFETIMES/log
\definemath{\hflavTAUBDerr}{\pm0.004}% from ../LIFETIMES/log
\definemath{\hflavTAUBUval}{1.638}% from ../LIFETIMES/log
\definemath{\hflavTAUBUerr}{\pm0.004}% from ../LIFETIMES/log
\definemath{\hflavRTAUBUval}{1.076}% from ../LIFETIMES/log
\definemath{\hflavRTAUBUerr}{\pm0.004}% from ../LIFETIMES/log
\definemath{\hflavTAUBSval}{1.505}% from ../LIFETIMES/log
\definemath{\hflavTAUBSerr}{\pm0.005}% from ../LIFETIMES/log
\definemath{\hflavRTAUBSval}{0.990}% from ../LIFETIMES/log
\definemath{\hflavRTAUBSerr}{\pm0.004}% from ../LIFETIMES/log
\definemath{\hflavTAULBval}{1.470}% from ../LIFETIMES/log
\definemath{\hflavTAULBerr}{\pm0.010}% from ../LIFETIMES/log
\definemath{\hflavTAULBSval}{1.247}% from ../LIFETIMES/log
\definemath{\hflavTAULBSerp}{^{+0.071}}% from ../LIFETIMES/log
\definemath{\hflavTAULBSern}{_{-0.069}}% from ../LIFETIMES/log
\definemath{\hflavTAULBEval}{1.470}% from ../LIFETIMES/log
\definemath{\hflavTAULBEerr}{\pm0.010}% from ../LIFETIMES/log
\definemath{\hflavTAUXBDval}{1.571}% from ../LIFETIMES/log
\definemath{\hflavTAUXBDerr}{\pm0.040}% from ../LIFETIMES/log
\definemath{\hflavTAUXBUval}{1.479}% from ../LIFETIMES/log
\definemath{\hflavTAUXBUerr}{\pm0.031}% from ../LIFETIMES/log
\definemath{\hflavTAUOBval}{1.64}% from ../LIFETIMES/log
\definemath{\hflavTAUOBerp}{^{+0.18}}% from ../LIFETIMES/log
\definemath{\hflavTAUOBern}{_{-0.17}}% from ../LIFETIMES/log
\definemath{\hflavTAUBCval}{0.507}% from ../LIFETIMES/log
\definemath{\hflavTAUBCerr}{\pm0.009}% from ../LIFETIMES/log
\definemath{\hflavTAUBSSLval}{1.516}% from ../LIFETIMES/log
\definemath{\hflavTAUBSSLerr}{\pm0.014}% from ../LIFETIMES/log
\definemath{\hflavTAUBSMEANCval}{1.505}% from ../LIFETIMES/log
\definemath{\hflavTAUBSMEANCerr}{\pm0.005}% from ../LIFETIMES/log
\definemath{\hflavTAUBSJFval}{1.479}% from ../LIFETIMES/log
\definemath{\hflavTAUBSJFerr}{\pm0.012}% from ../LIFETIMES/log
\definemath{\hflavRTAUBSSLval}{0.997}% from ../LIFETIMES/log
\definemath{\hflavRTAUBSSLerr}{\pm0.010}% from ../LIFETIMES/log
\definemath{\hflavRTAUBSMEANCval}{0.990}% from ../LIFETIMES/log
\definemath{\hflavRTAUBSMEANCerr}{\pm0.004}% from ../LIFETIMES/log
\definemath{\hflavRTAUBSMEANCsig}{2.2}% from ../LIFETIMES/log
\definemath{\hflavONEMINUSRTAUBSMEANCpercent}{(1.0\pm0.4)\%}% from ../LIFETIMES/log
\definemath{\hflavRTAULBval}{0.967}% from ../LIFETIMES/log
\definemath{\hflavRTAULBerr}{\pm0.007}% from ../LIFETIMES/log
\definemath{\hflavTAUBVTXval}{1.572}% from ../LIFETIMES/log
\definemath{\hflavTAUBVTXerr}{\pm0.009}% from ../LIFETIMES/log
\definemath{\hflavTAUBLEPval}{1.537}% from ../LIFETIMES/log
\definemath{\hflavTAUBLEPerr}{\pm0.020}% from ../LIFETIMES/log
\definemath{\hflavTAUBJPval}{1.533}% from ../LIFETIMES/log
\definemath{\hflavTAUBJPerr}{\pm0.036}% from ../LIFETIMES/log
\definemath{\hflavNSIGMATAULBCDFTWO}{2.4}% from ../LIFETIMES/log
\definemath{\hflavRTAUXBUXBDval}{0.929}% from ../LIFETIMES/log
\definemath{\hflavRTAUXBUXBDerr}{\pm0.028}% from ../LIFETIMES/log
\definemath{\hflavSDGDGDval}{-0.002}% from ../DGD/log
\definemath{\hflavSDGDGDerr}{\pm0.010}% from ../DGD/log
\definemath{\hflavTAUBSJPSIPIPIval}{1.658}% from ../DGS_PHIS/log
\definemath{\hflavTAUBSJPSIPIPIerr}{\pm0.032}% from ../DGS_PHIS/log
\definemath{\hflavTAUBSJPSIKSHORTval}{1.75}% from ../DGS_PHIS/log
\definemath{\hflavTAUBSJPSIKSHORTerr}{\pm0.14}% from ../DGS_PHIS/log
\definemath{\hflavTAUBSLONGval}{1.658}% from ../DGS_PHIS/log
\definemath{\hflavTAUBSLONGerr}{\pm0.032}% from ../DGS_PHIS/log
\definemath{\hflavTAUBSKKval}{1.408}% from ../DGS_PHIS/log
\definemath{\hflavTAUBSKKerr}{\pm0.017}% from ../DGS_PHIS/log
\definemath{\hflavTAUBSDSDSval}{1.379}% from ../DGS_PHIS/log
\definemath{\hflavTAUBSDSDSerr}{\pm0.031}% from ../DGS_PHIS/log
\definemath{\hflavTAUBSJPSIETAval}{1.479}% from ../DGS_PHIS/log
\definemath{\hflavTAUBSJPSIETAerr}{\pm0.036}% from ../DGS_PHIS/log
\definemath{\hflavTAUBSSHORTval}{1.422}% from ../DGS_PHIS/log
\definemath{\hflavTAUBSSHORTerr}{\pm0.023}% from ../DGS_PHIS/log
\definemath{\hflavGSval}{0.6654}% from ../DGS_PHIS/log
\definemath{\hflavGSerr}{\pm0.0022}% from ../DGS_PHIS/log
\definemath{\hflavTAUBSMEANval}{1.503}% from ../DGS_PHIS/log
\definemath{\hflavTAUBSMEANerr}{\pm0.005}% from ../DGS_PHIS/log
\definemath{\hflavDGSGSval}{+0.126}% from ../DGS_PHIS/log
\definemath{\hflavDGSGSerr}{\pm0.010}% from ../DGS_PHIS/log
\definemath{\hflavDGSval}{+0.084}% from ../DGS_PHIS/log
\definemath{\hflavDGSerr}{\pm0.007}% from ../DGS_PHIS/log
\definemath{\hflavRHOGSDGS}{-0.286}% from ../DGS_PHIS/log
\definemath{\hflavTAUBSLval}{1.414}% from ../DGS_PHIS/log
\definemath{\hflavTAUBSLerr}{\pm0.007}% from ../DGS_PHIS/log
\definemath{\hflavTAUBSHval}{1.604}% from ../DGS_PHIS/log
\definemath{\hflavTAUBSHerr}{\pm0.011}% from ../DGS_PHIS/log
\definemath{\hflavGSCOval}{0.6645}% from ../DGS_PHIS/log
\definemath{\hflavGSCOerr}{\pm0.0021}% from ../DGS_PHIS/log
\definemath{\hflavTAUBSMEANCOval}{1.505}% from ../DGS_PHIS/log
\definemath{\hflavTAUBSMEANCOerr}{\pm0.005}% from ../DGS_PHIS/log
\definemath{\hflavDGSGSCOval}{+0.130}% from ../DGS_PHIS/log
\definemath{\hflavDGSGSCOerr}{\pm0.009}% from ../DGS_PHIS/log
\definemath{\hflavDGSCOval}{+0.086}% from ../DGS_PHIS/log
\definemath{\hflavDGSCOerr}{\pm0.006}% from ../DGS_PHIS/log
\definemath{\hflavRHOGSDGSCO}{-0.267}% from ../DGS_PHIS/log
\definemath{\hflavTAUBSLCOval}{1.413}% from ../DGS_PHIS/log
\definemath{\hflavTAUBSLCOerr}{\pm0.006}% from ../DGS_PHIS/log
\definemath{\hflavTAUBSHCOval}{1.610}% from ../DGS_PHIS/log
\definemath{\hflavTAUBSHCOerr}{\pm0.011}% from ../DGS_PHIS/log
\definemath{\hflavGSCONval}{0.6646}% from ../DGS_PHIS/log
\definemath{\hflavGSCONerr}{\pm0.0020}% from ../DGS_PHIS/log
\definemath{\hflavTAUBSMEANCONval}{1.505}% from ../DGS_PHIS/log
\definemath{\hflavTAUBSMEANCONerr}{\pm0.005}% from ../DGS_PHIS/log
\definemath{\hflavDGSGSCONval}{+0.130}% from ../DGS_PHIS/log
\definemath{\hflavDGSGSCONerr}{\pm0.009}% from ../DGS_PHIS/log
\definemath{\hflavDGSCONval}{+0.086}% from ../DGS_PHIS/log
\definemath{\hflavDGSCONerr}{\pm0.006}% from ../DGS_PHIS/log
\definemath{\hflavRHOGSDGSCON}{-0.210}% from ../DGS_PHIS/log
\definemath{\hflavTAUBSLCONval}{1.413}% from ../DGS_PHIS/log
\definemath{\hflavTAUBSLCONerr}{\pm0.006}% from ../DGS_PHIS/log
\definemath{\hflavTAUBSHCONval}{1.609}% from ../DGS_PHIS/log
\definemath{\hflavTAUBSHCONerr}{\pm0.010}% from ../DGS_PHIS/log
\definemath{\hflavBETASCOMBval}{+0.015}% from ../DGS_PHIS/log
\definemath{\hflavBETASCOMBerr}{\pm0.016}% from ../DGS_PHIS/log
\definemath{\hflavPHISCOMBval}{-0.030}% from ../DGS_PHIS/log
\definemath{\hflavPHISCOMBerr}{\pm0.033}% from ../DGS_PHIS/log
\definemath{\hflavDGSCOMBval}{+0.085}% from ../DGS_PHIS/log
\definemath{\hflavDGSCOMBerr}{\pm0.007}% from ../DGS_PHIS/log
\definemath{\hflavPHISSMval}{-0.0370}% from ../DGS_PHIS/log
\definemath{\hflavPHISSMerr}{\pm0.0006}% from ../DGS_PHIS/log
\definemath{\hflavPHISTWELVESMval}{0.0046}% from ../DGS_PHIS/log
\definemath{\hflavPHISTWELVESMerr}{\pm0.0012}% from ../DGS_PHIS/log
\definemath{\hflavPHISTWELVEval}{0.012}% from ../DGS_PHIS/log
\definemath{\hflavPHISTWELVEerr}{\pm0.033}% from ../DGS_PHIS/log
\definemath{\hflavFCWval}{0.514}% from ../FRAC_4S/log
\definemath{\hflavFCWerr}{\pm0.006}% from ../FRAC_4S/log
\definemath{\hflavFNWval}{0.486}% from ../FRAC_4S/log
\definemath{\hflavFNWerr}{\pm0.006}% from ../FRAC_4S/log
\definemath{\hflavFFWval}{1.058}% from ../FRAC_4S/log
\definemath{\hflavFFWerr}{\pm0.024}% from ../FRAC_4S/log
\definemath{\hflavNSIGMAFFW}{2.4}% from ../FRAC_4S/log
\definemath{\hflavFCNval}{0.513}% from ../FRAC_4S/log
\definemath{\hflavFCNerr}{\pm0.013}% from ../FRAC_4S/log
\definemath{\hflavFNNval}{0.487}% from ../FRAC_4S/log
\definemath{\hflavFNNerr}{\pm0.013}% from ../FRAC_4S/log
\definemath{\hflavFFNval}{1.053}% from ../FRAC_4S/log
\definemath{\hflavFFNerr}{\pm0.054}% from ../FRAC_4S/log
\definemath{\hflavFCval}{0.514}% from ../FRAC_4S/log
\definemath{\hflavFCerr}{\pm0.006}% from ../FRAC_4S/log
\definemath{\hflavFNval}{0.486}% from ../FRAC_4S/log
\definemath{\hflavFNerr}{\pm0.006}% from ../FRAC_4S/log
\definemath{\hflavFFval}{1.059}% from ../FRAC_4S/log
\definemath{\hflavFFerr}{\pm0.027}% from ../FRAC_4S/log
\definemath{\hflavNSIGMAFF}{2.2}% from ../FRAC_4S/log
\definemath{\hflavFPRODval}{0.516}% from ../FRAC_4S/log
\definemath{\hflavFPRODerr}{\pm0.019}% from ../FRAC_4S/log
\definemath{\hflavFSUMval}{1.003}% from ../FRAC_4S/log
\definemath{\hflavFSUMerr}{\pm0.029}% from ../FRAC_4S/log
\definemath{\hflavFSFIVEOSval}{0.206}% from ../FRAC_5S/log
\definemath{\hflavFSFIVEOSsta}{\pm0.010}% from ../FRAC_5S/log
\definemath{\hflavFSFIVEOSsys}{\pm0.024}% from ../FRAC_5S/log
\definemath{\hflavFSFIVEOSerr}{\pm0.027}% from ../FRAC_5S/log
\definemath{\hflavFSFIVERLval}{0.215}% from ../FRAC_5S/log
\definemath{\hflavFSFIVERLerr}{\pm0.031}% from ../FRAC_5S/log
\definemath{\hflavFUDFIVEval}{0.761}% from ../FRAC_5S/log
\definemath{\hflavFUDFIVEerp}{^{+0.027}}% from ../FRAC_5S/log
\definemath{\hflavFUDFIVEern}{_{-0.042}}% from ../FRAC_5S/log
\definemath{\hflavFSFIVEval}{0.200}% from ../FRAC_5S/log
\definemath{\hflavFSFIVEerp}{^{+0.030}}% from ../FRAC_5S/log
\definemath{\hflavFSFIVEern}{_{-0.031}}% from ../FRAC_5S/log
\definemath{\hflavFSFUDFIVEval}{0.263}% from ../FRAC_5S/log
\definemath{\hflavFSFUDFIVEerp}{^{+0.052}}% from ../FRAC_5S/log
\definemath{\hflavFSFUDFIVEern}{_{-0.044}}% from ../FRAC_5S/log
\definemath{\hflavFNBFIVEval}{0.039}% from ../FRAC_5S/log
\definemath{\hflavFNBFIVEerp}{^{+0.050}}% from ../FRAC_5S/log
\definemath{\hflavFNBFIVEern}{_{-0.004}}% from ../FRAC_5S/log
\definemath{\hflavZFSFACTOR}{}% from ../FRAC_NOMIX/log
\definemath{\hflavZFBSNOMIXval}{0.088}% from ../FRAC_NOMIX/log
\definemath{\hflavZFBSNOMIXerr}{\pm0.013}% from ../FRAC_NOMIX/log
\definemath{\hflavZFBBNOMIXval}{0.089}% from ../FRAC_NOMIX/log
\definemath{\hflavZFBBNOMIXerr}{\pm0.012}% from ../FRAC_NOMIX/log
\definemath{\hflavZFBDNOMIXval}{0.412}% from ../FRAC_NOMIX/log
\definemath{\hflavZFBDNOMIXerr}{\pm0.008}% from ../FRAC_NOMIX/log
\definemath{\hflavWFSFACTOR}{1.1}% from ../FRAC_NOMIX/log
\definemath{\hflavWFBSNOMIXval}{0.102}% from ../FRAC_NOMIX/log
\definemath{\hflavWFBSNOMIXerr}{\pm0.005}% from ../FRAC_NOMIX/log
\definemath{\hflavWFBBNOMIXval}{0.090}% from ../FRAC_NOMIX/log
\definemath{\hflavWFBBNOMIXerr}{\pm0.012}% from ../FRAC_NOMIX/log
\definemath{\hflavWFBDNOMIXval}{0.404}% from ../FRAC_NOMIX/log
\definemath{\hflavWFBDNOMIXerr}{\pm0.006}% from ../FRAC_NOMIX/log
\definemath{\hflavTFSFACTOR}{}% from ../FRAC_NOMIX/log
\definemath{\hflavTFBSNOMIXval}{0.101}% from ../FRAC_NOMIX/log
\definemath{\hflavTFBSNOMIXerr}{\pm0.015}% from ../FRAC_NOMIX/log
\definemath{\hflavTFBBNOMIXval}{0.218}% from ../FRAC_NOMIX/log
\definemath{\hflavTFBBNOMIXerr}{\pm0.047}% from ../FRAC_NOMIX/log
\definemath{\hflavTFBDNOMIXval}{0.340}% from ../FRAC_NOMIX/log
\definemath{\hflavTFBDNOMIXerr}{\pm0.021}% from ../FRAC_NOMIX/log
\definemath{\hflavLFSFACTOR}{}% from ../FRAC_NOMIX/log
\definemath{\hflavLFBSNOMIXval}{0.087}% from ../FRAC_NOMIX/log
\definemath{\hflavLFBSNOMIXerr}{\pm0.005}% from ../FRAC_NOMIX/log
\definemath{\hflavLFBBNOMIXval}{0.230}% from ../FRAC_NOMIX/log
\definemath{\hflavLFBBNOMIXerr}{\pm0.021}% from ../FRAC_NOMIX/log
\definemath{\hflavLFBDNOMIXval}{0.342}% from ../FRAC_NOMIX/log
\definemath{\hflavLFBDNOMIXerr}{\pm0.009}% from ../FRAC_NOMIX/log
\definemath{\hflavCHIBARTEVval}{0.147}% from ../CHIBAR/log
\definemath{\hflavCHIBARTEVerr}{\pm0.011}% from ../CHIBAR/log
\definemath{\hflavCHIBARSFACTOR}{1.8}% from ../CHIBAR/log
\definemath{\hflavCHIBARval}{0.1284}% from ../CHIBAR/log
\definemath{\hflavCHIBARerr}{\pm0.0069}% from ../CHIBAR/log
\definemath{\hflavWFBSMIXval}{0.118}% from ../DMD/log
\definemath{\hflavWFBSMIXerr}{\pm0.018}% from ../DMD/log
\definemath{\hflavTFBSMIXval}{0.166}% from ../DMD/log
\definemath{\hflavTFBSMIXerr}{\pm0.029}% from ../DMD/log
\definemath{\hflavZFBSMIXval}{0.111}% from ../DMD/log
\definemath{\hflavZFBSMIXerr}{\pm0.011}% from ../DMD/log
\definemath{\hflavCHIDUval}{0.182}% from ../DMD/log
\definemath{\hflavCHIDUerr}{\pm0.015}% from ../DMD/log
\definemath{\hflavCHIDWUval}{0.1860}% from ../DMD/log
\definemath{\hflavCHIDWUerr}{\pm0.0011}% from ../DMD/log
\definemath{\hflavXDWval}{0.770}% from ../DMD/log
\definemath{\hflavXDWerr}{\pm0.004}% from ../DMD/log
\definemath{\hflavXDWUval}{0.770}% from ../DMD/log
\definemath{\hflavXDWUerr}{\pm0.004}% from ../DMD/log
\definemath{\hflavDMDWval}{0.5065}% from ../DMD/log
\definemath{\hflavDMDWsta}{\pm0.0016}% from ../DMD/log
\definemath{\hflavDMDWsys}{\pm0.0011}% from ../DMD/log
\definemath{\hflavDMDWerr}{\pm0.0019}% from ../DMD/log
\definemath{\hflavDMDWUval}{0.5064}% from ../DMD/log
\definemath{\hflavDMDWUerr}{\pm0.0019}% from ../DMD/log
\definemath{\hflavDMDLHCbval}{0.5063}% from ../DMD/log
\definemath{\hflavDMDLHCbsta}{\pm0.0019}% from ../DMD/log
\definemath{\hflavDMDLHCbsys}{\pm0.0010}% from ../DMD/log
\definemath{\hflavDMDLHCberr}{\pm0.0022}% from ../DMD/log
\definemath{\hflavZFBSval}{0.101}% from ../FRAC_MIX/log
\definemath{\hflavZFBSerr}{\pm0.008}% from ../FRAC_MIX/log
\definemath{\hflavZFBBval}{0.084}% from ../FRAC_MIX/log
\definemath{\hflavZFBBerr}{\pm0.011}% from ../FRAC_MIX/log
\definemath{\hflavZFBDval}{0.407}% from ../FRAC_MIX/log
\definemath{\hflavZFBDerr}{\pm0.007}% from ../FRAC_MIX/log
\definemath{\hflavZRHOFBBFBS}{+0.074}% from ../FRAC_MIX/log
\definemath{\hflavZRHOFBDFBS}{-0.629}% from ../FRAC_MIX/log
\definemath{\hflavZRHOFBDFBB}{-0.822}% from ../FRAC_MIX/log
\definemath{\hflavWFBSval}{0.103}% from ../FRAC_MIX/log
\definemath{\hflavWFBSerr}{\pm0.005}% from ../FRAC_MIX/log
\definemath{\hflavWFBBval}{0.088}% from ../FRAC_MIX/log
\definemath{\hflavWFBBerr}{\pm0.012}% from ../FRAC_MIX/log
\definemath{\hflavWFBDval}{0.404}% from ../FRAC_MIX/log
\definemath{\hflavWFBDerr}{\pm0.006}% from ../FRAC_MIX/log
\definemath{\hflavWRHOFBBFBS}{-0.254}% from ../FRAC_MIX/log
\definemath{\hflavWRHOFBDFBS}{-0.143}% from ../FRAC_MIX/log
\definemath{\hflavWRHOFBDFBB}{-0.921}% from ../FRAC_MIX/log
\definemath{\hflavTFBSval}{0.115}% from ../FRAC_MIX/log
\definemath{\hflavTFBSerr}{\pm0.013}% from ../FRAC_MIX/log
\definemath{\hflavTFBBval}{0.196}% from ../FRAC_MIX/log
\definemath{\hflavTFBBerr}{\pm0.046}% from ../FRAC_MIX/log
\definemath{\hflavTFBDval}{0.344}% from ../FRAC_MIX/log
\definemath{\hflavTFBDerr}{\pm0.021}% from ../FRAC_MIX/log
\definemath{\hflavTRHOFBBFBS}{-0.426}% from ../FRAC_MIX/log
\definemath{\hflavTRHOFBDFBS}{+0.153}% from ../FRAC_MIX/log
\definemath{\hflavTRHOFBDFBB}{-0.959}% from ../FRAC_MIX/log
\definemath{\hflavLFBSval}{0.090}% from ../FRAC_MIX/log
\definemath{\hflavLFBSerr}{\pm0.005}% from ../FRAC_MIX/log
\definemath{\hflavLFBBval}{0.222}% from ../FRAC_MIX/log
\definemath{\hflavLFBBerr}{\pm0.020}% from ../FRAC_MIX/log
\definemath{\hflavLFBDval}{0.344}% from ../FRAC_MIX/log
\definemath{\hflavLFBDerr}{\pm0.009}% from ../FRAC_MIX/log
\definemath{\hflavLRHOFBBFBS}{-0.509}% from ../FRAC_MIX/log
\definemath{\hflavLRHOFBDFBS}{+0.304}% from ../FRAC_MIX/log
\definemath{\hflavLRHOFBDFBB}{-0.975}% from ../FRAC_MIX/log
\definemath{\hflavZFBSBDval}{0.249}% from ../FRAC_MIX/log
\definemath{\hflavZFBSBDerr}{\pm0.023}% from ../FRAC_MIX/log
\definemath{\hflavWFBSBDval}{0.256}% from ../FRAC_MIX/log
\definemath{\hflavWFBSBDerr}{\pm0.013}% from ../FRAC_MIX/log
\definemath{\hflavTFBSBDval}{0.333}% from ../FRAC_MIX/log
\definemath{\hflavTFBSBDerr}{\pm0.041}% from ../FRAC_MIX/log
\definemath{\hflavLFBSBDval}{0.260}% from ../FRAC_MIX/log
\definemath{\hflavLFBSBDerr}{\pm0.013}% from ../FRAC_MIX/log
\definemath{\hflavDMDLval}{0.493}% from ../MORE_DMD/log
\definemath{\hflavDMDLsta}{\pm0.011}% from ../MORE_DMD/log
\definemath{\hflavDMDLsys}{\pm0.009}% from ../MORE_DMD/log
\definemath{\hflavDMDLerr}{\pm0.014}% from ../MORE_DMD/log
\definemath{\hflavDMDTval}{0.509}% from ../MORE_DMD/log
\definemath{\hflavDMDTsta}{\pm0.017}% from ../MORE_DMD/log
\definemath{\hflavDMDTsys}{\pm0.013}% from ../MORE_DMD/log
\definemath{\hflavDMDTerr}{\pm0.022}% from ../MORE_DMD/log
\definemath{\hflavDMDBval}{0.509}% from ../MORE_DMD/log
\definemath{\hflavDMDBsta}{\pm0.003}% from ../MORE_DMD/log
\definemath{\hflavDMDBsys}{\pm0.003}% from ../MORE_DMD/log
\definemath{\hflavDMDBerr}{\pm0.005}% from ../MORE_DMD/log
\definemath{\hflavDMDTWODval}{0.509}% from ../2D/log
\definemath{\hflavDMDTWODsta}{\pm0.004}% from ../2D/log
\definemath{\hflavDMDTWODsys}{\pm0.004}% from ../2D/log
\definemath{\hflavDMDTWODerr}{\pm0.006}% from ../2D/log
\definemath{\hflavTAUBDTWODval}{1.527}% from ../2D/log
\definemath{\hflavTAUBDTWODsta}{\pm0.006}% from ../2D/log
\definemath{\hflavTAUBDTWODsys}{\pm0.008}% from ../2D/log
\definemath{\hflavTAUBDTWODerr}{\pm0.010}% from ../2D/log
\definemath{\hflavRHOstaDMDTAUBD}{-0.19}% from ../2D/log
\definemath{\hflavRHOsysDMDTAUBD}{-0.25}% from ../2D/log
\definemath{\hflavRHODMDTAUBD}{-0.23}% from ../2D/log
\definemath{\hflavZRHOTAUHTAUL}{-0.398}% from ../TAUB/log
\definemath{\hflavTAUBZCALCval}{1.566}% from ../TAUB/log
\definemath{\hflavTAUBZCALCerr}{\pm0.003}% from ../TAUB/log
\definemath{\hflavQPDBval}{1.0009}% from ../CPV_MIX_BD/log
\definemath{\hflavQPDBerr}{\pm0.0013}% from ../CPV_MIX_BD/log
\definemath{\hflavQPDDval}{1.0000}% from ../CPV_MIX_BD/log
\definemath{\hflavQPDDerr}{\pm0.0010}% from ../CPV_MIX_BD/log
\definemath{\hflavQPDWval}{1.0005}% from ../CPV_MIX_BD/log
\definemath{\hflavQPDWerr}{\pm0.0009}% from ../CPV_MIX_BD/log
\definemath{\hflavQPDAval}{1.0005}% from ../CPV_MIX_BD/log
\definemath{\hflavQPDAerr}{\pm0.0009}% from ../CPV_MIX_BD/log
\definemath{\hflavASLDBval}{-0.0019}% from ../CPV_MIX_BD/log
\definemath{\hflavASLDBerr}{\pm0.0027}% from ../CPV_MIX_BD/log
\definemath{\hflavASLDDval}{+0.0001}% from ../CPV_MIX_BD/log
\definemath{\hflavASLDDerr}{\pm0.0020}% from ../CPV_MIX_BD/log
\definemath{\hflavASLDWval}{-0.0010}% from ../CPV_MIX_BD/log
\definemath{\hflavASLDWerr}{\pm0.0018}% from ../CPV_MIX_BD/log
\definemath{\hflavASLDAval}{-0.0010}% from ../CPV_MIX_BD/log
\definemath{\hflavASLDAerr}{\pm0.0018}% from ../CPV_MIX_BD/log
\definemath{\hflavREBDBval}{-0.0005}% from ../CPV_MIX_BD/log
\definemath{\hflavREBDBerr}{\pm0.0007}% from ../CPV_MIX_BD/log
\definemath{\hflavREBDDval}{+0.0000}% from ../CPV_MIX_BD/log
\definemath{\hflavREBDDerr}{\pm0.0005}% from ../CPV_MIX_BD/log
\definemath{\hflavREBDWval}{-0.0002}% from ../CPV_MIX_BD/log
\definemath{\hflavREBDWerr}{\pm0.0004}% from ../CPV_MIX_BD/log
\definemath{\hflavREBDAval}{-0.0002}% from ../CPV_MIX_BD/log
\definemath{\hflavREBDAerr}{\pm0.0004}% from ../CPV_MIX_BD/log
\definemath{\hflavASLLHCBDZERONSIGMA}{2.2}% from ../CPV_MIX_BS/log
\definemath{\hflavASLLHCBDZEROPVALPERCENT}{3.1}% from ../CPV_MIX_BS/log
\definemath{\hflavASLSval}{-0.0006}% from ../CPV_MIX_BS/log
\definemath{\hflavASLSerr}{\pm0.0028}% from ../CPV_MIX_BS/log
\definemath{\hflavQPSval}{1.0003}% from ../CPV_MIX_BS/log
\definemath{\hflavQPSerr}{\pm0.0014}% from ../CPV_MIX_BS/log
\definemath{\hflavASLDval}{-0.0021}% from ../CPV_MIX_BS/log
\definemath{\hflavASLDerr}{\pm0.0017}% from ../CPV_MIX_BS/log
\definemath{\hflavQPDval}{1.0010}% from ../CPV_MIX_BS/log
\definemath{\hflavQPDerr}{\pm0.0008}% from ../CPV_MIX_BS/log
\definemath{\hflavRHOASLSASLD}{-0.054}% from ../CPV_MIX_BS/log
\definemath{\hflavCLPERCENTASLSASLD}{4.5}% from ../CPV_MIX_BS/log
\definemath{\hflavREBDval}{-0.0005}% from ../CPV_MIX_BS/log
\definemath{\hflavREBDerr}{\pm0.0004}% from ../CPV_MIX_BS/log
\definemath{\hflavASLDNOMUval}{+0.0000}% from ../CPV_MIX_BS/log
\definemath{\hflavASLDNOMUerr}{\pm0.0019}% from ../CPV_MIX_BS/log
\definemath{\hflavASLSNOMUval}{+0.0016}% from ../CPV_MIX_BS/log
\definemath{\hflavASLSNOMUerr}{\pm0.0030}% from ../CPV_MIX_BS/log
\definemath{\hflavRHOASLSASLDNOMU}{+0.066}% from ../CPV_MIX_BS/log
\definemath{\hflavASLDASLSNSIGMA}{0.5}% from ../CPV_MIX_BS/log
\definemath{\hflavASLDASLSPVALPERCENT}{61.3}% from ../CPV_MIX_BS/log
\definemath{\hflavTANPHIval}{-0.1}% from ../CPV_MIX_BS/log
\definemath{\hflavTANPHIerr}{\pm0.6}% from ../CPV_MIX_BS/log
\definemath{\hflavDMSval}{17.757}% from ../DMS/log
\definemath{\hflavDMSsta}{\pm0.020}% from ../DMS/log
\definemath{\hflavDMSsys}{\pm0.007}% from ../DMS/log
\definemath{\hflavDMSerr}{\pm0.021}% from ../DMS/log
\definemath{\hflavXSval}{26.72}% from ../DMS/log
\definemath{\hflavXSerr}{\pm0.09}% from ../DMS/log
\definemath{\hflavCHISval}{0.499304}% from ../DMS/log
\definemath{\hflavCHISerr}{\pm0.000005}% from ../DMS/log
\definemath{\hflavRATIODGSDMSval}{0.00486}% from ../DMS/log
\definemath{\hflavRATIODGSDMSerr}{\pm0.00034}% from ../DMS/log
\definemath{\hflavRATIODMDDMSval}{0.02852}% from ../DMS/log
\definemath{\hflavRATIODMDDMSerr}{\pm0.00011}% from ../DMS/log
\definemath{\hflavVTDVTSval}{0.2053}% from ../DMS/log
\definemath{\hflavVTDVTSexx}{\pm0.0004}% from ../DMS/log
\definemath{\hflavVTDVTSthe}{\pm0.0032}% from ../DMS/log
\definemath{\hflavVTDVTSerr}{\pm0.0033}% from ../DMS/log
\definemath{\hflavXIval}{1.206}% from ../DMS/log
\definemath{\hflavXIerr}{\pm0.019}% from ../DMS/log
\definemath{\hflavXIsta}{\pm0.018}% from ../DMS/log
\definemath{\hflavXIsys}{\pm0.006}% from ../DMS/log

% Units
\newcommand{\unit}[1]{~\ensuremath{\rm #1}\xspace}
\renewcommand{\ps}{\unit{ps}}
\newcommand{\invps}{\unit{ps^{-1}}}
\newcommand{\TeV}{\unit{TeV}}
\newcommand{\MeVcc}{\unit{MeV/\mbox{$c$}^2}}
\newcommand{\MeV}{\unit{MeV}}

% HFLAV results (complete)
\definemath{\hflavCHIBARLEP}{\hflavCHIBARLEPval\hflavCHIBARLEPerr}
\definemath{\hflavTAUBD}{\hflavTAUBDval\hflavTAUBDerr\ps}
\definemath{\hflavTAUBDnounit}{\hflavTAUBDval\hflavTAUBDerr}
\definemath{\hflavTAUBU}{\hflavTAUBUval\hflavTAUBUerr\ps}
\definemath{\hflavTAUBUnounit}{\hflavTAUBUval\hflavTAUBUerr}
\definemath{\hflavRTAUBU}{\hflavRTAUBUval\hflavRTAUBUerr}
\definemath{\hflavTAUBS}{\hflavTAUBSval\hflavTAUBSerr\ps}
\definemath{\hflavTAUBSnounit}{\hflavTAUBSval\hflavTAUBSerr}
\definemath{\hflavRTAUBS}{\hflavRTAUBSval\hflavRTAUBSerr}
\definemath{\hflavTAULB}{\hflavTAULBval\hflavTAULBerr\ps}
\definemath{\hflavTAULBnounit}{\hflavTAULBval\hflavTAULBerr}
\definemath{\hflavTAULBSerr}{\hflavTAULBSerp\hflavTAULBSern}
\definemath{\hflavTAULBS}{\hflavTAULBSval\hflavTAULBSerr\ps}
\definemath{\hflavTAULBSnounit}{\hflavTAULBSval\hflavTAULBSerr}
\definemath{\hflavTAULBE}{\hflavTAULBEval\hflavTAULBEerr\ps}
\definemath{\hflavTAULBEnounit}{\hflavTAULBEval\hflavTAULBEerr}
\definemath{\hflavTAUXBD}{\hflavTAUXBDval\hflavTAUXBDerr\ps}
\definemath{\hflavTAUXBDnounit}{\hflavTAUXBDval\hflavTAUXBDerr}
\definemath{\hflavTAUXBU}{\hflavTAUXBUval\hflavTAUXBUerr\ps}
\definemath{\hflavTAUXBUnounit}{\hflavTAUXBUval\hflavTAUXBUerr}
\definemath{\hflavTAUOBerr}{\hflavTAUOBerp\hflavTAUOBern}
\definemath{\hflavTAUOB}{\hflavTAUOBval\hflavTAUOBerr\ps}
\definemath{\hflavTAUOBnounit}{\hflavTAUOBval\hflavTAUOBerr}
\definemath{\hflavTAUBC}{\hflavTAUBCval\hflavTAUBCerr\ps}
\definemath{\hflavTAUBCnounit}{\hflavTAUBCval\hflavTAUBCerr}
\definemath{\hflavTAUBSSL}{\hflavTAUBSSLval\hflavTAUBSSLerr\ps}
\definemath{\hflavTAUBSSLnounit}{\hflavTAUBSSLval\hflavTAUBSSLerr}
\definemath{\hflavTAUBSMEANC}{\hflavTAUBSMEANCval\hflavTAUBSMEANCerr\ps}
\definemath{\hflavTAUBSMEANCnounit}{\hflavTAUBSMEANCval\hflavTAUBSMEANCerr}
\definemath{\hflavTAUBSJF}{\hflavTAUBSJFval\hflavTAUBSJFerr\ps}
\definemath{\hflavTAUBSJFnounit}{\hflavTAUBSJFval\hflavTAUBSJFerr}
\definemath{\hflavRTAUBSSL}{\hflavRTAUBSSLval\hflavRTAUBSSLerr}
\definemath{\hflavRTAUBSMEANC}{\hflavRTAUBSMEANCval\hflavRTAUBSMEANCerr}
\definemath{\hflavRTAULB}{\hflavRTAULBval\hflavRTAULBerr}
\definemath{\hflavTAUBVTX}{\hflavTAUBVTXval\hflavTAUBVTXerr\ps}
\definemath{\hflavTAUBVTXnounit}{\hflavTAUBVTXval\hflavTAUBVTXerr}
\definemath{\hflavTAUBLEP}{\hflavTAUBLEPval\hflavTAUBLEPerr\ps}
\definemath{\hflavTAUBLEPnounit}{\hflavTAUBLEPval\hflavTAUBLEPerr}
\definemath{\hflavTAUBJP}{\hflavTAUBJPval\hflavTAUBJPerr\ps}
\definemath{\hflavTAUBJPnounit}{\hflavTAUBJPval\hflavTAUBJPerr}
\definemath{\hflavRTAUXBUXBD}{\hflavRTAUXBUXBDval\hflavRTAUXBUXBDerr}
\definemath{\hflavSDGDGD}{\hflavSDGDGDval\hflavSDGDGDerr}
\definemath{\hflavTAUBSJPSIPIPI}{\hflavTAUBSJPSIPIPIval\hflavTAUBSJPSIPIPIerr\ps}
\definemath{\hflavTAUBSJPSIPIPInounit}{\hflavTAUBSJPSIPIPIval\hflavTAUBSJPSIPIPIerr}
\definemath{\hflavTAUBSJPSIKSHORT}{\hflavTAUBSJPSIKSHORTval\hflavTAUBSJPSIKSHORTerr\ps}
\definemath{\hflavTAUBSJPSIKSHORTnounit}{\hflavTAUBSJPSIKSHORTval\hflavTAUBSJPSIKSHORTerr}
\definemath{\hflavTAUBSLONG}{\hflavTAUBSLONGval\hflavTAUBSLONGerr\ps}
\definemath{\hflavTAUBSLONGnounit}{\hflavTAUBSLONGval\hflavTAUBSLONGerr}
\definemath{\hflavTAUBSKK}{\hflavTAUBSKKval\hflavTAUBSKKerr\ps}
\definemath{\hflavTAUBSKKnounit}{\hflavTAUBSKKval\hflavTAUBSKKerr}
\definemath{\hflavTAUBSDSDS}{\hflavTAUBSDSDSval\hflavTAUBSDSDSerr\ps}
\definemath{\hflavTAUBSDSDSnounit}{\hflavTAUBSDSDSval\hflavTAUBSDSDSerr}
\definemath{\hflavTAUBSJPSIETA}{\hflavTAUBSJPSIETAval\hflavTAUBSJPSIETAerr\ps}
\definemath{\hflavTAUBSJPSIETAnounit}{\hflavTAUBSJPSIETAval\hflavTAUBSJPSIETAerr}
\definemath{\hflavTAUBSSHORT}{\hflavTAUBSSHORTval\hflavTAUBSSHORTerr\ps}
\definemath{\hflavTAUBSSHORTnounit}{\hflavTAUBSSHORTval\hflavTAUBSSHORTerr}
\definemath{\hflavGS}{\hflavGSval\hflavGSerr\invps}
\definemath{\hflavGSnounit}{\hflavGSval\hflavGSerr}
\definemath{\hflavTAUBSMEAN}{\hflavTAUBSMEANval\hflavTAUBSMEANerr\ps}
\definemath{\hflavTAUBSMEANnounit}{\hflavTAUBSMEANval\hflavTAUBSMEANerr}
\definemath{\hflavDGSGS}{\hflavDGSGSval\hflavDGSGSerr}
\definemath{\hflavDGS}{\hflavDGSval\hflavDGSerr\invps}
\definemath{\hflavDGSnounit}{\hflavDGSval\hflavDGSerr}
\definemath{\hflavTAUBSL}{\hflavTAUBSLval\hflavTAUBSLerr\ps}
\definemath{\hflavTAUBSLnounit}{\hflavTAUBSLval\hflavTAUBSLerr}
\definemath{\hflavTAUBSH}{\hflavTAUBSHval\hflavTAUBSHerr\ps}
\definemath{\hflavTAUBSHnounit}{\hflavTAUBSHval\hflavTAUBSHerr}
\definemath{\hflavGSCO}{\hflavGSCOval\hflavGSCOerr\invps}
\definemath{\hflavGSCOnounit}{\hflavGSCOval\hflavGSCOerr}
\definemath{\hflavTAUBSMEANCO}{\hflavTAUBSMEANCOval\hflavTAUBSMEANCOerr\ps}
\definemath{\hflavTAUBSMEANCOnounit}{\hflavTAUBSMEANCOval\hflavTAUBSMEANCOerr}
\definemath{\hflavDGSGSCO}{\hflavDGSGSCOval\hflavDGSGSCOerr}
\definemath{\hflavDGSCO}{\hflavDGSCOval\hflavDGSCOerr\invps}
\definemath{\hflavDGSCOnounit}{\hflavDGSCOval\hflavDGSCOerr}
\definemath{\hflavTAUBSLCO}{\hflavTAUBSLCOval\hflavTAUBSLCOerr\ps}
\definemath{\hflavTAUBSLCOnounit}{\hflavTAUBSLCOval\hflavTAUBSLCOerr}
\definemath{\hflavTAUBSHCO}{\hflavTAUBSHCOval\hflavTAUBSHCOerr\ps}
\definemath{\hflavTAUBSHCOnounit}{\hflavTAUBSHCOval\hflavTAUBSHCOerr}
\definemath{\hflavGSCON}{\hflavGSCONval\hflavGSCONerr\invps}
\definemath{\hflavGSCONnounit}{\hflavGSCONval\hflavGSCONerr}
\definemath{\hflavTAUBSMEANCON}{\hflavTAUBSMEANCONval\hflavTAUBSMEANCONerr\ps}
\definemath{\hflavTAUBSMEANCONnounit}{\hflavTAUBSMEANCONval\hflavTAUBSMEANCONerr}
\definemath{\hflavDGSGSCON}{\hflavDGSGSCONval\hflavDGSGSCONerr}
\definemath{\hflavDGSCON}{\hflavDGSCONval\hflavDGSCONerr\invps}
\definemath{\hflavDGSCONnounit}{\hflavDGSCONval\hflavDGSCONerr}
\definemath{\hflavTAUBSLCON}{\hflavTAUBSLCONval\hflavTAUBSLCONerr\ps}
\definemath{\hflavTAUBSLCONnounit}{\hflavTAUBSLCONval\hflavTAUBSLCONerr}
\definemath{\hflavTAUBSHCON}{\hflavTAUBSHCONval\hflavTAUBSHCONerr\ps}
\definemath{\hflavTAUBSHCONnounit}{\hflavTAUBSHCONval\hflavTAUBSHCONerr}
\definemath{\hflavBETASCOMB}{\hflavBETASCOMBval\hflavBETASCOMBerr}
\definemath{\hflavPHISCOMB}{\hflavPHISCOMBval\hflavPHISCOMBerr}
\definemath{\hflavDGSCOMB}{\hflavDGSCOMBval\hflavDGSCOMBerr\invps}
\definemath{\hflavDGSCOMBnounit}{\hflavDGSCOMBval\hflavDGSCOMBerr}
\definemath{\hflavPHISSM}{\hflavPHISSMval\hflavPHISSMerr}
\definemath{\hflavPHISTWELVESM}{\hflavPHISTWELVESMval\hflavPHISTWELVESMerr}
\definemath{\hflavPHISTWELVE}{\hflavPHISTWELVEval\hflavPHISTWELVEerr}
\definemath{\hflavFCW}{\hflavFCWval\hflavFCWerr}
\definemath{\hflavFNW}{\hflavFNWval\hflavFNWerr}
\definemath{\hflavFFW}{\hflavFFWval\hflavFFWerr}
\definemath{\hflavFCN}{\hflavFCNval\hflavFCNerr}
\definemath{\hflavFNN}{\hflavFNNval\hflavFNNerr}
\definemath{\hflavFFN}{\hflavFFNval\hflavFFNerr}
\definemath{\hflavFC}{\hflavFCval\hflavFCerr}
\definemath{\hflavFN}{\hflavFNval\hflavFNerr}
\definemath{\hflavFF}{\hflavFFval\hflavFFerr}
\definemath{\hflavFPROD}{\hflavFPRODval\hflavFPRODerr}
\definemath{\hflavFSUM}{\hflavFSUMval\hflavFSUMerr}
\definemath{\hflavFSFIVEOS}{\hflavFSFIVEOSval\hflavFSFIVEOSerr}
\definemath{\hflavFSFIVEOSfull}{\hflavFSFIVEOSval\hflavFSFIVEOSsta\hflavFSFIVEOSsys}
\definemath{\hflavFSFIVERL}{\hflavFSFIVERLval\hflavFSFIVERLerr}
\definemath{\hflavFUDFIVEerr}{\hflavFUDFIVEerp\hflavFUDFIVEern}
\definemath{\hflavFUDFIVE}{\hflavFUDFIVEval\hflavFUDFIVEerr}
\definemath{\hflavFSFIVEerr}{\hflavFSFIVEerp\hflavFSFIVEern}
\definemath{\hflavFSFIVE}{\hflavFSFIVEval\hflavFSFIVEerr}
\definemath{\hflavFSFUDFIVEerr}{\hflavFSFUDFIVEerp\hflavFSFUDFIVEern}
\definemath{\hflavFSFUDFIVE}{\hflavFSFUDFIVEval\hflavFSFUDFIVEerr}
\definemath{\hflavFNBFIVEerr}{\hflavFNBFIVEerp\hflavFNBFIVEern}
\definemath{\hflavFNBFIVE}{\hflavFNBFIVEval\hflavFNBFIVEerr}
\definemath{\hflavZFBSNOMIX}{\hflavZFBSNOMIXval\hflavZFBSNOMIXerr}
\definemath{\hflavZFBBNOMIX}{\hflavZFBBNOMIXval\hflavZFBBNOMIXerr}
\definemath{\hflavZFBDNOMIX}{\hflavZFBDNOMIXval\hflavZFBDNOMIXerr}
\definemath{\hflavWFBSNOMIX}{\hflavWFBSNOMIXval\hflavWFBSNOMIXerr}
\definemath{\hflavWFBBNOMIX}{\hflavWFBBNOMIXval\hflavWFBBNOMIXerr}
\definemath{\hflavWFBDNOMIX}{\hflavWFBDNOMIXval\hflavWFBDNOMIXerr}
\definemath{\hflavTFBSNOMIX}{\hflavTFBSNOMIXval\hflavTFBSNOMIXerr}
\definemath{\hflavTFBBNOMIX}{\hflavTFBBNOMIXval\hflavTFBBNOMIXerr}
\definemath{\hflavTFBDNOMIX}{\hflavTFBDNOMIXval\hflavTFBDNOMIXerr}
\definemath{\hflavLFBSNOMIX}{\hflavLFBSNOMIXval\hflavLFBSNOMIXerr}
\definemath{\hflavLFBBNOMIX}{\hflavLFBBNOMIXval\hflavLFBBNOMIXerr}
\definemath{\hflavLFBDNOMIX}{\hflavLFBDNOMIXval\hflavLFBDNOMIXerr}
\definemath{\hflavCHIBARTEV}{\hflavCHIBARTEVval\hflavCHIBARTEVerr}
\definemath{\hflavCHIBAR}{\hflavCHIBARval\hflavCHIBARerr}
\definemath{\hflavWFBSMIX}{\hflavWFBSMIXval\hflavWFBSMIXerr}
\definemath{\hflavTFBSMIX}{\hflavTFBSMIXval\hflavTFBSMIXerr}
\definemath{\hflavZFBSMIX}{\hflavZFBSMIXval\hflavZFBSMIXerr}
\definemath{\hflavCHIDU}{\hflavCHIDUval\hflavCHIDUerr}
\definemath{\hflavCHIDWU}{\hflavCHIDWUval\hflavCHIDWUerr}
\definemath{\hflavXDW}{\hflavXDWval\hflavXDWerr}
\definemath{\hflavXDWU}{\hflavXDWUval\hflavXDWUerr}
\definemath{\hflavDMDW}{\hflavDMDWval\hflavDMDWerr\invps}
\definemath{\hflavDMDWnounit}{\hflavDMDWval\hflavDMDWerr}
\definemath{\hflavDMDWfull}{\hflavDMDWval\hflavDMDWsta\hflavDMDWsys\invps}
\definemath{\hflavDMDWnounitfull}{\hflavDMDWval\hflavDMDWsta\hflavDMDWsys}
\definemath{\hflavDMDWU}{\hflavDMDWUval\hflavDMDWUerr\invps}
\definemath{\hflavDMDWUnounit}{\hflavDMDWUval\hflavDMDWUerr}
\definemath{\hflavDMDLHCb}{\hflavDMDLHCbval\hflavDMDLHCberr\invps}
\definemath{\hflavDMDLHCbnounit}{\hflavDMDLHCbval\hflavDMDLHCberr}
\definemath{\hflavDMDLHCbfull}{\hflavDMDLHCbval\hflavDMDLHCbsta\hflavDMDLHCbsys\invps}
\definemath{\hflavDMDLHCbnounitfull}{\hflavDMDLHCbval\hflavDMDLHCbsta\hflavDMDLHCbsys}
\definemath{\hflavZFBS}{\hflavZFBSval\hflavZFBSerr}
\definemath{\hflavZFBB}{\hflavZFBBval\hflavZFBBerr}
\definemath{\hflavZFBD}{\hflavZFBDval\hflavZFBDerr}
\definemath{\hflavWFBS}{\hflavWFBSval\hflavWFBSerr}
\definemath{\hflavWFBB}{\hflavWFBBval\hflavWFBBerr}
\definemath{\hflavWFBD}{\hflavWFBDval\hflavWFBDerr}
\definemath{\hflavTFBS}{\hflavTFBSval\hflavTFBSerr}
\definemath{\hflavTFBB}{\hflavTFBBval\hflavTFBBerr}
\definemath{\hflavTFBD}{\hflavTFBDval\hflavTFBDerr}
\definemath{\hflavLFBS}{\hflavLFBSval\hflavLFBSerr}
\definemath{\hflavLFBB}{\hflavLFBBval\hflavLFBBerr}
\definemath{\hflavLFBD}{\hflavLFBDval\hflavLFBDerr}
\definemath{\hflavZFBSBD}{\hflavZFBSBDval\hflavZFBSBDerr}
\definemath{\hflavWFBSBD}{\hflavWFBSBDval\hflavWFBSBDerr}
\definemath{\hflavTFBSBD}{\hflavTFBSBDval\hflavTFBSBDerr}
\definemath{\hflavLFBSBD}{\hflavLFBSBDval\hflavLFBSBDerr}
\definemath{\hflavDMDL}{\hflavDMDLval\hflavDMDLerr\invps}
\definemath{\hflavDMDLnounit}{\hflavDMDLval\hflavDMDLerr}
\definemath{\hflavDMDLfull}{\hflavDMDLval\hflavDMDLsta\hflavDMDLsys\invps}
\definemath{\hflavDMDLnounitfull}{\hflavDMDLval\hflavDMDLsta\hflavDMDLsys}
\definemath{\hflavDMDT}{\hflavDMDTval\hflavDMDTerr\invps}
\definemath{\hflavDMDTnounit}{\hflavDMDTval\hflavDMDTerr}
\definemath{\hflavDMDTfull}{\hflavDMDTval\hflavDMDTsta\hflavDMDTsys\invps}
\definemath{\hflavDMDTnounitfull}{\hflavDMDTval\hflavDMDTsta\hflavDMDTsys}
\definemath{\hflavDMDB}{\hflavDMDBval\hflavDMDBerr\invps}
\definemath{\hflavDMDBnounit}{\hflavDMDBval\hflavDMDBerr}
\definemath{\hflavDMDBfull}{\hflavDMDBval\hflavDMDBsta\hflavDMDBsys\invps}
\definemath{\hflavDMDBnounitfull}{\hflavDMDBval\hflavDMDBsta\hflavDMDBsys}
\definemath{\hflavDMDTWOD}{\hflavDMDTWODval\hflavDMDTWODerr\invps}
\definemath{\hflavDMDTWODnounit}{\hflavDMDTWODval\hflavDMDTWODerr}
\definemath{\hflavDMDTWODfull}{\hflavDMDTWODval\hflavDMDTWODsta\hflavDMDTWODsys\invps}
\definemath{\hflavDMDTWODnounitfull}{\hflavDMDTWODval\hflavDMDTWODsta\hflavDMDTWODsys}
\definemath{\hflavTAUBDTWOD}{\hflavTAUBDTWODval\hflavTAUBDTWODerr\ps}
\definemath{\hflavTAUBDTWODnounit}{\hflavTAUBDTWODval\hflavTAUBDTWODerr}
\definemath{\hflavTAUBDTWODfull}{\hflavTAUBDTWODval\hflavTAUBDTWODsta\hflavTAUBDTWODsys\ps}
\definemath{\hflavTAUBDTWODnounitfull}{\hflavTAUBDTWODval\hflavTAUBDTWODsta\hflavTAUBDTWODsys}
\definemath{\hflavTAUBZCALC}{\hflavTAUBZCALCval\hflavTAUBZCALCerr\ps}
\definemath{\hflavTAUBZCALCnounit}{\hflavTAUBZCALCval\hflavTAUBZCALCerr}
\definemath{\hflavQPDB}{\hflavQPDBval\hflavQPDBerr}
\definemath{\hflavQPDD}{\hflavQPDDval\hflavQPDDerr}
\definemath{\hflavQPDW}{\hflavQPDWval\hflavQPDWerr}
\definemath{\hflavQPDA}{\hflavQPDAval\hflavQPDAerr}
\definemath{\hflavASLDB}{\hflavASLDBval\hflavASLDBerr}
\definemath{\hflavASLDD}{\hflavASLDDval\hflavASLDDerr}
\definemath{\hflavASLDW}{\hflavASLDWval\hflavASLDWerr}
\definemath{\hflavASLDA}{\hflavASLDAval\hflavASLDAerr}
\definemath{\hflavREBDB}{\hflavREBDBval\hflavREBDBerr}
\definemath{\hflavREBDD}{\hflavREBDDval\hflavREBDDerr}
\definemath{\hflavREBDW}{\hflavREBDWval\hflavREBDWerr}
\definemath{\hflavREBDA}{\hflavREBDAval\hflavREBDAerr}
\definemath{\hflavASLS}{\hflavASLSval\hflavASLSerr}
\definemath{\hflavQPS}{\hflavQPSval\hflavQPSerr}
\definemath{\hflavASLD}{\hflavASLDval\hflavASLDerr}
\definemath{\hflavQPD}{\hflavQPDval\hflavQPDerr}
\definemath{\hflavREBD}{\hflavREBDval\hflavREBDerr}
\definemath{\hflavASLDNOMU}{\hflavASLDNOMUval\hflavASLDNOMUerr}
\definemath{\hflavASLSNOMU}{\hflavASLSNOMUval\hflavASLSNOMUerr}
\definemath{\hflavTANPHI}{\hflavTANPHIval\hflavTANPHIerr}
\definemath{\hflavDMS}{\hflavDMSval\hflavDMSerr\invps}
\definemath{\hflavDMSnounit}{\hflavDMSval\hflavDMSerr}
\definemath{\hflavDMSfull}{\hflavDMSval\hflavDMSsta\hflavDMSsys\invps}
\definemath{\hflavDMSnounitfull}{\hflavDMSval\hflavDMSsta\hflavDMSsys}
\definemath{\hflavXS}{\hflavXSval\hflavXSerr}
\definemath{\hflavCHIS}{\hflavCHISval\hflavCHISerr}
\definemath{\hflavRATIODGSDMS}{\hflavRATIODGSDMSval\hflavRATIODGSDMSerr}
\definemath{\hflavRATIODMDDMS}{\hflavRATIODMDDMSval\hflavRATIODMDDMSerr}
\definemath{\hflavVTDVTS}{\hflavVTDVTSval\hflavVTDVTSerr}
\definemath{\hflavVTDVTSfull}{\hflavVTDVTSval\hflavVTDVTSexx\hflavVTDVTSthe}
\definemath{\hflavXI}{\hflavXIval\hflavXIerr}
\definemath{\hflavXIfull}{\hflavXIval\hflavXIsta\hflavXIsys}

%%%%%%%%%%%%%%%%%%%%%%%%%%%%%%%%%%%%%%%%%%%%%%%

\renewcommand{\floatpagefraction}{0.8}
\renewcommand{\topfraction}{0.9}

\newcommand{\comment}[1]{}

% Fractions
\newcommand{\fBs}{\ensuremath{f_{\particle{s}}}\xspace}
\newcommand{\fBd}{\ensuremath{f_{\particle{d}}}\xspace}
\newcommand{\fBu}{\ensuremath{f_{\particle{u}}}\xspace}
\newcommand{\fbb}{\ensuremath{f_{\rm baryon}}\xspace}
\newcommand{\fLb}{\ensuremath{f_{\Lb}}\xspace}
\newcommand{\fXib}{\ensuremath{f_{\Xi_{b}}}\xspace}
\newcommand{\fOb}{\ensuremath{f_{\Omega_{b}}}\xspace}

% Mixing
\newcommand{\dmd}{\ensuremath{\Delta m_{\particle{d}}}\xspace}
\newcommand{\dms}{\ensuremath{\Delta m_{\particle{s}}}\xspace}
\newcommand{\xd}{\ensuremath{x_{\particle{d}}}\xspace}
\newcommand{\xs}{\ensuremath{x_{\particle{s}}}\xspace}
\newcommand{\yd}{\ensuremath{y_{\particle{d}}}\xspace}
\newcommand{\ys}{\ensuremath{y_{\particle{s}}}\xspace}
\newcommand{\chibar}{\ensuremath{\overline{\chi}}\xspace}
\newcommand{\chid}{\ensuremath{\chi_{\particle{d}}}\xspace}
\newcommand{\chis}{\ensuremath{\chi_{\particle{s}}}\xspace}
\newcommand{\Gd}{\ensuremath{\Gamma_{\particle{d}}}\xspace}
\newcommand{\DGd}{\ensuremath{\Delta\Gd}\xspace}
\newcommand{\DGGd}{\ensuremath{\DGd/\Gd}\xspace}
\newcommand{\Gs}{\ensuremath{\Gamma_{\particle{s}}}\xspace}
\newcommand{\DGs}{\ensuremath{\Delta\Gs}\xspace}
\newcommand{\DGGs}{\ensuremath{\Delta\Gs/\Gs}\xspace}
\newcommand{\ASLd}{\ensuremath{{\cal A}_{\rm SL}^\particle{d}}\xspace}
\newcommand{\ASLs}{\ensuremath{{\cal A}_{\rm SL}^\particle{s}}\xspace}
\newcommand{\ASLb}{\ensuremath{{\cal A}_{\rm SL}^\particle{b}}\xspace}

\newcommand{\DG}{\ensuremath{\Delta\Gamma}\xspace}
\newcommand{\phiccbars}{\ensuremath{\phi_s^{c\bar{c}s}}\xspace}

% Branching ratios, CL, ... and miscellaneous
\renewcommand{\BR}[1]{\particle{{\cal B}(#1)}}
\newcommand{\CL}[1]{#1\%~\mbox{CL}}
\newcommand{\Qjet}{\ensuremath{Q_{\rm jet}}\xspace}

% Labels, and references (equations, figures, tables, sections, ...)
\newcommand{\labe}[1]{\label{equ:#1}}
\newcommand{\labs}[1]{\label{sec:#1}}
\newcommand{\labf}[1]{\label{fig:#1}}
\newcommand{\labt}[1]{\label{tab:#1}}
\newcommand{\refe}[1]{\ref{equ:#1}}
\newcommand{\refs}[1]{\ref{sec:#1}}
\newcommand{\reff}[1]{\ref{fig:#1}}
\newcommand{\reft}[1]{\ref{tab:#1}}
\newcommand{\Ref}[1]{Ref.~\cite{#1}}
\newcommand{\Refs}[1]{Refs.~\cite{#1}}
\newcommand{\Refss}[2]{Refs.~\cite{#1} and \cite{#2}}
\newcommand{\Refsss}[3]{Refs.~\cite{#1}, \cite{#2} and \cite{#3}}
\newcommand{\eq}[1]{(\refe{#1})}
\newcommand{\Eq}[1]{Eq.~(\refe{#1})}
\newcommand{\Eqs}[1]{Eqs.~(\refe{#1})}
\newcommand{\Eqss}[2]{Eqs.~(\refe{#1}) and (\refe{#2})}
\newcommand{\Eqssor}[2]{Eqs.~(\refe{#1}) or (\refe{#2})}
\newcommand{\Eqsss}[3]{Eqs.~(\refe{#1}), (\refe{#2}), and (\refe{#3})}
\newcommand{\Figure}[1]{Figure~\reff{#1}}
\newcommand{\Figuress}[2]{Figures~\reff{#1} and \reff{#2}}
\newcommand{\Fig}[1]{Fig.~\reff{#1}}
\newcommand{\Figs}[1]{Figs.~\reff{#1}}
\newcommand{\Figss}[2]{Figs.~\reff{#1} and \reff{#2}}
\newcommand{\Figsss}[3]{Figs.~\reff{#1}, \reff{#2}, and \reff{#3}}
\newcommand{\Section}[1]{Section~\refs{#1}}
\newcommand{\Sec}[1]{Sec.~\refs{#1}}
\newcommand{\Secs}[1]{Secs.~\refs{#1}}
\newcommand{\Secss}[2]{Secs.~\refs{#1} and \refs{#2}}
\newcommand{\Secsss}[3]{Secs.~\refs{#1}, \refs{#2}, and \refs{#3}}
\newcommand{\Table}[1]{Table~\reft{#1}}
\newcommand{\Tables}[1]{Tables~\reft{#1}}
\newcommand{\Tabless}[2]{Tables~\reft{#1} and \reft{#2}}
\newcommand{\Tablesss}[3]{Tables~\reft{#1}, \reft{#2}, and \reft{#3}}

\newcommand{\subsubsubsection}[1]{\vspace{2ex}\par\noindent {\bf\boldmath\em #1} \vspace{2ex}\par}

% ---------------------
% Title of this chapter
% ---------------------

\mysection{Production fractions, lifetimes and mixing parameters of \b hadrons}
\labs{life_mix}

% ----------------------------
% Introduction to this chapter
% ----------------------------

Quantities such as \b-hadron production fractions, \b-hadron lifetimes, 
and neutral \B-meson oscillation frequencies have been studied
in the nineties at LEP and SLC (\ee colliders at $\sqrt{s}=m_{\particle{Z}}$) 
as well as at the first version of the Tevatron
(\particle{p\bar{p}} collider at $\sqrt{s}=1.8\TeV$). 
This was followed by precise measurements of the \Bd and \Bu mesons
performed at the asymmetric \B factories, KEKB and PEPII
(\ee colliders at $\sqrt{s}=m_{\Ups}$), as well as measurements related 
to the other \b hadrons, in particular \Bs, \Bc and \Lb, 
performed at the upgraded Tevatron ($\sqrt{s}=1.96\TeV$).
Since a few years, the most precise measurements are coming from the 
LHC ($pp$ collider at $\sqrt{s}=7$ and $8\TeV$),
in particular the LHCb experiment. 

In most cases, these basic quantities, although interesting by themselves,
became necessary ingredients for the more refined measurements,
such as those of decay-time dependent \CP-violating asymmetries.
It is therefore important that the best experimental
values of these quantities continue to be kept up-to-date and improved. 

In several cases, the averages presented in this section are 
needed and used as input for the results given in the subsequent sections. 
Within this section, some averages need the knowledge of other 
averages in a circular way. This coupling, which appears through the 
\b-hadron fractions whenever inclusive or semi-exclusive measurements 
have to be considered, has been reduced drastically in the past several years 
with increasingly precise exclusive measurements becoming available
and dominating practically all averages. 

In addition to \b-hadron fractions, lifetimes and 
oscillation frequencies, this section also deals with \CP violation
in the \Bd and \Bs mixing amplitudes, as well as the
\CP-violating phase $\phiccbars\simeq -2\beta_s$, which is the phase 
difference between the \Bs mixing amplitude and the 
$b\to c\bar{c}s$ decay amplitude.
The angle $\beta$, which is the equivalent of $\beta_s$ for the \Bd 
system, is discussed in Section~\ref{sec:cp_uta}. 

Throughout this section published results that have been superseded 
by subsequent publications are ignored (\ie, excluded from the averages)
and are only referred to if necessary.

% ----------------------------------------
\mysubsection{\b-hadron production fractions}
% ----------------------------------------
\labs{fractions}
 
We consider here the relative fractions of the different \b-hadron 
species found in an unbiased sample of weakly decaying \b hadrons 
produced under some specific conditions. The knowledge of these fractions
is useful to characterize the signal composition in inclusive \b-hadron 
analyses, to predict the background composition in exclusive analyses, 
or to convert (relative) observed rates into (relative) branching fraction 
measurements. 
% Many \B-physics analyses need these fractions as input.
We distinguish 
here the following three conditions: \Ups decays, \Upsfive decays, and 
high-energy collisions (including \Z decays). 

% -------------------------------------------------------------
\mysubsubsection{\b-hadron production fractions in \Ups decays}
% -------------------------------------------------------------
\labs{fraction_Ups4S}

Only pairs of the two lightest (charged and neutral) \B mesons 
can be produced in \Ups decays. 
Therefore only the following two branching fractions must be considered: 
\begin{eqnarray}
f^{+-} & = & \Gamma(\Ups \to \particle{B^+B^-})/
             \Gamma_{\rm tot}(\Ups)  \,, \\
f^{00} & = & \Gamma(\Ups \to \particle{B^0\bar{B}^0})/
             \Gamma_{\rm tot}(\Ups) \,.
\end{eqnarray}
In practice, most analyses measure their ratio
\begin{equation}
R^{+-/00} = f^{+-}/f^{00} = \Gamma(\Ups \to \particle{B^+B^-})/
             \Gamma(\Ups \to \particle{B^0\bar{B}^0}) \,,
\end{equation}
which is easier to access experimentally.
Since an inclusive (but separate) reconstruction of 
\Bu and \Bd is difficult, exclusive decay modes to specific final states $f$, 
${\Bu} \to f^+$ and ${\Bd} \to f^0$, are usually considered to perform 
a measurement of $R^{+-/00}$, whenever they can be related by 
isospin symmetry (for example \particle{\Bu \to \jpsi K^+} and 
\particle{\Bd \to \jpsi K^0}).
Under the assumption that $\Gamma(\Bu \to f^+) = \Gamma(\Bd \to f^0)$, 
\ie, that isospin invariance holds in these \B decays,
the ratio of the number of reconstructed
$\Bu \to f^+$ and $\Bd \to f^0$ mesons, after correcting for efficiency, is
proportional to
\begin{equation}
\frac{f^{+-}\, \BR{\Bu\to f^+}}{f^{00}\, \BR{\Bd\to f^0}}
= \frac{f^{+-}\, \Gamma({\Bu}\to f^+)\, \tau(\Bu)}%
{f^{00}\, \Gamma({\Bd}\to f^0)\,\tau(\Bd)}
= \frac{f^{+-}}{f^{00}} \, \frac{\tau(\Bu)}{\tau(\Bd)}  \,, 
\end{equation} 
where $\tau(\Bu)$ and $\tau(\Bd)$ are the \Bu and \Bd 
lifetimes respectively.
Hence the primary quantity measured in these analyses 
is $R^{+-/00} \, \tau(\Bu)/\tau(\Bd)$, 
and the extraction of $R^{+-/00}$ with this method therefore 
requires the knowledge of the $\tau(\Bu)/\tau(\Bd)$ lifetime ratio. 

\begin{table}
\caption{Published measurements of the $\Bu/\Bd$ production ratio
in \Ups decays, together with their average (see text).
Systematic uncertainties due to the imperfect knowledge of 
$\tau(\Bu)/\tau(\Bd)$ are included. 
%% The latest \babar result~\cite{Aubert:2004rz}
%% supersedes the earlier \babar measurements~\cite{Aubert:2001xs,Aubert:2004ur}.
}
\labt{R_data}
\begin{center}
\begin{tabular}{lccll}
\hline
Experiment, year  & Ref. & Decay modes & Published value of & Assumed value \\
& & or method & $R^{+-/00}=f^{+-}/f^{00}$ & of $\tau(\Bu)/\tau(\Bd)$ \\
\hline
CLEO,   2001 & \cite{Alexander:2000tb}  & \particle{\jpsi K^{(*)}} 
             & $1.04 \pm0.07 \pm0.04$ & $1.066 \pm0.024$ \\
%superseded% \babar, 2002 & \cite{Aubert:2001xs} & \particle{(c\bar{c})K^{(*)}}
%superseded%              & $1.10 \pm0.06 \pm0.05$ & $1.062 \pm0.029$\\ 
CLEO,   2002 & \cite{Athar:2002mr}  & \particle{D^*\ell\nu}
             & $1.058 \pm0.084 \pm0.136$ & $1.074 \pm0.028$\\
\belle, 2003 & \citehistory{Hastings:2002ff}{Hastings:2002ff,*Abe:2000yh_hist} & Dilepton events 
             & $1.01 \pm0.03 \pm0.09$ & $1.083 \pm0.017$\\
%superseded% \babar, 2004 & \cite{Aubert:2004ur} & \particle{\jpsi K}
%superseded%              & $1.006 \pm0.036 \pm0.031$ & $1.083 \pm0.017$ \\
\babar, 2005 & \citehistory{Aubert:2004rz}{Aubert:2004rz,*Aubert:2001xs_hist,*Aubert:2004ur_hist} & \particle{(c\bar{c})K^{(*)}}
             & $1.06 \pm0.02 \pm0.03$ & $1.086 \pm0.017$\\ 
\hline
Average      & & & \hflavFF~(tot) & \hflavRTAUBU \\
\hline
\end{tabular}
\end{center}
\end{table}

The published measurements of $R^{+-/00}$ are listed 
in \Table{R_data}\footnote{An old and imprecise measurement from
CLEO~\cite{Barish:1994mu} is not 
included in \Table{R_data} nor in the average.}
together with the corresponding assumed values of 
$\tau(\Bu)/\tau(\Bd)$.
All measurements are based on the above-mentioned method, 
except the one from \belle, which is a by-product of the 
\Bd mixing frequency analysis using dilepton events
(but note that it also assumes isospin invariance, 
namely $\Gamma(\Bu \to \ell^+{\rm X}) = \Gamma(\Bd \to \ell^+{\rm X})$).
The latter is therefore treated in a slightly different 
manner in the following procedure used to combine 
these measurements:
\begin{itemize} 
\item each published value of $R^{+-/00}$ from CLEO and \babar
      is first converted back to the original measurement of 
      $R^{+-/00} \, \tau(\Bu)/\tau(\Bd)$, using the value of the 
      lifetime ratio assumed in the corresponding analysis;
\item a simple weighted average of these original
      measurements of $R^{+-/00} \, \tau(\Bu)/\tau(\Bd)$ from 
      CLEO and \babar
      % (which do not depend on the assumed value of the lifetime ratio)
      is then computed, assuming no 
      statistical or systematic correlations between them;

% {\em ***** this may not be true in the case of \babar;
% waiting for more information 
% from David about averaging of the two \babar results ****}
% \marginpar{David}

\item the weighted average of $R^{+-/00} \, \tau(\Bu)/\tau(\Bd)$ 
      is converted into a value of $R^{+-/00}$, using the latest 
      average of the lifetime ratios, $\tau(\Bu)/\tau(\Bd)=\hflavRTAUBU$ 
      (see \Sec{lifetime_ratio});
\item the \belle measurement of $R^{+-/00}$ is adjusted to the 
      current values of $\tau(\Bd)=\hflavTAUBD$ and 
      $\tau(\Bu)/\tau(\Bd)=\hflavRTAUBU$ (see \Sec{lifetime_ratio}),
      using the quoted systematic uncertainties due to these parameters;
\item the combined value of $R^{+-/00}$ from CLEO and \babar is averaged 
      with the adjusted value of $R^{+-/00}$ from \belle, assuming a 100\% 
      correlation of the systematic uncertainty due to the limited 
      knowledge on $\tau(\Bu)/\tau(\Bd)$; no other correlation is considered. 
\end{itemize} 
The resulting global average, 
\begin{equation}
R^{+-/00} = \frac{f^{+-}}{f^{00}} =  \hflavFF \,,
\labe{Rplusminus}
\end{equation}
is consistent with equal production rate of charged and neutral \B mesons, 
although only at the $\hflavNSIGMAFF\,\sigma$ level.

On the other hand, the \babar collaboration has 
performed a direct measurement of the $f^{00}$ fraction 
using an original method, which neither relies on isospin symmetry nor requires 
the knowledge of $\tau(\Bu)/\tau(\Bd)$. Its analysis, 
based on a comparison between the number of events where a single 
$B^0 \to D^{*-} \ell^+ \nu$ decay could be reconstructed and the number 
of events where two such decays could be reconstructed, yields~\cite{Aubert:2005bq}
\begin{equation}
f^{00}= 0.487 \pm 0.010\,\mbox{(stat)} \pm 0.008\,\mbox{(syst)} \,.
% f^{00}= \hflavFNN \,.
\labe{fzerozero}
\end{equation}

The two results of \Eqss{Rplusminus}{fzerozero} are of very different natures 
and completely independent of each other. 
Their product is equal to $f^{+-} = \hflavFPROD$, 
while another combination of them gives $f^{+-} + f^{00}= \hflavFSUM$, 
compatible with unity.
Assuming\footnote{A few non-$\B\bar{B}$
decay modes of the $\Upsilon(4S)$ 
($\Upsilon(1S)\pi^+\pi^-$,
$\Upsilon(2S)\pi^+\pi^-$, $\Upsilon(1S)\eta$) 
have been observed with branching fractions
of the order of $10^{-4}$~\cite{Aubert:2006bm,Sokolov:2006sd,Aubert:2008az},
corresponding to a partial
width several times larger than that in the \ee channel.
However, this can still be
neglected and the assumption $f^{+-}+f^{00}=1$ remains valid
in the present context of the determination of $f^{+-}$ and $f^{00}$.}
 $f^{+-}+f^{00}= 1$, also consistent with 
CLEO's observation that the fraction of \Ups decays 
to \BB pairs is larger than 0.96 at \CL{95}~\cite{Barish:1995cx},
the results of \Eqss{Rplusminus}{fzerozero}
can be averaged (first converting \Eq{Rplusminus} 
into a value of $f^{00}=1/(R^{+-/00}+1)$) 
to yield the following more precise estimates:
\begin{equation}
f^{00} = \hflavFNW  \,,~~~ f^{+-} = 1 -f^{00} =  \hflavFCW \,,~~~
\frac{f^{+-}}{f^{00}} =  \hflavFFW \,.
\end{equation}
The latter ratio differs from one by $\hflavNSIGMAFFW\,\sigma$.

%%%%%%%%%%%%%%%%%%%%%%%%%%%%%%%%%%%%%%%%%%%%%%%%%%%%%%%%%%%%%%%%%%%%%%%%%%%%
%%%%% NB: it is a circular argument to assume isospin invariance in B decays
%%%%%     to test isopsin invariance in Ups(4S) decays, see for example
%%%%%     https://arxiv.org/abs/1510.03423 by Martin Jung
%%%%%%%%%%%%%%%%%%%%%%%%%%%%%%%%%%%%%%%%%%%%%%%%%%%%%%%%%%%%%%%%%%%%%%%%%%%%

% -------------------------------------------------------------
\mysubsubsection{\b-hadron production fractions in \Upsfive decays}
% -------------------------------------------------------------
\labs{fraction_Ups5S}

\newcommand{\fsfive}{\ensuremath{f^{\Upsfive}_{s}}}
\newcommand{\fudfive}{\ensuremath{f^{\Upsfive}_{u,d}}}
\newcommand{\fnBfive}{\ensuremath{f^{\Upsfive}_{B\!\!\!\!/}}}

Hadronic events produced in $e^+e^-$ collisions at the \Upsfive (also known as
$\Upsilon(10860)$) energy can be classified into three categories: 
light-quark ($u$, $d$, $s$, $c$) continuum events, $b\bar{b}$ continuum events,
and \Upsfive events. The latter two cannot be distinguished and will be called
$b\bar{b}$ events in the following. These $b\bar{b}$ events, which also include 
$b\bar{b}\gamma$ events because of possible initial-state radiation, 
can hadronize in different final states.
We define \fudfive\ as
the fraction of $b\bar{b}$ events with a pair of non-strange 
bottom mesons 
($B\bar{B}$, $B\bar{B}^*$, $B^*\bar{B}$, $B^*\bar{B}^*$,
$B\bar{B}\pi$, $B\bar{B}^*\pi$, $B^*\bar{B}\pi$,
$B^*\bar{B}^*\pi$, and $B\bar{B}\pi\pi$ final states, 
where
$B$ denotes a $B^0$ or $B^+$ meson and 
$\bar{B}$ denotes a $\bar{B}^0$ or $B^-$ meson), \fsfive\ as
the fraction of $b\bar{b}$ events with a pair of strange bottom mesons
($B_s^0\bar{B}_s^0$, $B_s^0\bar{B}_s^{*0}$, $B_s^{*0}\bar{B}_s^0$, and
$B_s^{*0}\bar{B}_s^{*0}$ final states), and 
\fnBfive\ as the fraction of $b\bar{b}$ events without 
any bottom meson in the final state.
Note that the excited bottom-meson states decay via $B^* \to B \gamma$ and
$B_s^{*0} \to B_s^0 \gamma$.
These fractions satisfy
\begin{equation}
\fudfive + \fsfive + \fnBfive = 1 \,.
\labe{sum_frac_five}
\end{equation} 

\begin{table}
\caption{Published measurements of \fsfive, obtained 
assuming $\fnBfive=0$ and 
quoted as in the original publications, except for the 2010
Belle measurement, which is quoted as 
$1-\fudfive$ with \fudfive\ from \Ref{Drutskoy:2010an}.
Our average of \fsfive\ assuming $\fnBfive=0$, given on the 
penultimate line, does not include the most recent Belle result
quoted on the last line.\footref{foot:life_mix:Esen:2012yz}}
\labt{fsFiveS}
\begin{center}
\begin{tabular}{lll}
\hline
Experiment, year, dataset                 & Decay mode or method    & Value of \fsfive\  \\
\hline
CLEO, 2006, 0.42\invfb~\citehistory{Huang:2006em}{Huang:2006em_hist} & $\Upsfive\to D_{s}X$     & $0.168 \pm 0.026^{+0.067}_{-0.034}$  \\
             & $\Upsfive \to \phi X$    & $0.246 \pm 0.029^{+0.110}_{-0.053}$ \\
             & $\Upsfive \to B\bar{B}X$ & $0.411 \pm 0.100 \pm 0.092$ \\  
             & CLEO average of above 3  & $0.21^{+0.06}_{-0.03}$      \\  \hline
Belle, 2006, 1.86\invfb~\cite{Drutskoy:2006fg} & $\Upsfive \to D_s X$     & $0.179 \pm 0.014 \pm 0.041$ \\
             & $\Upsfive \to D^0 X$     & $0.181 \pm 0.036 \pm 0.075$ \\  
             & Belle average of above 2 & $0.180 \pm 0.013 \pm 0.032$ \\  \hline 
Belle, 2010, 23.6\invfb~\cite{Drutskoy:2010an} & $\Upsfive \to B\bar{B}X$ & $0.263 \pm 0.032 \pm 0.051$ % \,$^a$
\\ \hline
\multicolumn{2}{l}{Average of all above %%RL
after adjustments to inputs of \Table{fsFiveS_external}} & %%RL
\hflavFSFIVERL             \\  \hline 
Belle, 2012, 121.4\invfb~\cite{Esen:2012yz} & $\Upsfive \to D_sX, D^0X$ & $0.172 \pm 0.030$ \\ \hline
%%%\multicolumn{4}{l}{$^a$ {\footnotesize 
%%%We quote here $1-\fudfive$, with \fudfive\ from \Ref{Drutskoy:2010an}.}}
\end{tabular}
\end{center}
\end{table}

\begin{table}
\caption{External inputs on which the \fsfive\ averages are based.}
\labt{fsFiveS_external}
\begin{center}
\begin{tabular}{lcl}
\hline
Branching fraction   & Value     & Explanation and reference \\
\hline
${\cal B}(B\to D_s X)\times {\cal B}(D_s \to \phi\pi)$ & 
$0.00374\pm 0.00014$ & Derived from~\cite{PDG_2016} \\
${\cal B}(B^0_s \to D_s X)$ & 
$0.92\pm0.11$ & Model-dependent estimate~\cite{Artuso:2005xw} \\
${\cal B}(D_s \to \phi\pi)$ & 
$0.045\pm0.004$ &\!\!\cite{PDG_2016} \\
${\cal B}(B\to D^0 X)\times {\cal B}(D^0 \to K\pi)$ & 
$0.0243\pm0.0011$ & Derived from~\cite{PDG_2016} \\
${\cal B}(B^0_s \to D^0 X)$ & 
$0.08\pm0.07$ & Model-dependent estimate~\cite{Drutskoy:2006fg,Artuso:2005xw} \\
${\cal B}(D^0 \to K\pi)$ & 
$0.0393\pm0.0004$ & \!\!\cite{PDG_2016} \\
${\cal B}(B \to \phi X)$ & 
$0.0343\pm0.0012$ &\!\!\cite{PDG_2016} \\
${\cal B}(B^0_s \to \phi X)$ &
$0.161\pm0.024$ & Model-dependent estimate~\citehistory{Huang:2006em}{Huang:2006em_hist} \\
\hline
\end{tabular}
\end{center}
\end{table}

The CLEO and Belle collaborations have published % in 2006
measurements of several inclusive \Upsfive branching fractions, 
${\cal B}(\Upsfive\to D_s X)$, 
${\cal B}(\Upsfive\to \phi X)$ and 
${\cal B}(\Upsfive\to D^0 X)$, %and%%RL 
%${\cal B}(\Upsfive\to B\bar{B} X)$, %%RL
from which they extracted the
model-dependent estimates of \fsfive\ reported in \Table{fsFiveS}.
This extraction was performed under the implicit assumption  
$\fnBfive=0$, using the relation 
\begin{equation}
\frac12{\cal B}(\Upsfive\to D_s X)=\fsfive\times{\cal B}(B_s^0\to D_s X) + 
\left(1-\fsfive-\fnBfive\right)\times{\cal B}(B\to D_s X) \,,
\labe{Ds_correct}
\end{equation}
and similar relations for
${\cal B}(\Upsfive\to D^0 X)$ and ${\cal B}(\Upsfive\to \phi X)$.
%%For completeness, %RL
In \Table{fsFiveS} we list also
the values of \fsfive\ derived from measurements of
$\fudfive={\cal B}(\Upsfive\to B\bar BX)$~\citehistory{Huang:2006em,Drutskoy:2010an}{Huang:2006em_hist,Drutskoy:2010an}, 
as well as our average value of  \fsfive, all obtained under the assumption $\fnBfive=0$.

However, the assumption $\fnBfive=0$ is known to be invalid since the observation of
the following final states in $e^+e^-$ collisions at the \Upsfive\ energy:
$\Upsilon(1S)\pi^+\pi^-$,
$\Upsilon(2S)\pi^+\pi^-$,
$\Upsilon(3S)\pi^+\pi^-$
and
$\Upsilon(1S)K^+K^-$~\citehistory{Abe:2007tk,Garmash:2014dhx}{Abe:2007tk,Garmash:2014dhx_hist},
$h_b(1P)\pi^+\pi^-$ and 
$h_b(2P)\pi^+\pi^-$~\cite{Adachi:2011ji},
and more recently 
$\Upsilon(1S)\pi^0\pi^0$,
$\Upsilon(2S)\pi^0\pi^0$ 
and
$\Upsilon(3S)\pi^0\pi^0$~\cite{Krokovny:2013mgx}.
The sum of the measurements of the corresponding visible cross-sections,
adding also the contributions of the unmeasured
$\Upsilon(1S)K^0\bar{K}^0$, $h_b(1P)\pi^0\pi^0$ and $h_b(2P)\pi^0\pi^0$ final states
assuming isospin conservation, amounts to
$$
\sigma^{\rm vis}(e^+e^-\to (\b\bar{\b})hh) = 13.2\pm1.4~{\rm pb} \,,
~~\mbox{for $(\b\bar{\b})=\Upsilon(1S,2S,3S),h_b(1P,2P)$ and $hh=\pi\pi,KK$}\,.
$$
We divide this by the $\b\bar{\b}$ production cross section, 
$\sigma(e^+e^- \to \b\bar{\b} X) = 337 \pm 15$~pb, obtained as the average of the 
CLEO~\cite{Artuso:2005xw} and Belle~\cite{Esen:2012yz}\footref{foot:life_mix:Esen:2012yz}
measurements, to obtain
$$
{\cal B}(\Upsfive\to (\b\bar{\b})hh) = 0.039\pm0.004 \,,
~~\mbox{for $(\b\bar{\b})=\Upsilon(1S,2S,3S),h_b(1P,2P)$ and $hh=\pi\pi,KK$}\,,
$$
which is to be considered as a lower bound for \fnBfive. 

%old% However, the assumption $\fnBfive=0$ is no longer valid since the 
%old% observation of \Upsfive\ decays to $\Upsilon(1S)\pi^+\pi^-$,
%old% $\Upsilon(2S)\pi^+\pi^-$,
%old% $\Upsilon(3S)\pi^+\pi^-$ and
%old% $\Upsilon(1S)K^+K^-$~\cite{Abe:2007tk},
%old% and more recently to 
%old% $h_b(1P)\pi^+\pi^-$ and 
%old% $h_b(2P)\pi^+\pi^-$~\cite{Adachi:2011ji}.
%old% The sum of these measured branching fractions, adding also the 
%old% contributions of the 
%old% $\Upsilon(1S)\pi^0\pi^0$, $\Upsilon(2S)\pi^0\pi^0$, $\Upsilon(3S)\pi^0\pi^0$,
%old% $\Upsilon(1S)K^0\bar{K}^0$, $h_b(1P)\pi^0\pi^0$ and $h_b(2P)\pi^0\pi^0$ final states
%old% assuming isospin conservation, amounts to
%old% $$
%old% {\cal B}(\Upsfive\to (\b\bar{\b})hh) = 0.042\pm0.006 \,,
%old% ~~~\mbox{for $(\b\bar{\b})=\Upsilon(1S,2S,3S),h_b(1P,2P)$ and $hh=\pi\pi,KK$}\,,
%old% $$
%old% which is to be considered as a lower bound for \fnBfive. 

%% Since the observation of \Upsfive\ decays to final states without 
%% bottom hadrons~\cite{Abe:2007tk},
%% the assumption $\fnBfive=0$ is no longer valid. The measured \Upsfive\ decays
%% to final states without bottom meson are summarized in \Table{fnonB}, together with 
%% our estimate of the $\Upsfive \to (\b\bar{\b}) hh$ branching fraction, where 
%% $(\b\bar{\b})$ is a $\Upsilon(1S)$, $\Upsilon(2S)$,
%% $\Upsilon(3S)$, $h_b(1P)$ or $h_b(2P)$ mesons, 
%% and $hh$ is a pair of neutral or oppositely-charged pions or kaons. 
Following the method described in \Ref{thesis_Louvot}, 
we perform a $\chi^2$ fit of the original 
measurements of the \Upsfive\ branching fractions of Refs.~%
\citehistory{Huang:2006em,Drutskoy:2006fg,Drutskoy:2010an}{Huang:2006em_hist,Drutskoy:2006fg,Drutskoy:2010an},\footnote{%
   \label{foot:life_mix:Esen:2012yz}
   Belle updated the analysis of \Ref{Drutskoy:2006fg} with the full \Upsfive dataset.
The resulting measurements of $\sigma(e^+e^- \to \b\bar{\b} X)$ and \fsfive,
   which supersede those of \Ref{Drutskoy:2006fg}, 
   are quoted and used in \Ref{Esen:2012yz}. However, no details are given.
   Because of the lack of relevant information, this measurement of \fsfive\
   cannot be included in the averages presented here.
} % end footnote
using the inputs of \Table{fsFiveS_external},
the relations of \Eqss{sum_frac_five}{Ds_correct} and the
one-sided Gaussian constraint $\fnBfive \ge {\cal B}(\Upsfive \to (\b\bar{\b}) hh)$,
to simultaneously extract \fudfive, \fsfive\ and \fnBfive. Taking all known 
correlations into account, the best fit values are
\begin{eqnarray}
\fudfive &=& \hflavFUDFIVE \,, \labe{fudfive} \\
\fsfive  &=& \hflavFSFIVE  \,, \labe{fsfive}  \\
\fnBfive &=& \hflavFNBFIVE \,, \labe{fnBfive}
\end{eqnarray}
where the strongly asymmetric uncertainty on \fnBfive\ is due to the one-sided constraint
from the observed $(\b\bar{\b}) hh$ decays. These results, together with their correlation, 
imply
\begin{eqnarray}
\fsfive/\fudfive  &=& \hflavFSFUDFIVE  \,, \labe{fsfudfive} 
\end{eqnarray}
in fair agreement with the results of a \babar
analysis~\cite{Lees:2011ji}, performed as a function 
of centre-of-mass energy.\footnote{%
\label{foot:life_mix:Lees:2011ji}
The results of \Ref{Lees:2011ji} are not included in the average 
since no numerical value is given for $\fsfive/\fudfive$.
%%This has not been included in the average, since 
%%no numerical value is given for $\fsfive/\fudfive$ in 
%%\Ref{Lees:2011ji}.
}

The production of $B^0_s$ mesons at the \Upsfive
is observed to be dominated by the $B_s^{*0}\bar{B}_s^{*0}$
channel, %~\cite{Bonvicini:2005ci,Drutskoy:2006xc,Louvot:2008sc},
with $\sigma(e^+e^- \to B_s^{*0}\bar{B}_s^{*0})/%
\sigma(e^+e^- \to B_s^{(*)0}\bar{B}_s^{(*)0})
= (87.0\pm 1.7)\%$~\cite{Li:2011pg,Louvot:2008sc}.
% = (90.1^{+3.8}_{-4.0}\pm 0.2)\%$~\cite{Louvot:2008sc}.
The proportions of the various production channels 
for non-strange $B$ mesons have also been measured~\cite{Drutskoy:2010an}.

%--------------------------------------------------------------
\mysubsubsection{\b-hadron production fractions at high energy}
%--------------------------------------------------------------
\labs{fractions_high_energy}
\labs{chibar}

At high energy, all species of weakly decaying \b hadrons 
may be produced, either directly or in strong and electromagnetic 
decays of excited \b hadrons. It is often assumed that the fractions 
of these different species are the same in unbiased samples of 
high-$p_{\rm T}$ \b jets originating from \particle{Z^0} decays, 
from \particle{p\bar{p}} collisions at the Tevatron, or from 
\particle{p p} collisions at the LHC.
This hypothesis is plausible under the condition that the square of
the momentum transfer to the produced \b quarks, $Q^2$, is large compared 
with the square of the hadronization energy scale, 
$Q^2 \gg \Lambda_{\rm QCD}^2$.
On the other hand, there is no strong argument that the
fractions at different machines should be strictly equal, so 
this assumption should be checked experimentally. 
%  RJT 10/16/16 New text
The available data show that the fractions depend on the kinematics 
of the produced \b hadron.
A simple phenomenological model appears to agree with
all data and indicates that the fractions are constant if the \b hadron
is produced with sufficiently high transverse momentum from any collider.
%Although the 
%available data is not sufficient at this time to perform a definitive
%check, it is expected that more refined 
%or new analyses from LHC 
%experiments may improve this situation and allow one to confirm or 
%disprove this assumption with reasonable confidence. 
Unless otherwise indicated, these fractions are assumed to be equal at 
all high-energy colliders until demonstrated otherwise by experiment.
%%% \footnote{Both CDF and 
%%% LHCb report a $p_{\rm T}$ dependence for \Lb production relative to
%%% ${\Bu}$ and ${\Bd}$.  The reported $p_{\rm T}$ dependence enhances the 
%%% number of \Lb baryons observed at low-$p_{\rm T}$ compared with the
%%% LEP results.}
%%% However, as explained below, the measurements performed at LEP, at 
%%% the Tevatron and from LHCb show discrepancies.
Both CDF and LHCb report a $p_{\rm T}$ dependence for \Lb
production relative to \Bu and \Bd; the number of \Lb baryons
observed at low $p_{\rm T}$ is enhanced with respect to that 
seen at LEP's higher $p_{\rm T}$.
Therefore we present 
three sets of complete averages: one set including only measurements 
performed at LEP, a second set including only measurements performed 
at the Tevatron, a third  set including measurements performed at LEP, 
Tevatron and LHC.  The LHCb production fractions results by themselves 
are still incomplete, lacking measurements of the production of 
weakly-decaying baryons heavier than \Lb.
%2015% other weakly decaying heavy-flavour baryons, $\Xi_b$ and $\Omega_b$, and a measurement of 
%%% $\overline{\chi}$ giving an extra constraint between \fBd and \fBs.
%2015% the average mixing probability \chibar defined in Eq.~(\refe{chibar}).

Contrary to what happens in the charm sector where the fractions of 
\particle{D^+} and \particle{D^0} are different, the relative amount 
of \Bu and \Bd is not affected by the electromagnetic decays of 
excited $B^{*+}$ and $B^{*0}$ states and strong decays of excited 
$B^{**+}$ and $B^{**0}$ states. Decays of the type 
\particle{B_s^{**0} \to B^{(*)}K} also contribute to the \Bu and \Bd rates, 
but with the same magnitude if mass effects can be neglected.  
We therefore assume equal production of \Bu and \Bd mesons. We also  
neglect the production of weakly decaying states
made of several heavy quarks (like \Bc and doubly heavy baryons) 
which is known to be very small. Hence, for the purpose of determining 
the \b-hadron fractions, we use the constraints
\begin{equation}
\fBu = \fBd ~~~~\mbox{and}~~~ \fBu + \fBd + \fBs + \fbb = 1 \,,
\labe{constraints}
\end{equation}
where \fBu, \fBd, \fBs and \fbb
are the unbiased fractions of \Bu, \Bd, \Bs and \b baryons, respectively.

We note that there are many measurements of the production cross-sections of
different species of \b hadrons.
In principle these could be included in a global fit to determine the
production fractions.
We do not perform such a fit at the current time, and instead average only the
explicit measurements of the production fractions.

The LEP experiments have measured
$\fBs \times \BR{\Bs\to\particle{D_s^-} \ell^+ \nu_\ell \mbox{$X$}}$~\cite{Abreu:1992rv,Acton:1992zq,Buskulic:1995bd}, 
$\BR{\b\to\Lb} \times \BR{\Lb\to\Lc\ell^-\bar{\nu}_\ell \mbox{$X$}}$~\cite{Abreu:1995me,Barate:1997if}
and $\BR{\b\to\Xib^-} \times \BR{\Xi_b^- \to \Xi^-\ell^-\overline\nu_\ell 
\mbox{$X$}}$~\citehistory{Buskulic:1996sm,Abdallah:2005cw}{Buskulic:1996sm,Abdallah:2005cw_hist}
from partially reconstructed final states including a lepton, \fbb
from protons identified in \b events~\cite{Barate:1997ty}, and the 
production rate of charged \b hadrons~\cite{Abdallah:2003xp}. 
Ratios of \b-hadron fractions have been measured by CDF using 
lepton+charm final 
states~\cite{Affolder:1999iq,Aaltonen:2008zd,Aaltonen:2008eu}\footnote{
  \label{foot:life_mix:Affolder:1999iq}
  CDF updated their measurement of \fLb/\fBd~\cite{Affolder:1999iq} to account 
  for a measured $p_{\rm T}$ dependence between exclusively reconstructed 
  \Lb and $B^0$~\cite{Aaltonen:2008eu}.
}\unpublished{ and}{,} double semileptonic decays 
with \particle{K^*\mu\mu} and \particle{\phi\mu\mu}
final states~\cite{Abe:1999ta}\unpublished{.}{,
and fully reconstructed $\Bs\to\jpsi\phi$ decays~\cite{CDFnote10795:2012}.}
Measurements of the production of other heavy 
flavour baryons at the Tevatron are included in the determination of 
\fbb~\cite{Abazov:2007am,Abazov:2008qm,Aaltonen:2009ny}\footnote{
  \label{foot:life_mix:Abazov:2008qm}
  \dzero reports $f_{\Omega_b^-}/f_{\Xi_b^-}$.  We use the CDF+\dzero average of 
  $f_{\Xi_b^-}/f_{\Lb}$ to obtain $f_{\Omega_b^-}/f_{\Lb}$ and then 
  combine it with the CDF result.
} using the constraint
\begin{eqnarray}
\fbb & = & f_{\Lb} + f_{\Xi_b^0} + f_{\Xi_b^-} + f_{\Omega_b^-} 
     \nonumber \\
     & = & f_{\Lb}\left(1 + 2\frac{f_{\Xi_b^-}}{f_{\Lb}} 
           + \frac{f_{\Omega_b^-}}{f_{\Lb}}\right),
\end{eqnarray}
where isospin invariance is assumed in the production of $\Xi_b^0$ and 
$\Xi_b^-$. Other \b baryons are expected to decay strongly or 
electromagnetically to those baryons listed. For the production 
measurements, both CDF and \dzero\ reconstruct their \b baryons exclusively 
to final states which include a $\jpsi$ and a hyperon 
($\Lb\to \jpsi \Lambda$, 
$\Xi_b^- \rightarrow \jpsi \Xi^-$ and 
$\Omega_b^- \rightarrow \jpsi \Omega^-$).  
We assume that the partial decay width of a \b baryon to a $\jpsi$ and the 
corresponding hyperon is equal to the partial width of any other \b baryon to 
a $\jpsi$ and the corresponding hyperon.  LHCb has also measured
ratios of \b-hadron fractions in charm+lepton final states~\cite{Aaij:2011jp} 
and in fully reconstructed hadronic two-body decays $\Bd \to D^-\pi^+$, $\Bs \to D_s^- \pi^+$ and 
$\Lb \to \Lc \pi^-$~\citehistory{Aaij:2013qqa,Aaij:2014jyk}{Aaij:2013qqa,*Aaij:2011hi_hist,Aaij:2014jyk}.

Both CDF and LHCb observe that the ratio $\fLb/\fBd$ depends on the $p_{\rm T}$
of the charm+lepton system~\cite{Aaltonen:2008eu,Aaij:2011jp}.%
\footnote{
  \label{foot:life_mix:Aaltonen:2008eu}
  CDF compares the $p_{\rm T}$ distribution of fully reconstructed 
  $\Lb \to \Lc \pi^-$ 
  with $\Bzb\rightarrow D^+\pi^-$, which 
  gives $\fLb/\fBd$ up to a scale factor. LHCb compares the $p_{\rm T}$ 
  in the charm+lepton system between \Lb and \Bd and \Bu, giving
  $R_{\Lb}/2 = \fLb/(\fBu+\fBd) = \fLb/2\fBd$.}
CDF chose to correct an older result to account for the $p_{\rm T}$ dependence.
In a second result, CDF binned their data in $p_{\rm T}$ of the charm+electron 
system~\cite{Aaltonen:2008zd}.
The more recent LHCb measurement using hadronic decays~\cite{Aaij:2014jyk} 
obtains the scale for $R_{\Lb} = \fLb/\fBd$ from their previous 
charm + lepton data~\cite{Aaij:2011jp}.  The LHCb measurement using hadronic
data also bins the same data in pseudorapidity ($\eta$) and sees a 
linear dependence of $R_{\Lb}$.  Since $\eta$ is not entirely
independent of $p_{\rm T}$ it is impossible to tell at this time whether 
this dependence is just an artifact of the $p_{\rm T}$ dependence.
\Figure{rlb_comb} shows the ratio $R_{\Lb}$ as a function of 
$p_{\rm T}$ for the \b hadron, as measured by LHCb.  LHCb fits their
scaled hadronic data to obtain
\begin{equation}
R_{\Lb} = (0.151\pm 0.030) + 
  \exp{\left\{-(0.57\pm 0.11) - 
  (0.095\pm 0.016)[\gevc]^{-1} \times p_{\rm T}\right\}}.
\end{equation}
A value of
$R_{\Lb}$ is also calculated for LEP and placed at the approximate $p_{\rm T}$ for the charm+lepton
system, but this value does not participate in any fit.\footnote{
  \label{foot:life_mix:Aaltonen:2008zd}
  The CDF semileptonic data would require significant corrections to obtain the $p_{\rm T}$ of the \b hadron and be included on the same plot with the LHCb data.
  We do not have these corrections at this time.}
Because the two LHCb results for $R_{\Lb}$ are not 
independent, we use only their semileptonic data for the averages.
Note that the $p_{\rm T}$ dependence
of $R_{\Lb}$ combined with the constraint from \Eq{constraints} implies
a compensating $p_{\rm T}$ dependence in one or more of the production fractions, \fBu, \fBd,
or \fBs.

\begin{figure}
 \begin{center}
  \includegraphics[width=\textwidth]{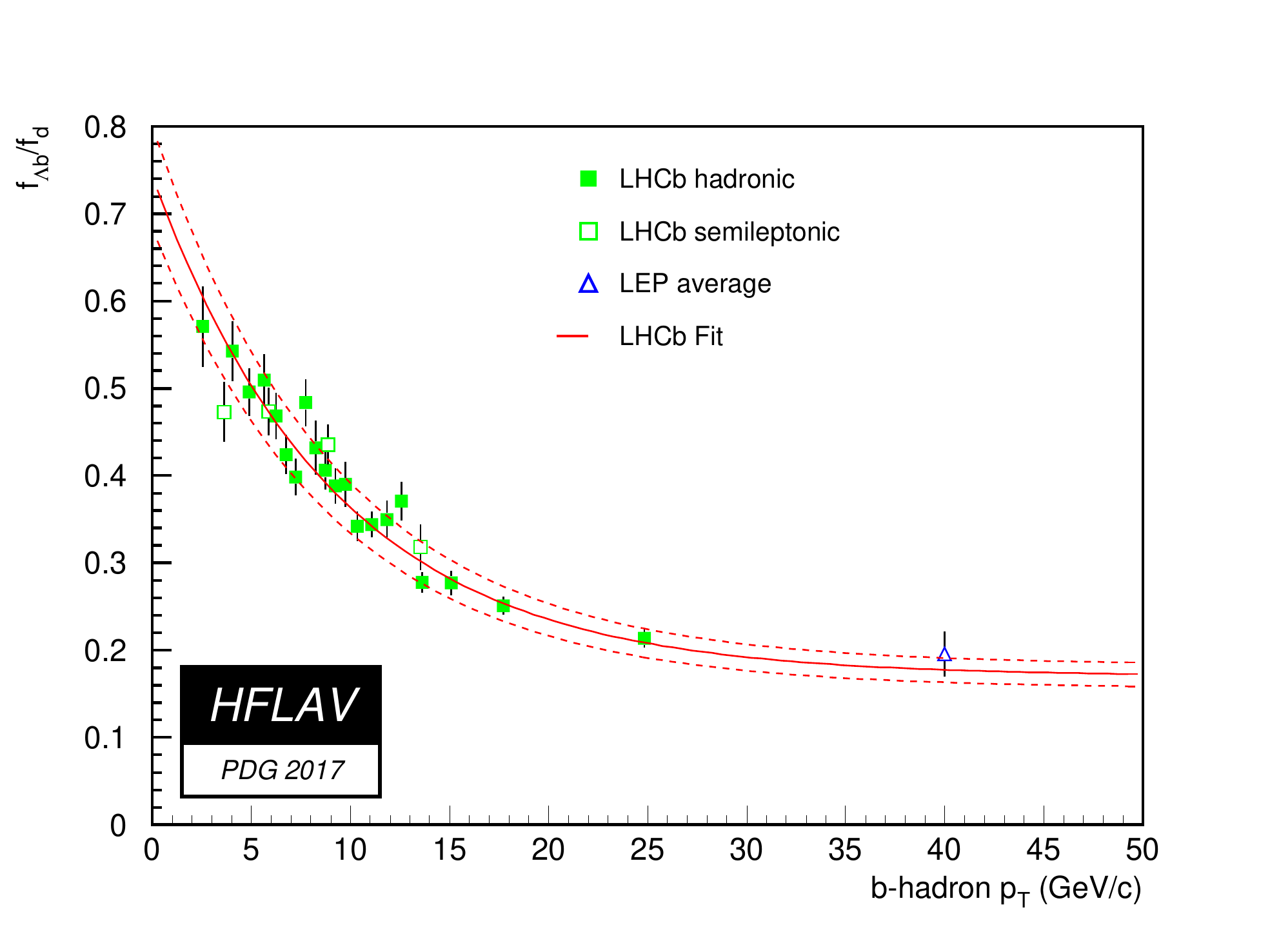}
  \caption{Ratio of production fractions $\fLb/\fBd$ as 
   a function of $p_{\rm T}$ of the \b hadron from
   LHCb data for \b hadrons decaying semileptonically~\cite{Aaij:2011jp}
   and fully reconstructed in hadronic decays~\cite{Aaij:2014jyk}. 
%   A scale uncertainty due to the common systematic uncertainty 
%   from the $\Lc \rightarrow pK^-\pi^+$ branching fraction
%   is omitted.
   The curve represents a fit to the LHCb hadronic data~\cite{Aaij:2014jyk}.
   The computed LEP ratio is included at an approximate $p_{\rm T}$ 
   in $Z$ decays, but does not participate in any fit.}
  \labf{rlb_comb}
 \end{center}
\end{figure}

\unpublished{}{CDF\footnote{
The analysis of \Ref{CDFnote10795:2012} is unpublished, 
therefore not further discussed here nor included in the averages.},}
LHCb and ATLAS have investigated the $p_{\rm T}$ dependence of $\fBs/\fBd$
using fully reconstructed $\Bs$ and $\Bd$ decays.
LHCb reported $3\sigma$ evidence that the ratio $R_s = \fBs/\fBd$ decreases with 
$p_{\rm T}$ using fully reconstructed $\Bs$ and $\Bd$ decays and theoretical predictions 
for branching ratios~\citehistory{Aaij:2013qqa}{Aaij:2013qqa,*Aaij:2011hi_hist}.
Data from the
ATLAS experiment~\cite{Aad:2015cda} using decays of $\Bs$ and $\Bd$ to $J/\psi$ final states 
and using theoretical predictions for branching ratios~\cite{Liu:2013nea} indicates 
that $R_s$ is consistent with no $p_T$ dependence.
\Figure{rs_comb} shows 
the ratio $R_s$ as a function of $p_{\rm T}$ measured by LHCb and ATLAS.  
Two fits are performed. % similar to the result for $R_{\Lb}$ above.r
The first fit, using a linear parameterization, yields
$R_s = (0.2701\pm 0.0058) - (0.00139\pm 0.00044)[\gevc]^{-1} \times p_{\rm T}$.  
A second fit, using a simple exponential, yields
$R_s = \exp\left\{(-1.304\pm 0.024) - (0.0058\pm 0.0019)[\gevc]^{-1} \times p_{\rm T}\right\}$.  
The two fits are nearly indistinguishable over the $p_{\rm T}$ range of the results,
but the second gives a physical value for all $p_{\rm T}$.  $R_s$ is also calculated
for LEP and placed at the approximate $p_{\rm T}$ for the \b hadron, though the LEP result
doesn't participate in the fit.  Our world average for $R_s$ is also included in the
figure for reference.  

\begin{figure}
 \begin{center}
  \includegraphics[width=\textwidth]{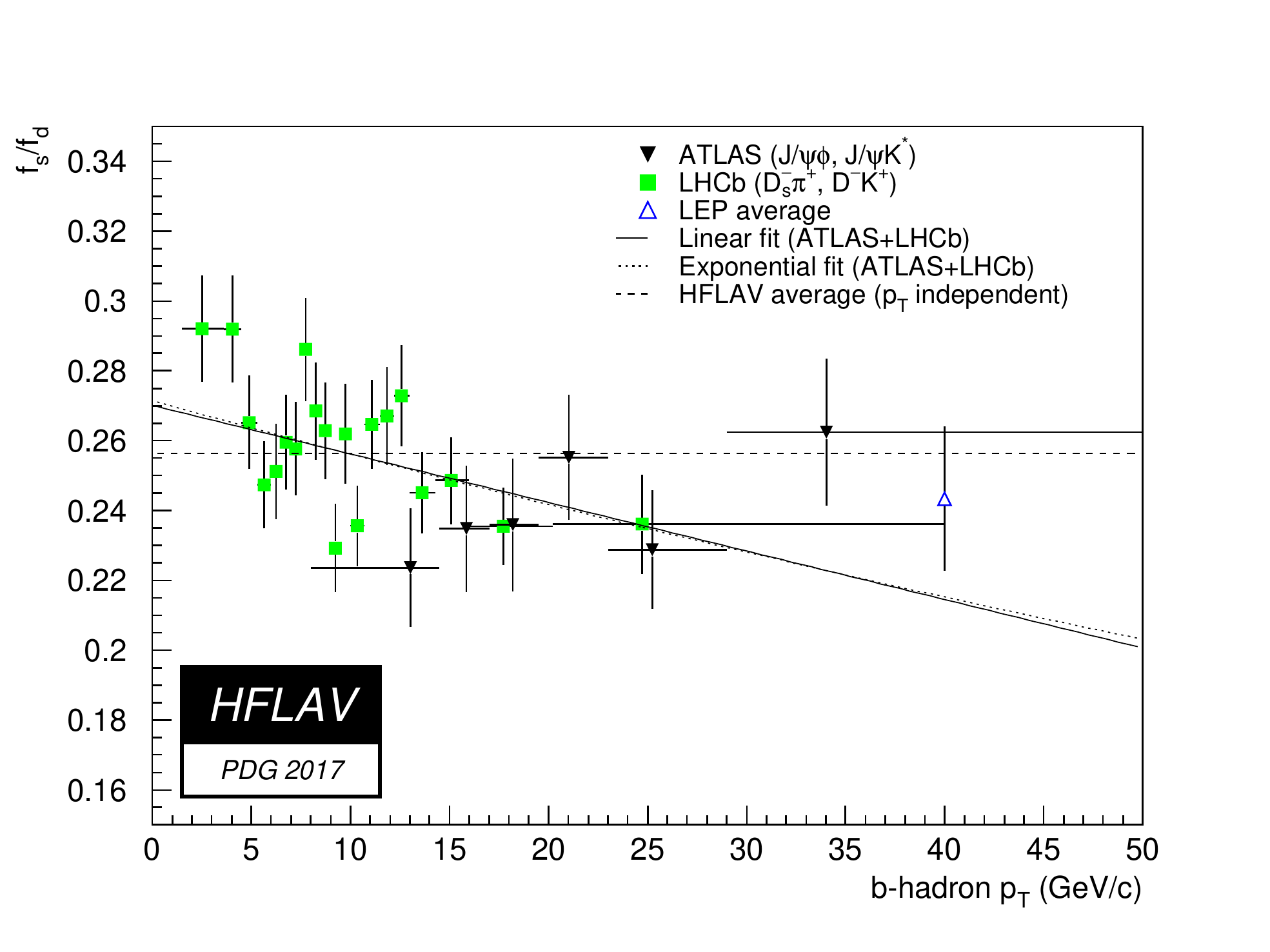}
  \caption{Ratio of production fractions $\fBs/\fBd$ as 
   a function of $p_{\rm T}$ of the reconstructed \b hadrons for the 
   %%CDF~\cite{CDFnote10795:2012} and 
   LHCb~\protect\citehistory{Aaij:2013qqa}{Aaij:2013qqa,*Aaij:2011hi_hist}
   %\footref{foot:life_mix:Aaij:2013qqa} 
   and ATLAS~\cite{Aad:2015cda}
   data. Note the suppressed zero for the vertical axis.
   The curves represent fits to the data:
   a linear fit (solid), and an exponential fit described in the text (dotted).
   The $p_{\rm T}$ independent value average of $R_s$ (dashed) is shown for 
   comparison.
   The computed LEP ratio is included at an 
   approximate $p_{\rm T}$ in $Z$ decays, but does not participate in any fit.}
  \labf{rs_comb}
 \end{center}
\end{figure}

In order to combine or compare LHCb results with other experiments,
the $p_{\rm T}$-dependent $\fLb/(\fBu + \fBd)$ is weighted by the $p_{\rm T}$ spectrum.\footnote{
  \label{foot:life_mix:Aaij:2011jp}
  In practice the LHCb data are given in 14 bins in $p_{\rm T}$ and $\eta$ with a full covariance matrix~\cite{Aaij:2011jp}. 
  The weighted average is calculated as
  $D^T C^{-1} M/\sigma$, where $\sigma = D^T C^{-1} D$, $M$ is a vector 
  of measurements, $C^{-1}$ is the inverse covariance matrix and $D^T$ is the 
  transpose of the design matrix (vector of 1's).}
\Table{LHCBcomp} compares 
the $p_{\rm T}$-weighted LHCb data with comparable averages from CDF. 
The average CDF and LHCb data are in agreement despite the 
\b hadrons being produced in different kinematic regimes.

\begin{table}
 \caption{Comparison of average production fraction ratios from 
 CDF~\cite{Aaltonen:2008eu,Aaltonen:2008zd} and LHCb~\cite{Aaij:2011jp}.
 The kinematic regime of the charm+lepton system reconstructed in each
 experiment is also shown.}
 \labt{LHCBcomp}
 \begin{center}
  \begin{tabular}{lccc}
   \hline
   Quantity                         & CDF               & LHCb \\
   \hline
%   $\fBs/(\fBu + \fBd)$             & \hflavRBSTEVNOCON  & \hflavRBSLHCBNOCON   \\
%   $\fLb/(\fBu + \fBd)$             & \hflavRLBTEVNOCON  & \hflavRLBLHCBNOCON   \\
   $\fBs/(\fBu + \fBd)$             & $0.224\pm 0.057$  & $0.134\pm 0.009$   \\
   $\fLb/(\fBu + \fBd)$             & $0.229\pm 0.062$  & $0.240\pm 0.022$    \\
   Average charm+lepton $p_{\rm T}$ & $\sim 13~\gevc$ & $\sim 7~\gevc$ \\
   Pseudorapidity range             & $-1 < \eta < 1$   & $2 < \eta < 5$      \\
   \hline
  \end{tabular}
 \end{center}
\end{table}

Ignoring $p_{\rm T}$ dependence, all these published results have been 
adjusted to the latest branching fraction averages~\cite{PDG_2016} and combined
%\footnote{%
%The latest preliminary results from CDF using $\Bs\to\jpsi\phi$ decays~\cite{CDFnote10795:2012}
%have not been included yet in our averages.
%}
following the procedure and 
assumptions described in \Ref{Abbaneo:2000ej_mod,*Abbaneo:2001bv_mod_cont},
to yield $\fBu=\fBd=\hflavWFBDNOMIX$, 
$\fBs=\hflavWFBSNOMIX$ and $\fbb=\hflavWFBBNOMIX$
under the constraints of \Eq{constraints}.  
%Following the PDG prescription, we have scaled the combined uncertainties 
%on these fractions by \hflavWFSFACTOR to account for slight discrepancies 
%in the input data. 
Repeating the combinations for LEP and the Tevatron, we obtain 
$\fBu=\fBd=\hflavZFBDNOMIX$,
$\fBs=\hflavZFBSNOMIX$ and $\fbb=\hflavZFBBNOMIX$ when using the LEP data only, and
$\fBu=\fBd=\hflavTFBDNOMIX$, $\fBs=\hflavTFBSNOMIX$ and
$\fbb = \hflavTFBBNOMIX$ when using the Tevatron data only.  
As noted previously,
the LHCb data are insufficient to determine a complete set of \b-hadron production
fractions. The world averages (LEP, Tevatron and LHC) for the various fractions 
are presented here for comparison with previous averages.  Significant differences
exist between the LEP and Tevatron fractions, therefore use of the world averages
should be taken with some care.
%When the Tevatron, 
%LHCb and LEP data are separated, we find no need to scale the uncertainties of 
%any combination.  
For these combinations other external inputs are used, 
\eg, the branching ratios of \B mesons to final states with a
\particle{D} or \particle{D^*} % or \particle{D^{**}}
in semileptonic decays, which are needed 
to evaluate the fraction of semileptonic \Bs decays with a \particle{D_s^-} 
in the final state.

Time-integrated mixing analyses performed with lepton pairs 
from \particle{b\bar{b}} 
events produced at high-energy colliders measure the quantity 
\begin{equation}
\chibar = f'_{\particle{d}} \,\chid + f'_{\particle{s}} \,\chis \,,
\labe{chibar}
\end{equation}
where $f'_{\particle{d}}$ and $f'_{\particle{s}}$ are 
the fractions of \Bd and \Bs hadrons 
in a sample of semileptonic \b-hadron decays, and where \chid and \chis 
are the \Bd and \Bs time-integrated mixing probabilities.
Assuming that all \b hadrons have the same semileptonic decay width implies 
$f'_i = f_i R_i$, where $R_i = \tau_i/\tau_{\b}$ is the ratio of the lifetime 
$\tau_i$ of species $i$ to the average \b-hadron lifetime 
$\tau_{\b} = \sum_i f_i \tau_i$.
Hence measurements of the mixing probabilities
\chibar, \chid and \chis can be used to improve our 
knowledge of \fBu, \fBd, \fBs and \fbb.
In practice, the above relations yield another determination of 
\fBs obtained from \fbb and mixing information, 
\begin{equation}
\fBs = \frac{1}{R_{\particle{s}}}
\frac{(1+r)\overline{\chi}-(1-\fbb R_{\rm baryon}) \chid}{(1+r)\chis - \chid} \,,
\labe{fBs-mixing}
\end{equation}
where $r=R_{\particle{u}}/R_{\particle{d}} = \tau(\Bu)/\tau(\Bd)$.

%%%%%%%%%%%%%%%%%%%%%%%%%%%%%%%%%%%%%%%%%%%%yy
%%%%% The paragraph below was used between 2012 and 2014, when the 
%%%%% unpublished Run II measurement of CDF was included in the chibar average.
%%%%%%%%%%%%%%%%%%%%%%%%%%%%%%%%%%%%%%%%%%%%yy
%old% The published measurements of \chibar performed by the LEP
%old% experiments have been combined by the LEP Electroweak Working Group to yield 
%old% $\chibar = \hflavCHIBARLEP$~\cite{ALEPH:2005ab}.
%old% This can be compared with the Tevatron average, $\chibar = \hflavCHIBARTEV$,
%old% obtained from \dzero~\cite{Abazov:2006qw} and CDF~\cite{CDFnote10335:2011}
%old% measurements with Run~II data.\footnote{
%old%   \label{foot:life_mix:Acosta:2003ie}
%old%   As explained in \Ref{CDFnote10335:2011}, a previous CDF analysis~\citehistory{Acosta:2003ie}{Acosta:2003ie_hist}
%old%   performed with Run~I data overlooked a background component, so the corresponding result is not 
%old%   included in the average.}
%old% The two averages agree, showing no evidence that the production fractions
%old% of \Bd and \Bs mesons at the \particle{Z} peak or at the Tevatron are different.
%old% We combine these two results in a simple weighted average,
%old% assuming no correlations, and obtain 
%old% $\chibar = \hflavCHIBAR$.

The published measurements of \chibar performed by the LEP
experiments have been combined by the LEP Electroweak Working Group to yield 
$\chibar = \hflavCHIBARLEP$~\cite{ALEPH:2005ab}.%
\footnote{We use the $\bar{\chi}$ average of Eq.~5.39 in \Ref{ALEPH:2005ab}, 
obtained from a 10-parameter global fit of all electroweak data where the
asymmetry measurements have been excluded}
This can be compared with the Tevatron average, $\chibar = \hflavCHIBARTEV$,
obtained from \dzero~\cite{Abazov:2006qw} and CDF~\citehistory{Acosta:2003ie}{Acosta:2003ie_hist}.%
\unpublished{}{\footnote{ \label{foot:life_mix:Acosta:2003ie}
The CDF result of Ref.~\citehistory{Acosta:2003ie}{Acosta:2003ie_hist} is from Run~I data.
A preliminary CDF measurement based on Run~II data~\cite{CDFnote10335:2011}
is still unpublished and therefore no longer included in our averages.
%  Reference~\cite{CDFnote10335:2011} claims that the Run~I analysis
%  overlooked a background component.
}}
The two averages deviate
from each other by $\hflavCHIBARSFACTOR\,\sigma$; 
this could be an indication that the production fractions of \b hadrons 
at the \particle{Z} peak or at the Tevatron are not the same. 
%Although this discrepancy 
%is not very significant it should be carefully monitored in the future. 
We choose to combine these two results in a simple weighted average,
assuming no correlations, and, following the PDG prescription, we 
multiply the combined uncertainty by \hflavCHIBARSFACTOR to account 
for the discrepancy. Our world average is then $\chibar = \hflavCHIBAR$.

\begin{table}
\centering
\caption{Time-integrated mixing probability \chibar (defined in \Eq{chibar}), 
and fractions of the different \b-hadron species in an unbiased sample of 
weakly decaying \b hadrons, obtained from both direct
and mixing measurements. The correlation coefficients between the fractions are 
also given.
The last column includes measurements performed at LEP, Tevatron and LHC.}
\labt{fractions}
\resizebox{\textwidth}{!}{
\begin{tabular}{@{}l@{~}l@{~}cccc@{}}
\hline
Quantity            &                      & $Z$ decays      & Tevatron       & LHCb~\citehistory{Aaij:2013qqa}{Aaij:2013qqa,*Aaij:2011hi_hist}
 & all    \\
\hline
Mixing probability  & $\overline{\chi}$    & \hflavCHIBARLEP  & \hflavCHIBARTEV &         & \hflavCHIBAR \\
\Bu or \Bd fraction & $\fBu = \fBd$        & \hflavZFBD       & \hflavTFBD      &         & \hflavWFBD   \\
\Bs fraction        & $\fBs$               & \hflavZFBS       & \hflavTFBS      &         & \hflavWFBS   \\
\b-baryon fraction  & $\fbb$               & \hflavZFBB       & \hflavTFBB      &         & \hflavWFBB   \\
$\Bs/\Bd$ ratio     & $\fBs/\fBd$          & \hflavZFBSBD     & \hflavTFBSBD    & $0.256 \pm 0.020^u$ & \hflavWFBSBD \\
\multicolumn{2}{@{}l}{$\rho(\fBs,\fBu) = \rho(\fBs,\fBd)$} & \hflavZRHOFBDFBS & \hflavTRHOFBDFBS &         & \hflavWRHOFBDFBS \\
\multicolumn{2}{@{}l}{$\rho(\fbb,\fBu) = \rho(\fbb,\fBd)$} & \hflavZRHOFBDFBB & \hflavTRHOFBDFBB &         & \hflavWRHOFBDFBB \\
\multicolumn{2}{@{}l}{$\rho(\fbb,\fBs)$}                   & \hflavZRHOFBBFBS & \hflavTRHOFBBFBS &         & \hflavWRHOFBBFBS \\
\hline
\multicolumn{6}{l}{$^u$ \footnotesize This value has been updated with new inputs by LHCb to yield $0.259 \pm 0.015$~\cite{LHCb-CONF-2013-011}.} 
\end{tabular}
}
\end{table}

Introducing the \chibar average in \Eq{fBs-mixing}, together with our world average 
$\chid = \hflavCHIDWU$ (see \Eq{chid} of \Sec{dmd}), the assumption $\chis= 1/2$ 
(justified by \Eq{chis} in \Sec{dms}), the 
best knowledge of the lifetimes (see \Sec{lifetimes}) and the estimate of \fbb given above, 
yields $\fBs = \hflavWFBSMIX$ 
(or $\fBs = \hflavZFBSMIX$ using only LEP data, 
or $\fBs = \hflavTFBSMIX$ using only Tevatron data),
an estimate dominated by the mixing information. 
Taking into account all known correlations (including that introduced by \fbb), 
this result is then combined with the set of fractions obtained from direct measurements 
(given above), to yield the % following improved estimates, 
improved estimates of \Table{fractions}, 
still under the constraints of \Eq{constraints}.
%%% \footnote{%
%%% The combined value of \fbb is smaller than the  
%%% results from either LEP, the Tevatron and LHCb separately.
%%% This seemingly surprising result  
%%% arises from the smaller uncertainties on the other fractions 
%%% and the application of the unitarity constraint of \Eq{constraints}.}
As can be seen, our knowledge on the mixing parameters % substantially
reduces the uncertainty on \fBs, quite substantially in the case of LEP data.
% , and this even in the case of the world averages where a rather strong deweighting was introduced in the computation of \chibar.
It should be noted that the results % of \Eqsss{fBd}{fBs}{fbb} 
are correlated, as indicated in \Table{fractions}.

%{\bf Need to include in this section the latest fractions from LHCb of \Ref{LHCb-CONF-2013-011}.}
%\marginpar{XXXX}

%2014 Although no recent measurements of the fractions have become available, 
%2014 the averages of \Table{fractions} (and most notably the \b-baryon fraction) 
%2014 have significantly improved in precision as compared to those given in our 
%2014 previous report~\cite{Amhis:2012bh}. This is mostly due to a new and precise 
%2014 model-independent measurement of the $\Lc \to p K^-\pi^+$
%2014 branching fraction from Belle~\cite{Zupanc:2013iki}, 
%2014 which has been used to adjust the fractions obtained from direct measurements. 

%------------------------------------------------
%\mysubsection{\b-hadron lifetimes}
%------------------------------------------------

% Introduction, b-hadron lifetime, B0 lifetime, B+ lifetime, B+/B0 lifetime ratio
%%%%%%%%%%%%%%%%%%%%%%%%%%%%%%%%%%%%%%%%%%%%%%%%%%
%
% This is file life_mix_tau1.tex containing the
% first part of the chapter on the b-hadron life-
% times: introduction, average b-hadron, B0 and B+
% lifetimes as well as the B+/B0 lifetime ratio
%
%%%%%%%%%%%%%%%%%%%%%%%%%%%%%%%%%%%%%%%%%%%%%%%%%

%------------------------------------------------
\mysubsection{\b-hadron lifetimes}
%------------------------------------------------
\labs{lifetimes}

In the spectator model the decay of \b hadrons $H_b$ is
governed entirely by the flavour changing \particle{b\to Wq} transition
($\particle{q}=\particle{c,u}$).  For this very reason, lifetimes of all
\b hadrons are the same in the spectator approximation
regardless of the (spectator) quark content of the $H_b$.  In the early
1990's experiments became sophisticated enough to start seeing the
differences of the lifetimes among various $H_b$ species.  The first
theoretical calculations of the spectator quark effects on $H_b$
lifetime emerged only few years earlier~\cite{Shifman:1986mx}.

Since then, such calculations are performed in the framework of
the Heavy Quark Expansion
(HQE)~\cite{Shifman:1986mx,Chay:1990da,Bigi:1992su}, using 
as most important assumption that 
of quark-hadron duality~\cite{Shifman:2000jv,Bigi:2001ys}.
Since a few years, possible quark-hadron duality violating effects are 
severely constrained by experiments~\cite{Jubb:2016mvq}.
% In the HQE, under certain assumptions
% (the most important of which is that of
% quark-hadron duality~\cite{Shifman:2000jv}\cite{Bigi:2001ys}),
In these calculations, 
the total decay rate of an $H_b$ 
is expressed as the sum
of a series of expectation values of operators of increasing dimension,
multiplied by the correspondingly higher powers of $\Lambda_{\rm
QCD}/m_b$:
\begin{equation}
\Gamma_{H_b} = |{\rm CKM}|^2 \sum_n c_n
\left(\frac{\Lambda_{\rm QCD}}{m_b}\right)^n \langle H_b|O_n|H_b\rangle \,,
\labe{hqe}
\end{equation}
where $|{\rm CKM}|^2$ is the relevant combination of CKM matrix elements.
The coefficients $c_n$ of this expansion, known as the Operator Product
Expansion~\cite{Wilson:1969zs},
can be calculated perturbatively. Hence, the HQE
predicts $\Gamma_{H_b}$ in the form of an expansion in both
$\Lambda_{\rm QCD}/m_{\b}$ and $\alpha_s(m_{\b})$.  The precision of
current experiments requires an expansion up to % makes it mandatory to go to
the next-to-leading order in QCD, \ie, the inclusion of corrections of the order of
$\alpha_s(m_{\b})$ to the $c_n$ terms. 
% All non-perturbative physics is shifted 
The non-perturbative parts of the calculation are grouped
into the expectation values $\langle H_b|O_n|H_b\rangle$ of
operators $O_n$.  These can be calculated using lattice QCD or QCD sum
rules, or can be related to other observables via the HQE. % ~\cite{Bigi:1995jr,Bellini:1996ra}.
One may reasonably expect that powers of
$\Lambda_{\rm QCD}/m_{\b}$ provide enough suppression that only the
first few terms of the sum in \Eq{hqe} matter.

Theoretical predictions are usually made for the ratios of the lifetimes
(with $\tau(\Bd)$ often chosen as the common denominator) rather than for the
individual lifetimes, for this allows several uncertainties to cancel.
The precision of the HQE calculations (see
\Refs{Ciuchini:2001vx,Beneke:2002rj,Franco:2002fc,Tarantino:2003qw,Gabbiani:2003pq,Gabbiani:2004tp}, and \Ref{Lenz:2015dra,Lenz:2014jha} for the latest updates)
is in some instances already surpassed by the measurements,
\eg, in the case of $\tau(\Bu)/\tau(\Bd)$.  
More accurate predictions are now a matter of progress
in the evaluation of the non-perturbative hadronic matrix
elements, in particular using lattice QCD where significant advances were made in the last decade.
% However, the HQE, even in its present shape, draws a number of important conclusions,
However, the following important conclusions can be drawn from the HQE, even in its present state, 
which are in agreement with experimental observations:
\begin{itemize}
\item The heavier the mass of the heavy quark, the smaller is the
  variation in the lifetimes among different hadrons containing this
  quark, which is to say that as $m_{\b}\to\infty$ we retrieve the
  spectator picture in which the lifetimes of all $H_b$ states are the same.
%OS  This is well illustrated by the fact that lifetimes in the $b$ sector
%OS  are all rather similar, while in the $c$ sector
%OS  ($m_{\particle{c}}<m_{\b}$) lifetimes differ by large factors.
   This is well illustrated by the fact that lifetimes are rather
   similar in the \b sector, while they differ by large factors
   in the charm sector ($m_{\particle{c}}<m_{\b}$).
\item The non-perturbative corrections arise only at the order of
  $\Lambda_{\rm QCD}^2/m_{\b}^2$, which translates into 
  differences among $H_b$ lifetimes of only a few percent.
\item It is only the difference between meson and baryon lifetimes that
  appears at the $\Lambda_{\rm QCD}^2/m_{\b}^2$ level.  The splitting of the
  meson lifetimes occurs at the $\Lambda_{\rm QCD}^3/m_{\b}^3$ level, yet it is
  enhanced by a phase space factor $16\pi^2$ with respect to the leading
  free \b decay.
\end{itemize}

To ensure that certain sources of systematic uncertainty cancel, 
lifetime analyses are sometimes designed to measure
ratios of lifetimes.  However, because of the differences in decay
topologies, abundance (or lack thereof) of decays of a certain kind,
{\em etc.}, measurements of the individual lifetimes are also 
common.  In the following section we review the most common
types of lifetime measurements.  This discussion is followed by the
presentation of the averaging of the various lifetime measurements, each
with a brief description of its particularities.

%% Experimental measurements too often benefit from a partial systematic
%% uncertainty cancellation if a measurement is that of the ratio of two
%% quantities of the same kind, which are affected similarly by one or
%% more systematic effect(s).  For this reason, rather often the lifetime
%% measurements are being designed to be those of the ratio of the
%% lifetimes.  However, because of the differences in decay topologies,
%% abundance (or lack thereof) of decays of a certain kind, {\em etc.}\
%% measurements of the individual lifetimes are not particularly rare.  In
%% the following section we review the most common types of the lifetime
%% measurements.  This discussion is followed by the presentation of the
%% averaging of the various lifetime measurements, each with a brief
%% description of its particularities.

%% Details of procedures used to combine the different measurements can be
%% found in \Ref{LEPBOSC:1996}. {\sc do we want this? HERE?}

\mysubsubsection{Lifetime measurements, uncertainties and correlations}

In most cases, the lifetime of an $H_b$ state is estimated from a flight
distance measurement
and a $\beta\gamma$ factor which is used to convert the geometrical
distance into the proper decay time.  Methods of accessing lifetime
information can roughly be divided in the following five categories:
\begin{enumerate}
\item {\bf\em Inclusive (flavour-blind) measurements}.  These early
  measurements were aimed at extracting the lifetime from a mixture of
  \b-hadron decays, without distinguishing the decaying species.  Often
  the knowledge of the mixture composition was limited, which made these
  measurements experiment-specific.  Also, these
  measurements had to rely on Monte Carlo simulation for estimating the
  $\beta\gamma$ factor, because the decaying hadrons are not fully
  reconstructed.  These were usually the largest
  statistics \b-hadron lifetime measurements accessible to a
  given experiment, and could therefore serve as an important
  performance benchmark.
\item {\bf\em Measurements in semileptonic decays of a specific
  {\boldmath $H_b$\unboldmath}}.  The \particle{W} boson from \particle{\b\to Wc}
  produces a $\ell\nu_l$ pair (\particle{\ell=e,\mu}) in about 21\% of the
  cases.  The electron or muon from such decays provides a clean and efficient
  trigger signature.
  The \particle{c} quark from the \particle{b\to Wc} transition and the other
  quark(s) making up the decaying $H_b$ combine into a charm hadron,
  which is reconstructed in one or more exclusive decay channels.
  Knowing what this charmed hadron is allows one to separate, at least
  statistically, different $H_b$ species.  The advantage of these
  measurements is in the sample size,
  which is usually larger than in the case of
  exclusively reconstructed $H_b$ decays.  Some of the main
  disadvantages are related to the difficulty of estimating the lepton+charm
  sample composition and to the Monte Carlo reliance for
  the momentum (and hence $\beta\gamma$ factor) estimate.
\item {\bf\em Measurements in exclusively reconstructed hadronic decays}.
  These
  have the advantage of complete reconstruction of the decaying $H_b$ state, 
  which allows one to infer the decaying species as well as to perform precise
  measurement of the $\beta\gamma$ factor.  Both lead to generally
  smaller systematic uncertainties than in the above two categories.
  The downsides are smaller branching ratios and larger combinatorial
  backgrounds, especially in $H_b\rightarrow H_c\pi(\pi\pi)$ and
  multi-body $H_c$ decays, or in a hadron collider environment with
  non-trivial underlying event.  Decays of the type $H_b\to \jpsi H_s$ are
  relatively clean and easy to trigger, due to the $\jpsi\to \ell^+\ell^-$
  signature, but their branching fraction is only about 1\%.
\item {\bf\em Measurements at asymmetric B factories}. 
  In the $\Ups\rightarrow B \bar{B}$ decay, the \B mesons (\Bu or \Bd) are
essentially at rest in the \Ups frame.  This makes direct lifetime
measurements impossible in experiments at symmetric colliders producing 
\Ups at rest. 
%Romulus% However, most time-integrated measurements that related to
%Romulus% directly measure the overall mixing probability, \chid, have been measured
%Romulus% by the CLEO and ARGUS. It is almost impossible to measure the \B-meson lifetime measurement
%Romulus% with time-integrated technique. The best approach for measuring the \B-meson lifetime 
%Romulus% is using the time dependent measurement at asymmetric \B factories such as \babar and
%Romulus% Belle. 
At asymmetric \B factories the \Ups meson is boosted
resulting in \B and \particle{\bar{B}} moving nearly parallel to each 
other with the same boost. The lifetime is inferred from the distance $\Delta z$        
separating the \B and \particle{\bar{B}} decay vertices along the beam axis 
%Romulus% (see \Fig{Ups_geometry})
and from the \Ups boost known from the beam energies. This boost is equal to 
$\beta \gamma \approx 0.55$ (0.43) in the \babar (\belle) experiment,
resulting in an average \B decay length of approximately 250~(190)~$\mu$m. 
In order to determine the charge of the \B mesons in each event, one of them is
fully reconstructed in a semileptonic or hadronic decay mode.
The other \B is typically not fully reconstructed, only the position
of its decay vertex is determined from the remaining tracks in the event.
These measurements benefit from large sample sizes, but suffer from poor proper time 
resolution, comparable to the \B lifetime itself. This resolution is dominated by the 
uncertainty on the decay vertices, which is typically 50~(100)~$\mu$m for a
fully (partially) reconstructed \B meson. 
%Romulus% $\Delta z$ resolution.
%Romulus% At asymmetric \B factories, the decay time different resolution is dominated by the uncertainty 
%Romulus% of the vertex location of both \B mesons. The decay distance and the momenta of a \B meson 
%Romulus% determine its lifetime. In order to determine the charge of the \B mesons in each event,
%Romulus% one of the them is fully reconstructed in semileptonic or fully hadronic decay modes.
%Romulus% The other \B is typically not fully reconstructed, only the position
%Romulus% of its decay vertex is determined from the remaining tracks in the event.
%Romulus% These measurements benefit from very large statistics, but suffer from
%Romulus% poor $\Delta z$ resolution (the distance between the two \B-meson decay
%Romulus% vertexes projected on the beam axis.
%Romulus% Alternatively one could apply a completely 
%Romulus% reconstructed events where both \B are fully reconstructed for getting the best
%Romulus% precision on the lifetime measurement, however, a price has to be paid due to 
%Romulus% a low statistics. 
With much larger samples in the future, 
the resolution and purity could be improved (and hence the systematics reduced)
by fully reconstructing both \B mesons in the event. 
 
\item {\bf\em Direct measurement of lifetime ratios}.  This method, 
  initially applied 
  in the measurement of $\tau(\Bu)/\tau(\Bd)$, is now also used for other 
  \b-hadron species at the LHC. 
  The ratio of the lifetimes is extracted from the proper time 
  dependence of the ratio of the observed yields of 
  of two different \b-hadron species, 
  both reconstructed in decay modes with similar topologies. 
  The advantage of this method is that subtle efficiency effects
  (partially) cancel in the ratio. 
\end{enumerate}

In some of the latest analyses, measurements of two (\eg, $\tau(\Bu)$ and
$\tau(\Bu)/\tau(\Bd)$) or three (\eg\ $\tau(\Bu)$,
$\tau(\Bu)/\tau(\Bd)$, and \dmd) quantities are combined.  This
introduces correlations among measurements.  Another source of
correlations among the measurements are the systematic effects, which
could be common to an experiment or to an analysis technique across the
experiments.  When calculating the averages, such known correlations are taken
into account.
%  following the general procedure described in \Ref{LEPBOSC:1996}.

%% ====================================================================
\mysubsubsection{Inclusive \b-hadron lifetimes}
%% ====================================================================

The inclusive \b-hadron lifetime is defined as $\tau_{\b} = \sum_i f_i
\tau_i$ where $\tau_i$ are the individual species lifetimes and $f_i$ are
the fractions of the various species present in an unbiased sample of
weakly decaying \b hadrons produced at a high-energy
collider.\footnote{In principle such a quantity could be slightly
different in \particle{Z} decays, at the Tevatron or at the LHC, 
% (it has not been measured at the LHC),
in case the
fractions of \b-hadron species are not exactly the same; see the
discussion in \Sec{fractions_high_energy}.}  This quantity is certainly
less fundamental than the lifetimes of the individual species, the
latter being much more useful in comparisons of the measurements with
the theoretical predictions.  Nonetheless, we perform the averaging of
the inclusive lifetime measurements for completeness and because
they might be of interest as ``technical numbers.''

In practice, an unbiased measurement of the inclusive lifetime is
difficult to achieve, because it would imply an efficiency which is
guaranteed to be the same across species.  So most of the measurements
are biased.  In an attempt to group analyses that are expected to
select the same mixture of \b hadrons, the available results (given in
\Table{lifeincl}) are divided into the following three sets:
\begin{enumerate}
\item measurements at LEP and SLD that include any \b-hadron decay, based 
      on topological reconstruction (secondary vertex or track impact
      parameters);
\item measurements at LEP based on the identification
      of a lepton from a \b decay; and
\item measurements at hadron colliders based on inclusive 
      \particle{H_b\to \jpsi X} reconstruction, where the
      \particle{\jpsi} is fully reconstructed.
\end{enumerate}

\begin{table}[t]
\caption{Measurements of average \b-hadron lifetimes.}
\labt{lifeincl}
\begin{center}
\begin{tabular}{lcccl} \hline
Experiment &Method           &Data set & $\tau_{\b}$ (ps)       &Ref.\\
\hline
ALEPH  &Dipole               &91     &$1.511\pm 0.022\pm 0.078$ &\cite{Buskulic:1993gj}\\
DELPHI &All track i.p.\ (2D) &91--92 &$1.542\pm 0.021\pm 0.045$ &\cite{Abreu:1994dr}$^a$\\
DELPHI &Sec.\ vtx            &91--93 &$1.582\pm 0.011\pm 0.027$ &\cite{Abreu:1996hv}$^a$\\
DELPHI &Sec.\ vtx            &94--95 &$1.570\pm 0.005\pm 0.008$ &\cite{Abdallah:2003sb}\\
L3     &Sec.\ vtx + i.p.     &91--94 &$1.556\pm 0.010\pm 0.017$ &\cite{Acciarri:1997tt}$^b$\\
OPAL   &Sec.\ vtx            &91--94 &$1.611\pm 0.010\pm 0.027$ &\cite{Ackerstaff:1996as}\\
SLD    &Sec.\ vtx            &93     &$1.564\pm 0.030\pm 0.036$ &\cite{Abe:1995rm}\\ 
\hline
\multicolumn{2}{l}{Average set 1 (\b vertex)} && \hflavTAUBVTXnounit &\\
\hline\hline
ALEPH  &Lepton i.p.\ (3D)    &91--93 &$1.533\pm 0.013\pm 0.022$ &\cite{Buskulic:1995rw}\\
L3     &Lepton i.p.\ (2D)    &91--94 &$1.544\pm 0.016\pm 0.021$ &\cite{Acciarri:1997tt}$^b$\\
OPAL   &Lepton i.p.\ (2D)    &90--91 &$1.523\pm 0.034\pm 0.038$ &\cite{Acton:1993xk}\\ 
\hline
\multicolumn{2}{l}{Average set 2 ($\b\to\ell$)} && \hflavTAUBLEPnounit &\\
\hline\hline
CDF1   &\particle{\jpsi} vtx&92--95 &$1.533\pm 0.015^{+0.035}_{-0.031}$ &\cite{Abe:1997bd} \\ 
%% CDF2       & \particle{\jpsi} vtx
%%                                &  02--03 & $1.526 \pm 0.034 \pm 0.035$ & \cite{CDFnote9203:2008,*CDFnote9203:2008_cont} \\  WARNING: the meaning of CDFnote9203:2008 has changed !!!
%ATLAS &\particle{\jpsi} vtx& 2010 & $1.489\pm 0.016 \pm 0.043$ & \cite{ATLAS-CONF-2011-145}$^p$ \\
% The above measurement has beeen removed from the average on Oct 15, 2016, since it has remained unpublished for 4 years
\hline
\multicolumn{2}{l}{Average set 3 (\particle{\b\to \jpsi})} && \hflavTAUBJPnounit & \\ 
%\hline\hline
%\multicolumn{2}{l}{Average of all above} && \hflavTAUBnounit & \\
\hline
\multicolumn{5}{l}{$^a$ \footnotesize The combined DELPHI result quoted in
\cite{Abreu:1996hv} is 1.575 $\pm$ 0.010 $\pm$ 0.026 ps.} \\[-0.5ex]
\multicolumn{5}{l}{$^b$ \footnotesize The combined L3 result quoted in \cite{Acciarri:1997tt} 
is 1.549 $\pm$ 0.009 $\pm$ 0.015 ps.} \\[-0.5ex]
%\multicolumn{5}{l}{$^p$ \footnotesize Preliminary.}
\end{tabular}
\end{center}
\end{table}

The measurements of the first set are generally considered as estimates
of $\tau_{\b}$, although the efficiency to reconstruct a secondary
vertex most probably depends, in an analysis-specific way, on the number
of tracks coming from the vertex, thereby depending on the type of the
$H_b$.  Even though these efficiency variations can in principle be
accounted for using Monte Carlo simulations (which inevitably contain
assumptions on branching fractions), the $H_b$ mixture in that case can
remain somewhat ill-defined and could be slightly different among
analyses in this set.

On the contrary, the mixtures corresponding to the other two sets of
measurements are better defined in the limit where the reconstruction
and selection efficiency of a lepton or a \particle{\jpsi} from an
$H_b$ does not depend on the decaying hadron type.  These mixtures are
given by the production fractions and the inclusive branching fractions
for each $H_b$ species to give a lepton or a \particle{\jpsi}.  In
particular, under the assumption that all \b hadrons have the same
semileptonic decay width, the analyses of the second set should measure
$\tau(\b\to\ell) = (\sum_i f_i \tau_i^3) /(\sum_i f_i \tau_i^2)$ which is
necessarily larger than $\tau_{\b}$ if lifetime differences exist.
Given the present knowledge on $\tau_i$ and $f_i$,
$\tau(\b\to\ell)-\tau_{\b}$ is expected to be of the order of 0.003\ps.
On the other hand, the third set measuring $\tau(\b\to\particle{\jpsi})$
is expected to give an average smaller than $\tau_{\b}$ because 
of the \Bc meson, which has a significantly
larger probability to decay to a \particle{\jpsi}
than other \b-hadron species. 

Measurements by SLC and LEP experiments are subject to a number of
common systematic uncertainties, such as those due to (lack of knowledge
of) \b and \particle{c} fragmentation, \b and \particle{c} decay models,
\BR{B\to\ell}, \BR{B\to c\to\ell}, \BR{c\to\ell}, $\tau_{\particle{c}}$,
and $H_b$ decay multiplicity.  In the averaging, these systematic
uncertainties are assumed to be 100\% correlated.  The averages for the
sets defined above (also given in \Table{lifeincl})%
\unpublished{}{\footnote{We do not include here an unpublished measurement from ATLAS~\cite{ATLAS-CONF-2011-145}.}}
are
\begin{eqnarray}
\tau(\b~\mbox{vertex}) &=& \hflavTAUBVTX \,, \labe{TAUBVTX} \\
\tau(\b\to\ell) &=& \hflavTAUBLEP  \,, \\
\tau(\b\to\particle{\jpsi}) &=& \hflavTAUBJP\,.
\end{eqnarray}
% whereas an average of all measurements, ignoring mixture differences, 
% yields \hflavTAUB.
The differences between these averages are consistent both with zero
and with expectations within less than $2\,\sigma$.

%% ====================================================================
\mysubsubsection{\Bd and \Bu lifetimes and their ratio}
%% ====================================================================
\labs{taubd}
\labs{taubu}
\labs{lifetime_ratio}

After a number of years of dominating these averages the LEP experiments
yielded the scene to the asymmetric \B~factories and
the Tevatron experiments. The \B~factories have been very successful in
utilizing their potential -- in only a few years of running, \babar and,
to a greater extent, \belle, have struck a balance between the
statistical and the systematic uncertainties, with both being close to
(or even better than) an impressive 1\% level.  In the meanwhile, CDF and
\dzero have emerged as significant contributors to the field as the
Tevatron Run~II data flowed in. Recently, the LHCb experiment reached 
a further step in precision, improving by a factor $\sim 2 $ 
over the previous best measurements. 

% Both appear to enjoy relatively small
% systematic effects, and while current statistical uncertainties of their
% measurements are factors of 2 to 4 larger than those of their \B-factory
% counterparts, both Tevatron experiments stand to increase their samples
% by almost an order of magnitude.

At the present time we are in an interesting position of having three sets
of measurements (from LEP/SLC, \B factories and Tevatron/LHC) that
originate from different environments, are obtained using substantially
different techniques and are precise enough for incisive comparison.

% While individual lifetimes are often of interest to experiments, \eg\ in
% extraction of CKM matrix elements, the ratios of the lifetimes are more
% interesting from the theoretical perspective as they are predicted more
% precisely.

\begin{table}[!t]
\caption{Measurements of the \Bd lifetime.}
\labt{lifebd}
\begin{center}
\begin{tabular}{lcccl} \hline
Experiment &Method                    &Data set &$\tau(\Bd)$ (ps)                  &Ref.\\
\hline
ALEPH  &\particle{D^{(*)} \ell}       &91--95 &$1.518\pm 0.053\pm 0.034$          &\cite{Barate:2000bs}\\
ALEPH  &Exclusive                     &91--94 &$1.25^{+0.15}_{-0.13}\pm 0.05$     &\cite{Buskulic:1996hy}\\
ALEPH  &Partial rec.\ $\pi^+\pi^-$    &91--94 &$1.49^{+0.17+0.08}_{-0.15-0.06}$   &\cite{Buskulic:1996hy}\\
DELPHI &\particle{D^{(*)} \ell}       &91--93 &$1.61^{+0.14}_{-0.13}\pm 0.08$     &\cite{Abreu:1995mc}\\
DELPHI &Charge sec.\ vtx              &91--93 &$1.63 \pm 0.14 \pm 0.13$           &\cite{Adam:1995mb}\\
DELPHI &Inclusive \particle{D^* \ell} &91--93 &$1.532\pm 0.041\pm 0.040$          &\cite{Abreu:1996gb}\\
DELPHI &Charge sec.\ vtx              &94--95 &$1.531 \pm 0.021\pm0.031$          &\cite{Abdallah:2003sb}\\
L3     &Charge sec.\ vtx              &94--95 &$1.52 \pm 0.06 \pm 0.04$           &\cite{Acciarri:1998uv}\\
OPAL   &\particle{D^{(*)} \ell}       &91--93 &$1.53 \pm 0.12 \pm 0.08$           &\cite{Akers:1995pa}\\
OPAL   &Charge sec.\ vtx              &93--95 &$1.523\pm 0.057\pm 0.053$          &\cite{Abbiendi:1998av}\\
OPAL   &Inclusive \particle{D^* \ell} &91--00 &$1.541\pm 0.028\pm 0.023$          &\cite{Abbiendi:2000ec}\\
SLD    &Charge sec.\ vtx $\ell$       &93--95 &$1.56^{+0.14}_{-0.13} \pm 0.10$    &\cite{Abe:1997ys}$^a$\\
SLD    &Charge sec.\ vtx              &93--95 &$1.66 \pm 0.08 \pm 0.08$           &\cite{Abe:1997ys}$^a$\\
CDF1   &\particle{D^{(*)} \ell}       &92--95 &$1.474\pm 0.039^{+0.052}_{-0.051}$ &\cite{Abe:1998wt}\\
CDF1  &Excl.\ \particle{\jpsi K^{*0}}&92--95 &$1.497\pm 0.073\pm 0.032$          &\cite{Acosta:2002nd}\\
% CDF2  &Excl.\ \particle{\jpsi K^{*0}}&02--04 &$1.541\pm 0.050\pm0.020$           &\cite{Aaltonen:2007gf}\\ %%% Published result superseded by preliminary result of \citehistory{Aaltonen:2010pj}{Aaltonen:2010pj,*Abulencia:2006dr_hist}
%%% CDF2   &Incl.\ \particle{D^{(*)} \ell}&02--04 &$1.473\pm 0.036\pm0.054$           &\cite{CDFnote7514:2005}$^p$\\
%%% CDF2   &Excl.\ \particle{D^-(3)\pi}   &02--04 &$1.511\pm 0.023\pm0.013$           &\cite{CDFnote7386:2005}$^p$\\
CDF2   &Excl.\ \particle {\jpsi K_S^0}, \particle{\jpsi K^{*0}} &02--09 &$1.507\pm 0.010\pm0.008$           &\citehistory{Aaltonen:2010pj}{Aaltonen:2010pj,*Abulencia:2006dr_hist} \\
%%%%%\dzero &Excl. \particle{\jpsi K^{*0}}&02--04 &$1.473^{+0.052}_{-0.050}\pm0.023$  &\cite{Abazov:2004ce}\\ % superseded by \citehistory{Abazov:2008jz}{Abazov:2008jz,*Abazov:2005sa_hist}
%%%%%\dzero &Excl.\ \particle{\jpsi K^{*0}}&02--05 &$1.530\pm0.043\pm0.023$ &\citehistory{Abazov:2008jz,Abazov:2004ce}{Abazov:2008jz,*Abazov:2005sa_hist,Abazov:2004ce} \\ % replaces 1.473+0.052-0.050 +-0.023 of \cite{Abazov:2004ce}  % superseded by \citehistory{Abazov:2008jz}{Abazov:2008jz,*Abazov:2005sa_hist}
\dzero &Excl.\ \particle{\jpsi K^{*0}}&03--07 &$1.414\pm0.018\pm0.034$ &\citehistory{Abazov:2008jz}{Abazov:2008jz,*Abazov:2005sa_hist}\\ % replaces 1.530+-0.043+-0.023 of above line
\dzero &Excl.\ \particle {\jpsi K_S^0} &02--11 &$1.508 \pm0.025 \pm0.043$  &\citehistory{Abazov:2012iy}{Abazov:2012iy,*Abazov:2007sf_hist,*Abazov:2004bn_hist} \\
\dzero &Inclusive \particle {D^-\mu^+} &02--11 &$1.534 \pm0.019 \pm0.021$  & \citehistory{Abazov:2014rua}{Abazov:2014rua,*Abazov:2006cb_hist} \\ % RUN II; 10.4 fb-1
\babar &Exclusive                     &99--00 &$1.546\pm 0.032\pm 0.022$          &\cite{Aubert:2001uw}\\
\babar &Inclusive \particle{D^* \ell} &99--01 &$1.529\pm 0.012\pm 0.029$          &\cite{Aubert:2002gi}\\
\babar &Exclusive \particle{D^* \ell} &99--02 &$1.523^{+0.024}_{-0.023}\pm 0.022$ &\cite{Aubert:2002sh}\\
\babar &Incl.\ \particle{D^*\pi}, \particle{D^*\rho} 
                                      &99--01 &$1.533\pm 0.034 \pm 0.038$         &\cite{Aubert:2002ms}\\
\babar &Inclusive \particle{D^* \ell}
&99--04 &$1.504\pm0.013^{+0.018}_{-0.013}$  &\cite{Aubert:2005kf} \\ 
%% 99 in the above line needs to be verified
%% 04 also. 81/fb, \ie\ by summer02 (to be confirmed by David) though reported in summer04
%% lastly, this may actually supersede or have large correlation to \cite{Aubert:2002gi}
%%\belle & Exclusive                     & 00--01 & $1.554\pm 0.030 \pm 0.019$      & \cite{BELLE1}\\
\belle & Exclusive                     & 00--03 & $1.534\pm 0.008\pm0.010$        &  \citehistory{Abe:2004mz}{Abe:2004mz,*Abe:2002id_hist,*Tomura:2002qs_hist,*Hara:2002mq_hist} \\
%% in the above Belle not use 99 data, for 140/fb by sum'03
% ATLAS & Excl.\ \particle{\jpsi K^{*0}} & 2010 & $1.51 \pm0.04 \pm0.04$ & \cite{ATLAS-CONF-2011-092}$^p$ \\
% above result removed from average on Oct 15, 2016 because it remained unpublished for 4 years
ATLAS & Excl.\ \particle {\jpsi K_S^0} & 2011 & $1.509 \pm 0.012 \pm 0.018$ & \cite{Aad:2012bpa} \\
%%% LHCb  & Excl.\ \particle{\jpsi K^{*0}} & 2010 & $1.512 \pm0.032 \pm 0.042$ & \cite{LHCb-CONF-2011-001}$^p$ \\
LHCb  & Excl.\ \particle{\jpsi K^{*0}} & 2011 & $1.524 \pm0.006 \pm 0.004$ & \cite{Aaij:2014owa} \\
%%% LHCb  & Excl.\ \particle {\jpsi K_S^0}   & 2010 & $1.558 \pm0.056 \pm 0.022$ & \cite{LHCb-CONF-2011-001}$^p$ \\
LHCb  & Excl.\ \particle {\jpsi K_S^0}   & 2011 & $1.499 \pm0.013 \pm 0.005$ & \cite{Aaij:2014owa} \\
LHCb    & \particle{K^+\pi^-}   & 2011 & $1.524 \pm 0.011 \pm 0.004$ & \citehistory{Aaij:2014fia}{Aaij:2014fia,*Aaij:2012ns_hist} \\
\hline
Average&                               &        & \hflavTAUBDnounit & \\
\hline\hline           
\multicolumn{5}{l}{$^a$ \footnotesize The combined SLD result 
quoted in \cite{Abe:1997ys} is 1.64 $\pm$ 0.08 $\pm$ 0.08 ps.}\\[-0.5ex]
% \multicolumn{5}{l}{$^p$ {\footnotesize Preliminary.}}
\end{tabular}
\end{center}
\end{table}

% \afterpage{\clearpage}

\begin{table}[p]
\centering
\caption{Measurements of the \Bu lifetime.}
\labt{lifebu}
\begin{tabular}{lcccl} \hline
Experiment &Method                 &Data set &$\tau(\Bu)$ (ps)                 &Ref.\\
\hline
ALEPH  &\particle{D^{(*)} \ell}    &91--95 &$1.648\pm 0.049\pm 0.035$          &\cite{Barate:2000bs}\\
ALEPH  &Exclusive                  &91--94 &$1.58^{+0.21+0.04}_{-0.18-0.03}$   &\cite{Buskulic:1996hy}\\
DELPHI &\particle{D^{(*)} \ell}    &91--93 &$1.61\pm 0.16\pm 0.12$             &\cite{Abreu:1995mc}$^a$\\
DELPHI &Charge sec.\ vtx           &91--93 &$1.72\pm 0.08\pm 0.06$             &\cite{Adam:1995mb}$^a$\\
DELPHI &Charge sec.\ vtx           &94--95 &$1.624\pm 0.014\pm 0.018$          &\cite{Abdallah:2003sb}\\
L3     &Charge sec.\ vtx           &94--95 &$1.66\pm  0.06\pm 0.03$            &\cite{Acciarri:1998uv}\\
OPAL   &\particle{D^{(*)} \ell}    &91--93 &$1.52 \pm 0.14\pm 0.09$            &\cite{Akers:1995pa}\\
OPAL   &Charge sec.\ vtx           &93--95 &$1.643\pm 0.037\pm 0.025$          &\cite{Abbiendi:1998av}\\
SLD    &Charge sec.\ vtx $\ell$    &93--95 &$1.61^{+0.13}_{-0.12}\pm 0.07$     &\cite{Abe:1997ys}$^b$\\
SLD    &Charge sec.\ vtx           &93--95 &$1.67\pm 0.07\pm 0.06$             &\cite{Abe:1997ys}$^b$\\
CDF1   &\particle{D^{(*)} \ell}    &92--95 &$1.637\pm 0.058^{+0.045}_{-0.043}$ &\cite{Abe:1998wt}\\
CDF1   &Excl.\ \particle{\jpsi K} &92--95 &$1.636\pm 0.058\pm 0.025$          &\cite{Acosta:2002nd}\\
CDF2   &Excl.\ \particle{\jpsi K} &02--09 &$1.639\pm 0.009\pm 0.009$          &\citehistory{Aaltonen:2010pj}{Aaltonen:2010pj,*Abulencia:2006dr_hist}\\ 
%%% CDF2   &Incl.\ \particle{D^0 \ell} &02--04 &$1.653\pm 0.029^{+0.033}_{-0.031}$ &\cite{CDFnote7514:2005}$^p$\\
CDF2   &Excl.\ \particle{D^0 \pi}  &02--06 &$1.663\pm 0.023\pm0.015$           &\cite{Aaltonen:2010ta}\\
\babar &Exclusive                  &99--00 &$1.673\pm 0.032\pm 0.023$          &\cite{Aubert:2001uw}\\
\belle &Exclusive                  &00--03 &$1.635\pm 0.011\pm 0.011$          &\citehistory{Abe:2004mz}{Abe:2004mz,*Abe:2002id_hist,*Tomura:2002qs_hist,*Hara:2002mq_hist} \\
%%% LHCb  & Excl.\ \particle{\jpsi K} & 2010 & $1.689 \pm0.022 \pm 0.047$ & \cite{LHCb-CONF-2011-001}$^p$ \\
LHCb  & Excl.\ \particle{\jpsi K} & 2011 & $1.637 \pm0.004 \pm 0.003$ & \cite{Aaij:2014owa} \\
\hline
Average&                           &       &\hflavTAUBUnounit &\\
\hline\hline
\multicolumn{5}{l}{$^a$ \footnotesize The combined DELPHI result quoted 
in~\cite{Adam:1995mb} is $1.70 \pm 0.09$ ps.} \\[-0.5ex]
\multicolumn{5}{l}{$^b$ \footnotesize The combined SLD result 
quoted in~\cite{Abe:1997ys} is $1.66 \pm 0.06 \pm 0.05$ ps.}\\[-0.5ex]
% \multicolumn{5}{l}{$^p$ {\footnotesize Preliminary.}}
\end{tabular}
\end{table}
\begin{table}[t]
\centering
\caption{Measurements of the ratio $\tau(\Bu)/\tau(\Bd)$.}
\labt{liferatioBuBd}
\begin{tabular}{lcccl} 
\hline
Experiment &Method                 &Data set &Ratio $\tau(\Bu)/\tau(\Bd)$      &Ref.\\
\hline
ALEPH  &\particle{D^{(*)} \ell}    &91--95 &$1.085\pm 0.059\pm 0.018$          &\cite{Barate:2000bs}\\
ALEPH  &Exclusive                  &91--94 &$1.27^{+0.23+0.03}_{-0.19-0.02}$   &\cite{Buskulic:1996hy}\\
DELPHI &\particle{D^{(*)} \ell}    &91--93 &$1.00^{+0.17}_{-0.15}\pm 0.10$     &\cite{Abreu:1995mc}\\
DELPHI &Charge sec.\ vtx           &91--93 &$1.06^{+0.13}_{-0.11}\pm 0.10$     &\cite{Adam:1995mb}\\
DELPHI &Charge sec.\ vtx           &94--95 &$1.060\pm 0.021 \pm 0.024$         &\cite{Abdallah:2003sb}\\
L3     &Charge sec.\ vtx           &94--95 &$1.09\pm 0.07  \pm 0.03$           &\cite{Acciarri:1998uv}\\
OPAL   &\particle{D^{(*)} \ell}    &91--93 &$0.99\pm 0.14^{+0.05}_{-0.04}$     &\cite{Akers:1995pa}\\
OPAL   &Charge sec.\ vtx           &93--95 &$1.079\pm 0.064 \pm 0.041$         &\cite{Abbiendi:1998av}\\
SLD    &Charge sec.\ vtx $\ell$    &93--95 &$1.03^{+0.16}_{-0.14} \pm 0.09$    &\cite{Abe:1997ys}$^a$\\
SLD    &Charge sec.\ vtx           &93--95 &$1.01^{+0.09}_{-0.08} \pm0.05$     &\cite{Abe:1997ys}$^a$\\
CDF1   &\particle{D^{(*)} \ell}    &92--95 &$1.110\pm 0.056^{+0.033}_{-0.030}$ &\cite{Abe:1998wt}\\
CDF1   &Excl.\ \particle{\jpsi K} &92--95 &$1.093\pm 0.066 \pm 0.028$         &\cite{Acosta:2002nd}\\
CDF2   &Excl.\ \particle{\jpsi K^{(*)}} &02--09 &$1.088\pm 0.009 \pm 0.004$   &\citehistory{Aaltonen:2010pj}{Aaltonen:2010pj,*Abulencia:2006dr_hist}\\ 
%%% CDF2   &Incl.\ \particle{D \ell}   &02--04 &$1.123\pm0.040^{+0.041}_{-0.039}$  &\cite{CDFnote7514:2005}$^p$\\
%%% CDF2   &Excl.\ \particle{D \pi}    &02--04 &$1.10\pm 0.02\pm 0.01$             &\cite{CDFnote7386:2005}$^p$\\
\dzero &\particle{D^{*+} \mu} \particle{D^0 \mu} ratio
	                           &02--04 &$1.080\pm 0.016\pm 0.014$          &\cite{Abazov:2004sa}\\
\babar &Exclusive                  &99--00 &$1.082\pm 0.026\pm 0.012$          &\cite{Aubert:2001uw}\\
\belle &Exclusive                  &00--03 &$1.066\pm 0.008\pm 0.008$          &\citehistory{Abe:2004mz}{Abe:2004mz,*Abe:2002id_hist,*Tomura:2002qs_hist,*Hara:2002mq_hist} \\
LHCb  & Excl.\ \particle{\jpsi K^{(*)}} & 2011 & $1.074 \pm0.005 \pm 0.003$ & \cite{Aaij:2014owa} \\
\hline
Average&                           &       & \hflavRTAUBU & \\   
\hline\hline
\multicolumn{5}{l}{$^a$ \footnotesize The combined SLD result quoted
	   in~\cite{Abe:1997ys} is $1.01 \pm 0.07 \pm 0.06$.}
%%% \\[-0.5ex] \multicolumn{5}{l}{$^p$ {\footnotesize Preliminary.}}
\end{tabular}
\end{table}

The averaging of $\tau(\Bu)$, $\tau(\Bd)$ and $\tau(\Bu)/\tau(\Bd)$
measurements is summarized\unpublished{}{\footnote{%
We do not include the old unpublished measurements of Refs.~\cite{CDFnote7514:2005,CDFnote7386:2005,ATLAS-CONF-2011-092}.}}
in \Tablesss{lifebd}{lifebu}{liferatioBuBd}.
For $\tau(\Bu)/\tau(\Bd)$ we average only the measurements of this
quantity provided by experiments rather than using all available
knowledge, which would have included, for example, $\tau(\Bu)$ and
$\tau(\Bd)$ measurements which did not contribute to any of the ratio
measurements.

The following sources of correlated (within experiment/machine)
systematic uncertainties have been considered:
% (central values and errors scaled accordingly):
\begin{itemize}
\item for SLC/LEP measurements -- \particle{D^{**}} branching ratio uncertainties~\cite{Abbaneo:2000ej_mod,*Abbaneo:2001bv_mod_cont},
momentum estimation of \b mesons from \particle{Z^0} decays
(\b-quark fragmentation parameter $\langle X_E \rangle = 0.702 \pm 0.008$~\cite{Abbaneo:2000ej_mod,*Abbaneo:2001bv_mod_cont}),
\Bs and \b-baryon lifetimes (see \Secss{taubs}{taulb}),
and \b-hadron fractions at high energy (see \Table{fractions}); 
\item for \B-factory measurements -- alignment, $z$ scale, machine boost,
sample composition (where applicable);
\item for Tevatron/LHC measurements -- alignment (separately
within each experiment).
\end{itemize}
The resultant averages are:
\begin{eqnarray}
\tau(\Bd) & = & \hflavTAUBD \,, \\
\tau(\Bu) & = & \hflavTAUBU \,, \\
\tau(\Bu)/\tau(\Bd) & = & \hflavRTAUBU \,.
\end{eqnarray}
  % from Konstantin
% Bs lifetime, Bc lifetime, \Lb and \b-baryon lifetimes, theoretical predictions
%%%%%%%%%%%%%%%%%%%%%%%%%%%%%%%%%%%%%%%%%%%%%%%%
%
% This is file life_mix_tau2.tex containing
% the second part of the chapter on the b-hadron lifetimes: 
% Bs, Bc, lambda_b and b-baryon lifetimes
% as well as theroretical predictions for all b-hadron lifetimes.
%
%%%%%%%%%%%%%%%%%%%%%%%%%%%%%%%%%%%%%%%%%%%%%%%
%

\mysubsubsection{\Bs lifetimes}
\labs{taubs}

Like neutral kaons, neutral \B mesons contain
short- and long-lived components, since the
light (L) and heavy (H)
eigenstates %, $\B_{q\rm L}$ and $\B_{q\rm H}$, 
differ not only
in their masses, but also in their total decay widths. 
% ,  with a decay width difference defined as $\DG_q = \Gamma_{q\rm L} - \Gamma_{q\rm H}$. 
Neglecting \CP violation in $\Bs-\Bsbar$ mixing, 
which is expected to be very
small~\citehistory{Jubb:2016mvq,Artuso:2015swg,Laplace:2002ik,Ciuchini:2003ww,Beneke:2003az}{Jubb:2016mvq,Artuso:2015swg,*Lenz_hist,Laplace:2002ik,Ciuchini:2003ww,Beneke:2003az}
(see also \Sec{qpd}), the mass eigenstates are also \CP eigenstates,
with the light % $\B_{q\rm L}$ 
state being \CP-even and the heavy % $\B_{q\rm H}$ 
state being \CP-odd. 
While the decay width difference \DGd can be neglected in the \Bd system, 
the \Bs system exhibits a significant value of 
$\DGs = \Gamma_{s\rm L} - \Gamma_{s\rm H}$, where $\Gamma_{s\rm L}$ and $\Gamma_{s\rm H}$
are the total decay widths of the light eigenstate $\B^0_{s\rm L}$ and the heavy eigenstate $\B^0_{s\rm H}$, respectively.
The sign of \DGs is known to be positive~\cite{Aaij:2012eq}, \ie,
$\B^0_{s\rm H}$ lives longer than $\B^0_{s\rm L}$. 
Specific measurements of \DGs and 
$\Gs = (\Gamma_{s\rm L} + \Gamma_{s\rm H})/2$ are explained
and averaged in \Sec{DGs}, but the results for
$1/\Gamma_{s\rm L} = 1/(\Gs+\DGs/2)$, $1/\Gamma_{s\rm H}= 1/(\Gs-\DGs/2)$
and the mean \Bs lifetime, defined as $\tau(\Bs) = 1/\Gs$, are also quoted at the end of this section. 

Many \Bs lifetime analyses, in particular the early 
ones performed before the non-zero value of \DGs was 
firmly established, ignore \DGs and fit the proper time 
distribution of a sample of \Bs candidates 
reconstructed in a certain final state $f$
with a model assuming a single exponential function 
for the signal. We denote such {\rm effective lifetime}
measurements~\cite{Fleischer:2011cw} as $\tau_{\rm single}(\Bs\to f)$; 
their true values may lie {\em a priori} anywhere
between $1/\Gamma_{s\rm L}$ % $ = 1/(\Gs+\DGs/2)$ 
and $1/\Gamma_{s,\rm H}$, % $ = 1/(\Gs-\DGs/2)$, 
depending on the proportion of $B^0_{s\rm L}$ and $B^0_{s\rm H}$
in the final state $f$. 
More recent determinations of effective lifetimes may be interpreted as
measurements of the relative composition of 
$B^0_{s\rm L}$ and $B^0_{s\rm H}$
decaying to the final state $f$. 
\Table{lifebs} summarizes the effective 
lifetime measurements.

Averaging measurements of $\tau_{\rm single}(\Bs\to f)$
over several final states $f$ will yield a result 
corresponding to an ill-defined observable
when the proportions of $B^0_{s\rm L}$ and $B^0_{s\rm H}$
differ. 
Therefore, the effective \Bs lifetime measurements are broken down into
several categories and averaged separately.

\begin{table}[t]
\centering
\caption{Measurements of the effective \Bs lifetimes obtained from single exponential fits.}
% without attempting to separate the short and long components.} % \CP-even and \CP-odd 
\labt{lifebs}
\resizebox{\textwidth}{!}{
\begin{tabular}{l@{}c@{}c@{}c@{}rc@{}l} \hline
Experiment & \multicolumn{2}{c}{Final state $f$}           & \multicolumn{2}{c}{Data set} & $\tau_{\rm single}(\Bs\to f)$ (ps) & Ref. \\
\hline \hline
ALEPH  & \particle{D_s h}     & ill-defined & 91--95 & & $1.47\pm 0.14\pm 0.08$           & \cite{Barate:1997ua}          \\
DELPHI & \particle{D_s h}     & ill-defined & 91--95 & & $1.53^{+0.16}_{-0.15}\pm 0.07$   & \citehistory{Abreu:2000ev}{Abreu:2000ev,*Abreu:1996ep_hist} \\
%%OS 23apr2005: this is superseded by \citehistory{Abreu:2000ev}{Abreu:2000ev,*Abreu:1996ep_hist} %% DELPHI & \particle{D_s} incl. & mixture & 91--94 & $1.60\pm 0.26^{+0.13}_{-0.15}$   & \cite{DELBS2}          \\
OPAL   & \particle{D_s} incl. & ill-defined & 90--95 & & $1.72^{+0.20+0.18}_{-0.19-0.17}$ & \cite{Ackerstaff:1997ne}          \\ 
%% ALEPH    & \particle{D_s^{(*)+}D_s^{(*)-}} & \CP-even ? & 91--95 & 4M \particle{Z\to q\bar{q}} & $1.27 \pm 0.33 \pm 0.08$ & \cite{Barate:2000kd} \\
\hline
ALEPH  & \particle{D_s^- \ell^+}  & flavour-specific & 91--95 & & $1.54^{+0.14}_{-0.13}\pm 0.04$   & \cite{Buskulic:1996ei}          \\
CDF1   & \particle{D_s^- \ell^+}  & flavour-specific & 92--96 & & $1.36\pm 0.09 ^{+0.06}_{-0.05}$  & \cite{Abe:1998cj}           \\
DELPHI & \particle{D_s^- \ell^+}  & flavour-specific & 92--95 & & $1.42^{+0.14}_{-0.13}\pm 0.03$   & \cite{Abreu:2000sh}          \\
OPAL   & \particle{D_s^- \ell^+}  & flavour-specific & 90--95 & & $1.50^{+0.16}_{-0.15}\pm 0.04$   & \cite{Ackerstaff:1997qi}  \\
%% superseded by line below: \dzero & \particle{D_s^- \mu^+}   & flavour-specific & 02--04 & 0.4 fb$^{-1}$ & $1.398 \pm 0.044 ^{+0.028}_{-0.025}   $   & \cite{Abazov:2006cb}       \\ 
\dzero & \particle{D_s^-\mu^+X}   & flavour-specific & Run~II & 10.4 fb$^{-1}$ & $1.479 \pm 0.010 \pm 0.021$   & \citehistory{Abazov:2014rua}{Abazov:2014rua,*Abazov:2006cb_hist} \\
%%% CDF2   & \particle{D_s^- \ell^+}  & flavour-specific & 02--04 & & $1.381 \pm 0.055 ^{+0.052}_{-0.046} $ & \cite{CDFnote7757:2005}$^p$ \\
%%%%% CDF2   & \particle{D_s^- \pi^+, D_s^- \pi^+ \pi^- \pi^+} 
CDF2   & \particle{D_s^- \pi^+ (X)} 
                              & flavour-specific & 02--06 & 1.3 fb$^{-1}$ & $1.518 \pm 0.041 \pm 0.027     $   & \unpublished{\cite{Aaltonen:2011qsa}}{\citehistory{Aaltonen:2011qsa}{Aaltonen:2011qsa,*Aaltonen:2011qsa_hist}} \\ %was \cite{CDFnote9203:2008,*CDFnote9203:2008_cont}$^p$      \\
LHCb   &  \particle{D_s^- D^+} & flavour-specific & 11--12 & 3 fb$^{-1}$ & $1.52 \pm 0.15 \pm 0.01$ & \cite{Aaij:2013bvd} \\
LHCb   &  \particle{D_s^- \pi^+} & flavour-specific & 11 & 1 fb$^{-1}$ & $1.535 \pm 0.015 \pm 0.014$ & \cite{Aaij:2014sua} \\
LHCb    & \particle{\pi^+K^-}   &  flavour-specific & 11 & 1.0 fb$^{-1}$ & $1.60 \pm 0.06 \pm 0.01$ & \citehistory{Aaij:2014fia}{Aaij:2014fia,*Aaij:2012ns_hist} \\
\multicolumn{5}{l}{Average of above 9 flavour-specific lifetime measurements} &  \hflavTAUBSSLnounit & \\  
\hline\hline
CDF1     & \particle{\jpsi\phi} & \CP even+odd & 92--95 &  & $1.34^{+0.23}_{-0.19}    \pm 0.05$ & \cite{Abe:1997bd} \\
%%% CDF2     & \particle{\jpsi\phi} & \CP even+odd & 02--06 &  & $1.494 \pm 0.054 \pm 0.009$ &  \citehistory{CDFnote8524:2007}{CDFnote8524:2007,*CDFnote8524:2007_hist}$^p$ \\
\dzero   & \particle{\jpsi\phi} & \CP even+odd & 02--04 &  & $1.444^{+0.098}_{-0.090} \pm 0.02$ & \cite{Abazov:2004ce}  \\
% ATLAS & \particle{\jpsi\phi} & \CP even+odd & 10 & 40 pb$^{-1}$ & $1.41 \pm0.08 \pm0.05$ & \cite{ATLAS-CONF-2011-092}$^p$ \\
% above measurement removed from average on Oct 15, 2016 because if remained unpublished for 4 years
%%% LHCb  & \particle{\jpsi\phi} & \CP even+odd & 10 & 36 pb$^{-1}$ & $1.447 \pm0.064 \pm 0.056$ & \cite{LHCb-CONF-2011-001}$^p$ \\
LHCb  & \particle{\jpsi\phi} & \CP even+odd & 11 & 1 fb$^{-1}$ & $1.480 \pm0.011 \pm 0.005$ & \cite{Aaij:2014owa} \\
\multicolumn{5}{l}{Average of above 3 \particle{\jpsi \phi} lifetime measurements} &  \hflavTAUBSJFnounit & \\ 
\hline\hline
ALEPH    & \particle{D_s^{(*)+}D_s^{(*)-}} & mostly \CP even & 91--95 & & $1.27 \pm 0.33 \pm 0.08$ & \cite{Barate:2000kd} \\
\hline
%%% CDF2 measurement below removed from the average because it remained unpublished for two long
%%% CDF2     & \particle{K^+K^-}   & \CP-even & 02--04 & 0.36 fb$^{-1}$ & $1.53 \pm 0.18 \pm 0.02$ & \cite{Tonelli:2006np}$^p$ \\
LHCb    & \particle{K^+K^-}   &  \CP-even & 10 & 0.037 fb$^{-1}$ & $1.440 \pm 0.096 \pm 0.009$ & \cite{Aaij:2012kn} \\
%%% superseded by next line LHCb    & \particle{K^+K^-}   &  \CP-even & 11 & 1.0 fb$^{-1}$ & $1.455 \pm 0.046 \pm 0.006$ & \cite{Aaij:2012ns} \\
LHCb    & \particle{K^+K^-}   &  \CP-even & 11 & 1.0 fb$^{-1}$ & $1.407 \pm 0.016 \pm 0.007$ & \citehistory{Aaij:2014fia}{Aaij:2014fia,*Aaij:2012ns_hist} \\
\multicolumn{5}{l}{Average of above 2 \particle{K^+K^-} lifetime measurements} &  \hflavTAUBSKKnounit & \\ 
\hline
LHCb   &  \particle{D_s^+ D_s^-} & \CP-even & 11--12 & 3 fb$^{-1}$ & $1.379 \pm 0.026 \pm 0.017$ & \cite{Aaij:2013bvd} \\
LHCb   &  \particle{\jpsi\eta} & \CP-even & 11--12 & 3 fb$^{-1}$ & $1.479 \pm 0.034 \pm 0.011$ &\cite{Aaij:2016dzn} \\
\multicolumn{5}{l}{Average of above 2 measurements of $1/\Gamma_{s\rm L}$} &  \hflavTAUBSSHORTnounit & \\ \hline \hline
LHCb     & \particle{\jpsi K^0_{\rm S}} & \CP-odd & 11   & 1.0 fb$^{-1}$ & $1.75 \pm 0.12 \pm 0.07$ & \cite{Aaij:2013eia} \\
\hline
CDF2     & \particle{\jpsi f_0(980)} & \CP-odd & 02--08 & 3.8 fb$^{-1}$ & $1.70^{+0.12}_{-0.11} \pm 0.03$ & \cite{Aaltonen:2011nk} \\
\dzero       & \particle{\jpsi f_0(980)} & \CP-odd & Run~II & 10.4 fb$^{-1}$ & $1.70\pm 0.14 \pm 0.05$ & \cite{Abazov:2016oqi} \\
%%LHCb     & \particle{\jpsi f_0(980)} & \CP-odd & 11   & 1.0 fb$^{-1}$ & $1.700 \pm 0.040 \pm 0.026$ & \cite{Aaij:2012nta} \\
LHCb     & \particle{\jpsi \pi^+\pi^-} & \CP-odd & 11   & 1.0 fb$^{-1}$ & $1.652 \pm 0.024 \pm 0.024$ & \citehistory{Aaij:2013oba}{Aaij:2013oba,*LHCb:2011aa_hist,*LHCb:2012ad_hist,*LHCb:2011ab_hist,*Aaij:2012nta_hist} \\
% \multicolumn{5}{l}{Average of above 3 \particle{\jpsi f_0(980)}, \particle{\jpsi \pi^+\pi^-} measurements} &  \hflavTAUBSJPSIPIPInounit & \\ \hline 
\multicolumn{5}{l}{Average of above 3 measurements of $1/\Gamma_{s\rm H}$} &  \hflavTAUBSLONGnounit & \\ \hline \hline
%\multicolumn{5}{l}{$^p$ \footnotesize Preliminary.}
\end{tabular}
}
\end{table}

\afterpage{\clearpage}

\begin{itemize}

\item
{\bf\em \boldmath $\Bs\to D_s^{\mp} X$ decays}
include mostly flavour-specific decays but also decays 
with an unknown mixture of light and heavy components. 
Measurements performed with such inclusive states are
no longer used in averages. 
%OLD% The corresponding effective lifetime average,
%OLD% \begin{equation}
%OLD% \tau_{\rm single}(\Bs\to D_s^{\mp} X) = \hflavTAUBSwaschanged \,,
%OLD% \end{equation}
%OLD% can still be a useful input
%OLD% for analyses examining an inclusive $D_s$ sample.
%OLD% The following correlated systematic errors were considered:
%OLD% average \B lifetime used in backgrounds,
%OLD% \Bs decay multiplicity, and branching ratios used to determine 
%OLD% backgrounds (\eg\ \BR{B\to D_s D}).
%OLD% A knowledge of the multiplicity of \Bs decays is important for
%OLD% measurements that partially reconstruct the final state such as 
%OLD% \particle{\B\to D_s \mbox{$X$}} (where $X$ is not a lepton). 
%OLD% The boost deduced from Monte Carlo simulation depends on the multiplicity used.
%OLD% Since this is not well known, the multiplicity in the simulation is
%OLD% varied and this range of values observed is taken to be a systematic.
%OLD% Similarly not all the branching ratios for the potential background
%OLD% processes are measured. Where they are available, the PDG values are
%OLD% used for the error estimate. Where no measurements are available
%OLD% estimates can usually be made by using measured branching ratios of
%OLD% related processes and using some reasonable extrapolation.

\item 
{\bf\em Decays to flavour-specific final states}, 
\ie, decays to final states $f$ with decay amplitudes satisfying 
$A(\Bs\to f) \ne 0$, $A(\Bsbar\to \bar{f}) \ne 0$, 
$A(\Bs\to \bar{f}) = 0$ and $A(\Bsbar\to f)=0$, 
have equal 
fractions of $B^0_{s\rm L}$ and $B^0_{s\rm H}$ at time zero.
%2015% \footnote{%
%2015% The assumption that such decays are flavour-specific is valid to an excellent approximation in the SM.
%2015% However, there are few experimental tests of it.}
% , where $\tau_{s\rm L} = 1/\Gamma_{s\rm L}$ 
% is expected to be the shorter-lived component and
% $\tau_{s\rm H} = 1/\Gamma_{s\rm H}$ 
% expected to be the longer-lived component. 
%2015% If the resulting superposition of two exponential distributions
%2015% is fitted with a single exponential function, 
%2015% one obtains a measure of the so-called {\em flavour-specific lifetime}~\cite{Hartkorn:1999ga}:
% A superposition of two exponentials thus results with decay
% widths $\Gs \pm \DGs /2$.  Fitting to a single exponential one obtains a
% measure of the flavour-specific lifetime~\cite{Hartkorn:1999ga}:
Their total untagged time-dependent decay rates $\Gamma_s(t)$
have a mean value $\int_0^\infty t\Gamma_s(t)dt/\int_0^\infty \Gamma_s(t)dt$,
called the {\em flavour-specific lifetime}, equal to~\cite{Hartkorn:1999ga}
\begin{equation}
\tau_{\rm single}(\Bs\to \mbox{flavour specific})
 =  \frac{1/\Gamma_{s\rm L}^2+1/\Gamma_{s\rm H}^2}{1/\Gamma_{s\rm L}+1/\Gamma_{s\rm H}}
% = \frac{\frac{1}{\Gamma_{s\rm L}^2}+\frac{1}{\Gamma_{s\rm H}^2}}{\frac{1}{\Gamma_{s\rm L}}+\frac{1}{\Gamma_{s\rm H}}}
 = \frac{1}{\Gs} \,
\frac{{1+\left(\frac{\DGs}{2\Gs}\right)^2}}{{1-\left(\frac{\DGs}{2\Gs}\right)^2}
}\,.
\labe{fslife}
\end{equation}
%Estimates of the flavour-specific lifetime are obtained by fitting the untagged proper-time distribution
%%%, which consist of the superposition of two exponentials functions, 
%with a single exponential function. 

Because of the fast $\Bs-\Bsbar$ oscillations, 
possible biases of the flavour-specific lifetime due to a
combination of $\Bs/\Bsbar$ production asymmetry,
\CP violation in the decay amplitudes ($|A(\Bs\to f)| \ne |A(\Bsbar\to \bar{f})|$), 
and \CP violation in $\Bs-\Bsbar$ mixing
($|q_{\particle{s}}/p_{\particle{s}}| \ne 1$) 
are strongly suppressed, by a factor $\sim x_s^2$ (given in \Eq{xs}).
The $\Bs/\Bsbar$ production asymmetry at LHCb and the \CP asymmetry due to mixing 
have been measured to be compatible with zero with a precision below 3\%~\cite{Aaij:2014bba} 
and 0.3\% (see \Eq{ASLS}), respectively. The corresponding effects on the flavour-specific lifetime, which therefore have a relative size of the order of $10^{-5}$ or smaller, can be neglected at the current level of experimental precision.
Under the assumption of no production asymmetry 
and no \CP violation in mixing, \Eq{fslife} is exact even for a flavour-specific decay with 
\CP violation in the decay amplitudes. Hence any flavour-specific decay 
mode can be used to measure the flavour-specific lifetime. 
%2015% This average does not include an effective lifetime measurement of 
%2015% $\Bs \to \pi^+K^-$ decays~\citehistory{Aaij:2014fia}{Aaij:2014fia,*Aaij:2012ns_hist}.
%2015% % where tree and penguin amplitudes may interfere. 

%2015% such as $\Bs \to \particle{D_s^- \ell^+ \nu}$
%2015% or $\Bs\to \particle {D_s^- \pi^+}$, have equal 
%2015% fractions of $\B_{\rm L}$ and $\B_{\rm H}$ at time zero.

The average of all flavour-specific 
\Bs lifetime measurements%
\unpublished{\citehistory{Buskulic:1996ei,Abe:1998cj,Abreu:2000sh,Ackerstaff:1997qi,Abazov:2014rua,Aaltonen:2011qsa,Aaij:2013bvd,Aaij:2014sua,Aaij:2014fia}{Buskulic:1996ei,Abe:1998cj,Abreu:2000sh,Ackerstaff:1997qi,Abazov:2014rua,*Abazov:2006cb_hist,Aaltonen:2011qsa,Aaij:2013bvd,Aaij:2014sua,Aaij:2014fia,*Aaij:2012ns_hist}}{\citehistory{Buskulic:1996ei,Abe:1998cj,Abreu:2000sh,Ackerstaff:1997qi,Abazov:2014rua,Aaltonen:2011qsa,Aaij:2013bvd,Aaij:2014sua,Aaij:2014fia}{Buskulic:1996ei,Abe:1998cj,Abreu:2000sh,Ackerstaff:1997qi,Abazov:2014rua,*Abazov:2006cb_hist,Aaltonen:2011qsa,*Aaltonen:2011qsa_hist,Aaij:2013bvd,Aaij:2014sua,Aaij:2014fia,*Aaij:2012ns_hist}}%
\unpublished{}{\footnote{%
An old unpublished measurement~\cite{CDFnote7757:2005} is not included.}}
is
\begin{equation}
\tau_{\rm single}(\Bs\to \mbox{flavour specific}) = \hflavTAUBSSL \,.
\labe{tau_fs}
\end{equation}
% is used in \Sec{DGs} as one of the ingredients 
% to determine $\tau(\Bs) = 1/\Gs$ and \DGs.

\item
{\bf\em 
{\boldmath $\Bs \to \jpsi\phi$ \unboldmath}decays}
contain a well-measured mixture of \CP-even and \CP-odd states.
% are expected to be
% dominated by the \CP-even state and its lifetime.
There are no known correlations
between the existing 
\particle{\Bs\to \jpsi\phi}
effective lifetime measurements; these are combined  
into the average\unpublished{}{\footnote{%
The old unpublished measurements of Refs.~\citehistory{CDFnote8524:2007,ATLAS-CONF-2011-092}{CDFnote8524:2007,*CDFnote8524:2007_hist,ATLAS-CONF-2011-092} are not included.}}
% \begin{equation}
$\tau_{\rm single}(\Bs\to \jpsi \phi) = \hflavTAUBSJF$. % \,.
% \end{equation}
A caveat is that different experimental acceptances
may lead to different admixtures of the 
\CP-even and \CP-odd states, and simple fits to a single
exponential may result in inherently different 
values of $\tau_{\rm single}(\Bs\to \jpsi \phi)$.
Analyses that separate the \CP-even and \CP-odd components in
this decay through a full angular study, outlined in \Sec{DGs},
provide directly precise measurements of $1/\Gs$ and $\DGs$ (see \Table{phisDGsGs}).

\item
{\bf\em Decays to \boldmath\CP eigenstates} have also 
been measured, in the \CP-even modes 
$\Bs \to D_s^{(*)+}D_s^{(*)-}$ by ALEPH~\cite{Barate:2000kd},
$\Bs \to K^+ K^-$ by LHCb~\citehistory{Aaij:2012kn,Aaij:2014fia}{Aaij:2012kn,Aaij:2014fia,*Aaij:2012ns_hist}%
\unpublished{}{\footnote{An old unpublished measurement of the $\Bs \to K^+ K^-$
effective lifetime by CDF~\cite{Tonelli:2006np} is no longer considered.}},
$\Bs \to D_s^+D_s^-$ by LHCb~\cite{Aaij:2013bvd}
and $\Bs \to J/\psi \eta$ by LHCb~\cite{Aaij:2016dzn}, as well as in the \CP-odd modes 
$\Bs \to \jpsi f_0(980)$ by CDF~\cite{Aaltonen:2011nk}
and \dzero~\cite{Abazov:2016oqi},
$\Bs \to \jpsi \pi^+\pi^-$ by LHCb~\citehistory{Aaij:2013oba}{Aaij:2013oba,*LHCb:2011aa_hist,*LHCb:2012ad_hist,*LHCb:2011ab_hist,*Aaij:2012nta_hist}
and $\Bs \to \jpsi K^0_{\rm S}$ by LHCb~\cite{Aaij:2013eia}.
If these 
decays are dominated by a single weak phase and if \CP violation 
can be neglected, then $\tau_{\rm single}(\Bs \to \mbox{\CP-even}) = 1/\Gamma_{s\rm L}$ 
and  $\tau_{\rm single}(\Bs \to \mbox{\CP-odd}) = 1/\Gamma_{s\rm H}$ 
(see \Eqss{tau_KK_approx}{tau_Jpsif0_approx} for approximate relations in presence of mixing-induced
\CP violation). 
However, not all these modes can be considered as pure \CP eigenstates:
a small \CP-odd component is most probably present
in $\Bs \to D_s^{(*)+}D_s^{(*)-}$ decays. Furthermore the decays
$\Bs \to K^+ K^-$ and $\Bs \to \jpsi K^0_{\rm S}$ %to pure \CP eigenstates 
may suffer from direct \CP violation due to interfering tree and loop amplitudes. 
The averages for the effective lifetimes obtained for decays to
pure \CP-even ($D_s^+D_s^-$, $\jpsi\eta$) and \CP-odd ($\jpsi f_0(980)$, $\jpsi \pi^+\pi^-$)
final states, where \CP conservation can be assumed, are
\begin{eqnarray}
\tau_{\rm single}(\Bs \to \mbox{\CP-even}) & = & \hflavTAUBSSHORT \,,
\labe{tau_KK}
\\
\tau_{\rm single}(\Bs \to \mbox{\CP-odd}) & = & \hflavTAUBSLONG \,.
\labe{tau_Jpsif0}
\end{eqnarray}
% A measurement of the effective lifetime of $\Bs \to D_s^{(*)+}D_s^{(*)-}$ decays by ALEPH~\cite{Barate:2000kd}
% is not included in the above \CP-even average, since a small \CP-odd component is most probably present. 

\end{itemize}

As described in \Sec{DGs}, 
the effective lifetime averages of \Eqsss{tau_fs}{tau_KK}{tau_Jpsif0}
are used as ingredients to improve the 
determination of $1/\Gs$ and \DGs obtained from the full angular analyses
of $\Bs\to \jpsi\phi$ and $\Bs\to \jpsi K^+K^-$ decays. 
The resulting world averages for the \Bs lifetimes are
\begin{eqnarray}
\tau(B^0_{s\rm L}) = \frac{1}{\Gamma_{s\rm L}}
 = \frac{1}{\Gs+\DGs/2} & = & \hflavTAUBSLCON \,, \\
\tau(B^0_{s\rm H}) = \frac{1}{\Gamma_{s\rm H}}
 = \frac{1}{\Gs-\DGs/2} & = & \hflavTAUBSHCON \,, \\
\tau(\Bs) = \frac{1}{\Gs} = \frac{2}{\Gamma_{s\rm L}+\Gamma_{s\rm H}} & = & \hflavTAUBSMEANCON \,.
\labe{oneoverGs}
\end{eqnarray}

\mysubsubsection{\Bc lifetime}
\labs{taubc}

Early measurements of the \Bc meson lifetime,
from CDF~\unpublished{\cite{Abe:1998wi,Abulencia:2006zu}}{\cite{Abe:1998wi,CDFnote9294:2008,Abulencia:2006zu}} and \dzero~\cite{Abazov:2008rba},
use the semileptonic decay mode \particle{\Bc \to \jpsi \ell^+ \nu} 
and are based on a 
simultaneous fit to the mass and lifetime using the vertex formed
with the leptons from the decay of the \particle{\jpsi} and
the third lepton. Correction factors
to estimate the boost due to the missing neutrino are used.
%OS% In the analysis of the CDF Run~I data~\cite{Abe:1998wi},
%OS% a mass value of 
%OS% $6.40 \pm 0.39 \pm 0.13~\gevcc$ 
%OS% is found by fitting
%OS% to the tri-lepton invariant mass spectrum. 
%OS% %%% START WARNING
%OS% %%% Text below is valid when CDFnote9294:2008,*Abulencia:2006zu_mod_cont is the published result from 2006
%OS% In the CDF Run~II result~\cite{Abulencia:2006zu}, the mass is fixed
%OS% to 6.271~\gevcc$, but then varied between 
%OS% 6.2 and 6.4~\gevcc$ to assess the systematic error on the
%OS% lifetime due to the \Bc mass value.
%OS% Finally, in the \dzero Run~II result~\cite{Abazov:2008rba}, 
%OS% %%% Text below valid when CDFnote9294:2008,*Abulencia:2006zu_mod_cont is the new CDF prel. result from CD note 9294
%OS% % In the CDF and \dzero Run~II results~\cite{CDFnote9294:2008,*Abulencia:2006zu_mod_cont,Abazov:2008rba}, 
%OS% %%% END WARNING
%OS% the \Bc mass is assumed to be 
%OS% $6285.7 \pm 5.3 \pm 1.2$~MeV/$c^2$, taken from a 
%OS% CDF result~\cite{Abulencia:2005usa}. 
%OS% These mass measurements
%OS% are consistent within uncertainties, and also consistent with the
%OS% most recent precision determination from CDF of 
%OS% $6275.6 \pm 2.9 \pm 2.5$~MeV/$c^2$~\cite{Aaltonen:2007gv}.
Correlated systematic errors include the impact
of the uncertainty of the \Bc $p_T$ spectrum on the correction
factors, the level of feed-down from $\psi(2S)$ decays, 
Monte Carlo modeling of the decay model varying from phase space
to the ISGW model, and mass variations.
With more statistics, CDF2 was able to perform the first \Bc lifetime 
based on fully reconstructed
$\Bc \to J/\psi \pi^+$ decays~\cite{Aaltonen:2012yb},
which does not suffer from a missing neutrino. Recent measurements from 
LHCb, both with  
\particle{\Bc \to \jpsi \mu^+ \nu}~\cite{Aaij:2014bva} and 
\particle{\Bc \to \jpsi \pi^+}~\cite{Aaij:2014gka} decays, achieve the 
highest level of precision. 

%OS% The latest determination of the \Bc lifetime from CDF2~\cite{Aaltonen:2012yb} is based on fully reconstructed 
%OS% $\Bc \to J/\psi \pi^+$ decays and does not suffer from a missing neutrino. 
All the measurements\unpublished{}{\footnote{We do not list (nor include in the average) an unpublished result from CDF2~\cite{CDFnote9294:2008}.}}
are summarized in 
\Table{lifebc} and the world average, dominated by the LHCb measurements, is
determined to be
\begin{equation}
\tau(\Bc) = \hflavTAUBC \,.
\end{equation}

\begin{table}[tb]
\centering
\caption{Measurements of the \Bc lifetime.}
\labt{lifebc}
\begin{tabular}{lccrcl} \hline
Experiment & Method                    & \multicolumn{2}{c}{Data set}  & $\tau(\Bc)$ (ps)
      & Ref.\\   \hline
CDF1       & \particle{\jpsi \ell} & 92--95 & 0.11 fb$^{-1}$ & $0.46^{+0.18}_{-0.16} \pm
 0.03$   & \cite{Abe:1998wi}  \\ 
CDF2       & \particle{\jpsi e} & 02--04 & 0.36 fb$^{-1}$ & $0.463^{+0.073}_{-0.065} \pm 0.036$   & \cite{Abulencia:2006zu} \\
%%unpublished%% CDF2       & \particle{\jpsi \ell} & 02--06 & 1.0 fb$^{-1}$ & $0.475^{+0.053}_{-0.049} \pm 0.018$   & \cite{CDFnote9294:2008,*Abulencia:2006zu_mod_cont}$^p$ \\
 \dzero & \particle{\jpsi \mu} & 02--06 & 1.3 fb$^{-1}$  & $0.448^{+0.038}_{-0.036} \pm 0.032$
   & \cite{Abazov:2008rba}  \\
CDF2       & \particle{\jpsi \pi} & & 6.7 fb$^{-1}$ & $0.452 \pm 0.048 \pm 0.027$  & \cite{Aaltonen:2012yb} \\
LHCb & \particle{\jpsi \mu} & 12 & 2 fb$^{-1}$  & $0.509 \pm 0.008 \pm 0.012$ & \cite{Aaij:2014bva}  \\
LHCb & \particle{\jpsi \pi} & 11--12 & 3 fb$^{-1}$  & $0.5134 \pm 0.0110 \pm 0.0057$ & \cite{Aaij:2014gka} \\
\hline
  \multicolumn{2}{l}{Average} & &  &  \hflavTAUBCnounit
                 &    \\   \hline
% \multicolumn{5}{l}{$^p$ \footnotesize Preliminary.}
\end{tabular}
\end{table}

\mysubsubsection{\Lb and \b-baryon lifetimes}
\labs{taulb}

The first measurements of \b-baryon lifetimes, performed at LEP,
originate from two classes of partially reconstructed decays.
In the first class, decays with an exclusively 
reconstructed \Lc baryon
and a lepton of opposite charge are used. These products are
more likely to occur in the decay of \Lb baryons.
In the second class, more inclusive final states with a baryon
(\particle{p}, \particle{\bar{p}}, $\Lambda$, or $\bar{\Lambda}$) 
and a lepton have been used, and these final states can generally
arise from any \b baryon.  With the large \b-hadron samples available
at the Tevatron and the LHC, the most precise measurements of \b baryons now
come from fully reconstructed exclusive decays.

The following sources of correlated systematic uncertainties have 
been considered:
experimental time resolution within a given experiment, \b-quark
fragmentation distribution into weakly decaying \b baryons,
\Lb polarisation, decay model,
and evaluation of the \b-baryon purity in the selected event samples.
In computing the averages
the central values of the masses are scaled to 
$M(\Lb) = 5619.51 \pm 0.23\MeVcc$~\cite{PDG_2016}.
% and $M(\mbox{\b-baryon}) = 5670 \pm 100\MeVcc$.

%For the semi-inclusive lifetime measurements, 
For measurements with partially reconstructed decays,
the meaning of the decay model
systematic uncertainties
and the correlation of these uncertainties between measurements
are not always clear.
Uncertainties related to the decay model are dominated by
assumptions on the fraction of $n$-body semileptonic decays.
To be conservative, it is assumed
that these are 100\%  correlated whenever given as an error.
DELPHI varies the fraction of four-body decays from 0.0 to 0.3. 
In computing the average, the DELPHI
result is corrected to a value of  $0.2 \pm 0.2$ for this fraction.
Furthermore
the semileptonic decay results from LEP are corrected for a
$\Lb$ polarisation of 
$-0.45^{+0.19}_{-0.17}$~\cite{Abbaneo:2000ej_mod,*Abbaneo:2001bv_mod_cont}
and a $b$ fragmentation parameter
$\langle x_E \rangle_b =0.702\pm 0.008$~\cite{ALEPH:2005ab}.

\begin{table}[!p]
\centering
\caption{Measurements of the \b-baryon lifetimes.
%Measurements of the \b-baryon and \Lb lifetime.
%The DELPHI and ALEPH $\Xi \ell$ results are not included 
%in the quoted average since the selected data samples
%contain mostly \Xib while 
%the data samples in the other measurements contain mostly \Lb.
}
\labt{lifelb}
\begin{tabular}{lcccl} 
\hline
Experiment&Method                &Data set& Lifetime (ps) & Ref. \\\hline\hline
ALEPH  &$\Lambda\ell$         & 91--95 &$1.20 \pm 0.08 \pm 0.06$ & \cite{Barate:1997if}\\
DELPHI &$\Lambda\ell\pi$ vtx  & 91--94 &$1.16 \pm 0.20 \pm 0.08$        & \cite{Abreu:1999hu}$^b$\\
DELPHI &$\Lambda\mu$ i.p.     & 91--94 &$1.10^{+0.19}_{-0.17} \pm 0.09$ & \cite{Abreu:1996nt}$^b$ \\
DELPHI &\particle{p\ell}      & 91--94 &$1.19 \pm 0.14 \pm 0.07$        & \cite{Abreu:1999hu}$^b$\\
OPAL   &$\Lambda\ell$ i.p.    & 90--94 &$1.21^{+0.15}_{-0.13} \pm 0.10$ & \cite{Akers:1995ui}$^c$  \\
OPAL   &$\Lambda\ell$ vtx     & 90--94 &$1.15 \pm 0.12 \pm 0.06$        & \cite{Akers:1995ui}$^c$ \\ 
%OS% It does no longer make sense to quote the mean \b-baryon lifetime, since this has anyway the 
%OS% same value as the \Lb lifetime (the above measurements are old and imprecise)
%OS% \multicolumn{3}{l}{Average of above 19: \hfill mean \b-baryon lifetime $=$} & \hflavTAUBBnounit & \\  
\hline
ALEPH  &$\Lc\ell$             & 91--95 &$1.18^{+0.13}_{-0.12} \pm 0.03$ & \cite{Barate:1997if}$^a$\\
ALEPH  &$\Lambda\ell^-\ell^+$ & 91--95 &$1.30^{+0.26}_{-0.21} \pm 0.04$ & \cite{Barate:1997if}$^a$\\
DELPHI &$\Lc\ell$             & 91--94 &$1.11^{+0.19}_{-0.18} \pm 0.05$ & \cite{Abreu:1999hu}$^b$\\
OPAL   &$\Lc\ell$, $\Lambda\ell^-\ell^+$ 
                                 & 90--95 & $1.29^{+0.24}_{-0.22} \pm 0.06$ & \cite{Ackerstaff:1997qi}\\ 
CDF1   &$\Lc\ell$             & 91--95 &$1.32 \pm 0.15        \pm 0.07$ & \cite{Abe:1996df}\\
\dzero &$\Lc\mu$              & 02--06 &$1.290^{+0.119+0.087}_{-0.110-0.091}$ & \cite{Abazov:2007al} \\
%\multicolumn{3}{l}{Average of above 6 (semileptonic \Lb decays)} & \hflavTAULBSnounit & \\
\multicolumn{3}{l}{Average of above 6} & \hflavTAULBSnounit & \\
\hline
CDF2   &$\Lc\pi$              & 02--06 &$1.401 \pm 0.046 \pm 0.035$ & \cite{Aaltonen:2009zn} \\
%%CDF2   &$\jpsi \Lambda$      & 02--09 &$1.537 \pm 0.045 \pm 0.014$ & \citehistory{Aaltonen:2010pj}{Aaltonen:2010pj,*Abulencia:2006dr_hist}\\
CDF2   &$\jpsi \Lambda$      & 02--11 &$1.565 \pm 0.035 \pm 0.020$ & \citehistory{Aaltonen:2014wfa}{Aaltonen:2014wfa,*Aaltonen:2014wfa_hist} \\
\dzero &$\jpsi \Lambda$      & 02--11 &$1.303 \pm 0.075 \pm 0.035$ & \citehistory{Abazov:2012iy}{Abazov:2012iy,*Abazov:2007sf_hist,*Abazov:2004bn_hist} \\
ATLAS  &$\jpsi \Lambda$      & 2011   &$1.449 \pm 0.036 \pm 0.017$ & \cite{Aad:2012bpa} \\
CMS    &$\jpsi \Lambda$      & 2011   &$1.503 \pm 0.052 \pm 0.031$ & \cite{Chatrchyan:2013sxa} \\ % 5 fb-1
%%% LHCb   &$\jpsi \Lambda$      & 2010   &$1.353 \pm 0.108 \pm 0.035$ & \cite{LHCb-CONF-2011-001}$^p$ \\
LHCb   &$\jpsi \Lambda$      & 2011   &$1.415 \pm 0.027 \pm 0.006$ & \cite{Aaij:2014owa} \\
LHCb   &$\jpsi pK$ (w.r.t.\ $B^0$)  & 11--12 &$1.479 \pm 0.009 \pm 0.010$ & \citehistory{Aaij:2014zyy}{Aaij:2014zyy,*Aaij:2013oha_hist} \\ % 3 fb-1
%\multicolumn{3}{l}{Average of above 7 (fully reconstructed \Lb decays)} & \hflavTAULBEnounit & \\
\multicolumn{3}{l}{Average of above 7: \hfill \Lb lifetime $=$} & \hflavTAULBnounit & \\
\hline\hline
ALEPH  &$\Xi^-\ell^-X$        & 90--95 &$1.35^{+0.37+0.15}_{-0.28-0.17}$ & \cite{Buskulic:1996sm}\\
DELPHI &$\Xi^-\ell^-X$        & 91--93 &$1.5 ^{+0.7}_{-0.4} \pm 0.3$     & \cite{Abreu:1995kt}$^d$ \\
DELPHI &$\Xi^-\ell^-X$        & 92--95 &$1.45 ^{+0.55}_{-0.43} \pm 0.13$     & \cite{Abdallah:2005cw}$^d$ \\
%OS% It does no longer make sense to quote the mean Xib lifetime, since it would merely be the average 
%OS% of the Xib- and Xib0 lifetimes (the above measurements are old and imprecise)
%OS% \multicolumn{3}{l}{Average of above 7: \hfill mean \Xib lifetime $=$} & \hflavTAUXBnounit & \\
\hline
%%CDF2   &$\jpsi \Xi^-$        & 02--09 &$1.56 ^{+0.27}_{-0.25} \pm 0.02$ & \cite{Aaltonen:2009ny} \\
CDF2   &$\jpsi \Xi^-$        & 02--11 &$1.32 \pm 0.14 \pm 0.02$ & \citehistory{Aaltonen:2014wfa}{Aaltonen:2014wfa,*Aaltonen:2014wfa_hist} \\ % full Run II data set = 9.6 fb-1
LHCb   &$\jpsi \Xi^-$         & 11--12 &$1.55 ^{+0.10}_{-0.09} \pm 0.03$ & \cite{Aaij:2014sia} \\ 
LHCb   &$\Xi_c^0\pi^-$ (w.r.t.\ $\Lb$)  & 11--12 &$1.599 \pm 0.041 \pm 0.022$ & \cite{Aaij:2014lxa} \\ 
\multicolumn{3}{l}{Average of above 3: \hfill \Xibd lifetime $=$} & \hflavTAUXBDnounit & \\
\hline\hline
LHCb   &$\Xi_c^+\pi^-$  (w.r.t.\ $\Lb$) & 11--12 &$1.477 \pm 0.026 \pm 0.019$ & \cite{Aaij:2014esa} \\ 
\multicolumn{3}{l}{Average of above 1: \hfill \Xibu lifetime $=$} & \hflavTAUXBUnounit & \\
\hline\hline
%%CDF2   &$\jpsi \Omega^-$     & 02--09 & $1.13 ^{+0.53}_{-0.40} \pm 0.02$ & \cite{Aaltonen:2009ny} \\
CDF2   &$\jpsi \Omega^-$     & 02--11 & $1.66 ^{+0.53}_{-0.40} \pm 0.02$ & \citehistory{Aaltonen:2014wfa}{Aaltonen:2014wfa,*Aaltonen:2014wfa_hist} \\ % full Run II data set = 9.6 fb-1
LHCb   &$\jpsi \Omega^-$     & 11--12 &$1.54 ^{+0.26}_{-0.21} \pm 0.05$ & \cite{Aaij:2014sia} \\ 
LHCb   &$\Omega_c^0 \pi^-$ (w.r.t.\ $\Xi_b^-$)  & 11--12 &$1.78 \pm 0.26 \pm 0.05 \pm 0.06$ & \cite{Aaij:2016dls} \\
\multicolumn{3}{l}{Average of above 3: \hfill \Omegab lifetime $=$} & \hflavTAUOBnounit & \\
\hline\hline
\multicolumn{5}{l}{$^a$ \footnotesize The combined ALEPH result quoted 
in \cite{Barate:1997if} is $1.21 \pm 0.11$ ps.} \\[-0.5ex]
\multicolumn{5}{l}{$^b$ \footnotesize The combined DELPHI result quoted 
in \cite{Abreu:1999hu} is $1.14 \pm 0.08 \pm 0.04$ ps.} \\[-0.5ex]
\multicolumn{5}{l}{$^c$ \footnotesize The combined OPAL result quoted 
in \cite{Akers:1995ui} is $1.16 \pm 0.11 \pm 0.06$ ps.} \\[-0.5ex]
\multicolumn{5}{l}{$^d$ \footnotesize The combined DELPHI result quoted 
in \cite{Abdallah:2005cw} is $1.48 ^{+0.40}_{-0.31} \pm 0.12$ ps.}
%%%\\[-0.5ex] \multicolumn{5}{l}{$^p$ \footnotesize Preliminary.}
\end{tabular}
\end{table}

The list of all measurements are given in \Table{lifelb}.
We do not attempt to average measurements performed with $p\ell$ or 
$\Lambda\ell$ correlations, which select unknown mixtures of $b$ baryons. 
Measurements performed with $\Lc\ell$ or $\Lambda\ell^+\ell^-$
correlations can be assumed to correspond to semileptonic \Lb decays. 
Their average (\hflavTAULBS) is significantly different 
from the average using only measurements performed with
exclusively reconstructed hadronic \Lb decays (\hflavTAULBE). 
The latter is much more precise
and less prone to potential biases than the former. 
The discrepancy between the two averages is at the level of
$\hflavNSIGMATAULBEXCLSEMI\,\sigma$ 
and assumed to be due to an experimental systematic effect in the 
semileptonic measurements or to a rare statistical fluctuation.
The best estimate of the \Lb lifetime is therefore taken 
as the average of the exclusive  measurements only. 
The CDF $\Lb \to \jpsi \Lambda$
lifetime result~\citehistory{Aaltonen:2014wfa}{Aaltonen:2014wfa,*Aaltonen:2014wfa_hist} 
is larger than the average of all other exclusive measurements
by $\hflavNSIGMATAULBCDFTWO\,\sigma$. 
It is nonetheless kept in the average
without adjustment of input errors.
The world average \Lb lifetime is then
\begin{equation}
\tau(\Lb) = \hflavTAULB \,. 
\end{equation}
%2015% For the \Lb lifetime average, we only include measurements obtained
%2015% with inclusive \particle{\Lambda^{\pm}_c \ell^{\mp}}, inclusive
%2015% $\Lambda \ell^- \ell^+$, and fully exclusive final states.
%2015% The CDF $\Lb \to \jpsi \Lambda$
%2015% lifetime result~\citehistory{Aaltonen:2014wfa}{Aaltonen:2014wfa,*Aaltonen:2014wfa_hist} 
%2015% is larger than the world average computed excluding this result
%2015% by $\hflavNSIGMATAULBCDFTWO\,\sigma$. 
%2015% It is nonetheless combined with the rest 
%2015% without adjustment of input errors.
%2015% The world average \Lb lifetime is then
%2015% \begin{equation}
%2015% \tau(\Lb) = \hflavTAULB \,. 
%2015% \end{equation}
%2015% % Adding also the measurements with more inclusive baryon final states yields the 
%2015% % following world average of \b baryons:
%2015% % \begin{equation}
%2015% % \langle\tau(\mbox{\b-baryon})\rangle = \hflavTAUBB \,.
%2015% % \end{equation}
%2015% It turns out that the average obtained using only measurements performed 
%2015% with semileptonic \Lb decays (\hflavTAULBS) is significantly different 
%2015% from the one using only measurements performed with exclusively reconstructed 
%2015% \Lb decays (\hflavTAULBE). The latter is much more precise
%2015% (and less prone to systematic uncertainties) than the former. 
%2015% This discrepancy can only be attributed to a systematic experimental effect
%2015% or to a statistical fluctuation. 

For the strange \b baryons, we do not include the measurements based on
inclusive $\Xi^{\mp} \ell^{\mp}$~\cite{Buskulic:1996sm,Abdallah:2005cw,Abreu:1995kt} 
final states, which consist of a mixture of 
$\Xibd$ and $\Xibu$ baryons. Instead we only average results obtained with 
fully reconstructed $\Xibd$, $\Xibu$ and $\Omegab$ baryons, and obtain
% \begin{equation}
% \langle\tau(\Xib)\rangle = \hflavTAUXB \,.
% \end{equation}
%old% First measurements of fully reconstructed 
%old% $\Xibd \to \jpsi\Xi^-$ and $\Omegab \to \jpsi\Omega^-$
%old% baryons yield~\citehistory{Aaltonen:2014wfa}{Aaltonen:2014wfa,*Aaltonen:2014wfa_hist}
\begin{eqnarray}
\tau(\Xibd) &=& \hflavTAUXBD \,, \\
\tau(\Xibu) &=& \hflavTAUXBU \,, \\
\tau(\Omegab) &=& \hflavTAUOB \,. 
\end{eqnarray}
It should be noted that several $b$-baryon lifetime measurements from LHCb~%
\citehistory{Aaij:2014zyy,Aaij:2014lxa,Aaij:2014esa,Aaij:2016dls}{Aaij:2014zyy,*Aaij:2013oha_hist,Aaij:2014lxa,Aaij:2014esa,Aaij:2016dls}
were made with respect to the lifetime of another $b$ hadron
(\ie, the original measurement is that of a decay width difference).
Before these measurements are included in the averages quoted above, we 
rescale them according to our latest lifetime average
of that reference $b$ hadron. This introduces correlations between 
our averages, in particular between the $\Xibd$ and $\Xibu$ lifetimes. 
Taking this correlation into account leads to 
\begin{equation}
\tau(\Xibu) / \tau(\Xibd) = \hflavRTAUXBUXBD \,.
\end{equation}

\mysubsubsection{Summary and comparison with theoretical predictions}
\labs{lifesummary}

Averages of lifetimes of specific \b-hadron species are collected
in \Table{sumlife}.
\begin{table}[t]
\centering
\caption{Summary of the lifetime averages for the different \b-hadron species.}
\labt{sumlife}
\begin{tabular}{lrc} \hline
\multicolumn{2}{l}{\b-hadron species} & Measured lifetime \\ \hline
\Bu &                       & \hflavTAUBU   \\
\Bd &                       & \hflavTAUBD   \\
% \Bs ($\to$ flavour specific) & \hflavTAUBSSL \\
% \Bs ($\to \jpsi\phi$)      & \hflavTAUBSJF \\
\Bs & $1/\Gs~\, =$               & \hflavTAUBSMEANC \\
~~ $B^0_{s\rm L}$ & $1/\Gamma_{s\rm L}=$  & \hflavTAUBSLCON \\
~~ $B^0_{s\rm H}$ & $1/\Gamma_{s\rm H}=$  & \hflavTAUBSHCON \\
\Bc     &                   & \hflavTAUBC   \\ 
\Lb     &                   & \hflavTAULB   \\
\Xibd   &                   & \hflavTAUXBD  \\
\Xibu   &                   & \hflavTAUXBU  \\
\Omegab &                   & \hflavTAUOB   \\
\hline
%\multicolumn{2}{l}{\b-hadron mixture}  & \hflavTAUB    \\
%OS% \b-baryon mixture           & \hflavTAUBB   \\
%OS% \Xib mixture                & \hflavTAUXB   \\
%\hline
\end{tabular}
\end{table}
%
%
% Predictions for tau(Omega_b-)/tau(B0) < 1.10 
% (quoted in D0 observation paper)
% 
% X. Liu et al., Phys. Rev. D 77, 014031 (2008);
% M. Karliner et al., arXiv:0804.1575;
% E. E. Jenkins, Phys. Rev. D 77, 034012 (2008);
% R. Roncaglia, D. B. Lichtenberg, and E. Predazzi, Phys. Rev. D 52, 1722 (1995);
% N. Mathur, R. Lewis, and R. M. Woloshyn, Phys. Rev. D 66, 014502 (2002);
% D. Ebert, R. N. Faustov, and V. O. Galkin, Phys. Rev. D 72, 034026 (2005);
% T. Ito, M. Matsuda, and Y. Matsui, Prog. Theor. Phys. 99, 271 (1998).
%
%
\begin{table}[t]
\centering
\caption{Experimental averages of \b-hadron lifetime ratios and
Heavy-Quark Expansion (HQE) predictions~\cite{Lenz:2015dra,Lenz:2014jha}.}
\labt{liferatio}
\begin{tabular}{lcc} \hline
Lifetime ratio & Experimental average & HQE prediction \\ \hline
$\tau(\Bu)/\tau(\Bd)$ & \hflavRTAUBU & $1.04 ^{+0.05}_{-0.01} \pm 0.02 \pm 0.01$ \\
$\tau(\Bs)/\tau(\Bd)$ & \hflavRTAUBSMEANC & $1.001 \pm 0.002$ \\
$\tau(\Lb)/\tau(\Bd)$ & \hflavRTAULB & $0.935 \pm 0.054$ \\
$\tau(\Xibu)/\tau(\Xibd)$ & \hflavRTAUXBUXBD & $0.95 \pm 0.06$ \\
\hline
\end{tabular}
\end{table}
As described in the introduction to \Sec{lifetimes},
the HQE can be employed to explain the hierarchy of
$\tau(\Bc) \ll \tau(\Lb) < \tau(\Bs) \approx \tau(\Bd) < \tau(\Bu)$,
and used to predict the ratios between lifetimes.
Recent predictions are compared to the measured 
lifetime ratios in \Table{liferatio}.

The predictions of the ratio between the \Bu and \Bd lifetimes,
$1.06 \pm 0.02$~\cite{Tarantino:2003qw,Gabbiani:2003pq} or
$1.04 ^{+0.05}_{-0.01} \pm 0.02 \pm 0.01$~\cite{Lenz:2015dra,Lenz:2014jha},
are in good agreement with experiment. 

The total widths of the \Bs and \Bd mesons
are expected to be very close and differ by at most 
1\%~\cite{Beneke:1996gn,Keum:1998fd,Gabbiani:2004tp,Lenz:2015dra,Lenz:2014jha}.
This prediction is consistent with the
experimental ratio $\tau(\Bs)/\tau(\Bd)=\Gd/\Gs$,
which is smaller than 1 by 
% $\hflavRTAUBSMEANCsig\,\sigma$ 
\hflavONEMINUSRTAUBSMEANCpercent. 
% at deviation with respect to the prediction. 
The authors of Ref.~\citehistory{Jubb:2016mvq,Artuso:2015swg}{Jubb:2016mvq,Artuso:2015swg,*Lenz_hist} predict
$\tau(\Bs)/\tau(\Bd) = 1.00050 \pm 0.00108 \pm 0.0225\times \delta$,
where $\delta$ quantifies a possible breaking of the quark-hadron duality.
In this context, they interpret the $2.5\sigma$ difference 
between theory and experiment as being due to either new physics
or a sizable duality violation.
The key message is that improved experimental precision
on this ratio is very welcome.

The ratio $\tau(\Lb)/\tau(\Bd)$ has particularly been the source of theoretical
scrutiny since earlier calculations using the HQE~\cite{Shifman:1986mx,Chay:1990da,Bigi:1992su,Voloshin:1999pz,*Guberina:1999bw,*Neubert:1996we,*Bigi:1997fj}
predicted a value larger than 0.90, almost $2\,\sigma$ 
above the world average at the time. 
Many predictions cluster around a most likely central value
of 0.94~\cite{Uraltsev:1996ta,*Pirjol:1998ur,*Colangelo:1996ta,*DiPierro:1999tb}.
Calculations
of this ratio that include higher-order effects predict a
lower ratio between the
\Lb and \Bd lifetimes~\cite{Tarantino:2003qw,Gabbiani:2003pq,Gabbiani:2004tp}
and reduce this difference.
%2014 References~\cite{Tarantino:2003qw,Gabbiani:2003pq,Gabbiani:2004tp} present probability density functions
%2014 of their predictions with a variation of theoretical inputs, and the
%2014 indicated ranges in \Table{liferatio}
%2014 are the RMS of the distributions from the most probable values, and for 
%2014 $\tau(\Lb)/\tau(\Bd)$, also encompass the earlier theoretical predictions%
%2014 ~\cite{Shifman:1986mx,Chay:1990da,Bigi:1992su,Voloshin:1999pz,*Guberina:1999bw,*Neubert:1996we,*Bigi:1997fj,Uraltsev:1996ta,*Pirjol:1998ur,*Colangelo:1996ta,*DiPierro:1999tb}.
%2014 % Next sentence added on Feb 18, 2011 (following a comment from A. Lenz)
%2014 Note that in contrast to the $B$ mesons, complete NLO QCD
%2014 corrections and fully reliable lattice
%2014 determinations of the matrix elements for $\Lb$ are not yet available.
Since then the experimental average is now definitely settling at a value 
significantly larger than initially, in agreement with the latest theoretical 
predictions. 
A recent review~\cite{Lenz:2015dra,Lenz:2014jha} concludes that 
the long-standing $\Lb$ lifetime puzzle is resolved, with a
nice agreement between the precise experimental determination
of $\tau(\Lb)/\tau(\Bd)$ and the less precise HQE prediction
which needs new lattice calculations.
There is also good agreement for the 
$\tau(\Xibu)/\tau(\Xibd)$ ratio.
%
% The paragraph below has been commented out on June 4, 2017, on the 
% suggestion of Diego Tonelli: 
%  "I appreciate that this comment is kind toward CDF but I also think it's
%   maybe overly politically correct. While there is no proof that something
%   was wrong in that measurement, it's clear that the subsequent, and more
%   precise, LHCb measurements have clarified a lot the LambdaB lifetime
%   issue and suggest that probably this outlier from CDF might be due to
%   some poorly understood systematic effect. I would drop this piece of
%   text altogether."
%
%2017% As already mentioned, the CDF measurement of the \Lb lifetime
%2017% in the exclusive decay mode $\jpsi \Lambda$~\citehistory{Aaltonen:2014wfa}{Aaltonen:2014wfa,*Aaltonen:2014wfa_hist} 
%2017% is significantly 
%2017% higher than the world average before inclusion, with a ratio
%2017% to the $\tau(\Bd)$ world average of 
%2017% $\tau(\Lb)/\tau(\Bd) = 1.030 \pm 0.027$, 
%2017% % OS, Jul 13, 2016:
%2017% %     CDF2 Lambda_b -> J/psi Lambda lifetime result: 1.565 +-0.035 +-0.020 ps
%2017% %     World average of B0 lifetime:                  1.5197 +-0.0044 ps
%2017% %     Ratio tau(Lambda_b CDF2)/tau(B0 world) = 1.030 +- 0.027
%2017% %          central value = 1.565/1.5197 = 1.0298
%2017% %          error = 1.565/1.5197*sqrt((0.035**2+0.020**2)/1.565**2+(0.0044/1.5197)**2) = 0.02669
%2017% %
%2017% resulting in continued interest in lifetimes of \b baryons.

The lifetimes of the most abundant \b-hadron species are now all known to sub-percent precision. Neglecting the 
contributions of the rarer species (\Bc meson and \b baryons other than the \Lb), one can compute the average 
\b-hadron lifetime from the individual lifetimes and production fractions as 
\begin{equation}
\tau_b = \frac%
{\fBd \tau(\Bd)^2+ \fBu \tau(\Bu)^2+0.5 \fBs \tau(B^0_{s\rm H})^2+0.5 \fBs \tau(B^0_{s\rm L})^2+ \fbb \tau(\Lb)^2}%
{\fBd \tau(\Bd)  + \fBu \tau(\Bu)  +0.5 \fBs \tau(B^0_{s\rm H})  +0.5 \fBs \tau(B^0_{s\rm L})  + \fbb \tau(\Lb)  } \,.
\end{equation}
Using the lifetimes of \Table{sumlife} and the fractions in $Z$ decays of \Table{fractions},
taking into account the correlations between the fractions (\Table{fractions}) as well as the correlation 
between $\tau(B_{s\rm H})$ and $\tau(B_{s\rm L})$ (\hflavZRHOTAUHTAUL), one obtains
\begin{equation}
\tau_b(Z) = \hflavTAUBZCALC \,.
\end{equation}
This is in very good agreement with (and three times more precise than)
the average of \Eq{TAUBVTX} for the inclusive measurements performed at LEP. 
  % from Rick

%------------------------------------------------
\mysubsection{Neutral \B-meson mixing}
%------------------------------------------------
\labs{mixing}

The $\Bd-\Bdbar$ and $\Bs-\Bsbar$ systems
both exhibit the phenomenon of particle-antiparticle mixing. For each of them, 
there are two mass eigenstates which are linear combinations of the two flavour states,
$B^0_q$ and $\bar{B}^0_q$, 
\begin{eqnarray}
| B^0_{q\rm L}\rangle &=& p_q |B^0_q \rangle +  q_q |\bar{B}^0_q \rangle \,, \\
| B^0_{q\rm H}\rangle &=& p_q |B^0_q \rangle -  q_q |\bar{B}^0_q \rangle  \,,
\end{eqnarray}
where the subscript $q=d$ is used for the  $B^0_d$ ($=\Bd$) meson and $q=s$ for the \Bs meson.
The heaviest (lightest) of these mass states is denoted
$B^0_{q\rm H}$ ($B^0_{q\rm L}$),
with mass $m_{q\rm H}$ ($m_{q\rm L}$)
and total decay width $\Gamma_{q\rm H}$ ($\Gamma_{q\rm L}$). We define
\begin{eqnarray}
\Delta m_q = m_{q\rm H} - m_{q\rm L} \,, &~~~~&  x_q = \Delta m_q/\Gamma_q \,, \labe{dm} \\
\Delta \Gamma_q \, = \Gamma_{q\rm L} - \Gamma_{q\rm H} \,, ~ &~~~~&  y_q= \Delta\Gamma_q/(2\Gamma_q) \,, \labe{dg}
\end{eqnarray}
where 
$\Gamma_q = (\Gamma_{q\rm H} + \Gamma_{q\rm L})/2 =1/\bar{\tau}(B^0_q)$ 
is the average decay width.
$\Delta m_q$ is positive by definition, and 
$\Delta \Gamma_q$ is expected to be positive within
the Standard Model.\footnote{
  \label{foot:life_mix:Eqdg}
  For reasons of symmetry in \Eqss{dm}{dg}, 
  $\Delta \Gamma$ is sometimes defined with the opposite sign. 
  The definition adopted in \Eq{dg} is the one used
  by most experimentalists and many phenomenologists in \B physics.}

There are four different time-dependent probabilities describing the 
case of a neutral \B meson produced as a flavour state and decaying without
\CP violation to a flavour-specific final state. 
If \CPT is conserved (which  
will be assumed throughout), they can be written as 
\begin{equation}
% \frac{\Delta\Gamma_q}{2}
% \frac{1}{2}\Delta\Gamma_q
\left\{
\begin{array}{rcl}
{\cal P}(B^0_q \to B^0_q) & = &  \frac{1}{2} e^{-\Gamma_q t} 
\left[ \cosh\!\left(\frac{1}{2}\Delta\Gamma_q t\right) + \cos\!\left(\Delta m_q t\right)\right]  \\
{\cal P}(B^0_q \to \bar{B}^0_q) & = &   \frac{1}{2} e^{-\Gamma_q t} 
\left[ \cosh\!\left(\frac{1}{2}\Delta\Gamma_q t\right) - \cos\!\left(\Delta m_q t\right)\right] 
\left|q_q/p_q\right|^2 \\
{\cal P}(\bar{B}^0_q \to B^0_q) & = &  \frac{1}{2} e^{-\Gamma_q t} 
\left[ \cosh\!\left(\frac{1}{2}\Delta\Gamma_q t\right) - \cos\!\left(\Delta m_q t\right)\right] 
\left|p_q/q_q\right|^2 \\
{\cal P}(\bar{B}^0_q \to\bar{B}^0_q) & = &  \frac{1}{2} e^{-\Gamma_q t}  
\left[ \cosh\!\left(\frac{1}{2}\Delta\Gamma_q t\right) + \cos\!\left(\Delta m_q t\right)\right] 
\end{array} \right. \,,
\labe{oscillations}
\end{equation}
where $t$ is the proper time of the system (\ie, the time interval between the production 
and the decay in the rest frame of the \B meson). 
At the \B factories, only the proper-time difference $\Delta t$ between the decays
of the two neutral \B mesons from the \Ups can be determined, but, 
because the two \B mesons evolve coherently (keeping opposite flavours as long
as neither of them has decayed), the 
above formulae remain valid 
if $t$ is replaced with $\Delta t$ and the production flavour is replaced by the flavour 
at the time of the decay of the accompanying \B meson in a flavour-specific state.
As can be seen in the above expressions,
the mixing probabilities 
depend on three mixing observables:
$\Delta m_q$, $\Delta\Gamma_q$,
and $|q_q/p_q|^2$, which signals \CP violation in the mixing if $|q_q/p_q|^2 \ne 1$.
Another (non independent)  observable often used to characterize \CP violation in the mixing 
is the so-called semileptonic asymmetry, defined as
\begin{equation} 
{\cal A}_{\rm SL}^q = 
\frac{|p_{\particle{q}}/q_{\particle{q}}|^2 - |q_{\particle{q}}/p_{\particle{q}}|^2}%
{|p_{\particle{q}}/q_{\particle{q}}|^2 + |q_{\particle{q}}/p_{\particle{q}}|^2} \,.
\labe{ASLq}
\end{equation} 
All  mixing observables depend on two complex numbers, $M^q_{12}$ and $\Gamma^q_{12}$, which are the off-diagonal elements of the mass and decay $2\times 2$ matrices describing the evolution of the $B^0_q-\bar{B}^0_q$ system. In the Standard Model the quantity $|\Gamma^q_{12}/M^q_{12}|$ is small, of the order of $(m_b/m_t)^2$ where $m_b$ and $m_t$ are the bottom and top quark masses. The following relations hold, to first order in $|\Gamma^q_{12}/M^q_{12}|$:
\begin{eqnarray}
\Delta m_q & = & 2 |M^q_{12}| \left[1 + {\cal O} \left(|\Gamma^q_{12}/M^q_{12}|^2 \right) \right] \,, \\
\Delta\Gamma_q & = & 2 |\Gamma^q_{12}| \cos\phi^q_{12} \left[1 + {\cal O} \left(|\Gamma^q_{12}/M^q_{12}|^2 \right) \right]   \,, \\
{\cal A}_{\rm SL}^q & = &  \Im \left(\Gamma^q_{12}/M^q_{12} \right) +
{\cal O} \left(|\Gamma^q_{12}/M^q_{12}|^2 \right) =
\frac{\DGs}{\dms}\tan\phi^q_{12} +
{\cal O} \left(|\Gamma^q_{12}/M^q_{12}|^2 \right)  \,,
\labe{ALSq_tanphi2}
\end{eqnarray}
where 
\begin{equation}
\phi^q_{12} = \arg \left( -{M^q_{12}}/{\Gamma^q_{12}} \right)
\labe{phi12}
\end{equation}
is the observable phase difference between $-M^q_{12}$ and $\Gamma^q_{12}$ (often called the mixing phase). 
It should be noted that the theoretical predictions for $\Gamma^q_{12}$ are based on the same HQE as the lifetime predictions. 

In the next sections we review in turn the experimental knowledge
on the \Bd decay-width and mass differences, 
the \Bs decay-width and mass differences,  
\CP violation in \Bd and \Bs mixing, and mixing-induced \CP violation in \Bs decays. 

%------------------------------------------------
\mysubsubsection{\Bd mixing parameters \DGd and \dmd}
%\mysubsubsection{Mass and decay width differences \dmd and \DGd}
%------------------------------------------------
\labs{DGd} \labs{dmd}

\begin{table}
\centering
\caption{Time-dependent measurements included in the \dmd average.
The results obtained from multi-dimensional fits involving also 
the \Bd (and \Bu) lifetimes
as free parameter(s)~\protect\citehistory{Aubert:2002sh,Aubert:2005kf,Abe:2004mz}{Aubert:2002sh,Aubert:2005kf,Abe:2004mz,*Abe:2002id_hist,*Tomura:2002qs_hist,*Hara:2002mq_hist} 
have been converted into one-dimensional measurements of \dmd.
All the measurements have then been adjusted to a common set of physics
parameters before being combined.}
\labt{dmd}
\begin{tabular}{@{}rc@{}cc@{}c@{}cc@{}c@{}c@{}}
\hline
%Experiment & \multicolumn{2}{c}{Method} & \multicolumn{3}{l}{Published value}   
%                                        & \multicolumn{3}{l}{Adjusted value}     \\
%and Ref.   &  rec. & tag                & \multicolumn{3}{l}{of \dmd in\invps} 
%                                        & \multicolumn{3}{l}{of \dmd in\invps} \\
Experiment & \multicolumn{2}{c}{Method} & \multicolumn{3}{l}{\dmd in\invps}   
                                        & \multicolumn{3}{l}{\dmd in\invps}     \\
and Ref.   &  rec. & tag                & \multicolumn{3}{l}{before adjustment} 
                                        & \multicolumn{3}{l}{after adjustment} \\
\hline
%    0.404 +-0.045 +-0.027      from ALEPH LEPTON/QJET PUBLISHED 1 l/Qjet (91-94)
%    0.452 +-0.039 +-0.044      from ALEPH LEPTON/LEPTON PUBLISHED 1 l/l (91-94)
 ALEPH~\cite{Buskulic:1996qt}  & \particle{ \ell  } & \particle{ \Qjet  } & $  0.404 $ & $ \pm  0.045 $ & $ \pm  0.027 $ & & & \\
 ALEPH~\cite{Buskulic:1996qt}  & \particle{ \ell  } & \particle{ \ell  } & $  0.452 $ & $ \pm  0.039 $ & $ \pm  0.044 $ & & & \\
 ALEPH~\cite{Buskulic:1996qt}  & \multicolumn{2}{c}{above two combined} & $  0.422 $ & $ \pm  0.032 $ & $ \pm  0.026 $ & $  0.440 $ & $ \pm  0.032 $ & $ ^{+  0.020 }_{-  0.019 } $ \\
 ALEPH~\cite{Buskulic:1996qt}  & \particle{ D^*  } & \particle{ \ell,\Qjet  } & $  0.482 $ & $ \pm  0.044 $ & $ \pm  0.024 $ & $  0.482 $ & $ \pm  0.044 $ & $ \pm  0.024 $ \\
 DELPHI~\cite{Abreu:1997xq}  & \particle{ \ell  } & \particle{ \Qjet  } & $  0.493 $ & $ \pm  0.042 $ & $ \pm  0.027 $ & $  0.499 $ & $ \pm  0.042 $ & $ \pm  0.024 $ \\
 DELPHI~\cite{Abreu:1997xq}  & \particle{ \pi^*\ell  } & \particle{ \Qjet  } & $  0.499 $ & $ \pm  0.053 $ & $ \pm  0.015 $ & $  0.500 $ & $ \pm  0.053 $ & $ \pm  0.015 $ \\
 DELPHI~\cite{Abreu:1997xq}  & \particle{ \ell  } & \particle{ \ell  } & $  0.480 $ & $ \pm  0.040 $ & $ \pm  0.051 $ & $  0.495 $ & $ \pm  0.040 $ & $ ^{+  0.042 }_{-  0.040 } $ \\
 DELPHI~\cite{Abreu:1997xq}  & \particle{ D^*  } & \particle{ \Qjet  } & $  0.523 $ & $ \pm  0.072 $ & $ \pm  0.043 $ & $  0.518 $ & $ \pm  0.072 $ & $ \pm  0.043 $ \\
 DELPHI~\cite{Abdallah:2002mr}  & \particle{ \mbox{vtx}  } & \particle{ \mbox{comb}  } & $  0.531 $ & $ \pm  0.025 $ & $ \pm  0.007 $ & $  0.525 $ & $ \pm  0.025 $ & $ \pm  0.006 $ \\
 L3~\citehistory{Acciarri:1998pq}{Acciarri:1998pq,*Acciarri:1996ia_hist}  & \particle{ \ell  } & \particle{ \ell  } & $  0.458 $ & $ \pm  0.046 $ & $ \pm  0.032 $ & $  0.466 $ & $ \pm  0.046 $ & $ \pm  0.028 $ \\
 L3~\citehistory{Acciarri:1998pq}{Acciarri:1998pq,*Acciarri:1996ia_hist}  & \particle{ \ell  } & \particle{ \Qjet  } & $  0.427 $ & $ \pm  0.044 $ & $ \pm  0.044 $ & $  0.439 $ & $ \pm  0.044 $ & $ \pm  0.042 $ \\
 L3~\citehistory{Acciarri:1998pq}{Acciarri:1998pq,*Acciarri:1996ia_hist}  & \particle{ \ell  } & \particle{ \ell\mbox{(IP)}  } & $  0.462 $ & $ \pm  0.063 $ & $ \pm  0.053 $ & $  0.470 $ & $ \pm  0.063 $ & $ \pm  0.044 $ \\
 OPAL~\cite{Ackerstaff:1997iw}  & \particle{ \ell  } & \particle{ \ell  } & $  0.430 $ & $ \pm  0.043 $ & $ ^{+  0.028 }_{-  0.030 } $ & $  0.466 $ & $ \pm  0.043 $ & $ ^{+  0.017 }_{-  0.016 } $ \\
 OPAL~\cite{Ackerstaff:1997vd}  & \particle{ \ell  } & \particle{ \Qjet  } & $  0.444 $ & $ \pm  0.029 $ & $ ^{+  0.020 }_{-  0.017 } $ & $  0.481 $ & $ \pm  0.029 $ & $ \pm  0.013 $ \\
 OPAL~\cite{Alexander:1996id}  & \particle{ D^*\ell  } & \particle{ \Qjet  } & $  0.539 $ & $ \pm  0.060 $ & $ \pm  0.024 $ & $  0.544 $ & $ \pm  0.060 $ & $ \pm  0.023 $ \\
 OPAL~\cite{Alexander:1996id}  & \particle{ D^*  } & \particle{ \ell  } & $  0.567 $ & $ \pm  0.089 $ & $ ^{+  0.029 }_{-  0.023 } $ & $  0.572 $ & $ \pm  0.089 $ & $ ^{+  0.028 }_{-  0.022 } $ \\
 OPAL~\cite{Abbiendi:2000ec}  & \particle{ \pi^*\ell  } & \particle{ \Qjet  } & $  0.497 $ & $ \pm  0.024 $ & $ \pm  0.025 $ & $  0.496 $ & $ \pm  0.024 $ & $ \pm  0.025 $ \\
 CDF1~\cite{Abe:1997qf,*Abe:1998sq}  & \particle{ D\ell  } & \particle{ \mbox{SST}  } & $  0.471 $ & $ ^{+  0.078 }_{-  0.068 } $ & $ ^{+  0.033 }_{-  0.034 } $ & $  0.470 $ & $ ^{+  0.078 }_{-  0.068 } $ & $ ^{+  0.033 }_{-  0.034 } $ \\
 CDF1~\cite{Abe:1999pv}  & \particle{ \mu  } & \particle{ \mu  } & $  0.503 $ & $ \pm  0.064 $ & $ \pm  0.071 $ & $  0.514 $ & $ \pm  0.064 $ & $ ^{+  0.070 }_{-  0.069 } $ \\
 CDF1~\cite{Abe:1999ds}  & \particle{ \ell  } & \particle{ \ell,\Qjet  } & $  0.500 $ & $ \pm  0.052 $ & $ \pm  0.043 $ & $  0.546 $ & $ \pm  0.052 $ & $ \pm  0.036 $ \\
 CDF1~\cite{Affolder:1999cn}  & \particle{ D^*\ell  } & \particle{ \ell  } & $  0.516 $ & $ \pm  0.099 $ & $ ^{+  0.029 }_{-  0.035 } $ & $  0.523 $ & $ \pm  0.099 $ & $ ^{+  0.028 }_{-  0.035 } $ \\
 \dzero~\cite{Abazov:2006qp}  & \particle{ D^{(*)}\mu  } & \particle{ \mbox{OST}  } & $  0.506 $ & $ \pm  0.020 $ & $ \pm  0.016 $ & $  0.506 $ & $ \pm  0.020 $ & $ \pm  0.016 $ \\
 \babar~\cite{Aubert:2001te,*Aubert:2002rg}  & \particle{ \Bd  } & \particle{ \ell,K,\mbox{NN}  } & $  0.516 $ & $ \pm  0.016 $ & $ \pm  0.010 $ & $  0.521 $ & $ \pm  0.016 $ & $ \pm  0.008 $ \\
 \babar~\cite{Aubert:2001tf}  & \particle{ \ell  } & \particle{ \ell  } & $  0.493 $ & $ \pm  0.012 $ & $ \pm  0.009 $ & $  0.487 $ & $ \pm  0.012 $ & $ \pm  0.006 $ \\
 \babar~\cite{Aubert:2002sh}  & \particle{ D^*\ell\nu  } & \particle{ \ell,K,\mbox{NN}  } & $  0.492 $ & $ \pm  0.018 $ & $ \pm  0.014 $ & $  0.493 $ & $ \pm  0.018 $ & $ \pm  0.013 $ \\
 \babar~\cite{Aubert:2005kf}  & \particle{ D^*\ell\nu\mbox{(part)}  } & \particle{ \ell  } & $  0.511 $ & $ \pm  0.007 $ & $ \pm  0.007 $ & $  0.513 $ & $ \pm  0.007 $ & $ \pm  0.007 $ \\
 \belle~\citehistory{Abe:2004mz}{Abe:2004mz,*Abe:2002id_hist,*Tomura:2002qs_hist,*Hara:2002mq_hist}  & \particle{ \Bd,D^*\ell\nu  } & \particle{ \mbox{comb}  } & $  0.511 $ & $ \pm  0.005 $ & $ \pm  0.006 $ & $  0.513 $ & $ \pm  0.005 $ & $ \pm  0.006 $ \\
 \belle~\cite{Zheng:2002jv}  & \particle{ D^*\pi\mbox{(part)}  } & \particle{ \ell  } & $  0.509 $ & $ \pm  0.017 $ & $ \pm  0.020 $ & $  0.513 $ & $ \pm  0.017 $ & $ \pm  0.019 $ \\
 \belle~\citehistory{Hastings:2002ff}{Hastings:2002ff,*Abe:2000yh_hist}  & \particle{ \ell  } & \particle{ \ell  } & $  0.503 $ & $ \pm  0.008 $ & $ \pm  0.010 $ & $  0.506 $ & $ \pm  0.008 $ & $ \pm  0.008 $ \\
 LHCb~\cite{LHCb-CONF-2011-010_published}  & \particle{ \Bd  } & \particle{ \mbox{OST}  } & $  0.499 $ & $ \pm  0.032 $ & $ \pm  0.003 $ & $  0.499 $ & $ \pm  0.032 $ & $ \pm  0.003 $ \\
 LHCb~\cite{Aaij:2012nt}  & \particle{ \Bd  } & \particle{ \mbox{OST,SST}  } & $  0.5156 $ & $ \pm  0.0051 $ & $ \pm  0.0033 $ & $  0.5156 $ & $ \pm  0.0051 $ & $ \pm  0.0033 $ \\
 LHCb~\cite{Aaij:2013gja}  & \particle{ D\mu  } & \particle{ \mbox{OST,SST}  } & $  0.503 $ & $ \pm  0.011 $ & $ \pm  0.013 $ & $  0.503 $ & $ \pm  0.011 $ & $ \pm  0.013 $ \\
 LHCb~\cite{Aaij:2016fdk}  & \particle{ D^{(*)}\mu  } & \particle{ \mbox{OST}  } & $  0.5050 $ & $ \pm  0.0021 $ & $ \pm  0.0010 $ & $  0.5050 $ & $ \pm  0.0021 $ & $ \pm  0.0010 $ \\
 \hline \\[-2.0ex]
 \multicolumn{6}{l}{World average (all above measurements included):} & $  0.5065 $ & $ \pm  0.0016 $ & $ \pm  0.0011 $ \\

%\hline  \\[-2.0ex]
%\multicolumn{6}{l}{World average (all above measurements included):}
%    & \hflavDMDWval & \hflavDMDWsta & \hflavDMDWsys \\
\\[-2.0ex]
\multicolumn{6}{l}{~~~ -- ALEPH, DELPHI, L3 and OPAL only:}
     & \hflavDMDLval & \hflavDMDLsta & \hflavDMDLsys \\
\multicolumn{6}{l}{~~~ -- CDF and \dzero only:}
     & \hflavDMDTval & \hflavDMDTsta & \hflavDMDTsys \\
%%% \multicolumn{6}{l}{~~~ -- ALEPH, DELPHI, L3, OPAL and CDF1 only:}
%%%      & \hflavDMDHval & \hflavDMDHsta & \hflavDMDHsys \\
\multicolumn{6}{l}{~~~ -- \babar and \belle only:}
     & \hflavDMDBval & \hflavDMDBsta & \hflavDMDBsys \\
\multicolumn{6}{l}{~~~ -- LHCb only:} & \hflavDMDLHCbval & \hflavDMDLHCbsta & \hflavDMDLHCbsys \\
\hline
\end{tabular}
\end{table}

A large number of time-dependent \Bd--\Bdbar oscillation analyses
have been performed in the past 20 years by the 
ALEPH, DELPHI, L3, OPAL, CDF, \dzero, \babar, \belle and  LHCb collaborations. 
The corresponding measurements of \dmd are summarized in 
\Table{dmd}\history{.}{,
where only the most recent results
are listed (\ie\ measurements superseded by more recent ones are omitted\unpublished{}{\footnote{
  \label{foot:life_mix:CDFnote8235:2006}
  Two old unpublished CDF2 measurements~\cite{CDFnote8235:2006,CDFnote7920:2005}
  are also omitted from our averages, \Table{dmd} and \Fig{dmd}.}}).}
Although a variety of different techniques have been used, the 
individual \dmd
results obtained at different colliders have remarkably similar precision.
The systematic uncertainties are comparable to the statistical uncertainties;
they are often dominated by sample composition, mistag probability,
or \b-hadron lifetime contributions.
%OS% Their average is compatible with the recent and more precise measurements 
%OS% from the asymmetric \B factories and the LHCb experiment.
Before being combined, the measurements are adjusted on the basis of a 
common set of input values, including the averages of the 
\b-hadron fractions and lifetimes given in this report 
(see \Secss{fractions}{lifetimes}).
Some measurements are statistically correlated. 
Systematic correlations arise both from common physics sources 
(fractions, lifetimes, branching ratios of \b hadrons), and from purely 
experimental or algorithmic effects (efficiency, resolution, flavour tagging, 
background description). Combining all published measurements
listed in \Table{dmd}
and accounting for all identified correlations
as described in \Ref{Abbaneo:2000ej_mod,*Abbaneo:2001bv_mod_cont} yields $\dmd = \hflavDMDWfull$.

On the other hand, ARGUS and CLEO have published 
measurements of the time-integrated mixing probability 
\chid~\cite{Albrecht:1992yd,*Albrecht:1993gr,Bartelt:1993cf,Behrens:2000qu}, 
which average to $\chid =\hflavCHIDU$.
Following \Ref{Behrens:2000qu}, 
the decay width difference \DGd could 
in principle be extracted from the
measured value of $\Gd=1/\tau(\Bd)$ and the above averages for 
\dmd and \chid 
(provided that \DGd has a negligible impact on 
the \dmd and $\tau(\Bd)$ analyses that have assumed $\DGd=0$), 
using the relation
\begin{equation}
\chid = \frac{\xd^2+\yd^2}{2(\xd^2+1)}
% ~~~ \mbox{with} ~~ \xd=\frac{\dmd}{\Gd}  ~~ \mbox{and} ~~ \yd=\frac{\DGd}{2\Gd}
\,.
\labe{chid_definition}
\end{equation}
However, direct time-dependent studies provide much stronger constraints: 
%%% DELPHI obtained
%%% $|\DGd|/\Gd < 18\%$ at \CL{95}~\cite{Abdallah:2002mr}, 
%%% while \babar obtained 
%%% % $-8.4\% < -{\rm sign}({\rm Re} \lambda_{\CP}) \DGGd < 6.8\%$ % closer to what \babar quoted
%%% $-6.8\% < {\rm sign}({\rm Re} \lambda_{\CP}) \DGGd < 8.4\%$
%%% at \CL{90}~\cite{Aubert:2003hd,*Aubert:2004xga}, 
$|\DGd|/\Gd < 18\%$ at \CL{95} from DELPHI~\cite{Abdallah:2002mr},
$-6.8\% < {\rm sign}({\rm Re} \lambda_{\CP}) \DGGd < 8.4\%$
at \CL{90} from \babar~\cite{Aubert:2003hd,*Aubert:2004xga},
and ${\rm sign}({\rm Re} \lambda_{\CP})\DGGd = (1.7 \pm 1.8 \pm 1.1)\%$~\cite{Higuchi:2012kx}
from Belle, 
where $\lambda_{\CP} = (q/p)_{\particle{d}} (\bar{A}_{\CP}/A_{\CP})$
is defined for a \CP-even final state 
%%% and where \DGd is defined as\footnote{This sign convention for 
%%% \DGd, taken from \Ref{Aubert:2003hd,*Aubert:2004xga},
%%% is opposite to that used for \DGs in \Secss{taubs}{DGs}.}
%%% $\DGd = \Gamma(\Bd_{\rm H})-\Gamma(\Bd_{\rm L})$
(the sensitivity to the overall sign of 
${\rm sign}({\rm Re} \lambda_{\CP}) \DGGd$ comes
from the use of \Bd decays to \CP final states).
In addition
LHCb has obtained $\DGGd=(-4.4 \pm 2.5 \pm 1.1)\%$~\cite{Aaij:2014owa}
by comparing measurements of the $\Bd \to \jpsi K^{*0}$ and $\Bd \to \jpsi K^0_{\rm S}$
decays, following the method of Ref.~\cite{Gershon:2010wx}.
More recently 
ATLAS has measured $\DGGd=(-0.1 \pm 1.1 \pm 0.9)\%$~\cite{Aaboud:2016bro} using a similar method. 
Assuming ${\rm Re} \lambda_{\CP} > 0$, as expected from the global fits
of the Unitarity Triangle within the Standard Model~\cite{Charles:2011va_mod,*Bona:2006ah_mod},
a combination of these five results (after adjusting the DELPHI and \babar results to  
$1/\Gd=\tau(\Bd)=\hflavTAUBD$) yields
\begin{equation}
%%% {\rm sign}({\rm Re} \lambda_{\CP}) 
\DGGd  = \hflavSDGDGD \,,
\end{equation}
%%% The sign of ${\rm Re} \lambda_{\CP}$ is not measured,
%%% but expected to be positive from the global fits
%%% of the Unitarity Triangle within the Standard Model~\cite{Charles:2011va_mod,*Bona:2006ah_mod}.
an average consistent with zero and with the Standard Model prediction of $(3.97\pm0.90)\times 10^{-3}$~\cite{Artuso:2015swg}. 
An independent result, 
$\DGGd=(0.50 \pm 1.38)\%$\citehistory{Abazov:2013uma}{Abazov:2013uma,*Abazov:2011yk_hist,*Abazov:2010hv_hist,*Abazov:2010hj_hist,*Abazov:dimuon_hist},
was obtained by the \dzero collaboration 
from their measurements of the single muon and same-sign dimuon charge asymmetries,
under the interpretation that 
the observed asymmetries are due to \CP violation in neutral $B$-meson mixing and interference.
This indirect determination was called into question~\cite{Nierste_CKM2014}
and is therefore not included in the above average, 
as explained in \Sec{qpd}.\footref{foot:life_mix:Abazov:2013uma}

Assuming $\DGd=0$ 
% and no \CP violation in mixing, and using the measured \Bd lifetime,
and using $1/\Gd=\tau(\Bd)=\hflavTAUBD$,
the \dmd and \chid results are combined through \Eq{chid_definition} 
to yield the 
world average
\begin{equation} 
\dmd = \hflavDMDWU \,,
\labe{dmd}
\end{equation} 
or, equivalently,
\begin{equation} 
\xd= \hflavXDWU ~~~ \mbox{and} ~~~ \chid=\hflavCHIDWU \,.  
\labe{chid}
\end{equation}
\Figure{dmd} compares the \dmd values obtained by the different experiments.

\begin{figure}
\begin{center}
\includegraphics[width=\textwidth,viewport=0 100 610 700,clip=true]{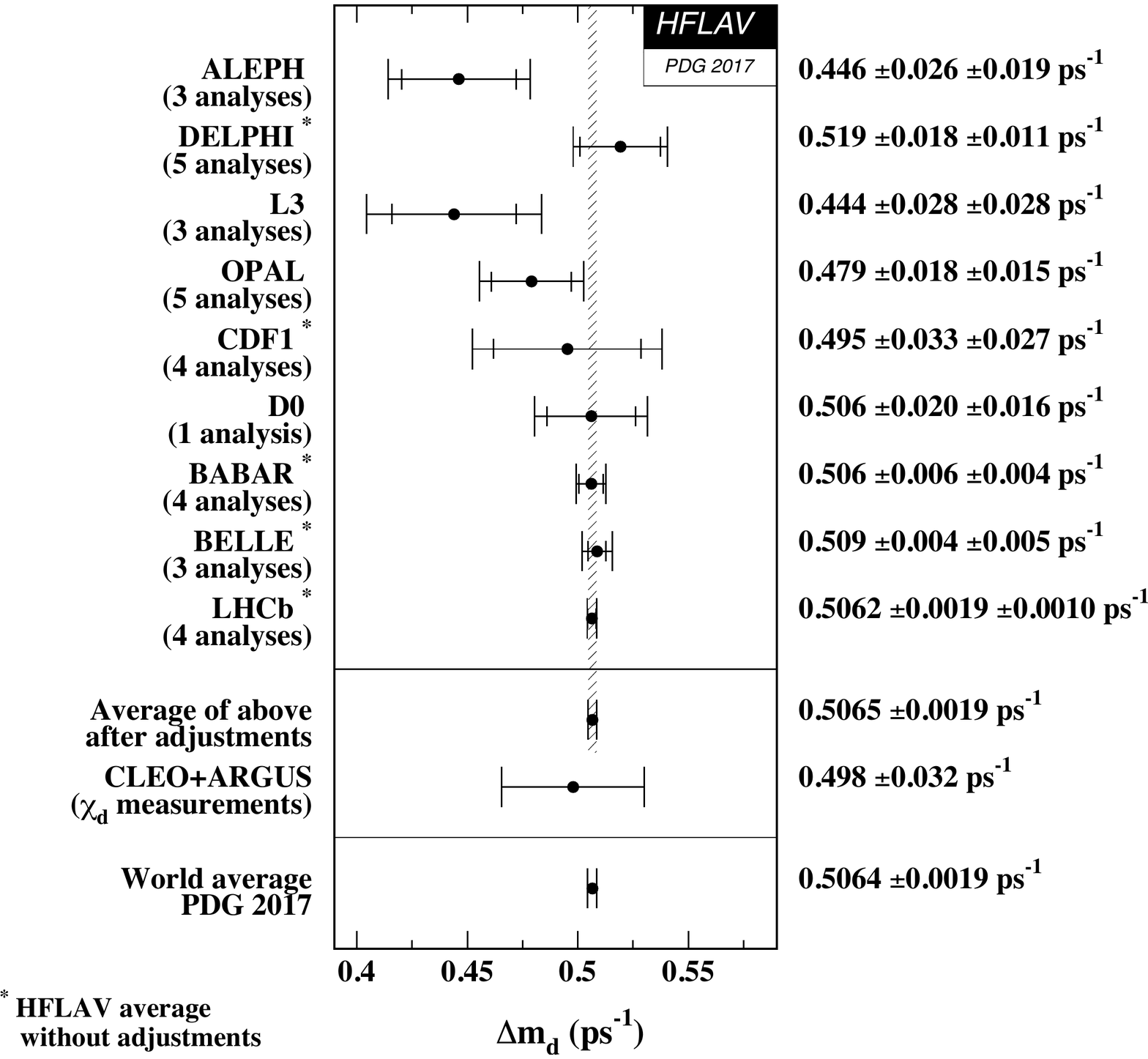}
\caption{The \Bd--\Bdbar oscillation frequency \dmd as measured by the different experiments. 
The averages quoted for ALEPH, L3 and OPAL are taken from the original publications, while the 
ones for DELPHI, CDF, \babar, \belle and LHCb are computed from the individual results 
listed in \Table{dmd} without performing any adjustments. The time-integrated measurements 
of \chid from the symmetric \B factory experiments ARGUS and CLEO are converted 
to a \dmd value using $\tau(\Bd)=\hflavTAUBD$. The two global averages are obtained 
after adjustments of all the individual \dmd results of \Table{dmd} (see text).}
\labf{dmd}
\end{center}
\end{figure}

The \Bd mixing averages given in \Eqss{dmd}{chid}
and the \b-hadron fractions of \Table{fractions} have been obtained in a fully 
consistent way, taking into account the fact that the fractions are computed using 
the \chid value of \Eq{chid} and that many individual measurements of \dmd
at high energy depend on the assumed values for the \b-hadron fractions.
Furthermore, this set of averages is consistent with the lifetime averages 
of \Sec{lifetimes}.

\mysubsubsection{\Bs mixing parameters \DGs and \dms}
%\mysubsubsection{Mass and decay width differences \dms and \DGs}
%------------------------------------------------
\labs{DGs} \labs{dms}

The best sensitivity to \DGs is currently achieved 
by the recent time-dependent measurements
of the $\Bs\to\jpsi\phi$ (or more generally $\Bs\to (c\bar{c}) K^+K^-$) decay rates performed at
CDF~\citehistory{Aaltonen:2012ie}{Aaltonen:2012ie,*CDF:2011af_hist,*Aaltonen:2007he_hist,*Aaltonen:2007gf_hist},
\dzero~\citehistory{Abazov:2011ry}{Abazov:2011ry,*Abazov:2008af_hist,*Abazov:2007tx_hist},
ATLAS~\citehistory{Aad:2014cqa,Aad:2016tdj}{Aad:2014cqa,*Aad:2012kba_hist,Aad:2016tdj}
CMS~\unpublished{\cite{Khachatryan:2015nza}}{\cite{CMS-PAS-BPH-11-006,Khachatryan:2015nza}}
and LHCb~\citehistory{Aaij:2014zsa,Aaij:2016ohx}{Aaij:2014zsa,*Aaij:2013oba_supersede2,Aaij:2016ohx},
where the \CP-even and \CP-odd
amplitudes are statistically separated through a full angular analysis.
%(see last two columns of \Table{phisDGsGs}). 
%
%OS% In addition, 
%OS% LHCb~\citehistory{Aaij:2013oba}{Aaij:2013oba,*LHCb:2011aa_hist,*LHCb:2012ad_hist,*LHCb:2011ab_hist,*Aaij:2012nta_hist}
%OS% has analyzed $\Bs\to\jpsi \pi^+\pi^-$ decays, for which no angular analysis is needed. 
\unpublished{These}{With the exception of the first CMS analysis~\cite{CMS-PAS-BPH-11-006}%
\footnote{The CMS result of \Ref{CMS-PAS-BPH-11-006}
is statistically independent of that of
Ref.~\cite{Khachatryan:2015nza} but, since it has not
been published, it is not included in \Table{GsDGs} nor in our averages.},
these}
studies use both untagged and tagged \Bs\ candidates and 
are optimized for the measurement of the \CP-violating 
phase \phiccbars, defined later in \Sec{phasebs}.
The LHCb collaboration analyzed the $\Bs \to \jpsi K^+K^-$
decay, considering that the $K^+K^-$ system can be in a $P$-wave or $S$-wave state, 
and measured the dependence of the strong phase difference between the 
$P$-wave and $S$-wave amplitudes as a function of the $K^+K^-$ invariant
mass~\cite{Aaij:2012eq}. 
This allowed, for the first time, the unambiguous determination of the sign of 
$\DGs$, which was found to be positive at the $4.7\,\sigma$ level. 
The following averages present only the $\DGs > 0$ solutions.

%wrong% The combined fit procedure used to extract simultaneously \DGs\ and \phiccbars
%wrong% is described in \Sec{phasebs}. 
%wrong% The results, displayed as the red contours labelled ``$\Bs \to \jpsi\phi$ measurements'' in the 
%wrong% plots of \Fig{DGs}, are given in the first column of numbers of \Table{tabtauLH}.
%wrong% In those averages, the correlation between \DGs and \Gs has been neglected. 

The published results~\citehistory%
{Aaltonen:2012ie,Abazov:2011ry,Aad:2014cqa,Aad:2016tdj,Khachatryan:2015nza,Aaij:2014zsa,Aaij:2016ohx}%
{Aaltonen:2012ie,*CDF:2011af_hist,*Aaltonen:2007he_hist,*Aaltonen:2007gf_hist,Abazov:2011ry,*Abazov:2008af_hist,*Abazov:2007tx_hist,Aad:2014cqa,*Aad:2012kba_hist,Aad:2016tdj,Khachatryan:2015nza,Aaij:2014zsa,*Aaij:2013oba_supersede2,Aaij:2016ohx}
are shown in \Table{GsDGs}. They are combined taking into account, in each analysis, the correlation between \DGs and \Gs.
The results, displayed as the red contours labelled ``$\Bs \to (c\bar{c}) KK$ measurements'' in the
plots of \Fig{DGs}, are given in the first column of numbers of \Table{tabtauLH}.

\begin{table}
\caption{Measurements of \DGs and \Gs using
$\Bs\to\jpsi\phi$, $\Bs\to\jpsi K^+K^-$ and $\Bs\to\psi(2S)\phi$ decays.
Only the solution with $\DGs > 0$ is shown, since the two-fold ambiguity has been
resolved in \Ref{Aaij:2012eq}. The first error is due to 
statistics, the second one to systematics. The last line gives our average.}
\labt{GsDGs}
\begin{center}
%\begin{tabular}{l@{\,}l@{\,}l@{\,}|@{\,}l@{\,}|@{\,}l@{\,}|@{\,}l} 
\begin{tabular}{ll@{\,}rll@{\,}l@{\,}} 
\hline
% Exp.\ & Mode & Dataset & \multicolumn{1}{c@{\,}|@{\,}}{\phiccbars}
%                      & \multicolumn{1}{c@{}}{\DGs (\!\!\invps)} & Ref.\ \\
Exp.\ & Mode & Dataset
      & \multicolumn{1}{c}{\DGs (\!\!\invps)}
      & \multicolumn{1}{c}{\Gs  (\!\!\invps)}
      & Ref.\ \\
\hline
CDF    & $\jpsi\phi$ & $9.6\invfb$
       & $+0.068\pm0.026\pm0.009$
       % syst error was +-0.007 instead of +-0.009 in CDF note 10778
       % & $0.6545\pm0.0081\pm0.0039$
       & $0.654\pm0.008\pm0.004$ % quoted in paper as tau(Bs) = 1/Gamma_s = 1.528 +-0.019 +-0.009
% 1./1.528 = 0.6544502617801047, 0.019/1.528**2 = 0.008137797757736905, 0.009//1.528**2 = 0.003854746306296428
       & \citehistory{Aaltonen:2012ie}{Aaltonen:2012ie,*CDF:2011af_hist,*Aaltonen:2007he_hist,*Aaltonen:2007gf_hist} \\
\dzero & $\jpsi\phi$ & $8.0\invfb$
       & $+0.163^{+0.065}_{-0.064}$ 
       & $0.693^{+0.018}_{-0.017}$
       & \citehistory{Abazov:2011ry}{Abazov:2011ry,*Abazov:2008af_hist,*Abazov:2007tx_hist} \\
ATLAS  & $\jpsi\phi$ & $4.9\invfb$
       & $+0.053 \pm0.021 \pm0.010$
       & $0.677 \pm0.007 \pm0.004$
       & \citehistory{Aad:2014cqa}{Aad:2014cqa,*Aad:2012kba_hist}  \\
ATLAS  & $\jpsi\phi$ & $14.3\invfb$
       & $+0.101 \pm0.013 \pm0.007$
       & $0.676 \pm0.004 \pm0.004$
       & \cite{Aad:2016tdj} \\
ATLAS  & \multicolumn{2}{r}{above 2 combined}
       & $+0.085 \pm0.011 \pm0.007$
       & $0.675 \pm0.003 \pm0.003$
       & \cite{Aad:2016tdj} \\
% CMS    & $\jpsi\phi$ & $5.0\invfb$ 
%        & $+0.048\pm0.024\pm0.003$
%        & $0.655\pm0.008\pm0.003$
% %%% the CMS note CMS-PAS-BPH-11-006 quotes a "mean Bs lifetime" of
% %%%    ctau = 458.0 +-5.9 +-2.2 microns = 1.528 +-0.020 +-0.007 ps
% %%%    corresponding to Gamma_s = 0.6546 +-0.0084 +-0.0031 ps-1
%        & \cite{CMS-PAS-BPH-11-006}$^p$ \\
% The above result is statisticall independent of the result below, 
% but it is not inlcuded here because it is unpublished
CMS    & $\jpsi\phi$ & $19.7\invfb$ 
       & $+0.095\pm0.013\pm0.007$
       & $0.6704 \pm0.0043 \pm0.0055$
%%% the CMS paper quotes a "mean Bs lifetime" of
%%%    ctau = 447.2 +-2.9 +-3.7 microns = 1.492 +-0.010 +-0.012 ps
%%%    corresponding to Gamma_s = 0.6704 +-0.0043 +-0.0055 ps-1
       & \cite{Khachatryan:2015nza} \\
LHCb   & $\jpsi K^+K^-$ & $3.0\invfb$
       & $+0.0805\pm0.0091\pm0.0032$
       & $0.6603\pm0.0027\pm0.0015$
       & \citehistory{Aaij:2014zsa}{Aaij:2014zsa,*Aaij:2013oba_supersede2} \\
LHCb   & $\psi(2S) \phi$ & $3.0\invfb$
       & $+0.066 ^{+0.041}_{-0.044} \pm 0.007$
       & $0.668 \pm0.011 \pm0.006$
       & \cite{Aaij:2016ohx} \\
\hline
\multicolumn{3}{l}{All combined} & \hflavDGSnounit & \hflavGSnounit & \\ 
\hline
% \multicolumn{6}{l}{$^p$ {\footnotesize Preliminary.}}
\end{tabular}
\end{center}
\end{table}

\begin{figure}
\begin{center}
\includegraphics[width=0.49\textwidth]{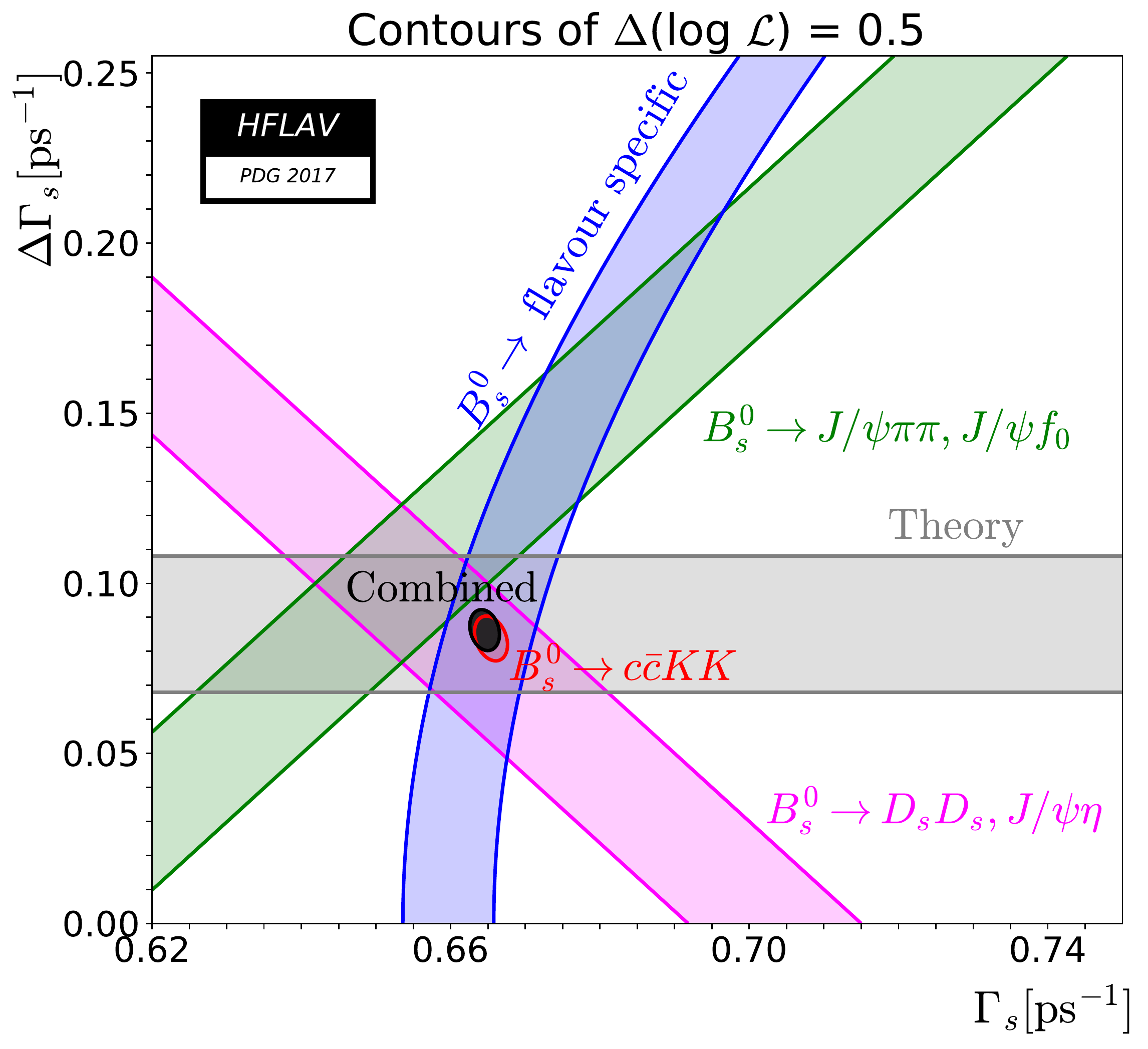}
\hfill
\includegraphics[width=0.49\textwidth]{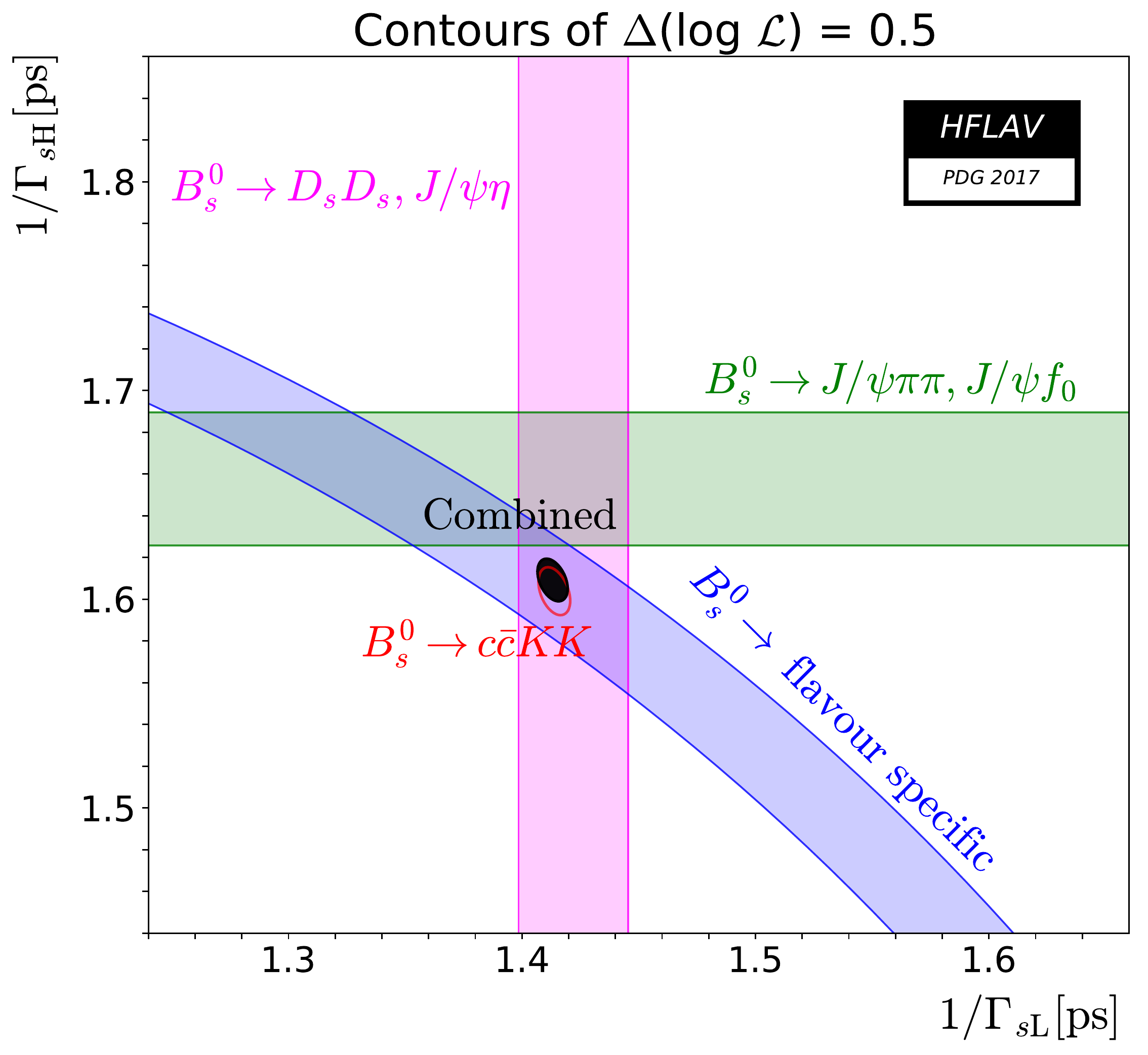}
\caption{Contours of $\Delta \ln L = 0.5$ (39\% CL for the enclosed 2D regions, 68\% CL for the bands)
shown in the $(\Gs,\,\DGs)$ plane on the left
and in the $(1/\Gamma_{s\rm L},\,1/\Gamma_{s\rm H})$ plane on the right. 
The average of all the $\Bs \to \jpsi\phi$, $\Bs\to \jpsi K^+K^-$ and
$\Bs \to \psi(2S)\phi$
% and $\jpsi\pi^+\pi^-$
results is shown as the red contour,
and the constraints given by the effective lifetime measurements of
\Bs\ to flavour-specific, pure \CP-odd and pure \CP-even final states
are shown as the blue, green and purple bands, 
respectively. The average taking all constraints into account is shown as the grey-filled contour.
The yellow band is a theory prediction
$\DGs = 0.088 \pm 0.020~\hbox{ps}^{-1}$~\protect\citehistory{Jubb:2016mvq,Artuso:2015swg}{Jubb:2016mvq,Artuso:2015swg,*Lenz_hist}
that assumes no new physics in \Bs\ mixing.}
% ,!!!!!!!!TO BE UPDATED !!!!!!\DGs combination results with one-sigma contours
% ($\Delta\log\mathcal{L} = 0.5$) shown for (a) \DGs versus
% $\bar{\tau}(\Bs) = 1/\Gs$  and (b)
% $\tau_{\rm H} = 1/\Gamma_{s\rm H}$ versus $\tau_{\rm L} = 1/\Gamma_{s\rm L}$.
% The red contours labelled ``Direct" are the result of the combination of
% last two measurements of \Table{dgammat}, the blue bands are the one-sigma
% contours due to the world average of flavour-specific 
% \Bs lifetime measurements,
% the green bands are the one-sigma contour of the $\Bs \to K^+K^-$ 
% lifetime measurement, 
% and the solid and dashed-outlined shaded regions result using
% the combination constraints described in the text.}
\labf{DGs}
\end{center}
\end{figure}

\begin{table}
\caption{Averages of \DGs, $\Gs$ and related quantities, obtained from
$\Bs\to\jpsi\phi$, $\Bs\to\jpsi K^+K^-$ and $\Bs\to\psi(2S)\phi$ alone (first column),
adding the constraints from the effective lifetimes measured in pure \CP modes
%OL2014 $\Bs\to K^+ K^-, \, D_s^+D_s^-, \, \jpsi f_0(980), \jpsi K_{\rm S}^0$ (second column),
$\Bs\to D_s^+D_s^-,J/\psi\eta$ and $\Bs \to \jpsi f_0(980), \jpsi \pi^+\pi^-$ (second column),
and adding the constraint from the effective lifetime measured in flavour-specific modes
$\Bs\to D_s^-\ell^+\nu X, \, D_s^-\pi^+, \, D_s^-D^+$ (third column, recommended world averages).}
\labt{tabtauLH}
\begin{center}
\begin{tabular}{c|c|c|c}
\hline
% & $\jpsi hh$
% & $\jpsi hh, \mbox{\CP-even}, \mbox{\CP-odd}$
% & $\jpsi hh, \mbox{\CP-even}, \mbox{\CP-odd}, \mbox{flavour-specific}$ \\
& $\Bs\to (c\bar{c}) K^+K^-$ modes & $\Bs\to (c\bar{c}) K^+K^-$ modes & $\Bs\to (c\bar{c}) K^+K^-$ modes \\
& only (see \Table{GsDGs}) & + pure \CP modes & + pure \CP modes \\
&                          &                  & + flavour-specific modes \\
\hline
\Gs                & \hflavGS        &  \hflavGSCO        &  \hflavGSCON        \\
$1/\Gs$            & \hflavTAUBSMEAN &  \hflavTAUBSMEANCO &  \hflavTAUBSMEANCON \\
$1/\Gamma_{s\rm L}$ & \hflavTAUBSL    &  \hflavTAUBSLCO    &  \hflavTAUBSLCON    \\
$1/\Gamma_{s\rm H}$ & \hflavTAUBSH    &  \hflavTAUBSHCO    &  \hflavTAUBSHCON    \\
\DGs               & \hflavDGS       &  \hflavDGSCO       &  \hflavDGSCON       \\
\DGs/\Gs           & \hflavDGSGS     &  \hflavDGSGSCO     &  \hflavDGSGSCON     \\
$\rho(\Gs,\DGs)$   & \hflavRHOGSDGS  &  \hflavRHOGSDGSCO  &  \hflavRHOGSDGSCON  \\
\hline
\end{tabular}
\end{center}
\end{table}

%The positive sign of $\DGs$ is due to the constraint applied 
%on $\phi_s$. In absence of such constraint, there would be two 
%mirror solutions related by the transformation
%$(\DGs, \phi_s)  \to (-\DGs, \pi-\phi_s)$.

An alternative approach, which is directly sensitive to first order in 
$\DGs/\Gs$, 
is to determine the effective lifetime of untagged \Bs\ candidates
decaying to %fairly
pure \CP eigenstates; we use here measurements with
%OL 2014 $\Bs \to K^+K^-$~\citehistory{Aaij:2012kn,Aaij:2014fia}{Aaij:2012kn,Aaij:2014fia,*Aaij:2012ns_hist}%
%\unpublished{}{\footnote{An old unpublished measurement of the $\Bs \to K^+ K^-$
%effective lifetime by CDF~\cite{Tonelli:2006np} is no longer considered.}},
$\Bs \to D_s^+D_s^-$~\cite{Aaij:2013bvd}, 
$\Bs \to J/\psi \eta$~\cite{Aaij:2016dzn}, 
$\Bs \to \jpsi f_0(980)$~\cite{Aaltonen:2011nk,Abazov:2016oqi}
and $\Bs\to \jpsi \pi^+\pi^-$~\citehistory{Aaij:2013oba}{Aaij:2013oba,*LHCb:2011aa_hist,*LHCb:2012ad_hist,*LHCb:2011ab_hist,*Aaij:2012nta_hist} decays.
% OL2014 and $\Bs \to \jpsi K_{\rm S}^0$~\cite{Aaij:2013eia}.
The precise extraction of $1/\Gs$ and $\DGs$
from such measurements, discussed in detail in \Ref{Fleischer:2011cw}, 
requires additional information 
in the form of theoretical assumptions or
external inputs on weak phases and hadronic parameters. 
If $f$ designates a final state in which both \Bs and \Bsbar can decay,
the ratio of the effective \Bs lifetime decaying to $f$ relative to the mean
\Bs lifetime is~\cite{Fleischer:2011cw}%
\footnote{%
\label{foot:life_mix:ADG-def}
The definition of $A_f^{\DG}$ given in \Eq{ADG} has the sign opposite to that given in \Ref{Fleischer:2011cw}.}
\begin{equation}
  \frac{\tau_{\rm single}(\Bs \to f)}{\tau(\Bs)} = \frac{1}{1-y_s^2} \left[ \frac{1 - 2A_f^{\DG} y_s + y_s^2}{1 - A_f^{\DG} y_s}\right ] \,,
%   = 1 - A_f^{\DG} y_s + 2 y_s^2\left [ 2 - (A_f^{\DG})^2 \right ] + \mathcal{O}(y_s^3),
\labe{tauf_fleisch}
\end{equation}
where
\begin{equation}
A_f^{\DG} = -\frac{2 \Re(\lambda_f)} {1+|\lambda_f|^2} \,.
\labe{ADG}
\end{equation}
To include the measurements of the effective
%OL2014 $\Bs\to K^+ K^-$ (\CP-even),
$\Bs \to D_s^+D_s^-$ (\CP-even), $\Bs \to \jpsi f_0(980)$ (\CP-odd) and
$\Bs \to \jpsi\pi^+\pi^-$ (\CP-odd) 
% $\Bs \to \jpsi K_{\rm S}^0$ (\CP-odd)
lifetimes as constraints in the \DGs fit,\footnote{%
The effective lifetimes measured in $\Bs\to K^+ K^-$ (mostly \CP-even) and  $\Bs \to \jpsi K_{\rm S}^0$ (mostly \CP-odd) are not used because we can not quantify the penguin contributions in those modes.}
we neglect sub-leading penguin contributions and possible direct \CP violation. 
Explicitly, in \Eq{ADG}, we set
$A_{\mbox{\scriptsize \CP-even}}^{\DG} = \cos \phiccbars$
and $A_{\mbox{\scriptsize \CP-odd}}^{\DG} = -\cos \phiccbars$.
Given the small value of $\phiccbars$, we have, to first order in $y_s$:
\begin{eqnarray}
\tau_{\rm single}(\Bs \to \mbox{\CP-even})
& \approx & \frac{1}{\Gamma_{s\rm L}} \left(1 + \frac{(\phiccbars)^2 y_s}{2} \right) \,,
\labe{tau_KK_approx}
\\
\tau_{\rm single}(\Bs \to \mbox{\CP-odd})
& \approx & \frac{1}{\Gamma_{s\rm H}} \left(1 - \frac{(\phiccbars)^2 y_s}{2} \right) \,.
\labe{tau_Jpsif0_approx}
\end{eqnarray}
The numerical inputs are taken from \Eqss{tau_KK}{tau_Jpsif0}
%OL2014 \footnote{The LHCb effective lifetime measurement obtained with 
%$\Bs \to\jpsi\pi^+\pi^-$ decays~\citehistory{Aaij:2013oba}{Aaij:2013oba,*LHCb:2011aa_hist,*LHCb:2012ad_hist,*LHCb:2011ab_hist,*Aaij:2012nta_hist}
%is first removed from the average of \Eq{tau_Jpsif0}, 
%yielding 
%$\tau_{\rm single}(\Bs \to \mbox{\CP-odd}) = \hflavTAUBSLONGCON$, because this 
%information is already contained in the $\Bs \to \jpsi hh$ analysis.}
and the resulting averages, combined with the $\Bs\to\jpsi K^+K^-$ information,
are indicated in the second column of numbers of \Table{tabtauLH}. 
These averages assume $\phiccbars = 0$, which is compatible with
%OS% the central value of
the \phiccbars average presented in \Sec{phasebs}.

Information on \DGs can also be obtained from the study of the
proper time distribution of untagged samples
of flavour-specific \Bs decays~\cite{Hartkorn:1999ga}, where
the flavour (\ie, \Bs or \Bsbar) at the time of decay can be determined by
the decay products. In such decays,
\eg\ semileptonic \Bs decays, there is
an equal mix of the heavy and light mass eigenstates at time zero.
The proper time distribution is then a superposition 
of two exponential functions with decay constants
$\Gamma_{s\rm L}$ and $\Gamma_{s\rm H}$. % $\Gamma_{s\rm L,H} = \Gs \pm \DGs/2$.
This provides sensitivity to both $1/\Gs$ and 
$(\DGs/\Gs)^2$. Ignoring \DGs and fitting for 
a single exponential leads to an estimate of \Gs with a 
relative bias proportional to $(\DGs/\Gs)^2$, as shown in \Eq{fslife}. 
Including the constraint from the world-average flavour-specific \Bs 
lifetime, given in \Eq{tau_fs}, leads to the results shown in the last column 
of \Table{tabtauLH}.
These world averages are displayed as the grey contours labelled ``Combined'' in the
plots of \Fig{DGs}. 
They correspond to the lifetime averages
$1/\Gs=\hflavTAUBSMEANCON$,
$1/\Gamma_{s\rm L}=\hflavTAUBSLCON$,
$1/\Gamma_{s\rm H}=\hflavTAUBSHCON$,
and to the decay-width difference
\begin{equation}
\DGs = \hflavDGSCON ~~~~\mbox{and} ~~~~~ \DGs/\Gs = \hflavDGSGSCON \,.
\labe{DGs_DGsGs}
\end{equation}
The good agreement with the Standard Model prediction 
$\DGs = 0.088 \pm 0.020~\hbox{ps}^{-1}$~\citehistory{Jubb:2016mvq,Artuso:2015swg}{Jubb:2016mvq,Artuso:2015swg,*Lenz_hist}
excludes significant quark-hadron duality violation in the HQE~\cite{Lenz:2012mb}. 

Estimates of $\DGs/\Gs$ obtained from measurements of the 
$\Bs \to D_s^{(*)+} D_s^{(*)-}$ branching fraction~\citehistory{Barate:2000kd,Esen:2010jq,Abazov:2008ig,Abulencia:2007zz}{Barate:2000kd,Esen:2010jq,Abazov:2008ig,*Abazov:2007rb_hist,Abulencia:2007zz}
have not been used,
%OS% \footnote{%
%OS% A new average is being prepared.},
%XXXXX%OS%%Our average is ${\cal B} = \hflavBRDSDS$, from which one would get 
%XXXXX%OS%%$\DGs/\Gs \sim 2{\cal B}/(1-{\cal B}) = \hflavDGSGSBRDSDS$.},
since they are based on the questionable~\cite{Lenz:2011ti,*Lenz:2006hd}
assumption that these decays account for all \CP-even final states.
The results of early lifetime analyses attempting
to measure $\DGs/\Gs$~\citehistory{Acciarri:1998uv,Abreu:2000sh,Abreu:2000ev,Abe:1997bd}{Acciarri:1998uv,Abreu:2000sh,Abreu:2000ev,*Abreu:1996ep_hist,Abe:1997bd}
have not been used either.

The strength of \Bs mixing has been known to be large for more than 20 years. 
Indeed the time-integrated measurements of \chibar (see \Sec{chibar}),
when compared to our knowledge
of \chid and the \b-hadron fractions, indicated that 
\chis should be close to its maximal possible value of $1/2$.
Many searches of the time dependence of this mixing 
were performed by ALEPH~\cite{Heister:2002gk}, % no history here, \ie\ we do not quote the papers superseded by \cite{Heister:2002gk}
DELPHI~\citehistory{Abreu:2000sh,Abreu:2000ev,Abdallah:2002mr,Abdallah:2003qga}{Abreu:2000sh,Abreu:2000ev,*Abreu:1996ep_hist,Abdallah:2002mr,Abdallah:2003qga}, % here we must keep some history because Abreu:2000ev,*Abreu:1996ep_hist contain also Bs lifetime measurements reported elsewhere
OPAL~\cite{Abbiendi:1999gm,Abbiendi:2000bh},
SLD~\unpublished{\cite{Abe:2002ua,Abe:2002wfa}}{\cite{Abe:2002ua,Abe:2002wfa,Abe:2000gp}},
CDF (Run~I)~\cite{Abe:1998qj} and
\dzero~\cite{Abazov:2006dm}
%(we omit references to searches that have been superseded
%by more recent ones).
but did not have enough statistical power
and proper time resolution to resolve 
the small period of the \Bs\ oscillations.

\Bs oscillations have been observed for the first time in 2006
by the CDF collaboration~\citehistory{Abulencia:2006ze}{Abulencia:2006ze,*Abulencia:2006mq_hist},
based on samples of flavour-tagged hadronic and semileptonic \Bs decays
(in flavour-specific final states), partially or fully reconstructed in 
$1\invfb$ of data collected during Tevatron's Run~II. 
% From the proper-time dependence of these \Bs candidates, CDF
% observe \Bs oscillations with a significance of at least $5\,\sigma$ 
% and measure $\dms = 17.77 \pm 0.10 \pm 0.07\invps$~\citehistory{Abulencia:2006ze}{Abulencia:2006ze,*Abulencia:2006mq_hist}.
\unpublished{}{
This was shortly followed by independent unpublished evidence obtained by the \dzero collaboration
with $2.4\invfb$ of
data~\cite{D0note5618:2008,*D0note5474:2007,*D0note5254:2006}.}
More recently the LHCb collaboration obtained the most precise results using fully reconstructed 
$\Bs \to D_s^-\pi^+$ and $\Bs \to D_s^-\pi^+\pi^-\pi^+$ decays at the 
LHC~\cite{Aaij:2011qx,Aaij:2013mpa}.
LHCb has also observed \Bs oscillations with 
$\Bs\to\jpsi K^+K^-$ decays~\citehistory{Aaij:2014zsa}{Aaij:2014zsa,*Aaij:2013oba_supersede2}
and with semileptonic $\Bs \to D_s^-\mu^+ X$ decays~\cite{Aaij:2013gja}.
The measurements of \dms are summarized in \Table{dms}. 

\begin{table}[t]
\caption{Measurements of \dms.}
\labt{dms}
\begin{center}
%%%% \resizebox{\textwidth}{!}{
\begin{tabular}{ll@{}crl@{\,}l@{\,}ll} \hline
Experiment & Method           & \multicolumn{2}{c}{Data set} & \multicolumn{3}{c}{\dms (\!\!\invps)} & Ref. \\
\hline
CDF2   & \multicolumn{2}{l}{\particle{D_s^{(*)-} \ell^+ \nu}, \particle{D_s^{(*)-} \pi^+}, \particle{D_s^{-} \rho^+}}
       & 1 \invfb & $17.77$ & $\pm 0.10$ & $\pm 0.07~$
       & \citehistory{Abulencia:2006ze}{Abulencia:2006ze,*Abulencia:2006mq_hist} \\
\unpublished{}{%
\dzero & \particle{D_s^- \ell^+ X}, \particle{D_s^- \pi^+ X}
       &  & 2.4 \invfb & $18.53$ & $\pm 0.93$ & $\pm 0.30~$ 
       & \cite{D0note5618:2008,*D0note5474:2007,*D0note5254:2006}$^u$ \\
}%
LHCb   & \particle{D_s^- \pi^+}, \particle{D_s^- \pi^+\pi^-\pi^+}
       & 2010 & 0.034 \invfb & $17.63$ & $\pm 0.11$ & $\pm 0.02~$   
       & \cite{Aaij:2011qx} \\
LHCb   & \particle{D_s^- \mu^+ X}
       & 2011 & 1.0 \invfb & $17.93$ & $\pm 0.22$ & $\pm 0.15$ 
       & \cite{Aaij:2013gja}  \\
LHCb   & \particle{D_s^- \pi^+}
       & 2011 & 1.0 \invfb & $17.768$ & $\pm 0.023$ & $\pm 0.006$ 
       & \cite{Aaij:2013mpa}  \\
LHCb   & \particle{\jpsi K^+K^-}
       & 2011--2012 & 3.0 \invfb & $17.711$ & $^{+0.055}_{-0.057}$ & $\pm 0.011$ 
       & \citehistory{Aaij:2014zsa}{Aaij:2014zsa,*Aaij:2013oba_supersede2}  \\
\hline
\multicolumn{4}{l}{Average \unpublished{}{of CDF and LHCb measurements}} & $\hflavDMSval$ & $\hflavDMSsta$ & $\hflavDMSsys$ & \\  
\hline
\unpublished{}{$^u$ \footnotesize Unpublished. }
\end{tabular}
%%%% }
\end{center}
\end{table}

\begin{figure}
\begin{center}
\includegraphics[width=0.8\textwidth]{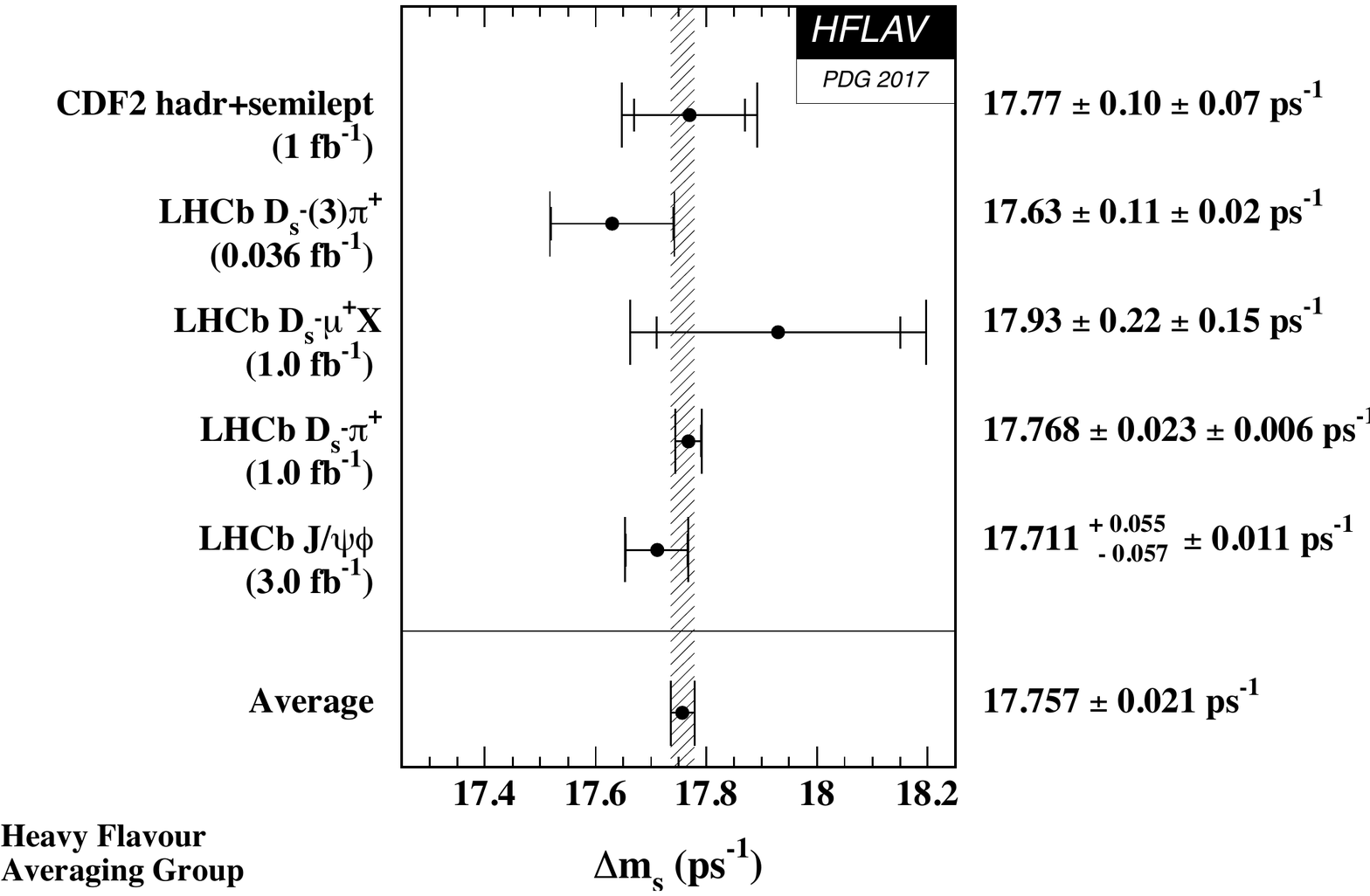}
\caption{Published % and recent preliminary
measurements of \dms, together with their average.} 
\labf{dms}
\end{center}
\end{figure}

A simple average of the CDF and LHCb results\unpublished{}{\footnote{
  \label{foot:life_mix:D0note5618:2008}
  We do not include the unpublished
  \dzero~\cite{D0note5618:2008,*D0note5474:2007,*D0note5254:2006} result in the average.}},
taking into account the correlated systematic uncertainties between the three 
LHCb measurements, yields 
\begin{equation}
\dms = \hflavDMSfull = \hflavDMS \labe{dms}
\end{equation}
and is illustrated in \Figure{dms}.
The Standard Model prediction 
$\dms = 18.3 \pm 2.7~\hbox{ps}^{-1}$~\citehistory{Jubb:2016mvq,Artuso:2015swg}{Jubb:2016mvq,Artuso:2015swg,*Lenz_hist} is consistent with the experimental value, but has a much larger error dominated by the uncertainty on the hadronic matrix elements.
The ratio $\DGs/\dms$ can be predicted more accurately, $0.0048 \pm 0.0008$~\citehistory{Jubb:2016mvq,Artuso:2015swg}{Jubb:2016mvq,Artuso:2015swg,*Lenz_hist},
and is in good agreement with the experimental determination of 
\begin{equation}
\DGs/\dms= \hflavRATIODGSDMS \,.
\end{equation}

Multiplying the \dms result of \Eq{dms} with the 
mean \Bs lifetime of \Eq{oneoverGs}, $1/\Gs=\hflavTAUBSMEANCON$,
yields
\begin{equation}
\xs % = \frac{\dms}{\Gs} 
= \hflavXS \,. \labe{xs}
\end{equation}
With $2\ys %= \DGGs
=\hflavDGSGSCON$ 
%(see \Eqss{DGGs_ave}{corrDGGs})
%(see \Eq{DGGs_ave})
(see \Eq{DGs_DGsGs})
and under the assumption of no \CP violation in \Bs mixing,
this corresponds to
\begin{equation}
\chis = \frac{\xs^2+\ys^2}{2(\xs^2+1)} = \hflavCHIS \,. \labe{chis}
\end{equation}
The ratio of the \Bd and \Bs oscillation frequencies, 
obtained from \Eqss{dmd}{dms}, 
\begin{equation}
\frac{\dmd}{\dms} = \hflavRATIODMDDMS \,, \labe{dmd_over_dms}
\end{equation}
can be used to extract the following magnitude of the ratio of CKM matrix elements, 
\begin{equation}
\left|\frac{V_{td}}{V_{ts}}\right| =
\xi \sqrt{\frac{\dmd}{\dms}\frac{m(\Bs)}{m(\Bd)}} = 
\hflavVTDVTSfull \,, \labe{Vtd_over_Vts}
\end{equation}
where the first quoted error is from experimental uncertainties 
(with the masses $m(\Bs)$ and $m(\Bd)$ taken from \Ref{PDG_2016}),
and where the second quoted error is from theoretical uncertainties 
in the estimation of the SU(3) flavour-symmetry breaking factor
$\xi %= (f_{B_s} \sqrt{B_{B_s}})/(f_{B_d} \sqrt{B_{B_d}})
= \hflavXIfull$,
obtained from 
%2016% unquenched 
recent three-flavour
lattice QCD calculations~\cite{Bazavov:2016nty,Aoki:2016frl}.
%2016% \cite{Aoki:2013ldr_mod}.
%2014% \cite{Laiho:2009eu_mod,*Evans:2008zzg_mod,*Gamiz:2009ku,*Albertus:2010nm}.
Note that \Eq{Vtd_over_Vts} assumes that \dms and \dmd only receive 
Standard Model contributions.
% An alternative approach would be to 
% take $V_{td}/V_{ts}$ from global fits to predict $\dmd/\dms$, and then 
% compare the prediction with the measurement
% of \Eq{dmd_over_dms} to set limits on 
% new physics effects. 

%------------------------------------------------
\mysubsubsection{\CP violation in \Bd and \Bs mixing}
%------------------------------------------------
\labs{qpd} \labs{qps}

Evidence for \CP violation in \Bd mixing
%, which is predicted to be very small in the Standard Model,
has been searched for,
both with flavour-specific and inclusive \Bd decays, 
in samples where the initial 
flavour state is tagged. In the case of semileptonic 
(or other flavour-specific) decays, 
where the final state tag is 
also available, the asymmetry
\begin{equation} 
\ASLd = \frac{
N(\hbox{\Bdbar}(t) \to \ell^+      \nu_{\ell} X) -
N(\hbox{\Bd}(t)    \to \ell^- \bar{\nu}_{\ell} X) }{
N(\hbox{\Bdbar}(t) \to \ell^+      \nu_{\ell} X) +
N(\hbox{\Bd}(t)    \to \ell^- \bar{\nu}_{\ell} X) } 
% = \frac{|p_{\particle{d}}/q_{\particle{d}}|^2 - |q_{\particle{d}}/p_{\particle{d}}|^2}{|p_{\particle{d}}/q_{\particle{d}}|^2 - |q_{\particle{d}}/p_{\particle{d}}|^2}
% \simeq 1 - |q_{\particle{d}}/p_{\particle{d}}|^2
\labe{ASLd}
\end{equation} 
has been measured, either in decay-time-integrated analyses at 
CLEO~\citehistory{Behrens:2000qu,Jaffe:2001hz}{Behrens:2000qu,Jaffe:2001hz,*Jaffe:2001hz_hist},
\babar~\citehistory{Lees:2014kep}{Lees:2014kep,*Lees:2014kep_hist},
CDF~\unpublished{\cite{Abe:1996zt}}{\cite{Abe:1996zt,CDFnote9015:2007}}
and \dzero~\citehistory{Abazov:2013uma}{Abazov:2013uma,*Abazov:2011yk_hist,*Abazov:2010hv_hist,*Abazov:2010hj_hist,*Abazov:dimuon_hist},
or in decay-time-dependent analyses at 
OPAL~\cite{Ackerstaff:1997vd}, ALEPH~\cite{Barate:2000uk}, 
\babar~\unpublished{%
\citehistory%
{Aubert:2003hd,*Aubert:2004xga,Lees:2013sua,Aubert:2006nf}%
{Aubert:2003hd,*Aubert:2004xga,Lees:2013sua,Aubert:2006nf,*Aubert:2002mn_hist}}{%
\citehistory%
{Aubert:2003hd,*Aubert:2004xga,Lees:2013sua,Aubert:2006nf}%
{Aubert:2003hd,*Aubert:2004xga,Lees:2013sua,*Margoni:2013qx_hist,*Aubert:2006sa_hist,Aubert:2006nf,*Aubert:2002mn_hist}}
and \belle~\cite{Nakano:2005jb}.
Note that the asymmetry of time-dependent decay rates in \Eq{ASLd} is 
related to  $|q_d/p_d|$ through \Eq{ASLq} and is therefore time-independent.
In the inclusive case, also investigated and published
% at LEP~\cite{DELPHIconf:1997,Barate:2000uk,Abbiendi:1998av},
by ALEPH~\cite{Barate:2000uk} and OPAL~\cite{Abbiendi:1998av},
no final state tag is used, and the asymmetry~\cite{Beneke:1996hv,*Dunietz:1998av}
\begin{equation} 
\frac{
N(\hbox{\Bd}(t) \to {\rm all}) -
N(\hbox{\Bdbar}(t) \to {\rm all}) }{
N(\hbox{\Bd}(t) \to {\rm all}) +
N(\hbox{\Bdbar}(t) \to {\rm all}) } 
\simeq
\ASLd \left[ \frac{\dmd}{2\Gd} \sin(\dmd \,t) - 
\sin^2\left(\frac{\dmd \,t}{2}\right)\right] 
\labe{ASLincl}
\end{equation} 
must be measured as a function of the proper time to extract information 
on \CP violation.

On the other hand, \dzero~\cite{Abazov:2012hha} and
LHCb~\cite{Aaij:2014nxa} have studied the time-dependence of the 
charge asymmetry of $B^0 \to D^{(*)-}\mu^+\nu_{\mu}X$ decays
without tagging the initial state,
which would be equal to 
\begin{equation} 
%TimGershon% \frac{N(\hbox{\Bd}(t) \to D^{(*)-}\mu^+\nu_{\mu}X)-N(\hbox{\Bdbar}(t) \to D^{(*)+}\mu^-\bar{\nu}_{\mu}X)}%
%TimGershon% {N(\hbox{\Bd}(t) \to D^{(*)-}\mu^+\nu_{\mu}X)+N(\hbox{\Bdbar}(t) \to D^{(*)+}\mu^-\bar{\nu}_{\mu}X)} =
% The above two lines were introduced instead of the two lines below, but I think this was wrong (\ie\ the two lines below are correct, since the initial state is untagged):
\frac{N(D^{(*)-}\mu^+\nu_{\mu}X)-N(D^{(*)+}\mu^-\bar{\nu}_{\mu}X)}%
{N(D^{(*)-}\mu^+\nu_{\mu}X)+N(D^{(*)+}\mu^-\bar{\nu}_{\mu}X)} =
%\ASLd \left[ 1- \cos(\dmd \,t)\right]
\ASLd \frac{1- \cos(\dmd \,t)}{2}
\label{eq:untagged_ASL}
\end{equation}
in absence of detection and production asymmetries.

\Table{qoverp} summarizes the different measurements%
\footnote{
\label{foot:life_mix:Abe:1996zt}
A low-statistics result published by CDF using the Run~I data~\cite{Abe:1996zt}
\unpublished{is}{and an unpublished result by CDF using Run~II data~\cite{CDFnote9015:2007} are}
not included in our averages, nor in \Table{qoverp}.
}
of \ASLd and $|q_{\particle{d}}/p_{\particle{d}}|$: 
in all cases asymmetries compatible with zero have been found,  
with a precision limited by the available statistics. 
\begin{table}
\caption{Measurements\footref{foot:life_mix:Abe:1996zt}
%\addtocounter{footnote}{0}\protect\footnotemark\addtocounter{footnote}{-1}
of \CP violation in \Bd mixing and their average
in terms of both \ASLd and $|q_{\particle{d}}/p_{\particle{d}}|$.
The individual results are listed as quoted in the original publications, 
or converted\footref{foot:life_mix:epsilon_B}
%\addtocounter{footnote}{4}\protect\footnotemark\addtocounter{footnote}{-5}
to an \ASLd value.
% (except in the case of CDF2, where 
% the quoted value of \ASLd has been derived from the original measurement 
% assuming that $\ASLs = 0$). 
When two errors are quoted, the first one is statistical and the 
second one systematic. The ALEPH and OPAL % and CDF2
results assume no \CP violation in \Bs mixing.}
% \ie\ $|q_{\particle{s}}/p_{\particle{s}}|=1$.}
\labt{qoverp}
\begin{center}
\resizebox{\textwidth}{!}{
\begin{tabular}{@{}rcl@{$\,\pm$}l@{$\pm$}ll@{$\,\pm$}l@{$\pm$}l@{}}
\hline
%Experiment & {Method} & \multicolumn{3}{c}{\ASLd} 
%                      & \multicolumn{3}{c}{$|q_{\particle{d}}/p_{\particle{d}}|$} \\
%Exp.\ \& Ref. & Method & ~~~\ASLd & stat & syst
%                       & $|q_{\particle{d}}/p_{\particle{d}}|$ & stat & syst \\
Exp.\ \& Ref. & Method & \multicolumn{3}{c}{Measured \ASLd} 
                       & \multicolumn{3}{c}{Measured $|q_{\particle{d}}/p_{\particle{d}}|$} \\
\hline
% CLEO   \cite{Behrens:2000qu} & $D^{*\pm}\pi^{\mp}$, $D^{*\pm}\rho^{\mp}$ (part.\ rec.) 
CLEO   \cite{Behrens:2000qu} & Partial hadronic rec. 
                             & $+0.017$ & 0.070 & 0.014 
                             & \multicolumn{3}{c}{} \\
CLEO   \citehistory{Jaffe:2001hz}{Jaffe:2001hz,*Jaffe:2001hz_hist}   & Dileptons 
                             & $+0.013$ & 0.050 & 0.005 
                             & \multicolumn{3}{c}{} \\
CLEO   \citehistory{Jaffe:2001hz}{Jaffe:2001hz,*Jaffe:2001hz_hist}   & Average of above two 
                             & $+0.014$ & 0.041 & 0.006 
                             & \multicolumn{3}{c}{} \\
\babar \cite{Aubert:2003hd,*Aubert:2004xga}   & Full hadronic rec. 
                             & \multicolumn{3}{c}{}  
                             & 1.029 & 0.013 & 0.011  \\
\babar \unpublished{\cite{Lees:2013sua}}{\citehistory{Lees:2013sua}{Lees:2013sua,*Margoni:2013qx_hist,*Aubert:2006sa_hist}} & Part.\ rec.\ $D^{*}X\ell\nu$ 
                             & $+0.0006$ & \multicolumn{2}{@{}l}{$0.0017 ^{+0.0038}_{-0.0032}$} 
                             & $0.99971$ & $0.00084$ & $0.00175$ \\ 
% \babar \cite{Aubert:2002mn}  & dileptons                 % superseded by \citehistory{Aubert:2006nf}{Aubert:2006nf,*Aubert:2002mn_hist}
%                              & $+0.005$ & 0.012 & 0.014  % superseded by \citehistory{Aubert:2006nf}{Aubert:2006nf,*Aubert:2002mn_hist}
%                              & 0.998 & 0.006 & 0.007 \\  % superseded by \citehistory{Aubert:2006nf}{Aubert:2006nf,*Aubert:2002mn_hist}
% \babar \citehistory{Aubert:2006nf}{Aubert:2006nf,*Aubert:2002mn_hist}  & dileptons % superseded by \citehistory{Lees:2014kep}{Lees:2014kep,*Lees:2014kep_hist}
%                              & \multicolumn{3}{c}{} % superseded by \citehistory{Lees:2014kep}{Lees:2014kep,*Lees:2014kep_hist}
%                              & 0.9992 & 0.0027 & 0.0019 \\  % superseded by \citehistory{Lees:2014kep}{Lees:2014kep,*Lees:2014kep_hist}
\babar \citehistory{Lees:2014kep}{Lees:2014kep,*Lees:2014kep_hist}  & Dileptons
                             & $-0.0039$ & 0.0035 & 0.0019 
                             & \multicolumn{3}{c}{} \\
\belle \cite{Nakano:2005jb}  & Dileptons 
                             & $-0.0011$ & 0.0079 & 0.0085 
                             & 1.0005 & 0.0040 & 0.0043 \\
%\hline
\multicolumn{2}{l}{Average of above 6 \B-factory results} & \multicolumn{3}{l}{\hflavASLDB\ (tot)} 
                             & \multicolumn{3}{l}{\hflavQPDB\  (tot)} \\ 
\hline
\dzero \cite{Abazov:2012hha} & $B^0 \to D^{(*)-}\mu^+\nu X$
                            & $+0.0068$ & 0.0045 & 0.0014 & \multicolumn{3}{c}{} \\
LHCb \cite{Aaij:2014nxa} & $B^0 \to D^{(*)-}\mu^+\nu X$
                            & $-0.0002$ & 0.0019 & 0.0030 & \multicolumn{3}{c}{} \\
\multicolumn{2}{l}{Average of above 8 pure $B^0$ results} & \multicolumn{3}{l}{\hflavASLDD\ (tot)}
                             & \multicolumn{3}{l}{\hflavQPDD\  (tot)} \\
\hline
\dzero  \citehistory{Abazov:2013uma}{Abazov:2013uma,*Abazov:2011yk_hist,*Abazov:2010hv_hist,*Abazov:2010hj_hist,*Abazov:dimuon_hist}  & Muons  \& dimuons
                             & $-0.0062$ & \multicolumn{2}{@{\hspace{0.26em}}l}{0.0043 (tot)}
                             & \multicolumn{3}{c}{} \\
%\hline
\multicolumn{2}{l}{Average of above 9 direct measurements} & \multicolumn{3}{l}{\hflavASLDW\ (tot)} 
                             & \multicolumn{3}{l}{\hflavQPDW\  (tot)} \\ 
\hline
OPAL   \cite{Ackerstaff:1997vd}   & Leptons     
                             & $+0.008$ & 0.028 & 0.012 
                             & \multicolumn{3}{c}{} \\
OPAL   \cite{Abbiendi:1998av}   & Inclusive (\Eq{ASLincl}) 
                             & $+0.005$ & 0.055 & 0.013 
                             & \multicolumn{3}{c}{} \\
ALEPH  \cite{Barate:2000uk}       & Leptons 
                             & $-0.037$ & 0.032 & 0.007 
                             & \multicolumn{3}{c}{} \\
ALEPH  \cite{Barate:2000uk}       & Inclusive (\Eq{ASLincl}) 
                             & $+0.016$ & 0.034 & 0.009 
                             & \multicolumn{3}{c}{} \\
ALEPH  \cite{Barate:2000uk}       & Average of above two 
                             & $-0.013$ & \multicolumn{2}{@{\hspace{0.26em}}l}{0.026 (tot)} 
                             & \multicolumn{3}{c}{} \\
%CDF2    \cite{CDFnote9015:2007}$^p$ & dimuons  
%                             & $+0.0136$ & 0.0151 & 0.0115
%                             & \multicolumn{3}{c}{} \\
\multicolumn{2}{l}{Average of above 13 results} & \multicolumn{3}{l}{\hflavASLDA\ (tot)} 
                             & \multicolumn{3}{l}{\hflavQPDA\  (tot)} \\ 
\hline
\multicolumn{5}{l}{Best fit value from 2D combination of} \\
\multicolumn{2}{l}{\ASLd and \ASLs results (see \Eq{ASLD})} & \multicolumn{3}{l}{\hflavASLD\ (tot)} 
                             & \multicolumn{3}{l}{\hflavQPD\  (tot)} \\ 
\hline
% \multicolumn{8}{l}{$^p$ {\footnotesize Preliminary.}}
\end{tabular}
}
\end{center}
\end{table}
A simple average of all measurements performed at the
%pre-2012% \B factories~\citehistory%
%{Behrens:2000qu,Jaffe:2001hz,Aubert:2003hd,*Aubert:2004xga,Aubert:2006nf,Aubert:2006sa,Nakano:2005jb}%
%{Behrens:2000qu,Jaffe:2001hz,*Jaffe:2001hz_hist,Aubert:2003hd,*Aubert:2004xga,Aubert:2006nf,*Aubert:2002mn_hist,Aubert:2006sa,Nakano:2005jb}
\B factories~\unpublished%
{\citehistory%
{Behrens:2000qu,Jaffe:2001hz,Aubert:2003hd,*Aubert:2004xga,Lees:2013sua,Lees:2014kep,Nakano:2005jb}%
{Behrens:2000qu,Jaffe:2001hz,*Jaffe:2001hz_hist,Aubert:2003hd,*Aubert:2004xga,Lees:2013sua,Lees:2014kep,*Lees:2014kep_hist,Nakano:2005jb}}%
{\citehistory%
{Behrens:2000qu,Jaffe:2001hz,Aubert:2003hd,*Aubert:2004xga,Lees:2013sua,Lees:2014kep,Nakano:2005jb}%
{Behrens:2000qu,Jaffe:2001hz,*Jaffe:2001hz_hist,Aubert:2003hd,*Aubert:2004xga,Lees:2013sua,*Margoni:2013qx_hist,*Aubert:2006sa_hist,Lees:2014kep,*Lees:2014kep_hist,Nakano:2005jb}}
yields
%2015% \begin{equation}
%2015% \ASLd = \hflavASLDB  ~~~ \Longleftrightarrow ~~~ |q_{\particle{d}}/p_{\particle{d}}| = \hflavQPDB \,
%2015% \labe{ASLDB}
%2015% \end{equation}
%2015% where the relation between \ASLd and $|q_{\particle{d}}/p_{\particle{d}}|$ is given in \Eq{ASLq}.
$\ASLd = \hflavASLDB$.
Adding also the \dzero~\cite{Abazov:2012hha}
and LHCb~\cite{Aaij:2014nxa} measurements obtained with reconstructed 
semileptonic \Bd decays yields $\ASLd = \hflavASLDD$.
%%%\begin{equation}
%%%\ASLd = \hflavASLDD  ~~~ \Longleftrightarrow ~~~ |q_{\particle{d}}/p_{\particle{d}}| = \hflavQPDD \,,
%%%\labe{ASLDD}
%%%\end{equation}
%%%where the relation between \ASLd and $|q_{\particle{d}}/p_{\particle{d}}|$ is given in \Eq{ASLq}.
As discussed in more detail later in this section, 
the \dzero analysis with single muons and like-sign dimuons~\citehistory{Abazov:2013uma}{Abazov:2013uma,*Abazov:2011yk_hist,*Abazov:2010hv_hist,*Abazov:2010hj_hist,*Abazov:dimuon_hist}
separates the \Bd and \Bs contributions by exploiting the dependence on the muon impact parameter cut; including the 
\ASLd result quoted by \dzero in the average yields
%%% \begin{equation}
%%%\ASLd = \hflavASLDW ~~~ \Longleftrightarrow ~~~ |q_{\particle{d}}/p_{\particle{d}}| = \hflavQPDW \,.
%%% \labe{ASLDW}
%%% \end{equation}
$\ASLd = \hflavASLDW$. % and $|q_{\particle{d}}/p_{\particle{d}}| = \hflavQPDW$.
All the other \Bd analyses performed at high energy, either at LEP or at the Tevatron,
did not separate the contributions from the \Bd and \Bs mesons.
Under the assumption of no \CP violation in \Bs mixing ($\ASLs =0$),
a number of these early analyses~\cite{Abazov:2006qw,Ackerstaff:1997vd,Barate:2000uk,Abbiendi:1998av}
quote a measurement of $\ASLd$ or $|q_{\particle{d}}/p_{\particle{d}}|$ for the \Bd meson.
However, these imprecise determinations no longer improve the world average of 
\ASLd.
The latter assumption makes sense within the Standard Model, 
since \ASLs is predicted to be much smaller than \ASLd~\citehistory{Jubb:2016mvq,Artuso:2015swg}{Jubb:2016mvq,Artuso:2015swg,*Lenz_hist},
but may not be suitable in the presence of new physics. 

The Tevatron experiments
have measured linear combinations of 
\ASLd and \ASLs
% (or equivalently $|q_{\particle{d}}/p_{\particle{d}}|$ and $|q_{\particle{s}}/p_{\particle{s}}|$)
using inclusive semileptonic decays of \b hadrons, 
$\ASLb = +0.0015 \pm 0.0038 \mbox{(stat)} \pm 0.0020 \mbox{(syst)}$~\cite{Abe:1996zt}
and $\ASLb = -0.00496 \pm 0.00153 \mbox{(stat)} \pm 0.00072 \mbox{(syst)}$~\citehistory{Abazov:2013uma}{Abazov:2013uma,*Abazov:2011yk_hist,*Abazov:2010hv_hist,*Abazov:2010hj_hist,*Abazov:dimuon_hist}, at CDF1 and \dzero respectively.
While the imprecise CDF1 result is compatible with no \CP violation%
\unpublished{}{\footnote{
  \label{foot:life_mix:CDFnote9015:2007}
  An unpublished measurement from CDF2, 
  $\ASLb = +0.0080 \pm 0.0090 \mbox{(stat)} \pm 0.0068 \mbox{(syst)}$~\cite{CDFnote9015:2007},
  more precise than the \dzero measurement,
  is also compatible with no \CP violation.
% but since it is unpublished since 2007 
% we no longer include it in our averages, nor in \Fig{ASLs}.
}},
the \dzero result, obtained by measuring
the single muon and like-sign dimuon charge asymmetries,
differs by 2.8 standard
deviations from the Standard Model expectation of
%\begin{eqnarray}
%\mbox{\hspace{3cm}}
${\cal A}_{\rm SL}^{b,\rm SM} = (-2.3\pm 0.4) \times 10^{-4}$%
~\citehistory{Abazov:2013uma,Lenz:2011ti,*Lenz:2006hd}{Abazov:2013uma,*Abazov:2011yk_hist,*Abazov:2010hv_hist,*Abazov:2010hj_hist,*Abazov:dimuon_hist,Lenz:2011ti,*Lenz:2006hd}
%&~~& \mbox{\citehistory{Abazov:2013uma,Lenz:2011ti,*Lenz:2006hd}{Abazov:2013uma,*Abazov:2011yk_hist,*Abazov:2010hv_hist,*Abazov:2010hj_hist,*Abazov:dimuon_hist,Lenz:2011ti,*Lenz:2006hd}}\,.
%\end{eqnarray}
With a more sophisticated analysis in bins of the
muon impact parameters, \dzero conclude that the overall deviation of 
their measurements from the SM is at the level of $3.6\,\sigma$.
% As mentioned above, the \dzero
% single muon and like-sign dimuon analysis investigates 
% the dependence of the charge asymmetries
% as a function of the muon impact parameters. 
Interpreting the observed asymmetries
in bins of the muon impact parameters
in terms of \CP violation in $B$-meson mixing and interference, 
and using 
the mixing parameters and the world \b-hadron fractions 
of \Ref{Amhis:2012bh}, the \dzero collaboration
extracts~\citehistory{Abazov:2013uma}{Abazov:2013uma,*Abazov:2011yk_hist,*Abazov:2010hv_hist,*Abazov:2010hj_hist,*Abazov:dimuon_hist}
values for \ASLd and \ASLs and their correlation
coefficient\footnote{
\label{foot:life_mix:Abazov:2013uma}
In each impact parameter bin $i$ the measured same-sign dimuon asymmetry is interpreted as  
$A_i = K^s_i \ASLs + K^d_i \ASLd + \lambda K^{\rm int}_i \DGd/\Gd$, where the factors  $K^s_i$, $K^d_i$ and $K^{\rm int_i}$ are obtained by \dzero from Monte Carlo simulation. The \dzero publication~\citehistory{Abazov:2013uma}{Abazov:2013uma,*Abazov:2011yk_hist,*Abazov:2010hv_hist,*Abazov:2010hj_hist,*Abazov:dimuon_hist} assumes $\lambda=1$, but it has been demonstrated 
subsequently that $\lambda \le 0.49$~\cite{Nierste_CKM2014}. This particular point invalidates the $\DGd/\Gd$ result published by \dzero, but not the \ASLd and \ASLs results. As stated by \dzero, their \ASLd and \ASLs results assume the above expression for $A_i$, \ie\ that the observed asymmetries are due to \CP violation in $B$ mixing. As long as this assumption is not shown to be wrong (or withdrawn by \dzero), we include the \ASLd and \ASLs results in our world average.},
%They also extract at the same time a value for \DGGd (see \Sec{DGd}).}, 
as shown in \Table{ASLs_ASLd}.
However, the various 
contributions to the total quoted errors from this analysis and from the
external inputs are not given, so the adjustment of these results to different
or more recent values of the external inputs cannot (easily) be done. 

Finally, direct determinations of \ASLs,
% and hence $|q_{\particle{s}}/p_{\particle{s}}|$ 
also shown in \Table{ASLs_ASLd},
are obtained by \dzero~\citehistory{Abazov:2012zz}{Abazov:2012zz,*Abazov:2009wg_hist,*Abazov:2007nw_hist}
and LHCb~\citehistory{Aaij:2016yze}{Aaij:2016yze,*Aaij:2013gta_hist}
from the time-integrated charge asymmetry of
untagged $\Bs \to D_s^- \mu^+\nu X$ decays.

Using a two-dimensional fit, all measurements of \ASLs and \ASLd obtained by 
\dzero and LHCb are combined with the 
\B-factory average of \Table{qoverp}. Correlations are taken into account as 
shown in \Table{ASLs_ASLd}.
The results, displayed graphically in \Fig{ASLs_ASLd}, are
\begin{table}
\caption{Measurements of \CP violation in \Bs and \Bd mixing, together 
with their correlations $\rho(\ASLs,\ASLd)$
and their two-dimensional average. Only total errors are quoted.}
\labt{ASLs_ASLd}
\begin{center}
\begin{tabular}{rcccl}
\hline
Exp.\ \& Ref.\ & Method & Measured \ASLs & Measured \ASLd & $\rho(\ASLs,\ASLd)$ \\
\hline
\multicolumn{2}{l}{\B-factory average  of \Table{qoverp}}
       & & \hflavASLDB & \\ 
\dzero  \citehistory{Abazov:2012zz,Abazov:2012hha}{Abazov:2012zz,*Abazov:2009wg_hist,*Abazov:2007nw_hist,Abazov:2012hha} & $B_{(s)}^0 \to D_{(s)}^{(*)-} \mu^+\nu X$
       & $-0.0112 \pm 0.0076$ % ASLs = -0.0112 +-0.0074(stat) +-0.0017(syst)
       & $+0.0068 \pm 0.0047$ % ASLd = +0.0068 +-0.0045(stat) +-0.0014(syst) 
       & ~~~$+0.$ \\
LHCb \citehistory{Aaij:2016yze,Aaij:2014nxa}{Aaij:2016yze,*Aaij:2013gta_hist,Aaij:2014nxa} & $B_{(s)}^0 \to D_{(s)}^{(*)-} \mu^+\nu X$
       & $+0.0039 \pm 0.0033$ % ASLs = +0.0039 +-0.0026(stat) +-0.0020(syst)
       & $-0.0002 \pm 0.0036$ % ASLd = -0.0002 +-0.0019(stat) +-0.0030(syst) 
       & ~~~$+0.13$ \\
\hline
\multicolumn{2}{l}{Average of above}
       & \hflavASLSNOMU & \hflavASLDNOMU & ~~~\hflavRHOASLSASLDNOMU \\ 
\dzero  \citehistory{Abazov:2013uma}{Abazov:2013uma,*Abazov:2011yk_hist,*Abazov:2010hv_hist,*Abazov:2010hj_hist,*Abazov:dimuon_hist}  & muons \& dimuons
       & $-0.0082 \pm 0.0099$ % ASLs
       & $-0.0062 \pm 0.0043$ % ASLd
       & ~~~$-0.61$ \\          % rho
\hline
\multicolumn{2}{l}{Average of all above}
       & \hflavASLS & \hflavASLD & ~~~\hflavRHOASLSASLD \\ 
\hline
%\multicolumn{5}{l}{$^p$ {\footnotesize Preliminary.}}
\end{tabular}
\end{center}
\end{table}
\begin{figure}
\begin{center}
\includegraphics[width=0.6\textwidth]{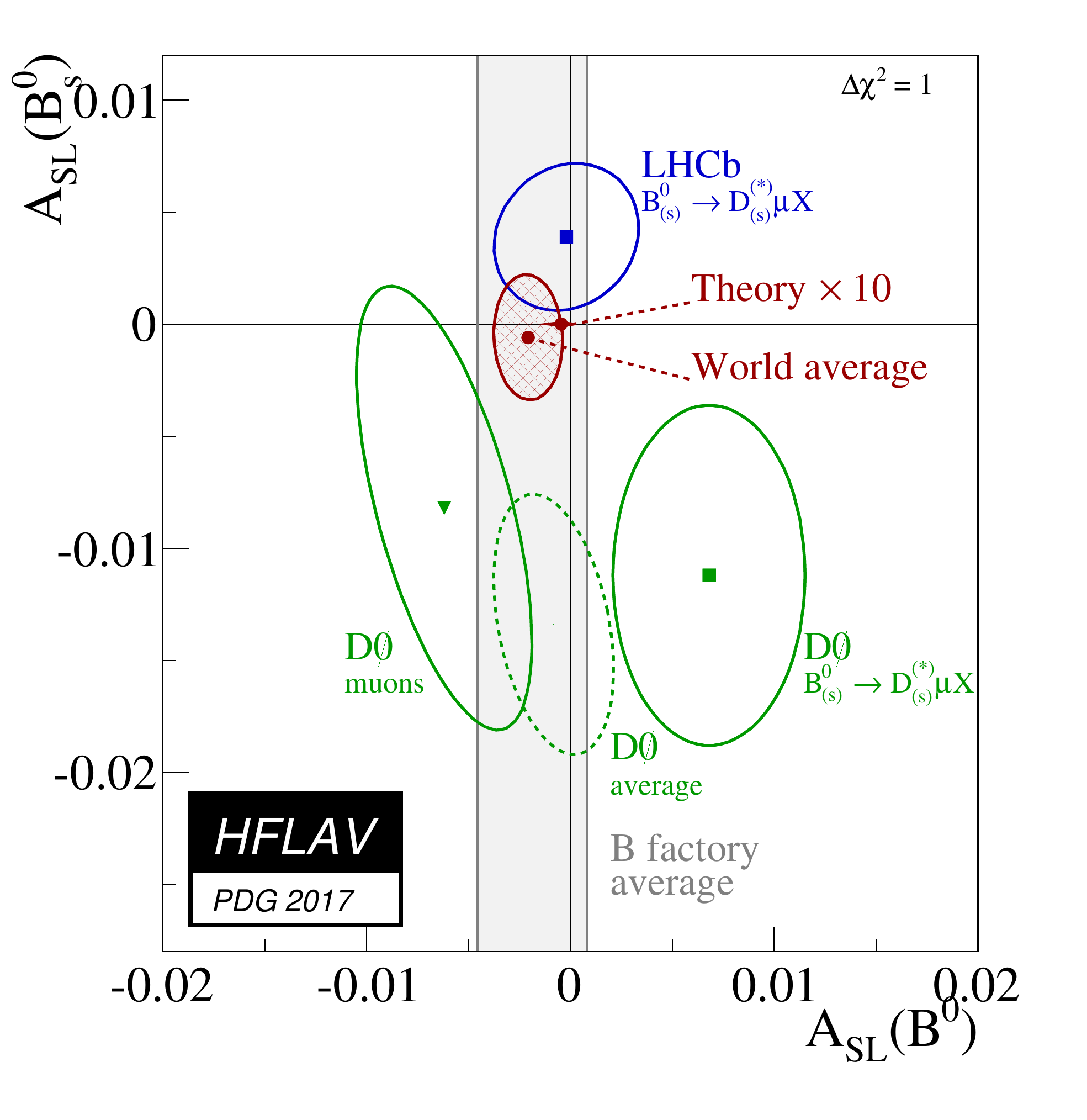}
\end{center}
\vspace{-5mm}
\caption{
Measurements of \ASLs and \ASLd listed in \Table{ASLs_ASLd}
(\B-factory average as the grey band, \dzero measurements as the green ellipses, LHCb measurements as the blue ellipse) 
together with their two-dimensional average (red hatched ellipse).
%%% Direct measurements of \ASLs and \ASLd listed in \Table{ASLs_ASLd}
%%% (\B-factory average as the vertical blue dotted band, \dzero measurements as the horizontal green dotted band and as the 
%%% green ellipse), together with their two-dimensional average (red hatched ellipse). 
The red point close to $(0,0)$ is the Standard Model prediction
of Ref.~\protect\citehistory{Jubb:2016mvq,Artuso:2015swg}{Jubb:2016mvq,Artuso:2015swg,*Lenz_hist} with error bars multiplied by 10.
The prediction and the experimental world average deviate from each other by $\hflavASLDASLSNSIGMA\,\sigma$.}
\labf{ASLs_ASLd}
\end{figure}
\begin{eqnarray}
\ASLd & = & \hflavASLD ~~~ \Longleftrightarrow ~~~ |q_{\particle{d}}/p_{\particle{d}}| = \hflavQPD \,,
\labe{ASLD}
\\
\ASLs & = & \hflavASLS ~~~ \Longleftrightarrow ~~~ |q_{\particle{s}}/p_{\particle{s}}| = \hflavQPS \,,
\labe{ASLS}
\\
\rho(\ASLd , \ASLs) & = & \hflavRHOASLSASLD \,,
\labe{rhoASLDASLS}
\end{eqnarray}
%%% and has a correlation of \hflavRHOASLSASLD with the \ASLd value already quoted in \Eq{ASLD}.
%2015% and represent our best averages of the available data.\footnote{
where the relation between ${\cal A}_{\rm SL}^q$ and $|q_{\particle{q}}/p_{\particle{q}}|$ is given in \Eq{ASLq}.%
\footnote{
  \label{foot:life_mix:epsilon_B}
  Early analyses and % (perhaps hence)
  the PDG use the complex
  parameter $\epsilon_{\B} = (p_q-q_q)/(p_q+q_q)$ for the \Bd; if \CP violation in the mixing is small,
  $\ASLd \cong 4 {\rm Re}(\epsilon_{\B})/(1+|\epsilon_{\B}|^2)$ and the 
%2015% averages of \Eqss{ASLDB}{ASLD} % \Eqsss{ASLDB}{ASLDW}{ASLD} correspond to
  average of \Eq{ASLD} corresponds to
  ${\rm Re}(\epsilon_{\B})/(1+|\epsilon_{\B}|^2)= \hflavREBD$.
% \hflavREBDB$ and %, $\hflavREBDW$ % , respectively.
}
However, the fit $\chi^2$ probability % confidence level of the fit
is only $\hflavCLPERCENTASLSASLD\%$.
This is mostly due to an overall discrepancy between the \dzero and 
LHCb averages at the level of $\hflavASLLHCBDZERONSIGMA\,\sigma$.
Since the assumptions underlying the inclusion of the \dzero muon results 
in the average\footref{foot:life_mix:Abazov:2013uma}
are somewhat controversial~\cite{Lenz_private_communication}, 
we also provide in  \Table{ASLs_ASLd} an average excluding these results.

The above averages show no evidence of \CP violation in \Bd or \Bs mixing.
They deviate by $\hflavASLDASLSNSIGMA\,\sigma$ from the very small predictions of the Standard Model (SM), 
${\cal A}_{\rm SL}^{d,\rm SM} = -(4.7\pm 0.6)\times 10^{-4}$ and 
${\cal A}_{\rm SL}^{s,\rm SM} = +(2.22\pm 0.27)\times 10^{-5}$~\citehistory{Jubb:2016mvq,Artuso:2015swg}{Jubb:2016mvq,Artuso:2015swg,*Lenz_hist}.
Given the current size of the experimental uncertainties,
there is still significant room for a possible new physics contribution, in particular in the \Bs system. 
In this respect, the deviation of the \dzero dimuon
asymmetry~\citehistory{Abazov:2013uma}{Abazov:2013uma,*Abazov:2011yk_hist,*Abazov:2010hv_hist,*Abazov:2010hj_hist,*Abazov:dimuon_hist}
from expectation has generated a lot of excitement.
However, the recent \ASLs and \ASLd results from LHCb 
are not precise enough yet to settle the issue.
It was pointed out~\cite{DescotesGenon:2012kr}
that the \dzero dimuon result can be reconciled with the SM expectations
of \ASLs and \ASLd if there were non-SM sources of \CP violation
in the semileptonic decays of the $b$ and $c$ quarks. 
A recent Run~1 ATLAS study~\cite{Aaboud:2016bmk} of charge asymmetries
in muon+jets $t\bar{t}$ events, % in 20.3\invfb of Run 1 data,
in which a \b-hadron decays semileptonically to a soft muon,
yields results with limited statistical precision, 
compatible both with the D0 dimuon asymmetry and with the SM predictions. 
More experimental data, especially from Run~2 of LHC, is awaited eagerly.

At the more fundamental level, \CP violation in \Bs
mixing is caused by the weak phase difference $\phi^s_{12}$ defined in \Eq{phi12}.
% \begin{equation}
% \phi_{12} = \arg \left[ -{M_{12}}/{\Gamma_{12}} \right], 
% \end{equation}
% where $M_{12}$ and $\Gamma_{12}$ are the off-diagonal
% elements of the mass and decay matrices of the $\Bs-\Bsbar$ system.
% This is related to the observed decay-width difference through the relation
% \begin{equation}
% \DGs = 2|\Gamma_{12}|\cos\phi_{12}+
% {\cal O} \left( \left|\frac{\Gamma_{12}}{M_{12}}\right|^2 \right) \,,
% \end{equation}
% where quadratic (or higher-order) terms in the small quantity
% $|\Gamma_{12}/M_{12}| \sim {\cal O}(m_b^2/m_t^2)$ can be neglected. 
The SM prediction for this phase is tiny~\citehistory{Jubb:2016mvq,Artuso:2015swg}{Jubb:2016mvq,Artuso:2015swg,*Lenz_hist},
\begin{equation}
\phi_{12}^{s,\rm SM} = \hflavPHISTWELVESM \,;
\labe{phis12SM}
\end{equation}
however, new physics in \Bs mixing could change this observed phase to
\begin{equation}
\phi^s_{12} = \phi_{12}^{s,\rm SM} + \phi_{12}^{s,\rm NP} \,.
\labe{phi12NP}
\end{equation}
% The \Bs semileptonic asymmetry can be expressed as~\cite{Beneke:2003az}
% \begin{equation}
% \ASLs = 
% \Im \left(\frac{\Gamma_{12}}{M_{12}} \right) +
% {\cal O} \left( \left|\frac{\Gamma_{12}}{M_{12}}\right|^2 \right) =
% \frac{\DGs}{\dms}\tan\phi_{12} +
% {\cal O} \left( \left|\frac{\Gamma_{12}}{M_{12}}\right|^2 \right) \,.
% \labe{ASLS_tanphi12}
% \end{equation}
Using \Eq{ALSq_tanphi2}, the current knowledge of \ASLs, \DGs and \dms, 
given in \Eqsss{ASLS}{DGs_DGsGs}{dms} respectively, yields an
experimental determination of $ \phi^s_{12}$,
\begin{equation}
\tan\phi^s_{12} = \ASLs \frac{\dms}{\DGs} = \hflavTANPHI \,,
\labe{tanphi12}
\comment{ % start python
from math import *
asls = -0.010460 ; easls = 0.006400
dms = 17.719032  ; edms = 0.042701
dgs = 0.0951919  ; edgs = 0.01362480381803716 
tanphi12 = asls*dms/dgs
etanphi12 = tanphi12*sqrt((easls/asls)**2+(edms/dms)**2+(edgs/dgs)**2)
print tanphi12, etanphi12
} % end python
\end{equation}
which represents only a very weak constraint at present.

%%% In order to better exploit the available data, a two-dimensional fit of all
%%% direct measurements of \ASLd and \ASLs can be performed. This fit, described 
%%% in \Sec{qps} and shown in \Fig{ASLs_ASLd}, gives 
%%% \begin{equation}
%%% \ASLd = \hflavASLD ~~~ \Longleftrightarrow ~~~ |q_{\particle{d}}/p_{\particle{d}}| = \hflavQPD \,,
%%% \labe{ASLD}
%%% \end{equation}
%%% which we take as our final average for \ASLd and $|q_{\particle{d}}/p_{\particle{d}}|$.
%%% All these results\footnote{Early analyses and (perhaps hence) the PDG use the complex
%%% parameter $\epsilon_{\B} = (p-q)/(p+q)$; if \CP violation in the mixing in small, 
%%% $\ASLd \cong 4 {\rm Re}(\epsilon_{\B})/(1+|\epsilon_{\B}|^2)$ and the averages of 
%%% \Eqss{ASLDD}{ASLD} % \Eqsss{ASLDD}{ASLDW}{ASLD}
%%% correspond to ${\rm Re}(\epsilon_{\B})/(1+|\epsilon_{\B}|^2)=\hflavREBDD$ %, $\hflavREBDW$
%%% and $\hflavREBD$, respectively.},
%%% summarized in \Table{qoverp},
%%% are compatible with no \CP violation in \Bd and \Bs mixing. % , an assumption we make for the rest of this section.

%------------------------------------------------
\mysubsubsection{Mixing-induced \CP violation in \Bs decays}
%------------------------------------------------

%%%%%%%%%%%%%%%%%%%%%%%%%%%%%%%%%%%%%%%%%%%%%%%%
%
% This is file life_mix_phis.tex containing
% the subsection about the phi_s
%
%%%%%%%%%%%%%%%%%%%%%%%%%%%%%%%%%%%%%%%%%%%%%%%
%

%---------------------------------------------
% \subsubsubsection{Weak phase in \Bs mixing}
%---------------------------------------------
\labs{phasebs}

\CP violation induced by $\Bs-\Bsbar$ mixing
% in $b \to c\bar{c}s$ decays
has been a field of 
very active study and fast experimental progress 
in the past few years.
%OS% Similarly to what has happened at the \B factories 
%OS% a decade ago, when the \Bd mixing-induced phase $2\beta$
%OS% was measured, the Tevatron and LHC experiments are 
%OS% now obtaining point estimates
%OS% of the \Bs mixing-induced phase \phiccbars.
The main observable is the 
\CP-violating phase \phiccbars, defined as 
the weak phase difference between
the $\Bs-\Bsbar$ mixing amplitude $M^s_{12}$
and the $b \to c\bar{c}s$ decay amplitude.

The golden mode for such studies is 
$\Bs \to \jpsi\phi$, followed by $\jpsi \to \mu^+\mu^-$ and 
$\phi\to K^+K^-$, for which a full angular 
analysis of the decay products is performed to 
separate statistically the \CP-even and \CP-odd
contributions in the final state. As already mentioned in 
\Sec{DGs},
CDF~\citehistory{Aaltonen:2012ie}{Aaltonen:2012ie,*CDF:2011af_hist,*Aaltonen:2007he_hist,*Aaltonen:2007gf_hist},
\dzero~\citehistory{Abazov:2011ry}{Abazov:2011ry,*Abazov:2008af_hist,*Abazov:2007tx_hist},
ATLAS~\citehistory{Aad:2014cqa,Aad:2016tdj}{Aad:2014cqa,*Aad:2012kba_hist,Aad:2016tdj},
CMS~\cite{Khachatryan:2015nza}
and LHCb~\citehistory{Aaij:2014zsa,Aaij:2016ohx}{Aaij:2014zsa,*Aaij:2013oba_supersede2,Aaij:2016ohx}
have used both untagged and tagged $\Bs \to \jpsi\phi$ (and more generally $\Bs \to (c\bar{c}) K^+K^-$) events 
for the measurement of \phiccbars.
LHCb~\citehistory{Aaij:2014dka}{Aaij:2014dka,*Aaij:2013oba_supersede}
has used $\Bs \to \jpsi \pi^+\pi^-$ events, 
analyzed with a full amplitude model
including several $\pi^+\pi^-$ resonances (\eg, $f_0(980)$),
although the
$\jpsi \pi^+\pi^-$ final state had already been shown
to be almost \CP pure with a \CP-odd fraction
larger than 0.977 at 95\% CL~\cite{LHCb:2012ae}. 
In addition, LHCb has used the $\Bs \to \Dsp\Dsm$ channel~\cite{Aaij:2014ywt} to measure \phiccbars.

All CDF, \dzero, ATLAS and CMS analyses provide 
two mirror solutions related by the transformation 
$(\DGs, \phiccbars) \to (-\DGs, \pi-\phiccbars)$. However, the
LHCb analysis of $\Bs \to \jpsi K^+K^-$ resolves this ambiguity and 
rules out the solution with negative \DGs~\cite{Aaij:2012eq},
a result in agreement with the Standard Model expectation.
Therefore, in what follows, we only consider the solution with $\DGs > 0$.

\begin{table}
\caption{Direct experimental measurements of \phiccbars, \DGs and \Gs using
$\Bs\to\jpsi\phi$, $\jpsi K^+K^-$, $\psi(2S)\phi$, $\jpsi\pi^+\pi^-$ and $D_s^+D_s^-$ decays.
Only the solution with $\DGs > 0$ is shown, since the two-fold ambiguity has been
resolved in \Ref{Aaij:2012eq}. The first error is due to 
statistics, the second one to systematics. The last line gives our average.}
\labt{phisDGsGs}
\begin{center}
%\begin{tabular}{l@{\,}l@{\,}l@{\,}|@{\,}l@{\,}|@{\,}l@{\,}|@{\,}l} 
\begin{tabular}{ll@{\,}rll@{\,}l} 
\hline
% Exp.\ & Mode & Dataset & \multicolumn{1}{c@{\,}|@{\,}}{\phiccbars}
%                      & \multicolumn{1}{c@{}}{\DGs (\!\!\invps)} & Ref.\ \\
Exp.\ & Mode & Dataset & \multicolumn{1}{c}{\phiccbars}
                     & \multicolumn{1}{c}{\DGs (\!\!\invps)} & Ref.\ \\
\hline
CDF    & $\jpsi\phi$ & $9.6\invfb$
       & $[-0.60,\, +0.12]$, 68\% CL & $+0.068\pm0.026\pm0.009$
       & \protect\citehistory{Aaltonen:2012ie}{Aaltonen:2012ie,*CDF:2011af_hist,*Aaltonen:2007he_hist,*Aaltonen:2007gf_hist} \\
       % syst error was +-0.007 instead of +-0.009 in CDF note 10778
       % & $0.654\pm0.008\pm0.004$ \\ % quoted in paper as tau(Bs) = 1/Gamma_s = 1.528 +-0.019 +-0.009
% 1./1.528 = 0.6544502617801047, 0.019/1.528**2 = 0.008137797757736905, 0.009//1.528**2 = 0.003854746306296428
\dzero & $\jpsi\phi$ & $8.0\invfb$
       & $-0.55^{+0.38}_{-0.36}$ & $+0.163^{+0.065}_{-0.064}$ % & $0.693^{+0.018}_{-0.017}$  \\
       & \citehistory{Abazov:2011ry}{Abazov:2011ry,*Abazov:2008af_hist,*Abazov:2007tx_hist} \\
ATLAS  & $\jpsi\phi$ & $4.9\invfb$
       & $+0.12 \pm 0.25 \pm 0.05$ & $+0.053 \pm0.021 \pm0.010$ % & $0.677 \pm0.007 \pm0.004$ \\
       & \citehistory{Aad:2014cqa}{Aad:2014cqa,*Aad:2012kba_hist} \\
ATLAS  & $\jpsi\phi$ & $14.3\invfb$
       & $-0.110 \pm 0.082 \pm 0.042$ & $+0.101 \pm0.013 \pm0.007$ % & $0.676 \pm0.004 \pm0.004 \\
       & \cite{Aad:2016tdj} \\
ATLAS  & \multicolumn{2}{r}{above 2 combined}
       & $-0.090 \pm 0.078 \pm 0.041$ & $+0.085 \pm0.011 \pm0.007$ % & $0.675 \pm0.003 \pm0.003$ \\
       & \cite{Aad:2016tdj} \\
%%% CMS    & $\jpsi\phi$ & $X.X\invfb$ % this CMS result not included in the (phi_s, DGs) averaging !
%%%        & (set to $0$) & $0.048\pm0.024\pm0.003$ % & --- \\
%%%        & \cite{CMS-PAS-BPH-11-006}$^p$ \\
CMS    & $\jpsi\phi$ & $19.7\invfb$ 
       & $-0.075 \pm 0.097 \pm 0.031$ & $+0.095\pm0.013\pm0.007$ % & $0.6704 \pm0.0043 \pm0.0051$ \\
       & \cite{Khachatryan:2015nza} \\ 
LHCb   & $\jpsi K^+K^-$ & $3.0\invfb$
       & $-0.058\pm0.049\pm0.006$ & $+0.0805\pm0.0091\pm0.0032$ % & $0.6603\pm0.0027\pm0.0015$  \\
       & \citehistory{Aaij:2014zsa}{Aaij:2014zsa,*Aaij:2013oba_supersede2} \\
LHCb   & $\jpsi\pi^+\pi^-$ & $3.0\invfb$
       & $+0.070 \pm0.068 \pm 0.008$ & --- % & --- \\
       & \citehistory{Aaij:2014dka}{Aaij:2014dka,*Aaij:2013oba_supersede} \\
LHCb   & \multicolumn{2}{r}{above 2 combined}
       & $-0.010\pm0.039(\rm tot)$ & --- % & --- \\
       & \citehistory{Aaij:2014zsa}{Aaij:2014zsa,*Aaij:2013oba_supersede2} \\
LHCb   & $\psi(2S)\phi$ & $3.0\invfb$
       & $+0.23 ^{+0.29}_{-0.28} \pm0.02$ & $+0.066 ^{+0.41}_{-0.44} \pm0.007$ % & $0.668 \pm0.011 \pm0.006$  \\
       & \cite{Aaij:2016ohx} \\
LHCb   & $D_s^+D_s^-$ & $3.0\invfb$
       & $+0.02 \pm0.17 \pm 0.02$ & --- % & --- \\
       & \cite{Aaij:2014ywt} \\
%old%LHCb   & $\jpsi KK$ & \citehistory{Aaij:2013oba}{Aaij:2013oba,*LHCb:2011aa_hist,*LHCb:2012ad_hist,*LHCb:2011ab_hist,*Aaij:2012nta_hist}
%old%       & $+0.07\pm0.09\pm0.01$ & $0.100\pm0.016\pm0.003$ & $0.663\pm0.005\pm0.006$  \\
%old%LHCb   & $\jpsi\pi\pi$ & \citehistory{Aaij:2013oba}{Aaij:2013oba,*LHCb:2011aa_hist,*LHCb:2012ad_hist,*LHCb:2011ab_hist,*Aaij:2012nta_hist}
%old%       & $-0.014 ^{+0.17}_{-0.16}\pm 0.01$ & --- & --- \\
%old%LHCb   & combined & \citehistory{Aaij:2013oba}{Aaij:2013oba,*LHCb:2011aa_hist,*LHCb:2012ad_hist,*LHCb:2011ab_hist,*Aaij:2012nta_hist}
%old%       & $+0.01\pm0.07\pm0.01$ & $0.106\pm0.011\pm0.007$ & $0.661\pm0.004\pm0.006$  \\
\hline
%\multicolumn{3}{@{}l@{\,}|@{\,}}{All combined} & \hflavPHISCOMB & \hflavDGSCOMBnounit & \\ % & \hflavGSnounit \\
\multicolumn{3}{l}{All combined} & \hflavPHISCOMB & \hflavDGSCOMBnounit & \\ % & \hflavGSnounit \\
\hline
%\multicolumn{6}{l}{$^a$ {\footnotesize LHCb combination of $\jpsi K^+K^-$~\citehistory{Aaij:2014zsa}{Aaij:2014zsa,*Aaij:2013oba_supersede2} and $\jpsi\pi^+\pi^-$~\citehistory{Aaij:2014dka}{Aaij:2014dka,*Aaij:2013oba_supersede}.}}\\[-0.8ex]
%\multicolumn{6}{l}{$^p$ {\footnotesize Preliminary.}}
\end{tabular}
\end{center}
\end{table}

\begin{figure}
\begin{center}
\includegraphics[width=0.65\textwidth]{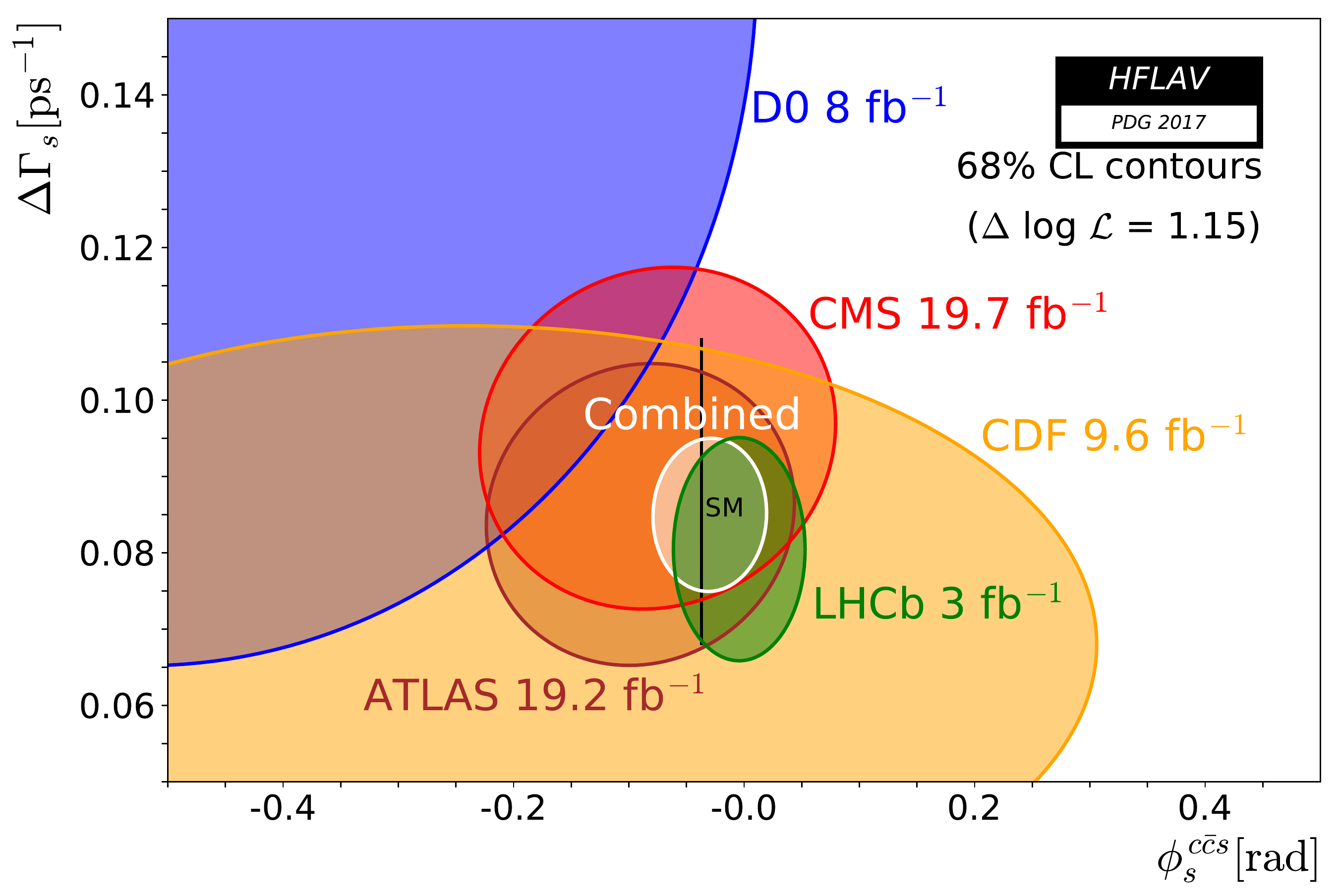}
\caption{
%2014% Left:
68\% CL regions in \Bs width difference \DGs and weak phase \phiccbars
obtained from individual and combined CDF~\protect\citehistory{Aaltonen:2012ie}{Aaltonen:2012ie,*CDF:2011af_hist,*Aaltonen:2007he_hist,*Aaltonen:2007gf_hist},
\dzero~\protect\citehistory{Abazov:2011ry}{Abazov:2011ry,*Abazov:2008af_hist,*Abazov:2007tx_hist},
ATLAS~\protect\citehistory{Aad:2014cqa,Aad:2016tdj}{Aad:2014cqa,*Aad:2012kba_hist,Aad:2016tdj}, 
CMS~\cite{Khachatryan:2015nza}
and LHCb~\protect\citehistory%
{Aaij:2014zsa,Aaij:2016ohx,Aaij:2014dka,Aaij:2014ywt}%
{Aaij:2014zsa,*Aaij:2013oba_supersede2,Aaij:2016ohx,Aaij:2014dka,*Aaij:2013oba_supersede,Aaij:2014ywt}
likelihoods of 
$\Bs\to \jpsi\phi$, $\Bs\to \jpsi K^+K^-$, $\Bs\to \psi(2S) \phi$, $\Bs\to\jpsi\pi^+\pi^-$ and 
$\Bs\to D_s^+D_s^-$ samples. 
The expectation within the Standard Model~\protect\citehistory{Charles:2011va_mod,Jubb:2016mvq,Artuso:2015swg}{Charles:2011va_mod,Jubb:2016mvq,Artuso:2015swg,*Lenz_hist}
is shown as the black rectangle.
%2014% Right: same, but zoomed on the region of interest.
}
\labf{DGs_phase}
\end{center}
\end{figure}

We perform a combination of the CDF~\citehistory{Aaltonen:2012ie}{Aaltonen:2012ie,*CDF:2011af_hist,*Aaltonen:2007he_hist,*Aaltonen:2007gf_hist},
\dzero~\citehistory{Abazov:2011ry}{Abazov:2011ry,*Abazov:2008af_hist,*Abazov:2007tx_hist},
ATLAS~\citehistory{Aad:2014cqa,Aad:2016tdj}{Aad:2014cqa,*Aad:2012kba_hist,Aad:2016tdj},
CMS~\cite{Khachatryan:2015nza}
and LHCb~\citehistory%
{Aaij:2014zsa,Aaij:2016ohx,Aaij:2014dka}%
{Aaij:2014zsa,*Aaij:2013oba_supersede2,Aaij:2016ohx,Aaij:2014dka,*Aaij:2013oba_supersede}
results summarized in \Table{phisDGsGs}.
%OL2014
%LHCb has performed an internal combination of \phis\ measured using $\Bs \to \jpsi KK$ and $\Bs \to \jpsi\pi\pi$ events~\citehistory{Aaij:2014zsa}{Aaij:2014zsa,*Aaij:2013oba_supersede2} and obtain $-0.010\pm0.040(\rm tot)$. 
% BEWARE HARD-CODED!
%We average this with the \phis\ value measured using using $\Bs \to D_s^+ D_s^-$ events~\cite{Aaij:2014ywt}. 
This is done by adding the two-dimensional log profile-likelihood scans of
\DGs and \phiccbars from all $\Bs\to\ (c\bar{c})  K^+K^-$ analyses and 
a one-dimensional log profile-likelihood of \phiccbars
from the $\Bs\to\jpsi\pi^+\pi^-$ and $\Bs \to D_s^+ D_s^-$ analyses; 
the combined likelihood is then maximized with respect to \DGs and \phiccbars.

In the $\Bs\to\jpsi\phi$ and $\Bs\to\jpsi K^+K^-$ analyses, \phiccbars and \DGs 
come from a simultaneous fit that determines also the \Bs lifetime,
the polarisation amplitudes and strong phases.
While the correlation between \phiccbars and all other parameters is small,
the correlations between \DGs and the polarisation amplitudes are sizable.
However, since the various experiments use different conventions
for the amplitudes and phases, a full combination including all
correlations is not performed. Instead, our average only takes
into account the correlation between \phiccbars and \DGs.

In the recent LHCb $\Bs \to \jpsi K^+K^-$ analysis~\citehistory{Aaij:2014zsa}{Aaij:2014zsa,*Aaij:2013oba_supersede2}, the \phiccbars values are measured for the first time for each polarisation of the final state. Since those values are compatible within each other, we still use the unique value of \phiccbars for our world average, corresponding to the one measured by the other-than-LHCb analyses. 
In the same analysis, the statistical correlation coefficient between \phiccbars and $|\lambda|$
(which signals \CP violation in the decay if different from unity) 
is measured to be very small ($-0.02$). We neglect this correlation in our average. 
Furthermore, the statistical correlation coefficient between \phiccbars and \DGs\ is measured to be small $(-0.08)$. When averaging LHCb results of 
$\Bs \to \jpsi K^+K^-$,  $\Bs \to \jpsi \pi^+\pi^-$ and $\Bs \to D_s^+ D_s^-$, we neglect this correlation coefficient (putting it to zero). 
Given the increasing experimental precision, we have also stopped using the two-dimensional $\DGs-\phiccbars$ histograms provided by the CDF and \dzero collaborations: we are now approximating those with two-dimensional Gaussian likelihoods. 

We obtain the individual and combined contours shown in \Fig{DGs_phase}. % (left).
Maximizing the likelihood, we find, as summarized in \Table{phisDGsGs}:  
\begin{eqnarray}
\DGs &=& \hflavDGSCOMB \,, \\    
\phiccbars &=& \hflavPHISCOMB \,.
\labe{phis}
\end{eqnarray}
The above \DGs average is consistent, but highly correlated with the average
of \Eq{DGs_DGsGs}. Our
final recommended average for \DGs is the one of \Eq{DGs_DGsGs}, which 
includes all available information on \DGs. 

In the Standard Model and ignoring sub-leading penguin contributions, 
\phiccbars is expected to be equal to $-2\beta_s$, 
%%% \begin{equation}
%%% (\phiccbars)^{\rm SM} = -2\beta_s \,,
%%% \end{equation}
where
%%% \begin{equation}
$\beta_s = \arg\left[-\left(V_{ts}V^*_{tb}\right)/\left(V_{cs}V^*_{cb}\right)\right]$ % \,.
%%% %= 0.036 \pm 0.002 \approx 0.04.
%%% \end{equation}
is a phase analogous to the angle $\beta$ of the usual CKM
unitarity triangle (aside from a sign change). % the negative sign.
% (resulting in a positive angle in the Standard Model).
An indirect determination via global fits to experimental data
gives~\cite{Charles:2011va_mod}
\begin{equation}
%(\phiccbars)^{\rm SM} = -2\beta_s = -0.0363^{+0.0012}_{-0.0014} \,.
(\phiccbars)^{\rm SM} = -2\beta_s = \hflavPHISSM \,.
\labe{phisSM}
\end{equation}
The average value of \phiccbars from \Eq{phis} is consistent with this
Standard Model expectation.

From its measurements of time-dependent \CP violation in $\Bs \to K^+K^-$ decays, the LHCb collaboration has determined the 
\Bs mixing phase to be $-2\beta_s = -0.12^{+0.14}_{-0.12}$~\cite{Aaij:2014xba},
assuming a U-spin relation (with up to 50\% breaking effects) between the decay amplitudes of $\Bs \to K^+K^-$ 
and $\Bd \to \pi^+\pi^-$, and a value of the CKM angle $\gamma$  
of $(70.1 \pm7.1)^{\circ}$. This determination is compatible with, 
and less precise than, the world average of \phiccbars from \Eq{phis}.

New physics could contribute to \phiccbars. Assuming that new physics only 
enters in $M^s_{12}$ (rather than in $\Gamma^s_{12}$),
one can write~\cite{Lenz:2011ti,*Lenz:2006hd}
%%% The same additional contribution due to new physics would show up in this
%%% observed phase~\cite{Lenz:2011ti,*Lenz:2006hd}, \ie\
\begin{equation}
\phiccbars = -  2\beta_s + \phi_{12}^{s,\rm NP} \,,
\end{equation}
where the new physics phase $\phi_{12}^{s,\rm NP}$ is the same as that appearing in \Eq{phi12NP}.
In this case
\begin{equation}
\phi^s_{12} = % \phi_{12}^{\rm SM}+ \phi_{12}^{\rm NP} =
\phi_{12}^{s,\rm SM} +2\beta_s + \phiccbars = \hflavPHISTWELVE \,,
\end{equation}
where the numerical estimation was performed with the values of \Eqsss{phis12SM}{phisSM}{phis}.
% and \Eq{ASLS_tanphi12} then provides a relation between \DGs and \phiccbars, 
% based on the measured values of \ASLs and \dms (\Eqss{ASLS}{dms}) 
% as well as the expectations
% for $\phi_{12}^{\rm SM}$ and $-2\beta_s$.
% The allowed region in the (\DGs, \phiccbars) plane is shown in 
% \Fig{DGs_phase_old} (right), where it is compared both with the
% direct measurement of \DGs and \phiccbars,
% and with the Standard Model expectations. 
% No inconsistency is observed between all these data.
Keeping in mind the approximation and assumption mentioned above,
this can serve as a reference value to which the measurement of \Eq{tanphi12} can be compared. 

%%% In \Fig{DGs_phase} (right), this result is compared to the 
%%% constraint provided by the world-average \Bs semileptonic asymmetry of
%%% \Eq{ASLS} and the world-average mass difference \dms of \Eq{dms} through~\cite{Beneke:2003az}:
%%% \begin{equation}
%%% \ASLs = 
%%% \Im \left(\frac{\Gamma^{12}_s}{M^{12}_s} \right) + {\cal O} \left( \frac{\Gamma_{12}}{M_{12}} \right)^2 =
%%% \frac{\DGs}{\dms}\tan\phi_{12} + {\cal O} \left( \frac{\Gamma_{12}}{M_{12}} \right)^2 \,,
%%% \end{equation}
%%% where, ignoring penguin contributions and using the Standard Model values of $\phi_{12}^{\rm SM}$ and $-2\beta_s$, we have
%%% \begin{equation}
%%% \phi_{12} = % \phi_{12}^{\rm SM}+ \phi_{12}^{\rm NP} =
%%% \phi_{12}^{\rm SM} +2\beta_s + \phiccbars \,.
%%% \end{equation}

% allowing the extraction of \phiccbars\ from \ASLs\ given the SM
% values of $\phi_{12}^{\mathrm SM}$ and $2\beta_s$. 
%
% And the result are
% citation to be update to should be  Lenz, arXiv:1102.4274
% \begin{eqnarray}
% \DGs &=& \hflavDGSCOMBCON \,, \\    
% \phiccbars =  &=& \hflavPHISCOMBCON \,.
% \end{eqnarray}

% UNDER DISCUSSION:
% In addition, we introduce the constraints due to the effective lifetime measured in $\hbox{\Bs} \to K^+K^-$~\cite{Aaij:2012kn},
% and $\hbox{\Bs} \to \jpsi f_0(980)$~\cite{Aaltonen:2011nk}, using \Eqss{tauf_fleisch,ADG}.
% We find: 
% \begin{eqnarray}
% \DGs &=&   ~~TO~BE~UPDATED \\
% \phiccbars =  &=&  ~~TO~BE~UPDATED 
% \end{eqnarray}.

\clearpage
% %% % Measurements related to the Unitarity Triangle
\mysection{Measurements related to Unitarity Triangle angles
}
\label{sec:cp_uta}

The charge of the ``$\CP(t)$ and Unitarity Triangle angles'' group is to provide averages of measurements obtained from analyses of decay-time-dependent asymmetries, and other quantities that are related to the angles of the Unitarity Triangle (UT).
In cases where considerable theoretical input is required to extract the fundamental quantities, no attempt to do so is made. 
However, straightforward interpretations of the averages are given, where possible.

In Sec.~\ref{sec:cp_uta:introduction} 
a brief introduction to the relevant phenomenology is given.
In Sec.~\ref{sec:cp_uta:notations}
an attempt is made to clarify the various different notations in use.
In Sec.~\ref{sec:cp_uta:common_inputs}
the common inputs to which experimental results are rescaled in the
averaging procedure are listed. 
We also briefly introduce the treatment of experimental errors. 
In the remainder of this section,
the experimental results and their averages are given,
divided into subsections based on the underlying quark-level decays.
All the measurements reported are quantities determined from decay-time-dependent analyses, with the exception of several in Sec.~\ref{sec:cp_uta:cus}, which are related to the UT angle $\gamma$ and are obtained from decay-time-integrated analyses.
In the compilations of measurements, indications of the sizes of the data samples used by each experiment are given.
For the $\epem$ $B$ factory experiments, this is quoted in terms of the number of $B\bar{B}$ pairs in the data sample, while the integrated luminosity is given for experiments at hadron colliders.

% \afterpage{\clearpage}
\mysubsection{Introduction
}
\label{sec:cp_uta:introduction}

The Standard Model Cabibbo-Kobayashi-Maskawa (CKM) quark mixing matrix $\VCKM$ must be unitary. 
% A $3 \times 3$ unitary matrix has four free parameters,\footnote{
%   In the general case there are nine free parameters,
%   but five of these are absorbed into unobservable quark phases.}
The CKM matrix has four free parameters
and these are conventionally written by the product
of three (complex) rotation matrices~\cite{Chau:1984fp}, 
where the rotations are characterised by the Euler mixing angles between the generations, $\theta_{12}$, $\theta_{13}$ and $\theta_{23}$, and one overall phase $\delta$,
\begin{equation}
\label{eq:ckmPdg}
\VCKM =
        \left(
          \begin{array}{ccc}
            V_{ud} & V_{us} & V_{ub} \\
            V_{cd} & V_{cs} & V_{cb} \\
            V_{td} & V_{ts} & V_{tb} \\
          \end{array}
        \right)
        =
        \left(
        \begin{array}{ccc}
        c_{12}c_{13}    
                &    s_{12}c_{13}   
                        &   s_{13}e^{-i\delta}  \\
        -s_{12}c_{23}-c_{12}s_{23}s_{13}e^{i\delta} 
                &  c_{12}c_{23}-s_{12}s_{23}s_{13}e^{i\delta} 
                        & s_{23}c_{13} \\
        s_{12}s_{23}-c_{12}c_{23}s_{13}e^{i\delta}  
                &  -c_{12}s_{23}-s_{12}c_{23}s_{13}e^{i\delta} 
                        & c_{23}c_{13} 
        \end{array}
        \right)
\end{equation}
where $c_{ij}=\cos\theta_{ij}$, $s_{ij}=\sin\theta_{ij}$ for 
$i<j=1,2,3$. 

Following the observation of a hierarchy between the different
matrix elements, the Wolfenstein parametrisation~\cite{Wolfenstein:1983yz}
is an expansion of $\VCKM$ in terms of the four real parameters $\lambda$
(the expansion parameter), $A$, $\rho$ and $\eta$. Defining to 
all orders in $\lambda$~\cite{Buras:1994ec}
\begin{eqnarray}
  \label{eq:burasdef}
  s_{12}             &\equiv& \lambda\,,\nonumber \\ 
  s_{23}             &\equiv& A\lambda^2\,, \\
  s_{13}e^{-i\delta} &\equiv& A\lambda^3(\rho -i\eta)\,,\nonumber
\end{eqnarray}
and inserting these into the representation of Eq.~(\ref{eq:ckmPdg}), 
unitarity of the CKM matrix is achieved to all orders.
A Taylor expansion of $\VCKM$ leads to the familiar approximation
\begin{equation}
  \label{eq:cp_uta:ckm}
  \VCKM
  = 
  \left(
    \begin{array}{ccc}
      1 - \lambda^2/2 & \lambda & A \lambda^3 ( \rho - i \eta ) \\
      - \lambda & 1 - \lambda^2/2 & A \lambda^2 \\
      A \lambda^3 ( 1 - \rho - i \eta ) & - A \lambda^2 & 1 \\
    \end{array}
  \right) + {\cal O}\left( \lambda^4 \right) \, .
\end{equation}
At order $\lambda^{5}$, the obtained CKM matrix in this extended
Wolfenstein parametrisation is:
{\small
  \begin{equation}
    \label{eq:cp_uta:ckm_lambda5}
    \VCKM
    =
    \left(
      \begin{array}{ccc}
        1 - \frac{1}{2}\lambda^{2} - \frac{1}{8}\lambda^4 &
        \lambda &
        A \lambda^{3} (\rho - i \eta) \\
        - \lambda + \frac{1}{2} A^2 \lambda^5 \left[ 1 - 2 (\rho + i \eta) \right] &
        1 - \frac{1}{2}\lambda^{2} - \frac{1}{8}\lambda^4 (1+4A^2) &
        A \lambda^{2} \\
        A \lambda^{3} \left[ 1 - (1-\frac{1}{2}\lambda^2)(\rho + i \eta) \right] &
%        -A \lambda^{2}(1-\frac{1}{2}\lambda^2) \left[ 1 + \lambda^{2}(\rho + i \eta) \right] &
        -A \lambda^{2} + \frac{1}{2}A\lambda^4 \left[ 1 - 2(\rho + i \eta) \right] &
        1 - \frac{1}{2}A^2 \lambda^4
      \end{array} 
    \right) + {\cal O}\left( \lambda^{6} \right)\,.
  \end{equation}
}

\vspace{-5mm}
\noindent
A non-zero value of $\eta$ implies that the CKM matrix is not purely real in this, or any, parametrisation, and is the origin of $\CP$ violation in the Standard Model.
This is encapsulated in a parametrisation-invariant way through the Jarlskog parameter $J = \Im\left(V_{us}V_{cb}V^*_{ub}V^*_{cs}\right)$~\cite{Jarlskog:1985ht}.

% By definition, the expression for $V_{ub}$ remains unchanged relative
% to the original Wolfenstein parametrization and the
% corrections to $V_{us}$ and $V_{cb}$ appear only at ${\cal O}(\lambda7)$
% and
% ${\cal O}(\lambda8)$, respectively.

%%% TJG this expression for V_{td} is only approximate
%%%     leave it out here and discuss it later
% It can be noted that the element $V_{td}$ can be re-expressed as:
% \begin{equation}
%   V_{td} = A \lambda^{3} (1-\overline{\rho} -i \overline{\eta})
% \end{equation}
% where~\cite{Buras:1994ec}
% \begin{equation}
%   \overline{\rho}=\rho (1-\frac{\lambda^2}{2}) + {\cal O}\left( \lambda^{4} \right),
%   \qquad
%   \overline{\eta}=\eta (1-\frac{\lambda^2}{2}) + {\cal O}\left( \lambda^{4} \right).
% \end{equation}
% This elegant change in $V_{td}$, 
% with respect to the original Wolfenstein parametrization,
% allows a simple generalization of the so-called unitarity triangle to
% higher orders in $\lambda$ as discussed below.

The unitarity relation $\VCKM^\dagger\VCKM = {\mathit 1}$
results in a total of nine expressions, that can be written as
$\sum_{i=u,c,t} V_{ij}^*V_{ik} = \delta_{jk}$,
where $\delta_{jk}$ is the Kronecker symbol.
Of the off-diagonal expressions ($j \neq k$),
three can be transformed into the other three 
(under $j \leftrightarrow k$, corresponding to complex conjugation).
This leaves three relations in which three complex numbers sum to zero,
which therefore can be expressed as triangles in the complex plane, together with three relations in which the squares of the elements in each column of the CKM matrix sum to unity.
Similar relations are obtained for the rows of the matrix from $\VCKM\VCKM^\dagger = {\mathit 1}$, so there are in total six triangle relations and six sums to unity.
More details about unitarity triangles can be found in Refs.~\cite{Jarlskog:2005uq,Harrison:2009bz,Frampton:2010ii,Frampton:2010uq}.

\begin{figure}[t]
  \begin{center}
    \resizebox{0.55\textwidth}{!}{\includegraphics{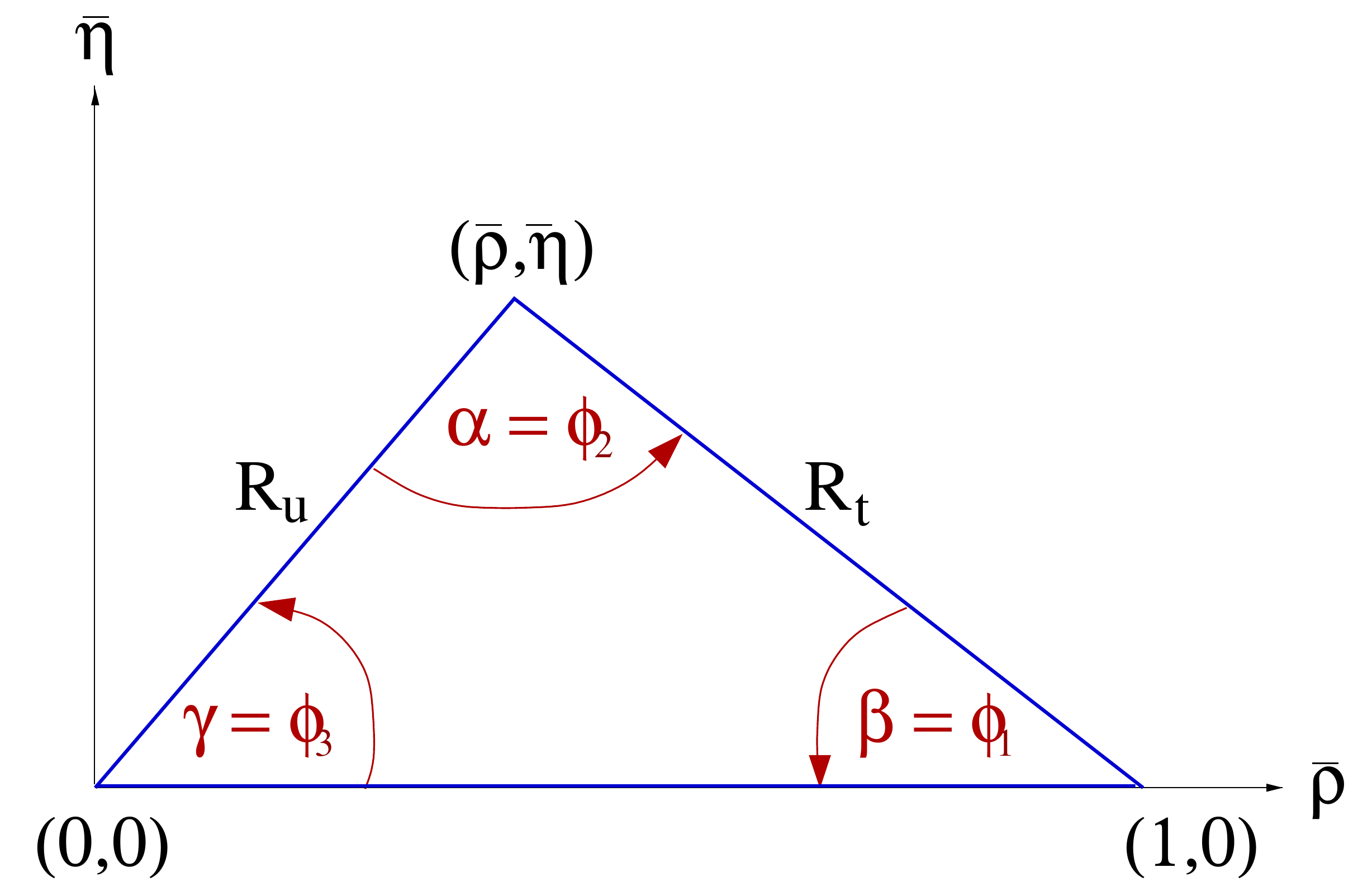}}
    \caption{The Unitarity Triangle.}
    \label{fig:cp_uta:ut}
  \end{center}
\end{figure}

One of the triangle relations,
\begin{equation}
  \label{eq:cp_uta:ut}
  V_{ud}V_{ub}^* + V_{cd}V_{cb}^* + V_{td}V_{tb}^* = 0\,,
\end{equation}
is of particular importance to the $\B$ system, 
being specifically related to flavour-changing neutral-current $b \to d$ transitions.
The three terms in Eq.~(\ref{eq:cp_uta:ut}) are of the same order,
${\cal O}\left( \lambda^3 \right)$,
and this relation is commonly known as the Unitarity Triangle.
For presentational purposes,
it is convenient to rescale the triangle by $(V_{cd}V_{cb}^*)^{-1}$,
as shown in Fig.~\ref{fig:cp_uta:ut}.

Two popular naming conventions for the UT angles exist in the literature:
\begin{equation}
  \label{eq:cp_uta:abc}
  \alpha  \equiv  \phi_2  = 
  \arg\left[ - \frac{V_{td}V_{tb}^*}{V_{ud}V_{ub}^*} \right]\,,
  \hspace{0.5cm}
  \beta   \equiv   \phi_1 =  
  \arg\left[ - \frac{V_{cd}V_{cb}^*}{V_{td}V_{tb}^*} \right]\,,
  \hspace{0.5cm}
  \gamma  \equiv   \phi_3  =  
  \arg\left[ - \frac{V_{ud}V_{ub}^*}{V_{cd}V_{cb}^*} \right]\,.
%  \nonumber
\end{equation}
In this document the $\left( \alpha, \beta, \gamma \right)$ set is used.
The sides $R_u$ and $R_t$ of the Unitarity Triangle 
(the third side being normalised to unity) 
are given by
%% read to all orders 
\begin{equation}
  \label{eq:ru_rt}
  R_u =
  \left|\frac{V_{ud}V_{ub}^*}{V_{cd}V_{cb}^*} \right|
  = \sqrt{\rhobar^2+\etabar^2} \,,
  \hspace{0.5cm}
  R_t = 
  \left|\frac{V_{td}V_{tb}^*}{V_{cd}V_{cb}^*}\right| 
  = \sqrt{(1-\rhobar)^2+\etabar^2} \,.
\end{equation} 
where $\rhobar$ and $\etabar$ define 
the apex of the Unitarity Triangle~\cite{Buras:1994ec} 
\begin{equation}
  \label{eq:rhoetabar}
  \rhobar + i\etabar
  \equiv -\frac{V_{ud}V_{ub}^*}{V_{cd}V_{cb}^*}
  \equiv 1 + \frac{V_{td}V_{tb}^*}{V_{cd}V_{cb}^*}
  = \frac{\sqrt{1-\lambda^2}\,(\rho + i \eta)}{\sqrt{1-A^2\lambda^4}+\sqrt{1-\lambda^2}A^2\lambda^4(\rho+i\eta)} \, .
%   = (\rho + i\eta) (1 - \frac{1}{2}\lambda^{2}) + {\cal O}(\lambda^4).
\end{equation}
The exact relation between $\left( \rho, \eta \right)$ and 
$\left( \rhobar, \etabar \right)$ is
\begin{equation}
  \label{eq:rhoetabarinv}
  \rho + i\eta \;=\; 
  \frac{ 
    \sqrt{ 1-A^2\lambda^4 }(\rhobar+i\etabar) 
  }{
    \sqrt{ 1-\lambda^2 } \left[ 1-A^2\lambda^4(\rhobar+i\etabar) \right]
  } \, .
\end{equation}

By expanding in powers of $\lambda$, several useful approximate expressions
can be obtained, including
\begin{equation}
  \label{eq:rhoeta_approx}
  \rhobar = \rho (1 - \frac{1}{2}\lambda^{2}) + {\cal O}(\lambda^4) \, ,
  \hspace{0.5cm}
  \etabar = \eta (1 - \frac{1}{2}\lambda^{2}) + {\cal O}(\lambda^4) \, ,
  \hspace{0.5cm}
  V_{td} = A \lambda^{3} (1-\rhobar -i\etabar) + {\cal O}(\lambda^6) \, .
\end{equation}
Recent world average values for the Wolfenstein parameters, evaluated using many of the measurements reported in this document, are~\cite{Charles:2004jd}
\begin{equation}
  A = 0.8227 \,^{+0.0066}_{-0.0136}\, , \quad
  \lambda = 0.22543 \,^{+0.00042}_{-0.00031} \, , \quad
  \rhobar = 0.1504 \,^{+0.0121}_{-0.0062} \, , \quad
  \etabar = 0.3540 \,^{+0.0069}_{-0.0076} \, .
\end{equation}

The relevant unitarity triangle for the $b \to s$ transition is obtained 
by replacing $d \leftrightarrow s$ in Eq.~(\ref{eq:cp_uta:ut}).
Definitions of the set of angles $( \alpha_s, \beta_s, \gamma_s )$ 
can be obtained using equivalent relations to those of Eq.~(\ref{eq:cp_uta:abc}).
However, this gives a value of $\beta_s$ that is negative in the Standard Model, so that the sign is usually flipped in the literature; this convention, \ie\ $\beta_s = \arg\left[ - (V_{ts}V_{tb}^*)/(V_{cs}V_{cb}^*) \right]$, is also followed here and in Sec.~\ref{sec:life_mix}.
Since the sides of the $b \to s$ unitarity triangle are not all of the same order of $\lambda$, the triangle is squashed and $\beta_s \sim \lambda^2\eta$.

% \afterpage{\clearpage}
\mysubsection{Notations
}
\label{sec:cp_uta:notations}

Several different notations for $\CP$ violation parameters
are commonly used.
This section reviews those found in the experimental literature,
in the hope of reducing the potential for confusion, 
and to define the frame that is used for the averages.

In some cases, when $\B$ mesons decay into 
multibody final states via broad resonances ($\rho$, $\Kstar$, \etc),
the experimental analyses ignore the effects of interference 
between the overlapping structures.
%% DP is only for 3body, but Q2B is also true for \rho\rho \etc 
% in the underlying multidimensional Dalitz plots.
%%%% --> that was meant by 'multidimensional' 
This is referred to as the quasi-two-body (Q2B) approximation
in the following.

\mysubsubsection{$\CP$ asymmetries
}
\label{sec:cp_uta:notations:pra}

The $\CP$ asymmetry is defined as the difference between the rate 
involving a $b$ quark and that involving a $\bar b$ quark, divided 
by the sum. For example, the partial rate (or charge) asymmetry for 
a charged $\B$ decay would be given as 
\begin{equation}
  \label{eq:cp_uta:pra}
  \Acp_{f} \;\equiv\; 
  \frac{\Gamma(\Bm \to f)-\Gamma(\Bp \to \bar{f})}{\Gamma(\Bm \to f)+\Gamma(\Bp \to \bar{f})}.
\end{equation}

\mysubsubsection{Time-dependent \CP asymmetries in decays to $\CP$ eigenstates
}
\label{sec:cp_uta:notations:cp_eigenstate}

If the amplitudes for $\Bz$ and $\Bzb$ to decay to a final state $f$, 
which is a $\CP$ eigenstate with eigenvalue $\etacpf$,
are given by $\Af$ and $\Abarf$, respectively, 
then the decay distributions for neutral $\B$ mesons, 
with known (\ie\ ``tagged'') flavour at time $\Delta t =0$,
are given by
\begin{eqnarray}
  \label{eq:cp_uta:td_cp_asp1}
  \Gamma_{\Bzb \to f} (\Delta t) & = &
  \frac{e^{-| \Delta t | / \tau(\Bz)}}{4\tau(\Bz)}
  \left[ 
    1 +
%%    \left\{ 
    \frac{2\, \Im(\lambda_f)}{1 + |\lambda_f|^2} \sin(\Delta m \Delta t) -
    \frac{1 - |\lambda_f|^2}{1 + |\lambda_f|^2} \cos(\Delta m \Delta t)
%%    \right\}  
  \right], \\
  \label{eq:cp_uta:td_cp_asp2}
  \Gamma_{\Bz \to f} (\Delta t) & = &
  \frac{e^{-| \Delta t | / \tau(\Bz)}}{4\tau(\Bz)}
  \left[ 
    1 -
%%    \left\{ 
    \frac{2\, \Im(\lambda_f)}{1 + |\lambda_f|^2} \sin(\Delta m \Delta t) +
    \frac{1 - |\lambda_f|^2}{1 + |\lambda_f|^2} \cos(\Delta m \Delta t)
%%    \right\}  
  \right].
\end{eqnarray}
Here $\lambda_f = \frac{q}{p} \frac{\Abarf}{\Af}$ 
contains terms related to $\Bz$--$\Bzb$ mixing and to the decay amplitude
(the eigenstates of the $\BzBzb$ system with physical masses and lifetimes
are $\left| B_\pm \right> = p \left| \Bz \right> \pm q \left| \Bzb \right>$).
This formulation assumes $\CPT$ invariance, 
and neglects possible lifetime differences between the two physical states (see Sec.~\ref{sec:mixing} where the mass difference $\Delta m$ is also defined)
in the neutral $\B$ meson system.
The case where non-zero lifetime differences are taken into account is 
discussed in Sec.~\ref{sec:cp_uta:notations:Bs}.

The notation and normalisation used in Eqs.~(\ref{eq:cp_uta:td_cp_asp1}) and~(\ref{eq:cp_uta:td_cp_asp2}) are relevant for the $e^+e^-$ $B$ factory experiments.
In this case, neutral $B$ mesons are produced via the $e^+e^- \to \Upsilon(4S) \to \B\Bbar$ process, and the wavefunction of the produced $\B\Bbar$ pair evolves coherently until one meson decays.  
When one of the pair decays into a final state that tags its flavour, the flavour of the other at that instant is known.  
The evolution of the other neutral $B$ meson is therefore described in terms of $\Delta t$, the difference between the decay times of the two mesons in the pair.
At hadron collider experiments, $t$ is usually used in place of $\Delta t$ since the flavour tagging is done at production ($t = 0$); due to the nature of the production in hadron colliders (incoherent $b\bar{b}$ quark pair production with many additional associated particles), very different methods are used for tagging compared to those in $e^+e^-$ experiments.
Moreover, since negative values of $t$ are not possible, the normalisation is such that 
$\int_0^{+\infty} \left( 
\Gamma_{\Bzb \to f} (t) + \Gamma_{\Bz \to f} (t) \right) dt = 1$,
rather than 
$\int_{-\infty}^{+\infty} \left( 
\Gamma_{\Bzb \to f} (\Delta t) + \Gamma_{\Bz \to f} (\Delta t) \right) d(\Delta t) = 1$,
as in Eqs.~(\ref{eq:cp_uta:td_cp_asp1}) and~(\ref{eq:cp_uta:td_cp_asp2}).

The time-dependent $\CP$ asymmetry,
again defined as the difference between the rate 
involving a $b$ quark and that involving a $\bar b$ quark,
is then given by
%% don't use {\cal A}_f since this will be used for A (cosine term)
\begin{equation}
  \label{eq:cp_uta:td_cp_asp}
  \Acp_{f} \left(\Delta t\right) \; \equiv \;
  \frac{
    \Gamma_{\Bzb \to f} (\Delta t) - \Gamma_{\Bz \to f} (\Delta t)
  }{
    \Gamma_{\Bzb \to f} (\Delta t) + \Gamma_{\Bz \to f} (\Delta t)
  } \; = \;
  \frac{2\, \Im(\lambda_f)}{1 + |\lambda_f|^2} \sin(\Delta m \Delta t) -
  \frac{1 - |\lambda_f|^2}{1 + |\lambda_f|^2} \cos(\Delta m \Delta t).
\end{equation}
While the coefficient of the $\sin(\Delta m \Delta t)$ term in 
Eq.~(\ref{eq:cp_uta:td_cp_asp}) is customarily\footnote
{
%  Actually, not quite everywhere.  
  Occasionally one also finds Eq.~(\ref{eq:cp_uta:td_cp_asp}) written as
  $\Acp_{f} \left(\Delta t\right) = 
  {\cal A}^{\rm mix}_f \sin(\Delta m \Delta t) + {\cal A}^{\rm dir}_f \cos(\Delta m \Delta t)$,
  or similar.
} denoted $S_f$:
\begin{equation}
  \label{eq:cp_uta:s_def}
  S_f \;\equiv\; \frac{2\, \Im(\lambda_f)}{1 + \left|\lambda_f\right|^2},
\end{equation}
different notations are in use for the
coefficient of the $\cos(\Delta m \Delta t)$ term:
%% cannot use A_f here as this is already used for the amplitude
%% use {\cal A}_f and clarify in the text
\begin{equation}
  \label{eq:cp_uta:c_def}
  C_f \;\equiv\; - A_f \;\equiv\; \frac{1 - \left|\lambda_f\right|^2}{1 + \left|\lambda_f\right|^2}.
\end{equation}
The $C$ notation has been used by the \babar\ collaboration 
(see \eg\ Ref.~\cite{Aubert:2001sp}), 
and is also adopted in this document.
The $A$ notation has been used by the \belle\ collaboration
(see \eg\ Ref.~\cite{Abe:2001xe}).
When the final state is a \CP eigenstate, the notation $S_{\CP}$ and $C_{\CP}$ is widely used, including in this document, instead of specifying the final state $f$.
In addition, particularly when grouping together measurements with different final states mediated by the same quark-level transition, the $S$, $C$ notation with a subscript indicating the transition is used.

Neglecting effects due to $\CP$ violation in mixing 
(by taking $|q/p| = 1$),
if the decay amplitude contains terms with 
a single weak (\ie\ $\CP$ violating) phase
then $\left|\lambda_f\right| = 1$ and one finds
$S_f = -\etacpf \sin(\phi_{\rm mix} + \phi_{\rm dec})$, $C_f = 0$,
where $\phi_{\rm mix}=\arg(q/p)$ and $\phi_{\rm dec}=\arg(\Abarf/\Af)$.
Note that the $\Bz$--$\Bzb$ mixing phase $\phi_{\rm mix}$ is approximately equal to $2\beta$ in the Standard Model (in the usual phase convention)~\cite{Carter:1980tk,Bigi:1981qs}. 

If amplitudes with different weak phases contribute to the decay, 
no clean interpretation of $S_f$ is possible without further input. 
In this document, only the theoretically cleanest channels are interpreted as measurements of the weak phase (\eg\ $b \to c\bar{c}s$ transitions for $\sin(2\beta)$), though even in these cases some care is necessary.
In channels in which the second amplitude is expected to be suppressed, the concept of an effective weak phase difference is sometimes used, \eg\ $\sin(2\beta^{\rm eff})$ in $b \to q\bar{q}s$ transition. % (see Eq.~(\ref{eq:cp_uta:sin2betaeff-def})).  

If, in addition to having a weak phase difference, the decay amplitudes have different $\CP$ conserving strong phases, then $\left| \lambda_f \right| \neq 1$.
Additional input is required for interpretation of the results.
The coefficient of the cosine term becomes non-zero, indicating $\CP$ violation in decay.
% The sign of $A_f$ as defined above is consistent with that of $\Acp_{f}$ in Eq.~(\ref{eq:cp_uta:pra}).

Due to the fact that $\sin(\Delta m \Delta t)$ and $\cos(\Delta m \Delta t)$
are respectively odd and even functions of $\Delta t$, only small correlations
(that can be induced by backgrounds, for example) between $S_f$ and $C_f$ are
expected at an $e^+e^-$ $B$ factory experiment, where the range of $\Delta t$
is $-\infty < \Delta t < +\infty$. 
The situation is different for measurements at hadron collider experiments, where the range of the time variable is $0 < t < +\infty$, so that more sizable correlations can be expected.  
We include the correlations in the averages where available.

Frequently, we are interested in combining measurements 
governed by similar or identical short-distance physics,
but with different final states
(\eg, $\Bz \to \jpsi \KS$ and $\Bz \to \jpsi \KL$).
In this case, we remove the dependence on the $\CP$ eigenvalue 
of the final state by quoting $-\etacp S_f$.
In cases where the final state is not a $\CP$ eigenstate but has
an effective $\CP$ content (see below),
the reported $-\etacp S$ is corrected by the effective $\CP$.

\mysubsubsection{Time-dependent distributions with non-zero decay width difference}
\label{sec:cp_uta:notations:Bs}

A complete analysis of the time-dependent decay rates of neutral $B$ mesons must also take into account the difference in lifetimes, denoted $\Delta \Gamma$, between the mass eigenstates.
This is particularly important in the $\Bs$ system, since a non-negligible value of $\Delta \Gamma_s$ has been established (see Sec.~\ref{sec:mixing} for the latest experimental constraints).
The formalism given here is therefore appropriate for measurements of $\Bs$ decays to a $\CP$ eigenstate $f$ as studied at hadron colliders, but appropriate modifications for $\Bz$ mesons or for the $e^+e^-$ environment are straightforward to make.

Neglecting $\CP$ violation in mixing, the relevant replacements for 
Eqs.~(\ref{eq:cp_uta:td_cp_asp1})~and~(\ref{eq:cp_uta:td_cp_asp2}) 
are~\cite{Dunietz:2000cr}
\begin{equation}
  \label{eq:cp_uta:td_cp_bs_asp1}
  \begin{array}{l@{\hspace{50mm}}cr}
    \mc{2}{l}{
      \Gamma_{\Bsbar \to f} (t) = 
      {\cal N} %need normalization factor due to cosh term (even)
%      (1 - (\frac{\Delta \Gamma_s}{2\Gamma_s})^2)
      \frac{e^{-t/\tau(\Bs)}}{2\tau(\Bs)}
      \Big[ 
      \cosh(\frac{\Delta \Gamma_s t}{2}) + {}
    } & \hspace{40mm} \\
    \hspace{40mm} &
    \mc{2}{r}{
%%    \left\{ 
%       \frac{2\, \Im(\lambda_f)}{1 + |\lambda_f|^2} \sin(\Delta m_s t) -
%       \frac{1 - |\lambda_f|^2}{1 + |\lambda_f|^2} \cos(\Delta m_s t) -
%       \frac{2\, \Re(\lambda_f)}{1 + |\lambda_f|^2} \sinh(\frac{\Delta \Gamma_s t}{2})
      S_f \sin(\Delta m_s t) - C_f \cos(\Delta m_s t) + 
      A^{\Delta \Gamma}_f \sinh(\frac{\Delta \Gamma_s t}{2})
%%    \right\}  
      \Big]\,,
    } \\
  \end{array}
\end{equation}
and
\begin{equation}
  \label{eq:cp_uta:td_cp_bs_asp2}
  \begin{array}{l@{\hspace{50mm}}cr}
    \mc{2}{l}{
      \Gamma_{\Bs \to f} (t) =
      {\cal N} %need normalization factor due to cosh term (even)
%      (1 - (\frac{\Delta \Gamma_s}{2\Gamma_s})^2)
      \frac{e^{-t/\tau(\Bs)}}{2\tau(\Bs)}
      \Big[ 
      \cosh(\frac{\Delta \Gamma_s t}{2}) - {}
    } & \hspace{40mm} \\
    \hspace{40mm} & 
    \mc{2}{r}{
%%    \left\{ 
%       \frac{2\, \Im(\lambda_f)}{1 + |\lambda_f|^2} \sin(\Delta m_s t) +
%       \frac{1 - |\lambda_f|^2}{1 + |\lambda_f|^2} \cos(\Delta m_s t) -
%       \frac{2\, \Re(\lambda_f)}{1 + |\lambda_f|^2} \sinh(\frac{\Delta \Gamma_s t}{2})
      S_f \sin(\Delta m_s t) + C_f \cos(\Delta m_s t) + 
      A^{\Delta \Gamma}_f \sinh(\frac{\Delta \Gamma_s t}{2})
%%    \right\}  
      \Big]\,,
    } \\
  \end{array}
\end{equation}
where $S_f$ and $C_f$ are as defined in Eqs.~\ref{eq:cp_uta:s_def} and~\ref{eq:cp_uta:c_def}, respectively, $\tau(\Bs) = 1/\Gamma_s$ is defined in Sec.~\ref{sec:taubs}, and the coefficient of the $\sinh$ term is\footnote{
  As ever, alternative and conflicting notations appear in the literature.
  One popular alternative notation for this parameter is 
  ${\cal A}_{\Delta \Gamma}$.
  Particular care must be taken over the signs.
}
\begin{equation}
  A^{\Delta \Gamma}_f = - \frac{2\, \Re(\lambda_f)}{1 + |\lambda_f|^2} \, .
\end{equation}
With the requirement
$\int_{0}^{+\infty} \left[ \Gamma_{\Bsbar \to f} (t) + \Gamma_{\Bs \to f} (t) \right] dt = 1$,
the normalisation factor is fixed to ${\cal N} = \left(1 - (\frac{\Delta \Gamma_s}{2\Gamma_s})^2\right)/\left(1 +\frac{A^{\Delta \Gamma}_f \Delta \Gamma_s}{2\Gamma_s}\right)$.\footnote{
  The prefactor of ${\cal N}/2\tau(\Bs)$ in Eqs.~\ref{eq:cp_uta:s_def} and~\ref{eq:cp_uta:c_def} has been chosen so that ${\cal N} = 1$ in the limit $\Delta \Gamma_s = 0$.
  In the $e^+e^-$ environment, where the range is $-\infty < \Delta t < \infty$, the prefactor should be ${\cal N}/4\tau(\Bs)$ and ${\cal N} = 1 - (\frac{\Delta \Gamma_s}{2\Gamma_s})^2$.
}

A time-dependent analysis of \CP asymmetries in flavour-tagged
$\Bs$ decays to a \CP eigenstate $f$ can thus determine the parameters 
$S_f$, $C_f$ and $A^{\Delta \Gamma}_f$.
Note that, by definition, 
\begin{equation}
  \left( S_f \right)^2 + \left( C_f \right)^2 + \left( A^{\Delta \Gamma}_f \right)^2 = 1 \, ,
\end{equation}
and this constraint can be imposed or not in the fits.
Since these parameters have sensitivity to both
$\Im(\lambda_f)$ and $\Re(\lambda_f)$,
alternative choices of parametrisation, 
including those directly involving \CP violating phases (such as $\beta_s$), 
are possible.
These can also be adopted for vector-vector final states.

The {\it untagged} time-dependent decay rate is given by
\begin{equation}
  \Gamma_{\Bsbar \to f} (t) + \Gamma_{\Bs \to f} (t)
  = 
  {\cal N} %need normalization factor due to cosh term (even)
%      (1 - (\frac{\Delta \Gamma_s}{2\Gamma_s})^2)
  \frac{e^{-t/\tau(\Bs)}}{\tau(\Bs)}
  \Big[ 
  \cosh\left(\frac{\Delta \Gamma_s t}{2}\right)
%  - \frac{2\, \Re(\lambda_f)}{1 + |\lambda_f|^2} \sinh\left(\frac{\Delta \Gamma_s t}{2}\right)
  + A^{\Delta \Gamma}_f \sinh\left(\frac{\Delta \Gamma_s t}{2}\right)
  \Big] \, .
\end{equation}
Thus, an untagged time-dependent analysis can probe $\lambda_f$, through the dependence of $A^{\Delta \Gamma}_f $ on $\Re(\lambda_f)$, when $\Delta \Gamma_s \neq 0$.
This is equivalent to determining the ``{\it effective lifetime}''~\cite{Fleischer:2011cw}, as discussed in Sec.~\ref{sec:taubs}.
The analysis of flavour-tagged \Bs\ mesons is, of course, more sensitive.

The discussion in this and the previous section is relevant for decays to \CP\ eigenstates.
In the following sections, various cases of time-dependent \CP\ asymmetries in decays to non-\CP eigenstates are considered.
For brevity, these will be given assuming that the decay width difference $\Delta \Gamma$ is negligible.
Modifications similar to those described here can be made to take into account a non-zero decay width difference.

% Note that when the final state contains 
% a mixture of $\CP$-even and $\CP$-odd states
% (as, for example, for vector-vector or multibody self-conjugate states),
% that $\Re(\lambda_f)$ contains terms proportional to 
% both the sine and cosine of the weak phase difference, 
% albeit with rather different sensitivities.

\mysubsubsection{Time-dependent \CP asymmetries in decays to vector-vector final states
}
\label{sec:cp_uta:notations:vv}

Consider \B decays to states consisting of two spin-1 particles,
such as $\jpsi K^{*0}(\to\KS\piz)$, $\jpsi\phi$, $D^{*+}D^{*-}$ and $\rho^+\rho^-$,
which are eigenstates of charge conjugation but not of parity.\footnote{
  \noindent
  This is not true for all vector-vector final states,
  \eg, $D^{*\pm}\rho^{\mp}$ is clearly not an eigenstate of 
  charge conjugation.
}
For such a system, there are three possible final states:
in the helicity basis these can be written $h_{-1}, h_0, h_{+1}$.
The $h_0$ state is an eigenstate of parity, and hence of $\CP$, however $\CP$ transforms $h_{+1} \leftrightarrow h_{-1}$ (up to an unobservable phase). 
In the transversity basis, these states are transformed into  $h_\parallel =  (h_{+1} + h_{-1})/2$ and $h_\perp = (h_{+1} - h_{-1})/2$.
In this basis all three states are $\CP$ eigenstates, and $h_\perp$ has the opposite $\CP$ to the others.

The amplitude for decays to the transversity basis states are usually given by $A_{0,\perp,\parallel}$, with normalisation such that $| A_0 |^2 + | A_\perp |^2 + | A_\parallel |^2 = 1$.
Given the relation between the $\CP$ eigenvalues of the states, the effective $\CP$ content of the vector-vector state is known if $| A_\perp |^2$ is measured.
An alternative strategy is to measure just the longitudinally polarised component,  $| A_0 |^2$ (sometimes denoted by $f_{\rm long}$), which allows a limit to be set on the effective $\CP$ since $| A_\perp |^2 \leq | A_\perp |^2 + | A_\parallel |^2 = 1 - | A_0 |^2$.
The value of the effective $\CP$ content can be used to treat the decay with the same formalism as for $\CP$ eigenstates.
The most complete treatment for neutral $\B$ decays to vector-vector final states is, however, time-dependent angular analysis (also known as time-dependent transversity analysis).
In such an analysis, interference between $\CP$-even and $\CP$-odd states provides additional sensitivity to the weak and strong phases involved.
% When non-negligible, the finite widths of the decaying vector particles should also be taken into account.

In most analyses of time-dependent \CP asymmetries in decays to vector-vector final states carried out to date, an assumption has been made that each helicity (or transversity) amplitude has the same weak phase.
This is a good approximation for decays that are dominated by 
amplitudes with a single weak phase, such $\Bz \to \jpsi K^{*0}$,
and is a reasonable approximation in any mode for which only 
very limited statistics are available.
However, for modes that have contributions from amplitudes with different 
weak phases, the relative size of these contributions can be different 
for each helicity (or transversity) amplitude,
and therefore the time-dependent \CP asymmetry parameters can also differ.
The most generic analysis, suitable for modes with sufficient statistics,
allows for this effect; such an analysis has been carried out by LHCb for the $\Bz \to \jpsi \rhoz$ decay~\cite{Aaij:2014vda}.
An intermediate analysis can allow different parameters for the $\CP$-even and $\CP$-odd components; such an analysis has been carried out by \babar\ for the decay $\Bz \to D^{*+}D^{*-}$~\cite{Aubert:2008ah}.
The independent treatment of each helicity (or transversity) amplitude, as in the latest result on $\Bs \to \jpsi\phi$~\cite{Aaij:2014zsa} (discussed in Sec.~\ref{sec:life_mix}), becomes increasingly important for high precision measurements.

\mysubsubsection{Time-dependent asymmetries: self-conjugate multiparticle final states
}
\label{sec:cp_uta:notations:dalitz}

Amplitudes for neutral \B decays into 
self-conjugate multiparticle final states
such as $\pi^+\pi^-\pi^0$, $K^+K^-\KS$, $\pi^+\pi^-\KS$,
$\jpsi \pi^+\pi^-$ or $D\pi^0$ with $D \to \KS\pi^+\pi^-$
may be written in terms of \CP-even and \CP-odd amplitudes.
As above, the interference between these terms 
provides additional sensitivity to the weak and strong phases
involved in the decay,
and the time-dependence depends on both the sine and cosine
of the weak phase difference.
In order to perform unbinned maximum likelihood fits,
and thereby extract as much information as possible from the distributions,
it is necessary to choose a model for the multiparticle decay,
and therefore the results acquire some model dependence.
In certain cases, model-independent methods are also possible, but the resulting need to bin the Dalitz plot leads to some loss of statistical precision.
The number of observables depends on the final state (and on the model used);
the key feature is that as long as there are regions where both
\CP-even and \CP-odd amplitudes contribute,
the interference terms will be sensitive to the cosine 
of the weak phase difference.
Therefore, these measurements allow distinction between multiple solutions
for, \eg, the two values of $2\beta$ from the measurement of $\sin(2\beta)$.

We now consider the various notations that have been used in experimental studies of
time-dependent asymmetries in decays to self-conjugate multiparticle final states.

% \newpage %% ugly hack to prevent pagebreak just after subsubsubsection heading
\mysubsubsubsection{$\Bz \to D^{(*)}h^0$ with $D \to \KS\pi^+\pi^-$
}
\label{sec:cp_uta:notations:dalitz:dh0}

The states $D\pi^0$, $D^*\pi^0$, $D\eta$, $D^*\eta$, $D\omega$
are collectively denoted $D^{(*)}h^0$.
When the $D$ decay model is fixed,
fits to the time-dependent decay distributions can be performed
to extract the weak phase difference.
However, it is experimentally advantageous to use the sine and cosine of 
this phase as fit parameters, since these behave as essentially 
independent parameters, with low correlations and (potentially)
rather different uncertainties.
A parameter representing $\CP$ violation in the $B$ decay 
can be simultaneously determined.  
For consistency with other analyses, this could be chosen to be $C_f$,
but could equally well be $\left| \lambda_f \right|$, or other possibilities.

\belle\ performed an analysis of these channels
with $\sin(2\phi_1)$ and $\cos(2\phi_1)$ as free parameters~\cite{Krokovny:2006sv}.
\babar\ has performed an analysis in which $\left| \lambda_f \right|$ was also determined~\cite{Aubert:2007rp}.
% (and, of course, replacing $\phi_1$ with $\beta$).
\belle\ has in addition performed a model-independent analysis~\cite{Vorobyev:2016npn} using as input information about the average strong phase difference between symmetric bins of the Dalitz plot determined by CLEO-c~\cite{Libby:2010nu}.\footnote{
  The external input needed for this analysis is the same as in the model-independent analysis of $\Bp \to D\Kp$ with $D \to \KS\pip\pim$, discussed in Sec.~\ref{sec:cp_uta:cus:dalitz:modInd}.
}
The results of this analysis are measurements of $\sin(2\phi_1)$ and $\cos(2\phi_1)$.

\mysubsubsubsection{$\Bz \to D^{*+}D^{*-}\KS$
}
\label{sec:cp_uta:notations:dalitz:dstardstarks}

The hadronic structure of the $\Bz \to D^{*+}D^{*-}\KS$ decay
is not sufficiently well understood to perform a full 
time-dependent Dalitz plot analysis.
Instead, following Ref.~\cite{Browder:1999ng},
\babar~\cite{Aubert:2006fh} and \belle~\cite{Dalseno:2007hx} divide the Dalitz plane in two regions:
$m(D^{*+}\KS)^2 > m(D^{*-}\KS)^2$ $(\eta_y = +1)$ and 
$m(D^{*+}\KS)^2 < m(D^{*-}\KS)^2$ $(\eta_y = -1)$;
and then fit to a decay time distribution with asymmetry given by
\begin{equation}
  \Acp_{f} \left(\Delta t\right) =
  \eta_y \frac{J_c}{J_0} \cos(\Delta m \Delta t) -  
  \left[ 
    \frac{2J_{s1}}{J_0} \sin(2\beta) + \eta_y \frac{2J_{s2}}{J_0} \cos(2\beta) 
  \right] \sin(\Delta m \Delta t) \, .
\end{equation}
% A similar analysis has also been carried out by \belle~\cite{Dalseno:2007hx}.
The fitted observables are $\frac{J_c}{J_0}$, $\frac{2J_{s1}}{J_0} \sin(2\beta)$ and $\frac{2J_{s2}}{J_0} \cos(2\beta)$, 
where the parameters $J_0$, $J_c$, $J_{s1}$ and $J_{s2}$ are the integrals 
over the half Dalitz plane $m(D^{*+}\KS)^2 < m(D^{*-}\KS)^2$ 
of the functions $|a|^2 + |\bar{a}|^2$, $|a|^2 - |\bar{a}|^2$, 
$\Re(\bar{a}a^*)$ and $\Im(\bar{a}a^*)$ respectively, 
where $a$ and $\bar{a}$ are the decay amplitudes of 
$\Bz \to D^{*+}D^{*-}\KS$ and $\Bzb \to D^{*+}D^{*-}\KS$ respectively. 
The parameter $J_{s2}$ (and hence $J_{s2}/J_0$) is predicted to be positive;
assuming this prediction to be correct, it is possible to determine the sign of $\cos(2\beta)$.

\mysubsubsubsection{$\Bz \to \jpsi \pip\pim$
}
\label{sec:cp_uta:notations:dalitz:jpsipipi}

Amplitude analyses of $\Bz \to \jpsi \pip\pim$ decays~\cite{Aaij:2014siy,Aaij:2014vda} show large contributions from the $\rho(770)^0$ and $f_0(500)$ states, together with smaller contributions from higher resonances.
Since modelling the $f_0(500)$ structure is challenging~\cite{Pelaez:2015qba}, it is difficult to determine reliably its associated \CP violation parameters.
Corresponding parameters for the $\jpsi\rhoz$ decay can, however, be determined.
In the LHCb analysis~\cite{Aaij:2014vda}, $2\beta^{\rm eff}$ is determined from the fit; results are then converted into values for $S_{\CP}$ and $C_{\CP}$ to allow comparison with other modes.
Here, the notation $S_{\CP}$ and $C_{\CP}$ denotes parameters obtained for the $\jpsi\rhoz$ final state accounting for the composition of \CP-even and \CP-odd amplitudes (while assuming that all amplitudes involve the same phases), so that no dilution occurs.
Possible \CP violation effects in the other amplitudes contributing to the Dalitz plot are treated as a source of systematic uncertainty.

Amplitude analyses have also been done for the $\Bs \to \jpsi\pip\pim$ decay, where the final state is dominated by scalar resonances including the $f_0(980)$~\cite{LHCb:2012ae,Aaij:2014dka}.
Time-dependent analyses of this \Bs decay allow a determination of $2\beta_s$, as discussed in Sec.~\ref{sec:life_mix}. 

\mysubsubsubsection{$\Bz \to K^+K^-\Kz$
}
\label{sec:cp_uta:notations:dalitz:kkk0}

Studies of $\Bz \to K^+K^-\Kz$~\cite{Aubert:2007sd,Nakahama:2010nj,Lees:2012kxa} 
and of the related decay 
$\Bp \to K^+K^-K^+$~\cite{Garmash:2004wa,Aubert:2006nu,Lees:2012kxa},
show that the decay is dominated by a large nonresonant contribution
with significant components from the 
intermediate $K^+K^-$ resonances $\phi(1020)$, $f_0(980)$,
and other higher resonances,
% \footnote{
%   The broad structure that peaks near 
%   $m(K^+K^-) \sim 1550 \ {\rm MeV}/c^2$ and was denoted $X_0(1550)$ 
%   is now believed to originate from interference effects.
% }
as well as a contribution from $\chi_{c0}$.

The full time-dependent Dalitz plot analysis allows 
the complex amplitudes of each contributing term to be determined from data,
including $\CP$ violation effects
(\ie\ allowing the complex amplitude for the $\Bz$ decay to be independent
from that for $\Bzb$ decay), although one amplitude must be fixed 
to serve as a reference.
There are several choices for parametrisation of the complex amplitudes 
(\eg\ real and imaginary part, or magnitude and phase).
Similarly, there are various approaches to include $\CP$ violation effects.
Note that positive definite parameters such as magnitudes are
disfavoured in certain circumstances 
(they inevitably lead to biases for small values).
In order to compare results between analyses,
it is useful for each experiment to present results in terms of the 
parameters that can be measured in a Q2B analysis
(such as $\Acp_{f}$, $S_f$, $C_f$, 
$\sin(2\beta^{\rm eff})$, $\cos(2\beta^{\rm eff})$, \etc)

In the \babar\ analysis of the $\Bz \to K^+K^-\Kz$ decay~\cite{Lees:2012kxa},
the complex amplitude for each resonant contribution is written as
\begin{equation}
  A_f = c_f ( 1 + b_f ) e^{i ( \phi_f + \delta_f )} 
  \ , \ \ \ \ 
  \bar{A}_f = c_f ( 1 - b_f ) e^{i ( \phi_f - \delta_f )} \, ,
\end{equation}
where $b_f$ and $\delta_f$ introduce $\CP$ violation in the magnitude 
and phase respectively.
Belle~\cite{Nakahama:2010nj} use the same parametrisation but with a different notation for the parameters.\footnote{
  $(c, b, \phi, \delta) \leftrightarrow (a, c, b, d)$.
  See Eq.~(\ref{eq:cp_uta:BelleDPCPparam}).
}
(The weak phase in $B^0$--$\bar{B}^0$ mixing ($2\beta$) also appears 
in the full formula for the time-dependent decay distribution.)
The Q2B parameter of $\CP$ violation in decay is directly related to $b_f$,
\begin{equation}
  \Acp_{f} = \frac{-2b_f}{1+b_f^2} \approx C_f \, ,
\end{equation}
and the mixing-induced $\CP$ violation parameter can be used to obtain
$\sin(2\beta^{\rm eff})$,
\begin{equation}
  \label{eq:cp_uta:sin2betaeff-def}
  -\eta_f S_f \approx \frac{1-b_f^2}{1+b_f^2}\sin(2\beta^{\rm eff}_f) \, ,
\end{equation}
where the approximations are exact in the case that $\left| q/p \right| = 1$.

Both \babar~\cite{Lees:2012kxa} and \belle~\cite{Nakahama:2010nj} present results for $c_f$ and $\phi_f$,
for each resonant contribution,
and in addition present results for $\Acp_{f}$ and $\beta^{\rm eff}_{f}$ for $\phi(1020) \Kz$, $f_0(980) \Kz$ and for the remainder of the contributions to the $K^+K^-\Kz$ Dalitz plot combined.
\babar also present results for the Q2B parameter $S_{f}$ for these channels.
The models used to describe the resonant structure of the Dalitz plot differ, however.  
Both analyses suffer from symmetries in the likelihood that lead to multiple solutions, from which we select only one for averaging.

\mysubsubsubsection{$\Bz \to \pi^+\pi^-\KS$
}
\label{sec:cp_uta:notations:dalitz:pipik0}

Studies of $\Bz \to \pi^+\pi^-\KS$~\cite{Aubert:2009me,Dalseno:2008wwa}
and of the related decay
$\Bp \to \pi^+\pi^-K^+$~\cite{Garmash:2004wa,Garmash:2005rv,Aubert:2005ce,Aubert:2008bj}
show that the decay is dominated by components from intermediate resonances 
in the $K\pi$ ($K^*(892)$, $K^*_0(1430)$) 
and $\pi\pi$ ($\rho(770)$, $f_0(980)$, $f_2(1270)$) spectra,
together with a poorly understood scalar structure that peaks near 
$m(\pi\pi) \sim 1300 \ {\rm MeV}/c^2$ and is denoted $f_X$\footnote{
  The $f_X$ component may originate from either the $f_0(1370)$ or $f_0(1500)$ resonances, or from interference between those or other states and nonresonant amplitudes in this region.
}
and a large nonresonant component.
There is also a contribution from the $\chi_{c0}$ state.

The full time-dependent Dalitz plot analysis allows 
the complex amplitudes of each contributing term to be determined from data,
including $\CP$ violation effects.
In the \babar\ analysis~\cite{Aubert:2009me}, 
the magnitude and phase of each component (for both $\Bz$ and $\Bzb$ decays) 
are measured relative to $\Bz \to f_0(980)\KS$, using the following
parametrisation
\begin{equation}
  A_f = \left| A_f \right| e^{i\,{\rm arg}(A_f)}
  \ , \ \ \ \ 
  \bar{A}_f = \left| \bar{A}_f \right| e^{i\,{\rm arg}(\bar{A}_f)} \, .
\end{equation}
In the \belle\ analysis~\cite{Dalseno:2008wwa}, the $\Bz \to K^{*+}\pi^-$ amplitude
is chosen as the reference, and the amplitudes are parametrised as 
\begin{equation}
  \label{eq:cp_uta:BelleDPCPparam}
  A_f = a_f ( 1 + c_f ) e^{i ( b_f + d_f )} 
  \ , \ \ \ \ 
  \bar{A}_f = a_f ( 1 - c_f ) e^{i ( b_f - d_f )} \, .
\end{equation}
In both cases, the results are translated into Q2B parameters 
such as $2\beta^{\rm eff}_f$, $S_f$, $C_f$ for each \CP\ eigenstate $f$,
and parameters of \CP\ violation in decay for each flavour-specific state.
Relative phase differences between resonant terms are also extracted.

\mysubsubsubsection{$\Bz \to \pi^+\pi^-\pi^0$
}
\label{sec:cp_uta:notations:dalitz:pipipi0}

The $\Bz \to \pi^+\pi^-\pi^0$ decay is dominated by 
intermediate $\rho$ resonances.
Though it is possible, as above, 
to determine directly the complex amplitudes for each component,
an alternative approach~\cite{Snyder:1993mx,Quinn:2000by}
has been used by both \babar~\cite{Aubert:2007jn,Lees:2013nwa}
and \belle~\cite{Kusaka:2007dv,Kusaka:2007mj}.
The amplitudes for $\Bz$ and $\Bzb$ decays to $\pi^+\pi^-\pi^0$ are written as
\begin{equation}
  A_{3\pi} = f_+ A_+ + f_- A_- + f_0 A_0
  \, , \ \ \ 
  \bar{A}_{3\pi} = f_+ \bar{A}_+ + f_- \bar{A}_- + f_0 \bar{A}_0 \, ,
\end{equation}
respectively.
The symbols $A_+$, $A_-$ and $A_0$
represent the complex decay amplitudes for 
$\Bz \to \rho^+\pi^-$, $\Bz \to \rho^-\pi^+$ and $\Bz \to \rho^0\pi^0$
while 
$\bar{A}_+$, $\bar{A}_-$ and $\bar{A}_0$
represent those for 
$\Bzb \to \rho^+\pi^-$, $\Bzb \to \rho^-\pi^+$ and $\Bzb \to \rho^0\pi^0$
respectively.
The terms $f_+$, $f_-$ and $f_0$ incorporate kinematic and dynamical factors
and depend on the Dalitz plot coordinates.
The full time-dependent decay distribution can then be written 
in terms of 27 free parameters,
one for each coefficient of the form factor bilinears,
as listed in Table~\ref{tab:cp_uta:pipipi0:uandi}.
These parameters are sometimes referred to as ``the $U$s and $I$s'',
and can be expressed in terms of 
$A_+$, $A_-$, $A_0$, $\bar{A}_+$, $\bar{A}_-$ and $\bar{A}_0$.
If the full set of parameters is determined,
together with their correlations,
other parameters, such as weak and strong phases,
parameters of $\CP$ violation in decay, \etc, 
can be subsequently extracted.
Note that one of the parameters (typically $U_+^+$) is often fixed to unity to provide a reference; this does not affect the analysis.

%Note that the $U$ parameters are $\CP$ conserving,
%while the $I$ parameters are $\CP$ violating.

\begin{table}[htbp]
  \begin{center}
    \caption{
      Definitions of the $U$ and $I$ coefficients.
      Modified from Ref.~\cite{Aubert:2007jn}.
    }
    \label{tab:cp_uta:pipipi0:uandi}
    \setlength{\tabcolsep}{0.3pc}
    \begin{tabular}{l@{\extracolsep{5mm}}l}
      \hline
      Parameter   & Description \\
      \hline
      $U_+^+$          & Coefficient of $|f_+|^2$ \\
      $U_0^+$          & Coefficient of $|f_0|^2$ \\
      $U_-^+$          & Coefficient of $|f_-|^2$ \\
      [0.15cm]
      $U_0^-$          & Coefficient of $|f_0|^2\cos(\Delta m\Delta t)$ \\
      $U_-^-$          & Coefficient of $|f_-|^2\cos(\Delta m\Delta t)$ \\
      $U_+^-$          & Coefficient of $|f_+|^2\cos(\Delta m\Delta t)$ \\
      [0.15cm]
      $I_0$            & Coefficient of $|f_0|^2\sin(\Delta m\Delta t)$ \\
      $I_-$            & Coefficient of $|f_-|^2\sin(\Delta m\Delta t)$ \\
      $I_+$            & Coefficient of $|f_+|^2\sin(\Delta m\Delta t)$ \\
      [0.15cm]
      $U_{+-}^{+,\Im}$ & Coefficient of $\Im[f_+f_-^*]$ \\
      $U_{+-}^{+,\Re}$ & Coefficient of $\Re[f_+f_-^*]$ \\
      $U_{+-}^{-,\Im}$ & Coefficient of $\Im[f_+f_-^*]\cos(\Delta m\Delta t)$ \\
      $U_{+-}^{-,\Re}$ & Coefficient of $\Re[f_+f_-^*]\cos(\Delta m\Delta t)$ \\
      $I_{+-}^{\Im}$   & Coefficient of $\Im[f_+f_-^*]\sin(\Delta m\Delta t)$ \\
      $I_{+-}^{\Re}$   & Coefficient of $\Re[f_+f_-^*]\sin(\Delta m\Delta t)$ \\
      [0.15cm]
      $U_{+0}^{+,\Im}$ & Coefficient of $\Im[f_+f_0^*]$ \\
      $U_{+0}^{+,\Re}$ & Coefficient of $\Re[f_+f_0^*]$ \\
      $U_{+0}^{-,\Im}$ & Coefficient of $\Im[f_+f_0^*]\cos(\Delta m\Delta t)$ \\
      $U_{+0}^{-,\Re}$ & Coefficient of $\Re[f_+f_0^*]\cos(\Delta m\Delta t)$ \\
      $I_{+0}^{\Im}$   & Coefficient of $\Im[f_+f_0^*]\sin(\Delta m\Delta t)$ \\
      $I_{+0}^{\Re}$   & Coefficient of $\Re[f_+f_0^*]\sin(\Delta m\Delta t)$ \\
      [0.15cm]
      $U_{-0}^{+,\Im}$ & Coefficient of $\Im[f_-f_0^*]$ \\
      $U_{-0}^{+,\Re}$ & Coefficient of $\Re[f_-f_0^*]$ \\
      $U_{-0}^{-,\Im}$ & Coefficient of $\Im[f_-f_0^*]\cos(\Delta m\Delta t)$ \\
      $U_{-0}^{-,\Re}$ & Coefficient of $\Re[f_-f_0^*]\cos(\Delta m\Delta t)$ \\
      $I_{-0}^{\Im}$   & Coefficient of $\Im[f_-f_0^*]\sin(\Delta m\Delta t)$ \\
      $I_{-0}^{\Re}$   & Coefficient of $\Re[f_-f_0^*]\sin(\Delta m\Delta t)$ \\     
      \hline
    \end{tabular}
  \end{center}
\end{table}

\mysubsubsection{Time-dependent \CP asymmetries in decays to non-$\CP$ eigenstates
}
\label{sec:cp_uta:notations:non_cp}

Consider a non-$\CP$ eigenstate $f$, and its conjugate $\bar{f}$. 
For neutral $\B$ decays to these final states,
there are four amplitudes to consider:
those for $\Bz$ to decay to $f$ and $\bar{f}$
($\Af$ and $\Afbar$, respectively),
and the equivalents for $\Bzb$
($\Abarf$ and $\Abarfbar$).
% $\CP$ invariance in the decay requires 
If $\CP$ is conserved in the decay, then
$\Af = \Abarfbar$ and $\Afbar = \Abarf$.

%% make definition Bbar - B for f then fbar
%% define so that asymmetry is C cos DmDt - S sin DmDt for both f and fbar

The time-dependent decay distributions can be written in many different ways.
Here, we follow Sec.~\ref{sec:cp_uta:notations:cp_eigenstate}
and define $\lambda_f = \frac{q}{p}\frac{\Abarf}{\Af}$ and
$\lambda_{\bar f} = \frac{q}{p}\frac{\Abarfbar}{\Afbar}$.
The time-dependent \CP asymmetries that are sensitive to mixing-induced
$\CP$ violation effects then follow Eq.~(\ref{eq:cp_uta:td_cp_asp}):
\begin{eqnarray}
\label{eq:cp_uta:non-cp-obs}
  {\cal A}_f (\Delta t) \; \equiv \;
  \frac{
    \Gamma_{\Bzb \to f} (\Delta t) - \Gamma_{\Bz \to f} (\Delta t)
  }{
    \Gamma_{\Bzb \to f} (\Delta t) + \Gamma_{\Bz \to f} (\Delta t)
  } & = & S_f \sin(\Delta m \Delta t) - C_f \cos(\Delta m \Delta t), \\
  {\cal A}_{\bar{f}} (\Delta t) \; \equiv \;
  \frac{
    \Gamma_{\Bzb \to \bar{f}} (\Delta t) - \Gamma_{\Bz \to \bar{f}} (\Delta t)
  }{
    \Gamma_{\Bzb \to \bar{f}} (\Delta t) + \Gamma_{\Bz \to \bar{f}} (\Delta t)
  } & = & S_{\bar{f}} \sin(\Delta m \Delta t) - C_{\bar{f}} \cos(\Delta m \Delta t),
\end{eqnarray}
with the definitions of the parameters 
$C_f$, $S_f$, $C_{\bar{f}}$ and $S_{\bar{f}}$,
following Eqs.~(\ref{eq:cp_uta:s_def}) and~(\ref{eq:cp_uta:c_def}).

The time-dependent decay rates are given by
\begin{eqnarray}
  \label{eq:cp_uta:non-CP-TD1}
  \Gamma_{\Bzb \to f} (\Delta t) & = &
  \frac{e^{-\left| \Delta t \right| / \tau(\Bz)}}{8\tau(\Bz)} 
  ( 1 + \Adirnoncp ) 
  \left[ 
    1 + S_f \sin(\Delta m \Delta t) - C_f \cos(\Delta m \Delta t) 
  \right],
  \\
  \label{eq:cp_uta:non-CP-TD2}
  \Gamma_{\Bz \to f} (\Delta t) & = &
  \frac{e^{-\left| \Delta t \right| / \tau(\Bz)}}{8\tau(\Bz)} 
  ( 1 + \Adirnoncp ) 
  \left[ 
    1 - S_f \sin(\Delta m \Delta t) + C_f \cos(\Delta m \Delta t) 
  \right],
  \\
  \label{eq:cp_uta:non-CP-TD3}
  \Gamma_{\Bzb \to \bar{f}} (\Delta t) & = &
  \frac{e^{-\left| \Delta t \right| / \tau(\Bz)}}{8\tau(\Bz)} 
  ( 1 - \Adirnoncp ) 
  \left[ 
    1 + S_{\bar{f}} \sin(\Delta m \Delta t) - C_{\bar{f}} \cos(\Delta m \Delta t) 
  \right],
  \\
  \label{eq:cp_uta:non-CP-TD4}
  \Gamma_{\Bz \to \bar{f}} (\Delta t) & = &
    \frac{e^{-\left| \Delta t \right| / \tau(\Bz)}}{8\tau(\Bz)} 
  ( 1 - \Adirnoncp ) 
  \left[ 
    1 - S_{\bar{f}} \sin(\Delta m \Delta t) + C_{\bar{f}} \cos(\Delta m \Delta t) 
  \right],
\end{eqnarray}
where the time-independent parameter \Adirnoncp
represents an overall asymmetry in the production of the 
$f$ and $\bar{f}$ final states,\footnote{
  This parameter is often denoted ${\cal A}_f$ (or ${\cal A}_{\CP}$),
  but here we avoid this notation to prevent confusion with the
  time-dependent $\CP$ asymmetry.
}
\begin{equation}
  \Adirnoncp = 
  \frac{
    \left( 
      \left| \Af \right|^2 + \left| \Abarf \right|^2
    \right) - 
    \left( 
      \left| \Afbar \right|^2 + \left| \Abarfbar \right|^2
    \right)
  }{
    \left( 
      \left| \Af \right|^2 + \left| \Abarf \right|^2
    \right) +
    \left( 
      \left| \Afbar \right|^2 + \left| \Abarfbar \right|^2
    \right)
  }.
\end{equation}
Assuming $|q/p| = 1$, \ie\ absence of \CP violation in mixing,
the parameters $C_f$ and $C_{\bar{f}}$
can also be written in terms of the decay amplitudes as follows:
\begin{equation}
  C_f = 
  \frac{
    \left| \Af \right|^2 - \left| \Abarf \right|^2 
  }{
    \left| \Af \right|^2 + \left| \Abarf \right|^2
  }
  \hspace{5mm}
  {\rm and}
  \hspace{5mm}
  C_{\bar{f}} = 
  \frac{
    \left| \Afbar \right|^2 - \left| \Abarfbar \right|^2
  }{
    \left| \Afbar \right|^2 + \left| \Abarfbar \right|^2
  },
\end{equation}
giving asymmetries in the decay amplitudes of $\Bz$ and $\Bzb$
to the final states $f$ and $\bar{f}$ respectively.
In this notation, the conditions for absence of $\CP$ violation in decay are
$\Adirnoncp = 0$ and $C_f = - C_{\bar{f}}$.
Note that $C_f$ and $C_{\bar{f}}$ are typically non-zero;
\eg, for a flavour-specific final state, 
$\Abarf = \Afbar = 0$ ($\Af = \Abarfbar = 0$), they take the values
$C_f = - C_{\bar{f}} = 1$ ($C_f = - C_{\bar{f}} = -1$).

The coefficients of the sine terms contain information about the weak phase. 
% Assuming $\left| \frac{q}{p} \right| = 1$, then 
In the case that each decay amplitude contains only a single weak phase
(\ie, no $\CP$ violation in decay as well as none in mixing),
these terms can be written as
%% should include angular momentum factor here (hep-ph/0304027)
%% or just absorb it into the strong phase difference
\begin{equation}
  S_f = 
  \frac{ 
    - 2 \left| \Af \right| \left| \Abarf \right| 
    \sin( \phi_{\rm mix} + \phi_{\rm dec} - \delta_f )
  }{
    \left| \Af \right|^2 + \left| \Abarf \right|^2
  } 
  \hspace{5mm}
  {\rm and}
  \hspace{5mm}
  S_{\bar{f}} = 
  \frac{
    - 2 \left| \Afbar \right| \left| \Abarfbar \right| 
    \sin( \phi_{\rm mix} + \phi_{\rm dec} + \delta_f )
  }{
    \left| \Afbar \right|^2 + \left| \Abarfbar \right|^2
  },
\end{equation}
where $\delta_f$ is the strong phase difference between the decay amplitudes.
If there is no $\CP$ violation, the condition $S_f = - S_{\bar{f}}$ holds.
If decay amplitudes with different weak and strong phases contribute,
no clean interpretation of $S_f$ and $S_{\bar{f}}$ is possible.

The conditions for $\CP$ invariance $C_f = - C_{\bar{f}}$ and $S_f = - S_{\bar{f}}$ motivate a rotation of the parameters:
\begin{equation}
\label{eq:cp_uta:non-cp-s_and_deltas}
  S_{f\bar{f}} = \frac{S_{f} + S_{\bar{f}}}{2},
  \hspace{4mm}
  \Delta S_{f\bar{f}} = \frac{S_{f} - S_{\bar{f}}}{2},
  \hspace{4mm}
  C_{f\bar{f}} = \frac{C_{f} + C_{\bar{f}}}{2},
  \hspace{4mm}
  \Delta C_{f\bar{f}} = \frac{C_{f} - C_{\bar{f}}}{2}.
\end{equation}
With these parameters, the $\CP$ invariance conditions become
$S_{f\bar{f}} = 0$ and $C_{f\bar{f}} = 0$. 
The parameter $\Delta C_{f\bar{f}}$ gives a measure of the ``flavour-specificity''
of the decay:
$\Delta C_{f\bar{f}}=\pm1$ corresponds to a completely flavour-specific decay,
in which no interference between decays with and without mixing can occur,
while $\Delta C_{f\bar{f}} = 0$ results in 
maximum sensitivity to mixing-induced $\CP$ violation.
% describes the ``flavour-eigenstateness'' 
% of  the decay: maximum sensitivity to mixing-induced $\CP$ violation is 
% achieved for $\Delta C_{f\bar{f}}=0$, while for $\Delta C_{f\bar{f}}=\pm1$
% (maximum dilution) no interference between decays with and without 
% mixing can occur. 
The parameter $\Delta S_{f\bar{f}}$ is related to the strong phase difference 
between the decay amplitudes of the $\Bz$ meson to the $f$ and to $\bar f$ final states. 
We note that the observables of Eq.~(\ref{eq:cp_uta:non-cp-s_and_deltas})
exhibit experimental correlations 
(typically of $\sim 20\%$, depending on the tagging purity, and other effects)
between $S_{f\bar{f}}$ and  $\Delta S_{f\bar{f}}$, 
and between $C_{f\bar{f}}$ and $\Delta C_{f\bar{f}}$. 
%% TJG try to clarify
% This is not the case for the final state
% specific observables~(\ref{eq:cp_uta:non-cp-obs}). 
On the other hand, 
the final state specific observables of Eqs.~(\ref{eq:cp_uta:non-CP-TD1})--(\ref{eq:cp_uta:non-CP-TD4}) tend to have low correlations. 
% when determined in an $\epem$ collider.
% since they are obtained from essentially independent data.
%% TJG not sure this is helpful
% Since the transformation is linear,
% both sets of observables are approximately Gaussian distributed.

Alternatively, if we recall that the $\CP$ invariance
conditions at the decay amplitude level are
$\Af = \Abarfbar$ and $\Afbar = \Abarf$,
% we are led to consider the parameters~\cite{ref:cp_uta:uud:charles}
we are led to consider the parameters~\cite{Charles:2004jd}
\begin{equation}
  \label{eq:cp_uta:non-cp-directcp}
  {\cal A}_{f\bar{f}} = 
  \frac{
    \left| \Abarfbar \right|^2 - \left| \Af \right|^2 
  }{
    \left| \Abarfbar \right|^2 + \left| \Af \right|^2
  }
  \hspace{5mm}
  {\rm and}
  \hspace{5mm}
  {\cal A}_{\bar{f}f} = 
  \frac{
    \left| \Abarf \right|^2 - \left| \Afbar \right|^2
  }{
    \left| \Abarf \right|^2 + \left| \Afbar \right|^2
  }.
\end{equation}
These are sometimes considered more physically intuitive parameters
since they characterise $\CP$ violation in decay
in decays with particular topologies.
For example, in the case of $\Bz \to \rho^\pm\pi^\mp$
(choosing $f =  \rho^+\pi^-$ and $\bar{f} = \rho^-\pi^+$),
${\cal A}_{f\bar{f}}$ (also denoted ${\cal A}^{+-}_{\rho\pi}$)
parametrises $\CP$ violation
in decays in which the produced $\rho$ meson does not contain the 
spectator quark,
while ${\cal A}_{\bar{f}f}$ (also denoted ${\cal A}^{-+}_{\rho\pi}$)
parametrises $\CP$ violation in decays in which it does.
Note that we have again followed the sign convention that the asymmetry 
is the difference between the rate involving a $b$ quark and that
involving a $\bar{b}$ quark, \cf\ Eq.~(\ref{eq:cp_uta:pra}). 
Of course, these parameters are not independent of the 
other sets of parameters given above, and can be written
\begin{equation}
  {\cal A}_{f\bar{f}} =
  - \frac{
    \Adirnoncp + C_{f\bar{f}} + \Adirnoncp \Delta C_{f\bar{f}} 
  }{
    1 + \Delta C_{f\bar{f}} + \Adirnoncp C_{f\bar{f}} 
  }
  \hspace{5mm}
  {\rm and}
  \hspace{5mm}
  {\cal A}_{\bar{f}f} =
  \frac{
    - \Adirnoncp + C_{f\bar{f}} + \Adirnoncp \Delta C_{f\bar{f}} 
  }{
    - 1 + \Delta C_{f\bar{f}} + \Adirnoncp C_{f\bar{f}}  
  }.
\end{equation}
They usually exhibit strong correlations.

We now consider the various notations used in experimental studies of
time-dependent $\CP$ asymmetries in decays to non-$\CP$ eigenstates.

\mysubsubsubsection{$\Bz \to D^{*\pm}D^\mp$
}
\label{sec:cp_uta:notations:non_cp:dstard}

The ($\Adirnoncp$, $C_f$, $S_f$, $C_{\bar{f}}$, $S_{\bar{f}}$)
set of parameters was used in early publications by both \babar~\cite{Aubert:2007pa} and \belle~\cite{Aushev:2004uc} (albeit with slightly different notations) in the $D^{*\pm}D^{\mp}$ system ($f = D^{*+}D^-$, $\bar{f} = D^{*-}D^+$).
In their most recent paper on this topic \belle~\cite{Rohrken:2012ta} instead used the parametrisation ($A_{D^*D}$, $S_{D^*D}$, $\Delta S_{D^*D}$, $C_{D^*D}$, $\Delta C_{D^*D}$), while \babar~\cite{Aubert:2008ah} give results in both sets of parameters.
We therefore use the ($A_{D^*D}$, $S_{D^*D}$, $\Delta S_{D^*D}$, $C_{D^*D}$, $\Delta C_{D^*D}$) set.

\mysubsubsubsection{$\Bz \to \rho^{\pm}\pi^\mp$
}
\label{sec:cp_uta:notations:non_cp:rhopi}

In the $\rho^\pm\pi^\mp$ system, the 
($\Adirnoncp$, $C_{f\bar{f}}$, $S_{f\bar{f}}$, $\Delta C_{f\bar{f}}$, 
$\Delta S_{f\bar{f}}$)
set of parameters has been used 
originally by \babar~\cite{Aubert:2003wr} and \belle~\cite{Wang:2004va}, 
in the Q2B approximation; 
the exact names\footnote{
  \babar\ has used the notations
  $A_{\CP}^{\rho\pi}$~\cite{Aubert:2003wr} and 
  ${\cal A}_{\rho\pi}$~\cite{Aubert:2007jn}
  in place of ${\cal A}_{\CP}^{\rho\pi}$.
}
used in this case are
$\left( 
  {\cal A}_{\CP}^{\rho\pi}, C_{\rho\pi}, S_{\rho\pi}, \Delta C_{\rho\pi}, \Delta S_{\rho\pi}
\right)$,
and these names are also used in this document.

Since $\rho^\pm\pi^\mp$ is reconstructed in the final state $\pi^+\pi^-\pi^0$,
the interference between the $\rho$ resonances
can provide additional information about the phases 
(see Sec.~\ref{sec:cp_uta:notations:dalitz}).
Both \babar~\cite{Aubert:2007jn} 
and \belle~\cite{Kusaka:2007dv,Kusaka:2007mj}
have performed time-dependent Dalitz plot analyses, 
from which the weak phase $\alpha$ is directly extracted.
In such an analysis, the measured Q2B parameters are 
also naturally corrected for interference effects.
% See Sec.~\ref{sec:cp_uta:notations:dalitz:pipipi0}.

\mysubsubsubsection{$\Bz \to D^{\mp}\pi^{\pm}, D^{*\mp}\pi^{\pm}, D^{\mp}\rho^{\pm}$
}
\label{sec:cp_uta:notations:non_cp:dstarpi}

Time-dependent $\CP$ analyses have also been performed for the
final states $D^{\mp}\pi^{\pm}$, $D^{*\mp}\pi^{\pm}$ and $D^{\mp}\rho^{\pm}$.
In these theoretically clean cases, no penguin contributions are possible,
so there is no $\CP$ violation in decay.
Furthermore, due to the smallness of the ratio of the magnitudes of the 
suppressed ($b \to u$) and favoured ($b \to c$) amplitudes (denoted $R_f$),
to a very good approximation, $C_f = - C_{\bar{f}} = 1$
(using $f = D^{(*)-}h^+$, $\bar{f} = D^{(*)+}h^-$ $h = \pi,\rho$),
and the coefficients of the sine terms are given by
\begin{equation}
  S_f = - 2 R_f \sin( \phi_{\rm mix} + \phi_{\rm dec} - \delta_f )
  \quad
  \text{and}
  \quad
  S_{\bar{f}} = - 2 R_f \sin( \phi_{\rm mix} + \phi_{\rm dec} + \delta_f ).
\end{equation}
Thus weak phase information can be cleanly obtained from measurements
of $S_f$ and $S_{\bar{f}}$, 
although external information on at least one of $R_f$ or $\delta_f$ is necessary.
(Note that $\phi_{\rm mix} + \phi_{\rm dec} = 2\beta + \gamma \equiv 2\phi_1 + \phi_3$ for all the decay modes 
in question, while $R_f$ and $\delta_f$ depend on the decay mode.)

Again, different notations have been used in the literature.
\babar~\cite{Aubert:2006tw,Aubert:2005yf}
defines the time-dependent probability function by
\begin{equation}
  f^\pm (\eta, \Delta t) = \frac{e^{-|\Delta t|/\tau}}{4\tau} 
  \left[  
    1 \mp S_\zeta \sin (\Delta m \Delta t) \mp \eta C_\zeta \cos(\Delta m \Delta t) 
  \right],
\end{equation} 
where the upper (lower) sign corresponds to the tagging meson being a $\Bz$ ($\Bzb$). 
%Note here that a tagging $\Bz$ ($\Bzb$) corresponds to $-S_\zeta$ ($+S_\zeta$).
The parameters $\eta$ and $\zeta$ take the values $+1$ and $+$ ($-1$ and $-$) 
when the final state is, \eg, $D^-\pi^+$ ($D^+\pi^-$). 
However, in the fit, the substitutions $C_\zeta = 1$ and 
$S_\zeta = a \mp \eta b_i - \eta c_i$ are made, where the subscript $i$ denotes tagging category.
Neglecting $b$ terms, 
\begin{equation}
  \label{eq:cp_uta:aandc}
    S_+ = a - c \quad \text{and} \quad S_- = a + c \ \Leftrightarrow \ 
    a = (S_+ + S_-)/2 \quad \text{and} \quad c = (S_- - S_+)/2 \, ,
\end{equation}
in analogy to the parameters of Eq.~(\ref{eq:cp_uta:non-cp-s_and_deltas}).
% The subscript $i$ denotes the tagging category. 
These are motivated by the possibility of 
$\CP$ violation on the tag side~\cite{Long:2003wq}, 
which is absent for semileptonic $\B$ decays (mostly lepton tags). 
The parameter $a$ is not affected by tag side $\CP$ violation. 
The parameter $b$ only depends on tag side $\CP$ violation parameters 
and is not directly useful for determining UT angles.
A clean interpretation of the $c$ parameter is only possible for 
lepton-tagged events,
so the \babar\ measurements report $c$ measured with those events only.

The parameters used by \belle\ in the analysis using 
partially reconstructed $\B$ decays~\cite{Bahinipati:2011yq}, 
are similar to the $S_\zeta$ parameters defined above. 
However, in the \belle\ convention, 
a tagging $\Bz$ corresponds to a $+$ sign in front of the sine coefficient; 
furthermore the correspondence between the super/subscript 
and the final state is opposite, so that $S_\pm$ (\babar) = $- S^\mp$ (\belle). 
In this analysis, only lepton tags are used, 
so there is no effect from tag side $\CP$ violation. 
In the \belle\ analysis using 
fully reconstructed $\B$ decays~\cite{Ronga:2006hv}, 
this effect is measured and taken into account using $\Dstar \ell \nu$ decays; 
in neither \belle\ analysis are the $a$, $b$ and $c$ parameters used. 
In the latter case, the measured parameters are 
$2 R_{D^{(*)}\pi} \sin( 2\phi_1 + \phi_3 \pm \delta_{D^{(*)}\pi} )$; 
the definition is such that 
$S^\pm$ (\belle) = $- 2 R_{\Dstar \pi} \sin( 2\phi_1 + \phi_3 \pm \delta_{\Dstar \pi} )$. 
However, the definition includes an 
angular momentum factor $(-1)^L$~\cite{Fleischer:2003yb}, 
and so for the results in the $D\pi$ system, 
there is an additional factor of $-1$ in the conversion.

Explicitly, the conversion then reads as given in 
Table~\ref{tab:cp_uta:notations:non_cp:dstarpi}, 
where we have neglected the $b_i$ terms used by \babar
(which are zero in the absence of tag side $\CP$ violation).
For the averages in this document,
we use the $a$ and $c$ parameters,
and give the explicit translations used in 
Table~\ref{tab:cp_uta:notations:non_cp:dstarpi2}.
It is to be fervently hoped that the experiments will
converge on a common notation in future.

\begin{table}
  \begin{center} 
    \caption{
      Conversion between the various notations used for 
      $\CP$ violation parameters in the 
      $D^{\pm}\pi^{\mp}$, $D^{*\pm}\pi^{\mp}$ and $D^{\pm}\rho^{\mp}$ systems.
      The $b_i$ terms used by \babar\ have been neglected.
      Recall that $\left( \alpha, \beta, \gamma \right) \equiv \left( \phi_2, \phi_1, \phi_3 \right)$.
    }
    \vspace{0.2cm}
    \setlength{\tabcolsep}{0.0pc}
    \begin{tabular*}{\textwidth}{@{\extracolsep{\fill}}cccc} \hline 
      & \babar\ & \belle\ partial rec. & \belle\ full rec. \\
      \hline
      $S_{D^+\pi^-}$    & $- S_- = - (a + c_i)$ &  ---  &
      $\phantom{-}2 R_{D\pi} \sin( 2\phi_1 + \phi_3 + \delta_{D\pi} )$ \\
      $S_{D^-\pi^+}$    & $- S_+ = - (a - c_i)$ &  ---  &
      $\phantom{-}2 R_{D\pi} \sin( 2\phi_1 + \phi_3 - \delta_{D\pi} )$ \\
      $S_{D^{*+}\pi^-}$ & $- S_- = - (a + c_i)$ & $S^+$ &   
      $- 2 R_{\Dstar \pi} \sin( 2\phi_1 + \phi_3 + \delta_{\Dstar \pi} )$ \\
      $S_{D^{*-}\pi^+}$ & $- S_+ = - (a - c_i)$ & $S^-$ &
      $- 2 R_{\Dstar \pi} \sin( 2\phi_1 + \phi_3 - \delta_{\Dstar \pi} )$ \\
      $S_{D^+\rho^-}$    & $- S_- = - (a + c_i)$ &  ---  &  ---  \\
      $S_{D^-\rho^+}$    & $- S_+ = - (a - c_i)$ &  ---  &  ---  \\
      \hline 
    \end{tabular*}
    \label{tab:cp_uta:notations:non_cp:dstarpi}
  \end{center}
\end{table}
   
\begin{table}
  \begin{center} 
    \caption{
      Translations used to convert the parameters measured by \belle
      to the parameters used for averaging in this document.
      The angular momentum factor $L$ is $-1$ for $\Dstar\pi$ and $+1$ for $D\pi$.
      Recall that $\left( \alpha, \beta, \gamma \right) \equiv \left( \phi_2, \phi_1, \phi_3 \right)$.
    }
    \vspace{0.2cm}
    \setlength{\tabcolsep}{0.0pc}
    \begin{tabular*}{\textwidth}{@{\extracolsep{\fill}}ccc} \hline 
        & $\Dstar\pi$ partial rec. & $D^{(*)}\pi$ full rec. \\
        \hline
        $a$ & $- (S^+ + S^-)$ &
        $\frac{1}{2} (-1)^{L+1}
        \left(
          2 R_{D^{(*)}\pi} \sin( 2\phi_1 + \phi_3 + \delta_{D^{(*)}\pi} ) + 
          2 R_{D^{(*)}\pi} \sin( 2\phi_1 + \phi_3 - \delta_{D^{(*)}\pi} )
        \right)$ \\
        $c$ & $- (S^+ - S^-)$ & 
        $\frac{1}{2} (-1)^{L+1}
        \left(
          2 R_{D^{(*)}\pi} \sin( 2\phi_1 + \phi_3 + \delta_{D^{(*)}\pi} ) -
          2 R_{D^{(*)}\pi} \sin( 2\phi_1 + \phi_3 - \delta_{D^{(*)}\pi} )
        \right)$ \\
        \hline 
      \end{tabular*}
    \label{tab:cp_uta:notations:non_cp:dstarpi2}
  \end{center}
\end{table}

\mysubsubsubsection{$\Bs \to D_s^{\mp}K^\pm$}
\label{sec:cp_uta:notations:non_cp:dsk}

The phenomenology of $\Bs \to D_s^{\mp}K^\pm$ decays is similar to that for $\Bz \to D^{\mp}\pi^{\pm}$, with some important caveats.
The two amplitudes $b \to u$ and $b \to c$ amplitudes have the same level of Cabibbo-suppression (\ie\ are of the same order in $\lambda$) though the former is suppressed by $\sqrt{\rho^2+\eta^2}$.
The large value of the ratio $R$ of their magnitudes allows it to be determined from data, as the deviation of $C_f$ and $C_{\bar{f}}$ from unity (in magnitude) can be observed.
Moreover, the non-zero value of $\Delta \Gamma_s$ allows the determination of additional terms, $A^{\Delta\Gamma}_f$ and $A^{\Delta\Gamma}_{\bar{f}}$ (see Sec.~\ref{sec:cp_uta:notations:Bs}), that break ambiguities in the solutions for $\phi_{\rm mix} + \phi_{\rm dec}$, which for $\Bs \to D_s^{\mp}K^\pm$ decays is equal to $\gamma-2\beta_s$.

LHCb~\cite{Aaij:2014fba} has performed such an analysis with $\Bs \to D_s^{\mp}K^\pm$ decays.
The absence of \CP violation in decay is assumed, and the parameters that are determined from the fit are labelled $C$, $A^{\Delta\Gamma}$, $\bar{A}{}^{\Delta\Gamma}$, $S$, $\bar{S}$.
These are trivially related to the definitions used in this section.

\mysubsubsubsection{Time-dependent asymmetries in radiative $\B$ decays
}
\label{sec:cp_uta:notations:non_cp:radiative}

As a special case of decays to non-$\CP$ eigenstates,
let us consider radiative $\B$ decays.
Here, the emitted photon has a distinct helicity,
which is in principle observable, but in practice is not usually measured.
Thus the measured time-dependent decay rates for \Bz\ decays 
are given by~\cite{Atwood:1997zr,Atwood:2004jj}
\begin{eqnarray}
  \label{eq:cp_uta:non-cp-radiative1}
  \Gamma_{\Bzb \to X \gamma} (\Delta t) & = &
  \Gamma_{\Bzb \to X \gamma_L} (\Delta t) + \Gamma_{\Bzb \to X \gamma_R} (\Delta t) \\ \nonumber
  & = &
  \frac{e^{-\left| \Delta t \right| / \tau(\Bz)}}{4\tau(\Bz)} 
  \left[ 
    1 + 
    \left( S_L + S_R \right) \sin(\Delta m \Delta t) - 
    \left( C_L + C_R \right) \cos(\Delta m \Delta t) 
  \right],
  \\
  \label{eq:cp_uta:non-cp-radiative2}
  \Gamma_{\Bz \to X \gamma} (\Delta t) & = & 
  \Gamma_{\Bz \to X \gamma_L} (\Delta t) + \Gamma_{\Bz \to X \gamma_R} (\Delta t) \\ \nonumber 
  & = &
  \frac{e^{-\left| \Delta t \right| / \tau(\Bz)}}{4\tau(\Bz)} 
  \left[ 
    1 - 
    \left( S_L + S_R \right) \sin(\Delta m \Delta t) + 
    \left( C_L + C_R \right) \cos(\Delta m \Delta t) 
  \right],
\end{eqnarray}
where in place of the subscripts $f$ and $\bar{f}$ we have used $L$ and $R$
to indicate the photon helicity.
In order for interference between decays with and without $\Bz$-$\Bzb$ mixing
to occur, the $X$ system must not be flavour-specific,
\eg, in case of $\Bz \to K^{*0}\gamma$, the final state must be $\KS \pi^0 \gamma$.
The sign of the sine term depends on the $C$ eigenvalue of the $X$ system.
At leading order, the photons from 
$b \to q \gamma$ ($\bar{b} \to \bar{q} \gamma$) are predominantly
left (right) polarised, with corrections of order of $m_q/m_b$,
thus interference effects are suppressed.
Higher-order effects can lead to corrections of order 
$\Lambda_{\rm QCD}/m_b$~\cite{Grinstein:2004uu,Grinstein:2005nu},
though explicit calculations indicate that such corrections may be small for exclusive final states~\cite{Matsumori:2005ax,Ball:2006cva}.
% In the case of $b \to s \gamma$, a single weak phase dominates,
% so one expects that $C_L + C_R \approx 0$ and
% $\left| S_L + S_R \right| \lesssim \frac{2 m_s}{m_b} \sin \left( 2\beta \right)$.
% In the case of $b \to d \gamma$, more than one weak phase is possible,
% so direct $\CP$ violation can occur, 
% but the $S$ term should be vanishingly small.
The predicted smallness of the $S$ terms in the Standard Model
results in sensitivity to new physics contributions.

The formalism discussed above is valid for any radiative decay to a final state where the hadronic system is an eigenstate of $C$.
In addition to $\KS\piz\gamma$, experiments have presented results using $\Bz$ decays to $\KS\eta\gamma$, $\KS\rho^0\gamma$ and $\KS\phi\gamma$.
For the case of the $\KS\rho^0\gamma$ final state, particular care is needed, as due to the non-negligible width of the $\rho^0$ meson, decays selected as $\Bz \to \KS\rho^0\gamma$ can include a significant contribution from $K^{*\pm}\pimp\gamma$ decays, which are flavour-specific and do not have the same oscillation phenomenology. 
It is therefore necessary to correct the fitted asymmetry parameter for a ``dilution factor''.

In the case of radiative \Bs\ decays, the time-dependent decay rates of Eqs.~(\ref{eq:cp_uta:non-cp-radiative1}) and~(\ref{eq:cp_uta:non-cp-radiative2}) must be modified, in a similar way as discussed in Sec.~\ref{sec:cp_uta:notations:Bs}, to account for the non-zero value of \DGs.
Thus, for decays such as $\Bs\to\phi\gamma$, there is an additional observable, $A^{\Delta \Gamma}_{\phi\gamma}$, which can be determined from an untagged effective lifetime measurement~\cite{Muheim:2008vu}.

\mysubsubsection{Asymmetries in $\B \to \DorDstar K^{(*)}$ decays
}
\label{sec:cp_uta:notations:cus}

$\CP$ asymmetries in $\B \to \DorDstar K^{(*)}$ decays are sensitive to $\gamma$.
The neutral $D^{(*)}$ meson produced 
% in the decay $\Bm \to \DorDstar K^{(*)-}$
is an admixture of $\DorDstarz$ (produced by a $b \to c$ transition) and 
$\DorDstarzb$ (produced by a colour-suppressed $b \to u$ transition) states.
If the final state is chosen so that both $\DorDstarz$ and $\DorDstarzb$ 
can contribute, the two amplitudes interfere,
and the resulting observables are sensitive to $\gamma$, 
the relative weak phase between 
the two $\B$ decay amplitudes~\cite{Bigi:1988ym}.
% \footnote{
%   The same is true for $\Bzb \to \DorDstar \bar{K}^{(*)0}$ decays,
%   where both $b \to c$ and $b \to u$ transitions are colour-suppressed.
% }
Various methods have been proposed to exploit this interference,
including those where the neutral $D$ meson is reconstructed 
as a $\CP$ eigenstate (GLW)~\cite{Gronau:1990ra,Gronau:1991dp},
in a suppressed final state (ADS)~\cite{Atwood:1996ci,Atwood:2000ck},
or in a self-conjugate three-body final state, 
such as $\KS \pi^+\pi^-$ (GGSZ or Dalitz)~\cite{Giri:2003ty,Poluektov:2004mf}.
It should be emphasised that while each method 
differs in the choice of $D$ decay,
they are all sensitive to the same parameters of the $B$ decay,
and can be considered as variations of the same technique.
% Each of these approaches, while theoretically clean,
% has some difficulty for $\gamma$ extraction due to intrinsic ambiguities
% or model dependence;
% these can be overcome by combining the results from different techniques,
% and including other modes, 
% such as $\Bmp \to \Dstar \Kmp$ and $\Bmp \to D \Kstarmp$.

Consider the case of $\Bmp \to D \Kmp$, with $D$ decaying to a final state $f$, which is accessible from both $\Dz$ and $\Dzb$.
We can write the decay rates for $\Bm$ and $\Bp$ ($\Gamma_\mp$), 
the charge averaged rate ($\Gamma = (\Gamma_- + \Gamma_+)/2$)
and the charge asymmetry 
($A = (\Gamma_- - \Gamma_+)/(\Gamma_- + \Gamma_+)$, see Eq.~(\ref{eq:cp_uta:pra})) as 
\begin{eqnarray}
  \label{eq:cp_uta:dk:rate_def}
  \Gamma_\mp  & \propto & 
  r_B^2 + r_D^2 + 2 r_B r_D \cos \left( \delta_B + \delta_D \mp \gamma \right), \\
  \label{eq:cp_uta:dk:av_rate_def}
  \Gamma & \propto &  
  r_B^2 + r_D^2 + 2 r_B r_D \cos \left( \delta_B + \delta_D \right) \cos \left( \gamma \right), \\
  \label{eq:cp_uta:dk:acp_def}
  A & = & 
  \frac{
    2 r_B r_D \sin \left( \delta_B + \delta_D \right) \sin \left( \gamma \right)
  }{
    r_B^2 + r_D^2 + 2 r_B r_D \cos \left( \delta_B + \delta_D \right) \cos \left( \gamma \right)  
  },
\end{eqnarray}
where the ratio of $\B$ decay amplitudes\footnote{
  Note that here we use the notation $r_B$ to denote the ratio
  of $\B$ decay amplitudes, 
  whereas in Sec.~\ref{sec:cp_uta:notations:non_cp:dstarpi} 
  we used, \eg, $R_{D\pi}$, for a rather similar quantity.
  The reason is that here we need to be concerned also with 
  $D$ decay amplitudes,
  and so it is convenient to use the subscript to denote the decaying particle.
  Hopefully, using $r$ in place of $R$ will reduce the potential for confusion.
} 
is usually defined to be less than one,
\begin{equation}
  \label{eq:cp_uta:dk:rb_def}
  r_B = 
  \left|   
    \frac{A\left( \Bm \to \Dzb K^- \right)}{A\left( \Bm \to \Dz  K^- \right)}
  \right| = 
  \left|   
    \frac{A\left( \Bp \to \Dz K^+ \right)}{A\left( \Bp \to \Dzb  K^+ \right)}
  \right| ,
\end{equation}
and the ratio of $D$ decay amplitudes is correspondingly defined by
\begin{equation}
  \label{eq:cp_uta:dk:rd_def}
  r_D = 
  \left| 
    \frac{A\left( \Dz  \to f \right)}{A\left( \Dzb \to f \right)}
  \right| .
\end{equation}
The relation between $\Bm$ and $\Bp$ amplitudes given in Eq.~(\ref{eq:cp_uta:dk:rb_def}) is a result of their being only one weak phase contributing to each amplitude in the Standard Model, which is the source of the theoretical cleanliness of this approach to measure $\gamma$~\cite{Brod:2013sga}.
The strong phase differences between the $\B$ and $D$ decay amplitudes 
are given by $\delta_B$ and $\delta_D$, respectively.
% Note that $r_B$ and $\delta_B$ take different values for different $\B$ decays;
% the values for $\Bm \to D \Km$ and $\Bm \to \Dstar \Km$ are not the same.
% On the other hand, the value of $r_{D^{(*)}}$ depends only on the final state of
% the $D$ decay, since the amplitudes for $D^{*0}$ and $\bar{D}^{*0}$ decays
% to $D^{0}$ and $\bar{D}^{0}$, respectively,
% via emission of either a pion or a photon, will cancel in the ratio.
The values of $r_D$ and $\delta_D$ depend on the final state $f$:
for the GLW analysis, $r_D = 1$ and $\delta_D$ is trivial (either zero or $\pi$);
for other modes, values of $r_D$ and $\delta_D$ are not trivial and for multibody final states they vary across the phase space.
This can be quantified either by an explicit $D$ decay amplitude model or by model-independent information.
In the case that the multibody final state is treated inclusively, the formalism is modified by the inclusion of a coherence factor, usually denoted $\kappa$, while $r_D$ and $\delta_D$ become effectively parameters corresponding to amplitude-weighted averages across the phase space.

Note that, for given values of $r_B$ and $r_D$, 
the maximum size of $A$ (at $\sin \left( \delta_B + \delta_D \right) = 1$)
is $2 r_B r_D \sin \left( \gamma \right) / \left( r_B^2 + r_D^2 \right)$.
Thus even for $D$ decay modes with small $r_D$, 
large asymmetries, and hence sensitivity to $\gamma$, 
may occur for $B$ decay modes with similar values of $r_B$.
For this reason, the ADS analysis of the decay $B^\mp \to D \pi^\mp$ is also of interest.

The expressions of Eq.~(\ref{eq:cp_uta:dk:rate_def})--(\ref{eq:cp_uta:dk:rd_def}) are for a specific point in phase space, and therefore are relevant where both $B$ and $D$ decays are to two-body final states.
Additional coherence factors enter the expressions when the $B$ decay is to a multibody final state (further discussion of multibody $D$ decays can be found below).
In particular, experiments have studied $B^+ \to DK^*(892)^+$, $B^0 \to DK^*(892)^0$ and $B^+ \to DK^+\pi^+\pi^-$ decays.
Considering, for concreteness, the $B \to DK^*(892)$ case, the non-negligible width of the $K^*(892)$ resonance implies that contributions from other $B \to DK\pi$ decays can pass the selection requirements.
Their effect on the Q2B analysis can be accounted for with a coherence factor~\cite{Gronau:2002mu}, usually denoted $\kappa$, which tends to unity in the limit that the $K^*(892)$ resonance is the only signal amplitude contributing in the selected region of phase space.
In this case, the hadronic parameters $r_B$ and $\delta_B$ become effectively weighted averages across the selected phase space of the magnitude ratio and relative strong phase between the CKM-suppressed and -favoured amplitudes; these effective parameters are denoted $\bar{r}_B$ and $\bar{\delta}_B$ (the notations $r_s$, $\delta_s$ and $r_S$, $\delta_S$ are also found in the literature).
An alternative, and in certain cases more advantageous, approach is Dalitz plot analysis of the full $B \to DK\pi$ phase space~\cite{Gershon:2008pe,Gershon:2009qc}.

\newpage
\mysubsubsubsection{$\B \to \DorDstar K^{(*)}$ with $D \to$ \CP\ eigenstate decays
}
\label{sec:cp_uta:notations:cus:glw}

In the GLW analysis, the measured quantities are the 
partial rate asymmetry and the charge averaged rate,
which are measured both for $\CP$-even and $\CP$-odd $D$ decays.
The latter is defined as 
\begin{equation}
  \label{eq:cp_uta:dk:glw-rdef}
  R_{\CP} = 
  \frac{2 \, \Gamma \left( \Bp \to D_{\CP} \Kp  \right)}
  {\Gamma\left( \Bp \to \Dzb \Kp \right)} \, .
\end{equation}
It is experimentally convenient to measure $R_{\CP}$ using a double ratio,
\begin{equation}
  \label{eq:cp_uta:dk:double_ratio}
  R_{\CP} = 
  \frac{
    \Gamma\left( \Bp \to D_{\CP} \Kp  \right) \, / \, \Gamma\left( \Bp \to \Dzb \Kp \right)
  }{
    \Gamma\left( \Bp \to D_{\CP} \pip \right) \, / \, \Gamma\left( \Bp \to \Dzb \pip \right)
  }
\end{equation}
that is normalised both to the rate for the favoured $\Dzb \to \Kp\pim$ decay, 
and to the equivalent quantities for $\Bp \to D\pip$ decays
(charge conjugate processes are implicitly included in 
Eqs.~(\ref{eq:cp_uta:dk:glw-rdef}) and~(\ref{eq:cp_uta:dk:double_ratio})).
In this way the constant of proportionality drops out of 
Eq.~(\ref{eq:cp_uta:dk:av_rate_def}).
Eq.~(\ref{eq:cp_uta:dk:double_ratio}) is exact in the limit that the
contribution of the $b \to u$ decay amplitude to $\Bp \to D \pip$ vanishes and
when the flavour-specific rates $\Gamma\left( \Bp \to \Dzb h^+ \right)$ ($h =
\pi,K$) are determined using appropriately flavour-specific $D$ decays.
In reality, the decay $D \to K\pi$ is used, leading to a small source of systematic uncertainty.
The \CP\ asymmetry is defined as
\begin{equation}
  \label{eq:cp_uta:dk:glw-adef}
  A_{\CP} = \frac{
    \Gamma\left(\Bm\to D_{\CP}\Km\right) - \Gamma\left(\Bp\to D_{\CP}\Kp\right)
  }{
    \Gamma\left(\Bm\to D_{\CP}\Km\right) + \Gamma\left(\Bp\to D_{\CP}\Kp\right)
  } \, .
\end{equation}

\mysubsubsubsection{$\B \to \DorDstar K^{(*)}$ with $D \to$ non-\CP\ eigenstate two-body decays
}
\label{sec:cp_uta:notations:cus:ads}

For the ADS analysis, based on a suppressed $D \to f$ decay,
the measured quantities are again the partial rate asymmetry, 
and the charge averaged rate.
In this case it is sufficient to measure the rate in a single ratio
(normalised to the favoured $D \to \bar{f}$ decay)
since potential systematic uncertainties related to detection cancel naturally;
the observed quantity is then
\begin{equation}
  \label{eq:cp_uta:dk:r_ads}
  R_{\rm ADS} = 
  \frac{
    \Gamma \left( \Bm \to \left[\,f\,\right]_D \Km \right) + 
    \Gamma \left( \Bp \to \left[\,\bar{f}\,\right]_D \Kp \right)
  }{
    \Gamma \left( \Bm \to \left[\,\bar{f}\,\right]_D \Km \right) +
    \Gamma \left( \Bp \to \left[\,f\,\right]_D \Kp \right)
  } \, ,
\end{equation}
where the inclusion of charge-conjugate modes has been made explicit.
The \CP\ asymmetry is defined as
\begin{equation}
  \label{eq:cp_uta:dk:a_ads}
  A_{\rm ADS} = 
  \frac{
    \Gamma\left(\Bm\to\left[\,f\,\right]_D\Km\right)-
    \Gamma\left(\Bp\to\left[\,f\,\right]_D\Kp\right)
  }{
    \Gamma\left(\Bm\to\left[\,f\,\right]_D\Km\right)+
    \Gamma\left(\Bp\to\left[\,f\,\right]_D\Kp\right)
  } \, .
\end{equation}
Since the uncertainty of $A_{\rm ADS}$ depends on the central value of $R_{\rm ADS}$, for some statistical treatments it is preferable to use an alternative pair of parameters~\cite{Bondar:2004bi}
\begin{equation}
  R_- = \frac{
    \Gamma \left( \Bm \to \left[\,f\,\right]_D \Km \right)
  }{
    \Gamma \left( \Bm \to \left[\,\bar{f}\,\right]_D \Km \right)
  } \, 
  \hspace{5mm}
  R_+ = \frac{
    \Gamma \left( \Bp \to \left[\,\bar{f}\,\right]_D \Kp \right)
  }{
    \Gamma \left( \Bp \to \left[\,f\,\right]_D \Kp \right)
  } \, ,
\end{equation}
where there is no implied inclusion of charge-conjugate processes.
These parameters are statistically uncorrelated but may be affected by common sources of systematic uncertainty.
We use the $(R_{\rm ADS}, A_{\rm ADS})$ set in our compilation where available.

In the ADS analysis, there are two additional unknowns ($r_D$ and $\delta_D$) compared to the GLW case.  
However, the value of $r_D$ can be measured using decays of $D$ mesons of known flavour, and $\delta_D$ can be measured from interference effects in decays of quantum-correlated $D\bar{D}$ pairs produced at the $\psi(3770)$ resonance.
More generally, one needs access to two different linear admixtures of $D^0$ and $\bar{D}{}^0$ states in order to determine the relative phase: one such sample can be flavour tagged $D$ mesons, which are available in abundant quantities in many experiments; the other can be \CP-tagged $D$ mesons from $\psi(3770)$ decays or could be mixed $D$ mesons (or could be the combination of $D^0$ and $\bar{D}{}^0$ that is found in $B \to DK$ decays).
In fact, the most precise information on both $r_D$ and $\delta_D$ currently comes from global fits on charm mixing parameters, as discussed in Sec.~\ref{sec:charm:mixcpv}.

The relation of $A_{\rm ADS}$ to the underlying parameters given in Eq.~(\ref{eq:cp_uta:dk:acp_def}) and Table~\ref{tab:cp_uta:notations:dk} is exact for a two-body $D$ decay.  
For multibody decays, a similar formalism can be used with the introduction of a coherence factor~\cite{Atwood:2003mj}.
This is most appropriate for doubly-Cabibbo-suppressed decays to non-self-conjugate final states, but can also be modified for use with singly-Cabibbo-suppressed decays~\cite{Grossman:2002aq}.
For multibody self-conjugate final states, such as $\KS\pi^+\pi^-$, a Dalitz plot analysis (discussed below) is often more appropriate.
However, in certain cases where the final state can be approximated as a \CP\ eigenstate, a modified version of the GLW formalism can be used~\cite{Nayak:2014tea}.
In such cases the observables are denoted $A_{\rm qGLW}$ and $R_{\rm qGLW}$ to indicate that the final state is not a pure \CP eigenstate.

\mysubsubsubsection{$\B \to \DorDstar K^{(*)}$ with $D \to$ multibody final state decays
}
\label{sec:cp_uta:notations:cus:ggsz}

In the Dalitz plot (or GGSZ) analysis of $D$ decays to multibody self-conjugate final states, once a model is assumed for the $D$ decay, 
which gives the values of $r_D$ and $\delta_D$ across the Dalitz plot,
it is possible to perform a simultaneous fit to the $B^+$ and $B^-$ samples 
and directly extract $\gamma$, $r_B$ and $\delta_B$.
However, the uncertainties on the phases depend approximately inversely on $r_B$.
Furthermore, $r_B$ is positive definite and therefore tends to be overestimated (unless $\sigma(r_B) \ll r_B$),
which leads to an underestimation of the uncertainty on $\gamma$ that must be
corrected statistically. % treatment is necessary to correct for this bias.
An alternative approach is to extract from the data the ``Cartesian''
variables
\begin{equation}
  \label{eq:cp_uta:cartesian}
  \left( x_\pm, y_\pm \right) = 
  \left( \Re(r_B e^{i(\delta_B\pm\gamma)}), \Im(r_B e^{i(\delta_B\pm\gamma)}) \right) = 
  \left( r_B \cos(\delta_B\pm\gamma), r_B \sin(\delta_B\pm\gamma) \right).
\end{equation}
These variables tend to be statistically well-behaved, and are therefore appropriate for combination of results.
The pairs of variables $\left( x_\pm, y_\pm \right)$ can be extracted
from independent fits of the $B^\pm$ data samples.

The assumption of a model for the $D$ decay leads to a non-negligible, and hard to quantify, source of uncertainty.
To obviate this, it is possible to use instead a model-independent approach, in which the Dalitz plot (or, more generally, the phase space) is binned~\cite{Giri:2003ty,Bondar:2005ki,Bondar:2008hh}.
In this case, hadronic parameters describing the average strong phase difference in each bin between the suppressed and favoured decay amplitudes enter the equations.
These parameters can be determined from interference effects in decays of quantum-correlated $D\bar{D}$ pairs produced at the $\psi(3770)$ resonance.
Measurements of such parameters have been made for various different hadronic $D$ decays by CLEO-c and BESIII.

If a multibody decay is dominated by one $\CP$ state, there will be additional sensitivity to $\gamma$ in the numbers of events in the $B^\pm$ data samples.
This can be taken into account in various ways.
One possibility is to perform a GLW-like analysis, as mentioned above.
An alternative approach proceeds by defining $z_\pm = x_\pm + i y_\pm$
and \mbox{$x_0 = - \int \Re \left[ f(s_1,s_2)f^*(s_2,s_1) \right] ds_1ds_2$},
where $s_1, s_2$ are the coordinates of invariant mass squared that
define the Dalitz plot and $f$ is the complex amplitude for $D$ decay
as a function of the Dalitz plot coordinates.\footnote{
  The $x_0$ parameter gives a model-dependent measure of the net \CP content of the final state~\cite{Nayak:2014tea,Gershon:2015xra}.
  It is closely related to the $c_i$ parameters of the model dependent Dalitz plot analysis~\cite{Giri:2003ty,Bondar:2005ki,Bondar:2008hh},
  and the coherence factor of inclusive ADS-type analyses~\cite{Atwood:2003mj}, integrated over the entire Dalitz plot.
}
The fitted parameters ($\rho^\pm, \theta^\pm$) are then defined by
\begin{equation}
  \rho^\pm e^{i \theta^\pm} = z_\pm - x_0 \, .
\end{equation}
Note that the yields of $B^\pm$ decays are proportional 
to $1 + (\rho^\pm)^2 - (x_0)^2$. 
This choice of variables has been used by \babar\ in the analysis of
$\Bp \to D\Kp$ with $D \to \pi^+\pi^-\pi^0$~\cite{Aubert:2007ii};
for this $D$ decay, and with the assumed amplitude model, a value of $x_0 = 0.850$ is obtained.
%More recently, it has been noted that $D \to \pi^+\pi^-\pi^0$ can be used in a GLW-like analysis~\cite{Nayak:2014tea}.

The relations between the measured quantities and the
underlying parameters are summarised in Table~\ref{tab:cp_uta:notations:dk}.
It must be emphasised that the hadronic factors $r_B$ and $\delta_B$ 
are different, in general, for each $\B$ decay mode.

\begin{table}[htbp]
  \begin{center} 
    \caption{
      Summary of relations between measured and physical parameters 
      in GLW, ADS and Dalitz analyses of $\B \to \DorDstar K^{(*)}$ decays.
    }
    \vspace{0.2cm}
    \setlength{\tabcolsep}{1.0pc}
    \begin{tabular}{cc} \hline 
      \mc{2}{l}{GLW analysis} \\
      $R_{\CP\pm}$ & $1 + r_B^2 \pm 2 r_B \cos \left( \delta_B \right) \cos \left( \gamma \right)$ \\
      $A_{\CP\pm}$ & $\pm 2 r_B \sin \left( \delta_B \right) \sin \left( \gamma \right) / R_{\CP\pm}$ \\
      \hline
      \mc{2}{l}{ADS analysis} \\
      $R_{\rm ADS}$ & $r_B^2 + r_D^2 + 2 r_B r_D \cos \left( \delta_B + \delta_D \right) \cos \left( \gamma \right)$ \\
      $A_{\rm ADS}$ & $2 r_B r_D \sin \left( \delta_B + \delta_D \right) \sin \left( \gamma \right) / R_{\rm ADS}$ \\
      \hline
      \mc{2}{l}{GGSZ Dalitz analysis ($D \to \KS \pi^+\pi^-$)} \\
      $x_\pm$ & $r_B \cos(\delta_B\pm\gamma)$ \\
      $y_\pm$ & $r_B \sin(\delta_B\pm\gamma)$ \\
      \hline
      \mc{2}{l}{Dalitz analysis ($D \to \pi^+\pi^-\pi^0$)} \\
      $\rho^\pm$ & $|z_\pm - x_0|$ \\
      $\theta^\pm$ & $\tan^{-1}(\Im(z_\pm)/(\Re(z_\pm) - x_0))$ \\
      \hline
    \end{tabular}
    \label{tab:cp_uta:notations:dk}
  \end{center}
\end{table}

% Results from model-dependent Dalitz plot fits tend to suffer from significant uncertainties due to the choice of model to describe hadronic effects.
% This can be obviated by a model-independent analysis, in which the Dalitz plot is binned~\cite{Giri:2003ty,Bondar:2005ki,Bondar:2008hh}.  It is then necessary to gain information on effective parameters which describe the average strong phase difference between a certain bin and its conjugate (found by reflecting in the symmetry axis of the Dalitz plot\footnote{Here we restrict the discussion to three-body self conjugate final states such as $\KS\pip\pim$ and $\KS\Kp\Km$, though it can be extended to other modes, including four-body final states.}).
% Such information can be obtained from interference effects in decays of quantum-correlated $D\bar{D}$ pairs produced at the $\psi(3770)$ resonance.

% \afterpage{\clearpage}
\mysubsection{Common inputs and error treatment
}
\label{sec:cp_uta:common_inputs}

The common inputs used for rescaling are listed in 
Table~\ref{tab:cp_uta:common_inputs}.
The $\Bz$ lifetime ($\tau(\Bz)$), mixing parameter ($\Delta m_d$) and relative width difference ($\Delta\Gamma_d / \Gamma_d$)
averages are provided by the HFLAV Lifetimes and Oscillations subgroup (Sec.~\ref{sec:life_mix}).
The fraction of the perpendicularly polarised component 
($\left| A_{\perp} \right|^2$) in $\B \to \jpsi \Kstar(892)$ decays,
which determines the $\CP$ composition in these decays, 
is averaged from results by 
\babar~\cite{Aubert:2007hz}, \belle~\cite{Itoh:2005ks}, CDF~\cite{Acosta:2004gt}, D0~\cite{Abazov:2008jz} and LHCb~\cite{Aaij:2013cma}.
See also the HFLAV $B$ to Charm Decay Parameters subgroup (Sec.~\ref{sec:b2c}).

At present, we only rescale to a common set of input parameters
for modes with reasonably small statistical errors
% ($b \to c\bar{c}s$ and $b \to q\bar{q}s$ transitions).
($b \to c\bar{c}s$ transitions of \Bz\ mesons).
Correlated systematic errors are taken into account in these modes as well.
For all other modes, the effect of such a procedure is currently negligible.

\begin{table}[htbp]
  \begin{center}
    \caption{
      Common inputs used in calculating the averages.
    }
    \vspace{0.2cm}
    \setlength{\tabcolsep}{1.0pc}
    \begin{tabular}{cr@{$\,\pm\,$}l} \hline 
      $\tau(\Bz)$ $({\rm ps})$  & $1.520$ & $0.004$  \\
      $\Delta m_d$ $({\rm ps}^{-1})$ & $0.5064$ & $0.0019$ \\
      $\Delta\Gamma_d / \Gamma_d$ & $-0.002$ & $0.010$ \\
      $\left| A_{\perp} \right|^2 (\jpsi \Kstar)$ & $0.209$ & $0.006$ \\
      \hline
    \end{tabular}
    \label{tab:cp_uta:common_inputs}
  \end{center}
\end{table}

As explained in Sec.~\ref{sec:intro},
we do not apply a rescaling factor on the error of an average
that has $\chi^2/\dof > 1$ 
(unlike the procedure currently used by the PDG~\cite{PDG_2014}).
We provide a confidence level of the fit so that
one can know the consistency of the measurements included in the average,
and attach comments in case some care needs to be taken in the interpretation.
Note that, in general, results obtained from data samples with low statistics
will exhibit some non-Gaussian behaviour.
We average measurements with asymmetric errors 
using the PDG~\cite{PDG_2014} prescription.
In cases where several measurements are correlated
(\eg\ $S_f$ and $C_f$ in measurements of time-dependent $\CP$ violation
in $B$ decays to a particular $\CP$ eigenstate)
we take these into account in the averaging procedure
if the uncertainties are sufficiently Gaussian.
For measurements where one error is given, 
it represents the total error, 
where statistical and systematic uncertainties have been added in quadrature.
If two errors are given, the first is statistical and the second systematic.
If more than two errors are given,
the origin of the additional uncertainty will be explained in the text.

%%%%%%%%
%%%
%%% ccs
%%%
%%%%%%%%
% \afterpage{\clearpage}
\mysubsection{Time-dependent asymmetries in $b \to c\bar{c}s$ transitions
}
\label{sec:cp_uta:ccs}

\mysubsubsection{Time-dependent $\CP$ asymmetries in $b \to c\bar{c}s$ decays to $\CP$ eigenstates
}
\label{sec:cp_uta:ccs:cp_eigen}

\begin{figure}[htbp]
  \begin{center}
    \resizebox{0.51\textwidth}{!}{
      \includegraphics{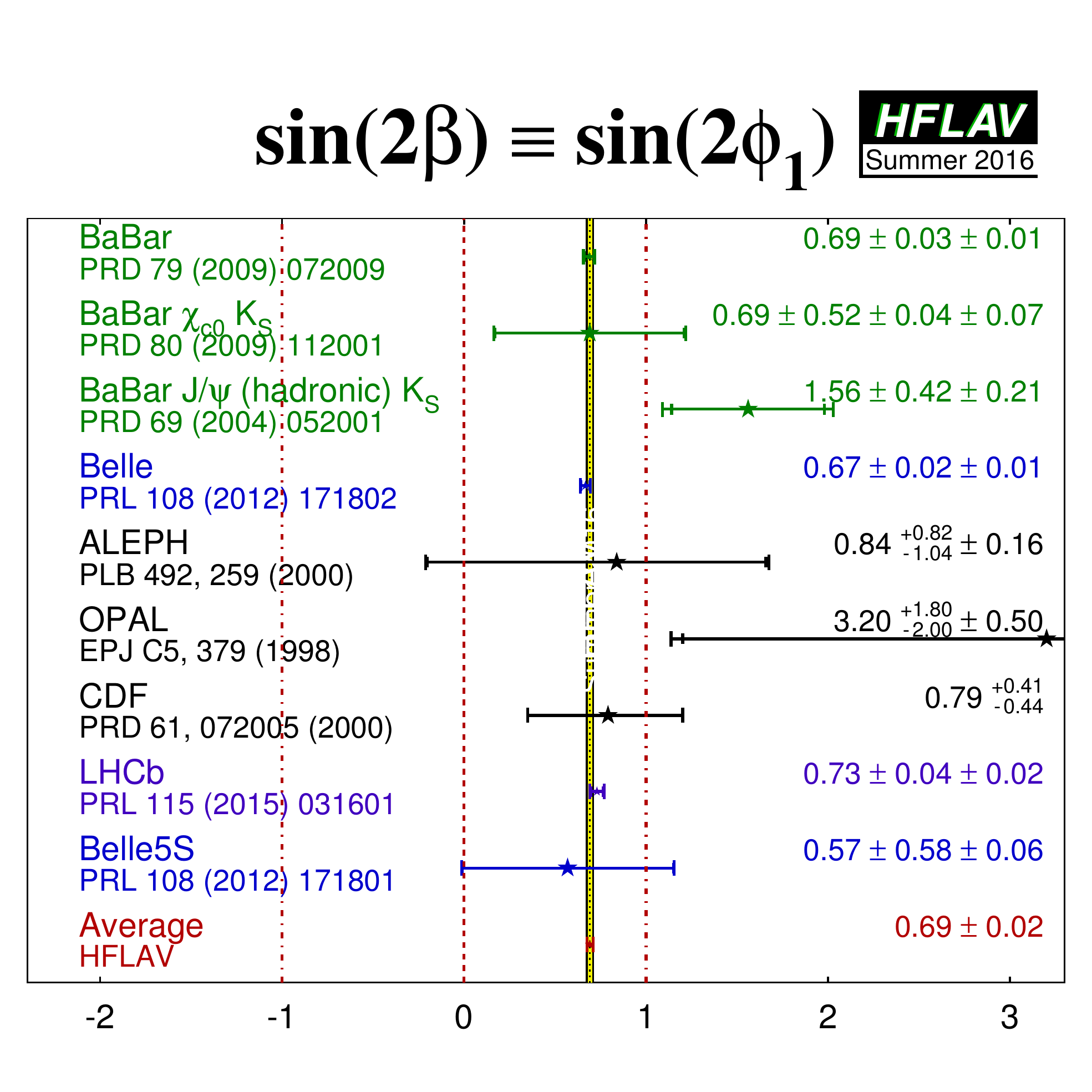}
    }
    \hfill
    \resizebox{0.48\textwidth}{!}{
      \includegraphics{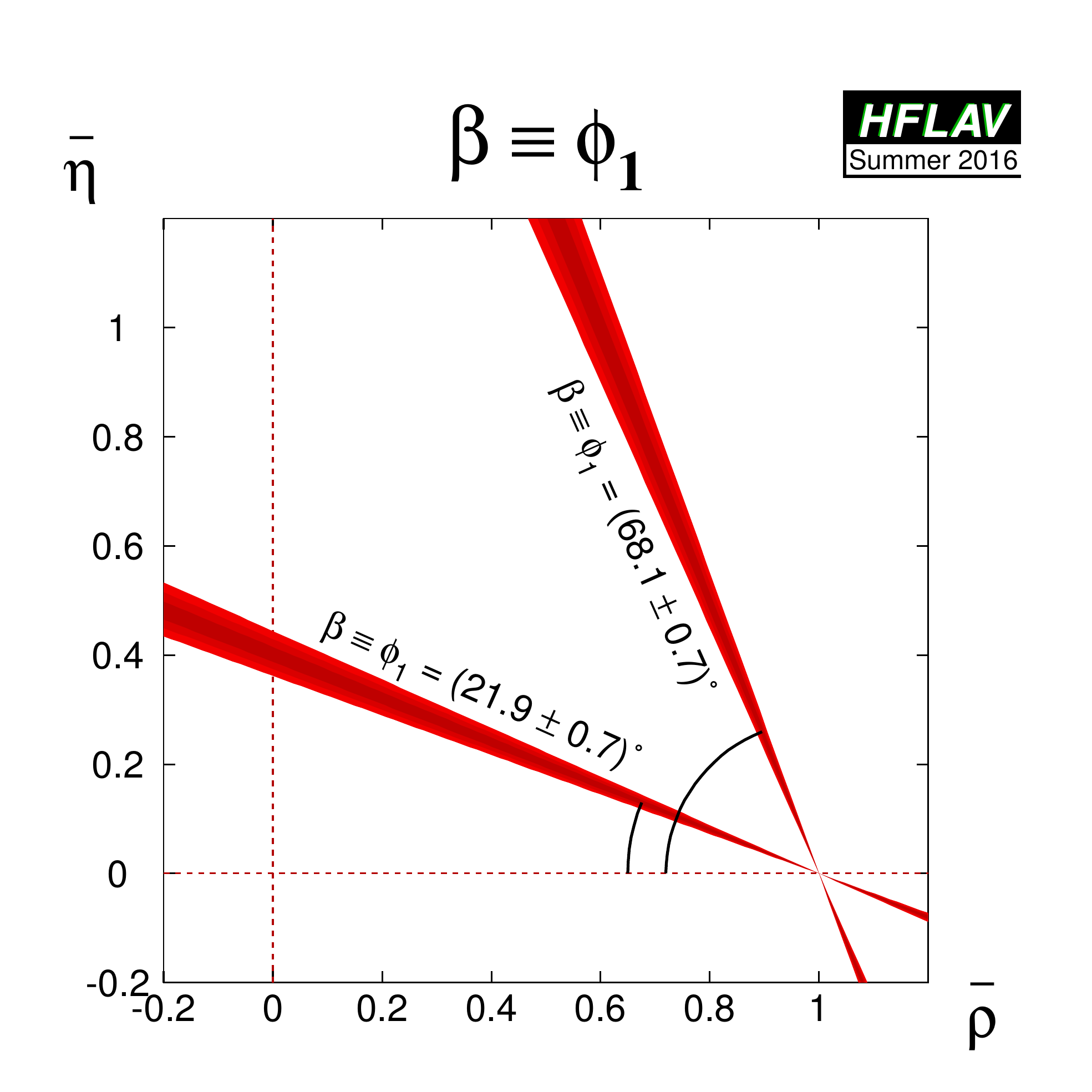}
    }
  \end{center}
  \vspace{-0.5cm}
  \caption{
    (Left) Average of measurements of $S_{b \to c\bar c s}$, interpreted as $\sin(2\beta)$.
    (Right) Constraints on the $(\rhobar,\etabar)$ plane,
    obtained from the average of $-\etacp S_{b \to c\bar c s}$ 
    and Eq.~(\ref{eq:cp_uta:sin2beta}).
    Note that the solution with the smaller (larger) value of $\beta$ has $\cos(2\beta)>0$ ($<0$).
  }
  \label{fig:cp_uta:ccs}
\end{figure}

In the Standard Model, the time-dependent parameters for $\Bz$ decays governed by $b \to c\bar c s$ transitions are predicted to be
$S_{b \to c\bar c s} = - \etacp \sin(2\beta)$ and $C_{b \to c\bar c s} = 0$ to very good accuracy.
Deviations from this relation are currently limited to the level of $\lsim 1\degrees$ on $2\beta$~\cite{Jung:2012mp,DeBruyn:2014oga,Frings:2015eva}.
The averages for $-\etacp S_{b \to c\bar c s}$ and $C_{b \to c\bar c s}$
are provided in Table~\ref{tab:cp_uta:ccs}.
The averages for $-\etacp S_{b \to c\bar c s}$ 
are shown in Fig.~\ref{fig:cp_uta:ccs}.

Both \babar\  and \belle\ have used the $\etacp = -1$ modes
$\jpsi \KS$, $\psi(2S) \KS$, $\chi_{c1} \KS$ and $\eta_c \KS$, 
as well as $\jpsi \KL$, which has $\etacp = +1$
and $\jpsi K^{*0}(892)$, which is found to have $\etacp$ close to $+1$
based on the measurement of $\left| A_\perp \right|$ 
(see Sec.~\ref{sec:cp_uta:common_inputs}).
The most recent \belle\ result does not use $\eta_c \KS$ or $\jpsi K^{*0}(892)$ decays.\footnote{
  Previous analyses from \belle\ did include these channels~\cite{Abe:2004mz},
  but it is not possible to obtain separate results for those modes from the
  published information.
}
ALEPH, OPAL, CDF and LHCb have used only the $\jpsi \KS$ final state.
%% In the latest result from \belle~\cite{Chen:2006nk}, 
%% only $\jpsi \KS$ and $\jpsi \KL$ are used,
%% while results from $\psi(2S) \KS$ have been presented
%% separately~\cite{Abe:2007gj}.
\babar\ has also determined the \CP violation parameters of the
$\Bz\to\chi_{c0} \KS$ decay from the time-dependent Dalitz plot analysis of
the $\Bz \to \pi^+\pi^-\KS$ mode (see Sec.~\ref{sec:cp_uta:qqs:dp}).
In addition, \belle\ has performed a measurement with data accumulated at the $\Upsilon(5S)$ resonance, using the $\jpsi\KS$ final state -- this involves a different flavour tagging method compared to the measurements performed with data accumulated at the $\Upsilon(4S)$ resonance.
A breakdown of results in each charmonium-kaon final state is given in 
Table~\ref{tab:cp_uta:ccs-BF}.

\begin{table}[htb]
	\begin{center}
		\caption{
                        Results and averages for $S_{b \to c\bar c s}$ and $C_{b \to c\bar c s}$.
                        The result marked ($^{*}$) uses ``{\it hadronic and previously unused muonic decays of the $J/\psi$}''. 
                        We neglect a small possible correlation of this result with the main \babar\ result~\cite{:2009yr} that could be caused by reprocessing of the data.
                }
		\vspace{0.2cm}
%		\setlength{\tabcolsep}{0.0pc}
% make this tabular (not tabular*) and resize down to \textwidth
% change @{\extracolsep{\fill}} to @{\extracolsep{2mm}}
    \resizebox{\textwidth}{!}{
\renewcommand{\arraystretch}{1.2}
		\begin{tabular}{@{\extracolsep{2mm}}lrccc} \hline
      \mc{2}{l}{Experiment} & Sample size & $- \etacp S_{b \to c\bar c s}$ & $C_{b \to c\bar c s}$ \\
      \hline
	\babar & \cite{:2009yr} & $N(B\bar{B})$ = 465M & $0.687 \pm 0.028 \pm 0.012$ & $0.024 \pm 0.020 \pm 0.016$ \\
	\babar\ $\chi_{c0} \KS$ & \cite{Aubert:2009me} & $N(B\bar{B})$ = 383M & $0.69 \pm 0.52 \pm 0.04 \pm 0.07$ & $-0.29 \,^{+0.53}_{-0.44} \pm 0.03 \pm 0.05$ \\
	\babar\ $J/\psi \KS$ ($^{*}$) & \cite{Aubert:2003xn} & $N(B\bar{B})$ = 88M & $1.56 \pm 0.42 \pm 0.21$ &  \textendash{} \\
	\belle & \cite{Adachi:2012et} & $N(B\bar{B})$ = 772M & $0.667 \pm 0.023 \pm 0.012$ & $-0.006 \pm 0.016 \pm 0.012$ \\
%	\hline
	\mc{3}{l}{\bf \boldmath $\B$ factory average} & $0.679 \pm 0.020$ & $0.005 \pm 0.017$ \\
	\mc{3}{l}{\small Confidence level} & {\small $0.28~(1.2\sigma)$} & {\small $0.47~(0.5\sigma)$} \\
        \hline
        ALEPH & \cite{Barate:2000tf} &  $N(Z \to \text{hadrons})$ = 4M & $0.84 \, ^{+0.82}_{-1.04} \pm 0.16$ &  \textendash{} \\
        OPAL  & \cite{Ackerstaff:1998xz} & $N(Z \to \text{hadrons})$ = 4.4M & $3.2 \, ^{+1.8}_{-2.0} \pm 0.5$ &  \textendash{} \\
        CDF   & \cite{Affolder:1999gg} & $\int {\cal L} \, dt = 110\ {\rm pb}^{-1}$ & $0.79 \, ^{+0.41}_{-0.44}$ &  \textendash{} \\
        LHCb & \cite{Aaij:2015vza} & $\int {\cal L} \, dt = 3\ {\rm fb}^{-1}$ & $0.731 \pm 0.035 \pm 0.020$ & $-0.038 \pm 0
.032 \pm 0.005$ \\
	Belle $\Upsilon(5S)$ & \cite{Sato:2012hu} & $\int {\cal L} \, dt = 121\ {\rm fb}^{-1}$ & $0.57 \pm 0.58 \pm 0.06$ &  \textendash{} \\
%        \hline
        \mc{3}{l}{\bf Average} & $0.691 \pm 0.017$ & $-0.004 \pm 0.015$ \\
%	\mc{3}{l}{\small Confidence level} & {\small $0.25$} & {\small $0.47$} \\
		\hline
		\end{tabular}
}
                \label{tab:cp_uta:ccs}
        \end{center}
\end{table}

\begin{table}[htb]
	\begin{center}
		\caption{
                        Breakdown of $B$ factory results on $S_{b \to c\bar c s}$ and $C_{b \to c\bar c s}$.
                }
		\vspace{0.2cm}
		\setlength{\tabcolsep}{0.0pc}
\renewcommand{\arraystretch}{1.1}
		\begin{tabular*}{\textwidth}{@{\extracolsep{\fill}}lrccc} \hline
        \mc{2}{l}{Mode} & $N(B\bar{B})$ & $- \etacp S_{b \to c\bar c s}$ & $C_{b \to c\bar c s}$ \\
        \hline
        \mc{5}{c}{\babar} \\
        $J/\psi \KS$ & \cite{:2009yr} & 465M & $0.657 \pm 0.036 \pm 0.012$ & $\phantom{-}0.026 \pm 0.025 \pm 0.016$ \\
        $J/\psi \KL$ & \cite{:2009yr} & 465M & $0.694 \pm 0.061 \pm 0.031$ & $-0.033 \pm 0.050 \pm 0.027$ \\
        {\bf \boldmath $J/\psi K^0$} & \cite{:2009yr} & 465M & $0.666 \pm 0.031 \pm 0.013$ & $\phantom{-}0.016 \pm 0.023 \pm 0.018$ \\
        $\psi(2S) \KS$ & \cite{:2009yr} & 465M & $0.897 \pm 0.100 \pm 0.036$ & $\phantom{-}0.089 \pm 0.076 \pm 0.020$ \\
        $\chi_{c1} \KS$ & \cite{:2009yr} & 465M & $0.614 \pm 0.160 \pm 0.040$ & $\phantom{-}0.129 \pm 0.109 \pm 0.025$ \\
        $\eta_c \KS$ & \cite{:2009yr} & 465M & $0.925 \pm 0.160 \pm 0.057$ & $\phantom{-}0.080 \pm 0.124 \pm 0.029$ \\
        $\jpsi K^{*0}(892)$ & \cite{:2009yr} & 465M & $0.601 \pm 0.239 \pm 0.087$ & $\phantom{-}0.025 \pm 0.083 \pm 0.054$ \\
        {\bf All} & \cite{:2009yr} & 465M & $0.687 \pm 0.028 \pm 0.012$ & $\phantom{-}0.024 \pm 0.020 \pm 0.016$ \\
	\hline
	\mc{5}{c}{\bf \belle} \\
        $J/\psi \KS$ & \cite{Adachi:2012et} & 772M & $0.670 \pm 0.029 \pm 0.013$ & $\phantom{-}0.015 \pm 0.021 \,^{+0.023}_{-0.045}$ \\
        $J/\psi \KL$ & \cite{Adachi:2012et} & 772M & $0.642 \pm 0.047 \pm 0.021$ & $-0.019 \pm 0.026 \,^{+0.041}_{-0.017}$ \\
%        {\bf \boldmath $J/\psi K^0$} & \cite{Chen:2006nk} & 535M & $0.642 \pm 0.031 \pm 0.017$ & $-0.018 \pm 0.021 \pm 0.014$ \\
	$\psi(2S) \KS$ & \cite{Adachi:2012et} & 772M & $0.738 \pm 0.079 \pm 0.036$ & $-0.104 \pm 0.055 \,^{+0.027}_{-0.047}$ \\
	$\chi_{c1} \KS$ & \cite{Adachi:2012et} & 772M & $0.640 \pm 0.117 \pm 0.040$ & $\phantom{-}0.017 \pm 0.083 \,^{+0.026}_{-0.046}$ \\
        {\bf All} & \cite{Adachi:2012et} & 772M & $0.667 \pm 0.023 \pm 0.012$ & $-0.006 \pm 0.016 \pm 0.012$ \\
	\hline
	\mc{5}{c}{\bf Averages} \\
        \mc{3}{l}{$J/\psi \KS$} & $0.665 \pm 0.024$ & $\phantom{-}0.024 \pm 0.026$ \\
        \mc{3}{l}{$J/\psi \KL$} & $0.663 \pm 0.041$ & $-0.023 \pm 0.030$ \\
        \mc{3}{l}{$\psi(2S) \KS$} & $0.807 \pm 0.067$ & $-0.009 \pm 0.055$ \\
        \mc{3}{l}{$\chi_{c1} \KS$} & $0.632 \pm 0.099$ & $\phantom{-}0.066 \pm 0.074$ \\
		\hline
		\end{tabular*}
                \label{tab:cp_uta:ccs-BF}
        \end{center}
\end{table}

It should be noted that, while the uncertainty in the average for 
$-\etacp S_{b \to c\bar c s}$ is still limited by statistics,
the precision for $C_{b \to c\bar c s}$ is close to being dominated by the systematic uncertainty, particularly for measurements from the $\epem$ $B$ factory experiments.
This occurs due to the possible effect of tag side interference~\cite{Long:2003wq} on the $C_{b \to c\bar c s}$ measurement, an effect which is correlated between different $e^+e^- \to \Upsilon(4S) \to B\bar{B}$ experiments.
Understanding of this effect may continue to improve in future, allowing the uncertainty to reduce.

%% straightforward interpretation
From the average for $-\etacp S_{b \to c\bar c s}$ above, 
we obtain the following solutions for $\beta$
(in $\left[ 0, \pi \right]$):
\begin{equation}
  \beta = \left( 21.9 \pm 0.7 \right)^\circ
  \hspace{5mm}
  {\rm or}
  \hspace{5mm}
  \beta = \left( 68.1 \pm 0.7 \right)^\circ \, .
  \label{eq:cp_uta:sin2beta}
\end{equation}
In radians, these values are 
$\beta = \left( 0.382 \pm 0.012 \right)$, 
$\beta = \left( 1.189 \pm 0.012 \right)$.

This result gives a precise constraint on the $(\rhobar,\etabar)$ plane,
as shown in Fig.~\ref{fig:cp_uta:ccs}.
The measurement is in remarkable agreement with other constraints from 
$\CP$ conserving quantities, 
and with $\CP$ violation in the kaon system, in the form of the parameter $\epsilon_K$.
Such comparisons have been performed by various phenomenological groups,
such as CKMfitter~\cite{Charles:2004jd} and UTFit~\cite{Bona:2005vz} (see also Refs.~\cite{Lunghi:2008aa,Eigen:2013cv}).

%%%%%%%%
%%%
%%% J/psiK*
%%%
%%%%%%%%
\mysubsubsection{Time-dependent transversity analysis of $\Bz \to J/\psi K^{*0}$ decays
}
\label{sec:cp_uta:ccs:vv}

\begin{figure}[htbp]
  \begin{center}
    \resizebox{0.46\textwidth}{!}{
      \includegraphics{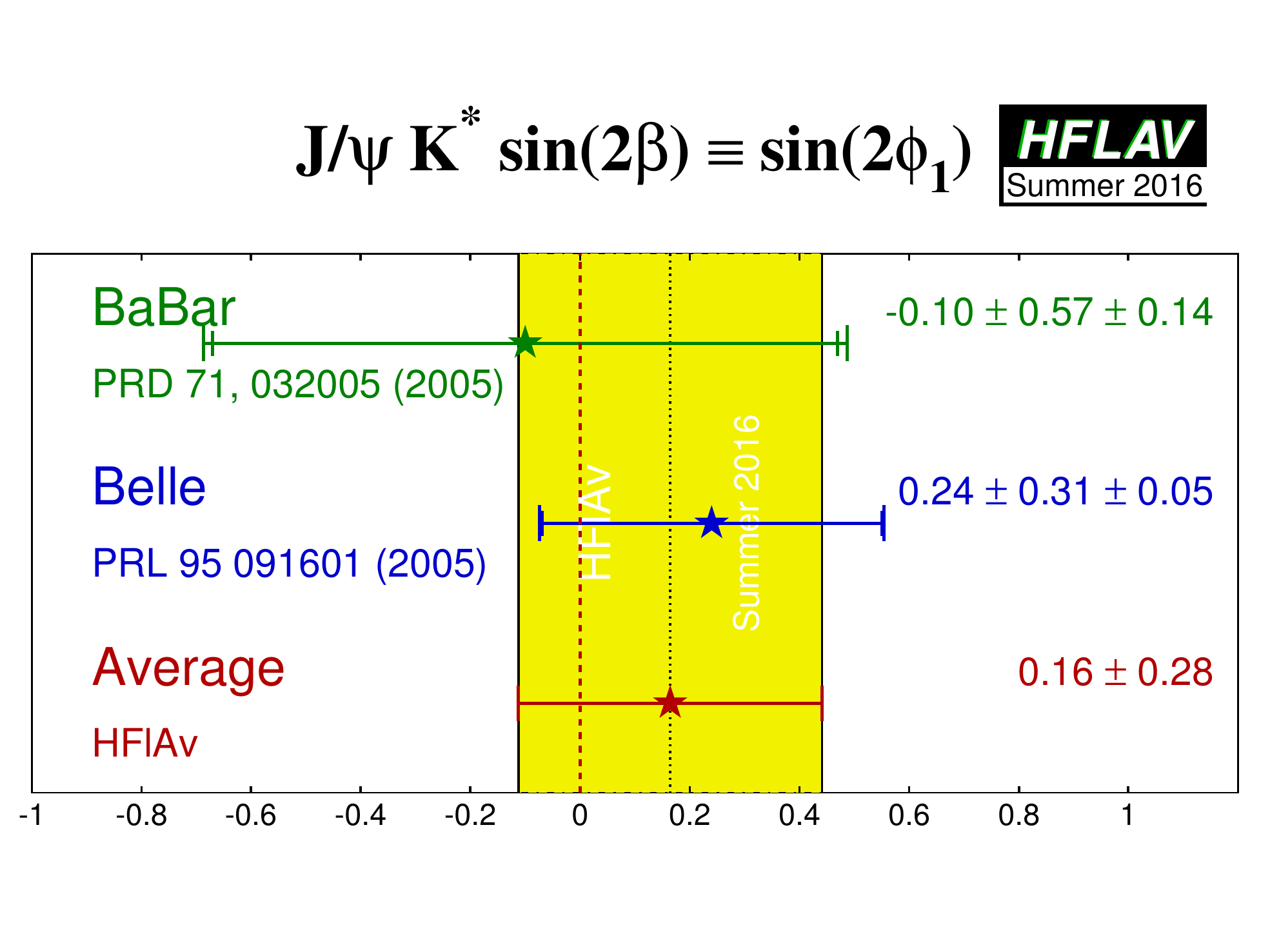}
    }
    \hfill
    \resizebox{0.46\textwidth}{!}{
      \includegraphics{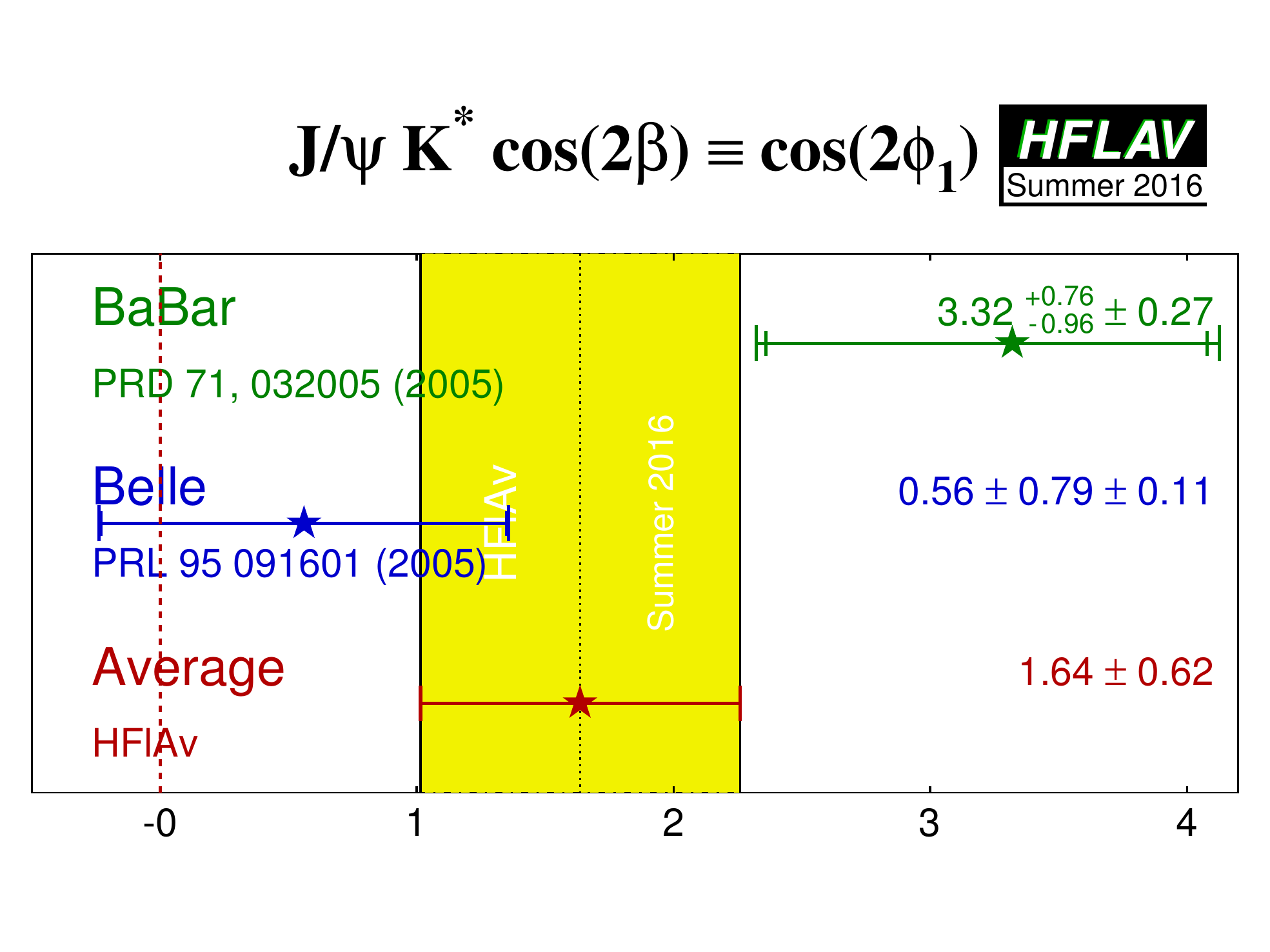}
    }
  \end{center}
  \vspace{-0.5cm}
  \caption{
    Averages of 
    (left) $\sin(2\beta) \equiv \sin(2\phi_1)$ and
    (right) $\cos(2\beta) \equiv \cos(2\phi_1)$
    from time-dependent analyses of $\Bz \to \jpsi K^{*0}$ decays.
  }
  \label{fig:cp_uta:JpsiKstar}
\end{figure}

$\B$ meson decays to the vector-vector final state $J/\psi K^{*0}$
are also mediated by the $b \to c \bar c s$ transition.
When a final state that is not flavour-specific ($K^{*0} \to \KS \pi^0$) is used,
a time-dependent transversity analysis can be performed 
allowing sensitivity to both 
$\sin(2\beta)$ and $\cos(2\beta)$~\cite{Dunietz:1990cj}.
Such analyses have been performed by both $\B$ factory experiments.
In principle, the strong phases between the transversity amplitudes
are not uniquely determined by such an analysis, 
leading to a discrete ambiguity in the sign of $\cos(2\beta)$.
The \babar\ collaboration resolves 
this ambiguity using the known variation~\cite{Aston:1987ir}
of the P-wave phase (fast) relative to the S-wave phase (slow) 
with the invariant mass of the $K\pi$ system 
in the vicinity of the $K^*(892)$ resonance. 
The result is in agreement with the prediction from 
$s$ quark helicity conservation,
and corresponds to Solution II defined by Suzuki~\cite{Suzuki:2001za}.
We include only the solutions consistent with this phase variation in 
Table~\ref{tab:cp_uta:ccs:psi_kstar} and Fig.~\ref{fig:cp_uta:JpsiKstar}.

\begin{table}[htb]
	\begin{center}
		\caption{
			Averages from $\Bz \to J/\psi K^{*0}$ transversity analyses.
		}
		\vspace{0.2cm}
		\setlength{\tabcolsep}{0.0pc}
\renewcommand{\arraystretch}{1.1}
		\begin{tabular*}{\textwidth}{@{\extracolsep{\fill}}lrcccc} \hline
		\mc{2}{l}{Experiment} & $N(B\bar{B})$ & $\sin 2\beta$ & $\cos 2\beta$ & Correlation \\
		\hline
	\babar & \cite{Aubert:2004cp} & 88M & $-0.10 \pm 0.57 \pm 0.14$ & $3.32 ^{+0.76}_{-0.96} \pm 0.27$ & $-0.37$ \\
	\belle & \cite{Itoh:2005ks} & 275M & $0.24 \pm 0.31 \pm 0.05$ & $0.56 \pm 0.79 \pm 0.11$ & $0.22$ \\
%	\hline
	\mc{3}{l}{\bf Average} & $0.16 \pm 0.28$ & $1.64 \pm 0.62$ &  \hspace{-8mm} {\small uncorrelated averages}  \\
        \mc{3}{l}{\small Confidence level} & {\small $0.61~(0.5\sigma)$} & {\small $0.03~(2.2\sigma)$} & \\
		\hline
		\end{tabular*}
		\label{tab:cp_uta:ccs:psi_kstar}
	\end{center}
\end{table}

At present the results are dominated by 
large and non-Gaussian statistical errors,
and exhibit significant correlations.
We perform uncorrelated averages, 
the interpretation of which has to be done with the greatest care. 
Nonetheless, it is clear that $\cos(2\beta)>0$ is preferred 
by the experimental data in $J/\psi \Kstarz$ 
(for example, \babar~\cite{Aubert:2004cp} 
find a confidence level for $\cos(2\beta)>0$ of $89\%$).

%%%%%%%%
%%%
%%% D*D*Ks
%%%
%%%%%%%%
\mysubsubsection{Time-dependent $\CP$ asymmetries in $\Bz \to \Dstarp \Dstarm \KS$ decays
}
\label{sec:cp_uta:ccs:DstarDstarKs}

\begin{figure}[htbp]
  \begin{center}
    \begin{tabular}{c@{\hspace{-1mm}}c@{\hspace{-1mm}}c}
      \resizebox{0.32\textwidth}{!}{
        \includegraphics{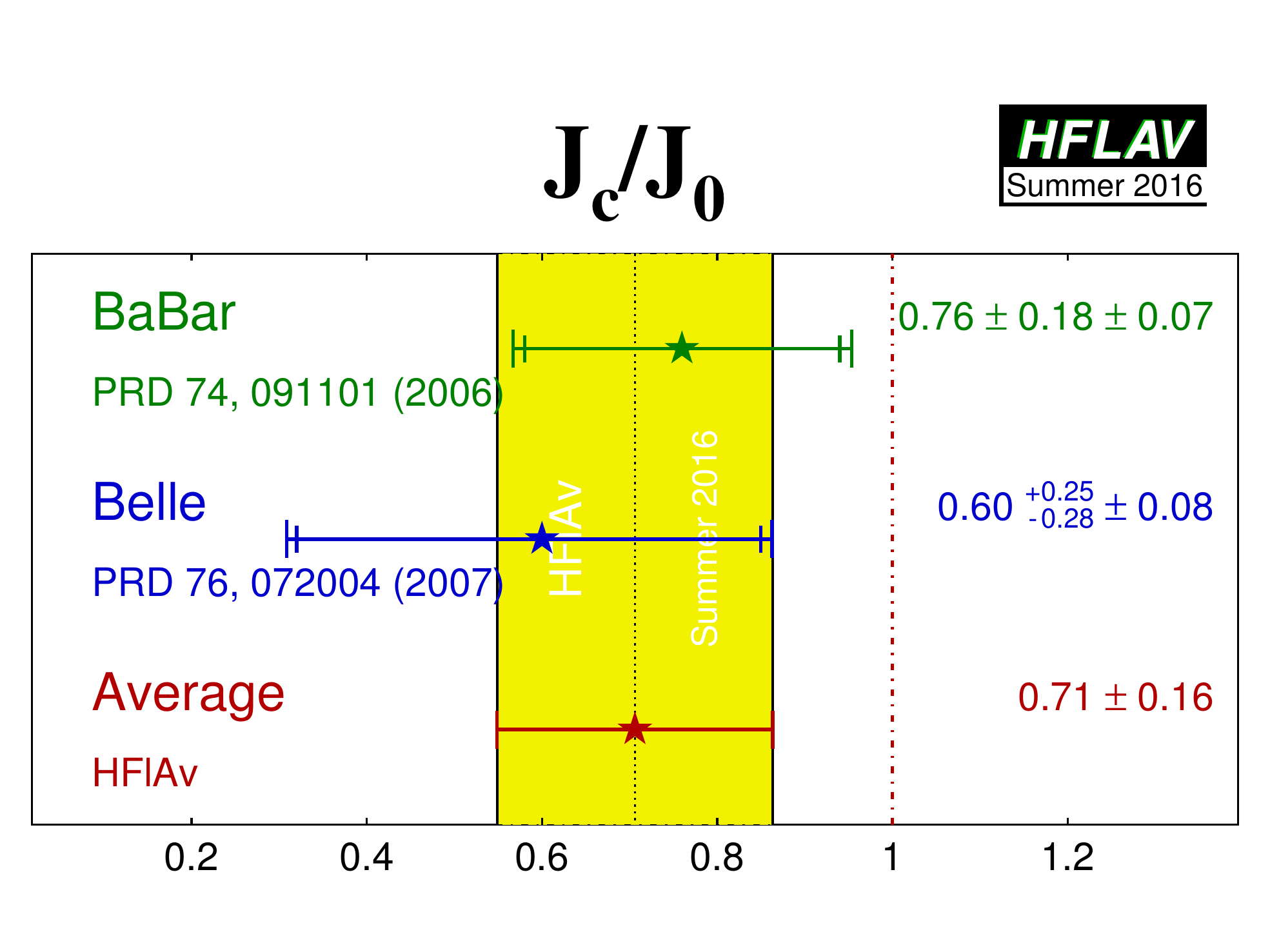}
      }
      &
      \resizebox{0.32\textwidth}{!}{
        \includegraphics{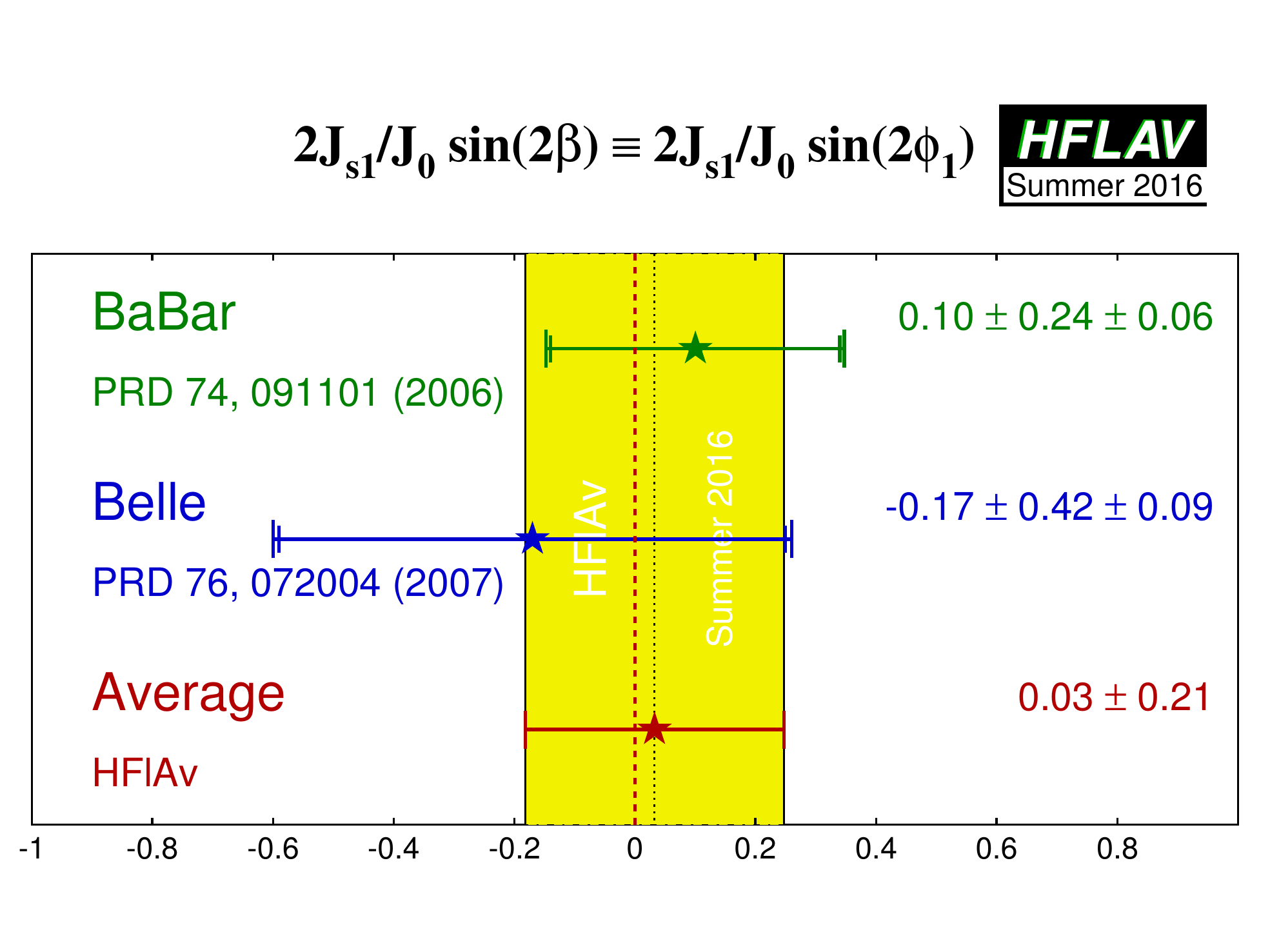}
      }
      &
      \resizebox{0.32\textwidth}{!}{
        \includegraphics{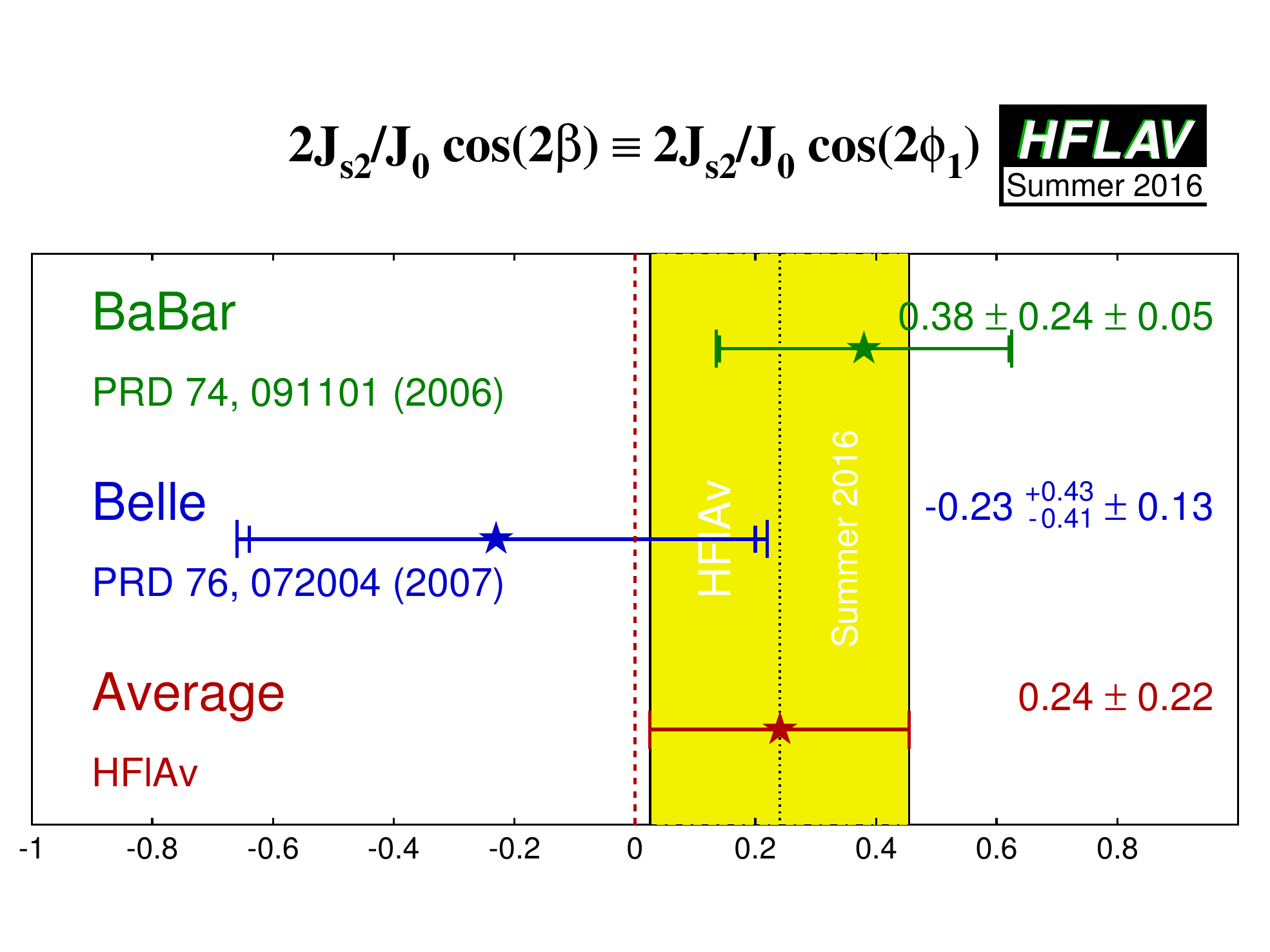}
      }
    \end{tabular}
  \end{center}
  \vspace{-0.8cm}
  \caption{
    Averages of 
    (left) $(J_c/J_0)$, (middle) $(2J_{s1}/J_0) \sin(2\beta)$ and 
    (right) $(2J_{s2}/J_0) \cos(2\beta)$
    from time-dependent analyses of $\Bz \to \Dstarp \Dstarm \KS$ decays.
  }
  \label{fig:cp_uta:ccs:dstardstarks}
\end{figure}

Both \babar~\cite{Aubert:2006fh} and \belle~\cite{Dalseno:2007hx} have performed
time-dependent analyses of the $\Bz \to \Dstarp \Dstarm \KS$ decay,
to obtain information on the sign of $\cos(2\beta)$.
More information can be found in 
Sec.~\ref{sec:cp_uta:notations:dalitz:dstardstarks}.
The results are given in Table~\ref{tab:cp_uta:ccs:dstardstarks}, 
and shown in Fig.~\ref{fig:cp_uta:ccs:dstardstarks}.

\begin{table}[htb]
	\begin{center}
		\caption{
                        Results from time-dependent analysis of $\Bz \to \Dstarp \Dstarm \KS$.
		}
		\vspace{0.2cm}
		\setlength{\tabcolsep}{0.0pc}
\renewcommand{\arraystretch}{1.2}
		\begin{tabular*}{\textwidth}{@{\extracolsep{\fill}}lrcccc} \hline
                \mc{2}{l}{Experiment} & $N(B\bar{B})$ & $\frac{J_c}{J_0}$ & $\frac{2J_{s1}}{J_0} \sin(2\beta)$ &  $\frac{2J_{s2}}{J_0} \cos(2\beta)$ \\
		\hline
	\babar & \cite{Aubert:2006fh} & 230M & $0.76 \pm 0.18 \pm 0.07$ & $0.10 \pm 0.24 \pm 0.06$ & $0.38 \pm 0.24 \pm 0.05$ \\
	\belle & \cite{Dalseno:2007hx} & 449M & $0.60 \,^{+0.25}_{-0.28} \pm 0.08$ & $-0.17 \pm 0.42 \pm 0.09$ & $-0.23 \,^{+0.43}_{-0.41} \pm 0.13$ \\
%	\hline
	\mc{3}{l}{\bf Average} & $0.71 \pm 0.16$ & $0.03 \pm 0.21$ & $0.24 \pm 0.22$ \\
	\mc{3}{l}{\small Confidence level} & {\small $0.63~(0.5\sigma)$} & {\small $0.59~(0.5\sigma)$} & {\small $0.23~(1.2\sigma)$} \\
		\hline
		\end{tabular*}
		\label{tab:cp_uta:ccs:dstardstarks}
	\end{center}
\end{table}

From the above result and the assumption that $J_{s2}>0$, 
\babar\ infer that $\cos(2\beta)>0$ at the $94\%$ confidence level~\cite{Aubert:2006fh}.

%%%%%%%%
%%%
%%% Bs -> J/psi phi
%%%
%%%%%%%%
\mysubsubsection{Time-dependent analysis of $\Bs$ decays through the $b \to c\bar{c}s$ transition
}
\label{sec:cp_uta:ccs:jpsiphi}

As described in Sec.~\ref{sec:cp_uta:notations:Bs},
time-dependent analysis of decays such as $\Bs \to J/\psi \phi$ probes the 
$\CP$ violating phase of $\Bs$--$\Bsbar$ oscillations, $\phi_s$.\footnote{
  We use $\phi_s$ here to denote the same quantity labelled $\phi_s^{c\bar{c}s}$ in Sec.~\ref{sec:life_mix}.
  It should not be confused with the parameter $\phi_{12} \equiv \arg\left[ -M_{12}/\Gamma_{12} \right]$, which historically was also often referred to as $\phi_s$.
}
%    We make the approximation $\phi_s = -2 \beta_s$, 
%    where $\phi_s \equiv \arg\left[ -M_{12}/\Gamma_{12} \right]$ 
%    and $2\beta_s \equiv 2 \arg\left[ -(V_{ts}V_{tb}^*)/(V_{cs}V_{cb}^*) \right]$
%    (see Sec.~\ref{sec:cp_uta:introduction}). 
%    This is a reasonable approximation since, 
%    although the equality does not hold in the Standard Model~\cite{Lenz:2011ti,*Lenz:2006hd}, 
%    both are much smaller than the current experimental resolution, 
%    whereas new physics contributions add a phase $\phi_{\rm NP}$ to $\phi_s$
%    and subtract the same phase from $2\beta_s$, 
%    such that the approximation remains valid.
%
%Within the Standard Model, this parameter is predicted to be small.
The combination of results on $\Bs \to \jpsi \phi$ decays, including also results from $\Bs \to \jpsi \pi^+\pi^-$ and $\Bs \to D_s^+D_s^-$ decays, is performed by the HFLAV Lifetimes and Oscillations subgroup, see Sec.~\ref{sec:life_mix}.

%%%%%%%%
%%%
%%% Dh0
%%%
%%%%%%%%
% \afterpage{\clearpage}
\mysubsection{Time-dependent $\CP$ asymmetries in colour-suppressed $b \to c\bar{u}d$ transitions
}
\label{sec:cp_uta:cud_beta}

\mysubsubsection{Time-dependent $\CP$ asymmetries: $b \to c\bar{u}d$ decays to
  \CP eigenstates
}
\label{sec:cp_uta:cud_beta:cp}

Decays of $\B$ mesons to final states such as $D\pi^0$ are 
governed by $b \to c\bar{u}d$ transitions. 
If the final state is a $\CP$ eigenstate, \eg\ $D_{\CP}\pi^0$, 
the usual time-dependence formulae are recovered, 
with the sine coefficient sensitive to $\sin(2\beta)$. 
Since there is no penguin contribution to these decays, 
there is even less associated theoretical uncertainty 
than for $b \to c\bar{c}s$ decays such as $\B \to \jpsi \KS$.
Such measurements therefore allow to test the Standard Model prediction
that the $\CP$ violation parameters in $b \to c\bar{u}d$ transitions
are the same as those in $b \to c\bar{c}s$~\cite{Grossman:1996ke}.
Although there is an additional contribution from CKM suppressed $b \to u \bar{c} d$ amplitudes, which have a different weak phase compared to the leading $b \to c\bar{u}d$ transition, the effect is small and can be taken into account in the analysis~\cite{Fleischer:2003ai,Fleischer:2003aj}.

% Results of such an analysis are available from \babar~\cite{Aubert:2007mn}.
% The decays $\Bz \to D\pi^0$, $\Bz \to D\eta$, $\Bz \to D\omega$,
% $\Bz \to D^*\pi^0$ and $\Bz \to D^*\eta$ are used.
% In the latter two modes, the daughter decay $D^* \to D\pi^0$ is used.
% The $\CP$-even $D$ decay to $K^+K^-$ is used for all decay modes,
% with the $\CP$-odd $D$ decay to $\KS\omega$ also used in $\Bz \to D^{(*)}\pi^0$
% and the additional $\CP$-odd $D$ decay to $\KS\pi^0$ 
% also used in $\Bz \to D\omega$.
% Results are presented separately for $\CP$-even and $\CP$-odd 
% $D^{(*)}$ decays (denoted $D^{(*)}_+ h^0$ and $D^{(*)}_- h^0$ respectively),
% and for both combined, taking into account the different $\CP$ factors
% (denoted $D^{(*)}_{\CP} h^0$).

Results are available from a joint analysis of \babar\ and \belle\ data~\cite{Abdesselam:2015gha}.
The following \CP-even final states are included: $D\piz$ and $D\eta$ with $D \to \KS\piz$ and $D \to \KS\omega$; $D\omega$ with $D \to \KS\piz$; $\Dstar\piz$ and $\Dstar\eta$ with $\Dstar \to D\piz$ and $D \to \Kp\Km$. 
The following \CP-odd final states are included: $D\piz$, $D\eta$ and $D\omega$ with $D \to \Kp\Km$, $\Dstar\piz$ and $\Dstar\eta$ with $\Dstar \to D\piz$ and $D \to \KS\piz$. 
All $\Bd \to \DorDstar h^0$ decays are analysed together, taking into account the different $\CP$ factors (denoted $\DorDstar_{\CP} h^0$). 
The results are summarised in Table~\ref{tab:cp_uta:cud_cp_beta}.

\begin{table}[htb]
	\begin{center}
		\caption{
			Results from analyses of $\Bz \to D^{(*)}h^0$, $D \to \CP$ eigenstates decays.
		}
		\vspace{0.2cm}
		\setlength{\tabcolsep}{0.0pc}
\renewcommand{\arraystretch}{1.1}
		\begin{tabular*}{\textwidth}{@{\extracolsep{\fill}}lrcccc} \hline
	\mc{2}{l}{Experiment} & $N(B\bar{B})$ & $S_{\CP}$ & $C_{\CP}$ & Correlation \\
	\hline
%        \mc{6}{c}{$D^{(*)}_{\CP} h^0$}  \\ 
	\babar\ \& \belle & \cite{Abdesselam:2015gha} & 1243M & $0.66 \pm 0.10 \pm 0.06$ & $-0.02 \pm 0.07 \pm 0.03$ & $-0.05$ \\
	\hline
		\end{tabular*}
		\label{tab:cp_uta:cud_cp_beta}
	\end{center}
\end{table}

\mysubsubsection{Time-dependent Dalitz plot analyses of $b \to c\bar{u}d$ decays
}
\label{sec:cp_uta:cud_beta:dalitz}

When multibody $D$ decays, such as $D \to \KS\pi^+\pi^-$ are used, 
a time-dependent analysis of the Dalitz plot of the neutral $D$ decay 
allows for a direct determination of the weak phase $2\beta$. 
(Equivalently, both $\sin(2\beta)$ and $\cos(2\beta)$ can be measured.)
This information can be used to resolve the ambiguity in the 
measurement of $2\beta$ from $\sin(2\beta)$~\cite{Bondar:2005gk}.

\begin{figure}[htbp]
  \begin{center}
    \begin{tabular}{cc}
      \resizebox{0.46\textwidth}{!}{
        \includegraphics{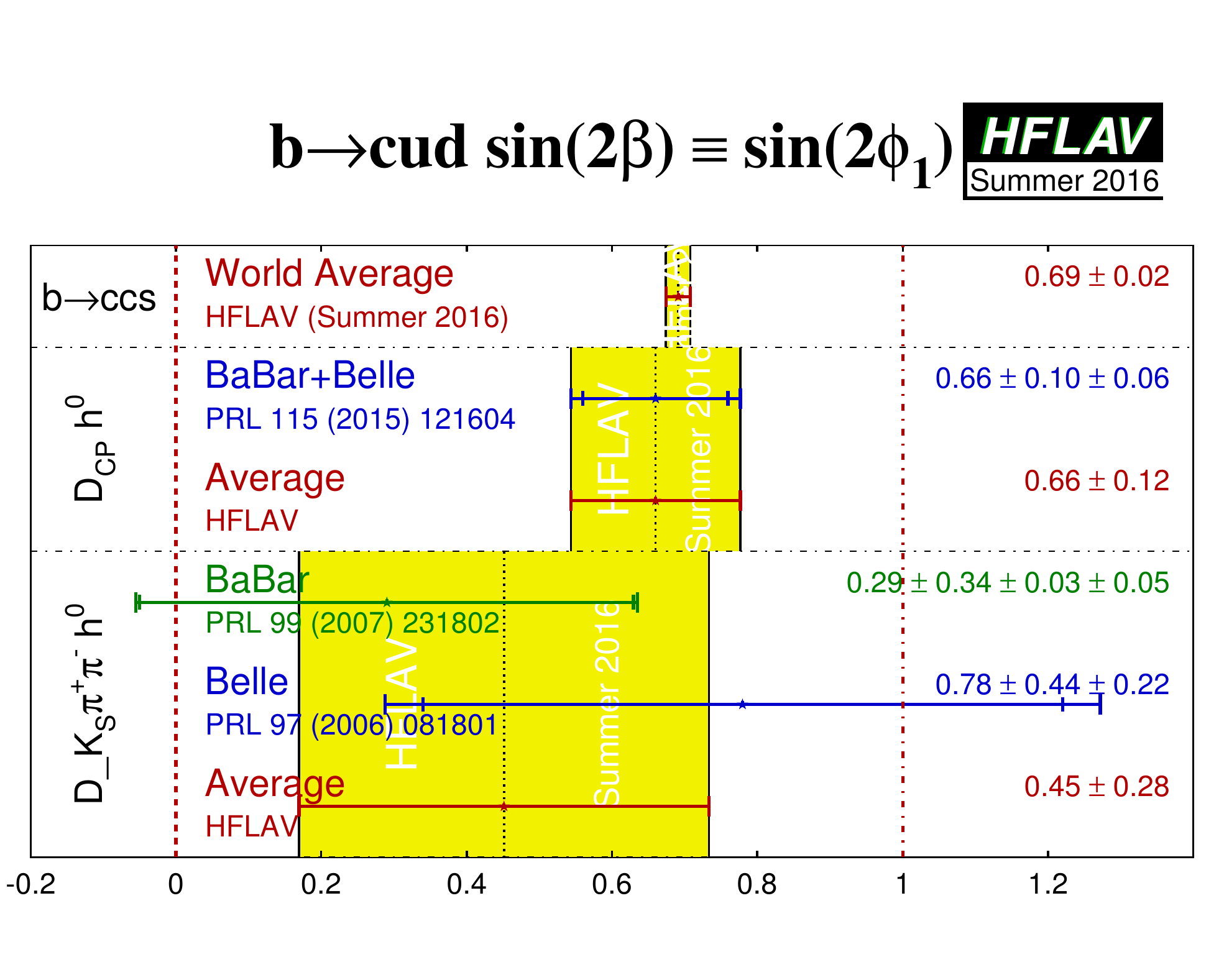} %figures/cp_uta/Dh0sin2beta}
      }
      &
      \resizebox{0.46\textwidth}{!}{
        \includegraphics{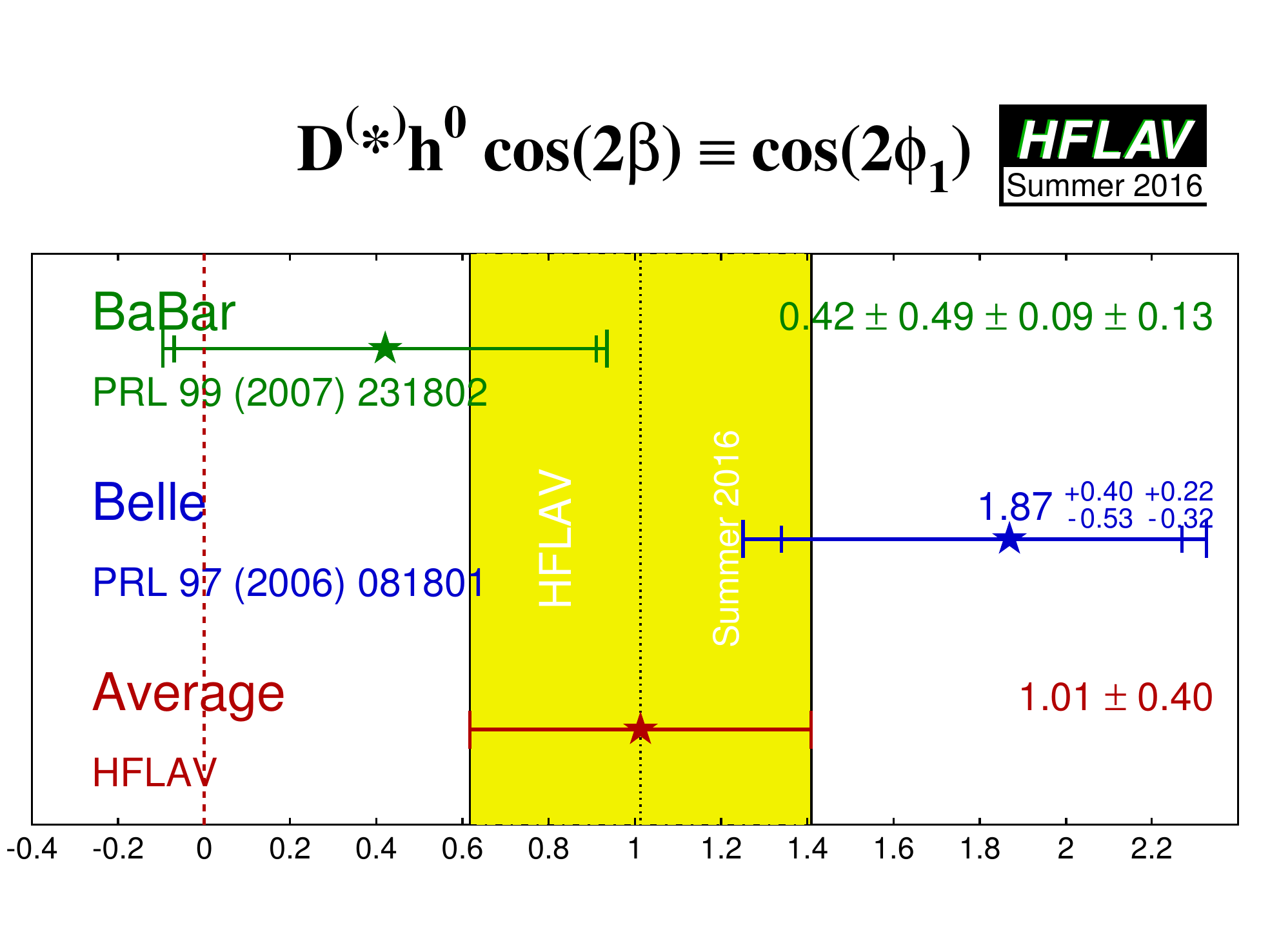}
      }
    \end{tabular}
  \end{center}
  \vspace{-0.8cm}
  \caption{
    Averages of 
    (left) $\sin(2\beta)$ and (right) $\cos(2\beta)$
    measured in colour-suppressed $b \to c\bar{u}d$ transitions.
  }
  \label{fig:cp_uta:cud_beta}
\end{figure}

Results of such analyses are available from both 
\belle~\cite{Krokovny:2006sv} and \babar~\cite{Aubert:2007rp}.
The decays $\B \to D\pi^0$, $\B \to D\eta$, $\B \to D\omega$, 
$\B \to D^*\pi^0$ and $\B \to D^*\eta$ are used. 
(This collection of states is denoted by $D^{(*)}h^0$.)
The daughter decays are $D^* \to D\pi^0$ and $D \to \KS\pi^+\pi^-$.
The results are given in Table~\ref{tab:cp_uta:cud_beta},
and shown in Fig.~\ref{fig:cp_uta:cud_beta}.
Note that \babar\ quote uncertainties due to the $D$ decay model 
separately from other systematic errors as a third source of uncertainty, while \belle\ do not.

\begin{table}[htb]
	\begin{center}
		\caption{
			Averages from $\Bz \to D^{(*)}h^0$, $D \to \KS\pi^+\pi^-$ analyses.
		}
		\vspace{0.2cm}
		\setlength{\tabcolsep}{0.0pc}
    \resizebox{\textwidth}{!}{
\renewcommand{\arraystretch}{1.1}
      		\begin{tabular*}{\textwidth}{@{\extracolsep{\fill}}lrcccc} \hline
	\mc{2}{l}{Experiment} & $N(B\bar{B})$ & $\sin 2\beta$ & $\cos 2\beta$ & $|\lambda|$ \\
		\hline
                & & & \mc{3}{c}{Model dependent} \\
	\babar & \cite{Aubert:2007rp} & 383M & $0.29 \pm 0.34 \pm 0.03 \pm 0.05$ & $0.42 \pm 0.49 \pm 0.09 \pm 0.13$ & $1.01 \pm 0.08 \pm 0.02$ \\
	\belle & \cite{Krokovny:2006sv} & 386M & $0.78 \pm 0.44 \pm 0.22$ & $1.87 \,^{+0.40}_{-0.53} \,^{+0.22}_{-0.32}$ & \textendash{} \\
%	\hline
	\mc{3}{l}{\bf Average} & $0.45 \pm 0.28$ & $1.01 \pm 0.40$ & $1.01 \pm 0.08$ \\
	\mc{3}{l}{\small Confidence level} & {\small $0.59~(0.5\sigma)$} & {\small $0.12~(1.6\sigma)$} & \textendash{} \\
		\hline
                & & & \mc{3}{c}{Model independent} \\
	\belle & \cite{Vorobyev:2016npn} & 772M & $0.43 \pm 0.27 \pm 0.08$ & $1.06 \pm 0.33 \,^{+0.21}_{-0.15}$ & \textendash{} \\
		\hline        
		\end{tabular*}
    }
		\label{tab:cp_uta:cud_beta}
	\end{center}
\end{table}

Again, it is clear that the data prefer $\cos(2\beta)>0$.
Indeed, \belle~\cite{Krokovny:2006sv} 
determine the sign of $\cos(2\phi_1)$ to be positive at $98.3\%$ confidence level,
while \babar~\cite{Aubert:2007rp} 
favour the solution of $\beta$ with $\cos(2\beta)>0$ at $87\%$ confidence level.
Note, however, that the Belle measurement has strongly non-Gaussian behaviour. 
Therefore, we perform uncorrelated averages, 
from which any interpretation has to be done with the greatest care. 

A model-independent time-dependent analysis of $\Bd \to \DorDstar h^0$ decays, with $D \to \KS\pip\pim$, has been performed by \belle~\cite{Vorobyev:2016npn}. 
The decays $\Bd \to D\piz$, $\Bd \to D\eta$, $\Bd \to D\etapr$, $\Bd \to D\omega$, $\Bd \to \Dstar\piz$ and $\Bd \to \Dstar\eta$ are used. 
The results are also included in Table~\ref{tab:cp_uta:cud_beta}.
From these results, \belle\ disfavour the solution with the value of $\sin(2\phi_1)$ from $b \to c\bar{c}s$ transitions but a negative value for $\cos(2\phi_1)$, at $5.1\,\sigma$ significance. 
The solution with the $b \to c\bar{c}s$ value of $\sin(2\phi_1)$ and positive $\cos(2\phi_1)$ is consistent with the data at the level of $1.3\,\sigma$. 
Note that due to the strong statistical and systematic correlations, model-dependent results and model-independent results from the same experiment cannot be combined.

%%%%%%%%
%%%
%%% ccd
%%%
%%%%%%%%
% \afterpage{\clearpage}
\mysubsection{Time-dependent $\CP$ asymmetries in $b \to c\bar{c}d$ transitions
}
\label{sec:cp_uta:ccd}

The transition $b \to c\bar c d$ can occur via either a $b \to c$ tree
or a $b \to d$ penguin amplitude.  
The flavour changing neutral current $b \to d$ penguin can be mediated by any up-type quark in the loop, and hence the amplitude can be written as
\begin{equation}
  \label{eq:cp_uta:b_to_d}
  \begin{array}{ccccc}
    A_{b \to d} & = & 
    \mc{3}{l}{F_u V_{ub}V^*_{ud} + F_c V_{cb}V^*_{cd} + F_t V_{tb}V^*_{td}} \\
    & = & (F_u - F_c) V_{ub}V^*_{ud} & + & (F_t - F_c) V_{tb}V^*_{td} \, , \\
%    & = & {\cal O}(\lambda^3) & + & {\cal O}(\lambda^3) \, , \\
  \end{array}
\end{equation}
where $F_{u,c,t}$ describe all factors except CKM suppression entering each quark loop diagram.
In the last line, both terms are ${\cal O}(\lambda^3)$, so it can be seen that the $b \to d$ penguin amplitude contains terms with different weak phases at the same order of CKM suppression.

In the above, we have chosen to eliminate the $F_c$ term using unitarity.
However, we could equally well write
\begin{equation}
  \label{eq:cp_uta:b_to_d_alt}
  \begin{array}{ccccc}
    A_{b \to d} 
    & = & (F_u - F_t) V_{ub}V^*_{ud} & + & (F_c - F_t) V_{cb}V^*_{cd} \\
    & = & (F_c - F_u) V_{cb}V^*_{cd} & + & (F_t - F_u) V_{tb}V^*_{td} \, . \\
  \end{array}
\end{equation}
Since the $b \to c\bar{c}d$ tree amplitude has the weak phase of $V_{cb}V^*_{cd}$,
either of the above expressions allow the penguin amplitude to be decomposed into a part with weak phase the same as the tree amplitude and another part with a different weak phase, which can be chosen to be either $\beta$ or $\gamma$.
The choice of parametrisation cannot, of course, affect the physics~\cite{Botella:2005ks}.
In any case, if the tree amplitude dominates, there is little sensitivity to any phase other than that from $\Bz$--$\Bzb$ mixing.

\begin{figure}[htbp]
  \begin{center}
    \begin{tabular}{cc}
      \resizebox{0.46\textwidth}{!}{
        \includegraphics{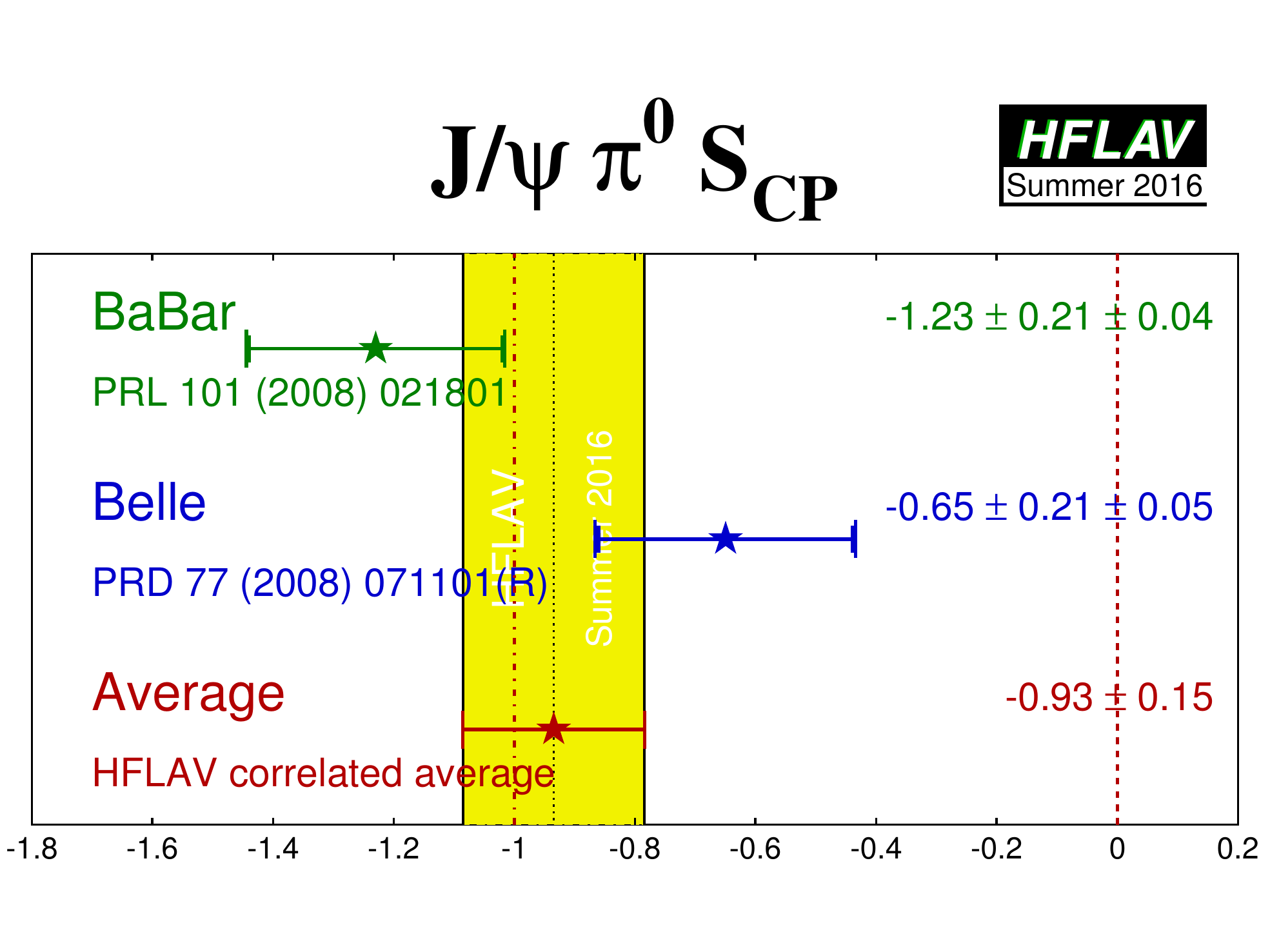}
      }
      &
      \resizebox{0.46\textwidth}{!}{
        \includegraphics{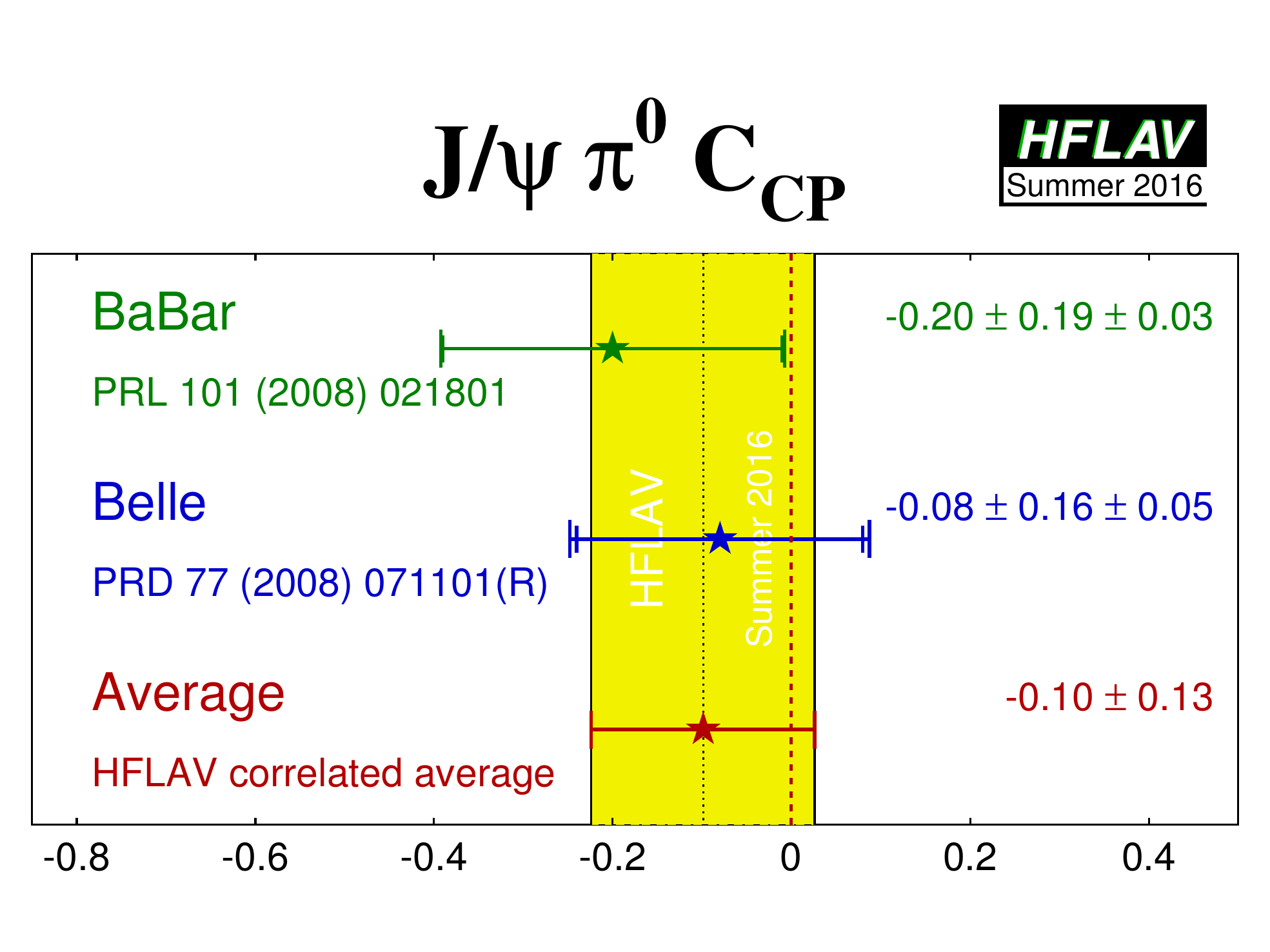}
      }
    \end{tabular}
  \end{center}
  \vspace{-0.8cm}
  \caption{
    Averages of 
    (left) $S_{b \to c\bar c d}$ and (right) $C_{b \to c\bar c d}$ 
    for the mode $\Bz \to J/ \psi \pi^0$.
  }
  \label{fig:cp_uta:ccd:psipi0}
\end{figure}

\begin{figure}[htbp]
  \begin{center}
    \begin{tabular}{cc}
      \resizebox{0.46\textwidth}{!}{
        \includegraphics{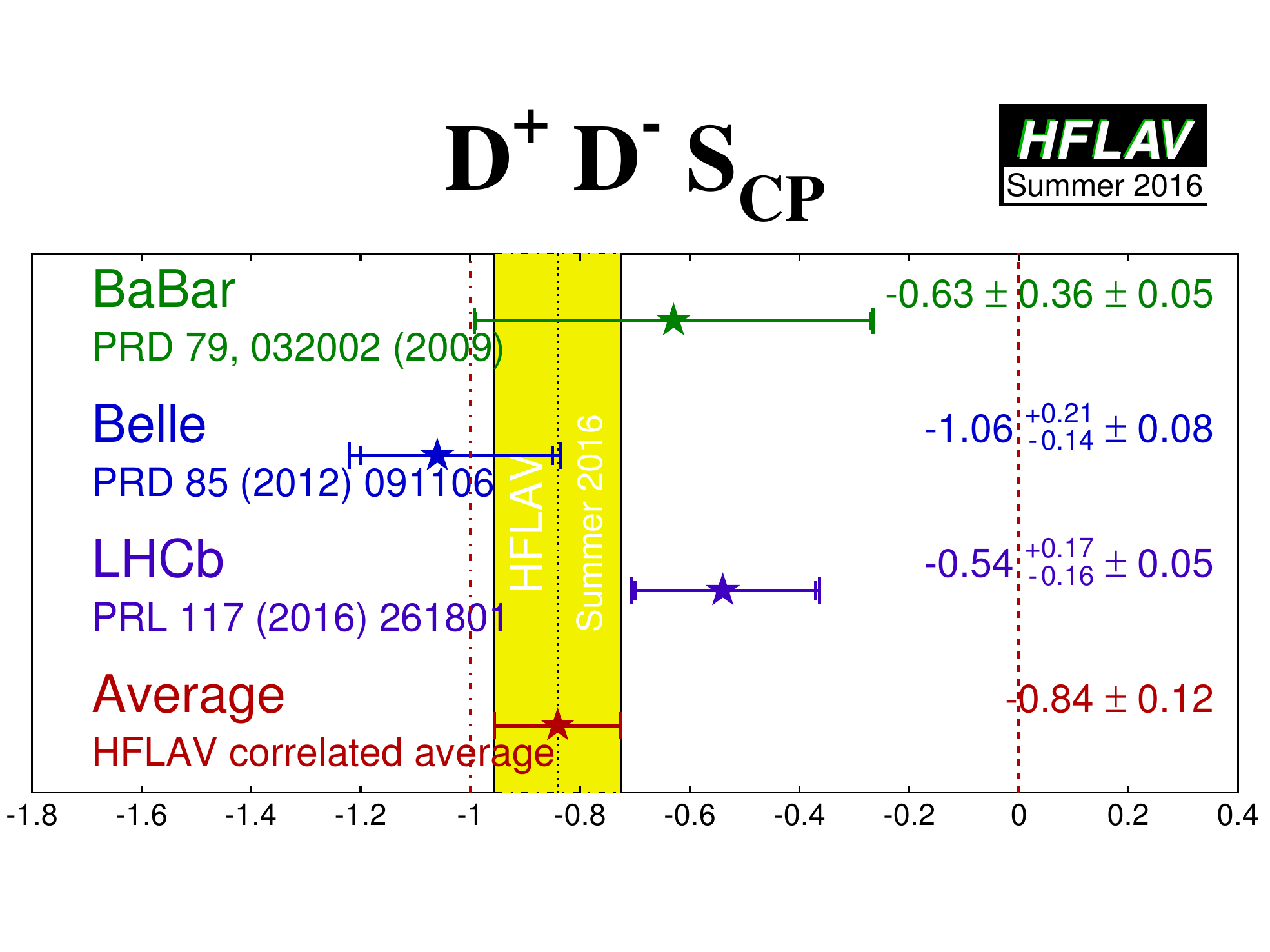}
      }
      &
      \resizebox{0.46\textwidth}{!}{
        \includegraphics{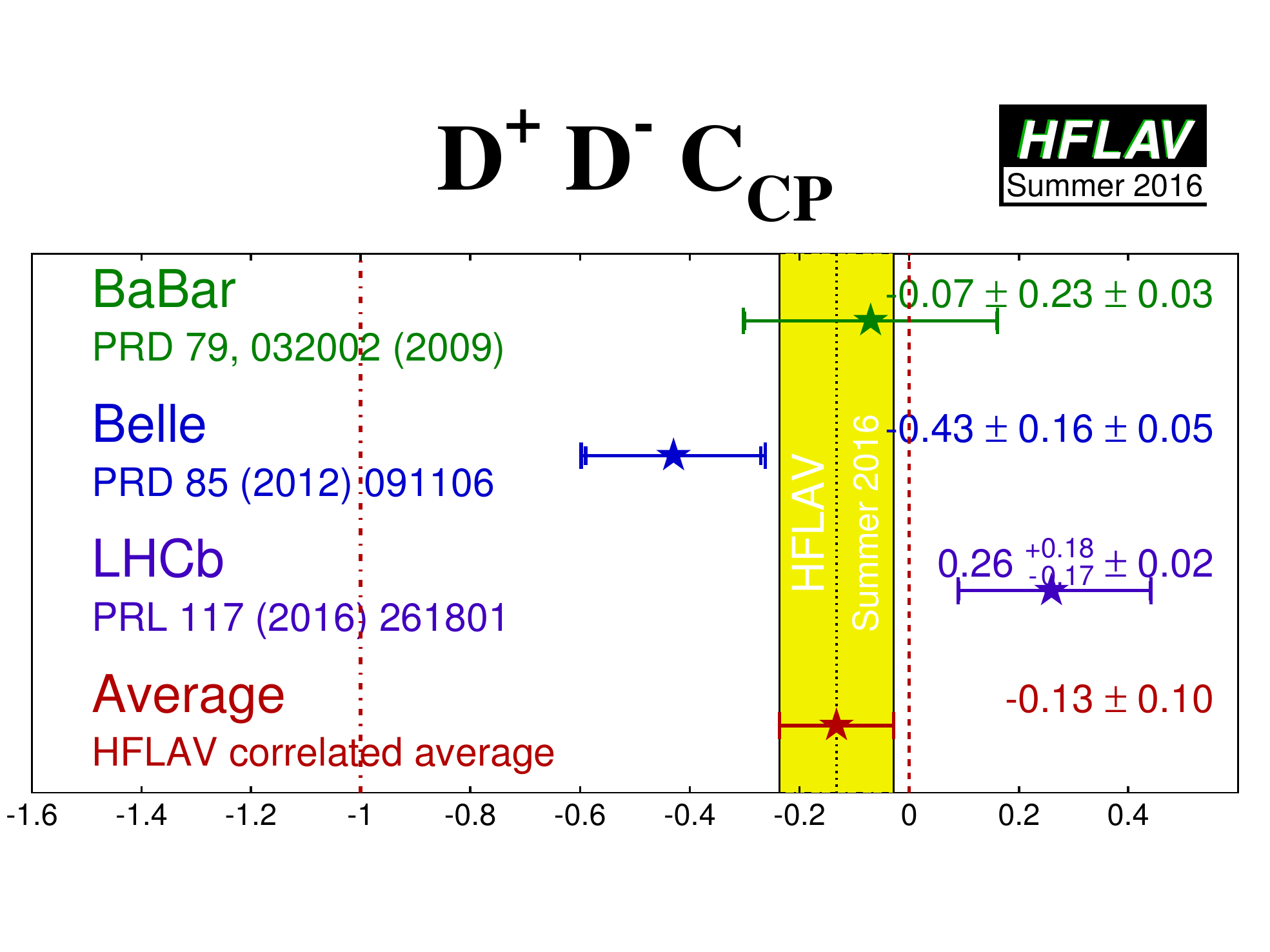}
      }
    \end{tabular}
  \end{center}
  \vspace{-0.8cm}
  \caption{
    Averages of 
    (left) $S_{b \to c\bar c d}$ and (right) $C_{b \to c\bar c d}$ 
    for the mode $\Bz \to D^+D^-$.
  }
  \label{fig:cp_uta:ccd:dd}
\end{figure}

\begin{figure}[htbp]
  \begin{center}
    \begin{tabular}{cc}
      \resizebox{0.46\textwidth}{!}{
        \includegraphics{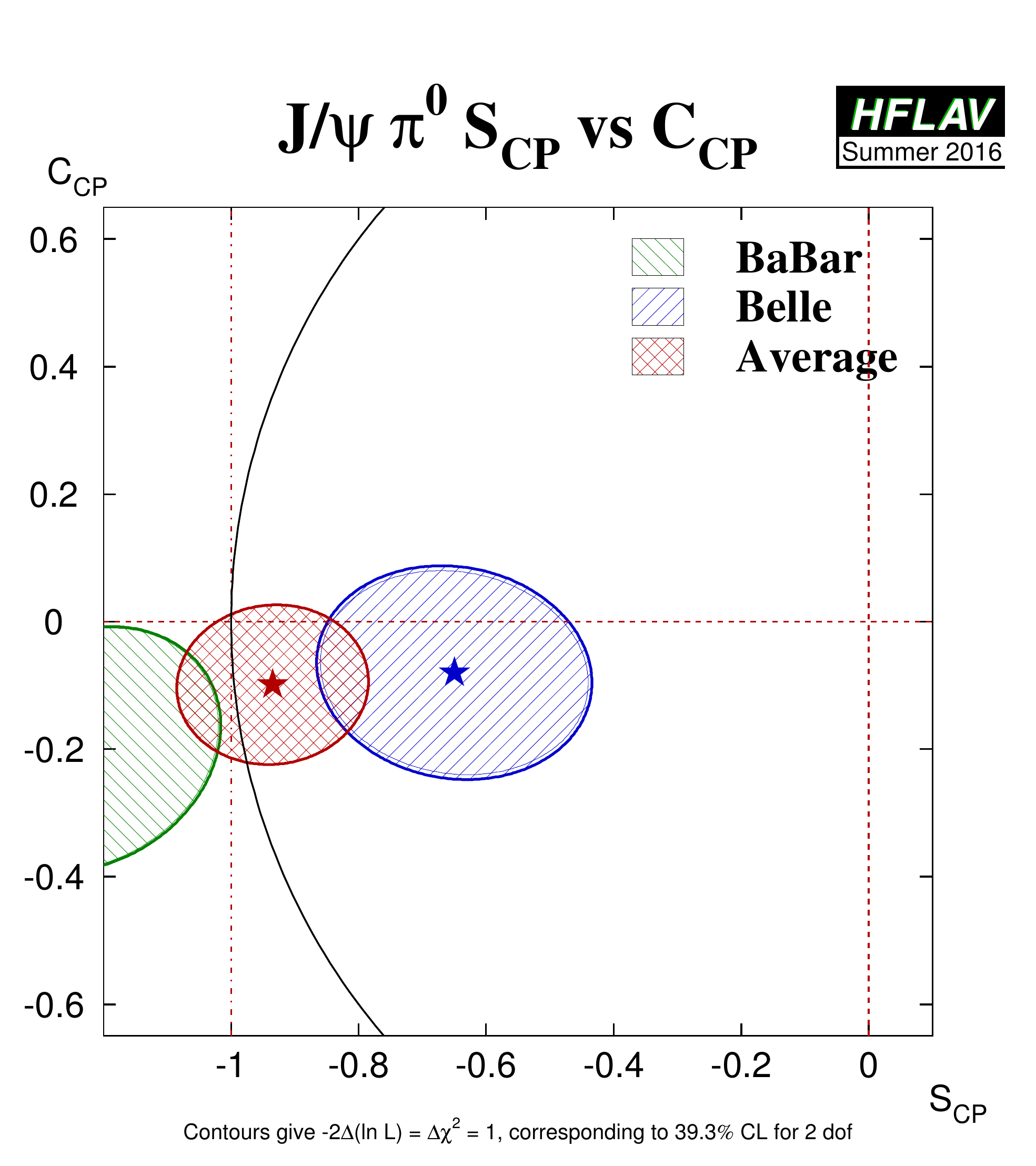}
      }
      &
      \resizebox{0.46\textwidth}{!}{
        \includegraphics{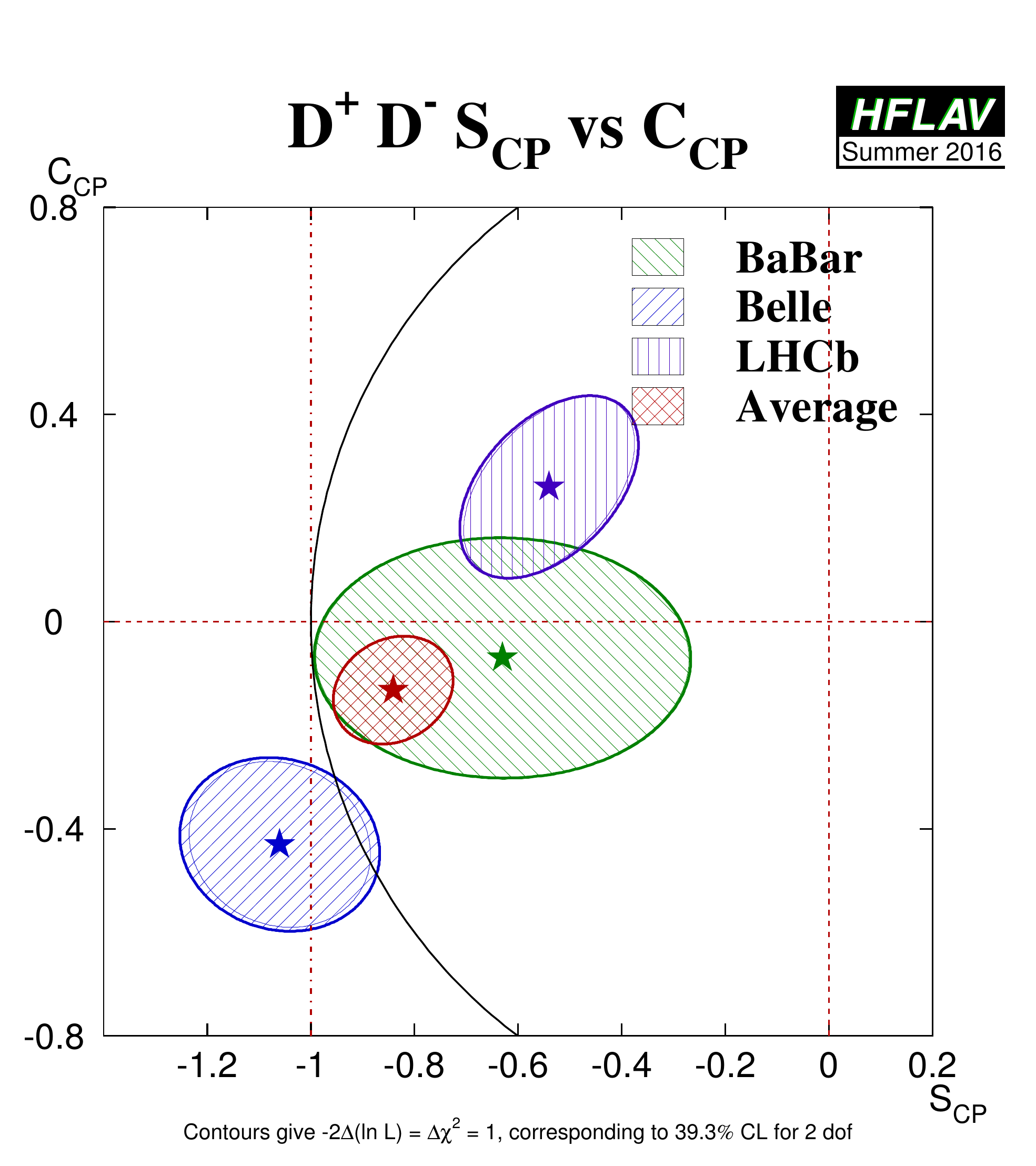}
      }
    \end{tabular}
  \end{center}
  \vspace{-0.8cm}
  \caption{
    Averages of two $b \to c\bar c d$ dominated channels,
    for which correlated averages are performed,
    in the $S_{\CP}$ \vs\ $C_{\CP}$ plane.
    (Left) $\Bz \to J/ \psi \pi^0$ and (right) $\Bz \to D^+D^-$.
  }
  \label{fig:cp_uta:ccd_SvsC}
\end{figure}

The $b \to c\bar{c}d$ transitions can be investigated with studies 
of various different final states. 
Results are available from both \babar\  and \belle\ 
using the final states $\jpsi \pi^0$, $D^+D^-$, $D^{*+}D^{*-}$ and $D^{*\pm}D^{\mp}$,
and from LHCb using the final states $\jpsi \rho^0$ and $D^+D^-$;
the averages of these results are given in Tables~\ref{tab:cp_uta:ccd1} and~\ref{tab:cp_uta:ccd2}.
The results using the $\CP$ eigenstate ($\etacp = +1$) modes
$\jpsi \pi^0$ and $D^+D^-$
are shown in Fig.~\ref{fig:cp_uta:ccd:psipi0} and 
Fig.~\ref{fig:cp_uta:ccd:dd} respectively,
with two-dimensional constraints shown in Fig.~\ref{fig:cp_uta:ccd_SvsC}.

Results for the vector-vector mode $\jpsi \rho^0$ are obtained from a full time-dependent amplitude analysis of $\Bz \to \jpsi \pip\pim$ decays.
LHCb~\cite{Aaij:2014vda} find a $\jpsi \rhoz$ fit fraction of $65.6 \pm 1.9\%$ and a longitudinal polarisation fraction of $56.7 \pm 1.8\%$ (uncertainties are statistical only; both results are consistent with those from a time-integrated amplitude analysis~\cite{Aaij:2014siy} where systematic uncertainties were also evaluated). 
Fits are performed to obtain $2\beta^{\rm eff}$ in the cases that all transversity amplitudes are assumed to have the same \CP violation parameter.
A separate fit is performed allowing different parameters. 
The results in the former case are presented in terms of $S_{\CP}$ and $C_{\CP}$ in Table~\ref{tab:cp_uta:ccd2}. 

\begin{figure}[htbp]
  \begin{center}
    \begin{tabular}{cc}
      \resizebox{0.46\textwidth}{!}{
        \includegraphics{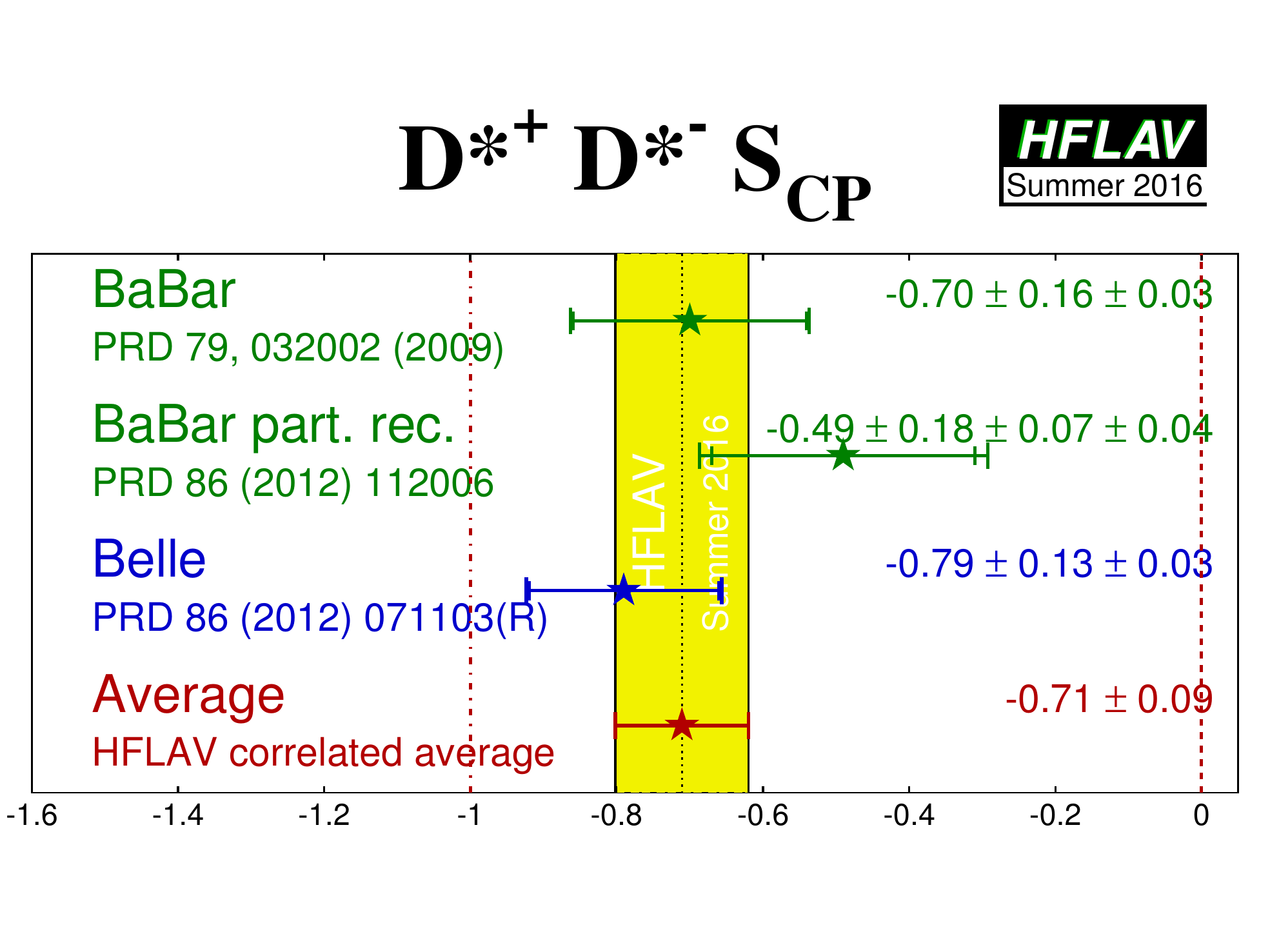}
      }
      &
      \resizebox{0.46\textwidth}{!}{
        \includegraphics{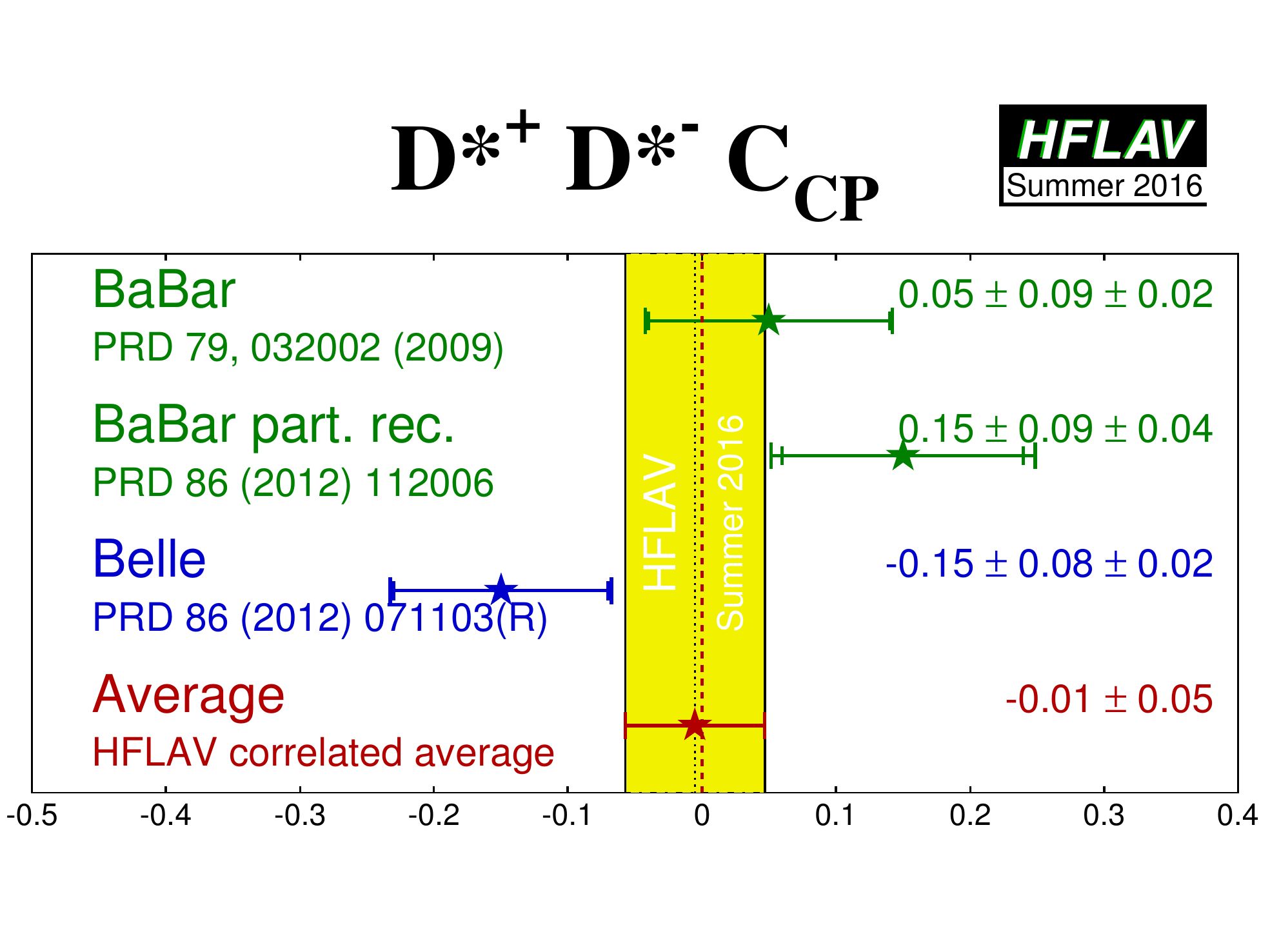}
      }
    \end{tabular}
  \end{center}
  \vspace{-0.8cm}
  \caption{
    Averages of 
    (left) $S_{b \to c\bar c d}$ and (right) $C_{b \to c\bar c d}$ 
    for the mode $\Bz \to D^{*+}D^{*-}$.
  }
  \label{fig:cp_uta:ccd:dstardstar}
\end{figure}

The vector-vector mode $D^{*+}D^{*-}$ 
is found to be dominated by the $\CP$-even longitudinally polarised component;
\babar\ measures a $\CP$-odd fraction of 
$0.158 \pm 0.028 \pm 0.006$~\cite{Aubert:2008ah} while
\belle\ measures a $\CP$-odd fraction of 
$0.138 \pm 0.024 \pm 0.006$~\cite{Kronenbitter:2012ha}.
These values, listed as $R_\perp$, are included in the averages, which ensures
that the correlations are taken into account.\footnote{
  Note that the \babar\ value given in Table~\ref{tab:cp_uta:ccd2} differs from
  the value quoted here, since that in the table is not corrected for efficiency.
}
\babar\ has also performed an additional fit in which the 
$\CP$-even and $\CP$-odd components are allowed to have different 
$\CP$ violation parameters $S$ and $C$.  
These results are included in Table~\ref{tab:cp_uta:ccd2}.
Results using $D^{*+}D^{*-}$ are shown in Fig.~\ref{fig:cp_uta:ccd:dstardstar}.

%% For the non-$\CP$ eigenstate mode $D^{*\pm}D^{\mp}$
%% \babar\ uses fully reconstructed events while 
%% \belle\ combines both fully and partially reconstructed samples.
%% At present we perform uncorrelated averages of the parameters in the 
%% $D^{*\pm}D^{\mp}$ system.

As discussed in Sec.~\ref{sec:cp_uta:notations:non_cp}, the most recent papers on the non-$\CP$ eigenstate mode $D^{*\pm}D^{\mp}$ use the ($A$, $S$, $\Delta S$, $C$, $\Delta C$) set of parameters, and we therefore perform the averages with this choice.

\begin{table}[htb]
	\begin{center}
		\caption{
     Averages for the $b \to c\bar{c}d$ modes,
     $\Bz \to J/\psi \pi^{0}$ and $D^+D^-$.
%			Averages for $J/\psi \pi^{0}$.
		}
		\vspace{0.2cm}
		\setlength{\tabcolsep}{0.0pc}
\renewcommand{\arraystretch}{1.1}
		\begin{tabular*}{\textwidth}{@{\extracolsep{\fill}}lrcccc} \hline
	\mc{2}{l}{Experiment} & Sample size & $S_{\CP}$ & $C_{\CP}$ & Correlation \\
	\hline
        \mc{6}{c}{$J/\psi \pi^{0}$} \\
	\babar & \cite{Aubert:2008bs} & $N(B\bar{B})$ = 466M & $-1.23 \pm 0.21 \pm 0.04$ & $-0.20 \pm 0.19 \pm 0.03$ & $0.20$ \\
	\belle & \cite{:2007wd} & $N(B\bar{B})$ = 535M & $-0.65 \pm 0.21 \pm 0.05$ & $-0.08 \pm 0.16 \pm 0.05$ & $-0.10$ \\
%	\hline
	\mc{3}{l}{\bf Average} & $-0.93 \pm 0.15$ & $-0.10 \pm 0.13$ & $0.04$ \\
	\mc{3}{l}{\small Confidence level} & \mc{2}{c}{\small $0.15~(1.4\sigma)$} & \\
		\hline
% 		\end{tabular*}
% 		\label{tab:cp_uta:yyy}
% 	\end{center}
% \end{table}

% \begin{table}[htb]
% 	\begin{center}
% 		\caption{
% 			Averages for $D^{+} D^{-}$.
% 		}
% 		\vspace{0.2cm}
% 		\setlength{\tabcolsep}{0.0pc}
%		\begin{tabular*}{\textwidth}{@{\extracolsep{\fill}}lrcccc} \hline
% 		\mc{2}{l}{Experiment} & $N(B\bar{B})$ & $S_{\CP}$ & $C_{\CP}$ & Correlation \\
% 		\hline
        \mc{6}{c}{$D^{+} D^{-}$} \\
	\babar & \cite{Aubert:2008ah} & $N(B\bar{B})$ = 467M & $-0.65 \pm 0.36 \pm 0.05$ & $-0.07 \pm 0.23 \pm 0.03$ & $-0.01$ \\
	\belle & \cite{Rohrken:2012ta} & $N(B\bar{B})$ = 772M & $-1.06 \,^{+0.21}_{-0.14} \pm 0.08$ & $-0.43 \pm 0.16 \pm 0.05$ & $-0.12$ \\
%	\hline
	LHCb & \cite{Aaij:2016yip} & $\int {\cal L}\,dt = 3 \, {\rm fb}^{-1}$ & $-0.54 \,^{+0.17}_{-0.16} \pm 0.05$ & $0.26 \,^{+0.18}_{-0.17} \pm 0.02$ & $0.48$ \\
	\mc{3}{l}{\bf Average} & $-0.84 \pm 0.12$ & $-0.13 \pm 0.10$ & $0.18$ \\
	\mc{3}{l}{\small Confidence level} & \mc{2}{c}{\small $0.027~(2.2\sigma)$} & \\
		\hline
 		\end{tabular*}
 		\label{tab:cp_uta:ccd1}
 	\end{center}
 \end{table}

% \begin{table}[htb]
\begin{sidewaystable}
 	\begin{center}
 		\caption{
      Averages for the $b \to c\bar{c}d$ modes,
      $\jpsi\rho^0$, $D^{*+} D^{*-}$ and $D^{*\pm}D^\mp$.
 		}
% 		\vspace{0.2cm}
% 		\setlength{\tabcolsep}{0.0pc}

\renewcommand{\arraystretch}{1.1}
 		\begin{tabular*}{\textwidth}{@{\extracolsep{\fill}}lrcccc} \hline
 		\mc{2}{l}{Experiment} & $N(B\bar{B})$ & $S_{\CP}$ & $C_{\CP}$ & $R_\perp$ \\
	\hline
        \mc{6}{c}{$\jpsi\rho^0$} \\
	LHCb & \cite{Aaij:2014vda} & 3 ${\rm fb}^{-1}$ & $-0.66 \,^{+0.13}_{-0.12} \,^{+0.09}_{-0.03}$ & $-0.06 \pm 0.06 \,^{+0.02}_{-0.01}$ & $0.198 \pm 0.017$ \\
		\hline
        \mc{6}{c}{$D^{*+} D^{*-}$} \\
	\babar & \cite{Aubert:2008ah} & 467M & $-0.70 \pm 0.16 \pm 0.03$ & $0.05 \pm 0.09 \pm 0.02$ & $0.17 \pm 0.03$ \\
	\babar part. rec. & \cite{Lees:2012px} & 471M & $-0.49 \pm 0.18 \pm 0.07 \pm 0.04$ & $0.15 \pm 0.09 \pm 0.04$ & \textemdash{} \\
	\belle & \cite{Kronenbitter:2012ha} & 772M & $-0.79 \pm 0.13 \pm 0.03$ & $-0.15 \pm 0.08 \pm 0.02$ & $0.14 \pm 0.02 \pm 0.01$ \\
%	\hline
	\mc{3}{l}{\bf Average} & $-0.71 \pm 0.09$ & $-0.01 \pm 0.05$ & $0.15 \pm 0.02$ \\
	\mc{3}{l}{\small Confidence level} & \mc{3}{c}{\small $0.72~(0.4\sigma)$} \\
		\hline
		\end{tabular*}
% 		\label{tab:cp_uta:yyy}
% 	\end{center}
% \end{table}

                \vspace{2ex}

%\begin{table}[htb]
% 	\begin{center}
% 		\caption{
% 			Averages for $D*^{+} D*^{-} 2$.
% 		}
% 		\vspace{0.2cm}
% 		\setlength{\tabcolsep}{0.0pc}
% make this tabular (not tabular*) and resize down to \textwidth
% change @{\extracolsep{\fill}} to @{\extracolsep{2mm}}
    \resizebox{\textwidth}{!}{
		\begin{tabular}{@{\extracolsep{2mm}}lrcccccc} \hline
	\mc{2}{l}{Experiment} & $N(B\bar{B})$ & $S_{\CP+}$ & $C_{\CP+}$ & $S_{\CP-}$ & $C_{\CP-}$ & $R_\perp$ \\
	\hline
        \mc{8}{c}{$D^{*+} D^{*-}$} \\
	\babar & \cite{Aubert:2008ah} & 467M & $-0.76 \pm 0.16 \pm 0.04$ & $0.02 \pm 0.12 \pm 0.02$ & $-1.81 \pm 0.71 \pm 0.16$ & $0.41 \pm 0.50 \pm 0.08$ & $0.15 \pm 0.03$ \\
%	\hline
%	\mc{3}{l}{\bf Average} & $-0.76 \pm 0.16$ & $0.02 \pm 0.12$ & $-1.81 \pm 0.73$ & $0.41 \pm 0.51$ & $0.15 \pm 0.03$ & \textendash{} \\
%	\mc{3}{l}{\small Confidence level} & \mc{5}{c}{\small $0.xx~(y.y\sigma)$} & \\
		\hline
		\end{tabular}
    }
% 		\label{tab:cp_uta:yyy}
% 	\end{center}
% \end{table}

                \vspace{2ex}

% \begin{table}[htb]
% 	\begin{center}
% 		\caption{
% 			Averages for $D*^{\pm} D^{\mp}$.
% 		}
% 		\vspace{0.2cm}
% 		\setlength{\tabcolsep}{0.0pc}
% make this tabular (not tabular*) and resize down to \textwidth
% change @{\extracolsep{\fill}} to @{\extracolsep{2mm}}
    \resizebox{\textwidth}{!}{
		\begin{tabular}{@{\extracolsep{2mm}}lrcccccc} \hline
	\mc{2}{l}{Experiment} & $N(B\bar{B})$ & $S$ & $C$ & $\Delta S$ & $\Delta C$ & ${\cal A}$ \\
        \hline
        \mc{8}{c}{$D^{*\pm} D^{\mp}$} \\
	\babar & \cite{Aubert:2008ah} & 467M & $-0.68 \pm 0.15 \pm 0.04$ & $0.04 \pm 0.12 \pm 0.03$ & $0.05 \pm 0.15 \pm 0.02$ & $0.04 \pm 0.12 \pm 0.03$ & $0.01 \pm 0.05 \pm 0.01$ \\
	\belle & \cite{Rohrken:2012ta} & 772M & $-0.78 \pm 0.15 \pm 0.05$ & $-0.01 \pm 0.11 \pm 0.04$ & $-0.13 \pm 0.15 \pm 0.04$ & $0.12 \pm 0.11 \pm 0.03$ & $0.06 \pm 0.05 \pm 0.02$ \\
%	\hline
	\mc{3}{l}{\bf Average} & $-0.73 \pm 0.11$ & $0.01 \pm 0.09$ & $-0.04 \pm 0.11$ & $0.08 \pm 0.08$ & $0.03 \pm 0.04$ \\
	\mc{3}{l}{\small Confidence level} & {\small $0.65~(0.5\sigma)$} & {\small $0.77~(0.3\sigma)$} & {\small $0.41~(0.8\sigma)$} & {\small $0.63~(0.5\sigma)$} & {\small $0.48~(0.7\sigma)$} \\
        \hline
                \end{tabular}
    }
		\label{tab:cp_uta:ccd2}
	\end{center}
\end{sidewaystable}
% \end{table}

In the absence of the penguin contribution (tree dominance),
the time-dependent parameters would be given by
$S_{b \to c\bar c d} = - \etacp \sin(2\beta)$,
$C_{b \to c\bar c d} = 0$,
$S_{+-} = \sin(2\beta + \delta)$,
$S_{-+} = \sin(2\beta - \delta)$,
$C_{+-} = - C_{-+}$ and 
${\cal A} = 0$,
where $\delta$ is the strong phase difference between the 
$D^{*+}D^-$ and $D^{*-}D^+$ decay amplitudes.
In the presence of the penguin contribution,
there is no clean interpretation in terms of CKM parameters;
however,
direct $\CP$ violation may be observed through any of
$C_{b \to c\bar c d} \neq 0$, $C_{+-} \neq - C_{-+}$ or $A_{+-} \neq 0$.

\begin{figure}[htbp]
  \begin{center}
    \begin{tabular}{cc}
      \resizebox{0.46\textwidth}{!}{
        \includegraphics{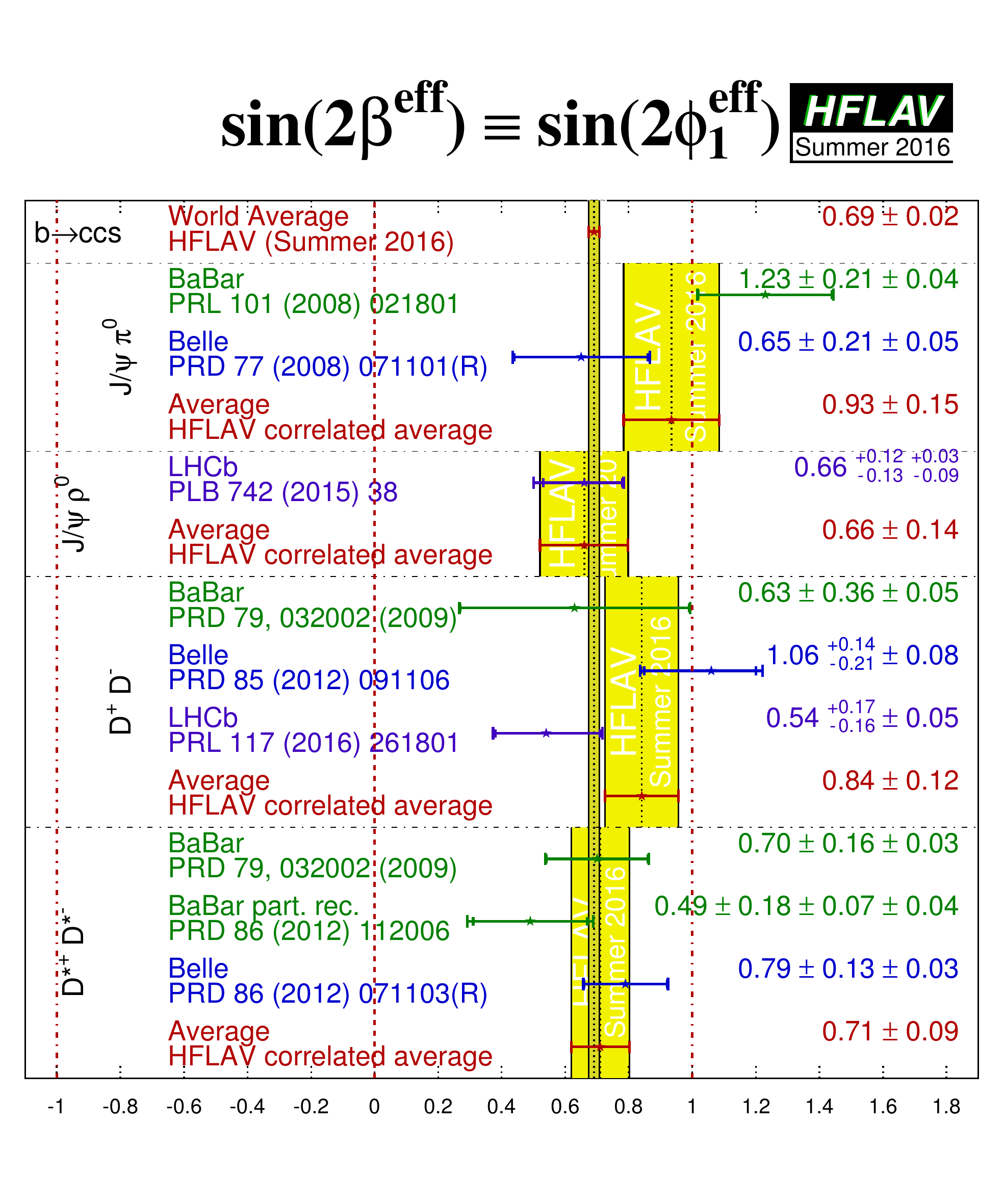}
      }
      &
      \resizebox{0.46\textwidth}{!}{
        \includegraphics{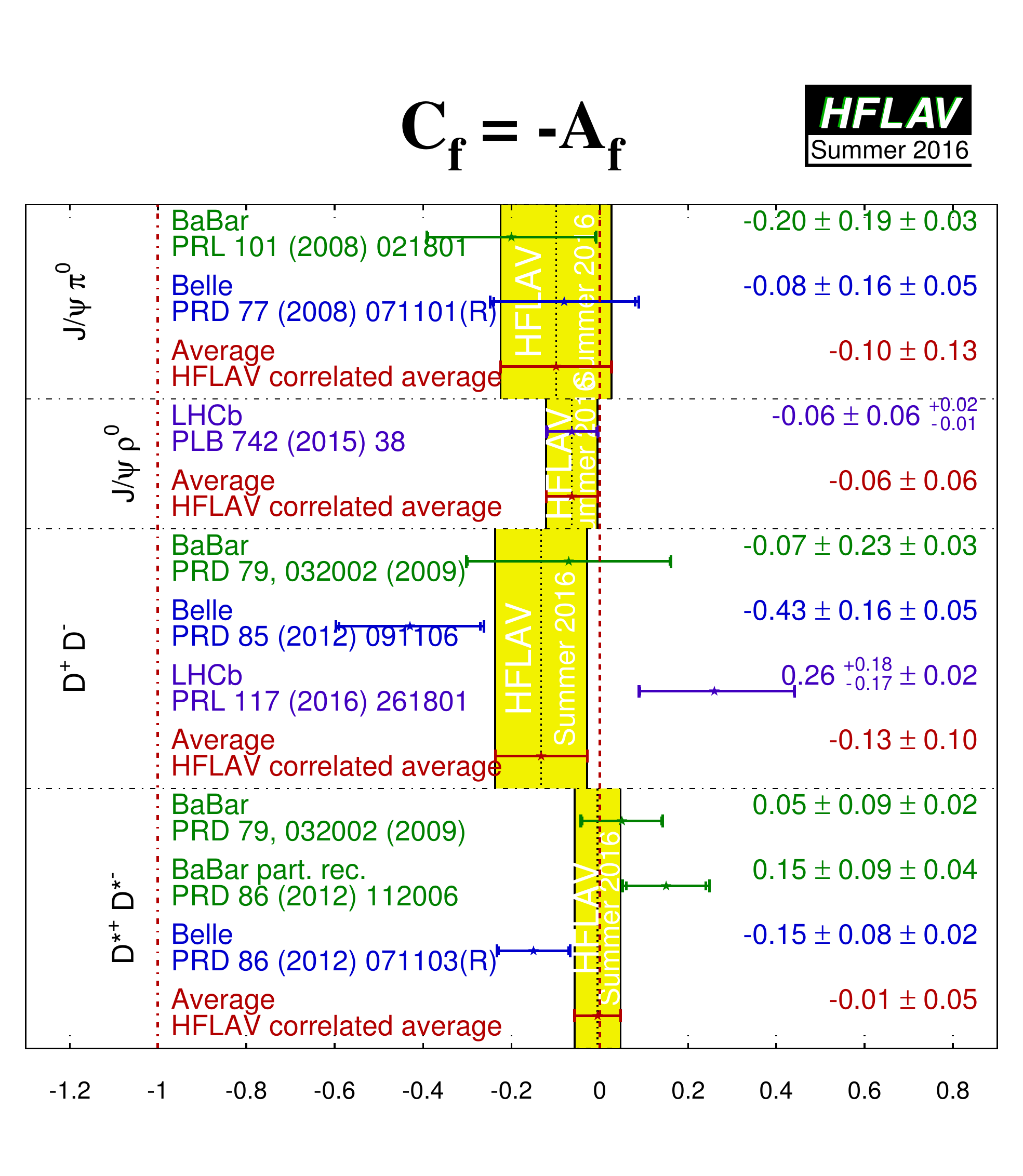}
      }
    \end{tabular}
  \end{center}
  \vspace{-0.8cm}
  \caption{
    Averages of 
    (left) $-\etacp S_{b \to c\bar c d}$ interpreted as $\sin(2\beta^{\rm eff})$ and (right) $C_{b \to c\bar c d}$.
    The $-\etacp S_{b \to c\bar c d}$ figure compares the results to 
    the world average 
    for $-\etacp S_{b \to c\bar c s}$ (see Sec.~\ref{sec:cp_uta:ccs:cp_eigen}).
  }
  \label{fig:cp_uta:ccd}
\end{figure}

\begin{figure}[htbp]
  \begin{center}
    \resizebox{0.66\textwidth}{!}{
      \includegraphics{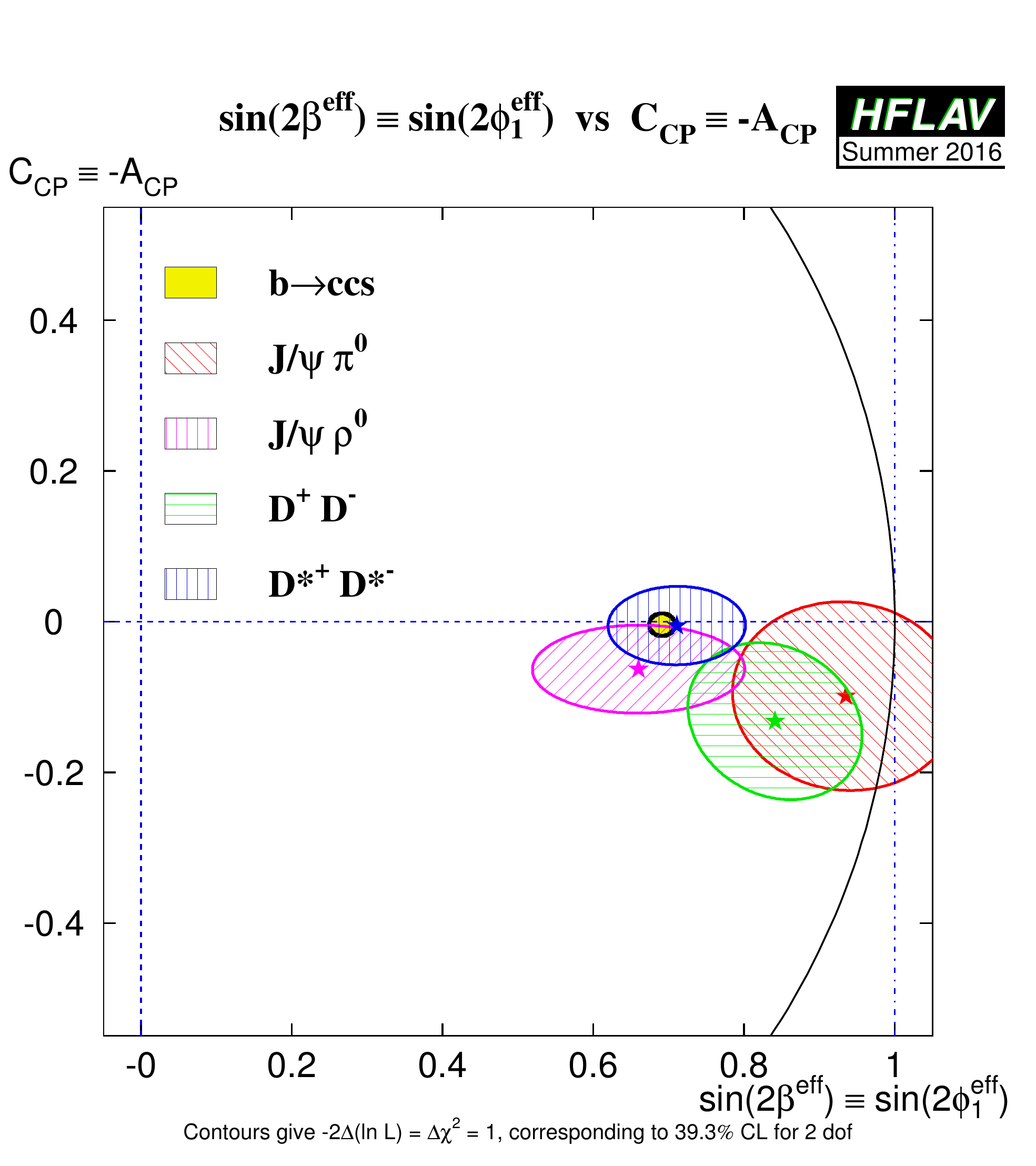}
    }
  \end{center}
  \vspace{-0.8cm}
  \caption{
    Compilation of constraints in the 
    $-\etacp S_{b \to c\bar c d}$, interpreted as $\sin(2\beta^{\rm eff})$, \vs\ $C_{b \to c\bar c d}$ plane.
  }
  \label{fig:cp_uta:ccd_SvsC-all}
\end{figure}

The averages for the $b \to c\bar c d$ modes 
are shown in Figs.~\ref{fig:cp_uta:ccd} and~\ref{fig:cp_uta:ccd_SvsC-all}.
Results are consistent with tree dominance,
and with the Standard Model,
though the \belle\ results in $\Bz \to D^+D^-$~\cite{Fratina:2007zk}
show an indication of $\CP$ violation in decay,
and hence a non-zero penguin contribution.
The average of $S_{b \to c\bar c d}$ in each of the $J/\psi \pi^{0}$, $\Dp\Dm$ and $D^{*+}D^{*-}$ final states is more than $5\sigma$ from zero, corresponding to observations of \CP violation in these decay channels.
% That in the $D^+D^-$ final state is more than $3\sigma$ from zero;
% however, due to the large uncertainty and possible non-Gaussian effects,
% any strong conclusion should be deferred.
Possible non-Gaussian effects due to some of the inputs measurements being outside the physical region ($S_{\CP}^2 + C_{\CP}^2 \leq 1$) should, however, be borne in mind.

% Comparisons of the results for the $b \to c\bar c d$ modes 
% to the $b \to c\bar c s$ and $b \to q\bar q s$ modes,
% can be seen in Fig.~\ref{fig:cp_uta:qqs_ccd}.

\mysubsubsection{Time-dependent $\CP$ asymmetries in $\Bs$ decays mediated by $b \to c\bar{c}d$ transitions
}
\label{sec:cp_uta:ccd:Bs}

Time-dependent \CP asymmetries in \Bs\ decays mediated by $b \to c\bar{c}d$ transitions provide a determination of $2\beta_s^{\rm eff}$ where possible effects from penguin amplitudes may cause a shift from the value of $2\beta_s$ seen in $b \to c\bar{c}s$ transitions. 
Results in the $b \to c\bar{c}d$ case, with larger penguin effects, can be used together with flavour symmetries to derive limits on the possible size of penguin effects in the $b \to c\bar{c}s$ transitions~\cite{Fleischer:1999nz,DeBruyn:2010hh}. 
% If the penguin effect is large, it may also be possible to determine $\gamma \equiv \phi_3$.

The parameters have been measured in $\Bs \to \jpsi\KS$ decays by LHCb, as summarised in Table~\ref{tab:cp_uta:ccd:Bs}.
The results supersede an earlier measurement of the effective lifetime, which is directly related to $A^{\Delta\Gamma}$, in the same mode~\cite{Aaij:2013eia}, which is discussed in Sec.~\ref{sec:life_mix}.

\begin{table}[!htb]
	\begin{center}
		\caption{
     Measurements of \CP violation parameters from $\Bs \to \jpsi\KS$.
%			Averages for $J/\psi K_{S}$.
		}
		\vspace{0.2cm}
		\setlength{\tabcolsep}{0.0pc}
\renewcommand{\arraystretch}{1.1}
		\begin{tabular*}{\textwidth}{@{\extracolsep{\fill}}lrcccc} \hline
	\mc{2}{l}{Experiment} & $\int {\cal L}\,dt$ & $S_{\CP}$ & $C_{\CP}$ & $A^{\Delta\Gamma}$ \\
	\hline
	LHCb & \cite{Aaij:2015tza} & $3 \ {\rm fb}^{-1}$ & $0.49 \,^{+0.77}_{-0.65} \pm 0.06$ & $-0.28 \pm 0.41 \pm 0.08$ & $-0.08 \pm 0.40 \pm 0.08$ \\
%	\hline
%	\mc{3}{l}{\bf Average} & $0.49 \pm 0.71$ & $-0.28 \pm 0.42$ & $-0.08 \pm 0.41$ & \textendash{} \\
%	\mc{3}{l}{\small Confidence level} & \mc{3}{c}{\small $0.xx~(y.y\sigma)$} & \\
		\hline
		\end{tabular*}
		\label{tab:cp_uta:ccd:Bs}
	\end{center}
\end{table}

%% straightforward interpretation

%%%%%%%%
%%%
%%% qqs
%%%
%%%%%%%%
% \afterpage{\clearpage}
\mysubsection{Time-dependent $\CP$ asymmetries in charmless $b \to q\bar{q}s$ transitions
}
\label{sec:cp_uta:qqs}

Similarly to Eq.~(\ref{eq:cp_uta:b_to_d}), the $b \to s$ penguin amplitude can be written as
\begin{equation}
  \label{eq:cp_uta:b_to_s}
  \begin{array}{ccccc}
    A_{b \to s} & = & 
    \mc{3}{l}{F_u V_{ub}V^*_{us} + F_c V_{cb}V^*_{cs} + F_t V_{tb}V^*_{ts}} \\
    & = & (F_u - F_c) V_{ub}V^*_{us} & + & (F_t - F_c) V_{tb}V^*_{ts} \, , \\
%    & = & {\cal O}(\lambda^4) & + & {\cal O}(\lambda^2) \, , \\
  \end{array}
\end{equation}
using the unitarity of the CKM matrix to eliminate the $F_c$ term.
In this case, the first term in the last line is ${\cal O}(\lambda^4)$ while the second is ${\cal O}(\lambda^2)$.
Therefore, in the Standard Model, this amplitude is dominated by $V_{tb}V^*_{ts}$, and to within a few degrees ($\left| \delta\beta^{\rm eff} \right| \equiv \left| \beta^{\rm eff}-\beta \right| \lesssim 2^\circ$ for $\beta \approx 20^\circ$) the time-dependent parameters can be written\footnote{
  The presence of a small (${\cal O}(\lambda^2)$) weak phase in the dominant amplitude of the $s$ penguin decays introduces a phase shift given by
  $S_{b \to q\bar q s} = -\eta\sin(2\beta)(1 + \Delta)$. 
  Using the CKMfitter results for the Wolfenstein parameters~\cite{Charles:2004jd}, one finds $\Delta \simeq 0.033$, which corresponds to a shift of $2\beta$ of $+2.1^\circ$. 
  Nonperturbative contributions can alter this result.
}
$S_{b \to q\bar q s} \approx - \etacp \sin(2\beta)$,
$C_{b \to q\bar q s} \approx 0$,
assuming $b \to s$ penguin contributions only ($q = u,d,s$).

Due to the suppression of the Standard Model amplitude, contributions of additional diagrams from physics beyond the Standard Model,
with heavy virtual particles in the penguin loops, may have observable effects.
In general, these contributions will affect the values of 
$S_{b \to q\bar q s}$ and $C_{b \to q\bar q s}$.
A discrepancy between the values of 
$S_{b \to c\bar c s}$ and $S_{b \to q\bar q s}$
can therefore provide a clean indication of new physics~\cite{Grossman:1996ke,Fleischer:1996bv,London:1997zk,Ciuchini:1997zp}.

However, there is an additional consideration to take into account.
The above argument assumes that only the $b \to s$ penguin contributes
to the $b \to q\bar q s$ transition.
For $q = s$ this is a good assumption, which neglects only rescattering effects.
However, for $q = u$ there is a colour-suppressed $b \to u$ tree diagram
(of order ${\cal O}(\lambda^4)$), 
which has a different weak (and possibly strong) phase.
In the case $q = d$, any light neutral meson that is formed from $d \bar{d}$ 
also has a $u \bar{u}$ component, and so again there is ``tree pollution''. 
The \Bz decays to $\piz\KS$, $\rho^0\KS$ and $\omega\KS$ belong to this category.
The mesons $\phi$, $f_0$ and $\etapr$ are expected to have predominant
$s\bar{s}$ composition, which reduces the relative size of the possible tree
pollution. 
If the inclusive decay $\Bz\to\Kp\Km\Kz$ (excluding $\phi\Kz$) is dominated by
a nonresonant three-body transition, 
an Okubo-Zweig-Iizuka-suppressed~\cite{Okubo:1963fa,Zweig:1964jf,Iizuka:1966fk} tree-level diagram can occur through insertion of an $s\sbar$ pair. 
The corresponding penguin-type transition 
proceeds via insertion of a $u\ubar$ pair, which is expected
to be favoured over the $s\sbar$ insertion by fragmentation models.
Neglecting rescattering, the final state $\Kz\Kzb\Kz$ 
(reconstructed as $\KS\KS\KS$) has no tree pollution~\cite{Gershon:2004tk}.
%% this is repeated below
% Note that the calculation of an average between different modes
% implicitly neglects contributions with different weak phases 
% to the $b \to s$ penguin amplitude.
Various estimates, using different theoretical approaches,
of the values of $\Delta S = S_{b \to q\bar q s} - S_{b \to c\bar c s}$
exist in the literature~\cite{Grossman:2003qp,Gronau:2003ep,Gronau:2003kx,Gronau:2004hp,Cheng:2005bg,Gronau:2005gz,Buchalla:2005us,Beneke:2005pu,Engelhard:2005hu,Cheng:2005ug,Engelhard:2005ky,Gronau:2006qh,Silvestrini:2007yf,Dutta:2008xw}.
In general, there is agreement that the modes
$\phi\Kz$, $\etapr\Kz$ and $\Kz\Kzb\Kz$ are the cleanest,
with values of $\left| \Delta S \right|$ at or below the few percent level 
($\Delta S$ is usually predicted to be positive).
Nonetheless, the uncertainty is sufficient that interpretation is given in terms of $\sin(2\beta^{\rm eff})$.

\mysubsubsection{Time-dependent $\CP$ asymmetries: $b \to q\bar{q}s$ decays to $\CP$ eigenstates
}
\label{sec:cp_uta:qqs:cp_eigen}

\begin{figure}[htbp]
  \begin{center}
    \resizebox{0.45\textwidth}{!}{
      \includegraphics{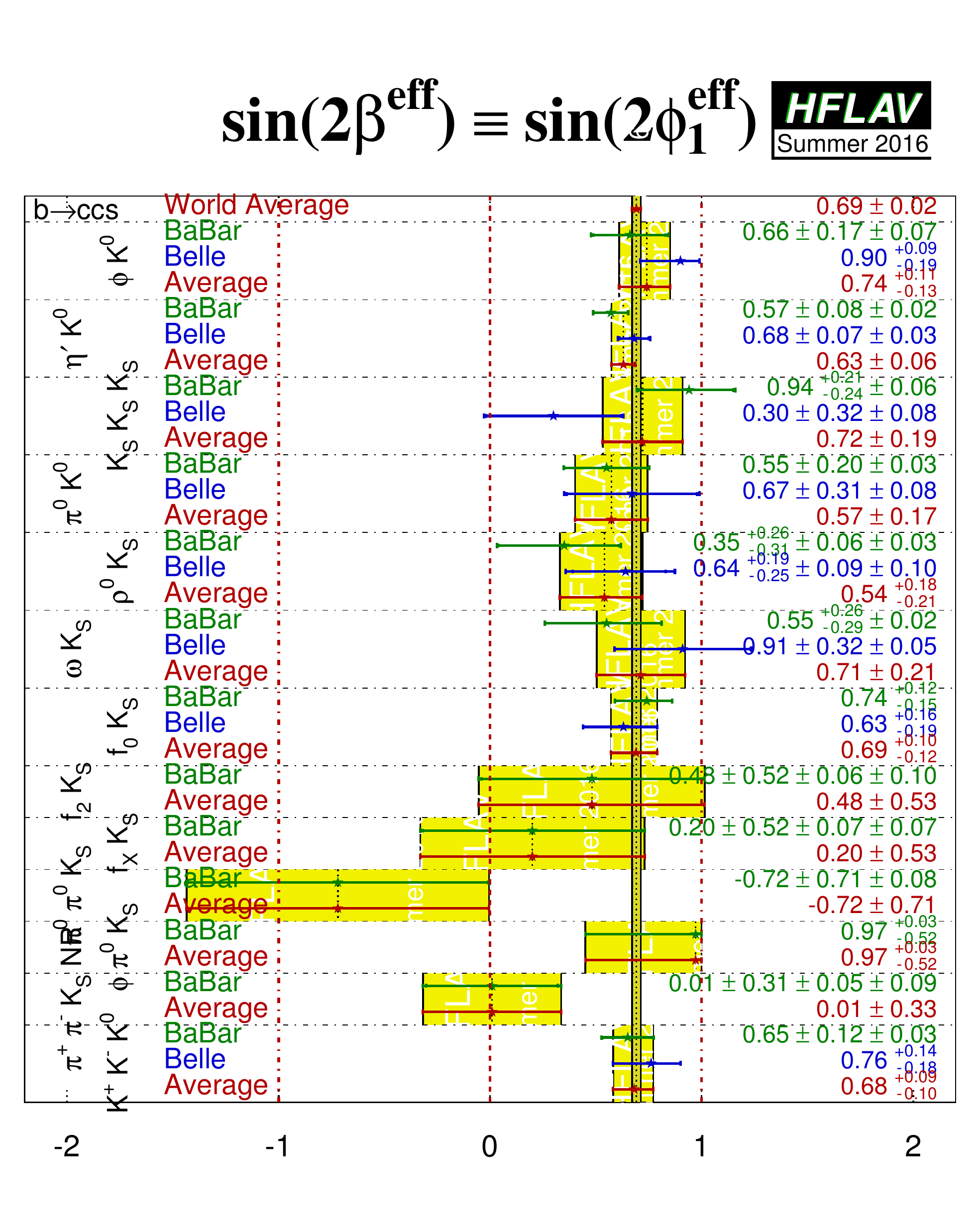}
    }
    \hfill
    \resizebox{0.45\textwidth}{!}{
      \includegraphics{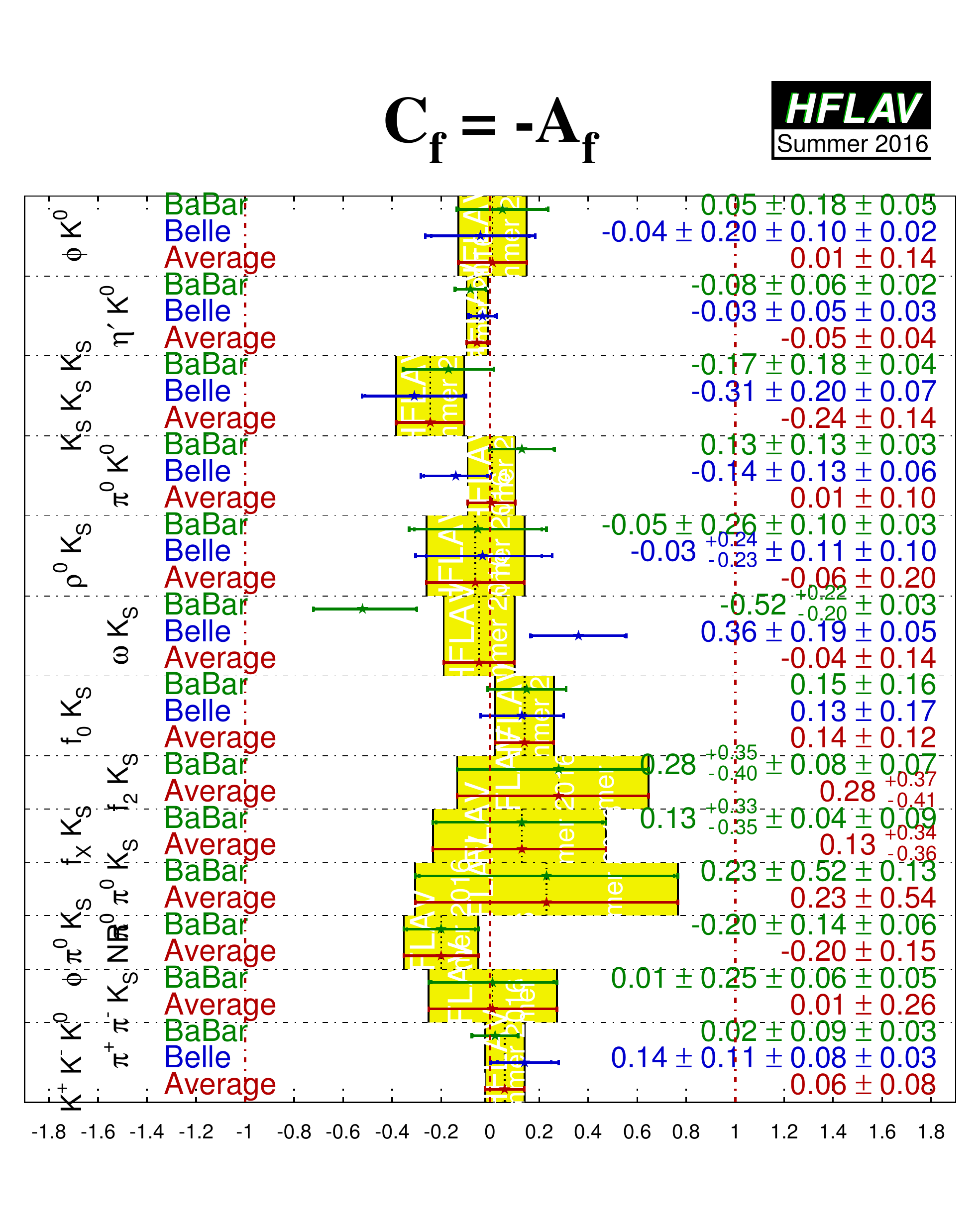}
    }
    \\
    \resizebox{0.45\textwidth}{!}{
      \includegraphics{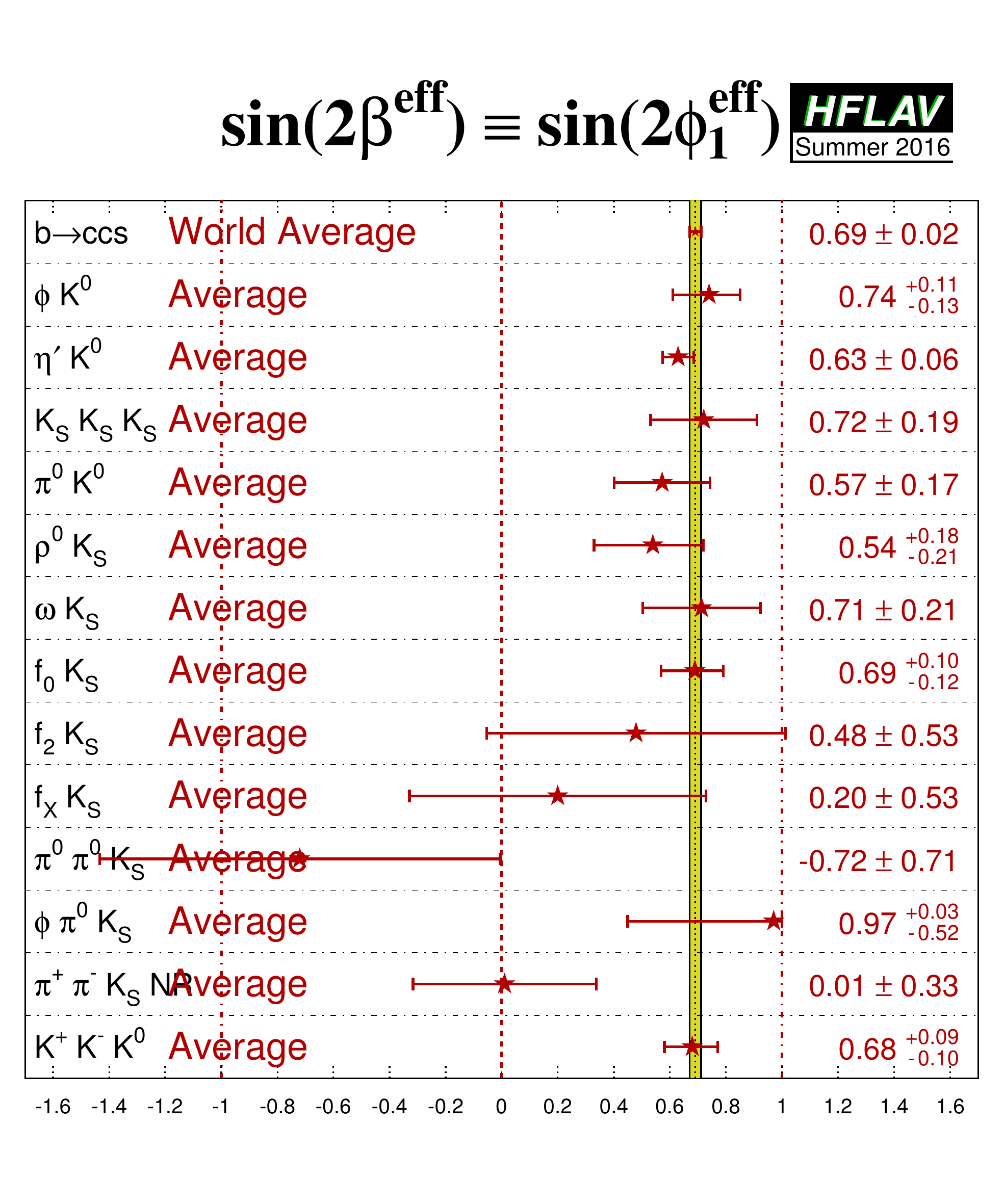}
    }
    \hfill
    \resizebox{0.45\textwidth}{!}{
      \includegraphics{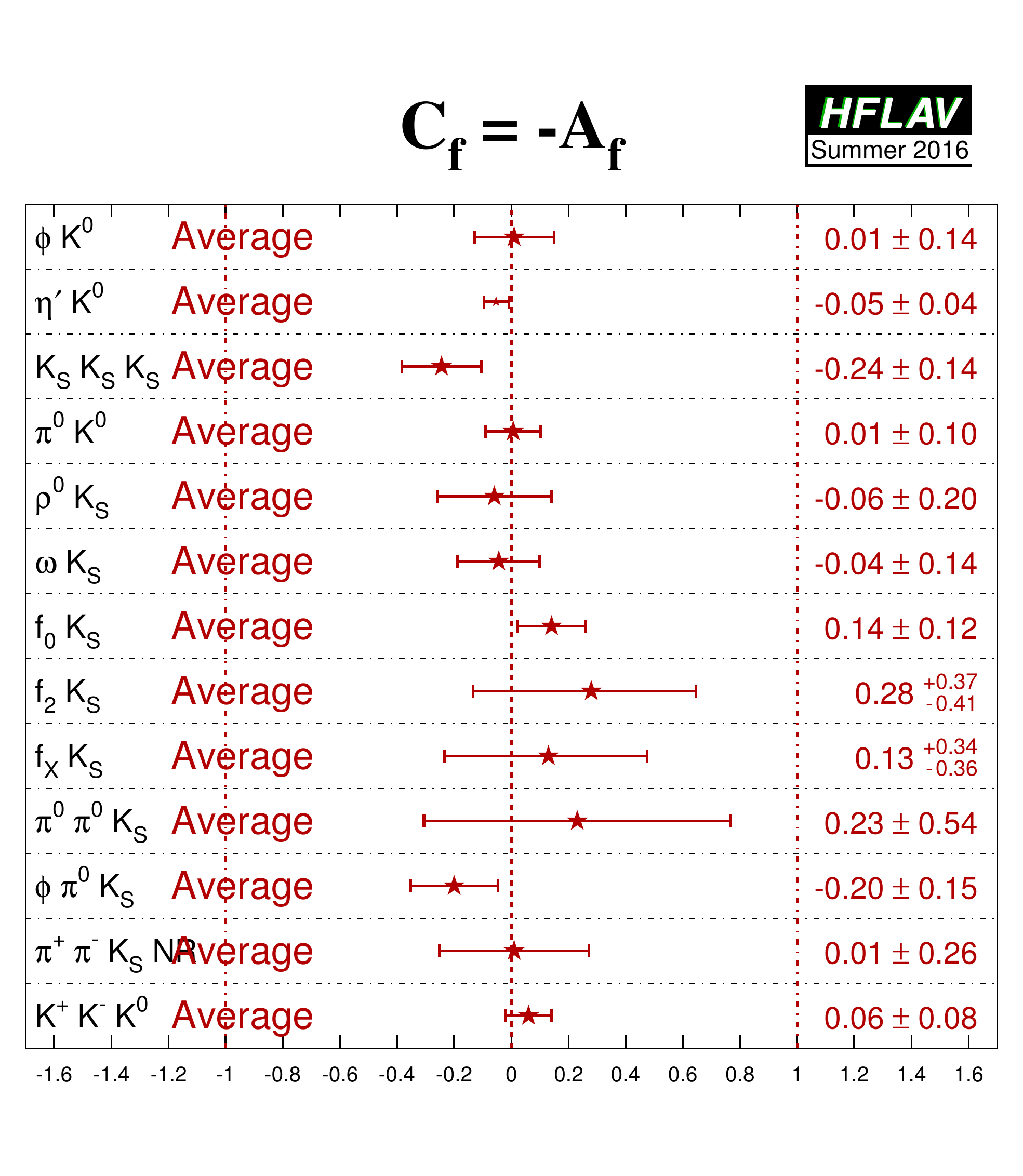}
    }
  \end{center}
  \vspace{-0.8cm}
  \caption{
    (Top)
    Averages of 
    (left) $-\etacp S_{b \to q\bar q s}$, interpreted as $\sin(2\beta^{\rm eff}$ and (right) $C_{b \to q\bar q s}$.
    The $-\etacp S_{b \to q\bar q s}$ figure compares the results to 
    the world average 
    for $-\etacp S_{b \to c\bar c s}$ (see Sec.~\ref{sec:cp_uta:ccs:cp_eigen}).
    (Bottom) Same, but only averages for each mode are shown.
    More figures are available from the HFLAV web pages.
  }
  \label{fig:cp_uta:qqs}
\end{figure}

\begin{figure}[htbp]
  \begin{center}
    \resizebox{0.38\textwidth}{!}{
      \includegraphics{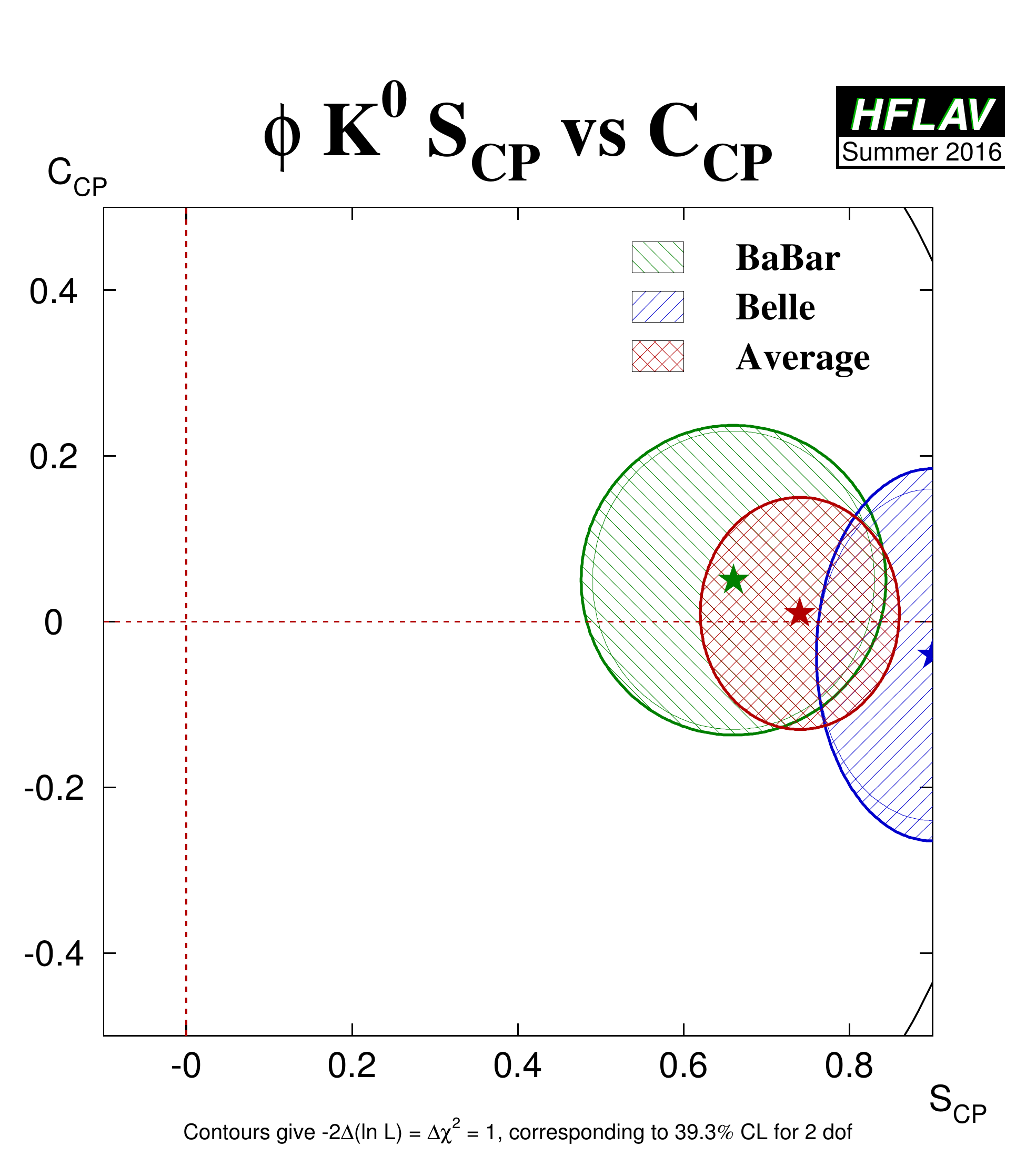}
    }
    \hspace{0.04\textwidth}
    \resizebox{0.38\textwidth}{!}{
      \includegraphics{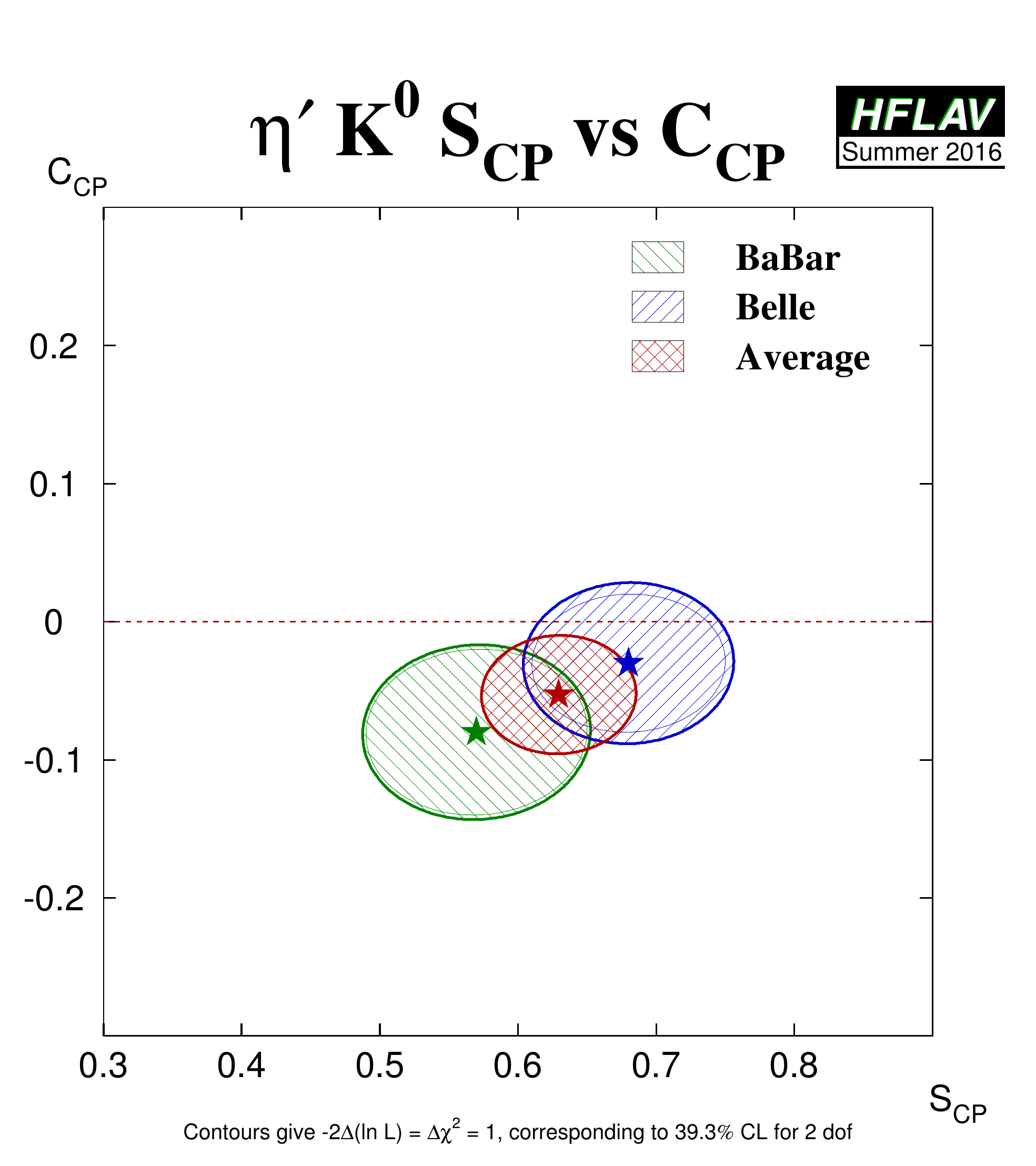}
    }
    \\
    \resizebox{0.38\textwidth}{!}{
      \includegraphics{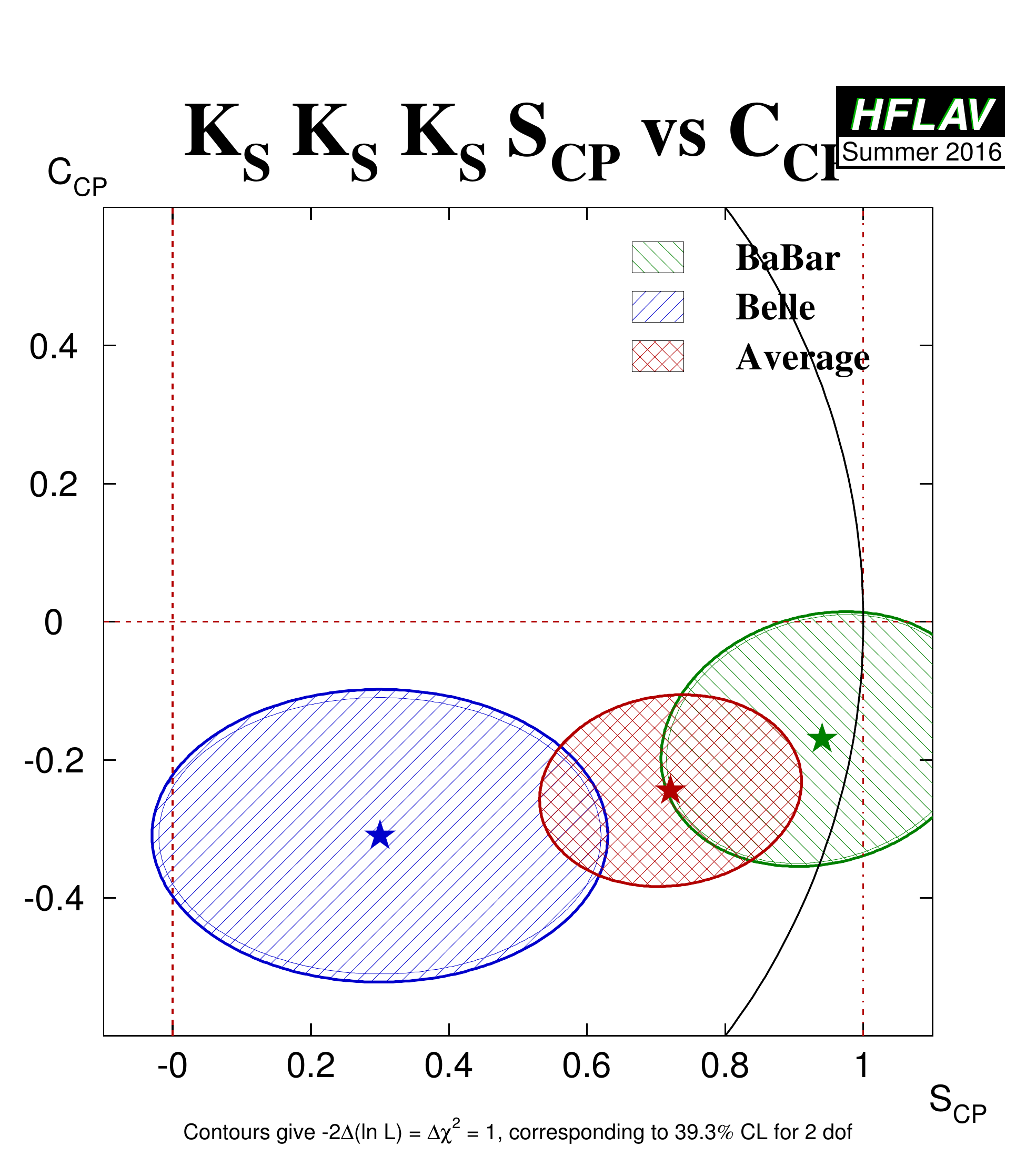}
    }
    \hspace{0.04\textwidth}
    \resizebox{0.38\textwidth}{!}{
      \includegraphics{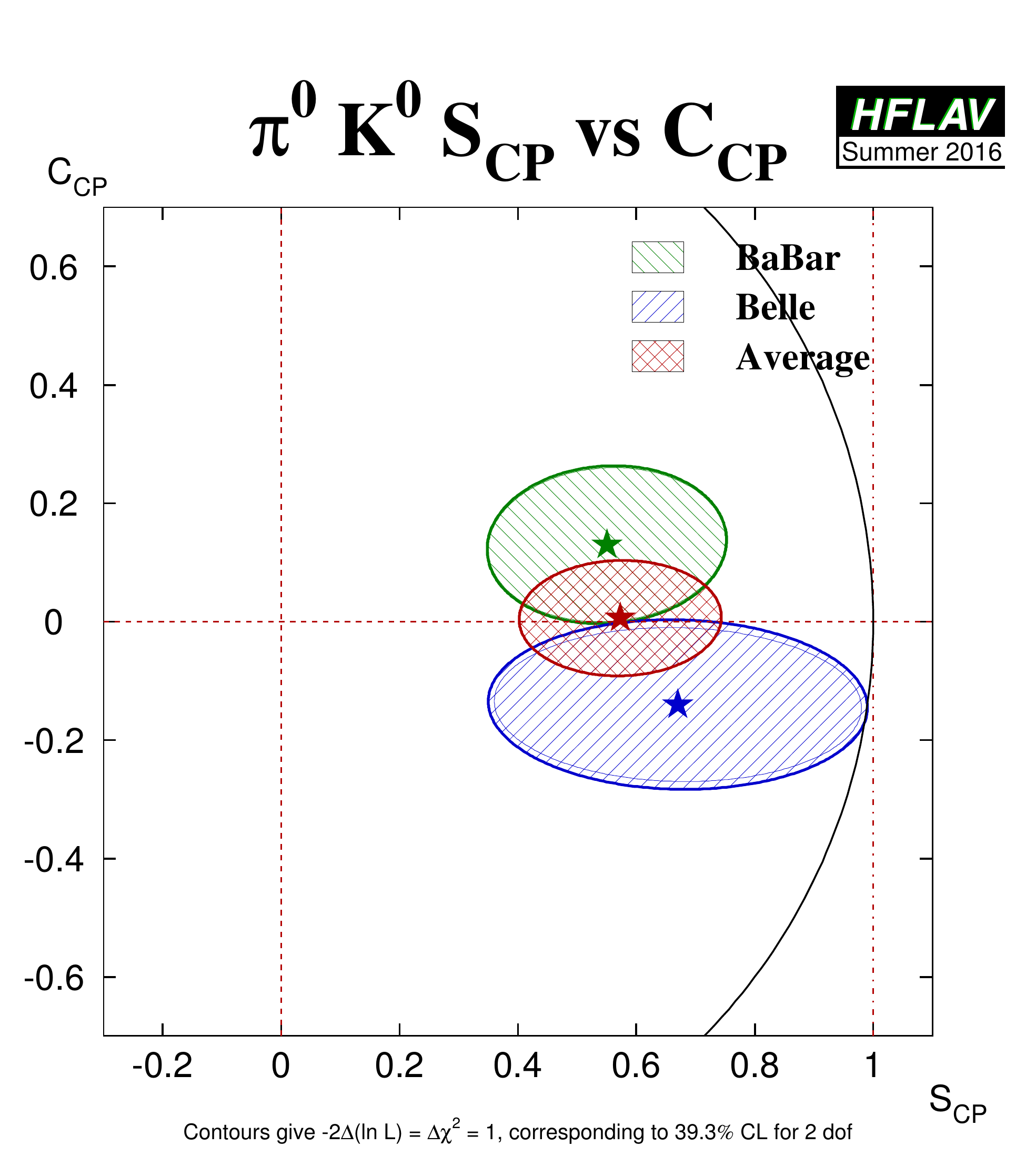}
    }
  \end{center}
  \vspace{-0.5cm}
  \caption{
    Averages of four $b \to q\bar q s$ dominated channels,
    for which correlated averages are performed,
    in the $S_{\CP}$ \vs\ $C_{\CP}$ plane,
    where $S_{\CP}$ has been corrected by the $\CP$ eigenvalue to give
    $\sin(2\beta^{\rm eff})$.
    (Top left) $\Bz \to \phi\Kz$,
    (top right) $\Bz \to \eta^\prime\Kz$,
    (bottom left) $\Bz \to \KS\KS\KS$,
    (bottom right) $\Bz \to \pi^0\KS$.
    More figures are available from the HFLAV web pages.
  }
  \label{fig:cp_uta:qqs_SvsC}
\end{figure}

\begin{figure}[htbp]
  \begin{center}
    \resizebox{0.66\textwidth}{!}{
      \includegraphics{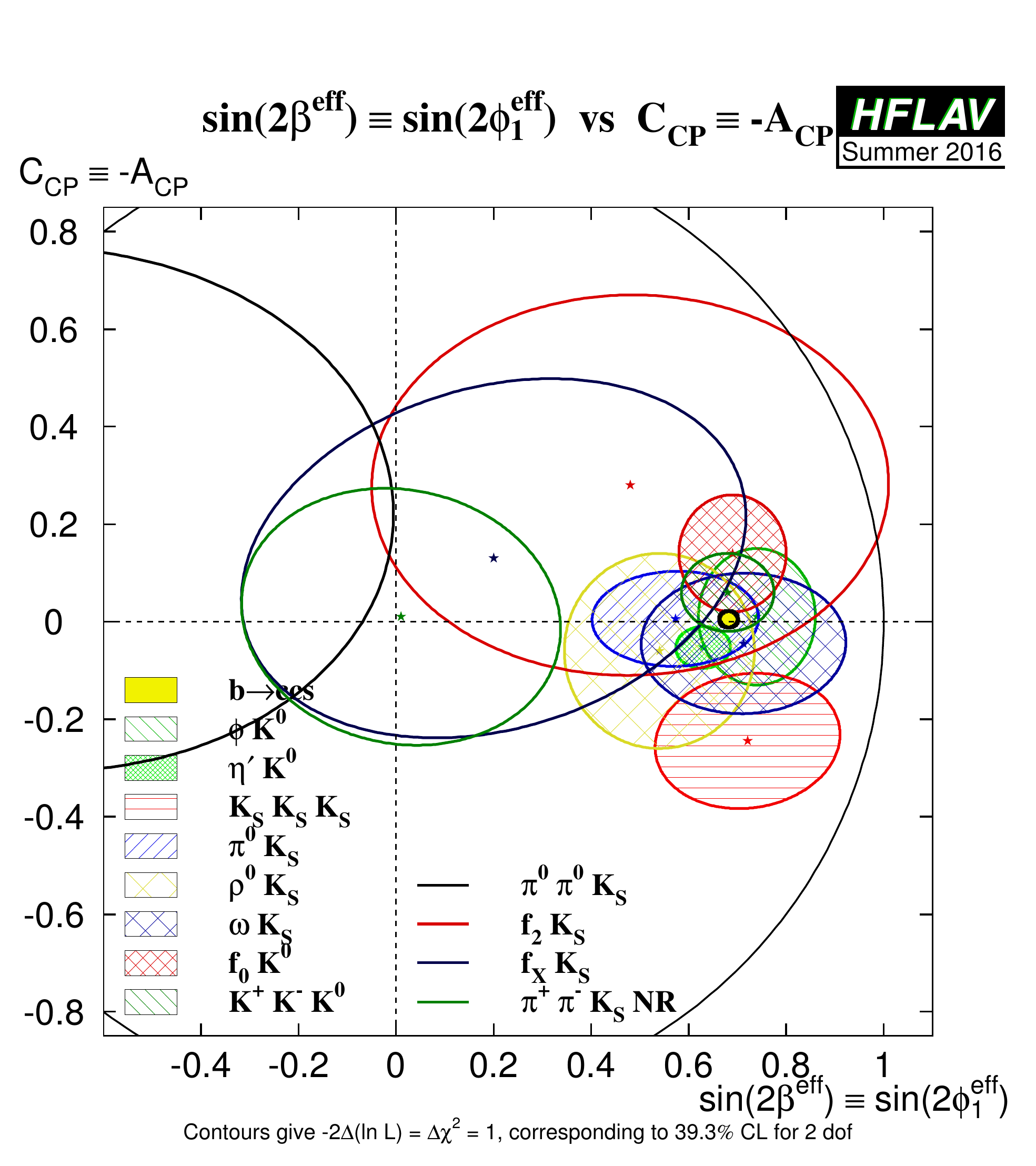}
    }
  \end{center}
  \vspace{-0.8cm}
  \caption{
    Compilation of constraints in the $-\etacp S_{b \to q\bar q s}$, interpreted as $\sin(2\beta^{\rm eff})$, \vs\ $C_{b \to q\bar q s}$ plane.
  }
  \label{fig:cp_uta:qqs_SvsC-all}
\end{figure}

The averages for $-\etacp S_{b \to q\bar q s}$ and $C_{b \to q\bar q s}$
can be found in Tables~\ref{tab:cp_uta:qqs} and~\ref{tab:cp_uta:qqs2},
and are shown in Figs.~\ref{fig:cp_uta:qqs},~\ref{fig:cp_uta:qqs_SvsC} 
and~\ref{fig:cp_uta:qqs_SvsC-all}.
Results from both \babar\  and \belle\ are averaged for the modes
$\etapr\Kz$ ($\Kz$ indicates that both $\KS$ and $\KL$ are used)
$\KS\KS\KS$, $\pi^0 \KS$ and $\omega\KS$.\footnote{
  \belle~\cite{Fujikawa:2008pk} include the $\pi^0\KL$ final state together with $\pi^0 \KS$ in order to improve the constraint on the parameter of \CP\ violation in decay; these events cannot be used for time-dependent analysis.
}
Results on $\phi\KS$ and $\Kp\Km\KS$ (implicitly excluding $\phi\KS$ and $f_0\KS$) are taken from time-dependent Dalitz plot analyses of $\Kp\Km\KS$;
results on $\rho^0\KS$, $f_2\KS$, $f_X\KS$ and $\pip\pim\KS$ nonresonant are taken from time-dependent Dalitz plot analyses of $\pip\pim\KS$ (see Sec.~\ref{sec:cp_uta:qqs:dp}).
The results on $f_0\KS$ are from combinations of both Dalitz plot analyses.
\babar\ has also presented results with the final states
$\pi^0\pi^0\KS$ and $\phi \KS \pi^0$. 
% \footnote{
%   We do not include a preliminary result from \belle~\cite{:2007xd}, which
%   remains unpublished after more than two years.
% }

Of these final states,
$\phi\KS$, $\etapr\KS$, $\pi^0 \KS$, $\rho^0\KS$, $\omega\KS$ and $f_0\KL$
have $\CP$ eigenvalue $\etacp = -1$, 
while $\phi\KL$, $\etapr\KL$, $\KS\KS\KS$, $f_0\KS$, $f_2\KS$, $f_X\KS$, $\pi^0\pi^0\KS$ and $\pi^+ \pi^- \KS$ nonresonant have $\etacp = +1$.
The final state $K^+K^-\KS$ (with $\phi\KS$ and $f_0\KS$ implicitly excluded)
is not a $\CP$ eigenstate, but the \CP-content can be absorbed in the amplitude analysis to allow the determination of a single effective $S$ parameter.
(In earlier analyses of the $K^+K^-\Kz$ final state,
its $\CP$ composition was determined using an isospin argument~\cite{Abe:2006gy}
and a moments analysis~\cite{Aubert:2005ja}.)
Throughout this section, $f_0 \equiv f_0(980)$ and $f_2 \equiv f_2(1270)$.
Details of the assumed lineshapes of these states, and of the $f_X$ (which is taken to have even spin), can be found in the relevant experimental papers~\cite{Lees:2012kxa,Aubert:2009me,Nakahama:2010nj,Dalseno:2008wwa}.

% consider using longtable next time
\begin{table}[!htb]
	\begin{center}
		\caption{
      Averages of $-\etacp S_{b \to q\bar q s}$ and $C_{b \to q\bar q s}$.
      Where a third source of uncertainty is given, it is due to model
      uncertainties arising in Dalitz plot analyses.
%			Averages for $\phi K^{0}$.
		}
		\vspace{0.2cm}
% make this tabular (not tabular*) and resize down to \textwidth
% change @{\extracolsep{\fill}} to @{\extracolsep{2mm}}
    \resizebox{\textwidth}{!}{
\renewcommand{\arraystretch}{1.1}
		\begin{tabular}{@{\extracolsep{2mm}}lrccc@{\hspace{-3pt}}c} \hline
%		\setlength{\tabcolsep}{0.0pc}
%		\begin{tabular*}{\textwidth}{@{\extracolsep{\fill}}lrccc@{\hspace{-3pt}}c} \hline
        \mc{2}{l}{Experiment} & $N(B\bar{B})$ & $- \etacp S_{b \to q\bar q s}$ & $C_{b \to q\bar q s}$ & Correlation \\
	\hline
      \mc{6}{c}{$\phi \Kz$} \\
	\babar & \cite{Lees:2012kxa} & 470M & $0.66 \pm 0.17 \pm 0.07$ & $0.05 \pm 0.18 \pm 0.05$ & \textendash{} \\
	\belle & \cite{Nakahama:2010nj} & 657M & $0.90 \,^{+0.09}_{-0.19}$ & $-0.04 \pm 0.20 \pm 0.10 \pm 0.02$ & \textendash{} \\
%	\hline
	\mc{3}{l}{\bf Average} & $0.74 \,^{+0.11}_{-0.13}$ & $0.01 \pm 0.14$ & {\small uncorrelated averages} \\
%	\mc{3}{l}{\small Confidence level} & {\small $0.xx~(y.y\sigma)$} & {\small $0.xx~(y.y\sigma)$} & \\
		\hline
% 		\end{tabular*}
% 		\label{tab:cp_uta:yyy}
% 	\end{center}
% \end{table}

% \begin{table}[htb]
% 	\begin{center}
% 		\caption{
% 			Averages for $\eta^{\prime} K^{0}$.
% 		}
% 		\vspace{0.2cm}
% 		\setlength{\tabcolsep}{0.0pc}
% 		\begin{tabular*}{\textwidth}{@{\extracolsep{\fill}}lrcccc} \hline
% 		\mc{2}{l}{Experiment} & $N(B\bar{B})$ & $S_{\CP}$ & $C_{\CP}$ & Correlation \\
% 		\hline
      \mc{6}{c}{$\etapr \Kz$} \\
	\babar & \cite{:2008se} & 467M & $0.57 \pm 0.08 \pm 0.02$ & $-0.08 \pm 0.06 \pm 0.02$ & $0.03$ \\
	\belle & \cite{Santelj:2014sja} & 772M & $0.68 \pm 0.07 \pm 0.03$ & $-0.03 \pm 0.05 \pm 0.03$ & $0.03$ \\
%	\hline
	\mc{3}{l}{\bf Average} & $0.63 \pm 0.06$ & $-0.05 \pm 0.04$ & $0.02$ \\
	\mc{3}{l}{\small Confidence level} & \mc{2}{c}{\small $0.53~(0.6\sigma)$} & \\
		\hline
% 		\end{tabular*}
% 		\label{tab:cp_uta:yyy}
% 	\end{center}
% \end{table}

% \begin{table}[htb]
% 	\begin{center}
% 		\caption{
% 			Averages for $K_{S} K_{S} K_{S}$.
% 		}
% 		\vspace{0.2cm}
% 		\setlength{\tabcolsep}{0.0pc}
% 		\begin{tabular*}{\textwidth}{@{\extracolsep{\fill}}lrcccc} \hline
% 		\mc{2}{l}{Experiment} & $N(B\bar{B})$ & $S_{\CP}$ & $C_{\CP}$ & Correlation \\
% 		\hline
      \mc{6}{c}{$\KS\KS\KS$} \\
	\babar & \cite{Lees:2011nf} & 468M & $0.94 \,^{+0.21}_{-0.24} \pm 0.06$ & $-0.17 \pm 0.18 \pm 0.04$ & $0.16$ \\
	\belle & \cite{Chen:2006nk} & 535M & $0.30 \pm 0.32 \pm 0.08$ & $-0.31 \pm 0.20 \pm 0.07$ & \textendash{} \\
%	\hline
	\mc{3}{l}{\bf Average} & $0.72 \pm 0.19$ & $-0.24 \pm 0.14$ & $0.09$ \\
	\mc{3}{l}{\small Confidence level} & \mc{2}{c}{\small $0.26~(1.1\sigma)$} & \\
		\hline
% 		\end{tabular*}
% 		\label{tab:cp_uta:yyy}
% 	\end{center}
% \end{table}

% \begin{table}[htb]
% 	\begin{center}
% 		\caption{
% 			Averages for $\pi^{0} K_{S}$.
% 		}
% 		\vspace{0.2cm}
% 		\setlength{\tabcolsep}{0.0pc}
% 		\begin{tabular*}{\textwidth}{@{\extracolsep{\fill}}lrcccc} \hline
% 		\mc{2}{l}{Experiment} & $N(B\bar{B})$ & $S_{\CP}$ & $C_{\CP}$ & Correlation \\
% 		\hline
      \mc{6}{c}{$\pi^0 K^0$} \\
	\babar & \cite{:2008se} & 467M & $0.55 \pm 0.20 \pm 0.03$ & $0.13 \pm 0.13 \pm 0.03$ & $0.06$ \\
	\belle & \cite{Fujikawa:2008pk} & 657M & $0.67 \pm 0.31 \pm 0.08$ & $-0.14 \pm 0.13 \pm 0.06$ & $-0.04$ \\
%	\hline
	\mc{3}{l}{\bf Average} & $0.57 \pm 0.17$ & $0.01 \pm 0.10$ & $0.02$ \\
	\mc{3}{l}{\small Confidence level} & \mc{2}{c}{\small $0.37~(0.9\sigma)$} & \\
		\hline
% 		\end{tabular*}
% 		\label{tab:cp_uta:yyy}
% 	\end{center}
% \end{table}

% \begin{table}[htb]
% 	\begin{center}
% 		\caption{
% 			Averages for $\rho^{0} K_{S}$.
% 		}
% 		\vspace{0.2cm}
% 		\setlength{\tabcolsep}{0.0pc}
% 		\begin{tabular*}{\textwidth}{@{\extracolsep{\fill}}lrcccc} \hline
% 		\mc{2}{l}{Experiment} & $N(B\bar{B})$ & $S_{\CP}$ & $C_{\CP}$ & Correlation \\
		\hline
      \mc{6}{c}{$\rho^0 \KS$} \\
	\babar & \cite{Aubert:2009me} & 383M & $0.35 \,^{+0.26}_{-0.31} \pm 0.06 \pm 0.03$ & $-0.05 \pm 0.26 \pm 0.10 \pm 0.03$ & \textendash{} \\
	\belle & \cite{Dalseno:2008wwa} & 657M & $0.64 \,^{+0.19}_{-0.25} \pm 0.09 \pm 0.10$ & $-0.03 \,^{+0.24}_{-0.23} \pm 0.11 \pm 0.10$ & \textendash{} \\
% 	\hline
	\mc{3}{l}{\bf Average} & $0.54 \,^{+0.18}_{-0.21}$ & $-0.06 \pm 0.20$ & {\small uncorrelated averages} \\
%	\mc{3}{l}{\small Confidence level} & {\small $0.xx~(y.y\sigma)$} & {\small $0.xx~(y.y\sigma)$} & \\
		\hline
% 		\end{tabular*}
% 		\label{tab:cp_uta:yyy}
% 	\end{center}
% \end{table}

% \begin{table}[htb]
% 	\begin{center}
% 		\caption{
% 			Averages for $\omega K_{S}$.
% 		}
% 		\vspace{0.2cm}
% 		\setlength{\tabcolsep}{0.0pc}
% 		\begin{tabular*}{\textwidth}{@{\extracolsep{\fill}}lrcccc} \hline
% 		\mc{2}{l}{Experiment} & $N(B\bar{B})$ & $S_{\CP}$ & $C_{\CP}$ & Correlation \\
% 		\hline
      \mc{6}{c}{$\omega \KS$} \\
	\babar & \cite{:2008se} & 467M & $0.55 \,^{+0.26}_{-0.29} \pm 0.02$ & $-0.52 \,^{+0.22}_{-0.20} \pm 0.03$ & $0.03$ \\
	\belle & \cite{Chobanova:2013ddr} & 772M & $0.91 \pm 0.32 \pm 0.05$ & $0.36 \pm 0.19 \pm 0.05$ & $-0.00$ \\
%	\hline
	\mc{3}{l}{\bf Average} & $0.71 \pm 0.21$ & $-0.04 \pm 0.14$ & $0.01$ \\
	\mc{3}{l}{\small Confidence level} & \mc{2}{c}{\small $0.007~(2.7\sigma)$} & \\
		\hline
% 		\end{tabular*}
% 		\label{tab:cp_uta:yyy}
% 	\end{center}
% \end{table}

% \begin{table}[htb]
% 	\begin{center}
% 		\caption{
% 			Averages for $f_{0} K^{0}$.
% 		}
% 		\vspace{0.2cm}
% 		\setlength{\tabcolsep}{0.0pc}
% 		\begin{tabular*}{\textwidth}{@{\extracolsep{\fill}}lrcccc} \hline
% 		\mc{2}{l}{Experiment} & $N(B\bar{B})$ & $S_{\CP}$ & $C_{\CP}$ & Correlation \\
% 		\hline
      \mc{6}{c}{$f_0 \Kz$} \\
	\babar & \cite{Lees:2012kxa,Aubert:2009me} & \textendash{} & $0.74 \,^{+0.12}_{-0.15}$ & $0.15 \pm 0.16$ & \textendash{} \\
	\belle & \cite{Nakahama:2010nj,Dalseno:2008wwa} & \textendash{} & $0.63 \,^{+0.16}_{-0.19}$ & $0.13 \pm 0.17$ & \textendash{} \\
%	\hline
	\mc{3}{l}{\bf Average} & $0.69 \,^{+0.10}_{-0.12}$ & $0.14 \pm 0.12$ & {\small uncorrelated averages} \\
%	\mc{3}{l}{\small Confidence level} & {\small $0.xx~(y.y\sigma)$} & {\small $0.xx~(y.y\sigma)$} & \\
		\hline
% 		\end{tabular*}
% 		\label{tab:cp_uta:yyy}
% 	\end{center}
% \end{table}

% \begin{table}[htb]
% 	\begin{center}
% 		\caption{
% 			Averages for $f_{2} \KS$.
% 		}
% 		\vspace{0.2cm}
% 		\setlength{\tabcolsep}{0.0pc}
% 		\begin{tabular*}{\textwidth}{@{\extracolsep{\fill}}lrcccc} \hline
% 		\mc{2}{l}{Experiment} & $N(B\bar{B})$ & $S_{\CP}$ & $C_{\CP}$ & Correlation \\
% 		\hline
      \mc{6}{c}{$f_2 \KS$} \\
	\babar & \cite{Aubert:2009me} & 383M & $0.48 \pm 0.52 \pm 0.06 \pm 0.10$ & $0.28 \,^{+0.35}_{-0.40} \pm 0.08 \pm 0.07$ & \textendash{} \\
%% 	\hline
%% 	\mc{3}{l}{\bf Average} & $0.48 \pm 0.53$ & $0.28 \,^{+0.37}_{-0.41}$ & {\small uncorrelated averages} \\
%% 	\mc{3}{l}{\small Confidence level} & {\small $0.xx~(y.y\sigma)$} & {\small $0.xx~(y.y\sigma)$} & \\
		\hline
% 		\end{tabular*}
% 		\label{tab:cp_uta:yyy}
% 	\end{center}
% \end{table}

% \begin{table}[htb]
% 	\begin{center}
% 		\caption{
% 			Averages for $f_{\rm X} \KS$.
% 		}
% 		\vspace{0.2cm}
% 		\setlength{\tabcolsep}{0.0pc}
% 		\begin{tabular*}{\textwidth}{@{\extracolsep{\fill}}lrcccc} \hline
% 		\mc{2}{l}{Experiment} & $N(B\bar{B})$ & $S_{\CP}$ & $C_{\CP}$ & Correlation \\
% 		\hline
      \mc{6}{c}{$f_{X} \KS$} \\
	\babar & \cite{Aubert:2009me} & 383M & $0.20 \pm 0.52 \pm 0.07 \pm 0.07$ & $0.13 \,^{+0.33}_{-0.35} \pm 0.04 \pm 0.09$ & \textendash{} \\
%% 	\hline
%% 	\mc{3}{l}{\bf Average} & $0.20 \pm 0.53$ & $0.13 \,^{+0.34}_{-0.36}$ & {\small uncorrelated averages} \\
%% 	\mc{3}{l}{\small Confidence level} & {\small $0.xx~(y.y\sigma)$} & {\small $0.xx~(y.y\sigma)$} & \\
		\hline
% 		\end{tabular*}
% 		\label{tab:cp_uta:yyy}
% 	\end{center}
% \end{table}

% 		\end{tabular*}
 		\end{tabular}
}
		\label{tab:cp_uta:qqs}
	\end{center}
\end{table}

\begin{table}[!htb]
	\begin{center}
		\caption{
      Averages of $-\etacp S_{b \to q\bar q s}$ and $C_{b \to q\bar q s}$ (continued).
      Where a third source of uncertainty is given, it is due to model
      uncertainties arising in Dalitz plot analyses.
		}
		\vspace{0.2cm}
		\setlength{\tabcolsep}{0.0pc}
\renewcommand{\arraystretch}{1.1}
		\begin{tabular*}{\textwidth}{@{\extracolsep{\fill}}lrccc@{\hspace{-3pt}}c} \hline
        \mc{2}{l}{Experiment} & $N(B\bar{B})$ & $- \etacp S_{b \to q\bar q s}$ & $C_{b \to q\bar q s}$ & Correlation \\
	\hline
% \begin{table}[htb]
% 	\begin{center}
% 		\caption{
% 			Averages for $\pi^{0} \pi^{0} K_{S}$.
% 		}
% 		\vspace{0.2cm}
% 		\setlength{\tabcolsep}{0.0pc}
% 		\begin{tabular*}{\textwidth}{@{\extracolsep{\fill}}lrcccc} \hline
% 		\mc{2}{l}{Experiment} & $N(B\bar{B})$ & $S_{\CP}$ & $C_{\CP}$ & Correlation \\
% 		\hline
      \mc{6}{c}{$\pi^0 \pi^0 \KS$} \\
	\babar & \cite{Aubert:2007ub} & 227M & $-0.72 \pm 0.71 \pm 0.08$ & $0.23 \pm 0.52 \pm 0.13$ & $-0.02$ \\
%	\belle & \cite{:2007xd} & 657M & $-0.43 \pm 0.49 \pm 0.09$ & $0.17 \pm 0.24 \pm 0.06$ & $0.09$ \\
%	\hline
%	\mc{3}{l}{\bf Average} & $-0.72 \pm 0.71$ & $0.23 \pm 0.54$ & $-0.02$ \\
%	\mc{3}{l}{\small Confidence level} & \mc{2}{c}{\small $0.94~(0.1\sigma)$} & \\
		\hline
% 		\end{tabular*}
% 		\label{tab:cp_uta:yyy}
% 	\end{center}
% \end{table}

% \begin{table}[htb]
% 	\begin{center}
% 		\caption{
% 			Averages for $\phi K_{S} \pi^0$.
% 		}
% 		\vspace{0.2cm}
% 		\setlength{\tabcolsep}{0.0pc}
% 		\begin{tabular*}{\textwidth}{@{\extracolsep{\fill}}lrcccc} \hline
% 		\mc{2}{l}{Experiment} & $N(B\bar{B})$ & $S_{\CP}$ & $C_{\CP}$ & Correlation \\
% 		\hline
      \mc{6}{c}{$\phi \KS \pi^0$} \\
	\babar & \cite{Aubert:2008zza} & 465M & $0.97 \,^{+0.03}_{-0.52}$ & $-0.20 \pm 0.14 \pm 0.06$ & \textendash{} \\
%% 	\hline
%% 	\mc{3}{l}{\bf Average} & $0.97 \,^{+0.03}_{-0.52}$ & $-0.20 \pm 0.15$ & {\small uncorrelated averages} \\
%% 	\mc{3}{l}{\small Confidence level} & {\small $0.xx~(y.y\sigma)$} & {\small $0.xx~(y.y\sigma)$} & \\
 		\hline
% 		\end{tabular*}
% 		\label{tab:cp_uta:yyy}
% 	\end{center}
% \end{table}

% \begin{table}[htb]
% 	\begin{center}
% 		\caption{
% 			Averages for $\pi^{+} \pi^{-} K_{S} nonresonant$.
% 		}
% 		\vspace{0.2cm}
% 		\setlength{\tabcolsep}{0.0pc}
% 		\begin{tabular*}{\textwidth}{@{\extracolsep{\fill}}lrcccc} \hline
% 		\mc{2}{l}{Experiment} & $N(B\bar{B})$ & $S_{\CP}$ & $C_{\CP}$ & Correlation \\
% 		\hline
      \mc{6}{c}{$\pi^+ \pi^- \KS$ nonresonant} \\
	\babar & \cite{Aubert:2009me} & 383M & $0.01 \pm 0.31 \pm 0.05 \pm 0.09$ & $0.01 \pm 0.25 \pm 0.06 \pm 0.05$ & \textendash{} \\
%% 	\hline
%% 	\mc{3}{l}{\bf Average} & $0.01 \pm 0.33$ & $0.01 \pm 0.26$ & {\small uncorrelated averages} \\
%% 	\mc{3}{l}{\small Confidence level} & {\small $0.xx~(y.y\sigma)$} & {\small $0.xx~(y.y\sigma)$} & \\
 		\hline
% 		\end{tabular*}
% 		\label{tab:cp_uta:yyy}
% 	\end{center}
% \end{table}

% \begin{table}[htb]
% 	\begin{center}
% 		\caption{
% 			Averages for $K^{+} K^{-} K^{0}$.
% 		}
% 		\vspace{0.2cm}
% 		\setlength{\tabcolsep}{0.0pc}
% 		\begin{tabular*}{\textwidth}{@{\extracolsep{\fill}}lrcccc} \hline
% 		\mc{2}{l}{Experiment} & $N(B\bar{B})$ & $S_{\CP}$ & $C_{\CP}$ & Correlation \\
% 		\hline
      \mc{6}{c}{$K^+K^- \Kz$} \\
	\babar & \cite{Lees:2012kxa} & 470M & $0.65 \pm 0.12 \pm 0.03$ & $0.02 \pm 0.09 \pm 0.03$ & \textendash{} \\
	\belle & \cite{Nakahama:2010nj} & 657M & $0.76 \,^{+0.14}_{-0.18}$ & $0.14 \pm 0.11 \pm 0.08 \pm 0.03$ & \textendash{} \\
%	\hline
	\mc{3}{l}{\bf Average} & $0.68 \,^{+0.09}_{-0.10}$ & $0.06 \pm 0.08$ & {\small uncorrelated averages} \\
%	\mc{3}{l}{\small Confidence level} & {\small $0.xx~(y.y\sigma)$} & {\small $0.xx~(y.y\sigma)$} & \\
		\hline
% 		\end{tabular*}
% 		\label{tab:cp_uta:yyy}
% 	\end{center}
% \end{table}

% \begin{table}[htb]
% 	\begin{center}
% 		\caption{
% 			Averages for $b\rightarrow qqs$.
% 		}
% 		\vspace{0.2cm}
% 		\setlength{\tabcolsep}{0.0pc}
% 		\begin{tabular*}{\textwidth}{@{\extracolsep{\fill}}lrcccc} \hline
% 		\mc{2}{l}{Experiment} & $N(B\bar{B})$ & $S_{\CP}$ & $C_{\CP}$ & Correlation \\
% 		\hline
% 	\babar & \cite{} & 0M & $0.21 \pm 0.26 \pm 0.11$ & $0.08 \pm 0.18 \pm 0.04$ & \textendash{} \\

% 	\belle & \cite{} & 0M & $0.50 \pm 0.21 \pm 0.06$ & $-0.07 \pm 0.15 \pm 0.05$ & \textendash{} \\
% % 	\hline
% 	\mc{3}{l}{\bf Average} & $0.53 \pm 0.05$ & $-0.01 \pm 0.04$ &  -  \\

		\hline
		\end{tabular*}
		\label{tab:cp_uta:qqs2}
	\end{center}
\end{table}

The final state $\phi \KS \pi^0$ is also not a \CP eigenstate but its
\CP-composition can be determined from an angular analysis.
Since the parameters are common to the $\Bz\to\phi \KS \pi^0$ and
$\Bz\to \phi \Kp\pim$ decays (because only $K\pi$ resonances contribute),
\babar\ perform a simultaneous analysis of the two final
states~\cite{Aubert:2008zza} (see Sec.~\ref{sec:cp_uta:qqs:vv}).

It must be noted that Q2B parameters extracted from Dalitz plot analyses 
are constrained to lie within the physical boundary ($S_{\CP}^2 + C_{\CP}^2 < 1$)
and consequently the obtained errors are highly non-Gaussian when
the central value is close to the boundary.  
This is particularly evident in the \babar\ results for 
$\Bz \to f_0\Kz$ with $f_0 \to \pi^+\pi^-$~\cite{Aubert:2009me}.
These results must be treated with extreme caution.

%% straightforward interpretation
As explained above, each of the modes listed in Tables~\ref{tab:cp_uta:qqs} and~\ref{tab:cp_uta:qqs2} has potentially different subleading contributions within the Standard Model,
and thus each may have a different value of $-\etacp S_{b \to q\bar q s}$.
Therefore, there is no strong motivation to make a combined average
over the different modes.
We refer to such an average as a ``na\"\i ve $s$-penguin average.''
It is na\"\i ve not only because the theoretical uncertainties are neglected,
but also since possible correlations of systematic effects 
between different modes are not included.
In spite of these caveats there remains interest in the value of this quantity
and therefore it is given here:
$\langle -\etacp S_{b \to q\bar q s} \rangle = 0.655 \pm 0.032$,
with confidence level $0.77~(0.3\sigma)$.
This value is in agreement with the average 
$-\etacp S_{b \to c\bar c s}$ given in Sec.~\ref{sec:cp_uta:ccs:cp_eigen}.
(The average for $C_{b \to q\bar q s}$ is 
$\langle C_{b \to q\bar q s} \rangle = -0.006 \pm 0.026$
with confidence level $0.53~(0.6\sigma)$.)
% We emphasise again that we do not advocate the use of these averages.
% and that the values should be treated with {\it extreme caution}, if at all.
% What is unambiguous (although only qualitative) 
% is that there is a trend that the values 
% of $-\etacp S_{b \to q\bar q s}$ in different modes 
% are below the average for $-\etacp S_{b \to c\bar c s}$.

From Table~\ref{tab:cp_uta:qqs} it may be noted 
that the averages for $-\etacp S_{b \to q\bar q s}$ in 
$\phi\KS$, $\etapr \Kz$, $f_0\KS$ and $\Kp\Km\KS$
are all now more than $5\sigma$ away from zero, 
so that $\CP$ violation in these modes can be considered well established.
There is no evidence (above $2\sigma$) for $\CP$ violation in any $b \to q \bar q s$ decay.

%%%%%%%%
%%%
%%% qqs
%%%
%%%%%%%%
\mysubsubsection{Time-dependent Dalitz plot analyses: $\Bz \to K^+K^-\Kz$ and $\Bz \to \pi^+\pi^-\KS$}
\label{sec:cp_uta:qqs:dp}

\begin{figure}[htbp]
  \begin{center}
    \resizebox{0.45\textwidth}{!}{
      \includegraphics{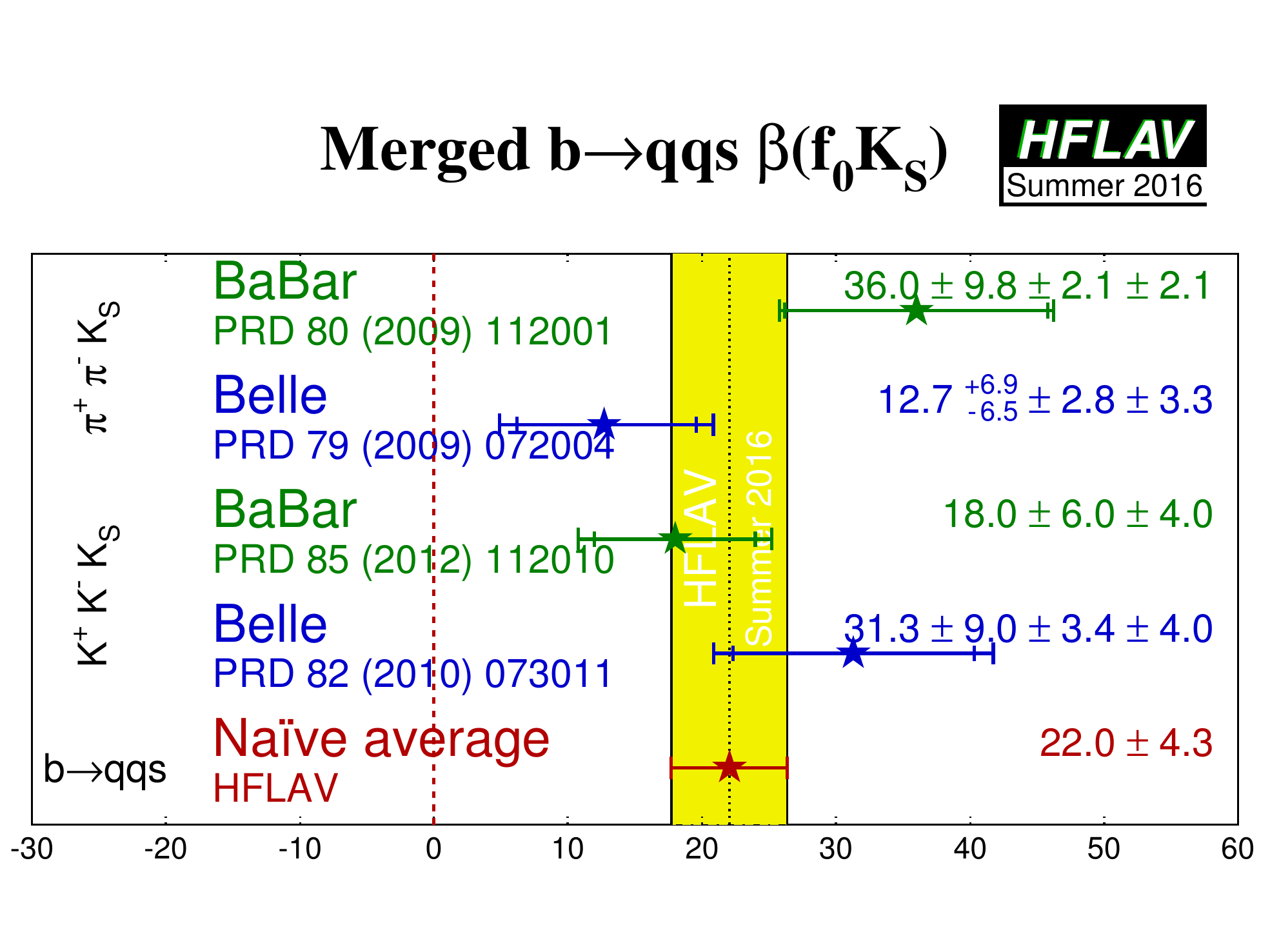}
    }
    \hfill
    \resizebox{0.45\textwidth}{!}{
      \includegraphics{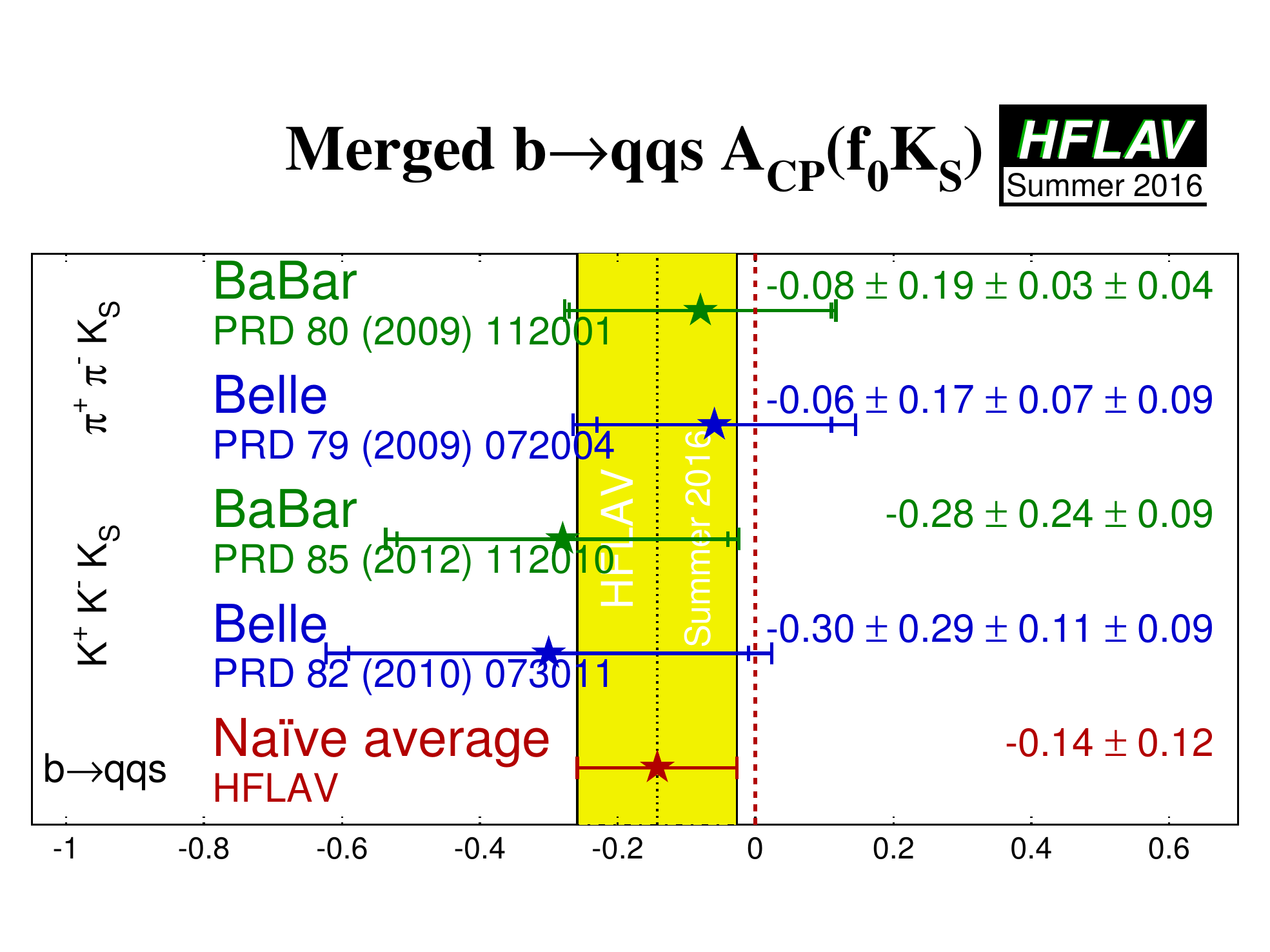}
    }
  \end{center}
  \vspace{-0.8cm}
  \caption{
    Averages of 
    (left) $\beta^{\rm eff} \equiv \phi_1^{\rm eff}$ and (right) $A_{\CP}$
    for the $\Bz\to f_0\KS$ decay including measurements from Dalitz plot analyses of both $\Bz\to K^+K^-\KS$ and $\Bz\to \pi^+\pi^-\KS$.
  }
  \label{fig:cp_uta:qqs:f0KS}
\end{figure}

As mentioned in Sec.~\ref{sec:cp_uta:notations:dalitz:kkk0} and above,
both \babar\ and \belle\ have performed time-dependent Dalitz plot analysis of
$\Bz \to K^+K^-\Kz$ and $\Bz \to \pi^+\pi^-\KS$ decays.
The results are summarised in Tables~\ref{tab:cp_uta:kkk0_tddp} 
and~\ref{tab:cp_uta:pipik0_tddp}.
Averages for the $\Bz\to f_0 \KS$ decay, which contributes to both Dalitz
plots, are shown in Fig.~\ref{fig:cp_uta:qqs:f0KS}.
Results are presented in terms of the effective weak phase (from mixing and
decay) difference $\beta^{\rm eff}$ and the parameter of $\CP$ violation in decay
$\Acp$ ($\Acp = -C$) for each of the resonant contributions.
Note that Dalitz plot analyses, including all those included in these
averages, often suffer from ambiguous solutions -- we quote the results
corresponding to those presented as solution 1 in all cases.
Results on flavour specific amplitudes that may contribute to these Dalitz
plots (such as $K^{*+}\pi^-$) are averaged by the HFLAV Rare Decays subgroup 
(Sec.~\ref{sec:rare}).

For the $\Bz \to K^+K^-\Kz$ decay, both \babar\ and \belle\ measure the \CP
violation parameters for the $\phi\Kz$, $f_0\Kz$ and ``other $\Kp\Km\Kz$''
amplitudes, where the latter includes all remaining resonant and nonresonant
contributions to the charmless three-body decay.
For the $\Bz \to \pi^+\pi^-\KS$ decay, \babar\ reports \CP violation parameters for all of the \CP eigenstate components in the Dalitz plot model ($\rhoz\KS$, $f_0\KS$, $f_2\KS$, $f_X\KS$ and nonresonant decays; see Sec.~\ref{sec:cp_uta:notations:dalitz:pipik0}), 
while \belle\ reports the \CP violation parameters for only the $\rhoz\KS$ and $f_0\KS$ amplitudes, although the used Dalitz plot model is rather similar.

% \begin{table}[htb]
\begin{sidewaystable}
  \begin{center}
    \caption{
      Results from time-dependent Dalitz plot analyses of 
      the $\Bz \to K^+K^-\Kz$ decay.
      Correlations (not shown) are taken into account in the average.
    }
    \vspace{0.2cm}
    \setlength{\tabcolsep}{0.0pc}
% make this tabular (not tabular*) and resize down to \textwidth
% change @{\extracolsep{\fill}} to @{\extracolsep{2mm}}
    \resizebox{\textwidth}{!}{
\renewcommand{\arraystretch}{1.2}
      \begin{tabular}{l@{\hspace{2mm}}r@{\hspace{2mm}}c@{\hspace{2mm}}|@{\hspace{2mm}}c@{\hspace{2mm}}c@{\hspace{2mm}}|@{\hspace{2mm}}c@{\hspace{2mm}}c@{\hspace{2mm}}|@{\hspace{2mm}}c@{\hspace{2mm}}c} 
        \hline 
        \mc{2}{l}{Experiment} & $N(B\bar{B})$ &
        \mc{2}{c}{$\phi\KS$} & \mc{2}{c}{$f_0\KS$} & \mc{2}{c}{$K^+K^-\KS$} \\
        & & & $\beta^{\rm eff}\,(^\circ)$ & $\Acp$ & $\beta^{\rm eff}\,(^\circ)$ & $\Acp$ & $\beta^{\rm eff}\,(^\circ)$ & $\Acp$ \\
	\babar & \cite{Lees:2012kxa} & 470M & $21 \pm 6 \pm 2$ & $-0.05 \pm 0.18 \pm 0.05$ & $18 \pm 6 \pm 4$ & $-0.28 \pm 0.24 \pm 0.09$ & $20.3 \pm 4.3 \pm 1.2$ & $-0.02 \pm 0.09 \pm 0.03$ \\
	\belle & \cite{Nakahama:2010nj} & 657M & $32.2 \pm 9.0 \pm 2.6 \pm 1.4$ & $0.04 \pm 0.20 \pm 0.10 \pm 0.02$ & $31.3 \pm 9.0 \pm 3.4 \pm 4.0$ & $-0.30 \pm 0.29 \pm 0.11 \pm 0.09$ & $24.9 \pm 6.4 \pm 2.1 \pm 2.5$ & $-0.14 \pm 0.11 \pm 0.08 \pm 0.03$ \\
	\mc{2}{l}{\bf Average} & & $24 \pm 5$ & $-0.01 \pm 0.14$ & $22 \pm 6$ & $-0.29 \pm 0.20$ & $21.6 \pm 3.7$ & $-0.06 \pm 0.08$ \\
	\mc{3}{l}{\small Confidence level} & \mc{6}{c}{\small $0.93~(0.1\sigma)$} \\
        \hline
      \end{tabular}
    }
    
    \label{tab:cp_uta:kkk0_tddp}
  \end{center}
\end{sidewaystable}
% \end{table}

% \begin{table}[htb]
\begin{sidewaystable}
  \begin{center}
    \caption{
      Results from time-dependent Dalitz plot analysis of 
      the $\Bz \to \pi^+\pi^-\KS$ decay.
      Correlations (not shown) are taken into account in the average.
    }
    \vspace{0.2cm}
    \setlength{\tabcolsep}{0.0pc}
% make this tabular (not tabular*) and resize down to \textwidth
% change @{\extracolsep{\fill}} to @{\extracolsep{2mm}}
    \resizebox{\textwidth}{!}{
\renewcommand{\arraystretch}{1.2}
      \begin{tabular}{l@{\hspace{2mm}}r@{\hspace{2mm}}c@{\hspace{2mm}}|@{\hspace{2mm}}c@{\hspace{2mm}}c@{\hspace{2mm}}|@{\hspace{2mm}}c@{\hspace{2mm}}c} 
        \hline 
        \mc{2}{l}{Experiment} & $N(B\bar{B})$ & 
        \mc{2}{c}{$\rho^0\KS$} & \mc{2}{c}{$f_0\KS$} \\
        & & & $\beta^{\rm eff}$ & $\Acp$ & $\beta^{\rm eff}$ & $\Acp$ \\
        \hline
        \babar & \cite{Aubert:2009me} & 383M & $(10.2 \pm 8.9 \pm 3.0 \pm 1.9)^\circ$ & $0.05 \pm 0.26 \pm 0.10 \pm 0.03$ & $(36.0 \pm 9.8 \pm 2.1 \pm 2.1)^\circ$ & $-0.08 \pm 0.19 \pm 0.03 \pm 0.04$ \\
        \belle & \cite{Dalseno:2008wwa} & 657M & $(20.0 \,^{+8.6}_{-8.5} \pm 3.2 \pm 3.5)^\circ$ & $0.03 \,^{+0.23}_{-0.24} \pm 0.11 \pm 0.10$ & $(12.7 \,^{+6.9}_{-6.5} \pm 2.8 \pm 3.3)^\circ$ & $-0.06 \pm 0.17 \pm 0.07 \pm 0.09$ \\
        \hline
        \mc{2}{l}{\bf Average} & & $16.4 \pm 6.8$ & $0.06 \pm 0.20$ & $20.6 \pm 6.2$ & $-0.07 \pm 0.14$  \\
        \mc{3}{l}{\small Confidence level} & \mc{4}{c}{\small $0.39~(0.9\sigma)$} \\
        \hline
      \end{tabular}
    }

    \vspace{2ex}

    \setlength{\tabcolsep}{0.0pc}
% make this tabular (not tabular*) and resize down to \textwidth
% change @{\extracolsep{\fill}} to @{\extracolsep{2mm}}
    \resizebox{\textwidth}{!}{
\renewcommand{\arraystretch}{1.2}
      \begin{tabular}{l@{\hspace{2mm}}r@{\hspace{2mm}}c@{\hspace{2mm}}|@{\hspace{2mm}}c@{\hspace{2mm}}c@{\hspace{2mm}}|@{\hspace{2mm}}c@{\hspace{2mm}}c} 
        \hline 
        \mc{2}{l}{Experiment} & $N(B\bar{B})$ & 
        \mc{2}{c}{$f_2\KS$} & \mc{2}{c}{$f_{\rm X}\KS$} \\
        & & & $\beta^{\rm eff}$ & $\Acp$ & $\beta^{\rm eff}$ & $\Acp$ \\
        \babar & \cite{Aubert:2009me} & 383M & $(14.9 \pm 17.9 \pm 3.1 \pm 5.2)^\circ$ & $-0.28 \,^{+0.40}_{-0.35} \pm 0.08 \pm 0.07$ & $(5.8 \pm 15.2 \pm 2.2 \pm 2.3)^\circ$ & $-0.13 \,^{+0.35}_{-0.33} \pm 0.04 \pm 0.09$ \\
        \hline
      \end{tabular}
    }

    \vspace{2ex}

    \setlength{\tabcolsep}{0.0pc}
% make this tabular (not tabular*) and resize down to \textwidth
% change @{\extracolsep{\fill}} to @{\extracolsep{2mm}}
    \resizebox{\textwidth}{!}{
\renewcommand{\arraystretch}{1.1}
      \begin{tabular}{l@{\hspace{2mm}}r@{\hspace{2mm}}c@{\hspace{2mm}}|@{\hspace{2mm}}c@{\hspace{2mm}}c@{\hspace{2mm}}|@{\hspace{2mm}}c@{\hspace{2mm}}c} 
        \hline 
        \mc{2}{l}{Experiment} & $N(B\bar{B})$ & 
        \mc{2}{c}{$\Bz \to \pi^+\pi^-\KS$ nonresonant} & \mc{2}{c}{$\chi_{c0}\KS$} \\
        & & & $\beta^{\rm eff}$ & $\Acp$ & $\beta^{\rm eff}$ & $\Acp$ \\
        \babar & \cite{Aubert:2009me} & 383M & $(0.4 \pm 8.8 \pm 1.9 \pm 3.8)^\circ$ & $-0.01 \pm 0.25 \pm 0.06 \pm 0.05$ & $(23.2 \pm 22.4 \pm 2.3 \pm 4.2)^\circ$ & $0.29 \,^{+0.44}_{-0.53} \pm 0.03 \pm 0.05$ \\
        \hline
      \end{tabular}
    }

    \label{tab:cp_uta:pipik0_tddp}
  \end{center}
\end{sidewaystable}
% \end{table}

%% From the results in Table~\ref{tab:cp_uta:pipik0_tddp},
%% \babar\ infer that the trigonometric reflection 
%% at $\pi/2 - \beta^{\rm eff}$ in $\Bz \to K^+K^-\Kz$,
%% which is inconsistent with the Standard Model expectation,
%% is disfavoured at $4.8\sigma$.

%%%%%%%%
%%%
%%% qqs
%%%
%%%%%%%%
\mysubsubsection{Time-dependent analyses of $\Bz \to \phi \KS \pi^0$}
\label{sec:cp_uta:qqs:vv}

The final state in the decay $\Bz \to \phi \KS \pi^0$ is a mixture of \CP-even
and \CP-odd amplitudes. However, since only $\phi K^{*0}$ resonant states
contribute (in particular, $\phi K^{*0}(892)$, $\phi K^{*0}_0(1430)$ and $\phi
K^{*0}_2(1430)$ are seen), the composition can be determined from the analysis
of $B \to \phi K^+ \pi^-$ decays, assuming only that the ratio of branching fractions ${\cal B}(K^{*0} \to \KS \pi^0)/{\cal B}(K^{*0} \to K^+ \pi^-)$ is the same
for each excited kaon state. 

\babar~\cite{Aubert:2008zza} has performed a simultaneous analysis of 
$\Bz \to \phi \KS \pi^0$ and $\Bz \to \phi K^+ \pi^-$ decays that is time-dependent for the former mode and time-integrated for the latter. 
Such an analysis allows, in principle, all parameters of the $\Bz \to \phi K^{*0}$ system to be determined, including mixing-induced \CP violation effects. 
The latter is determined to be $\Delta\phi_{00} = 0.28 \pm 0.42 \pm 0.04$, where $\Delta\phi_{00}$ is half the weak phase difference between $\Bz$ and $\Bzb$ decays to the $\phi K^{*0}_0(1430)$ final state. 
As discussed above, this can also be presented in terms of the Q2B parameter $\sin(2\beta^{\rm eff}_{00}) = \sin(2\beta+2\Delta\phi_{00}) = 0.97 \,^{+0.03}_{-0.52}$. 
The highly asymmetric uncertainty arises due to the conversion from the phase to the sine of the phase, and the proximity of the physical boundary. 

Similar $\sin(2\beta^{\rm eff})$ parameters can be defined for each of the helicity amplitudes for both $\phi K^{*0}(892)$ and $\phi K^{*0}_2(1430)$. 
However, the relative phases between these decays are constrained due to the nature of the simultaneous analysis of $\Bz \to \phi \KS \pi^0$ and $\Bz \to \phi K^+ \pi^-$, decays and therefore these measurements are highly correlated. 
Instead of quoting all these results, \babar provide an illustration of their measurements with the following differences: 
\begin{eqnarray}
  \sin(2\beta - 2\Delta\delta_{01}) - \sin(2\beta) & = & -0.42\,^{+0.26}_{-0.34} \, , \\
  \sin(2\beta - 2\Delta\phi_{\parallel1}) - \sin(2\beta) & = & -0.32\,^{+0.22}_{-0.30} \, , \\
  \sin(2\beta - 2\Delta\phi_{\perp1}) - \sin(2\beta) & = & -0.30\,^{+0.23}_{-0.32} \, , \\
  \sin(2\beta - 2\Delta\phi_{\perp1}) - \sin(2\beta - 2\Delta\phi_{\parallel1}) & = & 0.02 \pm 0.23 \, , \\
  \sin(2\beta - 2\Delta\delta_{02}) - \sin(2\beta) & = & -0.10\,^{+0.18}_{-0.29} \, ,
\end{eqnarray}
where the first subscript indicates the helicity amplitude and the second
indicates the spin of the kaon resonance. 
For the complete definitions of the
$\Delta\delta$ and $\Delta\phi$ parameters, please refer to the \babar\ paper~\cite{Aubert:2008zza}.

Parameters of \CP violation in decay for each of the contributing helicity amplitudes can also be measured. 
Again, these are determined from a simultaneous fit of $\Bz \to \phi \KS \pi^0$ and $\Bz \to \phi K^+ \pi^-$ decays, with the precision being dominated by the statistics of the latter mode. 
Measurements of \CP violation in decay, obtained from decay-time-integrated analyses, are tabulated by the HFLAV Rare Decays subgroup (Sec.~\ref{sec:rare}). 

%%%%%%%%
%%%
%%% qqs
%%%
%%%%%%%%
\mysubsubsection{Time-dependent \CP asymmetries in $\Bs \to \Kp\Km$}
\label{sec:cp_uta:qqs:BstoKK}

The decay $\Bs \to \Kp\Km$ involves a $b \to u\bar{u}s$ transition, and hence has both penguin and tree contributions. 
Both mixing-induced and \CP violation in decay effects may arise, and additional input is needed to disentangle the contributions and determine $\gamma$ and $\beta_s^{\rm eff}$. 
For example, the observables in $\Bd \to \pip\pim$ can be related using U-spin, as proposed in Refs.~\cite{Dunietz:1993rm,Fleischer:1999pa}.

The observables are $A_{\rm mix} = S_{\CP}$, $A_{\rm dir} = -C_{\CP}$, and $A_{\Delta\Gamma}$. 
They can all be treated as free parameters, but are physically constrained to satisfy $A_{\rm mix}^2 + A_{\rm dir}^2 + A_{\Delta\Gamma}^2 = 1$. 
Note that the untagged decay distribution, from which an ``effective lifetime'' can be measured, retains sensitivity to $A_{\Delta\Gamma}$; measurements of the $\Bs \to \Kp\Km$ effective lifetime have been made by LHCb~\cite{Aaij:2012kn,Aaij:2014fia}.
Compilations and averages of effective lifetimes are performed by the HFLAV Lifetimes and Oscillations subgroup, see Sec.~\ref{sec:life_mix}.

The observables in $\Bs \to \Kp\Km$ have been measured by LHCb~\cite{LHCb-CONF-2016-018}, who do not impose the constraint mentioned above to eliminate $A_{\rm \Delta\Gamma}$. 
The results are shown in Table~\ref{tab:cp_uta:BstoKK}, and correspond to evidence for \CP violation both in the interference between mixing and decay, and in the $\Bs \to \Kp\Km$ decay.

\begin{table}[!htb]
	\begin{center}
		\caption{
      Results from time-dependent analysis of the $\Bs \to K^{+} K^{-}$ decay.
		}
		\vspace{0.2cm}
		\setlength{\tabcolsep}{0.0pc}
\renewcommand{\arraystretch}{1.1}
		\begin{tabular*}{\textwidth}{@{\extracolsep{\fill}}lrcccc} \hline
	\mc{2}{l}{Experiment} & Sample size & $S_{\CP}$ & $C_{\CP}$ & $A^{\Delta\Gamma}$ \\
	\hline
	LHCb & \cite{LHCb-CONF-2016-018} & $\int {\cal L} \, dt = 3.0 \ {\rm fb}^{-1}$ & $0.22 \pm 0.06 \pm 0.02$ & $0.24 \pm 0.06 \pm 0.02$ & $-0.75 \pm 0.07 \pm 0.11$ \\
%	LHCb & \cite{Aaij:2013tna} & $\int {\cal L} \, dt = 1.0 \ {\rm fb}^{-1}$ & $0.30 \pm 0.12 \pm 0.04$ & $0.14 \pm 0.11 \pm 0.03$ & 0.02 \\
	\hline
%	\mc{3}{l}{\bf Average} & $0.17 \pm 0.19$ & $0.02 \pm 0.18$ & {\small uncorrelated averages} \\
%	\mc{3}{l}{\small Confidence level} & {\small $0.xx~(y.y\sigma)$} & {\small $0.xx~(y.y\sigma)$} & \\
%		\hline
		\end{tabular*}
		\label{tab:cp_uta:BstoKK}
	\end{center}
\end{table}

Interpretations of an earlier set of results~\cite{Aaij:2013tna}, in terms of constraints on $\gamma$ and $2\beta_s$, have been separately published by LHCb~\cite{Aaij:2014xba}.

\mysubsubsection{Time-dependent \CP asymmetries in $\Bs \to \phi\phi$}
\label{sec:cp_uta:qqs:Bstophiphi}

 The decay $\Bs \to \phi\phi$ involves a $b \to s\bar{s}s$ transition, and hence is a ``pure penguin'' mode (in the limit that the $\phi$ meson is considered a pure $s\bar{s}$ state). 
Since the mixing phase and the decay phase are expected to cancel in the Standard Model, the prediction for the phase from the interference of mixing and decay is predicted to be $\phi_s(\phi\phi) = 0$ with low uncertainty~\cite{Raidal:2002ph}. Due to the vector-vector nature of the final state, angular analysis is needed to separate the \CP-even and \CP-odd contributions. Such an analysis also makes it possible to fit directly for $\phi_s(\phi\phi)$.

A constraint on $\phi_s(\phi\phi)$ has been obtained by LHCb using $3.0 \,{\rm fb}^{-1}$ of data~\cite{Aaij:2014kxa}.
The result is $\phi_s(\phi\phi) = -0.17 \pm 0.15 \pm 0.03 \, {\rm rad}$ where the first uncertainty is statistical and the second is systematic.

%%%%%%%%
%%%
%%% qqd
%%%
%%%%%%%%
% \afterpage{\clearpage}
\mysubsection{Time-dependent $\CP$ asymmetries in $b \to q\bar{q}d$ transitions
}
\label{sec:cp_uta:qqd}

Decays such as $\Bz\to\KS\KS$ are pure $b \to q\bar{q}d$ penguin transitions.
As shown in Eq.~(\ref{eq:cp_uta:b_to_d}),
this diagram has different contributing weak phases,
and therefore the observables are sensitive to their difference 
(which can be chosen to be either $\beta$ or $\gamma$).
Note that if the contribution with the top quark in the loop dominates,
the weak phase from the decay amplitudes should cancel that from mixing,
so that no $\CP$ violation (neither mixing-induced nor in decay) occurs.
Non-zero contributions from loops with intermediate up and charm quarks
can result in both types of effect 
(as usual, a strong phase difference is required for $\CP$ violation in decay
to occur).

\begin{figure}[htbp]
  \begin{center}
    \begin{tabular}{cc}
      \resizebox{0.46\textwidth}{!}{
        \includegraphics{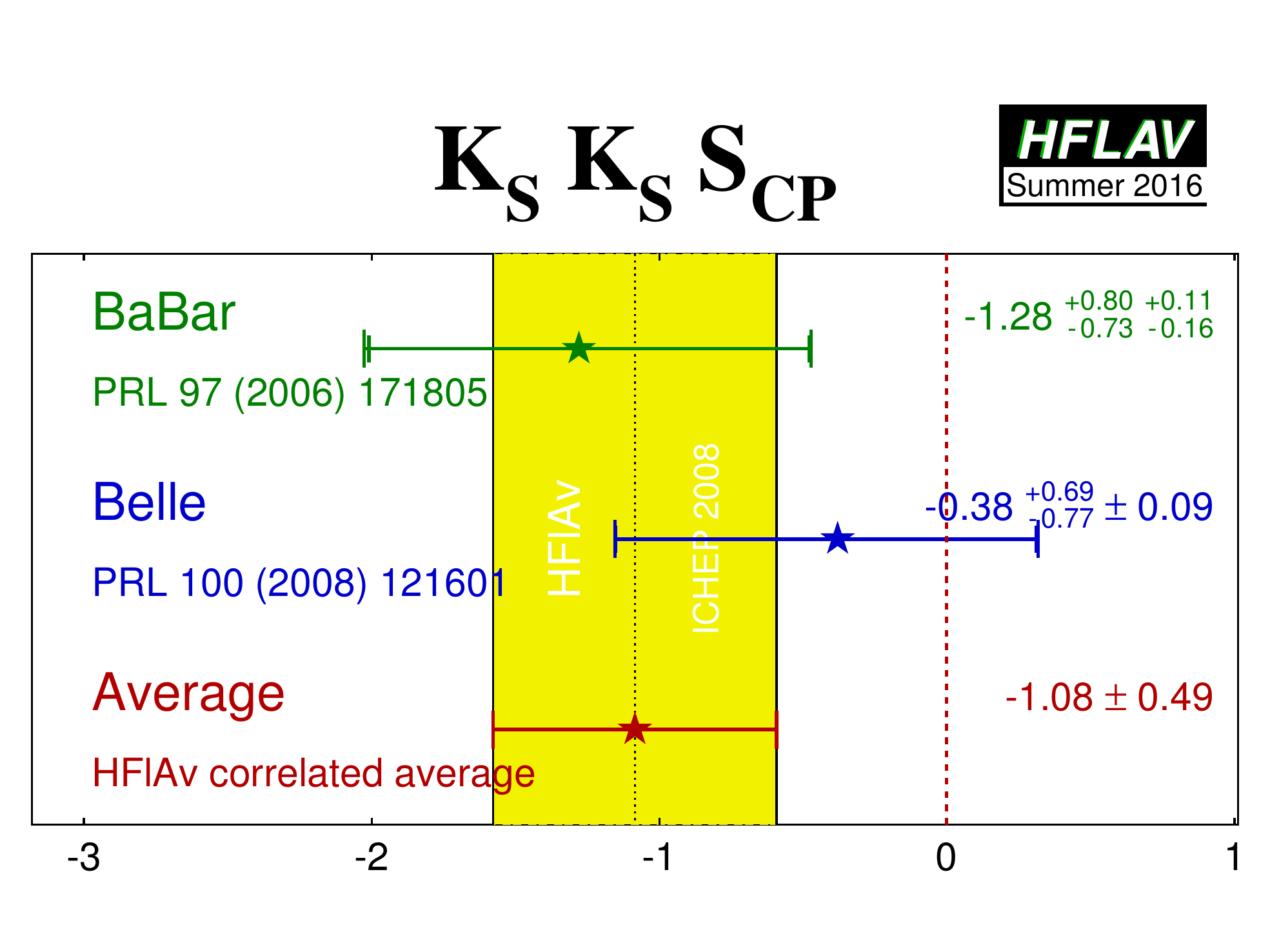}
      }
      &
      \resizebox{0.46\textwidth}{!}{
        \includegraphics{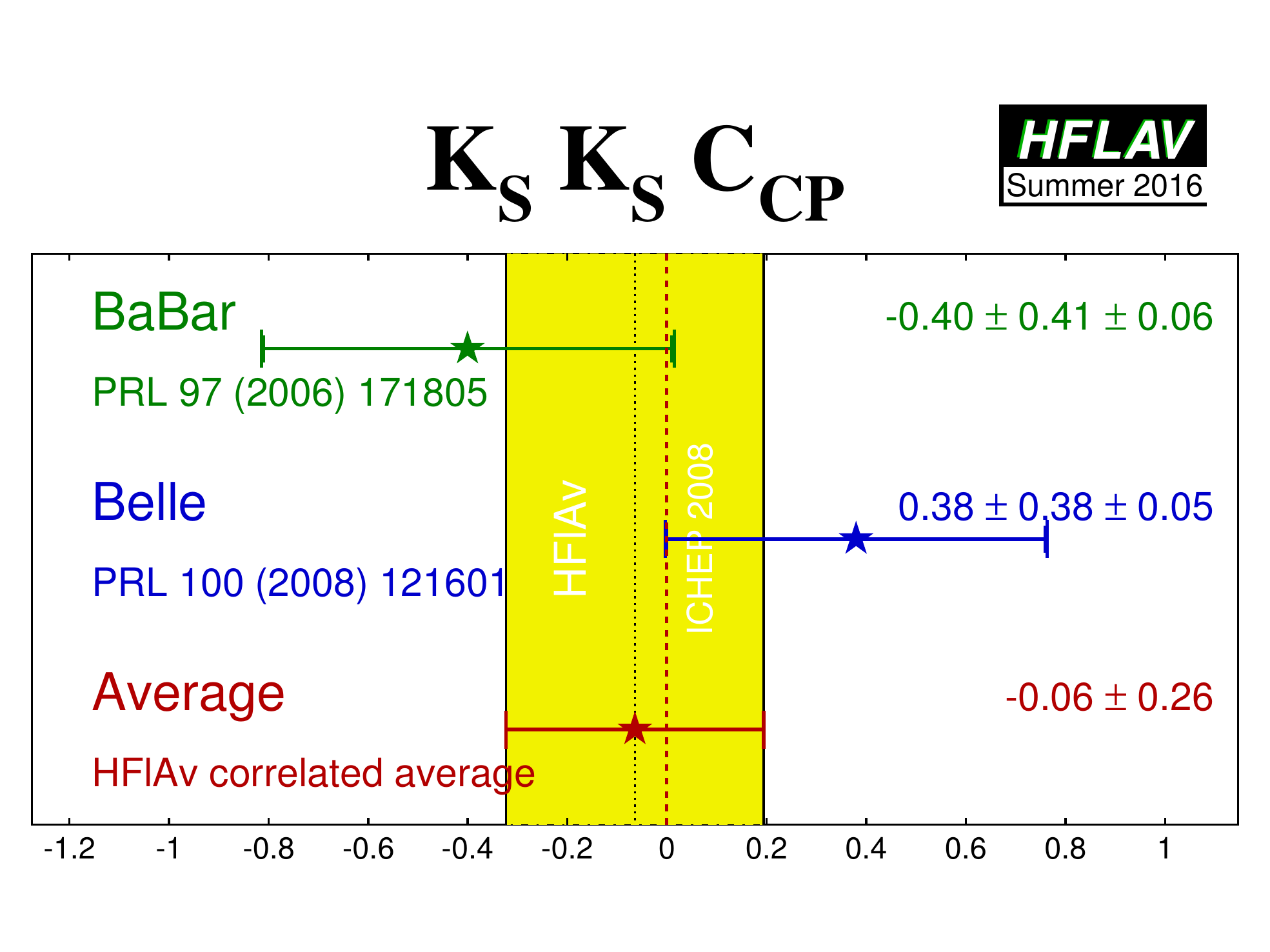}
      }
    \end{tabular}
  \end{center}
  \vspace{-0.8cm}
  \caption{
    Averages of (left) $S_{\CP}$ and (right) $C_{\CP}$ for the mode $\Bz \to \KS\KS$.
  }
  \label{fig:cp_uta:qqd:ksks}
\end{figure}

Both \babar~\cite{Aubert:2006gm} and \belle~\cite{Nakahama:2007dg}
have performed time-dependent analyses of $\Bz\to\KS\KS$ decays.
The results are given in Table~\ref{tab:cp_uta:qqd}
and shown in Fig.~\ref{fig:cp_uta:qqd:ksks}.

\begin{table}[htb]
	\begin{center}
		\caption{
			Results for $\Bz \to \KS\KS$.
		}
		\vspace{0.2cm}
		\setlength{\tabcolsep}{0.0pc}
\renewcommand{\arraystretch}{1.1}
		\begin{tabular*}{\textwidth}{@{\extracolsep{\fill}}lrcccc} \hline
	\mc{2}{l}{Experiment} & $N(B\bar{B})$ & $S_{\CP}$ & $C_{\CP}$ & Correlation \\
	\hline
	\babar & \cite{Aubert:2006gm} & 350M & $-1.28 \,^{+0.80}_{-0.73} \,^{+0.11}_{-0.16}$ & $-0.40 \pm 0.41 \pm 0.06$ & $-0.32$ \\
	\belle & \cite{Nakahama:2007dg} & 657M & $-0.38 \,^{+0.69}_{-0.77} \pm 0.09$ & $0.38 \pm 0.38 \pm 0.05$ & $0.48$ \\
	\hline
	\mc{3}{l}{\bf Average} & $-1.08 \pm 0.49$ & $-0.06 \pm 0.26$ & $0.14$ \\
	\mc{3}{l}{\small Confidence level} & \mc{2}{c}{\small $0.29~(1.1\sigma)$} & \\
		\hline
		\end{tabular*}
		\label{tab:cp_uta:qqd}
	\end{center}
\end{table}

%%%%%%%%
%%%
%%% b -> sg
%%%
%%%%%%%%
% \afterpage{\clearpage}
\mysubsection{Time-dependent asymmetries in $b \to s\gamma$ transitions
}
\label{sec:cp_uta:bsg}

The radiative decays $b \to s\gamma$ produce photons 
which are highly polarised in the Standard Model.
The decays $\Bz \to F \gamma$ and $\Bzb \to F \gamma$ 
produce photons with opposite helicities, 
and since the polarisation is, in principle, observable,
these final states cannot interfere.
The finite mass of the $s$ quark introduces small corrections
to the limit of maximum polarisation,
but any large mixing-induced $\CP$ violation would be a signal for new physics.
Since a single weak phase dominates the $b \to s \gamma$ transition in the 
Standard Model, the cosine term is also expected to be small.

Atwood {\it et al.}~\cite{Atwood:2004jj} have shown that 
an inclusive analysis of $\KS\pi^0\gamma$ can be performed,
since the properties of the decay amplitudes 
are independent of the angular momentum of the $\KS\pi^0$ system. 
However, if non-dipole operators contribute significantly to the amplitudes, 
then the Standard Model mixing-induced $\CP$ violation could be larger 
than the na\"\i ve expectation 
$S \simeq -2 (m_s/m_b) \sin \left(2\beta\right)$~\cite{Grinstein:2004uu,Grinstein:2005nu}.
In this case, 
the $\CP$ parameters may vary over the $\KS\pi^0\gamma$ Dalitz plot, 
for example as a function of the $\KS\pi^0$ invariant mass.
% Explicit calculations indicate such corrections are small for exclusive final states~\cite{Matsumori:2005ax,Ball:2006cva}.

With the above in mind, 
we quote two averages: one for $K^*(892)$ candidates only, 
and the other one for the inclusive $\KS\pi^0\gamma$ decay (including the $K^*(892)$).
If the Standard Model dipole operator is dominant, 
both should give the same quantities 
(the latter naturally with smaller statistical error). 
If not, care needs to be taken in interpretation of the inclusive parameters, 
while the results on the $K^*(892)$ resonance remain relatively clean.
Results from \babar\ and \belle\ are
used for both averages; both experiments use the invariant mass range 
$0.60 < M_{\KS\pi^0} < 1.80~\gevcc$
in the inclusive analysis.

In addition to the $\KS\pi^0\gamma$ decay, both \babar\ and \belle\ have presented results using the $\KS\rho\gamma$ mode, while \babar\ (\belle) has in addition presented results using the $\KS\eta\gamma$ ($\KS\phi\gamma$) channel.
For the $\KS\rho\gamma$ case, due to the non-negligible width of the $\rho^0$ meson, decays selected as $\Bz \to \KS\rho^0\gamma$ can include a significant contribution from $K^{*\pm}\pi^\mp\gamma$ decays, which are flavour-specific and do not have the same oscillation phenomenology. 
Both \babar\ and \belle\ measure $S_{\rm eff}$ for all \B decay candidates with the $\rho^0$ selection being $0.6 < m(\pip\pim) < 0.9~\gevcc$, obtaining $0.14 \pm 0.25 \,^{+0.04}_{-0.03}$ (\babar) and $0.09 \pm 0.27 \,^{+0.04}_{-0.07}$ (\belle). 
These values are then corrected for a ``dilution factor'', that is evaluated with different methods in the two experiments: \babar~\cite{Akar:2013ima,Sanchez:2015pxu} obtains a dilution factor of $-0.78 \,^{+0.19}_{-0.17}$ while \belle~\cite{Li:2008qma} obtains $+0.83 \,^{+0.19}_{-0.03}$. 
Until the discrepancy between these values is understood, the average of the results should be treated with caution.

\begin{table}[htb]
	\begin{center}
		\caption{
      Averages for $b \to s \gamma$ modes.
		}
		\vspace{0.2cm}
		\setlength{\tabcolsep}{0.0pc}
\renewcommand{\arraystretch}{1.1}
		\begin{tabular*}{\textwidth}{@{\extracolsep{\fill}}lrcccc} \hline
	\mc{2}{l}{Experiment} & $N(B\bar{B})$ & $S_{\CP} (b \to s \gamma)$ & $C_{\CP} (b \to s \gamma)$ & Correlation \\
        \hline
        \mc{6}{c}{$\Kstar(892)\gamma$} \\
	\babar & \cite{Aubert:2008gy} & 467M & $-0.03 \pm 0.29 \pm 0.03$ & $-0.14 \pm 0.16 \pm 0.03$ & $0.05$ \\
	\belle & \cite{Ushiroda:2006fi} & 535M & $-0.32 \,^{+0.36}_{-0.33} \pm 0.05$ & $0.20 \pm 0.24 \pm 0.05$ & $0.08$ \\
%	\hline
	\mc{3}{l}{\bf Average} & $-0.16 \pm 0.22$ & $-0.04 \pm 0.14$ & $0.06$ \\
	\mc{3}{l}{\small Confidence level} & \mc{2}{c}{\small $0.40~(0.9\sigma)$} & \\
		\hline
% 		\end{tabular*}
% 		\label{tab:cp_uta:yyy}
% 	\end{center}
% \end{table}

% \begin{table}
% 	\begin{center}
% 		\caption{
% 			Averages for $K_{S} \\pi^{0} \\gammama$.
% 		}
% 		\vspace{0.2cm}
% 		\setlength{\tabcolsep}{0.0pc}
% 		\begin{tabular*}{\textwidth}{@{\extracolsep{\fill}}lrcccc} \hline
% 		\mc{2}{l}{Experiment} & $N(B\bar{B})$ & $S_{\CP}$ & $C_{\CP}$ & Correlation \\
% 		\hline
        \mc{6}{c}{$\KS \pi^0 \gamma$ (including $\Kstar(892)\gamma$)} \\
	\babar & \cite{Aubert:2008gy} & 467M & $-0.17 \pm 0.26 \pm 0.03$ & $-0.19 \pm 0.14 \pm 0.03$ & $0.04$ \\
	\belle & \cite{Ushiroda:2006fi} & 535M & $-0.10 \pm 0.31 \pm 0.07$ & $0.20 \pm 0.20 \pm 0.06$ & $0.08$ \\
%	\hline
	\mc{3}{l}{\bf Average} & $-0.15 \pm 0.20$ & $-0.07 \pm 0.12$ & $0.05$ \\
        \mc{3}{l}{\small Confidence level} & \mc{2}{c}{\small $0.30~(1.0\sigma)$} & \\

		\hline
%% 		\end{tabular*}
%% 		\label{tab:cp_uta:bsg}
%% 	\end{center}
%% \end{table}

%% \begin{table}[!htb]
%% 	\begin{center}
%% 		\caption{
%% 			Averages for $K_{S} eta \\gammama$.
%% 		}
%% 		\vspace{0.2cm}
%% 		\setlength{\tabcolsep}{0.0pc}
%% 		\begin{tabular*}{\textwidth}{@{\extracolsep{\fill}}lrcccc} \hline
%% 	\mc{2}{l}{Experiment} & $N(B\bar{B})$ & $S_{\CP}$ & $C_{\CP}$ & Correlation \\
%% 	\hline
        \mc{6}{c}{$\KS \eta \gamma$} \\
	\babar & \cite{Aubert:2008js} & 465M & $-0.18 \,^{+0.49}_{-0.46} \pm 0.12$ & $-0.32 \,^{+0.40}_{-0.39} \pm 0.07$ & $-0.17$ \\
%	\belle & \cite{Belle-KSetagamma} & 772M & $-1.32 \pm 0.77 \pm 0.36$ & $0.48 \pm 0.41 \pm 0.07$ & $-0.14$ \\
	\hline
%	\mc{3}{l}{\bf Average} & $-0.49 \pm 0.42$ & $0.06 \pm 0.29$ & $-0.15$ \\
%	\mc{3}{l}{\small Confidence level} & \mc{2}{c}{\small $0.24~(1.2\sigma)$} & \\
%% 		\hline
%% 		\end{tabular*}
%% 		\label{tab:cp_uta:yyy}
%% 	\end{center}
%% \end{table}

%% \begin{table}[!htb]
%% 	\begin{center}
%% 		\caption{
%% 			Averages for $K_{S} \\rho^{0} \\gammama$.
%% 		}
%% 		\vspace{0.2cm}
%% 		\setlength{\tabcolsep}{0.0pc}
%% 		\begin{tabular*}{\textwidth}{@{\extracolsep{\fill}}lrcccc} \hline
%% 	\mc{2}{l}{Experiment} & $N(B\bar{B})$ & $S_{\CP}$ & $C_{\CP}$ & Correlation \\
%% 	\hline
        \mc{6}{c}{$\KS \rho^0 \gamma$} \\
	\babar & \cite{Sanchez:2015pxu} & 471M & $-0.18 \pm 0.32 \,^{+0.06}_{-0.05}$ & $-0.39 \pm 0.20 \,^{+0.03}_{-0.02}$ & $-0.09$ \\
	\belle & \cite{Li:2008qma} & 657M & $0.11 \pm 0.33 \,^{+0.05}_{-0.09}$ & $-0.05 \pm 0.18 \pm 0.06$ & $\phantom{-}0.04$ \\
%	\hline
	\mc{3}{l}{\bf Average} & $-0.06 \pm 0.23$ & $-0.22 \pm 0.14$ & $-0.02$ \\
	\mc{3}{l}{\small Confidence level} & \mc{2}{c}{\small $0.38~(0.9\sigma)$} & \\
 		\hline
%% 		\end{tabular*}
%% 		\label{tab:cp_uta:bsg}
%% 	\end{center}
%% \end{table}

%% \begin{table}[!htb]
%% 	\begin{center}
%% 		\caption{
%% 			Averages for $K_{S} \phi \\gammama$.
%% 		}
%% 		\vspace{0.2cm}
%% 		\setlength{\tabcolsep}{0.0pc}
%% 		\begin{tabular*}{\textwidth}{@{\extracolsep{\fill}}lrcccc} \hline
%% 	\mc{2}{l}{Experiment} & $N(B\bar{B})$ & $S_{\CP}$ & $C_{\CP}$ & Correlation \\
	%% \hline
        \mc{6}{c}{$\KS \phi \gamma$} \\
	\belle & \cite{Sahoo:2011zd} & 772M & $0.74 \,^{+0.72}_{-1.05} \,^{+0.10}_{-0.24}$ & $-0.35 \pm 0.58 \,^{+0.10}_{-0.23}$ & \textendash{} \\
	\hline
	%% \mc{3}{l}{\bf Average} & $0.74 \pm 0.90$ & $-0.35 \pm 0.60$ & $0.00$ \\
	%% \mc{3}{l}{\small Confidence level} & \mc{2}{c}{\small $0.xx~(y.y\sigma)$} & \\
	%% 	\hline
		\end{tabular*}
		\label{tab:cp_uta:bsg}
	\end{center}
\end{table}

\begin{figure}[htbp]
  \begin{center}
    \begin{tabular}{cc}
      \resizebox{0.46\textwidth}{!}{
        \includegraphics{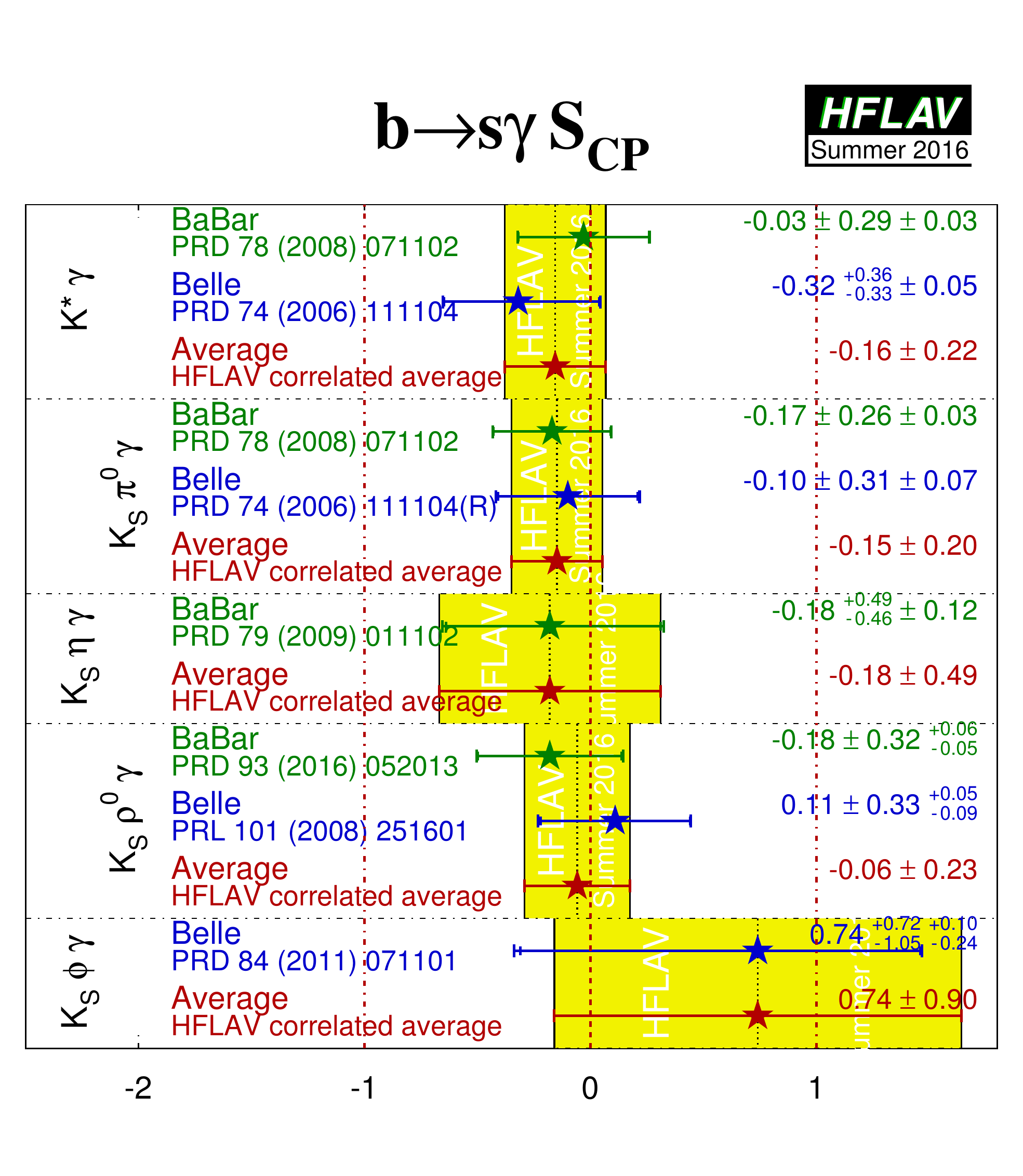}
      }
      &
      \resizebox{0.46\textwidth}{!}{
        \includegraphics{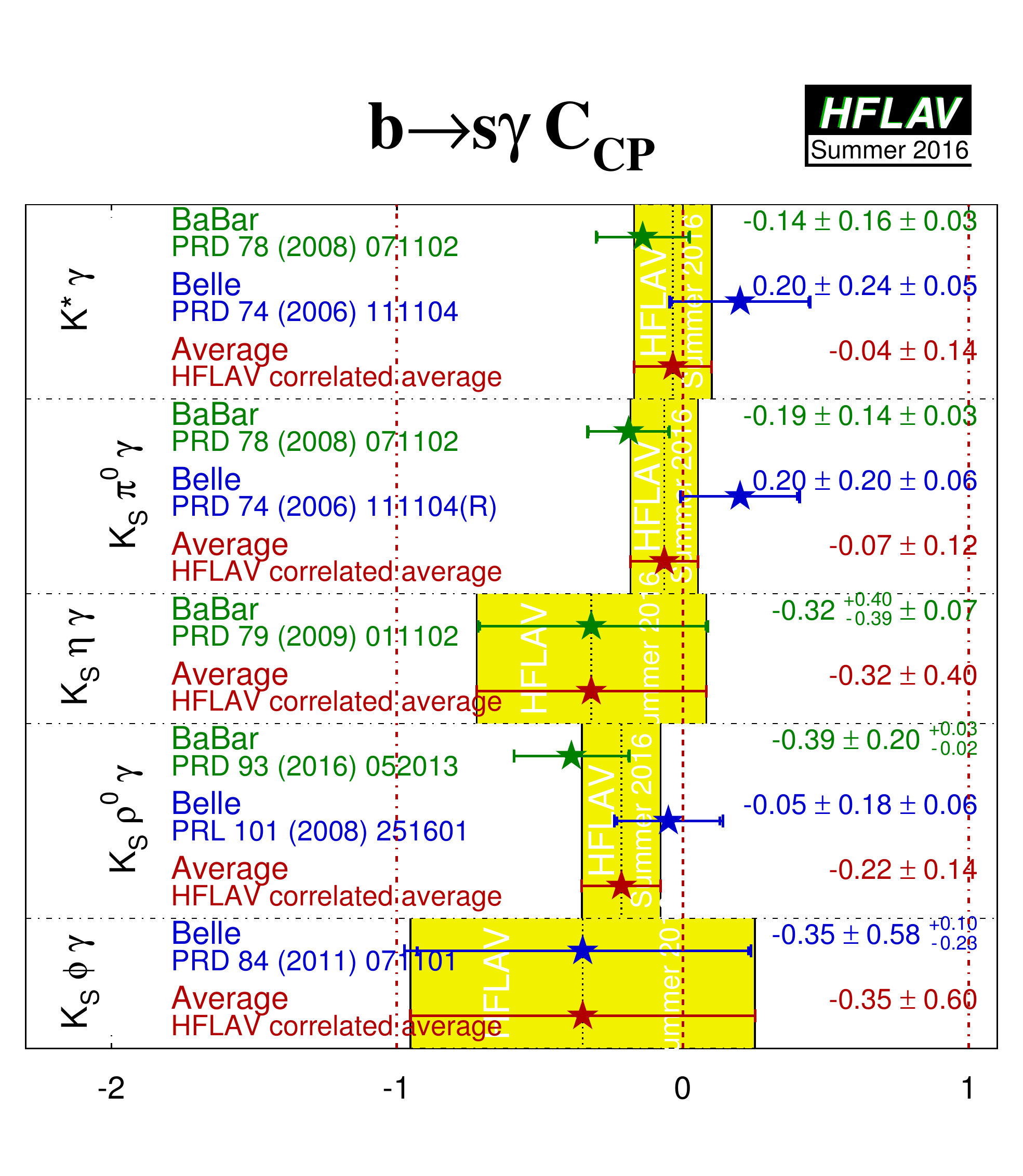}
      }
    \end{tabular}
  \end{center}
  \vspace{-0.8cm}
  \caption{
    Averages of (left) $S_{b \to s \gamma}$ and (right) $C_{b \to s \gamma}$.
    Recall that the data for $K^*\gamma$ is a subset of that for $\KS\pi^0\gamma$.
  }
  \label{fig:cp_uta:bsg}
\end{figure}

\begin{figure}[htbp]
  \centering
    \resizebox{0.32\textwidth}{!}{
      \includegraphics{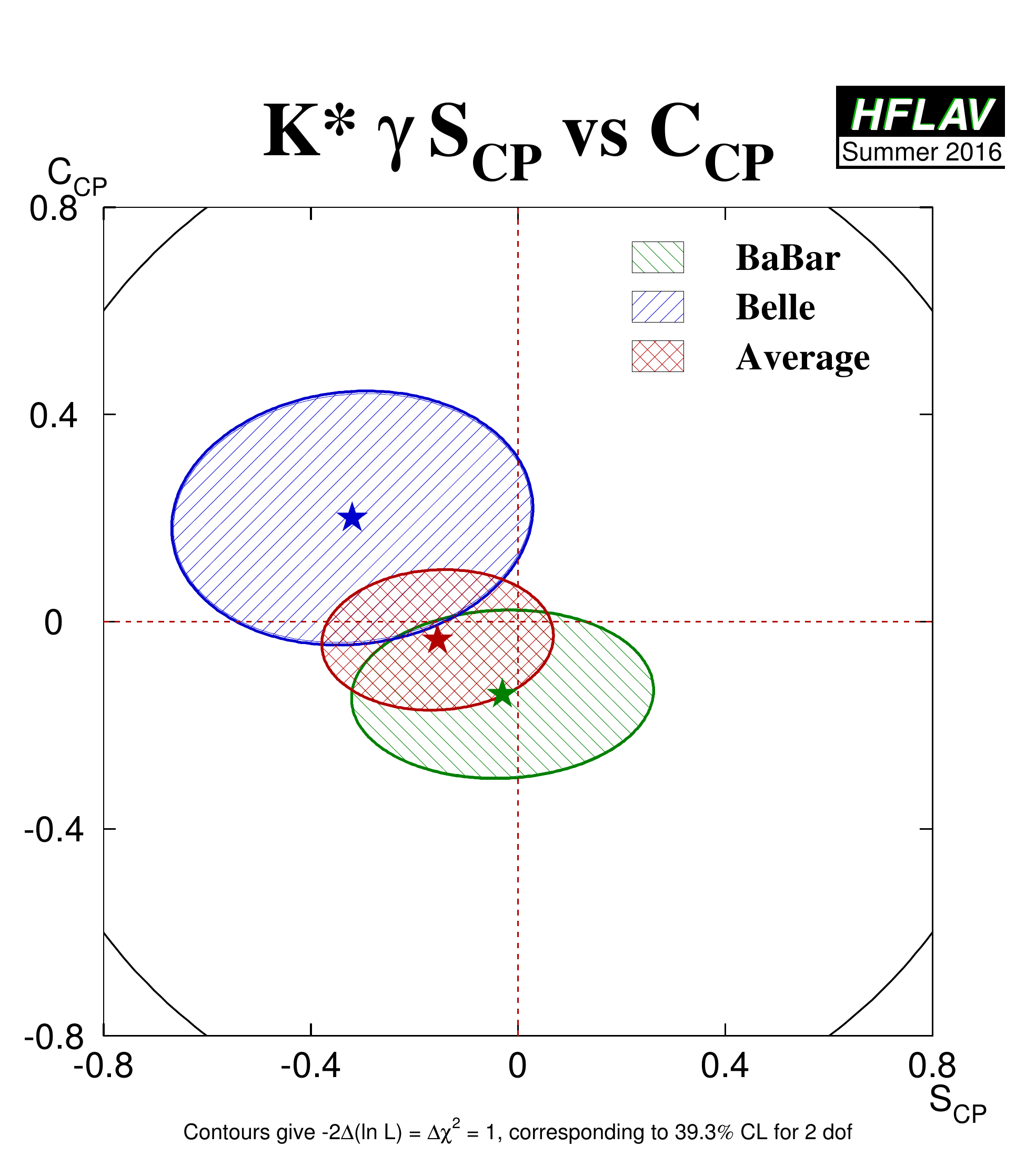}
    }
    \resizebox{0.32\textwidth}{!}{
      \includegraphics{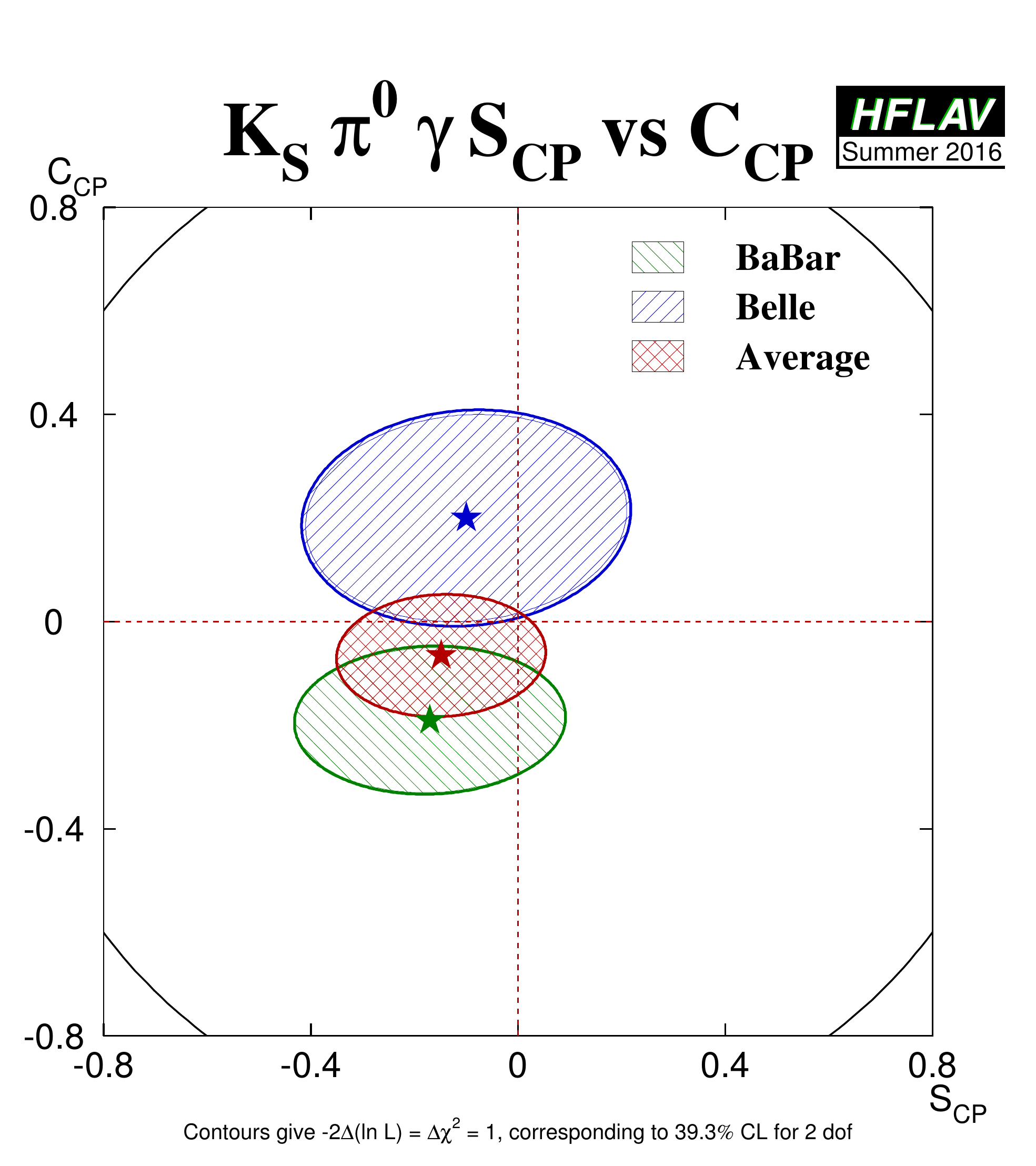}
    }
    \resizebox{0.32\textwidth}{!}{
      \includegraphics{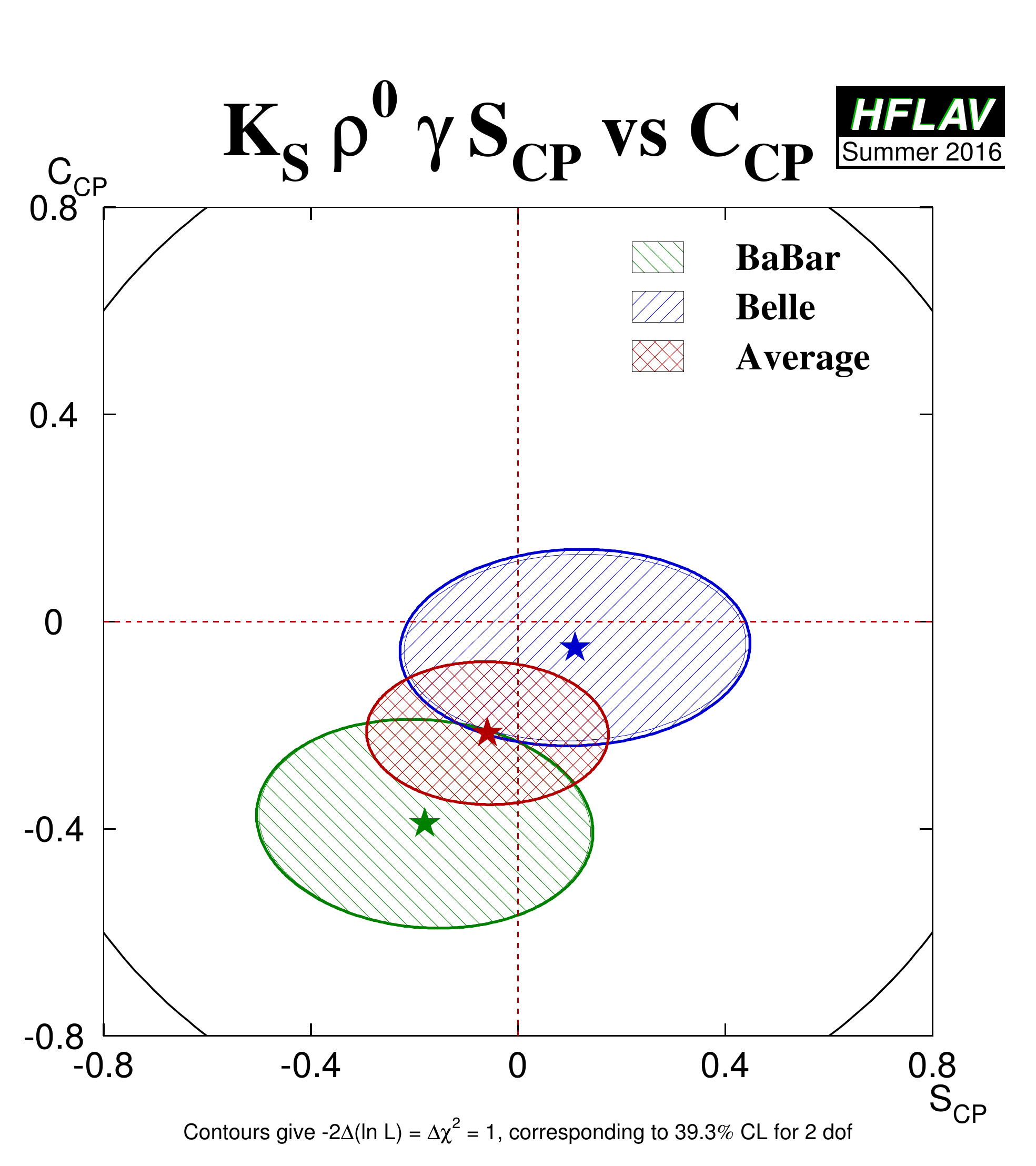}
    }
%  \vspace{-0.5cm}
  \caption{
    Averages of three $b \to s\gamma$ dominated channels,
    for which correlated averages are performed,
    in the $S_{\CP}$ \vs\ $C_{\CP}$ plane.
    (Left) $\Bz \to K^*\gamma$, (middle) $\Bz \to \KS\pi^0\gamma$ (including $K^*\gamma$), (right) $\Bz \to \KS\rho\gamma$.
  }
  \label{fig:cp_uta:bsg_SvsC}
\end{figure}

The results are given in Table~\ref{tab:cp_uta:bsg},
and shown in Figs.~\ref{fig:cp_uta:bsg} and~~\ref{fig:cp_uta:bsg_SvsC}.
No significant $\CP$ violation results are seen;
the results are consistent with the Standard Model
and with other measurements in the $b \to s\gamma$ system (see Sec.~\ref{sec:rare}).

A similar analysis can be performed for radiative \Bs decays to, for example, the $\phi\gamma$ final state. 
As for other observables determined with self-conjugate final states produced in \Bs decays, the effective lifetime also provides sensitivity, and can be determined without tagging the initial flavour of the decaying meson. 
The LHCb collaboration has determined the associated parameter $A_{\Delta\Gamma}(\phi\gamma) = -0.98 \,^{+0.46}_{-0.52}\,^{+0.23}_{-0.20}$~\cite{Aaij:2016ofv}. 

%%%%%%%%
%%%
%%% b -> dg
%%%
%%%%%%%%
% \afterpage{\clearpage}
\mysubsection{Time-dependent asymmetries in $b \to d\gamma$ transitions
}
\label{sec:cp_uta:bdg}

The formalism for the radiative decays $b \to d\gamma$ is much the same
as that for $b \to s\gamma$ discussed above.
Assuming dominance of the top quark in the loop,
the weak phase in decay should cancel with that from mixing,
so that the mixing-induced \CP\ violation parameter $S_{\CP}$ 
should be very small.
Corrections due to the finite light quark mass are smaller compared to $b \to s\gamma$, since $m_d < m_s$, but QCD corrections of ${\cal O}\left(\Lambda_{\rm QCD}/m_b\right)$ may be sizable~\cite{Grinstein:2004uu}.
% they cannot significantly affect the prediction $S_{b \to d \gamma} \simeq 0$.
Large \CP\ violation effects could be seen through a non-zero value of $C_{b \to d \gamma}$ since the top loop is not the only contribution.

Results using the mode $\Bz \to \rho^0\gamma$ are available from 
\belle\ and are given in Table~\ref{tab:cp_uta:bdg}.

\begin{table}[htb]
	\begin{center}
		\caption{
			Averages for $\Bz \to \rho^{0} \gamma$.
		}
		\vspace{0.2cm}
		\setlength{\tabcolsep}{0.0pc}
\renewcommand{\arraystretch}{1.1}
		\begin{tabular*}{\textwidth}{@{\extracolsep{\fill}}lrcccc} \hline
	\mc{2}{l}{Experiment} & $N(B\bar{B})$ & $S_{\CP}$ & $C_{\CP}$ & Correlation \\
	\hline
	\belle & \cite{Ushiroda:2007jf} & 657M & $-0.83 \pm 0.65 \pm 0.18$ & $0.44 \pm 0.49 \pm 0.14$ & $-0.08$ \\
		\hline
		\end{tabular*}
		\label{tab:cp_uta:bdg}
	\end{center}
\end{table}

%%%%%%%%
%%%
%%% b -> uud
%%%
%%%%%%%%
% \afterpage{\clearpage}
\mysubsection{Time-dependent $\CP$ asymmetries in $b \to u\bar{u}d$ transitions
}
\label{sec:cp_uta:uud}

The $b \to u \bar u d$ transition can be mediated by either 
a $b \to u$ tree amplitude or a $b \to d$ penguin amplitude.
These transitions can be investigated using 
the time dependence of $\Bz$ decays to final states containing light mesons.
Results are available from both \babar\ and \belle\ for the 
$\CP$ eigenstate ($\etacp = +1$) $\pi^+\pi^-$ final state
and for the vector-vector final state $\rho^+\rho^-$,
which is found to be dominated by the $\CP$-even
longitudinally polarised component
(\babar\ measures $f_{\rm long} = 
0.992 \pm 0.024 \, ^{+0.026}_{-0.013}$~\cite{Aubert:2007nua}
while \belle\ measures $f_{\rm long} = 
0.988 \pm 0.012 \pm 0.023$~\cite{Vanhoefer:2015ijw}).
\babar\ has also performed a time-dependent analysis of the 
vector-vector final state $\rho^0\rho^0$~\cite{Aubert:2008au},
in which they measure  $f_{\rm long} = 0.70 \pm 0.14 \pm 0.05$;
\belle\ measure a smaller branching fraction than \babar\ for
$\Bz\to\rho^0\rho^0$~\cite{Adachi:2012cz} with corresponding signal yields too small to perform a time-dependent analysis; for the longitudinal polarisation they measure $f_{\rm long} = 0.21 \,^{+0.18}_{-0.22} \pm 0.13$.
LHCb has measured the branching fraction and longitudinal polarisation for $\Bz\to\rho^0\rho^0$, and for the latter finds $f_{\rm long} = 0.745 \,^{+0.048}_{-0.058} \pm 0.034$~\cite{Aaij:2015ria}, but has not yet performed a time-dependent analysis of this decay.
The \belle\ measurement for $f_{\rm long}$ is thus in some tension with the other results.
\babar\ has furthermore performed a time-dependent analysis of the 
$\Bz \to a_1^\pm \pi^\mp$ decay~\cite{Aubert:2006gb}; further experimental
input for the extraction of $\alpha$ from this channel is reported in a later
publication~\cite{Aubert:2009ab}.

\begin{figure}[htbp]
  \begin{center}
    \begin{tabular}{cc}
      \resizebox{0.46\textwidth}{!}{
        \includegraphics{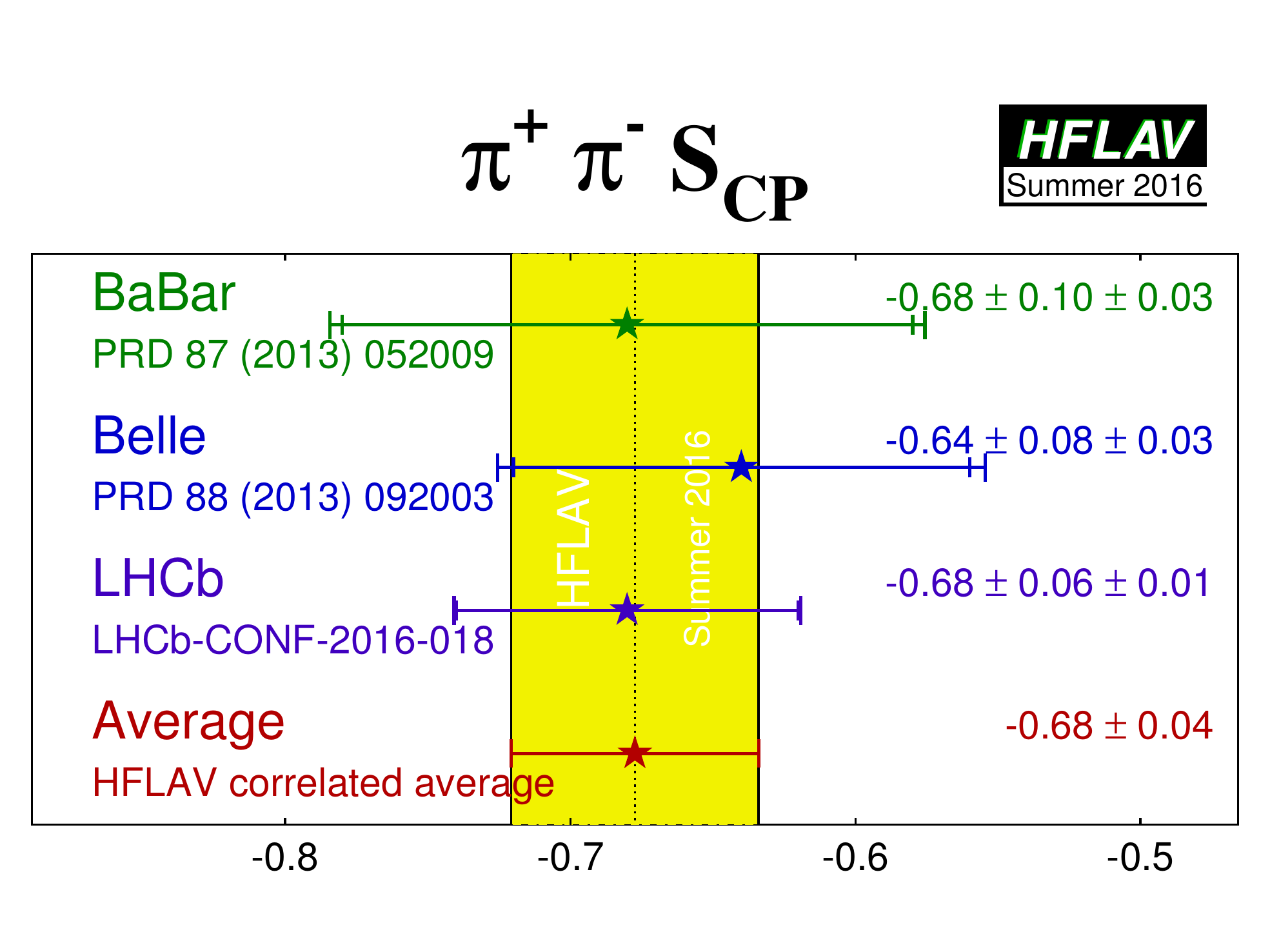}
      }
      &
      \resizebox{0.46\textwidth}{!}{
        \includegraphics{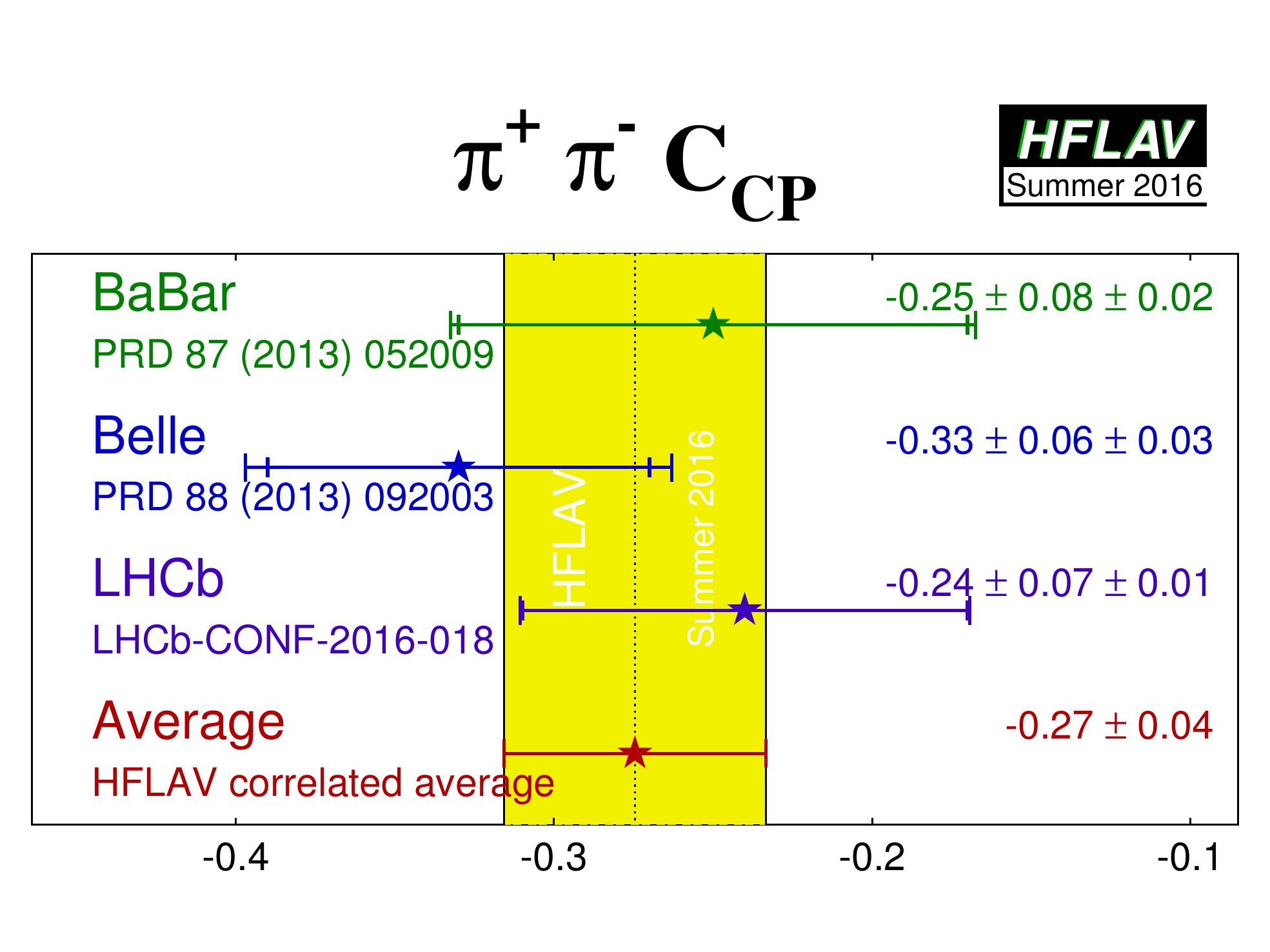}
      }
    \end{tabular}
  \end{center}
  \vspace{-0.8cm}
  \caption{
    Averages of (left) $S_{\CP}$ and (right) $C_{\CP}$ for the mode $\Bz \to \pi^+\pi^-$.
  }
  \label{fig:cp_uta:uud:pipi}
\end{figure}

\begin{figure}[htbp]
  \begin{center}
    \begin{tabular}{cc}
      \resizebox{0.46\textwidth}{!}{
        \includegraphics{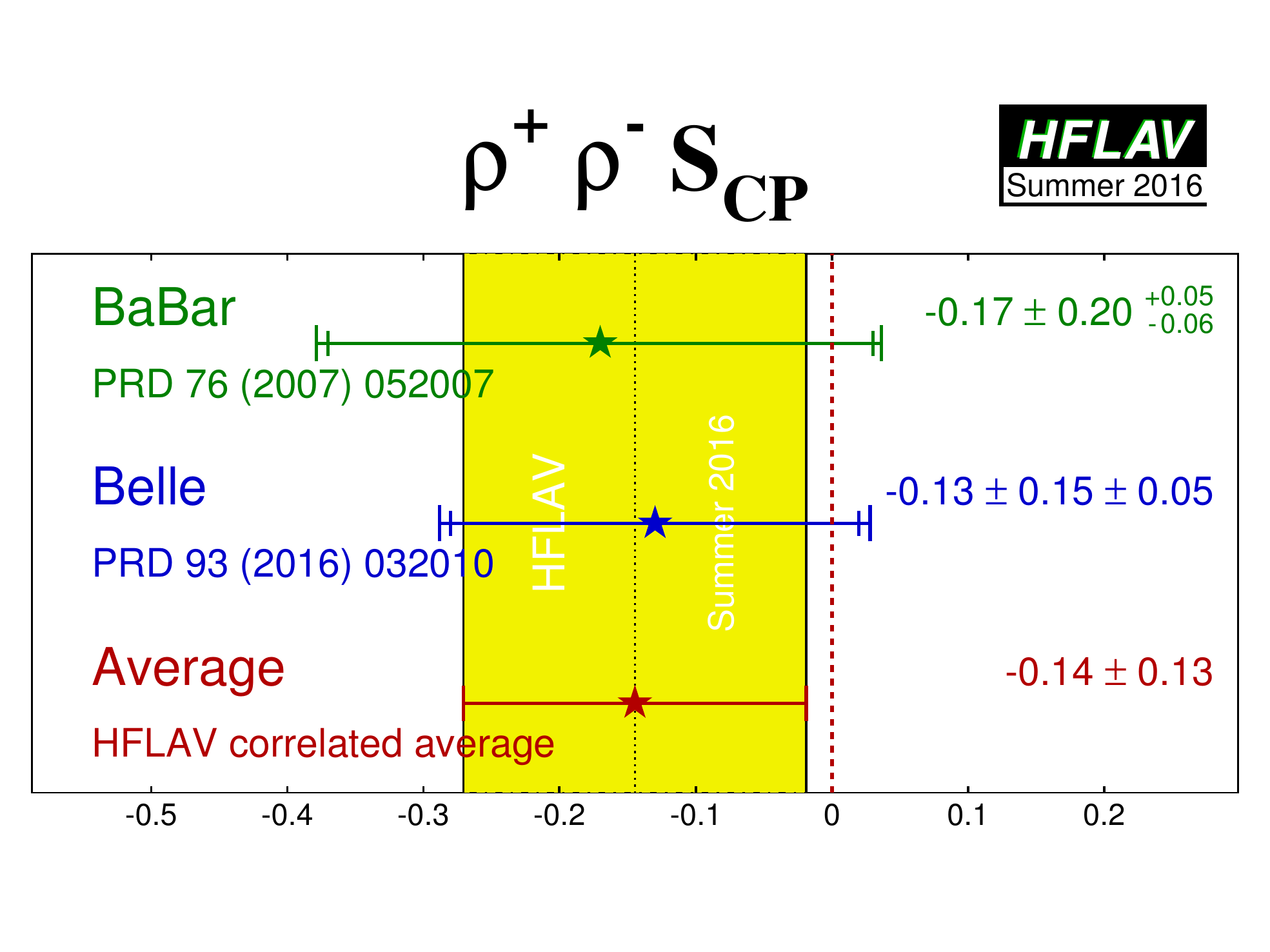}
      }
      &
      \resizebox{0.46\textwidth}{!}{
        \includegraphics{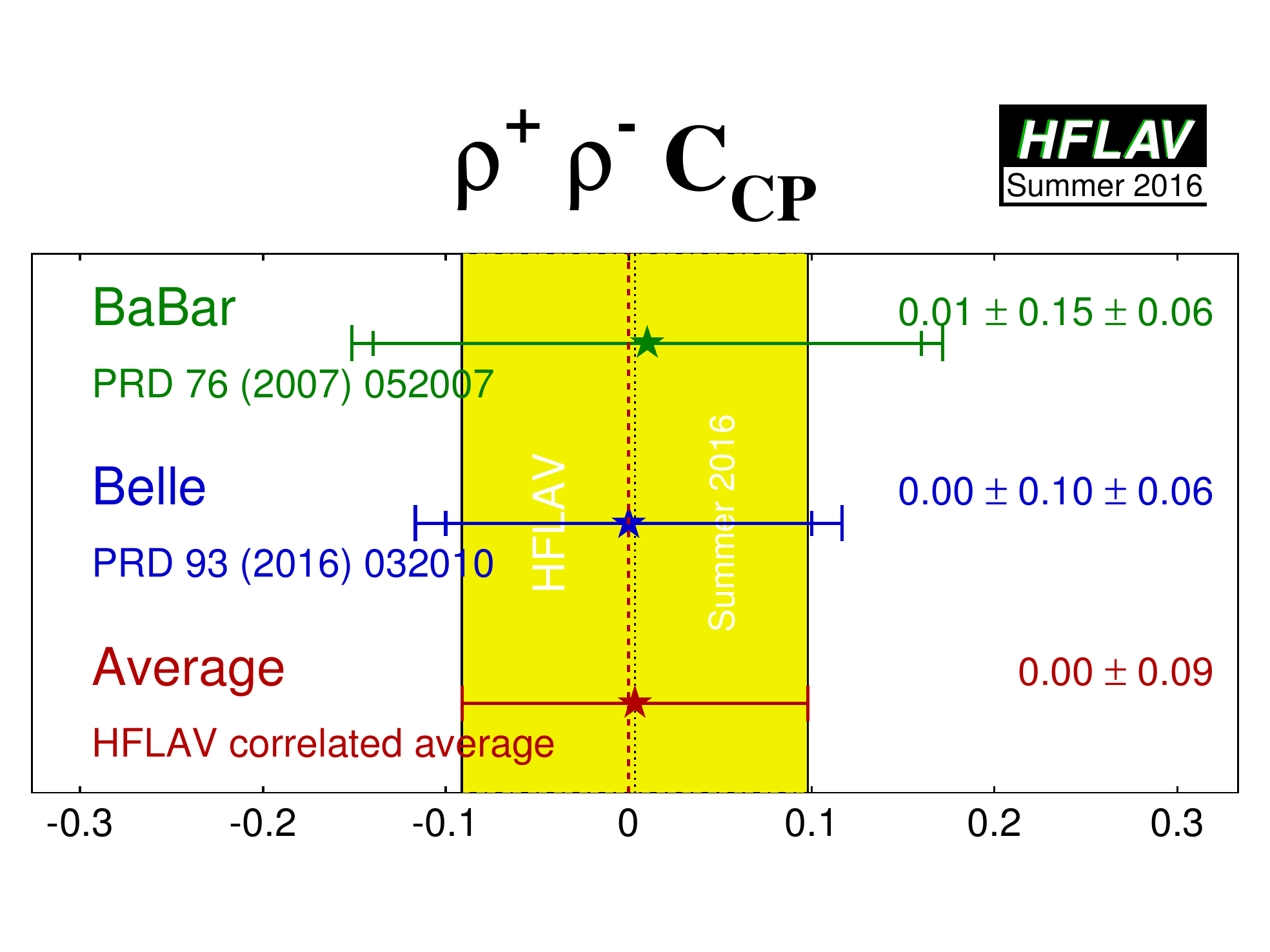}
      }
    \end{tabular}
  \end{center}
  \vspace{-0.8cm}
  \caption{
    Averages of (left) $S_{\CP}$ and (right) $C_{\CP}$ for the mode $\Bz \to \rho^+\rho^-$.
  }
  \label{fig:cp_uta:uud:rhorho}
\end{figure}

\begin{figure}[htbp]
  \begin{center}
    \begin{tabular}{cc}
      \resizebox{0.46\textwidth}{!}{
        \includegraphics{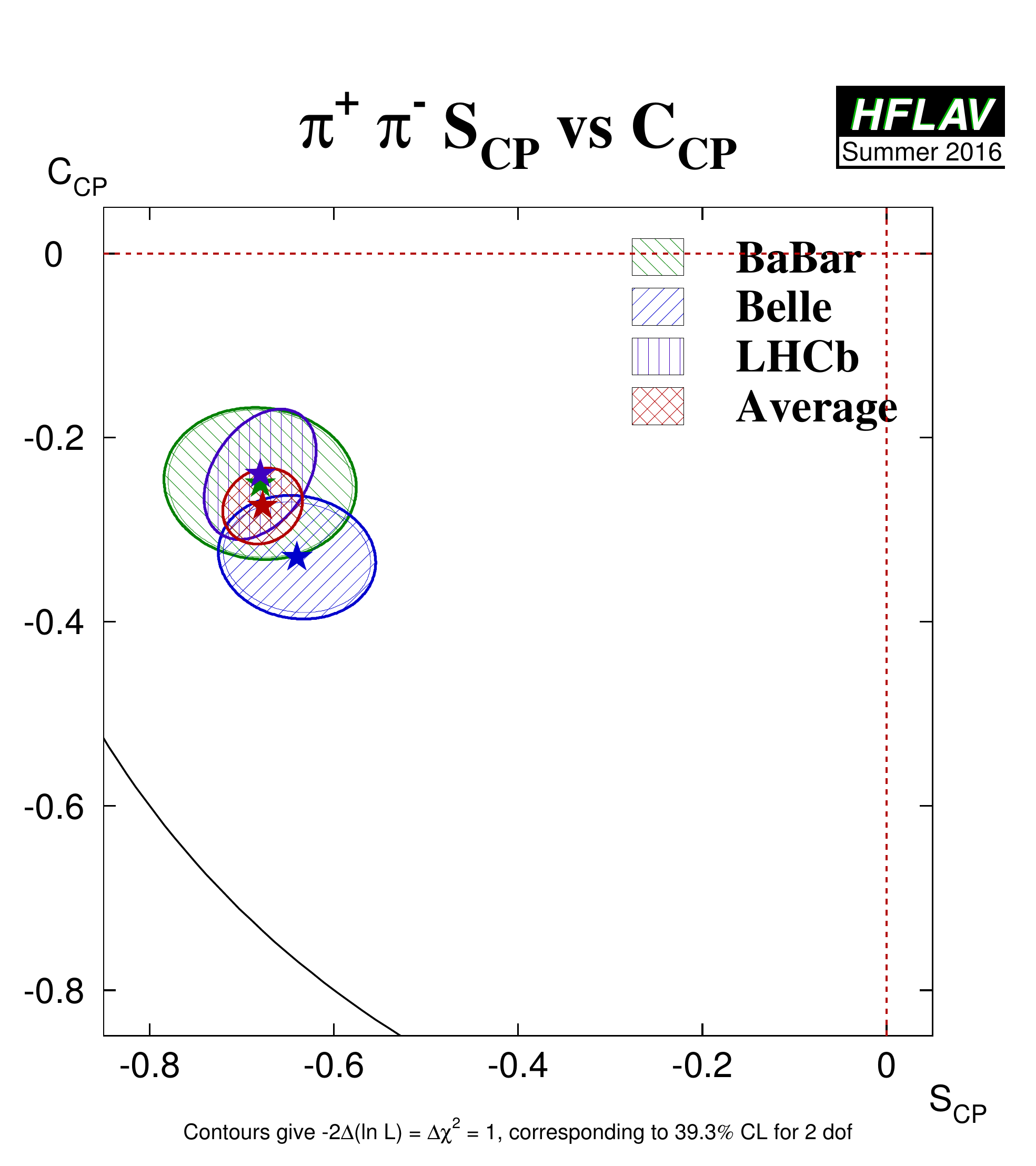}
      }      
      &
      \resizebox{0.46\textwidth}{!}{
        \includegraphics{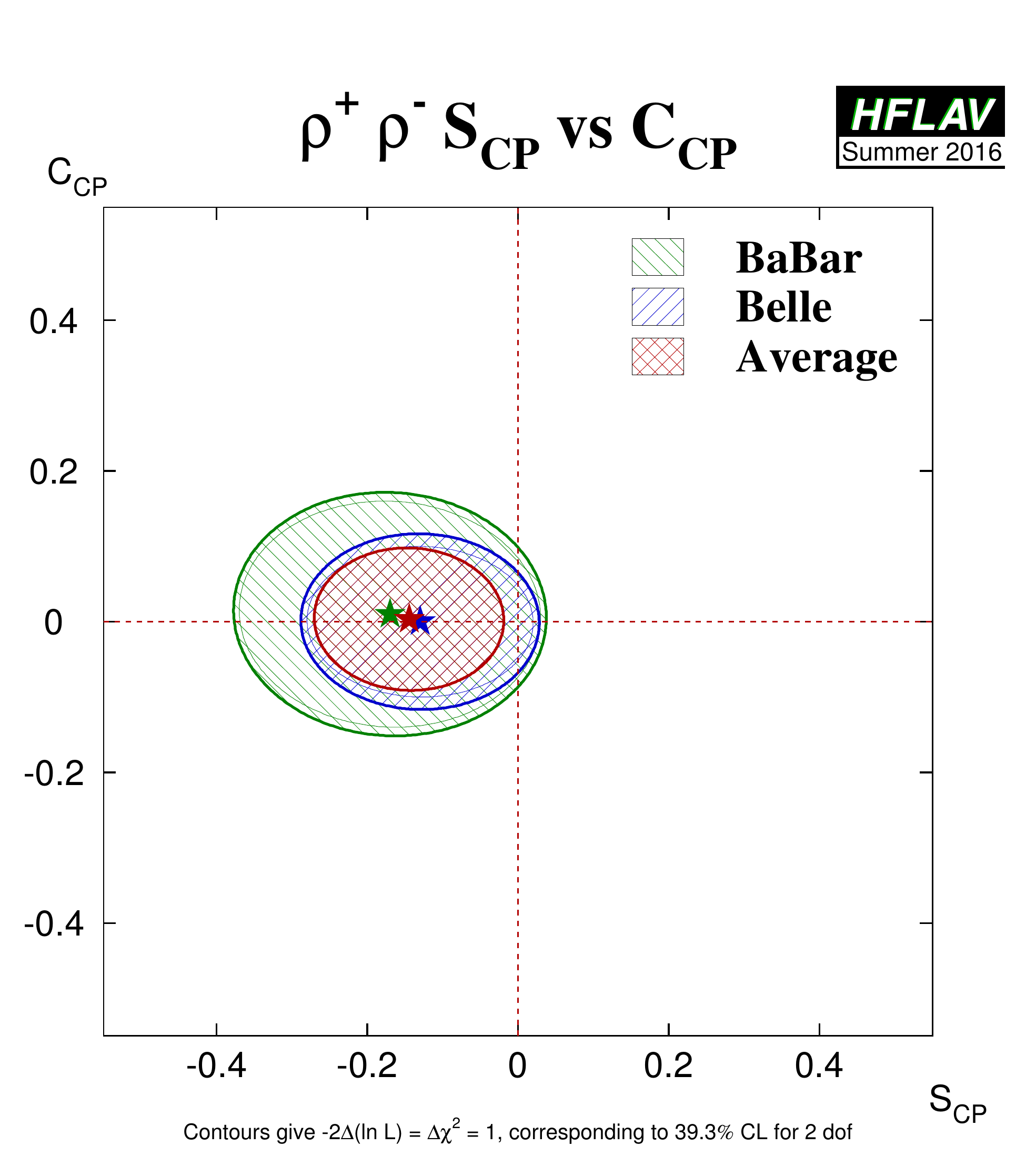}
      }
    \end{tabular}
  \end{center}
  \vspace{-0.8cm}
  \caption{
    Averages of $b \to u\bar u d$ dominated channels,
    for which correlated averages are performed,
    in the $S_{\CP}$ \vs\ $C_{\CP}$ plane.
    (Left) $\Bz \to \pi^+\pi^-$ and (right) $\Bz \to \rho^+\rho^-$.
  }
  \label{fig:cp_uta:uud_SvsC}
\end{figure}

\begin{figure}[htbp]
  \begin{center}
    \resizebox{0.46\textwidth}{!}{
      \includegraphics{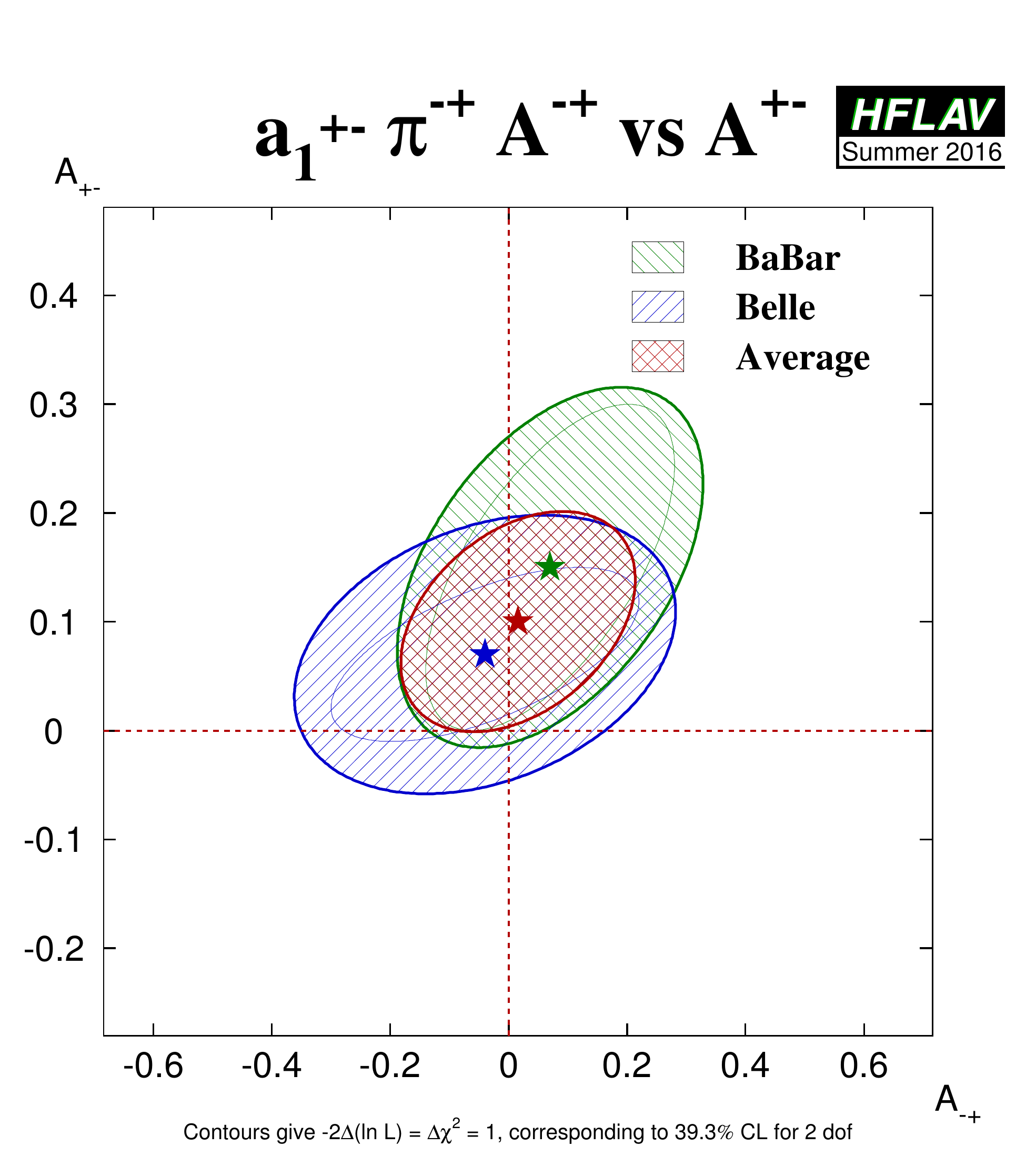}
    }
    \vspace{-0.3cm}
    \caption{
      Averages of \CP violation parameters in $\Bz \to a_1^\pm\pi^\mp$ in
      ${\cal A}^{-+}_{a_1\pi}$ \vs\ ${\cal A}^{+-}_{a_1\pi}$ space.
    }
    \label{fig:cp_uta:a1pi}
  \end{center}
\end{figure}

Results, and averages, of time-dependent \CP violation parameters in 
$b \to u \bar u d$ transitions are listed in Table~\ref{tab:cp_uta:uud}.
The averages for $\pi^+\pi^-$ are shown in Fig.~\ref{fig:cp_uta:uud:pipi},
and those for $\rho^+\rho^-$ are shown in Fig.~\ref{fig:cp_uta:uud:rhorho},
with the averages in the $S_{\CP}$ \vs\ $C_{\CP}$ plane 
shown in Fig.~\ref{fig:cp_uta:uud_SvsC} and
averages of \CP violation parameters in $\Bz \to a_1^\pm \pi^\mp$ decay shown in Fig.~\ref{fig:cp_uta:a1pi}.

% \begin{table}[htb]
\begin{sidewaystable}
	\begin{center}
		\caption{
      Averages for $b \to u \bar u d$ modes.
		}
		\vspace{0.2cm}
		\setlength{\tabcolsep}{0.0pc}
\renewcommand{\arraystretch}{1.1}
		\begin{tabular*}{\textwidth}{@{\extracolsep{\fill}}lrcccc} \hline
	\mc{2}{l}{Experiment} & Sample size & $S_{\CP}$ & $C_{\CP}$ & Correlation \\
	\hline
      \mc{6}{c}{$\pi^{+} \pi^{-}$} \\
	\babar & \cite{Lees:2012mma} & $N(B\bar{B})$ = 467M & $-0.68 \pm 0.10 \pm 0.03$ & $-0.25 \pm 0.08 \pm 0.02$ & $-0.06$ \\
	\belle & \cite{Adachi:2013mae} & $N(B\bar{B})$ = 772M & $-0.64 \pm 0.08 \pm 0.03$ & $-0.33 \pm 0.06 \pm 0.03$ & $-0.10$ \\
%	LHCb & \cite{Aaij:2013tna} & $\int {\cal L}\,dt = 1.0 \, {\rm fb}^{-1}$ & $-0.71 \pm 0.13 \pm 0.02$ & $-0.38 \pm 0.15 \pm 0.02$ & $0.38$ \\
	LHCb & \cite{LHCb-CONF-2016-018} & $\int {\cal L}\,dt = 3.0 \, {\rm fb}^{-1}$ & $-0.68 \pm 0.06 \pm 0.01$ & $-0.24 \pm 0.07 \pm 0.01$ & $0.38$ \\
%	\hline
	\mc{3}{l}{\bf Average} & $-0.68 \pm 0.04$ & $-0.27 \pm 0.04$ & $0.14$ \\
	\mc{3}{l}{\small Confidence level} & \mc{2}{c}{\small $0.88~(0.2\sigma)$} & \\
		\hline
% 		\end{tabular*}
% 		\label{tab:cp_uta:yyy}
% 	\end{center}
% \end{table}

% \begin{table}[htb]
% 	\begin{center}
% 		\caption{
% 			Averages for $\\rho^{+} \\rho^{-}$.
% 		}
% 		\vspace{0.2cm}
% 		\setlength{\tabcolsep}{0.0pc}
% 		\begin{tabular*}{\textwidth}{@{\extracolsep{\fill}}lrcccc} \hline
% 		\mc{2}{l}{Experiment} & $N(B\bar{B})$ & $S_{\CP}$ & $C_{\CP}$ & Correlation \\
% 		\hline
      \mc{6}{c}{$\rho^{+} \rho^{-}$} \\
	\babar & \cite{Aubert:2007nua} & $N(B\bar{B})$ = 387M & $-0.17 \pm 0.20 \,^{+0.05}_{-0.06}$ & $0.01 \pm 0.15 \pm 0.06$ & $-0.04$ \\
	\belle & \cite{Vanhoefer:2015ijw} & $N(B\bar{B})$ = 772M & $-0.13 \pm 0.15 \pm 0.05$ & $0.00 \pm 0.10 \pm 0.06$ & $-0.02$ \\
%	\hline
	\mc{3}{l}{\bf Average} & $-0.14 \pm 0.13$ & $0.00 \pm 0.09$ & $-0.02$ \\
	\mc{3}{l}{\small Confidence level} & \mc{2}{c}{\small $0.99~(0.02\sigma)$} & \\
		\hline
%		\end{tabular*}
% 		\label{tab:cp_uta:yyy}
% 	\end{center}
% \end{table}

% \begin{table}[htb]
% 	\begin{center}
% 		\caption{
% 			Averages for $\\rho^{0} \\rho^{0}$.
% 		}
% 		\vspace{0.2cm}
% 		\setlength{\tabcolsep}{0.0pc}
% 		\begin{tabular*}{\textwidth}{@{\extracolsep{\fill}}lrcccc} \hline
% 	\mc{2}{l}{Experiment} & $N(B\bar{B})$ & $S_{\CP}$ & $C_{\CP}$ & Correlation \\
% 	\hline
      \mc{6}{c}{$\rho^{0} \rho^{0}$} \\
	\babar & \cite{Aubert:2008au} & $N(B\bar{B}) =$ 465M & $0.3 \pm 0.7 \pm 0.2$ & $0.2 \pm 0.8 \pm 0.3$ & $-0.04$ \\
%	\hline
% 	\mc{3}{l}{\bf Average} & $0.50 \pm 0.92$ & $0.40 \pm 0.92$ & $0.00$ \\
% 	\mc{3}{l}{\small Confidence level} & \mc{2}{c}{\small $0.xx~(y.y\sigma)$} & \\
 		\hline
 		\end{tabular*}
% 		\label{tab:cp_uta:yyy}
% 	\end{center}
% \end{table}

                \vspace{2ex}

% \begin{table}[htb]
% 	\begin{center}
% 		\caption{
% 			Averages for $a_{1}^{\pm} \\pi^{\mp}$.
% 		}
% 		\vspace{0.2cm}
% 		\setlength{\tabcolsep}{0.0pc}
% make this tabular (not tabular*) and resize down to \textwidth
% change @{\extracolsep{\fill}} to @{\extracolsep{2mm}}
    \resizebox{\textwidth}{!}{
 		\begin{tabular}{@{\extracolsep{2mm}}lrcccccc} \hline
 		\mc{2}{l}{Experiment} & $N(B\bar{B})$ & $A_{\CP}^{a_1\pi}$ & $C_{a_1\pi}$ & $S_{a_1\pi}$ & $\Delta C_{a_1\pi}$ & $\Delta S_{a_1\pi}$ \\
 		\hline
      \mc{8}{c}{$a_1^{\pm} \pi^{\mp}$} \\
	\babar & \cite{Aubert:2006gb} & 384M & $-0.07 \pm 0.07 \pm 0.02$ & $-0.10 \pm 0.15 \pm 0.09$ & $0.37 \pm 0.21 \pm 0.07$ & $0.26 \pm 0.15 \pm 0.07$ & $-0.14 \pm 0.21 \pm 0.06$ \\
	\belle & \cite{Dalseno:2012hp} & 772M & $-0.06 \pm 0.05 \pm 0.07$ & $-0.01 \pm 0.11 \pm 0.09$ & $-0.51 \pm 0.14 \pm 0.08$ & $0.54 \pm 0.11 \pm 0.07$ & $-0.09 \pm 0.14 \pm 0.06$ \\ 
 	\hline
	\mc{3}{l}{\bf Average} & $-0.06 \pm 0.06$ & $-0.05 \pm 0.11$ & $-0.20 \pm 0.13$ & $0.43 \pm 0.10$ & $-0.10 \pm 0.12$ \\
	\mc{3}{l}{\small Confidence level} & \mc{5}{c}{\small $0.03~(2.1\sigma)$} \\
        \hline
		\end{tabular}
              }
% 		\label{tab:cp_uta:yyy}
% 	\end{center}
% \end{table}

                \vspace{2ex}

% \begin{table}
% 	\begin{center}
% 		\caption{
% 			Averages for $a_{1}^{\pm} \\pi^{\mp}.DCPV$.
% 		}
% 		\vspace{0.2cm}
% 		\setlength{\tabcolsep}{0.0pc}
		\begin{tabular*}{\textwidth}{@{\extracolsep{\fill}}lrcccc} \hline
		\mc{2}{l}{Experiment} & $N(B\bar{B})$ & ${\cal A}^{-+}_{a_1\pi}$ & ${\cal A}^{+-}_{a_1\pi}$ & Correlation \\
		\hline
	\babar & \cite{Aubert:2006gb} & 384M & $0.07 \pm 0.21 \pm 0.15$ & $0.15 \pm 0.15 \pm 0.07$ & 0.63 \\
	\belle & \cite{Dalseno:2012hp} & 772M & $-0.04 \pm 0.26 \pm 0.19$ & $0.07 \pm 0.08 \pm 0.10$ & 0.61 \\
	\mc{3}{l}{\bf Average} & $0.02 \pm 0.20$ & $0.10 \pm 0.10$ & 0.38 \\
        \mc{3}{l}{\small Confidence level} & \mc{2}{c}{\small $0.92~(0.1\sigma)$} \\
		\hline
		\end{tabular*}

		\label{tab:cp_uta:uud}
	\end{center}
\end{sidewaystable}
% \end{table}

If the penguin contribution is negligible, 
the time-dependent parameters for $\Bz \to \pi^+\pi^-$ 
and $\Bz \to \rho^+\rho^-$ are given by
$S_{b \to u\bar u d} = \etacp \sin(2\alpha)$ and
$C_{b \to u\bar u d} = 0$.
In the presence of the penguin contribution, 
$\CP$ violation in decay may arise, 
and there is no straightforward interpretation 
of $S_{b \to u\bar u d}$ and $C_{b \to u\bar u d}$.
An isospin analysis~\cite{Gronau:1990ka} 
can be used to disentangle the contributions and extract $\alpha$.

\begin{figure}[htbp]
  \begin{center}
    \begin{tabular}{cc}
      \resizebox{0.46\textwidth}{!}{
        \includegraphics{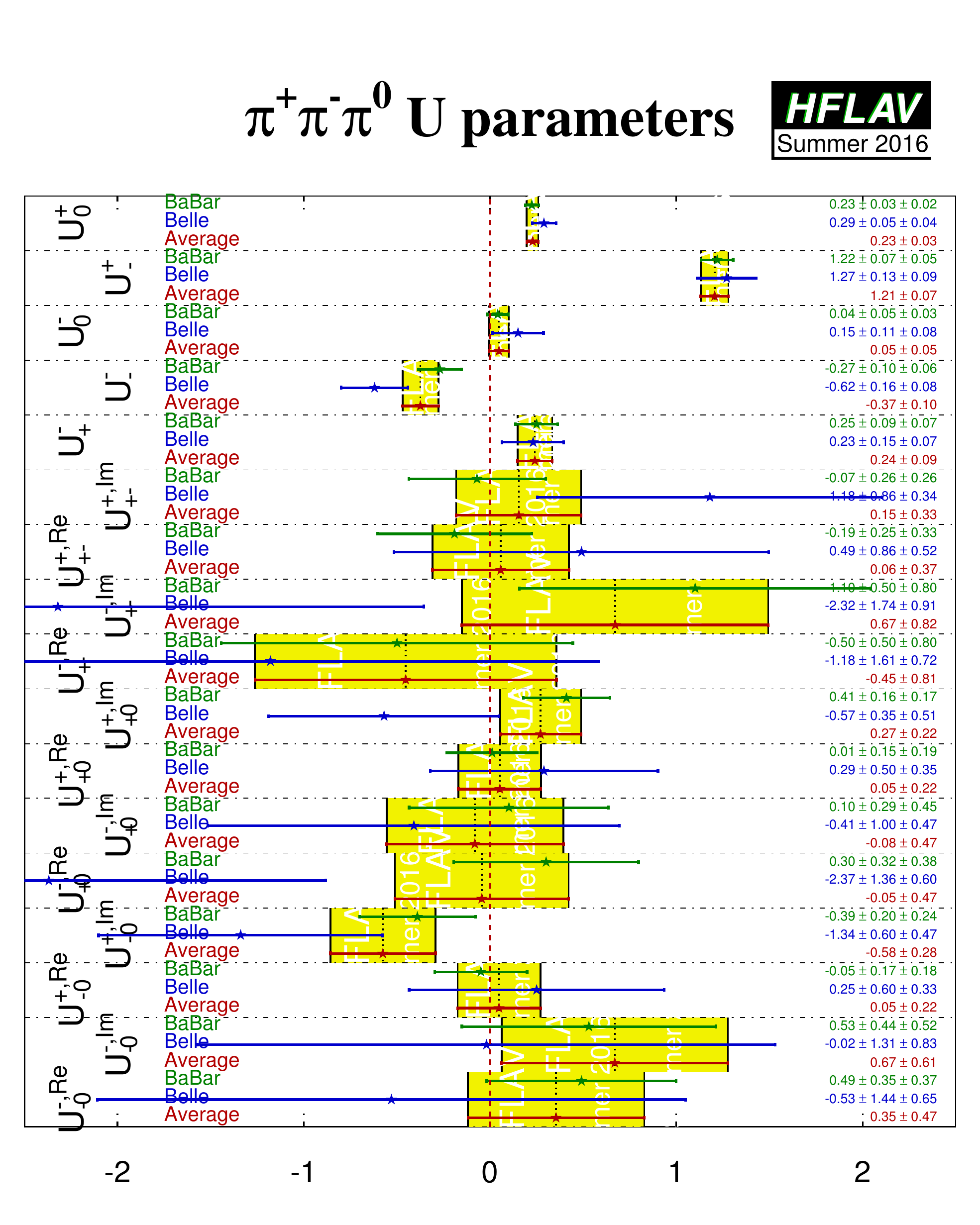}
      }
      &
      \resizebox{0.46\textwidth}{!}{
        \includegraphics{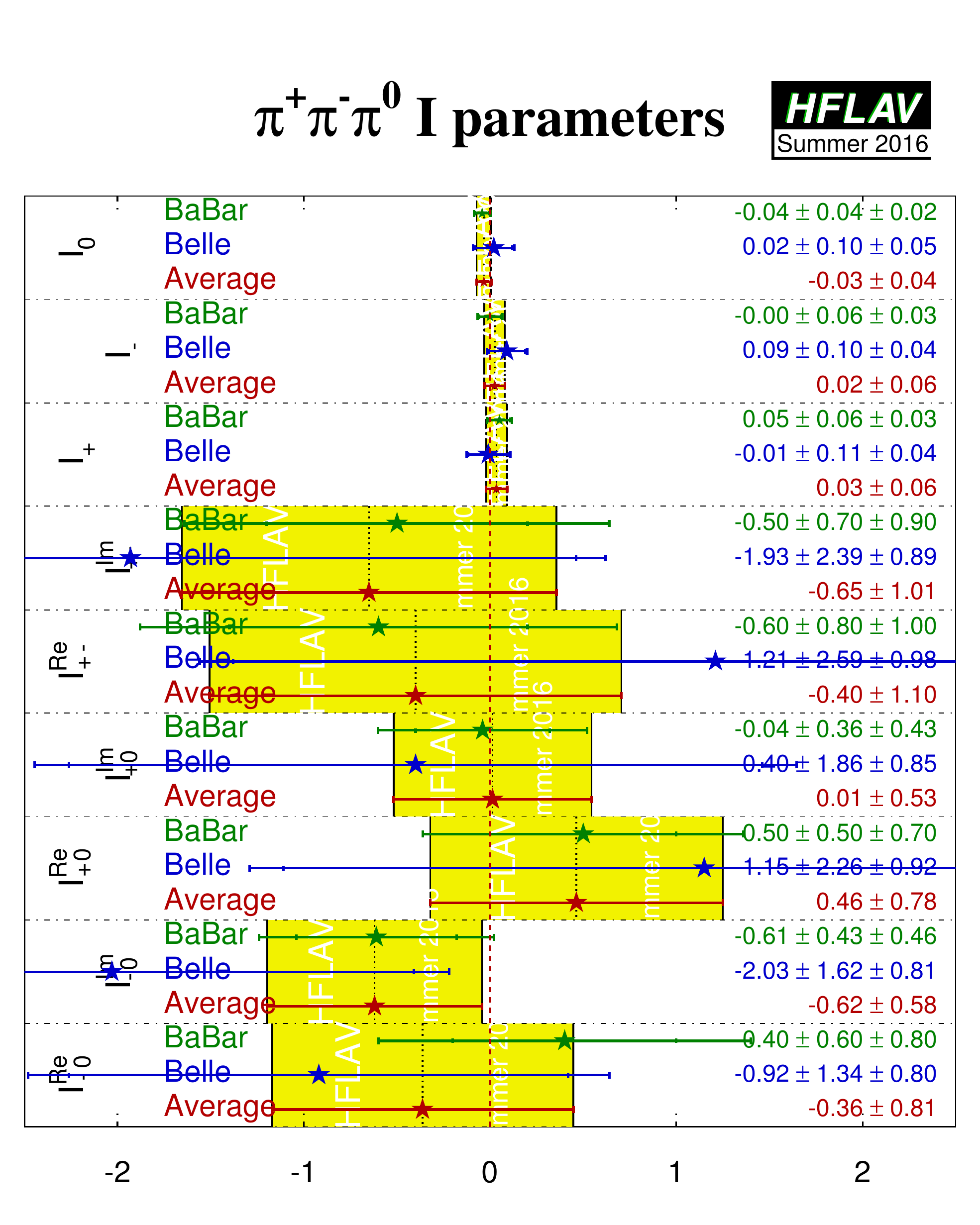}
      }
    \end{tabular}
  \end{center}
  \vspace{-0.8cm}
  \caption{
    Summary of the $U$ and $I$ parameters measured in the 
    time-dependent $\Bz \to \pi^+\pi^-\pi^0$ Dalitz plot analysis.
  }
  \label{fig:cp_uta:uud:uandi}
\end{figure}

For the non-$\CP$ eigenstate $\rho^{\pm}\pi^{\mp}$, 
both \babar~\cite{Aubert:2007jn} 
and \belle~\cite{Kusaka:2007dv,Kusaka:2007mj} have performed 
time-dependent Dalitz plot analyses
of the $\pi^+\pi^-\pi^0$ final state~\cite{Snyder:1993mx};
such analyses allow direct measurements of the phases.
Both experiments have measured the $U$ and $I$ parameters discussed in 
Sec.~\ref{sec:cp_uta:notations:dalitz:pipipi0} and defined in 
Table~\ref{tab:cp_uta:pipipi0:uandi}.
We have performed a full correlated average of these parameters,
the results of which are summarised in Fig.~\ref{fig:cp_uta:uud:uandi}.

\begin{figure}[htbp]
  \begin{center}
    \begin{tabular}{cc}
      \resizebox{0.46\textwidth}{!}{
        \includegraphics{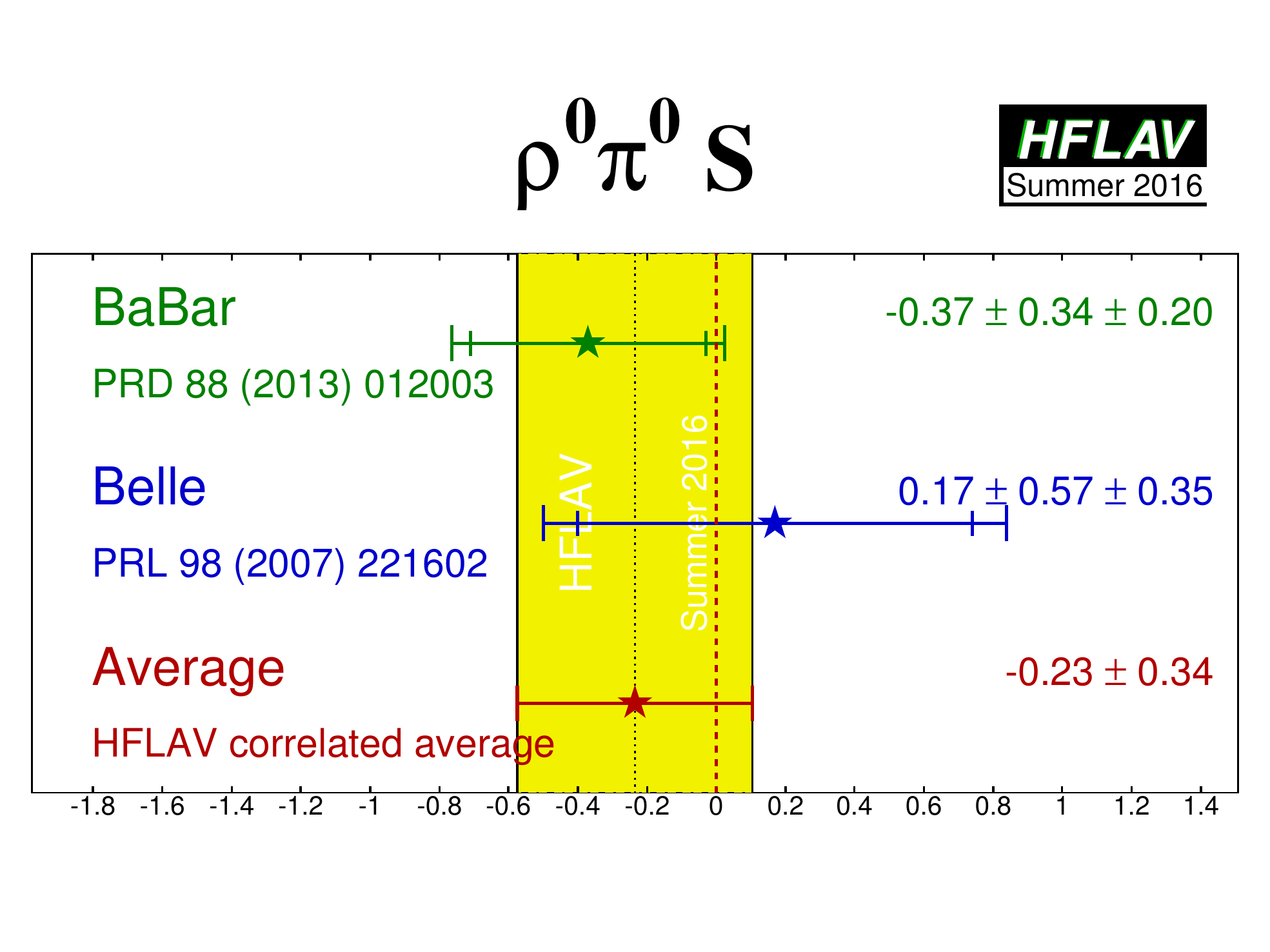}
      }
      &
      \resizebox{0.46\textwidth}{!}{
        \includegraphics{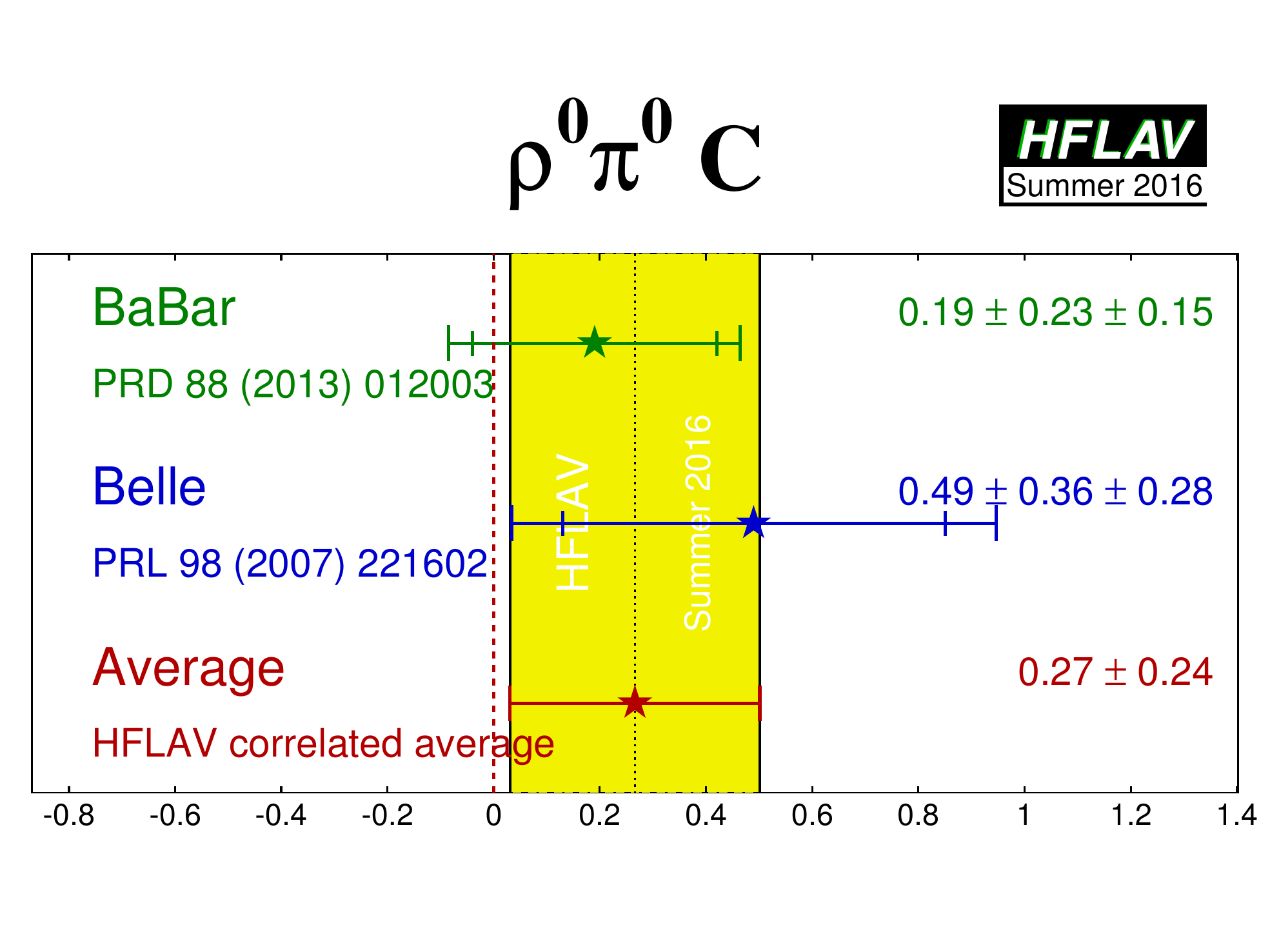}
      }
    \end{tabular}
  \end{center}
  \vspace{-0.8cm}
  \caption{
    Averages of (left) $S_{b \to u\bar u d}$ and (right) $C_{b \to u\bar u d}$
    for the mode $\Bz \to \rho^0\pi^0$.
  }
  \label{fig:cp_uta:uud:rho0pi0}
\end{figure}

\begin{figure}[htbp]
  \begin{center}
    \resizebox{0.46\textwidth}{!}{
      \includegraphics{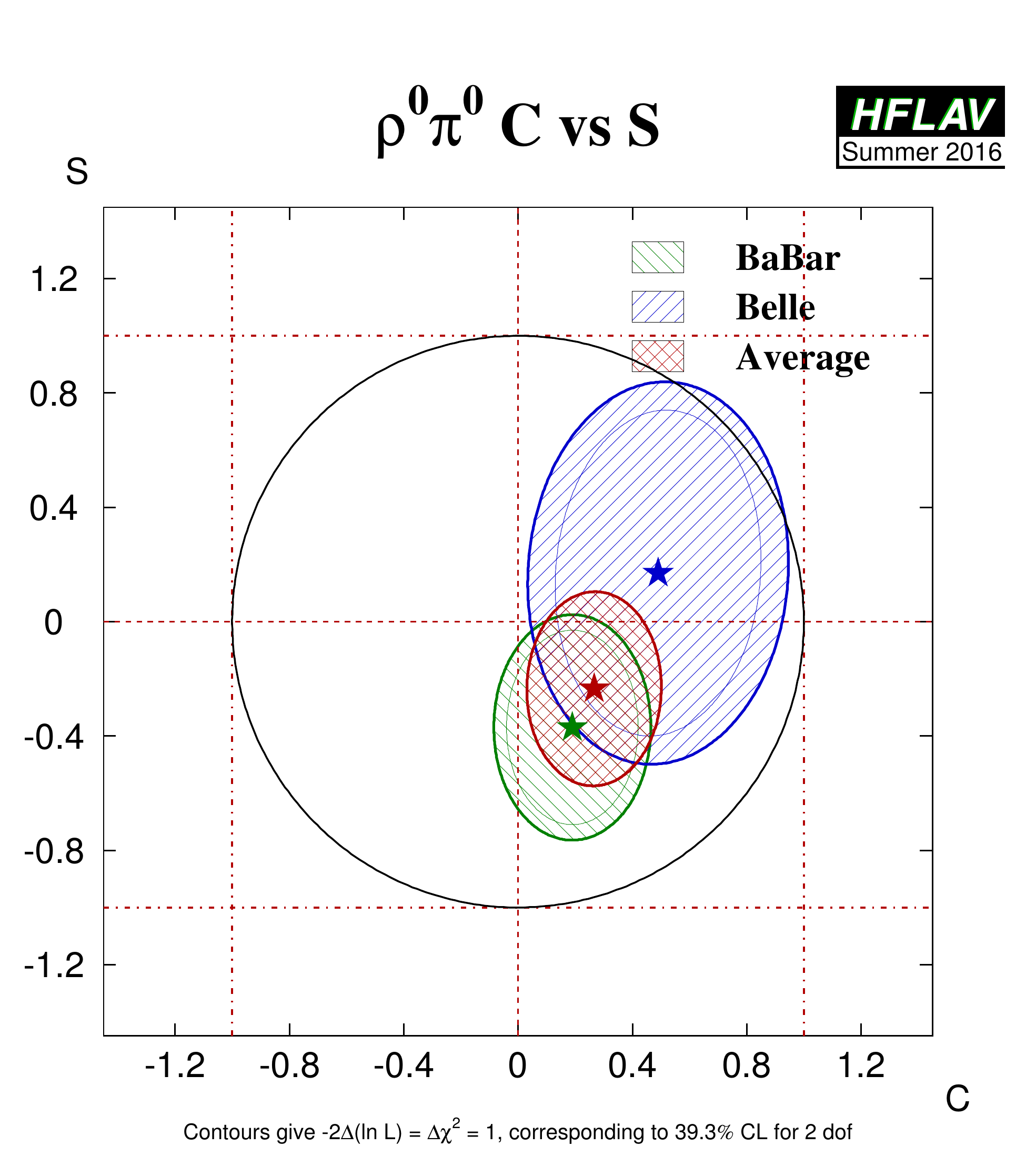}
    }      
  \end{center}
  \vspace{-0.8cm}
  \caption{
    Averages of $b \to u\bar u d$ dominated channels,
    for the mode $\Bz \to \rho^0\pi^0$
    in the $S_{\CP}$ \vs\ $C_{\CP}$ plane.
  }
  \label{fig:cp_uta:uud:rho0pi0_SvsC}
\end{figure}

Both experiments have also extracted the Q2B parameters.
We have performed a full correlated average of these parameters,
which is equivalent to determining the values from the 
averaged $U$ and $I$ parameters.
The results are given in Table~\ref{tab:cp_uta:uud:rhopi_q2b}.\footnote{
  The $\Bz \to \rho^\pm \pi^\mp$ Q2B parameters are comparable to the
  parameters used for $\Bz \to a_1^\pm \pi^\mp$ decays, reported in
  Table~\ref{tab:cp_uta:uud}.   
  For the $\Bz \to a_1^\pm \pi^\mp$ case there has not yet been a full
  amplitude analysis of $\Bz \to \pi^+\pi^-\pi^+\pi^-$ and therefore only the
  Q2B parameters are available.
}
Averages of the $\Bz \to \rho^0\pi^0$ Q2B parameters are shown in 
Figs.~\ref{fig:cp_uta:uud:rho0pi0} and~\ref{fig:cp_uta:uud:rho0pi0_SvsC}.

% \begin{table}[htb]
\begin{sidewaystable}
	\begin{center}
		\caption{
                  Averages of quasi-two-body parameters extracted
                  from time-dependent Dalitz plot analysis of 
                  $\Bz \to \pi^+\pi^-\pi^0$.
%			Averages for $\\rho^{\pm}\\pi^{\mp}$.
		}
		\vspace{0.2cm}
		\setlength{\tabcolsep}{0.0pc}
% make this tabular (not tabular*) and resize down to \textwidth
% change @{\extracolsep{\fill}} to @{\extracolsep{2mm}}
    \resizebox{\textwidth}{!}{
\renewcommand{\arraystretch}{1.1}
		\begin{tabular}{@{\extracolsep{2mm}}lrcccccc} \hline
		\mc{2}{l}{Experiment} & $N(B\bar{B})$ & ${\cal A}_{\CP}^{\rho\pi}$ & $C_{\rho\pi}$ & $S_{\rho\pi}$ & $\Delta C_{\rho\pi}$ & $\Delta S_{\rho\pi}$ \\
	\hline
	\babar & \cite{Lees:2013nwa} & 471M & $-0.10 \pm 0.03 \pm 0.02$ & $0.02 \pm 0.06 \pm 0.04$ & $0.05 \pm 0.08 \pm 0.03$ & $0.23 \pm 0.06 \pm 0.05$ & $0.05 \pm 0.08 \pm 0.04$ \\
	\belle & \cite{Kusaka:2007dv,Kusaka:2007mj} & 449M & $-0.12 \pm 0.05 \pm 0.04$ & $-0.13 \pm 0.09 \pm 0.05$ & $0.06 \pm 0.13 \pm 0.05$ & $0.36 \pm 0.10 \pm 0.05$ & $-0.08 \pm 0.13 \pm 0.05$ \\
%	\hline
	\mc{3}{l}{\bf Average} & $-0.11 \pm 0.03$ & $-0.03 \pm 0.06$ & $0.06 \pm 0.07$ & $0.27 \pm 0.06$ & $0.01 \pm 0.08$ \\
	\mc{3}{l}{\small Confidence level} & \mc{5}{c}{\small $0.63~(0.5\sigma)$} \\
        \hline
		\end{tabular}
              }
% 		\label{tab:cp_uta:yyy}
% 	\end{center}
% \end{table}

                \vspace{2ex}

% \begin{table}
% 	\begin{center}
% 		\caption{
% 			Averages for $\\rho^{\pm}\\pi^{\mp}.DCPV$.
% 		}
% 		\vspace{0.2cm}
% 		\setlength{\tabcolsep}{0.0pc}
		\begin{tabular*}{\textwidth}{@{\extracolsep{\fill}}lrcccc} \hline
		\mc{2}{l}{Experiment} & $N(B\bar{B})$ & ${\cal A}^{-+}_{\rho\pi}$ & ${\cal A}^{+-}_{\rho\pi}$ & Correlation \\
		\hline
	\babar & \cite{Lees:2013nwa} & 471M & $-0.12 \pm 0.08 \,^{+0.04}_{-0.05}$ & $0.09 \,^{+0.05}_{-0.06} \pm 0.04$ & $0.55$ \\
	\belle & \cite{Kusaka:2007dv,Kusaka:2007mj} & 449M & $0.08 \pm 0.16 \pm 0.11$ & $0.21 \pm 0.08 \pm 0.04$ & $0.47$ \\
%	\hline
	\mc{3}{l}{\bf Average} & $-0.08 \pm 0.08$ & $0.13 \pm 0.05$ & $0.37$ \\
        \mc{3}{l}{\small Confidence level} & \mc{2}{c}{\small $0.47~(0.7\sigma)$} \\
		\hline
		\end{tabular*}
% 		\label{tab:cp_uta:yyy}
% 	\end{center}
% \end{table}

                \vspace{2ex}

% \begin{table}
% 	\begin{center}
% 		\caption{
% 			Averages for $\\rho^{0}\\pi^{0}$.
% 		}
% 		\vspace{0.2cm}
% 		\setlength{\tabcolsep}{0.0pc}
		\begin{tabular*}{\textwidth}{@{\extracolsep{\fill}}lrcccc} \hline
		\mc{2}{l}{Experiment} & $N(B\bar{B})$ & $C_{\rho^0\pi^0}$ & $S_{\rho^0\pi^0}$ & Correlation \\
		\hline
	\babar & \cite{Lees:2013nwa} & 471M & $0.19 \pm 0.23 \pm 0.15$ & $-0.37 \pm 0.34 \pm 0.20$ & $0.00$ \\
	\belle & \cite{Kusaka:2007dv,Kusaka:2007mj} & 449M & $0.49 \pm 0.36 \pm 0.28$ & $0.17 \pm 0.57 \pm 0.35$ & $0.08$ \\
%	\hline
	\mc{3}{l}{\bf Average} & $0.27 \pm 0.24$ & $-0.23 \pm 0.34$ & $0.02$ \\
	\mc{3}{l}{\small Confidence level} & \mc{2}{c}{\small $0.68~(0.4\sigma)$} \\
		\hline
		\end{tabular*}
		\label{tab:cp_uta:uud:rhopi_q2b}
	\end{center}
\end{sidewaystable}
% \end{table}

\begin{figure}[htbp]
  \begin{center}
    \resizebox{0.46\textwidth}{!}{
      \includegraphics{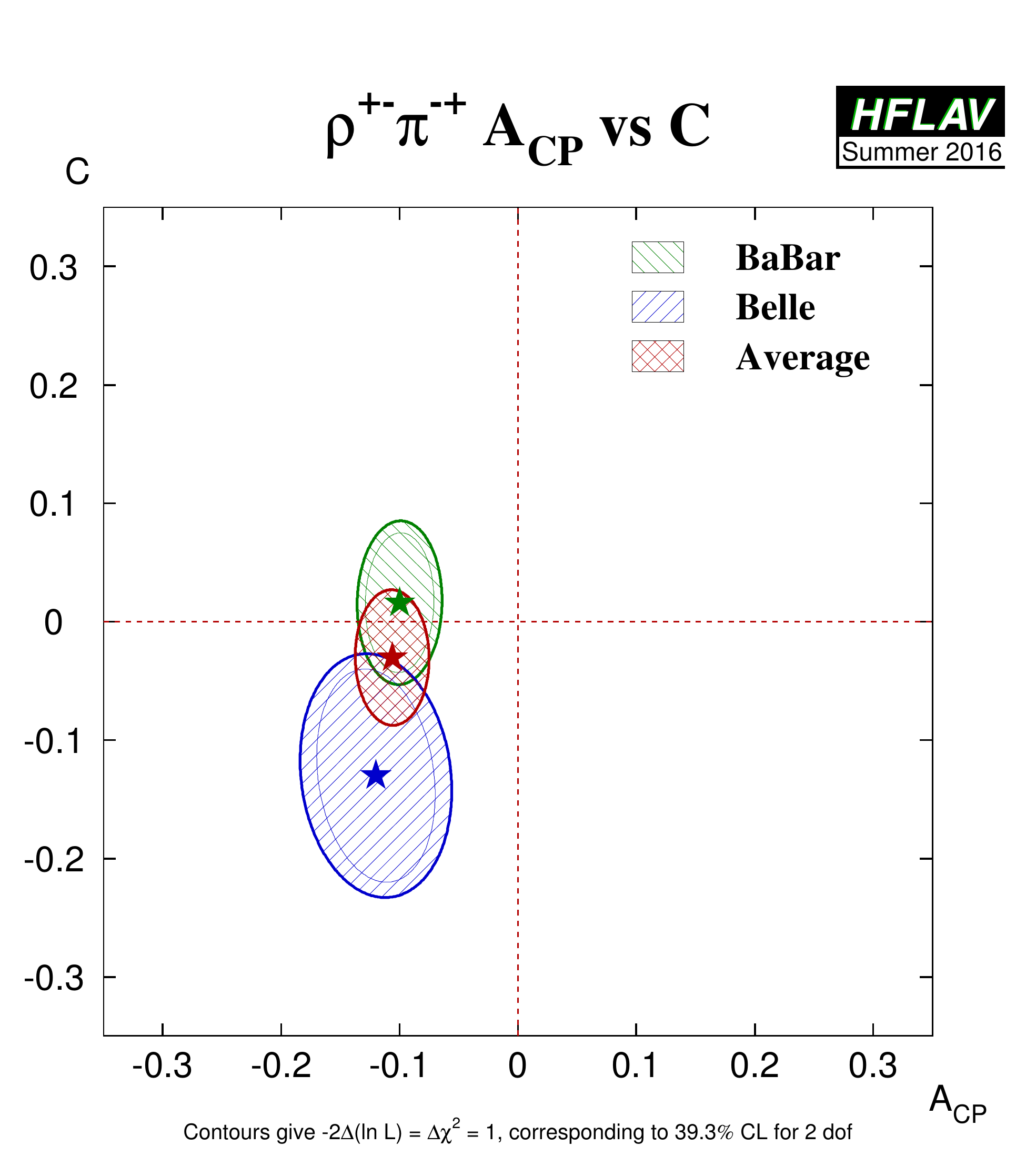}
    }
    \hfill
    \resizebox{0.46\textwidth}{!}{
      \includegraphics{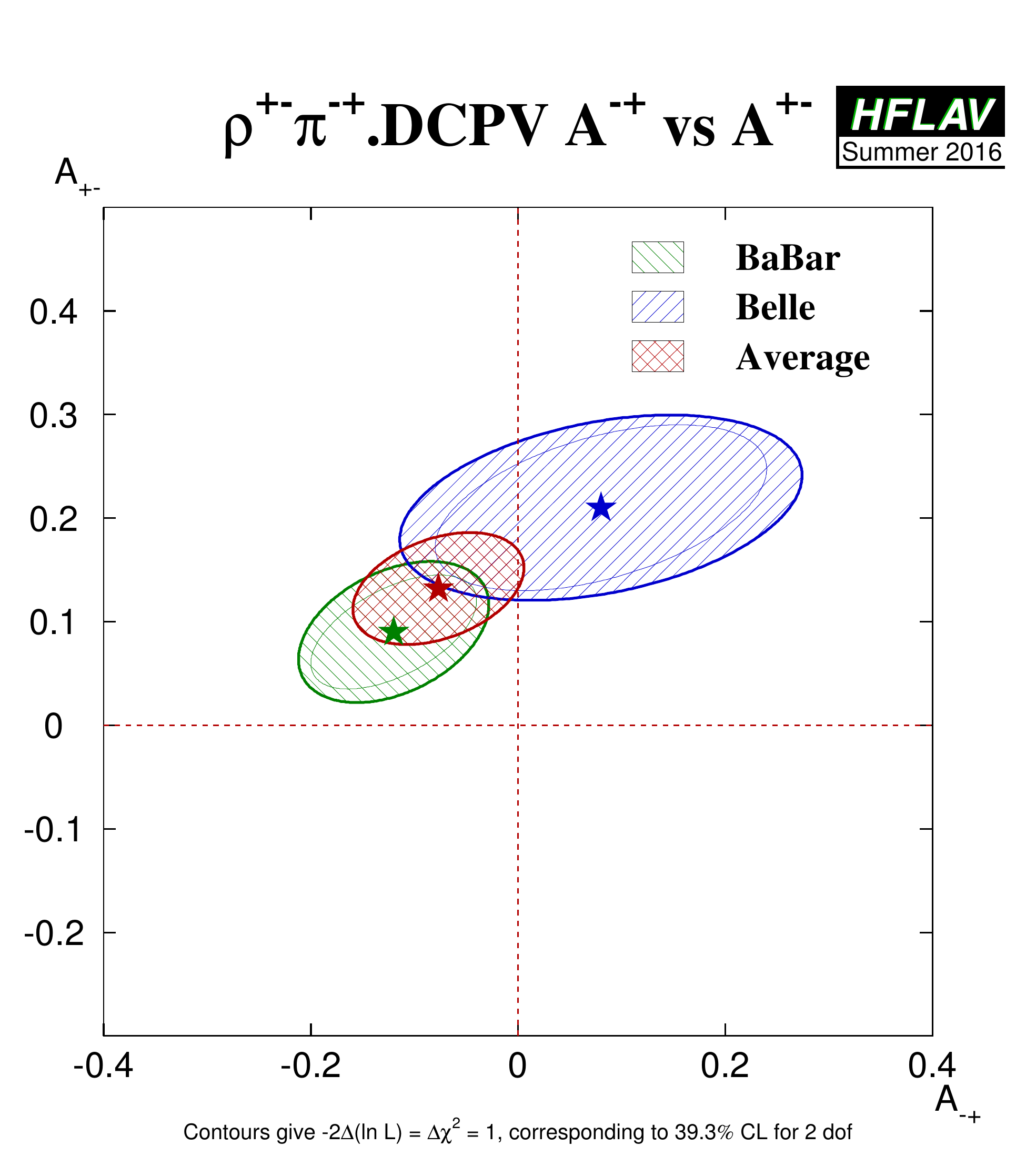}
    }
  \end{center}
  \vspace{-0.8cm}
  \caption{
    $\CP$ violation in $\Bz\to\rho^\pm\pi^\mp$ decays.
    (Left) ${\cal A}^{\rho\pi}_{\CP}$ \vs\ $C_{\rho\pi}$ space,
    (right) ${\cal A}^{-+}_{\rho\pi}$ \vs\ ${\cal A}^{+-}_{\rho\pi}$ space.
  }
  \label{fig:cp_uta:uud:rhopi-dircp}
\end{figure}

With the notation described in Sec.~\ref{sec:cp_uta:notations}
(Eq.~(\ref{eq:cp_uta:non-cp-s_and_deltas})), 
the time-dependent parameters for the Q2B $\Bz \to \rho^\pm\pi^\mp$ analysis are,
neglecting penguin contributions, given by
\begin{equation}
  S_{\rho\pi} = 
  \sqrt{1 - \left(\frac{\Delta C}{2}\right)^2}\sin(2\alpha)\cos(\delta)
  \ , \ \ \ 
  \Delta S_{\rho\pi} = 
  \sqrt{1 - \left(\frac{\Delta C}{2}\right)^2}\cos(2\alpha)\sin(\delta)
\end{equation} 
and $C_{\rho\pi} = {\cal A}_{\CP}^{\rho\pi} = 0$,
where $\delta=\arg(A_{-+}A^*_{+-})$ is the strong phase difference 
between the $\rho^-\pi^+$ and $\rho^+\pi^-$ decay amplitudes.
In the presence of the penguin contribution, there is no straightforward 
interpretation of the Q2B observables in the $\Bz \to \rho^\pm\pi^\mp$ system
in terms of CKM parameters.
However, $\CP$ violation in decay may arise,
resulting in either or both of $C_{\rho\pi} \neq 0$ and ${\cal A}_{\CP}^{\rho\pi} \neq 0$.
Equivalently,
$\CP$ violation in decay may be seen by either of
the decay-type-specific observables ${\cal A}^{+-}_{\rho\pi}$ 
and ${\cal A}^{-+}_{\rho\pi}$, defined in Eq.~(\ref{eq:cp_uta:non-cp-directcp}), 
deviating from zero.
Results and averages for these parameters
are also given in Table~\ref{tab:cp_uta:uud:rhopi_q2b}.
Averages of $\CP$ violation parameters in $\Bz \to \rho^\pm\pi^\mp$ decays
are shown in Fig.~\ref{fig:cp_uta:uud:rhopi-dircp},
both in 
${\cal A}^{\rho\pi}_{\CP}$ \vs\ $C_{\rho\pi}$ space and in 
${\cal A}^{-+}_{\rho\pi}$ \vs\ ${\cal A}^{+-}_{\rho\pi}$ space.

%% straightforward interpretation
The averages for $S_{b \to u\bar u d}$ and $C_{b \to u\bar u d}$ 
in $\Bz \to \pi^+\pi^-$ decays are both more than $5\sigma$ away from zero,
suggesting that both mixing-induced and $\CP$ violation in decay
are well-established in this channel.
The discrepancy between results from \babar\ and Belle that used to exist in
this channel (see, for example, Ref.~\cite{Asner:2010qj}) is no longer
apparent, and the results from LHCb are also fully consistent with other
measurements.  
Some difference is, however, seen between the \babar\ and \belle\ measurements
in the $a_1^\pm\pi^\mp$ system. 
The confidence level of the five-dimensional average is $0.03$,
which corresponds to a $2.1\sigma$ discrepancy.  
As seen in Table~\ref{tab:cp_uta:uud}, this discrepancy is primarily in the
values of $S_{a_1\pi}$, and is not evident in the ${\cal A}^{-+}_{a_1\pi}$
\vs\ ${\cal A}^{+-}_{a_1\pi}$ projection shown in Fig.~\ref{fig:cp_uta:a1pi}.
Since there is no
evidence of systematic problems in either analysis,
we do not rescale the errors of the averages.
% Nonetheless, due to the possible discrepancy mentioned above,
% a slightly cautious interpretation should be made 
% with regard to the significance of $\CP$ violation in decay.

In $\Bz \to \rho^\pm\pi^\mp$ decays,
both experiments see an indication of $\CP$ violation in the 
${\cal A}^{\rho\pi}_{\CP}$ parameter 
(as seen in Fig.~\ref{fig:cp_uta:uud:rhopi-dircp}).
The average is more than $3\sigma$ from zero,
providing evidence of direct $\CP$ violation in this channel.
In $\Bz \to \rho^+\rho^-$ decays there is no evidence for $\CP$ violation,
either mixing-induced or in decay.
The absence of evidence of penguin contributions in this mode leads to
strong constraints on $\alpha \equiv \phi_2$.

\mysubsubsection{Constraints on $\alpha \equiv \phi_2$}
\label{sec:cp_uta:uud:alpha}

The precision of the measured $\CP$ violation parameters in
$b \to u\bar{u}d$ transitions allows 
constraints to be set on the UT angle $\alpha \equiv \phi_2$. 
Constraints have been obtained with various methods:
\begin{itemize}\setlength{\itemsep}{0.5ex}
\item 
  Both \babar~\cite{Lees:2012mma}
  and  \belle~\cite{Adachi:2013mae} have performed 
  isospin analyses in the $\pi\pi$ system.
  \belle\ exclude $23.8^\circ < \phi_2 < 66.8^\circ$ at 68\% CL while
  \babar\ give a confidence level interpretation for $\alpha$, and constrain
  $\alpha \in \left[ 71^\circ, 109^\circ \right]$ at 68\% CL.
  % considering only the solution consistent with the Standard Model.
  Values in the range $\left[ 23^\circ, 67^\circ \right]$ are excluded at 90\% CL.
  In both cases, only solutions in $0^\circ$--$180^\circ$ are quoted.

\item
  Both experiments have also performed isospin analyses in the $\rho\rho$
  system. 
  The most recent result from \babar\ is given in an update of the
  measurements of the $B^+\to\rho^+\rho^0$ decay~\cite{Aubert:2009it}, and
  sets the constraint $\alpha = \left( 92.4 \,^{+6.0}_{-6.5}\right)^\circ$.
  The most recent result from \belle\ is given in their paper on time-dependent \CP violation parameters in $\Bz \to \rho^+\rho^-$ decays, and sets the constraint
  $\phi_2 = \left( 93.7 \pm 10.6 \right)^\circ$~\cite{Vanhoefer:2015ijw}.

\item
  The time-dependent Dalitz plot analysis of the $\Bz \to \pi^+\pi^-\pi^0$
  decay allows a determination of $\alpha$ without input from any other 
  channels.
  \babar~\cite{Lees:2013nwa} present a scan, but not an interval, for $\alpha$, since
  their studies indicate that the scan is not statistically robust and cannot
  be interpreted as 1$-$CL.
  \belle~\cite{Kusaka:2007dv,Kusaka:2007mj} has obtained a constraint on $\alpha$
  using additional information from the SU(2) partners of 
  $B \to \rho\pi$, which can be used to constrain $\alpha$
  via an isospin pentagon relation~\cite{Lipkin:1991st}. 
  With this analysis,
  \belle\ obtains the constraint $\phi_2 = (83 \, ^{+12}_{-23})^\circ$
  (where the errors correspond to $1\sigma$, \ie\ $68.3\%$ confidence level).

\item 
  The results from \babar\ on $\Bz \to a_1^\pm \pi^\mp$~\cite{Aubert:2006gb} can be
  combined with results from modes related by flavour symmetries ($a_1K$ and $K_1\pi$)~\cite{Gronau:2005kw}.
  This has been done by \babar~\cite{Aubert:2009ab}, resulting in the constraint
  $\alpha = \left( 79 \pm 7 \pm 11 \right)^\circ$, where the first uncertainty is from the analysis of $\Bz \to a_1^\pm \pi^\mp$ that obtains $\alpha^{\rm eff}$, and the second is due to the constraint on $\left| \alpha^{\rm eff} - \alpha \right|$.
  This approach gives a result with several ambiguous solutions; that consistent with other determinations of $\alpha$ and with global fits to the CKM matrix parameters is quoted here.

% \item 
%   Each experiment has obtained a value of $\alpha$ from combining its 
%   results in the different $b \to u \bar{u} d$ modes 
%   (with some input also from HFLAV).
%   These values have appeared in talks, but not in publications,
%   and are not listed here.

\item 
  The CKMfitter~\cite{Charles:2004jd} and 
  UTFit~\cite{Bona:2005vz} groups use the measurements 
  from \belle\ and \babar\ given above
  with other branching fractions and \CP asymmetries in 
  $\B\to\pi\pi$, $\rho\pi$ and $\rho\rho$ modes, 
  to perform isospin analyses for each system, 
  and to obtain combined constraints on $\alpha$.

\item
  The \babar\ and \belle\ collaborations have combined their results on $B \to \pi\pi$, $\pi\pi\pi^0$ and $\rho\rho$ decays to obtain~\cite{Bevan:2014iga}
  \begin{equation}
    \alpha \equiv \phi_2 = (88 \pm 5)^\circ \, .
  \end{equation}
  The above solution is that consistent with the Standard Model (an ambiguous solution shifted by $180^\circ$ exists). The strongest constraint currently comes from the $B \to \rho\rho$ system. The inclusion of results from $\Bz \to a_1^\pm \pi^\mp$ does not significantly affect the average. 

\end{itemize}

Note that methods based on isospin symmetry make extensive use of 
measurements of branching fractions and $\CP$ asymmetries,
as averaged by the HFLAV Rare Decays subgroup (Sec.~\ref{sec:rare}).
Note also that each method suffers from discrete ambiguities in the solutions.
The model assumption in the $\Bz \to \pi^+\pi^-\pi^0$ analysis 
helps resolve some of the multiple solutions, 
and results in a single preferred value for $\alpha$ in $\left[ 0, \pi \right]$.
All the above measurements correspond to the choice
that is in agreement with the global CKM fit.

At present we make no attempt to provide an HFLAV average for $\alpha \equiv \phi_2$.
More details on procedures to calculate a best fit value for $\alpha$ 
can be found in Refs.~\cite{Charles:2004jd,Bona:2005vz}.

%%%%%%%%
%%%
%%% b -> cud
%%%
%%%%%%%%
% \afterpage{\clearpage}
\mysubsection{Time-dependent $\CP$ asymmetries in $b \to c\bar{u}d / u\bar{c}d$ transitions
}
\label{sec:cp_uta:cud}

Non-$\CP$ eigenstates such as $D^\mp\pi^\pm$, $D^{*\mp}\pi^\pm$ and $D^\mp\rho^\pm$ can be produced 
in decays of $\Bz$ mesons either via Cabibbo favoured ($b \to c$) or
doubly Cabibbo suppressed ($b \to u$) tree amplitudes. 
Since no penguin contribution is possible,
these modes are theoretically clean.
The ratio of the magnitudes of the suppressed and favoured amplitudes, $R$,
is sufficiently small (predicted to be about $0.02$), that ${\cal O}(R^2)$ terms can be neglected, 
and the sine terms give sensitivity to the combination of UT angles $2\beta+\gamma$.

\begin{figure}[htbp]
  \begin{center}
    \begin{tabular}{cc}
      \resizebox{0.46\textwidth}{!}{
        \includegraphics{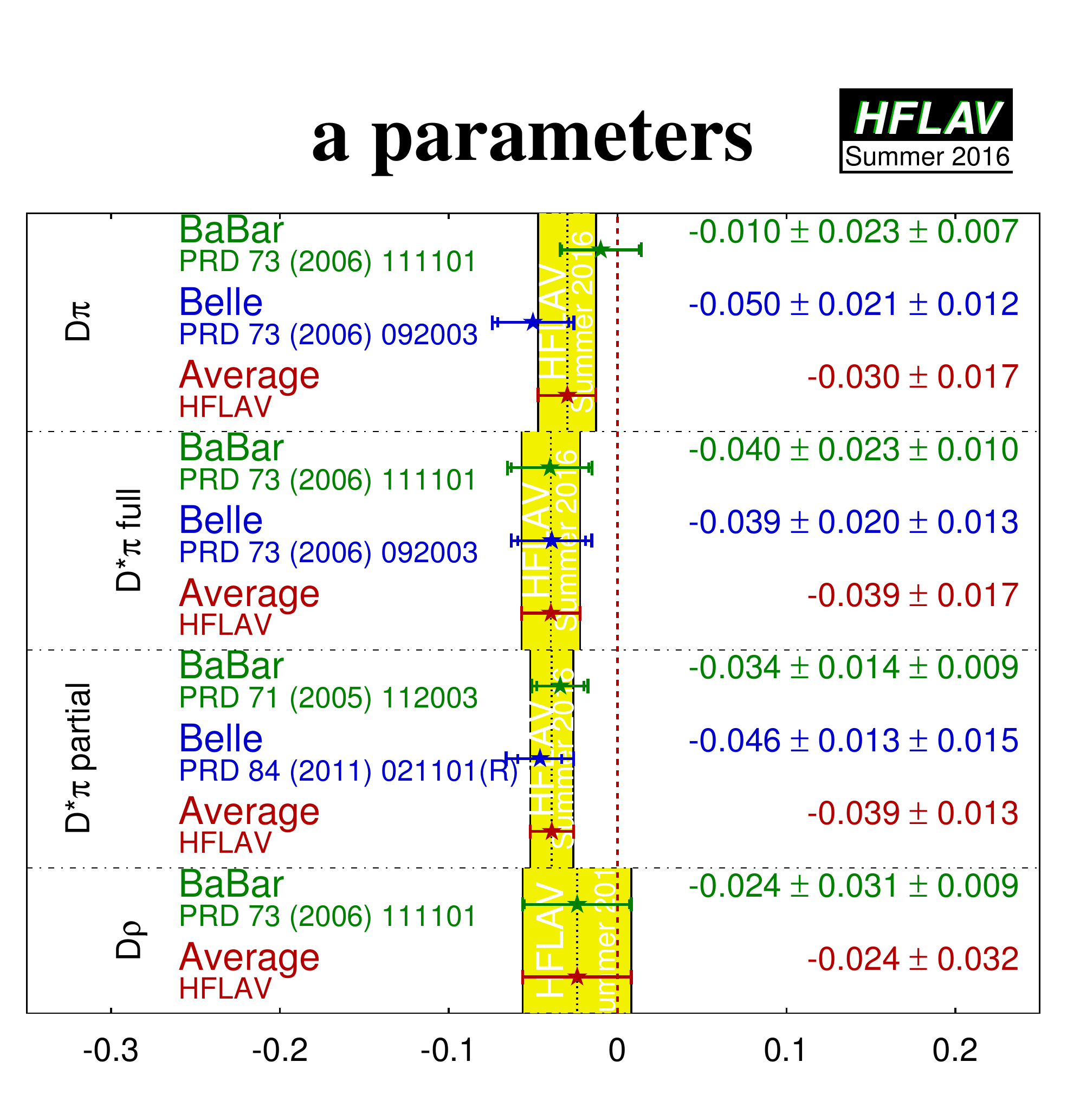}
      }
      &
      \resizebox{0.46\textwidth}{!}{
        \includegraphics{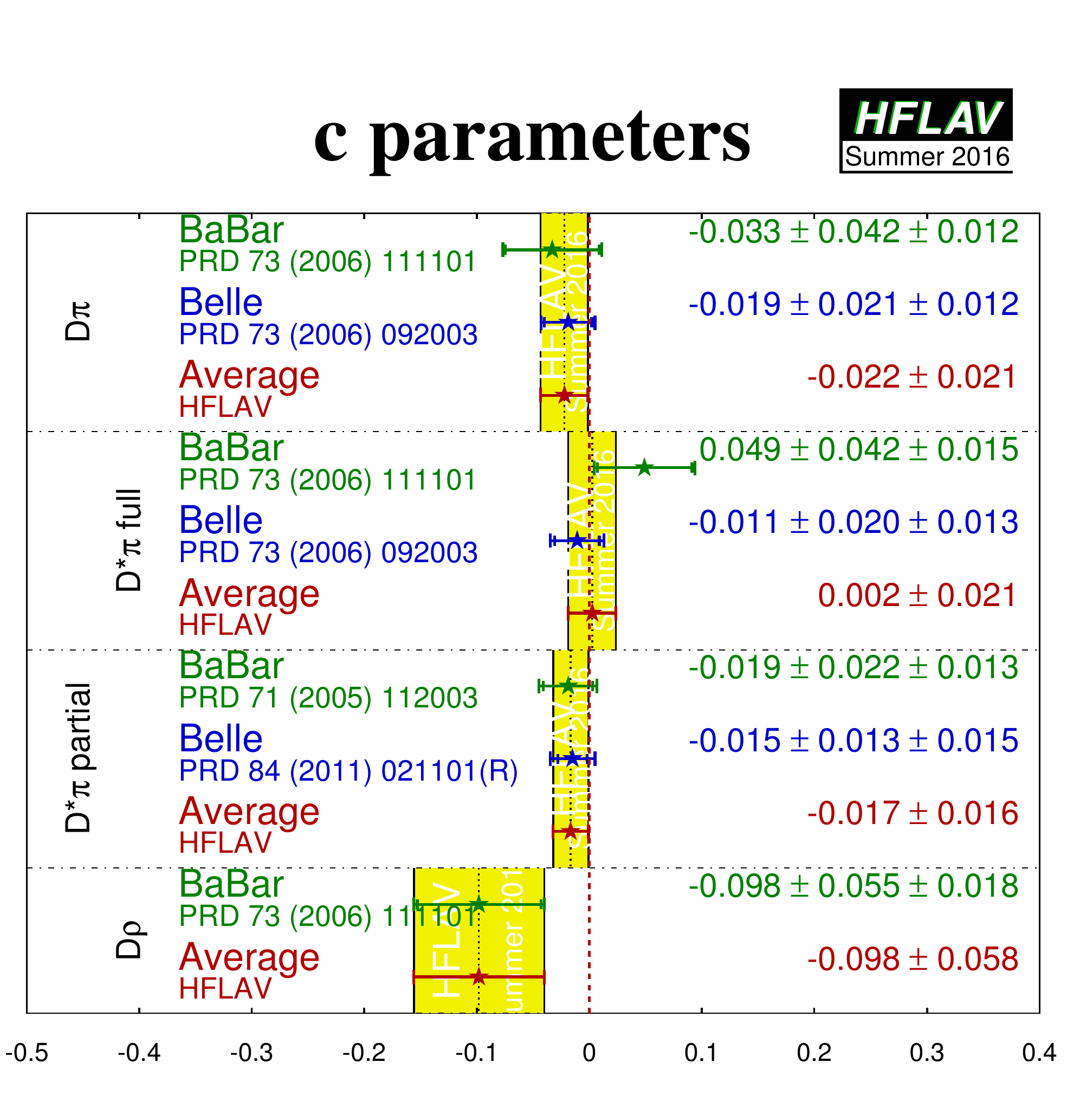}
      }
    \end{tabular}
  \end{center}
  \vspace{-0.8cm}
  \caption{
    Averages for $b \to c\bar{u}d / u\bar{c}d$ modes.
  }
  \label{fig:cp_uta:cud}
\end{figure}

\begin{figure}[htbp]
  \begin{center}
    \begin{tabular}{cc}
      \resizebox{0.46\textwidth}{!}{
        \includegraphics{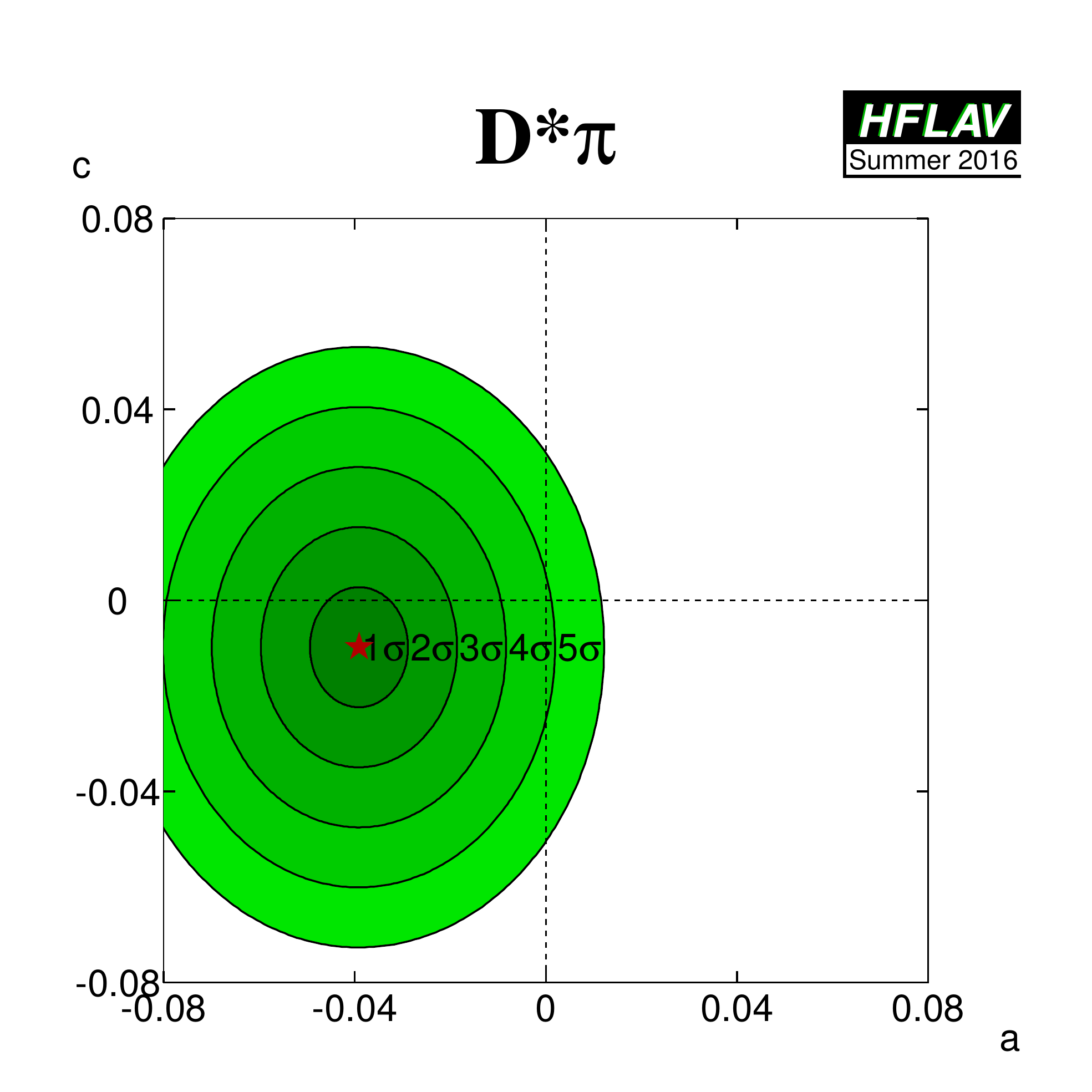}
      }
      &
      \resizebox{0.46\textwidth}{!}{
        \includegraphics{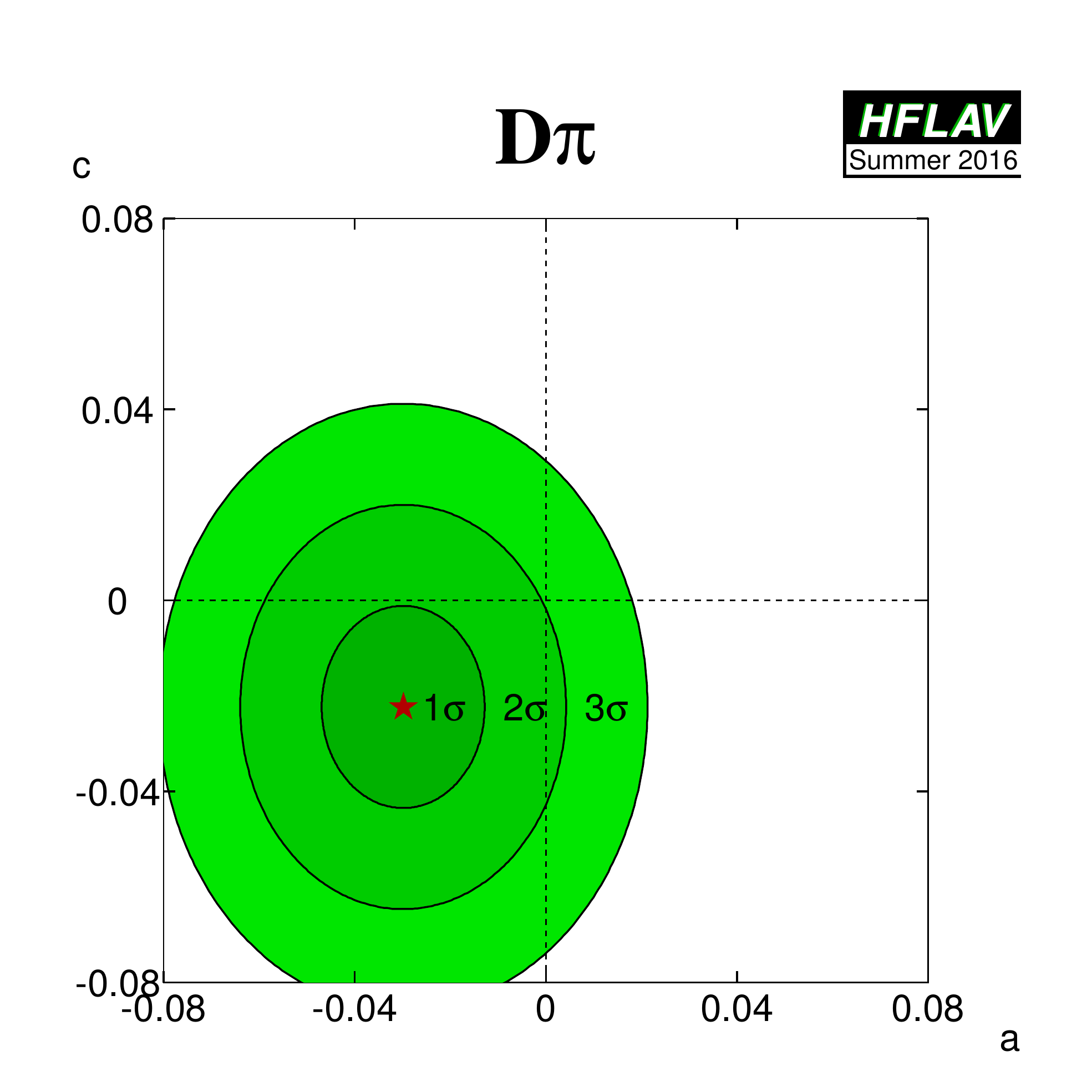}
      } \\
      \resizebox{0.46\textwidth}{!}{
        \includegraphics{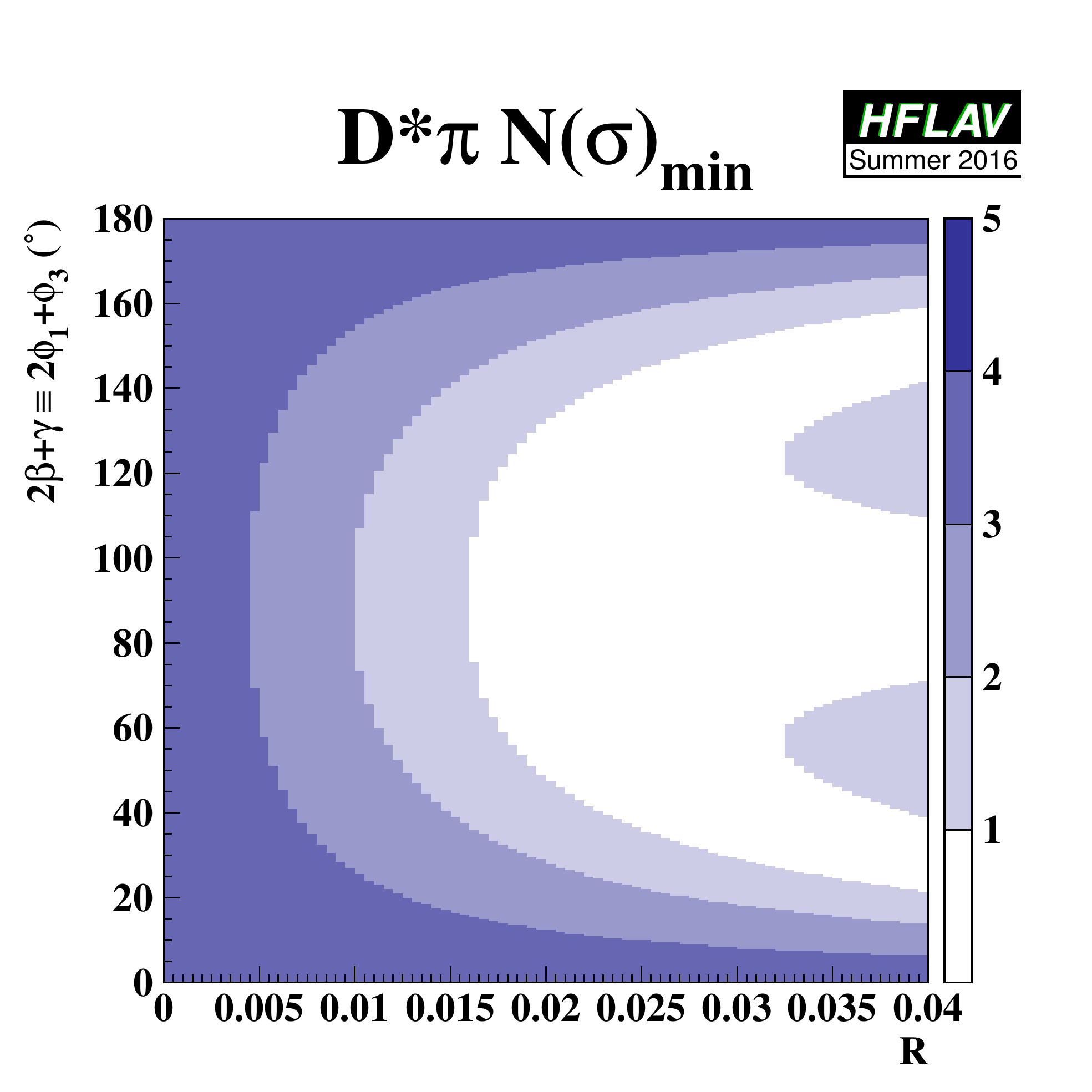}
      }
      &
      \resizebox{0.46\textwidth}{!}{
        \includegraphics{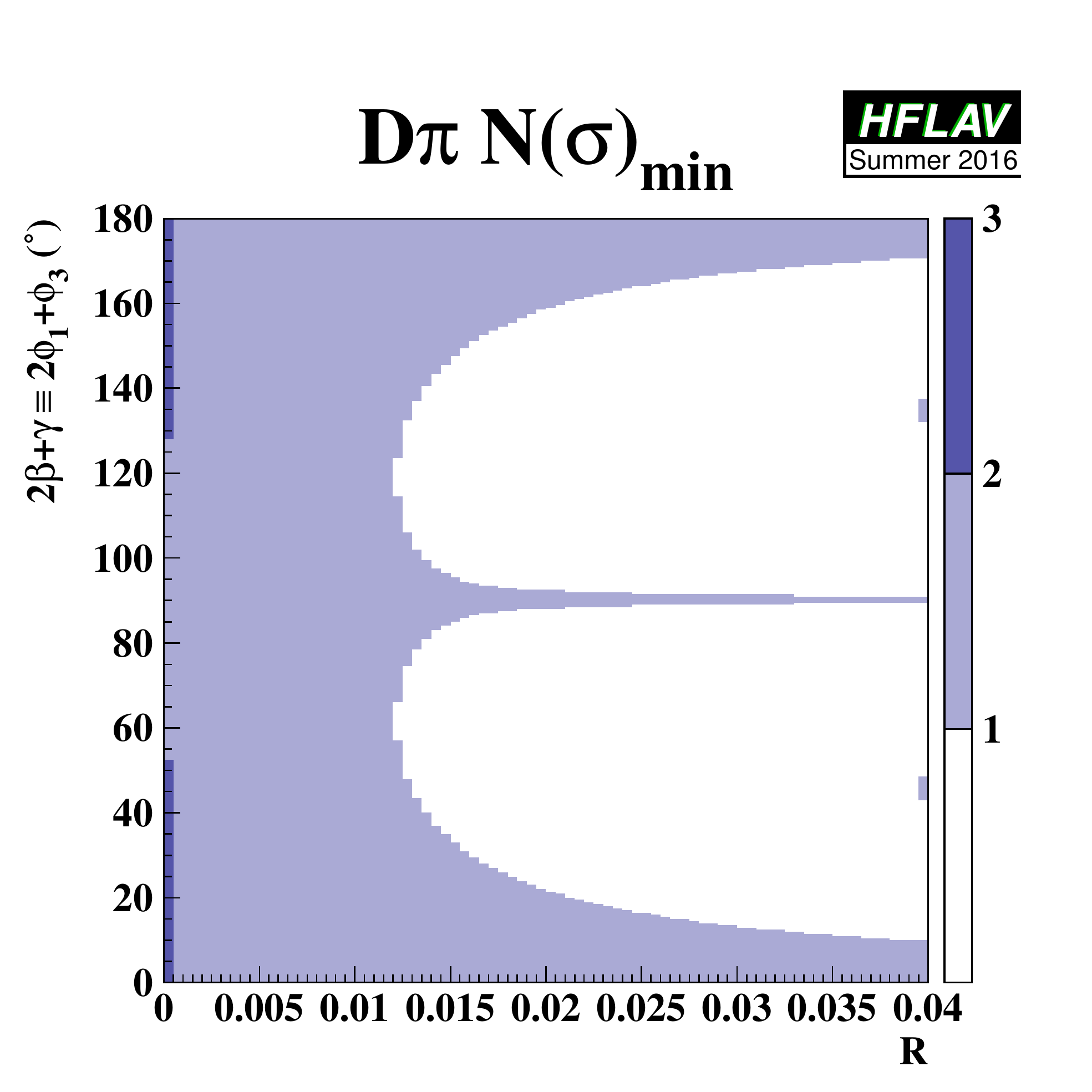}
      }          
    \end{tabular}
  \end{center}
  \vspace{-0.8cm}
  \caption{
    Results from $b \to c\bar{u}d / u\bar{c}d$ modes.
    (Top)~Constraints in $c$ {\it vs.}\ $a$ space.
    (Bottom)~Constraints in $2\beta+\gamma$ {\it vs.}\ $R$ space.
    (Left)~$D^*\pi$ and (right)~$D\pi$ modes.
  }
  \label{fig:cp_uta:cud_constraints}
\end{figure}

As described in Sec.~\ref{sec:cp_uta:notations:non_cp:dstarpi},
the averages are given in terms of the parameters $a$ and $c$ of Eq.~(\ref{eq:cp_uta:aandc}).
$\CP$ violation would appear as $a \neq 0$.
Results are available from both \babar\ and \belle\ in the modes
$D^\mp\pi^\pm$ and $D^{*\mp}\pi^\pm$; for the latter mode both experiments 
have used both full and partial reconstruction techniques.
% (\babar\ has provided separate results with each technique,
% while \belle\ has in addition provided a combined result.)
Results are also available from \babar\ using $D^\mp\rho^\pm$.
These results, and their averages, are listed in Table~\ref{tab:cp_uta:cud},
and are shown in Fig.~\ref{fig:cp_uta:cud}.
The constraints in $c$ \vs\ $a$ space for the $D\pi$ and $D^*\pi$ modes
are shown in Fig.~\ref{fig:cp_uta:cud_constraints}.
It is notable that the average value of $a$ from $D^*\pi$ is more than
$3\sigma$ from zero, providing evidence of $\CP$ violation in this channel.

\begin{table}[htb]
	\begin{center}
		\caption{
      Averages for $b \to c\bar{u}d / u\bar{c}d$ modes.
%       Note that the ``\belle (combined)'' result for $D^{*\pm}\pi^{\mp}$
%       is a combination of the 
%       ``\belle (full rec.)'' and ``\belle (partial rec.)'' results.
                }
                \vspace{0.2cm}
                \setlength{\tabcolsep}{0.0pc}
\renewcommand{\arraystretch}{1.1}
                \begin{tabular*}{\textwidth}{@{\extracolsep{\fill}}lrccc} \hline 
	\mc{2}{l}{Experiment} & $N(B\bar{B})$ & $a$ & $c$ \\
	\hline
      \mc{5}{c}{$D^{\mp}\pi^{\pm}$} \\
	\babar (full rec.) & \cite{Aubert:2006tw} & 232M & $-0.010 \pm 0.023 \pm 0.007$ & $-0.033 \pm 0.042 \pm 0.012$ \\
	\belle (full rec.) & \cite{Ronga:2006hv} & 386M & $-0.050 \pm 0.021 \pm 0.012$ & $-0.019 \pm 0.021 \pm 0.012$ \\
%	\hline
        \mc{3}{l}{\bf Average} & $ -0.030 \pm 0.017$ & $ -0.022 \pm 0.021 $ \\
        \mc{3}{l}{\small Confidence level} & {\small $0.24~(1.2\sigma)$} & {\small $0.78~(0.3\sigma)$} \\
        \hline
      \mc{5}{c}{$D^{*\mp}\pi^{\pm}$} \\
      \babar (full rec.) & \cite{Aubert:2006tw} & 232M & $-0.040 \pm 0.023 \pm 0.010$ & $0.049 \pm 0.042 \pm 0.015$ \\
      \babar (partial rec.)  & \cite{Aubert:2005yf} & 232M & $-0.034 \pm 0.014 \pm 0.009$ & $-0.019 \pm 0.022 \pm 0.013$ \\
      \belle (full rec.) & \cite{Ronga:2006hv} & 386M & $-0.039 \pm 0.020 \pm 0.013$ & $-0.011 \pm 0.020 \pm 0.013$ \\
      \belle (partial rec.) & \cite{Bahinipati:2011yq} & 657M & $-0.046 \pm 0.013 \pm 0.015$ & $-0.015 \pm 0.013 \pm 0.015$ \\
%      {\small \belle (combined)} & {\small \cite{Ronga:2006hv}} & {\small 386M} & {\small $-0.040 \pm 0.014 \pm 0.011$} & {\small $-0.009 \pm 0.014 \pm 0.011$} \\
	\mc{3}{l}{\bf Average} & $-0.039 \pm 0.010$ & $-0.010 \pm 0.013$ \\
      \mc{3}{l}{\small Confidence level} & {\small $0.97~(0.03\sigma)$} & {\small $0.59~(0.6\sigma)$} \\
      \hline
      \mc{5}{c}{$D^{\mp}\rho^{\pm}$} \\
      \babar (full rec.) & \cite{Aubert:2006tw} & 232M & $-0.024 \pm 0.031 \pm 0.009$ & $-0.098 \pm 0.055 \pm 0.018$ \\
%       \mc{3}{l}{\bf Average} & $ -0.024 \pm 0.033 $ & $ -0.098 \pm 0.058$ \\
      \hline 
    \end{tabular*}
    \label{tab:cp_uta:cud}
  \end{center}
\end{table}

For each mode, $D\pi$, $D^*\pi$ and $D\rho$, 
there are two measurements ($a$ and $c$, or $S^+$ and $S^-$) 
that depend on three unknowns ($R$, $\delta$ and $2\beta+\gamma$), 
of which two are different for each decay mode. 
Therefore, there is not enough information to solve directly for $2\beta+\gamma$. 
However, for each choice of $R$ and $2\beta+\gamma$, 
one can find the value of $\delta$ that allows $a$ and $c$ to be closest 
to their measured values, 
and calculate the separation in terms of numbers of standard deviations.
(We currently neglect experimental correlations in this analysis.) 
These values of $N(\sigma)_{\rm min}$ can then be displayed
as a function of $R$ and $2\beta+\gamma$
(and can trivially be converted to confidence levels). 
These plots are given for the $D\pi$ and $D^*\pi$ modes 
in Fig.~\ref{fig:cp_uta:cud_constraints}; 
the uncertainties in the $D\rho$ mode are currently too large 
to give any meaningful constraint.

The constraints can be tightened if one is willing 
to use theoretical input on the values of $R$ and/or $\delta$. 
One popular choice is the use of SU(3) symmetry to obtain 
$R$ by relating the suppressed decay mode to $\B$ decays 
involving $D_s$ mesons. 
More details can be found in Refs.~\cite{Dunietz:1997in,Fleischer:2003yb,Baak:2007gp,DeBruyn:2012jp,Kenzie:2016yee}.

%%%%%%%%
%%%
%%% b -> cus
%%%
%%%%%%%%
% \afterpage{\clearpage}
\mysubsection{Time-dependent $\CP$ asymmetries in $b \to c\bar{u}s / u\bar{c}s$ transitions
}
\label{sec:cp_uta:cus-td}

\mysubsubsection{Time-dependent $\CP$ asymmetries in $\Bz \to D^\mp \KS \pi^\pm$}
\label{sec:cp_uta:cus-td-DKSpi}

Time-dependent analyses of transitions such as $\Bz \to D^\mp \KS \pi^\pm$ can
be used to probe $\sin(2\beta+\gamma)$ in a similar way to that discussed
above (Sec.~\ref{sec:cp_uta:cud}). Since the final state contains three
particles, a Dalitz plot analysis is necessary to maximise the sensitivity. 
\babar~\cite{Aubert:2007qe} has carried out such an analysis. 
They obtain $2\beta+\gamma = \left( 83 \pm 53 \pm 20 \right)^\circ$
(with an ambiguity $2\beta+\gamma \leftrightarrow 2\beta+\gamma+\pi$) assuming
the ratio of the $b \to u$ and $b \to c$ amplitude to be constant across the
Dalitz plot at 0.3.

\mysubsubsection{Time-dependent $\CP$ asymmetries in $\Bs \to D_s^\mp K^\pm$}
\label{sec:cp_uta:cus-td-DsK}

Time-dependent analysis of $\Bs \to D_s^\mp K^\pm$ decays can be used to determine $\gamma-2\beta_s$~\cite{Dunietz:1987bv,Aleksan:1991nh}.
Compared to the situation for $\Bz \to D^{(*)\mp} \pi^\pm$ decays discussed in Sec.~\ref{sec:cp_uta:cud}, the larger value of the ratio $R$ of the magnitudes of the suppressed and favoured amplitudes allows it to be determined from the data.  
Moreover, the non-zero value of $\Delta \Gamma_s$ allows the determination of additional terms, labelled $A^{\Delta\Gamma}$ and $\bar{A}{}^{\Delta\Gamma}$, that break ambiguities in the solutions for $\gamma-2\beta_s$.

LHCb~\cite{Aaij:2014fba,LHCb-CONF-2016-015} has measured the time-dependent \CP violation parameters in $\Bs \to D_s^\mp K^\pm$ decays, using $3.0 \ {\rm fb}^{-1}$ of data.  
The results are given in Table~\ref{tab:cp_uta:DsK}, and correspond to $3.6\,\sigma$ evidence for \CP violation in the interference between mixing and $\Bs \to D_s^\mp K^\pm$ decays.
From these results, and a constraint on $2\beta_s$ from independent LHCb measurements~\cite{Aaij:2014zsa}, LHCb determine $\gamma = (127 \,^{+17}_{-22})^\circ$, $\delta_{D_sK} = (358 \,^{+15}_{-16})^\circ$ and $R_{D_sK} = 0.37 \,^{+0.10}_{-0.09}$. 

\begin{table}[!htb]
	\begin{center}
		\caption{
			Results for $\Bs \to D_s^\mp K^\pm$.
		}
%		\vspace{0.2cm}
%		\setlength{\tabcolsep}{0.0pc}
% make this tabular (not tabular*) and resize down to \textwidth
% change @{\extracolsep{\fill}} to @{\extracolsep{2mm}}
    \resizebox{\textwidth}{!}{
\renewcommand{\arraystretch}{1.2}
		\begin{tabular}{@{\extracolsep{2mm}}lrcccccc} \hline
	\mc{2}{l}{Experiment} & $\int {\cal L}\,dt$ & $C$ & $A^{\Delta\Gamma}$ & $\bar{A}{}^{\Delta\Gamma}$ & $S$ & $\bar{S}$ \\
	\hline
	LHCb & \cite{LHCb-CONF-2016-015} & 3 ${\rm fb}^{-1}$ & $0.74 \pm 0.14 \pm 0.05$ & $0.40 \pm 0.28 \pm 0.12$ & $0.31 \pm 0.27 \pm 0.11$ & $-0.52 \pm 0.20 \pm 0.07$ & $-0.50 \pm 0.20 \pm 0.07$ \\
%	LHCb & \cite{Aaij:2014fba} & 1 ${\rm fb}^{-1}$ & $0.53 \pm 0.25 \pm 0.04$ & $0.37 \pm 0.42 \pm 0.20$ & $0.20 \pm 0.41 \pm 0.20$ & $-1.09 \pm 0.33 \pm 0.08$ & $-0.36 \pm 0.34 \pm 0.08$ \\
	%% \hline
	%% \mc{3}{l}{\bf Average} & $0.53 \pm 0.25$ & $0.37 \pm 0.47$ & $0.20 \pm 0.46$ & $-1.09 \pm 0.34$ & $-0.36 \pm 0.35$ & {\small uncorrelated averages} \\
	%% \mc{3}{l}{\small Confidence level} & {\small $0.xx~(y.y\sigma)$} & {\small $0.xx~(y.y\sigma)$} & {\small $0.xx~(y.y\sigma)$} & {\small $0.xx~(y.y\sigma)$} & {\small $0.xx~(y.y\sigma)$} & \\
		\hline
		\end{tabular}
    }
		\label{tab:cp_uta:DsK}
	\end{center}
\end{table}

%%%%%%%%
%%%
%%% b -> cus
%%%
%%%%%%%%
% \afterpage{\clearpage}
\mysubsection{Rates and asymmetries in $\B \to \DorDstar K^{(*)}$ decays
}
\label{sec:cp_uta:cus}

As explained in Sec.~\ref{sec:cp_uta:notations:cus},
rates and asymmetries in $\Bp \to \DorDstar K^{(*)+}$ decays
are sensitive to $\gamma$, and have negligible theoretical uncertainty~\cite{Brod:2013sga}.
Various methods using different $\DorDstar$ final states have been used.

\mysubsubsection{$D$ decays to $\CP$ eigenstates}
\label{sec:cp_uta:cus:glw}

\begin{figure}[htbp]
  \begin{center}
    \begin{tabular}{cc}
      \resizebox{0.46\textwidth}{!}{
        \includegraphics{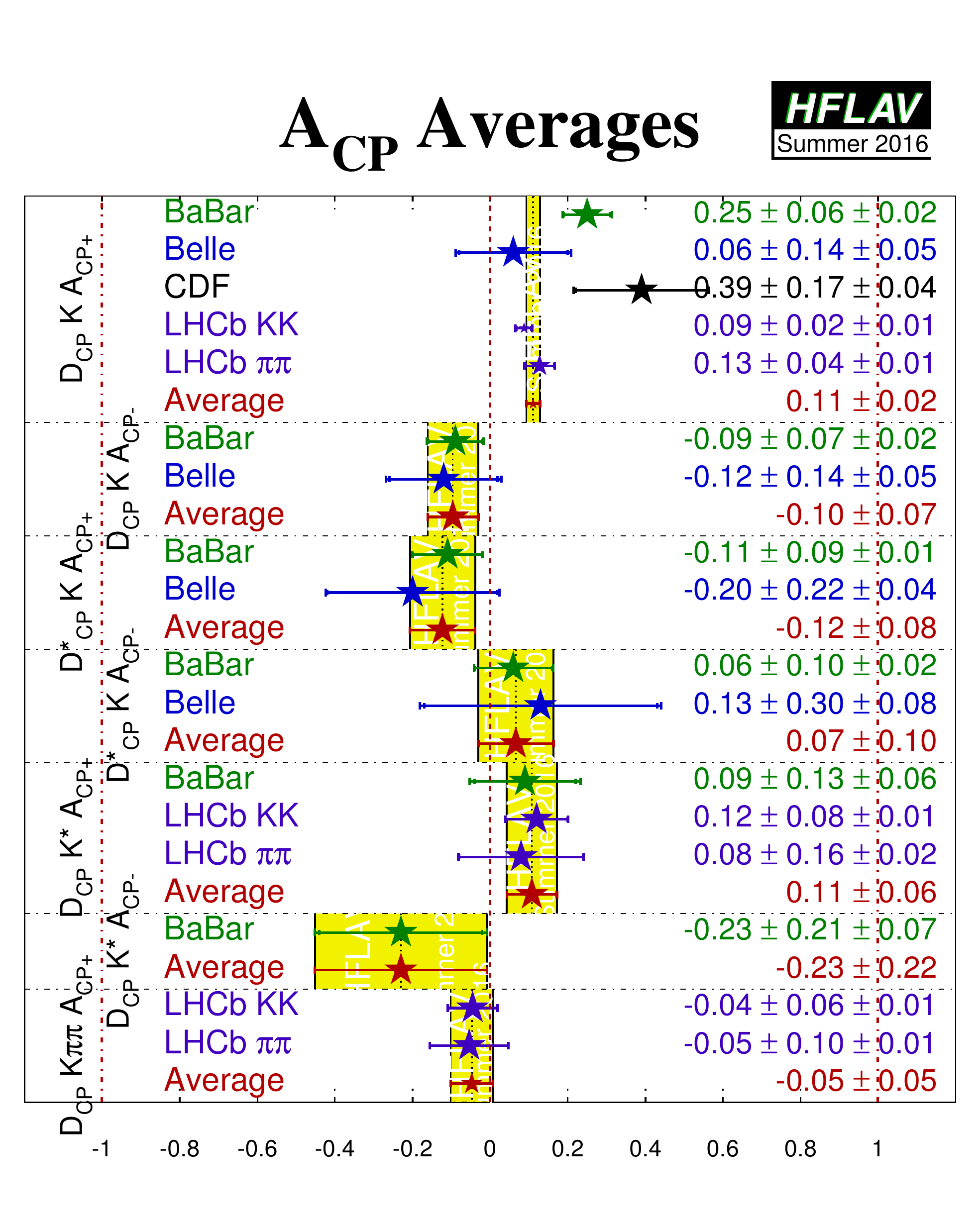}
      }
      &
      \resizebox{0.46\textwidth}{!}{
        \includegraphics{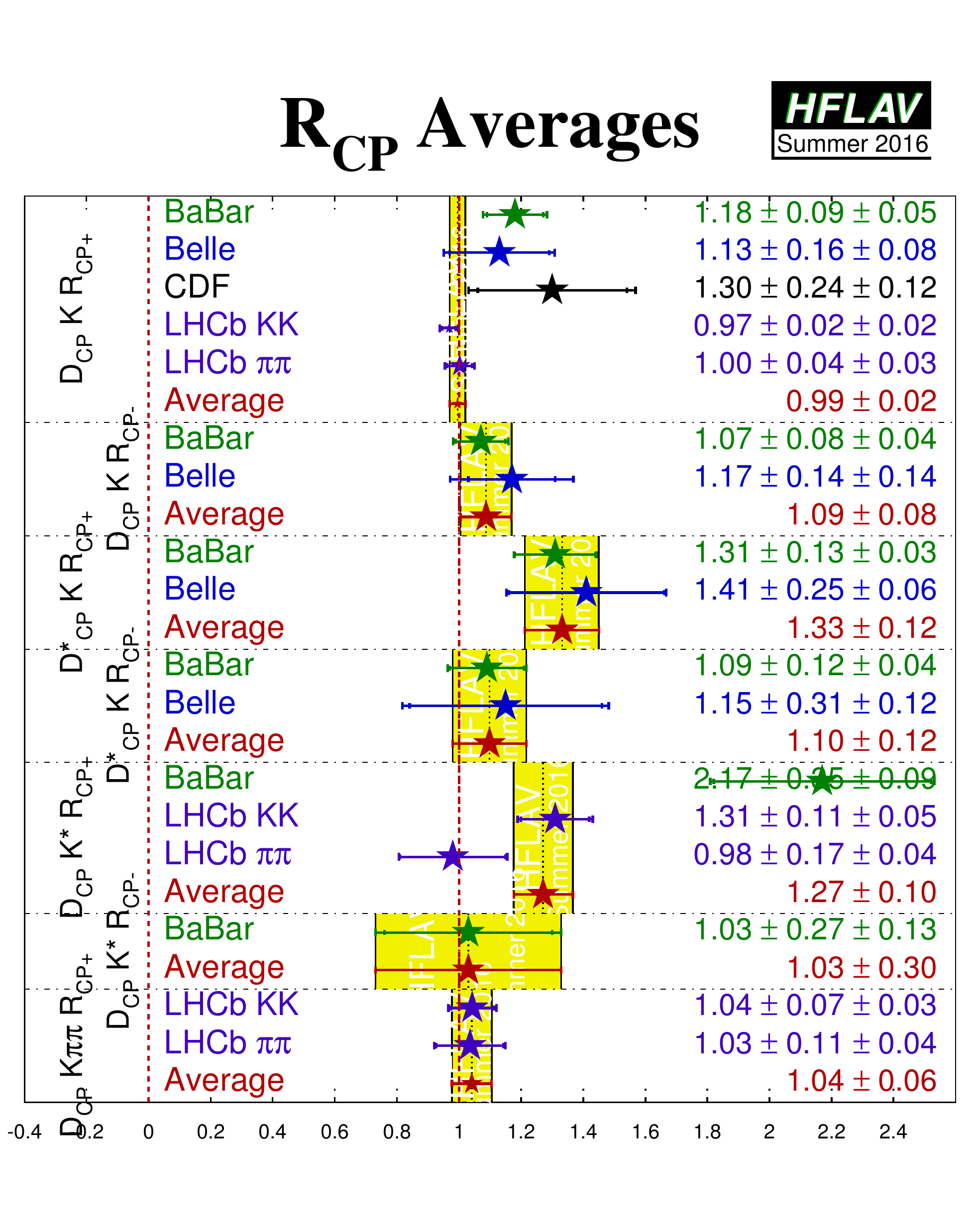}
      }
    \end{tabular}
 \end{center}
  \vspace{-0.8cm}
  \caption{
    Averages of $A_{\CP}$ and $R_{\CP}$ from GLW analyses.
  }
  \label{fig:cp_uta:cus:glw}
\end{figure}

Results are available from \babar, \belle, CDF and LHCb on GLW analyses in the
decay mode $\Bp \to D\Kp$.
All experiments use the $\CP$-even $D$ decay final states $K^+K^-$ and
$\pi^+\pi^-$; \babar\ and \belle\ in addition use the \CP-odd
decay modes $\KS\pi^0$, $\KS\omega$ and $\KS\phi$, though care is taken to
avoid statistical overlap with the $\KS K^+K^-$ sample used for Dalitz plot
analyses (see Sec.~\ref{sec:cp_uta:cus:dalitz}). 
%and asymmetric systematic errors are assigned due to $\CP$-even pollution under the $\KS\omega$ and $\KS\phi$ signals.
\babar\ and \belle\ also have results in the decay mode $\Bp \to \Dstar\Kp$,
using both the $\Dstar \to D\pi^0$ decay, which gives $\CP(\Dstar) = \CP(D)$,
and the $\Dstar \to D\gamma$ decays, which gives $\CP(\Dstar) = -\CP(D)$.
In addition, \babar\ and LHCb have results in the decay mode $\Bp \to D\Kstarp$,
and LHCb has results in the decay mode $\Bp \to D\Kp\pi^+\pi^-$.
% for $\CP$-even final states ($K^+K^-$ and $\pipi$) only.
The results and averages are given in Table~\ref{tab:cp_uta:cus:glw}
and shown in Fig.~\ref{fig:cp_uta:cus:glw}.

\begin{table}[htb]
	\begin{center}
		\caption{
%			Averages for $D_{\CP} K$.
                        Averages from GLW analyses of $b \to c\bar{u}s / u\bar{c}s$ modes.
                        The sample size is given in terms of number of $B\bar{B}$ pairs, $N(B\bar{B})$, for the $\epem$ $B$ factory experiments \babar\ and \belle, and in terms of integrated luminosity, $\int {\cal L}\,dt$, for the hadron collider experiments CDF and LHCb.
                }
                \vspace{0.2cm}
% make this tabular (not tabular*) and resize down to \textwidth
% change @{\extracolsep{\fill}} to @{\extracolsep{2mm}}
    \resizebox{\textwidth}{!}{
 \renewcommand{\arraystretch}{1.2}
     \setlength{\tabcolsep}{0.0pc}
      \begin{tabular}{@{\extracolsep{2mm}}lrccccc} \hline 
        \mc{2}{l}{Experiment} & Sample size & $A_{\CP+}$ & $A_{\CP-}$ & $R_{\CP+}$ & $R_{\CP-}$ \\
        \hline
        \mc{7}{c}{$D_{\CP} K^+$} \\
	\babar & \cite{delAmoSanchez:2010ji} & 467M & $0.25 \pm 0.06 \pm 0.02$ & $-0.09 \pm 0.07 \pm 0.02$ & $1.18 \pm 0.09 \pm 0.05$ & $1.07 \pm 0.08 \pm 0.04$ \\
	\belle & \cite{Abe:2006hc} & 275M & $0.06 \pm 0.14 \pm 0.05$ & $-0.12 \pm 0.14 \pm 0.05$ & $1.13 \pm 0.16 \pm 0.08$ & $1.17 \pm 0.14 \pm 0.14$ \\
	CDF & \cite{Aaltonen:2009hz} & $1 \, {\rm fb}^{-1}$ & $0.39 \pm 0.17 \pm 0.04$ & \textendash{} & $1.30 \pm 0.24 \pm 0.12$ &  \textendash{} \\
	LHCb $KK$ & \cite{Aaij:2016oso} & $3 \, {\rm fb}^{-1}$ & $0.087 \pm 0.020 \pm 0.008$ &  \textendash{} & $0.968 \pm 0.022 \pm 0.021$ &  \textendash{} \\
	LHCb $\pi\pi$ & \cite{Aaij:2016oso} & $3 \, {\rm fb}^{-1}$ & $0.128 \pm 0.037 \pm 0.012$ &  \textendash{} & $1.002 \pm 0.040 \pm 0.026$ &  \textendash{} \\
	LHCb average & \cite{Aaij:2016oso} & $3 \, {\rm fb}^{-1}$ & $0.097 \pm 0.018 \pm 0.009$ & \textendash{} & $0.978 \pm 0.019 \pm 0.018$ & \textendash{} \\
%	\hline
	\mc{3}{l}{\bf Average} & $0.111 \pm 0.018$ & $-0.10 \pm 0.07$ & $0.995 \pm 0.025$ & $1.09 \pm 0.08$ \\
	\mc{3}{l}{\small Confidence level} & {\small $0.063~(1.9\sigma)$} & {\small $0.86~(0.2\sigma)$} & {\small $0.21~(1.3\sigma)$} & {\small $0.65~(0.5\sigma)$} \\
		\hline
% 		\end{tabular*}
% 		\label{tab:cp_uta:yyy}
% 	\end{center}
% \end{table}

% \begin{table}[htb]
% 	\begin{center}
% 		\caption{
% 			Averages for $D*_{\CP} K$.
% 		}
% 		\vspace{0.2cm}
% 		\setlength{\tabcolsep}{0.0pc}
% 		\begin{tabular*}{\textwidth}{@{\extracolsep{\fill}}lrcccccc} \hline
% 	\mc{2}{l}{Experiment} & $N(B\bar{B})$ & $A_{\CP+}$ & $A_{\CP-}$ & $R_{\CP+}$ & $R_{\CP-}$ & Correlation \\
% 	\hline
        \mc{7}{c}{$\Dstar_{\CP} K^+$} \\
	\babar & \cite{:2008jd} & 383M & $-0.11 \pm 0.09 \pm 0.01$ & $0.06 \pm 0.10 \pm 0.02$ & $1.31 \pm 0.13 \pm 0.03$ & $1.09 \pm 0.12 \pm 0.04$ \\
	\belle & \cite{Abe:2006hc} & 275M & $-0.20 \pm 0.22 \pm 0.04$ & $0.13 \pm 0.30 \pm 0.08$ & $1.41 \pm 0.25 \pm 0.06$ & $1.15 \pm 0.31 \pm 0.12$ \\
%	\hline
	\mc{3}{l}{\bf Average} & $-0.12 \pm 0.07$ & $0.13 \pm 0.07$ & $1.25 \pm 0.09$ & $1.06 \pm 0.09$ \\
	\mc{3}{l}{\small Confidence level} & {\small $0.82~(0.2\sigma)$} & {\small $0.29~(1.1\sigma)$} & {\small $0.52~(0.6\sigma)$} & {\small $0.74~(0.3\sigma)$} \\
		\hline
%		\end{tabular*}
% 		\label{tab:cp_uta:yyy}
% 	\end{center}
% \end{table}

% \begin{table}[htb]
% 	\begin{center}
% 		\caption{
% 			Averages for $D_{\CP} K*$.
% 		}
% 		\vspace{0.2cm}
% 		\setlength{\tabcolsep}{0.0pc}
% 		\begin{tabular*}{\textwidth}{@{\extracolsep{\fill}}lrcccccc} \hline
%	\mc{2}{l}{Experiment} & $N(B\bar{B})$ & $A_{\CP+}$ & $A_{\CP-}$ & $R_{\CP+}$ & $R_{\CP-}$ & Correlation \\
%	\hline
        \mc{7}{c}{$D_{\CP} K^{*+}$} \\
	\babar & \cite{Aubert:2009yw} & 379M & $0.09 \pm 0.13 \pm 0.06$ & $-0.23 \pm 0.21 \pm 0.07$ & $2.17 \pm 0.35 \pm 0.09$ & $1.03 \pm 0.27 \pm 0.13$ \\
        LHCb $KK$ & \cite{LHCb-CONF-2016-014} & $4 \, {\rm fb}^{-1}$ & $0.12 \pm 0.08 \pm 0.01$ & \textendash{} & $1.31 \pm 0.11 \pm 0.05$ & \textendash{} \\
	LHCb $\pi\pi$ & \cite{LHCb-CONF-2016-014} & $4 \, {\rm fb}^{-1}$ & $0.08 \pm 0.16 \pm 0.02$ & \textendash{} & $0.98 \pm 0.17 \pm 0.04$ & \textendash{} \\
        LHCb average & \cite{LHCb-CONF-2016-014} & $4 \, {\rm fb}^{-1}$ & $0.11 \pm 0.07$ & \textendash{} & $1.21 \pm 0.10$ & \textendash{} \\
	\mc{3}{l}{\bf Average} & $0.11 \pm 0.06$ & $-0.23 \pm 0.22$ & $1.27 \pm 0.10$ & $1.03 \pm 0.30$ \\
	\mc{3}{l}{\small Confidence level} & {\small $0.97~(0.04\sigma)$} & & {\small $0.01~(2.6\sigma)$} \\
%	\hline
%	\mc{3}{l}{\bf Average} & $-0.08 \pm 0.21$ & $-0.26 \pm 0.42$ & $1.96 \pm 0.41$ & $0.65 \pm 0.27$ & {\small uncorrelated averages} \\
%	\mc{3}{l}{\small Confidence level} & {\small $0.xx~(y.y\sigma)$} & {\small $0.xx~(y.y\sigma)$} & {\small $0.xx~(y.y\sigma)$} & {\small $0.xx~(y.y\sigma)$} & \\
		\hline
% 		\end{tabular*}
% 		\label{tab:cp_uta:yyy}

	%% 	\caption{
	%% 		Averages for $D_{\CP} K\pi\pi$.
	%% 	}
	%% 	\vspace{0.2cm}
	%% 	\setlength{\tabcolsep}{0.0pc}
	%% 	\begin{tabular*}{\textwidth}{@{\extracolsep{\fill}}lrcccc} \hline
	%% \mc{2}{l}{Experiment} & $N(B\bar{B})$ & $A_{\CP+}$ & $R_{\CP+}$ & Correlation \\
	%% \hline
        \mc{7}{c}{$D_{\CP} K^+\pi^+\pi^-$} \\
	LHCb $KK$ & \cite{Aaij:2015ina} & $3 \, {\rm fb}^{-1}$ & $-0.045 \pm 0.064 \pm 0.011$ & \textendash{} & $1.043 \pm 0.069 \pm 0.034$ & \textendash{} \\
	LHCb $\pi\pi$ & \cite{Aaij:2015ina} & $3 \, {\rm fb}^{-1}$ & $-0.054 \pm 0.101 \pm 0.011$ & \textendash{} & $1.035 \pm 0.108 \pm 0.038$ & \textendash{} \\
        LHCb average & \cite{Aaij:2015ina} & $3 \, {\rm fb}^{-1}$ & $-0.048 \pm 0.055$ & \textendash{} & $1.040 \pm 0.064$ & \textendash{} \\
	\hline
	%% \mc{3}{l}{\bf Average} & $-0.14 \pm 0.10$ & $0.95 \pm 0.11$ & {\small uncorrelated averages} \\
	%% \mc{3}{l}{\small Confidence level} & {\small $0.xx~(y.y\sigma)$} & {\small $0.xx~(y.y\sigma)$} & \\
	%% 	\hline
	%% 	\end{tabular*}
	%% 	\label{tab:cp_uta:yyy}

      \end{tabular}
    }
    \label{tab:cp_uta:cus:glw}
	\end{center}
\end{table}

LHCb has performed a GLW analysis using the $B^0 \to DK^{*0}$ decay with the \CP-even $D \to K^+K^-$ and $D \to \pi^+\pi^-$ channels~\cite{Aaij:2014eha}.
The results are presented separately to allow for possible \CP violation effects in the charm decays, which are, however, known to be small.
The results are given in Table~\ref{tab:cp_uta:glw-DKstar} where an average is also reported.

\begin{table}[!htb]
        \begin{center}
                \caption{
%                        Averages for $D_K\pi K*$.
      Results from GLW analysis of $\Bz \to D\Kstarz$.
                }
                \vspace{0.2cm}
                \setlength{\tabcolsep}{0.0pc}
\renewcommand{\arraystretch}{1.1}
                \begin{tabular*}{\textwidth}{@{\extracolsep{\fill}}lrccc} \hline
        \mc{2}{l}{Experiment} & Sample size & $A_{\CP+}$ & $R_{\CP+}$ \\
        \hline
        LHCb $KK$ & \cite{Aaij:2014eha} & $\int {\cal L}\,dt = 3 \, {\rm fb}^{-1}$ & $-0.20 \pm 0.15 \pm 0.02$ & $1.05 \,^{+0.17}_{-0.15} \pm 0.04$ \\
        LHCb $\pi\pi$ & \cite{Aaij:2014eha} & $\int {\cal L}\,dt = 3 \, {\rm fb}^{-1}$ & $-0.09 \pm 0.22 \pm 0.02$ & $1.21 \,^{+0.28}_{-0.25} \pm 0.05$ \\
%        \hline
        \mc{3}{l}{\bf Average} & $-0.16 \pm 0.12$ & $1.10 \pm 0.14$ \\
        %% \mc{3}{l}{\small Confidence level} & {\small $0.xx~(y.y\sigma)$} & {\small $0.xx~(y.y\sigma)$} & \\
                \hline
                \end{tabular*}
                \label{tab:cp_uta:glw-DKstar}
        \end{center}
\end{table}

As pointed out in Refs.~\cite{Gershon:2008pe,Gershon:2009qc}, a Dalitz plot analysis of $B^0 \to DK^+\pi^-$ decays provides more sensitivity to $\gamma \equiv \phi_3$ than the quasi-two-body $DK^{*0}$ approach. 
The analysis provides direct sensitivity to the hadronic parameters $r_B$ and $\delta_B$ associated with the $B^0 \to DK^{*0}$ decay amplitudes, rather than effective hadronic parameters averaged over the $K^{*0}$ selection window as in the quasi-two-body case. 

Such an analysis has been performed by LHCb. 
A simultaneous fit is performed to the $B^0 \to DK^+\pi^-$ Dalitz plots with the neutral $D$ meson reconstructed in the $K^+\pi^-$, $K^+K^-$ and $\pi^+\pi^-$ final states. 
The reported results in Table~\ref{tab:cp_uta:cus:DKpiDalitz} are for the Cartesian parameters, defined in Eq.~(\ref{eq:cp_uta:cartesian}) associated with the $B^0 \to DK^*(892)^0$ decay. 
Note that, since the measurements use overlapping data samples, these results cannot be combined with the LHCb results for GLW observables in $B^0 \to DK^*(892)^0$ decays reported in Table~\ref{tab:cp_uta:glw-DKstar}.

\begin{table}[!htb]
	\begin{center}
		\caption{
			Results from Dalitz plot analysis of $\B^0 \to DK^+\pi^-$ decays with $D \to \Kp\Km$ and $\pip\pim$.
		}
		\vspace{0.2cm}
		\setlength{\tabcolsep}{0.0pc}
% make this tabular (not tabular*) and resize down to \textwidth
% change @{\extracolsep{\fill}} to @{\extracolsep{2mm}}
    \resizebox{\textwidth}{!}{
\renewcommand{\arraystretch}{1.2}
		\begin{tabular}{@{\extracolsep{2mm}}lrccccc} \hline
	\mc{2}{l}{Experiment} & $\int {\cal L}\,dt$ & $x_+$ & $y_+$ & $x_-$ & $y_-$ \\
	\hline
	LHCb & \cite{Aaij:2016bqv} & $3 \, {\rm fb}^{-1}$ & $0.04 \pm 0.16 \pm 0.11$ & $-0.47 \pm 0.28 \pm 0.22$ & $-0.02 \pm 0.13 \pm 0.14$ & $-0.35 \pm 0.26 \pm 0.41$ \\
% 	\hline
% 	\mc{3}{l}{\bf Average} & $0.04 \pm 0.19$ & $-0.47 \pm 0.36$ & $-0.02 \pm 0.19$ & $-0.35 \pm 0.49$ & \textendash{} \\
% 	\mc{3}{l}{\small Confidence level} & \mc{4}{c}{\small $0.xx~(y.y\sigma)$} & \\
		\hline
		\end{tabular}
}
		\label{tab:cp_uta:cus:DKpiDalitz}
	\end{center}
\end{table}

LHCb use these results to obtain confidence levels for $\gamma$, $r_B(DK^{*0})$ and $\delta_B(DK^{*0})$. 
In addition, results are reported for the hadronic parameters needed to relate these results to quasi-two-body measurements of $B^0 \to DK^*(892)^0$ decays, where a selection window of $m(K^+\pi^-)$ within $50~\mevcc$ of the pole mass and helicity angle satisfying $\left|\cos(\theta_{K^{*0}})\right|>0.4$ is assumed. 
These parameters are the coherence factor $\kappa$, the ratio of quasi-two-body and amplitude level $r_B$ values, $\bar{R}_B = \bar{r}_B/r_B$, and the difference between quasi-two-body and amplitude level $\delta_B$ values, $\Delta \bar{\delta}_B = \bar{\delta}_B-\delta_B$.
LHCb~\cite{Aaij:2016bqv} obtain
\begin{equation}
  \gamma = 0.958 \,^{+0.005}_{-0.010}\,^{+0.002}_{-0.045}\,,\quad 
  \bar{R}_B = 1.02 \,^{+0.03}_{-0.01} \pm 0.06\,,\quad
  \Delta \bar{\delta}_B = 0.02 \,^{+0.03}_{-0.02} \pm 0.11\,.
\end{equation}

\mysubsubsection{$D$ decays to quasi-$\CP$ eigenstates}
\label{sec:cp_uta:cus:quasi-glw}

As discussed in Sec.~\ref{sec:cp_uta:notations:cus}, if a multibody neutral $D$ meson decay can be shown to be dominated by one $\CP$ eigenstate, it can be used in a ``GLW-like'' (sometimes called ``quasi-GLW'') analysis~\cite{Nayak:2014tea}. 
The same observables $R_{\CP}$, $A_{\CP}$ as for the GLW case are measured, but an additional factor of $(2F_+-1)$, where $F_+$ is the fractional $\CP$-even content, enters the expressions relating these observables to $\gamma \equiv \phi_3$. 
The $F_+$ factors have been measured using CLEO-c data to be $F_+(\pi^+\pi^-\pi^0) = 0.973 \pm 0.017$, $F_+(K^+K^-\pi^0) = 0.732 \pm 0.055$, $F_+(\pi^+\pi^-\pi^+\pi^-) = 0.737 \pm 0.028$~\cite{Malde:2015mha}. 

The GLW-like observables for $D\to\pi^+\pi^-\pi^0$, $K^+K^-\pi^0$ and $D\to\pi^+\pi^-\pi^+\pi^-$ have been measured by LHCb. 
The $A_{\rm qGLW}$ observable for $D\to\pi^+\pi^-\pi^0$ was measured in an earlier analysis by \babar, from which additional observables, discussed in Sec.~\ref{sec:cp_uta:notations:cus} and reported in Table~\ref{tab:cp_uta:cus:dalitz} below, were reported.
The results are given in Table~\ref{tab:cp_uta:cus:glwLike}.

\begin{table}[!htb]
	\begin{center}
		\caption{
%			Averages for $D_\pi\pi\\pi^{0} K$.
                        Averages from GLW-like analyses of $b \to c\bar{u}s / u\bar{c}s$ modes.
		}
		\vspace{0.2cm}
		\setlength{\tabcolsep}{0.0pc}
\renewcommand{\arraystretch}{1.1}
		\begin{tabular*}{\textwidth}{@{\extracolsep{\fill}}lrccc} \hline
	\mc{2}{l}{Experiment} & Sample size & $A_{\rm qGLW}$ & $R_{\rm qGLW}$ \\
	\hline
        \mc{5}{c}{$D_{\pi^+\pi^-\pi^0} K^+$} \\
	LHCb & \cite{Aaij:2015jna} & $\int {\cal L}\,dt = 3 \, {\rm fb}^{-1}$ & $0.05 \pm 0.09 \pm 0.01$ & $0.98 \pm 0.11 \pm 0.05$ \\
	\babar & \cite{Aubert:2007ii} & $N(B\bar{B}) =$ 324M & $-0.02 \pm 0.15 \pm 0.03$ &  \textendash{} \\
%	\hline
	\mc{3}{l}{\bf Average} & $0.03 \pm 0.08$ & $0.98 \pm 0.12$ \\
	\mc{3}{l}{\small Confidence level} & {\small $0.68~(0.4\sigma)$} & \textendash{} \\
		\hline
        \mc{5}{c}{$D_{K^+K^-\pi^0} K^+$} \\
	LHCb & \cite{Aaij:2015jna} & $\int {\cal L}\,dt = 3 \, {\rm fb}^{-1}$ & $0.30 \pm 0.20 \pm 0.02$ & $0.95 \pm 0.22 \pm 0.04$ \\
		\hline
        \mc{5}{c}{$D_{\pi^+\pi^-\pi^+\pi^-} K^+$} \\
	LHCb & \cite{Aaij:2016oso} & $\int {\cal L}\,dt = 3 \, {\rm fb}^{-1}$ & $0.10 \pm 0.03 \pm 0.02$ & $0.97 \pm 0.04 \pm 0.02$ \\
		\hline
		\end{tabular*}
		\label{tab:cp_uta:cus:glwLike}
	\end{center}
\end{table}

\mysubsubsection{$D$ decays to suppressed final states}
\label{sec:cp_uta:cus:ads}

\begin{figure}[htbp]
  \begin{center}
    \begin{tabular}{cc}
      \resizebox{0.46\textwidth}{!}{
        \includegraphics{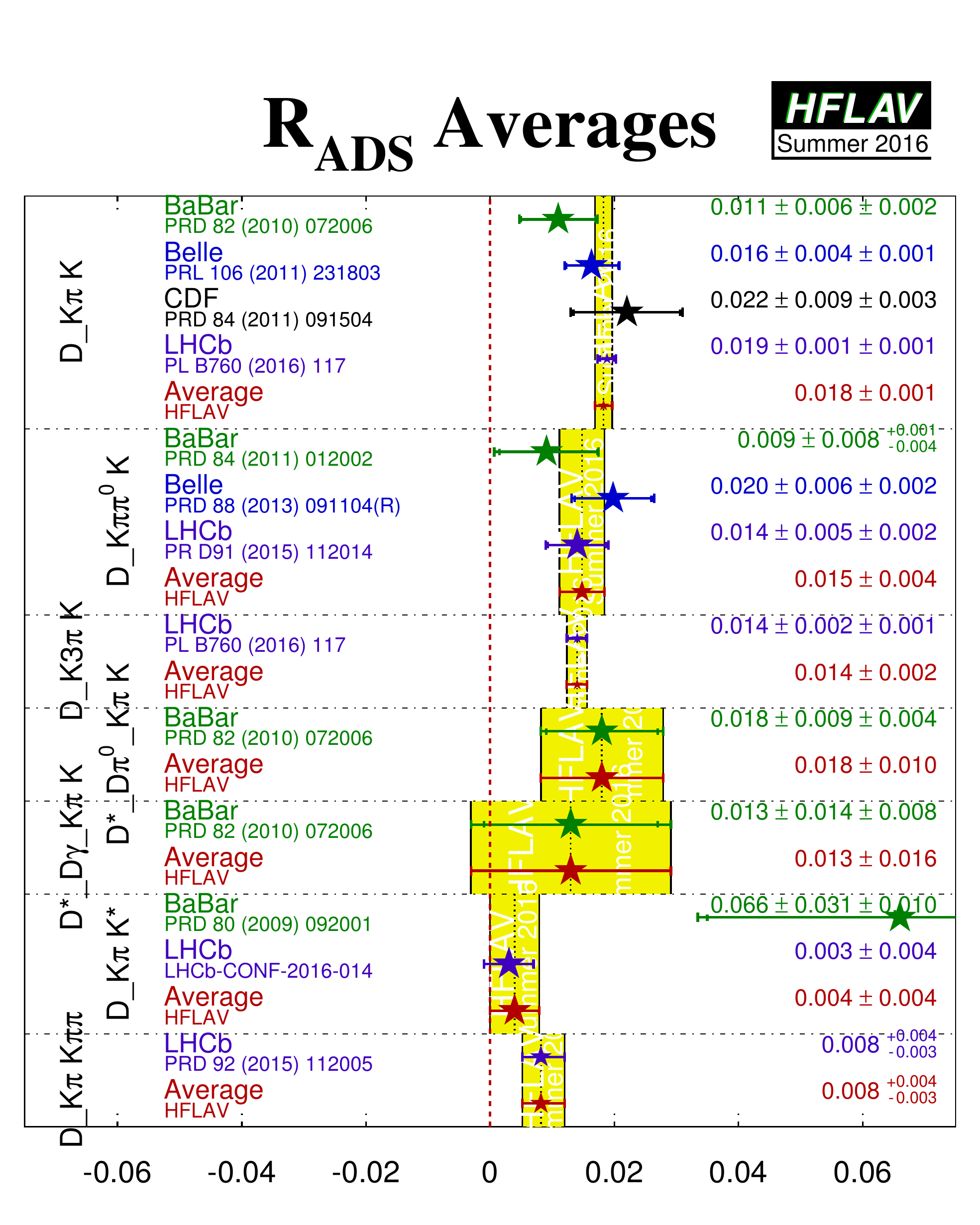}
      }
      &
      \resizebox{0.46\textwidth}{!}{
        \includegraphics{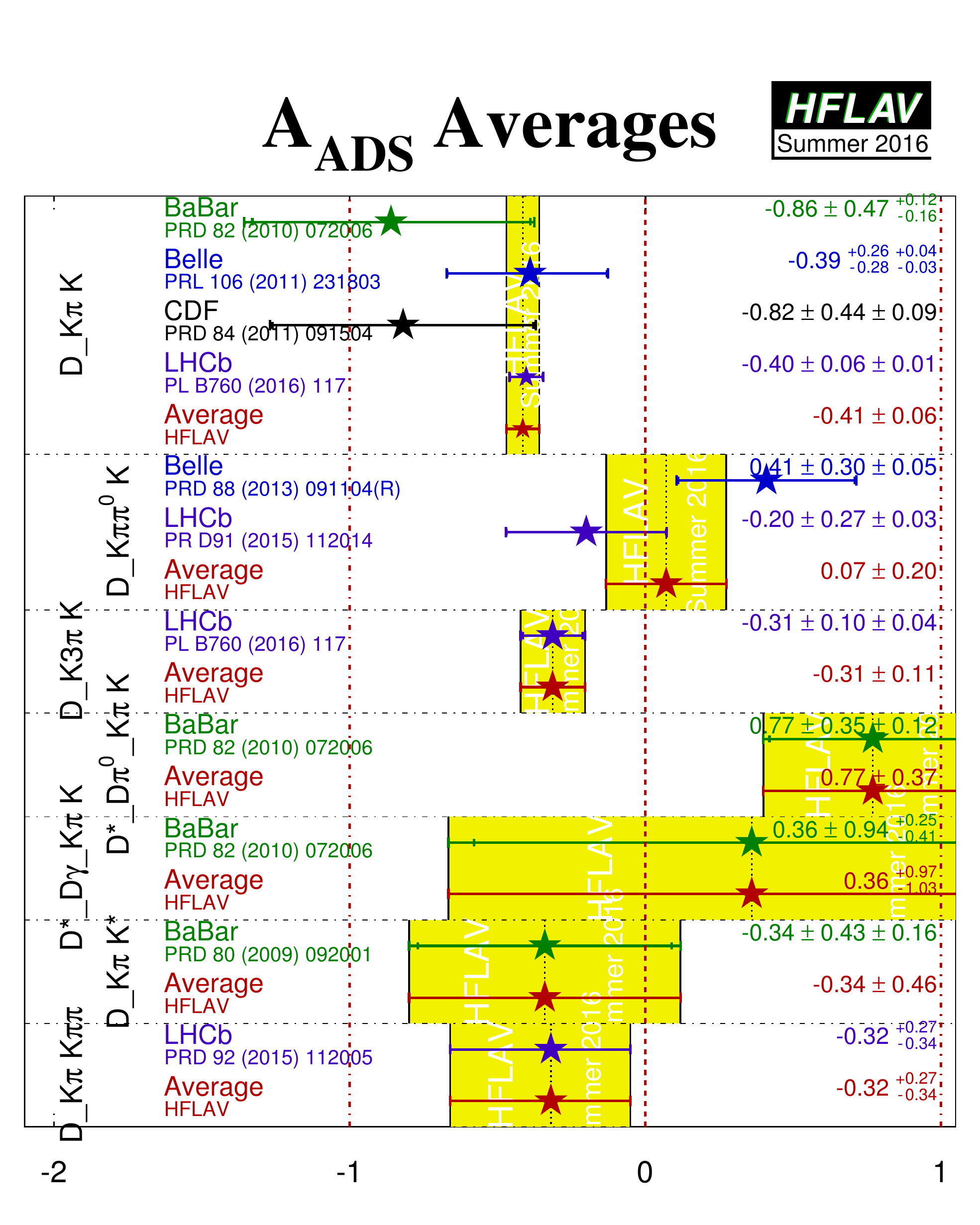}
      }
    \end{tabular}
  \end{center}
  \vspace{-0.8cm}
  \caption{
    Averages of $R_{\rm ADS}$ and $A_{\rm ADS}$ for $B \to D^{(*)}K^{(*)}$ decays.
  }
  \label{fig:cp_uta:cus:ads}
\end{figure}

For ADS analyses, all of \babar, \belle, CDF and LHCb have studied the modes 
$\Bp \to D\Kp$ and $\Bp \to D\pip$. 
\babar\ has also analysed the $\Bp \to \Dstar\Kp$ mode.
There is an effective shift of $\pi$ in the strong phase difference between
the cases that the $\Dstar$ is reconstructed as $D\pi^0$ and
$D\gamma$~\cite{Bondar:2004bi}, therefore these modes are studied separately.
In addition, \babar\ has studied the $\Bp \to D\Kstarp$ mode, 
where $\Kstarp$ is reconstructed as $\KS\pip$,
and LHCb has studied the $\Bp \to D\Kp\pip\pim$ mode.
In all the above cases the suppressed decay $D \to K^-\pi^+$ has been used.
\babar, \belle\ and LHCb also have results using $\Bp \to D\Kp$ with $D \to K^-\pi^+\pi^0$,
while LHCb has results using $\Bp \to D\Kp$ with $D \to K^-\pi^+\pi^+\pi^-$.
The results and averages are given in Table~\ref{tab:cp_uta:cus:ads}
and shown in Fig.~\ref{fig:cp_uta:cus:ads}.

\begin{figure}[htbp]
  \begin{center}
    \begin{tabular}{cc}
      \resizebox{0.46\textwidth}{!}{
        \includegraphics{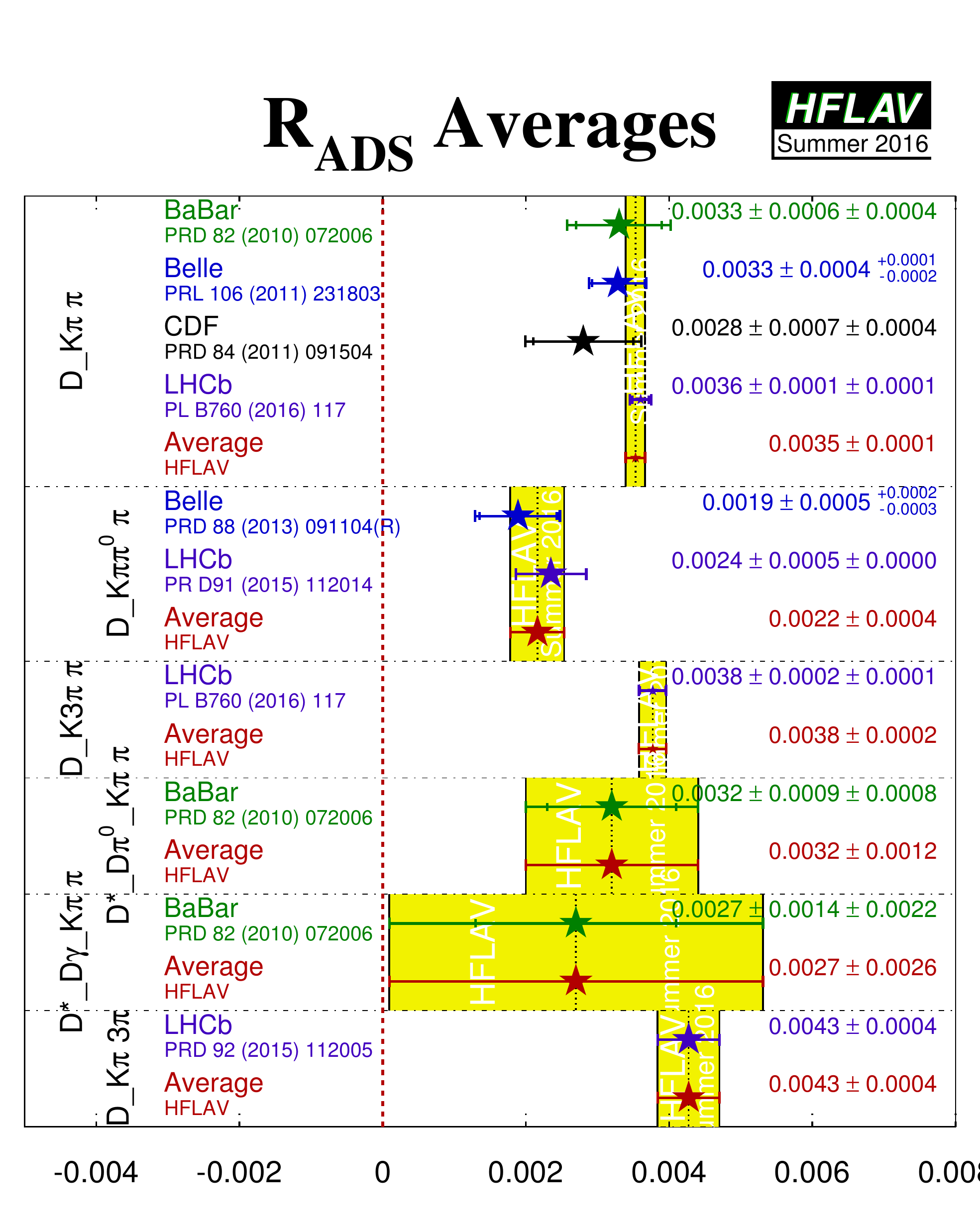}
      }
      &
      \resizebox{0.46\textwidth}{!}{
        \includegraphics{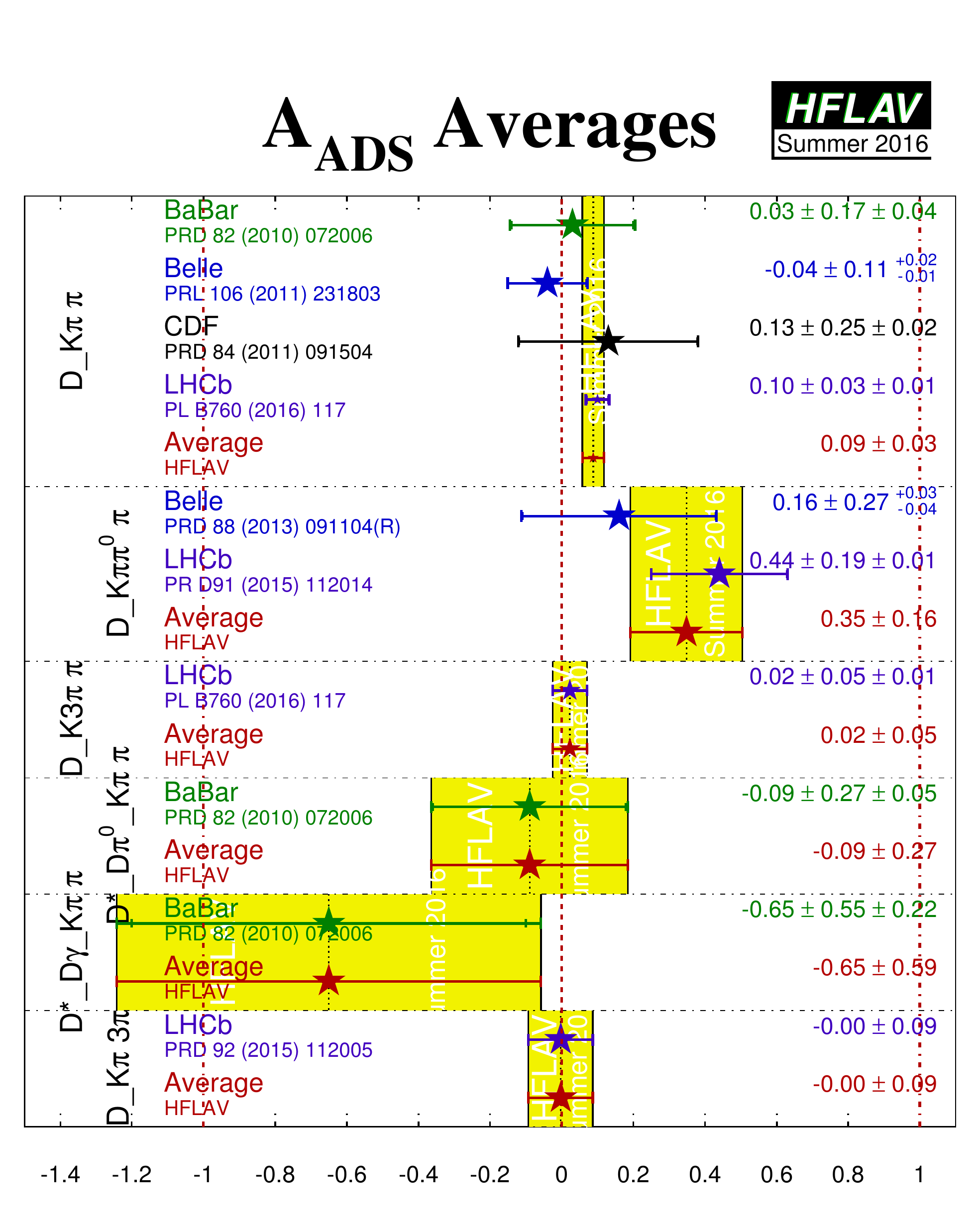}
      }
    \end{tabular}
  \end{center}
  \vspace{-0.8cm}
  \caption{
    Averages of $R_{\rm ADS}$ and $A_{\rm ADS}$ for $B \to D^{(*)}\pi$ decays.
  }
  \label{fig:cp_uta:cus:ads-Dpi}
\end{figure}

Similar phenomenology as for $B \to DK$ decays holds for $B \to D\pi$ decays, though in this case the interference is between $b \to c\bar{u}d$ and $b \to u\bar{c}d$ transitions, and the ratio of suppressed to favoured amplitudes is expected to be much smaller, ${\cal O}(1\%)$.
For most $D$ meson final states this implies that the interference effect is too small to be of interest, but in the case of ADS analysis it is possible that effects due to $\gamma$ may be observable.
Accordingly, the experiments now measure the corresponding observables in the $D\pi$ final states.
The results and averages are given in Table~\ref{tab:cp_uta:cus:ads2}
and shown in Fig.~\ref{fig:cp_uta:cus:ads-Dpi}.

\begin{table}[htb]
	\begin{center}
		\caption{
%			Averages for $D_K\pi K$.
      Averages from ADS analyses of $b \to c\bar{u}s / u\bar{c}s$ modes.
                }
                \vspace{0.2cm}
                \setlength{\tabcolsep}{0.0pc}
\renewcommand{\arraystretch}{1.1}
                \begin{tabular*}{\textwidth}{@{\extracolsep{\fill}}lrccc} \hline 
        \mc{2}{l}{Experiment} & Sample size & $A_{\rm ADS}$ & $R_{\rm ADS}$ \\
	\hline
        \mc{5}{c}{$D K^+$, $D \to K^-\pi^+$} \\
	\babar & \cite{delAmoSanchez:2010dz} & $N(B\bar{B}) =$ 467M & $-0.86 \pm 0.47 \,^{+0.12}_{-0.16}$ & $0.011 \pm 0.006 \pm 0.002$ \\
	\belle & \cite{Belle:2011ac} & $N(B\bar{B}) =$ 772M & $-0.39 \,^{+0.26}_{-0.28} \,^{+0.04}_{-0.03}$ & $0.0163 \,^{+0.0044}_{-0.0041} \,^{+0.0007}_{-0.0013}$ \\
	CDF & \cite{Aaltonen:2011uu} & $\int {\cal L}\,dt = 7 \, {\rm fb}^{-1}$ & $-0.82 \pm 0.44 \pm 0.09$ & $0.0220 \pm 0.0086 \pm 0.0026$ \\
	LHCb & \cite{Aaij:2016oso} & $\int {\cal L}\,dt = 3 \, {\rm fb}^{-1}$ & $-0.403 \pm 0.056 \pm 0.011$ & $0.0188 \pm 0.0011 \pm 0.0010$ \\
%	\hline
	\mc{3}{l}{\bf Average} & $-0.415 \pm 0.055$ & $0.0183 \pm 0.0014$ \\
	\mc{3}{l}{\small Confidence level} & {\small $0.64~(0.5\sigma)$} & {\small $0.61~(0.5\sigma)$} \\
		\hline
% 		\end{tabular*}
% 		\label{tab:cp_uta:yyy}
% 	\end{center}
% \end{table}

% \begin{table}[htb]
% 	\begin{center}
% 		\caption{
% 			Averages for $D_K\pi\\pi^{0} K$.
% 		}
% 		\vspace{0.2cm}
% 		\setlength{\tabcolsep}{0.0pc}
% 		\begin{tabular*}{\textwidth}{@{\extracolsep{\fill}}lrccc} \hline
% 	\mc{2}{l}{Experiment} & $N(B\bar{B})$ & $R_{\rm ADS}$ & Correlation \\
% 	\hline
        \mc{5}{c}{$D K^+$, $D \to K^-\pi^+\pi^0$} \\
	\babar & \cite{Lees:2011up} & 474M & \textendash{} & $0.0091 \,^{+0.0082}_{-0.0076} \,^{+0.0014}_{-0.0037}$ \\
	\belle & \cite{Nayak:2013tgg} & 772M & $0.41 \pm 0.30 \pm 0.05$ & $0.0198 \pm 0.0062 \pm 0.0024$ \\
	LHCb & \cite{Aaij:2015jna} & $\int {\cal L}\,dt = 3 \, {\rm fb}^{-1}$ & $-0.20 \pm 0.27 \pm 0.03$ & $0.0140 \pm 0.0047 \pm 0.0019$ \\
%        \hline 
	\mc{3}{l}{\bf Average} & $0.07 \pm 0.20$ & $0.0148 \pm 0.0036$ \\
 	\mc{3}{l}{\small Confidence level} & {\small $0.13~(1.5\sigma)$} & {\small $0.59~(0.5\sigma)$} \\
 	\hline
% 		\end{tabular*}
% 		\label{tab:cp_uta:yyy}
% 	\end{center}
% \end{table}

% \begin{table}[htb]
% 	\begin{center}
% 		\caption{
% 			Averages for $D_K\pi\pi\pi K$.
% 		}
% 		\vspace{0.2cm}
% 		\setlength{\tabcolsep}{0.0pc}
% 		\begin{tabular*}{\textwidth}{@{\extracolsep{\fill}}lrccc} \hline
% 	\mc{2}{l}{Experiment} & $N(B\bar{B})$ & $R_{\rm ADS}$ & Correlation \\
% 	\hline
        \mc{5}{c}{$D K^+$, $D \to K^-\pi^+\pi^+\pi^-$} \\
	LHCb & \cite{Aaij:2016oso} & $\int {\cal L}\,dt = 3 \, {\rm fb}^{-1}$ & $-0.313 \pm 0.102 \pm 0.038$ & $0.0140 \pm 0.0015 \pm 0.0006$ \\
        \hline
% 		\end{tabular*}
% 		\label{tab:cp_uta:yyy}
% 	\end{center}
% \end{table}

% \begin{table}[htb]
% 	\begin{center}
% 		\caption{
% 			Averages for $D*_D\\pi^{0}_K\pi K$.
% 		}
% 		\vspace{0.2cm}
% 		\setlength{\tabcolsep}{0.0pc}
% 		\begin{tabular*}{\textwidth}{@{\extracolsep{\fill}}lrccc} \hline
%        \mc{2}{l}{Experiment} & $N(B\bar{B})$ & $A_{\rm ADS}$ & $R_{\rm ADS}$ \\
% 	\hline
        \mc{5}{c}{$\Dstar K^+$, $\Dstar \to D\pi^0$, $D \to K^-\pi^+$} \\
	\babar & \cite{delAmoSanchez:2010dz} & $N(B\bar{B}) =$ 467M & $0.77 \pm 0.35 \pm 0.12$ & $0.018 \pm 0.009 \pm 0.004$ \\
%	\belle & \cite{belle:glwads:prelim} & 772M & $0.4 \,^{+1.1}_{-0.7} \,^{+0.2}_{-0.1}$ & $0.010 \,^{+0.008}_{-0.007} \,^{+0.001}_{-0.002}$ \\
%        \hline
%	\mc{3}{l}{\bf Average} & $0.72 \pm 0.34$ & $0.013 \pm 0.006$ \\
%	\mc{3}{l}{\small Confidence level} & {\small $0.71~(0.4\sigma)$} & {\small $0.52~(0.6\sigma)$} \\
 	\hline
% 		\end{tabular*}
% 		\label{tab:cp_uta:yyy}
% 	\end{center}
% \end{table}

% \begin{table}[htb]
% 	\begin{center}
% 		\caption{
% 			Averages for $D*_D\gamma_K\pi K$.
% 		}
% 		\vspace{0.2cm}
% 		\setlength{\tabcolsep}{0.0pc}
% 		\begin{tabular*}{\textwidth}{@{\extracolsep{\fill}}lrccc} \hline
% 	\mc{2}{l}{Experiment} & $N(B\bar{B})$ & $R_{\rm ADS}$ & Correlation \\
% 	\hline
        \mc{5}{c}{$\Dstar K^+$, $\Dstar \to D\gamma$, $D \to K^-\pi^+$} \\
	\babar & \cite{delAmoSanchez:2010dz} & $N(B\bar{B}) =$ 467M & $0.36 \pm 0.94 \,^{+0.25}_{-0.41}$ & $0.013 \pm 0.014 \pm 0.008$ \\
%	\belle & \cite{belle:glwads:prelim} & 772M & $-0.51 \,^{+0.33}_{-0.29} \pm 0.08$ & $0.036 \,^{+0.014}_{-0.012} \pm 0.002$ \\
%        \hline
%	\mc{3}{l}{\bf Average} & $-0.43 \pm 0.31$ & $0.027 \pm 0.010$ \\
%	\mc{3}{l}{\small Confidence level} & {\small $0.42~(0.8\sigma)$} & {\small $0.26~(1.1\sigma)$} \\
		\hline
% 		\end{tabular*}
% 		\label{tab:cp_uta:yyy}
% 	\end{center}
% \end{table}

% \begin{table}[htb]
% 	\begin{center}
% 		\caption{
% 			Averages for $D_K\pi K*$.
% 		}
% 		\vspace{0.2cm}
% 		\setlength{\tabcolsep}{0.0pc}
% 		\begin{tabular*}{\textwidth}{@{\extracolsep{\fill}}lrcccc} \hline
% 	\mc{2}{l}{Experiment} & $N(B\bar{B})$ & $A_{\rm ADS}$ & $R_{\rm ADS}$ & Correlation \\
% 	\hline
        \mc{5}{c}{$D K^{*+}$, $D \to K^-\pi^+$, $K^{*+} \to \KS \pi^+$} \\
	\babar & \cite{Aubert:2009yw} & $N(B\bar{B}) =$ 379M & $-0.34 \pm 0.43 \pm 0.16$ & $0.066 \pm 0.031 \pm 0.010$ \\
        LHCb   & \cite{LHCb-CONF-2016-014} & $\int {\cal L}\,dt = 4 \, {\rm fb}^{-1}$ & \textendash{} & $0.003 \pm 0.004$ \\
%        \hline
 	\mc{3}{l}{\bf Average} & $-0.34 \pm 0.46$ & $0.004 \pm 0.004$ \\
 	\mc{3}{l}{\small Confidence level} & \textendash{} & {\small $0.06~(1.9\sigma)$} \\
 		\hline
% 		\end{tabular*}
% 		\label{tab:cp_uta:yyy}
% 	\end{center}
% \end{table}

% \begin{table}[htb]
% 	\begin{center}
% 		\caption{
% 			Averages for $D_K\pi K\pi\pi$.
% 		}
% 		\vspace{0.2cm}
% 		\setlength{\tabcolsep}{0.0pc}
% 		\begin{tabular*}{\textwidth}{@{\extracolsep{\fill}}lrccc} \hline
% 	\mc{2}{l}{Experiment} & $N(B\bar{B})$ & $R_{\rm ADS}$ & Correlation \\
% 	\hline
        \mc{5}{c}{$D K^+\pi^+\pi^-$, $D \to K^-\pi^+$} \\
	LHCb & \cite{Aaij:2015ina} & $\int {\cal L}\,dt = 3 \, {\rm fb}^{-1}$ & $-0.32 \,^{+0.27}_{-0.34}$ & $0.0082 \,^{+0.0038}_{-0.0030}$ \\
        \hline
 		\end{tabular*}
                \label{tab:cp_uta:cus:ads}
 	\end{center}
 \end{table}

\begin{table}[htb]
	\begin{center}
		\caption{
%			Averages for $D_K\pi K$.
      Averages from ADS analyses of $b \to c\bar{u}d / u\bar{c}d$ modes.
                }
                \vspace{0.2cm}
                \setlength{\tabcolsep}{0.0pc}
\renewcommand{\arraystretch}{1.1}
                \begin{tabular*}{\textwidth}{@{\extracolsep{\fill}}lrccc} \hline 
        \mc{2}{l}{Experiment} & Sample size & $A_{\rm ADS}$ & $R_{\rm ADS}$ \\
        \hline
       \mc{5}{c}{$D \pi^+$, $D \to K^-\pi^+$} \\
	\babar & \cite{delAmoSanchez:2010dz} & $N(B\bar{B}) =$ 467M & $0.03 \pm 0.17 \pm 0.04$ & $0.0033 \pm 0.0006 \pm 0.0004$ \\
	\belle & \cite{Belle:2011ac} & $N(B\bar{B}) =$ 772M & $-0.04 \pm 0.11 \,^{+0.02}_{-0.01}$ & $0.00328 \,^{+0.00038}_{-0.00036} \,^{+0.00012}_{-0.00018}$ \\
	CDF & \cite{Aaltonen:2011uu} & $\int {\cal L}\,dt = 7 \, {\rm fb}^{-1}$ & $0.13 \pm 0.25 \pm 0.02$ & $0.0028 \pm 0.0007 \pm 0.0004$ \\
	LHCb & \cite{Aaij:2016oso} & $\int {\cal L}\,dt = 3 \, {\rm fb}^{-1}$ & $0.100 \pm 0.031 \pm 0.009$ & $0.00360 \pm 0.00012 \pm 0.00009$ \\
%	\hline
	\mc{3}{l}{\bf Average} & $0.088 \pm 0.030$ & $0.00353 \pm 0.00014$ \\
	\mc{3}{l}{\small Confidence level} & {\small $0.66~(0.4\sigma)$} & {\small $0.68~(0.4\sigma)$} \\
        \hline 
        \mc{5}{c}{$D \pi^+$, $D \to K^-\pi^+\pi^0$} \\
	\belle & \cite{Nayak:2013tgg} & 772M & $0.16 \pm 0.27 \,^{+0.03}_{-0.04}$ & $0.00189 \pm 0.00054 \,^{+0.00022}_{-0.00025}$ \\
	LHCb & \cite{Aaij:2015jna} & $\int {\cal L}\,dt = 3 \, {\rm fb}^{-1}$ & $0.44 \pm 0.19 \pm 0.01$ & $0.00235 \pm 0.00049 \pm 0.00004$ \\
	\mc{3}{l}{\bf Average} & $0.35 \pm 0.16$ & $0.00216 \pm 0.00038$ \\
	\mc{3}{l}{\small Confidence level} & {\small $0.40~(0.8\sigma)$} & {\small $0.55~(0.6\sigma)$} \\
        \hline 
        \mc{5}{c}{$D \pi^+$, $D \to K^-\pi^+\pi^+\pi^-$} \\
 	LHCb & \cite{Aaij:2016oso} & $\int {\cal L}\,dt = 3 \, {\rm fb}^{-1}$ & $0.023 \pm 0.048 \pm 0.005$ & $0.00377 \pm 0.00018 \pm 0.00006$ \\
        \hline
       \mc{5}{c}{$\Dstar \pi^+$, $\Dstar \to D\pi^0$, $D \to K^-\pi^+$} \\
	\babar & \cite{delAmoSanchez:2010dz} & 467M & $-0.09 \pm 0.27 \pm 0.05$ & $0.0032 \pm 0.0009 \pm 0.0008$ \\
        \hline 
        \mc{5}{c}{$\Dstar \pi^+$, $\Dstar \to D\gamma$, $D \to K^-\pi^+$} \\
	\babar & \cite{delAmoSanchez:2010dz} & 467M & $-0.65 \pm 0.55 \pm 0.22$ & $0.0027 \pm 0.0014 \pm 0.0022$ \\
        \hline
        \mc{5}{c}{$D \pi^+\pi^+\pi^-$, $D \to K^-\pi^+$} \\
        LHCb & \cite{Aaij:2015ina} & $\int {\cal L}\,dt = 3 \, {\rm fb}^{-1}$ & $-0.003 \pm 0.090$ & $0.00427 \pm 0.00043$ \\
        \hline
 		\end{tabular*}
                \label{tab:cp_uta:cus:ads2}
	\end{center}
\end{table}

\babar, \belle\ and LHCb have also presented results from a similar analysis method with self-tagging neutral $B$ decays: 
$\Bz \to DK^{*0}$ with $D \to K^-\pi^+$ (all), $D \to K^-\pi^+\pi^0$ and $D \to K^-\pi^+\pi^+\pi^-$ (\babar\ only).
All these results are obtained with the $K^{*0} \to K^+\pi^-$ decay. 
Effects due to the natural width of the $K^{*0}$ are
handled using the parametrisation suggested by Gronau~\cite{Gronau:2002mu}. 

The following 95\% CL limits are set by \babar~\cite{:2009au}:
\begin{equation}
  R_{\rm ADS}(K\pi) < 0.244 \hspace{5mm}
  R_{\rm ADS}(K\pi\pi^0) < 0.181 \hspace{5mm}
  R_{\rm ADS}(K\pi\pi\pi) < 0.391 \, ,
\end{equation}
while \belle~\cite{Negishi:2012uxa} obtain
\begin{equation}
  R_{\rm ADS}(K\pi) < 0.16 \, .
\end{equation}
The results from LHCb, which are presented in terms of the parameters $R_+$ and $R_-$ instead of $R_{\rm ADS}$ and $A_{\rm ADS}$, are given in Table~\ref{tab:cp_uta:ads-DKstar}.

\begin{table}[!htb]
        \begin{center}
                \caption{
%                        Averages for $D_K\pi K*$.
      Results from ADS analysis of $\Bz \to D\Kstarz$, $D \to K^-\pi^+$.
                }
                \vspace{0.2cm}
                \setlength{\tabcolsep}{0.0pc}
\renewcommand{\arraystretch}{1.1}
                \begin{tabular*}{\textwidth}{@{\extracolsep{\fill}}lrccc} \hline
        \mc{2}{l}{Experiment} & Sample size & $R_{+}$ & $R_{-}$ \\
        \hline
        LHCb & \cite{Aaij:2014eha} & $\int {\cal L}\,dt = 3 {\rm fb}^{-1}$ & $0.06 \pm 0.03 \pm 0.01$ & $0.06 \pm 0.03 \pm 0.01$ \\
        %% \hline
        %% \mc{3}{l}{\bf Average} & $0.06 \pm 0.03$ & $0.06 \pm 0.03$ & {\small uncorrelated averages} \\
        %% \mc{3}{l}{\small Confidence level} & {\small $0.xx~(y.y\sigma)$} & {\small $0.xx~(y.y\sigma)$} & \\
                \hline
                \end{tabular*}
                \label{tab:cp_uta:ads-DKstar}
        \end{center}
\end{table}

Combining the results and using additional input from
CLEO-c~\cite{Asner:2008ft,Lowery:2009id} a limit on the ratio between the 
$b \to u$ and $b \to c$ amplitudes of $\bar{r}_B(DK^{*0}) \in \left[ 0.07,0.41 \right]$ 
at 95\% CL limit is set by \babar.
Belle set a limit of $\bar{r}_B < 0.4$ at 95\% CL. 
LHCb take input from Sec.~\ref{sec:charm_physics} and obtain $\bar{r}_B = 0.240 \,^{+0.055}_{-0.048}$ (different from zero with $2.7\sigma$ significance). 

\mysubsubsection{$D$ decays to multiparticle self-conjugate final states (model-dependent analysis)}
\label{sec:cp_uta:cus:dalitz}

\begin{figure}[htbp]
  \begin{center}
    \resizebox{0.30\textwidth}{!}{
      \includegraphics{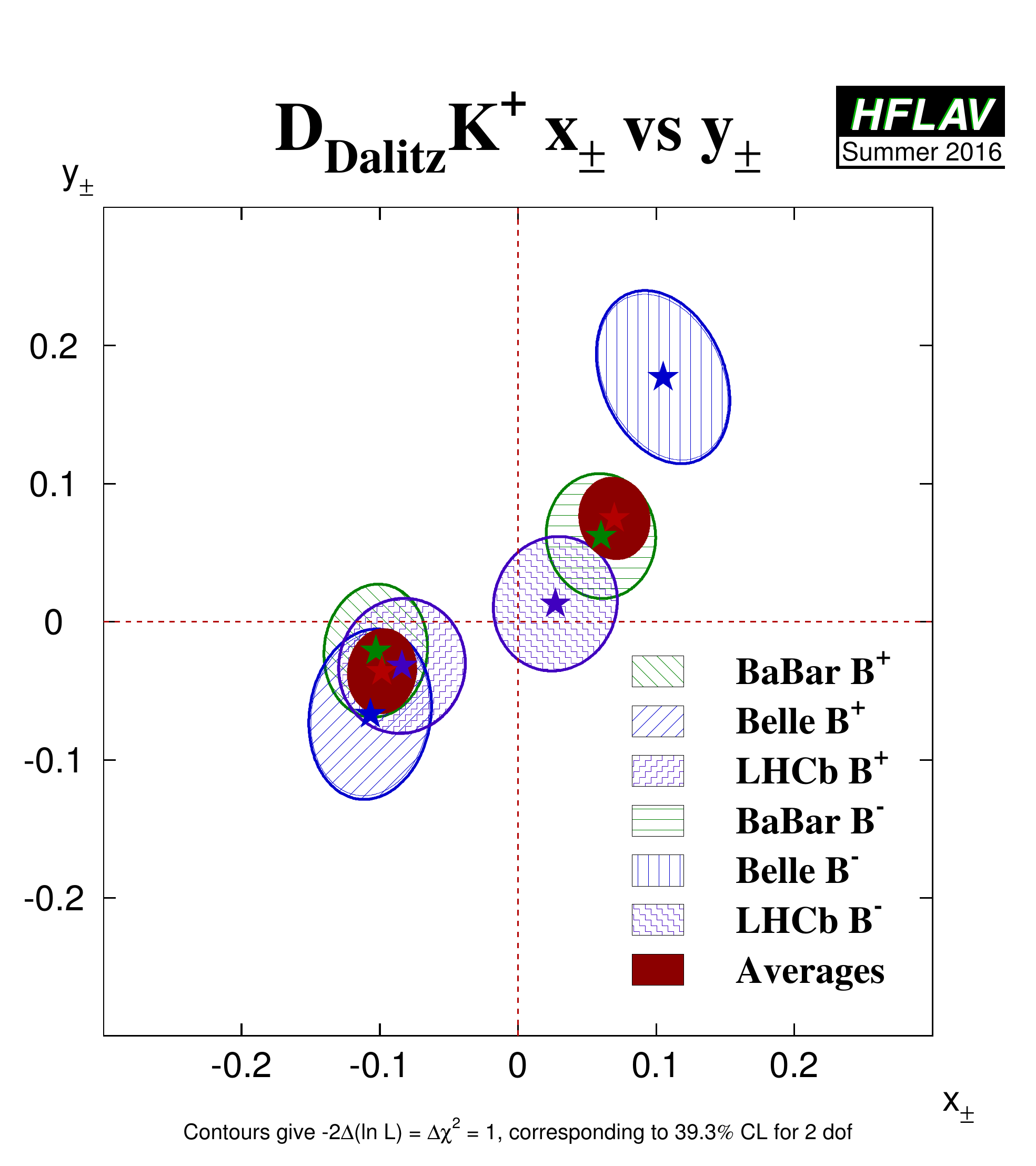}
    }
    \hfill
    \resizebox{0.30\textwidth}{!}{
      \includegraphics{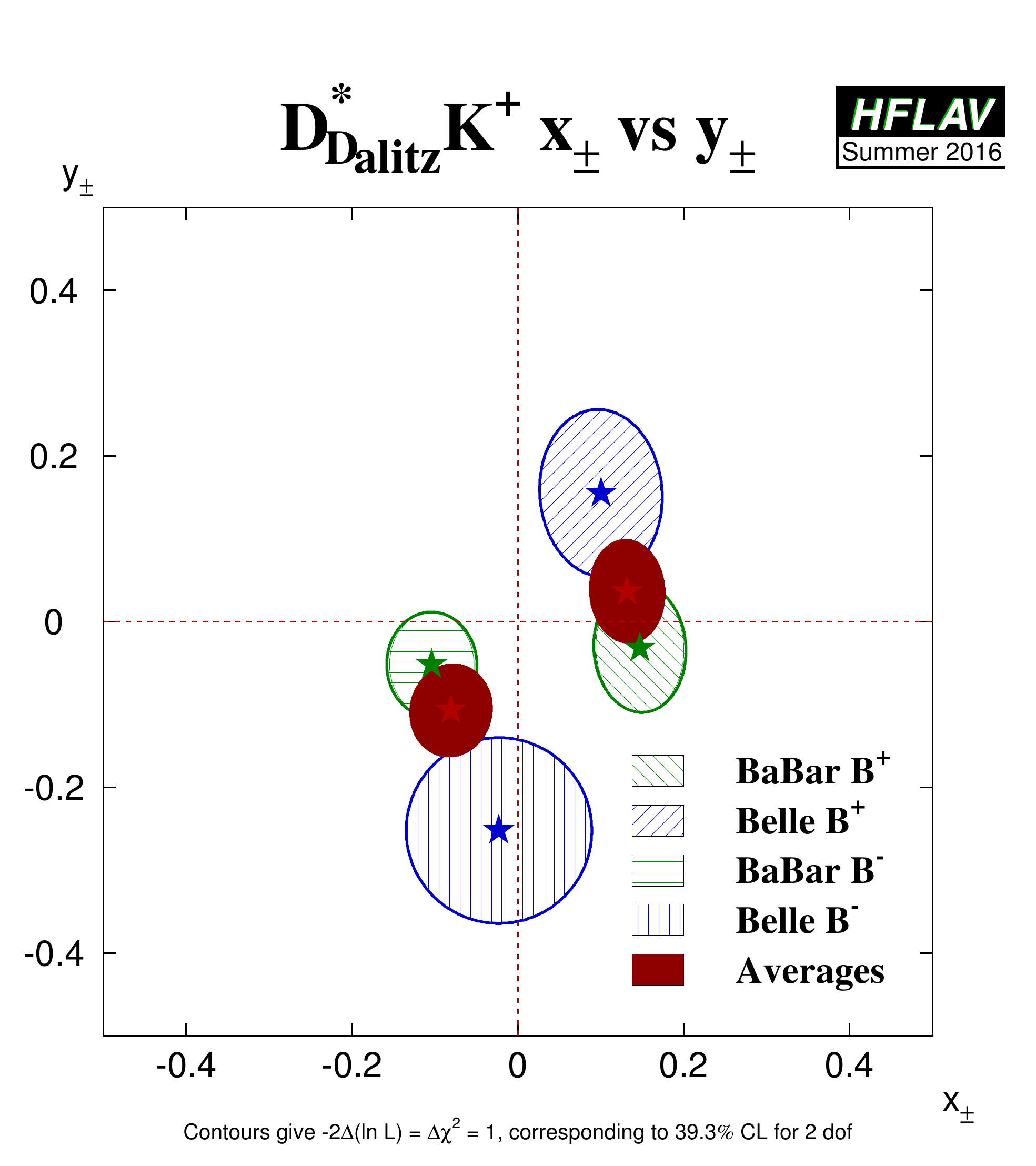}
    }
    \hfill
    \resizebox{0.30\textwidth}{!}{
      \includegraphics{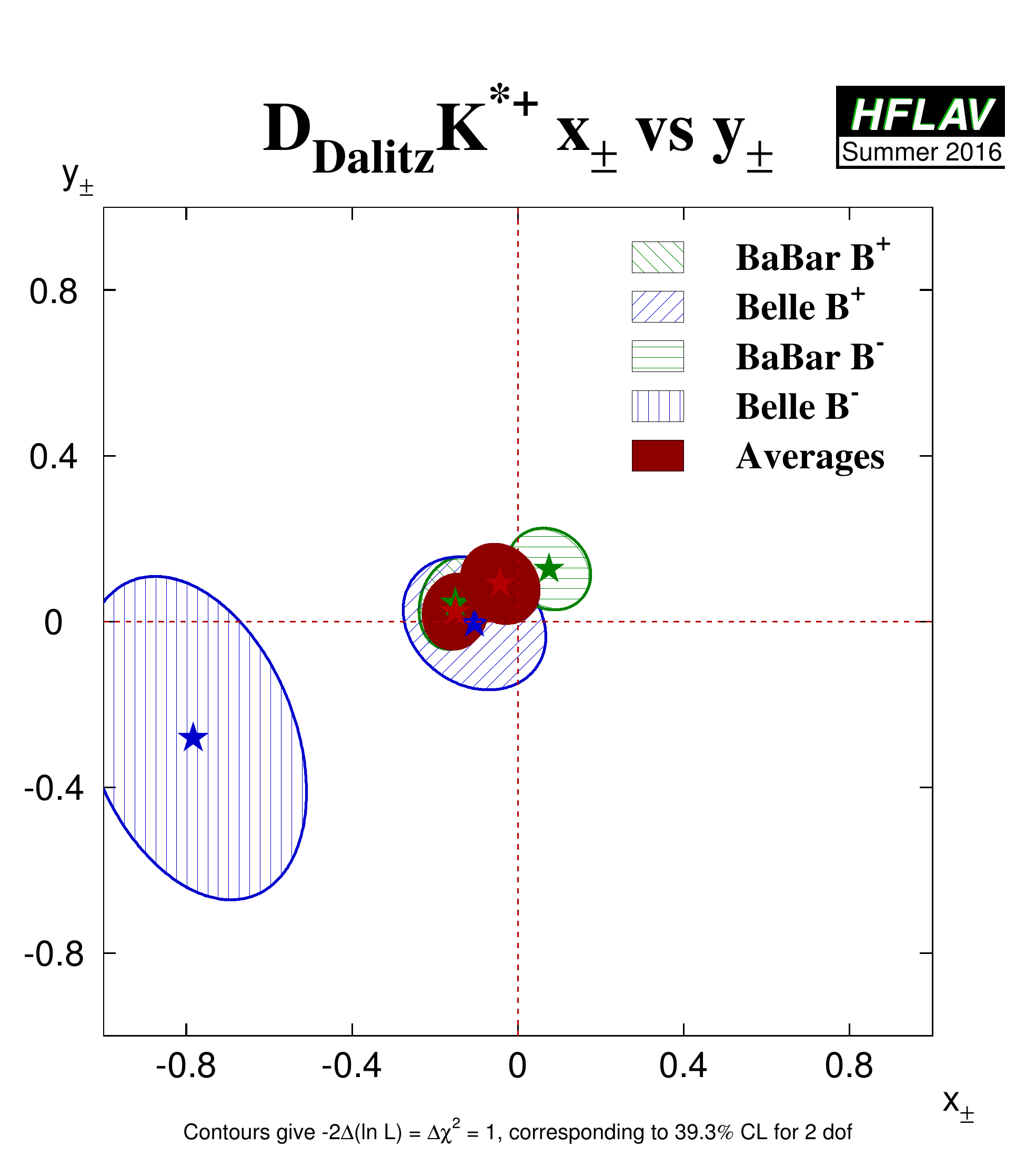}
    }
  \end{center}
  \vspace{-0.5cm}
  \caption{
    Contours in the $(x_\pm, y_\pm)$ from model-dependent analysis of $\Bp \to D^{(*)}K^{(*)+}$, $D \to \KS h^+ h^-$ ($h = \pi,K$).
    (Left) $\Bp \to D\Kp$, 
    (middle) $\Bp \to \Dstar\Kp$,
    (right) $\Bp \to D\Kstarp$.
    Note that the uncertainties assigned to the averages given in these plots
    do not include model errors.        
  }
  \label{fig:cp_uta:cus:dalitz_2d}
\end{figure}

\begin{figure}[htbp]
  \begin{center}
    \resizebox{0.40\textwidth}{!}{
      \includegraphics{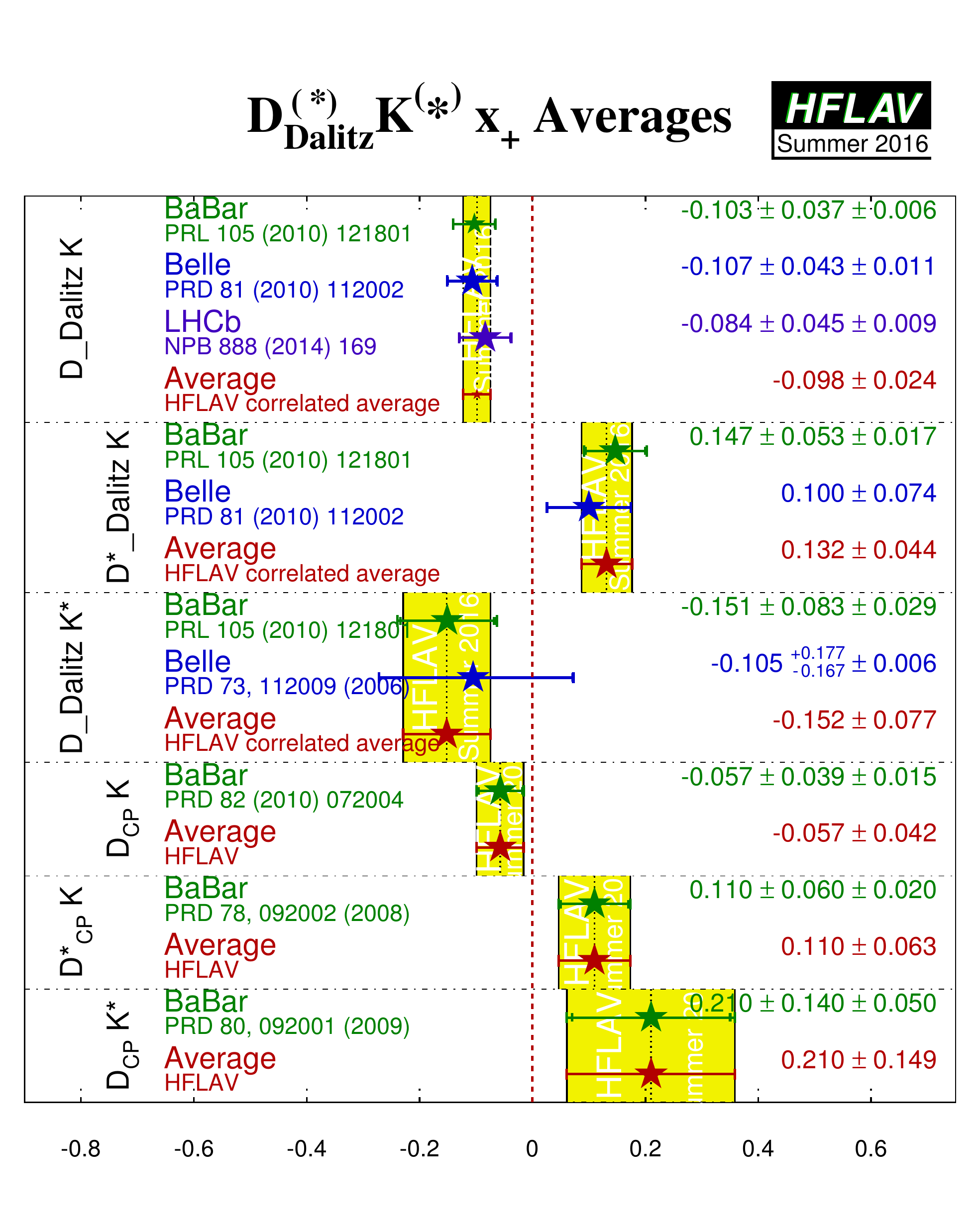}
    }
    \hspace{0.1\textwidth}
    \resizebox{0.40\textwidth}{!}{
      \includegraphics{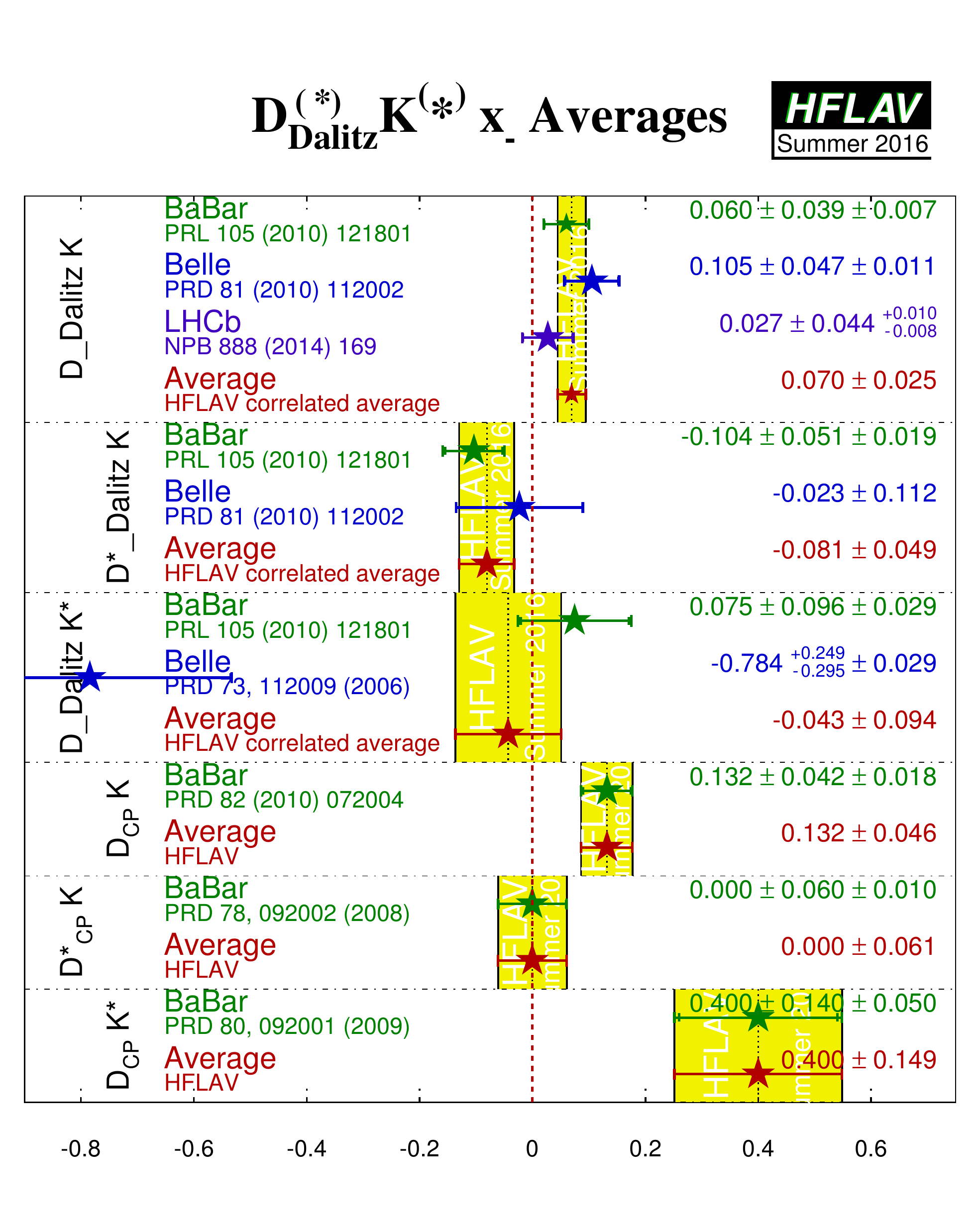}
    }
    \\
    \resizebox{0.40\textwidth}{!}{
      \includegraphics{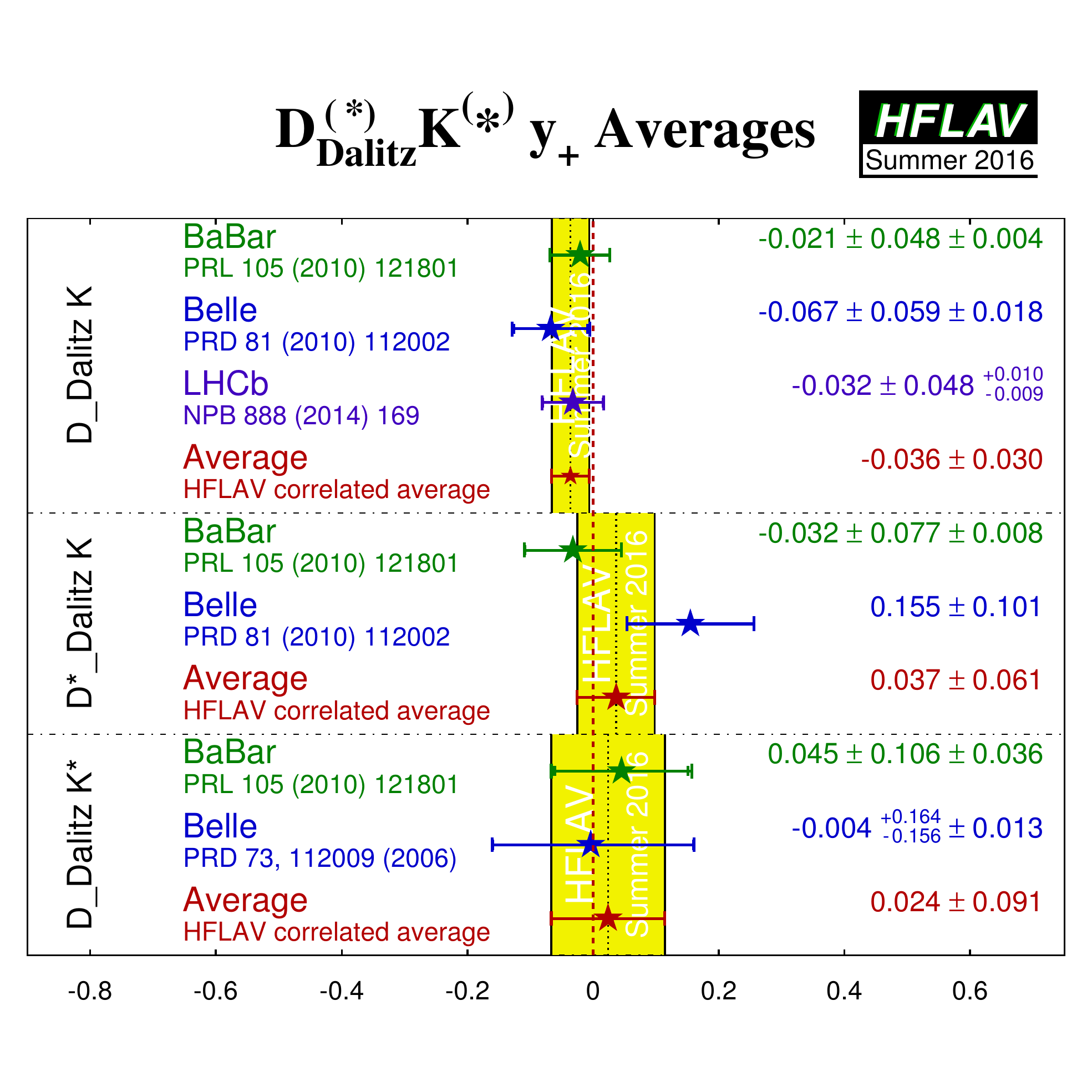}
    }
    \hspace{0.1\textwidth}
    \resizebox{0.40\textwidth}{!}{
      \includegraphics{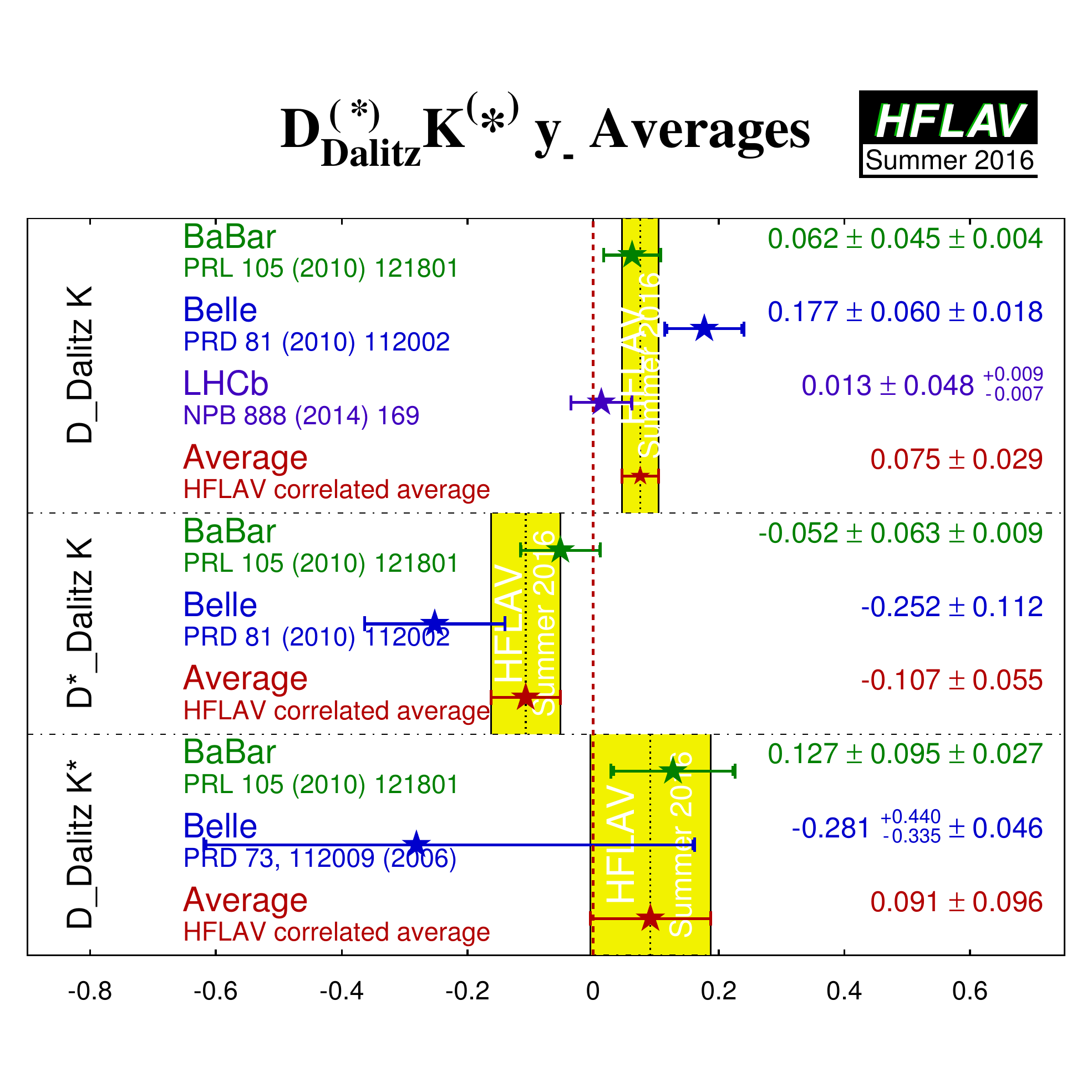}
    }
  \end{center}
  \vspace{-0.8cm}
  \caption{
    Averages of $(x_\pm, y_\pm)$ from model-dependent analyses of $\Bp \to D^{(*)}K^{(*)+}$ with $D \to \KS h^+h^-$ ($h=\pi,K$).
    (Top left) $x_+$, (top right) $x_-$,
    (bottom left) $y_+$, (bottom right) $y_-$.
    The top plots include constraints on $x_{\pm}$ obtained from GLW analyses (see Sec.~\ref{sec:cp_uta:cus:glw}).
    Note that the uncertainties assigned to the averages given in these plots
    do not include model errors.        
  }
  \label{fig:cp_uta:cus:dalitz_1d}
\end{figure}

For the model-dependent Dalitz plot analysis, both \babar\ and \belle\ have studied the modes 
$\Bp \to D\Kp$, $\Bp \to \Dstar\Kp$ and $\Bp \to D\Kstarp$.
For $\Bp \to \Dstar\Kp$,
both experiments have used both $\Dstar$ decay modes, $\Dstar \to D\pi^0$ and
$\Dstar \to D\gamma$, taking the effective shift in the strong phase
difference into account.\footnote{
  \belle~\cite{Poluektov:2010wz} quote separate results for $\Bp \to \Dstar\Kp$ with  $\Dstar \to D\pi^0$ and $\Dstar \to D\gamma$.
  The results quoted in Table~\ref{tab:cp_uta:cus:dalitz} are from our average, performed using the statistical correlations provided, and neglecting all systematic correlations; model uncertainties are not included. 
  The first uncertainty on the quoted results is combined statistical and systematic, the second is the model error (taken from the Belle results on $\Bp \to \Dstar\Kp$ with  $\Dstar \to D\pi^0$). 
}
In all cases the decay $D \to \KS\pi^+\pi^-$ has been used.
\babar\ also used the decay $D \to \KS K^+K^-$.
LHCb has also studied $\Bp \to D\Kp$ decays with $D \to \KS\pi^+\pi^-$.
\babar\ has also performed an analysis of $\Bp \to D\Kp$ with $D \to \pi^+\pi^-\pi^0$.
Results and averages are given in Table~\ref{tab:cp_uta:cus:dalitz}, and shown in Figs.~\ref{fig:cp_uta:cus:dalitz_2d} and~\ref{fig:cp_uta:cus:dalitz_1d}.
The third error on each measurement is due to $D$ decay model uncertainty.

The parameters measured in the analyses are explained in
Sec.~\ref{sec:cp_uta:notations:cus}.
All experiments measure the Cartesian variables, defined in Eq.~(\ref{eq:cp_uta:cartesian}), and perform frequentist statistical procedures, to convert these into measurements of $\gamma$, $r_B$ and $\delta_B$.
In the $\Bp \to D\Kp$ with $D \to \pi^+\pi^-\pi^0$ analysis,
the parameters $(\rho^{\pm}, \theta^\pm)$ are used instead.

Both experiments reconstruct $\Kstarp$ as $\KS\pip$,
but the treatment of possible nonresonant $\KS\pip$ differs:
\belle\ assign an additional model uncertainty,
while \babar\ use a parametrisation suggested by Gronau~\cite{Gronau:2002mu}.
The parameters $r_B$ and $\delta_B$ are replaced with 
effective parameters $\kappa \bar{r}_B$ and $\bar{\delta}_B$;
no attempt is made to extract the true hadronic parameters 
of the $\Bp \to D\Kstarp$ decay.

We perform averages using the following procedure, which is based on a set of
reasonable, though imperfect, assumptions. 

\begin{itemize}\setlength{\itemsep}{0.5ex}
\item 
  It is assumed that effects due to the different $D$ decay models 
  used by the two experiments are negligible. 
  Therefore, we do not rescale the results to a common model.
\item 
  It is further assumed that the model uncertainty is $100\%$ 
  correlated between experiments, 
  and therefore this source of error is not used in the averaging procedure.
  (This approximation is compromised by the fact that the \babar\ results
  include $D \to \KS K^+K^-$ decays in addition to $D \to \KS\pi^+\pi^-$.)
\item 
  We include in the average the effect of correlations 
  within each experiment's set of measurements.
\item 
  At present it is unclear how to assign an average model uncertainty. 
  We have not attempted to do so. 
  Our average includes only statistical and systematic errors. 
  An unknown amount of model uncertainty should be added to the final error.
\item 
  We follow the suggestion of Gronau~\cite{Gronau:2002mu} 
  in making the $DK^*$ averages. 
  Explicitly, we assume that the selection of $K^{*+} \to \KS\pip$
  is the same in both experiments 
  (so that $\kappa$, $\bar{r}_B$ and $\bar{\delta}_B$ are the same), 
  and drop the additional source of model uncertainty 
  assigned by Belle due to possible nonresonant decays.
\item 
  We do not consider common systematic errors, 
  other than the $D$ decay model. 
\end{itemize}

% \begin{table}[htb]
\begin{sidewaystable}
	\begin{center}
		\caption{
      Averages from model-dependent Dalitz plot analyses of $b \to c\bar{u}s / u\bar{c}s$ modes.
      Note that the uncertainities assigned to the averages do not include model errors.	
%			Averages for $D_Dalitz K$.
		}
		\vspace{0.2cm}
		\setlength{\tabcolsep}{0.0pc}
% make this tabular (not tabular*) and resize down to \textwidth
% change @{\extracolsep{\fill}} to @{\extracolsep{2mm}}
    \resizebox{\textwidth}{!}{
\renewcommand{\arraystretch}{1.2}
		\begin{tabular}{@{\extracolsep{2mm}}lrccccc} \hline
	\mc{2}{l}{Experiment} & Sample size & $x_+$ & $y_+$ & $x_-$ & $y_-$ \\
	\hline
        \mc{7}{c}{$D K^+$, $D \to \KS \pi^+\pi^-$} \\
	\babar & \cite{delAmoSanchez:2010rq} & $N(B\bar{B}) =$ 468M & $-0.103 \pm 0.037 \pm 0.006 \pm 0.007$ & $-0.021 \pm 0.048 \pm 0.004 \pm 0.009$ & $0.060 \pm 0.039 \pm 0.007 \pm 0.006$ & $0.062 \pm 0.045 \pm 0.004 \pm 0.006$ \\
	\belle & \cite{Poluektov:2010wz} & $N(B\bar{B}) =$ 657M & $-0.107 \pm 0.043 \pm 0.011 \pm 0.055$ & $-0.067 \pm 0.059 \pm 0.018 \pm 0.063$ & $0.105 \pm 0.047 \pm 0.011 \pm 0.064$ & $0.177 \pm 0.060 \pm 0.018 \pm 0.054$ \\
	LHCb & \cite{Aaij:2014iba} & $\int {\cal L}\,dt = 1 {\rm fb}^{-1}$ & $-0.084 \pm 0.045 \pm 0.009 \pm 0.005$ & $-0.032 \pm 0.048 \,^{+0.010}_{-0.009} \pm 0.008$ & $0.027 \pm 0.044 \,^{+0.010}_{-0.008} \pm 0.001$ & $0.013 \pm 0.048 \,^{+0.009}_{-0.007} \pm 0.003$ \\
% 	\hline
	\mc{3}{l}{\bf Average} & $-0.098 \pm 0.024$ & $-0.036 \pm 0.030$ & $0.070 \pm 0.025$ & $0.075 \pm 0.029$ \\
        \mc{3}{l}{\small Confidence level} &  \mc{4}{c}{\small $0.52~(0.7\sigma)$} \\
 		\hline
% 		\end{tabular*}
% 		\label{tab:cp_uta:yyy}
% 	\end{center}
% \end{table}

% \begin{table}[htb]
% 	\begin{center}
% 		\caption{
% 			Averages for $D*_Dalitz K$.
% 		}
% 		\vspace{0.2cm}
% 		\setlength{\tabcolsep}{0.0pc}
% 		\begin{tabular*}{\textwidth}{@{\extracolsep{\fill}}lrcccccc} \hline
% 		\mc{2}{l}{Experiment} & $N(B\bar{B})$ & $x+$ & $y+$ & $x-$ & $y-$ \\
% 		\hline
                \mc{7}{c}{$\Dstar K^+$, $\Dstar \to D\pi^0$ or $D\gamma$, $D \to \KS \pi^+\pi^-$} \\
	\babar & \cite{delAmoSanchez:2010rq} & $N(B\bar{B}) =$ 468M & $0.147 \pm 0.053 \pm 0.017 \pm 0.003$ & $-0.032 \pm 0.077 \pm 0.008 \pm 0.006$ & $-0.104 \pm 0.051 \pm 0.019 \pm 0.002$ & $-0.052 \pm 0.063 \pm 0.009 \pm 0.007$ \\
	\belle & \cite{Poluektov:2010wz} & $N(B\bar{B}) =$ 657M & $0.100 \pm 0.074 \pm 0.081$ & $0.155 \pm 0.101 \pm 0.063$ & $-0.023 \pm 0.112 \pm 0.090$ & $-0.252 \pm 0.112 \pm 0.049$ \\
% 	\hline
	\mc{3}{l}{\bf Average} & $0.132 \pm 0.044$ & $0.037 \pm 0.061$ & $-0.081 \pm 0.049$ & $-0.107 \pm 0.055$ \\
        \mc{3}{l}{\small Confidence level} & \mc{4}{c}{\small $0.22~(1.2\sigma)$} \\
 		\hline
% 		\end{tabular*}
% 		\label{tab:cp_uta:yyy}
% 	\end{center}
% \end{table}

% \begin{table}[htb]
% 	\begin{center}
% 		\caption{
% 			Averages for $D_Dalitz K*$.
% 		}
% 		\vspace{0.2cm}
% 		\setlength{\tabcolsep}{0.0pc}
% 		\begin{tabular*}{\textwidth}{@{\extracolsep{\fill}}lrcccccc} \hline
% 		\mc{2}{l}{Experiment} & $N(B\bar{B})$ & $x+$ & $y+$ & $x-$ & $y-$ \\
% 		\hline
                \mc{7}{c}{$D K^{*+}$, $D \to \KS \pi^+\pi^-$} \\
	\babar & \cite{delAmoSanchez:2010rq} & $N(B\bar{B}) =$ 468M & $-0.151 \pm 0.083 \pm 0.029 \pm 0.006$ & $0.045 \pm 0.106 \pm 0.036 \pm 0.008$ & $0.075 \pm 0.096 \pm 0.029 \pm 0.007$ & $0.127 \pm 0.095 \pm 0.027 \pm 0.006$ \\
 	\belle & \cite{Poluektov:2006ia} & $N(B\bar{B}) =$ 386M & $-0.105 \,^{+0.177}_{-0.167} \pm 0.006 \pm 0.088$ & $-0.004 \,^{+0.164}_{-0.156} \pm 0.013 \pm 0.095$ & $-0.784 \,^{+0.249}_{-0.295} \pm 0.029 \pm 0.097$ & $-0.281 \,^{+0.440}_{-0.335} \pm 0.046 \pm 0.086$ \\
% 	\hline
	\mc{3}{l}{\bf Average} & $-0.152 \pm 0.077$ & $0.024 \pm 0.091$ & $-0.043 \pm 0.094$ & $0.091 \pm 0.096$ \\
        \mc{3}{l}{\small Confidence level} & \mc{4}{c}{\small $0.011~(2.5\sigma)$} \\
 		\hline
%		\end{tabular}
% 		\label{tab:cp_uta:yyy}
% 	\end{center}
% \end{table}

% \begin{table}[!htb]
% 	\begin{center}
% 		\caption{
% 			Averages for $D_Dalitz K*0$.
% 		}
% 		\vspace{0.2cm}
% 		\setlength{\tabcolsep}{0.0pc}
% 		\begin{tabular*}{\textwidth}{@{\extracolsep{\fill}}lrcccccc} \hline
% 	\mc{2}{l}{Experiment} & $N(B\bar{B})$ & $x-$ & $y-$ & $x+$ & $y+$ & Correlation \\
% 	\hline
                \mc{7}{c}{$D K^{*0}$, $D \to \KS \pi^+\pi^-$, $K^{*0} \to \Kp\pim$} \\
	LHCb & \cite{Aaij:2016zlt} & $\int {\cal L}\,dt = 3 \, {\rm fb}^{-1}$ & $0.05 \pm 0.24 \pm 0.04 \pm 0.01$ & $-0.65 \,^{+0.24}_{-0.23} \pm 0.08 \pm 0.01$ & $-0.15 \pm 0.14 \pm 0.03 \pm 0.01$ & $0.25 \pm 0.15 \pm 0.06 \pm 0.01$ \\
% 	\hline
% 	\mc{3}{l}{\bf Average} & $-0.15 \pm 0.14$ & $0.25 \pm 0.16$ & $0.05 \pm 0.24$ & $-0.65 \pm 0.25$ & \textendash{} \\
% 	\mc{3}{l}{\small Confidence level} & \mc{4}{c}{\small $0.xx~(y.y\sigma)$} & \\
		\hline
% 		\end{tabular*}
% 		\label{tab:cp_uta:yyy}
% 	\end{center}
% \end{table}

                \vspace{1ex} \\

% \begin{table}[htb]
% 	\begin{center}
% 		\caption{
% 			Averages for $D_\pi\pi\\pi^{0} K$.
% 		}
% 		\vspace{0.2cm}
% 		\setlength{\tabcolsep}{0.0pc}
% 		\begin{tabular*}{\textwidth}{@{\extracolsep{\fill}}lrcccccc} \hline
	\hline
	\mc{2}{l}{Experiment} & $N(B\bar{B})$ & $\rho^{+}$ & $\theta^+$ & $\rho^{-}$ & $\theta^-$ \\
	\hline
        \mc{7}{c}{$D K^+$, $D \to \pi^+\pi^-\pi^0$} \\
	\babar & \cite{Aubert:2007ii} & 324M & $0.75 \pm 0.11 \pm 0.04$ & $147 \pm 23 \pm 1$ & $0.72 \pm 0.11 \pm 0.04$ & $173 \pm 42 \pm 2$ \\
	\hline
%	\mc{3}{l}{\bf Average} & $0.750 \pm 0.117$ & $147.000 \pm 23.022$ & $0.720 \pm 0.117$ & $173.000 \pm 42.048$ & \textendash{} \\
%	\mc{3}{l}{\small Confidence level} & \mc{4}{c}{\small $0.xx~(y.y\sigma)$} & \\
%		\hline
		\end{tabular}
%		\label{tab:cp_uta:yyy}
              }
		\label{tab:cp_uta:cus:dalitz}
	\end{center}
\end{sidewaystable}
% \end{table}

\vspace{3ex}

\noindent
\underline{Constraints on $\gamma \equiv \phi_3$}

The measurements of $(x_\pm, y_\pm)$ can be used to obtain constraints on 
$\gamma \equiv \phi_3$, as well as the hadronic parameters $r_B$ and $\delta_B$.
\babar~\cite{delAmoSanchez:2010rq},
\belle~\cite{Poluektov:2010wz,Poluektov:2006ia} and
LHCb~\cite{Aaij:2014iba} have all done so using a frequentist procedure 
(there are some differences in the details of the techniques used).

\begin{itemize}\setlength{\itemsep}{0.5ex}

\item 
  \babar\ obtain $\gamma = (68 \,^{+15}_{-14} \pm 4 \pm 3)^\circ$
  from $D\Kp$, $\Dstar\Kp$ and $D\Kstarp$.

\item
  \belle\ obtain $\phi_3 = (78 \,^{+11}_{-12} \pm 4 \pm 9)^\circ$
  from $D\Kp$ and $\Dstar\Kp$.

\item 
  LHCb obtain $\gamma = (84 \,^{+49}_{-42})^\circ$
  from $D\Kp$ using 1 ${\rm fb}^{-1}$ of data (a more precise result using 3 ${\rm fb}^{-1}$ and the model-independent method is reported below).

\item
  The experiments also obtain values for the hadronic parameters as detailed
  in Table~\ref{tab:cp_uta:rBdeltaB_summary}.

%% \item 
%%   Improved constraints can be achieved combining the information from
%%   $\Bp \to D\Kp$ analysis with different $D$ decay modes.
%%   The experiments have not yet published such results,
%%   and none are listed here.

% \item 
%   The CKMfitter~\cite{Charles:2004jd} and 
%   UTFit~\cite{Bona:2005vz} groups use the measurements 
%   from \belle\ and \babar\ given above
%   to obtain combined constraints on $\gamma$.

\item 
  In the \babar\ analysis of $\Bp \to D\Kp$ with 
  $D \to \pi^+\pi^-\pi^0$ decays~\cite{Aubert:2007ii},
  a constraint of $-30^\circ < \gamma < 76^\circ$ is obtained 
  at the 68\% confidence level.

\item
  The results discussed here are included in the HFLAV combination to obtain a world average value for $\gamma \equiv \phi_3$, as discussed in Sec.~\ref{sec:cp_uta:cus:gamma}.

\end{itemize}

\begin{table}
  \begin{center}
  \caption{
    Summary of constraints on hadronic parameters from model-dependent analyses of $\Bp \to \DorDstar\KorKstarp$ and $\Bz \to D\Kstarz$ decays.
    Note the alternative parametrisation of the hadronic parameters used by \babar\ in the $D\Kstarp$ mode.
  }
  \label{tab:cp_uta:rBdeltaB_summary}
  \renewcommand{\arraystretch}{1.1}
  \begin{tabular}{lcc}
    \hline
    & $r_B$ & $\delta_B$ \\
    \hline
    \multicolumn{3}{c}{In $D\Kp$} \\
    \babar & $0.096 \pm 0.029 \pm 0.005 \pm 0.004$ & $(119 \,^{+19}_{-20} \pm 3 \pm 3)^\circ$ \\
    \belle & $0.160 \,^{+0.040}_{-0.038} \pm 0.011 \,^{+0.05}_{-0.010}$ & 
    $(138 \,^{+13}_{-16} \pm 4 \pm 23)^\circ$ \\
    LHCb & $0.06 \pm 0.04$ & $(115 \,^{+41}_{-51})^\circ$ \\
    \hline
    \multicolumn{3}{c}{In $\Dstar\Kp$} \\
    \babar & $0.133 \,^{+0.042}_{-0.039} \pm 0.014 \pm 0.003$ & $(-82 \pm 21 \pm 5 \pm 3)^\circ$ \\
    \belle & $0.196 \,^{+0.072}_{-0.069} \pm 0.012 \,^{+0.062}_{-0.012}$ &
    $(342 \,^{+19}_{-21} \pm 3 \pm 23)^\circ$ \\
    \hline \\ [-2.4ex]
    & $\bar{r}_B$ & $\bar{\delta}_B$ \\
    \hline
    \multicolumn{3}{c}{In $D\Kstarp$} \\
    \babar & $\kappa \bar{r}_B = 0.149 \,^{+0.066}_{-0.062} \pm 0.026 \pm 0.006$ & $(111 \pm 32 \pm 11 \pm 3)^\circ$ \\
    \belle & $0.56 \,^{+0.22}_{-0.16} \pm 0.04 \pm 0.08$ & 
    $(243 \,^{+20}_{-23} \pm 3 \pm 50)^\circ$ \\
    \hline
    \multicolumn{3}{c}{In $D\Kstarz$} \\
    \babar & $< 0.55$ at 95\% probability & $(62 \pm 57)^\circ$ \\
    LHCb & $0.39 \pm 0.13$ & $(197 \,^{+24}_{-20})^\circ$ \\
    \hline
  \end{tabular}
  \end{center}
\end{table}

% At present we make no attempt to provide an HFLAV average for $\gamma$,
% nor indeed for the hadronic parameters.
% More details on procedures to calculate a best fit value for $\gamma$ 
% can be found in Refs.~\cite{Charles:2004jd,Bona:2005vz}.

\babar\ and LHCb have performed a similar analysis using the self-tagging neutral $B$ decay $\Bz \to DK^{*0}$ (with $K^{*0} \to K^+\pi^-$). 
Effects due to the natural width of the $K^{*0}$ are handled using the parametrisation suggested by Gronau~\cite{Gronau:2002mu}.
LHCb give results in terms of the Cartesian parameters, as shown in Table~\ref{tab:cp_uta:cus:dalitz}.
\babar~\cite{Aubert:2008yn} present results only in terms of $\gamma$ and the hadronic parameters.
The obtained constraints on $\gamma \equiv \phi_3$ are 
\begin{itemize}\setlength{\itemsep}{0.5ex}
\item 
  \babar\ obtain $\gamma = (162 \pm 56)^\circ$
\item
  LHCb obtain $\gamma = (80 \,^{+21}_{-22})^\circ$
\item
  Values for the hadronic parameters are given in Table~\ref{tab:cp_uta:rBdeltaB_summary}.
\end{itemize}
% Note that there is an ambiguity in the solutions 
% $\left( \gamma, \bar{\delta}_B \leftrightarrow \gamma+\pi, \bar{\delta}_B+\pi \right)$.

\mysubsubsection{$D$ decays to multiparticle self-conjugate final states (model-independent analysis)}
\label{sec:cp_uta:cus:dalitz:modInd}

A model-independent approach to the analysis of $\Bp \to \DorDstar \Kp$ with multibody $D$ decays was proposed by Giri, Grossman, Soffer and Zupan~\cite{Giri:2003ty}, and further developed by Bondar and Poluektov~\cite{Bondar:2005ki,Bondar:2008hh}. 
The method relies on information on the average strong phase difference between $\Dz$ and $\Dzb$ decays in bins of Dalitz plot position that can be obtained from quantum-correlated $\psi(3770) \to \Dz\Dzb$ events. 
This information is measured in the form of parameters $c_i$ and $s_i$ that are the amplitude weighted averages of the cosine and sine of the strong phase difference in a Dalitz plot bin labelled by $i$, respectively. 
These quantities have been obtained for $D \to \KS \pi^+\pi^-$ (and $D \to \KS K^+K^-$) decays by CLEO-c~\cite{Briere:2009aa,Libby:2010nu}.  
% (Preliminary results from BESIII are also available.)

\belle~\cite{Aihara:2012aw} and LHCb~\cite{Aaij:2014uva} have used the model-independent Dalitz plot analysis approach to study the mode $\Bp \to D\Kp$.
Both \belle~\cite{Negishi:2015vqa} and LHCb~\cite{Aaij:2016nao} have also used this approach to study $\Bz \to DK^*(892)^0$ decays.
In both cases, the experiments use $D \to \KS\pi^+\pi^-$ decays while LHCb has also included the $D \to \KS K^+K^-$ decay.
The Cartesian variables $(x_\pm, y_\pm)$, defined in Eq.~(\ref{eq:cp_uta:cartesian}), are determined from the data. 
Note that due to the strong statistical and systematic correlations with the model-dependent results given in Sec.~\ref{sec:cp_uta:cus:dalitz}, these results cannot be combined. 

\begin{figure}[htbp]
  \begin{center}
    \resizebox{0.45\textwidth}{!}{
      \includegraphics{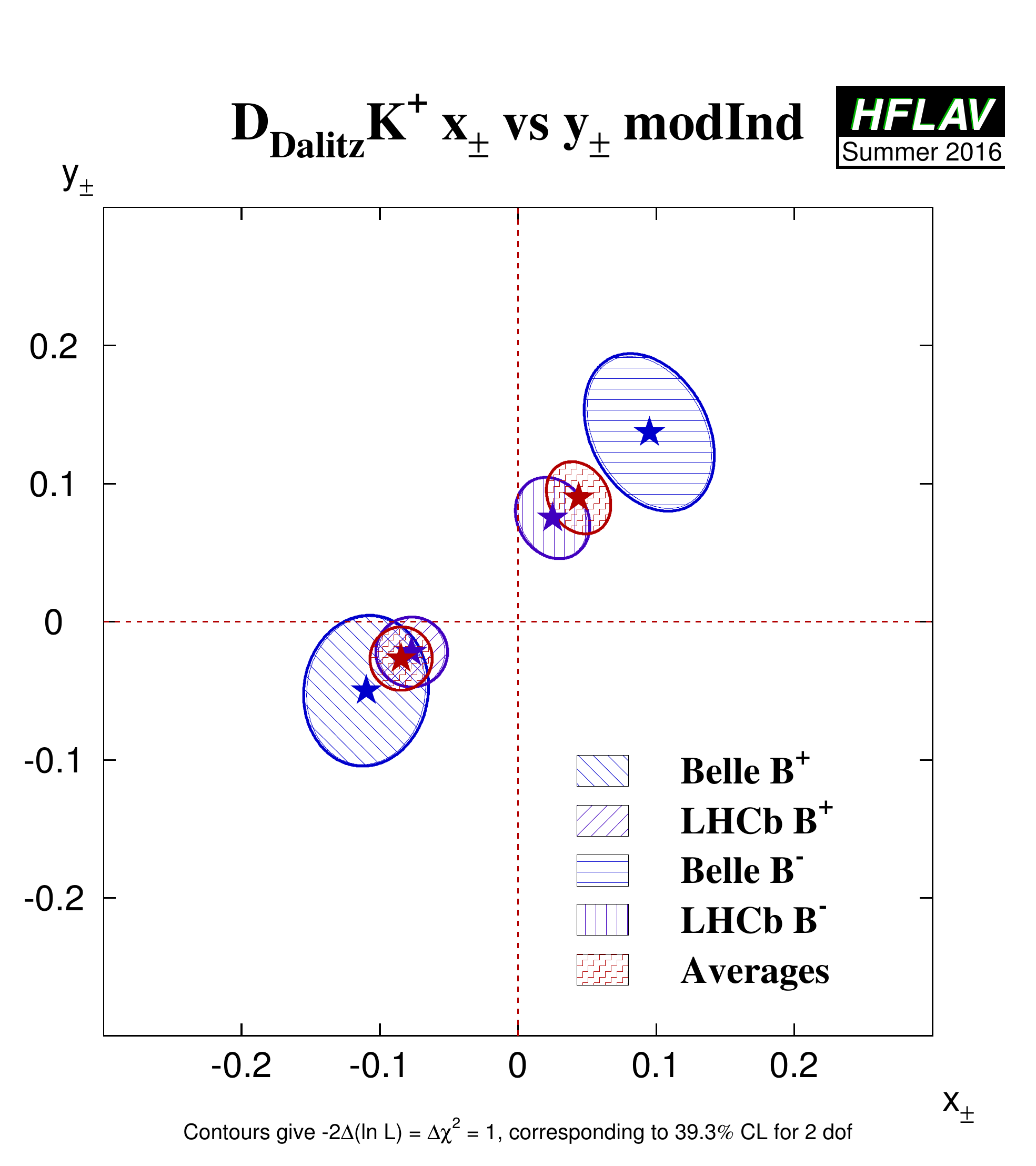}
    }
%     \hfill
%     \resizebox{0.30\textwidth}{!}{
%       \includegraphics{figures/cp_uta/D_DalitzKstarzx+vsy+}
%     }
%     \hfill
%     \resizebox{0.30\textwidth}{!}{
%       \includegraphics{figures/cp_uta/D_DalitzKstarzx-vsy-}
%     }
  \end{center}
  \vspace{-0.5cm}
  \caption{
    Contours in the $(x_\pm, y_\pm)$ plane from model-independent analysis of 
    $\Bp \to D\Kp$ with $D \to \KS h^+ h^-$ ($h = \pi,K$).
%     (left) $\Bp \to D\Kp$, (middle and right) $\Bz \to DK^*(892)^0$,
%     all with $D \to \KS h^+ h^-$,($h = \pi,K$).
  }
  \label{fig:cp_uta:cus:dalitz-modInd_2d}
\end{figure}

The results and averages are given in Table~\ref{tab:cp_uta:cus:dalitz-modInd}, and shown in Figs.~\ref{fig:cp_uta:cus:dalitz-modInd_2d}.
Most results have three sets of errors, which are statistical, systematic, and uncertainty coming from the knowledge of $c_i$ and $s_i$ respectively. 
To perform the average, we remove the last uncertainty, which should be 100\% correlated between the measurements. 
Since the size of the uncertainty from $c_i$ and $s_i$ is found to depend on the size of the $B \to DK$ data sample, we assign the LHCb uncertainties (which are mostly the smaller of the Belle and LHCb values) to the averaged result. 
This procedure should be conservative. 
In the LHCb $\Bz \to DK^*(892)^0$ results~\cite{Aaij:2016nao}, the values of $c_i$ and $s_i$ are constrained to their measured values within uncertainties in the fit to data, and hence the effect is absorbed in their statistical uncertainties.
The $\Bz \to DK^*(892)^0$ average is performed neglecting the model uncertainties on the Belle results.

% \begin{table}[htb]
\begin{sidewaystable}
	\begin{center}
		\caption{
      Averages from model-independent Dalitz plot analyses of $b \to c\bar{u}s / u\bar{c}s$ modes.
%			Averages for $D_Dalitz K$.
		}
		\vspace{0.2cm}
		\setlength{\tabcolsep}{0.0pc}
% make this tabular (not tabular*) and resize down to \textwidth
% change @{\extracolsep{\fill}} to @{\extracolsep{2mm}}
    \resizebox{\textwidth}{!}{
\renewcommand{\arraystretch}{1.2}
		\begin{tabular}{@{\extracolsep{2mm}}lrccccc} \hline
	\mc{2}{l}{Experiment} & Sample size & $x_+$ & $y_+$ & $x_-$ & $y_-$ \\
	\hline
        \mc{7}{c}{$D K^+$, $D \to \KS \pi^+\pi^-$} \\
	\belle & \cite{Aihara:2012aw} & $N(B\bar{B}) =$ 772M & $-0.110 \pm 0.043 \pm 0.014 \pm 0.007$ & $-0.050 \,^{+0.052}_{-0.055} \pm 0.011 \pm 0.007$ & $0.095 \pm 0.045 \pm 0.014 \pm 0.010$ & $0.137 \,^{+0.053}_{-0.057} \pm 0.015 \pm 0.023$ \\
	LHCb & \cite{Aaij:2014uva} & $\int {\cal L}\,dt = 3 {\rm fb}^{-1}$ & $-0.077 \pm 0.024 \pm 0.010 \pm 0.004$ & $-0.022 \pm 0.025 \pm 0.004 \pm 0.010$ & $0.025 \pm 0.025 \pm 0.010 \pm 0.005$ & $0.075 \pm 0.029 \pm 0.005 \pm 0.014$ \\
% 	\hline
	\mc{3}{l}{\bf Average} & $-0.085 \pm 0.023 \pm 0.04$ & $-0.027 \pm 0.023 \pm 0.010$ & $0.044 \pm 0.023 \pm 0.005$ & $0.090 \pm 0.026 \pm 0.014$ \\
        \mc{3}{l}{\small Confidence level} &  \mc{4}{c}{\small $0.39~(0.9\sigma)$} \\
 		\hline
        \mc{7}{c}{$D K^{*0}$, $D \to \KS \pi^+\pi^-$} \\
	\belle & \cite{Negishi:2015vqa} & $N(B\bar{B}) =$ 772M & $0.1 \,^{+0.7}_{-0.4} \,^{+0.0}_{-0.1} \pm 0.1$ & $0.3 \,^{+0.5}_{-0.8} \,^{+0.0}_{-0.1} \pm 0.1$ & $0.4 \,^{+1.0}_{-0.6} \,^{+0.0}_{-0.1} \pm 0.0$ & $-0.6 \,^{+0.8}_{-1.0} \,^{+0.1}_{-0.0} \pm 0.1$ \\
	LHCb & \cite{Aaij:2016nao} & $\int {\cal L}\,dt = 3 {\rm fb}^{-1}$ & $0.05 \pm 0.35 \pm 0.02$ & $-0.81 \pm 0.28 \pm 0.06$ & $-0.31 \pm 0.20 \pm 0.04$ & $0.31 \pm 0.21 \pm 0.05$ \\
%	\hline
	\mc{3}{l}{\bf Average} & $0.10 \pm 0.30$ & $-0.63 \pm 0.26$ & $-0.27 \pm 0.20$ & $0.27 \pm 0.21$ \\
	\mc{3}{l}{\small Confidence level} & \mc{4}{c}{\small $0.38~(0.9\sigma)$} \\
 		\hline
		\end{tabular}
              }
		\label{tab:cp_uta:cus:dalitz-modInd}
	\end{center}
\end{sidewaystable}
% \end{table}

\begin{sidewaystable}
	\begin{center}
		\caption{
      Results from model-independent Dalitz plot analysis of $B^+ \to DK^+$, $D \to \KS\Kpm\pimp$.
%			Averages for $D_K_{S}K\pi K modInd$.
		}
		\setlength{\tabcolsep}{0.0pc}
% make this tabular (not tabular*) and resize down to \textwidth
% change @{\extracolsep{\fill}} to @{\extracolsep{2mm}}
    \resizebox{\textwidth}{!}{
\renewcommand{\arraystretch}{1.2}
		\begin{tabular}{@{\extracolsep{4mm}}lrccccccc} \hline
	\mc{2}{l}{Experiment} & $\int {\cal L}\,dt$ & $R_{\rm SS}$ & $R_{\rm OS}$ & $A_{{\rm SS},DK}$ & $A_{{\rm OS},DK}$ & $A_{{\rm SS},D\pi}$ & $A_{{\rm OS},D\pi}$ \\
	\hline
        \mc{9}{c}{$D \to \KS\Kpm\pimp$ (whole Dalitz plot)} \\
	LHCb & \cite{Aaij:2014dia} & $3 \ {\rm fb}^{-1}$ & $0.092 \pm 0.009 \pm 0.004$ & $0.066 \pm 0.009 \pm 0.002$ & $0.040 \pm 0.091 \pm 0.018$ & $0.233 \pm 0.129 \pm 0.024$ & $-0.025 \pm 0.024 \pm 0.010$ & $-0.052 \pm 0.029 \pm 0.017$ \\
	\hline
        \mc{9}{c}{$D \to K^*(892)^\pm\Kmp$} \\
	LHCb & \cite{Aaij:2014dia} & $3 \ {\rm fb}^{-1}$ & $0.084 \pm 0.011 \pm 0.003$ & $0.056 \pm 0.013 \pm 0.002$ & $0.026 \pm 0.109 \pm 0.029$ & $0.336 \pm 0.208 \pm 0.026$ & $-0.012 \pm 0.028 \pm 0.010$ & $-0.054 \pm 0.043 \pm 0.017$ \\
        \hline
                \end{tabular}
    }
		\label{tab:cp_uta:cus:KSKpi-modInd}
	\end{center}
\end{sidewaystable}

\vspace{3ex}

\noindent
\underline{Constraints on $\gamma \equiv \phi_3$}

The measurements of $(x_\pm, y_\pm)$ can be used to obtain constraints on 
$\gamma$, as well as the hadronic parameters $r_B$ and $\delta_B$.
The experiments have done so using frequentist procedures (there are some differences in the details of the techniques used).

\begin{itemize}\setlength{\itemsep}{0.5ex}

\item 
  From $\Bp \to D\Kp$, \belle~\cite{Aihara:2012aw} obtain
  $\phi_3 = (77.3 \,^{+15.1}_{-14.9} \pm 4.1 \pm 4.3)^\circ$.

\item
  From $\Bp \to D\Kp$, LHCb~\cite{Aaij:2014uva} obtain
  $\gamma = (62 \,^{+15}_{-14})^\circ$.

\item 
  From $\Bz \to DK^*(892)^0$, LHCb~\cite{Aaij:2016nao} obtain 
  $\gamma = (71 \pm 20)^\circ$.

\item
  The experiments also obtain values for the hadronic parameters as detailed
  in Table~\ref{tab:cp_uta:rBdeltaB_summary-modInd}.

\item
  The results discussed here are included in the HFLAV combination to obtain a world average value for $\gamma \equiv \phi_3$, as discussed in Sec.~\ref{sec:cp_uta:cus:gamma}.

\end{itemize}

\begin{table}
  \begin{center}
  \caption{
    Summary of constraints on hadronic parameters from model-independent analyses of $\Bp \to D\Kp$ and $\Bz \to D\Kstarz$, $D \to \KS h^+h^-$ ($h=\pi,K$) decays.
  }
  \label{tab:cp_uta:rBdeltaB_summary-modInd}
  \renewcommand{\arraystretch}{1.1}
  \begin{tabular}{lcc}
    \hline
    & $r_B(D\Kp)$ & $\delta_B(D\Kp)$ \\
    \hline
    \belle & $0.145 \pm 0.030 \pm 0.010 \pm 0.011$ & $(129.9 \pm 15.0 \pm 3.8 \pm 4.7)^\circ$ \\
    LHCb & $0.080 \,^{+0.019}_{-0.021}$ & $(134 \,^{+14}_{-15})^\circ$ \\
    \hline
    & $\bar{r}_B(DK^{*0})$ & $\bar{\delta}_B(DK^{*0})$ \\
    \belle & $< 0.87 \ \text{at 68\% confidence level}$ & \\
    LHCb & $0.56 \pm 0.17$ & $(204 \,^{+21}_{-20})^\circ$ \\
    \hline
  \end{tabular}
  \end{center}
\end{table}

% At present we make no attempt to provide an HFLAV average for $\gamma$,
% nor indeed for the hadronic parameters.
% More details on procedures to calculate a best fit value for $\gamma$ 
% can be found in Refs.~\cite{Charles:2004jd,Bona:2005vz}.

\mysubsubsection{$D$ decays to multiparticle non-self-conjugate final states (model-independent analysis)}
\label{sec:cp_uta:cus:dalitz:KsKpi}

Following the original suggestion of Grossman, Ligeti and Soffer~\cite{Grossman:2002aq}, decays of $D$ mesons to $\KS\Kpm\pimp$ can be used in a similar approach to that discussed above to determine $\gamma \equiv \phi_3$. 
Since these decays are less abundant, the event samples available to date have not been sufficient for a fine binning of the Dalitz plots, but the analysis can be performed using only an overall coherence factor and related strong phase difference for the decay. 
These quantities have been determined by CLEO-c~\cite{Insler:2012pm} both for the full Dalitz plots and in a restricted region $\pm 100 \ {\rm MeV}/c^2$ around the peak of the $K^*(892)^\pm$ resonance.

LHCb~\cite{Aaij:2014dia} has reported results of an analysis of $B^+\to D K^+$ and $B^+ \to D \pi^+$ decays with $D \to \KS\Kpm\pimp$. 
The decays with different final states of the $D$ meson are distinguished by the charge of the kaon from the decay of the $D$ meson relative to the charge of the $B$ meson, and are labelled ``same sign'' (SS) and ``opposite sign'' (OS). 
Six observables potentially sensitive to $\gamma \equiv \phi_3$ are measured: two ratios of rates for $DK$ and $D\pi$ decays (one each for SS and OS) and four asymmetries (for $DK$ and $D\pi$, SS and OS). 
This is done both for the full Dalitz plot of the $D$ decay and for the $K^*(892)^\pm$-dominated region (with the same boundaries as used by CLEO-c). 
Note that there is a significant overlap of events between the two samples. 
The results, shown in Table~\ref{tab:cp_uta:cus:KSKpi-modInd} do not yet have sufficient precision to set significant constraints on $\gamma \equiv \phi_3$. 

\mysubsubsection{Combinations of results on rates and asymmetries in $B \to \DorDstar K^{(*)}$ decays to obtain constraints on $\gamma \equiv \phi_3$}
\label{sec:cp_uta:cus:gamma}

\babar\ and LHCb have both produced constraints on $\gamma \equiv \phi_3$ from combinations of their results on $B^+ \to DK^+$ and related processes.
The experiments use a frequentist procedure (there are some differences in the details of the techniques used).

\begin{itemize}\setlength{\itemsep}{0.5ex}

\item 
  \babar~\cite{Lees:2013nha} use results from $DK$, $D^*K$ and $DK^*$ modes with GLW, ADS and GGSZ analyses, to obtain $\gamma = (69 \,^{+17}_{-16})^\circ$.

% \item 
%   \belle~\cite{Trabelsi:2013uj} use results from $DK$ and $D^*K$ modes with GLW, ADS and GGSZ analyses, to obtain $\phi_3 = (68 \,^{+15}_{-14})^\circ$.

% @ArxivOnly{Trabelsi:2013uj,
%       author         = "Trabelsi, K.",
%       title          = "{Study of direct CP in charmed B decays and measurement
%                         of the CKM angle gamma at Belle}",
%       collaboration  = "Belle",
%       year           = "2013",
%       eprint         = "1301.2033",
%       archivePrefix  = "arXiv",
%       primaryClass   = "hep-ex",
%       SLACcitation   = "%%CITATION = ARXIV:1301.2033;%%",
%       note           = "{Proceedings of CKM2012}",
% }

\item
  LHCb~\cite{Aaij:2016kjh} use results from the $D\Kp$ mode with GLW, GLW-like, ADS, GGSZ ($\KS h^+h^-$) and GLS ($\KS\Kpm\pimp$) analyses, as well as $DK^{*0}$ with GLW, ADS and GGSZ analyses, $D\Kp\pim$ GLW Dalitz plot analysis, $D\Kp\pim\pip$ with GLW and ADS analyses and $\Bs \to D_s^\mp\Kpm$ decays. 
%  LHCb have in addition obtained a constraint (not quoted here) including results from $B \to D\pi$.
  The LHCb combination takes into account subleading effects due to charm mixing and \CP violation~\cite{Rama:2013voa}.  
  The result is $\gamma = (72.2 \,^{+6.8}_{-7.3})^\circ$.

\item
  All the combinations use inputs determined from $\psi(3770)\to \Dz\Dzb$ data samples (and/or from the HFLAV Charm Physics subgroup global fits on charm mixing parameters; see Sec.~\ref{sec:charm:mixcpv}) to constrain the hadronic parameters in the charm system. 

\item 
  Constraints are also obtained on the hadronic parameters involved in the decays.
  A summary of these is given in Table~\ref{tab:cp_uta:rBdeltaB_combination}.

\item 
  The CKMfitter~\cite{Charles:2004jd} and 
  UTFit~\cite{Bona:2005vz} groups perform similar combinations of all available results to obtain combined constraints on $\gamma \equiv \phi_3$.

\end{itemize}

\begin{table}
  \begin{center}
  \caption{
    Summary of constraints on hadronic parameters obtained from global combinations of results in $\Bp \to \DorDstar\KorKstarp$ and $\Bz \to D\Kstarz$ decays.
  }
  \label{tab:cp_uta:rBdeltaB_combination}
  \renewcommand{\arraystretch}{1.1}
  \begin{tabular}{l@{\hspace{5mm}}c@{\hspace{5mm}}c}
    \hline
    & $r_B(D\Kp)$ & $\delta_B(D\Kp)$ \\
    \hline
    \babar & $0.092 \,^{+0.013}_{-0.012}$ & $(105 \,^{+16}_{-17})^\circ$ \\
%    \belle & $ 0.112 \,^{+0.014}_{-0.015}$ & $(116 \,^{+18}_{-21})^\circ$ \\
    LHCb & $0.1019 \pm 0.0056$ & $(142.6 \,^{+5.7}_{-6.6})^\circ$ \\
    \hline
    & $r_B(DK^{*0})$ & $\delta_B(DK^{*0})$ \\
    LHCb & $0.218 \,^{+0.045}_{-0.047}$ & $(189 \,^{+23}_{-20})^\circ$ \\
    \hline
  \end{tabular}
  \end{center}
\end{table}

% symbols
\newcommand{\Dsmp}{\ensuremath{D_{s}^{\mp}}\xspace}
\newcommand{\phis}{\ensuremath{\phi_{s}}\xspace}

% beauty params
\newcommand{\rb}{\ensuremath{r_{B}}\xspace}
\newcommand{\db}{\ensuremath{\delta_{B}}\xspace}
\newcommand{\rbdk}{\ensuremath{r_{B}(D\Kp)}\xspace}
\newcommand{\dbdk}{\ensuremath{\delta_{B}(D\Kp)}\xspace}
\newcommand{\rbdstk}{\ensuremath{r_{B}(\Dstar\Kp)}\xspace}
\newcommand{\dbdstk}{\ensuremath{\delta_{B}(\Dstar\Kp)}\xspace}
\newcommand{\kbdstk}{\ensuremath{\kappa_{B}(\Dstar\Kp)}\xspace}
\newcommand{\rbdkst}{\ensuremath{r_{B}(D\Kstarp)}\xspace}
\newcommand{\dbdkst}{\ensuremath{\delta_{B}(D\Kstarp)}\xspace}
\newcommand{\kbdkst}{\ensuremath{\kappa_{B}(D\Kstarp)}\xspace}
\newcommand{\rbdkstz}{\ensuremath{r_{B}(D\Kstarz)}\xspace}
\newcommand{\dbdkstz}{\ensuremath{\delta_{B}(D\Kstarz)}\xspace}
\newcommand{\kbdkstz}{\ensuremath{\kappa_{B}(D\Kstarz)}\xspace}
\newcommand{\RbDKstz}{\ensuremath{\bar{R}_{B}^{DK^{*0}}}\xspace}
\newcommand{\DbDKstz}{\ensuremath{\bar{\Delta}_{B}^{DK^{*0}}}\xspace}

% charm params
\newcommand{\rD}{\ensuremath{r_{D}}\xspace}
\newcommand{\dD}{\ensuremath{\delta_{D}}\xspace}
\newcommand{\kD}{\ensuremath{\kappa_{D}}\xspace}
\newcommand{\rdKpi}  {\ensuremath{r_D^{K\pi}}\xspace}
\newcommand{\ddKpi}  {\ensuremath{\delta_D^{K\pi}}\xspace}
\newcommand{\Fp}{\texorpdfstring{\ensuremath{F_{+}}}{F+}\xspace}
\newcommand{\rDKpi}{\texorpdfstring{\ensuremath{r_{D}^{K\pi}}}{rDKPi}\xspace}
\newcommand{\dDKpi}{\texorpdfstring{\ensuremath{\delta_{D}^{K\pi}}}{dDKPi}\xspace}
\newcommand{\rdKpp}{\ensuremath{r_{D}^{K2\pi}}\xspace}
\newcommand{\rdKppsq}{\ensuremath{(r_{D}^{K2\pi})^{2}}\xspace}
\newcommand{\ddKpp}{\ensuremath{\delta_{D}^{K2\pi}}\xspace}
\newcommand{\kdKpp}{\ensuremath{\kappa_{D}^{K2\pi}}\xspace}
\newcommand{\Fppp}{\ensuremath{F_+(\pip\pim\piz)}\xspace}
\newcommand{\FKKp}{\ensuremath{F_+(\Kp\Km\piz)}\xspace}
\newcommand{\kdppp}{\ensuremath{\kappa_{\pi\pi\piz}}\xspace}
\newcommand{\kdkkp}{\ensuremath{\kappa_{KK\piz}}\xspace}
\newcommand{\Aprod}{\ensuremath{A_{B}^{\rm{prod}}}\xspace}
\newcommand{\rdKskpi}{\ensuremath{r_D^{K_SK\pi}}\xspace}
\newcommand{\rdKskpisq}{\ensuremath{(r_D^{K_SK\pi})^2}\xspace}
\newcommand{\ddKskpi}{\ensuremath{\delta_D^{K_SK\pi}}\xspace}
\newcommand{\kdKskpi}{\ensuremath{\kappa_D^{K_SK\pi}}\xspace}
\newcommand{\RdKskpi}{\ensuremath{R_D^{K_SK\pi}}\xspace}
\newcommand{\rdKppp}{\texorpdfstring{\ensuremath{r_D^{K3\pi}}}{rD(K3pi)}\xspace}
\newcommand{\rdKpppsq}{\ensuremath{(r_D^{K3\pi})^2}\xspace}
\newcommand{\ddKppp}{\ensuremath{\delta_D^{K3\pi}}\xspace}
\newcommand{\kdKppp}{\ensuremath{\kappa_D^{K3\pi}}\xspace}
\newcommand{\Fpppp}{\ensuremath{F_+(\pip\pim\pip\pim)}\xspace}
\newcommand{\kdpppp}{\ensuremath{\kappa_{\pi\pi\pi\pi}}\xspace}

% decays
\newcommand{\BuDK}    {\ensuremath{\Bp\to D \Kp}\xspace}
\newcommand{\BuDstK}  {\ensuremath{\Bp\to \Dstar \Kp}\xspace}
\newcommand{\BuDKst}  {\ensuremath{\Bp\to D \Kstarp}\xspace}
\newcommand{\BuDKpipi}{\ensuremath{\Bp\to D \Kp\pip\pim}\xspace}
\newcommand{\BuDhhpizK}    {\ensuremath{\Bp\to D_{hh\piz} \Kp}\xspace}
\newcommand{\BuDppppK}    {\ensuremath{\Bp\to D_{\pi\pi\pi\pi} \Kp}\xspace}
\newcommand{\BdDKstz} {\ensuremath{\Bd\to D \Kstarz}\xspace}
\newcommand{\BdDKpi}  {\ensuremath{\Bd\to D \Kp\pim} }
\newcommand{\BsDsK}	  {\ensuremath{\Bs\to \Dsmp \Kpm}\xspace}
\newcommand{\DKpi}     {\ensuremath{D\to K^{\pm}\pi^{\mp}}\xspace}
\newcommand{\Dhh}      {\ensuremath{D\to h^+h^-}\xspace}
\newcommand{\Dhhh}     {\ensuremath{D\to hhhh}\xspace}
\newcommand{\DKpipipi} {\texorpdfstring{\ensuremath{D\to K^{\pm}\pi^{\mp}\pip\pim}}{D -> K3pi}\xspace}
\newcommand{\DKpipiz}  {\texorpdfstring{\ensuremath{D\to K^{\pm}\pi^{\mp}\piz}}{D -> K2pi}\xspace}
\newcommand{\Dhpipipi} {\ensuremath{D\to h^+\pi^-\pi^+\pi^-}\xspace}
\newcommand{\Dpipipipi}{\ensuremath{D\to \pi^+\pi^-\pi^+\pi^-}\xspace}
\newcommand{\Dhhpiz}   {\ensuremath{D\to h^+h^-\piz}\xspace}
\newcommand{\Dzhh}     {\ensuremath{\Dz\to h^+h^-}\xspace}
\newcommand{\DzKK}     {\ensuremath{\Dz\to K^+K^-}\xspace}
\newcommand{\Dzpipi}   {\ensuremath{\Dz\to\pi^+\pi^-}\xspace}
\newcommand{\DzKShh}   {\ensuremath{\Dz\to\KS h^+h^-}\xspace}
\newcommand{\DKSpipi}  {\ensuremath{D\to\KS\pi^+\pi^-}\xspace}
\newcommand{\DKSKK}    {\ensuremath{D\to\KS K^+K^-}\xspace}
\newcommand{\DKShh}    {\ensuremath{D\to\KS h^+h^-}\xspace}
\newcommand{\DKSKpi}   {\ensuremath{D\to \KS K^+\pi^-}\xspace}
\newcommand{\Dzpipipiz}{\ensuremath{\Dz\to\pi^+\pi^-\piz}\xspace}
\newcommand{\DzKKpiz}  {\ensuremath{\Dz\to K^+ K^-\piz}\xspace}
\newcommand{\Dpipipiz} {\ensuremath{D\to\pi^+\pi^-\piz}\xspace}
\newcommand{\DKKpiz}   {\ensuremath{D\to K^+ K^-\piz}\xspace}
\newcommand{\Dshhh}    {\ensuremath{\Ds\to h^+h^-\pip}}
\newcommand{\DstD}     {\ensuremath{\Dstar\to D\piz(\g)}\xspace}
\newcommand{\DstDg}    {\ensuremath{\Dstar\to D\g}\xspace}
\newcommand{\DstDpiz}  {\ensuremath{\Dstar\to D\piz}\xspace}
\newcommand{\DKK}      {\ensuremath{D\to \Kp\Km}\xspace}
\newcommand{\Dpipi}      {\ensuremath{D\to \pip\pim}\xspace}
\newcommand{\DKSpi}      {\ensuremath{D\to \KS\piz}\xspace}
\newcommand{\DKSw}      {\ensuremath{D\to \KS\omega}\xspace}
\newcommand{\DKSphi}      {\ensuremath{D\to \KS\phi}\xspace}

Independently from the constraints on $\gamma \equiv \phi_3$ obtained by the
experiments, the results summarised in Sec.~\ref{sec:cp_uta:cus} are statistically combined to produce world average constraints on $\gamma \equiv \phi_3$ and the hadronic parameters involved.
The combination is performed with the \textsc{GammaCombo} framework~\cite{gammacombo} and follows a frequentist procedure, similar to those used by the experiments~\cite{Lees:2013nha,Aaij:2013zfa,Aaij:2016kjh}.

The input measurements used in the combination are listed in Table~\ref{tab:cp_uta:gamma:inputs}.
Individual measurements are used as inputs, rather than the averages presented in Sec.~\ref{sec:cp_uta:cus}, in order to facilitate cross-checks and to ensure the most appropriate treatment of correlations.
A combination based on our averages for each of the quantities measured by experiments gives consistent results.

All results from GLW and GLW-like analyses of $\Bp\to \DorDstar \KorKstarp$ modes, as listed in Tables~\ref{tab:cp_uta:cus:glw} and~\ref{tab:cp_uta:cus:glwLike}, are used.
All results from ADS analyses of $\Bp\to \DorDstar \KorKstarp$ as listed in Table~\ref{tab:cp_uta:cus:ads} are also used.
Regarding $\Bd\to D\Kstarz$ decays, the results of the $\Bd\to D\Kp\pim$ GLW-Dalitz analysis (Table~\ref{tab:cp_uta:cus:DKpiDalitz}) are included, as are the LHCb results of the ADS analysis of $\Bd\to D\Kstarz$ (Table~\ref{tab:cp_uta:ads-DKstar}). 
Concerning results of GGSZ analyses of $\Bp\to \DorDstar \KorKstarp$ with $D\to\KS h^{+}h^{-}$, the model-dependent results, as listed in Table~\ref{tab:cp_uta:cus:dalitz}, are used for the \babar\ and \belle\ experiments, whilst the model-independent results, as listed in Table~\ref{tab:cp_uta:cus:dalitz-modInd}, are used for LHCb. 
This choice is made in order to maintain consistency of the approach across experiments whilst maximising the size of the samples used to obtain inputs for the combination. 
For GGSZ analyses of $\Bd\to D\Kstarz$ with $D\to\KS h^{+}h^{-}$ the model-independent result from LHCb (given in Table~\ref{tab:cp_uta:cus:dalitz-modInd}) is used for consistency with the treatment of the LHCb $\Bp \to D\Kp$ GGSZ result; the model-independent result by Belle is also included. 
The result of the GLS analysis of $\Bp \to D \Kp$ with $D\to\Kstarpm\Kmp$ from LHCb (Table~\ref{tab:cp_uta:cus:KSKpi-modInd}) are used. 
Finally, results from the time-dependent analysis of $\Bs\to\Dsmp\Kpm$ from LHCb (Table~\ref{tab:cp_uta:DsK}) are used.

Several results with sensitivity to $\gamma$ are not included in the combination.
Results from time-dependent analyses of $\Bz \to D^{(*)\mp}\pi^\pm$ and $D^\mp\rho^\pm$ (Table~\ref{tab:cp_uta:cud}) are not used as there are insufficient constraints on the associated hadronic parameters.
Similarly, results from $\Bz \to \Dmp\KS\pipm$ (Sec.~\ref{sec:cp_uta:cus-td-DKSpi}) are not used.
Results from the LHCb $\Bz \to D\Kstarz$ GLW analysis (Table~\ref{tab:cp_uta:glw-DKstar}) are not used because of the statistical overlap with the GLW-Dalitz analysis which is used instead. 
Limits on ADS parameters reported in Sec.~\ref{sec:cp_uta:cus:ads} are not used.
Results on $\Bp \to D\pip$ decays, given in Table~\ref{tab:cp_uta:cus:ads2}, are not used since the small value of $r_B(D\pip)$ means these channels have less sensitivity to $\gamma$ and are more vulnerable to biases due to subleading effects~\cite{Aaij:2016kjh}.
Results from the \babar\ Dalitz plot analysis of $\Bp \to D\Kp$ with $D \to \pip\pim\piz$ (given in Table~\ref{tab:cp_uta:cus:dalitz}) are not included due to their limited sensitivity.  
Results from the $\Bp \to D\Kp$, $D \to \KS \pip\pim$ GGSZ model-dependent analysis by LHCb (given in Table~\ref{tab:cp_uta:cus:dalitz}), and of the model-independent analysis of the same decay by Belle (given in Table~\ref{tab:cp_uta:cus:dalitz-modInd}) are not included due to the statistical overlap with results from model-(in)dependent analyses of the same data.

% \begin{table}[h!]
%   \caption{List of the measurements used in the combinations.}
%   \label{tab:cp_uta:gamma:inputs}
%   \centering
%   \begin{tabular}{l l l l l}
\begin{longtable}{l l l l l}
  \caption{List of measurements used in the $\gamma$ combination.}
  \label{tab:cp_uta:gamma:inputs}
\endfirsthead
\multicolumn{5}{c}{List of measurements used in the $\gamma$ combination -- continued from previous page.}
\endhead
\endfoot
\endlastfoot
       \hline
        $B$ decay & $D$ decay & Method & Experiment & Ref. \\
        \hline
        \BuDK     & \DKK, \Dpipi,                        & GLW         & \babar & \cite{delAmoSanchez:2010ji} \\
                  & \DKSpi, \DKSw, \DKSphi               &             &       &         \\
        \BuDK     & \DKK, \Dpipi,                        & GLW         & \belle & \cite{Abe:2006hc} \\
                  & \DKSpi, \DKSw, \DKSphi               &             &       &         \\
        \BuDK     & \DKK, \Dpipi & GLW         & CDF   & \cite{Aaltonen:2009hz} \\
        \BuDK     & \DKK, \Dpipi &GLW         & LHCb  & \cite{Aaij:2016oso} \\
        \hline
        \BuDstK   & \DKK, \Dpipi,                        & GLW         & \babar & \cite{:2008jd} \\
        $\;\;$\DstDg(\piz)  & \DKSpi, \DKSw, \DKSphi               &             &       &         \\
        \BuDstK   & \DKK, \Dpipi,                        & GLW         & \belle & \cite{Abe:2006hc} \\
        $\;\;$\DstDg(\piz)  & \DKSpi, \DKSw, \DKSphi               &             &       &         \\
        \hline
        \BuDKst   & \DKK, \Dpipi,                        & GLW         & \babar & \cite{Aubert:2009yw} \\
                  & \DKSpi, \DKSw, \DKSphi               &             &       &         \\
        \BuDKst   & \DKK, \Dpipi,                        & GLW         & LHCb & \cite{LHCb-CONF-2016-014} \\
        \hline
        \BuDKpipi & \DKK, \Dpipi & GLW         & LHCb  & \cite{Aaij:2015ina} \\
        \hline
        \BuDK     & \Dpipipiz  & GLW-like & \babar & \cite{Aubert:2007ii} \\
        \BuDK     & \Dhhpiz    & GLW-like & LHCb & \cite{Aaij:2015jna} \\
        \BuDK     & \Dpipipipi & GLW-like & LHCb & \cite{Aaij:2016oso} \\
        \hline
        \BuDK  & \DKpi & ADS & \babar & \cite{delAmoSanchez:2010dz} \\
        \BuDK  & \DKpi & ADS & \belle & \cite{Belle:2011ac} \\
        \BuDK  & \DKpi & ADS & CDF & \cite{Aaltonen:2011uu} \\
        \BuDK  & \DKpi & ADS & LHCb & \cite{Aaij:2016oso} \\
        \hline
        \BuDK  & \DKpipiz & ADS & \babar & \cite{Lees:2011up} \\
        \BuDK  & \DKpipiz & ADS & \belle & \cite{Nayak:2013tgg} \\
        \BuDK  & \DKpipiz & ADS & LHCb & \cite{Aaij:2015jna} \\
        \hline
        \BuDK  & \DKpipipi & ADS & LHCb & \cite{Aaij:2016oso} \\
        \hline
        \BuDstK  & \DKpi & ADS & \babar & \cite{delAmoSanchez:2010dz} \\
        $\;\;$\DstDg  & & & &  \\
        \hline
        \BuDstK  & \DKpi & ADS & \babar & \cite{delAmoSanchez:2010dz} \\
        $\;\;$\DstDpiz  & & & &  \\
        \hline
        \BuDKst  & \DKpi  & ADS & \babar & \cite{Aubert:2009yw} \\
        \BuDKst  & \DKpi  & ADS & LHCb & \cite{LHCb-CONF-2016-014} \\
        \hline
        \BuDKpipi  & \DKpi & ADS & LHCb & \cite{Aaij:2015ina} \\
        \hline
        \BuDK  & \DKSpipi & GGSZ MD & \babar & \cite{delAmoSanchez:2010rq} \\
        \BuDK  & \DKSpipi & GGSZ MD & \belle & \cite{Poluektov:2010wz} \\
        \hline
        \BuDstK  & \DKSpipi & GGSZ MD & \babar & \cite{delAmoSanchez:2010rq} \\
        $\;\;$\DstDg(\piz)& &         &       &         \\
        \BuDstK  & \DKSpipi & GGSZ MD & \belle & \cite{Poluektov:2010wz} \\
        $\;\;$\DstDg(\piz)& &         &       &         \\
        \hline
        \BuDKst  & \DKSpipi & GGSZ MD & \babar & \cite{delAmoSanchez:2010rq} \\
        \BuDKst  & \DKSpipi & GGSZ MD & \belle & \cite{Poluektov:2006ia} \\
        \hline
        \BuDK  & \DKSpipi & GGSZ MI & LHCb & \cite{Aaij:2014uva} \\
        \hline
        \BuDK  & \DKSKpi & GLS & LHCb & \cite{Aaij:2014dia} \\
        \hline
        \BdDKstz  & \DKpi & ADS & LHCb & \cite{Aaij:2014eha} \\
        \hline
        \BdDKpi  & \Dhh & GLW-Dalitz & LHCb & \cite{Aaij:2016bqv} \\
        \hline
        \BdDKstz  & \DKShh & GGSZ MI & Belle & \cite{Negishi:2015vqa} \\
        \BdDKstz  & \DKShh & GGSZ MI & LHCb & \cite{Aaij:2016nao} \\
        \hline
        \BsDsK  & \Dshhh & TD & LHCb & \cite{LHCb-CONF-2016-015} \\
        \hline
\end{longtable}
%       \end{tabular}
% \end{table}

\begin{table}[b]
  \caption{List of the auxiliary inputs used in the combinations.}
  \label{tab:cp_uta:gamma:inputs_aux}
  \centering
  \renewcommand{\arraystretch}{1.1}
      \begin{tabular}{l l l l }
        \hline
        Decay      & Parameters                  & Source & Ref. \\
        \hline \\[-2.5ex]
         \DKpi              & \rdKpi, \ddKpi                       & HFLAV       &  Sec.~\ref{sec:charm_physics}       \\
         \DKpipipi          & \ddKppp, \kdKppp, \rdKppp            & CLEO+LHCb  &  \cite{Evans:2016tlp}       \\
         \Dpipipipi         & \Fpppp                               & CLEO       &  \cite{Malde:2015mha}       \\
         \DKpipiz           & \ddKpp, \kdKpp, \rdKpp               & CLEO+LHCb  &  \cite{Evans:2016tlp}       \\
         \Dhhpiz            & \Fppp, \FKKp                         & CLEO       &  \cite{Malde:2015mha}       \\
         \multirow{2}{*}{\DKSKpi}            & \ddKskpi, \kdKskpi, \rdKskpi         & CLEO       &  \cite{Insler:2012pm}       \\
                     & \rdKskpi                             & LHCb       &  \cite{Aaij:2015lsa} \\
         \BdDKstz           & \kbdkstz, \RbDKstz, \DbDKstz         & LHCb       &  \cite{Aaij:2016bqv} \\
         \BsDsK             & \phis                                & HFLAV       &  Sec.~\ref{sec:life_mix} \\
        \hline
      \end{tabular}
\end{table}

Auxiliary inputs are used in the combination in order to constrain the $D$ system parameters and subsequently improve the determination of $\gamma \equiv \phi_3$. 
These include the ratio of suppressed to favoured decay amplitudes and the strong phase difference for $D\to\Kpm\pimp$ decays, taken from the HFLAV Charm Physics subgroup global fits (see Sec.~\ref{sec:charm_physics}).
The amplitude ratios, strong phase differences and coherence factors of $D\to\Kpm\pimp\piz$, $D\to\Kpm\pimp\pip\pim$ and $D\to\KS\Kpm\pipm$ decays are taken from CLEO-c and LHCb measurements~\cite{Evans:2016tlp,Insler:2012pm,Aaij:2015lsa}.
The fraction of \CP-even content for quasi-GLW $D\to\pip\pim\pip\pim$, $D\to\Kp\Km\piz$ and $D\to\pip\pim\piz$ decays are taken from CLEO-c measurements~\cite{Malde:2015mha}.
Constraints required to relate the hadronic parameters of the $\Bd\to D\Kstarz$ GLW-Dalitz analysis to the effective hadronic parameters of the quasi-two-body approaches are taken from LHCb measurements~\cite{Aaij:2016bqv}.
Finally, the value of $-2\beta_{s}$ is taken from the HFLAV Lifetimes and Oscillations subgroup (see Sec.~\ref{sec:life_mix}); this is required to obtain sensitivity to $\gamma \equiv \phi_3$ from the time-dependent analysis of $\Bs\to\Dsmp\Kpm$ decays.
A summary of the auxiliary constraints is given in Table~\ref{tab:cp_uta:gamma:inputs_aux}.

The following reasonable, although imperfect, assumptions are made when performing the averages.
\begin{itemize}
  \item{\CP violation in \DKK and \Dpipi decays is assumed to be zero. The results of Sec.~\ref{sec:charm_physics} anyhow suggest such effects to be negligible.}
  \item{The combination is potentially sensitive to subleading effects from \Dz--\Dzb mixing which is not accounted for~\cite{Silva:1999bd,Grossman:2005rp,Rama:2013voa}. The effect is expected to be small given that $\rb\gsim0.1$ (for all included modes) whilst $\rD\approx 0.05$.}
  \item{All \BuDKst modes are treated as two-body decays. In other words any dilution caused by non-\Kstarp\ contributions in the selected regions of the $D\KS\pip$ or $D\Kp\piz$ Dalitz plots is assumed to be negligible.  As a check of this assumption, it was found that including a coherence factor for \BuDKst modes, $\kbdkst = 0.9$, had negligible impact on the results.}
  \item{All of the inputs are assumed to be completely uncorrelated. Whilst this is true of the statistical uncertainties, it is not necessarily the case for systematic uncertainties. In particular, the model uncertainties for different model-dependent GGSZ analyses are fully correlated (when the same model is used) and similarly the model-independent GGSZ analyses have correlated systematic uncertainties originating from the knowledge of the strong phase variation across the Dalitz plot. The effect of including these correlations is estimated to be $<1\degrees$.}
\end{itemize}

\begin{figure}
    \centering
    \includegraphics[width=0.6\textwidth]{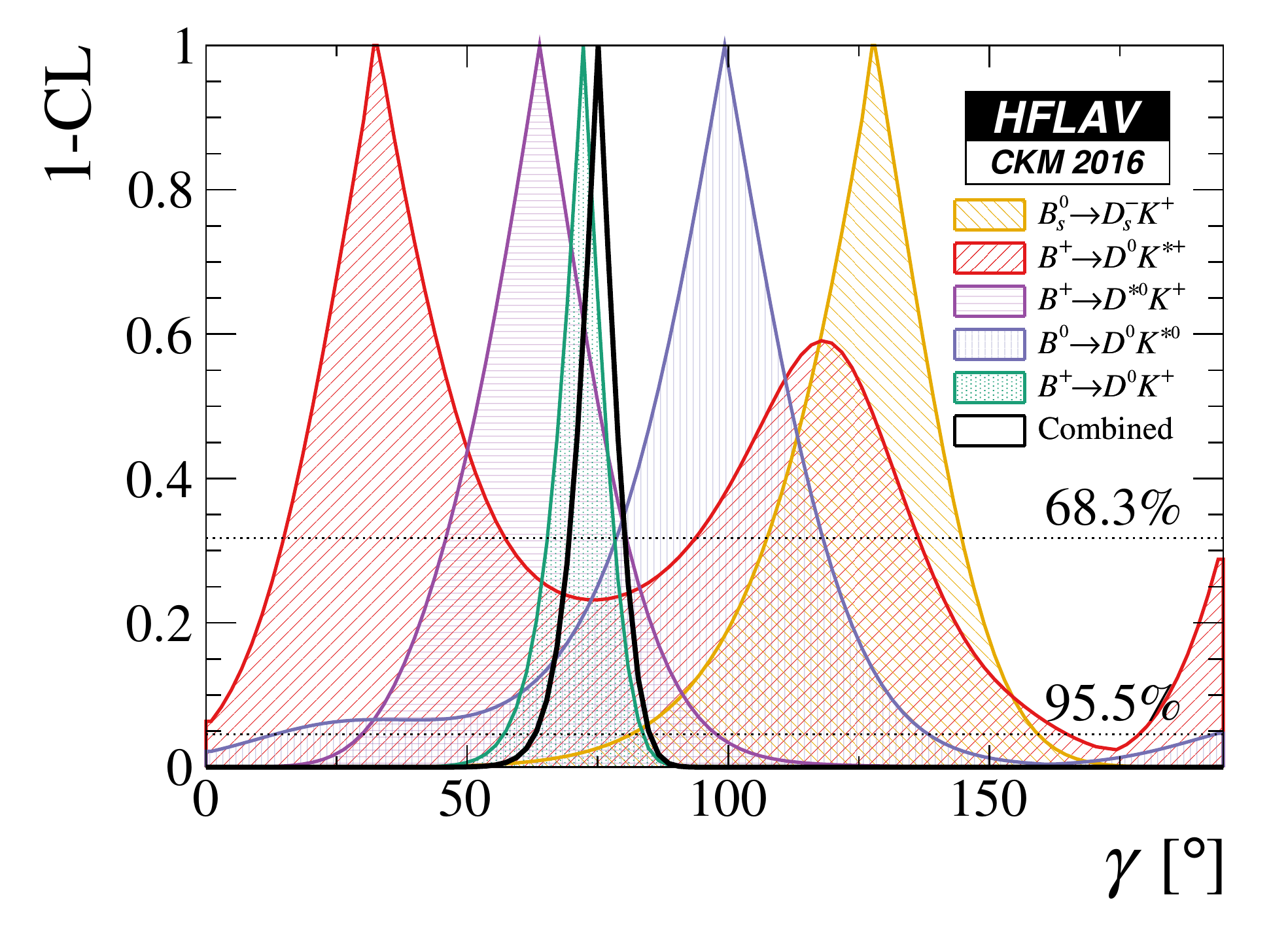}
    \caption{World average of $\gamma\equiv\phi_{3}$, in terms of 1$-$CL, split by decay mode.}
    \label{fig:cp_uta:gamma:results_mode}
\end{figure}

\begin{figure}
    \centering
    \includegraphics[width=0.6\textwidth]{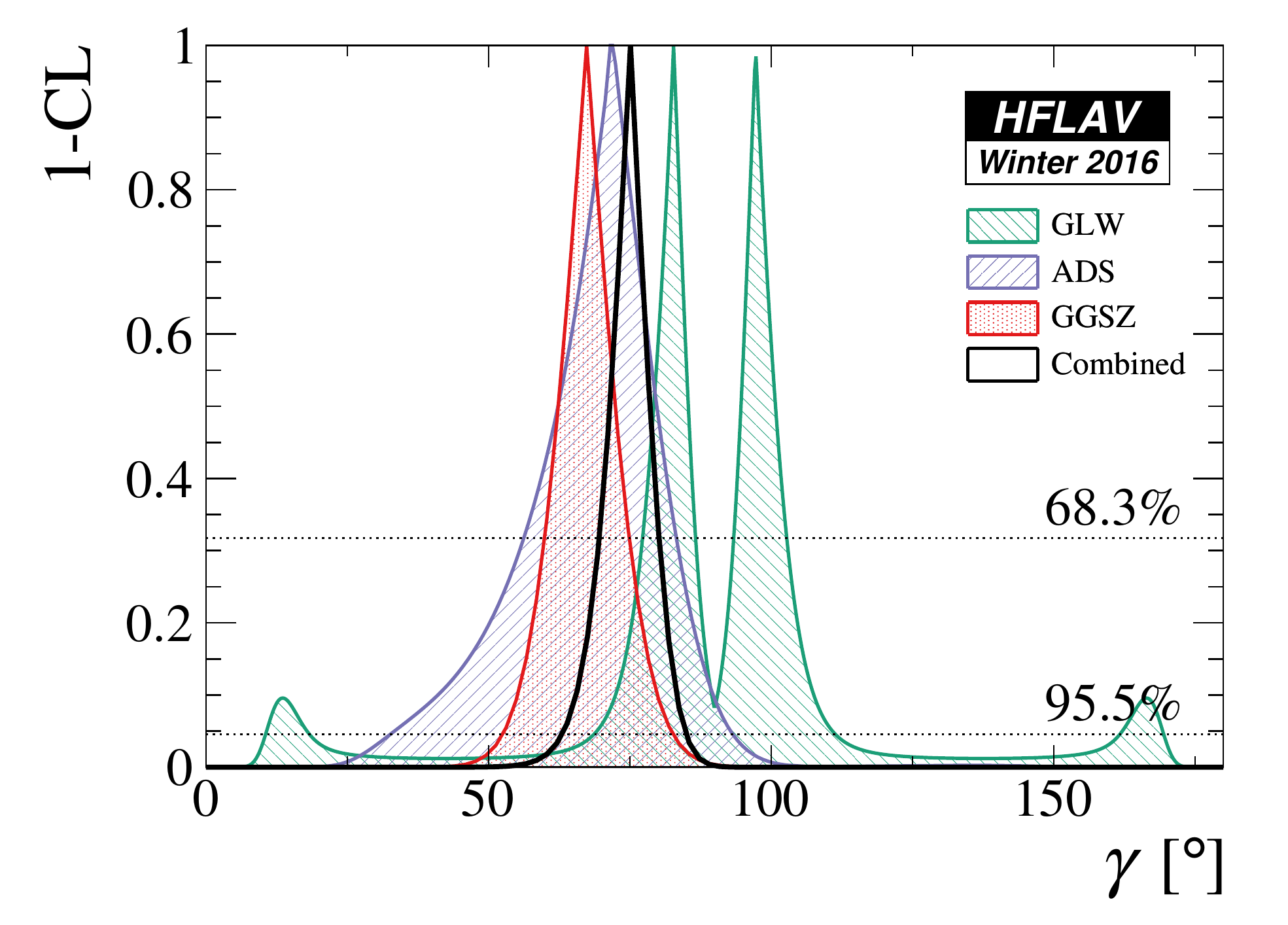}
    \caption{World average of $\gamma\equiv\phi_{3}$, in terms of 1$-$CL, split by analysis method.}
    \label{fig:cp_uta:gamma:results_method}
\end{figure}

\begin{figure}
    \centering
    \includegraphics[width=0.6\textwidth]{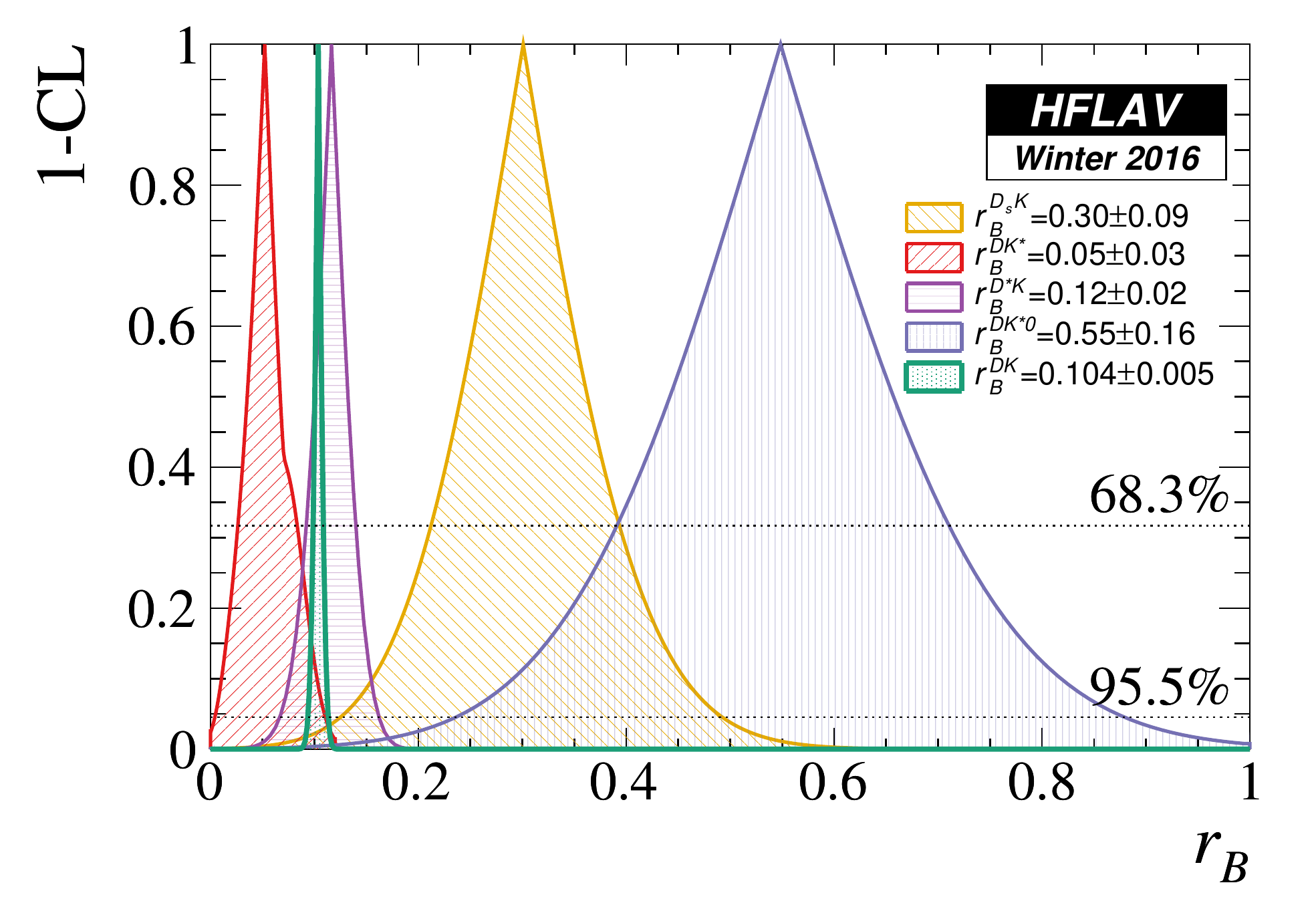}
    \caption{World averages for the hadronic parameters \rb\ in the different decay modes, in terms of 1$-$CL.}
    \label{fig:cp_uta:gamma:results_rb}
\end{figure}

\begin{figure}
    \centering
    \includegraphics[width=0.48\textwidth]{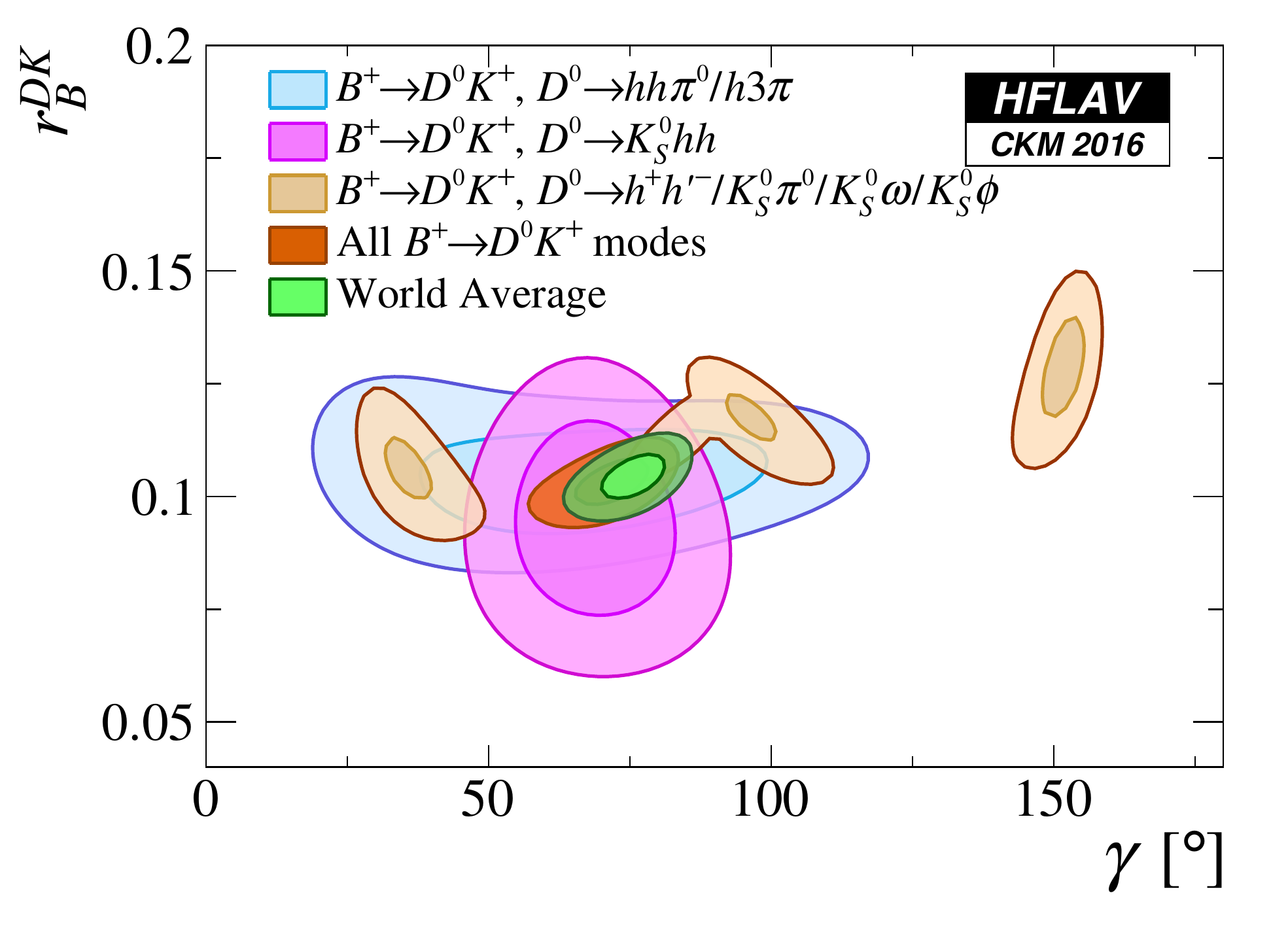}
    \includegraphics[width=0.48\textwidth]{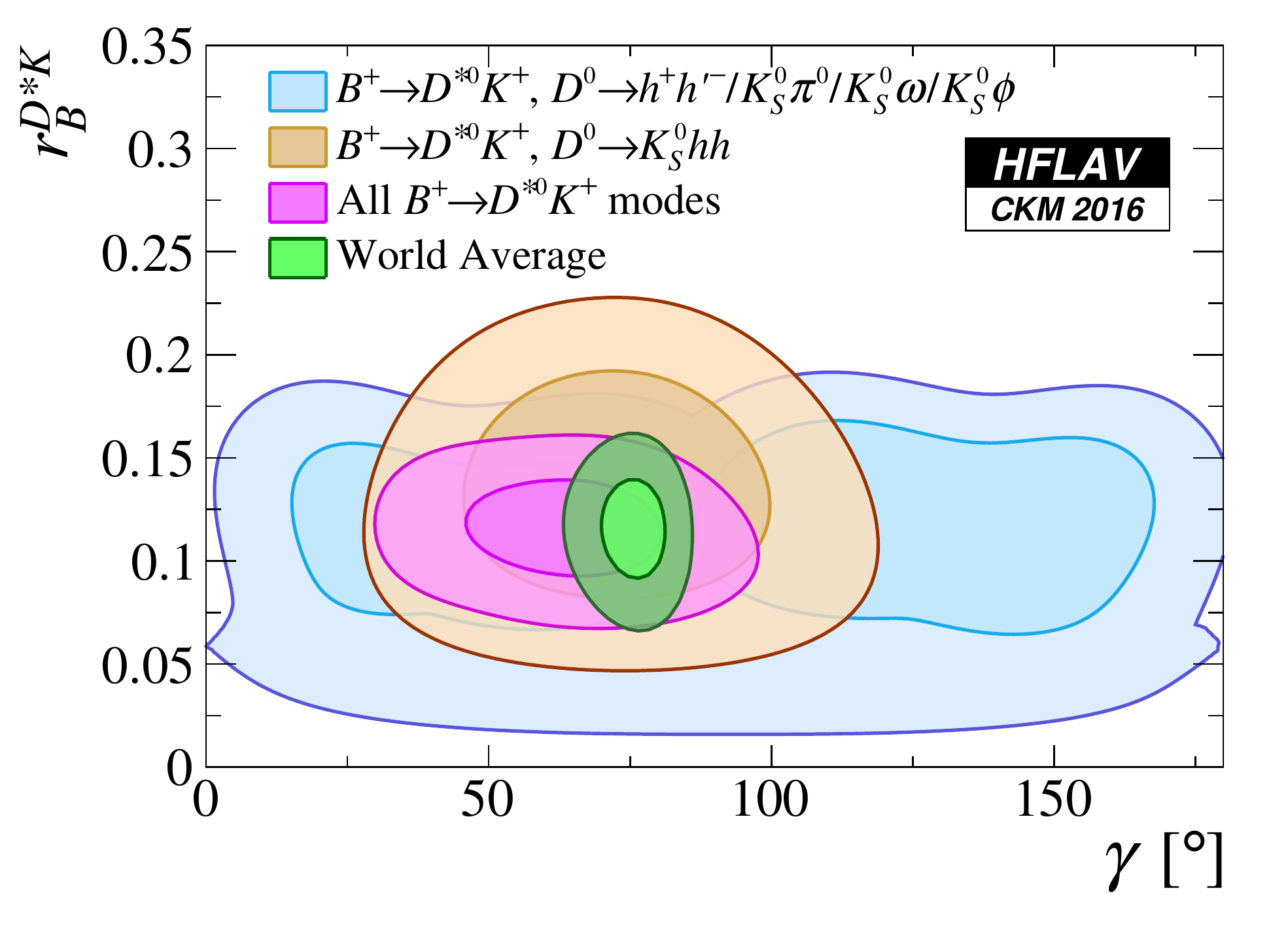} \\
    \includegraphics[width=0.48\textwidth]{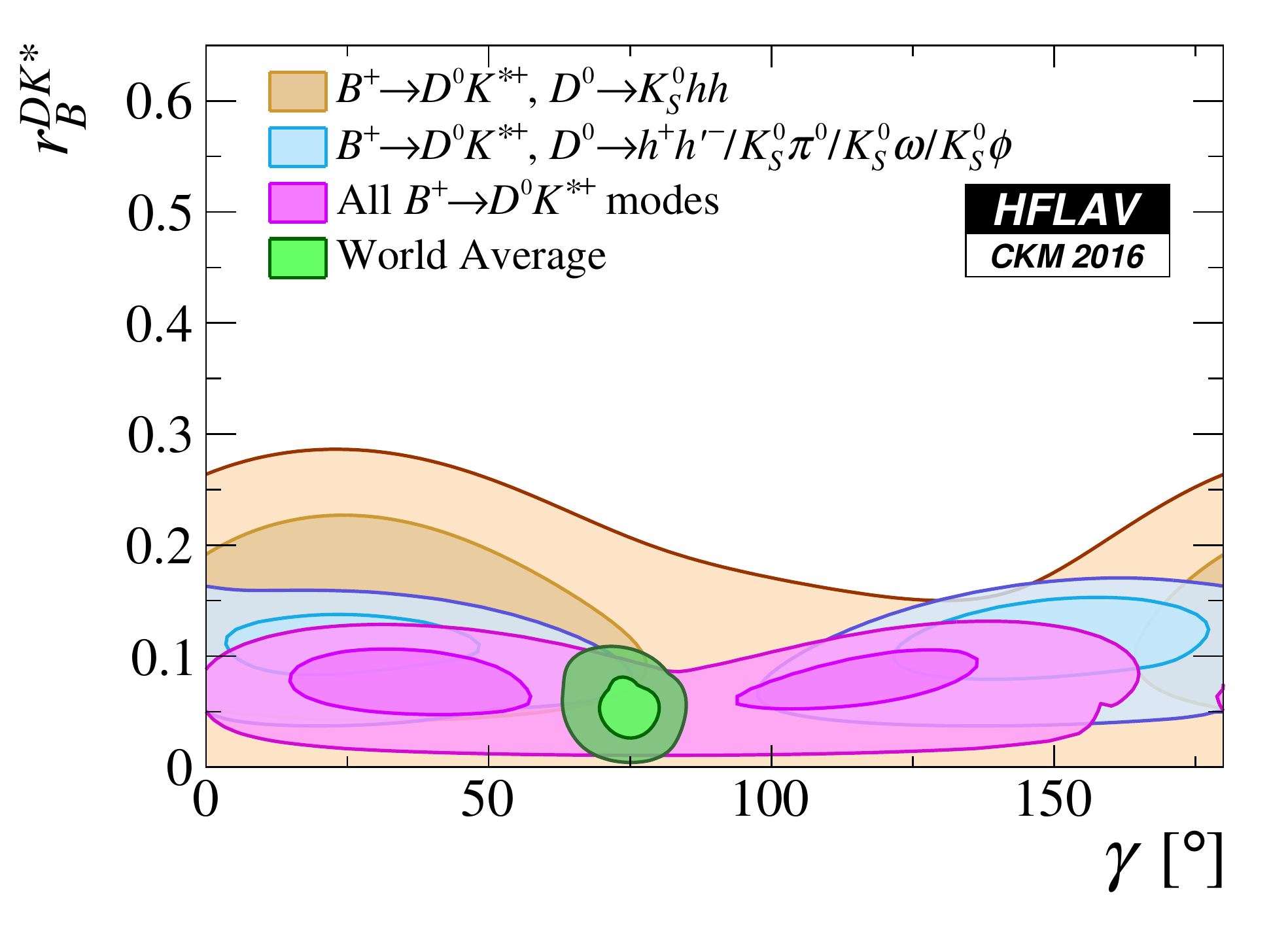}
    \includegraphics[width=0.48\textwidth]{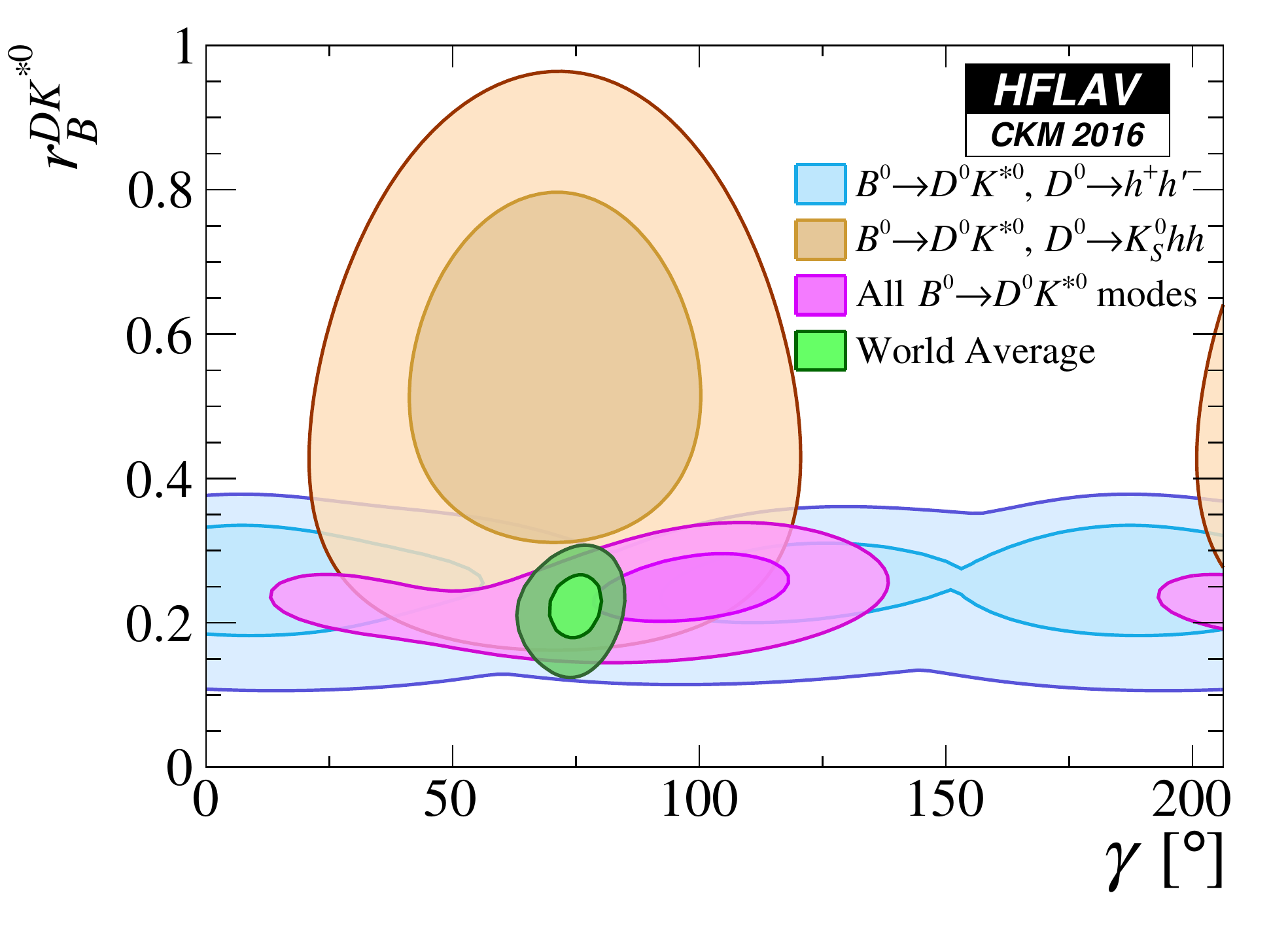}
  \caption{Contributions to the combination from different input measurements, shown in the plane of the relevant \rb parameter \vs\ $\gamma \equiv \phi_3$. 
    From left to right, top to bottom: \BuDK, \BuDstK, \BuDKst\ and \BdDKstz.
    Contours show the two-dimensional $68\,\%$ and $95\,\%$ CL regions.
}
  \label{fig:cp_uta:gamma:results_contribs}
\end{figure}

In total, there are 116 observables and 33 free parameters. 
The combination has a $\chi^2$ value of 95.5, which corresponds to a global p-value of 0.164.
The obtained world average for the Unitarity Triangle angle $\gamma \equiv \phi_3$ is
\begin{equation}
  \gamma \equiv \phi_3 = (74.0\,^{+5.8}_{-6.4})\degrees \, .
\end{equation}
An ambiguous solution at $\gamma \equiv \phi_3 \longrightarrow \gamma \equiv \phi_3+\pi$ also exists.
The results for the hadronic parameters are listed in Table~\ref{tab:cp_uta:gamma:results}. 
Results for input analyses as split by \B meson decay mode are shown in Table~\ref{tab:cp_uta:gamma:results_mode} and Fig.~\ref{fig:cp_uta:gamma:results_mode}. 
Results for input analyses as split by the method are shown in Table~\ref{tab:cp_uta:gamma:results_method} and Fig.~\ref{fig:cp_uta:gamma:results_method}. 
Results for the hadronic ratios, \rb, are shown in Fig.~\ref{fig:cp_uta:gamma:results_rb}. 
A demonstration of how the various analyses contribute to the combination is shown in Fig.~\ref{fig:cp_uta:gamma:results_contribs}.

\begin{table}
  \caption{Averages values obtained for the hadronic parameters in $\B \to \DorDstar\KorKstar$ decays.}
  \label{tab:cp_uta:gamma:results}
  \centering
  \renewcommand{\arraystretch}{1.1}
  \begin{tabular}{l c}
    \hline
    Parameter & Value \\
    \hline
%    $\gamma \equiv \phi_3$        &  $(74.0\,^{+5.8}_{-6.4})\degrees$ \\
    \rbdk     &  $0.104 \pm 0.005$ \\
    \rbdstk   &  $0.12 \pm 0.02$ \\
    \rbdkst   &  $0.05 \pm 0.03$ \\
    \rbdkstz  &  $0.55 \pm 0.16$ \\
    \dbdk     &  $(137.7 \,^{+5.1}_{-6.0})\degrees$ \\
    \dbdstk   &  $(311\,^{+13}_{-17})\degrees$ \\
    \dbdkst   &  $(108 \,^{+33}_{-74})\degrees$ \\
    \dbdkstz  &  $(203\,^{+22}_{-20})\degrees$ \\
    \hline
  \end{tabular}
\end{table}

\begin{table}
  \caption{Averages of $\gamma \equiv \phi_3$ split by \B meson decay mode.}
  \label{tab:cp_uta:gamma:results_mode}
  \centering
  \renewcommand{\arraystretch}{1.1}
  \begin{tabular}{l c}
    \hline
    Decay Mode & Value \\
    \hline
    \BsDsK   &  $(128 \,^{+18}_{-22})\degrees$ \\
    \BuDKst  &  $(33 \,^{+30}_{-20})\degrees$ \\
    \BuDstK  &  $(64 \,^{+18}_{-19})\degrees$ \\
    \BdDKstz &  $(92 \,^{+23}_{-21})\degrees$ \\
    \BuDK    &  $(72.2 \,^{+5.9}_{-7.0})\degrees$ \\
    \hline
  \end{tabular}
\end{table}

\begin{table}
  \caption{Averages of $\gamma \equiv \phi_3$ split by method. For GLW method only the solution nearest the combined average is shown.}
  \label{tab:cp_uta:gamma:results_method}
  \centering
  \renewcommand{\arraystretch}{1.1}
  \begin{tabular}{l c}
    \hline
    Method & Value \\
    \hline
    GLW  &  $(82.7\,^{+5.5}_{-6.9})\degrees$ \\
    ADS  &  $(72\,^{+12}_{-18})\degrees$ \\
    GGSZ &  $(67.3\,^{+8.1}_{-7.8})\degrees$ \\
    \hline
  \end{tabular}
\end{table}

\clearpage
%% % Semileptonic B decays
% -- \include{slbdecays.tex}
% ======================================================================

%\cleardoublepage

\section{Semileptonic $B$ decays}
\label{sec:slbdecays}

This section contains our averages for semileptonic $B$~meson decays,
\ie\ decays of the type $B\to X\ell\nu_\ell$, where $X$ refers to one
or more hadrons, $\ell$ to a charged lepton and $\nu_\ell$ to its associated
neutrino. Unless otherwise stated, $\ell$ stands for an electron
\emph{or} a muon, lepton universality is assumed, and both charge
conjugate states are combined. Some averages assume isospin symmetry,
this will be explicitly mentioned at every instance.

Averages are presented separately for CKM favored $b\to c$ quark transitions
and CKM suppressed $b\to u$ transitions. Among these transitions we
distinguish \emph{exclusive} decays involving a specific meson ($X=D, D^*, \pi,
\rho,\dots$) from \emph{inclusive} decay modes, \ie\ the sum over all possible
hadronic states, one or more mesons and baryons. Semileptonic decays proceed
via first order weak interactions and are well described in the framework of
the standard model (SM). Their decay rates are sensitive to the magnitude
squared of the CKM elements $V_{cb}$ and $V_{ub}$, and their determination is
one of the primary goals for the study of these decays. Semileptonic decays
involving the $\tau$~lepton might be sensitive to beyond SM processes because
of the high $\tau$~mass, which might result in enhanced couplings 
to a hypothetical charged Higgs boson or leptoquarks.

The technique for obtaining the averages follows the general HFLAV
procedure (Sec.~\ref{sec:method}) unless otherwise stated. More
information on the averages, in particular on the common input parameters
is available on the HFLAV semileptonic webpage.
%
%\centerline{\tt http://www.slac.stanford.edu/xorg/hfag/semi/summer16/main.shtml}

% ======================================================================

%% \clearpage

% ======================================================================
% Common set of input parameters

%%%commented out 2-Aug-2010 by AJS
%%%\input{slbdecays/common.tex}

% ======================================================================
% Exclusive CKM-favoured decays
% -- \include{b2cexcl.tex}
% ======================================================================
\subsection{Exclusive CKM-favoured decays}
\label{slbdecays_b2cexcl}
% -------------------------------------------
%This section is organized as follows: First, we present averages for
%the decays $\bar B\to D^*\ell^-\bar\nu_\ell$ and $\bar B\to
%D\ell^-\bar\nu_\ell$. In addition to the branching fractions, the CKM
%element $|V_{cb}|$ is extracted. We then provide
%averages for the inclusive branching fractions $\cbf(\bar B\to
%D^{(*)}\pi \ell^-\bar\nu_\ell)$ and for $B$ semileptonic decays into
%orbitally-excited $P$-wave charm mesons ($D^{**}$). As the $D^{**}$
%branching fraction is poorly known, we report the averages for the products 
%$\cbf(B^-\to D^{**}(D^{(*)}\pi)\ell^-\bar\nu_\ell)\times
%\cbf(D^{**}\to D^{(*)}\pi)$.

%===================================================================
% D and D* 
%===================================================================

\mysubsubsection{$\bar B\to D^*\ell^-\bar\nu_\ell$}
\label{slbdecays_dstarlnu}

The recoil variable $w$ used to describe $\bar B\to D^*\ell^-\bar\nu_\ell$
decays is the product of the four-velocities of the initial and final state
mesons, $w=v_B\cdot v_{D^{(*)}}$. The differential decay rate for massless
fermions as a function of $w$ is given by (see, \eg,~\cite{Neubert:1993mb})
\begin{equation}
  \frac{d\Gamma(\bar B\to D^*\ell^-\bar\nu_\ell)}{dw} = \frac{G^2_\mathrm{F} m^3_{D^*}}{48\pi^3}(m_B-m_{D^*})^2\chi(w)\eta_\mathrm{EW}^2\mathcal{F}^2(w)\vcb^2~,
\end{equation}
where $G_\mathrm{F}$ is Fermi's constant, $m_B$ and $m_{D^*}$ are the $B$ and
$D^*$ meson masses, $\chi(w)$ is a known expression of $w$ and
$\eta_\mathrm{EW}$ is a small electroweak correction~\cite{Sirlin:1981ie}.
Some authors also include a long-distance EM radiation effect (Coulomb
correction) in this factor.
The form factor $\mathcal{F}(w)$ for the $\bar B\to D^*\ell^-\bar\nu_\ell$
decay contains three indepedent functions, $h_{A_1}(w)$, $R_1(w)$ and $R_2(w)$,
\begin{eqnarray}
  &&\chi(w)\mathcal{F}^2(w)=\\
  && \phantom{\mathcal{F}^2(w)} h_{A_1}^2(w)\sqrt{w^2-1}(w+1)^2 \left\{2\left[\frac{1-2wr+r^2}{(1-r)^2}\right]\left[1+R^2_1(w)\frac{w^2-1}{w+1}\right]+\right. \nonumber\\
  && \phantom{\mathcal{F}^2(w)} \left.\left[1+(1-R_2(w))\frac{w-1}{1-r}\right]^2\right\}~, \nonumber
\end{eqnarray}
where $r=m_{D^*}/m_B$.

To extract $\vcb$, the experimental analyses we consider in this section use
the parametrization of these form factor functions by Caprini, Lellouch and
Neubert (CLN)~\cite{CLN},
\begin{eqnarray}
  h_{A_1}(w) & = &
  h_{A_1}(1)\big[1-8\rho^2z+(53\rho^2-15)z^2-(231\rho^2-91)z^3\big]~, \\
  R_1(w) & = & R_1(1)-0.12(w-1)+0.05(w-1)^2~, \\ 
  R_2(w) & = & R_2(1)+0.11(w-1)-0.06(w-1)^{2}~,
\end{eqnarray}
where $z=(\sqrt{w+1}-\sqrt{2})/(\sqrt{w+1}+\sqrt{2})$. The form factor
${\cal F}(w)$ is thus described by the slope $\rho^2$ and the ratios $R_1(1)$
and $R_2(1)$.

We use the measurements of these form factor parameters shown in
Table~\ref{tab:vcbf1} and rescale them to the latest values of the
input parameters (mainly branching fractions of charmed
mesons)~\cite{HFLAV_sl:inputparams}. Most of the measurements in
Table~\ref{tab:vcbf1} are based on the decay $\bar B^0\to
D^{*+}\ell^-\bar\nu_\ell$. Some
measurements~\cite{Adam:2002uw,Aubert:2009_1} are sensitive also to
the $B^-\to D^{*0}\ell^-\bar\nu_\ell$, and one
measurement~\cite{Aubert:2009_3} is based on the decay
$B^-\to D^{*0}\ell^-\bar\nu_\ell$. Isospin symmetry is assumed in this average.
The earlier results for the LEP experiments and CLEO have significantly
rescaled results, and significantly larger uncertainties than the recent
measurements by the B-factories Belle and \babar.
% ----------------------------------------------------------------------
\begin{table}[!htb]
\caption{Measurements of the Caprini, Lellouch and Neubert
  (CLN)~\cite{CLN} form factor parameters in $\bar B\to
  D^*\ell^-\bar\nu_\ell$ before and after rescaling. Most analyses
  (except \cite{Dungel:2010uk,Aubert:2006mb}) measure only
  $\eta_\mathrm{EW}{\cal F}(1)\vcb$, and $\rho^2$, so only these two
  parameters are shown here.}
\begin{center}
\resizebox{0.99\textwidth}{!}{
\begin{tabular}{|l|c|c|}
  \hline
  Experiment
  & $\eta_\mathrm{EW}{\cal F}(1)\vcb [10^{-3}]$ (rescaled)
  & $\rho^2$ (rescaled)\\
  & $\eta_\mathrm{EW}{\cal F}(1)\vcb [10^{-3}]$ (published)
  & $\rho^2$ (published)\\
  \hline\hline
  ALEPH~\cite{Buskulic:1996yq}
  & $30.97\pm 1.78_{\rm stat}\pm 1.29_{\rm syst}$
  & $0.491\pm 0.227_{\rm stat}\pm 0.146_{\rm syst}$\\
  & $31.9\pm 1.8_{\rm stat}\pm 1.9_{\rm syst}$
  & $0.37\pm 0.26_{\rm stat}\pm 0.14_{\rm syst}$\\
  \hline
  CLEO~\cite{Adam:2002uw}
  & $39.67\pm 1.22_{\rm stat}\pm 1.62_{\rm syst}$
  & $1.366\pm 0.085_{\rm stat}\pm 0.087_{\rm syst}$\\
  & $43.1\pm 1.3_{\rm stat}\pm 1.8_{\rm syst}$
  & $1.61\pm 0.09_{\rm stat}\pm 0.21_{\rm syst}$\\
  \hline
  OPAL excl~\cite{Abbiendi:2000hk}
  & $35.81\pm 1.57_{\rm stat}\pm 1.62_{\rm syst}$
  & $1.205\pm 0.207_{\rm stat}\pm 0.153_{\rm syst}$\\
  & $36.8\pm 1.6_{\rm stat}\pm 2.0_{\rm syst}$
  & $1.31\pm 0.21_{\rm stat}\pm 0.16_{\rm syst}$\\
  \hline
  OPAL partial reco~\cite{Abbiendi:2000hk}
  & $36.98\pm 1.19_{\rm stat}\pm 2.32_{\rm syst}$
  & $1.149\pm 0.145_{\rm stat}\pm 0.296_{\rm syst}$\\
  & $37.5\pm 1.2_{\rm stat}\pm 2.5_{\rm syst}$
  & $1.12\pm 0.14_{\rm stat}\pm 0.29_{\rm syst}$\\
  \hline
  DELPHI partial reco~\cite{Abreu:2001ic}
  & $35.15\pm 1.39_{\rm stat}\pm 2.30_{\rm syst}$
  & $1.168\pm 0.126_{\rm stat} \pm 0.381_{\rm syst}$\\
  & $35.5\pm 1.4_{\rm stat}\ {}^{+2.3}_{-2.4}{}_{\rm syst}$
  & $1.34\pm 0.14_{\rm stat}\ {}^{+0.24}_{-0.22}{}_{\rm syst}$\\
  \hline
  DELPHI excl~\cite{Abdallah:2004rz}
  & $35.85\pm 1.68_{\rm stat}\pm 1.98_{\rm syst}$
  & $1.084\pm 0.143_{\rm stat} \pm 0.151_{\rm syst}$\\
  & $39.2\pm 1.8_{\rm stat}\pm 2.3_{\rm syst}$
  & $1.32\pm 0.15_{\rm stat}\pm 0.33_{\rm syst}$\\
  \hline
  \belle~\cite{Dungel:2010uk}
  & $34.39\pm 0.17_{\rm stat}\pm 1.01_{\rm syst}$
  & $1.213\pm 0.034_{\rm stat}\pm 0.008_{\rm syst}$\\
  & $34.6\pm 0.2_{\rm stat}\pm 1.0_{\rm syst}$
  & $1.214\pm 0.034_{\rm stat} \pm 0.009_{\rm syst}$\\
  \hline
  \babar\ excl~\cite{Aubert:2006mb}
  & $33.59\pm 0.29_{\rm stat}\pm 1.03_{\rm syst}$
  & $1.184\pm 0.048_{\rm stat}\pm 0.029_{\rm syst}$\\
  & $34.7\pm 0.3_{\rm stat}\pm 1.1_{\rm syst}$
  & $1.18\pm 0.05_{\rm stat}\pm 0.03_{\rm syst}$\\
  \hline
  \babar\ $D^{*0}$~\cite{Aubert:2009_3}
  & $34.96\pm 0.58_{\rm stat}\pm 1.32_{\rm syst}$
  & $1.126\pm 0.058_{\rm stat}\pm 0.055_{\rm syst}$\\
  & $35.9\pm 0.6_{\rm stat}\pm 1.4_{\rm syst}$
  & $1.16\pm 0.06_{\rm stat}\pm 0.08_{\rm syst}$\\
  \hline
  \babar\ global fit~\cite{Aubert:2009_1}
  & $35.49\pm 0.20_{\rm stat}\pm 1.09_{\rm syst}$
  & $1.185\pm 0.020_{\rm stat}\pm 0.061_{\rm syst}$\\
  & $35.7\pm 0.2_{\rm stat}\pm 1.2_{\rm syst}$
  & $1.21\pm 0.02_{\rm stat}\pm 0.07_{\rm syst}$\\
  \hline
  {\bf Average}
  & \mathversion{bold} $35.61\pm 0.11_{\rm stat}\pm 0.41_{\rm syst}$ &
  \mathversion{bold} $1.205\pm 0.015_{\rm stat}\pm 0.021_{\rm syst}$\\
  \hline 
\end{tabular}
}
\end{center}
\label{tab:vcbf1}
\end{table}
% ----------------------------------------------------------------------

\begin{figure}[!ht]
  \begin{center}
  \unitlength 1.0cm % coordinates in cm
  \begin{picture}(14.,11.0)
    \put(  7.5,-0.2){\includegraphics[width=9.5cm]{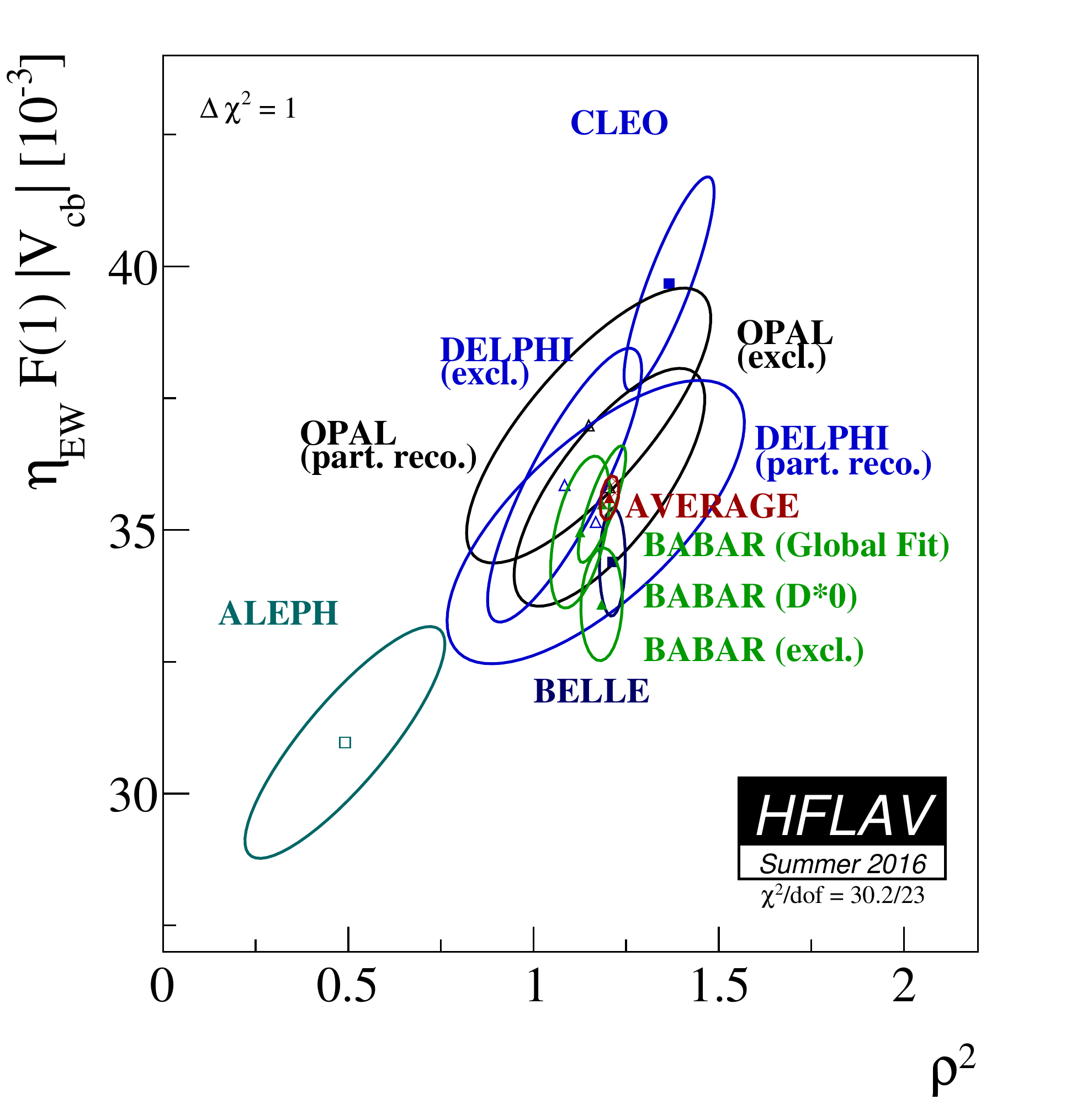}
    }
    \put( -1.5, 0.0){\includegraphics[width=9.0cm]{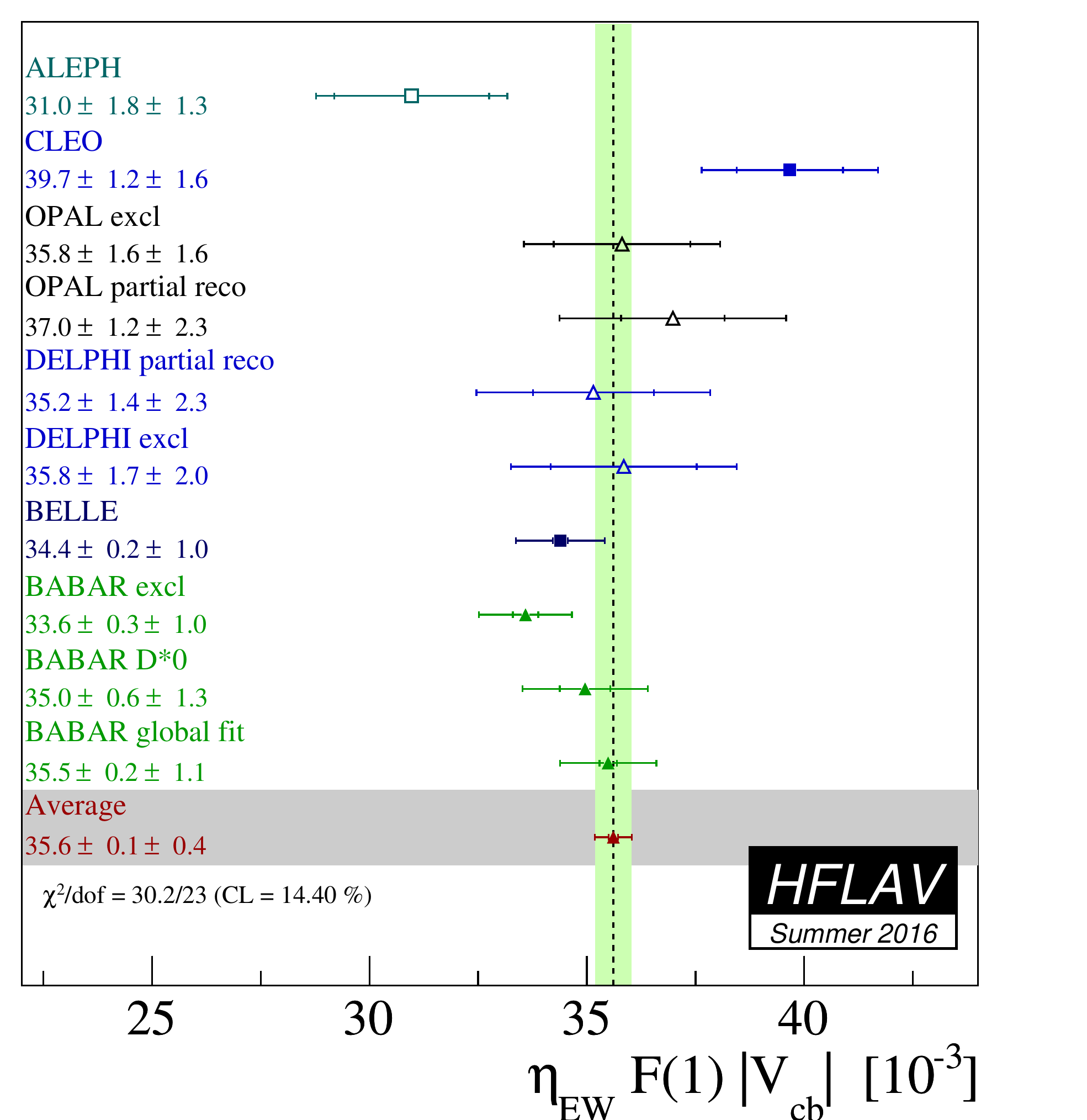}
    }
    \put(  5.8,  8.4){{\large\bf a)}}  
    \put( 14.7,  8.4){{\large\bf b)}}
  \end{picture}
  \caption{Illustration of the (a) the average and (b) the dependence of
    $\eta_\mathrm{EW}{\cal F}(1)\vcb$ on $\rho^2$. The error ellipses
    correspond to $\Delta\chi^2 = 1$ (CL=39\%).} \label{fig:vcbf1}
  \end{center}
\end{figure}

In the next step, we perform a four parameter fit of
$\eta_\mathrm{EW}{\cal F}(1)\vcb$, $\rho^2$, $R_1(1)$ and $R_2(1)$
to the rescaled measurements, taking into account correlated
statistical and systematic uncertainties. Only two measurements
constrain all four parameters~\cite{Dungel:2010uk,Aubert:2006mb}, the remaining
measurements determine only the normalization $\eta_\mathrm{EW}{\cal
  F}(1)\vcb$ and the slope $\rho^2$. The result of the fit is
\begin{eqnarray}
  \eta_\mathrm{EW}{\cal F}(1)\vcb & = & (35.61\pm 0.43)\times
  10^{-3}~, \label{eq:vcbf1} \\
  \rho^2 & = & 1.205\pm 0.026~,\\
  R_1(1) & = & 1.404\pm 0.032~, \label{eq:r1} \\
  R_2(1) & = & 0.854\pm 0.020~, \label{eq:r2}
\end{eqnarray}
and the correlation coefficients are
\begin{eqnarray}
  \rho_{\eta_\mathrm{EW}{\cal F}(1)\vcb,\rho^2} & = & 0.338~,\\
  \rho_{\eta_\mathrm{EW}{\cal F}(1)\vcb,R_1(1)} & = & -0.104~,\\
  \rho_{\eta_\mathrm{EW}{\cal F}(1)\vcb,R_2(1)} & = & -0.071~,\\
  \rho_{\rho^2,R_1(1)} & = & 0.570~,\\
  \rho_{\rho^2,R_2(1)} & = & -0.810~,\\
  \rho_{R_1(1),R_2(1)} & = & -0.758~.
\end{eqnarray}
The uncertainties and correlations quoted here include both
statistical and systematic contributions. The $\chi^2$ of the fit is
30.2 for 23 degrees of freedom, which corresponds to a confidence
level of 14.4\%. An illustration of this fit result is given in
Fig.~\ref{fig:vcbf1}.

Using the lastest update from the Fermilab Lattice and MILC
Collaborations~\cite{Bailey:2014tva}, the form factor normalization
$\eta_\mathrm{EW}{\cal F}(1)$ is
\begin{equation}
  \eta_\mathrm{EW}{\cal F}(1) = 0.912\pm 0.014~,
\end{equation}
where $\eta_\mathrm{EW}=1.0066\pm 0.0050$ has been used. The central value of
this number corresponds to the electroweak correction only. The uncertainty
has been increased to accomodate the Coulomb effect. Based on Eq.~(\ref{eq:vcbf1}), this results in
\begin{equation}
  \vcb = (39.05\pm 0.47_{\rm exp}\pm 0.58_{\rm th})\times
  10^{-3}~, \label{eq:vcbdstar}
\end{equation}
where the first uncertainty is experimental and the second error is
theoretical (lattice QCD calculation and electro-weak correction).

From each rescaled measurements in Table~\ref{tab:vcbf1}, we
calculate the $\bar B\to D^*\ell^-\bar\nu_\ell$ form factor
$\eta_\mathrm{EW}{\cal F}(w)$ and, by numerical integration, the
branching ratio of the decay $\bar B^0\to
D^{*+}\ell^-\bar\nu_\ell$. For measurements that do not determine the
parameters $R_1(1)$ and $R_2(1)$ we assume the average
values given in Eqs.~(\ref{eq:r1}) and~(\ref{eq:r2}). 
The results are quoted in Table~\ref{tab:dstarlnu}. The branching ratio is
\begin{equation}
  \cbf(\BzbDstarlnu)=(4.88\pm 0.10)\%~. \label{eq:br_dstarlnu}
\end{equation}

We have also performed a one-dimensional average of measurements of the
decay $B^-\to D^{*0}\ell^-\bar\nu_\ell$, which is shown in
Table~\ref{tab:dstar0lnu}. The result of this average,
$(5.59\pm 0.02\pm 0.19)$\%, is consistent with the average including
both charge states given in Eq.~(\ref{eq:br_dstarlnu}) rescaled by the
lifetime ratio $\tau(B^+)/\tau(B^0)$, $(5.26\pm 0.11)$\%.

% ----------------------------------------------------------------------
\begin{table}[!htb]
\caption{$\BzbDstarlnu$ branching fractions calculated from the
  rescaled CLN pameters in Table~\ref{tab:vcbf1}. For
  Ref.~\cite{Aubert:2009_3} the published value of ${\cal B}(B^-\to
  D^{*0}\ell^-\bar\nu_\ell)$ has been rescaled by the factor
  $\tau(B^0)/\tau(B^+)$ for comparison to the other measurements.}
\begin{center} 
\resizebox{0.99\textwidth}{!}{
\begin{tabular}{|l|c|c|}\hline
  Experiment & $\cbf(\BzbDstarlnu)$ [\%] (calculated) &
  $\cbf(\BzbDstarlnu)$ [\%] (published)\\
  \hline\hline
  ALEPH~\cite{Buskulic:1996yq}
  & $5.26\pm 0.25_{\rm stat} \pm 0.30_{\rm syst}$
  & $5.53\pm 0.26_{\rm stat} \pm 0.52_{\rm syst}$\\
  CLEO~\cite{Adam:2002uw}
  & $5.55\pm 0.17_{\rm stat} \pm 0.24_{\rm syst}$
  & $6.09\pm 0.19_{\rm stat} \pm 0.40_{\rm syst}$\\
  OPAL excl~\cite{Abbiendi:2000hk}
  & $4.93\pm 0.18_{\rm stat} \pm 0.43_{\rm syst}$
  & $5.11\pm 0.19_{\rm stat} \pm 0.49_{\rm syst}$\\
  OPAL partial reco~\cite{Abbiendi:2000hk}
  & $5.42\pm 0.25_{\rm stat} \pm 0.52_{\rm syst}$
  & $5.92\pm 0.27_{\rm stat} \pm 0.68_{\rm syst}$\\
  DELPHI partial reco~\cite{Abreu:2001ic}
  & $4.85\pm 0.13_{\rm stat} \pm 0.72_{\rm syst}$
  & $4.70\pm 0.13_{\rm stat} \ {}^{+0.36}_{-0.31}\ {}_{\rm syst}$\\
  DELPHI excl~\cite{Abdallah:2004rz}
  & $5.27\pm 0.20_{\rm stat} \pm 0.37_{\rm syst}$
  & $5.90\pm 0.22_{\rm stat} \pm 0.50_{\rm syst}$\\
  \belle~\cite{Dungel:2010uk}
  & $4.51\pm 0.03_{\rm stat} \pm 0.26_{\rm syst}$
  & $4.58\pm 0.03_{\rm stat} \pm 0.26_{\rm syst}$\\
  \babar\ excl~\cite{Aubert:2006mb}
  & $4.45\pm 0.04_{\rm stat}\pm 0.26_{\rm syst}$
  & $4.69\pm 0.04_{\rm stat} \pm 0.34_{\rm syst}$\\
  \babar\ $D^{*0}$~\cite{Aubert:2009_3}
  & $4.90\pm 0.07_{\rm stat}\pm 0.34_{\rm syst}$
  & $5.15\pm 0.07_{\rm stat} \pm 0.38_{\rm syst}$\\
  \babar\ global fit~\cite{Aubert:2009_1}
  & $4.90\pm 0.02_{\rm stat}\pm 0.19_{\rm syst}$
  & $5.00\pm 0.02_{\rm stat} \pm 0.19_{\rm syst}$\\
  \hline 
  {\bf Average} & \mathversion{bold}$4.88\pm 0.01_{\rm stat}\pm
  0.10_{\rm syst}$ & \mathversion{bold}$\chi^2/\dof = 30.2/23$ (CL=$14.4\%$)\\
  \hline 
\end{tabular}
}
\end{center}
\label{tab:dstarlnu}
\end{table}
% ----------------------------------------------------------------------

% ----------------------------------------------------------------------0
\begin{table}[!htb]
\caption{Average of the $B^-\to D^{*0}\ell^-\bar\nu_\ell$ branching
  fraction measurements.}
\begin{center}
\begin{tabular}{|l|c|c|}
  \hline
  Experiment & $\cbf(B^-\to D^{*0}\ell^-\bar\nu_\ell)$ [\%] (rescaled) &
  $\cbf(B^-\to D^{*0}\ell^-\bar\nu_\ell)$ [\%] (published)\\
  \hline \hline
  CLEO~\cite{Adam:2002uw}
  & $6.52\pm 0.20_{\rm stat}\pm 0.39_{\rm syst}$
  & $6.50\pm 0.20_{\rm stat}\pm 0.43_{\rm syst}$\\
  \babar tagged~\cite{Aubert:vcbExcl}
  & $5.48\pm 0.15_{\rm stat}\pm 0.35_{\rm syst}$
  & $5.83\pm 0.15_{\rm stat}\pm 0.30_{\rm syst}$\\
  \babar~\cite{Aubert:2009_3}
  & $5.28\pm 0.08_{\rm stat}\pm 0.40_{\rm syst}$
  & $5.56\pm 0.08_{\rm stat}\pm 0.41_{\rm syst}$\\
  \babar~\cite{Aubert:2009_1}
  & $5.36\pm 0.02_{\rm stat}\pm 0.21_{\rm syst}$
  & $5.40\pm 0.02_{\rm stat}\pm 0.21_{\rm syst}$\\
  \hline
  {\bf Average} & \mathversion{bold}$5.59\pm 0.02_{\rm stat}\pm
  0.19_{\rm syst}$ & \mathversion{bold}$\chi^2/\dof = 8.3/3$ (CL=$3.94\%$)\\
  \hline 
\end{tabular}
\end{center}
\label{tab:dstar0lnu}
\end{table}
% ----------------------------------------------------------------------

\begin{figure}[!ht]
  \begin{center}
  \unitlength1.0cm % coordinates in cm
  \begin{picture}(14.,9.0)  %ys(25.,6.0)
    \put( -1.5, 0.0){\includegraphics[width=9.0cm]{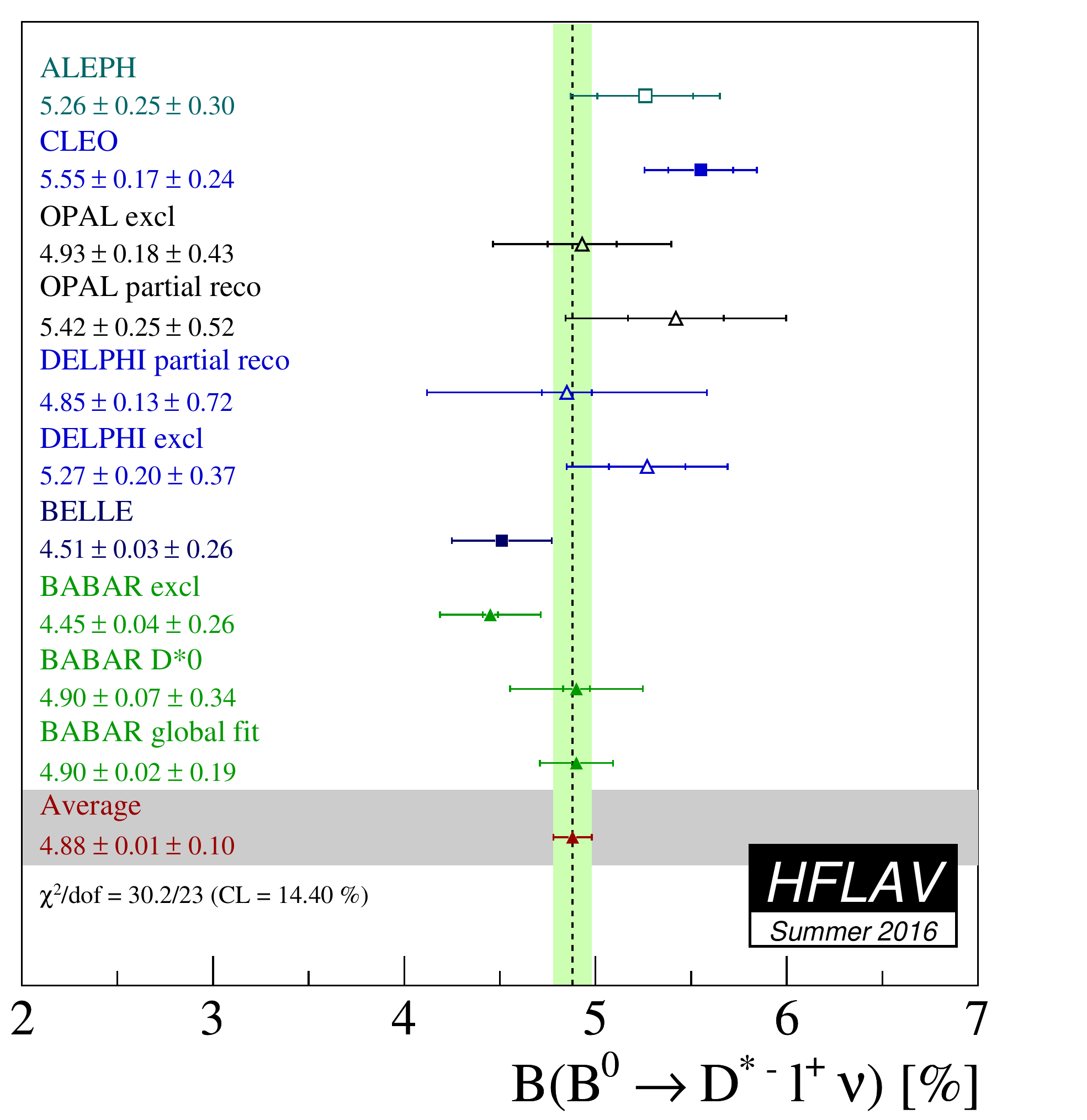}
    }
    \put(  7.5, 0.0){\includegraphics[width=9.0cm]{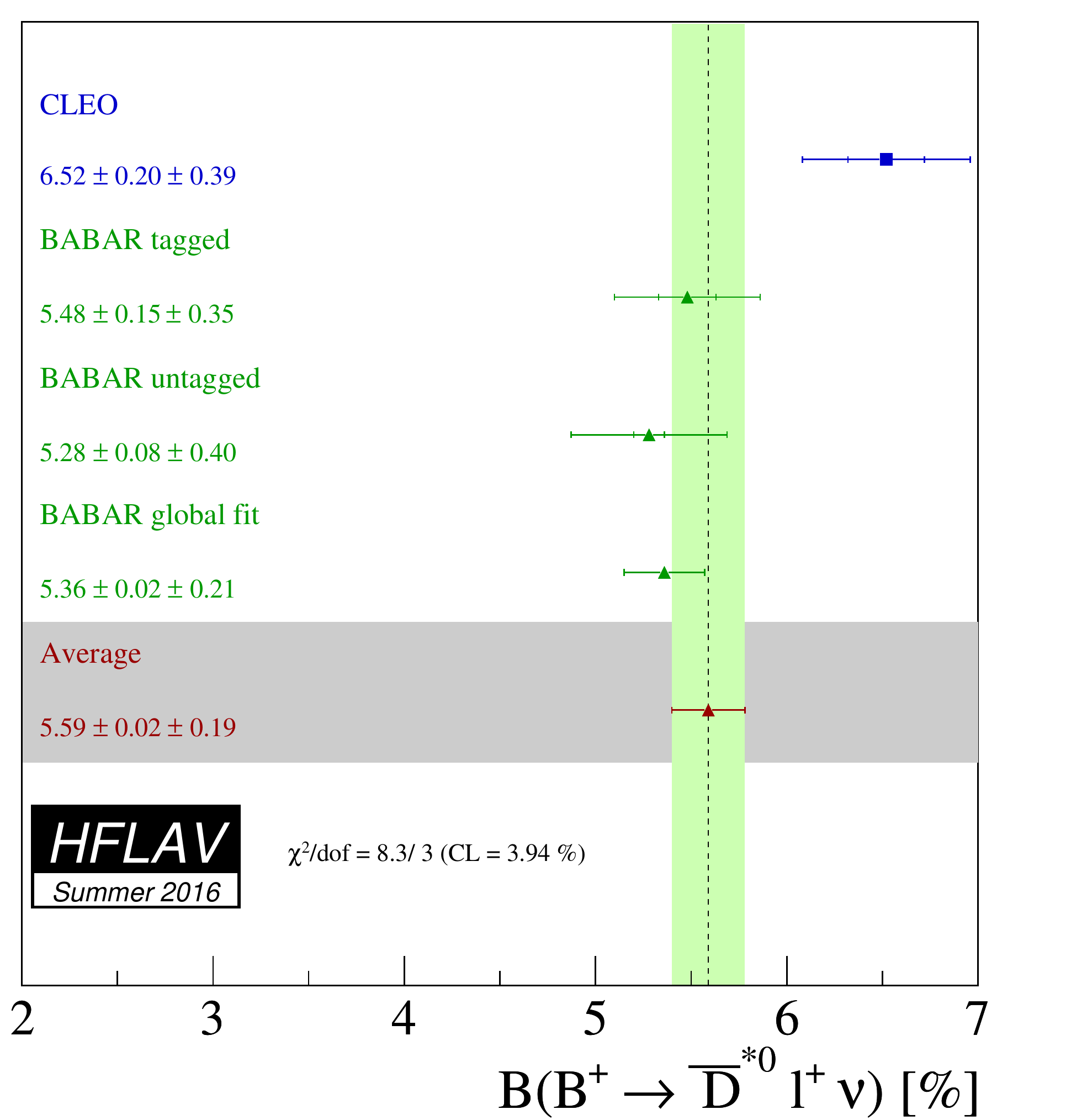}
    }
    \put(  5.8, 8.5){{\large\bf a)}}
    \put( 14.8, 8.5){{\large\bf b)}}
  \end{picture}
  \caption{Branching fractions of exclusive semileptonic
    $B$ decays: (a) $\BzbDstarlnu$
    (Table~\ref{tab:dstarlnu}) and (b) $B^-\to
    D^{*0}\ell^-\bar\nu_\ell$ (Table~\ref{tab:dstar0lnu}).} \label{fig:brdsl}
  \end{center}
\end{figure}

\mysubsubsection{$\bar B\to D\ell^-\bar\nu_\ell$}
\label{slbdecays_dlnu}

The differential decay rate for massless fermions as a function of $w$
(introduced in the previous section) is given by (see, \eg,~\cite{Neubert:1993mb})
\begin{equation}
  \frac{\bar B\to D\ell^-\bar\nu_\ell}{dw} = \frac{G^2_\mathrm{F} m^3_D}{48\pi^3}(m_B+m_D)^2(w^2-1)^{3/2}\eta_\mathrm{EW}^2\mathcal{G}^2(w)|V_{cb}|^2~,
\end{equation}
where $G_\mathrm{F}$ is Fermi's constant, and $m_B$ and $m_D$ are the $B$ and $D$
meson masses. Again, $\eta_\mathrm{EW}$ is the electroweak correction
introduced in the previous section. In contrast to
$\bar B\to D^*\ell^-\bar\nu_\ell$, $\mathcal{G}(w)$ contains a single
form-factor function $f_+(w)$,
\begin{equation}
  \mathcal{G}^2(w) = \frac{4r}{(1+r)^2} f^2_+(w)~,
\end{equation}
where $r=m_D/m_B$.

As for $\bar B\to D^*\ell^-\bar\nu_\ell$ decays, we adopt the prescription by
Caprini, Lellouch and Neubert~\cite{CLN}, which describes the shape and
normalization of the measured decay distributions in terms of two parameters:
the normalization ${\cal G}(1)$ and the slope $\rho^2$,
\begin{equation}
  \mathcal{G}(z)= \mathcal{G}(1)\big[1 - 8 \rho^2 z + (51 \rho^2 - 10 )
  z^2 - (252 \rho^2 - 84 ) z^3\big]~,
\end{equation}
where $z=(\sqrt{w+1}-\sqrt{2})/(\sqrt{w+1}+\sqrt{2})$.

Table~\ref{tab:vcbg1} shows experimental measurements of the two CLN
parameters, which are corrected to match the latest values of the input
parameters~\cite{HFLAV_sl:inputparams}. Both measurements of $\BzbDplnu$ and
$B^-\to D^0\ell^-\bar\nu_\ell$ are used and isospin symmetry is assumed in the
analysis.
\begin{table}[!htb]
\caption{Measurements of the Caprini, Lellouch and Neubert
  (CLN)~\cite{CLN} form factor parameters in $\bar B\to
  D\ell^-\bar\nu_\ell$ before and after rescaling.}
\begin{center}
\begin{tabular}{|l|c|c|}
  \hline
  Experiment
  & $\eta_\mathrm{EW}{\cal G}(1)\vcb$ [10$^{-3}$] (rescaled)
  & $\rho^2$ (rescaled)\\
  & $\eta_\mathrm{EW}{\cal G}(1)\vcb$ [10$^{-3}$] (published)
  & $\rho^2$ (published)\\
  \hline \hline
  ALEPH~\cite{Buskulic:1996yq}
  & $36.67\pm 10.05_{\rm stat}\pm 7.33_{\rm syst}$
  & $0.845\pm 0.879_{\rm stat}\pm 0.448_{\rm syst}$\\
  & $31.1\pm 9.9_{\rm stat}\pm 8.6_{\rm syst}$
  & $0.70\pm 0.98_{\rm stat}\pm 0.50_{\rm syst}$\\
  \hline
  CLEO~\cite{Bartelt:1998dq}
  & $44.18\pm 5.70_{\rm stat}\pm 3.47_{\rm syst}$
  & $1.270\pm 0.215_{\rm stat}\pm 0.121_{\rm syst}$\\
  & $44.8\pm 6.1_{\rm stat}\pm 3.7_{\rm syst}$
  & $1.30\pm 0.27_{\rm stat}\pm 0.14_{\rm syst}$\\
  \hline
  \belle~\cite{Glattauer:2015teq}
  & $41.94\pm 0.60_{\rm stat}\pm 1.21_{\rm syst}$
  & $1.090\pm 0.036_{\rm stat}\pm 0.019_{\rm syst}$\\
  & $42.29\pm 1.37$ & $1.09\pm 0.05$\\
  \hline
  \babar global fit~\cite{Aubert:2009_1}
  & $42.23\pm 0.74_{\rm stat}\pm 2.14_{\rm syst}$
  & $1.186\pm 0.035_{\rm stat}\pm 0.062_{\rm syst}$\\
  & $43.1\pm 0.8_{\rm stat}\pm 2.3_{\rm syst}$
  & $1.20\pm 0.04_{\rm stat}\pm 0.07_{\rm syst}$\\
  \hline
  \babar tagged~\cite{Aubert:2009_2}
  & $42.60\pm 1.71_{\rm stat}\pm 1.26_{\rm syst}$
  & $1.200\pm 0.088_{\rm stat}\pm 0.043_{\rm syst}$\\
  & $42.3\pm 1.9_{\rm stat}\pm 1.0_{\rm syst}$
  & $1.20\pm 0.09_{\rm stat}\pm 0.04_{\rm syst}$\\
  \hline 
  {\bf Average }
  & \mathversion{bold}$41.57\pm 0.45_{\rm stat}\pm 0.89_{\rm syst}$
  & \mathversion{bold}$1.128\pm 0.024_{\rm stat}\pm 0.023_{\rm syst}$\\
  \hline 
\end{tabular}
\end{center}
\label{tab:vcbg1}
\end{table}

\begin{figure}[!ht]
  \begin{center}
  \unitlength1.0cm % coordinates in cm
  \begin{picture}(14.,10.) %ys(25.,6.)
    \put(  7.5, -0.2){\includegraphics[width=9.5cm]{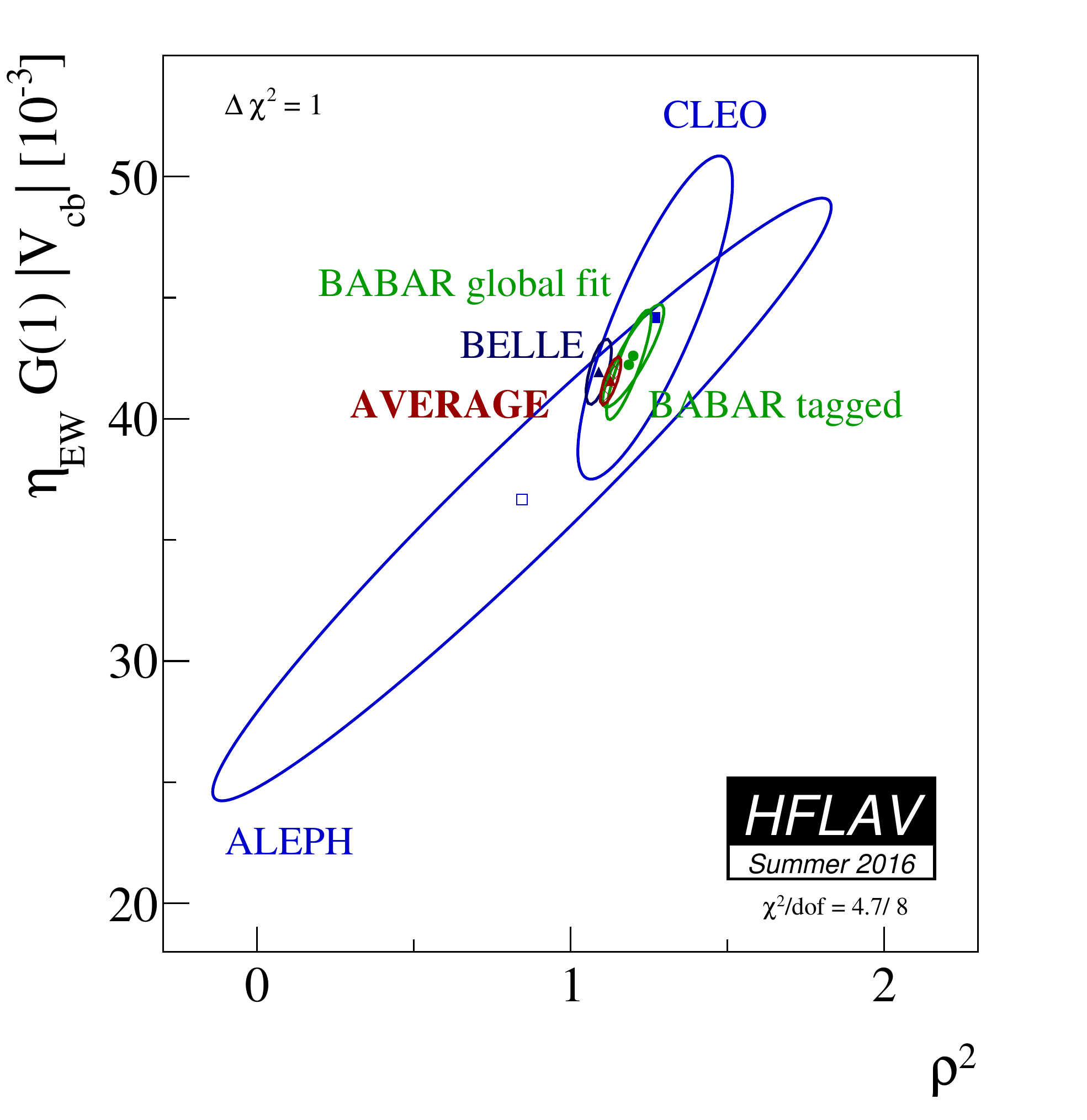}
    }
    \put( -1.5,  0.0){\includegraphics[width=9.0cm]{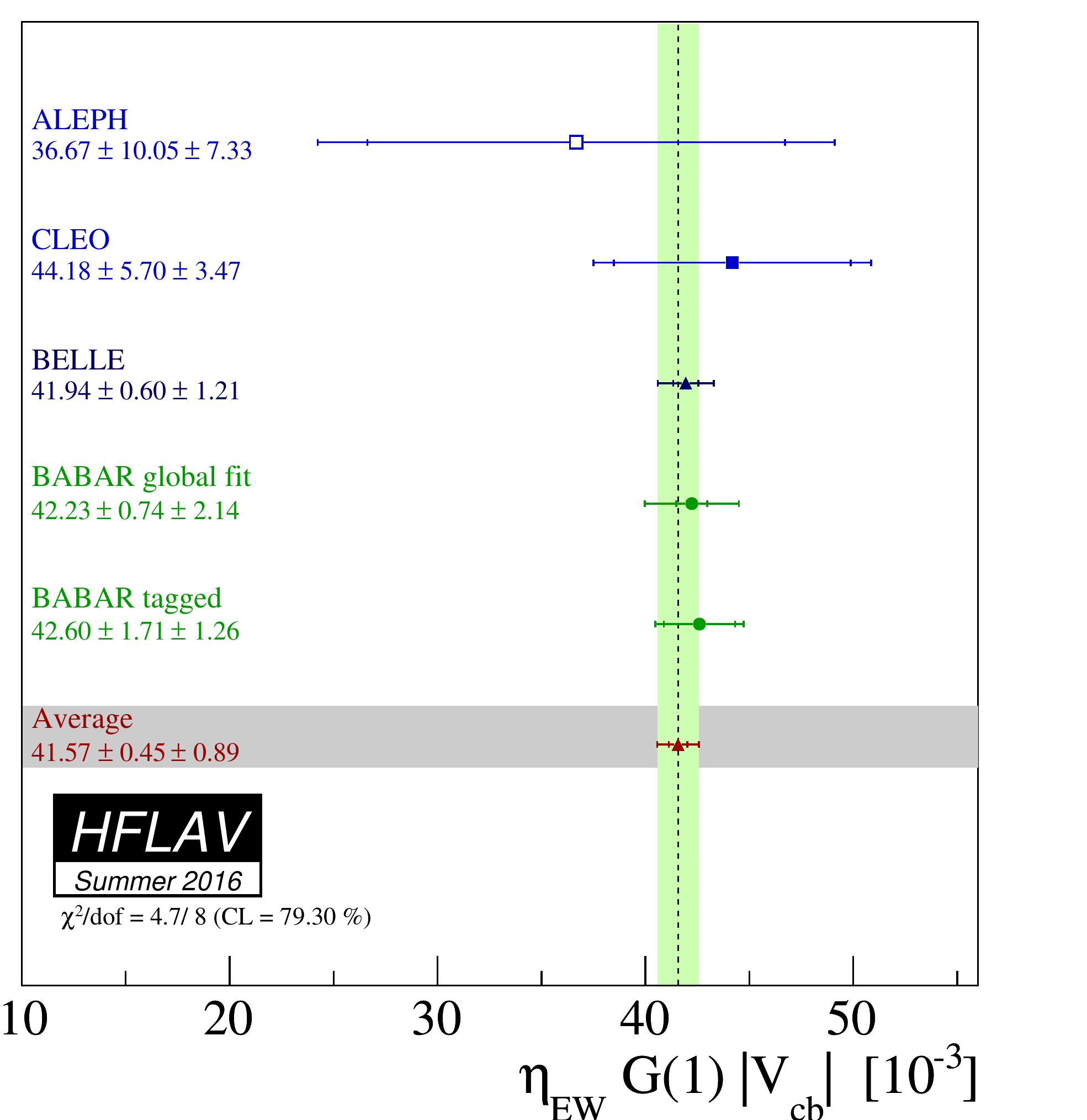}
    }
    \put(  5.8, 8.5){{\large\bf a)}}
    \put( 15.1, 8.5){{\large\bf b)}}
  \end{picture}
  \caption{Illustration of the (a) the average and (b) dependence of
    $\eta_\mathrm{EW}{\cal G}(w)\vcb$ on $\rho^2$. The error ellipses correspond
    to $\Delta\chi^2 = 1$ (CL=39\%).}
  \label{fig:vcbg1}
  \end{center}
\end{figure}

The form factor parameters are extracted by a two-parameter fit to
the rescaled measurements of $\eta_\mathrm{EW}{\cal G}(1)\vcb$ and
$\rho^2$ taking into account correlated statistical and systematic
uncertainties. The result of the fit is
\begin{eqnarray}
  \eta_\mathrm{EW}{\cal G}(1)\vcb & = & (41.57\pm 1.00)\times
  10^{-3}~, \label{eq:vcbg1} \\
  \rho^2 & = & 1.128 \pm 0.033~,
\end{eqnarray}
with a correlation of
\begin{equation}
  \rho_{\eta_\mathrm{EW}{\cal G}(1)\vcb,\rho^2} = 0.751~.
\end{equation}
The uncertainties and the correlation coefficient include both
statistical and systematic contributions. The $\chi^2$ of the fit is
4.7 for 8 degrees of freedom, which corresponds to a probability of 
79.3\%. An illustration of this fit result is given in
Fig.~\ref{fig:vcbg1}.

The most recent lattice QCD result obtained for the form factor
normalization is~\cite{Lattice:2015rga}
\begin{equation}
  {\cal G}(1) = 1.0541\pm 0.0083~.
\end{equation}
Using again $\eta_\mathrm{EW}=1.0066\pm 0.0050$, we determine $\vcb$ from
Eq.~(\ref{eq:vcbg1}),
\begin{equation}
  \vcb = (39.18\pm 0.94_{\rm exp}\pm 0.36_{\rm th})\times 10^{-3}~,
\end{equation}
where the first error is experimental and the second theoretical. This
number is in excellent agreement with $\vcb$ obtained from 
$\bar B\to D^*\ell^-\bar\nu_\ell$ decays given in Eq.~(\ref{eq:vcbdstar}).

\begin{figure}[!ht]
  \begin{center}
    \includegraphics[width=9.5cm]{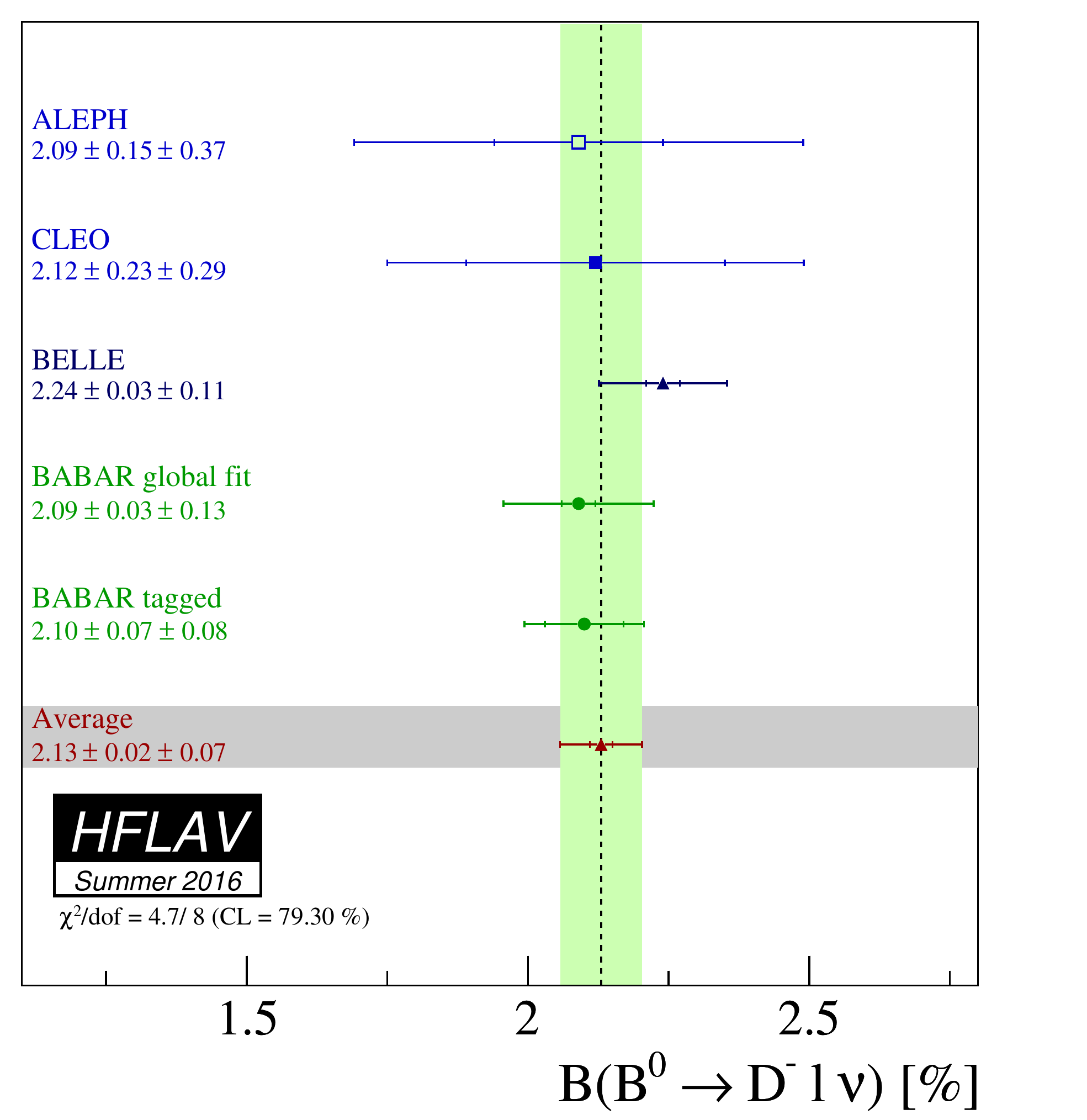}
    \caption{Illustration of Table~\ref{tab:dlnuIso}.} \label{fig:brdlIso}
  \end{center}
\end{figure}

From each rescaled measurement in Table~\ref{tab:vcbg1}, we have
calculated the $\bar B\to D\ell^-\bar\nu_\ell$ form factor ${\cal
  G}(w)$ and, by numerical integration, the branching ratio of the
decay $\BzbDplnu$. The results are quoted in Table~\ref{tab:dlnuIso} and
illustrated in Fig.~\ref{fig:brdlIso}. The branching ratio for
the average values of $\eta_\mathrm{EW}{\cal G}(1)\vcb$ and $\rho^2$ is
\begin{equation}
  \cbf(\BzbDplnu)=(2.13\pm 0.07)\%~. \label{eq:br_dlnu}
\end{equation}
% ----------------------------------------------------------------------0
\begin{table}[!htb]
\caption{$\bar B^0\to D^+\ell^-\bar\nu_\ell$ branching fractions
  calculated from the rescaled CLN parameters in Table~\ref{tab:vcbg1},
  which are based on both charged and neutral $B$ decays, combined under
  the assumption of isospin symmetry.}
\begin{center}
\resizebox{0.99\textwidth}{!}{
\begin{tabular}{|l|c|c|}
  \hline
  Experiment
  & $\cbf(\bar B^0\to D^+\ell^-\bar\nu_\ell)$ [\%] (calculated)
  & $\cbf(\bar B^0\to D^+\ell^-\bar\nu_\ell)$ [\%] (published)\\
  \hline \hline
  ALEPH~\cite{Buskulic:1996yq}
  & $2.09\pm 0.15_{\rm stat}\pm 0.37_{\rm syst}$
  & $2.35\pm 0.20_{\rm stat}\pm 0.44_{\rm syst}$\\
  CLEO~\cite{Bartelt:1998dq}
  & $2.12\pm 0.23_{\rm stat}\pm 0.29_{\rm syst}$
  & $2.20\pm 0.16_{\rm stat}\pm 0.19_{\rm syst}$\\
  \belle~\cite{Glattauer:2015teq}
  & $2.24\pm 0.03_{\rm stat}\pm 0.11_{\rm syst}$
  & $2.31\pm 0.03_{\rm stat}\pm 0.11_{\rm syst}$\\
  \babar global fit~\cite{Aubert:2009_1}
  & $2.09\pm 0.03_{\rm stat}\pm 0.13_{\rm syst}$
  & $2.34\pm 0.03_{\rm stat}\pm 0.13_{\rm syst}$\\
  \babar tagged~\cite{Aubert:2009_2}
  & $2.10\pm 0.07_{\rm stat}\pm 0.08_{\rm syst}$
  & $2.23\pm 0.11_{\rm stat}\pm 0.11_{\rm syst}$\\
  \hline 
  {\bf Average}
  & \mathversion{bold}$2.13\pm 0.02_{\rm stat}\pm 0.07_{\rm syst}$
  & \mathversion{bold}$\chi^2/\dof = 4.7/8$ (CL=$79.3\%$)\\
  \hline 
\end{tabular}
}
\end{center}
\label{tab:dlnuIso}
\end{table}
% ----------------------------------------------------------------------

We have also performed one-dimensional averages of measurements
of $\BzbDplnu$ and $B^-\to D^0\ell^-\bar\nu_\ell$ decays. The results
are shown in Tables~\ref{tab:dlnu} and \ref{tab:d0lnu}. The $\BzbDplnu$
average, $(2.20\pm 0.04\pm 0.09)$\%, is consistent with the result in
Eq.~(\ref{eq:br_dlnu}), $(2.13\pm 0.07)$\%. The $B^-\to D^0\ell^-\bar\nu_\ell$
average, $(2.33\pm 0.04\pm 0.09)$\%, also compares well to the result in
Eq.~(\ref{eq:br_dlnu}), rescaled by the lifetime ratio
$\tau(B^+)/\tau(B^0)$, $(2.30\pm 0.07)$\%.
% ----------------------------------------------------------------------0
\begin{table}[!htb]
\caption{Average of $\BzbDplnu$ branching fraction
  measurements.}
\begin{center}
\begin{tabular}{|l|c|c|}
  \hline
  Experiment
  & $\cbf(\BzbDplnu)$ [\%] (rescaled)
  & $\cbf(\BzbDplnu)$ [\%] (published)\\
  \hline \hline
  ALEPH~\cite{Buskulic:1996yq}
  & $2.14\pm 0.18_{\rm stat}\pm 0.36_{\rm syst}$
  & $2.35\pm 0.20_{\rm stat}\pm 0.44_{\rm syst}$\\
  CLEO~\cite{Bartelt:1998dq}
  & $2.09\pm 0.13_{\rm stat}\pm 0.16_{\rm syst}$
  & $2.20\pm 0.16_{\rm stat}\pm 0.19_{\rm syst}$\\
  \belle~\cite{Glattauer:2015teq}
  & $2.30\pm 0.04_{\rm stat}\pm 0.12_{\rm syst}$
  & $2.39\pm 0.04_{\rm stat}\pm 0.11_{\rm syst}$\\
  \babar~\cite{Aubert:vcbExcl}
  & $2.08\pm 0.11_{\rm stat}\pm 0.14_{\rm syst}$
  & $2.21\pm 0.11_{\rm stat}\pm 0.12_{\rm syst}$\\
  \hline 
  {\bf Average}
  & \mathversion{bold}$2.20\pm 0.04_{\rm stat}\pm 0.09_{\rm syst}$
  & \mathversion{bold}$\chi^2/\dof = 1.7/3$ (CL=$63.9\%$)\\
  \hline 
\end{tabular}
\end{center}
\label{tab:dlnu}
\end{table}
% ----------------------------------------------------------------------

% ----------------------------------------------------------------------0
\begin{table}[!htb]
\caption{Average of $B^-\to D^0\ell^-\bar\nu_\ell$ branching fraction
  measurements.}
\begin{center}
\begin{tabular}{|l|c|c|}
  \hline
  Experiment
  & $\cbf(B^-\to D^0\ell^-\bar\nu_\ell)$ [\%] (rescaled)
  & $\cbf(B^-\to D^0\ell^-\bar\nu_\ell)$ [\%] (published)\\
  \hline \hline
  CLEO~\cite{Bartelt:1998dq}
  & $2.16\pm 0.13_{\rm stat}\pm 0.17_{\rm syst}$
  & $2.32\pm 0.17_{\rm stat}\pm 0.20_{\rm syst}$\\
  \babar~\cite{Aubert:vcbExcl}
  & $2.21\pm 0.09_{\rm stat}\pm 0.12_{\rm syst}$
  & $2.33\pm 0.09_{\rm stat}\pm 0.09_{\rm syst}$\\
  \belle~\cite{Glattauer:2015teq}
  & $2.48\pm 0.04_{\rm stat}\pm 0.12_{\rm syst}$
  & $2.54\pm 0.04_{\rm stat}\pm 0.13_{\rm syst}$\\
  \hline
  {\bf Average}
  & \mathversion{bold}$2.33\pm 0.04_{\rm stat}\pm 0.09_{\rm syst}$
  & \mathversion{bold}$\chi^2/\dof = 2.8/2$ (CL=$25.2\%$)\\
  \hline
\end{tabular}
\end{center}
\label{tab:d0lnu}
\end{table}
% ----------------------------------------------------------------------

%===================================================================
% D** 
%===================================================================

\mysubsubsection{$\bar{B} \to D^{(*)}\pi \ell^-\bar{\nu}_{\ell}$}
\label{slbdecays_dpilnu}
% --------------------

\begin{figure}[!ht]
 \begin{center}
  \unitlength1.0cm % coordinates in cm
  \begin{picture}(14.,9.5)  %ys(25.,6.0)
   \put( -1.5,  0.0){\includegraphics[width=8.7cm]{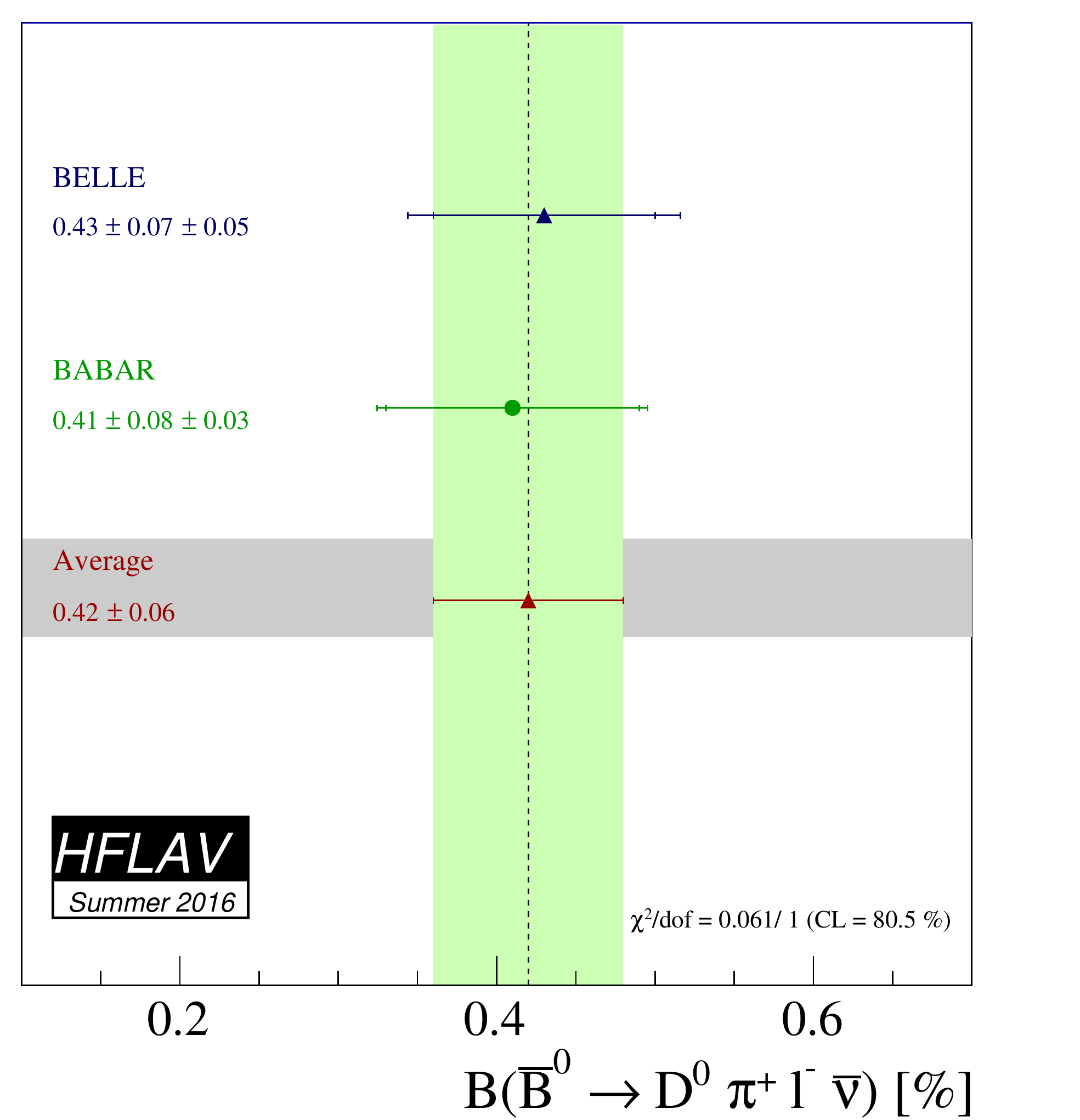}
   }
   \put(  7.5,  0.0){\includegraphics[width=8.7cm]{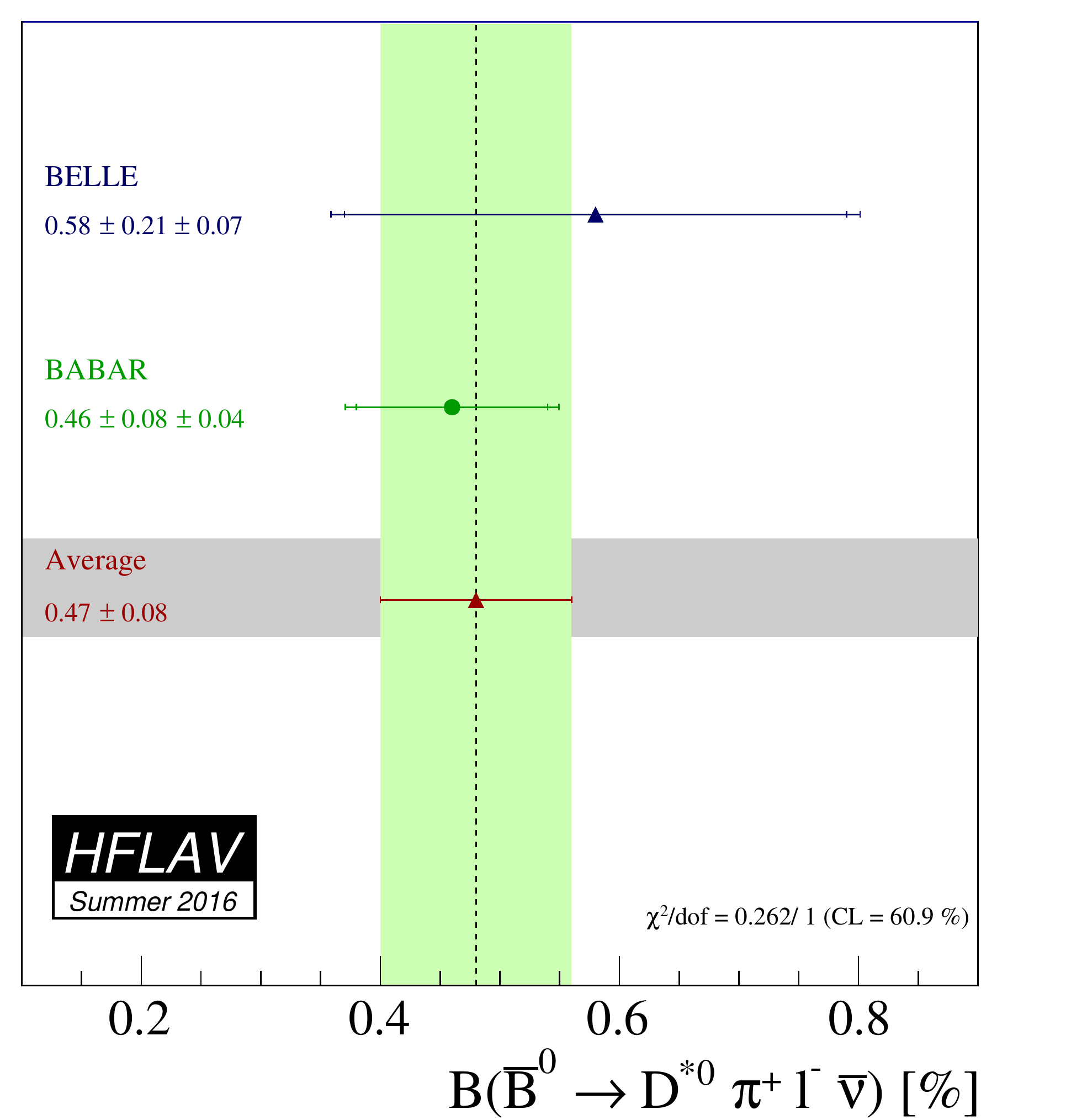}
   }
   \put(  5.5,  8.2){{\large\bf a)}}
   \put( 14.5,  8.2){{\large\bf b)}}
  \end{picture}
  \begin{picture}(14.,9.5)  %ys(25.,6.0)
   \put( -1.5,  0.0){\includegraphics[width=8.7cm]{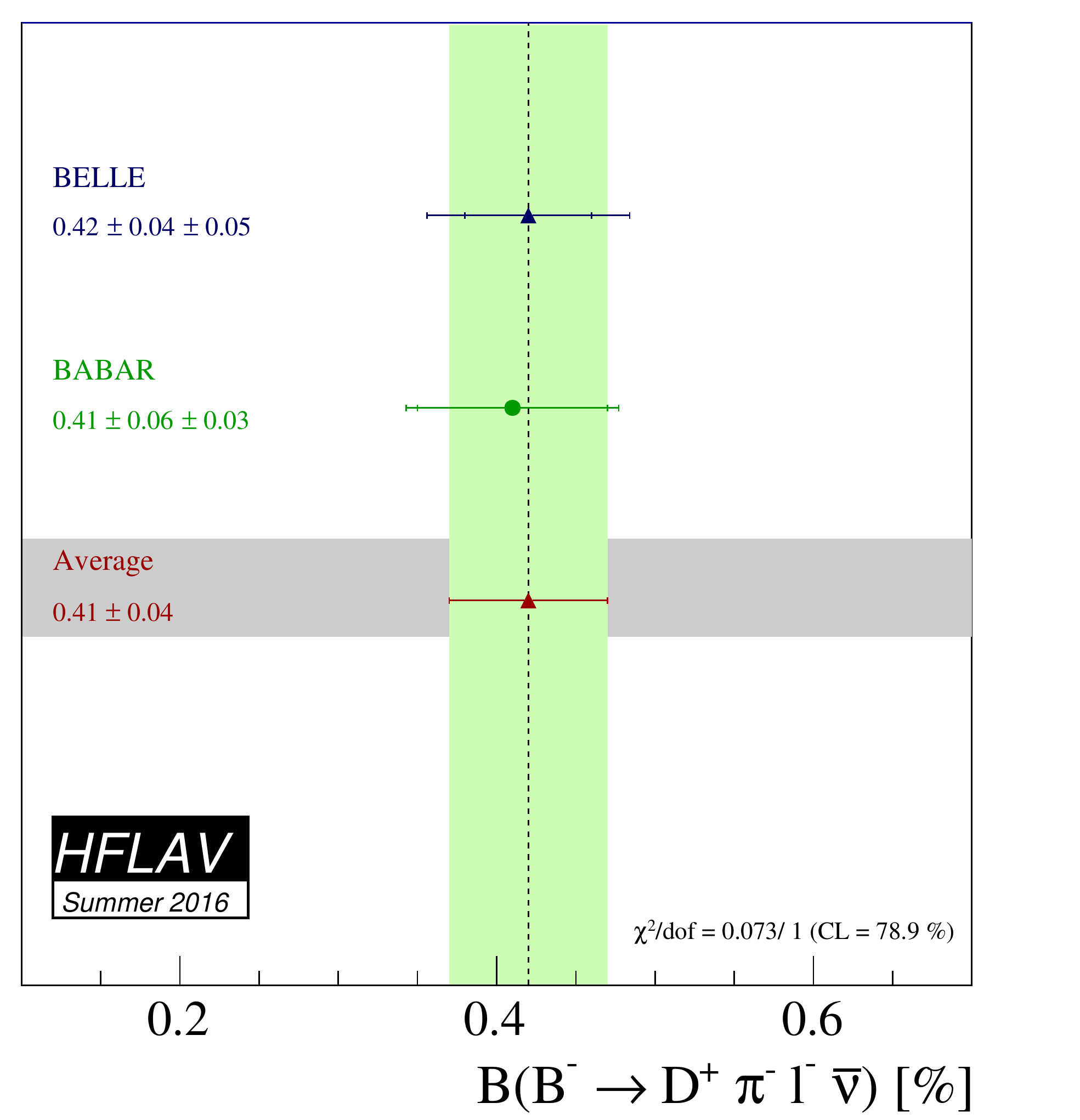}
   }
   \put(  7.5,  0.0){\includegraphics[width=8.7cm]{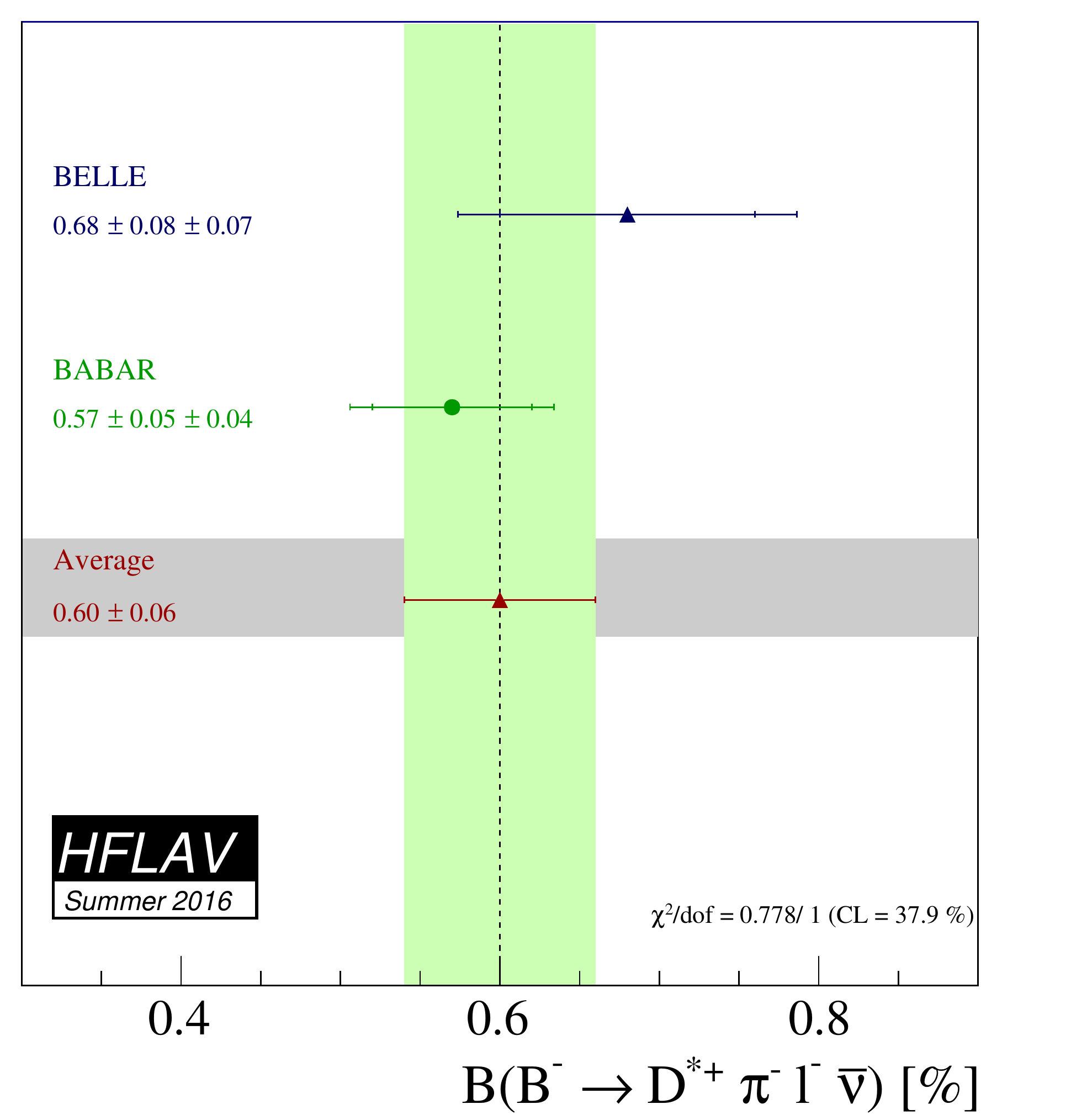}
   }
   \put(  5.5,  8.2){{\large\bf c)}}
   \put( 14.5,  8.2){{\large\bf d)}}
  \end{picture}
  \caption{Average branching fraction  of exclusive semileptonic $B$ decays
(a) $\bar{B}^0 \to D^0 \pi^+ \ell^-\bar{\nu}_{\ell}$, (b) $\bar{B}^0 \to D^{*0} \pi^+
\ell^-\bar{\nu}_{\ell}$, (c) $B^- \to D^+ \pi^-
\ell^-\bar{\nu}_{\ell}$, and (d) $B^- \to D^{*+} \pi^- \ell^-\bar{\nu}_{\ell}$.
The corresponding individual
  results are also shown.}
  \label{fig:brdpil}
 \end{center}
\end{figure}

The average inclusive branching fractions for $\bar{B} \to D^{(*)}\pi\ell^-\bar{\nu}_{\ell}$ 
decays, where no constraint is applied to the $D^{(*)}\pi$ system, are determined by the
combination of the results provided in Table~\ref{tab:dpilnu} for 
$\bar{B}^0 \to D^0 \pi^+ \ell^-\bar{\nu}_{\ell}$, $\bar{B}^0 \to D^{*0} \pi^+
\ell^-\bar{\nu}_{\ell}$, 
$B^- \to D^+ \pi^-
\ell^-\bar{\nu}_{\ell}$, and $B^- \to D^{*+} \pi^-
\ell^-\bar{\nu}_{\ell}$ decays.
The measurements included in the average 
are scaled to a consistent set of input
parameters and their uncertainties~\cite{HFLAV_sl:inputparams}.
For both the \babar\ and Belle results, the $B$ semileptonic signal yields are
 extracted from a fit to the missing mass squared distribution for a sample of fully
 reconstructed \BB\ events.  
Figure~\ref{fig:brdpil} shows the measurements and the resulting average for the 
four decay modes.

% ----------------------------------------------------------------------
\begin{table}[!htb]
\caption{Averages of the $B \to D^{(*)} \pi^- \ell^-\bar{\nu}_{\ell}$  branching fractions and individual results.}
\begin{center}
\begin{tabular}{|l|c c|}\hline
Experiment                                 &$\cbf(B^- \to D^+ \pi^- \ell^-\bar{\nu}_{\ell}) [\%]$ (rescaled) & $\cbf(B^- \to D^+ \pi^- \ell^-\bar{\nu}_{\ell}) [\%]$ (published)\\
\hline
\belle  ~\cite{Live:Dss}             &$0.42 \pm0.04_{\rm stat} \pm0.05_{\rm syst}$  & $0.40 \pm0.04_{\rm stat} \pm0.06_{\rm syst}$\\
\babar  ~\cite{Aubert:vcbExcl}       &$0.40 \pm0.06_{\rm stat} \pm0.03_{\rm syst}$ & $0.42 \pm0.06_{\rm stat} \pm0.03_{\rm syst}$  \\
\hline 
{\bf Average}                              &\mathversion{bold}$0.41 \pm0.04$ &\mathversion{bold}$\chi^2/\dof = 0.073$ (CL=$78.9\%$) \\
\hline\hline

Experiment                                 &$\cbf(B^- \to D^{*+} \pi^- \ell^-\bar{\nu}_{\ell}) [\%]$ (rescaled) & $\cbf(B^- \to D^{*+} \pi^- \ell^-\bar{\nu}_{\ell}) [\%]$ (published) \\
\hline 
\belle  ~\cite{Live:Dss}           &$0.68 \pm0.08_{\rm stat} \pm0.07_{\rm syst}$   & $0.64 \pm0.08_{\rm stat} \pm0.09_{\rm syst}$  \\
\babar  ~\cite{Aubert:vcbExcl}       &$0.57 \pm0.05_{\rm stat} \pm0.04_{\rm syst}$   & $0.59 \pm0.05_{\rm stat} \pm0.04_{\rm syst}$ \\
\hline 
{\bf Average}                              &\mathversion{bold}$0.60 \pm0.06$   & \mathversion{bold}$\chi^2/\dof = 0.778$ (CL=$37.9\%$) \\
\hline \hline

Experiment                               &$\cbf(\bar{B}^0 \to D^0 \pi^+ \ell^-\bar{\nu}_{\ell}) [\%]$ (rescaled) & $\cbf(\bar{B}^0 \to D^0 \pi^+ \ell^-\bar{\nu}_{\ell}) [\%]$ (published)\\
\hline 
\belle  ~\cite{Live:Dss}           &$0.43 \pm0.07_{\rm stat} \pm0.05_{\rm syst}$ & $0.42 \pm0.07_{\rm stat} \pm0.06_{\rm syst}$ \\
\babar  ~\cite{Aubert:vcbExcl}     &$0.40 \pm0.08_{\rm stat} \pm0.03_{\rm syst}$ & $0.43 \pm0.08_{\rm stat} \pm0.03_{\rm syst}$ \\
\hline 
{\bf Average}                              &\mathversion{bold}$0.42 \pm0.06$  &\mathversion{bold}$\chi^2/\dof = 0.061$ (CL=$80.5\%$) \\
\hline\hline

Experiment                                 &$\cbf(\bar{B}^0 \to D^{*0} \pi^+\ell^-\bar{\nu}_{\ell}) [\%]$ (rescaled) & $\cbf(\bar{B}^0 \to D^{*0} \pi^+\ell^-\bar{\nu}_{\ell}) [\%]$ (published) \\
\hline 
\belle  ~\cite{Live:Dss}           &$0.58 \pm0.21_{\rm stat} \pm0.07_{\rm syst}$  & $0.56 \pm0.21_{\rm stat} \pm0.08_{\rm syst}$ \\
\babar  ~\cite{Aubert:vcbExcl}       &$0.46 \pm0.08_{\rm stat} \pm0.04_{\rm syst}$ &$0.48 \pm0.08_{\rm stat} \pm0.04_{\rm syst}$ \\ 
\hline 
{\bf Average}                              &\mathversion{bold}$0.47 \pm0.08$ &\mathversion{bold}$\chi^2/\dof = 0.262$ (CL=$60.9\%$) \\
\hline

\end{tabular}
\end{center}
\label{tab:dpilnu}
\end{table}
% ----------------------------------------------------------------------

\mysubsubsection{$\bar{B} \to D^{**} \ell^-\bar{\nu}_{\ell}$}
\label{slbdecays_dsslnu}
% -------------------

$D^{**}$ mesons contain one charm quark and one light anti-quark with relative angular momentum $L=1$. 
According to Heavy Quark Symmetry (HQS)~\cite{Isgur:1991wq}, they form one doublet of states with 
angular momentum $j \equiv s_q + L= 3/2$  $\left[D_1(2420), D_2^*(2460)\right]$ and another doublet 
with $j=1/2$ $\left[D^*_0(2400), D_1'(2430)\right]$, where $s_q$ is the light quark spin. 
Parity and angular momentum conservation constrain the decays allowed for each state. The $D_1$ and $D_2^*$ 
states decay via a D-wave to $D^*\pi$ and $D^{(*)}\pi$, respectively, and have small decay widths, 
while the $D_0^*$ and $D_1'$  states decay via an S-wave to $D\pi$ and $D^*\pi$ and are very broad.
For the narrow states, the averages are determined by the
combination of the results provided in Table~\ref{tab:dss1lnu} and \ref{tab:dss2lnu} for 
$\cbf(B^- \to D_1^0\ell^-\bar{\nu}_{\ell})
\times \cbf(D_1^0 \to D^{*+}\pi^-)$ and $\cbf(B^- \to D_2^0\ell^-\bar{\nu}_{\ell})
\times \cbf(D_2^0 \to D^{*+}\pi^-)$. 
For the broad states, the averages are determined by the
combination of the results provided in Table~\ref{tab:dss1plnu} and \ref{tab:dss0lnu} for 
$\cbf(B^- \to D_1'^0\ell^-\bar{\nu}_{\ell})
\times \cbf(D_1'^0 \to D^{*+}\pi^-)$ and $\cbf(B^- \to D_0^{*0}\ell^-\bar{\nu}_{\ell})
\times \cbf(D_0^{*0} \to D^{+}\pi^-)$. 
The measurements are scaled to a consistent set of input
parameters and their uncertainties~\cite{HFLAV_sl:inputparams}.  

\begin{figure}[!ht]
 \begin{center}
  \unitlength1.0cm % coordinates in cm
  \begin{picture}(14.,9.0)  %ys(25.,6.0)
   \put( -1.5,  0.0){\includegraphics[width=8.7cm]{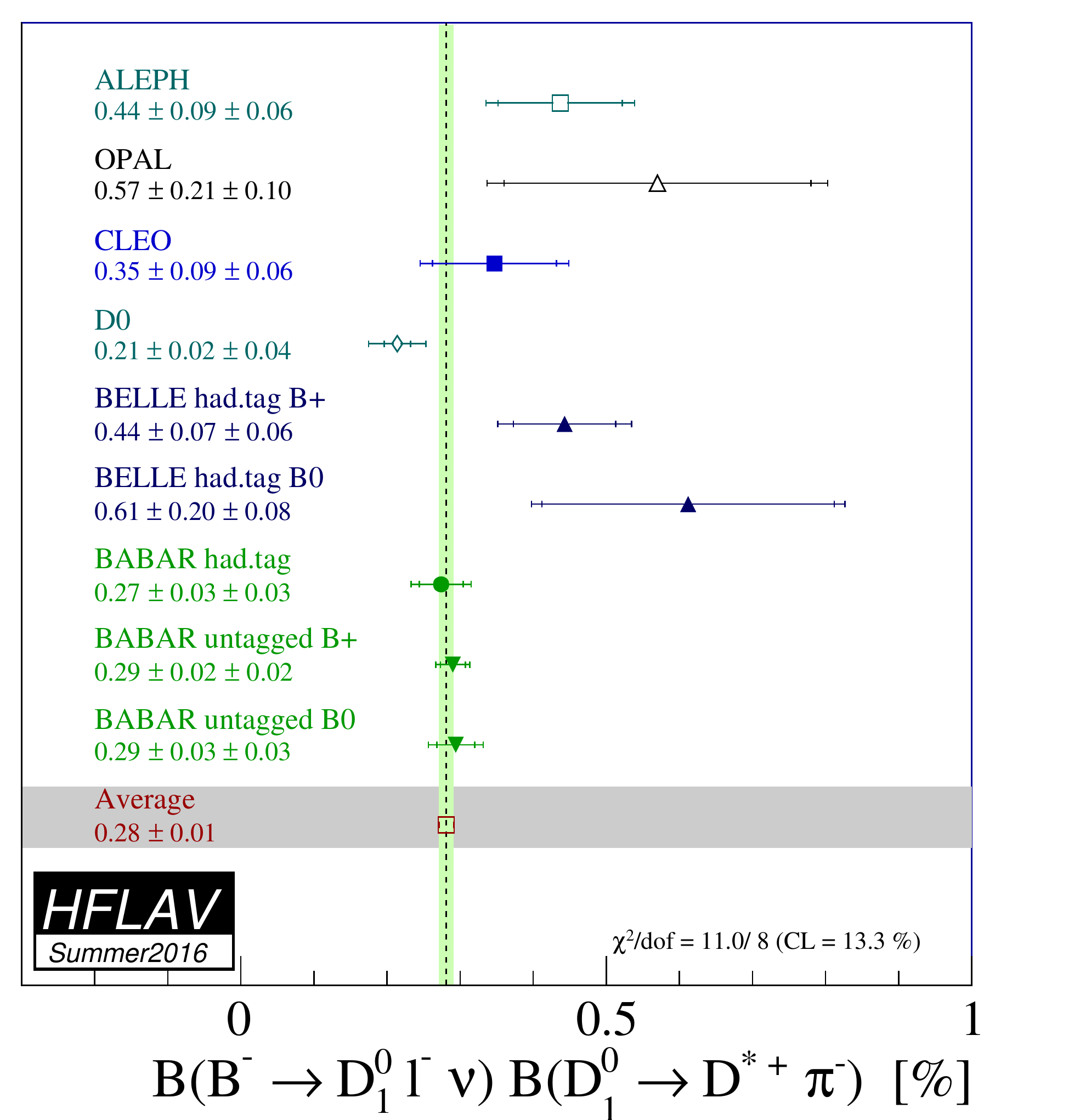}
   }
   \put(  7.5,  0.0){\includegraphics[width=8.7cm]{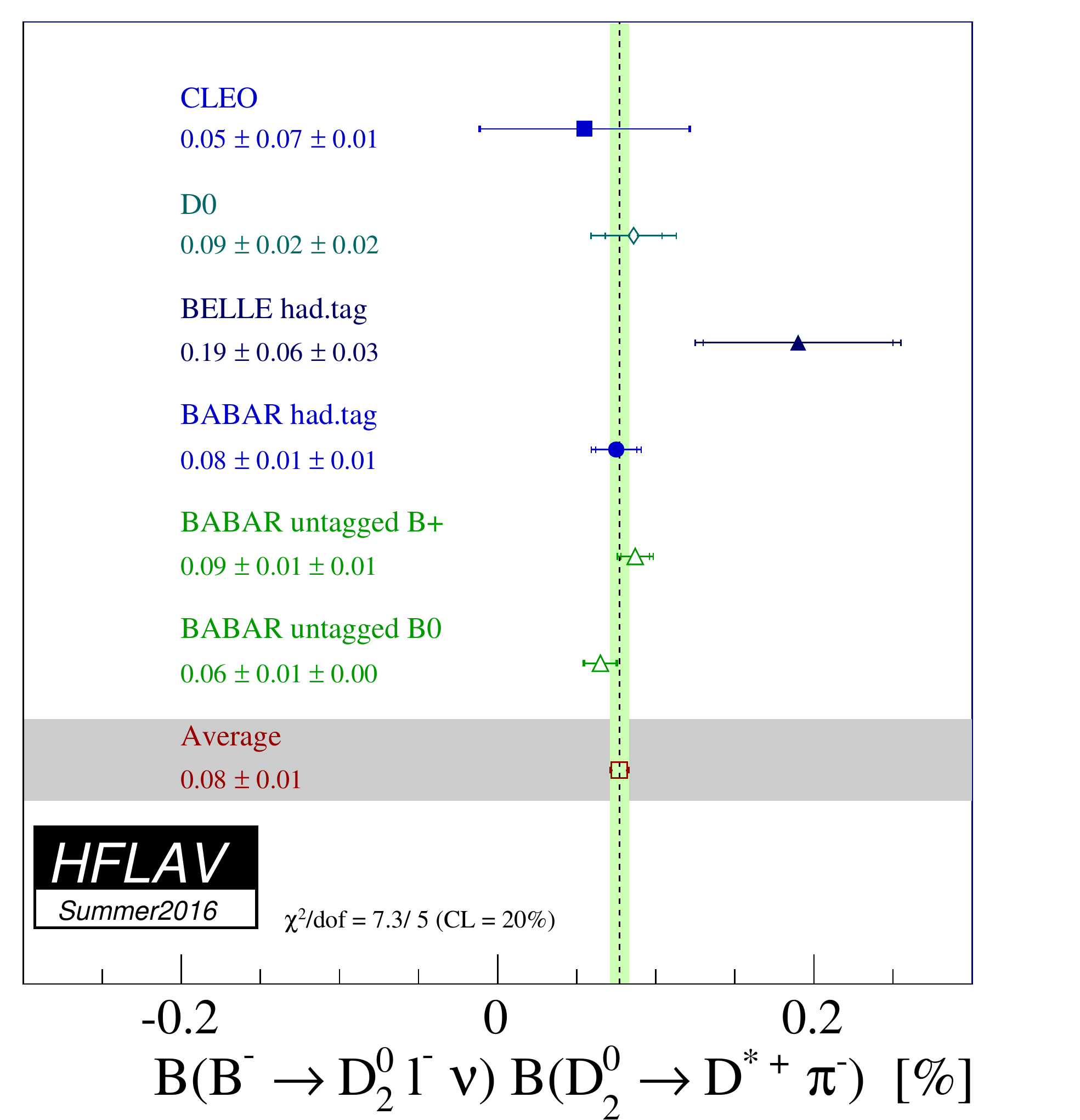}
   }
   \put(  5.6,  8.0){{\large\bf a)}}
   \put( 14.2,  8.0){{\large\bf b)}}
  \end{picture}
  \caption{Rescaled individual measurements and their averages for (a) 
  $\cbf(B^- \to D_1^0\ell^-\bar{\nu}_{\ell})
\times \cbf(D_1^0 \to D^{*+}\pi^-)$ and (b) $\cbf(B^- \to D_2^0\ell^-\bar{\nu}_{\ell})
\times \cbf(D_2^0 \to D^{*+}\pi^-)$.}
  \label{fig:brdssl}
 \end{center}
\end{figure}

\begin{figure}[!ht]
 \begin{center}
  \unitlength1.0cm % coordinates in cm
  \begin{picture}(14.,9.0)  %ys(25.,6.0)
   \put( -1.5,  0.0){\includegraphics[width=8.7cm]{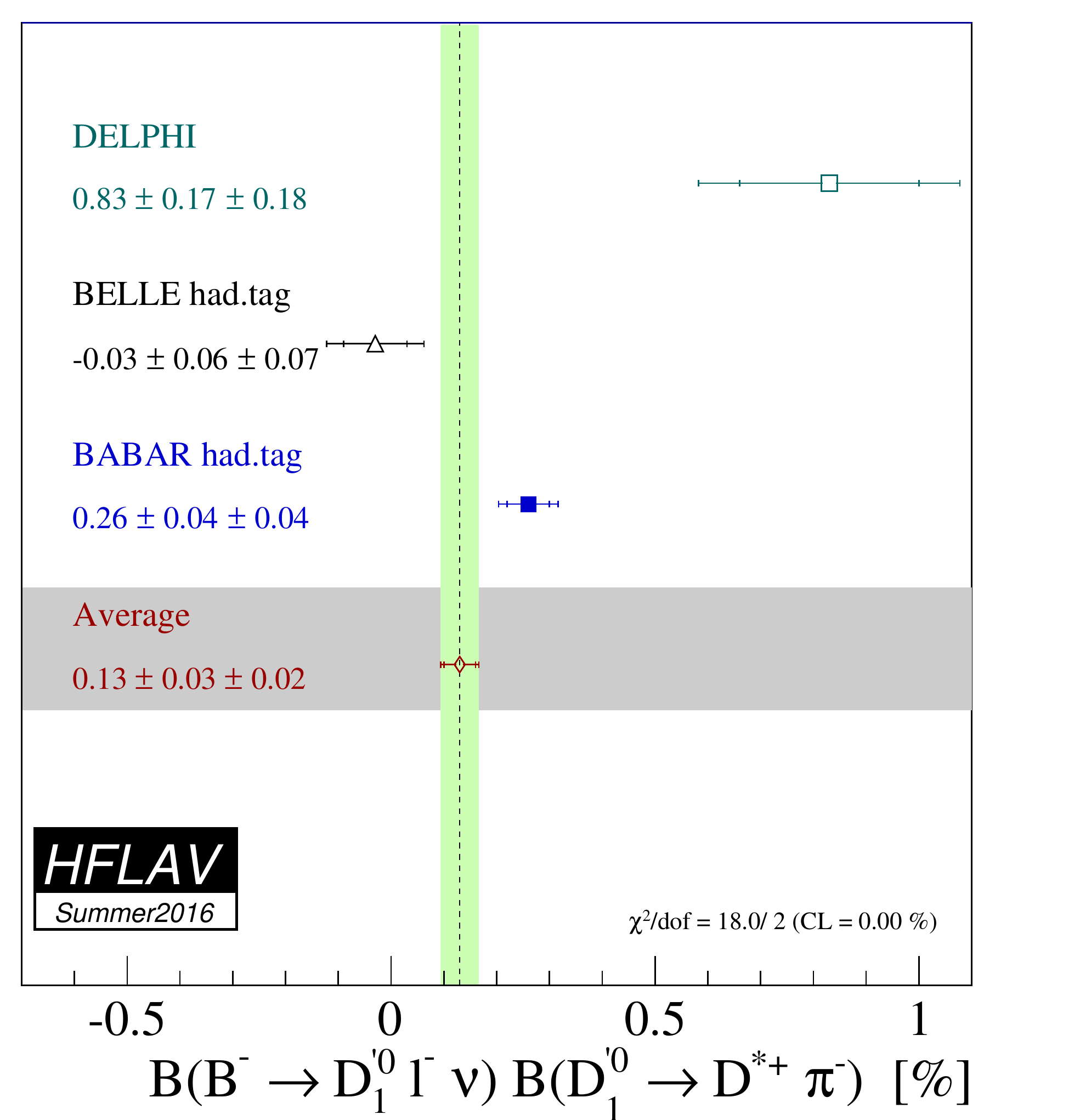}
   }
   \put(  7.5,  0.0){\includegraphics[width=8.7cm]{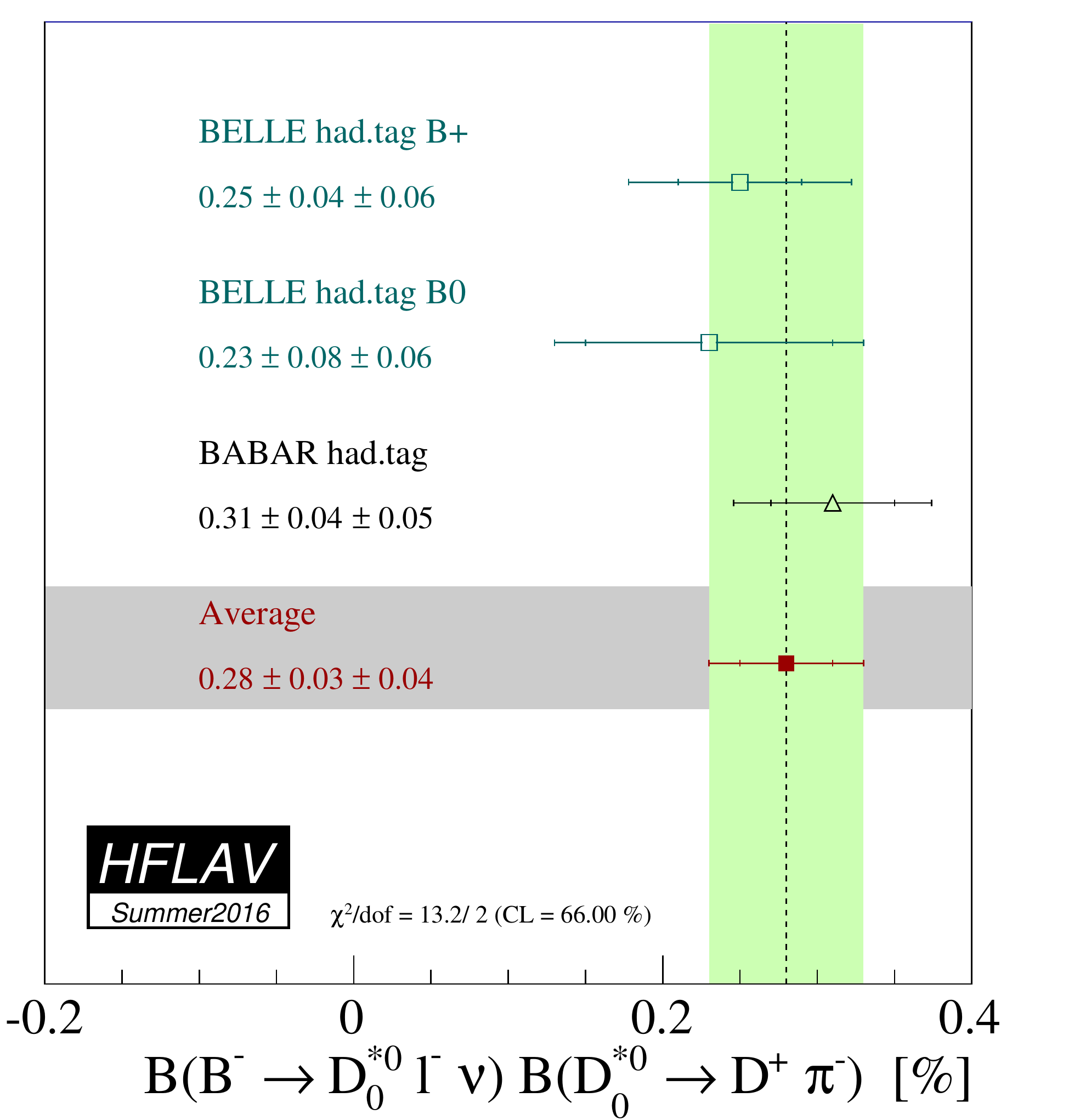}
   }
   \put(  5.8,  8.3){{\large\bf a)}}
   \put( 14.7,  8.3){{\large\bf b)}}
  \end{picture}
  \caption{Rescaled individual measurements and their averages for (a) 
  $\cbf(B^- \to D_1'^0\ell^-\bar{\nu}_{\ell})
\times \cbf(D_1'^0 \to D^{*+}\pi^-)$ and (b) $\cbf(B^- \to D_0^{*0}\ell^-\bar{\nu}_{\ell})
\times \cbf(D_0^{*0} \to D^{+}\pi^-)$.}
  \label{fig:brdssl2}
 \end{center}
\end{figure}

For both the B-factory and the LEP and Tevatron results, the $B$ semileptonic 
signal yields are extracted from a fit to the invariant mass distribution of the $D^{(*)+}\pi^-$ system.
 Apart for the CLEO, \belle and \babar results, the other measurements 
 are for the $\bar{B} \to D^{**}(D^*\pi^-)X \ell^- \bar{\nu}_{\ell}$ final state and 
 we assume that no particles are left in the $X$ system. 
 The \babar tagged $\bar{B} \to D_2^* \ell^- \bar{\nu}_{\ell}$ has been measured 
 selecting $D_s^*\to D\pi$ final state and it has been translated in 
 a result on $D_2^*\to D^*\pi$ decay mode, assuming 
 ${\cal B}(D_2^*\to D\pi)/{\cal B}(D_2^*\to D^*\pi)=1.54\pm 0.15$~\cite{PDG_2014}. 
Figure~\ref{fig:brdssl} and ~\ref{fig:brdssl2} show the measurements and the
resulting averages.

% ----------------------------------------------------------------------0
\begin{table}[!htb]
\caption{Published and rescaled individual measurements and their averages for
of the branching fraction $\cbf(B^- \to D_1^0\ell^-\bar{\nu}_{\ell})\times \cbf(D_1^0 \to D^{*+}\pi^-)$. 
%The ALEPH, OPAL and D0 measurements are for the 
%$D_1(D^*\pi)X$ final state and we assume that no particles are left in the $X$ system.
}
\begin{center}
\resizebox{0.99\textwidth}{!}{
\begin{tabular}{|l|c|c|}\hline
Experiment                                 &$\cbf(B^- \to D_1^0(D^{*+}\pi^-)\ell^-\bar{\nu}_{\ell})
 [\%]$  &$\cbf(B^- \to D_1^0(D^{*+}\pi^-)\ell^-\bar{\nu}_{\ell})
 [\%]$  \\
                                                & (rescaled) & (published) \\

\hline\hline 
ALEPH ~\cite{Aleph:Dss}        &$0.437 \pm0.085_{\rm stat} \pm0.056_{\rm syst}$ 
 &$0.47 \pm0.10_{\rm stat} \pm0.07_{\rm syst}$ \\
OPAL  ~\cite{opal:Dss}         &$0.570 \pm0.210_{\rm stat} \pm0.101_{\rm syst}$  
&$0.70 \pm0.21_{\rm stat} \pm0.10_{\rm syst}$ \\
CLEO  ~\cite{cleo:Dss}         &$0.347 \pm0.085_{\rm stat} \pm0.056_{\rm syst}$ 
 &$0.373 \pm0.085_{\rm stat} \pm0.057_{\rm syst}$ \\
D0  ~\cite{D0:Dss}         &$0.214 \pm0.018_{\rm stat} \pm0.035_{\rm syst}$  
&$0.219 \pm0.018_{\rm stat} \pm0.035_{\rm syst}$ \\
\belle Tagged $B^-$ ~\cite{Live:Dss}           &$0.443 \pm0.070_{\rm stat} \pm0.059_{\rm syst}$  
&$0.42 \pm0.07_{\rm stat} \pm0.07_{\rm syst}$ \\
\belle Tagged $B^0$ ~\cite{Live:Dss}           &$0.612 \pm0.200_{\rm stat} \pm0.077_{\rm syst}$  
&$0.42 \pm0.07_{\rm stat} \pm0.07_{\rm syst}$ \\ 
\babar Tagged ~\cite{Aubert:2009_4}           &$0.274 \pm0.030_{\rm stat} \pm0.029_{\rm syst}$
&$0.29 \pm0.03_{\rm stat} \pm0.03_{\rm syst}$ \\
\babar Untagged $B^-$ ~\cite{Aubert:2008zc}           &$0.290 \pm0.017_{\rm stat} \pm0.016_{\rm syst}$
&$0.30 \pm0.02_{\rm stat} \pm0.02_{\rm syst}$ \\
\babar Untagged $B^0$ ~\cite{Aubert:2008zc}           &$0.294 \pm0.026_{\rm stat} \pm0.027_{\rm syst}$
&$0.30 \pm0.02_{\rm stat} \pm0.02_{\rm syst}$ \\
\hline
{\bf Average}                              &\mathversion{bold}$0.281 \pm0.010 \pm 0.015$ 
    &\mathversion{bold}$\chi^2/\dof = 12.7/8$ (CL=$11.7\%$)  \\
\hline 
\end{tabular}
}
\end{center}
\label{tab:dss1lnu}
\end{table}
% ----------------------------------------------------------------------

% ----------------------------------------------------------------------0
\begin{table}[!htb]
\caption{Published and rescaled individual measurements and their averages for 
$\cbf(B^- \to D_2^0\ell^-\bar{\nu}_{\ell})\times \cbf(D_2^0 \to D^{*+}\pi^-)$. 
%The D0 measurement is for the $D_2^*(D^*\pi)X$
%final state and we assume that no particles are left in the X system.
%The \babar tagged measurement
%has been translated in a result on $D_2^*\to D^*\pi$ decay mode, assuming 
%${\cal B}(D_2^*\to D\pi)/{\cal B}(D_2^*\to D^*\pi)=1.54\pm
%0.15$~\cite{PDG_2014}.
}
\begin{center}
\resizebox{0.99\textwidth}{!}{
\begin{tabular}{|l|c|c|}\hline
Experiment                                 &$\cbf(B^- \to D_2^0(D^{*+}\pi^-)\ell^-\bar{\nu}_{\ell})
 [\%]$  &$\cbf(B^- \to D_2^0(D^{*+}\pi^-)\ell^-\bar{\nu}_{\ell})
 [\%]$  \\
                                                & (rescaled) & (published) \\
\hline\hline 
CLEO  ~\cite{cleo:Dss}         &$0.055 \pm0.066_{\rm stat} \pm0.011_{\rm syst}$ 
 &$0.059 \pm0.066_{\rm stat} \pm0.011_{\rm syst}$ \\
D0  ~\cite{D0:Dss}         &$0.086 \pm0.018_{\rm stat} \pm0.020_{\rm syst}$  
&$0.088 \pm0.018_{\rm stat} \pm0.020_{\rm syst}$ \\
\belle  ~\cite{Live:Dss}           &$0.190 \pm0.060_{\rm stat} \pm0.025_{\rm syst}$  
&$0.18 \pm0.06_{\rm stat} \pm0.03_{\rm syst}$ \\
\babar tagged ~\cite{Aubert:2009_4}           &$0.075 \pm0.013_{\rm stat} \pm0.009_{\rm syst}$
&$0.078 \pm0.013_{\rm stat} \pm0.010_{\rm syst}$ \\
\babar untagged $B^-$ ~\cite{Aubert:2008zc}           &$0.087 \pm0.009_{\rm stat} \pm0.007_{\rm syst}$
&$0.087 \pm0.013_{\rm stat} \pm0.007_{\rm syst}$ \\
\babar untagged $B^0$ ~\cite{Aubert:2008zc}           &$0.065 \pm0.010_{\rm stat} \pm0.004_{\rm syst}$
&$0.087 \pm0.013_{\rm stat} \pm0.007_{\rm syst}$ \\
\hline
{\bf Average}                              &\mathversion{bold}$0.077 \pm0.006 \pm 0.004$ 
    &\mathversion{bold}$\chi^2/\dof = 5.3/5$ (CL=$37.7\%$)  \\
\hline 
\end{tabular}
}
\end{center}
\label{tab:dss2lnu}
\end{table}
% ----------------------------------------------------------------------

% ----------------------------------------------------------------------
\begin{table}[!htb]
\caption{
Published and rescaled individual measurements and their averages for 
$\cbf(B^- \to D_1^{'0}\ell^-\bar{\nu}_{\ell})\times \cbf(D_1^{'0} \to D^{*+}\pi^-)$. 
%The DELPHI measurement 
%is for the final state $D_1'(D^*\pi)X$ and we assume that no particles are left in the X system.
}
\begin{center}
\begin{tabular}{|l|c|c|}\hline
Experiment                                 &$\cbf(B^- \to D_1^{'0}(D^{*+}\pi^-)\ell^-\bar{\nu}_{\ell})
 [\%]$  &$\cbf(B^- \to D_1^{'0}(D^{*+}\pi^-)\ell^-\bar{\nu}_{\ell})
 [\%]$  \\
                                                & (rescaled) & (published) \\
\hline\hline 
DELPHI ~\cite{Abdallah:2005cx}        &$0.71 \pm0.17_{\rm stat} \pm0.18_{\rm syst}$ 
 &$0.83 \pm0.17_{\rm stat} \pm0.18_{\rm syst}$ \\
\belle  ~\cite{Live:Dss}           &$-0.03 \pm0.06_{\rm stat} \pm0.07_{\rm syst}$  
&$-0.03 \pm0.06_{\rm stat} \pm0.07_{\rm syst}$ \\
\babar  ~\cite{Aubert:2009_4}           &$0.26 \pm0.04_{\rm stat} \pm0.04_{\rm syst}$
&$0.27 \pm0.04_{\rm stat} \pm0.05_{\rm syst}$ \\
\hline
{\bf Average}                              &\mathversion{bold}$0.13 \pm 0.03 \pm0.02$ 
    &\mathversion{bold}$\chi^2/\dof = 18./2$ (CL=$0.0001\%$)  \\
\hline 
\end{tabular}
\end{center}
\label{tab:dss1plnu}
\end{table}
% ----------------------------------------------------------------------

% ----------------------------------------------------------------------0
\begin{table}[!htb]
\caption{Published and rescaled individual measurements and their averages for  
$\cbf(B^- \to D_0^{*0}\ell^-\bar{\nu}_{\ell})\times \cbf(D_0^{*0} \to D^{+}\pi^-)$. }
\begin{center}
\begin{tabular}{|l|c|c|}\hline
Experiment                                 &$\cbf(B^- \to D_0^{*0}(D^{+}\pi^-)\ell^-\bar{\nu}_{\ell})
 [\%]$  &$\cbf(B^- \to D_0^{*0}(D^{+}\pi^-)\ell^-\bar{\nu}_{\ell})
 [\%]$ \\
						& (rescaled) & (published) \\
\hline\hline 
\belle Tagged $B^-$ ~\hfill\cite{Live:Dss}           &$0.25 \pm0.04_{\rm stat} \pm0.06_{\rm syst}$  
&$0.24 \pm0.04_{\rm stat} \pm0.06_{\rm syst}$ \\
\belle Tagged $B^0$ ~\hfill\cite{Live:Dss}           &$0.23 \pm0.08_{\rm stat} \pm0.06_{\rm syst}$  
&$0.24 \pm0.04_{\rm stat} \pm0.06_{\rm syst}$ \\
\babar Tagged ~\hfill\cite{Aubert:2009_4}            &$0.31 \pm0.04_{\rm stat} \pm0.05_{\rm syst}$
&$0.26 \pm0.05_{\rm stat} \pm0.04_{\rm syst}$ \\
\hline
{\bf Average}                              &\mathversion{bold}$0.28 \pm 0.03 \pm0.04$ 
    &\mathversion{bold}$\chi^2/\dof = 0.49/2$ (CL=$78.0\%$)  \\
\hline 
\end{tabular}
\end{center}
\label{tab:dss0lnu}
\end{table}
% ----------------------------------------------------------------------

%
% ======================================================================
% Inclusive CKM-favoured decays
% -- \include{b2cincl.tex}
% ======================================================================
\subsection{Inclusive CKM-favored decays}
\label{slbdecays_b2cincl}
% -------------------------------------------

\subsubsection{Global analysis of $\bar B\to X_c\ell^-\bar\nu_\ell$}

The semileptonic decay width $\Gamma(\bar B\to X_c\ell^-\bar\nu_\ell)$ has
been calculated in the framework of the operator production expansion
(OPE)~\cite{Shifman:1986mx,Chay:1990da,Bigi:1992su}.
The result is a double-expansion in $\Lambda_{\rm QCD}/m_b$ and
$\alpha_s$, which depends on a number of non-perturbative
parameters. These parameters describe the dynamics of the
$b$-quark inside the $B$~hadron and can be measured using
observables in $\bar B\to X_c\ell^-\bar\nu_\ell$ decays, such as the
moments of the lepton energy and the hadronic mass spectrum.

Two renormalization schemes are commonly used to defined the $b$-quark mass
and other theoretical quantities: the
kinetic~\cite{Benson:2003kp,Gambino:2004qm,Gambino:2011cq,Alberti:2014yda}
and the 1S~\cite{Bauer:2004ve} schemes. An independent set of theoretical
expressions is available for each, with several non-perturbative parameters.
The non-perturbative parameters in the kinetic scheme
are: the quark masses $m_b$ and $m_c$, $\mu^2_\pi$ and
$\mu^2_G$ at $O(1/m^2_b)$, and $\rho^3_D$ and $\rho^3_{LS}$ at
$O(1/m^3_b)$. In the 1S scheme, the parameters are: $m_b$, $\lambda_1$
at $O(1/m^2_b)$, and $\rho_1$, $\tau_1$, $\tau_2$ and $\tau_3$ at
$O(1/m^3_b)$.
Note that the numerical values of the kinetic and 1S $b$-quark masses cannot
be compared without converting one or the other, or both, to the same
renormalization scheme.

We used two kinematic distributions for
$\bar B\to X_c\ell^-\bar\nu_\ell$ decays, the hadron effective mass to derive
moments $\langle M^n_X\rangle$ of order $n=2,4,6$, and the charged lepton
momentum to derive moments $\langle E^n_\ell\rangle$ of order $n=0,1,2,3$.
Moments are determined for different values of $E_\mathrm{cut}$, the lower
limit on the minimum lepton momentum. The moments derived from the same
distributions with different value of $E_\mathrm{cut}$ are highly correlated.
The list of measurements is given in Table~\ref{tab:gf_input}. The only
input is the average lifetime~$\tau_B$ of neutral and charged $B$~mesons,
taken to be $(1.579\pm 0.004)$~ps (Sec.~\ref{sec:life_mix}).
\begin{table}[!htb]
\caption{Experimental inputs used in the global analysis of $\bar B\to
  X_c\ell^-\bar\nu_\ell$. $n$ is the order of the moment, $c$ is the
  threshold value of the lepton momentum in GeV. In total, there are
  23 measurements from \babar, 15 measurements from Belle and 12 from
  other experiments.} \label{tab:gf_input}
\begin{center}
\begin{tabular}{|l|l|l|}
  \hline
  Experiment
  & Hadron moments $\langle M^n_X\rangle$
  & Lepton moments $\langle E^n_\ell\rangle$\\
  \hline \hline
  \babar & $n=2$, $c=0.9,1.1,1.3,1.5$ & $n=0$, $c=0.6,1.2,1.5$\\
  & $n=4$, $c=0.8,1.0,1.2,1.4$ & $n=1$, $c=0.6,0.8,1.0,1.2,1.5$\\
  & $n=6$, $c=0.9,1.3$~\cite{Aubert:2009qda} & $n=2$, $c=0.6,1.0,1.5$\\
  & & $n=3$, $c=0.8,1.2$~\cite{Aubert:2009qda,Aubert:2004td}\\
  \hline
  Belle & $n=2$, $c=0.7,1.1,1.3,1.5$ & $n=0$, $c=0.6,1.4$\\
  & $n=4$, $c=0.7,0.9,1.3$~\cite{Schwanda:2006nf} & $n=1$,
  $c=1.0,1.4$\\
  & & $n=2$, $c=0.6,1.4$\\
  & & $n=3$, $c=0.8,1.2$~\cite{Urquijo:2006wd}\\
  \hline
  CDF & $n=2$, $c=0.7$ & \\
  & $n=4$, $c=0.7$~\cite{Acosta:2005qh} & \\
  \hline
  CLEO & $n=2$, $c=1.0,1.5$ & \\
  & $n=4$, $c=1.0,1.5$~\cite{Csorna:2004kp} & \\
  \hline
  DELPHI & $n=2$, $c=0.0$ & $n=1$, $c=0.0$ \\
  & $n=4$, $c=0.0$ & $n=2$, $c=0.0$ \\
  & $n=6$, $c=0.0$~\cite{Abdallah:2005cx} & $n=3$,
  $c=0.0$~\cite{Abdallah:2005cx}\\
  \hline
\end{tabular}
\end{center}
\end{table}

In the kinetic and 1S schemes, the moments in $\bar B\to
X_c\ell^-\bar\nu_\ell$ are not sufficient to determine the $b$-quark
mass precisely. In the kinetic scheme analysis we constrain the $c$-quark
mass (defined in the $\overline{\rm MS}$ scheme) to the value of
Ref.~\cite{Chetyrkin:2009fv},
\begin{equation}
  m_c^{\overline{\rm MS}}(3~{\rm GeV})=0.986\pm 0.013~{\rm GeV}~.
\end{equation}
In the 1S~scheme analysis, the $b$-quark mass is constrained by
measurements of the photon energy moments in $B\to
X_s\gamma$~\cite{Aubert:2005cua,Aubert:2006gg,Limosani:2009qg,Chen:2001fja}.

\subsubsection{Analysis in the kinetic scheme}
\label{globalfitsKinetic}

The fit relies on the calculations of the lepton energy and hadron mass
moments in $\bar B\to X_c\ell^-\bar\nu_\ell$~decays described in
Ref.~\cite{Gambino:2011cq,Alberti:2014yda} and closely follows the
procedure of Ref.~\cite{Gambino:2013rza}. The analysis determines
$\vcb$ and the six non-perturbative parameters mentioned above.

\begin{figure}
\begin{center}
  \includegraphics[width=8.2cm]{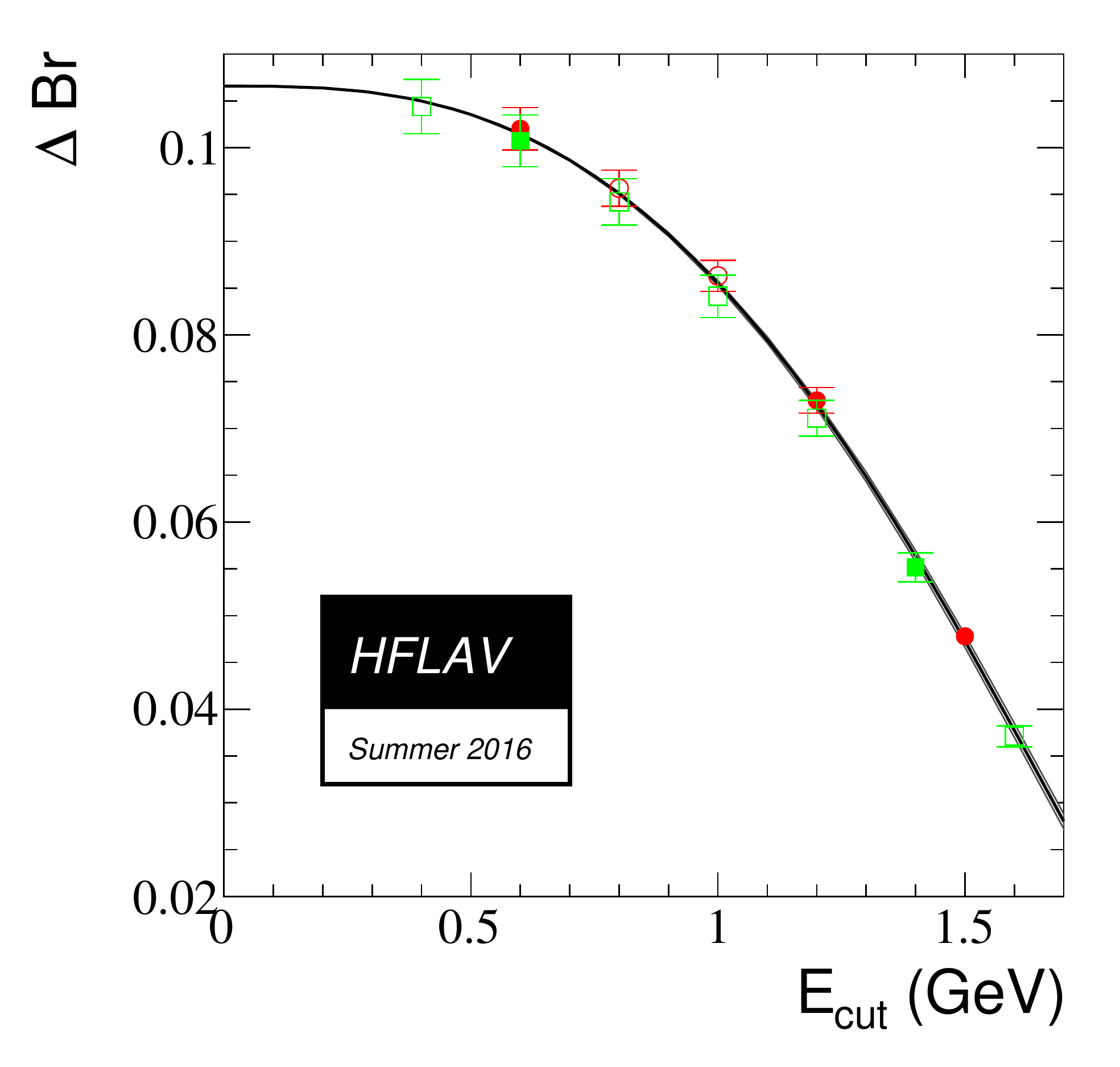}
  \includegraphics[width=8.2cm]{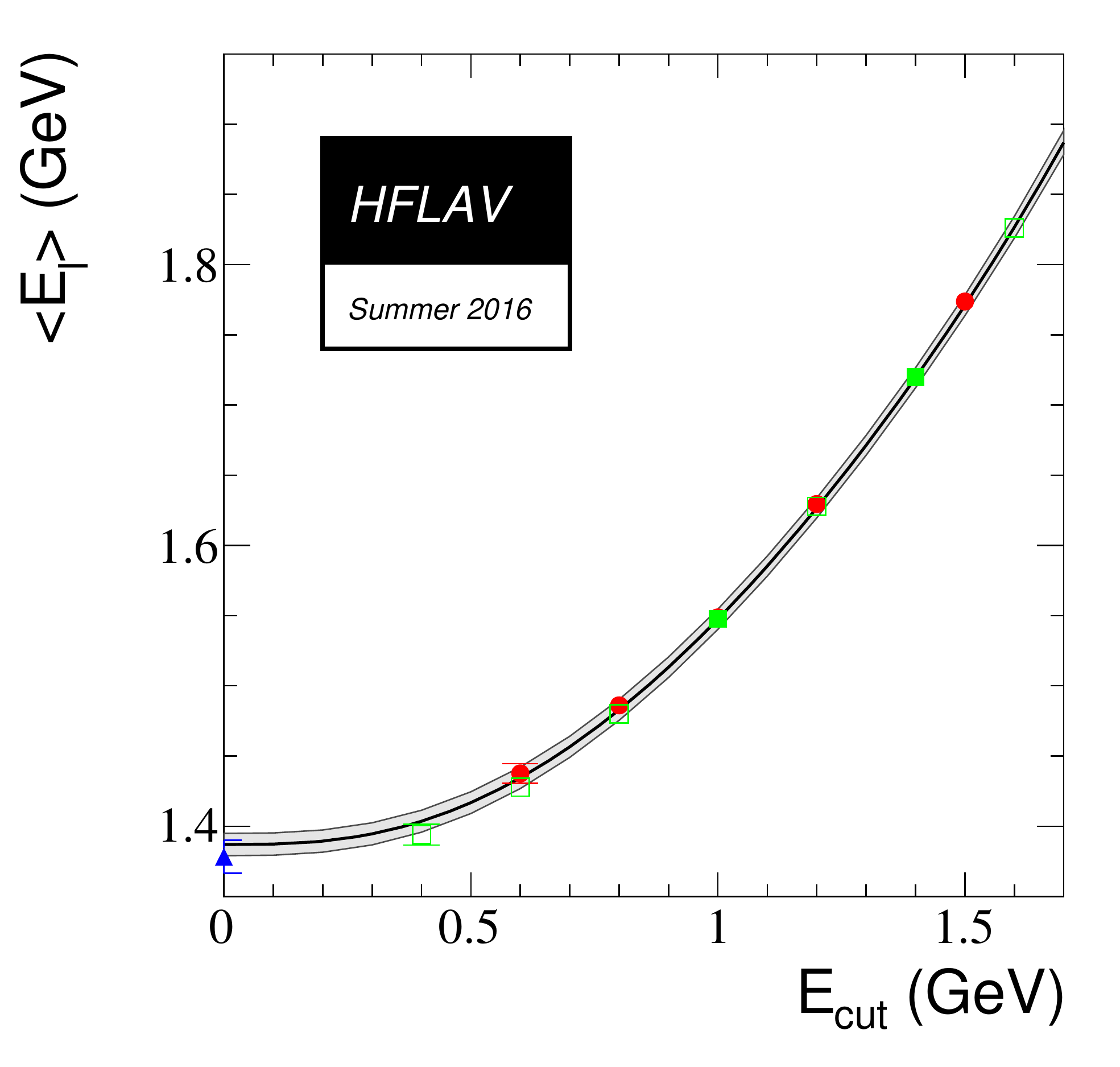}\\
  \includegraphics[width=8.2cm]{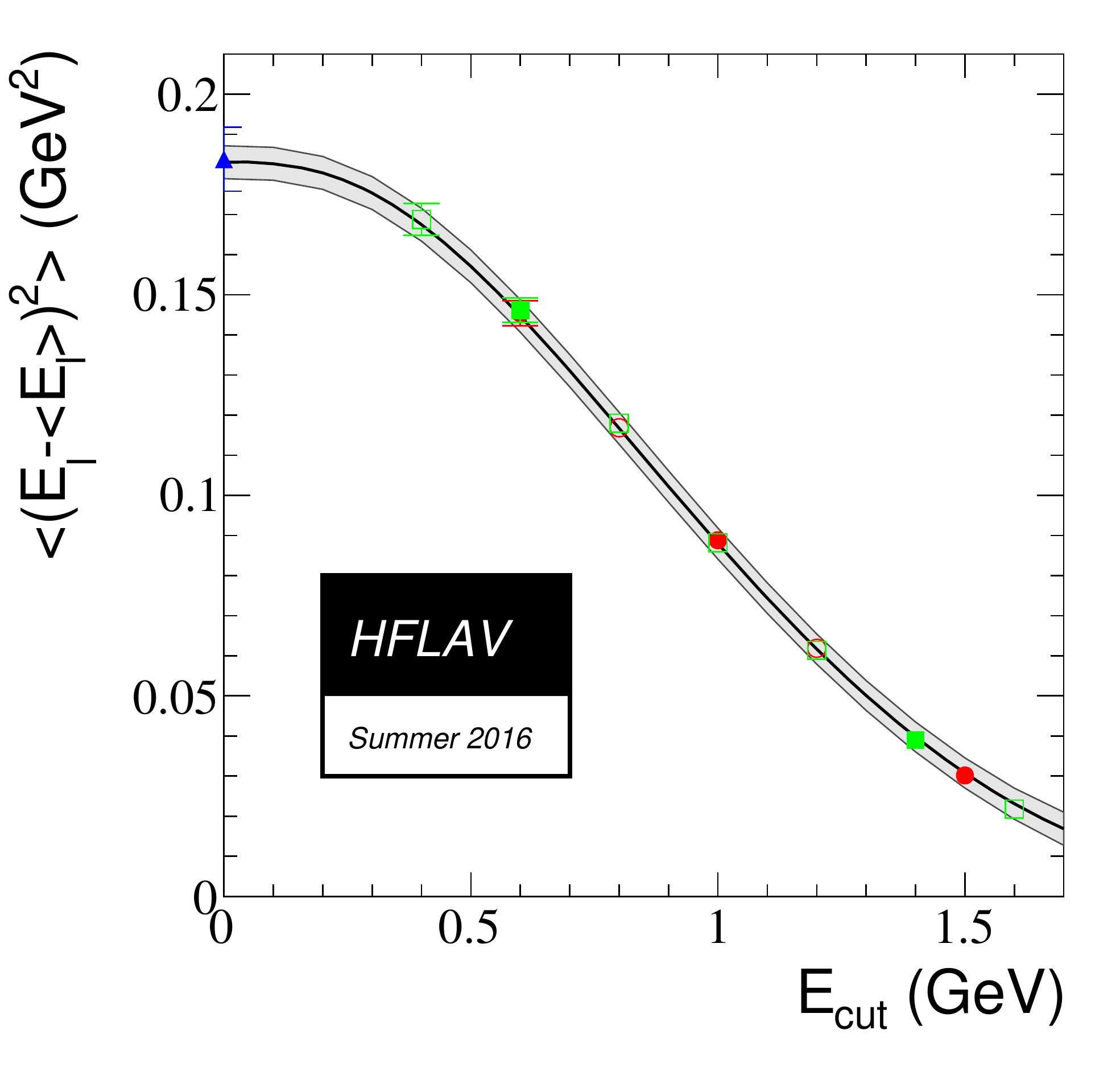}
  \includegraphics[width=8.2cm]{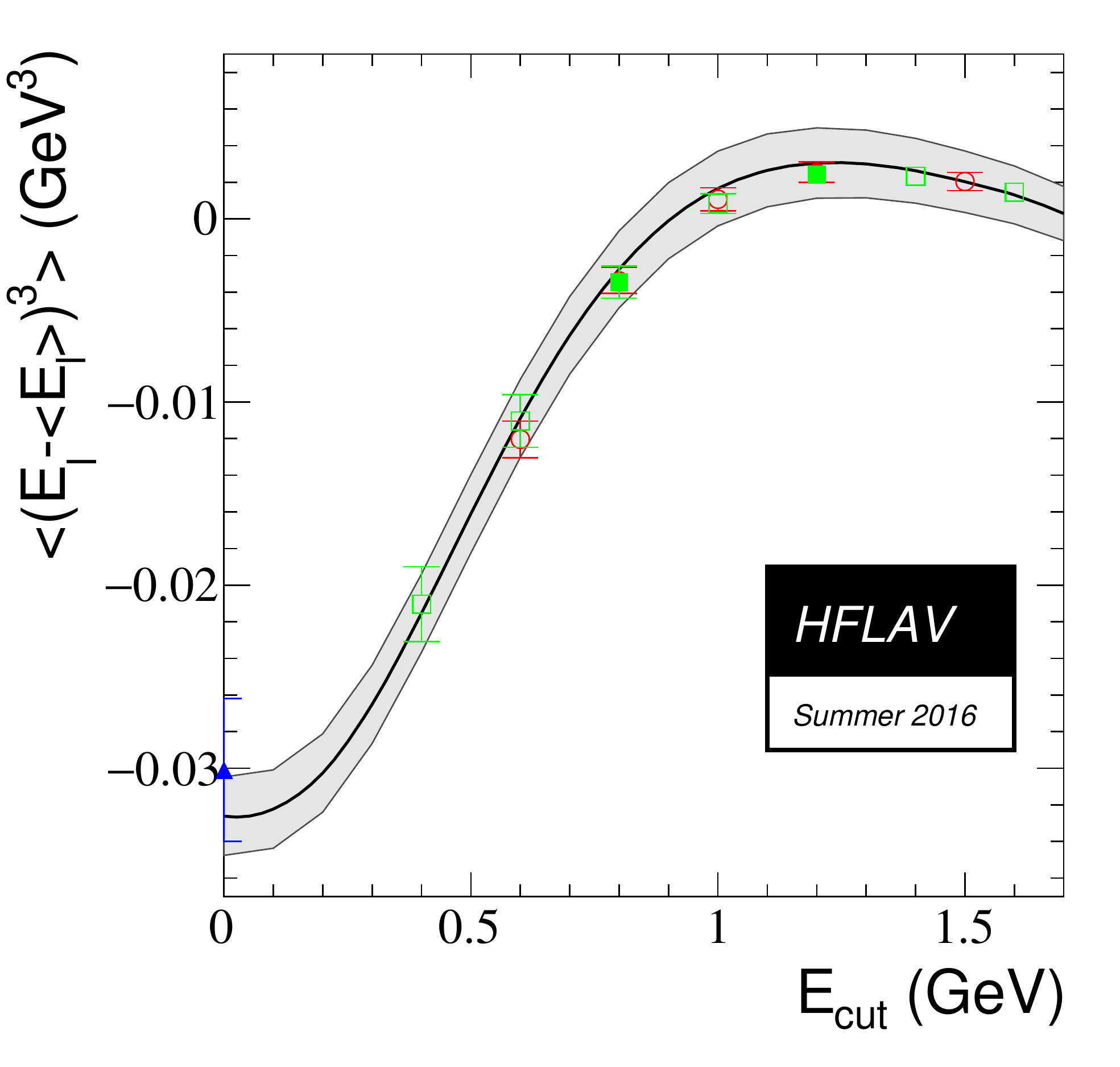}
\end{center}
\caption{Fit to the inclusive partial semileptonic branching ratios and to
  the lepton energy moments in the kinetic mass scheme. In all plots, the
  grey band is the theory prediction with total theory error. \babar
  data are shown by circles, Belle by squares and other experiments
  (DELPHI, CDF, CLEO) by triangles. Filled symbols mean that the point
  was used in the fit. Open symbols are measurements that were not
  used in the fit.} \label{fig:gf_res_kin_el}
\end{figure}

\begin{figure}
\begin{center}
  \includegraphics[width=8.2cm]{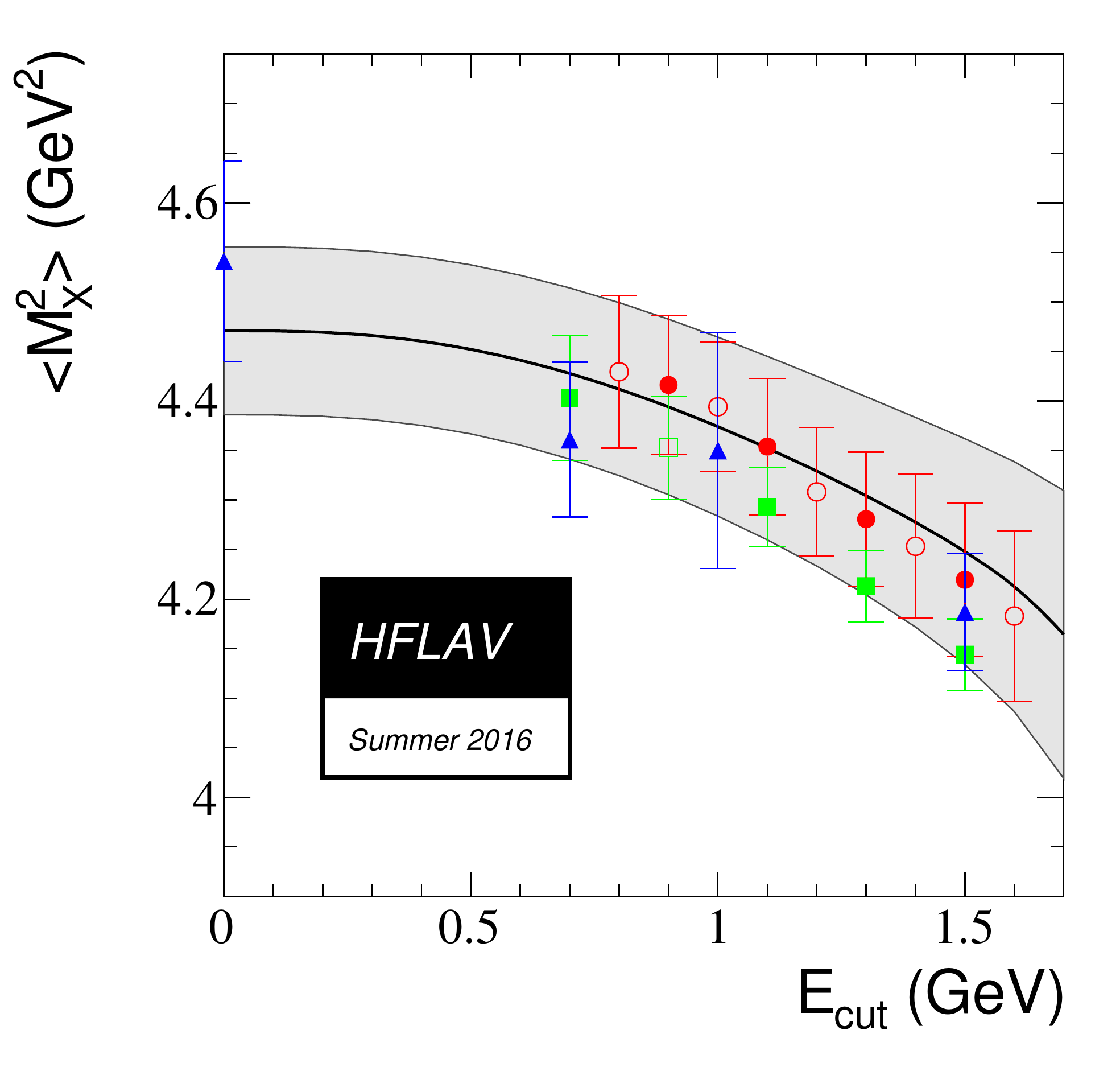}
  \includegraphics[width=8.2cm]{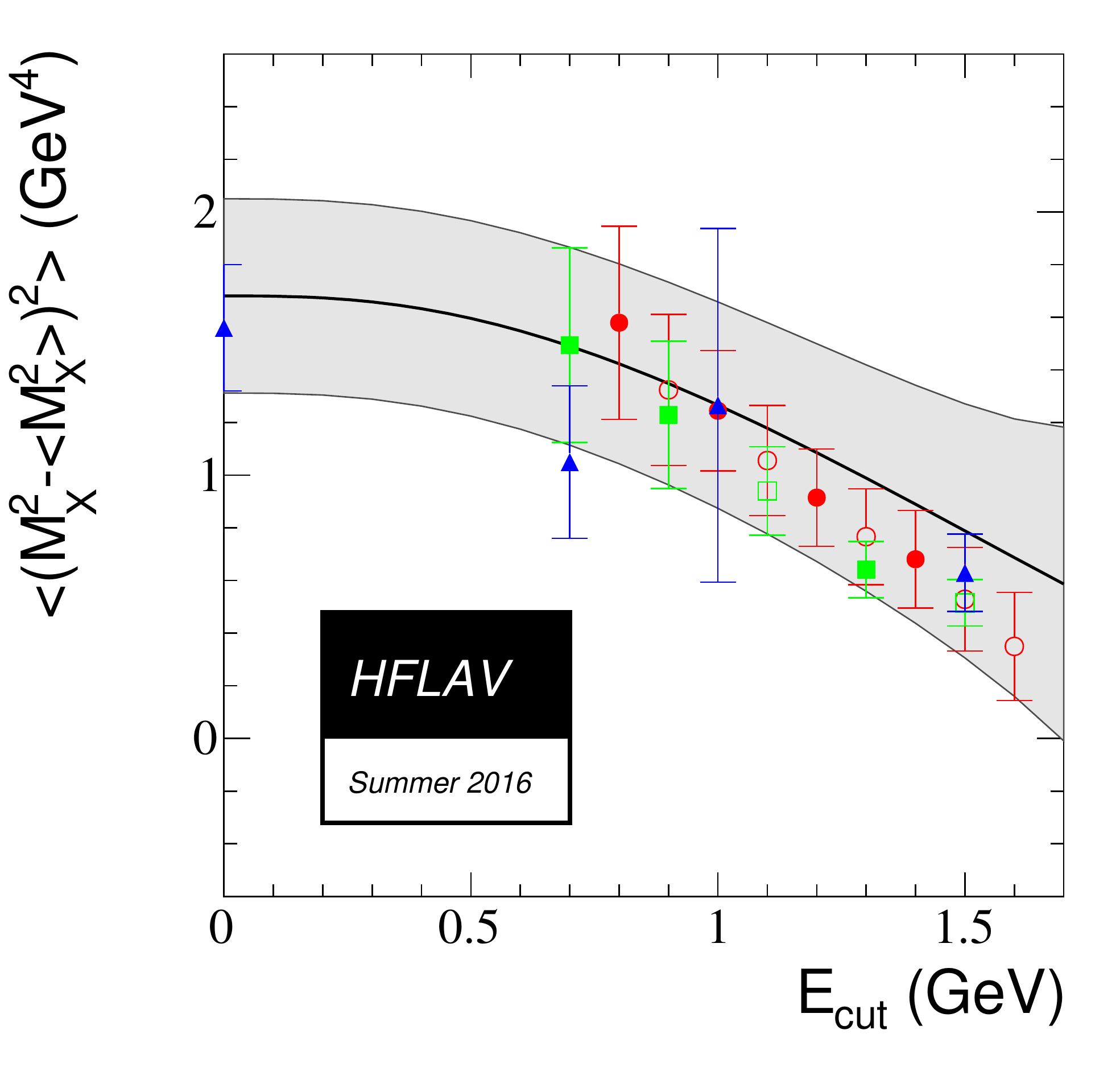}\\
  \includegraphics[width=8.2cm]{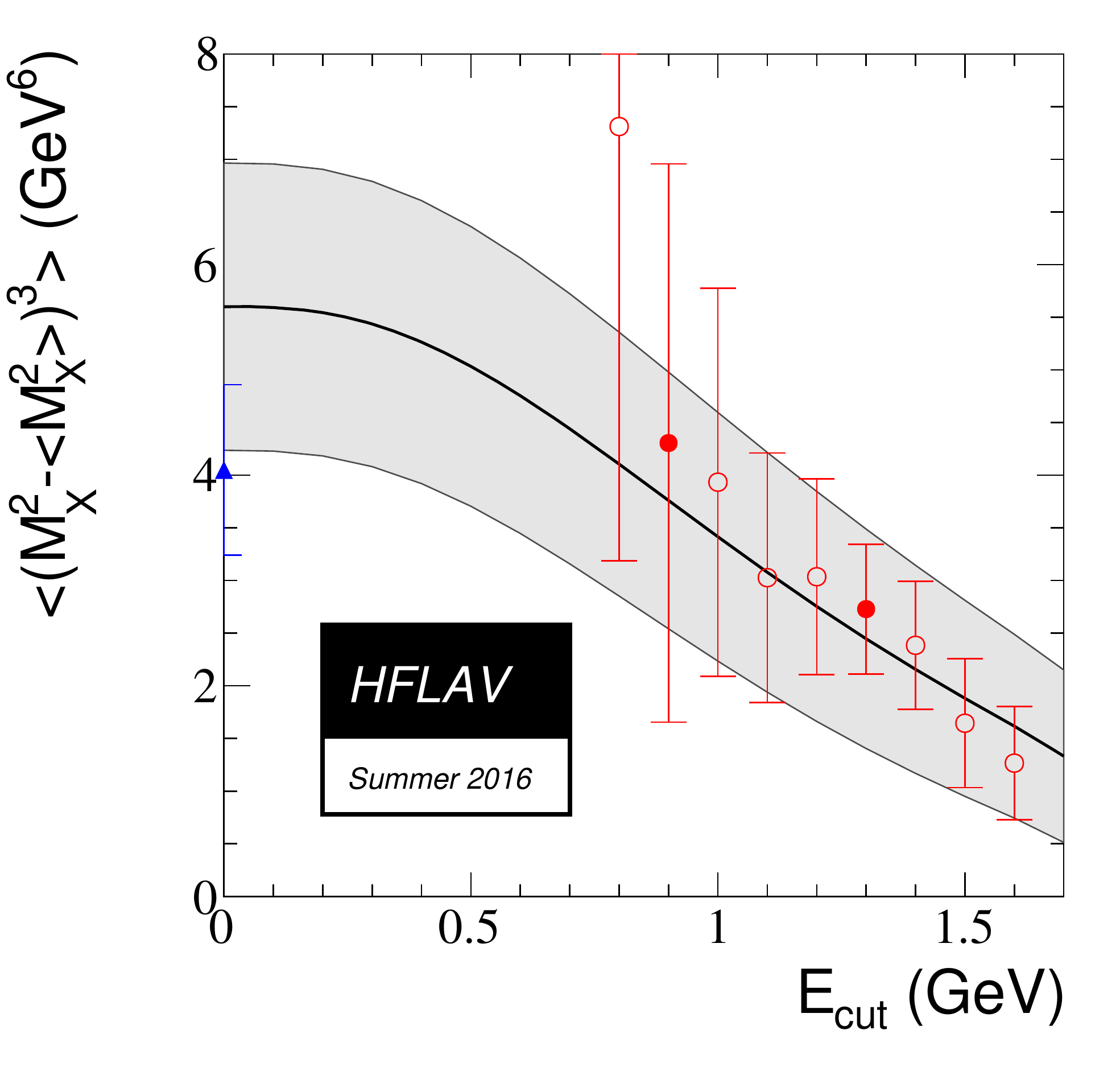}
\end{center}
\caption{Same as Fig.~\ref{fig:gf_res_kin_el} for the fit to the
  hadronic mass moments in the kinetic mass
  scheme.} \label{fig:gf_res_kin_mx}
\end{figure}

The detailed fit result and the matrix of the correlation coefficients is
given in Table~\ref{tab:gf_res_mc_kin}. The fit to the lepton energy and
hadronic mass moments is shown in Figs.~\ref{fig:gf_res_kin_el} and
\ref{fig:gf_res_kin_mx}, respectively. The result in terms of the main
parameters is
\begin{eqnarray}
  \vcb & = & (42.19\pm 0.78)\times 10^{-3}~, \\
  m_b^{\rm kin} & = & 4.554\pm 0.018~{\rm GeV}~, \\
  \mu^2_\pi & = & 0.464\pm 0.076~{\rm GeV^2}~,
\end{eqnarray}
with a $\chi^2$ of 15.6 for $43$ degrees of freedom. The scale $\mu$ of the
quantities in the kinematic scheme is 1~GeV.
\begin{table}[!htb]
\caption{Fit result in the kinetic scheme, using a precise $c$-quark
  mass constraint. The error matrix of the fit contains
  experimental and theoretical contributions. In the lower part of the
  table, the correlation matrix of the parameters is
  given. The scale $\mu$ of the quantities in the kinematic scheme is 1~GeV.}
\label{tab:gf_res_mc_kin}
\begin{center}
\resizebox{0.99\textwidth}{!}{
\begin{tabular}{|l|ccccccc|}
  \hline
  & \vcb\ [10$^{-3}$] & $m_b^{\rm kin}$ [GeV] &
  $m_c^{\overline{\rm MS}}$ [GeV] & $\mu^2_\pi$ [GeV$^2$]
  & $\rho^3_D$ [GeV$^3$] & $\mu^2_G$ [GeV$^2$] & $\rho^3_{LS}$ [GeV$^3$]\\
  \hline \hline
  value & 42.19 & \phantom{$-$}4.554 & \phantom{$-$}0.987 &
  \phantom{$-$}0.464 & \phantom{$-$}0.169 & \phantom{$-$}0.333 &
  $-$0.153\\
  error & 0.78 & \phantom{$-$}0.018 &
  \phantom{$-$}0.015 & \phantom{$-$}0.076 & \phantom{$-$}0.043 &
  \phantom{$-$}0.053 & \phantom{$-$}0.096\\
  \hline
  $|V_{cb}|$ & 1.000 & $-$0.257 & $-$0.078 &
  \phantom{$-$}0.354 & \phantom{$-$}0.289 & $-$0.080 &
  $-$0.051\\
  $m_b^{\rm kin}$ & & \phantom{$-$}1.000 & \phantom{$-$}0.769 &
  $-$0.054 & \phantom{$-$}0.097 & \phantom{$-$}0.360 & $-$0.087\\
  $m_c^{\overline{\rm MS}}$ & & & \phantom{$-$}1.000
  & $-$0.021 & \phantom{$-$}0.027 & \phantom{$-$}0.059 & $-$0.013\\
  $\mu^2_\pi$ & & & & \phantom{$-$}1.000 & \phantom{$-$}0.732 &
  \phantom{$-$}0.012 & \phantom{$-$}0.020\\
  $\rho^3_D$ & & & & & \phantom{$-$}1.000 & $-$0.173 & $-$0.123\\
  $\mu^2_G$ & & & & & & \phantom{$-$}1.000 & \phantom{$-$}0.066\\
  $\rho^3_{LS}$ & & & & & & & \phantom{$-$}1.000\\
  \hline
\end{tabular}
}
\end{center}
\end{table}

The inclusive $\bar B\to X_c\ell^-\bar\nu_\ell$ branching fraction
determined by this analysis is
\begin{equation}
  \cbf(\bar B\to X_c\ell^-\bar\nu_\ell)=(10.65\pm 0.16)\%~.
\end{equation}
Including the rate of charmless semileptonic decays
(Sec.~\ref{slbdecays_b2uincl}), $\cbf(\bar B\to
X_u\ell^-\bar\nu_\ell)=(2.13\pm 0.31)\times 10^{-3}$, we obtain the
semileptonic branching fraction,
\begin{equation}
  \cbf(\bar B\to X\ell^-\bar\nu_\ell)=(10.86\pm 0.16)\%~.
\end{equation}

\subsubsection{Analysis in the 1S scheme}
\label{globalfits1S}

The fit relies on the same set of moment measurements and the calculations of
the spectral moments described in
Ref.~\cite{Bauer:2004ve}. The theoretical uncertainties are estimated
as explained in Ref.~\cite{Schwanda:2008kw}. Only trivial theory
correlations, \ie\ between the same moment at the same
threshold are included in the analysis. The fit determines $\vcb$ and
the six non-perturbative parameters mentioned above.

The detailed result of the fit using the $B\to X_s\gamma$ constraint is given
in Table~\ref{tab:gf_res_xsgamma_1s}. The result in terms of the main
parameters is
\begin{eqnarray}
  \vcb & = & (41.98\pm 0.45)\times 10^{-3}~, \\
  m_b^{1S} & = & 4.691\pm 0.037~{\rm GeV}~, \\
  \lambda_1 & = & -0.362\pm 0.067~{\rm GeV^2}~,
\end{eqnarray}
with a $\chi^2$ of 23.0 for $59$ degrees of freedom. We find a good agreement
in the central values of \vcb\ between the kinetic and 1S scheme analyses. No
conclusion should, however, been drawn regarding the uncertainties in \vcb\ as
the two approaches are not equivalent in the number of higher-order corrections
included.

\begin{table}[!htb]
\caption{Fit result in the 1S scheme, using $B\to X_s\gamma$~moments
  as a constraint. In the lower part of the table, the correlation
  matrix of the parameters is given.} \label{tab:gf_res_xsgamma_1s}
\begin{center}
\begin{tabular}{|l|ccccccc|}
  \hline
  & $m_b^{1S}$ [GeV] & $\lambda_1$ [GeV$^2$] & $\rho_1$ [GeV$^3$] &
  $\tau_1$ [GeV$^3$] & $\tau_2$ [GeV$^3$] & $\tau_3$ [GeV$^3$] &
  $\vcb$ [10$^{-3}$]\\
  \hline \hline
  value & 4.691 & $-0.362$ & \phantom{$-$}0.043 &
  \phantom{$-$}0.161 & $-0.017$ & \phantom{$-$}0.213 &
  \phantom{$-$}41.98\\
  error & 0.037 & \phantom{$-$}0.067 & \phantom{$-$}0.048 &
  \phantom{$-$}0.122 & \phantom{$-$}0.062 & \phantom{$-$}0.102 &
  \phantom{$-$}0.45\\
  \hline
  $m_b^{1S}$ & 1.000 & \phantom{$-$}0.434 & \phantom{$-$}0.213 &
  $-0.058$ & $-0.629$ & $-0.019$ & $-0.215$\\
  $\lambda_1$ & & \phantom{$-$}1.000 & $-0.467$ & $-0.602$ & $-0.239$
  & $-0.547$ & $-0.403$\\
  $\rho_1$ & & & \phantom{$-$}1.000 & \phantom{$-$}0.129 & $-0.624$ &
  \phantom{$-$}0.494 & \phantom{$-$}0.286\\
  $\tau_1$ & & & & \phantom{$-$}1.000 & \phantom{$-$}0.062 & $-0.148$ &
  \phantom{$-$}0.194\\
  $\tau_2$ & & & & & \phantom{$-$}1.000 & $-0.009$ & $-0.145$\\
  $\tau_3$ & & & & & & \phantom{$-$}1.000 & \phantom{$-$}0.376\\
  $\vcb$ & & & & & & & \phantom{$-$}1.000\\
  \hline
\end{tabular}
\end{center}
\end{table}

% ======================================================================
% Exclusive CKM-suppressed decays
% ======================================================================
\subsection{Exclusive CKM-suppressed decays}
\label{slbdecays_b2uexcl}
% ----------------------------------------------
In this section, we give results on exclusive charmless semileptonic branching fractions
and the determination of $\Vub$ based on \Btopilnu\ decays.
The measurements are based on two different event selections: tagged
events, in which the second $B$ meson in the event is fully (or partially)
reconstructed, and untagged events, for which the momentum
of the undetected neutrino is inferred from measurements of the total 
momentum sum of the detected particles and the knowledge of the initial state.
The LHCb experiment recently reported a direct measurement of 
$|V_{ub}|/|V_{cb}|$ \cite{Aaij:2015bfa} reconstructing the 
$\Lb\to p\mu\nu$ decays and normalizing the branching fraction to the 
$\Lb\to\Lc(\to pK\pi)\mu\nu$ decays. 
We show a combination of $\Vub$-$\Vcb$ using the LHCb constraint on $|V_{ub}|/|V_{cb}|$, 
the exclusive determination of $\Vub$ from \Btopilnu\ and $\Vcb$ from both $B\to D^*\ell\nu$ 
and $B\to D\ell\nu$. 
We also present branching fraction averages for 
$\Bz\to\rho\ell^+\nu$, $\Bp\to\omega\ell^+\nu$, $\Bp\to\eta\ell^+\nu$ and $\Bp\to\etapr\ell^+\nu$.

\subsubsection{\Btopilnu\ branching fraction and $q^2$ spectrum}

\begin{figure} 
 \centering
 \includegraphics[width=0.8\textwidth,page=1]{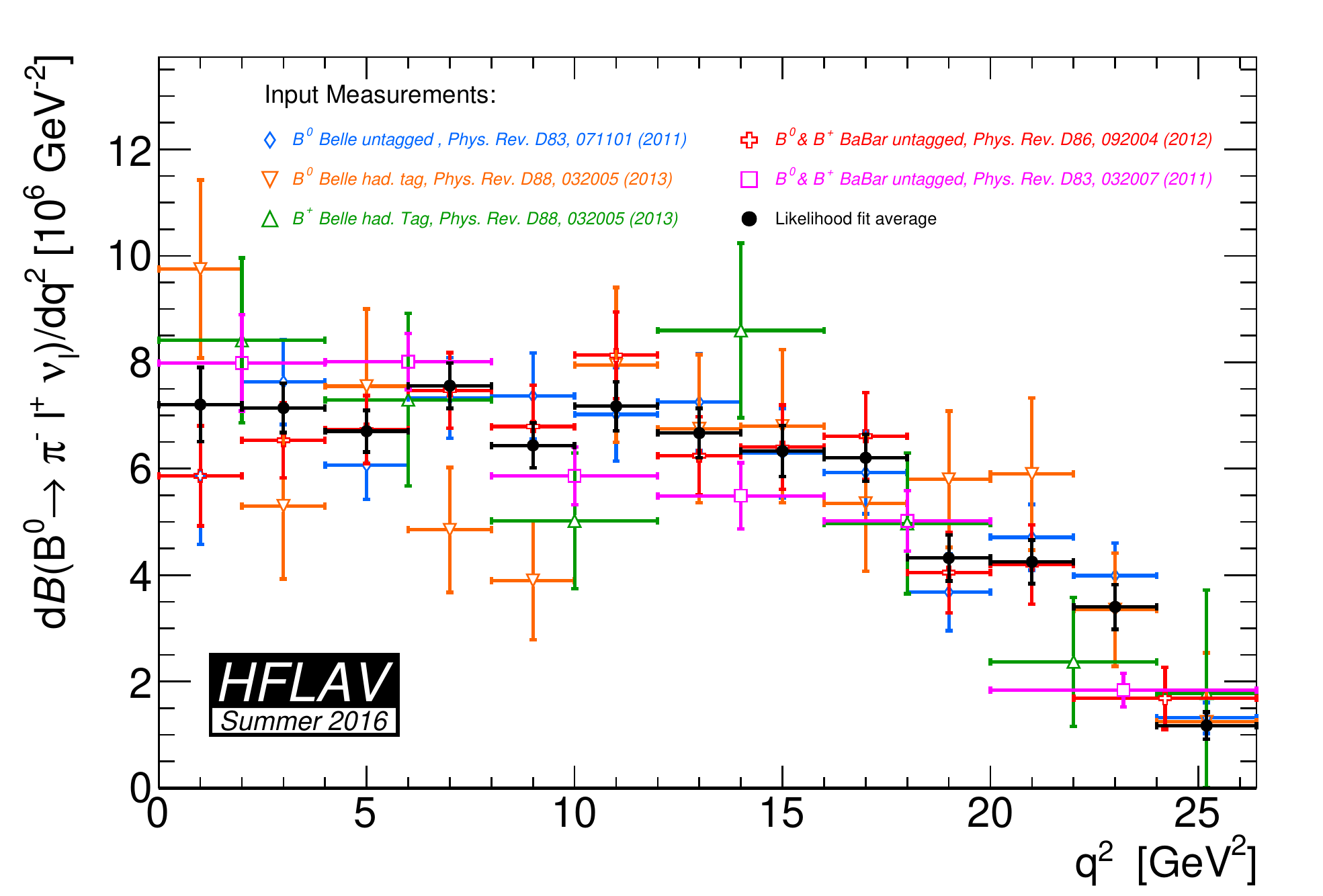}
\caption{The \Btopilnu\ $q^2$ spectrum measurements and the average spectrum obtained 
from the likelihood combination (shown in black). \label{fig:avg}}
\end{figure}

Currently, the four most precise measurements of the differential \Btopilnu\ decay rate as a function of the four-momentum transfer squared, $q^2$,
from \babar and Belle~\cite{Ha:2010rf,Sibidanov:2013rkk,delAmoSanchez:2010af,Lees:2012vv}
are used to obtain an average $q^2$ spectrum and an average for the total branching fraction. 
The measurements are presented in Fig.~\ref{fig:avg}.
From the two untagged \babar\ analyses~\cite{delAmoSanchez:2010af,Lees:2012vv},
the combined results for $B^0 \to \pi^- \ell^+ \nu$ and $B^+ \to \pi^0 \ell^+ \nu$ decays based on isospin symmetry are used.
The hadronic-tag analysis by Belle~\cite{Sibidanov:2013rkk} provides results for $B^0 \to \pi^- \ell^+ \nu$ and
$B^+ \to \pi^0 \ell^+ \nu$ separately, but not for the combination of both channels.
In the untagged analysis by Belle~\cite{Ha:2010rf}, only $B^0 \to \pi^- \ell^+ \nu$ decays were measured.
The experimental measurements use different binnings in $q^2$, but have matching bin edges, which allows
them to be easily combined.

To arrive at an average $q^2$ spectrum, a binned maximum-likelihood fit to determine the average partial branching fraction
in each $q^2$ interval is performed differentiating between common and individual uncertainties and correlations for the
various measurements.
Shared sources of systematic uncertainty of all measurements are included in the likelihood
as nuisance parameters constrained using standard normal distributions.
The most important shared sources of uncertainty are due to 
continuum subtraction, 
branching fractions, 
the number of $B$-meson pairs (only correlated among measurement by the same experiment), 
tracking efficiency (only correlated among measurements by the same experiment), 
uncertainties from modelling the $b \to u \, \ell \, \bar\nu_\ell$ contamination,
modelling of final state radiation, 
and contamination from $b \to c \, \ell \bar \nu_\ell$ decays. 

The averaged $q^2$ spectrum is shown in Fig.~\ref{fig:avg}. 
The probability of the average is computed as the $\chi^2$ probability quantifying the agreement between
the input spectra and the averaged spectrum and amounts to $6\%$.
The partial branching fractions and the full covariance matrix obtained from the likelihood fit 
are given in Tables \ref{tab:average} and ~\ref{tab:Cov}.
The average for the total $B^0 \to \pi^- \ell^+ \nu_\ell$ branching fraction is obtained by summing up the
partial branching fractions:
\begin{align}
{\cal B}(B^0 \to \pi^- \ell^+ \nu_\ell) = (1.50 \pm 0.02_{\rm stat} \pm 0.06_{\rm syst}) \times 10^{-4}.
\end{align}

\begin{table}
\centering
\small
\caption{Partial $B^0 \to \pi^- \ell^+ \nu_\ell$ branching fractions per GeV$^2$ for the input measurements and the average
obtained from the likelihood fit. The uncertainties are the combined statistical and systematic uncertainties.\label{tab:average}}
\begin{tabular}{|c | c | c | c | c | c || c|}
\hline
\multirow{2}{2cm}{$\Delta q^2$ [GeV${}^2$] }&\multicolumn{6}{ c |}{ $\Delta \mathcal{B}(B^0 \to \pi^- \ell^+ \nu_\ell) / \Delta q^2 $\quad$ [10^{-7}]$}\\\cline{2-7}
& Belle & Belle & Belle & \babar & \babar & Average\\
& untagged & tagged & tagged & untagged & untagged & \\
& ($B^0$) & ($B^0$) & ($B^+$) & ($B^{0,+}$, 12 bins) & ($B^{0,+}$, 6 bins) & \\\hline
$ 0 - 2 $ & $ 58.7 \pm 12.9 $ & $ 97.5 \pm 16.7 $ &\multirow{2}{2cm}{ $ 84.1 \pm 15.5 $ }& $ 58.7 \pm 9.4 $ &\multirow{2}{2cm}{ $ 79.9 \pm 9.1 $} & $ 72.0 \pm 7.0 $\\
$ 2 - 4 $ & $ 76.3 \pm 8.0 $ & $ 53.0 \pm 13.8 $ & & $ 65.3 \pm 7.1 $ & & $ 71.4 \pm 4.6 $\\
$ 4 - 6 $ & $ 60.6 \pm 6.4 $ & $ 75.5 \pm 14.5 $ &\multirow{2}{2cm}{ $ 73.0 \pm 16.2 $} & $ 67.3 \pm 6.4 $ &\multirow{2}{2cm}{ $ 80.1 \pm 5.3 $ }& $ 67.0 \pm 3.9 $\\
$ 6 - 8 $ & $ 73.3 \pm 7.6 $ & $ 48.5 \pm 11.8 $ & & $ 74.7 \pm 7.1 $ & & $ 75.6 \pm 4.3 $\\
$ 8 - 10 $ & $ 73.7 \pm 8.1 $ & $ 39.0 \pm 11.2 $ &\multirow{2}{2cm}{ $ 50.2 \pm 12.8 $ }& $ 67.9 \pm 7.8 $ &\multirow{2}{2cm}{ $ 58.7 \pm 5.5 $} & $ 64.4 \pm 4.3 $\\
$ 10 - 12 $ & $ 70.2 \pm 8.8 $ & $ 79.5 \pm 14.6 $ & & $ 81.3 \pm 8.2 $ & & $ 71.7 \pm 4.6 $\\
$ 12 - 14 $ & $ 72.5 \pm 9.1 $ & $ 67.5 \pm 13.9 $ &\multirow{2}{2cm}{ $ 86.0 \pm 16.4 $ }& $ 62.4 \pm 7.4 $ &\multirow{2}{2cm}{ $ 54.9 \pm 6.2 $} & $ 66.7 \pm 4.7 $\\
$ 14 - 16 $ & $ 63.0 \pm 8.4 $ & $ 68.0 \pm 14.4 $ & & $ 64.0 \pm 7.9 $ & & $ 63.3 \pm 4.8 $\\
$ 16 - 18 $ & $ 59.3 \pm 7.8 $ & $ 53.5 \pm 12.8 $ &\multirow{2}{2cm}{ $ 49.7 \pm 13.3 $} & $ 66.1 \pm 8.2 $ &\multirow{2}{2cm}{ $ 50.2 \pm 5.7 $ }& $ 62.0 \pm 4.4 $\\
$ 18 - 20 $ & $ 36.8 \pm 7.2 $ & $ 58.0 \pm 12.8 $ & & $ 40.5 \pm 7.6 $ & & $ 43.2 \pm 4.3 $\\
$ 20 - 22 $ & $ 47.1 \pm 6.2 $ & $ 59.0 \pm 14.3 $ &\multirow{2}{2cm}{ $ 23.7 \pm 12.1 $ }& $ 42.0 \pm 7.5 $ &\multirow{3}{2cm}{ $ 18.4 \pm 3.2 $ }& $ 42.5 \pm 4.1 $\\
$ 22 - 24 $ & $ 39.9 \pm 6.2 $ & $ 33.5 \pm 10.6 $ & & \multirow{2}{2cm}{ $ 16.8 \pm 5.9 $ }& & $ 34.0 \pm 4.2 $\\
$ 24 - 26.4 $ & $ 13.2 \pm 2.9 $ & $ 12.4 \pm 13.0 $ & $ 17.8 \pm 19.4 $ & & & $ 11.7 \pm 2.6 $\\\hline
\end{tabular}
\end{table}

\begin{table}
\caption{Covariance matrix of the averaged partial branching fractions per GeV$^2$ in units of $10^{-14}$.\label{tab:Cov}}
\tiny
\begin{tabular}{c | c c c c c c c c c c c c c c }
$\Delta q^2$ [GeV${}^{2}$] &$0-2$ &$ 2-4$ &$4-6$ &$6-8 $&$8-10$ &$10-12$ &$12-14$ &$14-16$ &$16-18 $&$18-20$ &$20-22$ &$22-24$ &$24-26.4$ \\\hline
$0-2$&$49.091$&$1.164$&$8.461$&$7.996$&$7.755$&$9.484$&$7.604$&$9.680$&$8.868$&$7.677$&$7.374$&$7.717$&$2.877$\\
$2-4$&        &$21.487$&$-0.0971$&$7.155$&$4.411$&$5.413$&$4.531$&$4.768$&$4.410$&$3.442$&$3.597$&$3.388$&$1.430$\\
$4-6$&        &        &$15.489$&$-0.563$&$5.818$&$4.449$&$4.392$&$4.157$&$4.024$&$3.185$&$3.169$&$3.013$&$1.343$\\
$6-8$&        &        &        &$18.2$&$2.377$&$7.889$&$6.014$&$5.938$&$5.429$&$4.096$&$3.781$&$3.863$&$1.428$\\
$8-10$&       &        &        &      &$18.124$&$1.540$&$7.496$&$5.224$&$5.441$&$4.197$&$3.848$&$4.094$&$1.673$\\
$10-12$&      &        &        &      &        &$21.340$&$4.213$&$7.696$&$6.493$&$5.170$&$4.686$&$4.888$&$1.950$\\
$12-14$&      &        &        &      &        &        &$21.875$&$0.719$&$6.144$&$3.846$&$3.939$&$3.922$&$1.500$\\
$14-16$&      &        &        &      &        &        &        &$23.040$&$5.219$&$6.123$&$4.045$&$4.681$&$1.807$\\
$16-18$&      &        &        &      &        &        &        &        &$19.798$&$1.662$&$4.362$&$4.140$&$1.690$\\
$18-20$&      &        &        &      &        &        &        &        &        &$18.0629$&$2.621$&$3.957$&$1.438$\\
$20-22$&      &        &        &      &        &        &        &        &        &         &$16.990$&$1.670$&$1.127$\\
$22-24$&      &        &        &      &        &        &        &        &        &         &        &$17.774$&$-0.293$\\
$24-26.4$&    &        &        &      &        &        &        &        &        &         &        &        &$6.516$\\
\end{tabular}
\end{table}

\subsubsection{\Vub from \Btopilnu}

The \Vub average can be determined from the averaged $q^2$ spectrum in combination with a prediction for
the normalization of the $\B \to \pi$ form factor.
The differential decay rate for light leptons ($e$, $\mu$) is given by
\begin{align}
 \Delta \Gamma =  \Delta \Gamma(q^2_{\rm low}, q^2_{\rm high})  = \int_{q^2_{\rm low}}^{q^2_{\rm high}} \text{d} q^2 \bigg[ \frac{8 \left| \vec p_\pi \right|  }{3} \frac{ G_F^2 \, \left| V_{ub} \right|^2 q^2 }{256 \, \pi^3 \, m_B^2}  H_0^2(q^2) \bigg]  \, ,
\end{align}
where $G_F$ is Fermi's constant, $\left| \vec p_\pi \right|$ is the absolute four-momentum of the 
final state $\pi$ (a function of $q^2$), $m_B$ the $B^0$-meson mass, 
and $H_0(q^2)$ the only non-zero helicity amplitude. 
The helicity amplitude is a function of the form factor $f_+$, 
\begin{align}
 H_0 = \frac{2 m_B \, \left| \vec p_\pi \right| }{\sqrt{q^2}}\, f_+(q^2) .
\end{align} 
The form factor $f_{+}$ can be calculated with non-perturbative methods, but its general form can be constrained by the differential \Btopilnu\ spectrum. 
Here, we parametrize the form factor using the BCL parametrization~\cite{Bourrely:2008za}.

The decay rate is proportional to $\Vub^2 |f_+(q^2)|^2$. Thus to extract \Vub one needs to determine $f_+(q^2)$
(at least at one value of $q^2$). In order to enhance the precision, a binned $\chi^2$ fit is performed
using a $\chi^2$ function of the form
\begin{align} \label{eq:chi2}
 \chi^2 & = \left( \vec{\cal B} - \Delta \vec{\Gamma} \, \tau \right)^T 
            C^{-1} 
            \left( \vec{\cal B} - \Delta \vec{\Gamma} \, \tau \right) + \chi^2_{\rm LQCD} + \chi^2_{\rm LCSR}
\end{align}
with $C$ denoting the covariance matrix given in Table~\ref{tab:Cov}, $\vec{\cal B}$ the vector of 
averaged branching fractions and $\Delta \vec{\Gamma} \, \tau$ the product of the vector of 
theoretical predictions of the partial decay rates and the $B^0$-meson lifetime. 
The form factor normalization is included in the fit by the two extra terms in Eq.~(\ref{eq:chi2}): 
$\chi_{\rm LQCD}$ uses the latest FLAG lattice average~\cite{Aoki:2016frl} from 
two state-of-the-art unquenched lattice QCD calculations~\cite{Lattice:2015tia, Flynn:2015mha}. 
The resulting constraints are quoted directly in terms of the coefficients $b_j$ of the BCL parameterization 
and enter Eq.~(\ref{eq:chi2}) as
\begin{align}
 \chi^2_{\rm LQCD} & = \left( {\vec{b}} - {\vec{b}_{\rm LQCD}} \right)^T \, C_{\rm LQCD}^{-1} \,  \left( {\vec{b}} - {\vec{b}_{\rm LQCD}} \right) \, ,
\end{align}
with ${\vec{b}}$ the vector containing the free parameters of the $\chi^2$ fit 
constraining the form factor, 
${\vec{b}_{\rm LQCD}}$ the averaged values from Ref.~\cite{Aoki:2016frl}, 
and $C_{\rm LQCD}$ their covariance matrix. 
Additional information about the form factor can be obtained from light-cone sum rule calculations. 
The state-of-the-art calculation includes up to two-loop contributions~\cite{Bharucha:2012wy}. 
It is included in Eq.~(\ref{eq:chi2}) via
\begin{align}
\chi^2_{\rm LQCR} & = \left( f_+^{\rm LCSR} - f_+(q^2 = 0; {\vec{b}}) \right)^2 / \sigma_{f_+^{\rm LCSR} }^2 \, .
\end{align}

\begin{figure} 
\centering
 \includegraphics[width=0.8\textwidth]{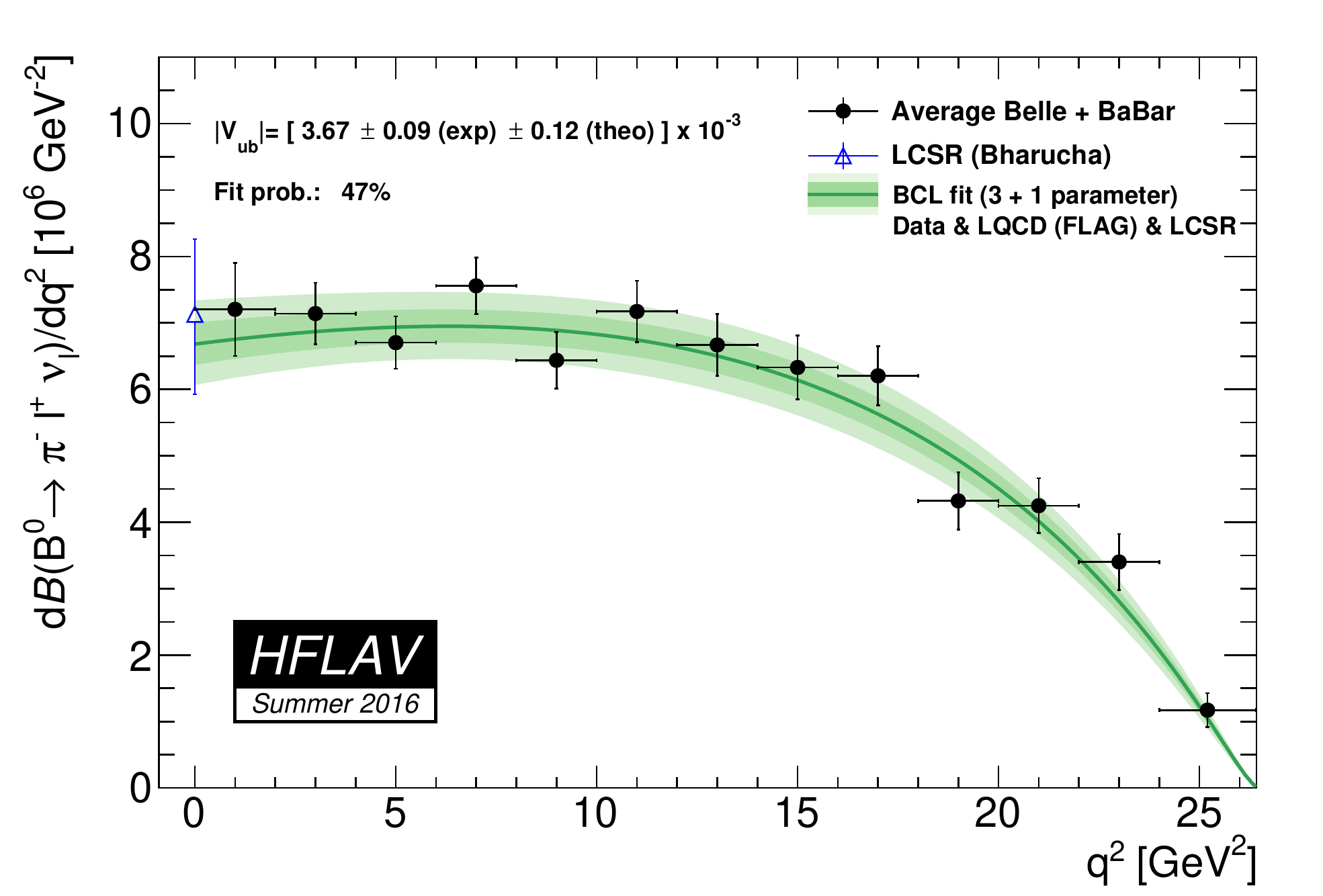}
\caption{Fit of the BCL parametrization to the averaged $q^2$ spectrum from \babar and Belle and the LQCD and LCSR calculations. 
The error bands represent the $1~\sigma$ (dark green) and $2~\sigma$ (light green) uncertainties 
of the fitted spectrum. \label{fig:vub}}
\end{figure}

The \Vub average is obtained for two versions: the first combines the data with the LQCD constraints  
and the second additionally includes the information from the LCSR calculation. 
The resulting values for $\Vub$ are
\begin{align}
 \Vub & = \left( 3.70 \pm 0.10 \, (\text{exp}) \pm 0.12 \, (\text{theo}) \right) \times 10^{-3} \, \ \rm (data+LQCD), \\
 \Vub & = \left( 3.67 \pm 0.09 \, (\text{exp}) \pm 0.12 \, (\text{theo}) \right) \times 10^{-3} \, \ \rm (data+LQCD+LCSR),
\end{align}
for the first and second fit version, respectively. 
The result of the fit including both LQCD and LCSR is shown in Figure~\ref{fig:vub}. 
The $\chi^2$ probability of the fit is $47\%$.
We quote the result of the fit including both LQCD and LCSR calculations as our average for \Vub. 
The best fit values for \Vub and the BCL parameters and their covariance matrix 
are given in Tables~\ref{tab:fitres2} and ~\ref{tab:fitcov2}. 

\begin{table}
\centering
\caption{Best fit values and uncertainties for the combined fit to data, LQCD and LCSR results. \label{tab:fitres2}}
\begin{tabular}{c | c }
Parameter & Value\\\hline
$\Vub  $&$(3.67 \pm 0.15)\times 10^{-3}$\\
$b_0$   & $0.418 \pm 0.012$\\
$b_1$   & $-0.399 \pm 0.033$\\
$b_2$   &$-0.578 \pm 0.130$\\
\end{tabular}
\end{table}

\begin{table}
\centering
\caption{Covariance matrix for the combined fit to data, LQCD and LCSR results. \label{tab:fitcov2}}

\begin{tabular}{c | c c c c}
Parameter & $\Vub$ & $b_0$ & $b_1$ & $b_2$\\\hline
$\Vub$    &$ 2.064\times 10^{-8} $&$ -1.321\times 10^{-6}$&$ -1.881\times 10^{-6}$&$ 7.454\times 10^{-6}$ \\
$b_0$   &                       &$ 1.390\times 10^{-4} $&$ 8.074\times 10^{-5}$ &$ -8.953\times 10^{-4} $\\
$b_1$   &                       &                       &$ 1.053\times 10^{-3}$ &$ -2.879\times 10^{-3}$ \\
$b_2$   &                       &                       &                       &$ 1.673\times 10^{-2}$\\
\end{tabular}

\end{table}

\subsubsection{Combined extraction of $\Vub$ and $\Vcb$}

The LHCb experiment reported the first observation of the CKM suppressed decay $\Lb\to p\mu\nu$
\cite{Aaij:2015bfa} and the measurement of the ratio of partial branching fractions at high $q^2$
for $\Lb\to p\mu\nu$ and $\Lb\to \Lc(\to pK\pi)\mu\nu$ decays

\begin{align}
R = \dfrac{{\cal B}(\Lb\to p\mu\nu)_{q^2>15~GeV^2} }{{\cal B}(\Lb\to \Lc\mu\nu)_{q^2>7~GeV^2} }=(1.00\pm 0.04\pm 0.08)\times 10^{-2}.
\end{align}

\noindent The ratio $R$ is proportional to $(|V_{ub}|/|V_{cb}|)^2$ and sensitive to the form factors 
of $\Lb\to p$ and $\Lb\to \Lc$ transitions that have to be computed with non-perturbative
methods, like lattice QCD.
%The measured ratio $R$ depends on the branching fraction of the $\Lc$ in the $p K \pi$ decay mode used to reconstruct the normalization decay. 
The uncertainty on ${\cal B}(\Lc\to p K \pi)$ is the largest source of systematic uncertainties
on $R$.
Using the recent HFLAV average ${\cal B}(\Lc\to p K \pi)=(6.46\pm 0.24)\%$ reported in Table~\ref{tab:Lc:br-fit}, 
which includes the recent BESIII measurements ~\cite{Ablikim:2015flg}, the rescaled value for $R$ is 

\begin{align}
R = (0.95\pm 0.04\pm 0.07)\times 10^{-2}
\end{align}

\begin{figure} 
\centering
\includegraphics[width=0.8\textwidth]{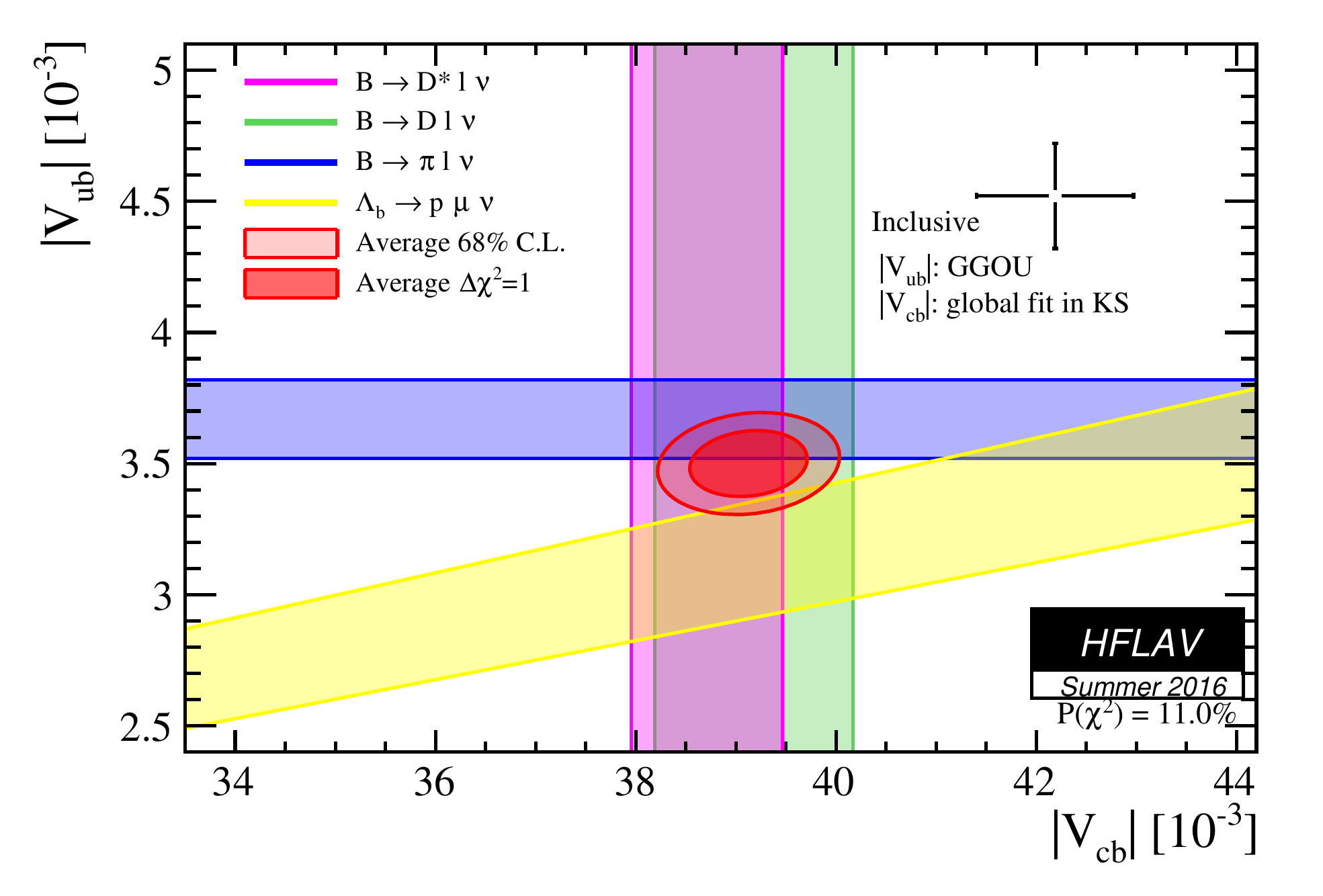}
 \caption{\Vub-\Vcb combined average including the LHCb measurement of $\Vub/\Vcb$, the exclusive $\Vub$ measurement from \Btopilnu, and \Vcb measurements from
both  $B\to D^*\ell\nu$ and $B\to D\ell\nu$. The point with the error bars corresponds to the inclusive \Vcb 
 from the kinetic scheme \ref{globalfitsKinetic}, and the 
 inclusive \Vub from GGOU calculation \ref{subsec:ggou}. \label{fig:vubvc}}
\end{figure}

\noindent With the precise lattice QCD prediction \cite{Detmold:2015aaa} of the form factors in the 
experimentally interesting $q^2$ region considered, results in

\begin{align}
\dfrac{|V_{ub}|}{|V_{cb}|} = 0.080\pm 0.004_{Exp.} \pm 0.004_{F.F.}
\end{align}

\noindent where the first uncertainty is the total experimental error and the second one is due to the 
knowledge of the form factors. A combined fit for \Vub and \Vcb that includes the constraint from LHCb, 
and the determination of \Vub and \Vcb from exclusive $B$ meson decays, results in 

\begin{align}
\left| V_{ub} \right| & = \left( 3.50 \pm 0.13 \right) \times 10^{-3}\,  \\
\left| V_{cb} \right| & = \left( 39.13 \pm 0.59 \right) \times 10^{-3} \, \\
\rho(\Vub|,\Vcb) & =0.14\,
\end{align}

\noindent where the uncertainties are considered uncorrelated. 
The $\chi^2$ of the fit is $4.4$ for $2$ d.o.f corresponding 
to a $P(\chi^2)$ of 11.0\%. The fit result is shown in Fig.~\ref{fig:vubvc}, where both 
the $\Delta\chi^2$ and the two-dimensional $68\%$ C.L. contours are indicated. 
The $\Vub/\Vcb$ value extracted from $R$ is more compatible with the exclusive determinations
of $\Vub$. Another recent calculation, by Faustov and Galkin \cite{Faustov:2016pal}, 
based on a relativistic quark model,
gives a value of $\Vub/\Vcb$  closer to the inclusive determinations. 
%More calculations of the relevant form factors for $\Lb\to p\ell\nu$ and $\Lb\to \Lc\ell\nu$ are highly desirable.

\subsubsection{Other exclusive charmless semileptonic \B decays}

\begin{figure}[!ht]
 \begin{center}
  \unitlength1.0cm % coordinates in cm
  \begin{picture}(14.,10.0)  %ys(25.,6.)
   \put( -1.5,  0.0){\includegraphics[width=9.0cm]{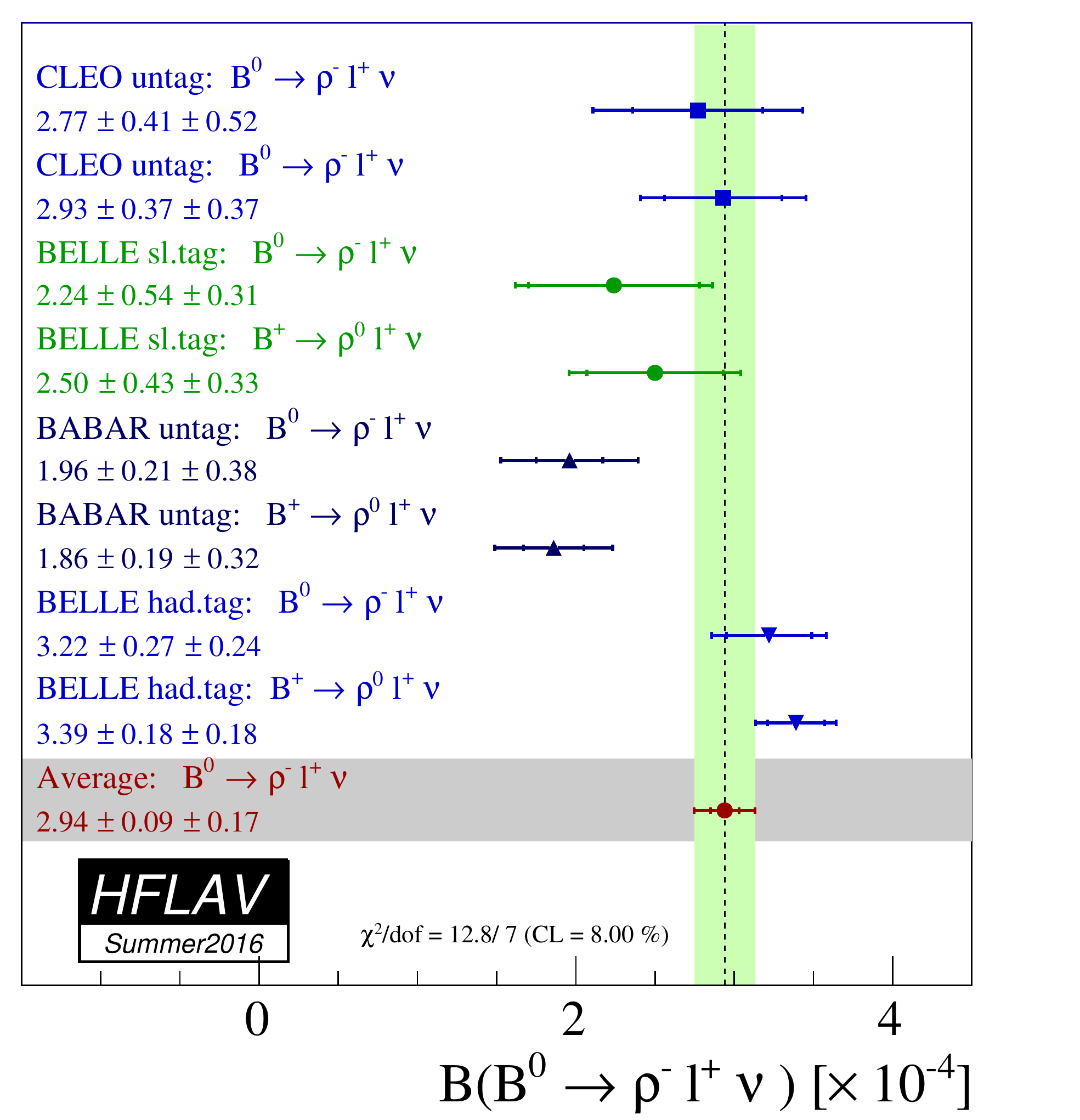}}
   \put( 7.5,  0.0){\includegraphics[width=9.0cm]{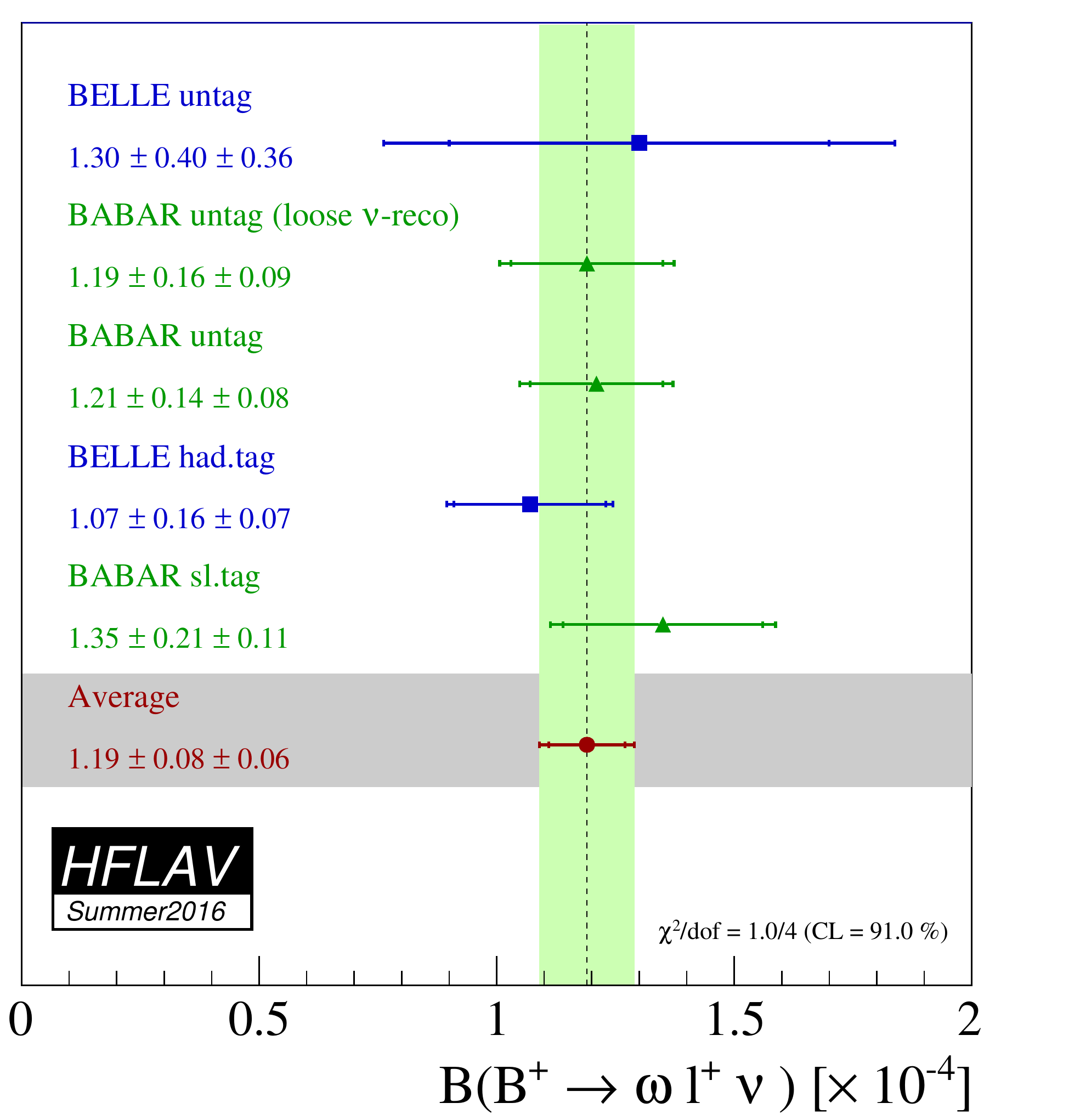}} 
   \put(  5.8,  8.5){{\large\bf a)}}  
   \put( 14.6,  8.5){{\large\bf b)}}
   \end{picture} \caption{
 (a) Summary of exclusive determinations of $\cbf(\Bz\to\rho\ell^+\nu)$ and their average. Measurements
 of $B^+ \to \rho^0\ell^+\nu$ branching fractions have been multiplied by $2\tau_{B^0}/\tau_{B^+}$ 
 in accordance with isospin symmetry.    
(b) Summary of exclusive determinations of $\Bp\to\omega\ell^+\nu$ and their average.
}
\label{fig:xulnu1}
\end{center}
\end{figure}

\begin{figure}[!ht]
 \begin{center}
  \unitlength1.0cm % coordinates in cm
  \begin{picture}(14.,10.0)  %ys(25.,6.)
   \put( -1.5,  0.0){\includegraphics[width=9.0cm]{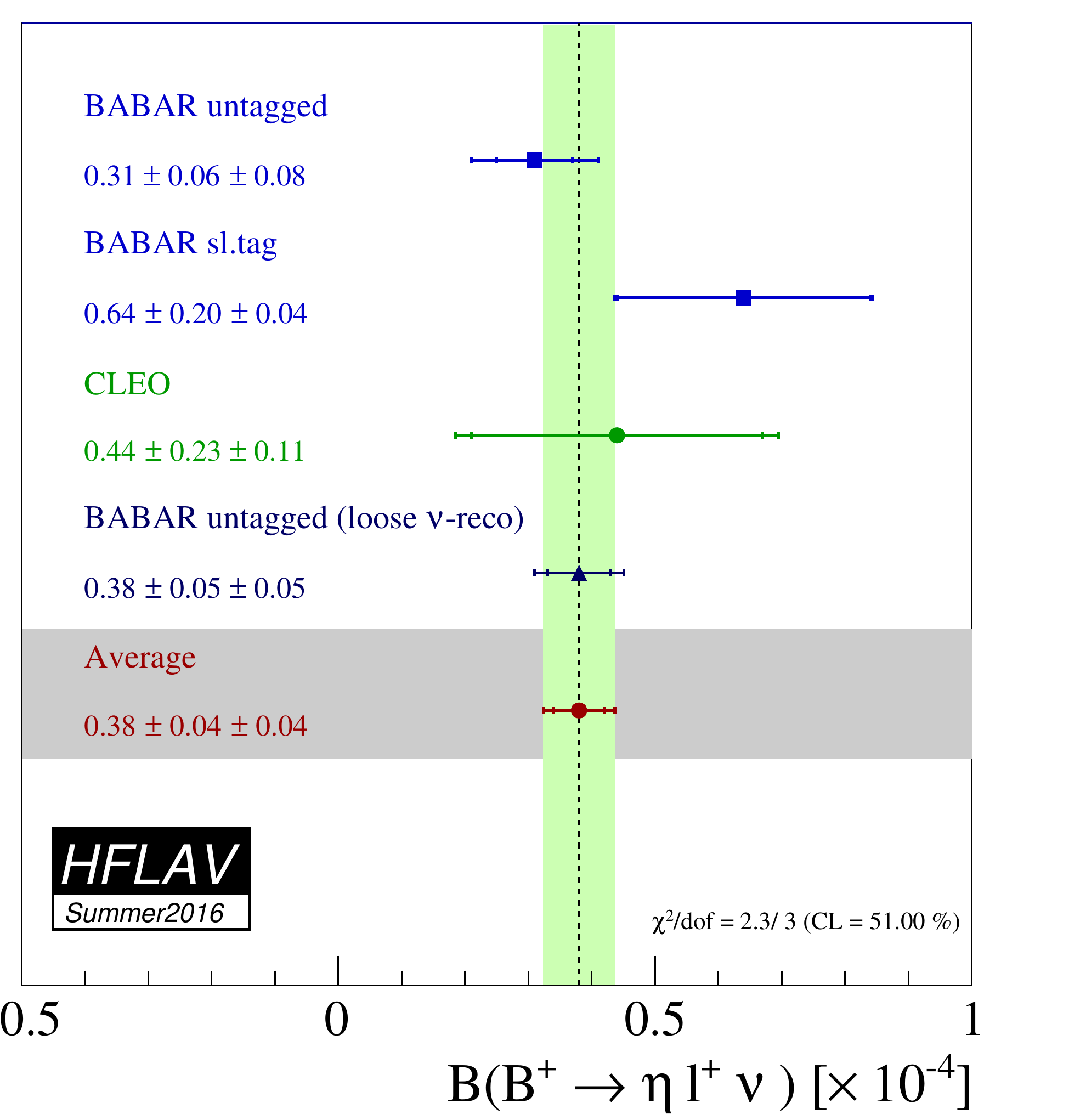}}
   \put( 7.5,  0.0){\includegraphics[width=9.0cm]{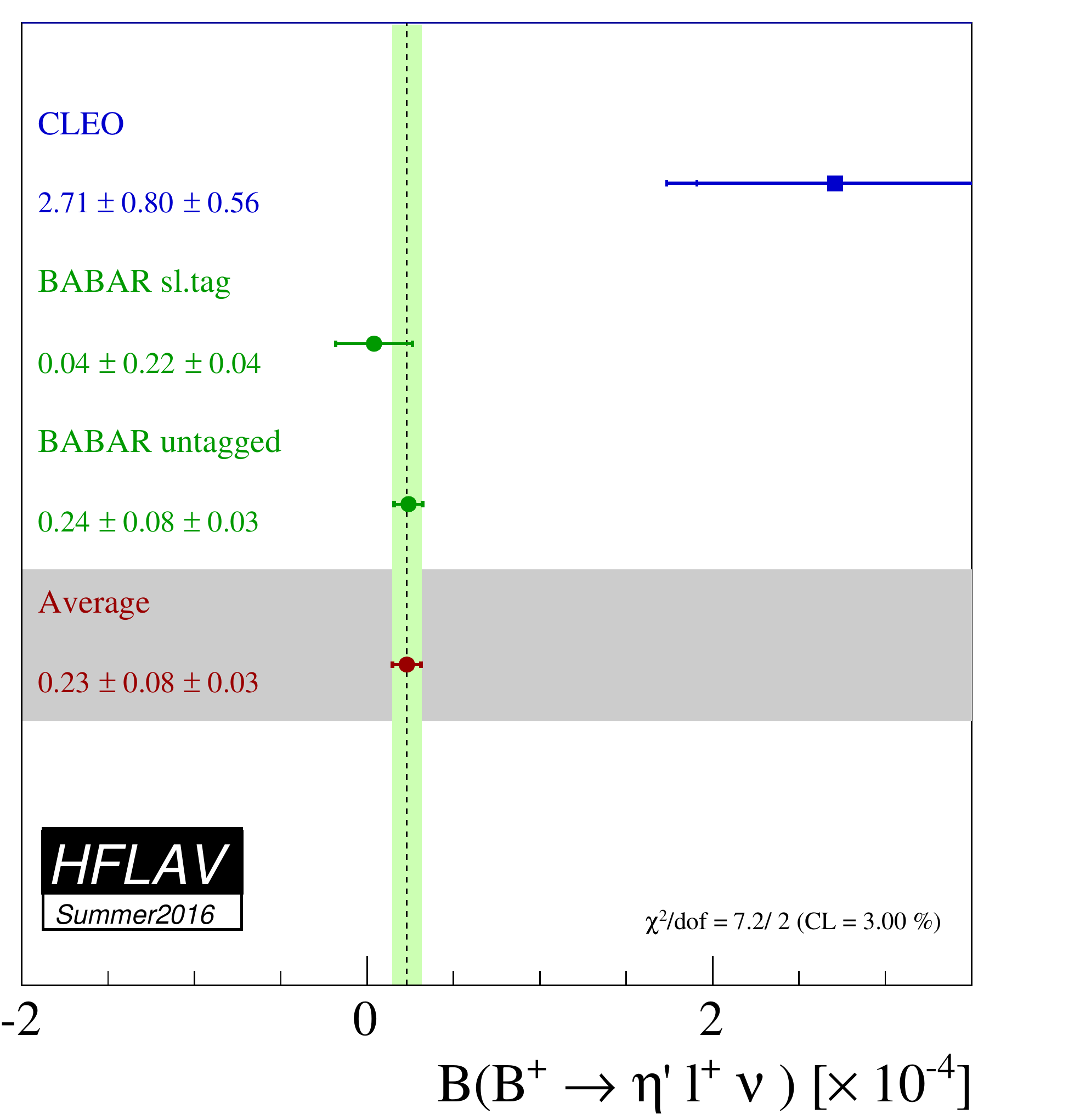}} 
   \put(  5.8,  8.5){{\large\bf a)}}     
   \put( 14.6,  8.5){{\large\bf b)}}
   
   \end{picture} \caption{
(a) Summary of exclusive determinations of $\cbf(\Bp\to\eta\ell^+\nu)$ and their average.
(b) Summary of exclusive determinations of $\cbf(\Bp\to\etapr\ell^+\nu)$ and their average.
}
\label{fig:xulnu2}
\end{center}
\end{figure}

We report the branching fraction average for $\Bz\to\rho\ell^+\nu$, $\Bp\to\omega\ell^+\nu$, $\Bp\to\eta\ell^+\nu$ 
and $\Bp\to\etapr\ell^+\nu$ decays. The measurements and their averages are listed in 
Tables~\ref{tab:rholnu},~\ref{tab:omegalnu},~\ref{tab:etalnu},~\ref{tab:etaprimelnu}, 
and presented in Figures ~\ref{fig:xulnu1} and ~\ref{fig:xulnu2}. 
%The measurements for $\Bz\to\rho\ell^+\nu$ are reported in  Table~\ref{tab:rholnu} and shown in Fig.~\ref{fig:xulnu1}~(a).
In the $\Bz\to\rho^-\ell^+\nu$ average, both the $\Bz\to\rho^-\ell^+\nu$ and $\Bp\to\rho^0\ell^+\nu$ decays are used, 
where the $\Bp\to\rho^0\ell^+\nu$ are rescaled by $2\tau_{B^0}/\tau_{B^+}$ assuming the isospin symmetry.
For  $\Bp\to\omega\ell^+\nu$ and $\Bp\to\eta\ell^+\nu$ decays, the agreement between the different measurements 
is good. $\Bp\to\etapr\ell^+\nu$ shows a discrepancy between the old CLEO measurement and the \babar untagged 
analysis, but the statistical uncertainties of the CLEO measuement are large.
The $\Bz\to\rho\ell^+\nu$ results, instead, show significant differences, in particular the \babar untagged analysis 
gives a branching fraction significantly lower (by about 2$\sigma$) that the Belle measurement based on the hadronic-tag. 
A possible reason for such discrepancy could be the broad nature of the $\rho$ resonance that makes the 
control of the background under the  $\rho$ mass peak more difficult in the untagged analysis than in the hadronic-tag
analysis.

We do not report \vub~ for these exclusive charmless decays, because the form factor calculations have not yet reached 
the precision achieved for $B\to\pi\ell\nu$ decays. 
Unquenched lattice QCD calculations of the form factors are not available for these decays, 
but LCSR calculations exist for all these decay modes. The most recent of these calculations for the  $B\to\rho\ell\nu$ 
and  $B\to\omega\ell\nu$ decays are reported in Ref.\cite{Ball:2004ye} and \cite{Straub:2015ica}.

%The measurement for $\Bp\to\omega\ell^+\nu$ and their average, are reported in Table~\ref{tab:omegalnu} 
%and shown in Fig.~\ref{fig:xulnu1}~(b), while the ones for $\Bp\to\eta\ell^+\nu$ and  $\Bp\to\etapr\ell^+\nu$ are reported in 
%Table~\ref{tab:etalnu} and~\ref{tab:etaprimelnu},  and are shown in Fig.~\ref{fig:xulnu2}. 

\begin{table}[!htb]
\begin{center}
\caption{Summary of exclusive determinations of $\Bz\to\rho\ell^+\nu$. The errors quoted
correspond to statistical and systematic uncertainties, respectively.}
\label{tab:rholnu}
\begin{small}
\begin{tabular}{|lc|}
\hline
& $\cbf [10^{-4}]$
\\
\hline\hline
CLEO (Untagged) $\rho^+$~\cite{Behrens:1999vv}
& $2.77\pm 0.41\pm 0.52\ $ 
\\ 
CLEO (Untagged) $\rho^+$~\cite{Adam:2007pv}
& $2.93\pm 0.37\pm 0.37\ $ 
\\ 
%\babar\ $\rho^+$~\cite{Aubert:2005cd}
%& $2.16\pm 0.21\pm 0.57\ $
% Belle Breco
Belle (Hadronic Tag) $\rho^+$~\cite{Sibidanov:2013rkk}
& $3.22\pm 0.27\pm 0.24\ $
\\
Belle (Hadronic Tag) $\rho^0$~\cite{Sibidanov:2013rkk}
& $3.39\pm 0.18\pm 0.18\ $
\\
%Belle SL
Belle (Semileptonic Tag) $\rho^+$~\cite{Hokuue:2006nr}
& $2.24\pm 0.54\pm 0.31\ $
\\
Belle (Semileptonic Tag) $\rho^0$~\cite{Hokuue:2006nr}
& $2.50\pm 0.43\pm 0.33\ $
\\
\babar (Untagged) $\rho^+$~\cite{delAmoSanchez:2010af}
& $1.96\pm 0.21\pm 0.38\ $
\\
\babar (Untagged) $\rho^0$~\cite{delAmoSanchez:2010af}
& $1.86\pm 0.19\pm 0.32\ $

\\  \hline
{\bf Average}
& \mathversion{bold}$2.94 \pm 0.09\pm 0.17 $
%\hline
%{\bf Average of published results}
%& \mathversion{bold}$2.34 \pm 0.15\pm 0.24 $
\\ 
\hline
\end{tabular}\\
\end{small}
\end{center}
\end{table}

\begin{table}[!htb]
\begin{center}
\caption{Summary of exclusive determinations of $\Bp\to\omega\ell^+\nu$. The errors quoted
correspond to statistical and systematic uncertainties, respectively.}
\label{tab:omegalnu}
\begin{small}
\begin{tabular}{|lc|}
\hline
& $\cbf [10^{-4}]$
\\
\hline\hline
Belle (Untagged) ~\cite{Schwanda:2004fa}
& $1.30\pm 0.40\pm 0.36\ $
\\
\babar (Loose $\nu$ reco.) ~\cite{Lees:2012vv}
& $1.19\pm 0.16\pm 0.09\ $
\\  
\babar (Untagged) ~\cite{Lees:2012mq}
& $1.21\pm 0.14\pm 0.08\ $
\\  
Belle (Hadronic Tag) ~\cite{Sibidanov:2013rkk}
& $1.07\pm 0.16\pm 0.07 $
\\
\babar (Semileptonic Tag) ~\cite{Lees:2013gja}
& $1.35\pm 0.21\pm 0.11\ $
\\  

\hline

{\bf Average}
& \mathversion{bold}$1.19 \pm 0.08 \pm 0.06\ $
\\ 
\hline
\end{tabular}\\
\end{small}
\end{center}
\end{table}

\begin{table}[!htb]
\begin{center}
\caption{Summary of exclusive determinations of $\Bp\to\eta\ell^+\nu$. 
The errors quoted correspond to statistical and systematic uncertainties, respectively.}
\label{tab:etalnu}
\begin{small}
\begin{tabular}{|lc|}
\hline
& $\cbf [10^{-4}]$
\\
\hline\hline
CLEO ~\cite{Gray:2007pw}
& $0.45\pm 0.23\pm 0.11\ $
\\
\babar\ (Untagged) ~\cite{Aubert:2008ct}
& $0.31\pm 0.06\pm 0.08\ $
\\ 
\babar\ (Semileptonic Tag) ~\cite{Aubert:2008bf}
& $0.64\pm 0.20\pm 0.04\ $
\\
\babar\ (Loose $\nu$-reco.) ~\cite{Lees:2012vv}
& $0.38\pm 0.05\pm 0.05\ $
\\  
 \hline
{\bf Average}
& \mathversion{bold}$0.38 \pm 0.04 \pm 0.04 $
\\ 
\hline
\end{tabular}\\
\end{small}
\end{center}
\end{table}

\begin{table}[!htb]
\begin{center}
\caption{Summary of exclusive determinations of  $\Bp\to\eta'\ell^+\nu$. The errors quoted
correspond to statistical and systematic uncertainties, respectively.}
\label{tab:etaprimelnu}
\begin{small}
\begin{tabular}{|lc|}
\hline
& $\cbf [10^{-4}]$
\\
\hline\hline
CLEO ~\cite{Gray:2007pw} & $2.71\pm 0.80\pm 0.56\ $
\\
\babar\ (Semileptonic Tag) ~\cite{Aubert:2008bf}
& $0.04\pm 0.22\pm 0.04$, $(<0.47 ~~@~ 90\% C.L.)$
\\ 
\babar\ (Untagged) ~\cite{Lees:2012vv} & $0.24\pm 0.08\pm 0.03$
\\  
 \hline
{\bf Average}
& \mathversion{bold}$0.23 \pm 0.08 \pm 0.03 $
\\ 
\hline
\end{tabular}\\
\end{small}
\end{center}
\end{table}

%Branching fractions for other $\Bb\to X_u\ell\nub$ decays are given in
%Table~\ref{tab:xslother}. 
%\input{tables/slb/xslother.tex}

% ----------------------------------------------------------------------

\clearpage

%
% ======================================================================
% Inclusive CKM-suppressed decays
% -- \include{b2uincl.tex}
% ======================================================================
\subsection{Inclusive CKM-suppressed decays}
\label{slbdecays_b2uincl}
% ----------------------------------------------
%% The large background from $\B\to X_c\ell^+\nul$ decays is the chief
%% experimental limitation in determinations of $\vub$.  Cuts designed to
%% reject this background limit the acceptance for $\B\to X_u\ell^+\nul$
%% decays. The calculation of partial rates for these restricted
%% acceptances is more complicated and requires substantial theoretical machinery.
%% VERA
Measurements of $B \to X_u \ell^+ \nu$  decays are very challenging because of the fifty times larger rates Cabibbo-favoured 
$B \to X_c \ell^+ \nu$ decays.  Cuts designed to suppress this dominant background severely complicate the perturbative 
QCD calculations required to extract $\vub$.  For strict phase space limitations, parameterizations of the so-called 
shape functions are required to describe the unmeasured regions of the phase space.  
In this update, we use several theoretical calculations to extract \vub~ and  
do not advocate the use of one method over another.
The authors of the different calculations have provided 
codes to compute the partial rates in limited regions of phase space covered by the measurements. 
Latest results by Belle~\cite{ref:belle-multivariate} and \babar~\cite{Lees:2011fv} 
explore large portions of phase space, with a consequent reduction of the theoretical 
uncertainties. 

In the averages, the systematic errors associated with the
modeling of $\B\to X_c\ell^+\nul$ and $\B\to X_u\ell^+\nul$ decays and the theoretical
uncertainties are taken as fully correlated among all measurements.
Reconstruction-related uncertainties are taken as fully correlated within a given experiment.
Measurements of partial branching fractions for $\B\to X_u\ell^+\nul$
transitions from $\Upsilon(4S)$ decays, together with the corresponding selected region, 
are given in Table~\ref{tab:BFbulnu}.  
The signal yields for all the measurements shown in Table~\ref{tab:BFbulnu}
are not rescaled to common input values of the $B$ meson lifetime (see
Sec.~\ref{sec:life_mix}) and the semileptonic width~\cite{PDG_2014}.
We use all results published by \babar\ in Ref.~\cite{Lees:2011fv}, since the 
statistical correlations are given. 
To make use of the theoretical calculations of Ref.~\cite{ref:BLL}, we restrict the
kinematic range of the invariant mass of the hadronic system, $M_X$, 
and the square of the invariant mass of the lepton pair, $q^2$. This reduces the size of the data
sample significantly, but also the theoretical uncertainty, as stated by the
authors~\cite{ref:BLL}.
The dependence of the quoted error on the measured value for each source of uncertainty 
is taken into account in the calculation of the averages.

It has been first suggested by Neubert~\cite{Neubert:1993um} and later detailed by Leibovich, 
Low, and Rothstein (LLR)~\cite{Leibovich:1999xf} and Lange, Neubert and Paz (LNP)~\cite{Lange:2005qn}, 
that the uncertainty of
the leading shape functions can be eliminated by comparing inclusive rates for
$\B\to X_u\ell^+\nul$ decays with the inclusive photon spectrum in $\B\to X_s\gamma$,
based on the assumption that the shape functions for transitions to light
quarks, $u$ or $s$, are the same at first order.
However, shape function uncertainties are only eliminated at the leading order
and they still enter via the signal models used for the determination of efficiency. 

%% For completeness, we provide at the end of this section a comparison of the results using 
%% calculations with reduced dependence on the shape function, as just
%% introduced, with our averages based on different theoretical approaches.
%% Results are presented by \babar\ in Ref.\cite{Aubert:2006qi} using the LLR prescription. 
%% In another work (Ref.~\cite{Golubev:2007cs}), \vub\ was extracted from the 
%% endpoint spectrum of $\B\to X_u\ell^+\nul$ from \babar~\cite{ref:babar-endpoint}, 
%% using several theoretical approaches with reduced dependence on the shape function.
%% In both cases, the photon energy spectrum in the 
%% rest frame of the $B$-meson by \babar~\cite{Aubert:2005cua} has been used.

In the following, the different theoretical methods and the resulting averages are described.
A recent \babar measurement of the inclusive electron spectrum~\cite{TheBABAR:2016lja} was released at the time of this writing and could 
not be included in the averages. 

\begin{table}[!htb]
\caption{\label{tab:BFbulnu}
Summary of measurements  of partial branching
fractions for $B\rightarrow X_u \ell^+ \nu_{\ell}$ decays.
The errors quoted on $\Delta\cbf$ correspond to
statistical and systematic uncertainties.
$E_e$ is the electron
energy in the $B$~rest frame, $p^*$ the lepton momentum in the
$B$~frame and $m_X$ is the invariant mass of the hadronic system. The
light-cone momentum $P_+$ is defined in the $B$ rest frame as
$P_+=E_X-|\vec p_X|$.
%%%%The statistical correlations between the analysis are given where applicable. 
The $s_\mathrm{h}^{\mathrm{max}}$ variable is described in Refs.~\cite{ref:shmax,ref:babar-elq2}. }
\begin{center}
\begin{small}
\begin{tabular}{|llcl|}
\hline
Measurement & Accepted region &  $\Delta\cbf [10^{-4}]$ & Notes\\
\hline\hline
CLEO~\cite{ref:cleo-endpoint}
& $E_e>2.1\,\gev$ & $3.3\pm 0.2\pm 0.7$ &  \\ 
\babar~\cite{ref:babar-elq2}
%%%%CB PUT NEW PREL BABAR& $E_e>2.0\,\gev$, $s_\mathrm{h}^{\mathrm{max}}<3.5\,\mathrm{GeV^2}$ & $4.4\pm 0.4\pm 0.4$ & \\
& $E_e>2.0~\gev$, $s_\mathrm{h}^{\mathrm{max}}<3.5\,\mathrm{GeV}^2$ & $4.4\pm 0.4\pm 0.4$ & \\
\babar~\cite{ref:babar-endpoint}
& $E_e>2.0\,\gev$  & $5.7\pm 0.4\pm 0.5$ & \\
Belle~\cite{ref:belle-endpoint}
& $E_e>1.9\,\gev$  & $8.5\pm 0.4\pm 1.5$ & \\
\babar~\cite{Lees:2011fv}
& $M_X<1.7\,\gevcc, q^2>8\,\gevgevcccc$ & $6.9\pm 0.6\pm 0.4$ & 
%(52\%,46\%) correlation with \babar\ ($p^*_{\ell} > 1~\gev/c$, $p^*_{\ell} > 1.3~\gev/c$) analyses
\\
Belle~\cite{ref:belle-mxq2Anneal}
& $M_X<1.7\,\gevcc, q^2>8\,\gevgevcccc$ & $7.4\pm 0.9\pm 1.3$ & \\
Belle~\cite{ref:belle-mx}
& $M_X<1.7\,\gevcc, q^2>8\,\gevgevcccc$ & $8.5\pm 0.9\pm 1.0$ & used only in BLL average\\
\babar~\cite{Lees:2011fv}
& $P_+<0.66\,\gev$  & $9.9\pm 0.9\pm 0.8 $ & 
%%(46\%, 78\%, 61\%) correlations with \babar\ ($(M_X-q^2)$, $p^*_{\ell} > 1~\gev/c$, $p^*_{\ell} > 1.3~\gev/c$) 
%% analyses
\\
%%%%BELLE~\cite{ref:belle-mx}
%%%%& $P_+<0.66\,\gev$  & $11.0\pm 1.0\pm 1.6$ & not used in averages\\ 
\babar~\cite{Lees:2011fv}
& $M_X<1.7\,\gevcc$ & $11.6\pm 1.0\pm 0.8 $ &
%%(86\%, 55\%, 94\%, 73\%) correlations with \babar\ ($P_+$, $(M_X-q^2)$,
%%$p^*_{\ell} > 1~\gev/c$, $p^*_{\ell} > 1.3~\gev/c$)  analyses 
\\ 
\babar~\cite{Lees:2011fv}
& $M_X<1.55\,\gevcc$ & $10.9\pm 0.8\pm 0.6 $ & 
%%(74\%, 77\%, 50\%, 72\%, 57\%) correlations with \babar\ ($P_+$, $M_X<1.7\,\gevcc$, $(M_X-q^2)$, 
%%$p^*_{\ell} > 1~\gev/c$, $p^*_{\ell} > 1.3~\gev/c$)  analyses 
\\ 
%%%%BELLE~\cite{ref:belle-mx}
%%%%& $M_X<1.7\,\gevcc$ & $12.3\pm 1.1\pm 1.2$ & not used in averages\\ 
Belle~\cite{ref:belle-multivariate}
& ($M_X, q^2$) fit, $p^*_{\ell} > 1~\gev/c$ & $19.6\pm 1.7\pm 1.6$ & \\
\babar~\cite{Lees:2011fv}
& ($M_X, q^2$) fit, $p^*_{\ell} > 1~\gev/c$  & $18.2\pm 1.3\pm 1.5$ & 
%%74\% correlation with $p^*_{\ell} > 1.3~\gev/c$ analysis\\ \hline
\\ 
\babar~\cite{Lees:2011fv}
& $p^*_{\ell} > 1.3~\gev/c$  & $15.5\pm 1.3\pm 1.4$ & 
%%67\% correlation with \babar\ $P_+$ analysis 
\\ \hline
\end{tabular}\\
\end{small}
\end{center}
\end{table}

\subsubsection{BLNP}
\begin{figure}[!ht]
 \begin{center}
  \unitlength1.0cm % coordinates in cm
  \begin{picture}(14.,10.0)  %ys(25.,6.0)
   \put( -1.5,  0.0){\includegraphics[width=9.4cm]{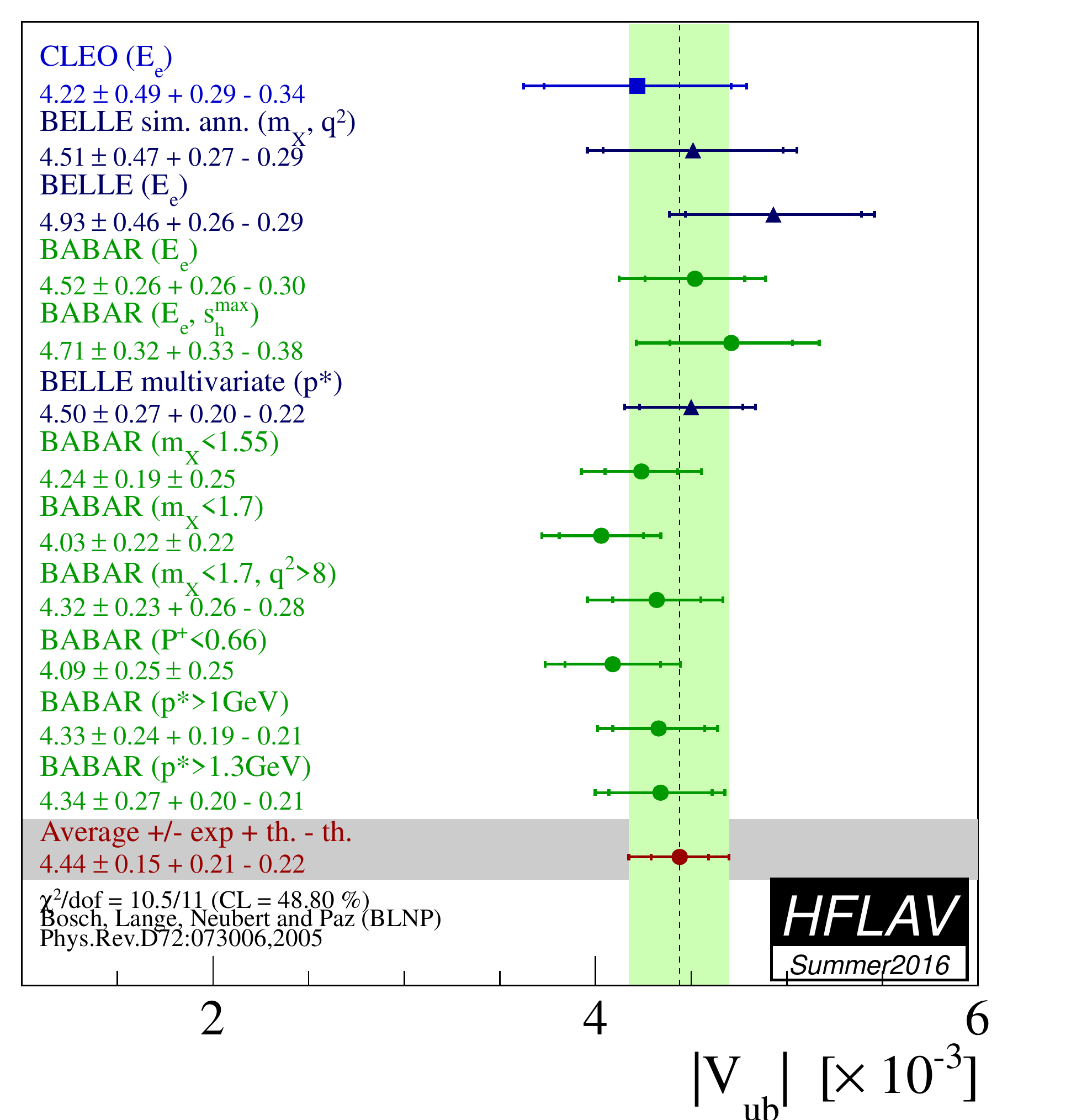}
   }
   \put(  7.4,  0.0){\includegraphics[width=9.4cm]{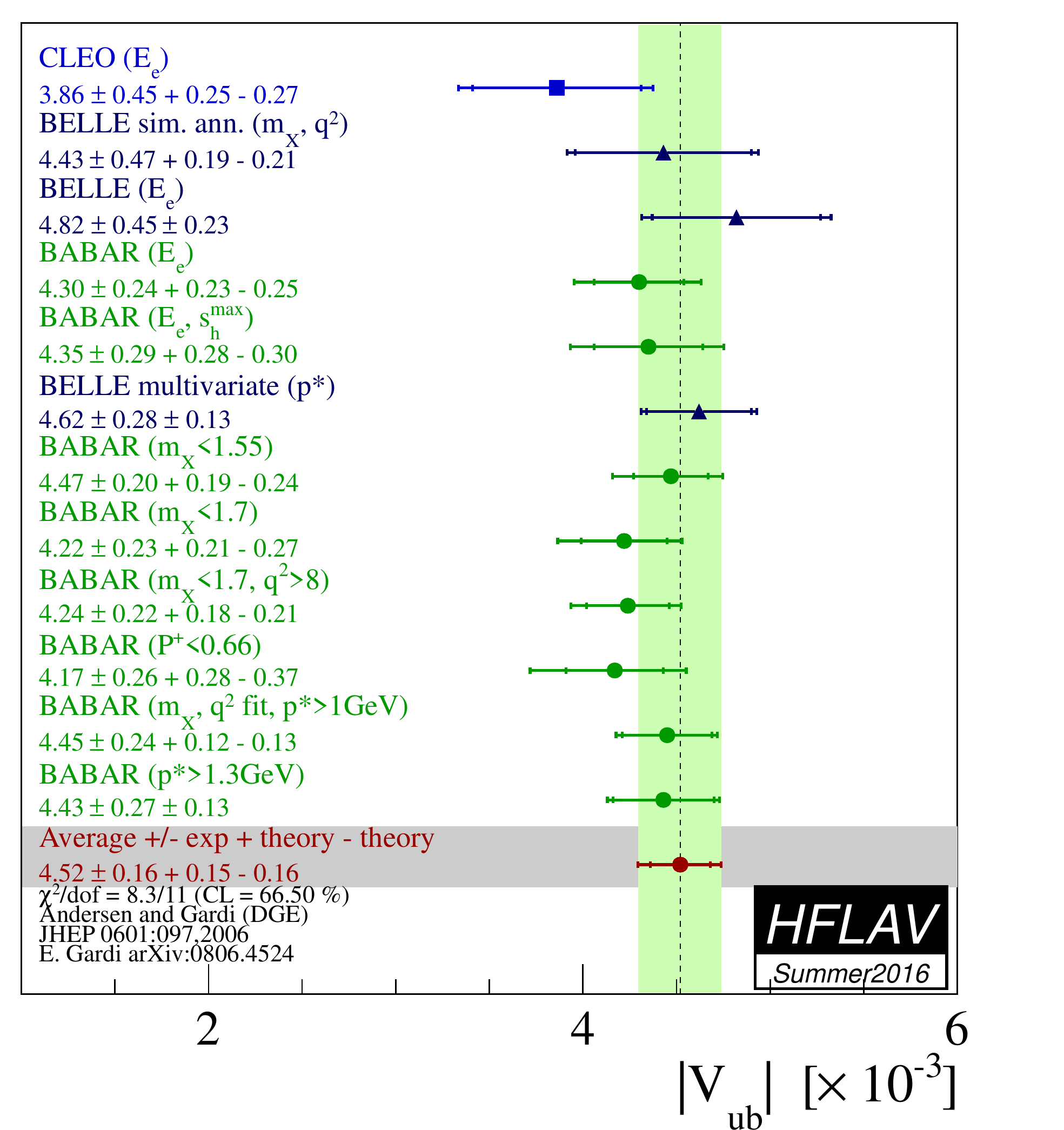}
   }
   \put(  6.0,  8.7){{\large\bf a)}}
   \put( 14.8,  8.7){{\large\bf b)}}
  \end{picture}
  \caption{Measurements of $\vub$ from inclusive semileptonic decays 
and their average based on the BLNP (a) and DGE (b) prescription. The
labels indicate the variabless and selections used to define the
signal regions in the different analyses.
%, where $E_e$ is the electron
%energy in the $B$~rest frame, $p^*$ the lepton momentum in the
%$B$~frame and $m_X$ is the invariant mass of the hadronic system. The
%light-cone momentum $P_+$ is defined in the $B$ rest frame as
%$P_+=E_X-|\vec p_X|$.
} \label{fig:BLNP_DGE}
 \end{center}
\end{figure}

Bosch, Lange, Neubert and Paz (BLNP)~\cite{ref:BLNP,
% removed missing reference TJG 13/5/2012
%  ref:Neubert-new-1,ref:Neubert-new-2,ref:Neubert-new-3,ref:Neubert-new-4}
  ref:Neubert-new-1,ref:Neubert-new-2,ref:Neubert-new-3}
provide theoretical expressions for the triple
differential decay rate for $B\to X_u \ell^+ \nul$ events, incorporating all known
contributions, whilst smoothly interpolating between the 
``shape-function region'' of large hadronic
energy and small invariant mass, and the ``OPE region'' in which all
hadronic kinematical variables scale with the $b$-quark mass. BLNP assign
uncertainties to the $b$-quark mass, which enters through the leading shape function, 
to sub-leading shape function forms, to possible weak annihilation
contribution, and to matching scales. 
The BLNP calculation uses the shape function renormalization scheme; the heavy quark parameters determined  
from the global fit in the kinetic scheme, described in \ref{globalfitsKinetic}, were therefore 
translated into the shape function scheme by using a prescription by Neubert 
\cite{Neubert:2004sp,Neubert:2005nt}. The resulting parameters are 
$m_b({\rm SF})=(4.582 \pm 0.023 \pm 0.018)~\gev$, 
$\mu_\pi^2({\rm SF})=(0.202 \pm 0.089 ^{+0.020}_{-0.040})~\gevcc$, 
where the second uncertainty is due to the scheme translation. 
The extracted values of \vub\, for each measurement along with their average are given in
Table~\ref{tab:bulnu} and illustrated in Fig.~\ref{fig:BLNP_DGE}(a). 
The total uncertainty is $^{+5.8}_{-6.0}\%$ and is due to:
statistics ($^{+2.1}_{-2.1}\%$),
detector effects ($^{+1.7}_{-1.8}\%$),
$B\to X_c \ell^+ \nul$ model ($^{+1.2}_{-1.2}\%$),
$B\to X_u \ell^+ \nul$ model ($^{+1.8}_{-1.7}\%$),
heavy quark parameters ($^{+2.6}_{-2.6}\%$),
SF functional form ($^{+0.2}_{-0.3}\%$),
sub-leading shape functions ($^{+0.6}_{-0.7}\%$),
BLNP theory: matching scales $\mu,\mu_i,\mu_h$ ($^{+3.8}_{-3.7}\%$), and
weak annihilation ($^{+0.0}_{-1.4}\%$).
The error assigned to the matching scales 
is the source of the largest uncertainty, while the
uncertainty due to HQE parameters ($b$-quark mass and $\mu_\pi^2)$ is second. The uncertainty due to 
weak annihilation has been assumed to be asymmetric, \ie\ it only tends to decrease \vub.

\begin{table}[!htb]
\caption{\label{tab:bulnu}
Summary of input parameters used by the different theory calculations,
corresponding inclusive determinations of $\vub$ and their average.
The errors quoted on \vub\ correspond to
experimental and theoretical uncertainties, respectively.}
\begin{center}
\resizebox{0.99\textwidth}{!}{
\begin{tabular}{|lccccc|}
\hline
 & BLNP &DGE & GGOU & ADFR &BLL \\
\hline\hline
\multicolumn{6}{|c|}{Input parameters}\\ \hline
scheme & SF           & $\overline{MS}$ & kinetic &  $\overline{MS}$ & $1S$ \\ 
Ref.       & \cite{Neubert:2004sp,Neubert:2005nt} & Ref.~\cite{ref:DGE} & 
see Sec.~\ref{globalfitsKinetic}  & Ref.~\cite{Aglietti:2006yb} & Ref.~\cite{ref:BLL} \\
%%%%%       & (only $b\to c \ell\nu$ & & ($b\to c \ell\nu$ + $b\to s\gamma$ &  & \\
%%%%%       & moments) & & moments) & &  \\
$m_b$ (GeV)           & 4.582 $\pm$ 0.026 & 4.188 $\pm 0.043$ & 4.554 $\pm 0.018$ & 4.188 $\pm 0.043$ & 4.704 $\pm 0.029$ \\
$\mu_\pi^2$ (GeV$^2$) & 0.145 $^{+0.091}_{-0.097}$ & -                 & 0.414 $\pm 0.078$ & - &  - \\
\hline\hline
Ref. & \multicolumn{5}{c|}{$|V_{ub}|$ values $[10^{-3}]$}\\ 
\hline
CLEO $E_e$~\cite{ref:cleo-endpoint} &
$4.22\pm 0.49 ^{+0.29}_{-0.34}$ &
$3.86\pm 0.45 ^{+0.25}_{-0.27}$ &
$4.23\pm 0.49 ^{+0.22}_{-0.31}$ &
$3.42\pm 0.40 ^{+0.17}_{-0.17}$ &
- \\

Belle $M_X, q^2$~\cite{ref:belle-mxq2Anneal}&
$4.51\pm 0.47 ^{+0.27}_{-0.29}$ &
$4.43\pm 0.47 ^{+0.19}_{-0.21}$ &
$4.52\pm 0.48 ^{+0.25}_{-0.28}$ &
$3.93\pm 0.41 ^{+0.18}_{-0.17}$ &
$4.68\pm 0.49 ^{+0.30}_{-0.30}$ \\

Belle $E_e$~\cite{ref:belle-endpoint}&
$4.93\pm 0.46 ^{+0.26}_{-0.29}$ &
$4.82\pm 0.45 ^{+0.23}_{-0.23}$ &
$4.95\pm 0.46 ^{+0.16}_{-0.21}$ &
$4.48\pm 0.42 ^{+0.20}_{-0.20}$ &
-\\

\babar $E_e$~\cite{ref:babar-endpoint}&
$4.52\pm 0.26 ^{+0.26}_{-0.30}$ &
$4.30\pm 0.24 ^{+0.23}_{-0.25}$ &
$4.52\pm 0.26 ^{+0.17}_{-0.24}$ &
$3.93\pm 0.22 ^{+0.20}_{-0.20}$ &
-\\

\babar $E_e,s_\mathrm{h}^{\mathrm{max}}$~\cite{ref:babar-elq2}&
$4.71\pm 0.32 ^{+0.33}_{-0.38}$ &
$4.35\pm 0.29 ^{+0.28}_{-0.30}$ &
- &
$3.81\pm 0.19 ^{+0.19}_{-0.18}$ &
%%%%%%%?!?!?!?!?$4.71\pm 0.50 ^{+0.35}_{-0.35}$ \\
 \\ 

Belle $p^*_{\ell}$, $(M_X,q^2)$ fit~\cite{ref:belle-multivariate}&
$4.50\pm 0.27 ^{+0.20}_{-0.22}$ &
$4.62\pm 0.28 ^{+0.13}_{-0.13}$ &
$4.62\pm 0.28 ^{+0.09}_{-0.10}$ &
$4.50\pm 0.30 ^{+0.20}_{-0.20}$ &
- \\

\babar $M_X$~\cite{Lees:2011fv}&
$4.24\pm 0.19 ^{+0.25}_{-0.25}$ &
$4.47\pm 0.20 ^{+0.19}_{-0.24}$ &
$4.30\pm 0.20 ^{+0.20}_{-0.21}$ &
$3.83\pm 0.18 ^{+0.20}_{-0.19}$ &
- \\
\babar $M_X$~\cite{Lees:2011fv}&
$4.03\pm 0.22 ^{+0.22}_{-0.22}$ &
$4.22\pm 0.23 ^{+0.21}_{-0.27}$ &
$4.10\pm 0.23 ^{+0.16}_{-0.17}$ &
$3.75\pm 0.21 ^{+0.18}_{-0.18}$ &
- \\

\babar $M_X,q^2$~\cite{Lees:2011fv}&
$4.32\pm 0.23 ^{+0.26}_{-0.28}$  &
$4.24\pm 0.22 ^{+0.18}_{-0.21}$  &
$4.33\pm 0.23 ^{+0.24}_{-0.27}$  &
$3.75\pm 0.20 ^{+0.17}_{-0.17}$  &
$4.50\pm 0.24 ^{+0.29}_{-0.29}$ \\

\babar $P_+$~\cite{Lees:2011fv}&
$4.09\pm 0.25 ^{+0.25}_{-0.25}$  &
$4.17\pm 0.25 ^{+0.28}_{-0.37}$  &
$4.25\pm 0.26 ^{+0.26}_{-0.27}$  &
$3.57\pm 0.22 ^{+0.19}_{-0.18}$  &
- \\

\babar $p^*_{\ell}$, $(M_X,q^2)$ fit~\cite{Lees:2011fv}&
$4.33\pm 0.24 ^{+0.19}_{-0.21}$  &
$4.45\pm 0.24 ^{+0.12}_{-0.13}$  &
$4.44\pm 0.24 ^{+0.09}_{-0.10}$  &
$4.33\pm 0.24 ^{+0.19}_{-0.19}$  &
- \\

\babar $p^*_{\ell}$~\cite{Lees:2011fv}&
$4.34\pm 0.27 ^{+0.20}_{-0.21}$  &
$4.43\pm 0.27 ^{+0.13}_{-0.13}$  &
$4.43\pm 0.27 ^{+0.09}_{-0.11}$  &
$4.28\pm 0.27 ^{+0.19}_{-0.19}$  &
- \\

Belle $M_X,q^2$~\cite{ref:belle-mx}&
- &
- &
- &
- &
$5.01\pm 0.39 ^{+0.32}_{-0.32}$ \\
\hline
Average &
$4.44\pm 0.15 ^{+0.21}_{-0.22}$ &
$4.52\pm 0.16 ^{+0.15}_{-0.16}$ &
$4.52\pm 0.15 ^{+0.11}_{-0.14}$ &
$4.08\pm 0.13 ^{+0.18}_{-0.12}$ &
$4.62\pm 0.20 ^{+0.29}_{-0.29}$ \\
\hline
\end{tabular}
}
\end{center}
\end{table}

%\begin{figure}
%\begin{center}
%\includegraphics[width=0.48\textwidth]{figures/slb/vub_clnu_mc_twomu_asym_BLNP.pdf}
%\end{center}
%\caption{Measurements of $\vub$ from inclusive semileptonic decays 
%and their average based on the BLNP prescription.
%``$E_e$'', ``$M_X$'', ``$(M_X,q^2)$'', ``$P^+$'', ``$p^*$ and ``($E_e,s^{max}_h$)'' indicate the 
%distributions and cuts used for the measurement of the partial decay rates.}
%\label{fig:BLNP}
%\end{figure}

\begin{figure}[!ht]
 \begin{center}
  \unitlength1.0cm % coordinates in cm
  \begin{picture}(14.,10.0)  %ys(25.,6.0)
   \put( -1.5,  0.0){\includegraphics[width=9.4cm]{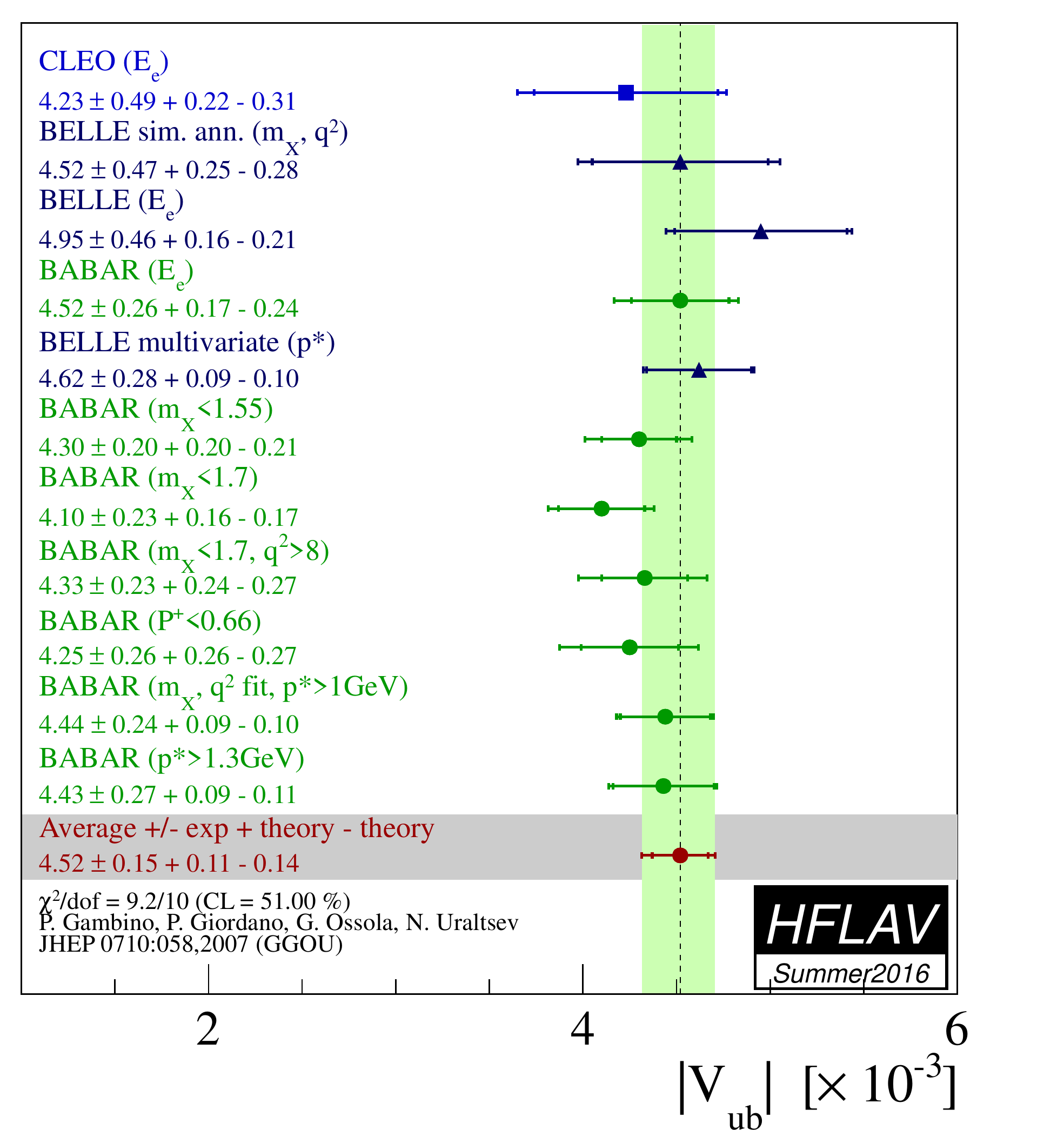}
   }
   \put(  7.4,  0.0){\includegraphics[width=9.4cm]{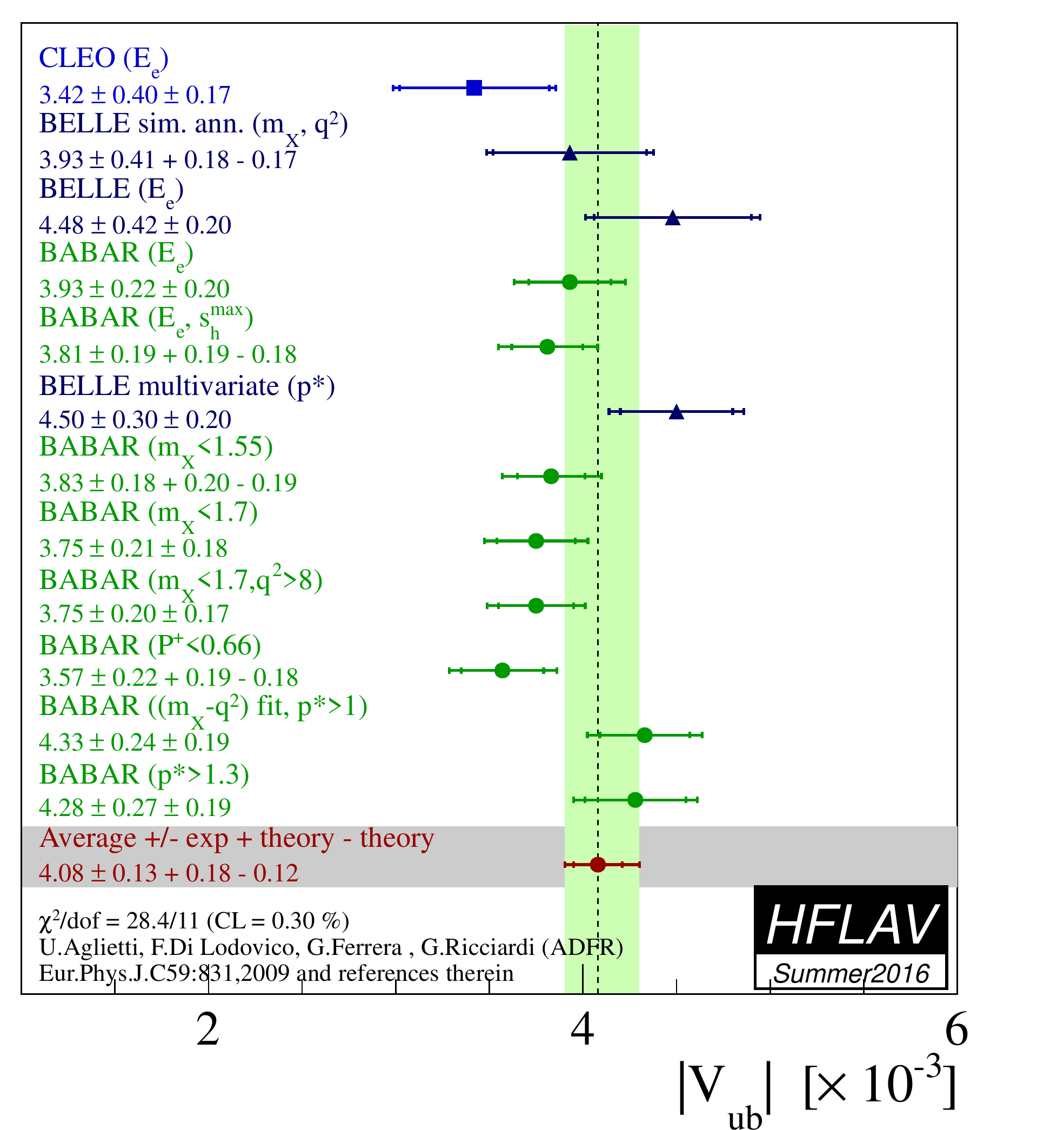}
   }
   \put(  6.0,  8.7){{\large\bf a)}}
   \put( 14.8,  8.7){{\large\bf b)}}
  \end{picture}
  \caption{Measurements of $\vub$ from inclusive semileptonic decays 
and their average based on the GGOU (a) and ADFR (b) prescription. The
labels indicate the variables and selections used to define the
signal regions in the different analyses
%, where $E_e$ is the electron
%energy in the $B$~rest frame, $p^*$ the lepton momentum in the
%$B$~frame and $m_X$ is the invariant mass of the hadronic system. The
%light-cone momentum $P_+$ is defined in the $B$ rest frame as
%$P_+=E_X-|\vec p_X|$
.} \label{fig:GGOU_ADFR}
 \end{center}
\end{figure}

\subsubsection{DGE}
Andersen and Gardi (Dressed Gluon Exponentiation, DGE)~\cite{ref:DGE} provide
a framework where the on-shell $b$-quark calculation, converted into hadronic variables, is
directly used as an approximation to the meson decay spectrum without
the use of a leading-power non-perturbative function (or, in other words,
a shape function). The on-shell mass of the $b$-quark within the $B$-meson ($m_b$) is
required as input. 
The DGE calculation uses the $\overline{MS}$ renormalization scheme. The heavy quark parameters determined  
from the global fit in the kinetic scheme, described in \ref{globalfitsKinetic}, were therefore 
translated into the $\overline{MS}$ scheme by using a calculation by Gardi, giving 
$m_b({\overline{MS}})=(4.188 \pm 0.043)~\gev$.
The extracted values
of \vub\, for each measurement along with their average are given in
Table~\ref{tab:bulnu} and illustrated in Fig.~\ref{fig:BLNP_DGE}(b).
The total error is $^{+4.8}_{-4.8}\%$, whose breakdown is:
statistics ($^{+1.9}_{-1.9}\%$),
detector effects ($^{+1.7}_{-1.7}\%$),
$B\to X_c \ell^+ \nul$ model ($^{+1.3}_{-1.3}\%$),
$B\to X_u \ell^+ \nul$ model ($^{+2.1}_{-1.7}\%$),
strong coupling $\alpha_s$ ($^{+0.5}_{-0.5}\%$),
$m_b$ ($^{+3.2}_{-2.9}\%$),
%%%%%%%%%%%%%%?!?!?!?spectral fraction ($m_b$) ($^{+3.0}_{-3.3}\%$),
%%%%%%%%%%%%%%?!?!?!?!total semileptonic width ($m_b$) ($^{+3.0}_{-3.0}\%$),
weak annihilation ($^{+0.0}_{-1.8}\%$),
matching scales in DGE ($^{+0.5}_{-0.4}\%$).
The largest contribution to the total error is due to the effect of the uncertainty 
on $m_b$. 
%%%%%%%%%%%%%%on the prediction of the event rate, closely followed by the 
%%%%%%%%%%%%%%specific theory error on overall DGE and the total semileptonic decay width.
The uncertainty due to 
weak annihilation has been assumed to be asymmetric, \ie\ it only tends to decrease \vub.

%\begin{figure}
%\begin{center}
%\includegraphics[width=0.48\textwidth]{figures/slb/vub_clnu_mc_asym_DGE.pdf}
%\end{center}
%\caption{Measurements of $\vub$ from inclusive semileptonic decays 
%and their average based on the DGE prescription.
%``$E_e$'', ``$M_X$'', ``$(M_X,q^2)$'' `$P^+$'', ``$p^*$ and ``($E_e,s^{max}_h$)'' indicate the 
%analysis type and applied cut.}
%\label{fig:DGE}
%\end{figure}

\subsubsection{GGOU}
\label{subsec:ggou}
Gambino, Giordano, Ossola and Uraltsev (GGOU)~\cite{Gambino:2007rp} 
compute the triple differential decay rates of $B \to X_u \ell^+ \nul$, 
including all perturbative and non--perturbative effects through $O(\alphas^2 \beta_0)$ 
and $O(1/m_b^3)$. 
The Fermi motion is parameterized in terms of a single light--cone function 
for each structure function and for any value of $q^2$, accounting for all subleading effects. 
The calculations are performed in the kinetic scheme, a framework characterized by a Wilsonian 
treatment with a hard cutoff $\mu \sim 1~\gev$.
GGOU have not included calculations for the ``($E_e,s^{max}_h$)'' analysis~\cite{ref:babar-elq2}. 
The heavy quark parameters determined  
from the global fit in the kinetic scheme, described in \ref{globalfitsKinetic}, are used as inputs: 
$m_b^{kin}=(4.554 \pm 0.018)~\gev$, 
$\mu_\pi^2=(0.464 \pm 0.076)~\gevcc$. 
The extracted values
of \vub\, for each measurement along with their average are given in
Table~\ref{tab:bulnu} and illustrated in Fig.~\ref{fig:GGOU_ADFR}(a).
The total error is $^{+4.2}_{-4.6}\%$ whose breakdown is:
statistics ($^{+2.0}_{-2.0}\%$),
detector effects ($^{+1.7}_{-1.7}\%$),
$B\to X_c \ell^+ \nul$ model ($^{+1.3}_{-1.3}\%$),
$B\to X_u \ell^+ \nul$ model ($^{+1.8}_{-1.8}\%$),
$\alpha_s$, $m_b$ and other non--perturbative parameters ($^{+1.4}_{-1.4}\%$), 
higher order perturbative and non--perturbative corrections ($^{+1.5}_{-1.5}\%$), 
modelling of the $q^2$ tail
%and choice of the scale $q^{2*}$
($^{+1.2}_{-1.2}\%$), 
weak annihilations matrix element ($^{+0.0}_{-1.9}\%$), 
functional form of the distribution functions ($^{+0.2}_{-0.2}\%$).  
The leading uncertainties
on  \vub\ are both from theory, and are due to perturbative and non--perturbative
parameters and the modelling of the $q^2$ tail.
% and choice of the scale $q^{2*}$. 
The uncertainty due to 
weak annihilation has been assumed to be asymmetric, \ie\ it only tends to decrease \vub.

%\begin{figure}
%\begin{center}
%\includegraphics[width=0.48\textwidth]{figures/slb/vub_clnu_mc_GGOU.pdf}
%\end{center}
%\caption{Measurements of $\vub$ from inclusive semileptonic decays 
%and their average based on the GGOU prescription.
%``$E_e$'', ``$M_X$'', ``$(M_X,q^2)$'' `$P^+$'', ``$p^*$ and ``($E_e,s^{max}_h$)''  indicate the
%analysis type and applied cut.}
%\label{fig:GGOU}
%\end{figure}

\subsubsection{ADFR}
Aglietti, Di Lodovico, Ferrera and Ricciardi (ADFR)~\cite{Aglietti:2007ik}
use an approach to extract \vub, that makes use of the ratio
of the  $B \to X_c \ell^+ \nul$ and $B \to X_u \ell^+ \nul$ widths. 
The normalized triple differential decay rate for 
$B \to X_u \ell^+ \nul$~\cite{Aglietti:2006yb,Aglietti:2005mb, Aglietti:2005bm, Aglietti:2005eq}
is calculated with a model based on (i) soft--gluon resummation 
to next--to--next--leading order and (ii) an effective QCD coupling without
Landau pole. This coupling is constructed by means of an extrapolation to low
energy of the high--energy behaviour of the standard coupling. More technically,
an analyticity principle is used.
The lower cut on the electron energy for the endpoint analyses is 2.3~GeV~\cite{Aglietti:2006yb}.
The ADFR calculation uses the $\overline{MS}$ renormalization scheme; the heavy quark parameters determined  
from the global fit in the kinetic scheme, described in \ref{globalfitsKinetic}, were therefore 
translated into the $\overline{MS}$ scheme by using a calculation by Gardi, giving 
$m_b({\overline{MS}})=(4.188 \pm 0.043)~\gev$.
The extracted values
of \vub\, for each measurement along with their average are given in
Table~\ref{tab:bulnu} and illustrated in Fig.~\ref{fig:GGOU_ADFR}(b).
The total error is $^{+5.5}_{-5.5}\%$ whose breakdown is:
statistics ($^{+1.9}_{-1.9}\%$),
detector effects ($^{+1.7}_{-1.7}\%$),
$B\to X_c \ell^+ \nul$ model ($^{+1.3}_{-1.3}\%$),
$B\to X_u \ell^+ \nul$ model ($^{+1.3}_{-1.3}\%$),
$\alpha_s$ ($^{+1.1}_{-1.0}\%$), 
$|V_{cb}|$ ($^{+1.9}_{-1.9}\%$), 
$m_b$ ($^{+0.7}_{-0.7}\%$), 
$m_c$ ($^{+1.3}_{-1.3}\%$), 
semileptonic branching fraction ($^{+0.8}_{-0.7}\%$), 
theory model ($^{+3.6}_{-3.6}\%$).
The leading uncertainty is due to the theory model.

%\begin{figure}
%\begin{center}
%\includegraphics[width=0.48\textwidth]{figures/slb/vub_clnu_mc_ADFR.pdf}
%\end{center}
%\caption{Measurements of $\vub$ from inclusive semileptonic decays 
%and their average based on the ADFR prescription.
%``$E_e$'', ``$M_X$'', ``$(M_X,q^2)$'' `$P^+$'', ``$p^*$ and ``($E_e,s^{max}_h$)'' indicate the 
%analysis type and applied cut.}
%\label{fig:AC}
%\end{figure}

\subsubsection{BLL}
\begin{figure}
\begin{center}
\includegraphics[width=0.52\textwidth]{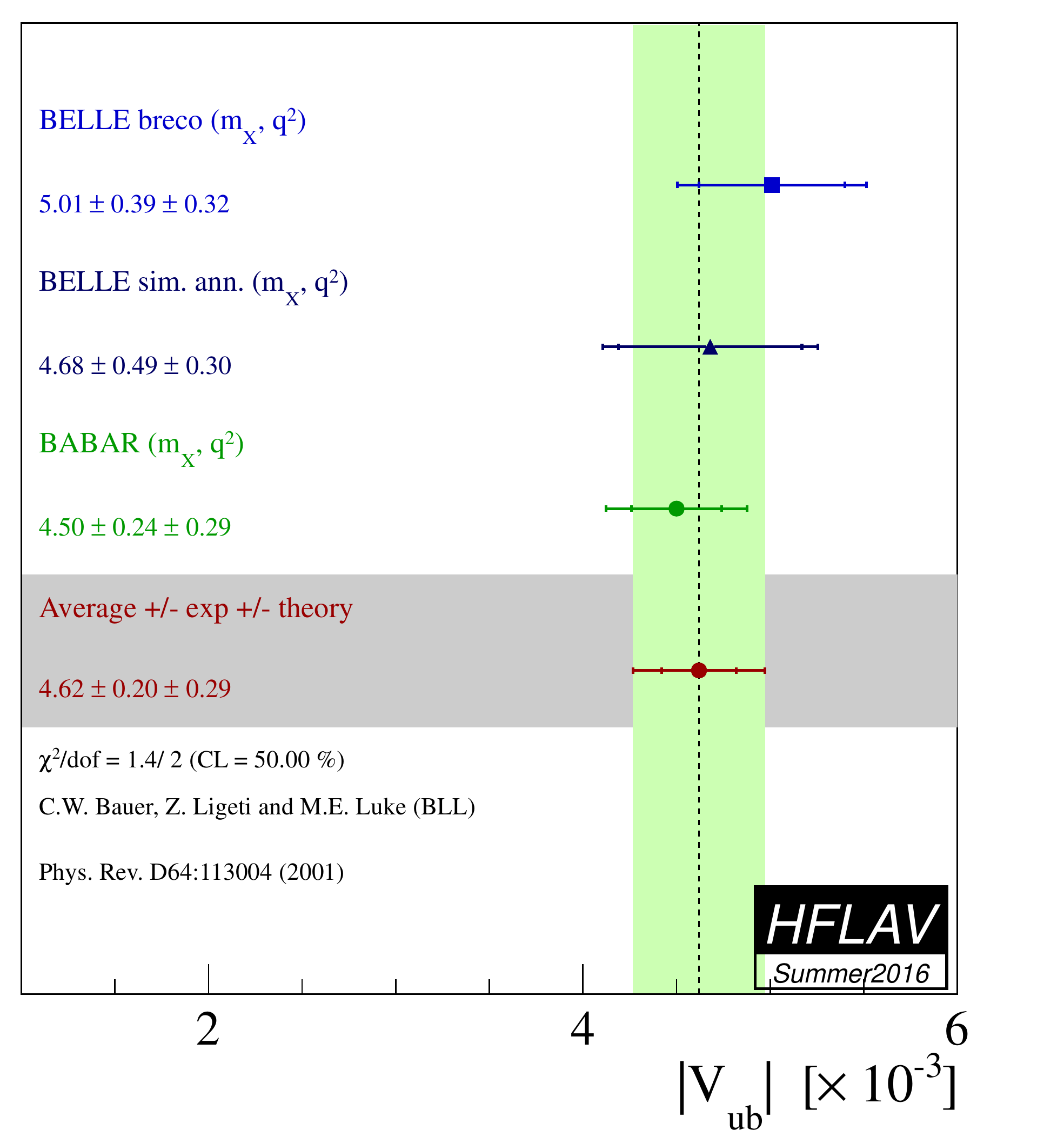}
\end{center}
\caption{Measurements of $\vub$ from inclusive semileptonic decays 
and their average in the BLL prescription.
%``$(M_X, q^2)$'' indicates the analysis type.
}
\label{fig:BLL}
\end{figure}

Bauer, Ligeti, and Luke (BLL)~\cite{ref:BLL} give a
HQET-based prescription that advocates combined cuts on the dilepton invariant mass, $q^2$,
and hadronic mass, $m_X$, to minimise the overall uncertainty on \vub.
In their reckoning a cut on $m_X$ only, although most efficient at
preserving phase space ($\sim$80\%), makes the calculation of the partial
rate untenable due to uncalculable corrections
to the $b$-quark distribution function or shape function. These corrections are
suppressed if events in the low $q^2$ region are removed. The cut combination used
in measurements is $M_x<1.7~\gevcc$ and $q^2 > 8~\gevgevcccc$.  
The extracted values
of \vub\, for each measurement along with their average are given in
Table~\ref{tab:bulnu} and illustrated in Fig.~\ref{fig:BLL}.
The total error is $^{+7.7}_{-7.7}\%$ whose breakdown is:
statistics ($^{+3.3}_{-3.3}\%$),
detector effects ($^{+3.0}_{-3.0}\%$),
$B\to X_c \ell^+ \nul$ model ($^{+1.6}_{-1.6}\%$),
$B\to X_u \ell^+ \nul$ model ($^{+1.1}_{-1.1}\%$),
spectral fraction ($m_b$) ($^{+3.0}_{-3.0}\%$),
perturbative approach: strong coupling $\alpha_s$ ($^{+3.0}_{-3.0}\%$),
residual shape function ($^{+2.5}_{-2.5}\%$),
third order terms in the OPE ($^{+4.0}_{-4.0}\%$). 
The leading
uncertainties, both from theory, are due to residual shape function
effects and third order terms in the OPE expansion. The leading
experimental uncertainty is due to statistics. 

\subsubsection{Summary}
The averages presented in several different
frameworks %and results by Golubev, Luth and Skovpen~\cite{Golubev:2007cs},
%% based on prescriptions by LLR~\cite{Leibovich:1999xf} and LNP~\cite{Lange:2005qn} 
%% to reduce the leading shape function uncertainties 
are presented in 
Table~\ref{tab:vubcomparison}.
In summary, we recognize that the experimental and theoretical uncertainties play out
differently between the schemes and the theoretical assumptions for the
theory calculations are different. Therefore, it is difficult to perform an average 
between the various determinations of \vub. 
Since the methodology is similar to that used to determine the 
inclusive \vcb\ average, we choose to quote as reference value the average determined 
by the GGOU calculation, which gives \vub $= (4.52 \pm 0.15 ^{+0.11}_{-0.14}) \times 10^{-3}$. 

\begin{table}[!htb]
\caption{\label{tab:vubcomparison}
Summary of inclusive determinations of $\vub$.
The errors quoted on \vub\ correspond to experimental and theoretical uncertainties.
%% , except for the last two 
%% measurements where the errors are due to the \babar\ endpoint analysis, the \babar $b\to s\gamma$ analysis~\cite{Aubert:2006qi}, 
%% the theoretical errors. For the next-to-last measurement, the fourth error is due to the uncertainty on $V_{ts}$.
}
\begin{center}
\begin{small}
\begin{tabular}{|lc|}
\hline
Framework
&  $\Vub [10^{-3}]$\\
\hline\hline
BLNP
& $4.44 \pm 0.15 ^{+0.21}_{-0.22}$ \\ 
DGE
& $4.52 \pm 0.16 ^{+0.15}_{-0.16}$ \\
GGOU
& $4.52 \pm 0.15 ^{+0.11}_{-0.14}$ \\
ADFR
& $4.08 \pm 0.13 ^{+0.18}_{-0.12}$ \\
BLL ($m_X/q^2$ only)
& $4.62 \pm 0.20 \pm 0.29$ \\ 
%% LLR (\babar)~\cite{Aubert:2006qi}
%% & $4.43 \pm 0.45 \pm 0.29$ \\
%% LLR (\babar)~\cite{Golubev:2007cs}
%% & $4.28 \pm 0.29 \pm 0.29 \pm 0.26 \pm0.28$ \\
%% LNP (\babar)~\cite{Golubev:2007cs}
%% & $4.40 \pm 0.30 \pm 0.41 \pm 0.23$ \\
\hline
\end{tabular}\\
\end{small}
\end{center}
\end{table}

%
% ======================================================================
% B --> D(*) tau nu_tau
% =====================================================================
\subsection{$B\to D^{(*)}\tau \nu_\tau$ decays}
\label{slbdecays_b2dtaunu}
% -------------------------------------------
%This section contains a summary of the existing measurements
%of $\bar B\to D^*\tau^-\bar\nu_\tau$ and $\bar B\to D\tau^-\bar\nu_\tau$
%branching ratios. 

%The leptonic and semileptonic decays with $\tau$ in the final state 
%are probes of physics beyond the SM. In the SM these decays proceed via the $W$ emission
%diagrams. In models with extended Higgs sectors, such as the Two Higgs Doublet Models (2HDM)
%or the MSSM, charged Higgs can contribute to the decay amplitude at the tree level. 
%Compared to $B^+\to\tau\nu_\tau$, the $B\to D^{(*)}\tau \nu_\tau$ decay
%has advantages: the branching fraction is relatively high, because it is not Cabibbo-suppressed, and 
%it is a three-body decay allowing access to many observables beside the branching fraction, such as the $D^*$
%and the $\tau$ polarisation, or the $q^2$ distribution (see Ref.~\cite{Duraisamy:2014sna} and reference therein for recent 
%calculations).
%

In the SM the semileptonic decay are tree level processes which proceed via coupling to the $W^{\pm}$ boson.
These couplings are assumed to be universal for all leptons and are well understood theoretically, (see Section 5.1 and 5.2.).
This universality has be tested in purely leptonic and semileptonic $B$ meson decays involving a $\tau$ lepton, which might 
be sensitive to a hypothetical charged Higgs boson or other non-SM processes.

Compared to $B^+\to\tau\nu_\tau$, the $B\to D^{(*)}\tau \nu_\tau$ decay has advantages: the branching fraction is 
relatively high, because it is not Cabibbo-suppressed, and it is a three-body decay allowing access to many 
observables besides the branching fraction, such as $D^{(*)}$ momentum, $q^2$ distributions, and measurements of the 
$D^*$ and $\tau$ polarisations (see Ref.~\cite{Duraisamy:2014sna} and references therein for recent calculations).

Experiments have measured two ratios of branching fractions defined as 
\begin{eqnarray}
{\cal R}(D)&=&\dfrac{ {\cal B}(B\to D\tau\nu_\tau) }{ {\cal B}(B\to D\ell\nu_\ell) },\\
{\cal R}(D^*)&=&\dfrac{ {\cal B}(B\to D^*\tau\nu_\tau) }{ {\cal B}(B\to D^*\ell\nu_\ell) } %
\end{eqnarray}
where $\ell$ refers either to electron or $\mu$. These ratios are independent of  $|V_{cb}|$ and to a large extent, also of 
the $B\to D^{(*)}$ form factors. As a consequences the SM predictions for these ratios are quite precise:
\begin{itemize}
\item ${\cal R}(D)=0.300\pm 0.008$, which is  
an average obtained by FLAG~\cite{Aoki:2016frl} by combining the most recent lattice
calculations of the $B\to D\ell\nu$ form factors~\cite{Lattice:2015rga,Na:2015kha}; 
\item ${\cal R}(D^*)=0.252\pm 0.003$, which is a prediction, \cite{Lees:2012xj,Lees:2013uzd} 
that  updates recent QCD calculations ~\cite{Kamenik:2008tj,Fajfer:2012vx}
 based on the recent $B\to D^{*}$ measurements from the B-Factories.
\end{itemize}

\noindent Recently, in Ref.~\cite{Bigi:2016mdz} Bigi and Gambino re-analysed the recent experimental results 
and theoretical calculation of $B\to D\ell\nu$ obtaining ${\cal R}(D)=0.299\pm 0.003$, compatible with the predictions reported 
above but with a total error reduced by a factor three.

From the experimental side, in the case of the leptonic $\tau$ decay, the ratios  ${\cal R}(D^{(*)})$ can be 
directly measured, and many systematic uncertainties cancel in the measurement.
The $B^0\to D^{*+}\tau\nu_\tau$ decay was first observed by Belle~\cite{Matyja:2007kt} performing 
an "inclusive" reconstruction, which is based on the reconstruction of the $B_{tag}$ from all the particles
of the events, other than the $D^{(*)}$ and the lepton candidate, without looking for any specific $B_{tag}$ decay chain.  
Since then, both \babar and Belle have published improved measurements 
and have found evidence for the $B\to D\tau\nu_\tau$ decays~\cite{Aubert:2007dsa,Bozek:2010xy}. %comment out Adachi:2009qg,

The most powerful way to study these decays at the B-Factories exploits the hadronic $B_{tag}$.
Using the full dataset and an improved $B_{tag}$ selection, \babar measured~\cite{Lees:2012xj}:  
\begin{equation}
{\cal R}(D)=0.440\pm 0.058\pm 0.042, ~ {\cal R}(D^*)=0.332\pm 0.024\pm 0.018 %
\end{equation}
where decays to both $e^\pm$ and $\mu^\pm$ were summed and $B^0$ and $B^-$ were combined in a 
isospin-constrained fit. The fact that the \babar result exceeded SM predictions by $3.4\sigma$, raised considerable interest.

Belle published various measurements using different techniques, and LHCb also joined the effort with a measurement of $R(D^*)$. 
The most important sources of systematic uncertainties correlated for the different measurement  
is due to the $B\to D^{**}$ background components that are difficult to disentangle from the signal. 
In the average the systematic uncertainties 
due to the $B\to D^{**}$ composition and kinematics are considered fully correlated among the measurements. 

\begin{figure}
\centering
\includegraphics[width=0.9\textwidth]{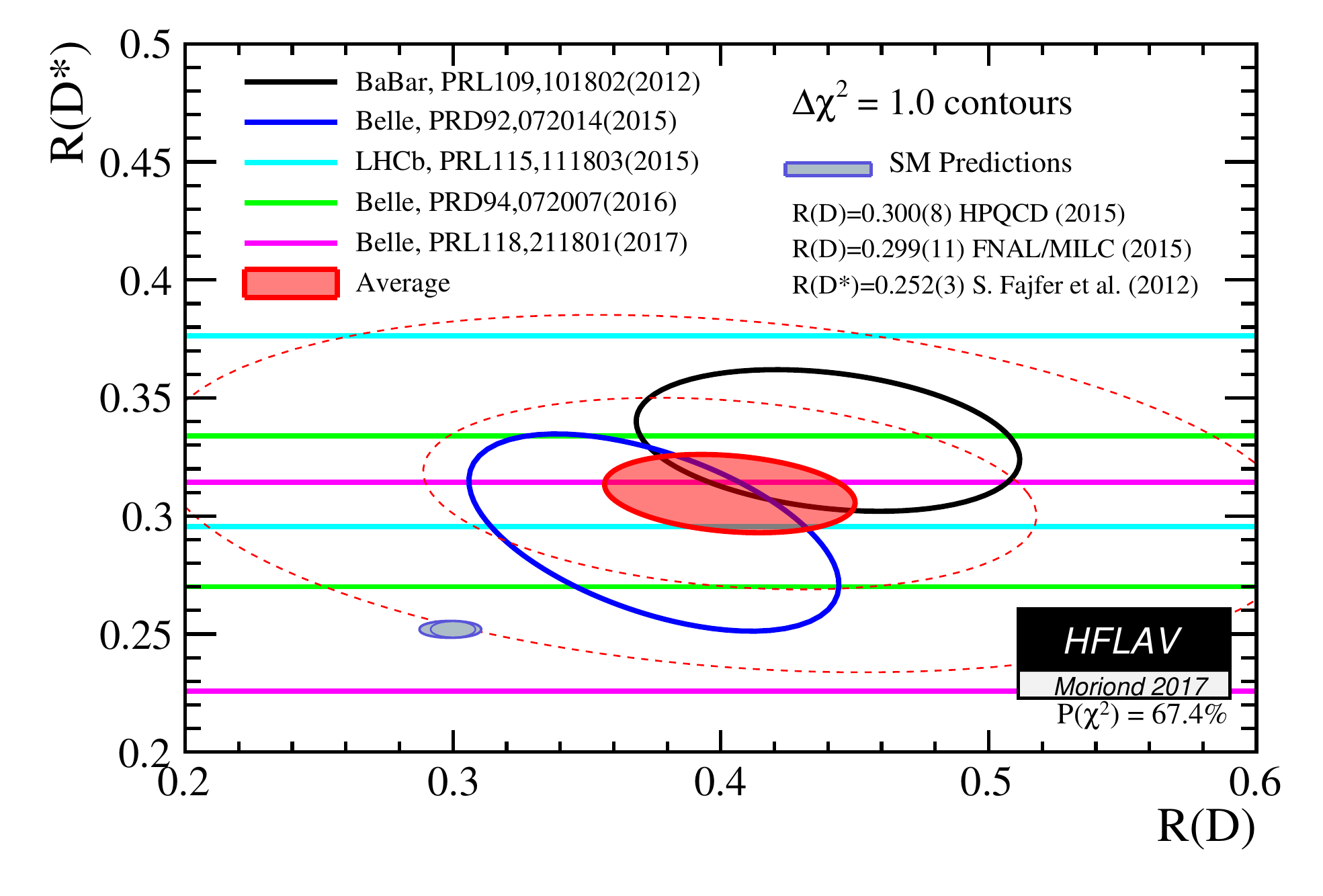}
\caption{
Measurement of ${\cal R}(D)$ and ${\cal R}(D^*)$ and their average compared with the prediction for ${\cal R}(D^*)$ ~\cite{Fajfer:2012vx}
and ${\cal R}(D)$ ~\cite{Lattice:2015rga,Na:2015kha}.
The dashed ellipses corresponds to the $2$ and $4$ $\sigma$ contours.
\label{fig:rds}}
\end{figure}

\begin{figure}[!ht]
  \begin{center}
  \unitlength 1.0cm % coordinates in cm
  \begin{picture}(14.,11.0)
    \put(  -1.5, 0.0){\includegraphics[width=9.2cm]{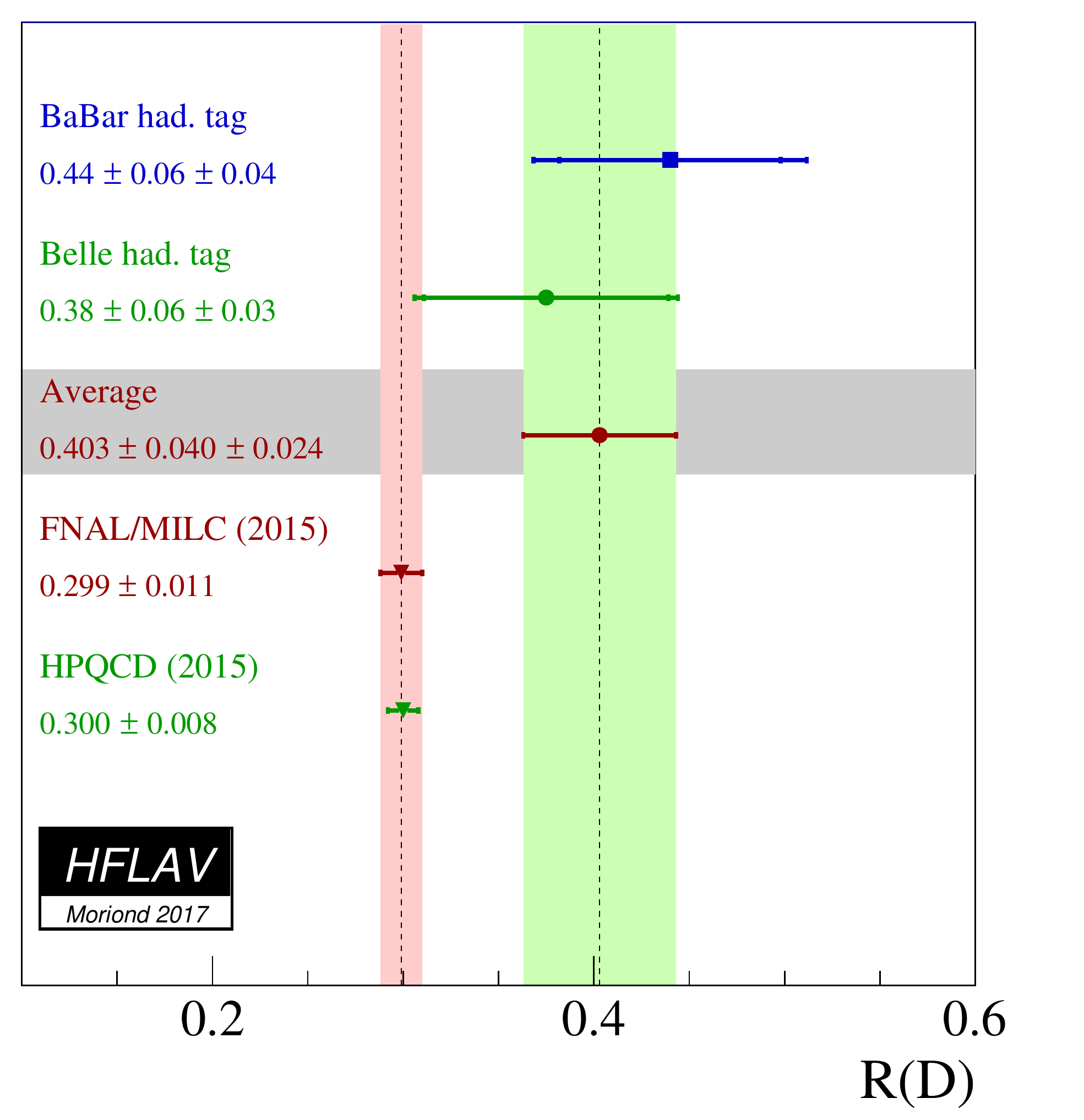}
    }
    \put(  7.5, 0.0){\includegraphics[width=9.2cm]{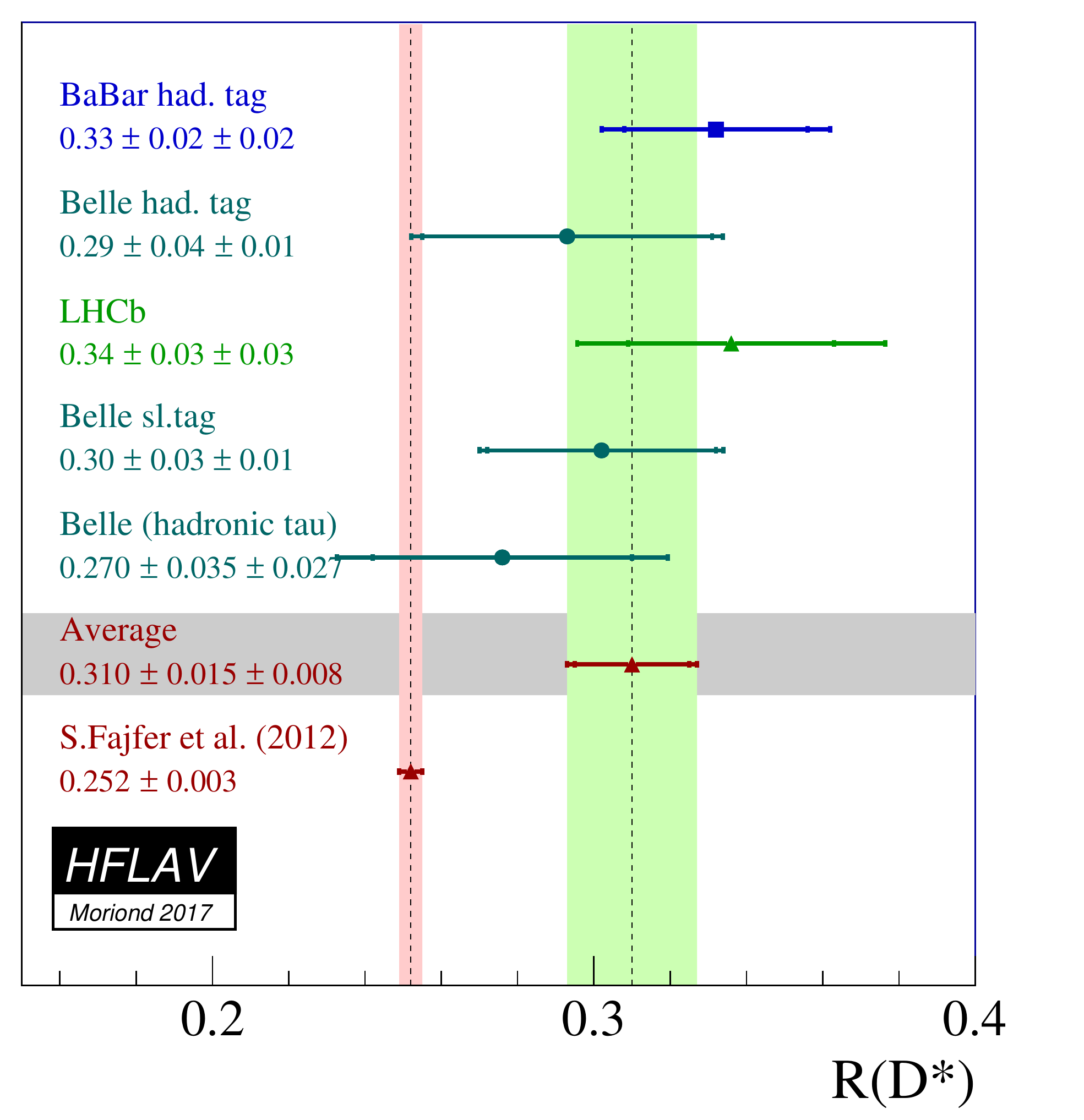}
    }
    \put(  5.8,  8.4){{\large\bf a)}}  
    \put( 14.7,  8.4){{\large\bf b)}}
  \end{picture}
  \caption{(a) Measurement of ${\cal R}(D)$ and (b) ${\cal R}(D^*)$. The average is the projection of the average
  obtained from the combined fit.} 
 \label{fig:rd}
 \end{center}
\end{figure}

The results of the individual measurements, their averages and correlations are presented in Table \ref{tab:dtaunu} and Fig.\ref{fig:rds}. 
The combined results, projected separately on ${\cal R}(D)$ 
and ${\cal R}(D^*)$, are reported in Figs.\ref{fig:rd}(a) and  Figs.\ref{fig:rd}(b) respectively. 

The averaged ${\cal R}(D)$ and ${\cal R}(D^*)$ exceed the SM predictions by 2.2$\sigma$ and 3.4$\sigma$ respectively. 
Considering the ${\cal R}(D)$ and ${\cal R}(D^*)$ total correlation of $-0.23$, the difference with respect to the 
SM is about 3.9~$\sigma$, the combined $\chi^2=18.83$ for 2 degrees of freedom corresponds to a $p$-value 
of $8.3\times 10^{-5}$, assuming Gaussian error distributions. 

% ----------------------------------------------------------------------0
\begin{table}[!htb]
\caption{Measurements of ${\cal R}(D^*)$ and ${\cal R}(D)$, their correlations and the combined average.}
\begin{center}
\resizebox{0.99\textwidth}{!}{
\begin{tabular}{l|c|c|c}\hline
Experiment  &${\cal R}(D^*)$ & ${\cal R}(D)$ & $\rho$ \\

\hline\hline 
\babar ~\cite{Lees:2012xj,Lees:2013uzd} &$0.332 \pm0.024_{\rm stat} \pm0.018_{\rm syst}$  &$0.440 \pm0.058_{\rm stat} \pm0.042_{\rm syst}$ & $-0.27$\\
Belle  ~\cite{Huschle:2015rga}         &$0.293 \pm0.038_{\rm stat} \pm0.015_{\rm syst}$  &$0.375 \pm0.064_{\rm stat} \pm0.026_{\rm syst}$ & $-0.49$ \\
LHCb   ~\cite{Aaij:2015yra}            &$0.336 \pm0.027_{\rm stat} \pm0.030_{\rm syst}$  &   &\\
Belle  ~\cite{Sato:2016svk}            &$0.302 \pm0.030_{\rm stat} \pm0.011_{\rm syst}$  &   &\\
Belle  ~\cite{Hirose:2016wfn}          &$0.270 \pm0.035_{\rm stat} {^{+0.028} _{-0.025}}_{\rm syst}$  & & \\
\hline
{\bf Average} &\mathversion{bold}$0.310 \pm0.015 \pm 0.008$ & \mathversion{bold}$0.403 \pm0.040 \pm 0.024$ & $-0.23$  \\
\hline 
\end{tabular}
}
\end{center}
\label{tab:dtaunu}
\end{table}
% ----------------------------------------------------------------------

%\begin{table}[!htb]
%\begin{center}
%\caption{Summary of the results on ${\cal R}(D)$ and ${\cal R}(D^*)$. The errors quoted
%correspond to statistical and systematic uncertainties, respectively.}
%\label{tab:dtaunu}
%\begin{small}
%\begin{tabular}{|lcc|}
%\hline
 % & Belle~\cite{Adachi:2009qg}   & \babar~\cite{Lees:2012xj} \\
%\hline\hline
%${\cal R}(D^0) $    &  $0.70^{+0.19}_{-0.18}~^{+0.11}_{-0.09}$ & $0.99\pm 0.19\pm 0.13$\\
%${\cal R}(D^{*0})$  &  $0.47^{+0.11}_{-0.10}~^{+0.06}_{-0.07}$ & $1.71\pm 0.17\pm 0.13$\\
%${\cal R}(D^+) $    &  $0.48^{+0.22}_{-0.19}~^{+0.06}_{-0.05}$ & $1.01\pm 0.18\pm 0.12$\\
%${\cal R}(D^{*+})$  &  $0.48^{+0.14}_{-0.12}~^{+0.06}_{-0.04}$ & $1.74\pm 0.19\pm 0.12$\\
%\hline
%\end{tabular}\\
%\end{small}
%\end{center}
%\end{table}

%
% ======================================================================
%

\clearpage
%% % Branching fractions for B decays to charm
\section{Decays of $b$-hadrons into open or hidden charm hadrons}
\label{sec:b2c}
Ground state $B$ mesons and $b$ baryons dominantly decay to particles containing a charm quark via the $b \rightarrow c$ quark transition.
%Therefore these decays are sensitive to the CKM matrix element $|V_{cb}|$.
%Usually semileptonic modes are used for $|V_{cb}|$ measurements, as discussed in Section~\ref{sec:slbdecays}.
In this section, measurements of such decays to hadronic final states are considered; semileptonic decay modes, which are usually used to determine the strength of the $b \rightarrow c$ transition as quantified in the magnitude of the CKM matrix element $|V_{cb}|$, are discussed in Section~\ref{sec:slbdecays}.
Some $B$ meson decays to open or hidden charm hadrons that are fundamental for the measurements of $\CP$-violation phases -- like $\phi_s^{c\bar{c}s}$ (Section~\ref{sec:life_mix}), $\beta \equiv \phi_1$ and $\gamma \equiv \phi_3$ (Section~\ref{sec:cp_uta}) -- are discussed elsewhere in this report. 
Similarly, the use of $b \rightarrow c$ decay modes for the determination of important properties of $b$-hadrons, like their masses or (absolute, relative or effective) lifetimes, is discussed in Section~\ref{sec:life_mix}.
% In particular the small $Q$ value of decays to two charmed hadrons allows the minimization of systematic uncertainties in mass measurements.
% The similar topology of decays of different $b$-hadrons can be exploited 
% to make very precise relative measurements.
The properties of certain $b$ hadron decays to open or hidden charm hadrons, such as small $Q$ values and similar topologies for different modes, allow the minimization of systematic uncertainties in these measurements.

The fact that decays to final states containing open or hidden charm hadrons dominate the $b$-hadron widths makes them a very important part of the experimental programme in heavy flavor physics.
Understanding the rate of charm production in $b$-hadron decays is crucial to
validate the HQE that underpins much of the theoretical framework for $b$
physics (see, for example, Ref.~\cite{Lenz:2014nka} for a review).
Moreover, such decays are often used as normalization modes for
measurements of rarer decays. 
In addition, they are the dominant background in many analyses.
To model accurately such backgrounds with simulated data, it is essential to have precise knowledge of the contributing decay modes.
In particular, with the expected increase in the data samples at LHCb and
Belle~II, the enhanced statistical sensitivity has to be matched by low
systematic uncertainties due to knowledge of the dominant $b$-hadron decay
modes. 
For multibody decays, knowledge of the distribution of decays across the
phase-space (\eg,\ the Dalitz plot density for three-body decays or the
polarization amplitudes for vector-vector final states) is required in
addition to the total branching fraction.

The large yields of $b \to c$ decays to multibody final states 
make them ideal to study the spectroscopy of both open and hidden charm hadrons.  
In particular, they have been used to both discover, and measure the properties
of, exotic particles such as the $X(3872)$~\cite{Choi:2003ue,Aaij:2013zoa},
$Z(4430)^+$~\cite{Choi:2007wga,Aaij:2014jqa} and $P_c(4450)^+$~\cite{Aaij:2015tga} states.
The large yields available similarly make decays involving $b \to c$ transitions very useful to study baryon-antibaryon pair production.

In addition to the dominant $b$-hadron decays to final states containing
charmed hadrons, there are several decays in this category that are expected
to be highly suppressed in the Standard Model.
These are of interest to probe particular decay topologies (\eg,\ the $B^- \to
\Dsm \phi$ decay, which is dominated by the annihilation diagram)
and thereby constrain effects in other hadronic decays or to search for new physics.
There are also other decays involving $b \to c$ transitions, such as $\Bzb \to \Dsm \pip$, that are
mediated by the $W$ emission involving the $|V_{ub}|$ CKM matrix element.
Finally, some $b \to c$ decays involving lepton flavour or number violation are
extremely suppressed in the Standard Model, and therefore provide highly
sensitive null tests.

In this section, we give an exhaustive list of measured branching ratios of decay modes to
hadrons containing charm quarks.
The averaging procedure follows the methodology described in Section~\ref{sec:method}.
Where available, correlations between measurements are taken into account.
If an insignificant measurement and a limit for the same parameter are provided 
the former is taken so that it can be included in averages.
The confidence level of an average is quoted if it is below 1\%.
We provide averages of the polarization amplitudes of $B$ meson decays to
vector-vector states, but we do not currently provide detailed averages of
quantities obtained from Dalitz plot analyses, due to the complications
arising from the dependence on the model used.

The results are presented in subsections organized according to the type of decaying bottom
hadron: $\Bzb$ (Sec.~\ref{sec:b2c:Bd}), $B^-$ (Sec.~\ref{sec:b2c:Bu}), $\Bzb/B^-$ admixture (Sec.~\ref{sec:b2c:B}), $\Bsb$ (Sec.~\ref{sec:b2c:Bs}), $B_c^-$ (Sec.~\ref{sec:b2c:Bc}), $b$ baryons (Sec.~\ref{sec:b2c:Bbaryon}).
For each subsection the measurements are arranged according to the final state
into the following groups: a single charmed meson, two charmed mesons, a
charmonium state, a charm baryon, or other states, like for example the
$X(3872)$ meson.
The individual measurements and averages are shown as numerical values in tables followed by a graphical representation of the averages.
The symbol $\mathcal{B}$ is used for branching ratios, $f$ for production fractions (see Section~\ref{sec:life_mix}), and $\sigma$ for cross sections.
The decay amplitudes for longitudinal, parallel, and perpendicular transverse polarization in pseudoscalar to vector-vector decays are denoted ${\cal{A}}_0$, ${\cal{A}}_\parallel$, and ${\cal{A}}_\perp$, respectively, and the definitions $\delta_\parallel = \arg({\cal{A}}_\parallel/{\cal{A}}_0)$ and $\delta_\perp = \arg({\cal{A}}_\perp/{\cal{A}}_0)$ are used for their relative phases.
The inclusion of charge conjugate modes is always implied.

Following the approach used by the PDG~\cite{PDG_2014}, for decays that involve
neutral kaons we mainly quote results in terms of final states including
either a $\Kz$ or $\Kzb$ meson (instead of a \KS or \KL).
In some cases where the decay is not flavour-specific and the final state is
not self-conjugate, the inclusion of the conjugate final state neutral kaon is implied --
% (\eg\ $\Dp\Kzb\pip$ should be read as the sum of $\Dp\Kzb\pip$ and
% $\Dm\Kz\pip$).
in fact, the flavour of the neutral kaon is never determined experimentally,
and so the specification as \Kz or \Kzb simply follows the quark model
expectation for the dominant decay.
An exception occurs for some \Bs decays, specifically those to \CP
eigenstates, where the width difference between the mass eigenstates (see
Sec.~\ref{sec:life_mix}) means that the measured branching fraction,
integrated over decay time, is specific to the studied final
state~\cite{DeBruyn:2012wj}. 
Therefore it is appropriate to quote the branching fraction for, \eg, $\Bsb \to
\jpsi \KS$ instead of $\Bsb \to \jpsi \Kzb$.

Several measurements assume $\Gamma(\Upsilon(4S) \to B^+B^-) = \Gamma(\Upsilon(4S) \to B^0\bar{B}^0)$.
While there is no evidence for isospin violation in $\Upsilon(4S)$ decays,
deviations from this assumptions can be of the order of a few percent,
see Section~\ref{sec:fraction_Ups4S} and Ref.~\cite{Jung:2015yma}.
As the effect is negligible for many averages, we do not apply a correction 
or additional systematic uncertainty, but we point out that it can be relevant
for averages with a percent level uncertainty.

% \clearpage

\newenvironment{btocharmtab}[2]{\begin{table}[H]\begin{center}\caption{#2.}\label{tab:b2c:#1}\begin{tabular}{|l l l |}}{\end{tabular}\end{center}\end{table}}
\newenvironment{widebtocharmtab}[2]{\begin{table}[H]\begin{center}\caption{#2.}\label{tab:b2c:#1}\begin{adjustbox}{width=\textwidth,center}\begin{tabular}{|l l l |}}{\end{tabular}\end{adjustbox}\end{center}\end{table}}
\newcommand{\btocharmfig}[1]{\begin{figure}[H]\begin{center}\includegraphics[width=0.99\textwidth]{b2charm/figs/#1}\caption{Summary of the averages from Table~\ref{tab:b2c:#1}.}\label{fig:b2c:#1}\end{center}\end{figure}}
\newcommand{\btocharm}[1]{\input{b2charm/#1.tex}}

\subsection{Decays of $\bar{B}^0$ mesons}
\label{sec:b2c:Bd}
Measurements of $\bar{B}^0$ decays to charmed hadrons are summarized in Sections~\ref{sec:b2c:Bd_D} to~\ref{sec:b2c:Bd_other}.

\subsubsection{Decays to a single open charm meson}
\label{sec:b2c:Bd_D}
Averages of $\bar{B}^0$ decays to a single open charm meson are shown in Tables~\ref{tab:b2c:Bd_D_1}--\ref{tab:b2c:Bd_D_13} and Figs.~\ref{fig:b2c:Bd_D_1}--\ref{fig:b2c:Bd_D_13}.
In this section $D^{**}$ refers to the sum of all the non-strange charm meson
states with masses in the range $2.2-2.8~\gevcc$.
\begin{btocharmtab}{Bd_D_1}{Decays to a $D^{(*)}$ meson and one or more pions I $[10^{-3}]$}
\hline
\textbf{Parameter} &
\begin{tabular}{l}\textbf{Measurements}\end{tabular} & \textbf{Average} \\
\hline
\hline
${\cal{B}} ( \bar{B}^{0} \to D^{+} \pi^{-} )$ & \begin{tabular}{l} \babar \cite{Aubert:2006cd}: $2.55 \pm 0.05 \pm 0.16$ \\ \babar \cite{Aubert:2006jc}: $3.03 \pm 0.23 \pm 0.23$ \\ \end{tabular} & $2.65 \pm 0.15$ \\
\hline
${\cal{B}} ( \bar{B}^{0} \to D^{*}(2010)^{+} \pi^{-} )$ & \begin{tabular}{l} \babar \cite{Aubert:2006cd}: $2.79 \pm 0.08 \pm 0.17$ \\ \babar \cite{Aubert:2006jc}: $2.99 \pm 0.23 \pm 0.24$ \\ \end{tabular} & $2.84 \pm 0.16$ \\
\hline
${\cal{B}} ( \bar{B}^{0} \to D^{*}(2010)^{+} \pi^{-} \pi^{+} \pi^{-} )$ & \begin{tabular}{l} Belle \cite{Majumder:2004su}: $6.81 \pm 0.23 \pm 0.72$ \\ \babar \cite{TheBABAR:2016vzj}: $7.26 \pm 0.11 \pm 0.31$ \\ \end{tabular} & $7.19 \pm 0.30$ \\
\hline
\multicolumn{3}{|l|}{${\cal{B}} ( \bar{B}^{0} \to D^{*}(2007)^{0} \pi^{-} \pi^{+} \pi^{-} \pi^{+} )$}\\
 & \begin{tabular}{l} Belle \cite{Majumder:2004su}: $2.60 \pm 0.47 \pm 0.37$ \\ \end{tabular} & $2.60 \pm 0.60$ \\
\hline
\multicolumn{3}{|l|}{${\cal{B}} ( \bar{B}^{0} \to D^{*}(2010)^{+} \pi^{-} \pi^{+} \pi^{-} \pi^{+} \pi^{-} )$}\\
 & \begin{tabular}{l} Belle \cite{Majumder:2004su}: $4.72 \pm 0.59 \pm 0.71$ \\ \end{tabular} & $4.72 \pm 0.92$ \\
\hline
${\cal{B}} ( \bar{B}^{0} \to D^{*}(2010)^{+} \omega(782) \pi^{-} )$ & \begin{tabular}{l} Belle \cite{Matvienko:2015gqa}: $2.31 \pm 0.11 \pm 0.14$ \\ \babar \cite{Aubert:2006zb}: $2.88 \pm 0.21 \pm 0.31$ \\ \end{tabular} & $2.41 \pm 0.16$ \\
\hline
\end{btocharmtab}
\btocharmfig{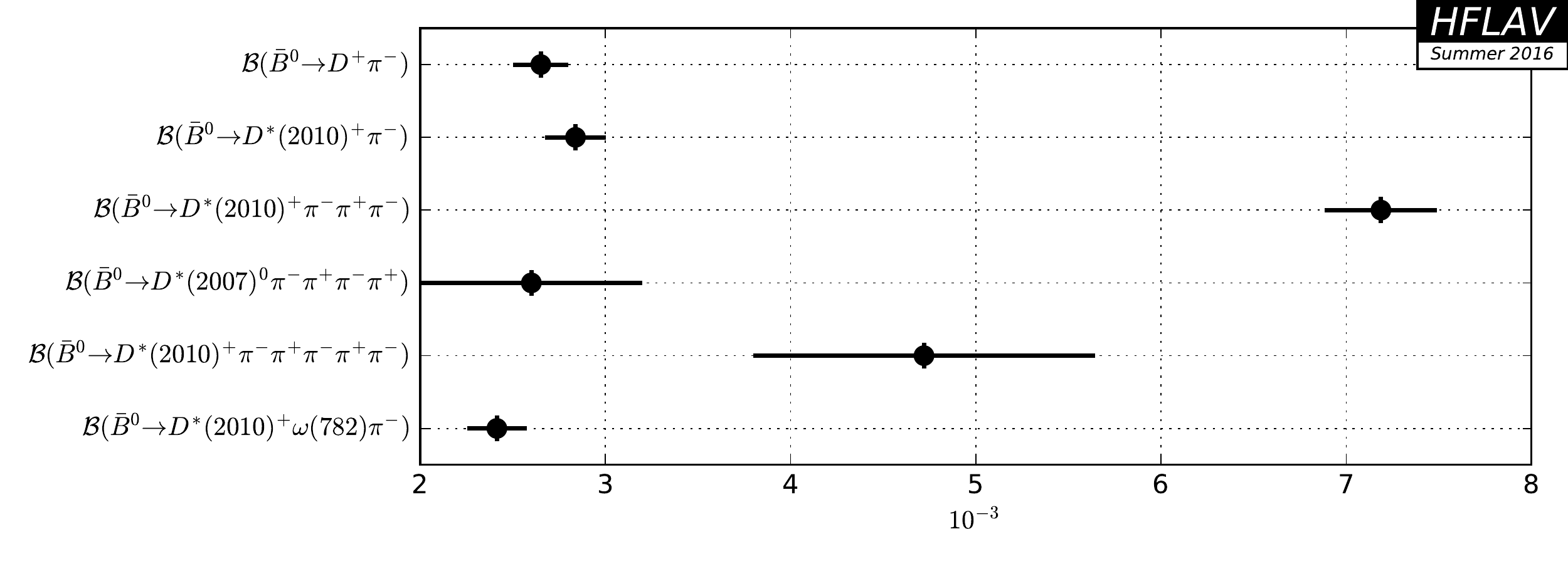}

\begin{btocharmtab}{Bd_D_2}{Decays to a $D^{(*)}$ meson and one or more pions II $[10^{-4}]$}
\hline
\textbf{Parameter} & \begin{tabular}{l}\textbf{Measurements}\end{tabular} & \textbf{Average} \\
\hline
\hline
${\cal{B}} ( \bar{B}^{0} \to D^{0} \pi^{0} )$ & \begin{tabular}{l} Belle \cite{Blyth:2006at}: $2.25 \pm 0.14 \pm 0.35$ \\ \babar \cite{Lees:2011gw}: $2.69 \pm 0.09 \pm 0.13$ \\ \end{tabular} & $2.62 \pm 0.15$ \\
\hline
${\cal{B}} ( \bar{B}^{0} \to D^{*}(2007)^{0} \pi^{0} )$ & \begin{tabular}{l} Belle \cite{Blyth:2006at}: $1.39 \pm 0.18 \pm 0.26$ \\ \babar \cite{Lees:2011gw}: $3.05 \pm 0.14 \pm 0.28$ \\ \end{tabular} & $2.23 \pm 0.22$ {\tiny CL=0.2\permil} \\
\hline
${\cal{B}} ( \bar{B}^{0} \to D^{0} \pi^{+} \pi^{-} )$ & \begin{tabular}{l} LHCb \cite{Aaij:2015sqa}: $8.46 \pm 0.14 \pm 0.49$ \\ Belle \cite{Satpathy:2002js}: $8.0 \pm 0.6 \pm 1.5$ \\ \end{tabular} & $8.42 \pm 0.49$ \\
\hline
${\cal{B}} ( \bar{B}^{0} \to D^{*}(2007)^{0} \pi^{+} \pi^{-} )$ & \begin{tabular}{l} Belle \cite{Satpathy:2002js}: $6.2 \pm 1.2 \pm 1.8$ \\ \end{tabular} & $6.2 \pm 2.2$ \\
\hline
\end{btocharmtab}
\btocharmfig{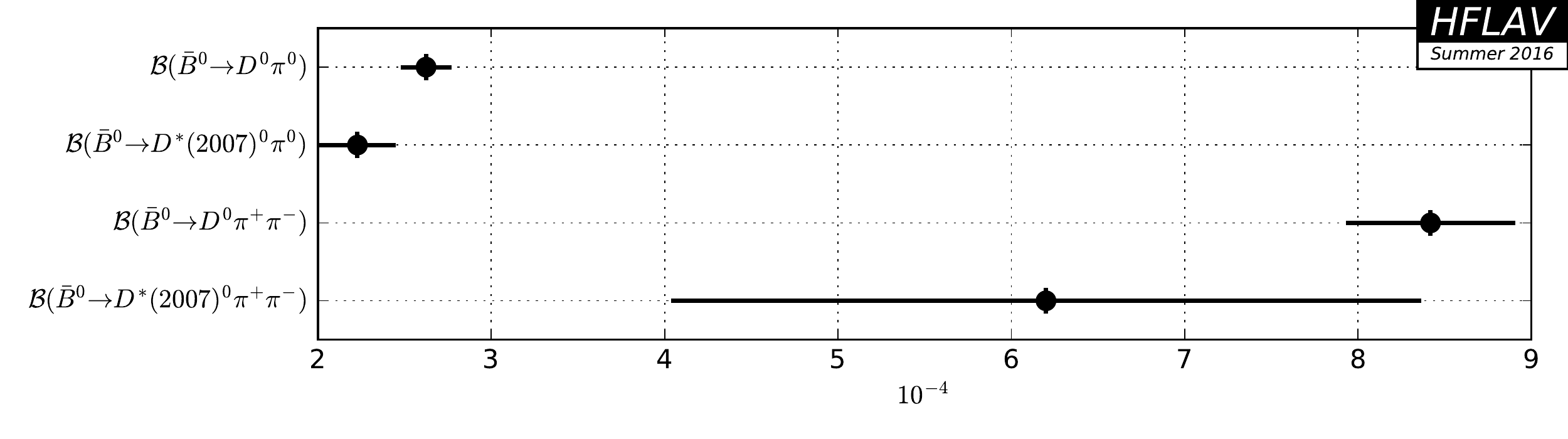}

\begin{btocharmtab}{Bd_D_3}{Decays to a $D^{(*)0}$ meson and a light meson $[10^{-4}]$}
\hline
\textbf{Parameter} & \begin{tabular}{l}\textbf{Measurements}\end{tabular} & \textbf{Average} \\
\hline
\hline
${\cal{B}} ( \bar{B}^{0} \to D^{0} \rho(770)^{0} )$ & \begin{tabular}{l} Belle \cite{Satpathy:2002js}: $2.9 \pm 1.0 \pm 0.4$ \\ \end{tabular} & $2.9 \pm 1.1$ \\
\hline
${\cal{B}} ( \bar{B}^{0} \to D^{*}(2007)^{0} \rho(770)^{0} )$ & \begin{tabular}{l} Belle \cite{Satpathy:2002js}: $< 5.1$ \\ \end{tabular} & $< 5.1$ \\
\hline
${\cal{B}} ( \bar{B}^{0} \to D^{0} \eta )$ & \begin{tabular}{l} Belle \cite{Blyth:2006at}: $1.77 \pm 0.16 \pm 0.21$ \\ \babar \cite{Lees:2011gw}: $2.53 \pm 0.09 \pm 0.11$ \\ \end{tabular} & $2.36 \pm 0.13$ \\
\hline
${\cal{B}} ( \bar{B}^{0} \to D^{*}(2007)^{0} \eta )$ & \begin{tabular}{l} Belle \cite{Blyth:2006at}: $1.40 \pm 0.28 \pm 0.26$ \\ \babar \cite{Lees:2011gw}: $2.69 \pm 0.14 \pm 0.23$ \\ \end{tabular} & $2.26 \pm 0.22$ {\tiny CL=5.8\permil} \\
\hline
${\cal{B}} ( \bar{B}^{0} \to D^{0} \eta^\prime(958) )$ & \begin{tabular}{l} Belle \cite{Schumann:2005ej}: $1.14 \pm 0.20 \,^{+0.10}_{-0.13}$ \\ \babar \cite{Lees:2011gw}: $1.48 \pm 0.13 \pm 0.07$ \\ \end{tabular} & $1.38 \pm 0.12$ \\
\hline
${\cal{B}} ( \bar{B}^{0} \to D^{*}(2007)^{0} \eta^\prime(958) )$ & \begin{tabular}{l} Belle \cite{Schumann:2005ej}: $1.21 \pm 0.34 \pm 0.22$ \\ \babar \cite{Lees:2011gw}: $1.48 \pm 0.22 \pm 0.13$ \\ \end{tabular} & $1.40 \pm 0.22$ \\
\hline
${\cal{B}} ( \bar{B}^{0} \to D^{0} \omega(782) )$ & \begin{tabular}{l} LHCb \cite{Aaij:2015sqa}: $2.81 \pm 0.72 \,^{+0.30}_{-0.33}$ \\ Belle \cite{Blyth:2006at}: $2.37 \pm 0.23 \pm 0.28$ \\ \babar \cite{Lees:2011gw}: $2.57 \pm 0.11 \pm 0.14$ \\ \end{tabular} & $2.54 \pm 0.16$ \\
\hline
${\cal{B}} ( \bar{B}^{0} \to D^{*}(2007)^{0} \omega(782) )$ & \begin{tabular}{l} Belle \cite{Blyth:2006at}: $2.29 \pm 0.39 \pm 0.40$ \\ \babar \cite{Lees:2011gw}: $4.55 \pm 0.24 \pm 0.39$ \\ \end{tabular} & $3.64 \pm 0.35$ {\tiny CL=1.8\permil} \\
\hline
${\cal{B}} ( \bar{B}^{0} \to D^{0} f_2(1270) )$ & \begin{tabular}{l} LHCb \cite{Aaij:2015sqa}: $1.61 \pm 0.11 \,^{+0.19}_{-0.18}$ \\ \end{tabular} & $1.61 \,^{+0.22}_{-0.21}$ \\
\hline
\end{btocharmtab}
\btocharmfig{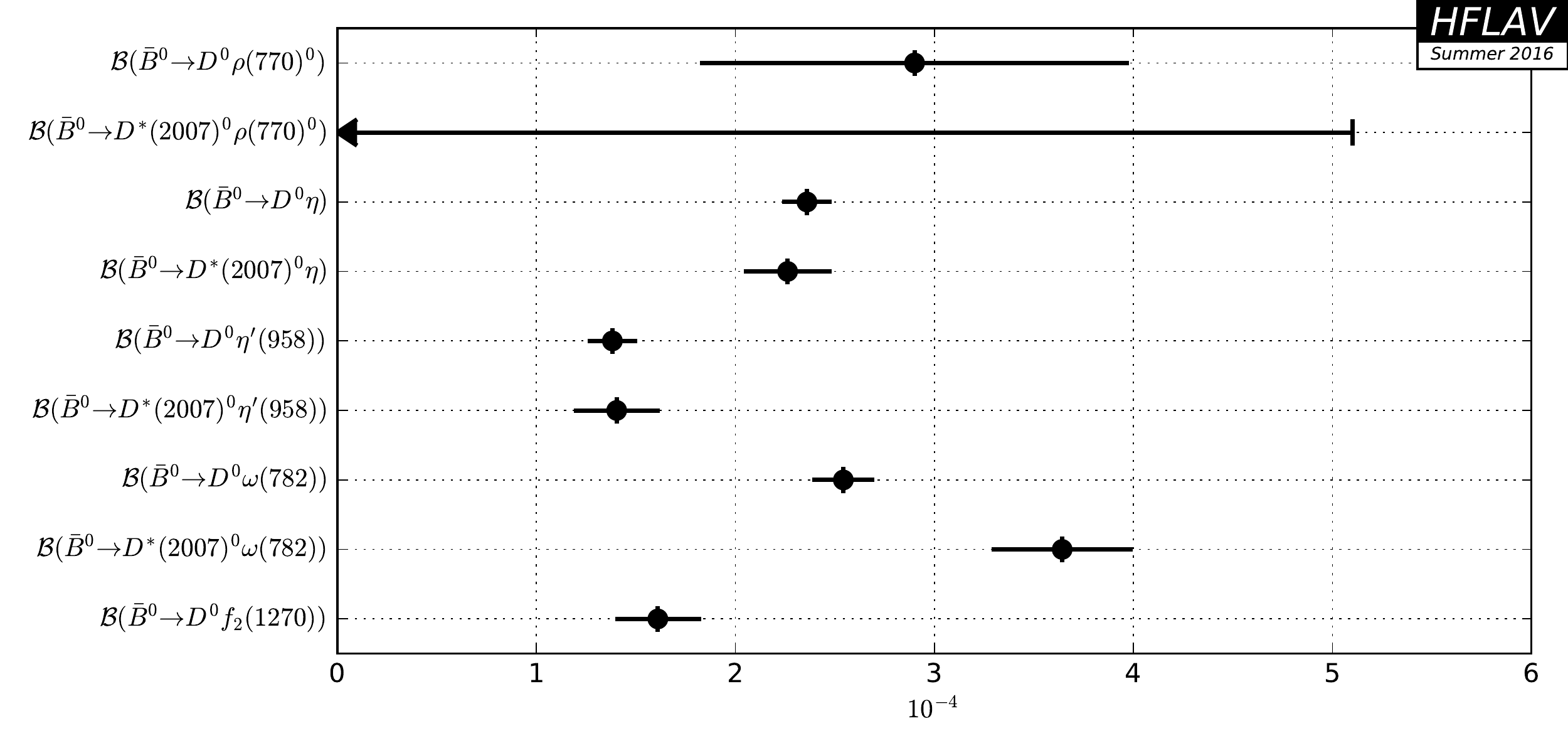}

\begin{btocharmtab}{Bd_D_4}{Decays to a $D^{(*)+}$ meson and one or more kaons $[10^{-3}]$}
\hline
\textbf{Parameter} & \begin{tabular}{l}\textbf{Measurements}\end{tabular} & \textbf{Average} \\
\hline
\hline
${\cal{B}} ( \bar{B}^{0} \to D^{+} K^{-} )$ & \begin{tabular}{l} LHCb \cite{Aaij:2013qqa}: $0.220 \pm 0.003 \pm 0.013$ \\ Belle \cite{Abe:2001waa}: $0.204 \pm 0.045 \pm 0.034$ \\ \end{tabular} & $0.219 \pm 0.013$ \\
\hline
${\cal{B}} ( \bar{B}^{0} \to D^{*}(2010)^{+} K^{-} )$ & \begin{tabular}{l} Belle \cite{Abe:2001waa}: $0.204 \pm 0.041 \pm 0.023$ \\ \end{tabular} & $0.204 \pm 0.047$ \\
\hline
${\cal{B}} ( \bar{B}^{0} \to D^{+} K^{*}(892)^{-} )$ & \begin{tabular}{l} \babar \cite{Aubert:2004at}: $0.46 \pm 0.06 \pm 0.05$ \\ \end{tabular} & $0.46 \pm 0.08$ \\
\hline
${\cal{B}} ( \bar{B}^{0} \to D^{*}(2010)^{+} K^{*}(892)^{-} )$ & \begin{tabular}{l} \babar \cite{Aubert:2004at}: $0.32 \pm 0.06 \pm 0.03$ \\ \end{tabular} & $0.32 \pm 0.07$ \\
\hline
${\cal{B}} ( \bar{B}^{0} \to D^{+} K^{0} \pi^{-} )$ & \begin{tabular}{l} \babar \cite{Aubert:2004at}: $0.49 \pm 0.07 \pm 0.05$ \\ \end{tabular} & $0.49 \pm 0.09$ \\
\hline
${\cal{B}} ( \bar{B}^{0} \to D^{*}(2010)^{+} K^{0} \pi^{-} )$ & \begin{tabular}{l} \babar \cite{Aubert:2004at}: $0.30 \pm 0.07 \pm 0.03$ \\ \end{tabular} & $0.30 \pm 0.08$ \\
\hline
${\cal{B}} ( \bar{B}^{0} \to D^{+} K^{-} K^{0} )$ & \begin{tabular}{l} Belle \cite{Drutskoy:2002ib}: $< 0.31$ \\ \end{tabular} & $< 0.31$ \\
\hline
${\cal{B}} ( \bar{B}^{0} \to D^{*}(2010)^{+} K^{-} K^{0} )$ & \begin{tabular}{l} Belle \cite{Drutskoy:2002ib}: $< 0.47$ \\ \end{tabular} & $< 0.47$ \\
\hline
${\cal{B}} ( \bar{B}^{0} \to D^{+} K^{-} K^{*}(892)^{0} )$ & \begin{tabular}{l} Belle \cite{Drutskoy:2002ib}: $0.88 \pm 0.11 \pm 0.15$ \\ \end{tabular} & $0.88 \pm 0.19$ \\
\hline
${\cal{B}} ( \bar{B}^{0} \to D^{*}(2010)^{+} K^{-} K^{*}(892)^{0} )$ & \begin{tabular}{l} Belle \cite{Drutskoy:2002ib}: $1.29 \pm 0.22 \pm 0.25$ \\ \end{tabular} & $1.29 \pm 0.33$ \\
\hline
\end{btocharmtab}
\btocharmfig{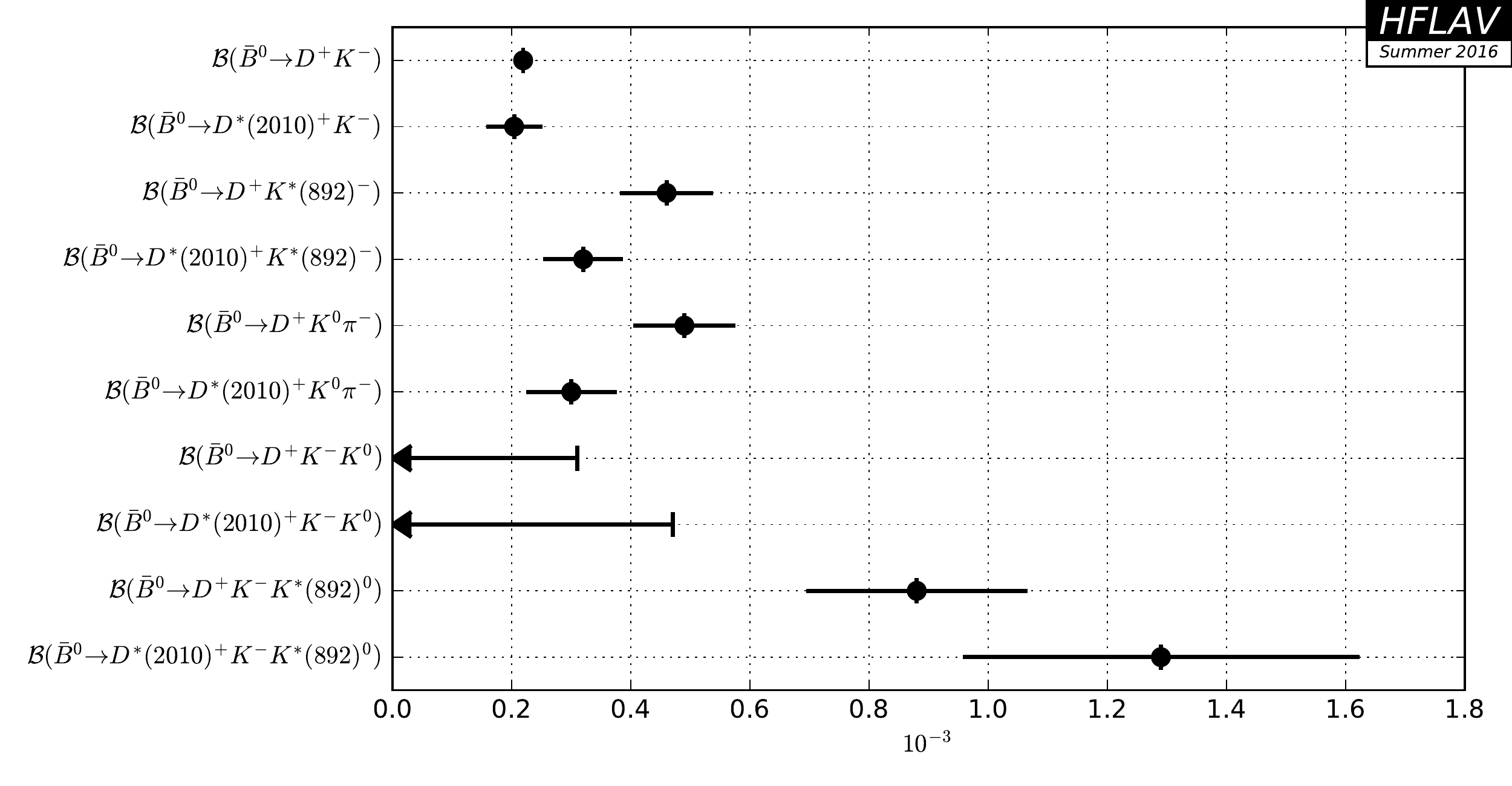}

\begin{btocharmtab}{Bd_D_5}{Decays to a $D^{(*)0}$ meson and a kaon $[10^{-4}]$}
\hline
\textbf{Parameter} & \begin{tabular}{l}\textbf{Measurements}\end{tabular} & \textbf{Average} \\
\hline
\hline
${\cal{B}} ( \bar{B}^{0} \to D^{0} \bar{K}^{0} )$ & \begin{tabular}{l} Belle \cite{Krokovny:2002ua}: $0.50 \,^{+0.13}_{-0.12} \pm 0.06$ \\ \babar \cite{Aubert:2006qn}: $0.53 \pm 0.07 \pm 0.03$ \\ \end{tabular} & $0.52 \pm 0.07$ \\
\hline
${\cal{B}} ( \bar{B}^{0} \to D^{*}(2007)^{0} \bar{K}^{0} )$ & \begin{tabular}{l} Belle \cite{Krokovny:2002ua}: $< 0.66$ \\ \babar \cite{Aubert:2006qn}: $0.36 \pm 0.12 \pm 0.03$ \\ \end{tabular} & $0.36 \pm 0.12$ \\
\hline
${\cal{B}} ( \bar{B}^{0} \to D^{0} K^{-} \pi^{+} )$ & \begin{tabular}{l} \babar \cite{Aubert:2005yt}: $0.88 \pm 0.15 \pm 0.09$ \\ \end{tabular} & $0.88 \pm 0.17$ \\
\hline
${\cal{B}} ( \bar{B}^{0} \to D^{0} \bar{K}^{*}(892)^{0} )$ & \begin{tabular}{l} Belle \cite{Krokovny:2002ua}: $0.48 \,^{+0.11}_{-0.10} \pm 0.05$ \\ \babar \cite{Aubert:2006qn}: $0.40 \pm 0.07 \pm 0.03$ \\ \end{tabular} & $0.42 \pm 0.06$ \\
\hline
\multicolumn{3}{|l|}{${{\cal{B}} ( \bar{B}^{0} \to D^{0} \bar{K}^{*}(892)^{0} )\times {\cal{B}} ( \bar{K}^{*}(892)^{0} \to K^{-} \pi^{+} )}$}\\
 & \begin{tabular}{l} \babar \cite{Aubert:2005yt}: $0.38 \pm 0.06 \pm 0.04$ \\ \end{tabular} & $0.38 \pm 0.07$ \\
\hline
${\cal{B}} ( \bar{B}^{0} \to D^{*}(2007)^{0} \bar{K}^{*}(892)^{0} )$ & \begin{tabular}{l} Belle \cite{Krokovny:2002ua}: $< 0.69$ \\ \end{tabular} & $< 0.69$ \\
\hline
${\cal{B}} ( \bar{B}^{0} \to \bar{D}^{*}(2007)^{0} \bar{K}^{*}(892)^{0} )$ & \begin{tabular}{l} Belle \cite{Krokovny:2002ua}: $< 0.40$ \\ \end{tabular} & $< 0.40$ \\
\hline
${\cal{B}} ( \bar{B}^{0} \to \bar{D}^{0} K^{-} \pi^{+} )$ & \begin{tabular}{l} \babar \cite{Aubert:2005yt}: $< 0.19$ \\ \end{tabular} & $< 0.19$ \\
\hline
${\cal{B}} ( \bar{B}^{0} \to \bar{D}^{0} \bar{K}^{*}(892)^{0} )$ & \begin{tabular}{l} Belle \cite{Krokovny:2002ua}: $< 0.18$ \\ \babar \cite{Aubert:2006qn}: $0.00 \pm 0.05 \pm 0.03$ \\ \end{tabular} & $0.00 \pm 0.06$ \\
\hline
\end{btocharmtab}
\btocharmfig{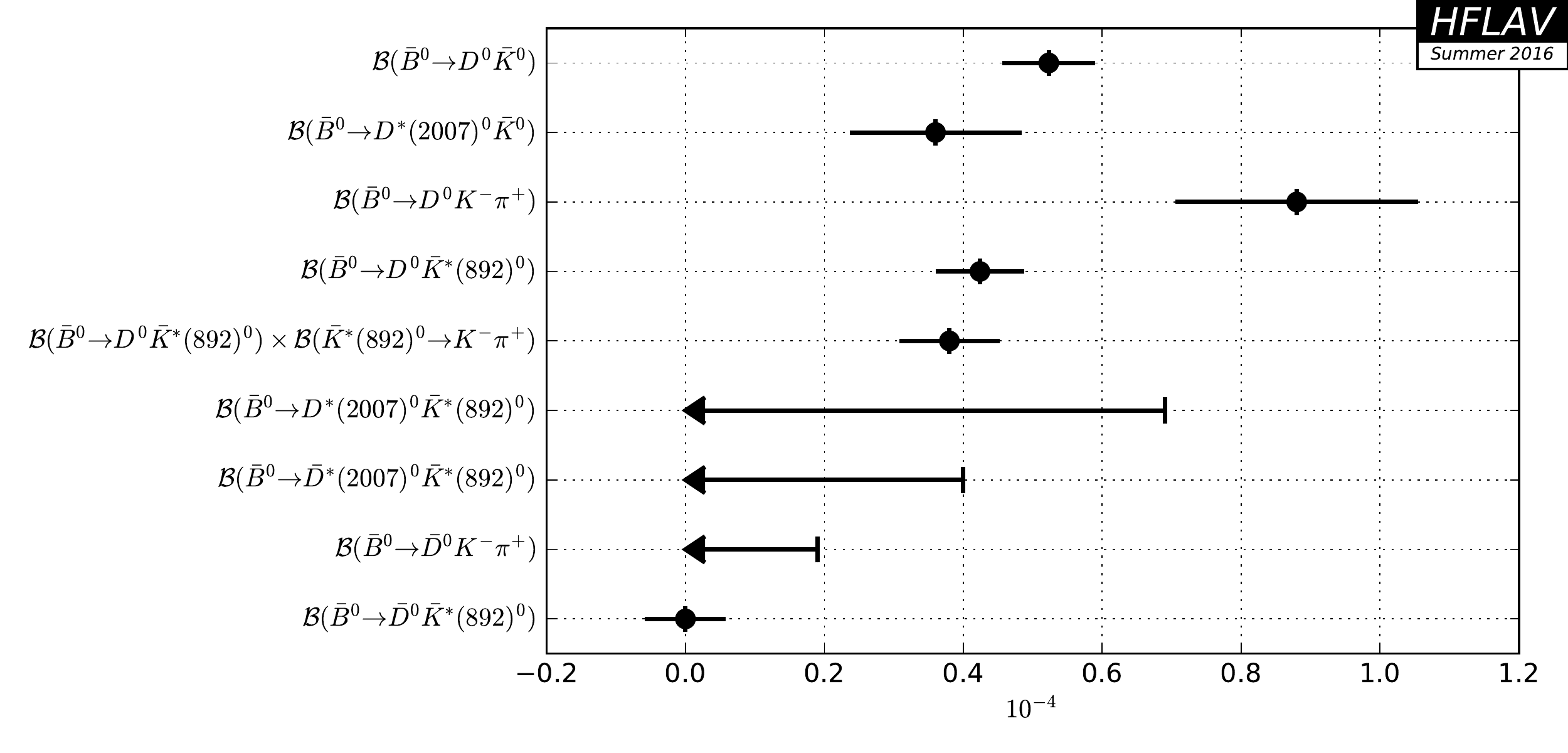}

\begin{btocharmtab}{Bd_D_6}{Decays to a $D_s^{(*)}$ meson $[10^{-4}]$}
\hline
\textbf{Parameter} & \begin{tabular}{l}\textbf{Measurements}\end{tabular} & \textbf{Average} \\
\hline
\hline
${\cal{B}} ( \bar{B}^{0} \to D_s^{-} \pi^{+} )$ & \begin{tabular}{l} Belle \cite{Das:2010be}: $0.199 \pm 0.026 \pm 0.018$ \\ \babar \cite{Aubert:2008zi}: $0.25 \pm 0.04 \pm 0.02$ \\ \end{tabular} & $0.216 \pm 0.026$ \\
\hline
${\cal{B}} ( \bar{B}^{0} \to D_s^{*-} \pi^{+} )$ & \begin{tabular}{l} Belle \cite{Joshi:2009yv}: $0.175 \pm 0.034 \pm 0.020$ \\ \babar \cite{Aubert:2008zi}: $0.26 \,^{+0.05}_{-0.04} \pm 0.02$ \\ \end{tabular} & $0.207 \pm 0.032$ \\
\hline
${\cal{B}} ( \bar{B}^{0} \to D_s^{-} \rho(770)^{+} )$ & \begin{tabular}{l} \babar \cite{Aubert:2008zi}: $0.11 \,^{+0.09}_{-0.08} \pm 0.03$ \\ \end{tabular} & $0.11 \,^{+0.09}_{-0.09}$ \\
\hline
${\cal{B}} ( \bar{B}^{0} \to D_s^{*-} \rho(770)^{+} )$ & \begin{tabular}{l} \babar \cite{Aubert:2008zi}: $0.41 \,^{+0.13}_{-0.12} \pm 0.04$ \\ \end{tabular} & $0.41 \,^{+0.14}_{-0.13}$ \\
\hline
${\cal{B}} ( \bar{B}^{0} \to D_s^{-} a_0(980)^{+} )$ & \begin{tabular}{l} \babar \cite{Aubert:2005qt}: $0.06 \,^{+0.14}_{-0.11} \pm 0.01$ \\ \end{tabular} & $0.06 \,^{+0.14}_{-0.11}$ \\
\hline
${\cal{B}} ( \bar{B}^{0} \to D_s^{*-} a_0(980)^{+} )$ & \begin{tabular}{l} \babar \cite{Aubert:2005qt}: $0.14 \,^{+0.21}_{-0.16} \pm 0.03$ \\ \end{tabular} & $0.14 \,^{+0.21}_{-0.16}$ \\
\hline
${\cal{B}} ( \bar{B}^{0} \to D_s^{-} a_2(1320)^{+} )$ & \begin{tabular}{l} \babar \cite{Aubert:2005qt}: $0.64 \,^{+1.04}_{-0.57} \pm 0.15$ \\ \end{tabular} & $0.64 \,^{+1.05}_{-0.59}$ \\
\hline
${\cal{B}} ( \bar{B}^{0} \to D_s^{*-} a_2(1320)^{+} )$ & \begin{tabular}{l} \babar \cite{Aubert:2005qt}: $< 2.0$ \\ \end{tabular} & $< 2.0$ \\
\hline
${\cal{B}} ( \bar{B}^{0} \to D_s^{+} K^{-} )$ & \begin{tabular}{l} Belle \cite{Das:2010be}: $0.191 \pm 0.024 \pm 0.017$ \\ \babar \cite{Aubert:2008zi}: $0.29 \pm 0.04 \pm 0.02$ \\ \end{tabular} & $0.221 \pm 0.025$ \\
\hline
${\cal{B}} ( \bar{B}^{0} \to D_s^{*+} K^{-} )$ & \begin{tabular}{l} Belle \cite{Joshi:2009yv}: $0.202 \pm 0.033 \pm 0.022$ \\ \babar \cite{Aubert:2008zi}: $0.24 \pm 0.04 \pm 0.02$ \\ \end{tabular} & $0.219 \pm 0.031$ \\
\hline
${\cal{B}} ( \bar{B}^{0} \to D_s^{+} K^{*}(892)^{-} )$ & \begin{tabular}{l} \babar \cite{Aubert:2008zi}: $0.35 \,^{+0.10}_{-0.09} \pm 0.04$ \\ \end{tabular} & $0.35 \,^{+0.11}_{-0.10}$ \\
\hline
${\cal{B}} ( \bar{B}^{0} \to D_s^{*+} K^{*}(892)^{-} )$ & \begin{tabular}{l} \babar \cite{Aubert:2008zi}: $0.32 \,^{+0.14}_{-0.12} \pm 0.04$ \\ \end{tabular} & $0.32 \,^{+0.15}_{-0.13}$ \\
\hline
${\cal{B}} ( \bar{B}^{0} \to D_s^{+} K_S^{0} \pi^{-} )$ & \begin{tabular}{l} \babar \cite{Aubert:2007xma}: $0.55 \pm 0.13 \pm 0.10$ \\ \end{tabular} & $0.55 \pm 0.17$ \\
\hline
${\cal{B}} ( \bar{B}^{0} \to D_s^{*+} K^{0} \pi^{-} )$ & \begin{tabular}{l} \babar \cite{Aubert:2007xma}: $< 0.55$ \\ \end{tabular} & $< 0.55$ \\
\hline
\end{btocharmtab}
\btocharmfig{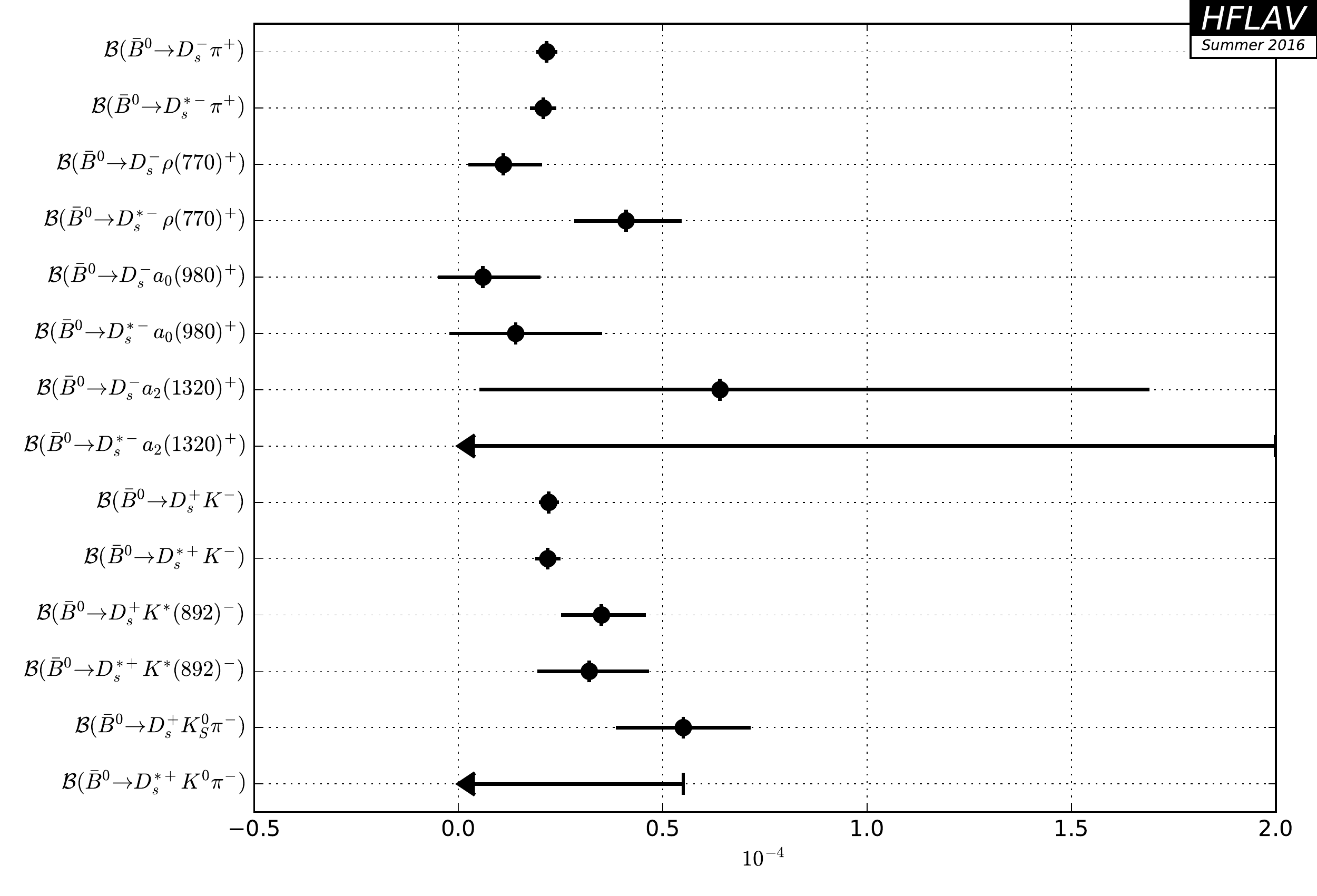}

\begin{btocharmtab}{Bd_D_7}{Relative decay rates I}
\hline
\textbf{Parameter} & \begin{tabular}{l}\textbf{Measurements}\end{tabular} & \textbf{Average} \\
\hline
\hline
\multicolumn{3}{|l|}{${{\cal{B}} ( \bar{B}^{0} \to D^{0} \rho(770)^{0} )}/{{\cal{B}} ( \bar{B}^{0} \to D^{0} \omega(782) )}$}\\
 & \begin{tabular}{l} Belle \cite{Satpathy:2002js}: $1.6 \pm 0.8$ \\ \end{tabular} & $1.6 \pm 0.8$ \\
\hline
\multicolumn{3}{|l|}{${{\cal{B}} ( \bar{B}^{0} \to D^{+} \pi^{+} \pi^{-} \pi^{-} ) }/{ {\cal{B}} ( \bar{B}^{0} \to D^{+} \pi^{-} )}$}\\
 & \begin{tabular}{l} LHCb \cite{Aaij:2011rj}: $2.38 \pm 0.11 \pm 0.21$ \\ \end{tabular} & $2.38 \pm 0.24$ \\
\hline
\multicolumn{3}{|l|}{${{\cal{B}} ( \bar{B}^{0} \to D^{*}(2010)^{+} \pi^{-} )}/{{\cal{B}} ( \bar{B}^{0} \to D^{+} \pi^{-} )}$}\\
 & \begin{tabular}{l} \babar \cite{Aubert:2006jc}: $0.99 \pm 0.11 \pm 0.08$ \\ \end{tabular} & $0.99 \pm 0.14$ \\
\hline
\multicolumn{3}{|l|}{${{\cal{B}} ( \bar{B}^{0} \to D^{**+} \pi^{-} )}/{{\cal{B}} ( \bar{B}^{0} \to D^{+} \pi^{-} )}$}\\
 & \begin{tabular}{l} \babar \cite{Aubert:2006jc}: $0.77 \pm 0.22 \pm 0.29$ \\ \end{tabular} & $0.77 \pm 0.36$ \\
\hline
\multicolumn{3}{|l|}{${{\cal{B}} ( \bar{B}^{0} \to D_s^{+} K^{-}  \pi^{+}  \pi^{-} )}/{{\cal{B}} ( \bar{B}_s^{0} \to D_s^{+} K^{-}  \pi^{+}  \pi^{-} )}$}\\
 & \begin{tabular}{l} LHCb \cite{Aaij:2012mra}: $0.54 \pm 0.07 \pm 0.07$ \\ \end{tabular} & $0.54 \pm 0.10$ \\
\hline
\end{btocharmtab}
\btocharmfig{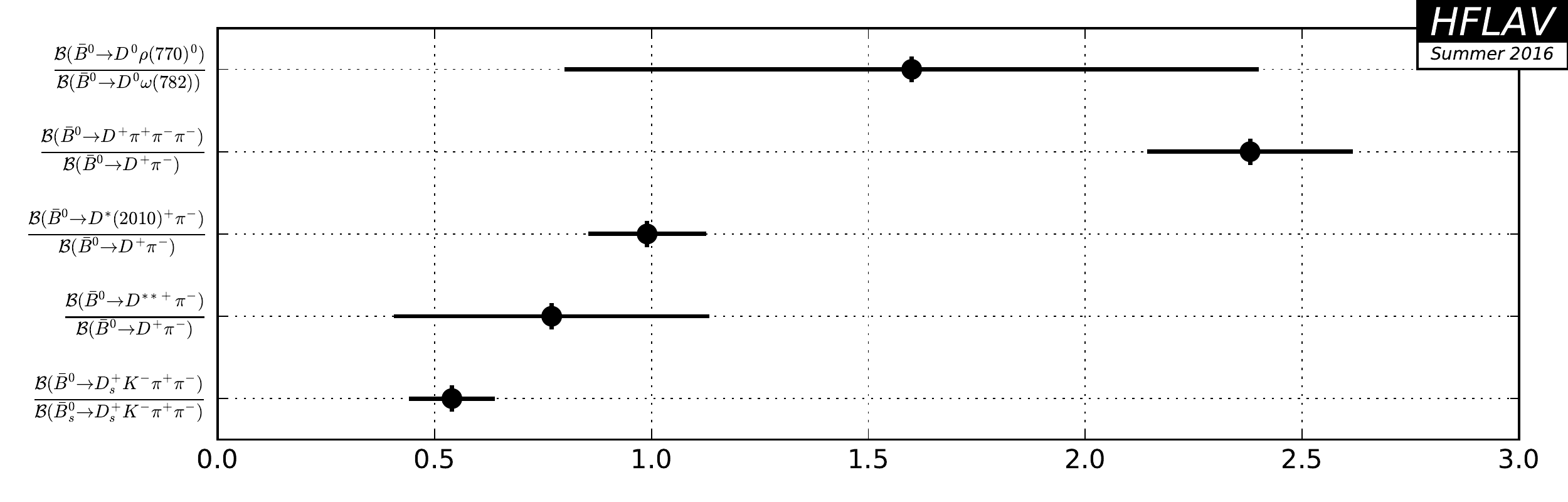}

\begin{btocharmtab}{Bd_D_8}{Relative decay rates II}
\hline
\textbf{Parameter} & \begin{tabular}{l}\textbf{Measurements}\end{tabular} & \textbf{Average} \\
\hline
\hline
\multicolumn{3}{|l|}{${ {\cal{B}} ( \bar{B}^{0} \to D^{0} K^{-} \pi^{+} )}/  {{\cal{B}} ( \bar{B}^{0} \to D^{0} \pi^{-} \pi^{+} )}$}\\
 & \begin{tabular}{l} LHCb \cite{Aaij:2013pua}: $0.106 \pm 0.007 \pm 0.008$ \\ \end{tabular} & $0.106 \pm 0.011$ \\
\hline
\multicolumn{3}{|l|}{${ {\cal{B}} ( \bar{B}^{0} \to \bar{D}^{0} K^{-} K^{+} )}/  {{\cal{B}} ( \bar{B}^{0} \to \bar{D}^{0} \pi^{-} \pi^{+} )}$}\\
 & \begin{tabular}{l} LHCb \cite{Aaij:2012zka}: $0.056 \pm 0.110 \pm 0.007$ \\ \end{tabular} & $0.056 \pm 0.110$ \\
\hline
\multicolumn{3}{|l|}{${{\cal{B}} ( \bar{B}^{0} \to D^{+} K^{-} )}/{{\cal{B}} ( \bar{B}^{0} \to D^{+} \pi^{-} )}$}\\
 & \begin{tabular}{l} LHCb \cite{Aaij:2013qqa}: $0.0822 \pm 0.0011 \pm 0.0025$ \\ Belle \cite{Abe:2001waa}: $0.068 \pm 0.015 \pm 0.007$ \\ \end{tabular} & $0.0818 \pm 0.0027$ \\
\hline
\multicolumn{3}{|l|}{${{\cal{B}} ( \bar{B}^{0} \to D^{+} K^{-} \pi^{+} \pi^{-} ) }/{ {\cal{B}} ( \bar{B}^{0} \to D^{+} \pi^{+} \pi^{-} \pi^{-} )}$}\\
 & \begin{tabular}{l} LHCb \cite{Aaij:2012bw}: $0.059 \pm 0.011 \pm 0.005$ \\ \end{tabular} & $0.059 \pm 0.012$ \\
\hline
\multicolumn{3}{|l|}{${{\cal{B}} ( \bar{B}^{0} \to D^{*}(2010)^{+} K^{-} )}/{{\cal{B}} ( \bar{B}^{0} \to D^{*}(2010)^{+} \pi^{-} )}$}\\
 & \begin{tabular}{l} Belle \cite{Abe:2001waa}: $0.074 \pm 0.015 \pm 0.006$ \\ \babar \cite{Aubert:2005yt}: $0.0776 \pm 0.0034 \pm 0.0029$ \\ \end{tabular} & $0.0773 \pm 0.0043$ \\
\hline
\multicolumn{3}{|l|}{${{\cal{B}} ( \bar{B}^{0} \to D_s^{+} K^{-} )}/{{\cal{B}} ( \bar{B}^{0} \to D^{+} \pi^{-} )}$}\\
 & \begin{tabular}{l} LHCb \cite{Aaij:2014jpa}: $0.0129 \pm 0.0005 \pm 0.0008$ \\ \end{tabular} & $0.0129 \pm 0.0009$ \\
\hline
\end{btocharmtab}
\btocharmfig{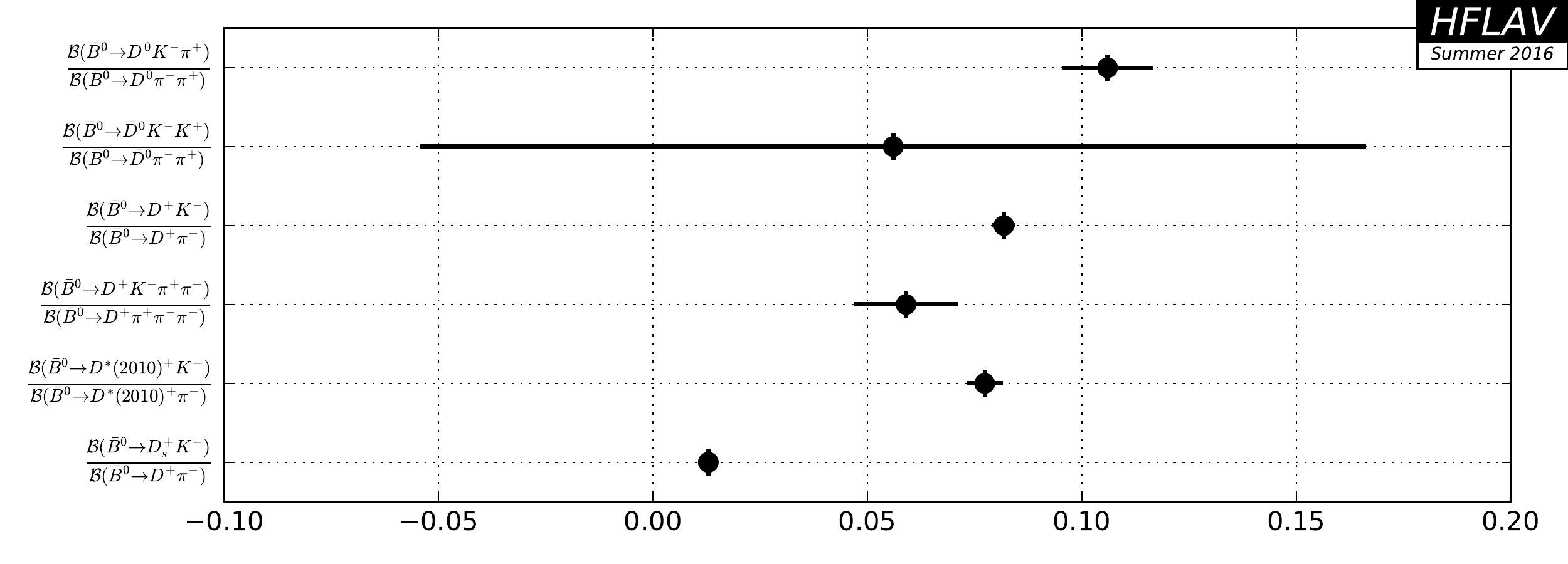}

\begin{btocharmtab}{Bd_D_9}{Absolute product decay rates to excited $D$ mesons I $[10^{-4}]$}
\hline
\textbf{Parameter} & \begin{tabular}{l}\textbf{Measurements}\end{tabular} & \textbf{Average} \\
\hline
\hline
\multicolumn{3}{|l|}{${{\cal{B}} ( \bar{B}^{0} \to D_1(2420)^{+} \pi^{-} )\times {\cal{B}} ( D_1(2420)^{+} \to D^{+} \pi^{-} \pi^{+} )}$}\\
 & \begin{tabular}{l} Belle \cite{Abe:2004sm}: $0.89 \pm 0.15 \,^{+0.17}_{-0.31}$ \\ \end{tabular} & $0.89 \,^{+0.23}_{-0.34}$ \\
\hline
\multicolumn{3}{|l|}{${{\cal{B}} ( \bar{B}^{0} \to D_1^{0}(H) \omega(782) )\times {\cal{B}} ( D_1^{0}(H) \to D^{*}(2010)^{+} \pi^{-} )}$}\\
 & \begin{tabular}{l} \babar \cite{Aubert:2006zb}: $4.1 \pm 1.2 \pm 1.1$ \\ \end{tabular} & $4.1 \pm 1.6$ \\
\hline
\end{btocharmtab}
\btocharmfig{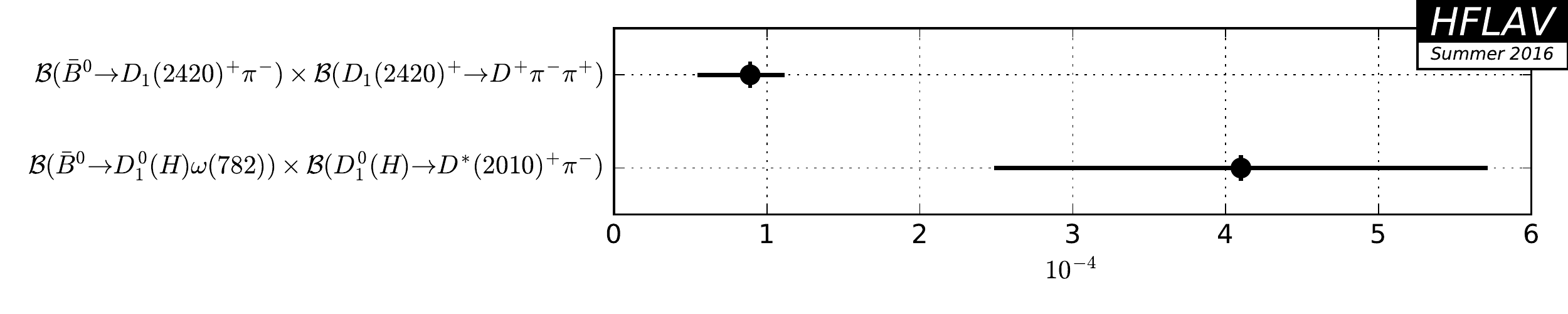}

\begin{btocharmtab}{Bd_D_10}{Absolute product decay rates to excited $D$ mesons II $[10^{-5}]$}
\hline
\textbf{Parameter} & \begin{tabular}{l}\textbf{Measurements}\end{tabular} & \textbf{Average} \\
\hline
\hline
\multicolumn{3}{|l|}{${{\cal{B}} ( \bar{B}^{0} \to D_1(2420)^{+} \pi^{-} )\times {\cal{B}} ( D_1(2420)^{+} \to D^{*}(2010)^{+} \pi^{-} \pi^{+} )}$}\\
 & \begin{tabular}{l} Belle \cite{Abe:2004sm}: $< 3.3$ \\ \end{tabular} & $< 3.3$ \\
\hline
\multicolumn{3}{|l|}{${{\cal{B}} ( \bar{B}^{0} \to D_2^{*}(2460)^{+} \pi^{-} )\times {\cal{B}} ( D_2^{*}(2460)^{+} \to D^{*}(2010)^{+} \pi^{-} \pi^{+} )}$}\\
 & \begin{tabular}{l} Belle \cite{Abe:2004sm}: $< 2.4$ \\ \end{tabular} & $< 2.4$ \\
\hline
\multicolumn{3}{|l|}{${{\cal{B}} ( \bar{B}^{0} \to D_2^{*}(2460)^{+} K^{-} )\times {\cal{B}} ( D_2^{*}(2460)^{+} \to D^{0} \pi^{+} )}$}\\
 & \begin{tabular}{l} \babar \cite{Aubert:2005yt}: $1.83 \pm 0.40 \pm 0.31$ \\ \end{tabular} & $1.83 \pm 0.51$ \\
\hline
\multicolumn{3}{|l|}{${{\cal{B}} ( \bar{B}^{0} \to D_{sJ}(2460)^{-} \pi^{+} )\times {\cal{B}} ( D_{sJ}(2460)^{-} \to D_s^{-} \gamma )}$}\\
 & \begin{tabular}{l} Belle \cite{Abe:2004wz}: $< 0.40$ \\ \end{tabular} & $< 0.40$ \\
\hline
\multicolumn{3}{|l|}{${{\cal{B}} ( \bar{B}^{0} \to D_{sJ}^{+}(2460) K^{-} )\times {\cal{B}} ( D_{sJ}^{+}(2460) \to D_s^{+} \gamma )}$}\\
 & \begin{tabular}{l} Belle \cite{Abe:2004wz}: $< 0.94$ \\ \end{tabular} & $< 0.94$ \\
\hline
\multicolumn{3}{|l|}{${{\cal{B}} ( \bar{B}^{0} \to D_{sJ}^{*}(2317)^{-} \pi^{+} )\times {\cal{B}} ( D_{sJ}^{*}(2317)^{-} \to D_s^{-} \pi^{0} )}$}\\
 & \begin{tabular}{l} Belle \cite{Abe:2004wz}: $< 2.5$ \\ \end{tabular} & $< 2.5$ \\
\hline
\multicolumn{3}{|l|}{${{\cal{B}} ( \bar{B}^{0} \to D_{sJ}^{*}(2317)^{+} K^{-} )\times {\cal{B}} ( D_{sJ}^{*}(2317)^{+} \to D_s^{+} \pi^{0} )}$}\\
 & \begin{tabular}{l} Belle \cite{Abe:2004wz}: $5.3 \,^{+1.5}_{-1.3} \pm 1.6$ \\ \end{tabular} & $5.3 \,^{+2.2}_{-2.0}$ \\
\hline
\end{btocharmtab}
\btocharmfig{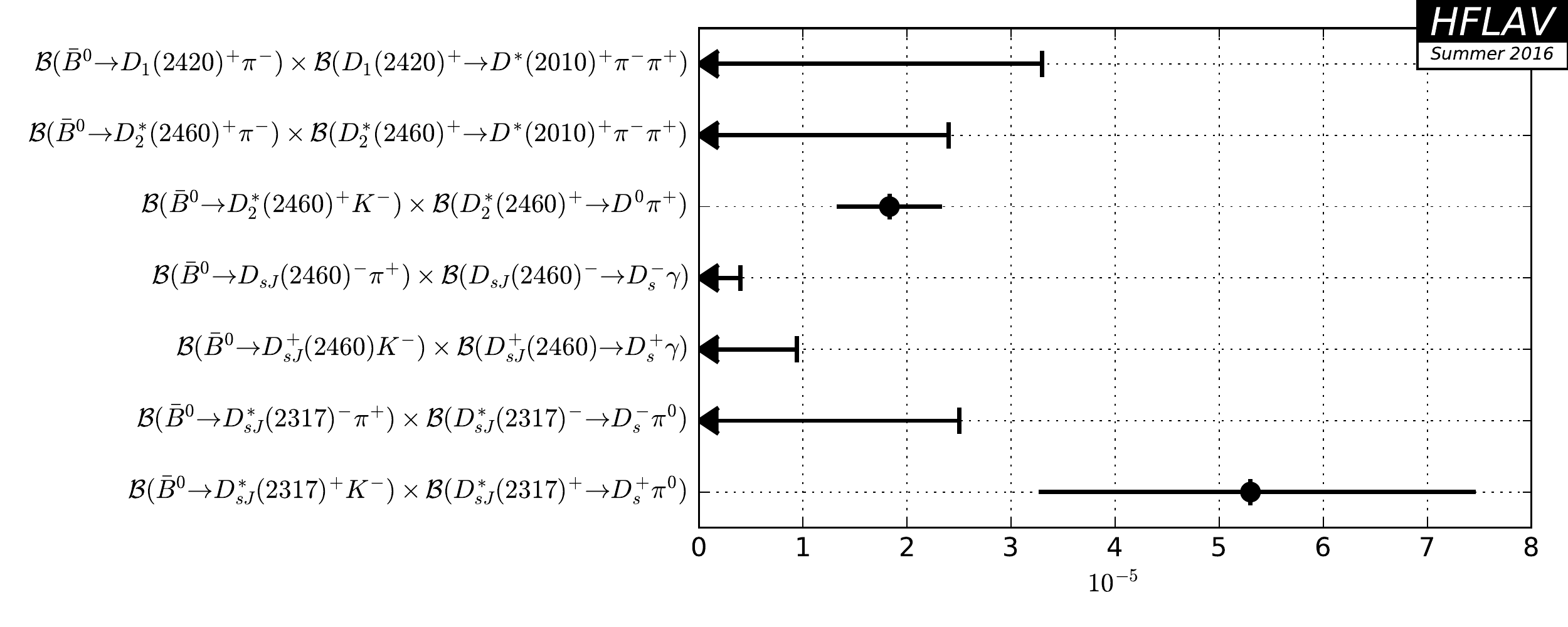}

\begin{btocharmtab}{Bd_D_11}{Absolute and relative decay rates to excited $D$ mesons $[10^{-2}]$}
\hline
\textbf{Parameter} & \begin{tabular}{l}\textbf{Measurements}\end{tabular} & \textbf{Average} \\
\hline
\hline
${\cal{B}} ( \bar{B}^{0} \to D^{**+} \pi^{-} )$ & \begin{tabular}{l} \babar \cite{Aubert:2006jc}: $0.234 \pm 0.065 \pm 0.088$ \\ \end{tabular} & $0.234 \pm 0.109$ \\
\hline
\multicolumn{3}{|l|}{$[{{\cal{B}} ( \bar{B}^{0} \to D_1^{+} \pi^{-} ) \times {\cal{B}} ( D_1^{+} \to D^{+} \pi^{+} \pi^{-} ) }]/{ {\cal{B}} ( \bar{B}^{0} \to D^{+} \pi^{+} \pi^{-} \pi^{-} )}$}\\
 & \begin{tabular}{l} LHCb \cite{Aaij:2011rj}: $2.1 \pm 0.5 \,^{+0.3}_{-0.5}$ \\ \end{tabular} & $2.1 \,^{+0.6}_{-0.7}$ \\
\hline
\end{btocharmtab}
\btocharmfig{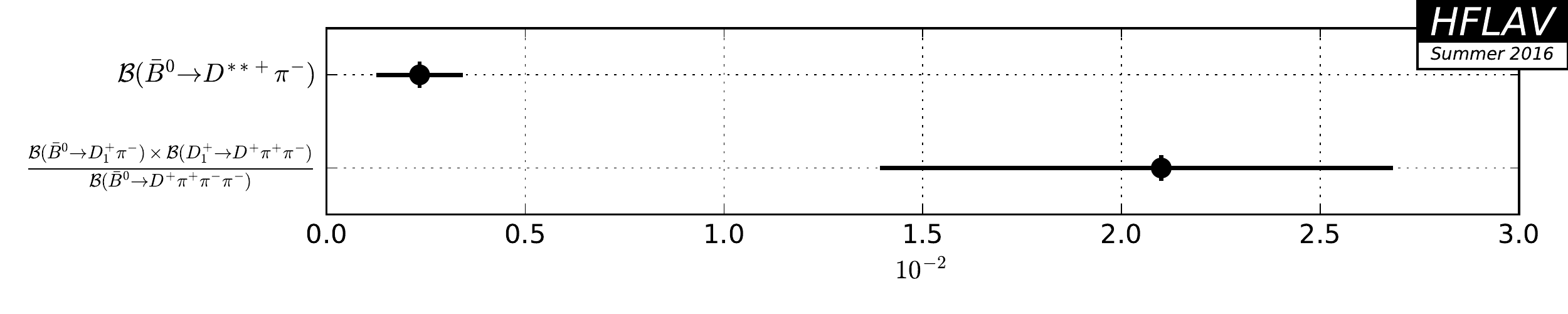}

\begin{btocharmtab}{Bd_D_12}{Baryonic decays I $[10^{-4}]$}
\hline
\textbf{Parameter} & \begin{tabular}{l}\textbf{Measurements}\end{tabular} & \textbf{Average} \\
\hline
\hline
${\cal{B}} ( \bar{B}^{0} \to D^{+} p \bar{p} \pi^{-} )$ & \begin{tabular}{l} \babar \cite{delAmoSanchez:2011gi}: $3.32 \pm 0.10 \pm 0.29$ \\ \end{tabular} & $3.32 \pm 0.31$ \\
\hline
${\cal{B}} ( \bar{B}^{0} \to D^{*}(2010)^{+} p \bar{p} \pi^{-} )$ & \begin{tabular}{l} \babar \cite{delAmoSanchez:2011gi}: $4.55 \pm 0.16 \pm 0.39$ \\ \end{tabular} & $4.55 \pm 0.42$ \\
\hline
${\cal{B}} ( \bar{B}^{0} \to D^{0} p \bar{p} \pi^{-} \pi^{+} )$ & \begin{tabular}{l} \babar \cite{delAmoSanchez:2011gi}: $2.99 \pm 0.21 \pm 0.45$ \\ \end{tabular} & $2.99 \pm 0.50$ \\
\hline
${\cal{B}} ( \bar{B}^{0} \to D^{*}(2007)^{0} p \bar{p} \pi^{-} \pi^{+} )$ & \begin{tabular}{l} \babar \cite{delAmoSanchez:2011gi}: $1.91 \pm 0.36 \pm 0.29$ \\ \end{tabular} & $1.91 \pm 0.46$ \\
\hline
\end{btocharmtab}
\btocharmfig{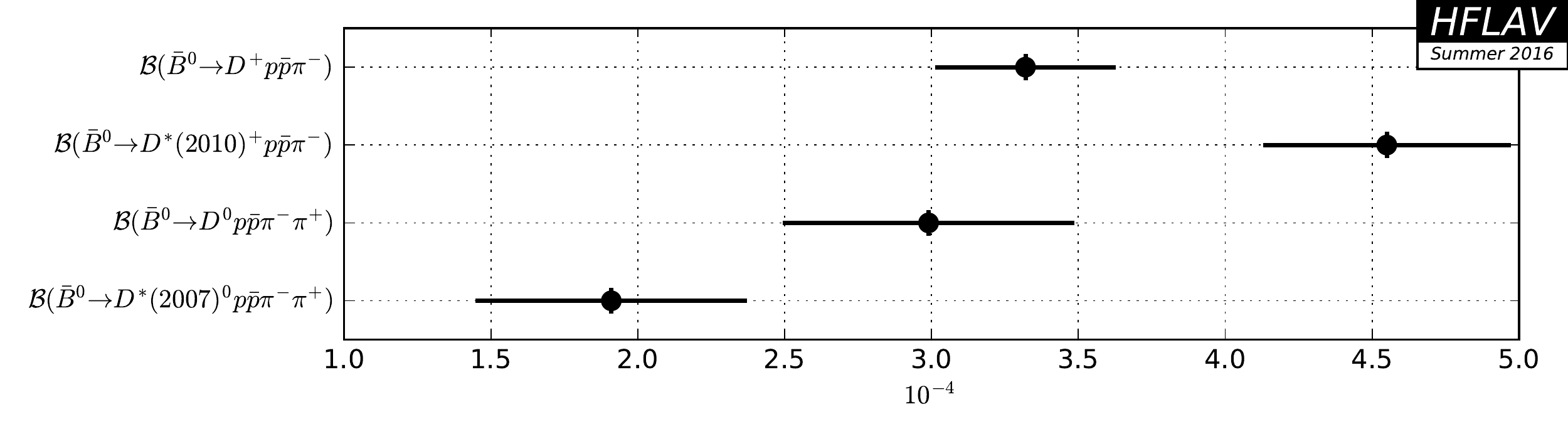}

\begin{btocharmtab}{Bd_D_13}{Baryonic decays II $[10^{-5}]$}
\hline
\textbf{Parameter} & \begin{tabular}{l}\textbf{Measurements}\end{tabular} & \textbf{Average} \\
\hline
\hline
${\cal{B}} ( \bar{B}^{0} \to D^{0} p \bar{p} )$ & \begin{tabular}{l} Belle \cite{Abe:2002tw}: $11.8 \pm 1.5 \pm 1.6$ \\ \babar \cite{delAmoSanchez:2011gi}: $10.2 \pm 0.4 \pm 0.6$ \\ \end{tabular} & $10.4 \pm 0.7$ \\
\hline
${\cal{B}} ( \bar{B}^{0} \to D^{*}(2007)^{0} p \bar{p} )$ & \begin{tabular}{l} Belle \cite{Abe:2002tw}: $12.0 \,^{+3.3}_{-2.9} \pm 2.1$ \\ \babar \cite{delAmoSanchez:2011gi}: $9.7 \pm 0.7 \pm 0.9$ \\ \end{tabular} & $9.9 \pm 1.1$ \\
\hline
${\cal{B}} ( \bar{B}^{0} \to D_s^{+} \Lambda \bar{p} )$ & \begin{tabular}{l} Belle \cite{Medvedeva:2007af}: $2.9 \pm 0.7 \pm 0.6$ \\ \end{tabular} & $2.9 \pm 0.9$ \\
\hline
${\cal{B}} ( \bar{B}^{0} \to D^{0} \Lambda^{0} \bar{\Lambda}^{0} )$ & \begin{tabular}{l} Belle \cite{Chang:2008yw}: $1.05 \,^{+0.57}_{-0.44} \pm 0.14$ \\ \babar \cite{Lees:2014mka}: $0.98 \,^{+0.29}_{-0.26} \pm 0.19$ \\ \end{tabular} & $1.00 \pm 0.28$ \\
\hline
\multicolumn{3}{|l|}{${\cal{B}}(\bar{B}^{0}\to D^{0}\Sigma^{0}\bar\Lambda+\bar{B}^{0}\to D^{0}\Lambda\bar\Sigma^{0})$}\\
 & \begin{tabular}{l} \babar \cite{Lees:2014mka}: $1.5 \,^{+0.9}_{-0.8} \pm 0.3$ \\ \end{tabular} & $1.5 \,^{+0.9}_{-0.9}$ \\
\hline
${\cal{B}} ( \bar{B}^{0} \to D^{+} \Lambda \bar{p} )$ & \begin{tabular}{l} Belle \cite{Chang:2015fja}: $3.36 \pm 0.63 \pm 0.44$ \\ \end{tabular} & $3.36 \pm 0.77$ \\
\hline
${\cal{B}} ( \bar{B}^{0} \to D^{*+} \Lambda \bar{p} )$ & \begin{tabular}{l} Belle \cite{Chang:2015fja}: $2.51 \pm 0.26 \pm 0.35$ \\ \end{tabular} & $2.51 \pm 0.44$ \\
\hline
\end{btocharmtab}
\btocharmfig{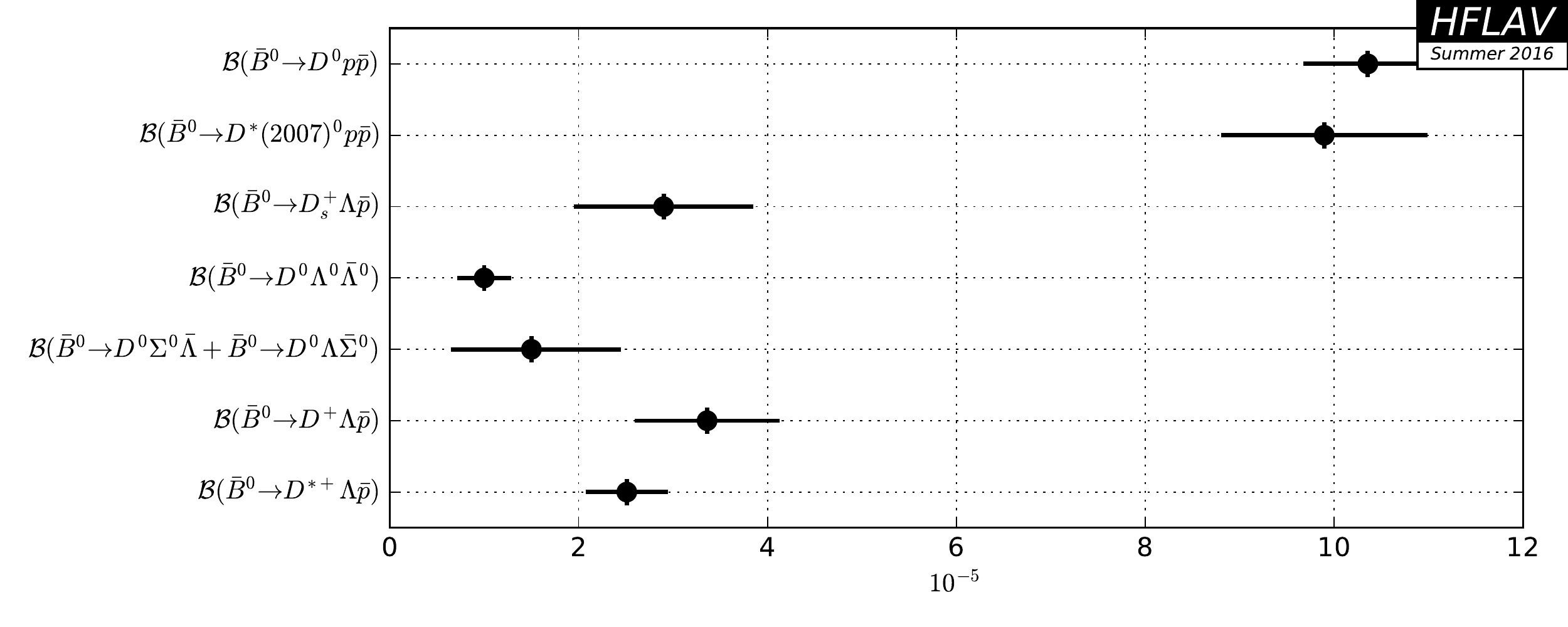}

\subsubsection{Decays to two open charm mesons}
\label{sec:b2c:Bd_DD}
Averages of $\bar{B}^0$ decays to two open charm mesons are shown in Tables~\ref{tab:b2c:Bd_DD_1}--\ref{tab:b2c:Bd_DD_10} and Figs.~\ref{fig:b2c:Bd_DD_1}--\ref{fig:b2c:Bd_DD_10}.
\begin{btocharmtab}{Bd_DD_1}{Decays to $D^{(*)+}D^{(*)-}$ $[10^{-3}]$}
\hline
\textbf{Parameter} & \begin{tabular}{l}\textbf{Measurements}\end{tabular} & \textbf{Average} \\
\hline
\hline
${\cal{B}} ( \bar{B}^{0} \to D^{+} D^{-} )$ & \begin{tabular}{l} Belle \cite{Rohrken:2012ta}: $0.212 \pm 0.016 \pm 0.018$ \\ \babar \cite{Aubert:2006ia}: $0.28 \pm 0.04 \pm 0.05$ \\ \end{tabular} & $0.220 \pm 0.023$ \\
\hline
${\cal{B}} ( \bar{B}^{0} \to D^{*}(2010)^{-} D^{+} )$ & \begin{tabular}{l} Belle \cite{Rohrken:2012ta}: $0.614 \pm 0.029 \pm 0.050$ \\ \babar \cite{Aubert:2006ia}: $0.57 \pm 0.07 \pm 0.07$ \\ \end{tabular} & $0.603 \pm 0.050$ \\
\hline
${\cal{B}} ( \bar{B}^{0} \to D^{*}(2010)^{-} D^{*}(2010)^{+} )$ & \begin{tabular}{l} Belle \cite{Kronenbitter:2012ha}: $0.782 \pm 0.038 \pm 0.060$ \\ \babar \cite{Aubert:2006ia}: $0.81 \pm 0.06 \pm 0.10$ \\ \end{tabular} & $0.790 \pm 0.061$ \\
\hline
${\cal{B}} ( \bar{B}^{0} \to D^{0} \bar{D}^{0} )$ & \begin{tabular}{l} Belle \cite{Adachi:2008cj}: $< 0.043$ \\ \babar \cite{Aubert:2006ia}: $< 0.06$ \\ \end{tabular} & $< 0.043$ \\
\hline
${\cal{B}} ( \bar{B}^{0} \to D^{0} \bar{D}^{*}(2007)^{0} )$ & \begin{tabular}{l} \babar \cite{Aubert:2006ia}: $< 0.29$ \\ \end{tabular} & $< 0.29$ \\
\hline
${\cal{B}} ( \bar{B}^{0} \to D^{*}(2007)^{0} \bar{D}^{*}(2007)^{0} )$ & \begin{tabular}{l} \babar \cite{Aubert:2006ia}: $< 0.09$ \\ \end{tabular} & $< 0.09$ \\
\hline
\end{btocharmtab}
\btocharmfig{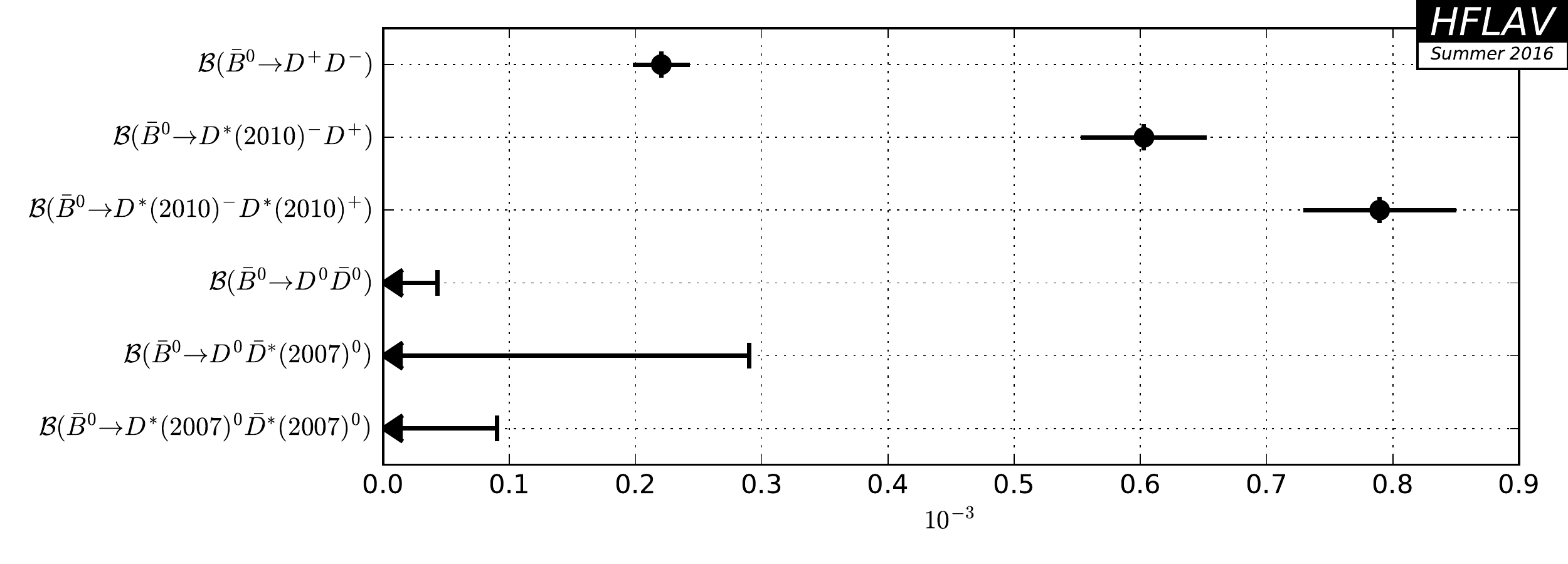}

\begin{btocharmtab}{Bd_DD_2}{Decays to two $D$ mesons and a kaon I $[10^{-3}]$}
\hline
\textbf{Parameter} & \begin{tabular}{l}\textbf{Measurements}\end{tabular} & \textbf{Average} \\
\hline
\hline
${\cal{B}} ( \bar{B}^{0} \to D^{*}(2010)^{+} D^{-} \bar{K}^{0} )$ & \begin{tabular}{l} \babar \cite{delAmoSanchez:2010pg}: $6.41 \pm 0.36 \pm 0.39$ \\ \end{tabular} & $6.41 \pm 0.53$ \\
\hline
${\cal{B}} ( \bar{B}^{0} \to D^{*}(2010)^{-} D^{*}(2010)^{+} \bar{K}^{0} )$ & \begin{tabular}{l} \babar \cite{delAmoSanchez:2010pg}: $8.26 \pm 0.43 \pm 0.67$ \\ \end{tabular} & $8.26 \pm 0.80$ \\
\hline
${\cal{B}} ( \bar{B}^{0} \to D^{*}(2010)^{+} D^{*}(2010)^{-} K_S^{0} )$ & \begin{tabular}{l} Belle \cite{Dalseno:2007hx}: $3.4 \pm 0.4 \pm 0.7$ \\ \babar \cite{Aubert:2006fh}: $4.4 \pm 0.4 \pm 0.7$ \\ \end{tabular} & $3.9 \pm 0.6$ \\
\hline
${\cal{B}} ( \bar{B}^{0} \to D^{*}(2010)^{+} \bar{D}^{0} K^{-} )$ & \begin{tabular}{l} \babar \cite{delAmoSanchez:2010pg}: $2.47 \pm 0.10 \pm 0.18$ \\ \end{tabular} & $2.47 \pm 0.21$ \\
\hline
${\cal{B}} ( \bar{B}^{0} \to D^{+} \bar{D}^{*}(2007)^{0} K^{-} )$ & \begin{tabular}{l} \babar \cite{delAmoSanchez:2010pg}: $3.46 \pm 0.18 \pm 0.37$ \\ \end{tabular} & $3.46 \pm 0.41$ \\
\hline
${\cal{B}} ( \bar{B}^{0} \to D^{*}(2010)^{+} \bar{D}^{*}(2007)^{0} K^{-} )$ & \begin{tabular}{l} \babar \cite{delAmoSanchez:2010pg}: $10.6 \pm 0.3 \pm 0.9$ \\ \end{tabular} & $10.6 \pm 0.9$ \\
\hline
${\cal{B}} ( \bar{B}^{0} \to D^{0} \bar{D}^{*}(2007)^{0} \bar{K}^{0} )$ & \begin{tabular}{l} \babar \cite{delAmoSanchez:2010pg}: $1.08 \pm 0.32 \pm 0.36$ \\ \end{tabular} & $1.08 \pm 0.48$ \\
\hline
${\cal{B}} ( \bar{B}^{0} \to D^{*}(2007)^{0} \bar{D}^{*}(2007)^{0} \bar{K}^{0} )$ & \begin{tabular}{l} \babar \cite{delAmoSanchez:2010pg}: $2.40 \pm 0.55 \pm 0.67$ \\ \end{tabular} & $2.40 \pm 0.87$ \\
\hline
\end{btocharmtab}
\btocharmfig{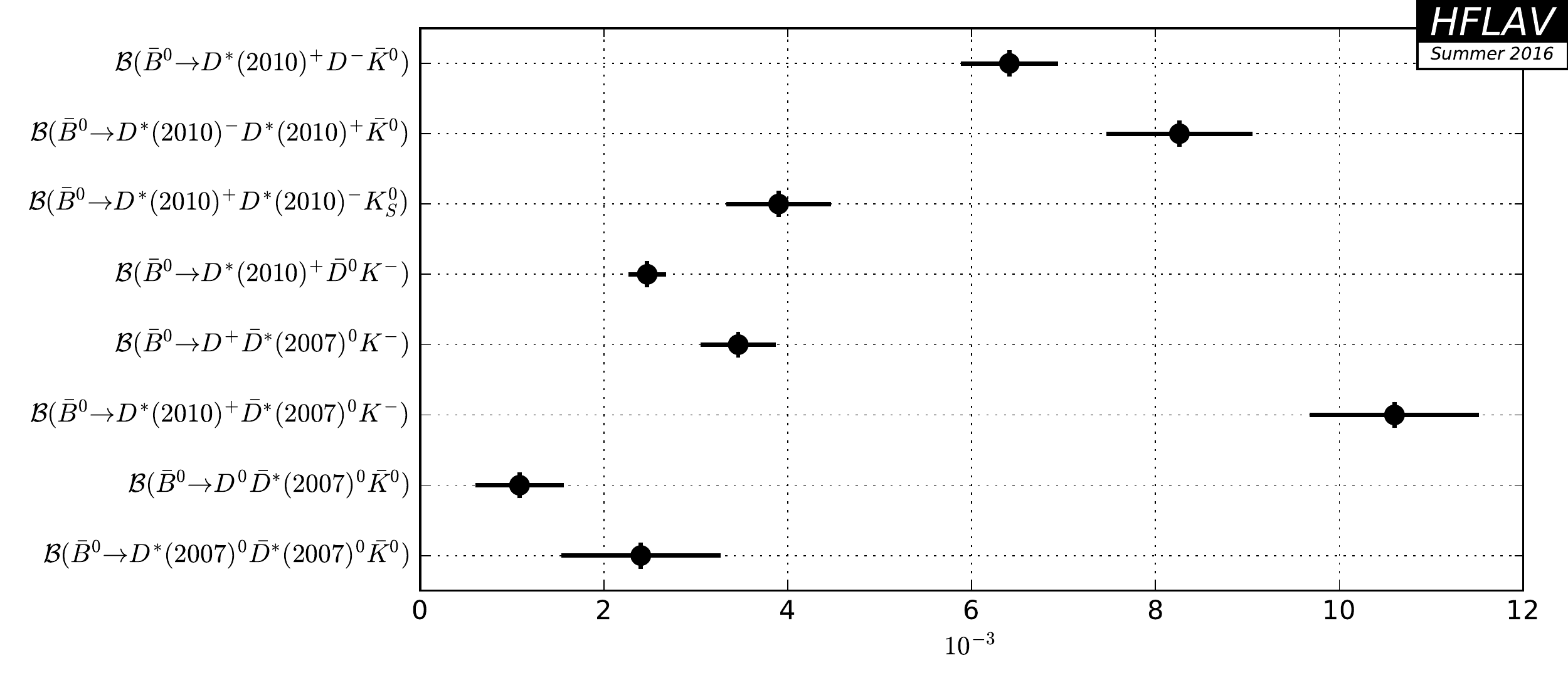}

\begin{btocharmtab}{Bd_DD_3}{Decays to two $D$ mesons and a kaon II $[10^{-4}]$}
\hline
\textbf{Parameter} & \begin{tabular}{l}\textbf{Measurements}\end{tabular} & \textbf{Average} \\
\hline
\hline
${\cal{B}} ( \bar{B}^{0} \to D^{+} D^{-} \bar{K}^{0} )$ & \begin{tabular}{l} \babar \cite{delAmoSanchez:2010pg}: $7.5 \pm 1.2 \pm 1.2$ \\ \end{tabular} & $7.5 \pm 1.7$ \\
\hline
${\cal{B}} ( \bar{B}^{0} \to D^{+} \bar{D}^{0} K^{-} )$ & \begin{tabular}{l} \babar \cite{delAmoSanchez:2010pg}: $10.7 \pm 0.7 \pm 0.9$ \\ \end{tabular} & $10.7 \pm 1.1$ \\
\hline
${\cal{B}} ( \bar{B}^{0} \to D^{0} \bar{D}^{0} \bar{K}^{0} )$ & \begin{tabular}{l} \babar \cite{delAmoSanchez:2010pg}: $2.7 \pm 1.0 \pm 0.5$ \\ \end{tabular} & $2.7 \pm 1.1$ \\
\hline
${\cal{B}} ( \bar{B}^{0} \to D^{0} \bar{D}^{0} \pi^{0} \bar{K}^{0} )$ & \begin{tabular}{l} Belle \cite{Gokhroo:2006bt}: $1.73 \pm 0.70 \,^{+0.31}_{-0.53}$ \\ \end{tabular} & $1.73 \,^{+0.77}_{-0.88}$ \\
\hline
\end{btocharmtab}
\btocharmfig{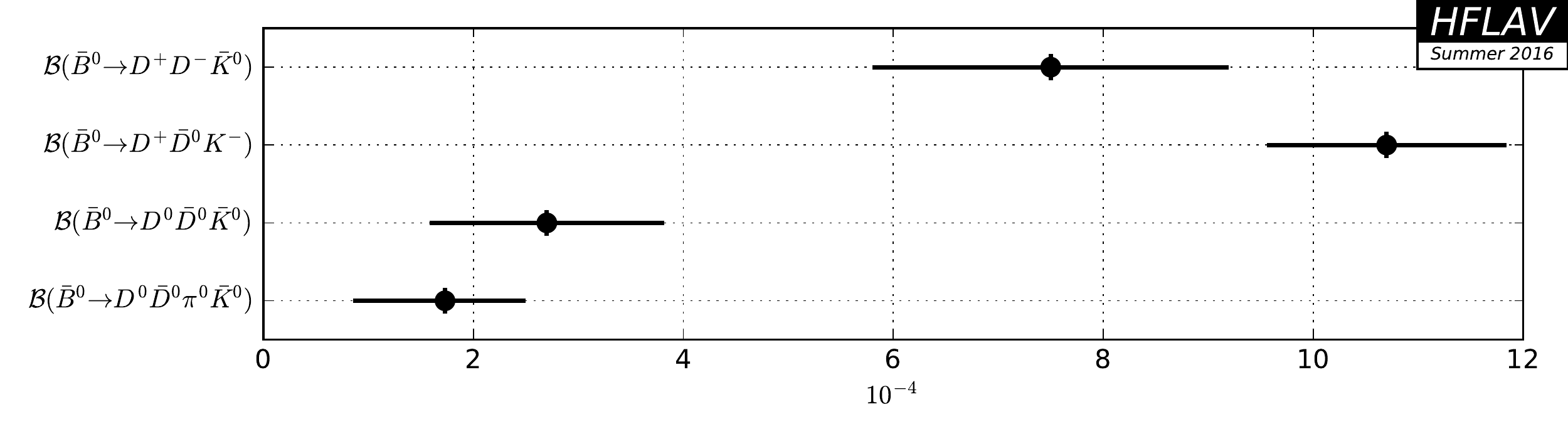}

\begin{btocharmtab}{Bd_DD_4}{Decays to $D_s^{(*)-}D^{(*)+}$ I $[10^{-3}]$}
\hline
\textbf{Parameter} & \begin{tabular}{l}\textbf{Measurements}\end{tabular} & \textbf{Average} \\
\hline
\hline
${\cal{B}} ( \bar{B}^{0} \to D_s^{-} D^{+} )$ & \begin{tabular}{l} Belle \cite{Zupanc:2007pu}: $7.5 \pm 0.2 \pm 1.1$ \\ \babar \cite{Aubert:2006nm}: $9.0 \pm 1.8 \pm 1.4$ \\ \end{tabular} & $7.8 \pm 1.0$ \\
\hline
${\cal{B}} ( \bar{B}^{0} \to D_s^{-} D^{*}(2010)^{+} )$ & \begin{tabular}{l} \babar \cite{Aubert:2006nm}: $5.7 \pm 1.6 \pm 0.9$ \\ \babar \cite{Aubert:2003jj}: $10.3 \pm 1.4 \pm 2.9$ \\ \end{tabular} & $6.8 \pm 1.6$ \\
\hline
${\cal{B}} ( \bar{B}^{0} \to D_s^{*-} D^{*}(2010)^{+} )$ & \begin{tabular}{l} \babar \cite{Aubert:2006nm}: $16.5 \pm 2.3 \pm 1.9$ \\ \babar \cite{Aubert:2005xu}: $18.8 \pm 0.9 \pm 1.7$ \\ \babar \cite{Aubert:2003jj}: $19.7 \pm 1.5 \pm 5.7$ \\ \end{tabular} & $18.1 \pm 1.6$ \\
\hline
${\cal{B}} ( \bar{B}^{0} \to D_s^{*-} D^{+} )$ & \begin{tabular}{l} \babar \cite{Aubert:2006nm}: $6.7 \pm 2.0 \pm 1.1$ \\ \end{tabular} & $6.7 \pm 2.3$ \\
\hline
\end{btocharmtab}
\btocharmfig{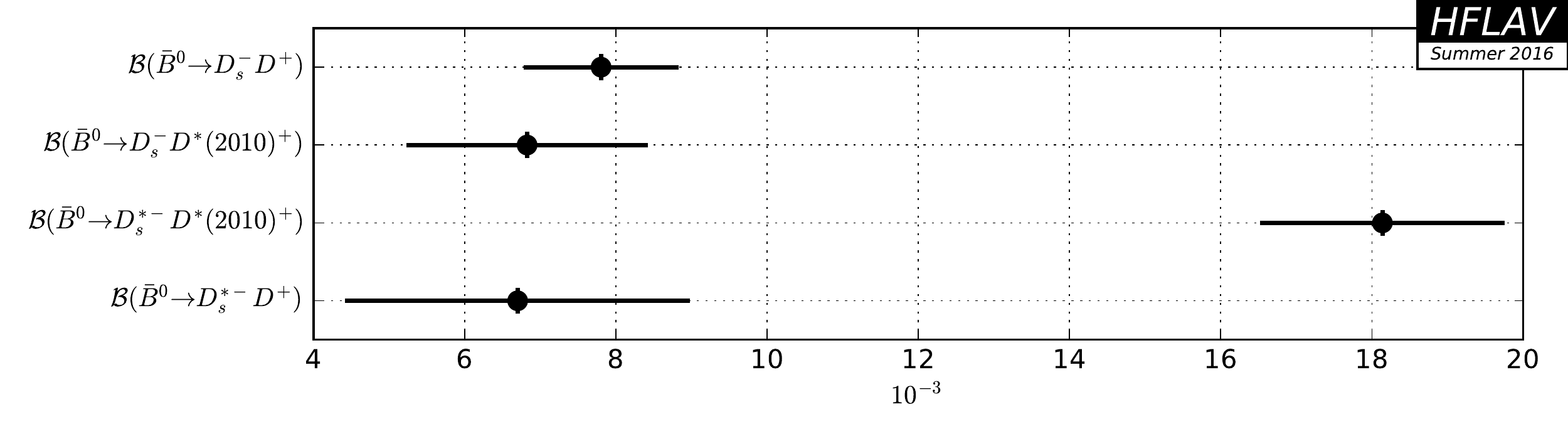}

\begin{btocharmtab}{Bd_DD_5}{Decays to $D_s^{(*)-}D^{(*)+}$ II $[10^{-4}]$}
\hline
\textbf{Parameter} & \begin{tabular}{l}\textbf{Measurements}\end{tabular} & \textbf{Average} \\
\hline
\hline
\multicolumn{3}{|l|}{${{\cal{B}} ( \bar{B}^{0} \to D_s^{-} D^{+} )\times {\cal{B}} ( D_s^{-} \to \phi(1020) \pi^{-} )}$}\\
 & \begin{tabular}{l} \babar \cite{Aubert:2006nm}: $2.67 \pm 0.61 \pm 0.47$ \\ \end{tabular} & $2.67 \pm 0.77$ \\
\hline
\multicolumn{3}{|l|}{${{\cal{B}} ( \bar{B}^{0} \to D_s^{-} D^{*}(2010)^{+} )\times {\cal{B}} ( D_s^{-} \to \phi(1020) \pi^{-} )}$}\\
 & \begin{tabular}{l} \babar \cite{Aubert:2006nm}: $5.11 \pm 0.94 \pm 0.72$ \\ \end{tabular} & $5.11 \pm 1.18$ \\
\hline
\multicolumn{3}{|l|}{${{\cal{B}} ( \bar{B}^{0} \to D_s^{*-} D^{+} )\times {\cal{B}} ( D_s^{-} \to \phi(1020) \pi^{-} )}$}\\
 & \begin{tabular}{l} \babar \cite{Aubert:2006nm}: $4.14 \pm 1.19 \pm 0.94$ \\ \end{tabular} & $4.14 \pm 1.52$ \\
\hline
\multicolumn{3}{|l|}{${{\cal{B}} ( \bar{B}^{0} \to D_s^{*-} D^{*}(2010)^{+} )\times {\cal{B}} ( D_s^{-} \to \phi(1020) \pi^{-} )}$}\\
 & \begin{tabular}{l} \babar \cite{Aubert:2006nm}: $12.2 \pm 2.2 \pm 2.2$ \\ \end{tabular} & $12.2 \pm 3.1$ \\
\hline
\end{btocharmtab}
\btocharmfig{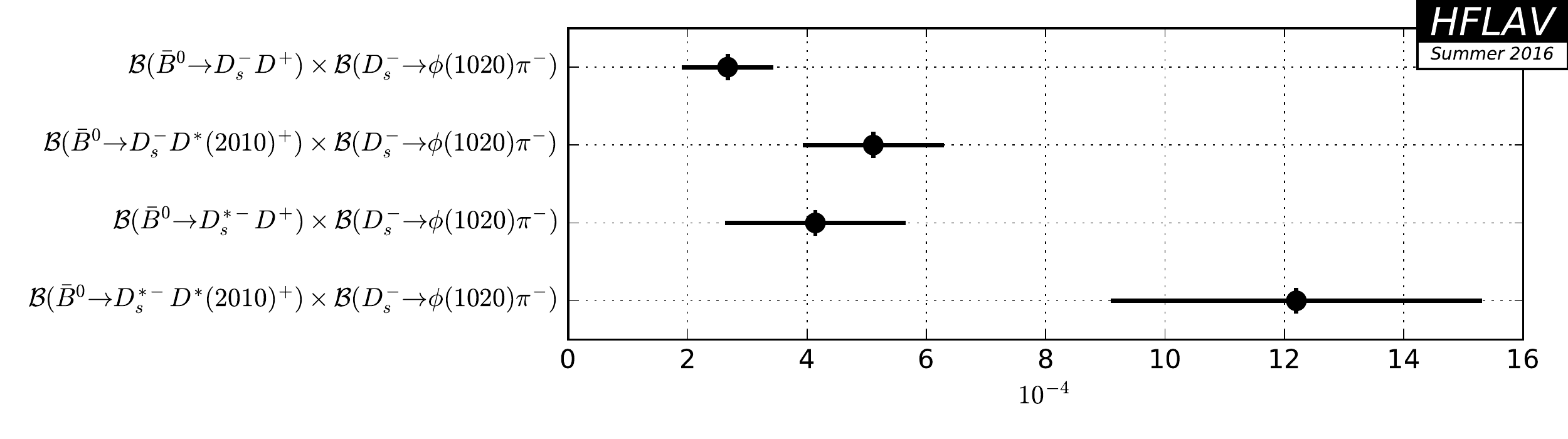}

\begin{btocharmtab}{Bd_DD_6}{Decays to $D_s^{(*)+}D_s^{(*)-}$ $[10^{-3}]$}
\hline
\textbf{Parameter} & \begin{tabular}{l}\textbf{Measurements}\end{tabular} & \textbf{Average} \\
\hline
\hline
${\cal{B}} ( \bar{B}^{0} \to D_s^{-} D_s^{+} )$ & \begin{tabular}{l} Belle \cite{Zupanc:2007pu}: $< 0.036$ \\ \babar \cite{Aubert:2005jv}: $< 0.10$ \\ \end{tabular} & $< 0.036$ \\
\hline
${\cal{B}} ( \bar{B}^{0} \to D_s^{-} D_s^{*+} )$ & \begin{tabular}{l} \babar \cite{Aubert:2005jv}: $< 0.13$ \\ \end{tabular} & $< 0.13$ \\
\hline
${\cal{B}} ( \bar{B}^{0} \to D_s^{*+} D_s^{*-} )$ & \begin{tabular}{l} \babar \cite{Aubert:2005jv}: $< 0.24$ \\ \end{tabular} & $< 0.24$ \\
\hline
\end{btocharmtab}
\btocharmfig{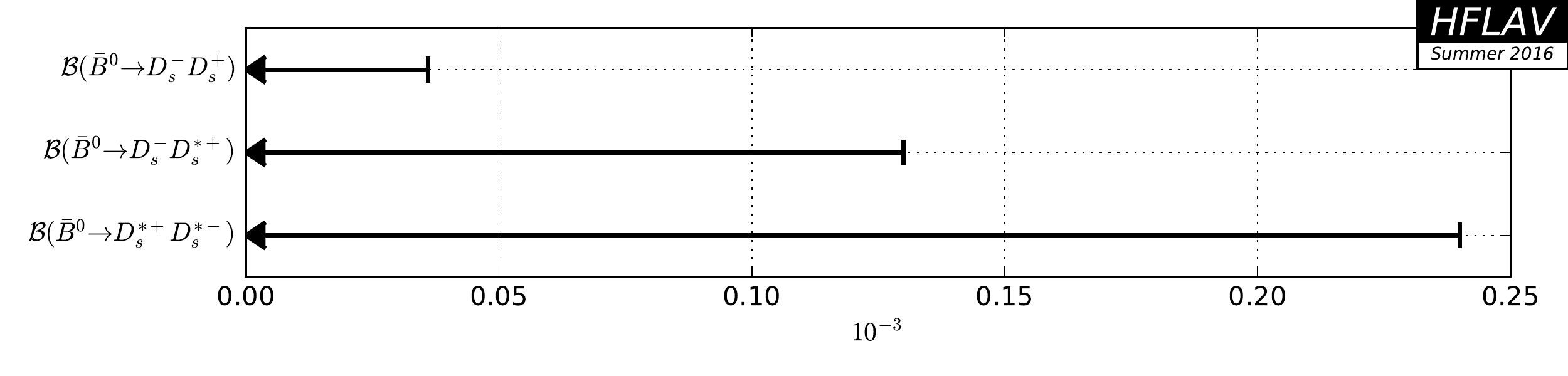}

\begin{btocharmtab}{Bd_DD_7}{Relative decay rates $[10^{-3}]$}
\hline
\textbf{Parameter} & \begin{tabular}{l}\textbf{Measurements}\end{tabular} & \textbf{Average} \\
\hline
\hline
\multicolumn{3}{|l|}{${{\cal{B}} ( \bar{B}^{0} \to D^{0} \bar{D}^{0} )}/{ {\cal{B}} ( \bar{B}^{-} \to D^{0} \bar{D}_s^{-} )}$}\\
 & \begin{tabular}{l} LHCb \cite{Aaij:2013fha}: $1.4 \pm 0.6 \pm 0.2$ \\ \end{tabular} & $1.4 \pm 0.6$ \\
\hline
\end{btocharmtab}

\begin{btocharmtab}{Bd_DD_8}{Absolute decay rates to excited $D_s$ mesons $[10^{-3}]$}
\hline
\textbf{Parameter} & \begin{tabular}{l}\textbf{Measurements}\end{tabular} & \textbf{Average} \\
\hline
\hline
${\cal{B}} ( \bar{B}^{0} \to D_{sJ}(2460)^{-} D^{+} )$ & \begin{tabular}{l} \babar \cite{Aubert:2006nm}: $2.6 \pm 1.5 \pm 0.7$ \\ \end{tabular} & $2.6 \pm 1.7$ \\
\hline
${\cal{B}} ( \bar{B}^{0} \to D_{sJ}(2460)^{-} D^{*}(2010)^{+} )$ & \begin{tabular}{l} \babar \cite{Aubert:2006nm}: $8.8 \pm 2.0 \pm 1.4$ \\ \end{tabular} & $8.8 \pm 2.4$ \\
\hline
${\cal{B}} ( \bar{B}^{0} \to D_{s1}(2536)^{-} D^{*}(2010)^{+} )$ & \begin{tabular}{l} \babar \cite{Aubert:2006fh}: $92 \pm 24 \pm 1$ \\ \end{tabular} & $92 \pm 24$ \\
\hline
\end{btocharmtab}
\btocharmfig{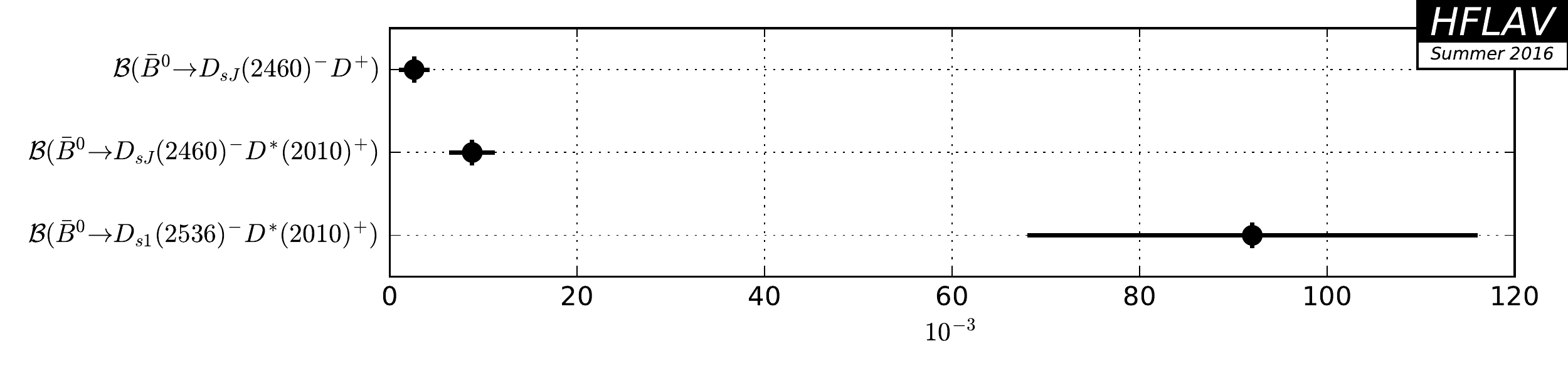}

\begin{btocharmtab}{Bd_DD_9}{Product decay rates to excited $D_s$ mesons I $[10^{-3}]$}
\hline
\textbf{Parameter} & \begin{tabular}{l}\textbf{Measurements}\end{tabular} & \textbf{Average} \\
\hline
\hline
\multicolumn{3}{|l|}{${{\cal{B}} ( \bar{B}^{0} \to D^{+} D_{sJ}(2460)^{-} )\times {\cal{B}} ( D_{sJ}(2460)^{-} \to D_s^{-} \gamma )}$}\\
 & \begin{tabular}{l} Belle \cite{Krokovny:2003zq}: $0.82 \,^{+0.22}_{-0.19} \pm 0.25$ \\ \babar \cite{Aubert:2004pw}: $0.8 \pm 0.2 \,^{+0.3}_{-0.2}$ \\ \end{tabular} & $0.81 \pm 0.23$ \\
\hline
\multicolumn{3}{|l|}{${{\cal{B}} ( \bar{B}^{0} \to D^{+} D_{sJ}(2460)^{-} )\times {\cal{B}} ( D_{sJ}(2460)^{-} \to D_s^{*-} \pi^{0} )}$}\\
 & \begin{tabular}{l} Belle \cite{Krokovny:2003zq}: $2.27 \,^{+0.73}_{-0.62} \pm 0.68$ \\ \babar \cite{Aubert:2004pw}: $2.8 \pm 0.8 \,^{+1.1}_{-0.8}$ \\ \end{tabular} & $2.47 \pm 0.76$ \\
\hline
\multicolumn{3}{|l|}{${{\cal{B}} ( \bar{B}^{0} \to D_{sJ}(2460)^{-} D^{*}(2010)^{+} )\times {\cal{B}} ( D_{sJ}(2460)^{-} \to D_s^{*-} \pi^{0} )}$}\\
 & \begin{tabular}{l} \babar \cite{Aubert:2004pw}: $5.5 \pm 1.2 \,^{+2.1}_{-1.6}$ \\ \end{tabular} & $5.5 \,^{+2.5}_{-2.0}$ \\
\hline
\multicolumn{3}{|l|}{${{\cal{B}} ( \bar{B}^{0} \to D_{sJ}(2460)^{-} D^{*}(2010)^{+} )\times {\cal{B}} ( D_{sJ}(2460)^{-} \to D_s^{-} \gamma )}$}\\
 & \begin{tabular}{l} \babar \cite{Aubert:2004pw}: $2.3 \pm 0.3 \,^{+0.9}_{-0.6}$ \\ \end{tabular} & $2.3 \,^{+0.9}_{-0.7}$ \\
\hline
\multicolumn{3}{|l|}{${{\cal{B}} ( \bar{B}^{0} \to D_{sJ}^{*}(2317)^{-} D^{*}(2010)^{+} )\times {\cal{B}} ( D_{sJ}^{*}(2317)^{-} \to D_s^{-} \pi^{0} )}$}\\
 & \begin{tabular}{l} \babar \cite{Aubert:2004pw}: $1.5 \pm 0.4 \,^{+0.5}_{-0.4}$ \\ \end{tabular} & $1.5 \,^{+0.7}_{-0.5}$ \\
\hline
\multicolumn{3}{|l|}{${{\cal{B}} ( \bar{B}^{0} \to D^{+} D_{sJ}^{*}(2317)^{-} )\times {\cal{B}} ( D_{sJ}^{*}(2317)^{-} \to D_s^{-} \pi^{0} )}$}\\
 & \begin{tabular}{l} Belle \cite{Choi:2015lpc}: $1.02 \,^{+0.13}_{-0.12} \pm 0.11$ \\ \babar \cite{Aubert:2004pw}: $1.8 \pm 0.4 \,^{+0.7}_{-0.5}$ \\ \end{tabular} & $1.03 \pm 0.16$ \\
\hline
\multicolumn{3}{|l|}{${{\cal{B}} ( \bar{B}^{0} \to D^{+} D_{sJ}^{*}(2317)^{-} )\times {\cal{B}} ( D_{sJ}^{*}(2317)^{-} \to D_s^{*-} \gamma )}$}\\
 & \begin{tabular}{l} Belle \cite{Krokovny:2003zq}: $< 0.95$ \\ \end{tabular} & $< 0.95$ \\
\hline
\end{btocharmtab}
\btocharmfig{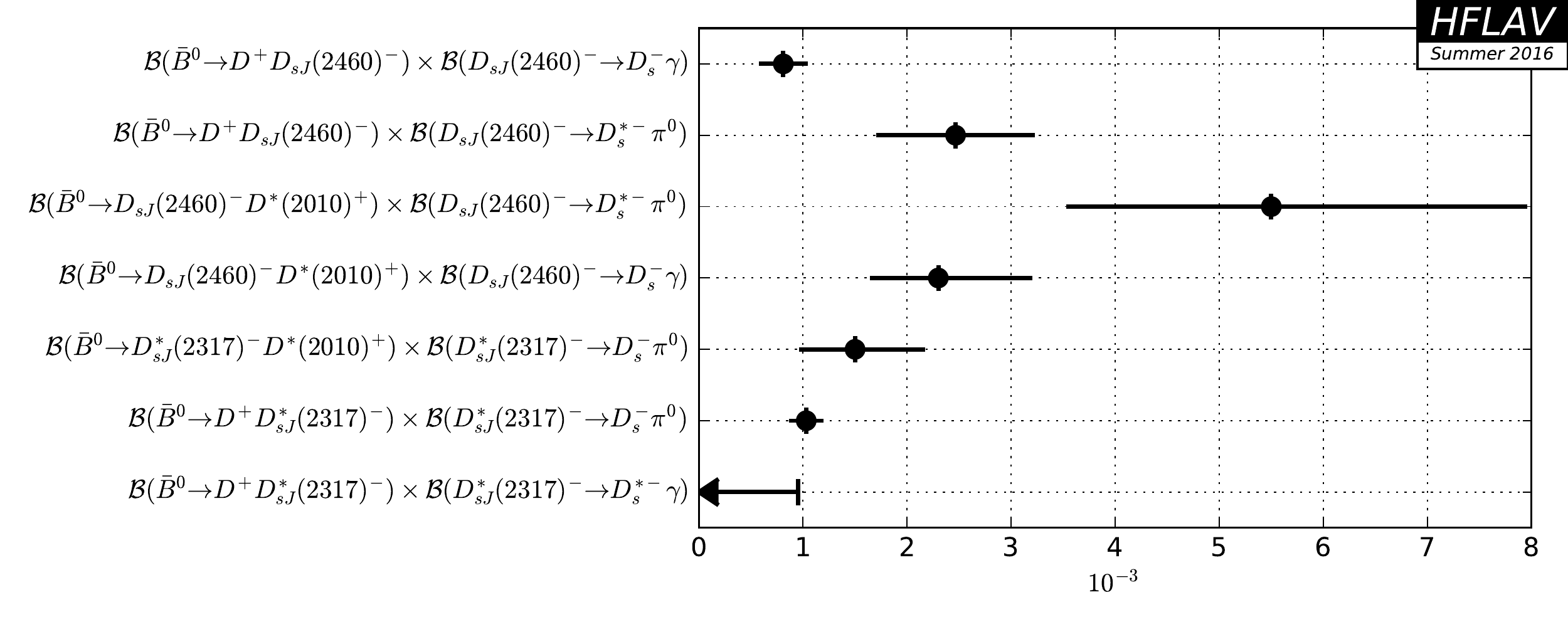}

\begin{btocharmtab}{Bd_DD_10}{Product decay rates to excited $D_s$ mesons II $[10^{-4}]$}
\hline
\textbf{Parameter} & \begin{tabular}{l}\textbf{Measurements}\end{tabular} & \textbf{Average} \\
\hline
\hline
\multicolumn{3}{|l|}{${{\cal{B}} ( \bar{B}^{0} \to D_{s1}(2536)^{-} D^{+} )\times {\cal{B}} ( D_{s1}(2536)^{-} \to D^{*}(2010)^{-} \bar{K}^{0} )}$}\\
 & \begin{tabular}{l} \babar \cite{Aubert:2007rva}: $2.61 \pm 1.03 \pm 0.31$ \\ \end{tabular} & $2.61 \pm 1.08$ \\
\hline
\multicolumn{3}{|l|}{${{\cal{B}} ( \bar{B}^{0} \to D_{s1}(2536)^{-} D^{+} )\times {\cal{B}} ( D_{s1}(2536)^{-} \to K^{-} \bar{D}^{*}(2007)^{0} )}$}\\
 & \begin{tabular}{l} \babar \cite{Aubert:2007rva}: $1.71 \pm 0.48 \pm 0.32$ \\ \end{tabular} & $1.71 \pm 0.58$ \\
\hline
\multicolumn{3}{|l|}{${{\cal{B}} ( \bar{B}^{0} \to D_{s1}(2536)^{-} D^{*}(2010)^{+} )\times {\cal{B}} ( D_{s1}(2536)^{-} \to D^{*}(2010)^{-} \bar{K}^{0} )}$}\\
 & \begin{tabular}{l} \babar \cite{Aubert:2007rva}: $5.00 \pm 1.51 \pm 0.67$ \\ \end{tabular} & $5.00 \pm 1.65$ \\
\hline
\multicolumn{3}{|l|}{${{\cal{B}} ( \bar{B}^{0} \to D_{s1}(2536)^{-} D^{*}(2010)^{+} )\times {\cal{B}} ( D_{s1}(2536)^{-} \to \bar{D}^{*}(2007)^{0} K^{+} )}$}\\
 & \begin{tabular}{l} \babar \cite{Aubert:2007rva}: $3.32 \pm 0.88 \pm 0.66$ \\ \end{tabular} & $3.32 \pm 1.10$ \\
\hline
\multicolumn{3}{|l|}{${{\cal{B}} ( \bar{B}^{0} \to D_{s1}(2536)^{+} D^{-} )\times {\cal{B}} ( D_{s1}(2536)^{+} \to D^{*}(2007)^{0} K^{+} + D^{*}(2010)^{+} K^{0} )}$}\\
 & \begin{tabular}{l} Belle \cite{Belle:2011ad}: $2.75 \pm 0.62 \pm 0.36$ \\ \end{tabular} & $2.75 \pm 0.72$ \\
\hline
\multicolumn{3}{|l|}{${{\cal{B}} ( \bar{B}^{0} \to D_{s1}(2536)^{+} D^{*}(2010)^{-} )\times {\cal{B}} ( D_{s1}(2536)^{+} \to D^{*}(2007)^{0} K^{+} + D^{*}(2010)^{+} K^{0} )}$}\\
 & \begin{tabular}{l} Belle \cite{Belle:2011ad}: $5.01 \pm 1.21 \pm 0.70$ \\ \end{tabular} & $5.01 \pm 1.40$ \\
\hline
\multicolumn{3}{|l|}{${{\cal{B}} ( \bar{B}^{0} \to D_{s1}(2536)^{+} D^{*}(2010)^{-} )\times {\cal{B}} ( D_{s1}(2536)^{+} \to D^{*}(2010)^{+} K_S^{0} )}$}\\
 & \begin{tabular}{l} Belle \cite{Dalseno:2007hx}: $< 6.0$ \\ \end{tabular} & $< 6.0$ \\
\hline
\multicolumn{3}{|l|}{${{\cal{B}} ( \bar{B}^{0} \to D^{+} D_{sJ}(2460)^{-} )\times {\cal{B}} ( D_{sJ}(2460)^{-} \to D_s^{-} \pi^{+} \pi^{-} )}$}\\
 & \begin{tabular}{l} Belle \cite{Krokovny:2003zq}: $< 2.0$ \\ \end{tabular} & $< 2.0$ \\
\hline
\multicolumn{3}{|l|}{${{\cal{B}} ( \bar{B}^{0} \to D^{+} D_{sJ}(2460)^{-} )\times {\cal{B}} ( D_{sJ}(2460)^{-} \to D_s^{-} \pi^{0} )}$}\\
 & \begin{tabular}{l} Belle \cite{Krokovny:2003zq}: $< 3.6$ \\ \end{tabular} & $< 3.6$ \\
\hline
\multicolumn{3}{|l|}{${{\cal{B}} ( \bar{B}^{0} \to D^{+} D_{sJ}(2460)^{-} )\times {\cal{B}} ( D_{sJ}(2460)^{-} \to D_s^{*-} \gamma )}$}\\
 & \begin{tabular}{l} Belle \cite{Krokovny:2003zq}: $< 6.0$ \\ \end{tabular} & $< 6.0$ \\
\hline
\end{btocharmtab}
\btocharmfig{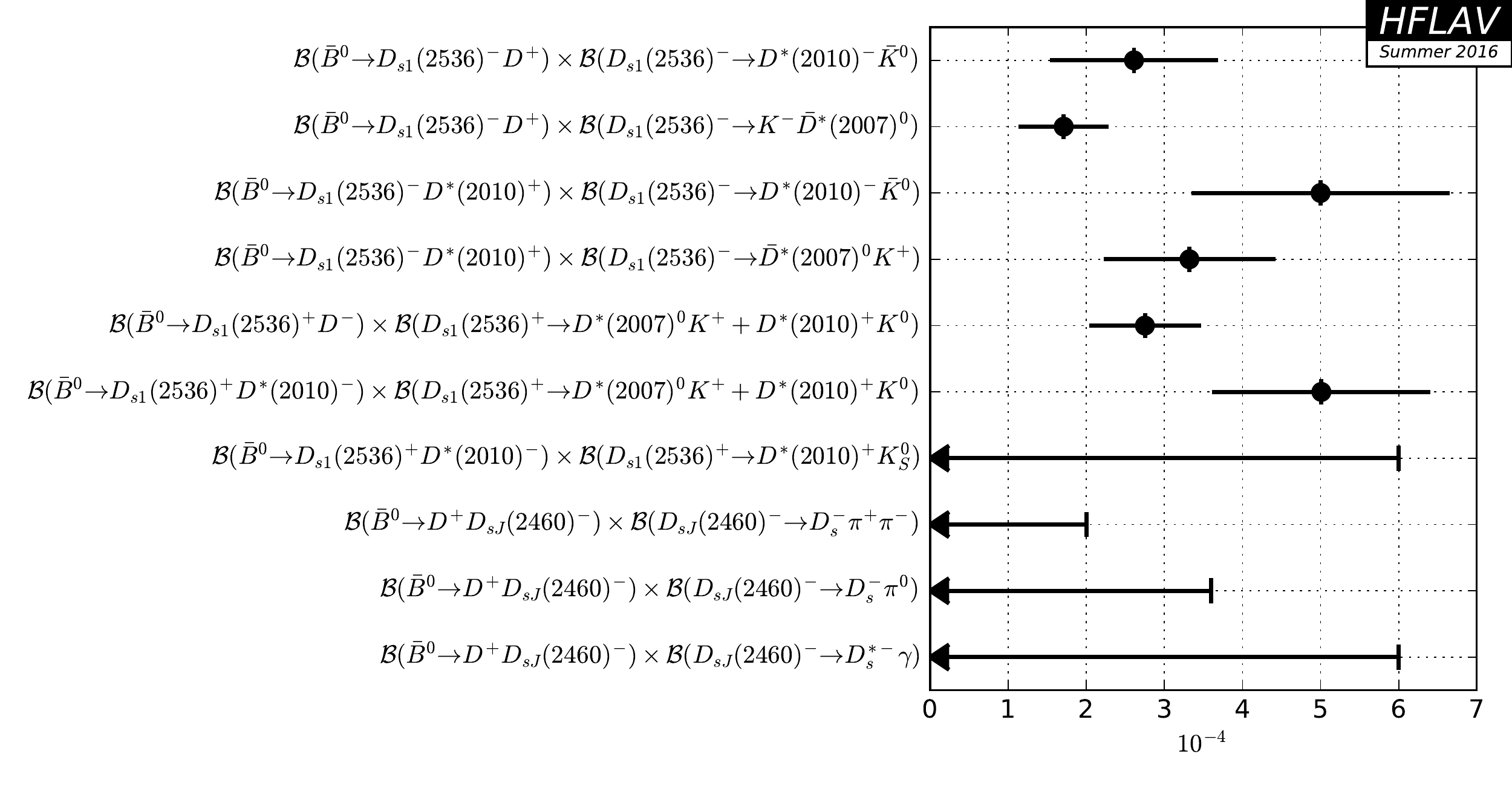}

\subsubsection{Decays to charmonium states}
\label{sec:b2c:Bd_cc}
Averages of $\bar{B}^0$ decays to charmonium states are shown in Tables~\ref{tab:b2c:Bd_cc_1}--\ref{tab:b2c:Bd_cc_9} and Figs.~\ref{fig:b2c:Bd_cc_1}--\ref{fig:b2c:Bd_cc_9}.
\begin{btocharmtab}{Bd_cc_1}{Decays to $J/\psi$ and one kaon $[10^{-3}]$}
\hline
\textbf{Parameter} & \begin{tabular}{l}\textbf{Measurements}\end{tabular} & \textbf{Average} \\
\hline
\hline
${\cal{B}} ( \bar{B}^{0} \to J/\psi \bar{K}^{0} )$ & \begin{tabular}{l} CDF \cite{Abe:1995aw}: $1.15 \pm 0.23 \pm 0.17$ \\ Belle \cite{Abe:2002rc}: $0.79 \pm 0.04 \pm 0.09$ \\ \babar \cite{Aubert:2004rz}: $0.869 \pm 0.022 \pm 0.030$ \\ \end{tabular} & $0.863 \pm 0.035$ \\
\hline
${\cal{B}} ( \bar{B}^{0} \to J/\psi K^{-} \pi^{+}  )$ & \begin{tabular}{l} Belle \cite{Chilikin:2014bkk}: $1.15 \pm 0.01 \pm 0.05$ \\ \end{tabular} & $1.15 \pm 0.05$ \\
\hline
${\cal{B}} ( \bar{B}^{0} \to J/\psi \bar{K}^{*}(892)^{0} )$ & \begin{tabular}{l} CDF \cite{Abe:1998yu}: $1.74 \pm 0.20 \pm 0.18$ \\ Belle \cite{Chilikin:2014bkk}: $1.19 \pm 0.01 \pm 0.08$ \\ \babar \cite{Aubert:2004rz}: $1.309 \pm 0.026 \pm 0.077$ \\ \end{tabular} & $1.270 \pm 0.056$ \\
\hline
${\cal{B}} ( \bar{B}^{0} \to J/\psi \bar{K}^{0} \pi^{+} \pi^{-} )$ & \begin{tabular}{l} LHCb \cite{Aaij:2014naa}: $0.430 \pm 0.030 \pm 0.037$ \\ CDF \cite{Affolder:2001qi}: $1.03 \pm 0.33 \pm 0.15$ \\ \end{tabular} & $0.440 \pm 0.047$ \\
\hline
${\cal{B}} ( \bar{B}^{0} \to J/\psi \bar{K}^{0} \rho(770)^{0} )$ & \begin{tabular}{l} CDF \cite{Affolder:2001qi}: $0.54 \pm 0.29 \pm 0.09$ \\ \end{tabular} & $0.54 \pm 0.30$ \\
\hline
${\cal{B}} ( \bar{B}^{0} \to J/\psi K^{*}(892)^{-} \pi^{+} )$ & \begin{tabular}{l} CDF \cite{Affolder:2001qi}: $0.77 \pm 0.41 \pm 0.13$ \\ \end{tabular} & $0.77 \pm 0.43$ \\
\hline
${\cal{B}} ( \bar{B}^{0} \to J/\psi \omega(782) \bar{K}^{0} )$ & \begin{tabular}{l} \babar \cite{delAmoSanchez:2010jr}: $0.23 \pm 0.03 \pm 0.03$ \\ \end{tabular} & $0.23 \pm 0.04$ \\
\hline
${\cal{B}} ( \bar{B}^{0} \to J/\psi \phi(1020) \bar{K}^{0} )$ & \begin{tabular}{l} \babar \cite{Aubert:2003ii}: $0.102 \pm 0.038 \pm 0.010$ \\ \end{tabular} & $0.102 \pm 0.039$ \\
\hline
${\cal{B}} ( \bar{B}^{0} \to J/\psi \bar{K}_1^{0}(1270) )$ & \begin{tabular}{l} Belle \cite{Abe:2001wa}: $1.30 \pm 0.34 \pm 0.31$ \\ \end{tabular} & $1.30 \pm 0.46$ \\
\hline
${\cal{B}} ( \bar{B}^{0} \to J/\psi \eta K_S^{0} )$ & \begin{tabular}{l} Belle \cite{Iwashita:2013wnn}: $0.0522 \pm 0.0078 \pm 0.0049$ \\ \babar \cite{Aubert:2004fc}: $0.084 \pm 0.026 \pm 0.027$ \\ \end{tabular} & $0.0540 \pm 0.0089$ \\
\hline
${\cal{B}} ( \bar{B}^{0} \to J/\psi \bar{K}^{*}(892)^{0} \pi^{+} \pi^{-} )$ & \begin{tabular}{l} CDF \cite{Affolder:2001qi}: $0.66 \pm 0.19 \pm 0.11$ \\ \end{tabular} & $0.66 \pm 0.22$ \\
\hline
\end{btocharmtab}
\btocharmfig{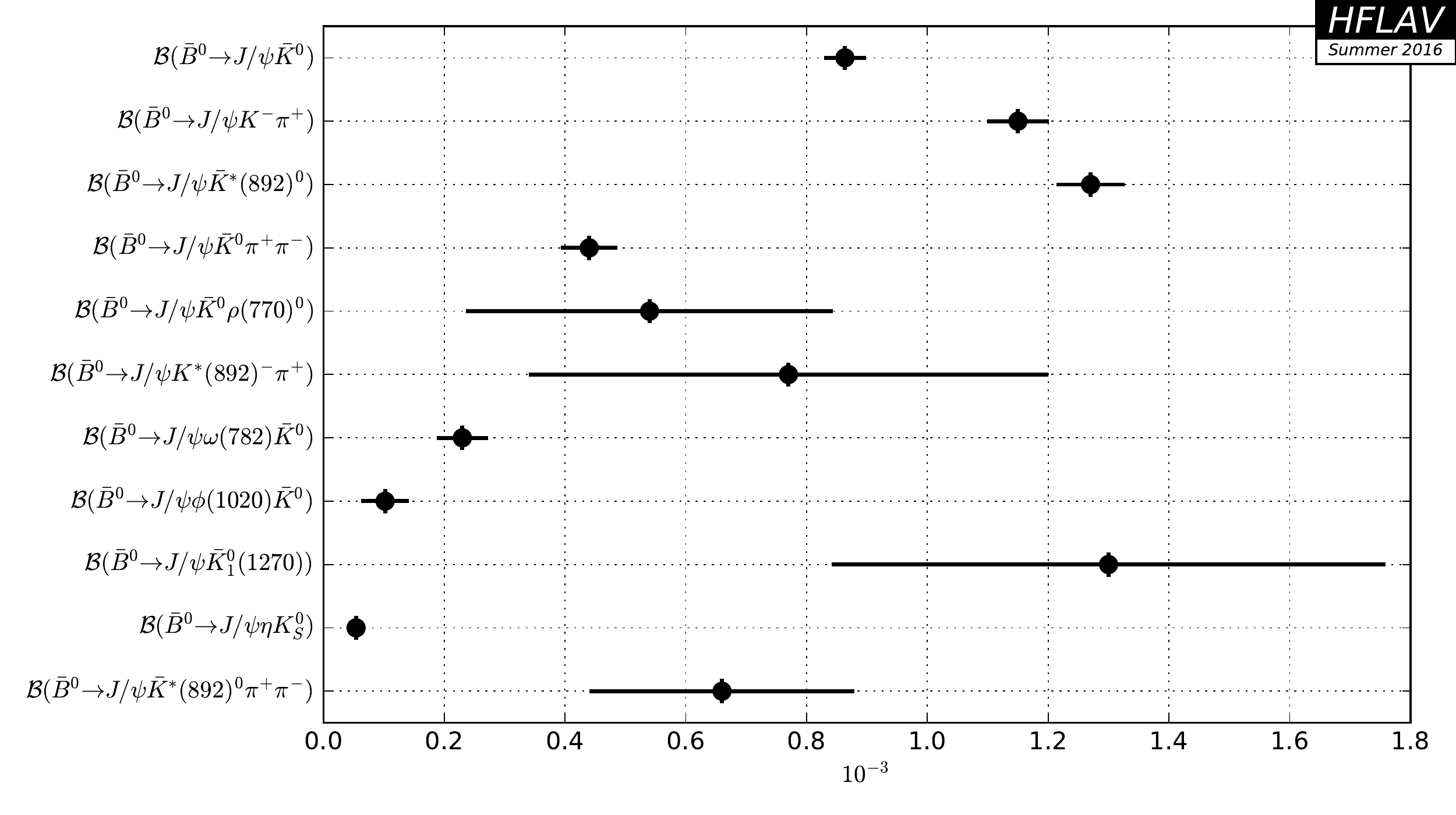}

\begin{btocharmtab}{Bd_cc_2}{Decays to charmonium other than $J/\psi$ and one kaon I $[10^{-3}]$}
\hline
\textbf{Parameter} & \begin{tabular}{l}\textbf{Measurements}\end{tabular} & \textbf{Average} \\
\hline
\hline
${\cal{B}} ( \bar{B}^{0} \to \psi(2S) \bar{K}^{0} )$ & \begin{tabular}{l} Belle \cite{Abe:2002rc}: $0.67 \pm 0.11$ \\ \babar \cite{Aubert:2004rz}: $0.646 \pm 0.065 \pm 0.051$ \\ \end{tabular} & $0.655 \pm 0.066$ \\
\hline
${\cal{B}} ( \bar{B}^{0} \to \psi(2S) \bar{K}^{*}(892)^{0} )$ & \begin{tabular}{l} CDF \cite{Abe:1998yu}: $0.90 \pm 0.22 \pm 0.09$ \\ \babar \cite{Aubert:2004rz}: $0.649 \pm 0.059 \pm 0.097$ \\ \end{tabular} & $0.696 \pm 0.103$ \\
\hline
${\cal{B}} ( \bar{B}^{0} \to \psi(2S) K^{0}  )$ & \begin{tabular}{l} LHCb \cite{Aaij:2014naa}: $0.47 \pm 0.07 \pm 0.07$ \\ \end{tabular} & $0.47 \pm 0.10$ \\
\hline
${\cal{B}} ( \bar{B}^{0} \to K^{*}(892)^{0} \psi(2S) )$ & \begin{tabular}{l} Belle \cite{Mizuk:2009da}: $0.552 \,^{+0.035}_{-0.032} \,^{+0.053}_{-0.058}$ \\ \end{tabular} & $0.552 \,^{+0.064}_{-0.066}$ \\
\hline
\multicolumn{3}{|l|}{${\cal{B}} ( \bar{B}^{0} \to \psi(2S) K^{0} )\times {\cal{B}} ( \psi(2S) \to \chi_{c1} \gamma )$}\\
 & \begin{tabular}{l} Belle \cite{Bhardwaj:2013rmw}: $0.68 \pm 0.10 \pm 0.07$ \\ \end{tabular} & $0.68 \pm 0.12$ \\
\hline
\multicolumn{3}{|l|}{${\cal{B}} ( \bar{B}^{0} \to \psi(2S) K^{0} )\times {\cal{B}} ( \psi(2S) \to \chi_{c2} \gamma )$}\\
 & \begin{tabular}{l} Belle \cite{Bhardwaj:2013rmw}: $0.47 \pm 0.16 \pm 0.08$ \\ \end{tabular} & $0.47 \pm 0.18$ \\
\hline
${\cal{B}} ( \bar{B}^{0} \to \chi_{c0} \bar{K}^{0} )$ & \begin{tabular}{l} \babar \cite{Aubert:2005vwa}: $< 1.24$ \\ \end{tabular} & $< 1.24$ \\
\hline
${\cal{B}} ( \bar{B}^{0} \to \chi_{c0} \bar{K}^{*}(892)^{0} )$ & \begin{tabular}{l} \babar \cite{Aubert:2005vwa}: $< 0.77$ \\ \babar \cite{Aubert:2008ak}: $0.17 \pm 0.03 \pm 0.02$ \\ \end{tabular} & $0.17 \pm 0.04$ \\
\hline
${\cal{B}} ( \bar{B}^{0} \to \chi_{c1} \bar{K}^{0} )$ & \begin{tabular}{l} Belle \cite{Bhardwaj:2011dj}: $0.378 \,^{+0.017}_{-0.016} \pm 0.033$ \\ \babar \cite{Aubert:2008ae}: $0.42 \pm 0.03 \pm 0.03$ \\ \end{tabular} & $0.396 \pm 0.028$ \\
\hline
${\cal{B}} ( \bar{B}^{0} \to \chi_{c1} K^{-} \pi^{+} )$ & \begin{tabular}{l} Belle \cite{Bhardwaj:2015rju}: $0.497 \pm 0.012 \pm 0.028$ \\ \babar \cite{Lees:2011ik}: $0.511 \pm 0.014 \pm 0.058$ \\ \end{tabular} & $0.500 \pm 0.027$ \\
\hline
${\cal{B}} ( \bar{B}^{0} \to \chi_{c1} \bar{K}^{*}(892)^{0} )$ & \begin{tabular}{l} Belle \cite{Soni:2005fw}: $0.31 \pm 0.03 \pm 0.07$ \\ \babar \cite{Aubert:2008ae}: $0.25 \pm 0.02 \pm 0.02$ \\ \end{tabular} & $0.26 \pm 0.03$ \\
\hline
${\cal{B}} ( \bar{B}^{0} \to \chi_{c1} K^{-} \pi^{+}  \pi^{0} )$ & \begin{tabular}{l} Belle \cite{Bhardwaj:2015rju}: $0.352 \pm 0.052 \pm 0.024$ \\ \end{tabular} & $0.352 \pm 0.057$ \\
\hline
${\cal{B}} ( \bar{B}^{0} \to \chi_{c1} \bar{K}^{0} \pi^{+}  \pi^{-} )$ & \begin{tabular}{l} Belle \cite{Bhardwaj:2015rju}: $0.316 \pm 0.035 \pm 0.032$ \\ \end{tabular} & $0.316 \pm 0.047$ \\
\hline
${\cal{B}} ( \bar{B}^{0} \to \eta_c \bar{K}^{0} )$ & \begin{tabular}{l} Belle \cite{Fang:2002gi}: $1.23 \pm 0.23 \,^{+0.40}_{-0.41}$ \\ \babar \cite{Aubert:2007qea}: $0.64 \,^{+0.22}_{-0.20} \,^{+0.28}_{-0.16}$ \\ \babar \cite{Aubert:2004gc}: $1.14 \pm 0.15 \pm 0.34$ \\ \end{tabular} & $0.85 \pm 0.24$ \\
\hline
${\cal{B}} ( \bar{B}^{0} \to \eta_c \bar{K}^{*}(892)^{0} )$ & \begin{tabular}{l} Belle \cite{Fang:2002gi}: $1.62 \pm 0.32 \,^{+0.55}_{-0.60}$ \\ \babar \cite{Aubert:2008kp}: $0.57 \pm 0.06 \pm 0.09$ \\ \babar \cite{Aubert:2007qea}: $0.80 \,^{+0.21}_{-0.19} \,^{+0.37}_{-0.23}$ \\ \end{tabular} & $0.61 \pm 0.10$ \\
\hline
${\cal{B}} ( \bar{B}^{0} \to \eta_c(2S) \bar{K}^{*}(892)^{0} )$ & \begin{tabular}{l} \babar \cite{Aubert:2008kp}: $< 0.39$ \\ \end{tabular} & $< 0.39$ \\
\hline
\multicolumn{3}{|l|}{${{\cal{B}} ( \bar{B}^{0} \to h_c(1P) \bar{K}^{*}(892)^{0} )\times {\cal{B}} ( h_c(1P) \to \eta_c \gamma )}$}\\
 & \begin{tabular}{l} \babar \cite{Aubert:2008kp}: $< 0.22$ \\ \end{tabular} & $< 0.22$ \\
\hline
\end{btocharmtab}
\btocharmfig{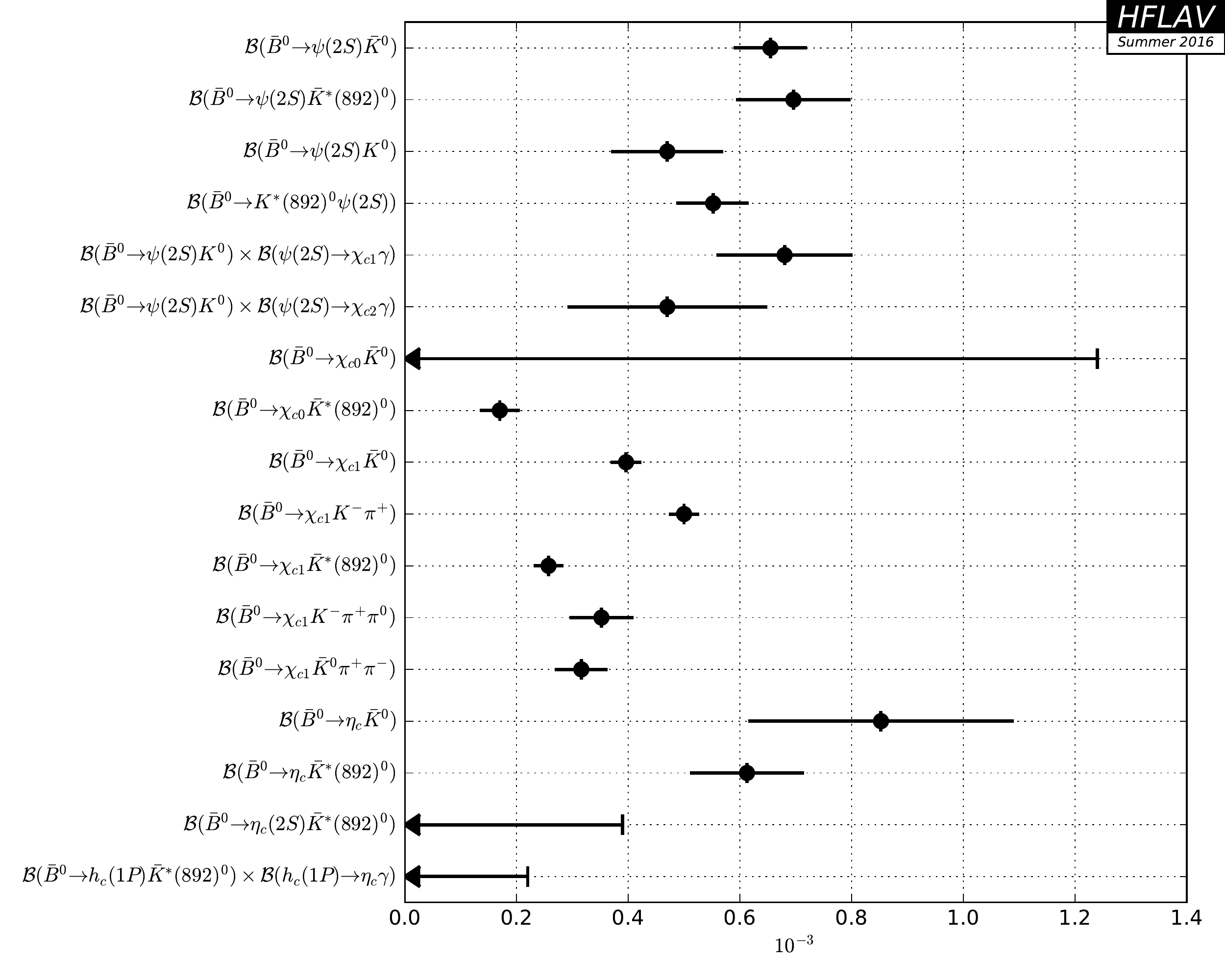}

\begin{btocharmtab}{Bd_cc_3}{Decays to charmonium other than $J/\psi$ and one kaon II $[10^{-4}]$}
\hline
\textbf{Parameter} & \begin{tabular}{l}\textbf{Measurements}\end{tabular} & \textbf{Average} \\
\hline
\hline
\multicolumn{3}{|l|}{${{\cal{B}} ( \bar{B}^{0} \to \psi(3770) \bar{K}^{0} )\times {\cal{B}} ( \psi(3770) \to D^{0} \bar{D}^{0} )}$}\\
 & \begin{tabular}{l} \babar \cite{Aubert:2007rva}: $< 1.23$ \\ \end{tabular} & $< 1.23$ \\
\hline
\multicolumn{3}{|l|}{${{\cal{B}} ( \bar{B}^{0} \to \psi(3770) \bar{K}^{0} )\times {\cal{B}} ( \psi(3770) \to D^{+} D^{-} )}$}\\
 & \begin{tabular}{l} \babar \cite{Aubert:2007rva}: $< 1.88$ \\ \end{tabular} & $< 1.88$ \\
\hline
${\cal{B}} ( \bar{B}^{0} \to \chi_{c2} \bar{K}^{0} )$ & \begin{tabular}{l} Belle \cite{Bhardwaj:2011dj}: $< 0.15$ \\ \babar \cite{Aubert:2008ae}: $0.15 \pm 0.09 \pm 0.03$ \\ \end{tabular} & $0.15 \pm 0.09$ \\
\hline
${\cal{B}} ( \bar{B}^{0} \to \chi_{c2} \bar{K}^{*}(892)^{0} )$ & \begin{tabular}{l} \babar \cite{Aubert:2008ae}: $0.66 \pm 0.18 \pm 0.05$ \\ \end{tabular} & $0.66 \pm 0.19$ \\
\hline
${\cal{B}} ( \bar{B}^{0} \to \chi_{c2} K^{-} \pi^{+} )$ & \begin{tabular}{l} Belle \cite{Bhardwaj:2015rju}: $0.72 \pm 0.09 \pm 0.05$ \\ \end{tabular} & $0.72 \pm 0.10$ \\
\hline
\end{btocharmtab}
\btocharmfig{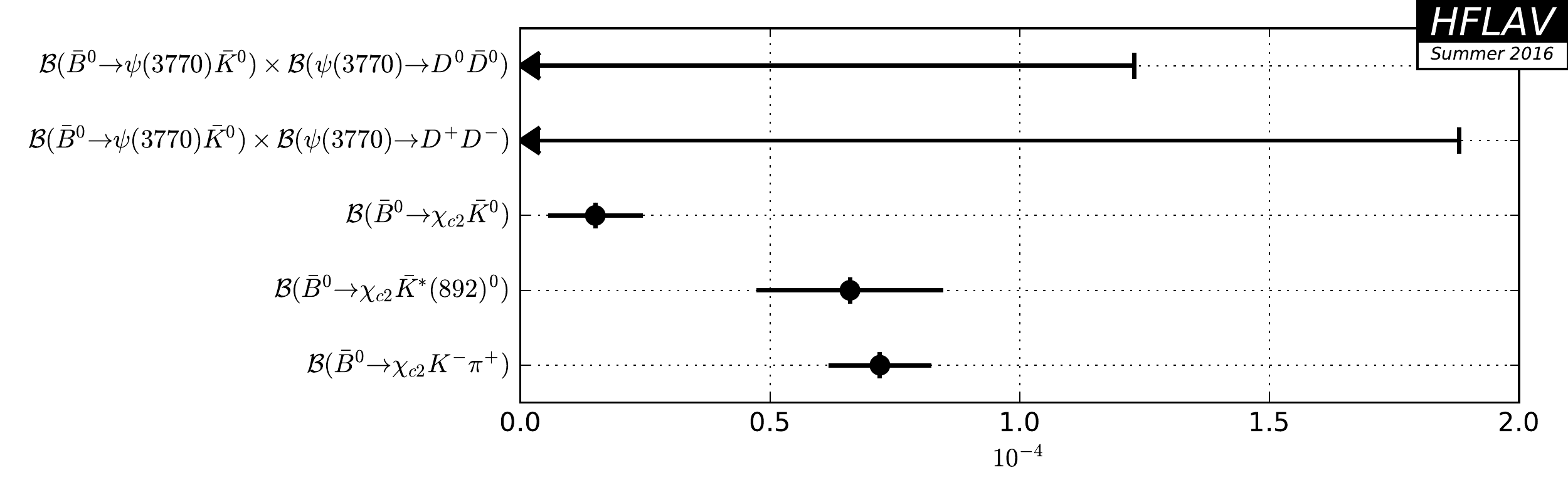}

\begin{btocharmtab}{Bd_cc_4}{Decays to charmonium and light mesons I $[10^{-5}]$}
\hline
\textbf{Parameter} & \begin{tabular}{l}\textbf{Measurements}\end{tabular} & \textbf{Average} \\
\hline
\hline
${\cal{B}} ( \bar{B}^{0} \to J/\psi \pi^{0} )$ & \begin{tabular}{l} Belle \cite{Abe:2002rc}: $2.3 \pm 0.5 \pm 0.2$ \\ \babar \cite{Aubert:2008bs}: $1.69 \pm 0.14 \pm 0.07$ \\ \end{tabular} & $1.74 \pm 0.15$ \\
\hline
${\cal{B}} ( \bar{B}^{0} \to J/\psi \pi^{+} \pi^{-} )$ & \begin{tabular}{l} \babar \cite{Aubert:2007xw}: $< 1.2$ \\ \end{tabular} & $< 1.2$ \\
\hline
${\cal{B}} ( \bar{B}^{0} \to J/\psi \rho(770)^{0} )$ & \begin{tabular}{l} \babar \cite{Aubert:2007xw}: $2.7 \pm 0.3 \pm 0.2$ \\ \end{tabular} & $2.7 \pm 0.4$ \\
\hline
${\cal{B}} ( \bar{B}^{0} \to J/\psi \eta )$ & \begin{tabular}{l} Belle \cite{Chang:2012gnb}: $1.23 \,^{+0.18}_{-0.17} \pm 0.07$ \\ \babar \cite{Aubert:2003ii}: $< 2.7$ \\ \end{tabular} & $1.23 \,^{+0.19}_{-0.18}$ \\
\hline
${\cal{B}} ( \bar{B}^{0} \to J/\psi \eta^\prime(958) )$ & \begin{tabular}{l} Belle \cite{Chang:2012gnb}: $< 0.74$ \\ \babar \cite{Aubert:2003ii}: $< 6.3$ \\ \end{tabular} & $< 0.74$ \\
\hline
${\cal{B}} ( \bar{B}^{0} \to J/\psi f_2(1270) )$ & \begin{tabular}{l} \babar \cite{Aubert:2007xw}: $< 0.46$ \\ \end{tabular} & $< 0.46$ \\
\hline
${\cal{B}} ( \bar{B}^{0} \to J/\psi f_1(1285) )$ & \begin{tabular}{l} LHCb \cite{Aaij:2013rja}: $0.837 \pm 0.195 \,^{+0.079}_{-0.075}$ \\ \end{tabular} & $0.837 \,^{+0.210}_{-0.209}$ \\
\hline
${\cal{B}} ( \bar{B}^{0} \to J/\psi K^{0} K^{\pm} \pi^{\mp} )$ & \begin{tabular}{l} LHCb \cite{Aaij:2014naa}: $< 2.1$ \\ \end{tabular} & $< 2.1$ \\
\hline
${\cal{B}} ( \bar{B}^{0} \to J/\psi K^{0} K^{+} K^{-} )$ & \begin{tabular}{l} LHCb \cite{Aaij:2014naa}: $2.02 \pm 0.43 \pm 0.19$ \\ \end{tabular} & $2.02 \pm 0.47$ \\
\hline
${\cal{B}} ( \bar{B}^{0} \to \chi_{c1} \pi^{0} )$ & \begin{tabular}{l} Belle \cite{Kumar:2008ir}: $1.12 \pm 0.25 \pm 0.12$ \\ \end{tabular} & $1.12 \pm 0.28$ \\
\hline
\end{btocharmtab}
\btocharmfig{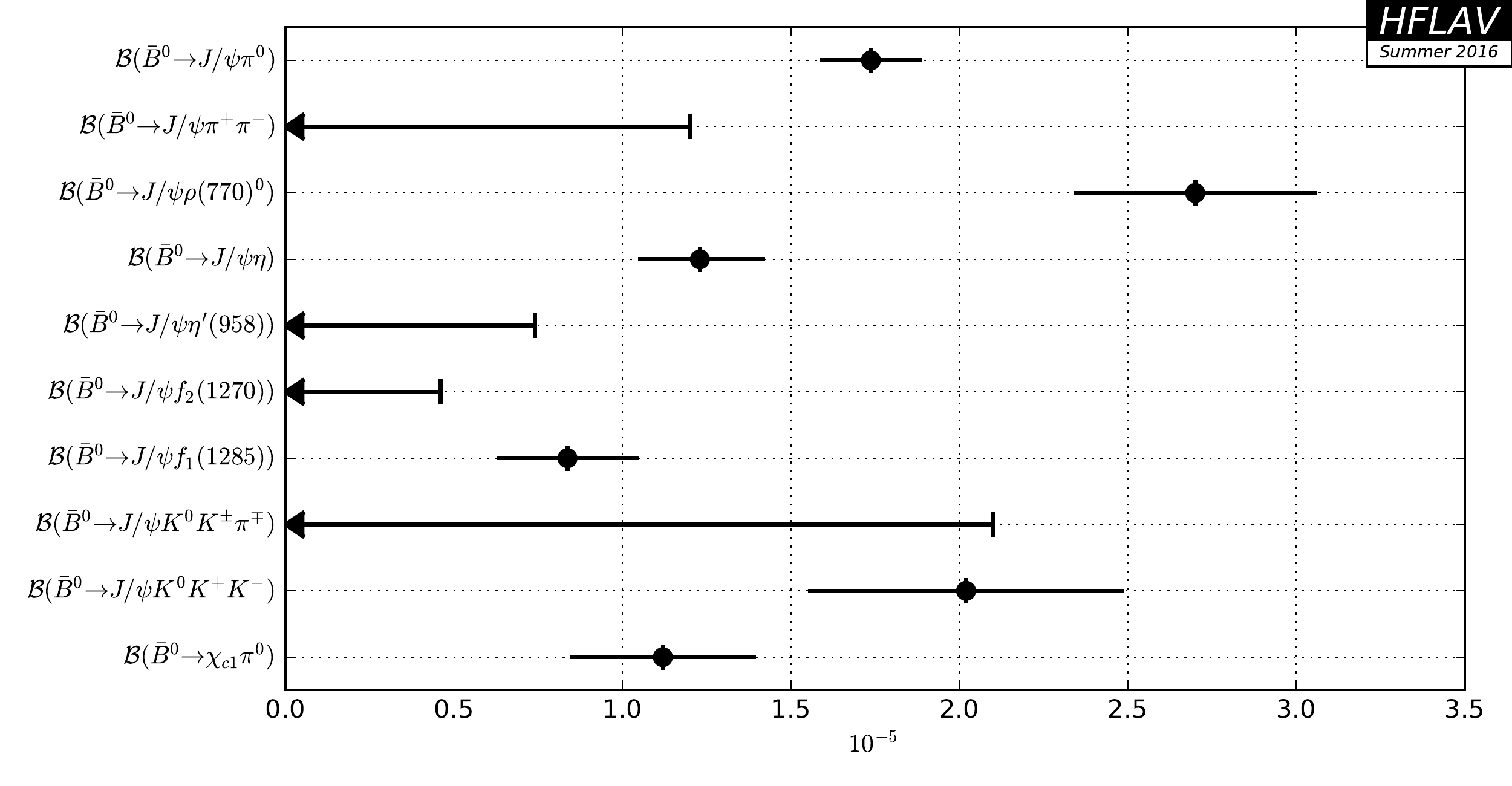}

\begin{btocharmtab}{Bd_cc_5}{Decays to charmonium and light mesons II $[10^{-5}]$}
\hline
\textbf{Parameter} & \begin{tabular}{l}\textbf{Measurements}\end{tabular} & \textbf{Average} \\
\hline
\hline
\multicolumn{3}{|l|}{${{\cal{B}} ( \bar{B}^{0} \to J/\psi a_0(980) )\times {\cal{B}} ( a_0(980) \to K^{+} K^{-} )}$}\\
 & \begin{tabular}{l} LHCb \cite{Aaij:2013mtm}: $< 0.090$ \\ \end{tabular} & $< 0.090$ \\
\hline
\multicolumn{3}{|l|}{${{\cal{B}} ( \bar{B}^{0} \to J/\psi f_0(980) )\times {\cal{B}} ( f_0(980) \to \pi^{+} \pi^{-} )}$}\\
 & \begin{tabular}{l} LHCb \cite{Aaij:2013zpt}: $< 0.11$ \\ \end{tabular} & $< 0.11$ \\
\hline
\multicolumn{3}{|l|}{${{\cal{B}} ( \bar{B}^{0} \to J/\psi f_1(1285) )\times {\cal{B}} ( f_1(1285) \to \pi^{+} \pi^{-} \pi^{+} \pi^{-} )}$}\\
 & \begin{tabular}{l} LHCb \cite{Aaij:2013rja}: $0.0921 \pm 0.0214 \pm 0.0064$ \\ \end{tabular} & $0.0921 \pm 0.0223$ \\
\hline
${\cal{B}} ( \bar{B}^{0} \to J/\psi K^{+} K^{-} )$ & \begin{tabular}{l} LHCb \cite{Aaij:2013mtm}: $0.253 \pm 0.031 \pm 0.019$ \\ \end{tabular} & $0.253 \pm 0.036$ \\
\hline
${\cal{B}} ( \bar{B}^{0} \to J/\psi \phi(1020) )$ & \begin{tabular}{l} LHCb \cite{Aaij:2013mtm}: $< 0.019$ \\ Belle \cite{Liu:2008bta}: $< 0.094$ \\ \babar \cite{Aubert:2003ii}: $< 0.9$ \\ \end{tabular} & $< 0.019$ \\
\hline
\end{btocharmtab}
\btocharmfig{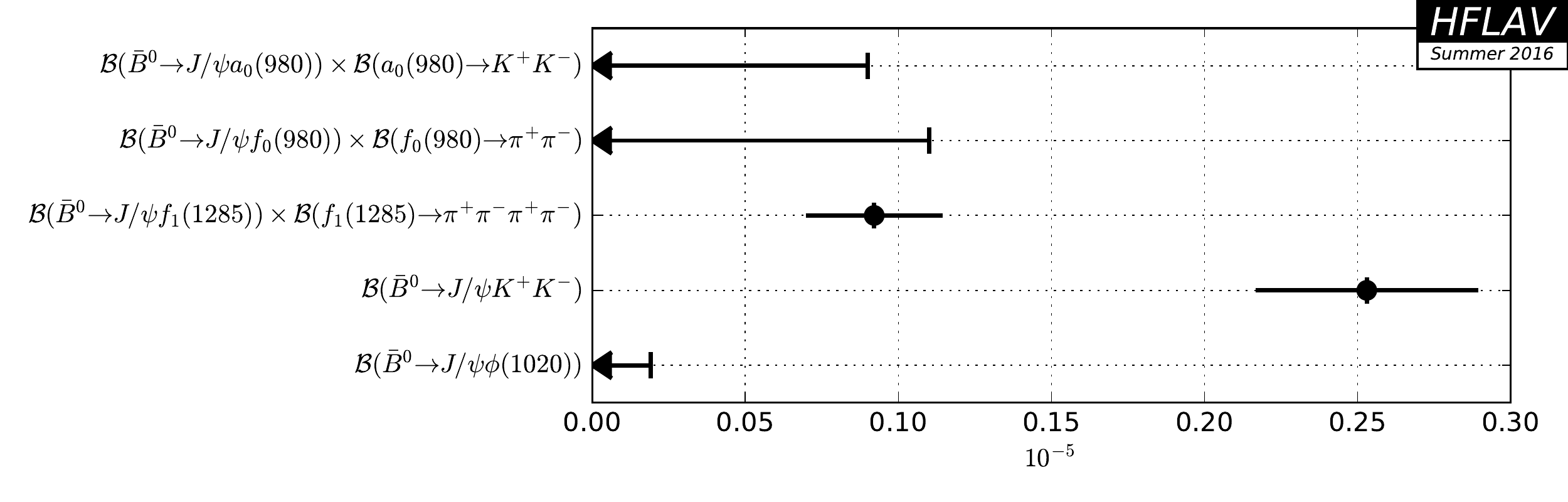}

\begin{btocharmtab}{Bd_cc_6}{Decays to  $J/\psi$ and photons, baryons, or heavy mesons $[10^{-5}]$}
\hline
\textbf{Parameter} & \begin{tabular}{l}\textbf{Measurements}\end{tabular} & \textbf{Average} \\
\hline
\hline
${\cal{B}} ( \bar{B}^{0} \to J/\psi \gamma )$ & \begin{tabular}{l} LHCb \cite{Aaij:2015uoa}: $< 0.15$ \\ \babar \cite{Aubert:2004xd}: $< 0.16$ \\ \end{tabular} & $< 0.15$ \\
\hline
${\cal{B}} ( \bar{B}^{0} \to J/\psi \bar{p} p )$ & \begin{tabular}{l} LHCb \cite{Aaij:2013yba}: $< 0.052$ \\ Belle \cite{Xie:2005tf}: $< 0.083$ \\ \babar \cite{Aubert:2003ww}: $< 0.19$ \\ \end{tabular} & $< 0.052$ \\
\hline
${\cal{B}} ( \bar{B}^{0} \to J/\psi D^{0} )$ & \begin{tabular}{l} Belle \cite{Zhang:2005bs}: $< 2.0$ \\ \babar \cite{Aubert:2005tr}: $< 1.3$ \\ \end{tabular} & $< 1.3$ \\
\hline
\end{btocharmtab}
\btocharmfig{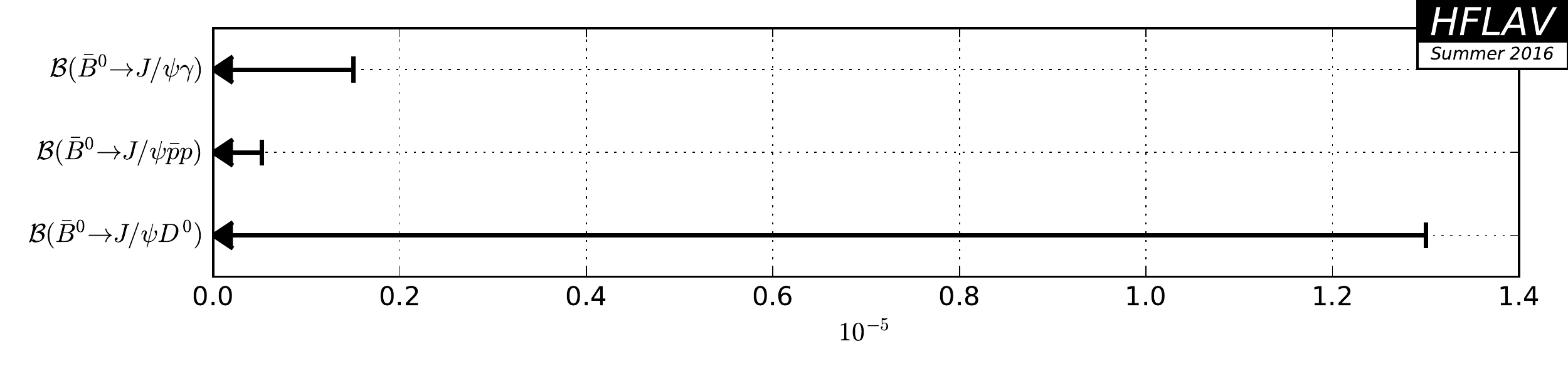}

\begin{btocharmtab}{Bd_cc_7}{Relative decay rates I}
\hline
\textbf{Parameter} & \begin{tabular}{l}\textbf{Measurements}\end{tabular} & \textbf{Average} \\
\hline
\hline
\multicolumn{3}{|l|}{${{\cal{B}} ( \bar{B}^{0} \to J/\psi \bar{K}_1^{0}(1270) )}/{{\cal{B}} ( B^{-} \to J/\psi K^{-} )}$}\\
 & \begin{tabular}{l} Belle \cite{Abe:2001wa}: $1.30 \pm 0.34 \pm 0.28$ \\ \end{tabular} & $1.30 \pm 0.44$ \\
\hline
\multicolumn{3}{|l|}{${{\cal{B}} ( \bar{B}^{0} \to J/\psi \bar{K}^{*}(892)^{0} )}/{{\cal{B}} ( \bar{B}^{0} \to J/\psi \bar{K}^{0} )}$}\\
 & \begin{tabular}{l} CDF \cite{Abe:1996kc}: $1.39 \pm 0.36 \pm 0.10$ \\ \babar \cite{Aubert:2004rz}: $1.51 \pm 0.05 \pm 0.08$ \\ \end{tabular} & $1.50 \pm 0.09$ \\
\hline
\multicolumn{3}{|l|}{${{\cal{B}} ( \bar{B}^{0} \to J/\psi \omega(782)) }/{ {\cal{B}} ( \bar{B}^{0} \to J/\psi \rho )}$}\\
 & \begin{tabular}{l} LHCb \cite{LHCb:2012cw}: $0.89 \pm 0.19 \,^{+0.07}_{-0.13}$ \\ \end{tabular} & $0.89 \,^{+0.20}_{-0.23}$ \\
\hline
\multicolumn{3}{|l|}{${{\cal{B}} ( \bar{B}^{0} \to J/\psi \omega(782) \bar{K}^{0} ) }/{ {\cal{B}} ( B^{-} \to J/\psi \omega(782) K^{-} )}$}\\
 & \begin{tabular}{l} \babar \cite{delAmoSanchez:2010jr}: $0.7 \pm 0.1 \pm 0.1$ \\ \end{tabular} & $0.7 \pm 0.1$ \\
\hline
\multicolumn{3}{|l|}{${{\cal{B}} ( \bar{B}^{0} \to J/\psi K_S^{0} \pi^{-} \pi^{+} )}/{{\cal{B}} ( \bar{B}^{0} \to J/\psi K_S^{0} )}$}\\
 & \begin{tabular}{l} LHCb \cite{Aaij:2014naa}: $0.493 \pm 0.034 \pm 0.027$ \\ \end{tabular} & $0.493 \pm 0.043$ \\
\hline
\multicolumn{3}{|l|}{$[{   {\cal{B}}  ( \bar{B}^{0} \to \psi(2S) K_S^{0} ) \times {\cal{B}}( \psi(2S)\to  J/\psi \pi^{-} \pi^{+})   }]/ { {\cal{B}} ( \bar{B}^{0} \to J/\psi K_S^{0} )}$}\\
 & \begin{tabular}{l} LHCb \cite{Aaij:2014naa}: $0.183 \pm 0.027 \pm 0.015$ \\ \end{tabular} & $0.183 \pm 0.031$ \\
\hline
\multicolumn{3}{|l|}{${{\cal{B}} ( \bar{B}^{0} \to \psi(2S) \bar{K}^{*}(892)^{0} )}/{{\cal{B}} ( \bar{B}^{0} \to \psi(2S) \bar{K}^{0} )}$}\\
 & \begin{tabular}{l} \babar \cite{Aubert:2004rz}: $1.00 \pm 0.14 \pm 0.09$ \\ \end{tabular} & $1.00 \pm 0.17$ \\
\hline
\multicolumn{3}{|l|}{${{\cal{B}} ( \bar{B}^{0} \to \psi(2S) K(892)^{*0} )}/{{\cal{B}} ( \bar{B}^{0} \to J/\psi  K(892)^{*0} )}$}\\
 & \begin{tabular}{l} LHCb \cite{Aaij:2012dda}: $0.476 \pm 0.014 \pm 0.016$ \\ \end{tabular} & $0.476 \pm 0.021$ \\
\hline
\multicolumn{3}{|l|}{${{\cal{B}} \bar{B}^{0} \to \psi(2S) \pi^{+}  \pi^{-} }/{{\cal{B}} \bar{B}^{0} \to J/\psi \pi^{+} \pi^{-}  }$}\\
 & \begin{tabular}{l} LHCb \cite{Aaij:2013cpa}: $0.56 \pm 0.07 \pm 0.05$ \\ \end{tabular} & $0.56 \pm 0.09$ \\
\hline
\multicolumn{3}{|l|}{${{\cal{B}} ( \bar{B}^{0} \to \eta_c \bar{K}^{0} )}/{{\cal{B}} ( B^{-} \to \eta_c K^{-} )}$}\\
 & \begin{tabular}{l} \babar \cite{Aubert:2004gc}: $0.87 \pm 0.13 \pm 0.07$ \\ \end{tabular} & $0.87 \pm 0.15$ \\
\hline
\multicolumn{3}{|l|}{${{\cal{B}} ( \bar{B}^{0} \to \eta_c \bar{K}^{0} )}/{{\cal{B}} ( \bar{B}^{0} \to J/\psi \bar{K}^{0} )}$}\\
 & \begin{tabular}{l} \babar \cite{Aubert:2004gc}: $1.34 \pm 0.19 \pm 0.40$ \\ \end{tabular} & $1.34 \pm 0.44$ \\
\hline
\multicolumn{3}{|l|}{${{\cal{B}} ( \bar{B}^{0} \to \eta_c \bar{K}^{*}(892)^{0} )}/{{\cal{B}} ( B^{-} \to \eta_c K^{-} )}$}\\
 & \begin{tabular}{l} \babar \cite{Aubert:2008kp}: $0.62 \pm 0.06 \pm 0.05$ \\ \end{tabular} & $0.62 \pm 0.08$ \\
\hline
\multicolumn{3}{|l|}{${{\cal{B}} ( \bar{B}^{0} \to \eta_c \bar{K}^{*}(892)^{0} )}/{{\cal{B}} ( \bar{B}^{0} \to \eta_c \bar{K}^{0} )}$}\\
 & \begin{tabular}{l} Belle \cite{Fang:2002gi}: $1.33 \pm 0.36 \,^{+0.24}_{-0.33}$ \\ \end{tabular} & $1.33 \,^{+0.43}_{-0.49}$ \\
\hline
\multicolumn{3}{|l|}{${{\cal{B}} ( \bar{B}^{0} \to \chi_{c1} K^{-} \pi^{+} )}/{{\cal{B}} ( \bar{B}^{0} \to J/\psi K^{-} \pi^{+} )}$}\\
 & \begin{tabular}{l} \babar \cite{Lees:2011ik}: $0.474 \pm 0.013 \pm 0.054$ \\ \end{tabular} & $0.474 \pm 0.056$ \\
\hline
\multicolumn{3}{|l|}{${{\cal{B}} ( \bar{B}^{0} \to \chi_{c1} \bar{K}^{*}(892)^{0} )}/{{\cal{B}} ( \bar{B}^{0} \to \chi_{c1} \bar{K}^{0} )}$}\\
 & \begin{tabular}{l} \babar \cite{Aubert:2004rz}: $0.72 \pm 0.11 \pm 0.12$ \\ \end{tabular} & $0.72 \pm 0.16$ \\
\hline
\multicolumn{3}{|l|}{$[{{\cal{B}} ( \bar{B}^{0} \to h_c(1P) \bar{K}^{*}(892)^{0} )\times {\cal{B}} ( h_c(1P) \to \eta_c \gamma )}]/{{\cal{B}} ( B^{-} \to \eta_c K^{-} )}$}\\
 & \begin{tabular}{l} \babar \cite{Aubert:2008kp}: $< 0.236$ \\ \end{tabular} & $< 0.236$ \\
\hline
\end{btocharmtab}
\btocharmfig{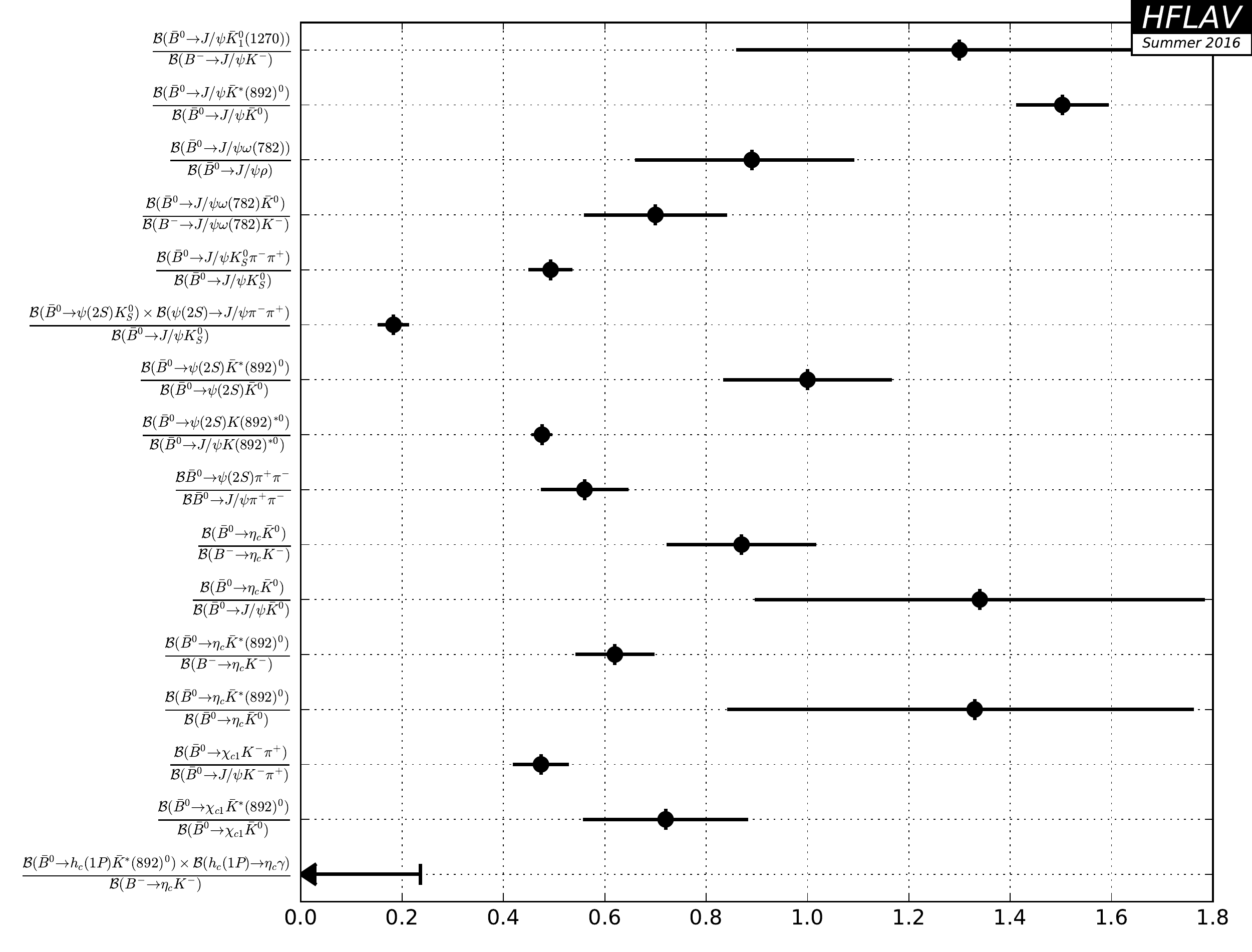}

\begin{btocharmtab}{Bd_cc_8}{Relative decay rates II}
\hline
\textbf{Parameter} & \begin{tabular}{l}\textbf{Measurements}\end{tabular} & \textbf{Average} \\
\hline
\hline
\multicolumn{3}{|l|}{${{\cal{B}} ( \bar{B}^{0} \to J/\psi \eta )}/  {{\cal{B}} ( \bar{B}_{s}^{0} \to J/\psi \eta )}$}\\
 & \begin{tabular}{l} LHCb \cite{Aaij:2014jna}: $0.0185 \pm 0.0061 \pm 0.0014$ \\ \end{tabular} & $0.0185 \pm 0.0063$ \\
\hline
\multicolumn{3}{|l|}{${{\cal{B}} ( \bar{B}^{0} \to J/\psi \eta^{\prime })}/  {{\cal{B}} ( \bar{B}_{s}^{0} \to J/\psi \eta^{\prime } )}$}\\
 & \begin{tabular}{l} LHCb \cite{Aaij:2014jna}: $0.0228 \pm 0.0065 \pm 0.0016$ \\ \end{tabular} & $0.0228 \pm 0.0067$ \\
\hline
\multicolumn{3}{|l|}{${{\cal{B}} ( \bar{B}^{0} \to J/\psi K_S^{0} K^{\pm} \pi^{\mp} )}/{{\cal{B}} ( \bar{B}^{0} \to J/\psi K_S^{0} \pi^{+} \pi^{-} )}$}\\
 & \begin{tabular}{l} LHCb \cite{Aaij:2014naa}: $< 0.048$ \\ \end{tabular} & $< 0.048$ \\
\hline
\multicolumn{3}{|l|}{${{\cal{B}} ( \bar{B}^{0} \to J/\psi K_S^{0} K^{+} K^{-} )}/{{\cal{B}} ( \bar{B}^{0} \to J/\psi K_S^{0} \pi^{+} \pi^{-} )}$}\\
 & \begin{tabular}{l} LHCb \cite{Aaij:2014naa}: $0.047 \pm 0.010 \pm 0.004$ \\ \end{tabular} & $0.047 \pm 0.011$ \\
\hline
\end{btocharmtab}
\btocharmfig{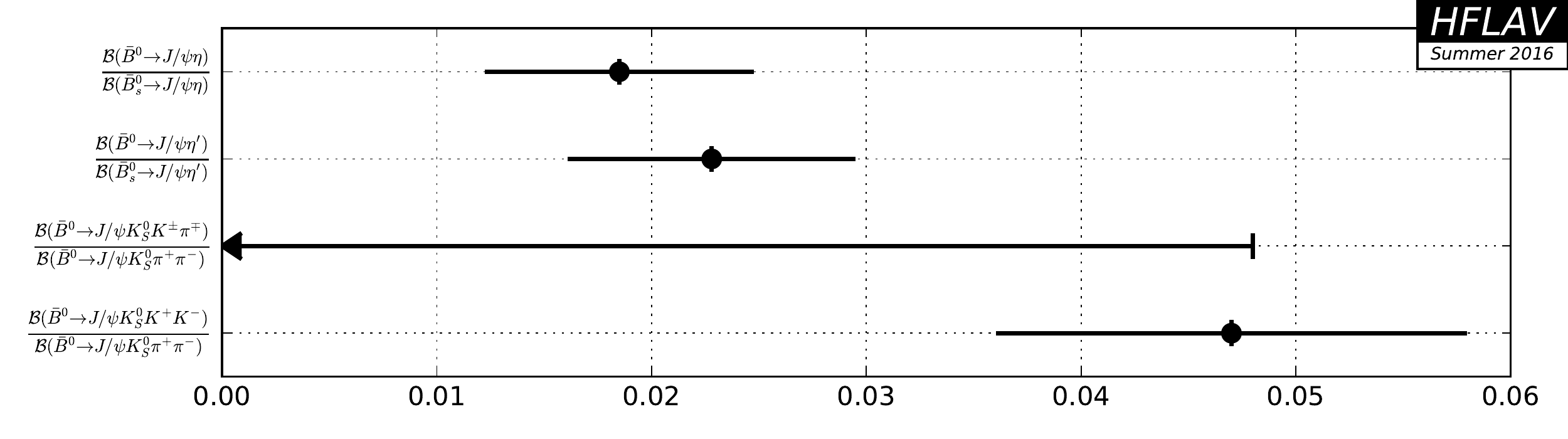}

\begin{btocharmtab}{Bd_cc_9}{Polarization fractions}
\hline
\textbf{Parameter} & \begin{tabular}{l}\textbf{Measurements}\end{tabular} & \textbf{Average} \\
\hline
\hline
\multicolumn{3}{|l|}{${\vert{\cal{A}}_{0}\vert^{2} ( B^{0} \to J/\psi \bar{K}^{*}(892)^{0} )}/{\vert{\cal{A}}_{0}\vert^{2} ( \bar{B}^{0} \to J/\psi \bar{K}^{*}(892)^{0} )}$}\\
 & \begin{tabular}{l} \babar \cite{Aubert:2004ei}: $< 0.32$ \\ \end{tabular} & $< 0.32$ \\
\hline
\multicolumn{3}{|l|}{${\vert{\cal{A}}_{0}\vert^{2} ( \bar{B}^{0} \to J/\psi K^{*}(892)^{0} )}/{\vert{\cal{A}}_{0}\vert^{2} ( B^{0} \to J/\psi K^{*}(892)^{0} )}$}\\
 & \begin{tabular}{l} \babar \cite{Aubert:2004ei}: $< 0.26$ \\ \end{tabular} & $< 0.26$ \\
\hline
\end{btocharmtab}
\btocharmfig{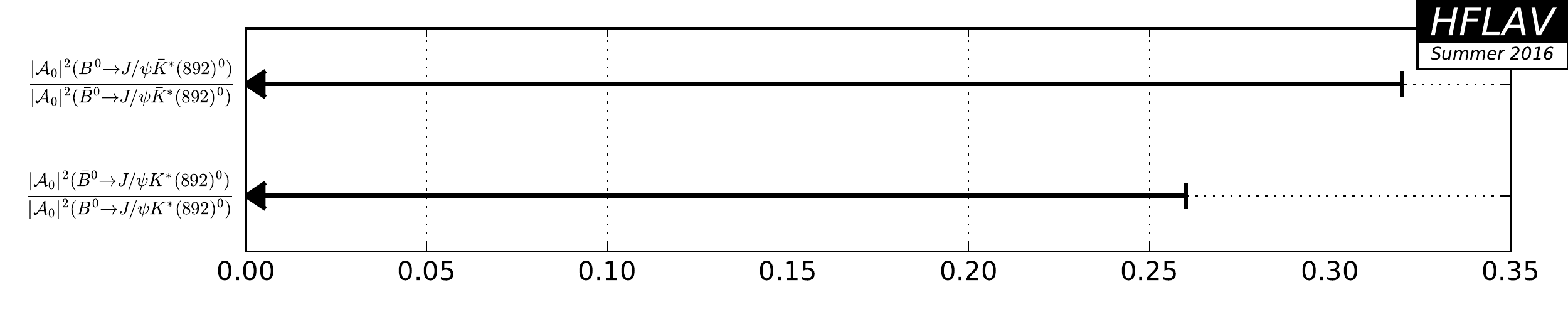}

\subsubsection{Decays to charm baryons}
\label{sec:b2c:Bd_baryon}
Averages of $\bar{B}^0$ decays to charm baryons are shown in Tables~\ref{tab:b2c:Bd_baryon_1}--\ref{tab:b2c:Bd_baryon_4} and Figs.~\ref{fig:b2c:Bd_baryon_1}--\ref{fig:b2c:Bd_baryon_3}.
\begin{btocharmtab}{Bd_baryon_1}{Absolute decay rates to charm baryons I $[10^{-4}]$}
\hline
\textbf{Parameter} & %
 & $7.9 \pm 2.1$ \\
\hline
\end{btocharmtab}
\btocharmfig{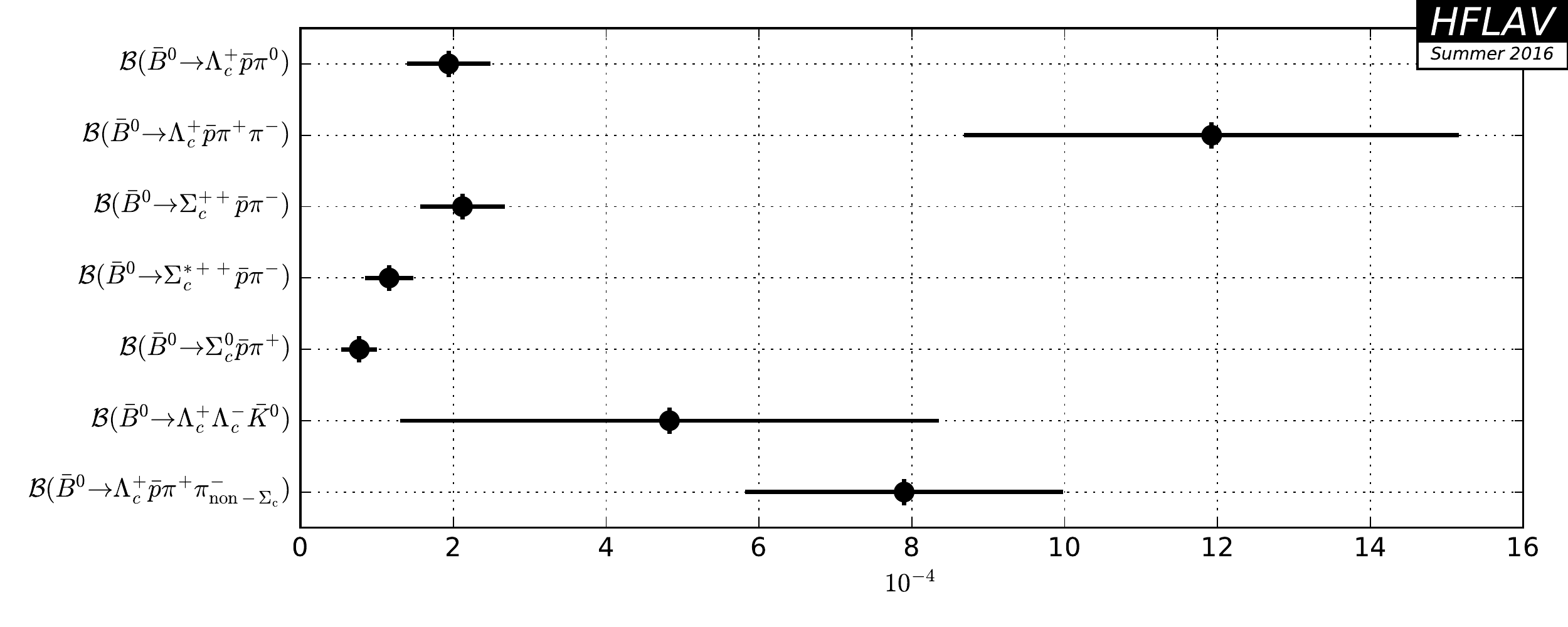}

\begin{btocharmtab}{Bd_baryon_2}{Absolute decay rates to charm baryons II $[10^{-5}]$}
\hline
\textbf{Parameter} & %
 & $4.33 \pm 1.43$ \\
\hline
\end{btocharmtab}
\btocharmfig{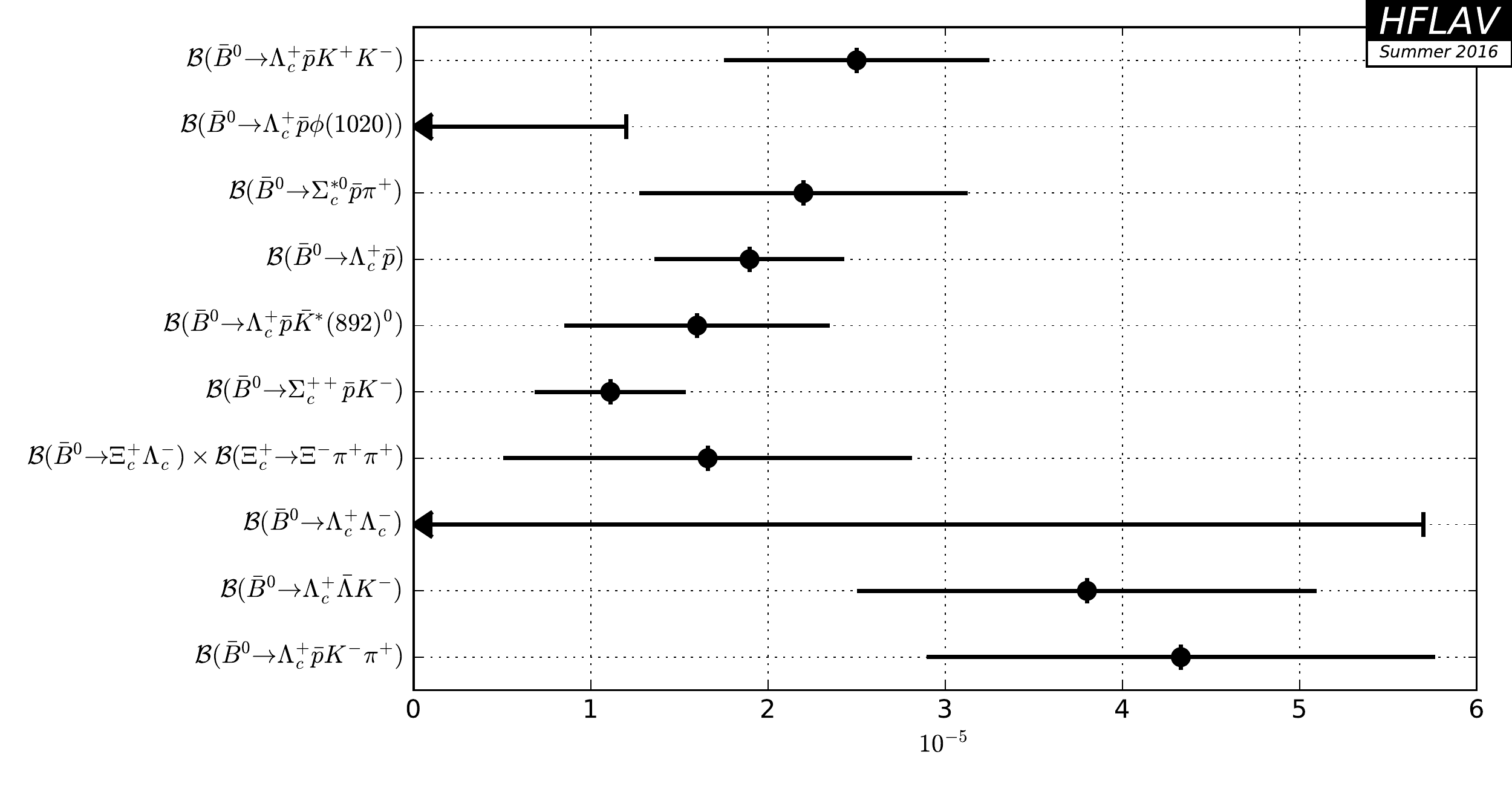}

\begin{btocharmtab}{Bd_baryon_3}{Absolute decay rates to charm baryons III $[10^{-6}]$}
\hline
\textbf{Parameter} & %
 & $< 0.14$ \\
\hline
\end{btocharmtab}
\btocharmfig{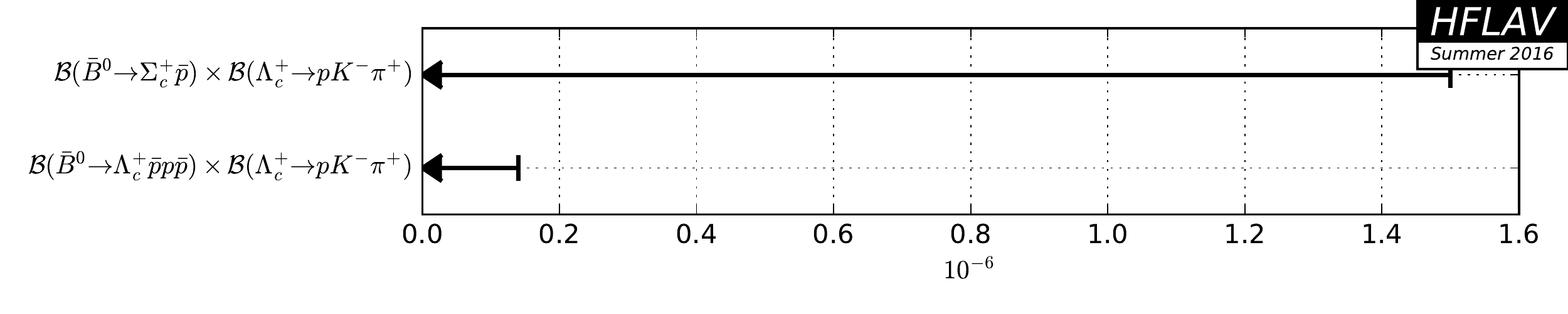}

\begin{btocharmtab}{Bd_baryon_4}{Relative decay rates to charm baryons $[10^{-3}]$}
\hline
\textbf{Parameter} & %
 & $< 2$ \\
\hline
\end{btocharmtab}

\subsubsection{Decays to other ($XYZ$) states}
\label{sec:b2c:Bd_other}
Averages of $\bar{B}^0$ decays to other ($XYZ$) states are shown in Tables~\ref{tab:b2c:Bd_other_1}--\ref{tab:b2c:Bd_other_6} and Figs.~\ref{fig:b2c:Bd_other_1}--\ref{fig:b2c:Bd_other_6}.
\begin{btocharmtab}{Bd_other_1}{Decays to $X(3872)$ $[10^{-5}]$}
\hline
\textbf{Parameter} & %
 & $< 4.37$ \\
\hline
\end{btocharmtab}

\begin{btocharmtab}{Bd_other_3}{Decays to neutral states other than $X(3872)$ $[10^{-4}]$}
\hline
\textbf{Parameter} & %
 & $0.40 \,^{+1.98}_{-0.10}$ \\
\hline
\end{btocharmtab}
\btocharmfig{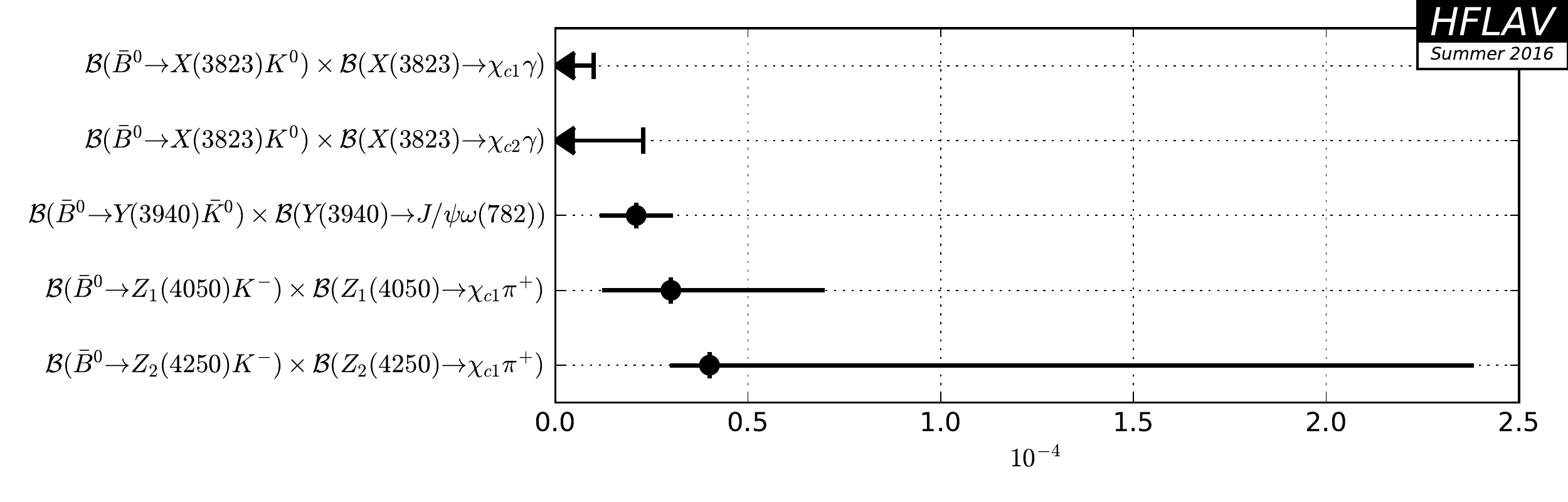}

\begin{btocharmtab}{Bd_other_4}{Decays to charged states I $[10^{-4}]$}
\hline
\textbf{Parameter} & %
 & $< 5.0$ \\
\hline
\end{btocharmtab}

\begin{btocharmtab}{Bd_other_5}{Decays to charged states II $[10^{-5}]$}
\hline
\textbf{Parameter} & %
 & $2.2 \,^{+1.3}_{-0.8}$ \\
\hline
\end{btocharmtab}
\btocharmfig{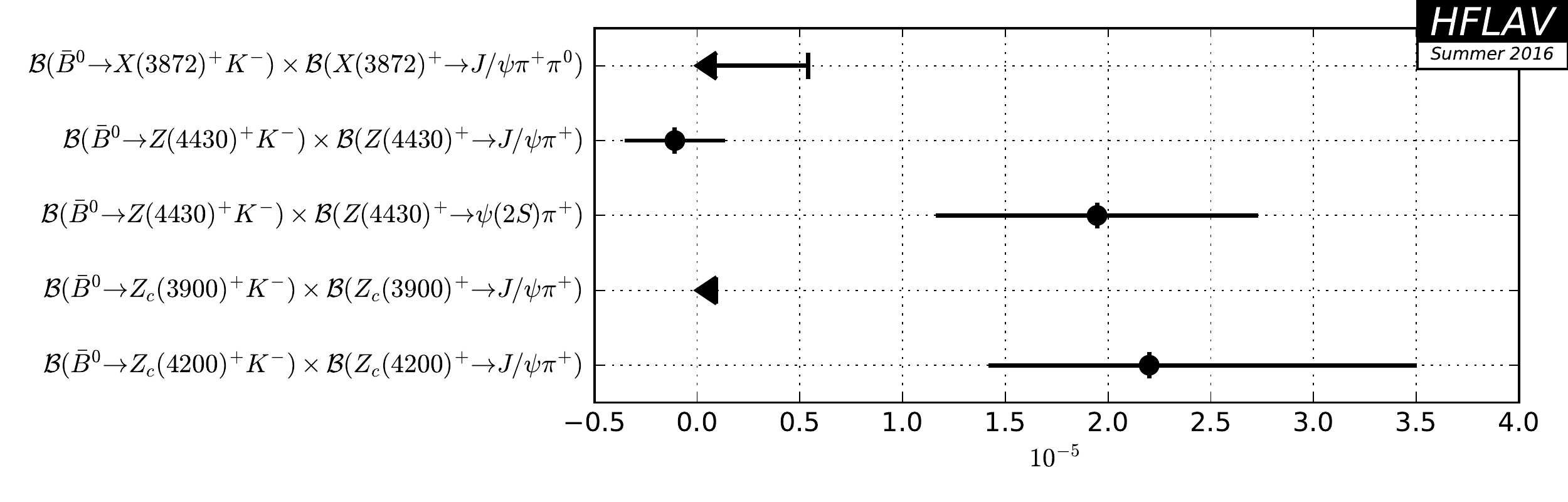}

\begin{btocharmtab}{Bd_other_6}{Relative decay rates}
\hline
\textbf{Parameter} & %
 & $0.7 \,^{+0.4}_{-0.3}$ \\
\hline
\end{btocharmtab}
\btocharmfig{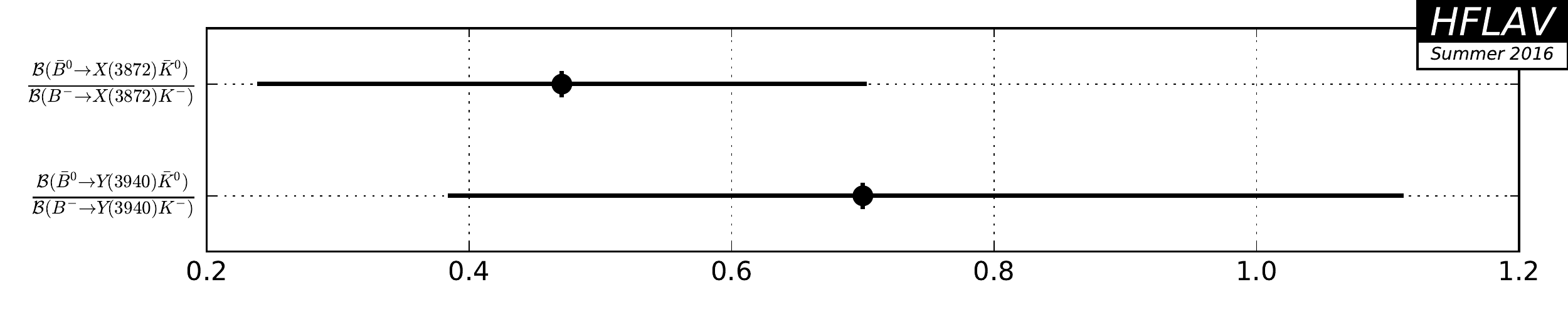}

\subsection{Decays of $B^-$ mesons}
\label{sec:b2c:Bu}
Measurements of $B^-$ decays to charmed hadrons are summarized in Sections~\ref{sec:b2c:Bu_D} to~\ref{sec:b2c:Bu_other}.

\subsubsection{Decays to a single open charm meson}
\label{sec:b2c:Bu_D}
Averages of $B^-$ decays to a single open charm meson are shown in Tables~\ref{tab:b2c:Bu_D_1}--\ref{tab:b2c:Bu_D_15} and Figs.~\ref{fig:b2c:Bu_D_1}--\ref{fig:b2c:Bu_D_15}.
In this section $D^{**}$ refers to the sum of all the non-strange charm meson
states with masses in the range $2.2-2.8~\gevcc$.
\begin{btocharmtab}{Bu_D_1}{Decays to a $D^{(*)}$ meson and one or more pions $[10^{-2}]$}
\hline
\textbf{Parameter} & & $0.567 \pm 0.125$ \\
\hline
\end{btocharmtab}
\btocharmfig{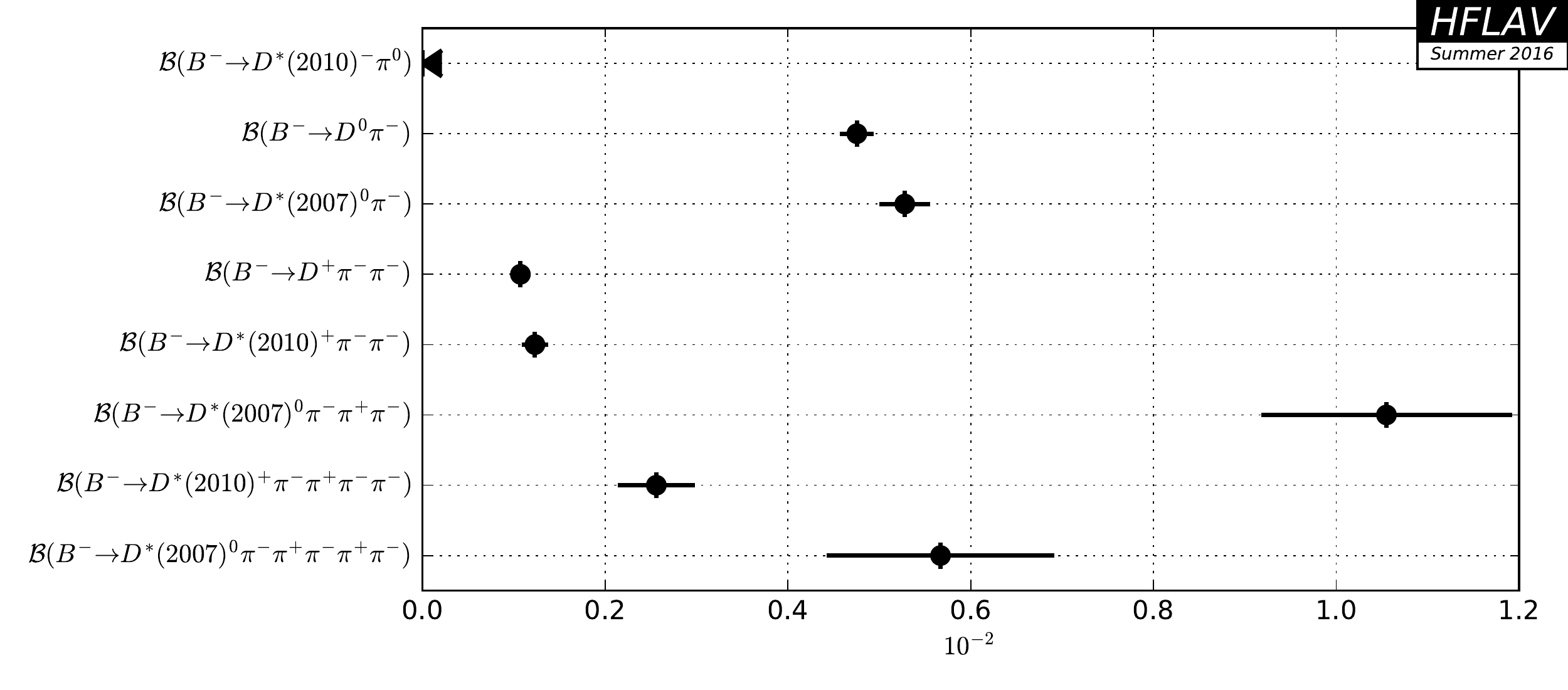}

\begin{btocharmtab}{Bu_D_2}{Decays to a $D^{(*)0}$ meson and one or more kaons $[10^{-3}]$}
\hline
\textbf{Parameter} & %
 & $0.0731 \pm 0.0049$ \\
\hline
\end{btocharmtab}
\btocharmfig{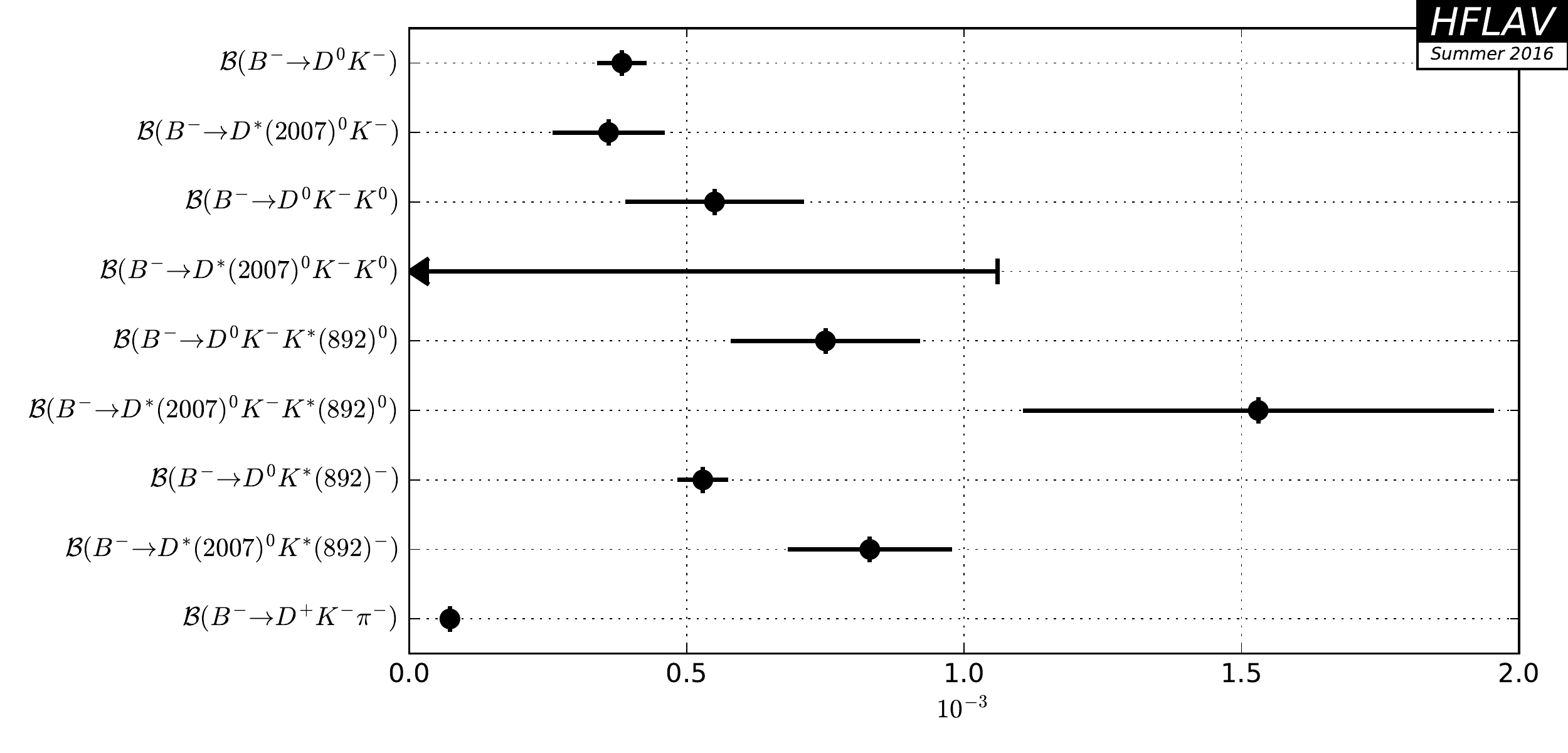}

\begin{btocharmtab}{Bu_D_3}{Decays to a $D^{(*)-}$ meson and a neutral kaon or a kaon and a pion $[10^{-6}]$}
\hline
\textbf{Parameter} & %
 & $< 9$ \\
\hline
\end{btocharmtab}
\btocharmfig{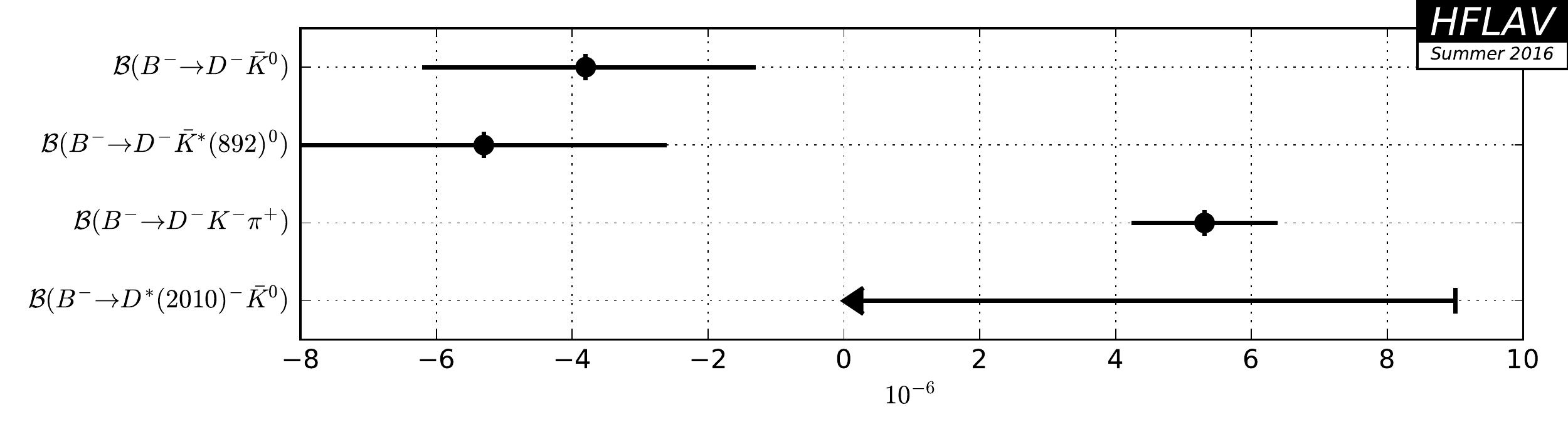}

\begin{btocharmtab}{Bu_D_4}{Relative decay rates to $D^{0}$ mesons I}
\hline
\textbf{Parameter} & %
 & $1.27 \pm 0.13$ \\
\hline
\end{btocharmtab}
\btocharmfig{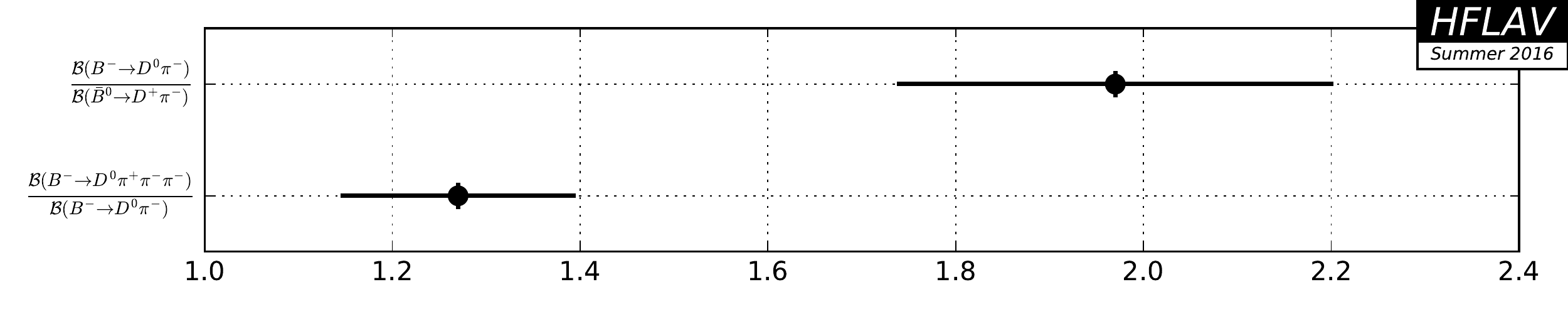}

\begin{btocharmtab}{Bu_D_5}{Relative decay rates to $D^{0}$ mesons II $[10^{-2}]$}
\hline
\textbf{Parameter} & %
 & $9.4 \pm 1.6$ \\
\hline
\end{btocharmtab}
\btocharmfig{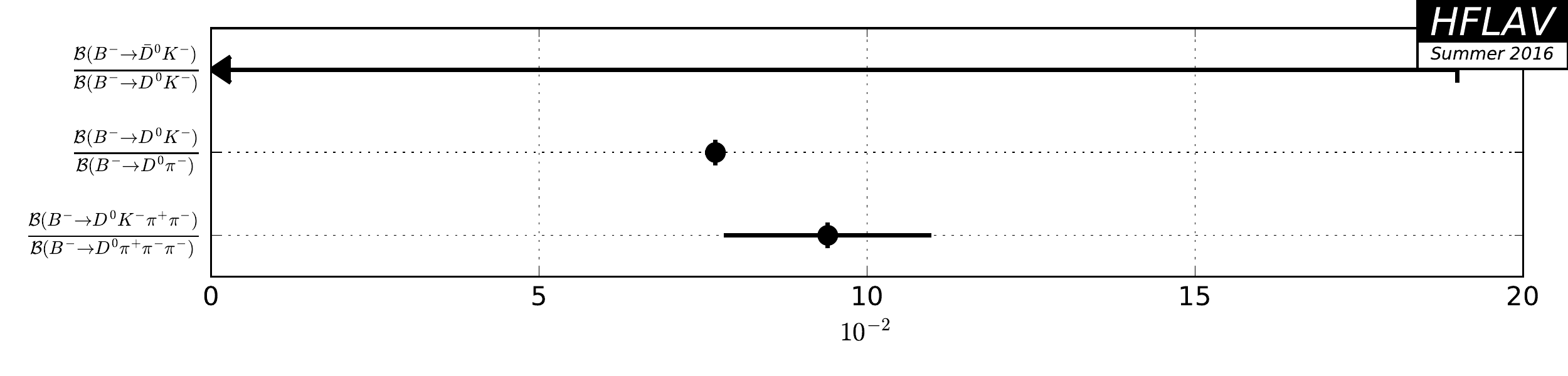}

\begin{btocharmtab}{Bu_D_6}{Absolute decay rates to excited $D$ mesons $[10^{-3}]$}
\hline
\textbf{Parameter} & %
 & $5.50 \pm 1.16$ \\
\hline
\end{btocharmtab}

\begin{btocharmtab}{Bu_D_7}{Absolute product decay rates to excited $D$ mesons I $[10^{-3}]$}
\hline
\textbf{Parameter} & %
 & $0.35 \pm 0.05$ \\
\hline
\end{btocharmtab}
\btocharmfig{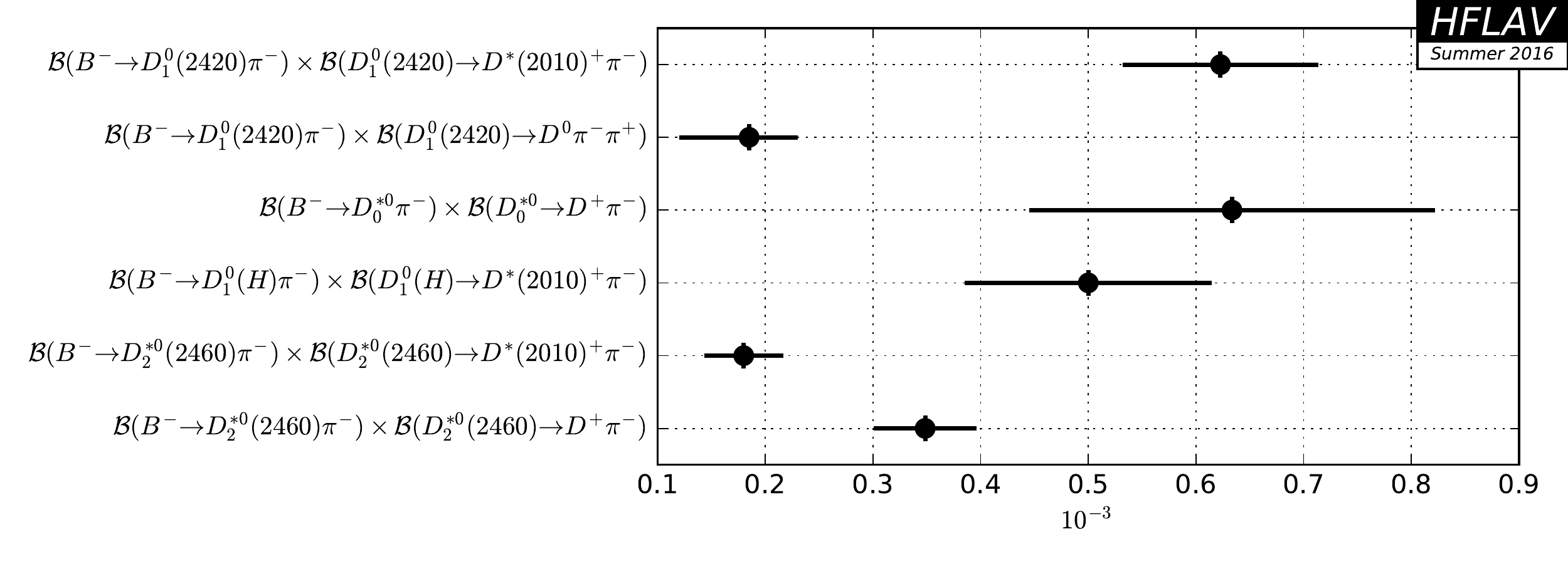}

\begin{btocharmtab}{Bu_D_8}{Absolute product decay rates to excited $D$ mesons II $[10^{-5}]$}
\hline
\textbf{Parameter} & %
 & $< 2.2$ \\
\hline
\end{btocharmtab}
\btocharmfig{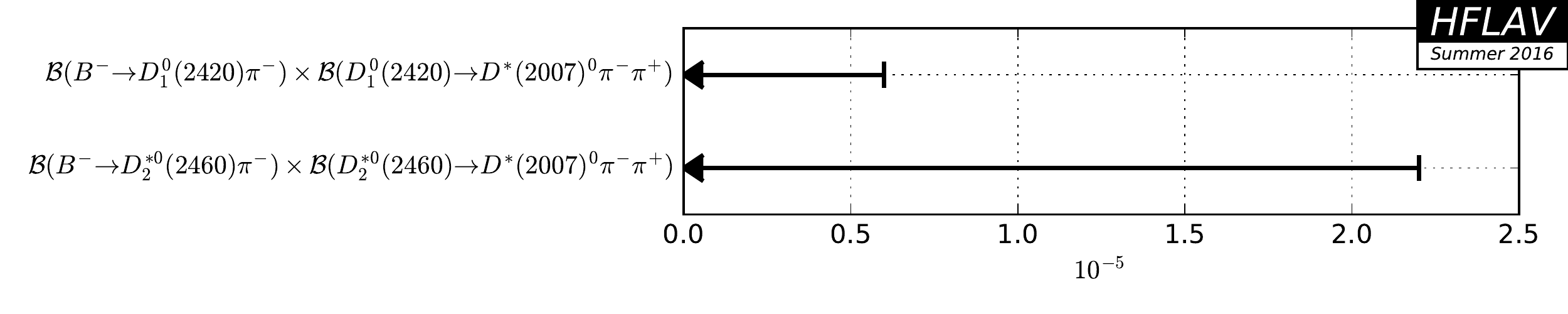}

\begin{btocharmtab}{Bu_D_9}{Relative decay rates to excited $D$ mesons}
\hline
\textbf{Parameter} & %
 & $0.0811 \pm 0.0053$ \\
\hline
\end{btocharmtab}
\btocharmfig{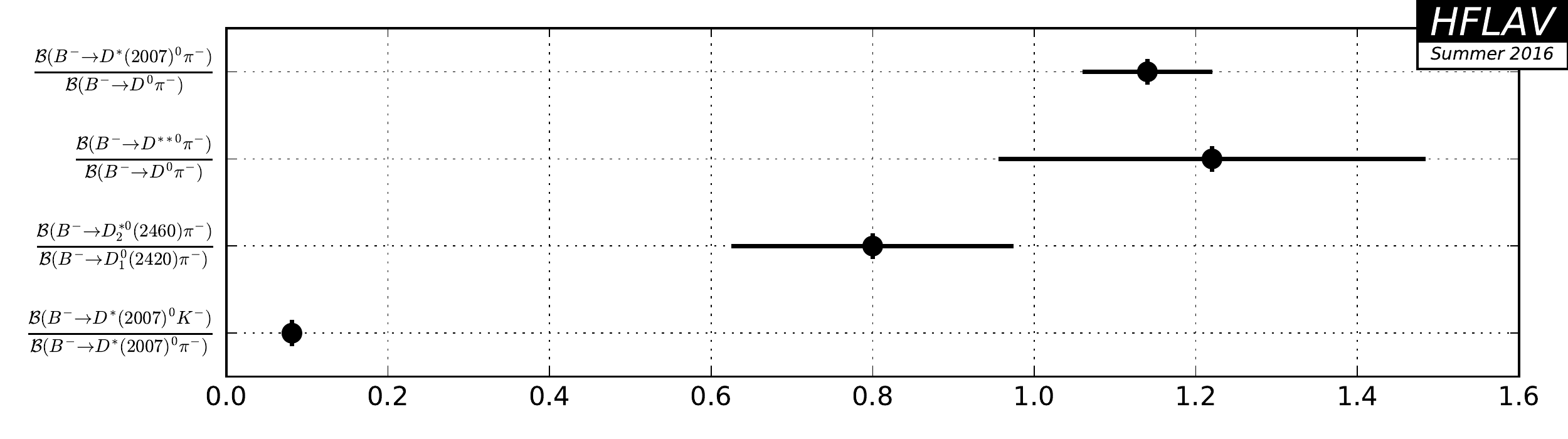}

\begin{btocharmtab}{Bu_D_10}{Relative product decay rates to excited $D$ mesons}
\hline
\textbf{Parameter} & %
 & $1.86 \pm 0.25$ \\
\hline
\end{btocharmtab}
\btocharmfig{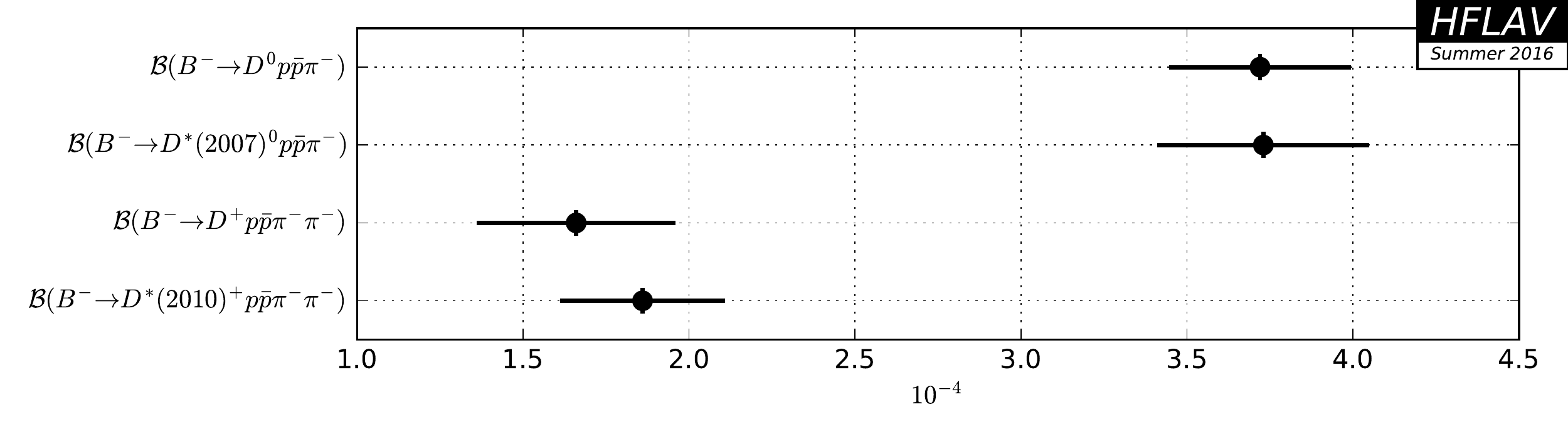}

\begin{btocharmtab}{Bu_D_14}{Baryonic decays II $[10^{-5}]$}
\hline
\textbf{Parameter} & %
 & $< 1.5$ \\
\hline
\end{btocharmtab}
\btocharmfig{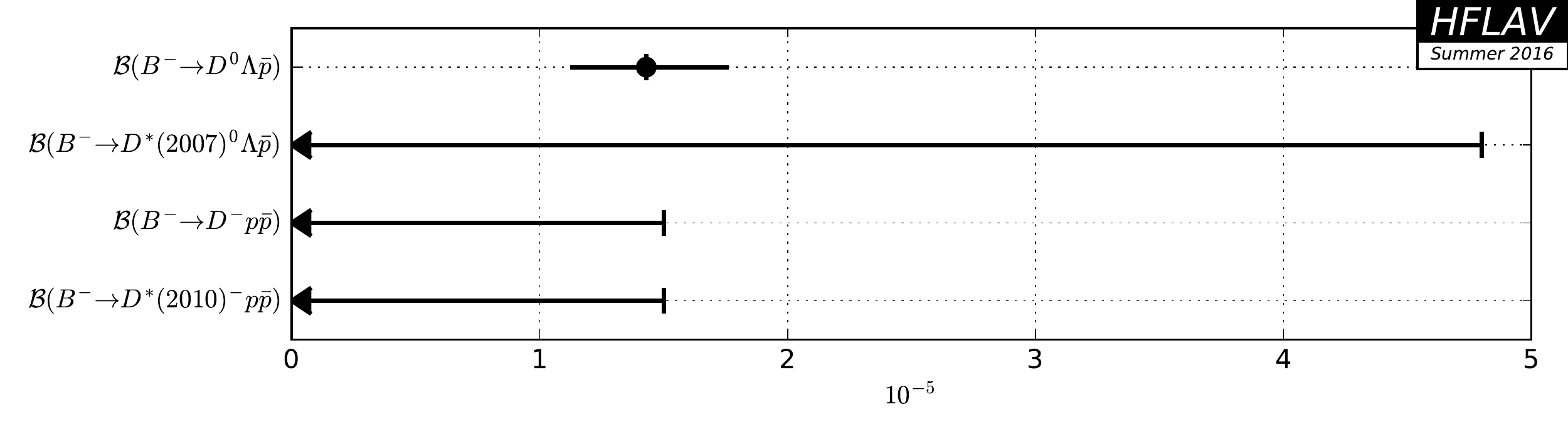}

\begin{btocharmtab}{Bu_D_15}{Lepton number violating decays $[10^{-6}]$}
\hline
\textbf{Parameter} & %
 & $< 1.0$ \\
\hline
\end{btocharmtab}
\btocharmfig{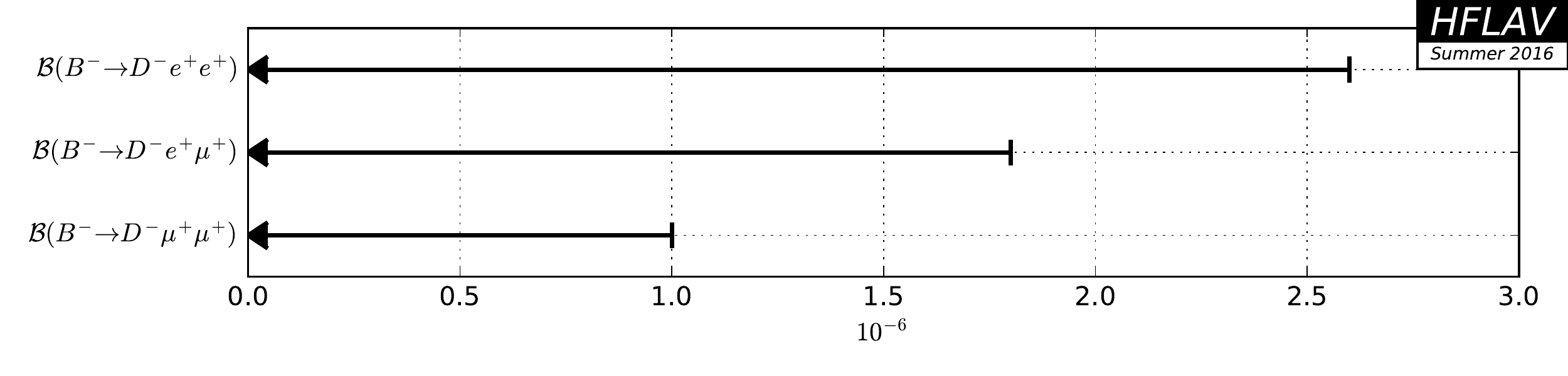}

\subsubsection{Decays to two open charm mesons}
\label{sec:b2c:Bu_DD}
Averages of $B^-$ decays to two open charm mesons are shown in Tables~\ref{tab:b2c:Bu_DD_1}--\ref{tab:b2c:Bu_DD_9} and Figs.~\ref{fig:b2c:Bu_DD_1}--\ref{fig:b2c:Bu_DD_9}.
\begin{btocharmtab}{Bu_DD_1}{Decays to $D^{(*)-}D^{(*)0}$ $[10^{-3}]$}
\hline
\textbf{Parameter} & %
 & $0.81 \pm 0.17$ \\
\hline
\end{btocharmtab}
\btocharmfig{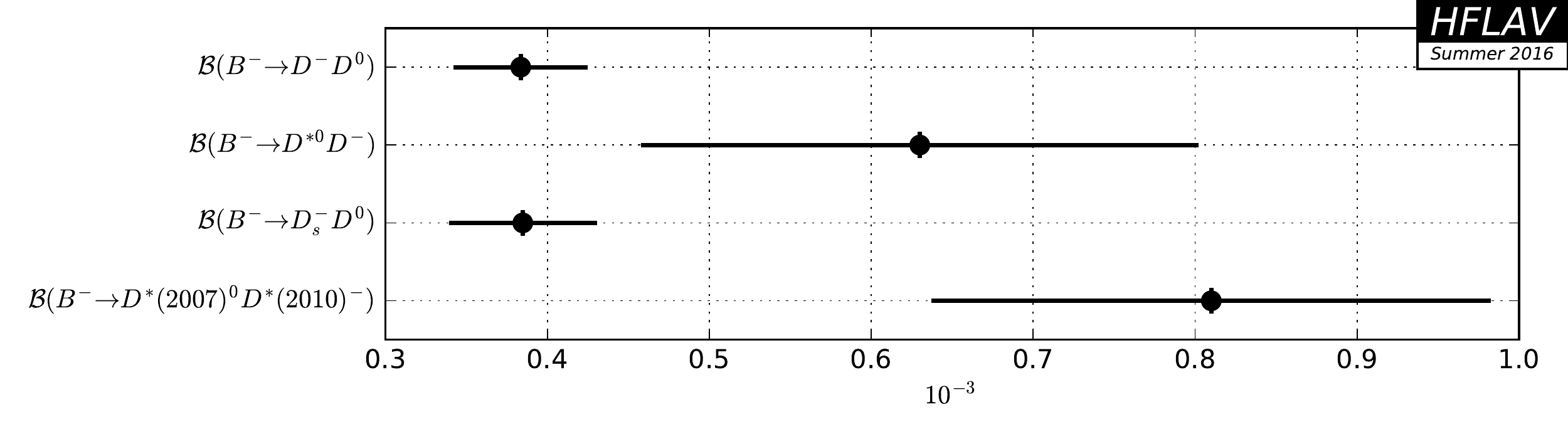}

\begin{btocharmtab}{Bu_DD_2}{Decays to two $D$ mesons and a kaon I $[10^{-2}]$}
\hline
\textbf{Parameter} & %
 & $0.917 \pm 0.122$ \\
\hline
\end{btocharmtab}
\btocharmfig{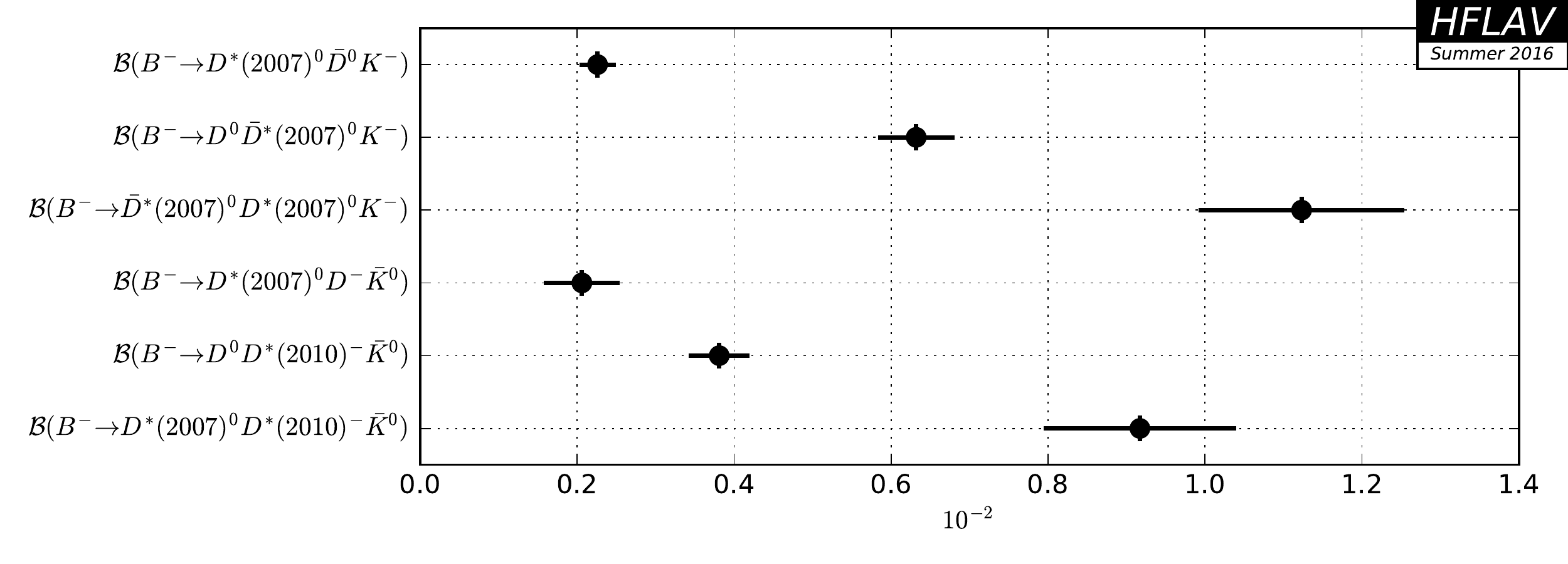}

\begin{btocharmtab}{Bu_DD_3}{Decays to two $D$ mesons and a kaon II $[10^{-3}]$}
\hline
\textbf{Parameter} & %
 & $1.70 \pm 0.35$ \\
\hline
\end{btocharmtab}
\btocharmfig{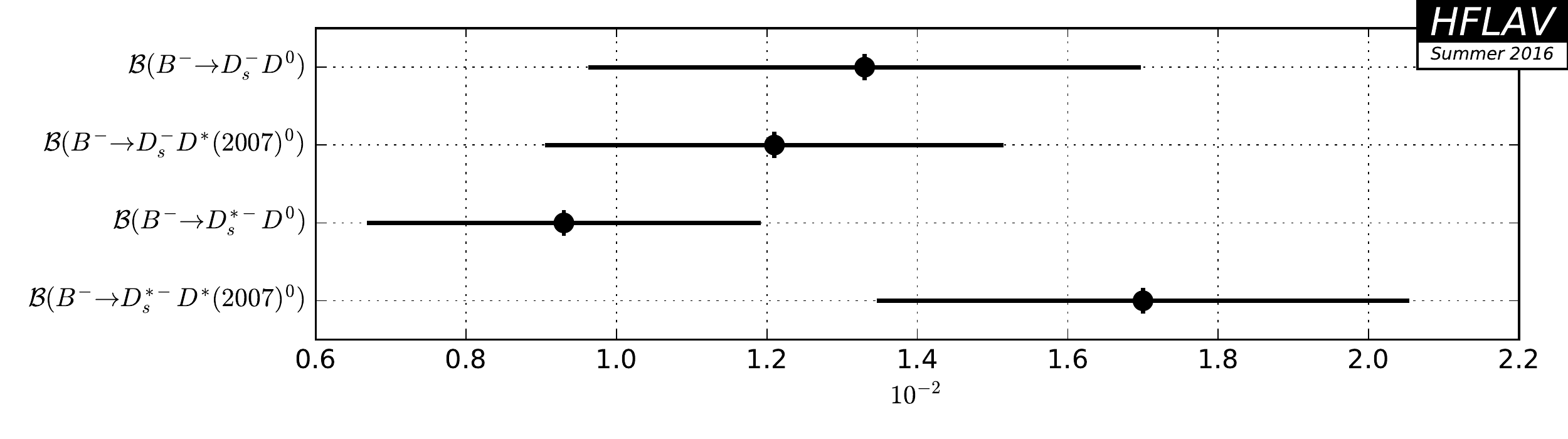}

\begin{btocharmtab}{Bu_DD_5}{Product decays rates to $D_s^{(*)-}D^{(*)+}$ $[10^{-4}]$}
\hline
\textbf{Parameter} & %
 & $8.57 \pm 1.86$ \\
\hline
\end{btocharmtab}
\btocharmfig{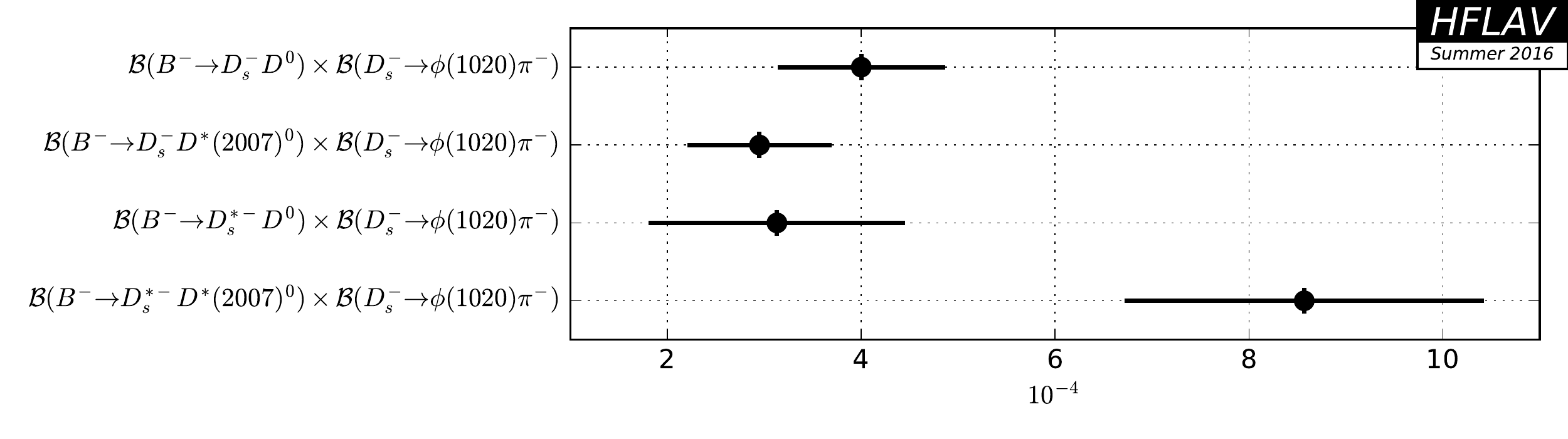}

\begin{btocharmtab}{Bu_DD_6}{Relative decay rates}
\hline
\textbf{Parameter} & %
 & $1.22 \pm 0.07$ \\
\hline
\end{btocharmtab}

\begin{btocharmtab}{Bu_DD_7}{Absolute decays rates to excited $D_s$ mesons $[10^{-3}]$}
\hline
\textbf{Parameter} & %
 & $11.2 \pm 3.3$ \\
\hline
\end{btocharmtab}
\btocharmfig{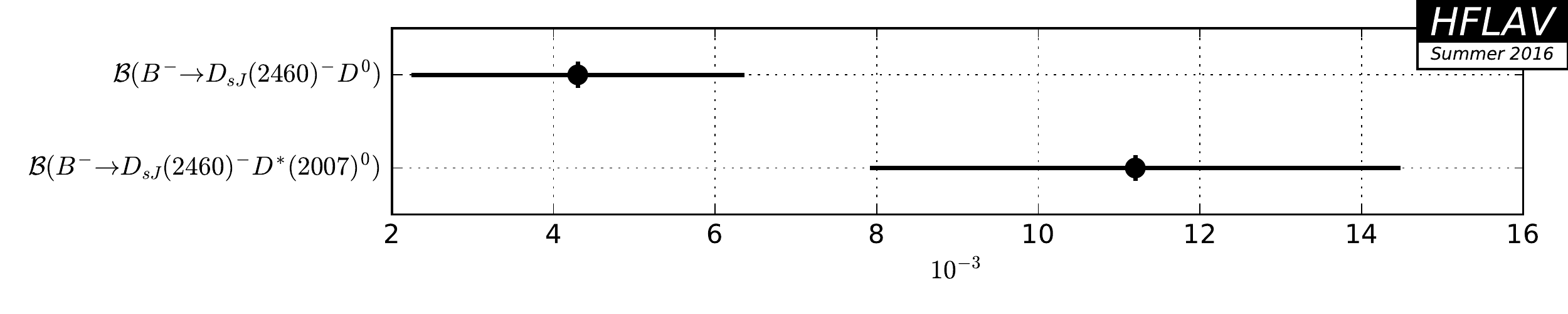}

\begin{btocharmtab}{Bu_DD_8}{Product decays rates to excited $D_s$ mesons I $[10^{-3}]$}
\hline
\textbf{Parameter} & %
 & $7.6 \,^{+3.6}_{-2.9}$ \\
\hline
\end{btocharmtab}
\btocharmfig{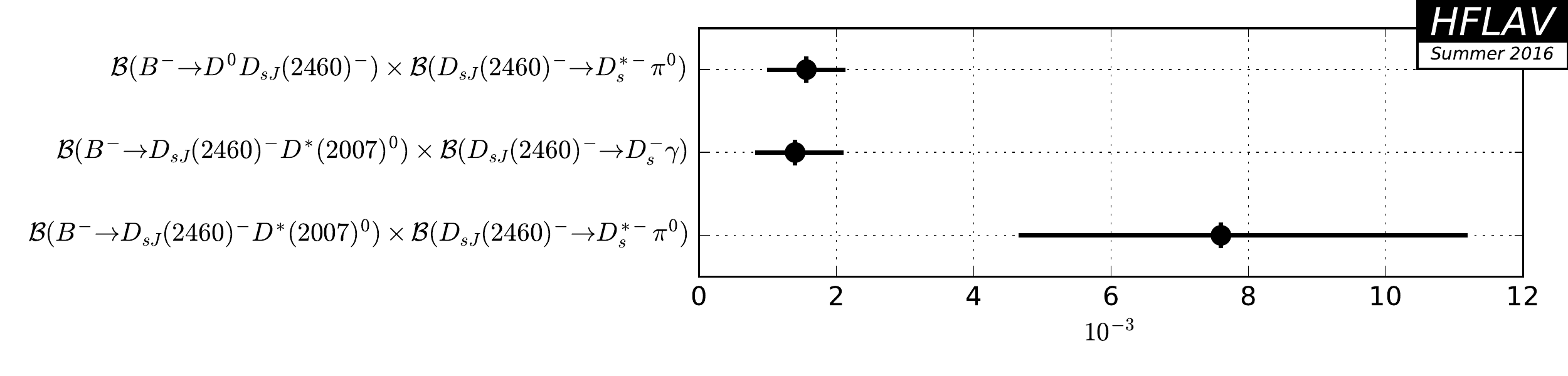}

\begin{btocharmtab}{Bu_DD_9}{Product decays rates to excited $D_s$ mesons II $[10^{-3}]$}
\hline
\textbf{Parameter} & %
 & $< 1.069$ \\
\hline
\end{btocharmtab}
\btocharmfig{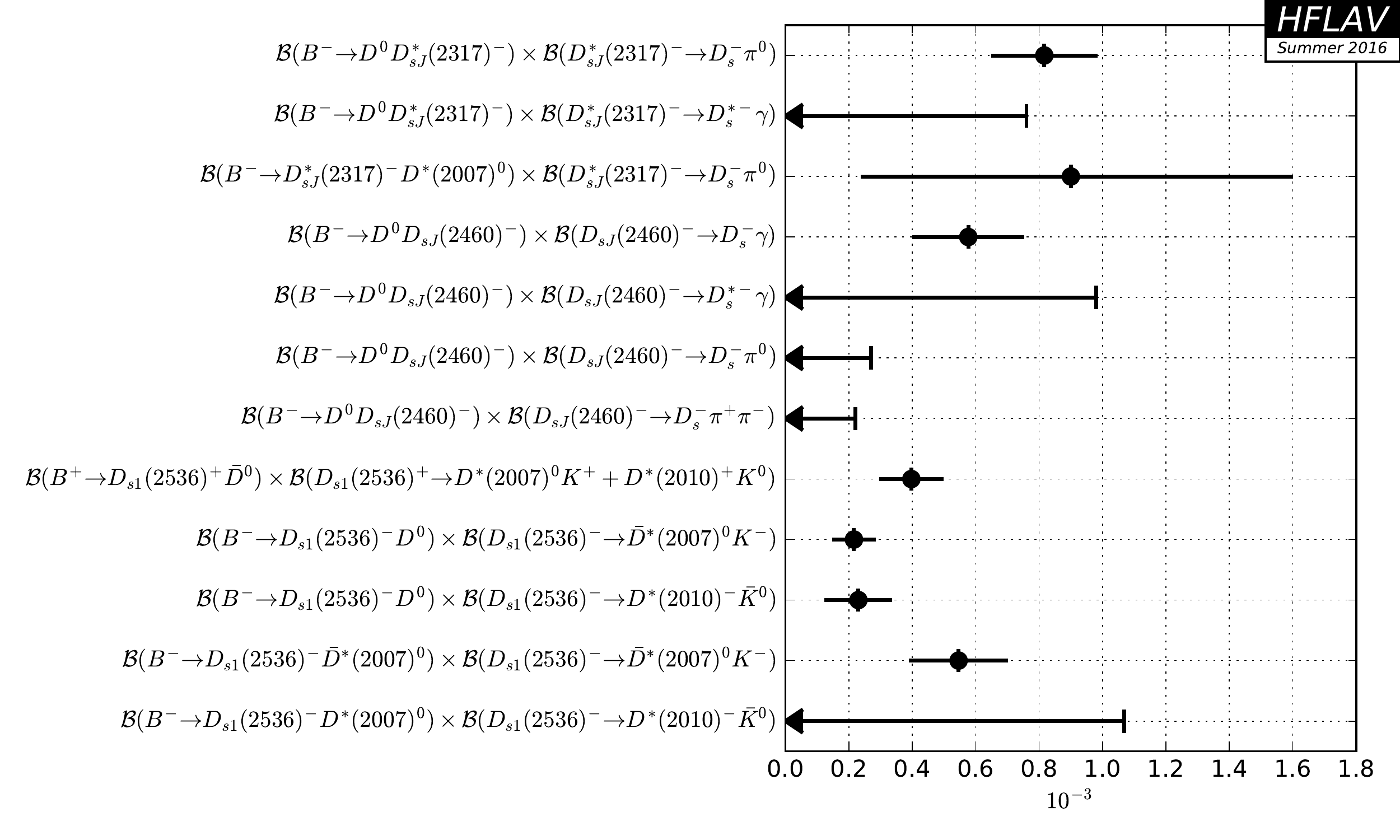}

\subsubsection{Decays to charmonium states}
\label{sec:b2c:Bu_cc}
Averages of $B^-$ decays to charmonium states are shown in Tables~\ref{tab:b2c:Bu_cc_1}--\ref{tab:b2c:Bu_cc_11} and Figs.~\ref{fig:b2c:Bu_cc_1}--\ref{fig:b2c:Bu_cc_11}.
\begin{btocharmtab}{Bu_cc_1}{Decays to $J/\psi$ and one kaon I $[10^{-3}]$}
\hline
\textbf{Parameter} & %
 & $0.807 \pm 0.052$ {\tiny CL=2.3\permil} \\
\hline
\end{btocharmtab}
\btocharmfig{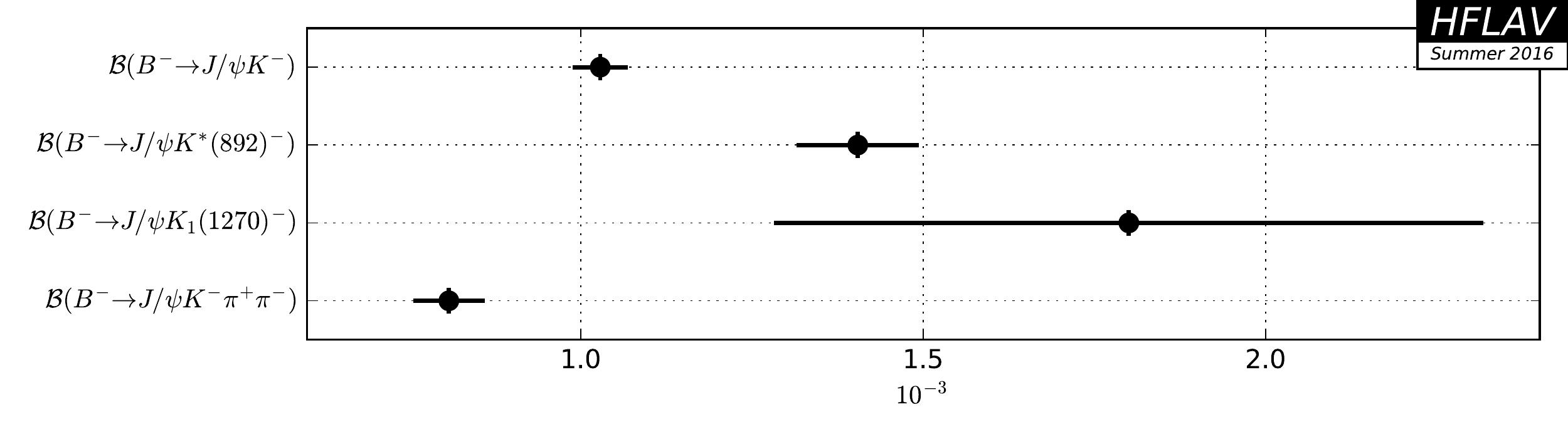}

\begin{btocharmtab}{Bu_cc_2}{Decays to $J/\psi$ and one kaon II $[10^{-4}]$}
\hline
\textbf{Parameter} & %
 & $0.44 \pm 0.15$ \\
\hline
\end{btocharmtab}
\btocharmfig{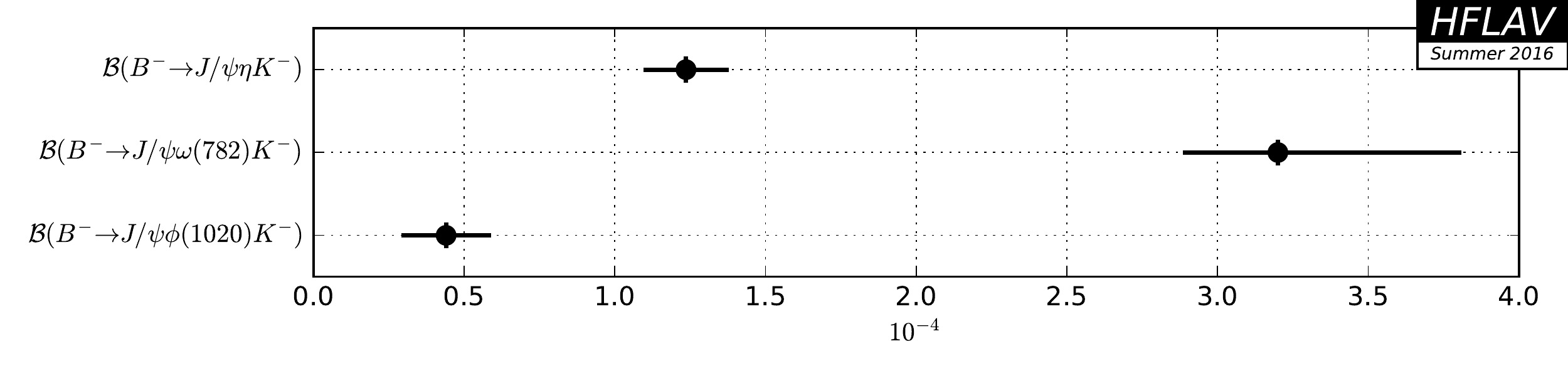}

\begin{btocharmtab}{Bu_cc_3}{Decays to charmonium other than $J/\psi$ and one kaon I $[10^{-3}]$}
\hline
\textbf{Parameter} & %
 & $0.34 \pm 0.18$ \\
\hline
\end{btocharmtab}
\btocharmfig{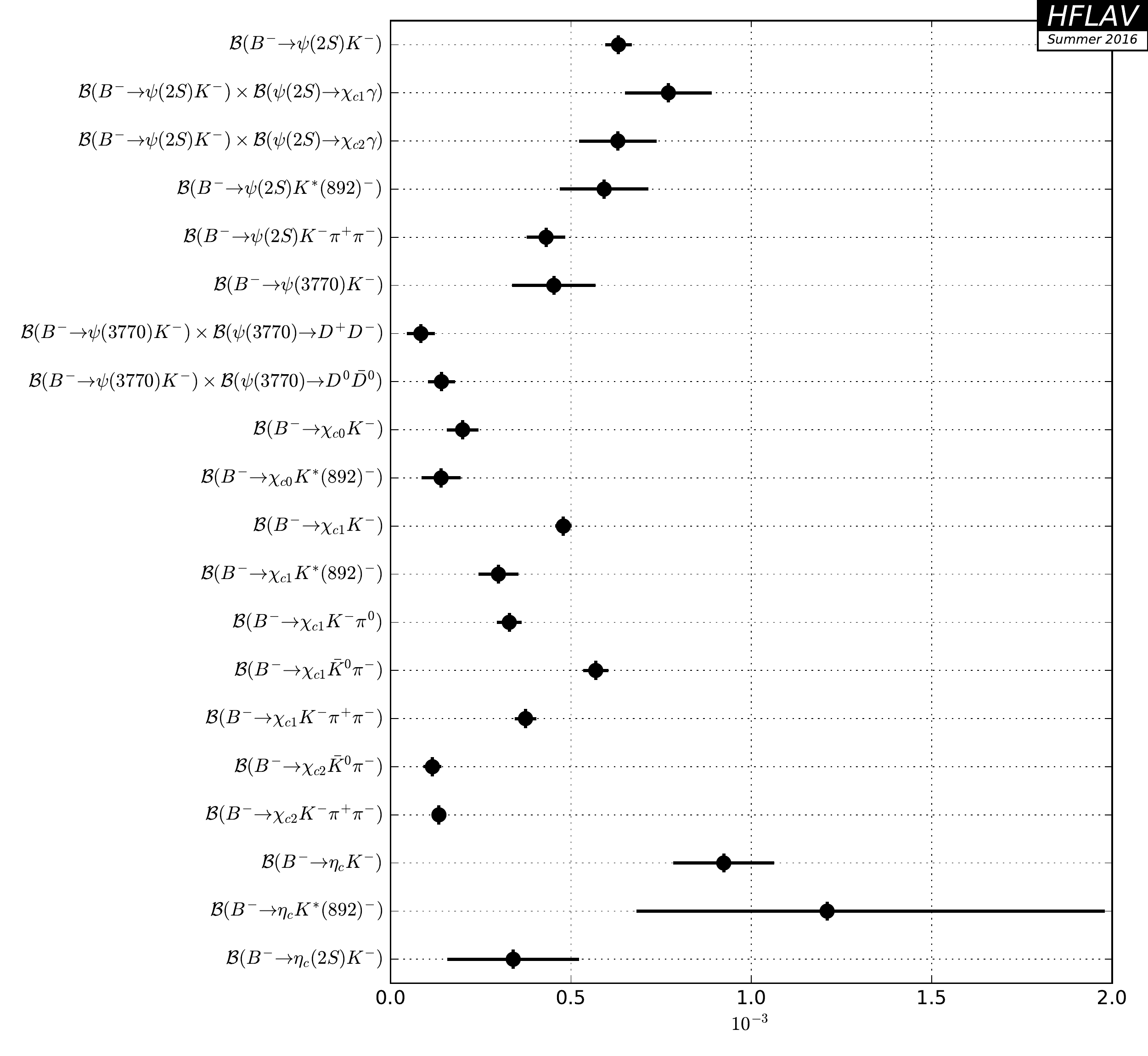}

\begin{btocharmtab}{Bu_cc_4}{Decays to charmonium other than $J/\psi$ and one kaon II $[10^{-5}]$}
\hline
\textbf{Parameter} & %
 & $< 4.8$ \\
\hline
\end{btocharmtab}
\btocharmfig{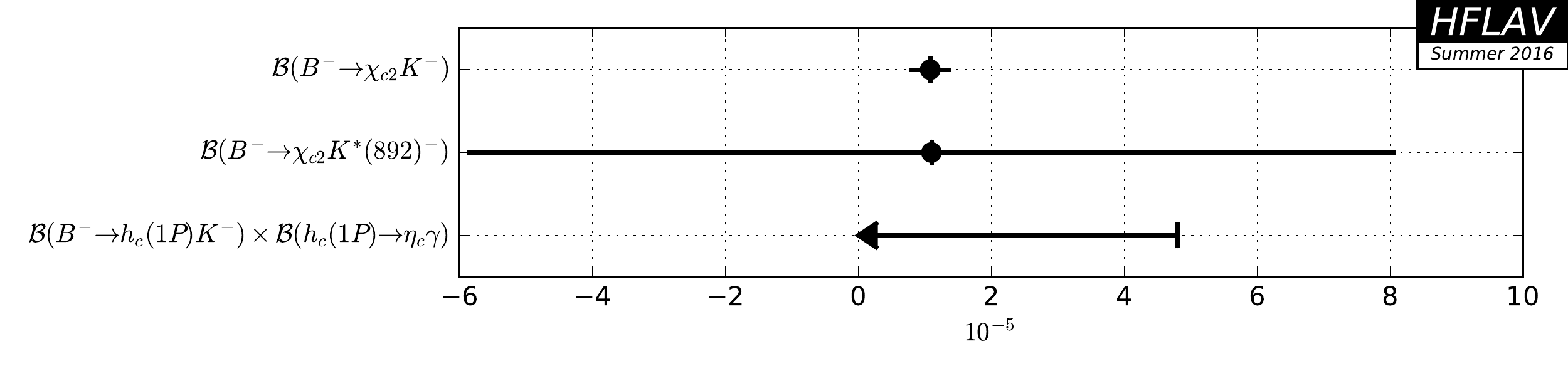}

\begin{btocharmtab}{Bu_cc_5}{Decays to charmonium other than $J/\psi$ and one kaon III $[10^{-6}]$}
\hline
\textbf{Parameter} & %
 & $< 3.4$ \\
\hline
\end{btocharmtab}
\btocharmfig{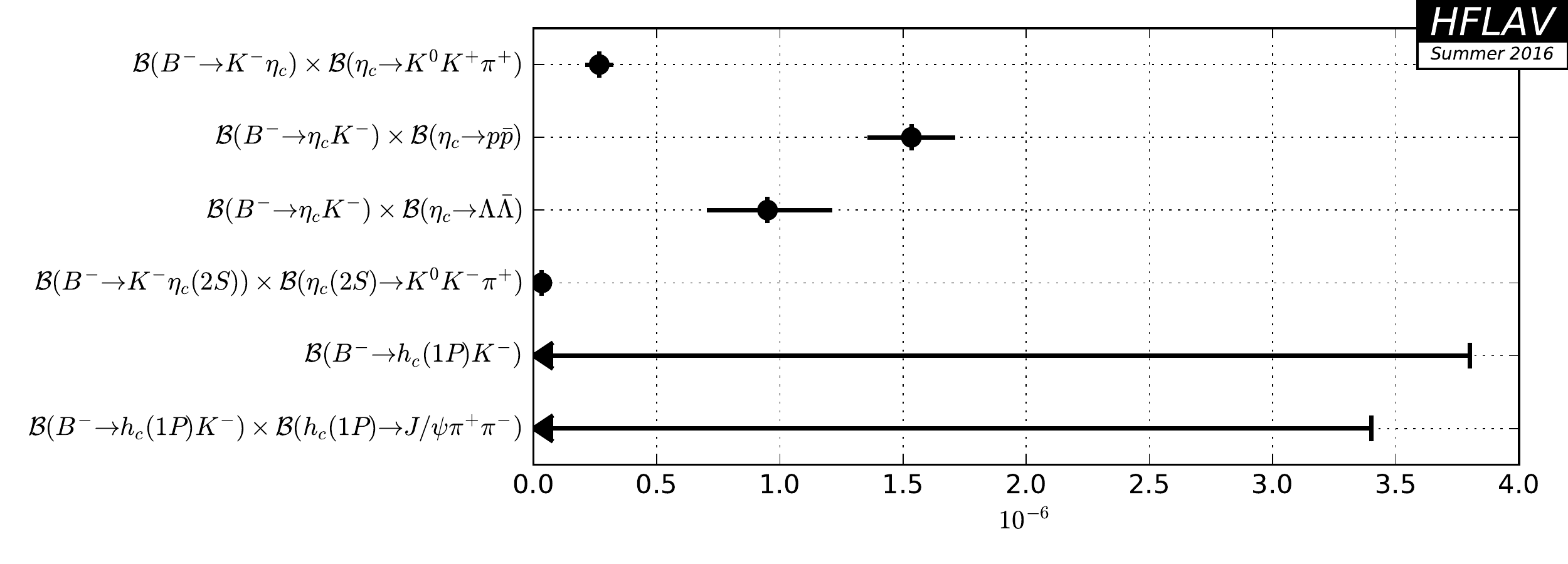}

\begin{btocharmtab}{Bu_cc_6}{Decays to charmonium and light mesons $[10^{-5}]$}
\hline
\textbf{Parameter} & %
 & $< 0.25$ \\
\hline
\end{btocharmtab}
\btocharmfig{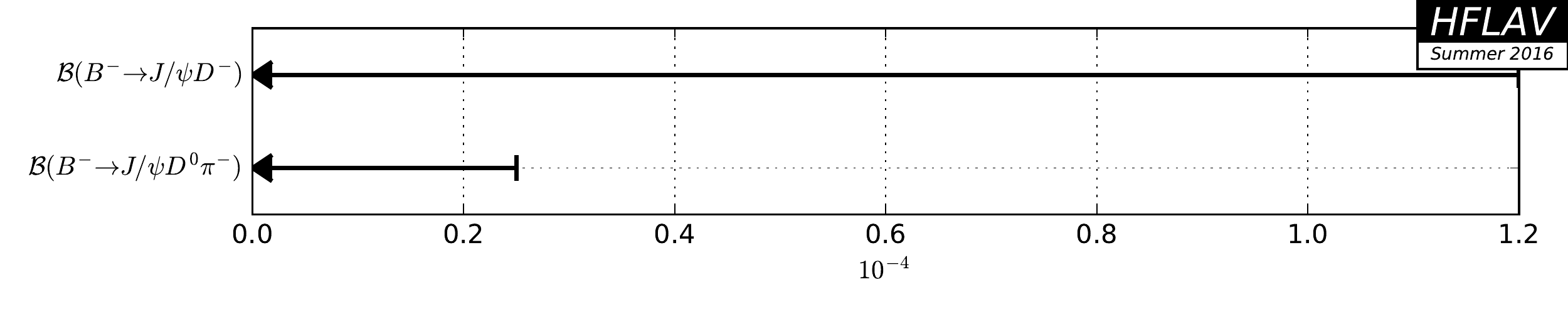}

\begin{btocharmtab}{Bu_cc_8}{Decays with baryons I $[10^{-5}]$}
\hline
\textbf{Parameter} & %
 & $< 1.1$ \\
\hline
\end{btocharmtab}
\btocharmfig{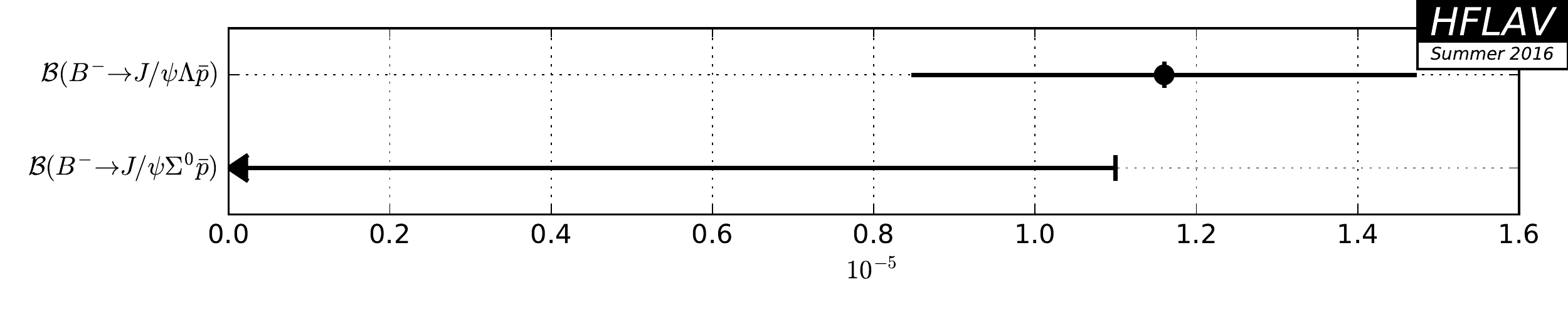}

\begin{btocharmtab}{Bu_cc_9}{Decays with baryons II $[10^{-6}]$}
\hline
\textbf{Parameter} & %
 & $2.21 \pm 0.13$ \\
\hline
\end{btocharmtab}
\btocharmfig{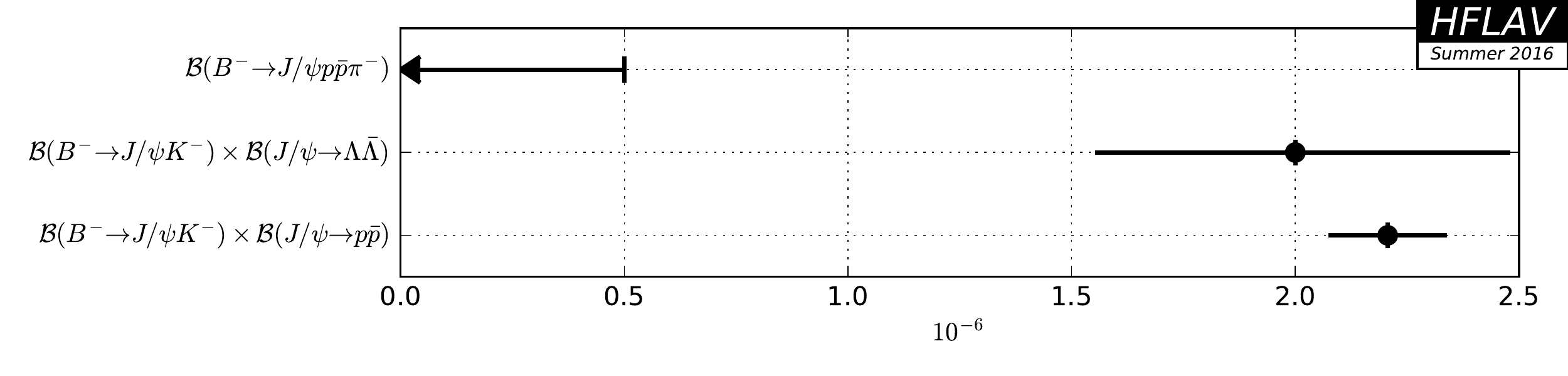}

\begin{btocharmtab}{Bu_cc_10}{Relative decay rates I}
\hline
\textbf{Parameter} & %
 & $0.578 \pm 0.044$ \\
\hline
\end{btocharmtab}
\btocharmfig{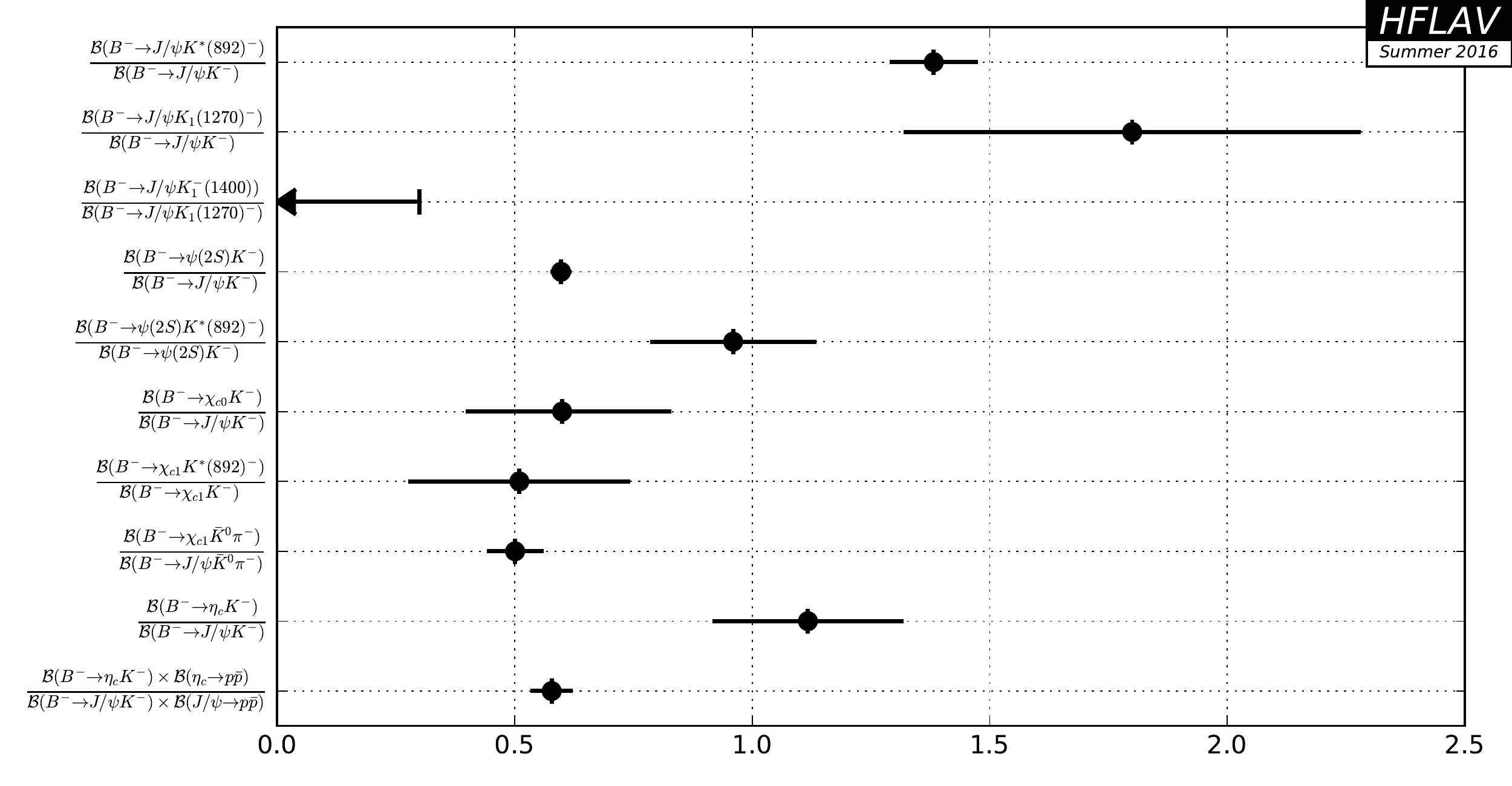}

\begin{btocharmtab}{Bu_cc_11}{Relative decay rates II}
\hline
\textbf{Parameter} & %
 & $< 0.052$ \\
\hline
\end{btocharmtab}
\btocharmfig{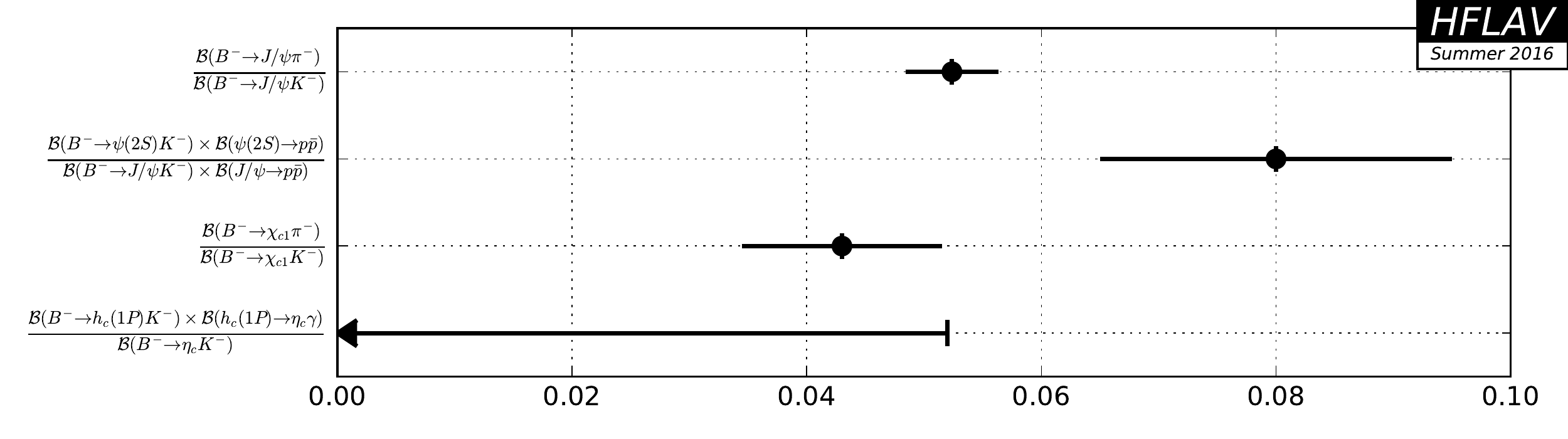}

%\btocharm{Bu_cc_12}

\subsubsection{Decays to charm baryons}
\label{sec:b2c:Bu_baryon}
Averages of $B^-$ decays to charm baryons are shown in Tables~\ref{tab:b2c:Bu_baryon_1}--\ref{tab:b2c:Bu_baryon_3} and Figs.~\ref{fig:b2c:Bu_baryon_1}--\ref{fig:b2c:Bu_baryon_3}.
\begin{btocharmtab}{Bu_baryon_1}{Absolute (product) decay rates $[10^{-4}]$}
\hline
\textbf{Parameter} & %
 & $2.98 \pm 0.80$ \\
\hline
\end{btocharmtab}
\btocharmfig{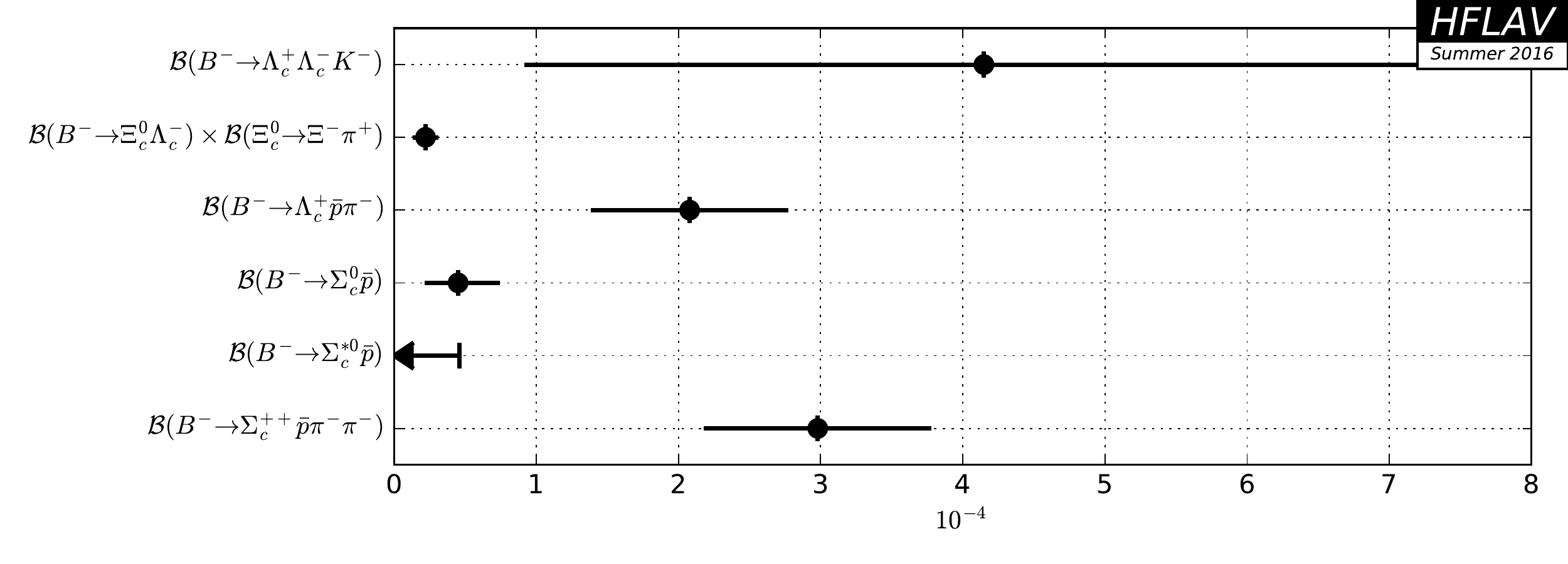}

\begin{btocharmtab}{Bu_baryon_2}{Relative decay rates I}
\hline
\textbf{Parameter} & %
 & $15.4 \pm 1.8$ \\
\hline
\end{btocharmtab}

\begin{btocharmtab}{Bu_baryon_3}{Relative decay rates II}
\hline
\textbf{Parameter} & %
 & $0.117 \pm 0.033$ \\
\hline
\end{btocharmtab}
\btocharmfig{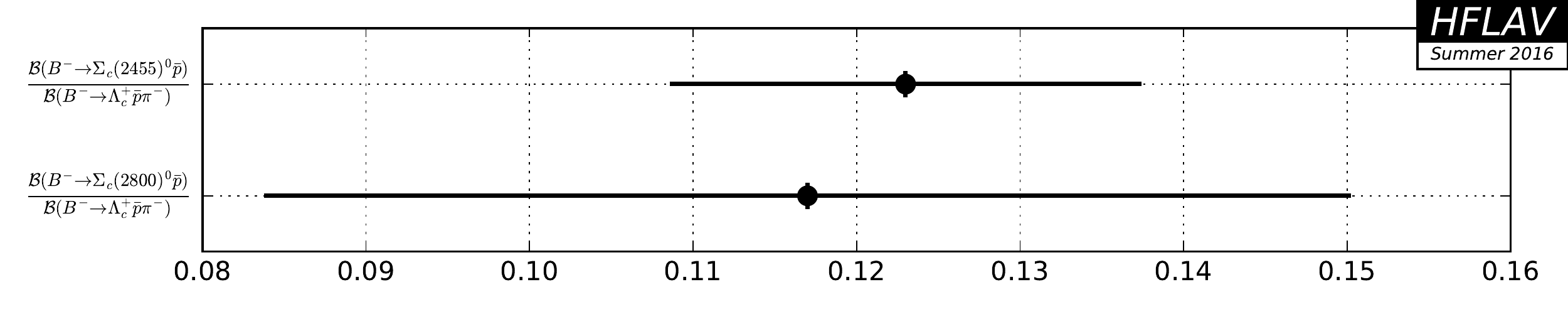}

\subsubsection{Decays to other ($XYZ$) states}
\label{sec:b2c:Bu_other}
Averages of $B^-$ decays to other ($XYZ$) states are shown in Tables~\ref{tab:b2c:Bu_other_1}--\ref{tab:b2c:Bu_other_6} and Figs.~\ref{fig:b2c:Bu_other_2}--\ref{fig:b2c:Bu_other_6}.
\begin{btocharmtab}{Bu_other_1}{Absolute decay rates $[10^{-4}]$}
\hline
\textbf{Parameter} & %
 & $< 3.2$ \\
\hline
\end{btocharmtab}

\begin{btocharmtab}{Bu_other_2}{Product decay rates to $X(3872)$ I $[10^{-4}]$}
\hline
\textbf{Parameter} & %
 & $< 0.4$ \\
\hline
\end{btocharmtab}
\btocharmfig{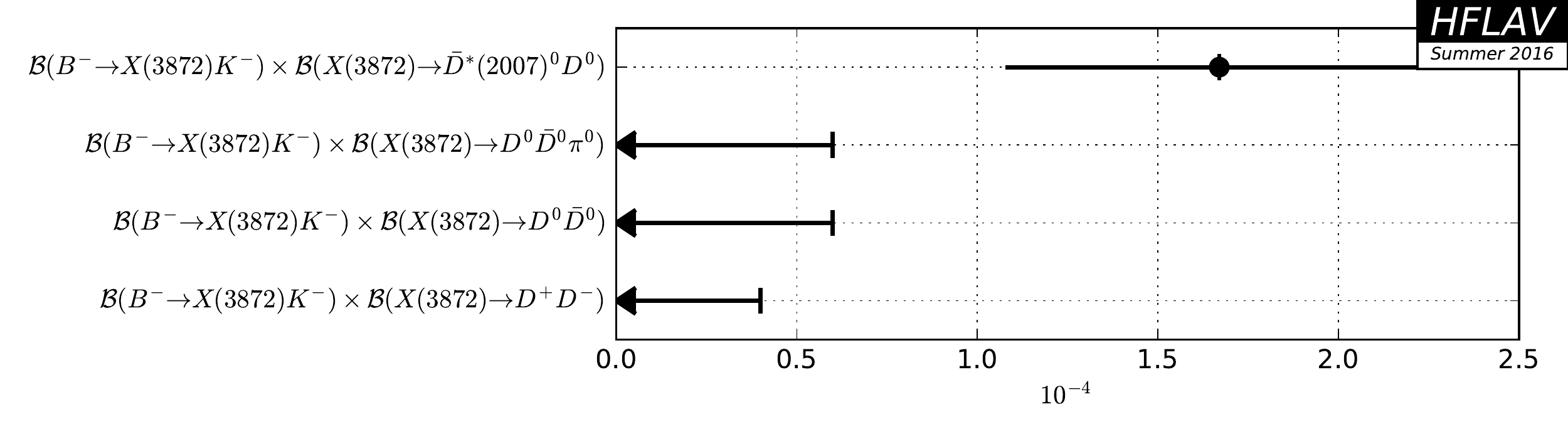}

\begin{btocharmtab}{Bu_other_3}{Product decay rates to $X(3872)$ II $[10^{-5}]$}
\hline
\textbf{Parameter} & %
 & $< 0.67$ \\
\hline
\end{btocharmtab}
\btocharmfig{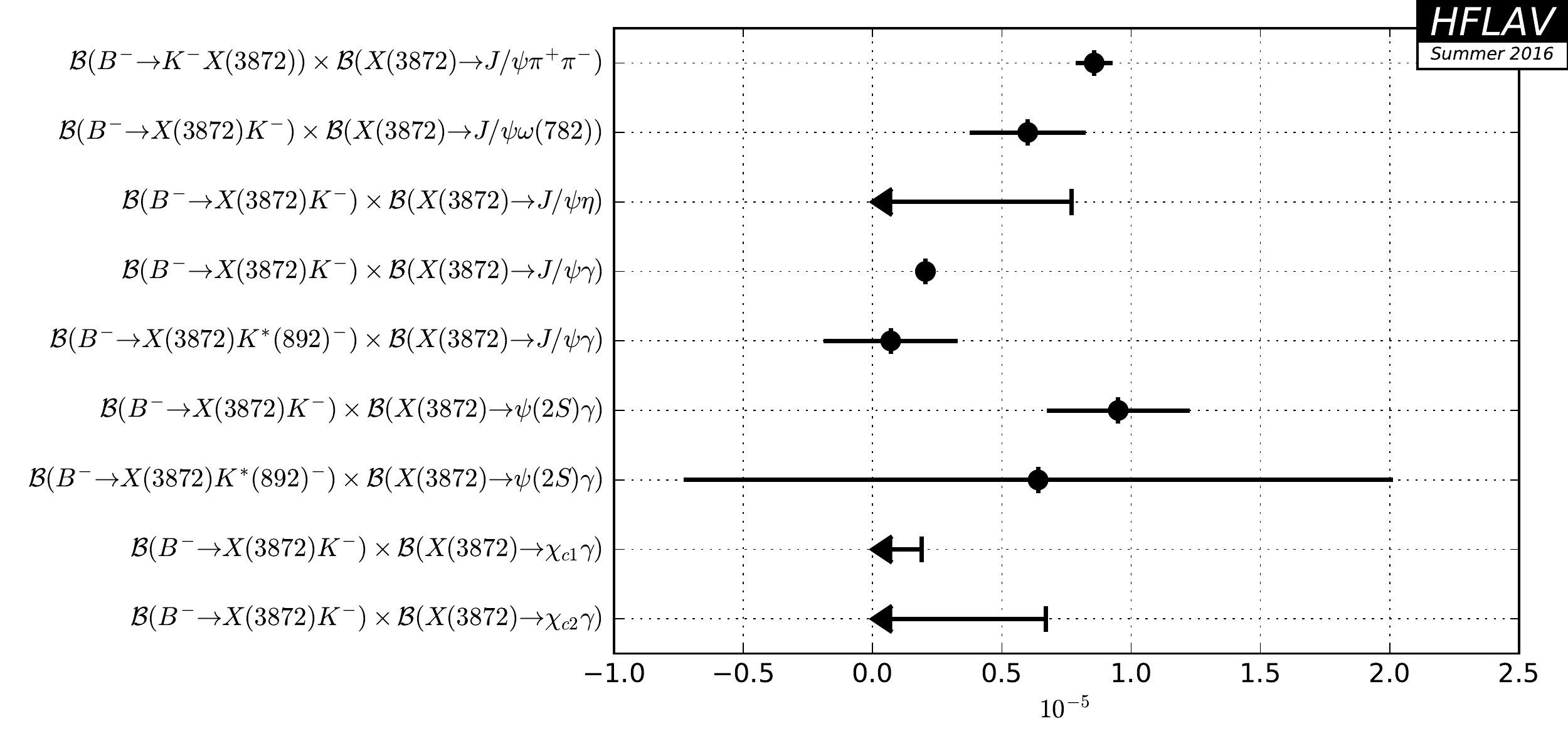}

\begin{btocharmtab}{Bu_other_4}{Product decay rates to neutral states other than $X(3872)$ $[10^{-5}]$}
\hline
\textbf{Parameter} & %
 & $2.0 \pm 0.7$ \\
\hline
\end{btocharmtab}
\btocharmfig{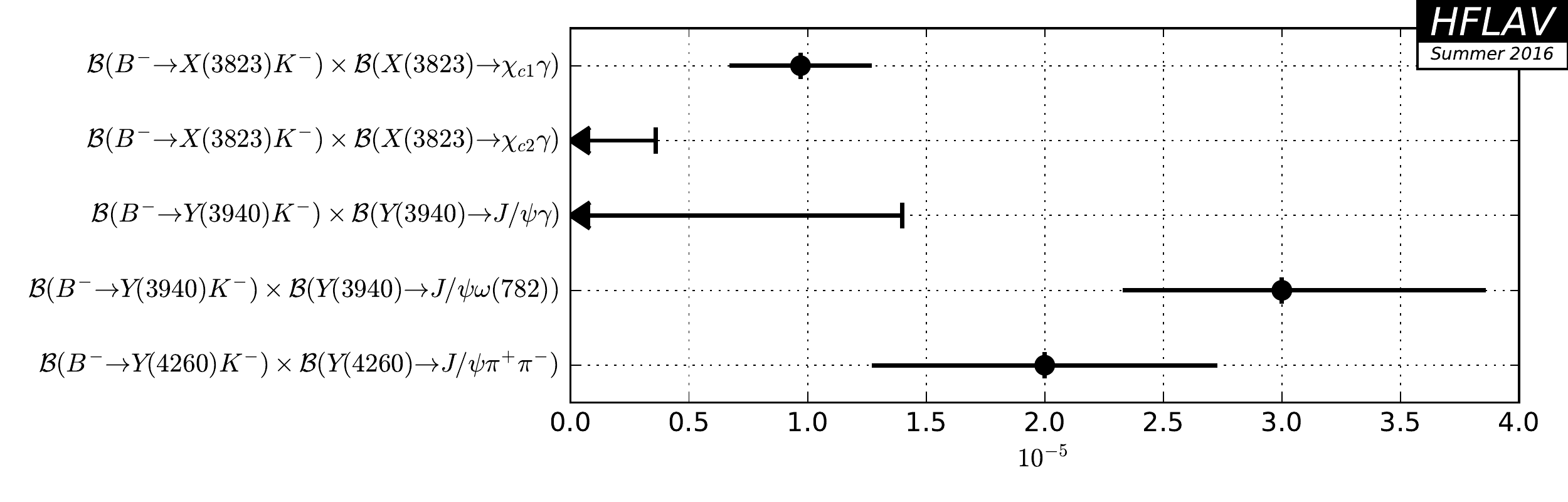}

\begin{btocharmtab}{Bu_other_5}{Relative product decay rates to states with $s\bar{s}$ component}
\hline
\textbf{Parameter} & %
 & $0.071 \,^{+0.043}_{-0.035}$ \\
\hline
\end{btocharmtab}
\btocharmfig{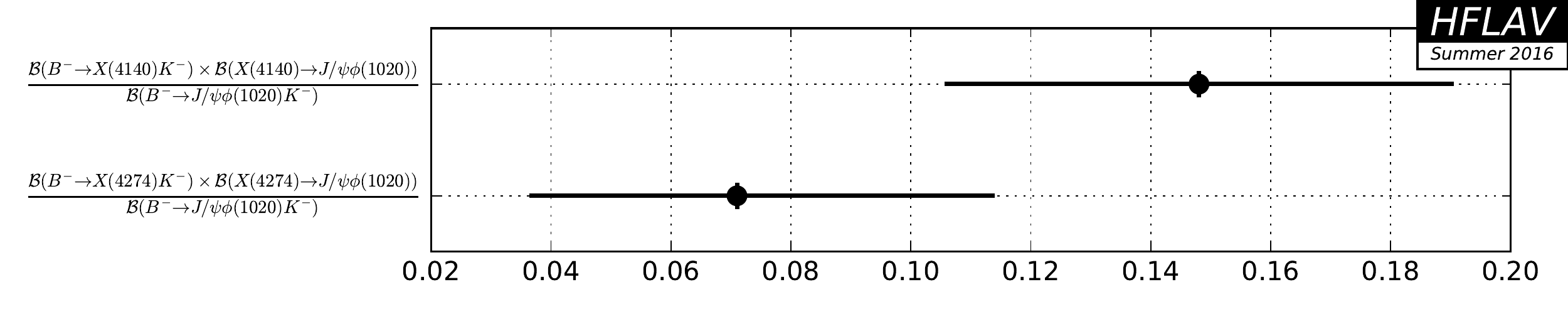}

\begin{btocharmtab}{Bu_other_6}{Product decay rates to charged states $[10^{-5}]$}
\hline
\textbf{Parameter} & %
 & $2.0 \pm 1.7$ \\
\hline
\end{btocharmtab}
\btocharmfig{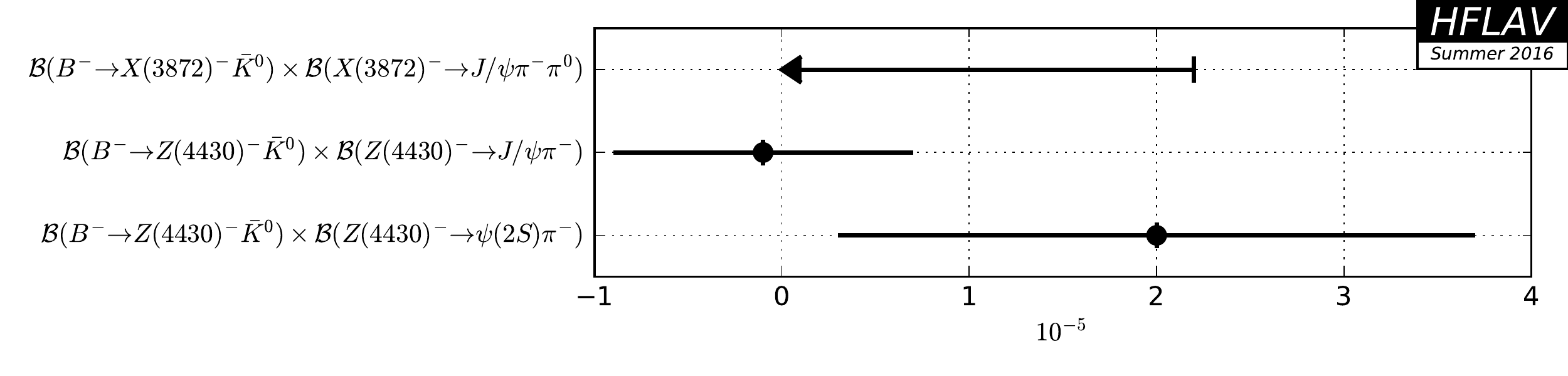}

\subsection{Decays of admixtures of $\bar{B}^0$ / $B^-$ mesons}
\label{sec:b2c:B}
Measurements of $\bar{B}^0$ / $B^-$ decays to charmed hadrons are summarized in Sections~\ref{sec:b2c:B_DD} to~\ref{sec:b2c:B_other}.

\subsubsection{Decays to two open charm mesons}
\label{sec:b2c:B_DD}
Averages of $\bar{B}^0$ / $B^-$ decays to two open charm mesons are shown in Table~\ref{tab:b2c:B_DD_1}.
\begin{btocharmtab}{B_DD_1}{ B decays to double charm $[10^{-4}]$}
\hline
\textbf{Parameter} & %
 & $1.27 \,^{+0.38}_{-0.50}$ \\
\hline
\end{btocharmtab}

\subsubsection{Decays to charmonium states}
\label{sec:b2c:B_cc}
Averages of $\bar{B}^0$ / $B^-$ decays to charmonium states are shown in Tables~\ref{tab:b2c:B_cc_1}--\ref{tab:b2c:B_cc_5} and Figs.~\ref{fig:b2c:B_cc_1}--\ref{fig:b2c:B_cc_5}.
\begin{btocharmtab}{B_cc_1}{Decay amplitudes for parallel transverse polarization}
\hline
\textbf{Parameter} & %
 & $0.22 \pm 0.06$ \\
\hline
\end{btocharmtab}
\btocharmfig{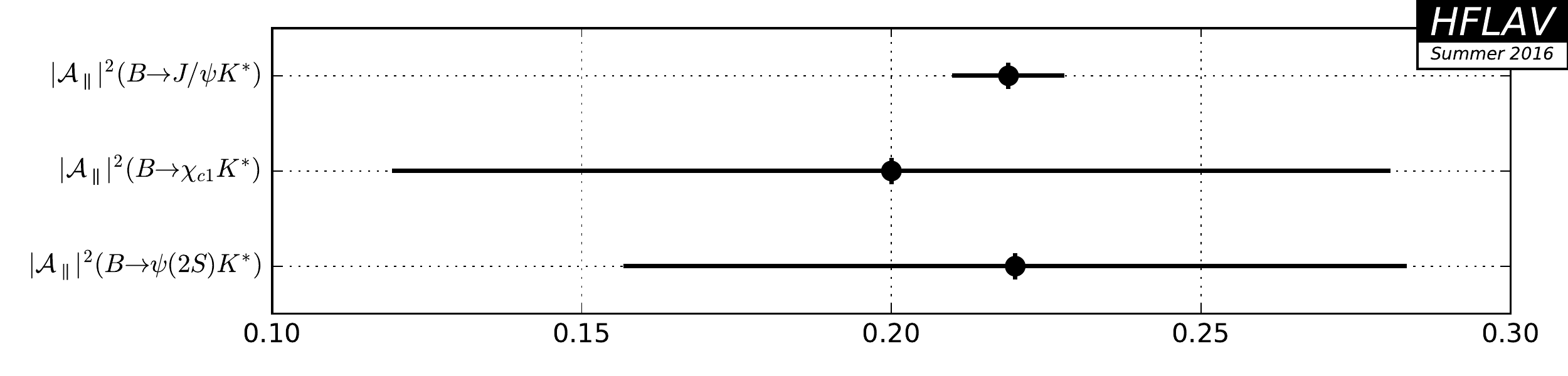}

\begin{btocharmtab}{B_cc_2}{Decay amplitudes for perpendicular transverse polarization}
\hline
\textbf{Parameter} & %
 & $0.30 \pm 0.06$ \\
\hline
\end{btocharmtab}
\btocharmfig{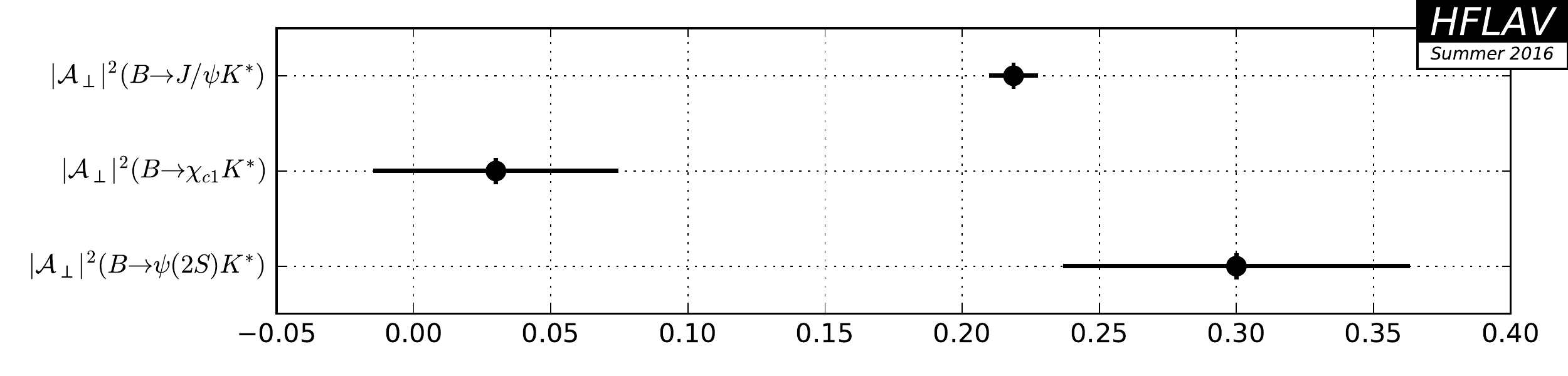}

\begin{btocharmtab}{B_cc_3}{Decay amplitudes for longitudinal polarization}
\hline
\textbf{Parameter} & %
 & $0.48 \pm 0.05$ \\
\hline
\end{btocharmtab}
\btocharmfig{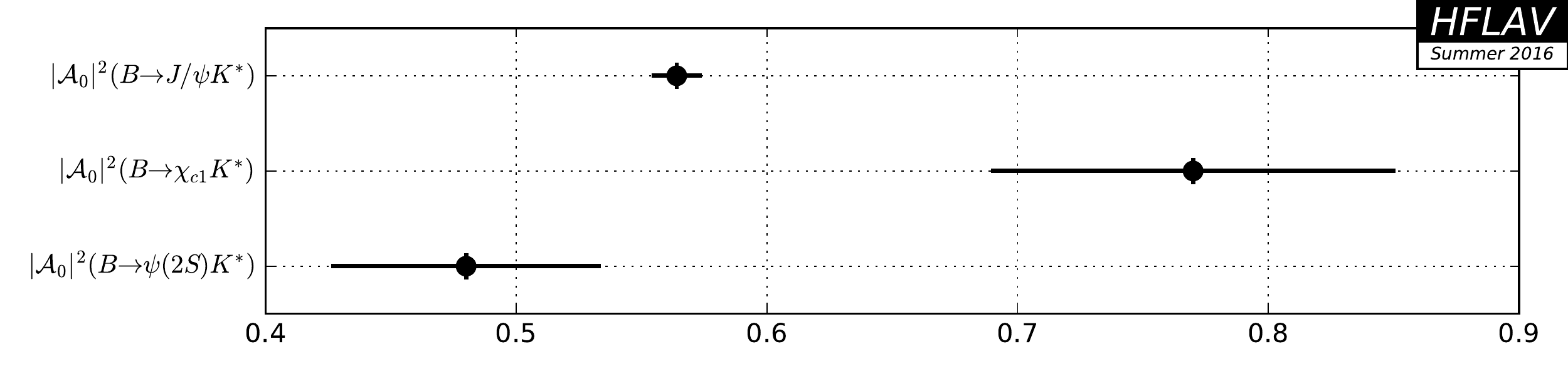}

\begin{btocharmtab}{B_cc_4}{Relative phases of parallel transverse polarization decay amplitudes}
\hline
\textbf{Parameter} & %
 & $-2.8 \pm 0.4$ \\
\hline
\end{btocharmtab}
\btocharmfig{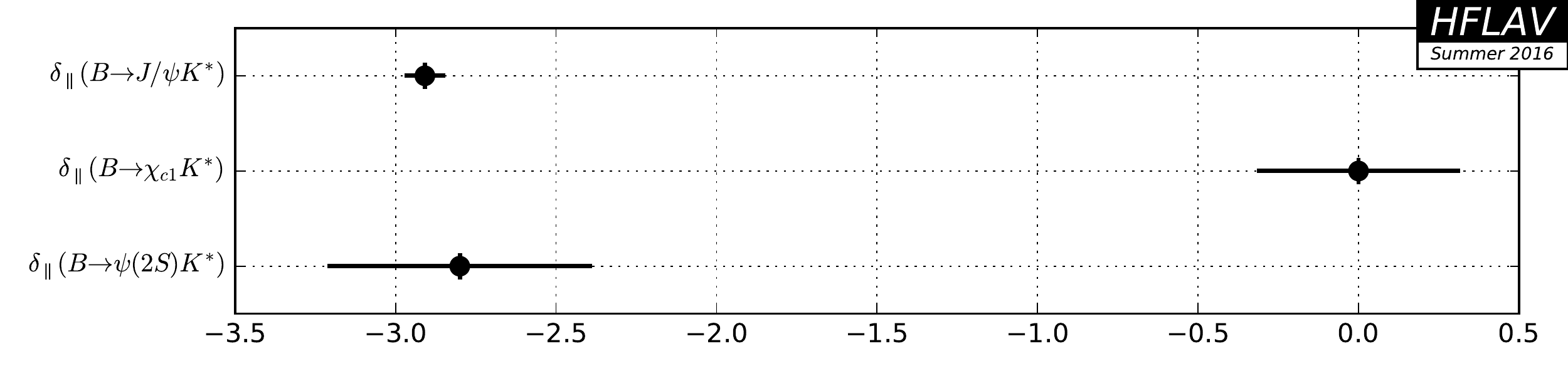}

\begin{btocharmtab}{B_cc_5}{Relative phases of perpendicular transverse polarization decay amplitudes}
\hline
\textbf{Parameter} & %
 & $2.8 \pm 0.3$ \\
\hline
\end{btocharmtab}
\btocharmfig{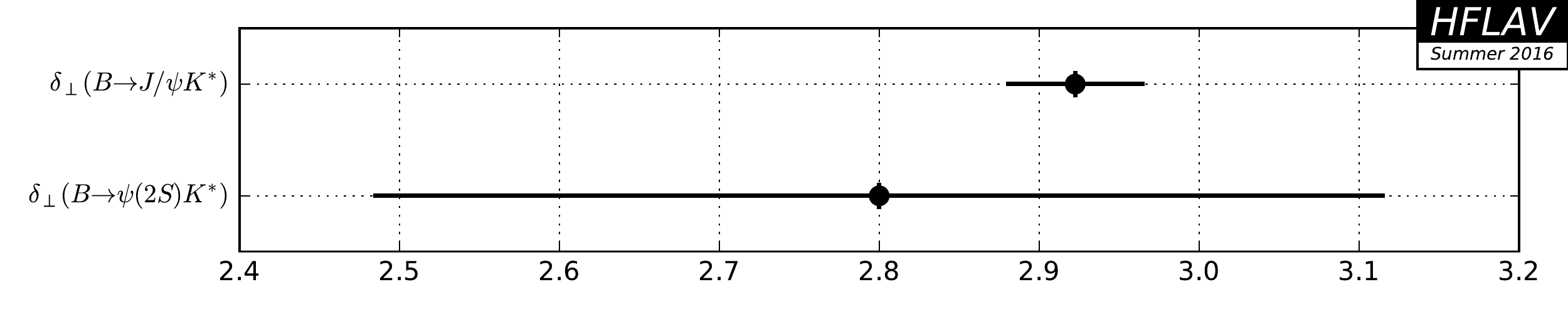}

\subsubsection{Decays to other ($XYZ$) states}
\label{sec:b2c:B_other}
Averages of $\bar{B}^0$ / $B^-$ decays to other ($XYZ$) states are shown in Table~\ref{tab:b2c:B_other_1} and Fig.~\ref{fig:b2c:B_other_1}.
\begin{btocharmtab}{B_other_1}{Absolute decay rates to $X$/$Y$ states $[10^{-4}]$}
\hline
\textbf{Parameter} & %
 & $0.71 \pm 0.34$ \\
\hline
\end{btocharmtab}
\btocharmfig{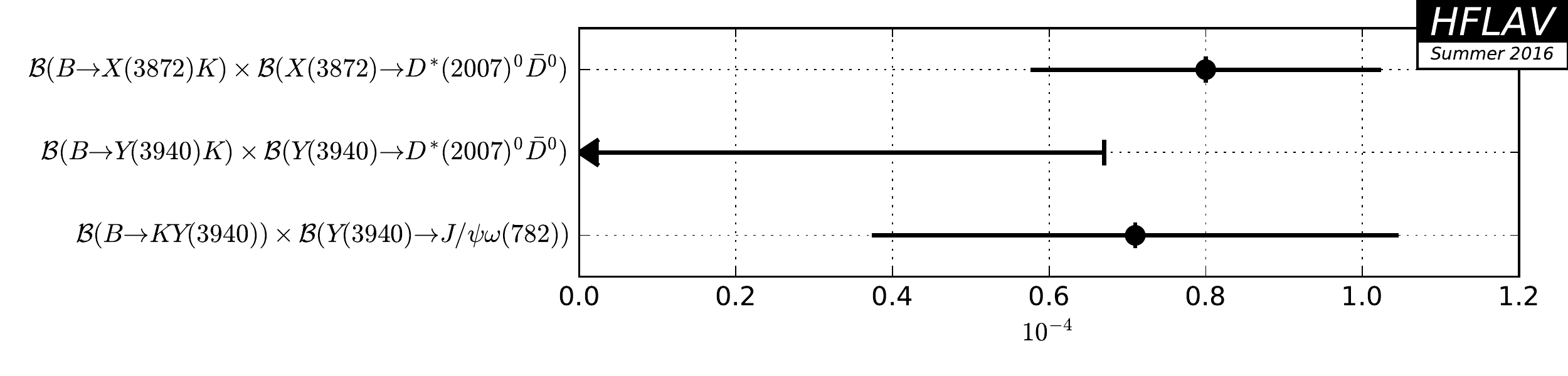}

\subsection{Decays of $\bar{B}_s^0$ mesons}
\label{sec:b2c:Bs}
Measurements of $\bar{B}_s^0$ decays to charmed hadrons are summarized in Sections~\ref{sec:b2c:Bs_D} to~\ref{sec:b2c:Bs_baryon}.
These measurements require knowledge of the production rates of $\bar{B}_s^0$ mesons, usually measured relative to those of $\bar{B}^0$ and $B^-$ mesons, in the appropriate experimental environment. 
Since these production fractions are reasonably well known, see Sec.~\ref{sec:fractions}, they can be corrected for allowing the results to be presented in terms of the absolute $\bar{B}_s^0$ branching fraction, or the relative branching fraction to a lighter $B$ meson decay mode.  
This is usually done in the publications; we do not make any attempt to rescale results according to more recent determinations of the relative production fractions. 
Ratios of branching fractions of two decays of the same hadron do not require any such correction.

\subsubsection{Decays to a single open charm meson}
\label{sec:b2c:Bs_D}
Averages of $\bar{B}_s^0$ decays to a single open charm meson are shown in Tables~\ref{tab:b2c:Bs_D_1}--\ref{tab:b2c:Bs_D_6} and Figs.~\ref{fig:b2c:Bs_D_1}--\ref{fig:b2c:Bs_D_6}.
\begin{btocharmtab}{Bs_D_1}{Decays to a $D_s^{(*)}$ and a light meson I $[10^{-3}]$}
\hline
\textbf{Parameter} & \begin{tabular}{l}\textbf{Measurements}\end{tabular} & \textbf{Average} \\
\hline
\hline
${\cal{B}} ( \bar{B}_s^{0} \to D_s^{+} \pi^{-} )$ & \begin{tabular}{l} LHCb \cite{Aaij:2012zz}: $2.95 \pm 0.05 \,^{+0.25}_{-0.28}$ \\ Belle \cite{Louvot:2008sc}: $3.67 \,^{+0.35}_{-0.33} \,^{+0.65}_{-0.65}$ \\ \end{tabular} & $3.03 \pm 0.25$ \\
\hline
${\cal{B}} ( \bar{B}_s^{0} \to D_s^{*+} \pi^{-} )$ & \begin{tabular}{l} Belle \cite{Louvot:2010rd}: $2.4 \,^{+0.5}_{-0.4} \pm 0.4$ \\ \end{tabular} & $2.4 \,^{+0.7}_{-0.6}$ \\
\hline
${\cal{B}} ( \bar{B}_s^{0} \to D_s^{+} \rho^{-}(770) )$ & \begin{tabular}{l} Belle \cite{Louvot:2010rd}: $8.5 \,^{+1.3}_{-1.2} \pm 1.7$ \\ \end{tabular} & $8.5 \,^{+2.1}_{-2.1}$ \\
\hline
${\cal{B}} ( \bar{B}_s^{0} \to D_s^{*+} \rho^{-}(770) )$ & \begin{tabular}{l} Belle \cite{Louvot:2010rd}: $11.8 \,^{+2.2}_{-2.0} \pm 2.5$ \\ \end{tabular} & $11.8 \,^{+3.3}_{-3.2}$ \\
\hline
\end{btocharmtab}
\btocharmfig{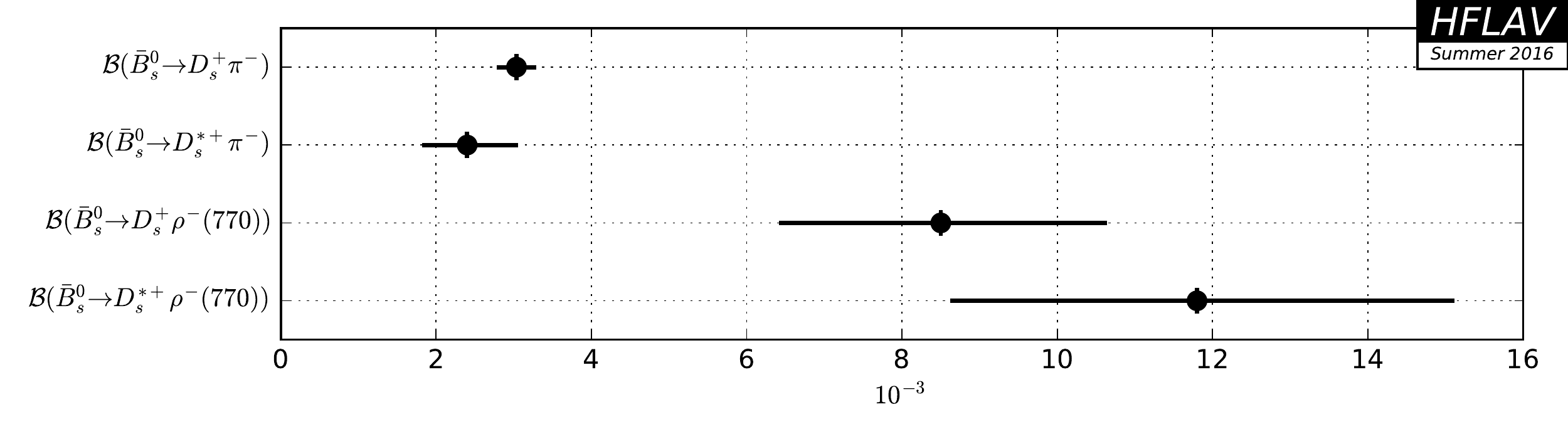}

\begin{btocharmtab}{Bs_D_2}{Decays to a $D_s^{(*)}$ and a light meson II $[10^{-4}]$}
\hline
\textbf{Parameter} & \begin{tabular}{l}\textbf{Measurements}\end{tabular} & \textbf{Average} \\
\hline
\hline
${\cal{B}} ( \bar{B}_s^{0} \to D_s^{+} K^{-} )$ & \begin{tabular}{l} LHCb \cite{Aaij:2012zz}: $1.90 \pm 0.12 \,^{+0.18}_{-0.19}$ \\ Belle \cite{Louvot:2008sc}: $2.4 \,^{+1.2}_{-1.0} \pm 0.4$ \\ \end{tabular} & $1.92 \pm 0.22$ \\
\hline
${\cal{B}} ( \bar{B}_s^{0} \to D_s^{*+} K^{-} )$ & \begin{tabular}{l} LHCb \cite{Aaij:2015dsa}: $1.63 \pm 0.12 \,^{+0.49}_{-0.48}$ \\ \end{tabular} & $1.63 \,^{+0.50}_{-0.50}$ \\
\hline
\end{btocharmtab}
\btocharmfig{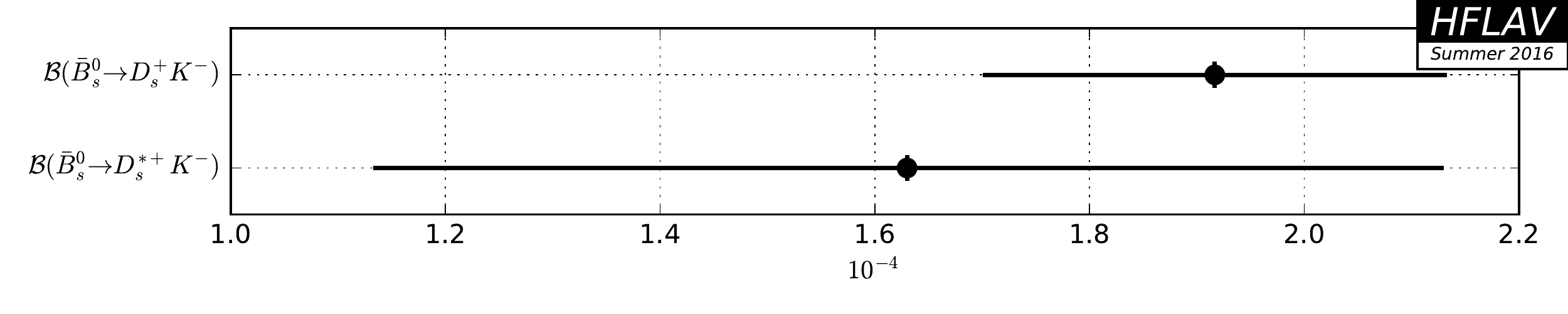}

\begin{btocharmtab}{Bs_D_3}{Decays to a $D^{(*)}$ and a light meson I $[10^{-4}]$}
\hline
\textbf{Parameter} & \begin{tabular}{l}\textbf{Measurements}\end{tabular} & \textbf{Average} \\
\hline
\hline
${\cal{B}} ( \bar{B}_s^{0} \to D^{0} K^{0} )$ & \begin{tabular}{l} LHCb \cite{Aaij:2016amk}: $4.3 \pm 0.5 \pm 0.8$ \\ \end{tabular} & $4.3 \pm 0.9$ \\
\hline
${\cal{B}} ( \bar{B}_s^{0} \to D^{*0} K^{0} )$ & \begin{tabular}{l} LHCb \cite{Aaij:2016amk}: $2.8 \pm 1.0 \pm 0.5$ \\ \end{tabular} & $2.8 \pm 1.1$ \\
\hline
${\cal{B}} ( \bar{B}^{0}_s \to D^{0} K^{*0} )$ & \begin{tabular}{l} LHCb \cite{Aaij:2011tz}: $4.72 \pm 1.07 \pm 0.96$ \\ \end{tabular} & $4.72 \pm 1.44$ \\
\hline
\end{btocharmtab}
\btocharmfig{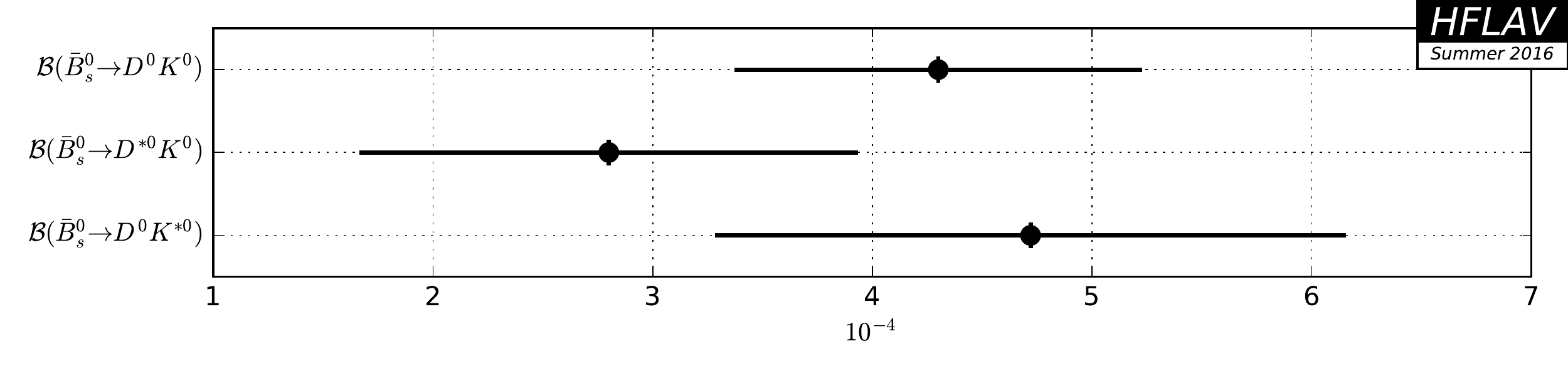}

\begin{btocharmtab}{Bs_D_4}{Decays to a $D^{(*)}$ and a light meson II $[10^{-6}]$}
\hline
\textbf{Parameter} & \begin{tabular}{l}\textbf{Measurements}\end{tabular} & \textbf{Average} \\
\hline
\hline
${\cal{B}} ( \bar{B}_s^{0} \to D^{*}(2010)^{\pm} \pi^{\mp} )$ & \begin{tabular}{l} LHCb \cite{Aaij:2013fpa}: $< 6.1$ \\ \end{tabular} & $< 6.1$ \\
\hline
${\cal{B}} ( \bar{B}^{0}_s \to D^{0} f_0(980))$ & \begin{tabular}{l} LHCb \cite{Aaij:2015rqa}: $< 3.1$ \\ \end{tabular} & $< 3.1$ \\
\hline
\end{btocharmtab}
\btocharmfig{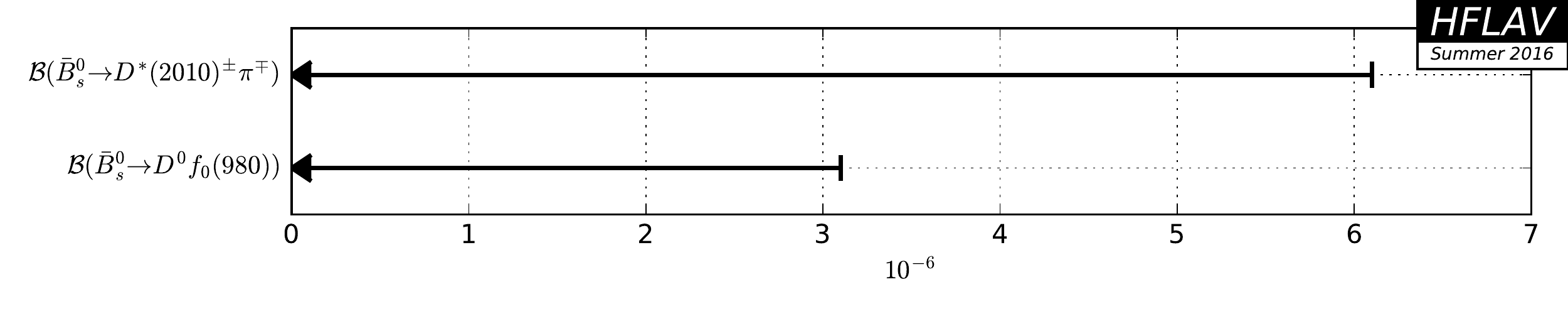}

\begin{btocharmtab}{Bs_D_5}{Relative decay rates I}
\hline
\textbf{Parameter} & \begin{tabular}{l}\textbf{Measurements}\end{tabular} & \textbf{Average} \\
\hline
\hline
\multicolumn{3}{|l|}{${{\cal{B}} ( \bar{B}_s^{0} \to D_s^{+} \pi^{-} )}/{{\cal{B}} ( \bar{B}^{0} \to D^{+} \pi^{-} )}$}\\
 & \begin{tabular}{l} CDF \cite{Abulencia:2006aa}: $1.13 \pm 0.08 \pm 0.23$ \\ \end{tabular} & $1.13 \pm 0.25$ \\
\hline
\multicolumn{3}{|l|}{${{\cal{B}} ( \bar{B}^{0}_s \to D_s^{+} \pi^{+} \pi^{-} \pi^{-} ) }/{ {\cal{B}} ( \bar{B}^{0}_s \to D_s^{+} \pi^{-} )}$}\\
 & \begin{tabular}{l} LHCb \cite{Aaij:2011rj}: $2.01 \pm 0.37 \pm 0.20$ \\ \end{tabular} & $2.01 \pm 0.42$ \\
\hline
\multicolumn{3}{|l|}{${{\cal{B}} ( \bar{B}_s^{0} \to D_s^{+} \pi^{+} \pi^{-} \pi^{-} )}/{{\cal{B}} ( \bar{B}^{0} \to D^{+} \pi^{+} \pi^{-} \pi^{-} )}$}\\
 & \begin{tabular}{l} CDF \cite{Abulencia:2006aa}: $1.05 \pm 0.10 \pm 0.22$ \\ \end{tabular} & $1.05 \pm 0.24$ \\
\hline
\multicolumn{3}{|l|}{${{\cal{B}} ( \bar{B}^{0}_s \to D^{0} K^{*0} ) }/{ {\cal{B}} ( \bar{B}^{0} \to D^{0} \rho^{0} )}$}\\
 & \begin{tabular}{l} LHCb \cite{Aaij:2011tz}: $1.48 \pm 0.34 \pm 0.19$ \\ \end{tabular} & $1.48 \pm 0.39$ \\
\hline
\multicolumn{3}{|l|}{${{\cal{B}} ( \bar{B}^{0}_s \to D^{0} K^{*0} ) }/{ {\cal{B}} ( \bar{B}^{0} \to D^{0} \bar{K}^{*0} )}$}\\
 & \begin{tabular}{l} LHCb \cite{Aaij:2013dda}: $7.8 \pm 0.7 \pm 0.7$ \\ \end{tabular} & $7.8 \pm 1.0$ \\
\hline
\multicolumn{3}{|l|}{${ {\cal{B}} ( \bar{B}_s^{0} \to D^{0} K^{+} \pi^{-} )}/  {{\cal{B}} ( \bar{B}^{0} \to D^{0} \pi^{-} \pi^{+} )}$}\\
 & \begin{tabular}{l} LHCb \cite{Aaij:2013pua}: $1.18 \pm 0.05 \pm 0.12$ \\ \end{tabular} & $1.18 \pm 0.13$ \\
\hline
\end{btocharmtab}
\btocharmfig{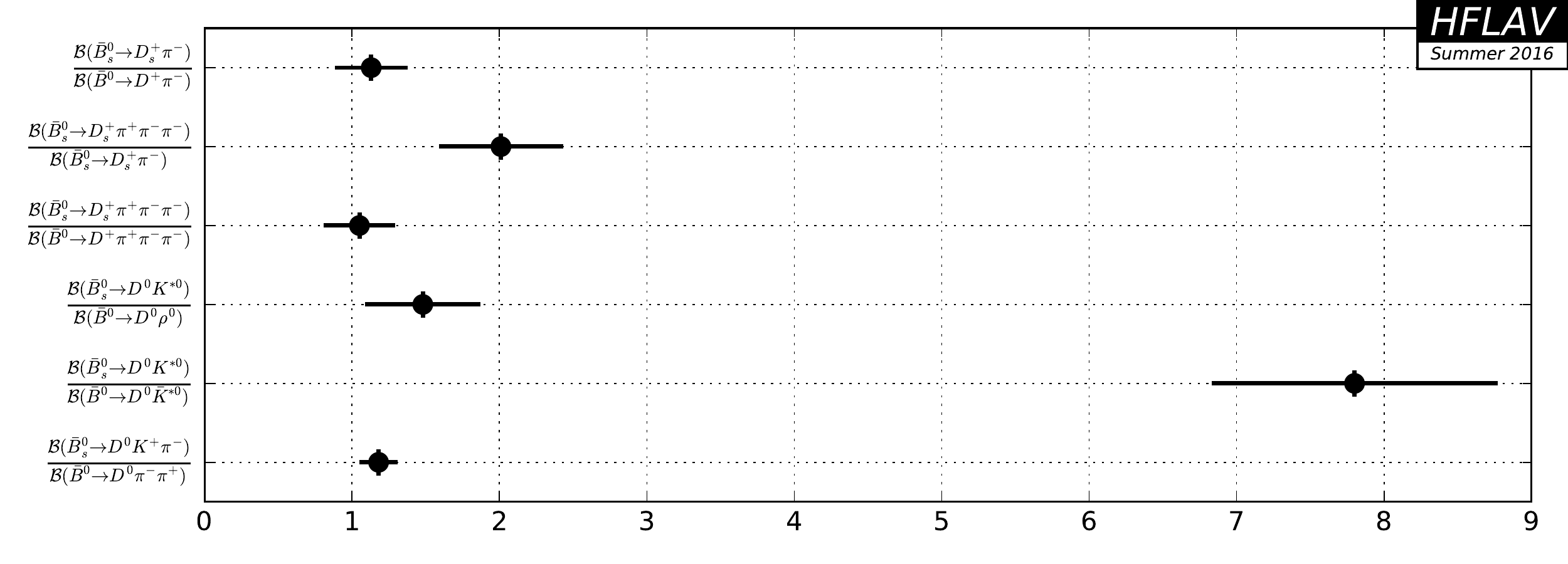}

\begin{btocharmtab}{Bs_D_6}{Relative decay rates II $[10^{-2}]$}
\hline
\textbf{Parameter} & \begin{tabular}{l}\textbf{Measurements}\end{tabular} & \textbf{Average} \\
\hline
\hline
\multicolumn{3}{|l|}{${{\cal{B}} ( \bar{B}_s^{0} \to D_s^{+} K^{-} )}/{{\cal{B}} ( \bar{B}_s^{0} \to D_s^{+} \pi^{-} )}$}\\
 & \begin{tabular}{l} LHCb \cite{Aaij:2014jpa}: $7.52 \pm 0.15 \pm 0.19$ \\ CDF \cite{Aaltonen:2008ab}: $9.7 \pm 1.8 \pm 0.9$ \\ \end{tabular} & $7.55 \pm 0.24$ \\
\hline
\multicolumn{3}{|l|}{${  {\cal{B}} ( \bar{B}_s^{0} \to D_s^{*+} K^{-} )}/{    {\cal{B}} ( \bar{B}_s^{0} \to D_s^{*+} \pi^{-} ) }$}\\
 & \begin{tabular}{l} LHCb \cite{Aaij:2015dsa}: $6.8 \pm 0.5 \,^{+0.3}_{-0.2}$ \\ \end{tabular} & $6.8 \,^{+0.6}_{-0.5}$ \\
\hline
\multicolumn{3}{|l|}{${{\cal{B}} ( \bar{B}_s^{0} \to D_s^{+} K^{-}  \pi^{+}  \pi^{-} )}/{{\cal{B}} ( \bar{B}^{0} \to D_s^{+} \pi^{-}  \pi^{+}  \pi^{-} )}$}\\
 & \begin{tabular}{l} LHCb \cite{Aaij:2012mra}: $5.2 \pm 0.5 \pm 0.3$ \\ \end{tabular} & $5.2 \pm 0.6$ \\
\hline
\multicolumn{3}{|l|}{${{\cal{B}} ( \bar{B}^{0}_s \to D^{0} \phi(1020) ) }/{ {\cal{B}} ( \bar{B}^{0}_s \to D^{0} K^{*0} )}$}\\
 & \begin{tabular}{l} LHCb \cite{Aaij:2013dda}: $6.9 \pm 1.3 \pm 0.7$ \\ \end{tabular} & $6.9 \pm 1.5$ \\
\hline
\multicolumn{3}{|l|}{$[{{\cal{B}} ( \bar{B}_s^{0} \to D_{s1}^{+} \pi^{-} ) \times {\cal{B}}(D_{s1}^{+} \to  D_s^{+} \pi^{-}  \pi^{+})  }]/{{\cal{B}} ( \bar{B}^{0} \to D_s^{+} \pi^{-}  \pi^{+}  \pi^{-} )}$}\\
 & \begin{tabular}{l} LHCb \cite{Aaij:2012mra}: $0.40 \pm 0.10 \pm 0.04$ \\ \end{tabular} & $0.40 \pm 0.11$ \\
\hline
\end{btocharmtab}
\btocharmfig{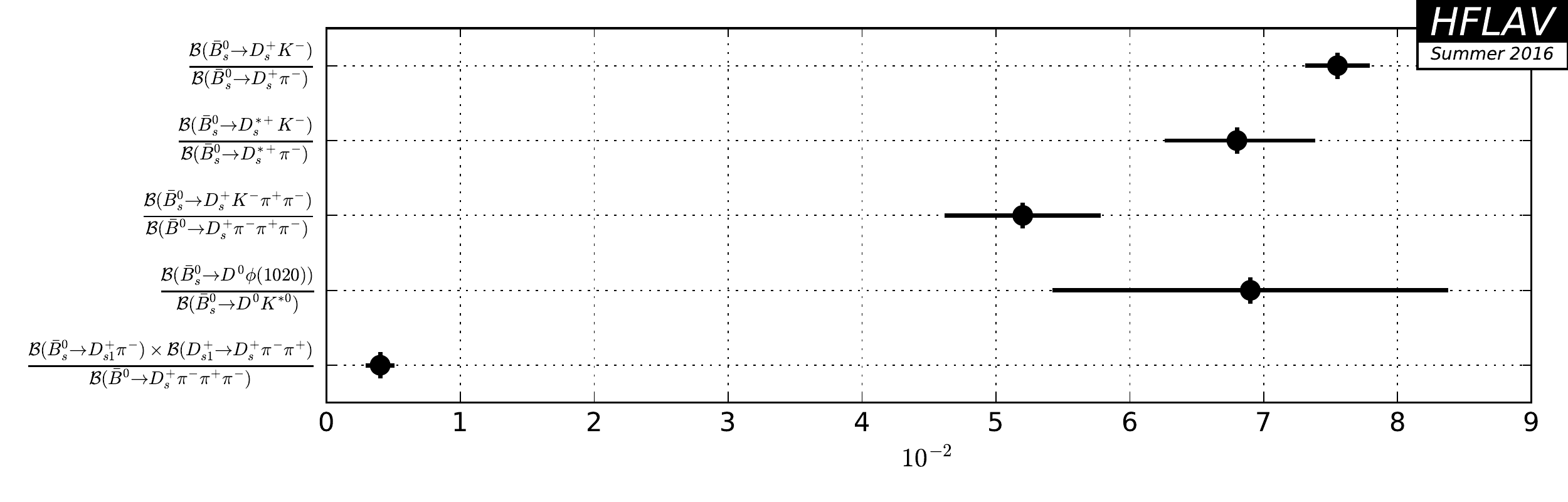}

\subsubsection{Decays to two open charm mesons}
\label{sec:b2c:Bs_DD}
Averages of $\bar{B}_s^0$ decays to two open charm mesons are shown in Tables~\ref{tab:b2c:Bs_DD_1}--\ref{tab:b2c:Bs_DD_3} and Figs.~\ref{fig:b2c:Bs_DD_1}--\ref{fig:b2c:Bs_DD_3}.
\begin{btocharmtab}{Bs_DD_1}{Absolute decay rates $[10^{-2}]$}
\hline
\textbf{Parameter} & \begin{tabular}{l}\textbf{Measurements}\end{tabular} & \textbf{Average} \\
\hline
\hline
${\cal{B}} ( \bar{B}_s^{0} \to D_s^{+} D_s^{-} )$ & \begin{tabular}{l} CDF \cite{Aaltonen:2012mg}: $0.49 \pm 0.06 \pm 0.09$ \\ Belle \cite{Esen:2012yz}: $0.58 \,^{+0.11}_{-0.09} \pm 0.13$ \\ \end{tabular} & $0.52 \pm 0.09$ \\
\hline
${\cal{B}} ( \bar{B}_s^{0} \to D_s^{+} D_s^{*-} )$ & \begin{tabular}{l} LHCb \cite{Aaij:2016rja}: $1.35 \pm 0.06 \pm 0.17$ \\ CDF \cite{Aaltonen:2012mg}: $1.13 \pm 0.12 \pm 0.21$ \\ Belle \cite{Esen:2012yz}: $1.76 \,^{+0.23}_{-0.22} \pm 0.40$ \\ \end{tabular} & $1.38 \pm 0.17$ \\
\hline
${\cal{B}} ( \bar{B}_s^{0} \to D_s^{*+} D_s^{*-} )$ & \begin{tabular}{l} LHCb \cite{Aaij:2016rja}: $1.27 \pm 0.08 \pm 0.17$ \\ CDF \cite{Aaltonen:2012mg}: $1.75 \pm 0.19 \pm 0.34$ \\ Belle \cite{Esen:2012yz}: $1.98 \,^{+0.33}_{-0.31} \,^{+0.51}_{-0.50}$ \\ \end{tabular} & $1.32 \pm 0.18$ \\
\hline
${\cal{B}} ( \bar{B}_s^{0} \to D_s^{(*)+} D_s^{(*)-} )$ & \begin{tabular}{l} LHCb \cite{Aaij:2016rja}: $3.05 \pm 0.10 \pm 0.39$ \\ \dzero \cite{Abazov:2008ig}: $3.5 \pm 1.0 \pm 1.1$ \\ CDF \cite{Aaltonen:2012mg}: $3.38 \pm 0.25 \pm 0.64$ \\ Belle \cite{Esen:2012yz}: $4.32 \,^{+0.42}_{-0.39} \,^{+1.04}_{-1.03}$ \\ \end{tabular} & $3.19 \pm 0.37$ \\
\hline
\end{btocharmtab}
\btocharmfig{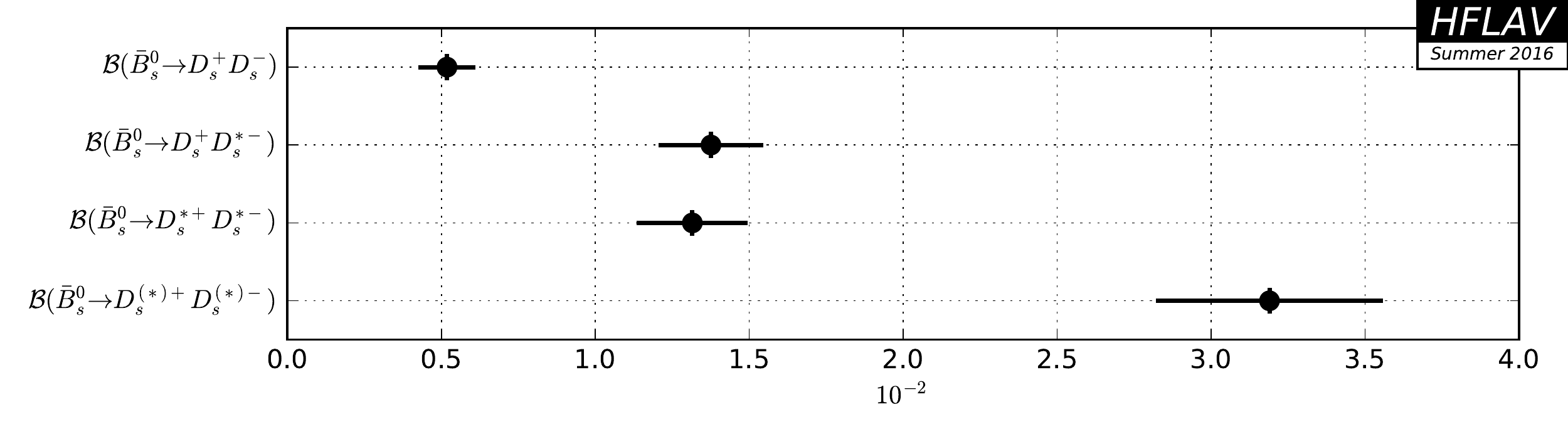}

\begin{btocharmtab}{Bs_DD_2}{Relative decay rates I}
\hline
\textbf{Parameter} & \begin{tabular}{l}\textbf{Measurements}\end{tabular} & \textbf{Average} \\
\hline
\hline
\multicolumn{3}{|l|}{${{\cal{B}} ( \bar{B}^{0}_s \to D^{-} D^{+} )}/{{\cal{B}} ( \bar{B}^{0} \to D^{-} D^{+} )}$}\\
 & \begin{tabular}{l} LHCb \cite{Aaij:2013fha}: $1.08 \pm 0.20 \pm 0.10$ \\ \end{tabular} & $1.08 \pm 0.22$ \\
\hline
\multicolumn{3}{|l|}{${{\cal{B}} ( \bar{B}_s^{0} \to D_s^{-} D_s^{+} )}/{{\cal{B}} ( \bar{B}^{0} \to D_s^{-} D^{+} )}$}\\
 & \begin{tabular}{l} LHCb \cite{Aaij:2013fha}: $0.56 \pm 0.03 \pm 0.04$ \\ \end{tabular} & $0.56 \pm 0.05$ \\
\hline
\end{btocharmtab}
\btocharmfig{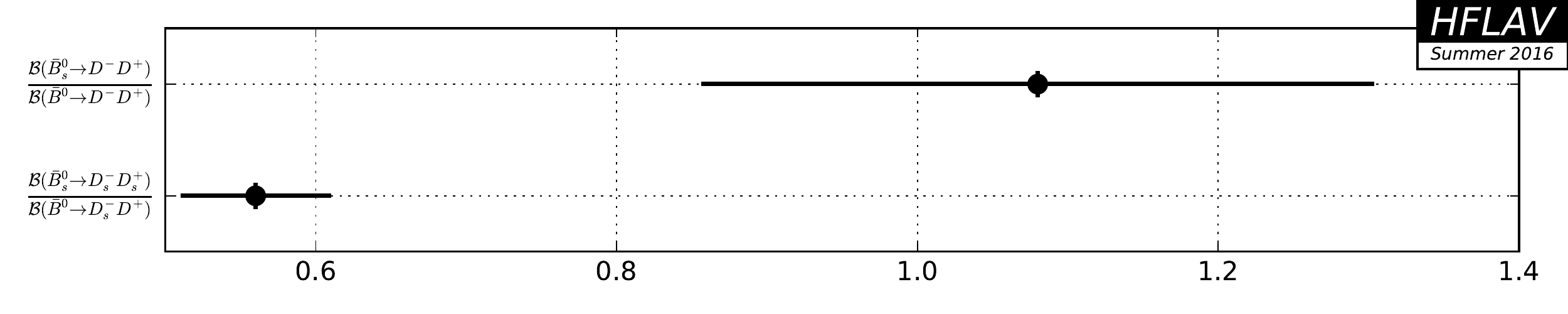}

\begin{btocharmtab}{Bs_DD_3}{Relative decay rates II $[10^{-2}]$}
\hline
\textbf{Parameter} & \begin{tabular}{l}\textbf{Measurements}\end{tabular} & \textbf{Average} \\
\hline
\hline
\multicolumn{3}{|l|}{${ {\cal{B}} ( \bar{B}^{0}_s \to D_s^{+} D^{-} )}/{ {\cal{B}} ( B^{0} \to D_s^{+} D^{-} )}$}\\
 & \begin{tabular}{l} LHCb \cite{Aaij:2013fha}: $5.0 \pm 0.8 \pm 0.4$ \\ \end{tabular} & $5.0 \pm 0.9$ \\
\hline
\multicolumn{3}{|l|}{${{\cal{B}} ( \bar{B}^{0}_s \to \bar{D}^{0} D^{0} )}/{{\cal{B}} ( B^{-} \to D^{0} D_s^{-} )}$}\\
 & \begin{tabular}{l} LHCb \cite{Aaij:2013fha}: $1.9 \pm 0.3 \pm 0.3$ \\ \end{tabular} & $1.9 \pm 0.4$ \\
\hline
\end{btocharmtab}
\btocharmfig{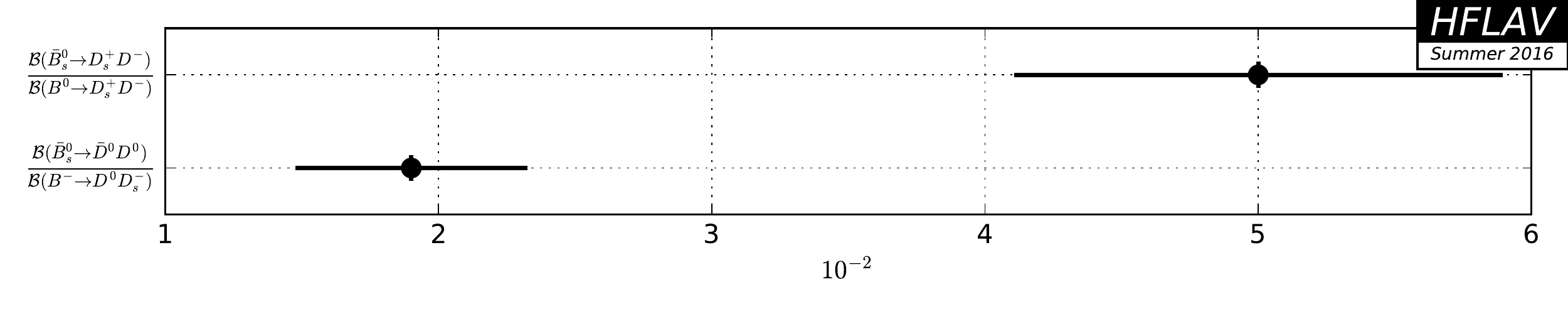}

\subsubsection{Decays to charmonium states}
\label{sec:b2c:Bs_cc}
Averages of $\bar{B}_s^0$ decays to charmonium states are shown in Tables~\ref{tab:b2c:Bs_cc_1}--\ref{tab:b2c:Bs_cc_5} and Figs.~\ref{fig:b2c:Bs_cc_1}--\ref{fig:b2c:Bs_cc_5}.
\begin{btocharmtab}{Bs_cc_1}{Absolute decay rates I $[10^{-4}]$}
\hline
\textbf{Parameter} & \begin{tabular}{l}\textbf{Measurements}\end{tabular} & \textbf{Average} \\
\hline
\hline
${\cal{B}} ( \bar{B}_s^{0} \to J/\psi \eta )$ & \begin{tabular}{l} Belle \cite{Belle:2012aa}: $5.10 \pm 0.50 \,^{+1.17}_{-0.83}$ \\ \end{tabular} & $5.10 \,^{+1.27}_{-0.97}$ \\
\hline
${\cal{B}} ( \bar{B}_s^{0} \to J/\psi \eta^{\prime } )$ & \begin{tabular}{l} Belle \cite{Belle:2012aa}: $3.71 \pm 0.61 \,^{+0.85}_{-0.60}$ \\ \end{tabular} & $3.71 \,^{+1.05}_{-0.85}$ \\
\hline
${\cal{B}} ( \bar{B}_s^{0} \to J/\psi \phi(1020) )$ & \begin{tabular}{l} LHCb \cite{Aaij:2013orb}: $10.5 \pm 0.1 \pm 1.0$ \\ CDF \cite{Abe:1996kc}: $9.3 \pm 2.8 \pm 1.7$ \\ Belle \cite{Thorne:2013llu}: $12.5 \pm 0.7 \pm 2.3$ \\ \end{tabular} & $10.0 \pm 0.9$ \\
\hline
${\cal{B}} ( \bar{B}_s^{0} \to J/\psi K^{0} K^{\pm} \pi^{\mp} )$ & \begin{tabular}{l} LHCb \cite{Aaij:2014naa}: $9.1 \pm 0.6 \pm 0.7$ \\ \end{tabular} & $9.1 \pm 0.9$ \\
\hline
\multicolumn{3}{|l|}{${{\cal{B}} ( \bar{B}_s^{0} \to J/\psi f_0(980) )\times {\cal{B}} ( f_0(980) \to \pi^{+} \pi^{-} )}$}\\
 & \begin{tabular}{l} Belle \cite{Li:2011pg}: $1.16 \,^{+0.31}_{-0.19} \,^{+0.30}_{-0.25}$ \\ \end{tabular} & $1.16 \,^{+0.43}_{-0.32}$ \\
\hline
\end{btocharmtab}
\btocharmfig{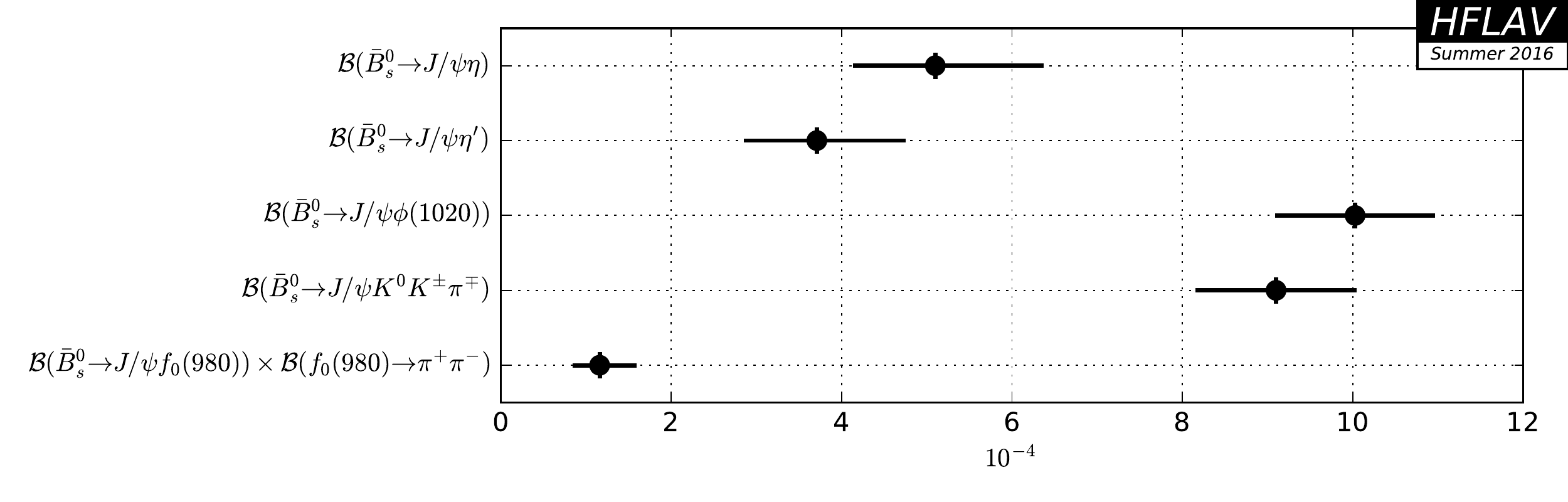}

\begin{btocharmtab}{Bs_cc_2}{Absolute decay rates II $[10^{-5}]$}
\hline
\textbf{Parameter} & \begin{tabular}{l}\textbf{Measurements}\end{tabular} & \textbf{Average} \\
\hline
\hline
${\cal{B}} ( \bar{B}^{0}_s \to J/\psi \bar{K}^{0} )$ & \begin{tabular}{l} LHCb \cite{Aaij:2012di}: $3.66 \pm 0.42 \pm 0.37$ \\ CDF \cite{Aaltonen:2011sy}: $3.5 \pm 0.6 \pm 0.6$ \\ \end{tabular} & $3.61 \pm 0.46$ \\
\hline
${\cal{B}} ( \bar{B}^{0}_s \to J/\psi K^{*0} )$ & \begin{tabular}{l} LHCb \cite{Aaij:2015mea}: $4.17 \pm 0.18 \pm 0.35$ \\ CDF \cite{Aaltonen:2011sy}: $8.3 \pm 1.2 \pm 3.6$ \\ \end{tabular} & $4.15 \pm 0.40$ \\
\hline
${\cal{B}} ( \bar{B}_s^{0} \to J/\psi p \bar{p} )$ & \begin{tabular}{l} LHCb \cite{Aaij:2013yba}: $< 0.48$ \\ \end{tabular} & $< 0.48$ \\
\hline
${\cal{B}} ( \bar{B}_s^{0} \to J/\psi f_1(1285) )$ & \begin{tabular}{l} LHCb \cite{Aaij:2013rja}: $7.14 \pm 0.99 \,^{+0.93}_{-1.00}$ \\ \end{tabular} & $7.14 \,^{+1.36}_{-1.41}$ \\
\hline
${\cal{B}} ( \bar{B}_s^{0} \to J/\psi K^{0} \pi^{+} \pi^{-} )$ & \begin{tabular}{l} LHCb \cite{Aaij:2014naa}: $< 4.4$ \\ \end{tabular} & $< 4.4$ \\
\hline
${\cal{B}} ( \bar{B}_s^{0} \to J/\psi K^{0} K^{+} K^{-} )$ & \begin{tabular}{l} LHCb \cite{Aaij:2014naa}: $< 1.2$ \\ \end{tabular} & $< 1.2$ \\
\hline
\multicolumn{3}{|l|}{${{\cal{B}} ( \bar{B}_s^{0} \to J/\psi f_0(1370) )\times {\cal{B}} ( f_0(1370) \to \pi^{+} \pi^{-} )}$}\\
 & \begin{tabular}{l} Belle \cite{Li:2011pg}: $3.4 \,^{+1.1}_{-1.4} \,^{+0.9}_{-0.5}$ \\ \end{tabular} & $3.4 \,^{+1.4}_{-1.5}$ \\
\hline
\multicolumn{3}{|l|}{${{\cal{B}} ( \bar{B}_s^{0} \to J/\psi f_1(1285) )\times {\cal{B}} ( f_1(1285) \to  \pi^{+} \pi^{-} \pi^{+} \pi^{-} )}$}\\
 & \begin{tabular}{l} LHCb \cite{Aaij:2013rja}: $0.785 \pm 0.109 \,^{+0.089}_{-0.101}$ \\ \end{tabular} & $0.785 \,^{+0.141}_{-0.149}$ \\
\hline
${\cal{B}} ( \bar{B}^{0}_s \to J/\psi \gamma )$ & \begin{tabular}{l} LHCb \cite{Aaij:2015uoa}: $< 0.73$ \\ \end{tabular} & $< 0.73$ \\
\hline
\end{btocharmtab}
\btocharmfig{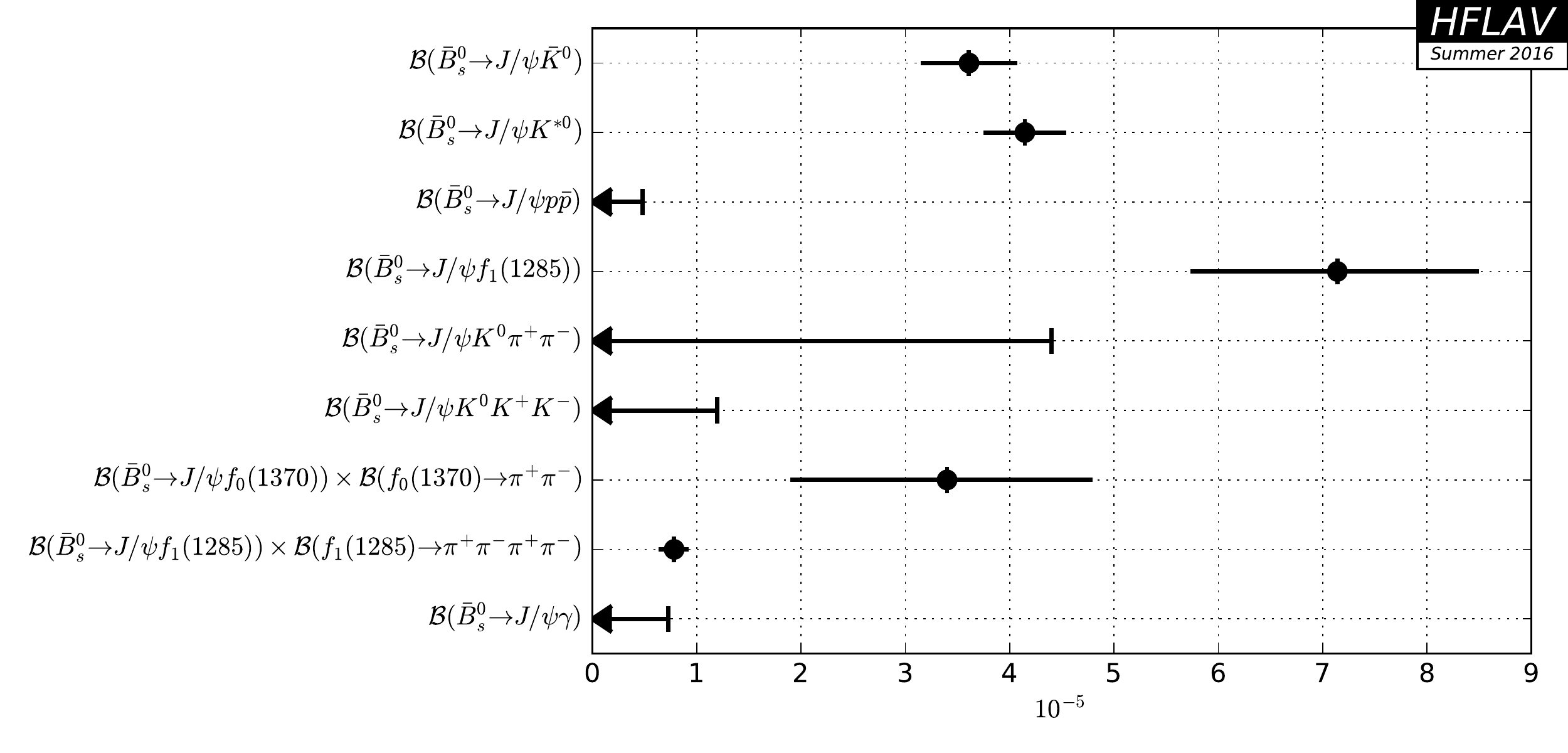}

\begin{btocharmtab}{Bs_cc_3}{Relative decay rates I}
\hline
\textbf{Parameter} & \begin{tabular}{l}\textbf{Measurements}\end{tabular} & \textbf{Average} \\
\hline
\hline
\multicolumn{3}{|l|}{${{\cal{B}} ( \bar{B}^{0}_s \to J/\psi \eta)}/ { {\cal{B}} ( \bar{B}^{0} \to J/\psi \rho )}$}\\
 & \begin{tabular}{l} LHCb \cite{LHCb:2012cw}: $14.0 \pm 1.2 \,^{+1.6}_{-1.8}$ \\ \end{tabular} & $14.0 \,^{+2.0}_{-2.2}$ \\
\hline
\multicolumn{3}{|l|}{${ {\cal{B}} ( \bar{B}^{0}_s \to J/\psi \eta^{\prime })}/ {{\cal{B}} ( \bar{B}^{0} \to J/\psi \rho )}$}\\
 & \begin{tabular}{l} LHCb \cite{LHCb:2012cw}: $12.7 \pm 1.1 \,^{+1.1}_{-0.9}$ \\ \end{tabular} & $12.7 \,^{+1.6}_{-1.4}$ \\
\hline
\multicolumn{3}{|l|}{${{\cal{B}} ( \bar{B}_s^{0} \to J/\psi K_S^{0} K^{\pm} \pi^{\mp} )}/{{\cal{B}} ( \bar{B}^{0} \to J/\psi \pi^{+} \pi^{-} )}$}\\
 & \begin{tabular}{l} LHCb \cite{Aaij:2014naa}: $2.12 \pm 0.15 \pm 0.18$ \\ \end{tabular} & $2.12 \pm 0.23$ \\
\hline
\end{btocharmtab}
\btocharmfig{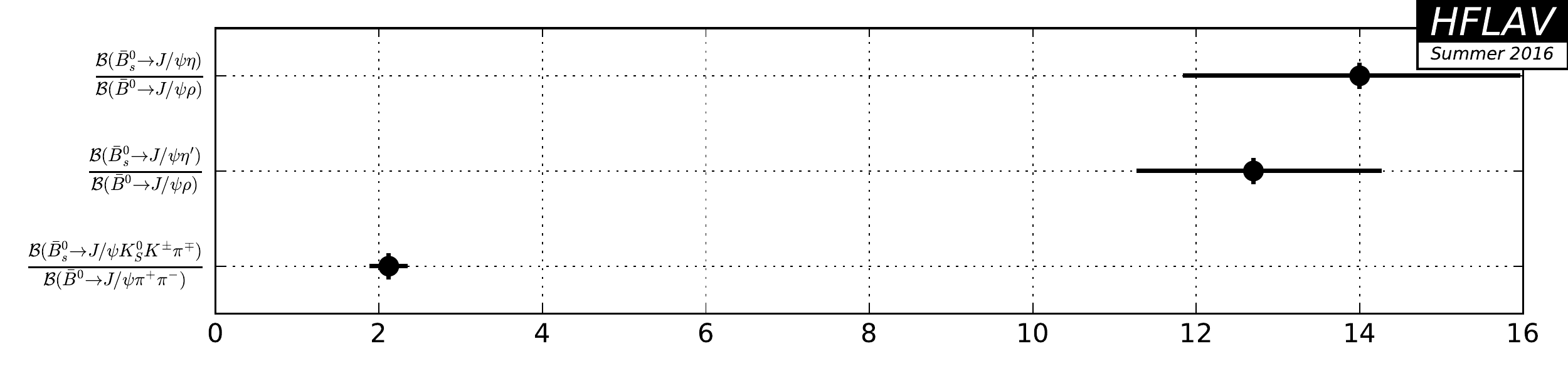}

\begin{btocharmtab}{Bs_cc_4}{Relative decay rates II}
\hline
\textbf{Parameter} & \begin{tabular}{l}\textbf{Measurements}\end{tabular} & \textbf{Average} \\
\hline
\hline
\multicolumn{3}{|l|}{${ {\cal{B}} ( \bar{B}^{0}_s \to J/\psi \eta)}/ {{\cal{B}} ( \bar{B}^{0} \to J/\psi \eta^{\prime} ) }$}\\
 & \begin{tabular}{l} Belle \cite{Belle:2012aa}: $0.73 \pm 0.14 \pm 0.02$ \\ \end{tabular} & $0.73 \pm 0.14$ \\
\hline
\multicolumn{3}{|l|}{${{\cal{B}} ( \bar{B}^{0}_s \to J/\psi \eta^{\prime })}/  {{\cal{B}} ( \bar{B}_{s}^{0} \to J/\psi \eta )}$}\\
 & \begin{tabular}{l} LHCb \cite{LHCb:2012cw}: $0.90 \pm 0.09 \,^{+0.06}_{-0.02}$ \\ \end{tabular} & $0.90 \,^{+0.11}_{-0.09}$ \\
\hline
\multicolumn{3}{|l|}{${{\cal{B}} ( \bar{B}^{0}_s \to J/\psi f^\prime_2 )}/ { {\cal{B}} ( \bar{B}^{0}_s \to J/\psi \phi(1020) )}$}\\
 & \begin{tabular}{l} LHCb \cite{Aaij:2011ac}: $0.264 \pm 0.027 \pm 0.024$ \\ \dzero \cite{Abazov:2012dz}: $0.19 \pm 0.05 \pm 0.04$ \\ \end{tabular} & $0.246 \pm 0.031$ \\
\hline
\multicolumn{3}{|l|}{${{\cal{B}} ( \bar{B}^{0}_s \to J/\psi \pi^{+} \pi^{-} )}/ { {\cal{B}} ( \bar{B}^{0}_s \to J/\psi \phi(1020) )}$}\\
 & \begin{tabular}{l} LHCb \cite{Aaij:2011fx}: $0.162 \pm 0.022 \pm 0.016$ \\ \end{tabular} & $0.162 \pm 0.027$ \\
\hline
\multicolumn{3}{|l|}{${{\cal{B}} \bar{B}_s^{0} \to \psi(2S) \pi^{+}  \pi^{-} }/{{\cal{B}} \bar{B}_s^{0} \to J/\psi \pi^{+} \pi^{-}  }$}\\
 & \begin{tabular}{l} LHCb \cite{Aaij:2013cpa}: $0.34 \pm 0.04 \pm 0.03$ \\ \end{tabular} & $0.34 \pm 0.05$ \\
\hline
\multicolumn{3}{|l|}{${{\cal{B}} ( \bar{B}_s^{0} \to \psi(2S) \phi(1020) )}/{{\cal{B}} ( \bar{B}_s^{0} \to J/\psi \phi(1020) )}$}\\
 & \begin{tabular}{l} LHCb \cite{Aaij:2012dda}: $0.489 \pm 0.026 \pm 0.024$ \\ \dzero \cite{Abazov:2008jk}: $0.55 \pm 0.11 \pm 0.09$ \\ CDF \cite{Abulencia:2006jp}: $0.52 \pm 0.13 \pm 0.07$ \\ \end{tabular} & $0.494 \pm 0.034$ \\
\hline
\multicolumn{3}{|l|}{${{\cal{B}} ( \bar{B}_s^{0} \to J/\psi K_S^{0} \pi^{+} \pi^{-} )}/{{\cal{B}} ( \bar{B}^{0} \to J/\psi \pi^{+} \pi^{-} )}$}\\
 & \begin{tabular}{l} LHCb \cite{Aaij:2014naa}: $< 0.10$ \\ \end{tabular} & $< 0.10$ \\
\hline
\multicolumn{3}{|l|}{$[{{\cal{B}} ( \bar{B}^{0}_s \to J/\psi f_0(980) ) \times {\cal{B}} ( f_0(980) \to \pi^{+} \pi^{-} )}]/[ { {\cal{B}} ( \bar{B}^{0}_s \to J/\psi \phi(1020) )) \times {\cal{B}} ( \phi \to K^{+} K^{-} )}]$}\\
 & \begin{tabular}{l} LHCb \cite{Aaij:2011fx}: $0.252 \,^{+0.046}_{-0.032} \,^{+0.027}_{-0.033}$ \\ \dzero \cite{Abazov:2011hv}: $0.275 \pm 0.041 \pm 0.061$ \\ CMS \cite{Khachatryan:2015lua}: $0.140 \pm 0.008 \pm 0.023$ \\ CDF \cite{Aaltonen:2011nk}: $0.257 \pm 0.020 \pm 0.014$ \\ \end{tabular} & $0.207 \pm 0.016$ {\tiny CL=3.8\permil} \\
\hline
\end{btocharmtab}
\btocharmfig{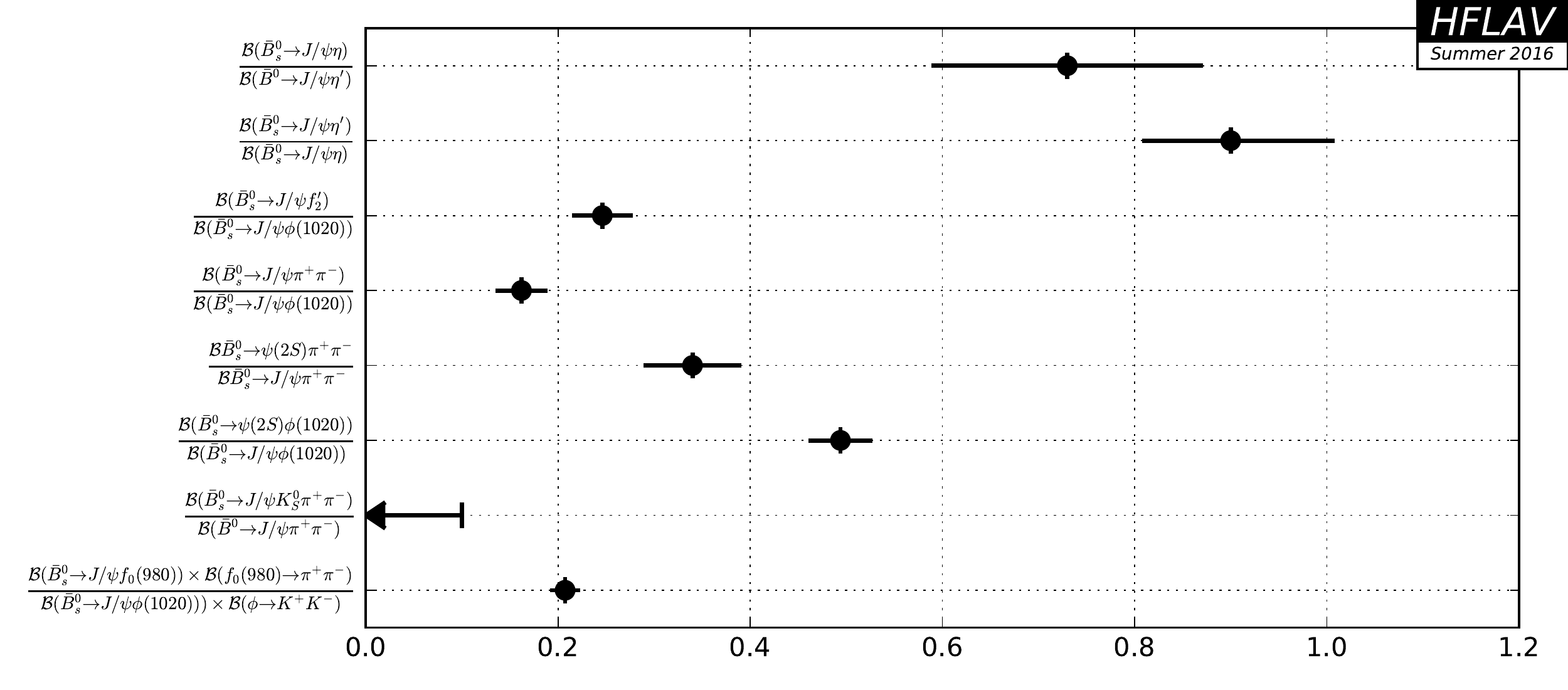}

\begin{btocharmtab}{Bs_cc_5}{Relative decay rates III $[10^{-2}]$}
\hline
\textbf{Parameter} & \begin{tabular}{l}\textbf{Measurements}\end{tabular} & \textbf{Average} \\
\hline
\hline
\multicolumn{3}{|l|}{${{\cal{B}} ( \bar{B}^{0}_s \to J/\psi K^{0}_S )}/ {{\cal{B}} ( \bar{B}^{0} \to J/\psi K^{0}_S )}$}\\
 & \begin{tabular}{l} LHCb \cite{Aaij:2012di}: $4.20 \pm 0.49 \pm 0.40$ \\ \end{tabular} & $4.20 \pm 0.63$ \\
\hline
\multicolumn{3}{|l|}{${{\cal{B}} ( \bar{B}^{0}_s \to J/\psi \phi(1020) \phi(1020) )}/ { {\cal{B}} ( \bar{B}^{0}_s \to J/\psi \phi(1020) )}$}\\
 & \begin{tabular}{l} LHCb \cite{Aaij:2016qim}: $1.15 \pm 0.12 \,^{+0.05}_{-0.09}$ \\ \end{tabular} & $1.15 \,^{+0.13}_{-0.15}$ \\
\hline
\multicolumn{3}{|l|}{${{\cal{B}} ( \bar{B}_s^{0} \to \psi(2S) K^{+} \pi^{-} )}/{{\cal{B}} ( \bar{B}^{0} \to \psi(2S)  K^{+} \pi^{-}  )}$}\\
 & \begin{tabular}{l} LHCb \cite{Aaij:2015wza}: $5.38 \pm 0.36 \pm 0.38$ \\ \end{tabular} & $5.38 \pm 0.52$ \\
\hline
\multicolumn{3}{|l|}{${{\cal{B}} ( \bar{B}_s^{0} \to \psi(2S)  K^{*0}  )}/{{\cal{B}} ( \bar{B}^{0} \to \psi(2S)  K^{*0}   )}$}\\
 & \begin{tabular}{l} LHCb \cite{Aaij:2015wza}: $5.38 \pm 0.57 \pm 0.51$ \\ \end{tabular} & $5.38 \pm 0.77$ \\
\hline
\multicolumn{3}{|l|}{${{\cal{B}} ( \bar{B}_s^{0} \to J/\psi K_S^{0} K^{+} K^{-} )}/{{\cal{B}} ( \bar{B}^{0} \to J/\psi \pi^{+} \pi^{-} )}$}\\
 & \begin{tabular}{l} LHCb \cite{Aaij:2014naa}: $< 2.7$ \\ \end{tabular} & $< 2.7$ \\
\hline
\multicolumn{3}{|l|}{$[{{\cal{B}} ( \bar{B}^{0}_s \to J/\psi f_0(500) ) \times {\cal{B}} ( f_0(500) \to \pi^{+} \pi^{-} )}]/[  { {\cal{B}} ( \bar{B}^{0}_s \to J/\psi f_0(980)  )) \times {\cal{B}} ( f_0(500)  \to \pi^{+} \pi^{-} )}]$}\\
 & \begin{tabular}{l} LHCb \cite{Aaij:2014emv}: $< 3.4$ \\ \end{tabular} & $< 3.4$ \\
\hline
\end{btocharmtab}
\btocharmfig{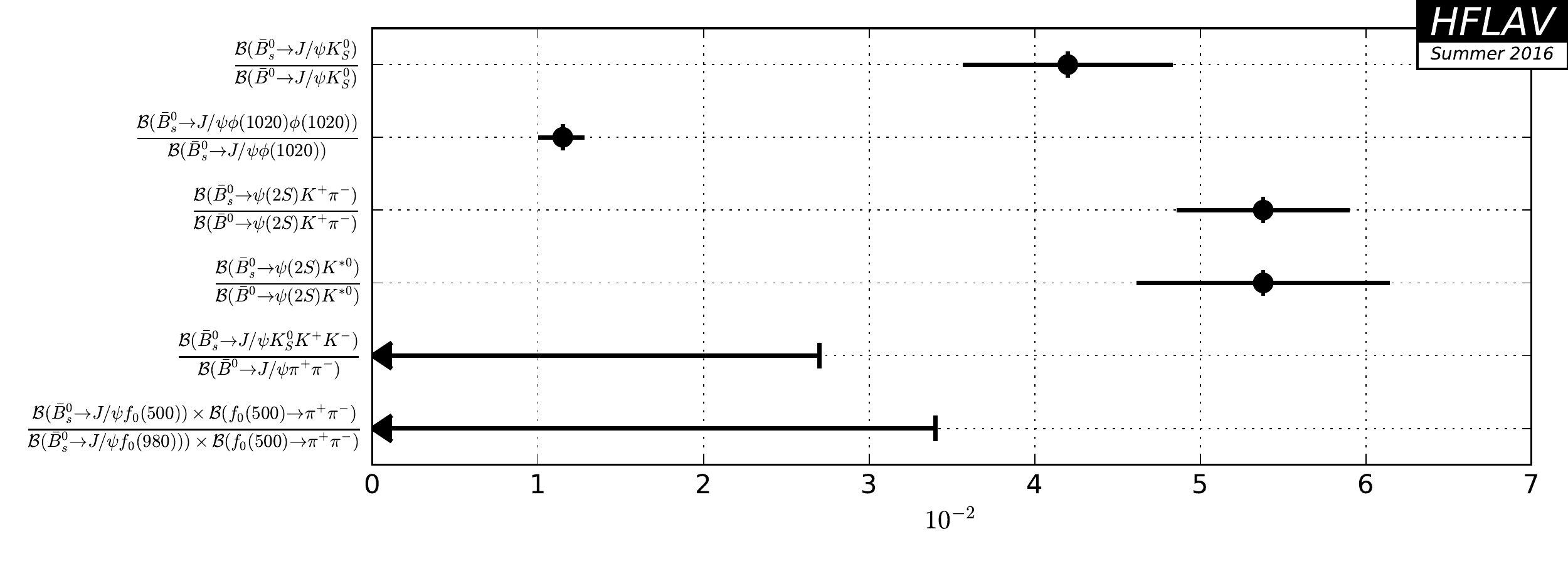}

\subsubsection{Decays to charm baryons}
\label{sec:b2c:Bs_baryon}
Averages of $\bar{B}_s^0$ decays to charm baryons are shown in Tables~\ref{tab:b2c:Bs_baryon_1}--\ref{tab:b2c:Bs_baryon_2}.
\begin{btocharmtab}{Bs_baryon_1}{Decays to one charm baryon $[10^{-4}]$}
\hline
\textbf{Parameter} & \begin{tabular}{l}\textbf{Measurements}\end{tabular} & \textbf{Average} \\
\hline
\hline
${\cal{B}} ( \bar{B}_s^{0} \to \Lambda_c^{+} \bar{\Lambda} \pi^{-} )$ & \begin{tabular}{l} Belle \cite{Solovieva:2013rhq}: $3.6 \pm 1.1 \,^{+1.2}_{-1.2}$ \\ \end{tabular} & $3.6 \,^{+1.6}_{-1.7}$ \\
\hline
\end{btocharmtab}

\begin{btocharmtab}{Bs_baryon_2}{Decays to two charm baryons}
\hline
\textbf{Parameter} & \begin{tabular}{l}\textbf{Measurements}\end{tabular} & \textbf{Average} \\
\hline
\hline
\multicolumn{3}{|l|}{${{\cal{B}} ( \bar{B}^{0}_s \to \Lambda_c^{-} \Lambda_c^{+} )}/  { {\cal{B}} ( \bar{B}_s^{0} \to D^{-} D_s^{+} )}$}\\
 & \begin{tabular}{l} LHCb \cite{Aaij:2014pha}: $< 0.30$ \\ \end{tabular} & $< 0.30$ \\
\hline
\end{btocharmtab}

\subsection{Decays of $B_c^-$ mesons}
\label{sec:b2c:Bc}
Measurements of $B_c^-$ decays to charmed hadrons are summarized in Sections~\ref{sec:b2c:Bc_cc} to~\ref{sec:b2c:Bc_B}.
Since the absolute cross-section for $B_c^-$ meson production in any production environment is currently not known, it is not possible to determine absolute branching fractions. 
Instead, results are presented either as ratios of branching fractions of different $B_c^-$ decays, or are normalised to the branching fraction of the decay of a lighter $B$ meson (usually $B^-$).
In the latter case the measured quantity is the absolute or relative $B_c^-$ branching fraction multiplied by the ratio of cross-sections (or, equivalently, production fractions) of the $B_c^-$ and the lighter $B$ meson.

It should be noted that the ratio of cross-sections for different $b$ hadron species can depend on production environment, and on the fiducial region accessed by each experiment.
While this has been studied for certain $b$ hadron species (see Sec.~\ref{sec:fractions}), there is currently little published data that would allow to investigate the effect for $B_c^-$ mesons.
Therefore, we do not attempt to apply any correction for this effect.

\subsubsection{Decays to charmonium states}
\label{sec:b2c:Bc_cc}
Averages of $B_c^-$ decays to charmonium states are shown in Tables~\ref{tab:b2c:Bc_cc_1}--\ref{tab:b2c:Bc_cc_4} and Figs.~\ref{fig:b2c:Bc_cc_1}--\ref{fig:b2c:Bc_cc_2}.
\begin{btocharmtab}{Bc_cc_1}{Relative decay rates I}
\hline
\textbf{Parameter} & \begin{tabular}{l}\textbf{Measurements}\end{tabular} & \textbf{Average} \\
\hline
\hline
\multicolumn{3}{|l|}{${{\cal{B}} (B_c^{-} \to J/\psi D_s^{-})}/ {{\cal{B}} (B_c^{-} \to J/\psi \pi^{-})}$}\\
 & \begin{tabular}{l} LHCb \cite{Aaij:2013gia}: $2.90 \pm 0.57 \pm 0.24$ \\ ATLAS \cite{Aad:2015eza}: $3.8 \pm 1.1 \pm 0.4$ \\ \end{tabular} & $3.09 \pm 0.55$ \\
\hline
\multicolumn{3}{|l|}{${{\cal{B}} (B_c^{-} \to J/\psi D_s^{*-}}/ {{\cal{B}} (B_c^{-} \to J/\psi D_s^{-})}$}\\
 & \begin{tabular}{l} ATLAS \cite{Aad:2015eza}: $2.8 \,^{+1.2}_{-0.8} \pm 0.3$ \\ \end{tabular} & $2.8 \,^{+1.2}_{-0.9}$ \\
\hline
\multicolumn{3}{|l|}{${{\cal{B}} (B_c^{-} \to J/\psi D_s^{*-}}/ {{\cal{B}} (B_c^{-} \to J/\psi \pi^{-})}$}\\
 & \begin{tabular}{l} ATLAS \cite{Aad:2015eza}: $10.4 \pm 3.1 \pm 1.6$ \\ \end{tabular} & $10.4 \pm 3.5$ \\
\hline
\multicolumn{3}{|l|}{${{\cal{B}} ( B_c^{-} \to J/\psi \pi^{+} \pi^{-} \pi^{-} )}/ {{\cal{B}} ( B_c^{-} \to J/\psi \pi^{-} )}$}\\
 & \begin{tabular}{l} LHCb \cite{LHCb:2012ag}: $2.41 \pm 0.30 \pm 0.33$ \\ CMS \cite{Khachatryan:2014nfa}: $2.55 \pm 0.80 \,^{+0.33}_{-0.33}$ \\ \end{tabular} & $2.44 \pm 0.40$ \\
\hline
\end{btocharmtab}
\btocharmfig{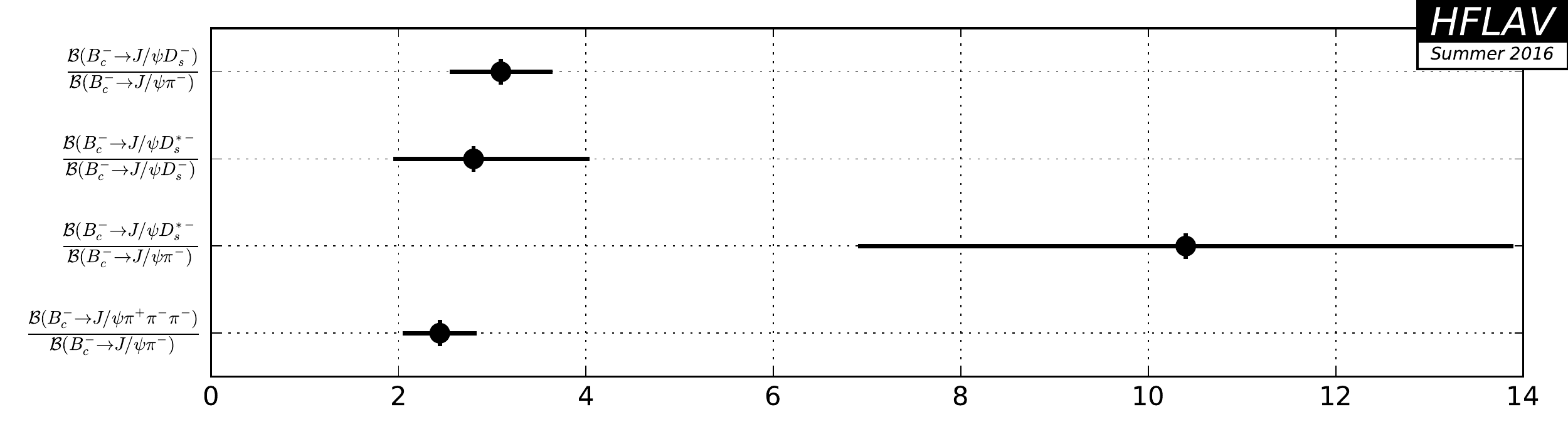}

\begin{btocharmtab}{Bc_cc_2}{Relative decay rates II}
\hline
\textbf{Parameter} & \begin{tabular}{l}\textbf{Measurements}\end{tabular} & \textbf{Average} \\
\hline
\hline
\multicolumn{3}{|l|}{${{\cal{B}}( B_c^{-} \to J/\psi K^{-})}/ { {\cal{B}}( B_c^{-} \to J/\psi \pi^{-})}$}\\
 & \begin{tabular}{l} LHCb \cite{Aaij:2013vcx}: $0.069 \pm 0.019 \pm 0.005$ \\ \end{tabular} & $0.069 \pm 0.020$ \\
\hline
\multicolumn{3}{|l|}{${{\cal{B}} ( B_c^{-} \to J/\psi K^{-} K^{+} \pi^{-} )}/ { {\cal{B}} ( B_c^{-} \to J/\psi \pi^{-} )}$}\\
 & \begin{tabular}{l} LHCb \cite{Aaij:2013gxa}: $0.53 \pm 0.10 \pm 0.05$ \\ \end{tabular} & $0.53 \pm 0.11$ \\
\hline
\multicolumn{3}{|l|}{${{\cal{B}} ( B_c^{-} \to \psi(2S) \pi^{-} ) }/  {{\cal{B}} ( B_c^{-} \to J/\psi \pi^{-} ) }$}\\
 & \begin{tabular}{l} LHCb \cite{Aaij:2015xga}: $0.268 \pm 0.032 \pm 0.009$ \\ \end{tabular} & $0.268 \pm 0.033$ \\
\hline
\end{btocharmtab}
\btocharmfig{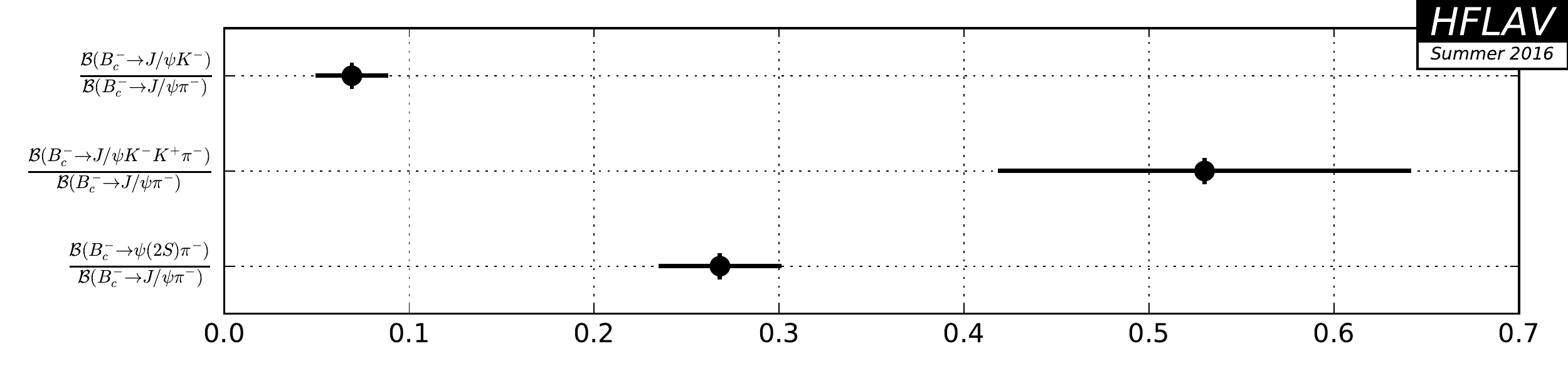}

\begin{btocharmtab}{Bc_cc_3}{Relative production times decay rates $[10^{-3}]$}
\hline
\textbf{Parameter} & \begin{tabular}{l}\textbf{Measurements}\end{tabular} & \textbf{Average} \\
\hline
\hline
\multicolumn{3}{|l|}{$[{{\sigma(B_c^{-}) \times \cal{B}} ( B_c^{-} \to J/\psi \pi^{-} )}]/[  {\sigma(B^{-}) \times {\cal{B}} ( B^{-} \to J/\psi K^{-} )}]$}\\
 & \begin{tabular}{l} LHCb \cite{Aaij:2014ija}: $6.83 \pm 0.18 \pm 0.09$ \\ LHCb \cite{Aaij:2012dd}: $6.8 \pm 1.0 \pm 0.6$ \\ CMS \cite{Khachatryan:2014nfa}: $4.8 \pm 0.5 \pm 0.6$ \\ \end{tabular} & $6.72 \pm 0.19$ \\
\hline
\end{btocharmtab}

\begin{btocharmtab}{Bc_cc_4}{Decay rates times relative production rates $[10^{-6}]$}
\hline
\textbf{Parameter} & \begin{tabular}{l}\textbf{Measurements}\end{tabular} & \textbf{Average} \\
\hline
\hline
\multicolumn{3}{|l|}{$[{{\sigma(B_c^{-})}/{\sigma(B^{-})}}] \times {\cal{B}} ( B_c^{-} \to \chi_{c0} \pi^{-} )$}\\
 & \begin{tabular}{l} LHCb \cite{Aaij:2016xas}: $9.8 \,^{+3.4}_{-3.0} \pm 0.8$ \\ \end{tabular} & $9.8 \,^{+3.5}_{-3.1}$ \\
\hline
\end{btocharmtab}

\subsubsection{Decays to a $B$ meson}
\label{sec:b2c:Bc_B}
Averages of $B_c^-$ decays to a $B$ meson are shown in Table~\ref{tab:b2c:Bc_B_1}.
\begin{btocharmtab}{Bc_B_1}{Decays to $B_s^{0}$ meson $[10^{-3}]$}
\hline
\textbf{Parameter} & \begin{tabular}{l}\textbf{Measurements}\end{tabular} & \textbf{Average} \\
\hline
\hline
\multicolumn{3}{|l|}{${ [\sigma(B_c^{+})}/{ \sigma(B_s^{0})} ] \times {\cal B} ( B_c^{+} \to  B_s^{0}\pi^{+} )$}\\
 & \begin{tabular}{l} LHCb \cite{Aaij:2013cda}: $2.37 \pm 0.31 \,^{+0.20}_{-0.17}$ \\ \end{tabular} & $2.37 \,^{+0.37}_{-0.35}$ \\
\hline
\end{btocharmtab}

\subsection{Decays of $b$ baryons}
\label{sec:b2c:Bbaryon}
Measurements of $b$ baryons decays to charmed hadrons are summarized in Sections~\ref{sec:b2c:Bbaryon_D} to~\ref{sec:b2c:Bbaryon_baryon}.
Comments regarding the production rates of $\bar{B}_s^0$ and $B_c^-$ mesons relative to lighter $B$ mesons, in Sec.~\ref{sec:b2c:Bs} and Sec.~\ref{sec:b2c:Bc} respectively, are also appropriate here.
Specifically, since the cross-section for production of $\Lb$ baryons is reasonably well-known, it is possible to determine absolute or relative branching fractions for its decays (although some older measurements are presented as products involving the cross-section).
The cross-sections for production of heavier $b$ baryons are not known, and therefore measured quantities are presented as absolute or relative branching fraction multiplied by a ratio of cross-sections (or, equivalently, production fractions).

\subsubsection{Decays to a single open charm meson}
\label{sec:b2c:Bbaryon_D}
Averages of $b$ baryons decays to a single open charm meson are shown in Table~\ref{tab:b2c:Bbaryon_D_1} and Fig.~\ref{fig:b2c:Bbaryon_D_1}.
\begin{btocharmtab}{Bbaryon_D_1}{Relative decay rates to $D^{0}$ mesons}
\hline
\textbf{Parameter} & \begin{tabular}{l}\textbf{Measurements}\end{tabular} & \textbf{Average} \\
\hline
\hline
\multicolumn{3}{|l|}{${{\cal{B}} ( \Lambda_b^{0} \to D^{0} p K^{-} ) }/{ {\cal{B}} ( \Lambda_b^{0} \to D^{0} p \pi^{-} )}$}\\
 & \begin{tabular}{l} LHCb \cite{Aaij:2013pka}: $0.073 \pm 0.008 \,^{+0.005}_{-0.006}$ \\ \end{tabular} & $0.073 \,^{+0.009}_{-0.010}$ \\
\hline
\multicolumn{3}{|l|}{$[{ {\cal{B}} ( \Lambda_b^{0} \to D^{0} p \pi^{-})  \times {\cal{B}} ( D^{0} \to K^{+} \pi^{-} )}]/[ {  {\cal{B}} ( \Lambda_b^{0} \to \Lambda_c^{+} \pi^{-} )   \times   {\cal{B}}  (\Lambda_c^{+} \to  p K^{-} \pi^{+}) }]$}\\
 & \begin{tabular}{l} LHCb \cite{Aaij:2013pka}: $0.0806 \pm 0.0023 \pm 0.0035$ \\ \end{tabular} & $0.0806 \pm 0.0042$ \\
\hline
\multicolumn{3}{|l|}{$[{ f_{\Xi_b^{0}} \times{\cal{B}} ( \Xi_b^{0} \to D^{0} p K^{-} )   }]/[ { f_{\Lambda_b^{0}} \times{\cal{B}} ( \Lambda_b^{0} \to D^{0} p K^{-})  }]$}\\
 & \begin{tabular}{l} LHCb \cite{Aaij:2013pka}: $0.44 \pm 0.09 \pm 0.06$ \\ \end{tabular} & $0.44 \pm 0.11$ \\
\hline
\end{btocharmtab}
\btocharmfig{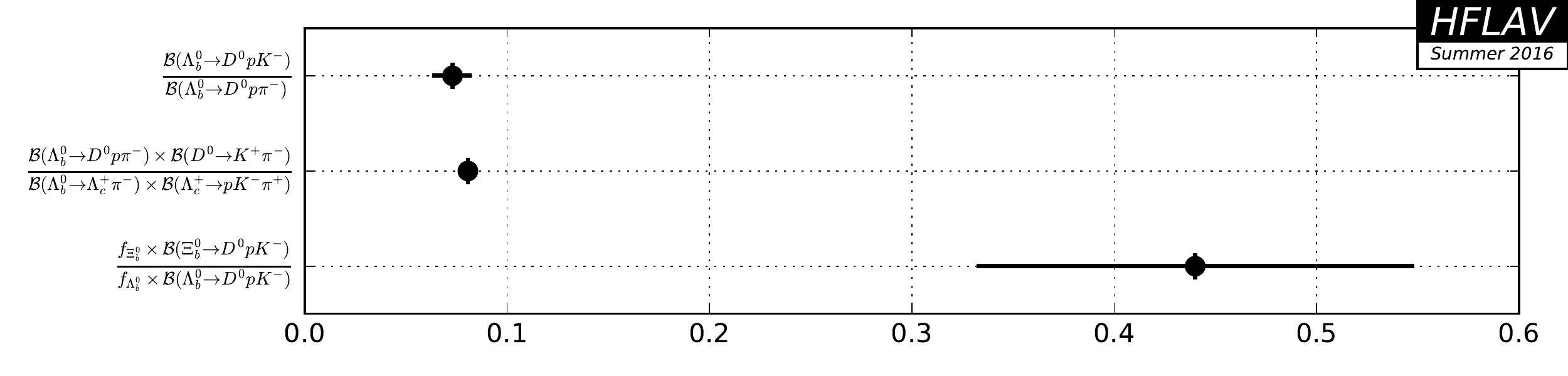}

\subsubsection{Decays to charmonium states}
\label{sec:b2c:Bbaryon_cc}
Averages of $b$ baryons decays to charmonium states are shown in Tables~\ref{tab:b2c:Bbaryon_cc_1}--\ref{tab:b2c:Bbaryon_cc_5} and Figs.~\ref{fig:b2c:Bbaryon_cc_1}--\ref{fig:b2c:Bbaryon_cc_4}.
\begin{btocharmtab}{Bbaryon_cc_1}{$\Lambda_b^{0}$ decays to charmonium $[10^{-4}]$}
\hline
\textbf{Parameter} & \begin{tabular}{l}\textbf{Measurements}\end{tabular} & \textbf{Average} \\
\hline
\hline
${\cal{B}} ( \Lambda_b^{0} \to J/\psi p K^{-} )$ & \begin{tabular}{l} LHCb \cite{Aaij:2015fea}: $3.17 \pm 0.04 \,^{+0.46}_{-0.29}$ \\ \end{tabular} & $3.17 \,^{+0.46}_{-0.29}$ \\
\hline
${\cal{B}} ( \Lambda_b^{0} \to J/\psi \Lambda )$ & \begin{tabular}{l} CDF \cite{Abe:1996tr}: $4.7 \pm 2.1 \pm 1.9$ \\ \end{tabular} & $4.7 \pm 2.8$ \\
\hline
\end{btocharmtab}
\btocharmfig{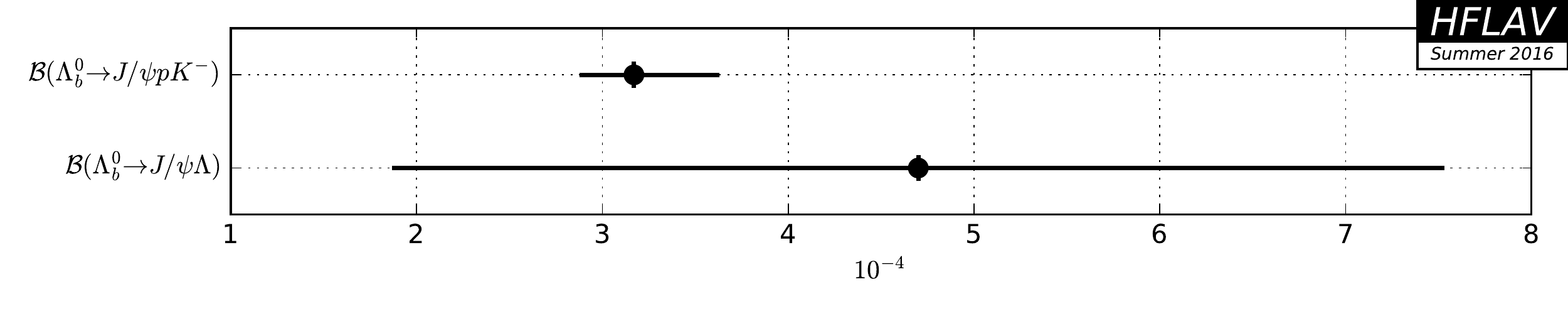}

\begin{btocharmtab}{Bbaryon_cc_2}{$f_b$ times $\Lambda_b^{0}$ decay to charmonium $[10^{-5}]$}
\hline
\textbf{Parameter} & \begin{tabular}{l}\textbf{Measurements}\end{tabular} & \textbf{Average} \\
\hline
\hline
${f_{\Lambda_b} \times \cal{B}} ( \Lambda_b^{0} \to J/\psi \Lambda )$ & \begin{tabular}{l} \dzero \cite{Abazov:2011wt}: $6.01 \pm 0.60 \pm 0.64$ \\ \end{tabular} & $6.01 \pm 0.88$ \\
\hline
\end{btocharmtab}

\begin{btocharmtab}{Bbaryon_cc_3}{Relative $\Lambda_b^{0}$ decay rates}
\hline
\textbf{Parameter} & \begin{tabular}{l}\textbf{Measurements}\end{tabular} & \textbf{Average} \\
\hline
\hline
\multicolumn{3}{|l|}{${{\cal{B}} ( \Lambda_b^{0} \to \psi(2S) \Lambda )}/{{\cal{B}} ( \Lambda_b^{0} \to J/\psi \Lambda )}$}\\
 & \begin{tabular}{l} ATLAS \cite{Aad:2015msa}: $0.501 \pm 0.033 \pm 0.019$ \\ \end{tabular} & $0.501 \pm 0.038$ \\
\hline
\multicolumn{3}{|l|}{${{\cal{B}} ( \Lambda_b^{0} \to J/\psi p \pi^{-} )}/{{\cal{B}} ( \Lambda_b^{0} \to J/\psi p K^{-} )}$}\\
 & \begin{tabular}{l} LHCb \cite{Aaij:2014zoa}: $0.0824 \pm 0.0025 \pm 0.0042$ \\ \end{tabular} & $0.0824 \pm 0.0049$ \\
\hline
\multicolumn{3}{|l|}{${{\cal{B}} ( \Lambda_b^{0} \to J/\psi \pi^{+} \pi^{-} p K^{-} )}/{{\cal{B}} ( \Lambda_b^{0} \to J/\psi p K^{-} )}$}\\
 & \begin{tabular}{l} LHCb \cite{Aaij:2016wxd}: $0.2086 \pm 0.0096 \pm 0.0134$ \\ \end{tabular} & $0.2086 \pm 0.0165$ \\
\hline
\multicolumn{3}{|l|}{${{\cal{B}} ( \Lambda_b^{0} \to \psi(2S)  p K^{-} )}/{{\cal{B}} ( \Lambda_b^{0} \to J/\psi p K^{-} )}$}\\
 & \begin{tabular}{l} LHCb \cite{Aaij:2016wxd}: $0.2070 \pm 0.0076 \pm 0.0059$ \\ \end{tabular} & $0.2070 \pm 0.0096$ \\
\hline
\end{btocharmtab}
\btocharmfig{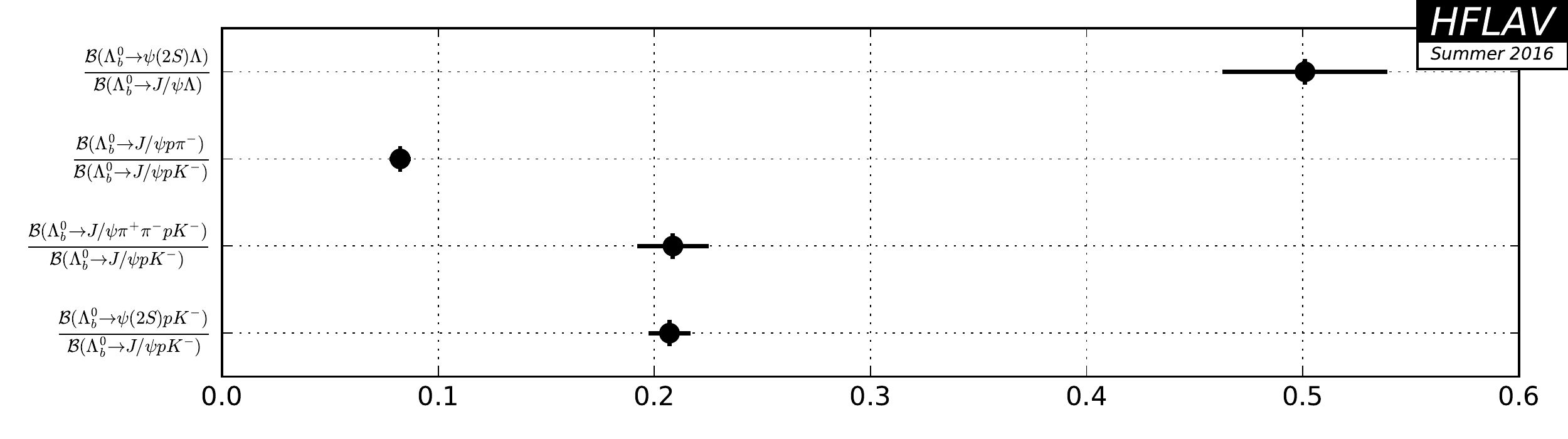}

\begin{btocharmtab}{Bbaryon_cc_4}{$\Xi_b^{-}$ and $\Omega_b^{-}$ decays to charmonium}
\hline
\textbf{Parameter} & \begin{tabular}{l}\textbf{Measurements}\end{tabular} & \textbf{Average} \\
\hline
\hline
\multicolumn{3}{|l|}{$[{\sigma(\Xi_b^{-}) \times {\cal{B}} ( \Xi_b^{-} \to J/\psi \Xi^{-} ) }]/[{ \sigma(\Lambda_b^{0}) \times {\cal{B}} ( \Lambda_b^{0} \to J/\psi \Lambda )}]$}\\
 & \begin{tabular}{l} CDF \cite{Aaltonen:2009ny}: $0.167 \,^{+0.037}_{-0.025} \pm 0.012$ \\ \end{tabular} & $0.167 \,^{+0.039}_{-0.028}$ \\
\hline
\multicolumn{3}{|l|}{$[{\sigma(\Omega_b^{-}) \times {\cal{B}} ( \Omega_b^{-} \to J/\psi \Omega^{-} ) }]/[{ \sigma(\Lambda_b^{0})  \times   {\cal{B}} ( \Lambda_b^{0} \to J/\psi \Lambda )}]$}\\
 & \begin{tabular}{l} CDF \cite{Aaltonen:2009ny}: $0.045 \,^{+0.017}_{-0.012} \pm 0.004$ \\ \end{tabular} & $0.045 \,^{+0.017}_{-0.013}$ \\
\hline
\end{btocharmtab}
\btocharmfig{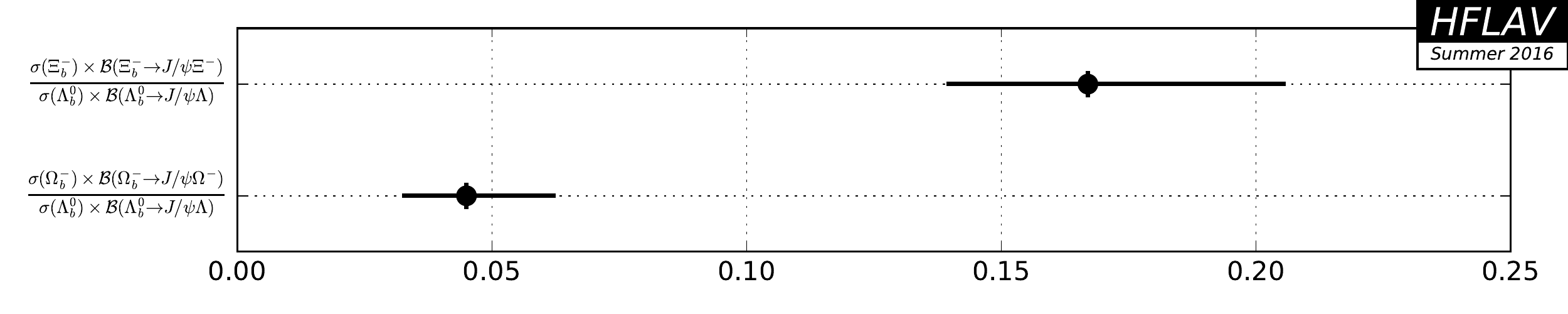}

\begin{btocharmtab}{Bbaryon_cc_5}{Parity violation in $\Lambda_b^{0}$ decays to charmonium}
\hline
\textbf{Parameter} & \begin{tabular}{l}\textbf{Measurements}\end{tabular} & \textbf{Average} \\
\hline
\hline
$\alpha_b ( \Lambda_b^{0} \to J/\psi \Lambda )$ & \begin{tabular}{l} ATLAS \cite{Aad:2014iba}: $0.30 \pm 0.16 \pm 0.06$ \\ \end{tabular} & $0.30 \pm 0.17$ \\
\hline
\end{btocharmtab}

\subsubsection{Decays to charm baryons}
\label{sec:b2c:Bbaryon_baryon}
Averages of $b$ baryons decays to charm baryons are shown in Tables~\ref{tab:b2c:Bbaryon_baryon_1}--\ref{tab:b2c:Bbaryon_baryon_5} and Figs.~\ref{fig:b2c:Bbaryon_baryon_1}--\ref{fig:b2c:Bbaryon_baryon_4}.
\begin{btocharmtab}{Bbaryon_baryon_1}{Absolute decay rates $[10^{-2}]$}
\hline
\textbf{Parameter} & \begin{tabular}{l}\textbf{Measurements}\end{tabular} & \textbf{Average} \\
\hline
\hline
${\cal{B}} ( \Lambda_b^{0} \to \Lambda_c^{+} \pi^{-} )$ & \begin{tabular}{l} LHCb \cite{Aaij:2014jyk}: $0.430 \pm 0.003 \,^{+0.036}_{-0.035}$ \\ \end{tabular} & $0.430 \,^{+0.036}_{-0.035}$ \\
\hline
${\cal{B}} ( \Lambda_b^{0} \to \Lambda_c^{+} \pi^{+} \pi^{-} \pi^{-} )$ & \begin{tabular}{l} CDF \cite{CDF:2011aa}: $2.68 \pm 0.29 \,^{+1.15}_{-1.09}$ \\ \end{tabular} & $2.68 \,^{+1.19}_{-1.12}$ \\
\hline
\end{btocharmtab}
\btocharmfig{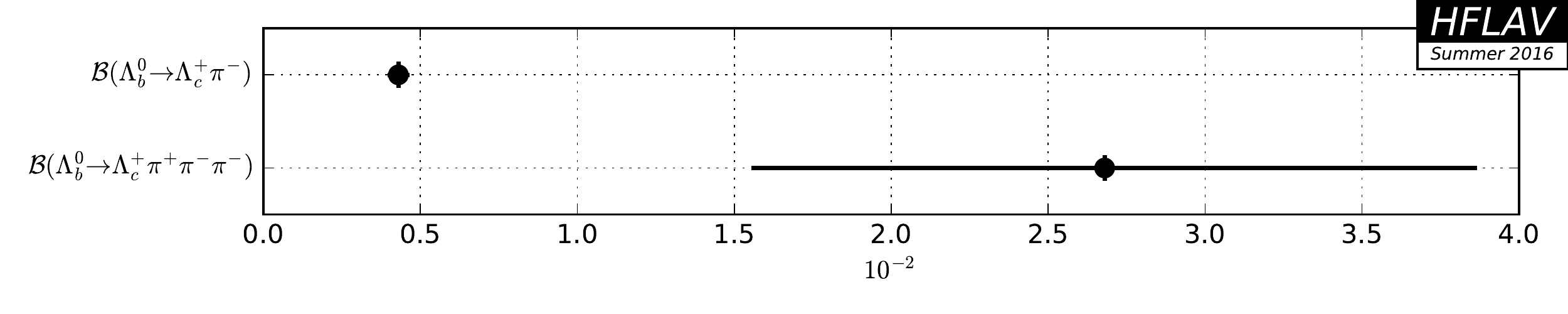}

\begin{btocharmtab}{Bbaryon_baryon_2}{Relative decay rates to $\Lambda_c$ I}
\hline
\textbf{Parameter} & \begin{tabular}{l}\textbf{Measurements}\end{tabular} & \textbf{Average} \\
\hline
\hline
\multicolumn{3}{|l|}{${{\cal{B}} ( \Lambda_b^{0} \to \Lambda_c^{+} \pi^{-} )}/{{\cal{B}} ( \bar{B}^{0} \to D^{+} \pi^{-} )}$}\\
 & \begin{tabular}{l} CDF \cite{Abulencia:2006df}: $3.3 \pm 0.3 \pm 1.2$ \\ \end{tabular} & $3.3 \pm 1.2$ \\
\hline
\multicolumn{3}{|l|}{${{\cal{B}} ( \Lambda_b^{0} \to \Lambda_c^{+} \pi^{+} \pi^{-} \pi^{-} ) }/{ {\cal{B}} ( \Lambda_b^{0} \to \Lambda_c^{+} \pi^{-} )}$}\\
 & \begin{tabular}{l} LHCb \cite{Aaij:2011rj}: $1.43 \pm 0.16 \pm 0.13$ \\ CDF \cite{CDF:2011aa}: $3.04 \pm 0.33 \,^{+0.70}_{-0.55}$ \\ \end{tabular} & $1.55 \pm 0.20$ \\
\hline
\multicolumn{3}{|l|}{$[{ {\cal{B}} ( \Xi_b^{0} \to \Lambda_c^{+} K^{-} )   \times   {\cal{B}}  (\Lambda_c^{+} \to  p K^{-} \pi^{+}) }]/[ { {\cal{B}} ( \Xi_b^{0} \to D^{0} p K^{-})  \times {\cal{B}} ( D^{0} \to K^{+} \pi^{-} ) }]$}\\
 & \begin{tabular}{l} LHCb \cite{Aaij:2013pka}: $0.57 \pm 0.22 \pm 0.21$ \\ \end{tabular} & $0.57 \pm 0.30$ \\
\hline
\end{btocharmtab}
\btocharmfig{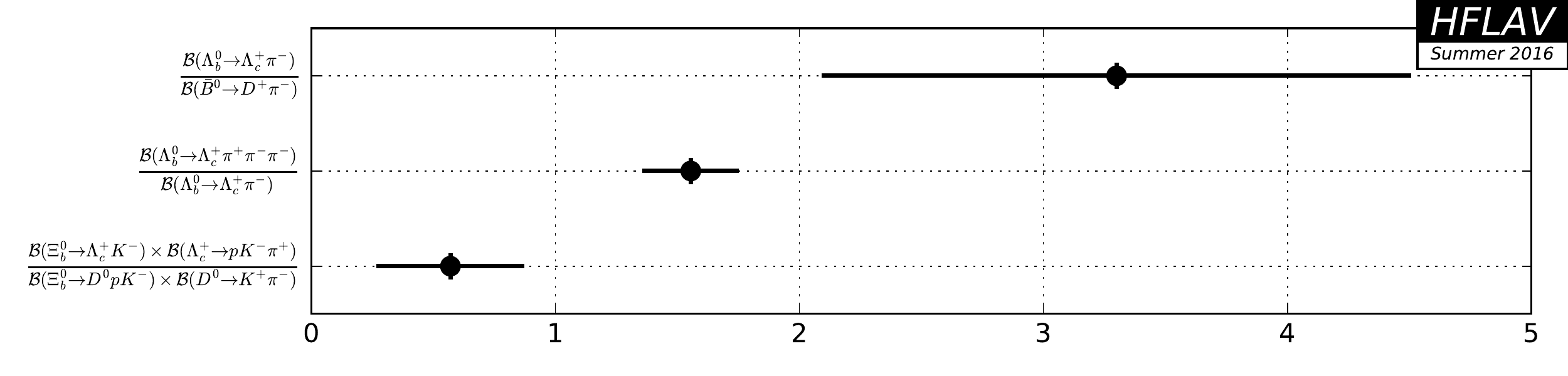}

\begin{btocharmtab}{Bbaryon_baryon_3}{Relative decay rates to $\Lambda_c$ II $[10^{-2}]$}
\hline
\textbf{Parameter} & \begin{tabular}{l}\textbf{Measurements}\end{tabular} & \textbf{Average} \\
\hline
\hline
\multicolumn{3}{|l|}{${ {\cal{B}} ( \Lambda_b^{0} \to \Lambda_c^{+} K^{-} )  }/{ {\cal{B}} ( \Lambda_b^{0} \to \Lambda_c^{+} \pi^{-} )  }$}\\
 & \begin{tabular}{l} LHCb \cite{Aaij:2013pka}: $7.31 \pm 0.16 \pm 0.16$ \\ \end{tabular} & $7.31 \pm 0.23$ \\
\hline
\multicolumn{3}{|l|}{${ {\cal{B}} ( \Lambda_b^{0} \to \Lambda_c^{+} D^{-} )  }/{ {\cal{B}} ( \Lambda_b^{0} \to \Lambda_c^{+} D_s^{-} )  }$}\\
 & \begin{tabular}{l} LHCb \cite{Aaij:2014pha}: $4.2 \pm 0.3 \pm 0.3$ \\ \end{tabular} & $4.2 \pm 0.4$ \\
\hline
\end{btocharmtab}
\btocharmfig{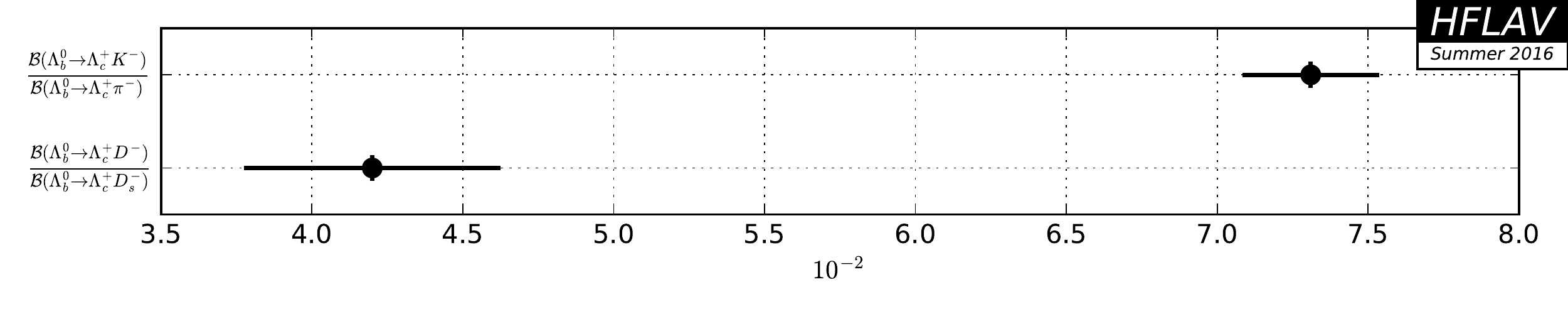}

\begin{btocharmtab}{Bbaryon_baryon_4}{Relative decay rates to excited or $\Sigma_c$ states}
\hline
\textbf{Parameter} & \begin{tabular}{l}\textbf{Measurements}\end{tabular} & \textbf{Average} \\
\hline
\hline
\multicolumn{3}{|l|}{$[{{\cal{B}} ( \Lambda_b^{0} \to \Lambda_c(2595)^{+} \pi^{-} ) \times {\cal{B}} ( \Lambda_c(2595)^{+} \to \Lambda_c^{+} \pi^{+} \pi^{-} ) }]/{ {\cal{B}} ( \Lambda_b^{0} \to \Lambda_c^{+} \pi^{+} \pi^{-} \pi^{-} )}$}\\
 & \begin{tabular}{l} LHCb \cite{Aaij:2011rj}: $0.044 \pm 0.017 \,^{+0.006}_{-0.004}$ \\ \end{tabular} & $0.044 \,^{+0.018}_{-0.017}$ \\
\hline
\multicolumn{3}{|l|}{$[{{\cal{B}} ( \Lambda_b^{0} \to \Lambda_c(2625)^{+} \pi^{-} ) \times {\cal{B}} ( \Lambda_c(2625)^{+} \to \Lambda_c^{+} \pi^{+} \pi^{-} ) }]/{ {\cal{B}} ( \Lambda_b^{0} \to \Lambda_c^{+} \pi^{+} \pi^{-} \pi^{-} )}$}\\
 & \begin{tabular}{l} LHCb \cite{Aaij:2011rj}: $0.043 \pm 0.015 \pm 0.004$ \\ \end{tabular} & $0.043 \pm 0.016$ \\
\hline
\multicolumn{3}{|l|}{$[{{\cal{B}} ( \Lambda_b^{0} \to \Sigma_c^{0} \pi^{+} \pi^{-} ) \times {\cal{B}} ( \Sigma_c^{0} \to \Lambda_c^{+} \pi^{-} ) }]/{ {\cal{B}} ( \Lambda_b^{0} \to \Lambda_c^{+} \pi^{+} \pi^{-} \pi^{-} )}$}\\
 & \begin{tabular}{l} LHCb \cite{Aaij:2011rj}: $0.074 \pm 0.024 \pm 0.012$ \\ \end{tabular} & $0.074 \pm 0.027$ \\
\hline
\multicolumn{3}{|l|}{$[{{\cal{B}} ( \Lambda_b^{0} \to \Sigma_c^{++} \pi^{-} \pi^{-} ) \times {\cal{B}} ( \Sigma_c^{++} \to \Lambda_c^{+} \pi^{+} ) }]/{ {\cal{B}} ( \Lambda_b^{0} \to \Lambda_c^{+} \pi^{+} \pi^{-} \pi^{-} )}$}\\
 & \begin{tabular}{l} LHCb \cite{Aaij:2011rj}: $0.042 \pm 0.018 \pm 0.007$ \\ \end{tabular} & $0.042 \pm 0.019$ \\
\hline
\end{btocharmtab}
\btocharmfig{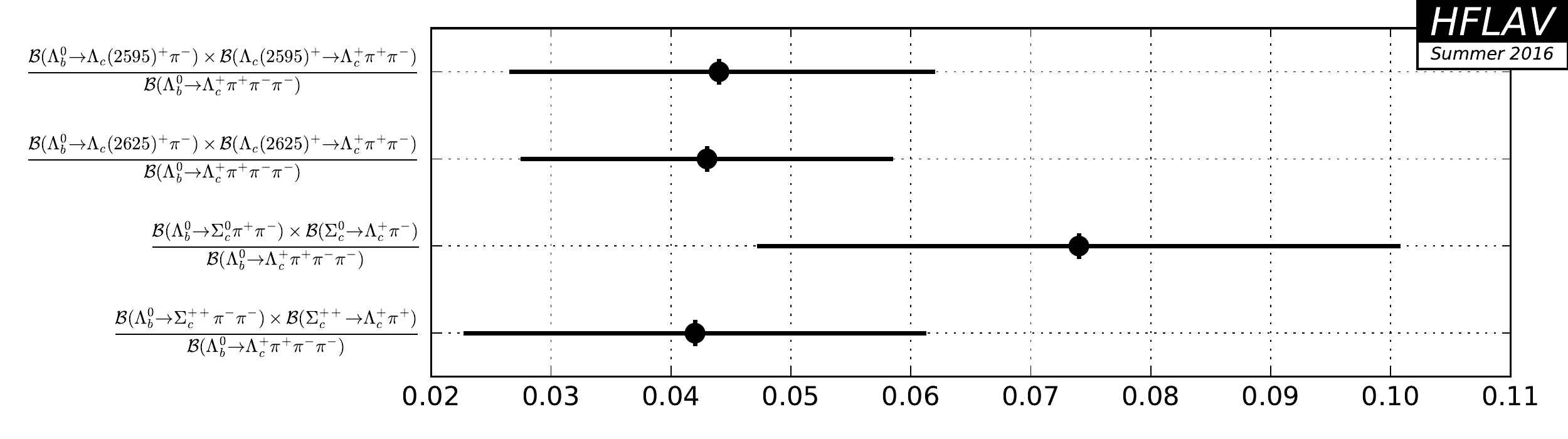}

\begin{btocharmtab}{Bbaryon_baryon_5}{$\Xi_b$ decay rates $[10^{-4}]$}
\hline
\textbf{Parameter} & \begin{tabular}{l}\textbf{Measurements}\end{tabular} & \textbf{Average} \\
\hline
\hline
\multicolumn{3}{|l|}{$[{f_{\Xi_b^{-}}}/{f_{\Lambda_b^{0}}}] \times {\cal{B}} ( \Xi_b^{-} \to \Lambda_b^{0}\pi^{-})$}\\
 & \begin{tabular}{l} LHCb \cite{Aaij:2015yoy}: $5.7 \pm 1.8 \,^{+0.8}_{-0.9}$ \\ \end{tabular} & $5.7 \,^{+2.0}_{-2.0}$ \\
\hline
\end{btocharmtab}

\clearpage
% Charmless B-decay branching fractions and their asymmetries
\mysection{$B$ decays to charmless final states}

\label{sec:rare}

%The aim of this section is to provide the branching fractions, polarization 
This section provides branching fractions (BF), polarization 
fractions, partial rate asymmetries ($A_{\CP}$) and other observables of 
% charmless $B$ decays.
$B$ decays to final states that do not contain charm hadrons or charmonia mesons.
The order of entries in the tables corresponds to that in PDG2014~\cite{PDG_2014}, and the quoted RPP numbers are the PDG numbers of the corresponding branching fractions.
The asymmetry is defined as
%\begin{equation}
%	A_{\CP} = \frac{N_{\Bbar} -N_B}{N_{\Bbar} +N_B},
%\end{equation}
%where $N_{\Bbar}$ and $N_B$ are, respectively,
%the number of $\Bzb/\Bm$ mesons (or of heavier mesons containing a $b$ quark)
%and $\Bz/\Bp$ mesons (containing a $\bar{b}$ quark) decaying into a specific final state.
\begin{equation}
	A_{\CP} = \frac{N_b - N_{\bbar}}{N_b + N_{\bbar}},
\end{equation}
where $N_b$ ($N_{\bbar}$) is the number of hadrons containing a $b$ ($\bbar$) quark 
decaying into a specific final state.
This definition is consistent with that of Eq.~(\ref{eq:cp_uta:pra}) in Sec.~\ref{sec:cp_uta:notations:pra}.
Four different $\Bz$ and $\Bp$ decay categories are considered: 
charmless mesonic (\ie, final states containing only mesons), baryonic (only hadrons, but including a baryon-antibaryon pair), radiative (including a photon or a lepton-antilepton pair) and semileptonic/leptonic (including/only leptons).
We also include measurements of $\Bs$, $\Bc$ and $b$-baryon decays.
Measurements supported with  written documents are accepted in  
the averages; written documents include journal papers, 
conference contributed papers, preprints or conference proceedings.
In all the tables of this section, values in red (blue) are new published (preliminary) results since PDG2014.  
Results from  $A_{\CP}$ measurements  obtained from time-dependent analyses 
are listed and described in Sec.~\ref{sec:cp_uta}.

%So far all
Most of the branching fractions from \babar\ and Belle assume equal production 
of charged and neutral $B$ pairs.  The best measurements to date show that this
is still a reasonable approximation (see Sec.~\ref{sec:life_mix}).
For branching fractions, we provide either averages or the most stringent
% 90\% confidence level (CL) 
upper limits. If one or more experiments have
measurements with $>\!4 \sigma$ for a decay channel, all available central values
for that channel are used in the averaging.  We also give central values
and errors for cases where the significance of the average value is at
least $3 \sigma$, even if no single measurement is above $4 \sigma$. 
%Since a few decay modes are sensitive to the contribution of
%new physics and the current experimental upper limits are not far from the 
%Standard Model expectation, we provide the combined upper limits or
%averages in these cases.
%Their upper limits can be estimated assuming that the errors are 
%Gaussian.
For $A_{\CP}$ we provide averages in all cases. 
At the end of some of the tables we give a list of results that were not
included. Typical cases are the measurements of distributions, such as differential
branching fractions or longitudinal polarizations, which are measured in different
binning schemes by the different collaborations, and thus cannot be directly
used to obtain averages.

Our averaging is performed by maximizing the likelihood,
%\begin{eqnarray}
   $\displaystyle {\mathcal L} = \prod_i {\mathcal P}_i(x),$  
%\end{eqnarray}
where ${\mathcal P_i}$ is the probability density function (PDF) of the
$i^{\rm th}$  measurement, and $x$ is, \eg, the branching fraction or $A_{\CP}$.
The PDF is modelled by an asymmetric Gaussian function with the measured
central value as its most probable value and the quadratic sum of the statistical
and systematic errors as the standard deviation. The experimental
uncertainties are considered to be uncorrelated with each other when the 
averaging is performed. As mentioned in Sec.~\ref{sec:method}, no error scaling is applied when the fit $\chi^2$ is 
greater than 1,
%since that tends to overestimate the errors
except for cases of extreme disagreement (at present we have no such cases).
% EB: we need to decide what to do with the B -> Xs gamma measurement. In the previous paper there was a dedicated section
%%One exception to consider the correlated systematic errors is the inclusive
%%$B\to X_s\gamma$ mode, which is sensitive to physics beyond the Standard Model.
%%In this update, we have included new measurements from both Belle and \babar\
%%to perform the average. The detail is  
%%described  in Sec. ~\ref{sec:btosg}. 

The largest improvement since the last report has come from the inclusion of a
variety of new measurements from the LHC, especially LHCb.  The
measurements of $\Bs$ decays are particularly noteworthy.

Sections \ref{sec:rare-charmless} and \ref{sec:rare-bary} provide compilations of branching fractions of $\Bz$ and $\Bp$ to mesonic and baryonic charmless final states, respectively, 
while Sec.~\ref{sec:rare-lb} gives branching fractions of $b$-baryon decays.
In Secs.~\ref{sec:rare-bs} and \ref{sec:rare-radll} various observables of interest are given in addition to branching fractions: in the former,
%observables related to
branching fractions of 
\Bs-meson charmless decays, and in the latter observables related to leptonic and radiative \Bz\ and \Bp\ meson decays, including processes in which the photon yields a pair of charged or neutral leptons. Section~\ref{sec:rare-radll} also reports limits from searches for lepton-flavor/number-violating decays.
Sections~\ref{sec:rare-acp} and \ref{sec:rare-polar} give \CP\ asymmetries and results of polarization measurements, respectively, in various $b$-hadron charmless decays.
Finally, Sec.~\ref{sec:rare-bc} gives branching fractions of $\Bc$ meson decays to charmless final states.
%Finally, Sec.~\ref{sec:rare-plots} shows a graphic representation of a selection of results in the chapter.

\mysubsection{Mesonic decays of \Bz and \Bp\ mesons}
\label{sec:rare-charmless}

This section provides branching fractions of charmless mesonic decays:
Tables~\ref{tab:charmless_BpFirst} to \ref{tab:charmless_BpLast} for \Bp\
and Tables~\ref{tab:charmless_BdFirst} to
\ref{tab:charmless_BdLast}
for \Bz\ mesons.
The tables are separated according to the presence or absence of kaons in the final state. 
Finally, Table~\ref{tab:charmless_Ratio} details several relative branching fractions of \Bz decays.

\begin{figure}[htbp!]
\centering
\includegraphics[width=0.40\textwidth]{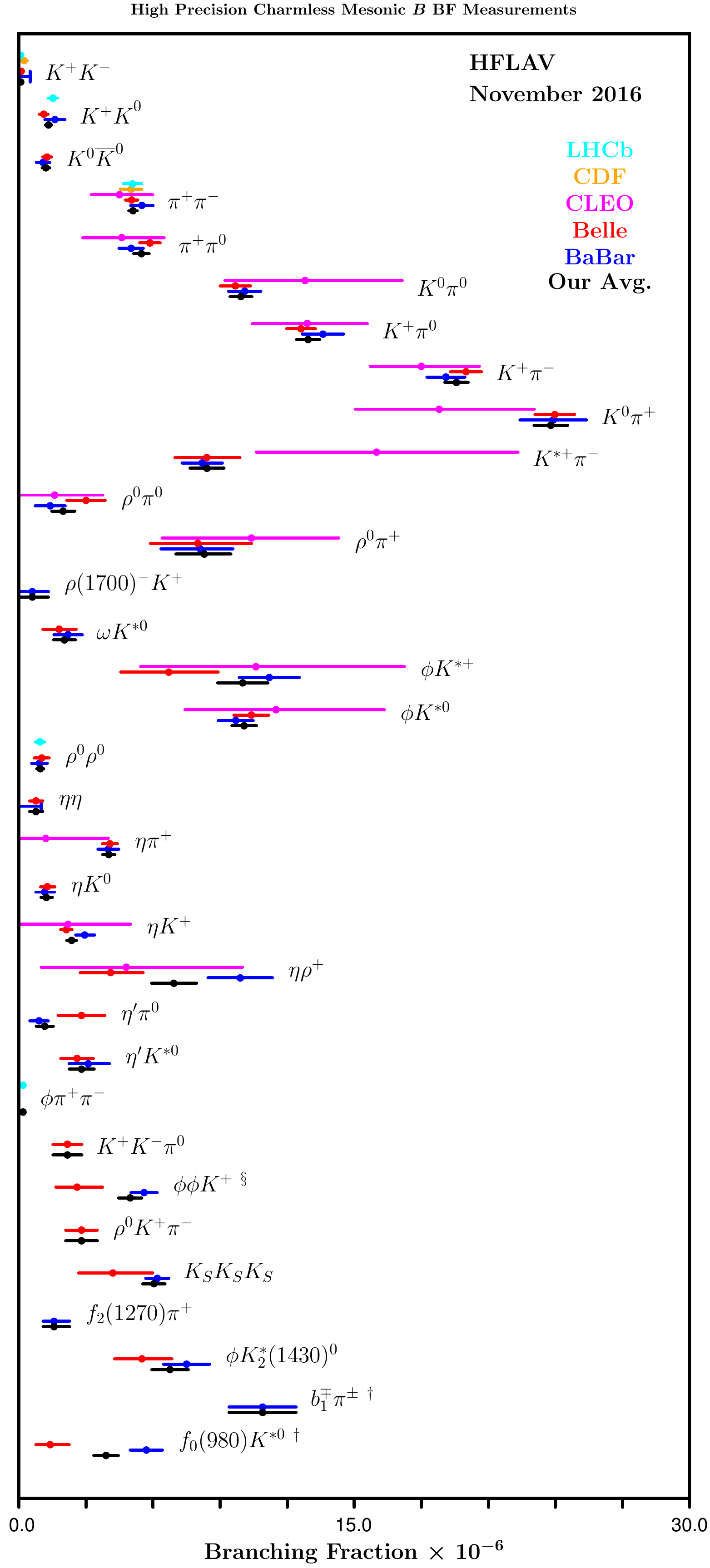}
\caption{Selection of high-precision charmless mesonic $B$ meson branching fraction measurements.}
\label{fig:rare-mostprec}
\end{figure}

Figure~\ref{fig:rare-mostprec} gives a graphic representation of a selection of high-precision branching fractions given in this section.
Footnote symbols indicate that the footnote in the corresponding table should be consulted.
%For comments in the plot, marked with a symbol or a number, refer to %the corresponding table.

\input{rare/charmless}

\clearpage

\mysubsection{Baryonic decays of \Bp\ and \Bz mesons}
\label{sec:rare-bary}

This section provides branching fractions of charmless baryonic decays of \Bp\ and \Bz mesons in Tables~\ref{tab:bary_Bp} and~\ref{tab:bary_Bz}, respectively. Relative branching fractions are given in Table~\ref{tab:bary_Ratio}.

\begin{figure}[htbp!]
\centering
\includegraphics[width=0.5\textwidth]{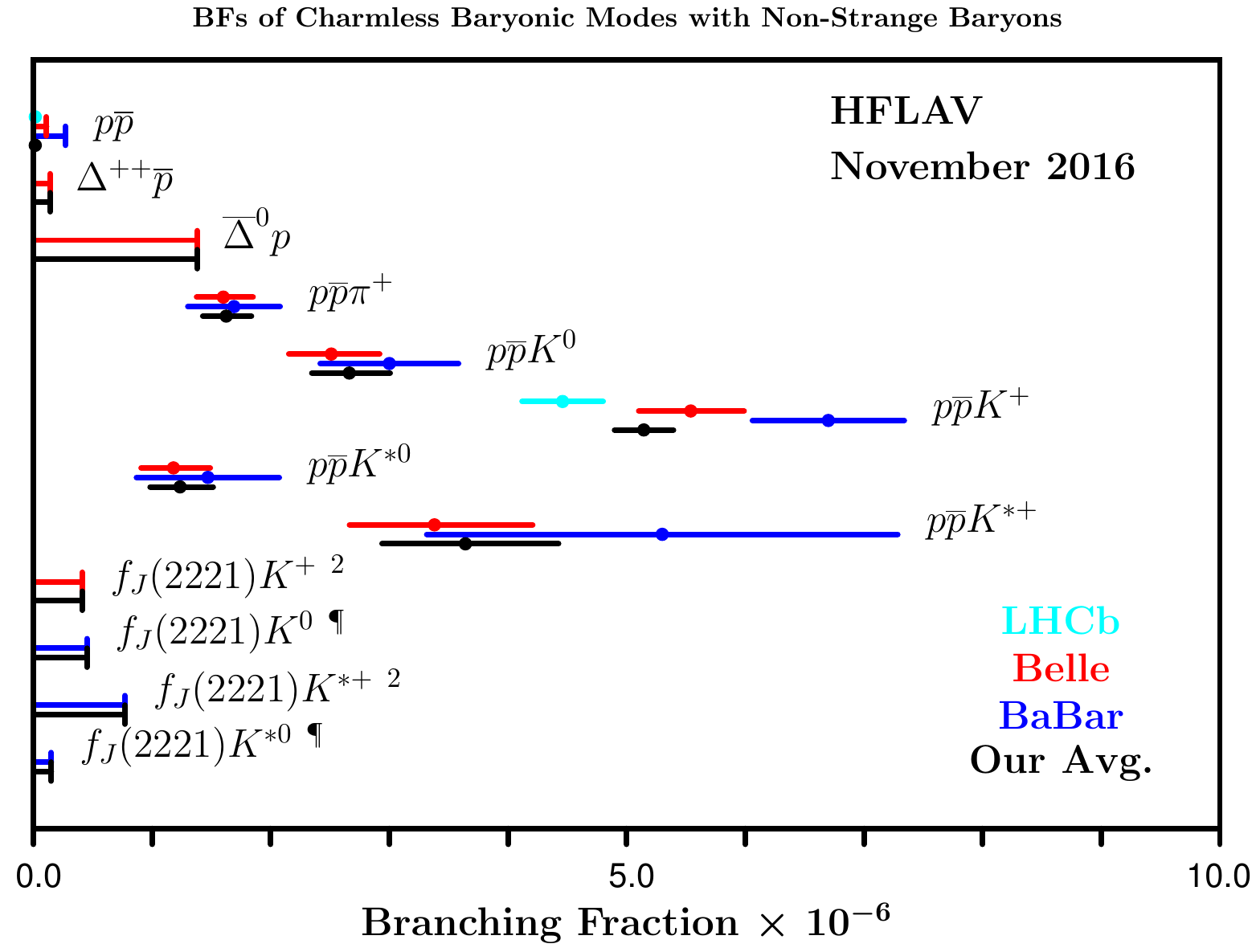}
\caption{Branching fractions of charmless baryonic modes with non-strange baryons.}
\label{fig:rare-baryns}
\end{figure}

\begin{figure}[htbp!]
\centering
\includegraphics[width=0.5\textwidth]{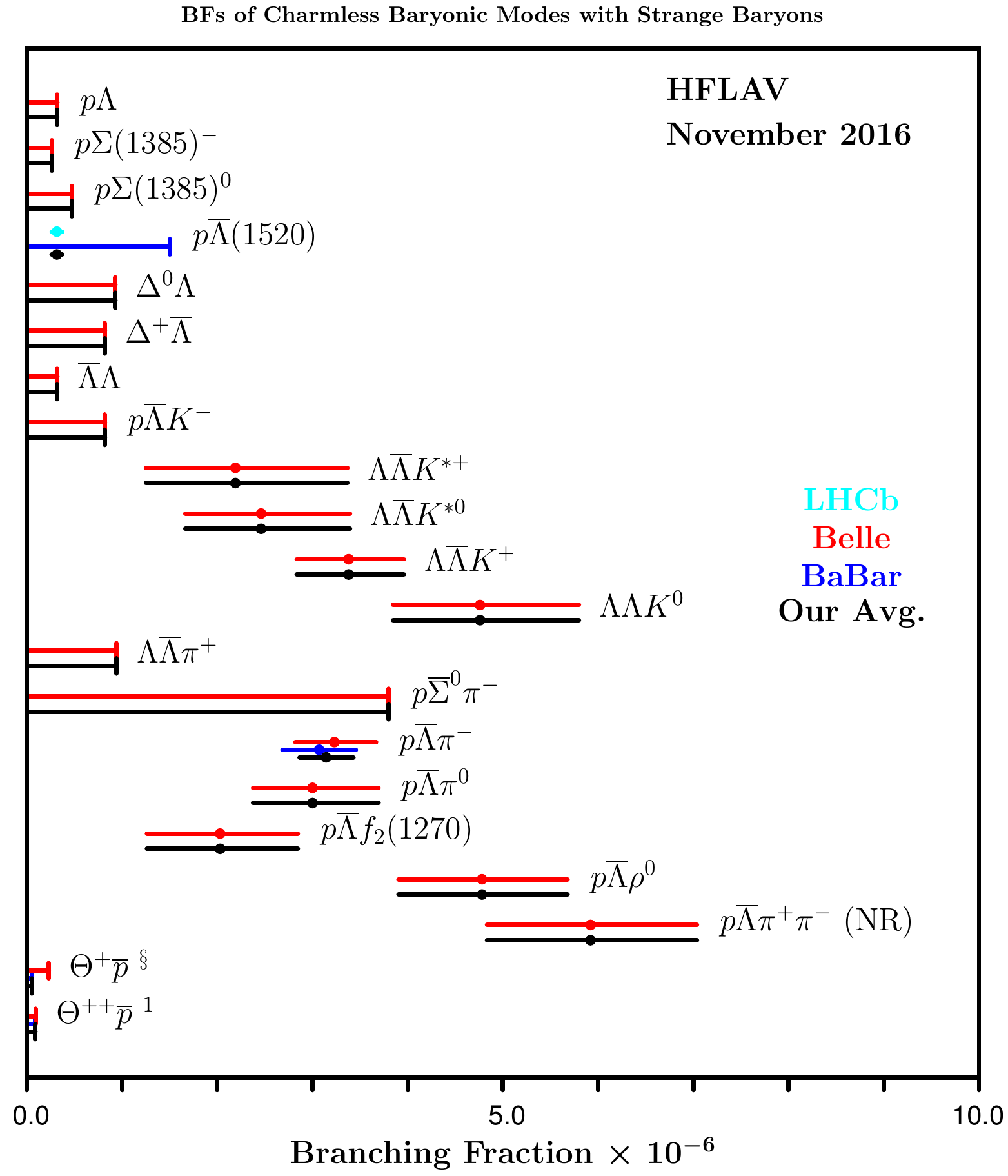}
\caption{Branching fractions of charmless baryonic modes with strange baryons.}
\label{fig:rare-bary}
\end{figure}

Figures~\ref{fig:rare-baryns} and~\ref{fig:rare-bary} show graphic representations of a selection of results given in this section.
Footnote symbols indicate that the footnote in the corresponding table should be consulted.
%For comments in the plots, marked with a symbol or a number, refer to the corresponding table.

%------------------------------------------------------------------
%Article Name Definitions
%-----------------------------------------------------
%\newcommand{\ppbar}             {\mbox{${p\bar p}$}}
%\newcommand{\pL}                {\mbox{${p \bar\Lambda}$}}
%\newcommand{\LL}                {\mbox{${\Lambda\bar\Lambda}$}}
%\newcommand{\pLpi}              {\mbox{${p\bar\Lambda\pi^-}$}}

%NEW_TABLE

\begin{table}[!htbp]
\begin{center}
\caption{Branching fractions of charmless baryonic  
$\Bp$ decays in units of $\times 10^{-6}$. Upper limits are
at 90\% CL.
Where values are shown in \red{red} (\blue{blue}), this indicates that
they are new \red{published} (\blue{preliminary}) results since PDG2014.}
%Values in \red{red} (\blue{blue}) are new \red{published}
%(\blue{preliminary}) results since PDG2014.}
\label{tab:bary_Bp}
\resizebox{\textwidth}{!}{
\begin{tabular}{|lccc @{}c c @{}c c @{}c c|}
\sgline
RPP\#   & Mode & PDG2014 Avg. & \babar & & Belle & & LHCb & & Our Avg. \\
\sgline

%TABLE_BODY
%-----------------------------------------------------
% ppbar pi
%-----------------------------------------------------
417                                               & %         RPP\#
$p \overline p \pi^+$                             & %         MODE @baryns
$1.62\pm0.20$                                     & %         PDG2014 AVG.
{$\err{1.69}{0.29}{0.26}~^\dag$}                  & %   [14]  BABAR
\ifref {\cite{Aubert:2007qea}} \fi \phantom{.}& %   [0]  
{$\aerr{1.60}{0.22}{0.19}{0.12}$}                 & %   [5]  BELLE
\ifref {\cite{Wei:2007fg}} \fi \phantom{.}& %   [0]  
\nodata                                           & %   [0]  LHCB
\phantom{.}                                       & %   [0]  
$\cerr{1.62}{0.21}{0.20}$                         \\

%-----------------------------------------------------
% ppbar pi, m(ppbar)<2.85 GeV/c^2
%-----------------------------------------------------
417                                               & %         RPP\#
$p \overline p \pi^+~^\S$                         & %         MODE @baryns
\nodata                                           & %         PDG2014 AVG.
\nodata                                           & %   [0]  BABAR
\phantom{.}                                       & %   [0]  
\nodata                                           & %   [0]  BELLE
\phantom{.}                                       & %   [0]  
\red{$\err{1.07}{0.11}{0.11}$}                    & %   [17]  LHCB
\ifref {\cite{Aaij:2014tua}} \fi \phantom{.}& %   [0]  
$1.07 \pm 0.16$                                   \\

%-----------------------------------------------------
% ppK
%-----------------------------------------------------
420                                               & %         RPP\#
$p \overline p K^+$                               & %         MODE @baryns
$5.9\pm0.5$                                       & %         PDG2014 AVG.
{$\err{6.7}{0.5}{0.4}~^\dag$}                     & %   [12]  BABAR
\ifref {\cite{Aubert:2005gw}} \fi \phantom{.}& %   [0]  
{$\aerr{5.54}{0.27}{0.25}{0.36}$}                 & %   [5]  BELLE
\ifref {\cite{Wei:2007fg}} \fi \phantom{.}& %   [0]  
\red{$\err{4.46}{0.21}{0.27}$}~$^\P$              & %   [16]  LHCB
\ifref {\cite{Aaij:2013rha}} \fi \phantom{.}& %   [0]  
$5.14 \pm 0.25$                                   \\

%-----------------------------------------------------
% B-> theta ++ pbar , theta ++ -> p K
%-----------------------------------------------------
421                                               & %         RPP\#
$\Theta^{++} \overline p$ $^1$                    & %         MODE @bary
$<0.091$                                          & %         PDG2014 AVG.
{$<0.09$}                                         & %   [12]  BABAR
\ifref {\cite{Aubert:2005gw}} \fi \phantom{.}& %   [0]  
{$<0.091$}                                        & %   [13]  BELLE
\ifref {\cite{Wang:2005fc}} \fi \phantom{.}& %   [0]  
\nodata                                           & %   [0]  LHCB
\phantom{.}                                       & %   [0]  
{$<0.09$}                                         \\

%-----------------------------------------------------
% B-> f_J(2221) K+, f_J(2221) -> p overline p
%-----------------------------------------------------
422                                               & %         RPP\#
$f_J(2221) K^+$ $^2$                              & %         MODE @baryns
$<0.41$                                           & %         PDG2014 AVG.
\nodata                                           & %   [0]  BABAR
\phantom{.}                                       & %   [0]  
{$<0.41$}                                         & %   [13]  BELLE
\ifref {\cite{Wang:2005fc}} \fi \phantom{.}& %   [0]  
\nodata                                           & %   [0]  LHCB
\phantom{.}                                       & %   [0]  
{$<0.41$}                                         \\

%-----------------------------------------------------
% p Lambdabar(1520)
%-----------------------------------------------------
423                                               & %         RPP\#
$p \overline\Lambda(1520)$                        & %         MODE @bary
$< 1.5$                                           & %         PDG2014 AVG.
{$< 1.5$}                                         & %   [12]  BABAR
\ifref {\cite{Aubert:2005gw}} \fi \phantom{.}& %   [0]  
\nodata                                           & %   [0]  BELLE
\phantom{.}                                       & %   [0]  
\red{\err{0.315}{0.048}{0.027}}                   & %   [17]  LHCB
\ifref {\cite{Aaij:2014tua}} \fi \phantom{.}& %   [0]  
$0.315 \pm 0.055$                                 \\

%-----------------------------------------------------
% ppK^*
%-----------------------------------------------------
425                                               & %         RPP\#
$p \overline p K^{*+}$                            & %         MODE @baryns
$\cerr{3.6}{0.8}{0.7}$                            & %         PDG2014 AVG.
{$\err{5.3}{1.5}{1.3}~^\dag$}                     & %   [14]  BABAR
\ifref {\cite{Aubert:2007qea}} \fi \phantom{.}& %   [0]  
{$\aerr{3.38}{0.73}{0.60}{0.39}~^\ddag$}          & %   [2]  BELLE
\ifref {\cite{Chen:2008jy}} \fi \phantom{.}& %   [0]  
\nodata                                           & %   [0]  LHCB
\phantom{.}                                       & %   [0]  
$\cerr{3.64}{0.79}{0.70}$                         \\

%-----------------------------------------------------
% B-> f_J(2221) K*+, f_J(2221) -> p overline p
%-----------------------------------------------------
426                                               & %         RPP\#
$f_J(2221) K^{*+}$ $^2$                           & %         MODE @baryns
$<0.77$                                           & %         PDG2014 AVG.
{$<0.77$}                                         & %   [14]  BABAR
\ifref {\cite{Aubert:2007qea}} \fi \phantom{.}& %   [0]  
\nodata                                           & %   [0]  BELLE
\phantom{.}                                       & %   [0]  
\nodata                                           & %   [0]  LHCB
\phantom{.}                                       & %   [0]  
{$<0.77$}                                         \\

%-----------------------------------------------------
% p Lambdabar
%-----------------------------------------------------
427                                               & %         RPP\#
$p \overline\Lambda$                              & %         MODE @bary
$<0.32$                                           & %         PDG2014 AVG.
\nodata                                           & %   [0]  BABAR
\phantom{.}                                       & %   [0]  
{$< 0.32$}                                        & %   [1]  BELLE
\ifref {\cite{Tsai:2007pp}} \fi \phantom{.}& %   [0]  
\nodata                                           & %   [0]  LHCB
\phantom{.}                                       & %   [0]  
{$< 0.32$}                                        \\

%-----------------------------------------------------
% p Lambdabar pi0
%-----------------------------------------------------
429                                               & %         RPP\#
$p \overline\Lambda \pi^0$                        & %         MODE @bary
$\cerr{3.00}{0.7}{0.6}$                           & %         PDG2014 AVG.
\nodata                                           & %   [0]  BABAR
\phantom{.}                                       & %   [0]  
{$\aerr{3.00}{0.61}{0.53}{0.33}$}                 & %   [9]  BELLE
\ifref {\cite{Wang:2007as}} \fi \phantom{.}& %   [0]  
\nodata                                           & %   [0]  LHCB
\phantom{.}                                       & %   [0]  
$\cerr{3.00}{0.69}{0.62}$                         \\

%-----------------------------------------------------
% p Sigmabar(1385)^0
%-----------------------------------------------------
430                                               & %         RPP\#
$p \overline\Sigma(1385)^0$                       & %         MODE @bary
$<0.47$                                           & %         PDG2014 AVG.
\nodata                                           & %   [0]  BABAR
\phantom{.}                                       & %   [0]  
$<0.47$                                           & %   [9]  BELLE
\ifref {\cite{Wang:2007as}} \fi \phantom{.}& %   [0]  
\nodata                                           & %   [0]  LHCB
\phantom{.}                                       & %   [0]  
$<0.47$                                           \\

%-----------------------------------------------------
% Delta^+ Lambdabar
%-----------------------------------------------------
431                                               & %         RPP\#
$\Delta^+\overline \Lambda$                       & %         MODE @bary
$<0.82$                                           & %         PDG2014 AVG.
\nodata                                           & %   [0]  BABAR
\phantom{.}                                       & %   [0]  
$<0.82$                                           & %   [9]  BELLE
\ifref {\cite{Wang:2007as}} \fi \phantom{.}& %   [0]  
\nodata                                           & %   [0]  LHCB
\phantom{.}                                       & %   [0]  
$<0.82$                                           \\

%-----------------------------------------------------
% p Lambda pi pi (n.r))
%-----------------------------------------------------
433                                               & %         RPP\#
$p \overline{\Lambda} \pi^+\pi^-$ (NR)            & %         MODE @bary
$5.9\pm1.1$                                       & %         PDG2014 AVG.
\nodata                                           & %   [0]  BABAR
\phantom{.}                                       & %   [0]  
$\aerr{5.92}{0.88}{0.84}{0.69}$                   & %   [15]  BELLE
\ifref {\cite{Chen:2009xg}} \fi \phantom{.}& %   [0]  
\nodata                                           & %   [0]  LHCB
\phantom{.}                                       & %   [0]  
$\cerr{5.92}{1.12}{1.09}$                         \\

%-----------------------------------------------------
% p Lambda rho
%-----------------------------------------------------
434                                               & %         RPP\#
$p \overline{\Lambda} \rho^0$                     & %         MODE @bary
$4.8\pm0.9$                                       & %         PDG2014 AVG.
\nodata                                           & %   [0]  BABAR
\phantom{.}                                       & %   [0]  
$\aerr{4.78}{0.67}{0.64}{0.60}$                   & %   [15]  BELLE
\ifref {\cite{Chen:2009xg}} \fi \phantom{.}& %   [0]  
\nodata                                           & %   [0]  LHCB
\phantom{.}                                       & %   [0]  
$\cerr{4.78}{0.90}{0.88}$                         \\

%-----------------------------------------------------
% p Lambda f2
%-----------------------------------------------------
435                                               & %         RPP\#
$p \overline{\Lambda} f_2(1270)$                  & %         MODE @bary
$2.0\pm0.8$                                       & %         PDG2014 AVG.
\nodata                                           & %   [0]  BABAR
\phantom{.}                                       & %   [0]  
$\aerr{2.03}{0.77}{0.72}{0.27}$                   & %   [15]  BELLE
\ifref {\cite{Chen:2009xg}} \fi \phantom{.}& %   [0]  
\nodata                                           & %   [0]  LHCB
\phantom{.}                                       & %   [0]  
$\cerr{2.03}{0.82}{0.77}$                         \\

%-----------------------------------------------------
% Lambda Lambdabar pi+
%-----------------------------------------------------
436                                               & %         RPP\#
$\Lambda \overline{\Lambda} \pi^+$                & %         MODE @bary
$<0.94$                                           & %         PDG2014 AVG.
\nodata                                           & %   [0]  BABAR
\phantom{.}                                       & %   [0]  
$<0.94~\S$                                        & %   [11]  BELLE
\ifref {\cite{Chang:2008yw}} \fi \phantom{.}& %   [0]  
\nodata                                           & %   [0]  LHCB
\phantom{.}                                       & %   [0]  
$<0.94~\S$                                        \\

%-----------------------------------------------------
% Lambda Lambdabar K+
%-----------------------------------------------------
437                                               & %         RPP\#
$\Lambda \overline{\Lambda} K^+$                  & %         MODE @bary
$3.4\pm0.6$                                       & %         PDG2014 AVG.
\nodata                                           & %   [0]  BABAR
\phantom{.}                                       & %   [0]  
$\aerr{3.38}{0.41}{0.36}{0.41}~^\ddag$            & %   [11]  BELLE
\ifref {\cite{Chang:2008yw}} \fi \phantom{.}& %   [0]  
\nodata                                           & %   [0]  LHCB
\phantom{.}                                       & %   [0]  
$\cerr{3.38}{0.58}{0.55}$                         \\

%-----------------------------------------------------
% Lambda Lambdabar K*+
%-----------------------------------------------------
438                                               & %         RPP\#
$\Lambda \overline{\Lambda} K^{*+}$               & %         MODE @bary
$\cerr{2.2}{1.2}{0.9}$                            & %         PDG2014 AVG.
\nodata                                           & %   [0]  BABAR
\phantom{.}                                       & %   [0]  
$\aerr{2.19}{1.13}{0.88}{0.33}~^\S$               & %   [11]  BELLE
\ifref {\cite{Chang:2008yw}} \fi \phantom{.}& %   [0]  
\nodata                                           & %   [0]  LHCB
\phantom{.}                                       & %   [0]  
$\cerr{2.19}{1.18}{0.94}$                         \\

%-------------------------------------------------------------------------
%p Delta \#
%-----------------------------------------------------
439                                               & %         RPP\#
$\overline{\Delta}^0 p$                           & %         MODE @baryns
$<1.38$                                           & %         PDG2014 AVG.
\nodata                                           & %   [0]  BABAR
\phantom{.}                                       & %   [0]  
{$<1.38$} $^\S$                                   & %   [5]  BELLE
\ifref {\cite{Wei:2007fg}} \fi \phantom{.}& %   [0]  
\nodata                                           & %   [0]  LHCB
\phantom{.}                                       & %   [0]  
{$<1.38$} $^\S$                                   \\

%--------------------------------------------------------------------------
%pbar Delta++ \#
%-----------------------------------------------------
440                                               & %         RPP\#
$\Delta^{++} \overline p$                         & %         MODE @baryns
$<0.14$                                           & %         PDG2014 AVG.
\nodata                                           & %   [0]  BABAR
\phantom{.}                                       & %   [0]  
{$<0.14$} $^\S$                                   & %   [5]  BELLE
\ifref {\cite{Wei:2007fg}} \fi \phantom{.}& %   [0]  
\nodata                                           & %   [0]  LHCB
\phantom{.}                                       & %   [0]  
{$<0.14$} $^\S$                                   \\

%TABLE_BODY

\hline
\end{tabular}
}
\end{center}
\scriptsize
Results for LHCb are relative BFs converted to absolute BFs.\\    % FOOTNOTE
$^\dag$ Charmonium decays to $p\bar p$ have been statistically subtracted.\\    % FOOTNOTE
$^\ddag$ The charmonium mass region has been vetoed.\\    % FOOTNOTE
$^\S$~Di-baryon mass is less than $2.85~\gevcc$.\\    % FOOTNOTE
$^\P$~Includes contribution where $p \bar{p}$ is produced in charmonia decays.\\    % FOOTNOTE
$^1$~$\Theta(1540)^{++}\to K^+p$ (pentaquark candidate). \\    % FOOTNOTE
$^2$~In this product of BFs, all daughter BFs not shown are set to 100\%.     % FOOTNOTE
%$^2$ Product BFs --- daughter BFs taken to be 100\%.     % FOOTNOTE
\end{table}
\clearpage

%NEW_TABLE

\begin{table}[!htbp]
\begin{center}
\caption{Branching fractions of charmless baryonic  
$\Bz$ decays in units of $\times 10^{-6}$. Upper limits are
at 90\% CL. 
Where values are shown in \red{red} (\blue{blue}), this indicates that
they are new \red{published} (\blue{preliminary}) results since PDG2014.}
%Values in \red{red} (\blue{blue}) are new \red{published}
%(\blue{preliminary}) results since PDG2014.}
\label{tab:bary_Bz}
\resizebox{\textwidth}{!}{
\begin{tabular}{|lccc @{}c c @{}c c @{}c c|}
\sgline
RPP\#   & Mode & PDG2014 Avg. & \babar & & Belle & & LHCb & & Our Avg. \\
\sgline

%TABLE_BODY
%-----------------------------------------------------
% p pbar
%-----------------------------------------------------
407                                               & %         RPP\#
$p \overline{p}$                                  & %         MODE @baryns
$\cerr{0.015}{0.007}{0.005}$                      & %         PDG2014 AVG.
{$<0.27$}                                         & %   [10]  BABAR
\ifref {\cite{Aubert:2004fy}} \fi \phantom{.}& %   [0]  
{$<0.11$}                                         & %   [1]  BELLE
\ifref {\cite{Tsai:2007pp}} \fi \phantom{.}& %   [0]  
$\aerrsy{0.0147}{0.0062}{0.0051}{0.0035}{0.0014}$ & %   [18]  LHCB
\ifref {\cite{Aaij:2013fta}} \fi \phantom{.}& %   [0]  
$\cerr{0.0150}{0.0070}{0.0050}$                   \\

%-----------------------------------------------------
% p p K^0
%-----------------------------------------------------
409                                               & %         RPP\#
$p \overline{p} K^0$                              & %         MODE @baryns
$2.66\pm0.32$                                     & %         PDG2014 AVG.
{$\err{3.0}{0.5}{0.3}~^\dag$}                     & %   [14]  BABAR
\ifref {\cite{Aubert:2007qea}} \fi \phantom{.}& %   [0]  
$\aerr{2.51}{0.35}{0.29}{0.21}~^\ddag$            & %   [2]  BELLE
\ifref {\cite{Chen:2008jy}} \fi \phantom{.}& %   [0]  
\nodata                                           & %   [0]  LHCB
\phantom{.}                                       & %   [0]  
$\cerr{2.66}{0.34}{0.32}$                         \\

%-----------------------------------------------------
% \theta+ pbar
%-----------------------------------------------------
410                                               & %         RPP\#
$\Theta^+ \overline{p}$~$^\S$                     & %         MODE @bary
$<0.05$                                           & %         PDG2014 AVG.
{$<0.05$}                                         & %   [14]  BABAR
\ifref {\cite{Aubert:2007qea}} \fi \phantom{.}& %   [0]  
{$<0.23$}                                         & %   [13]  BELLE
\ifref {\cite{Wang:2005fc}} \fi \phantom{.}& %   [0]  
\nodata                                           & %   [0]  LHCB
\phantom{.}                                       & %   [0]  
{$<0.05$}                                         \\

%-----------------------------------------------------
% B-> f_J(2221) K0, f_J(2221) -> p overline p
%-----------------------------------------------------
411                                               & %         RPP\#
$f_J(2221) K^0$~$^\P$                             & %         MODE @baryns
$<0.45$                                           & %         PDG2014 AVG.
{$<0.45$}                                         & %   [14]  BABAR
\ifref {\cite{Aubert:2007qea}} \fi \phantom{.}& %   [0]  
\nodata                                           & %   [0]  BELLE
\phantom{.}                                       & %   [0]  
\nodata                                           & %   [0]  LHCB
\phantom{.}                                       & %   [0]  
{$<0.45$}                                         \\

%-----------------------------------------------------
% p p K^*0
%-----------------------------------------------------
412                                               & %         RPP\#
$p \overline{p} K^{*0}$                           & %         MODE @baryns
$\cerr{1.24}{0.28}{0.25}$                         & %         PDG2014 AVG.
{$\err{1.47}{0.45}{0.40}~^\dag$}                  & %   [14]  BABAR
\ifref {\cite{Aubert:2007qea}} \fi \phantom{.}& %   [0]  
$\aerr{1.18}{0.29}{0.25}{0.11}~^\ddag$            & %   [2]  BELLE
\ifref {\cite{Chen:2008jy}} \fi \phantom{.}& %   [0]  
\nodata                                           & %   [0]  LHCB
\phantom{.}                                       & %   [0]  
$\cerr{1.24}{0.28}{0.25}$                         \\

%-----------------------------------------------------
% B-> f_J(2221) K*0, f_J(2221) -> p overline p
%-----------------------------------------------------
413                                               & %         RPP\#
$f_J(2221) K^{*0}$~$^\P$                          & %         MODE @baryns
$<0.15$                                           & %         PDG2014 AVG.
{$<0.15$}                                         & %   [14]  BABAR
\ifref {\cite{Aubert:2007qea}} \fi \phantom{.}& %   [0]  
\nodata                                           & %   [0]  BELLE
\phantom{.}                                       & %   [0]  
\nodata                                           & %   [0]  LHCB
\phantom{.}                                       & %   [0]  
{$<0.15$}                                         \\

%-----------------------------------------------------
% p Lambdabar pi-
%-----------------------------------------------------
414                                               & %         RPP\#
$p \overline\Lambda \pi^-$                        & %         MODE @bary
$3.14\pm0.29$                                     & %         PDG2014 AVG.
$\err{3.07}{0.31}{0.23}$                          & %   [4]  BABAR
\ifref {\cite{Aubert:2009am}} \fi \phantom{.}& %   [0]  
{$\aerr{3.23}{0.33}{0.29}{0.29}$}                 & %   [9]  BELLE
\ifref {\cite{Wang:2007as}} \fi \phantom{.}& %   [0]  
\nodata                                           & %   [0]  LHCB
\phantom{.}                                       & %   [0]  
$\cerr{3.14}{0.29}{0.28}$                         \\

%-----------------------------------------------------
% p Sigmabar(1385)^-
%-----------------------------------------------------
415                                               & %         RPP\#
$p \overline\Sigma(1385)^-$                       & %         MODE @bary
$<0.26$                                           & %         PDG2014 AVG.
\nodata                                           & %   [0]  BABAR
\phantom{.}                                       & %   [0]  
$<0.26$                                           & %   [9]  BELLE
\ifref {\cite{Wang:2007as}} \fi \phantom{.}& %   [0]  
\nodata                                           & %   [0]  LHCB
\phantom{.}                                       & %   [0]  
$<0.26$                                           \\

%-----------------------------------------------------
% Delta^0 Lambdabar
%-----------------------------------------------------
416                                               & %         RPP\#
$\Delta^0 \overline\Lambda$                       & %         MODE @bary
$<0.93$                                           & %         PDG2014 AVG.
\nodata                                           & %   [0]  BABAR
\phantom{.}                                       & %   [0]  
$<0.93$                                           & %   [9]  BELLE
\ifref {\cite{Wang:2007as}} \fi \phantom{.}& %   [0]  
\nodata                                           & %   [0]  LHCB
\phantom{.}                                       & %   [0]  
$<0.93$                                           \\

%-----------------------------------------------------
% p Lambdabar K-
%-----------------------------------------------------
417                                               & %         RPP\#
$p \overline\Lambda K^-$                          & %         MODE @bary
$<0.82$                                           & %         PDG2014 AVG.
\nodata                                           & %   [0]  BABAR
\phantom{.}                                       & %   [0]  
$< 0.82$                                          & %   [3]  BELLE
\ifref {\cite{Wang:2003yi}} \fi \phantom{.}& %   [0]  
\nodata                                           & %   [0]  LHCB
\phantom{.}                                       & %   [0]  
$< 0.82$                                          \\

%-----------------------------------------------------
% p Sigmabar pi-
%-----------------------------------------------------
418                                               & %         RPP\#
$p \overline\Sigma^0 \pi^-$                       & %         MODE @bary
$<3.8$                                            & %         PDG2014 AVG.
\nodata                                           & %   [0]  BABAR
\phantom{.}                                       & %   [0]  
$< 3.8$                                           & %   [3]  BELLE
\ifref {\cite{Wang:2003yi}} \fi \phantom{.}& %   [0]  
\nodata                                           & %   [0]  LHCB
\phantom{.}                                       & %   [0]  
$< 3.8$                                           \\

%-----------------------------------------------------
% Lambda Lambdabar
%-----------------------------------------------------
419                                               & %         RPP\#
$\overline\Lambda \Lambda$                        & %         MODE @bary
$<0.32$                                           & %         PDG2014 AVG.
\nodata                                           & %   [0]  BABAR
\phantom{.}                                       & %   [0]  
{$<0.32$}                                         & %   [1]  BELLE
\ifref {\cite{Tsai:2007pp}} \fi \phantom{.}& %   [0]  
\nodata                                           & %   [0]  LHCB
\phantom{.}                                       & %   [0]  
{$<0.32$}                                         \\

%-----------------------------------------------------
% Lambda Lambdabar K^0
%-----------------------------------------------------
420                                               & %         RPP\#
$\overline\Lambda \Lambda K^0$                    & %         MODE @bary
$\cerr{4.8}{1.0}{0.9}$                            & %         PDG2014 AVG.
\nodata                                           & %   [0]  BABAR
\phantom{.}                                       & %   [0]  
$\aerr{4.76}{0.84}{0.68}{0.61}~^\ddag$            & %   [11]  BELLE
\ifref {\cite{Chang:2008yw}} \fi \phantom{.}& %   [0]  
\nodata                                           & %   [0]  LHCB
\phantom{.}                                       & %   [0]  
$\cerr{4.76}{1.04}{0.91}$                         \\

%-----------------------------------------------------
% Lambda Lambdabar K^*0
%-----------------------------------------------------
421                                               & %         RPP\#
$\Lambda \overline{\Lambda} K^{*0}$               & %         MODE @bary
$\cerr{2.5}{0.9}{0.8}$                            & %         PDG2014 AVG.
\nodata                                           & %   [0]  BABAR
\phantom{.}                                       & %   [0]  
$\aerr{2.46}{0.87}{0.72}{0.34}~^\ddag$            & %   [11]  BELLE
\ifref {\cite{Chang:2008yw}} \fi \phantom{.}& %   [0]  
\nodata                                           & %   [0]  LHCB
\phantom{.}                                       & %   [0]  
$\cerr{2.46}{0.93}{0.80}$                         \\

%TABLE_BODY

\hline
\end{tabular}
}
\end{center}
\scriptsize
Results for LHCb are relative BFs converted to absolute BFs.\\    % FOOTNOTE
$^\dag$ Charmonium decays to $p\bar p$ have been statistically subtracted.\\    % FOOTNOTE
$^\ddag$ The charmonium mass region has been vetoed.\\    % FOOTNOTE
$^\S~\Theta(1540)^+\to p K^0$ (pentaquark candidate).\\    % FOOTNOTE
$^\P$~In this product of BFs, all daughter BFs not shown are set to 100\%.     % FOOTNOTE
%$^\P$ Product BF --- daughter BFs taken to be 100\%.    % FOOTNOTE
\end{table}

%NEW_TABLE

\begin{table}[!htbp]
\begin{center}
\caption{Relative branching fractions of charmless baryonic  
$\B$ decays.
Where values are shown in \red{red} (\blue{blue}), this indicates that
they are new \red{published} (\blue{preliminary}) results since PDG2014.}
%Values in \red{red} (\blue{blue}) are new \red{published}
%(\blue{preliminary}) results since PDG2014.}
\label{tab:bary_Ratio}
\resizebox{\textwidth}{!}{
\begin{tabular}{|lccc @{}c c|} \hline
RPP\# & Mode & PDG2014 Avg. & LHCb & & Our Avg.  \\ \sglinespb
%TABLE_BODY
$417$                                             & %         RPP\#
$\mathcal{B}(B^+\rightarrow p \overline p \pi^+, m_{p \overline p}<2.85~\gevcc)/\mathcal{B}(B^+\rightarrow J/\psi(\to p \bar{p})\pi^+)$& %         MODE
\nodata                                           & %         PDG2014 AVG.
\red{$\err{12.0}{1.2}{0.3}$}                      & %   [17]  LHCB
\ifref {\cite{Aaij:2014tua}} \fi \phantom{.}& %   [0]  
$12.0 \pm 1.2$                                    \\

$420$                                             & %         RPP\#
$\mathcal{B}(B^+\rightarrow p \overline p K^+)/\mathcal{B}(B^+\rightarrow J/\psi(\to p \bar{p})K^+)$& %         MODE
\nodata                                           & %         PDG2014 AVG.
$\err{4.91}{0.19}{0.14}~^\dag$                    & %   [16]  LHCB
\ifref {\cite{Aaij:2013rha}} \fi \phantom{.}& %   [0]  
$4.91 \pm 0.24$                                   \\

$420$                                             & %         RPP\#
$\mathcal{B}(B^+\rightarrow p \overline p K^+)/\mathcal{B}(B^+\rightarrow J/\psi K^+)$& %         MODE
$\err{0.0104}{0.0005}{0.0001}$                    & %         PDG2014 AVG.
$\err{0.0104}{0.0005}{0.0001}$~$^{\dag\ddag}$     & %   [16]  LHCB
\ifref {\cite{Aaij:2013rha}} \fi \phantom{.}& %   [0]  
$0.0104 \pm 0.0005$                    \\

$423$                                             & %         RPP\#
$\mathcal{B}(B^+\rightarrow \overline\Lambda(1520)(\to K^+\bar{p}) p)/\mathcal{B}(B^+\rightarrow J/\psi(\to p \bar{p})\pi^+)$& %         MODE
\nodata                                           & %         PDG2014 AVG.
\red{$\err{0.033}{0.005}{0.007}$}                 & %   [17]  LHCB
\ifref {\cite{Aaij:2014tua}} \fi \phantom{.}& %   [0]  
$0.033 \pm 0.009$                                 \\

%TABLE_BODY

\sglinespt
\end{tabular}
}
\end{center}
\scriptsize
$^\dag$~Includes contribution where $p \bar{p}$ is produced in charmonia decays.\\
$^\ddag$~Original experimental relative BF multiplied by the best values (PDG2014) of certain reference BFs. The first error is experimental, and the second is from the reference BFs.\\
\end{table}

%\newpage
\clearpage

\mysubsection{Decays of \b baryons}
\label{sec:rare-lb}

A compilation of branching fractions of $\Lb$ baryon decays is given in Table~\ref{tab:bbaryons_Lb}. Table~\ref{tab:bbaryons_LbPartialBF} provides the partial branching fractions of $\Lb\to\Lambda\mup\mun$ decays. A compilation of branching fractions of $\Xi^{0}_{b}$  baryon decays is given in Table~\ref{tab:bbaryons_Xhib}.

\begin{figure}[htbp!]
\centering
\includegraphics[width=0.45\textwidth]{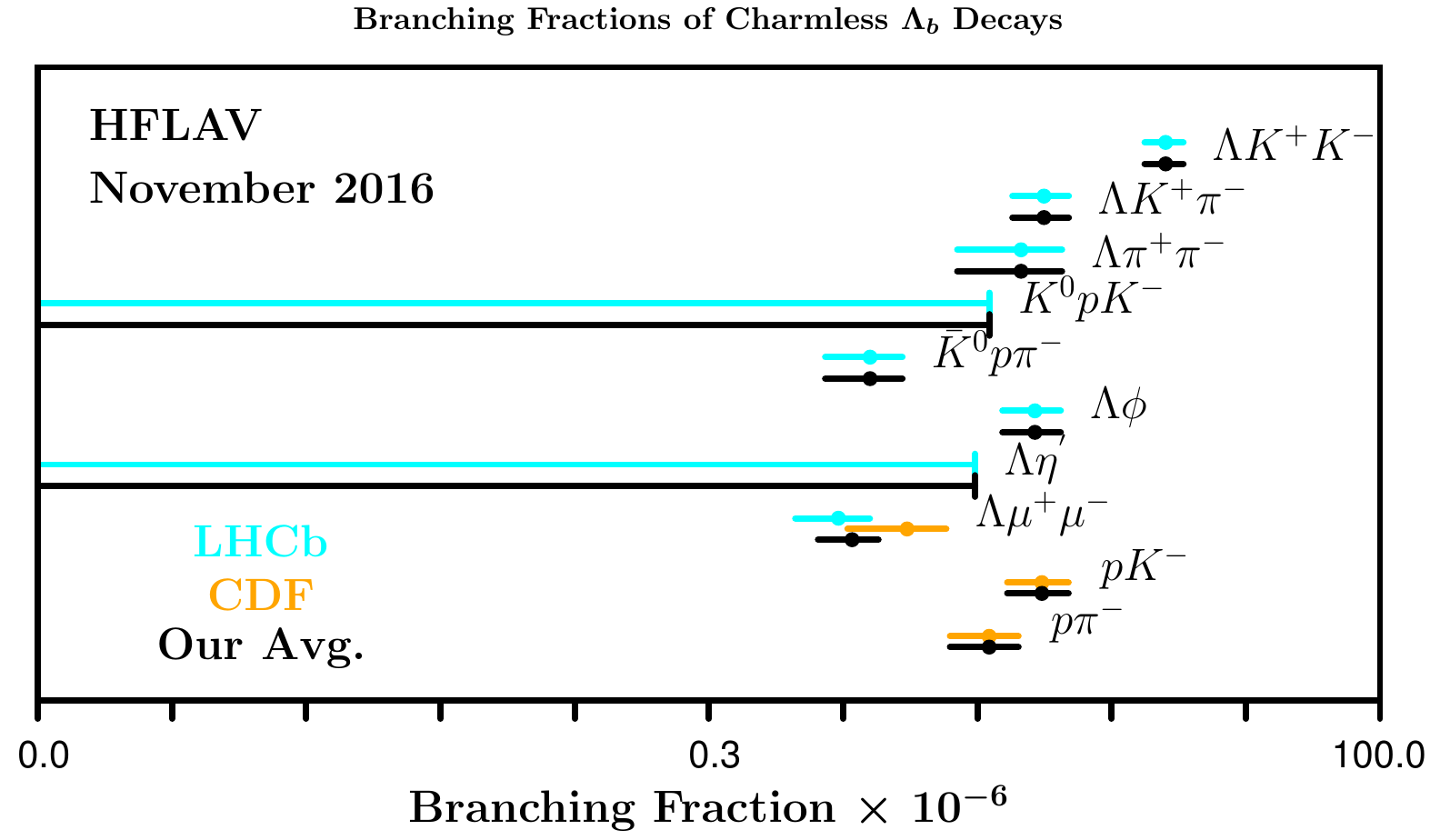}
\caption{Branching fractions of charmless $\Lb$ decays.}
\label{fig:rare-lb}
\end{figure}

Figure~\ref{fig:rare-lb} shows a graphic representation of branching fractions of $\Lb$ decays.
Footnote symbols indicate that the footnote in the corresponding table should be consulted.
%For comments in the plot, marked with a symbol or a number, refer to the corresponding table.

%NEW_TABLE

\begin{table}[h!]
\begin{center}
\caption{Branching fractions of charmless $\Lb$ decays in units of
$\times 10^{-6}$.
Upper limits are at 90\% CL.
Where values are shown in \red{red} (\blue{blue}), this indicates that
they are new \red{published} (\blue{preliminary}) results since PDG2014.}
%Values in \red{red} (\blue{blue}) are new \red{published}
%(\blue{preliminary}) results since PDG2014.}
\label{tab:bbaryons_Lb}
\resizebox{\textwidth}{!}{
\begin{tabular}{|lccc @{}c c @{}c c|} \hline
RPP\# &Mode & PDG2014 Avg. & CDF & & LHCb & & Our Avg.  \\ \sglinespb
%TABLE_BODY
$~19$                                             & %         RPP\#
$p\pi^-$                                          & %         MODE @lb
$\err{3.5}{0.8}{0.6}$                             & %         PDG2014 AVG.
$\err{3.5}{0.8}{0.6}$                             & %   [1]  CDF
\ifref {\cite{Aaltonen:2008hg}} \fi \phantom{.}& %   [0]  
\nodata                                           & %   [0]  LHCB
\phantom{.}                                       & %   [0]  
$3.5 \pm 1.0$                                     \\

$~20$                                             & %         RPP\#
$p K^-$                                           & %         MODE @lb
$\err{5.5}{1.0}{1.0}$                             & %         PDG2014 AVG.
$\err{5.5}{1.0}{1.0}$                             & %   [1]  CDF
\ifref {\cite{Aaltonen:2008hg}} \fi \phantom{.}& %   [0]  
\nodata                                           & %   [0]  LHCB
\phantom{.}                                       & %   [0]  
$5.5 \pm 1.4$                                     \\

$~21$                                             & %         RPP\#
$\Lambda \mu^+\mu^-$                              & %         MODE @lb
$\err{1.73}{0.42}{0.55}$                          & %         PDG2014 AVG.
$\err{1.73}{0.42}{0.55}$                          & %   [2]  CDF
\ifref {\cite{Aaltonen:2011qs}} \fi \phantom{.}& %   [0]  
\red{$\err{0.96}{0.16}{0.25}$}                    & %   [3]  LHCB
\ifref {\cite{Aaij:2013mna}} \fi \phantom{.}& %   [0]  
$1.08 \pm 0.27$                                   \\

\nodata                                           & %         RPP\#
$\Lambda \eta$                                    & %         MODE @lb
\nodata                                           & %         PDG2014 AVG.
\nodata                                           & %   [2]  CDF
\phantom{.}                                       & %   [0]  
\red{$\cerr{9.3}{7.3}{5.3}$}~$^\P$                & %   [6]  LHCB
\ifref {\cite{Aaij:2015eqa}} \fi \phantom{.}& %   [0]  
$\cerr{9.3}{7.3}{5.3}$                            \\

\nodata                                           & %         RPP\#
$\Lambda \etapr$                                & %         MODE @lb
\nodata                                           & %         PDG2014 AVG.
\nodata                                           & %   [2]  CDF
\phantom{.}                                       & %   [0]  
\red{$<3.1$}                                      & %   [6]  LHCB
\ifref {\cite{Aaij:2015eqa}} \fi \phantom{.}& %   [0]  
{$<3.1$}                                          \\

\nodata                                           & %         RPP\#
$\Lambda \phi$                                    & %         MODE @lb
\nodata                                           & %         PDG2014 AVG.
\nodata                                           & %   [2]  CDF
\phantom{.}                                       & %   [0]  
\red{$\derrsyt{5.18}{1.04}{0.35}{0.67}{0.62}$}~$^\ddag$& %   [7]  LHCB
\ifref {\cite{Aaij:2016zhm}} \fi \phantom{.}& %   [0]  
$\cerr{5.18}{1.29}{1.26}$                         \\

\nodata                                           & %         RPP\#
$\bar{K}^0 p \pi^{-}$                             & %         MODE @lb
\nodata                                           & %         PDG2014 AVG.
\nodata                                           & %   [2]  CDF
\phantom{.}                                       & %   [0]  
\red{$\ferrsyt{1.26}{0.19}{0.09}{0.34}{0.05}$}~$^\S$& %   [5]  LHCB
\ifref {\cite{Aaij:2014lpa}} \fi \phantom{.}& %   [0]  
$1.26 \pm 0.40$                                   \\

%ADD CP ASYMMETRY TOO!!!!
\nodata                                           & %         RPP\#
$K^0 p K^{-}$                                     & %         MODE @lb
\nodata                                           & %         PDG2014 AVG.
\nodata                                           & %   [2]  CDF
\phantom{.}                                       & %   [0]  
\red{$<3.5$}                                      & %   [5]  LHCB
\ifref {\cite{Aaij:2014lpa}} \fi \phantom{.}& %   [0]  
{$<3.5$}                                          \\

\nodata                                           & %         RPP\#
$\Lambda \pi^+\pi^-$                              & %         MODE @lb
\nodata                                           & %         PDG2014 AVG.
\nodata                                           & %   [2]  CDF
\phantom{.}                                       & %   [0]  
\red{$\gerrsyt{4.6}{1.2}{1.4}{0.6}$}~$^\dag$      & %   [4]  LHCB
\ifref {\cite{Aaij:2016nrq}} \fi \phantom{.}& %   [0]  
$4.6 \pm 1.9$                                     \\

\nodata                                           & %         RPP\#
$\Lambda K^+\pi^-$                                & %         MODE @lb
\nodata                                           & %         PDG2014 AVG.
\nodata                                           & %   [2]  CDF
\phantom{.}                                       & %   [0]  
\red{$\gerrsyt{5.6}{0.8}{0.8}{0.7}$}~$^\dag$      & %   [4]  LHCB
\ifref {\cite{Aaij:2016nrq}} \fi \phantom{.}& %   [0]  
$5.6 \pm 1.3$                                     \\

\nodata                                           & %         RPP\#
$\Lambda K^+K^-$                                  & %         MODE @lb
\nodata                                           & %         PDG2014 AVG.
\nodata                                           & %   [2]  CDF
\phantom{.}                                       & %   [0]  
\red{$\gerrsyt{15.9}{1.2}{1.2}{2.0}$}~$^\dag$     & %   [4]  LHCB
\ifref {\cite{Aaij:2016nrq}} \fi \phantom{.}& %   [0]  
$15.9 \pm 2.6$                                    \\

%TABLE_BODY

\hline
\end{tabular}
}
\end{center}
\scriptsize
Results for CDF and LHCb are relative BFs converted to absolute BFs.\\   %FOOTNOTE
$^\dag$~Last quoted uncertainty is due to the precision with which the normalization channel branching fraction is known. \\     % FOOTNOTE
$^\ddag$~Third uncertainty is related to external inputs. \\     % FOOTNOTE
$^\S$~Third uncertainty is from the ratio of fragmentation fractions $f_{\Lambda^{0}_{b}}/f_d$, and the fourth is due to the uncertainty on ${\cal B}(B^{0}\rightarrow K^{0} \pi^+ \pi^-)$. \\     % FOOTNOTE
$^\P$~Result at 68\% CL.     % FOOTNOTE
\end{table}

%NEW_TABLE

\begin{table}[h!]
\begin{center}
\caption{Partial branching fractions of $\Lb \to \mu^+\mu^-$
decays in intervals of $q^2=m^2(\mu^+\mu^-)$ in units of
$\times 10^{-6}$. 
Where values are shown in \red{red} (\blue{blue}), this indicates that
they are new \red{published} (\blue{preliminary}) results since PDG2014.}
%Values in \red{red} (\blue{blue}) are new \red{published} (\blue{preliminary}) results since PDG2014.}
\label{tab:bbaryons_LbPartialBF}
\resizebox{\textwidth}{!}{
\begin{tabular}{|llccc @{}c c @{}c c|}
\sgline
RPP\# & Mode & $q^2~[\gevgevcccc]$~$^\dag$ & PDG2014 Avg. & CDF & & LHCb & & Our Avg. \\
\sglinespb
%TABLE_BODY
21                                                & %         RPP\#
$\Lambda\mu^+\mu^-$~$^\ddag$                      & %         MODE @lb
$< 2.0$                                           & %         $Q^2~[(\MATHRM{GEV}/C^2)^2]$~$^\DAG$
{$\err{0.15}{2.01}{0.05}$}                        & %   [0]  PDG2014 AVG.
{$\err{0.15}{2.01}{0.05}$}                        & %   [2]  CDF
\ifref {\cite{Aaltonen:2011qs}} \fi \phantom{.}& %   [0]  
\red{$\err{0.56}{0.76}{0.80}$}                    & %   [3]  LHCB
\ifref {\cite{Aaij:2013mna}} \fi \phantom{.}& %   [0]  
$0.41 \pm 0.87$                                   \\

\nodata                                           & %         RPP\#
$\Lambda\mu^+\mu^-$                               & %         MODE @lb
$[2.0,4.3]$                                       & %         $Q^2~[(\MATHRM{GEV}/C^2)^2]$~$^\DAG$
{$\err{1.8}{1.7}{0.6}$}                           & %   [0]  PDG2014 AVG.
{$\err{1.8}{1.7}{0.6}$}                           & %   [2]  CDF
\phantom{.}                                       & %   [0]  
\red{$\err{0.71}{0.60}{0.10}$}                    & %   [3]  LHCB
\phantom{.}                                       & %   [0]  
$0.91 \pm 0.55$                                   \\

\nodata                                           & %         RPP\#
$\Lambda\mu^+\mu^-$                               & %         MODE @lb
$[4.3,8.68]$                                      & %         $Q^2~[(\MATHRM{GEV}/C^2)^2]$~$^\DAG$
{$\err{-0.2}{1.6}{0.1}$}                          & %   [0]  PDG2014 AVG.
{$\err{-0.2}{1.6}{0.1}$}                          & %   [2]  CDF
\phantom{.}                                       & %   [0]  
\red{$\err{0.66}{0.72}{0.16}$}                    & %   [3]  LHCB
\phantom{.}                                       & %   [0]  
$0.40 \pm 0.62$                                   \\

\nodata                                           & %         RPP\#
$\Lambda\mu^+\mu^-$                               & %         MODE @lb
$[10.09,12.86]$                                   & %         $Q^2~[(\MATHRM{GEV}/C^2)^2]$~$^\DAG$
{$\err{3.0}{1.5}{1.0}$}                           & %   [0]  PDG2014 AVG.
{$\err{3.0}{1.5}{1.0}$}                           & %   [2]  CDF
\phantom{.}                                       & %   [0]  
\red{$\err{1.55}{0.58}{0.55}$}                    & %   [3]  LHCB
\phantom{.}                                       & %   [0]  
$1.96 \pm 0.68$                                   \\

\nodata                                           & %         RPP\#
$\Lambda\mu^+\mu^-$                               & %         MODE @lb
$[14.18,16.00]$                                   & %         $Q^2~[(\MATHRM{GEV}/C^2)^2]$~$^\DAG$
{$\err{1.0}{0.7}{0.3}$}                           & %   [0]  PDG2014 AVG.
{$\err{1.0}{0.7}{0.3}$}                           & %   [2]  CDF
\phantom{.}                                       & %   [0]  
\red{$\err{1.44}{0.44}{0.42}$}                    & %   [3]  LHCB
\phantom{.}                                       & %   [0]  
$1.19 \pm 0.40$                                   \\

\nodata                                           & %         RPP\#
$\Lambda\mu^+\mu^-$                               & %         MODE @lb
$>16.00$                                          & %         $Q^2~[(\MATHRM{GEV}/C^2)^2]$~$^\DAG$
{$\err{7.0}{1.9}{2.2}$}                           & %   [0]  PDG2014 AVG.
{$\err{7.0}{1.9}{2.2}$}                           & %   [2]  CDF
\phantom{.}                                       & %   [0]  
\red{$\err{4.7}{0.8}{1.2}$}                       & %   [3]  LHCB
\phantom{.}                                       & %   [0]  
$5.5 \pm 1.2$                                     \\

%TABLE_BODY
\sglinespt
\end{tabular}
}
\end{center}
\scriptsize
Results for CDF and LHCb are relative BFs converted to absolute BFs.\\    %FOOTNOTE
$^\dag$ ~See the original paper for the exact $m^2(\mu^+\mu^-)$ selection.\\     % FOOTNOTE
$^\ddag$~The LHCb measurement was superseded with a more accurate result in different $m^2(\mu^+\mu^-)$ bins (see list of not-included results).     % FOOTNOTE
\end{table}

%NEW_TABLE

\begin{table}
\begin{center}
\caption{Branching fractions of charmless $\Xi^{0}_{b}$ decays in units of
$\times 10^{-6}$.
Upper limits are at 90\% CL.
Where values are shown in \red{red} (\blue{blue}), this indicates that
they are new \red{published} (\blue{preliminary}) results since PDG2014.}
%Values in \red{red} (\blue{blue}) are new \red{published}
%(\blue{preliminary}) results since PDG2014.}
\label{tab:bbaryons_Xhib}
\begin{tabular}{|lccc @{}c c|} \hline
RPP\# &Mode & PDG2014 Avg. & LHCb & & Our Avg.  \\ \sglinespb
%%
%% -------------- IMPORTANT -------------- (PG 11/08/2016)     % COMMENT
%% These modes have the same final state as the Lambda_B     % COMMENT
%% decays in the first table above. The plotting macro     % COMMENT
%% distinguishes the modes via the decay string. I added an     % COMMENT
%% extra space between the first two particles in each of     % COMMENT
%% the Xi_b modes to distinguish them from the Lambda_B modes.     % COMMENT
%% This is matched in the xib plot defined in mode-ordering.txt.     % COMMENT
%%
%TABLE_BODY
\nodata                                           & %         RPP\#
$\Lambda  \pi^+\pi^-$                             & %         MODE @xib
\nodata                                           & %         PDG2014 AVG.
\red{$<1.7$}                                      & %   [4]  LHCB
\ifref {\cite{Aaij:2016nrq}} \fi \phantom{.}& %   [0]  
{$<1.7$}                                          \\

\nodata                                           & %         RPP\#
$\Lambda  K^+\pi^-$                               & %         MODE @xib
\nodata                                           & %         PDG2014 AVG.
\red{$<0.8$}                                      & %   [4]  LHCB
\ifref {\cite{Aaij:2016nrq}} \fi \phantom{.}& %   [0]  
{$<0.8$}                                          \\

\nodata                                           & %         RPP\#
$\Lambda  K^+K^-$                                 & %         MODE @xib
\nodata                                           & %         PDG2014 AVG.
\red{$<0.3$}                                      & %   [4]  LHCB
\ifref {\cite{Aaij:2016nrq}} \fi \phantom{.}& %   [0]  
{$<0.3$}                                          \\

\nodata                                           & %         RPP\#
$\bar{K}^0  p \pi^{-}$                            & %         MODE @xib
\nodata                                           & %         PDG2014 AVG.
\red{$<1.6$}                                      & %   [5]  LHCB
\ifref {\cite{Aaij:2014lpa}} \fi \phantom{.}& %   [0]  
{$<1.6$}                                          \\

\nodata                                           & %         RPP\#
$\bar{K}^0  p K^{-}$                              & %         MODE @xib
\nodata                                           & %         PDG2014 AVG.
\red{$<1.1$}                                      & %   [5]  LHCB
\ifref {\cite{Aaij:2014lpa}} \fi \phantom{.}& %   [0]  
{$<1.1$}                                          \\

%TABLE_BODY
%\\
\hline
\end{tabular}
\end{center}
\scriptsize
Results for LHCb are relative BFs converted to absolute BFs.\\ \\    %FOOTNOTE
\end{table}

\newpage
 List of other measurements that are not included in the tables:
\begin{itemize}
\item In Ref.~%[8],
\cite{Aaij:2015xza},
LHCb provides a measurement of the differential $\Lb \to \Lambda \mu^+\mu^-$ branching fraction. It is given in bins of $m^2(\mu^+\mu^-)$ that are different from those used in the past by the LHCb and CDF collaborations (see table of differential branching fractions).
\item In the paper %Phys. Rev. Lett. 114, 062004,
\cite{Aaij:2014yka},
LHCb measures the ratios
$$
\frac{\sigma(pp\rightarrow \Xi^{\prime -}_{b}X) {\cal B}(\Xi^{\prime -}_{b}\rightarrow\Xi^{0}_{b}\pi^-)}{\sigma(pp\rightarrow \Xi^{0}_{b}X)},
\frac{\sigma(pp\rightarrow \Xi^{\prime -}_{b}X) {\cal B}(\Xi^{\ast -}_{b}\rightarrow\Xi^{0}_{b}\pi^-)}{\sigma(pp\rightarrow \Xi^{\prime -}_{b}X) {\cal B}(\Xi^{\prime -}_{b}\rightarrow\Xi^{0}_{b}\pi^-)}.
$$
\item In the paper %JHEP 05 (2016) 161,
\cite{Aaij:2016jnn},
LHCb measures the ratio $$
\frac{\sigma(pp\rightarrow \Xi^{\ast -}_{b}X) {\cal B}(\Xi^{\ast -}_{b}\rightarrow\Xi^{0}_{b}\pi^-)}{\sigma(pp\rightarrow \Xi^{0}_{b}X)}.
$$

\end{itemize}

\vspace*{-5mm}

\mysubsection{Decays of \Bs mesons}
\label{sec:rare-bs}

Tables~\ref{tab:Bs_BF} and~\ref{tab:Bs_BF_rel} detail branching fractions and relative branching fractions of \Bs meson decays, respectively. 
%Tables~\ref{tab:Bs_PartBF} to~\ref{tab:Bs_A9} give different observables for $\Bs\to\phi\mup\mun$ decays in bins of the dimuon invariant mass.

\begin{figure}[htbp!]
\centering
\includegraphics[width=0.5\textwidth]{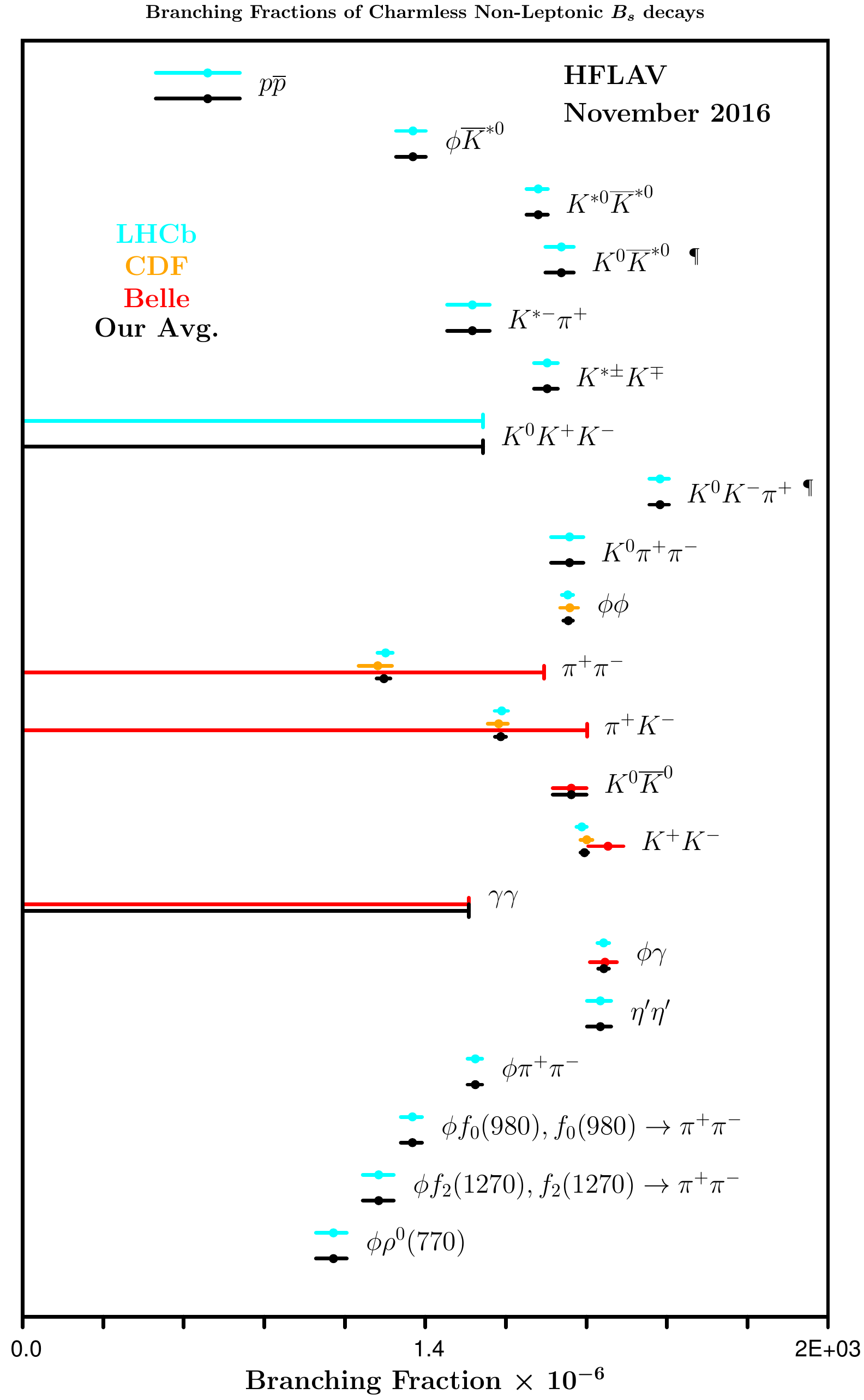}
\caption{Branching fractions of charmless non-leptonic $\Bs$ decays.}
\label{fig:rare-bs}
\end{figure}

\begin{figure}[htbp!]
\centering
\includegraphics[width=0.5\textwidth]{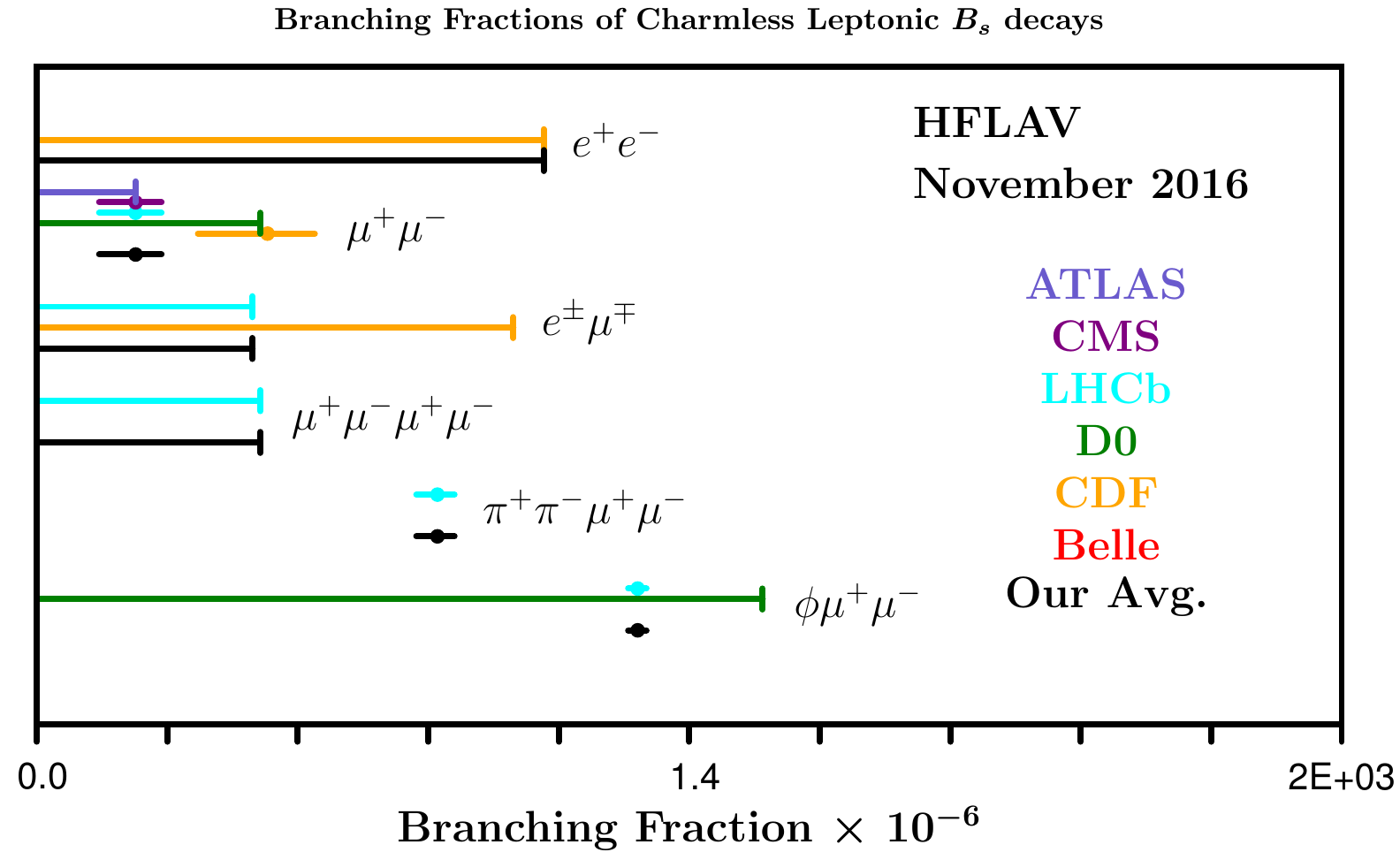}
\caption{Branching fractions of charmless leptonic $\Bs$ decays.}
\label{fig:rare-bsleptonic}
\end{figure}

Figures~\ref{fig:rare-bs} and~\ref{fig:rare-bsleptonic} show graphic representations of a selection of results given in this section.
Footnote symbols indicate that the footnote in the corresponding table should be consulted.
%For comments in the plots, marked with a symbol or a number, refer to the corresponding table.

%NEW_TABLE

%\begin{sidewaystable}
\begin{sidewaystable}[htbp!]
\begin{center}
\caption{Branching fractions of charmless 
$\Bs$ decays in units of $\times 10^{-6}$. Upper limits are
at 90\% CL.
Where values are shown in \red{red} (\blue{blue}), this indicates that
they are new \red{published} (\blue{preliminary}) results since PDG2014.}
%Values in \red{red} (\blue{blue}) are new \red{published} (\blue{preliminary}) results since PDG2014.}
\label{tab:Bs_BF}
\resizebox{\textwidth}{!}{

}
\end{center}
\scriptsize{
Results for CDF, D0, LHCb, CMS and ATLAS are relative BFs converted to absolute BFs.\\     %FOOTNOTE
$^\dag$~The first error is experimental, and the second is from the reference BF.\\[0.1cm]     %FOOTNOTE
$^\ddag$~Last error represents the uncertainty due to the total number of $\Bs \Bsb$ pairs.\\     %FOOTNOTE
$^\S$~Last error takes into account ${\cal B}(B^0 \to \phi K^{*0})$ and $f_d/f_s$.\\     %FOOTNOTE
\quad $^\P$~Includes two distinct decay processes: ${\cal B}(\Bs \to f)+{\cal B}(\Bs \to \bar{f})$.\\    % FOOTNOTE
%$^\P$~Sum of charge-conjugate states.\\     %FOOTNOTE
$^1$ UL at 95\% CL.\\     %FOOTNOTE
$^2$ Muon pairs do not originate from resonances and $0.5<m(\pi^+\pi^-)<1.3~\gevcc$.\\     %FOOTNOTE
$^3$ The average is done between the combined LHCb and CMS result, $\cerr{0.0028}{0.0007}{0.0006}$ (Ref.\cite{CMS:2014xfa}) and CDF.\\     %FOOTNOTE
$^4$ In the mass range $400<m(\pi^+\pi^-)<1600~\gevcc$.\\     %FOOTNOTE
}
\end{sidewaystable}
\clearpage

%NEW_TABLE

\begin{table}
\begin{center}
\caption{Relative branching fractions of charmless 
$\Bs$ decays. Upper limits are
at 90\% CL.
Where values are shown in \red{red} (\blue{blue}), this indicates that
they are new \red{published} (\blue{preliminary}) results since PDG2014.}
%Values in \red{red} (\blue{blue}) are new \red{published}
%(\blue{preliminary}) results since PDG2014.}
\label{tab:Bs_BF_rel}
\resizebox{\textwidth}{!}{
\begin{tabular}{|lccc @{}c c @{}c c|} \hline
RPP\# & Mode & PDG2014 Avg. & CDF & & LHCb & & Our Avg.  \\ \sglinespb
%TABLE_BODY
$45$                                              & %         RPP\#
$\it{f}_s\mathcal{B}(B^0_s\rightarrow\pi^+\pi^-)/\it{f}_d\mathcal{B}(B^0\rightarrow K^+\pi^-)$& %         MODE
\nodata                                           & %         PDG2014 AVG.
$0.008\pm0.002\pm0.001$                           & %   [6]  CDF
\ifref {\cite{Aaltonen:2011jv}} \fi \phantom{.}& %   [0]  
\red{$\err{0.00915}{0.00071}{0.00083}$}           & %   [36]  LHCB
\ifref {\cite{Aaij:2016elb}} \fi \phantom{.}& %   [0]  
$0.00880 \pm 0.00090$                             \\

$45$                                              & %         RPP\#
$\it{f}_s\mathcal{B}(B^0_s\rightarrow\pi^+\pi^-)/\it{f}_d\mathcal{B}(B^0\rightarrow \pi^+\pi^-)$& %         MODE
\nodata                                           & %         PDG2014 AVG.
\nodata                                           & %   [0]  CDF
\phantom{.}                                       & %   [0]  
$\aerr{0.050}{0.011}{0.009}{0.004}$               & %   [22]  LHCB
\ifref {\cite{Aaij:2012as}} \fi \phantom{.}& %   [0]  
$\cerr{0.050}{0.012}{0.010}$                      \\

$51$                                              & %         RPP\#
$\mathcal{B}(B^0_s\rightarrow\phi\phi)/\mathcal{B}(B^0_s\rightarrow J/\psi\phi)$& %         MODE
\nodata                                           & %         PDG2014 AVG.
$\err{0.0178}{0.0014}{0.0020}$                    & %   [5]  CDF
\ifref {\cite{Aaltonen:2011rs}} \fi \phantom{.}& %   [0]  
\nodata                                           & %   [0]  LHCB
\phantom{.}                                       & %   [0]  
$0.0180 \pm 0.0020$                               \\

\nodata                                           & %         RPP\#
$\mathcal{B}(B^0_s\rightarrow\phi\phi)/\mathcal{B}(B^0\rightarrow \phi K^*)$& %         MODE
\nodata                                           & %         PDG2014 AVG.
\nodata                                           & %   [0]  CDF
\phantom{.}                                       & %   [0]  
\red{$1.84 \pm 0.05 \pm 0.13$}                    & %   [34]  LHCB
\ifref {\cite{Aaij:2014lba}} \fi \phantom{.}& %   [0]  
$1.84 \pm 0.14$                                   \\

$52$                                              & %         RPP\#
$\it{f}_s\mathcal{B}(B^0_s\rightarrow K^+\pi^-)/\it{f}_d\mathcal{B}(B^0_d\rightarrow K^+\pi^-)$& %         MODE
\nodata                                           & %         PDG2014 AVG.
$0.071\pm0.010\pm0.007 $                          & %   [2]  CDF
\ifref {\cite{Aaltonen:2008hg}} \fi \phantom{.}& %   [0]  
$0.074\pm0.006\pm0.006$                           & %   [22]  LHCB
\ifref {\cite{Aaij:2012as}} \fi \phantom{.}& %   [0]  
$0.073 \pm 0.007$                                 \\

$53$                                              & %         RPP\#
$\it{f}_s\mathcal{B}(B^0_s\rightarrow K^+K^-)/\it{f}_d\mathcal{B}(B^0_d\rightarrow K^+\pi^-)$& %         MODE
\nodata                                           & %         PDG2014 AVG.
$0.347\pm0.020\pm0.021 $                          & %   [7]  CDF
\ifref {\cite{Aaltonen:2011qt}} \fi \phantom{.}& %   [0]  
\err{0.316}{0.009}{0.019}                         & %   [22]  LHCB
\ifref {\cite{Aaij:2012as}} \fi \phantom{.}& %   [0]  
$0.327 \pm 0.017$                                 \\

$55$                                              & %         RPP\#
$\mathcal{B}(B^0_s\to K^0\pi^+\pi^-)/\mathcal{B}(B^0 \to K^0\pi^+\pi^-)$& %         MODE
\nodata                                           & %         PDG2014 AVG.
\nodata                                           & %   [0]  CDF
\phantom{.}                                       & %   [0]  
$\err{0.29}{0.06}{0.04}$                          & %   [26]  LHCB
\ifref {\cite{Aaij:2013uta}} \fi \phantom{.}& %   [0]  
$0.29 \pm 0.07$                                   \\

$56$                                              & %         RPP\#
$\mathcal{B}(B^0_s\to K^0 K^- \pi^+)/\it\mathcal{B}(B^0 \to K^0 K^- \pi^+)$~$^\dag$& %         MODE
\nodata                                           & %         PDG2014 AVG.
\nodata                                           & %   [0]  CDF
\phantom{.}                                       & %   [0]  
$\err{1.48}{0.12}{0.14}$                          & %   [26]  LHCB
\ifref {\cite{Aaij:2013uta}} \fi \phantom{.}& %   [0]  
$1.48 \pm 0.18$                                   \\

$57$                                              & %         RPP\#
$\mathcal{B}(B^0_s\to K^0 K^+ K^-)/\mathcal{B}(B^0 \to K^0 K^+ K^-)$& %         MODE
\nodata                                           & %         PDG2014 AVG.
\nodata                                           & %   [0]  CDF
\phantom{.}                                       & %   [0]  
$<0.068$                                          & %   [26]  LHCB
\ifref {\cite{Aaij:2013uta}} \fi \phantom{.}& %   [0]  
$<0.068$                                          \\

\nodata                                           & %         RPP\#
${\cal B}(B^0_s\to K^{*-}K^+)/{\cal B}(B^0 \to K^{*+}\pi^-)$& %         MODE
\nodata                                           & %         PDG2014 AVG.
\nodata                                           & %   [0]  CDF
\phantom{.}                                       & %   [0]  
\red{$\err{1.49}{0.22}{0.18}$}                    & %   [30]  LHCB
\ifref {\cite{Aaij:2014aaa}} \fi \phantom{.}& %   [0]  
$1.49 \pm 0.28$                                   \\

\nodata                                           & %         RPP\#
${\cal B}(B^0_s\to K^{*-}\pi^+)/{\cal B}(B^0 \to K^{*+}\pi^-)$& %         MODE
\nodata                                           & %         PDG2014 AVG.
\nodata                                           & %   [0]  CDF
\phantom{.}                                       & %   [0]  
\red{$\err{0.39}{0.13}{0.05}$}                    & %   [30]  LHCB
\ifref {\cite{Aaij:2014aaa}} \fi \phantom{.}& %   [0]  
$0.39 \pm 0.14$                                   \\

$59$                                              & %         RPP\#
${\cal B}(B^0_s\to K^{*0}\overline K^{*0})/{\cal B}(B^0 \to K^{*+}\pi^-)$& %         MODE
\nodata                                           & %         PDG2014 AVG.
\nodata                                           & %   [0]  CDF
\phantom{.}                                       & %   [0]  
\red{$\err{1.11}{0.22}{0.13}$}                    & %   [21]  LHCB
\ifref {\cite{Aaij:2015kba}} \fi \phantom{.}& %   [0]  
$1.11 \pm 0.26$                                   \\

$60$                                              & %         RPP\#
${\cal B}(B^0_s\to \phi \overline{K}^{*0})/{\cal B}(B^0 \to \phi K^{*0})$& %         MODE
\nodata                                           & %         PDG2014 AVG.
\nodata                                           & %   [0]  CDF
\phantom{.}                                       & %   [0]  
$\err{0.113}{0.024}{0.016}$                       & %   [27]  LHCB
\ifref {\cite{Aaij:2013lla}} \fi \phantom{.}& %   [0]  
$0.113 \pm 0.029$                                 \\

$64$                                              & %         RPP\#
${\cal B}(B^0_s\to \phi \gamma)/{\cal B}(B^0 \to K^{*0}\gamma)$& %         MODE @bs
\nodata                                           & %         PDG2014 AVG.
\nodata                                           & %   [0]  CDF
\phantom{.}                                       & %   [0]  
\err{0.81}{0.04}{0.07}                            & %   [23]  LHCB
\ifref {\cite{Aaij:2012ita}} \fi \phantom{.}& %   [0]  
$0.81 \pm 0.08$                                   \\

$70$                                              & %         RPP\#
$\mathcal{B}(B^0_s\to \phi\mu^+\mu^-)/\mathcal{B}(B^0_s\to J/\psi\phi)\times10^4$& %         MODE
$7.1 \pm 1.3$                                     & %         PDG2014 AVG.
\nodata                                           & %   [0]  CDF
\phantom{.}                                       & %   [0]  
\red{$\aerr{7.41}{0.42}{0.40}{0.29}$}             & %   [24]  LHCB
\ifref {\cite{Aaij:2015esa}} \fi \phantom{.}& %   [0]  
$\cerr{7.41}{0.51}{0.49}$                         \\

\nodata                                           & %         RPP\#
$\mathcal{B}(B^0_s\to K^0_S K^{*0})/\mathcal{B}(B^0\to K^0_S \pi^+ \pi^-)$~$^\dag$& %         MODE
\nodata                                           & %         PDG2014 AVG.
\nodata                                           & %   [0]  CDF
\phantom{.}                                       & %   [0]  
\red{$\err{0.33}{0.07}{0.04}$}                    & %   [33]  LHCB
\ifref {\cite{Aaij:2015asa}} \fi \phantom{.}& %   [0]  
$0.33 \pm 0.08$                                   \\

%TABLE_BODY
%\\
\sglinespt
\end{tabular}
}
\end{center}
\scriptsize{
\quad $^\dag$~Numerator includes two distinct decay processes: ${\cal B}(\Bs \to f)+{\cal B}(\Bs \to \bar{f})$.\\    % FOOTNOTE
%$^\dag$~Sum of charge-conjugate states in the numerator.\\     %FOOTNOTE
%$^\ddag$~Sum of charge-conjugate states in the numerator and denominator.     %FOOTNOTE
}
\end{table}

\newpage

List of other measurements that are not included in the tables:
\begin{itemize}
\item  $\Bs \to \phi\mu^+\mu^-$ : LHCb measures the differential BF in bins of $m^2(\mu^+\mu^-)$. It also performs an angular analysis and measures $F_L$, $S_3$, $S_4$, $S_7$, $A_5$, $A_6$, $A_8$ and $A_9$ in bins of $m^2(\mu^+\mu^-)$~%[24].
\cite{Aaij:2015esa}.
\item  $\Bs \to \phi \gamma$ : LHCb has measured the photon polarization~%[37].
\cite{Aaij:2016ofv}.
\end{itemize}

%\newpage

%\newpage 
\clearpage

\mysubsection{Radiative and leptonic decays of \Bz and \Bp\ mesons.}
\label{sec:rare-radll}

This section reports different observables for leptonic and radiative \Bz\ and \Bp\ meson decays, including processes in which the photon yields a pair of charged or neutral leptons.
Tables~\ref{tab:radll_Bp_1} and \ref{tab:radll_Bp_2} provide compilations of branching fractions of \Bp mesons to radiative, and lepton-flavor/number-violating final states, respectively. Tables~\ref{tab:radll_Bz} and~\ref{tab:radll_B} provide compilations of branching fractions of \Bz mesons, and $\Bpm/\Bz$ meson admixture, respectively.
Table~\ref{tab:radll_lep} contains branching fractions of leptonic and radiative-leptonic $\Bp$ and $\Bz$ decays. It is followed by Tables~\ref{tab:radll_rel} and~\ref{tab:radll_gluon}, which give relative branching fractions of $\Bp$ decays and a compilations of inclusive decays, respectively. Table~\ref{tab:radll_AI} contains isospin asymmetry measurements.

\begin{figure}[htbp!]
\centering
\includegraphics[width=0.5\textwidth]{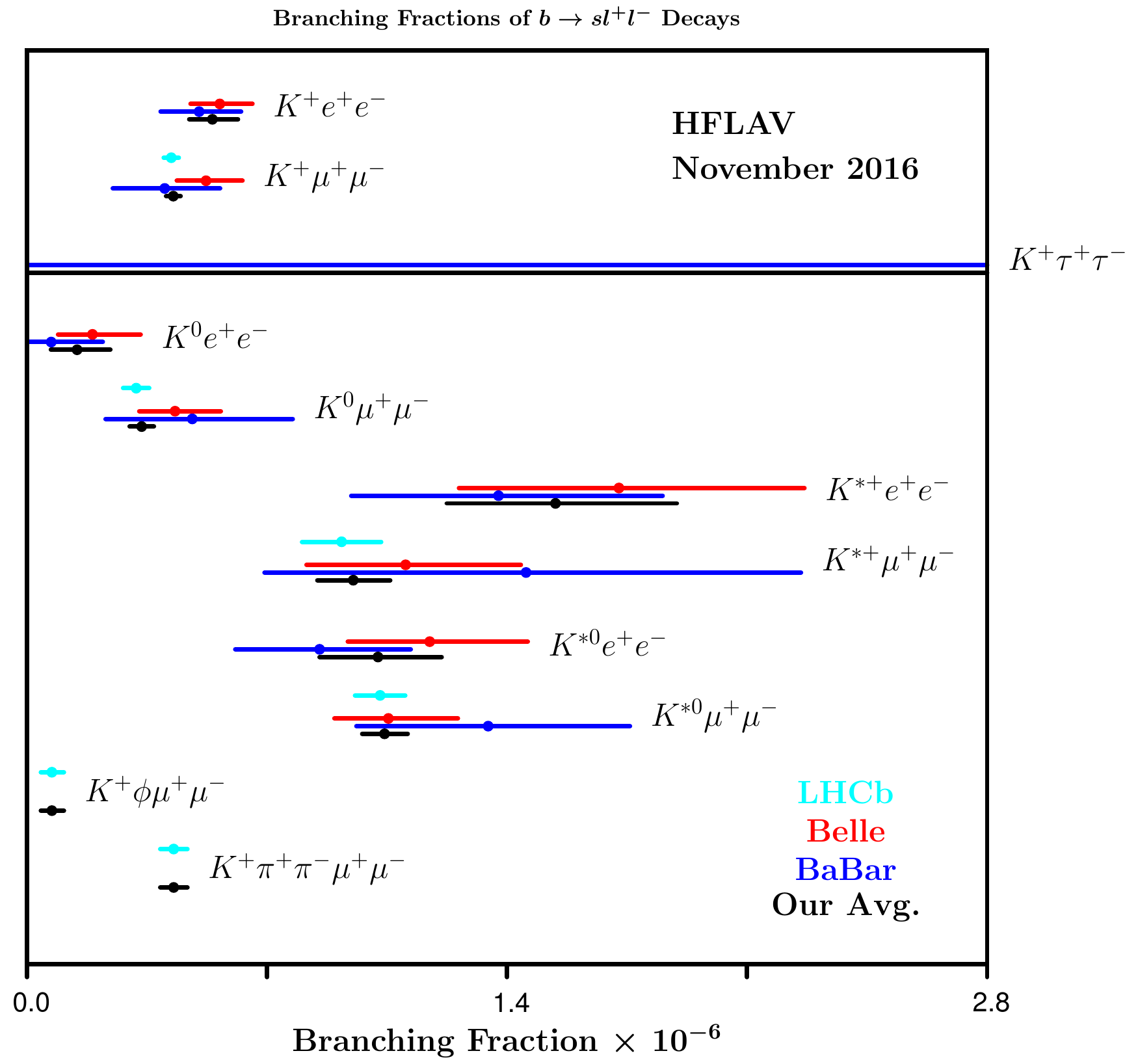}
\caption{Branching fractions of $b\to s\ell^{+}\ell^{-}$ decays.}
\label{fig:rare-btosll}
\end{figure}

\begin{figure}[htbp!]
\centering
\includegraphics[width=0.5\textwidth]{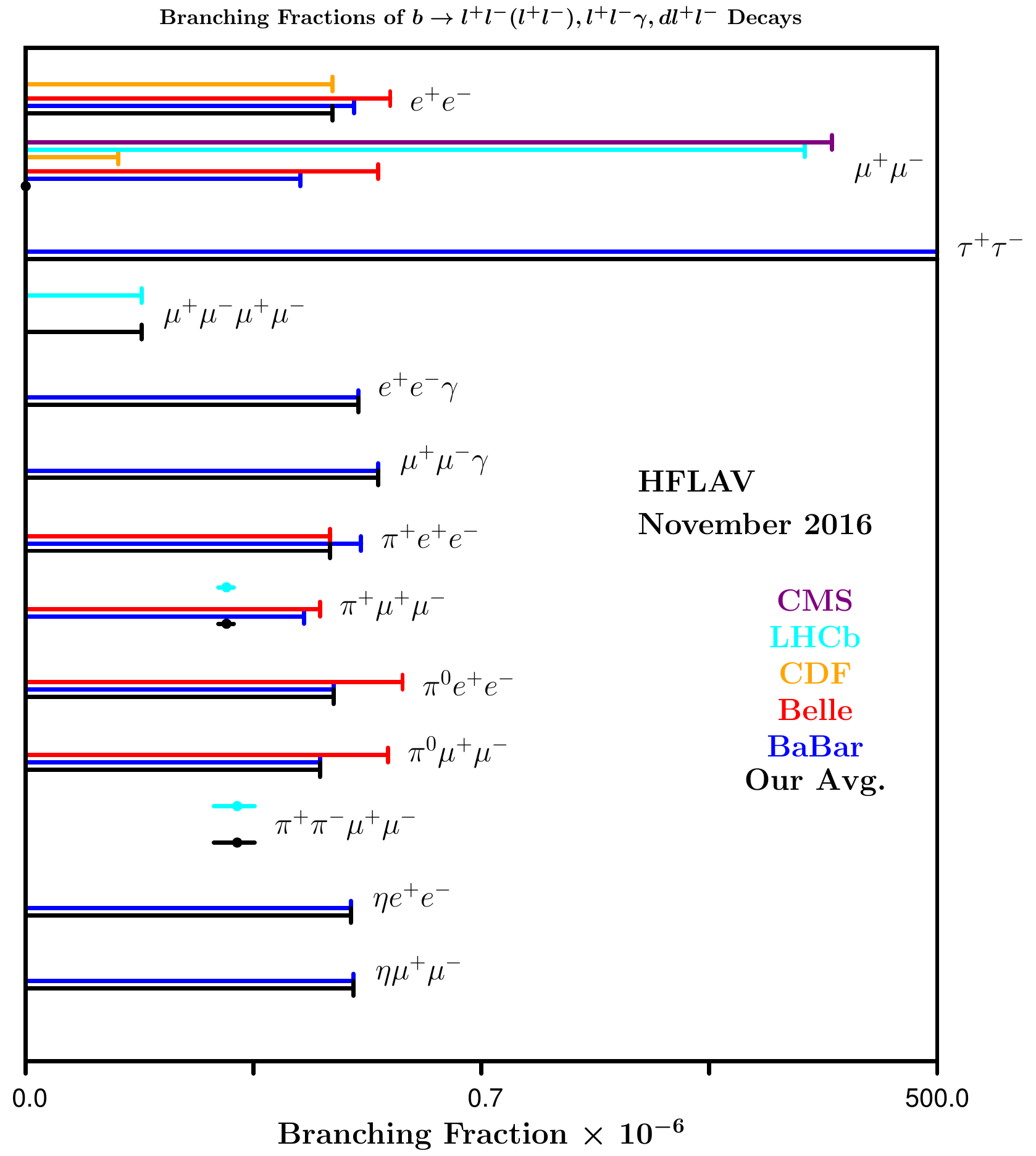}
\caption{Branching fractions of $b\to \ell^{+}\ell^{-}(\ell^{+}\ell^{-}), \ell^{+}\ell^{-}\gamma$ and $b\to d \ell^{+}\ell^{-}$ decays.}
\label{fig:rare-btodllg}
\end{figure}

\begin{figure}[htbp!]
\centering
\includegraphics[width=0.5\textwidth]{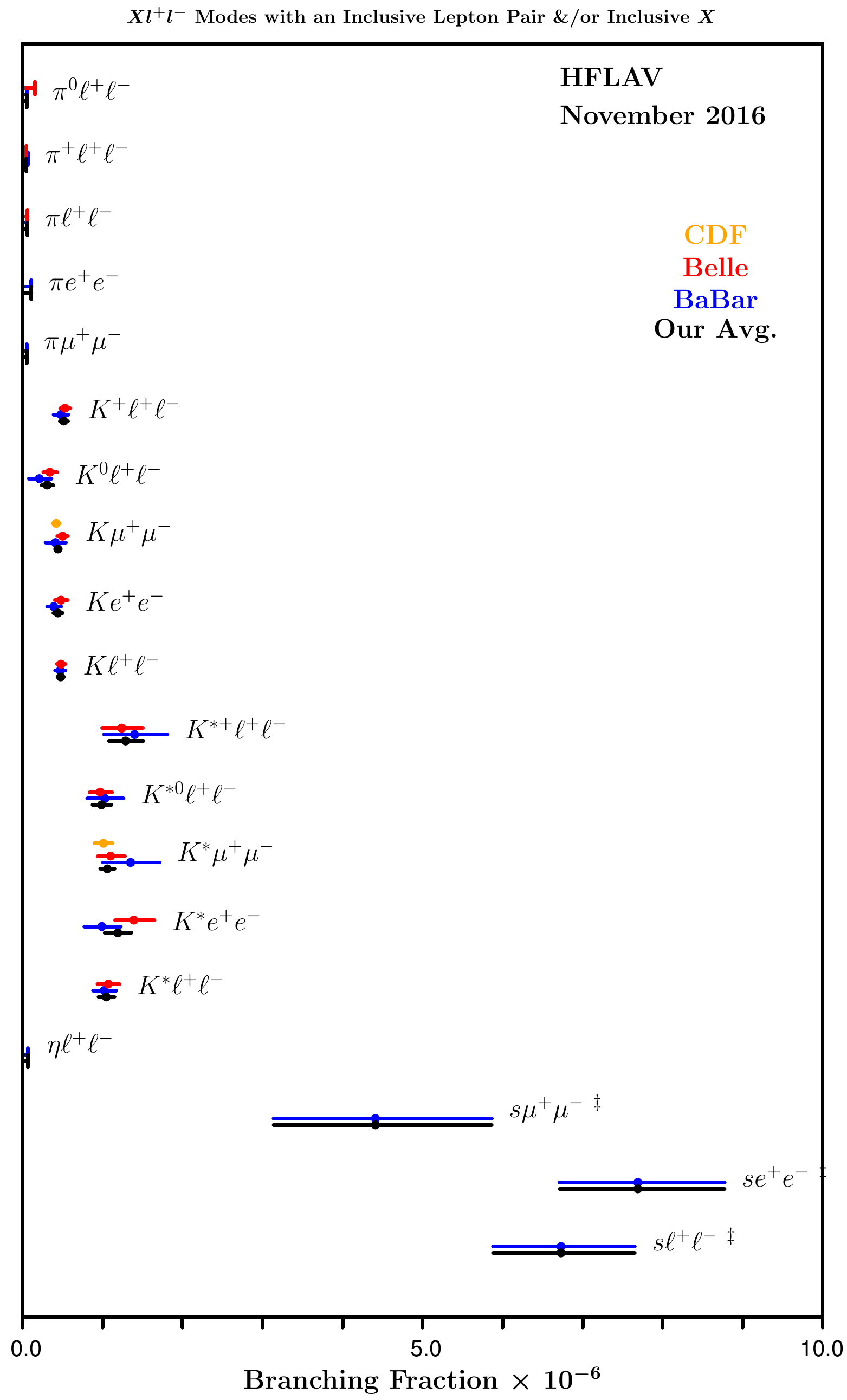}
\caption{$X \ell^+ \ell^-$ modes with an inclusive lepton pair and/or inclusive $X$.}
\label{fig:rare-kllsummary}
\end{figure}

\begin{figure}[htbp!]
\centering
\includegraphics[width=0.5\textwidth]{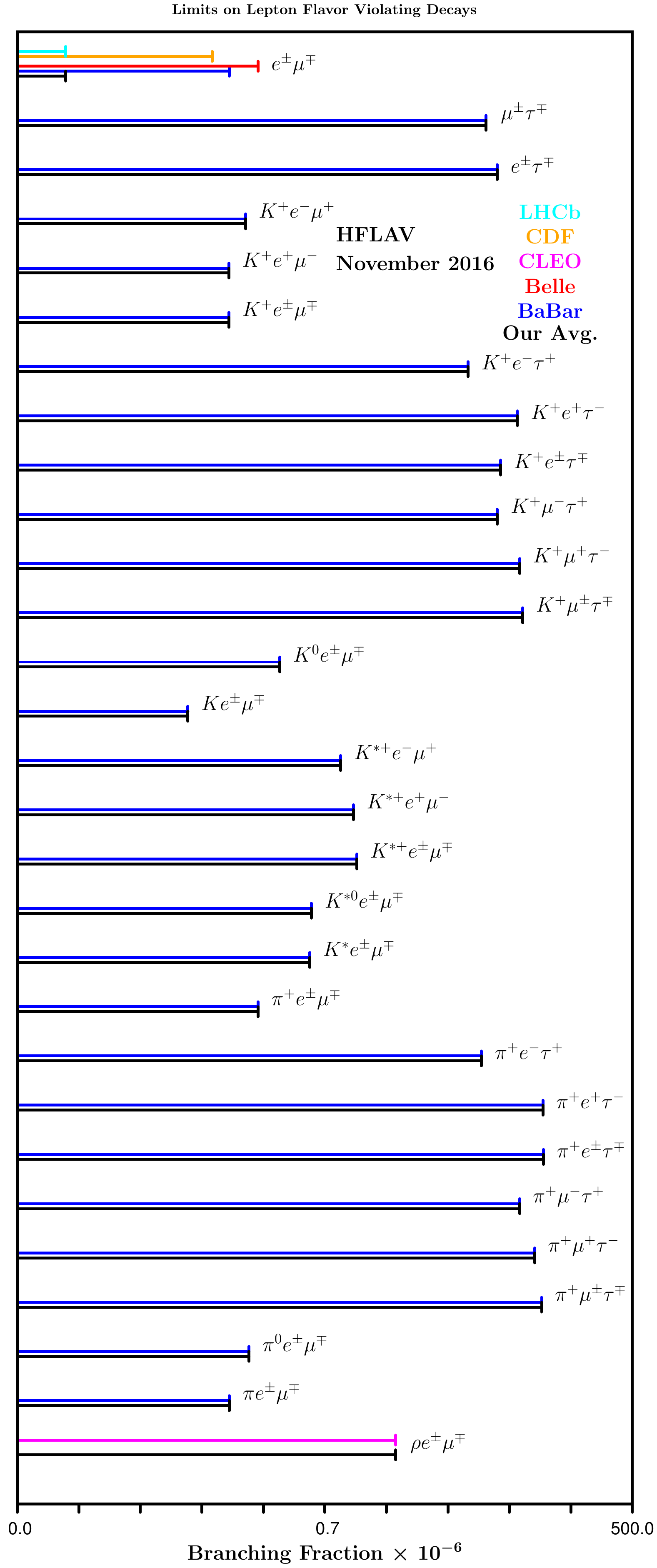}
\caption{Limits on lepton-flavor-violating decays.}
\label{fig:rare-leptonflavorviol}
\end{figure}

\begin{figure}[htbp!]
\centering
\includegraphics[width=0.5\textwidth]{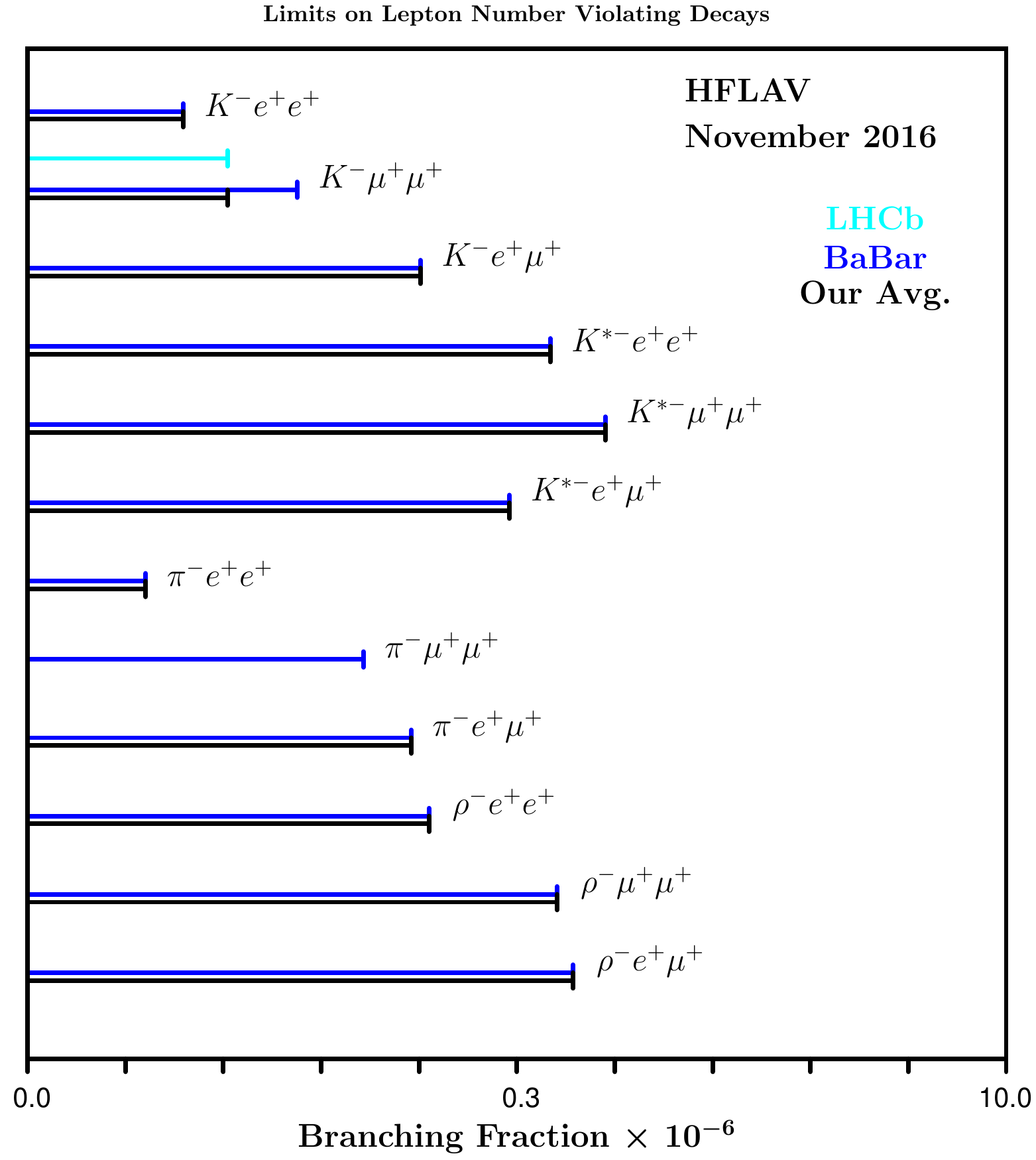}
\caption{Limits on lepton-number-violating decays.}
\label{fig:rare-leptonnumberviol}
\end{figure}

\begin{figure}[htbp!]
\centering
\includegraphics[width=0.5\textwidth]{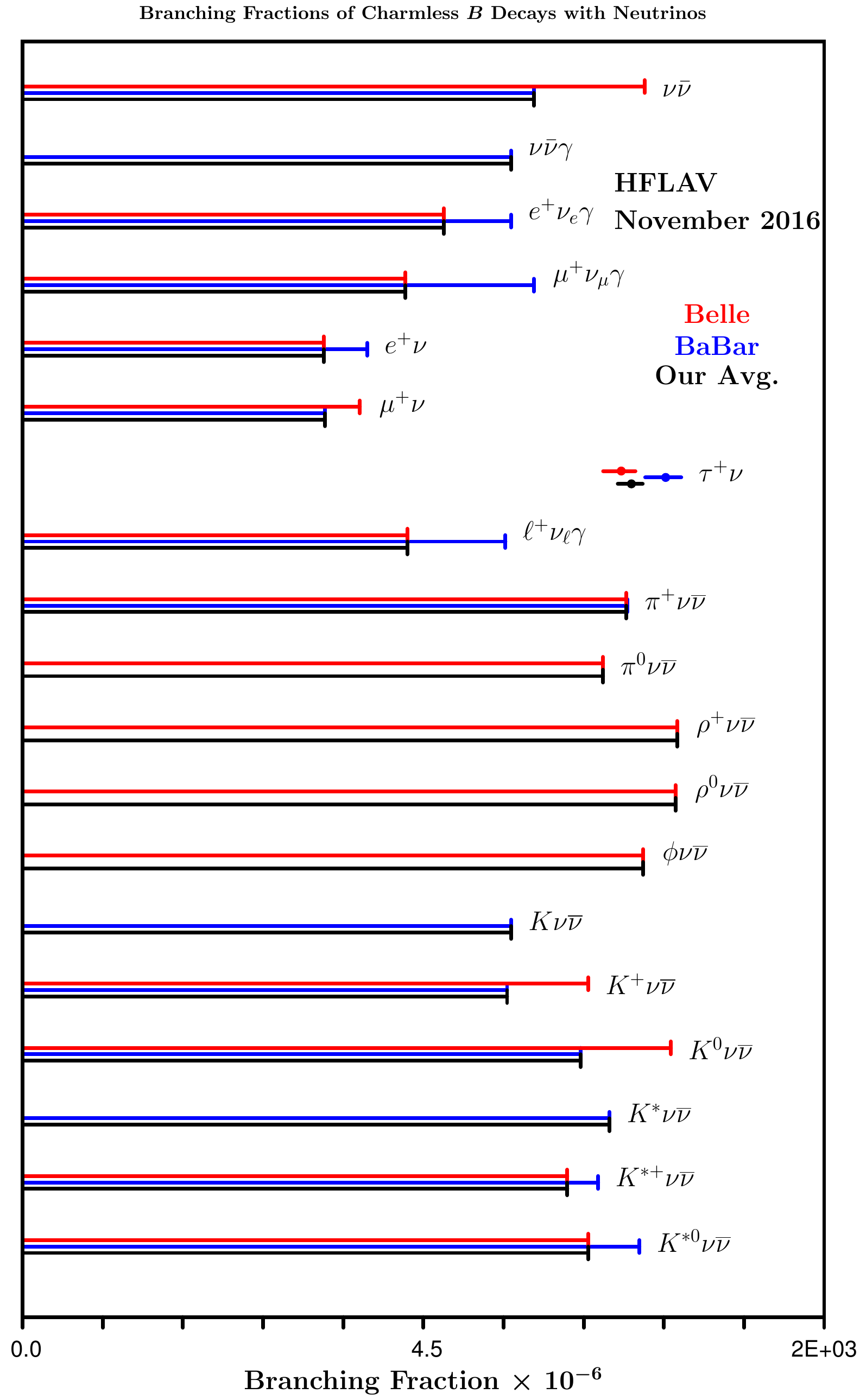}
\caption{Branching fractions of charmless $B$ decays with neutrinos.}
\label{fig:rare-neutrino}
\end{figure}

Figures \ref{fig:rare-btosll} to~\ref{fig:rare-neutrino} show graphic representations of a selection of results given in this section.
Footnote symbols indicate that the footnote in the corresponding table should be consulted.
%For comments in the plots, marked with a symbol or a number, refer to the corresponding table.

%NEW_TABLE

\begin{table}[!htbp]
\begin{center}
\caption{Branching fractions of charmless semileptonic and radiative
$B^+$ decays in units of $\times 10^{-6}$. Upper limits are
at 90\% CL. 
Where values are shown in \red{red} (\blue{blue}), this indicates that
they are new \red{published} (\blue{preliminary}) results since PDG2014.}
%Values in \red{red} (\blue{blue}) are new \red{published}
%(\blue{preliminary}) results since PDG2014.}
\label{tab:radll_Bp_1}
\resizebox{\textwidth}{!}{

}
\end{center}
\scriptsize{
Results for LHCb are relative BFs converted to absolute BFs.\\     %FOOTNOTE
CLEO upper limits that have been greatly superseded are not shown. \\     % FOOTNOTE
$^\dag$ $M_{K\pi\pi} < 1.8~\gevcc$.\\     % FOOTNOTE
$^\ddag$~$1.0 < M_{K\pi\pi} < 2.0~\gevcc$.\\     % FOOTNOTE
$^\S$~$M_{K\pi\pi} < 2.4~\gevcc$.\\     % FOOTNOTE
$^1$~PDG2014 cites only the measurement: $\mathcal{B}(\pi^+\mu^+\mu^-)/\mathcal{B}(K^+\mu^+\mu^-)=\err{0.053}{0.014}{0.01}$.\\     % FOOTNOTE
$^2$~Differential BF in bins of $m(\mu^+\mu^-)$ is also available.\\     % FOOTNOTE
%$^4$~PDG considers here the BF measured in $B^+ \to K^+\mu^+\mu^-$.     % FOOTNOTE
}
\end{table}

%NEW_TABLE

\begin{table}[!htbp]
\begin{center}
\caption{Branching fractions of charmless semileptonic
$B^+$ decays to LFV and LNV final states in units of $\times 10^{-6}$. Upper limits are
at 90\% CL. 
Where values are shown in \red{red} (\blue{blue}), this indicates that
they are new \red{published} (\blue{preliminary}) results since PDG2014.}
%Values in \red{red} (\blue{blue}) are new \red{published}
%(\blue{preliminary}) results since PDG2014.}
\label{tab:radll_Bp_2}
%\resizebox{\textwidth}{!}{

%}
\end{center}
\scriptsize{
Results for LHCb are relative BFs converted to absolute BFs.\\     %FOOTNOTE
CLEO upper limits that have been greatly superseded are not shown. \\     % FOOTNOTE
$^\dag$~UL at $95\%$ CL.     % FOOTNOTE
%$^4$~PDG considers here the BF measured in $B^+ \to K^+\mu^+\mu^-$.     % FOOTNOTE
}
\end{table}

%NEW_TABLE

\begin{table}[!htbp]
\begin{center}
\caption{Branching fractions of charmless semileptonic and radiative 
$\Bz$ decays in units of $\times 10^{-6}$. Upper limits are
at 90\% CL. 
Where values are shown in \red{red} (\blue{blue}), this indicates that
they are new \red{published} (\blue{preliminary}) results since PDG2014.}
%Values in \red{red} (\blue{blue}) are new \red{published}
%(\blue{preliminary}) results since PDG2014.}
\label{tab:radll_Bz}
\resizebox{\textwidth}{!}{

}
\end{center}
\scriptsize{
Results for LHCb are relative BFs converted to absolute BFs.\\     %FOOTNOTE
CLEO upper limits that have been greatly superseded are not shown. \\     % FOOTNOTE
$^\dag$ $M_{K\pi\pi} < 1.8~\gevcc$.\\     % FOOTNOTE
$^\ddag$~$1.0 < M_{K\pi\pi} < 2.0~\gevcc$.\\     % FOOTNOTE
$^\S$~$1.25 < M_{K\pi} < 1.60~\gevcc$.\\     % FOOTNOTE
$^\P$~This result takes into account the S-wave fraction in the $K\pi$ system.\\     % FOOTNOTE
$^1$~Muon pairs do not originate from resonances and $0.5<m(\pi^+\pi^-)<1.3~\gevcc$.     % FOOTNOTE
}
\end{table}

%NEW_TABLE

\begin{table}[!htbp]
\begin{center}
\caption{Branching fractions of charmless semileptonic and radiative 
decays of $\Bpm/\Bz$ admixture in units of $\times 10^{-6}$. Upper limits are
at 90\% CL. 
Where values are shown in \red{red} (\blue{blue}), this indicates that
they are new \red{published} (\blue{preliminary}) results since PDG2014.}
%Values in \red{red} (\blue{blue}) are new \red{published}
%(\blue{preliminary}) results since PDG2014.}
\label{tab:radll_B}
\resizebox{\textwidth}{!}{

}
\end{center}
\scriptsize{
Results for CDF are relative BFs converted to absolute BFs.\\     %FOOTNOTE
CLEO upper limits that have been greatly superseded are not shown. \\     % FOOTNOTE
$^\dag$~Results extrapolated to $E_{\gamma}>1.6~\gev$, using the method of Ref.~\cite{Buchmuller:2005zv}.\\ %O.-L. Buchmuller \etal, \prd{73}, 073008 (2006). \\     % FOOTNOTE
$^\ddag$~Belle: $m(\ell^+\ell^-)>0.2~\gevcc$, \babar: $m^2(\ell^+\ell^-)>0.1~\gevgevcccc$.\\     % FOOTNOTE
$^\S$~The value quoted is   ${\cal B}(B \to K^*_3 \gamma) \times {\cal B}(K^*_3 \to K\eta)$. PDG gives the BF assuming ${\cal B}(K^*_3 \to K\eta)=\cerr{11}{5}{4}\%$.\\     % FOOTNOTE
%$^\S$~Product BF ($\times {\cal B}(K^*_3 \to K\eta)$). PDG gives the BF assuming ${\cal B}(K^*_3 \to K\eta)=\cerr{11}{5}{4}\%$.\\     % FOOTNOTE
%$^\P$~$E_\gamma > 2.0~\gev$.\\     % FOOTNOTE
$^1$~Average of several results, obtained with different methods.\\     % FOOTNOTE
$^2$~Only results originally measured in the interval $E_\gamma > 1.9~\gev$ (also taken into account in the previous line).\\     % FOOTNOTE
}
\end{table}

%NEW_TABLE

\begin{table}[!htbp]
\begin{center}
\caption{Branching fractions of leptonic and radiative-leptonic 
$\Bp$ and $\Bz$ decays in units of $\times 10^{-6}$. Upper limits are
at 90\% CL. 
Where values are shown in \red{red} (\blue{blue}), this indicates that
they are new \red{published} (\blue{preliminary}) results since PDG2014.}
%Values in \red{red} (\blue{blue}) are new \red{published}
%(\blue{preliminary}) results since PDG2014.}
\label{tab:radll_lep}
\resizebox{\textwidth}{!}{
\begin{tabular}{|lccc @{}c c @{}c c @{}c c @{}c c @{}c c @{}c c|} \hline
RPP\# &Mode & PDG2014 Avg. & \babar & & Belle & & CDF & & LHCb & & CMS & & ATLAS & & Our Avg.  \\ \sglinespb
%TABLE_BODY
~29                                               & %         RPP\#
$e^+ \nu$                                         & %         MODE @neutrino
$<0.98$                                           & %         PDG2014 AVG.
{$<1.9$}                                          & %   [23]  BABAR
\ifref {\cite{Aubert:2009ar}} \fi \phantom{.}& %   [0]  
$<0.98$~$^\dag$                                   & %   [45]  BELLE
\ifref {\cite{Satoyama:2006xn}} \fi \phantom{.}& %   [0]  
\nodata                                           & %   [0]  CDF
\phantom{.}                                       & %   [0]  
\nodata                                           & %   [0]  LHCB
\phantom{.}                                       & %   [0]  
\nodata                                           & %   [0]  CMS
\phantom{.}                                       & %   [0]  
\nodata                                           & %   [0]  ATLAS
\phantom{.}                                       & %   [0]  
$<0.98$~$^\dag$                                   \\

~30                                               & %         RPP\#
$\mu^+ \nu$                                       & %         MODE @neutrino
$<1.0$                                            & %         PDG2014 AVG.
$<1.0$                                            & %   [23]  BABAR
\ifref {\cite{Aubert:2009ar}} \fi \phantom{.}& %   [0]  
$<1.7$~$^\dag$                                    & %   [45]  BELLE
\ifref {\cite{Satoyama:2006xn}} \fi \phantom{.}& %   [0]  
\nodata                                           & %   [0]  CDF
\phantom{.}                                       & %   [0]  
\nodata                                           & %   [0]  LHCB
\phantom{.}                                       & %   [0]  
\nodata                                           & %   [0]  CMS
\phantom{.}                                       & %   [0]  
\nodata                                           & %   [0]  ATLAS
\phantom{.}                                       & %   [0]  
$<1.0$                                            \\

~31                                               & %         RPP\#
$\tau^+ \nu$                                      & %         MODE @neutrino
$114\pm27$                                        & %         PDG2014 AVG.
$179\pm48$~$^\ddag$                               & %   [9]  BABAR
\ifref {\cite{Lees:2012ju}} \fi \phantom{.}& %   [0]  
\red{$\err{91}{19}{11}$}~$^\ddag$                 & %   [44]  BELLE
\ifref {\cite{Kronenbitter:2015kls}} \fi \phantom{.}& %   [0]  
\nodata                                           & %   [0]  CDF
\phantom{.}                                       & %   [0]  
\nodata                                           & %   [0]  LHCB
\phantom{.}                                       & %   [0]  
\nodata                                           & %   [0]  CMS
\phantom{.}                                       & %   [0]  
\nodata                                           & %   [0]  ATLAS
\phantom{.}                                       & %   [0]  
$106 \pm 19$                                      \\

~32                                               & %         RPP\#
$\ell^+ \nu_{\ell} \gamma$                        & %         MODE @neutrino
$<15.6$                                           & %         PDG2014 AVG.
$<15.6$                                           & %   [26]  BABAR
\ifref {\cite{Aubert:2009ya}} \fi \phantom{.}& %   [0]  
\red{$<3.5$}                                      & %   [62]  BELLE
\ifref {\cite{Heller:2015vvm}} \fi \phantom{.}& %   [0]  
\nodata                                           & %   [0]  CDF
\phantom{.}                                       & %   [0]  
\nodata                                           & %   [0]  LHCB
\phantom{.}                                       & %   [0]  
\nodata                                           & %   [0]  CMS
\phantom{.}                                       & %   [0]  
\nodata                                           & %   [0]  ATLAS
\phantom{.}                                       & %   [0]  
{$<3.5$}                                          \\

~33                                               & %         RPP\#
$e^+ \nu_e \gamma$                                & %         MODE @neutrino
$<17$                                             & %         PDG2014 AVG.
{$<17$}                                           & %   [26]  BABAR
\ifref {\cite{Aubert:2009ya}} \fi \phantom{.}& %   [0]  
\red{$<6.1$}                                      & %   [62]  BELLE
\ifref {\cite{Heller:2015vvm}} \fi \phantom{.}& %   [0]  
\nodata                                           & %   [0]  CDF
\phantom{.}                                       & %   [0]  
\nodata                                           & %   [0]  LHCB
\phantom{.}                                       & %   [0]  
\nodata                                           & %   [0]  CMS
\phantom{.}                                       & %   [0]  
\nodata                                           & %   [0]  ATLAS
\phantom{.}                                       & %   [0]  
{$<6.1$}                                          \\

~34                                               & %         RPP\#
$\mu^+ \nu_{\mu} \gamma$                          & %         MODE @neutrino
$<24$                                             & %         PDG2014 AVG.
{$<24$}                                           & %   [26]  BABAR
\ifref {\cite{Aubert:2009ya}} \fi \phantom{.}& %   [0]  
\red{$<3.4$}                                      & %   [62]  BELLE
\ifref {\cite{Heller:2015vvm}} \fi \phantom{.}& %   [0]  
\nodata                                           & %   [0]  CDF
\phantom{.}                                       & %   [0]  
\nodata                                           & %   [0]  LHCB
\phantom{.}                                       & %   [0]  
\nodata                                           & %   [0]  CMS
\phantom{.}                                       & %   [0]  
\nodata                                           & %   [0]  ATLAS
\phantom{.}                                       & %   [0]  
{$<3.4$}                                          \\

457                                               & %         RPP\#
$\gamma \gamma$                                   & %         MODE @bsgamma
$<0.32$                                           & %         PDG2014 AVG.
$<0.32$                                           & %   [14]  BABAR
\ifref {\cite{delAmoSanchez:2010bx}} \fi \phantom{.}& %   [0]  
{$<0.62$}                                         & %   [58]  BELLE
\ifref {\cite{Abe:2005bs}} \fi \phantom{.}& %   [0]  
\nodata                                           & %   [0]  CDF
\phantom{.}                                       & %   [0]  
\nodata                                           & %   [0]  LHCB
\phantom{.}                                       & %   [0]  
\nodata                                           & %   [0]  CMS
\phantom{.}                                       & %   [0]  
\nodata                                           & %   [0]  ATLAS
\phantom{.}                                       & %   [0]  
$<0.32$                                           \\

458                                               & %         RPP\#
$e^+ e^-$                                         & %         MODE @kll
$<0.083$                                          & %         PDG2014 AVG.
{$<0.113$}                                        & %   [5]  BABAR
\ifref {\cite{Aubert:2007hb}} \fi \phantom{.}& %   [0]  
$<0.19$                                           & %   [46]  BELLE
\ifref {\cite{Chang:2003yy}} \fi \phantom{.}& %   [0]  
$<0.083$                                          & %   [83]  CDF
\ifref {\cite{Aaltonen:2009vr}} \fi \phantom{.}& %   [0]  
\nodata                                           & %   [0]  LHCB
\phantom{.}                                       & %   [0]  
\nodata                                           & %   [0]  CMS
\phantom{.}                                       & %   [0]  
\nodata                                           & %   [0]  ATLAS
\phantom{.}                                       & %   [0]  
$<0.083$                                          \\

459                                               & %         RPP\#
$e^+ e^- \gamma$                                  & %         MODE @kll
$<0.12$                                           & %         PDG2014 AVG.
{$<0.12$}                                         & %   [20]  BABAR
\ifref {\cite{Aubert:2007up}} \fi \phantom{.}& %   [0]  
\nodata                                           & %   [0]  BELLE
\phantom{.}                                       & %   [0]  
\nodata                                           & %   [0]  CDF
\phantom{.}                                       & %   [0]  
\nodata                                           & %   [0]  LHCB
\phantom{.}                                       & %   [0]  
\nodata                                           & %   [0]  CMS
\phantom{.}                                       & %   [0]  
\nodata                                           & %   [0]  ATLAS
\phantom{.}                                       & %   [0]  
{$<0.12$}                                         \\

460                                               & %         RPP\#
$\mu^+ \mu^-$                                     & %         MODE NoAv  @kll
$<0.00063$                                        & %         PDG2014 AVG.
{$<0.052$}                                        & %   [5]  BABAR
\ifref {\cite{Aubert:2007hb}} \fi \phantom{.}& %   [0]  
$<0.16$                                           & %   [46]  BELLE
\ifref {\cite{Chang:2003yy}} \fi \phantom{.}& %   [0]  
$<0.0038$                                         & %   [82]  CDF
\ifref {\cite{Aaltonen:2013as}} \fi \phantom{.}& %   [0]  
$<00074$ $^\P$                                    & %   [87]  LHCB
\ifref {\cite{Aaij:2013aka}} \fi \phantom{.}& %   [0]  
$<00110$ $^\P$                                    & %   [111]  CMS
\ifref {\cite{Chatrchyan:2013bka}} \fi \phantom{.}& %   [0]  
\red{$<0.00042$} $^\P$                            & %   [113]  ATLAS
\ifref {\cite{Aaboud:2016ire}} \fi \phantom{.}& %   [0]  
{$\cerr{0.00039}{0.00016}{0.00014}$}~$^\S$        \\

461                                               & %         RPP\#
$\mu^+ \mu^- \gamma$                              & %         MODE @kll
$<0.16$                                           & %         PDG2014 AVG.
{$<0.16$}                                         & %   [20]  BABAR
\ifref {\cite{Aubert:2007up}} \fi \phantom{.}& %   [0]  
\nodata                                           & %   [0]  BELLE
\phantom{.}                                       & %   [0]  
\nodata                                           & %   [0]  CDF
\phantom{.}                                       & %   [0]  
\nodata                                           & %   [0]  LHCB
\phantom{.}                                       & %   [0]  
\nodata                                           & %   [0]  CMS
\phantom{.}                                       & %   [0]  
\nodata                                           & %   [0]  ATLAS
\phantom{.}                                       & %   [0]  
{$<0.16$}                                         \\

462                                               & %         RPP\#
$\mu^+ \mu^- \mu^+ \mu^-$                         & %         MODE @kll
$<0.0053$                                         & %         PDG2014 AVG.
\nodata                                           & %   [0]  BABAR
\phantom{.}                                       & %   [0]  
\nodata                                           & %   [0]  BELLE
\phantom{.}                                       & %   [0]  
\nodata                                           & %   [0]  CDF
\phantom{.}                                       & %   [0]  
$<0.0053$                                         & %   [90]  LHCB
\ifref {\cite{Aaij:2013lla}} \fi \phantom{.}& %   [0]  
\nodata                                           & %   [0]  CMS
\phantom{.}                                       & %   [0]  
\nodata                                           & %   [0]  ATLAS
\phantom{.}                                       & %   [0]  
$<0.0053$                                         \\

464                                               & %         RPP\#
$\tau^+ \tau^-$                                   & %         MODE @kll
$<4100$                                           & %         PDG2014 AVG.
{$<4100$}                                         & %   [16]  BABAR
\ifref {\cite{Aubert:2005qw}} \fi \phantom{.}& %   [0]  
\nodata                                           & %   [0]  BELLE
\phantom{.}                                       & %   [0]  
\nodata                                           & %   [0]  CDF
\phantom{.}                                       & %   [0]  
\nodata                                           & %   [0]  LHCB
\phantom{.}                                       & %   [0]  
\nodata                                           & %   [0]  CMS
\phantom{.}                                       & %   [0]  
\nodata                                           & %   [0]  ATLAS
\phantom{.}                                       & %   [0]  
{$<4100$}                                         \\

482                                               & %         RPP\#
$e^\pm \mu^\mp$                                   & %         MODE @leptonviol
$<0.0028$                                         & %         PDG2014 AVG.
{$<0.092$}                                        & %   [5]  BABAR
\ifref {\cite{Aubert:2007hb}} \fi \phantom{.}& %   [0]  
$<0.17$                                           & %   [46]  BELLE
\ifref {\cite{Chang:2003yy}} \fi \phantom{.}& %   [0]  
$<0.064$                                          & %   [83]  CDF
\ifref {\cite{Aaltonen:2009vr}} \fi \phantom{.}& %   [0]  
$<0.0028$                                         & %   [88]  LHCB
\ifref {\cite{Aaij:2013cby}} \fi \phantom{.}& %   [0]  
\nodata                                           & %   [0]  CMS
\phantom{.}                                       & %   [0]  
\nodata                                           & %   [0]  ATLAS
\phantom{.}                                       & %   [0]  
$<0.0028$                                         \\

488                                               & %         RPP\#
$e^\pm \tau^\mp$                                  & %         MODE @leptonviol
$<28$                                             & %         PDG2014 AVG.
{$<28$}                                           & %   [10]  BABAR
\ifref {\cite{Aubert:2008cu}} \fi \phantom{.}& %   [0]  
\nodata                                           & %   [0]  BELLE
\phantom{.}                                       & %   [0]  
\nodata                                           & %   [0]  CDF
\phantom{.}                                       & %   [0]  
\nodata                                           & %   [0]  LHCB
\phantom{.}                                       & %   [0]  
\nodata                                           & %   [0]  CMS
\phantom{.}                                       & %   [0]  
\nodata                                           & %   [0]  ATLAS
\phantom{.}                                       & %   [0]  
{$<28$}                                           \\

489                                               & %         RPP\#
$\mu^\pm \tau^\mp$                                & %         MODE @leptonviol
$<22$                                             & %         PDG2014 AVG.
{$<22$}                                           & %   [10]  BABAR
\ifref {\cite{Aubert:2008cu}} \fi \phantom{.}& %   [0]  
\nodata                                           & %   [0]  BELLE
\phantom{.}                                       & %   [0]  
\nodata                                           & %   [0]  CDF
\phantom{.}                                       & %   [0]  
\nodata                                           & %   [0]  LHCB
\phantom{.}                                       & %   [0]  
\nodata                                           & %   [0]  CMS
\phantom{.}                                       & %   [0]  
\nodata                                           & %   [0]  ATLAS
\phantom{.}                                       & %   [0]  
{$<22$}                                           \\

490                                               & %         RPP\#
$\nu \bar\nu$                                     & %         MODE @neutrino
$<24$                                             & %         PDG2014 AVG.
$<24$                                             & %   [12]  BABAR
\ifref {\cite{Lees:2012wv}} \fi \phantom{.}& %   [0]  
$<130$                                            & %   [60]  BELLE
\ifref {\cite{Hsu:2012uh}} \fi \phantom{.}& %   [0]  
\nodata                                           & %   [0]  CDF
\phantom{.}                                       & %   [0]  
\nodata                                           & %   [0]  LHCB
\phantom{.}                                       & %   [0]  
\nodata                                           & %   [0]  CMS
\phantom{.}                                       & %   [0]  
\nodata                                           & %   [0]  ATLAS
\phantom{.}                                       & %   [0]  
$<24$                                             \\

491                                               & %         RPP\#
$\nu \bar\nu \gamma$                              & %         MODE @neutrino
$<17$                                             & %         PDG2014 AVG.
$<17$                                             & %   [12]  BABAR
\ifref {\cite{Lees:2012wv}} \fi \phantom{.}& %   [0]  
\nodata                                           & %   [0]  BELLE
\phantom{.}                                       & %   [0]  
\nodata                                           & %   [0]  CDF
\phantom{.}                                       & %   [0]  
\nodata                                           & %   [0]  LHCB
\phantom{.}                                       & %   [0]  
\nodata                                           & %   [0]  CMS
\phantom{.}                                       & %   [0]  
\nodata                                           & %   [0]  ATLAS
\phantom{.}                                       & %   [0]  
$<17$                                             \\

%TABLE_BODY

\sglinespt
\end{tabular}
}
\end{center}
\scriptsize{
Results for CDF, LHCb, CMS and ATLAS are relative BFs converted to absolute BFs.\\     %FOOTNOTE
$^\dag$~More recent results exist, with hadronic tagging~\cite{Yook:2014kga},
which do not improve the limits ({$<3.5$} and {$<2.7$}) for $e^+\nu$ and $\mu^+\nu$, respectively).\\ %(Ref.~\cite{39})     % FOOTNOTE
%$^\dag$~More recent results exist, with hadronic tagging (PRD 91, 052016 (Belle)). It does not improve the limits ({\blue{$<3.5$} and \blue{$<2.7$}} for $e^+\nu$ and $\mu^+\nu$, respectively).\\ %(Ref.~\cite{39})     % FOOTNOTE
$^\ddag$~The authors make the average with their previous results, derived from statistically independent samples~\cite{Aubert:2009wt,Adachi:2012mm}.\\      % FOOTNOTE
%$^\ddag$~The authors make the average with their previous results, derived from statistically independent samples. BABAR: PRD 81, 051101(R) (2010), Belle: PRL 110, 131801 (2013).      % FOOTNOTE
$^\S$~This is the combined result obtained by the LHCb and CMS collaborations~\cite{CMS:2014xfa}.\\% (Ref. [109]) .\\      % FOOTNOTE
$^\P$~UL at 95\% CL.      % FOOTNOTE
}
\end{table}

%NEW_TABLE

\begin{table}[!htbp]
\begin{center}
\caption{Relative branching fractions of semileptonic and radiative 
$\Bp$ decays. 
Where values are shown in \red{red} (\blue{blue}), this indicates that
they are new \red{published} (\blue{preliminary}) results since PDG2014.}
%Values in \red{red} (\blue{blue}) are new \red{published}
%(\blue{preliminary}) results since PDG2014.}
\label{tab:radll_rel}
\resizebox{\textwidth}{!}{
\begin{tabular}{|lccc @{}c c @{}c c @{}c c|} \hline
RPP\# &Mode & PDG2014 AVG. & Belle & & \babar & & LHCb & & Our Avg.  \\ \sglinespb
%TABLE_BODY
$~$                                               & %         RPP\#
$10^4\times\mathcal{B}(K^+\pi^+\pi^-\mu^+\mu^-)/\mathcal{B}(\psi(2S)K^+)$& %         MODE
\nodata                                           & %         PDG2014 AVG.
\nodata                                           & %   [0]  BELLE
\phantom{.}                                       & %   [0]  
\nodata                                           & %   [0]  BABAR
\phantom{.}                                       & %   [0]  
\red{$\aerr{6.95}{0.46}{0.43}{0.34}$}             & %   [98]  LHCB
\ifref {\cite{Aaij:2014kwa}} \fi \phantom{.}& %   [0]  
$\cerr{6.95}{0.57}{0.55}$                         \\

\nodata                                           & %         RPP\#
$10^4\times\mathcal{B}(K^+\phi\mu^+\mu^-)/\mathcal{B}(\psi(2S)K^+)$& %         MODE
\nodata                                           & %         PDG2014 AVG.
\nodata                                           & %   [0]  BELLE
\phantom{.}                                       & %   [0]  
\nodata                                           & %   [0]  BABAR
\phantom{.}                                       & %   [0]  
\red{$\aerrsy{1.58}{0.36}{0.32}{0.19}{0.07}$}     & %   [98]  LHCB
\ifref {\cite{Aaij:2014kwa}} \fi \phantom{.}& %   [0]  
$\cerr{1.58}{0.41}{0.33}$                         \\

469                                               & %         RPP\#
$\mathcal{B}(\pi^+\mu^+\mu^-)/\mathcal{B}(K^+\mu^+\mu^-)$~$^\dag$& %         MODE
$\err{0.053}{0.014}{0.01}$                        & %         PDG2014 AVG.
\nodata                                           & %   [0]  BELLE
\phantom{.}                                       & %   [0]  
\nodata                                           & %   [0]  BABAR
\phantom{.}                                       & %   [0]  
\red{$\err{0.038}{0.009}{0.001}$}                 & %   [92]  LHCB
\ifref {\cite{Aaij:2015nea}} \fi \phantom{.}& %   [0]  
$0.038 \pm 0.009$                                 \\

473                                               & %         RPP\#
$\mathcal{B}(K^+\mu^+\mu^-)/\mathcal{B}(K^+e^+e^-)$~$^\ddag$& %         MODE
\nodata                                           & %         PDG2014 AVG.
\nodata                                           & %   [0]  BELLE
\phantom{.}                                       & %   [0]  
\nodata                                           & %   [0]  BABAR
\phantom{.}                                       & %   [0]  
\red{\aerr{0.745}{0.090}{0.074}{0.036}}           & %   [100]  LHCB
\ifref {\cite{Aaij:2014ora}} \fi \phantom{.}& %   [0]  
$\cerr{0.745}{0.097}{0.082}$                      \\

473                                               & %         RPP\#
$\mathcal{B}(K^+\mu^+\mu^-)/\mathcal{B}(K^+e^+e^-)$~$^\S$& %         MODE
\nodata                                           & %         PDG2014 AVG.
$\err{1.03}{0.19}{0.06}$                          & %   [41]  BELLE
\ifref {\cite{Wei:2009zv}} \fi \phantom{.}& %   [0]  
\nodata                                           & %   [0]  BABAR
\phantom{.}                                       & %   [0]  
\nodata                                           & %   [0]  LHCB
\phantom{.}                                       & %   [0]  
$1.03 \pm 0.20$                                   \\

473                                               & %         RPP\#
$\mathcal{B}(K^+\mu^+\mu^-)/\mathcal{B}(K^+e^+e^-)$~$^\P$& %         MODE
\nodata                                           & %         PDG2014 AVG.
\nodata                                           & %   [0]  BELLE
\phantom{.}                                       & %   [0]  
$\aerr{1.00}{0.31}{0.25}{0.07}$                   & %   [31]  BABAR
\ifref {\cite{Lees:2012tva}} \fi \phantom{.}& %   [0]  
\nodata                                           & %   [0]  LHCB
\phantom{.}                                       & %   [0]  
$\cerr{1.00}{0.32}{0.26}$                         \\

\nodata                                           & %         RPP\#
$\mathcal{B}(K^*\mu^+\mu^-)/\mathcal{B}(K^*e^+e^-)$~$^\S$& %         MODE
\nodata                                           & %         PDG2014 AVG.
$\err{0.83}{0.17}{0.08}$                          & %   [41]  BELLE
\ifref {\cite{Wei:2009zv}} \fi \phantom{.}& %   [0]  
\nodata                                           & %   [0]  BABAR
\phantom{.}                                       & %   [0]  
\nodata                                           & %   [0]  LHCB
\phantom{.}                                       & %   [0]  
$0.83 \pm 0.19$                                   \\

\nodata                                           & %         RPP\#
$\mathcal{B}(K^*\mu^+\mu^-)/\mathcal{B}(K^*e^+e^-)$~$^\P$& %         MODE
\nodata                                           & %         PDG2014 AVG.
\nodata                                           & %   [0]  BELLE
\phantom{.}                                       & %   [0]  
$\aerr{1.013}{0.34}{0.26}{0.010}$                 & %   [31]  BABAR
\ifref {\cite{Lees:2012tva}} \fi \phantom{.}& %   [0]  
\nodata                                           & %   [0]  LHCB
\phantom{.}                                       & %   [0]  
$\cerr{1.013}{0.340}{0.260}$                      \\

%TABLE_BODY
%\\
\hline
\end{tabular}
}
\end{center}
\scriptsize{
$^\dag$~For $0.1<m^2(\ell^+\ell^-)<6.0~\gevgevcccc$.\\      % FOOTNOTE
$^\ddag$~For $1.0<m^2(\ell^+\ell^-)<6.0~\gevgevcccc$.\\      % FOOTNOTE
$^\S$~For the full $m^2(\ell^+\ell^-)$ range.\\      % FOOTNOTE
$^\P$~For $0.10<m^2(\ell^+\ell^-)<8.12~\gevgevcccc$ and $m^2(\ell^+\ell^-)>10.11~\gevgevcccc$.\\      % FOOTNOTE
}
\end{table}

%NEW_TABLE

\begin{table}[!htbp]
\begin{center}
\caption{Branching fractions of $\Bp/\Bz\to\bar{q}$ gluon decays in units of $\times 10^{-6}$. Upper limits are
at 90\% CL. 
Where values are shown in \red{red} (\blue{blue}), this indicates that
they are new \red{published} (\blue{preliminary}) results since PDG2014.}
%Values in \red{red} (\blue{blue}) are new \red{published}
%(\blue{preliminary}) results since PDG2014.}
\label{tab:radll_gluon}
\resizebox{\textwidth}{!}{
\begin{tabular}{|lccc @{}c c @{}c c @{}c c|} \hline
RPP\# &Mode & PDG2014 Avg. & \babar & & Belle & & CLEO & & Our Avg. \\
\sglinespb
%TABLE_BODY
$~80$                                             & %         RPP\#
$\eta X$                                          & %         MODE
$\cerr{260}{50}{80}$                              & %         PDG2014 AVG.
\nodata                                           & %   [0]  BABAR
\phantom{.}                                       & %   [0]  
{\berr{261}{30}{44}{74}}~$^\S$                    & %   [59]  BELLE
\ifref {\fontsize{11}{11}\selectfont \cite{Nishimura:2009ae}} \fi \phantom{.}& %   [0]  
$<440$                                            & %   [81]  CLEO
\ifref {\fontsize{11}{11}\selectfont \cite{Browder:1998yb}} \fi \phantom{.}& %   [0]  
$\cerr{261}{53}{79}$                              \\

$~81$                                             & %         RPP\#
$\etapr X$                                         & %         MODE
$420 \pm 90$                                      & %         PDG2014 AVG.
\err{390}{80}{90}~$^\dag$                         & %   [30]  BABAR
\ifref {\fontsize{11}{11}\selectfont \cite{Aubert:2004eq}} \fi \phantom{.}& %   [0]  
\nodata                                           & %   [0]  BELLE
\phantom{.}                                       & %   [0]  
\err{460}{110}{60}~$^\dag$                        & %   [79]  CLEO
\ifref {\fontsize{11}{11}\selectfont \cite{Bonvicini:2003aw}} \fi \phantom{.}& %   [0]  
$423 \pm 86$                                      \\

$~82$                                             & %         RPP\#
$K^+ X$                                           & %         MODE
$<187$                                            & %         PDG2014 AVG.
$<187$~$^\ddag$                                   & %   [19]  BABAR
\ifref {\fontsize{11}{11}\selectfont \cite{delAmoSanchez:2010gx}} \fi \phantom{.}& %   [0]  
\nodata                                           & %   [0]  BELLE
\phantom{.}                                       & %   [0]  
\nodata                                           & %   [0]  CLEO
\phantom{.}                                       & %   [0]  
$<187$                                            \\

$~83$                                             & %         RPP\#
$K^0 X$                                           & %         MODE
\cerr{195}{71}{67}                                & %         PDG2014 AVG.
\aerr{195}{51}{45}{50}~$^\ddag$                   & %   [19]  BABAR
\ifref {\fontsize{11}{11}\selectfont \cite{delAmoSanchez:2010gx}} \fi \phantom{.}& %   [0]  
\nodata                                           & %   [0]  BELLE
\phantom{.}                                       & %   [0]  
\nodata                                           & %   [0]  CLEO
\phantom{.}                                       & %   [0]  
$\cerr{195}{71}{67}$                              \\

$~94$                                             & %         RPP\#
$\pi^+ X$                                         & %         MODE
$370\pm80$                                        & %         PDG2014 AVG.
{\aerr{372}{50}{47}{59}}~$^\P$                    & %   [19]  BABAR
\ifref {\fontsize{11}{11}\selectfont \cite{delAmoSanchez:2010gx}} \fi \phantom{.}& %   [0]  
\nodata                                           & %   [0]  BELLE
\phantom{.}                                       & %   [0]  
\nodata                                           & %   [0]  CLEO
\phantom{.}                                       & %   [0]  
$\cerr{372}{77}{75}$                              \\

%TABLE_BODY

\hline
\end{tabular}
}
\end{center}
\scriptsize{
$^\dag$~$2.0 < p^*(\etapr) < 2.7~\gevc$. \\      % FOOTNOTE
$^\ddag$~$m_{X} < 1.69~\gevcc$. \\      % FOOTNOTE
$^\S$~$0.4 < m_{X} < 2.6~\gevcc$. \\      % FOOTNOTE
$^\P$~$m_{X} < 1.71~\gevcc$.      % FOOTNOTE
}
\end{table}

%NEW_TABLE

\begin{table}[!htbp]
\begin{center}
\caption{Isospin asymmetry in radiative and semileptonic $B$ meson decays.
The notations are those adopted by the PDG.
Where values are shown in \red{red} (\blue{blue}), this indicates that
they are new \red{published} (\blue{preliminary}) results since PDG2014.}
%Values in \red{red} (\blue{blue}) are new \red{published}
%(\blue{preliminary}) results since PDG2014.}
\label{tab:radll_AI}
\resizebox{\textwidth}{!}{
\begin{tabular}{|cccc @{}c c @{}c c @{}c c|}
\sgline
Parameter & & PDG2014 Avg. & \babar & & Belle & & LHCb & & Our Avg. \\
\sglinespb
%TABLE_BODY
$\Delta_{0^-}(X_s\gamma)$                         & %         PARAMETER
\phantom{.}                                       & %         
$-0.01 \pm 0.06$                                  & %         PDG2014 AVG.
$-0.01 \pm 0.06$~$^\ddag$                         & %   [1]  BABAR
\ifref {\fontsize{8}{11}\selectfont \cite{Aubert:2005cua,Aubert:2007my}} \fi \phantom{.}& %   [0]  
\nodata                                           & %   [0]  BELLE
\phantom{.}                                       & %   [0]  
\nodata                                           & %   [0]  LHCB
\phantom{.}                                       & %   [0]  
$-0.01 \pm 0.06$                                  \\

$\Delta_{0^+}(K^* \gamma)$                        & %         PARAMETER
\phantom{.}                                       & %         
$0.052 \pm 0.026$                                 & %         PDG2014 AVG.
\err{0.066}{0.021}{0.022}                         & %   [1]  BABAR
\ifref {\fontsize{8}{11}\selectfont \cite{Aubert:2009ak}} \fi \phantom{.}& %   [0]  
\err{0.012}{0.044}{0.026}                         & %   [42]  BELLE
\ifref {\fontsize{8}{11}\selectfont \cite{Nakao:2004th}} \fi \phantom{.}& %   [0]  
\nodata                                           & %   [0]  LHCB
\phantom{.}                                       & %   [0]  
$0.052 \pm 0.026$                                 \\

$\Delta_{\rho \gamma}$                            & %         PARAMETER
\phantom{.}                                       & %         
$-0.46 \pm 0.17$                                  & %         PDG2014 AVG.
{$\aerr{-0.43}{0.25}{0.22}{0.10}$}                & %   [1]  BABAR
\ifref {\fontsize{8}{11}\selectfont \cite{Aubert:2008al}} \fi \phantom{.}& %   [0]  
{$\aerrsy{-0.48}{0.21}{0.19}{0.08}{0.09}$}        & %   [54]  BELLE
\ifref {\fontsize{8}{11}\selectfont \cite{Taniguchi:2008ty}} \fi \phantom{.}& %   [0]  
\nodata                                           & %   [0]  LHCB
\phantom{.}                                       & %   [0]  
$\cerr{-0.46}{0.17}{0.16}$                        \\

$\Delta_{0-}(K\ell\ell)~^\dag$                    & %         PARAMETER
\phantom{.}                                       & %         
$-0.37 \pm 0.13$                                  & %         PDG2014 AVG.
{$\err{-0.41}{0.25}{0.01}$}                       & %   [1]  BABAR
\ifref {\fontsize{8}{11}\selectfont \cite{Lees:2012tva}} \fi \phantom{.}& %   [0]  
{$\aerr{-0.41}{0.25}{0.20}{0.07}$}                & %   [41]  BELLE
\ifref {\fontsize{8}{11}\selectfont \cite{Wei:2009zv}} \fi \phantom{.}& %   [0]  
\red{$\aerr{-0.10}{0.08}{0.09}{0.02}$}~$^\S$      & %   [95]  LHCB
\ifref {\fontsize{8}{11}\selectfont \cite{Aaij:2014pli}} \fi \phantom{.}& %   [0]  
$-0.16 \pm 0.08$                                  \\

$\Delta_{0-}(K^*\ell\ell)~^\dag$                  & %         PARAMETER
\phantom{.}                                       & %         
$-0.22 \pm 0.10$                                  & %         PDG2014 AVG.
$\aerr{-0.20}{0.30}{0.23}{0.03}$                  & %   [1]  BABAR
\ifref {\fontsize{8}{11}\selectfont \cite{Lees:2012tva}} \fi \phantom{.}& %   [0]  
$\aerr{0.33}{0.37}{0.43}{0.08}$                   & %   [41]  BELLE
\ifref {\fontsize{8}{11}\selectfont \cite{Wei:2009zv}} \fi \phantom{.}& %   [0]  
\red{$\aerr{0.00}{0.12}{0.10}{0.02}$}~$^\S$       & %   [95]  LHCB
\ifref {\fontsize{8}{11}\selectfont \cite{Aaij:2014pli}} \fi \phantom{.}& %   [0]  
$\cerr{-0.01}{0.11}{0.09}$                        \\

%TABLE_BODY

\sglinespt
\end{tabular}
}
\end{center}
\scriptsize{
In some of the $B$-factory results it is assumed that
$\mathcal{B}(\Upsilon(4S) \to B^+ B^-) = \mathcal{B}(\Upsilon(4S) \to B^0 \overline{B}{}^0)$,
and in others a measured value of the ratio of branching fractions is used.
See original papers for details.
The averages quoted above are computed naively and should be treated with caution.\\
$^\dag$~Results given for the bin $1<m^2(\ell^+\ell^-)<6~\gevgevcccc$, see references for the other bins.\\      % FOOTNOTE
$^\ddag$~Average of two independent measurements from \babar.\\      % FOOTNOTE
$^\S$~Only muons are used. \\      % FOOTNOTE
%% THERE ARE NO BABAR HYPERLINK REFERENCES FOR THIS TABLE. PUT [1] FOR NOW SO AVERAGES RUN OK   % FOOTNOTE
}
\end{table}

%\newpage
\clearpage
 List of other measurements that are not included in the tables:
\begin{itemize}
\item  $B^+ \to K^+ \pi^- \pi^+ \gamma$ : LHCb has measured the up-down asymmetries in bins of the $K\pi\pi\gamma$ mass~%[97].
\cite{Aaij:2014wgo}.
\item  In Ref.~%[99], 
\cite{Aaij:2013hha},
LHCb has also measured the branching fraction of $B^+ \to K^+ e^- e^+$  in the $m^2(\ell \ell)$ bin $[1, 6]~\gevgevcccc$.
\item  In the $B^+ \to \pi^+ \mu^+ \mu^-$ paper~%[92],
\cite{Aaij:2015nea},
LHCb has also measured the differential branching fraction in bins of  $m^2(\ell \ell)$.
\item For $B \to K \ell^- \ell^+$, LHCb has measured $F_H$ and $A_{\rm FB}$ in 17 (5) bins of $m^2(\ell \ell)$  for the $K^+$ (\ks) final state~%[105].
\cite{Aaij:2014tfa}.
Belle has measured $F_L$ and $A_{\rm FB}$ in 6 $m^2(\ell \ell)$ bins~%[64].% no babar results ?
\cite{Abdesselam:2016llu}. % no babar results?
\item For the  $B \to K^{*} \ell^- \ell^+$ analyses, partial branching fractions and angular observables in bins of $m^2(\ell\ell)$ are also available:  
\begin{itemize}
\item   $B^0 \to K^{*0} e^- e^+$ : LHCb has measured $F_L$, $A_T^{(2)}$, $A_T^{\rm Im}$, $A_T^{\rm Re}$ in the $[0.002, 1.120]~\gevgevcccc$ bin of $m^2(\ell \ell)$~%[101],
\cite{Aaij:2015dea},
and has also determined the branching fraction in the dilepton mass region $[10,1000]~\mevcc$~%[103].
\cite{Aaij:2013hha}.
\item   $B \to K^{*} \ell^- \ell^+$ : Belle has measured $F_L$, $A_{\rm FB}$, isospin asymmetry in 6 $m^2(\ell \ell)$ bins~%[41]
\cite{Wei:2009zv}
and $P_4'$, $P_5'$, $P_6'$, $P_8'$ in 4 $m^2(\ell \ell)$ bins~%[64].
\cite{Abdesselam:2016llu}.
\babar\ has measured $F_L$, $A_{\rm FB}$, $P_2$ in 5 $m^2(\ell \ell)$ bins~%[37].
\cite{Lees:2015ymt}.
\item   $B^0 \to K^{*0} \mu^- \mu^+$ : LHCb has measured $F_L$, $A_{\rm FB}$, $S_3-S_9$, $A_3-A_9$, $P_1-P_3$, $P_4'-P_8'$ in 8 $m^2(\ell \ell)$ bins~%[104].
\cite{Aaij:2015oid}.
CMS has  measured $F_L$ and $A_{\rm FB}$ in 7 $m^2(\ell \ell)$ bins~%[115].
\cite{Khachatryan:2015isa}.
%ATLAS has measured  $F_L$ and $A_{\rm FB}$ in 5 $m^2(\ell \ell)$ bins~[114]
%%\cite{ATLAS:2013ola}
%% --> conference note that was withdrawn by ATLAS. Should be replaced by ATLAS-CONF-2017-023.

\end{itemize}
\item For $B \to X_s \ell^- \ell^+$ ($X_s$ is a hadronic system with an $s$ quark), Belle has measured $A_{\rm FB}$ in bins of $m^2(\ell \ell)$ with a sum of 10 exclusive final states~%[61].
\cite{Sato:2014pjr}.

\item  $B^0 \to K^+ \pi^- \mu^+ \mu^-$, with $1330 < m(K^+ \pi^-) < 1530~\gevcc$: LHCb has measured the partial branching fraction in bins of  $m^2(\mu^- \mu^+)$ in the range $[0.1,8.0]~\gevgevcccc$, and has also determined angular moments~%[107]. 
\cite{Aaij:2016kqt}.

\end{itemize}

\clearpage

\mysubsection{Charge asymmetries in \b-hadron decays}
\label{sec:rare-acp}

This section contains, in Tables~\ref{tab:acp_Bp1} to~\ref{tab:acp_Lb},
compilations of \CP\ asymmetries in decays of various \b-hadrons: \Bp, \Bz
mesons, $\Bpm/\Bz$ admixtures, \Bs mesons and finally \Lb baryons.
Measurements of time-dependent \CP\ asymmetries are not listed here but are
discussed in Sec.~\ref{sec:cp_uta}.

\begin{figure}[htbp!]
\centering
\includegraphics[width=0.5\textwidth]{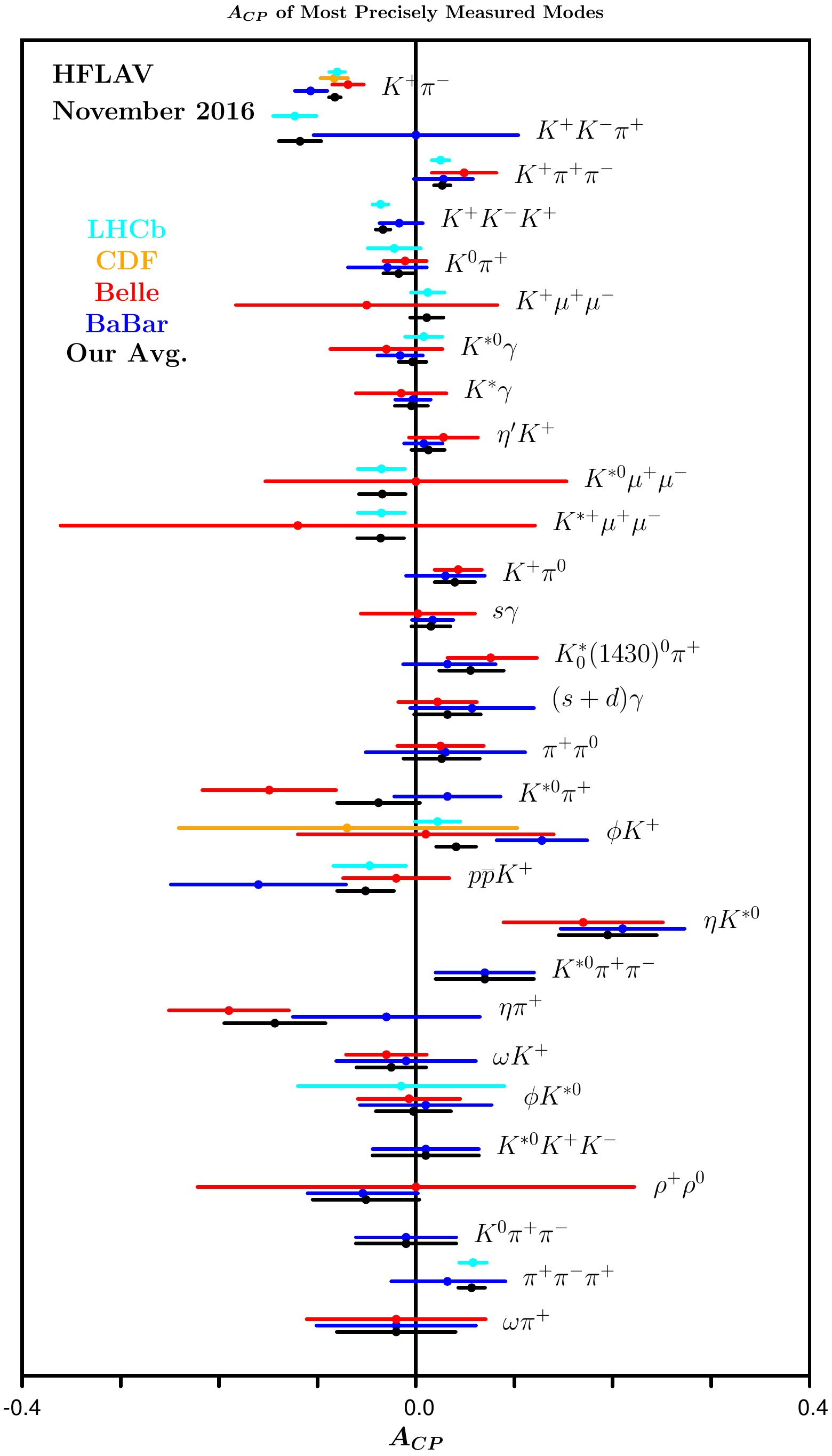}
\caption{$A_{\CP}$ of most precisely measured modes.}
\label{fig:rare-acpselect}
\end{figure}

Figure~\ref{fig:rare-acpselect} shows a graphic representation of a selection of results given in this section.
Footnote symbols indicate that the footnote in the corresponding table should be consulted.
%For comments in the plot, marked with a symbol or a number, refer to the corresponding table.

\newpage

%NEW_TABLE

\begin{sidewaystable}[!htbp]
\begin{center}
\caption{\CP\ asymmetries of charmless hadronic $\Bp$ decays (part 1).
Where values are shown in \red{red} (\blue{blue}), this indicates that
they are new \red{published} (\blue{preliminary}) results since PDG2014.}
%Values in \red{red} (\blue{blue}) are new \red{published}
%(\blue{preliminary}) results since PDG2014.}
\label{tab:acp_Bp1}
\resizebox{0.92\textwidth}{!}{

}
\end{center}
\vspace{-0.35cm}
\tiny $^\dag$~PDG takes the value from the \babar\ amplitude analysis of $B^+ \to K^+ K^- K^+$, while our numbers are from amplitude analyses of $B^+ \to K^+ \pi^- \pi^+$.\\     %FOOTNOTE
\tiny $^\ddag$~PDG uses also a result from CLEO.     %FOOTNOTE
\end{sidewaystable}
\clearpage

%NEW_TABLE

\begin{sidewaystable}[!htbp]
\begin{center}
\caption{\CP\ asymmetries of charmless hadronic $\Bp$ decays (part 2).
Where values are shown in \red{red} (\blue{blue}), this indicates that
they are new \red{published} (\blue{preliminary}) results since PDG2014.}
%Values in \red{red} (\blue{blue}) are new \red{published}
%(\blue{preliminary}) results since PDG2014.}
\label{tab:acp_Bp2}
\resizebox{0.92\textwidth}{!}{

}
\end{center}
\vspace{-0.35cm}
\tiny
$^\dag$~PDG uses also a result from CLEO.\\     %FOOTNOTE
$^\ddag$~PDG swaps the Belle results corresponding to $A_{C\!P}(p\bar{p}\pi^+)$ and $A_{C\!P}(p\bar{p}K^+)$.\\
$^\S$~PDG uses also a previous result from \babar\ (\cite{Aubert:2008ps}).\\%(\cite{12})     %FOOTNOTE
$^\P$~LHCb also quotes results in bins of $m(\ell^+\ell^-)^2$.     %FOOTNOTE
\end{sidewaystable}
\clearpage

%NEW_TABLE

\begin{sidewaystable}[!htbp]
\begin{center}
\caption{\CP\ asymmetries of charmless hadronic $\Bz$ decays.
Where values are shown in \red{red} (\blue{blue}), this indicates that
they are new \red{published} (\blue{preliminary}) results since PDG2014.}
%Values in \red{red} (\blue{blue}) are new \red{published}
%(\blue{preliminary}) results since PDG2014.}
\label{tab:acp_Bz}
\resizebox{\textwidth}{!}{

}
\end{center}
\scriptsize
Measurements of time-dependent $\CP$ asymmetries are listed in     %FOOTNOTE
%the Unitarity Triangle home page. (http://www.slac.stanford.edu/xorg/hfag/triangle/index.html)\\ \\     %FOOTNOTE
Sec.~\ref{sec:cp_uta}.\\ \\     %FOOTNOTE
\scriptsize
$^\dag$~PDG uses also a result from CLEO.\\     %FOOTNOTE
$^\ddag$ Average of \babar\ results from     %FOOTNOTE
$B^0 \to K^+ \pi^- \pi^0$ and $B^0 \to K^0 \pi^+ \pi^-$.\\     %FOOTNOTE
$^\S$ PDG quotes the opposite asymmetry. \\     %FOOTNOTE
$^\P$~Extracted from measured $\Delta A_{\CP}=A_{\CP}     %FOOTNOTE
(\phi K^{*0})-A_{\CP}(J/\psi K^{*0})=     %FOOTNOTE
\red{\err{0.015}{0.032}{0.005}}$.\\     %FOOTNOTE
$^\diamond$~LHCb also quotes results in bins of $m(\ell^+\ell^-)^2$.     %FOOTNOTE
%{\normalsize All entries are time-integrated except where     %FOOTNOTE
%indicated by labelled coefficients $S_{xy}$ and $C_{xy}$.     %FOOTNOTE
%Note     %FOOTNOTE
%$$S_{xy}\equiv     %FOOTNOTE
%\frac{2Im(\lambda)}{1+|\lambda|^2}$$     %FOOTNOTE
%and     %FOOTNOTE
%$$C_{xy}\equiv     %FOOTNOTE
%\frac{1-|\lambda|^2}{1+|\lambda|^2}     %FOOTNOTE
%\equiv -A_{xy}$$     %FOOTNOTE
%where $\lambda = {\cal A}(\bar B\to f_{\CP})/{\cal A}(B\to f_{\CP})$.}     %FOOTNOTE

\end{sidewaystable}

\clearpage

%NEW_TABLE

\begin{table}[!htbp]
\begin{center}
\caption{\CP\ asymmetries of charmless hadronic decays of $B^\pm/B^0$ admixture.
Where values are shown in \red{red} (\blue{blue}), this indicates that
they are new \red{published} (\blue{preliminary}) results since PDG2014.}
%Values in \red{red} (\blue{blue}) are new \red{published}
%(\blue{preliminary}) results since PDG2014.}
\label{tab:acp_B}
\resizebox{\textwidth}{!}{
\begin{tabular}{|lccc @{}c c @{}c c|}
\hline
RPP\#    & Mode & PDG2014 Avg. & \babar & & Belle & & Our Avg. \\
\sglinespb
%TABLE_BODY
~65                                               & %         RPP\#
$K^* \gamma$                                      & %         MODE @acp4,acpselect
$-0.003\pm 0.017$~$^\dag$                         & %         PDG2014 AVG.
{$-0.003\pm 0.017\pm 0.007$}                      & %   [41]  BABAR
\ifref {\cite{Aubert:2009ak}} \fi \phantom{.}& %   [0]  
{$-0.015\pm 0.044\pm 0.012$}                      & %   [56]  BELLE
\ifref {\cite{Nakao:2004th}} \fi \phantom{.}& %   [0]  
$-0.005 \pm 0.017$                                \\

~77                                               & %         RPP\#
$s \gamma$                                        & %         MODE @acp4,acpselect
$-0.008\pm0.029$                                  & %         PDG2014 AVG.
\red{$\err{0.017}{0.019}{0.010}$}~$^\ddag$        & %   [2]  BABAR
\ifref {\cite{Lees:2014uoa}} \fi \phantom{.}& %   [0]  
$\err{0.002}{0.050}{0.030}$                       & %   [61]  BELLE
\ifref {\cite{Nishida:2003paa}} \fi \phantom{.}& %   [0]  
$0.015 \pm 0.020$                                 \\

\nodata                                           & %         RPP\#
$(s+d) \gamma$                                    & %         MODE @acp4,acpselect
$-0.01\pm0.05$                                    & %         PDG2014 AVG.
$\err{0.057}{0.060}{0.018}$~$^\S$                 & %   [46]  BABAR
\ifref {\cite{Lees:2012ym}} \fi \phantom{.}& %   [0]  
\red{$\err{0.022}{0.039}{0.009}$}~$^\diamond$     & %   [77]  BELLE
\ifref {\cite{Pesantez:2015fza}} \fi \phantom{.}& %   [0]  
$0.032 \pm 0.034$                                 \\

~80                                               & %         RPP\#
$s \eta$                                          & %         MODE @acp2,acpselect
$\cerr{-0.13}{0.04}{0.05}$                        & %         PDG2014 AVG.
\nodata                                           & %   [0]  BABAR
\phantom{.}                                       & %   [0]  
{\berr{-0.13}{0.04}{0.02}{0.03}}                  & %   [72]  BELLE
\ifref {\cite{Nishimura:2009ae}} \fi \phantom{.}& %   [0]  
$\cerr{-0.13}{0.04}{0.05}$                        \\

~86                                               & %         RPP\#
$\pi^+ X$                                         & %         MODE @acp5
$0.10 \pm 0.17$                                   & %         PDG2014 AVG.
{\err{0.10}{0.16}{0.05}}                          & %   [42]  BABAR
\ifref {\cite{delAmoSanchez:2010gx}} \fi \phantom{.}& %   [0]  
\nodata                                           & %   [0]  BELLE
\phantom{.}                                       & %   [0]  
$0.10 \pm 0.17$                                   \\

121                                               & %         RPP\#
$s \ell \ell$                                     & %         MODE @acp4
$-0.22\pm0.26$                                    & %         PDG2014 AVG.
\red{$\err{0.04}{0.11}{0.01}$}                    & %   [14]  BABAR
\ifref {\cite{Lees:2013nxa}} \fi \phantom{.}& %   [0]  
\nodata                                           & %   [0]  BELLE
\phantom{.}                                       & %   [0]  
$0.04 \pm 0.11$                                   \\

126                                               & %         RPP\#
$K^*e^+e^-$                                       & %         MODE @acp4
$-0.18 \pm 0.15$                                  & %         PDG2014 AVG.
\nodata                                           & %   [0]  BABAR
\phantom{.}                                       & %   [0]  
$\err{-0.18}{0.15}{0.01}$                         & %   [71]  BELLE
\ifref {\cite{Wei:2009zv}} \fi \phantom{.}& %   [0]  
$-0.18 \pm 0.15$                                  \\

128                                               & %         RPP\#
$K^*\mu^+\mu^-$                                   & %         MODE @acp4
$-0.03 \pm 0.13$                                  & %         PDG2014 AVG.
\nodata                                           & %   [0]  BABAR
\phantom{.}                                       & %   [0]  
$\err{-0.03}{0.13}{0.02}$                         & %   [71]  BELLE
\ifref {\cite{Wei:2009zv}} \fi \phantom{.}& %   [0]  
$-0.03 \pm 0.13$                                  \\

129                                               & %         RPP\#
$K \ell \ell$                                     & %         MODE @acp4
\nodata                                           & %         PDG2014 AVG.
\red{$\err{-0.03}{0.14}{0.01}$}                   & %   [19]  BABAR
\ifref {\cite{Lees:2012tva}} \fi \phantom{.}& %   [0]  
\nodata                                           & %   [71]  BELLE
\phantom{.}                                       & %   [0]  
$-0.03 \pm 0.14$                                  \\

130                                               & %         RPP\#
$K^* \ell \ell$                                   & %         MODE @acp4
$-0.04 \pm 0.07$                                  & %         PDG2014 AVG.
$\err{0.03}{0.13}{0.01}$~$^\P$                    & %   [19]  BABAR
\ifref {\cite{Lees:2012tva}} \fi \phantom{.}& %   [0]  
$\err{-0.10}{0.10}{0.01}$                         & %   [71]  BELLE
\ifref {\cite{Wei:2009zv}} \fi \phantom{.}& %   [0]  
$-0.05 \pm 0.08$                                  \\

%TABLE_BODY
\hline
\end{tabular}
}
\end{center}
%~~~~~~~~~~~~~~~~~~\dag~$p^* > 2.34$ GeV;     %FOOTNOTE
%~\S~$0.4 < M_{X_s} < 2.6$ GeV;     %FOOTNOTE
\scriptsize $^\dag$~PDG includes also a result from CLEO.\\     %FOOTNOTE
\scriptsize $^\ddag$~\babar\ also measures the difference in direct $C\!P$ asymmetry for charged and neutral $B$ mesons: $\Delta A_{C\!P}= +(5.0\pm3.9\pm1.5)\%$.\\     %FOOTNOTE
\scriptsize $^\S$~There is another \babar\ result using the recoil method %(Phys. Rev. D 77, 051103),     %FOOTNOTE
\cite{Aubert:2007my},     %FOOTNOTE
and a CLEO result %(Phys. Rev. Lett. 86, 5661)     %FOOTNOTE
\cite{Coan:2000pu}     %FOOTNOTE
that are used in the PDG average.\\     %FOOTNOTE
\scriptsize $^\P$~Previous \babar\ result is also included in the PDG Average.\\     %FOOTNOTE
\scriptsize $^\diamond$~Requires $E_\gamma >2.1$~GeV. \\     %FOOTNOTE
\end{table}

%NEW_TABLE

\begin{table}[!htbp]
\begin{center}
\caption{\CP\ asymmetries of charmless hadronic $\Bs$ decays.
Where values are shown in \red{red} (\blue{blue}), this indicates that
they are new \red{published} (\blue{preliminary}) results since PDG2014.}
%Values in \red{red} (\blue{blue}) are new \red{published}
%(\blue{preliminary}) results since PDG2014.}
\label{tab:acp_Bs}
%\hspace*{-1.2cm}
%\small
\resizebox{\textwidth}{!}{
\begin{tabular}{|lccc @{}c c @{}c c|}
\hline
RPP\#    & Mode & PDG2014 Avg. & CDF & & LHCb & & Our Avg. \\
\sglinespb
%TABLE_BODY
~52                                               & %         RPP\#
$\pi^+ K^-$                                       & %         MODE
$0.28 \pm 0.04$                                   & %         PDG2014 AVG.
\red{$\err{0.22}{0.07}{0.02}$}                    & %   [82]  CDF
\ifref {\cite{Aaltonen:2014vra}} \fi \phantom{.}& %   [0]  
$\err{0.27}{0.04}{0.01}$                          & %   [91]  LHCB
\ifref {\cite{Aaij:2013iua}} \fi \phantom{.}& %   [0]  
$0.26 \pm 0.04$                                   \\

%TABLE_BODY
\hline
\end{tabular}
}
\end{center}
\end{table}

%NEW_TABLE

\begin{table}[!htbp]
\begin{center}
\caption{\CP\ asymmetries of charmless hadronic $\Lb$ decays.
Where values are shown in \red{red} (\blue{blue}), this indicates that
they are new \red{published} (\blue{preliminary}) results since PDG2014.}
%Values in \red{red} (\blue{blue}) are new \red{published}
%(\blue{preliminary}) results since PDG2014.}
\label{tab:acp_Lb}
\resizebox{\textwidth}{!}{
\begin{tabular}{|lccc @{}c c @{}c c|}
\sgline
RPP\#   & Mode & PDG2014 Avg. & CDF & & LHCb & & Our Avg. \\
\hline
%TABLE_BODY
$~21$                                             & %         RPP\#
$p\pi^-$                                          & %         MODE @Lb
$0.03 \pm 0.18$                                   & %         PDG2014 AVG.
\red{$\err{0.06}{0.07}{0.03}$}                    & %   [82]  CDF
\ifref {\cite{Aaltonen:2014vra}} \fi \phantom{.}& %   [0]  
\nodata                                           & %   [0]  LHCB
\phantom{.}                                       & %   [0]  
$0.06 \pm 0.08$                                   \\

$~22$                                             & %         RPP\#
$p K^-$                                           & %         MODE @Lb
$0.37 \pm 0.17$                                   & %         PDG2014 AVG.
\red{$\err{-0.10}{0.08}{0.04}$}                   & %   [82]  CDF
\ifref {\cite{Aaltonen:2014vra}} \fi \phantom{.}& %   [0]  
\nodata                                           & %   [0]  LHCB
\phantom{.}                                       & %   [0]  
$-0.10 \pm 0.09$                                  \\

\nodata                                           & %         RPP\#
$\kzb p \pi^-$                                    & %         MODE @Lb
\nodata                                           & %         PDG2014 AVG.
\nodata                                           & %   [0]  CDF
\phantom{.}                                       & %   [0]  
\red{$\err{0.22}{0.13}{0.03}$}                    & %   [101]  LHCB
\ifref {\cite{Aaij:2014lpa}} \fi \phantom{.}& %   [0]  
$0.22 \pm 0.13$                                   \\

\nodata                                           & %         RPP\#
$\Lambda K^+\pi^-$                                & %         MODE @Lb
\nodata                                           & %         PDG2014 AVG.
\nodata                                           & %   [0]  CDF
\phantom{.}                                       & %   [0]  
\red{$\err{-0.53}{0.23}{0.11}$}                   & %   [102]  LHCB
\ifref {\cite{Aaij:2016nrq}} \fi \phantom{.}& %   [0]  
$-0.53 \pm 0.26$                                  \\

\nodata                                           & %         RPP\#
$\Lambda K^+ K^-$                                 & %         MODE @Lb
\nodata                                           & %         PDG2014 AVG.
\nodata                                           & %   [0]  CDF
\phantom{.}                                       & %   [0]  
\red{$\err{-0.28}{0.10}{0.07}$}                   & %   [102]  LHCB
\ifref {\cite{Aaij:2016nrq}} \fi \phantom{.}& %   [0]  
$-0.28 \pm 0.12$                                  \\

%TABLE_BODY
\sglinespt
\end{tabular}
}
\end{center}
\end{table}

\clearpage

List of other measurements that are not included in the tables:
\begin{itemize}

\item In the paper %LHCB-PAPER-2016-030,
\cite{Aaij:2016cla},
LHCb has measured the triple-product asymmetries for the decays $\Lb \to p\pi^-\pi^+\pi^-$ and $\Lb \to p\pi^- K^+ K^-$.
\end{itemize}

\clearpage

\mysubsection{Polarization measurements in \b-hadron decays}
\label{sec:rare-polar}

In this section, compilations of polarization measurements in \b-hadron decays
are given. Table~\ref{tab:polar_Bp} (\ref{tab:polar_Bz}) details measurements
of the longitudinal fraction, $f_L$, in \Bp\ (\Bz) decays, and
Table~\ref{tab:polar_BpAng} (\ref{tab:polar_BzAng}) the results of the full
angular analyses of $\Bp$ (\Bz) $\to\phi\Kstar$ decays. 
Table~\ref{tab:polar_BzAng1430} gives results of the full angular analysis of
$\Bz\to\phi K_2^{\ast 0}(1430)$ decays. 
Tables~\ref{tab:polar_Bs} to~\ref{tab:polar_BsAng2} detail quantities of \Bs decays: $f_L$ measurements, and observables from full angular analyses of decays to $\phi\phi$ and $\phi\Kstarzb$.

\begin{figure}[p!]
\centering
\includegraphics[width=0.5\textwidth]{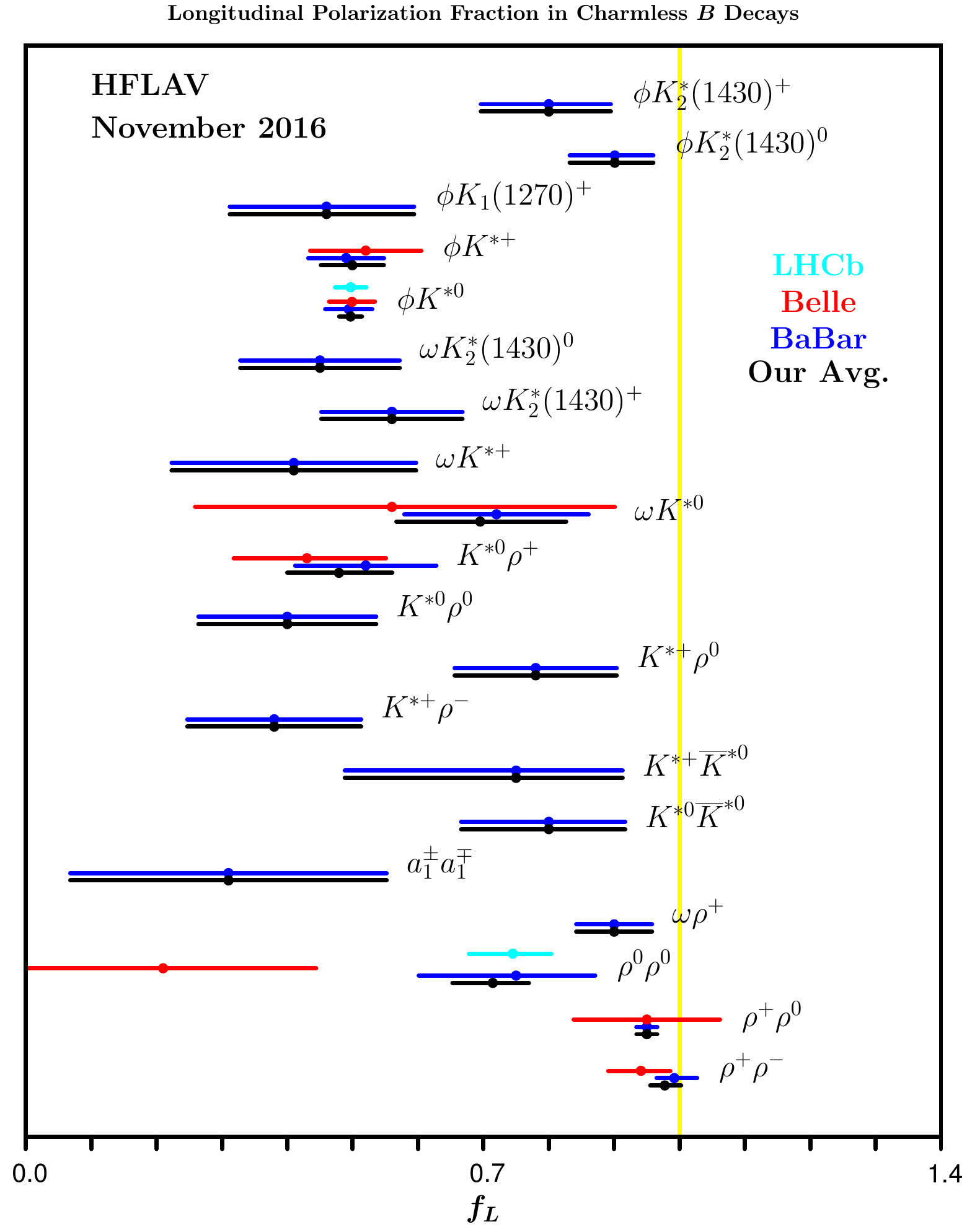}
\caption{Longitudinal polarization fraction in charmless $B$ decays.}
\label{fig:rare-polar}
\end{figure}

\begin{figure}[ht!]
\centering
\includegraphics[width=0.5\textwidth]{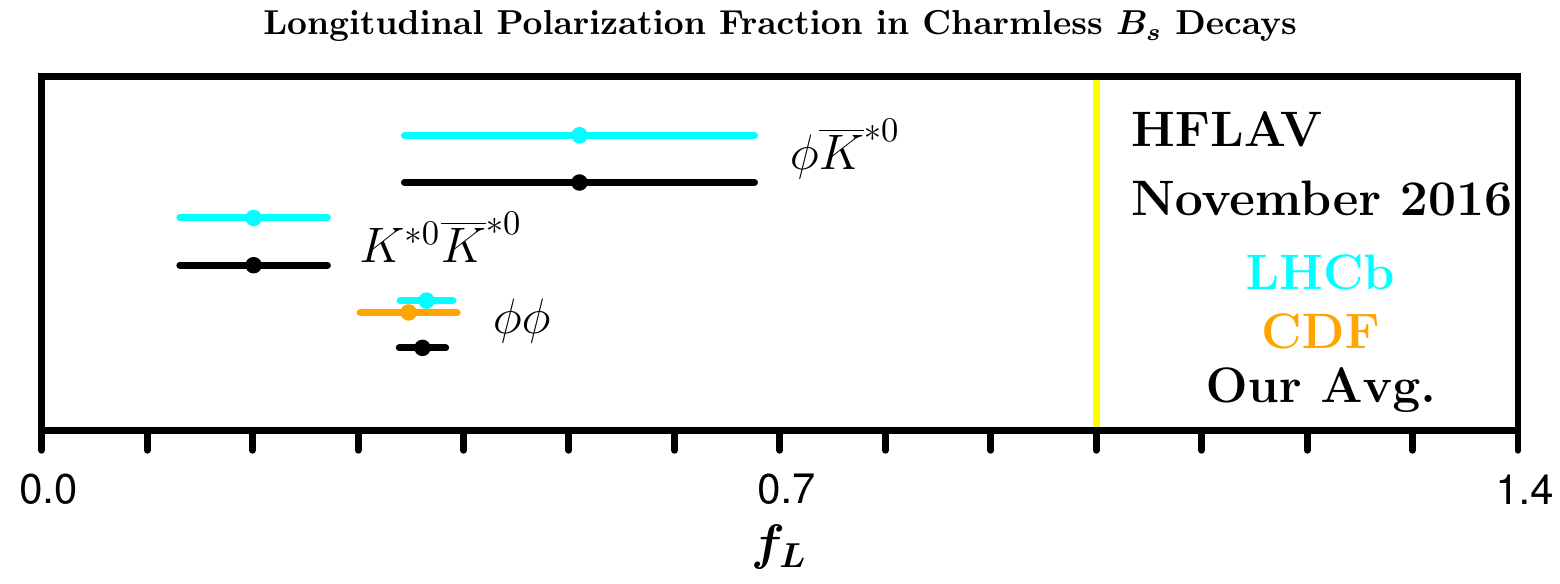}
\caption{Longitudinal polarization fraction in charmless $\Bs$ decays.}
\label{fig:rare-polarbs}
\end{figure}

Figures~\ref{fig:rare-polar} and~\ref{fig:rare-polarbs} show graphic representations of a selection of results shown in this section.
Footnote symbols indicate that the footnote in the corresponding table should be consulted.
%For comments in the plots, marked with a symbol or a number, refer to the corresponding table.

%NEW_TABLE

\begin{table}[!htbp]
\begin{center}
\caption{Longitudinal polarization fraction $f_L$ for $B^+$ decays. 
Where values are shown in \red{red} (\blue{blue}), this indicates that
they are new \red{published} (\blue{preliminary}) results since PDG2014.}
%Values in \red{red} (\blue{blue}) are new \red{published}
%(\blue{preliminary}) results since PDG2014.}
\label{tab:polar_Bp}
\resizebox{\textwidth}{!}{
\begin{tabular}{|lccc @{}c c @{}c c|}
\sgline
RPP\#   & Mode & PDG2014 Avg. & \babar & & Belle & & Our Avg. \\
\sglinespb
%TABLE_BODY
282                                               & %         RPP\#
$\omega K^{*+}$                                   & %         MODE @polar
$\err{0.41}{0.18}{0.05}$                          & %         PDG2014 AVG.
$\err{0.41}{0.18}{0.05}$                          & %   [5]  BABAR
\ifref {\cite{Aubert:2009sx}} \fi \phantom{.}& %   [0]  
\nodata                                           & %   [0]  BELLE
\phantom{.}                                       & %   [0]  
$0.41 \pm 0.19$                                   \\

285                                               & %         RPP\#
$\omega K_2^*(1430)^+$                            & %         MODE @polar
$\err{0.56}{0.10}{0.04}$                          & %         PDG2014 AVG.
$\err{0.56}{0.10}{0.04}$                          & %   [5]  BABAR
\ifref {\cite{Aubert:2009sx}} \fi \phantom{.}& %   [0]  
\nodata                                           & %   [0]  BELLE
\phantom{.}                                       & %   [0]  
$0.56 \pm 0.11$                                   \\

312                                               & %         RPP\#
$K^{*+} \rho^0$                                   & %         MODE @polar
{$\err{0.78}{0.12}{0.03}$}                        & %         PDG2014 AVG.
{$\err{0.78}{0.12}{0.03}$}                        & %   [1]  BABAR
\ifref {\cite{delAmoSanchez:2010mz}} \fi \phantom{.}& %   [0]  
\nodata                                           & %   [0]  BELLE
\phantom{.}                                       & %   [0]  
$0.78 \pm 0.12$                                   \\

316                                               & %         RPP\#
$K^{*0} \rho^+$                                   & %         MODE @polar
$0.48\pm0.08$                                     & %         PDG2014 AVG.
{$\err{0.52}{0.10}{0.04}$}                        & %   [6]  BABAR
\ifref {\cite{Aubert:2006fs}} \fi \phantom{.}& %   [0]  
{$\berr{0.43}{0.11}{0.05}{0.02}$}                 & %   [23]  BELLE
\ifref {\cite{Abe:2004mq}} \fi \phantom{.}& %   [0]  
$0.48 \pm 0.08$                                   \\

338                                               & %         RPP\#
$K^{*+} \overline{K}^{*0}$                        & %         MODE @polar
$\aerr{0.75}{0.16}{0.26}{0.03}$                   & %         PDG2014 AVG.
$\aerr{0.75}{0.16}{0.26}{0.03}$                   & %   [11]  BABAR
\ifref {\cite{Aubert:2009ax}} \fi \phantom{.}& %   [0]  
\nodata                                           & %   [0]  BELLE
\phantom{.}                                       & %   [0]  
$\cerr{0.75}{0.16}{0.26}$                         \\

349                                               & %         RPP\#
$\phi K^{*+}$                                     & %         MODE @polar
$0.50\pm0.05$                                     & %         PDG2014 AVG.
{$\err{0.49}{0.05}{0.03}$}                        & %   [8]  BABAR
\ifref {\cite{Aubert:2007ac}} \fi \phantom{.}& %   [0]  
{$\err{0.52}{0.08}{0.03}$}                        & %   [22]  BELLE
\ifref {\cite{Chen:2005zv}} \fi \phantom{.}& %   [0]  
$0.50 \pm 0.05$                                   \\

351                                               & %         RPP\#
$\phi K_1(1270)^+$                                & %         MODE @polar
$\aerrsy{0.46}{0.12}{0.13}{0.06}{0.07}$           & %         PDG2014 AVG.
$\aerrsy{0.46}{0.12}{0.13}{0.06}{0.07}$           & %   [10]  BABAR
\ifref {\cite{Aubert:2008bc}} \fi \phantom{.}& %   [0]  
\nodata                                           & %   [0]  BELLE
\phantom{.}                                       & %   [0]  
$\cerr{0.46}{0.13}{0.15}$                         \\

355                                               & %         RPP\#
$\phi K_2^*(1430)^+$                              & %         MODE @polar
$\aerr{0.80}{0.09}{0.10}{0.03}$                   & %         PDG2014 AVG.
$\aerr{0.80}{0.09}{0.10}{0.03}$                   & %   [10]  BABAR
\ifref {\cite{Aubert:2008bc}} \fi \phantom{.}& %   [0]  
\nodata                                           & %   [0]  BELLE
\phantom{.}                                       & %   [0]  
$0.80 \pm 0.10$                                   \\

391                                               & %         RPP\#
$\rho^+ \rho^0$                                   & %         MODE @polar
$0.950\pm0.016$                                   & %         PDG2014 AVG.
$\err{0.950}{0.015}{0.006}$                       & %   [7]  BABAR
\ifref {\cite{Aubert:2009it}} \fi \phantom{.}& %   [0]  
$\err{0.95}{0.11}{0.02}$                          & %   [21]  BELLE
\ifref {\cite{Zhang:2003up}} \fi \phantom{.}& %   [0]  
$0.950 \pm 0.016$                                 \\

396                                               & %         RPP\#
$\omega \rho^+$                                   & %         MODE @polar
$\err{0.90}{0.05}{0.03}$                          & %         PDG2014 AVG.
$\err{0.90}{0.05}{0.03}$                          & %   [5]  BABAR
\ifref {\cite{Aubert:2009sx}} \fi \phantom{.}& %   [0]  
\nodata                                           & %   [0]  BELLE
\phantom{.}                                       & %   [0]  
$0.90 \pm 0.06$                                   \\

%TABLE_BODY
\sglinespt
\end{tabular}
}
\end{center}
\end{table}

%NEW_TABLE

\begin{table}[!htbp]
\begin{center}
\caption{Longitudinal polarization fraction $f_L$ for $\Bz$ decays.
Where values are shown in \red{red} (\blue{blue}), this indicates that
they are new \red{published} (\blue{preliminary}) results since PDG2014.}
%Values in \red{red} (\blue{blue}) are new \red{published}
%(\blue{preliminary}) results since PDG2014.}
\label{tab:polar_Bz}
\resizebox{\textwidth}{!}{
\begin{tabular}{|lccc @{}c c @{}c c @{}c c|}
\sgline
RPP\#   & Mode & PDG2014 Avg. & \babar & & Belle & & LHCb & & Our Avg. \\
\hline
%TABLE_BODY
246                                               & %         RPP\#
$\omega K^{*0}$                                   & %         MODE @polar
$0.69\pm0.13$                                     & %         PDG2014 AVG.
$\err{0.72}{0.14}{0.02}$                          & %   [5]  BABAR
\ifref {\cite{Aubert:2009sx}} \fi \phantom{.}& %   [0]  
$\berr{0.56}{0.29}{0.18}{0.08}$                   & %   [25]  BELLE
\ifref {\cite{Goldenzweig:2008sz}} \fi \phantom{.}& %   [0]  
\nodata                                           & %   [0]  LHCB
\phantom{.}                                       & %   [0]  
$0.70 \pm 0.13$                                   \\

249                                               & %         RPP\#
$\omega K_2^*(1430)^0$                            & %         MODE @polar
$\err{0.45}{0.12}{0.02}$                          & %         PDG2014 AVG.
$\err{0.45}{0.12}{0.02}$                          & %   [5]  BABAR
\ifref {\cite{Aubert:2009sx}} \fi \phantom{.}& %   [0]  
\nodata                                           & %   [0]  BELLE
\phantom{.}                                       & %   [0]  
\nodata                                           & %   [0]  LHCB
\phantom{.}                                       & %   [0]  
$0.45 \pm 0.12$                                   \\

279                                               & %         RPP\#
$K^{*0} \rho^0$                                   & %         MODE @polar
$\err{0.40}{0.08}{0.11}$                          & %         PDG2014 AVG.
$\err{0.40}{0.08}{0.11}$                          & %   [13]  BABAR
\ifref {\cite{Lees:2011dq}} \fi \phantom{.}& %   [0]  
\nodata                                           & %   [0]  BELLE
\phantom{.}                                       & %   [0]  
\nodata                                           & %   [0]  LHCB
\phantom{.}                                       & %   [0]  
$0.40 \pm 0.14$                                   \\

284                                               & %         RPP\#
$K^{*+} \rho^-$                                   & %         MODE @polar
$\err{0.38}{0.13}{0.03}$                          & %         PDG2014 AVG.
$\err{0.38}{0.13}{0.03}$                          & %   [13]  BABAR
\ifref {\cite{Lees:2011dq}} \fi \phantom{.}& %   [0]  
\nodata                                           & %   [0]  BELLE
\phantom{.}                                       & %   [0]  
\nodata                                           & %   [0]  LHCB
\phantom{.}                                       & %   [0]  
$0.38 \pm 0.13$                                   \\

312                                               & %         RPP\#
$\phi K^{*0}$                                     & %         MODE @polar
$0.497\pm0.025$                                   & %         PDG2014 AVG.
$\err{0.494}{0.034}{0.013}$                       & %   [2]  BABAR
\ifref {\cite{Aubert:2008zza}} \fi \phantom{.}& %   [0]  
$\err{0.499}{0.030}{0.018}$                       & %   [26]  BELLE
\ifref {\cite{Prim:2013nmy}} \fi \phantom{.}& %   [0]  
\red{$\err{0.497}{0.019}{0.015}$}                 & %   [34]  LHCB
\ifref {\cite{Aaij:2014tpa}} \fi \phantom{.}& %   [0]  
$0.497 \pm 0.017$                                 \\

315                                               & %         RPP\#
$K^{*0} \overline{K}^{*0}$                        & %         MODE @polar
$\aerr{0.80}{0.10}{0.12}{0.06}$                   & %         PDG2014 AVG.
{$\aerr{0.80}{0.10}{0.12}{0.06}$}                 & %   [9]  BABAR
\ifref {\cite{Aubert:2007xc}} \fi \phantom{.}& %   [0]  
\nodata                                           & %   [0]  BELLE
\phantom{.}                                       & %   [0]  
\nodata                                           & %   [0]  LHCB
\phantom{.}                                       & %   [0]  
$\cerr{0.80}{0.12}{0.13}$                         \\

333                                               & %         RPP\#
$\phi K_2^*(1430)^0$                              & %         MODE @polar
$\aerr{0.901}{0.046}{0.058}{0.037}$               & %         PDG2014 AVG.
$\aerr{0.901}{0.046}{0.058}{0.037}$               & %   [2]  BABAR
\ifref {\cite{Aubert:2008zza}} \fi \phantom{.}& %   [0]  
\nodata                                           & %   [0]  BELLE
\phantom{.}                                       & %   [0]  
\nodata                                           & %   [0]  LHCB
\phantom{.}                                       & %   [0]  
$\cerr{0.901}{0.059}{0.069}$                      \\

386                                               & %         RPP\#
$\rho^0 \rho^0$                                   & %         MODE @polar
$\aerr{0.75}{0.11}{0.14}{0.05}$                   & %         PDG2014 AVG.
{$\aerr{0.75}{0.11}{0.14}{0.05}$}                 & %   [4]  BABAR
\ifref {\cite{Aubert:2008au}} \fi \phantom{.}& %   [0]  
\red{$\aerr{0.21}{0.18}{0.22}{0.15}$}             & %   [27]  BELLE
\ifref {\cite{Adachi:2012cz}} \fi \phantom{.}& %   [0]  
\red{$\aerr{0.745}{0.048}{0.058}{0.034}$}         & %   [35]  LHCB
\ifref {\cite{Aaij:2015ria}} \fi \phantom{.}& %   [0]  
$\cerr{0.714}{0.055}{0.062}$                      \\

394                                               & %         RPP\#
$\rho^+ \rho^-$                                   & %         MODE @polar
$\cerr{0.977}{0.028}{0.024}$                      & %         PDG2014 AVG.
{$\berr{0.992}{0.024}{0.026}{0.013}$}             & %   [3]  BABAR
\ifref {\cite{Aubert:2007nua}} \fi \phantom{.}& %   [0]  
{$\aerr{0.941}{0.034}{0.040}{0.030}$}             & %   [24]  BELLE
\ifref {\cite{Somov:2006sg}} \fi \phantom{.}& %   [0]  
\nodata                                           & %   [0]  LHCB
\phantom{.}                                       & %   [0]  
$\cerr{0.978}{0.025}{0.022}$                      \\

405                                               & %         RPP\#
$a_1^\pm a_1^\mp$                                 & %         MODE @polar
$\err{0.31}{0.22}{0.10}$                          & %         PDG2014 AVG.
{$\err{0.31}{0.22}{0.10}$}                        & %   [12]  BABAR
\ifref {\cite{Aubert:2009zr}} \fi \phantom{.}& %   [0]  
\nodata                                           & %   [0]  BELLE
\phantom{.}                                       & %   [0]  
\nodata                                           & %   [0]  LHCB
\phantom{.}                                       & %   [0]  
$0.31 \pm 0.24$                                   \\

%TABLE_BODY
\sglinespt
\end{tabular}
}
\end{center}
\end{table}

%NEW_TABLE

\begin{table}[!p]
%\vspace*{-1.0cm}
\begin{center}
\caption{
Results of the full angular analyses of $B^+ \to \phi K^{*+}$ decays.
Where values are shown in \red{red} (\blue{blue}), this indicates that
they are new \red{published} (\blue{preliminary}) results since PDG2014.}
%Values in \red{red} (\blue{blue}) are new \red{published}
%(\blue{preliminary}) results since PDG2014.}
\label{tab:polar_BpAng}
\resizebox{\textwidth}{!}{
\begin{tabular}{|cc@{}cc @{}c c @{}c c|}
\sgline
Parameter & & PDG2014 Avg. & \babar & & Belle & & Our Avg. \\
\sglinespb
%TABLE_BODY
$f_\perp = \Lambda_{\perp\perp}$                  & %         PARAMETER
\phantom{.}                                       & %         
$0.20\pm0.05$                                     & %         PDG2014 AVG.
{$\err{0.21}{0.05}{0.02}$}                        & %   [8]  BABAR
\ifref {\cite{Aubert:2007ac}} \fi \phantom{.}& %   [0]  
{$\err{0.19}{0.08}{0.02}$}                        & %   [22]  BELLE
\ifref {\cite{Chen:2005zv}} \fi \phantom{.}& %   [0]  
$0.20 \pm 0.05$                                   \\

$\phi_\parallel$                                  & %         PARAMETER
\phantom{.}                                       & %         
$2.34\pm0.18$                                     & %         PDG2014 AVG.
{$\err{2.47}{0.20}{0.07}$}                        & %   [8]  BABAR
\phantom{.}                                       & %   [0]  
{$\err{2.10}{0.28}{0.04}$}                        & %   [22]  BELLE
\phantom{.}                                       & %   [0]  
$2.34 \pm 0.17$                                   \\

$\phi_\perp$                                      & %         PARAMETER
\phantom{.}                                       & %         
$2.58\pm0.17$                                     & %         PDG2014 AVG.
{$\err{2.69}{0.20}{0.03}$}                        & %   [8]  BABAR
\phantom{.}                                       & %   [0]  
{$\err{2.31}{0.30}{0.07}$}                        & %   [22]  BELLE
\phantom{.}                                       & %   [0]  
$2.58 \pm 0.17$                                   \\

$\delta_0$                                        & %         PARAMETER
\phantom{.}                                       & %         
{$\err{3.07}{0.18}{0.06}$}                        & %         PDG2014 AVG.
{$\err{3.07}{0.18}{0.06}$}                        & %   [8]  BABAR
\phantom{.}                                       & %   [0]  
\nodata                                           & %   [0]  BELLE
\phantom{.}                                       & %   [0]  
$3.07 \pm 0.19$                                   \\

$A_{\CP}^0$                                        & %         PARAMETER
\phantom{.}                                       & %         
$\err{0.17}{0.11}{0.02}$                          & %         PDG2014 AVG.
{$\err{0.17}{0.11}{0.02}$}                        & %   [8]  BABAR
\phantom{.}                                       & %   [0]  
\nodata                                           & %   [0]  BELLE
\phantom{.}                                       & %   [0]  
$0.17 \pm 0.11$                                   \\

$A_{\CP}^\perp$                                    & %         PARAMETER
\phantom{.}                                       & %         
$\err{0.22}{0.24}{0.08}$                          & %         PDG2014 AVG.
{$\err{0.22}{0.24}{0.08}$}                        & %   [8]  BABAR
\phantom{.}                                       & %   [0]  
\nodata                                           & %   [0]  BELLE
\phantom{.}                                       & %   [0]  
$0.22 \pm 0.25$                                   \\

$\Delta\phi_\parallel$                            & %         PARAMETER
\phantom{.}                                       & %         
$\err{0.07}{0.20}{0.05}$                          & %         PDG2014 AVG.
{$\err{0.07}{0.20}{0.05}$}                        & %   [8]  BABAR
\phantom{.}                                       & %   [0]  
\nodata                                           & %   [0]  BELLE
\phantom{.}                                       & %   [0]  
$0.07 \pm 0.21$                                   \\

$\Delta\phi_\perp$                                & %         PARAMETER
\phantom{.}                                       & %         
$\err{0.19}{0.20}{0.07}$                          & %         PDG2014 AVG.
{$\err{0.19}{0.20}{0.07}$}                        & %   [8]  BABAR
\phantom{.}                                       & %   [0]  
\nodata                                           & %   [0]  BELLE
\phantom{.}                                       & %   [0]  
$0.19 \pm 0.21$                                   \\

$\Delta\delta_0$                                  & %         PARAMETER
\phantom{.}                                       & %         
{$\err{0.20}{0.18}{0.03}$}                        & %         PDG2014 AVG.
{$\err{0.20}{0.18}{0.03}$}                        & %   [8]  BABAR
\phantom{.}                                       & %   [0]  
\nodata                                           & %   [0]  BELLE
\phantom{.}                                       & %   [0]  
$0.20 \pm 0.18$                                   \\

%TABLE_BODY
\sglinespt
\end{tabular}
}
\end{center}
\scriptsize
Angles ($\phi$, $\delta$) are in radians. BF, $f_L$ and $A_{\CP}$ are tabulated separately.    % FOOTNOTE
\end{table}

%NEW_TABLE

\begin{table}[!p]
\begin{center}
\caption{
Results of the full angular analyses of $B^0 \to \phi K^{*0}$ decays.
Where values are shown in \red{red} (\blue{blue}), this indicates that
they are new \red{published} (\blue{preliminary}) results since PDG2014.}
%Values in \red{red} (\blue{blue}) are new \red{published}
%(\blue{preliminary}) results since PDG2014.}
\label{tab:polar_BzAng}
\resizebox{\textwidth}{!}{
\begin{tabular}{|cc@{}cc @{}c c @{}c c @{}c c|}
\sgline
Parameter & & PDG2014 Avg. & \babar & & Belle & & LHCb & & Our Avg. \\
\sglinespb
%TABLE_BODY
$f_\perp = \Lambda_{\perp\perp}$                  & %         PARAMETER
\phantom{.}                                       & %         
$0.228 \pm 0.021$                                 & %         PDG2014 AVG.
{$\err{0.212}{0.032}{0.013}$}                     & %   [2]  BABAR
\ifref {\cite{Aubert:2008zza}} \fi \phantom{.}& %   [0]  
$\err{0.238}{0.026}{0.008}$                       & %   [26]  BELLE
\ifref {\cite{Prim:2013nmy}} \fi \phantom{.}& %   [0]  
\red{$\err{0.221}{0.016}{0.013}$}                 & %   [34]  LHCB
\ifref {\cite{Aaij:2014tpa}} \fi \phantom{.}& %   [0]  
$0.225 \pm 0.015$                                 \\

$f_S(K\pi)$                                       & %         PARAMETER
\phantom{.}                                       & %         
\nodata                                           & %         PDG2014 AVG.
\nodata                                           & %   [0]  BABAR
\phantom{.}                                       & %   [0]  
\nodata                                           & %   [0]  BELLE
\phantom{.}                                       & %   [0]  
\red{$\err{0.143}{0.013}{0.012}$}                 & %   [34]  LHCB
\phantom{.}                                       & %   [0]  
$0.143 \pm 0.018$                                 \\

$f_S(KK)$                                         & %         PARAMETER
\phantom{.}                                       & %         
\nodata                                           & %         PDG2014 AVG.
\nodata                                           & %   [0]  BABAR
\phantom{.}                                       & %   [0]  
\nodata                                           & %   [0]  BELLE
\phantom{.}                                       & %   [0]  
\red{$\err{0.122}{0.013}{0.008}$}                 & %   [34]  LHCB
\phantom{.}                                       & %   [0]  
$0.122 \pm 0.015$                                 \\

$\phi_\parallel$                                  & %         PARAMETER
\phantom{.}                                       & %         
$2.28 \pm 0.08$                                   & %         PDG2014 AVG.
{$\err{2.40}{0.13}{0.08}$}                        & %   [2]  BABAR
\phantom{.}                                       & %   [0]  
$\err{2.23}{0.10}{0.02}$                          & %   [26]  BELLE
\phantom{.}                                       & %   [0]  
\red{$\err{2.562}{0.069}{0.040}$}                 & %   [34]  LHCB
\phantom{.}                                       & %   [0]  
$2.430 \pm 0.058$                                 \\

$\phi_\perp$                                      & %         PARAMETER
\phantom{.}                                       & %         
$2.36 \pm 0.09$                                   & %         PDG2014 AVG.
{$\err{2.35}{0.13}{0.09}$}                        & %   [2]  BABAR
\phantom{.}                                       & %   [0]  
$\err{2.37}{0.10}{0.04}$                          & %   [26]  BELLE
\phantom{.}                                       & %   [0]  
\red{$\err{2.633}{0.062}{0.037}$}                 & %   [34]  LHCB
\phantom{.}                                       & %   [0]  
$2.527 \pm 0.056$                                 \\

$\delta_0$                                        & %         PARAMETER
\phantom{.}                                       & %         
$2.88 \pm 0.10$                                   & %         PDG2014 AVG.
{$\err{2.82}{0.15}{0.09}$}                        & %   [2]  BABAR
\phantom{.}                                       & %   [0]  
$\err{2.91}{0.10}{0.08}$                          & %   [26]  BELLE
\phantom{.}                                       & %   [0]  
\nodata                                           & %   [34]  LHCB
\phantom{.}                                       & %   [0]  
$2.88 \pm 0.10$                                   \\

$\phi_S(K\pi)$~$^\dag$                            & %         PARAMETER
\phantom{.}                                       & %         
\nodata                                           & %         PDG2014 AVG.
\nodata                                           & %   [0]  BABAR
\phantom{.}                                       & %   [0]  
\nodata                                           & %   [0]  BELLE
\phantom{.}                                       & %   [0]  
\red{$\err{2.222}{0.063}{0.081}$}                 & %   [34]  LHCB
\phantom{.}                                       & %   [0]  
$2.222 \pm 0.103$                                 \\

$\phi_S(KK)$~$^\dag$                              & %         PARAMETER
\phantom{.}                                       & %         
\nodata                                           & %         PDG2014 AVG.
\nodata                                           & %   [0]  BABAR
\phantom{.}                                       & %   [0]  
\nodata                                           & %   [0]  BELLE
\phantom{.}                                       & %   [0]  
\red{$\err{2.481}{0.072}{0.048}$}                 & %   [34]  LHCB
\phantom{.}                                       & %   [0]  
$2.481 \pm 0.087$                                 \\

$A_{\CP}^0$                                        & %         PARAMETER
\phantom{.}                                       & %         
$-0.01 \pm 0.05$                                  & %         PDG2014 AVG.
{$\err{0.01}{0.07}{0.02}$}                        & %   [2]  BABAR
\phantom{.}                                       & %   [0]  
$\err{-0.03}{0.06}{0.01}$                         & %   [26]  BELLE
\phantom{.}                                       & %   [0]  
\red{$\err{-0.003}{0.038}{0.005}$}                & %   [34]  LHCB
\phantom{.}                                       & %   [0]  
$-0.007 \pm 0.030$                                \\

$A_{\CP}^\perp$                                    & %         PARAMETER
\phantom{.}                                       & %         
$-0.11 \pm 0.09$                                  & %         PDG2014 AVG.
{$\err{-0.04}{0.15}{0.06}$}                       & %   [2]  BABAR
\phantom{.}                                       & %   [0]  
$\err{-0.14}{0.11}{0.01}$                         & %   [26]  BELLE
\phantom{.}                                       & %   [0]  
\red{$\err{0.047}{0.072}{0.009}$}                 & %   [34]  LHCB
\phantom{.}                                       & %   [0]  
$-0.014 \pm 0.057$                                \\

${\cal A}_{\CP}^S(K\pi)$                           & %         PARAMETER
\phantom{.}                                       & %         
\nodata                                           & %         PDG2014 AVG.
\nodata                                           & %   [0]  BABAR
\phantom{.}                                       & %   [0]  
\nodata                                           & %   [0]  BELLE
\phantom{.}                                       & %   [0]  
\red{$\err{0.073}{0.091}{0.035}$}                 & %   [34]  LHCB
\phantom{.}                                       & %   [0]  
$0.073 \pm 0.097$                                 \\

${\cal A}_{\CP}^S(KK)$                             & %         PARAMETER
\phantom{.}                                       & %         
\nodata                                           & %         PDG2014 AVG.
\nodata                                           & %   [0]  BABAR
\phantom{.}                                       & %   [0]  
\nodata                                           & %   [0]  BELLE
\phantom{.}                                       & %   [0]  
\red{$\err{-0.209}{0.105}{0.012}$}                & %   [34]  LHCB
\phantom{.}                                       & %   [0]  
$-0.209 \pm 0.106$                                \\

$\Delta\phi_\parallel$                            & %         PARAMETER
\phantom{.}                                       & %         
$0.06 \pm 0.11$                                   & %         PDG2014 AVG.
{$\err{0.22}{0.12}{0.08}$}                        & %   [2]  BABAR
\phantom{.}                                       & %   [0]  
$\err{-0.02}{0.10}{0.01}$                         & %   [26]  BELLE
\phantom{.}                                       & %   [0]  
\red{$\err{0.045}{0.068}{0.015}$}                 & %   [34]  LHCB
\phantom{.}                                       & %   [0]  
$0.051 \pm 0.053$                                 \\

$\Delta\phi_\perp$                                & %         PARAMETER
\phantom{.}                                       & %         
$0.10 \pm 0.08$                                   & %         PDG2014 AVG.
{$\err{0.21}{0.13}{0.08}$}                        & %   [2]  BABAR
\phantom{.}                                       & %   [0]  
$\err{0.05}{0.10}{0.02}$                          & %   [26]  BELLE
\phantom{.}                                       & %   [0]  
\red{$\err{0.062}{0.062}{0.006}$}                 & %   [34]  LHCB
\phantom{.}                                       & %   [0]  
$0.075 \pm 0.050$                                 \\

$\Delta\delta_0$                                  & %         PARAMETER
\phantom{.}                                       & %         
$0.13 \pm 0.09$                                   & %         PDG2014 AVG.
{$\err{0.27}{0.14}{0.08}$}                        & %   [2]  BABAR
\phantom{.}                                       & %   [0]  
$\err{0.08}{0.10}{0.01}$                          & %   [26]  BELLE
\phantom{.}                                       & %   [0]  
\nodata                                           & %   [34]  LHCB
\phantom{.}                                       & %   [0]  
$0.13 \pm 0.08$                                   \\

$\Delta \phi_S(K\pi)$~$^\dag$                     & %         PARAMETER
\phantom{.}                                       & %         
\nodata                                           & %         PDG2014 AVG.
\nodata                                           & %   [0]  BABAR
\phantom{.}                                       & %   [0]  
\nodata                                           & %   [0]  BELLE
\phantom{.}                                       & %   [0]  
\red{$\err{0.062}{0.062}{0.022}$}                 & %   [34]  LHCB
\phantom{.}                                       & %   [0]  
$0.062 \pm 0.066$                                 \\

$\Delta \phi_S(KK)$~$^\dag$                       & %         PARAMETER
\phantom{.}                                       & %         
\nodata                                           & %         PDG2014 AVG.
\nodata                                           & %   [0]  BABAR
\phantom{.}                                       & %   [0]  
\nodata                                           & %   [0]  BELLE
\phantom{.}                                       & %   [0]  
\red{$\err{0.022}{0.072}{0.004}$}                 & %   [34]  LHCB
\phantom{.}                                       & %   [0]  
$0.022 \pm 0.072$                                 \\

%TABLE_BODY
\sglinespt
\end{tabular}
}
\end{center}
\scriptsize
Angles ($\phi$, $\delta$) are in radians. BF, $f_L$ and $A_{\CP}$ are tabulated separately.\\     % FOOTNOTE
$^\dag$~Original LHCb notation adapted to match similar existing quantities.     % FOOTNOTE
\end{table}

%NEW_TABLE

\begin{table}[!p]
\begin{center}
\caption{
Results of the full angular analyses of $B^0 \to \phi K_2^{*0}(1430)$ decays.
Where values are shown in \red{red} (\blue{blue}), this indicates that
they are new \red{published} (\blue{preliminary}) results since PDG2014.}
%Values in \red{red} (\blue{blue}) are new \red{published}
%(\blue{preliminary}) results since PDG2014.}
\label{tab:polar_BzAng1430}
\resizebox{\textwidth}{!}{
\begin{tabular}{|c@{}ccc @{}c c @{}c c|}
\sgline
Parameter & & PDG2014 Avg. & \babar & & Belle & & Our Avg. \\
\sglinespb
%TABLE_BODY
$f_\perp = \Lambda_{\perp\perp}$                  & %         PARAMETER
\phantom{.}                                       & %         
$\cerr{0.027}{0.031}{0.025}$                      & %         PDG2014 AVG.
{$\aerr{0.002}{0.018}{0.002}{0.031}$}             & %   [2]  BABAR
\ifref {\cite{Aubert:2008zza}} \fi \phantom{.}& %   [0]  
$\aerr{0.056}{0.050}{0.035}{0.009}$               & %   [26]  BELLE
\ifref {\cite{Prim:2013nmy}} \fi \phantom{.}& %   [0]  
$\cerr{0.027}{0.027}{0.024}$                      \\

$\phi_\parallel$                                  & %         PARAMETER
\phantom{.}                                       & %         
$4.0 \pm 0.4$                                     & %         PDG2014 AVG.
{$\err{3.96}{0.38}{0.06}$}                        & %   [2]  BABAR
\phantom{.}                                       & %   [0]  
$\err{3.76}{2.88}{1.32}$                          & %   [26]  BELLE
\phantom{.}                                       & %   [0]  
$3.96 \pm 0.38$                                   \\

$\phi_\perp$                                      & %         PARAMETER
\phantom{.}                                       & %         
$4.5 \pm 0.4$                                     & %         PDG2014 AVG.
\nodata                                           & %   [2]  BABAR
\phantom{.}                                       & %   [0]  
$\aerr{4.45}{0.43}{0.38}{0.13}$                   & %   [26]  BELLE
\phantom{.}                                       & %   [0]  
$\cerr{4.45}{0.45}{0.40}$                         \\

$\delta_0$                                        & %         PARAMETER
\phantom{.}                                       & %         
$3.46 \pm 0.14$                                   & %         PDG2014 AVG.
{$\err{3.41}{0.13}{0.13}$}                        & %   [2]  BABAR
\phantom{.}                                       & %   [0]  
$\err{3.53}{0.11}{0.19}$                          & %   [26]  BELLE
\phantom{.}                                       & %   [0]  
$3.46 \pm 0.14$                                   \\

$A_{\CP}^0$                                        & %         PARAMETER
\phantom{.}                                       & %         
$-0.03 \pm 0.04$                                  & %         PDG2014 AVG.
{$\err{-0.05}{0.06}{0.01}$}                       & %   [2]  BABAR
\phantom{.}                                       & %   [0]  
$\aerr{-0.016}{0.066}{0.051}{0.008}$              & %   [26]  BELLE
\phantom{.}                                       & %   [0]  
$\cerr{-0.032}{0.043}{0.038}$                     \\

$A_{\CP}^\perp$                                    & %         PARAMETER
\phantom{.}                                       & %         
$\cerr{0.0}{0.9}{0.7}$                            & %         PDG2014 AVG.
\nodata                                           & %   [2]  BABAR
\phantom{.}                                       & %   [0]  
$\aerr{-0.01}{0.85}{0.67}{0.09}$                  & %   [26]  BELLE
\phantom{.}                                       & %   [0]  
$\cerr{-0.01}{0.85}{0.68}$                        \\

$\Delta\phi_\parallel$                            & %         PARAMETER
\phantom{.}                                       & %         
$-0.9 \pm 0.4$                                    & %         PDG2014 AVG.
{$\err{-1.00}{0.38}{0.09}$}                       & %   [2]  BABAR
\phantom{.}                                       & %   [0]  
$\err{-0.02}{1.08}{1.01}$                         & %   [26]  BELLE
\phantom{.}                                       & %   [0]  
$-0.94 \pm 0.38$                                  \\

$\Delta\phi_\perp$                                & %         PARAMETER
\phantom{.}                                       & %         
$-0.2 \pm 0.4$                                    & %         PDG2014 AVG.
\nodata                                           & %   [2]  BABAR
\phantom{.}                                       & %   [0]  
$\err{-0.19}{0.42}{0.11}$                         & %   [26]  BELLE
\phantom{.}                                       & %   [0]  
$-0.19 \pm 0.43$                                  \\

$\Delta\delta_0$                                  & %         PARAMETER
\phantom{.}                                       & %         
$0.08 \pm 0.09$                                   & %         PDG2014 AVG.
{$\err{0.11}{0.13}{0.06}$}                        & %   [2]  BABAR
\phantom{.}                                       & %   [0]  
$\err{0.06}{0.11}{0.02}$                          & %   [26]  BELLE
\phantom{.}                                       & %   [0]  
$0.08 \pm 0.09$                                   \\

%TABLE_BODY
\sglinespt
\end{tabular}
}
\end{center}
\scriptsize
Angles ($\phi$, $\delta$) are in radians. BF, $f_L$ and $A_{\CP}$ are tabulated separately.     % FOOTNOTE
\end{table}

%NEW_TABLE

\begin{table}[!p]
\begin{center}
\caption{Longitudinal polarization fraction $f_L$ for $\Bs$ decays.
Where values are shown in \red{red} (\blue{blue}), this indicates that
they are new \red{published} (\blue{preliminary}) results since PDG2014.}
%Values in \red{red} (\blue{blue}) are new \red{published}
%(\blue{preliminary}) results since PDG2014.}
\label{tab:polar_Bs}
\resizebox{\textwidth}{!}{
\begin{tabular}{|lccc @{}c c @{}c c|}
\sgline
RPP\#   & Mode & PDG2014 Avg. & CDF & & LHCb & & Our Avg. \\
\hline
%TABLE_BODY
$51$                                              & %         RPP\#
$\phi  \phi$                                      & %         MODE @polar
$0.361 \pm 0.022$                                 & %         PDG2014 AVG.
$\err{0.348}{0.041}{0.021}$                       & %   [30]  CDF
\ifref {\cite{Aaltonen:2011rs}} \fi \phantom{.}& %   [0]  
$\err{0.365}{0.022}{0.012}$                       & %   [32]  LHCB
\ifref {\cite{Aaij:2012ud}} \fi \phantom{.}& %   [0]  
$0.361 \pm 0.022$                                 \\

$59$                                              & %         RPP\#
$K^{*0}  \overline{K}^{*0}$                       & %         MODE @polar
$0.31 \pm 0.13$                                   & %         PDG2014 AVG.
\nodata                                           & %   [0]  CDF
\phantom{.}                                       & %   [0]  
\red{$\err{0.201}{0.057}{0.040}$}                 & %   [31]  LHCB
\ifref {\cite{Aaij:2015kba}} \fi \phantom{.}& %   [0]  
$0.201 \pm 0.070$                                 \\

$60$                                              & %         RPP\#
$\phi  \overline{K}^{*0}$                         & %         MODE @polar
$0.51 \pm 0.17$                                   & %         PDG2014 AVG.
\nodata                                           & %   [0]  CDF
\phantom{.}                                       & %   [0]  
$\err{0.51}{0.15}{0.07}$                          & %   [33]  LHCB
\ifref {\cite{Aaij:2013gga}} \fi \phantom{.}& %   [0]  
$0.51 \pm 0.17$                                   \\

%TABLE_BODY
\sglinespt
\end{tabular}
}
\end{center}
\end{table}

%NEW_TABLE

\begin{table}[!p]
\begin{center}
\caption{
Results of the full angular analyses of $\Bs \to \phi\phi$ decays.
Where values are shown in \red{red} (\blue{blue}), this indicates that
they are new \red{published} (\blue{preliminary}) results since PDG2014.}
%Values in \red{red} (\blue{blue}) are new \red{published}
%(\blue{preliminary}) results since PDG2014.}
\label{tab:polar_BsAng1}
\resizebox{\textwidth}{!}{
\begin{tabular}{|cc@{}cc @{}c c @{}c c|}
\sgline
Parameter & & PDG2014 Avg. & CDF & & LHCb & & Our Avg. \\
\sglinespb
%TABLE_BODY
$f_\perp = \Lambda_{\perp\perp}$                  & %         PARAMETER
\phantom{.}                                       & %         
$0.306 \pm 0.030$                                 & %         PDG2014 AVG.
$\err{0.365}{0.044}{0.027}$                       & %   [30]  CDF
\ifref {\cite{Aaltonen:2011rs}} \fi \phantom{.}& %   [0]  
$\err{0.291}{0.024}{0.010}$                       & %   [32]  LHCB
\ifref {\cite{Aaij:2012ud}} \fi \phantom{.}& %   [0]  
$0.306 \pm 0.023$                                 \\

$\phi_\parallel$                                  & %         PARAMETER
\phantom{.}                                       & %         
$2.59 \pm 0.15$                                   & %         PDG2014 AVG.
$\aerr{2.71}{0.31}{0.36}{0.22}$                   & %   [30]  CDF
\phantom{.}                                       & %   [0]  
$\err{2.57}{0.15}{0.06}$                          & %   [32]  LHCB
\phantom{.}                                       & %   [0]  
$2.59 \pm 0.15$                                   \\

%TABLE_BODY
\sglinespt
\end{tabular}
}
\end{center}
\scriptsize
The parameter $\phi$ is in radians. BF, $f_L$ and $A_{\CP}$ are tabulated separately.     % FOOTNOTE
\end{table}

%NEW_TABLE

\begin{table}[!p]
\begin{center}
\caption{
Results of the full angular analyses of $\Bs \to \phi\overline{K}^{*0}$ decays.
Where values are shown in \red{red} (\blue{blue}), this indicates that
they are new \red{published} (\blue{preliminary}) results since PDG2014.}
%Values in \red{red} (\blue{blue}) are new \red{published}
%(\blue{preliminary}) results since PDG2014.}
\label{tab:polar_BsAng2}
\begin{tabular}{|cc@{}cc @{}c c|}
\sgline
Parameter & & PDG2014 Avg. & LHCb & & Our Avg. \\
\sglinespb
%TABLE_BODY
$f_\perp = \Lambda_{\perp\perp}$                  & %         PARAMETER
\phantom{.}                                       & %         
\nodata                                           & %         PDG2014 AVG.
$\err{0.28}{0.12}{0.03}$                          & %   [33]  LHCB
\ifref {\cite{Aaij:2013gga}} \fi \phantom{.}& %   [0]  
$0.28 \pm 0.12$                                   \\

$f_0$                                             & %         PARAMETER
\phantom{.}                                       & %         
\nodata                                           & %         PDG2014 AVG.
$\err{0.51}{0.15}{0.07}$                          & %   [33]  LHCB
\phantom{.}                                       & %   [0]  
$0.51 \pm 0.17$                                   \\

$f_\parallel$                                     & %         PARAMETER
\phantom{.}                                       & %         
$0.21 \pm 0.11$                                   & %         PDG2014 AVG.
$\err{0.21}{0.11}{0.02}$                          & %   [33]  LHCB
\phantom{.}                                       & %   [0]  
$0.21 \pm 0.11$                                   \\

$\phi_\parallel$~$^\dag$                          & %         PARAMETER
\phantom{.}                                       & %         
$1.75 \pm 0.53 \pm 0.29$                          & %         PDG2014 AVG.
$\aerrsy{1.75}{0.59}{0.53}{0.38}{0.30}$           & %   [33]  LHCB
\phantom{.}                                       & %   [0]  
$\cerr{1.75}{0.70}{0.61}$                         \\

%TABLE_BODY
\sglinespt
\end{tabular}
\end{center}
\scriptsize
The parameter $\phi$ is in radians. BF, $f_L$ and $A_{\CP}$ are tabulated separately.\\     % FOOTNOTE
$^\dag$~Converted from the measurement of $\cos(\phi_\parallel)$. PDG takes the smallest resulting asymmetric error as parabolic.     % FOOTNOTE
\end{table}

%NEW_TABLE

\begin{table}[!p]
\begin{center}
\caption{
Results of the full angular analyses of $\Bs \to {K}^{*0}\overline{K}^{*0}$ decays.
Where values are shown in \red{red} (\blue{blue}), this indicates that
they are new \red{published} (\blue{preliminary}) results since PDG2014.}
%Values in \red{red} (\blue{blue}) are new \red{published}
%(\blue{preliminary}) results since PDG2014.}
\label{tab:polar_BsAng3}
\begin{tabular}{|cc@{}cc @{}c c|}
\sgline
Parameter & & PDG2014 Avg. & LHCb & & Our Avg. \\
\sglinespb
%TABLE_BODY
$f_L$                                             & %         PARAMETER
\phantom{.}                                       & %         
$\err{0.31}{0.12}{0.04}$                          & %         PDG2014 AVG.
\red{$\err{0.201}{0.057}{0.040}$}                 & %   [31]  LHCB
\ifref {\cite{Aaij:2015kba}} \fi \phantom{.}& %   [0]  
$0.201 \pm 0.070$                                 \\

$f_\parallel$                                     & %         PARAMETER
\phantom{.}                                       & %         
\nodata                                           & %         PDG2014 AVG.
\red{$\err{0.215}{0.046}{0.015}$}                 & %   [31]  LHCB
\phantom{.}                                       & %   [0]  
$0.215 \pm 0.048$                                 \\

$|A_s^+|^2$                                       & %         PARAMETER
\phantom{.}                                       & %         
\nodata                                           & %         PDG2014 AVG.
\red{$\err{0.114}{0.037}{0.023}$}                 & %   [31]  LHCB
\phantom{.}                                       & %   [0]  
$0.114 \pm 0.044$                                 \\

$|A_s^-|^2$                                       & %         PARAMETER
\phantom{.}                                       & %         
\nodata                                           & %         PDG2014 AVG.
\red{$\err{0.485}{0.051}{0.019}$}                 & %   [31]  LHCB
\phantom{.}                                       & %   [0]  
$0.485 \pm 0.054$                                 \\

$|A_{ss}|^2$                                      & %         PARAMETER
\phantom{.}                                       & %         
\nodata                                           & %         PDG2014 AVG.
\red{$\err{0.066}{0.022}{0.007}$}                 & %   [31]  LHCB
\phantom{.}                                       & %   [0]  
$0.066 \pm 0.023$                                 \\

$\delta_\parallel$                                & %         PARAMETER
\phantom{.}                                       & %         
\nodata                                           & %         PDG2014 AVG.
\red{$\err{5.31}{0.24}{0.14}$}                    & %   [31]  LHCB
\phantom{.}                                       & %   [0]  
$5.31 \pm 0.28$                                   \\

$\delta_\perp - \delta_s^+$                       & %         PARAMETER
\phantom{.}                                       & %         
\nodata                                           & %         PDG2014 AVG.
\red{$\err{1.95}{0.21}{0.04}$}                    & %   [31]  LHCB
\phantom{.}                                       & %   [0]  
$1.95 \pm 0.21$                                   \\

$\delta_s^-$                                      & %         PARAMETER
\phantom{.}                                       & %         
\nodata                                           & %         PDG2014 AVG.
\red{$\err{1.79}{0.19}{0.19}$}                    & %   [31]  LHCB
\phantom{.}                                       & %   [0]  
$1.79 \pm 0.27$                                   \\

$\delta_{ss}$                                     & %         PARAMETER
\phantom{.}                                       & %         
\nodata                                           & %         PDG2014 AVG.
\red{$\err{1.06}{0.27}{0.23}$}                    & %   [31]  LHCB
\phantom{.}                                       & %   [0]  
$1.06 \pm 0.35$                                   \\

%TABLE_BODY
\sglinespt
\end{tabular}
\end{center}
\end{table}

\clearpage

\mysubsection{Decays of $\Bc$ mesons}
\label{sec:rare-bc}

Table~\ref{tab:Bc_BF} details branching fractions of $\Bc$ meson decays to
charmless hadronic final states.

%NEW_TABLE

\begin{table}[ht!]
\begin{center}
\caption{Relative branching fractions of
$\Bc$ decays.
Where values are shown in \red{red} (\blue{blue}), this indicates that
they are new \red{published} (\blue{preliminary}) results since PDG2014.}
%Values in \red{red} (\blue{blue}) are new \red{published}
%(\blue{preliminary}) results since PDG2014.}
\label{tab:Bc_BF}
\resizebox{\textwidth}{!}{
\begin{tabular}{|lccc @{}c c|} \hline
RPP\# & Mode & PDG2014 AVG. & LHCb & & Our Avg.  \\ \sglinespb
%TABLE_BODY
\nodata                                           & %         RPP\#
$\it{f}_c\mathcal{B}(\Bc\rightarrow K^+ K^0)/\it{f}_u\mathcal{B}(B^+\rightarrow K_S^0\pi^+)$& %         MODE @bc
$^\ddag$                                          & %         PDG2014 AVG.
\red{$<5.8\times 10^{-2}$}                        & %   [1]  LHCB
\ifref {\cite{Aaij:2013fja}} \fi \phantom{.}& %   [0]  
{$<5.8\times 10^{-2}$}                            \\

\nodata                                           & %         RPP\#
$\it{f}_c\mathcal{B}(\Bc\rightarrow p \bar{p}\pi^+)/\it{f}_u$& %         MODE @bc
\nodata                                           & %         PDG2014 AVG.
\red{$<2.8 \times 10^{-8}$}                       & %   [2]  LHCB
\ifref {\cite{Aaij:2016xxs}} \fi \phantom{.}& %   [0]  
{$<2.8 \times 10^{-8}$}                           \\

\nodata                                           & %         RPP\#
$\it{\sigma(\Bc)}\mathcal{B}(\Bc\rightarrow K^+ K^- \pi^+)/\it{\sigma(B^+)}$~$^\dag$& %         MODE @bc
\nodata                                           & %         PDG2014 AVG.
\red{$<15 \times 10^{-8}$}                        & %   [3]  LHCB
\ifref {\cite{Aaij:2016xas}} \fi \phantom{.}& %   [0]  
{$<15 \times 10^{-8}$}                            \\

%TABLE_BODY
%\\
\sglinespt
\hline
\end{tabular}
}
\end{center}
\scriptsize
$^\dag$~Measured in the annihilation region m($K^-\pi^+)<1.834 $GeV$/c^2$.\\
$^\ddag$~PDG converts the LHCb result to ${f}_c\mathcal{B}(\Bc\rightarrow K^+ K^0)<4.6\times 10^{-7}$.
\end{table}

\clearpage
%% %Charm decays
\section{Charm physics}
\label{sec:charm_physics}

\def\kbar{\overline{K}{}^{\,0}}
\def\dbar{\overline{D}{}^{\,0}}
\def\bbar{\overline{B}{}^{\,0}}
\def\cp{$\CP$}
\def\cpv{$CPV$}
\def\ra{\!\rightarrow\!}
\def\ddbar{$D^0$-$\dbar$}
\def\ycp{$y^{}_{\CP}$}

\def\dklnu{$D^0\ra K^+\ell^-\bar{\nu}$}
\def\dkpi{$D^0\ra K^+\pi^-$}
\def\dkk{$D^0\ra K^+K^-$}
\def\dpipi{$D^0\ra\pi^+\pi^-$}
\def\dkkpp{$D^0\ra K^+K^-/\pi^+\pi^-$}
\def\dkspp{$D^0\ra K^0_S\,\pi^+\pi^-$}
\def\dppp{$D^0\ra \pi^0\,\pi^+\pi^-$}
\def\dkskk{$D^0\ra K^0_S\,K^+ K^-$}
\def\dkppp{$D^0\ra K^+\pi^-\pi^+\pi^-$}

\def\dsphipi{$D^+_s\ra\phi\,\pi^+$}

\def\lcp{\Lambda_c^+}

% \def\gevm{~GeV/$c^2$}
% \def\gevp{~GeV/$c$}
% \def\geve{~GeV}
% \def\mevm{~MeV/$c^2$}
% \def\meve{~MeV}

%% defined properly elsewhere
% \def\babar{Babar}

\def\simge{\mathrel{%
   \rlap{\raise 0.511ex \hbox{$>$}}{\lower 0.511ex \hbox{$\sim$}}}}
\def\simle{\mathrel{
   \rlap{\raise 0.511ex \hbox{$<$}}{\lower 0.511ex \hbox{$\sim$}}}}

\newcommand{\Dnan}{\ensuremath{D_0^\ast(2400)^0}}
\newcommand{\Dtan}{\ensuremath{D_2^\ast(2460)^0}}
\newcommand{\Don}{\ensuremath{D_1(2420)^{0}}}
\newcommand{\Dopn}{\ensuremath{D_1(2430)^{0}}}
\newcommand{\Dnap}{\ensuremath{D_0^\ast(2400)^\pm}}
\newcommand{\Dtap}{\ensuremath{D_2^\ast(2460)^\pm}}
\newcommand{\Dop}{\ensuremath{D_1(2420)^{\pm}}}
\newcommand{\Dopp}{\ensuremath{D_1(2430)^{\pm}}}

\newcommand{\Dsa}{\ensuremath{D_s^{\ast\pm}}}
\newcommand{\Dsna}{\ensuremath{D_{s0}^\ast(2317)^{\pm}}}
\newcommand{\Dsop}{\ensuremath{D_{s1}(2460)^{\pm}}}
\newcommand{\Dso}{\ensuremath{D_{s1}(2536)^{\pm}}}
\newcommand{\Dst}{\ensuremath{D_{s2}(2573)^{\pm}}}
\newcommand{\Dsts}{\ensuremath{D_{sJ}(2700)^{\pm}}}
\newcommand{\Dste}{\ensuremath{D_{sJ}(2860)^{\pm}}}
\newcommand{\Dstsi}{\ensuremath{D_{sJ}(2632)^{\pm}}}

\newcommand{\citep}{\cite}

\newcommand{\kst}{K^*(892)^0}
\newcommand{\akst}{\overline{K}^*(892)^0}
\newcommand{\kstp}{K^*(1410)^0}
\newcommand{\akstp}{\overline{K}^*(1410)^0}
\newcommand{\kstd}{K^*_2(1430)^0}
\newcommand{\akstd}{\overline{K}^*_2(1430)^0}

\newcommand{\ksts}{K^*_0(1430)^0}
\newcommand{\aksts}{\overline{K}^*_0(1430)^0}

\newcommand{\ds}{D_{s}}
\newcommand{\dsp}{D_{s}^+}
\newcommand{\dsm}{D_{s}^-}
\newcommand{\dspm}{D_{s}^{\pm}}
\newcommand{\dsmunu}{\ds^+\to\mu^+\nu_{\mu}}
\newcommand{\dsellnu}{\ds^+\to\ell^+\nu_{\ell}}
\newcommand{\br}{{\cal B}}
\newcommand{\ellnu}{\ell^+\nu_{\ell}}
\newcommand{\enu}{e^+\nu_{e}}
\newcommand{\munu}{\mu^+\nu_{\mu}}
\newcommand{\taunu}{\tau^+\nu_{\tau}}
\newcommand{\taumunu}{\tau^+(\mu^+)\nu_{\tau}}
\newcommand{\tauenu}{\tau^+(e^+)\nu_{\tau}}
\newcommand{\taupinuCharm}{\tau^+(\pi^+)\nu_{\tau}}
\newcommand{\taurhonu}{\tau^+(\rho^+)\nu_{\tau}}

% Mixing and CPV
\subsection{\emph{$D^0$-$\dbar$} mixing and \emph{\cp}\ violation}
\label{sec:charm:mixcpv}

\subsubsection{Introduction}

In 2007 Belle~\cite{Staric:2007dt} and \babar~\cite{Aubert:2007wf} 
obtained the first evidence of $D^0$-$\dbar$ mixing, for which 
experiments had searched for more than two decades. 
These results were later confirmed by CDF~\cite{Aaltonen:2007uc}
and more recently by LHCb~\cite{Aaij:2013wda}.
There are now numerous measurements of $D^0$-$\dbar$ mixing 
with various levels of sensitivity. All measurements are
input into a global fit to determine world average values 
of mixing parameters, \cp-violation (\cpv) parameters, 
and strong phase differences.

Our notation is as follows.
The mass eigenstates are denoted
\begin{eqnarray} 
D^{}_1 & = & p|D^0\rangle-q|\dbar\rangle \\ 
D^{}_2 & = & p|D^0\rangle+q|\dbar\rangle\,,
\end{eqnarray}
where we use the convention~\cite{Bergmann:2000id} 
$\CP|D^0\rangle=-|\dbar\rangle$ and 
$\CP|\dbar\rangle=-|D^0\rangle$. Thus in the absence of 
\cp\ violation, $D^{}_1$ is \cp-even and $D^{}_2$ is \cp-odd.
The weak phase $\phi$ is defined as ${\rm Arg}(q/p)$.
The mixing parameters are defined as 
$x\equiv(m^{}_1-m^{}_2)/\Gamma$ and 
$y\equiv (\Gamma^{}_1-\Gamma^{}_2)/(2\Gamma)$, where 
$m^{}_1,\,m^{}_2$ and $\Gamma^{}_1,\,\Gamma^{}_2$ are
the masses and decay widths for the mass eigenstates,
and $\Gamma\equiv (\Gamma^{}_1+\Gamma^{}_2)/2$.

The global fit determines central values and errors
for ten underlying parameters. These consist of the
mixing parameters $x$ and $y$; 
indirect \cpv\ parameters $|q/p|$ and $\phi$; 
the ratio of decay rates
%$R^{}_D\equiv\left|{\cal A}(D^0\ra K^+\pi^-)/
%              {\cal A}(\dbar\ra K^+\pi^-)\right|^2$;
$R^{}_D\equiv
[\Gamma(D^0\ra K^+\pi^-)+\Gamma(\dbar\ra K^-\pi^+)]/
[\Gamma(D^0\ra K^-\pi^+)+\Gamma(\dbar\ra K^+\pi^-)]$;
direct \cpv\ asymmetries $A^{}_D$, $A^{}_K$, and $A^{}_\pi$
in $D^0\ra K^+\pi^-$, $K^+K^-$, and $\pi^+\pi^-$ decays,
respectively; 
%(see Table~\ref{tab:relationships}), and
%$A^{}_D =(R^+_D-R^-_D)/(R^+_D+R^-_D)$, where the $+\,(-)$
%superscript corresponds to $D^0\,(\dbar)$ decays;
the strong phase difference
$\delta$ between $\dbar\ra K^-\pi^+$ and 
$D^0\ra K^-\pi^+$ amplitudes; and 
the strong phase difference $\delta^{}_{K\pi\pi}$ between 
$\dbar\ra K^-\rho^+$ and $D^0\ra K^-\rho^+$ amplitudes. 

The fit uses 50 observables from 
measurements of \dklnu, \dkk, \dpipi, \dkpi, 
$D^0\ra K^+\pi^-\pi^0$, %$D^0\ra K^+\pi^-\pi^+\pi^-$, 
\dkspp, \dppp, \dkskk, and \dkppp\ decays\footnote{Charge-conjugate 
modes are implicitly included.}, and from double-tagged branching 
fractions measured at the $\psi(3770)$ resonance. 
The relationships between the measured observables and the
fitted parameters are given in Table~\ref{tab:relationships}.
Correlations among observables are accounted for by using covariance 
matrices provided by the experimental collaborations. Errors are 
assumed to be Gaussian, and systematic errors among different 
experiments are assumed to be uncorrelated unless specific 
correlations have been identified.
We have checked this method with a second method that adds
together three-dimensional log-likelihood functions 
for $x$, $y$, and $\delta$ obtained from several analyses;
this combination accounts for non-Gaussian errors.
When both methods are applied to the same set of 
measurements, equivalent results are obtained. 

\begin{table}
\renewcommand{\arraycolsep}{0.02in}
\renewcommand{\arraystretch}{1.3}
\begin{center}
\caption{\label{tab:relationships}
Left: decay modes used to determine fitted parameters 
$x,\,y,\,\delta,\,\delta^{}_{K\pi\pi},\,R^{}_D,\,A^{}_D,\,A^{}_K,\,A^{}_\pi,\,|q/p|$, 
and $\phi$.
Middle: the measured observables for each decay mode. Right: the 
relationships between the measured observables and the fitted
parameters. $\langle t\rangle$ is the mean reconstructed decay
time for $D^0\ra K^+K^-$ or $D^0\ra\pi^+\pi^-$ decays.}
\vspace*{6pt}
\footnotesize
\resizebox{0.99\textwidth}{!}{
\begin{tabular}{l|c|l}
\hline
\textbf{Decay Mode} & \textbf{Observables} & \textbf{Relationship} \\
\hline
$D^0\ra K^+K^-/\pi^+\pi^-$  & 
\begin{tabular}{c}
 $y^{}_{\CP}$  \\
 $A^{}_{\Gamma}$
\end{tabular} & 
$\begin{array}{c}
2y^{}_{\CP} = 
\left(\left|q/p\right|+\left|p/q\right|\right)y\cos\phi - \\
\hskip0.50in \left(\left|q/p\right|-\left|p/q\right|\right)x\sin\phi \\
2A^{}_\Gamma = 
\left(\left|q/p\right|-\left|p/q\right|\right)y\cos\phi - \\
\hskip0.50in \left(\left|q/p\right|+\left|p/q\right|\right)x\sin\phi
\end{array}$   \\
\hline
$D^0\ra K^0_S\,\pi^+\pi^-$ & 
$\begin{array}{c}
x \\ 
y \\ 
|q/p| \\ 
\phi
\end{array}$ &   \\ 
\hline
$D^0\ra K^+\ell^-\bar{\nu}$ & $R^{}_M$  & $R^{}_M = (x^2 + y^2)/2$ \\
\hline \hskip-0.10in
\begin{tabular}{l}
$D^0\ra K^+\pi^-\pi^0$ \\
(Dalitz plot analysis)
\end{tabular} & 
$\begin{array}{c}
x'' \\ 
y''
\end{array}$ &
$\begin{array}{l}
x'' = x\cos\delta^{}_{K\pi\pi} + y\sin\delta^{}_{K\pi\pi} \\ 
y'' = y\cos\delta^{}_{K\pi\pi} - x\sin\delta^{}_{K\pi\pi}
\end{array}$ \\
\hline\hskip-0.10in
\begin{tabular}{l}
``Double-tagged'' \\
branching fractions \\
measured in \\
$\psi(3770)\ra DD$ decays
\end{tabular} & 
$\begin{array}{c}
R^{}_M \\
y \\
R^{}_D \\
\sqrt{R^{}_D}\cos\delta
\end{array}$ &   $R^{}_M = (x^2 + y^2)/2$ \\
\hline
$D^0\ra K^+\pi^-$ &
$\begin{array}{c}
%R^+_D,\ R^-_D \\
x'^2,\ y' \\
x'^{2+},\ x'^{2-} \\
y'^+,\ y'^-
\end{array}$ & 
$\begin{array}{l}
%R^{}_D = (R^+_D + R^-_D)/2 \\
%A^{}_D = (R^+_D - R^-_D)/(R^+_D + R^-_D)  \\ \\
%R^\pm_M=(x'^{\pm 2}+y'^{\pm 2})/2 \\
%(|q/p|^4-1)/(|q/p|^4+1)=(R^+_M-R^-_M)/(R^+_M+R^-_M)\equiv A^{}_M \\ \\
x' = x\cos\delta + y\sin\delta \\ 
y' = y\cos\delta - x\sin\delta \\
A^{}_M\equiv (|q/p|^4-1)/(|q/p|^4+1) \\
x'^\pm = [(1\pm A^{}_M)/(1\mp A^{}_M)]^{1/4} \times \\
\hskip0.50in (x'\cos\phi\pm y'\sin\phi) \\
y'^\pm = [(1\pm A^{}_M)/(1\mp A^{}_M)]^{1/4} \times \\
\hskip0.50in (y'\cos\phi\mp x'\sin\phi) \\
%x'^\pm = |q/p|^{\pm 1}(x'\cos\phi\pm y'\sin\phi) \\
%y'^\pm = |q/p|^{\pm 1}(y'\cos\phi\mp x'\sin\phi) \\
\end{array}$ \\
\hline\hskip-0.10in
\begin{tabular}{l}
$D^0\ra K^+\pi^-/K^-\pi^+$ \\
(time-integrated)
\end{tabular} & 
\begin{tabular}{c}
$\frac{\displaystyle \Gamma(D^0\ra K^+\pi^-)+\Gamma(\dbar\ra K^-\pi^+)}
{\displaystyle \Gamma(D^0\ra K^-\pi^+)+\Gamma(\dbar\ra K^+\pi^-)}$  \\ \\
$\frac{\displaystyle \Gamma(D^0\ra K^+\pi^-)-\Gamma(\dbar\ra K^-\pi^+)}
{\displaystyle \Gamma(D^0\ra K^+\pi^-)+\Gamma(\dbar\ra K^-\pi^+)}$ 
\end{tabular} & 
\begin{tabular}{c}
$R^{}_D$ \\ \\ \\
$A^{}_D$ 
\end{tabular} \\
\hline\hskip-0.10in
\begin{tabular}{l}
$D^0\ra K^+K^-/\pi^+\pi^-$ \\
(time-integrated)
\end{tabular} & 
\begin{tabular}{c}
$\frac{\displaystyle \Gamma(D^0\ra K^+K^-)-\Gamma(\dbar\ra K^+K^-)}
{\displaystyle \Gamma(D^0\ra K^+K^-)+\Gamma(\dbar\ra K^+K^-)}$    \\ \\
$\frac{\displaystyle \Gamma(D^0\ra\pi^+\pi^-)-\Gamma(\dbar\ra\pi^+\pi^-)}
{\displaystyle \Gamma(D^0\ra\pi^+\pi^-)+\Gamma(\dbar\ra\pi^+\pi^-)}$ 
\end{tabular} & 
\begin{tabular}{c}
$A^{}_K  + \frac{\displaystyle \langle t\rangle}
{\displaystyle \tau^{}_D}\,{\cal A}_{\CP}^{\rm indirect}$ 
\ \ (${\cal A}_{\CP}^{\rm indirect}\approx -A^{}_\Gamma$)
\\ \\ \\
$A^{}_\pi + \frac{\displaystyle \langle t\rangle}
{\displaystyle \tau^{}_D}\,{\cal A}_{\CP}^{\rm indirect}$ 
\ \ (${\cal A}_{\CP}^{\rm indirect}\approx -A^{}_\Gamma$)
\end{tabular} \\
\hline
%2{\cal A}_{\CP}^{\rm indirect} & = & 
%\Big(\left|q/p\right| + \left|p/q\right|\Big) x \sin\phi\ -\ 
%\Big(\left|q/p\right| - \left|p/q\right|\Big) y \cos\phi \\
\end{tabular}
}
\end{center}
\end{table}

Mixing in the $D^0$, $B^0$, and $B^0_s$ heavy flavor systems
is governed by a short-distance box diagram. In the $D^0$ 
system, this diagram is doubly-Cabibbo-suppressed 
{\it relative to amplitudes dominating the decay width}.
In addition, because the $d$ and $s$ quark masses are 
sufficiently close, this diagram is also GIM-suppressed.
Thus the short-distance mixing rate is extremely small, and 
$D^0$-$\dbar$ mixing is expected to be dominated by long-distance 
processes. These are difficult to calculate, and theoretical 
estimates for $x$ and $y$ range over three orders of 
magnitude (up to the percent 
level)~\cite{Bigi:2000wn,Petrov:2003un,Petrov:2004rf,Falk:2004wg}.

Almost all methods besides that of the $\psi(3770)\ra \overline{D}D$
measurements~\cite{Asner:2012xb} identify the flavor of the
$D^0$ or $\dbar$ when produced by reconstructing the decay
$D^{*+}\ra D^0\pi^+$ or $D^{*-}\ra\dbar\pi^-$. The charge
of the pion, which has low momentum relative to that of the 
$D^0$ and is often referred to as the ``soft'' pion, 
identifies the $D$ flavor. For this 
decay $M^{}_{D^*}-M^{}_{D^0}-M^{}_{\pi^+}\equiv Q\approx 6~\mev$, 
which is close to the kinematic threshold; thus analyses typically
require that the reconstructed $Q$ be small to suppress backgrounds. 
An LHCb measurement~\cite{Aaij:2014gsa} of the difference
between time-integrated \cp\ asymmetries
$A_{\CP}(K^+K^-) - A_{\CP}(\pi^+\pi^-)$ identifies the flavor of
the $D^0$ by partially reconstructing $\overline{B}\ra D^0\mu^- X$ 
decays (and charge-conjugates); in this case the charge of
the muon originating from the $B$ decay identifies the flavor
of the $D^0$.

For time-dependent measurements, the $D^0$ decay time is 
calculated as 
$M^{}_{D^0}\times (\vec{\bf d}\cdot\vec{\bf p})/(cp^2)$, 
where $\vec{\bf d}$ is the displacement vector between the
$D^*$ and $D^0$ decay vertices, $\vec{\bf p}$ is the
reconstructed $D^0$ momentum, and $p$ and $M^{}_{D^0}$ 
are in GeV. The $D^*$ vertex position is 
taken to be the intersection of the $D^0$ momentum vector 
with the beamspot profile for $e^+e^-$ experiments, and 
at the primary interaction vertex for $\bar{p}p$ and $pp$
experiments~\cite{Aaltonen:2007uc,Aaij:2013wda}.

\subsubsection{Input observables}

The global fit determines central values and errors for
ten underlying parameters using a $\chi^2$ statistic.
The fitted parameters are $x$, $y$, $R^{}_D$, $A^{}_D$,
$|q/p|$, $\phi$, $\delta$, $\delta^{}_{K\pi\pi}$,
$A^{}_K$, and $A^{}_\pi$. In the $D\ra K^+\pi^-\pi^0$ 
Dalitz plot analysis~\cite{Aubert:2008zh}, 
%that provides sensitivity to $x$ and $y$, 
the phases of intermediate resonances in the $\dbar\ra K^+\pi^-\pi^0$ 
decay amplitude are determined relative to the phase for
${\cal A}(\dbar\ra K^+\rho^-)$, and the phases of intermediate
resonances for $D^0\ra K^+\pi^-\pi^0$ are determined 
relative to the phase for ${\cal A}(D^0\ra K^+\rho^-)$. 
As the $\dbar$ and $D^0$ Dalitz plots are fitted independently, 
the phase difference $\delta^{}_{K\pi\pi}$ between the
two reference amplitudes cannot be determined from 
these fits. However, the phase difference can be constrained
in the global fit and thus is included as a fitted parameter.

\begin{figure}
\begin{center}
\includegraphics[width=4.2in]{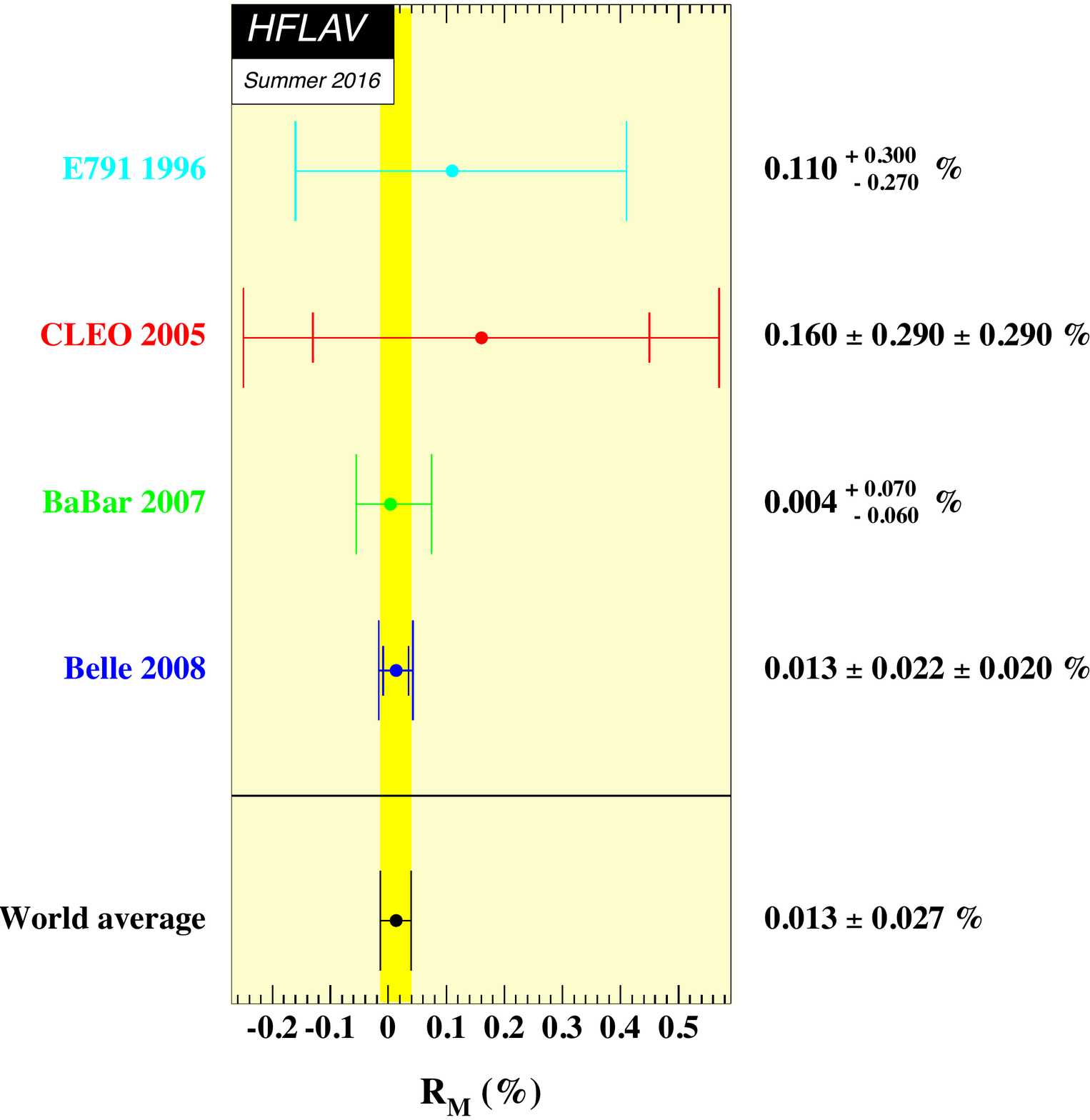}
\end{center}
\vskip-0.20in
\caption{\label{fig:rm_semi}
World average value of $R^{}_M=(x^2+y^2)/2$ from 
Ref.~\cite{HFLAV_charm:webpage},
as calculated from $D^0\ra K^+\ell^-\bar{\nu}$ 
measurements~\cite{Aitala:1996vz,Cawlfield:2005ze,Aubert:2007aa,Bitenc:2008bk}. }
\end{figure}

\begin{figure}
\vskip-0.20in
\begin{center}
\includegraphics[width=4.2in]{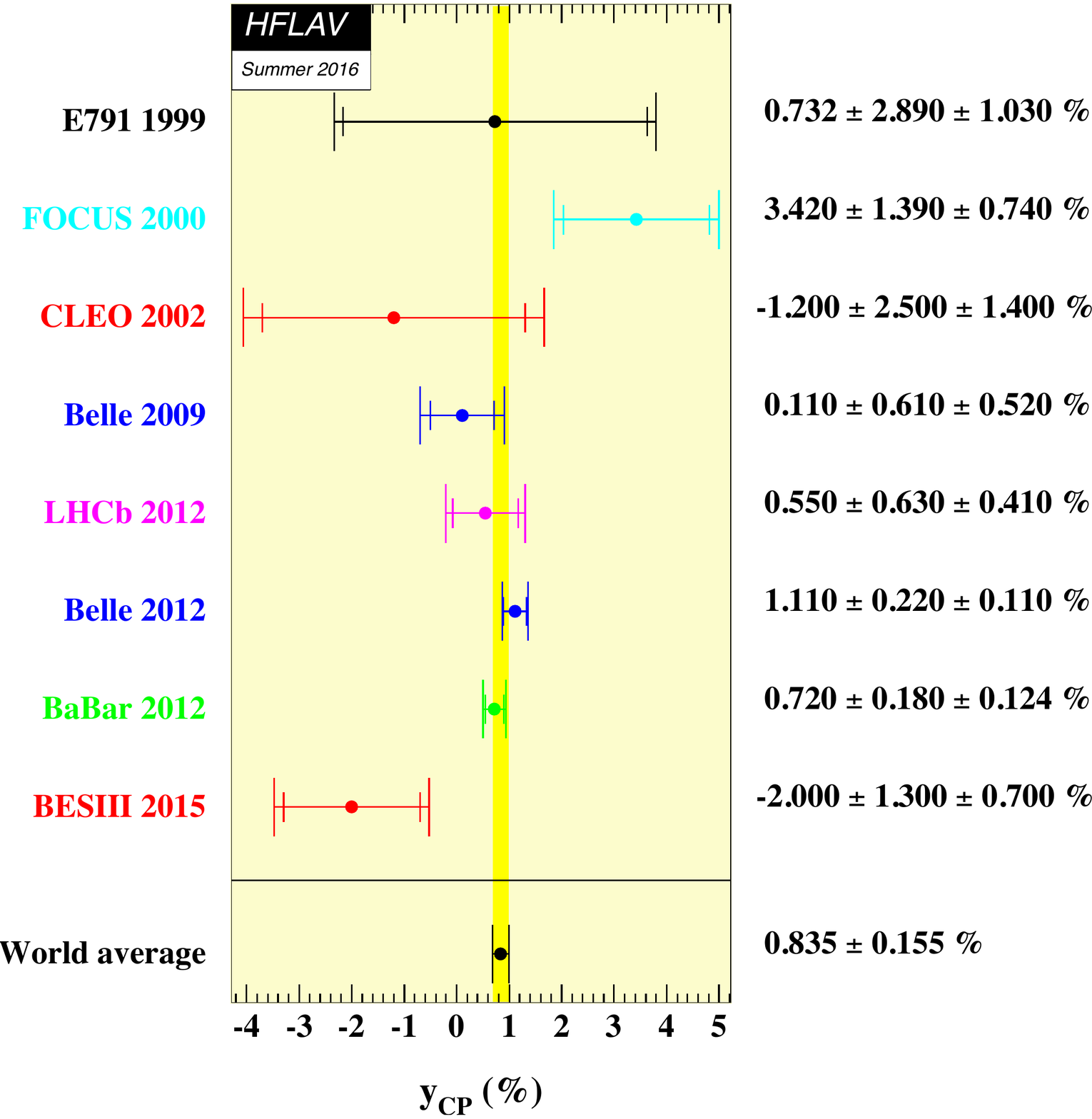}
\end{center}
\vskip-0.80in
\caption{\label{fig:ycp}
World average value of $y^{}_{\CP}$ from Ref.~\cite{HFLAV_charm:webpage}, 
as calculated from $D^0\ra K^+ K^-\!\!,\,\pi^+\pi^-$
measurements~\cite{Aitala:1999dt,Link:2000cu,Csorna:2001ww,
Zupanc:2009sy,Aaij:2011ad,Staric:2015sta,Lees:2012qh,Ablikim:2015hih}.  }
\end{figure}

\begin{figure}
\vskip-0.20in
\begin{center}
\includegraphics[width=4.4in]{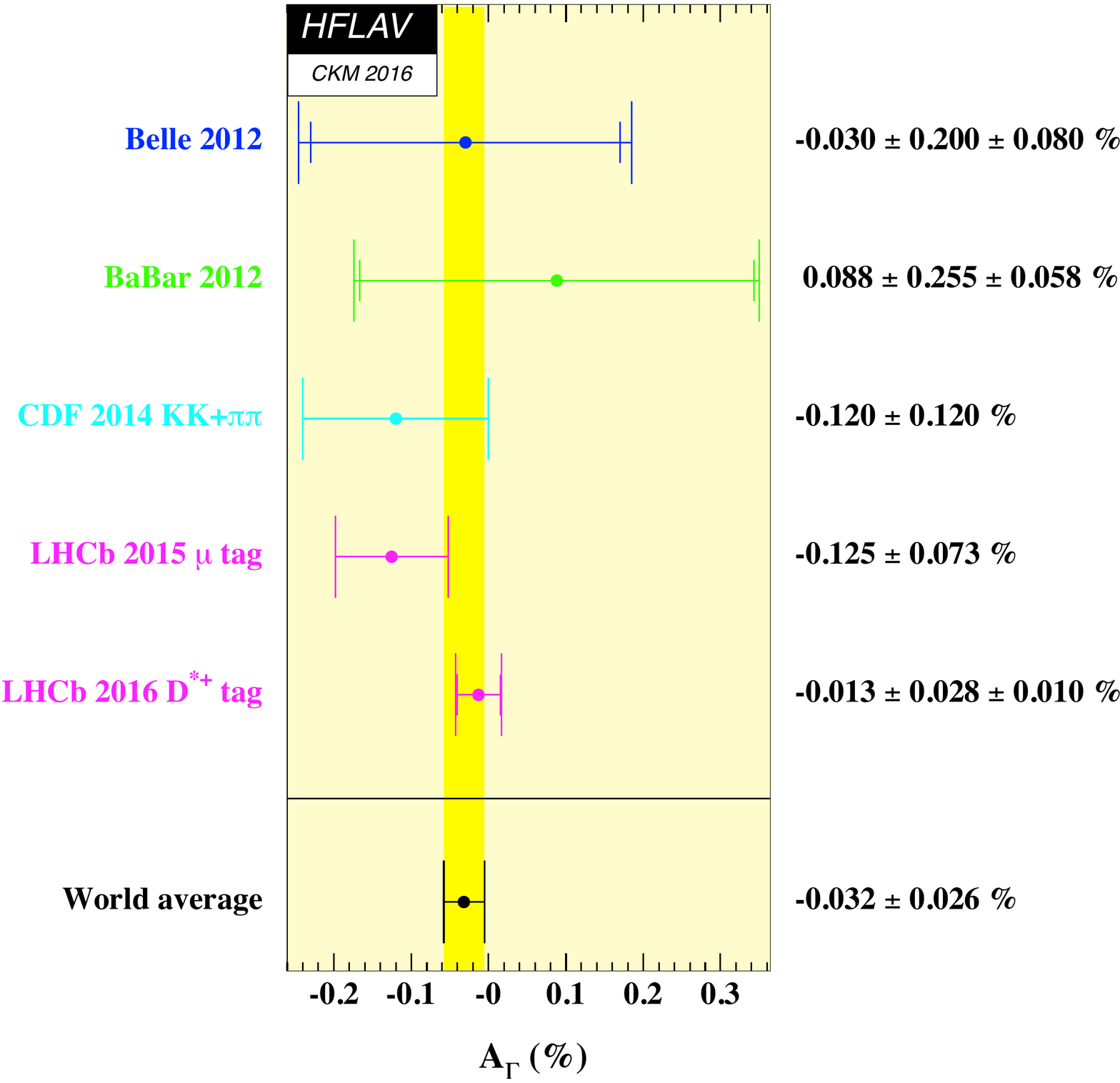}
\end{center}
\vskip-1.0in
\caption{\label{fig:Agamma}
World average value of $A^{}_\Gamma$ from Ref.~\cite{HFLAV_charm:webpage}, 
as calculated from $D^0\ra K^+ K^-\!\!,\,\pi^+\pi^-$
measurements~\cite{Staric:2015sta,Lees:2012qh,
Aaltonen:2014efa,Aaij:2015yda,Aaij:2017idz}.}
\end{figure}

All input measurements are listed in 
Tables~\ref{tab:observables1}-\ref{tab:observables3}. 
The observable $R^{}_M=(x^2+y^2)/2$ is measured in both
$D^0\ra K^+\pi^-\pi^+\pi^-$~\cite{Aaij:2016rhq} and
\dklnu\ decays. In the case of the latter, the HFLAV 
world average~\cite{HFLAV_charm:webpage} is used in 
the global fit. The inputs used for this 
average~\cite{Aitala:1996vz,Cawlfield:2005ze,Aubert:2007aa,Bitenc:2008bk}
are plotted in Fig.~\ref{fig:rm_semi}. The observables 
\begin{eqnarray}
y^{}_{\CP} & = & 
\frac{1}{2}\left(\left|\frac{q}{p}\right| + \left|\frac{p}{q}\right|\right)
y\cos\phi - 
\frac{1}{2}\left(\left|\frac{q}{p}\right| - \left|\frac{p}{q}\right|\right)
x\sin\phi \\
 & & \nonumber \\
A^{}_\Gamma & = & 
\frac{1}{2}\left(\left|\frac{q}{p}\right| - \left|\frac{p}{q}\right|\right)
y\cos\phi - 
\frac{1}{2}\left(\left|\frac{q}{p}\right| + \left|\frac{p}{q}\right|\right)
x\sin\phi 
\end{eqnarray} 
are also HFLAV world average values~\cite{HFLAV_charm:webpage}; 
the inputs used for these averages are plotted in
Figs.~\ref{fig:ycp} and \ref{fig:Agamma}, respectively.
The \dkpi\ measurements used are from 
Belle~\cite{Zhang:2006dp,Ko:2014qvu}, 
\babar~\cite{Aubert:2007wf}, 
CDF~\cite{Aaltonen:2013pja}, and more recently
LHCb~\cite{Aaij:2013wda,Aaij:2016roz};
earlier measurements have much less precision and are not used.
The observables from \dkspp\ decays are measured in two ways:
assuming \cp\ conservation ($D^0$ and $\dbar$ decays combined),
and allowing for \cp\ violation ($D^0$ and $\dbar$ decays
fitted separately). The no-\cpv\ measurements are from 
Belle~\cite{Peng:2014oda}, \babar~\cite{delAmoSanchez:2010xz},
and LHCb~\cite{Aaij:2015xoa}, but for the \cpv-allowed case only 
Belle measurements~\cite{Peng:2014oda} are available. The 
$D^0\ra K^+\pi^-\pi^0$, 
$D^0\ra K^0_S K^+ K^-$, 
and $D^0\ra \pi^0\,\pi^+\pi^-$ results 
are from \babar~\cite{Aubert:2008zh,Lees:2016gom}, the 
$D^0\ra K^+\pi^-\pi^+\pi^-$ results are from LHCb~\cite{Aaij:2016rhq},
and the $\psi(3770)\ra\overline{D}D$ results are from 
CLEOc~\cite{Asner:2012xb}.

\begin{table}
\renewcommand{\arraystretch}{1.4}
\renewcommand{\arraycolsep}{0.02in}
\renewcommand{\tabcolsep}{0.05in}
\caption{\label{tab:observables1}
Observables used in the global fit except those from
time-dependent \dkpi\ measurements, and those from direct 
\cpv\ measurements. The $D^0\ra K^+\pi^-\pi^0$ observables are
$x'' = x\cos\delta^{}_{K\pi\pi} + y\sin\delta^{}_{K\pi\pi}$ and 
$y'' = -x\sin\delta^{}_{K\pi\pi} + y\cos\delta^{}_{K\pi\pi}$.}
\vspace*{6pt}
\footnotesize
\resizebox{0.99\textwidth}{!}{
\begin{tabular}{l|ccc}
\hline
{\bf Mode} & \textbf{Observable} & {\bf Values} & {\bf Correlation coefficients} \\
\hline
\begin{tabular}{l}  
$D^0\ra K^+K^-/\pi^+\pi^-$, \\
\hskip0.30in $\phi\,K^0_S$~\cite{HFLAV_charm:webpage} 
\end{tabular}
&
\begin{tabular}{c}
 $y^{}_{\CP}$  \\
 $A^{}_{\Gamma}$
\end{tabular} & 
$\begin{array}{c}
(0.835\pm 0.155)\% \\
(-0.032\pm 0.026)\% 
\end{array}$   & \\ 
\hline
\begin{tabular}{l}  
$D^0\ra K^0_S\,\pi^+\pi^-$~\cite{Peng:2014oda} \\
\ (Belle: no \cpv)
\end{tabular}
&
\begin{tabular}{c}
$x$ \\
$y$ 
\end{tabular} & 
\begin{tabular}{c}
 $(0.56\pm 0.19\,^{+0.067}_{-0.127})\%$ \\
 $(0.30\pm 0.15\,^{+0.050}_{-0.078})\%$ 
\end{tabular} & $+0.012$ \\ 
 & & \\
\begin{tabular}{l}  
$D^0\ra K^0_S\,\pi^+\pi^-$~\cite{Peng:2014oda} \\
\ (Belle: no direct \cpv)
\end{tabular}
&
\begin{tabular}{c}
$|q/p|$ \\
$\phi$  
\end{tabular} & 
\begin{tabular}{c}
 $0.90\,^{+0.16}_{-0.15}{}^{+0.078}_{-0.064}$ \\
 $(-6\pm 11\,^{+4.2}_{-5.0})$ degrees
\end{tabular} & \\
 & & \\
\begin{tabular}{l}  
$D^0\ra K^0_S\,\pi^+\pi^-$~\cite{Peng:2014oda} \\
\ (Belle: direct \cpv\ allowed)
\end{tabular}
&
\begin{tabular}{c}
$x$ \\
$y$ \\
$|q/p|$ \\
$\phi$  
\end{tabular} & 
\begin{tabular}{c}
 $(0.58\pm 0.19^{+0.0734}_{-0.1177})\%$ \\
 $(0.27\pm 0.16^{+0.0546}_{-0.0854})\%$ \\
 $0.82\,^{+0.20}_{-0.18}{}^{+0.0807}_{-0.0645}$ \\
 $(-13\,^{+12}_{-13}\,^{+4.15}_{-4.77})$ degrees
\end{tabular} & 
$\left\{ \begin{array}{cccc}
 1 &  0.054 & -0.074 & -0.031  \\
 0.054 &  1 & 0.034 & -0.019 \\
 -0.074 &  0.034 & 1 & 0.044  \\
 -0.031 &  -0.019 & 0.044 & 1 
\end{array} \right\}$  \\
 & & \\
\begin{tabular}{l}  
$D^0\ra K^0_S\,\pi^+\pi^-$~\cite{Aaij:2015xoa} \\
\ (LHCb: no \cpv)
\end{tabular}
&
\begin{tabular}{c}
$x$ \\
$y$ 
\end{tabular} & 
\begin{tabular}{c}
 $(-0.86\,\pm 0.53\,\pm 0.17)\%$ \\
 $(0.03\,\pm 0.46\,\pm 0.13)\%$ 
\end{tabular} & $+0.37$ \\ 
 & & \\
\begin{tabular}{l}  
$D^0\ra K^0_S\,\pi^+\pi^-$~\cite{delAmoSanchez:2010xz} \\
\hskip0.30in $K^0_S\,K^+ K^-$ \\
\ (\babar: no \cpv) 
\end{tabular}
&
\begin{tabular}{c}
$x$ \\
$y$ 
\end{tabular} & 
\begin{tabular}{c}
 $(0.16\pm 0.23\pm 0.12\pm 0.08)\%$ \\
 $(0.57\pm 0.20\pm 0.13\pm 0.07)\%$ 
\end{tabular} &  $+0.0615$ \\ 
 & & \\
\begin{tabular}{l}  
$D^0\ra \pi^0\,\pi^+\pi^-$~\cite{Lees:2016gom} \\
\ (\babar: no \cpv) 
\end{tabular}
&
\begin{tabular}{c}
$x$ \\
$y$ 
\end{tabular} & 
\begin{tabular}{c}
 $(1.5\pm 1.2\pm 0.6)\%$ \\
 $(0.2\pm 0.9\pm 0.5)\%$ 
\end{tabular} &  $-0.006$ \\ 
\hline
\begin{tabular}{l}  
$D^0\ra K^+\ell^-\bar{\nu}$~\cite{HFLAV_charm:webpage}
\end{tabular} 
  & $R^{}_M =(x^2+y^2)/2$ & $(0.0130\pm 0.0269)\%$  &  \\ 
\hline
\begin{tabular}{l}  
$D^0\ra K^+\pi^-\pi^0$~\cite{Aubert:2008zh}
\end{tabular} 
&
\begin{tabular}{c}
$x''$ \\ 
$y''$ 
\end{tabular} &
\begin{tabular}{c}
$(2.61\,^{+0.57}_{-0.68}\,\pm 0.39)\%$ \\ 
$(-0.06\,^{+0.55}_{-0.64}\,\pm 0.34)\%$ 
\end{tabular} & $-0.75$ \\
\hline
\begin{tabular}{l}  
$D^0\ra K^+\pi^-\pi^+\pi^-$~\cite{Aaij:2016rhq}
\end{tabular} 
  & $R^{}_M/2$  & $(4.8\pm 1.8)\times 10^{-5}$  &  \\ 
\hline
\begin{tabular}{c}  
$\psi(3770)\ra\overline{D}D$~\cite{Asner:2012xb} \\
(CLEOc)
\end{tabular}
&
\begin{tabular}{c}
$R^{}_D$ \\
$x^2$ \\
$y$ \\
$\cos\delta$ \\
$\sin\delta$ 
\end{tabular} & 
\begin{tabular}{c}
$(0.533 \pm 0.107 \pm 0.045)\%$ \\
$(0.06 \pm 0.23 \pm 0.11)\%$ \\
$(4.2 \pm 2.0 \pm 1.0)\%$ \\
$0.81\,^{+0.22}_{-0.18}\,^{+0.07}_{-0.05}$ \\
$-0.01\pm 0.41\pm 0.04$
\end{tabular} &
$\left\{ \begin{array}{ccccc}
1 & 0 &  0    & -0.42 &  0.01 \\
  & 1 & -0.73 &  0.39 &  0.02 \\
  &   &  1    & -0.53 & -0.03 \\
  &   &       &  1    &  0.04 \\
  &   &       &       &  1    
\end{array} \right\}$ \\
\hline
\end{tabular}
}
\end{table}

\begin{table}
\renewcommand{\arraystretch}{1.3}
\renewcommand{\arraycolsep}{0.02in}
\caption{\label{tab:observables2}
Time-dependent \dkpi\ observables used for the global fit.
The observables $R^+_D$ and $R^-_D$ are related to parameters 
$R^{}_D$ and $A^{}_D$ via $R^\pm_D = R^{}_D (1\pm A^{}_D)$.}
\vspace*{6pt}
\footnotesize
\begin{center}
\begin{tabular}{l|ccc}
\hline
{\bf Mode} & \textbf{Observable} & {\bf Values} & {\bf Correlation coefficients} \\
\hline
\begin{tabular}{l}  
$D^0\ra K^+\pi^-$~\cite{Aubert:2007wf} \\
(\babar~384~fb$^{-1}$)
\end{tabular}
&
\begin{tabular}{c}
$R^{}_D$ \\
$x'^{2+}$ \\
$y'^+$ 
\end{tabular} & 
\begin{tabular}{c}
 $(0.303\pm 0.0189)\%$ \\
 $(-0.024\pm 0.052)\%$ \\
 $(0.98\pm 0.78)\%$ 
\end{tabular} &
$\left\{ \begin{array}{ccc}
 1 &  0.77 &  -0.87 \\
0.77 & 1 & -0.94 \\
-0.87 & -0.94 & 1 
\end{array} \right\}$ \\ \\
\begin{tabular}{l}  
$\dbar\ra K^-\pi^+$~\cite{Aubert:2007wf} \\
(\babar~384~fb$^{-1}$)
\end{tabular}
&
\begin{tabular}{c}
$A^{}_D$ \\
$x'^{2-}$ \\
$y'^-$ 
\end{tabular} & 
\begin{tabular}{c}
 $(-2.1\pm 5.4)\%$ \\
 $(-0.020\pm 0.050)\%$ \\
 $(0.96\pm 0.75)\%$ 
\end{tabular} & same as above \\
\hline
\begin{tabular}{l}  
$D^0\ra K^+\pi^-$~\cite{Ko:2014qvu} \\
(Belle 976~fb$^{-1}$ No \cpv)
\end{tabular}
&
\begin{tabular}{c}
$R^{}_D$ \\
$x'^{2}$ \\
$y'$ 
\end{tabular} & 
\begin{tabular}{c}
 $(0.353\pm 0.013)\%$ \\
 $(0.009\pm 0.022)\%$ \\
 $(0.46\pm 0.34)\%$ 
\end{tabular} &
$\left\{ \begin{array}{ccc}
 1 &  0.737 &  -0.865 \\
0.737 & 1 & -0.948 \\
-0.865 & -0.948 & 1 
\end{array} \right\}$ \\ \\
\begin{tabular}{l}  
$D^0\ra K^+\pi^-$~\cite{Zhang:2006dp} \\
(Belle 400~fb$^{-1}$ \cpv-allowed)
\end{tabular}
&
\begin{tabular}{c}
$R^{}_D$ \\
$x'^{2+}$ \\
$y'^+$ 
\end{tabular} & 
\begin{tabular}{c}
 $(0.364\pm 0.018)\%$ \\
 $(0.032\pm 0.037)\%$ \\
 $(-0.12\pm 0.58)\%$ 
\end{tabular} &
$\left\{ \begin{array}{ccc}
 1 &  0.655 &  -0.834 \\
0.655 & 1 & -0.909 \\
-0.834 & -0.909 & 1 
\end{array} \right\}$ \\ \\
\begin{tabular}{l}  
$\dbar\ra K^-\pi^+$~\cite{Zhang:2006dp} \\
(Belle 400~fb$^{-1}$ \cpv-allowed)
\end{tabular}
&
\begin{tabular}{c}
$A^{}_D$ \\
$x'^{2-}$ \\
$y'^-$ 
\end{tabular} & 
\begin{tabular}{c}
 $(+2.3\pm 4.7)\%$ \\
 $(0.006\pm 0.034)\%$ \\
 $(0.20\pm 0.54)\%$ 
\end{tabular} & same as above \\
\hline
\begin{tabular}{l}  
$D^0\ra K^+\pi^-$~\cite{Aaltonen:2013pja} \\
%\ \ \ \ \ + c.c. \\
(CDF 9.6~fb$^{-1}$ No \cpv)
\end{tabular}
&
\begin{tabular}{c}
$R^{}_D$ \\
$x'^{2}$ \\
$y'$ 
\end{tabular} & 
\begin{tabular}{c}
 $(0.351\pm 0.035)\%$ \\
 $(0.008\pm 0.018)\%$ \\
 $(0.43\pm 0.43)\%$ 
\end{tabular} & 
$\left\{ \begin{array}{ccc}
 1 &  0.90 &  -0.97 \\
0.90 & 1 & -0.98 \\
-0.97 & -0.98 & 1 
\end{array} \right\}$ \\ 
\hline
\begin{tabular}{l}  
$D^0\ra K^+\pi^-$~\cite{Aaij:2016roz} \\  %Aaij:2013wda
(LHCb 3.0~fb$^{-1}$ \cpv-allowed)
\end{tabular}
&
\begin{tabular}{c}
$R^{+}_D$ \\
$x'^{2+}$ \\
$y'^+$ 
\end{tabular} & 
\begin{tabular}{c}
 $(0.3474\pm 0.0081)\%$ \\
 $(0.0011\pm 0.0065)\%$ \\
 $(0.597\pm 0.125)\%$ 
\end{tabular} &
$\left\{ \begin{array}{ccc}
 1 &  0.823 &  -0.920 \\
0.823 & 1 & -0.962 \\
-0.920 & -0.962 & 1 
\end{array} \right\}$ \\ \\
\begin{tabular}{l}  
$\dbar\ra K^-\pi^+$~\cite{Aaij:2016roz} \\
(LHCb 3.0~fb$^{-1}$ \cpv-allowed)
\end{tabular}
&
\begin{tabular}{c}
$R^{-}_D$ \\
$x'^{2-}$ \\
$y'^-$ 
\end{tabular} & 
\begin{tabular}{c}
 $(0.3591\pm 0.0081)\%$ \\
 $(0.0061\pm 0.0061)\%$ \\
 $(0.450\pm 0.121)\%$ 
\end{tabular} & 
$\left\{ \begin{array}{ccc}
 1 &  0.812 &  -0.918 \\
0.812 & 1 & -0.956 \\
-0.918 & -0.956 & 1 
\end{array} \right\}$ \\
\hline
\end{tabular}
\end{center}
\end{table}

\begin{table}
\renewcommand{\arraystretch}{1.3}
\renewcommand{\arraycolsep}{0.02in}
\caption{\label{tab:observables3}
Measurements of time-integrated \cp\ asymmetries. The observable 
$A^{}_{\CP}(f)= [\Gamma(D^0\ra f)-\Gamma(\dbar\ra f)]/
[\Gamma(D^0\ra f)+\Gamma(\dbar\ra f)]$, and
$\Delta\langle t\rangle$ is the difference
between the mean reconstructed decay times 
for $D^0\ra K^+K^-$ and $D^0\ra\pi^+\pi^-$  
(due to different trigger and reconstruction efficiencies).}
\vspace*{6pt}
\footnotesize
\begin{center}
\resizebox{\textwidth}{!}{
\begin{tabular}{l|ccc}
\hline
{\bf Mode} & \textbf{Observable} & {\bf Values} & 
                  {\boldmath $\Delta\langle t\rangle/\tau^{}_D$} \\
\hline
\begin{tabular}{c}
$D^0\ra h^+ h^-$~\cite{Aubert:2007if} \\
(\babar\ 386 fb$^{-1}$)
\end{tabular} & 
\begin{tabular}{c}
$A^{}_{\CP}(K^+K^-)$ \\
$A^{}_{\CP}(\pi^+\pi^-)$ 
\end{tabular} & 
\begin{tabular}{c}
$(+0.00 \pm 0.34 \pm 0.13)\%$ \\
$(-0.24 \pm 0.52 \pm 0.22)\%$ 
\end{tabular} &
0 \\
\hline
\begin{tabular}{c}
$D^0\ra h^+ h^-$~\cite{Ko:2012jh} \\
(Belle 976~fb$^{-1}$)
\end{tabular} & 
\begin{tabular}{c}
$A^{}_{\CP}(K^+K^-)$ \\
$A^{}_{\CP}(\pi^+\pi^-)$ 
\end{tabular} & 
\begin{tabular}{c}
$(-0.32 \pm 0.21 \pm 0.09)\%$ \\
$(+0.55 \pm 0.36 \pm 0.09)\%$ 
\end{tabular} &
0 \\
\hline
\begin{tabular}{c}
$D^0\ra h^+ h^-$~\cite{cdf_public_note_10784,Collaboration:2012qw} \\
(CDF 9.7~fb$^{-1}$)
\end{tabular} & 
\begin{tabular}{c}
$A^{}_{\CP}(K^+K^-)-A^{}_{\CP}(\pi^+\pi^-)$ \\
$A^{}_{\CP}(K^+K^-)$ \\
$A^{}_{\CP}(\pi^+\pi^-)$ 
\end{tabular} & 
\begin{tabular}{c}
$(-0.62 \pm 0.21 \pm 0.10)\%$ \\
$(-0.32 \pm 0.21)\%$ \\
$(+0.31 \pm 0.22)\%$ 
\end{tabular} &
$0.27 \pm 0.01$ \\
\hline
\begin{tabular}{c}
$D^0\ra h^+ h^-$~\cite{Aaij:2016cfh} \\
(LHCb 3.0~fb$^{-1}$, \\
$D^{*+}\ra D^0\pi^+$ tag)
\end{tabular} & 
$A^{}_{\CP}(K^+K^-)-A^{}_{\CP}(\pi^+\pi^-)$ &
$(-0.10 \pm 0.08 \pm 0.03)\%$ &
$0.1153 \pm 0.0007 \pm 0.0018$ \\
\hline
\begin{tabular}{c}
$D^0\ra h^+ h^-$~\cite{Aaij:2014gsa} \\
(LHCb 3~fb$^{-1}$, \\
$\overline{B}\ra D^0\mu^- X$ tag)
\end{tabular} & 
$A^{}_{\CP}(K^+K^-)-A^{}_{\CP}(\pi^+\pi^-)$ &
$(+0.14 \pm 0.16 \pm 0.08)\%$ &
$0.014 \pm 0.004$ \\
\hline
\end{tabular}
}
\end{center}
\end{table}

The relationships between the observables and the fitted
parameters are listed in Table~\ref{tab:relationships}.
For each set of correlated observables we construct a 
difference vector $\vec{V}$ between the measured values 
and those calculated from fitted parameters using these 
relations;
% of Table~\ref{tab:relationships}; 
\eg, for $D^0\ra K^0_S\,\pi^+\pi^-$ decays,
$\vec{V}=(\Delta x,\Delta y,\Delta |q/p|,\Delta \phi)$.
%where $\Delta$ represents the difference between the 
%measured value and the fitted value.
The contribution of a set of observables to the fit $\chi^2$ 
is calculated as $\vec{V}\cdot (M^{-1})\cdot\vec{V}^T$, 
where $M^{-1}$ is the inverse of the covariance matrix 
for the measurement. Covariance matrices are constructed 
from the correlation coefficients among the measured observables.
These correlation coefficients are also listed in 
Tables~\ref{tab:observables1}-\ref{tab:observables3}.

\subsubsection{Fit results}

The global fit uses MINUIT with the MIGRAD minimizer, 
and all errors are obtained from MINOS~\cite{MINUIT:webpage}. 
Four separate fits are performed: 
\begin{enumerate}
\item 
assuming \cp\ conservation, \ie, fixing
$A^{}_D\!=\!0$, $A_K\!=\!0$, $A^{}_\pi\!=\!0$, $\phi\!=\!0$, 
and $|q/p|\!=\!1$;
\item  
assuming no direct \cpv\ in doubly Cabibbo-suppressed (DCS)
decays ($A^{}_D\!=\!0$) and fitting for parameters $(x,y,|q/p|)$ 
or $(x,y,\phi)$; 
\item  
assuming no direct \cpv\ in DCS decays and fitting 
for alternative parameters~\cite{Grossman:2009mn,Kagan:2009gb}
$x^{}_{12}= 2|M^{}_{12}|/\Gamma$, 
$y^{}_{12}= |\Gamma^{}_{12}|/\Gamma$, and 
$\phi^{}_{12}= {\rm Arg}(M^{}_{12}/\Gamma^{}_{12})$,
where $M^{}_{12}$ and $\Gamma^{}_{12}$ are the off-diagonal
elements of the $D^0$-$\dbar$ mass and decay matrices, respectively.
The parameter $\phi^{}_{12}$ is a weak phase that is responsible 
for $\CP$ violation in mixing.
\item  
allowing full \cpv\ (floating all parameters). 
\end{enumerate}

For fits (2) and (3) assuming no direct \cpv\ in DCS decays, in addition to
$A^{}_D\!=\!0$ we impose other constraints that reduce four independent 
parameters to three\footnote{One can also use Eq.~(15) of 
Ref.~\cite{Grossman:2009mn} to reduce four parameters to three.}. 
For fit (2)  we impose the 
relation~\cite{Ciuchini:2007cw,Kagan:2009gb}
$\tan\phi = (1-|q/p|^2)/(1+|q/p|^2)\times (x/y)$ in 
two ways: first we float parameters
$x$, $y$, and $\phi$ and from these derive $|q/p|$; we then repeat
the fit floating $x$, $y$, and $|q/p|$ and from these derive 
$\phi$. The central values returned by the two fits are identical,
but the first fit yields MINOS errors for $\phi$, while the second
fit yields MINOS errors for $|q/p|$. For no-direct-\cpv\ fit (3),
we fit for underlying parameters $x^{}_{12}$, $y^{}_{12}$, 
and $\phi^{}_{12}$, and from these calculate $x$, $y$, $|q/p|$, and 
$\phi$ to which measured observables are compared.
All fit results are listed in 
Table~\ref{tab:results}. For the \cpv-allowed fit,
individual contributions to the $\chi^2$ are listed 
in Table~\ref{tab:results_chi2}. The total $\chi^2$ 
is 76.8 for $50-10=40$ degrees of freedom.

\begin{figure}
\vskip-0.20in
\begin{center}
\includegraphics[width=4.0in]{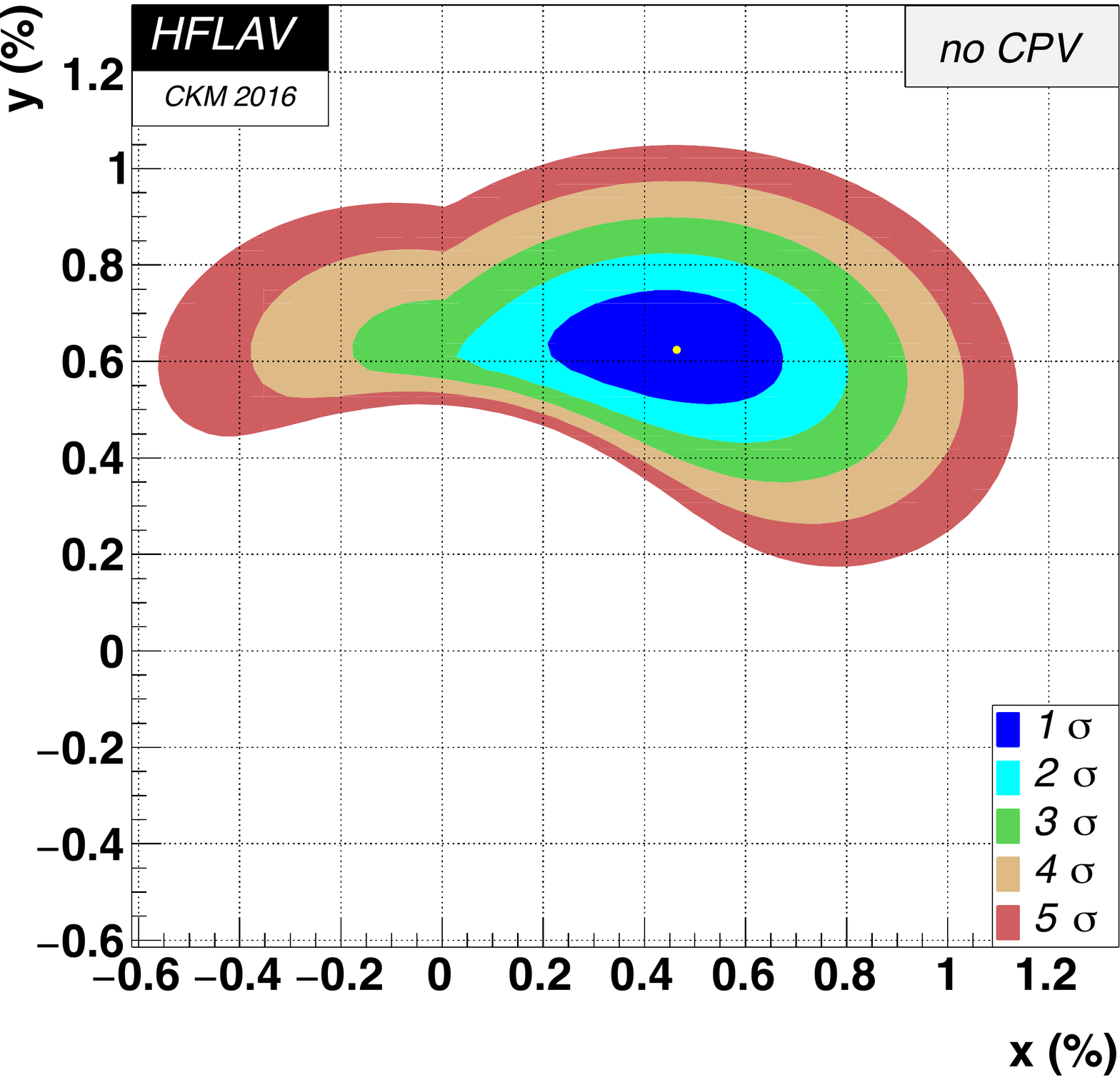}
\end{center}
\vskip-0.90in
\caption{\label{fig:contours_ncpv}
Two-dimensional contours for mixing parameters $(x,y)$, for no \cpv. }
\end{figure}

\begin{figure}
\begin{center}
\vbox{
\includegraphics[width=84mm]{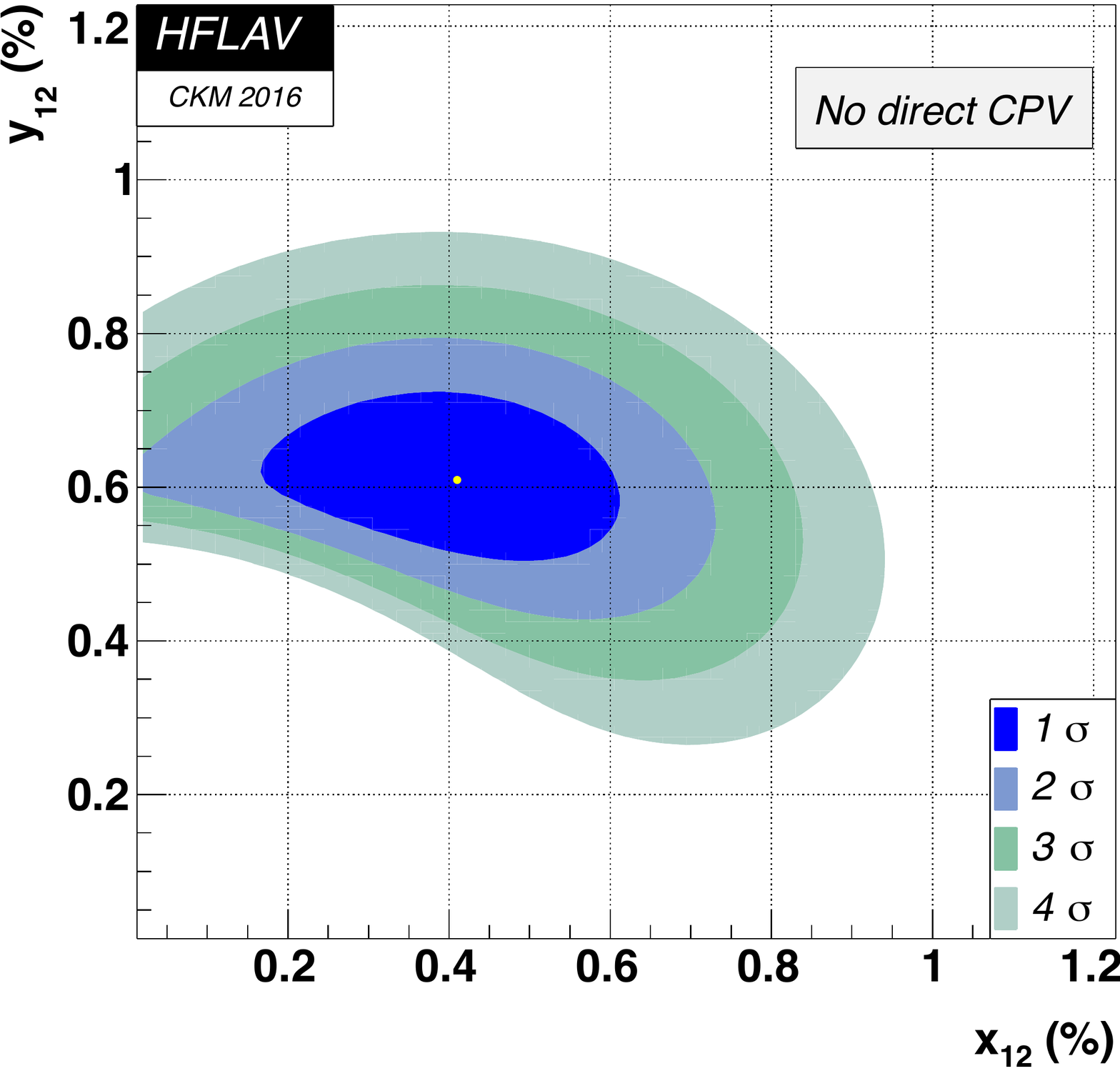}
%\vskip0.10in
\includegraphics[width=84mm]{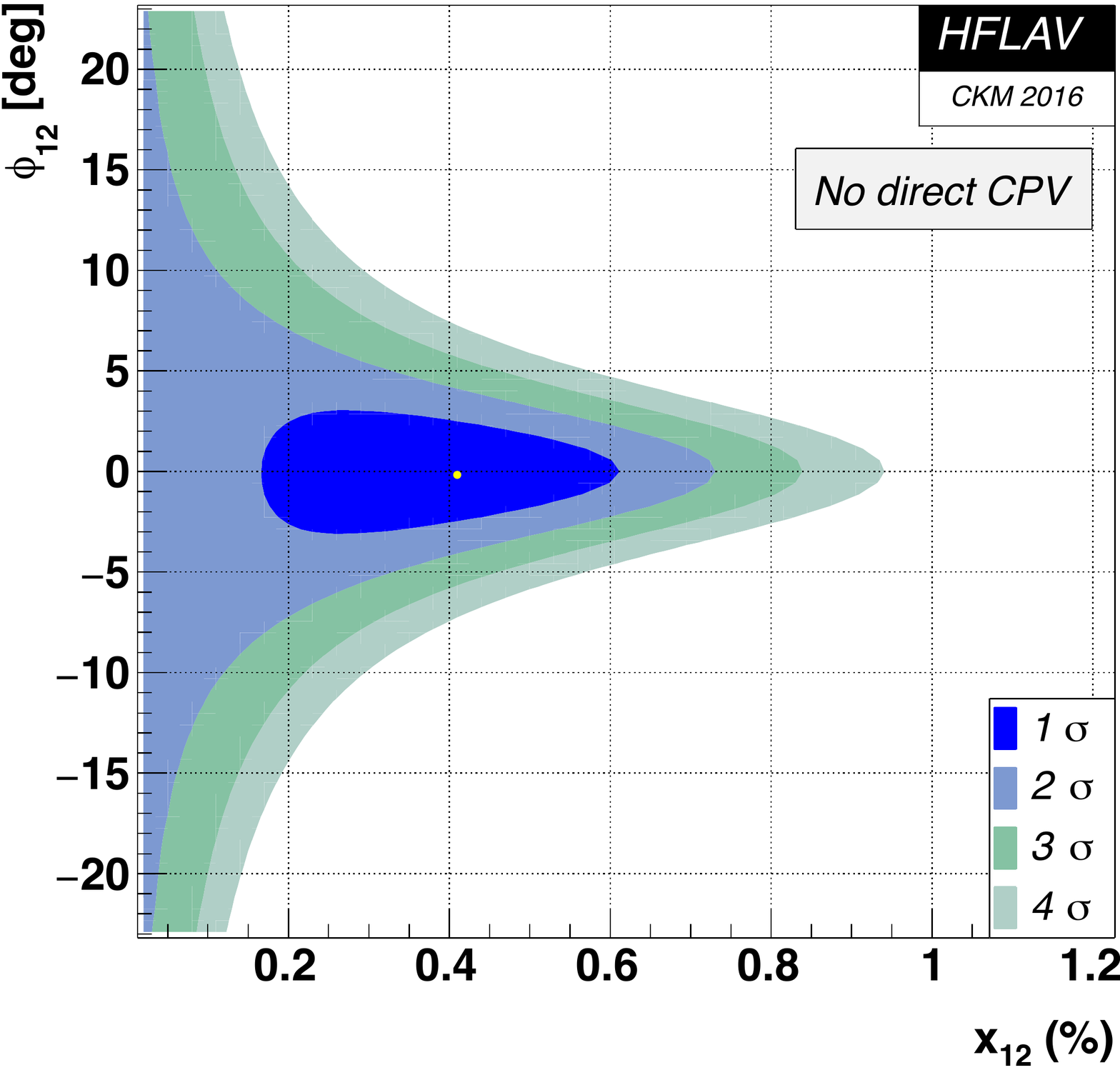}
\vskip-0.60in
\includegraphics[width=84mm]{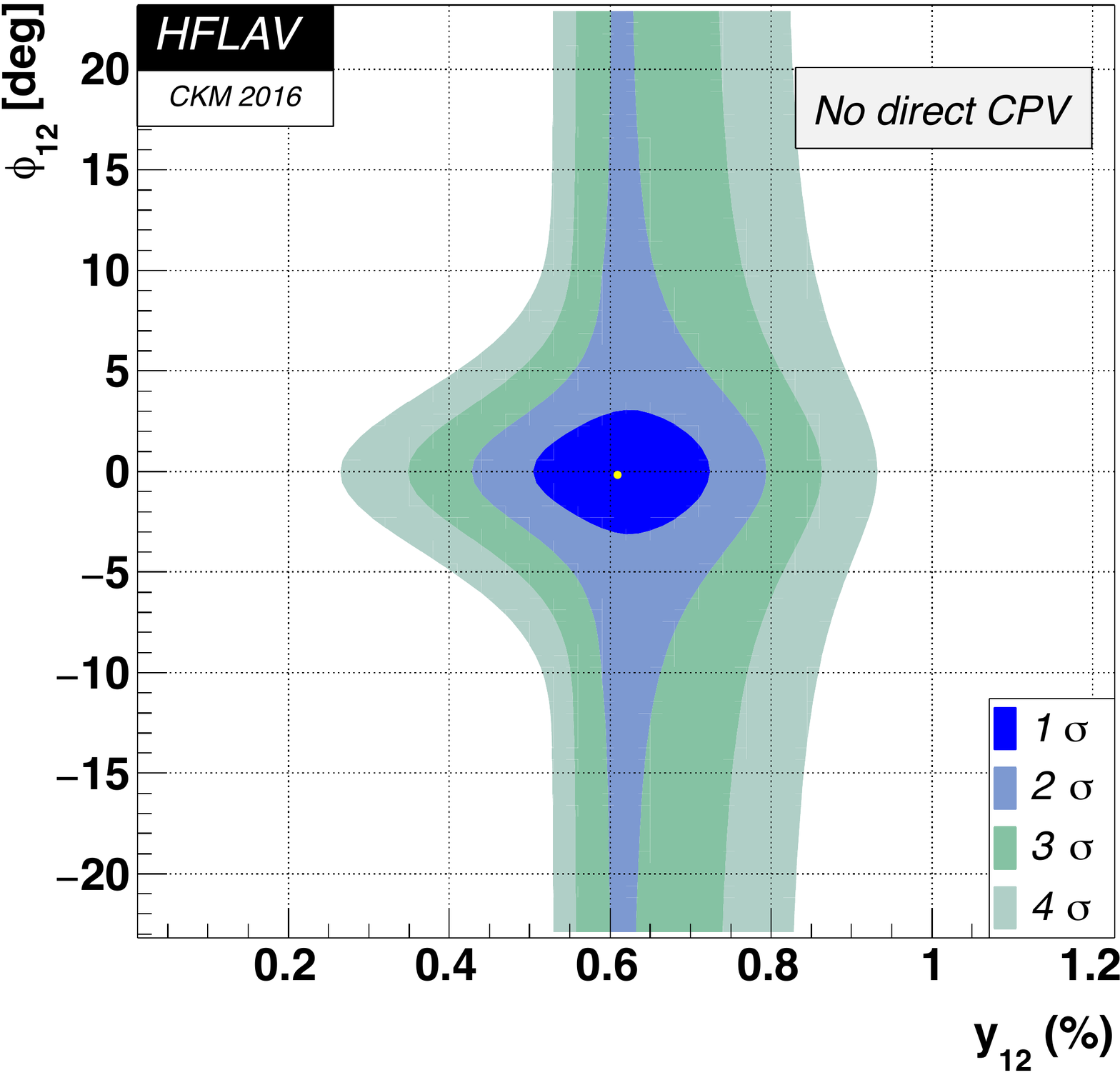}
}
\end{center}
\vskip-0.70in
\caption{\label{fig:contours_ndcpv}
Two-dimensional contours for theoretical parameters 
$(x^{}_{12},y^{}_{12})$ (top left), 
$(x^{}_{12},\phi^{}_{12})$ (top right), and 
$(y^{}_{12},\phi^{}_{12})$ (bottom), 
for no direct \cpv\ in DCS decays.}
\end{figure}

\begin{figure}
\vskip-0.40in
\begin{center}
\vbox{
\includegraphics[width=4.0in]{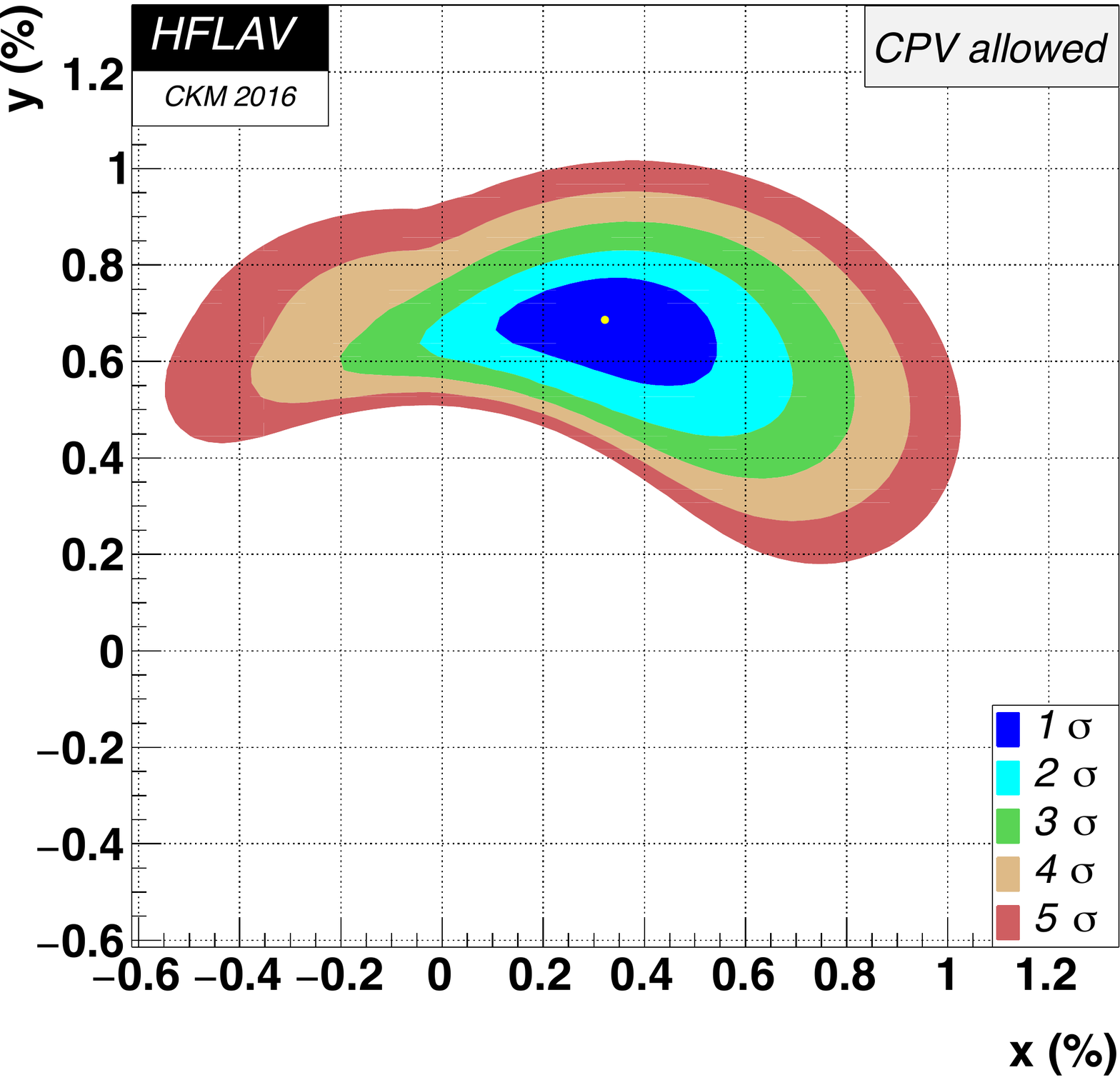}
\vskip-1.0in
\includegraphics[width=4.0in]{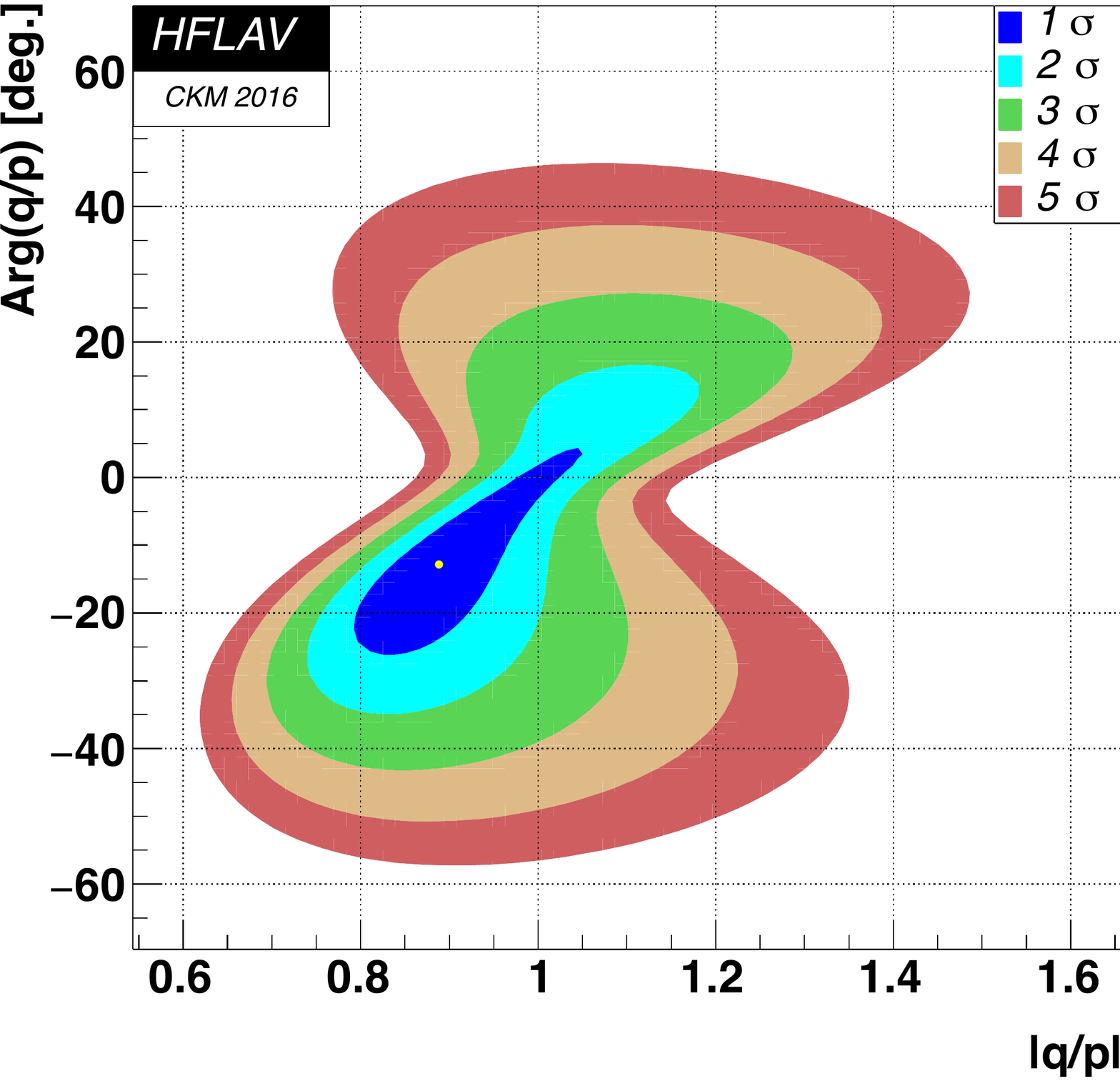}
}
\end{center}
\vskip-0.90in
\caption{\label{fig:contours_cpv}
Two-dimensional contours for parameters $(x,y)$ (top) 
and $(|q/p|,\phi)$ (bottom), allowing for \cpv.}
\end{figure}

Confidence contours in the two dimensions $(x,y)$ or 
$(|q/p|,\phi)$ are obtained by allowing, for any point in the
two-dimensional plane, all other fitted parameters to take their 
preferred values. The resulting $1\sigma$-$5\sigma$ contours 
are shown 
in Fig.~\ref{fig:contours_ncpv} for the \cp-conserving case, 
in Fig.~\ref{fig:contours_ndcpv} for the no-direct-\cpv\ case, 
and in Fig.~\ref{fig:contours_cpv} for the \cpv-allowed 
case. The contours are determined from the increase of the
$\chi^2$ above the minimum value.
One observes that the $(x,y)$ contours for the no-\cpv\ fit 
are very similar to those for the \cpv-allowed fit. In the latter
fit, the $\chi^2$ at the no-mixing point $(x,y)\!=\!(0,0)$ is 450
units above the minimum value, which, for two degrees of freedom,
corresponds to a confidence level\footnote{This is
the limit of the CERNLIB PROB routine~\cite{CERNLIB:webpage}
used for this calculation.}
(C.L.) $>11.5\sigma$.
Thus, no mixing is excluded at this high level. In the $(|q/p|,\phi)$
plot, the no-\cpv\ point $(1,0)$ is within the $1\sigma$ contour; thus 
the data is consistent with \cp\ conservation.

\begin{figure}
\centering
\vskip-0.40in
%\hbox{
\includegraphics[width=68mm]{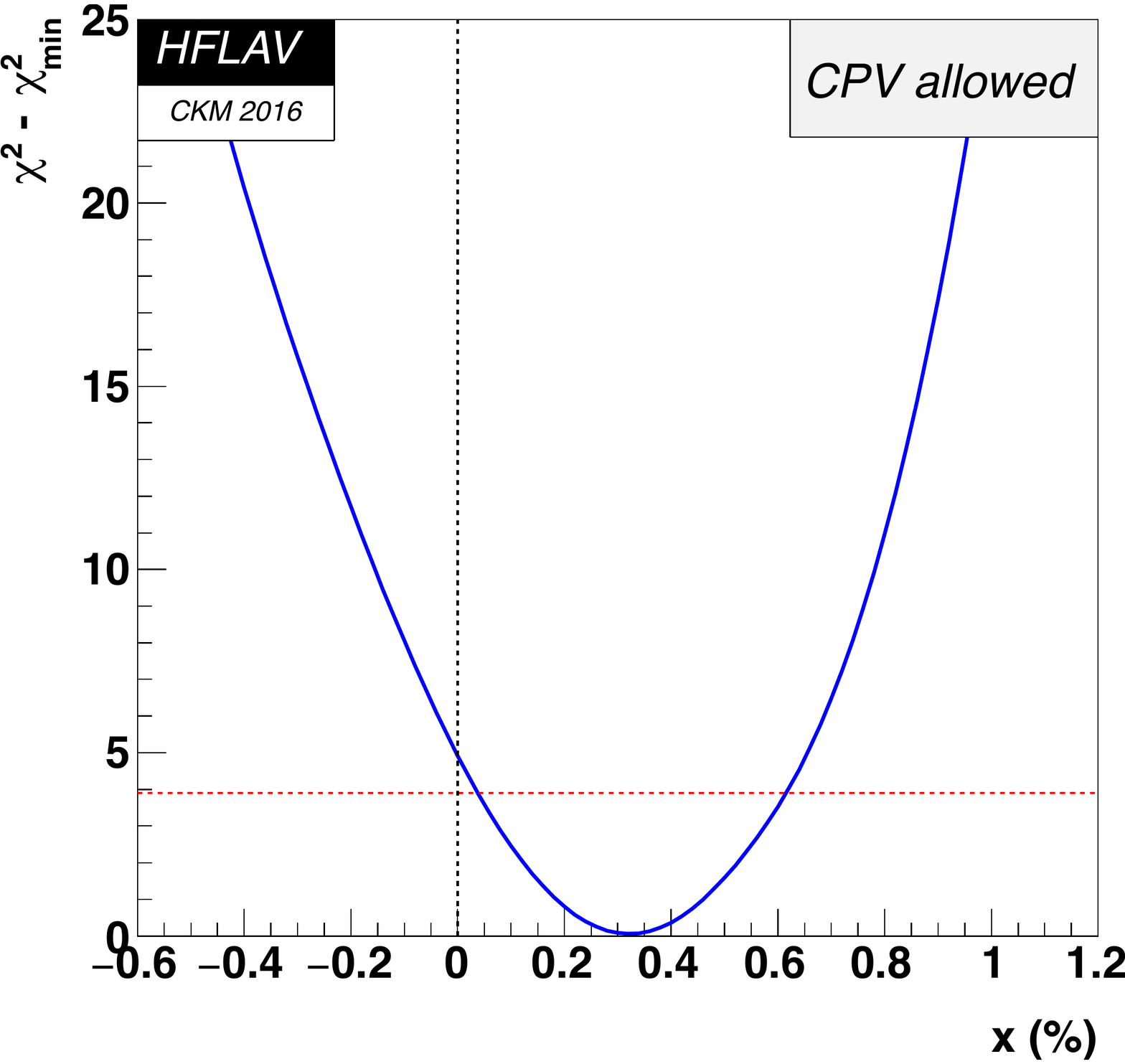}
\includegraphics[width=68mm]{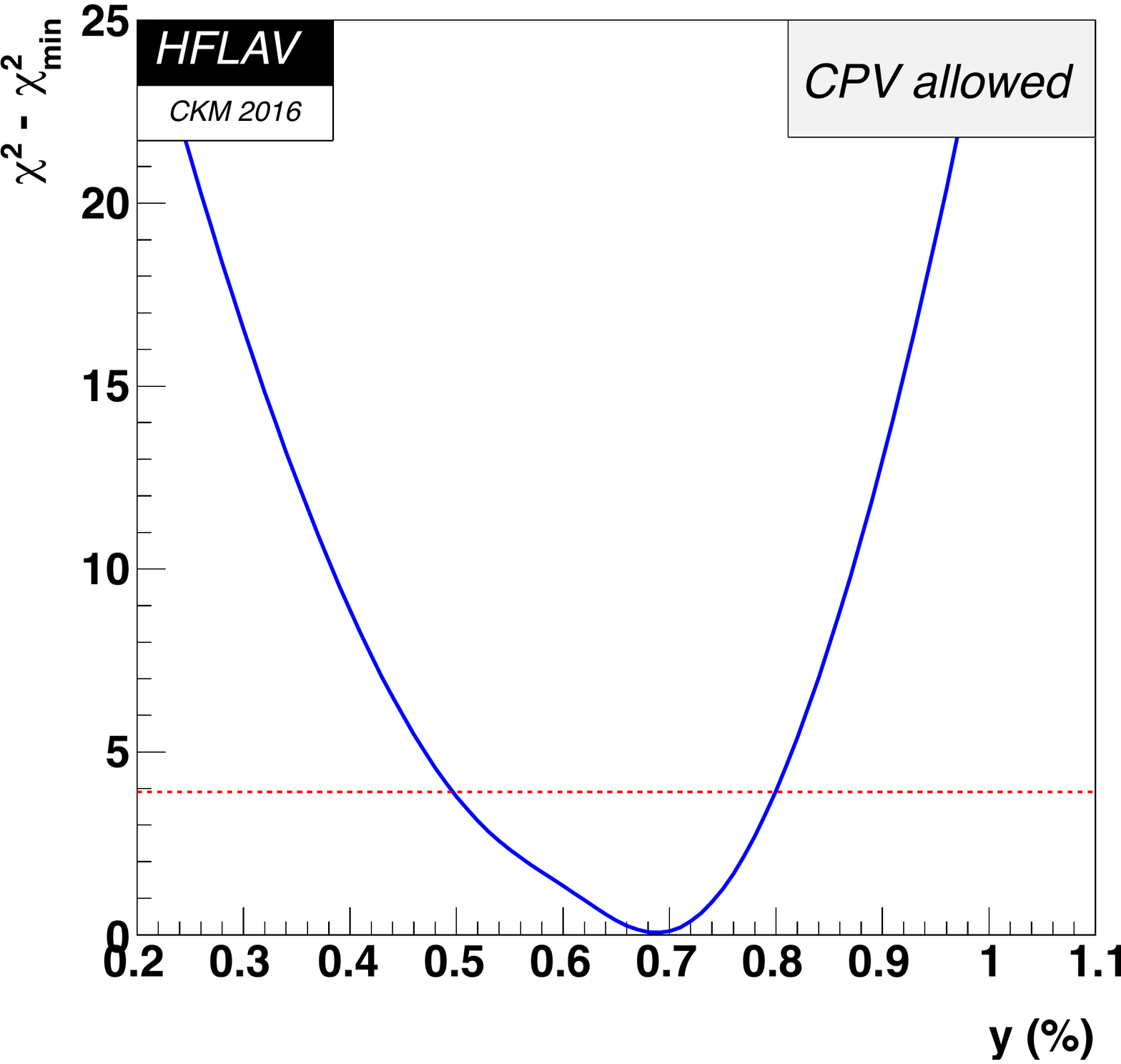}
%}
\vskip-0.70in
%\hbox{
\includegraphics[width=68mm]{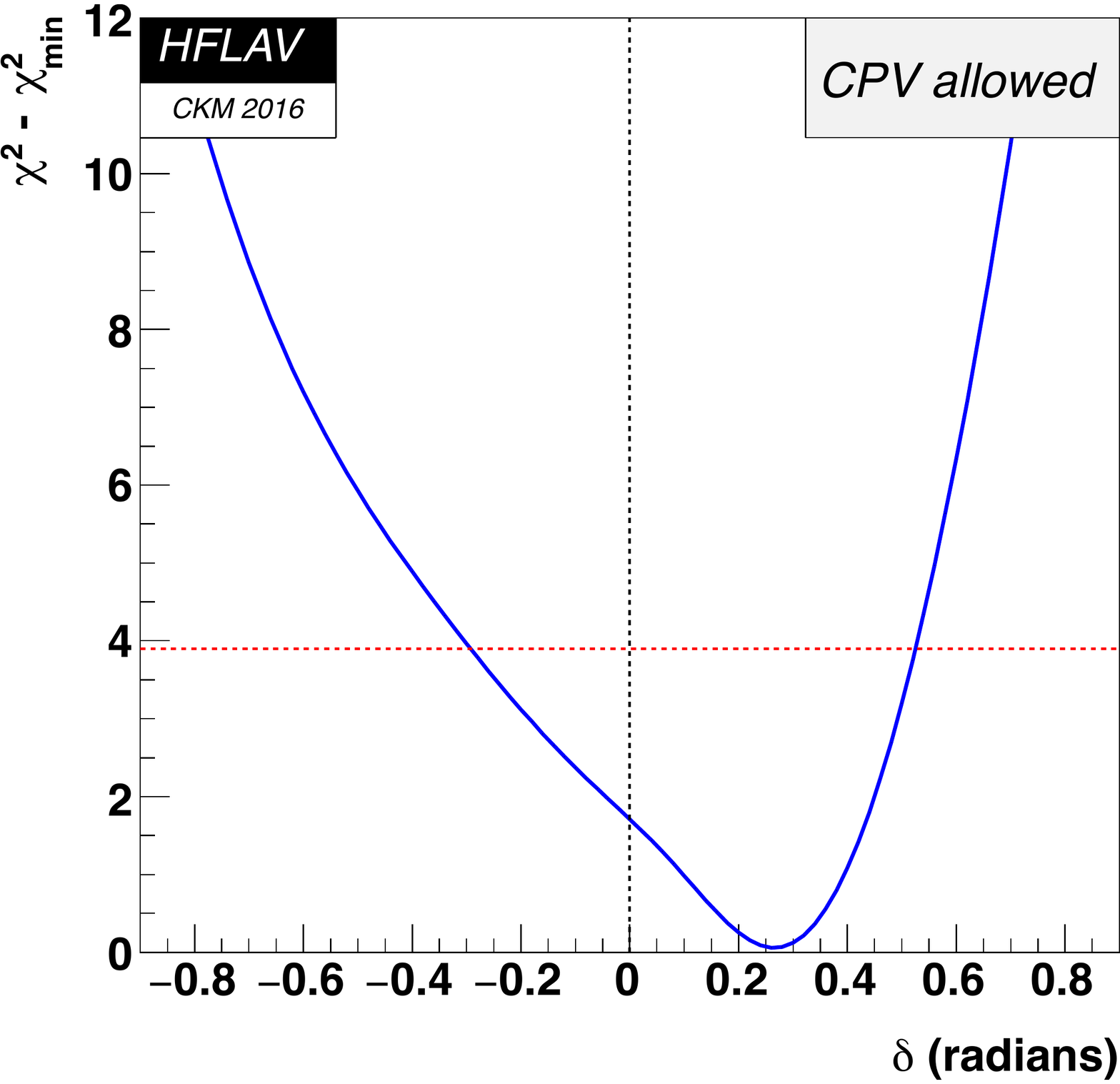}
\includegraphics[width=68mm]{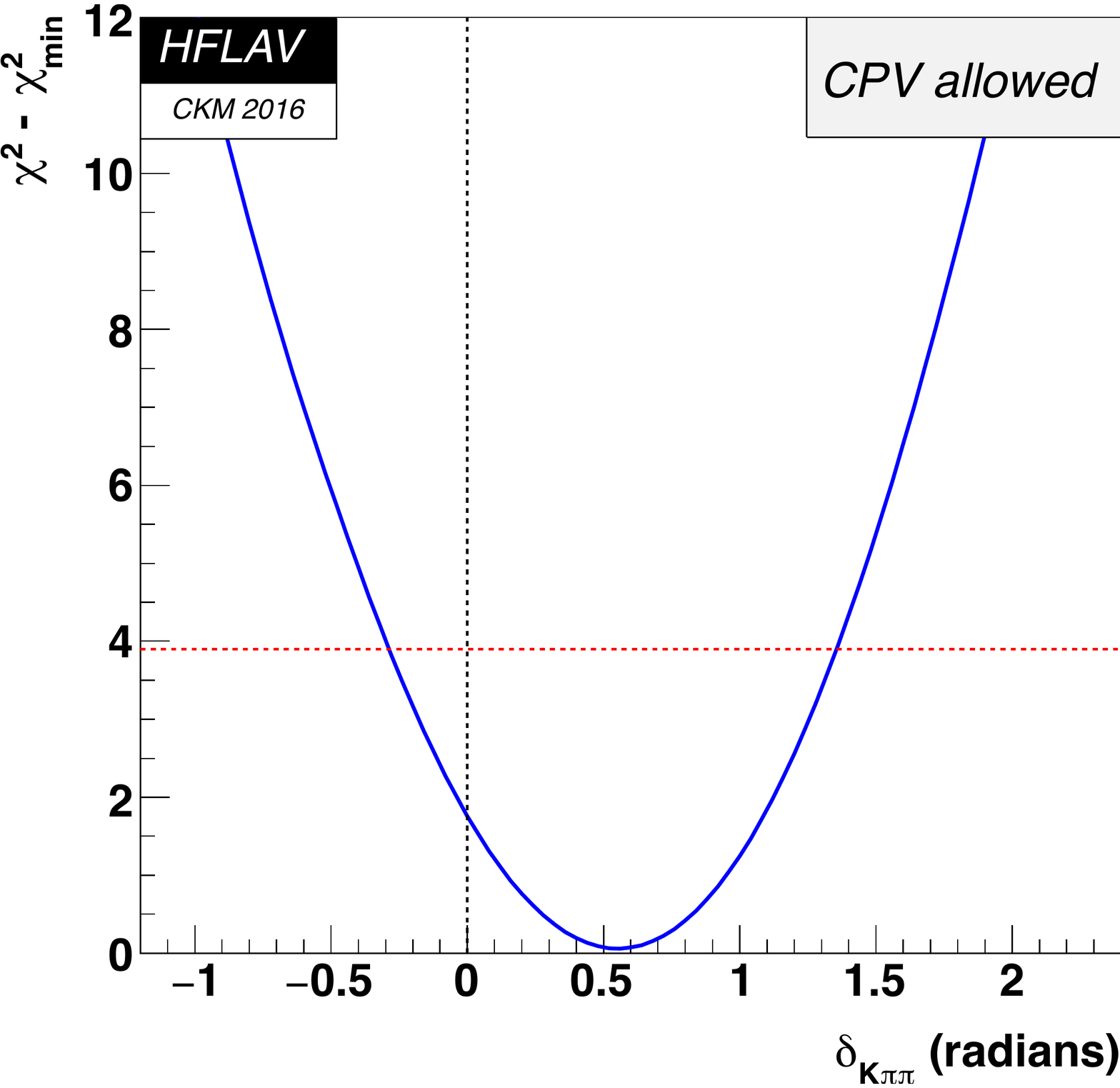}
%}
\vskip-0.70in
%\hbox{
\includegraphics[width=68mm]{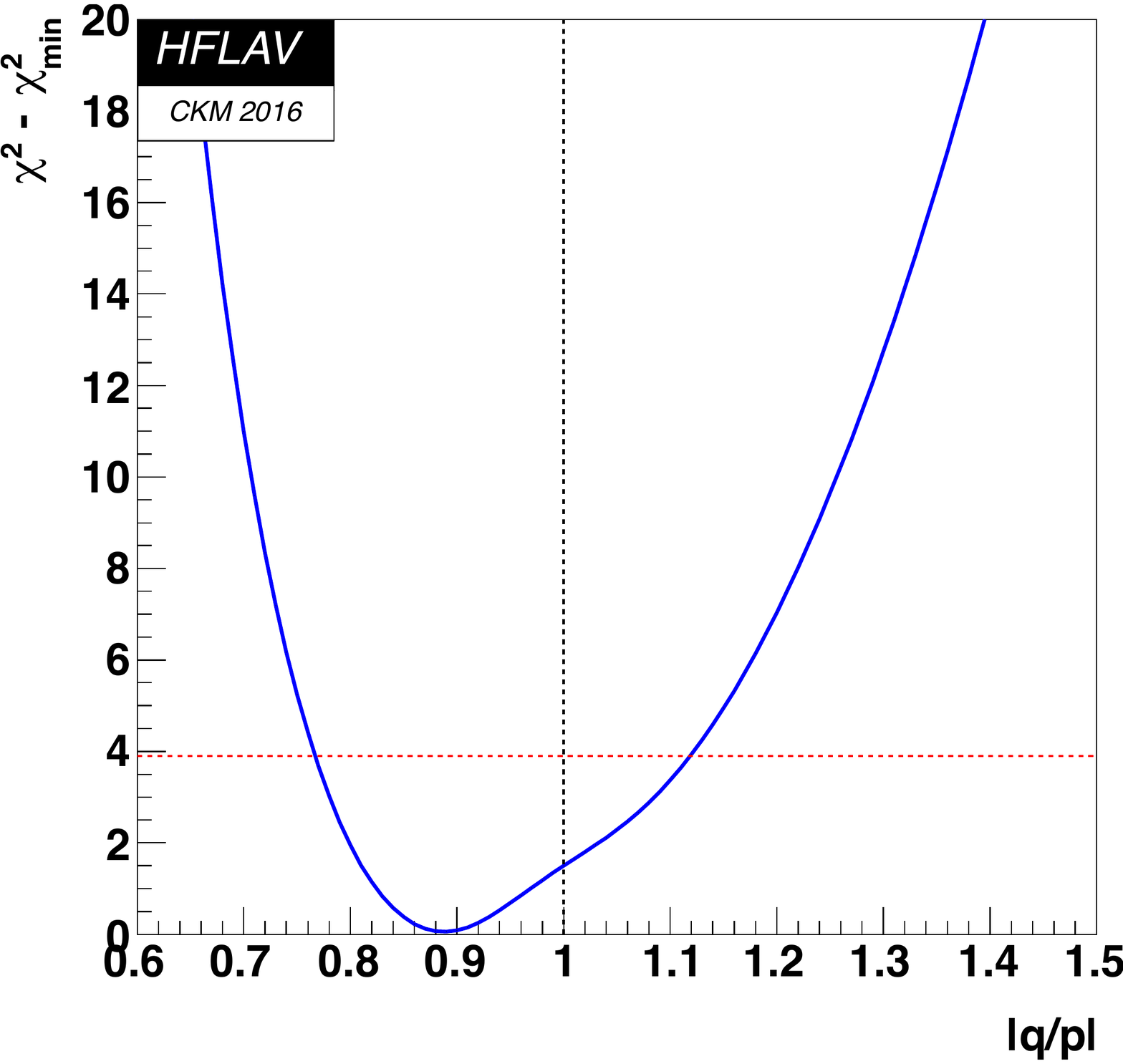}
\includegraphics[width=68mm]{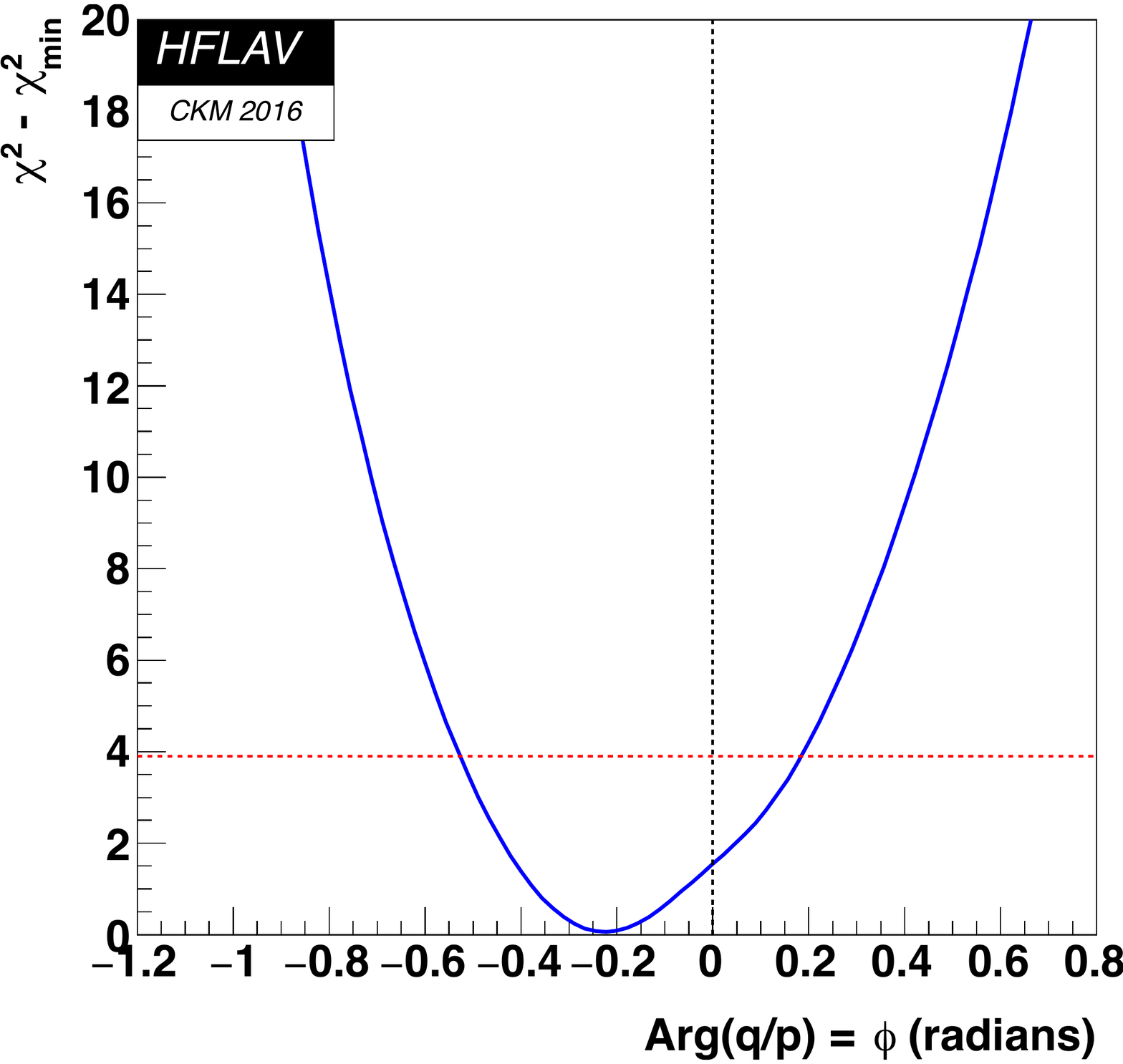}
%}
\vskip-0.40in
\caption{\label{fig:1dlikelihood}
The function $\Delta\chi^2=\chi^2-\chi^2_{\rm min}$ 
for fitted parameters
$x,\,y,\,\delta,\,\delta^{}_{K\pi\pi},\,|q/p|$, and $\phi$.
The points where $\Delta\chi^2=3.84$ (denoted by dashed 
horizontal lines) determine 95\% C.L. intervals. }
\end{figure}

One-dimensional likelihood curves for individual parameters 
are obtained by allowing, for a fixed value of a selected parameter, 
all other fitted parameters to take their preferred values. The resulting
functions $\Delta\chi^2=\chi^2-\chi^2_{\rm min}$ ($\chi^2_{\rm min}$
is the minimum value) are shown in Fig.~\ref{fig:1dlikelihood}.
The points where $\Delta\chi^2=3.84$ determine 95\% C.L. intervals 
for the parameters. These intervals are listed in Table~\ref{tab:results}.

\begin{table}
\renewcommand{\arraystretch}{1.4}
\begin{center}
\caption{\label{tab:results}
Results of the global fit for different assumptions concerning~\cpv.}
\vspace*{6pt}
\footnotesize
\begin{tabular}{c|cccc}
\hline
\textbf{Parameter} & \textbf{\boldmath No \cpv} & \textbf{\boldmath No direct \cpv} 
& \textbf{\boldmath \cpv-allowed} & \textbf{\boldmath \cpv-allowed} \\
 & & \textbf{\boldmath in DCS decays} & & \textbf{\boldmath 95\% C.L.\ Interval} \\
%\begin{tabular}{c|cc}
%\hline
%\textbf{Parameter} & \textbf{\boldmath No \cpv} & \textbf{\boldmath \cpv-allowed} \\
\hline
$\begin{array}{c}
x\ (\%) \\ 
y\ (\%) \\ 
\delta^{}_{K\pi}\ (^\circ) \\ 
R^{}_D\ (\%) \\ 
A^{}_D\ (\%) \\ 
|q/p| \\ 
\phi\ (^\circ) \\
\delta^{}_{K\pi\pi}\ (^\circ)  \\
A^{}_{\pi} (\%) \\
A^{}_K (\%) \\
x^{}_{12}\ (\%) \\ 
y^{}_{12}\ (\%) \\ 
\phi^{}_{12} (^\circ)
\end{array}$ & 
$\begin{array}{c}
0.46\,^{+0.14}_{-0.15} \\
0.62\,\pm 0.08 \\
8.0\,^{+9.7}_{-11.2} \\
0.348\,^{+0.004}_{-0.003} \\
- \\
- \\
- \\
20.4\,^{+23.3}_{-23.8} \\
- \\
- \\
- \\
- \\
- 
\end{array}$ &
$\begin{array}{c}
0.41\,^{+0.14}_{-0.15}\\
0.61\,\,\pm 0.07 \\
4.8\,^{+10.4}_{-12.3}\\
0.347\,\,^{+0.004}_{-0.003} \\
- \\
0.999\,\pm 0.014 \\
0.05\,\,^{+0.54}_{-0.53} \\ 
22.6\,^{+24.1}_{-24.4} \\ 
+0.02\,\pm 0.13 \\
-0.11\,\pm 0.13 \\
0.41\,^{+0.14}_{-0.15}\\
0.61\,\pm 0.07 \\
-0.17\,\pm 1.8 
\end{array}$ &
$\begin{array}{c}
0.32\,\pm 0.14 \\
0.69\,\,^{+0.06}_{-0.07}\\
15.2\,^{+7.6}_{-10.0} \\
0.349\,^{+0.004}_{-0.003} \\
-0.88\,\pm 0.99 \\
0.89\,^{+0.08}_{-0.07} \\ 
-12.9\,^{+9.9}_{-8.7} \\ 
31.7\,^{+23.5}_{-24.2} \\
+0.01\,\pm 0.14 \\
-0.11\,\pm 0.13 \\
 \\
 \\
 \\
\end{array}$ &
$\begin{array}{c}
\left[ 0.04,\, 0.62\right] \\
\left[ 0.50,\, 0.80\right] \\
\left[ -16.8,\, 30.1\right] \\
\left[ 0.342,\, 0.356\right] \\
\left[ -2.8,\, 1.0\right] \\
\left[ 0.77,\, 1.12\right] \\\
\left[ -30.2,\, 10.6\right] \\
\left[ -16.4,\, 77.7\right] \\
\left[ -0.25,\, 0.28\right] \\
\left[ -0.37,\, 0.14\right] \\
\left[ 0.10,\, 0.67\right] \\
\left[ 0.47,\, 0.75\right] \\
\left[ -5.3,\, 4.4\right] \\
\end{array}$ \\
\hline
\end{tabular}
\end{center}
\end{table}

\begin{table}
\renewcommand{\arraystretch}{1.4}
\begin{center}
\caption{\label{tab:results_chi2}
Individual contributions to the $\chi^2$ for the \cpv-allowed fit.}
\vspace*{6pt}
\footnotesize
\begin{tabular}{l|rr}
\hline
\textbf{Observable} & \textbf{\boldmath $\chi^2$} & \textbf{\boldmath $\sum\chi^2$} \\
\hline
$y^{}_{\CP}$                      & 1.19 & 1.19 \\
$A^{}_\Gamma$                    & 0.83 & 2.01 \\
\hline
$x^{}_{K^0_S\pi^+\pi^-}$ Belle       & 1.33 & 3.35 \\
$y^{}_{K^0_S\pi^+\pi^-}$ Belle       & 5.30 & 8.64 \\
$|q/p|^{}_{K^0_S\pi^+\pi^-}$ Belle   & 0.10 & 8.74 \\
$\phi^{}_{K^0_S\pi^+\pi^-}$  Belle   & 0.23 & 8.97 \\
\hline
$x^{}_{K^0_S\pi^+\pi^-}$ LHCb       & 4.51 & 13.48 \\
$y^{}_{K^0_S\pi^+\pi^-}$ LHCb       & 0.40 & 13.88 \\
\hline
$x^{}_{K^0_S h^+ h^-}$ \babar        & 0.36 & 14.24 \\
$y^{}_{K^0_S h^+ h^-}$ \babar        & 0.19 & 14.43 \\
\hline
$x^{}_{\pi^0\pi^+\pi^-}$ \babar    & 0.77 & 15.20 \\
$y^{}_{\pi^0\pi^+\pi^-}$ \babar    & 0.22 & 15.42 \\
\hline
$(x^2+y^2)^{}_{K^+\ell^-\bar{\nu}}$ & 0.14 & 15.56 \\
\hline
$x^{}_{K^+\pi^-\pi^0}$ \babar      & 7.10 & 22.67 \\
$y^{}_{K^+\pi^-\pi^0}$ \babar      & 3.92 & 26.58 \\
\hline
CLEOc                           &      &       \\
($x/y/R^{}_D/\cos\delta/\sin\delta$) 
                                & 10.53 & 37.12 \\
\hline
$R^+_D/x'{}^{2+}/y'{}^+$ \babar  & 11.13 & 48.25    \\
$R^-_D/x'{}^{2-}/y'{}^-$ \babar  &  6.04 & 54.29    \\
$R^+_D/x'{}^{2+}/y'{}^+$ Belle   &  2.08 & 56.36    \\
$R^-_D/x'{}^{2-}/y'{}^-$ Belle   &  3.22 & 59.58    \\
$R^{}_D/x'{}^{2}/y'$ CDF         &  1.29 & 60.87    \\
$R^+_D/x'{}^{2+}/y'{}^+$ LHCb    &  0.58 & 61.46    \\
$R^-_D/x'{}^{2-}/y'{}^-$ LHCb    &  1.65 & 63.11    \\
\hline
$A^{}_{KK}/A^{}_{\pi\pi}$  \babar & 0.30 & 63.41  \\
$A^{}_{KK}/A^{}_{\pi\pi}$  Belle  & 2.89 & 66.30  \\
$A^{}_{KK}/A^{}_{\pi\pi}$  CDF    & 4.63 & 70.94  \\
$A^{}_{KK}-A^{}_{\pi\pi}$  LHCb ($D^{*+}\ra D^0\pi^+$ tag)   
                                & 0.12 & 71.05  \\
$A^{}_{KK}-A^{}_{\pi\pi}$  LHCb ($\overline{B}\ra D^0\mu^-X$ tag)
                                & 2.24 & 73.30  \\
\hline
$(x^2+y^2)^{}_{K^+\pi^-\pi^+\pi^-}$ LHCb  & 3.48 & 76.78 \\
\hline
\end{tabular}
\end{center}
\end{table}

%\newpage

\subsubsection{Conclusions}

From the fit results listed in Table~\ref{tab:results}
and shown in Figs.~\ref{fig:contours_cpv} and \ref{fig:1dlikelihood},
we conclude that:
\begin{itemize}
\item 
the experimental data consistently indicate that 
$D^0$ mesons mix. The no-mixing point $x=y=0$
is excluded at $>11.5\sigma$. The parameter $x$ differs
from zero by $1.9\sigma$, and $y$ differs from zero by
$9.4\sigma$. This mixing is presumably dominated 
by long-distance processes, which are difficult to calculate.
Thus unless it turns out that $|x|\gg |y|$, which is not 
indicated, it will be difficult to identify new physics 
from $(x,y)$ alone~\cite{Bigi:2000wn}.
\item 
Since \ycp\ is positive, the \cp-even state is shorter-lived
as in the $K^0$-$\kbar$ system. However, since $x$ also appears
to be positive, the \cp-even state is heavier, unlike in the 
$K^0$-$\kbar$ system.
\item 
There is no evidence for \cpv\ arising from $D^0$-$\dbar$
mixing ($|q/p|\neq 1$) or from a phase difference between
the mixing amplitude and a direct decay amplitude ($\phi\neq 0$). 
The CDF experiment (and initially LHCb) measured a 
time-integrated asymmetry in $D^0\ra K^+K^-, \pi^+\pi^-$ 
decays that hints at direct \cpv\ (see Table~\ref{tab:observables3}); 
however, recent measurements from LHCb with higher statistics
disfavor this hypothesis and are consistent with zero.
\end{itemize}

\clearpage
% Direct CPV
\mysubsection{\CP\ asymmetries}\label{sec:cp_asym}

One way \CP\ violation manifests itself is if the decay rate for a particle 
differs from that of its \CP-conjugate\cite{Bigi:2000yz}. 
Such phenomena can be classified into two broad categories, 
%In general there are two classes of \CP\ violation, 
termed {\it direct\/} \CP\ violation and 
{\it indirect\/} \CP\ violation~\cite{Nir:1999mg}. 
Direct \CP\ violation refers to charm changing, $\Delta C\!=\!1$,
processes and can occur in both charged and neutral 
charm hadron decays. It results from interference between two different decay
amplitudes (\eg, a penguin and tree amplitude) that have
different weak (CKM) and strong phases.\footnote{The weak 
phase difference will have opposite signs for $D\ra f$ and 
$\overline{D}\ra\bar{f}$ decays, while the strong phase difference 
will have the same sign. As a result, squaring the total amplitudes 
to obtain the decay rates gives interference terms having 
opposite sign, \ie, non-identical decay rates.}
In the Standard Model a difference in strong phases may arise for example
due to final-state interactions (FSI)\cite{Buccella:1994nf}, different isospin 
amplitudes, intermediate resonance contributions, or different partial waves.
A difference in weak phases arises from different CKM vertex couplings, as 
is often the case for tree and penguin diagrams. Within the SM direct 
\CP\ violation is expected only in Singly Cabibbo Suppressed (SCS) charm 
decays, as only these decays receive a contribution from the penguin amplitude. 
This type of \CP\ violation depends on the decay mode, the SM asymmetries may 
reach a percent level. Indirect \CP\ violation refers to $\Delta C\!=\!2$ 
processes and arises in $D^0$ decays due to $D^0$-$\dbar$ mixing. 
It can occur as an asymmetry in the mixing itself, or it can 
result from interference between a decay amplitude following mixing and a 
non-mixed amplitude. Within the SM charm indirect \CP\ asymmetry is expected 
to be universal.

\vspace{0.6cm}
The \CP\ asymmetry is defined as the difference between 
$D$ and $\overline{D}$ partial widths divided by their sum:
\begin{eqnarray}  
A_{\CP} & = & \frac{\Gamma(D)-\Gamma(\overline{D})}
{\Gamma(D)+\Gamma(\overline{D})}\,.
\end{eqnarray}
In the case of $D^+$ and $D^+_s$ decays, $A^{}_{\CP}$ measures 
direct \CP\ violation; in the case of $D^0$ decays, $A^{}_{\CP}$ 
measures direct and indirect \CP\ violation combined 
(see also Sec.~\ref{sec:charm:cpvdir}).

In each experiment, care must be taken to correct for production 
and detection asymmetries.
To take into account differences in production rates between 
$D$ and $\overline{D}$ (which would affect the number of respective 
decays observed), some experiments (like FOCUS and E791) normalize 
to a Cabibbo-favored mode. In this case there is the additional benefit 
that most corrections due to reconstruction inefficiencies cancel out, 
reducing systematic uncertainties. An implicit assumption is that 
there is no measurable \CP\ violation in the Cabibbo-favored 
normalizing mode. The \CP\ asymmetry is calculated as
\begin{eqnarray}
A_{\CP} & = & \frac{\eta(D)-\eta(\overline{D})}{\eta(D)+\eta(\overline{D})}\,,
\end{eqnarray}
where (considering, for example, $D^0 \to K^-K^+$)
\begin{eqnarray}
 \eta(D) & = & \frac{N(D^0 \rightarrow K^-K^+)}{N(D^0 \rightarrow K^-\pi^+)}\,, \\
 \eta(\overline{D}) & = & \frac{N(\dbar\rightarrow K^-K^+)}
{N(\dbar\rightarrow K^+\pi^-)}\,.
\end{eqnarray}
Other experiments (like LHCb) determine $A_{\CP}$ through the relation:
\begin{eqnarray}
A_{\rm meas} & = & A_{\CP} + A_{\rm prod} + A_{\rm det}\,,
\end{eqnarray}
where $A_{\rm meas}$ is the measured asymmetry, $A_{\rm prod}$ is the 
asymmetry in the charm meson production, and $A_{\rm det}$ is due to difference 
in detection efficiencies between positevely and negatively charged hadrons.

Values of $A^{}_{\CP}$ for $D^+$, $D^0$ and $D_s^+$ decays are listed in
Tables~\ref{tab:cp_charged}, \ref{tab:cp_charged2}, \ref{tab:cp_neutral}, 
\ref{tab:cp_neutral2} and \ref{tab:cp_ds} respectively. In these tables 
we report asymmetries for the actual final state, \ie, resonant 
substructure is implicitly included but not considered separately. 
The high accuracy of these measurements allows one to see and 
correct for \CP\ violation due to the CPV in $K^0$-$\kbar$ 
mixing~\cite{Grossman:2012aa}. For example, the decay modes 
$D^+ \to (\kbar/K^0)K^+$ and $D_s^+ \to (\kbar/K^0)\pi^+$ 
(shown in Tables~\ref{tab:cp_charged} and 
\ref{tab:cp_ds}, respectively) are the modes 
$D^+ \to K^0_s K^+$ and $D_s^+ \to K^0_s \pi^+$ 
after subtracting for this effect. For multi-body decays some experiments 
use model independent techniques to reveal local \CP\ asymmetry.
The first technique (Miranda method)~\cite{Bediaga:2009tr} uses a binned 
$\chi^2$ approach to compare the relative density in a bin of phase 
space of a decay with that of its \CP\ conjugate. In the Energy Test 
technique~\cite{doi:10.1080/00949650410001661440} two event samples are 
compared and a test statistic variable (T) is used to determine the average 
distances of events in phase space. If the distributions of events 
in both samples are identical, T will randomly fluctuate around a value 
close to zero. 

\vspace{0.4cm}
Overall, \CP\ asymmetry measurements have been carried out 
for 49 charm decay modes, and in several modes the sensitivity 
is well below $5 \times 10^{-3}$. There is currently no evidence 
for \CP\ violation in the charm meson sector.
The \CP\ asymmetry observed in the mode $D^+ \to K^0_s\pi^+$ 
is consistent with what expected from the $K^0-\kbar$ 
system~\cite{Grossman:2012aa}, and thus it is not attributed 
to charm.

Neither in the charm baryon sector there is evidence of \CP\ asymmetry.
These are just two measurements on $\Lambda_C^+$, with limited sensitivity, 
done by FOCUS~\cite{ Link:2005ft} and by CLEO~\cite{Hinson:2004pj}.

Taken together, the limits obtained for \CP\ asymmetries 
in the charm sector pose tight constraints on new physics models.

\begin{table}[!htb]
\renewcommand{\arraystretch}{1.4}
\caption{\CP\ asymmetries 
$A^{}_{\CP}= [\Gamma(D^+)-\Gamma(D^-)]/[\Gamma(D^+)+\Gamma(D^-)]$
for two-body $D^\pm$ decays.
\label{tab:cp_charged}}
\footnotesize
\begin{center}
\begin{tabular}{|l|c|c|c|} 
\hline
{\bf Mode} & {\bf Year} & {\bf Collaboration} & {\boldmath $A^{}_{\CP}$} \\
\hline
{\boldmath $D^+ \to \mu^+ \nu$} &
  2008 & CLEO~\cite{Eisenstein:2008aa} &  $ +0.08  \pm 0.08 $ \\
\hline
{\boldmath $D^+ \to \pi^+ \pi^0$} &
  2010 & CLEO~\cite{Mendez:2009aa} &  $ +0.029  \pm 0.029 \pm 0.003 $ \\
\hline
{\boldmath $D^+ \to \pi^+ \eta$} &
  2011 & Belle~\cite{Won:2011ng}    &  $ +0.0174  \pm 0.0113 \pm 0.0019 $ \\
& 2010 & CLEO~\cite{Mendez:2009aa} &  $ -0.020   \pm 0.023  \pm 0.003 $ \\
&      & HFLAV average             &  $ +0.010   \pm 0.010 $ \\
\hline
{\boldmath $D^+ \to \pi^+ \eta^\prime$} &
  2011 & Belle~\cite{Won:2011ng}    &  $ -0.0012  \pm 0.0112 \pm 0.0017 $ \\
& 2010 & CLEO~\cite{Mendez:2009aa} &  $ -0.040   \pm 0.034  \pm 0.003  $ \\
&      & HFLAV average             &  $ -0.005   \pm 0.011 $ \\  
\hline
{\boldmath $D^+ \to K^+ \pi^0$} &
  2010 & CLEO~\cite{Mendez:2009aa} &  $ -0.035  \pm 0.107 \pm 0.009 $ \\
\hline
{\boldmath $D^+ \to K^0_s\pi^+$}   &
   2014 & CLEO~\cite{Bonvicini:2013vxi} &  $ -0.011   \pm 0.006   \pm 0.002   $ \\
&  2012 & Belle~\cite{Ko:2012pe}        &  $ -0.00363 \pm 0.00094 \pm 0.00067 $ \\
&  2011 & \babar~\cite{Amo:2011ab}      &  $ -0.0044  \pm 0.0013  \pm 0.0010  $ \\
&  2002 & FOCUS~\cite{Link:2001zj}      &  $ -0.016   \pm 0.015   \pm 0.009   $ \\
&       & HFLAV average                &  $ -0.0041  \pm 0.0009 $ \\
\hline
{\boldmath $D^+ \to K^0_sK^+$} &
  2013 & \babar~\cite{Lees:2013aa}      &  $ +0.0013 \pm 0.0036 \pm 0.0025 $ \\
& 2013 & Belle~\cite{Ko:2013aa}         &  $ -0.0025 \pm 0.0028 \pm 0.0014 $ \\
& 2010 & CLEO~\cite{Mendez:2009aa}      &  $ -0.002  \pm 0.015  \pm 0.009  $ \\  
& 2002 & FOCUS~\cite{Link:2001zj}       &  $ +0.071  \pm 0.061  \pm 0.012  $ \\
&      & HFLAV average                 &  $ -0.0011 \pm 0.0025 $ \\
\hline
{\boldmath $D^+ \to (\kbar/K^0)K^+$} &
  2014 & LHCb~\cite{Aaij:2014ac}        &  $ +0.0003 \pm 0.0017 \pm 0.0014 $ \\
& 2013 & \babar~\cite{Lees:2013aa}      &  $ +0.0046 \pm 0.0036 \pm 0.0025 $ \\
& 2013 & Belle~\cite{Ko:2013aa}         &  $ -0.0008 \pm 0.0028 \pm 0.0014 $ \\
&      & HFLAV average                 &  $ +0.0011 \pm 0.0017 $ \\
\hline
\end{tabular}
\end{center} 
\end{table}

\begin{table}[!htb]
\renewcommand{\arraystretch}{1.4}
\caption{\CP\ asymmetries 
$A^{}_{\CP}= [\Gamma(D^+)-\Gamma(D^-)]/[\Gamma(D^+)+\Gamma(D^-)]$
for three- and four-body $D^\pm$ decays.
\label{tab:cp_charged2}}
\footnotesize
\begin{center}
\begin{tabular}{|l|c|c|c|} 
\hline
{\bf Mode} & {\bf Year} & {\bf Collaboration} & {\boldmath $A^{}_{\CP}$} \\
\hline
{\boldmath $D^+ \to \pi^+\pi^-\pi^+$} &
  2014 & LHCb~\cite{Aaij:2014aa}        &  Model independent technique, no 
evidence for \CP\ violation \\
& 1997 & E791~\cite{Aitala:1996sh}      &  $ -0.017  \pm 0.042  $ (stat.) \\
\hline
{\boldmath $D^+ \to K^-\pi^+\pi^+$} &
   2014 & D0~\cite{Abazov:2014wga}       &  $ -0.0016 \pm 0.0015 \pm 0.0009 $ \\
&  2014 & CLEO~\cite{Bonvicini:2013vxi}  &  $ -0.003  \pm 0.002  \pm 0.004  $ \\
&       & HFLAV average                 &  $ -0.0018 \pm 0.0016            $ \\        
\hline
{\boldmath $D^+ \to K^0_s\pi^+\pi^0$} &
  2014 & CLEO~\cite{Bonvicini:2013vxi} &  $ -0.001  \pm 0.007 \pm 0.002  $ \\
\hline
{\boldmath $D^+ \to K^+K^-\pi^+$} &
   2014 & CLEO~\cite{Bonvicini:2013vxi}  &  $ -0.001  \pm 0.009  \pm 0.004  $ \\
&  2013 & \babar~\cite{Lees:2013ab}      &  $ +0.0037 \pm 0.0030 \pm 0.0015 $ \\
&  2008 & CLEO~\cite{Rubin:2008zi}     &  Dalitz plot analysis, no evidence 
for \CP\ violation\\
&  2000 & FOCUS~\cite{Link:2000aw}       &  $ +0.006  \pm 0.011  \pm 0.005  $ \\
&  1997 & E791~\cite{Aitala:1996sh}      &  $ -0.014  \pm 0.029  $ (stat.)    \\
&       & HFLAV average                 &  $ +0.0032 \pm 0.0031 $            \\
\hline
{\boldmath $D^+ \to K^-\pi^+\pi^+\pi^0$} &
  2014 & CLEO~\cite{Bonvicini:2013vxi}   &  $ -0.003  \pm 0.006  \pm 0.004  $ \\
\hline
{\boldmath $D^+ \to K^0_s\pi^+\pi^+\pi^-$} &
  2014 & CLEO~\cite{Bonvicini:2013vxi}   &  $ +0.000  \pm 0.012  \pm 0.003  $ \\
\hline
{\boldmath $D^+ \to K^0_sK^+\pi^+\pi^-$} &
  2005 & FOCUS~\cite{Link:2005th}  &  $ -0.042  \pm 0.064  \pm 0.022  $ \\
\hline 
\end{tabular}
\end{center} 
\end{table}

\begin{table}[!htb]
\renewcommand{\arraystretch}{1.3}
\caption{\CP\ asymmetries 
$A^{}_{\CP}=[\Gamma(D^0)-\Gamma(\dbar)]/[\Gamma(D^0)+\Gamma(\dbar)]$
for two-body $D^0,\dbar$ decays.
\label{tab:cp_neutral}}
\footnotesize
\begin{center}
\begin{tabular}{|l|c|c|c|} 
\hline
{\bf Mode} & {\bf Year} & {\bf Collaboration} & {\boldmath $A^{}_{\CP}$} \\
\hline
{\boldmath $D^0 \to \pi^+\pi^-$} &
  2014 & LHCb~\cite{Aaij:2014gsa}     & $ -0.0020 \pm 0.0019 \pm 0.0010  $ \\
& 2012 & CDF~\cite{Aaltonen:2012ab}  & $ +0.0022 \pm 0.0024 \pm 0.0011  $ \\
& 2008 & \babar~\cite{Aubert:2007if} & $ -0.0024 \pm 0.0052 \pm 0.0022  $ \\
& 2012 & Belle~\cite{Staric:2008rx}  & $ +0.0043 \pm 0.0052 \pm 0.0012  $ \\
& 2002 & CLEO~\cite{Csorna:2001ww}   & $ +0.019  \pm 0.032  \pm 0.008   $ \\
& 2000 & FOCUS~\cite{Link:2000aw}    & $ +0.048  \pm 0.039  \pm 0.025   $ \\
& 1998 & E791~\cite{Aitala:1997ff}   & $ -0.049  \pm 0.078  \pm 0.030   $ \\
&      & HFLAV average              & $ +0.0000 \pm 0.0015 $ \\
\hline
{\boldmath $D^0 \to \pi^0\pi^0$} &
  2014 & Belle~\cite{Nisar:2014fkc}     & $ -0.0003 \pm 0.0064 \pm 0.0010  $ \\
& 2001 & CLEO~\cite{Bonvicini:2000qm}  & $ +0.001  \pm 0.048 $ (stat. and syst. 
combined) \\
&      & HFLAV average                & $ -0.0003 \pm 0.0064 $ \\ 
\hline
{\boldmath $D^0 \to K_s^0\pi^0$} &
  2014 & Belle~\cite{Nisar:2014fkc}     & $ -0.0021 \pm 0.0016 \pm 0.0007 $ \\
& 2001 & CLEO~\cite{Bonvicini:2000qm}  & $ +0.001  \pm 0.013 $ (stat. and syst. 
combined) \\
&      & HFLAV average                & $ -0.0020 \pm 0.0017 $ \\
\hline
{\boldmath $D^0 \to K_s^0\eta$} &
  2011 & Belle~\cite{Ko:2011ab}        & $ +0.0054 \pm 0.0051 \pm 0.0016 $ \\
\hline
{\boldmath $D^0 \to K_s^0\eta^\prime$} &
  2011 & Belle~\cite{Ko:2011ab}        & $ +0.0098 \pm 0.0067 \pm 0.0014 $ \\  
\hline
{\boldmath $D^0 \to K^0_sK^0_s$} &
  2015 & LHCb~\cite{Aaij:2015fua}        & $ -0.029 \pm 0.052 \pm 0.022 $ \\
& 2001 & CLEO~\cite{Bonvicini:2000qm}   & $ -0.23  \pm 0.19  $ (stat. and syst. 
combined) \\
&      & HFLAV average                 & $ -0.046 \pm 0.054           $ \\ 
\hline
{\boldmath $D^0 \to K^-\pi^+$} &
 2014 & CLEO~\cite{Bonvicini:2013vxi} & $ +0.003  \pm 0.003  \pm 0.006 $  \\
\hline
{\boldmath $D^0 \to K^+K^-$} &
  2014 & LHCb~\cite{Aaij:2014gsa}     & $ -0.0006 \pm 0.0015 \pm 0.0010 $ \\
& 2012 & CDF~\cite{Aaltonen:2012ab}  & $ -0.0024 \pm 0.0022 \pm 0.0009 $ \\
& 2008 & \babar~\cite{Aubert:2007if} & $ +0.0000 \pm 0.0034 \pm 0.0013 $ \\
& 2012 & Belle~\cite{Staric:2008rx}  & $ -0.0043 \pm 0.0030 \pm 0.0011 $ \\ 
& 2002 & CLEO~\cite{Csorna:2001ww}   & $ +0.000  \pm 0.022  \pm 0.008  $ \\
& 2000 & FOCUS~\cite{Link:2000aw}    & $ -0.001  \pm 0.022  \pm 0.015  $ \\
& 1998 & E791~\cite{Aitala:1997ff}   & $ -0.010  \pm 0.049  \pm 0.012  $ \\
&      & HFLAV average              & $ -0.0016 \pm 0.0012            $ \\
\hline
\end{tabular}
\end{center} 
\end{table}

\begin{table}[!htb]
\renewcommand{\arraystretch}{1.3}
\caption{\CP\ asymmetries 
$A^{}_{\CP}=[\Gamma(D^0)-\Gamma(\dbar)]/[\Gamma(D^0)+\Gamma(\dbar)]$
for three- and four-body $D^0,\dbar$ decays.
\label{tab:cp_neutral2}}
\footnotesize
\begin{center}
\begin{tabular}{|l|c|c|c|} 
\hline
{\bf Mode} & {\bf Year} & {\bf Collaboration} & {\boldmath $A^{}_{\CP}$} \\
\hline
{\boldmath $D^0 \to \pi^+\pi^-\pi^0$} &
   2015 & LHCb~\cite{Aaij:2014afa}          & Model independent technique, no evidence for 
\CP violation \\
&  2008 & \babar~\cite{Aubert:2008yd}       & $ +0.0031 \pm  0.0041 \pm  0.0017$ \\
&  2008 & Belle~\cite{Arinstein:2008zh}     & $ +0.0043 \pm  0.0130 $ (stat. and 
syst. combined) \\
&  2005 & CLEO~\cite{CroninHennessy:2005sy} & $ +0.01^{+0.09}_{-0.07} \pm  0.05 $ \\
&       & HFLAV average                    & $ +0.0032 \pm 0.0042 $ \\
\hline
{\boldmath $D^0 \to K^-\pi^+\pi^0$} &
  2014 & CLEO~\cite{Bonvicini:2013vxi} & $ +0.001  \pm 0.003 \pm 0.004   $ \\
\hline   
{\boldmath $D^0 \to K^+\pi^-\pi^0$} &
  2005 & Belle~\cite{Tian:2005ik}        & $ -0.006  \pm 0.053  $ (stat.) \\
& 2001 & CLEO~\cite{Brandenburg:2001ze}  & $ +0.09^{+0.25}_{-0.22}  $ (stat.) \\
&      & HFLAV average                  & $ -0.0014 \pm 0.0517 $ \\
\hline
{\boldmath $D^0 \to K^0_s\pi^+\pi^-$} &
  2012 & CDF~\cite{Aaltonen:2012ac}  & $ -0.0005 \pm 0.0057 \pm 0.0054 $ \\
& 2004 & CLEO~\cite{Asner:2003uz}    & $ -0.009  \pm 0.021^{+0.016}_{-0.057} $ \\
&      & HFLAV average              & $ -0.0008 \pm 0.0077 $ \\
\hline
{\boldmath $D^0 \to K^0_s+ K^-\pi^+$} &
   2016 & LHCb~\cite{Aaij:2015lsa} &  Amplitude analysis, no evidence for 
\CP\ violation \\ 
\hline
{\boldmath $D^0 \to K^0_s+ K^+\pi^-$} &
   2016 & LHCb~\cite{Aaij:2015lsa} &  Amplitude analysis, no evidence for 
\CP\ violation \\ 
\hline
{\boldmath $D^0 \to K^+ K^-\pi^0$} &
   2008 & \babar~\cite{Aubert:2008yd} & $ -0.0100 \pm  0.0167 \pm  0.0025$ \\ 
\hline
{\boldmath $D^0 \to \pi^-\pi^-\pi^+\pi^+$} &
  2013 & LHCb~\cite{Aaij:2013aa}  & Model independent technique, no evidence for 
\CP violation \\
\hline
{\boldmath $D^0 \to K^-\pi^+\pi^+\pi^-$} &
  2014 & CLEO~\cite{Bonvicini:2013vxi} & $ +0.002  \pm 0.003 \pm 0.004   $ \\
\hline 
{\boldmath $D^0 \to K^+\pi^-\pi^+\pi^-$} &
  2005 & Belle~\cite{Tian:2005ik} & $ -0.018  \pm 0.044  $ (stat.) \\
\hline
{\boldmath $D^0 \to K^+K^-\pi^+\pi^-$} &
  2013 & LHCb~\cite{Aaij:2013aa}   & Model independent technique, no evidence for 
\CP\ violation \\
& 2012 & CLEO~\cite{Artuso:2012df} & Amplitude analysis, no evidence for 
\CP\ violation \\  
& 2005 & FOCUS~\cite{Link:2005th}  & $ -0.082  \pm 0.056  \pm 0.047  $ \\
\hline                   
\end{tabular}
\end{center} 
\end{table}

\begin{table}[!htb]
\renewcommand{\arraystretch}{1.4}
\caption{\CP\ asymmetries 
$A^{}_{\CP}= [\Gamma(D_s^+)-\Gamma(D_s^-)]/[\Gamma(D_s^+)+\Gamma(D_s^-)]$
for $D_s^\pm$ decays.
\label{tab:cp_ds}}
\footnotesize
\begin{center}
\begin{tabular}{|l|c|c|c|} 
\hline
{\bf Mode} & {\bf Year} & {\bf Collaboration} & {\boldmath $A^{}_{\CP}$} \\
\hline
{\boldmath $D_s^+ \to \mu^+ \nu$} &
  2009 & CLEO~\cite{Alexander:2009ux} & $ +0.048 \pm 0.061 $ \\
\hline
{\boldmath $D_s^+ \to \pi^+ \eta$} &
  2013 & CLEO~\cite{Onyisi:2013bjt}     & $ +0.011 \pm 0.030 \pm 0.008 $ \\
\hline
{\boldmath $D_s^+ \to \pi^+ \eta^\prime$} &
  2013 & CLEO~\cite{Onyisi:2013bjt}     & $ -0.022 \pm 0.022 \pm 0.006 $ \\
\hline
{\boldmath $D_s^+ \to K^0_s\pi^+$}  &
  2013 & \babar~\cite{Lees:2013aa}  & $ +0.006  \pm 0.020  \pm 0.003  $ \\  
& 2010 & Belle~\cite{Ko:2010ng}     & $ +0.0545 \pm 0.0250 \pm 0.0033 $ \\
& 2010 & CLEO~\cite{Mendez:2009aa}  & $ +0.163  \pm 0.073  \pm 0.003  $ \\
&      & HFLAV average             & $ +0.0311 \pm 0.0154 $            \\
\hline
{\boldmath $D_s^+ \to (\kbar/K^0)\pi^+$}  &
  2014 & LHCb~\cite{Aaij:2014ac}    & $ +0.0038 \pm 0.0046 \pm 0.0017 $ \\
& 2013 & \babar~\cite{Lees:2013aa}  & $ +0.003  \pm 0.020  \pm 0.003  $ \\  
&      & HFLAV average             & $ +0.0038 \pm 0.0048 $            \\
\hline
{\boldmath $D_s^+ \to K^0_s K^+$}   &
  2013 & CLEO~\cite{Onyisi:2013bjt}  & $ +0.026  \pm 0.015  \pm 0.006  $ \\
& 2013 & \babar~\cite{Lees:2013aa}  & $ -0.0005 \pm 0.0023 \pm 0.0024 $ \\  
& 2010 & Belle~\cite{Ko:2010ng}     & $ +0.0012 \pm 0.0036 \pm 0.0022 $ \\
&      & HFLAV average             & $ +0.0008 \pm 0.0026 $            \\
\hline
{\boldmath $D_s^+ \to K^+ \pi^0$}   &
  2010 & CLEO~\cite{Mendez:2009aa} &  $ +0.266 \pm 0.228 \pm 0.009 $ \\
\hline
{\boldmath $D_s^+ \to K^+ \eta$}    &
  2010 & CLEO~\cite{Mendez:2009aa} &  $ +0.093 \pm 0.152 \pm 0.009 $ \\
\hline
{\boldmath $D_s^+ \to K^+ \eta^\prime$}  &
  2010 & CLEO~\cite{Mendez:2009aa}      &  $ +0.060 \pm 0.189 \pm 0.009 $ \\
\hline
{\boldmath $D_s^+ \to \pi^+ \pi^+ \pi^-$} &
  2013 & CLEO~\cite{Onyisi:2013bjt}        & $ -0.007 \pm 0.030 \pm 0.006 $ \\
\hline
{\boldmath $D_s^+ \to \pi^+ \pi^0 \eta$}  &
  2013 & CLEO~\cite{Onyisi:2013bjt}        & $ -0.005 \pm 0.039 \pm 0.020 $ \\
\hline
{\boldmath $D_s^+ \to \pi^+ \pi^0 \eta^\prime$} &
  2013 & CLEO~\cite{Onyisi:2013bjt}        & $ -0.004 \pm 0.074 \pm 0.019 $ \\
\hline
{\boldmath $D_s^+ \to K^0_s K^+ \pi^0$}   &
  2013 & CLEO~\cite{Onyisi:2013bjt}        & $ -0.016 \pm 0.060 \pm 0.011 $ \\
\hline
{\boldmath $D_s^+ \to K^0_s K^0_s \pi^+$} &
  2013 & CLEO~\cite{Onyisi:2013bjt}        & $ +0.031 \pm 0.052 \pm 0.006 $ \\
\hline
{\boldmath $D_s^+ \to K^+ \pi^+ \pi^-$} &
  2013 & CLEO~\cite{Onyisi:2013bjt}        & $ +0.045 \pm 0.048 \pm 0.006 $ \\
\hline
{\boldmath $D_s^+ \to K^+ K^- \pi^+$} &
  2013 & CLEO~\cite{Onyisi:2013bjt}        & $ -0.005 \pm 0.008 \pm 0.004 $ \\
\hline
{\boldmath $D_s^+ \to K^0_s K^- \pi^+\pi^+$} &
  2013 & CLEO~\cite{Onyisi:2013bjt}        & $ +0.041 \pm 0.027 \pm 0.009 $ \\
\hline
{\boldmath $D_s^+ \to K^0_s K^+ \pi^+\pi^-$} &
  2013 & CLEO~\cite{Onyisi:2013bjt}        & $ -0.057 \pm 0.053 \pm 0.009 $ \\
\hline
{\boldmath $D_s^+ \to K^+ K^- \pi^+\pi^0$} &
  2013 & CLEO~\cite{Onyisi:2013bjt}        & $ +0.000 \pm 0.027 \pm 0.012 $ \\ 
\hline 
\end{tabular}
\end{center} 
\end{table}

\clearpage
% T-odd
\mysubsection{$T$-odd asymmetries}
                                               
Measuring $T$-odd asymmetries provides an alternative way to search 
for \CP\ violation in the charm sector, due to ${\CP}T$ invariance.
$T$-odd asymmetries are measured using triple-product correlations
of the form $\vec{a}\cdot(\vec{b}\times\vec{c})$, where $a$, $b$, 
and $c$ are spins or momenta; this combination is odd under time 
reversal~($T$).
If the triple-product is formed using {\it both\/} spin and 
momenta, \ie, 
\begin{eqnarray}
\vec{s_1} \cdot(\vec{p_2} \times \vec{p}_3)\,,
\end{eqnarray}
then it is even for $P$-conjugation. However, if only momenta
are used, then it is odd for  $P$-conjugation. In this case
the asymmetry allows one to probe \CP\ violation occuring 
via $P$-violation. 
This may arise in $P$-odd amplitudes, which are allowed 
in decays to final states with 4 spinless particles.

Taking as an example the decay mode $D^0 \to K^+K^-\pi^+\pi^-$, 
one forms the triple-product correlation using the momenta 
of the final state particles. We note that when using only momenta,
at least four daughter particles are required to give a nonzero
correlation (as three daughters decay in a plane).
%For a spinless decaying particle, such a correlation 
%Due to momentum conservation, a nonzero correlation 
%must necessarily involve at least four final-state particles.
Defining for $D^0$ the $T$-odd correlation
\begin{eqnarray} 
C_T \equiv \vec{p}^{}_{K^+}\cdot(\vec{p}_{\pi^+}\times \vec{p}_{\pi^-})\,,
\end{eqnarray}  
and the corresponding quantity for $\dbar$
\begin{eqnarray}
\overline{C}_T \equiv 
      \vec{p}^{}_{K^-}\cdot(\vec{p}_{\pi^-}\times \vec{p}_{\pi^+})\,,
\end{eqnarray}      
one constructs the asymmetry
\begin{eqnarray}
 A_{T} & = &
    \frac{\Gamma(C_T>0)-\Gamma(C_T<0)}{\Gamma(C_T>0)+\Gamma(C_T<0)}
\end{eqnarray}
for $D^0$ decays and
\begin{eqnarray}
\overline{A}_{T} & = & 
   \frac{\Gamma(-\overline{C}_T>0)-\Gamma(-\overline{C}_T<0)}
                        {\Gamma(-\overline{C}_T>0)+\Gamma(-\overline{C}_T<0)}
\end{eqnarray} 
for $\dbar$ 
%(\CP\ conjugate) 
decays. In these expressions, $\Gamma$ 
represents a partial width. The asymmetries $A_T$ and $\overline{A}_T$
%, being $P$-asymmetries for respectively particles and anti-particles, 
depend on the angular distribution of the daughter particles and may
be nonzero due to final state interactions or $P$-violation in weak decays.

Since $P(C_T)=-C_T$ and $C(C_T)=\overline{C_T}$, $C\!P(A_T) = \overline{A}_T$.
One can thus construct the \CP-odd (and $P$-odd, $T$-odd) quantity 
\begin{eqnarray}
\label{eqn:atodd}
%A^{}_{T-{\rm odd}} & \equiv & \frac{A_{T}-\overline{A}_{T}}{2}\,,
{\cal A}^{}_{T} & \equiv & \frac{A_{T}-\overline{A}_{T}}{2}\,;
\end{eqnarray}
a nonzero value indicates \CP\ violation
%can have a nonzero values because of FSI or because of $P$-violation 
%in weak decays.
%However, one can construct the $T$-odd observable
%and $P$-odd asymmetry
(see Refs.~\cite{Golowich:1988ig,Bigi:2001sg,Bensalem:2002ys,
Bensalem:2000hq,Bensalem:2002pz,Gronau:2011cf}).

Recently, this topic has been revisited (see 
Refs.~\cite{Bevan:2015xra,Durieux:2015zwa}) with the suggestion 
to use other asymmetries constructed from triple products in 
multi-body decays to probe $C$, $P$, and \CP\ symmetries. 
Up until now, experiments have measured only the asymmetry 
${\cal A}^{}_T$ defined in Eq.~(\ref{eqn:atodd}). 
(Note that this asymmetry is referred to in the literature by 
several names: $A^{}_{T\,{\rm viol}}$, $a^P_{\CP}$, and $a^{T-{\rm odd}}_{\CP}$.)

Values of ${\cal A}^{}_{T}$ for $D^+$, $D^+_s$, and
$D^0$ decay modes are listed in Table~\ref{tab:t_odd}. 
The first measurements were made by FOCUS, and subsequent 
\babar\ measurements reached a sensitivity of $\sim 1\%$. 
Currently the best sensitivity is from LHCb. 
%As also concluded in Sec.~\ref{sec:cp_asym}, 
However, despite relatively high precision ($<1\%$), 
there is no evidence for \CP\ violation.

\begin{table}[ht]
\renewcommand{\arraystretch}{1.4}
\caption{Measurements of the $T$-odd asymmetry 
${\cal A}^{}_{T} = (A_{T}-\overline{A}_{T})/2$.
\label{tab:t_odd}}
\footnotesize
\begin{center}
\begin{tabular}{|l|c|c|c|} 
\hline
{\bf Mode} & {\bf Year} & {\bf Collaboration} & {\boldmath ${\cal A}^{}_{T}$} \\
\hline
{\boldmath $D^0 \to K^+K^-\pi^+\pi^-$} &
   2014 & LHCb~\cite{Aaij:2014qwa}     &  $ +0.0018 \pm 0.0029 \pm 0.0004 $ \\
&  2010 & \babar~\cite{Sanchez:2010xj} &  $ +0.0010 \pm 0.0051 \pm 0.0044 $ \\
&  2005 & FOCUS~\cite{Link:2005th}     &  $ +0.010  \pm 0.057  \pm 0.037  $ \\
&       & HFLAV average               &  $ +0.0017 \pm 0.0027            $ \\  
\hline
{\boldmath $D^+ \to K^0_sK^+\pi^+\pi^-$} &
  2011 & \babar~\cite{Lees:2011ab} &  $ -0.0120 \pm 0.0100 \pm 0.0046 $ \\
& 2005 & FOCUS~\cite{Link:2005th}  &  $ +0.023  \pm 0.062  \pm 0.022  $ \\
&      & HFLAV average            &  $ -0.0110 \pm 0.0109            $ \\
\hline
{\boldmath $D^+_s \to K^0_sK^+\pi^+\pi^-$} &
  2011 & \babar~\cite{Lees:2011ab} &  $ -0.0136 \pm 0.0077 \pm 0.0034 $ \\
& 2005 & FOCUS~\cite{Link:2005th}  &  $ -0.036  \pm 0.067  \pm 0.023  $ \\
&      & HFLAV average            &  $ -0.0139 \pm 0.0084            $ \\
\hline                    
\end{tabular}
\end{center} 
\end{table}

% \vskip0.30in
% \begin{center}  ---------------  \end{center}
% \vskip0.30in

\clearpage
% Interplay of direct and indirect CPV
\subsection{Interplay of direct and indirect \cp\ violation}
\label{sec:charm:cpvdir}

In decays of $D^0$ mesons, \cp\ asymmetry measurements have contributions from 
both direct and indirect \cp\ violation as discussed in Sec.~\ref{sec:charm:mixcpv}.
The contribution from indirect \cp\ violation depends on the decay-time distribution 
of the data sample~\cite{Kagan:2009gb}. This section describes a combination of 
measurements that allows the extraction of the individual contributions of the 
two types of \cp\ violation.
At the same time, the level of agreement for a no-\cp-violation hypothesis is 
tested. The observables are: 
\begin{equation}
A_{\Gamma} \equiv \frac{\tau(\dbar \ra h^+ h^-) - \tau(D^0 \ra h^+ h^- )}
{\tau(\dbar \ra h^+ h^-) + \tau(D^0 \ra h^+ h^- )},
\end{equation}
where $h^+ h^-$ can be $K^+ K^-$ or $\pi^+\pi^-$, and 
\begin{equation}
\Delta A_{\CP}   \equiv A_{\CP}(K^+K^-) - A_{\CP}(\pi^+\pi^-),
\end{equation}
where $A_{\CP}$ are time-integrated \cp\ asymmetries. The underlying 
theoretical parameters are: 
\begin{eqnarray}
a_{\CP}^{\rm dir} & \equiv & 
\frac{|{\cal A}_{D^0\rightarrow f} |^2 - |{\cal A}_{\dbar\rightarrow f} |^2} 
{|{\cal A}_{D^0\rightarrow f} |^2 + |{\cal A}_{\dbar\rightarrow f} |^2} ,\nonumber\\ 
a_{\CP}^{\rm ind}  & \equiv & \frac{1}{2} 
\left[ \left(\left|\frac{q}{p}\right| + \left|\frac{p}{q}\right|\right) x \sin \phi - 
\left(\left|\frac{q}{p}\right| - \left|\frac{p}{q}\right|\right) y \cos \phi \right] ,
\end{eqnarray}
where ${\cal A}_{D\rightarrow f}$ is the amplitude for $D\ra f$~\cite{Grossman:2006jg}. 
We use the following relations 
between the observables and the underlying parameters~\cite{Gersabeck:2011xj}: 
\begin{eqnarray}
A_{\Gamma} & = & - a_{\CP}^{\rm ind} - a_{\CP}^{\rm dir} y_{\CP},\label{eqn:charm_MG_AGamma}\\ 
\Delta A_{\CP} & = &  \Delta a_{\CP}^{\rm dir} \left(1 + y_{\CP} 
\frac{\overline{\langle t\rangle}}{\tau} \right)   +   
   a_{\CP}^{\rm ind} \frac{\Delta\langle t\rangle}{\tau}   +   
  \overline{a_{\CP}^{\rm dir}} y_{\CP} \frac{\Delta\langle t\rangle}{\tau},\nonumber\\ 
& \approx & \Delta a_{\CP}^{\rm dir} \left(1 + y_{\CP} 
\frac{\overline{\langle t\rangle}}{\tau} \right)   +   a_{\CP}^{\rm ind} 
\frac{\Delta\langle t\rangle}{\tau}.\label{eqn:charm_MG_DACP}
\end{eqnarray}
Equation~(\ref{eqn:charm_MG_AGamma}) constrains mostly indirect \cp\ violation, and the 
direct \cp\ violation contribution can differ for different final states. 
In Eq.~(\ref{eqn:charm_MG_DACP}), $\langle t\rangle/\tau$ denotes the mean decay 
time in units of the $D^0$ lifetime; $\Delta X$ denotes the difference 
in quantity $X$ between $K^+K^-$ and $\pi^+\pi^-$ final states; and $\overline{X}$ 
denotes the average for quantity $X$. 
We neglect the last term in this relation as all three factors are 
$\mathcal{O}(10^{-2})$ or smaller, and thus this term is negligible 
with respect to the other two terms. 
Note that $\Delta\langle t\rangle/\tau \ll\langle t\rangle/\tau$, and 
it is expected that $|a_{\CP}^{\rm dir}| < |\Delta a_{\CP}^{\rm dir}|$ 
because $a_{\CP}^{\rm dir}(K^+K^-)$ and $a_{\CP}^{\rm dir}(\pi^+\pi^-)$ 
are expected to have opposite signs in the Standard Model~\cite{Grossman:2006jg}. 

A $\chi^2$ fit is performed in the plane $\Delta a_{\CP}^{\rm dir}$ 
vs. $a_{\CP}^{\rm ind}$. 
For the \babar result the difference of the quoted values for 
$A_{\CP}(K^+K^-)$ and $A_{\CP}(\pi^+\pi^-)$ is calculated, 
adding all uncertainties in quadrature. 
This may overestimate the systematic uncertainty for the difference 
as it neglects correlated errors; however, the result is conservative 
and the effect is small as all measurements are statistically limited. 
For all measurements, statistical and systematic uncertainties are added 
in quadrature when calculating the $\chi^2$. 
We use the current world average value $y_{\CP} = (0.835 \pm 0.155)\%$ 
(see Sec.~\ref{sec:charm:mixcpv}) and the measurements listed in 
Table~\ref{tab:charm:dir_indir_comb}. 

In this fit, $A_\Gamma(KK)$ and $A_\Gamma(\pi\pi)$ are assumed to be identical.
This assumption (expected to hold in the Standard Model) is supported by all measurements to date.
A significant relative shift due to final-state dependent $A_\Gamma$ values between $\Delta A_{\CP}$ measurements with different mean decay times is excluded by these measurements.

\begin{table}
\centering 
\caption{Inputs to the fit for direct and indirect \cp\ violation. 
The first uncertainty listed is statistical, and the second is systematic.}
\label{tab:charm:dir_indir_comb}
\vspace{3pt}
\begin{tabular}{ll|ccccc}
\hline \hline
Year & 	Experiment	& Results
& $\Delta \langle t\rangle/\tau$ & $\langle t\rangle/\tau$ & Reference\\
\hline
2012	& \babar	& $A_\Gamma = (+0.09 \pm 0.26 \pm 0.06 )\%$ &	-&	-&	 
\cite{Lees:2012qh}\\
2016	& LHCb	prompt & $A_\Gamma(KK) = (-0.030 \pm 0.032 \pm 0.010 )\%$ &	-&	-&	 
\cite{Aaij:2017idz}\\
    	&     	& $A_\Gamma(\pi\pi) = (+0.046 \pm 0.058 \pm 0.012 )\%$ &  -&	-&	 
                   \\
2014	& CDF & $A_\Gamma = (-0.12 \pm 0.12 )\%$ &	-&	-&	 
\cite{Aaltonen:2014efa}\\
2015	& LHCb SL & $A_\Gamma = (-0.125 \pm 0.073 )\%$ &	-&	-&	 
\cite{Aaij:2015yda}\\
2015	& Belle	& $A_\Gamma = (-0.03 \pm 0.20 \pm 0.07 )\%$ &	-&	-&	 
\cite{Staric:2015sta}\\
2008	& \babar	& $A_{\CP}(KK) = (+0.00 \pm 0.34 \pm 0.13 )\%$&&&\\ 
& & $A_{\CP}(\pi\pi) = (-0.24 \pm 0.52 \pm 0.22 )\%$ &	$0.00$ &	
$1.00$ &	 \cite{Aubert:2007if}\\
2012	& Belle	prel. & $\Delta A_{\CP} = (-0.87 \pm 0.41 \pm 0.06 )\%$ &	
$0.00$ &	$1.00$ &	 \cite{Ko:2012px}\\
2012	& CDF	 & $\Delta A_{\CP} = (-0.62 \pm 0.21 \pm 0.10 )\%$ &	
$0.25$ &	$2.58$ &	 \cite{Collaboration:2012qw}\\
2014	& LHCb	SL & $\Delta A_{\CP} = (+0.14 \pm 0.16 \pm 0.08 )\%$ &	
$0.01$ &	$1.07$ &	 \cite{Aaij:2014gsa}\\
2016	& LHCb	prompt & $\Delta A_{\CP} = (-0.10 \pm 0.08 \pm 0.03 )\%$ &	
$0.12$ &	$2.10$ &	 \cite{Aaij:2016cfh}\\
\hline
\end{tabular}
\end{table}

\begin{figure}
\begin{center}
\includegraphics[width=0.90\textwidth]{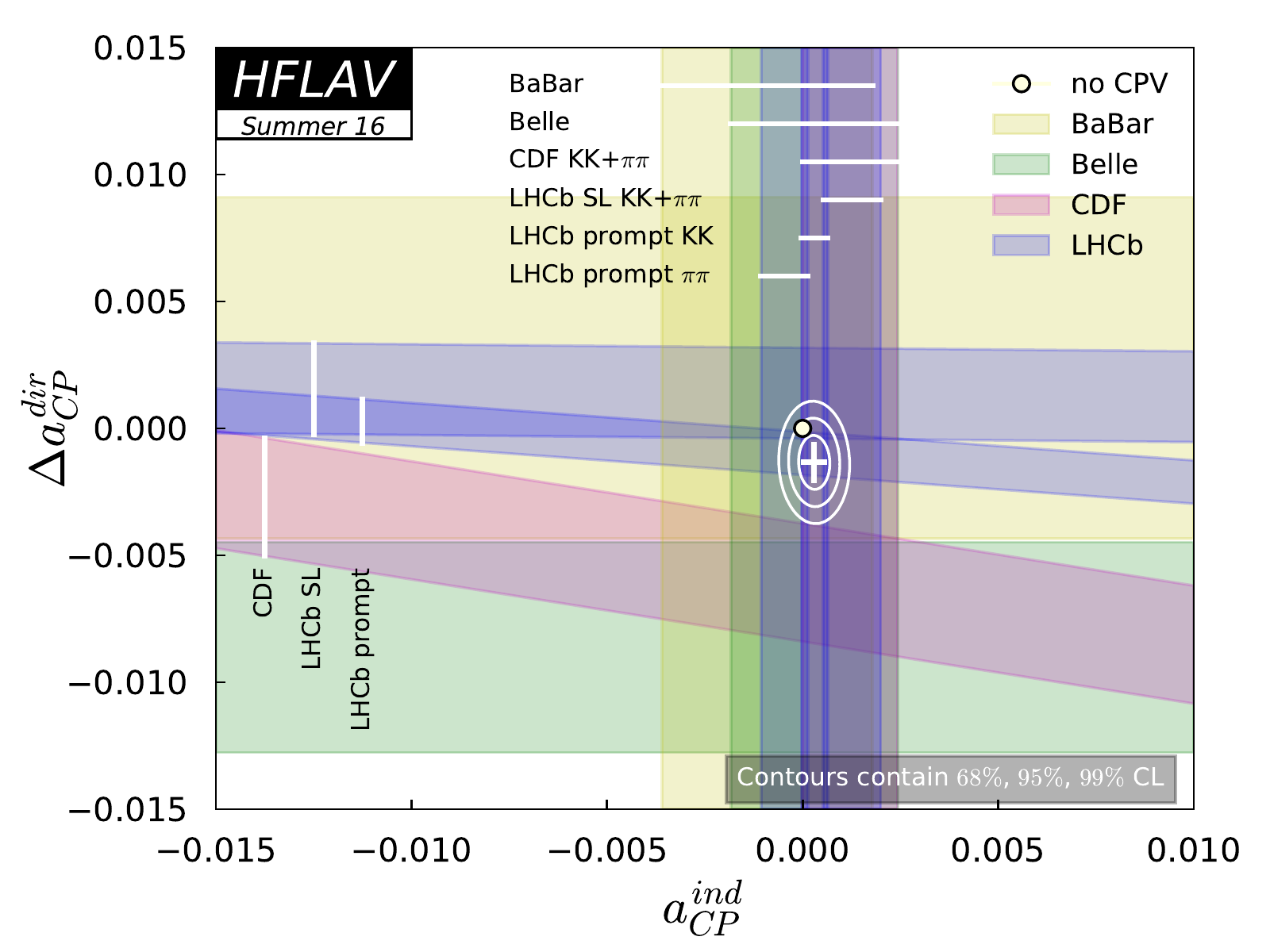}
\caption{Plot of all data and the fit result. Individual 
measurements are plotted as bands showing their $\pm1\sigma$ range. 
The no-\cpv\ point (0,0) is shown as a filled circle, and the best 
fit value is indicated by a cross showing the one-dimensional uncertainties. 
Two-dimensional $68\%$ C.L., $95\%$ C.L., and $99.7\%$ C.L.\ regions are 
plotted as ellipses. }
\label{fig:charm:dir_indir_comb}
\end{center}
\end{figure}

The combination plot (see Fig.~\ref{fig:charm:dir_indir_comb}) shows the measurements listed in 
Table~\ref{tab:charm:dir_indir_comb} for
$\Delta A_{\CP}$ and $A_\Gamma$.
%, where the bands represent $\pm1\sigma$ 
%intervals.  The point of no \cp\ violation (0,0) is shown as a filled circle, 
%and two-dimensional $68\%$ C.L., $95\%$ C.L., and $99.7\%$ C.L.\ regions are plotted 
%as ellipses. The best fit value is indicated by a cross showing the
%one-dimensional errors.
%
From the fit, the change in $\chi^2$ from the minimum value for the no-\cpv\ 
point (0,0) is $4.7$, which corresponds to a C.L.\ of $9.3\times 10^{-2}$ for 
two degrees of freedom. Thus the data are consistent with the no-\cp-violation 
hypothesis at $9.3\%$ C.L. This $p$-value corresponds to $1.7\sigma$. The central
values and $\pm1\sigma$ uncertainties for the individual parameters are
\begin{eqnarray}
a_{\CP}^{\rm ind} & = & (+0.030 \pm 0.026 )\% \nonumber\\
\Delta a_{\CP}^{\rm dir} & = & (-0.134 \pm 0.070 )\%.
\end{eqnarray}
Compared to the previous average, the tension in the difference between direct
\cp\ violation in the two final states is reduced, while the common indirect 
\cp\ violation moved away from the no-\cp-violation point by about one standard 
deviation.

\clearpage

% Semileptonic decays
\subsection{Semileptonic decays}
\label{sec:charm:semileptonic}

\subsubsection{Introduction}

Semileptonic decays of $D$ mesons involve the interaction of a leptonic
current with a hadronic current. The latter is nonperturbative
and cannot be calculated from first principles; thus it is usually
parameterized in terms of form factors. The transition matrix element 
is written
\begin{eqnarray}
  {\cal M} & = & -i\,\frac{G_F}{\sqrt{2}}\,V^{}_{cq}\,L^\mu H_\mu\,,
  \label{Melem}
\end{eqnarray}
where $G_F$ is the Fermi constant and $V^{}_{cq}$ is a CKM matrix element.
The leptonic current $L^\mu$ is evaluated directly from the lepton spinors 
and has a simple structure; this allows one to extract information about 
the form factors (in $H^{}_\mu$) from data on semileptonic decays~\cite{Becher:2005bg}.  
Conversely, because there are no strong final-state interactions between the
leptonic and hadronic systems, semileptonic decays for which the form 
factors can be calculated allow one to 
determine~$V^{}_{cq}$~\cite{Kobayashi:1973fv}.

\subsubsection{$D\ra P \overline \ell \nu_{\ell}$ decays}

When the final state hadron is a pseudoscalar, the hadronic 
current is given by
\begin{eqnarray}
\hspace{-1cm}
H_\mu & = & \left< P(p) | \bar{q}\gamma_\mu c | D(p') \right> \ =\  
f_+(q^2)\left[ (p' + p)_\mu -\frac{m_D^2-m_P^2}{q^2}q_\mu\right] + 
 f_0(q^2)\frac{m_D^2-m_P^2}{q^2}q_\mu\,,
\label{eq:hadronic}
\end{eqnarray}
where $m_D$ and $p'$ are the mass and four momentum of the 
parent $D$ meson, $m_P$ and $p$ are those of the daughter meson, 
$f_+(q^2)$ and $f_0(q^2)$ are form factors, and $q = p' - p$.  
Kinematics require that $f_+(0) = f_0(0)$.
The contraction $q_\mu L^\mu$ results in terms proportional 
to $m^{}_\ell$\cite{Gilman:1989uy}, and thus for $\ell=e $
the terms proportionals to $q_\mu$ in Eq.~(\ref{eq:hadronic}) are negligible. 
For light leptons only the $f_+(q^2)$ vector form factor 
is relevant and the differential partial width is
% integrated over various angular distributions is
\begin{eqnarray}
\frac{d\Gamma(D \to P \bar \ell \nu_\ell)}{dq^2\, d\cos\theta_\ell} & = & 
   \frac{G_F^2|V_{cq}|^2}{32\pi^3} p^{*\,3}|f_{+}(q^2)|^2\sin\theta^2_\ell\,,
\label{eq:dGamma}
\end{eqnarray}
where ${p^*}$ is the magnitude of the momentum of the final state hadron
in the $D$ rest frame, and $\theta_\ell$ is the angle of the lepton in the 
$\ell\nu$ rest frame with respect to the direction of the pseudoscalar meson 
in the $D$ rest frame.

%parameterization takes advantage of dispersion relations (see,
%\eg\ Ref.~\cite{Boyd:1994tt}), which allow expression of a form
%factor in terms of

\subsubsection{Form factor parameterizations} 

The form factor is traditionally parameterized with an explicit pole 
and a sum of effective poles:
\begin{eqnarray}
f_+(q^2) & = & \frac{f_+(0)}{(1-\alpha)}\Bigg [
\left(\frac{1}{1- q^2/m^2_{\rm pole}}\right)\ +\ 
\sum_{k=1}^{N}\frac{\rho_k}{1- q^2/(\gamma_k\,m^2_{\rm pole})}\Bigg ]\,,
\label{eqn:expansion}
\end{eqnarray}
where $\rho_k$ and $\gamma_k$ are expansion parameters and $\alpha$ is 
a parameter that normalizes the form factor at $q^2=0$, $f_+(0)$. 
The parameter $m_{{\rm pole}}$ is the mass of the lowest-lying $c\bar{q}$ 
resonance with the vector quantum numbers; this is expected to 
provide the largest contribution to the form factor for the $c\ra q$ 
transition. The sum over $N$ gives the contribution of higher mass states.  
For example, for $D\to\pi$ transitions the dominant resonance is
expected to be $D^*(2010)$, and thus $m^{}_{\rm pole}=m^{}_{D^*(2010)}$. 
For $D\to K$ transitions, the dominant contribution is expected from 
$D^{*}_s(2112)$, with $m^{}_{\rm pole}=m^{}_{D^{*}_{s}(2112)}$.

\subsubsection{Simple pole}

Equation~(\ref{eqn:expansion}) can be simplified by neglecting the 
sum over effective poles, leaving only the explicit vector meson pole. 
This approximation is referred to as ``nearest pole dominance'' or 
``vector-meson dominance.''  The resulting parameterization is
\begin{eqnarray}
  f_+(q^2) & = & \frac{f_+(0)}{(1-q^2/m^2_{\rm pole})}\,. 
\label{SimplePole}
\end{eqnarray}
However, values of $m_{{\rm pole}}$ that give a good fit to the data 
do not agree with the expected vector meson masses~\cite{Hill:2006ub}. 
To address this problem, the ``modified pole'' or Becirevic-Kaidalov~(BK) 
parameterization~\cite{Becirevic:1999kt} was introduced.
$m_{\rm pole} /\sqrt{\alpha_{\rm BK}}$
is interpreted as the mass of an effective pole, higher than 
$m_{\rm pole}$, thus it is expected that $\alpha_{\rm BK}<1$.
 
%This parametrization assumes that gluon 
%hard-scattering contributions ($\delta$) are near zero, and scaling
%violations ($\beta$) are near unity~\cite{Hill:2006ub}:
%\begin{eqnarray}
%1 + 1\slash \beta - \delta & \equiv & 
%\frac{\left(m_D^2 - m_{P}^2\right)}{f_+(0)}\ 
%\left.\frac{df_+}{dq^2}\right|_{q^2=0}\ \approx\ 2\,.
%\end{eqnarray}

The parameterization takes the form
\begin{eqnarray}
f_+(q^2) & = & \frac{f_+(0)}{(1-q^2/m^2_{\rm pole})}
\frac{1}{\left(1-\alpha^{}_{\rm BK}\frac{q^2}{m^2_{\rm pole}}\right)}\,.
\end{eqnarray}
%To be consistent with $1 + 1\slash \beta - \delta\approx 2$, the 
%parameter $\alpha^{}_{\rm BK}$ should be near the value~1.75.
These parameterizations are used by several experiments to 
determine form factor parameters.
Measured values of $m^{}_{\rm pole}$ and $\alpha^{}_{\rm BK}$ are
listed in Tables~\ref{kPseudoPole} and~\ref{piPseudoPole} for
$D\to K\ell\nu_{\ell}$ and $D\to\pi\ell\nu_{\ell}$ decays, respectively.
%These values are plotted in Figs.~\ref{Kpole} and~\ref{pipole}.
%Both tables show $\alpha^{}_{BK}$ to be substantially lower than
%the expected value of~$\sim$\,1.75.

%As is clear from the figures, the $K\ell\nu$ data yield pole masses that
%are significantly lower than the mass of the $D_s^*$ resonance that
%should dominate in the vector dominance (simple pole) model.  The
%$\pi\ell\nu_{\ell}$ data also show a trend toward a pole mass lower than the
%physical $D^*$ pole mass, with the recent CLEO-c measurement showing a
%significant discrepancy. The simple pole parameterization can typically
%provide a good fit to the data.  However, the unphysically low mass
%parameters indicate that higher mass resonances and the $DK$ and $D\pi$
%continuum contribute non-negligibly to the decay process.

\subsubsection{$z$ expansion}

An alternative series expansion around some value $q^2=t_0$ to parameterize 
$f^{}_+(q^2)$ can be used~\cite{Boyd:1994tt,Boyd:1997qw,Arnesen:2005ez,Becher:2005bg}. 
This parameterization is model independent and satisfies general QCD 
constraints, being suitable for fitting experimental data. The 
expansion is given in terms of a complex parameter $z$, which is 
the analytic continuation of $q^2$ into the complex plane:
\begin{eqnarray}
z(q^2,t_0) & = & \frac{\sqrt{t_+ - q^2} - \sqrt{t_+ - t_0}}{\sqrt{t_+ - q^2}
	  + \sqrt{t_+ - t_0}}\,, 
\end{eqnarray}
where $t_\pm \equiv (m_D \pm m_P)^2$ and $t_0$ is the (arbitrary) $q^2$ 
value corresponding to $z=0$. The physical region corresponds to 
$\pm|z|_{max} = \pm 0.051$ for $D\to K \ell \nu_\ell$ and $= \pm 0.17$ 
for  $D\to \pi \ell \nu_\ell$, using $t_{0}= t_{+} (1-\sqrt{1-t_{-}/t_{+}})$.  

The form factor is expressed as
\begin{eqnarray}
f_+(q^2) & = & \frac{1}{P(q^2)\,\phi(q^2,t_0)}\sum_{k=0}^\infty
a_k(t_0)[z(q^2,t_0)]^k\,,
\label{z_expansion}
\end{eqnarray}
where the $P(q^2)$ factor accommodates sub-threshold resonances via
\begin{eqnarray}
P(q^2) & \equiv & 
\begin{cases} 
1 & (D\to \pi) \\
z(q^2,M^2_{D^*_s}) & (D\to K)\,. 
\end{cases}
\end{eqnarray}
The ``outer'' function $\phi(t,t_0)$ can be any analytic function,
but a preferred choice (see, \eg\
Refs.~\cite{Boyd:1994tt,Boyd:1997qw,Bourrely:1980gp}) obtained
from the Operator Product Expansion (OPE) is
\begin{eqnarray}
\phi(q^2,t_0) & =  & \alpha 
\left(\sqrt{t_+ - q^2}+\sqrt{t_+ - t_0}\right) \times  \nonumber \\
 & & \hskip0.20in \frac{t_+ - q^2}{(t_+ - t_0)^{1/4}}\  
\frac{(\sqrt{t_+ - q^2}\ +\ \sqrt{t_+ - t_-})^{3/2}}
     {(\sqrt{t_+ - q^2}+\sqrt{t_+})^5}\,,
\label{eqn:outer}
\end{eqnarray}
with $\alpha = \sqrt{\pi m_c^2/3}$.
The OPE analysis provides a constraint upon the 
expansion coefficients, $\sum_{k=0}^{N}a_k^2 \leq 1$.
These coefficients receive $1/M_D$ corrections, and thus
the constraint is only approximate. However, the
expansion is expected to converge rapidly since 
$|z|<0.051\ (0.17)$ for $D\ra K$ ($D\ra\pi$) over 
the entire physical $q^2$ range, and Eq.~(\ref{z_expansion}) 
remains a useful parameterization. The main disadvantage as compared to 
phenomenological approaches is that there is no physical interpretation 
of the fitted coefficients $a_K$.

\subsubsection{Three-pole formalism}
An update of the vector pole dominance model has been developed for 
the $D \to \pi \ell \nu_\ell$ channel\cite{Becirevic:2014kaa}. It uses 
information of the residues of the semileptonic form factor at its first 
two poles, the $D^\ast(2010)$ and $D^{\ast '}(2600)$ resonances.  The form 
factor is expressed as an infinite sum of residues from $J^P =1^-$ states 
with masses $m_{D^\ast_n}$: 
\begin{eqnarray}
f_+(q^2) = \sum_{n=0}^\infty 
\frac{\displaystyle{\underset{q^2= m_{D^\ast_n}^2}{\rm Res}} f_+(q^2)}{m_{D^\ast_n}^2-q^2} \,,
\label{ThreePole}
\end{eqnarray}
with the residues given by 
\begin{eqnarray}
\displaystyle{\underset{q^2=m_{D_n^\ast}^2}{\rm Res}} 
f_+(q^2)= \frac{1}{2} m_{D_n^\ast} f_{D_n^\ast} g_{D_n^\ast D\pi}\,. 
\label{Residua}
\end{eqnarray}
Values of the $f_{D^\ast}$ and $f_{D^{\ast '}}$ decay constants have been 
obtained by lattice QCD calculations, relative to $f_{D}$, with 2$\%$ 
and 28$\%$ precision, respectively~\cite{Becirevic:2014kaa}. 
The couplings to the $D\pi$ state, $g_{D^\ast D\pi}$ and $g_{D^{\ast '} D\pi}$, 
are extracted from measurements of the $D^\ast(2010)$ and  $D^{\ast '}(2600)$ 
widths by \babar and LHCb 
experiments~\cite{Lees:2013uxa,delAmoSanchez:2010vq,Aaij:2013sza}. 
Thus the contribution from the first pole is known with a $3\%$ accuracy. 
The contribution from the $D^{\ast '}(2600)$ is determined with poorer 
accuracy, $\sim 30\%$, mainly due to lattice uncertainties.  
A {\it superconvergence} condition~\cite{Burdman:1996kr} is applied: 
\begin{eqnarray}
\sum_{n=0}^\infty 
{\displaystyle{\underset{q^2=m_{D^\ast_n}^2}{\rm Res}} f_+(q^2) }= 0 \,,
\label{superconvergence}
\end{eqnarray}
protecting the form factor behavior at large $q^2$. Within this model 
the first two poles are not sufficient to describe the data, and a third 
effective pole needs to be included. 

One of the advantages of this phenomenological model is that it can 
be extrapolated outside the charm physical region, providing a method 
to extract the CKM matrix element $V_{ub}$ using the ratio of the form 
factors of the $D\to \pi\ell \nu$ and $B\to \pi\ell \nu$ decay channels. 
It will be used once lattice calculations provide the form factor ratio 
$f^{+}_{B\pi}(q^2)/f^{+}_{D\pi}(q^2)$ at the same pion energy. 

This form factor description can be extended to the $D\to K \ell \nu$ 
decay channel, considering the contribution of several $c\bar s$ 
resonances with $J^P = 1^-$. The first two pole masses contributing 
to the form factor correspond to the $D^{*}_s(2112)$ and $D^{*}_{s1}(2700)$ 
resonant states \cite{PDG_2014}. A constraint on the first residue can be 
obtained using information of the $f_K$ decay constant~\cite{PDG_2014} and 
the $g$ coupling extracted from the $D^{\ast +}$ width~\cite{Lees:2013uxa}. 
The contribution from the second pole can be evaluated using the decay 
constants from~\cite{Becirevic:2012te}, the measured total width and 
the ratio of $D^{\ast} K$ and $D K$ decay branching fractions~\cite{PDG_2014}.

\subsubsection{Experimental techniques and results}

Different techniques by several experiments are used to measure 
$D$ meson semileptonic decays having a pseudoscalar particle in the 
final state. The most recent results are provided by the \babar~\cite{Lees:2014ihu} 
and BES III~\cite{Ablikim:2015ixa, Ablikim:2015qgt} collaborations.
Belle~\cite{Widhalm:2006wz}, \babar~\cite{Aubert:2007wg}, and 
CLEO-c~\cite{Besson:2009uv,Dobbs:2007aa} have all  
previously reported results. Belle fully 
reconstructs $e^+e^- \to D \bar D X$ events from the continuum 
under the $\Upsilon(4S)$ resonance, achieving very good $q^2$ 
resolution (15${\rm~MeV}^2$) and a low background level but with 
a low efficiency. Using 282~$\fb^{-1}$ of data, about 1300 and 115 
signal semileptonic decays are isolated for both lepton channels
together ($e+\mu$), for the Cabibbo-favored and Cabibbo-suppressed 
modes, respectively. The \babar experiment uses a partial 
reconstruction technique in which the semileptonic decays 
are tagged via $ D^{\ast +}\to D^0\pi^+$ decays. 
The $D$ direction and neutrino energy are obtained 
using information from the rest of the event. 
With 75~$\fb^{-1}$ of data, 74000 signal events in the 
$D^0 \to {K}^- e^+ \nu$ mode are obtained. This technique 
provides a large signal yield but also a high background level 
and a poor $q^2$ resolution (ranging from 66 to 219 MeV$^2$). In this 
case the measurement of the branching fraction is obtained by normalizing 
to the $D^0 \to K^- \pi^+$ decay channel; thus the measurement would
benefit from future improvements in the determination of this 
reference channel. The Cabibbo-suppressed mode has been recently 
measured using the same technique and 350~fb$^{-1}$ data. For
this measurement, 5000 $D^0 \to {\pi}^- e^+ \nu$ 
signal events were reconstructed~\cite{Lees:2014ihu}.  

The CLEO-c experiment uses two different methods to measure charm semileptonic 
decays. Tagged analyses~\cite{Besson:2009uv} rely on the full reconstruction of 
$\Psi(3770)\to D {\overline D}$ events. One of the $D$ mesons is reconstructed 
in a hadronic decay mode, the other in the semileptonic channel. The only missing 
particle is the neutrino so the $q^2$ resolution is very good and the background 
level very low.   
With the entire CLEO-c data sample, 818 $\pb^{-1}$, 14123 and 1374 signal events 
are reconstructed for 
the $D^0 \to K^{-} e^+\nu$ and $D^0\to \pi^{-} e^+\nu$ channels, and 8467 and 838 for 
the $D^+\to {\overline K}^{0} e^+\nu$ and $D^+\to \pi^{0} e^+\nu$ decays, respectively. 
Another technique without tagging the $D$ meson in a hadronic mode (``untagged'' in 
the following) has been also 
used by CLEO-c~\cite{Dobbs:2007aa}. In this method, the entire missing energy and momentum 
in an event are associated with the neutrino four momentum, with the penalty of 
larger background as compared to the tagged method. 
Using the ``tagged'' method the BES III experiment measures the $D^0 \to {K}^- e^+ \nu$ 
and $D^0 \to {\pi}^- e^+ \nu$ decay channels. With 2.9~fb$^{-1}$ they fully reconstruct 
70700 and 6300 signal events for each channel, respectively\cite{Ablikim:2015ixa}. In 
a separated analysis the BES III experiment measures also the $D^+$ decay mode into 
$D^+ \to K^{0}_{L} e^+ \nu$ \cite{Ablikim:2015qgt}. Using several tagged hadronic 
events they reconstruct 20100 semileptonic candidates. 

Previous measurements were also performed by several experiments. Events registered 
at the $\Upsilon (4S)$ energy corresponding to an integrated luminosity of 7 $\fb^{-1}$ 
were analyzed by CLEO~III~\cite{Huang:2004fra}. Fixed targed photo-production 
experiments performed also measurements of the normalized form factor distribution  
(FOCUS~\cite{Link:2004dh}) and total decay rates (Mark-III~\cite{Adler:1989rw}, 
E653~\cite{Kodama:1991ij, Kodama:1994aj}, E687\cite{Frabetti:1995xq,Frabetti:1993vz}, 
E691~\cite{Anjos:1988ue}, BES II~\cite{Ablikim:2004ej,Ablikim:2006bv}, 
CLEO II~\cite{Bean:1993zv}). In the FOCUS fixed target photo-production 
experiment, $D^0$ semimuonic events were obtained from the decay of a 
$D^{\ast +}$, with a kaon or a pion detected. 

%The $q^2$ resolution is 0.22~\gev2$. 
%They reconstructed 12840 signal events were obtained for the $\D^0 \to K^- \mu^+ \nu$ chnnel and 

Results of the hadronic form factor parameters, $m_{pole}$ and $\alpha_{BK}$, obtained from the measurements  
discussed above, are given in Tables \ref{kPseudoPole} and \ref{piPseudoPole}.
\begin{table}[htbp]
\centering
\caption{Results for $m_{\rm pole}$ and $\alpha_{\rm BK}$ from various 
experiments for $D^0\to K^-\ell^+\nu$ and $D^+\to \bar K^0\ell^+\nu$ decays. 
\label{kPseudoPole}}
\resizebox{\textwidth}{!}{
\begin{tabular}{ccccc}
\hline
\vspace*{-10pt} & \\
 $D\to K\ell\nu_\ell$ Expt. &  Mode & Ref.  & $m_{\rm pole}$ ($\gevcc$) 
& $\alpha^{}_{\rm BK}$       \\
\vspace*{-10pt} & \\
\hline
% \omit        & \omit   &  \omit               & \omit                                  & \omit                  \\
 CLEO III   &  ($D^0$; $\ell=e,\mu$) & \cite{Huang:2004fra}         & $1.89\pm0.05^{+0.04}_{-0.03}$          & $0.36\pm0.10^{+0.03}_{-0.07}$ \\
 FOCUS      &  ($D^0$; $\ell=\mu$)& \cite{Link:2004dh}            & $1.93\pm0.05\pm0.03$                   & $0.28\pm0.08\pm0.07$     \\
 Belle      &  ($D^0$; $\ell=e,\mu$)& \cite{Widhalm:2006wz}         & $1.82\pm0.04\pm0.03$                   & $0.52\pm0.08\pm0.06$     \\
 \babar     &  ($D^0$; $\ell=e$) & \cite{Aubert:2007wg}          & $1.889\pm0.012\pm0.015$                & $0.366\pm0.023\pm0.029$  \\

 CLEO-c (tagged)  & ($D^0,D^+$; $\ell=e$) &\cite{Besson:2009uv}      & $1.93\pm0.02\pm0.01$                   & $0.30\pm0.03\pm0.01$     \\
 CLEO-c (untagged) & ($D^0$; $\ell=e$) &\cite{Dobbs:2007aa}       & $1.97 \pm0.03 \pm 0.01 $ & $0.21 \pm 0.05 \pm 0.03 $  \\
 CLEO-c (untagged) & ($D^+$; $\ell=e$) &\cite{Dobbs:2007aa}       & $1.96 \pm0.04 \pm 0.02 $ & $0.22 \pm 0.08 \pm 0.03$  \\
% BESIII (prel)     &\cite{BESIII}                     & $1.943 \pm 0.025 \pm 0.003$ & $ 0.265 \pm 0.045 \pm 0.006$   \\ 0.923/fb
  BESIII      & ($D^0$; $\ell=e$)          &\cite{Ablikim:2015ixa}                & $1.921 \pm 0.010 \pm 0.007$ & $ 0.309 \pm 0.020 \pm 0.013$   \\ %3/fb
  BESIII      & ($D^+$; $\ell=e$)          &\cite{Ablikim:2015qgt}                & $1.953 \pm 0.044 \pm 0.036$ & $ 0.239 \pm 0.077 \pm 0.065$   \\ %3/fb
% CLEO-c ($D^0\to K^+$) & \cite{GaoICHEP06}    & $1.943^{+0.037}_{-0.033}\pm 0.011$ & $0.258^{+0.063}_{-0.065}\pm0.020$  \\
% CLEO-c ($D^0\to K^+$) & \cite{Dobbs:2007sm}  & $1.97\pm 0.03\pm 0.01$                 & $0.21\pm0.05\pm0.03$                  \\
% CLEO-c ($D^+\to K_S$) & \cite{GaoICHEP06}    & $2.02^{+0.07}_{-0.06}\pm 0.02$     & $0.127^{+0.099}_{-0.104}\pm0.031$  \\
% CLEO-c ($D^+\to K_S$) & \cite{Dobbs:2007sm}  & $1.96\pm 0.04\pm 0.02$                 & $0.22\pm0.08\pm0.03$                  \\
%\hline
% Fermilab lattice/MILC/HPQCD & \cite{Aubin:2004ej}            & --                                &  $0.50\pm0.04\pm0.07$         \\
\vspace*{-10pt} & \\
\hline
\end{tabular}
}
\end{table}

\begin{table}[htbp]
\centering
\caption{Results for $m_{\rm pole}$ and
  $\alpha_{\rm BK}$ from various experiments for 
  $D^0\to \pi^-\ell^+\nu$ and $D^+\to \pi^0\ell^+\nu$ decays.  
%The last entry is a lattice QCD prediction (errors have been increased as compared to the publication 
%to take into account remaining systematic uncertainties in Lattice calculations, as advised by the authors).
\label{piPseudoPole}}
\resizebox{\textwidth}{!}{
\begin{tabular}{ccccc}
\hline
\vspace*{-10pt} & \\
 $D\to \pi\ell\nu_\ell$ Expt. &  Mode & Ref.               & $m_{\rm pole}$ ($\gevcc$) & $\alpha_{\rm BK}$ \\
\vspace*{-10pt} & \\
\hline
 \omit        & \omit         & \omit                & \omit                                  & \omit                  \\
 CLEO III     &   ($D^0$; $\ell=e,\mu$) & \cite{Huang:2004fra}      & $1.86^{+0.10+0.07}_{-0.06-0.03}$       & $0.37^{+0.20}_{-0.31}\pm0.15$         \\
 FOCUS        &    ($D^0$; $\ell=\mu$)  & \cite{Link:2004dh}      & $1.91^{+0.30}_{-0.15}\pm0.07$          & --                                    \\
 Belle        &  ($D^0$; $\ell=e,\mu$)  & \cite{Widhalm:2006wz}      & $1.97\pm0.08\pm0.04$                   & $0.10\pm0.21\pm0.10$                  \\
 CLEO-c (tagged)  & ($D^0,D^+$; $\ell=e$)  &\cite{Besson:2009uv}   & $1.91\pm0.02\pm0.01$                   & $0.21\pm0.07\pm0.02$     \\
 CLEO-c (untagged) &  ($D^0$; $\ell=e$)  &\cite{Dobbs:2007aa}   & $1.87 \pm0.03 \pm 0.01 $ & $0.37 \pm 0.08 \pm 0.03 $  \\
 CLEO-c (untagged) & ($D^+$; $\ell=e$) &\cite{Dobbs:2007aa}     & $1.97 \pm0.07 \pm 0.02 $ & $0.14 \pm 0.16 \pm 0.04$  \\
% BESIII (prel)     &\cite{BESIII}                    & $1.876 \pm 0.023 \pm 0.004$ & $ 0.315 \pm 0.071 \pm 0.012$   \\ 0.9/fb
 BES III   &  ($D^0$; $\ell=e$) &\cite{Ablikim:2015ixa}       & $1.911 \pm 0.012 \pm 0.004$ & $ 0.279 \pm 0.035 \pm 0.011$   \\ %3/fb
  \babar  &($D^0$; $\ell=e$) &\cite{Lees:2014ihu}  & $1.906 \pm 0.029 \pm 0.023$ & $ 0.268 \pm 0.074 \pm 0.059$   \\
% CLEO-c ($D^0\to\pi^+$) & \cite{GaoICHEP06}   & $1.941^{+0.042}_{-0.034}\pm0.009$ & $0.20^{+0.10}_{-0.11}\pm0.03$  \\
% CLEO-c ($D^0\to\pi^+$) & \cite{Dobbs:2007sm} & $1.87\pm 0.03\pm 0.01$                 & $0.37\pm0.08\pm0.03$                  \\
% CLEO-c ($D^+\to\pi^0$) & \cite{GaoICHEP06}   & $1.99^{+0.11}_{-0.08}\pm 0.06$     & $0.05^{+0.19}_{-0.22}\pm0.13$  \\
% CLEO-c ($D^+\to\pi^0$) & \cite{Dobbs:2007sm} & $1.97\pm 0.07\pm 0.02$                 & $0.14\pm0.16\pm0.04$                  \\
%\hline
% Fermilab lattice/MILC/HPQCD & \cite{Aubin:2004ej}            & --                              & $0.44\pm 0.04\pm 0.07$         \\
\vspace*{-10pt} & \\
\hline
\end{tabular}
}
\end{table}
The $z$-expansion formalism has been used by \babar~\cite{Aubert:2007wg,Lees:2014ihu}, 
BES III\cite{BESIII-new} and CLEO-c~\cite{Besson:2009uv},~\cite{Dobbs:2007aa}.
Their fits use the first three terms of the expansion, %, where $a_0$ controls the
%absolute normalization of $f_+(q^2)$, and $a_1$ and $a_2$ control its %$q^2$ dependence. 
and the results for the ratios $r_1\equiv a_1/a_0$ and $r_2\equiv a_2/a_0$ are 
listed in Tables~\ref{KPseudoZ} and~\ref{piPseudoZ}. 
%The CLEO~III\cite{Huang:2004fra} and FOCUS\cite{Link:2004dh} results 
%listed are obtained by refitting their data using the full
%covariance matrix. 
% 
%
%These measurements correspond to using the standard 
%outer function $\phi(q^2,t_0)$ of Eq.~(\ref{eqn:outer}) and 
%$t_0=t_+\left(1-\sqrt{1-t_-/t_+}\right)$. This choice of $t^{}_0$
%constrains $|z|$ to vary between $\pm z_{max.}$

\begin{table}[htbp]
\caption{Results for $r_1$ and $r_2$ from various experiments for the $D\to K\ell\nu_{\ell}$ decay channel.
 The correlation coefficient between these parameters is larger than 0.9.} 
\label{KPseudoZ}
\begin{center}
\begin{tabular}{ccccc}
\hline
\vspace*{-10pt} & \\
Expt. $D\to K\ell\nu_{\ell}$    & Mode &  Ref.                         & $r_1$               & $r_2$   \\
\hline
 \omit    & \omit         & \omit                & \omit               & \omit         \\
 %CLEO III &  ($D^0$; $\ell=e,\mu$) & \cite{Huang:2004fra}              & $0.2^{+3.6}_{-3.0}$ & $-89^{+104}_{-120}$  \\
%  FOCUS             & ($D^0$; $\ell=\mu$) & \cite{Link:2004dh}      & $-2.54\pm0.75$  & $7\pm 13$ \\
 \babar              & ($D^0$; $\ell=e$)  & \cite{Aubert:2007wg}   & $-2.5\pm0.2\pm0.2$  & $0.6\pm6.0\pm5.0$    \\
 CLEO-c (tagged)     & ($D^0$; $\ell=e$)  & \cite{Besson:2009uv}   & $-2.65\pm0.34\pm0.08$  & $13\pm9\pm1$   \\
 CLEO-c (tagged)     & ($D^+$; $\ell=e$)  & \cite{Besson:2009uv}   & $-1.66\pm0.44\pm0.10$  & $-14\pm11\pm1$  \\
 CLEO-c (untagged)   & ($D^0$; $\ell=e$)  &\cite{Dobbs:2007aa}     & $-2.4\pm0.4\pm0.1$  & $21\pm11\pm2$      \\
 CLEO-c (untagged)   & ($D^+$; $\ell=e$)  & \cite{Dobbs:2007aa}    & $-2.8\pm6\pm2$      & $32\pm18\pm4$       \\
 % CLEO-c and FOCUS values obtained by Lawrence. 
 BES III             & ($D^0$; $\ell=e$)  & \cite{Ablikim:2015ixa}      & $-2.334\pm0.159\pm0.080$ & $3.42\pm 3.91\pm 2.41$  \\
 BES III             & ($D^+$; $\ell=e$) & \cite{Ablikim:2015qgt}     & $ -2.23\pm 0.42 \pm 0.53 $ & $ 11.3\pm 8.5 \pm 8.7$  \\
\hline
%\hline
% Combined (preliminary) & \omit         &  \omit               & $-2.39\pm0.17$       & $6.2\pm3.8$         & -0.82        \\ 
\end{tabular}
\end{center}
\end{table}

\begin{table}[htbp]
\caption{Results for $r_1$ and $r_2$ from various experiments, for $D\to \pi \ell\nu_{\ell}$. The correlation coefficient 
between these parameters is larger than 0.9.}
\label{piPseudoZ}
\begin{center}
\begin{tabular}{cccccc}
\hline
\vspace*{-10pt} & \\
Expt. $D\to \pi\ell\nu_{\ell}$     & Mode &  Ref.                         & $r_1$               & $r_2$    \\
\hline
 \omit    & \omit         & \omit                & \omit               & \omit              \\
 CLEO-c (tagged)     & $(D^0$; $\ell=e$) & \cite{Besson:2009uv}      &  $-2.80\pm0.49\pm0.04$ & 6 $\pm$ 3 $\pm$ 0\\            
 CLEO-c (tagged)     & $(D^+$; $\ell=e$) & \cite{Besson:2009uv}      &  $-1.37\pm0.88\pm0.24$ & -4 $\pm$ 5 $\pm$ 1\\            
 CLEO-c  (untagged)  & $(D^0$; $\ell=e$) & \cite{Dobbs:2007aa}  & $-2.1\pm0.7\pm0.3$      & $-1.2\pm4.8\pm1.7$ \\
 CLEO-c   (untagged) & $(D^+$; $\ell=e$) & \cite{Dobbs:2007aa}  & $-0.2\pm1.5\pm0.4$    & $-9.8\pm9.1\pm2.1$ \\
% Belle               & $(D^0$;$\ell=e$)  & \cite{Widhalm:2006wz} & $-1.84\pm 1.02 $ & $1.69\pm 6.5$ \\
 BES III             & $(D^0$; $\ell=e$)  & \cite{Ablikim:2015ixa}                    & $-1.85 \pm 0.22 \pm 0.07$ & $-1.4 \pm 1.5 \pm 0.5$ \\
 \babar              & $(D^0$; $\ell=e$)  & \cite{Lees:2014ihu}                   & $ -1.31 \pm 0.70 \pm 0.43 $ & $-4.2 \pm 4.0 \pm 1.9$ \\
\vspace*{-10pt} & \\
\hline
\end{tabular}
\end{center}
\end{table}

% KENU 2016 (NEW, A. Oyanguren and P. Roudeau) 

\subsubsection{Combined results for the $D\to K\ell\nu_\ell$ channel}
 
\begin{figure}[ph]
\begin{center}
\includegraphics[width=0.47\textwidth]{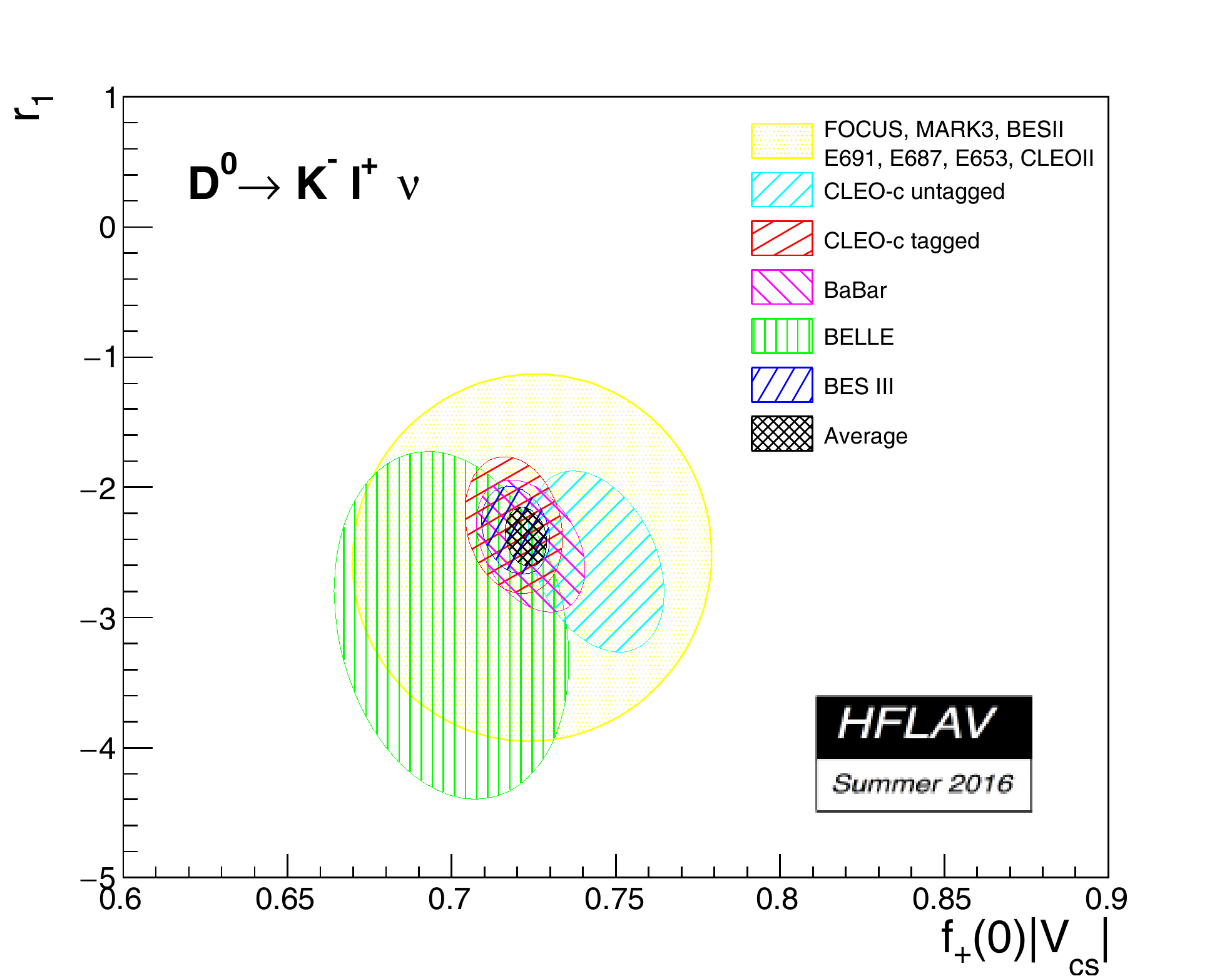}\hfill
\includegraphics[width=0.46\textwidth]{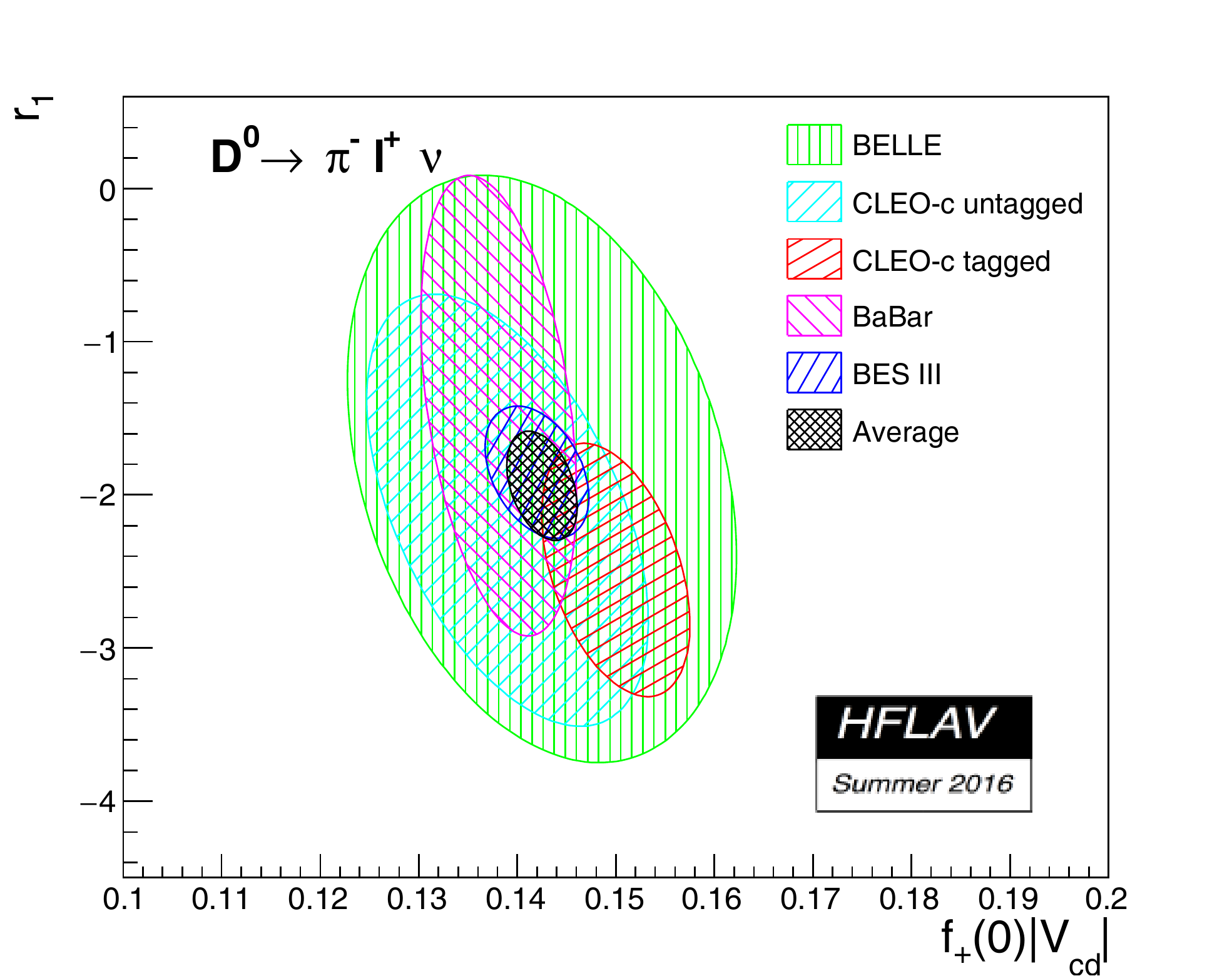}
\includegraphics[width=0.47\textwidth]{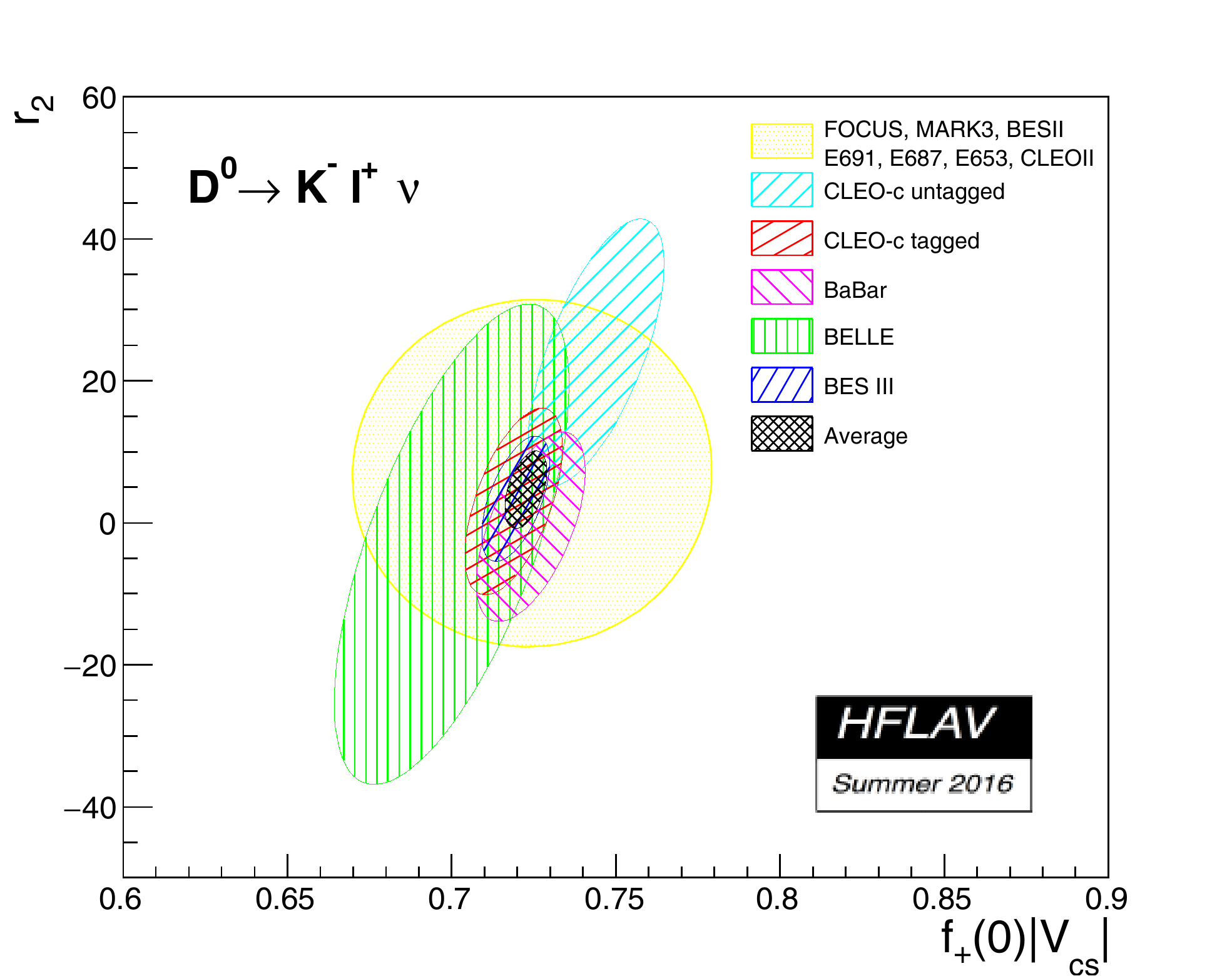}\hfill
\includegraphics[width=0.46\textwidth]{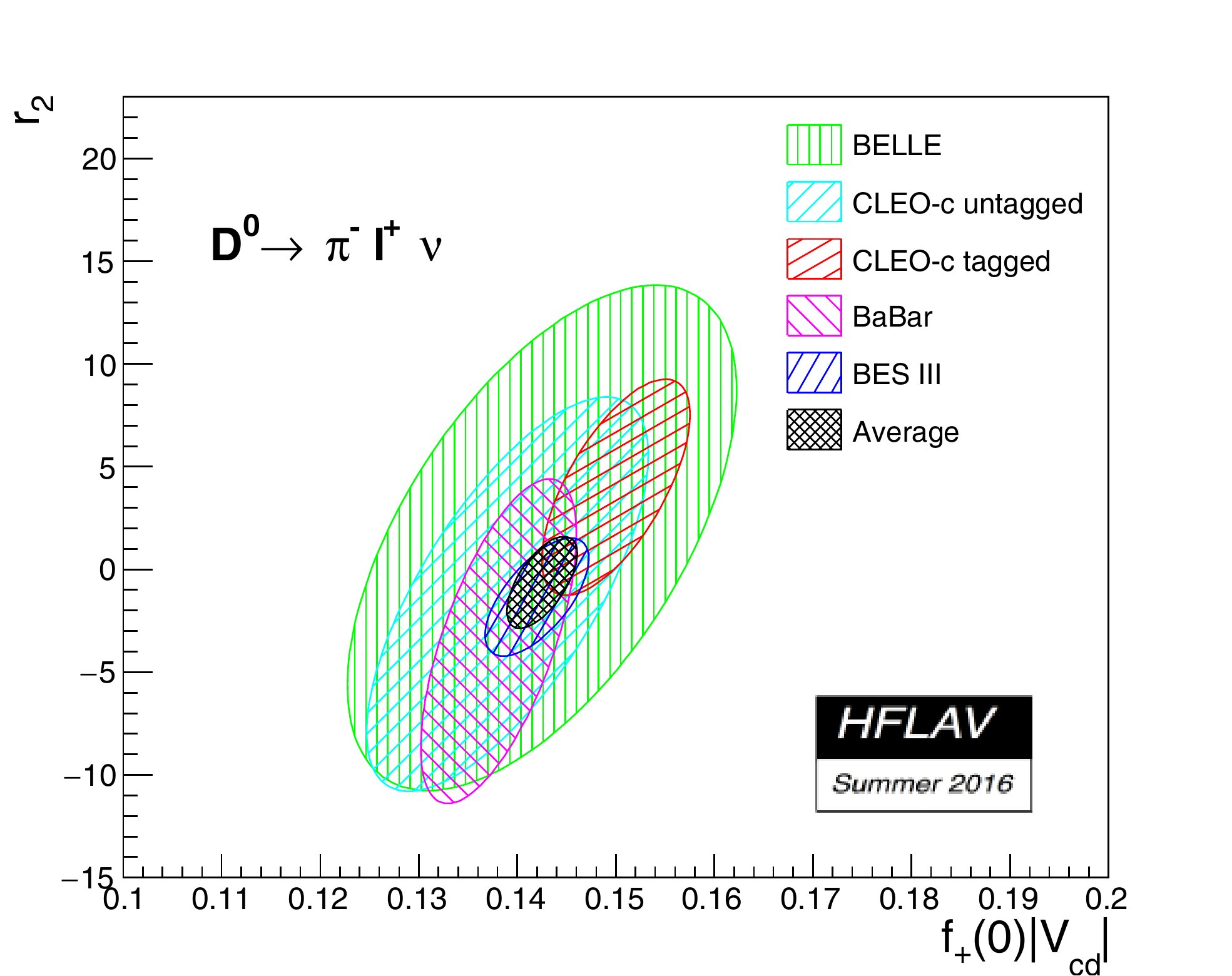}\hfill
\includegraphics[width=0.47\textwidth]{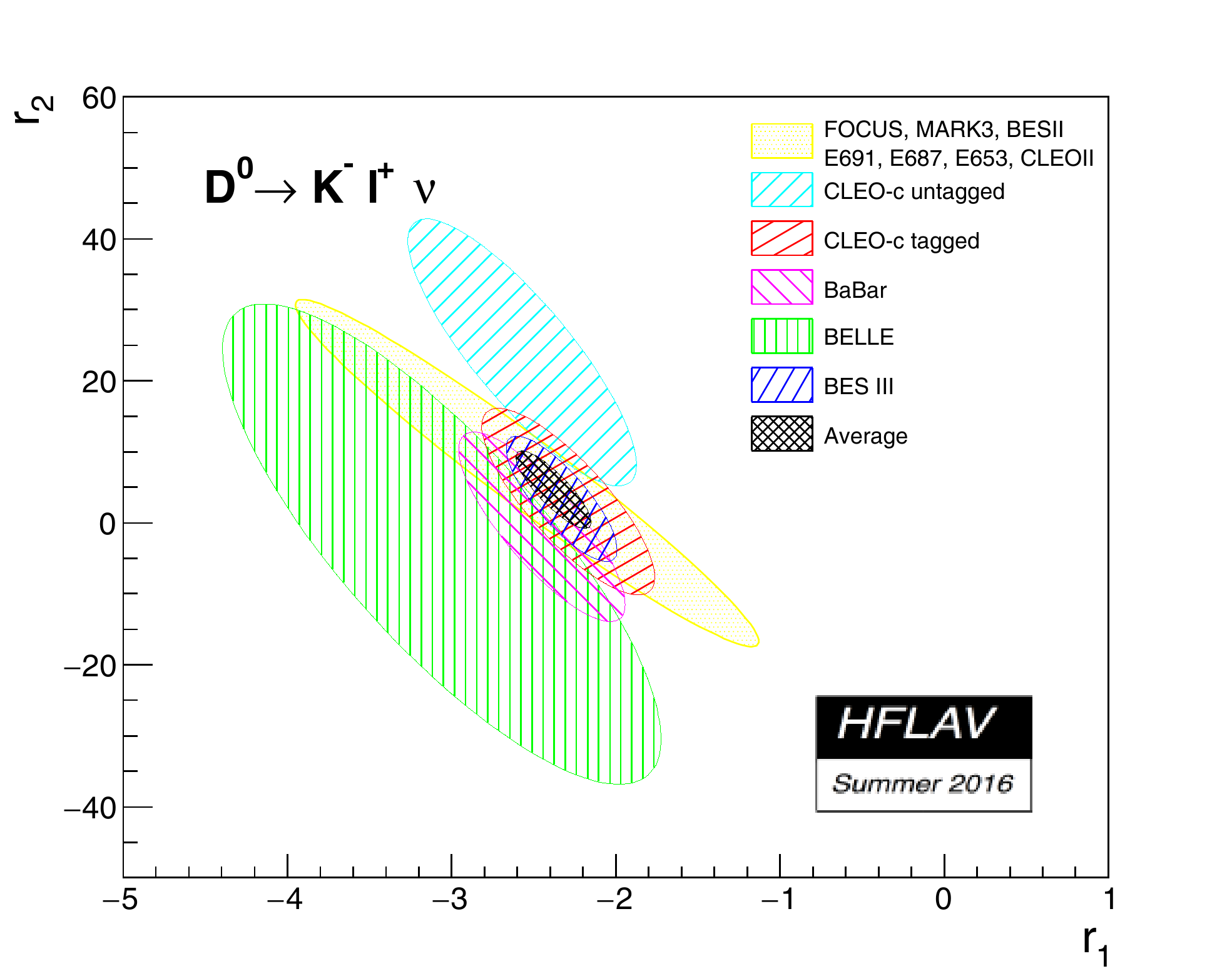}\hfill
\includegraphics[width=0.46\textwidth]{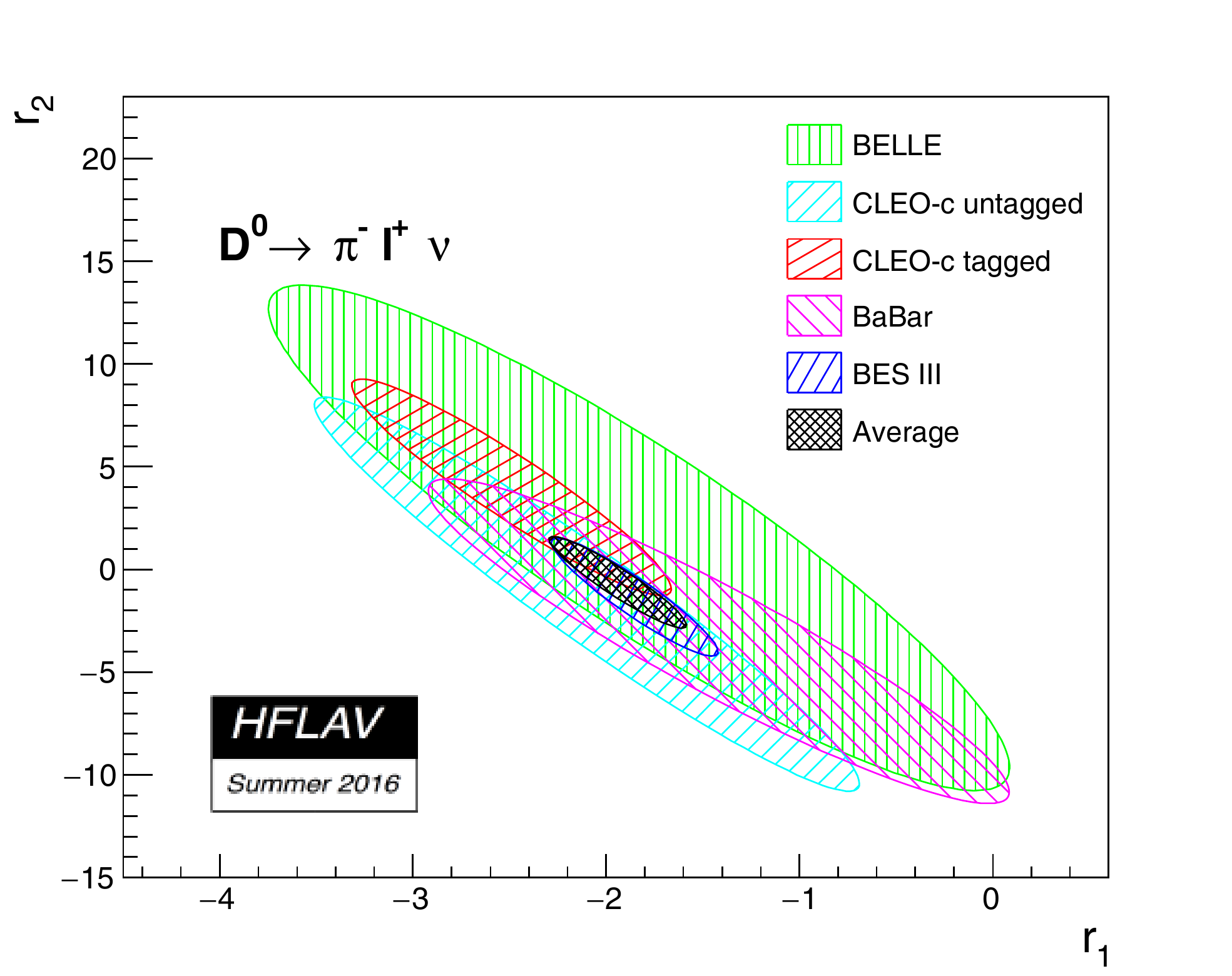}
\caption{The $D^0\to K^-\ell^+\nu$ (left) and $D^0\to \pi^-\ell^+\nu$ (right) 68\% C.L. error ellipses from the
average fit of the 3-parameter $z$-expansion results. 
\label{fig:fitellipse}}
\end{center}
\end{figure}

The $q^2$ distribution provided by each individual measurement is used to determine a combined result by performing a fit to the $z$-expansion 
formalism at second order. Results for the form factor normalization $f_+^K(0)|V_{cs}|$ and the shape parameters $r_1$ and $r_2$ for 
each individual measurement and for the combination are presented in Table \ref{krefitted}. 
Measurements have been corrected with respect to the original ones using recent values from PDG \cite{PDG_2014}. 
This includes updated branching fractions of normalization channels, corrected CKM matrix elements and the $D$ meson lifetime. 
The \babar measurement has been corrected accounting for final-state radiation. 
The result for the $D^+\to K_{L}^{0} e^+ \nu_e$ decay channel from BES III \cite{Ablikim:2015qgt} is included as a constraint in the combined result 
since correlation matrices are not provided. Correlation coefficients of the parameters are quoted in the last column of Table \ref{krefitted}. 
The $\chi^2$ per degree of freedom is 114.7/101. Results are 
shown in Figure \ref{fig:fitellipse}. 

\begin{figure}[ph]
\begin{center}
\includegraphics[width=0.47\textwidth]{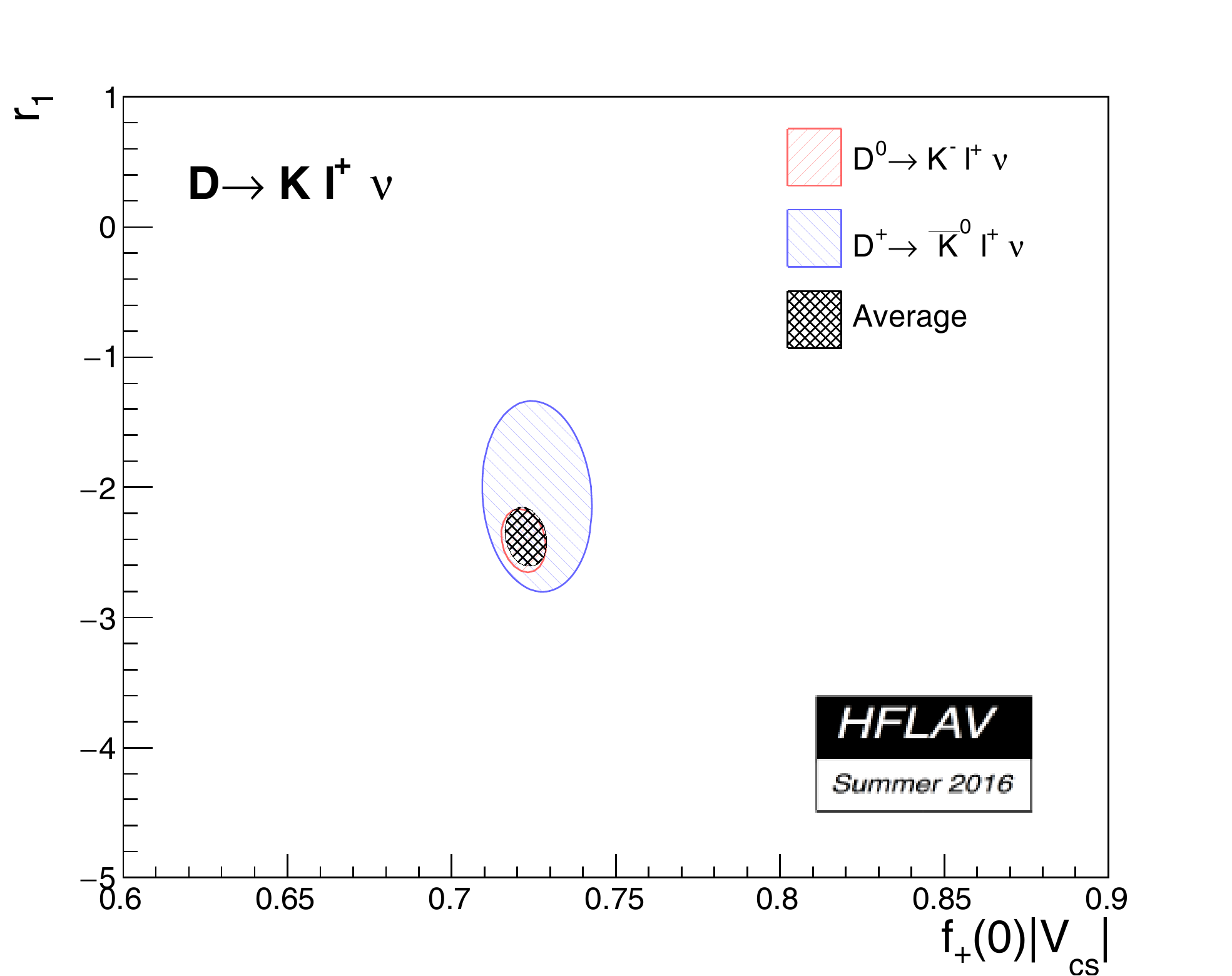}\hfill
\includegraphics[width=0.47\textwidth]{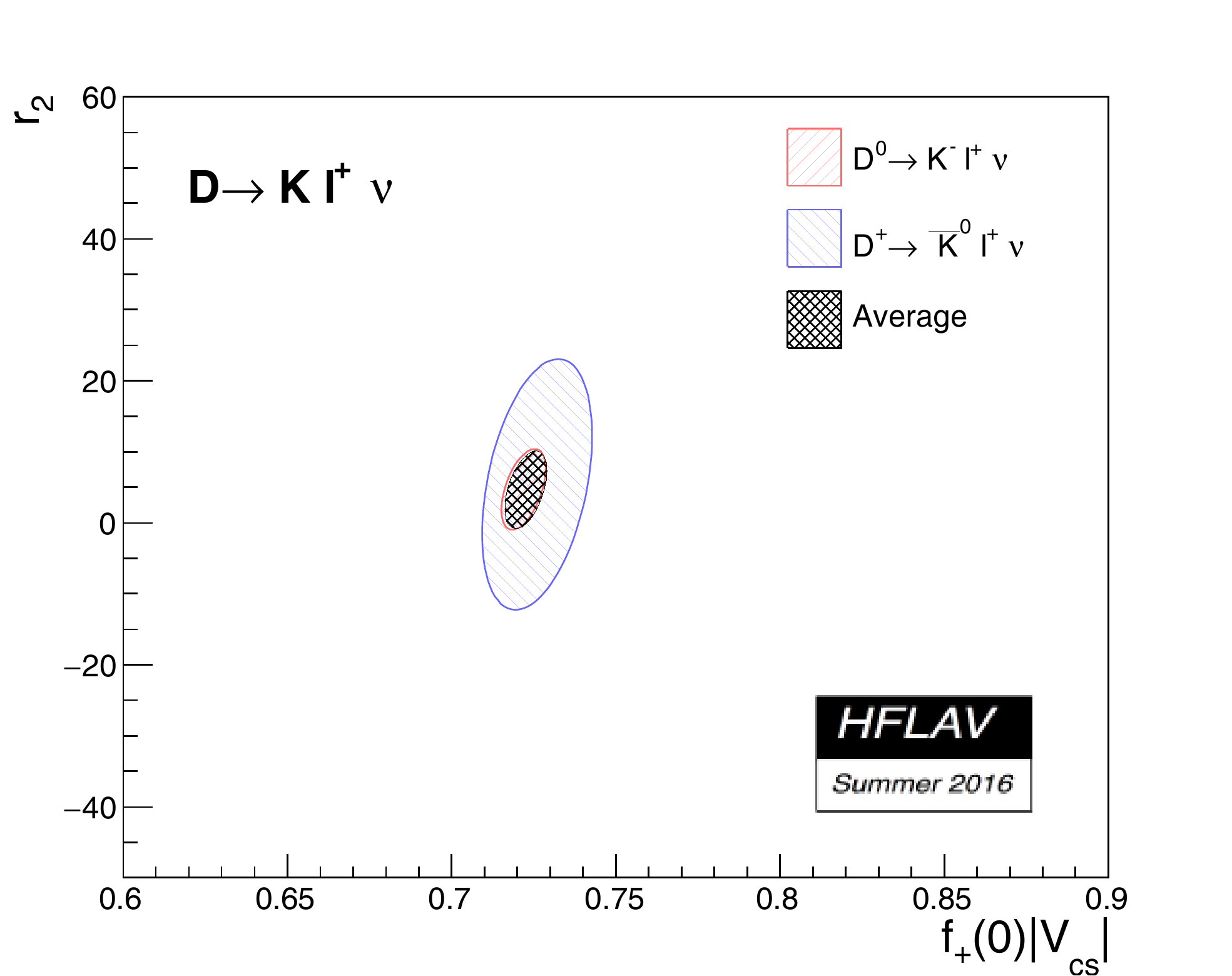}\hfill
\includegraphics[width=0.47\textwidth]{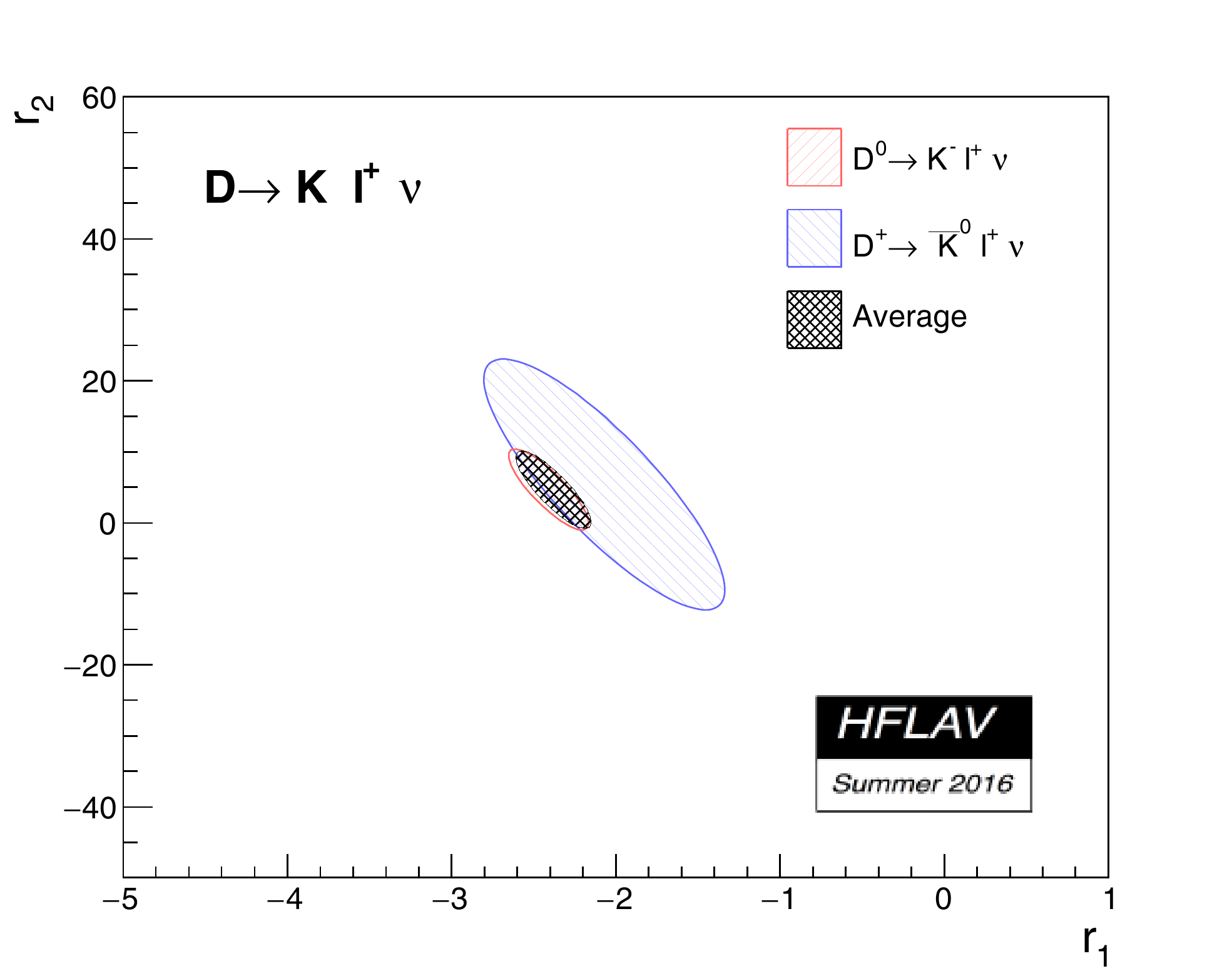}\hfill
\caption{Results of the combined fit shown separately for the $D^0\to K^-\ell^+\nu$ and 
$D^+\to \bar K^0 \ell^+\nu$ decay channels. Ellipses are shown for 68\% C.L.  
\label{fig:kiso}}
\end{center}
\end{figure}

In the combination of the electron and muon channels, the measurements with muons 
%are obtained in terms of $D \to K e \nu_e$, after having 
are corrected for the reduction of phase space and for the $f_0(q^2)$ contribution\cite{Korner:1989qb}. Channels with a $D^0$ or a $D^+$ are combined assuming isospin invariance and using physical meson and lepton masses. These combined results are noted as $D \to K \ell \nu_\ell$ in the following. 
Hadronic form factors are assumed to be the same for charged and neutral $D$ mesons. 
Separate results for the $D^0 \to K^- \ell^+ \nu_\ell$ and $D^+ \to \bar K^0 \ell^+ \nu_\ell$ decay channels are
shown in Table \ref{tab:kiso} and Figure \ref{fig:kiso}. 
 Using the fitted parameters and integrating over the full $q^2$ range, the combined semileptonic branching fraction, expressed in terms of the $D^0$ decay channel gives: 
 \begin{equation}
 \label{brpi}
 {\cal B}(D^0 \to K^- \ell^+ \nu_\ell) = (3.490\pm 0.011\pm0.020)\%
 \end{equation}

%%------------------------------------------------------------------------------------------------

\begin{table}[htbp]
\centering
\caption{Results of the fits to $D\to K \ell\nu_{\ell}$ measurements from several experiments, using the $z$-expansion. External inputs have been updated to PDG~\cite{PDG_2014}.
The correlation coefficients listed in the last column refer to 
$\rho_{12} \equiv \rho_{|V_{cs}| f_{+}^{K}(0) ,r_1}$, $\rho_{13} \equiv \rho_{|V_{cs}| f_{+}^{K}(0) ,r_2}$, and  $\rho_{23} \equiv \rho_{r_1,r_2}$ 
and are for the total uncertainties (statistical $\oplus$ systematic).
The result for the $D^+\to K_{L}^{0} e^+ \nu_e$ decay channel from BES III \cite{Ablikim:2015qgt} is included in the 
combined results as a constraint on the normalization,  $|V_{cs}| f_{+}^{K}(0)$.
The entry {\it others} refers to total decay rates measured by Mark-III~\cite{Adler:1989rw}, E653~\cite{Kodama:1991ij, Kodama:1994aj}, E687\cite{Frabetti:1995xq,Frabetti:1993vz}, E691~\cite{Anjos:1988ue}, BES II~\cite{Ablikim:2004ej,Ablikim:2006bv} and CLEO II~\cite{Bean:1993zv}.}

\label{krefitted}
\resizebox{\textwidth}{!}{
\begin{tabular}{ccc ccc}
\hline
\vspace*{-10pt} & \\
 Expt. $D\to K \ell\nu_{\ell}$   & Mode   &    $|V_{cs}| f_{+}^{K}(0)$  & $r_1$    & $r_2$        & $\rho_{12}/\rho_{13}/\rho_{23}$  \\
\hline
 \omit    & \omit         & \omit           & \omit     &\omit          & \omit         \\
  BES III (tagged)~\cite{Ablikim:2015ixa} & ($D^0$)      & 0.7195(35)(43)    & $-$2.33(16)(8)   & 3.4(4.0)(2.5)    & $-$0.21/0.58/$-$0.81 \\ 
  CLEO-c (tagged)~\cite{Besson:2009uv} & ($D^0$, $D^+$)  & 0.7189(64)(48)   & $-$2.29(28)(27)  & 3.0(7.0)(1.0)    & $-$0.19/0.58/$-$0.81 \\
  CLEO-c (untagged)~\cite{Dobbs:2007aa} & ($D^0$, $D^+$) & 0.7436(76)(79)   & $-$2.57(33)(18)  & 23.9(8.9)(4.3)   & $-$0.34/0.66/$-$0.84 \\
 \babar~\cite{Aubert:2007wg}  &  ($D^0$)                & 0.7241(64)(60)    & $-$2.45(20)(18)  &$-$0.6(6.0)(3.8)    & $-$0.36/0.59/$-$0.82\\     
  Belle~\cite{Widhalm:2006wz}  &  ($D^0$)               &  0.700(19)        & $-$3.06(71)      &$-$3.3(17.9)        & $-$0.20/0.66/$-$0.81\\ 
  FOCUS\cite{Link:2004dh} and others & \omit  & 0.724(29) & $-$2.54(75) & 7.0(12.8) & $-$0.02/0.02/$-$0.97 \\
 \hline 
\bf Combined & ($D^0$, $D^+$) & \bf 0.7226(22)(26) & \bf $-$2.38(11)(6) & \bf 4.7(2.6)(1.4) & \bf $-$0.19/0.51/$-$0.84\\   
\hline
\vspace*{-10pt} & \\
\end{tabular}
}
\end{table}

%----------------------------------------------------

\begin{table}[htbp]
\centering
\caption{Results for the $D^0 \to K^- \ell^+ \nu_\ell$ and $D^+ \to \bar K^0 \ell^+ \nu_\ell$ decays channels using the 
$z$-expansion formalism at second order.}
\label{tab:kiso}
\begin{tabular}{ccc}
\hline 
Fit value & $D^0 \to K^- \ell^+ \nu_\ell$ & $D^+ \to \bar K^0 \ell^+ \nu_\ell$ \\ 
\hline 
 $|V_{cs}| f_{+}^{K}(0)$  & 0.7219 $\pm$ 0.0024 $\pm$ 0.0027 & 0.726 $\pm$ 0.005 $\pm$ 0.007 \\
 $r_1$        & $-$2.41 $\pm$ 0.11 $\pm$ 0.07  & $-$2.07 $\pm$ 0.38 $\pm$ 0.10 \\
 $r_2$        & 4.7 $\pm$ 2.7 $\pm$ 1.4 & 5.4 $\pm$ 8.2 $\pm$ 4.6\\
 $\rho_{12}/\rho_{13}/\rho_{23}$ & $-$0.19/0.51/$-$0.84 & $-$0.10/0.39/$-$0.84 \\
 \hline
 \vspace*{-10pt} & \\
\end{tabular}
\end{table}

\begin{figure}[ph]
\hskip-1cm
\includegraphics[width=0.55\textwidth]{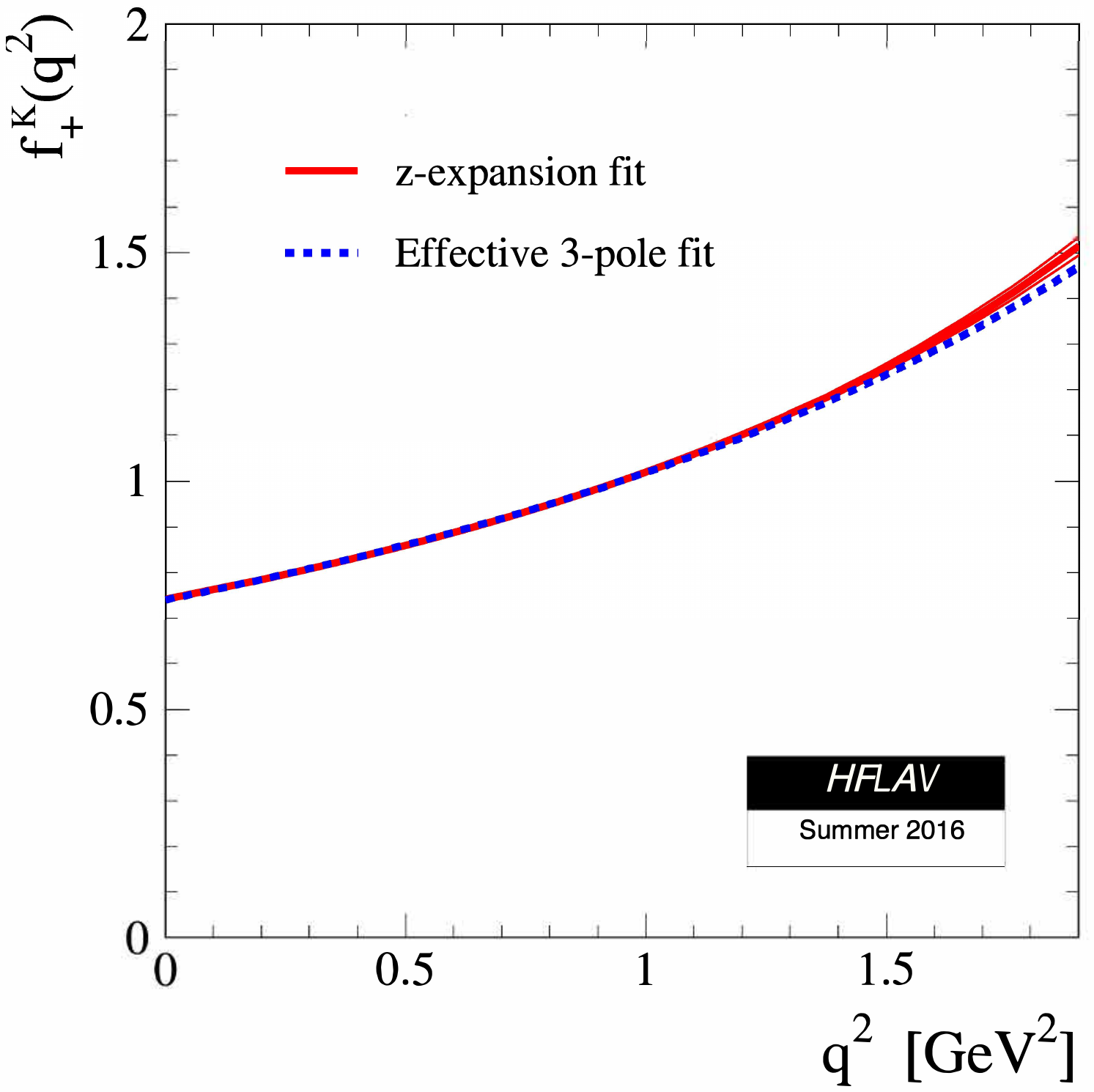}\hfill
\hskip-3cm
\includegraphics[width=0.56\textwidth]{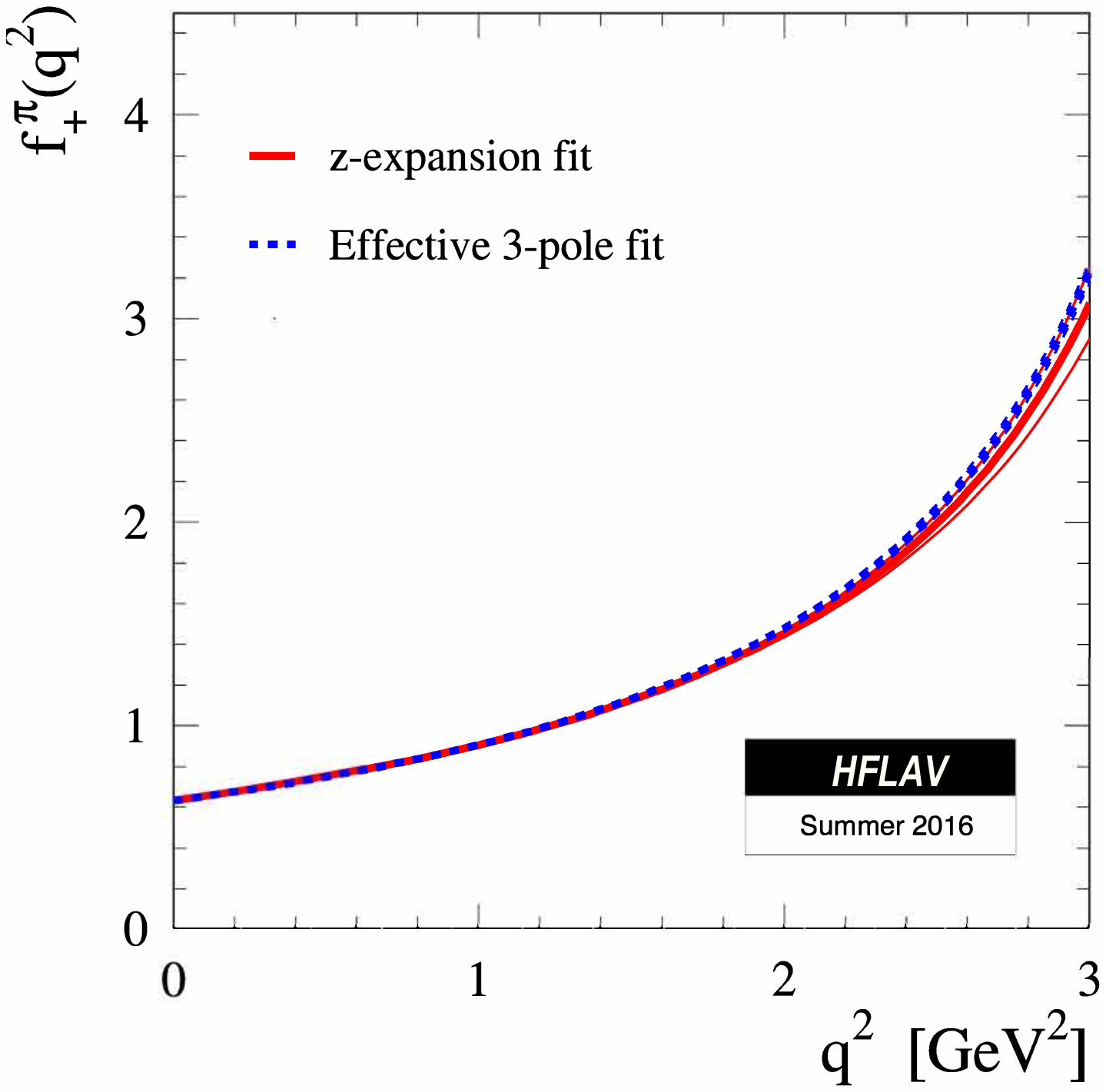}\hfill
\vskip-3.5cm
\caption{Form factors as function of $q^2$ for the $D \to K \ell \nu_\ell$ (left) 
and $D \to \pi \ell \nu_\ell$ (right) channels, obtained from a fit to all 
experimental data. Central values (central lines) and uncertainties (one $\sigma$ 
deviation) are shown for the z-expansion and the 3-pole parameterization.}
\label{fig:ffs}
\end{figure}

Data from the different experiments are also fitted within the three-pole form 
factor formalism. 
Constraints on the first and second poles are imposed using information of the 
$D^{*}_s(2112)$ and $D^{*}_{s1}(2700)$ resonances. 
%The superconvergence condition of Eq.~(\ref{superconvergence}) is applied. 
Results are presented in Table~\ref{tab:3Pole_k}. 
Fitted parameters are the first two residues 
$\gamma_{0}^{K}=\underset{ q^2=m_{D^{*}_s(2112)}^2} {\rm Res} f_{+}^{K}(q^2) $ 
and $\gamma^{\pi}_{1}=\underset{q^2=m_{D^{*}_{s1}(2700)}^2}{\rm Res} f_{+}^{K}(q^2)$ 
and an effective mass, $m_{D^{\ast ''}_{s\;{\rm eff}}}$, 
accounting for higher mass hadronic contributions.  
It is found that the fitted effective third pole mass is 
larger than the mass of the second radial excitation, 
around $3.2~\gevcc$, as expected.  
The contribution to the form factor by only the $D^{*}_{s}$ 
resonance is disfavoured by the data.   
Figure \ref{fig:ffs} (left) shows the result of the fitted form 
factors for the $z$-expansion and three-pole parametrizations. 

\begin{table}[htbp]
\caption{Results of the three-pole model form factors obtained from a fit to 
all measurements. Fitted parameters are the first two residues $\gamma_{0}^{K}$ 
and $\gamma_{1}^{K}$, which are constrained using present measurements of masses
and widths of the $D^{\ast}_{s}$ and $D^{\ast}_{s1}$ mesons, and lattice computations 
of decay constants, and the effective mass, $m_{D^{\ast ''}_{s\;{\rm eff}}}$, accounting for 
higher mass hadronic contributions. 
\label{tab:3Pole_k}}
\begin{center}
\begin{tabular}{cc}
\hline 
Parameter & Combined result ($D \to K \ell \nu_\ell $) \\ 
\hline 
\vspace*{-10pt} & \\
$\gamma_{0}^{K}$  & $4.85 \pm 0.08~\gev^2$ \\
$\gamma_{1}^{K}$  & $-1.2 \pm 0.30~\gev^2$  \\
$m_{D^{\ast ''}_{s\;{\rm eff}}}$  & $4.46 \pm 0.26~\gev$ \\
\hline
\vspace*{-10pt} & \\
\end{tabular}
\end{center}
\end{table}

%%%--------------------------------------------------------------------------------------------------

%%%--------------------------------------------------------------------------------------------------

\subsubsection{Combined results for the $D\to \pi \ell\nu_\ell$ channel}
%%%--------------------------------------------------------------------------------------------------
% PIENU 2016 (NEW, A. Oyanguren and P. Roudeau) 

The combined result for the $D\to \pi \ell\nu_{\ell}$ decay channel is obtained from a fit 
to \babar, Belle, BES III, and CLEO-c data,  with updated input values from \cite{PDG_2014}. 
The available measurements are fitted in bins of $q^2$ to the $z$-expansion model at 
second order. 
%Published values of $f^{D\pi}_{+} (q^2) \times |V_{cd}|$ Belle data~\cite{Widhalm:2006wz} are modified 
%by subtracting the uncertainty on $V_{cd}$ from the systematic error. Since the experimental $q^2$ resolution is very high, measurements at different $q^2$ are assumed uncorrelated. 
%$D\to \pi \ell\nu_{\ell}$ data from the Belle experiment~\cite{Widhalm:2006wz} is modified 
% Belle experiment in Table~\ref{piPseudoZ} and figure  
%\ref{fig:fitellipse} is obtained by a fit in the z-expansion formalism using their published values of 
%$f^{D\pi}_{+} (q^2) \times |V_{cd}|$, and removing the uncertainty on $V_{cd}$ from the systematic error.  
Results of the individual fits for each experiment and the combined result are shown 
in Table~\ref{pirefitted}. The $\chi^2$ per degree of freedom of the 
combined fit is $51/55$. 

%%------------------------------------------------------------------------------------------------

\begin{table}
\centering
\caption{Results of the fits to $D\to \pi \ell\nu_{\ell}$ measurements from several 
experiments, using the $z$-expansion. External inputs are updated to PDG~\cite{PDG_2014}. 
The correlation coefficients listed in the last column 
refer to $\rho_{12} \equiv \rho_{|V_{cd}| f_{+}^{\pi}(0) ,r_1}$, 
$\rho_{13} \equiv \rho_{|V_{cd}| f_{+}^{\pi}(0) ,r_2}$, and  $\rho_{23} \equiv \rho_{r_1,r_2}$ 
and are for the total uncertainties (statistical $\oplus$ systematic).}
\label{pirefitted}
\resizebox{\textwidth}{!}{
\begin{tabular}{cccccc}
\hline
\vspace*{-10pt} & \\
 Expt. $D\to \pi \ell\nu_{\ell}$   & mode   &    $|V_{cd}| f_{+}^{\pi}(0)$  & $r_1$    & $r_2$        & $\rho_{12}/\rho_{13}/\rho_{23}$  \\
\hline
 \omit    & \omit         & \omit           & \omit     &\omit          & \omit         \\
  BES III (tagged)~\cite{Ablikim:2015ixa}  & ($D^0$)         & 0.1422(25)(10)    & $-$1.86(23)(7)   & $-$1.24(1.51)(47)  & $-$0.37/0.64/$-$0.93 \\ 
  CLEO-c (tagged)~\cite{Besson:2009uv}     & ($D^0$, $D^+$)  & 0.1507(42)(11)   & $-$2.45(43)(9)   &  3.8(2.8)(6)     & $-$0.43/0.67/$-$0.94 \\
  CLEO-c (untagged)~\cite{Dobbs:2007aa}    & ($D^0$, $D^+$)  & 0.1394(58)(25)  & $-$1.71(62)(25)  & $-$2.8(4.0)(1.6)   & $-$0.50/0.69/$-$0.96 \\
  \babar~\cite{Aubert:2007wg}              &  ($D^0$)        & 0.1381(36)(22) & $-$1.42(66)(45)   & $-$3.5(3.7)(2.0)   & $-$0.40/0.57/$-$0.97 \\     
  Belle~\cite{Widhalm:2006wz}  &  ($D^0$)                    &  0.142(11)     &  $-$1.83(1.00)    & 1.5(6.5)         & $-$0.30/0.59/$-$0.91 \\ 
 \hline 
\bf Combined   & ($D^0$, $D^+$)  & \bf 0.1426(17)(8)   & \bf $-$1.95(18)(1)  & \bf $-$0.52(1.17)(32) & \bf $-$0.37/0.63/$-$0.94 \\   
\hline
\vspace*{-10pt} & \\
\end{tabular}
}
\end{table}

%----------------------------------------------------

Using the fitted parameters and integrating over the full $q^2$ range, the combined 
semileptonic branching fraction, expressed in terms of the $D^0$ decay channel gives: 
 \begin{equation}
 \label{brK}
 {\cal B}(D^0 \to \pi^- \ell^+ \nu_\ell) = (2.891\pm 0.030\pm0.022)\times10^{-3}
 \end{equation}

Results of the three-pole model to the $D \to \pi \ell \nu_\ell $ data are shown 
in Table \ref{3Pole_pi}. Fitted parameters are the first two residues 
$\gamma_{0}^{\pi}=\underset{ q^2=m_{D^{\ast}}^2} {\rm Res} f^{\pi}_{+}(q^2) $ 
and $\gamma^{\pi}_{1}=\underset{q^2=m_{D^{\ast '}}^2}{\rm Res} f^{\pi}_{+}(q^2)$ 
(which are constrained using present measurements 
of masses and widths of the $D^\ast(2010)$ and $D^{\ast '}(2600)$ mesons, 
and lattice computations of decay constants, following~\cite{Becirevic:2014kaa}), 
and an effective mass, $m_{D^{\ast ''}_{\rm eff}}$, accounting for higher mass hadronic 
contributions. 
The $V_{cd}$ value entering in the fit is given in Eq.~(\ref{eq:charm:ckm}).
The $\chi^2$ per degree of freedom of the combined fit is $57.5/57$. 

The effective mass $m_{D^{\ast ''}_{\rm eff}}$ is larger than the predicted mass 
of the second radially excited state with $J^P = 1^{-}$ ($\sim 3.11~\gev$), 
indicating that more contributions are needed to explain the form factor. 
 Figure \ref{fig:ffs} (right) shows the result of the combined form factor 
for the $z$-expansion and three-pole parameterizations. 

%Comparison of the combined fit with the individual data is shown in Figure \ref{plot3}.

%\begin{figure}[p]
%\begin{center}
%\includegraphics[width=0.7\textwidth]{figures/charm/sl_plot3_logo.pdf}
%\caption{Result of the three-pole model fit~\cite{Becirevic:2014kaa} to \babar~\cite{Lees:2014jka}, Belle\cite{Widhalm:2006wz}, 
%BES III\cite{BESIII-new} and CLEO-c~\cite{Besson:2009uv},~\cite{Dobbs:2007aa} $D \to \pi \ell \nu_\ell $ data. Points are the 
%measured data in $q^2$ bins and the black line correspond to the result of the combined fit.
%\label{plot3}
%}
%\end{center}
%\end{figure}

%For the Belle measurement published values of $f^{D\pi}_{+} (q^2) \times |V_{cd}|$ have been modifyied by subtracting the $V_{cd}$ 
%uncertainty from the total systematic error.

\begin{table}[htbp]
\caption{Results of the three-pole model to \babar, Belle, BES III and CLEO-c 
(tagged and untagged) data. Fitted parameters are the first two residues 
$\gamma_{0}^{\pi}$  and $\gamma_{1}^{\pi}$, which are constrained using present 
measurements of masses and widths of the $D^\ast$ and $D^{\ast '}$ mesons, and 
lattice computations of decay constants, and the effective mass, 
$m_{D^{\ast ''}_{\rm eff}}$, accounting for higher mass hadronic contributions. 
\label{3Pole_pi}}
\vskip0.5cm
\begin{center}
\begin{tabular}{cc}
\hline 
Parameter & Combined result ($D \to \pi \ell \nu_\ell $) \\ 
\hline 
\vspace*{-10pt} & \\
$\gamma_{0}^{\pi}$  & $3.881 \pm 0.093~\gev^2$ \\
$\gamma_{1}^{\pi}$  & $-1.18 \pm 0.30~\gev^2$  \\
$m_{D^{\ast ''}_{\rm eff}}$  & $4.17 \pm 0.42~\gev$ \\
%$V_{cd}$ & 0.22506 $\pm$ 0.00087\\
\hline
\vspace*{-10pt} & \\
\end{tabular}
\end{center}
\end{table}

\subsubsection{$V_{cs}$ and $V_{cd}$ determination}
Assuming unitarity of the CKM matrix, the values of the CKM matrix elements 
entering in charm semileptonic decays are evaluated from the $V_{ud}$, $V_{td}$ 
and $V_{cb}$ elements~\cite{PDG_2014}:
\begin{equation}
\label{eq:charm:ckm}
\begin{aligned}
|V_{cs}| & = 0.97343 \pm 0.00015 \, ,\\
|V_{cd}| & = 0.22521 \pm 0.00061 \, .
\end{aligned}
\end {equation}
Using the combined values of $f_+^K(0)|V_{cs}|$ and $f_+^{\pi}(0)|V_{cd}|$ in 
Tables \ref{krefitted} and \ref{pirefitted}, leads to the form factor values: 
\begin{equation}
%\label{ff_measured}
\begin{aligned}
 f_+^K(0) & = 0.7423 \pm 0.0035  \, , \\ 
 f_+^{\pi}(0) &= 0.6327 \pm 0.0086 \,, 
\end{aligned}\nonumber 
\end {equation}
which are in agreement with  present lattice QCD computations~\cite{Aoki:2016frl}: 
$f_+^K(0) = 0.747 \pm 0.019$ and $f_+^\pi(0) = 0.666 \pm 0.029$. 
The experimental accuracy is at present higher than the one from lattice calculations.  
If instead one assumes the lattice QCD form factor values, one obtains 
for the CKM matrix elements using the combined results in 
Tables~\ref{krefitted} and~\ref{pirefitted}:
\begin{equation}
%\label{ckm}
\begin{aligned}
|V_{cs}| &= 0.967 \pm 0.025  \, , \\ 
|V_{cd}| &= 0.2140 \pm 0.0097 \, , 
\end{aligned}\nonumber 
\end {equation} 
still compatible with unitarity of the CKM matrix.

\subsubsection{$D\ra V\overline \ell \nu_\ell$ decays}

\begin{figure}[htbp]
  \begin{center}
\includegraphics[width=2.5in, viewport=0 0 320 200]{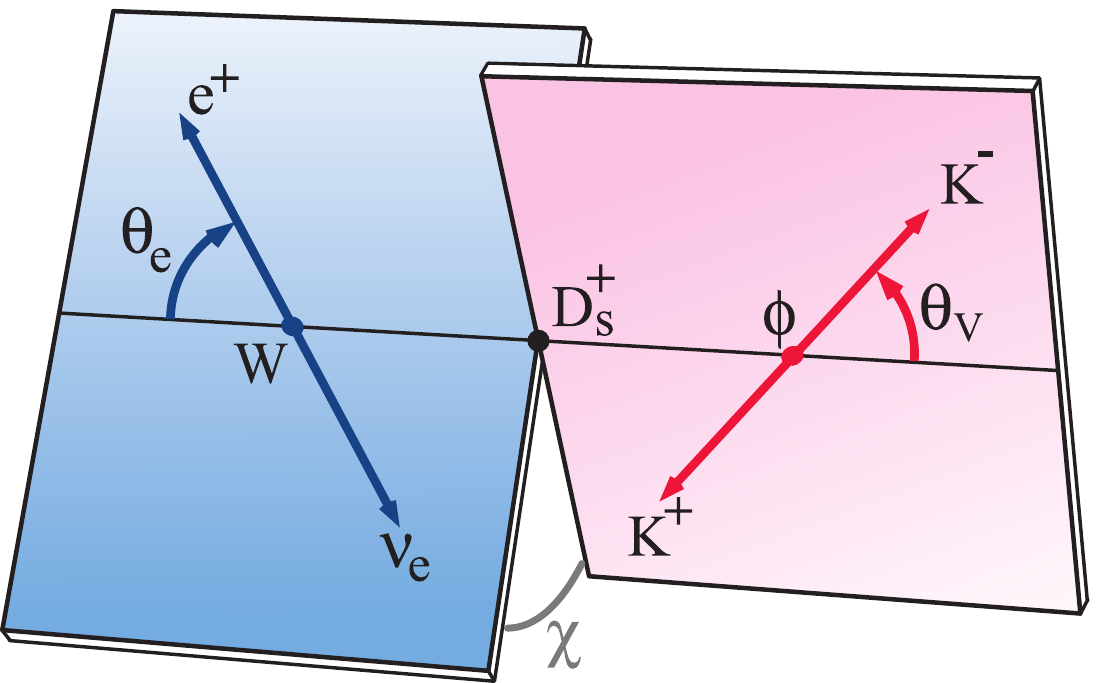}
  \end{center}
  \caption{
    Decay angles $\theta_V$, $\theta_\ell$ 
    and $\chi$. Note that the angle $\chi$ between the decay
    planes is defined in the $D$-meson reference frame, whereas
    the angles $\theta^{}_V$ and $\theta^{}_\ell$ are defined
    in the $V$ meson and $W$ reference frames, respectively.}
  \label{DecayAngles}
\end{figure}

When the final state hadron is a vector meson, the decay can proceed through
both vector and axial vector currents, and four form factors are needed.
The hadronic current is $H^{}_\mu = V^{}_\mu + A^{}_\mu$, 
where~\cite{Gilman:1989uy} 
\begin{eqnarray}
V_\mu & = & \left< V(p,\varepsilon) | \bar{q}\gamma_\mu c | D(p') \right> \ =\  
\frac{2V(q^2)}{m_D+m_V} 
\varepsilon_{\mu\nu\rho\sigma}\varepsilon^{*\nu}p^{\prime\rho}p^\sigma \\
 & & \nonumber\\
A_\mu & = & \left< V(p,\varepsilon) | -\bar{q}\gamma_\mu\gamma_5 c | D(p') \right> 
 \ =\  -i\,(m_D+m_V)A_1(q^2)\varepsilon^*_\mu \nonumber \\
 & & \hskip2.10in 
  +\ i \frac{A_2(q^2)}{m_D+m_V}(\varepsilon^*\cdot q)(p' + p)_\mu \\
 & & \hskip2.10in 
+\ i\,\frac{2m_V}{q^2}\left(A_3(q^2)-A_0(q^2)\right)[\varepsilon^*\cdot (p' +
p)] q_\mu\,. \nonumber 
\end{eqnarray}
In this expression, $m_V$ is the daughter meson mass and
\begin{equation}
  A_3(q^2) = \frac{m_D + m_V}{2m_V}A_1(q^2)\ -\ \frac{m_D - m_V}{2m_V}A_2(q^2)\,.
\end{equation}
Kinematics require that $A_3(0) = A_0(0)$. Terms proportional to $q_\mu$ are only important 
for the case of $\tau$ leptons. Thus, only three form factors are relevant in the decays involving $\mu$ or $e$: 
$A_1(q^2)$, $A_2(q^2)$ and $V(q^2)$. The differential partial width is
%integrated over various angular distributions is
\begin{eqnarray}
\frac{d\Gamma(D \to V \overline \ell \nu_\ell)}{dq^2\, d\cos\theta_\ell} & = & 
  \frac{G_F^2\,|V_{cq}|^2}{128\pi^3m_D^2}\,p^*\,q^2 \times \nonumber \\
 & &  
\left[\frac{(1-\cos\theta_\ell)^2}{2}|H_-|^2\ +\  
\frac{(1+\cos\theta_\ell)^2}{2}|H_+|^2\ +\ \sin^2\theta_\ell|H_0|^2\right]\,,
\end{eqnarray}
where $H^{}_\pm$ and $H^{}_0$ are helicity amplitudes, corresponding to helicities of the V meson or virtual $W$, given by
\begin{eqnarray}
H_\pm & = & \frac{1}{m_D + m_V}\left[(m_D+m_V)^2A_1(q^2)\ \mp\ 
      2m^{}_D\,p^* V(q^2)\right] \\
 & & \nonumber \\
H_0 & = & \frac{1}{|q|}\frac{m_D^2}{2m_V(m_D + m_V)}\ \times\ \nonumber \\
 & & \hskip0.01in \left[
    \left(1- \frac{m_V^2 - q^2}{m_D^2}\right)(m_D + m_V)^2 A_1(q^2) 
    \ -\ 4{p^*}^2 A_2(q^2) \right]\,.
\label{HelDef}
\end{eqnarray}
$p^*$ is the magnitude of the three-momentum of the $V$ system, measured in 
the $D$ rest frame, and $\theta_\ell$ is the angle of the lepton momentum, in the W rest frame, with respect to 
the opposite direction of the $D$ meson (see Figure \ref{DecayAngles} for the electron case ($\theta_e$)).
The left-handed nature of the quark current manifests itself as
$|H_-|>|H_+|$. The differential decay rate for $D\ra V\ell\nu$ 
followed by the vector meson decaying into two pseudoscalars is

\begin{eqnarray}
\frac{d\Gamma(D\ra V \overline \ell\nu, V\ra P_1P_2)}{dq^2 d\cos\theta_V d\cos\theta_\ell d\chi} 
 &  = & \frac{3G_F^2}{2048\pi^4}
       |V_{cq}|^2 \frac{p^*(q^2)q^2}{m_D^2} {\cal B}(V\to P_1P_2)\ \times \nonumber \\ 
 & & \hskip0.10in \Big\{ (1 + \cos\theta_\ell)^2 \sin^2\theta_V |H_+(q^2)|^2 \nonumber \\
 & & \hskip0.20in +\ (1 - \cos\theta_\ell)^2 \sin^2\theta_V |H_-(q^2)|^2 \nonumber \\
 & & \hskip0.30in +\ 4\sin^2\theta_\ell\cos^2\theta_V|H_0(q^2)|^2 \nonumber \\
 & & \hskip0.40in -\ 4\sin\theta_\ell (1 + \cos\theta_\ell) 
             \sin\theta_V \cos\theta_V \cos\chi H_+(q^2) H_0(q^2) \nonumber \\
 & & \hskip0.50in +\ 4\sin\theta_\ell (1 - \cos\theta_\ell) 
          \sin\theta_V \cos\theta_V \cos\chi H_-(q^2) H_0(q^2) \nonumber \\
 & & \hskip0.60in -\ 2\sin^2\theta_\ell \sin^2\theta_V 
                \cos 2\chi H_+(q^2) H_-(q^2) \Big\}\,,
\label{eq:dGammaVector}
\end{eqnarray}
where the helicity angles $\theta^{}_\ell$, $\theta^{}_V$, and acoplanarity angle $\chi$ are defined
in Fig.~\ref{DecayAngles}. 

%Assuming that the simple pole form of Eq.~(\ref{SimplePole}) describes 
%the $q^2$-dependence of the form factors, the %four-dimensional 
%distribution of Eq.~(\ref{eq:dGammaVector}) will depend only on the parameters
Ratios between the values of the hadronic form factors expressed at $q^2=0$ are 
usually introduced:
\begin{eqnarray}
r_V \equiv V(0) / A_1(0), & &  r_2 \equiv A_2(0) / A_1(0) \label{rVr2_eq}\,.
\end{eqnarray}

\subsubsection{Form factor measurements}

In 2002 FOCUS reported~\cite{Link:2002ev} an asymmetry in 
the observed $\cos(\theta_V)$ distribution in $D^+\to K^-\pi^+\mu^+\nu$ decays. This is interpreted as
evidence for an $S$-wave $K^-\pi^+$ component in the decay amplitude. 
Since $H_0$ typically dominates over $H_{\pm}$, the distribution given 
by Eq.~(\ref{eq:dGammaVector}) is, after integration over $\chi$,
roughly proportional to $\cos^2\theta_V$. 
Inclusion of a constant $S$-wave amplitude of the form $A\,e^{i\delta}$ 
leads to an interference term proportional to 
$|A H_0 \sin\theta_\ell \cos\theta_V|$; this term causes an asymmetry 
in $\cos(\theta_V)$.
When FOCUS fit their data including this $S$-wave amplitude, 
they obtained $A = 0.330 \pm 0.022 \pm 0.015~\gev^{-1}$ and 
$\delta = 0.68 \pm 0.07 \pm 0.05$~\cite{Link:2002wg}. 
Both \babar~\cite{Aubert:2008rs} and CLEO-c~\cite{Ecklund:2009fia} 
have also found evidence for an $f^{}_0 \to K^+ K^-$ component in semileptonic $D^{}_s$ decays.

The CLEO-c collaboration extracted the form factors $H_+(q^2)$, $H_-(q^2)$, 
and $H_0(q^2)$ from 11000 $D^+ \rightarrow K^- \pi^+ \ell^+ \nu_\ell$ events 
in a model-independent fashion directly as functions of $q^2$~\cite{Briere:2010zc}. 
They also determined the $S$-wave form factor $h_0(q^2)$ via the interference term, despite the
fact that the $K\pi$ mass distribution appears dominated by the vector
$K^*(892)$ state. 
%Their results are shown in Figs.~\ref{fig:cleoc_h0} and \ref{fig:cleoc_H}.  
%Plots in Fig.~\ref{fig:cleoc_H} clearly show that
It is observed that $H_0(q^2)$ dominates over a wide range of $q^2$, especially at 
low $q^2$. The transverse form factor $H_t(q^2)$, which can be related 
to $A_3(q^2)$, is small compared to lattice gauge theory calculations 
and suggests that the form factor ratio $r_3 \equiv A_3(0) / A_1(0)$ is large and negative.

\babar~\cite{delAmoSanchez:2010fd} selected a large sample of 
$244\times 10^3$ $D^+ \rightarrow K^- \pi^+ e^+ \nu_e$ candidates with a ratio $S/B\sim 2.3$ from an analyzed 
integrated luminosity of $347~\fb^{-1}$. With four particles emitted in the 
final state, the differential decay rate depends on five variables.
In addition to the four variables defined in previous sections there is also
$m^2$, the mass squared of the $K\pi$ system.
%Apart from this last variable, the reconstruction algorithm does not provide 
%a high resolution on the other measured quantities 
%(see the similar measurement of the $D^0 \rightarrow K^- e^+ \nu_e$ decay channel)
%and a multi-dimensional unfolding procedure
%is not used to correct for efficiency and resolution effects. However, these
%limitations still allow an essentially model independent measurement of
%the differential decay rate. This is because, apart from the $q^2$
%and mass dependence of the form factors, angular distributions are fixed by
%kinematics. In addition, present accurate measurements of 
%$D \rightarrow P \overline{\ell}\nu_{\ell}$ decays have shown that the 
%$q^2$ dependence of the form factors can be well described by several models
%as long as the corresponding model parameter(s) are fitted from data.
%This is even more true in $D \rightarrow V \overline{\ell}\nu_{\ell}$ decays
%because the $q^2$ range is reduced. 
To analyze the $D^+ \rightarrow K^- \pi^+ e^+ \nu_e$ decay channel it is assumed
that all form factors have a $q^2$ variation given by the simple pole model and the effective pole mass value,
$m_A=(2.63 \pm 0.10 \pm 0.13)~\gevcc$,
is fitted for the axial vector form factors. This value is compatible
with expectations when comparing with the mass of $J^P=1^+$ charm mesons. 
%Data are not sensitive to the effective mass
%of the vector form factor for which $m_V=(2.1 \pm 0.1)~\gevcc$ is used,
%nor to the effective pole mass of the scalar component for which $m_A$ is used.
 For the mass dependence of the form factors, a Breit-Wigner with a mass dependent width and a Blatt-Weisskopf damping factor is used. For the S-wave amplitude, 
considering what was measured in $D^+ \rightarrow K^- \pi^+\pi^+$ decays,
a polynomial variation below the $\overline{K}^*_0(1430)$ and a Breit-Wigner 
distribution, above are assumed. For the polynomial part, a linear term is sufficient to fit data.
It is verified that the variation of the S-wave phase is compatible 
with expectations from elastic $K\pi$ ~\cite{Estabrooks:1977xe,Aston:1987ir} (after correcting for $\delta^{3/2}$) according to the Watson theorem. At variance with elastic scattering, a negative relative sign between the 
S- and P-waves is measured; this is compatible with the previous theorem. 
Contributions from other spin-1 and spin-2 resonances decaying into $K^-\pi^+$ are considered.

\begin{figure}[htbp!]
 \begin{center}
\includegraphics[width=.60\textwidth]{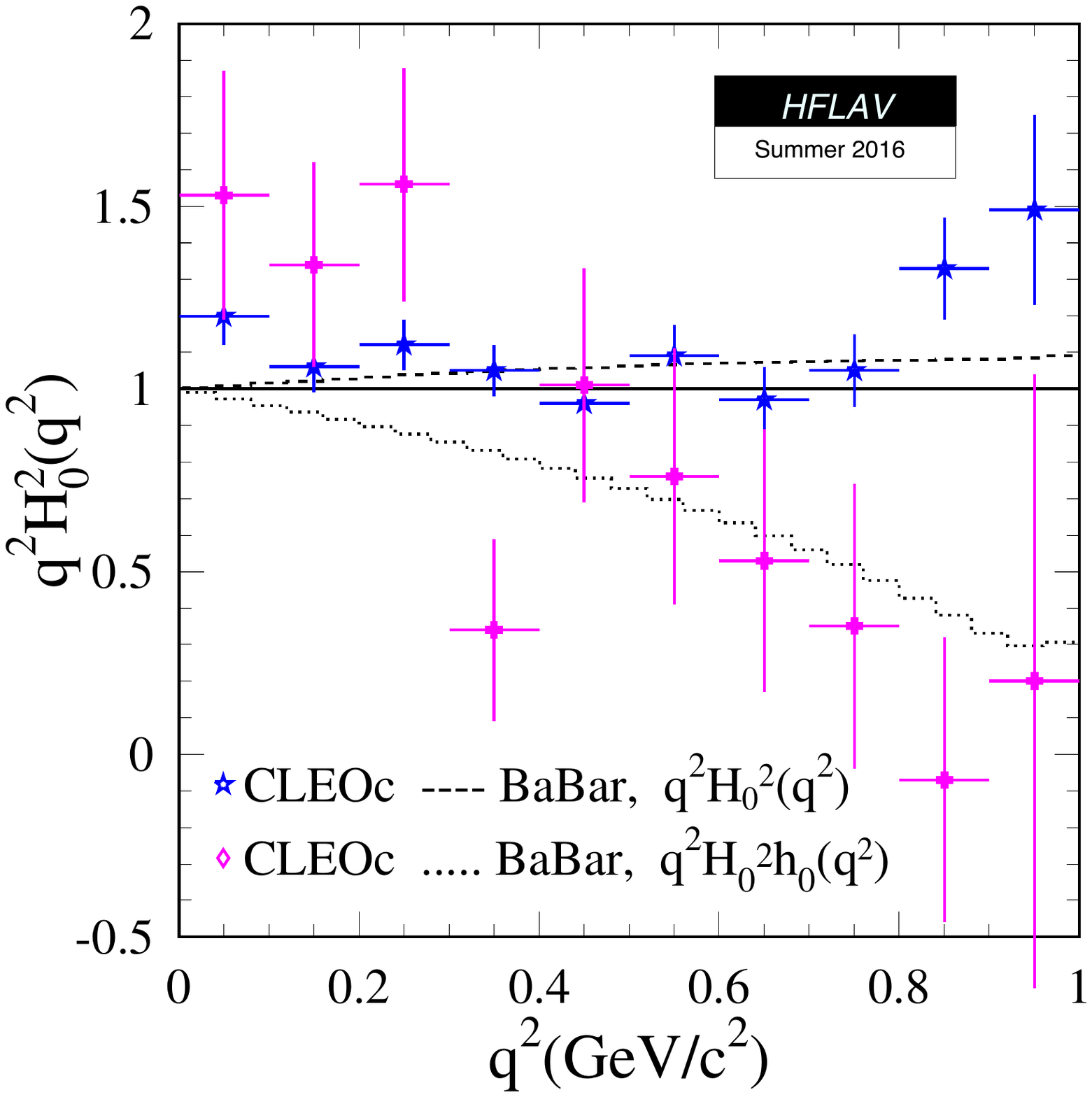}
 \end{center}
\vskip-4.4cm
\caption[]{{Comparison between CLEO-c and \babar results for 
the quantities $q^2 H_0^2(q^2)$ and $q^2 h_0(q^2) H_0(q^2)$.}
   \label{fig:h0FF}}
\end{figure}

In Fig. \ref{fig:h0FF}, measured values from CLEO-c
of the products $q^2H_0^2(q^2)$ and $q^2h_0(q^2)H_0(q^2)$ are compared with 
corresponding results from \babar illustrating the difference in behavior
of the scalar $h_0$ component and the $H_0$ form factor.
For this comparison, the plotted values from \babar for the two distributions
are fixed to 1 at $q^2=0$. The different behavior of $h_0(q^2)$
and $H_0(q^2)$ can be explained by their different dependence in the 
$p^*$ variable.

\begin{figure}[htbp]
  \begin{center}
    \includegraphics[width=1.2\textwidth,angle=90]{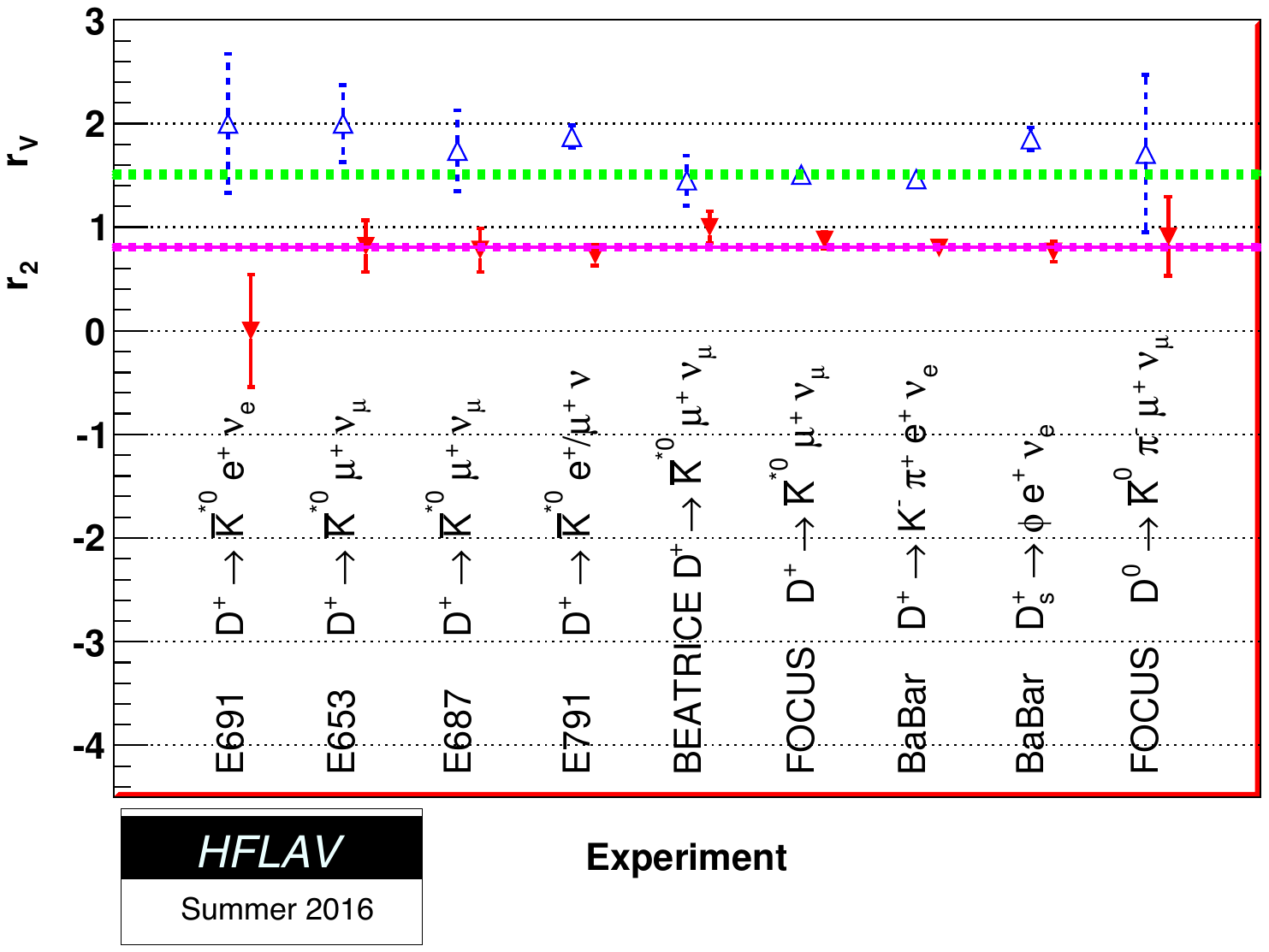}
  \end{center}
\vskip-9.2cm
  \caption{A comparison of $r_2$ (filled inverted triangles) 
    and $r_V$ (open triangles) values 
    from various experiments. The first seven measurements are for 
    $D^+\to K^-\pi^+ l^+\nu_l$ decays. Also shown as a line with
    1-$\sigma$ limits is the average of these. The last two points are
    $D_s^+$ and $D^0$ decays.  
  \label{fig:r2rv}}
\end{figure}

Table \ref{Table1} lists measurements of $r_V$ and $r_2$ from several
experiments. Most of the measurements assume that the $q^2$ dependence of 
hadronic form factors is given by the simple pole ansatz. Some of these 
measurements do not consider a S-wave contribution and it is 
included in the measured values. 
The measurements are plotted in Fig.~\ref{fig:r2rv}, which shows that 
they are all consistent.

\begin{table}[htbp]
\caption{Results for $r_V$ and $r_2$ from various experiments. 
\label{Table1}}
\begin{center}
\begin{tabular}{cccc}
\hline
\vspace*{-10pt} & \\
Experiment & Ref. & $r_V$ & $r_2$ \\
\vspace*{-10pt} & \\
\hline
\vspace*{-10pt} & \\
$D^+\to \overline{K}^{*0}l^+\nu$ & \omit & \omit & \omit         \\
E691         & \cite{Anjos:1990pn}     & 2.0$\pm$  0.6$\pm$  0.3  & 0.0$\pm$  0.5$\pm$  0.2    \\
E653         & \cite{Kodama:1992tn}     & 2.00$\pm$ 0.33$\pm$ 0.16 & 0.82$\pm$ 0.22$\pm$ 0.11   \\
E687         & \cite{Frabetti:1993jq}     & 1.74$\pm$ 0.27$\pm$ 0.28 & 0.78$\pm$ 0.18$\pm$ 0.11   \\
E791 (e)     & \cite{Aitala:1997cm}    & 1.90$\pm$ 0.11$\pm$ 0.09 & 0.71$\pm$ 0.08$\pm$ 0.09   \\
E791 ($\mu$) & \cite{Aitala:1998ey}    & 1.84$\pm$0.11$\pm$0.09   & 0.75$\pm$0.08$\pm$0.09     \\
Beatrice     & \cite{Adamovich:1998ia} & 1.45$\pm$ 0.23$\pm$ 0.07 & 1.00$\pm$ 0.15$\pm$ 0.03   \\
FOCUS        & \cite{Link:2002wg}   & 1.504$\pm$0.057$\pm$0.039& 0.875$\pm$0.049$\pm$0.064  \\
\hline
$D^0\to \overline{K}^0\pi^-\mu^+\nu$ & \omit & \omit & \omit         \\
FOCUS        & \cite{Link:2004uk}    & 1.706$\pm$0.677$\pm$0.342& 0.912$\pm$0.370$\pm$0.104 \\
\babar        & \cite{delAmoSanchez:2010fd} & $1.493 \pm 0.014 \pm 0.021$ & $0.775 \pm 0.011 \pm 0.011$ \\
\hline
%Average      & \omit              & x.xx $\pm$ x.xx          & x.xx $\pm$ x.xx             \\
%\hline
$D_s^+ \to \phi\,e^+ \nu$ &\omit  &\omit     & \omit                  \\
\babar        & \cite{Aubert:2008rs}    & 1.849$\pm$0.060$\pm$0.095& 0.763$\pm$0.071$\pm$0.065\\
\hline
$D^0, D^+\to \rho\,e \nu$ & \omit  & \omit    & \omit                 \\
CLEO         & \cite{Mahlke:2007uf}    & 1.40$\pm$0.25$\pm$0.03   & 0.57$\pm$0.18$\pm$0.06    \\
%& \babar    && 0.711$\pm$0.111$\pm$0.096&& 1.633$\pm$0.081$\pm$0.068 &&  2.53$_{-0.35}^{+0.54}$$\pm$0.54 && fixed       &\cr
%& $D_s^+ \to \phi e^+ \nu$ &&\omit && \omit                 && \omit && \omit        \dl
\hline
\end{tabular}
\end{center}
\end{table}

%%%%%%%%%%%%%%%%%%%%%%%%%%%%%%%%%%%%%%%%%%%%%%%%%%%%%%

\clearpage
% Leptonic decays
\subsection{Leptonic decays}

Purely leptonic decays of $\Dp$ and $\dsp$ mesons are among the simplest and 
theoretically cleanest
probes of $c\to d$ and $c\to s$ quark flavor-changing transitions. The 
branching fraction of leptonic 
decays that proceed via the annihilation of the initial quark-antiquark 
pair ($c\overline{d}$ or 
$c\overline{s}$) into a virtual $W^+$ that finally materializes as an 
antilepton-neutrino pair ($\ellnu$). 
Their Standard Model branching fraction is given by
\begin{equation}
  \br(D_{q}^+\to \ell^+\nu_{\ell})=
  \frac{G_F^2}{8\pi}\tau_{D_q}f_{D_{q}}^2|V_{cq}|^2m_{D_{q}}m_{\ell}^2
  \left(1-\frac{m_{\ell}^2}{m_{D_{q}}^2} \right)^2\,.
  \label{eq:brCharmLeptonicSM}
\end{equation}
Here, $m_{D_{q}}$ is the $D_{q}$ meson mass, $\tau_{D_q}$ is its lifetime, 
$m_{\ell}$ is the charged lepton mass, 
$|V_{cq}|$ is the magnitude of the relevant CKM matrix element, and 
$G_F$ is the Fermi coupling constant. The parameter 
$f_{D_{q}}$ is the $D_q$ meson decay constant and is related to the 
wave-function overlap of the meson's 
constituent quark and anti-quark. The decay constants 
have been predicted using several 
methods, the most accurate and robust being the lattice gauge theory (LQCD) 
calculations. The Flavor Lattice Averaging 
Group~\cite{Aoki:2016frl} combines all LQCD calculations and provides 
averaged values for $f_D$ and $f_{D_s}$ (see 
Table~\ref{tab:Lattice}) that are used within this section to extract 
the magnitudes of the $V_{cd}$ and $V_{cs}$ CKM
matrix elements from experimentally measured branching fractions of 
$D^+\to \ell^+\nu_{\ell}$ and 
$D_s^+\to \ell^+\nu_{\ell}$ decays, respectively.
\begin{table}[b!]
\caption{The LQCD average for $D$ and $D_s$ meson decay constants and 
their ratio from the Flavor Lattice Averaging 
Group~\cite{Aoki:2016frl}.
\label{tab:Lattice}}
\begin{center}
\begin{tabular}{ll}
\toprule
\rowcolor{Gray} Quantity & Value \\ 
\midrule
$f_D$ 		& $212.15\pm1.45$~MeV\\
$f_{D_s}$ 	& $248.83\pm1.27$~MeV\\
$f_{D_s}/f_D$	& $1.1716\pm0.0032$
\\ \bottomrule
\end{tabular}
\end{center}
\end{table}

The leptonic decays of pseudoscalar mesons 
are helicity-suppressed and their decay rates are thus 
proportional to the square of 
the charged lepton mass. Leptonic decays into electrons, with 
expected $\br\lesssim 10^{-7}$, are not experimentally 
observable yet, whereas decays to taus are favored over decays 
to muons. In particular, the ratio of the 
latter decays is equal to 
$R^{D_q}_{\tau/\mu}\equiv \br(D^+_q\to\tau^+\nu_{\tau})/\br(D^+_q\to\mu^+\nu_{\mu})=m_{\tau}^2/m_{\mu}^2\cdot(1-m^2_{\tau}/m^2_{D_q})^2/(1-m^2_{\mu}/m^2_{D_q})^2$,
and amounts to $9.76\pm0.03$ 
in the case of $D_s^+$ decays and to $2.67\pm0.01$ in the case 
of $D^+$ decays based on the world average values of masses of the 
muon, tau and $D_q$ meson given in Ref.~\cite{PDG_2016}. 
Any deviation from this expectation could only be interpreted as 
violation of lepton universality in charged 
currents and would hence point to NP effects~\cite{Filipuzzi:2012mg}.

Averages presented within this subsection are weighted averages, in which 
correlations between measurements and dependencies on input parameters
are taken into account. There is only one new experimental result on 
leptonic charm decays since our last report from 2014 -- the  
measurements of 
$\br(D_s^+\to\munu)$ and $\br(D_s^+\to\taunu)$ by BESIII collaboration~\cite{Ablikim:2016duz}.
The Lattice QCD calculations
of the $D$ and $D_s$ meson decay constants have improved significantly 
since our last report and we use the latest averages of $N_f=2+1+1$ 
calculations provided by the
Flavour Lattice Averaging Group~\cite{Aoki:2016frl} in our 
determinations of the CKM matrix elements $|V_{cd}|$ and $|V_{cs}|$.

\subsubsection{$D^+\to \ell^+\nu_{\ell}$ decays and $|V_{cd}|$}

\begin{figure}[hbt!]
\centering
\includegraphics[width=0.6\textwidth]{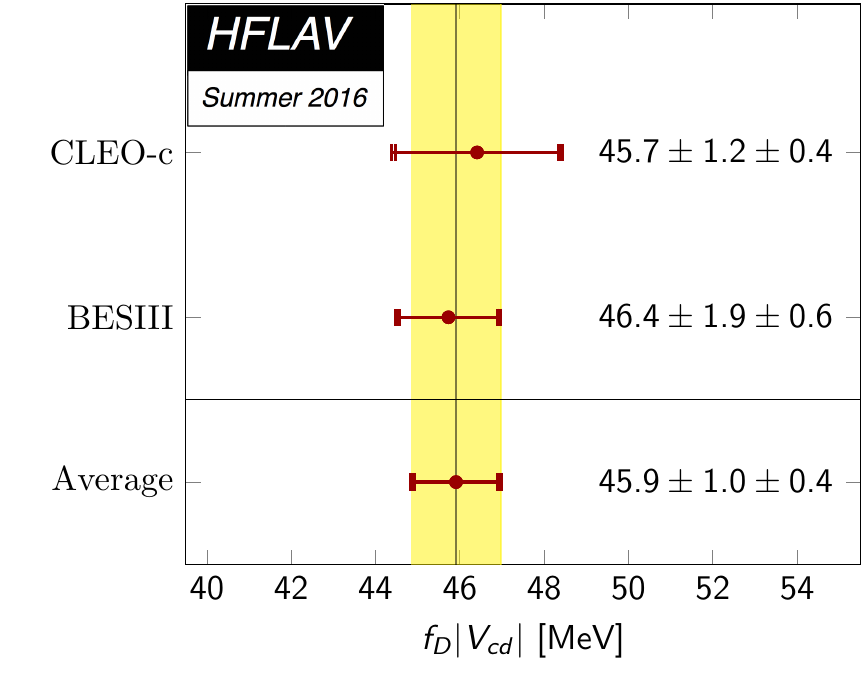}
\caption{
WA value for $f_{D}|V_{cd}|$. For each point, the first error listed is the 
statistical and the second error is the systematic error.
\label{fig:ExpDLeptonic}
}
\end{figure}

We use measurements of the branching fraction $\br(D^+\to\munu)$ 
from \mbox{CLEO-c}~\cite{Eisenstein:2008aa} and 
BESIII~\cite{Ablikim:2013uvu} to calculate the world average (WA) 
value. We obtain
\begin{equation}
 \br^{\rm WA}(D^+\to\munu) = (3.74\pm0.17)\times10^{-4},
 \label{eq:Br:WA:DtoMuNu}
\end{equation}
from which we determine the product of the decay constant and the 
CKM matrix element to be
\begin{equation}
 f_{D}|V_{cd}| = \left(45.9\pm1.1\right)~\mbox{MeV},
 \label{eq:expFDVCD}
\end{equation}
where the uncertainty includes the uncertainty on $\br^{\rm WA}(D^+\to\munu)$ 
and external inputs\footnote{These values (taken from the 
PDG 2014 edition~\cite{PDG_2014}) are 
$m_{\mu} = (0.1056583715\pm0.0000000035)~\gevcc$, $m_D = (1.86961\pm0.00009)~\gevcc$ 
and $\tau_D = (1040\pm7)\times 10^{-15}$~s.} 
needed to extract $f_{D}|V_{cd}|$ from the measured branching 
fraction using Eq.~(\ref{eq:brCharmLeptonicSM}). 
Using the LQCD value for $f_D$ from Table~\ref{tab:Lattice} we 
finally obtain the CKM matrix element $V_{cd}$ to be
\begin{equation}
 |V_{cd}| = 0.2164\pm0.0050(\rm exp.)\pm0.0015(\rm LQCD),
 \label{eq:Vcd:WA:Leptonic}
\end{equation}
where the uncertainties are from the experiments and lattice calculations, 
respectively. All input values
and the resulting world averages are summarized in Table~\ref{tab:DExpLeptonic} 
and plotted in Fig.~\ref{fig:ExpDLeptonic}.
\begin{table}[t!]
\caption{Experimental results and world averages for 
${\cal{B}}(D^+\to \ell^+\nu_{\ell})$ and $f_{D}|V_{cd}|$.
The first uncertainty is statistical and the second is experimental 
systematic. The third uncertainty in the case of $f_{D^+}|V_{cd}|$ is 
due to external inputs (dominated by the uncertainty of $\tau_D$).
\label{tab:DExpLeptonic}}
\begin{center}
\begin{tabular}{lccll}
\toprule
\rowcolor{Gray} Mode 	& ${\cal{B}}$ ($10^{-4}$)	& $f_{D}|V_{cd}|$ (MeV)		& Reference & \\ 
\midrule
% this is from floated taunu fit
%$\munu$		& $3.93\pm0.35\pm 0.09$ 	& $???.?\pm?.?\pm?.?$	& CLEO-c\hfill \cite{Eisenstein:2008aa}\\
% this is from fixed taunu contribution fit (same for BESIII)
\multirow{2}{*}{$\munu$} & $3.82\pm0.32\pm 0.09$ 	& $46.4\pm1.9\pm0.5\pm0.2$	& CLEO-c & \cite{Eisenstein:2008aa}\\ 
			& $3.71\pm0.19\pm 0.06$ 	& $45.7\pm1.2\pm0.4\pm0.2$	& BESIII & \cite{Ablikim:2013uvu}\\
\midrule
\rowcolor{Gray}$\munu$ 	& $3.74\pm0.16\pm 0.05$		& $45.9\pm1.0\pm0.3\pm0.2$	& Average & \\
\midrule
$\enu$	 		& {$<0.088$ at 90\% C.L.}	&& CLEO-c & \cite{Eisenstein:2008aa}\\
\midrule
$\taunu$ 		& {$<12$ at 90\% C.L.}		&& CLEO-c & \cite{Eisenstein:2008aa}
\\ \bottomrule
\end{tabular}
\end{center}
\end{table}
 
The upper limit on the ratio of branching fractions is found to be 
$R_{\tau/\mu}^D<3.2$ at 90\%~C.L., which is just slightly above the 
SM expected value.

\subsubsection{$D_s^+\to \ell^+\nu_{\ell}$ decays and $|V_{cs}|$}

We use measurements of the absolute branching fraction $\br(D_s^+\to\munu)$ 
from CLEO-c~\cite{Alexander:2009ux}, \babar~\cite{delAmoSanchez:2010jg},
Belle~\cite{Zupanc:2013byn}, and BESIII~\cite{Ablikim:2016duz}, and obtain 
a WA value of
\begin{equation}
 \br^{\rm WA}(\dsmunu) = (5.54\pm0.23)\times10^{-3}.
 \label{eq:Br:WA:DstoMuNu}
\end{equation}
The WA value for $\br(D_s^+\to\taunu)$ is also calculated from CLEO-c, 
\babar, Belle, and BESIII measurements. 
CLEO-c made separate measurements for 
$\tau^+\to e^+\nu_e\overline{\nu}{}_{\tau}$~\cite{Naik:2009tk},
$\tau^+\to\pi^+\overline{\nu}{}_{\tau}$~\cite{Alexander:2009ux}, and
$\tau^+\to\rho^+\overline{\nu}{}_{\tau}$~\cite{Onyisi:2009th},
\babar made separate measurements for 
$\tau^+\to e^+\nu_e\overline{\nu}{}_{\tau}$~\cite{delAmoSanchez:2010jg} 
and $\tau^+\to \mu^+\nu_{\mu}\overline{\nu}{}_{\tau}$, 
Belle made separate measurements for 
$\tau^+\to e^+\nu_e\overline{\nu}{}_{\tau}$, $\tau^+\to \mu^+\nu_{\mu}\overline{\nu}{}_{\tau}$, 
and $\tau^+\to\pi^+\overline{\nu}{}_{\tau}$~\cite{Zupanc:2013byn}, and 
BESIII made measurements using 
$\tau^+\to\pi^+\overline{\nu}{}_{\tau}$~\cite{Ablikim:2016duz} decays.
Combining all of them we obtain the WA value of
\begin{equation}
 \br^{\rm WA}(\dsp\to\taunu) = (5.51\pm0.24)\times10^{-2}.
 \label{eq:Br:WA:DstoTauNu}
\end{equation}

The ratio of branching fractions is found to be
\begin{equation}
R_{\tau/\mu}^{\ds} = 9.95\pm0.57,
\label{eq:R:WA:Leptonic}
\end{equation}
and is consistent with the value expected in the SM.

\begin{figure}[hbt!]
\centering
\includegraphics[width=0.8\textwidth]{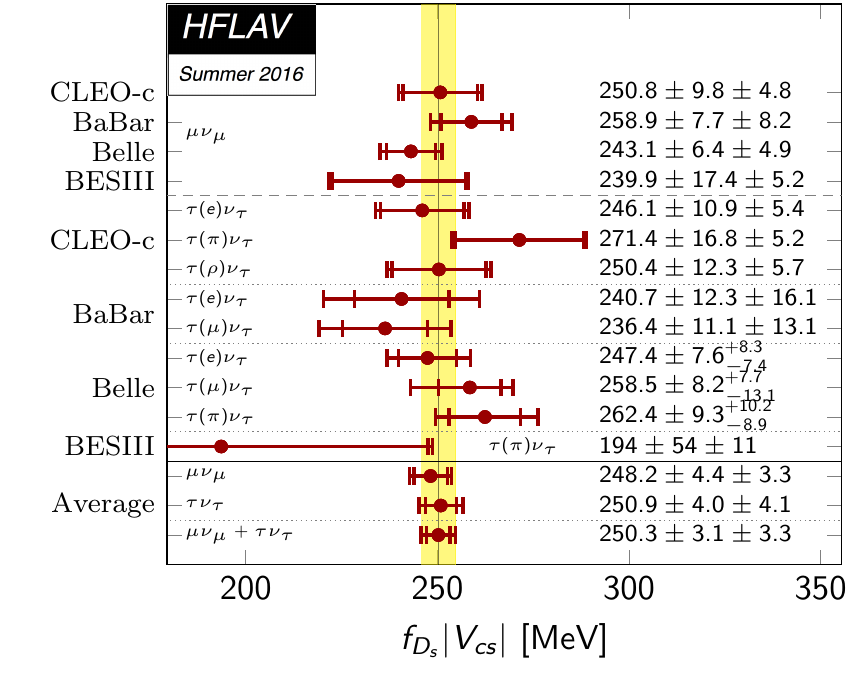}
\caption{
WA value for $f_{D_s}|V_{cs}|$. For each point, the first error listed is 
the statistical and the second error is the systematic error.
\label{fig:ExpDsLeptonic}
}
\end{figure}

From the average values of branching fractions of muonic and tauonic 
decays we determine\footnote{
We use the following values (taken from PDG 2014 edition~\cite{PDG_2014}) 
for external parameters entering Eq.~(\ref{eq:brCharmLeptonicSM}): 
$m_{\tau} = (1.77686\pm0.00012)~\gevcc$, 
$m_{D_s} = (1.96830\pm0.00010)~\gevcc$ 
and $\tau_{D_s} = (500\pm7)\times 10^{-15}$~s.} 
the product of $D_s$ meson decay constant 
and the $|V_{cs}|$ CKM matrix element to be
\begin{equation}
 \fds|V_{cs}|=\left(250.3\pm4.5\right)~\mbox{MeV},
 \label{eq:expFDSVCS}
\end{equation}
where the uncertainty is due to the uncertainties on $\br^{\rm WA}(D_s^+\to\munu)$ 
and $\br^{\rm WA}(D_s^+\to\taunu)$ and the external inputs. All input values and 
the resulting world averages are 
summarized in Table~\ref{tab:DsLeptonic} and plotted in Fig.~\ref{fig:ExpDsLeptonic}. 
To obtain the 
averages given within this subsection and in Table~\ref{tab:DsLeptonic} we have 
taken into account
the correlations within each experiment\footnote{In the case of \babar we use 
the covariance matrix from 
the errata of~Ref.\cite{delAmoSanchez:2010jg}.} for the uncertainties related 
to: normalization, tracking, particle identification, 
signal and background parameterizations, and peaking background contributions.
% \end{itemize}

Using the LQCD value for $\fds$ from Table~\ref{tab:Lattice}, we 
finally obtain the magnitude of the CKM matrix element $V_{cs}$ to be
\begin{equation}
 |V_{cs}| = 1.006\pm0.018(\rm exp.)\pm0.005(\rm LQCD),
 \label{eq:Vcs:WA:Leptonic}
\end{equation}
where the uncertainties are from the experiments and lattice calculations, respectively.

\begin{table}[t!]
\caption{Experimental results and world averages for ${\cal{B}}(\dsellnu)$ and 
$f_{D_s}|V_{cs}|$.
The first uncertainty is statistical and the second is experimental systematic. 
The third uncertainty 
in the case of $f_{D_s}|V_{cs}|$ is due to external inputs (dominated by the 
uncertainty of $\tau_{D_s}$).
We have recalculated $\br(\dsp\to\taunu)$ quoted by CLEO-c and \babar using the 
latest values for branching fractions of $\tau$ decays to electron, muon, or pion 
and neutrinos~\cite{PDG_2016}.
CLEO-c and \babar include statistical uncertainty of number of $\ds$ tags 
(denominator in the calculation of 
branching fraction) in the statistical uncertainty of measured $\br$. 
We subtract this uncertainty from the
statistical one and add it to the systematic uncertainty. 
\label{tab:DsLeptonic}}
\begin{center}
\begin{tabular}{lccll}
\toprule
\rowcolor{Gray}
Mode 		& ${\cal{B}}$ ($10^{-2}$) 	& $f_{D_s}|V_{cs}|$ (MeV) 		& Reference & 
\\ \midrule
\multirow{4}{*}{$\munu$}	& $0.565\pm0.044\pm 0.020$ 	& $250.8 \pm 9.8 \pm 4.4 \pm 1.8$	& CLEO-c &\cite{Alexander:2009ux}\\		
				& $0.602\pm0.037\pm 0.032$ 	& $258.9 \pm 8.0 \pm 6.9 \pm 1.8$	& \babar  &\cite{delAmoSanchez:2010jg}\\
				& $0.531\pm0.028\pm 0.020$ 	& $243.1 \pm 6.4 \pm 4.6 \pm 1.7$ 	& Belle  &\cite{Zupanc:2013byn}\\
                                & $0.517\pm0.075\pm 0.021$      & $239.9 \pm 17.4 \pm 4.9 \pm 1.7$      & BESIII &\cite{Ablikim:2016duz}\\
\midrule
\rowcolor{Gray}
$\munu$ 			& $0.554\pm0.020\pm0.013$ 	& $248.2 \pm 4.4 \pm 2.8 \pm 1.7$ 	& Average & \\
\midrule
$\tauenu$ 			& $5.31\pm0.47\pm0.22$ 		& $246.1 \pm 10.9 \pm 5.1 \pm 1.7$ 	& CLEO-c &\cite{Onyisi:2009th}\\
$\taupinuCharm$ 			& $6.46\pm0.80\pm0.23$ 		& $271.4 \pm 16.8 \pm 4.8 \pm 1.9$  & CLEO-c &\cite{Alexander:2009ux}\\
$\taurhonu$ 			& $5.50\pm0.54\pm0.24$ 		& $250.4 \pm 12.3 \pm 5.5 \pm 1.8$  & CLEO-c &\cite{Naik:2009tk}\\
\midrule
\rowcolor{LightGray}
$\taunu$			& $5.57\pm0.32\pm0.15$		& $252.0 \pm 7.2 \pm 3.4 \pm 1.8$   & CLEO-c & \\
\midrule
$\tauenu$ 			& $5.08\pm0.52\pm0.68$ 		& $240.7 \pm 12.3 \pm 16.1 \pm 1.7$	& \multirow{2}{*}{\babar} & \multirow{2}{*}{\cite{delAmoSanchez:2010jg}}\\
$\taumunu$ 			& $4.90\pm0.46\pm0.54$ 		& $236.4 \pm 11.1 \pm 13.0 \pm 1.7$	&  & \\
\midrule
\rowcolor{LightGray}
$\taunu$			& $4.95\pm0.36\pm0.58$		& $237.6 \pm 8.6 \pm 13.8 \pm 1.7$   & \babar &\\
\midrule
$\tauenu$  			& $5.37\pm0.33^{+0.35}_{-0.31}$ & $247.4 \pm 7.6^{+8.1}_{-7.1} \pm 1.7$  & \multirow{3}{*}{Belle} & \multirow{3}{*}{\cite{Zupanc:2013byn}} \\
$\taumunu$ 		 	& $5.86\pm0.37^{+0.34}_{-0.59}$ & $258.5 \pm 8.2^{+7.5}_{-13.0} \pm 1.8$  & &\\ 
$\taupinuCharm$  			& $6.04\pm0.43^{+0.46}_{-0.40}$ & $262.4 \pm 9.3^{+10.0}_{-8.7} \pm 1.8$  & &\\
\midrule
\rowcolor{LightGray}
$\taunu$			& $5.70\pm0.21\pm0.31$		& $254.9 \pm 4.7 \pm 6.9 \pm 1.8$   & Belle & \\
\midrule
$\taupinuCharm$ 	        & $3.28\pm1.83\pm0.37$ 		& $194 \pm 54 \pm 11 \pm 1$  & BESIII &\cite{Ablikim:2016duz}\\
\midrule
\rowcolor{Gray}
$\taunu$ 			& $5.51\pm0.18\pm0.16$ 		& $250.9 \pm 4.0 \pm 3.7 \pm 1.8$     & Average & \\
\midrule
\rowcolor{Gray}
$\munu$  			&		 		& \multirow{2}{*}{$250.3\pm 3.1\pm 2.7\pm1.8$}	& \multirow{2}{*}{Average} & \\
\rowcolor{Gray}
$\taunu$ 			&		 		& \multirow{-2}{*}{$250.3\pm 3.1\pm 2.7\pm1.8$}	& \multirow{-2}{*}{Average} & \\
\midrule
$\enu$				& $<0.0083$ at 90\% C.L.	& 					& Belle & \cite{Zupanc:2013byn} \\
\bottomrule
\end{tabular}
\end{center}
\end{table}

\subsubsection{Comparison with other determinations of $|V_{cd}|$ and $|V_{cs}|$}

\begin{figure}[hbt!]
\centering
\includegraphics[width=0.45\textwidth]{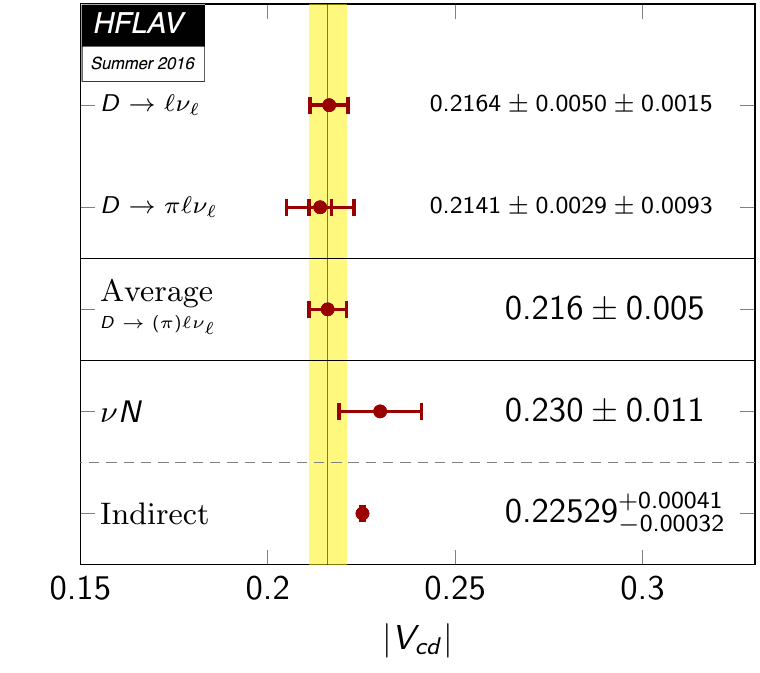}~
\includegraphics[width=0.45\textwidth]{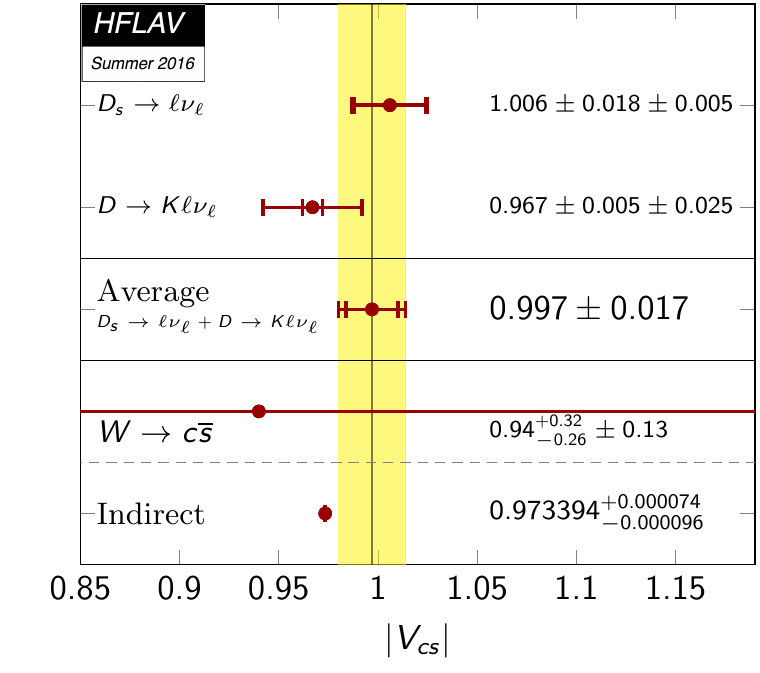}
\caption{
Comparison of magnitudes of the CKM matrix elements $|V_{cd}|$ (left) and 
$|V_{cs}|$ (right) determined from the (semi-)leptonic charm decays and from 
neutrino scattering data or $W$ decays and determination from the global fit 
assuming CKM unitarity~\cite{Charles:2004jd}.
\label{fig:VcdVcsComparions}
}
\end{figure}

Table~\ref{tab:CKMVcdVcs} summarizes and Fig.~\ref{fig:VcdVcsComparions} shows 
all determinations of the CKM matrix elements $|V_{cd}|$ and $|V_{cs}|$. As
can be seen, the most precise direct determinations of these CKM matrix elements 
are those from leptonic and semileptonic $D_{(s)}$ decays. The values are in agreement
within uncertainties with the values obtained from the global fit assuming CKM 
matrix unitarity.

\begin{table}[htb]
\centering
\caption{Average of the magnitudes of the CKM matrix elements $|V_{cd}|$ and 
$|V_{cs}|$ determined from the leptonic and semileptonic $D$ and $D_s$ decays.
In the calculation of average values we assume 100\% correlations in uncertainties 
due to LQCD.  The values determined from neutrino scattering or $W$ decays and 
determination from the global fit to the CKM matrix are given for comparison as well.
\label{tab:CKMVcdVcs}}
\begin{tabular}{lcc}
\toprule
\rowcolor{Gray} Method & Reference & Value \\ 
\midrule
&&{$|V_{cd}|$}\\
\cline{3-3}
$D\to\ell\nu_{\ell}$ 	 & This section	                 		& $0.2164\pm0.0050(\rm exp.)\pm0.0015(\rm LQCD)$\\
$D\to\pi\ell\nu_{\ell}$  & Section~\ref{sec:charm:semileptonic}		& $0.2141\pm0.0029(\rm exp.)\pm0.0093(\rm LQCD)$\\
\midrule
\rowcolor{Gray} $D\to\ell\nu_{\ell}$ 	& \multirow{2}{*}{Average}	& \multirow{2}{*}{$0.216\pm0.005$}\\
\rowcolor{Gray} $D\to\pi\ell\nu_{\ell}$ & \multirow{-2}{*}{Average}	& \multirow{-2}{*}{$0.216\pm0.005$}\\
\midrule
$\nu N$			& PDG~\cite{PDG_2016}	& $0.230\pm0.011$\\
Global CKM Fit		& CKMFitter~\cite{Charles:2004jd}		& $0.22529^{+0.00041}_{-0.00032}$\\
\midrule
\midrule
&&{$|V_{cs}|$}\\
\cline{3-3}
$D_s\to\ell\nu_{\ell}$ 	 & This section            			& $1.006\pm0.018(\rm exp.)\pm0.005(\rm LQCD)$\\
$D\to K\ell\nu_{\ell}$   & Section~\ref{sec:charm:semileptonic}		& $0.967\pm0.005(\rm exp.)\pm0.025(\rm LQCD)$\\
\midrule
\rowcolor{Gray} $D_s\to\ell\nu_{\ell}$ 	& \multirow{2}{*}{Average}	& \multirow{2}{*}{$0.997\pm0.017$}\\
\rowcolor{Gray} $D\to K\ell\nu_{\ell}$ & \multirow{-2}{*}{Average}	& \multirow{-2}{*}{$0.997\pm0.017$}\\
\midrule
$W\to c\overline{s}$	& PDG~\cite{PDG_2016}	& $0.94^{+0.32}_{-0.26}\pm0.13$\\
Global CKM Fit		& CKMFitter~\cite{Charles:2004jd}		& $0.973394^{+0.000074}_{-0.000096}$\\
\bottomrule
\end{tabular}
\end{table}

\subsubsection{Extraction of $D_{(s)}$ meson decay constants}

Assuming unitarity of the CKM matrix, the values of the elements relevant 
in the case of \mbox{(semi-)leptonic} charm decays are known from the global 
fit of the CKM matrix and are given in Table~\ref{tab:CKMVcdVcs}.
These values can be used to extract the $D$ and $D_s$ meson decay constants 
from the experimentally measured products $f_{D}|V_{cd}|$ and $f_{D_s}|V_{cs}|$ using Eq.~(\ref{eq:expFDVCD}) and Eq.~(\ref{eq:expFDSVCS}), respectively. 
This leads to the experimentally measured $D_{(s)}$ meson decay constants to be:
\begin{eqnarray}
f_D^{\rm exp} & = & (203.7\pm4.9)~{\rm MeV,}\\ 
f_{D_s}^{\rm exp} & = & (257.1\pm4.6)~{\rm MeV,}
\end{eqnarray}
and the ratio of the constants is determined to be
\begin{equation}
f_{D_s}^{\rm exp}/f_D^{\rm exp} = 1.262\pm0.037.
\label{eq:fDsfDRatio:WA}
\end{equation}
The values are in agreement with the LQCD determinations given in 
Table~\ref{tab:Lattice} within the uncertainties. The largest discrepancy 
is in the determinations of the ratio of the decay constants where the 
agreement is only at the level of $2.4\sigma$.

\clearpage
\subsection{Hadronic decays of $D_s$ mesons}

\begin{figure}[hbt!]
\centering
\includegraphics[width=0.50\textwidth]{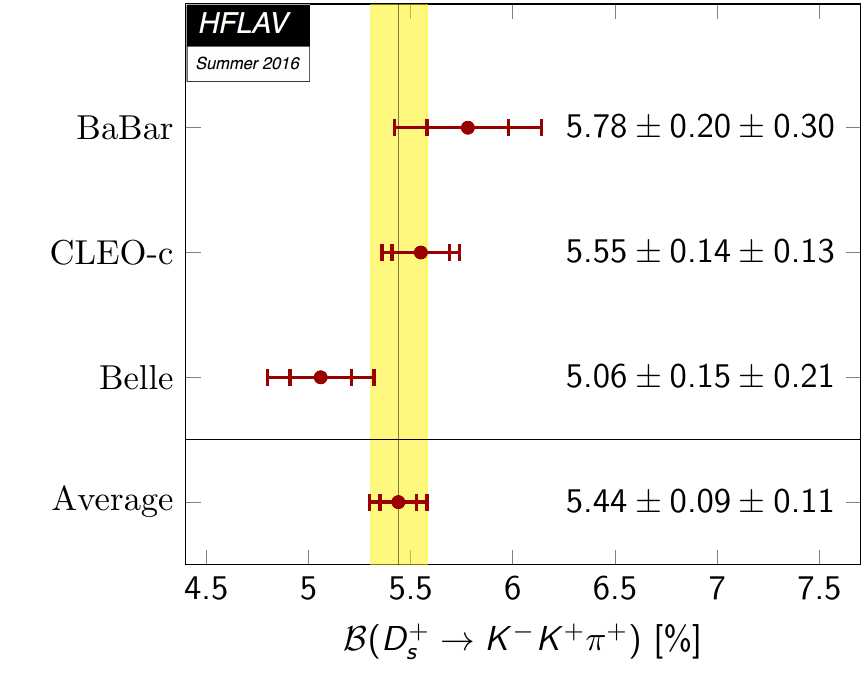}
\includegraphics[width=0.50\textwidth]{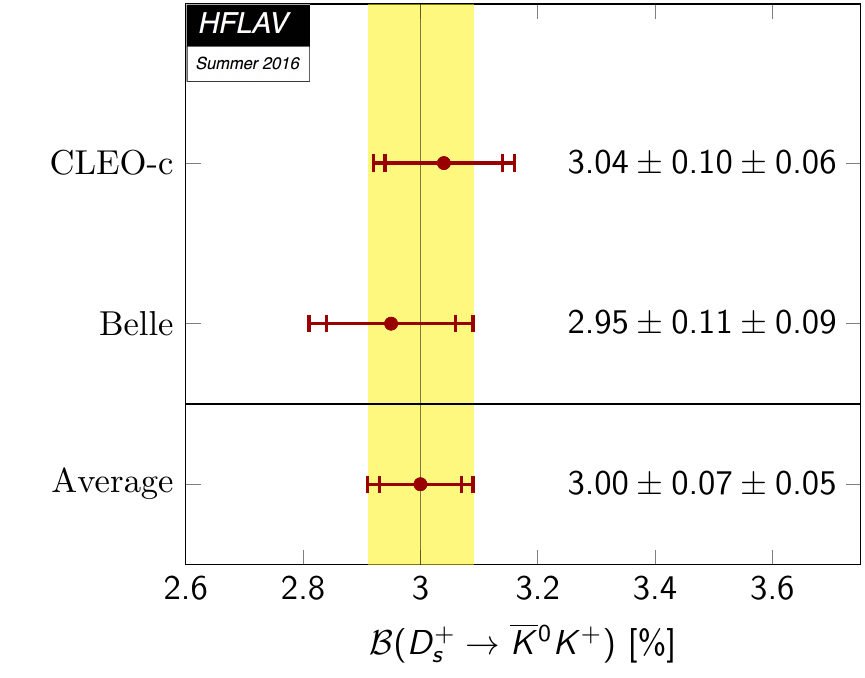}
\includegraphics[width=0.50\textwidth]{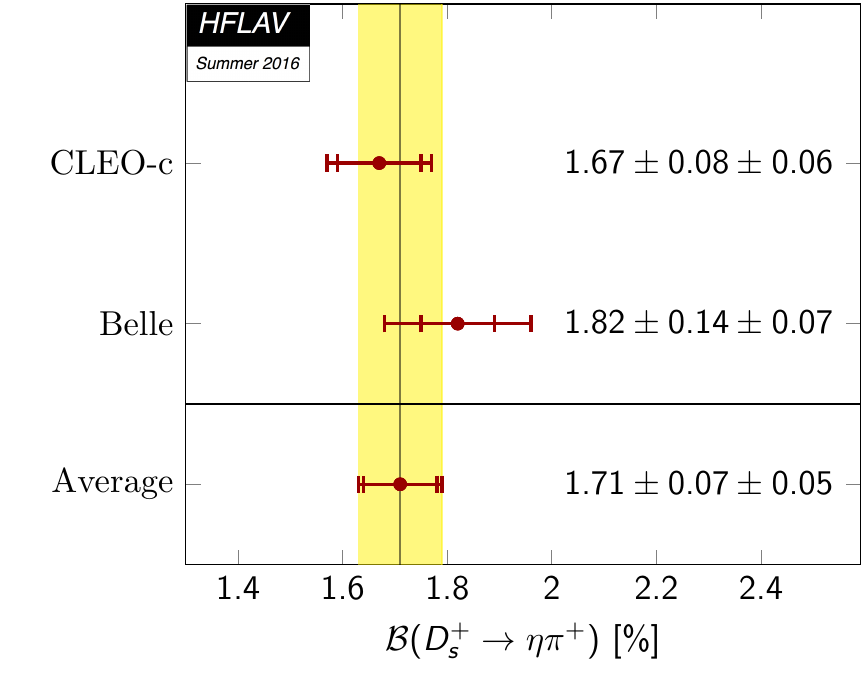}
\caption{
WA values for $\br(\dsp\to K^-K^+\pi^+)$ (top),
$\br(\dsp\to \overline{K}{}^0\pi^+)$ (middle), $\br(\dsp\to \eta\pi^+)$ (bottom).
For each point, the first error listed is the statistical and the second error
is the systematic error.
\label{fig:DSExpHadronic}
}
\end{figure}

\babar, CLEO-c and Belle collaborations have measured the absolute branching 
fractions of hadronic decays, $\dsp\to K^-K^+\pi^+$, $\dsp\to \overline{K}{}^0\pi^+$, 
and $\dsp \to \eta\pi^+$. The first two 
decay modes are the reference modes for the measurements of branching fractions of 
the $\dsp$ decays to any other final state. Table \ref{tab:DSExpHadronic} and 
Fig.~\ref{fig:DSExpHadronic} summarise the individual measurements and averaged 
values, which are found to be 
\begin{eqnarray}
\br^{\rm WA}(\dsp\to K^-K^+\pi^+) & = & (5.44\pm0.14)\%,\\
\br^{\rm WA}(\dsp\to \overline{K}{}^0\pi^+) & = & (3.00\pm0.09)\%,\\
\br^{\rm WA}(\dsp\to \eta\pi^+) & = & (1.71\pm0.08)\%,
\end{eqnarray}
where the uncertainties are total uncertainties. These averages are the same 
as in our previous report from 2014. The $\br(\dsp\to K^-K^+\pi^+)$ is for a phase space 
integrated decay and therfore includes all intermediate resonances.

\begin{table}[!htb]
\caption{Experimental results and world averages for branching fractions 
of $\dsp\to K^-K^+\pi^+$, $\dsp\to \overline{K}{}^0K^+$, and
$\dsp\to \eta\pi^+$ decays. The first uncertainty is statistical and the
second is experimental systematic. CLEO-c reports in Ref.~\cite{Onyisi:2013bjt}
$\br(\dsp\to K^0_SK^+)$. We include it in the average of 
$\br(\dsp\to \overline{K}{}^0K^+)$ by using the relation 
$\br(\dsp\to \overline{K}{}^0K^+)\equiv 2\br(\dsp\to K^0_SK^+)$.
\label{tab:DSExpHadronic}}
\begin{center}
\begin{tabular}{lcll}
\toprule
\rowcolor{Gray} Mode 	& ${\cal{B}}$ ($10^{-2}$)				& Reference 	& \\ 
\midrule
\multirow{3}{*}{$K^-K^+\pi^+$}  & $5.78\pm0.20\pm 0.30$ 		& \babar		& \cite{delAmoSanchez:2010jg}\\ 
						& $5.55\pm0.14\pm 0.13$ 		& CLEO-c		& \cite{Onyisi:2013bjt}\\ 
						& $5.06\pm0.15\pm 0.21$ 		& Belle   		& \cite{Zupanc:2013byn}\\
\midrule
\rowcolor{Gray}$K^-K^+\pi^+$	& $5.44\pm0.09\pm 0.11$			& Average & \\
\midrule
\multirow{2}{*}{$\overline{K}{}^0K^+$}		& $3.04\pm0.10\pm 0.06$ 		& CLEO-c		& \cite{Onyisi:2013bjt}\\ 
								& $2.95\pm0.11\pm 0.09$ 		& Belle   		& \cite{Zupanc:2013byn}\\
\midrule
\rowcolor{Gray}$\overline{K}{}^0K^+$		& $3.00\pm0.07\pm 0.05$			& Average & \\
\midrule
\multirow{2}{*}{$\eta\pi^+$}  	& $1.67\pm0.08\pm 0.06$ 		& CLEO-c		& \cite{Onyisi:2013bjt}\\ 
						& $1.82\pm0.14\pm 0.07$ 		& Belle   		& \cite{Zupanc:2013byn}\\
\midrule
\rowcolor{Gray}$\eta\pi^+$	& $1.71\pm0.07\pm 0.08$			& Average & 
\\ \bottomrule
\end{tabular}
\end{center}
\end{table}

\clearpage
% Hadronic decays
\subsection{Two-body hadronic $D^0$ decays and final state radiation}

\begin{figure}[bh]
\begin{center}
\includegraphics[width=0.48\textwidth,angle=0.]{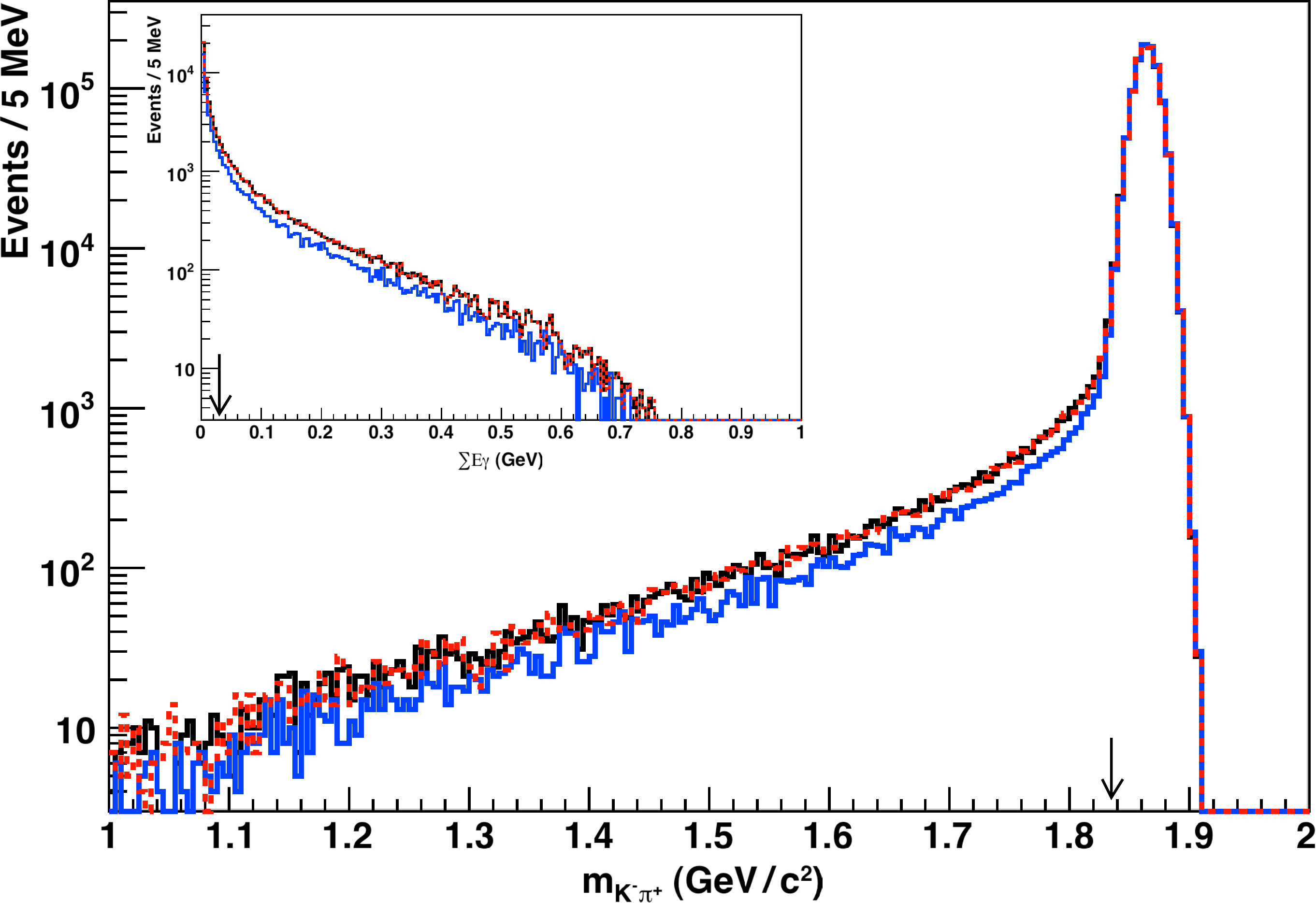}
\caption{The $K\pi$ invariant mass distribution for 
$D^0\to K^-\pi^+ (n\gamma)$ decays. The 3 curves correspond 
to three different configurations of PHOTOS for modeling FSR: 
version 2.02 without interference (blue/grey), version 2.02 with 
interference (red dashed) and version 2.15 with interference (black).  
The true invariant mass has been smeared with a typical experimental 
resolution of 10 MeV${}/c^2$.  Inset: The corresponding spectrum of 
total energy radiated per event.  The arrow indicates the $\sum E_\gamma$ 
value that begins to shift kinematic quantities outside of the range 
typically accepted in a measurement.}
\label{fig:FSR_mass_shift}
\end{center}
\end{figure}

Measurements of the branching fractions for the decays $D^0\to K^-\pi^+$,
$D^0\to \pi^+\pi^-$, and $D^0\to K^+ K^-$ have reached sufficient precision to
allow averages with ${\cal O}(1\%)$ relative uncertainties. 
At these precisions, Final 
State Radiation (FSR) must be treated correctly and consistently across 
the input measurements for the accuracy of the averages to match the 
precision.  The sensitivity of measurements to FSR arises because of 
a tail in the distribution of radiated energy that extends to the 
kinematic limit.  The tail beyond $\sum{E_\gamma} \approx 30$ MeV causes 
typical selection variables like the hadronic invariant mass to 
shift outside the selection range dictated by experimental 
resolution, as shown in Fig.~\ref{fig:FSR_mass_shift}.  While the 
differential rate for the tail is small, the integrated rate 
amounts to several percent of the total $h^+ h^-(n\gamma)$ 
rate because of the tail's extent.  The tail therefore 
translates directly into a several percent loss in 
experimental efficiency.

All measurements that include an FSR correction 
have a correction based on the use of 
PHOTOS~\cite{Barberio:1990ms,Barberio:1993qi,Golonka:2005pn,Golonka:2006tw} 
within the experiment's Monte Carlo simulation.  
PHOTOS itself, however, has evolved, over the period spanning the set of
measurements.  In particular, the incorporation of interference between
radiation off %of 
the two separate mesons has proceeded in stages: it was first
available for particle--antiparticle pairs in version 2.00 (1993), extended 
to any two-body, all-charged, final states in version 2.02 (1999), and 
further extended to multi-body final states in version 2.15 (2005).
The effects of interference are clearly visible, as shown in
Figure~\ref{fig:FSR_mass_shift}, and cause a 
roughly 30\% increase in the integrated rate into 
the high energy photon tail.  To evaluate the FSR 
correction incorporated into a given measurement, 
we must therefore note whether any correction was 
made, the version of PHOTOS used in correction, 
and whether the interference terms in PHOTOS were 
turned on.  

\subsubsection{Branching fraction corrections}

Before averaging the measured branching fractions, the published 
results are updated, as necessary, to the FSR prediction of 
PHOTOS~2.15 with interference included.  The correction will 
always shift a branching fraction to a higher value: with no 
FSR correction or with no interference term in the correction, 
the experimental efficiency determination will be biased high, 
and therefore the branching fraction will be biased low.

Most of the branching fraction analyses used the kinematic quantity 
sensitive to FSR in the candidate selection criteria.  For the 
analyses at the $\psi(3770)$, this variable was $\Delta E$, the 
difference between the candidate $D^0$ energy and the beam energy 
(\eg, $E_K + E_\pi - E_{\rm beam}$ for $D^0\to K^-\pi^+$).  
In the remainder of the analyses, the relevant quantity was the 
reconstructed hadronic two-body mass $m_{h^+h^-}$.  To make the 
correction, 
we  only need to evaluate the fraction of decays that FSR moves 
outside of the range accepted for the analysis.  

The corrections were evaluated using an event generator (EvtGen 
\cite{Ryd:2005zz}) that incorporates PHOTOS to simulate the 
portions of the decay process most relevant to the correction.  
We compared corrections determined both with and without smearing 
to account for experimental resolution.  The differences were 
negligible, typically of ${\cal O}(1\%)$ of the correction itself.  
The immunity of the correction to resolution effects comes about because 
most of the long FSR-induced tail in, for example, the $m_{h^+h^-}$ 
distribution resides well away from the selection boundaries.  The 
smearing from resolution, on the other hand, mainly affects the 
distribution of events right at the boundary.

For measurements incorporating an FSR correction that did not 
include interference, we update by assessing the FSR-induced 
efficiency loss for both the PHOTOS version and configuration 
used in the analysis and our nominal version 2.15 with interference.  
For measurements that published their sensitivity to FSR, our 
generator-level predictions for the original efficiency loss 
agreed to within a few percent of the correction. 
This agreement lends additional credence to the procedure.

Once the event loss from FSR in the most sensitive kinematic 
quantity is accounted for, the event loss in other quantities 
is very small. For example, analyses using $D^{*+}$ tags show 
little sensitivity to FSR in the reconstructed $D^{*+}\!-D^0$ 
mass difference, \ie, in $m^{}_{K^-\pi^+\pi^+}\!-m^{}_{K^-\pi^+}$. 
In this case the effect of FSR tends to cancel in the difference 
of reconstructed masses. In the $\psi(3770)$ analyses, the 
beam-constrained mass distributions 
($\sqrt{E_{\rm beam}^2 - |\vec{p}_K + \vec{p}_\pi|^2}$)  
also show much smaller sensitivity than does the two-body mass.
\begin{figure}
\begin{center}
\includegraphics[width=1.00\textwidth]{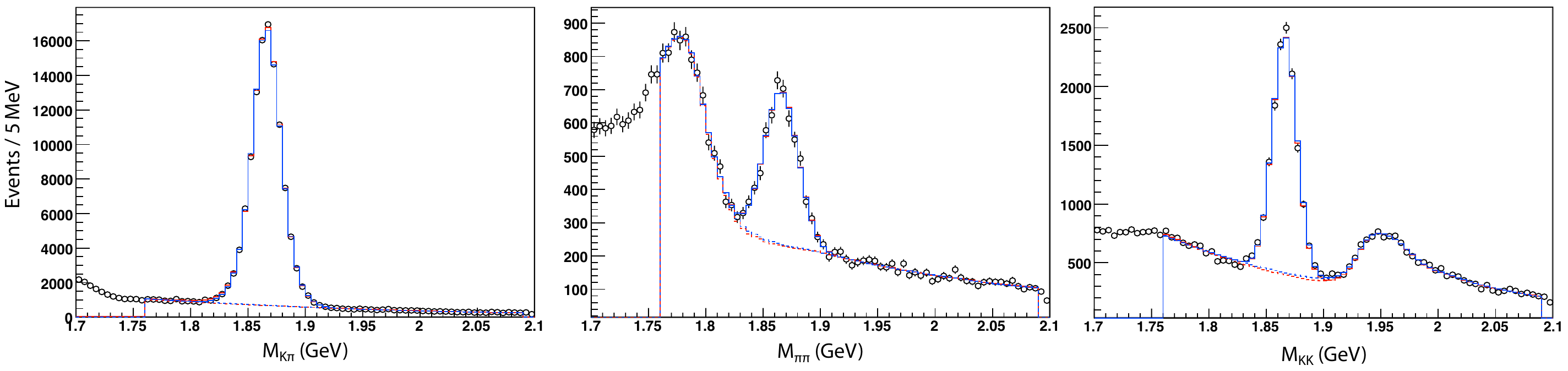}
\caption{FOCUS data (dots), original fits (blue) and 
toy MC parameterization (red) for $D^0\to K^-\pi^+$ (left), 
$D^0\to \pi^+\pi^-$ (center), and $D^0\to \pi^+\pi^-$ (right).}
\label{fig:FocusFits}
\end{center}
\end{figure}

The FOCUS~\cite{Link:2002hi} analysis of the branching fraction
ratios ${\cal B}(D^0\to \pi^+\pi^-)/{\cal B}(D^0\to K^-\pi^+)$ and 
${\cal B}(D^0\to K^+ K^-)/{\cal B}(D^0\to K^-\pi^+)$ obtained 
yields using fits to the two-body mass distributions.  FSR will 
both distort the low end of the signal mass peak, and will 
contribute a signal component to the low side tail used to 
estimate the background.  The fitting procedure is not sensitive 
to signal events out in the FSR tail, which would be counted as 
part of the background.

A more complex toy Monte Carlo procedure was required to analyze 
the effect of FSR on the fitted yields, which were published with 
no FSR corrections applied.  A detailed description of the procedure 
and results is available on the HFLAV web site, and a brief summary is provided
here.  Determining the correction involved an iterative procedure in which samples of similar size to the FOCUS sample were 
generated and then fit using the FOCUS signal and background 
parameterizations.  The MC parameterizations were tuned based 
on differences between the fits to the toy MC data and the FOCUS 
fits, and the procedure was repeated. These steps were iterated until 
the fit parameters matched the original FOCUS parameters.  

\begin{table}[!htb]
  \centering 
  \caption{The experimental measurements relating to ${\cal B}(D^0\to K^-\pi^+)$, ${\cal B}(D^0\to \pi^+\pi^-)$, and ${\cal B}(D^0\to K^+ K^-)$ after correcting to the common version and configuration of PHOTOS.  The uncertainties are statistical and total systematic, with the FSR-related systematic estimated in this procedure shown in parentheses.  Also listed are the percent shifts in the results from the correction, if any, applied here, as well as the original PHOTOS and interference configuration for each publication.}
  \label{tab:FSR_corrections}
\begin{tabular}{lccc}
\hline \hline
Experiment (acronym) & Result (rescaled) & Correction [\%] & PHOTOS \\ \hline
\multicolumn{4}{l}{$D^{0} \to K^{-} \pi^{+}$} \\
      CLEO-c 14  (CC14) \cite{Bonvicini:2013vxi} & $3.934 \pm 0.021 \pm 0.061(31)\%$ & --   & 2.15/Yes \\
%       CLEO-c 07  (CC07) \cite{Dobbs:2007zt}   & $3.891 \pm 0.035 \pm 0.065(27)\%$ & --   & 2.15/Yes \\      
      \babar 07  (BB07) \cite{Aubert:2007wn}   & $4.035 \pm 0.037 \pm 0.074(24)\%$ & 0.69 & 2.02/No \\
      CLEO II 98 (CL98) \cite{Artuso:1997mc}   & $3.920 \pm 0.154 \pm 0.168(32)\%$ & 2.80 & none \\
      ALEPH 97   (AL97) \cite{Barate:1997mm}   & $3.930 \pm 0.091 \pm 0.125(32)\%$ & 0.79 & 2.0/No \\
      ARGUS 94   (AR94) \cite{Albrecht:1994nb} & $3.490 \pm 0.123 \pm 0.288(24)\%$ & 2.33 & none \\
      CLEO II 93 (CL93) \cite{Akerib:1993pm}   & $3.960 \pm 0.080 \pm 0.171(15)\%$ & 0.38 & 2.0/No \\
      ALEPH 91   (AL91) \cite{Decamp:1991jw}   & $3.730 \pm 0.351 \pm 0.455(34)\%$ & 3.12 & none \\
\multicolumn{4}{l}{$D^{0} \to \pi^{+}\pi^{-} / D^{0} \to K^{-} \pi^{+}$} \\
      CLEO-c 10  (CC10) \cite{Mendez:2009aa}   & $0.0370  \pm 0.0006  \pm 0.0009(02)$  & --   & 2.15/Yes \\
      CDF 05     (CD05) \cite{Acosta:2004ts}   & $0.03594 \pm 0.00054 \pm 0.00043(15)$ & --   & 2.15/Yes \\
      FOCUS 02   (FO02) \cite{Link:2002hi}     & $0.0364  \pm 0.0012  \pm 0.0006(02)$  & 3.10 & none \\
\multicolumn{4}{l}{$D^{0} \to K^{+}K^{-} / D^{0} \to K^{-} \pi^{+}$} \\
      CLEO-c 10   \cite{Mendez:2009aa}         & $0.1041 \pm 0.0011 \pm 0.0012(03)$ & --    & 2.15/Yes \\ 
      CDF 05      \cite{Acosta:2004ts}         & $0.0992 \pm 0.0011 \pm 0.0012(01)$ & --    & 2.15/Yes \\
      FOCUS 02    \cite{Link:2002hi}           & $0.0982 \pm 0.0014 \pm 0.0014(01)$ & -1.12 & none \\ \hline
\end{tabular}
\end{table}

The toy MC samples for the first iteration were based on the generator-level 
distribution of $m_{K^-\pi^+}$, $m_{\pi^+\pi^-}$, and $m_{K^+K^-}$, including 
the effects of FSR, smeared according to the original FOCUS resolution 
function,  and on backgrounds 
generated 
%thrown 
using the parameterization from the final
FOCUS fits.  For each iteration, 400 to 1600 individual 
data-sized samples were 
generated 
%thrown 
and fit. The means of the parameters from these fits determined the 
corrections to the generator parameters for the following iteration.  The 
ratio between the number of signal events generated and the final signal 
yield provides the required FSR correction in the final iteration.  Only a 
few iterations were required in each mode.  Figure~\ref{fig:FocusFits} 
shows the FOCUS data, the published FOCUS fits, and the final toy MC 
parameterizations.  The toy MC provides an excellent description of the 
data.

The corrections obtained to the individual FOCUS yields were 
$1.0298\pm 0.0001$ for $K^-\pi^+$, $1.062 \pm 0.001$ for $\pi^+\pi^-$, 
and $1.0183 \pm 0.0003$ for $K^+K^-$.  These corrections tend to 
cancel in the branching ratios, leading to corrections of 
$1.031\pm 0.001$ for 
${\cal B}(D^0\to \pi^+\pi^-)/{\cal B}(D^0\to K^-\pi^+)$, and 
$0.9888\pm 0.0003$ for 
${\cal B}(D^0\to K^+ K^-)/{\cal B}(D^0\to K^-\pi^+)$.

Table~\ref{tab:FSR_corrections} summarizes the corrected branching fractions. 
The published FSR-related modeling uncertainties have been replaced by with a
new, common, estimate based on the assumption that the dominant uncertainty 
in the FSR corrections comes from the fact that the mesons are treated like 
structureless particles. No contributions from structure-dependent terms in 
the decay process (\eg, radiation off individual quarks) are included in PHOTOS. 
Internal studies done by various experiments have indicated that 
in $K\pi$ decays, 
the PHOTOS corrections agree with data at the 20-30\% level. 
We therefore attribute a 25\% uncertainty to the FSR prediction from potential 
structure-dependent contributions. For the other two modes, the only difference 
in structure is the final state valence quark content. While radiative corrections 
typically come in with a $1/M$ dependence, one would expect the additional 
contribution from the structure terms to come in on time scales shorter than 
the hadronization time scale. In this case, you might expect
$\rm \Lambda_{\rm QCD}$ to be the relevant scale, rather than the quark masses,
and therefore that the amplitude is the same for the three modes. In treating
the correlations among the measurements this is what we assume. We also assume
that the PHOTOS amplitudes and any missing structure amplitudes are relatively 
real with constructive interference.  The uncertainties largely cancel 
in the branching fraction ratios. For the final average branching 
fractions, the FSR uncertainty on $K\pi$ dominates. Note that because 
of the relative sizes of FSR in the different modes, the $\pi\pi/K\pi$ 
branching ratio uncertainty from FSR is 
positively correlated with that 
for the $K\pi$ branching fraction, while the $KK/K\pi$ branching ratio FSR
uncertainty is negatively correlated.

The ${\cal B}(D^0\to K^-\pi^+)$ measurement of reference~\cite{Coan:1997ye}, the  
${\cal B}(D^0\to \pi^+\pi^-)/{\cal B}(D^0\to K^-\pi^+)$ measurements of 
references~\cite{Aitala:1997ff} 
and~\cite{Csorna:2001ww}, and the 
${\cal B}(D^0\to K^+ K^-)/{\cal B}(D^0\to K^-\pi^+)$ measurement
of reference~\cite{Csorna:2001ww} are excluded from the branching 
fraction averages presented here.
These measurements appear not to have incorporated any FSR corrections, 
and insufficient information
is available to determine the 2-3\% corrections that would be required.

\begin{sidewaystable}[p]
  \centering 
  \caption{The correlation matrix corresponding to the full covariance matrix. 
  %from the sum of statistical,
  %systematic and FSR covariances. 
  Subscripts $h$ denote which of the $D^0 \to h^+ h^-$ decay results from a single experiment
  is represented in that row or column.}\label{tab:correlations}
  \small
\begin{tabular}{lr@{.}lr@{.}lr@{.}lr@{.}lr@{.}lr@{.}lr@{.}lr@{.}lr@{.}lr@{.}lr@{.}lr@{.}lr@{.}l}
\hline\hline
           & \multicolumn{2}{c}{CC14}
                   & \multicolumn{2}{c}{BB07}
                           & \multicolumn{2}{c}{CL98}
                                   & \multicolumn{2}{c}{AL97}
                                           & \multicolumn{2}{c}{AR94} 
                                                   & \multicolumn{2}{c}{CL93} 
                                                           & \multicolumn{2}{c}{AL91} 
                                                                   & \multicolumn{2}{c}{FO02$_\pi$} 
                                                                           & \multicolumn{2}{c}{CD05$_\pi$} 
                                                                                   & \multicolumn{2}{c}{CC10$_\pi$} 
                                                                                           & \multicolumn{2}{c}{FO02$_K$}
                                                                                                    & \multicolumn{2}{c}{CD05$_K$} 
                                                                                                            & \multicolumn{2}{c}{CC10$_K$} \\ \hline
% 2014 matrix
CC14 & 1&000 & 0&139 & 0&057 & 0&084 & 0&031 & 0&033 & 0&023 & 0&070 & 0&103 & 0&068 &-0&019 &-0&032 &-0&085 \\
BB07 & 0&139 & 1&000 & 0&035 & 0&051 & 0&019 & 0&020 & 0&014 & 0&042 & 0&062 & 0&041 &-0&012 &-0&019 &-0&051 \\
CL98 & 0&057 & 0&035 & 1&000 & 0&021 & 0&008 & 0&298 & 0&006 & 0&017 & 0&026 & 0&017 &-0&005 &-0&008 &-0&021 \\
AL97 & 0&084 & 0&051 & 0&021 & 1&000 & 0&011 & 0&012 & 0&116 & 0&025 & 0&038 & 0&025 &-0&007 &-0&012 &-0&031 \\
AR94 & 0&031 & 0&019 & 0&008 & 0&011 & 1&000 & 0&004 & 0&003 & 0&009 & 0&014 & 0&009 &-0&003 &-0&004 &-0&011 \\
CL93 & 0&033 & 0&020 & 0&298 & 0&012 & 0&004 & 1&000 & 0&003 & 0&010 & 0&015 & 0&010 &-0&003 &-0&005 &-0&012 \\
AL91 & 0&023 & 0&014 & 0&006 & 0&116 & 0&003 & 0&003 & 1&000 & 0&007 & 0&010 & 0&007 &-0&002 &-0&003 &-0&009 \\
FO02$_\pi$ & 0&070 & 0&042 & 0&017 & 0&025 & 0&009 & 0&010 & 0&007 & 1&000 & 0&031 & 0&021 &-0&006 &-0&010 &-0&026 \\
CD05$_\pi$ & 0&103 & 0&062 & 0&026 & 0&038 & 0&014 & 0&015 & 0&010 & 0&031 & 1&000 & 0&031 &-0&009 &-0&014 &-0&038 \\
CC10$_\pi$ & 0&068 & 0&041 & 0&017 & 0&025 & 0&009 & 0&010 & 0&007 & 0&021 & 0&031 & 1&000 &-0&006 &-0&010 &-0&025 \\
FO02$_K$ &-0&019 &-0&012 &-0&005 &-0&007 &-0&003 &-0&003 &-0&002 &-0&006 &-0&009 &-0&006 & 1&000 & 0&003 & 0&007 \\
CD05$_K$ &-0&032 &-0&019 &-0&008 &-0&012 &-0&004 &-0&005 &-0&003 &-0&010 &-0&014 &-0&010 & 0&003 & 1&000 & 0&012 \\
CC10$_K$ &-0&085 &-0&051 &-0&021 &-0&031 &-0&011 &-0&012 &-0&009 &-0&026 &-0&038 &-0&025 & 0&007 & 0&012 & 1&000 \\
\hline
\end{tabular}
\end{sidewaystable}

\subsubsection{Average branching fractions}
\begin{figure}
\begin{center}
\includegraphics[width=0.6\textwidth,angle=0.]{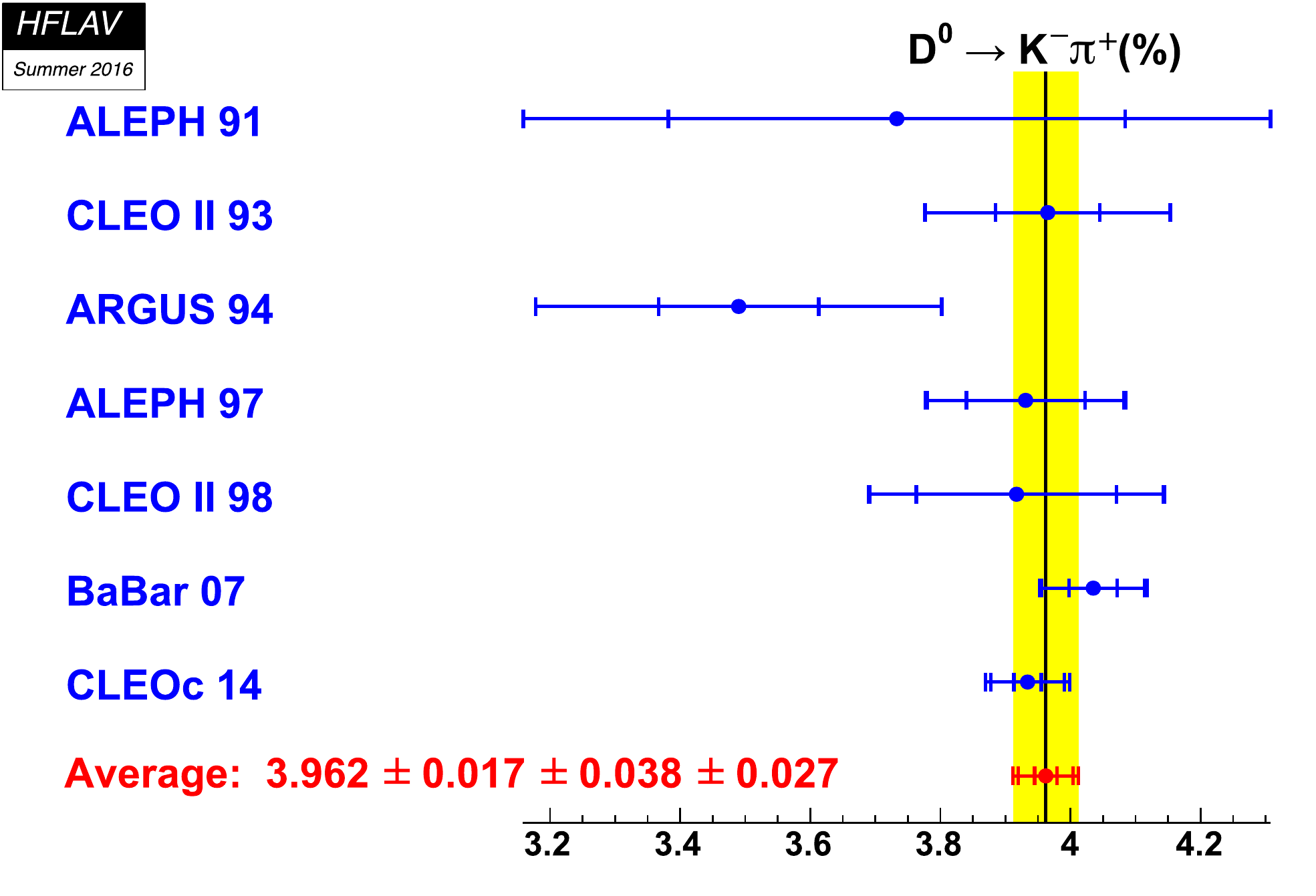}
\caption{Comparison of measurements of 
${\cal B}(D^0\to K^-\pi^+)$ (blue) with the average 
branching fraction obtained here (red, and yellow band).}
\label{D0bfs}
\end{center}
\end{figure}

%%%%
% OLD VERSION OF THIS PARAGRAPH - before 27 Nov 2014
%The average branching fractions for $D^0\to K^-\pi^+$, $D^0\to \pi^+\pi^-$ and 
%$D^0\to K^+ K^-$ are obtained from
%a single $\chi^2$ minimization procedure, in which the three branching 
%fractions are floating parameters. The
%central values derive from a fit in which the covariance matrix is the sum 
%of the covariance matrices for the
%statistical, systematic (excluding FSR) and FSR uncertainties.  
%The statistical uncertainties are obtained from
%a fit using only the statistical covariance matrix.  The systematic 
%uncertainties are obtained from the
%quadrature uncertainties from a fit with statistical-only and 
%statistical+systematic covariance matrices, and
%the FSR uncertainties on the averages from the quadrature 
%differences in the uncertainties obtained from the
%nominal fit and a fit excluding the FSR uncertainties.
%%%

%%%
% NEW VERSION OF THIS PARAGRAPH - 27 Nov 2014
The average branching fractions for 
$D^0\to K^-\pi^+$, $D^0\to \pi^+\pi^-$ and $D^0\to K^+ K^-$ decays
are obtained from a single $\chi^2$ minimization procedure, 
in which the three branching fractions are floating parameters. 
The central values are obtained from a fit in which the full covariance matrix
--\,accounting for all statistical, systematic (excluding FSR), and FSR 
measurement uncertainties\,-- is used.  
Table~\ref{tab:correlations} presents the correlation matrix for 
this nominal fit. % (stat.+syst.+FSR).
We then obtain the three reported uncertainties on those central values as follows:
The statistical uncertainties are obtained from a fit using only the statistical 
covariance matrix.  
The systematic uncertainties are obtained by subtracting (in quadrature) the 
statistical uncertainties 
from the uncertainties determined via a fit using a covariance matrix that 
accounts for both statistical and systematic measurement uncertainties.  
The FSR uncertainties are obtained by subtracting (in quadrature)
the uncertainties determined via a fit using a covariance matrix that accounts 
for both statistical and systematic measurement uncertainties
from the uncertainties determined via the fit using the full covariance matrix.
%%%

In forming the full covariance matrix, the FSR
uncertainties are treated as fully correlated (or anti-correlated) as 
described above.  % PN after 27 Nov
%In forming the covariance matrix for the FSR uncertainties, the FSR
%uncertainties are treated as fully correlated (or anti-correlated) as 
%described above.  PN before 27 Nov
%For the systematic covariance matrix, ALEPH's systematic
For the covariance matrices involving systematic measurement uncertainties, ALEPH's systematic % Nov 27
uncertainties in the $\theta_{D^*}$ parameter are treated
as fully correlated between the ALEPH 97 and ALEPH 91 measurements.  Similarly,
the tracking efficiency uncertainties in the CLEO II 98 and the
CLEO II 93 measurements are treated as fully correlated.  
%Finally, the CLEO-c 07 $D^0\to K^-\pi^+$ measurement and the CLEO-c 08 $D^0\to K^+ K^-$
%measurements have a significant statistical correlation.  
%The 2007 hadronic branching fraction analysis
%derives the number of $N_{D^0\bar{D}^0}$ pairs produced in CLEO-c, and that quantity is statistically
%correlated with the $D^0\to K^-\pi^+$ branching fraction in that analysis ($\rho=0.65$).  The 2008 
%$K^+K^-$ analysis in turn uses that value of $N_{D^0\bar{D}^0}$ as the normalization for its branching
%fraction.  

The averaging procedure results in a 
final $\chi^2$ of $11.0$ for $10$ ($13-3$) degrees 
of freedom.  The branching
fractions obtained are
\begin{eqnarray}
\label{DHad_results}
  {\cal B}(D^0\to K^-\pi^+)   & = & ( 3.962 \pm 0.017 \pm 0.038 \pm 0.027 )\,\%, \\
  {\cal B}(D^0\to \pi^+\pi^-) & = & ( 0.144 \pm 0.002 \pm 0.002 \pm 0.002 )\,\%, \\
  {\cal B}(D^0\to K^+ K^-)    & = & ( 0.399 \pm 0.003 \pm 0.005 \pm 0.002 )\,\%\,. 
\end{eqnarray}The uncertainties, estimated as described above, are statistical, 
systematic (excluding FSR), and
FSR modeling.  The correlation coefficients from the fit using the 
total uncertainties are
\begin{center}
\begin{tabular}{llll}
               & $K^-\pi^+$ & $\pi^+\pi^-$ & $K^+ K^-$ \\
$K^-\pi^+$     &  1.00 & 0.71 & 0.76  \\
$\pi^+\pi^-$   &  0.71 & 1.00 & 0.53  \\
$K^+ K^-$      &  0.76 & 0.53 & 1.00  \\
\end{tabular}
\end{center}

\begin{table}[!htb]
  \centering 
  \caption{Evolution of the $D^0\to K^-\pi^+$ branching fraction from a fit with
  no FSR corrections or correlations (similar to the average in the 
  PDG 2016 update~\cite{PDG_2016}) to the nominal fit presented
here.}\label{tab:fit_evolution}
\resizebox{\textwidth}{!}{
\begin{tabular}{cccll}
\hline\hline
Modes &  Description & ${\cal B}(D^0\to K^-\pi^+)$ (\%)  & $\chi^2$/(deg.~of freedom) \\
fit   &              &                                   &     \\ \hline
$K^-\pi^+$ & PDG 2016\,\cite{PDG_2016} equivalent    
     & $3.930 \pm 0.017 \pm 0.042$ & $4.5/(8-1)=0.64$ \\
$K^-\pi^+$ & drop Ref.~\cite{Coan:1997ye}  & $3.938 \pm 0.017 \pm 0.042$ & $4.5/(7-1)=0.75$ \\
% $K^-\pi^+$ & PDG summer 2014 equivalent    & $3.913 \pm 0.022 \pm 0.043 $ & $6.0/(8-1)$ \\
% $K^-\pi^+$ & drop Ref.~\cite{Coan:1997ye}  & $3.921 \pm 0.023 \pm 0.044$   & $4.8/(7-1)$ \\
% $K^-\pi^+$ & use Ref.~\cite{Bonvicini:2013vxi} instead of Ref.\cite{Dobbs:2007zt}  & $3.938 \pm 0.017 \pm 0.042$ & $4.5/(7-1)$  \\
$K^-\pi^+$ & add FSR corrections           & $3.955 \pm 0.017 \pm 0.038 \pm 0.018$ & $3.5/(7-1)=0.58$  \\
$K^-\pi^+$ & add FSR correlations          & $3.956 \pm 0.017 \pm 0.038 \pm 0.027$ & $3.6/(7-1)=0.60$  \\
all        & --   & $3.962 \pm 0.017 \pm 0.038 \pm 0.027$ &   $11.0/(13-3)=1.10$  \\
\hline
\end{tabular}  
}
\end{table}

\begin{figure}
\begin{center}
\includegraphics[width=0.47\textwidth,angle=0.]{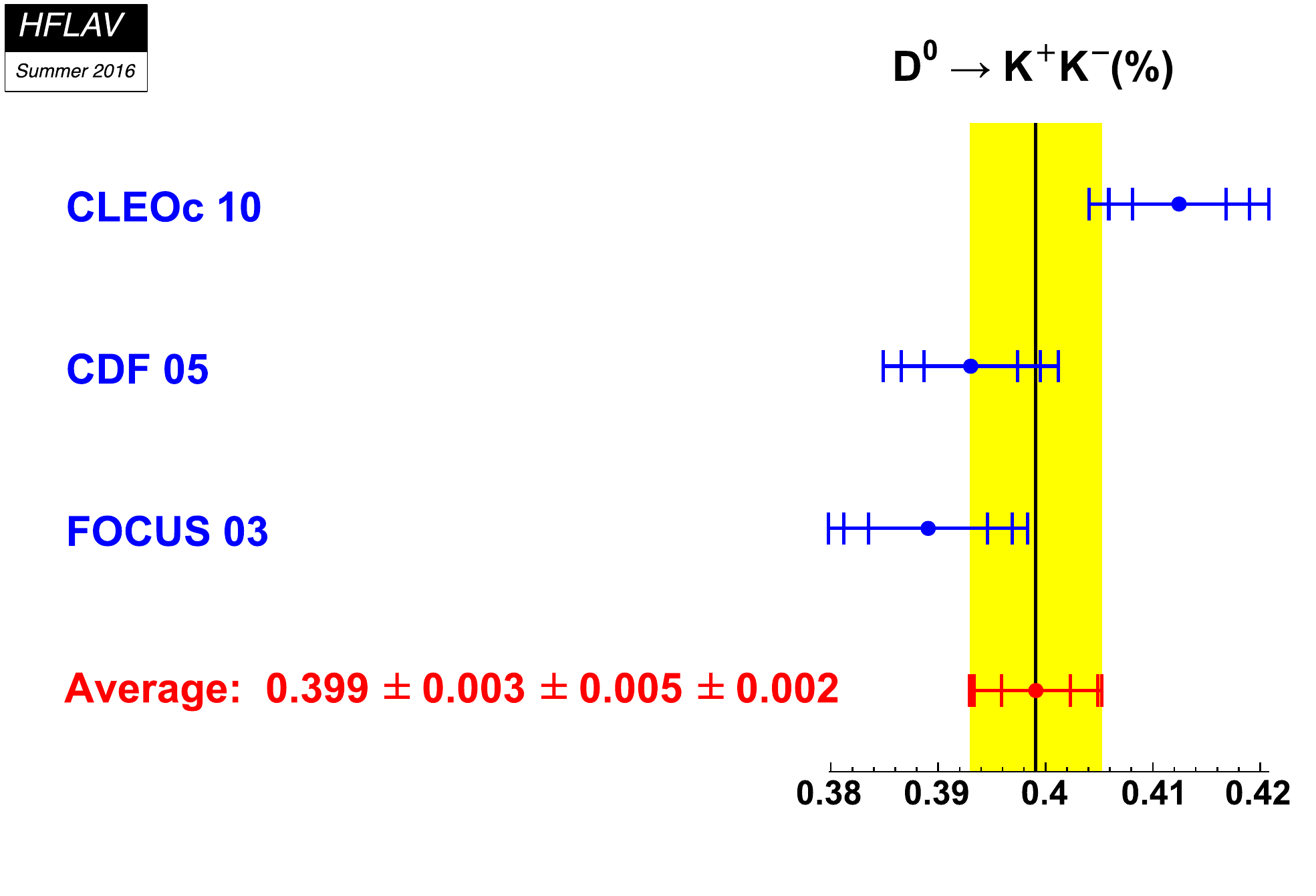}\hfill
\includegraphics[width=0.47\textwidth,angle=0.]{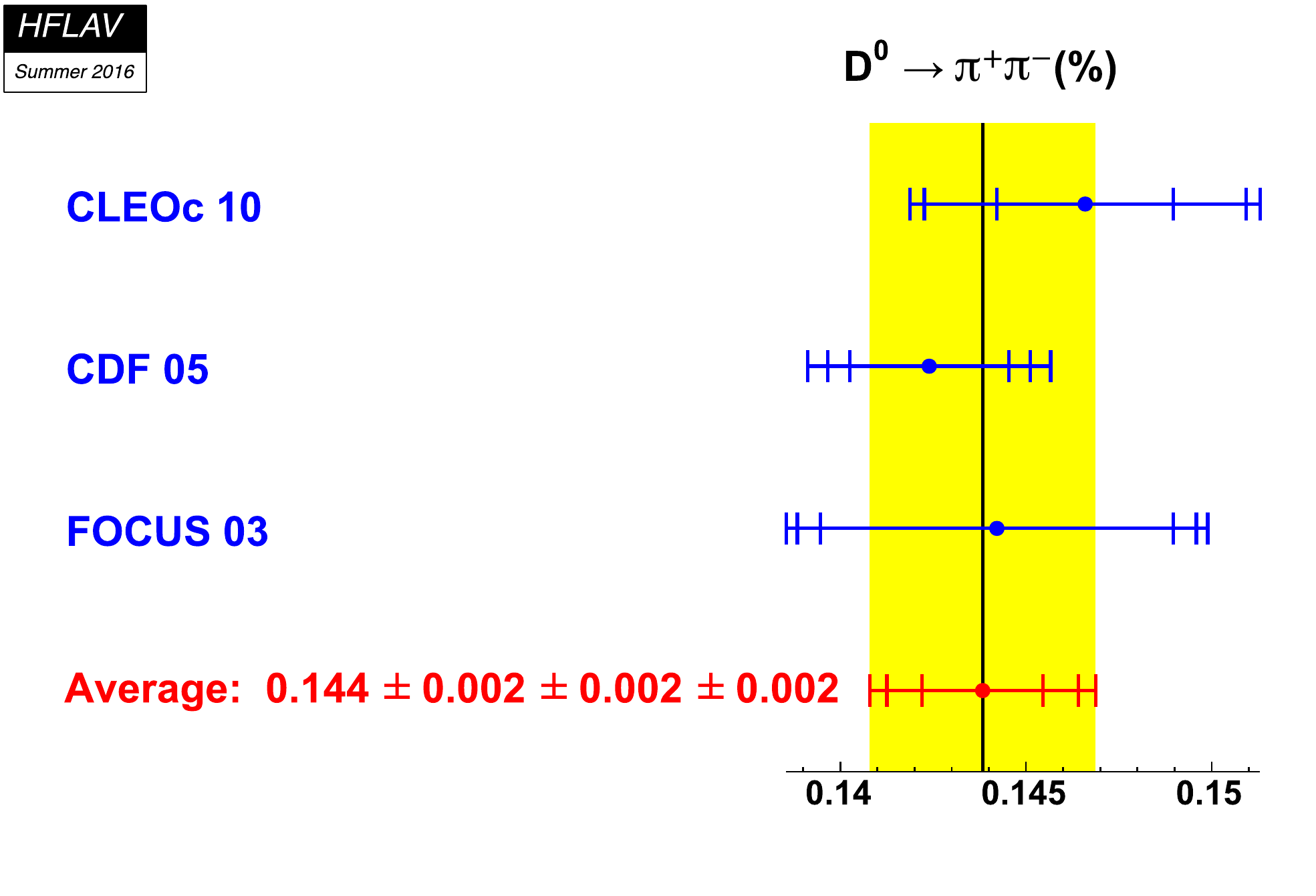}
\caption{The ${\cal B}(D^0\to K^+K^-)$ (left) and ${\cal B}(D^0\to \pi^+\pi^-)$ (right) 
values obtained by scaling the measured branching ratios with the ${\cal B}(D^0\to K^-\pi^+)$ branching fraction
average obtained here.  For the measurements (blue points), the error bars correspond to the statistical, systematic
and $K\pi$ normalization uncertainties.  The average obtained here (red point, yellow band) lists the statistical,
systematics excluding FSR, and the FSR systematic.
\label{fig:kkpipi}}
\end{center}
\end{figure}

As the $\chi^2$ would suggest and Fig.~\ref{D0bfs} shows, the average 
value for ${\cal B}(D^0\to K^-\pi^+)$ and
the input branching fractions agree very well.  With the estimated 
uncertainty in the FSR modeling used here,
the FSR uncertainty dominates the statistical uncertainty 
in the average, suggesting that experimental
work in the near future should focus on verification of FSR with 
$\sum E_\gamma \simge 100$ MeV.  Note that the systematic uncertainty 
excluding FSR
is still larger than the FSR uncertainty; in the most 
precise measurements of these branching fractions, the 
largest systematic
uncertainty is the uncertainty on the tracking efficiency. The ${\cal B}(D^0\to
K^+K^-)$ and ${\cal B}(D^0\to \pi^+\pi^-)$ measurements inferred
from the branching ratio measurements also agree well 
(Fig.~\ref{fig:kkpipi}). 

The ${\cal B}(D^0\to K^-\pi^+)$ average obtained here is 
approximately two statistical standard deviations higher
than the 2016 PDG update average~\cite{PDG_2016}. 
Table~\ref{tab:fit_evolution} shows the evolution from a
fit similar to the PDG's (no FSR corrections or correlations, 
reference~\cite{Coan:1997ye} 
included)
%, uses reference~\cite{Dobbs:2007zt} instead of reference~\cite{Bonvicini:2013vxi} [the latter being a recent, superseding result]) 
 to the average presented here.
There are two main contributions to the difference. The
branching fraction in reference~\cite{Coan:1997ye} is
low, and its exclusion shifts the result upwards. The dominant shift
($+0.017\%$) is due to the FSR corrections, which as
expected shift the result upwards.
%, and the more precise result from reference~\cite{Bonvicini:2013vxi}.
%Finally, including the CLEO-c
%absolute $D^0\to K^+ K^-$ branching fraction contributes the 
%final shift of $+0.009\%$.  As Fig.~\ref{fig:D0KK}
%shows, the $K^+ K^-$ branching fractions inferred from the 
%combining the CDF and FOCUS branching ratios and
%the average $K^-\pi^+$ branching fraction (excluding the 
%CLEO-c $K^+ K^-$ result) are both lower than
%the CLEO-c absolute measurement.  The fit, therefore, 
%exerts an upward pressure on the $K^-\pi^+$ result
%to improve the agreement in the $K^+ K^-$ sector.

There is no reason to presume that the effects of FSR should be
different in $D^0\to K^+\pi^-$ and $D^0\to K^-\pi^+$ decays, as both decay to
one charged kaon and one charged pion. Measurements of the relative branching
fraction ratio between the doubly Cabibbo-suppressed decay
$D^0\to K^+\pi^-$ and the Cabibbo-favored decay $D^0\to K^-\pi^+$ ($R_D$,
determined in Section~\ref{sec:charm:mixcpv}) are now approaching ${\cal O}(1\%)$ relative uncertainties. 
This makes it worthwhile to combine our $R_D$ average with the ${\cal B}(D^0\to K^-\pi^+)$ average obtained in Eq.~(\ref{DHad_results}), to
provide measurements of the branching fraction:
\begin{eqnarray}
  {\cal B}(D^0\to K^+\pi^-)   & = & ( 1.379 \pm 0.023 ) \times 10^{-4}~({{\rm assuming~no}~CPV}),  % previous R_D: 1.387 \pm 0.024 times 10^-4
  \\   
 {\cal B}(D^0\to K^+\pi^-)   & = & ( 1.383 \pm 0.023 ) \times 10^{-4}~({CPV~{\rm allowed}}).    % previous R_D, also: 1.383 \pm 0.023 times 10^-4
\end{eqnarray} 
%where the uncertainty is the total uncertainty. 
Note that, by
definition of $R_D$, these branching fractions do not include any contribution
from Cabibbo-favored $\Dzb \to K^+\pi^-$ decays. 
% PN - I think I will not bother mentioning the PDG result for DCS Kpi BF, as
% their calculation is based both off of their R_D and their Kpi BF. How our R_D
% and our Kpi BF calculations compare to the PDG should have already been
% discused in our paper, so there is no need to repeat such a discussion here.
% PDG 2015 result: 	(1.399 \pm 0.027) \times 10^{-4}

\clearpage
% Excited D
\mysubsection{Excited $D_{(s)}$ mesons}

\begin{figure}[htb!]
\begin{centering}
\includegraphics[width=0.49\textwidth]{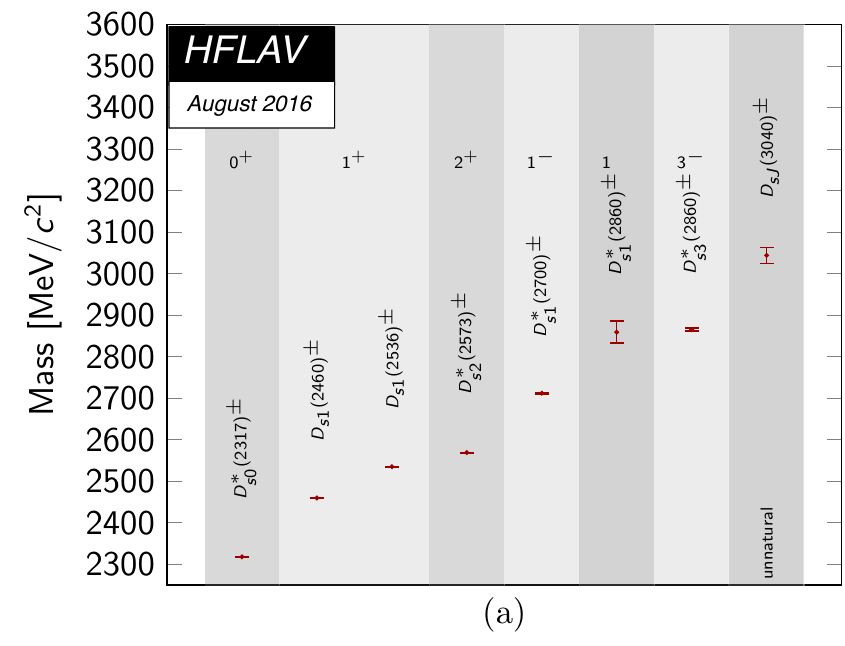}
\includegraphics[width=0.49\textwidth]{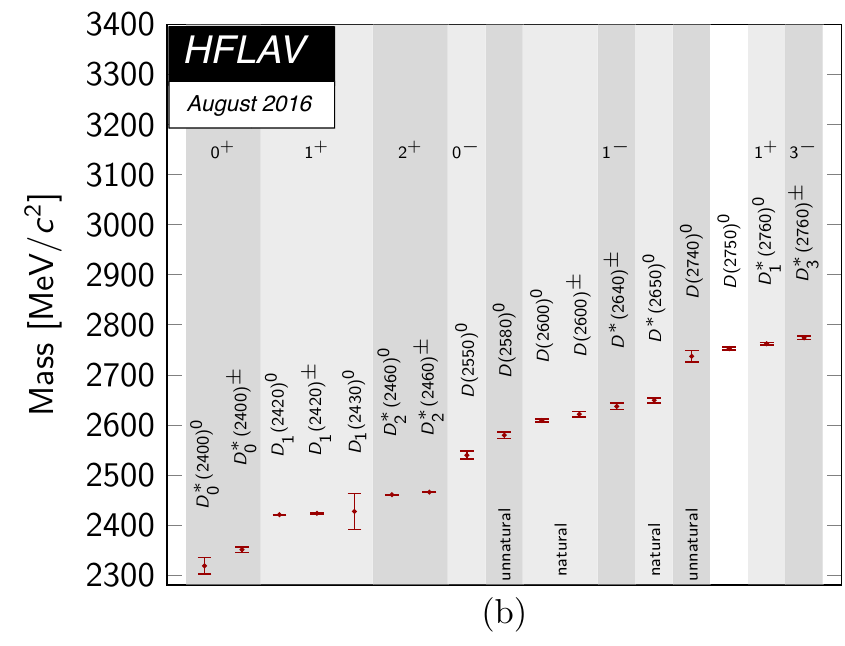}\\
\includegraphics[width=0.49\textwidth]{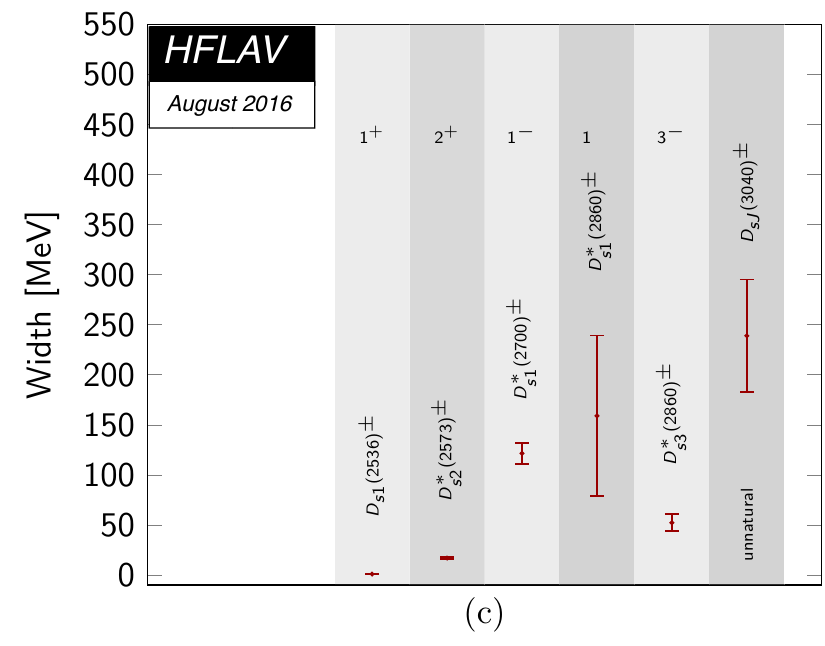}
\includegraphics[width=0.49\textwidth]{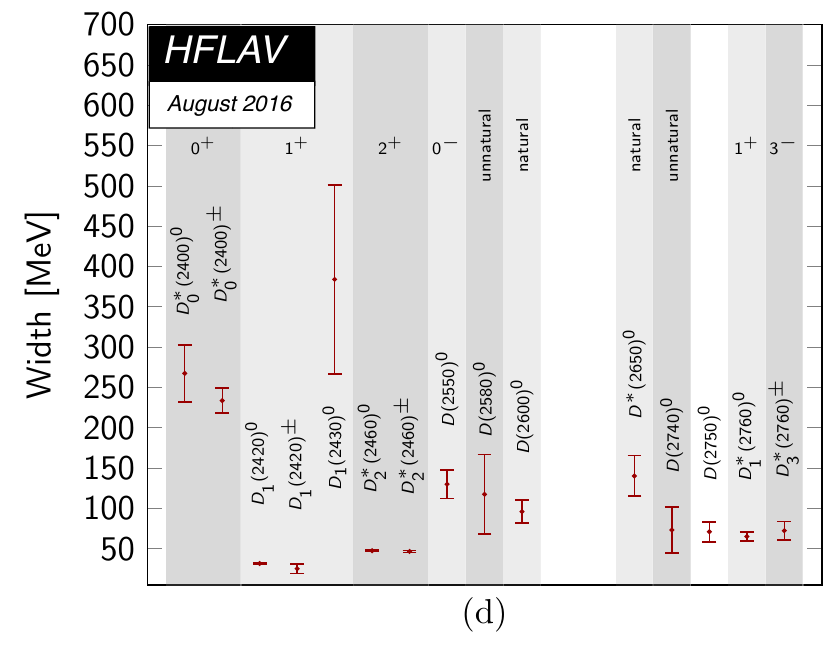}
\caption{\label{fig:charm:spect:1}  Averaged masses for excited $D_{s}$ mesons 
are shown in subfigure (a) and for $D$ mesons in subfigure (b). The 
average widths for excited $D_{s}$ mesons are shown in subfigure (c) and for excited 
$D$ mesons in subfigure (d). The vertical shaded regions distinguish 
between different spin parity states.}
\end{centering}
\end{figure}

Excited charm meson states have received increased attention since 
the first observation of states that could not be accommodated by QCD 
predictions~\cite{Aubert:2003fg,
Besson:2003cp,Abe:2003jk,Aubert:2003pe}. Tables \ref{table:charm:spect:1}, 
\ref{table:charm:spect:2a} and \ref{table:charm:spect:2} summarize recent 
measurements of the masses and widths of excited $D$ and $D_{s}$ mesons, 
respectively. If a preferred assignment of spin and parity was measured, 
it is listed in the column $J^{P}$, where the label natural denotes 
$J^{P}=0^{-},1^{+},2^{-}\ldots$ and unnatural $J^{P}=0^{+},1^{-},2^{+}\ldots$ 
If possible, an average mass and width are calculated; these are listed in 
the gray shaded row. The calculation of the averages assumes no correlation 
between individual measurements. A summary of the averaged masses and widths 
is shown in Figure~\ref{fig:charm:spect:1}. 

The masses and widths of narrow ($\Gamma<50$~MeV) orbitally excited 
$D$ mesons (1P states, denoted $D^{\ast\ast}$), both neutral and charged, are 
well-established. Measurements of broad states ($\Gamma\sim$ 200--400~MeV) 
are less abundant, as identifying the signal is more challenging. There is 
a slight discrepancy between the $D_0^\ast(2400)^0$ masses measured by the 
Belle~\cite{Abe:2003zm} and  FOCUS~\cite{Link:2003bd} experiments. No data 
exist yet for the $D_1(2430)^{\pm}$ state. Dalitz plot analyses of 
$B\to \overline{D}{}^{(\ast)}\pi\pi$ decays strongly favor the assignments $0^+$ 
and $1^+$ for the spin-parity quantum numbers of the 
$D_0^\ast(2400)^0/D_0^\ast(2400)^\pm$ and $D_1(2430)^{0}$ states, 
respectively. The measured masses and widths, as well as the $J^P$ 
values, are in agreement with theoretical predictions based on potential 
models~\cite{Godfrey:1985xj, Godfrey:1986wj, Isgur:1991wq, Schweitzer:2002nm}.

Tables~\ref{table:charm:spect:3} and \ref{table:charm:spect:4} summarize 
the branching fractions of $B$ meson decays to excited $D$ and $D_{s}$ 
states, respectively. It is notable that the branching fractions for 
$B$ mesons decaying to a narrow $D^{\ast\ast}$ state and a pion are similar 
for charged and neutral $B$ initial states, while the branching fractions 
to a broad $D^{\ast\ast}$ state and $\pi^+$ are much larger for $B^+$ than 
for $B^0$. This may be due to the fact that color-suppressed amplitudes 
contribute only to the $B^+$ decay and not to the $B^0$ decay (for a 
theoretical discussion, see Ref.~\cite{Jugeau:2005yr,Colangelo:2004vu}). 
Measurements of individual branching fractions of $D$ mesons are difficult 
due to the unknown fragmentation of a $c$ quark to $D^{\ast\ast}$ or due to 
the unknown $B \to D^{\ast\ast} X$ branching fractions.

The discoveries of the $D_{s0}^\ast(2317)^{\pm}$ and $D_{s1}(2460)^{\pm}$ 
have triggered increased interest in properties of, and searches for, 
excited $D_s$ mesons (here generically denoted $D_s^{\ast\ast}$). While 
the masses and widths of $D_{s1}(2536)^{\pm}$ and $D_{s2}^{\ast}(2573)^{\pm}$ 
states are in relatively good agreement with potential model predictions, 
the masses of $D_{s0}^\ast(2317)^{\pm}$ and $D_{s1}(2460)^{\pm}$ states are 
significantly lower than expected (see Ref.~\cite{Cahn:2003cw} for a 
discussion of $c\bar{s}$ models). Moreover, the mass splitting between 
these two states greatly exceeds that between the $D_{s1}(2536)^{\pm}$ 
and $D_{s2}(2573)^{\pm}$. These unexpected properties have led to 
interpretations of the $D_{s0}^\ast(2317)^{\pm}$ and $D_{s1}(2460)^{\pm}$ 
as exotic four-quark states~\cite{Barnes:2003dj,Lipkin:2003zk}.

While there are few measurements of the $J^P$ values of 
$D_{s0}^\ast(2317)^{\pm}$ and $D_{s1}(2460)^{\pm}$, the available data 
favor $0^+$ and $1^+$, respectively. A molecule-like ($DK$) interpretation 
of the $D_{s0}^\ast(2317)^{\pm}$ and 
$D_{s1}(2460)^{\pm}$~\cite{Barnes:2003dj,Lipkin:2003zk} that can account 
for their low masses and isospin-breaking decay modes is tested by searching 
for charged and neutral isospin partners of these states; thus far such 
searches have yielded negative results. Therefore the subset of models 
that predict equal production rates for different charged states is 
excluded. The molecular picture can also be tested by measuring the 
rates for the radiative processes 
$D_{s0}^\ast(2317)^{\pm}/D_{s1}(2460)^{\pm}\to D_s^{(\ast)}\gamma$ and 
comparing to theoretical predictions. The predicted rates, however, 
are below the sensitivity of current experiments. 

Another model successful in explaining the total widths and the 
$D_{s0}^\ast(2317)^{\pm}$ -- $D_{s1}(2460)^{\pm}$ mass splitting is based 
on the assumption that these states are chiral partners of the ground 
states $D_{s}^{+}$ and~$D_{s}^{*}$~\cite{Bardeen:2003kt}. While some 
measured branching fraction ratios agree with predicted values, further 
experimental tests with better sensitivity are needed to confirm or 
refute this scenario. A summary of the mass difference measurements 
is given in Table~\ref{table:charm:spect:5}.

Measurements by \babar{}~\cite{Aubert:2009ah} and LHCb~\cite{Aaij:2012pc} 
first indicated the existence of a strange-charm $D^{*}_{sJ}(2860)^{\pm}$ 
meson. An LHCb study of $B_{s}^{0}\to \overline{D}{}^{0}K^{-}\pi^{+}$ decays, 
in which they searched for excited $D_{s}$ mesons~\cite{Aaij:2014xza}, showed 
with $10\sigma$ significance that this state is comprised of two different 
particles,  one of spin 1 and one of spin 3. This represents the first 
measurement of a heavy flavored spin-3 particle, and the first observation 
of $B$ meson decays to spin 3 particles. A subsequent study of $D_{sJ}$ mesons 
by the LHCb collaboration~\cite{Aaij:2016utb} supports the natural parity 
assignment for this state ($J^P=3^-$). This study also shows weak evidence 
for a further structure at a mass around 3040 MeV$/c^2$ with unnatural 
parity, which was first hinted at by a \babar\ analysis~\cite{Aubert:2009ah}.

Recent evidence shows that the 1D family of charm resonances can be explored 
in the Dalitz plot analyses of $B$-meson decays in the same way as seen for 
the charm-strange resonances. The LHCb collaboration performed an analysis 
of $B^0 \to \overline{D}{}^0\pi^+\pi^-$ decays, in which they measured the 
spin-parity assignment of the state $D^*_3(2760)^{\pm}$, which was observed 
previously by \babar{}~\cite{delAmoSanchez:2010vq} and LHCb~\cite{Aaij:2013sza}, 
to be $J^P=3^-$. The measurement suggests a spectroscopic assignment of ${}^3D_3$. 
This is the second observation of a spin-3 charm meson. 

Other observed excited $D_s$ states include $D_{s1}^{\ast}(2700)^{\pm}$  and 
$D_{s2}^{\ast}(2573)^{\pm}$. The properties of both (mass, width, $J^P$) have 
been measured and determined in several analyses. A theoretical 
discussion~\cite{Matsuki:2006rz} investigates the possibility that 
the $D_{s1}(2700)^{\pm}$ could represent radial excitations of the 
$D_s^{\ast\pm}$. Similarly, the  $D_{s1}^{\ast}(2860)^{\pm}$ and $D_{sJ}(3040)^{\pm}$ 
could be excitations of $D_{s0}^\ast(2317)^{\pm}$ and $D_{s1}(2460)^{\pm}$ or 
$D_{s1}(2536)^{\pm}$, respectively.

Table~\ref{table:charm:spect:6} summarizes measurements of the helicity parameter
 $A_{D}$ (also referred to as polarization amplitude). In $D^{\ast\ast}$ meson 
decays to $D^{\ast\ast} \to D^{\ast}\pi$, $D^{\ast} \to D \pi$, the helicity distribution varies like $1 + A_{D}\cos^{2}\theta_{H}$, 
where $\theta_{H}$ is the angle in the $D^{\ast}$ rest frame between the two 
pions emitted by decay $D^{\ast\ast} \to D^{\ast}\pi$ and the $D^{\ast} \to D \pi$. 
The parameter is sensitive to possible S-wave contributions in the decay. 
In the case of a $D$ meson decay decaying purely via D-wave, 
the helicity parameter is predicted to give $A_{D}=3$. Studies of the 
$D_{1}(2420)^{0}$ meson by the ZEUS and \babar{} collaborations suggest 
that there is an S-wave admixture in the decay, which is  contrary to 
Heavy Quark Effective Theory calculations~\cite{Isgur:1989vq,Neubert:1993mb}.

\begin{table}[htb!]
\caption{\label{table:charm:spect:1} Recent measurements of mass and 
width for different excited $D_{s}$ mesons. The column $J^{P}$ list 
the most significant assignment of spin and parity. If possible an 
average mass or width is calculated.}
\begin{adjustbox}{width=\textwidth,center}
{\setlength\tabcolsep{0pt}

}
\end{adjustbox}

\end{table} 

\begin{table}[htb!]
\caption{\label{table:charm:spect:2a} Recent measurements of mass 
and width for different excited $D$ mesons. The column $J^{P}$ list 
the most significant assignment of spin and parity. If possible an 
average mass or width is calculated.}
\begin{adjustbox}{width=\textwidth,center}
{\setlength\tabcolsep{0pt}
 %
}
\end{adjustbox}

\end{table} 

\begin{table}[htb!]
\caption{\label{table:charm:spect:2} Recent measurements of mass and 
width for different excited $D$ mesons. The column $J^{P}$ list the 
most significant assignment of spin and parity. If possible an 
average mass or width is calculated.}
\begin{adjustbox}{width=\textwidth,center}
{\setlength\tabcolsep{0pt}
 %
}
\end{adjustbox}

\end{table}

\begin{table}[htb!]
\begin{center}
\caption{\label{table:charm:spect:3} 
Product of $B$ meson branching fraction and 
(daughter) excited $D$ meson branching fraction.}
\begin{adjustbox}{width=\textwidth,center}
{\setlength\tabcolsep{0pt}
 %
}
\end{adjustbox}

\end{center}
\end{table} 

\begin{table}[htb!]
\begin{center}
\caption{\label{table:charm:spect:4} 
Product of $B$ meson branching fraction and 
(daughter) excited $D^{}_s$ meson branching fraction.}
\begin{adjustbox}{width=\textwidth,center}
{\setlength\tabcolsep{0pt}
 %
}
\end{adjustbox}

\end{center}
\end{table}

\begin{table}[htb!]
\begin{center}
\caption{\label{table:charm:spect:5} 
Mass difference measurements for excited $D$ mesons.}
{\setlength\tabcolsep{0pt}
 %
}

\end{center}
\end{table} 

\begin{table}[htb!]
\begin{center}
\caption{\label{table:charm:spect:6}
Measurements of polarization amplitudes for excited $D$ mesons.}
{\setlength\tabcolsep{0pt}
 %
}

\end{center}
\end{table}

\clearpage
% Lambda_c+ decays
\subsection{$\lcp$ branching fractions}

Charmed baryon decays play an important role in studies of weak 
and strong interactions. For example, they provide crucial input
for measurements of exclusive and inclusive decay rates of $b$-flavored 
mesons and baryons, and also for measurements of fragmentation fractions 
of charm and bottom quarks. In spite of this importance, experimental 
data on $\lcp$ baryon decays was scarce until 2014, when Belle 
published the first model-independent measurement of the
branching fraction for $\lcp\to pK^-\pi^+$~\cite{Zupanc:2013iki}.
This measurement improved upon the precision of previous 
(model-dependent) measurements by a factor of five. Since then
the precision of other $\lcp$ branching fractions has improved 
due to measurements based on threshold data performed by 
BESIII~\cite{Ablikim:2015flg}. BESIII also reported the 
first measurement of the branching fraction for the
semileptonic decay
$\lcp\to \Lambda e^+ \nu_e$~\cite{Ablikim:2015prg}. Here we 
present a global fit for branching fractions of Cabbibo-favored 
$\lcp$ decays, taking into account all relevant experimental
measurements and their correlations. All measurements used 
assume unpolarised production of the $\lcp$.

The measurements listed in Table~\ref{tab:Lc:br-fit} are input
to a least-squares fit minimizing a $\chi^2$ statistic. The fitted 
quantities are the $\lcp$ branching fractions for twelve hadronic 
modes and one semileptonic mode. 
The measurements are labelled using the $\Gamma_n$ notation
employed by the Particle Data Group~\cite{PDG_2016}, where $n$ 
is an integer that specifies the decay mode. 
The fitted output consists of 13 quantities -- twelve hadronic and 
one semileptonic branching fraction. The advantage of our fit is that
it takes into account correlations among measurements from the same 
experiment, \ie, systematic uncertainties related to normalization, 
track-finding efficiency, particle identification efficiency, and
$\pi^0$, $K^0_S$, and $\Lambda$ reconstruction efficiencies. For the
twelve hadronic branching fractions measured by BESIII,
we use BESIII's published correlation matrix~\cite{Ablikim:2015flg}.

\begin{figure}[hbt!]
\centering
\includegraphics[width=\textwidth]{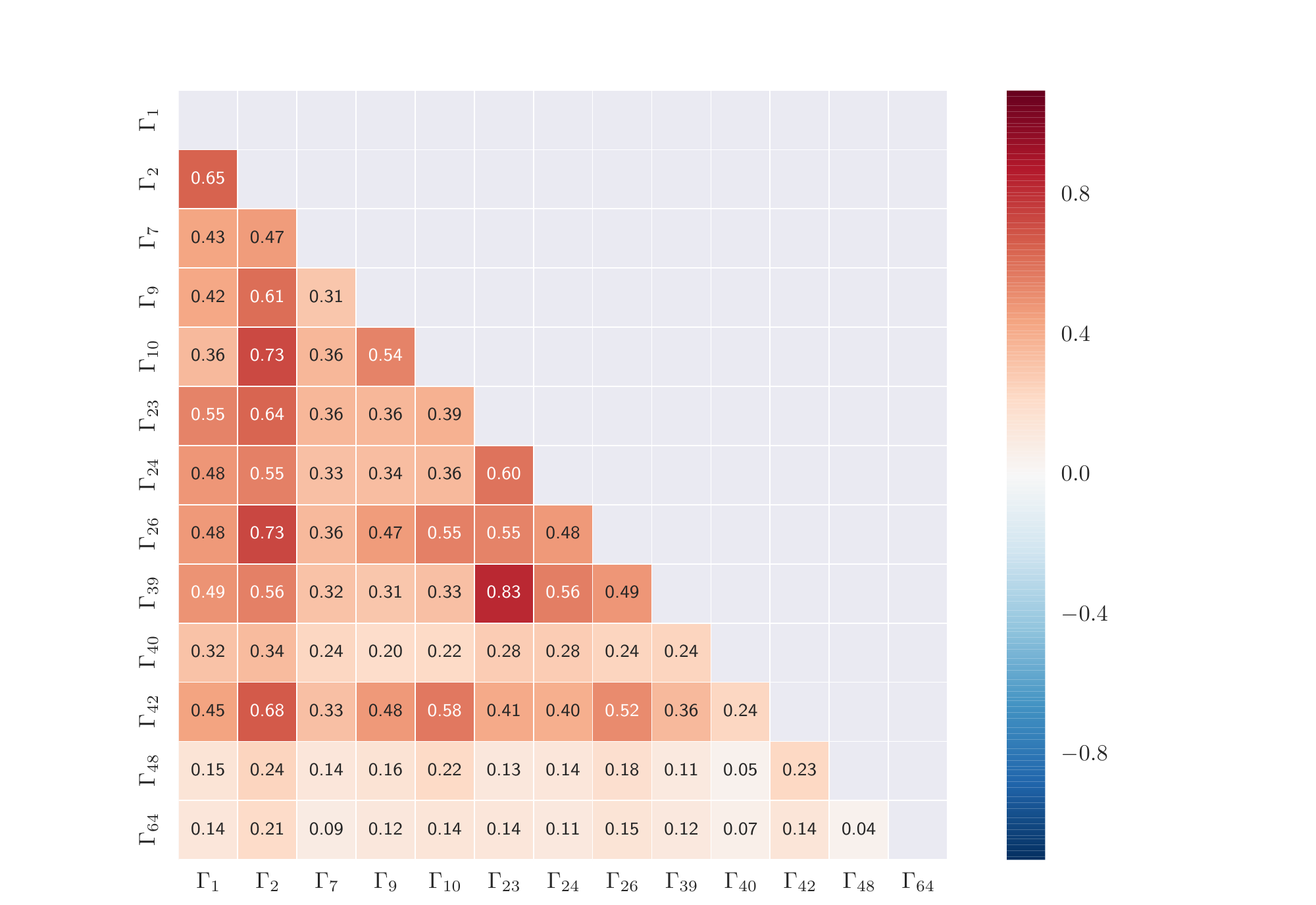}
\caption{Correlation coefficients between averaged $\lcp$ branching fractions.
\label{fig:Lc:corrM}}
\end{figure}

\begin{figure}[hbt!]
\includegraphics[width=0.32\textwidth]{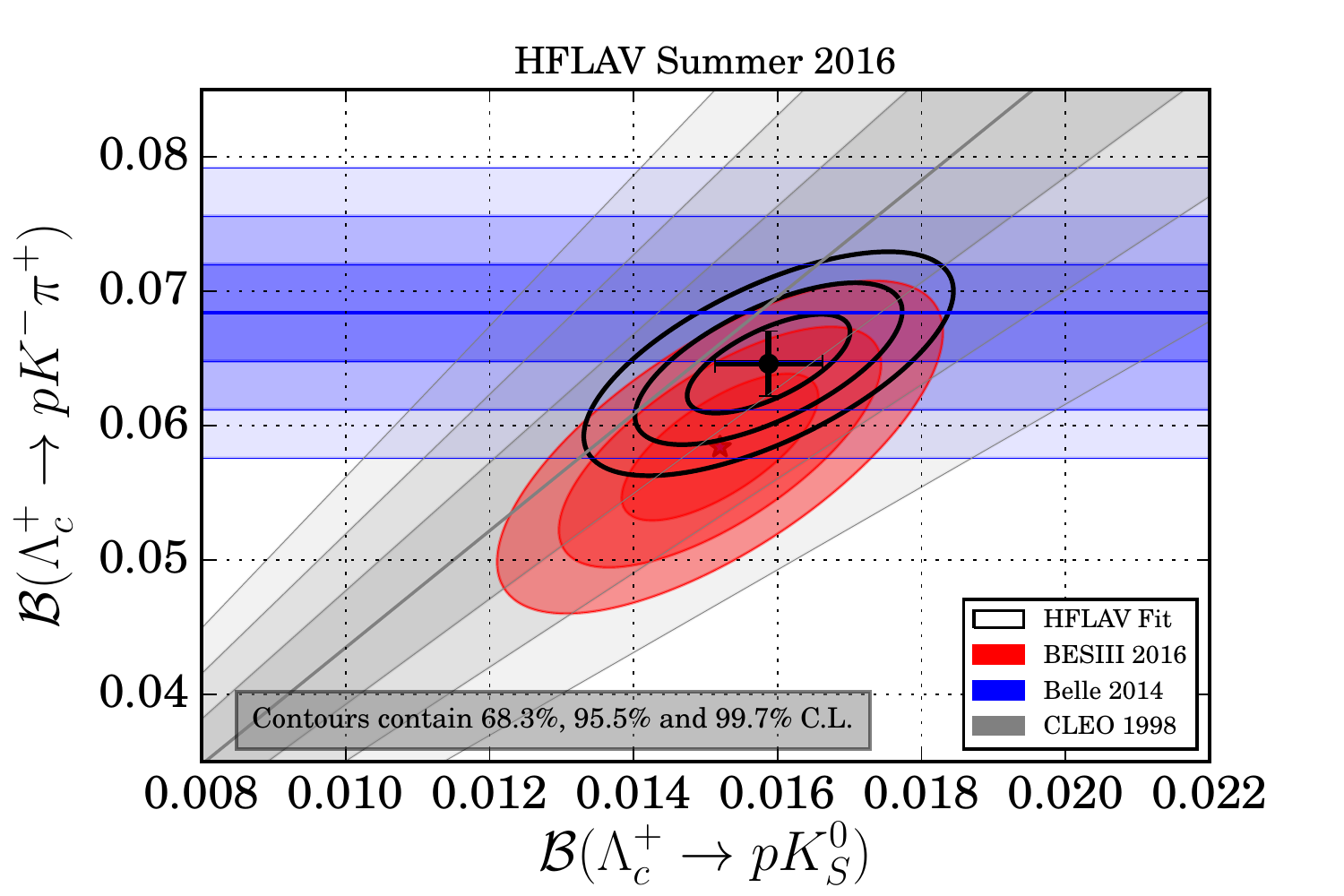}
\includegraphics[width=0.32\textwidth]{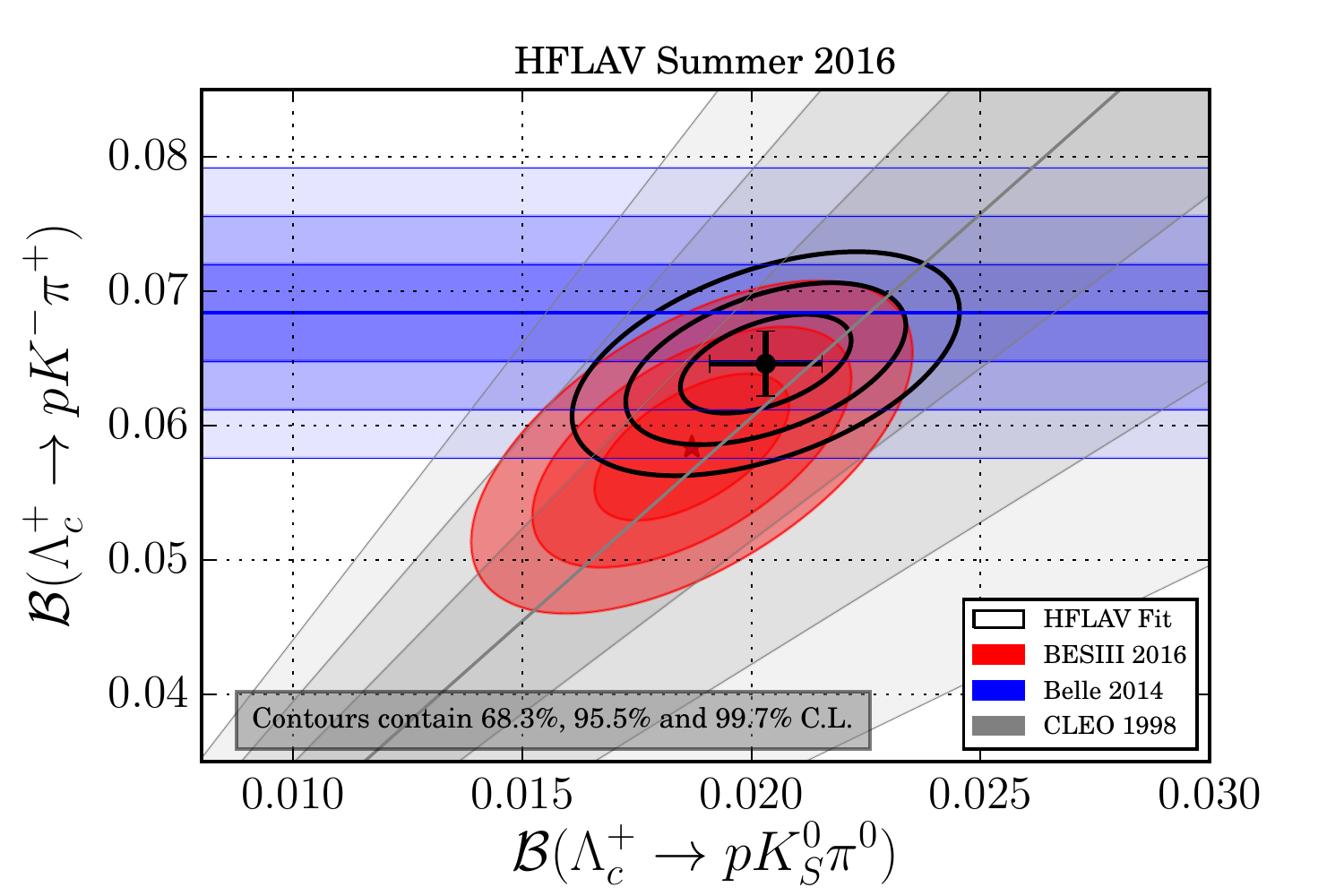}
\includegraphics[width=0.32\textwidth]{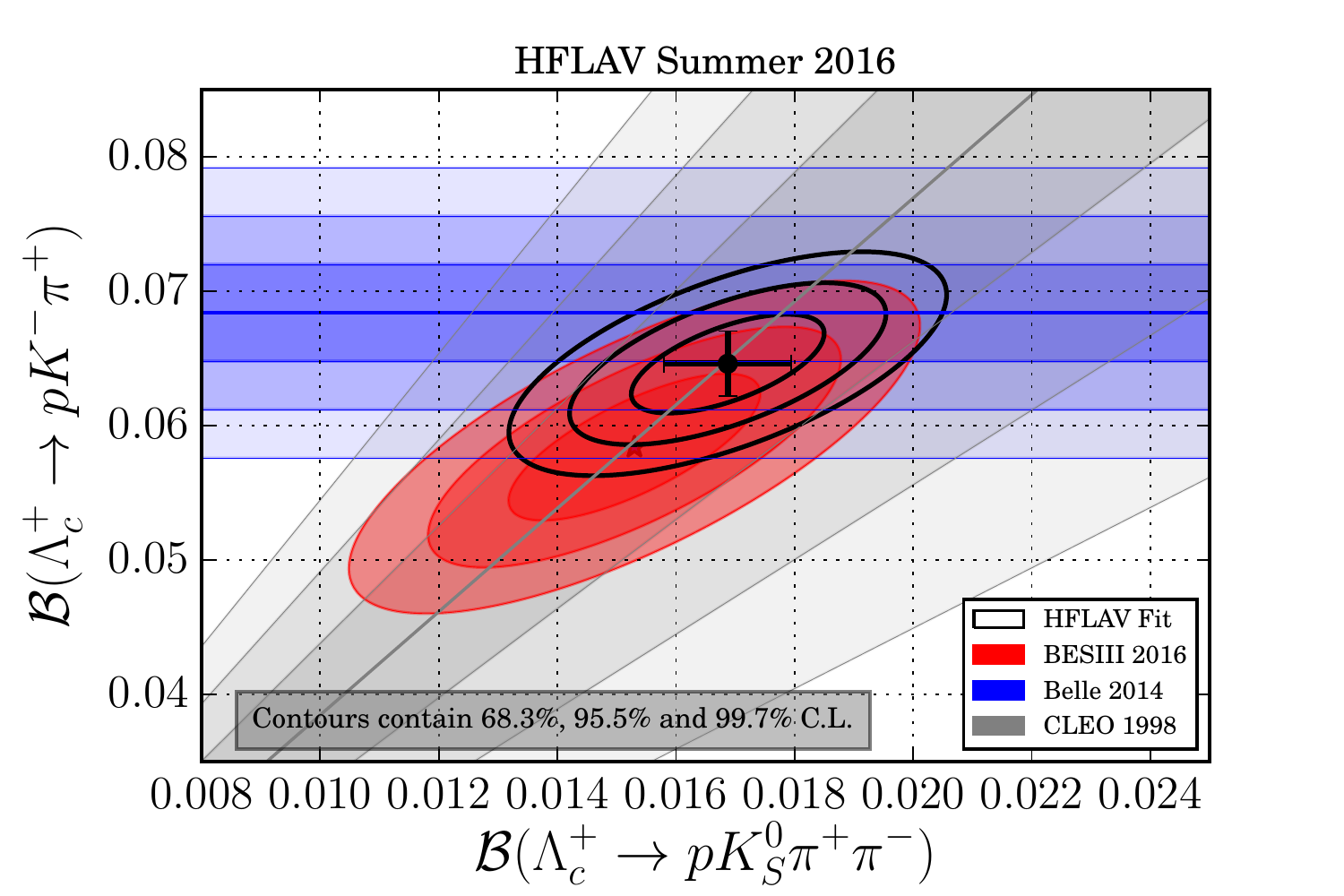}\\
\includegraphics[width=0.32\textwidth]{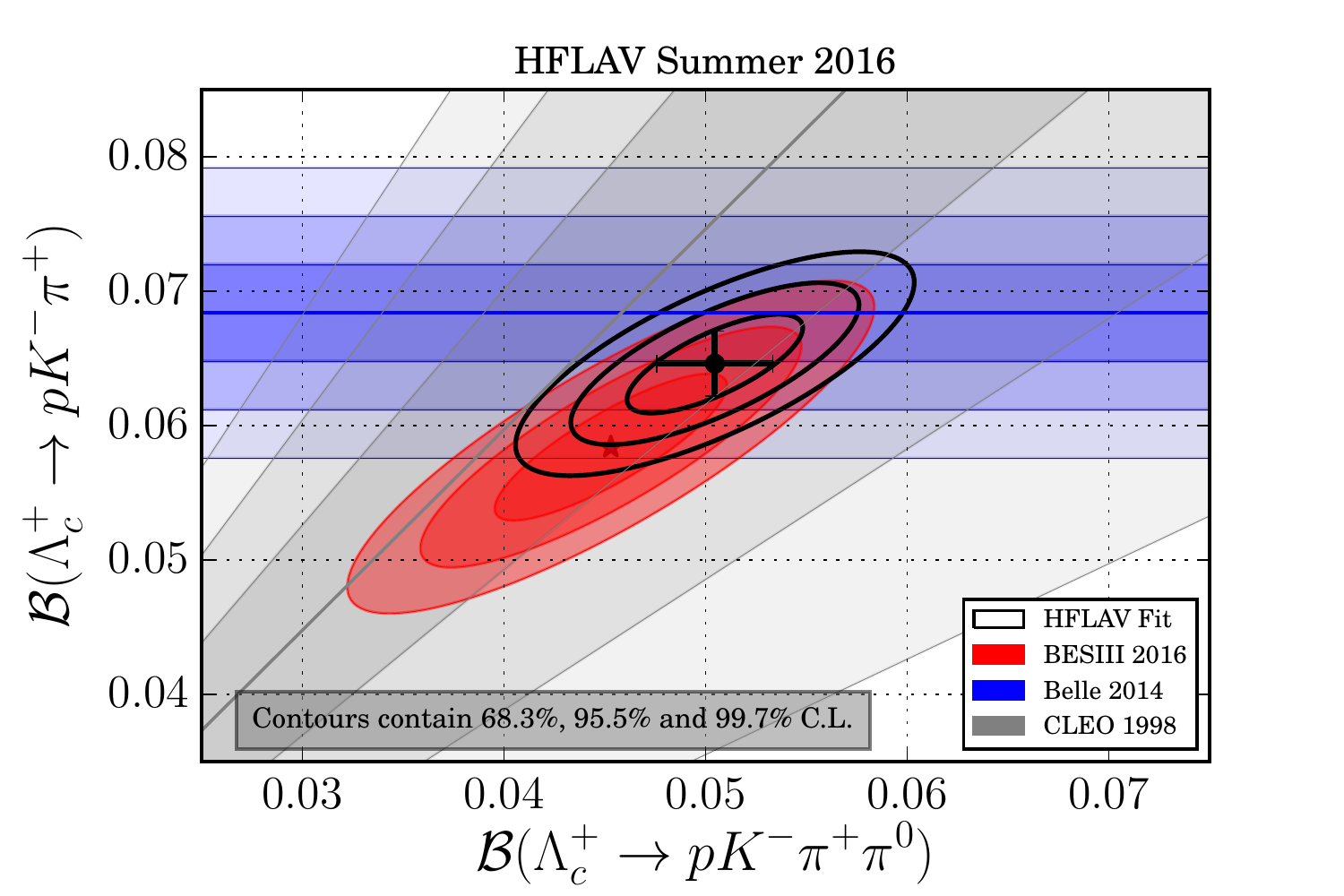}
\includegraphics[width=0.32\textwidth]{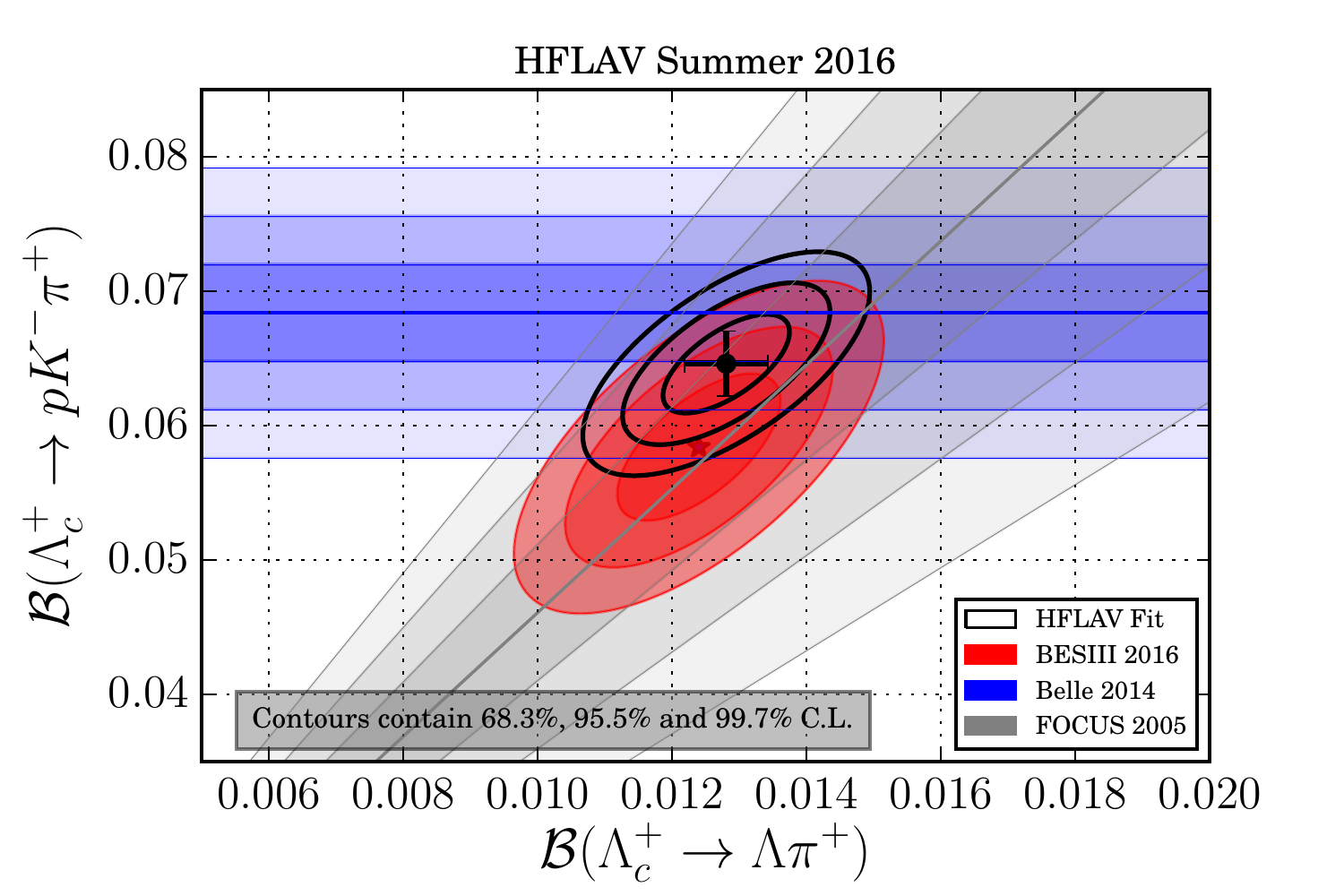}
\includegraphics[width=0.32\textwidth]{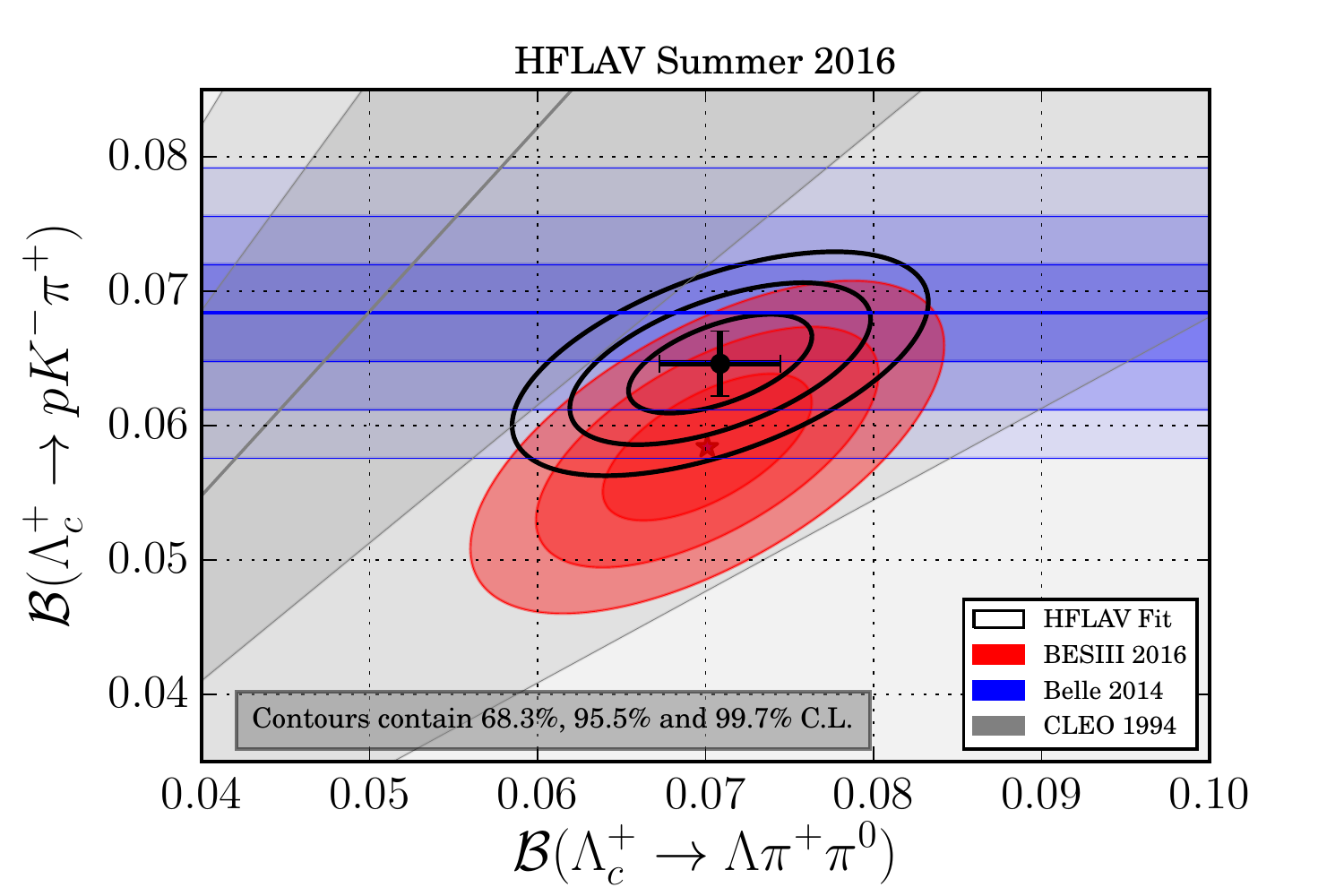}\\
\includegraphics[width=0.32\textwidth]{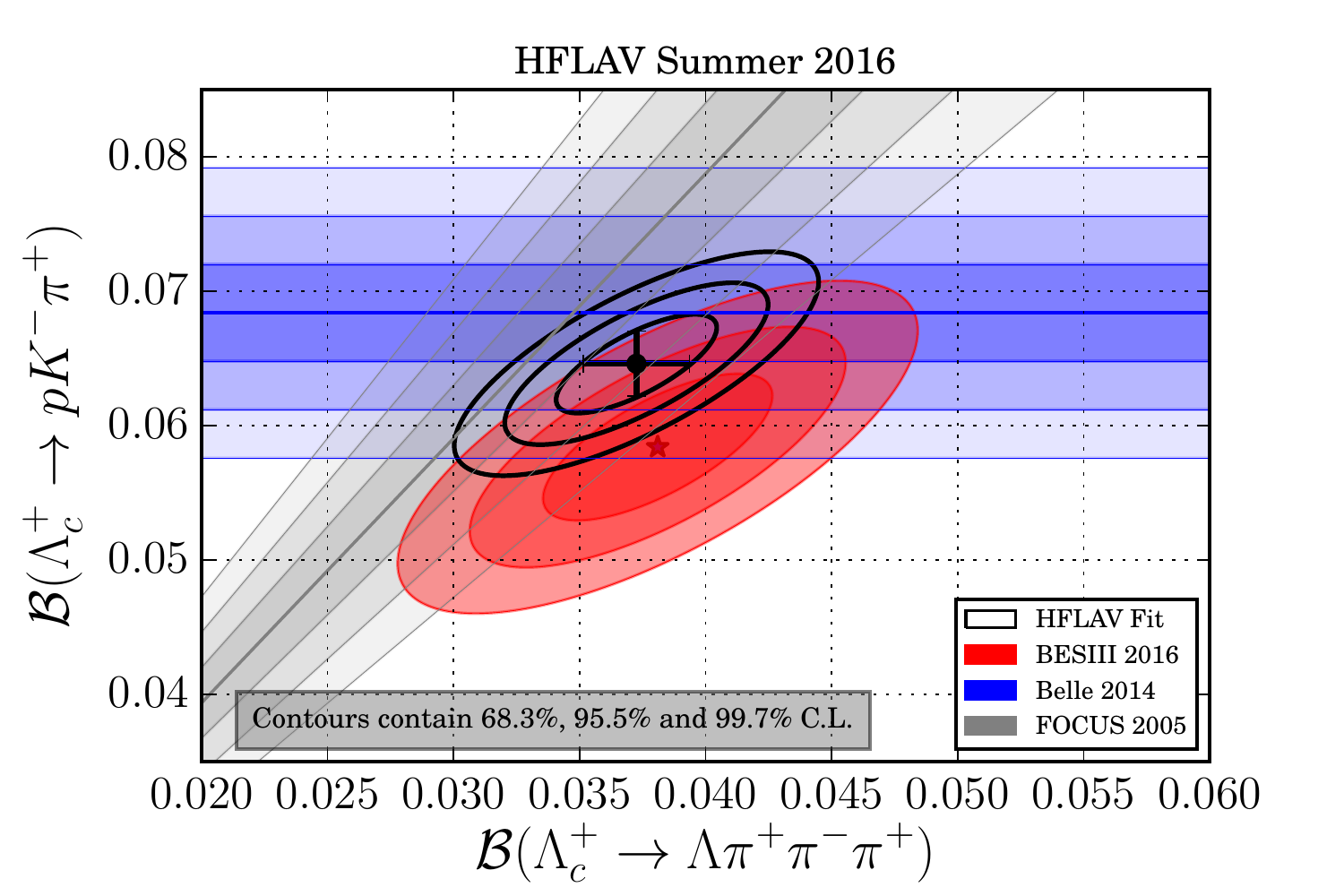}
\includegraphics[width=0.32\textwidth]{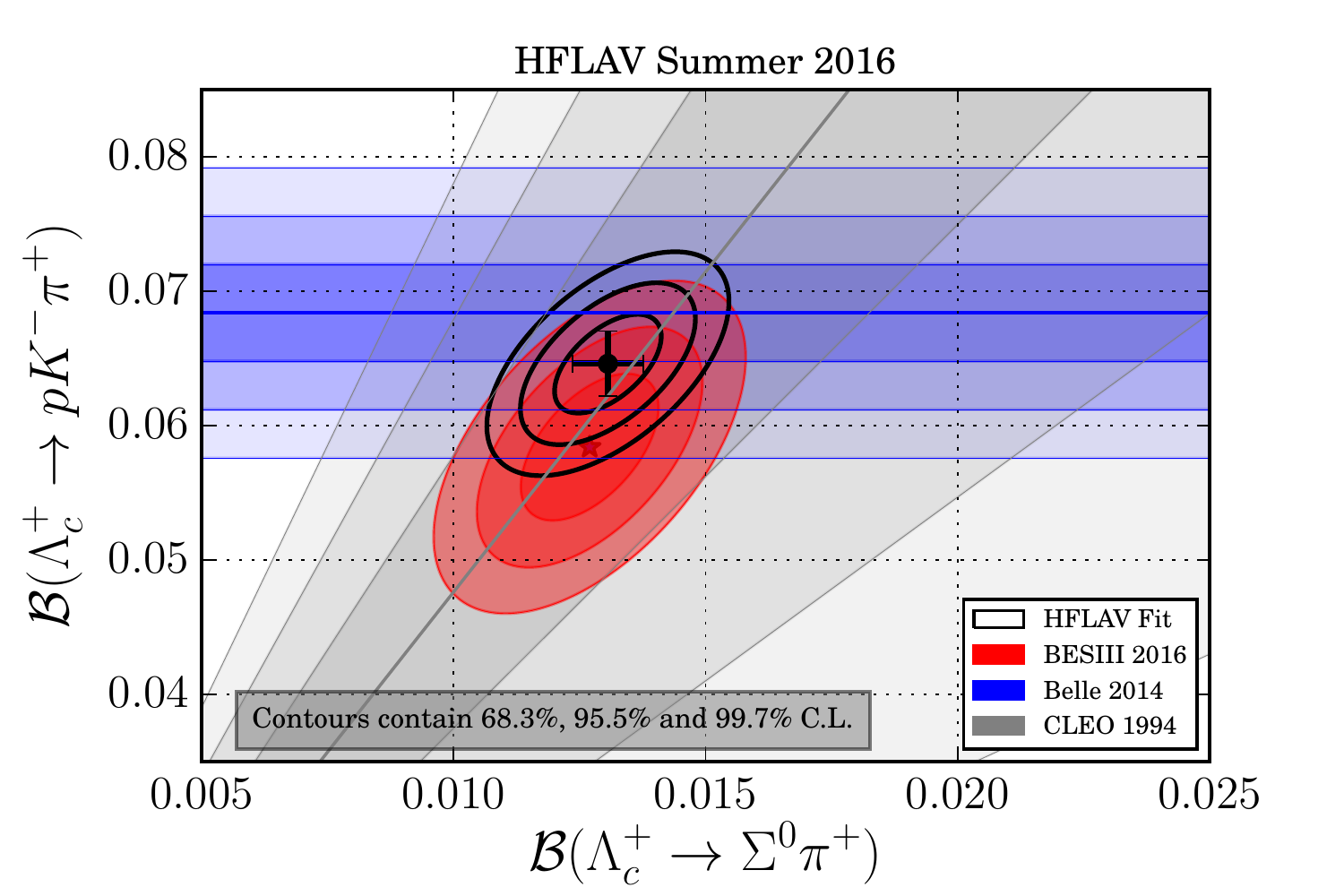}
\includegraphics[width=0.32\textwidth]{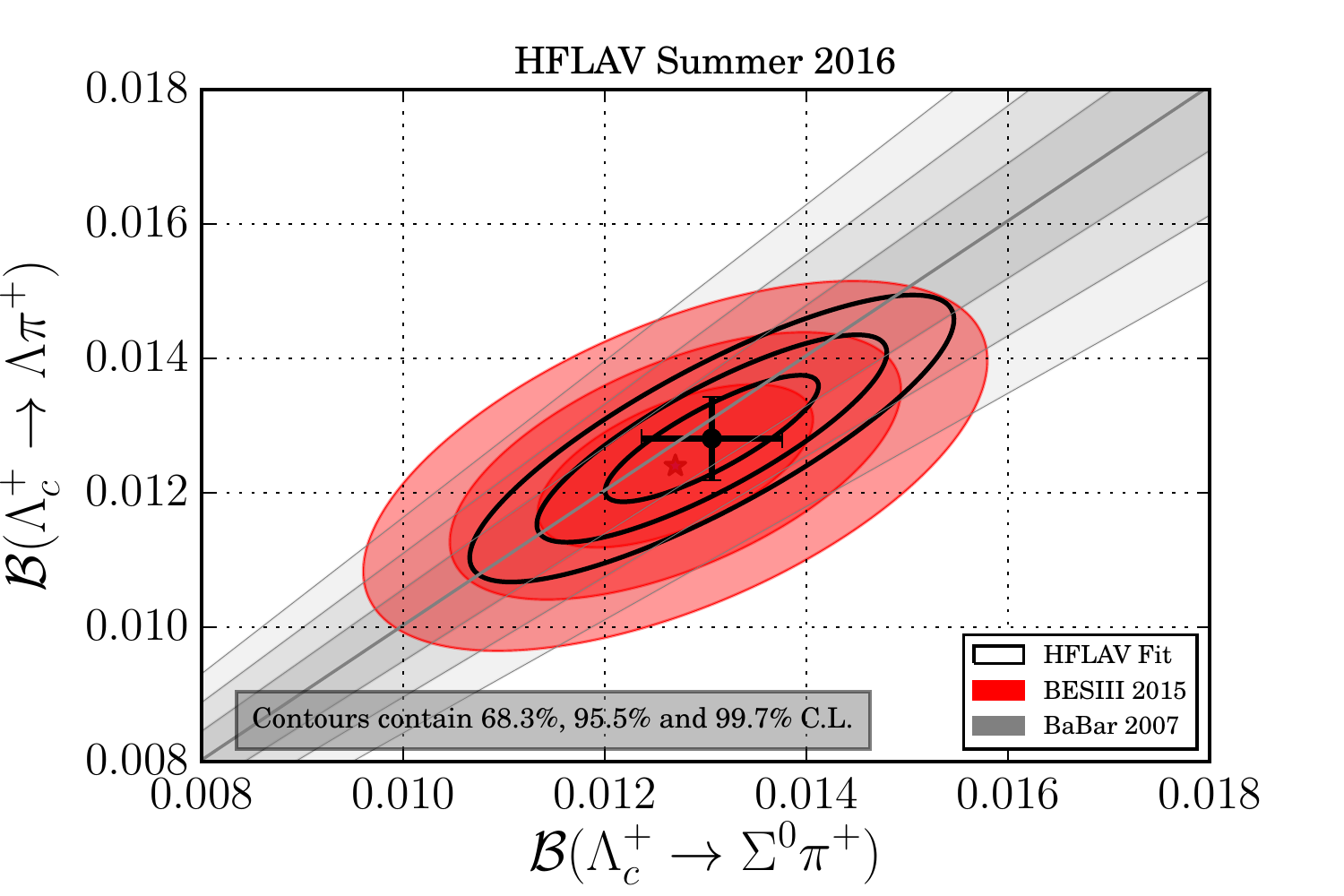}\\
\includegraphics[width=0.32\textwidth]{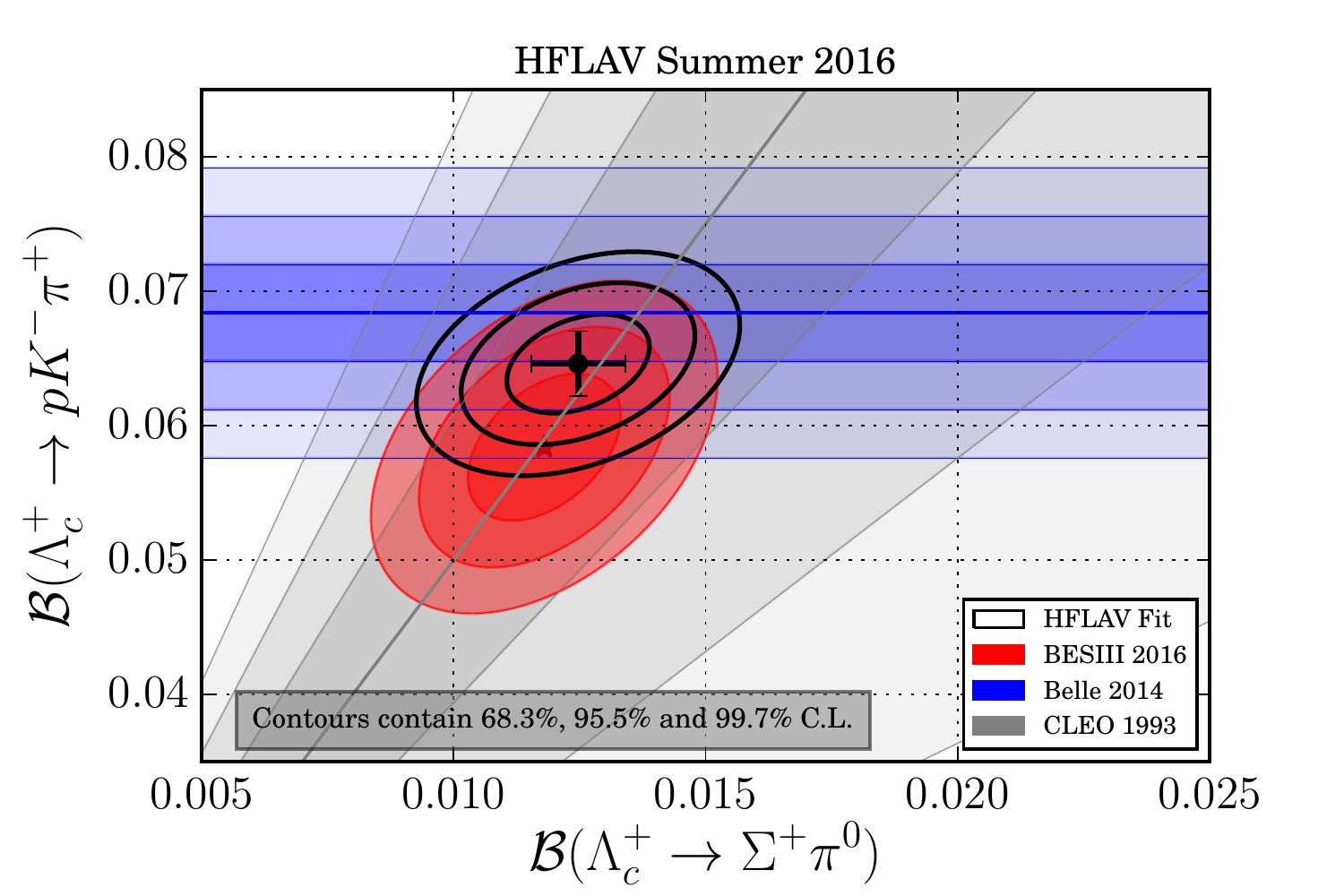}
\includegraphics[width=0.32\textwidth]{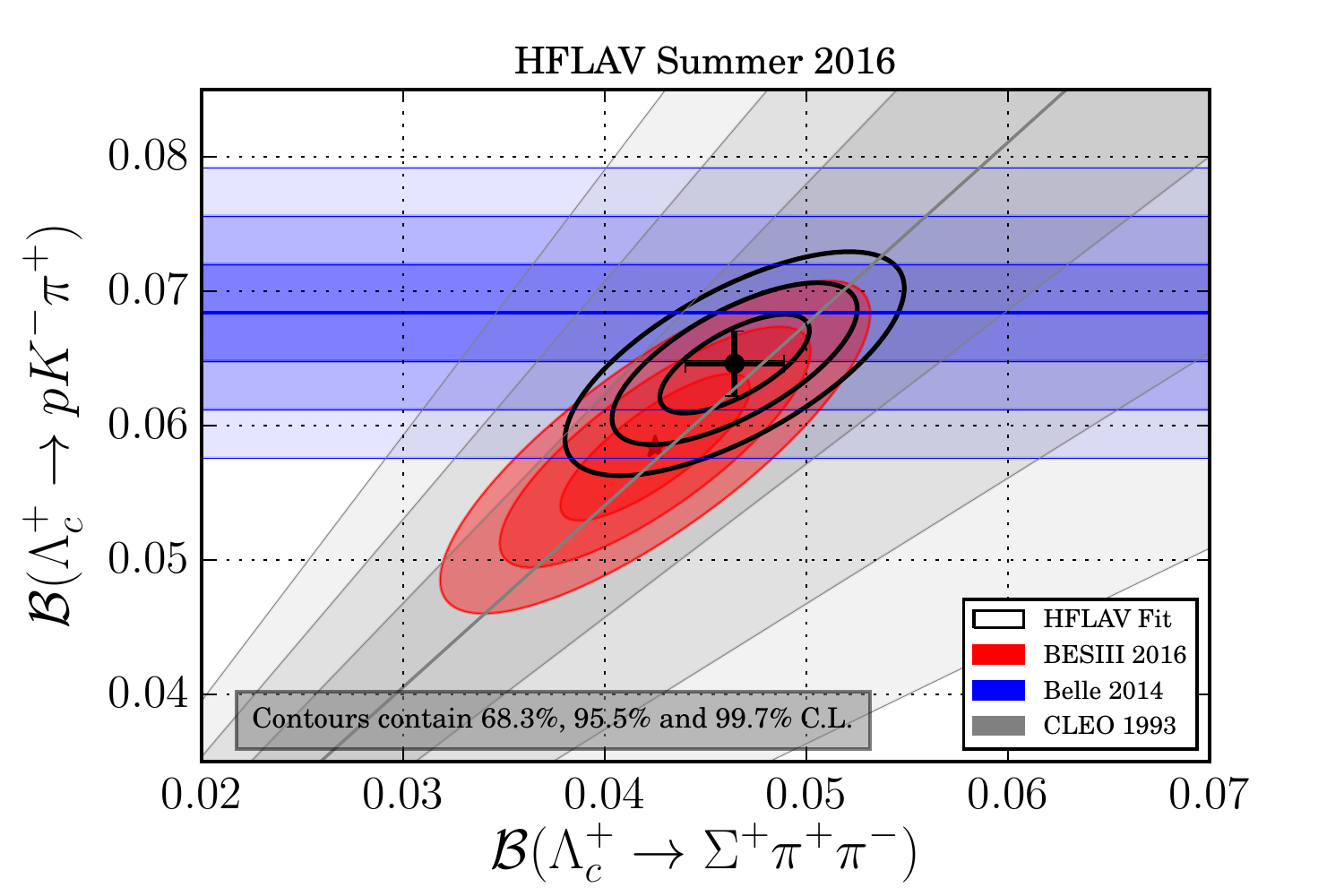}
\includegraphics[width=0.32\textwidth]{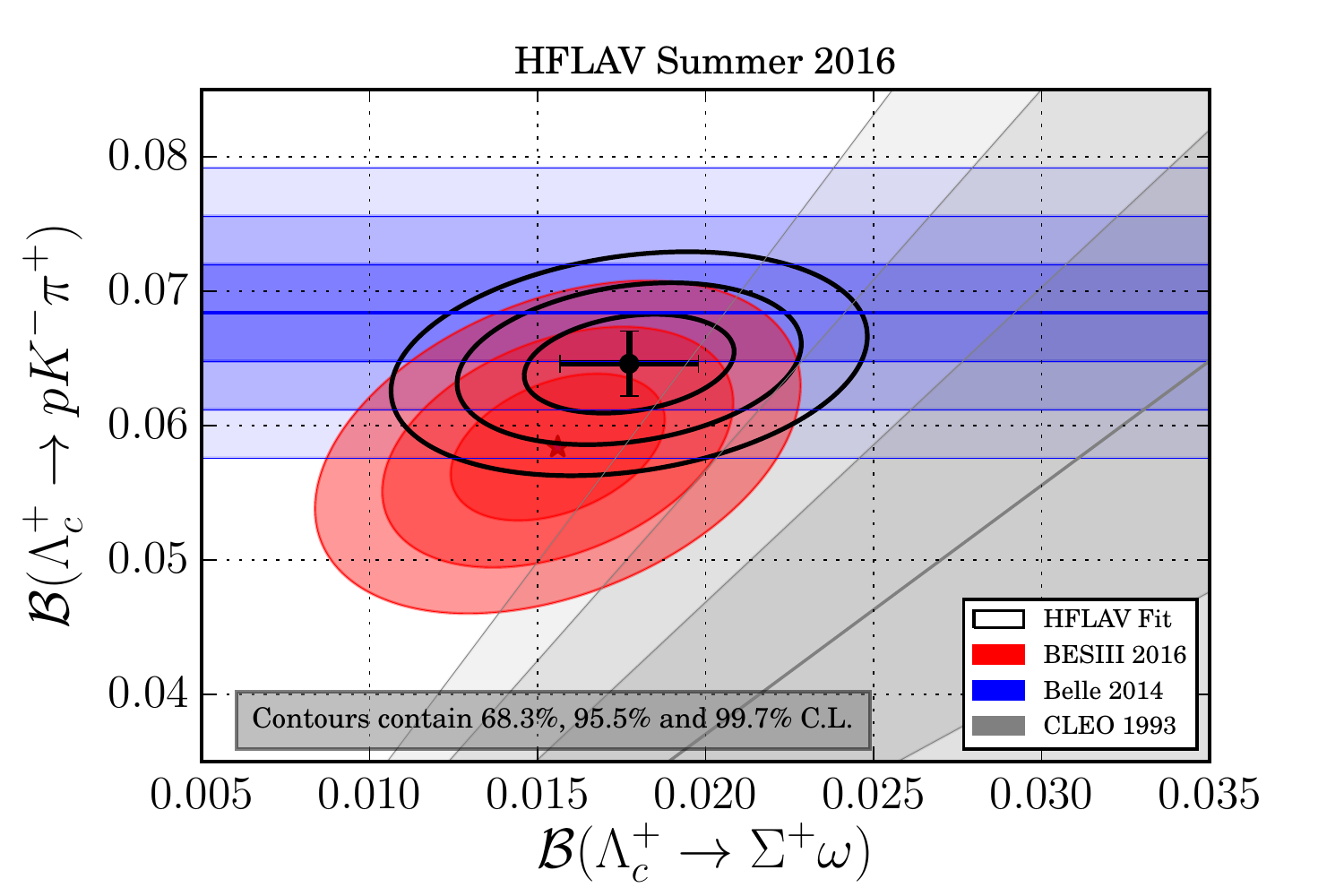}\\
\includegraphics[width=0.32\textwidth]{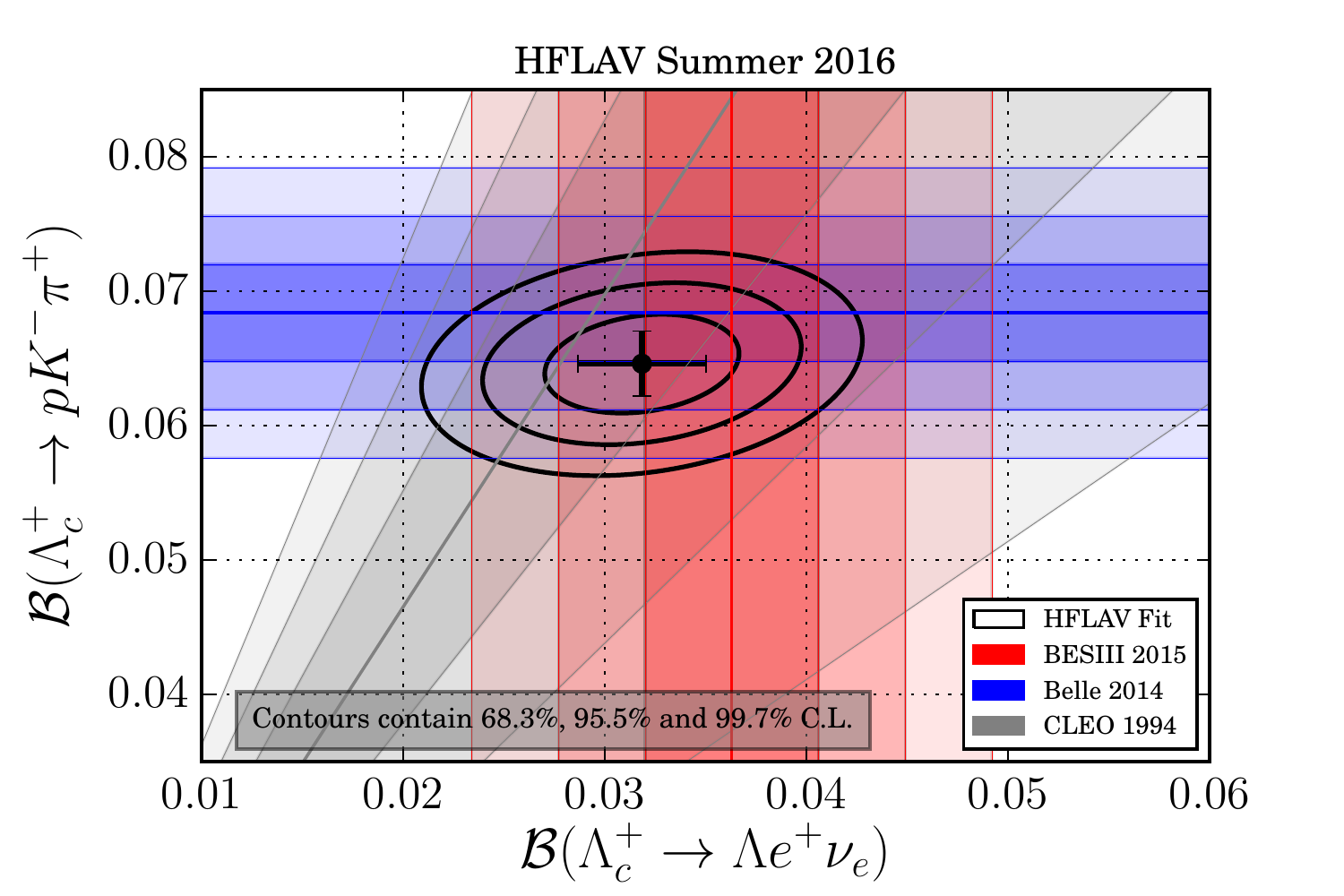}
\caption{Plots of all individual measurements and the fitted averages. 
Individual measurements are plotted as bands (ellipses) showing their 
$\pm1\sigma$, $\pm2\sigma$, and $\pm3\sigma$ ranges. The best fit value 
is indicated by a cross showing the one-dimensional errors. In cases 
where multiple ratio measurements exists ($\Gamma_i/\Gamma_j$), only 
the most precise one is plotted.
\label{fig:Lc:Brs}}
\end{figure}

The resulting fitted values for the 
%twelve hadronic and one semileptonic 
branching fractions are given in Table~\ref{tab:Lc:br-fit}. 
The overall $\chi^2$ of the fit is 30.0 for 23 degrees 
of freedom, which corresponds to a $p$ value of 0.149. The 
correlation matrix for the fitted branching fractions is shown 
in Fig.~\ref{fig:Lc:corrM}, and constraints from individual 
measurements for pairs of fitted branching fractions are 
shown in Fig.~\ref{fig:Lc:Brs}. The branching fraction 
of the normalisation decay $\lcp\to pK^-\pi^+$ is found 
to be
\[
{\cal B}(\lcp\to pK^-\pi^+)=(6.46\pm 0.24)\%.
\] 
%%
%% quantities and measurements
%%
\begin{center}
%\begin{envsmall}
%\setlength{\LTcapwidth}{0.85\linewidth}
\renewcommand*{\arraystretch}{1.3}%
\begin{longtable}{llll}
\caption{Experimental results and world averages for branching 
fractions of twelve hadronic and one semileptonic $\lcp$ decay. 
The first uncertainty is statistical and the second is systematic.
\label{tab:Lc:br-fit}}%
\\
\hline\hline
\rowcolor{Gray}
\multicolumn{1}{l}{\bfseries $\lcp$ branching fraction} &
\multicolumn{1}{l}{\bfseries Value} &
\multicolumn{1}{l}{\bfseries Reference} \\
\hline
\endfirsthead
\multicolumn{4}{c}{{\bfseries \tablename\ \thetable{} -- continued from previous page}} 
\\ \hline
\multicolumn{1}{l}{\bfseries  $\lcp$ branching fraction} &
\multicolumn{1}{l}{\bfseries Value} &
\multicolumn{1}{l}{\bfseries Reference} \\
\hline
\endhead
\endfoot
\endlastfoot
\rowcolor{LightGray}
\boldmath
$\Gamma_1=p K^0_S$ 					& \boldmath$(1.59\pm 0.07)\%$ 			& HFLAV Fit\\
BESIII								& $(1.52\pm0.08\pm0.03)\%$ 				& \cite{Ablikim:2015flg}\\
\hline
\rowcolor{LightGray}
\boldmath
$\frac{\Gamma_1}{\Gamma_2}=\frac{p K^0_S}{p K^- \pi^+}$ 					
									& \boldmath$0.246\pm0.009$ 				& HFLAV Fit\\
CLEO								& $0.22\pm0.04\pm0.03$ 					& \cite{Avery:1990bc}\\
CLEO								& $0.23\pm0.01\pm0.02$ 					& \cite{Alam:1998nb}\\
\hline
\rowcolor{LightGray}
\boldmath
$\Gamma_2=p K^- \pi^+$ 					& \boldmath$(6.46\pm0.24)\%$				& HFLAV Fit\\
Belle									& $(6.84\pm0.24{}^{+0.21}_{-0.27})\%$		& \cite{Zupanc:2013iki}\\
BESIII								& $(5.84\pm0.27\pm0.23)\%$ 				& \cite{Ablikim:2015flg}\\
\hline
\rowcolor{LightGray}
\boldmath
$\Gamma_7=p K^0_S\pi^0$	 			& \boldmath$(2.03\pm0.12)\%$ 			& HFLAV Fit\\
BESIII								& $(1.87\pm0.13\pm0.05)\%$ 				& \cite{Ablikim:2015flg}\\
\hline
\rowcolor{LightGray}
\boldmath
$\frac{\Gamma_7}{\Gamma_2}=\frac{p K^0_S\pi^0}{p K^- \pi^+}$ 					
									& \boldmath$0.314\pm0.017$ 				& HFLAV Fit\\
CLEO								& $0.33\pm0.03\pm0.04$ 					& \cite{Alam:1998nb}\\
\hline
\rowcolor{LightGray}
\boldmath
$\Gamma_9=p K^0_S\pi^+\pi^-$			& \boldmath$(1.69\pm0.11)\%$ 			& HFLAV Fit\\
BESIII								& $(1.53\pm0.11\pm0.09)\%$ 				& \cite{Ablikim:2015flg}\\
\hline
\rowcolor{LightGray}
\boldmath
$\frac{\Gamma_9}{\Gamma_2}=\frac{p K^0_S\pi^+\pi^-}{p K^- \pi^+}$ 					
									& \boldmath$0.261\pm0.013$ 				& HFLAV Fit\\
CLEO								& $0.22\pm0.06\pm0.02$ 					& \cite{Avery:1990bc}\\
CLEO								& $0.26\pm0.02\pm0.03$ 					& \cite{Alam:1998nb}\\
\hline
\rowcolor{LightGray}
\boldmath
$\Gamma_{10}=p K^-\pi^+\pi^0$			& \boldmath$(5.05\pm0.29)\%$ 					& HFLAV Fit\\
BESIII								& $(4.53\pm0.23\pm0.30)\%$ 				& \cite{Ablikim:2015flg}\\
\hline
\rowcolor{LightGray}
\boldmath
$\frac{\Gamma_{10}}{\Gamma_2}=\frac{p K^- \pi^+ \pi^0}{p K^- \pi^+}$ 					
									& \boldmath$0.781\pm0.031$ 				& HFLAV Fit\\
CLEO								& $0.67\pm0.04\pm0.11$ 					& \cite{Alam:1998nb}\\
\hline
\rowcolor{LightGray}
\boldmath
$\Gamma_{23}=\Lambda \pi^+$			& \boldmath$(1.28\pm0.06)\%$ 			& HFLAV Fit\\
BESIII								& $(1.24\pm0.07\pm0.03)\%$ 				& \cite{Ablikim:2015flg}\\
\hline
\rowcolor{LightGray}
\boldmath
$\frac{\Gamma_{23}}{\Gamma_2}=\frac{\Lambda \pi^+}{pK^-\pi^+}$			
									& \boldmath$0.198\pm0.008$ 				& HFLAV Fit\\
CLEO								& $0.18\pm0.03\pm0.03$ 					& \cite{Avery:1990bc}\\
ARGUS								& $0.18\pm0.03\pm0.04$					& \cite{Albrecht:1991vs}\\
FOCUS								& $0.217\pm0.013\pm0.020$ 				& \cite{Link:2005ut}\\
\hline
\rowcolor{LightGray}
\boldmath
$\Gamma_{24}=\Lambda \pi^+\pi^0$			& \boldmath$(7.09\pm0.36)\%$ 			& HFLAV Fit\\
BESIII								& $(7.01\pm0.37\pm0.19)\%$ 				& \cite{Ablikim:2015flg}\\
\hline
\rowcolor{LightGray}
\boldmath
$\frac{\Gamma_{24}}{\Gamma_2}=\frac{\Lambda \pi^+\pi^0}{pK^-\pi^+}$			
									& \boldmath$1.10\pm0.05$ 				& HFLAV Fit\\
CLEO								& $0.73\pm0.09\pm0.16$ 					& \cite{Avery:1993ri}\\
\hline
\rowcolor{LightGray}
\boldmath
$\Gamma_{26}=\Lambda \pi^+\pi^-\pi^+$		& \boldmath$(3.73\pm0.21)\%$ 			& HFLAV Fit\\
BESIII								& $(3.81\pm0.24\pm0.18)\%$ 				& \cite{Ablikim:2015flg}\\
\hline
\rowcolor{LightGray}
\boldmath
$\frac{\Gamma_{26}}{\Gamma_2}=\frac{\Lambda \pi^+\pi^-\pi^+}{pK^-\pi^+}$			
									& \boldmath$0.577\pm0.022$ 				& HFLAV Fit\\
CLEO								& $0.65\pm0.11\pm0.12$ 					& \cite{Avery:1990bc}\\
FOCUS								& $0.508\pm0.024\pm0.024$ 				& \cite{Link:2005ut}\\
ARGUS								& $0.61\pm0.16\pm0.04$ 					& \cite{Albrecht:1988an}\\
\hline
\rowcolor{LightGray}
\boldmath
$\Gamma_{39}=\Sigma^0 \pi^+$			& \boldmath$(1.31\pm0.07)\%$ 			& HFLAV Fit\\
BESIII								& $(1.27\pm0.08\pm0.03)\%$ 				& \cite{Ablikim:2015flg}\\
\hline
\rowcolor{LightGray}
\boldmath
$\frac{\Gamma_{39}}{\Gamma_2}=\frac{\Sigma^0 \pi^+}{pK^-\pi^+}$			
									& \boldmath$0.202\pm0.009$ 				& HFLAV Fit\\
CLEO								& $0.21\pm0.02\pm0.04$ 					& \cite{Avery:1993ri}\\
ARGUS								& $0.17\pm0.06\pm0.04$ 					& \cite{Albrecht:1991vs}\\
\hline
\rowcolor{LightGray}
\boldmath
$\frac{\Gamma_{39}}{\Gamma_{23}}=\frac{\Sigma^0 \pi^+}{\Lambda \pi^+}$		
									& \boldmath$1.02\pm0.03$ 				& HFLAV Fit\\
FOCUS								& $1.09\pm0.11\pm0.19$ 					& \cite{Link:2005ut}\\
\babar								& $0.997\pm0.015\pm0.051$ 	 			& \cite{Aubert:2006wm}\\
\hline
\rowcolor{LightGray}
\boldmath
$\Gamma_{40}=\Sigma^+ \pi^0$			& \boldmath$(1.25\pm0.09)\%$ 			& HFLAV Fit\\
BESIII								& $(1.18\pm0.10\pm0.03)\%$ 				& \cite{Ablikim:2015flg}\\
\hline
\rowcolor{LightGray}
\boldmath
$\frac{\Gamma_{40}}{\Gamma_2}=\frac{\Sigma^+ \pi^0}{pK^-\pi^+}$			
									& \boldmath$0.193\pm0.014$ 				& HFLAV Fit\\
CLEO								& $0.20\pm0.03\pm0.03$ 					& \cite{Kubota:1993pw}\\
\hline
\rowcolor{LightGray}
\boldmath
$\Gamma_{42}=\Sigma^+ \pi^+\pi^-$			& \boldmath$(4.64\pm0.24)\%$ 			& HFLAV Fit\\
BESIII								& $(4.25\pm0.24\pm0.20)\%$ 				& \cite{Ablikim:2015flg}\\
\hline
\rowcolor{LightGray}
\boldmath
$\frac{\Gamma_{42}}{\Gamma_2}=\frac{\Sigma^+ \pi^+\pi^-}{pK^-\pi^+}$			
									& \boldmath$0.719\pm0.028$ 				& HFLAV Fit\\
CLEO								& $0.74\pm0.07\pm0.09$ 					& \cite{Kubota:1993pw}\\
\hline
\rowcolor{LightGray}
\boldmath
$\Gamma_{48}=\Sigma^+ \omega$			& \boldmath$(1.77\pm0.21)\%$ 			& HFLAV Fit\\
BESIII								& $(1.56\pm0.20\pm0.07)\%$ 				& \cite{Ablikim:2015flg}\\
\hline
\rowcolor{LightGray}
\boldmath
$\frac{\Gamma_{48}}{\Gamma_2}=\frac{\Sigma^+ \omega}{pK^-\pi^+}$			
									& \boldmath$0.274\pm0.031$ 				& HFLAV Fit\\
CLEO								& $0.54\pm0.13\pm0.06$		 			& \cite{Kubota:1993pw}\\
\hline
\rowcolor{LightGray}
\boldmath
$\Gamma_{64}=\Lambda e^+ \nu_e$ 		& \boldmath$(3.18\pm0.32)\%$ 			& HFLAV Fit\\
BESIII								& $(3.63\pm0.38\pm0.20)\%$ 				& \cite{Ablikim:2015prg}\\
\hline
\rowcolor{LightGray}
\boldmath
$\frac{\Gamma_{64}}{\Gamma_2}=\frac{\Lambda e^+ \nu_e}{pK^-\pi^+}$ 		
									& \boldmath$0.492\pm0.049$ 				& HFLAV Fit\\
CLEO								& $0.43\pm0.08$ 						& \cite{Bergfeld:1994gt}\\
ARGUS								& $0.36\pm0.14$						& \cite{Albrecht:1991bu}\\
\hline\hline
\end{longtable}
%\end{envsmall}
\end{center}

\clearpage
% Excited baryons
\subsection{Excited charm baryons}

In this section we summarize the present status of excited charmed
baryons, decaying strongly or electromagnetically. We list their
masses (or the mass difference between the excited baryon and the
corresponding ground state), natural widths, decay modes, and 
assigned quantum numbers. 
The present ground-state measurements are: $M(\Lambda_c^+)=2286.46\pm0.14$~MeV/$c^2$
measured by \babar~\cite{Aubert:2005gt},
$M(\Xi_c^0)=(2470.85^{+0.28}_{-0.04}$)~MeV/$c^2$ and
$M(\Xi_c^+)=(2467.93^{+0.28}_{-0.40}$)~MeV/$c^2$, both 
dominated by CDF~\cite{Aaltonen:2014wfa}, and
$M(\Omega_c^0)=(2695.2\pm1.7$)~MeV/$c^2$, dominated 
by Belle~\cite{Solovieva:2008fw}. Should these values 
change, so will some of the values for the masses of the excited states.

Table~\ref{sumtable1} summarizes the excited $\Lambda_c^+$ baryons.  
The first two states listed, namely the $\Lambda_c(2595)^+$ and $\Lambda_c(2625)^+$,
are well-established. 
The measured masses and decay patterns suggest that 
they are orbitally 
excited $\Lambda_c^+$ baryons with total angular momentum of the
light quarks $L=1$. Thus their quantum numbers are assigned to be 
$J^P=(\frac{1}{2})^-$ and $J^P=(\frac{3}{2})^-$, respectively. 
Their mass measurements are  
%and widths 
dominated by CDF~\cite{Aaltonen:2011sf}: 
$M(\Lambda_c(2595)^+)=(2592.25\pm 0.24\pm 0.14$)~MeV/$c^2$ and
$M(\Lambda_c(2625)^+)=(2628.11\pm 0.13\pm 0.14$)~MeV/$c^2$. 
Earlier measurements did not fully take into account the restricted
phase-space of the $\Lambda_c(2595)^+$ decays.

The next two states, $\Lambda_c(2765)^+$ and $\Lambda_c(2880)^+$, 
were discovered by CLEO~\cite{Artuso:2000xy} in the $\Lambda_c^+\pi^+\pi^-$ 
final state. CLEO found that a significant fraction of the $\Lambda_c(2880)^+$ decays 
proceeds via an intermediate $\Sigma_c(2445)^{++/0}\pi^{-/+}$.  
Later, \babar~\cite{Aubert:2006sp} 
observed that this state has also a $D^0 p$ decay mode. This was the 
first example of an excited charmed baryon decaying into a charm meson 
plus a baryon; previously all excited charmed baryon were found in their 
hadronic transitions into lower lying charmed 
baryons. In the same analysis, \babar observed for the
first time an additional state, $\Lambda_c(2940)^+$, 
decaying into $D^0 p$. Studying the $D^+ p$ final state,
\babar found no signal; this implies that the $\Lambda_c(2880)^+$ 
and $\Lambda_c(2940)^+$ are $\Lambda_c^+$ excited states
rather than $\Sigma_c$ excitations. 
Belle reported the result of an angular analysis that favors
$5/2$ for the $\Lambda_c(2880)^+$ spin hypothesis. 
Moreover, the measured ratio of branching fractions 
${\cal B}(\Lambda_c(2880)^+\rightarrow \Sigma_c(2520)\pi^{\pm})/{\cal B}(\Lambda_c(2880)^+\rightarrow \Sigma_c(2455)\pi^{\pm})=(0.225\pm 0.062\pm 0.025)$, combined 
with theoretical predictions based on HQS~\cite{Isgur:1991wq,Cheng:2006dk}, 
favor even parity. However this prediction is only valid if the P-wave 
portion of $\Sigma_c(2520)\pi$ is suppressed.     
The current open questions in the excited $\Lambda_c^+$ family include
the determination of quantum numbers for the other states, and 
the nature of the $\Lambda_c(2765)^+$ state, in particular whether it is
an excited $\Sigma_c^+$ or $\Lambda_c^+$. However, there is no doubt that 
the state exists, as it is clearly visible in Belle data.

%\begin{sidewaystable}
\begin{table}[htb]
\caption{Summary of excited $\Lambda_c^+$ baryons.} 
\vskip0.15in
\begin{center}
\renewcommand{\arraystretch}{1.2}
\begin{tabular}{c|c|c|c|c}
\hline
Charmed baryon   & Mode  & Mass & Natural width  & $J^P$  \\
excited state &  &  (MeV/$c^2$) & (MeV)  \\
\hline
$\Lambda_c(2595)^+$ & $\Lambda_c^+\pi^+\pi^-$, $\Sigma_c(2455)\pi$ &  $2592.25\pm 0.28$ & $2.59\pm 0.30 \pm 0.47$  & $1/2^-$  \\
\hline
$\Lambda_c(2625)^+$ & $\Lambda_c^+\pi^+\pi^-$   & $2628.11\pm 0.19$ & $<0.97$ & $3/2^-$  \\
\hline
$\Lambda_c(2765)^+$ & $\Lambda_c^+\pi^+\pi^-$, $\Sigma_c(2455)\pi$ & $2766.6\pm 2.4$ & $50$ & ?  \\
\hline
$\Lambda_c(2880)^+$ & $\Lambda_c^+\pi^+\pi^-$, $\Sigma_c(2455)\pi$,  &$2881.53\pm 0.35$ & $5.8\pm 1.1$ & $5/2^+$ \\
 &  $\Sigma_c(2520)\pi$, $D^0p$     & & &  \\
\hline
$\Lambda_c(2940)^+$ & $D^0p$, $\Sigma_c(2455)\pi$ & $2939.3^{+1.4}_{-1.5}$ & $17^{+8}_{-6}$  & ?  \\
\hline 
\end{tabular}
\end{center}
\label{sumtable1} 
\end{table}
%\end{landscape}
%\end{sidewaystable}

Table~\ref{sumtable2} summarizes the excited $\Sigma_c^{++,+,0}$ baryons.
The ground iso-triplets of $\Sigma_c(2455)^{++,+,0}$ and
$\Sigma_c(2520)^{++,+,0}$ baryons are well-established. 
Belle~\cite{Lee:2014htd} 
%%%%%%%%%%%%% CDF~\cite{Aaltonen:2011sf}
precisely measured the mass differences  
and widths of the doubly charged and neutral members of this triplet.
%The results are
%\begin{eqnarray}
%\Delta M(\Sigma_c(2520)^{++}) & = & (231.99\pm 0.10\pm 0.02){\rm\ MeV}/c^2 \\ 
%\Gamma(\Sigma_c(2520)^{++}) & = & (14.77\pm 0.25^{+0.18}_{-0.30}){\rm\ MeV} \\ 
%\Delta M(\Sigma_c(2520)^{0}) & = & (231.98\pm 0.11\pm 0.04){\rm\ MeV}/c^2 \\ 
%\Gamma(\Sigma_c(2520)^{0}) & = & (15.41\pm 0.41^{+0.20}_{-0.32}){\rm\ MeV}\,.
%\end{eqnarray} 
%\Delta M(\Sigma_c(2520)^{++})& =& (231.99\pm 0.10\pm 0.02)$~MeV/$c^2$, 
%$\Gamma(\Sigma_c(2520)^{++})=(14.77\pm 0.25^{+0.18}_{-0.30})$~MeV/$c^2$ and 
%$\Delta M(\Sigma_c(2520)^{0})=(231.98\pm 0.11\pm 0.04)$~MeV/$c^2$, 
%$\Gamma(\Sigma_c(2520)^{0})=(15.41\pm 0.41^{+0.20}_{-0.32})$~MeV/$c^2$, respectively. 
The short list of excited $\Sigma_c$ baryons is completed by the triplet 
of $\Sigma_c(2800)$ states observed by Belle~\cite{Mizuk:2004yu}. Based 
on the measured masses and theoretical predictions~\cite{Copley:1979wj,Pirjol:1997nh}, 
these states are assumed to be members of the predicted $\Sigma_{c2}$ $3/2^-$
triplet. From a study of resonant substructure 
in $B^-\rightarrow \Lambda_c^+\bar{p}\pi^-$ decays, \babar found 
a significant signal in the $\Lambda_c^+\pi^-$ final state with a mean value 
higher than measured for the $\Sigma_c(2800)$ by Belle by about $3\sigma$
(Table~\ref{sumtable2}). The decay widths measured by
Belle and \babar are consistent, but it is an open question if the 
observed state is the same as the Belle state.

%\begin{sidewaystable}
\begin{table}[!htb]
\caption{Summary of the excited $\Sigma_c^{++,+,0}$ baryon family.} 
\vskip0.15in
\begin{center}
\renewcommand{\arraystretch}{1.2}
\begin{tabular}{c|c|c|c|c}
\hline
Charmed baryon   & Mode  & $\Delta M$ & Natural width  & $J^P$  \\
excited state &  &  (MeV/$c^2$) & (MeV)  \\
\hline
$\Sigma_c(2455)^{++}$ &$\Lambda_c^+\pi^+$  & $167.510 \pm 0.17$ & $1.89\,^{+0.09}_{-0.18}$ & $1/2^+$   \\
$\Sigma_c(2455)^{+}$ &$\Lambda_c^+\pi^0$  & $166.4\pm 0.4$ & $<4.6$~@~90$\%$~C.L. & $1/2^+$ \\
$\Sigma_c(2455)^{0}$ &$\Lambda_c^+\pi^-$  & $167.29\pm 0.17$ & $1.83\,^{+0.11}_{-0.19}$ & $1/2^+$    \\
\hline
$\Sigma_c(2520)^{++}$ &$\Lambda_c^+\pi^+$  & $231.95\,^{+0.17}_{-0.12}$ & $14.78\,^{+0.30}_{-0.40}$ & $3/2^+$   \\
$\Sigma_c(2520)^{+}$ &$\Lambda_c^+\pi^0$  & $231.0\pm 2.3$ & $<17$~@~90$\%$~C.L. & $3/2^+$ \\
$\Sigma_c(2520)^{0}$ &$\Lambda_c^+\pi^-$  & $232.02\,^{+0.15}_{-0.14}$ & $15.3\,^{+0.4}_{-0.5}$ & $3/2^+$    \\
\hline
$\Sigma_c(2800)^{++}$ & $\Lambda_c^+\pi^{+}$ & $514\,^{+4}_{-6}$ & $75\,^{+18+12}_{-13-11}$ & $3/2^-$?     \\
$\Sigma_c(2800)^{+}$ & $\Lambda_c^+\pi^{0}$&$505\,^{+15}_{-5}$ &$62\,^{+37+52}_{-23-38}$ &   \\
$\Sigma_c(2800)^{0}$ & $\Lambda_c^+\pi^{-}$&$519\,^{+5}_{-7}$ & $72\,^{+22}_{-15}$ &   \\
 & $\Lambda_c^+\pi^{-}$ & $560\pm 8\pm 10$ & $86\,^{+33}_{-22}$  \\

\hline 
\end{tabular}
\end{center}
\label{sumtable2} 
\end{table}
%\end{landscape}
%\end{sidewaystable}
% =====================================================================================

Table~\ref{sumtable3} summarizes the excited $\Xi_c^{+,0}$ and $\Omega_c^0$ 
baryons. The list of excited $\Xi_c$ baryons has several states, of unknown quantum
numbers, having masses 
above 2900~MeV/$c^2$ and decaying into three different types of decay modes:
$\Lambda_c/\Sigma_c n\pi$, $\Xi_c n\pi$ and the most recently observed $\Lambda D$.  
Some of these states ($\Xi_c(2970)^+$, $\Xi_c(3055)$ and $\Xi_c(3080)^{+,0}$) have been
observed by
both Belle~\cite{Chistov:2006zj,YKato:2014,YKato:2016} 
and \babar~\cite{Aubert:2007eb} and are considered well-established.
The $\Xi_c(2930)^0$ state decaying into $\Lambda_c^+ K^-$ is seen only
by \babar~\cite{Aubert:2007bd} and needs confirmation.  
The $\Xi_c(3123)^+$ observed by \babar~\cite{Aubert:2007eb}
in the $\Sigma_c(2520)^{++}\pi^-$ final state has not been
confirmed by Belle~\cite{YKato:2014} with twice the statistics; 
thus its existence in in doubt and it is omitted from Tab.~\ref{sumtable3}.

Several of the width and mass measurements for the $\Xi_c(3055)$ and $\Xi_c(3080)$ 
iso-doublets are only in marginal agreement between experiments and 
decay modes. However, there seems little doubt that the differing 
measurements are of the same particle.

Belle~\cite{Yelton:2016fqw} has recently analyzed large samples of 
$\Xi_c^\prime$, $\Xi_c(2645)$, $\Xi_c(2790)$, $\Xi_c(2815)$ and 
$\Xi_c(2970)$ decays. From this analysis they obtain the most 
precise mass measurements of all five iso-doublets, and the first
significant width measurements of the $\Xi_c(2645)$, $\Xi_c(2790)$ and $\Xi_c(2815)$.
The level of agreement in the different measurements of the mass and width 
of the $\Xi_c(2970)$, formerly named by the PDG as the $\Xi_c(2980)$, is not 
satisfactory. This leaves open the possibility of there being other resonances 
nearby or that threshold effects have not been fully understood.
The present situation in the excited $\Xi_c$ sector is summarized in
in Table~\ref{sumtable3}. 

The excited $\Omega_c^0$ doubly strange charmed baryon has been seen by both 
\babar~\cite{Aubert:2006je} and Belle~\cite{Solovieva:2008fw}.
The mass differences $\delta M=M(\Omega_c^{*0})-M(\Omega_c^0)$ 
measured by the experiments are in good agreement
and are also consistent with most theoretical 
predictions~\cite{Rosner:1995yu,Glozman:1995xy,Jenkins:1996de,
Burakovsky:1997vm}. 
No higher mass $\Omega_c$ states have yet been observed.

%\begin{sidewaystable}
\begin{table}[b]
\caption{Summary of excited $\Xi_c^{+,0}$ and $\Omega_c^0$ baryon families. 
For the first four iso-doublets, the mass difference with respect to the 
ground state is given, as the uncertainties are dominated by the uncertainty
in the ground state mass. In the remaining cases, the uncertainty on the 
measurement of the excited state itself dominates.} 
\vskip0.15in
\resizebox{\textwidth}{!}{
\renewcommand{\arraystretch}{1.2}
\begin{tabular}{c|c|c|c|c}
\hline
Charmed baryon   & Mode  & Mass or & Natural width  & $J^P$  \\
excited state &  &  mass difference & (MeV)  \\
              &  &  (MeV/$c^2$)     &        \\
\hline
$\Xi_c'^+$ & $\Xi_c^+\gamma$ & $110.5 \pm 0.4$  &  & $1/2^+$    \\
$\Xi_c'^0$ & $\Xi_c^0\gamma$ & $108.3\pm 0.4$   &  & $1/2^+$   \\
\hline
$\Xi_c(2645)^+$ & $\Xi_c^0\pi^+$ & $178.5 \pm 0.1$  & $2.1 \pm 0.2 $ & $3/2^+$   \\
$\Xi_c(2645)^0$ & $\Xi_c^+\pi^-$ & $174.7 \pm 0.1$  & $2.4 \pm 0.2 $ & $3/2^+$   \\
\hline
$\Xi_c(2790)^+$ &$\Xi_c'^0\pi^+$ & $320.7\pm 0.5$ & $9  \pm 1$ & $1/2^-$   \\
$\Xi_c(2790)^0$ &$\Xi_c'^+\pi^-$ & $323.8\pm 0.5$ & $10 \pm 1$ & $1/2^-$   \\
\hline
$\Xi_c(2815)^+$ &$\Xi_c(2645)^0\pi^+$ & $348.8\pm 0.1$ & $2.43\pm0.23$ &  $3/2^-$  \\
$\Xi_c(2815)^0$ &$\Xi_c(2645)^+\pi^-$ & $349.4\pm 0.1$  & $2.54\pm0.23$  & $3/2^-$   \\
\hline
\hline
Charmed baryon   & Mode  & Mass  & Natural width  & $J^P$  \\
excited state &  &  (MeV/$c^2$) & (MeV)  \\
\hline
$\Xi_c(2930)^0$ & $\Lambda_c^+ K^-$ & $2931.6\pm 6$ & $36\pm 13$ & ?     \\
\hline
$\Xi_c(2970)^+$ & $\Lambda_c^+K^-\pi^+$, $\Sigma_c^{++}K^-$, $\Xi_c(2645)^0\pi^+$
%\footnote{this mode obs. by Belle~\cite{lesiak} but natural width is systematically lower than in other modes }
 &  $2967.2\pm 0.8$  & $21 \pm 3$ & ?     \\
$\Xi_c(2970)^0$ & $\Xi_c(2645)^+\pi^-$
%\footnote{this mode obs. by Belle~\cite{lesiak} (6.1$\sigma$) but natural width is systematically lower than in other modes } 
&  $2970.4\pm 0.8$ &$28\pm 3$ & ?       \\
\hline
$\Xi_c(3055)^+$ & $\Sigma_c^{++}K^-$, $\Lambda D$ & 	$3055.7\pm 0.4$  &     	$8.0 \pm 1.9 $  & ?   \\
$\Xi_c(3055)^0$ & $\Lambda D$ &                         $3059.0\pm 0.8$   &     $6.2 \pm 2.4$ & ?    \\
\hline
$\Xi_c(3080)^+$ & $\Lambda_c^+K^-\pi^+$, $\Sigma_c^{++}K^-$, $\Sigma_c(2520)^{++}K^-$ , $\Lambda D$ & $3077.8\pm 0.3$ & $3.6\pm 0.7$ & ?   \\
$\Xi_c(3080)^0$ &$\Lambda_c^+ K^0_S\pi^-$, $\Sigma_c^0K^0_S$, $\Sigma_c(2520)^{0}K^0_S$ & $3079.9\pm 1.0$ & $5.6\pm 2.2$ & ?   \\
\hline
%$\Xi_c(3123)^+$ &$\Sigma_c(2520)^{++}K^-$ & $3122.9\pm 1.3$ & $4\pm 4$ & ?   \\
%\hline
%\\
$\Omega_c(2770)^0$ & $\Omega_c^0\gamma$& $2765.9\pm 2.0$  & $70.7^{+0.8}_{-0.9}$ & $3/2^+$  \\
\hline 
\end{tabular}
}
\label{sumtable3} 
\end{table}
%\end{landscape}
%\end{sidewaystable}

\begin{figure}[!htb]
\includegraphics[width=1.0\textwidth]{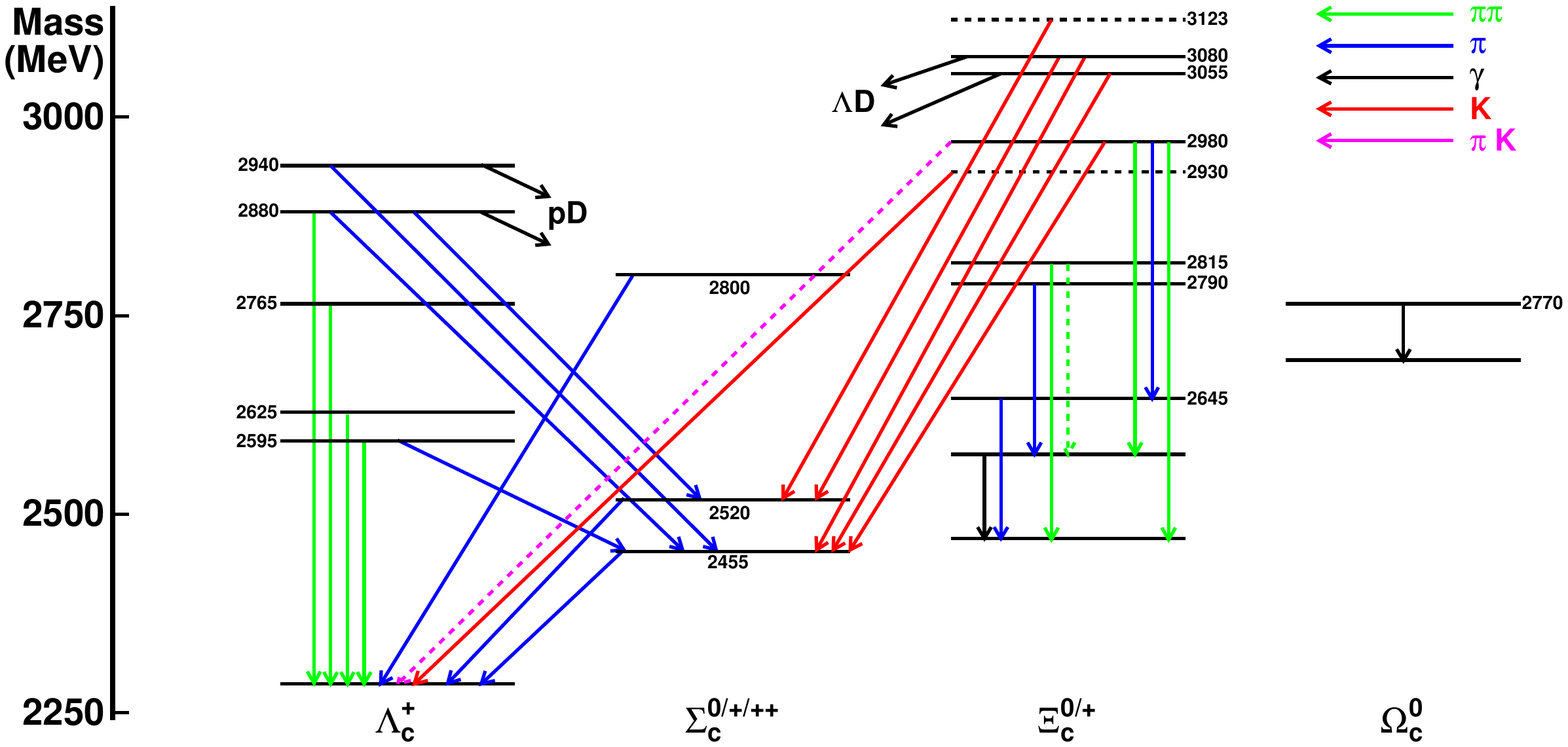}
\caption{Level diagram for multiplets and transitions for excited charm baryons.}
\label{charm:leveldiagram}
\end{figure}

Figure~\ref{charm:leveldiagram} shows the levels of excited charm
baryons along with corresponding transitions between them, and
also transitions to the ground states.
We note that Belle and \babar recently discovered
that transitions between families are possible, \ie, between 
the $\Xi_c$ and $\Lambda_c^+$ families of excited charmed 
baryons~\cite{Chistov:2006zj,Aubert:2007eb} and that 
highly excited states are found to decay into a
non-charmed baryons and a $D$ meson\cite{Aubert:2006sp,YKato:2016}.

\clearpage
% Rare and forbidden decays
\subsection{Rare and forbidden decays}
\label{sec:charm:rare}

\begin{figure}
\begin{center}
\includegraphics[width=5.00in]{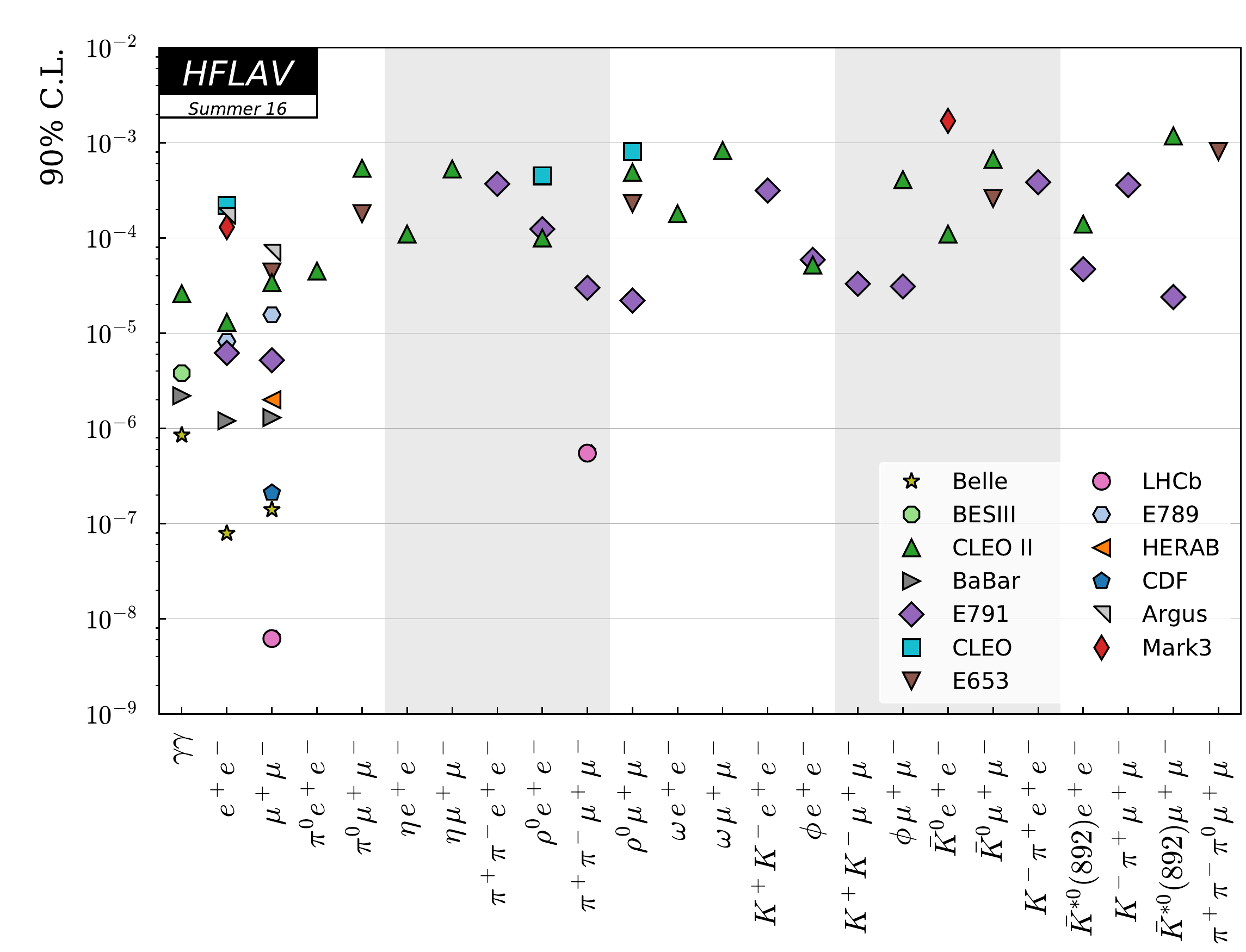}
\vskip0.10in
\includegraphics[width=5.00in]{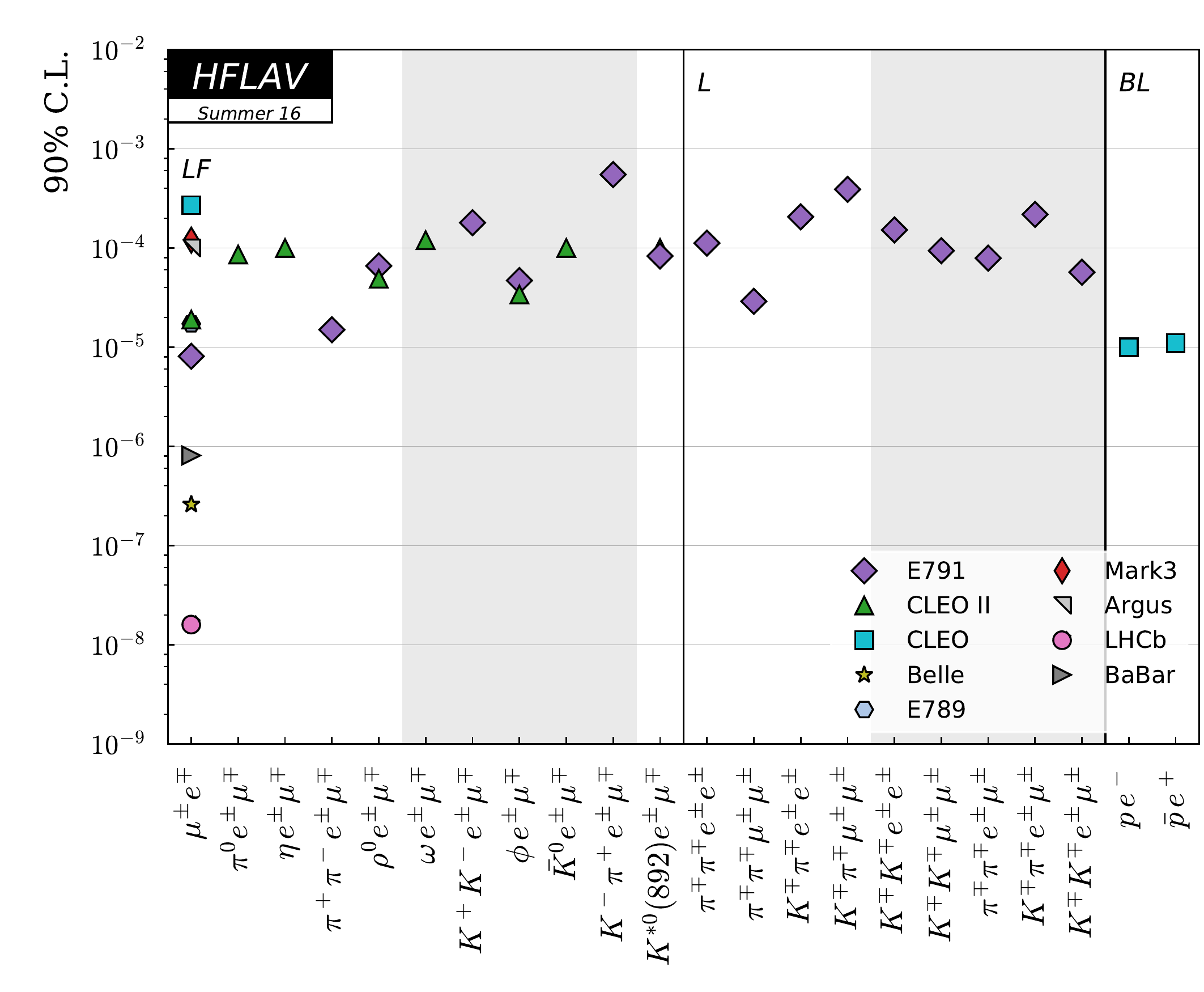}
\caption{Upper limits at $90\%$ C.L.\ for $D^0$ decays. The top plot
shows flavor-changing neutral current decays, and the bottom plot
shows lepton-flavor-changing (LF), lepton-number-changing (L), and 
both baryon- and lepton-number-changing (BL) decays.
}
\label{fig:charm:rare_d0}
\end{center}
\end{figure}

\begin{figure}
\begin{center}
\includegraphics[width=4.00in]{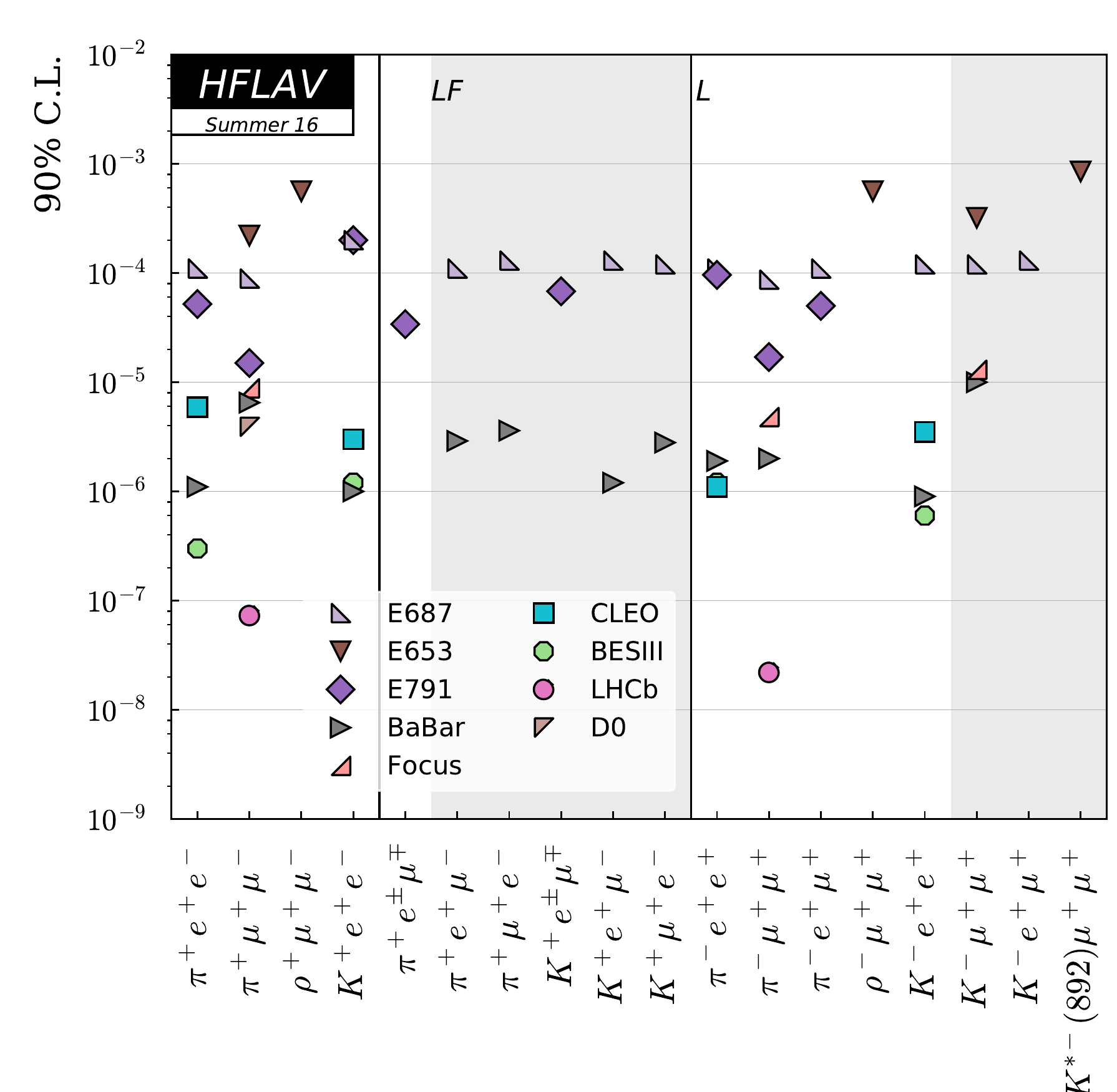}
\vskip0.10in
\includegraphics[width=4.00in]{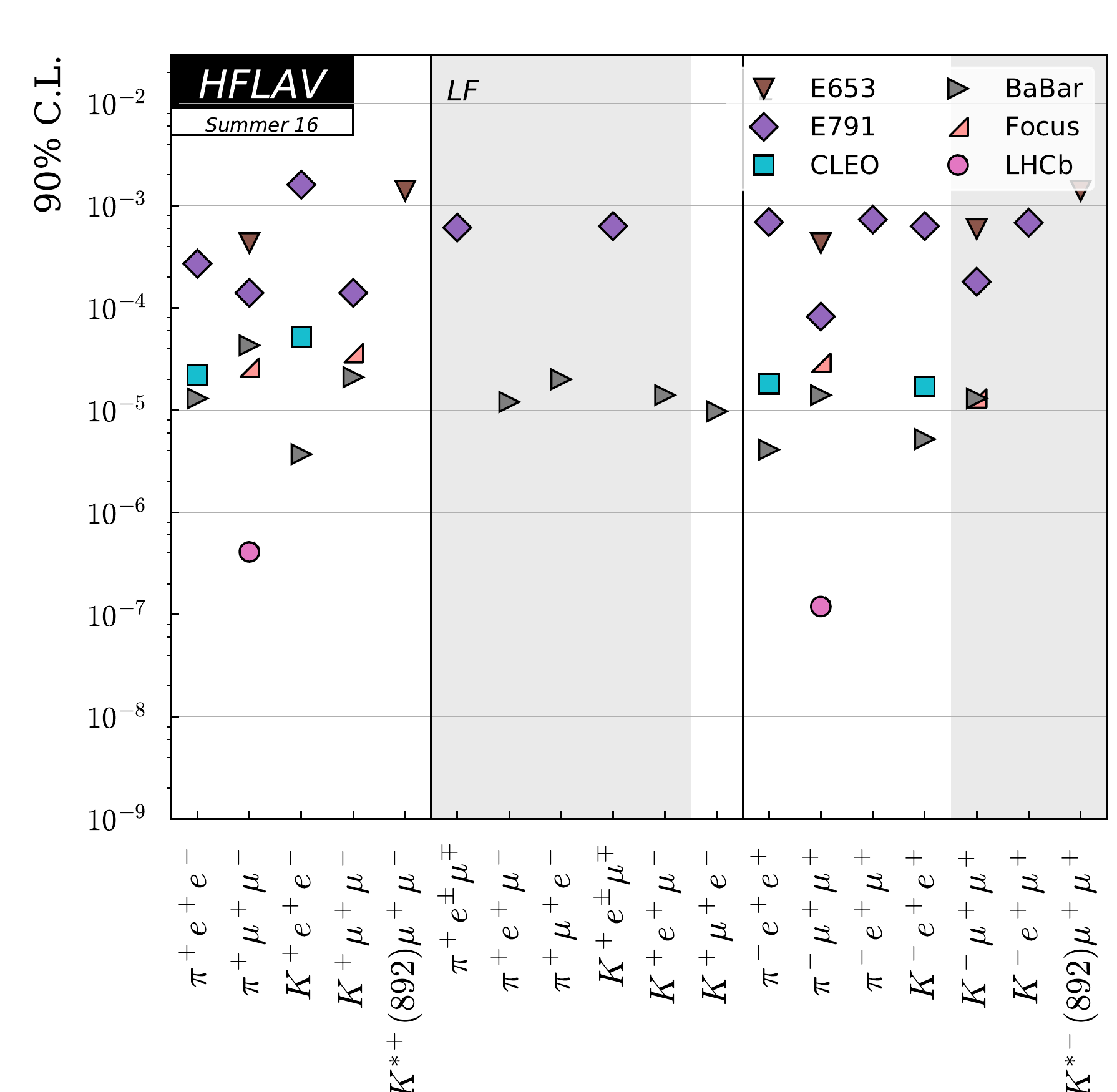}
\caption{Upper limits at $90\%$ C.L.\ for $D^+$ (top) and $D_s^+$ (bottom) 
decays. Each plot shows flavor-changing neutral current decays, 
lepton-flavor-changing decays (LF), and lepton-number-changing (L) decays. 
}
\label{fig:charm:rare_charged}
\end{center}
\end{figure}

\begin{figure}
\begin{center}
\includegraphics[width=3.0in]{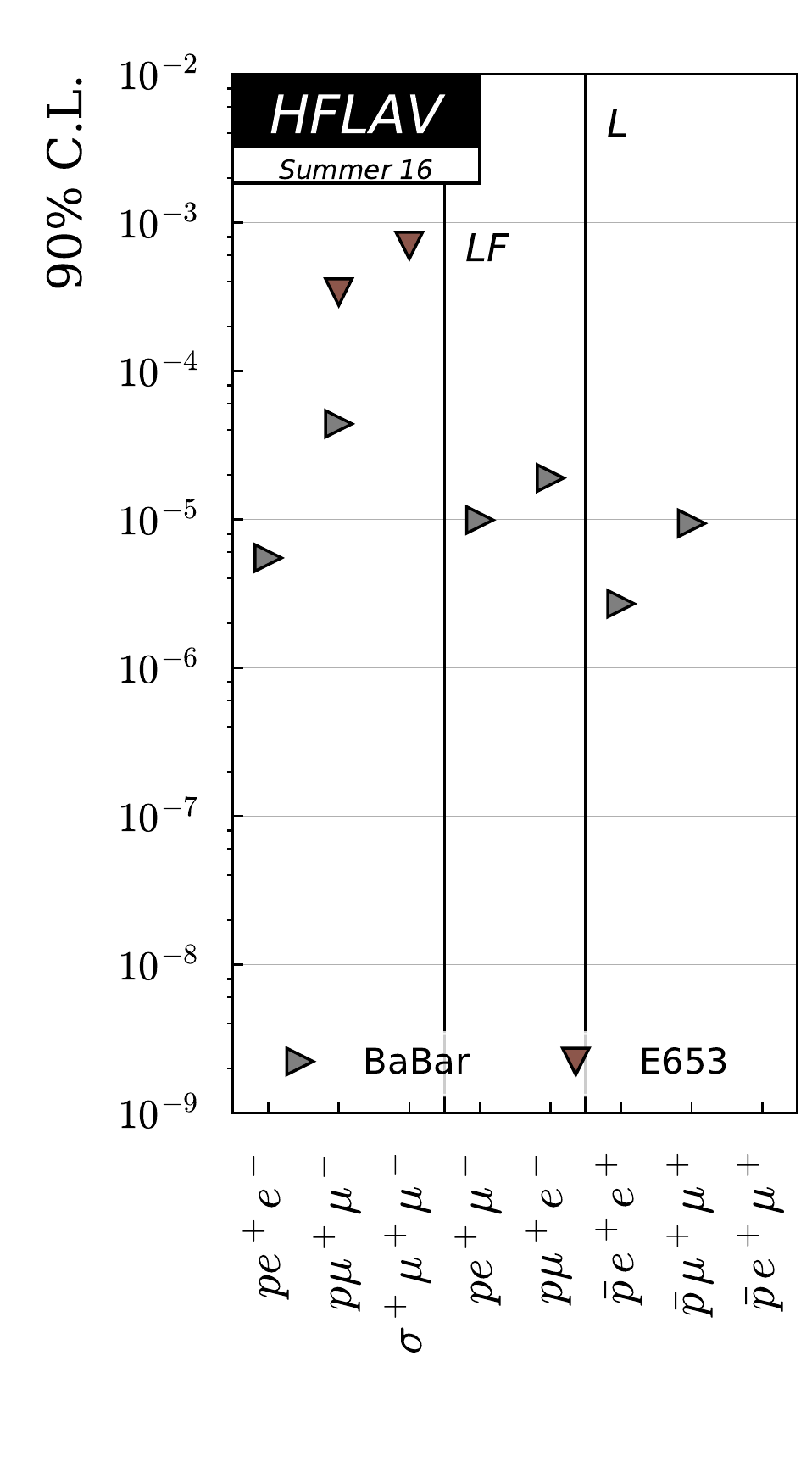}
\caption{Upper limits at $90\%$ C.L.\ for $\Lambda_c^+$ decays. Shown are 
flavor-changing neutral current decays, lepton-flavor-changing (LF) 
decays, and lepton-number-changing (L) decays. }
\label{fig:charm:lambdac}
\end{center}
\end{figure}

This section provides a summary of searches for rare and forbidden charm decays
in tabular form. The decay modes can be categorized as 
flavor-changing neutral currents, lepton-flavor-violating, 
lepton-number-violating, and both baryon- and lepton-number-violating decays.
Figures~\ref{fig:charm:rare_d0}-\ref{fig:charm:lambdac} plot the 
upper limits for $D^0$, $D^+$, $D_s^+$, and $\Lambda_c^+$ decays. 
Tables~\ref{tab:charm:rare_d0}-\ref{tab:charm:rare_lambdac} give the 
corresponding numerical results. Some theoretical predictions are given in 
Refs.~\cite{Burdman:2001tf,Fajfer:2002bu,Fajfer:2007dy,Golowich:2009ii,Paul:2010pq,Borisov:2011aa,Wang:2014dba,deBoer:2015boa}.

In several cases the rare-decay final states have been observed with 
the di-lepton pair being the decay product of a vector meson.
For these measurements the quoted limits are those expected for the 
non-resonant di-lepton spectrum.
For the extrapolation to the full spectrum a phase-space distribution 
of the non-resonant component has been assumed.
This applies to the CLEO measurement of the decays 
$D_{(s)}^+\to(K^+,\pi^+)e^+e^-$~\cite{Rubin:2010cq}, to the D0 measurements 
of the decays $D_{(s)}^+\to\pi^+\mu^+\mu^-$~\cite{Abazov:2007aj}, and to 
the \babar measurements of the decays $D_{(s)}^+\to(K^+,\pi^+)e^+e^-$ and 
$D_{(s)}^+\to(K^+,\pi^+)\mu^+\mu^-$, where the contribution from 
$\phi\to l^+l^-$ ($l=e,\mu$) has been excluded.
In the case of the LHCb measurements of the decays 
$D^0\to\pi^+\pi^-\mu^+\mu^-$~\cite{Aaij:2013uoa} as 
well as the decays $D_{(s)}^+\to\pi^+\mu^+\mu^-$~\cite{Aaij:2013sua} 
the contributions from $\phi\to l^+l^-$ as well as from 
$\rho,\omega\to l^+l^-$ ($l=e,\mu$) have been excluded. 

%\begin{table}
\begin{longtable}{l|ccc}
%\centering 
\caption{Upper limits at $90\%$ C.L.\ for $D^0$ decays.%}
\label{tab:charm:rare_d0}
}\\
%\vspace{3pt}
%\begin{tabular}{l|ccc}

\hline\hline%\\
Decay & Limit $\times10^6$ & Experiment & Reference\\
\endfirsthead
% \hline\hline%\\
\multicolumn{4}{c}{\tablename\ \thetable{} -- continued from previous page} \\ \hline
Decay & Limit $\times10^6$ & Experiment & Reference\\
\endhead
\hline
$\gamma{}\gamma{}$ & 26.0 & CLEO II & \cite{Coan:2002te}\\
& 3.8 & BESIII & \cite{Ablikim:2015djc}\\
& 2.2 & \babar & \cite{Lees:2011qz}\\
& 0.85 & Belle & \cite{Nisar:2015gvd}\\
\hline
$e^+e^-$ & 220.0 & CLEO & \cite{Haas:1988bh}\\
& 170.0 & Argus & \cite{Albrecht:1988ge}\\
& 130.0 & Mark3 & \cite{Adler:1987cp}\\
& 13.0 & CLEO II & \cite{Freyberger:1996it}\\
& 8.19 & E789 & \cite{Pripstein:1999tq}\\
& 6.2 & E791 & \cite{Aitala:1999db}\\
& 1.2 & \babar & \cite{Aubert:2004bs}\\
& 0.079 & Belle & \cite{Petric:2010yt}\\
\hline
$\mu{}^+\mu{}^-$ & 70.0 & Argus & \cite{Albrecht:1988ge}\\
& 44.0 & E653 & \cite{Kodama:1995ia}\\
& 34.0 & CLEO II & \cite{Freyberger:1996it}\\
& 15.6 & E789 & \cite{Pripstein:1999tq}\\
& 5.2 & E791 & \cite{Aitala:1999db}\\
& 2.0 & HERAb & \cite{Abt:2004hn}\\
& 1.3 & \babar & \cite{Aubert:2004bs}\\
& 0.21 & CDF & \cite{Aaltonen:2010hz}\\
& 0.14 & Belle & \cite{Petric:2010yt}\\
& 0.0062 & LHCb & \cite{Aaij:2013cza}\\
\hline
$\pi{}^0e^+e^-$ & 45.0 & CLEO II & \cite{Freyberger:1996it}\\
\hline
$\pi{}^0\mu{}^+\mu{}^-$ & 540.0 & CLEO II & \cite{Freyberger:1996it}\\
& 180.0 & E653 & \cite{Kodama:1995ia}\\
\hline
$\eta\,{}e^+e^-$ & 110.0 & CLEO II & \cite{Freyberger:1996it}\\
\hline
$\eta\,{}\mu{}^+\mu{}^-$ & 530.0 & CLEO II & \cite{Freyberger:1996it}\\
\hline
$\pi{}^+\pi{}^-e^+e^-$ & 370.0 & E791 & \cite{Aitala:2000kk}\\
\hline
$\rho{}^0e^+e^-$ & 450.0 & CLEO & \cite{Haas:1988bh}\\
& 124.0 & E791 & \cite{Aitala:2000kk}\\
& 100.0 & CLEO II & \cite{Freyberger:1996it}\\
\hline
$\pi{}^+\pi{}^-\mu{}^+\mu{}^-$ & 30.0 & E791 & \cite{Aitala:2000kk}\\
& 0.55 & LHCb & \cite{Aaij:2013uoa}\\
\hline
$\rho{}^0\mu{}^+\mu{}^-$ & 810.0 & CLEO & \cite{Haas:1988bh}\\
& 490.0 & CLEO II & \cite{Freyberger:1996it}\\
& 230.0 & E653 & \cite{Kodama:1995ia}\\
& 22.0 & E791 & \cite{Aitala:2000kk}\\
\hline
$\omega\,{}e^+e^-$ & 180.0 & CLEO II & \cite{Freyberger:1996it}\\
\hline
$\omega\,{}\mu{}^+\mu{}^-$ & 830.0 & CLEO II & \cite{Freyberger:1996it}\\
\hline
$K^+K^-e^+e^-$ & 315.0 & E791 & \cite{Aitala:2000kk}\\
\hline
$\phi\,{}e^+e^-$ & 59.0 & E791 & \cite{Aitala:2000kk}\\
& 52.0 & CLEO II & \cite{Freyberger:1996it}\\
\hline
$K^+K^-\mu{}^+\mu{}^-$ & 33.0 & E791 & \cite{Aitala:2000kk}\\
\hline
$\phi\,{}\mu{}^+\mu{}^-$ & 410.0 & CLEO II & \cite{Freyberger:1996it}\\
& 31.0 & E791 & \cite{Aitala:2000kk}\\
\hline
$\overline{K}^0e^+e^-$ & 1700.0 & Mark3 & \cite{Adler:1988es}\\
& 110.0 & CLEO II & \cite{Freyberger:1996it}\\
\hline
$\overline{K}^0\mu{}^+\mu{}^-$ & 670.0 & CLEO II & \cite{Freyberger:1996it}\\
& 260.0 & E653 & \cite{Kodama:1995ia}\\
\hline
$K^-\pi{}^+e^+e^-$ & 385.0 & E791 & \cite{Aitala:2000kk}\\
\hline
$\overline{K}^{*0}(892)e^+e^-$ & 140.0 & CLEO II & \cite{Freyberger:1996it}\\
& 47.0 & E791 & \cite{Aitala:2000kk}\\
\hline
$K^-\pi{}^+\mu{}^+\mu{}^-$ & 360.0 & E791 & \cite{Aitala:2000kk}\\
\hline
$\overline{K}^{*0}(892)\mu{}^+\mu{}^-$ & 1180.0 & CLEO II & \cite{Freyberger:1996it}\\
& 24.0 & E791 & \cite{Aitala:2000kk}\\
\hline
$\pi{}^+\pi{}^-\pi{}^0\mu{}^+\mu{}^-$ & 810.0 & E653 & \cite{Kodama:1995ia}\\
\hline
$\mu{}^{\pm}e^{\mp}$ & 270.0 & CLEO & \cite{Haas:1988bh}\\
& 120.0 & Mark3 & \cite{Becker:1987mu}\\
& 100.0 & Argus & \cite{Albrecht:1988ge}\\
& 19.0 & CLEO II & \cite{Freyberger:1996it}\\
& 17.2 & E789 & \cite{Pripstein:1999tq}\\
& 8.1 & E791 & \cite{Aitala:1999db}\\
& 0.81 & \babar & \cite{Aubert:2004bs}\\
& 0.26 & Belle & \cite{Petric:2010yt}\\
& 0.016 & LHCb & \cite{Aaij:2015qmj}\\
\hline
$\pi{}^0e^{\pm}\mu{}^{\mp}$ & 86.0 & CLEO II & \cite{Freyberger:1996it}\\
\hline
$\eta\,{}e^{\pm}\mu{}^{\mp}$ & 100.0 & CLEO II & \cite{Freyberger:1996it}\\
\hline
$\pi{}^+\pi{}^-e^{\pm}\mu{}^{\mp}$ & 15.0 & E791 & \cite{Aitala:2000kk}\\
\hline
$\rho{}^0e^{\pm}\mu{}^{\mp}$ & 66.0 & E791 & \cite{Aitala:2000kk}\\
& 49.0 & CLEO II & \cite{Freyberger:1996it}\\
\hline
$\omega\,{}e^{\pm}\mu{}^{\mp}$ & 120.0 & CLEO II & \cite{Freyberger:1996it}\\
\hline
$K^+K^-e^{\pm}\mu{}^{\mp}$ & 180.0 & E791 & \cite{Aitala:2000kk}\\
\hline
$\phi\,{}e^{\pm}\mu{}^{\mp}$ & 47.0 & E791 & \cite{Aitala:2000kk}\\
& 34.0 & CLEO II & \cite{Freyberger:1996it}\\
\hline
$\overline{K}^0e^{\pm}\mu{}^{\mp}$ & 100.0 & CLEO II & \cite{Freyberger:1996it}\\
\hline
$K^-\pi{}^+e^{\pm}\mu{}^{\mp}$ & 550.0 & E791 & \cite{Aitala:2000kk}\\
\hline
$K^{*0}(892)e^{\pm}\mu{}^{\mp}$ & 100.0 & CLEO II & \cite{Freyberger:1996it}\\
& 83.0 & E791 & \cite{Aitala:2000kk}\\
\hline
$\pi{}^{\mp}\pi{}^{\mp}e^{\pm}e^{\pm}$ & 112.0 & E791 & \cite{Aitala:2000kk}\\
\hline
$\pi{}^{\mp}\pi{}^{\mp}\mu{}^{\pm}\mu{}^{\pm}$ & 29.0 & E791 & \cite{Aitala:2000kk}\\
\hline
$K^{\mp}\pi{}^{\mp}e^{\pm}e^{\pm}$ & 206.0 & E791 & \cite{Aitala:2000kk}\\
\hline
$K^{\mp}\pi{}^{\mp}\mu{}^{\pm}\mu{}^{\pm}$ & 390.0 & E791 & \cite{Aitala:2000kk}\\
\hline
$K^{\mp}K^{\mp}e^{\pm}e^{\pm}$ & 152.0 & E791 & \cite{Aitala:2000kk}\\
\hline
$K^{\mp}K^{\mp}\mu{}^{\pm}\mu{}^{\pm}$ & 94.0 & E791 & \cite{Aitala:2000kk}\\
\hline
$\pi{}^{\mp}\pi{}^{\mp}e^{\pm}\mu{}^{\pm}$ & 79.0 & E791 & \cite{Aitala:2000kk}\\
\hline
$K^{\mp}\pi{}^{\mp}e^{\pm}\mu{}^{\pm}$ & 218.0 & E791 & \cite{Aitala:2000kk}\\
\hline
$K^{\mp}K^{\mp}e^{\pm}\mu{}^{\pm}$ & 57.0 & E791 & \cite{Aitala:2000kk}\\
\hline
$p\,e^-$ & 10.0 & CLEO & \cite{Rubin:2009aa}\\
\hline
$\overline{p}\,{}e^+$ & 11.0 & CLEO & \cite{Rubin:2009aa}\\
\hline
\end{longtable}
%\end{tabular}

%\end{table}
\pagebreak

\begin{longtable}{l|ccc}
\caption{Upper limits at $90\%$ C.L.\ for $D^+$ decays.\label{tab:charm:rare_dplus}}\\
\hline\hline
Decay & Limit $\times10^6$ & Experiment & Reference\\
\endfirsthead
%\hline\hline
\multicolumn{4}{c}{\tablename\ \thetable{} -- continued from previous page} \\ \hline
Decay & Limit $\times10^6$ & Experiment & Reference\\
\endhead

\hline
$\pi{}^+e^+e^-$ & 110.0 & E687 & \cite{Frabetti:1997wp}\\
& 52.0 & E791 & \cite{Aitala:1999db}\\
& 5.9 & CLEO & \cite{Rubin:2010cq}\\
& 1.1 & \babar & \cite{Lees:2011hb}\\
& 0.3 & BESIII & \cite{Zhao:2016jna}\\
\hline
$\pi{}^+\mu{}^+\mu{}^-$ & 220.0 & E653 & \cite{Kodama:1995ia}\\
& 89.0 & E687 & \cite{Frabetti:1997wp}\\
& 15.0 & E791 & \cite{Aitala:1999db}\\
& 8.8 & Focus & \cite{Link:2003qp}\\
& 6.5 & \babar & \cite{Lees:2011hb}\\
& 3.9 & D0 & \cite{Abazov:2007aj}\\
& 0.073 & LHCb & \cite{Aaij:2013sua}\\
\hline
$\rho{}^+\mu{}^+\mu{}^-$ & 560.0 & E653 & \cite{Kodama:1995ia}\\
\hline
$K^+e^+e^-$ & 200.0 & E687 & \cite{Frabetti:1997wp}\\
& 3.0 & CLEO & \cite{Rubin:2010cq}\\
& 1.2 & BESIII & \cite{Zhao:2016jna}\\
& 1.0 & \babar & \cite{Lees:2011hb}\\
\hline
$\pi{}^+e^{\pm}\mu{}^{\mp}$ & 34.0 & E791 & \cite{Aitala:1999db}\\
\hline
$\pi{}^+e^+\mu{}^-$ & 110.0 & E687 & \cite{Frabetti:1997wp}\\
& 2.9 & \babar & \cite{Lees:2011hb}\\
\hline
$\pi{}^+\mu{}^+e^-$ & 130.0 & E687 & \cite{Frabetti:1997wp}\\
& 3.6 & \babar & \cite{Lees:2011hb}\\
\hline
$K^+e^{\pm}\mu{}^{\mp}$ & 68.0 & E791 & \cite{Aitala:1999db}\\
\hline
$K^+e^+\mu{}^-$ & 130.0 & E687 & \cite{Frabetti:1997wp}\\
& 1.2 & \babar & \cite{Lees:2011hb}\\
\hline
$K^+\mu{}^+e^-$ & 120.0 & E687 & \cite{Frabetti:1997wp}\\
& 2.8 & \babar & \cite{Lees:2011hb}\\
\hline
$\pi{}^-e^+e^+$ & 110.0 & E687 & \cite{Frabetti:1997wp}\\
& 96.0 & E791 & \cite{Aitala:1999db}\\
& 1.9 & \babar & \cite{Lees:2011hb}\\
& 1.2 & BESIII & \cite{Zhao:2016jna}\\
& 1.1 & CLEO & \cite{Rubin:2010cq}\\
\hline
$\pi{}^-\mu{}^+\mu{}^+$ & 87.0 & E687 & \cite{Frabetti:1997wp}\\
& 17.0 & E791 & \cite{Aitala:1999db}\\
& 4.8 & Focus & \cite{Link:2003qp}\\
& 2.0 & \babar & \cite{Lees:2011hb}\\
& 0.022 & LHCb & \cite{Aaij:2013sua}\\
\hline
$\pi{}^-e^+\mu{}^+$ & 110.0 & E687 & \cite{Frabetti:1997wp}\\
& 50.0 & E791 & \cite{Aitala:1999db}\\
\hline
$\rho{}^-\mu{}^+\mu{}^+$ & 560.0 & E653 & \cite{Kodama:1995ia}\\
\hline
$K^-e^+e^+$ & 120.0 & E687 & \cite{Frabetti:1997wp}\\
& 3.5 & CLEO & \cite{Rubin:2010cq}\\
& 0.9 & \babar & \cite{Lees:2011hb}\\
& 0.6 & BESIII & \cite{Zhao:2016jna}\\
\hline
$K^-\mu{}^+\mu{}^+$ & 320.0 & E653 & \cite{Kodama:1995ia}\\
& 120.0 & E687 & \cite{Frabetti:1997wp}\\
& 13.0 & Focus & \cite{Link:2003qp}\\
& 10.0 & \babar & \cite{Lees:2011hb}\\
\hline
$K^-e^+\mu{}^+$ & 130.0 & E687 & \cite{Frabetti:1997wp}\\
\hline
$K^{*-}(892)\mu{}^+\mu{}^+$ & 850.0 & E653 & \cite{Kodama:1995ia}\\
\hline
\end{longtable}

\begin{longtable}{l|ccc}
\caption{Upper limits at $90\%$ C.L.\ for $D_s^+$ decays.\label{tab:charm:rare_dsplus}}\\
\hline\hline
Decay & Limit $\times10^6$ & Experiment & Reference\\
\endfirsthead
%\hline\hline
\multicolumn{4}{c}{\tablename\ \thetable{} -- continued from previous page} \\ \hline
Decay & Limit $\times10^6$ & Experiment & Reference\\
\endhead

\hline
$\pi{}^+e^+e^-$ & 270.0 & E791 & \cite{Aitala:1999db}\\
& 22.0 & CLEO & \cite{Rubin:2010cq}\\
& 13.0 & \babar & \cite{Lees:2011hb}\\
\hline
$\pi{}^+\mu{}^+\mu{}^-$ & 430.0 & E653 & \cite{Kodama:1995ia}\\
& 140.0 & E791 & \cite{Aitala:1999db}\\
& 43.0 & \babar & \cite{Lees:2011hb}\\
& 26.0 & Focus & \cite{Link:2003qp}\\
& 0.41 & LHCb & \cite{Aaij:2013sua}\\
\hline
$K^+e^+e^-$ & 1600.0 & E791 & \cite{Aitala:1999db}\\
& 52.0 & CLEO & \cite{Rubin:2010cq}\\
& 3.7 & \babar & \cite{Lees:2011hb}\\
\hline
$K^+\mu{}^+\mu{}^-$ & 140.0 & E791 & \cite{Aitala:1999db}\\
& 36.0 & Focus & \cite{Link:2003qp}\\
& 21.0 & \babar & \cite{Lees:2011hb}\\
\hline
$K^{*+}(892)\mu{}^+\mu{}^-$ & 1400.0 & E653 & \cite{Kodama:1995ia}\\
\hline
$\pi{}^+e^{\pm}\mu{}^{\mp}$ & 610.0 & E791 & \cite{Aitala:1999db}\\
\hline
$\pi{}^+e^+\mu{}^-$ & 12.0 & \babar & \cite{Lees:2011hb}\\
\hline
$\pi{}^+\mu{}^+e^-$ & 20.0 & \babar & \cite{Lees:2011hb}\\
\hline
$K^+e^{\pm}\mu{}^{\mp}$ & 630.0 & E791 & \cite{Aitala:1999db}\\
\hline
$K^+e^+\mu{}^-$ & 14.0 & \babar & \cite{Lees:2011hb}\\
\hline
$K^+\mu{}^+e^-$ & 9.7 & \babar & \cite{Lees:2011hb}\\
\hline
$\pi{}^-e^+e^+$ & 690.0 & E791 & \cite{Aitala:1999db}\\
& 18.0 & CLEO & \cite{Rubin:2010cq}\\
& 4.1 & \babar & \cite{Lees:2011hb}\\
\hline
$\pi{}^-\mu{}^+\mu{}^+$ & 430.0 & E653 & \cite{Kodama:1995ia}\\
& 82.0 & E791 & \cite{Aitala:1999db}\\
& 29.0 & Focus & \cite{Link:2003qp}\\
& 14.0 & \babar & \cite{Lees:2011hb}\\
& 0.12 & LHCb & \cite{Aaij:2013sua}\\
\hline
$\pi{}^-e^+\mu{}^+$ & 730.0 & E791 & \cite{Aitala:1999db}\\
\hline
$K^-e^+e^+$ & 630.0 & E791 & \cite{Aitala:1999db}\\
& 17.0 & CLEO & \cite{Rubin:2010cq}\\
& 5.2 & \babar & \cite{Lees:2011hb}\\
\hline
$K^-\mu{}^+\mu{}^+$ & 590.0 & E653 & \cite{Kodama:1995ia}\\
& 180.0 & E791 & \cite{Aitala:1999db}\\
& 13.0 & \babar & \cite{Lees:2011hb}\\
\hline
$K^-e^+\mu{}^+$ & 680.0 & E791 & \cite{Aitala:1999db}\\
\hline
$K^{*-}(892)\mu{}^+\mu{}^+$ & 1400.0 & E653 & \cite{Kodama:1995ia}\\
\hline
\end{longtable}

\begin{longtable}{l|ccc}
\caption{Upper limits at $90\%$ C.L.\ for $\Lambda_c^+$ decays.\label{tab:charm:rare_lambdac}}\\
\hline\hline
Decay & Limit $\times10^6$ & Experiment & Reference\\
\endfirsthead
%\hline\hline
\multicolumn{4}{c}{\tablename\ \thetable{} -- continued from previous page} \\ \hline
Decay & Limit $\times10^6$ & Experiment & Reference\\
\endhead

\hline
$pe^+e^-$ & 5.5 & \babar & \cite{Lees:2011hb}\\
\hline
$p\mu{}^+\mu{}^-$ & 340.0 & E653 & \cite{Kodama:1995ia}\\
& 44.0 & \babar & \cite{Lees:2011hb}\\
\hline
$\sigma{}^+\mu{}^+\mu{}^-$ & 700.0 & E653 & \cite{Kodama:1995ia}\\
\hline
$pe^+\mu{}^-$ & 9.9 & \babar & \cite{Lees:2011hb}\\
\hline
$p\mu{}^+e^-$ & 19.0 & \babar & \cite{Lees:2011hb}\\
\hline
$\overline{p}\,{}e^+e^+$ & 2.7 & \babar & \cite{Lees:2011hb}\\
\hline
$\overline{p}\,{}\mu{}^+\mu{}^+$ & 9.4 & \babar & \cite{Lees:2011hb}\\
\hline
\end{longtable}

\clearpage
%Tau decays
%% -*- mode: LaTeX; TeX-master: "../master.tex" -*-
%%
%% define report tags
%%
\newcommand{\hfagTauTag}{Spring 2017\xspace}
\newcommand{\hfagFitLabel}{HFLAV \hfagTauTag fit}

%%--- hevea flag
\newif\ifhevea\heveafalse
%%--- provide \cutname for non-hevea compiling
\providecommand{\cutname}[1]{}

%%--- env to get left aligned eqs
\makeatletter
\newenvironment*{fleqn}[1][\leftmargini minus\leftmargini]{\@fleqntrue
  \setlength\@mathmargin{#1}\ignorespaces
}{%
  \ignorespacesafterend
}

%%--- env to get center aligned eqs
\newenvironment{ceqn}{\@fleqnfalse
  \@mathmargin\@centering \ignorespaces
}{%
  \ignorespacesafterend
}

%%--- env to remove spaces before \cite{} from cite package
\newenvironment{citenoleadsp}{\@fleqnfalse
  \let\cite@adjust\@empty
}{%
  \ignorespacesafterend
}
\makeatother

%%--- be able to cite in math mode
\let\citeOld\cite
\renewcommand{\cite}[1]{\ifmmode\text{\citeOld{#1}}\else\citeOld{#1}\fi}

%%--- env for display math
\newenvironment{ensuredisplaymath}
  {\(\displaystyle}
  {\)}

%%--- bold math with \bfseries
\makeatletter
\DeclareRobustCommand\bfseries{%
  \not@math@alphabet\bfseries\mathbf
  \fontseries\bfdefault\selectfont\boldmath}
\DeclareRobustCommand*{\bm}[1]{%
    \mathchoice{\bmbox{\displaystyle}{#1}}%
               {\bmbox{\textstyle}{#1}}%
               {\bmbox{\scriptstyle}{#1}}%
               {\bmbox{\scriptscriptstyle}{#1}}}
\DeclareRobustCommand*{\bmbox}[2]{\mbox{\bfseries$#1 #2$}}
\makeatother

%%--- normal size captions in longtable
\makeatletter
\def\LT@makecaption#1#2#3{%
  \LT@mcol\LT@cols c{\hbox to\z@{\hss\parbox[t]\LTcapwidth{%
    \sbox\@tempboxa{\normalsize#1{#2: }#3}%
    \ifdim\wd\@tempboxa>\hsize
      \normalsize#1{#2: }#3%
    \else
      \hbox to\hsize{\hfil\box\@tempboxa\hfil}%
    \fi
    \endgraf\vskip\baselineskip}%
  \hss}}}
\makeatother

%%
%% defs for using LaTeX defs from HFLAV tau SW
%%
\makeatletter

%%--- define HFLAV tau quantity
\newcommand{\htdef}[2]{%
  \@namedef{hfagtau@#1}{#2}%
}

%%--- retrieve HFLAV tau quantity
\iffalse
\newcommand{\htuse}[1]{%
  \@nameuse{hfagtau@#1}}%
\else
\newcommand{\htuse}[1]{%
  \ifcsname hfagtau@#1\endcsname
  \@nameuse{hfagtau@#1}%
  \else
  \@latex@error{Undefined name hfagtau@#1}\@eha
  \fi}
\fi

%% \htconstrdef{name}{left}{right}
\newcommand{\htconstrdef}[4]{%
  \@namedef{hfagtau@#1.left}{\ensuremath{#2}}%
  \@namedef{hfagtau@#1.right}{\ensuremath{#3}}%
  \@namedef{hfagtau@#1.right.split}{\ensuremath{#4}}%
  \@namedef{hfagtau@#1.constr.eq}{\htuse{#1.left} ={}& \htuse{#1.right}}%
}

%% \htquantdef{name}{gammaname}{texdescr}{valerr}{val}{err}
\iffalse
\newcommand{\htquantdef}[6]{%
  \@namedef{hfagtau@#1.gn}{\ensuremath{#2}}%
  \@namedef{hfagtau@#1.td}{\ensuremath{#3}}%
  \@namedef{hfagtau@#1}{\ensuremath{#4}}%
  \@namedef{hfagtau@#1.v}{\ensuremath{#5}}%
  \@namedef{hfagtau@#1.e}{\ensuremath{#6}}%
}
\else
\newcommand{\htquantdef}[6]{%
  \ifx&#2&\else
  \@namedef{hfagtau@#1.gn}{\ensuremath{#2}}%
  \fi
  \ifx&#3&\else
  \@namedef{hfagtau@#1.td}{\ensuremath{#3}}%
  \fi
  \ifx&#6&%
    \@namedef{hfagtau@#1}{\ensuremath{#5}}%
  \else
    \ifthenelse{\equal{#6}{0}}{%
      \@namedef{hfagtau@#1}{\ensuremath{#5}}%
    }{%
      \@namedef{hfagtau@#1}{\ensuremath{#4}}%
      \@namedef{hfagtau@#1.v}{\ensuremath{#5}}%
      \@namedef{hfagtau@#1.e}{\ensuremath{#6}}%
    }%
  \fi
}
\fi

%% \htmeasdef{name}{quant}{exp}{ref}{valerr}{val}{err}{syst}
\newcommand{\htmeasdef}[8]{%
  \@namedef{hfagtau@#1,quant}{\ensuremath{#2}}%
  \@namedef{hfagtau@#1,exp}{#3}%
  \@namedef{hfagtau@#1,ref}{\cite{#4}}%
  \@namedef{hfagtau@#1}{\ensuremath{#5}}%
  \@namedef{hfagtau@#1,val}{\ensuremath{#6}}%
  \@namedef{hfagtau@#1,stat}{\ensuremath{#7}}%
  \@namedef{hfagtau@#1,syst}{\ensuremath{#8}}%
}

\newcommand{\htQuantLine}[3]{\ensuremath{\htuse{#1.td}}&\ensuremath{#2}\\}

\makeatother

\newcommand{\BRF}[2]{#2}
\renewcommand{\BR}{\ensuremath{B}\xspace}
\ifhevea
\renewcommand{\babar}{BaBar\xspace}
\else
\providecommand{\babar}{\mbox{\slshape B\kern-0.1em{\smaller A}\kern-0.1em
    B\kern-0.1em{\smaller A\kern-0.2em R}}\xspace}
\fi
%%
%%\newcommand{\chisq}{\ensuremath{\chi^2}\xspace}
%%\newcommand{\nub}{\ensuremath{\bar{\nu}}\xspace}
%%

%%--- higher and lower mass lepton for \eg\ tau leptonic partial widths
\newcommand{\lepth}{\lambda}
\newcommand{\leptl}{\rho}

%%--- symbol for multiplicative radiative correction, typically ~1 
\newcommand{\radRatio}{R}
%%--- symbol for additive radiative correction, typically <<1 
\newcommand{\radDelta}{\delta R}

\newcommand{\Rhad}{\ensuremath{R_{\text{had}}}\xspace}
\newcommand{\BRhad}{\ensuremath{\BR_{\text{had}}}\xspace}
\newcommand{\Gammahad}{\ensuremath{\Gamma_{\text{had}}}\xspace}
\newcommand{\Rstrange}{\ensuremath{R_s}\xspace}
\newcommand{\BRstrange}{\ensuremath{\BR_s}\xspace}
\newcommand{\Gammastrange}{\ensuremath{\Gamma_s}\xspace}
\newcommand{\Rnonstrange}{\ensuremath{R_{\text{VA}}}\xspace}
\newcommand{\BRnonstrange}{\ensuremath{\BR_{\text{VA}}}\xspace}
\newcommand{\Gammanonstrange}{\ensuremath{\Gamma_{\text{VA}}}\xspace}
\newcommand{\tauknu}{\ensuremath{\tau^{-} \to K^{-} \nut}\xspace}
\newcommand{\taupinu}{\ensuremath{\tau^{-} \to \pi^{-} \nut}\xspace}
\newcommand{\BFtautoknu}{\ensuremath{\BR(\tauknu)}\xspace}
\newcommand{\BFtautopinu}{\ensuremath{\BR(\taupinu)}\xspace}
\newcommand{\VusUni}{\ensuremath{\Vus_{\text{uni}}}\xspace}
\newcommand{\VusTauIncl}{\ensuremath{\Vus_{\tau s}}\xspace}
\newcommand{\VusTauKpi}{\ensuremath{\Vus_{\tau K/\pi}}\xspace}
\newcommand{\VusTauKnu}{\ensuremath{\Vus_{\tau K}}\xspace}
\newcommand{\VusTauKPinu}{\ensuremath{\Vus_{\tau K\pi}}\xspace}
\newcommand{\hfagtau}{HFLAV-Tau\xspace}
%% Decker:1994dd hadronic corrections
\newcommand{\dRradTauhHmu}{\ensuremath{\radDelta_{\tau/h}}\xspace}
\newcommand{\dRradTaupiPimu}{\ensuremath{\radDelta_{\tau/\pi}}\xspace}
\newcommand{\dRradTaukKmu}{\ensuremath{\radDelta_{\tau/K}}\xspace}
\newcommand{\RradKmuPimu}{\ensuremath{\radRatio_{K/\pi}}\xspace}
\providecommand{\slashLikeH}{%
  \raisebox{.9ex}{%
    \scalebox{.7}{%
      \rotatebox[origin=c]{18}{$-$}%
    }%
  }%
}
\providecommand{\hslash}{%
  {%
   \vphantom{h}%
   \ooalign{\kern.05em\smash{\slashLikeH}\hidewidth\cr$h$\cr}%
   \kern.05em
  }%
}

\htdef{UnitarityResid}{(0.0355 \pm 0.1031)\%}%
\htdef{MeasNum}{170}%
\htdef{QuantNum}{135}%
\htdef{QuantNumNonRatio}{118}%
\htdef{QuantNumRatio}{17}%
\htdef{QuantNumWithMeas}{84}%
\htdef{QuantNumNonRatioWithMeas}{71}%
\htdef{QuantNumRatioWithMeas}{13}%
\htdef{QuantNumPdg}{129}%
\htdef{QuantNumNonRatioPdg}{112}%
\htdef{QuantNumRatioPdg}{17}%
\htdef{QuantNumWithMeasPdg}{82}%
\htdef{QuantNumNonRatioWithMeasPdg}{69}%
\htdef{QuantNumRatioWithMeasPdg}{13}%
\htdef{IndepQuantNum}{47}%
\htdef{BaseQuantNum}{47}%
\htdef{UnitarityQuantNum}{48}%
\htdef{ConstrNum}{88}%
\htdef{ConstrNumPdg}{82}%
\htdef{Chisq}{137}%
\htdef{Dof}{123}%
\htdef{ChisqProb}{17.84\%}%
\htdef{ChisqProbRound}{18\%}%
\htmeasdef{ALEPH.Gamma10.pub.BARATE.99K}{Gamma10}{ALEPH}{Barate:1999hi}{0.00696 \pm 0.0002865}{0.00696}{\pm 0.0002865}{0}%
\htmeasdef{ALEPH.Gamma103.pub.SCHAEL.05C}{Gamma103}{ALEPH}{Schael:2005am}{0.00072 \pm 0.00015}{0.00072}{\pm 0.00015}{0}%
\htmeasdef{ALEPH.Gamma104.pub.SCHAEL.05C}{Gamma104}{ALEPH}{Schael:2005am}{( 0.021 \pm 0.007 \pm 0.009 ) \cdot 10^{ -2 }}{0.021e-2}{\pm 0.007e-2}{0.009e-2}%
\htmeasdef{ALEPH.Gamma126.pub.BUSKULIC.97C}{Gamma126}{ALEPH}{Buskulic:1996qs}{0.0018 \pm 0.0004472}{0.0018}{\pm 0.0004472}{0}%
\htmeasdef{ALEPH.Gamma128.pub.BUSKULIC.97C}{Gamma128}{ALEPH}{Buskulic:1996qs}{( 2.9 {}^{+1.3\cdot 10^{-4}}_{-1.2} \pm 0.7 ) \cdot 10^{ -4 }}{2.9e-4}{{}^{+1.3e-4}_{-1.2e-4}}{0.7e-4}%
\htmeasdef{ALEPH.Gamma13.pub.SCHAEL.05C}{Gamma13}{ALEPH}{Schael:2005am}{0.25924 \pm 0.00128973}{0.25924}{\pm 0.00128973}{0}%
\htmeasdef{ALEPH.Gamma150.pub.BUSKULIC.97C}{Gamma150}{ALEPH}{Buskulic:1996qs}{0.0191 \pm 0.000922}{0.0191}{\pm 0.000922}{0}%
\htmeasdef{ALEPH.Gamma150by66.pub.BUSKULIC.96}{Gamma150by66}{ALEPH}{Buskulic:1995ty}{0.431 \pm 0.033}{0.431}{\pm 0.033}{0}%
\htmeasdef{ALEPH.Gamma152.pub.BUSKULIC.97C}{Gamma152}{ALEPH}{Buskulic:1996qs}{0.0043 \pm 0.000781}{0.0043}{\pm 0.000781}{0}%
\htmeasdef{ALEPH.Gamma16.pub.BARATE.99K}{Gamma16}{ALEPH}{Barate:1999hi}{0.00444 \pm 0.0003538}{0.00444}{\pm 0.0003538}{0}%
\htmeasdef{ALEPH.Gamma19.pub.SCHAEL.05C}{Gamma19}{ALEPH}{Schael:2005am}{0.09295 \pm 0.00121655}{0.09295}{\pm 0.00121655}{0}%
\htmeasdef{ALEPH.Gamma23.pub.BARATE.99K}{Gamma23}{ALEPH}{Barate:1999hi}{0.00056 \pm 0.00025}{0.00056}{\pm 0.00025}{0}%
\htmeasdef{ALEPH.Gamma26.pub.SCHAEL.05C}{Gamma26}{ALEPH}{Schael:2005am}{0.01082 \pm 0.000925581}{0.01082}{\pm 0.000925581}{0}%
\htmeasdef{ALEPH.Gamma28.pub.BARATE.99K}{Gamma28}{ALEPH}{Barate:1999hi}{0.00037 \pm 0.0002371}{0.00037}{\pm 0.0002371}{0}%
\htmeasdef{ALEPH.Gamma3.pub.SCHAEL.05C}{Gamma3}{ALEPH}{Schael:2005am}{0.17319 \pm 0.000769675}{0.17319}{\pm 0.000769675}{0}%
\htmeasdef{ALEPH.Gamma30.pub.SCHAEL.05C}{Gamma30}{ALEPH}{Schael:2005am}{0.00112 \pm 0.000509313}{0.00112}{\pm 0.000509313}{0}%
\htmeasdef{ALEPH.Gamma33.pub.BARATE.98E}{Gamma33}{ALEPH}{Barate:1997tt}{0.0097 \pm 0.000849}{0.0097}{\pm 0.000849}{0}%
\htmeasdef{ALEPH.Gamma35.pub.BARATE.99K}{Gamma35}{ALEPH}{Barate:1999hi}{0.00928 \pm 0.000564}{0.00928}{\pm 0.000564}{0}%
\htmeasdef{ALEPH.Gamma37.pub.BARATE.98E}{Gamma37}{ALEPH}{Barate:1997tt}{0.00158 \pm 0.0004531}{0.00158}{\pm 0.0004531}{0}%
\htmeasdef{ALEPH.Gamma37.pub.BARATE.99K}{Gamma37}{ALEPH}{Barate:1999hi}{0.00162 \pm 0.0002371}{0.00162}{\pm 0.0002371}{0}%
\htmeasdef{ALEPH.Gamma40.pub.BARATE.98E}{Gamma40}{ALEPH}{Barate:1997tt}{0.00294 \pm 0.0008184}{0.00294}{\pm 0.0008184}{0}%
\htmeasdef{ALEPH.Gamma40.pub.BARATE.99K}{Gamma40}{ALEPH}{Barate:1999hi}{0.00347 \pm 0.0006464}{0.00347}{\pm 0.0006464}{0}%
\htmeasdef{ALEPH.Gamma42.pub.BARATE.98E}{Gamma42}{ALEPH}{Barate:1997tt}{0.00152 \pm 0.0007885}{0.00152}{\pm 0.0007885}{0}%
\htmeasdef{ALEPH.Gamma42.pub.BARATE.99K}{Gamma42}{ALEPH}{Barate:1999hi}{0.00143 \pm 0.0002915}{0.00143}{\pm 0.0002915}{0}%
\htmeasdef{ALEPH.Gamma44.pub.BARATE.99R}{Gamma44}{ALEPH}{Barate:1999hj}{0.00026 \pm 0.00024}{0.00026}{\pm 0.00024}{0}%
\htmeasdef{ALEPH.Gamma47.pub.BARATE.98E}{Gamma47}{ALEPH}{Barate:1997tt}{0.00026 \pm 0.0001118}{0.00026}{\pm 0.0001118}{0}%
\htmeasdef{ALEPH.Gamma48.pub.BARATE.98E}{Gamma48}{ALEPH}{Barate:1997tt}{0.00101 \pm 0.0002642}{0.00101}{\pm 0.0002642}{0}%
\htmeasdef{ALEPH.Gamma5.pub.SCHAEL.05C}{Gamma5}{ALEPH}{Schael:2005am}{0.17837 \pm 0.000804984}{0.17837}{\pm 0.000804984}{0}%
\htmeasdef{ALEPH.Gamma51.pub.BARATE.98E}{Gamma51}{ALEPH}{Barate:1997tt}{( 3.1 \pm 1.1 \pm 0.5 ) \cdot 10^{ -4 }}{3.1e-4}{\pm 1.1e-4}{0.5e-4}%
\htmeasdef{ALEPH.Gamma53.pub.BARATE.98E}{Gamma53}{ALEPH}{Barate:1997tt}{0.00023 \pm 0.000202485}{0.00023}{\pm 0.000202485}{0}%
\htmeasdef{ALEPH.Gamma58.pub.SCHAEL.05C}{Gamma58}{ALEPH}{Schael:2005am}{0.09469 \pm 0.000957758}{0.09469}{\pm 0.000957758}{0}%
\htmeasdef{ALEPH.Gamma66.pub.SCHAEL.05C}{Gamma66}{ALEPH}{Schael:2005am}{0.04734 \pm 0.000766942}{0.04734}{\pm 0.000766942}{0}%
\htmeasdef{ALEPH.Gamma76.pub.SCHAEL.05C}{Gamma76}{ALEPH}{Schael:2005am}{0.00435 \pm 0.000460977}{0.00435}{\pm 0.000460977}{0}%
\htmeasdef{ALEPH.Gamma8.pub.SCHAEL.05C}{Gamma8}{ALEPH}{Schael:2005am}{0.11524 \pm 0.00104805}{0.11524}{\pm 0.00104805}{0}%
\htmeasdef{ALEPH.Gamma805.pub.SCHAEL.05C}{Gamma805}{ALEPH}{Schael:2005am}{( 4 \pm 2 ) \cdot 10^{ -4 }}{4e-04}{\pm 2e-04}{0}%
\htmeasdef{ALEPH.Gamma85.pub.BARATE.98}{Gamma85}{ALEPH}{Barate:1997ma}{0.00214 \pm 0.0004701}{0.00214}{\pm 0.0004701}{0}%
\htmeasdef{ALEPH.Gamma88.pub.BARATE.98}{Gamma88}{ALEPH}{Barate:1997ma}{0.00061 \pm 0.0004295}{0.00061}{\pm 0.0004295}{0}%
\htmeasdef{ALEPH.Gamma93.pub.BARATE.98}{Gamma93}{ALEPH}{Barate:1997ma}{0.00163 \pm 0.0002702}{0.00163}{\pm 0.0002702}{0}%
\htmeasdef{ALEPH.Gamma94.pub.BARATE.98}{Gamma94}{ALEPH}{Barate:1997ma}{0.00075 \pm 0.0003265}{0.00075}{\pm 0.0003265}{0}%
\htmeasdef{ARGUS.Gamma103.pub.ALBRECHT.88B}{Gamma103}{ARGUS}{Albrecht:1987zf}{0.00064 \pm 0.00023 \pm 0.0001}{0.00064}{\pm 0.00023}{0.0001}%
\htmeasdef{ARGUS.Gamma3by5.pub.ALBRECHT.92D}{Gamma3by5}{ARGUS}{Albrecht:1991rh}{0.997 \pm 0.035 \pm 0.04}{0.997}{\pm 0.035}{0.04}%
\htmeasdef{BaBar.Gamma10by5.pub.AUBERT.10F}{Gamma10by5}{\babar}{Aubert:2009qj}{0.03882 \pm 0.00032 \pm 0.00057}{0.03882}{\pm 0.00032}{0.00057}%
\htmeasdef{BaBar.Gamma128.pub.DEL-AMO-SANCHEZ.11E}{Gamma128}{\babar}{delAmoSanchez:2010pc}{0.000142 \pm 1.1\cdot 10^{-5} \pm 7\cdot 10^{-6}}{0.000142}{\pm 1.1e-05}{7e-06}%
\htmeasdef{BaBar.Gamma16.pub.AUBERT.07AP}{Gamma16}{\babar}{Aubert:2007jh}{0.00416 \pm 3\cdot 10^{-5} \pm 0.00018}{0.00416}{\pm 3e-05}{0.00018}%
\htmeasdef{BaBar.Gamma3by5.pub.AUBERT.10F}{Gamma3by5}{\babar}{Aubert:2009qj}{0.9796 \pm 0.0016 \pm 0.0036}{0.9796}{\pm 0.0016}{0.0036}%
\htmeasdef{BaBar.Gamma47.pub.LEES.12Y}{Gamma47}{\babar}{Lees:2012de}{( 2.31 \pm 0.04 \pm 0.08 ) \cdot 10^{ -4 }}{2.31e-4}{\pm 0.04e-4}{0.08e-4}%
\htmeasdef{BaBar.Gamma50.pub.LEES.12Y}{Gamma50}{\babar}{Lees:2012de}{( 1.60 \pm 0.20 \pm 0.22 ) \cdot 10^{ -5 }}{1.60e-5}{\pm 0.20e-5}{0.22e-5}%
\htmeasdef{BaBar.Gamma60.pub.AUBERT.08}{Gamma60}{\babar}{Aubert:2007mh}{0.0883 \pm 0.0001 \pm 0.0013}{0.0883}{\pm 0.0001}{0.0013}%
\htmeasdef{BaBar.Gamma811.pub.LEES.12X}{Gamma811}{\babar}{Lees:2012ks}{( 7.3 \pm 1.2 \pm 1.2 ) \cdot 10^{ -5 }}{7.3e-5}{\pm 1.2e-5}{1.2e-5}%
\htmeasdef{BaBar.Gamma812.pub.LEES.12X}{Gamma812}{\babar}{Lees:2012ks}{( 0.1 \pm 0.08 \pm 0.30 ) \cdot 10^{ -4 }}{0.1e-4}{\pm 0.08e-4}{0.30e-4}%
\htmeasdef{BaBar.Gamma821.pub.LEES.12X}{Gamma821}{\babar}{Lees:2012ks}{( 7.68 \pm 0.04 \pm 0.40 ) \cdot 10^{ -4 }}{7.68e-4}{\pm 0.04e-4}{0.40e-4}%
\htmeasdef{BaBar.Gamma822.pub.LEES.12X}{Gamma822}{\babar}{Lees:2012ks}{( 0.6 \pm 0.5 \pm 1.1 ) \cdot 10^{ -6 }}{0.6e-06}{\pm 0.5e-06}{1.1e-06}%
\htmeasdef{BaBar.Gamma831.pub.LEES.12X}{Gamma831}{\babar}{Lees:2012ks}{( 8.4 \pm 0.4 \pm 0.6 ) \cdot 10^{ -5 }}{8.4e-5}{\pm 0.4e-5}{0.6e-5}%
\htmeasdef{BaBar.Gamma832.pub.LEES.12X}{Gamma832}{\babar}{Lees:2012ks}{( 0.36 \pm 0.03 \pm 0.09 ) \cdot 10^{ -4 }}{0.36e-4}{\pm 0.03e-4}{0.09e-4}%
\htmeasdef{BaBar.Gamma833.pub.LEES.12X}{Gamma833}{\babar}{Lees:2012ks}{( 1.1 \pm 0.4 \pm 0.4 ) \cdot 10^{ -6 }}{1.1e-6}{\pm 0.4e-6}{0.4e-6}%
\htmeasdef{BaBar.Gamma85.pub.AUBERT.08}{Gamma85}{\babar}{Aubert:2007mh}{0.00273 \pm 2\cdot 10^{-5} \pm 9\cdot 10^{-5}}{0.00273}{\pm 2e-05}{9e-05}%
\htmeasdef{BaBar.Gamma910.pub.LEES.12X}{Gamma910}{\babar}{Lees:2012ks}{( 8.27 \pm 0.88 \pm 0.81 ) \cdot 10^{ -5 }}{8.27e-5}{\pm 0.88e-5}{0.81e-5}%
\htmeasdef{BaBar.Gamma911.pub.LEES.12X}{Gamma911}{\babar}{Lees:2012ks}{( 4.57 \pm 0.77 \pm 0.50 ) \cdot 10^{ -5 }}{4.57e-5}{\pm 0.77e-5}{0.50e-5}%
\htmeasdef{BaBar.Gamma920.pub.LEES.12X}{Gamma920}{\babar}{Lees:2012ks}{( 5.20 \pm 0.31 \pm 0.37 ) \cdot 10^{ -5 }}{5.20e-5}{\pm 0.31e-5}{0.37e-5}%
\htmeasdef{BaBar.Gamma93.pub.AUBERT.08}{Gamma93}{\babar}{Aubert:2007mh}{0.001346 \pm 1\cdot 10^{-5} \pm 3.6\cdot 10^{-5}}{0.001346}{\pm 1e-05}{3.6e-05}%
\htmeasdef{BaBar.Gamma930.pub.LEES.12X}{Gamma930}{\babar}{Lees:2012ks}{( 5.39 \pm 0.27 \pm 0.41 ) \cdot 10^{ -5 }}{5.39e-5}{\pm 0.27e-5}{0.41e-5}%
\htmeasdef{BaBar.Gamma944.pub.LEES.12X}{Gamma944}{\babar}{Lees:2012ks}{( 8.26 \pm 0.35 \pm 0.51 ) \cdot 10^{ -5 }}{8.26e-5}{\pm 0.35e-5}{0.51e-5}%
\htmeasdef{BaBar.Gamma96.pub.AUBERT.08}{Gamma96}{\babar}{Aubert:2007mh}{1.5777\cdot 10^{-5} \pm 1.3\cdot 10^{-6} \pm 1.2308\cdot 10^{-6}}{1.5777e-05}{\pm 1.3e-06}{1.2308e-06}%
\htmeasdef{BaBar.Gamma9by5.pub.AUBERT.10F}{Gamma9by5}{\babar}{Aubert:2009qj}{0.5945 \pm 0.0014 \pm 0.0061}{0.5945}{\pm 0.0014}{0.0061}%
\htmeasdef{Belle.Gamma126.pub.INAMI.09}{Gamma126}{Belle}{Inami:2008ar}{0.00135 \pm 3\cdot 10^{-5} \pm 7\cdot 10^{-5}}{0.00135}{\pm 3e-05}{7e-05}%
\htmeasdef{Belle.Gamma128.pub.INAMI.09}{Gamma128}{Belle}{Inami:2008ar}{0.000158 \pm 5\cdot 10^{-6} \pm 9\cdot 10^{-6}}{0.000158}{\pm 5e-06}{9e-06}%
\htmeasdef{Belle.Gamma13.pub.FUJIKAWA.08}{Gamma13}{Belle}{Fujikawa:2008ma}{0.2567 \pm 1\cdot 10^{-4} \pm 0.0039}{0.2567}{\pm 1e-04}{0.0039}%
\htmeasdef{Belle.Gamma130.pub.INAMI.09}{Gamma130}{Belle}{Inami:2008ar}{4.6\cdot 10^{-5} \pm 1.1\cdot 10^{-5} \pm 4\cdot 10^{-6}}{4.6e-05}{\pm 1.1e-05}{4e-06}%
\htmeasdef{Belle.Gamma132.pub.INAMI.09}{Gamma132}{Belle}{Inami:2008ar}{8.8\cdot 10^{-5} \pm 1.4\cdot 10^{-5} \pm 6\cdot 10^{-6}}{8.8e-05}{\pm 1.4e-05}{6e-06}%
\htmeasdef{Belle.Gamma35.pub.RYU.14vpc}{Gamma35}{Belle}{Ryu:2014vpc}{8.32\cdot 10^{-3} \pm 0.3\% \pm 1.8\%}{8.32e-03}{\pm 0.3\%}{1.8\%}%
\htmeasdef{Belle.Gamma37.pub.RYU.14vpc}{Gamma37}{Belle}{Ryu:2014vpc}{14.8\cdot 10^{-4} \pm 0.9\% \pm 3.7\%}{14.8e-04}{\pm 0.9\%}{3.7\%}%
\htmeasdef{Belle.Gamma40.pub.RYU.14vpc}{Gamma40}{Belle}{Ryu:2014vpc}{3.86\cdot 10^{-3} \pm 0.8\% \pm 3.5\%}{3.86e-03}{\pm 0.8\%}{3.5\%}%
\htmeasdef{Belle.Gamma42.pub.RYU.14vpc}{Gamma42}{Belle}{Ryu:2014vpc}{14.96\cdot 10^{-4} \pm 1.3\% \pm 4.9\%}{14.96e-04}{\pm 1.3\%}{4.9\%}%
\htmeasdef{Belle.Gamma47.pub.RYU.14vpc}{Gamma47}{Belle}{Ryu:2014vpc}{2.33\cdot 10^{-4} \pm 1.4\% \pm 4.0\%}{2.33e-04}{\pm 1.4\%}{4.0\%}%
\htmeasdef{Belle.Gamma50.pub.RYU.14vpc}{Gamma50}{Belle}{Ryu:2014vpc}{2.00\cdot 10^{-5} \pm 10.8\% \pm 10.1\%}{2.00e-05}{\pm 10.8\%}{10.1\%}%
\htmeasdef{Belle.Gamma60.pub.LEE.10}{Gamma60}{Belle}{Lee:2010tc}{0.0842 \pm 0 {}^{+0.0026}_{-0.0025}}{0.0842}{\pm 0}{{}^{+0.0026}_{-0.0025}}%
\htmeasdef{Belle.Gamma85.pub.LEE.10}{Gamma85}{Belle}{Lee:2010tc}{0.0033 \pm 1\cdot 10^{-5} {}^{+0.00016}_{-0.00017}}{0.0033}{\pm 1e-05}{{}^{+0.00016}_{-0.00017}}%
\htmeasdef{Belle.Gamma93.pub.LEE.10}{Gamma93}{Belle}{Lee:2010tc}{0.00155 \pm 1\cdot 10^{-5} {}^{+6\cdot 10^{-5}}_{-5\cdot 10^{-5}}}{0.00155}{\pm 1e-05}{{}^{+6e-05}_{-5e-05}}%
\htmeasdef{Belle.Gamma96.pub.LEE.10}{Gamma96}{Belle}{Lee:2010tc}{3.29\cdot 10^{-5} \pm 1.7\cdot 10^{-6} {}^{+1.9\cdot 10^{-6}}_{-2.0\cdot 10^{-6}}}{3.29e-05}{\pm 1.7e-06}{{}^{+1.9e-06}_{-2.0e-06}}%
\htmeasdef{CELLO.Gamma54.pub.BEHREND.89B}{Gamma54}{CELLO}{Behrend:1989wc}{0.15 \pm 0.004 \pm 0.003}{0.15}{\pm 0.004}{0.003}%
\htmeasdef{CLEO.Gamma10.pub.BATTLE.94}{Gamma10}{CLEO}{Battle:1994by}{0.0066 \pm 0.0007 \pm 0.0009}{0.0066}{\pm 0.0007}{0.0009}%
\htmeasdef{CLEO.Gamma102.pub.GIBAUT.94B}{Gamma102}{CLEO}{Gibaut:1994ik}{0.00097 \pm 5\cdot 10^{-5} \pm 0.00011}{0.00097}{\pm 5e-05}{0.00011}%
\htmeasdef{CLEO.Gamma103.pub.GIBAUT.94B}{Gamma103}{CLEO}{Gibaut:1994ik}{0.00077 \pm 5\cdot 10^{-5} \pm 9\cdot 10^{-5}}{0.00077}{\pm 5e-05}{9e-05}%
\htmeasdef{CLEO.Gamma104.pub.ANASTASSOV.01}{Gamma104}{CLEO}{Anastassov:2000xu}{0.00017 \pm 2\cdot 10^{-5} \pm 2\cdot 10^{-5}}{0.00017}{\pm 2e-05}{2e-05}%
\htmeasdef{CLEO.Gamma126.pub.ARTUSO.92}{Gamma126}{CLEO}{Artuso:1992qu}{0.0017 \pm 0.0002 \pm 0.0002}{0.0017}{\pm 0.0002}{0.0002}%
\htmeasdef{CLEO.Gamma128.pub.BARTELT.96}{Gamma128}{CLEO}{Bartelt:1996iv}{( 2.6 \pm 0.5 \pm 0.5 ) \cdot 10^{ -4 }}{2.6e-4}{\pm 0.5e-4}{0.5e-4}%
\htmeasdef{CLEO.Gamma13.pub.ARTUSO.94}{Gamma13}{CLEO}{Artuso:1994ii}{0.2587 \pm 0.0012 \pm 0.0042}{0.2587}{\pm 0.0012}{0.0042}%
\htmeasdef{CLEO.Gamma130.pub.BISHAI.99}{Gamma130}{CLEO}{Bishai:1998gf}{( 1.77 \pm 0.56 \pm 0.71 ) \cdot 10^{ -4 }}{1.77e-4}{\pm 0.56e-4}{0.71e-4}%
\htmeasdef{CLEO.Gamma132.pub.BISHAI.99}{Gamma132}{CLEO}{Bishai:1998gf}{( 2.2 \pm 0.70 \pm 0.22 ) \cdot 10^{ -4 }}{2.2e-4}{\pm 0.70e-4}{0.22e-4}%
\htmeasdef{CLEO.Gamma150.pub.BARINGER.87}{Gamma150}{CLEO}{Baringer:1987tr}{0.016 \pm 0.0027 \pm 0.0041}{0.016}{\pm 0.0027}{0.0041}%
\htmeasdef{CLEO.Gamma150by66.pub.BALEST.95C}{Gamma150by66}{CLEO}{Balest:1995kq}{0.464 \pm 0.016 \pm 0.017}{0.464}{\pm 0.016}{0.017}%
\htmeasdef{CLEO.Gamma152by76.pub.BORTOLETTO.93}{Gamma152by76}{CLEO}{Bortoletto:1993px}{0.81 \pm 0.06 \pm 0.06}{0.81}{\pm 0.06}{0.06}%
\htmeasdef{CLEO.Gamma16.pub.BATTLE.94}{Gamma16}{CLEO}{Battle:1994by}{0.0051 \pm 0.001 \pm 0.0007}{0.0051}{\pm 0.001}{0.0007}%
\htmeasdef{CLEO.Gamma19by13.pub.PROCARIO.93}{Gamma19by13}{CLEO}{Procario:1992hd}{0.342 \pm 0.006 \pm 0.016}{0.342}{\pm 0.006}{0.016}%
\htmeasdef{CLEO.Gamma23.pub.BATTLE.94}{Gamma23}{CLEO}{Battle:1994by}{0.0009 \pm 0.001 \pm 0.0003}{0.0009}{\pm 0.001}{0.0003}%
\htmeasdef{CLEO.Gamma26by13.pub.PROCARIO.93}{Gamma26by13}{CLEO}{Procario:1992hd}{0.044 \pm 0.003 \pm 0.005}{0.044}{\pm 0.003}{0.005}%
\htmeasdef{CLEO.Gamma29.pub.PROCARIO.93}{Gamma29}{CLEO}{Procario:1992hd}{0.0016 \pm 0.0005 \pm 0.0005}{0.0016}{\pm 0.0005}{0.0005}%
\htmeasdef{CLEO.Gamma31.pub.BATTLE.94}{Gamma31}{CLEO}{Battle:1994by}{0.017 \pm 0.0012 \pm 0.0019}{0.017}{\pm 0.0012}{0.0019}%
\htmeasdef{CLEO.Gamma34.pub.COAN.96}{Gamma34}{CLEO}{Coan:1996iu}{0.00855 \pm 0.00036 \pm 0.00073}{0.00855}{\pm 0.00036}{0.00073}%
\htmeasdef{CLEO.Gamma37.pub.COAN.96}{Gamma37}{CLEO}{Coan:1996iu}{0.00151 \pm 0.00021 \pm 0.00022}{0.00151}{\pm 0.00021}{0.00022}%
\htmeasdef{CLEO.Gamma39.pub.COAN.96}{Gamma39}{CLEO}{Coan:1996iu}{0.00562 \pm 0.0005 \pm 0.00048}{0.00562}{\pm 0.0005}{0.00048}%
\htmeasdef{CLEO.Gamma3by5.pub.ANASTASSOV.97}{Gamma3by5}{CLEO}{Anastassov:1996tc}{0.9777 \pm 0.0063 \pm 0.0087}{0.9777}{\pm 0.0063}{0.0087}%
\htmeasdef{CLEO.Gamma42.pub.COAN.96}{Gamma42}{CLEO}{Coan:1996iu}{0.00145 \pm 0.00036 \pm 0.0002}{0.00145}{\pm 0.00036}{0.0002}%
\htmeasdef{CLEO.Gamma47.pub.COAN.96}{Gamma47}{CLEO}{Coan:1996iu}{0.00023 \pm 5\cdot 10^{-5} \pm 3\cdot 10^{-5}}{0.00023}{\pm 5e-05}{3e-05}%
\htmeasdef{CLEO.Gamma5.pub.ANASTASSOV.97}{Gamma5}{CLEO}{Anastassov:1996tc}{0.1776 \pm 0.0006 \pm 0.0017}{0.1776}{\pm 0.0006}{0.0017}%
\htmeasdef{CLEO.Gamma57.pub.BALEST.95C}{Gamma57}{CLEO}{Balest:1995kq}{0.0951 \pm 0.0007 \pm 0.002}{0.0951}{\pm 0.0007}{0.002}%
\htmeasdef{CLEO.Gamma66.pub.BALEST.95C}{Gamma66}{CLEO}{Balest:1995kq}{0.0423 \pm 0.0006 \pm 0.0022}{0.0423}{\pm 0.0006}{0.0022}%
\htmeasdef{CLEO.Gamma69.pub.EDWARDS.00A}{Gamma69}{CLEO}{Edwards:1999fj}{0.0419 \pm 0.001 \pm 0.0021}{0.0419}{\pm 0.001}{0.0021}%
\htmeasdef{CLEO.Gamma76by54.pub.BORTOLETTO.93}{Gamma76by54}{CLEO}{Bortoletto:1993px}{0.034 \pm 0.002 \pm 0.003}{0.034}{\pm 0.002}{0.003}%
\htmeasdef{CLEO.Gamma78.pub.ANASTASSOV.01}{Gamma78}{CLEO}{Anastassov:2000xu}{0.00022 \pm 3\cdot 10^{-5} \pm 4\cdot 10^{-5}}{0.00022}{\pm 3e-05}{4e-05}%
\htmeasdef{CLEO.Gamma8.pub.ANASTASSOV.97}{Gamma8}{CLEO}{Anastassov:1996tc}{0.1152 \pm 0.0005 \pm 0.0012}{0.1152}{\pm 0.0005}{0.0012}%
\htmeasdef{CLEO.Gamma80by60.pub.RICHICHI.99}{Gamma80by60}{CLEO}{Richichi:1998bc}{0.0544 \pm 0.0021 \pm 0.0053}{0.0544}{\pm 0.0021}{0.0053}%
\htmeasdef{CLEO.Gamma81by69.pub.RICHICHI.99}{Gamma81by69}{CLEO}{Richichi:1998bc}{0.0261 \pm 0.0045 \pm 0.0042}{0.0261}{\pm 0.0045}{0.0042}%
\htmeasdef{CLEO.Gamma93by60.pub.RICHICHI.99}{Gamma93by60}{CLEO}{Richichi:1998bc}{0.016 \pm 0.0015 \pm 0.003}{0.016}{\pm 0.0015}{0.003}%
\htmeasdef{CLEO.Gamma94by69.pub.RICHICHI.99}{Gamma94by69}{CLEO}{Richichi:1998bc}{0.0079 \pm 0.0044 \pm 0.0016}{0.0079}{\pm 0.0044}{0.0016}%
\htmeasdef{CLEO3.Gamma151.pub.ARMS.05}{Gamma151}{CLEO3}{Arms:2005qg}{( 4.1 \pm 0.6 \pm 0.7 ) \cdot 10^{ -4 }}{4.1e-4}{\pm 0.6e-4}{0.7e-4}%
\htmeasdef{CLEO3.Gamma60.pub.BRIERE.03}{Gamma60}{CLEO3}{Briere:2003fr}{0.0913 \pm 0.0005 \pm 0.0046}{0.0913}{\pm 0.0005}{0.0046}%
\htmeasdef{CLEO3.Gamma85.pub.BRIERE.03}{Gamma85}{CLEO3}{Briere:2003fr}{0.00384 \pm 0.00014 \pm 0.00038}{0.00384}{\pm 0.00014}{0.00038}%
\htmeasdef{CLEO3.Gamma88.pub.ARMS.05}{Gamma88}{CLEO3}{Arms:2005qg}{0.00074 \pm 8\cdot 10^{-5} \pm 0.00011}{0.00074}{\pm 8e-05}{0.00011}%
\htmeasdef{CLEO3.Gamma93.pub.BRIERE.03}{Gamma93}{CLEO3}{Briere:2003fr}{0.00155 \pm 6\cdot 10^{-5} \pm 9\cdot 10^{-5}}{0.00155}{\pm 6e-05}{9e-05}%
\htmeasdef{CLEO3.Gamma94.pub.ARMS.05}{Gamma94}{CLEO3}{Arms:2005qg}{( 5.5 \pm 1.4 \pm 1.2 ) \cdot 10^{ -5 }}{5.5e-05}{\pm 1.4e-05}{1.2e-05}%
\htmeasdef{DELPHI.Gamma10.pub.ABREU.94K}{Gamma10}{DELPHI}{Abreu:1994fi}{0.0085 \pm 0.0018}{0.0085}{\pm 0.0018}{0}%
\htmeasdef{DELPHI.Gamma103.pub.ABDALLAH.06A}{Gamma103}{DELPHI}{Abdallah:2003cw}{0.00097 \pm 0.00015 \pm 5\cdot 10^{-5}}{0.00097}{\pm 0.00015}{5e-05}%
\htmeasdef{DELPHI.Gamma104.pub.ABDALLAH.06A}{Gamma104}{DELPHI}{Abdallah:2003cw}{0.00016 \pm 0.00012 \pm 6\cdot 10^{-5}}{0.00016}{\pm 0.00012}{6e-05}%
\htmeasdef{DELPHI.Gamma13.pub.ABDALLAH.06A}{Gamma13}{DELPHI}{Abdallah:2003cw}{0.2574 \pm 0.00201 \pm 0.00138}{0.2574}{\pm 0.00201}{0.00138}%
\htmeasdef{DELPHI.Gamma19.pub.ABDALLAH.06A}{Gamma19}{DELPHI}{Abdallah:2003cw}{0.09498 \pm 0.0032 \pm 0.00275}{0.09498}{\pm 0.0032}{0.00275}%
\htmeasdef{DELPHI.Gamma25.pub.ABDALLAH.06A}{Gamma25}{DELPHI}{Abdallah:2003cw}{0.01403 \pm 0.00214 \pm 0.00224}{0.01403}{\pm 0.00214}{0.00224}%
\htmeasdef{DELPHI.Gamma3.pub.ABREU.99X}{Gamma3}{DELPHI}{Abreu:1999rb}{0.17325 \pm 0.00095 \pm 0.00077}{0.17325}{\pm 0.00095}{0.00077}%
\htmeasdef{DELPHI.Gamma31.pub.ABREU.94K}{Gamma31}{DELPHI}{Abreu:1994fi}{0.0154 \pm 0.0024}{0.0154}{\pm 0.0024}{0}%
\htmeasdef{DELPHI.Gamma5.pub.ABREU.99X}{Gamma5}{DELPHI}{Abreu:1999rb}{0.17877 \pm 0.00109 \pm 0.0011}{0.17877}{\pm 0.00109}{0.0011}%
\htmeasdef{DELPHI.Gamma57.pub.ABDALLAH.06A}{Gamma57}{DELPHI}{Abdallah:2003cw}{0.09317 \pm 0.0009 \pm 0.00082}{0.09317}{\pm 0.0009}{0.00082}%
\htmeasdef{DELPHI.Gamma66.pub.ABDALLAH.06A}{Gamma66}{DELPHI}{Abdallah:2003cw}{0.04545 \pm 0.00106 \pm 0.00103}{0.04545}{\pm 0.00106}{0.00103}%
\htmeasdef{DELPHI.Gamma7.pub.ABREU.92N}{Gamma7}{DELPHI}{Abreu:1992gn}{0.124 \pm 0.007 \pm 0.007}{0.124}{\pm 0.007}{0.007}%
\htmeasdef{DELPHI.Gamma74.pub.ABDALLAH.06A}{Gamma74}{DELPHI}{Abdallah:2003cw}{0.00561 \pm 0.00068 \pm 0.00095}{0.00561}{\pm 0.00068}{0.00095}%
\htmeasdef{DELPHI.Gamma8.pub.ABDALLAH.06A}{Gamma8}{DELPHI}{Abdallah:2003cw}{0.11571 \pm 0.0012 \pm 0.00114}{0.11571}{\pm 0.0012}{0.00114}%
\htmeasdef{HRS.Gamma102.pub.BYLSMA.87}{Gamma102}{HRS}{Bylsma:1986zy}{0.00102 \pm 0.00029}{0.00102}{\pm 0.00029}{0}%
\htmeasdef{HRS.Gamma103.pub.BYLSMA.87}{Gamma103}{HRS}{Bylsma:1986zy}{0.00051 \pm 0.0002}{0.00051}{\pm 0.0002}{0}%
\htmeasdef{L3.Gamma102.pub.ACHARD.01D}{Gamma102}{L3}{Achard:2001pk}{0.0017 \pm 0.00022 \pm 0.00026}{0.0017}{\pm 0.00022}{0.00026}%
\htmeasdef{L3.Gamma13.pub.ACCIARRI.95}{Gamma13}{L3}{Acciarri:1994vr}{0.2505 \pm 0.0035 \pm 0.005}{0.2505}{\pm 0.0035}{0.005}%
\htmeasdef{L3.Gamma19.pub.ACCIARRI.95}{Gamma19}{L3}{Acciarri:1994vr}{0.0888 \pm 0.0037 \pm 0.0042}{0.0888}{\pm 0.0037}{0.0042}%
\htmeasdef{L3.Gamma26.pub.ACCIARRI.95}{Gamma26}{L3}{Acciarri:1994vr}{0.017 \pm 0.0024 \pm 0.0038}{0.017}{\pm 0.0024}{0.0038}%
\htmeasdef{L3.Gamma3.pub.ACCIARRI.01F}{Gamma3}{L3}{Acciarri:2001sg}{0.17342 \pm 0.0011 \pm 0.00067}{0.17342}{\pm 0.0011}{0.00067}%
\htmeasdef{L3.Gamma35.pub.ACCIARRI.95F}{Gamma35}{L3}{Acciarri:1995kx}{0.0095 \pm 0.0015 \pm 0.0006}{0.0095}{\pm 0.0015}{0.0006}%
\htmeasdef{L3.Gamma40.pub.ACCIARRI.95F}{Gamma40}{L3}{Acciarri:1995kx}{0.0041 \pm 0.0012 \pm 0.0003}{0.0041}{\pm 0.0012}{0.0003}%
\htmeasdef{L3.Gamma5.pub.ACCIARRI.01F}{Gamma5}{L3}{Acciarri:2001sg}{0.17806 \pm 0.00104 \pm 0.00076}{0.17806}{\pm 0.00104}{0.00076}%
\htmeasdef{L3.Gamma54.pub.ADEVA.91F}{Gamma54}{L3}{Adeva:1991qq}{0.144 \pm 0.006 \pm 0.003}{0.144}{\pm 0.006}{0.003}%
\htmeasdef{L3.Gamma55.pub.ACHARD.01D}{Gamma55}{L3}{Achard:2001pk}{0.14556 \pm 0.00105 \pm 0.00076}{0.14556}{\pm 0.00105}{0.00076}%
\htmeasdef{L3.Gamma7.pub.ACCIARRI.95}{Gamma7}{L3}{Acciarri:1994vr}{0.1247 \pm 0.0026 \pm 0.0043}{0.1247}{\pm 0.0026}{0.0043}%
\htmeasdef{OPAL.Gamma10.pub.ABBIENDI.01J}{Gamma10}{OPAL}{Abbiendi:2000ee}{0.00658 \pm 0.00027 \pm 0.00029}{0.00658}{\pm 0.00027}{0.00029}%
\htmeasdef{OPAL.Gamma103.pub.ACKERSTAFF.99E}{Gamma103}{OPAL}{Ackerstaff:1998ia}{0.00091 \pm 0.00014 \pm 6\cdot 10^{-5}}{0.00091}{\pm 0.00014}{6e-05}%
\htmeasdef{OPAL.Gamma104.pub.ACKERSTAFF.99E}{Gamma104}{OPAL}{Ackerstaff:1998ia}{0.00027 \pm 0.00018 \pm 9\cdot 10^{-5}}{0.00027}{\pm 0.00018}{9e-05}%
\htmeasdef{OPAL.Gamma13.pub.ACKERSTAFF.98M}{Gamma13}{OPAL}{Ackerstaff:1997tx}{0.2589 \pm 0.0017 \pm 0.0029}{0.2589}{\pm 0.0017}{0.0029}%
\htmeasdef{OPAL.Gamma16.pub.ABBIENDI.04J}{Gamma16}{OPAL}{Abbiendi:2004xa}{0.00471 \pm 0.00059 \pm 0.00023}{0.00471}{\pm 0.00059}{0.00023}%
\htmeasdef{OPAL.Gamma17.pub.ACKERSTAFF.98M}{Gamma17}{OPAL}{Ackerstaff:1997tx}{0.0991 \pm 0.0031 \pm 0.0027}{0.0991}{\pm 0.0031}{0.0027}%
\htmeasdef{OPAL.Gamma3.pub.ABBIENDI.03}{Gamma3}{OPAL}{Abbiendi:2002jw}{0.1734 \pm 0.0009 \pm 0.0006}{0.1734}{\pm 0.0009}{0.0006}%
\htmeasdef{OPAL.Gamma31.pub.ABBIENDI.01J}{Gamma31}{OPAL}{Abbiendi:2000ee}{0.01528 \pm 0.00039 \pm 0.0004}{0.01528}{\pm 0.00039}{0.0004}%
\htmeasdef{OPAL.Gamma33.pub.AKERS.94G}{Gamma33}{OPAL}{Akers:1994td}{0.0097 \pm 0.0009 \pm 0.0006}{0.0097}{\pm 0.0009}{0.0006}%
\htmeasdef{OPAL.Gamma35.pub.ABBIENDI.00C}{Gamma35}{OPAL}{Abbiendi:1999pm}{0.00933 \pm 0.00068 \pm 0.00049}{0.00933}{\pm 0.00068}{0.00049}%
\htmeasdef{OPAL.Gamma38.pub.ABBIENDI.00C}{Gamma38}{OPAL}{Abbiendi:1999pm}{0.0033 \pm 0.00055 \pm 0.00039}{0.0033}{\pm 0.00055}{0.00039}%
\htmeasdef{OPAL.Gamma43.pub.ABBIENDI.00C}{Gamma43}{OPAL}{Abbiendi:1999pm}{0.00324 \pm 0.00074 \pm 0.00066}{0.00324}{\pm 0.00074}{0.00066}%
\htmeasdef{OPAL.Gamma5.pub.ABBIENDI.99H}{Gamma5}{OPAL}{Abbiendi:1998cx}{0.1781 \pm 0.0009 \pm 0.0006}{0.1781}{\pm 0.0009}{0.0006}%
\htmeasdef{OPAL.Gamma55.pub.AKERS.95Y}{Gamma55}{OPAL}{Akers:1995ry}{0.1496 \pm 0.0009 \pm 0.0022}{0.1496}{\pm 0.0009}{0.0022}%
\htmeasdef{OPAL.Gamma57by55.pub.AKERS.95Y}{Gamma57by55}{OPAL}{Akers:1995ry}{0.66 \pm 0.004 \pm 0.014}{0.66}{\pm 0.004}{0.014}%
\htmeasdef{OPAL.Gamma7.pub.ALEXANDER.91D}{Gamma7}{OPAL}{Alexander:1991am}{0.121 \pm 0.007 \pm 0.005}{0.121}{\pm 0.007}{0.005}%
\htmeasdef{OPAL.Gamma8.pub.ACKERSTAFF.98M}{Gamma8}{OPAL}{Ackerstaff:1997tx}{0.1198 \pm 0.0013 \pm 0.0016}{0.1198}{\pm 0.0013}{0.0016}%
\htmeasdef{OPAL.Gamma85.pub.ABBIENDI.04J}{Gamma85}{OPAL}{Abbiendi:2004xa}{0.00415 \pm 0.00053 \pm 0.0004}{0.00415}{\pm 0.00053}{0.0004}%
\htmeasdef{OPAL.Gamma92.pub.ABBIENDI.00D}{Gamma92}{OPAL}{Abbiendi:1999cq}{0.00159 \pm 0.00053 \pm 0.0002}{0.00159}{\pm 0.00053}{0.0002}%
\htmeasdef{TPC.Gamma54.pub.AIHARA.87B}{Gamma54}{TPC}{Aihara:1986mw}{0.151 \pm 0.008 \pm 0.006}{0.151}{\pm 0.008}{0.006}%
\htmeasdef{TPC.Gamma82.pub.BAUER.94}{Gamma82}{TPC}{Bauer:1993wn}{0.0058 {}^{+0.0015}_{-0.0013} \pm 0.0012}{0.0058}{{}^{+0.0015}_{-0.0013}}{0.0012}%
\htmeasdef{TPC.Gamma92.pub.BAUER.94}{Gamma92}{TPC}{Bauer:1993wn}{0.0015 {}^{+0.0009}_{-0.0007} \pm 0.0003}{0.0015}{{}^{+0.0009}_{-0.0007}}{0.0003}%
\htdef{Gamma1.qt}{\ensuremath{0.8519 \pm 0.0011}}% 
\htdef{Gamma2.qt}{\ensuremath{0.8453 \pm 0.0010}}% 
\htdef{Gamma3.qt}{\ensuremath{0.17392 \pm 0.00040}}% 
\htdef{ALEPH.Gamma3.pub.SCHAEL.05C,qt}{\ensuremath{0.17319 \pm 0.00077 \pm 0.00000}}%
\htdef{DELPHI.Gamma3.pub.ABREU.99X,qt}{\ensuremath{0.17325 \pm 0.00095 \pm 0.00077}}%
\htdef{L3.Gamma3.pub.ACCIARRI.01F,qt}{\ensuremath{0.17342 \pm 0.00110 \pm 0.00067}}%
\htdef{OPAL.Gamma3.pub.ABBIENDI.03,qt}{\ensuremath{0.17340 \pm 0.00090 \pm 0.00060}}% 
\htdef{Gamma3by5.qt}{\ensuremath{0.9762 \pm 0.0028}}% 
\htdef{ARGUS.Gamma3by5.pub.ALBRECHT.92D,qt}{\ensuremath{0.9970 \pm 0.0350 \pm 0.0400}}%
\htdef{BaBar.Gamma3by5.pub.AUBERT.10F,qt}{\ensuremath{0.9796 \pm 0.0016 \pm 0.0036}}%
\htdef{CLEO.Gamma3by5.pub.ANASTASSOV.97,qt}{\ensuremath{0.9777 \pm 0.0063 \pm 0.0087}}% 
\htdef{Gamma5.qt}{\ensuremath{0.17816 \pm 0.00041}}% 
\htdef{ALEPH.Gamma5.pub.SCHAEL.05C,qt}{\ensuremath{0.17837 \pm 0.00080 \pm 0.00000}}%
\htdef{CLEO.Gamma5.pub.ANASTASSOV.97,qt}{\ensuremath{0.17760 \pm 0.00060 \pm 0.00170}}%
\htdef{DELPHI.Gamma5.pub.ABREU.99X,qt}{\ensuremath{0.17877 \pm 0.00109 \pm 0.00110}}%
\htdef{L3.Gamma5.pub.ACCIARRI.01F,qt}{\ensuremath{0.17806 \pm 0.00104 \pm 0.00076}}%
\htdef{OPAL.Gamma5.pub.ABBIENDI.99H,qt}{\ensuremath{0.17810 \pm 0.00090 \pm 0.00060}}% 
\htdef{Gamma7.qt}{\ensuremath{0.12023 \pm 0.00054}}% 
\htdef{DELPHI.Gamma7.pub.ABREU.92N,qt}{\ensuremath{0.12400 \pm 0.00700 \pm 0.00700}}%
\htdef{L3.Gamma7.pub.ACCIARRI.95,qt}{\ensuremath{0.12470 \pm 0.00260 \pm 0.00430}}%
\htdef{OPAL.Gamma7.pub.ALEXANDER.91D,qt}{\ensuremath{0.12100 \pm 0.00700 \pm 0.00500}}% 
\htdef{Gamma8.qt}{\ensuremath{0.11506 \pm 0.00054}}% 
\htdef{ALEPH.Gamma8.pub.SCHAEL.05C,qt}{\ensuremath{0.11524 \pm 0.00105 \pm 0.00000}}%
\htdef{CLEO.Gamma8.pub.ANASTASSOV.97,qt}{\ensuremath{0.11520 \pm 0.00050 \pm 0.00120}}%
\htdef{DELPHI.Gamma8.pub.ABDALLAH.06A,qt}{\ensuremath{0.11571 \pm 0.00120 \pm 0.00114}}%
\htdef{OPAL.Gamma8.pub.ACKERSTAFF.98M,qt}{\ensuremath{0.11980 \pm 0.00130 \pm 0.00160}}% 
\htdef{Gamma8by5.qt}{\ensuremath{0.6458 \pm 0.0033}}% 
\htdef{Gamma9.qt}{\ensuremath{0.10810 \pm 0.00053}}% 
\htdef{Gamma9by5.qt}{\ensuremath{0.6068 \pm 0.0032}}% 
\htdef{BaBar.Gamma9by5.pub.AUBERT.10F,qt}{\ensuremath{0.5945 \pm 0.0014 \pm 0.0061}}% 
\htdef{Gamma10.qt}{\ensuremath{(0.6960 \pm 0.0096) \cdot 10^{-2}}}% 
\htdef{ALEPH.Gamma10.pub.BARATE.99K,qt}{\ensuremath{(0.6960 \pm 0.0287 \pm 0.0000) \cdot 10^{-2} }}%
\htdef{CLEO.Gamma10.pub.BATTLE.94,qt}{\ensuremath{(0.6600 \pm 0.0700 \pm 0.0900) \cdot 10^{-2} }}%
\htdef{DELPHI.Gamma10.pub.ABREU.94K,qt}{\ensuremath{(0.8500 \pm 0.1800 \pm 0.0000) \cdot 10^{-2} }}%
\htdef{OPAL.Gamma10.pub.ABBIENDI.01J,qt}{\ensuremath{(0.6580 \pm 0.0270 \pm 0.0290) \cdot 10^{-2} }}% 
\htdef{Gamma10by5.qt}{\ensuremath{(3.906 \pm 0.054) \cdot 10^{-2}}}% 
\htdef{BaBar.Gamma10by5.pub.AUBERT.10F,qt}{\ensuremath{(3.882 \pm 0.032 \pm 0.057) \cdot 10^{-2} }}% 
\htdef{Gamma10by9.qt}{\ensuremath{(6.438 \pm 0.094) \cdot 10^{-2}}}% 
\htdef{Gamma11.qt}{\ensuremath{0.36973 \pm 0.00097}}% 
\htdef{Gamma12.qt}{\ensuremath{0.36475 \pm 0.00097}}% 
\htdef{Gamma13.qt}{\ensuremath{0.25935 \pm 0.00091}}% 
\htdef{ALEPH.Gamma13.pub.SCHAEL.05C,qt}{\ensuremath{0.25924 \pm 0.00129 \pm 0.00000}}%
\htdef{Belle.Gamma13.pub.FUJIKAWA.08,qt}{\ensuremath{0.25670 \pm 0.00010 \pm 0.00390}}%
\htdef{CLEO.Gamma13.pub.ARTUSO.94,qt}{\ensuremath{0.25870 \pm 0.00120 \pm 0.00420}}%
\htdef{DELPHI.Gamma13.pub.ABDALLAH.06A,qt}{\ensuremath{0.25740 \pm 0.00201 \pm 0.00138}}%
\htdef{L3.Gamma13.pub.ACCIARRI.95,qt}{\ensuremath{0.25050 \pm 0.00350 \pm 0.00500}}%
\htdef{OPAL.Gamma13.pub.ACKERSTAFF.98M,qt}{\ensuremath{0.25890 \pm 0.00170 \pm 0.00290}}% 
\htdef{Gamma14.qt}{\ensuremath{0.25502 \pm 0.00092}}% 
\htdef{Gamma16.qt}{\ensuremath{(0.4327 \pm 0.0149) \cdot 10^{-2}}}% 
\htdef{ALEPH.Gamma16.pub.BARATE.99K,qt}{\ensuremath{(0.4440 \pm 0.0354 \pm 0.0000) \cdot 10^{-2} }}%
\htdef{BaBar.Gamma16.pub.AUBERT.07AP,qt}{\ensuremath{(0.4160 \pm 0.0030 \pm 0.0180) \cdot 10^{-2} }}%
\htdef{CLEO.Gamma16.pub.BATTLE.94,qt}{\ensuremath{(0.5100 \pm 0.1000 \pm 0.0700) \cdot 10^{-2} }}%
\htdef{OPAL.Gamma16.pub.ABBIENDI.04J,qt}{\ensuremath{(0.4710 \pm 0.0590 \pm 0.0230) \cdot 10^{-2} }}% 
\htdef{Gamma17.qt}{\ensuremath{0.10775 \pm 0.00095}}% 
\htdef{OPAL.Gamma17.pub.ACKERSTAFF.98M,qt}{\ensuremath{0.09910 \pm 0.00310 \pm 0.00270}}% 
\htdef{Gamma18.qt}{\ensuremath{(9.458 \pm 0.097) \cdot 10^{-2}}}% 
\htdef{Gamma19.qt}{\ensuremath{(9.306 \pm 0.097) \cdot 10^{-2}}}% 
\htdef{ALEPH.Gamma19.pub.SCHAEL.05C,qt}{\ensuremath{(9.295 \pm 0.122 \pm 0.000) \cdot 10^{-2} }}%
\htdef{DELPHI.Gamma19.pub.ABDALLAH.06A,qt}{\ensuremath{(9.498 \pm 0.320 \pm 0.275) \cdot 10^{-2} }}%
\htdef{L3.Gamma19.pub.ACCIARRI.95,qt}{\ensuremath{(8.880 \pm 0.370 \pm 0.420) \cdot 10^{-2} }}% 
\htdef{Gamma19by13.qt}{\ensuremath{0.3588 \pm 0.0044}}% 
\htdef{CLEO.Gamma19by13.pub.PROCARIO.93,qt}{\ensuremath{0.3420 \pm 0.0060 \pm 0.0160}}% 
\htdef{Gamma20.qt}{\ensuremath{(9.242 \pm 0.100) \cdot 10^{-2}}}% 
\htdef{Gamma23.qt}{\ensuremath{(0.0640 \pm 0.0220) \cdot 10^{-2}}}% 
\htdef{ALEPH.Gamma23.pub.BARATE.99K,qt}{\ensuremath{(0.0560 \pm 0.0250 \pm 0.0000) \cdot 10^{-2} }}%
\htdef{CLEO.Gamma23.pub.BATTLE.94,qt}{\ensuremath{(0.0900 \pm 0.1000 \pm 0.0300) \cdot 10^{-2} }}% 
\htdef{Gamma24.qt}{\ensuremath{(1.318 \pm 0.065) \cdot 10^{-2}}}% 
\htdef{Gamma25.qt}{\ensuremath{(1.233 \pm 0.065) \cdot 10^{-2}}}% 
\htdef{DELPHI.Gamma25.pub.ABDALLAH.06A,qt}{\ensuremath{(1.403 \pm 0.214 \pm 0.224) \cdot 10^{-2} }}% 
\htdef{Gamma26.qt}{\ensuremath{(1.158 \pm 0.072) \cdot 10^{-2}}}% 
\htdef{ALEPH.Gamma26.pub.SCHAEL.05C,qt}{\ensuremath{(1.082 \pm 0.093 \pm 0.000) \cdot 10^{-2} }}%
\htdef{L3.Gamma26.pub.ACCIARRI.95,qt}{\ensuremath{(1.700 \pm 0.240 \pm 0.380) \cdot 10^{-2} }}% 
\htdef{Gamma26by13.qt}{\ensuremath{(4.465 \pm 0.277) \cdot 10^{-2}}}% 
\htdef{CLEO.Gamma26by13.pub.PROCARIO.93,qt}{\ensuremath{(4.400 \pm 0.300 \pm 0.500) \cdot 10^{-2} }}% 
\htdef{Gamma27.qt}{\ensuremath{(1.029 \pm 0.075) \cdot 10^{-2}}}% 
\htdef{Gamma28.qt}{\ensuremath{(4.283 \pm 2.161) \cdot 10^{-4}}}% 
\htdef{ALEPH.Gamma28.pub.BARATE.99K,qt}{\ensuremath{(3.700 \pm 2.371 \pm 0.000) \cdot 10^{-4} }}% 
\htdef{Gamma29.qt}{\ensuremath{(0.1568 \pm 0.0391) \cdot 10^{-2}}}% 
\htdef{CLEO.Gamma29.pub.PROCARIO.93,qt}{\ensuremath{(0.1600 \pm 0.0500 \pm 0.0500) \cdot 10^{-2} }}% 
\htdef{Gamma30.qt}{\ensuremath{(0.1099 \pm 0.0391) \cdot 10^{-2}}}% 
\htdef{ALEPH.Gamma30.pub.SCHAEL.05C,qt}{\ensuremath{(0.1120 \pm 0.0509 \pm 0.0000) \cdot 10^{-2} }}% 
\htdef{Gamma31.qt}{\ensuremath{(1.545 \pm 0.030) \cdot 10^{-2}}}% 
\htdef{CLEO.Gamma31.pub.BATTLE.94,qt}{\ensuremath{(1.700 \pm 0.120 \pm 0.190) \cdot 10^{-2} }}%
\htdef{DELPHI.Gamma31.pub.ABREU.94K,qt}{\ensuremath{(1.540 \pm 0.240 \pm 0.000) \cdot 10^{-2} }}%
\htdef{OPAL.Gamma31.pub.ABBIENDI.01J,qt}{\ensuremath{(1.528 \pm 0.039 \pm 0.040) \cdot 10^{-2} }}% 
\htdef{Gamma32.qt}{\ensuremath{(0.8528 \pm 0.0286) \cdot 10^{-2}}}% 
\htdef{Gamma33.qt}{\ensuremath{(0.9372 \pm 0.0292) \cdot 10^{-2}}}% 
\htdef{ALEPH.Gamma33.pub.BARATE.98E,qt}{\ensuremath{(0.9700 \pm 0.0849 \pm 0.0000) \cdot 10^{-2} }}%
\htdef{OPAL.Gamma33.pub.AKERS.94G,qt}{\ensuremath{(0.9700 \pm 0.0900 \pm 0.0600) \cdot 10^{-2} }}% 
\htdef{Gamma34.qt}{\ensuremath{(0.9865 \pm 0.0139) \cdot 10^{-2}}}% 
\htdef{CLEO.Gamma34.pub.COAN.96,qt}{\ensuremath{(0.8550 \pm 0.0360 \pm 0.0730) \cdot 10^{-2} }}% 
\htdef{Gamma35.qt}{\ensuremath{(0.8386 \pm 0.0141) \cdot 10^{-2}}}% 
\htdef{ALEPH.Gamma35.pub.BARATE.99K,qt}{\ensuremath{(0.9280 \pm 0.0564 \pm 0.0000) \cdot 10^{-2} }}%
\htdef{Belle.Gamma35.pub.RYU.14vpc,qt}{\ensuremath{(0.8320 \pm 0.0025 \pm 0.0150) \cdot 10^{-2} }}%
\htdef{L3.Gamma35.pub.ACCIARRI.95F,qt}{\ensuremath{(0.9500 \pm 0.1500 \pm 0.0600) \cdot 10^{-2} }}%
\htdef{OPAL.Gamma35.pub.ABBIENDI.00C,qt}{\ensuremath{(0.9330 \pm 0.0680 \pm 0.0490) \cdot 10^{-2} }}% 
\htdef{Gamma37.qt}{\ensuremath{(0.1479 \pm 0.0053) \cdot 10^{-2}}}% 
\htdef{ALEPH.Gamma37.pub.BARATE.98E,qt}{\ensuremath{(0.1580 \pm 0.0453 \pm 0.0000) \cdot 10^{-2} }}%
\htdef{ALEPH.Gamma37.pub.BARATE.99K,qt}{\ensuremath{(0.1620 \pm 0.0237 \pm 0.0000) \cdot 10^{-2} }}%
\htdef{Belle.Gamma37.pub.RYU.14vpc,qt}{\ensuremath{(0.1480 \pm 0.0013 \pm 0.0055) \cdot 10^{-2} }}%
\htdef{CLEO.Gamma37.pub.COAN.96,qt}{\ensuremath{(0.1510 \pm 0.0210 \pm 0.0220) \cdot 10^{-2} }}% 
\htdef{Gamma38.qt}{\ensuremath{(0.2982 \pm 0.0079) \cdot 10^{-2}}}% 
\htdef{OPAL.Gamma38.pub.ABBIENDI.00C,qt}{\ensuremath{(0.3300 \pm 0.0550 \pm 0.0390) \cdot 10^{-2} }}% 
\htdef{Gamma39.qt}{\ensuremath{(0.5314 \pm 0.0134) \cdot 10^{-2}}}% 
\htdef{CLEO.Gamma39.pub.COAN.96,qt}{\ensuremath{(0.5620 \pm 0.0500 \pm 0.0480) \cdot 10^{-2} }}% 
\htdef{Gamma40.qt}{\ensuremath{(0.3812 \pm 0.0129) \cdot 10^{-2}}}% 
\htdef{ALEPH.Gamma40.pub.BARATE.98E,qt}{\ensuremath{(0.2940 \pm 0.0818 \pm 0.0000) \cdot 10^{-2} }}%
\htdef{ALEPH.Gamma40.pub.BARATE.99K,qt}{\ensuremath{(0.3470 \pm 0.0646 \pm 0.0000) \cdot 10^{-2} }}%
\htdef{Belle.Gamma40.pub.RYU.14vpc,qt}{\ensuremath{(0.3860 \pm 0.0031 \pm 0.0135) \cdot 10^{-2} }}%
\htdef{L3.Gamma40.pub.ACCIARRI.95F,qt}{\ensuremath{(0.4100 \pm 0.1200 \pm 0.0300) \cdot 10^{-2} }}% 
\htdef{Gamma42.qt}{\ensuremath{(0.1502 \pm 0.0071) \cdot 10^{-2}}}% 
\htdef{ALEPH.Gamma42.pub.BARATE.98E,qt}{\ensuremath{(0.1520 \pm 0.0789 \pm 0.0000) \cdot 10^{-2} }}%
\htdef{ALEPH.Gamma42.pub.BARATE.99K,qt}{\ensuremath{(0.1430 \pm 0.0291 \pm 0.0000) \cdot 10^{-2} }}%
\htdef{Belle.Gamma42.pub.RYU.14vpc,qt}{\ensuremath{(0.1496 \pm 0.0019 \pm 0.0073) \cdot 10^{-2} }}%
\htdef{CLEO.Gamma42.pub.COAN.96,qt}{\ensuremath{(0.1450 \pm 0.0360 \pm 0.0200) \cdot 10^{-2} }}% 
\htdef{Gamma43.qt}{\ensuremath{(0.4046 \pm 0.0260) \cdot 10^{-2}}}% 
\htdef{OPAL.Gamma43.pub.ABBIENDI.00C,qt}{\ensuremath{(0.3240 \pm 0.0740 \pm 0.0660) \cdot 10^{-2} }}% 
\htdef{Gamma44.qt}{\ensuremath{(2.340 \pm 2.306) \cdot 10^{-4}}}% 
\htdef{ALEPH.Gamma44.pub.BARATE.99R,qt}{\ensuremath{(2.600 \pm 2.400 \pm 0.000) \cdot 10^{-4} }}% 
\htdef{Gamma46.qt}{\ensuremath{(0.1513 \pm 0.0247) \cdot 10^{-2}}}% 
\htdef{Gamma47.qt}{\ensuremath{(2.332 \pm 0.065) \cdot 10^{-4}}}% 
\htdef{ALEPH.Gamma47.pub.BARATE.98E,qt}{\ensuremath{(2.600 \pm 1.118 \pm 0.000) \cdot 10^{-4} }}%
\htdef{BaBar.Gamma47.pub.LEES.12Y,qt}{\ensuremath{(2.310 \pm 0.040 \pm 0.080) \cdot 10^{-4} }}%
\htdef{Belle.Gamma47.pub.RYU.14vpc,qt}{\ensuremath{(2.330 \pm 0.033 \pm 0.093) \cdot 10^{-4} }}%
\htdef{CLEO.Gamma47.pub.COAN.96,qt}{\ensuremath{(2.300 \pm 0.500 \pm 0.300) \cdot 10^{-4} }}% 
\htdef{Gamma48.qt}{\ensuremath{(0.1047 \pm 0.0247) \cdot 10^{-2}}}% 
\htdef{ALEPH.Gamma48.pub.BARATE.98E,qt}{\ensuremath{(0.1010 \pm 0.0264 \pm 0.0000) \cdot 10^{-2} }}% 
\htdef{Gamma49.qt}{\ensuremath{(3.540 \pm 1.193) \cdot 10^{-4}}}% 
\htdef{Gamma50.qt}{\ensuremath{(1.815 \pm 0.207) \cdot 10^{-5}}}% 
\htdef{BaBar.Gamma50.pub.LEES.12Y,qt}{\ensuremath{(1.600 \pm 0.200 \pm 0.220) \cdot 10^{-5} }}%
\htdef{Belle.Gamma50.pub.RYU.14vpc,qt}{\ensuremath{(2.000 \pm 0.216 \pm 0.202) \cdot 10^{-5} }}% 
\htdef{Gamma51.qt}{\ensuremath{(3.177 \pm 1.192) \cdot 10^{-4}}}% 
\htdef{ALEPH.Gamma51.pub.BARATE.98E,qt}{\ensuremath{(3.100 \pm 1.100 \pm 0.500) \cdot 10^{-4} }}% 
\htdef{Gamma53.qt}{\ensuremath{(2.218 \pm 2.024) \cdot 10^{-4}}}% 
\htdef{ALEPH.Gamma53.pub.BARATE.98E,qt}{\ensuremath{(2.300 \pm 2.025 \pm 0.000) \cdot 10^{-4} }}% 
\htdef{Gamma54.qt}{\ensuremath{0.15215 \pm 0.00061}}% 
\htdef{CELLO.Gamma54.pub.BEHREND.89B,qt}{\ensuremath{0.15000 \pm 0.00400 \pm 0.00300}}%
\htdef{L3.Gamma54.pub.ADEVA.91F,qt}{\ensuremath{0.14400 \pm 0.00600 \pm 0.00300}}%
\htdef{TPC.Gamma54.pub.AIHARA.87B,qt}{\ensuremath{0.15100 \pm 0.00800 \pm 0.00600}}% 
\htdef{Gamma55.qt}{\ensuremath{0.14567 \pm 0.00057}}% 
\htdef{L3.Gamma55.pub.ACHARD.01D,qt}{\ensuremath{0.14556 \pm 0.00105 \pm 0.00076}}%
\htdef{OPAL.Gamma55.pub.AKERS.95Y,qt}{\ensuremath{0.14960 \pm 0.00090 \pm 0.00220}}% 
\htdef{Gamma56.qt}{\ensuremath{(9.780 \pm 0.054) \cdot 10^{-2}}}% 
\htdef{Gamma57.qt}{\ensuremath{(9.439 \pm 0.053) \cdot 10^{-2}}}% 
\htdef{CLEO.Gamma57.pub.BALEST.95C,qt}{\ensuremath{(9.510 \pm 0.070 \pm 0.200) \cdot 10^{-2} }}%
\htdef{DELPHI.Gamma57.pub.ABDALLAH.06A,qt}{\ensuremath{(9.317 \pm 0.090 \pm 0.082) \cdot 10^{-2} }}% 
\htdef{Gamma57by55.qt}{\ensuremath{0.6480 \pm 0.0030}}% 
\htdef{OPAL.Gamma57by55.pub.AKERS.95Y,qt}{\ensuremath{0.6600 \pm 0.0040 \pm 0.0140}}% 
\htdef{Gamma58.qt}{\ensuremath{(9.408 \pm 0.053) \cdot 10^{-2}}}% 
\htdef{ALEPH.Gamma58.pub.SCHAEL.05C,qt}{\ensuremath{(9.469 \pm 0.096 \pm 0.000) \cdot 10^{-2} }}% 
\htdef{Gamma59.qt}{\ensuremath{(9.290 \pm 0.052) \cdot 10^{-2}}}% 
\htdef{Gamma60.qt}{\ensuremath{(9.000 \pm 0.051) \cdot 10^{-2}}}% 
\htdef{BaBar.Gamma60.pub.AUBERT.08,qt}{\ensuremath{(8.830 \pm 0.010 \pm 0.130) \cdot 10^{-2} }}%
\htdef{Belle.Gamma60.pub.LEE.10,qt}{\ensuremath{(8.420 \pm 0.000 {}^{+0.260}_{-0.250}) \cdot 10^{-2} }}%
\htdef{CLEO3.Gamma60.pub.BRIERE.03,qt}{\ensuremath{(9.130 \pm 0.050 \pm 0.460) \cdot 10^{-2} }}% 
\htdef{Gamma62.qt}{\ensuremath{(8.970 \pm 0.052) \cdot 10^{-2}}}% 
\htdef{Gamma63.qt}{\ensuremath{(5.325 \pm 0.050) \cdot 10^{-2}}}% 
\htdef{Gamma64.qt}{\ensuremath{(5.120 \pm 0.049) \cdot 10^{-2}}}% 
\htdef{Gamma65.qt}{\ensuremath{(4.790 \pm 0.052) \cdot 10^{-2}}}% 
\htdef{Gamma66.qt}{\ensuremath{(4.606 \pm 0.051) \cdot 10^{-2}}}% 
\htdef{ALEPH.Gamma66.pub.SCHAEL.05C,qt}{\ensuremath{(4.734 \pm 0.077 \pm 0.000) \cdot 10^{-2} }}%
\htdef{CLEO.Gamma66.pub.BALEST.95C,qt}{\ensuremath{(4.230 \pm 0.060 \pm 0.220) \cdot 10^{-2} }}%
\htdef{DELPHI.Gamma66.pub.ABDALLAH.06A,qt}{\ensuremath{(4.545 \pm 0.106 \pm 0.103) \cdot 10^{-2} }}% 
\htdef{Gamma67.qt}{\ensuremath{(2.820 \pm 0.070) \cdot 10^{-2}}}% 
\htdef{Gamma68.qt}{\ensuremath{(4.651 \pm 0.053) \cdot 10^{-2}}}% 
\htdef{Gamma69.qt}{\ensuremath{(4.519 \pm 0.052) \cdot 10^{-2}}}% 
\htdef{CLEO.Gamma69.pub.EDWARDS.00A,qt}{\ensuremath{(4.190 \pm 0.100 \pm 0.210) \cdot 10^{-2} }}% 
\htdef{Gamma70.qt}{\ensuremath{(2.769 \pm 0.071) \cdot 10^{-2}}}% 
\htdef{Gamma74.qt}{\ensuremath{(0.5135 \pm 0.0312) \cdot 10^{-2}}}% 
\htdef{DELPHI.Gamma74.pub.ABDALLAH.06A,qt}{\ensuremath{(0.5610 \pm 0.0680 \pm 0.0950) \cdot 10^{-2} }}% 
\htdef{Gamma75.qt}{\ensuremath{(0.5024 \pm 0.0310) \cdot 10^{-2}}}% 
\htdef{Gamma76.qt}{\ensuremath{(0.4925 \pm 0.0310) \cdot 10^{-2}}}% 
\htdef{ALEPH.Gamma76.pub.SCHAEL.05C,qt}{\ensuremath{(0.4350 \pm 0.0461 \pm 0.0000) \cdot 10^{-2} }}% 
\htdef{Gamma76by54.qt}{\ensuremath{(3.237 \pm 0.202) \cdot 10^{-2}}}% 
\htdef{CLEO.Gamma76by54.pub.BORTOLETTO.93,qt}{\ensuremath{(3.400 \pm 0.200 \pm 0.300) \cdot 10^{-2} }}% 
\htdef{Gamma77.qt}{\ensuremath{(9.759 \pm 3.550) \cdot 10^{-4}}}% 
\htdef{Gamma78.qt}{\ensuremath{(2.107 \pm 0.299) \cdot 10^{-4}}}% 
\htdef{CLEO.Gamma78.pub.ANASTASSOV.01,qt}{\ensuremath{(2.200 \pm 0.300 \pm 0.400) \cdot 10^{-4} }}% 
\htdef{Gamma79.qt}{\ensuremath{(0.6297 \pm 0.0141) \cdot 10^{-2}}}% 
\htdef{Gamma80.qt}{\ensuremath{(0.4363 \pm 0.0073) \cdot 10^{-2}}}% 
\htdef{Gamma80by60.qt}{\ensuremath{(4.847 \pm 0.080) \cdot 10^{-2}}}% 
\htdef{CLEO.Gamma80by60.pub.RICHICHI.99,qt}{\ensuremath{(5.440 \pm 0.210 \pm 0.530) \cdot 10^{-2} }}% 
\htdef{Gamma81.qt}{\ensuremath{(8.726 \pm 1.177) \cdot 10^{-4}}}% 
\htdef{Gamma81by69.qt}{\ensuremath{(1.931 \pm 0.266) \cdot 10^{-2}}}% 
\htdef{CLEO.Gamma81by69.pub.RICHICHI.99,qt}{\ensuremath{(2.610 \pm 0.450 \pm 0.420) \cdot 10^{-2} }}% 
\htdef{Gamma82.qt}{\ensuremath{(0.4780 \pm 0.0137) \cdot 10^{-2}}}% 
\htdef{TPC.Gamma82.pub.BAUER.94,qt}{\ensuremath{(0.5800 {}^{+0.1500}_{-0.1300} \pm 0.1200) \cdot 10^{-2} }}% 
\htdef{Gamma83.qt}{\ensuremath{(0.3741 \pm 0.0135) \cdot 10^{-2}}}% 
\htdef{Gamma84.qt}{\ensuremath{(0.3441 \pm 0.0070) \cdot 10^{-2}}}% 
\htdef{Gamma85.qt}{\ensuremath{(0.2929 \pm 0.0067) \cdot 10^{-2}}}% 
\htdef{ALEPH.Gamma85.pub.BARATE.98,qt}{\ensuremath{(0.2140 \pm 0.0470 \pm 0.0000) \cdot 10^{-2} }}%
\htdef{BaBar.Gamma85.pub.AUBERT.08,qt}{\ensuremath{(0.2730 \pm 0.0020 \pm 0.0090) \cdot 10^{-2} }}%
\htdef{Belle.Gamma85.pub.LEE.10,qt}{\ensuremath{(0.3300 \pm 0.0010 {}^{+0.0160}_{-0.0170}) \cdot 10^{-2} }}%
\htdef{CLEO3.Gamma85.pub.BRIERE.03,qt}{\ensuremath{(0.3840 \pm 0.0140 \pm 0.0380) \cdot 10^{-2} }}%
\htdef{OPAL.Gamma85.pub.ABBIENDI.04J,qt}{\ensuremath{(0.4150 \pm 0.0530 \pm 0.0400) \cdot 10^{-2} }}% 
\htdef{Gamma85by60.qt}{\ensuremath{(3.254 \pm 0.074) \cdot 10^{-2}}}% 
\htdef{Gamma87.qt}{\ensuremath{(0.1331 \pm 0.0119) \cdot 10^{-2}}}% 
\htdef{Gamma88.qt}{\ensuremath{(8.115 \pm 1.168) \cdot 10^{-4}}}% 
\htdef{ALEPH.Gamma88.pub.BARATE.98,qt}{\ensuremath{(6.100 \pm 4.295 \pm 0.000) \cdot 10^{-4} }}%
\htdef{CLEO3.Gamma88.pub.ARMS.05,qt}{\ensuremath{(7.400 \pm 0.800 \pm 1.100) \cdot 10^{-4} }}% 
\htdef{Gamma89.qt}{\ensuremath{(7.761 \pm 1.168) \cdot 10^{-4}}}% 
\htdef{Gamma92.qt}{\ensuremath{(0.1495 \pm 0.0033) \cdot 10^{-2}}}% 
\htdef{OPAL.Gamma92.pub.ABBIENDI.00D,qt}{\ensuremath{(0.1590 \pm 0.0530 \pm 0.0200) \cdot 10^{-2} }}%
\htdef{TPC.Gamma92.pub.BAUER.94,qt}{\ensuremath{(0.1500 {}^{+0.0900}_{-0.0700} \pm 0.0300) \cdot 10^{-2} }}% 
\htdef{Gamma93.qt}{\ensuremath{(0.1434 \pm 0.0027) \cdot 10^{-2}}}% 
\htdef{ALEPH.Gamma93.pub.BARATE.98,qt}{\ensuremath{(0.1630 \pm 0.0270 \pm 0.0000) \cdot 10^{-2} }}%
\htdef{BaBar.Gamma93.pub.AUBERT.08,qt}{\ensuremath{(0.1346 \pm 0.0010 \pm 0.0036) \cdot 10^{-2} }}%
\htdef{Belle.Gamma93.pub.LEE.10,qt}{\ensuremath{(0.1550 \pm 0.0010 {}^{+0.0060}_{-0.0050}) \cdot 10^{-2} }}%
\htdef{CLEO3.Gamma93.pub.BRIERE.03,qt}{\ensuremath{(0.1550 \pm 0.0060 \pm 0.0090) \cdot 10^{-2} }}% 
\htdef{Gamma93by60.qt}{\ensuremath{(1.593 \pm 0.030) \cdot 10^{-2}}}% 
\htdef{CLEO.Gamma93by60.pub.RICHICHI.99,qt}{\ensuremath{(1.600 \pm 0.150 \pm 0.300) \cdot 10^{-2} }}% 
\htdef{Gamma94.qt}{\ensuremath{(0.611 \pm 0.183) \cdot 10^{-4}}}% 
\htdef{ALEPH.Gamma94.pub.BARATE.98,qt}{\ensuremath{(7.500 \pm 3.265 \pm 0.000) \cdot 10^{-4} }}%
\htdef{CLEO3.Gamma94.pub.ARMS.05,qt}{\ensuremath{(0.550 \pm 0.140 \pm 0.120) \cdot 10^{-4} }}% 
\htdef{Gamma94by69.qt}{\ensuremath{(0.1353 \pm 0.0405) \cdot 10^{-2}}}% 
\htdef{CLEO.Gamma94by69.pub.RICHICHI.99,qt}{\ensuremath{(0.7900 \pm 0.4400 \pm 0.1600) \cdot 10^{-2} }}% 
\htdef{Gamma96.qt}{\ensuremath{(2.174 \pm 0.800) \cdot 10^{-5}}}% 
\htdef{BaBar.Gamma96.pub.AUBERT.08,qt}{\ensuremath{(1.578 \pm 0.130 \pm 0.123) \cdot 10^{-5} }}%
\htdef{Belle.Gamma96.pub.LEE.10,qt}{\ensuremath{(3.290 \pm 0.170 {}^{+0.190}_{-0.200}) \cdot 10^{-5} }}% 
\htdef{Gamma102.qt}{\ensuremath{(0.0985 \pm 0.0037) \cdot 10^{-2}}}% 
\htdef{CLEO.Gamma102.pub.GIBAUT.94B,qt}{\ensuremath{(0.0970 \pm 0.0050 \pm 0.0110) \cdot 10^{-2} }}%
\htdef{HRS.Gamma102.pub.BYLSMA.87,qt}{\ensuremath{(0.1020 \pm 0.0290 \pm 0.0000) \cdot 10^{-2} }}%
\htdef{L3.Gamma102.pub.ACHARD.01D,qt}{\ensuremath{(0.1700 \pm 0.0220 \pm 0.0260) \cdot 10^{-2} }}% 
\htdef{Gamma103.qt}{\ensuremath{(8.216 \pm 0.316) \cdot 10^{-4}}}% 
\htdef{ALEPH.Gamma103.pub.SCHAEL.05C,qt}{\ensuremath{(7.200 \pm 1.500 \pm 0.000) \cdot 10^{-4} }}%
\htdef{ARGUS.Gamma103.pub.ALBRECHT.88B,qt}{\ensuremath{(6.400 \pm 2.300 \pm 1.000) \cdot 10^{-4} }}%
\htdef{CLEO.Gamma103.pub.GIBAUT.94B,qt}{\ensuremath{(7.700 \pm 0.500 \pm 0.900) \cdot 10^{-4} }}%
\htdef{DELPHI.Gamma103.pub.ABDALLAH.06A,qt}{\ensuremath{(9.700 \pm 1.500 \pm 0.500) \cdot 10^{-4} }}%
\htdef{HRS.Gamma103.pub.BYLSMA.87,qt}{\ensuremath{(5.100 \pm 2.000 \pm 0.000) \cdot 10^{-4} }}%
\htdef{OPAL.Gamma103.pub.ACKERSTAFF.99E,qt}{\ensuremath{(9.100 \pm 1.400 \pm 0.600) \cdot 10^{-4} }}% 
\htdef{Gamma104.qt}{\ensuremath{(1.634 \pm 0.114) \cdot 10^{-4}}}% 
\htdef{ALEPH.Gamma104.pub.SCHAEL.05C,qt}{\ensuremath{(2.100 \pm 0.700 \pm 0.900) \cdot 10^{-4} }}%
\htdef{CLEO.Gamma104.pub.ANASTASSOV.01,qt}{\ensuremath{(1.700 \pm 0.200 \pm 0.200) \cdot 10^{-4} }}%
\htdef{DELPHI.Gamma104.pub.ABDALLAH.06A,qt}{\ensuremath{(1.600 \pm 1.200 \pm 0.600) \cdot 10^{-4} }}%
\htdef{OPAL.Gamma104.pub.ACKERSTAFF.99E,qt}{\ensuremath{(2.700 \pm 1.800 \pm 0.900) \cdot 10^{-4} }}% 
\htdef{Gamma106.qt}{\ensuremath{(0.7748 \pm 0.0534) \cdot 10^{-2}}}% 
\htdef{Gamma110.qt}{\ensuremath{(2.909 \pm 0.048) \cdot 10^{-2}}}% 
\htdef{Gamma126.qt}{\ensuremath{(0.1386 \pm 0.0072) \cdot 10^{-2}}}% 
\htdef{ALEPH.Gamma126.pub.BUSKULIC.97C,qt}{\ensuremath{(0.1800 \pm 0.0447 \pm 0.0000) \cdot 10^{-2} }}%
\htdef{Belle.Gamma126.pub.INAMI.09,qt}{\ensuremath{(0.1350 \pm 0.0030 \pm 0.0070) \cdot 10^{-2} }}%
\htdef{CLEO.Gamma126.pub.ARTUSO.92,qt}{\ensuremath{(0.1700 \pm 0.0200 \pm 0.0200) \cdot 10^{-2} }}% 
\htdef{Gamma128.qt}{\ensuremath{(1.547 \pm 0.080) \cdot 10^{-4}}}% 
\htdef{ALEPH.Gamma128.pub.BUSKULIC.97C,qt}{\ensuremath{(2.900 {}^{+1.300}_{-1.200} \pm 0.700) \cdot 10^{-4} }}%
\htdef{BaBar.Gamma128.pub.DEL-AMO-SANCHEZ.11E,qt}{\ensuremath{(1.420 \pm 0.110 \pm 0.070) \cdot 10^{-4} }}%
\htdef{Belle.Gamma128.pub.INAMI.09,qt}{\ensuremath{(1.580 \pm 0.050 \pm 0.090) \cdot 10^{-4} }}%
\htdef{CLEO.Gamma128.pub.BARTELT.96,qt}{\ensuremath{(2.600 \pm 0.500 \pm 0.500) \cdot 10^{-4} }}% 
\htdef{Gamma130.qt}{\ensuremath{(0.483 \pm 0.116) \cdot 10^{-4}}}% 
\htdef{Belle.Gamma130.pub.INAMI.09,qt}{\ensuremath{(0.460 \pm 0.110 \pm 0.040) \cdot 10^{-4} }}%
\htdef{CLEO.Gamma130.pub.BISHAI.99,qt}{\ensuremath{(1.770 \pm 0.560 \pm 0.710) \cdot 10^{-4} }}% 
\htdef{Gamma132.qt}{\ensuremath{(0.937 \pm 0.149) \cdot 10^{-4}}}% 
\htdef{Belle.Gamma132.pub.INAMI.09,qt}{\ensuremath{(0.880 \pm 0.140 \pm 0.060) \cdot 10^{-4} }}%
\htdef{CLEO.Gamma132.pub.BISHAI.99,qt}{\ensuremath{(2.200 \pm 0.700 \pm 0.220) \cdot 10^{-4} }}% 
\htdef{Gamma136.qt}{\ensuremath{(2.184 \pm 0.130) \cdot 10^{-4}}}% 
\htdef{Gamma149.qt}{\ensuremath{(2.401 \pm 0.075) \cdot 10^{-2}}}% 
\htdef{Gamma150.qt}{\ensuremath{(1.995 \pm 0.064) \cdot 10^{-2}}}% 
\htdef{ALEPH.Gamma150.pub.BUSKULIC.97C,qt}{\ensuremath{(1.910 \pm 0.092 \pm 0.000) \cdot 10^{-2} }}%
\htdef{CLEO.Gamma150.pub.BARINGER.87,qt}{\ensuremath{(1.600 \pm 0.270 \pm 0.410) \cdot 10^{-2} }}% 
\htdef{Gamma150by66.qt}{\ensuremath{0.4332 \pm 0.0139}}% 
\htdef{ALEPH.Gamma150by66.pub.BUSKULIC.96,qt}{\ensuremath{0.4310 \pm 0.0330 \pm 0.0000}}%
\htdef{CLEO.Gamma150by66.pub.BALEST.95C,qt}{\ensuremath{0.4640 \pm 0.0160 \pm 0.0170}}% 
\htdef{Gamma151.qt}{\ensuremath{(4.100 \pm 0.922) \cdot 10^{-4}}}% 
\htdef{CLEO3.Gamma151.pub.ARMS.05,qt}{\ensuremath{(4.100 \pm 0.600 \pm 0.700) \cdot 10^{-4} }}% 
\htdef{Gamma152.qt}{\ensuremath{(0.4058 \pm 0.0419) \cdot 10^{-2}}}% 
\htdef{ALEPH.Gamma152.pub.BUSKULIC.97C,qt}{\ensuremath{(0.4300 \pm 0.0781 \pm 0.0000) \cdot 10^{-2} }}% 
\htdef{Gamma152by54.qt}{\ensuremath{(2.667 \pm 0.275) \cdot 10^{-2}}}% 
\htdef{Gamma152by76.qt}{\ensuremath{0.8241 \pm 0.0757}}% 
\htdef{CLEO.Gamma152by76.pub.BORTOLETTO.93,qt}{\ensuremath{0.8100 \pm 0.0600 \pm 0.0600}}% 
\htdef{Gamma167.qt}{\ensuremath{(4.445 \pm 1.636) \cdot 10^{-5}}}% 
\htdef{Gamma168.qt}{\ensuremath{(2.174 \pm 0.800) \cdot 10^{-5}}}% 
\htdef{Gamma169.qt}{\ensuremath{(1.520 \pm 0.560) \cdot 10^{-5}}}% 
\htdef{Gamma800.qt}{\ensuremath{(1.954 \pm 0.065) \cdot 10^{-2}}}% 
\htdef{Gamma802.qt}{\ensuremath{(0.2923 \pm 0.0067) \cdot 10^{-2}}}% 
\htdef{Gamma803.qt}{\ensuremath{(4.103 \pm 1.429) \cdot 10^{-4}}}% 
\htdef{Gamma804.qt}{\ensuremath{(2.332 \pm 0.065) \cdot 10^{-4}}}% 
\htdef{Gamma805.qt}{\ensuremath{(4.000 \pm 2.000) \cdot 10^{-4}}}% 
\htdef{ALEPH.Gamma805.pub.SCHAEL.05C,qt}{\ensuremath{(4.000 \pm 2.000 \pm 0.000) \cdot 10^{-4} }}% 
\htdef{Gamma806.qt}{\ensuremath{(1.815 \pm 0.207) \cdot 10^{-5}}}% 
\htdef{Gamma810.qt}{\ensuremath{(1.924 \pm 0.298) \cdot 10^{-4}}}% 
\htdef{Gamma811.qt}{\ensuremath{(7.105 \pm 1.586) \cdot 10^{-5}}}% 
\htdef{BaBar.Gamma811.pub.LEES.12X,qt}{\ensuremath{(7.300 \pm 1.200 \pm 1.200) \cdot 10^{-5} }}% 
\htdef{Gamma812.qt}{\ensuremath{(1.344 \pm 2.683) \cdot 10^{-5}}}% 
\htdef{BaBar.Gamma812.pub.LEES.12X,qt}{\ensuremath{(1.000 \pm 0.800 \pm 3.000) \cdot 10^{-5} }}% 
\htdef{Gamma820.qt}{\ensuremath{(8.197 \pm 0.315) \cdot 10^{-4}}}% 
\htdef{Gamma821.qt}{\ensuremath{(7.677 \pm 0.297) \cdot 10^{-4}}}% 
\htdef{BaBar.Gamma821.pub.LEES.12X,qt}{\ensuremath{(7.680 \pm 0.040 \pm 0.400) \cdot 10^{-4} }}% 
\htdef{Gamma822.qt}{\ensuremath{(0.596 \pm 1.208) \cdot 10^{-6}}}% 
\htdef{BaBar.Gamma822.pub.LEES.12X,qt}{\ensuremath{(0.600 \pm 0.500 \pm 1.100) \cdot 10^{-6} }}% 
\htdef{Gamma830.qt}{\ensuremath{(1.623 \pm 0.114) \cdot 10^{-4}}}% 
\htdef{Gamma831.qt}{\ensuremath{(8.359 \pm 0.626) \cdot 10^{-5}}}% 
\htdef{BaBar.Gamma831.pub.LEES.12X,qt}{\ensuremath{(8.400 \pm 0.400 \pm 0.600) \cdot 10^{-5} }}% 
\htdef{Gamma832.qt}{\ensuremath{(3.771 \pm 0.875) \cdot 10^{-5}}}% 
\htdef{BaBar.Gamma832.pub.LEES.12X,qt}{\ensuremath{(3.600 \pm 0.300 \pm 0.900) \cdot 10^{-5} }}% 
\htdef{Gamma833.qt}{\ensuremath{(1.108 \pm 0.566) \cdot 10^{-6}}}% 
\htdef{BaBar.Gamma833.pub.LEES.12X,qt}{\ensuremath{(1.100 \pm 0.400 \pm 0.400) \cdot 10^{-6} }}% 
\htdef{Gamma910.qt}{\ensuremath{(7.136 \pm 0.424) \cdot 10^{-5}}}% 
\htdef{BaBar.Gamma910.pub.LEES.12X,qt}{\ensuremath{(8.270 \pm 0.880 \pm 0.810) \cdot 10^{-5} }}% 
\htdef{Gamma911.qt}{\ensuremath{(4.420 \pm 0.867) \cdot 10^{-5}}}% 
\htdef{BaBar.Gamma911.pub.LEES.12X,qt}{\ensuremath{(4.570 \pm 0.770 \pm 0.500) \cdot 10^{-5} }}% 
\htdef{Gamma920.qt}{\ensuremath{(5.197 \pm 0.444) \cdot 10^{-5}}}% 
\htdef{BaBar.Gamma920.pub.LEES.12X,qt}{\ensuremath{(5.200 \pm 0.310 \pm 0.370) \cdot 10^{-5} }}% 
\htdef{Gamma930.qt}{\ensuremath{(5.005 \pm 0.297) \cdot 10^{-5}}}% 
\htdef{BaBar.Gamma930.pub.LEES.12X,qt}{\ensuremath{(5.390 \pm 0.270 \pm 0.410) \cdot 10^{-5} }}% 
\htdef{Gamma944.qt}{\ensuremath{(8.606 \pm 0.511) \cdot 10^{-5}}}% 
\htdef{BaBar.Gamma944.pub.LEES.12X,qt}{\ensuremath{(8.260 \pm 0.350 \pm 0.510) \cdot 10^{-5} }}% 
\htdef{Gamma945.qt}{\ensuremath{(1.929 \pm 0.378) \cdot 10^{-4}}}% 
\htdef{Gamma998.qt}{\ensuremath{(0.0355 \pm 0.1031) \cdot 10^{-2}}}%
\htdef{Gamma1.qm}{%
\begin{ensuredisplaymath}
\htuse{Gamma1.gn} = \htuse{Gamma1.td}
\end{ensuredisplaymath}
 & \htuse{Gamma1.qt} & \hfagFitLabel}% 
\htdef{Gamma2.qm}{%
\begin{ensuredisplaymath}
\htuse{Gamma2.gn} = \htuse{Gamma2.td}
\end{ensuredisplaymath}
 & \htuse{Gamma2.qt} & \hfagFitLabel}% 
\htdef{Gamma3.qm}{%
\begin{ensuredisplaymath}
\htuse{Gamma3.gn} = \htuse{Gamma3.td}
\end{ensuredisplaymath}
 & \htuse{Gamma3.qt} & \hfagFitLabel\\
\htuse{ALEPH.Gamma3.pub.SCHAEL.05C,qt} & \htuse{ALEPH.Gamma3.pub.SCHAEL.05C,exp} & \htuse{ALEPH.Gamma3.pub.SCHAEL.05C,ref} \\
\htuse{DELPHI.Gamma3.pub.ABREU.99X,qt} & \htuse{DELPHI.Gamma3.pub.ABREU.99X,exp} & \htuse{DELPHI.Gamma3.pub.ABREU.99X,ref} \\
\htuse{L3.Gamma3.pub.ACCIARRI.01F,qt} & \htuse{L3.Gamma3.pub.ACCIARRI.01F,exp} & \htuse{L3.Gamma3.pub.ACCIARRI.01F,ref} \\
\htuse{OPAL.Gamma3.pub.ABBIENDI.03,qt} & \htuse{OPAL.Gamma3.pub.ABBIENDI.03,exp} & \htuse{OPAL.Gamma3.pub.ABBIENDI.03,ref}
}% 
\htdef{Gamma3by5.qm}{%
\begin{ensuredisplaymath}
\htuse{Gamma3by5.gn} = \htuse{Gamma3by5.td}
\end{ensuredisplaymath}
 & \htuse{Gamma3by5.qt} & \hfagFitLabel\\
\htuse{ARGUS.Gamma3by5.pub.ALBRECHT.92D,qt} & \htuse{ARGUS.Gamma3by5.pub.ALBRECHT.92D,exp} & \htuse{ARGUS.Gamma3by5.pub.ALBRECHT.92D,ref} \\
\htuse{BaBar.Gamma3by5.pub.AUBERT.10F,qt} & \htuse{BaBar.Gamma3by5.pub.AUBERT.10F,exp} & \htuse{BaBar.Gamma3by5.pub.AUBERT.10F,ref} \\
\htuse{CLEO.Gamma3by5.pub.ANASTASSOV.97,qt} & \htuse{CLEO.Gamma3by5.pub.ANASTASSOV.97,exp} & \htuse{CLEO.Gamma3by5.pub.ANASTASSOV.97,ref}
}% 
\htdef{Gamma5.qm}{%
\begin{ensuredisplaymath}
\htuse{Gamma5.gn} = \htuse{Gamma5.td}
\end{ensuredisplaymath}
 & \htuse{Gamma5.qt} & \hfagFitLabel\\
\htuse{ALEPH.Gamma5.pub.SCHAEL.05C,qt} & \htuse{ALEPH.Gamma5.pub.SCHAEL.05C,exp} & \htuse{ALEPH.Gamma5.pub.SCHAEL.05C,ref} \\
\htuse{CLEO.Gamma5.pub.ANASTASSOV.97,qt} & \htuse{CLEO.Gamma5.pub.ANASTASSOV.97,exp} & \htuse{CLEO.Gamma5.pub.ANASTASSOV.97,ref} \\
\htuse{DELPHI.Gamma5.pub.ABREU.99X,qt} & \htuse{DELPHI.Gamma5.pub.ABREU.99X,exp} & \htuse{DELPHI.Gamma5.pub.ABREU.99X,ref} \\
\htuse{L3.Gamma5.pub.ACCIARRI.01F,qt} & \htuse{L3.Gamma5.pub.ACCIARRI.01F,exp} & \htuse{L3.Gamma5.pub.ACCIARRI.01F,ref} \\
\htuse{OPAL.Gamma5.pub.ABBIENDI.99H,qt} & \htuse{OPAL.Gamma5.pub.ABBIENDI.99H,exp} & \htuse{OPAL.Gamma5.pub.ABBIENDI.99H,ref}
}% 
\htdef{Gamma7.qm}{%
\begin{ensuredisplaymath}
\htuse{Gamma7.gn} = \htuse{Gamma7.td}
\end{ensuredisplaymath}
 & \htuse{Gamma7.qt} & \hfagFitLabel\\
\htuse{DELPHI.Gamma7.pub.ABREU.92N,qt} & \htuse{DELPHI.Gamma7.pub.ABREU.92N,exp} & \htuse{DELPHI.Gamma7.pub.ABREU.92N,ref} \\
\htuse{L3.Gamma7.pub.ACCIARRI.95,qt} & \htuse{L3.Gamma7.pub.ACCIARRI.95,exp} & \htuse{L3.Gamma7.pub.ACCIARRI.95,ref} \\
\htuse{OPAL.Gamma7.pub.ALEXANDER.91D,qt} & \htuse{OPAL.Gamma7.pub.ALEXANDER.91D,exp} & \htuse{OPAL.Gamma7.pub.ALEXANDER.91D,ref}
}% 
\htdef{Gamma8.qm}{%
\begin{ensuredisplaymath}
\htuse{Gamma8.gn} = \htuse{Gamma8.td}
\end{ensuredisplaymath}
 & \htuse{Gamma8.qt} & \hfagFitLabel\\
\htuse{ALEPH.Gamma8.pub.SCHAEL.05C,qt} & \htuse{ALEPH.Gamma8.pub.SCHAEL.05C,exp} & \htuse{ALEPH.Gamma8.pub.SCHAEL.05C,ref} \\
\htuse{CLEO.Gamma8.pub.ANASTASSOV.97,qt} & \htuse{CLEO.Gamma8.pub.ANASTASSOV.97,exp} & \htuse{CLEO.Gamma8.pub.ANASTASSOV.97,ref} \\
\htuse{DELPHI.Gamma8.pub.ABDALLAH.06A,qt} & \htuse{DELPHI.Gamma8.pub.ABDALLAH.06A,exp} & \htuse{DELPHI.Gamma8.pub.ABDALLAH.06A,ref} \\
\htuse{OPAL.Gamma8.pub.ACKERSTAFF.98M,qt} & \htuse{OPAL.Gamma8.pub.ACKERSTAFF.98M,exp} & \htuse{OPAL.Gamma8.pub.ACKERSTAFF.98M,ref}
}% 
\htdef{Gamma8by5.qm}{%
\begin{ensuredisplaymath}
\htuse{Gamma8by5.gn} = \htuse{Gamma8by5.td}
\end{ensuredisplaymath}
 & \htuse{Gamma8by5.qt} & \hfagFitLabel}% 
\htdef{Gamma9.qm}{%
\begin{ensuredisplaymath}
\htuse{Gamma9.gn} = \htuse{Gamma9.td}
\end{ensuredisplaymath}
 & \htuse{Gamma9.qt} & \hfagFitLabel}% 
\htdef{Gamma9by5.qm}{%
\begin{ensuredisplaymath}
\htuse{Gamma9by5.gn} = \htuse{Gamma9by5.td}
\end{ensuredisplaymath}
 & \htuse{Gamma9by5.qt} & \hfagFitLabel\\
\htuse{BaBar.Gamma9by5.pub.AUBERT.10F,qt} & \htuse{BaBar.Gamma9by5.pub.AUBERT.10F,exp} & \htuse{BaBar.Gamma9by5.pub.AUBERT.10F,ref}
}% 
\htdef{Gamma10.qm}{%
\begin{ensuredisplaymath}
\htuse{Gamma10.gn} = \htuse{Gamma10.td}
\end{ensuredisplaymath}
 & \htuse{Gamma10.qt} & \hfagFitLabel\\
\htuse{ALEPH.Gamma10.pub.BARATE.99K,qt} & \htuse{ALEPH.Gamma10.pub.BARATE.99K,exp} & \htuse{ALEPH.Gamma10.pub.BARATE.99K,ref} \\
\htuse{CLEO.Gamma10.pub.BATTLE.94,qt} & \htuse{CLEO.Gamma10.pub.BATTLE.94,exp} & \htuse{CLEO.Gamma10.pub.BATTLE.94,ref} \\
\htuse{DELPHI.Gamma10.pub.ABREU.94K,qt} & \htuse{DELPHI.Gamma10.pub.ABREU.94K,exp} & \htuse{DELPHI.Gamma10.pub.ABREU.94K,ref} \\
\htuse{OPAL.Gamma10.pub.ABBIENDI.01J,qt} & \htuse{OPAL.Gamma10.pub.ABBIENDI.01J,exp} & \htuse{OPAL.Gamma10.pub.ABBIENDI.01J,ref}
}% 
\htdef{Gamma10by5.qm}{%
\begin{ensuredisplaymath}
\htuse{Gamma10by5.gn} = \htuse{Gamma10by5.td}
\end{ensuredisplaymath}
 & \htuse{Gamma10by5.qt} & \hfagFitLabel\\
\htuse{BaBar.Gamma10by5.pub.AUBERT.10F,qt} & \htuse{BaBar.Gamma10by5.pub.AUBERT.10F,exp} & \htuse{BaBar.Gamma10by5.pub.AUBERT.10F,ref}
}% 
\htdef{Gamma10by9.qm}{%
\begin{ensuredisplaymath}
\htuse{Gamma10by9.gn} = \htuse{Gamma10by9.td}
\end{ensuredisplaymath}
 & \htuse{Gamma10by9.qt} & \hfagFitLabel}% 
\htdef{Gamma11.qm}{%
\begin{ensuredisplaymath}
\htuse{Gamma11.gn} = \htuse{Gamma11.td}
\end{ensuredisplaymath}
 & \htuse{Gamma11.qt} & \hfagFitLabel}% 
\htdef{Gamma12.qm}{%
\begin{ensuredisplaymath}
\htuse{Gamma12.gn} = \htuse{Gamma12.td}
\end{ensuredisplaymath}
 & \htuse{Gamma12.qt} & \hfagFitLabel}% 
\htdef{Gamma13.qm}{%
\begin{ensuredisplaymath}
\htuse{Gamma13.gn} = \htuse{Gamma13.td}
\end{ensuredisplaymath}
 & \htuse{Gamma13.qt} & \hfagFitLabel\\
\htuse{ALEPH.Gamma13.pub.SCHAEL.05C,qt} & \htuse{ALEPH.Gamma13.pub.SCHAEL.05C,exp} & \htuse{ALEPH.Gamma13.pub.SCHAEL.05C,ref} \\
\htuse{Belle.Gamma13.pub.FUJIKAWA.08,qt} & \htuse{Belle.Gamma13.pub.FUJIKAWA.08,exp} & \htuse{Belle.Gamma13.pub.FUJIKAWA.08,ref} \\
\htuse{CLEO.Gamma13.pub.ARTUSO.94,qt} & \htuse{CLEO.Gamma13.pub.ARTUSO.94,exp} & \htuse{CLEO.Gamma13.pub.ARTUSO.94,ref} \\
\htuse{DELPHI.Gamma13.pub.ABDALLAH.06A,qt} & \htuse{DELPHI.Gamma13.pub.ABDALLAH.06A,exp} & \htuse{DELPHI.Gamma13.pub.ABDALLAH.06A,ref} \\
\htuse{L3.Gamma13.pub.ACCIARRI.95,qt} & \htuse{L3.Gamma13.pub.ACCIARRI.95,exp} & \htuse{L3.Gamma13.pub.ACCIARRI.95,ref} \\
\htuse{OPAL.Gamma13.pub.ACKERSTAFF.98M,qt} & \htuse{OPAL.Gamma13.pub.ACKERSTAFF.98M,exp} & \htuse{OPAL.Gamma13.pub.ACKERSTAFF.98M,ref}
}% 
\htdef{Gamma14.qm}{%
\begin{ensuredisplaymath}
\htuse{Gamma14.gn} = \htuse{Gamma14.td}
\end{ensuredisplaymath}
 & \htuse{Gamma14.qt} & \hfagFitLabel}% 
\htdef{Gamma16.qm}{%
\begin{ensuredisplaymath}
\htuse{Gamma16.gn} = \htuse{Gamma16.td}
\end{ensuredisplaymath}
 & \htuse{Gamma16.qt} & \hfagFitLabel\\
\htuse{ALEPH.Gamma16.pub.BARATE.99K,qt} & \htuse{ALEPH.Gamma16.pub.BARATE.99K,exp} & \htuse{ALEPH.Gamma16.pub.BARATE.99K,ref} \\
\htuse{BaBar.Gamma16.pub.AUBERT.07AP,qt} & \htuse{BaBar.Gamma16.pub.AUBERT.07AP,exp} & \htuse{BaBar.Gamma16.pub.AUBERT.07AP,ref} \\
\htuse{CLEO.Gamma16.pub.BATTLE.94,qt} & \htuse{CLEO.Gamma16.pub.BATTLE.94,exp} & \htuse{CLEO.Gamma16.pub.BATTLE.94,ref} \\
\htuse{OPAL.Gamma16.pub.ABBIENDI.04J,qt} & \htuse{OPAL.Gamma16.pub.ABBIENDI.04J,exp} & \htuse{OPAL.Gamma16.pub.ABBIENDI.04J,ref}
}% 
\htdef{Gamma17.qm}{%
\begin{ensuredisplaymath}
\htuse{Gamma17.gn} = \htuse{Gamma17.td}
\end{ensuredisplaymath}
 & \htuse{Gamma17.qt} & \hfagFitLabel\\
\htuse{OPAL.Gamma17.pub.ACKERSTAFF.98M,qt} & \htuse{OPAL.Gamma17.pub.ACKERSTAFF.98M,exp} & \htuse{OPAL.Gamma17.pub.ACKERSTAFF.98M,ref}
}% 
\htdef{Gamma18.qm}{%
\begin{ensuredisplaymath}
\htuse{Gamma18.gn} = \htuse{Gamma18.td}
\end{ensuredisplaymath}
 & \htuse{Gamma18.qt} & \hfagFitLabel}% 
\htdef{Gamma19.qm}{%
\begin{ensuredisplaymath}
\htuse{Gamma19.gn} = \htuse{Gamma19.td}
\end{ensuredisplaymath}
 & \htuse{Gamma19.qt} & \hfagFitLabel\\
\htuse{ALEPH.Gamma19.pub.SCHAEL.05C,qt} & \htuse{ALEPH.Gamma19.pub.SCHAEL.05C,exp} & \htuse{ALEPH.Gamma19.pub.SCHAEL.05C,ref} \\
\htuse{DELPHI.Gamma19.pub.ABDALLAH.06A,qt} & \htuse{DELPHI.Gamma19.pub.ABDALLAH.06A,exp} & \htuse{DELPHI.Gamma19.pub.ABDALLAH.06A,ref} \\
\htuse{L3.Gamma19.pub.ACCIARRI.95,qt} & \htuse{L3.Gamma19.pub.ACCIARRI.95,exp} & \htuse{L3.Gamma19.pub.ACCIARRI.95,ref}
}% 
\htdef{Gamma19by13.qm}{%
\begin{ensuredisplaymath}
\htuse{Gamma19by13.gn} = \htuse{Gamma19by13.td}
\end{ensuredisplaymath}
 & \htuse{Gamma19by13.qt} & \hfagFitLabel\\
\htuse{CLEO.Gamma19by13.pub.PROCARIO.93,qt} & \htuse{CLEO.Gamma19by13.pub.PROCARIO.93,exp} & \htuse{CLEO.Gamma19by13.pub.PROCARIO.93,ref}
}% 
\htdef{Gamma20.qm}{%
\begin{ensuredisplaymath}
\htuse{Gamma20.gn} = \htuse{Gamma20.td}
\end{ensuredisplaymath}
 & \htuse{Gamma20.qt} & \hfagFitLabel}% 
\htdef{Gamma23.qm}{%
\begin{ensuredisplaymath}
\htuse{Gamma23.gn} = \htuse{Gamma23.td}
\end{ensuredisplaymath}
 & \htuse{Gamma23.qt} & \hfagFitLabel\\
\htuse{ALEPH.Gamma23.pub.BARATE.99K,qt} & \htuse{ALEPH.Gamma23.pub.BARATE.99K,exp} & \htuse{ALEPH.Gamma23.pub.BARATE.99K,ref} \\
\htuse{CLEO.Gamma23.pub.BATTLE.94,qt} & \htuse{CLEO.Gamma23.pub.BATTLE.94,exp} & \htuse{CLEO.Gamma23.pub.BATTLE.94,ref}
}% 
\htdef{Gamma24.qm}{%
\begin{ensuredisplaymath}
\htuse{Gamma24.gn} = \htuse{Gamma24.td}
\end{ensuredisplaymath}
 & \htuse{Gamma24.qt} & \hfagFitLabel}% 
\htdef{Gamma25.qm}{%
\begin{ensuredisplaymath}
\htuse{Gamma25.gn} = \htuse{Gamma25.td}
\end{ensuredisplaymath}
 & \htuse{Gamma25.qt} & \hfagFitLabel\\
\htuse{DELPHI.Gamma25.pub.ABDALLAH.06A,qt} & \htuse{DELPHI.Gamma25.pub.ABDALLAH.06A,exp} & \htuse{DELPHI.Gamma25.pub.ABDALLAH.06A,ref}
}% 
\htdef{Gamma26.qm}{%
\begin{ensuredisplaymath}
\htuse{Gamma26.gn} = \htuse{Gamma26.td}
\end{ensuredisplaymath}
 & \htuse{Gamma26.qt} & \hfagFitLabel\\
\htuse{ALEPH.Gamma26.pub.SCHAEL.05C,qt} & \htuse{ALEPH.Gamma26.pub.SCHAEL.05C,exp} & \htuse{ALEPH.Gamma26.pub.SCHAEL.05C,ref} \\
\htuse{L3.Gamma26.pub.ACCIARRI.95,qt} & \htuse{L3.Gamma26.pub.ACCIARRI.95,exp} & \htuse{L3.Gamma26.pub.ACCIARRI.95,ref}
}% 
\htdef{Gamma26by13.qm}{%
\begin{ensuredisplaymath}
\htuse{Gamma26by13.gn} = \htuse{Gamma26by13.td}
\end{ensuredisplaymath}
 & \htuse{Gamma26by13.qt} & \hfagFitLabel\\
\htuse{CLEO.Gamma26by13.pub.PROCARIO.93,qt} & \htuse{CLEO.Gamma26by13.pub.PROCARIO.93,exp} & \htuse{CLEO.Gamma26by13.pub.PROCARIO.93,ref}
}% 
\htdef{Gamma27.qm}{%
\begin{ensuredisplaymath}
\htuse{Gamma27.gn} = \htuse{Gamma27.td}
\end{ensuredisplaymath}
 & \htuse{Gamma27.qt} & \hfagFitLabel}% 
\htdef{Gamma28.qm}{%
\begin{ensuredisplaymath}
\htuse{Gamma28.gn} = \htuse{Gamma28.td}
\end{ensuredisplaymath}
 & \htuse{Gamma28.qt} & \hfagFitLabel\\
\htuse{ALEPH.Gamma28.pub.BARATE.99K,qt} & \htuse{ALEPH.Gamma28.pub.BARATE.99K,exp} & \htuse{ALEPH.Gamma28.pub.BARATE.99K,ref}
}% 
\htdef{Gamma29.qm}{%
\begin{ensuredisplaymath}
\htuse{Gamma29.gn} = \htuse{Gamma29.td}
\end{ensuredisplaymath}
 & \htuse{Gamma29.qt} & \hfagFitLabel\\
\htuse{CLEO.Gamma29.pub.PROCARIO.93,qt} & \htuse{CLEO.Gamma29.pub.PROCARIO.93,exp} & \htuse{CLEO.Gamma29.pub.PROCARIO.93,ref}
}% 
\htdef{Gamma30.qm}{%
\begin{ensuredisplaymath}
\htuse{Gamma30.gn} = \htuse{Gamma30.td}
\end{ensuredisplaymath}
 & \htuse{Gamma30.qt} & \hfagFitLabel\\
\htuse{ALEPH.Gamma30.pub.SCHAEL.05C,qt} & \htuse{ALEPH.Gamma30.pub.SCHAEL.05C,exp} & \htuse{ALEPH.Gamma30.pub.SCHAEL.05C,ref}
}% 
\htdef{Gamma31.qm}{%
\begin{ensuredisplaymath}
\htuse{Gamma31.gn} = \htuse{Gamma31.td}
\end{ensuredisplaymath}
 & \htuse{Gamma31.qt} & \hfagFitLabel\\
\htuse{CLEO.Gamma31.pub.BATTLE.94,qt} & \htuse{CLEO.Gamma31.pub.BATTLE.94,exp} & \htuse{CLEO.Gamma31.pub.BATTLE.94,ref} \\
\htuse{DELPHI.Gamma31.pub.ABREU.94K,qt} & \htuse{DELPHI.Gamma31.pub.ABREU.94K,exp} & \htuse{DELPHI.Gamma31.pub.ABREU.94K,ref} \\
\htuse{OPAL.Gamma31.pub.ABBIENDI.01J,qt} & \htuse{OPAL.Gamma31.pub.ABBIENDI.01J,exp} & \htuse{OPAL.Gamma31.pub.ABBIENDI.01J,ref}
}% 
\htdef{Gamma32.qm}{%
\begin{ensuredisplaymath}
\htuse{Gamma32.gn} = \htuse{Gamma32.td}
\end{ensuredisplaymath}
 & \htuse{Gamma32.qt} & \hfagFitLabel}% 
\htdef{Gamma33.qm}{%
\begin{ensuredisplaymath}
\htuse{Gamma33.gn} = \htuse{Gamma33.td}
\end{ensuredisplaymath}
 & \htuse{Gamma33.qt} & \hfagFitLabel\\
\htuse{ALEPH.Gamma33.pub.BARATE.98E,qt} & \htuse{ALEPH.Gamma33.pub.BARATE.98E,exp} & \htuse{ALEPH.Gamma33.pub.BARATE.98E,ref} \\
\htuse{OPAL.Gamma33.pub.AKERS.94G,qt} & \htuse{OPAL.Gamma33.pub.AKERS.94G,exp} & \htuse{OPAL.Gamma33.pub.AKERS.94G,ref}
}% 
\htdef{Gamma34.qm}{%
\begin{ensuredisplaymath}
\htuse{Gamma34.gn} = \htuse{Gamma34.td}
\end{ensuredisplaymath}
 & \htuse{Gamma34.qt} & \hfagFitLabel\\
\htuse{CLEO.Gamma34.pub.COAN.96,qt} & \htuse{CLEO.Gamma34.pub.COAN.96,exp} & \htuse{CLEO.Gamma34.pub.COAN.96,ref}
}% 
\htdef{Gamma35.qm}{%
\begin{ensuredisplaymath}
\htuse{Gamma35.gn} = \htuse{Gamma35.td}
\end{ensuredisplaymath}
 & \htuse{Gamma35.qt} & \hfagFitLabel\\
\htuse{ALEPH.Gamma35.pub.BARATE.99K,qt} & \htuse{ALEPH.Gamma35.pub.BARATE.99K,exp} & \htuse{ALEPH.Gamma35.pub.BARATE.99K,ref} \\
\htuse{Belle.Gamma35.pub.RYU.14vpc,qt} & \htuse{Belle.Gamma35.pub.RYU.14vpc,exp} & \htuse{Belle.Gamma35.pub.RYU.14vpc,ref} \\
\htuse{L3.Gamma35.pub.ACCIARRI.95F,qt} & \htuse{L3.Gamma35.pub.ACCIARRI.95F,exp} & \htuse{L3.Gamma35.pub.ACCIARRI.95F,ref} \\
\htuse{OPAL.Gamma35.pub.ABBIENDI.00C,qt} & \htuse{OPAL.Gamma35.pub.ABBIENDI.00C,exp} & \htuse{OPAL.Gamma35.pub.ABBIENDI.00C,ref}
}% 
\htdef{Gamma37.qm}{%
\begin{ensuredisplaymath}
\htuse{Gamma37.gn} = \htuse{Gamma37.td}
\end{ensuredisplaymath}
 & \htuse{Gamma37.qt} & \hfagFitLabel\\
\htuse{ALEPH.Gamma37.pub.BARATE.98E,qt} & \htuse{ALEPH.Gamma37.pub.BARATE.98E,exp} & \htuse{ALEPH.Gamma37.pub.BARATE.98E,ref} \\
\htuse{ALEPH.Gamma37.pub.BARATE.99K,qt} & \htuse{ALEPH.Gamma37.pub.BARATE.99K,exp} & \htuse{ALEPH.Gamma37.pub.BARATE.99K,ref} \\
\htuse{Belle.Gamma37.pub.RYU.14vpc,qt} & \htuse{Belle.Gamma37.pub.RYU.14vpc,exp} & \htuse{Belle.Gamma37.pub.RYU.14vpc,ref} \\
\htuse{CLEO.Gamma37.pub.COAN.96,qt} & \htuse{CLEO.Gamma37.pub.COAN.96,exp} & \htuse{CLEO.Gamma37.pub.COAN.96,ref}
}% 
\htdef{Gamma38.qm}{%
\begin{ensuredisplaymath}
\htuse{Gamma38.gn} = \htuse{Gamma38.td}
\end{ensuredisplaymath}
 & \htuse{Gamma38.qt} & \hfagFitLabel\\
\htuse{OPAL.Gamma38.pub.ABBIENDI.00C,qt} & \htuse{OPAL.Gamma38.pub.ABBIENDI.00C,exp} & \htuse{OPAL.Gamma38.pub.ABBIENDI.00C,ref}
}% 
\htdef{Gamma39.qm}{%
\begin{ensuredisplaymath}
\htuse{Gamma39.gn} = \htuse{Gamma39.td}
\end{ensuredisplaymath}
 & \htuse{Gamma39.qt} & \hfagFitLabel\\
\htuse{CLEO.Gamma39.pub.COAN.96,qt} & \htuse{CLEO.Gamma39.pub.COAN.96,exp} & \htuse{CLEO.Gamma39.pub.COAN.96,ref}
}% 
\htdef{Gamma40.qm}{%
\begin{ensuredisplaymath}
\htuse{Gamma40.gn} = \htuse{Gamma40.td}
\end{ensuredisplaymath}
 & \htuse{Gamma40.qt} & \hfagFitLabel\\
\htuse{ALEPH.Gamma40.pub.BARATE.98E,qt} & \htuse{ALEPH.Gamma40.pub.BARATE.98E,exp} & \htuse{ALEPH.Gamma40.pub.BARATE.98E,ref} \\
\htuse{ALEPH.Gamma40.pub.BARATE.99K,qt} & \htuse{ALEPH.Gamma40.pub.BARATE.99K,exp} & \htuse{ALEPH.Gamma40.pub.BARATE.99K,ref} \\
\htuse{Belle.Gamma40.pub.RYU.14vpc,qt} & \htuse{Belle.Gamma40.pub.RYU.14vpc,exp} & \htuse{Belle.Gamma40.pub.RYU.14vpc,ref} \\
\htuse{L3.Gamma40.pub.ACCIARRI.95F,qt} & \htuse{L3.Gamma40.pub.ACCIARRI.95F,exp} & \htuse{L3.Gamma40.pub.ACCIARRI.95F,ref}
}% 
\htdef{Gamma42.qm}{%
\begin{ensuredisplaymath}
\htuse{Gamma42.gn} = \htuse{Gamma42.td}
\end{ensuredisplaymath}
 & \htuse{Gamma42.qt} & \hfagFitLabel\\
\htuse{ALEPH.Gamma42.pub.BARATE.98E,qt} & \htuse{ALEPH.Gamma42.pub.BARATE.98E,exp} & \htuse{ALEPH.Gamma42.pub.BARATE.98E,ref} \\
\htuse{ALEPH.Gamma42.pub.BARATE.99K,qt} & \htuse{ALEPH.Gamma42.pub.BARATE.99K,exp} & \htuse{ALEPH.Gamma42.pub.BARATE.99K,ref} \\
\htuse{Belle.Gamma42.pub.RYU.14vpc,qt} & \htuse{Belle.Gamma42.pub.RYU.14vpc,exp} & \htuse{Belle.Gamma42.pub.RYU.14vpc,ref} \\
\htuse{CLEO.Gamma42.pub.COAN.96,qt} & \htuse{CLEO.Gamma42.pub.COAN.96,exp} & \htuse{CLEO.Gamma42.pub.COAN.96,ref}
}% 
\htdef{Gamma43.qm}{%
\begin{ensuredisplaymath}
\htuse{Gamma43.gn} = \htuse{Gamma43.td}
\end{ensuredisplaymath}
 & \htuse{Gamma43.qt} & \hfagFitLabel\\
\htuse{OPAL.Gamma43.pub.ABBIENDI.00C,qt} & \htuse{OPAL.Gamma43.pub.ABBIENDI.00C,exp} & \htuse{OPAL.Gamma43.pub.ABBIENDI.00C,ref}
}% 
\htdef{Gamma44.qm}{%
\begin{ensuredisplaymath}
\htuse{Gamma44.gn} = \htuse{Gamma44.td}
\end{ensuredisplaymath}
 & \htuse{Gamma44.qt} & \hfagFitLabel\\
\htuse{ALEPH.Gamma44.pub.BARATE.99R,qt} & \htuse{ALEPH.Gamma44.pub.BARATE.99R,exp} & \htuse{ALEPH.Gamma44.pub.BARATE.99R,ref}
}% 
\htdef{Gamma46.qm}{%
\begin{ensuredisplaymath}
\htuse{Gamma46.gn} = \htuse{Gamma46.td}
\end{ensuredisplaymath}
 & \htuse{Gamma46.qt} & \hfagFitLabel}% 
\htdef{Gamma47.qm}{%
\begin{ensuredisplaymath}
\htuse{Gamma47.gn} = \htuse{Gamma47.td}
\end{ensuredisplaymath}
 & \htuse{Gamma47.qt} & \hfagFitLabel\\
\htuse{ALEPH.Gamma47.pub.BARATE.98E,qt} & \htuse{ALEPH.Gamma47.pub.BARATE.98E,exp} & \htuse{ALEPH.Gamma47.pub.BARATE.98E,ref} \\
\htuse{BaBar.Gamma47.pub.LEES.12Y,qt} & \htuse{BaBar.Gamma47.pub.LEES.12Y,exp} & \htuse{BaBar.Gamma47.pub.LEES.12Y,ref} \\
\htuse{Belle.Gamma47.pub.RYU.14vpc,qt} & \htuse{Belle.Gamma47.pub.RYU.14vpc,exp} & \htuse{Belle.Gamma47.pub.RYU.14vpc,ref} \\
\htuse{CLEO.Gamma47.pub.COAN.96,qt} & \htuse{CLEO.Gamma47.pub.COAN.96,exp} & \htuse{CLEO.Gamma47.pub.COAN.96,ref}
}% 
\htdef{Gamma48.qm}{%
\begin{ensuredisplaymath}
\htuse{Gamma48.gn} = \htuse{Gamma48.td}
\end{ensuredisplaymath}
 & \htuse{Gamma48.qt} & \hfagFitLabel\\
\htuse{ALEPH.Gamma48.pub.BARATE.98E,qt} & \htuse{ALEPH.Gamma48.pub.BARATE.98E,exp} & \htuse{ALEPH.Gamma48.pub.BARATE.98E,ref}
}% 
\htdef{Gamma49.qm}{%
\begin{ensuredisplaymath}
\htuse{Gamma49.gn} = \htuse{Gamma49.td}
\end{ensuredisplaymath}
 & \htuse{Gamma49.qt} & \hfagFitLabel}% 
\htdef{Gamma50.qm}{%
\begin{ensuredisplaymath}
\htuse{Gamma50.gn} = \htuse{Gamma50.td}
\end{ensuredisplaymath}
 & \htuse{Gamma50.qt} & \hfagFitLabel\\
\htuse{BaBar.Gamma50.pub.LEES.12Y,qt} & \htuse{BaBar.Gamma50.pub.LEES.12Y,exp} & \htuse{BaBar.Gamma50.pub.LEES.12Y,ref} \\
\htuse{Belle.Gamma50.pub.RYU.14vpc,qt} & \htuse{Belle.Gamma50.pub.RYU.14vpc,exp} & \htuse{Belle.Gamma50.pub.RYU.14vpc,ref}
}% 
\htdef{Gamma51.qm}{%
\begin{ensuredisplaymath}
\htuse{Gamma51.gn} = \htuse{Gamma51.td}
\end{ensuredisplaymath}
 & \htuse{Gamma51.qt} & \hfagFitLabel\\
\htuse{ALEPH.Gamma51.pub.BARATE.98E,qt} & \htuse{ALEPH.Gamma51.pub.BARATE.98E,exp} & \htuse{ALEPH.Gamma51.pub.BARATE.98E,ref}
}% 
\htdef{Gamma53.qm}{%
\begin{ensuredisplaymath}
\htuse{Gamma53.gn} = \htuse{Gamma53.td}
\end{ensuredisplaymath}
 & \htuse{Gamma53.qt} & \hfagFitLabel\\
\htuse{ALEPH.Gamma53.pub.BARATE.98E,qt} & \htuse{ALEPH.Gamma53.pub.BARATE.98E,exp} & \htuse{ALEPH.Gamma53.pub.BARATE.98E,ref}
}% 
\htdef{Gamma54.qm}{%
\begin{ensuredisplaymath}
\htuse{Gamma54.gn} = \htuse{Gamma54.td}
\end{ensuredisplaymath}
 & \htuse{Gamma54.qt} & \hfagFitLabel\\
\htuse{CELLO.Gamma54.pub.BEHREND.89B,qt} & \htuse{CELLO.Gamma54.pub.BEHREND.89B,exp} & \htuse{CELLO.Gamma54.pub.BEHREND.89B,ref} \\
\htuse{L3.Gamma54.pub.ADEVA.91F,qt} & \htuse{L3.Gamma54.pub.ADEVA.91F,exp} & \htuse{L3.Gamma54.pub.ADEVA.91F,ref} \\
\htuse{TPC.Gamma54.pub.AIHARA.87B,qt} & \htuse{TPC.Gamma54.pub.AIHARA.87B,exp} & \htuse{TPC.Gamma54.pub.AIHARA.87B,ref}
}% 
\htdef{Gamma55.qm}{%
\begin{ensuredisplaymath}
\htuse{Gamma55.gn} = \htuse{Gamma55.td}
\end{ensuredisplaymath}
 & \htuse{Gamma55.qt} & \hfagFitLabel\\
\htuse{L3.Gamma55.pub.ACHARD.01D,qt} & \htuse{L3.Gamma55.pub.ACHARD.01D,exp} & \htuse{L3.Gamma55.pub.ACHARD.01D,ref} \\
\htuse{OPAL.Gamma55.pub.AKERS.95Y,qt} & \htuse{OPAL.Gamma55.pub.AKERS.95Y,exp} & \htuse{OPAL.Gamma55.pub.AKERS.95Y,ref}
}% 
\htdef{Gamma56.qm}{%
\begin{ensuredisplaymath}
\htuse{Gamma56.gn} = \htuse{Gamma56.td}
\end{ensuredisplaymath}
 & \htuse{Gamma56.qt} & \hfagFitLabel}% 
\htdef{Gamma57.qm}{%
\begin{ensuredisplaymath}
\htuse{Gamma57.gn} = \htuse{Gamma57.td}
\end{ensuredisplaymath}
 & \htuse{Gamma57.qt} & \hfagFitLabel\\
\htuse{CLEO.Gamma57.pub.BALEST.95C,qt} & \htuse{CLEO.Gamma57.pub.BALEST.95C,exp} & \htuse{CLEO.Gamma57.pub.BALEST.95C,ref} \\
\htuse{DELPHI.Gamma57.pub.ABDALLAH.06A,qt} & \htuse{DELPHI.Gamma57.pub.ABDALLAH.06A,exp} & \htuse{DELPHI.Gamma57.pub.ABDALLAH.06A,ref}
}% 
\htdef{Gamma57by55.qm}{%
\begin{ensuredisplaymath}
\htuse{Gamma57by55.gn} = \htuse{Gamma57by55.td}
\end{ensuredisplaymath}
 & \htuse{Gamma57by55.qt} & \hfagFitLabel\\
\htuse{OPAL.Gamma57by55.pub.AKERS.95Y,qt} & \htuse{OPAL.Gamma57by55.pub.AKERS.95Y,exp} & \htuse{OPAL.Gamma57by55.pub.AKERS.95Y,ref}
}% 
\htdef{Gamma58.qm}{%
\begin{ensuredisplaymath}
\htuse{Gamma58.gn} = \htuse{Gamma58.td}
\end{ensuredisplaymath}
 & \htuse{Gamma58.qt} & \hfagFitLabel\\
\htuse{ALEPH.Gamma58.pub.SCHAEL.05C,qt} & \htuse{ALEPH.Gamma58.pub.SCHAEL.05C,exp} & \htuse{ALEPH.Gamma58.pub.SCHAEL.05C,ref}
}% 
\htdef{Gamma59.qm}{%
\begin{ensuredisplaymath}
\htuse{Gamma59.gn} = \htuse{Gamma59.td}
\end{ensuredisplaymath}
 & \htuse{Gamma59.qt} & \hfagFitLabel}% 
\htdef{Gamma60.qm}{%
\begin{ensuredisplaymath}
\htuse{Gamma60.gn} = \htuse{Gamma60.td}
\end{ensuredisplaymath}
 & \htuse{Gamma60.qt} & \hfagFitLabel\\
\htuse{BaBar.Gamma60.pub.AUBERT.08,qt} & \htuse{BaBar.Gamma60.pub.AUBERT.08,exp} & \htuse{BaBar.Gamma60.pub.AUBERT.08,ref} \\
\htuse{Belle.Gamma60.pub.LEE.10,qt} & \htuse{Belle.Gamma60.pub.LEE.10,exp} & \htuse{Belle.Gamma60.pub.LEE.10,ref} \\
\htuse{CLEO3.Gamma60.pub.BRIERE.03,qt} & \htuse{CLEO3.Gamma60.pub.BRIERE.03,exp} & \htuse{CLEO3.Gamma60.pub.BRIERE.03,ref}
}% 
\htdef{Gamma62.qm}{%
\begin{ensuredisplaymath}
\htuse{Gamma62.gn} = \htuse{Gamma62.td}
\end{ensuredisplaymath}
 & \htuse{Gamma62.qt} & \hfagFitLabel}% 
\htdef{Gamma63.qm}{%
\begin{ensuredisplaymath}
\htuse{Gamma63.gn} = \htuse{Gamma63.td}
\end{ensuredisplaymath}
 & \htuse{Gamma63.qt} & \hfagFitLabel}% 
\htdef{Gamma64.qm}{%
\begin{ensuredisplaymath}
\htuse{Gamma64.gn} = \htuse{Gamma64.td}
\end{ensuredisplaymath}
 & \htuse{Gamma64.qt} & \hfagFitLabel}% 
\htdef{Gamma65.qm}{%
\begin{ensuredisplaymath}
\htuse{Gamma65.gn} = \htuse{Gamma65.td}
\end{ensuredisplaymath}
 & \htuse{Gamma65.qt} & \hfagFitLabel}% 
\htdef{Gamma66.qm}{%
\begin{ensuredisplaymath}
\htuse{Gamma66.gn} = \htuse{Gamma66.td}
\end{ensuredisplaymath}
 & \htuse{Gamma66.qt} & \hfagFitLabel\\
\htuse{ALEPH.Gamma66.pub.SCHAEL.05C,qt} & \htuse{ALEPH.Gamma66.pub.SCHAEL.05C,exp} & \htuse{ALEPH.Gamma66.pub.SCHAEL.05C,ref} \\
\htuse{CLEO.Gamma66.pub.BALEST.95C,qt} & \htuse{CLEO.Gamma66.pub.BALEST.95C,exp} & \htuse{CLEO.Gamma66.pub.BALEST.95C,ref} \\
\htuse{DELPHI.Gamma66.pub.ABDALLAH.06A,qt} & \htuse{DELPHI.Gamma66.pub.ABDALLAH.06A,exp} & \htuse{DELPHI.Gamma66.pub.ABDALLAH.06A,ref}
}% 
\htdef{Gamma67.qm}{%
\begin{ensuredisplaymath}
\htuse{Gamma67.gn} = \htuse{Gamma67.td}
\end{ensuredisplaymath}
 & \htuse{Gamma67.qt} & \hfagFitLabel}% 
\htdef{Gamma68.qm}{%
\begin{ensuredisplaymath}
\htuse{Gamma68.gn} = \htuse{Gamma68.td}
\end{ensuredisplaymath}
 & \htuse{Gamma68.qt} & \hfagFitLabel}% 
\htdef{Gamma69.qm}{%
\begin{ensuredisplaymath}
\htuse{Gamma69.gn} = \htuse{Gamma69.td}
\end{ensuredisplaymath}
 & \htuse{Gamma69.qt} & \hfagFitLabel\\
\htuse{CLEO.Gamma69.pub.EDWARDS.00A,qt} & \htuse{CLEO.Gamma69.pub.EDWARDS.00A,exp} & \htuse{CLEO.Gamma69.pub.EDWARDS.00A,ref}
}% 
\htdef{Gamma70.qm}{%
\begin{ensuredisplaymath}
\htuse{Gamma70.gn} = \htuse{Gamma70.td}
\end{ensuredisplaymath}
 & \htuse{Gamma70.qt} & \hfagFitLabel}% 
\htdef{Gamma74.qm}{%
\begin{ensuredisplaymath}
\htuse{Gamma74.gn} = \htuse{Gamma74.td}
\end{ensuredisplaymath}
 & \htuse{Gamma74.qt} & \hfagFitLabel\\
\htuse{DELPHI.Gamma74.pub.ABDALLAH.06A,qt} & \htuse{DELPHI.Gamma74.pub.ABDALLAH.06A,exp} & \htuse{DELPHI.Gamma74.pub.ABDALLAH.06A,ref}
}% 
\htdef{Gamma75.qm}{%
\begin{ensuredisplaymath}
\htuse{Gamma75.gn} = \htuse{Gamma75.td}
\end{ensuredisplaymath}
 & \htuse{Gamma75.qt} & \hfagFitLabel}% 
\htdef{Gamma76.qm}{%
\begin{ensuredisplaymath}
\htuse{Gamma76.gn} = \htuse{Gamma76.td}
\end{ensuredisplaymath}
 & \htuse{Gamma76.qt} & \hfagFitLabel\\
\htuse{ALEPH.Gamma76.pub.SCHAEL.05C,qt} & \htuse{ALEPH.Gamma76.pub.SCHAEL.05C,exp} & \htuse{ALEPH.Gamma76.pub.SCHAEL.05C,ref}
}% 
\htdef{Gamma76by54.qm}{%
\begin{ensuredisplaymath}
\htuse{Gamma76by54.gn} = \htuse{Gamma76by54.td}
\end{ensuredisplaymath}
 & \htuse{Gamma76by54.qt} & \hfagFitLabel\\
\htuse{CLEO.Gamma76by54.pub.BORTOLETTO.93,qt} & \htuse{CLEO.Gamma76by54.pub.BORTOLETTO.93,exp} & \htuse{CLEO.Gamma76by54.pub.BORTOLETTO.93,ref}
}% 
\htdef{Gamma77.qm}{%
\begin{ensuredisplaymath}
\htuse{Gamma77.gn} = \htuse{Gamma77.td}
\end{ensuredisplaymath}
 & \htuse{Gamma77.qt} & \hfagFitLabel}% 
\htdef{Gamma78.qm}{%
\begin{ensuredisplaymath}
\htuse{Gamma78.gn} = \htuse{Gamma78.td}
\end{ensuredisplaymath}
 & \htuse{Gamma78.qt} & \hfagFitLabel\\
\htuse{CLEO.Gamma78.pub.ANASTASSOV.01,qt} & \htuse{CLEO.Gamma78.pub.ANASTASSOV.01,exp} & \htuse{CLEO.Gamma78.pub.ANASTASSOV.01,ref}
}% 
\htdef{Gamma79.qm}{%
\begin{ensuredisplaymath}
\htuse{Gamma79.gn} = \htuse{Gamma79.td}
\end{ensuredisplaymath}
 & \htuse{Gamma79.qt} & \hfagFitLabel}% 
\htdef{Gamma80.qm}{%
\begin{ensuredisplaymath}
\htuse{Gamma80.gn} = \htuse{Gamma80.td}
\end{ensuredisplaymath}
 & \htuse{Gamma80.qt} & \hfagFitLabel}% 
\htdef{Gamma80by60.qm}{%
\begin{ensuredisplaymath}
\htuse{Gamma80by60.gn} = \htuse{Gamma80by60.td}
\end{ensuredisplaymath}
 & \htuse{Gamma80by60.qt} & \hfagFitLabel\\
\htuse{CLEO.Gamma80by60.pub.RICHICHI.99,qt} & \htuse{CLEO.Gamma80by60.pub.RICHICHI.99,exp} & \htuse{CLEO.Gamma80by60.pub.RICHICHI.99,ref}
}% 
\htdef{Gamma81.qm}{%
\begin{ensuredisplaymath}
\htuse{Gamma81.gn} = \htuse{Gamma81.td}
\end{ensuredisplaymath}
 & \htuse{Gamma81.qt} & \hfagFitLabel}% 
\htdef{Gamma81by69.qm}{%
\begin{ensuredisplaymath}
\htuse{Gamma81by69.gn} = \htuse{Gamma81by69.td}
\end{ensuredisplaymath}
 & \htuse{Gamma81by69.qt} & \hfagFitLabel\\
\htuse{CLEO.Gamma81by69.pub.RICHICHI.99,qt} & \htuse{CLEO.Gamma81by69.pub.RICHICHI.99,exp} & \htuse{CLEO.Gamma81by69.pub.RICHICHI.99,ref}
}% 
\htdef{Gamma82.qm}{%
\begin{ensuredisplaymath}
\htuse{Gamma82.gn} = \htuse{Gamma82.td}
\end{ensuredisplaymath}
 & \htuse{Gamma82.qt} & \hfagFitLabel\\
\htuse{TPC.Gamma82.pub.BAUER.94,qt} & \htuse{TPC.Gamma82.pub.BAUER.94,exp} & \htuse{TPC.Gamma82.pub.BAUER.94,ref}
}% 
\htdef{Gamma83.qm}{%
\begin{ensuredisplaymath}
\htuse{Gamma83.gn} = \htuse{Gamma83.td}
\end{ensuredisplaymath}
 & \htuse{Gamma83.qt} & \hfagFitLabel}% 
\htdef{Gamma84.qm}{%
\begin{ensuredisplaymath}
\htuse{Gamma84.gn} = \htuse{Gamma84.td}
\end{ensuredisplaymath}
 & \htuse{Gamma84.qt} & \hfagFitLabel}% 
\htdef{Gamma85.qm}{%
\begin{ensuredisplaymath}
\htuse{Gamma85.gn} = \htuse{Gamma85.td}
\end{ensuredisplaymath}
 & \htuse{Gamma85.qt} & \hfagFitLabel\\
\htuse{ALEPH.Gamma85.pub.BARATE.98,qt} & \htuse{ALEPH.Gamma85.pub.BARATE.98,exp} & \htuse{ALEPH.Gamma85.pub.BARATE.98,ref} \\
\htuse{BaBar.Gamma85.pub.AUBERT.08,qt} & \htuse{BaBar.Gamma85.pub.AUBERT.08,exp} & \htuse{BaBar.Gamma85.pub.AUBERT.08,ref} \\
\htuse{Belle.Gamma85.pub.LEE.10,qt} & \htuse{Belle.Gamma85.pub.LEE.10,exp} & \htuse{Belle.Gamma85.pub.LEE.10,ref} \\
\htuse{CLEO3.Gamma85.pub.BRIERE.03,qt} & \htuse{CLEO3.Gamma85.pub.BRIERE.03,exp} & \htuse{CLEO3.Gamma85.pub.BRIERE.03,ref} \\
\htuse{OPAL.Gamma85.pub.ABBIENDI.04J,qt} & \htuse{OPAL.Gamma85.pub.ABBIENDI.04J,exp} & \htuse{OPAL.Gamma85.pub.ABBIENDI.04J,ref}
}% 
\htdef{Gamma85by60.qm}{%
\begin{ensuredisplaymath}
\htuse{Gamma85by60.gn} = \htuse{Gamma85by60.td}
\end{ensuredisplaymath}
 & \htuse{Gamma85by60.qt} & \hfagFitLabel}% 
\htdef{Gamma87.qm}{%
\begin{ensuredisplaymath}
\htuse{Gamma87.gn} = \htuse{Gamma87.td}
\end{ensuredisplaymath}
 & \htuse{Gamma87.qt} & \hfagFitLabel}% 
\htdef{Gamma88.qm}{%
\begin{ensuredisplaymath}
\htuse{Gamma88.gn} = \htuse{Gamma88.td}
\end{ensuredisplaymath}
 & \htuse{Gamma88.qt} & \hfagFitLabel\\
\htuse{ALEPH.Gamma88.pub.BARATE.98,qt} & \htuse{ALEPH.Gamma88.pub.BARATE.98,exp} & \htuse{ALEPH.Gamma88.pub.BARATE.98,ref} \\
\htuse{CLEO3.Gamma88.pub.ARMS.05,qt} & \htuse{CLEO3.Gamma88.pub.ARMS.05,exp} & \htuse{CLEO3.Gamma88.pub.ARMS.05,ref}
}% 
\htdef{Gamma89.qm}{%
\begin{ensuredisplaymath}
\htuse{Gamma89.gn} = \htuse{Gamma89.td}
\end{ensuredisplaymath}
 & \htuse{Gamma89.qt} & \hfagFitLabel}% 
\htdef{Gamma92.qm}{%
\begin{ensuredisplaymath}
\htuse{Gamma92.gn} = \htuse{Gamma92.td}
\end{ensuredisplaymath}
 & \htuse{Gamma92.qt} & \hfagFitLabel\\
\htuse{OPAL.Gamma92.pub.ABBIENDI.00D,qt} & \htuse{OPAL.Gamma92.pub.ABBIENDI.00D,exp} & \htuse{OPAL.Gamma92.pub.ABBIENDI.00D,ref} \\
\htuse{TPC.Gamma92.pub.BAUER.94,qt} & \htuse{TPC.Gamma92.pub.BAUER.94,exp} & \htuse{TPC.Gamma92.pub.BAUER.94,ref}
}% 
\htdef{Gamma93.qm}{%
\begin{ensuredisplaymath}
\htuse{Gamma93.gn} = \htuse{Gamma93.td}
\end{ensuredisplaymath}
 & \htuse{Gamma93.qt} & \hfagFitLabel\\
\htuse{ALEPH.Gamma93.pub.BARATE.98,qt} & \htuse{ALEPH.Gamma93.pub.BARATE.98,exp} & \htuse{ALEPH.Gamma93.pub.BARATE.98,ref} \\
\htuse{BaBar.Gamma93.pub.AUBERT.08,qt} & \htuse{BaBar.Gamma93.pub.AUBERT.08,exp} & \htuse{BaBar.Gamma93.pub.AUBERT.08,ref} \\
\htuse{Belle.Gamma93.pub.LEE.10,qt} & \htuse{Belle.Gamma93.pub.LEE.10,exp} & \htuse{Belle.Gamma93.pub.LEE.10,ref} \\
\htuse{CLEO3.Gamma93.pub.BRIERE.03,qt} & \htuse{CLEO3.Gamma93.pub.BRIERE.03,exp} & \htuse{CLEO3.Gamma93.pub.BRIERE.03,ref}
}% 
\htdef{Gamma93by60.qm}{%
\begin{ensuredisplaymath}
\htuse{Gamma93by60.gn} = \htuse{Gamma93by60.td}
\end{ensuredisplaymath}
 & \htuse{Gamma93by60.qt} & \hfagFitLabel\\
\htuse{CLEO.Gamma93by60.pub.RICHICHI.99,qt} & \htuse{CLEO.Gamma93by60.pub.RICHICHI.99,exp} & \htuse{CLEO.Gamma93by60.pub.RICHICHI.99,ref}
}% 
\htdef{Gamma94.qm}{%
\begin{ensuredisplaymath}
\htuse{Gamma94.gn} = \htuse{Gamma94.td}
\end{ensuredisplaymath}
 & \htuse{Gamma94.qt} & \hfagFitLabel\\
\htuse{ALEPH.Gamma94.pub.BARATE.98,qt} & \htuse{ALEPH.Gamma94.pub.BARATE.98,exp} & \htuse{ALEPH.Gamma94.pub.BARATE.98,ref} \\
\htuse{CLEO3.Gamma94.pub.ARMS.05,qt} & \htuse{CLEO3.Gamma94.pub.ARMS.05,exp} & \htuse{CLEO3.Gamma94.pub.ARMS.05,ref}
}% 
\htdef{Gamma94by69.qm}{%
\begin{ensuredisplaymath}
\htuse{Gamma94by69.gn} = \htuse{Gamma94by69.td}
\end{ensuredisplaymath}
 & \htuse{Gamma94by69.qt} & \hfagFitLabel\\
\htuse{CLEO.Gamma94by69.pub.RICHICHI.99,qt} & \htuse{CLEO.Gamma94by69.pub.RICHICHI.99,exp} & \htuse{CLEO.Gamma94by69.pub.RICHICHI.99,ref}
}% 
\htdef{Gamma96.qm}{%
\begin{ensuredisplaymath}
\htuse{Gamma96.gn} = \htuse{Gamma96.td}
\end{ensuredisplaymath}
 & \htuse{Gamma96.qt} & \hfagFitLabel\\
\htuse{BaBar.Gamma96.pub.AUBERT.08,qt} & \htuse{BaBar.Gamma96.pub.AUBERT.08,exp} & \htuse{BaBar.Gamma96.pub.AUBERT.08,ref} \\
\htuse{Belle.Gamma96.pub.LEE.10,qt} & \htuse{Belle.Gamma96.pub.LEE.10,exp} & \htuse{Belle.Gamma96.pub.LEE.10,ref}
}% 
\htdef{Gamma102.qm}{%
\begin{ensuredisplaymath}
\htuse{Gamma102.gn} = \htuse{Gamma102.td}
\end{ensuredisplaymath}
 & \htuse{Gamma102.qt} & \hfagFitLabel\\
\htuse{CLEO.Gamma102.pub.GIBAUT.94B,qt} & \htuse{CLEO.Gamma102.pub.GIBAUT.94B,exp} & \htuse{CLEO.Gamma102.pub.GIBAUT.94B,ref} \\
\htuse{HRS.Gamma102.pub.BYLSMA.87,qt} & \htuse{HRS.Gamma102.pub.BYLSMA.87,exp} & \htuse{HRS.Gamma102.pub.BYLSMA.87,ref} \\
\htuse{L3.Gamma102.pub.ACHARD.01D,qt} & \htuse{L3.Gamma102.pub.ACHARD.01D,exp} & \htuse{L3.Gamma102.pub.ACHARD.01D,ref}
}% 
\htdef{Gamma103.qm}{%
\begin{ensuredisplaymath}
\htuse{Gamma103.gn} = \htuse{Gamma103.td}
\end{ensuredisplaymath}
 & \htuse{Gamma103.qt} & \hfagFitLabel\\
\htuse{ALEPH.Gamma103.pub.SCHAEL.05C,qt} & \htuse{ALEPH.Gamma103.pub.SCHAEL.05C,exp} & \htuse{ALEPH.Gamma103.pub.SCHAEL.05C,ref} \\
\htuse{ARGUS.Gamma103.pub.ALBRECHT.88B,qt} & \htuse{ARGUS.Gamma103.pub.ALBRECHT.88B,exp} & \htuse{ARGUS.Gamma103.pub.ALBRECHT.88B,ref} \\
\htuse{CLEO.Gamma103.pub.GIBAUT.94B,qt} & \htuse{CLEO.Gamma103.pub.GIBAUT.94B,exp} & \htuse{CLEO.Gamma103.pub.GIBAUT.94B,ref} \\
\htuse{DELPHI.Gamma103.pub.ABDALLAH.06A,qt} & \htuse{DELPHI.Gamma103.pub.ABDALLAH.06A,exp} & \htuse{DELPHI.Gamma103.pub.ABDALLAH.06A,ref} \\
\htuse{HRS.Gamma103.pub.BYLSMA.87,qt} & \htuse{HRS.Gamma103.pub.BYLSMA.87,exp} & \htuse{HRS.Gamma103.pub.BYLSMA.87,ref} \\
\htuse{OPAL.Gamma103.pub.ACKERSTAFF.99E,qt} & \htuse{OPAL.Gamma103.pub.ACKERSTAFF.99E,exp} & \htuse{OPAL.Gamma103.pub.ACKERSTAFF.99E,ref}
}% 
\htdef{Gamma104.qm}{%
\begin{ensuredisplaymath}
\htuse{Gamma104.gn} = \htuse{Gamma104.td}
\end{ensuredisplaymath}
 & \htuse{Gamma104.qt} & \hfagFitLabel\\
\htuse{ALEPH.Gamma104.pub.SCHAEL.05C,qt} & \htuse{ALEPH.Gamma104.pub.SCHAEL.05C,exp} & \htuse{ALEPH.Gamma104.pub.SCHAEL.05C,ref} \\
\htuse{CLEO.Gamma104.pub.ANASTASSOV.01,qt} & \htuse{CLEO.Gamma104.pub.ANASTASSOV.01,exp} & \htuse{CLEO.Gamma104.pub.ANASTASSOV.01,ref} \\
\htuse{DELPHI.Gamma104.pub.ABDALLAH.06A,qt} & \htuse{DELPHI.Gamma104.pub.ABDALLAH.06A,exp} & \htuse{DELPHI.Gamma104.pub.ABDALLAH.06A,ref} \\
\htuse{OPAL.Gamma104.pub.ACKERSTAFF.99E,qt} & \htuse{OPAL.Gamma104.pub.ACKERSTAFF.99E,exp} & \htuse{OPAL.Gamma104.pub.ACKERSTAFF.99E,ref}
}% 
\htdef{Gamma106.qm}{%
\begin{ensuredisplaymath}
\htuse{Gamma106.gn} = \htuse{Gamma106.td}
\end{ensuredisplaymath}
 & \htuse{Gamma106.qt} & \hfagFitLabel}% 
\htdef{Gamma110.qm}{%
\begin{ensuredisplaymath}
\htuse{Gamma110.gn} = \htuse{Gamma110.td}
\end{ensuredisplaymath}
 & \htuse{Gamma110.qt} & \hfagFitLabel}% 
\htdef{Gamma126.qm}{%
\begin{ensuredisplaymath}
\htuse{Gamma126.gn} = \htuse{Gamma126.td}
\end{ensuredisplaymath}
 & \htuse{Gamma126.qt} & \hfagFitLabel\\
\htuse{ALEPH.Gamma126.pub.BUSKULIC.97C,qt} & \htuse{ALEPH.Gamma126.pub.BUSKULIC.97C,exp} & \htuse{ALEPH.Gamma126.pub.BUSKULIC.97C,ref} \\
\htuse{Belle.Gamma126.pub.INAMI.09,qt} & \htuse{Belle.Gamma126.pub.INAMI.09,exp} & \htuse{Belle.Gamma126.pub.INAMI.09,ref} \\
\htuse{CLEO.Gamma126.pub.ARTUSO.92,qt} & \htuse{CLEO.Gamma126.pub.ARTUSO.92,exp} & \htuse{CLEO.Gamma126.pub.ARTUSO.92,ref}
}% 
\htdef{Gamma128.qm}{%
\begin{ensuredisplaymath}
\htuse{Gamma128.gn} = \htuse{Gamma128.td}
\end{ensuredisplaymath}
 & \htuse{Gamma128.qt} & \hfagFitLabel\\
\htuse{ALEPH.Gamma128.pub.BUSKULIC.97C,qt} & \htuse{ALEPH.Gamma128.pub.BUSKULIC.97C,exp} & \htuse{ALEPH.Gamma128.pub.BUSKULIC.97C,ref} \\
\htuse{BaBar.Gamma128.pub.DEL-AMO-SANCHEZ.11E,qt} & \htuse{BaBar.Gamma128.pub.DEL-AMO-SANCHEZ.11E,exp} & \htuse{BaBar.Gamma128.pub.DEL-AMO-SANCHEZ.11E,ref} \\
\htuse{Belle.Gamma128.pub.INAMI.09,qt} & \htuse{Belle.Gamma128.pub.INAMI.09,exp} & \htuse{Belle.Gamma128.pub.INAMI.09,ref} \\
\htuse{CLEO.Gamma128.pub.BARTELT.96,qt} & \htuse{CLEO.Gamma128.pub.BARTELT.96,exp} & \htuse{CLEO.Gamma128.pub.BARTELT.96,ref}
}% 
\htdef{Gamma130.qm}{%
\begin{ensuredisplaymath}
\htuse{Gamma130.gn} = \htuse{Gamma130.td}
\end{ensuredisplaymath}
 & \htuse{Gamma130.qt} & \hfagFitLabel\\
\htuse{Belle.Gamma130.pub.INAMI.09,qt} & \htuse{Belle.Gamma130.pub.INAMI.09,exp} & \htuse{Belle.Gamma130.pub.INAMI.09,ref} \\
\htuse{CLEO.Gamma130.pub.BISHAI.99,qt} & \htuse{CLEO.Gamma130.pub.BISHAI.99,exp} & \htuse{CLEO.Gamma130.pub.BISHAI.99,ref}
}% 
\htdef{Gamma132.qm}{%
\begin{ensuredisplaymath}
\htuse{Gamma132.gn} = \htuse{Gamma132.td}
\end{ensuredisplaymath}
 & \htuse{Gamma132.qt} & \hfagFitLabel\\
\htuse{Belle.Gamma132.pub.INAMI.09,qt} & \htuse{Belle.Gamma132.pub.INAMI.09,exp} & \htuse{Belle.Gamma132.pub.INAMI.09,ref} \\
\htuse{CLEO.Gamma132.pub.BISHAI.99,qt} & \htuse{CLEO.Gamma132.pub.BISHAI.99,exp} & \htuse{CLEO.Gamma132.pub.BISHAI.99,ref}
}% 
\htdef{Gamma136.qm}{%
\begin{ensuredisplaymath}
\htuse{Gamma136.gn} = \htuse{Gamma136.td}
\end{ensuredisplaymath}
 & \htuse{Gamma136.qt} & \hfagFitLabel}% 
\htdef{Gamma149.qm}{%
\begin{ensuredisplaymath}
\htuse{Gamma149.gn} = \htuse{Gamma149.td}
\end{ensuredisplaymath}
 & \htuse{Gamma149.qt} & \hfagFitLabel}% 
\htdef{Gamma150.qm}{%
\begin{ensuredisplaymath}
\htuse{Gamma150.gn} = \htuse{Gamma150.td}
\end{ensuredisplaymath}
 & \htuse{Gamma150.qt} & \hfagFitLabel\\
\htuse{ALEPH.Gamma150.pub.BUSKULIC.97C,qt} & \htuse{ALEPH.Gamma150.pub.BUSKULIC.97C,exp} & \htuse{ALEPH.Gamma150.pub.BUSKULIC.97C,ref} \\
\htuse{CLEO.Gamma150.pub.BARINGER.87,qt} & \htuse{CLEO.Gamma150.pub.BARINGER.87,exp} & \htuse{CLEO.Gamma150.pub.BARINGER.87,ref}
}% 
\htdef{Gamma150by66.qm}{%
\begin{ensuredisplaymath}
\htuse{Gamma150by66.gn} = \htuse{Gamma150by66.td}
\end{ensuredisplaymath}
 & \htuse{Gamma150by66.qt} & \hfagFitLabel\\
\htuse{ALEPH.Gamma150by66.pub.BUSKULIC.96,qt} & \htuse{ALEPH.Gamma150by66.pub.BUSKULIC.96,exp} & \htuse{ALEPH.Gamma150by66.pub.BUSKULIC.96,ref} \\
\htuse{CLEO.Gamma150by66.pub.BALEST.95C,qt} & \htuse{CLEO.Gamma150by66.pub.BALEST.95C,exp} & \htuse{CLEO.Gamma150by66.pub.BALEST.95C,ref}
}% 
\htdef{Gamma151.qm}{%
\begin{ensuredisplaymath}
\htuse{Gamma151.gn} = \htuse{Gamma151.td}
\end{ensuredisplaymath}
 & \htuse{Gamma151.qt} & \hfagFitLabel\\
\htuse{CLEO3.Gamma151.pub.ARMS.05,qt} & \htuse{CLEO3.Gamma151.pub.ARMS.05,exp} & \htuse{CLEO3.Gamma151.pub.ARMS.05,ref}
}% 
\htdef{Gamma152.qm}{%
\begin{ensuredisplaymath}
\htuse{Gamma152.gn} = \htuse{Gamma152.td}
\end{ensuredisplaymath}
 & \htuse{Gamma152.qt} & \hfagFitLabel\\
\htuse{ALEPH.Gamma152.pub.BUSKULIC.97C,qt} & \htuse{ALEPH.Gamma152.pub.BUSKULIC.97C,exp} & \htuse{ALEPH.Gamma152.pub.BUSKULIC.97C,ref}
}% 
\htdef{Gamma152by54.qm}{%
\begin{ensuredisplaymath}
\htuse{Gamma152by54.gn} = \htuse{Gamma152by54.td}
\end{ensuredisplaymath}
 & \htuse{Gamma152by54.qt} & \hfagFitLabel}% 
\htdef{Gamma152by76.qm}{%
\begin{ensuredisplaymath}
\htuse{Gamma152by76.gn} = \htuse{Gamma152by76.td}
\end{ensuredisplaymath}
 & \htuse{Gamma152by76.qt} & \hfagFitLabel\\
\htuse{CLEO.Gamma152by76.pub.BORTOLETTO.93,qt} & \htuse{CLEO.Gamma152by76.pub.BORTOLETTO.93,exp} & \htuse{CLEO.Gamma152by76.pub.BORTOLETTO.93,ref}
}% 
\htdef{Gamma167.qm}{%
\begin{ensuredisplaymath}
\htuse{Gamma167.gn} = \htuse{Gamma167.td}
\end{ensuredisplaymath}
 & \htuse{Gamma167.qt} & \hfagFitLabel}% 
\htdef{Gamma168.qm}{%
\begin{ensuredisplaymath}
\htuse{Gamma168.gn} = \htuse{Gamma168.td}
\end{ensuredisplaymath}
 & \htuse{Gamma168.qt} & \hfagFitLabel}% 
\htdef{Gamma169.qm}{%
\begin{ensuredisplaymath}
\htuse{Gamma169.gn} = \htuse{Gamma169.td}
\end{ensuredisplaymath}
 & \htuse{Gamma169.qt} & \hfagFitLabel}% 
\htdef{Gamma800.qm}{%
\begin{ensuredisplaymath}
\htuse{Gamma800.gn} = \htuse{Gamma800.td}
\end{ensuredisplaymath}
 & \htuse{Gamma800.qt} & \hfagFitLabel}% 
\htdef{Gamma802.qm}{%
\begin{ensuredisplaymath}
\htuse{Gamma802.gn} = \htuse{Gamma802.td}
\end{ensuredisplaymath}
 & \htuse{Gamma802.qt} & \hfagFitLabel}% 
\htdef{Gamma803.qm}{%
\begin{ensuredisplaymath}
\htuse{Gamma803.gn} = \htuse{Gamma803.td}
\end{ensuredisplaymath}
 & \htuse{Gamma803.qt} & \hfagFitLabel}% 
\htdef{Gamma804.qm}{%
\begin{ensuredisplaymath}
\htuse{Gamma804.gn} = \htuse{Gamma804.td}
\end{ensuredisplaymath}
 & \htuse{Gamma804.qt} & \hfagFitLabel}% 
\htdef{Gamma805.qm}{%
\begin{ensuredisplaymath}
\htuse{Gamma805.gn} = \htuse{Gamma805.td}
\end{ensuredisplaymath}
 & \htuse{Gamma805.qt} & \hfagFitLabel\\
\htuse{ALEPH.Gamma805.pub.SCHAEL.05C,qt} & \htuse{ALEPH.Gamma805.pub.SCHAEL.05C,exp} & \htuse{ALEPH.Gamma805.pub.SCHAEL.05C,ref}
}% 
\htdef{Gamma806.qm}{%
\begin{ensuredisplaymath}
\htuse{Gamma806.gn} = \htuse{Gamma806.td}
\end{ensuredisplaymath}
 & \htuse{Gamma806.qt} & \hfagFitLabel}% 
\htdef{Gamma810.qm}{%
\begin{ensuredisplaymath}
\htuse{Gamma810.gn} = \htuse{Gamma810.td}
\end{ensuredisplaymath}
 & \htuse{Gamma810.qt} & \hfagFitLabel}% 
\htdef{Gamma811.qm}{%
\begin{ensuredisplaymath}
\htuse{Gamma811.gn} = \htuse{Gamma811.td}
\end{ensuredisplaymath}
 & \htuse{Gamma811.qt} & \hfagFitLabel\\
\htuse{BaBar.Gamma811.pub.LEES.12X,qt} & \htuse{BaBar.Gamma811.pub.LEES.12X,exp} & \htuse{BaBar.Gamma811.pub.LEES.12X,ref}
}% 
\htdef{Gamma812.qm}{%
\begin{ensuredisplaymath}
\htuse{Gamma812.gn} = \htuse{Gamma812.td}
\end{ensuredisplaymath}
 & \htuse{Gamma812.qt} & \hfagFitLabel\\
\htuse{BaBar.Gamma812.pub.LEES.12X,qt} & \htuse{BaBar.Gamma812.pub.LEES.12X,exp} & \htuse{BaBar.Gamma812.pub.LEES.12X,ref}
}% 
\htdef{Gamma820.qm}{%
\begin{ensuredisplaymath}
\htuse{Gamma820.gn} = \htuse{Gamma820.td}
\end{ensuredisplaymath}
 & \htuse{Gamma820.qt} & \hfagFitLabel}% 
\htdef{Gamma821.qm}{%
\begin{ensuredisplaymath}
\htuse{Gamma821.gn} = \htuse{Gamma821.td}
\end{ensuredisplaymath}
 & \htuse{Gamma821.qt} & \hfagFitLabel\\
\htuse{BaBar.Gamma821.pub.LEES.12X,qt} & \htuse{BaBar.Gamma821.pub.LEES.12X,exp} & \htuse{BaBar.Gamma821.pub.LEES.12X,ref}
}% 
\htdef{Gamma822.qm}{%
\begin{ensuredisplaymath}
\htuse{Gamma822.gn} = \htuse{Gamma822.td}
\end{ensuredisplaymath}
 & \htuse{Gamma822.qt} & \hfagFitLabel\\
\htuse{BaBar.Gamma822.pub.LEES.12X,qt} & \htuse{BaBar.Gamma822.pub.LEES.12X,exp} & \htuse{BaBar.Gamma822.pub.LEES.12X,ref}
}% 
\htdef{Gamma830.qm}{%
\begin{ensuredisplaymath}
\htuse{Gamma830.gn} = \htuse{Gamma830.td}
\end{ensuredisplaymath}
 & \htuse{Gamma830.qt} & \hfagFitLabel}% 
\htdef{Gamma831.qm}{%
\begin{ensuredisplaymath}
\htuse{Gamma831.gn} = \htuse{Gamma831.td}
\end{ensuredisplaymath}
 & \htuse{Gamma831.qt} & \hfagFitLabel\\
\htuse{BaBar.Gamma831.pub.LEES.12X,qt} & \htuse{BaBar.Gamma831.pub.LEES.12X,exp} & \htuse{BaBar.Gamma831.pub.LEES.12X,ref}
}% 
\htdef{Gamma832.qm}{%
\begin{ensuredisplaymath}
\htuse{Gamma832.gn} = \htuse{Gamma832.td}
\end{ensuredisplaymath}
 & \htuse{Gamma832.qt} & \hfagFitLabel\\
\htuse{BaBar.Gamma832.pub.LEES.12X,qt} & \htuse{BaBar.Gamma832.pub.LEES.12X,exp} & \htuse{BaBar.Gamma832.pub.LEES.12X,ref}
}% 
\htdef{Gamma833.qm}{%
\begin{ensuredisplaymath}
\htuse{Gamma833.gn} = \htuse{Gamma833.td}
\end{ensuredisplaymath}
 & \htuse{Gamma833.qt} & \hfagFitLabel\\
\htuse{BaBar.Gamma833.pub.LEES.12X,qt} & \htuse{BaBar.Gamma833.pub.LEES.12X,exp} & \htuse{BaBar.Gamma833.pub.LEES.12X,ref}
}% 
\htdef{Gamma910.qm}{%
\begin{ensuredisplaymath}
\htuse{Gamma910.gn} = \htuse{Gamma910.td}
\end{ensuredisplaymath}
 & \htuse{Gamma910.qt} & \hfagFitLabel\\
\htuse{BaBar.Gamma910.pub.LEES.12X,qt} & \htuse{BaBar.Gamma910.pub.LEES.12X,exp} & \htuse{BaBar.Gamma910.pub.LEES.12X,ref}
}% 
\htdef{Gamma911.qm}{%
\begin{ensuredisplaymath}
\htuse{Gamma911.gn} = \htuse{Gamma911.td}
\end{ensuredisplaymath}
 & \htuse{Gamma911.qt} & \hfagFitLabel\\
\htuse{BaBar.Gamma911.pub.LEES.12X,qt} & \htuse{BaBar.Gamma911.pub.LEES.12X,exp} & \htuse{BaBar.Gamma911.pub.LEES.12X,ref}
}% 
\htdef{Gamma920.qm}{%
\begin{ensuredisplaymath}
\htuse{Gamma920.gn} = \htuse{Gamma920.td}
\end{ensuredisplaymath}
 & \htuse{Gamma920.qt} & \hfagFitLabel\\
\htuse{BaBar.Gamma920.pub.LEES.12X,qt} & \htuse{BaBar.Gamma920.pub.LEES.12X,exp} & \htuse{BaBar.Gamma920.pub.LEES.12X,ref}
}% 
\htdef{Gamma930.qm}{%
\begin{ensuredisplaymath}
\htuse{Gamma930.gn} = \htuse{Gamma930.td}
\end{ensuredisplaymath}
 & \htuse{Gamma930.qt} & \hfagFitLabel\\
\htuse{BaBar.Gamma930.pub.LEES.12X,qt} & \htuse{BaBar.Gamma930.pub.LEES.12X,exp} & \htuse{BaBar.Gamma930.pub.LEES.12X,ref}
}% 
\htdef{Gamma944.qm}{%
\begin{ensuredisplaymath}
\htuse{Gamma944.gn} = \htuse{Gamma944.td}
\end{ensuredisplaymath}
 & \htuse{Gamma944.qt} & \hfagFitLabel\\
\htuse{BaBar.Gamma944.pub.LEES.12X,qt} & \htuse{BaBar.Gamma944.pub.LEES.12X,exp} & \htuse{BaBar.Gamma944.pub.LEES.12X,ref}
}% 
\htdef{Gamma945.qm}{%
\begin{ensuredisplaymath}
\htuse{Gamma945.gn} = \htuse{Gamma945.td}
\end{ensuredisplaymath}
 & \htuse{Gamma945.qt} & \hfagFitLabel}% 
\htdef{Gamma998.qm}{%
\begin{ensuredisplaymath}
\htuse{Gamma998.gn} = \htuse{Gamma998.td}
\end{ensuredisplaymath}
 & \htuse{Gamma998.qt} & \hfagFitLabel}%
\htdef{BrVal}{%
\htuse{Gamma1.qm} \\
\midrule
\htuse{Gamma2.qm} \\
\midrule
\htuse{Gamma3.qm} \\
\midrule
\htuse{Gamma3by5.qm} \\
\midrule
\htuse{Gamma5.qm} \\
\midrule
\htuse{Gamma7.qm} \\
\midrule
\htuse{Gamma8.qm} \\
\midrule
\htuse{Gamma8by5.qm} \\
\midrule
\htuse{Gamma9.qm} \\
\midrule
\htuse{Gamma9by5.qm} \\
\midrule
\htuse{Gamma10.qm} \\
\midrule
\htuse{Gamma10by5.qm} \\
\midrule
\htuse{Gamma10by9.qm} \\
\midrule
\htuse{Gamma11.qm} \\
\midrule
\htuse{Gamma12.qm} \\
\midrule
\htuse{Gamma13.qm} \\
\midrule
\htuse{Gamma14.qm} \\
\midrule
\htuse{Gamma16.qm} \\
\midrule
\htuse{Gamma17.qm} \\
\midrule
\htuse{Gamma18.qm} \\
\midrule
\htuse{Gamma19.qm} \\
\midrule
\htuse{Gamma19by13.qm} \\
\midrule
\htuse{Gamma20.qm} \\
\midrule
\htuse{Gamma23.qm} \\
\midrule
\htuse{Gamma24.qm} \\
\midrule
\htuse{Gamma25.qm} \\
\midrule
\htuse{Gamma26.qm} \\
\midrule
\htuse{Gamma26by13.qm} \\
\midrule
\htuse{Gamma27.qm} \\
\midrule
\htuse{Gamma28.qm} \\
\midrule
\htuse{Gamma29.qm} \\
\midrule
\htuse{Gamma30.qm} \\
\midrule
\htuse{Gamma31.qm} \\
\midrule
\htuse{Gamma32.qm} \\
\midrule
\htuse{Gamma33.qm} \\
\midrule
\htuse{Gamma34.qm} \\
\midrule
\htuse{Gamma35.qm} \\
\midrule
\htuse{Gamma37.qm} \\
\midrule
\htuse{Gamma38.qm} \\
\midrule
\htuse{Gamma39.qm} \\
\midrule
\htuse{Gamma40.qm} \\
\midrule
\htuse{Gamma42.qm} \\
\midrule
\htuse{Gamma43.qm} \\
\midrule
\htuse{Gamma44.qm} \\
\midrule
\htuse{Gamma46.qm} \\
\midrule
\htuse{Gamma47.qm} \\
\midrule
\htuse{Gamma48.qm} \\
\midrule
\htuse{Gamma49.qm} \\
\midrule
\htuse{Gamma50.qm} \\
\midrule
\htuse{Gamma51.qm} \\
\midrule
\htuse{Gamma53.qm} \\
\midrule
\htuse{Gamma54.qm} \\
\midrule
\htuse{Gamma55.qm} \\
\midrule
\htuse{Gamma56.qm} \\
\midrule
\htuse{Gamma57.qm} \\
\midrule
\htuse{Gamma57by55.qm} \\
\midrule
\htuse{Gamma58.qm} \\
\midrule
\htuse{Gamma59.qm} \\
\midrule
\htuse{Gamma60.qm} \\
\midrule
\htuse{Gamma62.qm} \\
\midrule
\htuse{Gamma63.qm} \\
\midrule
\htuse{Gamma64.qm} \\
\midrule
\htuse{Gamma65.qm} \\
\midrule
\htuse{Gamma66.qm} \\
\midrule
\htuse{Gamma67.qm} \\
\midrule
\htuse{Gamma68.qm} \\
\midrule
\htuse{Gamma69.qm} \\
\midrule
\htuse{Gamma70.qm} \\
\midrule
\htuse{Gamma74.qm} \\
\midrule
\htuse{Gamma75.qm} \\
\midrule
\htuse{Gamma76.qm} \\
\midrule
\htuse{Gamma76by54.qm} \\
\midrule
\htuse{Gamma77.qm} \\
\midrule
\htuse{Gamma78.qm} \\
\midrule
\htuse{Gamma79.qm} \\
\midrule
\htuse{Gamma80.qm} \\
\midrule
\htuse{Gamma80by60.qm} \\
\midrule
\htuse{Gamma81.qm} \\
\midrule
\htuse{Gamma81by69.qm} \\
\midrule
\htuse{Gamma82.qm} \\
\midrule
\htuse{Gamma83.qm} \\
\midrule
\htuse{Gamma84.qm} \\
\midrule
\htuse{Gamma85.qm} \\
\midrule
\htuse{Gamma85by60.qm} \\
\midrule
\htuse{Gamma87.qm} \\
\midrule
\htuse{Gamma88.qm} \\
\midrule
\htuse{Gamma89.qm} \\
\midrule
\htuse{Gamma92.qm} \\
\midrule
\htuse{Gamma93.qm} \\
\midrule
\htuse{Gamma93by60.qm} \\
\midrule
\htuse{Gamma94.qm} \\
\midrule
\htuse{Gamma94by69.qm} \\
\midrule
\htuse{Gamma96.qm} \\
\midrule
\htuse{Gamma102.qm} \\
\midrule
\htuse{Gamma103.qm} \\
\midrule
\htuse{Gamma104.qm} \\
\midrule
\htuse{Gamma106.qm} \\
\midrule
\htuse{Gamma110.qm} \\
\midrule
\htuse{Gamma126.qm} \\
\midrule
\htuse{Gamma128.qm} \\
\midrule
\htuse{Gamma130.qm} \\
\midrule
\htuse{Gamma132.qm} \\
\midrule
\htuse{Gamma136.qm} \\
\midrule
\htuse{Gamma149.qm} \\
\midrule
\htuse{Gamma150.qm} \\
\midrule
\htuse{Gamma150by66.qm} \\
\midrule
\htuse{Gamma151.qm} \\
\midrule
\htuse{Gamma152.qm} \\
\midrule
\htuse{Gamma152by54.qm} \\
\midrule
\htuse{Gamma152by76.qm} \\
\midrule
\htuse{Gamma167.qm} \\
\midrule
\htuse{Gamma168.qm} \\
\midrule
\htuse{Gamma169.qm} \\
\midrule
\htuse{Gamma800.qm} \\
\midrule
\htuse{Gamma802.qm} \\
\midrule
\htuse{Gamma803.qm} \\
\midrule
\htuse{Gamma804.qm} \\
\midrule
\htuse{Gamma805.qm} \\
\midrule
\htuse{Gamma806.qm} \\
\midrule
\htuse{Gamma810.qm} \\
\midrule
\htuse{Gamma811.qm} \\
\midrule
\htuse{Gamma812.qm} \\
\midrule
\htuse{Gamma820.qm} \\
\midrule
\htuse{Gamma821.qm} \\
\midrule
\htuse{Gamma822.qm} \\
\midrule
\htuse{Gamma830.qm} \\
\midrule
\htuse{Gamma831.qm} \\
\midrule
\htuse{Gamma832.qm} \\
\midrule
\htuse{Gamma833.qm} \\
\midrule
\htuse{Gamma910.qm} \\
\midrule
\htuse{Gamma911.qm} \\
\midrule
\htuse{Gamma920.qm} \\
\midrule
\htuse{Gamma930.qm} \\
\midrule
\htuse{Gamma944.qm} \\
\midrule
\htuse{Gamma945.qm} \\
\midrule
\htuse{Gamma998.qm}}%
\htdef{BARATE 98.cite}{\cite{Barate:1997ma}}%
\htdef{BARATE 98.collab}{ALEPH}%
\htdef{BARATE 98.ref}{BARATE 98 (ALEPH) \cite{Barate:1997ma}}%
\htdef{BARATE 98.meas}{%
\begin{ensuredisplaymath}
\htuse{Gamma85.gn} = \htuse{Gamma85.td}
\end{ensuredisplaymath} & \htuse{ALEPH.Gamma85.pub.BARATE.98}
\\
\begin{ensuredisplaymath}
\htuse{Gamma88.gn} = \htuse{Gamma88.td}
\end{ensuredisplaymath} & \htuse{ALEPH.Gamma88.pub.BARATE.98}
\\
\begin{ensuredisplaymath}
\htuse{Gamma93.gn} = \htuse{Gamma93.td}
\end{ensuredisplaymath} & \htuse{ALEPH.Gamma93.pub.BARATE.98}
\\
\begin{ensuredisplaymath}
\htuse{Gamma94.gn} = \htuse{Gamma94.td}
\end{ensuredisplaymath} & \htuse{ALEPH.Gamma94.pub.BARATE.98}}%
\htdef{BARATE 98E.cite}{\cite{Barate:1997tt}}%
\htdef{BARATE 98E.collab}{ALEPH}%
\htdef{BARATE 98E.ref}{BARATE 98E (ALEPH) \cite{Barate:1997tt}}%
\htdef{BARATE 98E.meas}{%
\begin{ensuredisplaymath}
\htuse{Gamma33.gn} = \htuse{Gamma33.td}
\end{ensuredisplaymath} & \htuse{ALEPH.Gamma33.pub.BARATE.98E}
\\
\begin{ensuredisplaymath}
\htuse{Gamma37.gn} = \htuse{Gamma37.td}
\end{ensuredisplaymath} & \htuse{ALEPH.Gamma37.pub.BARATE.98E}
\\
\begin{ensuredisplaymath}
\htuse{Gamma40.gn} = \htuse{Gamma40.td}
\end{ensuredisplaymath} & \htuse{ALEPH.Gamma40.pub.BARATE.98E}
\\
\begin{ensuredisplaymath}
\htuse{Gamma42.gn} = \htuse{Gamma42.td}
\end{ensuredisplaymath} & \htuse{ALEPH.Gamma42.pub.BARATE.98E}
\\
\begin{ensuredisplaymath}
\htuse{Gamma47.gn} = \htuse{Gamma47.td}
\end{ensuredisplaymath} & \htuse{ALEPH.Gamma47.pub.BARATE.98E}
\\
\begin{ensuredisplaymath}
\htuse{Gamma48.gn} = \htuse{Gamma48.td}
\end{ensuredisplaymath} & \htuse{ALEPH.Gamma48.pub.BARATE.98E}
\\
\begin{ensuredisplaymath}
\htuse{Gamma51.gn} = \htuse{Gamma51.td}
\end{ensuredisplaymath} & \htuse{ALEPH.Gamma51.pub.BARATE.98E}
\\
\begin{ensuredisplaymath}
\htuse{Gamma53.gn} = \htuse{Gamma53.td}
\end{ensuredisplaymath} & \htuse{ALEPH.Gamma53.pub.BARATE.98E}}%
\htdef{BARATE 99K.cite}{\cite{Barate:1999hi}}%
\htdef{BARATE 99K.collab}{ALEPH}%
\htdef{BARATE 99K.ref}{BARATE 99K (ALEPH) \cite{Barate:1999hi}}%
\htdef{BARATE 99K.meas}{%
\begin{ensuredisplaymath}
\htuse{Gamma10.gn} = \htuse{Gamma10.td}
\end{ensuredisplaymath} & \htuse{ALEPH.Gamma10.pub.BARATE.99K}
\\
\begin{ensuredisplaymath}
\htuse{Gamma16.gn} = \htuse{Gamma16.td}
\end{ensuredisplaymath} & \htuse{ALEPH.Gamma16.pub.BARATE.99K}
\\
\begin{ensuredisplaymath}
\htuse{Gamma23.gn} = \htuse{Gamma23.td}
\end{ensuredisplaymath} & \htuse{ALEPH.Gamma23.pub.BARATE.99K}
\\
\begin{ensuredisplaymath}
\htuse{Gamma28.gn} = \htuse{Gamma28.td}
\end{ensuredisplaymath} & \htuse{ALEPH.Gamma28.pub.BARATE.99K}
\\
\begin{ensuredisplaymath}
\htuse{Gamma35.gn} = \htuse{Gamma35.td}
\end{ensuredisplaymath} & \htuse{ALEPH.Gamma35.pub.BARATE.99K}
\\
\begin{ensuredisplaymath}
\htuse{Gamma37.gn} = \htuse{Gamma37.td}
\end{ensuredisplaymath} & \htuse{ALEPH.Gamma37.pub.BARATE.99K}
\\
\begin{ensuredisplaymath}
\htuse{Gamma40.gn} = \htuse{Gamma40.td}
\end{ensuredisplaymath} & \htuse{ALEPH.Gamma40.pub.BARATE.99K}
\\
\begin{ensuredisplaymath}
\htuse{Gamma42.gn} = \htuse{Gamma42.td}
\end{ensuredisplaymath} & \htuse{ALEPH.Gamma42.pub.BARATE.99K}}%
\htdef{BARATE 99R.cite}{\cite{Barate:1999hj}}%
\htdef{BARATE 99R.collab}{ALEPH}%
\htdef{BARATE 99R.ref}{BARATE 99R (ALEPH) \cite{Barate:1999hj}}%
\htdef{BARATE 99R.meas}{%
\begin{ensuredisplaymath}
\htuse{Gamma44.gn} = \htuse{Gamma44.td}
\end{ensuredisplaymath} & \htuse{ALEPH.Gamma44.pub.BARATE.99R}}%
\htdef{BUSKULIC 96.cite}{\cite{Buskulic:1995ty}}%
\htdef{BUSKULIC 96.collab}{ALEPH}%
\htdef{BUSKULIC 96.ref}{BUSKULIC 96 (ALEPH) \cite{Buskulic:1995ty}}%
\htdef{BUSKULIC 96.meas}{%
\begin{ensuredisplaymath}
\htuse{Gamma150by66.gn} = \htuse{Gamma150by66.td}
\end{ensuredisplaymath} & \htuse{ALEPH.Gamma150by66.pub.BUSKULIC.96}}%
\htdef{BUSKULIC 97C.cite}{\cite{Buskulic:1996qs}}%
\htdef{BUSKULIC 97C.collab}{ALEPH}%
\htdef{BUSKULIC 97C.ref}{BUSKULIC 97C (ALEPH) \cite{Buskulic:1996qs}}%
\htdef{BUSKULIC 97C.meas}{%
\begin{ensuredisplaymath}
\htuse{Gamma126.gn} = \htuse{Gamma126.td}
\end{ensuredisplaymath} & \htuse{ALEPH.Gamma126.pub.BUSKULIC.97C}
\\
\begin{ensuredisplaymath}
\htuse{Gamma128.gn} = \htuse{Gamma128.td}
\end{ensuredisplaymath} & \htuse{ALEPH.Gamma128.pub.BUSKULIC.97C}
\\
\begin{ensuredisplaymath}
\htuse{Gamma150.gn} = \htuse{Gamma150.td}
\end{ensuredisplaymath} & \htuse{ALEPH.Gamma150.pub.BUSKULIC.97C}
\\
\begin{ensuredisplaymath}
\htuse{Gamma152.gn} = \htuse{Gamma152.td}
\end{ensuredisplaymath} & \htuse{ALEPH.Gamma152.pub.BUSKULIC.97C}}%
\htdef{SCHAEL 05C.cite}{\cite{Schael:2005am}}%
\htdef{SCHAEL 05C.collab}{ALEPH}%
\htdef{SCHAEL 05C.ref}{SCHAEL 05C (ALEPH) \cite{Schael:2005am}}%
\htdef{SCHAEL 05C.meas}{%
\begin{ensuredisplaymath}
\htuse{Gamma3.gn} = \htuse{Gamma3.td}
\end{ensuredisplaymath} & \htuse{ALEPH.Gamma3.pub.SCHAEL.05C}
\\
\begin{ensuredisplaymath}
\htuse{Gamma5.gn} = \htuse{Gamma5.td}
\end{ensuredisplaymath} & \htuse{ALEPH.Gamma5.pub.SCHAEL.05C}
\\
\begin{ensuredisplaymath}
\htuse{Gamma8.gn} = \htuse{Gamma8.td}
\end{ensuredisplaymath} & \htuse{ALEPH.Gamma8.pub.SCHAEL.05C}
\\
\begin{ensuredisplaymath}
\htuse{Gamma13.gn} = \htuse{Gamma13.td}
\end{ensuredisplaymath} & \htuse{ALEPH.Gamma13.pub.SCHAEL.05C}
\\
\begin{ensuredisplaymath}
\htuse{Gamma19.gn} = \htuse{Gamma19.td}
\end{ensuredisplaymath} & \htuse{ALEPH.Gamma19.pub.SCHAEL.05C}
\\
\begin{ensuredisplaymath}
\htuse{Gamma26.gn} = \htuse{Gamma26.td}
\end{ensuredisplaymath} & \htuse{ALEPH.Gamma26.pub.SCHAEL.05C}
\\
\begin{ensuredisplaymath}
\htuse{Gamma30.gn} = \htuse{Gamma30.td}
\end{ensuredisplaymath} & \htuse{ALEPH.Gamma30.pub.SCHAEL.05C}
\\
\begin{ensuredisplaymath}
\htuse{Gamma58.gn} = \htuse{Gamma58.td}
\end{ensuredisplaymath} & \htuse{ALEPH.Gamma58.pub.SCHAEL.05C}
\\
\begin{ensuredisplaymath}
\htuse{Gamma66.gn} = \htuse{Gamma66.td}
\end{ensuredisplaymath} & \htuse{ALEPH.Gamma66.pub.SCHAEL.05C}
\\
\begin{ensuredisplaymath}
\htuse{Gamma76.gn} = \htuse{Gamma76.td}
\end{ensuredisplaymath} & \htuse{ALEPH.Gamma76.pub.SCHAEL.05C}
\\
\begin{ensuredisplaymath}
\htuse{Gamma103.gn} = \htuse{Gamma103.td}
\end{ensuredisplaymath} & \htuse{ALEPH.Gamma103.pub.SCHAEL.05C}
\\
\begin{ensuredisplaymath}
\htuse{Gamma104.gn} = \htuse{Gamma104.td}
\end{ensuredisplaymath} & \htuse{ALEPH.Gamma104.pub.SCHAEL.05C}
\\
\begin{ensuredisplaymath}
\htuse{Gamma805.gn} = \htuse{Gamma805.td}
\end{ensuredisplaymath} & \htuse{ALEPH.Gamma805.pub.SCHAEL.05C}}%
\htdef{ALBRECHT 88B.cite}{\cite{Albrecht:1987zf}}%
\htdef{ALBRECHT 88B.collab}{ARGUS}%
\htdef{ALBRECHT 88B.ref}{ALBRECHT 88B (ARGUS) \cite{Albrecht:1987zf}}%
\htdef{ALBRECHT 88B.meas}{%
\begin{ensuredisplaymath}
\htuse{Gamma103.gn} = \htuse{Gamma103.td}
\end{ensuredisplaymath} & \htuse{ARGUS.Gamma103.pub.ALBRECHT.88B}}%
\htdef{ALBRECHT 92D.cite}{\cite{Albrecht:1991rh}}%
\htdef{ALBRECHT 92D.collab}{ARGUS}%
\htdef{ALBRECHT 92D.ref}{ALBRECHT 92D (ARGUS) \cite{Albrecht:1991rh}}%
\htdef{ALBRECHT 92D.meas}{%
\begin{ensuredisplaymath}
\htuse{Gamma3by5.gn} = \htuse{Gamma3by5.td}
\end{ensuredisplaymath} & \htuse{ARGUS.Gamma3by5.pub.ALBRECHT.92D}}%
\htdef{AUBERT 07AP.cite}{\cite{Aubert:2007jh}}%
\htdef{AUBERT 07AP.collab}{\babar}%
\htdef{AUBERT 07AP.ref}{AUBERT 07AP (\babar) \cite{Aubert:2007jh}}%
\htdef{AUBERT 07AP.meas}{%
\begin{ensuredisplaymath}
\htuse{Gamma16.gn} = \htuse{Gamma16.td}
\end{ensuredisplaymath} & \htuse{BaBar.Gamma16.pub.AUBERT.07AP}}%
\htdef{AUBERT 08.cite}{\cite{Aubert:2007mh}}%
\htdef{AUBERT 08.collab}{\babar}%
\htdef{AUBERT 08.ref}{AUBERT 08 (\babar) \cite{Aubert:2007mh}}%
\htdef{AUBERT 08.meas}{%
\begin{ensuredisplaymath}
\htuse{Gamma60.gn} = \htuse{Gamma60.td}
\end{ensuredisplaymath} & \htuse{BaBar.Gamma60.pub.AUBERT.08}
\\
\begin{ensuredisplaymath}
\htuse{Gamma85.gn} = \htuse{Gamma85.td}
\end{ensuredisplaymath} & \htuse{BaBar.Gamma85.pub.AUBERT.08}
\\
\begin{ensuredisplaymath}
\htuse{Gamma93.gn} = \htuse{Gamma93.td}
\end{ensuredisplaymath} & \htuse{BaBar.Gamma93.pub.AUBERT.08}
\\
\begin{ensuredisplaymath}
\htuse{Gamma96.gn} = \htuse{Gamma96.td}
\end{ensuredisplaymath} & \htuse{BaBar.Gamma96.pub.AUBERT.08}}%
\htdef{AUBERT 10F.cite}{\cite{Aubert:2009qj}}%
\htdef{AUBERT 10F.collab}{\babar}%
\htdef{AUBERT 10F.ref}{AUBERT 10F (\babar) \cite{Aubert:2009qj}}%
\htdef{AUBERT 10F.meas}{%
\begin{ensuredisplaymath}
\htuse{Gamma3by5.gn} = \htuse{Gamma3by5.td}
\end{ensuredisplaymath} & \htuse{BaBar.Gamma3by5.pub.AUBERT.10F}
\\
\begin{ensuredisplaymath}
\htuse{Gamma9by5.gn} = \htuse{Gamma9by5.td}
\end{ensuredisplaymath} & \htuse{BaBar.Gamma9by5.pub.AUBERT.10F}
\\
\begin{ensuredisplaymath}
\htuse{Gamma10by5.gn} = \htuse{Gamma10by5.td}
\end{ensuredisplaymath} & \htuse{BaBar.Gamma10by5.pub.AUBERT.10F}}%
\htdef{DEL-AMO-SANCHEZ 11E.cite}{\cite{delAmoSanchez:2010pc}}%
\htdef{DEL-AMO-SANCHEZ 11E.collab}{\babar}%
\htdef{DEL-AMO-SANCHEZ 11E.ref}{DEL-AMO-SANCHEZ 11E (\babar) \cite{delAmoSanchez:2010pc}}%
\htdef{DEL-AMO-SANCHEZ 11E.meas}{%
\begin{ensuredisplaymath}
\htuse{Gamma128.gn} = \htuse{Gamma128.td}
\end{ensuredisplaymath} & \htuse{BaBar.Gamma128.pub.DEL-AMO-SANCHEZ.11E}}%
\htdef{LEES 12X.cite}{\cite{Lees:2012ks}}%
\htdef{LEES 12X.collab}{\babar}%
\htdef{LEES 12X.ref}{LEES 12X (\babar) \cite{Lees:2012ks}}%
\htdef{LEES 12X.meas}{%
\begin{ensuredisplaymath}
\htuse{Gamma811.gn} = \htuse{Gamma811.td}
\end{ensuredisplaymath} & \htuse{BaBar.Gamma811.pub.LEES.12X}
\\
\begin{ensuredisplaymath}
\htuse{Gamma812.gn} = \htuse{Gamma812.td}
\end{ensuredisplaymath} & \htuse{BaBar.Gamma812.pub.LEES.12X}
\\
\begin{ensuredisplaymath}
\htuse{Gamma821.gn} = \htuse{Gamma821.td}
\end{ensuredisplaymath} & \htuse{BaBar.Gamma821.pub.LEES.12X}
\\
\begin{ensuredisplaymath}
\htuse{Gamma822.gn} = \htuse{Gamma822.td}
\end{ensuredisplaymath} & \htuse{BaBar.Gamma822.pub.LEES.12X}
\\
\begin{ensuredisplaymath}
\htuse{Gamma831.gn} = \htuse{Gamma831.td}
\end{ensuredisplaymath} & \htuse{BaBar.Gamma831.pub.LEES.12X}
\\
\begin{ensuredisplaymath}
\htuse{Gamma832.gn} = \htuse{Gamma832.td}
\end{ensuredisplaymath} & \htuse{BaBar.Gamma832.pub.LEES.12X}
\\
\begin{ensuredisplaymath}
\htuse{Gamma833.gn} = \htuse{Gamma833.td}
\end{ensuredisplaymath} & \htuse{BaBar.Gamma833.pub.LEES.12X}
\\
\begin{ensuredisplaymath}
\htuse{Gamma910.gn} = \htuse{Gamma910.td}
\end{ensuredisplaymath} & \htuse{BaBar.Gamma910.pub.LEES.12X}
\\
\begin{ensuredisplaymath}
\htuse{Gamma911.gn} = \htuse{Gamma911.td}
\end{ensuredisplaymath} & \htuse{BaBar.Gamma911.pub.LEES.12X}
\\
\begin{ensuredisplaymath}
\htuse{Gamma920.gn} = \htuse{Gamma920.td}
\end{ensuredisplaymath} & \htuse{BaBar.Gamma920.pub.LEES.12X}
\\
\begin{ensuredisplaymath}
\htuse{Gamma930.gn} = \htuse{Gamma930.td}
\end{ensuredisplaymath} & \htuse{BaBar.Gamma930.pub.LEES.12X}
\\
\begin{ensuredisplaymath}
\htuse{Gamma944.gn} = \htuse{Gamma944.td}
\end{ensuredisplaymath} & \htuse{BaBar.Gamma944.pub.LEES.12X}}%
\htdef{LEES 12Y.cite}{\cite{Lees:2012de}}%
\htdef{LEES 12Y.collab}{\babar}%
\htdef{LEES 12Y.ref}{LEES 12Y (\babar) \cite{Lees:2012de}}%
\htdef{LEES 12Y.meas}{%
\begin{ensuredisplaymath}
\htuse{Gamma47.gn} = \htuse{Gamma47.td}
\end{ensuredisplaymath} & \htuse{BaBar.Gamma47.pub.LEES.12Y}
\\
\begin{ensuredisplaymath}
\htuse{Gamma50.gn} = \htuse{Gamma50.td}
\end{ensuredisplaymath} & \htuse{BaBar.Gamma50.pub.LEES.12Y}}%
\htdef{FUJIKAWA 08.cite}{\cite{Fujikawa:2008ma}}%
\htdef{FUJIKAWA 08.collab}{Belle}%
\htdef{FUJIKAWA 08.ref}{FUJIKAWA 08 (Belle) \cite{Fujikawa:2008ma}}%
\htdef{FUJIKAWA 08.meas}{%
\begin{ensuredisplaymath}
\htuse{Gamma13.gn} = \htuse{Gamma13.td}
\end{ensuredisplaymath} & \htuse{Belle.Gamma13.pub.FUJIKAWA.08}}%
\htdef{INAMI 09.cite}{\cite{Inami:2008ar}}%
\htdef{INAMI 09.collab}{Belle}%
\htdef{INAMI 09.ref}{INAMI 09 (Belle) \cite{Inami:2008ar}}%
\htdef{INAMI 09.meas}{%
\begin{ensuredisplaymath}
\htuse{Gamma126.gn} = \htuse{Gamma126.td}
\end{ensuredisplaymath} & \htuse{Belle.Gamma126.pub.INAMI.09}
\\
\begin{ensuredisplaymath}
\htuse{Gamma128.gn} = \htuse{Gamma128.td}
\end{ensuredisplaymath} & \htuse{Belle.Gamma128.pub.INAMI.09}
\\
\begin{ensuredisplaymath}
\htuse{Gamma130.gn} = \htuse{Gamma130.td}
\end{ensuredisplaymath} & \htuse{Belle.Gamma130.pub.INAMI.09}
\\
\begin{ensuredisplaymath}
\htuse{Gamma132.gn} = \htuse{Gamma132.td}
\end{ensuredisplaymath} & \htuse{Belle.Gamma132.pub.INAMI.09}}%
\htdef{LEE 10.cite}{\cite{Lee:2010tc}}%
\htdef{LEE 10.collab}{Belle}%
\htdef{LEE 10.ref}{LEE 10 (Belle) \cite{Lee:2010tc}}%
\htdef{LEE 10.meas}{%
\begin{ensuredisplaymath}
\htuse{Gamma60.gn} = \htuse{Gamma60.td}
\end{ensuredisplaymath} & \htuse{Belle.Gamma60.pub.LEE.10}
\\
\begin{ensuredisplaymath}
\htuse{Gamma85.gn} = \htuse{Gamma85.td}
\end{ensuredisplaymath} & \htuse{Belle.Gamma85.pub.LEE.10}
\\
\begin{ensuredisplaymath}
\htuse{Gamma93.gn} = \htuse{Gamma93.td}
\end{ensuredisplaymath} & \htuse{Belle.Gamma93.pub.LEE.10}
\\
\begin{ensuredisplaymath}
\htuse{Gamma96.gn} = \htuse{Gamma96.td}
\end{ensuredisplaymath} & \htuse{Belle.Gamma96.pub.LEE.10}}%
\htdef{RYU 14vpc.cite}{\cite{Ryu:2014vpc}}%
\htdef{RYU 14vpc.collab}{Belle}%
\htdef{RYU 14vpc.ref}{RYU 14vpc (Belle) \cite{Ryu:2014vpc}}%
\htdef{RYU 14vpc.meas}{%
\begin{ensuredisplaymath}
\htuse{Gamma35.gn} = \htuse{Gamma35.td}
\end{ensuredisplaymath} & \htuse{Belle.Gamma35.pub.RYU.14vpc}
\\
\begin{ensuredisplaymath}
\htuse{Gamma37.gn} = \htuse{Gamma37.td}
\end{ensuredisplaymath} & \htuse{Belle.Gamma37.pub.RYU.14vpc}
\\
\begin{ensuredisplaymath}
\htuse{Gamma40.gn} = \htuse{Gamma40.td}
\end{ensuredisplaymath} & \htuse{Belle.Gamma40.pub.RYU.14vpc}
\\
\begin{ensuredisplaymath}
\htuse{Gamma42.gn} = \htuse{Gamma42.td}
\end{ensuredisplaymath} & \htuse{Belle.Gamma42.pub.RYU.14vpc}
\\
\begin{ensuredisplaymath}
\htuse{Gamma47.gn} = \htuse{Gamma47.td}
\end{ensuredisplaymath} & \htuse{Belle.Gamma47.pub.RYU.14vpc}
\\
\begin{ensuredisplaymath}
\htuse{Gamma50.gn} = \htuse{Gamma50.td}
\end{ensuredisplaymath} & \htuse{Belle.Gamma50.pub.RYU.14vpc}}%
\htdef{BEHREND 89B.cite}{\cite{Behrend:1989wc}}%
\htdef{BEHREND 89B.collab}{CELLO}%
\htdef{BEHREND 89B.ref}{BEHREND 89B (CELLO) \cite{Behrend:1989wc}}%
\htdef{BEHREND 89B.meas}{%
\begin{ensuredisplaymath}
\htuse{Gamma54.gn} = \htuse{Gamma54.td}
\end{ensuredisplaymath} & \htuse{CELLO.Gamma54.pub.BEHREND.89B}}%
\htdef{ANASTASSOV 01.cite}{\cite{Anastassov:2000xu}}%
\htdef{ANASTASSOV 01.collab}{CLEO}%
\htdef{ANASTASSOV 01.ref}{ANASTASSOV 01 (CLEO) \cite{Anastassov:2000xu}}%
\htdef{ANASTASSOV 01.meas}{%
\begin{ensuredisplaymath}
\htuse{Gamma78.gn} = \htuse{Gamma78.td}
\end{ensuredisplaymath} & \htuse{CLEO.Gamma78.pub.ANASTASSOV.01}
\\
\begin{ensuredisplaymath}
\htuse{Gamma104.gn} = \htuse{Gamma104.td}
\end{ensuredisplaymath} & \htuse{CLEO.Gamma104.pub.ANASTASSOV.01}}%
\htdef{ANASTASSOV 97.cite}{\cite{Anastassov:1996tc}}%
\htdef{ANASTASSOV 97.collab}{CLEO}%
\htdef{ANASTASSOV 97.ref}{ANASTASSOV 97 (CLEO) \cite{Anastassov:1996tc}}%
\htdef{ANASTASSOV 97.meas}{%
\begin{ensuredisplaymath}
\htuse{Gamma3by5.gn} = \htuse{Gamma3by5.td}
\end{ensuredisplaymath} & \htuse{CLEO.Gamma3by5.pub.ANASTASSOV.97}
\\
\begin{ensuredisplaymath}
\htuse{Gamma5.gn} = \htuse{Gamma5.td}
\end{ensuredisplaymath} & \htuse{CLEO.Gamma5.pub.ANASTASSOV.97}
\\
\begin{ensuredisplaymath}
\htuse{Gamma8.gn} = \htuse{Gamma8.td}
\end{ensuredisplaymath} & \htuse{CLEO.Gamma8.pub.ANASTASSOV.97}}%
\htdef{ARTUSO 92.cite}{\cite{Artuso:1992qu}}%
\htdef{ARTUSO 92.collab}{CLEO}%
\htdef{ARTUSO 92.ref}{ARTUSO 92 (CLEO) \cite{Artuso:1992qu}}%
\htdef{ARTUSO 92.meas}{%
\begin{ensuredisplaymath}
\htuse{Gamma126.gn} = \htuse{Gamma126.td}
\end{ensuredisplaymath} & \htuse{CLEO.Gamma126.pub.ARTUSO.92}}%
\htdef{ARTUSO 94.cite}{\cite{Artuso:1994ii}}%
\htdef{ARTUSO 94.collab}{CLEO}%
\htdef{ARTUSO 94.ref}{ARTUSO 94 (CLEO) \cite{Artuso:1994ii}}%
\htdef{ARTUSO 94.meas}{%
\begin{ensuredisplaymath}
\htuse{Gamma13.gn} = \htuse{Gamma13.td}
\end{ensuredisplaymath} & \htuse{CLEO.Gamma13.pub.ARTUSO.94}}%
\htdef{BALEST 95C.cite}{\cite{Balest:1995kq}}%
\htdef{BALEST 95C.collab}{CLEO}%
\htdef{BALEST 95C.ref}{BALEST 95C (CLEO) \cite{Balest:1995kq}}%
\htdef{BALEST 95C.meas}{%
\begin{ensuredisplaymath}
\htuse{Gamma57.gn} = \htuse{Gamma57.td}
\end{ensuredisplaymath} & \htuse{CLEO.Gamma57.pub.BALEST.95C}
\\
\begin{ensuredisplaymath}
\htuse{Gamma66.gn} = \htuse{Gamma66.td}
\end{ensuredisplaymath} & \htuse{CLEO.Gamma66.pub.BALEST.95C}
\\
\begin{ensuredisplaymath}
\htuse{Gamma150by66.gn} = \htuse{Gamma150by66.td}
\end{ensuredisplaymath} & \htuse{CLEO.Gamma150by66.pub.BALEST.95C}}%
\htdef{BARINGER 87.cite}{\cite{Baringer:1987tr}}%
\htdef{BARINGER 87.collab}{CLEO}%
\htdef{BARINGER 87.ref}{BARINGER 87 (CLEO) \cite{Baringer:1987tr}}%
\htdef{BARINGER 87.meas}{%
\begin{ensuredisplaymath}
\htuse{Gamma150.gn} = \htuse{Gamma150.td}
\end{ensuredisplaymath} & \htuse{CLEO.Gamma150.pub.BARINGER.87}}%
\htdef{BARTELT 96.cite}{\cite{Bartelt:1996iv}}%
\htdef{BARTELT 96.collab}{CLEO}%
\htdef{BARTELT 96.ref}{BARTELT 96 (CLEO) \cite{Bartelt:1996iv}}%
\htdef{BARTELT 96.meas}{%
\begin{ensuredisplaymath}
\htuse{Gamma128.gn} = \htuse{Gamma128.td}
\end{ensuredisplaymath} & \htuse{CLEO.Gamma128.pub.BARTELT.96}}%
\htdef{BATTLE 94.cite}{\cite{Battle:1994by}}%
\htdef{BATTLE 94.collab}{CLEO}%
\htdef{BATTLE 94.ref}{BATTLE 94 (CLEO) \cite{Battle:1994by}}%
\htdef{BATTLE 94.meas}{%
\begin{ensuredisplaymath}
\htuse{Gamma10.gn} = \htuse{Gamma10.td}
\end{ensuredisplaymath} & \htuse{CLEO.Gamma10.pub.BATTLE.94}
\\
\begin{ensuredisplaymath}
\htuse{Gamma16.gn} = \htuse{Gamma16.td}
\end{ensuredisplaymath} & \htuse{CLEO.Gamma16.pub.BATTLE.94}
\\
\begin{ensuredisplaymath}
\htuse{Gamma23.gn} = \htuse{Gamma23.td}
\end{ensuredisplaymath} & \htuse{CLEO.Gamma23.pub.BATTLE.94}
\\
\begin{ensuredisplaymath}
\htuse{Gamma31.gn} = \htuse{Gamma31.td}
\end{ensuredisplaymath} & \htuse{CLEO.Gamma31.pub.BATTLE.94}}%
\htdef{BISHAI 99.cite}{\cite{Bishai:1998gf}}%
\htdef{BISHAI 99.collab}{CLEO}%
\htdef{BISHAI 99.ref}{BISHAI 99 (CLEO) \cite{Bishai:1998gf}}%
\htdef{BISHAI 99.meas}{%
\begin{ensuredisplaymath}
\htuse{Gamma130.gn} = \htuse{Gamma130.td}
\end{ensuredisplaymath} & \htuse{CLEO.Gamma130.pub.BISHAI.99}
\\
\begin{ensuredisplaymath}
\htuse{Gamma132.gn} = \htuse{Gamma132.td}
\end{ensuredisplaymath} & \htuse{CLEO.Gamma132.pub.BISHAI.99}}%
\htdef{BORTOLETTO 93.cite}{\cite{Bortoletto:1993px}}%
\htdef{BORTOLETTO 93.collab}{CLEO}%
\htdef{BORTOLETTO 93.ref}{BORTOLETTO 93 (CLEO) \cite{Bortoletto:1993px}}%
\htdef{BORTOLETTO 93.meas}{%
\begin{ensuredisplaymath}
\htuse{Gamma76by54.gn} = \htuse{Gamma76by54.td}
\end{ensuredisplaymath} & \htuse{CLEO.Gamma76by54.pub.BORTOLETTO.93}
\\
\begin{ensuredisplaymath}
\htuse{Gamma152by76.gn} = \htuse{Gamma152by76.td}
\end{ensuredisplaymath} & \htuse{CLEO.Gamma152by76.pub.BORTOLETTO.93}}%
\htdef{COAN 96.cite}{\cite{Coan:1996iu}}%
\htdef{COAN 96.collab}{CLEO}%
\htdef{COAN 96.ref}{COAN 96 (CLEO) \cite{Coan:1996iu}}%
\htdef{COAN 96.meas}{%
\begin{ensuredisplaymath}
\htuse{Gamma34.gn} = \htuse{Gamma34.td}
\end{ensuredisplaymath} & \htuse{CLEO.Gamma34.pub.COAN.96}
\\
\begin{ensuredisplaymath}
\htuse{Gamma37.gn} = \htuse{Gamma37.td}
\end{ensuredisplaymath} & \htuse{CLEO.Gamma37.pub.COAN.96}
\\
\begin{ensuredisplaymath}
\htuse{Gamma39.gn} = \htuse{Gamma39.td}
\end{ensuredisplaymath} & \htuse{CLEO.Gamma39.pub.COAN.96}
\\
\begin{ensuredisplaymath}
\htuse{Gamma42.gn} = \htuse{Gamma42.td}
\end{ensuredisplaymath} & \htuse{CLEO.Gamma42.pub.COAN.96}
\\
\begin{ensuredisplaymath}
\htuse{Gamma47.gn} = \htuse{Gamma47.td}
\end{ensuredisplaymath} & \htuse{CLEO.Gamma47.pub.COAN.96}}%
\htdef{EDWARDS 00A.cite}{\cite{Edwards:1999fj}}%
\htdef{EDWARDS 00A.collab}{CLEO}%
\htdef{EDWARDS 00A.ref}{EDWARDS 00A (CLEO) \cite{Edwards:1999fj}}%
\htdef{EDWARDS 00A.meas}{%
\begin{ensuredisplaymath}
\htuse{Gamma69.gn} = \htuse{Gamma69.td}
\end{ensuredisplaymath} & \htuse{CLEO.Gamma69.pub.EDWARDS.00A}}%
\htdef{GIBAUT 94B.cite}{\cite{Gibaut:1994ik}}%
\htdef{GIBAUT 94B.collab}{CLEO}%
\htdef{GIBAUT 94B.ref}{GIBAUT 94B (CLEO) \cite{Gibaut:1994ik}}%
\htdef{GIBAUT 94B.meas}{%
\begin{ensuredisplaymath}
\htuse{Gamma102.gn} = \htuse{Gamma102.td}
\end{ensuredisplaymath} & \htuse{CLEO.Gamma102.pub.GIBAUT.94B}
\\
\begin{ensuredisplaymath}
\htuse{Gamma103.gn} = \htuse{Gamma103.td}
\end{ensuredisplaymath} & \htuse{CLEO.Gamma103.pub.GIBAUT.94B}}%
\htdef{PROCARIO 93.cite}{\cite{Procario:1992hd}}%
\htdef{PROCARIO 93.collab}{CLEO}%
\htdef{PROCARIO 93.ref}{PROCARIO 93 (CLEO) \cite{Procario:1992hd}}%
\htdef{PROCARIO 93.meas}{%
\begin{ensuredisplaymath}
\htuse{Gamma19by13.gn} = \htuse{Gamma19by13.td}
\end{ensuredisplaymath} & \htuse{CLEO.Gamma19by13.pub.PROCARIO.93}
\\
\begin{ensuredisplaymath}
\htuse{Gamma26by13.gn} = \htuse{Gamma26by13.td}
\end{ensuredisplaymath} & \htuse{CLEO.Gamma26by13.pub.PROCARIO.93}
\\
\begin{ensuredisplaymath}
\htuse{Gamma29.gn} = \htuse{Gamma29.td}
\end{ensuredisplaymath} & \htuse{CLEO.Gamma29.pub.PROCARIO.93}}%
\htdef{RICHICHI 99.cite}{\cite{Richichi:1998bc}}%
\htdef{RICHICHI 99.collab}{CLEO}%
\htdef{RICHICHI 99.ref}{RICHICHI 99 (CLEO) \cite{Richichi:1998bc}}%
\htdef{RICHICHI 99.meas}{%
\begin{ensuredisplaymath}
\htuse{Gamma80by60.gn} = \htuse{Gamma80by60.td}
\end{ensuredisplaymath} & \htuse{CLEO.Gamma80by60.pub.RICHICHI.99}
\\
\begin{ensuredisplaymath}
\htuse{Gamma81by69.gn} = \htuse{Gamma81by69.td}
\end{ensuredisplaymath} & \htuse{CLEO.Gamma81by69.pub.RICHICHI.99}
\\
\begin{ensuredisplaymath}
\htuse{Gamma93by60.gn} = \htuse{Gamma93by60.td}
\end{ensuredisplaymath} & \htuse{CLEO.Gamma93by60.pub.RICHICHI.99}
\\
\begin{ensuredisplaymath}
\htuse{Gamma94by69.gn} = \htuse{Gamma94by69.td}
\end{ensuredisplaymath} & \htuse{CLEO.Gamma94by69.pub.RICHICHI.99}}%
\htdef{ARMS 05.cite}{\cite{Arms:2005qg}}%
\htdef{ARMS 05.collab}{CLEO3}%
\htdef{ARMS 05.ref}{ARMS 05 (CLEO3) \cite{Arms:2005qg}}%
\htdef{ARMS 05.meas}{%
\begin{ensuredisplaymath}
\htuse{Gamma88.gn} = \htuse{Gamma88.td}
\end{ensuredisplaymath} & \htuse{CLEO3.Gamma88.pub.ARMS.05}
\\
\begin{ensuredisplaymath}
\htuse{Gamma94.gn} = \htuse{Gamma94.td}
\end{ensuredisplaymath} & \htuse{CLEO3.Gamma94.pub.ARMS.05}
\\
\begin{ensuredisplaymath}
\htuse{Gamma151.gn} = \htuse{Gamma151.td}
\end{ensuredisplaymath} & \htuse{CLEO3.Gamma151.pub.ARMS.05}}%
\htdef{BRIERE 03.cite}{\cite{Briere:2003fr}}%
\htdef{BRIERE 03.collab}{CLEO3}%
\htdef{BRIERE 03.ref}{BRIERE 03 (CLEO3) \cite{Briere:2003fr}}%
\htdef{BRIERE 03.meas}{%
\begin{ensuredisplaymath}
\htuse{Gamma60.gn} = \htuse{Gamma60.td}
\end{ensuredisplaymath} & \htuse{CLEO3.Gamma60.pub.BRIERE.03}
\\
\begin{ensuredisplaymath}
\htuse{Gamma85.gn} = \htuse{Gamma85.td}
\end{ensuredisplaymath} & \htuse{CLEO3.Gamma85.pub.BRIERE.03}
\\
\begin{ensuredisplaymath}
\htuse{Gamma93.gn} = \htuse{Gamma93.td}
\end{ensuredisplaymath} & \htuse{CLEO3.Gamma93.pub.BRIERE.03}}%
\htdef{ABDALLAH 06A.cite}{\cite{Abdallah:2003cw}}%
\htdef{ABDALLAH 06A.collab}{DELPHI}%
\htdef{ABDALLAH 06A.ref}{ABDALLAH 06A (DELPHI) \cite{Abdallah:2003cw}}%
\htdef{ABDALLAH 06A.meas}{%
\begin{ensuredisplaymath}
\htuse{Gamma8.gn} = \htuse{Gamma8.td}
\end{ensuredisplaymath} & \htuse{DELPHI.Gamma8.pub.ABDALLAH.06A}
\\
\begin{ensuredisplaymath}
\htuse{Gamma13.gn} = \htuse{Gamma13.td}
\end{ensuredisplaymath} & \htuse{DELPHI.Gamma13.pub.ABDALLAH.06A}
\\
\begin{ensuredisplaymath}
\htuse{Gamma19.gn} = \htuse{Gamma19.td}
\end{ensuredisplaymath} & \htuse{DELPHI.Gamma19.pub.ABDALLAH.06A}
\\
\begin{ensuredisplaymath}
\htuse{Gamma25.gn} = \htuse{Gamma25.td}
\end{ensuredisplaymath} & \htuse{DELPHI.Gamma25.pub.ABDALLAH.06A}
\\
\begin{ensuredisplaymath}
\htuse{Gamma57.gn} = \htuse{Gamma57.td}
\end{ensuredisplaymath} & \htuse{DELPHI.Gamma57.pub.ABDALLAH.06A}
\\
\begin{ensuredisplaymath}
\htuse{Gamma66.gn} = \htuse{Gamma66.td}
\end{ensuredisplaymath} & \htuse{DELPHI.Gamma66.pub.ABDALLAH.06A}
\\
\begin{ensuredisplaymath}
\htuse{Gamma74.gn} = \htuse{Gamma74.td}
\end{ensuredisplaymath} & \htuse{DELPHI.Gamma74.pub.ABDALLAH.06A}
\\
\begin{ensuredisplaymath}
\htuse{Gamma103.gn} = \htuse{Gamma103.td}
\end{ensuredisplaymath} & \htuse{DELPHI.Gamma103.pub.ABDALLAH.06A}
\\
\begin{ensuredisplaymath}
\htuse{Gamma104.gn} = \htuse{Gamma104.td}
\end{ensuredisplaymath} & \htuse{DELPHI.Gamma104.pub.ABDALLAH.06A}}%
\htdef{ABREU 92N.cite}{\cite{Abreu:1992gn}}%
\htdef{ABREU 92N.collab}{DELPHI}%
\htdef{ABREU 92N.ref}{ABREU 92N (DELPHI) \cite{Abreu:1992gn}}%
\htdef{ABREU 92N.meas}{%
\begin{ensuredisplaymath}
\htuse{Gamma7.gn} = \htuse{Gamma7.td}
\end{ensuredisplaymath} & \htuse{DELPHI.Gamma7.pub.ABREU.92N}}%
\htdef{ABREU 94K.cite}{\cite{Abreu:1994fi}}%
\htdef{ABREU 94K.collab}{DELPHI}%
\htdef{ABREU 94K.ref}{ABREU 94K (DELPHI) \cite{Abreu:1994fi}}%
\htdef{ABREU 94K.meas}{%
\begin{ensuredisplaymath}
\htuse{Gamma10.gn} = \htuse{Gamma10.td}
\end{ensuredisplaymath} & \htuse{DELPHI.Gamma10.pub.ABREU.94K}
\\
\begin{ensuredisplaymath}
\htuse{Gamma31.gn} = \htuse{Gamma31.td}
\end{ensuredisplaymath} & \htuse{DELPHI.Gamma31.pub.ABREU.94K}}%
\htdef{ABREU 99X.cite}{\cite{Abreu:1999rb}}%
\htdef{ABREU 99X.collab}{DELPHI}%
\htdef{ABREU 99X.ref}{ABREU 99X (DELPHI) \cite{Abreu:1999rb}}%
\htdef{ABREU 99X.meas}{%
\begin{ensuredisplaymath}
\htuse{Gamma3.gn} = \htuse{Gamma3.td}
\end{ensuredisplaymath} & \htuse{DELPHI.Gamma3.pub.ABREU.99X}
\\
\begin{ensuredisplaymath}
\htuse{Gamma5.gn} = \htuse{Gamma5.td}
\end{ensuredisplaymath} & \htuse{DELPHI.Gamma5.pub.ABREU.99X}}%
\htdef{BYLSMA 87.cite}{\cite{Bylsma:1986zy}}%
\htdef{BYLSMA 87.collab}{HRS}%
\htdef{BYLSMA 87.ref}{BYLSMA 87 (HRS) \cite{Bylsma:1986zy}}%
\htdef{BYLSMA 87.meas}{%
\begin{ensuredisplaymath}
\htuse{Gamma102.gn} = \htuse{Gamma102.td}
\end{ensuredisplaymath} & \htuse{HRS.Gamma102.pub.BYLSMA.87}
\\
\begin{ensuredisplaymath}
\htuse{Gamma103.gn} = \htuse{Gamma103.td}
\end{ensuredisplaymath} & \htuse{HRS.Gamma103.pub.BYLSMA.87}}%
\htdef{ACCIARRI 01F.cite}{\cite{Acciarri:2001sg}}%
\htdef{ACCIARRI 01F.collab}{L3}%
\htdef{ACCIARRI 01F.ref}{ACCIARRI 01F (L3) \cite{Acciarri:2001sg}}%
\htdef{ACCIARRI 01F.meas}{%
\begin{ensuredisplaymath}
\htuse{Gamma3.gn} = \htuse{Gamma3.td}
\end{ensuredisplaymath} & \htuse{L3.Gamma3.pub.ACCIARRI.01F}
\\
\begin{ensuredisplaymath}
\htuse{Gamma5.gn} = \htuse{Gamma5.td}
\end{ensuredisplaymath} & \htuse{L3.Gamma5.pub.ACCIARRI.01F}}%
\htdef{ACCIARRI 95.cite}{\cite{Acciarri:1994vr}}%
\htdef{ACCIARRI 95.collab}{L3}%
\htdef{ACCIARRI 95.ref}{ACCIARRI 95 (L3) \cite{Acciarri:1994vr}}%
\htdef{ACCIARRI 95.meas}{%
\begin{ensuredisplaymath}
\htuse{Gamma7.gn} = \htuse{Gamma7.td}
\end{ensuredisplaymath} & \htuse{L3.Gamma7.pub.ACCIARRI.95}
\\
\begin{ensuredisplaymath}
\htuse{Gamma13.gn} = \htuse{Gamma13.td}
\end{ensuredisplaymath} & \htuse{L3.Gamma13.pub.ACCIARRI.95}
\\
\begin{ensuredisplaymath}
\htuse{Gamma19.gn} = \htuse{Gamma19.td}
\end{ensuredisplaymath} & \htuse{L3.Gamma19.pub.ACCIARRI.95}
\\
\begin{ensuredisplaymath}
\htuse{Gamma26.gn} = \htuse{Gamma26.td}
\end{ensuredisplaymath} & \htuse{L3.Gamma26.pub.ACCIARRI.95}}%
\htdef{ACCIARRI 95F.cite}{\cite{Acciarri:1995kx}}%
\htdef{ACCIARRI 95F.collab}{L3}%
\htdef{ACCIARRI 95F.ref}{ACCIARRI 95F (L3) \cite{Acciarri:1995kx}}%
\htdef{ACCIARRI 95F.meas}{%
\begin{ensuredisplaymath}
\htuse{Gamma35.gn} = \htuse{Gamma35.td}
\end{ensuredisplaymath} & \htuse{L3.Gamma35.pub.ACCIARRI.95F}
\\
\begin{ensuredisplaymath}
\htuse{Gamma40.gn} = \htuse{Gamma40.td}
\end{ensuredisplaymath} & \htuse{L3.Gamma40.pub.ACCIARRI.95F}}%
\htdef{ACHARD 01D.cite}{\cite{Achard:2001pk}}%
\htdef{ACHARD 01D.collab}{L3}%
\htdef{ACHARD 01D.ref}{ACHARD 01D (L3) \cite{Achard:2001pk}}%
\htdef{ACHARD 01D.meas}{%
\begin{ensuredisplaymath}
\htuse{Gamma55.gn} = \htuse{Gamma55.td}
\end{ensuredisplaymath} & \htuse{L3.Gamma55.pub.ACHARD.01D}
\\
\begin{ensuredisplaymath}
\htuse{Gamma102.gn} = \htuse{Gamma102.td}
\end{ensuredisplaymath} & \htuse{L3.Gamma102.pub.ACHARD.01D}}%
\htdef{ADEVA 91F.cite}{\cite{Adeva:1991qq}}%
\htdef{ADEVA 91F.collab}{L3}%
\htdef{ADEVA 91F.ref}{ADEVA 91F (L3) \cite{Adeva:1991qq}}%
\htdef{ADEVA 91F.meas}{%
\begin{ensuredisplaymath}
\htuse{Gamma54.gn} = \htuse{Gamma54.td}
\end{ensuredisplaymath} & \htuse{L3.Gamma54.pub.ADEVA.91F}}%
\htdef{ABBIENDI 00C.cite}{\cite{Abbiendi:1999pm}}%
\htdef{ABBIENDI 00C.collab}{OPAL}%
\htdef{ABBIENDI 00C.ref}{ABBIENDI 00C (OPAL) \cite{Abbiendi:1999pm}}%
\htdef{ABBIENDI 00C.meas}{%
\begin{ensuredisplaymath}
\htuse{Gamma35.gn} = \htuse{Gamma35.td}
\end{ensuredisplaymath} & \htuse{OPAL.Gamma35.pub.ABBIENDI.00C}
\\
\begin{ensuredisplaymath}
\htuse{Gamma38.gn} = \htuse{Gamma38.td}
\end{ensuredisplaymath} & \htuse{OPAL.Gamma38.pub.ABBIENDI.00C}
\\
\begin{ensuredisplaymath}
\htuse{Gamma43.gn} = \htuse{Gamma43.td}
\end{ensuredisplaymath} & \htuse{OPAL.Gamma43.pub.ABBIENDI.00C}}%
\htdef{ABBIENDI 00D.cite}{\cite{Abbiendi:1999cq}}%
\htdef{ABBIENDI 00D.collab}{OPAL}%
\htdef{ABBIENDI 00D.ref}{ABBIENDI 00D (OPAL) \cite{Abbiendi:1999cq}}%
\htdef{ABBIENDI 00D.meas}{%
\begin{ensuredisplaymath}
\htuse{Gamma92.gn} = \htuse{Gamma92.td}
\end{ensuredisplaymath} & \htuse{OPAL.Gamma92.pub.ABBIENDI.00D}}%
\htdef{ABBIENDI 01J.cite}{\cite{Abbiendi:2000ee}}%
\htdef{ABBIENDI 01J.collab}{OPAL}%
\htdef{ABBIENDI 01J.ref}{ABBIENDI 01J (OPAL) \cite{Abbiendi:2000ee}}%
\htdef{ABBIENDI 01J.meas}{%
\begin{ensuredisplaymath}
\htuse{Gamma10.gn} = \htuse{Gamma10.td}
\end{ensuredisplaymath} & \htuse{OPAL.Gamma10.pub.ABBIENDI.01J}
\\
\begin{ensuredisplaymath}
\htuse{Gamma31.gn} = \htuse{Gamma31.td}
\end{ensuredisplaymath} & \htuse{OPAL.Gamma31.pub.ABBIENDI.01J}}%
\htdef{ABBIENDI 03.cite}{\cite{Abbiendi:2002jw}}%
\htdef{ABBIENDI 03.collab}{OPAL}%
\htdef{ABBIENDI 03.ref}{ABBIENDI 03 (OPAL) \cite{Abbiendi:2002jw}}%
\htdef{ABBIENDI 03.meas}{%
\begin{ensuredisplaymath}
\htuse{Gamma3.gn} = \htuse{Gamma3.td}
\end{ensuredisplaymath} & \htuse{OPAL.Gamma3.pub.ABBIENDI.03}}%
\htdef{ABBIENDI 04J.cite}{\cite{Abbiendi:2004xa}}%
\htdef{ABBIENDI 04J.collab}{OPAL}%
\htdef{ABBIENDI 04J.ref}{ABBIENDI 04J (OPAL) \cite{Abbiendi:2004xa}}%
\htdef{ABBIENDI 04J.meas}{%
\begin{ensuredisplaymath}
\htuse{Gamma16.gn} = \htuse{Gamma16.td}
\end{ensuredisplaymath} & \htuse{OPAL.Gamma16.pub.ABBIENDI.04J}
\\
\begin{ensuredisplaymath}
\htuse{Gamma85.gn} = \htuse{Gamma85.td}
\end{ensuredisplaymath} & \htuse{OPAL.Gamma85.pub.ABBIENDI.04J}}%
\htdef{ABBIENDI 99H.cite}{\cite{Abbiendi:1998cx}}%
\htdef{ABBIENDI 99H.collab}{OPAL}%
\htdef{ABBIENDI 99H.ref}{ABBIENDI 99H (OPAL) \cite{Abbiendi:1998cx}}%
\htdef{ABBIENDI 99H.meas}{%
\begin{ensuredisplaymath}
\htuse{Gamma5.gn} = \htuse{Gamma5.td}
\end{ensuredisplaymath} & \htuse{OPAL.Gamma5.pub.ABBIENDI.99H}}%
\htdef{ACKERSTAFF 98M.cite}{\cite{Ackerstaff:1997tx}}%
\htdef{ACKERSTAFF 98M.collab}{OPAL}%
\htdef{ACKERSTAFF 98M.ref}{ACKERSTAFF 98M (OPAL) \cite{Ackerstaff:1997tx}}%
\htdef{ACKERSTAFF 98M.meas}{%
\begin{ensuredisplaymath}
\htuse{Gamma8.gn} = \htuse{Gamma8.td}
\end{ensuredisplaymath} & \htuse{OPAL.Gamma8.pub.ACKERSTAFF.98M}
\\
\begin{ensuredisplaymath}
\htuse{Gamma13.gn} = \htuse{Gamma13.td}
\end{ensuredisplaymath} & \htuse{OPAL.Gamma13.pub.ACKERSTAFF.98M}
\\
\begin{ensuredisplaymath}
\htuse{Gamma17.gn} = \htuse{Gamma17.td}
\end{ensuredisplaymath} & \htuse{OPAL.Gamma17.pub.ACKERSTAFF.98M}}%
\htdef{ACKERSTAFF 99E.cite}{\cite{Ackerstaff:1998ia}}%
\htdef{ACKERSTAFF 99E.collab}{OPAL}%
\htdef{ACKERSTAFF 99E.ref}{ACKERSTAFF 99E (OPAL) \cite{Ackerstaff:1998ia}}%
\htdef{ACKERSTAFF 99E.meas}{%
\begin{ensuredisplaymath}
\htuse{Gamma103.gn} = \htuse{Gamma103.td}
\end{ensuredisplaymath} & \htuse{OPAL.Gamma103.pub.ACKERSTAFF.99E}
\\
\begin{ensuredisplaymath}
\htuse{Gamma104.gn} = \htuse{Gamma104.td}
\end{ensuredisplaymath} & \htuse{OPAL.Gamma104.pub.ACKERSTAFF.99E}}%
\htdef{AKERS 94G.cite}{\cite{Akers:1994td}}%
\htdef{AKERS 94G.collab}{OPAL}%
\htdef{AKERS 94G.ref}{AKERS 94G (OPAL) \cite{Akers:1994td}}%
\htdef{AKERS 94G.meas}{%
\begin{ensuredisplaymath}
\htuse{Gamma33.gn} = \htuse{Gamma33.td}
\end{ensuredisplaymath} & \htuse{OPAL.Gamma33.pub.AKERS.94G}}%
\htdef{AKERS 95Y.cite}{\cite{Akers:1995ry}}%
\htdef{AKERS 95Y.collab}{OPAL}%
\htdef{AKERS 95Y.ref}{AKERS 95Y (OPAL) \cite{Akers:1995ry}}%
\htdef{AKERS 95Y.meas}{%
\begin{ensuredisplaymath}
\htuse{Gamma55.gn} = \htuse{Gamma55.td}
\end{ensuredisplaymath} & \htuse{OPAL.Gamma55.pub.AKERS.95Y}
\\
\begin{ensuredisplaymath}
\htuse{Gamma57by55.gn} = \htuse{Gamma57by55.td}
\end{ensuredisplaymath} & \htuse{OPAL.Gamma57by55.pub.AKERS.95Y}}%
\htdef{ALEXANDER 91D.cite}{\cite{Alexander:1991am}}%
\htdef{ALEXANDER 91D.collab}{OPAL}%
\htdef{ALEXANDER 91D.ref}{ALEXANDER 91D (OPAL) \cite{Alexander:1991am}}%
\htdef{ALEXANDER 91D.meas}{%
\begin{ensuredisplaymath}
\htuse{Gamma7.gn} = \htuse{Gamma7.td}
\end{ensuredisplaymath} & \htuse{OPAL.Gamma7.pub.ALEXANDER.91D}}%
\htdef{AIHARA 87B.cite}{\cite{Aihara:1986mw}}%
\htdef{AIHARA 87B.collab}{TPC}%
\htdef{AIHARA 87B.ref}{AIHARA 87B (TPC) \cite{Aihara:1986mw}}%
\htdef{AIHARA 87B.meas}{%
\begin{ensuredisplaymath}
\htuse{Gamma54.gn} = \htuse{Gamma54.td}
\end{ensuredisplaymath} & \htuse{TPC.Gamma54.pub.AIHARA.87B}}%
\htdef{BAUER 94.cite}{\cite{Bauer:1993wn}}%
\htdef{BAUER 94.collab}{TPC}%
\htdef{BAUER 94.ref}{BAUER 94 (TPC) \cite{Bauer:1993wn}}%
\htdef{BAUER 94.meas}{%
\begin{ensuredisplaymath}
\htuse{Gamma82.gn} = \htuse{Gamma82.td}
\end{ensuredisplaymath} & \htuse{TPC.Gamma82.pub.BAUER.94}
\\
\begin{ensuredisplaymath}
\htuse{Gamma92.gn} = \htuse{Gamma92.td}
\end{ensuredisplaymath} & \htuse{TPC.Gamma92.pub.BAUER.94}}%
\htdef{MeasPaper}{%
\multicolumn{2}{l}{\htuse{BARATE 98.ref}} \\
\htuse{BARATE 98.meas} \\\hline
\multicolumn{2}{l}{\htuse{BARATE 98E.ref}} \\
\htuse{BARATE 98E.meas} \\\hline
\multicolumn{2}{l}{\htuse{BARATE 99K.ref}} \\
\htuse{BARATE 99K.meas} \\\hline
\multicolumn{2}{l}{\htuse{BARATE 99R.ref}} \\
\htuse{BARATE 99R.meas} \\\hline
\multicolumn{2}{l}{\htuse{BUSKULIC 96.ref}} \\
\htuse{BUSKULIC 96.meas} \\\hline
\multicolumn{2}{l}{\htuse{BUSKULIC 97C.ref}} \\
\htuse{BUSKULIC 97C.meas} \\\hline
\multicolumn{2}{l}{\htuse{SCHAEL 05C.ref}} \\
\htuse{SCHAEL 05C.meas} \\\hline
\multicolumn{2}{l}{\htuse{ALBRECHT 88B.ref}} \\
\htuse{ALBRECHT 88B.meas} \\\hline
\multicolumn{2}{l}{\htuse{ALBRECHT 92D.ref}} \\
\htuse{ALBRECHT 92D.meas} \\\hline
\multicolumn{2}{l}{\htuse{AUBERT 07AP.ref}} \\
\htuse{AUBERT 07AP.meas} \\\hline
\multicolumn{2}{l}{\htuse{AUBERT 08.ref}} \\
\htuse{AUBERT 08.meas} \\\hline
\multicolumn{2}{l}{\htuse{AUBERT 10F.ref}} \\
\htuse{AUBERT 10F.meas} \\\hline
\multicolumn{2}{l}{\htuse{DEL-AMO-SANCHEZ 11E.ref}} \\
\htuse{DEL-AMO-SANCHEZ 11E.meas} \\\hline
\multicolumn{2}{l}{\htuse{LEES 12X.ref}} \\
\htuse{LEES 12X.meas} \\\hline
\multicolumn{2}{l}{\htuse{LEES 12Y.ref}} \\
\htuse{LEES 12Y.meas} \\\hline
\multicolumn{2}{l}{\htuse{FUJIKAWA 08.ref}} \\
\htuse{FUJIKAWA 08.meas} \\\hline
\multicolumn{2}{l}{\htuse{INAMI 09.ref}} \\
\htuse{INAMI 09.meas} \\\hline
\multicolumn{2}{l}{\htuse{LEE 10.ref}} \\
\htuse{LEE 10.meas} \\\hline
\multicolumn{2}{l}{\htuse{RYU 14vpc.ref}} \\
\htuse{RYU 14vpc.meas} \\\hline
\multicolumn{2}{l}{\htuse{BEHREND 89B.ref}} \\
\htuse{BEHREND 89B.meas} \\\hline
\multicolumn{2}{l}{\htuse{ANASTASSOV 01.ref}} \\
\htuse{ANASTASSOV 01.meas} \\\hline
\multicolumn{2}{l}{\htuse{ANASTASSOV 97.ref}} \\
\htuse{ANASTASSOV 97.meas} \\\hline
\multicolumn{2}{l}{\htuse{ARTUSO 92.ref}} \\
\htuse{ARTUSO 92.meas} \\\hline
\multicolumn{2}{l}{\htuse{ARTUSO 94.ref}} \\
\htuse{ARTUSO 94.meas} \\\hline
\multicolumn{2}{l}{\htuse{BALEST 95C.ref}} \\
\htuse{BALEST 95C.meas} \\\hline
\multicolumn{2}{l}{\htuse{BARINGER 87.ref}} \\
\htuse{BARINGER 87.meas} \\\hline
\multicolumn{2}{l}{\htuse{BARTELT 96.ref}} \\
\htuse{BARTELT 96.meas} \\\hline
\multicolumn{2}{l}{\htuse{BATTLE 94.ref}} \\
\htuse{BATTLE 94.meas} \\\hline
\multicolumn{2}{l}{\htuse{BISHAI 99.ref}} \\
\htuse{BISHAI 99.meas} \\\hline
\multicolumn{2}{l}{\htuse{BORTOLETTO 93.ref}} \\
\htuse{BORTOLETTO 93.meas} \\\hline
\multicolumn{2}{l}{\htuse{COAN 96.ref}} \\
\htuse{COAN 96.meas} \\\hline
\multicolumn{2}{l}{\htuse{EDWARDS 00A.ref}} \\
\htuse{EDWARDS 00A.meas} \\\hline
\multicolumn{2}{l}{\htuse{GIBAUT 94B.ref}} \\
\htuse{GIBAUT 94B.meas} \\\hline
\multicolumn{2}{l}{\htuse{PROCARIO 93.ref}} \\
\htuse{PROCARIO 93.meas} \\\hline
\multicolumn{2}{l}{\htuse{RICHICHI 99.ref}} \\
\htuse{RICHICHI 99.meas} \\\hline
\multicolumn{2}{l}{\htuse{ARMS 05.ref}} \\
\htuse{ARMS 05.meas} \\\hline
\multicolumn{2}{l}{\htuse{BRIERE 03.ref}} \\
\htuse{BRIERE 03.meas} \\\hline
\multicolumn{2}{l}{\htuse{ABDALLAH 06A.ref}} \\
\htuse{ABDALLAH 06A.meas} \\\hline
\multicolumn{2}{l}{\htuse{ABREU 92N.ref}} \\
\htuse{ABREU 92N.meas} \\\hline
\multicolumn{2}{l}{\htuse{ABREU 94K.ref}} \\
\htuse{ABREU 94K.meas} \\\hline
\multicolumn{2}{l}{\htuse{ABREU 99X.ref}} \\
\htuse{ABREU 99X.meas} \\\hline
\multicolumn{2}{l}{\htuse{BYLSMA 87.ref}} \\
\htuse{BYLSMA 87.meas} \\\hline
\multicolumn{2}{l}{\htuse{ACCIARRI 01F.ref}} \\
\htuse{ACCIARRI 01F.meas} \\\hline
\multicolumn{2}{l}{\htuse{ACCIARRI 95.ref}} \\
\htuse{ACCIARRI 95.meas} \\\hline
\multicolumn{2}{l}{\htuse{ACCIARRI 95F.ref}} \\
\htuse{ACCIARRI 95F.meas} \\\hline
\multicolumn{2}{l}{\htuse{ACHARD 01D.ref}} \\
\htuse{ACHARD 01D.meas} \\\hline
\multicolumn{2}{l}{\htuse{ADEVA 91F.ref}} \\
\htuse{ADEVA 91F.meas} \\\hline
\multicolumn{2}{l}{\htuse{ABBIENDI 00C.ref}} \\
\htuse{ABBIENDI 00C.meas} \\\hline
\multicolumn{2}{l}{\htuse{ABBIENDI 00D.ref}} \\
\htuse{ABBIENDI 00D.meas} \\\hline
\multicolumn{2}{l}{\htuse{ABBIENDI 01J.ref}} \\
\htuse{ABBIENDI 01J.meas} \\\hline
\multicolumn{2}{l}{\htuse{ABBIENDI 03.ref}} \\
\htuse{ABBIENDI 03.meas} \\\hline
\multicolumn{2}{l}{\htuse{ABBIENDI 04J.ref}} \\
\htuse{ABBIENDI 04J.meas} \\\hline
\multicolumn{2}{l}{\htuse{ABBIENDI 99H.ref}} \\
\htuse{ABBIENDI 99H.meas} \\\hline
\multicolumn{2}{l}{\htuse{ACKERSTAFF 98M.ref}} \\
\htuse{ACKERSTAFF 98M.meas} \\\hline
\multicolumn{2}{l}{\htuse{ACKERSTAFF 99E.ref}} \\
\htuse{ACKERSTAFF 99E.meas} \\\hline
\multicolumn{2}{l}{\htuse{AKERS 94G.ref}} \\
\htuse{AKERS 94G.meas} \\\hline
\multicolumn{2}{l}{\htuse{AKERS 95Y.ref}} \\
\htuse{AKERS 95Y.meas} \\\hline
\multicolumn{2}{l}{\htuse{ALEXANDER 91D.ref}} \\
\htuse{ALEXANDER 91D.meas} \\\hline
\multicolumn{2}{l}{\htuse{AIHARA 87B.ref}} \\
\htuse{AIHARA 87B.meas} \\\hline
\multicolumn{2}{l}{\htuse{BAUER 94.ref}} \\
\htuse{BAUER 94.meas}}%
\htdef{BrStrangeVal}{%
\htQuantLine{Gamma10}{0.6960 \pm 0.0096}{-2} 
\htQuantLine{Gamma16}{0.4327 \pm 0.0149}{-2} 
\htQuantLine{Gamma23}{0.0640 \pm 0.0220}{-2} 
\htQuantLine{Gamma28}{0.0428 \pm 0.0216}{-2} 
\htQuantLine{Gamma35}{0.8386 \pm 0.0141}{-2} 
\htQuantLine{Gamma40}{0.3812 \pm 0.0129}{-2} 
\htQuantLine{Gamma44}{0.0234 \pm 0.0231}{-2} 
\htQuantLine{Gamma53}{0.0222 \pm 0.0202}{-2} 
\htQuantLine{Gamma128}{0.0155 \pm 0.0008}{-2} 
\htQuantLine{Gamma130}{0.0048 \pm 0.0012}{-2} 
\htQuantLine{Gamma132}{0.0094 \pm 0.0015}{-2} 
\htQuantLine{Gamma151}{0.0410 \pm 0.0092}{-2} 
\htQuantLine{Gamma168}{0.0022 \pm 0.0008}{-2} 
\htQuantLine{Gamma169}{0.0015 \pm 0.0006}{-2} 
\htQuantLine{Gamma802}{0.2923 \pm 0.0067}{-2} 
\htQuantLine{Gamma803}{0.0410 \pm 0.0143}{-2} 
\htQuantLine{Gamma822}{0.0001 \pm 0.0001}{-2} 
\htQuantLine{Gamma833}{0.0001 \pm 0.0001}{-2}}%
\htdef{BrStrangeTotVal}{%
\htQuantLine{Gamma110}{2.9087 \pm 0.0482}{-2}}%
\htdef{UnitarityQuants}{%
\htConstrLine{Gamma3}{17.3917 \pm 0.0396}{1.0000}{-2}{0} 
\htConstrLine{Gamma5}{17.8162 \pm 0.0410}{1.0000}{-2}{0} 
\htConstrLine{Gamma9}{10.8103 \pm 0.0526}{1.0000}{-2}{0} 
\htConstrLine{Gamma10}{0.6960 \pm 0.0096}{1.0000}{-2}{0} 
\htConstrLine{Gamma14}{25.5023 \pm 0.0918}{1.0000}{-2}{0} 
\htConstrLine{Gamma16}{0.4327 \pm 0.0149}{1.0000}{-2}{0} 
\htConstrLine{Gamma20}{9.2424 \pm 0.0997}{1.0000}{-2}{0} 
\htConstrLine{Gamma23}{0.0640 \pm 0.0220}{1.0000}{-2}{0} 
\htConstrLine{Gamma27}{1.0287 \pm 0.0749}{1.0000}{-2}{0} 
\htConstrLine{Gamma28}{0.0428 \pm 0.0216}{1.0000}{-2}{0} 
\htConstrLine{Gamma30}{0.1099 \pm 0.0391}{1.0000}{-2}{0} 
\htConstrLine{Gamma35}{0.8386 \pm 0.0141}{1.0000}{-2}{0} 
\htConstrLine{Gamma37}{0.1479 \pm 0.0053}{1.0000}{-2}{0} 
\htConstrLine{Gamma40}{0.3812 \pm 0.0129}{1.0000}{-2}{0} 
\htConstrLine{Gamma42}{0.1502 \pm 0.0071}{1.0000}{-2}{0} 
\htConstrLine{Gamma44}{0.0234 \pm 0.0231}{1.0000}{-2}{0} 
\htConstrLine{Gamma47}{0.0233 \pm 0.0007}{2.0000}{-2}{0} 
\htConstrLine{Gamma48}{0.1047 \pm 0.0247}{1.0000}{-2}{0} 
\htConstrLine{Gamma50}{0.0018 \pm 0.0002}{2.0000}{-2}{0} 
\htConstrLine{Gamma51}{0.0318 \pm 0.0119}{1.0000}{-2}{0} 
\htConstrLine{Gamma53}{0.0222 \pm 0.0202}{1.0000}{-2}{0} 
\htConstrLine{Gamma62}{8.9704 \pm 0.0515}{1.0000}{-2}{0} 
\htConstrLine{Gamma70}{2.7694 \pm 0.0711}{1.0000}{-2}{0} 
\htConstrLine{Gamma77}{0.0976 \pm 0.0355}{1.0000}{-2}{0} 
\htConstrLine{Gamma93}{0.1434 \pm 0.0027}{1.0000}{-2}{0} 
\htConstrLine{Gamma94}{0.0061 \pm 0.0018}{1.0000}{-2}{0} 
\htConstrLine{Gamma126}{0.1386 \pm 0.0072}{1.0000}{-2}{0} 
\htConstrLine{Gamma128}{0.0155 \pm 0.0008}{1.0000}{-2}{0} 
\htConstrLine{Gamma130}{0.0048 \pm 0.0012}{1.0000}{-2}{0} 
\htConstrLine{Gamma132}{0.0094 \pm 0.0015}{1.0000}{-2}{0} 
\htConstrLine{Gamma136}{0.0218 \pm 0.0013}{1.0000}{-2}{0} 
\htConstrLine{Gamma151}{0.0410 \pm 0.0092}{1.0000}{-2}{0} 
\htConstrLine{Gamma152}{0.4058 \pm 0.0419}{1.0000}{-2}{0} 
\htConstrLine{Gamma167}{0.0044 \pm 0.0016}{0.8310}{-2}{0} 
\htConstrLine{Gamma800}{1.9544 \pm 0.0647}{1.0000}{-2}{0} 
\htConstrLine{Gamma802}{0.2923 \pm 0.0067}{1.0000}{-2}{0} 
\htConstrLine{Gamma803}{0.0410 \pm 0.0143}{1.0000}{-2}{0} 
\htConstrLine{Gamma805}{0.0400 \pm 0.0200}{1.0000}{-2}{0} 
\htConstrLine{Gamma811}{0.0071 \pm 0.0016}{1.0000}{-2}{0} 
\htConstrLine{Gamma812}{0.0013 \pm 0.0027}{1.0000}{-2}{0} 
\htConstrLine{Gamma821}{0.0768 \pm 0.0030}{1.0000}{-2}{0} 
\htConstrLine{Gamma822}{0.0001 \pm 0.0001}{1.0000}{-2}{0} 
\htConstrLine{Gamma831}{0.0084 \pm 0.0006}{1.0000}{-2}{0} 
\htConstrLine{Gamma832}{0.0038 \pm 0.0009}{1.0000}{-2}{0} 
\htConstrLine{Gamma833}{0.0001 \pm 0.0001}{1.0000}{-2}{0} 
\htConstrLine{Gamma920}{0.0052 \pm 0.0004}{1.0000}{-2}{0} 
\htConstrLine{Gamma945}{0.0193 \pm 0.0038}{1.0000}{-2}{0} 
\htConstrLine{Gamma998}{0.0355 \pm 0.1031}{1.0000}{-2}{0}}%
\htdef{BaseQuants}{%
\htQuantLine{Gamma3}{17.3917 \pm 0.0396}{-2} 
\htQuantLine{Gamma5}{17.8162 \pm 0.0410}{-2} 
\htQuantLine{Gamma9}{10.8103 \pm 0.0526}{-2} 
\htQuantLine{Gamma10}{0.6960 \pm 0.0096}{-2} 
\htQuantLine{Gamma14}{25.5023 \pm 0.0918}{-2} 
\htQuantLine{Gamma16}{0.4327 \pm 0.0149}{-2} 
\htQuantLine{Gamma20}{9.2424 \pm 0.0997}{-2} 
\htQuantLine{Gamma23}{0.0640 \pm 0.0220}{-2} 
\htQuantLine{Gamma27}{1.0287 \pm 0.0749}{-2} 
\htQuantLine{Gamma28}{0.0428 \pm 0.0216}{-2} 
\htQuantLine{Gamma30}{0.1099 \pm 0.0391}{-2} 
\htQuantLine{Gamma35}{0.8386 \pm 0.0141}{-2} 
\htQuantLine{Gamma37}{0.1479 \pm 0.0053}{-2} 
\htQuantLine{Gamma40}{0.3812 \pm 0.0129}{-2} 
\htQuantLine{Gamma42}{0.1502 \pm 0.0071}{-2} 
\htQuantLine{Gamma44}{0.0234 \pm 0.0231}{-2} 
\htQuantLine{Gamma47}{0.0233 \pm 0.0007}{-2} 
\htQuantLine{Gamma48}{0.1047 \pm 0.0247}{-2} 
\htQuantLine{Gamma50}{0.0018 \pm 0.0002}{-2} 
\htQuantLine{Gamma51}{0.0318 \pm 0.0119}{-2} 
\htQuantLine{Gamma53}{0.0222 \pm 0.0202}{-2} 
\htQuantLine{Gamma62}{8.9704 \pm 0.0515}{-2} 
\htQuantLine{Gamma70}{2.7694 \pm 0.0711}{-2} 
\htQuantLine{Gamma77}{0.0976 \pm 0.0355}{-2} 
\htQuantLine{Gamma93}{0.1434 \pm 0.0027}{-2} 
\htQuantLine{Gamma94}{0.0061 \pm 0.0018}{-2} 
\htQuantLine{Gamma126}{0.1386 \pm 0.0072}{-2} 
\htQuantLine{Gamma128}{0.0155 \pm 0.0008}{-2} 
\htQuantLine{Gamma130}{0.0048 \pm 0.0012}{-2} 
\htQuantLine{Gamma132}{0.0094 \pm 0.0015}{-2} 
\htQuantLine{Gamma136}{0.0218 \pm 0.0013}{-2} 
\htQuantLine{Gamma151}{0.0410 \pm 0.0092}{-2} 
\htQuantLine{Gamma152}{0.4058 \pm 0.0419}{-2} 
\htQuantLine{Gamma167}{0.0044 \pm 0.0016}{-2} 
\htQuantLine{Gamma800}{1.9544 \pm 0.0647}{-2} 
\htQuantLine{Gamma802}{0.2923 \pm 0.0067}{-2} 
\htQuantLine{Gamma803}{0.0410 \pm 0.0143}{-2} 
\htQuantLine{Gamma805}{0.0400 \pm 0.0200}{-2} 
\htQuantLine{Gamma811}{0.0071 \pm 0.0016}{-2} 
\htQuantLine{Gamma812}{0.0013 \pm 0.0027}{-2} 
\htQuantLine{Gamma821}{0.0768 \pm 0.0030}{-2} 
\htQuantLine{Gamma822}{0.0001 \pm 0.0001}{-2} 
\htQuantLine{Gamma831}{0.0084 \pm 0.0006}{-2} 
\htQuantLine{Gamma832}{0.0038 \pm 0.0009}{-2} 
\htQuantLine{Gamma833}{0.0001 \pm 0.0001}{-2} 
\htQuantLine{Gamma920}{0.0052 \pm 0.0004}{-2} 
\htQuantLine{Gamma945}{0.0193 \pm 0.0038}{-2}}%
\htdef{BrCorr}{%
%%
%% basis quantities correlation, 1
%%
\ifhevea\begin{table}\fi%% otherwise cannot have normalsize caption
\begin{center}
\ifhevea
\caption{Basis quantities correlation coefficients in percent, subtable 1.\label{tab:tau:br-fit-corr1}}%
\else
\begin{minipage}{\linewidth}
\begin{center}
\captionof{table}{Basis quantities correlation coefficients in percent, subtable 1.}\label{tab:tau:br-fit-corr1}%
\fi
\begin{envsmall}
\begin{center}
\renewcommand*{\arraystretch}{1.1}%
\begin{tabular}{rrrrrrrrrrrrrrr}
\hline
\( \Gamma_{5} \) &   23 &  &  &  &  &  &  &  &  &  &  &  &  &  \\
\( \Gamma_{9} \) &    7 &    5 &  &  &  &  &  &  &  &  &  &  &  &  \\
\( \Gamma_{10} \) &    3 &    5 &    1 &  &  &  &  &  &  &  &  &  &  &  \\
\( \Gamma_{14} \) &  -13 &  -14 &  -12 &   -3 &  &  &  &  &  &  &  &  &  &  \\
\( \Gamma_{16} \) &    0 &   -1 &    2 &   -1 &  -16 &  &  &  &  &  &  &  &  &  \\
\( \Gamma_{20} \) &   -5 &   -5 &   -7 &   -1 &  -40 &    2 &  &  &  &  &  &  &  &  \\
\( \Gamma_{23} \) &    0 &    0 &    0 &   -2 &    2 &  -13 &  -22 &  &  &  &  &  &  &  \\
\( \Gamma_{27} \) &   -4 &   -3 &   -8 &   -1 &    0 &    3 &  -36 &    6 &  &  &  &  &  &  \\
\( \Gamma_{28} \) &    0 &    0 &    0 &   -2 &    2 &  -13 &    5 &  -21 &  -29 &  &  &  &  &  \\
\( \Gamma_{30} \) &   -5 &   -4 &  -11 &   -2 &   -9 &    0 &    6 &    0 &  -42 &    0 &  &  &  &  \\
\( \Gamma_{35} \) &    0 &    0 &    0 &    0 &    0 &    0 &    0 &    1 &    0 &    1 &    0 &  &  &  \\
\( \Gamma_{37} \) &    0 &    0 &    0 &    0 &    0 &   -2 &    1 &   -3 &    1 &   -3 &    0 &  -22 &  &  \\
\( \Gamma_{40} \) &    0 &    0 &    0 &    0 &    0 &    1 &    0 &    1 &   -2 &    1 &    0 &  -12 &    4 &  \\
 & \( \Gamma_{3} \) & \( \Gamma_{5} \) & \( \Gamma_{9} \) & \( \Gamma_{10} \) & \( \Gamma_{14} \) & \( \Gamma_{16} \) & \( \Gamma_{20} \) & \( \Gamma_{23} \) & \( \Gamma_{27} \) & \( \Gamma_{28} \) & \( \Gamma_{30} \) & \( \Gamma_{35} \) & \( \Gamma_{37} \) & \( \Gamma_{40} \)
\\\hline
\end{tabular}
\end{center}
\end{envsmall}
\ifhevea\else
\end{center}
\end{minipage}
\fi
\end{center}
\ifhevea\end{table}\fi
%%
%% basis quantities correlation, 2
%%
\ifhevea\begin{table}\fi%% otherwise cannot have normalsize caption
\begin{center}
\ifhevea
\caption{Basis quantities correlation coefficients in percent, subtable 2.\label{tab:tau:br-fit-corr2}}%
\else
\begin{minipage}{\linewidth}
\begin{center}
\captionof{table}{Basis quantities correlation coefficients in percent, subtable 2.}\label{tab:tau:br-fit-corr2}%
\fi
\begin{envsmall}
\begin{center}
\renewcommand*{\arraystretch}{1.1}%
\begin{tabular}{rrrrrrrrrrrrrrr}
\hline
\( \Gamma_{42} \) &    0 &    0 &    0 &    0 &    1 &   -3 &    1 &   -5 &    1 &   -5 &    0 &    2 &  -21 &  -20 \\
\( \Gamma_{44} \) &    0 &    0 &    0 &    0 &    0 &    0 &    0 &    0 &    0 &    0 &    0 &   -1 &    0 &   -4 \\
\( \Gamma_{47} \) &    0 &    0 &    0 &    0 &    0 &    0 &    0 &    0 &    0 &    0 &    0 &   -1 &    1 &   -4 \\
\( \Gamma_{48} \) &    0 &    0 &    0 &    0 &    0 &    0 &    0 &    0 &    0 &    0 &    0 &   -3 &    0 &   -2 \\
\( \Gamma_{50} \) &    0 &    0 &    0 &    0 &    0 &    0 &    0 &   -1 &    0 &   -1 &    0 &    0 &    7 &    0 \\
\( \Gamma_{51} \) &    0 &    0 &    0 &    0 &    0 &    0 &    0 &    0 &    0 &    0 &    0 &   -1 &    0 &   -1 \\
\( \Gamma_{53} \) &    0 &    0 &    0 &    0 &    0 &    0 &    0 &    0 &    0 &    0 &    0 &    0 &    0 &    0 \\
\( \Gamma_{62} \) &   -3 &   -5 &    8 &    0 &   -4 &    5 &   -7 &   -1 &   -5 &   -1 &   -5 &    0 &    0 &    0 \\
\( \Gamma_{70} \) &   -6 &   -6 &   -7 &   -1 &   -8 &   -1 &   -1 &    0 &   -1 &    0 &    3 &    0 &    0 &    0 \\
\( \Gamma_{77} \) &   -1 &    0 &   -3 &   -1 &   -2 &    0 &    0 &    0 &    2 &    0 &    2 &    0 &    0 &    0 \\
\( \Gamma_{93} \) &   -1 &   -1 &    3 &    0 &   -1 &    2 &   -1 &    0 &   -1 &    0 &   -1 &    0 &    0 &    0 \\
\( \Gamma_{94} \) &    0 &    0 &    0 &    0 &    0 &    0 &    0 &    0 &    0 &    0 &    0 &    0 &    0 &    0 \\
\( \Gamma_{126} \) &    0 &    0 &    0 &    0 &    0 &    0 &   -1 &    0 &    0 &    0 &   -2 &    0 &    0 &    0 \\
\( \Gamma_{128} \) &    0 &    0 &    1 &    0 &    0 &    1 &    0 &   -1 &    0 &   -1 &    0 &    0 &    0 &    0 \\
 & \( \Gamma_{3} \) & \( \Gamma_{5} \) & \( \Gamma_{9} \) & \( \Gamma_{10} \) & \( \Gamma_{14} \) & \( \Gamma_{16} \) & \( \Gamma_{20} \) & \( \Gamma_{23} \) & \( \Gamma_{27} \) & \( \Gamma_{28} \) & \( \Gamma_{30} \) & \( \Gamma_{35} \) & \( \Gamma_{37} \) & \( \Gamma_{40} \)
\\\hline
\end{tabular}
\end{center}
\end{envsmall}
\ifhevea\else
\end{center}
\end{minipage}
\fi
\end{center}
\ifhevea\end{table}\fi
%%
%% basis quantities correlation, 3
%%
\ifhevea\begin{table}\fi%% otherwise cannot have normalsize caption
\begin{center}
\ifhevea
\caption{Basis quantities correlation coefficients in percent, subtable 3.\label{tab:tau:br-fit-corr3}}%
\else
\begin{minipage}{\linewidth}
\begin{center}
\captionof{table}{Basis quantities correlation coefficients in percent, subtable 3.}\label{tab:tau:br-fit-corr3}%
\fi
\begin{envsmall}
\begin{center}
\renewcommand*{\arraystretch}{1.1}%
\begin{tabular}{rrrrrrrrrrrrrrr}
\hline
\( \Gamma_{130} \) &    0 &    0 &    0 &    0 &    0 &    0 &    0 &    0 &    0 &    0 &    0 &    0 &    0 &    0 \\
\( \Gamma_{132} \) &    0 &    0 &    0 &    0 &    0 &    0 &    0 &    0 &    0 &    0 &    0 &    0 &    0 &    0 \\
\( \Gamma_{136} \) &    0 &    0 &    0 &    0 &    0 &    0 &    0 &    0 &    0 &    0 &    0 &    0 &    0 &    0 \\
\( \Gamma_{151} \) &    0 &    0 &    0 &    0 &    0 &    0 &    0 &    0 &    0 &    0 &    0 &    0 &    0 &    0 \\
\( \Gamma_{152} \) &   -1 &    0 &   -3 &   -1 &   -2 &    0 &   -1 &    0 &    2 &    0 &    2 &    0 &    0 &    0 \\
\( \Gamma_{167} \) &    0 &    0 &    0 &    0 &    0 &    0 &    0 &    0 &    0 &    0 &    0 &    0 &    0 &    0 \\
\( \Gamma_{800} \) &   -2 &   -2 &   -2 &    0 &   -3 &    0 &    0 &    0 &    0 &    0 &    1 &    0 &    0 &    0 \\
\( \Gamma_{802} \) &   -1 &   -1 &    0 &    0 &   -1 &    0 &   -2 &    0 &   -2 &    0 &   -1 &    0 &    0 &    0 \\
\( \Gamma_{803} \) &    0 &    0 &    0 &    0 &    0 &    0 &    0 &    0 &    0 &    0 &    0 &    0 &    0 &    0 \\
\( \Gamma_{805} \) &    0 &    0 &    0 &    0 &    0 &    0 &    0 &    0 &    0 &    0 &    0 &    0 &    0 &    0 \\
\( \Gamma_{811} \) &    0 &    0 &    0 &    0 &    0 &    0 &    0 &    0 &    0 &    0 &    0 &    0 &    0 &    0 \\
\( \Gamma_{812} \) &    0 &    1 &    0 &    0 &    0 &    0 &    0 &    0 &    0 &    0 &    0 &    0 &    0 &    0 \\
\( \Gamma_{821} \) &    0 &    0 &    1 &    0 &    0 &    0 &   -1 &    0 &    0 &    0 &   -1 &    0 &    0 &    0 \\
\( \Gamma_{822} \) &    0 &    0 &    0 &    0 &    0 &    0 &    0 &    0 &    0 &    0 &    0 &    0 &    0 &    0 \\
 & \( \Gamma_{3} \) & \( \Gamma_{5} \) & \( \Gamma_{9} \) & \( \Gamma_{10} \) & \( \Gamma_{14} \) & \( \Gamma_{16} \) & \( \Gamma_{20} \) & \( \Gamma_{23} \) & \( \Gamma_{27} \) & \( \Gamma_{28} \) & \( \Gamma_{30} \) & \( \Gamma_{35} \) & \( \Gamma_{37} \) & \( \Gamma_{40} \)
\\\hline
\end{tabular}
\end{center}
\end{envsmall}
\ifhevea\else
\end{center}
\end{minipage}
\fi
\end{center}
\ifhevea\end{table}\fi
%%
%% basis quantities correlation, 4
%%
\ifhevea\begin{table}\fi%% otherwise cannot have normalsize caption
\begin{center}
\ifhevea
\caption{Basis quantities correlation coefficients in percent, subtable 4.\label{tab:tau:br-fit-corr4}}%
\else
\begin{minipage}{\linewidth}
\begin{center}
\captionof{table}{Basis quantities correlation coefficients in percent, subtable 4.}\label{tab:tau:br-fit-corr4}%
\fi
\begin{envsmall}
\begin{center}
\renewcommand*{\arraystretch}{1.1}%
\begin{tabular}{rrrrrrrrrrrrrrr}
\hline
\( \Gamma_{831} \) &    0 &    0 &    0 &    0 &    0 &    0 &    0 &    0 &    0 &    0 &    0 &    0 &    0 &    0 \\
\( \Gamma_{832} \) &    0 &    0 &    0 &    0 &    0 &    0 &    0 &    0 &    0 &    0 &    0 &    0 &    0 &    0 \\
\( \Gamma_{833} \) &    0 &    0 &    0 &    0 &    0 &    0 &    0 &    0 &    0 &    0 &    0 &    0 &    0 &    0 \\
\( \Gamma_{920} \) &    0 &    0 &    0 &    0 &    0 &    0 &    0 &    0 &    0 &    0 &    0 &    0 &    0 &    0 \\
\( \Gamma_{945} \) &    0 &    0 &    0 &    0 &    0 &    0 &    0 &    0 &    0 &    0 &    0 &    0 &    0 &    0 \\
 & \( \Gamma_{3} \) & \( \Gamma_{5} \) & \( \Gamma_{9} \) & \( \Gamma_{10} \) & \( \Gamma_{14} \) & \( \Gamma_{16} \) & \( \Gamma_{20} \) & \( \Gamma_{23} \) & \( \Gamma_{27} \) & \( \Gamma_{28} \) & \( \Gamma_{30} \) & \( \Gamma_{35} \) & \( \Gamma_{37} \) & \( \Gamma_{40} \)
\\\hline
\end{tabular}
\end{center}
\end{envsmall}
\ifhevea\else
\end{center}
\end{minipage}
\fi
\end{center}
\ifhevea\end{table}\fi
%%
%% basis quantities correlation, 5
%%
\ifhevea\begin{table}\fi%% otherwise cannot have normalsize caption
\begin{center}
\ifhevea
\caption{Basis quantities correlation coefficients in percent, subtable 5.\label{tab:tau:br-fit-corr5}}%
\else
\begin{minipage}{\linewidth}
\begin{center}
\captionof{table}{Basis quantities correlation coefficients in percent, subtable 5.}\label{tab:tau:br-fit-corr5}%
\fi
\begin{envsmall}
\begin{center}
\renewcommand*{\arraystretch}{1.1}%
\begin{tabular}{rrrrrrrrrrrrrrr}
\hline
\( \Gamma_{44} \) &    0 &  &  &  &  &  &  &  &  &  &  &  &  &  \\
\( \Gamma_{47} \) &    1 &    0 &  &  &  &  &  &  &  &  &  &  &  &  \\
\( \Gamma_{48} \) &   -1 &   -6 &    0 &  &  &  &  &  &  &  &  &  &  &  \\
\( \Gamma_{50} \) &    5 &    0 &   -7 &    0 &  &  &  &  &  &  &  &  &  &  \\
\( \Gamma_{51} \) &    0 &   -3 &    0 &   -6 &    0 &  &  &  &  &  &  &  &  &  \\
\( \Gamma_{53} \) &    0 &    0 &    0 &    0 &    0 &    0 &  &  &  &  &  &  &  &  \\
\( \Gamma_{62} \) &    0 &    0 &    1 &    0 &    0 &    0 &    0 &  &  &  &  &  &  &  \\
\( \Gamma_{70} \) &    0 &    0 &    0 &    0 &    0 &    0 &    0 &  -20 &  &  &  &  &  &  \\
\( \Gamma_{77} \) &    0 &    0 &    0 &    0 &    0 &    0 &    0 &   -1 &   -7 &  &  &  &  &  \\
\( \Gamma_{93} \) &    0 &    0 &    0 &    0 &    0 &    0 &    0 &   14 &   -4 &    0 &  &  &  &  \\
\( \Gamma_{94} \) &    0 &    0 &    0 &    0 &    0 &    0 &    0 &    0 &   -2 &    0 &    0 &  &  &  \\
\( \Gamma_{126} \) &    0 &    0 &    1 &    0 &    0 &    0 &    0 &    1 &    0 &   -5 &    0 &    0 &  &  \\
\( \Gamma_{128} \) &    0 &    0 &    1 &    0 &    0 &    0 &    0 &    2 &    0 &    0 &    1 &    0 &    4 &  \\
 & \( \Gamma_{42} \) & \( \Gamma_{44} \) & \( \Gamma_{47} \) & \( \Gamma_{48} \) & \( \Gamma_{50} \) & \( \Gamma_{51} \) & \( \Gamma_{53} \) & \( \Gamma_{62} \) & \( \Gamma_{70} \) & \( \Gamma_{77} \) & \( \Gamma_{93} \) & \( \Gamma_{94} \) & \( \Gamma_{126} \) & \( \Gamma_{128} \)
\\\hline
\end{tabular}
\end{center}
\end{envsmall}
\ifhevea\else
\end{center}
\end{minipage}
\fi
\end{center}
\ifhevea\end{table}\fi
%%
%% basis quantities correlation, 6
%%
\ifhevea\begin{table}\fi%% otherwise cannot have normalsize caption
\begin{center}
\ifhevea
\caption{Basis quantities correlation coefficients in percent, subtable 6.\label{tab:tau:br-fit-corr6}}%
\else
\begin{minipage}{\linewidth}
\begin{center}
\captionof{table}{Basis quantities correlation coefficients in percent, subtable 6.}\label{tab:tau:br-fit-corr6}%
\fi
\begin{envsmall}
\begin{center}
\renewcommand*{\arraystretch}{1.1}%
\begin{tabular}{rrrrrrrrrrrrrrr}
\hline
\( \Gamma_{130} \) &    0 &    0 &    0 &    0 &    0 &    0 &    0 &    0 &    0 &   -1 &    0 &    0 &    1 &    1 \\
\( \Gamma_{132} \) &    0 &    0 &    0 &    0 &    0 &    0 &    0 &    0 &    0 &    0 &    0 &    0 &    2 &    1 \\
\( \Gamma_{136} \) &    0 &    0 &    0 &    0 &    0 &    0 &    0 &    0 &   -1 &    0 &    0 &    0 &    0 &    0 \\
\( \Gamma_{151} \) &    0 &    0 &    0 &    0 &    0 &    0 &    0 &    0 &   12 &    0 &    0 &    0 &    0 &    0 \\
\( \Gamma_{152} \) &    0 &    0 &    0 &    0 &    0 &    0 &    0 &   -1 &  -11 &  -64 &    0 &    0 &    0 &    0 \\
\( \Gamma_{167} \) &    0 &    0 &    0 &    0 &    0 &    0 &    0 &   -1 &    0 &    0 &    1 &    0 &    0 &    0 \\
\( \Gamma_{800} \) &    0 &    0 &    0 &    0 &    0 &    0 &    0 &   -8 &  -69 &   -2 &   -1 &    0 &    0 &    0 \\
\( \Gamma_{802} \) &    0 &    0 &    0 &    0 &    0 &    0 &    0 &   16 &   -6 &    0 &    0 &    0 &    0 &    0 \\
\( \Gamma_{803} \) &    0 &    0 &    0 &    0 &    0 &    0 &    0 &   -1 &  -19 &    0 &    0 &   -2 &    0 &   -1 \\
\( \Gamma_{805} \) &    0 &    0 &    0 &    0 &    0 &    0 &    0 &    0 &    0 &    0 &    0 &    0 &    0 &    0 \\
\( \Gamma_{811} \) &    0 &    0 &    0 &    0 &    0 &    0 &    0 &    0 &    0 &    0 &    0 &    0 &    0 &    0 \\
\( \Gamma_{812} \) &    0 &    0 &    0 &    0 &   -1 &    0 &    0 &    0 &   -1 &    0 &    0 &    0 &    0 &    0 \\
\( \Gamma_{821} \) &    0 &    0 &    0 &    0 &    0 &    0 &    0 &    0 &   -1 &    0 &    0 &    0 &    0 &    0 \\
\( \Gamma_{822} \) &    0 &    0 &    0 &    0 &    0 &    0 &    0 &    0 &    0 &    0 &    0 &    0 &    0 &    0 \\
 & \( \Gamma_{42} \) & \( \Gamma_{44} \) & \( \Gamma_{47} \) & \( \Gamma_{48} \) & \( \Gamma_{50} \) & \( \Gamma_{51} \) & \( \Gamma_{53} \) & \( \Gamma_{62} \) & \( \Gamma_{70} \) & \( \Gamma_{77} \) & \( \Gamma_{93} \) & \( \Gamma_{94} \) & \( \Gamma_{126} \) & \( \Gamma_{128} \)
\\\hline
\end{tabular}
\end{center}
\end{envsmall}
\ifhevea\else
\end{center}
\end{minipage}
\fi
\end{center}
\ifhevea\end{table}\fi
%%
%% basis quantities correlation, 7
%%
\ifhevea\begin{table}\fi%% otherwise cannot have normalsize caption
\begin{center}
\ifhevea
\caption{Basis quantities correlation coefficients in percent, subtable 7.\label{tab:tau:br-fit-corr7}}%
\else
\begin{minipage}{\linewidth}
\begin{center}
\captionof{table}{Basis quantities correlation coefficients in percent, subtable 7.}\label{tab:tau:br-fit-corr7}%
\fi
\begin{envsmall}
\begin{center}
\renewcommand*{\arraystretch}{1.1}%
\begin{tabular}{rrrrrrrrrrrrrrr}
\hline
\( \Gamma_{831} \) &    0 &    0 &    0 &    0 &    0 &    0 &    0 &    0 &    0 &    0 &    0 &    0 &    0 &    0 \\
\( \Gamma_{832} \) &    0 &    0 &    0 &    0 &    0 &    0 &    0 &    0 &    0 &    0 &    0 &    0 &    0 &    0 \\
\( \Gamma_{833} \) &    0 &    0 &    0 &    0 &    0 &    0 &    0 &    0 &    0 &    0 &    0 &    0 &    0 &    0 \\
\( \Gamma_{920} \) &    0 &    0 &    0 &    0 &    0 &    0 &    0 &    0 &    0 &    0 &    0 &    0 &    0 &    0 \\
\( \Gamma_{945} \) &    0 &    0 &    0 &    0 &    0 &    0 &    0 &    0 &    0 &    0 &    0 &    0 &    0 &    0 \\
 & \( \Gamma_{42} \) & \( \Gamma_{44} \) & \( \Gamma_{47} \) & \( \Gamma_{48} \) & \( \Gamma_{50} \) & \( \Gamma_{51} \) & \( \Gamma_{53} \) & \( \Gamma_{62} \) & \( \Gamma_{70} \) & \( \Gamma_{77} \) & \( \Gamma_{93} \) & \( \Gamma_{94} \) & \( \Gamma_{126} \) & \( \Gamma_{128} \)
\\\hline
\end{tabular}
\end{center}
\end{envsmall}
\ifhevea\else
\end{center}
\end{minipage}
\fi
\end{center}
\ifhevea\end{table}\fi
%%
%% basis quantities correlation, 8
%%
\ifhevea\begin{table}\fi%% otherwise cannot have normalsize caption
\begin{center}
\ifhevea
\caption{Basis quantities correlation coefficients in percent, subtable 8.\label{tab:tau:br-fit-corr8}}%
\else
\begin{minipage}{\linewidth}
\begin{center}
\captionof{table}{Basis quantities correlation coefficients in percent, subtable 8.}\label{tab:tau:br-fit-corr8}%
\fi
\begin{envsmall}
\begin{center}
\renewcommand*{\arraystretch}{1.1}%
\begin{tabular}{rrrrrrrrrrrrrrr}
\hline
\( \Gamma_{132} \) &    0 &  &  &  &  &  &  &  &  &  &  &  &  &  \\
\( \Gamma_{136} \) &    0 &    0 &  &  &  &  &  &  &  &  &  &  &  &  \\
\( \Gamma_{151} \) &    0 &    0 &    0 &  &  &  &  &  &  &  &  &  &  &  \\
\( \Gamma_{152} \) &    0 &    0 &    0 &    0 &  &  &  &  &  &  &  &  &  &  \\
\( \Gamma_{167} \) &    0 &    0 &    0 &    0 &    0 &  &  &  &  &  &  &  &  &  \\
\( \Gamma_{800} \) &    0 &    0 &    0 &  -14 &   -3 &    0 &  &  &  &  &  &  &  &  \\
\( \Gamma_{802} \) &    0 &    0 &    0 &   -2 &    0 &    1 &   -1 &  &  &  &  &  &  &  \\
\( \Gamma_{803} \) &    0 &    0 &    0 &  -58 &    0 &    0 &    9 &    1 &  &  &  &  &  &  \\
\( \Gamma_{805} \) &    0 &    0 &    0 &    0 &    0 &    0 &    0 &    0 &    0 &  &  &  &  &  \\
\( \Gamma_{811} \) &    0 &   -1 &   20 &    0 &    0 &    0 &    0 &    0 &    0 &    0 &  &  &  &  \\
\( \Gamma_{812} \) &    0 &   -2 &   -8 &    0 &    0 &    0 &    0 &    0 &    0 &    0 &  -16 &  &  &  \\
\( \Gamma_{821} \) &    0 &    0 &   47 &    0 &    0 &    0 &    0 &    0 &    0 &    0 &    8 &   -4 &  &  \\
\( \Gamma_{822} \) &    0 &    0 &   -1 &    0 &    0 &    0 &    0 &    0 &    0 &    0 &    0 &    0 &   -1 &  \\
 & \( \Gamma_{130} \) & \( \Gamma_{132} \) & \( \Gamma_{136} \) & \( \Gamma_{151} \) & \( \Gamma_{152} \) & \( \Gamma_{167} \) & \( \Gamma_{800} \) & \( \Gamma_{802} \) & \( \Gamma_{803} \) & \( \Gamma_{805} \) & \( \Gamma_{811} \) & \( \Gamma_{812} \) & \( \Gamma_{821} \) & \( \Gamma_{822} \)
\\\hline
\end{tabular}
\end{center}
\end{envsmall}
\ifhevea\else
\end{center}
\end{minipage}
\fi
\end{center}
\ifhevea\end{table}\fi
%%
%% basis quantities correlation, 9
%%
\ifhevea\begin{table}\fi%% otherwise cannot have normalsize caption
\begin{center}
\ifhevea
\caption{Basis quantities correlation coefficients in percent, subtable 9.\label{tab:tau:br-fit-corr9}}%
\else
\begin{minipage}{\linewidth}
\begin{center}
\captionof{table}{Basis quantities correlation coefficients in percent, subtable 9.}\label{tab:tau:br-fit-corr9}%
\fi
\begin{envsmall}
\begin{center}
\renewcommand*{\arraystretch}{1.1}%
\begin{tabular}{rrrrrrrrrrrrrrr}
\hline
\( \Gamma_{831} \) &    0 &    0 &   39 &    0 &    0 &    0 &    0 &    0 &    0 &    0 &   14 &   -4 &   39 &   -1 \\
\( \Gamma_{832} \) &    0 &    0 &    3 &    0 &    0 &    0 &    0 &    0 &    0 &    0 &    2 &    0 &    3 &    0 \\
\( \Gamma_{833} \) &    0 &    0 &   -1 &    0 &    0 &    0 &    0 &    0 &    0 &    0 &    0 &    0 &   -1 &    0 \\
\( \Gamma_{920} \) &    0 &    0 &   21 &    0 &    0 &    0 &    0 &    0 &    0 &    0 &    3 &   -2 &   35 &   -1 \\
\( \Gamma_{945} \) &    0 &   -1 &   25 &    0 &    0 &    0 &    0 &    0 &    0 &    0 &   10 &  -11 &   10 &    0 \\
 & \( \Gamma_{130} \) & \( \Gamma_{132} \) & \( \Gamma_{136} \) & \( \Gamma_{151} \) & \( \Gamma_{152} \) & \( \Gamma_{167} \) & \( \Gamma_{800} \) & \( \Gamma_{802} \) & \( \Gamma_{803} \) & \( \Gamma_{805} \) & \( \Gamma_{811} \) & \( \Gamma_{812} \) & \( \Gamma_{821} \) & \( \Gamma_{822} \)
\\\hline
\end{tabular}
\end{center}
\end{envsmall}
\ifhevea\else
\end{center}
\end{minipage}
\fi
\end{center}
\ifhevea\end{table}\fi
%%
%% basis quantities correlation, 10
%%
\ifhevea\begin{table}\fi%% otherwise cannot have normalsize caption
\begin{center}
\ifhevea
\caption{Basis quantities correlation coefficients in percent, subtable 10.\label{tab:tau:br-fit-corr10}}%
\else
\begin{minipage}{\linewidth}
\begin{center}
\captionof{table}{Basis quantities correlation coefficients in percent, subtable 10.}\label{tab:tau:br-fit-corr10}%
\fi
\begin{envsmall}
\begin{center}
\renewcommand*{\arraystretch}{1.1}%
\begin{tabular}{rrrrrr}
\hline
\( \Gamma_{832} \) &   -2 &  &  &  &  \\
\( \Gamma_{833} \) &   -1 &   -1 &  &  &  \\
\( \Gamma_{920} \) &   17 &    1 &    0 &  &  \\
\( \Gamma_{945} \) &   17 &    2 &    0 &    4 &  \\
 & \( \Gamma_{831} \) & \( \Gamma_{832} \) & \( \Gamma_{833} \) & \( \Gamma_{920} \) & \( \Gamma_{945} \)
\\\hline
\end{tabular}
\end{center}
\end{envsmall}
\ifhevea\else
\end{center}
\end{minipage}
\fi
\end{center}
\ifhevea\end{table}\fi}%
\htconstrdef{Gamma1.c}{\Gamma_{1}}{\Gamma_{3} + \Gamma_{5} + \Gamma_{9} + \Gamma_{10} + \Gamma_{14} + \Gamma_{16} + \Gamma_{20} + \Gamma_{23} + \Gamma_{27} + \Gamma_{28} + \Gamma_{30} + \Gamma_{35} + \Gamma_{40} + \Gamma_{44} + \Gamma_{37} + \Gamma_{42} + \Gamma_{47} + \Gamma_{48} + \Gamma_{804} + \Gamma_{50} + \Gamma_{51} + \Gamma_{806} + \Gamma_{126}\cdot{}\Gamma_{\eta\to\text{neutral}} + \Gamma_{128}\cdot{}\Gamma_{\eta\to\text{neutral}} + \Gamma_{130}\cdot{}\Gamma_{\eta\to\text{neutral}} + \Gamma_{132}\cdot{}\Gamma_{\eta\to\text{neutral}} + \Gamma_{800}\cdot{}\Gamma_{\omega\to\pi^0\gamma} + \Gamma_{151}\cdot{}\Gamma_{\omega\to\pi^0\gamma} + \Gamma_{152}\cdot{}\Gamma_{\omega\to\pi^0\gamma} + \Gamma_{167}\cdot{}\Gamma_{\phi\to K_S K_L}}{\Gamma_{3} + \Gamma_{5} + \Gamma_{9} + \Gamma_{10} + \Gamma_{14} + \Gamma_{16}  \\ 
  {}& + \Gamma_{20} + \Gamma_{23} + \Gamma_{27} + \Gamma_{28} + \Gamma_{30} + \Gamma_{35}  \\ 
  {}& + \Gamma_{40} + \Gamma_{44} + \Gamma_{37} + \Gamma_{42} + \Gamma_{47} + \Gamma_{48}  \\ 
  {}& + \Gamma_{804} + \Gamma_{50} + \Gamma_{51} + \Gamma_{806} + \Gamma_{126}\cdot{}\Gamma_{\eta\to\text{neutral}}  \\ 
  {}& + \Gamma_{128}\cdot{}\Gamma_{\eta\to\text{neutral}} + \Gamma_{130}\cdot{}\Gamma_{\eta\to\text{neutral}} + \Gamma_{132}\cdot{}\Gamma_{\eta\to\text{neutral}}  \\ 
  {}& + \Gamma_{800}\cdot{}\Gamma_{\omega\to\pi^0\gamma} + \Gamma_{151}\cdot{}\Gamma_{\omega\to\pi^0\gamma} + \Gamma_{152}\cdot{}\Gamma_{\omega\to\pi^0\gamma}  \\ 
  {}& + \Gamma_{167}\cdot{}\Gamma_{\phi\to K_S K_L}}%
\htconstrdef{Gamma2.c}{\Gamma_{2}}{\Gamma_{3} + \Gamma_{5} + \Gamma_{9} + \Gamma_{10} + \Gamma_{14} + \Gamma_{16} + \Gamma_{20} + \Gamma_{23} + \Gamma_{27} + \Gamma_{28} + \Gamma_{30} + \Gamma_{35}\cdot{}(\Gamma_{<\bar{K}^0|K_S>}\cdot{}\Gamma_{K_S\to\pi^0\pi^0}+\Gamma_{<\bar{K}^0|K_L>}) + \Gamma_{40}\cdot{}(\Gamma_{<\bar{K}^0|K_S>}\cdot{}\Gamma_{K_S\to\pi^0\pi^0}+\Gamma_{<\bar{K}^0|K_L>}) + \Gamma_{44}\cdot{}(\Gamma_{<\bar{K}^0|K_S>}\cdot{}\Gamma_{K_S\to\pi^0\pi^0}+\Gamma_{<\bar{K}^0|K_L>}) + \Gamma_{37}\cdot{}(\Gamma_{<\bar{K}^0|K_S>}\cdot{}\Gamma_{K_S\to\pi^0\pi^0}+\Gamma_{<\bar{K}^0|K_L>}) + \Gamma_{42}\cdot{}(\Gamma_{<\bar{K}^0|K_S>}\cdot{}\Gamma_{K_S\to\pi^0\pi^0}+\Gamma_{<\bar{K}^0|K_L>}) + \Gamma_{47}\cdot{}(\Gamma_{K_S\to\pi^0\pi^0}\cdot{}\Gamma_{K_S\to\pi^0\pi^0}) + \Gamma_{48}\cdot{}\Gamma_{K_S\to\pi^0\pi^0} + \Gamma_{804} + \Gamma_{50}\cdot{}(\Gamma_{K_S\to\pi^0\pi^0}\cdot{}\Gamma_{K_S\to\pi^0\pi^0}) + \Gamma_{51}\cdot{}\Gamma_{K_S\to\pi^0\pi^0} + \Gamma_{806} + \Gamma_{126}\cdot{}\Gamma_{\eta\to\text{neutral}} + \Gamma_{128}\cdot{}\Gamma_{\eta\to\text{neutral}} + \Gamma_{130}\cdot{}\Gamma_{\eta\to\text{neutral}} + \Gamma_{132}\cdot{}(\Gamma_{\eta\to\text{neutral}}\cdot{}(\Gamma_{<\bar{K}^0|K_S>}\cdot{}\Gamma_{K_S\to\pi^0\pi^0}+\Gamma_{<\bar{K}^0|K_L>})) + \Gamma_{800}\cdot{}\Gamma_{\omega\to\pi^0\gamma} + \Gamma_{151}\cdot{}\Gamma_{\omega\to\pi^0\gamma} + \Gamma_{152}\cdot{}\Gamma_{\omega\to\pi^0\gamma} + \Gamma_{167}\cdot{}(\Gamma_{\phi\to K_S K_L}\cdot{}\Gamma_{K_S\to\pi^0\pi^0})}{\Gamma_{3} + \Gamma_{5} + \Gamma_{9} + \Gamma_{10} + \Gamma_{14} + \Gamma_{16}  \\ 
  {}& + \Gamma_{20} + \Gamma_{23} + \Gamma_{27} + \Gamma_{28} + \Gamma_{30} + \Gamma_{35}\cdot{}(\Gamma_{<\bar{K}^0|K_S>}\cdot{}\Gamma_{K_S\to\pi^0\pi^0} \\ 
  {}& +\Gamma_{<\bar{K}^0|K_L>}) + \Gamma_{40}\cdot{}(\Gamma_{<\bar{K}^0|K_S>}\cdot{}\Gamma_{K_S\to\pi^0\pi^0}+\Gamma_{<\bar{K}^0|K_L>}) + \Gamma_{44}\cdot{}(\Gamma_{<\bar{K}^0|K_S>}\cdot{}\Gamma_{K_S\to\pi^0\pi^0} \\ 
  {}& +\Gamma_{<\bar{K}^0|K_L>}) + \Gamma_{37}\cdot{}(\Gamma_{<\bar{K}^0|K_S>}\cdot{}\Gamma_{K_S\to\pi^0\pi^0}+\Gamma_{<\bar{K}^0|K_L>}) + \Gamma_{42}\cdot{}(\Gamma_{<\bar{K}^0|K_S>}\cdot{}\Gamma_{K_S\to\pi^0\pi^0} \\ 
  {}& +\Gamma_{<\bar{K}^0|K_L>}) + \Gamma_{47}\cdot{}(\Gamma_{K_S\to\pi^0\pi^0}\cdot{}\Gamma_{K_S\to\pi^0\pi^0}) + \Gamma_{48}\cdot{}\Gamma_{K_S\to\pi^0\pi^0}  \\ 
  {}& + \Gamma_{804} + \Gamma_{50}\cdot{}(\Gamma_{K_S\to\pi^0\pi^0}\cdot{}\Gamma_{K_S\to\pi^0\pi^0}) + \Gamma_{51}\cdot{}\Gamma_{K_S\to\pi^0\pi^0}  \\ 
  {}& + \Gamma_{806} + \Gamma_{126}\cdot{}\Gamma_{\eta\to\text{neutral}} + \Gamma_{128}\cdot{}\Gamma_{\eta\to\text{neutral}} + \Gamma_{130}\cdot{}\Gamma_{\eta\to\text{neutral}}  \\ 
  {}& + \Gamma_{132}\cdot{}(\Gamma_{\eta\to\text{neutral}}\cdot{}(\Gamma_{<\bar{K}^0|K_S>}\cdot{}\Gamma_{K_S\to\pi^0\pi^0}+\Gamma_{<\bar{K}^0|K_L>})) + \Gamma_{800}\cdot{}\Gamma_{\omega\to\pi^0\gamma}  \\ 
  {}& + \Gamma_{151}\cdot{}\Gamma_{\omega\to\pi^0\gamma} + \Gamma_{152}\cdot{}\Gamma_{\omega\to\pi^0\gamma} + \Gamma_{167}\cdot{}(\Gamma_{\phi\to K_S K_L}\cdot{}\Gamma_{K_S\to\pi^0\pi^0})}%
\htconstrdef{Gamma3by5.c}{\frac{\Gamma_{3}}{\Gamma_{5}}}{\frac{\Gamma_{3}}{\Gamma_{5}}}{\frac{\Gamma_{3}}{\Gamma_{5}}}%
\htconstrdef{Gamma7.c}{\Gamma_{7}}{\Gamma_{35}\cdot{}\Gamma_{<\bar{K}^0|K_L>} + \Gamma_{9} + \Gamma_{804} + \Gamma_{37}\cdot{}\Gamma_{<K^0|K_L>} + \Gamma_{10}}{\Gamma_{35}\cdot{}\Gamma_{<\bar{K}^0|K_L>} + \Gamma_{9} + \Gamma_{804} + \Gamma_{37}\cdot{}\Gamma_{<K^0|K_L>}  \\ 
  {}& + \Gamma_{10}}%
\htconstrdef{Gamma8.c}{\Gamma_{8}}{\Gamma_{9} + \Gamma_{10}}{\Gamma_{9} + \Gamma_{10}}%
\htconstrdef{Gamma8by5.c}{\frac{\Gamma_{8}}{\Gamma_{5}}}{\frac{\Gamma_{8}}{\Gamma_{5}}}{\frac{\Gamma_{8}}{\Gamma_{5}}}%
\htconstrdef{Gamma9by5.c}{\frac{\Gamma_{9}}{\Gamma_{5}}}{\frac{\Gamma_{9}}{\Gamma_{5}}}{\frac{\Gamma_{9}}{\Gamma_{5}}}%
\htconstrdef{Gamma10by5.c}{\frac{\Gamma_{10}}{\Gamma_{5}}}{\frac{\Gamma_{10}}{\Gamma_{5}}}{\frac{\Gamma_{10}}{\Gamma_{5}}}%
\htconstrdef{Gamma10by9.c}{\frac{\Gamma_{10}}{\Gamma_{9}}}{\frac{\Gamma_{10}}{\Gamma_{9}}}{\frac{\Gamma_{10}}{\Gamma_{9}}}%
\htconstrdef{Gamma11.c}{\Gamma_{11}}{\Gamma_{14} + \Gamma_{16} + \Gamma_{20} + \Gamma_{23} + \Gamma_{27} + \Gamma_{28} + \Gamma_{30} + \Gamma_{35}\cdot{}(\Gamma_{<K^0|K_S>}\cdot{}\Gamma_{K_S\to\pi^0\pi^0}) + \Gamma_{37}\cdot{}(\Gamma_{<K^0|K_S>}\cdot{}\Gamma_{K_S\to\pi^0\pi^0}) + \Gamma_{40}\cdot{}(\Gamma_{<K^0|K_S>}\cdot{}\Gamma_{K_S\to\pi^0\pi^0}) + \Gamma_{42}\cdot{}(\Gamma_{<K^0|K_S>}\cdot{}\Gamma_{K_S\to\pi^0\pi^0}) + \Gamma_{47}\cdot{}(\Gamma_{K_S\to\pi^0\pi^0}\cdot{}\Gamma_{K_S\to\pi^0\pi^0}) + \Gamma_{50}\cdot{}(\Gamma_{K_S\to\pi^0\pi^0}\cdot{}\Gamma_{K_S\to\pi^0\pi^0}) + \Gamma_{126}\cdot{}\Gamma_{\eta\to\text{neutral}} + \Gamma_{128}\cdot{}\Gamma_{\eta\to\text{neutral}} + \Gamma_{130}\cdot{}\Gamma_{\eta\to\text{neutral}} + \Gamma_{132}\cdot{}(\Gamma_{<K^0|K_S>}\cdot{}\Gamma_{K_S\to\pi^0\pi^0}\cdot{}\Gamma_{\eta\to\text{neutral}}) + \Gamma_{151}\cdot{}\Gamma_{\omega\to\pi^0\gamma} + \Gamma_{152}\cdot{}\Gamma_{\omega\to\pi^0\gamma} + \Gamma_{800}\cdot{}\Gamma_{\omega\to\pi^0\gamma}}{\Gamma_{14} + \Gamma_{16} + \Gamma_{20} + \Gamma_{23} + \Gamma_{27} + \Gamma_{28}  \\ 
  {}& + \Gamma_{30} + \Gamma_{35}\cdot{}(\Gamma_{<K^0|K_S>}\cdot{}\Gamma_{K_S\to\pi^0\pi^0}) + \Gamma_{37}\cdot{}(\Gamma_{<K^0|K_S>}\cdot{}\Gamma_{K_S\to\pi^0\pi^0})  \\ 
  {}& + \Gamma_{40}\cdot{}(\Gamma_{<K^0|K_S>}\cdot{}\Gamma_{K_S\to\pi^0\pi^0}) + \Gamma_{42}\cdot{}(\Gamma_{<K^0|K_S>}\cdot{}\Gamma_{K_S\to\pi^0\pi^0})  \\ 
  {}& + \Gamma_{47}\cdot{}(\Gamma_{K_S\to\pi^0\pi^0}\cdot{}\Gamma_{K_S\to\pi^0\pi^0}) + \Gamma_{50}\cdot{}(\Gamma_{K_S\to\pi^0\pi^0}\cdot{}\Gamma_{K_S\to\pi^0\pi^0})  \\ 
  {}& + \Gamma_{126}\cdot{}\Gamma_{\eta\to\text{neutral}} + \Gamma_{128}\cdot{}\Gamma_{\eta\to\text{neutral}} + \Gamma_{130}\cdot{}\Gamma_{\eta\to\text{neutral}}  \\ 
  {}& + \Gamma_{132}\cdot{}(\Gamma_{<K^0|K_S>}\cdot{}\Gamma_{K_S\to\pi^0\pi^0}\cdot{}\Gamma_{\eta\to\text{neutral}}) + \Gamma_{151}\cdot{}\Gamma_{\omega\to\pi^0\gamma}  \\ 
  {}& + \Gamma_{152}\cdot{}\Gamma_{\omega\to\pi^0\gamma} + \Gamma_{800}\cdot{}\Gamma_{\omega\to\pi^0\gamma}}%
\htconstrdef{Gamma12.c}{\Gamma_{12}}{\Gamma_{128}\cdot{}\Gamma_{\eta\to3\pi^0} + \Gamma_{30} + \Gamma_{23} + \Gamma_{28} + \Gamma_{14} + \Gamma_{16} + \Gamma_{20} + \Gamma_{27} + \Gamma_{126}\cdot{}\Gamma_{\eta\to3\pi^0} + \Gamma_{130}\cdot{}\Gamma_{\eta\to3\pi^0}}{\Gamma_{128}\cdot{}\Gamma_{\eta\to3\pi^0} + \Gamma_{30} + \Gamma_{23} + \Gamma_{28} + \Gamma_{14}  \\ 
  {}& + \Gamma_{16} + \Gamma_{20} + \Gamma_{27} + \Gamma_{126}\cdot{}\Gamma_{\eta\to3\pi^0} + \Gamma_{130}\cdot{}\Gamma_{\eta\to3\pi^0}}%
\htconstrdef{Gamma13.c}{\Gamma_{13}}{\Gamma_{14} + \Gamma_{16}}{\Gamma_{14} + \Gamma_{16}}%
\htconstrdef{Gamma17.c}{\Gamma_{17}}{\Gamma_{128}\cdot{}\Gamma_{\eta\to3\pi^0} + \Gamma_{30} + \Gamma_{23} + \Gamma_{28} + \Gamma_{35}\cdot{}(\Gamma_{<K^0|K_S>}\cdot{}\Gamma_{K_S\to\pi^0\pi^0}) + \Gamma_{40}\cdot{}(\Gamma_{<K^0|K_S>}\cdot{}\Gamma_{K_S\to\pi^0\pi^0}) + \Gamma_{42}\cdot{}(\Gamma_{<K^0|K_S>}\cdot{}\Gamma_{K_S\to\pi^0\pi^0}) + \Gamma_{20} + \Gamma_{27} + \Gamma_{47}\cdot{}(\Gamma_{K_S\to\pi^0\pi^0}\cdot{}\Gamma_{K_S\to\pi^0\pi^0}) + \Gamma_{50}\cdot{}(\Gamma_{K_S\to\pi^0\pi^0}\cdot{}\Gamma_{K_S\to\pi^0\pi^0}) + \Gamma_{126}\cdot{}\Gamma_{\eta\to3\pi^0} + \Gamma_{37}\cdot{}(\Gamma_{<K^0|K_S>}\cdot{}\Gamma_{K_S\to\pi^0\pi^0}) + \Gamma_{130}\cdot{}\Gamma_{\eta\to3\pi^0}}{\Gamma_{128}\cdot{}\Gamma_{\eta\to3\pi^0} + \Gamma_{30} + \Gamma_{23} + \Gamma_{28} + \Gamma_{35}\cdot{}(\Gamma_{<K^0|K_S>}\cdot{}\Gamma_{K_S\to\pi^0\pi^0})  \\ 
  {}& + \Gamma_{40}\cdot{}(\Gamma_{<K^0|K_S>}\cdot{}\Gamma_{K_S\to\pi^0\pi^0}) + \Gamma_{42}\cdot{}(\Gamma_{<K^0|K_S>}\cdot{}\Gamma_{K_S\to\pi^0\pi^0})  \\ 
  {}& + \Gamma_{20} + \Gamma_{27} + \Gamma_{47}\cdot{}(\Gamma_{K_S\to\pi^0\pi^0}\cdot{}\Gamma_{K_S\to\pi^0\pi^0}) + \Gamma_{50}\cdot{}(\Gamma_{K_S\to\pi^0\pi^0}\cdot{}\Gamma_{K_S\to\pi^0\pi^0})  \\ 
  {}& + \Gamma_{126}\cdot{}\Gamma_{\eta\to3\pi^0} + \Gamma_{37}\cdot{}(\Gamma_{<K^0|K_S>}\cdot{}\Gamma_{K_S\to\pi^0\pi^0}) + \Gamma_{130}\cdot{}\Gamma_{\eta\to3\pi^0}}%
\htconstrdef{Gamma18.c}{\Gamma_{18}}{\Gamma_{23} + \Gamma_{35}\cdot{}(\Gamma_{<K^0|K_S>}\cdot{}\Gamma_{K_S\to\pi^0\pi^0}) + \Gamma_{20} + \Gamma_{37}\cdot{}(\Gamma_{<K^0|K_S>}\cdot{}\Gamma_{K_S\to\pi^0\pi^0})}{\Gamma_{23} + \Gamma_{35}\cdot{}(\Gamma_{<K^0|K_S>}\cdot{}\Gamma_{K_S\to\pi^0\pi^0}) + \Gamma_{20} + \Gamma_{37}\cdot{}(\Gamma_{<K^0|K_S>}\cdot{}\Gamma_{K_S\to\pi^0\pi^0})}%
\htconstrdef{Gamma19.c}{\Gamma_{19}}{\Gamma_{23} + \Gamma_{20}}{\Gamma_{23} + \Gamma_{20}}%
\htconstrdef{Gamma19by13.c}{\frac{\Gamma_{19}}{\Gamma_{13}}}{\frac{\Gamma_{19}}{\Gamma_{13}}}{\frac{\Gamma_{19}}{\Gamma_{13}}}%
\htconstrdef{Gamma24.c}{\Gamma_{24}}{\Gamma_{27} + \Gamma_{28} + \Gamma_{30} + \Gamma_{40}\cdot{}(\Gamma_{<K^0|K_S>}\cdot{}\Gamma_{K_S\to\pi^0\pi^0}) + \Gamma_{42}\cdot{}(\Gamma_{<K^0|K_S>}\cdot{}\Gamma_{K_S\to\pi^0\pi^0}) + \Gamma_{47}\cdot{}(\Gamma_{K_S\to\pi^0\pi^0}\cdot{}\Gamma_{K_S\to\pi^0\pi^0}) + \Gamma_{50}\cdot{}(\Gamma_{K_S\to\pi^0\pi^0}\cdot{}\Gamma_{K_S\to\pi^0\pi^0}) + \Gamma_{126}\cdot{}\Gamma_{\eta\to3\pi^0} + \Gamma_{128}\cdot{}\Gamma_{\eta\to3\pi^0} + \Gamma_{130}\cdot{}\Gamma_{\eta\to3\pi^0} + \Gamma_{132}\cdot{}(\Gamma_{<K^0|K_S>}\cdot{}\Gamma_{K_S\to\pi^0\pi^0}\cdot{}\Gamma_{\eta\to3\pi^0})}{\Gamma_{27} + \Gamma_{28} + \Gamma_{30} + \Gamma_{40}\cdot{}(\Gamma_{<K^0|K_S>}\cdot{}\Gamma_{K_S\to\pi^0\pi^0})  \\ 
  {}& + \Gamma_{42}\cdot{}(\Gamma_{<K^0|K_S>}\cdot{}\Gamma_{K_S\to\pi^0\pi^0}) + \Gamma_{47}\cdot{}(\Gamma_{K_S\to\pi^0\pi^0}\cdot{}\Gamma_{K_S\to\pi^0\pi^0})  \\ 
  {}& + \Gamma_{50}\cdot{}(\Gamma_{K_S\to\pi^0\pi^0}\cdot{}\Gamma_{K_S\to\pi^0\pi^0}) + \Gamma_{126}\cdot{}\Gamma_{\eta\to3\pi^0} + \Gamma_{128}\cdot{}\Gamma_{\eta\to3\pi^0}  \\ 
  {}& + \Gamma_{130}\cdot{}\Gamma_{\eta\to3\pi^0} + \Gamma_{132}\cdot{}(\Gamma_{<K^0|K_S>}\cdot{}\Gamma_{K_S\to\pi^0\pi^0}\cdot{}\Gamma_{\eta\to3\pi^0})}%
\htconstrdef{Gamma25.c}{\Gamma_{25}}{\Gamma_{128}\cdot{}\Gamma_{\eta\to3\pi^0} + \Gamma_{30} + \Gamma_{28} + \Gamma_{27} + \Gamma_{126}\cdot{}\Gamma_{\eta\to3\pi^0} + \Gamma_{130}\cdot{}\Gamma_{\eta\to3\pi^0}}{\Gamma_{128}\cdot{}\Gamma_{\eta\to3\pi^0} + \Gamma_{30} + \Gamma_{28} + \Gamma_{27} + \Gamma_{126}\cdot{}\Gamma_{\eta\to3\pi^0}  \\ 
  {}& + \Gamma_{130}\cdot{}\Gamma_{\eta\to3\pi^0}}%
\htconstrdef{Gamma26.c}{\Gamma_{26}}{\Gamma_{128}\cdot{}\Gamma_{\eta\to3\pi^0} + \Gamma_{28} + \Gamma_{40}\cdot{}(\Gamma_{<K^0|K_S>}\cdot{}\Gamma_{K_S\to\pi^0\pi^0}) + \Gamma_{42}\cdot{}(\Gamma_{<K^0|K_S>}\cdot{}\Gamma_{K_S\to\pi^0\pi^0}) + \Gamma_{27}}{\Gamma_{128}\cdot{}\Gamma_{\eta\to3\pi^0} + \Gamma_{28} + \Gamma_{40}\cdot{}(\Gamma_{<K^0|K_S>}\cdot{}\Gamma_{K_S\to\pi^0\pi^0})  \\ 
  {}& + \Gamma_{42}\cdot{}(\Gamma_{<K^0|K_S>}\cdot{}\Gamma_{K_S\to\pi^0\pi^0}) + \Gamma_{27}}%
\htconstrdef{Gamma26by13.c}{\frac{\Gamma_{26}}{\Gamma_{13}}}{\frac{\Gamma_{26}}{\Gamma_{13}}}{\frac{\Gamma_{26}}{\Gamma_{13}}}%
\htconstrdef{Gamma29.c}{\Gamma_{29}}{\Gamma_{30} + \Gamma_{126}\cdot{}\Gamma_{\eta\to3\pi^0} + \Gamma_{130}\cdot{}\Gamma_{\eta\to3\pi^0}}{\Gamma_{30} + \Gamma_{126}\cdot{}\Gamma_{\eta\to3\pi^0} + \Gamma_{130}\cdot{}\Gamma_{\eta\to3\pi^0}}%
\htconstrdef{Gamma31.c}{\Gamma_{31}}{\Gamma_{128}\cdot{}\Gamma_{\eta\to\text{neutral}} + \Gamma_{23} + \Gamma_{28} + \Gamma_{42} + \Gamma_{16} + \Gamma_{37} + \Gamma_{10} + \Gamma_{167}\cdot{}(\Gamma_{\phi\to K_S K_L}\cdot{}\Gamma_{K_S\to\pi^0\pi^0})}{\Gamma_{128}\cdot{}\Gamma_{\eta\to\text{neutral}} + \Gamma_{23} + \Gamma_{28} + \Gamma_{42} + \Gamma_{16}  \\ 
  {}& + \Gamma_{37} + \Gamma_{10} + \Gamma_{167}\cdot{}(\Gamma_{\phi\to K_S K_L}\cdot{}\Gamma_{K_S\to\pi^0\pi^0})}%
\htconstrdef{Gamma32.c}{\Gamma_{32}}{\Gamma_{16} + \Gamma_{23} + \Gamma_{28} + \Gamma_{37} + \Gamma_{42} + \Gamma_{128}\cdot{}\Gamma_{\eta\to\text{neutral}} + \Gamma_{130}\cdot{}\Gamma_{\eta\to\text{neutral}} + \Gamma_{167}\cdot{}(\Gamma_{\phi\to K_S K_L}\cdot{}\Gamma_{K_S\to\pi^0\pi^0})}{\Gamma_{16} + \Gamma_{23} + \Gamma_{28} + \Gamma_{37} + \Gamma_{42} + \Gamma_{128}\cdot{}\Gamma_{\eta\to\text{neutral}}  \\ 
  {}& + \Gamma_{130}\cdot{}\Gamma_{\eta\to\text{neutral}} + \Gamma_{167}\cdot{}(\Gamma_{\phi\to K_S K_L}\cdot{}\Gamma_{K_S\to\pi^0\pi^0})}%
\htconstrdef{Gamma33.c}{\Gamma_{33}}{\Gamma_{35}\cdot{}\Gamma_{<\bar{K}^0|K_S>} + \Gamma_{40}\cdot{}\Gamma_{<\bar{K}^0|K_S>} + \Gamma_{42}\cdot{}\Gamma_{<K^0|K_S>} + \Gamma_{47} + \Gamma_{48} + \Gamma_{50} + \Gamma_{51} + \Gamma_{37}\cdot{}\Gamma_{<K^0|K_S>} + \Gamma_{132}\cdot{}(\Gamma_{<\bar{K}^0|K_S>}\cdot{}\Gamma_{\eta\to\text{neutral}}) + \Gamma_{44}\cdot{}\Gamma_{<\bar{K}^0|K_S>} + \Gamma_{167}\cdot{}\Gamma_{\phi\to K_S K_L}}{\Gamma_{35}\cdot{}\Gamma_{<\bar{K}^0|K_S>} + \Gamma_{40}\cdot{}\Gamma_{<\bar{K}^0|K_S>} + \Gamma_{42}\cdot{}\Gamma_{<K^0|K_S>}  \\ 
  {}& + \Gamma_{47} + \Gamma_{48} + \Gamma_{50} + \Gamma_{51} + \Gamma_{37}\cdot{}\Gamma_{<K^0|K_S>}  \\ 
  {}& + \Gamma_{132}\cdot{}(\Gamma_{<\bar{K}^0|K_S>}\cdot{}\Gamma_{\eta\to\text{neutral}}) + \Gamma_{44}\cdot{}\Gamma_{<\bar{K}^0|K_S>} + \Gamma_{167}\cdot{}\Gamma_{\phi\to K_S K_L}}%
\htconstrdef{Gamma34.c}{\Gamma_{34}}{\Gamma_{35} + \Gamma_{37}}{\Gamma_{35} + \Gamma_{37}}%
\htconstrdef{Gamma38.c}{\Gamma_{38}}{\Gamma_{42} + \Gamma_{37}}{\Gamma_{42} + \Gamma_{37}}%
\htconstrdef{Gamma39.c}{\Gamma_{39}}{\Gamma_{40} + \Gamma_{42}}{\Gamma_{40} + \Gamma_{42}}%
\htconstrdef{Gamma43.c}{\Gamma_{43}}{\Gamma_{40} + \Gamma_{44}}{\Gamma_{40} + \Gamma_{44}}%
\htconstrdef{Gamma46.c}{\Gamma_{46}}{\Gamma_{48} + \Gamma_{47} + \Gamma_{804}}{\Gamma_{48} + \Gamma_{47} + \Gamma_{804}}%
\htconstrdef{Gamma49.c}{\Gamma_{49}}{\Gamma_{50} + \Gamma_{51} + \Gamma_{806}}{\Gamma_{50} + \Gamma_{51} + \Gamma_{806}}%
\htconstrdef{Gamma54.c}{\Gamma_{54}}{\Gamma_{35}\cdot{}(\Gamma_{<K^0|K_S>}\cdot{}\Gamma_{K_S\to\pi^+\pi^-}) + \Gamma_{37}\cdot{}(\Gamma_{<K^0|K_S>}\cdot{}\Gamma_{K_S\to\pi^+\pi^-}) + \Gamma_{40}\cdot{}(\Gamma_{<K^0|K_S>}\cdot{}\Gamma_{K_S\to\pi^+\pi^-}) + \Gamma_{42}\cdot{}(\Gamma_{<K^0|K_S>}\cdot{}\Gamma_{K_S\to\pi^+\pi^-}) + \Gamma_{47}\cdot{}(2\cdot{}\Gamma_{K_S\to\pi^+\pi^-}\cdot{}\Gamma_{K_S\to\pi^0\pi^0}) + \Gamma_{48}\cdot{}\Gamma_{K_S\to\pi^+\pi^-} + \Gamma_{50}\cdot{}(2\cdot{}\Gamma_{K_S\to\pi^+\pi^-}\cdot{}\Gamma_{K_S\to\pi^0\pi^0}) + \Gamma_{51}\cdot{}\Gamma_{K_S\to\pi^+\pi^-} + \Gamma_{53}\cdot{}(\Gamma_{<\bar{K}^0|K_S>}\cdot{}\Gamma_{K_S\to\pi^0\pi^0}+\Gamma_{<\bar{K}^0|K_L>}) + \Gamma_{62} + \Gamma_{70} + \Gamma_{77} + \Gamma_{78} + \Gamma_{93} + \Gamma_{94} + \Gamma_{126}\cdot{}\Gamma_{\eta\to\text{charged}} + \Gamma_{128}\cdot{}\Gamma_{\eta\to\text{charged}} + \Gamma_{130}\cdot{}\Gamma_{\eta\to\text{charged}} + \Gamma_{132}\cdot{}(\Gamma_{<\bar{K}^0|K_L>}\cdot{}\Gamma_{\eta\to\pi^+\pi^-\pi^0} + \Gamma_{<\bar{K}^0|K_S>}\cdot{}\Gamma_{K_S\to\pi^0\pi^0}\cdot{}\Gamma_{\eta\to\pi^+\pi^-\pi^0} + \Gamma_{<\bar{K}^0|K_S>}\cdot{}\Gamma_{K_S\to\pi^+\pi^-}\cdot{}\Gamma_{\eta\to3\pi^0}) + \Gamma_{151}\cdot{}(\Gamma_{\omega\to\pi^+\pi^-\pi^0}+\Gamma_{\omega\to\pi^+\pi^-}) + \Gamma_{152}\cdot{}(\Gamma_{\omega\to\pi^+\pi^-\pi^0}+\Gamma_{\omega\to\pi^+\pi^-}) + \Gamma_{167}\cdot{}(\Gamma_{\phi\to K^+K^-} + \Gamma_{\phi\to K_S K_L}\cdot{}\Gamma_{K_S\to\pi^+\pi^-}) + \Gamma_{802} + \Gamma_{803} + \Gamma_{800}\cdot{}(\Gamma_{\omega\to\pi^+\pi^-\pi^0}+\Gamma_{\omega\to\pi^+\pi^-})}{\Gamma_{35}\cdot{}(\Gamma_{<K^0|K_S>}\cdot{}\Gamma_{K_S\to\pi^+\pi^-}) + \Gamma_{37}\cdot{}(\Gamma_{<K^0|K_S>}\cdot{}\Gamma_{K_S\to\pi^+\pi^-})  \\ 
  {}& + \Gamma_{40}\cdot{}(\Gamma_{<K^0|K_S>}\cdot{}\Gamma_{K_S\to\pi^+\pi^-}) + \Gamma_{42}\cdot{}(\Gamma_{<K^0|K_S>}\cdot{}\Gamma_{K_S\to\pi^+\pi^-})  \\ 
  {}& + \Gamma_{47}\cdot{}(2\cdot{}\Gamma_{K_S\to\pi^+\pi^-}\cdot{}\Gamma_{K_S\to\pi^0\pi^0}) + \Gamma_{48}\cdot{}\Gamma_{K_S\to\pi^+\pi^-}  \\ 
  {}& + \Gamma_{50}\cdot{}(2\cdot{}\Gamma_{K_S\to\pi^+\pi^-}\cdot{}\Gamma_{K_S\to\pi^0\pi^0}) + \Gamma_{51}\cdot{}\Gamma_{K_S\to\pi^+\pi^-}  \\ 
  {}& + \Gamma_{53}\cdot{}(\Gamma_{<\bar{K}^0|K_S>}\cdot{}\Gamma_{K_S\to\pi^0\pi^0}+\Gamma_{<\bar{K}^0|K_L>}) + \Gamma_{62} + \Gamma_{70}  \\ 
  {}& + \Gamma_{77} + \Gamma_{78} + \Gamma_{93} + \Gamma_{94} + \Gamma_{126}\cdot{}\Gamma_{\eta\to\text{charged}}  \\ 
  {}& + \Gamma_{128}\cdot{}\Gamma_{\eta\to\text{charged}} + \Gamma_{130}\cdot{}\Gamma_{\eta\to\text{charged}} + \Gamma_{132}\cdot{}(\Gamma_{<\bar{K}^0|K_L>}\cdot{}\Gamma_{\eta\to\pi^+\pi^-\pi^0}  \\ 
  {}& + \Gamma_{<\bar{K}^0|K_S>}\cdot{}\Gamma_{K_S\to\pi^0\pi^0}\cdot{}\Gamma_{\eta\to\pi^+\pi^-\pi^0} + \Gamma_{<\bar{K}^0|K_S>}\cdot{}\Gamma_{K_S\to\pi^+\pi^-}\cdot{}\Gamma_{\eta\to3\pi^0})  \\ 
  {}& + \Gamma_{151}\cdot{}(\Gamma_{\omega\to\pi^+\pi^-\pi^0}+\Gamma_{\omega\to\pi^+\pi^-}) + \Gamma_{152}\cdot{}(\Gamma_{\omega\to\pi^+\pi^-\pi^0}+\Gamma_{\omega\to\pi^+\pi^-})  \\ 
  {}& + \Gamma_{167}\cdot{}(\Gamma_{\phi\to K^+K^-} + \Gamma_{\phi\to K_S K_L}\cdot{}\Gamma_{K_S\to\pi^+\pi^-}) + \Gamma_{802} + \Gamma_{803}  \\ 
  {}& + \Gamma_{800}\cdot{}(\Gamma_{\omega\to\pi^+\pi^-\pi^0}+\Gamma_{\omega\to\pi^+\pi^-})}%
\htconstrdef{Gamma55.c}{\Gamma_{55}}{\Gamma_{128}\cdot{}\Gamma_{\eta\to\text{charged}} + \Gamma_{152}\cdot{}(\Gamma_{\omega\to\pi^+\pi^-\pi^0}+\Gamma_{\omega\to\pi^+\pi^-}) + \Gamma_{78} + \Gamma_{77} + \Gamma_{94} + \Gamma_{62} + \Gamma_{70} + \Gamma_{93} + \Gamma_{126}\cdot{}\Gamma_{\eta\to\text{charged}} + \Gamma_{802} + \Gamma_{803} + \Gamma_{800}\cdot{}(\Gamma_{\omega\to\pi^+\pi^-\pi^0}+\Gamma_{\omega\to\pi^+\pi^-}) + \Gamma_{151}\cdot{}(\Gamma_{\omega\to\pi^+\pi^-\pi^0}+\Gamma_{\omega\to\pi^+\pi^-}) + \Gamma_{130}\cdot{}\Gamma_{\eta\to\text{charged}} + \Gamma_{168}}{\Gamma_{128}\cdot{}\Gamma_{\eta\to\text{charged}} + \Gamma_{152}\cdot{}(\Gamma_{\omega\to\pi^+\pi^-\pi^0}+\Gamma_{\omega\to\pi^+\pi^-}) + \Gamma_{78}  \\ 
  {}& + \Gamma_{77} + \Gamma_{94} + \Gamma_{62} + \Gamma_{70} + \Gamma_{93} + \Gamma_{126}\cdot{}\Gamma_{\eta\to\text{charged}}  \\ 
  {}& + \Gamma_{802} + \Gamma_{803} + \Gamma_{800}\cdot{}(\Gamma_{\omega\to\pi^+\pi^-\pi^0}+\Gamma_{\omega\to\pi^+\pi^-}) + \Gamma_{151}\cdot{}(\Gamma_{\omega\to\pi^+\pi^-\pi^0} \\ 
  {}& +\Gamma_{\omega\to\pi^+\pi^-}) + \Gamma_{130}\cdot{}\Gamma_{\eta\to\text{charged}} + \Gamma_{168}}%
\htconstrdef{Gamma56.c}{\Gamma_{56}}{\Gamma_{35}\cdot{}(\Gamma_{<K^0|K_S>}\cdot{}\Gamma_{K_S\to\pi^+\pi^-}) + \Gamma_{62} + \Gamma_{93} + \Gamma_{37}\cdot{}(\Gamma_{<K^0|K_S>}\cdot{}\Gamma_{K_S\to\pi^+\pi^-}) + \Gamma_{802} + \Gamma_{800}\cdot{}\Gamma_{\omega\to\pi^+\pi^-} + \Gamma_{151}\cdot{}\Gamma_{\omega\to\pi^+\pi^-} + \Gamma_{168}}{\Gamma_{35}\cdot{}(\Gamma_{<K^0|K_S>}\cdot{}\Gamma_{K_S\to\pi^+\pi^-}) + \Gamma_{62} + \Gamma_{93} + \Gamma_{37}\cdot{}(\Gamma_{<K^0|K_S>}\cdot{}\Gamma_{K_S\to\pi^+\pi^-})  \\ 
  {}& + \Gamma_{802} + \Gamma_{800}\cdot{}\Gamma_{\omega\to\pi^+\pi^-} + \Gamma_{151}\cdot{}\Gamma_{\omega\to\pi^+\pi^-} + \Gamma_{168}}%
\htconstrdef{Gamma57.c}{\Gamma_{57}}{\Gamma_{62} + \Gamma_{93} + \Gamma_{802} + \Gamma_{800}\cdot{}\Gamma_{\omega\to\pi^+\pi^-} + \Gamma_{151}\cdot{}\Gamma_{\omega\to\pi^+\pi^-} + \Gamma_{167}\cdot{}\Gamma_{\phi\to K^+K^-}}{\Gamma_{62} + \Gamma_{93} + \Gamma_{802} + \Gamma_{800}\cdot{}\Gamma_{\omega\to\pi^+\pi^-} + \Gamma_{151}\cdot{}\Gamma_{\omega\to\pi^+\pi^-}  \\ 
  {}& + \Gamma_{167}\cdot{}\Gamma_{\phi\to K^+K^-}}%
\htconstrdef{Gamma57by55.c}{\frac{\Gamma_{57}}{\Gamma_{55}}}{\frac{\Gamma_{57}}{\Gamma_{55}}}{\frac{\Gamma_{57}}{\Gamma_{55}}}%
\htconstrdef{Gamma58.c}{\Gamma_{58}}{\Gamma_{62} + \Gamma_{93} + \Gamma_{802} + \Gamma_{167}\cdot{}\Gamma_{\phi\to K^+K^-}}{\Gamma_{62} + \Gamma_{93} + \Gamma_{802} + \Gamma_{167}\cdot{}\Gamma_{\phi\to K^+K^-}}%
\htconstrdef{Gamma59.c}{\Gamma_{59}}{\Gamma_{35}\cdot{}(\Gamma_{<K^0|K_S>}\cdot{}\Gamma_{K_S\to\pi^+\pi^-}) + \Gamma_{62} + \Gamma_{800}\cdot{}\Gamma_{\omega\to\pi^+\pi^-}}{\Gamma_{35}\cdot{}(\Gamma_{<K^0|K_S>}\cdot{}\Gamma_{K_S\to\pi^+\pi^-}) + \Gamma_{62} + \Gamma_{800}\cdot{}\Gamma_{\omega\to\pi^+\pi^-}}%
\htconstrdef{Gamma60.c}{\Gamma_{60}}{\Gamma_{62} + \Gamma_{800}\cdot{}\Gamma_{\omega\to\pi^+\pi^-}}{\Gamma_{62} + \Gamma_{800}\cdot{}\Gamma_{\omega\to\pi^+\pi^-}}%
\htconstrdef{Gamma63.c}{\Gamma_{63}}{\Gamma_{40}\cdot{}(\Gamma_{<K^0|K_S>}\cdot{}\Gamma_{K_S\to\pi^+\pi^-}) + \Gamma_{42}\cdot{}(\Gamma_{<K^0|K_S>}\cdot{}\Gamma_{K_S\to\pi^+\pi^-}) + \Gamma_{47}\cdot{}(2\cdot{}\Gamma_{K_S\to\pi^+\pi^-}\cdot{}\Gamma_{K_S\to\pi^0\pi^0}) + \Gamma_{50}\cdot{}(2\cdot{}\Gamma_{K_S\to\pi^+\pi^-}\cdot{}\Gamma_{K_S\to\pi^0\pi^0}) + \Gamma_{70} + \Gamma_{77} + \Gamma_{78} + \Gamma_{94} + \Gamma_{126}\cdot{}\Gamma_{\eta\to\text{charged}} + \Gamma_{128}\cdot{}\Gamma_{\eta\to\text{charged}} + \Gamma_{130}\cdot{}\Gamma_{\eta\to\text{charged}} + \Gamma_{132}\cdot{}(\Gamma_{<\bar{K}^0|K_S>}\cdot{}\Gamma_{K_S\to\pi^+\pi^-}\cdot{}\Gamma_{\eta\to\text{neutral}} + \Gamma_{<\bar{K}^0|K_S>}\cdot{}\Gamma_{K_S\to\pi^0\pi^0}\cdot{}\Gamma_{\eta\to\text{charged}}) + \Gamma_{151}\cdot{}\Gamma_{\omega\to\pi^+\pi^-\pi^0} + \Gamma_{152}\cdot{}(\Gamma_{\omega\to\pi^+\pi^-\pi^0}+\Gamma_{\omega\to\pi^+\pi^-}) + \Gamma_{800}\cdot{}\Gamma_{\omega\to\pi^+\pi^-\pi^0} + \Gamma_{803}}{\Gamma_{40}\cdot{}(\Gamma_{<K^0|K_S>}\cdot{}\Gamma_{K_S\to\pi^+\pi^-}) + \Gamma_{42}\cdot{}(\Gamma_{<K^0|K_S>}\cdot{}\Gamma_{K_S\to\pi^+\pi^-})  \\ 
  {}& + \Gamma_{47}\cdot{}(2\cdot{}\Gamma_{K_S\to\pi^+\pi^-}\cdot{}\Gamma_{K_S\to\pi^0\pi^0}) + \Gamma_{50}\cdot{}(2\cdot{}\Gamma_{K_S\to\pi^+\pi^-}\cdot{}\Gamma_{K_S\to\pi^0\pi^0})  \\ 
  {}& + \Gamma_{70} + \Gamma_{77} + \Gamma_{78} + \Gamma_{94} + \Gamma_{126}\cdot{}\Gamma_{\eta\to\text{charged}}  \\ 
  {}& + \Gamma_{128}\cdot{}\Gamma_{\eta\to\text{charged}} + \Gamma_{130}\cdot{}\Gamma_{\eta\to\text{charged}} + \Gamma_{132}\cdot{}(\Gamma_{<\bar{K}^0|K_S>}\cdot{}\Gamma_{K_S\to\pi^+\pi^-}\cdot{}\Gamma_{\eta\to\text{neutral}}  \\ 
  {}& + \Gamma_{<\bar{K}^0|K_S>}\cdot{}\Gamma_{K_S\to\pi^0\pi^0}\cdot{}\Gamma_{\eta\to\text{charged}}) + \Gamma_{151}\cdot{}\Gamma_{\omega\to\pi^+\pi^-\pi^0} + \Gamma_{152}\cdot{}(\Gamma_{\omega\to\pi^+\pi^-\pi^0} \\ 
  {}& +\Gamma_{\omega\to\pi^+\pi^-}) + \Gamma_{800}\cdot{}\Gamma_{\omega\to\pi^+\pi^-\pi^0} + \Gamma_{803}}%
\htconstrdef{Gamma64.c}{\Gamma_{64}}{\Gamma_{78} + \Gamma_{77} + \Gamma_{94} + \Gamma_{70} + \Gamma_{126}\cdot{}\Gamma_{\eta\to\pi^+\pi^-\pi^0} + \Gamma_{128}\cdot{}\Gamma_{\eta\to\pi^+\pi^-\pi^0} + \Gamma_{130}\cdot{}\Gamma_{\eta\to\pi^+\pi^-\pi^0} + \Gamma_{800}\cdot{}\Gamma_{\omega\to\pi^+\pi^-\pi^0} + \Gamma_{151}\cdot{}\Gamma_{\omega\to\pi^+\pi^-\pi^0} + \Gamma_{152}\cdot{}(\Gamma_{\omega\to\pi^+\pi^-\pi^0}+\Gamma_{\omega\to\pi^+\pi^-}) + \Gamma_{803}}{\Gamma_{78} + \Gamma_{77} + \Gamma_{94} + \Gamma_{70} + \Gamma_{126}\cdot{}\Gamma_{\eta\to\pi^+\pi^-\pi^0}  \\ 
  {}& + \Gamma_{128}\cdot{}\Gamma_{\eta\to\pi^+\pi^-\pi^0} + \Gamma_{130}\cdot{}\Gamma_{\eta\to\pi^+\pi^-\pi^0} + \Gamma_{800}\cdot{}\Gamma_{\omega\to\pi^+\pi^-\pi^0}  \\ 
  {}& + \Gamma_{151}\cdot{}\Gamma_{\omega\to\pi^+\pi^-\pi^0} + \Gamma_{152}\cdot{}(\Gamma_{\omega\to\pi^+\pi^-\pi^0}+\Gamma_{\omega\to\pi^+\pi^-}) + \Gamma_{803}}%
\htconstrdef{Gamma65.c}{\Gamma_{65}}{\Gamma_{40}\cdot{}(\Gamma_{<K^0|K_S>}\cdot{}\Gamma_{K_S\to\pi^+\pi^-}) + \Gamma_{42}\cdot{}(\Gamma_{<K^0|K_S>}\cdot{}\Gamma_{K_S\to\pi^+\pi^-}) + \Gamma_{70} + \Gamma_{94} + \Gamma_{128}\cdot{}\Gamma_{\eta\to\pi^+\pi^-\pi^0} + \Gamma_{151}\cdot{}\Gamma_{\omega\to\pi^+\pi^-\pi^0} + \Gamma_{152}\cdot{}\Gamma_{\omega\to\pi^+\pi^-} + \Gamma_{800}\cdot{}\Gamma_{\omega\to\pi^+\pi^-\pi^0} + \Gamma_{803}}{\Gamma_{40}\cdot{}(\Gamma_{<K^0|K_S>}\cdot{}\Gamma_{K_S\to\pi^+\pi^-}) + \Gamma_{42}\cdot{}(\Gamma_{<K^0|K_S>}\cdot{}\Gamma_{K_S\to\pi^+\pi^-})  \\ 
  {}& + \Gamma_{70} + \Gamma_{94} + \Gamma_{128}\cdot{}\Gamma_{\eta\to\pi^+\pi^-\pi^0} + \Gamma_{151}\cdot{}\Gamma_{\omega\to\pi^+\pi^-\pi^0}  \\ 
  {}& + \Gamma_{152}\cdot{}\Gamma_{\omega\to\pi^+\pi^-} + \Gamma_{800}\cdot{}\Gamma_{\omega\to\pi^+\pi^-\pi^0} + \Gamma_{803}}%
\htconstrdef{Gamma66.c}{\Gamma_{66}}{\Gamma_{70} + \Gamma_{94} + \Gamma_{128}\cdot{}\Gamma_{\eta\to\pi^+\pi^-\pi^0} + \Gamma_{151}\cdot{}\Gamma_{\omega\to\pi^+\pi^-\pi^0} + \Gamma_{152}\cdot{}\Gamma_{\omega\to\pi^+\pi^-} + \Gamma_{800}\cdot{}\Gamma_{\omega\to\pi^+\pi^-\pi^0} + \Gamma_{803}}{\Gamma_{70} + \Gamma_{94} + \Gamma_{128}\cdot{}\Gamma_{\eta\to\pi^+\pi^-\pi^0} + \Gamma_{151}\cdot{}\Gamma_{\omega\to\pi^+\pi^-\pi^0}  \\ 
  {}& + \Gamma_{152}\cdot{}\Gamma_{\omega\to\pi^+\pi^-} + \Gamma_{800}\cdot{}\Gamma_{\omega\to\pi^+\pi^-\pi^0} + \Gamma_{803}}%
\htconstrdef{Gamma67.c}{\Gamma_{67}}{\Gamma_{70} + \Gamma_{94} + \Gamma_{128}\cdot{}\Gamma_{\eta\to\pi^+\pi^-\pi^0} + \Gamma_{803}}{\Gamma_{70} + \Gamma_{94} + \Gamma_{128}\cdot{}\Gamma_{\eta\to\pi^+\pi^-\pi^0} + \Gamma_{803}}%
\htconstrdef{Gamma68.c}{\Gamma_{68}}{\Gamma_{40}\cdot{}(\Gamma_{<K^0|K_S>}\cdot{}\Gamma_{K_S\to\pi^+\pi^-}) + \Gamma_{70} + \Gamma_{152}\cdot{}\Gamma_{\omega\to\pi^+\pi^-} + \Gamma_{800}\cdot{}\Gamma_{\omega\to\pi^+\pi^-\pi^0}}{\Gamma_{40}\cdot{}(\Gamma_{<K^0|K_S>}\cdot{}\Gamma_{K_S\to\pi^+\pi^-}) + \Gamma_{70} + \Gamma_{152}\cdot{}\Gamma_{\omega\to\pi^+\pi^-}  \\ 
  {}& + \Gamma_{800}\cdot{}\Gamma_{\omega\to\pi^+\pi^-\pi^0}}%
\htconstrdef{Gamma69.c}{\Gamma_{69}}{\Gamma_{152}\cdot{}\Gamma_{\omega\to\pi^+\pi^-} + \Gamma_{70} + \Gamma_{800}\cdot{}\Gamma_{\omega\to\pi^+\pi^-\pi^0}}{\Gamma_{152}\cdot{}\Gamma_{\omega\to\pi^+\pi^-} + \Gamma_{70} + \Gamma_{800}\cdot{}\Gamma_{\omega\to\pi^+\pi^-\pi^0}}%
\htconstrdef{Gamma74.c}{\Gamma_{74}}{\Gamma_{152}\cdot{}\Gamma_{\omega\to\pi^+\pi^-\pi^0} + \Gamma_{78} + \Gamma_{77} + \Gamma_{126}\cdot{}\Gamma_{\eta\to\pi^+\pi^-\pi^0} + \Gamma_{130}\cdot{}\Gamma_{\eta\to\pi^+\pi^-\pi^0}}{\Gamma_{152}\cdot{}\Gamma_{\omega\to\pi^+\pi^-\pi^0} + \Gamma_{78} + \Gamma_{77} + \Gamma_{126}\cdot{}\Gamma_{\eta\to\pi^+\pi^-\pi^0}  \\ 
  {}& + \Gamma_{130}\cdot{}\Gamma_{\eta\to\pi^+\pi^-\pi^0}}%
\htconstrdef{Gamma75.c}{\Gamma_{75}}{\Gamma_{152}\cdot{}\Gamma_{\omega\to\pi^+\pi^-\pi^0} + \Gamma_{47}\cdot{}(2\cdot{}\Gamma_{K_S\to\pi^+\pi^-}\cdot{}\Gamma_{K_S\to\pi^0\pi^0}) + \Gamma_{77} + \Gamma_{126}\cdot{}\Gamma_{\eta\to\pi^+\pi^-\pi^0} + \Gamma_{130}\cdot{}\Gamma_{\eta\to\pi^+\pi^-\pi^0}}{\Gamma_{152}\cdot{}\Gamma_{\omega\to\pi^+\pi^-\pi^0} + \Gamma_{47}\cdot{}(2\cdot{}\Gamma_{K_S\to\pi^+\pi^-}\cdot{}\Gamma_{K_S\to\pi^0\pi^0})  \\ 
  {}& + \Gamma_{77} + \Gamma_{126}\cdot{}\Gamma_{\eta\to\pi^+\pi^-\pi^0} + \Gamma_{130}\cdot{}\Gamma_{\eta\to\pi^+\pi^-\pi^0}}%
\htconstrdef{Gamma76.c}{\Gamma_{76}}{\Gamma_{152}\cdot{}\Gamma_{\omega\to\pi^+\pi^-\pi^0} + \Gamma_{77} + \Gamma_{126}\cdot{}\Gamma_{\eta\to\pi^+\pi^-\pi^0} + \Gamma_{130}\cdot{}\Gamma_{\eta\to\pi^+\pi^-\pi^0}}{\Gamma_{152}\cdot{}\Gamma_{\omega\to\pi^+\pi^-\pi^0} + \Gamma_{77} + \Gamma_{126}\cdot{}\Gamma_{\eta\to\pi^+\pi^-\pi^0} + \Gamma_{130}\cdot{}\Gamma_{\eta\to\pi^+\pi^-\pi^0}}%
\htconstrdef{Gamma76by54.c}{\frac{\Gamma_{76}}{\Gamma_{54}}}{\frac{\Gamma_{76}}{\Gamma_{54}}}{\frac{\Gamma_{76}}{\Gamma_{54}}}%
\htconstrdef{Gamma78.c}{\Gamma_{78}}{\Gamma_{810} + \Gamma_{50}\cdot{}(2\cdot{}\Gamma_{K_S\to\pi^+\pi^-}\cdot{}\Gamma_{K_S\to\pi^0\pi^0}) + \Gamma_{132}\cdot{}(\Gamma_{<\bar{K}^0|K_S>}\cdot{}\Gamma_{K_S\to\pi^+\pi^-}\cdot{}\Gamma_{\eta\to3\pi^0})}{\Gamma_{810} + \Gamma_{50}\cdot{}(2\cdot{}\Gamma_{K_S\to\pi^+\pi^-}\cdot{}\Gamma_{K_S\to\pi^0\pi^0}) + \Gamma_{132}\cdot{}(\Gamma_{<\bar{K}^0|K_S>}\cdot{}\Gamma_{K_S\to\pi^+\pi^-}\cdot{}\Gamma_{\eta\to3\pi^0})}%
\htconstrdef{Gamma79.c}{\Gamma_{79}}{\Gamma_{37}\cdot{}(\Gamma_{<K^0|K_S>}\cdot{}\Gamma_{K_S\to\pi^+\pi^-}) + \Gamma_{42}\cdot{}(\Gamma_{<K^0|K_S>}\cdot{}\Gamma_{K_S\to\pi^+\pi^-}) + \Gamma_{93} + \Gamma_{94} + \Gamma_{128}\cdot{}\Gamma_{\eta\to\text{charged}} + \Gamma_{151}\cdot{}(\Gamma_{\omega\to\pi^+\pi^-\pi^0}+\Gamma_{\omega\to\pi^+\pi^-}) + \Gamma_{168} + \Gamma_{802} + \Gamma_{803}}{\Gamma_{37}\cdot{}(\Gamma_{<K^0|K_S>}\cdot{}\Gamma_{K_S\to\pi^+\pi^-}) + \Gamma_{42}\cdot{}(\Gamma_{<K^0|K_S>}\cdot{}\Gamma_{K_S\to\pi^+\pi^-})  \\ 
  {}& + \Gamma_{93} + \Gamma_{94} + \Gamma_{128}\cdot{}\Gamma_{\eta\to\text{charged}} + \Gamma_{151}\cdot{}(\Gamma_{\omega\to\pi^+\pi^-\pi^0} \\ 
  {}& +\Gamma_{\omega\to\pi^+\pi^-}) + \Gamma_{168} + \Gamma_{802} + \Gamma_{803}}%
\htconstrdef{Gamma80.c}{\Gamma_{80}}{\Gamma_{93} + \Gamma_{802} + \Gamma_{151}\cdot{}\Gamma_{\omega\to\pi^+\pi^-}}{\Gamma_{93} + \Gamma_{802} + \Gamma_{151}\cdot{}\Gamma_{\omega\to\pi^+\pi^-}}%
\htconstrdef{Gamma80by60.c}{\frac{\Gamma_{80}}{\Gamma_{60}}}{\frac{\Gamma_{80}}{\Gamma_{60}}}{\frac{\Gamma_{80}}{\Gamma_{60}}}%
\htconstrdef{Gamma81.c}{\Gamma_{81}}{\Gamma_{128}\cdot{}\Gamma_{\eta\to\pi^+\pi^-\pi^0} + \Gamma_{94} + \Gamma_{803} + \Gamma_{151}\cdot{}\Gamma_{\omega\to\pi^+\pi^-\pi^0}}{\Gamma_{128}\cdot{}\Gamma_{\eta\to\pi^+\pi^-\pi^0} + \Gamma_{94} + \Gamma_{803} + \Gamma_{151}\cdot{}\Gamma_{\omega\to\pi^+\pi^-\pi^0}}%
\htconstrdef{Gamma81by69.c}{\frac{\Gamma_{81}}{\Gamma_{69}}}{\frac{\Gamma_{81}}{\Gamma_{69}}}{\frac{\Gamma_{81}}{\Gamma_{69}}}%
\htconstrdef{Gamma82.c}{\Gamma_{82}}{\Gamma_{128}\cdot{}\Gamma_{\eta\to\text{charged}} + \Gamma_{42}\cdot{}(\Gamma_{<K^0|K_S>}\cdot{}\Gamma_{K_S\to\pi^+\pi^-}) + \Gamma_{802} + \Gamma_{803} + \Gamma_{151}\cdot{}(\Gamma_{\omega\to\pi^+\pi^-\pi^0}+\Gamma_{\omega\to\pi^+\pi^-}) + \Gamma_{37}\cdot{}(\Gamma_{<K^0|K_S>}\cdot{}\Gamma_{K_S\to\pi^+\pi^-})}{\Gamma_{128}\cdot{}\Gamma_{\eta\to\text{charged}} + \Gamma_{42}\cdot{}(\Gamma_{<K^0|K_S>}\cdot{}\Gamma_{K_S\to\pi^+\pi^-}) + \Gamma_{802}  \\ 
  {}& + \Gamma_{803} + \Gamma_{151}\cdot{}(\Gamma_{\omega\to\pi^+\pi^-\pi^0}+\Gamma_{\omega\to\pi^+\pi^-}) + \Gamma_{37}\cdot{}(\Gamma_{<K^0|K_S>}\cdot{}\Gamma_{K_S\to\pi^+\pi^-})}%
\htconstrdef{Gamma83.c}{\Gamma_{83}}{\Gamma_{128}\cdot{}\Gamma_{\eta\to\pi^+\pi^-\pi^0} + \Gamma_{802} + \Gamma_{803} + \Gamma_{151}\cdot{}(\Gamma_{\omega\to\pi^+\pi^-\pi^0}+\Gamma_{\omega\to\pi^+\pi^-})}{\Gamma_{128}\cdot{}\Gamma_{\eta\to\pi^+\pi^-\pi^0} + \Gamma_{802} + \Gamma_{803} + \Gamma_{151}\cdot{}(\Gamma_{\omega\to\pi^+\pi^-\pi^0} \\ 
  {}& +\Gamma_{\omega\to\pi^+\pi^-})}%
\htconstrdef{Gamma84.c}{\Gamma_{84}}{\Gamma_{802} + \Gamma_{151}\cdot{}\Gamma_{\omega\to\pi^+\pi^-} + \Gamma_{37}\cdot{}(\Gamma_{<K^0|K_S>}\cdot{}\Gamma_{K_S\to\pi^+\pi^-})}{\Gamma_{802} + \Gamma_{151}\cdot{}\Gamma_{\omega\to\pi^+\pi^-} + \Gamma_{37}\cdot{}(\Gamma_{<K^0|K_S>}\cdot{}\Gamma_{K_S\to\pi^+\pi^-})}%
\htconstrdef{Gamma85.c}{\Gamma_{85}}{\Gamma_{802} + \Gamma_{151}\cdot{}\Gamma_{\omega\to\pi^+\pi^-}}{\Gamma_{802} + \Gamma_{151}\cdot{}\Gamma_{\omega\to\pi^+\pi^-}}%
\htconstrdef{Gamma85by60.c}{\frac{\Gamma_{85}}{\Gamma_{60}}}{\frac{\Gamma_{85}}{\Gamma_{60}}}{\frac{\Gamma_{85}}{\Gamma_{60}}}%
\htconstrdef{Gamma87.c}{\Gamma_{87}}{\Gamma_{42}\cdot{}(\Gamma_{<K^0|K_S>}\cdot{}\Gamma_{K_S\to\pi^+\pi^-}) + \Gamma_{128}\cdot{}\Gamma_{\eta\to\pi^+\pi^-\pi^0} + \Gamma_{151}\cdot{}\Gamma_{\omega\to\pi^+\pi^-\pi^0} + \Gamma_{803}}{\Gamma_{42}\cdot{}(\Gamma_{<K^0|K_S>}\cdot{}\Gamma_{K_S\to\pi^+\pi^-}) + \Gamma_{128}\cdot{}\Gamma_{\eta\to\pi^+\pi^-\pi^0} + \Gamma_{151}\cdot{}\Gamma_{\omega\to\pi^+\pi^-\pi^0}  \\ 
  {}& + \Gamma_{803}}%
\htconstrdef{Gamma88.c}{\Gamma_{88}}{\Gamma_{128}\cdot{}\Gamma_{\eta\to\pi^+\pi^-\pi^0} + \Gamma_{803} + \Gamma_{151}\cdot{}\Gamma_{\omega\to\pi^+\pi^-\pi^0}}{\Gamma_{128}\cdot{}\Gamma_{\eta\to\pi^+\pi^-\pi^0} + \Gamma_{803} + \Gamma_{151}\cdot{}\Gamma_{\omega\to\pi^+\pi^-\pi^0}}%
\htconstrdef{Gamma89.c}{\Gamma_{89}}{\Gamma_{803} + \Gamma_{151}\cdot{}\Gamma_{\omega\to\pi^+\pi^-\pi^0}}{\Gamma_{803} + \Gamma_{151}\cdot{}\Gamma_{\omega\to\pi^+\pi^-\pi^0}}%
\htconstrdef{Gamma92.c}{\Gamma_{92}}{\Gamma_{94} + \Gamma_{93}}{\Gamma_{94} + \Gamma_{93}}%
\htconstrdef{Gamma93by60.c}{\frac{\Gamma_{93}}{\Gamma_{60}}}{\frac{\Gamma_{93}}{\Gamma_{60}}}{\frac{\Gamma_{93}}{\Gamma_{60}}}%
\htconstrdef{Gamma94by69.c}{\frac{\Gamma_{94}}{\Gamma_{69}}}{\frac{\Gamma_{94}}{\Gamma_{69}}}{\frac{\Gamma_{94}}{\Gamma_{69}}}%
\htconstrdef{Gamma96.c}{\Gamma_{96}}{\Gamma_{167}\cdot{}\Gamma_{\phi\to K^+K^-}}{\Gamma_{167}\cdot{}\Gamma_{\phi\to K^+K^-}}%
\htconstrdef{Gamma102.c}{\Gamma_{102}}{\Gamma_{103} + \Gamma_{104}}{\Gamma_{103} + \Gamma_{104}}%
\htconstrdef{Gamma103.c}{\Gamma_{103}}{\Gamma_{820} + \Gamma_{822} + \Gamma_{831}\cdot{}\Gamma_{\omega\to\pi^+\pi^-}}{\Gamma_{820} + \Gamma_{822} + \Gamma_{831}\cdot{}\Gamma_{\omega\to\pi^+\pi^-}}%
\htconstrdef{Gamma104.c}{\Gamma_{104}}{\Gamma_{830} + \Gamma_{833}}{\Gamma_{830} + \Gamma_{833}}%
\htconstrdef{Gamma106.c}{\Gamma_{106}}{\Gamma_{30} + \Gamma_{44}\cdot{}\Gamma_{<\bar{K}^0|K_S>} + \Gamma_{47} + \Gamma_{53}\cdot{}\Gamma_{<K^0|K_S>} + \Gamma_{77} + \Gamma_{103} + \Gamma_{126}\cdot{}(\Gamma_{\eta\to3\pi^0}+\Gamma_{\eta\to\pi^+\pi^-\pi^0}) + \Gamma_{152}\cdot{}\Gamma_{\omega\to\pi^+\pi^-\pi^0}}{\Gamma_{30} + \Gamma_{44}\cdot{}\Gamma_{<\bar{K}^0|K_S>} + \Gamma_{47} + \Gamma_{53}\cdot{}\Gamma_{<K^0|K_S>}  \\ 
  {}& + \Gamma_{77} + \Gamma_{103} + \Gamma_{126}\cdot{}(\Gamma_{\eta\to3\pi^0}+\Gamma_{\eta\to\pi^+\pi^-\pi^0}) + \Gamma_{152}\cdot{}\Gamma_{\omega\to\pi^+\pi^-\pi^0}}%
\htconstrdef{Gamma110.c}{\Gamma_{110}}{\Gamma_{10} + \Gamma_{16} + \Gamma_{23} + \Gamma_{28} + \Gamma_{35} + \Gamma_{40} + \Gamma_{128} + \Gamma_{802} + \Gamma_{803} + \Gamma_{151} + \Gamma_{130} + \Gamma_{132} + \Gamma_{44} + \Gamma_{53} + \Gamma_{168} + \Gamma_{169} + \Gamma_{822} + \Gamma_{833}}{\Gamma_{10} + \Gamma_{16} + \Gamma_{23} + \Gamma_{28} + \Gamma_{35} + \Gamma_{40}  \\ 
  {}& + \Gamma_{128} + \Gamma_{802} + \Gamma_{803} + \Gamma_{151} + \Gamma_{130} + \Gamma_{132}  \\ 
  {}& + \Gamma_{44} + \Gamma_{53} + \Gamma_{168} + \Gamma_{169} + \Gamma_{822} + \Gamma_{833}}%
\htconstrdef{Gamma149.c}{\Gamma_{149}}{\Gamma_{152} + \Gamma_{800} + \Gamma_{151}}{\Gamma_{152} + \Gamma_{800} + \Gamma_{151}}%
\htconstrdef{Gamma150.c}{\Gamma_{150}}{\Gamma_{800} + \Gamma_{151}}{\Gamma_{800} + \Gamma_{151}}%
\htconstrdef{Gamma150by66.c}{\frac{\Gamma_{150}}{\Gamma_{66}}}{\frac{\Gamma_{150}}{\Gamma_{66}}}{\frac{\Gamma_{150}}{\Gamma_{66}}}%
\htconstrdef{Gamma152by54.c}{\frac{\Gamma_{152}}{\Gamma_{54}}}{\frac{\Gamma_{152}}{\Gamma_{54}}}{\frac{\Gamma_{152}}{\Gamma_{54}}}%
\htconstrdef{Gamma152by76.c}{\frac{\Gamma_{152}}{\Gamma_{76}}}{\frac{\Gamma_{152}}{\Gamma_{76}}}{\frac{\Gamma_{152}}{\Gamma_{76}}}%
\htconstrdef{Gamma168.c}{\Gamma_{168}}{\Gamma_{167}\cdot{}\Gamma_{\phi\to K^+K^-}}{\Gamma_{167}\cdot{}\Gamma_{\phi\to K^+K^-}}%
\htconstrdef{Gamma169.c}{\Gamma_{169}}{\Gamma_{167}\cdot{}\Gamma_{\phi\to K_S K_L}}{\Gamma_{167}\cdot{}\Gamma_{\phi\to K_S K_L}}%
\htconstrdef{Gamma804.c}{\Gamma_{804}}{\Gamma_{47} \cdot{} ((\Gamma_{<K^0|K_L>}\cdot{}\Gamma_{<\bar{K}^0|K_L>}) / (\Gamma_{<K^0|K_S>}\cdot{}\Gamma_{<\bar{K}^0|K_S>}))}{\Gamma_{47} \cdot{} ((\Gamma_{<K^0|K_L>}\cdot{}\Gamma_{<\bar{K}^0|K_L>}) / (\Gamma_{<K^0|K_S>}\cdot{}\Gamma_{<\bar{K}^0|K_S>}))}%
\htconstrdef{Gamma806.c}{\Gamma_{806}}{\Gamma_{50} \cdot{} ((\Gamma_{<K^0|K_L>}\cdot{}\Gamma_{<\bar{K}^0|K_L>}) / (\Gamma_{<K^0|K_S>}\cdot{}\Gamma_{<\bar{K}^0|K_S>}))}{\Gamma_{50} \cdot{} ((\Gamma_{<K^0|K_L>}\cdot{}\Gamma_{<\bar{K}^0|K_L>}) / (\Gamma_{<K^0|K_S>}\cdot{}\Gamma_{<\bar{K}^0|K_S>}))}%
\htconstrdef{Gamma810.c}{\Gamma_{810}}{\Gamma_{910} + \Gamma_{911} + \Gamma_{811}\cdot{}\Gamma_{\omega\to\pi^+\pi^-\pi^0} + \Gamma_{812}}{\Gamma_{910} + \Gamma_{911} + \Gamma_{811}\cdot{}\Gamma_{\omega\to\pi^+\pi^-\pi^0} + \Gamma_{812}}%
\htconstrdef{Gamma820.c}{\Gamma_{820}}{\Gamma_{920} + \Gamma_{821}}{\Gamma_{920} + \Gamma_{821}}%
\htconstrdef{Gamma830.c}{\Gamma_{830}}{\Gamma_{930} + \Gamma_{831}\cdot{}\Gamma_{\omega\to\pi^+\pi^-\pi^0} + \Gamma_{832}}{\Gamma_{930} + \Gamma_{831}\cdot{}\Gamma_{\omega\to\pi^+\pi^-\pi^0} + \Gamma_{832}}%
\htconstrdef{Gamma910.c}{\Gamma_{910}}{\Gamma_{136}\cdot{}\Gamma_{\eta\to3\pi^0}}{\Gamma_{136}\cdot{}\Gamma_{\eta\to3\pi^0}}%
\htconstrdef{Gamma911.c}{\Gamma_{911}}{\Gamma_{945}\cdot{}\Gamma_{\eta\to\pi^+\pi^-\pi^0}}{\Gamma_{945}\cdot{}\Gamma_{\eta\to\pi^+\pi^-\pi^0}}%
\htconstrdef{Gamma930.c}{\Gamma_{930}}{\Gamma_{136}\cdot{}\Gamma_{\eta\to\pi^+\pi^-\pi^0}}{\Gamma_{136}\cdot{}\Gamma_{\eta\to\pi^+\pi^-\pi^0}}%
\htconstrdef{Gamma944.c}{\Gamma_{944}}{\Gamma_{136}\cdot{}\Gamma_{\eta\to\gamma\gamma}}{\Gamma_{136}\cdot{}\Gamma_{\eta\to\gamma\gamma}}%
\htconstrdef{GammaAll.c}{\Gamma_{\text{All}}}{\Gamma_{3} + \Gamma_{5} + \Gamma_{9} + \Gamma_{10} + \Gamma_{14} + \Gamma_{16} + \Gamma_{20} + \Gamma_{23} + \Gamma_{27} + \Gamma_{28} + \Gamma_{30} + \Gamma_{35} + \Gamma_{37} + \Gamma_{40} + \Gamma_{42} + \Gamma_{47}\cdot{}(1 + ((\Gamma_{<K^0|K_L>}\cdot{}\Gamma_{<\bar{K}^0|K_L>}) / (\Gamma_{<K^0|K_S>}\cdot{}\Gamma_{<\bar{K}^0|K_S>}))) + \Gamma_{48} + \Gamma_{62} + \Gamma_{70} + \Gamma_{77} + \Gamma_{811} + \Gamma_{812} + \Gamma_{93} + \Gamma_{94} + \Gamma_{832} + \Gamma_{833} + \Gamma_{126} + \Gamma_{128} + \Gamma_{802} + \Gamma_{803} + \Gamma_{800} + \Gamma_{151} + \Gamma_{130} + \Gamma_{132} + \Gamma_{44} + \Gamma_{53} + \Gamma_{50}\cdot{}(1 + ((\Gamma_{<K^0|K_L>}\cdot{}\Gamma_{<\bar{K}^0|K_L>}) / (\Gamma_{<K^0|K_S>}\cdot{}\Gamma_{<\bar{K}^0|K_S>}))) + \Gamma_{51} + \Gamma_{167}\cdot{}(\Gamma_{\phi\to K^+K^-}+\Gamma_{\phi\to K_S K_L}) + \Gamma_{152} + \Gamma_{920} + \Gamma_{821} + \Gamma_{822} + \Gamma_{831} + \Gamma_{136} + \Gamma_{945} + \Gamma_{805}}{\Gamma_{3} + \Gamma_{5} + \Gamma_{9} + \Gamma_{10} + \Gamma_{14} + \Gamma_{16}  \\ 
  {}& + \Gamma_{20} + \Gamma_{23} + \Gamma_{27} + \Gamma_{28} + \Gamma_{30} + \Gamma_{35}  \\ 
  {}& + \Gamma_{37} + \Gamma_{40} + \Gamma_{42} + \Gamma_{47}\cdot{}(1 + ((\Gamma_{<K^0|K_L>}\cdot{}\Gamma_{<\bar{K}^0|K_L>}) / (\Gamma_{<K^0|K_S>}\cdot{}\Gamma_{<\bar{K}^0|K_S>})))  \\ 
  {}& + \Gamma_{48} + \Gamma_{62} + \Gamma_{70} + \Gamma_{77} + \Gamma_{811} + \Gamma_{812}  \\ 
  {}& + \Gamma_{93} + \Gamma_{94} + \Gamma_{832} + \Gamma_{833} + \Gamma_{126} + \Gamma_{128}  \\ 
  {}& + \Gamma_{802} + \Gamma_{803} + \Gamma_{800} + \Gamma_{151} + \Gamma_{130} + \Gamma_{132}  \\ 
  {}& + \Gamma_{44} + \Gamma_{53} + \Gamma_{50}\cdot{}(1 + ((\Gamma_{<K^0|K_L>}\cdot{}\Gamma_{<\bar{K}^0|K_L>}) / (\Gamma_{<K^0|K_S>}\cdot{}\Gamma_{<\bar{K}^0|K_S>})))  \\ 
  {}& + \Gamma_{51} + \Gamma_{167}\cdot{}(\Gamma_{\phi\to K^+K^-}+\Gamma_{\phi\to K_S K_L}) + \Gamma_{152} + \Gamma_{920}  \\ 
  {}& + \Gamma_{821} + \Gamma_{822} + \Gamma_{831} + \Gamma_{136} + \Gamma_{945} + \Gamma_{805}}%
\htconstrdef{Unitarity}{1}{\Gamma_{\text{All}} + \Gamma_{998}}{\Gamma_{\text{All}} + \Gamma_{998}}%
\htdef{ConstrEqs}{%
\begin{align*}
\htuse{Gamma1.c.left} ={}& \htuse{Gamma1.c.right.split}
\end{align*}
\begin{align*}
\htuse{Gamma2.c.left} ={}& \htuse{Gamma2.c.right.split}
\end{align*}
\begin{align*}
\htuse{Gamma7.c.left} ={}& \htuse{Gamma7.c.right.split}
\end{align*}
\begin{align*}
\htuse{Gamma8.c.left} ={}& \htuse{Gamma8.c.right.split}
\end{align*}
\begin{align*}
\htuse{Gamma11.c.left} ={}& \htuse{Gamma11.c.right.split}
\end{align*}
\begin{align*}
\htuse{Gamma12.c.left} ={}& \htuse{Gamma12.c.right.split}
\end{align*}
\begin{align*}
\htuse{Gamma13.c.left} ={}& \htuse{Gamma13.c.right.split}
\end{align*}
\begin{align*}
\htuse{Gamma17.c.left} ={}& \htuse{Gamma17.c.right.split}
\end{align*}
\begin{align*}
\htuse{Gamma18.c.left} ={}& \htuse{Gamma18.c.right.split}
\end{align*}
\begin{align*}
\htuse{Gamma19.c.left} ={}& \htuse{Gamma19.c.right.split}
\end{align*}
\begin{align*}
\htuse{Gamma24.c.left} ={}& \htuse{Gamma24.c.right.split}
\end{align*}
\begin{align*}
\htuse{Gamma25.c.left} ={}& \htuse{Gamma25.c.right.split}
\end{align*}
\begin{align*}
\htuse{Gamma26.c.left} ={}& \htuse{Gamma26.c.right.split}
\end{align*}
\begin{align*}
\htuse{Gamma29.c.left} ={}& \htuse{Gamma29.c.right.split}
\end{align*}
\begin{align*}
\htuse{Gamma31.c.left} ={}& \htuse{Gamma31.c.right.split}
\end{align*}
\begin{align*}
\htuse{Gamma32.c.left} ={}& \htuse{Gamma32.c.right.split}
\end{align*}
\begin{align*}
\htuse{Gamma33.c.left} ={}& \htuse{Gamma33.c.right.split}
\end{align*}
\begin{align*}
\htuse{Gamma34.c.left} ={}& \htuse{Gamma34.c.right.split}
\end{align*}
\begin{align*}
\htuse{Gamma38.c.left} ={}& \htuse{Gamma38.c.right.split}
\end{align*}
\begin{align*}
\htuse{Gamma39.c.left} ={}& \htuse{Gamma39.c.right.split}
\end{align*}
\begin{align*}
\htuse{Gamma43.c.left} ={}& \htuse{Gamma43.c.right.split}
\end{align*}
\begin{align*}
\htuse{Gamma46.c.left} ={}& \htuse{Gamma46.c.right.split}
\end{align*}
\begin{align*}
\htuse{Gamma49.c.left} ={}& \htuse{Gamma49.c.right.split}
\end{align*}
\begin{align*}
\htuse{Gamma54.c.left} ={}& \htuse{Gamma54.c.right.split}
\end{align*}
\begin{align*}
\htuse{Gamma55.c.left} ={}& \htuse{Gamma55.c.right.split}
\end{align*}
\begin{align*}
\htuse{Gamma56.c.left} ={}& \htuse{Gamma56.c.right.split}
\end{align*}
\begin{align*}
\htuse{Gamma57.c.left} ={}& \htuse{Gamma57.c.right.split}
\end{align*}
\begin{align*}
\htuse{Gamma58.c.left} ={}& \htuse{Gamma58.c.right.split}
\end{align*}
\begin{align*}
\htuse{Gamma59.c.left} ={}& \htuse{Gamma59.c.right.split}
\end{align*}
\begin{align*}
\htuse{Gamma60.c.left} ={}& \htuse{Gamma60.c.right.split}
\end{align*}
\begin{align*}
\htuse{Gamma63.c.left} ={}& \htuse{Gamma63.c.right.split}
\end{align*}
\begin{align*}
\htuse{Gamma64.c.left} ={}& \htuse{Gamma64.c.right.split}
\end{align*}
\begin{align*}
\htuse{Gamma65.c.left} ={}& \htuse{Gamma65.c.right.split}
\end{align*}
\begin{align*}
\htuse{Gamma66.c.left} ={}& \htuse{Gamma66.c.right.split}
\end{align*}
\begin{align*}
\htuse{Gamma67.c.left} ={}& \htuse{Gamma67.c.right.split}
\end{align*}
\begin{align*}
\htuse{Gamma68.c.left} ={}& \htuse{Gamma68.c.right.split}
\end{align*}
\begin{align*}
\htuse{Gamma69.c.left} ={}& \htuse{Gamma69.c.right.split}
\end{align*}
\begin{align*}
\htuse{Gamma74.c.left} ={}& \htuse{Gamma74.c.right.split}
\end{align*}
\begin{align*}
\htuse{Gamma75.c.left} ={}& \htuse{Gamma75.c.right.split}
\end{align*}
\begin{align*}
\htuse{Gamma76.c.left} ={}& \htuse{Gamma76.c.right.split}
\end{align*}
\begin{align*}
\htuse{Gamma78.c.left} ={}& \htuse{Gamma78.c.right.split}
\end{align*}
\begin{align*}
\htuse{Gamma79.c.left} ={}& \htuse{Gamma79.c.right.split}
\end{align*}
\begin{align*}
\htuse{Gamma80.c.left} ={}& \htuse{Gamma80.c.right.split}
\end{align*}
\begin{align*}
\htuse{Gamma81.c.left} ={}& \htuse{Gamma81.c.right.split}
\end{align*}
\begin{align*}
\htuse{Gamma82.c.left} ={}& \htuse{Gamma82.c.right.split}
\end{align*}
\begin{align*}
\htuse{Gamma83.c.left} ={}& \htuse{Gamma83.c.right.split}
\end{align*}
\begin{align*}
\htuse{Gamma84.c.left} ={}& \htuse{Gamma84.c.right.split}
\end{align*}
\begin{align*}
\htuse{Gamma85.c.left} ={}& \htuse{Gamma85.c.right.split}
\end{align*}
\begin{align*}
\htuse{Gamma87.c.left} ={}& \htuse{Gamma87.c.right.split}
\end{align*}
\begin{align*}
\htuse{Gamma88.c.left} ={}& \htuse{Gamma88.c.right.split}
\end{align*}
\begin{align*}
\htuse{Gamma89.c.left} ={}& \htuse{Gamma89.c.right.split}
\end{align*}
\begin{align*}
\htuse{Gamma92.c.left} ={}& \htuse{Gamma92.c.right.split}
\end{align*}
\begin{align*}
\htuse{Gamma96.c.left} ={}& \htuse{Gamma96.c.right.split}
\end{align*}
\begin{align*}
\htuse{Gamma102.c.left} ={}& \htuse{Gamma102.c.right.split}
\end{align*}
\begin{align*}
\htuse{Gamma103.c.left} ={}& \htuse{Gamma103.c.right.split}
\end{align*}
\begin{align*}
\htuse{Gamma104.c.left} ={}& \htuse{Gamma104.c.right.split}
\end{align*}
\begin{align*}
\htuse{Gamma106.c.left} ={}& \htuse{Gamma106.c.right.split}
\end{align*}
\begin{align*}
\htuse{Gamma110.c.left} ={}& \htuse{Gamma110.c.right.split}
\end{align*}
\begin{align*}
\htuse{Gamma149.c.left} ={}& \htuse{Gamma149.c.right.split}
\end{align*}
\begin{align*}
\htuse{Gamma150.c.left} ={}& \htuse{Gamma150.c.right.split}
\end{align*}
\begin{align*}
\htuse{Gamma168.c.left} ={}& \htuse{Gamma168.c.right.split}
\end{align*}
\begin{align*}
\htuse{Gamma169.c.left} ={}& \htuse{Gamma169.c.right.split}
\end{align*}
\begin{align*}
\htuse{Gamma804.c.left} ={}& \htuse{Gamma804.c.right.split}
\end{align*}
\begin{align*}
\htuse{Gamma806.c.left} ={}& \htuse{Gamma806.c.right.split}
\end{align*}
\begin{align*}
\htuse{Gamma810.c.left} ={}& \htuse{Gamma810.c.right.split}
\end{align*}
\begin{align*}
\htuse{Gamma820.c.left} ={}& \htuse{Gamma820.c.right.split}
\end{align*}
\begin{align*}
\htuse{Gamma830.c.left} ={}& \htuse{Gamma830.c.right.split}
\end{align*}
\begin{align*}
\htuse{Gamma910.c.left} ={}& \htuse{Gamma910.c.right.split}
\end{align*}
\begin{align*}
\htuse{Gamma911.c.left} ={}& \htuse{Gamma911.c.right.split}
\end{align*}
\begin{align*}
\htuse{Gamma930.c.left} ={}& \htuse{Gamma930.c.right.split}
\end{align*}
\begin{align*}
\htuse{Gamma944.c.left} ={}& \htuse{Gamma944.c.right.split}
\end{align*}
\begin{align*}
\htuse{GammaAll.c.left} ={}& \htuse{GammaAll.c.right.split}
\end{align*}}%
\htdef{NumMeasALEPH}{39}%
\htdef{NumMeasARGUS}{2}%
\htdef{NumMeasBaBar}{23}%
\htdef{NumMeasBelle}{15}%
\htdef{NumMeasCELLO}{1}%
\htdef{NumMeasCLEO}{35}%
\htdef{NumMeasCLEO3}{6}%
\htdef{NumMeasDELPHI}{14}%
\htdef{NumMeasHRS}{2}%
\htdef{NumMeasL3}{11}%
\htdef{NumMeasOPAL}{19}%
\htdef{NumMeasTPC}{3}%

\htquantdef{B_tau_had_fit}{B_tau_had_fit}{}{64.76 \pm 0.10}{64.76}{0.10}%
\htquantdef{B_tau_s_fit}{B_tau_s_fit}{}{2.909 \pm 0.048}{2.909}{0.048}%
\htquantdef{B_tau_s_unitarity}{B_tau_s_unitarity}{}{(2.944 \pm 0.103) \cdot 10^{-2}}{2.944\cdot 10^{-2}}{0.103\cdot 10^{-2}}%
\htquantdef{B_tau_VA}{B_tau_VA}{}{0.6185 \pm 0.0010}{0.6185}{0.0010}%
\htquantdef{B_tau_VA_fit}{B_tau_VA_fit}{}{61.85 \pm 0.10}{61.85}{0.10}%
\htquantdef{B_tau_VA_unitarity}{B_tau_VA_unitarity}{}{0.61883 \pm 0.00080}{0.61883}{0.00080}%
\htquantdef{Be_fit}{Be_fit}{}{0.17816 \pm 0.00041}{0.17816}{0.00041}%
\htquantdef{Be_from_Bmu}{Be_from_Bmu}{}{0.17882 \pm 0.00041}{0.17882}{0.00041}%
\htquantdef{Be_from_taulife}{Be_from_taulife}{}{0.17780 \pm 0.00032}{0.17780}{0.00032}%
\htquantdef{Be_lept}{Be_lept}{}{17.850 \pm 0.032}{17.850}{0.032}%
\htquantdef{Be_unitarity}{Be_unitarity}{}{0.1785 \pm 0.0010}{0.1785}{0.0010}%
\htquantdef{Be_univ}{Be_univ}{}{17.815 \pm 0.023}{17.815}{0.023}%
\htquantdef{Bmu_by_Be_th}{Bmu_by_Be_th}{}{0.9725606 \pm 0.0000036}{0.9725606}{0.0000036}%
\htquantdef{Bmu_fit}{Bmu_fit}{}{0.17392 \pm 0.00040}{0.17392}{0.00040}%
\htquantdef{Bmu_from_taulife}{Bmu_from_taulife}{}{0.17292 \pm 0.00032}{0.17292}{0.00032}%
\htquantdef{Bmu_unitarity}{Bmu_unitarity}{}{0.1743 \pm 0.0010}{0.1743}{0.0010}%
\htquantdef{BR_a1_pigamma}{BR_a1_pigamma}{}{0.2100\cdot 10^{-2}}{0.2100\cdot 10^{-2}}{0}%
\htquantdef{BR_eta_2gam}{BR_eta_2gam}{}{0.3941}{0.3941}{0}%
\htquantdef{BR_eta_3piz}{BR_eta_3piz}{}{0.3268}{0.3268}{0}%
\htquantdef{BR_eta_charged}{BR_eta_charged}{}{0.2810}{0.2810}{0}%
\htquantdef{BR_eta_neutral}{BR_eta_neutral}{}{0.7212}{0.7212}{0}%
\htquantdef{BR_eta_pimpipgamma}{BR_eta_pimpipgamma}{}{4.220\cdot 10^{-2}}{4.220\cdot 10^{-2}}{0}%
\htquantdef{BR_eta_pimpippiz}{BR_eta_pimpippiz}{}{0.2292}{0.2292}{0}%
\htquantdef{BR_f1_2pip2pim}{BR_f1_2pip2pim}{}{0.1100}{0.1100}{0}%
\htquantdef{BR_f1_2pizpippim}{BR_f1_2pizpippim}{}{0.2200}{0.2200}{0}%
\htquantdef{BR_KS_2piz}{BR_KS_2piz}{}{0.3069}{0.3069}{0}%
\htquantdef{BR_KS_pimpip}{BR_KS_pimpip}{}{0.6920}{0.6920}{0}%
\htquantdef{BR_om_pimpip}{BR_om_pimpip}{}{1.530\cdot 10^{-2}}{1.530\cdot 10^{-2}}{0}%
\htquantdef{BR_om_pimpippiz}{BR_om_pimpippiz}{}{0.8920}{0.8920}{0}%
\htquantdef{BR_om_pizgamma}{BR_om_pizgamma}{}{8.280\cdot 10^{-2}}{8.280\cdot 10^{-2}}{0}%
\htquantdef{BR_phi_KmKp}{BR_phi_KmKp}{}{0.4890}{0.4890}{0}%
\htquantdef{BR_phi_KSKL}{BR_phi_KSKL}{}{0.3420}{0.3420}{0}%
\htquantdef{BRA_Kz_KL_KET}{BRA_Kz_KL_KET}{}{0.5000}{0.5000}{0}%
\htquantdef{BRA_Kz_KS_KET}{BRA_Kz_KS_KET}{}{0.5000}{0.5000}{0}%
\htquantdef{BRA_Kzbar_KL_KET}{BRA_Kzbar_KL_KET}{}{0.5000}{0.5000}{0}%
\htquantdef{BRA_Kzbar_KS_KET}{BRA_Kzbar_KS_KET}{}{0.5000}{0.5000}{0}%
\htquantdef{delta_mu_gamma}{delta_mu_gamma}{}{0.9958}{0.9958}{0}%
\htquantdef{delta_mu_W}{delta_mu_W}{}{1.00000103667 \pm 0.00000000039}{1.00000103667}{0.00000000039}%
\htquantdef{delta_tau_gamma}{delta_tau_gamma}{}{0.9957}{0.9957}{0}%
\htquantdef{delta_tau_W}{delta_tau_W}{}{1.00029627 \pm 0.00000012}{1.00029627}{0.00000012}%
\htquantdef{deltaR_su3break}{deltaR_su3break}{}{0.242 \pm 0.032}{0.242}{0.032}%
\htquantdef{deltaR_su3break_d2pert}{deltaR_su3break_d2pert}{}{9.300 \pm 3.400}{9.300}{3.400}%
\htquantdef{deltaR_su3break_pheno}{deltaR_su3break_pheno}{}{0.1544 \pm 0.0037}{0.1544}{0.0037}%
\htquantdef{deltaR_su3break_remain}{deltaR_su3break_remain}{}{(0.3400 \pm 0.2800) \cdot 10^{-2}}{0.3400\cdot 10^{-2}}{0.2800\cdot 10^{-2}}%
\htquantdef{dRrad_k_munu}{dRrad_k_munu}{}{1.30 \pm 0.20}{1.30}{0.20}%
\htquantdef{dRrad_kmunu_by_pimunu}{dRrad_kmunu_by_pimunu}{}{-1.13 \pm 0.23}{-1.13}{0.23}%
\htquantdef{dRrad_tauK_by_Kmu}{dRrad_tauK_by_Kmu}{}{0.90 \pm 0.22}{0.90}{0.22}%
\htquantdef{dRrad_taupi_by_pimu}{dRrad_taupi_by_pimu}{}{0.16 \pm 0.14}{0.16}{0.14}%
\htquantdef{EmNuNumb}{EmNuNumb}{}{0.1783}{0.1783}{0}%
\htquantdef{f_K}{f_K}{}{155.6 \pm 0.4}{155.6}{0.4}%
\htquantdef{f_K_by_f_pi}{f_K_by_f_pi}{}{1.1930 \pm 0.0030}{1.1930}{0.0030}%
\htquantdef{fp0_Kpi}{fp0_Kpi}{}{0.9677 \pm 0.0027}{0.9677}{0.0027}%
\htquantdef{G_F_by_hcut3_c3}{G_F_by_hcut3_c3}{}{(1.16637870 \pm 0.00000060) \cdot 10^{-11}}{1.16637870\cdot 10^{-11}}{0.00000060\cdot 10^{-11}}%
\htquantdef{Gamma1}{\Gamma_{1}}{\BRF{\tau^-}{(\text{particles})^- \ge{} 0\, \text{neutrals} \ge{} 0\,  K^0\, \nu_\tau}}{0.8519 \pm 0.0011}{0.8519}{0.0011}%
\htquantdef{Gamma10}{\Gamma_{10}}{\BRF{\tau^-}{K^- \nu_\tau}}{(0.6960 \pm 0.0096) \cdot 10^{-2}}{0.6960\cdot 10^{-2}}{0.0096\cdot 10^{-2}}%
\htquantdef{Gamma102}{\Gamma_{102}}{\BRF{\tau^-}{3h^- 2h^+ \ge{} 0\,  \text{neutrals}\, \nu_\tau\;(\text{ex.~} K^0)}}{(9.850 \pm 0.370) \cdot 10^{-4}}{9.850\cdot 10^{-4}}{0.370\cdot 10^{-4}}%
\htquantdef{Gamma103}{\Gamma_{103}}{\BRF{\tau^-}{3h^- 2h^+ \nu_\tau ~(\text{ex.~}K^0)}}{(8.216 \pm 0.316) \cdot 10^{-4}}{8.216\cdot 10^{-4}}{0.316\cdot 10^{-4}}%
\htquantdef{Gamma104}{\Gamma_{104}}{\BRF{\tau^-}{3h^- 2h^+ \pi^0 \nu_\tau ~(\text{ex.~}K^0)}}{(1.634 \pm 0.114) \cdot 10^{-4}}{1.634\cdot 10^{-4}}{0.114\cdot 10^{-4}}%
\htquantdef{Gamma106}{\Gamma_{106}}{\BRF{\tau^-}{(5\pi)^- \nu_\tau}}{(0.7748 \pm 0.0534) \cdot 10^{-2}}{0.7748\cdot 10^{-2}}{0.0534\cdot 10^{-2}}%
\htquantdef{Gamma10by5}{\frac{\Gamma_{10}}{\Gamma_{5}}}{\frac{\BRF{\tau^-}{K^- \nu_\tau}}{\BRF{\tau^-}{e^- \bar{\nu}_e \nu_\tau}}}{(3.906 \pm 0.054) \cdot 10^{-2}}{3.906\cdot 10^{-2}}{0.054\cdot 10^{-2}}%
\htquantdef{Gamma10by9}{\frac{\Gamma_{10}}{\Gamma_{9}}}{\frac{\BRF{\tau^-}{K^- \nu_\tau}}{\BRF{\tau^-}{\pi^- \nu_\tau}}}{(6.438 \pm 0.094) \cdot 10^{-2}}{6.438\cdot 10^{-2}}{0.094\cdot 10^{-2}}%
\htquantdef{Gamma11}{\Gamma_{11}}{\BRF{\tau^-}{h^- \ge{} 1\,  \text{neutrals}\, \nu_\tau}}{0.36973 \pm 0.00097}{0.36973}{0.00097}%
\htquantdef{Gamma110}{\Gamma_{110}}{\BRF{\tau^-}{X_s^- \nu_\tau}}{(2.909 \pm 0.048) \cdot 10^{-2}}{2.909\cdot 10^{-2}}{0.048\cdot 10^{-2}}%
\htquantdef{Gamma110_pdg09}{\Gamma_{110}_pdg09}{}{(2.841 \pm 0.038) \cdot 10^{-2}}{2.841\cdot 10^{-2}}{0.038\cdot 10^{-2}}%
\htquantdef{Gamma12}{\Gamma_{12}}{\BRF{\tau^-}{h^- \ge{} 1\, \pi^0\, \nu_\tau\;(\text{ex.~} K^0)}}{0.36475 \pm 0.00097}{0.36475}{0.00097}%
\htquantdef{Gamma126}{\Gamma_{126}}{\BRF{\tau^-}{\pi^- \pi^0 \eta \nu_\tau}}{(0.1386 \pm 0.0072) \cdot 10^{-2}}{0.1386\cdot 10^{-2}}{0.0072\cdot 10^{-2}}%
\htquantdef{Gamma128}{\Gamma_{128}}{\BRF{\tau^-}{K^- \eta \nu_\tau}}{(1.547 \pm 0.080) \cdot 10^{-4}}{1.547\cdot 10^{-4}}{0.080\cdot 10^{-4}}%
\htquantdef{Gamma13}{\Gamma_{13}}{\BRF{\tau^-}{h^- \pi^0 \nu_\tau}}{0.25935 \pm 0.00091}{0.25935}{0.00091}%
\htquantdef{Gamma130}{\Gamma_{130}}{\BRF{\tau^-}{K^- \pi^0 \eta \nu_\tau}}{(4.827 \pm 1.161) \cdot 10^{-5}}{4.827\cdot 10^{-5}}{1.161\cdot 10^{-5}}%
\htquantdef{Gamma132}{\Gamma_{132}}{\BRF{\tau^-}{\pi^- \bar{K}^0 \eta \nu_\tau}}{(9.371 \pm 1.491) \cdot 10^{-5}}{9.371\cdot 10^{-5}}{1.491\cdot 10^{-5}}%
\htquantdef{Gamma136}{\Gamma_{136}}{\BRF{\tau^-}{\pi^- \pi^+ \pi^- \eta \nu_\tau\;(\text{ex.~} K^0)}}{(2.184 \pm 0.130) \cdot 10^{-4}}{2.184\cdot 10^{-4}}{0.130\cdot 10^{-4}}%
\htquantdef{Gamma14}{\Gamma_{14}}{\BRF{\tau^-}{\pi^- \pi^0 \nu_\tau}}{0.25502 \pm 0.00092}{0.25502}{0.00092}%
\htquantdef{Gamma149}{\Gamma_{149}}{\BRF{\tau^-}{h^- \omega \ge{} 0\,  \text{neutrals}\, \nu_\tau}}{(2.401 \pm 0.075) \cdot 10^{-2}}{2.401\cdot 10^{-2}}{0.075\cdot 10^{-2}}%
\htquantdef{Gamma150}{\Gamma_{150}}{\BRF{\tau^-}{h^- \omega \nu_\tau}}{(1.995 \pm 0.064) \cdot 10^{-2}}{1.995\cdot 10^{-2}}{0.064\cdot 10^{-2}}%
\htquantdef{Gamma150by66}{\frac{\Gamma_{150}}{\Gamma_{66}}}{\frac{\BRF{\tau^-}{h^- \omega \nu_\tau}}{\BRF{\tau^-}{h^- h^- h^+ \pi^0 \nu_\tau\;(\text{ex.~} K^0)}}}{0.4332 \pm 0.0139}{0.4332}{0.0139}%
\htquantdef{Gamma151}{\Gamma_{151}}{\BRF{\tau^-}{K^- \omega \nu_\tau}}{(4.100 \pm 0.922) \cdot 10^{-4}}{4.100\cdot 10^{-4}}{0.922\cdot 10^{-4}}%
\htquantdef{Gamma152}{\Gamma_{152}}{\BRF{\tau^-}{h^- \pi^0 \omega \nu_\tau}}{(0.4058 \pm 0.0419) \cdot 10^{-2}}{0.4058\cdot 10^{-2}}{0.0419\cdot 10^{-2}}%
\htquantdef{Gamma152by54}{\frac{\Gamma_{152}}{\Gamma_{54}}}{\frac{\BRF{\tau^-}{h^- \omega \pi^0 \nu_\tau}}{\BRF{\tau^-}{h^- h^- h^+ \ge{} 0\, \text{neutrals} \ge{} 0\,  K_L^0\, \nu_\tau}}}{(2.667 \pm 0.275) \cdot 10^{-2}}{2.667\cdot 10^{-2}}{0.275\cdot 10^{-2}}%
\htquantdef{Gamma152by76}{\frac{\Gamma_{152}}{\Gamma_{76}}}{\frac{\BRF{\tau^-}{h^- \omega \pi^0 \nu_\tau}}{\BRF{\tau^-}{h^- h^- h^+ 2\pi^0 \nu_\tau\;(\text{ex.~} K^0)}}}{0.8241 \pm 0.0757}{0.8241}{0.0757}%
\htquantdef{Gamma16}{\Gamma_{16}}{\BRF{\tau^-}{K^- \pi^0 \nu_\tau}}{(0.4327 \pm 0.0149) \cdot 10^{-2}}{0.4327\cdot 10^{-2}}{0.0149\cdot 10^{-2}}%
\htquantdef{Gamma167}{\Gamma_{167}}{\BRF{\tau^-}{K^- \phi \nu_\tau}}{(4.445 \pm 1.636) \cdot 10^{-5}}{4.445\cdot 10^{-5}}{1.636\cdot 10^{-5}}%
\htquantdef{Gamma168}{\Gamma_{168}}{\BRF{\tau^-}{K^- \phi \nu_\tau ~(\phi \to K^+ K^-)}}{(2.174 \pm 0.800) \cdot 10^{-5}}{2.174\cdot 10^{-5}}{0.800\cdot 10^{-5}}%
\htquantdef{Gamma169}{\Gamma_{169}}{\BRF{\tau^-}{K^- \phi \nu_\tau ~(\phi \to K_S^0 K_L^0)}}{(1.520 \pm 0.560) \cdot 10^{-5}}{1.520\cdot 10^{-5}}{0.560\cdot 10^{-5}}%
\htquantdef{Gamma17}{\Gamma_{17}}{\BRF{\tau^-}{h^- \ge{} 2\,  \pi^0\, \nu_\tau}}{0.10775 \pm 0.00095}{0.10775}{0.00095}%
\htquantdef{Gamma18}{\Gamma_{18}}{\BRF{\tau^-}{h^- 2\pi^0 \nu_\tau}}{(9.458 \pm 0.097) \cdot 10^{-2}}{9.458\cdot 10^{-2}}{0.097\cdot 10^{-2}}%
\htquantdef{Gamma19}{\Gamma_{19}}{\BRF{\tau^-}{h^- 2\pi^0 \nu_\tau\;(\text{ex.~} K^0)}}{(9.306 \pm 0.097) \cdot 10^{-2}}{9.306\cdot 10^{-2}}{0.097\cdot 10^{-2}}%
\htquantdef{Gamma19by13}{\frac{\Gamma_{19}}{\Gamma_{13}}}{\frac{\BRF{\tau^-}{h^- 2\pi^0 \nu_\tau\;(\text{ex.~} K^0)}}{\BRF{\tau^-}{h^- \pi^0 \nu_\tau}}}{0.3588 \pm 0.0044}{0.3588}{0.0044}%
\htquantdef{Gamma2}{\Gamma_{2}}{\BRF{\tau^-}{(\text{particles})^- \ge{} 0\, \text{neutrals} \ge{} 0\,  K_L^0\, \nu_\tau}}{0.8453 \pm 0.0010}{0.8453}{0.0010}%
\htquantdef{Gamma20}{\Gamma_{20}}{\BRF{\tau^-}{\pi^- 2\pi^0 \nu_\tau ~(\text{ex.~}K^0)}}{(9.242 \pm 0.100) \cdot 10^{-2}}{9.242\cdot 10^{-2}}{0.100\cdot 10^{-2}}%
\htquantdef{Gamma23}{\Gamma_{23}}{\BRF{\tau^-}{K^- 2\pi^0 \nu_\tau ~(\text{ex.~}K^0)}}{(6.398 \pm 2.204) \cdot 10^{-4}}{6.398\cdot 10^{-4}}{2.204\cdot 10^{-4}}%
\htquantdef{Gamma24}{\Gamma_{24}}{\BRF{\tau^-}{h^- \ge{} 3\, \pi^0\, \nu_\tau}}{(1.318 \pm 0.065) \cdot 10^{-2}}{1.318\cdot 10^{-2}}{0.065\cdot 10^{-2}}%
\htquantdef{Gamma25}{\Gamma_{25}}{\BRF{\tau^-}{h^- \ge{} 3\, \pi^0\, \nu_\tau\;(\text{ex.~} K^0)}}{(1.233 \pm 0.065) \cdot 10^{-2}}{1.233\cdot 10^{-2}}{0.065\cdot 10^{-2}}%
\htquantdef{Gamma26}{\Gamma_{26}}{\BRF{\tau^-}{h^- 3\pi^0 \nu_\tau}}{(1.158 \pm 0.072) \cdot 10^{-2}}{1.158\cdot 10^{-2}}{0.072\cdot 10^{-2}}%
\htquantdef{Gamma26by13}{\frac{\Gamma_{26}}{\Gamma_{13}}}{\frac{\BRF{\tau^-}{h^- 3\pi^0 \nu_\tau}}{\BRF{\tau^-}{h^- \pi^0 \nu_\tau}}}{(4.465 \pm 0.277) \cdot 10^{-2}}{4.465\cdot 10^{-2}}{0.277\cdot 10^{-2}}%
\htquantdef{Gamma27}{\Gamma_{27}}{\BRF{\tau^-}{\pi^- 3\pi^0 \nu_\tau ~(\text{ex.~}K^0)}}{(1.029 \pm 0.075) \cdot 10^{-2}}{1.029\cdot 10^{-2}}{0.075\cdot 10^{-2}}%
\htquantdef{Gamma28}{\Gamma_{28}}{\BRF{\tau^-}{K^- 3\pi^0 \nu_\tau ~(\text{ex.~}K^0,\eta)}}{(4.283 \pm 2.161) \cdot 10^{-4}}{4.283\cdot 10^{-4}}{2.161\cdot 10^{-4}}%
\htquantdef{Gamma29}{\Gamma_{29}}{\BRF{\tau^-}{h^- 4\pi^0 \nu_\tau\;(\text{ex.~} K^0)}}{(0.1568 \pm 0.0391) \cdot 10^{-2}}{0.1568\cdot 10^{-2}}{0.0391\cdot 10^{-2}}%
\htquantdef{Gamma3}{\Gamma_{3}}{\BRF{\tau^-}{\mu^- \bar{\nu}_\mu \nu_\tau}}{0.17392 \pm 0.00040}{0.17392}{0.00040}%
\htquantdef{Gamma30}{\Gamma_{30}}{\BRF{\tau^-}{h^- 4\pi^0 \nu_\tau ~(\text{ex.~}K^0,\eta)}}{(0.1099 \pm 0.0391) \cdot 10^{-2}}{0.1099\cdot 10^{-2}}{0.0391\cdot 10^{-2}}%
\htquantdef{Gamma31}{\Gamma_{31}}{\BRF{\tau^-}{K^- \ge{} 0\, \pi^0 \ge{} 0\, K^0 \ge{} 0\, \gamma \nu_\tau}}{(1.545 \pm 0.030) \cdot 10^{-2}}{1.545\cdot 10^{-2}}{0.030\cdot 10^{-2}}%
\htquantdef{Gamma32}{\Gamma_{32}}{\BRF{\tau^-}{K^- \ge{} 1\, (\pi^0\,\text{or}\,K^0\,\text{or}\,\gamma) \nu_\tau}}{(0.8528 \pm 0.0286) \cdot 10^{-2}}{0.8528\cdot 10^{-2}}{0.0286\cdot 10^{-2}}%
\htquantdef{Gamma33}{\Gamma_{33}}{\BRF{\tau^-}{K_S^0 (\text{particles})^- \nu_\tau}}{(0.9372 \pm 0.0292) \cdot 10^{-2}}{0.9372\cdot 10^{-2}}{0.0292\cdot 10^{-2}}%
\htquantdef{Gamma34}{\Gamma_{34}}{\BRF{\tau^-}{h^- \bar{K}^0 \nu_\tau}}{(0.9865 \pm 0.0139) \cdot 10^{-2}}{0.9865\cdot 10^{-2}}{0.0139\cdot 10^{-2}}%
\htquantdef{Gamma35}{\Gamma_{35}}{\BRF{\tau^-}{\pi^- \bar{K}^0 \nu_\tau}}{(0.8386 \pm 0.0141) \cdot 10^{-2}}{0.8386\cdot 10^{-2}}{0.0141\cdot 10^{-2}}%
\htquantdef{Gamma37}{\Gamma_{37}}{\BRF{\tau^-}{K^- K^0 \nu_\tau}}{(0.1479 \pm 0.0053) \cdot 10^{-2}}{0.1479\cdot 10^{-2}}{0.0053\cdot 10^{-2}}%
\htquantdef{Gamma38}{\Gamma_{38}}{\BRF{\tau^-}{K^- K^0 \ge{} 0\,  \pi^0\, \nu_\tau}}{(0.2982 \pm 0.0079) \cdot 10^{-2}}{0.2982\cdot 10^{-2}}{0.0079\cdot 10^{-2}}%
\htquantdef{Gamma39}{\Gamma_{39}}{\BRF{\tau^-}{h^- \bar{K}^0 \pi^0 \nu_\tau}}{(0.5314 \pm 0.0134) \cdot 10^{-2}}{0.5314\cdot 10^{-2}}{0.0134\cdot 10^{-2}}%
\htquantdef{Gamma3by5}{\frac{\Gamma_{3}}{\Gamma_{5}}}{\frac{\BRF{\tau^-}{\mu^- \bar{\nu}_\mu \nu_\tau}}{\BRF{\tau^-}{e^- \bar{\nu}_e \nu_\tau}}}{0.9762 \pm 0.0028}{0.9762}{0.0028}%
\htquantdef{Gamma40}{\Gamma_{40}}{\BRF{\tau^-}{\pi^- \bar{K}^0 \pi^0 \nu_\tau}}{(0.3812 \pm 0.0129) \cdot 10^{-2}}{0.3812\cdot 10^{-2}}{0.0129\cdot 10^{-2}}%
\htquantdef{Gamma42}{\Gamma_{42}}{\BRF{\tau^-}{K^- \pi^0 K^0 \nu_\tau}}{(0.1502 \pm 0.0071) \cdot 10^{-2}}{0.1502\cdot 10^{-2}}{0.0071\cdot 10^{-2}}%
\htquantdef{Gamma43}{\Gamma_{43}}{\BRF{\tau^-}{\pi^- \bar{K}^0 \ge{} 1\,  \pi^0\, \nu_\tau}}{(0.4046 \pm 0.0260) \cdot 10^{-2}}{0.4046\cdot 10^{-2}}{0.0260\cdot 10^{-2}}%
\htquantdef{Gamma44}{\Gamma_{44}}{\BRF{\tau^-}{\pi^- \bar{K}^0 \pi^0 \pi^0 \nu_\tau ~(\text{ex.~}K^0)}}{(2.340 \pm 2.306) \cdot 10^{-4}}{2.340\cdot 10^{-4}}{2.306\cdot 10^{-4}}%
\htquantdef{Gamma46}{\Gamma_{46}}{\BRF{\tau^-}{\pi^- K^0 \bar{K}^0 \nu_\tau}}{(0.1513 \pm 0.0247) \cdot 10^{-2}}{0.1513\cdot 10^{-2}}{0.0247\cdot 10^{-2}}%
\htquantdef{Gamma47}{\Gamma_{47}}{\BRF{\tau^-}{\pi^- K_S^0 K_S^0 \nu_\tau}}{(2.332 \pm 0.065) \cdot 10^{-4}}{2.332\cdot 10^{-4}}{0.065\cdot 10^{-4}}%
\htquantdef{Gamma48}{\Gamma_{48}}{\BRF{\tau^-}{\pi^- K_S^0 K_L^0 \nu_\tau}}{(0.1047 \pm 0.0247) \cdot 10^{-2}}{0.1047\cdot 10^{-2}}{0.0247\cdot 10^{-2}}%
\htquantdef{Gamma49}{\Gamma_{49}}{\BRF{\tau^-}{\pi^- K^0 \bar{K}^0 \pi^0 \nu_\tau}}{(3.540 \pm 1.193) \cdot 10^{-4}}{3.540\cdot 10^{-4}}{1.193\cdot 10^{-4}}%
\htquantdef{Gamma5}{\Gamma_{5}}{\BRF{\tau^-}{e^- \bar{\nu}_e \nu_\tau}}{0.17816 \pm 0.00041}{0.17816}{0.00041}%
\htquantdef{Gamma50}{\Gamma_{50}}{\BRF{\tau^-}{\pi^- \pi^0 K_S^0 K_S^0 \nu_\tau}}{(1.815 \pm 0.207) \cdot 10^{-5}}{1.815\cdot 10^{-5}}{0.207\cdot 10^{-5}}%
\htquantdef{Gamma51}{\Gamma_{51}}{\BRF{\tau^-}{\pi^- \pi^0 K_S^0 K_L^0 \nu_\tau}}{(3.177 \pm 1.192) \cdot 10^{-4}}{3.177\cdot 10^{-4}}{1.192\cdot 10^{-4}}%
\htquantdef{Gamma53}{\Gamma_{53}}{\BRF{\tau^-}{\bar{K}^0 h^- h^- h^+ \nu_\tau}}{(2.218 \pm 2.024) \cdot 10^{-4}}{2.218\cdot 10^{-4}}{2.024\cdot 10^{-4}}%
\htquantdef{Gamma54}{\Gamma_{54}}{\BRF{\tau^-}{h^- h^- h^+ \ge{} 0\, \text{neutrals} \ge{} 0\,  K_L^0\, \nu_\tau}}{0.15215 \pm 0.00061}{0.15215}{0.00061}%
\htquantdef{Gamma55}{\Gamma_{55}}{\BRF{\tau^-}{h^- h^- h^+ \ge{} 0\,  \text{neutrals}\, \nu_\tau\;(\text{ex.~} K^0)}}{0.14567 \pm 0.00057}{0.14567}{0.00057}%
\htquantdef{Gamma56}{\Gamma_{56}}{\BRF{\tau^-}{h^- h^- h^+ \nu_\tau}}{(9.780 \pm 0.054) \cdot 10^{-2}}{9.780\cdot 10^{-2}}{0.054\cdot 10^{-2}}%
\htquantdef{Gamma57}{\Gamma_{57}}{\BRF{\tau^-}{h^- h^- h^+ \nu_\tau\;(\text{ex.~} K^0)}}{(9.439 \pm 0.053) \cdot 10^{-2}}{9.439\cdot 10^{-2}}{0.053\cdot 10^{-2}}%
\htquantdef{Gamma57by55}{\frac{\Gamma_{57}}{\Gamma_{55}}}{\frac{\BRF{\tau^-}{h^- h^- h^+ \nu_\tau\;(\text{ex.~} K^0)}}{\BRF{\tau^-}{h^- h^- h^+ \ge{} 0\,  \text{neutrals}\, \nu_\tau\;(\text{ex.~} K^0)}}}{0.6480 \pm 0.0030}{0.6480}{0.0030}%
\htquantdef{Gamma58}{\Gamma_{58}}{\BRF{\tau^-}{h^- h^- h^+ \nu_\tau\;(\text{ex.~} K^0, \omega)}}{(9.408 \pm 0.053) \cdot 10^{-2}}{9.408\cdot 10^{-2}}{0.053\cdot 10^{-2}}%
\htquantdef{Gamma59}{\Gamma_{59}}{\BRF{\tau^-}{\pi^- \pi^+ \pi^- \nu_\tau}}{(9.290 \pm 0.052) \cdot 10^{-2}}{9.290\cdot 10^{-2}}{0.052\cdot 10^{-2}}%
\htquantdef{Gamma60}{\Gamma_{60}}{\BRF{\tau^-}{\pi^- \pi^+ \pi^- \nu_\tau\;(\text{ex.~} K^0)}}{(9.000 \pm 0.051) \cdot 10^{-2}}{9.000\cdot 10^{-2}}{0.051\cdot 10^{-2}}%
\htquantdef{Gamma62}{\Gamma_{62}}{\BRF{\tau^-}{\pi^- \pi^- \pi^+ \nu_\tau ~(\text{ex.~}K^0,\omega)}}{(8.970 \pm 0.052) \cdot 10^{-2}}{8.970\cdot 10^{-2}}{0.052\cdot 10^{-2}}%
\htquantdef{Gamma63}{\Gamma_{63}}{\BRF{\tau^-}{h^- h^- h^+ \ge{} 1\,  \text{neutrals}\, \nu_\tau}}{(5.325 \pm 0.050) \cdot 10^{-2}}{5.325\cdot 10^{-2}}{0.050\cdot 10^{-2}}%
\htquantdef{Gamma64}{\Gamma_{64}}{\BRF{\tau^-}{h^- h^- h^+ \ge{} 1\,  \pi^0\, \nu_\tau\;(\text{ex.~} K^0)}}{(5.120 \pm 0.049) \cdot 10^{-2}}{5.120\cdot 10^{-2}}{0.049\cdot 10^{-2}}%
\htquantdef{Gamma65}{\Gamma_{65}}{\BRF{\tau^-}{h^- h^- h^+ \pi^0 \nu_\tau}}{(4.790 \pm 0.052) \cdot 10^{-2}}{4.790\cdot 10^{-2}}{0.052\cdot 10^{-2}}%
\htquantdef{Gamma66}{\Gamma_{66}}{\BRF{\tau^-}{h^- h^- h^+ \pi^0 \nu_\tau\;(\text{ex.~} K^0)}}{(4.606 \pm 0.051) \cdot 10^{-2}}{4.606\cdot 10^{-2}}{0.051\cdot 10^{-2}}%
\htquantdef{Gamma67}{\Gamma_{67}}{\BRF{\tau^-}{h^- h^- h^+ \pi^0 \nu_\tau\;(\text{ex.~} K^0, \omega)}}{(2.820 \pm 0.070) \cdot 10^{-2}}{2.820\cdot 10^{-2}}{0.070\cdot 10^{-2}}%
\htquantdef{Gamma68}{\Gamma_{68}}{\BRF{\tau^-}{\pi^- \pi^+ \pi^- \pi^0 \nu_\tau}}{(4.651 \pm 0.053) \cdot 10^{-2}}{4.651\cdot 10^{-2}}{0.053\cdot 10^{-2}}%
\htquantdef{Gamma69}{\Gamma_{69}}{\BRF{\tau^-}{\pi^- \pi^+ \pi^- \pi^0 \nu_\tau\;(\text{ex.~} K^0)}}{(4.519 \pm 0.052) \cdot 10^{-2}}{4.519\cdot 10^{-2}}{0.052\cdot 10^{-2}}%
\htquantdef{Gamma7}{\Gamma_{7}}{\BRF{\tau^-}{h^- \ge{} 0\,  K_L^0\, \nu_\tau}}{0.12023 \pm 0.00054}{0.12023}{0.00054}%
\htquantdef{Gamma70}{\Gamma_{70}}{\BRF{\tau^-}{\pi^- \pi^- \pi^+ \pi^0 \nu_\tau ~(\text{ex.~}K^0,\omega)}}{(2.769 \pm 0.071) \cdot 10^{-2}}{2.769\cdot 10^{-2}}{0.071\cdot 10^{-2}}%
\htquantdef{Gamma74}{\Gamma_{74}}{\BRF{\tau^-}{h^- h^- h^+ \ge{} 2\, \pi^0\, \nu_\tau\;(\text{ex.~} K^0)}}{(0.5135 \pm 0.0312) \cdot 10^{-2}}{0.5135\cdot 10^{-2}}{0.0312\cdot 10^{-2}}%
\htquantdef{Gamma75}{\Gamma_{75}}{\BRF{\tau^-}{h^- h^- h^+ 2\pi^0 \nu_\tau}}{(0.5024 \pm 0.0310) \cdot 10^{-2}}{0.5024\cdot 10^{-2}}{0.0310\cdot 10^{-2}}%
\htquantdef{Gamma76}{\Gamma_{76}}{\BRF{\tau^-}{h^- h^- h^+ 2\pi^0 \nu_\tau\;(\text{ex.~} K^0)}}{(0.4925 \pm 0.0310) \cdot 10^{-2}}{0.4925\cdot 10^{-2}}{0.0310\cdot 10^{-2}}%
\htquantdef{Gamma76by54}{\frac{\Gamma_{76}}{\Gamma_{54}}}{\frac{\BRF{\tau^-}{h^- h^- h^+ 2\pi^0 \nu_\tau\;(\text{ex.~} K^0)}}{\BRF{\tau^-}{h^- h^- h^+ \ge{} 0\, \text{neutrals} \ge{} 0\,  K_L^0\, \nu_\tau}}}{(3.237 \pm 0.202) \cdot 10^{-2}}{3.237\cdot 10^{-2}}{0.202\cdot 10^{-2}}%
\htquantdef{Gamma77}{\Gamma_{77}}{\BRF{\tau^-}{h^- h^- h^+ 2\pi^0 \nu_\tau ~(\text{ex.~}K^0,\omega,\eta)}}{(9.759 \pm 3.550) \cdot 10^{-4}}{9.759\cdot 10^{-4}}{3.550\cdot 10^{-4}}%
\htquantdef{Gamma78}{\Gamma_{78}}{\BRF{\tau^-}{h^- h^- h^+ 3\pi^0 \nu_\tau}}{(2.107 \pm 0.299) \cdot 10^{-4}}{2.107\cdot 10^{-4}}{0.299\cdot 10^{-4}}%
\htquantdef{Gamma79}{\Gamma_{79}}{\BRF{\tau^-}{K^- h^- h^+ \ge{} 0\,  \text{neutrals}\, \nu_\tau}}{(0.6297 \pm 0.0141) \cdot 10^{-2}}{0.6297\cdot 10^{-2}}{0.0141\cdot 10^{-2}}%
\htquantdef{Gamma8}{\Gamma_{8}}{\BRF{\tau^-}{h^- \nu_\tau}}{0.11506 \pm 0.00054}{0.11506}{0.00054}%
\htquantdef{Gamma80}{\Gamma_{80}}{\BRF{\tau^-}{K^- \pi^- h^+ \nu_\tau\;(\text{ex.~} K^0)}}{(0.4363 \pm 0.0073) \cdot 10^{-2}}{0.4363\cdot 10^{-2}}{0.0073\cdot 10^{-2}}%
\htquantdef{Gamma800}{\Gamma_{800}}{\BRF{\tau^-}{\pi^- \omega \nu_\tau}}{(1.954 \pm 0.065) \cdot 10^{-2}}{1.954\cdot 10^{-2}}{0.065\cdot 10^{-2}}%
\htquantdef{Gamma802}{\Gamma_{802}}{\BRF{\tau^-}{K^- \pi^- \pi^+ \nu_\tau ~(\text{ex.~}K^0,\omega)}}{(0.2923 \pm 0.0067) \cdot 10^{-2}}{0.2923\cdot 10^{-2}}{0.0067\cdot 10^{-2}}%
\htquantdef{Gamma803}{\Gamma_{803}}{\BRF{\tau^-}{K^- \pi^- \pi^+ \pi^0 \nu_\tau ~(\text{ex.~}K^0,\omega,\eta)}}{(4.103 \pm 1.429) \cdot 10^{-4}}{4.103\cdot 10^{-4}}{1.429\cdot 10^{-4}}%
\htquantdef{Gamma804}{\Gamma_{804}}{\BRF{\tau^-}{\pi^- K_L^0 K_L^0 \nu_\tau}}{(2.332 \pm 0.065) \cdot 10^{-4}}{2.332\cdot 10^{-4}}{0.065\cdot 10^{-4}}%
\htquantdef{Gamma805}{\Gamma_{805}}{\BRF{\tau^-}{a_1^- (\to \pi^- \gamma) \nu_\tau}}{(4.000 \pm 2.000) \cdot 10^{-4}}{4.000\cdot 10^{-4}}{2.000\cdot 10^{-4}}%
\htquantdef{Gamma806}{\Gamma_{806}}{\BRF{\tau^-}{\pi^- \pi^0 K_L^0 K_L^0 \nu_\tau}}{(1.815 \pm 0.207) \cdot 10^{-5}}{1.815\cdot 10^{-5}}{0.207\cdot 10^{-5}}%
\htquantdef{Gamma80by60}{\frac{\Gamma_{80}}{\Gamma_{60}}}{\frac{\BRF{\tau^-}{K^- \pi^- h^+ \nu_\tau\;(\text{ex.~} K^0)}}{\BRF{\tau^-}{\pi^- \pi^+ \pi^- \nu_\tau\;(\text{ex.~} K^0)}}}{(4.847 \pm 0.080) \cdot 10^{-2}}{4.847\cdot 10^{-2}}{0.080\cdot 10^{-2}}%
\htquantdef{Gamma81}{\Gamma_{81}}{\BRF{\tau^-}{K^- \pi^- h^+ \pi^0 \nu_\tau\;(\text{ex.~} K^0)}}{(8.726 \pm 1.177) \cdot 10^{-4}}{8.726\cdot 10^{-4}}{1.177\cdot 10^{-4}}%
\htquantdef{Gamma810}{\Gamma_{810}}{\BRF{\tau^-}{2\pi^- \pi^+ 3\pi^0 \nu_\tau ~(\text{ex.~}K^0)}}{(1.924 \pm 0.298) \cdot 10^{-4}}{1.924\cdot 10^{-4}}{0.298\cdot 10^{-4}}%
\htquantdef{Gamma811}{\Gamma_{811}}{\BRF{\tau^-}{\pi^- 2\pi^0 \omega \nu_\tau ~(\text{ex.~}K^0)}}{(7.105 \pm 1.586) \cdot 10^{-5}}{7.105\cdot 10^{-5}}{1.586\cdot 10^{-5}}%
\htquantdef{Gamma812}{\Gamma_{812}}{\BRF{\tau^-}{2\pi^- \pi^+ 3\pi^0 \nu_\tau ~(\text{ex.~}K^0, \eta, \omega, f_1)}}{(1.344 \pm 2.683) \cdot 10^{-5}}{1.344\cdot 10^{-5}}{2.683\cdot 10^{-5}}%
\htquantdef{Gamma81by69}{\frac{\Gamma_{81}}{\Gamma_{69}}}{\frac{\BRF{\tau^-}{K^- \pi^- h^+ \pi^0 \nu_\tau\;(\text{ex.~} K^0)}}{\BRF{\tau^-}{\pi^- \pi^+ \pi^- \pi^0 \nu_\tau\;(\text{ex.~} K^0)}}}{(1.931 \pm 0.266) \cdot 10^{-2}}{1.931\cdot 10^{-2}}{0.266\cdot 10^{-2}}%
\htquantdef{Gamma82}{\Gamma_{82}}{\BRF{\tau^-}{K^- \pi^- \pi^+ \ge{} 0\,  \text{neutrals}\, \nu_\tau}}{(0.4780 \pm 0.0137) \cdot 10^{-2}}{0.4780\cdot 10^{-2}}{0.0137\cdot 10^{-2}}%
\htquantdef{Gamma820}{\Gamma_{820}}{\BRF{\tau^-}{3\pi^- 2\pi^+ \nu_\tau ~(\text{ex.~}K^0, \omega)}}{(8.197 \pm 0.315) \cdot 10^{-4}}{8.197\cdot 10^{-4}}{0.315\cdot 10^{-4}}%
\htquantdef{Gamma821}{\Gamma_{821}}{\BRF{\tau^-}{3\pi^- 2\pi^+ \nu_\tau ~(\text{ex.~}K^0, \omega, f_1)}}{(7.677 \pm 0.297) \cdot 10^{-4}}{7.677\cdot 10^{-4}}{0.297\cdot 10^{-4}}%
\htquantdef{Gamma822}{\Gamma_{822}}{\BRF{\tau^-}{K^- 2\pi^- 2\pi^+ \nu_\tau ~(\text{ex.~}K^0)}}{(0.596 \pm 1.208) \cdot 10^{-6}}{0.596\cdot 10^{-6}}{1.208\cdot 10^{-6}}%
\htquantdef{Gamma83}{\Gamma_{83}}{\BRF{\tau^-}{K^- \pi^- \pi^+ \ge{} 0\,  \pi^0\, \nu_\tau\;(\text{ex.~} K^0)}}{(0.3741 \pm 0.0135) \cdot 10^{-2}}{0.3741\cdot 10^{-2}}{0.0135\cdot 10^{-2}}%
\htquantdef{Gamma830}{\Gamma_{830}}{\BRF{\tau^-}{3\pi^- 2\pi^+ \pi^0 \nu_\tau ~(\text{ex.~}K^0)}}{(1.623 \pm 0.114) \cdot 10^{-4}}{1.623\cdot 10^{-4}}{0.114\cdot 10^{-4}}%
\htquantdef{Gamma831}{\Gamma_{831}}{\BRF{\tau^-}{2\pi^- \pi^+ \omega \nu_\tau ~(\text{ex.~}K^0)}}{(8.359 \pm 0.626) \cdot 10^{-5}}{8.359\cdot 10^{-5}}{0.626\cdot 10^{-5}}%
\htquantdef{Gamma832}{\Gamma_{832}}{\BRF{\tau^-}{3\pi^- 2\pi^+ \pi^0 \nu_\tau ~(\text{ex.~}K^0, \eta, \omega, f_1)}}{(3.771 \pm 0.875) \cdot 10^{-5}}{3.771\cdot 10^{-5}}{0.875\cdot 10^{-5}}%
\htquantdef{Gamma833}{\Gamma_{833}}{\BRF{\tau^-}{K^- 2\pi^- 2\pi^+ \pi^0 \nu_\tau ~(\text{ex.~}K^0)}}{(1.108 \pm 0.566) \cdot 10^{-6}}{1.108\cdot 10^{-6}}{0.566\cdot 10^{-6}}%
\htquantdef{Gamma84}{\Gamma_{84}}{\BRF{\tau^-}{K^- \pi^- \pi^+ \nu_\tau}}{(0.3441 \pm 0.0070) \cdot 10^{-2}}{0.3441\cdot 10^{-2}}{0.0070\cdot 10^{-2}}%
\htquantdef{Gamma85}{\Gamma_{85}}{\BRF{\tau^-}{K^- \pi^+ \pi^- \nu_\tau\;(\text{ex.~} K^0)}}{(0.2929 \pm 0.0067) \cdot 10^{-2}}{0.2929\cdot 10^{-2}}{0.0067\cdot 10^{-2}}%
\htquantdef{Gamma85by60}{\frac{\Gamma_{85}}{\Gamma_{60}}}{\frac{\BRF{\tau^-}{K^- \pi^+ \pi^- \nu_\tau\;(\text{ex.~}K^0)}}{\BRF{\tau^-}{\pi^- \pi^+ \pi^- \nu_\tau\;(\text{ex.~}K^0)}}}{(3.254 \pm 0.074) \cdot 10^{-2}}{3.254\cdot 10^{-2}}{0.074\cdot 10^{-2}}%
\htquantdef{Gamma87}{\Gamma_{87}}{\BRF{\tau^-}{K^- \pi^- \pi^+ \pi^0 \nu_\tau}}{(0.1331 \pm 0.0119) \cdot 10^{-2}}{0.1331\cdot 10^{-2}}{0.0119\cdot 10^{-2}}%
\htquantdef{Gamma88}{\Gamma_{88}}{\BRF{\tau^-}{K^- \pi^- \pi^+ \pi^0 \nu_\tau\;(\text{ex.~} K^0)}}{(8.115 \pm 1.168) \cdot 10^{-4}}{8.115\cdot 10^{-4}}{1.168\cdot 10^{-4}}%
\htquantdef{Gamma89}{\Gamma_{89}}{\BRF{\tau^-}{K^- \pi^- \pi^+ \pi^0 \nu_\tau\;(\text{ex.~} K^0, \eta)}}{(7.761 \pm 1.168) \cdot 10^{-4}}{7.761\cdot 10^{-4}}{1.168\cdot 10^{-4}}%
\htquantdef{Gamma8by5}{\frac{\Gamma_{8}}{\Gamma_{5}}}{\frac{\BRF{\tau^-}{h^- \nu_\tau}}{\BRF{\tau^-}{e^- \bar{\nu}_e \nu_\tau}}}{0.6458 \pm 0.0033}{0.6458}{0.0033}%
\htquantdef{Gamma9}{\Gamma_{9}}{\BRF{\tau^-}{\pi^- \nu_\tau}}{0.10810 \pm 0.00053}{0.10810}{0.00053}%
\htquantdef{Gamma910}{\Gamma_{910}}{\BRF{\tau^-}{2\pi^- \pi^+ \eta \nu_\tau ~(\eta \to 3\pi^0) ~(\text{ex.~}K^0)}}{(7.136 \pm 0.424) \cdot 10^{-5}}{7.136\cdot 10^{-5}}{0.424\cdot 10^{-5}}%
\htquantdef{Gamma911}{\Gamma_{911}}{\BRF{\tau^-}{\pi^- 2\pi^0 \eta \nu_\tau ~(\eta \to \pi^+ \pi^- \pi^0) ~(\text{ex.~}K^0)}}{(4.420 \pm 0.867) \cdot 10^{-5}}{4.420\cdot 10^{-5}}{0.867\cdot 10^{-5}}%
\htquantdef{Gamma92}{\Gamma_{92}}{\BRF{\tau^-}{\pi^- K^- K^+ \ge{} 0\,  \text{neutrals}\, \nu_\tau}}{(0.1495 \pm 0.0033) \cdot 10^{-2}}{0.1495\cdot 10^{-2}}{0.0033\cdot 10^{-2}}%
\htquantdef{Gamma920}{\Gamma_{920}}{\BRF{\tau^-}{\pi^- f_1 \nu_\tau ~(f_1 \to 2\pi^- 2\pi^+)}}{(5.197 \pm 0.444) \cdot 10^{-5}}{5.197\cdot 10^{-5}}{0.444\cdot 10^{-5}}%
\htquantdef{Gamma93}{\Gamma_{93}}{\BRF{\tau^-}{\pi^- K^- K^+ \nu_\tau}}{(0.1434 \pm 0.0027) \cdot 10^{-2}}{0.1434\cdot 10^{-2}}{0.0027\cdot 10^{-2}}%
\htquantdef{Gamma930}{\Gamma_{930}}{\BRF{\tau^-}{2\pi^- \pi^+ \eta \nu_\tau ~(\eta \to \pi^+\pi^-\pi^0) ~(\text{ex.~}K^0)}}{(5.005 \pm 0.297) \cdot 10^{-5}}{5.005\cdot 10^{-5}}{0.297\cdot 10^{-5}}%
\htquantdef{Gamma93by60}{\frac{\Gamma_{93}}{\Gamma_{60}}}{\frac{\BRF{\tau^-}{\pi^- K^- K^+ \nu_\tau}}{\BRF{\tau^-}{\pi^- \pi^+ \pi^- \nu_\tau\;(\text{ex.~} K^0)}}}{(1.593 \pm 0.030) \cdot 10^{-2}}{1.593\cdot 10^{-2}}{0.030\cdot 10^{-2}}%
\htquantdef{Gamma94}{\Gamma_{94}}{\BRF{\tau^-}{\pi^- K^- K^+ \pi^0 \nu_\tau}}{(6.113 \pm 1.829) \cdot 10^{-5}}{6.113\cdot 10^{-5}}{1.829\cdot 10^{-5}}%
\htquantdef{Gamma944}{\Gamma_{944}}{\BRF{\tau^-}{2\pi^- \pi^+ \eta \nu_\tau ~(\eta \to \gamma\gamma) ~(\text{ex.~}K^0)}}{(8.606 \pm 0.511) \cdot 10^{-5}}{8.606\cdot 10^{-5}}{0.511\cdot 10^{-5}}%
\htquantdef{Gamma945}{\Gamma_{945}}{\BRF{\tau^-}{\pi^- 2\pi^0 \eta \nu_\tau}}{(1.929 \pm 0.378) \cdot 10^{-4}}{1.929\cdot 10^{-4}}{0.378\cdot 10^{-4}}%
\htquantdef{Gamma94by69}{\frac{\Gamma_{94}}{\Gamma_{69}}}{\frac{\BRF{\tau^-}{\pi^- K^- K^+ \pi^0 \nu_\tau}}{\BRF{\tau^-}{\pi^- \pi^+ \pi^- \pi^0 \nu_\tau\;(\text{ex.~} K^0)}}}{(0.1353 \pm 0.0405) \cdot 10^{-2}}{0.1353\cdot 10^{-2}}{0.0405\cdot 10^{-2}}%
\htquantdef{Gamma96}{\Gamma_{96}}{\BRF{\tau^-}{K^- K^- K^+ \nu_\tau}}{(2.174 \pm 0.800) \cdot 10^{-5}}{2.174\cdot 10^{-5}}{0.800\cdot 10^{-5}}%
\htquantdef{Gamma998}{\Gamma_{998}}{1 - \Gamma_{\text{All}}}{(0.0355 \pm 0.1031) \cdot 10^{-2}}{0.0355\cdot 10^{-2}}{0.1031\cdot 10^{-2}}%
\htquantdef{Gamma9by5}{\frac{\Gamma_{9}}{\Gamma_{5}}}{\frac{\BRF{\tau^-}{\pi^- \nu_\tau}}{\BRF{\tau^-}{e^- \bar{\nu}_e \nu_\tau}}}{0.6068 \pm 0.0032}{0.6068}{0.0032}%
\htquantdef{GammaAll}{\Gamma_{\text{All}}}{\Gamma_{\text{All}}}{0.9996 \pm 0.0010}{0.9996}{0.0010}%
\htquantdef{gmubyge_tau}{gmubyge_tau}{}{1.0019 \pm 0.0014}{1.0019}{0.0014}%
\htquantdef{gtaubyge_tau}{gtaubyge_tau}{}{1.0029 \pm 0.0015}{1.0029}{0.0015}%
\htquantdef{gtaubygmu_fit}{gtaubygmu_fit}{}{1.0000 \pm 0.0014}{1.0000}{0.0014}%
\htquantdef{gtaubygmu_K}{gtaubygmu_K}{}{0.9860 \pm 0.0070}{0.9860}{0.0070}%
\htquantdef{gtaubygmu_pi}{gtaubygmu_pi}{}{0.9961 \pm 0.0027}{0.9961}{0.0027}%
\htquantdef{gtaubygmu_tau}{gtaubygmu_tau}{}{1.0010 \pm 0.0015}{1.0010}{0.0015}%
\htquantdef{hcut}{hcut}{}{(6.582119514 \pm 0.000000040) \cdot 10^{-22}}{6.582119514\cdot 10^{-22}}{0.000000040\cdot 10^{-22}}%
\htquantdef{KmKzsNu}{KmKzsNu}{}{7.450\cdot 10^{-4}}{7.450\cdot 10^{-4}}{0}%
\htquantdef{KmPizKzsNu}{KmPizKzsNu}{}{7.550\cdot 10^{-4}}{7.550\cdot 10^{-4}}{0}%
\htquantdef{KtoENu}{KtoENu}{}{(1.5820 \pm 0.0070) \cdot 10^{-5}}{1.5820\cdot 10^{-5}}{0.0070\cdot 10^{-5}}%
\htquantdef{KtoMuNu}{KtoMuNu}{}{0.6356 \pm 0.0011}{0.6356}{0.0011}%
\htquantdef{m_e}{m_e}{}{0.510998928 \pm 0.000000011}{0.510998928}{0.000000011}%
\htquantdef{m_K}{m_K}{}{493.677 \pm 0.016}{493.677}{0.016}%
\htquantdef{m_mu}{m_mu}{}{105.6583715 \pm 0.0000035}{105.6583715}{0.0000035}%
\htquantdef{m_pi}{m_pi}{}{139.57018 \pm 0.00035}{139.57018}{0.00035}%
\htquantdef{m_s}{m_s}{}{95.00 \pm 5.00}{95.00}{5.00}%
\htquantdef{m_tau}{m_tau}{}{(1.77686 \pm 0.00012) \cdot 10^{3}}{1.77686\cdot 10^{3}}{0.00012\cdot 10^{3}}%
\htquantdef{m_W}{m_W}{}{(8.0385 \pm 0.0015) \cdot 10^{4}}{8.0385\cdot 10^{4}}{0.0015\cdot 10^{4}}%
\htquantdef{MumNuNumb}{MumNuNumb}{}{0.1741}{0.1741}{0}%
\htquantdef{phspf_mebymmu}{phspf_mebymmu}{}{0.999812949174 \pm 0.000000000015}{0.999812949174}{0.000000000015}%
\htquantdef{phspf_mebymtau}{phspf_mebymtau}{}{0.999999338359 \pm 0.000000000089}{0.999999338359}{0.000000000089}%
\htquantdef{phspf_mmubymtau}{phspf_mmubymtau}{}{0.9725600 \pm 0.0000036}{0.9725600}{0.0000036}%
\htquantdef{pi}{pi}{}{3.142}{3.142}{0}%
\htquantdef{PimKmKpNu}{PimKmKpNu}{}{0.1440\cdot 10^{-2}}{0.1440\cdot 10^{-2}}{0}%
\htquantdef{PimKmPipNu}{PimKmPipNu}{}{0.2940\cdot 10^{-2}}{0.2940\cdot 10^{-2}}{0}%
\htquantdef{PimKzsKzlNu}{PimKzsKzlNu}{}{0.1200\cdot 10^{-2}}{0.1200\cdot 10^{-2}}{0}%
\htquantdef{PimKzsKzsNu}{PimKzsKzsNu}{}{2.320\cdot 10^{-4}}{2.320\cdot 10^{-4}}{0}%
\htquantdef{PimPimPipNu}{PimPimPipNu}{}{8.990\cdot 10^{-2}}{8.990\cdot 10^{-2}}{0}%
\htquantdef{PimPimPipPizNu}{PimPimPipPizNu}{}{4.610\cdot 10^{-2}}{4.610\cdot 10^{-2}}{0}%
\htquantdef{PimPizKzsNu}{PimPizKzsNu}{}{0.1940\cdot 10^{-2}}{0.1940\cdot 10^{-2}}{0}%
\htquantdef{PimPizNu}{PimPizNu}{}{0.2552}{0.2552}{0}%
\htquantdef{pitoENu}{pitoENu}{}{(1.2300 \pm 0.0040) \cdot 10^{-4}}{1.2300\cdot 10^{-4}}{0.0040\cdot 10^{-4}}%
\htquantdef{pitoMuNu}{pitoMuNu}{}{0.99987700 \pm 0.00000040}{0.99987700}{0.00000040}%
\htquantdef{R_tau}{R_tau}{}{3.6349 \pm 0.0082}{3.6349}{0.0082}%
\htquantdef{R_tau_s}{R_tau_s}{}{0.1633 \pm 0.0027}{0.1633}{0.0027}%
\htquantdef{R_tau_VA}{R_tau_VA}{}{3.4717 \pm 0.0081}{3.4717}{0.0081}%
\htquantdef{Rrad_kmunu_by_pimunu}{Rrad_kmunu_by_pimunu}{}{0.9930 \pm 0.0035}{0.9930}{0.0035}%
\htquantdef{Rrad_SEW_tau_Knu}{Rrad_SEW_tau_Knu}{}{1.02010 \pm 0.00030}{1.02010}{0.00030}%
\htquantdef{Rrad_tauK_by_taupi}{Rrad_tauK_by_taupi}{}{1.00 \pm 0.00}{1.00}{0.00}%
\htquantdef{sigmataupmy4s}{sigmataupmy4s}{}{0.9190}{0.9190}{0}%
\htquantdef{tau_K}{tau_K}{}{(1.2380 \pm 0.0020) \cdot 10^{-8}}{1.2380\cdot 10^{-8}}{0.0020\cdot 10^{-8}}%
\htquantdef{tau_mu}{tau_mu}{}{(2.196981 \pm 0.000022) \cdot 10^{-6}}{2.196981\cdot 10^{-6}}{0.000022\cdot 10^{-6}}%
\htquantdef{tau_pi}{tau_pi}{}{(2.60330 \pm 0.00050) \cdot 10^{-8}}{2.60330\cdot 10^{-8}}{0.00050\cdot 10^{-8}}%
\htquantdef{tau_tau}{tau_tau}{}{290.3 \pm 0.5}{290.3}{0.5}%
\htquantdef{Vud}{Vud}{}{0.97417 \pm 0.00021}{0.97417}{0.00021}%
\htquantdef{Vud_moulson_ckm14}{Vud_moulson_ckm14}{}{0.97417 \pm 0.00021}{0.97417}{0.00021}%
\htquantdef{Vus}{Vus}{}{0.2186 \pm 0.0021}{0.2186}{0.0021}%
\htquantdef{Vus_err_exp}{Vus_err_exp}{}{0.1869\cdot 10^{-2}}{0.1869\cdot 10^{-2}}{0}%
\htquantdef{Vus_err_exp_perc}{Vus_err_exp_perc}{}{0.8547}{0.8547}{0}%
\htquantdef{Vus_err_perc}{Vus_err_perc}{}{0.9765}{0.9765}{0}%
\htquantdef{Vus_err_th}{Vus_err_th}{}{0.1032\cdot 10^{-2}}{0.1032\cdot 10^{-2}}{0}%
\htquantdef{Vus_err_th_perc}{Vus_err_th_perc}{}{0.47}{0.47}{0}%
\htquantdef{Vus_kl2_maltman_mainz16}{Vus_kl2_maltman_mainz16}{}{0.22500 \pm 0.00098}{0.22500}{0.00098}%
\htquantdef{Vus_kl2_moulson_ckm14}{Vus_kl2_moulson_ckm14}{}{0.22484 \pm 0.00059}{0.22484}{0.00059}%
\htquantdef{Vus_kl3_maltman_mainz16}{Vus_kl3_maltman_mainz16}{}{0.22310 \pm 0.00081}{0.22310}{0.00081}%
\htquantdef{Vus_kl3_moulson_ckm14}{Vus_kl3_moulson_ckm14}{}{0.22320 \pm 0.00090}{0.22320}{0.00090}%
\htquantdef{Vus_maltman_mainz16}{Vus_maltman_mainz16}{}{0.2228 \pm 0.0024}{0.2228}{0.0024}%
\htquantdef{Vus_mism}{Vus_mism}{}{(-0.7206 \pm 0.2334) \cdot 10^{-2}}{-0.7206\cdot 10^{-2}}{0.2334\cdot 10^{-2}}%
\htquantdef{Vus_mism_sigma}{Vus_mism_sigma}{}{-3.1}{-3.1}{0}%
\htquantdef{Vus_mism_sigma_abs}{Vus_mism_sigma_abs}{}{3.1}{3.1}{0}%
\htquantdef{Vus_tau}{Vus_tau}{}{0.2216 \pm 0.0015}{0.2216}{0.0015}%
\htquantdef{Vus_tau_mism}{Vus_tau_mism}{}{(-0.4171 \pm 0.1730) \cdot 10^{-2}}{-0.4171\cdot 10^{-2}}{0.1730\cdot 10^{-2}}%
\htquantdef{Vus_tau_mism_sigma}{Vus_tau_mism_sigma}{}{-2.4}{-2.4}{0}%
\htquantdef{Vus_tau_mism_sigma_abs}{Vus_tau_mism_sigma_abs}{}{2.4}{2.4}{0}%
\htquantdef{Vus_tauKnu}{Vus_tauKnu}{}{0.2221 \pm 0.0017}{0.2221}{0.0017}%
\htquantdef{Vus_tauKnu_err_th_perc}{Vus_tauKnu_err_th_perc}{}{0.2962}{0.2962}{0}%
\htquantdef{Vus_tauKnu_mism}{Vus_tauKnu_mism}{}{(-0.3736 \pm 0.1899) \cdot 10^{-2}}{-0.3736\cdot 10^{-2}}{0.1899\cdot 10^{-2}}%
\htquantdef{Vus_tauKnu_mism_sigma}{Vus_tauKnu_mism_sigma}{}{-2.0}{-2.0}{0}%
\htquantdef{Vus_tauKnu_mism_sigma_abs}{Vus_tauKnu_mism_sigma_abs}{}{2.0}{2.0}{0}%
\htquantdef{Vus_tauKpi}{Vus_tauKpi}{}{0.2236 \pm 0.0018}{0.2236}{0.0018}%
\htquantdef{Vus_tauKpi_err_th_perc}{Vus_tauKpi_err_th_perc}{}{0.3058}{0.3058}{0}%
\htquantdef{Vus_tauKpi_err_th_perc_dRrad_kmunu_by_pimunu}{Vus_tauKpi_err_th_perc_dRrad_kmunu_by_pimunu}{}{-0.1163}{-0.1163}{0}%
\htquantdef{Vus_tauKpi_err_th_perc_dRrad_tauK_by_Kmu}{Vus_tauKpi_err_th_perc_dRrad_tauK_by_Kmu}{}{-0.1090}{-0.1090}{0}%
\htquantdef{Vus_tauKpi_err_th_perc_dRrad_taupi_by_pimu}{Vus_tauKpi_err_th_perc_dRrad_taupi_by_pimu}{}{6.989\cdot 10^{-2}}{6.989\cdot 10^{-2}}{0}%
\htquantdef{Vus_tauKpi_err_th_perc_f_K_by_f_pi}{Vus_tauKpi_err_th_perc_f_K_by_f_pi}{}{-0.2515}{-0.2515}{0}%
\htquantdef{Vus_tauKpi_mism}{Vus_tauKpi_mism}{}{(-0.2231 \pm 0.1998) \cdot 10^{-2}}{-0.2231\cdot 10^{-2}}{0.1998\cdot 10^{-2}}%
\htquantdef{Vus_tauKpi_mism_sigma}{Vus_tauKpi_mism_sigma}{}{-1.1}{-1.1}{0}%
\htquantdef{Vus_tauKpi_mism_sigma_abs}{Vus_tauKpi_mism_sigma_abs}{}{1.1}{1.1}{0}%
\htquantdef{Vus_uni}{Vus_uni}{}{0.22582 \pm 0.00089}{0.22582}{0.00089}%
\htquantdef{VusbyVud_moulson_ckm14}{VusbyVud_moulson_ckm14}{}{0.23080 \pm 0.00060}{0.23080}{0.00060}%
\htdef{couplingsCorr}{%
$\left( \frac{g_\tau}{g_e} \right)$ &   53\\
$\left( \frac{g_\mu}{g_e} \right)$ &  -49 &   48\\
$\left( \frac{g_\tau}{g_\mu} \right)_\pi$ &   24 &   26 &    2\\
$\left( \frac{g_\tau}{g_\mu} \right)_K$ &   11 &   10 &   -1 &    6\\
 & $\left( \frac{g_\tau}{g_\mu} \right)$ & $\left( \frac{g_\tau}{g_e} \right)$ & $\left( \frac{g_\mu}{g_e} \right)$ & $\left( \frac{g_\tau}{g_\mu} \right)_\pi$}%

\htdef{g156.descr}{\ensuremath{e^- \gamma}}%
\htdef{g157.descr}{\ensuremath{\mu^- \gamma}}%
\htdef{g158.descr}{\ensuremath{e^- \pi^0}}%
\htdef{g159.descr}{\ensuremath{\mu^- \pi^0}}%
\htdef{g162.descr}{\ensuremath{e^- \eta}}%
\htdef{g163.descr}{\ensuremath{\mu^- \eta}}%
\htdef{g172.descr}{\ensuremath{e^- \eta^\prime(958)}}%
\htdef{g173.descr}{\ensuremath{\mu^- \eta^\prime(958)}}%
\htdef{g160.descr}{\ensuremath{e^- K^0_S}}%
\htdef{g161.descr}{\ensuremath{\mu^- K^0_S}}%
\htdef{g174.descr}{\ensuremath{e^- f_0(980)}}%
\htdef{g175.descr}{\ensuremath{\mu^- f_0(980)}}%
\htdef{g164.descr}{\ensuremath{e^- \rho^0}}%
\htdef{g165.descr}{\ensuremath{\mu^- \rho^0}}%
\htdef{g168.descr}{\ensuremath{e^- K^*(892)^0}}%
\htdef{g169.descr}{\ensuremath{\mu^- K^*(892)^0}}%
\htdef{g170.descr}{\ensuremath{e^- \bar{K}^*(892)^0}}%
\htdef{g171.descr}{\ensuremath{\mu^- \bar{K}^*(892)^0}}%
\htdef{g176.descr}{\ensuremath{e^- \phi}}%
\htdef{g177.descr}{\ensuremath{\mu^- \phi}}%
\htdef{g166.descr}{\ensuremath{e^- \omega}}%
\htdef{g167.descr}{\ensuremath{\mu^- \omega}}%
\htdef{g178.descr}{\ensuremath{e^- e^+ e^-}}%
\htdef{g181.descr}{\ensuremath{\mu^- e^+ e^-}}%
\htdef{g179.descr}{\ensuremath{e^- \mu^+ \mu^-}}%
\htdef{g183.descr}{\ensuremath{\mu^- \mu^+ \mu^-}}%
\htdef{g182.descr}{\ensuremath{e^- \mu^+ e^-}}%
\htdef{g180.descr}{\ensuremath{\mu^- e^+ \mu^-}}%
\htdef{g184.descr}{\ensuremath{e^- \pi^+ \pi^-}}%
\htdef{g186.descr}{\ensuremath{\mu^- \pi^+  \pi^-}}%
\htdef{g188.descr}{\ensuremath{e^- \pi^+ K^-}}%
\htdef{g194.descr}{\ensuremath{\mu^- \pi^+  K^-}}%
\htdef{g189.descr}{\ensuremath{e^- K^+ \pi^-}}%
\htdef{g195.descr}{\ensuremath{\mu^- K^+  \pi^-}}%
\htdef{g192.descr}{\ensuremath{e^- K^+ K^-}}%
\htdef{g198.descr}{\ensuremath{\mu^- K^+  K^-}}%
\htdef{g191.descr}{\ensuremath{e^- K^0_S K^0_S}}%
\htdef{g197.descr}{\ensuremath{\mu^- K^0_S  K^0_S}}%
\htdef{g185.descr}{\ensuremath{e^+ \pi^- \pi^-}}%
\htdef{g187.descr}{\ensuremath{\mu^+ \pi^- \pi^-}}%
\htdef{g190.descr}{\ensuremath{e^+ \pi^- K^-}}%
\htdef{g196.descr}{\ensuremath{\mu^+ \pi^- K^-}}%
\htdef{g193.descr}{\ensuremath{e^+ K^- K^-}}%
\htdef{g199.descr}{\ensuremath{\mu^+ K^- K^-}}%
\htdef{g211.descr}{\ensuremath{\pi^- \Lambda}}%
\htdef{g212.descr}{\ensuremath{\pi^- \bar{\Lambda}}}%
\htdef{g213.descr}{\ensuremath{K^- \Lambda}}%
\htdef{g214.descr}{\ensuremath{K^- \bar{\Lambda}}}%
\htdef{g215.descr}{\ensuremath{p \mu^- \mu^-}}%
\htdef{g216.descr}{\ensuremath{\bar{p} \mu^+ \mu^-}}%
\htdef{CombExtraLines}{%
\htCombExtraLine%
  {\ensuremath{\Gamma_{156} = e^- \gamma}}%
  {\babar}%
  {Aubert:2009ag}%
  {963}%
  {\ensuremath{3.90 \pm 0.30}}%
  {\ensuremath{1.60 \pm 0.40}}%
  {0}%
\htCombExtraLine%
  {\ensuremath{\Gamma_{156} = e^- \gamma}}%
  {\belle}%
  {Hayasaka:2007vc}%
  {983}%
  {\ensuremath{3.00 \pm 0.10}}%
  {\ensuremath{5.14 \pm 3.30}}%
  {5}%
\htCombExtraLine%
  {\ensuremath{\Gamma_{157} = \mu^- \gamma}}%
  {\babar}%
  {Aubert:2009ag}%
  {963}%
  {\ensuremath{6.10 \pm 0.50}}%
  {\ensuremath{3.60 \pm 0.70}}%
  {2}%
\htCombExtraLine%
  {\ensuremath{\Gamma_{157} = \mu^- \gamma}}%
  {\belle}%
  {Hayasaka:2007vc}%
  {983}%
  {\ensuremath{5.07 \pm 0.20}}%
  {\ensuremath{13.90 \pm 5.00}}%
  {10}%
\htCombExtraLine%
  {\ensuremath{\Gamma_{158} = e^- \pi^0}}%
  {\babar}%
  {Aubert:2006cz}%
  {339}%
  {\ensuremath{2.83 \pm 0.25}}%
  {\ensuremath{0.17 \pm 0.04}}%
  {0}%
\htCombExtraLine%
  {\ensuremath{\Gamma_{158} = e^- \pi^0}}%
  {\belle}%
  {Miyazaki:2007jp}%
  {401}%
  {\ensuremath{3.93 \pm 0.18}}%
  {\ensuremath{0.20 \pm 0.20}}%
  {0}%
\htCombExtraLine%
  {\ensuremath{\Gamma_{159} = \mu^- \pi^0}}%
  {\babar}%
  {Aubert:2006cz}%
  {339}%
  {\ensuremath{4.75 \pm 0.37}}%
  {\ensuremath{1.33 \pm 0.15}}%
  {1}%
\htCombExtraLine%
  {\ensuremath{\Gamma_{159} = \mu^- \pi^0}}%
  {\belle}%
  {Miyazaki:2007jp}%
  {401}%
  {\ensuremath{4.53 \pm 0.20}}%
  {\ensuremath{0.58 \pm 0.34}}%
  {1}%
\htCombExtraLine%
  {\ensuremath{\Gamma_{160} = e^- K^0_S}}%
  {\babar}%
  {Aubert:2009ys}%
  {862}%
  {\ensuremath{9.10 \pm 1.73}}%
  {\ensuremath{0.59 \pm 0.25}}%
  {1}%
\htCombExtraLine%
  {\ensuremath{\Gamma_{160} = e^- K^0_S}}%
  {\belle}%
  {Miyazaki:2010qb}%
  {1274}%
  {\ensuremath{10.20 \pm 0.67}}%
  {\ensuremath{0.18 \pm 0.18}}%
  {0}%
\htCombExtraLine%
  {\ensuremath{\Gamma_{161} = \mu^- K^0_S}}%
  {\babar}%
  {Aubert:2009ys}%
  {862}%
  {\ensuremath{6.14 \pm 0.20}}%
  {\ensuremath{0.30 \pm 0.18}}%
  {1}%
\htCombExtraLine%
  {\ensuremath{\Gamma_{161} = \mu^- K^0_S}}%
  {\belle}%
  {Miyazaki:2010qb}%
  {1274}%
  {\ensuremath{10.70 \pm 0.73}}%
  {\ensuremath{0.35 \pm 0.21}}%
  {0}%
\htCombExtraLine%
  {\ensuremath{\Gamma_{162} = e^- \eta}}%
  {\babar}%
  {Aubert:2006cz}%
  {339}%
  {\ensuremath{2.12 \pm 0.20}}%
  {\ensuremath{0.22 \pm 0.05}}%
  {0}%
\htCombExtraLine%
  {\ensuremath{\Gamma_{162} = e^- \eta}}%
  {\belle}%
  {Miyazaki:2007jp}%
  {401}%
  {\ensuremath{2.87 \pm 0.20}}%
  {\ensuremath{0.78 \pm 0.78}}%
  {0}%
\htCombExtraLine%
  {\ensuremath{\Gamma_{163} = \mu^- \eta}}%
  {\babar}%
  {Aubert:2006cz}%
  {339}%
  {\ensuremath{3.59 \pm 0.41}}%
  {\ensuremath{0.75 \pm 0.08}}%
  {1}%
\htCombExtraLine%
  {\ensuremath{\Gamma_{163} = \mu^- \eta}}%
  {\belle}%
  {Miyazaki:2007jp}%
  {401}%
  {\ensuremath{4.08 \pm 0.28}}%
  {\ensuremath{0.64 \pm 0.04}}%
  {0}%
\htCombExtraLine%
  {\ensuremath{\Gamma_{172} = e^- \eta^\prime(958)}}%
  {\babar}%
  {Aubert:2006cz}%
  {339}%
  {\ensuremath{1.53 \pm 0.16}}%
  {\ensuremath{0.12 \pm 0.03}}%
  {0}%
\htCombExtraLine%
  {\ensuremath{\Gamma_{172} = e^- \eta^\prime(958)}}%
  {\belle}%
  {Miyazaki:2007jp}%
  {401}%
  {\ensuremath{1.59 \pm 0.13}}%
  {\ensuremath{0.01 \pm 0.41}}%
  {0}%
\htCombExtraLine%
  {\ensuremath{\Gamma_{173} = \mu^- \eta^\prime(958)}}%
  {\babar}%
  {Aubert:2006cz}%
  {339}%
  {\ensuremath{2.18 \pm 0.26}}%
  {\ensuremath{0.49 \pm 0.26}}%
  {0}%
\htCombExtraLine%
  {\ensuremath{\Gamma_{173} = \mu^- \eta^\prime(958)}}%
  {\belle}%
  {Miyazaki:2007jp}%
  {401}%
  {\ensuremath{2.47 \pm 0.20}}%
  {\ensuremath{0.23 \pm 0.46}}%
  {0}%
\htCombExtraLine%
  {\ensuremath{\Gamma_{164} = e^- \rho^0}}%
  {\babar}%
  {Aubert:2009ap}%
  {829}%
  {\ensuremath{7.31 \pm 0.20}}%
  {\ensuremath{1.32 \pm 0.17}}%
  {1}%
\htCombExtraLine%
  {\ensuremath{\Gamma_{164} = e^- \rho^0}}%
  {\belle}%
  {Miyazaki:2011xe}%
  {1554}%
  {\ensuremath{7.58 \pm 0.41}}%
  {\ensuremath{0.29 \pm 0.15}}%
  {0}%
\htCombExtraLine%
  {\ensuremath{\Gamma_{165} = \mu^- \rho^0}}%
  {\babar}%
  {Aubert:2009ap}%
  {829}%
  {\ensuremath{4.52 \pm 0.40}}%
  {\ensuremath{2.04 \pm 0.19}}%
  {0}%
\htCombExtraLine%
  {\ensuremath{\Gamma_{165} = \mu^- \rho^0}}%
  {\belle}%
  {Miyazaki:2011xe}%
  {1554}%
  {\ensuremath{7.09 \pm 0.37}}%
  {\ensuremath{1.48 \pm 0.35}}%
  {0}%
\htCombExtraLine%
  {\ensuremath{\Gamma_{166} = e^- \omega}}%
  {\babar}%
  {Aubert:2007kx}%
  {829}%
  {\ensuremath{2.96 \pm 0.13}}%
  {\ensuremath{0.35 \pm 0.06}}%
  {0}%
\htCombExtraLine%
  {\ensuremath{\Gamma_{166} = e^- \omega}}%
  {\belle}%
  {Miyazaki:2011xe}%
  {1554}%
  {\ensuremath{2.92 \pm 0.18}}%
  {\ensuremath{0.30 \pm 0.14}}%
  {0}%
\htCombExtraLine%
  {\ensuremath{\Gamma_{167} = \mu^- \omega}}%
  {\babar}%
  {Aubert:2007kx}%
  {829}%
  {\ensuremath{2.56 \pm 0.16}}%
  {\ensuremath{0.73 \pm 0.03}}%
  {0}%
\htCombExtraLine%
  {\ensuremath{\Gamma_{167} = \mu^- \omega}}%
  {\belle}%
  {Miyazaki:2011xe}%
  {1554}%
  {\ensuremath{2.38 \pm 0.14}}%
  {\ensuremath{0.72 \pm 0.18}}%
  {0}%
\htCombExtraLine%
  {\ensuremath{\Gamma_{168} = e^- K^*(892)^0}}%
  {\babar}%
  {Aubert:2009ap}%
  {829}%
  {\ensuremath{8.00 \pm 0.20}}%
  {\ensuremath{1.65 \pm 0.23}}%
  {2}%
\htCombExtraLine%
  {\ensuremath{\Gamma_{168} = e^- K^*(892)^0}}%
  {\belle}%
  {Miyazaki:2011xe}%
  {1554}%
  {\ensuremath{4.37 \pm 0.24}}%
  {\ensuremath{0.29 \pm 0.14}}%
  {0}%
\htCombExtraLine%
  {\ensuremath{\Gamma_{169} = \mu^- K^*(892)^0}}%
  {\babar}%
  {Aubert:2009ap}%
  {829}%
  {\ensuremath{4.60 \pm 0.40}}%
  {\ensuremath{1.79 \pm 0.21}}%
  {4}%
\htCombExtraLine%
  {\ensuremath{\Gamma_{169} = \mu^- K^*(892)^0}}%
  {\belle}%
  {Miyazaki:2011xe}%
  {1554}%
  {\ensuremath{3.39 \pm 0.19}}%
  {\ensuremath{0.53 \pm 0.20}}%
  {1}%
\htCombExtraLine%
  {\ensuremath{\Gamma_{170} = e^- \bar{K}^*(892)^0}}%
  {\babar}%
  {Aubert:2009ap}%
  {829}%
  {\ensuremath{7.80 \pm 0.20}}%
  {\ensuremath{2.76 \pm 0.28}}%
  {2}%
\htCombExtraLine%
  {\ensuremath{\Gamma_{170} = e^- \bar{K}^*(892)^0}}%
  {\belle}%
  {Miyazaki:2011xe}%
  {1554}%
  {\ensuremath{4.41 \pm 0.25}}%
  {\ensuremath{0.08 \pm 0.08}}%
  {0}%
\htCombExtraLine%
  {\ensuremath{\Gamma_{171} = \mu^- \bar{K}^*(892)^0}}%
  {\babar}%
  {Aubert:2009ap}%
  {829}%
  {\ensuremath{4.10 \pm 0.30}}%
  {\ensuremath{1.72 \pm 0.17}}%
  {1}%
\htCombExtraLine%
  {\ensuremath{\Gamma_{171} = \mu^- \bar{K}^*(892)^0}}%
  {\belle}%
  {Miyazaki:2011xe}%
  {1554}%
  {\ensuremath{3.60 \pm 0.20}}%
  {\ensuremath{0.45 \pm 0.17}}%
  {1}%
\htCombExtraLine%
  {\ensuremath{\Gamma_{176} = e^- \phi}}%
  {\babar}%
  {Aubert:2009ap}%
  {829}%
  {\ensuremath{6.40 \pm 0.20}}%
  {\ensuremath{0.68 \pm 0.12}}%
  {0}%
\htCombExtraLine%
  {\ensuremath{\Gamma_{176} = e^- \phi}}%
  {\belle}%
  {Miyazaki:2011xe}%
  {1554}%
  {\ensuremath{4.18 \pm 0.25}}%
  {\ensuremath{0.47 \pm 0.19}}%
  {0}%
\htCombExtraLine%
  {\ensuremath{\Gamma_{177} = \mu^- \phi}}%
  {\babar}%
  {Aubert:2009ap}%
  {829}%
  {\ensuremath{5.20 \pm 0.30}}%
  {\ensuremath{2.76 \pm 0.16}}%
  {6}%
\htCombExtraLine%
  {\ensuremath{\Gamma_{177} = \mu^- \phi}}%
  {\belle}%
  {Miyazaki:2011xe}%
  {1554}%
  {\ensuremath{3.21 \pm 0.19}}%
  {\ensuremath{0.06 \pm 0.06}}%
  {1}%
\htCombExtraLine%
  {\ensuremath{\Gamma_{178} = e^- e^+ e^-}}%
  {\babar}%
  {Lees:2010ez}%
  {868}%
  {\ensuremath{8.60 \pm 0.20}}%
  {\ensuremath{0.12 \pm 0.02}}%
  {0}%
\htCombExtraLine%
  {\ensuremath{\Gamma_{178} = e^- e^+ e^-}}%
  {\belle}%
  {Hayasaka:2010np}%
  {1437}%
  {\ensuremath{6.00 \pm 0.59}}%
  {\ensuremath{0.21 \pm 0.15}}%
  {0}%
\htCombExtraLine%
  {\ensuremath{\Gamma_{179} = e^- \mu^+ \mu^-}}%
  {\babar}%
  {Lees:2010ez}%
  {868}%
  {\ensuremath{6.40 \pm 0.40}}%
  {\ensuremath{0.54 \pm 0.14}}%
  {0}%
\htCombExtraLine%
  {\ensuremath{\Gamma_{179} = e^- \mu^+ \mu^-}}%
  {\belle}%
  {Hayasaka:2010np}%
  {1437}%
  {\ensuremath{6.10 \pm 0.58}}%
  {\ensuremath{0.10 \pm 0.04}}%
  {0}%
\htCombExtraLine%
  {\ensuremath{\Gamma_{180} = \mu^- e^+ \mu^-}}%
  {\babar}%
  {Lees:2010ez}%
  {868}%
  {\ensuremath{10.20 \pm 0.60}}%
  {\ensuremath{0.03 \pm 0.02}}%
  {0}%
\htCombExtraLine%
  {\ensuremath{\Gamma_{180} = \mu^- e^+ \mu^-}}%
  {\belle}%
  {Hayasaka:2010np}%
  {1437}%
  {\ensuremath{10.10 \pm 0.77}}%
  {\ensuremath{0.02 \pm 0.02}}%
  {0}%
\htCombExtraLine%
  {\ensuremath{\Gamma_{181} = \mu^- e^+ e^-}}%
  {\babar}%
  {Lees:2010ez}%
  {868}%
  {\ensuremath{8.80 \pm 0.50}}%
  {\ensuremath{0.64 \pm 0.19}}%
  {0}%
\htCombExtraLine%
  {\ensuremath{\Gamma_{181} = \mu^- e^+ e^-}}%
  {\belle}%
  {Hayasaka:2010np}%
  {1437}%
  {\ensuremath{9.30 \pm 0.73}}%
  {\ensuremath{0.04 \pm 0.04}}%
  {0}%
\htCombExtraLine%
  {\ensuremath{\Gamma_{182} = e^- \mu^+ e^-}}%
  {\babar}%
  {Lees:2010ez}%
  {868}%
  {\ensuremath{12.70 \pm 0.70}}%
  {\ensuremath{0.34 \pm 0.12}}%
  {0}%
\htCombExtraLine%
  {\ensuremath{\Gamma_{182} = e^- \mu^+ e^-}}%
  {\belle}%
  {Hayasaka:2010np}%
  {1437}%
  {\ensuremath{11.50 \pm 0.89}}%
  {\ensuremath{0.01 \pm 0.01}}%
  {0}%
\htCombExtraLine%
  {\ensuremath{\Gamma_{183} = \mu^- \mu^+ \mu^-}}%
  {\babar}%
  {Lees:2010ez}%
  {868}%
  {\ensuremath{6.60 \pm 0.60}}%
  {\ensuremath{0.44 \pm 0.17}}%
  {0}%
\htCombExtraLine%
  {\ensuremath{\Gamma_{183} = \mu^- \mu^+ \mu^-}}%
  {\belle}%
  {Hayasaka:2010np}%
  {1437}%
  {\ensuremath{7.60 \pm 0.56}}%
  {\ensuremath{0.13 \pm 0.20}}%
  {0}}%
\htdef{LimitLines}{%
\htLimitLine%
  {\ensuremath{\Gamma_{156} = e^- \gamma}}%
  {\ensuremath{\ell\gamma}}%
  {\ensuremath{3.3 \cdot 10^{-8}}}%
  {\babar}%
  {Aubert:2009ag}%
\htLimitLine%
  {\ensuremath{\Gamma_{156} = e^- \gamma}}%
  {\ensuremath{}}%
  {\ensuremath{1.2 \cdot 10^{-7}}}%
  {\belle}%
  {Hayasaka:2007vc}%
\htLimitLine%
  {\ensuremath{\Gamma_{157} = \mu^- \gamma}}%
  {\ensuremath{}}%
  {\ensuremath{4.4 \cdot 10^{-8}}}%
  {\babar}%
  {Aubert:2009ag}%
\htLimitLine%
  {\ensuremath{\Gamma_{157} = \mu^- \gamma}}%
  {\ensuremath{}}%
  {\ensuremath{4.5 \cdot 10^{-8}}}%
  {\belle}%
  {Hayasaka:2007vc}%
\midrule%
\htLimitLine%
  {\ensuremath{\Gamma_{158} = e^- \pi^0}}%
  {\ensuremath{\ell P^0}}%
  {\ensuremath{1.3 \cdot 10^{-7}}}%
  {\babar}%
  {Aubert:2006cz}%
\htLimitLine%
  {\ensuremath{\Gamma_{158} = e^- \pi^0}}%
  {\ensuremath{}}%
  {\ensuremath{8.0 \cdot 10^{-8}}}%
  {\belle}%
  {Miyazaki:2007jp}%
\htLimitLine%
  {\ensuremath{\Gamma_{159} = \mu^- \pi^0}}%
  {\ensuremath{}}%
  {\ensuremath{1.1 \cdot 10^{-7}}}%
  {\babar}%
  {Aubert:2006cz}%
\htLimitLine%
  {\ensuremath{\Gamma_{159} = \mu^- \pi^0}}%
  {\ensuremath{}}%
  {\ensuremath{1.2 \cdot 10^{-7}}}%
  {\belle}%
  {Miyazaki:2007jp}%
\htLimitLine%
  {\ensuremath{\Gamma_{160} = e^- K^0_S}}%
  {\ensuremath{}}%
  {\ensuremath{3.3 \cdot 10^{-8}}}%
  {\babar}%
  {Aubert:2009ys}%
\htLimitLine%
  {\ensuremath{\Gamma_{160} = e^- K^0_S}}%
  {\ensuremath{}}%
  {\ensuremath{2.6 \cdot 10^{-8}}}%
  {\belle}%
  {Miyazaki:2010qb}%
\htLimitLine%
  {\ensuremath{\Gamma_{161} = \mu^- K^0_S}}%
  {\ensuremath{}}%
  {\ensuremath{4.0 \cdot 10^{-8}}}%
  {\babar}%
  {Aubert:2009ys}%
\htLimitLine%
  {\ensuremath{\Gamma_{161} = \mu^- K^0_S}}%
  {\ensuremath{}}%
  {\ensuremath{2.3 \cdot 10^{-8}}}%
  {\belle}%
  {Miyazaki:2010qb}%
\htLimitLine%
  {\ensuremath{\Gamma_{162} = e^- \eta}}%
  {\ensuremath{}}%
  {\ensuremath{1.6 \cdot 10^{-7}}}%
  {\babar}%
  {Aubert:2006cz}%
\htLimitLine%
  {\ensuremath{\Gamma_{162} = e^- \eta}}%
  {\ensuremath{}}%
  {\ensuremath{9.2 \cdot 10^{-8}}}%
  {\belle}%
  {Miyazaki:2007jp}%
\htLimitLine%
  {\ensuremath{\Gamma_{163} = \mu^- \eta}}%
  {\ensuremath{}}%
  {\ensuremath{1.5 \cdot 10^{-7}}}%
  {\babar}%
  {Aubert:2006cz}%
\htLimitLine%
  {\ensuremath{\Gamma_{163} = \mu^- \eta}}%
  {\ensuremath{}}%
  {\ensuremath{6.5 \cdot 10^{-8}}}%
  {\belle}%
  {Miyazaki:2007jp}%
\htLimitLine%
  {\ensuremath{\Gamma_{172} = e^- \eta^\prime(958)}}%
  {\ensuremath{}}%
  {\ensuremath{2.4 \cdot 10^{-7}}}%
  {\babar}%
  {Aubert:2006cz}%
\htLimitLine%
  {\ensuremath{\Gamma_{172} = e^- \eta^\prime(958)}}%
  {\ensuremath{}}%
  {\ensuremath{1.6 \cdot 10^{-7}}}%
  {\belle}%
  {Miyazaki:2007jp}%
\htLimitLine%
  {\ensuremath{\Gamma_{173} = \mu^- \eta^\prime(958)}}%
  {\ensuremath{}}%
  {\ensuremath{1.4 \cdot 10^{-7}}}%
  {\babar}%
  {Aubert:2006cz}%
\htLimitLine%
  {\ensuremath{\Gamma_{173} = \mu^- \eta^\prime(958)}}%
  {\ensuremath{}}%
  {\ensuremath{1.3 \cdot 10^{-7}}}%
  {\belle}%
  {Miyazaki:2007jp}%
\midrule%
\htLimitLine%
  {\ensuremath{\Gamma_{164} = e^- \rho^0}}%
  {\ensuremath{\ell V^0}}%
  {\ensuremath{4.6 \cdot 10^{-8}}}%
  {\babar}%
  {Aubert:2009ap}%
\htLimitLine%
  {\ensuremath{\Gamma_{164} = e^- \rho^0}}%
  {\ensuremath{}}%
  {\ensuremath{1.8 \cdot 10^{-8}}}%
  {\belle}%
  {Miyazaki:2011xe}%
\htLimitLine%
  {\ensuremath{\Gamma_{165} = \mu^- \rho^0}}%
  {\ensuremath{}}%
  {\ensuremath{2.6 \cdot 10^{-8}}}%
  {\babar}%
  {Aubert:2009ap}%
\htLimitLine%
  {\ensuremath{\Gamma_{165} = \mu^- \rho^0}}%
  {\ensuremath{}}%
  {\ensuremath{1.2 \cdot 10^{-8}}}%
  {\belle}%
  {Miyazaki:2011xe}%
\htLimitLine%
  {\ensuremath{\Gamma_{166} = e^- \omega}}%
  {\ensuremath{}}%
  {\ensuremath{1.1 \cdot 10^{-7}}}%
  {\babar}%
  {Aubert:2007kx}%
\htLimitLine%
  {\ensuremath{\Gamma_{166} = e^- \omega}}%
  {\ensuremath{}}%
  {\ensuremath{4.8 \cdot 10^{-8}}}%
  {\belle}%
  {Miyazaki:2011xe}%
\htLimitLine%
  {\ensuremath{\Gamma_{167} = \mu^- \omega}}%
  {\ensuremath{}}%
  {\ensuremath{1.0 \cdot 10^{-7}}}%
  {\babar}%
  {Aubert:2007kx}%
\htLimitLine%
  {\ensuremath{\Gamma_{167} = \mu^- \omega}}%
  {\ensuremath{}}%
  {\ensuremath{4.7 \cdot 10^{-8}}}%
  {\belle}%
  {Miyazaki:2011xe}%
\htLimitLine%
  {\ensuremath{\Gamma_{168} = e^- K^*(892)^0}}%
  {\ensuremath{}}%
  {\ensuremath{5.9 \cdot 10^{-8}}}%
  {\babar}%
  {Aubert:2009ap}%
\htLimitLine%
  {\ensuremath{\Gamma_{168} = e^- K^*(892)^0}}%
  {\ensuremath{}}%
  {\ensuremath{3.2 \cdot 10^{-8}}}%
  {\belle}%
  {Miyazaki:2011xe}%
\htLimitLine%
  {\ensuremath{\Gamma_{169} = \mu^- K^*(892)^0}}%
  {\ensuremath{}}%
  {\ensuremath{1.7 \cdot 10^{-7}}}%
  {\babar}%
  {Aubert:2009ap}%
\htLimitLine%
  {\ensuremath{\Gamma_{169} = \mu^- K^*(892)^0}}%
  {\ensuremath{}}%
  {\ensuremath{7.2 \cdot 10^{-8}}}%
  {\belle}%
  {Miyazaki:2011xe}%
\htLimitLine%
  {\ensuremath{\Gamma_{170} = e^- \bar{K}^*(892)^0}}%
  {\ensuremath{}}%
  {\ensuremath{4.6 \cdot 10^{-8}}}%
  {\babar}%
  {Aubert:2009ap}%
\htLimitLine%
  {\ensuremath{\Gamma_{170} = e^- \bar{K}^*(892)^0}}%
  {\ensuremath{}}%
  {\ensuremath{3.4 \cdot 10^{-8}}}%
  {\belle}%
  {Miyazaki:2011xe}%
\htLimitLine%
  {\ensuremath{\Gamma_{171} = \mu^- \bar{K}^*(892)^0}}%
  {\ensuremath{}}%
  {\ensuremath{7.3 \cdot 10^{-8}}}%
  {\babar}%
  {Aubert:2009ap}%
\htLimitLine%
  {\ensuremath{\Gamma_{171} = \mu^- \bar{K}^*(892)^0}}%
  {\ensuremath{}}%
  {\ensuremath{7.0 \cdot 10^{-8}}}%
  {\belle}%
  {Miyazaki:2011xe}%
\htLimitLine%
  {\ensuremath{\Gamma_{176} = e^- \phi}}%
  {\ensuremath{}}%
  {\ensuremath{3.1 \cdot 10^{-8}}}%
  {\babar}%
  {Aubert:2009ap}%
\htLimitLine%
  {\ensuremath{\Gamma_{176} = e^- \phi}}%
  {\ensuremath{}}%
  {\ensuremath{3.1 \cdot 10^{-8}}}%
  {\belle}%
  {Miyazaki:2011xe}%
\htLimitLine%
  {\ensuremath{\Gamma_{177} = \mu^- \phi}}%
  {\ensuremath{}}%
  {\ensuremath{1.9 \cdot 10^{-7}}}%
  {\babar}%
  {Aubert:2009ap}%
\htLimitLine%
  {\ensuremath{\Gamma_{177} = \mu^- \phi}}%
  {\ensuremath{}}%
  {\ensuremath{8.4 \cdot 10^{-8}}}%
  {\belle}%
  {Miyazaki:2011xe}%
\midrule%
\htLimitLine%
  {\ensuremath{\Gamma_{174} = e^- f_0(980)}}%
  {\ensuremath{\ell S^0}}%
  {\ensuremath{3.2 \cdot 10^{-8}}}%
  {\belle}%
  {Miyazaki:2008mw}%
\htLimitLine%
  {\ensuremath{\Gamma_{175} = \mu^- f_0(980)}}%
  {\ensuremath{}}%
  {\ensuremath{3.4 \cdot 10^{-8}}}%
  {\belle}%
  {Miyazaki:2008mw}%
\midrule%
\htLimitLine%
  {\ensuremath{\Gamma_{178} = e^- e^+ e^-}}%
  {\ensuremath{\ell\ell\ell}}%
  {\ensuremath{2.9 \cdot 10^{-8}}}%
  {\babar}%
  {Lees:2010ez}%
\htLimitLine%
  {\ensuremath{\Gamma_{178} = e^- e^+ e^-}}%
  {\ensuremath{}}%
  {\ensuremath{2.7 \cdot 10^{-8}}}%
  {\belle}%
  {Hayasaka:2010np}%
\htLimitLine%
  {\ensuremath{\Gamma_{179} = e^- \mu^+ \mu^-}}%
  {\ensuremath{}}%
  {\ensuremath{3.2 \cdot 10^{-8}}}%
  {\babar}%
  {Lees:2010ez}%
\htLimitLine%
  {\ensuremath{\Gamma_{179} = e^- \mu^+ \mu^-}}%
  {\ensuremath{}}%
  {\ensuremath{2.7 \cdot 10^{-8}}}%
  {\belle}%
  {Hayasaka:2010np}%
\htLimitLine%
  {\ensuremath{\Gamma_{180} = \mu^- e^+ \mu^-}}%
  {\ensuremath{}}%
  {\ensuremath{2.6 \cdot 10^{-8}}}%
  {\babar}%
  {Lees:2010ez}%
\htLimitLine%
  {\ensuremath{\Gamma_{180} = \mu^- e^+ \mu^-}}%
  {\ensuremath{}}%
  {\ensuremath{1.7 \cdot 10^{-8}}}%
  {\belle}%
  {Hayasaka:2010np}%
\htLimitLine%
  {\ensuremath{\Gamma_{181} = \mu^- e^+ e^-}}%
  {\ensuremath{}}%
  {\ensuremath{2.2 \cdot 10^{-8}}}%
  {\babar}%
  {Lees:2010ez}%
\htLimitLine%
  {\ensuremath{\Gamma_{181} = \mu^- e^+ e^-}}%
  {\ensuremath{}}%
  {\ensuremath{1.8 \cdot 10^{-8}}}%
  {\belle}%
  {Hayasaka:2010np}%
\htLimitLine%
  {\ensuremath{\Gamma_{182} = e^- \mu^+ e^-}}%
  {\ensuremath{}}%
  {\ensuremath{1.8 \cdot 10^{-8}}}%
  {\babar}%
  {Lees:2010ez}%
\htLimitLine%
  {\ensuremath{\Gamma_{182} = e^- \mu^+ e^-}}%
  {\ensuremath{}}%
  {\ensuremath{1.5 \cdot 10^{-8}}}%
  {\belle}%
  {Hayasaka:2010np}%
\htLimitLine%
  {\ensuremath{\Gamma_{183} = \mu^- \mu^+ \mu^-}}%
  {\ensuremath{}}%
  {\ensuremath{3.8 \cdot 10^{-7}}}%
  {ATLAS}%
  {Aad:2016wce}%
\htLimitLine%
  {\ensuremath{\Gamma_{183} = \mu^- \mu^+ \mu^-}}%
  {\ensuremath{}}%
  {\ensuremath{3.3 \cdot 10^{-8}}}%
  {\babar}%
  {Lees:2010ez}%
\htLimitLine%
  {\ensuremath{\Gamma_{183} = \mu^- \mu^+ \mu^-}}%
  {\ensuremath{}}%
  {\ensuremath{2.1 \cdot 10^{-8}}}%
  {\belle}%
  {Hayasaka:2010np}%
\htLimitLine%
  {\ensuremath{\Gamma_{183} = \mu^- \mu^+ \mu^-}}%
  {\ensuremath{}}%
  {\ensuremath{4.6 \cdot 10^{-8}}}%
  {LHCb}%
  {Aaij:2014azz}%
\midrule%
\htLimitLine%
  {\ensuremath{\Gamma_{184} = e^- \pi^+ \pi^-}}%
  {\ensuremath{\ell hh}}%
  {\ensuremath{1.2 \cdot 10^{-7}}}%
  {\babar}%
  {Aubert:2005tp}%
\htLimitLine%
  {\ensuremath{\Gamma_{184} = e^- \pi^+ \pi^-}}%
  {\ensuremath{}}%
  {\ensuremath{2.3 \cdot 10^{-8}}}%
  {\belle}%
  {Miyazaki:2012mx}%
\htLimitLine%
  {\ensuremath{\Gamma_{185} = e^+ \pi^- \pi^-}}%
  {\ensuremath{}}%
  {\ensuremath{2.7 \cdot 10^{-7}}}%
  {\babar}%
  {Aubert:2005tp}%
\htLimitLine%
  {\ensuremath{\Gamma_{185} = e^+ \pi^- \pi^-}}%
  {\ensuremath{}}%
  {\ensuremath{2.0 \cdot 10^{-8}}}%
  {\belle}%
  {Miyazaki:2012mx}%
\htLimitLine%
  {\ensuremath{\Gamma_{186} = \mu^- \pi^+  \pi^-}}%
  {\ensuremath{}}%
  {\ensuremath{2.9 \cdot 10^{-7}}}%
  {\babar}%
  {Aubert:2005tp}%
\htLimitLine%
  {\ensuremath{\Gamma_{186} = \mu^- \pi^+  \pi^-}}%
  {\ensuremath{}}%
  {\ensuremath{2.1 \cdot 10^{-8}}}%
  {\belle}%
  {Miyazaki:2012mx}%
\htLimitLine%
  {\ensuremath{\Gamma_{187} = \mu^+ \pi^- \pi^-}}%
  {\ensuremath{}}%
  {\ensuremath{7.0 \cdot 10^{-8}}}%
  {\babar}%
  {Aubert:2005tp}%
\htLimitLine%
  {\ensuremath{\Gamma_{187} = \mu^+ \pi^- \pi^-}}%
  {\ensuremath{}}%
  {\ensuremath{3.9 \cdot 10^{-8}}}%
  {\belle}%
  {Miyazaki:2012mx}%
\htLimitLine%
  {\ensuremath{\Gamma_{188} = e^- \pi^+ K^-}}%
  {\ensuremath{}}%
  {\ensuremath{3.2 \cdot 10^{-7}}}%
  {\babar}%
  {Aubert:2005tp}%
\htLimitLine%
  {\ensuremath{\Gamma_{188} = e^- \pi^+ K^-}}%
  {\ensuremath{}}%
  {\ensuremath{3.7 \cdot 10^{-8}}}%
  {\belle}%
  {Miyazaki:2012mx}%
\htLimitLine%
  {\ensuremath{\Gamma_{189} = e^- K^+ \pi^-}}%
  {\ensuremath{}}%
  {\ensuremath{1.7 \cdot 10^{-7}}}%
  {\babar}%
  {Aubert:2005tp}%
\htLimitLine%
  {\ensuremath{\Gamma_{189} = e^- K^+ \pi^-}}%
  {\ensuremath{}}%
  {\ensuremath{3.1 \cdot 10^{-8}}}%
  {\belle}%
  {Miyazaki:2012mx}%
\htLimitLine%
  {\ensuremath{\Gamma_{190} = e^+ \pi^- K^-}}%
  {\ensuremath{}}%
  {\ensuremath{1.8 \cdot 10^{-7}}}%
  {\babar}%
  {Aubert:2005tp}%
\htLimitLine%
  {\ensuremath{\Gamma_{190} = e^+ \pi^- K^-}}%
  {\ensuremath{}}%
  {\ensuremath{3.2 \cdot 10^{-8}}}%
  {\belle}%
  {Miyazaki:2012mx}%
\htLimitLine%
  {\ensuremath{\Gamma_{191} = e^- K^0_S K^0_S}}%
  {\ensuremath{}}%
  {\ensuremath{7.1 \cdot 10^{-8}}}%
  {\belle}%
  {Miyazaki:2010qb}%
\htLimitLine%
  {\ensuremath{\Gamma_{192} = e^- K^+ K^-}}%
  {\ensuremath{}}%
  {\ensuremath{1.4 \cdot 10^{-7}}}%
  {\babar}%
  {Aubert:2005tp}%
\htLimitLine%
  {\ensuremath{\Gamma_{192} = e^- K^+ K^-}}%
  {\ensuremath{}}%
  {\ensuremath{3.4 \cdot 10^{-8}}}%
  {\belle}%
  {Miyazaki:2012mx}%
\htLimitLine%
  {\ensuremath{\Gamma_{193} = e^+ K^- K^-}}%
  {\ensuremath{}}%
  {\ensuremath{1.5 \cdot 10^{-7}}}%
  {\babar}%
  {Aubert:2005tp}%
\htLimitLine%
  {\ensuremath{\Gamma_{193} = e^+ K^- K^-}}%
  {\ensuremath{}}%
  {\ensuremath{3.3 \cdot 10^{-8}}}%
  {\belle}%
  {Miyazaki:2012mx}%
\htLimitLine%
  {\ensuremath{\Gamma_{194} = \mu^- \pi^+  K^-}}%
  {\ensuremath{}}%
  {\ensuremath{2.6 \cdot 10^{-7}}}%
  {\babar}%
  {Aubert:2005tp}%
\htLimitLine%
  {\ensuremath{\Gamma_{194} = \mu^- \pi^+  K^-}}%
  {\ensuremath{}}%
  {\ensuremath{8.6 \cdot 10^{-8}}}%
  {\belle}%
  {Miyazaki:2012mx}%
\htLimitLine%
  {\ensuremath{\Gamma_{195} = \mu^- K^+  \pi^-}}%
  {\ensuremath{}}%
  {\ensuremath{3.2 \cdot 10^{-7}}}%
  {\babar}%
  {Aubert:2005tp}%
\htLimitLine%
  {\ensuremath{\Gamma_{195} = \mu^- K^+  \pi^-}}%
  {\ensuremath{}}%
  {\ensuremath{4.5 \cdot 10^{-8}}}%
  {\belle}%
  {Miyazaki:2012mx}%
\htLimitLine%
  {\ensuremath{\Gamma_{196} = \mu^+ \pi^- K^-}}%
  {\ensuremath{}}%
  {\ensuremath{2.2 \cdot 10^{-7}}}%
  {\babar}%
  {Aubert:2005tp}%
\htLimitLine%
  {\ensuremath{\Gamma_{196} = \mu^+ \pi^- K^-}}%
  {\ensuremath{}}%
  {\ensuremath{4.8 \cdot 10^{-8}}}%
  {\belle}%
  {Miyazaki:2012mx}%
\htLimitLine%
  {\ensuremath{\Gamma_{197} = \mu^- K^0_S  K^0_S}}%
  {\ensuremath{}}%
  {\ensuremath{8.0 \cdot 10^{-8}}}%
  {\belle}%
  {Miyazaki:2010qb}%
\htLimitLine%
  {\ensuremath{\Gamma_{198} = \mu^- K^+  K^-}}%
  {\ensuremath{}}%
  {\ensuremath{2.5 \cdot 10^{-7}}}%
  {\babar}%
  {Aubert:2005tp}%
\htLimitLine%
  {\ensuremath{\Gamma_{198} = \mu^- K^+  K^-}}%
  {\ensuremath{}}%
  {\ensuremath{4.4 \cdot 10^{-8}}}%
  {\belle}%
  {Miyazaki:2012mx}%
\htLimitLine%
  {\ensuremath{\Gamma_{199} = \mu^+ K^- K^-}}%
  {\ensuremath{}}%
  {\ensuremath{4.8 \cdot 10^{-7}}}%
  {\babar}%
  {Aubert:2005tp}%
\htLimitLine%
  {\ensuremath{\Gamma_{199} = \mu^+ K^- K^-}}%
  {\ensuremath{}}%
  {\ensuremath{4.7 \cdot 10^{-8}}}%
  {\belle}%
  {Miyazaki:2012mx}%
\midrule%
\htLimitLine%
  {\ensuremath{\Gamma_{211} = \pi^- \Lambda}}%
  {\ensuremath{\text{BNV}}}%
  {\ensuremath{7.2 \cdot 10^{-8}}}%
  {\belle}%
  {Miyazaki:2005ng}%
\htLimitLine%
  {\ensuremath{\Gamma_{212} = \pi^- \bar{\Lambda}}}%
  {\ensuremath{}}%
  {\ensuremath{1.4 \cdot 10^{-7}}}%
  {\belle}%
  {Miyazaki:2005ng}%
\htLimitLine%
  {\ensuremath{\Gamma_{215} = p \mu^- \mu^-}}%
  {\ensuremath{}}%
  {\ensuremath{4.4 \cdot 10^{-7}}}%
  {LHCb}%
  {Aaij:2013fia}%
\htLimitLine%
  {\ensuremath{\Gamma_{216} = \bar{p} \mu^+ \mu^-}}%
  {\ensuremath{}}%
  {\ensuremath{3.3 \cdot 10^{-7}}}%
  {LHCb}%
  {Aaij:2013fia}}%
\htdef{PrelimLimitLines}{%
\htLimitLine%
  {\ensuremath{\Gamma_{158} = e^- \pi^0}}%
  {\ensuremath{\ell P^0}}%
  {\ensuremath{2.2 \cdot 10^{-8}}}%
  {Belle}%
  {Hayasaka:2011zz}%
\htLimitLine%
  {\ensuremath{\Gamma_{159} = \mu^- \pi^0}}%
  {\ensuremath{}}%
  {\ensuremath{2.7 \cdot 10^{-8}}}%
  {Belle}%
  {Hayasaka:2011zz}%
\htLimitLine%
  {\ensuremath{\Gamma_{162} = e^- \eta}}%
  {\ensuremath{}}%
  {\ensuremath{4.4 \cdot 10^{-8}}}%
  {Belle}%
  {Hayasaka:2011zz}%
\htLimitLine%
  {\ensuremath{\Gamma_{163} = \mu^- \eta}}%
  {\ensuremath{}}%
  {\ensuremath{2.3 \cdot 10^{-8}}}%
  {Belle}%
  {Hayasaka:2011zz}%
\htLimitLine%
  {\ensuremath{\Gamma_{172} = e^- \eta^\prime(958)}}%
  {\ensuremath{}}%
  {\ensuremath{3.6 \cdot 10^{-8}}}%
  {Belle}%
  {Hayasaka:2011zz}%
\htLimitLine%
  {\ensuremath{\Gamma_{173} = \mu^- \eta^\prime(958)}}%
  {\ensuremath{}}%
  {\ensuremath{3.8 \cdot 10^{-8}}}%
  {Belle}%
  {Hayasaka:2011zz}%
\midrule%
\htLimitLine%
  {\ensuremath{\Gamma_{211} = \pi^- \Lambda}}%
  {\ensuremath{\text{BNV}}}%
  {\ensuremath{5.8 \cdot 10^{-8}}}%
  {BaBar}%
  {Lafferty:2007zz}%
\htLimitLine%
  {\ensuremath{\Gamma_{211} = \pi^- \Lambda}}%
  {\ensuremath{}}%
  {\ensuremath{3.0 \cdot 10^{-8}}}%
  {Belle}%
  {Hayasaka:2012pj}%
\htLimitLine%
  {\ensuremath{\Gamma_{212} = \pi^- \bar{\Lambda}}}%
  {\ensuremath{}}%
  {\ensuremath{5.9 \cdot 10^{-8}}}%
  {BaBar}%
  {Lafferty:2007zz}%
\htLimitLine%
  {\ensuremath{\Gamma_{212} = \pi^- \bar{\Lambda}}}%
  {\ensuremath{}}%
  {\ensuremath{2.8 \cdot 10^{-8}}}%
  {Belle}%
  {Hayasaka:2012pj}%
\htLimitLine%
  {\ensuremath{\Gamma_{213} = K^- \Lambda}}%
  {\ensuremath{}}%
  {\ensuremath{1.5 \cdot 10^{-7}}}%
  {BaBar}%
  {Lafferty:2007zz}%
\htLimitLine%
  {\ensuremath{\Gamma_{213} = K^- \Lambda}}%
  {\ensuremath{}}%
  {\ensuremath{4.2 \cdot 10^{-8}}}%
  {Belle}%
  {Hayasaka:2012pj}%
\htLimitLine%
  {\ensuremath{\Gamma_{214} = K^- \bar{\Lambda}}}%
  {\ensuremath{}}%
  {\ensuremath{7.2 \cdot 10^{-8}}}%
  {BaBar}%
  {Lafferty:2007zz}%
\htLimitLine%
  {\ensuremath{\Gamma_{214} = K^- \bar{\Lambda}}}%
  {\ensuremath{}}%
  {\ensuremath{3.1 \cdot 10^{-8}}}%
  {Belle}%
  {Hayasaka:2012pj}}%
\htdef{CombLines}{%
\htCombLimitLine%
  {\ensuremath{\Gamma_{156} = e^- \gamma}}%
  {\ensuremath{\ell\gamma}}%
  {\ensuremath{5.4 \cdot 10^{-8}}}%
  {\cite{Hayasaka:2007vc,Aubert:2009ag}}%
\htCombLimitLine%
  {\ensuremath{\Gamma_{157} = \mu^- \gamma}}%
  {\ensuremath{}}%
  {\ensuremath{5.0 \cdot 10^{-8}}}%
  {\cite{Hayasaka:2007vc,Aubert:2009ag}}%
\midrule%
\htCombLimitLine%
  {\ensuremath{\Gamma_{158} = e^- \pi^0}}%
  {\ensuremath{\ell P^0}}%
  {\ensuremath{4.9 \cdot 10^{-8}}}%
  {\cite{Miyazaki:2007jp,Aubert:2006cz}}%
\htCombLimitLine%
  {\ensuremath{\Gamma_{159} = \mu^- \pi^0}}%
  {\ensuremath{}}%
  {\ensuremath{3.6 \cdot 10^{-8}}}%
  {\cite{Miyazaki:2007jp,Aubert:2006cz}}%
\htCombLimitLine%
  {\ensuremath{\Gamma_{160} = e^- K^0_S}}%
  {\ensuremath{}}%
  {\ensuremath{1.4 \cdot 10^{-8}}}%
  {\cite{Miyazaki:2010qb,Aubert:2009ys}}%
\htCombLimitLine%
  {\ensuremath{\Gamma_{161} = \mu^- K^0_S}}%
  {\ensuremath{}}%
  {\ensuremath{1.5 \cdot 10^{-8}}}%
  {\cite{Miyazaki:2010qb,Aubert:2009ys}}%
\htCombLimitLine%
  {\ensuremath{\Gamma_{162} = e^- \eta}}%
  {\ensuremath{}}%
  {\ensuremath{5.5 \cdot 10^{-8}}}%
  {\cite{Miyazaki:2007jp,Aubert:2006cz}}%
\htCombLimitLine%
  {\ensuremath{\Gamma_{163} = \mu^- \eta}}%
  {\ensuremath{}}%
  {\ensuremath{3.8 \cdot 10^{-8}}}%
  {\cite{Miyazaki:2007jp,Aubert:2006cz}}%
\htCombLimitLine%
  {\ensuremath{\Gamma_{172} = e^- \eta^\prime(958)}}%
  {\ensuremath{}}%
  {\ensuremath{9.9 \cdot 10^{-8}}}%
  {\cite{Miyazaki:2007jp,Aubert:2006cz}}%
\htCombLimitLine%
  {\ensuremath{\Gamma_{173} = \mu^- \eta^\prime(958)}}%
  {\ensuremath{}}%
  {\ensuremath{6.3 \cdot 10^{-8}}}%
  {\cite{Miyazaki:2007jp,Aubert:2006cz}}%
\midrule%
\htCombLimitLine%
  {\ensuremath{\Gamma_{164} = e^- \rho^0}}%
  {\ensuremath{\ell V^0}}%
  {\ensuremath{1.5 \cdot 10^{-8}}}%
  {\cite{Miyazaki:2011xe,Aubert:2009ap}}%
\htCombLimitLine%
  {\ensuremath{\Gamma_{165} = \mu^- \rho^0}}%
  {\ensuremath{}}%
  {\ensuremath{1.5 \cdot 10^{-8}}}%
  {\cite{Miyazaki:2011xe,Aubert:2009ap}}%
\htCombLimitLine%
  {\ensuremath{\Gamma_{166} = e^- \omega}}%
  {\ensuremath{}}%
  {\ensuremath{3.3 \cdot 10^{-8}}}%
  {\cite{Miyazaki:2011xe,Aubert:2007kx}}%
\htCombLimitLine%
  {\ensuremath{\Gamma_{167} = \mu^- \omega}}%
  {\ensuremath{}}%
  {\ensuremath{4.0 \cdot 10^{-8}}}%
  {\cite{Miyazaki:2011xe,Aubert:2007kx}}%
\htCombLimitLine%
  {\ensuremath{\Gamma_{168} = e^- K^*(892)^0}}%
  {\ensuremath{}}%
  {\ensuremath{2.3 \cdot 10^{-8}}}%
  {\cite{Miyazaki:2011xe,Aubert:2009ap}}%
\htCombLimitLine%
  {\ensuremath{\Gamma_{169} = \mu^- K^*(892)^0}}%
  {\ensuremath{}}%
  {\ensuremath{6.0 \cdot 10^{-8}}}%
  {\cite{Miyazaki:2011xe,Aubert:2009ap}}%
\htCombLimitLine%
  {\ensuremath{\Gamma_{170} = e^- \bar{K}^*(892)^0}}%
  {\ensuremath{}}%
  {\ensuremath{2.2 \cdot 10^{-8}}}%
  {\cite{Miyazaki:2011xe,Aubert:2009ap}}%
\htCombLimitLine%
  {\ensuremath{\Gamma_{171} = \mu^- \bar{K}^*(892)^0}}%
  {\ensuremath{}}%
  {\ensuremath{4.2 \cdot 10^{-8}}}%
  {\cite{Miyazaki:2011xe,Aubert:2009ap}}%
\htCombLimitLine%
  {\ensuremath{\Gamma_{176} = e^- \phi}}%
  {\ensuremath{}}%
  {\ensuremath{2.0 \cdot 10^{-8}}}%
  {\cite{Miyazaki:2011xe,Aubert:2009ap}}%
\htCombLimitLine%
  {\ensuremath{\Gamma_{177} = \mu^- \phi}}%
  {\ensuremath{}}%
  {\ensuremath{6.8 \cdot 10^{-8}}}%
  {\cite{Miyazaki:2011xe,Aubert:2009ap}}%
\midrule%
\htCombLimitLine%
  {\ensuremath{\Gamma_{178} = e^- e^+ e^-}}%
  {\ensuremath{\ell\ell\ell}}%
  {\ensuremath{1.4 \cdot 10^{-8}}}%
  {\cite{Hayasaka:2010np,Lees:2010ez}}%
\htCombLimitLine%
  {\ensuremath{\Gamma_{179} = e^- \mu^+ \mu^-}}%
  {\ensuremath{}}%
  {\ensuremath{1.6 \cdot 10^{-8}}}%
  {\cite{Hayasaka:2010np,Lees:2010ez}}%
\htCombLimitLine%
  {\ensuremath{\Gamma_{180} = \mu^- e^+ \mu^-}}%
  {\ensuremath{}}%
  {\ensuremath{9.8 \cdot 10^{-9}}}%
  {\cite{Hayasaka:2010np,Lees:2010ez}}%
\htCombLimitLine%
  {\ensuremath{\Gamma_{181} = \mu^- e^+ e^-}}%
  {\ensuremath{}}%
  {\ensuremath{1.1 \cdot 10^{-8}}}%
  {\cite{Hayasaka:2010np,Lees:2010ez}}%
\htCombLimitLine%
  {\ensuremath{\Gamma_{182} = e^- \mu^+ e^-}}%
  {\ensuremath{}}%
  {\ensuremath{8.4 \cdot 10^{-9}}}%
  {\cite{Hayasaka:2010np,Lees:2010ez}}%
\htCombLimitLine%
  {\ensuremath{\Gamma_{183} = \mu^- \mu^+ \mu^-}}%
  {\ensuremath{}}%
  {\ensuremath{1.2 \cdot 10^{-8}}}%
  {\cite{Hayasaka:2010np,Lees:2010ez,Aaij:2014azz}}}%

\begin{fleqn}
\let\tausection\subsection
\let\tausubsection\subsubsection
\newenvironment{envsmall}%
  {\small}%
  {}
\section{Tau lepton properties}
\label{sec:tau}
We present averages of a selection of \mtau lepton quantities with
the goal to provide the best tests of the
universality of the charged-current weak interaction
(Section~\ref{sec:tau:leptonuniv}) and of the Cabibbo-Kobayashi-Maskawa
(CKM) matrix coefficient \Vus from \mtau decays
(Section~\ref{sec:tau:vus}).  We focus on the averages
that benefit most from the adoption of the HFLAV
methodology~\cite{Asner:2010qj}, namely a global fit of the \mtau branching
fractions that best exploits the available experimental information.
Since the 2016 edition,
the HFLAV-Tau group has collaborated to the determination of the
$\tau$-lepton branching fractions based on a global fit and to the
related mini-review that are included in the ``Review of particle
physics''~\cite{PDG_2016}. The differences between the PDG 2016 fit and the fit
presented here are detailed in Section~\ref{sec:tau:diffpdg}.

All relevant published statistical correlations are used, and a selection of
measurements, particularly the most precise and the most recent ones, was
studied to take into account the significant systematic dependencies from
external parameters and common sources of systematic uncertainty.

Finally, we report in Section~\ref{sec:tau:lfv-limits} the latest limits
on the lepton-flavour-violating \mtau branching fractions and in
Section~\ref{sec:tau:lfv-combs} we determine
the combined upper limits for the branching
fractions that have multiple experimental results.

The $\tau$ lepton results are obtained from inputs available through
summer 2016 and have been published on the web in 2016 with the label
``Summer 2016''. However, there have been minor revisions since then,
and we have updated tables and plots in this report with the label
``Spring 2017''.

%% ///////////////////////////////////////////////////////////////////////////
\tausection{Branching fraction fit}
\cutname{br-fit.html}
\label{sec:tau:br-fit}

A global fit of the available experimental measurements is used to
determine the \mtau branching fractions, together with their
uncertainties and statistical correlations.
The \mtau branching fractions provide a test for theory predictions based
on the Standard Model (SM) EW and QCD interactions and can be further
elaborated to test the EW charged-current universality for leptons, to
determine the CKM matrix coefficient \Vus (and the QCD coupling constant
$\alpha_s$ at the \mtau mass). 

The measurements used in the fit are listed in
Table~\ref{tab:tau:br-fit} and consist of either \mtau decay
branching fractions, labelled as $\Gamma_{i}$, or ratios of two \mtau
decay branching fractions, labelled as $\Gamma_{i}/\Gamma_{j}$. A
minimum \chisq fit is performed for all the measured quantities and for
some additional branching fractions and ratios of branching fractions, 
and all fit results are listed in Table~\ref{tab:tau:br-fit}. Some fitted
quantities are equal to the ratio of two other fitted quantities, as
documented with the notation $\Gamma_{i}/\Gamma_{j}$ in
Table~\ref{tab:tau:br-fit}. Some fitted quantities are sums of other
fitted quantities, for instance $\htuse{Gamma8.gn} =
\BR(\tau\to\htuse{Gamma8.td})$ is the sum of $\htuse{Gamma9.gn} =
\BR(\tau\to\htuse{Gamma9.td})$ and $\htuse{Gamma10.gn} =
\BR(\tau\to\htuse{Gamma10.td})$. The symbol $h$ is used to mean either a
$\pi$ or $K$. Section~\ref{sec:tau:constraints} lists all equations
relating one quantity to the sum of other quantities. In the following,
we refer to both types of relations between fitted quantities
collectively as constraint equations or constraints. The fit \chisq is
minimized subject to all the above mentioned constraints, listed in
Table~\ref{tab:tau:br-fit} and Section~\ref{sec:tau:constraints}. The fit
procedure is equivalent to that employed in the previous
HFLAV reports~\cite{Asner:2010qj,Amhis:2012bh,Amhis:2014hma}.

\tausubsection{Technical implementation of the fit procedure}
\label{sec:tau:fit-techn}%
\newcommand{\htvarE}{V}%
\newcommand{\htvarm}{x}%

The fit computes the quantities $q_i$ by minimizing a $\chi^2$
while respecting a series of equality constraints on the $q_i$.
The $\chi^2$ is computed using the measurements $\htvarm_i$ and their
covariance matrix $\htvarE_{ij}$ as
\begin{align}
\chi^2 = (\htvarm_i - A_{ik}q_k)^t \htvarE_{ij}^{-1} (\htvarm_j - A_{jl}q_l)~,\label{eq:tau:chisq}
\end{align}
where the model matrix $A_{ij}$ is used to get the
vector of the predicted measurements $\htvarm_i^\prime$ from the vector of the
fit parameters $q_j$ as $\htvarm_i^\prime = A_{ij}q_j$. In this
particular implementation, the measurements are grouped according to
the measured quantity, and all quantities with at least one measurement
correspond to a fit parameter. Therefore, the
matrix $A_{ij}$ has one row per
measurement $\htvarm_i$ and one column per fitted
quantity $q_j$, with unity coefficients for the rows and column that
identify a measurement $\htvarm_i$ of the quantity $q_j$.
In summary, the \chisq given in Eq.~(\ref{eq:tau:chisq}) is minimized subject 
to the constraints
\begin{align}
  f_r(q_s) - c_r = 0~, \label{eq1;b}
\end{align}
where Eq.~(\ref{eq1;b}) corresponds to the constraint equations, written
as a set of ``constraint expressions'' that are equated to zero.
Using the method of Lagrange multipliers, a set of equations is obtained by
taking the derivatives with respect to the fitted quantities $q_k$ and the
Lagrange multipliers $\lambda_r$ of the sum of the $\chi^2$ and the
constraint expressions multiplied by the
Lagrange multipliers $\lambda_r$, one for each constraint:
\begin{align}
&\text{min} \left[ (A_{ik}q_k {-} \htvarm_i)^t \htvarE_{ij}^{-1} (A_{jl}q_l {-} \htvarm_j) +
2\lambda_r(f_r(q_s) - c_r) \right] \label{eq2;a}
\\
&(\partial/\partial q_k, \partial/\partial \lambda_r) \, [\text{expression above}] = 0~. \label{eq2;b}
\end{align}
%%vector of unknowns:   g_j = (f_j, l_k)
%%vector of knowns:     v_i = (\htvarm_i, c_k)
Equation~(\ref{eq2;b}) defines a set of equations for the vector of the unknowns $(q_k,
\lambda_r)$, some of which may be non-linear, in case of non-linear
constraints. An iterative minimization procedure approximates at
each step the non-linear constraint expressions by their first order
Taylor expansion around the current values of the fitted quantities,
$\bar{q}_s$:
\begin{align}
f_r(q_s) - c_r \simeq
f_r(\bar{q}_s) + \left.\frac{\partial f_r(q_s)}{\partial q_s}\right|_{\bar{q}_s} (q_s - \bar{q}_s) - c_r~, \label{eq3;a}
\end{align}
which can be written as
\begin{align}
B_{rs} q_s - c^\prime_r~, \label{eq3;b}
\end{align}
where $c^\prime_r$ are the resulting constant known terms, independent
of $q_s$ at first order. After linearization, the differentiation by
$q_k$ and $\lambda_r$ is trivial and leads to a set of linear equations
\begin{align}
&A_{ki}^t \htvarE_{ij}^{-1} A_{jl} q_l + B^t_{kr} \lambda_r =  A_{ki}^t \htvarE_{ij}^{-1} \htvarm_j \label{eq4;a}
\\
&B_{rs} q_s = c^\prime_r~, \label{eq4;b}
\end{align}
which can be expressed as:
\begin{align}
F_{ij} u_j = v_i~, \label{eq5}
\end{align}
where $u_j = (q_k, \lambda_r)$ and $v_i$ is the vector of the known
constant terms running over the index $k$ and then $r$ in the right
terms of Eq.~(\ref{eq4;a}) and Eq.~(\ref{eq4;b}). Solving the equation
set in Eq.~(\ref{eq5}) gives the fitted quantities and
their covariance matrix, using the measurements and their
covariance matrix.
%%%Once all equations are linear, the unknowns are obtained with the
%%%straightforward linear algebra solution for $n$ equations in $n$
%%%unknowns as $(q_i, \lambda_j) = S (\htvarm_k, c_l$), where $c_l$ are the
%%%constant terms of the constraint equations. The variance and covariance
%%%matrix of the fitted quantities is obtained using the same matrix $S$
%%%and the measurements variance and covariance.
The fit procedure starts by computing the linear
approximation of the non-linear constraint
expressions around the quantities seed values. With an
iterative procedure, the unknowns are updated at each step by solving
the equations and the equations are then linearized around the updated
values, until the RMS average of relative variation of the
fitted unknowns is reduced below $10^{-12}$.

\tausubsection{Fit results}
\label{sec:tau:fit}

The fit output consists of \htuse{QuantNum} fitted quantities that correspond to
either branching fractions or ratios of branching
fractions. The fitted quantities values and uncertainties are listed in
Table~\ref{tab:tau:br-fit}. The off-diagonal statistical correlation terms
between a subset of \htuse{BaseQuantNum} ``basis quantities'' are listed
in Section~\ref{sec:tau:fitcorr}. All the remaining statistical
correlation terms can be obtained using the constraint equations listed
in Table~\ref{tab:tau:br-fit} and Section~\ref{sec:tau:constraints}.

\iffalse
Although the fit treats all quantities in the same way, for the
purpose of describing the results, we divide them in a set
of \htuse{BaseQuantNum} ``basis nodes'' that permit the definition of all
the remaining ones using the relations listed in Section~\ref{sec:tau:constraints} and
Table~\ref{tab:tau:br-fit}.

Furthermore we define (see Section~\ref{sec:tau:constraints}) $\Gamma_{110}
= \BR(\tau^- \to X_s^- \nu_\tau)$, the total branching fraction of the \mtau
decays to final states with the strangeness quantum number equal to one, and
$\Gamma_{\text{All}}$, the branching fraction of the \mtau into any
measured final state, which should be equal to $1$ within the
experimental uncertainty. We define the unitarity residual as $\Gamma_{998}
= 1 -\Gamma_{\text{All}}$.
\fi

The fit has $\chi^2/\text{d.o.f.} = \htuse{Chisq}/\htuse{Dof}$,
corresponding to a confidence level $\text{CL} = \htuse{ChisqProb}$. We use a total of
\htuse{MeasNum} measurements to fit the above mentioned
\htuse{QuantNum} quantities subjected to \htuse{ConstrNum} constraints.
Although the unitarity constraint is not applied, the fit is statistically
consistent with unitarity, where the residual is
\htuse{Gamma998.gn} =  \htuse{Gamma998.td} = \htuse{Gamma998}.

\label{ref:tau:kkk-error-scale-factor}%
A scale factor of 5.44 (as in the three previous
reports~\cite{Asner:2010qj,Amhis:2012bh,Amhis:2014hma}) has been applied
to the published uncertainties of the two severely inconsistent
measurements of \(\Gamma_{96} = \tau \to KKK\nu\) by \babar and
Belle. The scale factor has been determined using the PDG procedure,
\ie, to the proper size in order to obtain a reduced \chisq equal to $1$
when fitting just the two \(\Gamma_{96}\) measurements.

For several old results, for historical reasons, 
the table reports the total error (statistical plus systematic) in the
position of the statistical error and zero in the position of the
systematic error. Since the fit depends only on the total errors, the
results are unaffected.

%% ///////////////////////////////////////////////////////////////////////////
\tausubsection{Changes with respect to the previous report}

The following changes have been introduced with respect to the previous
HFLAV report~\cite{Amhis:2014hma}.

Two old preliminary results have been removed:
\begin{itemize}
\item $\Gamma_{35} = \BR(\tau\to\pi K_S\nu)$, \babar~\cite{Aubert:2008an},
\item $\Gamma_{40} = \BR(\tau\to\pi K_S\pi^0\nu)$, \babar~\cite{Paramesvaran:2009ec}.
\end{itemize}
They were announced in 2008 and 2009 but have not been published.

In the 2014 report, for several \babar and Belle experimental results we
used more precise numerical values than the published ones, using
internal information from the Collaborations.
We revert to the published figures in this
report, as the improvements in the fit results were negligible. In so
doing, we use in this report the same values that are used in the PDG
2016 fit.

The Belle result on $\tau^- \to \htuse{Gamma33.td}$~\cite{Ryu:2014vpc}
has been discarded, because it was determined that the published
information does not permit a reliable determination of the correlations
with the other results in the same paper. The correlations estimated
for the HFLAV 2014 report were inconsistent. As a result, both the
covariance matrix of the Belle results and the overall correlation
matrix for the branching ratio fit results were non-positive-definite.
It has been found that the inconsistency had negligible impact on lepton
universality tests and on the \Vus measurements.

\iffalse
The ALEPH measurement on \htuse{Gamma49.gn} ($\tau^- \to
\htuse{Gamma49.td}$)~\cite{Barate:1999hj} has been added. It is
computed by ALEPH as the sum of $\htuse{Gamma51.gn} =
\htuse{Gamma51.td}$, also measured by ALEPH~\cite{Barate:1999hj}, 
$\htuse{Gamma50.gn} = \htuse{Gamma50.td}$, which was measured by ALEPH
as not being significant above zero~\cite{Barate:1999hj}, and
$\htuse{Gamma806.gn} = \htuse{Gamma806.td}$. It is assumed that
\htuse{Gamma806.gn} and \htuse{Gamma50.gn} are equal.
By using in the fit the measurements of \htuse{Gamma49.gn} and
\htuse{Gamma51.gn} by ALEPH, with a proper statistical correlation
coefficient to account for the determination of \htuse{Gamma49.gn},
the two measurements contribute to determine also \htuse{Gamma50.gn}
and \htuse{Gamma806.gn}.
\fi

The ALEPH result on \htuse{Gamma46.gn} ($\tau^- \to
\htuse{Gamma46.td}$)~\htuse{ALEPH.Gamma47.pub.BARATE.98E,ref} has been
removed from the fit inputs, since it is simply the sum of twice
$\htuse{Gamma47.gn} = \htuse{Gamma47.td}$ and
$\htuse{Gamma48.gn} = \htuse{Gamma48.td}$ from the same paper, hence
100\% correlated with them.

Several minor corrections have been applied to the constraints. The list of
constraints included in the following fully documents the changes when
compared with the same list in the 2014 edition. In some cases the
relation equating one decay mode to a sum of modes included some minor
terms that did not match the mode definitions. In other cases, the sum
included modes with overlapping components. The effects on the 2014 fit
results have been found to be modest with respect to the quoted
uncertainties.
For instance, the definition of the total branching fraction has been
updated as follows:
\vspace{2.5ex}

\begin{tabularx}{\linewidth-\parindent}{@{}lX@{}}
  \htuse{GammaAll.c.constr.eq}~.
\end{tabularx}
\vspace{2.0ex}

\noindent In the 2014 definition, the term $\htuse{Gamma78.gn} =
\htuse{Gamma78.td}$ included the contributions of $\htuse{Gamma50.gn} =
\htuse{Gamma50.td}$ and $\htuse{Gamma132.gn} =
\htuse{Gamma132.td}$, which were already included explicitly in
\htuse{GammaAll.gn}. In the present
definition, \htuse{Gamma78.gn} has been replaced with modes whose sum
corresponds to 
\begin{align*}
  \htuse{Gamma810.gn} = \htuse{Gamma810.td}~.
\end{align*}
As in 2014, the total \mtau branching fraction \htuse{GammaAll.gn} definition
includes two modes that have overlapping final states, to a minor
extent, which we consider negligible:
\begin{align*}
  &\htuse{Gamma50.gn} =  \htuse{Gamma50.td} \\
  &\htuse{Gamma132.gn} =  \htuse{Gamma132.td}~.
\end{align*}

\noindent Finally, we updated to the PDG 2015 results~\cite{PDG_2014} all the
parameters corresponding to the measurements' systematic biases and
uncertainties and all the parameters appearing in the constraint
equations in Section~\ref{sec:tau:constraints} and
Table~\ref{tab:tau:br-fit}.

%% ///////////////////////////////////////////////////////////////////////////
\tausubsection{Differences between the HFLAV Spring 2017 fit and the PDG 2016
  fit}
\label{sec:tau:diffpdg}

As is standard for the PDG branching fraction fits, the PDG 2016 \mtau branching
fraction fit is unitarity constrained, while the HFLAV 2016 fit is
unconstrained.

The HFLAV-Tau fit uses the ALEPH measurements of branching
fractions defined according to the final state content of ``hadrons'' and
kaons, where a ``hadron'' corresponds to either a pion or a kaon, since
this set of results is closer to the actual experimental measurements
and facilitates a more
comprehensive treatment of the experimental results correlations~\cite{Asner:2010qj}.
The PDG 2016 fit on the other hand
continues to use -- as in the past editions -- the ALEPH measurements
of modes with pions and kaons, which correspond
to the final set of published measurements of the collaboration.
\iffalse
, but neglect some
correlation between the pions results in one paper and the kaon
results in another paper, because the pion branching fractions are
computed by subtracting the kaon contributions from the hadron modes.
\fi
It is planned eventually to update the PDG fit to use the same ALEPH
measurement set that is used by HFLAV.

\iffalse
The HFLAV Spring 2017 fit includes two preliminary results that are not
included in the PDG 2016 fit:
\begin{align*}
  \begin{tabular}{@{}lll}
    \htuse{BaBar prelim. ICHEP08.meas} & \htuse{BaBar prelim. ICHEP08.cite}~, \\
    \htuse{BaBar prelim. DPF09.meas} & \htuse{BaBar prelim. DPF09.cite}~.
  \end{tabular}
\end{align*}
\fi

The HFLAV Spring 2017 fit, as in 2014, uses the ALEPH estimate for
$\htuse{Gamma805.gn} =  \BR(\tau\to\htuse{Gamma805.td})$, which is not a direct
measurement. The PDG 2016 fit uses the PDG average of
$\BR(a_1\to\pi\gamma)$ as a parameter and defines $\htuse{Gamma805.gn}
= \BR(a_1\to\pi\gamma)\times\BR(\tau \to 3\pi \nu)$. As a consequence, the PDG fit
procedure does not take into account the large uncertainty on $\BR(a_1\to\pi\gamma)$,
resulting in an underestimated fit uncertainty on \htuse{Gamma805.gn}.
Therefore, in this case an appropriate correction has to be applied after the fit.

\iffalse
Unlike this edition of the HFLAV-Tau fit, the PDG 2016 fit does not
include the  \htuse{Gamma49.gn} ($\tau^- \to
\htuse{Gamma49.td}$)~\cite{Barate:1999hj} measurement. Its inclusion
would have required adding a proper correlation term in the PDG tables,
which has been postponed.
\fi

%% ///////////////////////////////////////////////////////////////////////////
\tausubsection{Branching ratio fit results and experimental inputs}
\label{sec:tau:br-fit-results}

Table~\ref{tab:tau:br-fit} reports the \mtau branching ratio fit results
and experimental inputs.

%%
%% quantities and measurements
%%
\begin{center}
\begin{envsmall}
\setlength{\LTcapwidth}{0.85\linewidth}
\renewcommand*{\arraystretch}{1.3}%
\ifhevea
\renewcommand{\bar}[1]{\textoverline{#1}}
\else
\begin{citenoleadsp}
\fi
\begin{longtable}{llll}
\caption{HFLAV \hfagTauTag branching fractions fit results.\label{tab:tau:br-fit}}%
\\
\toprule
\multicolumn{1}{l}{\bfseries \mtau lepton branching fraction} &
\multicolumn{1}{l}{\bfseries Fit value / Exp.} &
\multicolumn{1}{l}{\bfseries HFLAV Fit / Ref.} \\
\midrule
\endfirsthead
\multicolumn{4}{c}{{\bfseries \tablename\ \thetable{} -- continued from previous page}} \\ \midrule
\multicolumn{1}{l}{\bfseries \mtau lepton branching fraction} &
\multicolumn{1}{l}{\bfseries Fit value / Exp.} &
\multicolumn{1}{l}{\bfseries HFLAV Fit / Ref.} \\
\midrule
\endhead
\endfoot
\endlastfoot
\htuse{BrVal} \\
\bottomrule
\end{longtable}
\ifhevea\else
\end{citenoleadsp}
\fi
\end{envsmall}
\end{center}

\tausubsection{Correlation terms between basis branching fractions uncertainties}
\label{sec:tau:fitcorr}

The following tables report the correlation coefficients between basis quantities,
in percent.

%%
%% basis quantities correlations
%%
\htuse{BrCorr}

\tausubsection{Equality constraints}
\label{sec:tau:constraints}

We list in the following the equality constraints that relate a branching fraction to a sum of
branching fractions.
The constraint equations include as coefficients the
values of some non-tau branching fractions, denoted \eg, with the
self-describing notation $\Gamma_{K_S \to \pi^0\pi^0}$. Some coefficients
are probabilities corresponding to modulus square amplitudes describing quantum
mixtures of states such as $K^0$, $\bar{K}^0$, $K_S$, $K_L$, denoted with
\eg, $\Gamma_{<K^0|K_S>} = |{<}K^0|K_S{>}|^2$.
All non-tau quantities are taken from the PDG 2015~\cite{PDG_2014}
fits (when available) or averages, and are used without accounting for their
uncertainties, which are however in general small with respect
to the uncertainties on the \mtau branching fractions.

The following list does not include the constraints listed in
Table~\ref{tab:tau:br-fit}, where some measured ratios of branching
fractions are expressed as ratios of two branching fractions.

\begin{envsmall}
  \setlength\abovedisplayskip{0pt}
  \setlength\belowdisplayshortskip{0pt}
  \ifhevea\renewcommand{\bar}[1]{\textoverline{#1}}\fi
  %%
  %% after editing content of \htuse{ConstrVal} macro for better formatting
  %%
  \htuse{ConstrEqs}
\end{envsmall}

\iffalse
%% actually, values and uncertainties are listed
%% one needs the constraints only for the non diagonal terms
%% of the correlation matrix when one non-basis mode is involved
The \mtau branching fractions are listed in Table~\ref{tab:tau:br-fit}.
The equations in the following permit the computation of the values and
uncertainties for branching fractions that are not listed in
Table~\ref{tab:tau:br-fit}, once they are expressed as function of the
quantities that are listed there. 
\fi

%% ///////////////////////////////////////////////////////////////////////////
\tausection{Tests of lepton universality}
\cutname{lepton-univ.html}
\label{sec:tau:leptonuniv}

Lepton universality tests probe the Standard Model prediction that the
charged weak current interaction has the same coupling for all lepton generations.
The precision of such tests has been significantly improved since the
2014 edition by the
addition of the Belle \mtau lifetime
measurement~\cite{Belous:2013dba}, while improvements from the \mtau
branching fraction fit are negligible.
We compute the universality tests as in the previous report by using
ratios of the partial widths of a heavier lepton $\lepth$
decaying to a lighter lepton $\leptl$~\cite{Marciano:1988vm},
\begin{align*}
  \Gamma(\lepth \to \nu_{\lepth} \leptl \nub_{\leptl} (\gamma)) =
  \frac{\BR(\lepth \to \nu_{\lepth} \leptl \nub_{\leptl})}{\tau_{\lepth}} =
  \frac {G_{\lepth} G_{\leptl} m^5_{\lepth}}{192 \pi^3}\, f\left(\frac {m^2_{\leptl}}{m^2_{\lepth}}\right)
  \radRatio^{\lepth}_W \radRatio^{\lepth}_\gamma~,
\end{align*}
where
\begin{alignat*}{3}
 G_{\leptl} &= \frac {g^2_{\leptl}}{4 \sqrt{2} M^2_W}~, &\quad&&
 f(x) &= 1 -8x +8x^3 -x^4 -12x^2 \text{ln}x~, \\
 \radRatio^{\lepth}_W &= 1 + \frac {3}{5} \frac {m^2_{\lepth}}{M^2_W} + \frac {9}{5} \frac {m^2_{\leptl}}{M^2_W}~
\cite{Pich:2013lsa, Ferroglia:2013dga, Fael:2013pja}, &\quad\quad&&
 \radRatio^{\lepth}_\gamma &= 1+\frac {\alpha(m_{\lepth})}{2\pi} \left(\frac {25}{4}-\pi^2\right)~.
\end{alignat*}
We use $\radRatio^\tau_\gamma=1-43.2\cdot 10^{-4}$ and
$\radRatio^\mu_\gamma=1-42.4\cdot 10^{-4}$~\cite{Marciano:1988vm} and $M_W$
from PDG 2015~\cite{PDG_2014}.
We use HFLAV Spring 2017 averages and PDG 2015 for the other quantities.
Using pure leptonic processes we obtain
\begin{align*}
  \left( \frac{g_\tau}{g_\mu} \right) = \htuse{gtaubygmu_tau}~,
  && \left( \frac{g_\tau}{g_e} \right) = \htuse{gtaubyge_tau}~,
  && \left( \frac{g_\mu}{g_e} \right) = \htuse{gmubyge_tau}~.
\end{align*}
Using the expressions for the \mtau semi-hadronic partial
widths, we obtain
\begin{align*}
  \left( \frac{g_\tau}{g_\mu} \right)^2 =
  \frac{\BR({\tau \to h \nu_\tau})}{\BR({h \to \mu \bar{\nu}_\mu})}
  \frac{2m_h m^2_{\mu}\tau_h}{(1 + \dRradTauhHmu)m^3_{\tau}\tau_{\tau}}
  \left( \frac{1-m^2_{\mu}/m^2_h}{1-m^2_h/m^2_{\tau}} \right)^2~,
\end{align*}
where $h$ = $\pi$ or $K$ and the radiative corrections are
$\dRradTaupiPimu = (\htuse{dRrad_taupi_by_pimu})\%$ and
$\dRradTaukKmu = (\htuse{dRrad_tauK_by_Kmu})\%$~\cite{Marciano:1993sh,Decker:1994dd,Decker:1994ea,Decker:1994kw}.
We measure:
\begin{align*}
  \left( \frac{g_\tau}{g_\mu} \right)_\pi &= \htuse{gtaubygmu_pi}~,
  & \left( \frac{g_\tau}{g_\mu} \right)_K = \htuse{gtaubygmu_K}~.
\end{align*}
Similar tests could be performed with decays to electrons, however they are
less precise because the hadron two body decays to electrons are
helicity-suppressed.
Averaging the three \(g_\tau/g_\mu\) ratios we obtain
\begin{align*}
  \left( \frac{g_\tau}{g_\mu} \right)_{\tau{+}\pi{+}K} &= \htuse{gtaubygmu_fit}~,
\end{align*}
accounting for statistical correlations.
Table~\ref{tab:tau:univ-fit-corr} reports the statistical correlation coefficients for the fitted coupling ratios.
\ifhevea\begin{table}\fi%% otherwise cannot have normalsize caption
\begin{center}
\ifhevea
\caption{Universality coupling ratios correlation coefficients (\%).\label{tab:tau:univ-fit-corr}}%
\else
\begin{minipage}{\linewidth}
\begin{center}
\captionof{table}{Universality coupling ratios correlation coefficients (\%).}\label{tab:tau:univ-fit-corr}%
\fi
\begin{center}
\renewcommand*{\arraystretch}{1.1}%
\begin{tabular}{lcccc}
\toprule
\htuse{couplingsCorr}
\\\bottomrule
\end{tabular}
\end{center}
\ifhevea\else
\end{center}
\end{minipage}
\fi
\end{center}
\ifhevea\end{table}\fi
\noindent Since there is 100\% correlation between $g_\tau/g_\mu$,
$g_\tau/g_e$ and $g_\mu/g_e$, the correlation matrix is expected to be
positive semi-definite, with one eigenvalue equal to zero. Due to
numerical inaccuracies, one eigenvalue is expected to be close to zero
rather than exactly zero.

%% ///////////////////////////////////////////////////////////////////////////
\tausection{Universality improved $\BR(\tau \to e \nu \bar{\nu})$ and $\Rhad$}
\cutname{Be_univ_and_Rtau.html}
\label{sec:tau:be-univ-rtau}

We compute two quantities that are used in this report and that have been
traditionally used for further elaborations and
tests involving the \mtau branching fractions:
\begin{itemize}
  
\item the ``universality improved'' experimental
  determination of $\BR_e = \BR(\tau \to e \nu \bar{\nu})$, which relies on the assumption
  that the Standard Model and lepton universality hold;

\item the ratio \Rhad between the
total branching fraction of the \mtau to hadrons and the universality
improved $\BR_e$, which is the same as the ratio of the two respective
partial widths.
\end{itemize}

Following Ref.~\cite{Davier:2005xq}, we obtain a more precise experimental
determination of $\BR_e$ using the
\mtau branching fraction to $\mu \nu \bar{\nu}$, $\BR_\mu$, and the \mtau lifetime. We average:
\begin{itemize}

\item the $\BR_e$ fit value \htuse{Gamma5.gn},

\item
  the $\BR_e$ determination from the $\BR_\mu = \BR(\tau \to \mu \nu
  \bar{\nu})$ fit value \htuse{Gamma3.gn} assuming that $g_\mu/g_e = 1$,
  hence (see also Section~\ref{sec:tau:leptonuniv}) $\BR_e = \BR_\mu \cdot
  f(m^2_e/m^2_\tau)/f(m^2_\mu/m^2_\tau)$,

\item
  the $\BR_e$ determination from the \mtau lifetime assuming that
  $g_\tau/g_\mu =1$, hence $\BR_e = \BR(\mu \to e \bar{\nu}_e
  \nu_\mu)\cdot (\tau_\tau / \tau_\mu) \cdot (m_\tau/m_\mu)^5 \cdot
  f(m^2_e/m^2_\tau)/f(m^2_e/m^2_\mu) \cdot (\delta_\gamma^\tau
  \delta_W^\tau)/(\delta_\gamma^\mu \delta_W^\mu)$ where $\BR(\mu \to e
  \bar{\nu}_e \nu_\mu) = 1$.

\end{itemize}
Accounting for statistical correlations, we obtain
\begin{align*}
  \BR_e^{\text{uni}} = (\htuse{Be_univ})\%.
\end{align*}
We use $\BR_e^{\text{uni}}$ to obtain the ratio
\begin{align*}
  \Rhad = \frac{\Gamma(\tau \to \text{hadrons})}{\Gamma(\tau\to
    e\nu\bar{\nu})} = \frac{\Gamma_{\text{hadrons}}}{\BR_e^{\text{uni}}} =
  \htuse{R_tau},
\end{align*}
where $\Gamma(\tau \to \text{hadrons})$ and $\Gamma(\tau\to e\nu\bar{\nu})$
indicate the partial widths and $\Gamma_{\text{hadrons}}$ is the total
branching fraction of the \mtau to hadrons, or the total branching fraction
in any measured final state minus the leptonic branching fractions, \ie,
with our notation $\Gamma_{\text{hadrons}} = \htuse{GammaAll.gn} -
\htuse{Gamma3.gn} - \htuse{Gamma5.gn} = (\htuse{B_tau_had_fit})\%$ (see
Section~\ref{sec:tau:br-fit} and Table~\ref{tab:tau:br-fit} for the
definitions of \htuse{GammaAll.gn}, \htuse{Gamma3.gn}, \htuse{Gamma5.gn}).
We underline that this report's definition of $\Gamma_{\text{hadrons}}$
corresponds to summing all \mtau hadronic decay modes, like in the previous
report, rather than -- as done elsewhere -- subtracting the leptonic
branching fractions from unity, \ie, $\Gamma_{\text{hadrons}} = 1 -
\htuse{Gamma3.gn} - \htuse{Gamma5.gn}$.

%% ///////////////////////////////////////////////////////////////////////////
\tausection{$\Vus$ measurement}
\cutname{vus.html}
\label{sec:tau:vus}

The CKM matrix element magnitude \Vus is most precisely determined from kaon
decays~\cite{Antonelli:2010yf} (see Figure~\ref{fig:tau:vus-summary}),
and its precision is limited by the
uncertainties of the lattice QCD estimates of the meson decay constants $f_+^{K\pi}(0)$ and $f_K/f_\pi$.
Using the \mtau branching fractions, it is possible to determine \Vus in an
alternative way~\cite{Gamiz:2002nu,Gamiz:2004ar} that does not depend on lattice QCD and
has small theory uncertainties (as discussed in Section~\ref{sec:tau:vus:incl}).
Moreover, \Vus can be determined using the \mtau branching fractions
similarly to the kaon case, using the same meson decay constants from Lattice QCD.

\iffalse
In the following Sections~\ref{sec:tau:vus:incl},
\ref{sec:tau:vus:taukpi} and \ref{sec:tau:vus:tauk} we update the CKM
coefficient \Vus determinations
that were shown in the previous report using the 2015 determination of
\Vud~\cite{Hardy:2014qxa} and the updated averages from HFLAV Sprint 2017 and PDG
2015 for the other quantities.
\fi

%% ///////////////////////////////////////////////////////////////////////////
\tausubsection{\Vus from $\BR(\tau \to X_s\nu)$}
\label{sec:tau:vus:incl}

The \mtau hadronic partial width is the sum of the \mtau partial widths to
strange and to non-strange hadronic final states, $\Gammahad =
\Gammastrange + \Gammanonstrange$.  The suffix ``VA'' traditionally denotes
the sum of the \mtau partial widths to non-strange final states, which proceed
through either vector or axial-vector currents.

Dividing any partial width $\Gamma_x$
by the electronic partial width, $\Gamma_e$, we obtain partial width ratios
$R_x$ (which are equal to the respective branching fraction ratios
$\BR_x/\BR_e$) for which $\Rhad = \Rstrange + \Rnonstrange$. In terms of
such ratios, \Vus can be measured as~\cite{Gamiz:2002nu,Gamiz:2004ar}
\begin{align*}
  \VusTauIncl &= \sqrt{\Rstrange/\left[\frac{\Rnonstrange}{\Vud^2} -  \delta R_{\text{theory}}\right]}~,
\end{align*}
where $\delta R_{\text{theory}}$ can be determined in the context of low
energy QCD theory, partly relying on experimental low energy scattering
data. The literature reports several
calculations~\cite{Gamiz:2006xx,Gamiz:2007qs,Maltman:2010hb}. In this
report we use Ref.~\cite{Gamiz:2006xx}, whose estimated uncertainty size is
intermediate between the two other ones. We use the information in that paper and the
PDG 2015 value for the $s$-quark mass $m_s = \htuse{m_s}\,\text{MeV}$~\cite{PDG_2014}
to calculate $\delta R_{\text{theory}} = \htuse{deltaR_su3break}$.

We proceed following the same procedure of the 2012 HFLAV
report~\cite{Amhis:2012bh}. We sum the relevant \mtau branching
fractions to compute $\BRnonstrange$ and $\BRstrange$ and we use the
universality improved $\BR_e^{\text{uni}}$ (see
Section~\ref{sec:tau:be-univ-rtau}) to compute the $\Rnonstrange$ and $\Rstrange$  ratios.
In past determinations of \Vus, for example in the 2009 HFLAV
report~\cite{Asner:2010qj}, the total hadronic branching fraction
has been computed using unitarity as $\BRhad^{\text{uni}} = 1 - \BR_e
-\BR_\mu$, obtaining then $\BRstrange$ from the sum of the strange
branching fractions and $\BRnonstrange$ from $\BRhad^{\text{uni}} -
\BRstrange$. We prefer to use the more direct experimental determination
of $\BRnonstrange$ for two reasons.
First, both methods result in comparable uncertainties on \Vus,
since the better precision on  $\BRhad^{\text{uni}} = 1 - \BR_e
-\BR_\mu$ is vanified by increased statistical correlations in the
expressions $(1-\BR_e -\BR_\mu)/\BR_e^{\text{univ}}$ and
$\BRstrange/(\BRhad-\BRstrange)$ in the \Vus calculation. Second, if
there are unobserved \mtau hadronic decay modes, they would affect
\BRnonstrange and \BRstrange in a more asymmetric way when using
unitarity.

Using the \mtau branching fraction fit results with their uncertainties
and correlations (Section~\ref{sec:tau:br-fit}), we compute $\BRstrange =
(\htuse{B_tau_s_fit})\%$ (see also Table~\ref{tab:tau:vus}) and
$\BRnonstrange = \BR_{\text{hadrons}} - \BRstrange=
(\htuse{B_tau_VA_fit})\%$, where $\BR_{\text{hadrons}}$ is equal to
$\Gamma_{\text{hadrons}}$ defined in section~\ref{sec:tau:be-univ-rtau}. PDG 2015 averages
are used for non-\mtau quantities, and $\Vud =
\htuse{Vud}$~\cite{Hardy:2014qxa}.

We obtain $\VusTauIncl = \htuse{Vus}$, which
is $\htuse{Vus_mism_sigma_abs}\sigma$ lower than the unitarity CKM
prediction $\VusUni = \htuse{Vus_uni}$, from $(\VusUni)^2 = 1 -
\Vud^2$. The \VusTauIncl uncertainty includes a systematic error
contribution of \htuse{Vus_err_th_perc}\% from the theory uncertainty on
$\delta R_{\text{theory}}$. There is no significant change with respect to
the previous HFLAV report.

%%
%% Gamma110 quantities
%%
\begin{table}
\begin{center}
\renewcommand*{\arraystretch}{1.3}%
\caption{HFLAV \hfagTauTag \mtau branching fractions to strange final states.\label{tab:tau:vus}}%
\ifhevea\renewcommand{\bar}[1]{\textoverline{#1}}\fi
\begin{envsmall}
\begin{center}
\begin{tabular}{lc}
\toprule
\multicolumn{1}{l}{\bfseries Branching fraction} &
\multicolumn{1}{c}{\bfseries HFLAV \hfagTauTag fit (\%)} \\
\midrule
\htuse{BrStrangeVal}
\midrule
\htuse{BrStrangeTotVal}
\bottomrule
\end{tabular}
\end{center}
\end{envsmall}
\end{center}
\end{table}

\tausubsection{\Vus from $\BR(\tau \to K\nu) / \BR(\tau \to \pi\nu)$}
\label{sec:tau:vus:taukpi}

We compute \Vus from
the ratio of branching fractions $\BR(\tau \to \htuse{Gamma10.td}) / \BR(\tau \to \htuse{Gamma9.td}) =
\htuse{Gamma10by9}$
from the equation~\cite{Pich:2013lsa}:
\begin{align*}
\frac{\BR(\tau \to \htuse{Gamma10.td})}{\BR(\tau \to \htuse{Gamma9.td})} &=
\frac{f_K^2 \Vus^2}{f_\pi^2 \Vud^2} \frac{\left( m_\tau^2 - m_K^2 \right)^2}{\left( m_\tau^2 -  m_\pi^2 \right)^2}
\frac{1+\delta R_{\tau/K}}{1+\delta R_{\tau/\pi}}(1+\delta R_{K/\pi})
\end{align*}
We use $f_K/f_\pi = \htuse{f_K_by_f_pi}$ from the
FLAG 2016 Lattice averages with $N_f=2+1+1$~\cite{Aoki:2016frl},
\begin{align*}
  &\frac{1+\delta R_{\tau/K}}{1+\delta R_{\tau/\pi}} =
  \frac{1+(\htuse{dRrad_tauK_by_Kmu})\%}{1+(\htuse{dRrad_taupi_by_pimu})\%}%
  ~\cite{Marciano:1993sh,Decker:1994dd,Decker:1994ea,Decker:1994kw}~,\\
  &1+\delta R_{K/\pi} = 1 + (\htuse{dRrad_kmunu_by_pimunu})\%~\cite{Pich:2013lsa,Cirigliano:2011tm,Marciano:2004uf}~.
\end{align*}
We compute $\VusTauKpi = \htuse{Vus_tauKpi}$,
$\htuse{Vus_tauKpi_mism_sigma_abs}\sigma$ below the CKM unitarity prediction.

\iffalse
%%
%%
\tausubsection{\Vus from $\BR(\tau \to K\nu)$}
\label{sec:tau:vus:tauk}
%%
%%

We determine \Vus from the branching fraction
$\BFtautoknu$ using
\begin{align*}
  \BR(\tau^- \to K^-\nu_\tau) =
  \frac{G^2_F}{16\pi\hslash} f^2_K \Vus^2 \tau_{\tau}  m_{\tau}^3 \left(1 - \frac{m_K^2}{m_\tau^2}\right)^2
  \iffalse S_{EW} \fi
  (1+\delta R_{\tau/K}) (1+\delta R_{K\mu2})~.
\end{align*}
We use $f_K = \htuse{f_K}\,\mev$ from FLAG 2016 with
$N_f=2+1+1$~\cite{Aoki:2016frl},
\iffalse $S_{EW} = \htuse{Rrad_SEW_tau_Knu}$~\cite{Erler:2002mv}, \fi
$\delta R_{\tau/K} =
(\htuse{dRrad_tauK_by_Kmu})\%$~\cite{Marciano:1993sh,Decker:1994dd,Decker:1994ea,Decker:1994kw}
and $\delta R_{K\mu2} = (\htuse{dRrad_k_munu})\%$~\cite{Finkemeier:1995gi}.
We obtain $\VusTauKnu = \htuse{Vus_tauKnu}$,
which is $\htuse{Vus_tauKnu_mism_sigma_abs}\sigma$ below
the CKM unitarity prediction. The physical constants have been taken
from PDG 2015 (which uses CODATA 2014~\cite{Mohr:2015ccw}).
\fi

\tausubsection{\Vus from \mtau summary}

\begin{figure}[tb]
  \begin{center}
   \ifhevea
    \begin{tabular}{@{}cc@{}}
      \larger\bfseries\ahref{plot-vus-hfag16.png}{PNG format} &
      \larger\bfseries\ahref{plot-vus-hfag16.pdf}{PDF format} \\
      \multicolumn{2}{c}{\ahref{plot-vus-hfag16.png}{%
          \imgsrc[alt="Vus summary plot"]{plot-vus-hfag16.png}}}
    \end{tabular}
    \else
    {\fboxsep=2pt\fbox{%
        \includegraphics[width=0.75\linewidth,clip]{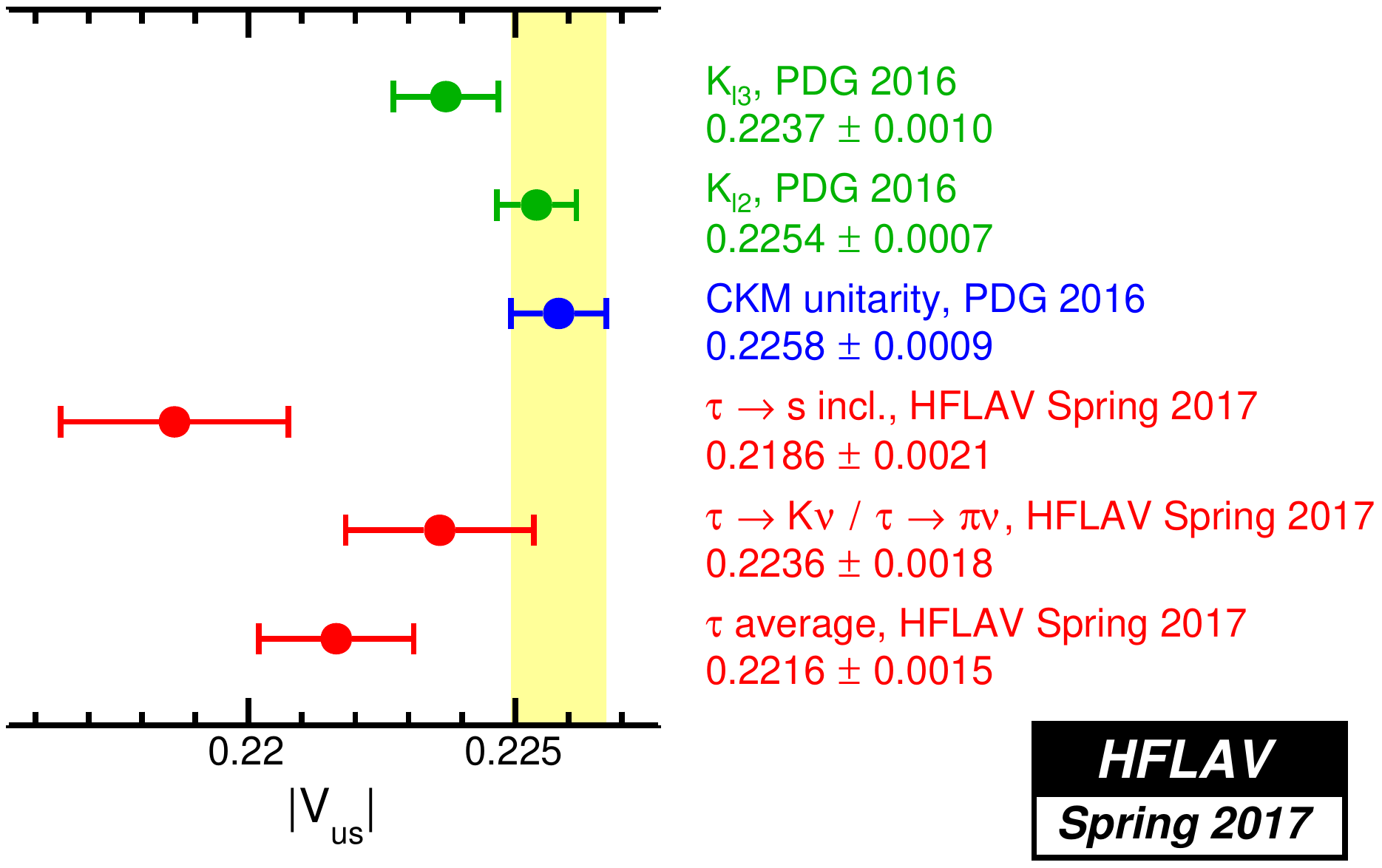}}}
    \fi
    \caption{\Vus averages.
      \iffalse
      The ``Maltman 2017'' \Vus
      determination~\cite{Hudspith:2017vew} reports the experimental
      uncertainty followed by the theoretical uncertainty.
      \fi
      \label{fig:tau:vus-summary}%
    }
  \end{center}
\end{figure}

We summarize the \Vus results reporting the values, the discrepancy with
respect to the \Vus determination from CKM unitarity, and an illustration
of the measurement method:
\begin{alignat*}{6}
  &\VusUni &&= \htuse{Vus_uni.v} &&\pm \htuse{Vus_uni.e} & \quad & & \quad
  & [\text{from } \sqrt{1 - \Vud^2} \quad\text{(CKM unitarity)}]~, \\
  &\VusTauIncl &&= \htuse{Vus.v} &&\pm \htuse{Vus.e} & \quad & \htuse{Vus_mism_sigma}\sigma &
  & [\text{from } \Gamma(\tau^- \to X_s^- \nut)]~, \\
  &\VusTauKpi &&= \htuse{Vus_tauKpi.v} &&\pm \htuse{Vus_tauKpi.e} & \quad & \htuse{Vus_tauKpi_mism_sigma}\sigma &
  & [\text{from } \Gamma(\tauknu)/\Gamma(\taupinu)]~.
%%   \iffalse
%%   %%
%%   &\VusTauKnu &&= \htuse{Vus_tauKnu.v} &&\pm \htuse{Vus_tauKnu.e} & \quad & \htuse{Vus_tauKnu_mism_sigma}\sigma &
%%   & [\text{from } \Gamma(\tauknu)]~.
%%   \fi
\end{alignat*}

Averaging the two above \Vus determinations that rely on the \mtau branching fractions (taking into account all
correlations due to the \mtau HFLAV and other mentioned inputs) we obtain, for \Vus and its discrepancy:
\begin{alignat*}{6}
  & \Vus_\tau &&= \htuse{Vus_tau} &\quad &\htuse{Vus_tau_mism_sigma}\sigma \quad
  & [\text{average of 2 \Vus \mtau measurements}]~.
\end{alignat*}
\iffalse
We could not find a published estimate of the correlation of the
uncertainties on $f_K$ and $f_K/f_\pi$, but even if we assume $\pm
100\%$ correlation, the uncertainty on $\Vus_\tau$ does not change by
more than $6\%$.
\fi

All \Vus determinations based on measured $\tau$ branching fractions are
lower than both the kaon and the CKM-unitarity determinations. This is
correlated with the fact that the direct measurements of the
three major $\tau$ branching fractions to kaons
[$\BR(\tau\to\htuse{Gamma10.td})$, $\BR(\tau\to\htuse{Gamma16.td})$ and
$\BR(\tau\to\htuse{Gamma35.td})$] are lower than their determinations
from the kaon branching fractions into final states with leptons within
the SM~\cite{Pich:2013lsa, Jamin:2008qg, Antonelli:2013usa}.

A recent determination of \Vus~\cite{Maltman:2015xwa,
Hudspith:2017vew} that relies on  the \mtau
spectral functions in addition to the inclusive $\tau \to X_s \nu$
branching fraction reports a \Vus value  about $1\sigma$ lower than the
CKM-unitarity determination. This determination uses inputs that are
partially different from the ones used in this report. Specifically, 
the HFLAV average of $\BR(\tau\to\htuse{Gamma10.td})$ has been replaced
with the SM prediction based on the measured $\BR(K^- \to \mu^-
\bar{\nu}_\mu)$ and the HFLAV average of
$\BR(\tau\to\htuse{Gamma16.td})$ has been replaced with an in-progress
\babar measurement that is published in a PhD thesis. Both changes
increase the resulting $\tau \to X_s \nu$ inclusive branching fraction.
This study claims that the newly proposed \Vus calculation has a more
stable and reliable theory uncertainty, which could possibly have been
underestimated in former studies, which are used for the HFLAV \Vus average.

In previous editions of the HFLAV report, we also computed \Vus using the
branching fraction $\BR(\tau \to K\nu)$ and without taking the ratio with
$\BR(\tau \to \pi\nu)$. We do not report this additional determination
because it did not include the long-distance radiative corrections in
addition to the short-distance contribution, and because it had a negligible
effect on the overall precision of the \Vus calculation
with \mtau data.

Figure~\ref{fig:tau:vus-summary} reports the HFLAV \Vus determinations
that use the \mtau branching fractions, compared to two \Vus
determinations based on kaon data~\cite{PDG_2016} and to \Vus obtained
from \Vud and the CKM matrix unitarity~\cite{PDG_2016}.

\iffalse
the above mentioned determination
of \Vus from inclusive $\tau \to X_s \nu$ decays and \mtau spectral
functions~\cite{Hudspith:2017vew}
\fi

%% -*- mode: LaTeX; TeX-master: "../master.tex" -*-
\tausection{Upper limits on \mtau lepton-flavour-violating branching fractions}
\cutname{lfv-limits.html}
\label{sec:tau:lfv-limits}
\label{sec:tau:lfv}
\newcommand{\htLimitLine}[5]{%
  #1 & #2 & #3 & #4 & \cite{#5} \\
}
The Standard Model predicts that the \mtau lepton-flavour-violating
(LFV) branching fractions are
too small to be measured with the available experimental precision.
We report in Table~\ref{tab:tau:lfv-upper-limits} and
Figure~\ref{fig:tau:lfv-limits-plot} the experimental upper
limits on these branching fractions that have been published by the
$B$-factories \babar and Belle and later experiments. We omit 
previous weaker upper limits (mainly from CLEO) and all preliminary
results presented several years ago. The previous
HFLAV report~\cite{Amhis:2014hma} still included a few preliminary
results, which have all been removed now.

\begin{center}

\begin{longtable}{lcclc}
\caption{Experimental upper limits on lepton flavour violating \mtau
  decays. The modes are grouped according to the properties of their final
  states. Modes with baryon number violation are labelled with ``BNV''.
  \label{tab:tau:lfv-upper-limits}}%
\\
\toprule
\multicolumn{1}{l}{\bfseries Decay mode} &
\multicolumn{1}{l}{\bfseries Category} &
\multicolumn{1}{c}{\bfseries \begin{tabular}{@{}c@{}}90\% CL\\Limit\end{tabular}} &
\multicolumn{1}{l}{\bfseries Exp.} &
\multicolumn{1}{l}{\bfseries Ref.} \\
\midrule
\endfirsthead
\multicolumn{5}{c}{{\bfseries \tablename\ \thetable{} -- continued from previous page}} \\ \midrule
\multicolumn{1}{l}{\bfseries Decay mode} &
\multicolumn{1}{l}{\bfseries Category} &
\multicolumn{1}{c}{\bfseries \begin{tabular}{@{}c@{}}90\% CL\\Limit\end{tabular}} &
\multicolumn{1}{l}{\bfseries Exp.} &
\multicolumn{1}{l}{\bfseries Ref.} \\
\midrule
\endhead
\htuse{LimitLines}
\bottomrule
\end{longtable}
\end{center}
 
%% -*- mode: LaTeX; TeX-master: "../master.tex" -*-
%% ///////////////////////////////////////////////////////////////////////////
\ifhevea
\tausection{Upper limits on \mtau lepton-flavour-violating branching fractions: summary plot}
\cutname{lfv-limits-plot.html}
\fi

\begin{figure}[tb]
  \begin{center}
    \ifhevea
    \begin{tabular}{@{}cc@{}}
      \larger\bfseries\ahref{tau-lfv-limits.png}{full size PNG} &
      \larger\bfseries\ahref{tau-lfv-limits.pdf}{PDF format} \\
      \multicolumn{2}{c}{\ahref{tau-lfv-limits.png}{%
          \imgsrc[alt="Tau LFV limits combinations plot" width=720]{tau-lfv-limits.png}}}
    \end{tabular}
    \else
    \includegraphics[angle=90,totalheight=0.9\textheight,clip]{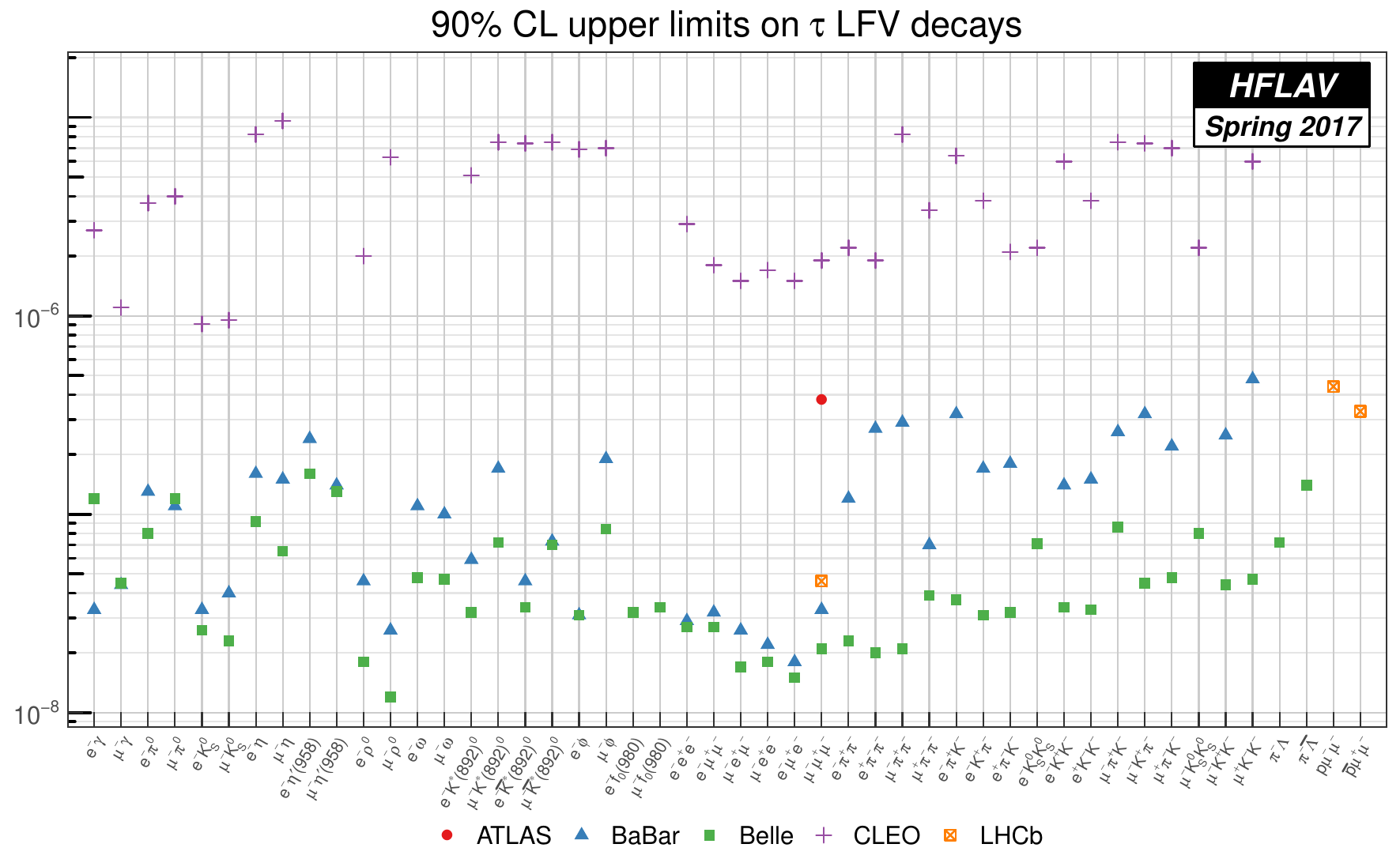}
    \fi
    \caption{Tau lepton-flavor-violating branching fraction upper
      limits summary plot. In order to appreciate the physics reach
      improvement over time, the plot includes also the CLEO upper
      limits reported by PDG 2016~\cite{PDG_2016}.
      \label{fig:tau:lfv-limits-plot}
    }
  \end{center}
\end{figure}

%% -*- mode: LaTeX; TeX-master: "../master.tex" -*-
\tausection{Combination of upper limits on \mtau lepton-flavour-violating branching fractions}
\cutname{lfv-combinations.html}
\label{sec:tau:lfv-combs}
\newcommand{\cls}{\ensuremath{\text{CL}_s}\xspace}
\newcommand{\clsb}{\ensuremath{\text{CL}_{s+b}}\xspace}
\newcommand{\clb}{\ensuremath{\text{CL}_b}\xspace}

Combining upper limits is a delicate issue, since there is no standard and
generally agreed procedure. Furthermore, the \mtau LFV searches
published limits are extracted from the data with a variety of
methods, and cannot be directly combined with a uniform procedure. It
is however possible to use a single and effective
upper limit combination procedure for all modes by re-computing the published upper
limits with just one extraction method, using the published
information that documents the upper limit determination:
number of observed candidates, expected background, signal efficiency and
number of analyzed \mtau decays.

We chose to use the \cls method~\cite{Read:2002hq} to re-compute the
\mtau LFV upper limits, since it is well known and widely used (see the
Statistics review of PDG 2013~\cite{PDG_2016}), and since the
limits computed with the \cls method can be combined in a straightforward
way (see below). The \cls method is based on two hypotheses: signal plus background and
background only. We calculate the observed confidence levels for the two
hypotheses:
\begin{align}
&\clsb = P_{s+b}(Q \leq Q_{obs}) = \int_{- \infty}^{Q_{obs}} \frac{dP_{s+b}}{dQ} dQ,
\label{eq:tau:clspdf1} \\
&\clb = P_{b}(Q \leq Q_{obs}) = \int_{- \infty}^{Q_{obs}} \frac{dP_{b}}{dQ} dQ,
\label{eq:tau:clspdf2}
\end{align}
where \clsb is the confidence level observed for the signal plus background
hypotheses, \clb is the confidence level observed for the background only
hypothesis, $\frac{dP_{s+b}}{dQ}$ and $\frac{dP_{b}}{dQ}$ are the probability
distribution functions (PDFs) for the two corresponding hypothesis and
$Q$ is called the test statistic. The \cls value is defined as the ratio
between the confidence level for the signal plus background hypothesis and
the confidence level for the background hypothesis:
\begin{align}
\cls = \dfrac{\clsb}{\clb}.
\end{align}
When multiple results are combined, the PDFs in
Eqs.~(\ref{eq:tau:clspdf1}) and~(\ref{eq:tau:clspdf2}) are the
product of the individual PDFs,
\begin{align}
\cls = \dfrac{\prod_{i=1}^{N}\sum_{n=0}^{n_i} \dfrac{e^{-(s_i+b_i)} (s_i+b_i)^{n}}{n!} }{\prod_{i=1}^{N}  \sum_{n=0}^{n_i} \dfrac{e^{-b_i} b_i^{n}}{n!}}    \dfrac{\prod_{j=1}^{N} \left[s_iS_i(x_{ij})+b_iB_i(x_{ij})\right]}{\prod_{j=1}^{N}B_i(x_{ij})}~,
\end{align}
where $N$ is the number of results (or channels), and, for each channel $i$,
$n_i$ is the number of observed candidates, $x_{ij}$ are the values of the
discriminating variables (with index $j$), $s_i$ and $b_i$ are the number
of signal and background events and $S_i$, $B_i$ are the probability
distribution functions of the discriminating variables. The
discriminating variables $x_{ij}$ are assumed to be uncorrelated.
The expected signal $s_i$ is related to the \mtau lepton branching
fraction $\BR(\tau \rightarrow f_i)$ into
the searched final state $f_i$ by $s_i = N_i\epsilon_i\BR(\tau \rightarrow
f_i)$, where $N_i$ is the number of produced \mtau leptons and
$\epsilon_i$ is the detection efficiency for observing the decay $\tau\to
f_i$. For $e^+ e^-$ experiments,
$N_i = 2\mathcal{L}_i\sigma_{\tau\tau}$, where $\mathcal{L}_i$ is the
integrated luminosity and $\sigma_{\tau\tau}$ is the
\mtau pair production cross section $\sigma(e^+ e^- \rightarrow \tau^+
\tau^-)$~\cite{Banerjee:2007is}.
In experiments where \mtau leptons are produced in more complex multiple
reactions, the effective $N_i$ is typically estimated with Monte Carlo simulations
calibrated with related data yields.

The extraction of the upper limits is performed using the code provided by
Tom Junk~\cite{junk:2007:cdfnote}. The systematic uncertainties are modeled
in the Monte Carlo toy experiments by convolving the $S_i$ and $B_i$
PDFs with Gaussian distributions corresponding to the nuisance
parameters.

Table~\ref{tab:tau:lfv-upper-limits-comb} reports the HFLAV combinations of the
\mtau LFV limits. Since there is negligible gain in combining limits of very
different strength, the combinations do not include the CLEO searches
and do not include results where the single event sensitivity is more
than a factor of 5 lower than the value for the search with the best limit.

Figure~\ref{fig:tau:lfv-limits-plot-average} reports a graphical
representation of the limits in Table~\ref{tab:tau:lfv-upper-limits-comb}.
The published information that has been used to obtain these limits is
reported in Table~\ref{tab:tau:lfv-extra-info-comb}.

\newcommand{\htCombLimitLine}[4]{%
  #1 & #2 & #3 & #4 \\
}
\newcommand{\htCombExtraLine}[7]{%
  #1 & #2 & \cite{#3} & #4 & #5 & #6 & #7 \\
}

\begin{center}
\begin{longtable}{lcrc}
\caption{Combinations of upper limits on lepton flavour violating \mtau decay
  modes. The modes are grouped according to the properties of their final
  states. Modes with baryon number violation are labelled with ``BNV''.
\label{tab:tau:lfv-upper-limits-comb}}%
\\
\toprule
\multicolumn{1}{l}{\bfseries Decay mode} &
\multicolumn{1}{c}{\bfseries Category} &
\multicolumn{1}{r}{\bfseries \begin{tabular}{@{}c@{}}90\% CL\\Limit\end{tabular}} &
\multicolumn{1}{c}{\bfseries Refs.} \\
\midrule
\endfirsthead
\multicolumn{3}{c}{{\bfseries \tablename\ \thetable{} -- continued from previous page}} \\
\midrule
\multicolumn{1}{l}{\bfseries Decay mode} &
\multicolumn{1}{c}{\bfseries Category} &
\multicolumn{1}{r}{\bfseries \begin{tabular}{@{}c@{}}90\% CL\\Limit\end{tabular}} &
\multicolumn{1}{c}{\bfseries Refs.} \\
\midrule
\endhead
\htuse{CombLines}
\bottomrule
\end{longtable}
\end{center}

\begin{center}\smaller
\begin{longtable}{llcrrrrrr}
\caption{
  % The table includes, for the \babar and Belle limits, the
  Published information that has been used to re-compute upper limits
  with the \cls method, \ie\ the number of \mtau leptons produced, the
  signal detection efficiency and its uncertainty, the number of
  expected background events and its uncertainty,
  and the number of observed events. The uncertainty on the efficiency
  includes the minor uncertainty contribution on the number of \mtau leptons
  (typically originating on the uncertainties on the integrated
  luminosity and on the production cross-section).
  The additional limit used in the
  combinations (from \lhcb) has been originally determined with the \cls
  method.
\label{tab:tau:lfv-extra-info-comb}}%
\\
\toprule
\multicolumn{1}{l}{\bfseries Decay mode} &
\multicolumn{1}{c}{\bfseries Exp.} &
\multicolumn{1}{c}{\bfseries Ref.} &
\multicolumn{1}{c}{\bfseries \begin{tabular}{@{}c@{}}$N_{\tau}$\\(millions)\end{tabular}} &
%% \multicolumn{1}{c}{\bfseries \begin{tabular}{@{}c@{}}{\cal L}\\($\text{fb}^{-1}$)\end{tabular}} &
%% \multicolumn{1}{c}{\bfseries \begin{tabular}{@{}c@{}}$\sigma_{\tau\tau}$\\(nb)\end{tabular}} &
\multicolumn{1}{c}{\bfseries \begin{tabular}{@{}c@{}}efficiency\\(\%)\end{tabular}} &
\multicolumn{1}{c}{\bfseries $N_{\text{bkg}}$} &
\multicolumn{1}{c}{\bfseries $N_{\text{obs}}$} \\
\midrule
\endfirsthead
\multicolumn{5}{c}{{\bfseries \tablename\ \thetable{} -- continued from previous page}} \\
\midrule
\multicolumn{1}{l}{\bfseries Decay mode} &
\multicolumn{1}{c}{\bfseries Exp.} &
\multicolumn{1}{c}{\bfseries Ref.} &
\multicolumn{1}{c}{\bfseries \begin{tabular}{@{}c@{}}$N_{\tau}$\\(millions)\end{tabular}} &
%% \multicolumn{1}{c}{\bfseries \begin{tabular}{@{}c@{}}{\cal L}\\($\text{fb}^{-1}$)\end{tabular}} &
%% \multicolumn{1}{c}{\bfseries \begin{tabular}{@{}c@{}}$\sigma_{\tau\tau}$\\(nb)\end{tabular}} &
\multicolumn{1}{c}{\bfseries \begin{tabular}{@{}c@{}}efficiency\\(\%)\end{tabular}} &
\multicolumn{1}{c}{\bfseries $N_{\text{bkg}}$} &
\multicolumn{1}{c}{\bfseries $N_{\text{obs}}$} \\
\midrule
\endhead
\htuse{CombExtraLines}
\bottomrule
\end{longtable}
\end{center}

%% -*- mode: LaTeX; TeX-master: "master.tex" -*-
%% ///////////////////////////////////////////////////////////////////////////
\ifhevea
\tausection{Combination of upper limits on \mtau lepton-flavour-violating branching fractions: summary plot}
\cutname{lfv-combinations-plot.html}
\fi

\begin{figure}[tb]
  \begin{center}
    \ifhevea
    \begin{tabular}{@{}cc@{}}
      \larger\bfseries\ahref{tau-lfv-combs.png}{full size PNG} &
      \larger\bfseries\ahref{tau-lfv-combs.pdf}{PDF format} \\
      \multicolumn{2}{c}{\ahref{tau-lfv-combs.png}{%
          \imgsrc[alt="Tau LFV limits combinations plot" width=720]{tau-lfv-combs.png}}}
    \end{tabular}
    \else
    \includegraphics[angle=90,totalheight=0.82\textheight,clip]{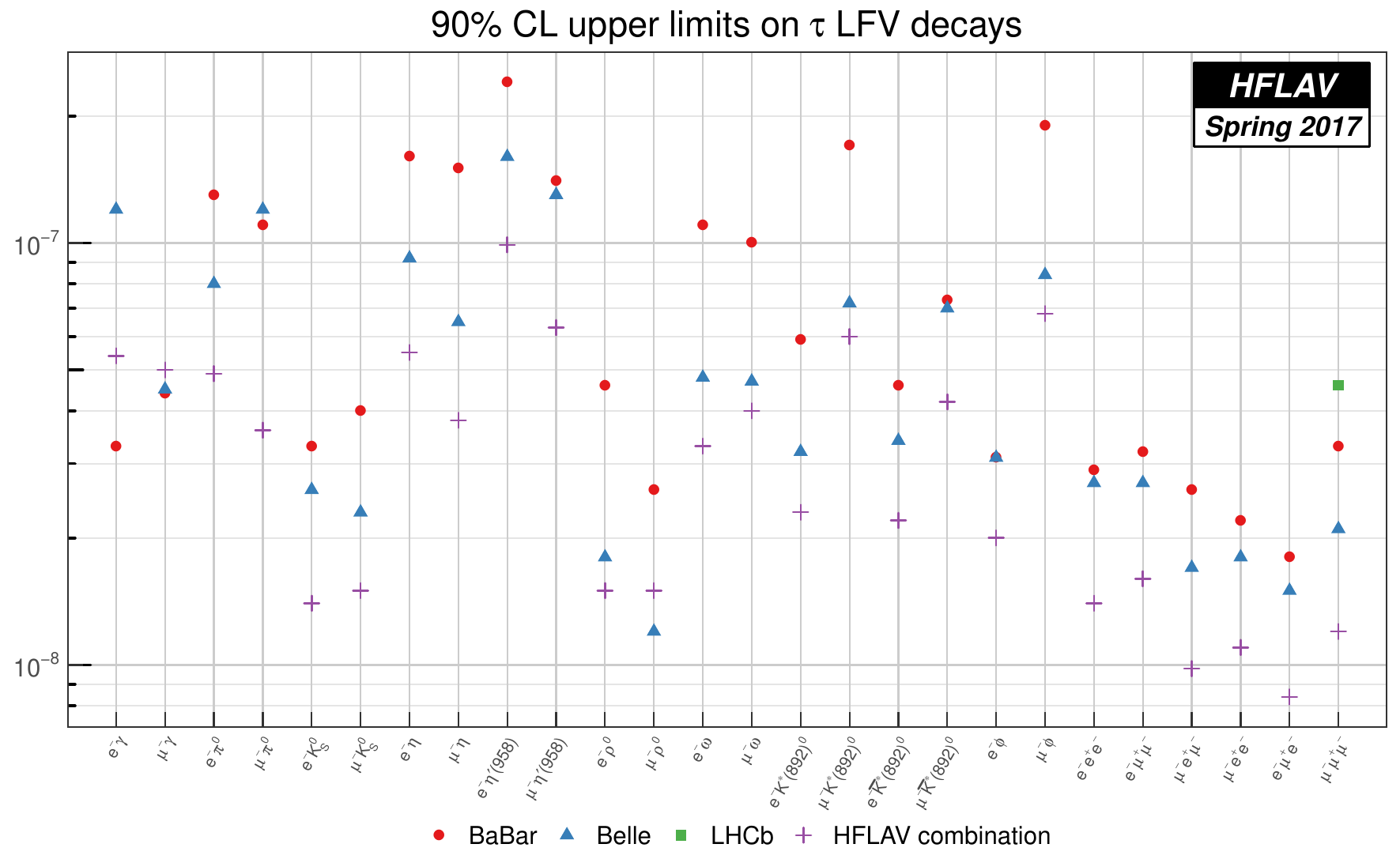}
    \fi
    \caption{Tau lepton-flavour-violating branching fraction upper limits
      combinations summary plot. For each channel we report the HFLAV
      combined limit, and the experimental published limits. In some cases,
      the combined limit is weaker than the limit published by a single
      experiment. This arises since the \cls method used in the
      combination can be more conservative compared to other legitimate
      methods, especially when the number of observed events fluctuates below the
      expected background. 
      \label{fig:tau:lfv-limits-plot-average}
    }
  \end{center}
\end{figure}

\let\tausection\subsection
%%--- end
\let\cite\citeOld
\end{fleqn}

% Summary and Acknowledgements

\clearpage
\section{Summary}
\label{sec:summary}

This article provides updated world averages of measurements of $b$-hadron, $c$-hadron, and $\tau$-lepton properties using results available through Summer 2016.
A small selection of highlights of the results described in 
Sections~\ref{sec:life_mix}--\ref{sec:tau} is given in Table~\ref{tab_summary}.
% Tables~\ref{tab_summary1},~\ref{tab_summary2} and~\ref{tab_summary3}.

\renewcommand*{\arraystretch}{1.08}
\begin{longtable}{|l|c|}
\caption{
  Selected world averages.
  Where two uncertainties are given the first is statistical and the second is systematic, except where indicated otherwise.
} % from Sections~\ref{sec:life_mix} and~\ref{sec:cp_uta}.}
\label{tab_summary}
\endfirsthead
\multicolumn{2}{c}{Selected world averages -- continued from previous page.}
\endhead
\endfoot
\endlastfoot
\hline
% {\bf\boldmath \b-hadron fractions} &   \\
% ~~$f^{+-}/f^{00}$ in \Ups decays  & \hflavFF \\ 
% ~~\fBs in \Upsfive decays & \hflavFSFIVE \\
% ~~\fBs, \fbb in $Z$ decays & \hflavZFBS, \hflavZFBB \\
% ~~\fBs, \fbb at Tevatron & \hflavTFBS, \hflavTFBB \\
% \hline
{\bf\boldmath \b-hadron lifetimes} &   \\
~~$\tau(\Bd)$         & \hflavTAUBD \\
~~$\tau(\Bu)$         & \hflavTAUBU \\
% ~~$\tau(\Bs\to~\mbox{flavour specific})$  & \hflavTAUBSSL \\
~~$\bar{\tau}(\Bs) = 1/\Gs$  & \hflavTAUBSMEANC \\
~~$\tau(B^0_{s\rm L})$ & \hflavTAUBSLCON \\
~~$\tau(B^0_{s\rm H})$  & \hflavTAUBSHCON \\
~~$\tau(\Bc)$         & \hflavTAUBC \\
~~$\tau(\Lb)$         & \hflavTAULB \\
~~$\tau(\Xibd)$       & \hflavTAUXBD  \\
~~$\tau(\Xibu)$       & \hflavTAUXBU  \\
~~$\tau(\Omegab)$     & \hflavTAUOB   \\
% ~~$\tau(\Xib)$ (mean) & \hflavTAUXB \\
% ~~$\tau(\Omegab)$ & \hflavTAUOB \\
\hline
\multicolumn{2}{|l|}{{\bf\boldmath \Bd\ and \Bs\ mixing / \CP violation parameters}}  \\
~~\dmd &  \hflavDMDWU \\
~~\DGGd  & \hflavSDGDGD \\
~~$|q_{\particle{d}}/p_{\particle{d}}|$ & \hflavQPDB  \\
~~\dms  &  \hflavDMS \\
%~~$\DGGs = (\Gamma_{s\rm L} - \Gamma_{s\rm H})/\Gs$ & \hflavDGSGSCON \\
~~\DGs & \hflavDGSCON \\
% ~~$\DGs = \Gamma_{s\rm L} - \Gamma_{s\rm H}$ & \hflavDGSCON \\
~~$|q_{\particle{s}}/p_{\particle{s}}|$ & \hflavQPS   \\
~~\phiccbars  & \hflavPHISCOMB \\
\hline
\multicolumn{2}{|l|}{{\bf Parameters related to Unitarity Triangle angles}} \\
 ~~ $\stwob \equiv \sin\! 2\phi_1$ & $\phantom{-}0.691 \pm 0.017$ \\
 ~~ $\beta \equiv \phi_1$          & $\phantom{\degrees}\left( 21.9 \pm 0.7 \right)\degrees$ \\
 ~~ $-\etacp S_{\phi \KS}$         & $0.74\,^{+0.11}_{-0.13}$ \\
 ~~ $-\etacp S_{\etapr \Kz}$       & $\phantom{-}0.63 \pm 0.06$ \\
 ~~ $-\etacp S_{\KS \KS \KS}$      & $\phantom{-}0.72 \pm 0.19$ \\
% ~~ $-\etacp S_{\Kp \Km \KS}$      & $0.68\,^{+0.09}_{-0.10}$ \\
 ~~ $\phi_s(\phi\phi)$             & $-0.17 \pm 0.15 \pm 0.03 \, {\rm rad}$ \\
 ~~ $-\etacp S_{\jpsi \piz}$       & $\phantom{-}0.93 \pm 0.15$ \\
 ~~ $-\etacp S_{\Dp\Dm}$           & $\phantom{-}0.84 \pm 0.12$ \\
 ~~ $-\etacp S_{\jpsi \rhoz}$      & $\phantom{-}0.66 \, ^{+0.13}_{-0.12}\,^{+0.09}_{-0.03}$ \\
 ~~ $S_{K^* \gamma}$               & $-0.16 \pm 0.22$ \\
 ~~ $\left( S_{\pi^+\pi^-}, C_{\pi^+\pi^-} \right)$ & $\left( -0.68 \pm 0.04, -0.27 \pm 0.04 \right)$ \\  
 ~~ $\left( S_{\rho^+\rho^-}, C_{\rho^+\rho^-} \right)$             & $\left( -0.14 \pm 0.13, \phantom{-}0.00 \pm 0.09\right)$ \\
 ~~ $a(D^{*\pm}\pi^{\mp})$         & $-0.039 \pm 0.010$ \\
 ~~ $A^{}_{\CP}(B\ra D^{}_{\CP+}K)$       & $\phantom{-}0.111 \pm 0.018$ \\
 ~~ $A_{\rm ADS}(B\ra D^{}_{K\pi}K)$     & $-0.415 \pm 0.055$ \\
 %~~ $R_{\rm ADS}(B\ra D^{}_{K\pi}K)$     & $\phantom{-}0.0183 \pm 0.0014$ \\
 ~~ $\gamma \equiv \phi_3$               & $\phantom{\degrees}(74.0\,^{+5.8}_{-6.4})\degrees$ \\
\hline
{\bf\boldmath Semileptonic \B decay parameters} & \\
 ~~${\cal B}(\Bzb\to D^{*+}\ell^-\nub_\ell)$ & $(4.88\pm 0.10)\%$\\
 ~~${\cal B}(\B^-\to D^{*0}\ell^-\nub_\ell)$ & $(5.59\pm 0.19)\%$\\
 ~~$\eta_{\rm EW}{\cal F}(1)\vcb$ & $(35.61\pm 0.43)\times 10^{-3}$\\
 ~~$\vcb$ from $\bar B\to D^*\ell^-\bar\nu_\ell$ & $(39.05\pm 0.47_{\rm exp}\pm 0.58_{\rm th})\times 10^{-3}$\\
\hline
 ~~${\cal B}(\Bzb\to D^+\ell^-\nub_\ell)$ & $(2.20\pm 0.10)\%$\\
 ~~${\cal B}(\B^-\to D^0\ell^-\nub_\ell)$ & $(2.33\pm 0.10)\%$\\
 ~~$\eta_{\rm EW}{\cal G}(1)\vcb$ & $(41.57 \pm 1.00)\times 10^{-3}$\\
 ~~$\vcb$ from $\bar B\to D\ell^-\bar\nu_\ell$ & $(39.18 \pm 0.94_{\rm exp}\pm 0.36_{\rm th})\times 10^{-3}$\\
\hline
 ~~${\cal B}(\bar B\to X_c\ell^-\bar\nu_\ell)$ & $(10.65\pm 0.16)\%$\\
 ~~${\cal B}(\bar B\to X\ell^-\bar\nu_\ell)$ & $(10.86\pm 0.16)\%$\\
 ~~$\vcb$ from $\bar B\to X\ell^-\bar\nu_\ell$ & $(42.19\pm 0.78)\times 10^{-3}$\\
\hline
 ~~${\cal B}(\Bb\to\pi\ell^-\nub_\ell)$ & $(1.50\pm 0.06)\times 10^{-4}$\\
 ~~$\vub$ from $\Bb\to\pi\ell^-\nub_\ell$ & $(3.67\pm 0.15)\times
 10^{-3}$\\
 ~~$\vub$ from $\Bb\to X_u\ell^-\nub_\ell$ & $(4.52\pm 0.15_{\rm exp}\pm 0.13_{\rm th})\times 10^{-3}$\\
%\hline
 ~~$\vub/\vcb$ from $\Lambda_b^0\to p\mu^-\nub_\mu/\Lambda_b^0\to \Lambda_c^+\mu^-\nub_\mu$ & $0.080\pm 0.004_{\rm exp}\pm 0.004_{\rm th}$\\

\hline
~~${\cal R}(D)={\cal B}(B\to D\tau\nu_\tau)/{\cal B}(B\to
  D\ell\nu_\ell)$ & $0.403\pm 0.047$\\
~~${\cal R}(D^*)={\cal B}(B\to D^*\tau\nu_\tau)/{\cal B}(B\to
  D^*\ell\nu_\ell)$ & $0.310\pm 0.017$\\
\hline
{\bf\boldmath \b-hadron to charmed hadron decays} & \\
 ~~ ${\cal B}(\Bzb \to D^+ \pi^-)$ & $(2.65\pm 0.15) \times 10^{-3}$ \\
 ~~ ${\cal B}(B^- \to D^0 \pi^-)$ & $(4.75 \pm 0.19) \times 10^{-3}$ \\
 ~~ ${\cal B}(\Bsb \to D_s^+ \pi^-)$ & $(3.03 \pm 0.25) \times 10^{-3}$ \\
 ~~ ${\cal B}(\Lambda_b^0 \to \Lambda_c^+ \pi^-)$ & $(4.30^{+0.36}_{-0.35}) \times 10^{-3}$ \\
 ~~ ${\cal B}(\Bzb \to J/\psi \bar{K}^0)$ & $(0.863 \pm 0.035) \times 10^{-3}$ \\
 ~~ ${\cal B}(B^- \to J/\psi K^-)$ & $(1.028 \pm 0.040) \times 10^{-3}$ \\
 ~~ ${\cal B}(\Bsb \to J/\psi \phi)$ & $(1.00 \pm 0.09) \times 10^{-3}$ \\
% ~~ ${\cal B}(\Lambda_b^0 \to J/\psi \Lambda^0)$ & $(0.47 \pm 0.28) \times 10^{-3}$ \\
\hline
{\bf\boldmath Rare \B decays} &   \\
 ~~ ${\cal B}(\Bs \to \mu^+\mu^-)$ & $\left( 2.8\,^{+0.07}_{-0.06} \right) \times 10^{-9}$ \\
 ~~ ${\cal B}(\Bz \to \mu^+\mu^-)$ & $\left( 0.39\,^{+0.16}_{-0.14} \right) \times 10^{-9}$ \\
 ~~ ${\cal B}(B \to X_s \gamma)$  ($E_{\gamma}>1.6~\gev$) & $(3.32 \pm 0.16) \times 10^{-4}$ \\
 ~~ ${\cal B}(\Bp \to \tau^+ \nu)$ & $(1.06 \pm 0.19) \times 10^{-4}$ \\
 ~~ $R_K = \mathcal{B}(\Bp \to K^+\mu^+\mu^-)/\mathcal{B}(\Bp \to K^+e^+e^-)$ &
 \multirow{2}{*} {$\aerr{0.745}{0.090}{0.074}{0.036}$}\\
 ~~ ~~~~ in $1.0<m^2_{\ell^+\ell^-}<6.0~{\mathrm{Ge\kern -0.1em V}^2/c^4}$ & \\ 
%  ~~ $A_{\CP}(\particle{\Bd\to K^+\pi^-})$ & $-0.082 \pm 0.006$ \\
%  ~~ $A_{\CP}(\particle{B^+\to K^+\pi^0})$ & $0.040 \pm 0.021$ \\
 ~~ $A_{\CP}(\Bd\to K^+\pi^-)$, $A_{\CP}(B^+\to K^+\pi^0)$  & $-0.082 \pm 0.006$, $0.040 \pm 0.021$ \\
 ~~ $A_{\CP}(\Bs\to K^-\pi^+)$ & $0.26 \pm 0.04$ \\
 ~~ Longitudinal polarisation of $\Bd \to \phi  \Kstarz$ & $0.497 \pm 0.017$ \\
 ~~ Longitudinal polarisation of $\Bs \to \phi  \phi$ & $0.361 \pm 0.022$ \\
%  ~~ Longitudinal polarisation of $\Bs \to K^{*0}  \Kstarzb$ & $0.201 \pm 0.070$ \\
%  ~~ Longitudinal polarisation of $\Bs \to \phi  \Kstarzb$ & $0.51 \pm 0.17$ \\
% ~~ Several new branching fractions of $b$-baryon decays & {See Sec.~\ref{sec:rare-lb}} \\ 
 ~~ Observables in $\Bz \to K^{*0}\mu^+\mu^-$ decays & \multirow{2}{*}{See Sec.~\ref{sec:rare-radll}} \\
 ~~ ~~~~ in bins of $q^2 = m^2(\mu^+\mu^-)$ & \\
% ~~ Several new branching fractions of \Bc meson decays & {See Sec.~\ref{sec:rare-bc}} \\
\hline
 {\bf\boldmath $D^0$ mixing and \CP violation parameters} &   \\
 ~~$x$ &  $(0.32\,\pm 0.14)\%$  \\
 ~~$y$ &  $(0.69\,^{+0.06}_{-0.07})\%$  \\
%~~$R^{}_D$ &  $(0.349\,^{+0.004}_{-0.003})\%$  \\
~~$\delta^{}_{K\pi}$ &  $(15.2\,^{+7.6}_{-10.0})^\circ$  \\
 ~~$A^{}_D$ &  $(-0.88\,\pm 0.99)\%$  \\
 ~~$|q/p|$ & $0.89\,^{+0.08}_{-0.07}$  \\
 ~~$\phi$ &  $(-12.9\,^{+9.9}_{-8.7})^\circ$  \\
\hline
 ~~$x^{}_{12}$ (no direct \CP violation) &  $(0.41\,^{+0.14}_{-0.15})\%$  \\
 ~~$y^{}_{12}$ (no direct \CP violation) &  $(0.61\,\pm 0.07)\%$  \\
 ~~$\phi^{}_{12}$ (no direct \CP violation) &  $(-0.17\,\pm 1.8)^\circ$  \\
\hline
~~$a^{\rm ind}_{\CP}$ & $(0.030 \pm 0.026)\%$ \\
~~$\Delta a^{\rm dir}_{\CP}$ & $(-0.134 \pm 0.070)\%$ \\
\hline
 {\bf\boldmath Leptonic $D$ decays} &   \\
 ~~$f^{}_D$     & $(203.7\,\pm 4.9)$~MeV  \\
 ~~$f^{}_{D_s}$  & $(257.1\,\pm 4.6)$~MeV  \\
 ~~$|V^{}_{cd}|$ & $0.2164\,\pm 0.0050_{\rm exp} \pm 0.0015_{\rm LQCD}$  \\
 ~~$|V^{}_{cs}|$ & $1.006\,\pm 0.018_{\rm exp} \pm 0.005_{\rm LQCD}$  \\
\hline
 {\bf\boldmath Benchmark charm branching fractions} &   \\
 ~~${\cal B}(\Lambda^+_c\ra pK^-\pi^+)$ & $(6.46\,\pm 0.24)\%$  \\
 ~~${\cal B}(D^0\ra K^-\pi^+)$   & 
$(3.962\,\pm 0.017\,\pm 0.038\,\pm 0.027_{\rm FSR})\%$ \\
 ~~${\cal B}(D^0\ra K^+\pi^-)/{\cal B}(D^0\ra K^-\pi^+)$    & 
$(0.349\,^{+0.004}_{-0.003})\%$  \\
 ~~${\cal B}(D^+_s\ra K^+K^-\pi^+)$   &
$(5.44\,\pm 0.09\,\pm 0.11)\%$ \\
% ~~${\cal B}(D^0\ra\mu^+\mu^-)$          & $<6.2\times 10^{-9}$ (90\% C.L.) \\
% ~~${\cal B}(D^0\ra\mu^\pm e^\mp)$          &  $<16\times 10^{-9}$ (90\% C.L.) \\
% ~~${\cal B}(D^0\ra e^+ e^-)$               & $<79\times 10^{-9}$ (90\% C.L.) \\
\hline
{\bf\boldmath $\tau$ parameters, lepton universality, and $|V_{us}|$} &   \\
%% ~~ $m^{}_\tau$ (MeV/$c^2$)                   & $1776.77 \pm 0.15$ \\
~~ $g^{}_{\tau}/g^{}_{\mu}$    & \htuse{gtaubygmu_tau} \\
~~ $g^{}_{\tau}/g^{}_{e}$      & \htuse{gtaubyge_tau} \\
~~ $g^{}_\mu/g^{}_e$           & \htuse{gmubyge_tau} \\
~~ ${\cal B}_e^{\text{uni}}$   & \htuse{Be_univ}\% \\
~~ $R_{\text{had}}$            & \htuse{R_tau} \\
~~ $|V_{us}|$ from sum of strange branching fractions                                   & \htuse{Vus} \\
~~ $|V_{us}|$ from ${\cal{B}}(\tau^- \to K^-\nu^{}_\tau)/ {\cal{B}}(\tau^- \to \pi^-\nu^{}_\tau)$ & \htuse{Vus_tauKpi} \\ 
%% ~~ $|V_{us}|$ from ${\cal{B}}(\tau^-\to K^-\nu^{}_\tau)$                                          & \htuse{Vus_tauKnu} \\
~~ $|V_{us}|$ \mtau average                                                                         & \htuse{Vus_tau} \\
\hline
%\end{tabular}
\end{longtable}

%%% lifetimes and mixing highlights
Since the previous version of this document~\cite{Amhis:2014hma}, 
the \b-hadron lifetime and mixing averages have mostly made gradual
progress in precision.
Notable exceptions with significant improvement are 
the averages for the mass difference in the $\Bd$--$\Bzb$ system (\dmd) and the 
\CP violation parameter in $\Bs$--$\Bsb$ system ($|q_{\particle{s}}/p_{\particle{s}}|$). 
In total eleven new results 
(of which ten from the LHC Run~1 data and one from the Tevatron data)
have been incorporated in these averages.
On the other hand, all results that remained unpublished and 
for which there is no publication plan,
have been removed from the averages. 
The lifetime hierarchy for the most abundant weakly decaying \b-hadron species
is well established, with impressive precisions of 5~fs or less
for the most common \Bd, \Bu and \Bs mesons,
and compatible with the expectations from the Heavy Quark Expansion. 
However, statistics are still lacking for \b baryons heavier 
than \Lb (\Xibd, \Xibu, $\Omega_b$, and all other yet-to-be-discovered 
\b baryons), but this will surely come from the LHC with sufficient time. 
A sizable value of the decay width difference in the $\Bs$--$\Bsb$ system 
is measured with a relative precision of 7\% and is well predicted by the 
Standard Model (SM). In contrast, 
the experimental results for the decay width difference in the
$\Bd$--$\Bzb$ system are not yet precise enough to distinguish
the small (expected) value from zero.
The mass differences in both systems are known very accurately, to the (few)
per mil level. On the other hand, \CP violation in the mixing of either system 
has not been observed yet, with asymmetries known within a couple per mil but
still consistent both with zero and their SM predictions. 
A similar conclusion holds for the \CP violation induced
by \Bs mixing in the $b\to c\bar{c}s$ transition, although in this case 
the experimental precision on the corresponding weak phase is an order
of magnitude larger, but now becoming just smaller than the SM central value. 
Many measurements are still dominated by statistical uncertainties and will improve once new results from the LHC Run~2 become available.

%% %%% cp(t) & ut highlights
The measurement of $\sin 2\beta \equiv \sin 2\phi_1$ from $b \to
c\bar{c}s$ transitions such as $\Bz \to \jpsi\KS$ has reached $<2.5\,\%$
precision: $\sin 2\beta \equiv \sin 2\phi_1 = 0.691 \pm 0.017$.
Measurements of the same parameter using different quark-level processes
provide a consistency test of the Standard Model and allow insight into
possible new physics.  
% Recent improvements include the use of time-dependent Dalitz plot analyses of $\Bz \to \KS\Kp\Km$ and $\Bz \to \KS\pip\pim$ to obtain \CP\ violation parameters for $\phi\KS$, $f_0(980)\KS$ and $\rho\KS$.  
All results among hadronic $b \to s$ penguin dominated decays of \Bz mesons are currently consistent with the Standard Model expectations.  
Measurements of \CP violation parameters in $\Bs \to \phi\phi$ allow a similar comparison to the value of $\phiccbars$; again, results are consistent with the SM expectation (which in this case is very close to zero).
Among measurements related to the Unitarity Triangle angle $\alpha \equiv \phi_2$, results from the $\rho\rho$ system allow constraints at the level of $\approx
6^\circ$.  
These remain the strongest constraints, although results from all of \babar, \belle and LHCb lead to good precision on the \CP violation parameters in $\Bz \to \pip\pim$ decays.
Knowledge of the third angle $\gamma \equiv \phi_3$ also continues to improve, with the current world average being $(74.0\,^{+5.8}_{-6.4})\degrees$.
The precision is expected to improve further as more data becomes available at LHCb and Belle~II.

%% %%% semileptonic highlights
In semileptonic $B$~meson decays, the anomalies reported in the last
version of the document have remained: 
The discrepancy between \vcb\ measured with inclusive and exclusive decays is of the order of $3\sigma$ ($3.2\sigma$ for \vcb\ from $\Bb\to D^*\ell^-\bar\nu_\ell$, $2.4\sigma$ for \vcb\ from $\Bb\to D\ell^-\bar\nu_\ell$). 
The difference between \vub\ measured with
inclusive decays $\Bb\to X_u\ell^-\nub_\ell$ and \vub\ from $\Bb\to\pi\ell^-\nub_\ell$ has risen to $3.6\sigma$. 
An important new contribution to the determination of the values of \vub\ and \vcb\ comes from exclusive $b$-baryon decays.
The largest anomaly however is observed in $B\to D^{(*)}\tau\nu_\tau$ decays: The combined discrepancy of the measured values of ${\cal R}(D^*)$ and ${\cal R}(D)$ to their standard model expectations is found to be $3.9\sigma$.

%% %%% rare decays highlights
The most important new measurements of rare $b$-hadron decays are coming from the LHC. 
Precision measurements of $\Bs$ decays are particularly noteworthy, including several measurements of the longitudinal polarisation fraction from LHCb. 
ATLAS, CMS and LHCb have significantly improved the sensitivity to the $B^0_{(s)}\to\mumu$ decays.
Recently, CMS and LHCb published a combined analysis that allowed the first observation of the $\Bs\to\mumu$ decay to be obtained, and provided three standard deviations evidence of the $\Bz\to\mumu$ decay. 
The results are compatible with the SM predictions, and yield constrains on the parameter space of new physics models.
CMS and LHCb have also performed angular analyses of the $\Bz\to\Kstarz\mumu$ decay, complementing, extending and improving on the precision of results from \babar and Belle. 
One of the observables measured by LHCb, $P_5^{\prime}$, differs from the SM prediction by $3.7\sigma$ in one of the $m^2_{\mumu}$ intervals; results from Belle on this observable are consistent but less precise.
Improved measurements from LHCb and other experiments are keenly anticipated.
A measurement of the ratio of branching fractions of $\Bp \to K^+\mu^+\mu^-$ and $\Bp \to K^+e^+e^-$ decays ($R_K$) has been made by LHCb. 
In the low $m^2_{\ell^+\ell^-}$ region, it differs from the standard model prediction by $2.6\sigma$.
Among the \CP violating observables in rare decays, the ``$K\pi$ puzzle'' persists, and important new results have appeared in three-body decays.
LHCb has produced many other results on a wide variety of decays, including $b$-baryon and \Bc-meson decays. %, as indicated in the tables in Sec.~\ref{sec:rare}.  
Belle and \babar\ continue to produce new results though their output rates are dwindling.  
It will still be some years before we see new results from the upgraded SuperKEKB $B$ factory and the Belle~II experiment.

%% %%% b to charm highlights
About 800 $b$ to charm results from \babar, Belle, CDF, D0, LHCb, CMS, and ATLAS reported in more than 200 papers
are compiled in a list of over 600 averages.
%Most new results are from LHCb, but also the B-factory, Tevatron, and general-purpose LHC experiments contributed to the update.
The huge samples of $b$ hadrons that are available in contemporary experiments allows measurements of decays to states with open or hidden charm content with unprecedented precision.
In addition to improvements in precision for branching fractions of $\Bzb$ and $B^-$ mesons, many new decay modes have been discovered.
In addition, there is a rapidly increasing set of measurements available for $\Bsb$ and $B_c^-$ mesons as well as for $b$ baryon decays.

%%% charm highlights

In the charm sector, $\Dz$--$\Dzb$ mixing is now well-established and
the emphasis has shifted to searching for \CP\ violation. Measurements 
of 49 observables from the E791, FOCUS, Belle, \babar, CLEO, BESIII, CDF, 
and LHCb experiments are input into a global fit for 10 underlying parameters, 
and the no-mixing hypothesis is excluded at a confidence level $>11.5\sigma$. 
The mixing parameters $x$ and $y$ individually differ from zero by 
$1.9\sigma$ and $9.4\sigma$, respectively. The world average value for 
the observable $y_{\CP}$ is positive, indicating that the \CP-even 
state is shorter-lived as in the $\Kz$--$\Kzb$ system. 
The \CP\ violation parameters $|q/p|$ and $\phi$ are consistent with the 
no-\CP\ violation hypothesis within~$1\sigma$. Thus there is no evidence
for \CP\ violation arising from mixing ($|q/p|\neq 1$) or 
from a phase difference between the mixing amplitude and 
a direct decay amplitude ($\phi\neq 0$). In addition,
the most recent data indicates no direct \CP\ violation in 
$D^0\ra K^+K^-/\pi^+\pi^-$ decays; performing a global 
fit to all relevant measurements gives 
$\Delta a^{\rm dir}_{\CP}= (-0.134\pm 0.070)\%$.
The world's most precise measurements of $|V^{}_{cd}|$ and $|V^{}_{cs}|$
are obtained from leptonic $D^+\ra \mu^+\nu$ and $D^+_s\ra\mu^+\nu/\tau^+\nu$ 
decays, respectively. 
These measurements have theoretical uncertainties arising from decay constants.
% ; the results are listed in Table~\ref{tab_summary}.
% The second uncertainty listed is due to form factors, which are calculated
% using lattice QCD. The 
However, calculations of decay constants within lattice QCD have improved such 
that the theory error is $<1/3$ the experimental errors of the measurements.

%% tau highlights

Since 2016, HFLAV provides the \mtau branching fraction fit averages for the
PDG Review of Particle Physics. For the PDG, a unitarity constrained
variant of the fit is performed, using only inputs that are published
and included in the PDG. Two preliminary results used in the HFLAV 2014
report have been removed both in the HFLAV and in the PDG variants of the
fit. A few minor imperfections of the 2014 fit have been corrected. There
are no non-negligible changes to the lepton universality tests and to
the \Vus determinations from the \mtau branching fractions. There is
still a large discrepancy between \Vus from \mtau, \Vus from kaons and
\Vus from \Vud and CKM matrix unitarity. On this topic, recent
studies~\cite{Maltman:2015xwa, Hudspith:2017vew}
claim to get a more reliable theory uncertainty on
\Vus with a revised calculation method that uses also the $\tau$
spectral functions.
Just one more \mtau lepton-flavour-violating branching
fraction upper limit has been published, which does not change the
computed combined related limit. The list of limits and their combinations has
been revised to remove old preliminary results.

\clearpage
\section{Acknowledgments}

We are grateful for the strong support of the 
ATLAS, \babar, \belle, BESIII, CLEO(c), CDF, CMS, 
\dzero\ and LHCb collaborations, without whom this 
compilation of results and world averages would not have  
been possible. The success of these experiments in turn would 
not have been possible without the excellent operations of the 
BEPC, CESR, KEKB, LHC, PEP-II, and Tevatron accelerators.
%, and fruitful collaborations between the accelerator 
% groups and the experiments.
We also recognise the interplay between theoretical and experimental
communities that has provided a stimulus for many of the measurements 
in this document.

Our averages and this compilation have benefitted greatly from 
contributions to the Heavy Flavor Averaging Group from numerous
individuals. We thank David Kirkby, Yoshihide Sakai, Simon Eidelman, 
Soeren Prell, and Gianluca Cavoto for their past leadership of HFLAV. 
We are grateful to Paolo Gambino and Bob Kowalewski for assistance with
averages that appear in Section~\ref{sec:slbdecays};
to Ruslan Chistov, Lawrence Gibbons, Bostjan Golob, Milind Purohit,
and Patrick Roudeau for significant contributions to 
Section~\ref{sec:charm_physics};
and to Michel Davier for providing valuable input to Section~\ref{sec:tau}.
We especially thank the following for their careful review of the 
text in preparing this paper for publication: 
%Alexander Lenz, Guennadi Borissov, Zoltan Ligeti, 
%Fernando Martinez Vidal, Paolo Gambino, Andreas Kronfeld, 
%Vera L\"{u}th, Martin Jung, Vanya Belyaev, David London, 
%Paoti Chang, Alexey Petrov, Jolanta Brodzicka, Antonio Pich, 
%George Lafferty.
Vanya Belyaev, 
Guennadi Borissov, 
Jolanta Brodzicka, 
Paoti Chang, 
Paolo Gambino, 
Martin Jung, 
Andreas Kronfeld, 
George Lafferty,
Alexander Lenz, 
Zoltan Ligeti, 
David London, 
Vera L\"{u}th, 
Alexey Petrov, 
Antonio Pich, and 
Fernando Martinez Vidal.
Finally, we thank the SLAC National Accelerator Laboratory for 
providing crucial computing resources and support to HFLAV.

\clearpage

\bibliographystyle{HFLAV}
\raggedright
\setlength{\parskip}{0pt}
\setlength{\itemsep}{0pt plus 0.3ex}
\begin{small}
\bibliography{Summer16,life_mix/life_mix,cp_uta,slbdecays/slb_ref,rare/RareDecaysBib,b2charm/b2charm,charm/charm_refs,tau/tau-refs,tau/tau-refs-pdg}

\ifx\mcitethebibliography\mciteundefinedmacro
\PackageError{unsrtM.bst}{mciteplus.sty has not been loaded}
{This bibstyle requires the use of the mciteplus package.}\fi
\begin{mcitethebibliography}{1000}

\bibitem{Cabibbo:1963yz}
N.~Cabibbo, \href{http://dx.doi.org/10.1103/PhysRevLett.10.531}{Phys.\ Rev.\
  Lett.\ {\bf 10},  531}  (1963)\relax
\mciteBstWouldAddEndPuncttrue
\mciteSetBstMidEndSepPunct{\mcitedefaultmidpunct}
{\mcitedefaultendpunct}{\mcitedefaultseppunct}\relax
\EndOfBibitem
\bibitem{Kobayashi:1973fv}
M.~Kobayashi and T.~Maskawa, \href{http://dx.doi.org/10.1143/PTP.49.652}{Prog.\
  Theor.\ Phys.\ {\bf 49},  652}  (1973)\relax
\mciteBstWouldAddEndPuncttrue
\mciteSetBstMidEndSepPunct{\mcitedefaultmidpunct}
{\mcitedefaultendpunct}{\mcitedefaultseppunct}\relax
\EndOfBibitem
\bibitem{Abbaneo:2000ej_mod}
D.~Abbaneo {\em et al.} ({ALEPH, CDF, DELPHI, L3, OPAL, and SLD}
  collaborations), \href{http://arxiv.org/abs/hep-ex/0009052}{{\tt
  arXiv:hep-ex/0009052}}  (2000), CERN-EP-2000-096\relax
\mciteBstWouldAddEndPuncttrue
\mciteSetBstMidEndSepPunct{\mcitedefaultmidpunct}
{\mcitedefaultendpunct}{\mcitedefaultseppunct}\relax
\EndOfBibitem
\bibitem{Abbaneo:2001bv_mod_cont}
\href{http://arxiv.org/abs/hep-ex/0112028}{{\tt arXiv:hep-ex/0112028}}
  (2001), CERN-EP-2001-050\relax
\mciteBstWouldAddEndPuncttrue
\mciteSetBstMidEndSepPunct{\mcitedefaultmidpunct}
{\mcitedefaultendpunct}{\mcitedefaultseppunct}\relax
\EndOfBibitem
\bibitem{Amhis:2014hma}
Y.~Amhis {\em et al.} ({Heavy Flavor Averaging Group}){,}
  \href{http://arxiv.org/abs/1412.7515}{{\tt arXiv:1412.7515 [hep-ex]}}
  (2014)\relax
\mciteBstWouldAddEndPuncttrue
\mciteSetBstMidEndSepPunct{\mcitedefaultmidpunct}
{\mcitedefaultendpunct}{\mcitedefaultseppunct}\relax
\EndOfBibitem
\bibitem{PDG_2016}
C.~Patrignani {\em et al.} ({Particle Data Group}){,}
  \href{http://dx.doi.org/10.1088/1674-1137/40/10/100001}{Chin.\ Phys.\ {\bf
  C40},  100001}  (2016)\relax
\mciteBstWouldAddEndPuncttrue
\mciteSetBstMidEndSepPunct{\mcitedefaultmidpunct}
{\mcitedefaultendpunct}{\mcitedefaultseppunct}\relax
\EndOfBibitem
\bibitem{Alexander:2000tb}
J.~P.\ Alexander {\em et al.} ({CLEO} collaboration){,}
  \href{http://dx.doi.org/10.1103/PhysRevLett.86.2737}{Phys.\ Rev.\ Lett.\ {\bf
  86},  2737}  (2001), \href{http://arxiv.org/abs/hep-ex/0006002}{{\tt
  arXiv:hep-ex/0006002 [hep-ex]}}\relax
\mciteBstWouldAddEndPuncttrue
\mciteSetBstMidEndSepPunct{\mcitedefaultmidpunct}
{\mcitedefaultendpunct}{\mcitedefaultseppunct}\relax
\EndOfBibitem
\bibitem{Athar:2002mr}
S.~B.\ Athar {\em et al.} ({CLEO} collaboration){,}
  \href{http://dx.doi.org/10.1103/PhysRevD.66.052003}{Phys.\ Rev.\ {\bf D66}{,}
  052003}  (2002), \href{http://arxiv.org/abs/hep-ex/0202033}{{\tt
  arXiv:hep-ex/0202033 [hep-ex]}}\relax
\mciteBstWouldAddEndPuncttrue
\mciteSetBstMidEndSepPunct{\mcitedefaultmidpunct}
{\mcitedefaultendpunct}{\mcitedefaultseppunct}\relax
\EndOfBibitem
\bibitem{Hastings:2002ff}
N.~C.\ Hastings {\em et al.} ({Belle} collaboration){,}
  \href{http://dx.doi.org/10.1103/PhysRevD.67.052004}{Phys.\ Rev.\ {\bf D67}{,}
  052004}  (2003), \href{http://arxiv.org/abs/hep-ex/0212033}{{\tt
  arXiv:hep-ex/0212033 [hep-ex]}}\relax
\mciteBstWouldAddEndPuncttrue
\mciteSetBstMidEndSepPunct{\mcitedefaultmidpunct}
{\mcitedefaultendpunct}{\mcitedefaultseppunct}\relax
\EndOfBibitem
\bibitem{Aubert:2004rz}
B.~Aubert {\em et al.} ({\babar} collaboration){,}
  \href{http://dx.doi.org/10.1103/PhysRevLett.94.141801}{Phys.\ Rev.\ Lett.\ {\bf
  94},  141801}  (2005), \href{http://arxiv.org/abs/hep-ex/0412062}{{\tt
  arXiv:hep-ex/0412062 [hep-ex]}}\relax
\mciteBstWouldAddEndPuncttrue
\mciteSetBstMidEndSepPunct{\mcitedefaultmidpunct}
{\mcitedefaultendpunct}{\mcitedefaultseppunct}\relax
\EndOfBibitem
\bibitem{Barish:1994mu}
B.~Barish {\em et al.} ({CLEO} collaboration){,}
  \href{http://dx.doi.org/10.1103/PhysRevD.51.1014}{Phys.\ Rev.\ {\bf D51}{,}
  1014}  (1995), \href{http://arxiv.org/abs/hep-ex/9406005}{{\tt
  arXiv:hep-ex/9406005 [hep-ex]}}\relax
\mciteBstWouldAddEndPuncttrue
\mciteSetBstMidEndSepPunct{\mcitedefaultmidpunct}
{\mcitedefaultendpunct}{\mcitedefaultseppunct}\relax
\EndOfBibitem
\bibitem{Aubert:2005bq}
B.~Aubert {\em et al.} ({\babar} collaboration){,}
  \href{http://dx.doi.org/10.1103/PhysRevLett.95.042001}{Phys.\ Rev.\ Lett.\ {\bf
  95},  042001}  (2005), \href{http://arxiv.org/abs/hep-ex/0504001}{{\tt
  arXiv:hep-ex/0504001 [hep-ex]}}\relax
\mciteBstWouldAddEndPuncttrue
\mciteSetBstMidEndSepPunct{\mcitedefaultmidpunct}
{\mcitedefaultendpunct}{\mcitedefaultseppunct}\relax
\EndOfBibitem
\bibitem{Aubert:2006bm}
B.~Aubert {\em et al.} ({\babar} collaboration){,}
  \href{http://dx.doi.org/10.1103/PhysRevLett.96.232001}{Phys.\ Rev.\ Lett.\ {\bf
  96},  232001}  (2006), \href{http://arxiv.org/abs/hep-ex/0604031}{{\tt
  arXiv:hep-ex/0604031 [hep-ex]}}\relax
\mciteBstWouldAddEndPuncttrue
\mciteSetBstMidEndSepPunct{\mcitedefaultmidpunct}
{\mcitedefaultendpunct}{\mcitedefaultseppunct}\relax
\EndOfBibitem
\bibitem{Sokolov:2006sd}
A.~Sokolov {\em et al.} ({Belle} collaboration){,}
  \href{http://dx.doi.org/10.1103/PhysRevD.75.071103}{Phys.\ Rev.\ {\bf D75}{,}
  071103}  (2007), \href{http://arxiv.org/abs/hep-ex/0611026}{{\tt
  arXiv:hep-ex/0611026 [hep-ex]}}\relax
\mciteBstWouldAddEndPuncttrue
\mciteSetBstMidEndSepPunct{\mcitedefaultmidpunct}
{\mcitedefaultendpunct}{\mcitedefaultseppunct}\relax
\EndOfBibitem
\bibitem{Aubert:2008az}
B.~Aubert {\em et al.} ({\babar} collaboration){,}
  \href{http://dx.doi.org/10.1103/PhysRevD.78.112002}{Phys.\ Rev.\ {\bf D78}{,}
  112002}  (2008), \href{http://arxiv.org/abs/0807.2014}{{\tt arXiv:0807.2014
  [hep-ex]}}\relax
\mciteBstWouldAddEndPuncttrue
\mciteSetBstMidEndSepPunct{\mcitedefaultmidpunct}
{\mcitedefaultendpunct}{\mcitedefaultseppunct}\relax
\EndOfBibitem
\bibitem{Barish:1995cx}
B.~Barish {\em et al.} ({CLEO} collaboration){,}
  \href{http://dx.doi.org/10.1103/PhysRevLett.76.1570}{Phys.\ Rev.\ Lett.\ {\bf
  76},  1570}  (1996)\relax
\mciteBstWouldAddEndPuncttrue
\mciteSetBstMidEndSepPunct{\mcitedefaultmidpunct}
{\mcitedefaultendpunct}{\mcitedefaultseppunct}\relax
\EndOfBibitem
\bibitem{Drutskoy:2010an}
A.~Drutskoy {\em et al.} ({Belle} collaboration){,}
  \href{http://dx.doi.org/10.1103/PhysRevD.81.112003}{Phys.\ Rev.\ {\bf D81}{,}
  112003}  (2010), \href{http://arxiv.org/abs/1003.5885}{{\tt arXiv:1003.5885
  [hep-ex]}}\relax
\mciteBstWouldAddEndPuncttrue
\mciteSetBstMidEndSepPunct{\mcitedefaultmidpunct}
{\mcitedefaultendpunct}{\mcitedefaultseppunct}\relax
\EndOfBibitem
\bibitem{Huang:2006em}
G.~S.\ Huang {\em et al.} ({CLEO} collaboration){,}
  \href{http://dx.doi.org/10.1103/PhysRevD.75.012002}{Phys.\ Rev.\ {\bf D75}{,}
  012002}  (2007), \href{http://arxiv.org/abs/hep-ex/0610035}{{\tt
  arXiv:hep-ex/0610035 [hep-ex]}}\relax
\mciteBstWouldAddEndPuncttrue
\mciteSetBstMidEndSepPunct{\mcitedefaultmidpunct}
{\mcitedefaultendpunct}{\mcitedefaultseppunct}\relax
\EndOfBibitem
\bibitem{Drutskoy:2006fg}
A.~Drutskoy {\em et al.} ({Belle} collaboration){,}
  \href{http://dx.doi.org/10.1103/PhysRevLett.98.052001}{Phys.\ Rev.\ Lett.\ {\bf
  98},  052001}  (2007), \href{http://arxiv.org/abs/hep-ex/0608015}{{\tt
  arXiv:hep-ex/0608015 [hep-ex]}}\relax
\mciteBstWouldAddEndPuncttrue
\mciteSetBstMidEndSepPunct{\mcitedefaultmidpunct}
{\mcitedefaultendpunct}{\mcitedefaultseppunct}\relax
\EndOfBibitem
\bibitem{Esen:2012yz}
S.~Esen {\em et al.} ({Belle} collaboration){,}
  \href{http://dx.doi.org/10.1103/PhysRevD.87.031101}{Phys.\ Rev.\ {\bf D87}{,}
  031101}  (2013), \href{http://arxiv.org/abs/1208.0323}{{\tt
  arXiv:1208.0323}}\relax
\mciteBstWouldAddEndPuncttrue
\mciteSetBstMidEndSepPunct{\mcitedefaultmidpunct}
{\mcitedefaultendpunct}{\mcitedefaultseppunct}\relax
\EndOfBibitem
\bibitem{Artuso:2005xw}
M.~Artuso {\em et al.} ({CLEO} collaboration){,}
  \href{http://dx.doi.org/10.1103/PhysRevLett.95.261801}{Phys.\ Rev.\ Lett.\ {\bf
  95},  261801}  (2005), \href{http://arxiv.org/abs/hep-ex/0508047}{{\tt
  arXiv:hep-ex/0508047}}\relax
\mciteBstWouldAddEndPuncttrue
\mciteSetBstMidEndSepPunct{\mcitedefaultmidpunct}
{\mcitedefaultendpunct}{\mcitedefaultseppunct}\relax
\EndOfBibitem
\bibitem{Abe:2007tk}
K.~F.\ Chen {\em et al.} ({Belle} collaboration){,}
  \href{http://dx.doi.org/10.1103/PhysRevLett.100.112001}{Phys.\ Rev.\ Lett.\ {\bf
  100},  112001}  (2008), \href{http://arxiv.org/abs/0710.2577}{{\tt
  arXiv:0710.2577 [hep-ex]}}\relax
\mciteBstWouldAddEndPuncttrue
\mciteSetBstMidEndSepPunct{\mcitedefaultmidpunct}
{\mcitedefaultendpunct}{\mcitedefaultseppunct}\relax
\EndOfBibitem
\bibitem{Garmash:2014dhx}
A.~Garmash {\em et al.} ({Belle} collaboration){,}
  \href{http://dx.doi.org/10.1103/PhysRevD.91.072003}{Phys.\ Rev.\ {\bf D91}{,}
  072003}  (2015), \href{http://arxiv.org/abs/1403.0992}{{\tt arXiv:1403.0992
  [hep-ex]}}\relax
\mciteBstWouldAddEndPuncttrue
\mciteSetBstMidEndSepPunct{\mcitedefaultmidpunct}
{\mcitedefaultendpunct}{\mcitedefaultseppunct}\relax
\EndOfBibitem
\bibitem{Adachi:2011ji}
I.~Adachi {\em et al.} ({Belle} collaboration){,}
  \href{http://dx.doi.org/10.1103/PhysRevLett.108.032001}{Phys.\ Rev.\ Lett.\ {\bf
  108},  032001}  (2012), \href{http://arxiv.org/abs/1103.3419}{{\tt
  arXiv:1103.3419 [hep-ex]}}\relax
\mciteBstWouldAddEndPuncttrue
\mciteSetBstMidEndSepPunct{\mcitedefaultmidpunct}
{\mcitedefaultendpunct}{\mcitedefaultseppunct}\relax
\EndOfBibitem
\bibitem{Krokovny:2013mgx}
P.~Krokovny {\em et al.} ({Belle} collaboration){,}
  \href{http://dx.doi.org/10.1103/PhysRevD.88.052016}{Phys.\ Rev.\ {\bf D88}{,}
  052016}  (2013), \href{http://arxiv.org/abs/1308.2646}{{\tt arXiv:1308.2646
  [hep-ex]}}\relax
\mciteBstWouldAddEndPuncttrue
\mciteSetBstMidEndSepPunct{\mcitedefaultmidpunct}
{\mcitedefaultendpunct}{\mcitedefaultseppunct}\relax
\EndOfBibitem
\bibitem{thesis_Louvot}
 R.~Louvot, PhD thesis \#5213, EPFL, Lausanne, 2012{,}
  \url{{http://dx.doi.org/10.5075/epfl-thesis-5213}}\relax
\mciteBstWouldAddEndPuncttrue
\mciteSetBstMidEndSepPunct{\mcitedefaultmidpunct}
{\mcitedefaultendpunct}{\mcitedefaultseppunct}\relax
\EndOfBibitem
\bibitem{Lees:2011ji}
J.~P.\ Lees {\em et al.} ({\babar} collaboration){,}
  \href{http://dx.doi.org/10.1103/PhysRevD.85.011101}{Phys.\ Rev.\ {\bf D85}{,}
  011101}  (2012), \href{http://arxiv.org/abs/1110.5600}{{\tt arXiv:1110.5600
  [hep-ex]}}\relax
\mciteBstWouldAddEndPuncttrue
\mciteSetBstMidEndSepPunct{\mcitedefaultmidpunct}
{\mcitedefaultendpunct}{\mcitedefaultseppunct}\relax
\EndOfBibitem
\bibitem{Li:2011pg}
J.~Li {\em et al.} ({Belle} collaboration){,}
  \href{http://dx.doi.org/10.1103/PhysRevLett.106.121802}{Phys.\ Rev.\ Lett.\ {\bf
  106},  121802}  (2011), \href{http://arxiv.org/abs/1102.2759}{{\tt
  arXiv:1102.2759 [hep-ex]}}\relax
\mciteBstWouldAddEndPuncttrue
\mciteSetBstMidEndSepPunct{\mcitedefaultmidpunct}
{\mcitedefaultendpunct}{\mcitedefaultseppunct}\relax
\EndOfBibitem
\bibitem{Louvot:2008sc}
R.~Louvot {\em et al.} ({Belle} collaboration){,}
  \href{http://dx.doi.org/10.1103/PhysRevLett.102.021801}{Phys.\ Rev.\ Lett.\ {\bf
  102},  021801}  (2009), \href{http://arxiv.org/abs/0809.2526}{{\tt
  arXiv:0809.2526 [hep-ex]}}\relax
\mciteBstWouldAddEndPuncttrue
\mciteSetBstMidEndSepPunct{\mcitedefaultmidpunct}
{\mcitedefaultendpunct}{\mcitedefaultseppunct}\relax
\EndOfBibitem
\bibitem{Abreu:1992rv}
P.~Abreu {\em et al.} ({DELPHI} collaboration){,}
  \href{http://dx.doi.org/10.1016/0370-2693(92)91385-M}{Phys.\ Lett.\ {\bf B289}{,}
   199}  (1992)\relax
\mciteBstWouldAddEndPuncttrue
\mciteSetBstMidEndSepPunct{\mcitedefaultmidpunct}
{\mcitedefaultendpunct}{\mcitedefaultseppunct}\relax
\EndOfBibitem
\bibitem{Acton:1992zq}
P.~D.\ Acton {\em et al.} ({OPAL} collaboration){,}
  \href{http://dx.doi.org/10.1016/0370-2693(92)91578-W}{Phys.\ Lett.\ {\bf B295}{,}
   357}  (1992)\relax
\mciteBstWouldAddEndPuncttrue
\mciteSetBstMidEndSepPunct{\mcitedefaultmidpunct}
{\mcitedefaultendpunct}{\mcitedefaultseppunct}\relax
\EndOfBibitem
\bibitem{Buskulic:1995bd}
D.~Buskulic {\em et al.} ({ALEPH} collaboration){,}
  \href{http://dx.doi.org/10.1016/0370-2693(95)01173-N}{Phys.\ Lett.\ {\bf B361}{,}
   221}  (1995)\relax
\mciteBstWouldAddEndPuncttrue
\mciteSetBstMidEndSepPunct{\mcitedefaultmidpunct}
{\mcitedefaultendpunct}{\mcitedefaultseppunct}\relax
\EndOfBibitem
\bibitem{Abreu:1995me}
P.~Abreu {\em et al.} ({DELPHI} collaboration){,}
  \href{http://dx.doi.org/10.1007/BF01620713}{Z.\ Phys.\ {\bf C68},  375}
  (1995)\relax
\mciteBstWouldAddEndPuncttrue
\mciteSetBstMidEndSepPunct{\mcitedefaultmidpunct}
{\mcitedefaultendpunct}{\mcitedefaultseppunct}\relax
\EndOfBibitem
\bibitem{Barate:1997if}
R.~Barate {\em et al.} ({ALEPH} collaboration){,}
  \href{http://dx.doi.org/10.1007/s100520050133}{Eur.\ Phys.\ J.\ {\bf C2},  197}
   (1998)\relax
\mciteBstWouldAddEndPuncttrue
\mciteSetBstMidEndSepPunct{\mcitedefaultmidpunct}
{\mcitedefaultendpunct}{\mcitedefaultseppunct}\relax
\EndOfBibitem
\bibitem{Buskulic:1996sm}
D.~Buskulic {\em et al.} ({ALEPH} collaboration){,}
  \href{http://dx.doi.org/10.1016/0370-2693(96)00925-2}{Phys.\ Lett.\ {\bf B384}{,}
   449}  (1996)\relax
\mciteBstWouldAddEndPuncttrue
\mciteSetBstMidEndSepPunct{\mcitedefaultmidpunct}
{\mcitedefaultendpunct}{\mcitedefaultseppunct}\relax
\EndOfBibitem
\bibitem{Abdallah:2005cw}
J.~Abdallah {\em et al.} ({DELPHI} collaboration){,}
  \href{http://dx.doi.org/10.1140/epjc/s2005-02388-4}{Eur.\ Phys.\ J.\ {\bf C44}{,}
  299}  (2005), \href{http://arxiv.org/abs/hep-ex/0510023}{{\tt
  arXiv:hep-ex/0510023 [hep-ex]}}\relax
\mciteBstWouldAddEndPuncttrue
\mciteSetBstMidEndSepPunct{\mcitedefaultmidpunct}
{\mcitedefaultendpunct}{\mcitedefaultseppunct}\relax
\EndOfBibitem
\bibitem{Barate:1997ty}
R.~Barate {\em et al.} ({ALEPH} collaboration){,}
  \href{http://dx.doi.org/10.1007/s100520050263}{Eur.\ Phys.\ J.\ {\bf C5},  205}
   (1998)\relax
\mciteBstWouldAddEndPuncttrue
\mciteSetBstMidEndSepPunct{\mcitedefaultmidpunct}
{\mcitedefaultendpunct}{\mcitedefaultseppunct}\relax
\EndOfBibitem
\bibitem{Abdallah:2003xp}
J.~Abdallah {\em et al.} ({DELPHI} collaboration){,}
  \href{http://dx.doi.org/10.1016/j.physletb.2003.09.070}{Phys.\ Lett.\ {\bf
  B576},  29}  (2003), \href{http://arxiv.org/abs/hep-ex/0311005}{{\tt
  arXiv:hep-ex/0311005 [hep-ex]}}\relax
\mciteBstWouldAddEndPuncttrue
\mciteSetBstMidEndSepPunct{\mcitedefaultmidpunct}
{\mcitedefaultendpunct}{\mcitedefaultseppunct}\relax
\EndOfBibitem
\bibitem{Affolder:1999iq}
T.~Affolder {\em et al.} ({CDF} collaboration){,}
  \href{http://dx.doi.org/10.1103/PhysRevLett.84.1663}{Phys.\ Rev.\ Lett.\ {\bf
  84},  1663}  (2000), \href{http://arxiv.org/abs/hep-ex/9909011}{{\tt
  arXiv:hep-ex/9909011 [hep-ex]}}\relax
\mciteBstWouldAddEndPuncttrue
\mciteSetBstMidEndSepPunct{\mcitedefaultmidpunct}
{\mcitedefaultendpunct}{\mcitedefaultseppunct}\relax
\EndOfBibitem
\bibitem{Aaltonen:2008zd}
T.~Aaltonen {\em et al.} ({CDF} collaboration){,}
  \href{http://dx.doi.org/10.1103/PhysRevD.77.072003}{Phys.\ Rev.\ {\bf D77}{,}
  072003}  (2008), \href{http://arxiv.org/abs/0801.4375}{{\tt arXiv:0801.4375
  [hep-ex]}}\relax
\mciteBstWouldAddEndPuncttrue
\mciteSetBstMidEndSepPunct{\mcitedefaultmidpunct}
{\mcitedefaultendpunct}{\mcitedefaultseppunct}\relax
\EndOfBibitem
\bibitem{Aaltonen:2008eu}
T.~Aaltonen {\em et al.} ({CDF} collaboration){,}
  \href{http://dx.doi.org/10.1103/PhysRevD.79.032001}{Phys.\ Rev.\ {\bf D79}{,}
  032001}  (2009), \href{http://arxiv.org/abs/0810.3213}{{\tt arXiv:0810.3213
  [hep-ex]}}\relax
\mciteBstWouldAddEndPuncttrue
\mciteSetBstMidEndSepPunct{\mcitedefaultmidpunct}
{\mcitedefaultendpunct}{\mcitedefaultseppunct}\relax
\EndOfBibitem
\bibitem{Abe:1999ta}
F.~Abe {\em et al.} ({CDF} collaboration){,}
  \href{http://dx.doi.org/10.1103/PhysRevD.60.092005}{Phys.\ Rev.\ {\bf D60}{,}
  092005}  (1999)\relax
\mciteBstWouldAddEndPuncttrue
\mciteSetBstMidEndSepPunct{\mcitedefaultmidpunct}
{\mcitedefaultendpunct}{\mcitedefaultseppunct}\relax
\EndOfBibitem
\bibitem{Abazov:2007am}
V.~M.\ Abazov {\em et al.} ({\dzero} collaboration){,}
  \href{http://dx.doi.org/10.1103/PhysRevLett.99.052001}{Phys.\ Rev.\ Lett.\ {\bf
  99},  052001}  (2007), \href{http://arxiv.org/abs/0706.1690}{{\tt
  arXiv:0706.1690 [hep-ex]}}\relax
\mciteBstWouldAddEndPuncttrue
\mciteSetBstMidEndSepPunct{\mcitedefaultmidpunct}
{\mcitedefaultendpunct}{\mcitedefaultseppunct}\relax
\EndOfBibitem
\bibitem{Abazov:2008qm}
V.~M.\ Abazov {\em et al.} ({\dzero} collaboration){,}
  \href{http://dx.doi.org/10.1103/PhysRevLett.101.232002}{Phys.\ Rev.\ Lett.\ {\bf
  101},  232002}  (2008), \href{http://arxiv.org/abs/0808.4142}{{\tt
  arXiv:0808.4142 [hep-ex]}}\relax
\mciteBstWouldAddEndPuncttrue
\mciteSetBstMidEndSepPunct{\mcitedefaultmidpunct}
{\mcitedefaultendpunct}{\mcitedefaultseppunct}\relax
\EndOfBibitem
\bibitem{Aaltonen:2009ny}
T.~Aaltonen {\em et al.} ({CDF} collaboration){,}
  \href{http://dx.doi.org/10.1103/PhysRevD.80.072003}{Phys.\ Rev.\ {\bf D80}{,}
  072003}  (2009), \href{http://arxiv.org/abs/0905.3123}{{\tt arXiv:0905.3123
  [hep-ex]}}\relax
\mciteBstWouldAddEndPuncttrue
\mciteSetBstMidEndSepPunct{\mcitedefaultmidpunct}
{\mcitedefaultendpunct}{\mcitedefaultseppunct}\relax
\EndOfBibitem
\bibitem{Aaij:2011jp}
R.~Aaij {\em et al.} ({LHCb} collaboration){,}
  \href{http://dx.doi.org/10.1103/PhysRevD.85.032008}{Phys.\ Rev.\ {\bf D85}{,}
  032008}  (2012), \href{http://arxiv.org/abs/1111.2357}{{\tt arXiv:1111.2357
  [hep-ex]}}, with numerical results and full covariance matrix available at
  \url{https://cdsweb.cern.ch/record/1390838}\relax
\mciteBstWouldAddEndPuncttrue
\mciteSetBstMidEndSepPunct{\mcitedefaultmidpunct}
{\mcitedefaultendpunct}{\mcitedefaultseppunct}\relax
\EndOfBibitem
\bibitem{Aaij:2013qqa}
R.~Aaij {\em et al.} ({LHCb} collaboration){,}
  \href{http://dx.doi.org/10.1007/JHEP04(2013)001}{JHEP {\bf 04},  001}
  (2013), \href{http://arxiv.org/abs/1301.5286}{{\tt arXiv:1301.5286
  [hep-ex]}}, with numerical results and full covariance matrix available at
  \url{http://cdsweb.cern.ch/record/1507868}\relax
\mciteBstWouldAddEndPuncttrue
\mciteSetBstMidEndSepPunct{\mcitedefaultmidpunct}
{\mcitedefaultendpunct}{\mcitedefaultseppunct}\relax
\EndOfBibitem
\bibitem{Aaij:2014jyk}
R.~Aaij {\em et al.} ({LHCb} collaboration){,}
  \href{http://dx.doi.org/10.1007/JHEP08(2014)143}{JHEP {\bf 08},  143}
  (2014), \href{http://arxiv.org/abs/1405.6842}{{\tt arXiv:1405.6842
  [hep-ex]}}\relax
\mciteBstWouldAddEndPuncttrue
\mciteSetBstMidEndSepPunct{\mcitedefaultmidpunct}
{\mcitedefaultendpunct}{\mcitedefaultseppunct}\relax
\EndOfBibitem
\bibitem{Aad:2015cda}
G.~Aad {\em et al.} ({ATLAS} collaboration){,}
  \href{http://dx.doi.org/10.1103/PhysRevLett.115.262001}{Phys.\ Rev.\ Lett.\ {\bf
  115},  262001}  (2015), \href{http://arxiv.org/abs/1507.08925}{{\tt
  arXiv:1507.08925 [hep-ex]}}\relax
\mciteBstWouldAddEndPuncttrue
\mciteSetBstMidEndSepPunct{\mcitedefaultmidpunct}
{\mcitedefaultendpunct}{\mcitedefaultseppunct}\relax
\EndOfBibitem
\bibitem{Liu:2013nea}
X.~Liu, W.~Wang, and Y.~Xie{,}
  \href{http://dx.doi.org/10.1103/PhysRevD.89.094010}{Phys.\ Rev.\ {\bf D89}{,}
  094010}  (2014), \href{http://arxiv.org/abs/1309.0313}{{\tt arXiv:1309.0313
  [hep-ph]}}\relax
\mciteBstWouldAddEndPuncttrue
\mciteSetBstMidEndSepPunct{\mcitedefaultmidpunct}
{\mcitedefaultendpunct}{\mcitedefaultseppunct}\relax
\EndOfBibitem
\bibitem{ALEPH:2005ab}
S.~Schael {\em et al.} ({ALEPH, CDF, DELPHI, L3, OPAL, and SLD collaborations{,}
  LEP electroweak working group, SLD electroweak and heavy flavour working
  groups}), \href{http://dx.doi.org/10.1016/j.physrep.2005.12.006}{Phys.\ Rept.\
  {\bf 427},  257}  (2006), \href{http://arxiv.org/abs/hep-ex/0509008}{{\tt
  arXiv:hep-ex/0509008 [hep-ex]}}\relax
\mciteBstWouldAddEndPuncttrue
\mciteSetBstMidEndSepPunct{\mcitedefaultmidpunct}
{\mcitedefaultendpunct}{\mcitedefaultseppunct}\relax
\EndOfBibitem
\bibitem{Abazov:2006qw}
V.~M.\ Abazov {\em et al.} ({\dzero} collaboration){,}
  \href{http://dx.doi.org/10.1103/PhysRevD.74.092001}{Phys.\ Rev.\ {\bf D74}{,}
  092001}  (2006), \href{http://arxiv.org/abs/hep-ex/0609014}{{\tt
  arXiv:hep-ex/0609014 [hep-ex]}}\relax
\mciteBstWouldAddEndPuncttrue
\mciteSetBstMidEndSepPunct{\mcitedefaultmidpunct}
{\mcitedefaultendpunct}{\mcitedefaultseppunct}\relax
\EndOfBibitem
\bibitem{Acosta:2003ie}
D.~Acosta {\em et al.} ({CDF} collaboration){,}
  \href{http://dx.doi.org/10.1103/PhysRevD.69.012002}{Phys.\ Rev.\ {\bf D69}{,}
  012002}  (2004), \href{http://arxiv.org/abs/hep-ex/0309030}{{\tt
  arXiv:hep-ex/0309030 [hep-ex]}}\relax
\mciteBstWouldAddEndPuncttrue
\mciteSetBstMidEndSepPunct{\mcitedefaultmidpunct}
{\mcitedefaultendpunct}{\mcitedefaultseppunct}\relax
\EndOfBibitem
\bibitem{LHCb-CONF-2013-011}
{LHCb} collaboration, LHCb-CONF-2013-011, 2013{,}
  \url{{https://cdsweb.cern.ch/record/1559262}}\relax
\mciteBstWouldAddEndPuncttrue
\mciteSetBstMidEndSepPunct{\mcitedefaultmidpunct}
{\mcitedefaultendpunct}{\mcitedefaultseppunct}\relax
\EndOfBibitem
\bibitem{Shifman:1986mx}
M.~A.\ Shifman and M.~B.\ Voloshin, Sov.\ Phys.\ JETP {\bf 64},  698  (1986)\relax
\mciteBstWouldAddEndPuncttrue
\mciteSetBstMidEndSepPunct{\mcitedefaultmidpunct}
{\mcitedefaultendpunct}{\mcitedefaultseppunct}\relax
\EndOfBibitem
\bibitem{Chay:1990da}
J.~Chay, H.~Georgi, and B.~Grinstein{,}
  \href{http://dx.doi.org/10.1016/0370-2693(90)90916-T}{Phys.\ Lett.\ {\bf B247}{,}
   399}  (1990)\relax
\mciteBstWouldAddEndPuncttrue
\mciteSetBstMidEndSepPunct{\mcitedefaultmidpunct}
{\mcitedefaultendpunct}{\mcitedefaultseppunct}\relax
\EndOfBibitem
\bibitem{Bigi:1992su}
I.~I.\ Bigi, N.~G.\ Uraltsev, and A.~I.\ Vainshtein{,}
  \href{http://dx.doi.org/10.1016/0370-2693(92)90908-M}{Phys.\ Lett.\ {\bf B293}{,}
   430}  (1992), \href{http://arxiv.org/abs/hep-ph/9207214}{{\tt
  arXiv:hep-ph/9207214 [hep-ph]}}, Erratum ibid.\
  \href{http://dx.doi.org/10.1016/0370-2693(92)91287-J}{{\bf B297}, 477}
  (1992)\relax
\mciteBstWouldAddEndPuncttrue
\mciteSetBstMidEndSepPunct{\mcitedefaultmidpunct}
{\mcitedefaultendpunct}{\mcitedefaultseppunct}\relax
\EndOfBibitem
\bibitem{Shifman:2000jv}
M.~A.\ Shifman, \href{http://arxiv.org/abs/hep-ph/0009131}{{\tt
  arXiv:hep-ph/0009131 [hep-ph]}}  (2000)\relax
\mciteBstWouldAddEndPuncttrue
\mciteSetBstMidEndSepPunct{\mcitedefaultmidpunct}
{\mcitedefaultendpunct}{\mcitedefaultseppunct}\relax
\EndOfBibitem
\bibitem{Bigi:2001ys}
I.~I.~Y.\ Bigi and N.~Uraltsev{,}
  \href{http://dx.doi.org/10.1142/S0217751X01005535}{Int.\ J.\ Mod.\ Phys.\ {\bf
  A16},  5201}  (2001), \href{http://arxiv.org/abs/hep-ph/0106346}{{\tt
  arXiv:hep-ph/0106346 [hep-ph]}}\relax
\mciteBstWouldAddEndPuncttrue
\mciteSetBstMidEndSepPunct{\mcitedefaultmidpunct}
{\mcitedefaultendpunct}{\mcitedefaultseppunct}\relax
\EndOfBibitem
\bibitem{Jubb:2016mvq}
T.~Jubb, M.~Kirk, A.~Lenz, and G.~Tetlalmatzi-Xolocotzi{,}
  \href{http://dx.doi.org/10.1016/j.nuclphysb.2016.12.020}{Nucl.\ Phys.\ {\bf
  B915},  431}  (2017), \href{http://arxiv.org/abs/1603.07770}{{\tt
  arXiv:1603.07770 [hep-ph]}}\relax
\mciteBstWouldAddEndPuncttrue
\mciteSetBstMidEndSepPunct{\mcitedefaultmidpunct}
{\mcitedefaultendpunct}{\mcitedefaultseppunct}\relax
\EndOfBibitem
\bibitem{Wilson:1969zs}
K.~G.\ Wilson, \href{http://dx.doi.org/10.1103/PhysRev.179.1499}{Phys.\ Rev.\ {\bf
  179},  1499}  (1969)\relax
\mciteBstWouldAddEndPuncttrue
\mciteSetBstMidEndSepPunct{\mcitedefaultmidpunct}
{\mcitedefaultendpunct}{\mcitedefaultseppunct}\relax
\EndOfBibitem
\bibitem{Ciuchini:2001vx}
M.~Ciuchini, E.~Franco, V.~Lubicz, and F.~Mescia{,}
  \href{http://dx.doi.org/10.1016/S0550-3213(02)00006-8}{Nucl.\ Phys.\ {\bf
  B625},  211}  (2002), \href{http://arxiv.org/abs/hep-ph/0110375}{{\tt
  arXiv:hep-ph/0110375 [hep-ph]}}\relax
\mciteBstWouldAddEndPuncttrue
\mciteSetBstMidEndSepPunct{\mcitedefaultmidpunct}
{\mcitedefaultendpunct}{\mcitedefaultseppunct}\relax
\EndOfBibitem
\bibitem{Beneke:2002rj}
M.~Beneke, G.~Buchalla, C.~Greub, A.~Lenz, and U.~Nierste{,}
  \href{http://dx.doi.org/10.1016/S0550-3213(02)00561-8}{Nucl.\ Phys.\ {\bf
  B639},  389}  (2002), \href{http://arxiv.org/abs/hep-ph/0202106}{{\tt
  arXiv:hep-ph/0202106 [hep-ph]}}\relax
\mciteBstWouldAddEndPuncttrue
\mciteSetBstMidEndSepPunct{\mcitedefaultmidpunct}
{\mcitedefaultendpunct}{\mcitedefaultseppunct}\relax
\EndOfBibitem
\bibitem{Franco:2002fc}
E.~Franco, V.~Lubicz, F.~Mescia, and C.~Tarantino{,}
  \href{http://dx.doi.org/10.1016/S0550-3213(02)00262-6}{Nucl.\ Phys.\ {\bf
  B633},  212}  (2002), \href{http://arxiv.org/abs/hep-ph/0203089}{{\tt
  arXiv:hep-ph/0203089 [hep-ph]}}\relax
\mciteBstWouldAddEndPuncttrue
\mciteSetBstMidEndSepPunct{\mcitedefaultmidpunct}
{\mcitedefaultendpunct}{\mcitedefaultseppunct}\relax
\EndOfBibitem
\bibitem{Tarantino:2003qw}
C.~Tarantino, \href{http://dx.doi.org/10.1140/epjcd/s2003-03-1006-y}{Eur.\ Phys.\
  J.\ {\bf C33},  S895}  (2004){,}
  \href{http://arxiv.org/abs/hep-ph/0310241}{{\tt arXiv:hep-ph/0310241
  [hep-ph]}}\relax
\mciteBstWouldAddEndPuncttrue
\mciteSetBstMidEndSepPunct{\mcitedefaultmidpunct}
{\mcitedefaultendpunct}{\mcitedefaultseppunct}\relax
\EndOfBibitem
\bibitem{Gabbiani:2003pq}
F.~Gabbiani, A.~I.\ Onishchenko, and A.~A.\ Petrov{,}
  \href{http://dx.doi.org/10.1103/PhysRevD.68.114006}{Phys.\ Rev.\ {\bf D68}{,}
  114006}  (2003), \href{http://arxiv.org/abs/hep-ph/0303235}{{\tt
  arXiv:hep-ph/0303235 [hep-ph]}}\relax
\mciteBstWouldAddEndPuncttrue
\mciteSetBstMidEndSepPunct{\mcitedefaultmidpunct}
{\mcitedefaultendpunct}{\mcitedefaultseppunct}\relax
\EndOfBibitem
\bibitem{Gabbiani:2004tp}
F.~Gabbiani, A.~I.\ Onishchenko, and A.~A.\ Petrov{,}
  \href{http://dx.doi.org/10.1103/PhysRevD.70.094031}{Phys.\ Rev.\ {\bf D70}{,}
  094031}  (2004), \href{http://arxiv.org/abs/hep-ph/0407004}{{\tt
  arXiv:hep-ph/0407004 [hep-ph]}}\relax
\mciteBstWouldAddEndPuncttrue
\mciteSetBstMidEndSepPunct{\mcitedefaultmidpunct}
{\mcitedefaultendpunct}{\mcitedefaultseppunct}\relax
\EndOfBibitem
\bibitem{Lenz:2015dra}
A.~Lenz, \href{http://dx.doi.org/10.1142/S0217751X15430058}{Int.\ J.\ Mod.\ Phys.\
  {\bf A30},  1543005}  (2015)\relax
\mciteBstWouldAddEndPuncttrue
\mciteSetBstMidEndSepPunct{\mcitedefaultmidpunct}
{\mcitedefaultendpunct}{\mcitedefaultseppunct}\relax
\EndOfBibitem
\bibitem{Lenz:2014jha}
A.~Lenz, \href{http://arxiv.org/abs/1405.3601}{{\tt arXiv:1405.3601 [hep-ph]}}
   (2014)\relax
\mciteBstWouldAddEndPuncttrue
\mciteSetBstMidEndSepPunct{\mcitedefaultmidpunct}
{\mcitedefaultendpunct}{\mcitedefaultseppunct}\relax
\EndOfBibitem
\bibitem{Buskulic:1993gj}
D.~Buskulic {\em et al.} ({ALEPH} collaboration){,}
  \href{http://dx.doi.org/10.1016/0370-2693(93)91265-O}{Phys.\ Lett.\ {\bf B314}{,}
   459}  (1993)\relax
\mciteBstWouldAddEndPuncttrue
\mciteSetBstMidEndSepPunct{\mcitedefaultmidpunct}
{\mcitedefaultendpunct}{\mcitedefaultseppunct}\relax
\EndOfBibitem
\bibitem{Abreu:1994dr}
P.~Abreu {\em et al.} ({DELPHI} collaboration){,}
  \href{http://dx.doi.org/10.1007/BF01577539}{Z.\ Phys.\ {\bf C63},  3}
  (1994)\relax
\mciteBstWouldAddEndPuncttrue
\mciteSetBstMidEndSepPunct{\mcitedefaultmidpunct}
{\mcitedefaultendpunct}{\mcitedefaultseppunct}\relax
\EndOfBibitem
\bibitem{Abreu:1996hv}
P.~Abreu {\em et al.} ({DELPHI} collaboration){,}
  \href{http://dx.doi.org/10.1016/0370-2693(96)00452-2}{Phys.\ Lett.\ {\bf B377}{,}
   195}  (1996)\relax
\mciteBstWouldAddEndPuncttrue
\mciteSetBstMidEndSepPunct{\mcitedefaultmidpunct}
{\mcitedefaultendpunct}{\mcitedefaultseppunct}\relax
\EndOfBibitem
\bibitem{Abdallah:2003sb}
J.~Abdallah {\em et al.} ({DELPHI} collaboration){,}
  \href{http://dx.doi.org/10.1140/epjc/s2004-01599-5}{Eur.\ Phys.\ J.\ {\bf C33}{,}
  307}  (2004), \href{http://arxiv.org/abs/hep-ex/0401025}{{\tt
  arXiv:hep-ex/0401025 [hep-ex]}}\relax
\mciteBstWouldAddEndPuncttrue
\mciteSetBstMidEndSepPunct{\mcitedefaultmidpunct}
{\mcitedefaultendpunct}{\mcitedefaultseppunct}\relax
\EndOfBibitem
\bibitem{Acciarri:1997tt}
M.~Acciarri {\em et al.} ({L3} collaboration){,}
  \href{http://dx.doi.org/10.1016/S0370-2693(97)01379-8}{Phys.\ Lett.\ {\bf
  B416},  220}  (1998)\relax
\mciteBstWouldAddEndPuncttrue
\mciteSetBstMidEndSepPunct{\mcitedefaultmidpunct}
{\mcitedefaultendpunct}{\mcitedefaultseppunct}\relax
\EndOfBibitem
\bibitem{Ackerstaff:1996as}
K.~Ackerstaff {\em et al.} ({OPAL} collaboration){,}
  \href{http://dx.doi.org/10.1007/s002880050329}{Z.\ Phys.\ {\bf C73},  397}
  (1997)\relax
\mciteBstWouldAddEndPuncttrue
\mciteSetBstMidEndSepPunct{\mcitedefaultmidpunct}
{\mcitedefaultendpunct}{\mcitedefaultseppunct}\relax
\EndOfBibitem
\bibitem{Abe:1995rm}
K.~Abe {\em et al.} ({SLD} collaboration){,}
  \href{http://dx.doi.org/10.1103/PhysRevLett.75.3624}{Phys.\ Rev.\ Lett.\ {\bf
  75},  3624}  (1995), \href{http://arxiv.org/abs/hep-ex/9511005}{{\tt
  arXiv:hep-ex/9511005 [hep-ex]}}\relax
\mciteBstWouldAddEndPuncttrue
\mciteSetBstMidEndSepPunct{\mcitedefaultmidpunct}
{\mcitedefaultendpunct}{\mcitedefaultseppunct}\relax
\EndOfBibitem
\bibitem{Buskulic:1995rw}
D.~Buskulic {\em et al.} ({ALEPH} collaboration){,}
  \href{http://dx.doi.org/10.1016/0370-2693(95)01586-8}{Phys.\ Lett.\ {\bf B369}{,}
   151}  (1996)\relax
\mciteBstWouldAddEndPuncttrue
\mciteSetBstMidEndSepPunct{\mcitedefaultmidpunct}
{\mcitedefaultendpunct}{\mcitedefaultseppunct}\relax
\EndOfBibitem
\bibitem{Acton:1993xk}
P.~D.\ Acton {\em et al.} ({OPAL} collaboration){,}
  \href{http://dx.doi.org/10.1007/BF01474617}{Z.\ Phys.\ {\bf C60},  217}
  (1993)\relax
\mciteBstWouldAddEndPuncttrue
\mciteSetBstMidEndSepPunct{\mcitedefaultmidpunct}
{\mcitedefaultendpunct}{\mcitedefaultseppunct}\relax
\EndOfBibitem
\bibitem{Abe:1997bd}
F.~Abe {\em et al.} ({CDF} collaboration){,}
  \href{http://dx.doi.org/10.1103/PhysRevD.57.5382}{Phys.\ Rev.\ {\bf D57}{,}
  5382}  (1998)\relax
\mciteBstWouldAddEndPuncttrue
\mciteSetBstMidEndSepPunct{\mcitedefaultmidpunct}
{\mcitedefaultendpunct}{\mcitedefaultseppunct}\relax
\EndOfBibitem
\bibitem{Barate:2000bs}
R.~Barate {\em et al.} ({ALEPH} collaboration){,}
  \href{http://dx.doi.org/10.1016/S0370-2693(00)01093-5}{Phys.\ Lett.\ {\bf
  B492},  275}  (2000), \href{http://arxiv.org/abs/hep-ex/0008016}{{\tt
  arXiv:hep-ex/0008016 [hep-ex]}}\relax
\mciteBstWouldAddEndPuncttrue
\mciteSetBstMidEndSepPunct{\mcitedefaultmidpunct}
{\mcitedefaultendpunct}{\mcitedefaultseppunct}\relax
\EndOfBibitem
\bibitem{Buskulic:1996hy}
D.~Buskulic {\em et al.} ({ALEPH} collaboration){,}
  \href{http://dx.doi.org/10.1007/s002880050145}{Z.\ Phys.\ {\bf C71},  31}
  (1996)\relax
\mciteBstWouldAddEndPuncttrue
\mciteSetBstMidEndSepPunct{\mcitedefaultmidpunct}
{\mcitedefaultendpunct}{\mcitedefaultseppunct}\relax
\EndOfBibitem
\bibitem{Abreu:1995mc}
P.~Abreu {\em et al.} ({DELPHI} collaboration){,}
  \href{http://dx.doi.org/10.1007/BF01579800}{Z.\ Phys.\ {\bf C68},  13}
  (1995)\relax
\mciteBstWouldAddEndPuncttrue
\mciteSetBstMidEndSepPunct{\mcitedefaultmidpunct}
{\mcitedefaultendpunct}{\mcitedefaultseppunct}\relax
\EndOfBibitem
\bibitem{Adam:1995mb}
W.~Adam {\em et al.} ({DELPHI} collaboration){,}
  \href{http://dx.doi.org/10.1007/BF01620712}{Z.\ Phys.\ {\bf C68},  363}
  (1995)\relax
\mciteBstWouldAddEndPuncttrue
\mciteSetBstMidEndSepPunct{\mcitedefaultmidpunct}
{\mcitedefaultendpunct}{\mcitedefaultseppunct}\relax
\EndOfBibitem
\bibitem{Abreu:1996gb}
P.~Abreu {\em et al.} ({DELPHI} collaboration){,}
  \href{http://dx.doi.org/10.1007/s002880050367}{Z.\ Phys.\ {\bf C74},  19}
  (1997)\relax
\mciteBstWouldAddEndPuncttrue
\mciteSetBstMidEndSepPunct{\mcitedefaultmidpunct}
{\mcitedefaultendpunct}{\mcitedefaultseppunct}\relax
\EndOfBibitem
\bibitem{Acciarri:1998uv}
M.~Acciarri {\em et al.} ({L3} collaboration){,}
  \href{http://dx.doi.org/10.1016/S0370-2693(98)01114-9}{Phys.\ Lett.\ {\bf
  B438},  417}  (1998)\relax
\mciteBstWouldAddEndPuncttrue
\mciteSetBstMidEndSepPunct{\mcitedefaultmidpunct}
{\mcitedefaultendpunct}{\mcitedefaultseppunct}\relax
\EndOfBibitem
\bibitem{Akers:1995pa}
R.~Akers {\em et al.} ({OPAL} collaboration){,}
  \href{http://dx.doi.org/10.1007/BF01624581}{Z.\ Phys.\ {\bf C67},  379}
  (1995)\relax
\mciteBstWouldAddEndPuncttrue
\mciteSetBstMidEndSepPunct{\mcitedefaultmidpunct}
{\mcitedefaultendpunct}{\mcitedefaultseppunct}\relax
\EndOfBibitem
\bibitem{Abbiendi:1998av}
G.~Abbiendi {\em et al.} ({OPAL} collaboration){,}
  \href{http://dx.doi.org/10.1007/s100520000322}{Eur.\ Phys.\ J.\ {\bf C12}{,}
  609}  (2000), \href{http://arxiv.org/abs/hep-ex/9901017}{{\tt
  arXiv:hep-ex/9901017 [hep-ex]}}\relax
\mciteBstWouldAddEndPuncttrue
\mciteSetBstMidEndSepPunct{\mcitedefaultmidpunct}
{\mcitedefaultendpunct}{\mcitedefaultseppunct}\relax
\EndOfBibitem
\bibitem{Abbiendi:2000ec}
G.~Abbiendi {\em et al.} ({OPAL} collaboration){,}
  \href{http://dx.doi.org/10.1016/S0370-2693(00)01145-X}{Phys.\ Lett.\ {\bf
  B493},  266}  (2000), \href{http://arxiv.org/abs/hep-ex/0010013}{{\tt
  arXiv:hep-ex/0010013 [hep-ex]}}\relax
\mciteBstWouldAddEndPuncttrue
\mciteSetBstMidEndSepPunct{\mcitedefaultmidpunct}
{\mcitedefaultendpunct}{\mcitedefaultseppunct}\relax
\EndOfBibitem
\bibitem{Abe:1997ys}
K.~Abe {\em et al.} ({SLD} collaboration){,}
  \href{http://dx.doi.org/10.1103/PhysRevLett.79.590}{Phys.\ Rev.\ Lett.\ {\bf
  79},  590}  (1997)\relax
\mciteBstWouldAddEndPuncttrue
\mciteSetBstMidEndSepPunct{\mcitedefaultmidpunct}
{\mcitedefaultendpunct}{\mcitedefaultseppunct}\relax
\EndOfBibitem
\bibitem{Abe:1998wt}
F.~Abe {\em et al.} ({CDF} collaboration){,}
  \href{http://dx.doi.org/10.1103/PhysRevD.58.092002}{Phys.\ Rev.\ {\bf D58}{,}
  092002}  (1998), \href{http://arxiv.org/abs/hep-ex/9806018}{{\tt
  arXiv:hep-ex/9806018 [hep-ex]}}\relax
\mciteBstWouldAddEndPuncttrue
\mciteSetBstMidEndSepPunct{\mcitedefaultmidpunct}
{\mcitedefaultendpunct}{\mcitedefaultseppunct}\relax
\EndOfBibitem
\bibitem{Acosta:2002nd}
D.~Acosta {\em et al.} ({CDF} collaboration){,}
  \href{http://dx.doi.org/10.1103/PhysRevD.65.092009}{Phys.\ Rev.\ {\bf D65}{,}
  092009}  (2002)\relax
\mciteBstWouldAddEndPuncttrue
\mciteSetBstMidEndSepPunct{\mcitedefaultmidpunct}
{\mcitedefaultendpunct}{\mcitedefaultseppunct}\relax
\EndOfBibitem
\bibitem{Aaltonen:2010pj}
T.~Aaltonen {\em et al.} ({CDF} collaboration){,}
  \href{http://dx.doi.org/10.1103/PhysRevLett.106.121804}{Phys.\ Rev.\ Lett.\ {\bf
  106},  121804}  (2011), \href{http://arxiv.org/abs/1012.3138}{{\tt
  arXiv:1012.3138 [hep-ex]}}\relax
\mciteBstWouldAddEndPuncttrue
\mciteSetBstMidEndSepPunct{\mcitedefaultmidpunct}
{\mcitedefaultendpunct}{\mcitedefaultseppunct}\relax
\EndOfBibitem
\bibitem{Abazov:2008jz}
V.~M.\ Abazov {\em et al.} ({\dzero} collaboration){,}
  \href{http://dx.doi.org/10.1103/PhysRevLett.102.032001}{Phys.\ Rev.\ Lett.\ {\bf
  102},  032001}  (2009), \href{http://arxiv.org/abs/0810.0037}{{\tt
  arXiv:0810.0037 [hep-ex]}}\relax
\mciteBstWouldAddEndPuncttrue
\mciteSetBstMidEndSepPunct{\mcitedefaultmidpunct}
{\mcitedefaultendpunct}{\mcitedefaultseppunct}\relax
\EndOfBibitem
\bibitem{Abazov:2012iy}
V.~M.\ Abazov {\em et al.} ({\dzero} collaboration){,}
  \href{http://dx.doi.org/10.1103/PhysRevD.85.112003}{Phys.\ Rev.\ {\bf D85}{,}
  112003}  (2012), \href{http://arxiv.org/abs/1204.2340}{{\tt arXiv:1204.2340
  [hep-ex]}}\relax
\mciteBstWouldAddEndPuncttrue
\mciteSetBstMidEndSepPunct{\mcitedefaultmidpunct}
{\mcitedefaultendpunct}{\mcitedefaultseppunct}\relax
\EndOfBibitem
\bibitem{Abazov:2014rua}
V.~M.\ Abazov {\em et al.} ({\dzero} collaboration){,}
  \href{http://dx.doi.org/10.1103/PhysRevLett.114.062001}{Phys.\ Rev.\ Lett.\ {\bf
  114},  062001}  (2015), \href{http://arxiv.org/abs/1410.1568}{{\tt
  arXiv:1410.1568 [hep-ex]}}\relax
\mciteBstWouldAddEndPuncttrue
\mciteSetBstMidEndSepPunct{\mcitedefaultmidpunct}
{\mcitedefaultendpunct}{\mcitedefaultseppunct}\relax
\EndOfBibitem
\bibitem{Aubert:2001uw}
B.~Aubert {\em et al.} ({\babar} collaboration){,}
  \href{http://dx.doi.org/10.1103/PhysRevLett.87.201803}{Phys.\ Rev.\ Lett.\ {\bf
  87},  201803}  (2001), \href{http://arxiv.org/abs/hep-ex/0107019}{{\tt
  arXiv:hep-ex/0107019 [hep-ex]}}\relax
\mciteBstWouldAddEndPuncttrue
\mciteSetBstMidEndSepPunct{\mcitedefaultmidpunct}
{\mcitedefaultendpunct}{\mcitedefaultseppunct}\relax
\EndOfBibitem
\bibitem{Aubert:2002gi}
B.~Aubert {\em et al.} ({\babar} collaboration){,}
  \href{http://dx.doi.org/10.1103/PhysRevLett.89.011802}{Phys.\ Rev.\ Lett.\ {\bf
  89},  011802}  (2002), \href{http://arxiv.org/abs/hep-ex/0202005}{{\tt
  arXiv:hep-ex/0202005 [hep-ex]}}, Erratum ibid.\
  \href{http://dx.doi.org/10.1103/PhysRevLett.89.169903}{{\bf 89}, 169903}
  (2002)\relax
\mciteBstWouldAddEndPuncttrue
\mciteSetBstMidEndSepPunct{\mcitedefaultmidpunct}
{\mcitedefaultendpunct}{\mcitedefaultseppunct}\relax
\EndOfBibitem
\bibitem{Aubert:2002sh}
B.~Aubert {\em et al.} ({\babar} collaboration){,}
  \href{http://dx.doi.org/10.1103/PhysRevD.67.072002}{Phys.\ Rev.\ {\bf D67}{,}
  072002}  (2003), \href{http://arxiv.org/abs/hep-ex/0212017}{{\tt
  arXiv:hep-ex/0212017 [hep-ex]}}\relax
\mciteBstWouldAddEndPuncttrue
\mciteSetBstMidEndSepPunct{\mcitedefaultmidpunct}
{\mcitedefaultendpunct}{\mcitedefaultseppunct}\relax
\EndOfBibitem
\bibitem{Aubert:2002ms}
B.~Aubert {\em et al.} ({\babar} collaboration){,}
  \href{http://dx.doi.org/10.1103/PhysRevD.67.091101}{Phys.\ Rev.\ {\bf D67}{,}
  091101}  (2003), \href{http://arxiv.org/abs/hep-ex/0212012}{{\tt
  arXiv:hep-ex/0212012 [hep-ex]}}\relax
\mciteBstWouldAddEndPuncttrue
\mciteSetBstMidEndSepPunct{\mcitedefaultmidpunct}
{\mcitedefaultendpunct}{\mcitedefaultseppunct}\relax
\EndOfBibitem
\bibitem{Aubert:2005kf}
B.~Aubert {\em et al.} ({\babar} collaboration){,}
  \href{http://dx.doi.org/10.1103/PhysRevD.73.012004}{Phys.\ Rev.\ {\bf D73}{,}
  012004}  (2006), \href{http://arxiv.org/abs/hep-ex/0507054}{{\tt
  arXiv:hep-ex/0507054 [hep-ex]}}\relax
\mciteBstWouldAddEndPuncttrue
\mciteSetBstMidEndSepPunct{\mcitedefaultmidpunct}
{\mcitedefaultendpunct}{\mcitedefaultseppunct}\relax
\EndOfBibitem
\bibitem{Abe:2004mz}
K.~Abe {\em et al.} ({Belle} collaboration){,}
  \href{http://dx.doi.org/10.1103/PhysRevD.71.072003}{Phys.\ Rev.\ {\bf D71}{,}
  072003}  (2005), \href{http://arxiv.org/abs/hep-ex/0408111}{{\tt
  arXiv:hep-ex/0408111 [hep-ex]}}, Erratum ibid.\
  \href{http://dx.doi.org/10.1103/PhysRevD.71.079903}{{\bf D71}, 079903}
  (2005)\relax
\mciteBstWouldAddEndPuncttrue
\mciteSetBstMidEndSepPunct{\mcitedefaultmidpunct}
{\mcitedefaultendpunct}{\mcitedefaultseppunct}\relax
\EndOfBibitem
\bibitem{Aad:2012bpa}
G.~Aad {\em et al.} ({ATLAS} collaboration){,}
  \href{http://dx.doi.org/10.1103/PhysRevD.87.032002}{Phys.\ Rev.\ {\bf D87}{,}
  032002}  (2013), \href{http://arxiv.org/abs/1207.2284}{{\tt arXiv:1207.2284
  [hep-ex]}}\relax
\mciteBstWouldAddEndPuncttrue
\mciteSetBstMidEndSepPunct{\mcitedefaultmidpunct}
{\mcitedefaultendpunct}{\mcitedefaultseppunct}\relax
\EndOfBibitem
\bibitem{Aaij:2014owa}
R.~Aaij {\em et al.} ({LHCb} collaboration){,}
  \href{http://dx.doi.org/10.1007/JHEP04(2014)114}{JHEP {\bf 04},  114}
  (2014), \href{http://arxiv.org/abs/1402.2554}{{\tt arXiv:1402.2554
  [hep-ex]}}\relax
\mciteBstWouldAddEndPuncttrue
\mciteSetBstMidEndSepPunct{\mcitedefaultmidpunct}
{\mcitedefaultendpunct}{\mcitedefaultseppunct}\relax
\EndOfBibitem
\bibitem{Aaij:2014fia}
R.~Aaij {\em et al.} ({LHCb} collaboration){,}
  \href{http://dx.doi.org/10.1016/j.physletb.2014.07.051}{Phys.\ Lett.\ {\bf
  B736},  446}  (2014), \href{http://arxiv.org/abs/1406.7204}{{\tt
  arXiv:1406.7204 [hep-ex]}}\relax
\mciteBstWouldAddEndPuncttrue
\mciteSetBstMidEndSepPunct{\mcitedefaultmidpunct}
{\mcitedefaultendpunct}{\mcitedefaultseppunct}\relax
\EndOfBibitem
\bibitem{Aaltonen:2010ta}
T.~Aaltonen {\em et al.} ({CDF} collaboration){,}
  \href{http://dx.doi.org/10.1103/PhysRevD.83.032008}{Phys.\ Rev.\ {\bf D83}{,}
  032008}  (2011), \href{http://arxiv.org/abs/1004.4855}{{\tt arXiv:1004.4855
  [hep-ex]}}\relax
\mciteBstWouldAddEndPuncttrue
\mciteSetBstMidEndSepPunct{\mcitedefaultmidpunct}
{\mcitedefaultendpunct}{\mcitedefaultseppunct}\relax
\EndOfBibitem
\bibitem{Abazov:2004sa}
V.~M.\ Abazov {\em et al.} ({\dzero} collaboration){,}
  \href{http://dx.doi.org/10.1103/PhysRevLett.94.182001}{Phys.\ Rev.\ Lett.\ {\bf
  94},  182001}  (2005), \href{http://arxiv.org/abs/hep-ex/0410052}{{\tt
  arXiv:hep-ex/0410052 [hep-ex]}}\relax
\mciteBstWouldAddEndPuncttrue
\mciteSetBstMidEndSepPunct{\mcitedefaultmidpunct}
{\mcitedefaultendpunct}{\mcitedefaultseppunct}\relax
\EndOfBibitem
\bibitem{Artuso:2015swg}
M.~Artuso, G.~Borissov, and A.~Lenz{,}
  \href{http://dx.doi.org/10.1103/RevModPhys.88.045002}{Rev.\ Mod.\ Phys.\ {\bf
  88},  045002}  (2016), \href{http://arxiv.org/abs/1511.09466}{{\tt
  arXiv:1511.09466 [hep-ph]}}\relax
\mciteBstWouldAddEndPuncttrue
\mciteSetBstMidEndSepPunct{\mcitedefaultmidpunct}
{\mcitedefaultendpunct}{\mcitedefaultseppunct}\relax
\EndOfBibitem
\bibitem{Laplace:2002ik}
S.~Laplace, Z.~Ligeti, Y.~Nir, and G.~Perez{,}
  \href{http://dx.doi.org/10.1103/PhysRevD.65.094040}{Phys.\ Rev.\ {\bf D65}{,}
  094040}  (2002), \href{http://arxiv.org/abs/hep-ph/0202010}{{\tt
  arXiv:hep-ph/0202010 [hep-ph]}}\relax
\mciteBstWouldAddEndPuncttrue
\mciteSetBstMidEndSepPunct{\mcitedefaultmidpunct}
{\mcitedefaultendpunct}{\mcitedefaultseppunct}\relax
\EndOfBibitem
\bibitem{Ciuchini:2003ww}
M.~Ciuchini, E.~Franco, V.~Lubicz, F.~Mescia, and C.~Tarantino{,}
  \href{http://dx.doi.org/10.1088/1126-6708/2003/08/031}{JHEP {\bf 08},  031}
  (2003), \href{http://arxiv.org/abs/hep-ph/0308029}{{\tt arXiv:hep-ph/0308029
  [hep-ph]}}\relax
\mciteBstWouldAddEndPuncttrue
\mciteSetBstMidEndSepPunct{\mcitedefaultmidpunct}
{\mcitedefaultendpunct}{\mcitedefaultseppunct}\relax
\EndOfBibitem
\bibitem{Beneke:2003az}
M.~Beneke, G.~Buchalla, A.~Lenz, and U.~Nierste{,}
  \href{http://dx.doi.org/10.1016/j.physletb.2003.09.089}{Phys.\ Lett.\ {\bf
  B576},  173}  (2003), \href{http://arxiv.org/abs/hep-ph/0307344}{{\tt
  arXiv:hep-ph/0307344 [hep-ph]}}\relax
\mciteBstWouldAddEndPuncttrue
\mciteSetBstMidEndSepPunct{\mcitedefaultmidpunct}
{\mcitedefaultendpunct}{\mcitedefaultseppunct}\relax
\EndOfBibitem
\bibitem{Aaij:2012eq}
R.~Aaij {\em et al.} ({LHCb} collaboration){,}
  \href{http://dx.doi.org/10.1103/PhysRevLett.108.241801}{Phys.\ Rev.\ Lett.\ {\bf
  108},  241801}  (2012), \href{http://arxiv.org/abs/1202.4717}{{\tt
  arXiv:1202.4717 [hep-ex]}}\relax
\mciteBstWouldAddEndPuncttrue
\mciteSetBstMidEndSepPunct{\mcitedefaultmidpunct}
{\mcitedefaultendpunct}{\mcitedefaultseppunct}\relax
\EndOfBibitem
\bibitem{Fleischer:2011cw}
R.~Fleischer and R.~Knegjens{,}
  \href{http://dx.doi.org/10.1140/epjc/s10052-011-1789-9}{Eur.\ Phys.\ J.\ {\bf
  C71},  1789}  (2011), \href{http://arxiv.org/abs/1109.5115}{{\tt
  arXiv:1109.5115 [hep-ph]}}\relax
\mciteBstWouldAddEndPuncttrue
\mciteSetBstMidEndSepPunct{\mcitedefaultmidpunct}
{\mcitedefaultendpunct}{\mcitedefaultseppunct}\relax
\EndOfBibitem
\bibitem{Barate:1997ua}
R.~Barate {\em et al.} ({ALEPH} collaboration){,}
  \href{http://dx.doi.org/10.1007/s100520050215}{Eur.\ Phys.\ J.\ {\bf C4},  367}
   (1998)\relax
\mciteBstWouldAddEndPuncttrue
\mciteSetBstMidEndSepPunct{\mcitedefaultmidpunct}
{\mcitedefaultendpunct}{\mcitedefaultseppunct}\relax
\EndOfBibitem
\bibitem{Abreu:2000ev}
P.~Abreu {\em et al.} ({DELPHI} collaboration){,}
  \href{http://dx.doi.org/10.1007/s100520000531}{Eur.\ Phys.\ J.\ {\bf C18}{,}
  229}  (2000), \href{http://arxiv.org/abs/hep-ex/0105077}{{\tt
  arXiv:hep-ex/0105077 [hep-ex]}}\relax
\mciteBstWouldAddEndPuncttrue
\mciteSetBstMidEndSepPunct{\mcitedefaultmidpunct}
{\mcitedefaultendpunct}{\mcitedefaultseppunct}\relax
\EndOfBibitem
\bibitem{Ackerstaff:1997ne}
K.~Ackerstaff {\em et al.} ({OPAL} collaboration){,}
  \href{http://dx.doi.org/10.1007/s100520050150}{Eur.\ Phys.\ J.\ {\bf C2},  407}
   (1998), \href{http://arxiv.org/abs/hep-ex/9708023}{{\tt arXiv:hep-ex/9708023
  [hep-ex]}}\relax
\mciteBstWouldAddEndPuncttrue
\mciteSetBstMidEndSepPunct{\mcitedefaultmidpunct}
{\mcitedefaultendpunct}{\mcitedefaultseppunct}\relax
\EndOfBibitem
\bibitem{Buskulic:1996ei}
D.~Buskulic {\em et al.} ({ALEPH} collaboration){,}
  \href{http://dx.doi.org/10.1016/0370-2693(96)00451-0}{Phys.\ Lett.\ {\bf B377}{,}
   205}  (1996)\relax
\mciteBstWouldAddEndPuncttrue
\mciteSetBstMidEndSepPunct{\mcitedefaultmidpunct}
{\mcitedefaultendpunct}{\mcitedefaultseppunct}\relax
\EndOfBibitem
\bibitem{Abe:1998cj}
F.~Abe {\em et al.} ({CDF} collaboration){,}
  \href{http://dx.doi.org/10.1103/PhysRevD.59.032004}{Phys.\ Rev.\ {\bf D59}{,}
  032004}  (1999), \href{http://arxiv.org/abs/hep-ex/9808003}{{\tt
  arXiv:hep-ex/9808003 [hep-ex]}}\relax
\mciteBstWouldAddEndPuncttrue
\mciteSetBstMidEndSepPunct{\mcitedefaultmidpunct}
{\mcitedefaultendpunct}{\mcitedefaultseppunct}\relax
\EndOfBibitem
\bibitem{Abreu:2000sh}
P.~Abreu {\em et al.} ({DELPHI} collaboration){,}
  \href{http://dx.doi.org/10.1007/s100520000415}{Eur.\ Phys.\ J.\ {\bf C16}{,}
  555}  (2000), \href{http://arxiv.org/abs/hep-ex/0107077}{{\tt
  arXiv:hep-ex/0107077 [hep-ex]}}\relax
\mciteBstWouldAddEndPuncttrue
\mciteSetBstMidEndSepPunct{\mcitedefaultmidpunct}
{\mcitedefaultendpunct}{\mcitedefaultseppunct}\relax
\EndOfBibitem
\bibitem{Ackerstaff:1997qi}
K.~Ackerstaff {\em et al.} ({OPAL} collaboration){,}
  \href{http://dx.doi.org/10.1016/S0370-2693(98)00289-5}{Phys.\ Lett.\ {\bf
  B426},  161}  (1998), \href{http://arxiv.org/abs/hep-ex/9802002}{{\tt
  arXiv:hep-ex/9802002 [hep-ex]}}\relax
\mciteBstWouldAddEndPuncttrue
\mciteSetBstMidEndSepPunct{\mcitedefaultmidpunct}
{\mcitedefaultendpunct}{\mcitedefaultseppunct}\relax
\EndOfBibitem
\bibitem{Aaltonen:2011qsa}
T.~Aaltonen {\em et al.} ({CDF} collaboration){,}
  \href{http://dx.doi.org/10.1103/PhysRevLett.107.272001}{Phys.\ Rev.\ Lett.\ {\bf
  107},  272001}  (2011), \href{http://arxiv.org/abs/1103.1864}{{\tt
  arXiv:1103.1864 [hep-ex]}}\relax
\mciteBstWouldAddEndPuncttrue
\mciteSetBstMidEndSepPunct{\mcitedefaultmidpunct}
{\mcitedefaultendpunct}{\mcitedefaultseppunct}\relax
\EndOfBibitem
\bibitem{Aaij:2013bvd}
R.~Aaij {\em et al.} ({LHCb} collaboration){,}
  \href{http://dx.doi.org/10.1103/PhysRevLett.112.111802}{Phys.\ Rev.\ Lett.\ {\bf
  112},  111802}  (2014), \href{http://arxiv.org/abs/1312.1217}{{\tt
  arXiv:1312.1217 [hep-ex]}}\relax
\mciteBstWouldAddEndPuncttrue
\mciteSetBstMidEndSepPunct{\mcitedefaultmidpunct}
{\mcitedefaultendpunct}{\mcitedefaultseppunct}\relax
\EndOfBibitem
\bibitem{Aaij:2014sua}
R.~Aaij {\em et al.} ({LHCb} collaboration){,}
  \href{http://dx.doi.org/10.1103/PhysRevLett.113.172001}{Phys.\ Rev.\ Lett.\ {\bf
  113},  172001}  (2014), \href{http://arxiv.org/abs/1407.5873}{{\tt
  arXiv:1407.5873 [hep-ex]}}\relax
\mciteBstWouldAddEndPuncttrue
\mciteSetBstMidEndSepPunct{\mcitedefaultmidpunct}
{\mcitedefaultendpunct}{\mcitedefaultseppunct}\relax
\EndOfBibitem
\bibitem{Abazov:2004ce}
V.~M.\ Abazov {\em et al.} ({\dzero} collaboration){,}
  \href{http://dx.doi.org/10.1103/PhysRevLett.94.042001}{Phys.\ Rev.\ Lett.\ {\bf
  94},  042001}  (2005), \href{http://arxiv.org/abs/hep-ex/0409043}{{\tt
  arXiv:hep-ex/0409043 [hep-ex]}}\relax
\mciteBstWouldAddEndPuncttrue
\mciteSetBstMidEndSepPunct{\mcitedefaultmidpunct}
{\mcitedefaultendpunct}{\mcitedefaultseppunct}\relax
\EndOfBibitem
\bibitem{Barate:2000kd}
R.~Barate {\em et al.} ({ALEPH} collaboration){,}
  \href{http://dx.doi.org/10.1016/S0370-2693(00)00750-4}{Phys.\ Lett.\ {\bf
  B486},  286}  (2000)\relax
\mciteBstWouldAddEndPuncttrue
\mciteSetBstMidEndSepPunct{\mcitedefaultmidpunct}
{\mcitedefaultendpunct}{\mcitedefaultseppunct}\relax
\EndOfBibitem
\bibitem{Aaij:2012kn}
R.~Aaij {\em et al.} ({LHCb} collaboration){,}
  \href{http://dx.doi.org/10.1016/j.physletb.2011.12.058}{Phys.\ Lett.\ {\bf
  B707},  349}  (2012), \href{http://arxiv.org/abs/1111.0521}{{\tt
  arXiv:1111.0521 [hep-ex]}}\relax
\mciteBstWouldAddEndPuncttrue
\mciteSetBstMidEndSepPunct{\mcitedefaultmidpunct}
{\mcitedefaultendpunct}{\mcitedefaultseppunct}\relax
\EndOfBibitem
\bibitem{Aaij:2016dzn}
R.~Aaij {\em et al.} ({LHCb} collaboration){,}
  \href{http://dx.doi.org/10.1016/j.physletb.2016.10.006}{Phys.\ Lett.\ {\bf
  B762},  484}  (2016), \href{http://arxiv.org/abs/1607.06314}{{\tt
  arXiv:1607.06314 [hep-ex]}}\relax
\mciteBstWouldAddEndPuncttrue
\mciteSetBstMidEndSepPunct{\mcitedefaultmidpunct}
{\mcitedefaultendpunct}{\mcitedefaultseppunct}\relax
\EndOfBibitem
\bibitem{Aaij:2013eia}
R.~Aaij {\em et al.} ({LHCb} collaboration){,}
  \href{http://dx.doi.org/10.1016/j.nuclphysb.2013.04.021}{Nucl.\ Phys.\ {\bf
  B873},  275}  (2013), \href{http://arxiv.org/abs/1304.4500}{{\tt
  arXiv:1304.4500 [hep-ex]}}\relax
\mciteBstWouldAddEndPuncttrue
\mciteSetBstMidEndSepPunct{\mcitedefaultmidpunct}
{\mcitedefaultendpunct}{\mcitedefaultseppunct}\relax
\EndOfBibitem
\bibitem{Aaltonen:2011nk}
T.~Aaltonen {\em et al.} ({CDF} collaboration){,}
  \href{http://dx.doi.org/10.1103/PhysRevD.84.052012}{Phys.\ Rev.\ {\bf D84}{,}
  052012}  (2011), \href{http://arxiv.org/abs/1106.3682}{{\tt arXiv:1106.3682
  [hep-ex]}}\relax
\mciteBstWouldAddEndPuncttrue
\mciteSetBstMidEndSepPunct{\mcitedefaultmidpunct}
{\mcitedefaultendpunct}{\mcitedefaultseppunct}\relax
\EndOfBibitem
\bibitem{Abazov:2016oqi}
V.~M.\ Abazov {\em et al.} ({\dzero} collaboration){,}
  \href{http://dx.doi.org/10.1103/PhysRevD.94.012001}{Phys.\ Rev.\ {\bf D94}{,}
  012001}  (2016), \href{http://arxiv.org/abs/1603.01302}{{\tt
  arXiv:1603.01302 [hep-ex]}}\relax
\mciteBstWouldAddEndPuncttrue
\mciteSetBstMidEndSepPunct{\mcitedefaultmidpunct}
{\mcitedefaultendpunct}{\mcitedefaultseppunct}\relax
\EndOfBibitem
\bibitem{Aaij:2013oba}
R.~Aaij {\em et al.} ({LHCb} collaboration){,}
  \href{http://dx.doi.org/10.1103/PhysRevD.87.112010}{Phys.\ Rev.\ {\bf D87}{,}
  112010}  (2013), \href{http://arxiv.org/abs/1304.2600}{{\tt arXiv:1304.2600
  [hep-ex]}}\relax
\mciteBstWouldAddEndPuncttrue
\mciteSetBstMidEndSepPunct{\mcitedefaultmidpunct}
{\mcitedefaultendpunct}{\mcitedefaultseppunct}\relax
\EndOfBibitem
\bibitem{Hartkorn:1999ga}
K.~Hartkorn and H.~G.\ Moser{,}
  \href{http://dx.doi.org/10.1007/s100520050472}{Eur.\ Phys.\ J.\ {\bf C8},  381}
   (1999)\relax
\mciteBstWouldAddEndPuncttrue
\mciteSetBstMidEndSepPunct{\mcitedefaultmidpunct}
{\mcitedefaultendpunct}{\mcitedefaultseppunct}\relax
\EndOfBibitem
\bibitem{Aaij:2014bba}
R.~Aaij {\em et al.} ({LHCb} collaboration){,}
  \href{http://dx.doi.org/10.1016/j.physletb.2014.10.005}{Phys.\ Lett.\ {\bf
  B739},  218}  (2014), \href{http://arxiv.org/abs/1408.0275}{{\tt
  arXiv:1408.0275 [hep-ex]}}\relax
\mciteBstWouldAddEndPuncttrue
\mciteSetBstMidEndSepPunct{\mcitedefaultmidpunct}
{\mcitedefaultendpunct}{\mcitedefaultseppunct}\relax
\EndOfBibitem
\bibitem{Abe:1998wi}
F.~Abe {\em et al.} ({CDF} collaboration){,}
  \href{http://dx.doi.org/10.1103/PhysRevLett.81.2432}{Phys.\ Rev.\ Lett.\ {\bf
  81},  2432}  (1998), \href{http://arxiv.org/abs/hep-ex/9805034}{{\tt
  arXiv:hep-ex/9805034 [hep-ex]}}\relax
\mciteBstWouldAddEndPuncttrue
\mciteSetBstMidEndSepPunct{\mcitedefaultmidpunct}
{\mcitedefaultendpunct}{\mcitedefaultseppunct}\relax
\EndOfBibitem
\bibitem{Abulencia:2006zu}
A.~Abulencia {\em et al.} ({CDF} collaboration){,}
  \href{http://dx.doi.org/10.1103/PhysRevLett.97.012002}{Phys.\ Rev.\ Lett.\ {\bf
  97},  012002}  (2006), \href{http://arxiv.org/abs/hep-ex/0603027}{{\tt
  arXiv:hep-ex/0603027 [hep-ex]}}\relax
\mciteBstWouldAddEndPuncttrue
\mciteSetBstMidEndSepPunct{\mcitedefaultmidpunct}
{\mcitedefaultendpunct}{\mcitedefaultseppunct}\relax
\EndOfBibitem
\bibitem{Abazov:2008rba}
V.~M.\ Abazov {\em et al.} ({\dzero} collaboration){,}
  \href{http://dx.doi.org/10.1103/PhysRevLett.102.092001}{Phys.\ Rev.\ Lett.\ {\bf
  102},  092001}  (2009), \href{http://arxiv.org/abs/0805.2614}{{\tt
  arXiv:0805.2614 [hep-ex]}}\relax
\mciteBstWouldAddEndPuncttrue
\mciteSetBstMidEndSepPunct{\mcitedefaultmidpunct}
{\mcitedefaultendpunct}{\mcitedefaultseppunct}\relax
\EndOfBibitem
\bibitem{Aaltonen:2012yb}
T.~Aaltonen {\em et al.} ({CDF} collaboration){,}
  \href{http://dx.doi.org/10.1103/PhysRevD.87.011101}{Phys.\ Rev.\ {\bf D87}{,}
  011101}  (2013), \href{http://arxiv.org/abs/1210.2366}{{\tt arXiv:1210.2366
  [hep-ex]}}\relax
\mciteBstWouldAddEndPuncttrue
\mciteSetBstMidEndSepPunct{\mcitedefaultmidpunct}
{\mcitedefaultendpunct}{\mcitedefaultseppunct}\relax
\EndOfBibitem
\bibitem{Aaij:2014bva}
R.~Aaij {\em et al.} ({LHCb} collaboration){,}
  \href{http://dx.doi.org/10.1140/epjc/s10052-014-2839-x}{Eur.\ Phys.\ J.\ {\bf
  C74},  2839}  (2014), \href{http://arxiv.org/abs/1401.6932}{{\tt
  arXiv:1401.6932 [hep-ex]}}\relax
\mciteBstWouldAddEndPuncttrue
\mciteSetBstMidEndSepPunct{\mcitedefaultmidpunct}
{\mcitedefaultendpunct}{\mcitedefaultseppunct}\relax
\EndOfBibitem
\bibitem{Aaij:2014gka}
R.~Aaij {\em et al.} ({LHCb} collaboration){,}
  \href{http://dx.doi.org/10.1016/j.physletb.2015.01.010}{Phys.\ Lett.\ {\bf
  B742},  29}  (2015), \href{http://arxiv.org/abs/1411.6899}{{\tt
  arXiv:1411.6899 [hep-ex]}}\relax
\mciteBstWouldAddEndPuncttrue
\mciteSetBstMidEndSepPunct{\mcitedefaultmidpunct}
{\mcitedefaultendpunct}{\mcitedefaultseppunct}\relax
\EndOfBibitem
\bibitem{Abreu:1999hu}
P.~Abreu {\em et al.} ({DELPHI} collaboration){,}
  \href{http://dx.doi.org/10.1007/s100520050582}{Eur.\ Phys.\ J.\ {\bf C10}{,}
  185}  (1999)\relax
\mciteBstWouldAddEndPuncttrue
\mciteSetBstMidEndSepPunct{\mcitedefaultmidpunct}
{\mcitedefaultendpunct}{\mcitedefaultseppunct}\relax
\EndOfBibitem
\bibitem{Abreu:1996nt}
P.~Abreu {\em et al.} ({DELPHI} collaboration){,}
  \href{http://dx.doi.org/10.1007/s002880050164}{Z.\ Phys.\ {\bf C71},  199}
  (1996)\relax
\mciteBstWouldAddEndPuncttrue
\mciteSetBstMidEndSepPunct{\mcitedefaultmidpunct}
{\mcitedefaultendpunct}{\mcitedefaultseppunct}\relax
\EndOfBibitem
\bibitem{Akers:1995ui}
R.~Akers {\em et al.} ({OPAL} collaboration){,}
  \href{http://dx.doi.org/10.1007/s002880050020}{Z.\ Phys.\ {\bf C69},  195}
  (1996)\relax
\mciteBstWouldAddEndPuncttrue
\mciteSetBstMidEndSepPunct{\mcitedefaultmidpunct}
{\mcitedefaultendpunct}{\mcitedefaultseppunct}\relax
\EndOfBibitem
\bibitem{Abe:1996df}
F.~Abe {\em et al.} ({CDF} collaboration){,}
  \href{http://dx.doi.org/10.1103/PhysRevLett.77.1439}{Phys.\ Rev.\ Lett.\ {\bf
  77},  1439}  (1996)\relax
\mciteBstWouldAddEndPuncttrue
\mciteSetBstMidEndSepPunct{\mcitedefaultmidpunct}
{\mcitedefaultendpunct}{\mcitedefaultseppunct}\relax
\EndOfBibitem
\bibitem{Abazov:2007al}
V.~M.\ Abazov {\em et al.} ({\dzero} collaboration){,}
  \href{http://dx.doi.org/10.1103/PhysRevLett.99.182001}{Phys.\ Rev.\ Lett.\ {\bf
  99},  182001}  (2007), \href{http://arxiv.org/abs/0706.2358}{{\tt
  arXiv:0706.2358 [hep-ex]}}\relax
\mciteBstWouldAddEndPuncttrue
\mciteSetBstMidEndSepPunct{\mcitedefaultmidpunct}
{\mcitedefaultendpunct}{\mcitedefaultseppunct}\relax
\EndOfBibitem
\bibitem{Aaltonen:2009zn}
T.~Aaltonen {\em et al.} ({CDF} collaboration){,}
  \href{http://dx.doi.org/10.1103/PhysRevLett.104.102002}{Phys.\ Rev.\ Lett.\ {\bf
  104},  102002}  (2010), \href{http://arxiv.org/abs/0912.3566}{{\tt
  arXiv:0912.3566 [hep-ex]}}\relax
\mciteBstWouldAddEndPuncttrue
\mciteSetBstMidEndSepPunct{\mcitedefaultmidpunct}
{\mcitedefaultendpunct}{\mcitedefaultseppunct}\relax
\EndOfBibitem
\bibitem{Aaltonen:2014wfa}
T.~A.\ Aaltonen {\em et al.} ({CDF} collaboration){,}
  \href{http://dx.doi.org/10.1103/PhysRevD.89.072014}{Phys.\ Rev.\ {\bf D89}{,}
  072014}  (2014), \href{http://arxiv.org/abs/1403.8126}{{\tt arXiv:1403.8126
  [hep-ex]}}\relax
\mciteBstWouldAddEndPuncttrue
\mciteSetBstMidEndSepPunct{\mcitedefaultmidpunct}
{\mcitedefaultendpunct}{\mcitedefaultseppunct}\relax
\EndOfBibitem
\bibitem{Chatrchyan:2013sxa}
S.~Chatrchyan {\em et al.} ({CMS} collaboration){,}
  \href{http://dx.doi.org/10.1007/JHEP07(2013)163}{JHEP {\bf 07},  163}
  (2013), \href{http://arxiv.org/abs/1304.7495}{{\tt arXiv:1304.7495
  [hep-ex]}}\relax
\mciteBstWouldAddEndPuncttrue
\mciteSetBstMidEndSepPunct{\mcitedefaultmidpunct}
{\mcitedefaultendpunct}{\mcitedefaultseppunct}\relax
\EndOfBibitem
\bibitem{Aaij:2014zyy}
R.~Aaij {\em et al.} ({LHCb} collaboration){,}
  \href{http://dx.doi.org/10.1016/j.physletb.2014.05.021}{Phys.\ Lett.\ {\bf
  B734},  122}  (2014), \href{http://arxiv.org/abs/1402.6242}{{\tt
  arXiv:1402.6242 [hep-ex]}}\relax
\mciteBstWouldAddEndPuncttrue
\mciteSetBstMidEndSepPunct{\mcitedefaultmidpunct}
{\mcitedefaultendpunct}{\mcitedefaultseppunct}\relax
\EndOfBibitem
\bibitem{Abreu:1995kt}
P.~Abreu {\em et al.} ({DELPHI} collaboration){,}
  \href{http://dx.doi.org/10.1007/BF01565255}{Z.\ Phys.\ {\bf C68},  541}
  (1995)\relax
\mciteBstWouldAddEndPuncttrue
\mciteSetBstMidEndSepPunct{\mcitedefaultmidpunct}
{\mcitedefaultendpunct}{\mcitedefaultseppunct}\relax
\EndOfBibitem
\bibitem{Aaij:2014sia}
R.~Aaij {\em et al.} ({LHCb} collaboration){,}
  \href{http://dx.doi.org/10.1016/j.physletb.2014.06.064}{Phys.\ Lett.\ {\bf
  B736},  154}  (2014), \href{http://arxiv.org/abs/1405.1543}{{\tt
  arXiv:1405.1543 [hep-ex]}}\relax
\mciteBstWouldAddEndPuncttrue
\mciteSetBstMidEndSepPunct{\mcitedefaultmidpunct}
{\mcitedefaultendpunct}{\mcitedefaultseppunct}\relax
\EndOfBibitem
\bibitem{Aaij:2014lxa}
R.~Aaij {\em et al.} ({LHCb} collaboration){,}
  \href{http://dx.doi.org/10.1103/PhysRevLett.113.242002}{Phys.\ Rev.\ Lett.\ {\bf
  113},  242002}  (2014), \href{http://arxiv.org/abs/1409.8568}{{\tt
  arXiv:1409.8568 [hep-ex]}}\relax
\mciteBstWouldAddEndPuncttrue
\mciteSetBstMidEndSepPunct{\mcitedefaultmidpunct}
{\mcitedefaultendpunct}{\mcitedefaultseppunct}\relax
\EndOfBibitem
\bibitem{Aaij:2014esa}
R.~Aaij {\em et al.} ({LHCb} collaboration){,}
  \href{http://dx.doi.org/10.1103/PhysRevLett.113.032001}{Phys.\ Rev.\ Lett.\ {\bf
  113},  032001}  (2014), \href{http://arxiv.org/abs/1405.7223}{{\tt
  arXiv:1405.7223 [hep-ex]}}\relax
\mciteBstWouldAddEndPuncttrue
\mciteSetBstMidEndSepPunct{\mcitedefaultmidpunct}
{\mcitedefaultendpunct}{\mcitedefaultseppunct}\relax
\EndOfBibitem
\bibitem{Aaij:2016dls}
R.~Aaij {\em et al.} ({LHCb} collaboration){,}
  \href{http://dx.doi.org/10.1103/PhysRevD.93.092007}{Phys.\ Rev.\ {\bf D93}{,}
  092007}  (2016), \href{http://arxiv.org/abs/1604.01412}{{\tt
  arXiv:1604.01412 [hep-ex]}}\relax
\mciteBstWouldAddEndPuncttrue
\mciteSetBstMidEndSepPunct{\mcitedefaultmidpunct}
{\mcitedefaultendpunct}{\mcitedefaultseppunct}\relax
\EndOfBibitem
\bibitem{Beneke:1996gn}
M.~Beneke, G.~Buchalla, and I.~Dunietz{,}
  \href{http://dx.doi.org/10.1103/PhysRevD.54.4419}{Phys.\ Rev.\ {\bf D54}{,}
  4419}  (1996), \href{http://arxiv.org/abs/hep-ph/9605259}{{\tt
  arXiv:hep-ph/9605259 [hep-ph]}}, Erratum ibid.\
  \href{http://dx.doi.org/10.1103/PhysRevD.83.119902}{{\bf D83}, 119902}
  (2011)\relax
\mciteBstWouldAddEndPuncttrue
\mciteSetBstMidEndSepPunct{\mcitedefaultmidpunct}
{\mcitedefaultendpunct}{\mcitedefaultseppunct}\relax
\EndOfBibitem
\bibitem{Keum:1998fd}
Y.-Y.\ Keum and U.~Nierste{,}
  \href{http://dx.doi.org/10.1103/PhysRevD.57.4282}{Phys.\ Rev.\ {\bf D57}{,}
  4282}  (1998), \href{http://arxiv.org/abs/hep-ph/9710512}{{\tt
  arXiv:hep-ph/9710512 [hep-ph]}}\relax
\mciteBstWouldAddEndPuncttrue
\mciteSetBstMidEndSepPunct{\mcitedefaultmidpunct}
{\mcitedefaultendpunct}{\mcitedefaultseppunct}\relax
\EndOfBibitem
\bibitem{Voloshin:1999pz}
M.~B.\ Voloshin, \href{http://dx.doi.org/10.1016/S0370-1573(99)00054-X}{Phys.\
  Rept.\ {\bf 320},  275}  (1999){,}
  \href{http://arxiv.org/abs/hep-ph/9901445}{{\tt arXiv:hep-ph/9901445
  [hep-ph]}}\relax
\mciteBstWouldAddEndPuncttrue
\mciteSetBstMidEndSepPunct{\mcitedefaultmidpunct}
{\mcitedefaultendpunct}{\mcitedefaultseppunct}\relax
\EndOfBibitem
\bibitem{Guberina:1999bw}
B.~Guberina, B.~Melic, and H.~Stefancic{,}
  \href{http://dx.doi.org/10.1016/S0370-2693(99)01198-3}{Phys.\ Lett.\ {\bf
  B469},  253}  (1999), \href{http://arxiv.org/abs/hep-ph/9907468}{{\tt
  arXiv:hep-ph/9907468 [hep-ph]}}\relax
\mciteBstWouldAddEndPuncttrue
\mciteSetBstMidEndSepPunct{\mcitedefaultmidpunct}
{\mcitedefaultendpunct}{\mcitedefaultseppunct}\relax
\EndOfBibitem
\bibitem{Neubert:1996we}
M.~Neubert and C.~T.\ Sachrajda{,}
  \href{http://dx.doi.org/10.1016/S0550-3213(96)00559-7}{Nucl.\ Phys.\ {\bf
  B483},  339}  (1997), \href{http://arxiv.org/abs/hep-ph/9603202}{{\tt
  arXiv:hep-ph/9603202 [hep-ph]}}\relax
\mciteBstWouldAddEndPuncttrue
\mciteSetBstMidEndSepPunct{\mcitedefaultmidpunct}
{\mcitedefaultendpunct}{\mcitedefaultseppunct}\relax
\EndOfBibitem
\bibitem{Bigi:1997fj}
I.~I.~Y.\ Bigi, M.~A.\ Shifman, and N.~Uraltsev{,}
  \href{http://dx.doi.org/10.1146/annurev.nucl.47.1.591}{Ann.\ Rev.\ Nucl.\ Part.\
  Sci.\ {\bf 47},  591}  (1997){,}
  \href{http://arxiv.org/abs/hep-ph/9703290}{{\tt arXiv:hep-ph/9703290
  [hep-ph]}}\relax
\mciteBstWouldAddEndPuncttrue
\mciteSetBstMidEndSepPunct{\mcitedefaultmidpunct}
{\mcitedefaultendpunct}{\mcitedefaultseppunct}\relax
\EndOfBibitem
\bibitem{Uraltsev:1996ta}
N.~G.\ Uraltsev, \href{http://dx.doi.org/10.1016/0370-2693(96)00305-X}{Phys.\
  Lett.\ {\bf B376},  303}  (1996){,}
  \href{http://arxiv.org/abs/hep-ph/9602324}{{\tt arXiv:hep-ph/9602324
  [hep-ph]}}\relax
\mciteBstWouldAddEndPuncttrue
\mciteSetBstMidEndSepPunct{\mcitedefaultmidpunct}
{\mcitedefaultendpunct}{\mcitedefaultseppunct}\relax
\EndOfBibitem
\bibitem{Pirjol:1998ur}
D.~Pirjol and N.~Uraltsev{,}
  \href{http://dx.doi.org/10.1103/PhysRevD.59.034012}{Phys.\ Rev.\ {\bf D59}{,}
  034012}  (1999), \href{http://arxiv.org/abs/hep-ph/9805488}{{\tt
  arXiv:hep-ph/9805488 [hep-ph]}}\relax
\mciteBstWouldAddEndPuncttrue
\mciteSetBstMidEndSepPunct{\mcitedefaultmidpunct}
{\mcitedefaultendpunct}{\mcitedefaultseppunct}\relax
\EndOfBibitem
\bibitem{Colangelo:1996ta}
P.~Colangelo and F.~De~Fazio{,}
  \href{http://dx.doi.org/10.1016/0370-2693(96)01049-0}{Phys.\ Lett.\ {\bf B387}{,}
   371}  (1996), \href{http://arxiv.org/abs/hep-ph/9604425}{{\tt
  arXiv:hep-ph/9604425 [hep-ph]}}\relax
\mciteBstWouldAddEndPuncttrue
\mciteSetBstMidEndSepPunct{\mcitedefaultmidpunct}
{\mcitedefaultendpunct}{\mcitedefaultseppunct}\relax
\EndOfBibitem
\bibitem{DiPierro:1999tb}
M.~Di~Pierro, C.~T.\ Sachrajda, and C.~Michael ({UKQCD} collaboration){,}
  \href{http://dx.doi.org/10.1016/S0370-2693(99)01166-1}{Phys.\ Lett.\ {\bf
  B468},  143}  (1999), \href{http://arxiv.org/abs/hep-lat/9906031}{{\tt
  arXiv:hep-lat/9906031 [hep-lat]}}, Erratum ibid.\
  \href{http://dx.doi.org/10.1016/S0370-2693(01)01458-7}{{\bf D525}, 360}
  (2002)\relax
\mciteBstWouldAddEndPuncttrue
\mciteSetBstMidEndSepPunct{\mcitedefaultmidpunct}
{\mcitedefaultendpunct}{\mcitedefaultseppunct}\relax
\EndOfBibitem
\bibitem{Buskulic:1996qt}
D.~Buskulic {\em et al.} ({ALEPH} collaboration){,}
  \href{http://dx.doi.org/10.1007/s002880050483}{Z.\ Phys.\ {\bf C75},  397}
  (1997)\relax
\mciteBstWouldAddEndPuncttrue
\mciteSetBstMidEndSepPunct{\mcitedefaultmidpunct}
{\mcitedefaultendpunct}{\mcitedefaultseppunct}\relax
\EndOfBibitem
\bibitem{Abreu:1997xq}
P.~Abreu {\em et al.} ({DELPHI} collaboration){,}
  \href{http://dx.doi.org/10.1007/s002880050582}{Z.\ Phys.\ {\bf C76},  579}
  (1997)\relax
\mciteBstWouldAddEndPuncttrue
\mciteSetBstMidEndSepPunct{\mcitedefaultmidpunct}
{\mcitedefaultendpunct}{\mcitedefaultseppunct}\relax
\EndOfBibitem
\bibitem{Abdallah:2002mr}
J.~Abdallah {\em et al.} ({DELPHI} collaboration){,}
  \href{http://dx.doi.org/10.1140/epjc/s2003-01183-7}{Eur.\ Phys.\ J.\ {\bf C28}{,}
  155}  (2003), \href{http://arxiv.org/abs/hep-ex/0303032}{{\tt
  arXiv:hep-ex/0303032 [hep-ex]}}\relax
\mciteBstWouldAddEndPuncttrue
\mciteSetBstMidEndSepPunct{\mcitedefaultmidpunct}
{\mcitedefaultendpunct}{\mcitedefaultseppunct}\relax
\EndOfBibitem
\bibitem{Acciarri:1998pq}
M.~Acciarri {\em et al.} ({L3} collaboration){,}
  \href{http://dx.doi.org/10.1007/s100520050262}{Eur.\ Phys.\ J.\ {\bf C5},  195}
   (1998)\relax
\mciteBstWouldAddEndPuncttrue
\mciteSetBstMidEndSepPunct{\mcitedefaultmidpunct}
{\mcitedefaultendpunct}{\mcitedefaultseppunct}\relax
\EndOfBibitem
\bibitem{Ackerstaff:1997iw}
K.~Ackerstaff {\em et al.} ({OPAL} collaboration){,}
  \href{http://dx.doi.org/10.1007/s002880050565}{Z.\ Phys.\ {\bf C76},  417}
  (1997), \href{http://arxiv.org/abs/hep-ex/9707010}{{\tt arXiv:hep-ex/9707010
  [hep-ex]}}\relax
\mciteBstWouldAddEndPuncttrue
\mciteSetBstMidEndSepPunct{\mcitedefaultmidpunct}
{\mcitedefaultendpunct}{\mcitedefaultseppunct}\relax
\EndOfBibitem
\bibitem{Ackerstaff:1997vd}
K.~Ackerstaff {\em et al.} ({OPAL} collaboration){,}
  \href{http://dx.doi.org/10.1007/s002880050564}{Z.\ Phys.\ {\bf C76},  401}
  (1997), \href{http://arxiv.org/abs/hep-ex/9707009}{{\tt arXiv:hep-ex/9707009
  [hep-ex]}}\relax
\mciteBstWouldAddEndPuncttrue
\mciteSetBstMidEndSepPunct{\mcitedefaultmidpunct}
{\mcitedefaultendpunct}{\mcitedefaultseppunct}\relax
\EndOfBibitem
\bibitem{Alexander:1996id}
G.~Alexander {\em et al.} ({OPAL} collaboration){,}
  \href{http://dx.doi.org/10.1007/s002880050258}{Z.\ Phys.\ {\bf C72},  377}
  (1996)\relax
\mciteBstWouldAddEndPuncttrue
\mciteSetBstMidEndSepPunct{\mcitedefaultmidpunct}
{\mcitedefaultendpunct}{\mcitedefaultseppunct}\relax
\EndOfBibitem
\bibitem{Abe:1997qf}
F.~Abe {\em et al.} ({CDF} collaboration){,}
  \href{http://dx.doi.org/10.1103/PhysRevLett.80.2057}{Phys.\ Rev.\ Lett.\ {\bf
  80},  2057}  (1998), \href{http://arxiv.org/abs/hep-ex/9712004}{{\tt
  arXiv:hep-ex/9712004 [hep-ex]}}\relax
\mciteBstWouldAddEndPuncttrue
\mciteSetBstMidEndSepPunct{\mcitedefaultmidpunct}
{\mcitedefaultendpunct}{\mcitedefaultseppunct}\relax
\EndOfBibitem
\bibitem{Abe:1998sq}
F.~Abe {\em et al.} ({CDF} collaboration){,}
  \href{http://dx.doi.org/10.1103/PhysRevD.59.032001}{Phys.\ Rev.\ {\bf D59}{,}
  032001}  (1999), \href{http://arxiv.org/abs/hep-ex/9806026}{{\tt
  arXiv:hep-ex/9806026 [hep-ex]}}\relax
\mciteBstWouldAddEndPuncttrue
\mciteSetBstMidEndSepPunct{\mcitedefaultmidpunct}
{\mcitedefaultendpunct}{\mcitedefaultseppunct}\relax
\EndOfBibitem
\bibitem{Abe:1999pv}
F.~Abe {\em et al.} ({CDF} collaboration){,}
  \href{http://dx.doi.org/10.1103/PhysRevD.60.051101}{Phys.\ Rev.\ {\bf D60}{,}
  051101}  (1999)\relax
\mciteBstWouldAddEndPuncttrue
\mciteSetBstMidEndSepPunct{\mcitedefaultmidpunct}
{\mcitedefaultendpunct}{\mcitedefaultseppunct}\relax
\EndOfBibitem
\bibitem{Abe:1999ds}
F.~Abe {\em et al.} ({CDF} collaboration){,}
  \href{http://dx.doi.org/10.1103/PhysRevD.60.072003}{Phys.\ Rev.\ {\bf D60}{,}
  072003}  (1999), \href{http://arxiv.org/abs/hep-ex/9903011}{{\tt
  arXiv:hep-ex/9903011 [hep-ex]}}\relax
\mciteBstWouldAddEndPuncttrue
\mciteSetBstMidEndSepPunct{\mcitedefaultmidpunct}
{\mcitedefaultendpunct}{\mcitedefaultseppunct}\relax
\EndOfBibitem
\bibitem{Affolder:1999cn}
T.~Affolder {\em et al.} ({CDF} collaboration){,}
  \href{http://dx.doi.org/10.1103/PhysRevD.60.112004}{Phys.\ Rev.\ {\bf D60}{,}
  112004}  (1999), \href{http://arxiv.org/abs/hep-ex/9907053}{{\tt
  arXiv:hep-ex/9907053 [hep-ex]}}\relax
\mciteBstWouldAddEndPuncttrue
\mciteSetBstMidEndSepPunct{\mcitedefaultmidpunct}
{\mcitedefaultendpunct}{\mcitedefaultseppunct}\relax
\EndOfBibitem
\bibitem{Abazov:2006qp}
V.~M.\ Abazov {\em et al.} ({\dzero} collaboration){,}
  \href{http://dx.doi.org/10.1103/PhysRevD.74.112002}{Phys.\ Rev.\ {\bf D74}{,}
  112002}  (2006), \href{http://arxiv.org/abs/hep-ex/0609034}{{\tt
  arXiv:hep-ex/0609034 [hep-ex]}}\relax
\mciteBstWouldAddEndPuncttrue
\mciteSetBstMidEndSepPunct{\mcitedefaultmidpunct}
{\mcitedefaultendpunct}{\mcitedefaultseppunct}\relax
\EndOfBibitem
\bibitem{Aubert:2001te}
B.~Aubert {\em et al.} ({\babar} collaboration){,}
  \href{http://dx.doi.org/10.1103/PhysRevLett.88.221802}{Phys.\ Rev.\ Lett.\ {\bf
  88},  221802}  (2002), \href{http://arxiv.org/abs/hep-ex/0112044}{{\tt
  arXiv:hep-ex/0112044 [hep-ex]}}\relax
\mciteBstWouldAddEndPuncttrue
\mciteSetBstMidEndSepPunct{\mcitedefaultmidpunct}
{\mcitedefaultendpunct}{\mcitedefaultseppunct}\relax
\EndOfBibitem
\bibitem{Aubert:2002rg}
B.~Aubert {\em et al.} ({\babar} collaboration){,}
  \href{http://dx.doi.org/10.1103/PhysRevD.66.032003}{Phys.\ Rev.\ {\bf D66}{,}
  032003}  (2002), \href{http://arxiv.org/abs/hep-ex/0201020}{{\tt
  arXiv:hep-ex/0201020 [hep-ex]}}\relax
\mciteBstWouldAddEndPuncttrue
\mciteSetBstMidEndSepPunct{\mcitedefaultmidpunct}
{\mcitedefaultendpunct}{\mcitedefaultseppunct}\relax
\EndOfBibitem
\bibitem{Aubert:2001tf}
B.~Aubert {\em et al.} ({\babar} collaboration){,}
  \href{http://dx.doi.org/10.1103/PhysRevLett.88.221803}{Phys.\ Rev.\ Lett.\ {\bf
  88},  221803}  (2002), \href{http://arxiv.org/abs/hep-ex/0112045}{{\tt
  arXiv:hep-ex/0112045 [hep-ex]}}\relax
\mciteBstWouldAddEndPuncttrue
\mciteSetBstMidEndSepPunct{\mcitedefaultmidpunct}
{\mcitedefaultendpunct}{\mcitedefaultseppunct}\relax
\EndOfBibitem
\bibitem{Zheng:2002jv}
Y.~Zheng {\em et al.} ({Belle} collaboration){,}
  \href{http://dx.doi.org/10.1103/PhysRevD.67.092004}{Phys.\ Rev.\ {\bf D67}{,}
  092004}  (2003), \href{http://arxiv.org/abs/hep-ex/0211065}{{\tt
  arXiv:hep-ex/0211065 [hep-ex]}}\relax
\mciteBstWouldAddEndPuncttrue
\mciteSetBstMidEndSepPunct{\mcitedefaultmidpunct}
{\mcitedefaultendpunct}{\mcitedefaultseppunct}\relax
\EndOfBibitem
\bibitem{LHCb-CONF-2011-010_published}
{LHCb} collaboration, LHCb-CONF-2011-010, 2011{,}
  \url{{https://cdsweb.cern.ch/record/1331124}}, this result has been published
  in Ref.~\cite{Aaij:2011qx}\relax
\mciteBstWouldAddEndPuncttrue
\mciteSetBstMidEndSepPunct{\mcitedefaultmidpunct}
{\mcitedefaultendpunct}{\mcitedefaultseppunct}\relax
\EndOfBibitem
\bibitem{Aaij:2012nt}
R.~Aaij {\em et al.} ({LHCb} collaboration){,}
  \href{http://dx.doi.org/10.1016/j.physletb.2013.01.019}{Phys.\ Lett.\ {\bf
  B719},  318}  (2013), \href{http://arxiv.org/abs/1210.6750}{{\tt
  arXiv:1210.6750 [hep-ex]}}\relax
\mciteBstWouldAddEndPuncttrue
\mciteSetBstMidEndSepPunct{\mcitedefaultmidpunct}
{\mcitedefaultendpunct}{\mcitedefaultseppunct}\relax
\EndOfBibitem
\bibitem{Aaij:2013gja}
R.~Aaij {\em et al.} ({LHCb} collaboration){,}
  \href{http://dx.doi.org/10.1140/epjc/s10052-013-2655-8}{Eur.\ Phys.\ J.\ {\bf
  C73},  2655}  (2013), \href{http://arxiv.org/abs/1308.1302}{{\tt
  arXiv:1308.1302 [hep-ex]}}\relax
\mciteBstWouldAddEndPuncttrue
\mciteSetBstMidEndSepPunct{\mcitedefaultmidpunct}
{\mcitedefaultendpunct}{\mcitedefaultseppunct}\relax
\EndOfBibitem
\bibitem{Aaij:2016fdk}
R.~Aaij {\em et al.} ({LHCb} collaboration){,}
  \href{http://dx.doi.org/10.1140/epjc/s10052-016-4250-2}{Eur.\ Phys.\ J.\ {\bf
  C76},  412}  (2016), \href{http://arxiv.org/abs/1604.03475}{{\tt
  arXiv:1604.03475 [hep-ex]}}\relax
\mciteBstWouldAddEndPuncttrue
\mciteSetBstMidEndSepPunct{\mcitedefaultmidpunct}
{\mcitedefaultendpunct}{\mcitedefaultseppunct}\relax
\EndOfBibitem
\bibitem{Albrecht:1992yd}
H.~Albrecht {\em et al.} ({ARGUS} collaboration){,}
  \href{http://dx.doi.org/10.1007/BF01565092}{Z.\ Phys.\ {\bf C55},  357}
  (1992)\relax
\mciteBstWouldAddEndPuncttrue
\mciteSetBstMidEndSepPunct{\mcitedefaultmidpunct}
{\mcitedefaultendpunct}{\mcitedefaultseppunct}\relax
\EndOfBibitem
\bibitem{Albrecht:1993gr}
H.~Albrecht {\em et al.} ({ARGUS} collaboration){,}
  \href{http://dx.doi.org/10.1016/0370-2693(94)90415-4}{Phys.\ Lett.\ {\bf B324}{,}
   249}  (1994)\relax
\mciteBstWouldAddEndPuncttrue
\mciteSetBstMidEndSepPunct{\mcitedefaultmidpunct}
{\mcitedefaultendpunct}{\mcitedefaultseppunct}\relax
\EndOfBibitem
\bibitem{Bartelt:1993cf}
J.~E.\ Bartelt {\em et al.} ({CLEO} collaboration){,}
  \href{http://dx.doi.org/10.1103/PhysRevLett.71.1680}{Phys.\ Rev.\ Lett.\ {\bf
  71},  1680}  (1993)\relax
\mciteBstWouldAddEndPuncttrue
\mciteSetBstMidEndSepPunct{\mcitedefaultmidpunct}
{\mcitedefaultendpunct}{\mcitedefaultseppunct}\relax
\EndOfBibitem
\bibitem{Behrens:2000qu}
B.~H.\ Behrens {\em et al.} ({CLEO} collaboration){,}
  \href{http://dx.doi.org/10.1016/S0370-2693(00)00990-4}{Phys.\ Lett.\ {\bf
  B490},  36}  (2000), \href{http://arxiv.org/abs/hep-ex/0005013}{{\tt
  arXiv:hep-ex/0005013 [hep-ex]}}\relax
\mciteBstWouldAddEndPuncttrue
\mciteSetBstMidEndSepPunct{\mcitedefaultmidpunct}
{\mcitedefaultendpunct}{\mcitedefaultseppunct}\relax
\EndOfBibitem
\bibitem{Aubert:2003hd}
B.~Aubert {\em et al.} ({\babar} collaboration){,}
  \href{http://dx.doi.org/10.1103/PhysRevLett.92.181801}{Phys.\ Rev.\ Lett.\ {\bf
  92},  181801}  (2004), \href{http://arxiv.org/abs/hep-ex/0311037}{{\tt
  arXiv:hep-ex/0311037 [hep-ex]}}\relax
\mciteBstWouldAddEndPuncttrue
\mciteSetBstMidEndSepPunct{\mcitedefaultmidpunct}
{\mcitedefaultendpunct}{\mcitedefaultseppunct}\relax
\EndOfBibitem
\bibitem{Aubert:2004xga}
B.~Aubert {\em et al.} ({\babar} collaboration){,}
  \href{http://dx.doi.org/10.1103/PhysRevD.70.012007}{Phys.\ Rev.\ {\bf D70}{,}
  012007}  (2004), \href{http://arxiv.org/abs/hep-ex/0403002}{{\tt
  arXiv:hep-ex/0403002 [hep-ex]}}\relax
\mciteBstWouldAddEndPuncttrue
\mciteSetBstMidEndSepPunct{\mcitedefaultmidpunct}
{\mcitedefaultendpunct}{\mcitedefaultseppunct}\relax
\EndOfBibitem
\bibitem{Higuchi:2012kx}
T.~Higuchi {\em et al.} ({Belle} collaboration){,}
  \href{http://dx.doi.org/10.1103/PhysRevD.85.071105}{Phys.\ Rev.\ {\bf D85}{,}
  071105}  (2012), \href{http://arxiv.org/abs/1203.0930}{{\tt arXiv:1203.0930
  [hep-ex]}}\relax
\mciteBstWouldAddEndPuncttrue
\mciteSetBstMidEndSepPunct{\mcitedefaultmidpunct}
{\mcitedefaultendpunct}{\mcitedefaultseppunct}\relax
\EndOfBibitem
\bibitem{Gershon:2010wx}
T.~Gershon, \href{http://dx.doi.org/10.1088/0954-3899/38/1/015007}{J.\ Phys.\
  {\bf G38},  015007}  (2011), \href{http://arxiv.org/abs/1007.5135}{{\tt
  arXiv:1007.5135 [hep-ph]}}, Erratum ibid.\
  \href{http://dx.doi.org/10.1088/0954-3899/42/11/119501}{{\bf G42}, 119501}
  (2015)\relax
\mciteBstWouldAddEndPuncttrue
\mciteSetBstMidEndSepPunct{\mcitedefaultmidpunct}
{\mcitedefaultendpunct}{\mcitedefaultseppunct}\relax
\EndOfBibitem
\bibitem{Aaboud:2016bro}
M.~Aaboud {\em et al.} ({ATLAS} collaboration){,}
  \href{http://dx.doi.org/10.1007/JHEP06(2016)081}{JHEP {\bf 06},  081}
  (2016), \href{http://arxiv.org/abs/1605.07485}{{\tt arXiv:1605.07485
  [hep-ex]}}\relax
\mciteBstWouldAddEndPuncttrue
\mciteSetBstMidEndSepPunct{\mcitedefaultmidpunct}
{\mcitedefaultendpunct}{\mcitedefaultseppunct}\relax
\EndOfBibitem
\bibitem{Charles:2011va_mod}
J.~Charles {\em et al.} ({CKMfitter group}){,}
  \href{http://dx.doi.org/10.1103/PhysRevD.84.033005}{Phys.\ Rev.\ {\bf D84}{,}
  033005}  (2011), \href{http://arxiv.org/abs/1106.4041}{{\tt arXiv:1106.4041
  [hep-ph]}}, with updated results and plots available at
  \url{http://ckmfitter.in2p3.fr}\relax
\mciteBstWouldAddEndPuncttrue
\mciteSetBstMidEndSepPunct{\mcitedefaultmidpunct}
{\mcitedefaultendpunct}{\mcitedefaultseppunct}\relax
\EndOfBibitem
\bibitem{Bona:2006ah_mod}
M.~Bona {\em et al.} ({UTfit} collaboration){,}
  \href{http://dx.doi.org/10.1088/1126-6708/2006/10/081}{JHEP {\bf 10},  081}
  (2006), \href{http://arxiv.org/abs/hep-ph/0606167}{{\tt arXiv:hep-ph/0606167
  [hep-ph]}}, with similar updated results and plots available at
  \url{http://www.utfit.org}\relax
\mciteBstWouldAddEndPuncttrue
\mciteSetBstMidEndSepPunct{\mcitedefaultmidpunct}
{\mcitedefaultendpunct}{\mcitedefaultseppunct}\relax
\EndOfBibitem
\bibitem{Abazov:2013uma}
V.~M.\ Abazov {\em et al.} ({\dzero} collaboration){,}
  \href{http://dx.doi.org/10.1103/PhysRevD.89.012002}{Phys.\ Rev.\ {\bf D89}{,}
  012002}  (2014), \href{http://arxiv.org/abs/1310.0447}{{\tt arXiv:1310.0447
  [hep-ex]}}\relax
\mciteBstWouldAddEndPuncttrue
\mciteSetBstMidEndSepPunct{\mcitedefaultmidpunct}
{\mcitedefaultendpunct}{\mcitedefaultseppunct}\relax
\EndOfBibitem
\bibitem{Nierste_CKM2014}
 {U.~Nierste, talk presented at the 8th International Workshop on the CKM
  unitarity Triangle (CKM 2014)}, 2014, {\small
  \url{{http://indico.cern.ch/event/253826/contributions/567426/}}}\relax
\mciteBstWouldAddEndPuncttrue
\mciteSetBstMidEndSepPunct{\mcitedefaultmidpunct}
{\mcitedefaultendpunct}{\mcitedefaultseppunct}\relax
\EndOfBibitem
\bibitem{Aaltonen:2012ie}
T.~Aaltonen {\em et al.} ({CDF} collaboration){,}
  \href{http://dx.doi.org/10.1103/PhysRevLett.109.171802}{Phys.\ Rev.\ Lett.\ {\bf
  109},  171802}  (2012), \href{http://arxiv.org/abs/1208.2967}{{\tt
  arXiv:1208.2967 [hep-ex]}}\relax
\mciteBstWouldAddEndPuncttrue
\mciteSetBstMidEndSepPunct{\mcitedefaultmidpunct}
{\mcitedefaultendpunct}{\mcitedefaultseppunct}\relax
\EndOfBibitem
\bibitem{Abazov:2011ry}
V.~M.\ Abazov {\em et al.} ({\dzero} collaboration){,}
  \href{http://dx.doi.org/10.1103/PhysRevD.85.032006}{Phys.\ Rev.\ {\bf D85}{,}
  032006}  (2012), \href{http://arxiv.org/abs/1109.3166}{{\tt arXiv:1109.3166
  [hep-ex]}}\relax
\mciteBstWouldAddEndPuncttrue
\mciteSetBstMidEndSepPunct{\mcitedefaultmidpunct}
{\mcitedefaultendpunct}{\mcitedefaultseppunct}\relax
\EndOfBibitem
\bibitem{Aad:2014cqa}
G.~Aad {\em et al.} ({ATLAS} collaboration){,}
  \href{http://dx.doi.org/10.1103/PhysRevD.90.052007}{Phys.\ Rev.\ {\bf D90}{,}
  052007}  (2014), \href{http://arxiv.org/abs/1407.1796}{{\tt arXiv:1407.1796
  [hep-ex]}}\relax
\mciteBstWouldAddEndPuncttrue
\mciteSetBstMidEndSepPunct{\mcitedefaultmidpunct}
{\mcitedefaultendpunct}{\mcitedefaultseppunct}\relax
\EndOfBibitem
\bibitem{Aad:2016tdj}
G.~Aad {\em et al.} ({ATLAS} collaboration){,}
  \href{http://dx.doi.org/10.1007/JHEP08(2016)147}{JHEP {\bf 08},  147}
  (2016), \href{http://arxiv.org/abs/1601.03297}{{\tt arXiv:1601.03297
  [hep-ex]}}\relax
\mciteBstWouldAddEndPuncttrue
\mciteSetBstMidEndSepPunct{\mcitedefaultmidpunct}
{\mcitedefaultendpunct}{\mcitedefaultseppunct}\relax
\EndOfBibitem
\bibitem{Khachatryan:2015nza}
V.~Khachatryan {\em et al.} ({CMS} collaboration){,}
  \href{http://dx.doi.org/10.1016/j.physletb.2016.03.046}{Phys.\ Lett.\ {\bf
  B757},  97}  (2016), \href{http://arxiv.org/abs/1507.07527}{{\tt
  arXiv:1507.07527 [hep-ex]}}\relax
\mciteBstWouldAddEndPuncttrue
\mciteSetBstMidEndSepPunct{\mcitedefaultmidpunct}
{\mcitedefaultendpunct}{\mcitedefaultseppunct}\relax
\EndOfBibitem
\bibitem{Aaij:2014zsa}
R.~Aaij {\em et al.} ({LHCb} collaboration){,}
  \href{http://dx.doi.org/10.1103/PhysRevLett.114.041801}{Phys.\ Rev.\ Lett.\ {\bf
  114},  041801}  (2015), \href{http://arxiv.org/abs/1411.3104}{{\tt
  arXiv:1411.3104 [hep-ex]}}\relax
\mciteBstWouldAddEndPuncttrue
\mciteSetBstMidEndSepPunct{\mcitedefaultmidpunct}
{\mcitedefaultendpunct}{\mcitedefaultseppunct}\relax
\EndOfBibitem
\bibitem{Aaij:2016ohx}
R.~Aaij {\em et al.} ({LHCb} collaboration){,}
  \href{http://dx.doi.org/10.1016/j.physletb.2016.09.028}{Phys.\ Lett.\ {\bf
  B762},  253}  (2016), \href{http://arxiv.org/abs/1608.04855}{{\tt
  arXiv:1608.04855 [hep-ex]}}\relax
\mciteBstWouldAddEndPuncttrue
\mciteSetBstMidEndSepPunct{\mcitedefaultmidpunct}
{\mcitedefaultendpunct}{\mcitedefaultseppunct}\relax
\EndOfBibitem
\bibitem{Lenz:2012mb}
A.~Lenz, \href{http://arxiv.org/abs/1205.1444}{{\tt arXiv:1205.1444 [hep-ph]}}
   (2012)\relax
\mciteBstWouldAddEndPuncttrue
\mciteSetBstMidEndSepPunct{\mcitedefaultmidpunct}
{\mcitedefaultendpunct}{\mcitedefaultseppunct}\relax
\EndOfBibitem
\bibitem{Esen:2010jq}
S.~Esen {\em et al.} ({Belle} collaboration){,}
  \href{http://dx.doi.org/10.1103/PhysRevLett.105.201802}{Phys.\ Rev.\ Lett.\ {\bf
  105},  201802}  (2010), \href{http://arxiv.org/abs/1005.5177}{{\tt
  arXiv:1005.5177 [hep-ex]}}\relax
\mciteBstWouldAddEndPuncttrue
\mciteSetBstMidEndSepPunct{\mcitedefaultmidpunct}
{\mcitedefaultendpunct}{\mcitedefaultseppunct}\relax
\EndOfBibitem
\bibitem{Abazov:2008ig}
V.~Abazov {\em et al.} ({\dzero} collaboration){,}
  \href{http://dx.doi.org/10.1103/PhysRevLett.102.091801}{Phys.\ Rev.\ Lett.\ {\bf
  102},  091801}  (2009), \href{http://arxiv.org/abs/0811.2173}{{\tt
  arXiv:0811.2173 [hep-ex]}}\relax
\mciteBstWouldAddEndPuncttrue
\mciteSetBstMidEndSepPunct{\mcitedefaultmidpunct}
{\mcitedefaultendpunct}{\mcitedefaultseppunct}\relax
\EndOfBibitem
\bibitem{Abulencia:2007zz}
T.~Aaltonen {\em et al.} ({CDF} collaboration){,}
  \href{http://dx.doi.org/10.1103/PhysRevLett.100.021803}{Phys.\ Rev.\ Lett.\ {\bf
  100},  021803}  (2008)\relax
\mciteBstWouldAddEndPuncttrue
\mciteSetBstMidEndSepPunct{\mcitedefaultmidpunct}
{\mcitedefaultendpunct}{\mcitedefaultseppunct}\relax
\EndOfBibitem
\bibitem{Lenz:2011ti}
A.~Lenz and U.~Nierste, \href{http://arxiv.org/abs/1102.4274}{{\tt
  arXiv:1102.4274 [hep-ph]}}  (2011)\relax
\mciteBstWouldAddEndPuncttrue
\mciteSetBstMidEndSepPunct{\mcitedefaultmidpunct}
{\mcitedefaultendpunct}{\mcitedefaultseppunct}\relax
\EndOfBibitem
\bibitem{Lenz:2006hd}
A.~Lenz and U.~Nierste{,}
  \href{http://dx.doi.org/10.1088/1126-6708/2007/06/072}{JHEP {\bf 06},  072}
  (2007), \href{http://arxiv.org/abs/hep-ph/0612167}{{\tt arXiv:hep-ph/0612167
  [hep-ph]}}\relax
\mciteBstWouldAddEndPuncttrue
\mciteSetBstMidEndSepPunct{\mcitedefaultmidpunct}
{\mcitedefaultendpunct}{\mcitedefaultseppunct}\relax
\EndOfBibitem
\bibitem{Heister:2002gk}
A.~Heister {\em et al.} ({ALEPH} collaboration){,}
  \href{http://dx.doi.org/10.1140/epjc/s2003-01230-5}{Eur.\ Phys.\ J.\ {\bf C29}{,}
  143}  (2003)\relax
\mciteBstWouldAddEndPuncttrue
\mciteSetBstMidEndSepPunct{\mcitedefaultmidpunct}
{\mcitedefaultendpunct}{\mcitedefaultseppunct}\relax
\EndOfBibitem
\bibitem{Abdallah:2003qga}
J.~Abdallah {\em et al.} ({DELPHI} collaboration){,}
  \href{http://dx.doi.org/10.1140/epjc/s2004-01827-0}{Eur.\ Phys.\ J.\ {\bf C35}{,}
  35}  (2004), \href{http://arxiv.org/abs/hep-ex/0404013}{{\tt
  arXiv:hep-ex/0404013 [hep-ex]}}\relax
\mciteBstWouldAddEndPuncttrue
\mciteSetBstMidEndSepPunct{\mcitedefaultmidpunct}
{\mcitedefaultendpunct}{\mcitedefaultseppunct}\relax
\EndOfBibitem
\bibitem{Abbiendi:1999gm}
G.~Abbiendi {\em et al.} ({OPAL} collaboration){,}
  \href{http://dx.doi.org/10.1007/s100520050658}{Eur.\ Phys.\ J.\ {\bf C11}{,}
  587}  (1999), \href{http://arxiv.org/abs/hep-ex/9907061}{{\tt
  arXiv:hep-ex/9907061 [hep-ex]}}\relax
\mciteBstWouldAddEndPuncttrue
\mciteSetBstMidEndSepPunct{\mcitedefaultmidpunct}
{\mcitedefaultendpunct}{\mcitedefaultseppunct}\relax
\EndOfBibitem
\bibitem{Abbiendi:2000bh}
G.~Abbiendi {\em et al.} ({OPAL} collaboration){,}
  \href{http://dx.doi.org/10.1007/s100520100591}{Eur.\ Phys.\ J.\ {\bf C19}{,}
  241}  (2001), \href{http://arxiv.org/abs/hep-ex/0011052}{{\tt
  arXiv:hep-ex/0011052 [hep-ex]}}\relax
\mciteBstWouldAddEndPuncttrue
\mciteSetBstMidEndSepPunct{\mcitedefaultmidpunct}
{\mcitedefaultendpunct}{\mcitedefaultseppunct}\relax
\EndOfBibitem
\bibitem{Abe:2002ua}
K.~Abe {\em et al.} ({SLD} collaboration){,}
  \href{http://dx.doi.org/10.1103/PhysRevD.67.012006}{Phys.\ Rev.\ {\bf D67}{,}
  012006}  (2003), \href{http://arxiv.org/abs/hep-ex/0209002}{{\tt
  arXiv:hep-ex/0209002 [hep-ex]}}\relax
\mciteBstWouldAddEndPuncttrue
\mciteSetBstMidEndSepPunct{\mcitedefaultmidpunct}
{\mcitedefaultendpunct}{\mcitedefaultseppunct}\relax
\EndOfBibitem
\bibitem{Abe:2002wfa}
K.~Abe {\em et al.} ({SLD} collaboration){,}
  \href{http://dx.doi.org/10.1103/PhysRevD.66.032009}{Phys.\ Rev.\ {\bf D66}{,}
  032009}  (2002), \href{http://arxiv.org/abs/hep-ex/0207048}{{\tt
  arXiv:hep-ex/0207048 [hep-ex]}}\relax
\mciteBstWouldAddEndPuncttrue
\mciteSetBstMidEndSepPunct{\mcitedefaultmidpunct}
{\mcitedefaultendpunct}{\mcitedefaultseppunct}\relax
\EndOfBibitem
\bibitem{Abe:1998qj}
F.~Abe {\em et al.} ({CDF} collaboration){,}
  \href{http://dx.doi.org/10.1103/PhysRevLett.82.3576}{Phys.\ Rev.\ Lett.\ {\bf
  82},  3576}  (1999)\relax
\mciteBstWouldAddEndPuncttrue
\mciteSetBstMidEndSepPunct{\mcitedefaultmidpunct}
{\mcitedefaultendpunct}{\mcitedefaultseppunct}\relax
\EndOfBibitem
\bibitem{Abazov:2006dm}
V.~M.\ Abazov {\em et al.} ({\dzero} collaboration){,}
  \href{http://dx.doi.org/10.1103/PhysRevLett.97.021802}{Phys.\ Rev.\ Lett.\ {\bf
  97},  021802}  (2006), \href{http://arxiv.org/abs/hep-ex/0603029}{{\tt
  arXiv:hep-ex/0603029 [hep-ex]}}\relax
\mciteBstWouldAddEndPuncttrue
\mciteSetBstMidEndSepPunct{\mcitedefaultmidpunct}
{\mcitedefaultendpunct}{\mcitedefaultseppunct}\relax
\EndOfBibitem
\bibitem{Abulencia:2006ze}
A.~Abulencia {\em et al.} ({CDF} collaboration){,}
  \href{http://dx.doi.org/10.1103/PhysRevLett.97.242003}{Phys.\ Rev.\ Lett.\ {\bf
  97},  242003}  (2006), \href{http://arxiv.org/abs/hep-ex/0609040}{{\tt
  arXiv:hep-ex/0609040 [hep-ex]}}\relax
\mciteBstWouldAddEndPuncttrue
\mciteSetBstMidEndSepPunct{\mcitedefaultmidpunct}
{\mcitedefaultendpunct}{\mcitedefaultseppunct}\relax
\EndOfBibitem
\bibitem{Aaij:2011qx}
R.~Aaij {\em et al.} ({LHCb} collaboration){,}
  \href{http://dx.doi.org/10.1016/j.physletb.2012.02.031}{Phys.\ Lett.\ {\bf
  B709},  177}  (2012), \href{http://arxiv.org/abs/1112.4311}{{\tt
  arXiv:1112.4311 [hep-ex]}}\relax
\mciteBstWouldAddEndPuncttrue
\mciteSetBstMidEndSepPunct{\mcitedefaultmidpunct}
{\mcitedefaultendpunct}{\mcitedefaultseppunct}\relax
\EndOfBibitem
\bibitem{Aaij:2013mpa}
R.~Aaij {\em et al.} ({LHCb} collaboration){,}
  \href{http://dx.doi.org/10.1088/1367-2630/15/5/053021}{New J.\ Phys.\ {\bf 15}{,}
   053021}  (2013), \href{http://arxiv.org/abs/1304.4741}{{\tt arXiv:1304.4741
  [hep-ex]}}\relax
\mciteBstWouldAddEndPuncttrue
\mciteSetBstMidEndSepPunct{\mcitedefaultmidpunct}
{\mcitedefaultendpunct}{\mcitedefaultseppunct}\relax
\EndOfBibitem
\bibitem{Bazavov:2016nty}
A.~Bazavov {\em et al.} ({Fermilab Lattice and MILC} collaborations){,}
  \href{http://dx.doi.org/10.1103/PhysRevD.93.113016}{Phys.\ Rev.\ {\bf D93}{,}
  113016}  (2016), \href{http://arxiv.org/abs/1602.03560}{{\tt
  arXiv:1602.03560 [hep-lat]}}\relax
\mciteBstWouldAddEndPuncttrue
\mciteSetBstMidEndSepPunct{\mcitedefaultmidpunct}
{\mcitedefaultendpunct}{\mcitedefaultseppunct}\relax
\EndOfBibitem
\bibitem{Aoki:2016frl}
S.~Aoki {\em et al.} ({FLAG working group}){,}
  \href{http://dx.doi.org/10.1140/epjc/s10052-016-4509-7}{Eur.\ Phys.\ J.\ {\bf
  C77},  112}  (2017), \href{http://arxiv.org/abs/1607.00299}{{\tt
  arXiv:1607.00299 [hep-lat]}}, see also
  \href{http://itpwiki.unibe.ch/flag/}{{\tt
  http://itpwiki.unibe.ch/flag/}}\relax
\mciteBstWouldAddEndPuncttrue
\mciteSetBstMidEndSepPunct{\mcitedefaultmidpunct}
{\mcitedefaultendpunct}{\mcitedefaultseppunct}\relax
\EndOfBibitem
\bibitem{Jaffe:2001hz}
D.~E.\ Jaffe {\em et al.} ({CLEO} collaboration){,}
  \href{http://dx.doi.org/10.1103/PhysRevLett.86.5000}{Phys.\ Rev.\ Lett.\ {\bf
  86},  5000}  (2001), \href{http://arxiv.org/abs/hep-ex/0101006}{{\tt
  arXiv:hep-ex/0101006 [hep-ex]}}\relax
\mciteBstWouldAddEndPuncttrue
\mciteSetBstMidEndSepPunct{\mcitedefaultmidpunct}
{\mcitedefaultendpunct}{\mcitedefaultseppunct}\relax
\EndOfBibitem
\bibitem{Lees:2014kep}
J.~P.\ Lees {\em et al.} ({\babar} collaboration){,}
  \href{http://dx.doi.org/10.1103/PhysRevLett.114.081801}{Phys.\ Rev.\ Lett.\ {\bf
  114},  081801}  (2015), \href{http://arxiv.org/abs/1411.1842}{{\tt
  arXiv:1411.1842 [hep-ex]}}\relax
\mciteBstWouldAddEndPuncttrue
\mciteSetBstMidEndSepPunct{\mcitedefaultmidpunct}
{\mcitedefaultendpunct}{\mcitedefaultseppunct}\relax
\EndOfBibitem
\bibitem{Abe:1996zt}
F.~Abe {\em et al.} ({CDF} collaboration){,}
  \href{http://dx.doi.org/10.1103/PhysRevD.55.2546}{Phys.\ Rev.\ {\bf D55}{,}
  2546}  (1997)\relax
\mciteBstWouldAddEndPuncttrue
\mciteSetBstMidEndSepPunct{\mcitedefaultmidpunct}
{\mcitedefaultendpunct}{\mcitedefaultseppunct}\relax
\EndOfBibitem
\bibitem{Barate:2000uk}
R.~Barate {\em et al.} ({ALEPH} collaboration){,}
  \href{http://dx.doi.org/10.1007/s100520100644}{Eur.\ Phys.\ J.\ {\bf C20}{,}
  431}  (2001)\relax
\mciteBstWouldAddEndPuncttrue
\mciteSetBstMidEndSepPunct{\mcitedefaultmidpunct}
{\mcitedefaultendpunct}{\mcitedefaultseppunct}\relax
\EndOfBibitem
\bibitem{Lees:2013sua}
J.~P.\ Lees {\em et al.} ({\babar} collaboration){,}
  \href{http://dx.doi.org/10.1103/PhysRevLett.111.101802}{Phys.\ Rev.\ Lett.\ {\bf
  111},  101802}  (2013), \href{http://arxiv.org/abs/1305.1575}{{\tt
  arXiv:1305.1575 [hep-ex]}}, Erratum ibid.\
  \href{http://dx.doi.org/10.1103/PhysRevLett.111.159901}{{\bf 111}, 159901}
  (2013)\relax
\mciteBstWouldAddEndPuncttrue
\mciteSetBstMidEndSepPunct{\mcitedefaultmidpunct}
{\mcitedefaultendpunct}{\mcitedefaultseppunct}\relax
\EndOfBibitem
\bibitem{Aubert:2006nf}
B.~Aubert {\em et al.} ({\babar} collaboration){,}
  \href{http://dx.doi.org/10.1103/PhysRevLett.96.251802}{Phys.\ Rev.\ Lett.\ {\bf
  96},  251802}  (2006), \href{http://arxiv.org/abs/hep-ex/0603053}{{\tt
  arXiv:hep-ex/0603053 [hep-ex]}}\relax
\mciteBstWouldAddEndPuncttrue
\mciteSetBstMidEndSepPunct{\mcitedefaultmidpunct}
{\mcitedefaultendpunct}{\mcitedefaultseppunct}\relax
\EndOfBibitem
\bibitem{Nakano:2005jb}
E.~Nakano {\em et al.} ({Belle} collaboration){,}
  \href{http://dx.doi.org/10.1103/PhysRevD.73.112002}{Phys.\ Rev.\ {\bf D73}{,}
  112002}  (2006), \href{http://arxiv.org/abs/hep-ex/0505017}{{\tt
  arXiv:hep-ex/0505017 [hep-ex]}}\relax
\mciteBstWouldAddEndPuncttrue
\mciteSetBstMidEndSepPunct{\mcitedefaultmidpunct}
{\mcitedefaultendpunct}{\mcitedefaultseppunct}\relax
\EndOfBibitem
\bibitem{Beneke:1996hv}
M.~Beneke, G.~Buchalla, and I.~Dunietz{,}
  \href{http://dx.doi.org/10.1016/S0370-2693(96)01648-6}{Phys.\ Lett.\ {\bf
  B393},  132}  (1997), \href{http://arxiv.org/abs/hep-ph/9609357}{{\tt
  arXiv:hep-ph/9609357 [hep-ph]}}\relax
\mciteBstWouldAddEndPuncttrue
\mciteSetBstMidEndSepPunct{\mcitedefaultmidpunct}
{\mcitedefaultendpunct}{\mcitedefaultseppunct}\relax
\EndOfBibitem
\bibitem{Dunietz:1998av}
I.~Dunietz, \href{http://dx.doi.org/10.1007/s100529801005}{Eur.\ Phys.\ J.\ {\bf
  C7},  197}  (1999), \href{http://arxiv.org/abs/hep-ph/9806521}{{\tt
  arXiv:hep-ph/9806521 [hep-ph]}}\relax
\mciteBstWouldAddEndPuncttrue
\mciteSetBstMidEndSepPunct{\mcitedefaultmidpunct}
{\mcitedefaultendpunct}{\mcitedefaultseppunct}\relax
\EndOfBibitem
\bibitem{Abazov:2012hha}
V.~M.\ Abazov {\em et al.} ({\dzero} collaboration){,}
  \href{http://dx.doi.org/10.1103/PhysRevD.86.072009}{Phys.\ Rev.\ {\bf D86}{,}
  072009}  (2012), \href{http://arxiv.org/abs/1208.5813}{{\tt arXiv:1208.5813
  [hep-ex]}}\relax
\mciteBstWouldAddEndPuncttrue
\mciteSetBstMidEndSepPunct{\mcitedefaultmidpunct}
{\mcitedefaultendpunct}{\mcitedefaultseppunct}\relax
\EndOfBibitem
\bibitem{Aaij:2014nxa}
R.~Aaij {\em et al.} ({LHCb} collaboration){,}
  \href{http://dx.doi.org/10.1103/PhysRevLett.114.041601}{Phys.\ Rev.\ Lett.\ {\bf
  114},  041601}  (2015), \href{http://arxiv.org/abs/1409.8586}{{\tt
  arXiv:1409.8586 [hep-ex]}}\relax
\mciteBstWouldAddEndPuncttrue
\mciteSetBstMidEndSepPunct{\mcitedefaultmidpunct}
{\mcitedefaultendpunct}{\mcitedefaultseppunct}\relax
\EndOfBibitem
\bibitem{Amhis:2012bh}
Y.~Amhis {\em et al.} ({Heavy Flavor Averaging Group}){,}
  \href{http://arxiv.org/abs/1207.1158}{{\tt arXiv:1207.1158 [hep-ex]}}
  (2012)\relax
\mciteBstWouldAddEndPuncttrue
\mciteSetBstMidEndSepPunct{\mcitedefaultmidpunct}
{\mcitedefaultendpunct}{\mcitedefaultseppunct}\relax
\EndOfBibitem
\bibitem{Abazov:2012zz}
V.~M.\ Abazov {\em et al.} ({\dzero} collaboration){,}
  \href{http://dx.doi.org/10.1103/PhysRevLett.110.011801}{Phys.\ Rev.\ Lett.\ {\bf
  110},  011801}  (2013), \href{http://arxiv.org/abs/1207.1769}{{\tt
  arXiv:1207.1769 [hep-ex]}}\relax
\mciteBstWouldAddEndPuncttrue
\mciteSetBstMidEndSepPunct{\mcitedefaultmidpunct}
{\mcitedefaultendpunct}{\mcitedefaultseppunct}\relax
\EndOfBibitem
\bibitem{Aaij:2016yze}
R.~Aaij {\em et al.} ({LHCb} collaboration){,}
  \href{http://dx.doi.org/10.1103/PhysRevLett.117.061803}{Phys.\ Rev.\ Lett.\ {\bf
  117},  061803}  (2016), \href{http://arxiv.org/abs/1605.09768}{{\tt
  arXiv:1605.09768 [hep-ex]}}\relax
\mciteBstWouldAddEndPuncttrue
\mciteSetBstMidEndSepPunct{\mcitedefaultmidpunct}
{\mcitedefaultendpunct}{\mcitedefaultseppunct}\relax
\EndOfBibitem
\bibitem{Lenz_private_communication}
 A.~Lenz, private communication, 2017\relax
\mciteBstWouldAddEndPuncttrue
\mciteSetBstMidEndSepPunct{\mcitedefaultmidpunct}
{\mcitedefaultendpunct}{\mcitedefaultseppunct}\relax
\EndOfBibitem
\bibitem{DescotesGenon:2012kr}
S.~Descotes-Genon and J.~F.\ Kamenik{,}
  \href{http://dx.doi.org/10.1103/PhysRevD.87.074036}{Phys.\ Rev.\ {\bf D87}{,}
  074036}  (2013), \href{http://arxiv.org/abs/1207.4483}{{\tt arXiv:1207.4483
  [hep-ph]}}, Erratum ibid.\
  \href{http://dx.doi.org/10.1103/PhysRevD.92.079903}{{\bf D92}, 079903}
  (2015)\relax
\mciteBstWouldAddEndPuncttrue
\mciteSetBstMidEndSepPunct{\mcitedefaultmidpunct}
{\mcitedefaultendpunct}{\mcitedefaultseppunct}\relax
\EndOfBibitem
\bibitem{Aaboud:2016bmk}
M.~Aaboud {\em et al.} ({ATLAS} collaboration){,}
  \href{http://dx.doi.org/10.1007/JHEP02(2017)071}{JHEP {\bf 02},  071}
  (2017), \href{http://arxiv.org/abs/1610.07869}{{\tt arXiv:1610.07869
  [hep-ex]}}\relax
\mciteBstWouldAddEndPuncttrue
\mciteSetBstMidEndSepPunct{\mcitedefaultmidpunct}
{\mcitedefaultendpunct}{\mcitedefaultseppunct}\relax
\EndOfBibitem
\bibitem{Aaij:2014dka}
R.~Aaij {\em et al.} ({LHCb} collaboration){,}
  \href{http://dx.doi.org/10.1016/j.physletb.2014.06.079}{Phys.\ Lett.\ {\bf
  B736},  186}  (2014), \href{http://arxiv.org/abs/1405.4140}{{\tt
  arXiv:1405.4140 [hep-ex]}}\relax
\mciteBstWouldAddEndPuncttrue
\mciteSetBstMidEndSepPunct{\mcitedefaultmidpunct}
{\mcitedefaultendpunct}{\mcitedefaultseppunct}\relax
\EndOfBibitem
\bibitem{LHCb:2012ae}
R.~Aaij {\em et al.} ({LHCb} collaboration){,}
  \href{http://dx.doi.org/10.1103/PhysRevD.86.052006}{Phys.\ Rev.\ {\bf D86}{,}
  052006}  (2012), \href{http://arxiv.org/abs/1204.5643}{{\tt arXiv:1204.5643
  [hep-ex]}}\relax
\mciteBstWouldAddEndPuncttrue
\mciteSetBstMidEndSepPunct{\mcitedefaultmidpunct}
{\mcitedefaultendpunct}{\mcitedefaultseppunct}\relax
\EndOfBibitem
\bibitem{Aaij:2014ywt}
R.~Aaij {\em et al.} ({LHCb} collaboration){,}
  \href{http://dx.doi.org/10.1103/PhysRevLett.113.211801}{Phys.\ Rev.\ Lett.\ {\bf
  113},  211801}  (2014), \href{http://arxiv.org/abs/1409.4619}{{\tt
  arXiv:1409.4619 [hep-ex]}}\relax
\mciteBstWouldAddEndPuncttrue
\mciteSetBstMidEndSepPunct{\mcitedefaultmidpunct}
{\mcitedefaultendpunct}{\mcitedefaultseppunct}\relax
\EndOfBibitem
\bibitem{Aaij:2014xba}
R.~Aaij {\em et al.} ({LHCb} collaboration){,}
  \href{http://dx.doi.org/10.1016/j.physletb.2014.12.015}{Phys.\ Lett.\ {\bf
  B741},  1}  (2015), \href{http://arxiv.org/abs/1408.4368}{{\tt
  arXiv:1408.4368 [hep-ex]}}\relax
\mciteBstWouldAddEndPuncttrue
\mciteSetBstMidEndSepPunct{\mcitedefaultmidpunct}
{\mcitedefaultendpunct}{\mcitedefaultseppunct}\relax
\EndOfBibitem
\bibitem{Chau:1984fp}
L.-L.\ Chau and W.-Y.\ Keung{,}
  \href{http://dx.doi.org/10.1103/PhysRevLett.53.1802}{Phys.\ Rev.\ Lett.\ {\bf
  53},  1802}  (1984)\relax
\mciteBstWouldAddEndPuncttrue
\mciteSetBstMidEndSepPunct{\mcitedefaultmidpunct}
{\mcitedefaultendpunct}{\mcitedefaultseppunct}\relax
\EndOfBibitem
\bibitem{Wolfenstein:1983yz}
L.~Wolfenstein, \href{http://dx.doi.org/10.1103/PhysRevLett.51.1945}{Phys.\ Rev.\
  Lett.\ {\bf 51},  1945}  (1983)\relax
\mciteBstWouldAddEndPuncttrue
\mciteSetBstMidEndSepPunct{\mcitedefaultmidpunct}
{\mcitedefaultendpunct}{\mcitedefaultseppunct}\relax
\EndOfBibitem
\bibitem{Buras:1994ec}
A.~J.\ Buras, M.~E.\ Lautenbacher, and G.~Ostermaier{,}
  \href{http://dx.doi.org/10.1103/PhysRevD.50.3433}{Phys.\ Rev.\ {\bf D50}{,}
  3433}  (1994), \href{http://arxiv.org/abs/hep-ph/9403384}{{\tt
  arXiv:hep-ph/9403384}}\relax
\mciteBstWouldAddEndPuncttrue
\mciteSetBstMidEndSepPunct{\mcitedefaultmidpunct}
{\mcitedefaultendpunct}{\mcitedefaultseppunct}\relax
\EndOfBibitem
\bibitem{Jarlskog:1985ht}
C.~Jarlskog, \href{http://dx.doi.org/10.1103/PhysRevLett.55.1039}{Phys.\ Rev.\
  Lett.\ {\bf 55},  1039}  (1985)\relax
\mciteBstWouldAddEndPuncttrue
\mciteSetBstMidEndSepPunct{\mcitedefaultmidpunct}
{\mcitedefaultendpunct}{\mcitedefaultseppunct}\relax
\EndOfBibitem
\bibitem{Jarlskog:2005uq}
C.~Jarlskog, \href{http://dx.doi.org/10.1016/j.physletb.2005.04.033}{Phys.\
  Lett.\ {\bf B615},  207}  (2005){,}
  \href{http://arxiv.org/abs/hep-ph/0503199}{{\tt arXiv:hep-ph/0503199}}\relax
\mciteBstWouldAddEndPuncttrue
\mciteSetBstMidEndSepPunct{\mcitedefaultmidpunct}
{\mcitedefaultendpunct}{\mcitedefaultseppunct}\relax
\EndOfBibitem
\bibitem{Harrison:2009bz}
P.~F.\ Harrison, S.~Dallison, and W.~G.\ Scott{,}
  \href{http://dx.doi.org/10.1016/j.physletb.2009.09.004}{Phys.\ Lett.\ {\bf
  B680},  328}  (2009), \href{http://arxiv.org/abs/0904.3077}{{\tt
  arXiv:0904.3077 [hep-ph]}}\relax
\mciteBstWouldAddEndPuncttrue
\mciteSetBstMidEndSepPunct{\mcitedefaultmidpunct}
{\mcitedefaultendpunct}{\mcitedefaultseppunct}\relax
\EndOfBibitem
\bibitem{Frampton:2010ii}
P.~H.\ Frampton and X.-G.\ He{,}
  \href{http://dx.doi.org/10.1016/j.physletb.2010.03.077}{Phys.\ Lett.\ {\bf
  B688},  67}  (2010), \href{http://arxiv.org/abs/1003.0310}{{\tt
  arXiv:1003.0310 [hep-ph]}}\relax
\mciteBstWouldAddEndPuncttrue
\mciteSetBstMidEndSepPunct{\mcitedefaultmidpunct}
{\mcitedefaultendpunct}{\mcitedefaultseppunct}\relax
\EndOfBibitem
\bibitem{Frampton:2010uq}
P.~H.\ Frampton and X.-G.\ He{,}
  \href{http://dx.doi.org/10.1103/PhysRevD.82.017301}{Phys.\ Rev.\ {\bf D82}{,}
  017301}  (2010), \href{http://arxiv.org/abs/1004.3679}{{\tt arXiv:1004.3679
  [hep-ph]}}\relax
\mciteBstWouldAddEndPuncttrue
\mciteSetBstMidEndSepPunct{\mcitedefaultmidpunct}
{\mcitedefaultendpunct}{\mcitedefaultseppunct}\relax
\EndOfBibitem
\bibitem{Charles:2004jd}
J.~Charles {\em et al.} ({CKMfitter group}){,}
  \href{http://dx.doi.org/10.1140/epjc/s2005-02169-1}{Eur.\ Phys.\ J.\ {\bf C41}{,}
  1}  (2005), \href{http://arxiv.org/abs/hep-ph/0406184}{{\tt
  arXiv:hep-ph/0406184}}, see also online updates{,}
  \url{http://ckmfitter.in2p3.fr/}\relax
\mciteBstWouldAddEndPuncttrue
\mciteSetBstMidEndSepPunct{\mcitedefaultmidpunct}
{\mcitedefaultendpunct}{\mcitedefaultseppunct}\relax
\EndOfBibitem
\bibitem{Aubert:2001sp}
B.~Aubert {\em et al.} ({\babar} collaboration){,}
  \href{http://dx.doi.org/10.1103/PhysRevLett.86.2515}{Phys.\ Rev.\ Lett.\ {\bf
  86},  2515}  (2001), \href{http://arxiv.org/abs/hep-ex/0102030}{{\tt
  arXiv:hep-ex/0102030}}\relax
\mciteBstWouldAddEndPuncttrue
\mciteSetBstMidEndSepPunct{\mcitedefaultmidpunct}
{\mcitedefaultendpunct}{\mcitedefaultseppunct}\relax
\EndOfBibitem
\bibitem{Abe:2001xe}
K.~Abe {\em et al.} ({Belle} collaboration){,}
  \href{http://dx.doi.org/10.1103/PhysRevLett.87.091802}{Phys.\ Rev.\ Lett.\ {\bf
  87},  091802}  (2001), \href{http://arxiv.org/abs/hep-ex/0107061}{{\tt
  arXiv:hep-ex/0107061}}\relax
\mciteBstWouldAddEndPuncttrue
\mciteSetBstMidEndSepPunct{\mcitedefaultmidpunct}
{\mcitedefaultendpunct}{\mcitedefaultseppunct}\relax
\EndOfBibitem
\bibitem{Carter:1980tk}
A.~B.\ Carter and A.~I.\ Sanda{,}
  \href{http://dx.doi.org/10.1103/PhysRevD.23.1567}{Phys.\ Rev.\ {\bf D23}{,}
  1567}  (1981)\relax
\mciteBstWouldAddEndPuncttrue
\mciteSetBstMidEndSepPunct{\mcitedefaultmidpunct}
{\mcitedefaultendpunct}{\mcitedefaultseppunct}\relax
\EndOfBibitem
\bibitem{Bigi:1981qs}
I.~I.~Y.\ Bigi and A.~I.\ Sanda{,}
  \href{http://dx.doi.org/10.1016/0550-3213(81)90519-8}{Nucl.\ Phys.\ {\bf B193}{,}
   85}  (1981)\relax
\mciteBstWouldAddEndPuncttrue
\mciteSetBstMidEndSepPunct{\mcitedefaultmidpunct}
{\mcitedefaultendpunct}{\mcitedefaultseppunct}\relax
\EndOfBibitem
\bibitem{Dunietz:2000cr}
I.~Dunietz, R.~Fleischer, and U.~Nierste{,}
  \href{http://dx.doi.org/10.1103/PhysRevD.63.114015}{Phys.\ Rev.\ {\bf D63}{,}
  114015}  (2001), \href{http://arxiv.org/abs/hep-ph/0012219}{{\tt
  arXiv:hep-ph/0012219 [hep-ph]}}\relax
\mciteBstWouldAddEndPuncttrue
\mciteSetBstMidEndSepPunct{\mcitedefaultmidpunct}
{\mcitedefaultendpunct}{\mcitedefaultseppunct}\relax
\EndOfBibitem
\bibitem{Aaij:2014vda}
R.~Aaij {\em et al.} ({LHCb} collaboration){,}
  \href{http://dx.doi.org/10.1016/j.physletb.2015.01.008}{Phys.\ Lett.\ {\bf
  B742},  38}  (2015), \href{http://arxiv.org/abs/1411.1634}{{\tt
  arXiv:1411.1634 [hep-ex]}}\relax
\mciteBstWouldAddEndPuncttrue
\mciteSetBstMidEndSepPunct{\mcitedefaultmidpunct}
{\mcitedefaultendpunct}{\mcitedefaultseppunct}\relax
\EndOfBibitem
\bibitem{Aubert:2008ah}
B.~Aubert {\em et al.} ({\babar} collaboration){,}
  \href{http://dx.doi.org/10.1103/PhysRevD.79.032002}{Phys.\ Rev.\ {\bf D79}{,}
  032002}  (2009), \href{http://arxiv.org/abs/0808.1866}{{\tt arXiv:0808.1866
  [hep-ex]}}\relax
\mciteBstWouldAddEndPuncttrue
\mciteSetBstMidEndSepPunct{\mcitedefaultmidpunct}
{\mcitedefaultendpunct}{\mcitedefaultseppunct}\relax
\EndOfBibitem
\bibitem{Krokovny:2006sv}
P.~Krokovny {\em et al.} ({Belle} collaboration){,}
  \href{http://dx.doi.org/10.1103/PhysRevLett.97.081801}{Phys.\ Rev.\ Lett.\ {\bf
  97},  081801}  (2006), \href{http://arxiv.org/abs/hep-ex/0605023}{{\tt
  arXiv:hep-ex/0605023}}\relax
\mciteBstWouldAddEndPuncttrue
\mciteSetBstMidEndSepPunct{\mcitedefaultmidpunct}
{\mcitedefaultendpunct}{\mcitedefaultseppunct}\relax
\EndOfBibitem
\bibitem{Aubert:2007rp}
B.~Aubert {\em et al.} ({\babar} collaboration){,}
  \href{http://dx.doi.org/10.1103/PhysRevLett.99.231802}{Phys.\ Rev.\ Lett.\ {\bf
  99},  231802}  (2007), \href{http://arxiv.org/abs/0708.1544}{{\tt
  arXiv:0708.1544 [hep-ex]}}\relax
\mciteBstWouldAddEndPuncttrue
\mciteSetBstMidEndSepPunct{\mcitedefaultmidpunct}
{\mcitedefaultendpunct}{\mcitedefaultseppunct}\relax
\EndOfBibitem
\bibitem{Vorobyev:2016npn}
V.~Vorobyev {\em et al.} ({Belle} collaboration){,}
  \href{http://dx.doi.org/10.1103/PhysRevD.94.052004}{Phys.\ Rev.\ {\bf D94}{,}
  052004}  (2016), \href{http://arxiv.org/abs/1607.05813}{{\tt
  arXiv:1607.05813 [hep-ex]}}\relax
\mciteBstWouldAddEndPuncttrue
\mciteSetBstMidEndSepPunct{\mcitedefaultmidpunct}
{\mcitedefaultendpunct}{\mcitedefaultseppunct}\relax
\EndOfBibitem
\bibitem{Libby:2010nu}
J.~Libby {\em et al.} ({CLEO} collaboration){,}
  \href{http://dx.doi.org/10.1103/PhysRevD.82.112006}{Phys.\ Rev.\ {\bf D82}{,}
  112006}  (2010), \href{http://arxiv.org/abs/1010.2817}{{\tt arXiv:1010.2817
  [hep-ex]}}\relax
\mciteBstWouldAddEndPuncttrue
\mciteSetBstMidEndSepPunct{\mcitedefaultmidpunct}
{\mcitedefaultendpunct}{\mcitedefaultseppunct}\relax
\EndOfBibitem
\bibitem{Browder:1999ng}
T.~E.\ Browder, A.~Datta, P.~J.\ O'Donnell, and S.~Pakvasa{,}
  \href{http://dx.doi.org/10.1103/PhysRevD.61.054009}{Phys.\ Rev.\ {\bf D61}{,}
  054009}  (2000), \href{http://arxiv.org/abs/hep-ph/9905425}{{\tt
  arXiv:hep-ph/9905425}}\relax
\mciteBstWouldAddEndPuncttrue
\mciteSetBstMidEndSepPunct{\mcitedefaultmidpunct}
{\mcitedefaultendpunct}{\mcitedefaultseppunct}\relax
\EndOfBibitem
\bibitem{Aubert:2006fh}
B.~Aubert {\em et al.} ({\babar} collaboration){,}
  \href{http://dx.doi.org/10.1103/PhysRevD.74.091101}{Phys.\ Rev.\ {\bf D74}{,}
  091101}  (2006), \href{http://arxiv.org/abs/hep-ex/0608016}{{\tt
  arXiv:hep-ex/0608016 [hep-ex]}}\relax
\mciteBstWouldAddEndPuncttrue
\mciteSetBstMidEndSepPunct{\mcitedefaultmidpunct}
{\mcitedefaultendpunct}{\mcitedefaultseppunct}\relax
\EndOfBibitem
\bibitem{Dalseno:2007hx}
J.~Dalseno {\em et al.} ({Belle} collaboration){,}
  \href{http://dx.doi.org/10.1103/PhysRevD.76.072004}{Phys.\ Rev.\ {\bf D76}{,}
  072004}  (2007), \href{http://arxiv.org/abs/0706.2045}{{\tt arXiv:0706.2045
  [hep-ex]}}\relax
\mciteBstWouldAddEndPuncttrue
\mciteSetBstMidEndSepPunct{\mcitedefaultmidpunct}
{\mcitedefaultendpunct}{\mcitedefaultseppunct}\relax
\EndOfBibitem
\bibitem{Aaij:2014siy}
R.~Aaij {\em et al.} ({LHCb} collaboration){,}
  \href{http://dx.doi.org/10.1103/PhysRevD.90.012003}{Phys.\ Rev.\ {\bf D90}{,}
  012003}  (2014), \href{http://arxiv.org/abs/1404.5673}{{\tt arXiv:1404.5673
  [hep-ex]}}\relax
\mciteBstWouldAddEndPuncttrue
\mciteSetBstMidEndSepPunct{\mcitedefaultmidpunct}
{\mcitedefaultendpunct}{\mcitedefaultseppunct}\relax
\EndOfBibitem
\bibitem{Pelaez:2015qba}
J.~R.\ Pelaez, \href{http://dx.doi.org/10.1016/j.physrep.2016.09.001}{Phys.\
  Rept.\ {\bf 658},  1}  (2016), \href{http://arxiv.org/abs/1510.00653}{{\tt
  arXiv:1510.00653 [hep-ph]}}\relax
\mciteBstWouldAddEndPuncttrue
\mciteSetBstMidEndSepPunct{\mcitedefaultmidpunct}
{\mcitedefaultendpunct}{\mcitedefaultseppunct}\relax
\EndOfBibitem
\bibitem{Aubert:2007sd}
B.~Aubert {\em et al.} ({\babar} collaboration){,}
  \href{http://dx.doi.org/10.1103/PhysRevLett.99.161802}{Phys.\ Rev.\ Lett.\ {\bf
  99},  161802}  (2007), \href{http://arxiv.org/abs/0706.3885}{{\tt
  arXiv:0706.3885 [hep-ex]}}\relax
\mciteBstWouldAddEndPuncttrue
\mciteSetBstMidEndSepPunct{\mcitedefaultmidpunct}
{\mcitedefaultendpunct}{\mcitedefaultseppunct}\relax
\EndOfBibitem
\bibitem{Nakahama:2010nj}
Y.~Nakahama {\em et al.} ({Belle} collaboration){,}
  \href{http://dx.doi.org/10.1103/PhysRevD.82.073011}{Phys.\ Rev.\ {\bf D82}{,}
  073011}  (2010), \href{http://arxiv.org/abs/1007.3848}{{\tt arXiv:1007.3848
  [hep-ex]}}\relax
\mciteBstWouldAddEndPuncttrue
\mciteSetBstMidEndSepPunct{\mcitedefaultmidpunct}
{\mcitedefaultendpunct}{\mcitedefaultseppunct}\relax
\EndOfBibitem
\bibitem{Lees:2012kxa}
J.~P.\ Lees {\em et al.} ({\babar} collaboration){,}
  \href{http://dx.doi.org/10.1103/PhysRevD.85.112010}{Phys.\ Rev.\ {\bf D85}{,}
  112010}  (2012), \href{http://arxiv.org/abs/1201.5897}{{\tt arXiv:1201.5897
  [hep-ex]}}\relax
\mciteBstWouldAddEndPuncttrue
\mciteSetBstMidEndSepPunct{\mcitedefaultmidpunct}
{\mcitedefaultendpunct}{\mcitedefaultseppunct}\relax
\EndOfBibitem
\bibitem{Garmash:2004wa}
A.~Garmash {\em et al.} ({Belle} collaboration){,}
  \href{http://dx.doi.org/10.1103/PhysRevD.71.092003}{Phys.\ Rev.\ {\bf D71}{,}
  092003}  (2005), \href{http://arxiv.org/abs/hep-ex/0412066}{{\tt
  arXiv:hep-ex/0412066 [hep-ex]}}\relax
\mciteBstWouldAddEndPuncttrue
\mciteSetBstMidEndSepPunct{\mcitedefaultmidpunct}
{\mcitedefaultendpunct}{\mcitedefaultseppunct}\relax
\EndOfBibitem
\bibitem{Aubert:2006nu}
B.~Aubert {\em et al.} ({\babar} collaboration){,}
  \href{http://dx.doi.org/10.1103/PhysRevD.74.032003}{Phys.\ Rev.\ {\bf D74}{,}
  032003}  (2006), \href{http://arxiv.org/abs/hep-ex/0605003}{{\tt
  arXiv:hep-ex/0605003 [hep-ex]}}\relax
\mciteBstWouldAddEndPuncttrue
\mciteSetBstMidEndSepPunct{\mcitedefaultmidpunct}
{\mcitedefaultendpunct}{\mcitedefaultseppunct}\relax
\EndOfBibitem
\bibitem{Aubert:2009me}
B.~Aubert {\em et al.} ({\babar} collaboration){,}
  \href{http://dx.doi.org/10.1103/PhysRevD.80.112001}{Phys.\ Rev.\ {\bf D80}{,}
  112001}  (2009), \href{http://arxiv.org/abs/0905.3615}{{\tt arXiv:0905.3615
  [hep-ex]}}\relax
\mciteBstWouldAddEndPuncttrue
\mciteSetBstMidEndSepPunct{\mcitedefaultmidpunct}
{\mcitedefaultendpunct}{\mcitedefaultseppunct}\relax
\EndOfBibitem
\bibitem{Dalseno:2008wwa}
J.~Dalseno {\em et al.} ({Belle} collaboration){,}
  \href{http://dx.doi.org/10.1103/PhysRevD.79.072004}{Phys.\ Rev.\ {\bf D79}{,}
  072004}  (2009), \href{http://arxiv.org/abs/0811.3665}{{\tt arXiv:0811.3665
  [hep-ex]}}\relax
\mciteBstWouldAddEndPuncttrue
\mciteSetBstMidEndSepPunct{\mcitedefaultmidpunct}
{\mcitedefaultendpunct}{\mcitedefaultseppunct}\relax
\EndOfBibitem
\bibitem{Garmash:2005rv}
A.~Garmash {\em et al.} ({Belle} collaboration){,}
  \href{http://dx.doi.org/10.1103/PhysRevLett.96.251803}{Phys.\ Rev.\ Lett.\ {\bf
  96},  251803}  (2006), \href{http://arxiv.org/abs/hep-ex/0512066}{{\tt
  arXiv:hep-ex/0512066 [hep-ex]}}\relax
\mciteBstWouldAddEndPuncttrue
\mciteSetBstMidEndSepPunct{\mcitedefaultmidpunct}
{\mcitedefaultendpunct}{\mcitedefaultseppunct}\relax
\EndOfBibitem
\bibitem{Aubert:2005ce}
B.~Aubert {\em et al.} ({\babar} collaboration){,}
  \href{http://dx.doi.org/10.1103/PhysRevD.72.072003}{Phys.\ Rev.\ {\bf D72}{,}
  072003}  (2005), \href{http://arxiv.org/abs/hep-ex/0507004}{{\tt
  arXiv:hep-ex/0507004 [hep-ex]}}, Erratum ibid.\
  \href{http://dx.doi.org/10.1103/PhysRevD.74.099903}{{\bf D74}, 099903}
  (2006)\relax
\mciteBstWouldAddEndPuncttrue
\mciteSetBstMidEndSepPunct{\mcitedefaultmidpunct}
{\mcitedefaultendpunct}{\mcitedefaultseppunct}\relax
\EndOfBibitem
\bibitem{Aubert:2008bj}
B.~Aubert {\em et al.} ({\babar} collaboration){,}
  \href{http://dx.doi.org/10.1103/PhysRevD.78.012004}{Phys.\ Rev.\ {\bf D78}{,}
  012004}  (2008), \href{http://arxiv.org/abs/0803.4451}{{\tt arXiv:0803.4451
  [hep-ex]}}\relax
\mciteBstWouldAddEndPuncttrue
\mciteSetBstMidEndSepPunct{\mcitedefaultmidpunct}
{\mcitedefaultendpunct}{\mcitedefaultseppunct}\relax
\EndOfBibitem
\bibitem{Snyder:1993mx}
A.~E.\ Snyder and H.~R.\ Quinn{,}
  \href{http://dx.doi.org/10.1103/PhysRevD.48.2139}{Phys.\ Rev.\ {\bf D48}{,}
  2139}  (1993)\relax
\mciteBstWouldAddEndPuncttrue
\mciteSetBstMidEndSepPunct{\mcitedefaultmidpunct}
{\mcitedefaultendpunct}{\mcitedefaultseppunct}\relax
\EndOfBibitem
\bibitem{Quinn:2000by}
H.~R.\ Quinn and J.~P.\ Silva{,}
  \href{http://dx.doi.org/10.1103/PhysRevD.62.054002}{Phys.\ Rev.\ {\bf D62}{,}
  054002}  (2000), \href{http://arxiv.org/abs/hep-ph/0001290}{{\tt
  arXiv:hep-ph/0001290}}\relax
\mciteBstWouldAddEndPuncttrue
\mciteSetBstMidEndSepPunct{\mcitedefaultmidpunct}
{\mcitedefaultendpunct}{\mcitedefaultseppunct}\relax
\EndOfBibitem
\bibitem{Aubert:2007jn}
B.~Aubert {\em et al.} ({\babar} collaboration){,}
  \href{http://dx.doi.org/10.1103/PhysRevD.76.012004}{Phys.\ Rev.\ {\bf D76}{,}
  012004}  (2007), \href{http://arxiv.org/abs/hep-ex/0703008}{{\tt
  arXiv:hep-ex/0703008}}\relax
\mciteBstWouldAddEndPuncttrue
\mciteSetBstMidEndSepPunct{\mcitedefaultmidpunct}
{\mcitedefaultendpunct}{\mcitedefaultseppunct}\relax
\EndOfBibitem
\bibitem{Lees:2013nwa}
J.~P.\ Lees {\em et al.} ({\babar} collaboration){,}
  \href{http://dx.doi.org/10.1103/PhysRevD.88.012003}{Phys.\ Rev.\ {\bf D88}{,}
  012003}  (2013), \href{http://arxiv.org/abs/1304.3503}{{\tt arXiv:1304.3503
  [hep-ex]}}\relax
\mciteBstWouldAddEndPuncttrue
\mciteSetBstMidEndSepPunct{\mcitedefaultmidpunct}
{\mcitedefaultendpunct}{\mcitedefaultseppunct}\relax
\EndOfBibitem
\bibitem{Kusaka:2007dv}
A.~Kusaka {\em et al.} ({Belle} collaboration){,}
  \href{http://dx.doi.org/10.1103/PhysRevLett.98.221602}{Phys.\ Rev.\ Lett.\ {\bf
  98},  221602}  (2007), \href{http://arxiv.org/abs/hep-ex/0701015}{{\tt
  arXiv:hep-ex/0701015}}\relax
\mciteBstWouldAddEndPuncttrue
\mciteSetBstMidEndSepPunct{\mcitedefaultmidpunct}
{\mcitedefaultendpunct}{\mcitedefaultseppunct}\relax
\EndOfBibitem
\bibitem{Kusaka:2007mj}
A.~Kusaka {\em et al.} ({Belle} collaboration){,}
  \href{http://dx.doi.org/10.1103/PhysRevD.77.072001}{Phys.\ Rev.\ {\bf D77}{,}
  072001}  (2008), \href{http://arxiv.org/abs/0710.4974}{{\tt arXiv:0710.4974
  [hep-ex]}}\relax
\mciteBstWouldAddEndPuncttrue
\mciteSetBstMidEndSepPunct{\mcitedefaultmidpunct}
{\mcitedefaultendpunct}{\mcitedefaultseppunct}\relax
\EndOfBibitem
\bibitem{Aubert:2007pa}
B.~Aubert {\em et al.} ({\babar} collaboration){,}
  \href{http://dx.doi.org/10.1103/PhysRevLett.99.071801}{Phys.\ Rev.\ Lett.\ {\bf
  99},  071801}  (2007), \href{http://arxiv.org/abs/0705.1190}{{\tt
  arXiv:0705.1190 [hep-ex]}}\relax
\mciteBstWouldAddEndPuncttrue
\mciteSetBstMidEndSepPunct{\mcitedefaultmidpunct}
{\mcitedefaultendpunct}{\mcitedefaultseppunct}\relax
\EndOfBibitem
\bibitem{Aushev:2004uc}
T.~Aushev {\em et al.} ({Belle} collaboration){,}
  \href{http://dx.doi.org/10.1103/PhysRevLett.93.201802}{Phys.\ Rev.\ Lett.\ {\bf
  93},  201802}  (2004), \href{http://arxiv.org/abs/hep-ex/0408051}{{\tt
  arXiv:hep-ex/0408051}}\relax
\mciteBstWouldAddEndPuncttrue
\mciteSetBstMidEndSepPunct{\mcitedefaultmidpunct}
{\mcitedefaultendpunct}{\mcitedefaultseppunct}\relax
\EndOfBibitem
\bibitem{Rohrken:2012ta}
M.~Rohrken {\em et al.} ({Belle} collaboration){,}
  \href{http://dx.doi.org/10.1103/PhysRevD.85.091106}{Phys.\ Rev.\ {\bf D85}{,}
  091106}  (2012), \href{http://arxiv.org/abs/1203.6647}{{\tt arXiv:1203.6647
  [hep-ex]}}\relax
\mciteBstWouldAddEndPuncttrue
\mciteSetBstMidEndSepPunct{\mcitedefaultmidpunct}
{\mcitedefaultendpunct}{\mcitedefaultseppunct}\relax
\EndOfBibitem
\bibitem{Aubert:2003wr}
B.~Aubert {\em et al.} ({\babar} collaboration){,}
  \href{http://dx.doi.org/10.1103/PhysRevLett.91.201802}{Phys.\ Rev.\ Lett.\ {\bf
  91},  201802}  (2003), \href{http://arxiv.org/abs/hep-ex/0306030}{{\tt
  arXiv:hep-ex/0306030 [hep-ex]}}\relax
\mciteBstWouldAddEndPuncttrue
\mciteSetBstMidEndSepPunct{\mcitedefaultmidpunct}
{\mcitedefaultendpunct}{\mcitedefaultseppunct}\relax
\EndOfBibitem
\bibitem{Wang:2004va}
C.~C.\ Wang {\em et al.} ({Belle} collaboration){,}
  \href{http://dx.doi.org/10.1103/PhysRevLett.94.121801}{Phys.\ Rev.\ Lett.\ {\bf
  94},  121801}  (2005), \href{http://arxiv.org/abs/hep-ex/0408003}{{\tt
  arXiv:hep-ex/0408003}}\relax
\mciteBstWouldAddEndPuncttrue
\mciteSetBstMidEndSepPunct{\mcitedefaultmidpunct}
{\mcitedefaultendpunct}{\mcitedefaultseppunct}\relax
\EndOfBibitem
\bibitem{Aubert:2006tw}
B.~Aubert {\em et al.} ({\babar} collaboration){,}
  \href{http://dx.doi.org/10.1103/PhysRevD.73.111101}{Phys.\ Rev.\ {\bf D73}{,}
  111101}  (2006), \href{http://arxiv.org/abs/hep-ex/0602049}{{\tt
  arXiv:hep-ex/0602049}}\relax
\mciteBstWouldAddEndPuncttrue
\mciteSetBstMidEndSepPunct{\mcitedefaultmidpunct}
{\mcitedefaultendpunct}{\mcitedefaultseppunct}\relax
\EndOfBibitem
\bibitem{Aubert:2005yf}
B.~Aubert {\em et al.} ({\babar} collaboration){,}
  \href{http://dx.doi.org/10.1103/PhysRevD.71.112003}{Phys.\ Rev.\ {\bf D71}{,}
  112003}  (2005), \href{http://arxiv.org/abs/hep-ex/0504035}{{\tt
  arXiv:hep-ex/0504035}}\relax
\mciteBstWouldAddEndPuncttrue
\mciteSetBstMidEndSepPunct{\mcitedefaultmidpunct}
{\mcitedefaultendpunct}{\mcitedefaultseppunct}\relax
\EndOfBibitem
\bibitem{Long:2003wq}
O.~Long, M.~Baak, R.~N.\ Cahn, and D.~Kirkby{,}
  \href{http://dx.doi.org/10.1103/PhysRevD.68.034010}{Phys.\ Rev.\ {\bf D68}{,}
  034010}  (2003), \href{http://arxiv.org/abs/hep-ex/0303030}{{\tt
  arXiv:hep-ex/0303030}}\relax
\mciteBstWouldAddEndPuncttrue
\mciteSetBstMidEndSepPunct{\mcitedefaultmidpunct}
{\mcitedefaultendpunct}{\mcitedefaultseppunct}\relax
\EndOfBibitem
\bibitem{Bahinipati:2011yq}
S.~Bahinipati {\em et al.} ({Belle} collaboration){,}
  \href{http://dx.doi.org/10.1103/PhysRevD.84.021101}{Phys.\ Rev.\ {\bf D84}{,}
  021101}  (2011), \href{http://arxiv.org/abs/1102.0888}{{\tt arXiv:1102.0888
  [hep-ex]}}\relax
\mciteBstWouldAddEndPuncttrue
\mciteSetBstMidEndSepPunct{\mcitedefaultmidpunct}
{\mcitedefaultendpunct}{\mcitedefaultseppunct}\relax
\EndOfBibitem
\bibitem{Ronga:2006hv}
F.~J.\ Ronga {\em et al.} ({Belle} collaboration){,}
  \href{http://dx.doi.org/10.1103/PhysRevD.73.092003}{Phys.\ Rev.\ {\bf D73}{,}
  092003}  (2006), \href{http://arxiv.org/abs/hep-ex/0604013}{{\tt
  arXiv:hep-ex/0604013}}\relax
\mciteBstWouldAddEndPuncttrue
\mciteSetBstMidEndSepPunct{\mcitedefaultmidpunct}
{\mcitedefaultendpunct}{\mcitedefaultseppunct}\relax
\EndOfBibitem
\bibitem{Fleischer:2003yb}
R.~Fleischer, \href{http://dx.doi.org/10.1016/j.nuclphysb.2003.08.010}{Nucl.\
  Phys.\ {\bf B671},  459}  (2003){,}
  \href{http://arxiv.org/abs/hep-ph/0304027}{{\tt arXiv:hep-ph/0304027}}\relax
\mciteBstWouldAddEndPuncttrue
\mciteSetBstMidEndSepPunct{\mcitedefaultmidpunct}
{\mcitedefaultendpunct}{\mcitedefaultseppunct}\relax
\EndOfBibitem
\bibitem{Aaij:2014fba}
R.~Aaij {\em et al.} ({LHCb} collaboration){,}
  \href{http://dx.doi.org/10.1007/JHEP11(2014)060}{JHEP {\bf 11},  060}
  (2014), \href{http://arxiv.org/abs/1407.6127}{{\tt arXiv:1407.6127
  [hep-ex]}}\relax
\mciteBstWouldAddEndPuncttrue
\mciteSetBstMidEndSepPunct{\mcitedefaultmidpunct}
{\mcitedefaultendpunct}{\mcitedefaultseppunct}\relax
\EndOfBibitem
\bibitem{Atwood:1997zr}
D.~Atwood, M.~Gronau, and A.~Soni{,}
  \href{http://dx.doi.org/10.1103/PhysRevLett.79.185}{Phys.\ Rev.\ Lett.\ {\bf
  79},  185}  (1997), \href{http://arxiv.org/abs/hep-ph/9704272}{{\tt
  arXiv:hep-ph/9704272}}\relax
\mciteBstWouldAddEndPuncttrue
\mciteSetBstMidEndSepPunct{\mcitedefaultmidpunct}
{\mcitedefaultendpunct}{\mcitedefaultseppunct}\relax
\EndOfBibitem
\bibitem{Atwood:2004jj}
D.~Atwood, T.~Gershon, M.~Hazumi, and A.~Soni{,}
  \href{http://dx.doi.org/10.1103/PhysRevD.71.076003}{Phys.\ Rev.\ {\bf D71}{,}
  076003}  (2005), \href{http://arxiv.org/abs/hep-ph/0410036}{{\tt
  arXiv:hep-ph/0410036}}\relax
\mciteBstWouldAddEndPuncttrue
\mciteSetBstMidEndSepPunct{\mcitedefaultmidpunct}
{\mcitedefaultendpunct}{\mcitedefaultseppunct}\relax
\EndOfBibitem
\bibitem{Grinstein:2004uu}
B.~Grinstein, Y.~Grossman, Z.~Ligeti, and D.~Pirjol{,}
  \href{http://dx.doi.org/10.1103/PhysRevD.71.011504}{Phys.\ Rev.\ {\bf D71}{,}
  011504}  (2005), \href{http://arxiv.org/abs/hep-ph/0412019}{{\tt
  arXiv:hep-ph/0412019}}\relax
\mciteBstWouldAddEndPuncttrue
\mciteSetBstMidEndSepPunct{\mcitedefaultmidpunct}
{\mcitedefaultendpunct}{\mcitedefaultseppunct}\relax
\EndOfBibitem
\bibitem{Grinstein:2005nu}
B.~Grinstein and D.~Pirjol{,}
  \href{http://dx.doi.org/10.1103/PhysRevD.73.014013}{Phys.\ Rev.\ {\bf D73}{,}
  014013}  (2006), \href{http://arxiv.org/abs/hep-ph/0510104}{{\tt
  arXiv:hep-ph/0510104}}\relax
\mciteBstWouldAddEndPuncttrue
\mciteSetBstMidEndSepPunct{\mcitedefaultmidpunct}
{\mcitedefaultendpunct}{\mcitedefaultseppunct}\relax
\EndOfBibitem
\bibitem{Matsumori:2005ax}
M.~Matsumori and A.~I.\ Sanda{,}
  \href{http://dx.doi.org/10.1103/PhysRevD.73.114022}{Phys.\ Rev.\ {\bf D73}{,}
  114022}  (2006), \href{http://arxiv.org/abs/hep-ph/0512175}{{\tt
  arXiv:hep-ph/0512175}}\relax
\mciteBstWouldAddEndPuncttrue
\mciteSetBstMidEndSepPunct{\mcitedefaultmidpunct}
{\mcitedefaultendpunct}{\mcitedefaultseppunct}\relax
\EndOfBibitem
\bibitem{Ball:2006cva}
P.~Ball and R.~Zwicky{,}
  \href{http://dx.doi.org/10.1016/j.physletb.2006.10.013}{Phys.\ Lett.\ {\bf
  B642},  478}  (2006), \href{http://arxiv.org/abs/hep-ph/0609037}{{\tt
  arXiv:hep-ph/0609037}}\relax
\mciteBstWouldAddEndPuncttrue
\mciteSetBstMidEndSepPunct{\mcitedefaultmidpunct}
{\mcitedefaultendpunct}{\mcitedefaultseppunct}\relax
\EndOfBibitem
\bibitem{Muheim:2008vu}
F.~Muheim, Y.~Xie, and R.~Zwicky{,}
  \href{http://dx.doi.org/10.1016/j.physletb.2008.05.032}{Phys.\ Lett.\ {\bf
  B664},  174}  (2008), \href{http://arxiv.org/abs/0802.0876}{{\tt
  arXiv:0802.0876 [hep-ph]}}\relax
\mciteBstWouldAddEndPuncttrue
\mciteSetBstMidEndSepPunct{\mcitedefaultmidpunct}
{\mcitedefaultendpunct}{\mcitedefaultseppunct}\relax
\EndOfBibitem
\bibitem{Bigi:1988ym}
I.~I.~Y.\ Bigi and A.~I.\ Sanda{,}
  \href{http://dx.doi.org/10.1016/0370-2693(88)90836-2}{Phys.\ Lett.\ {\bf B211}{,}
   213}  (1988)\relax
\mciteBstWouldAddEndPuncttrue
\mciteSetBstMidEndSepPunct{\mcitedefaultmidpunct}
{\mcitedefaultendpunct}{\mcitedefaultseppunct}\relax
\EndOfBibitem
\bibitem{Gronau:1990ra}
M.~Gronau and D.~London.{,}
  \href{http://dx.doi.org/10.1016/0370-2693(91)91756-L}{Phys.\ Lett.\ {\bf B253}{,}
   483}  (1991)\relax
\mciteBstWouldAddEndPuncttrue
\mciteSetBstMidEndSepPunct{\mcitedefaultmidpunct}
{\mcitedefaultendpunct}{\mcitedefaultseppunct}\relax
\EndOfBibitem
\bibitem{Gronau:1991dp}
M.~Gronau and D.~Wyler{,}
  \href{http://dx.doi.org/10.1016/0370-2693(91)90034-N}{Phys.\ Lett.\ {\bf B265}{,}
   172}  (1991)\relax
\mciteBstWouldAddEndPuncttrue
\mciteSetBstMidEndSepPunct{\mcitedefaultmidpunct}
{\mcitedefaultendpunct}{\mcitedefaultseppunct}\relax
\EndOfBibitem
\bibitem{Atwood:1996ci}
D.~Atwood, I.~Dunietz, and A.~Soni{,}
  \href{http://dx.doi.org/10.1103/PhysRevLett.78.3257}{Phys.\ Rev.\ Lett.\ {\bf
  78},  3257}  (1997), \href{http://arxiv.org/abs/hep-ph/9612433}{{\tt
  arXiv:hep-ph/9612433}}\relax
\mciteBstWouldAddEndPuncttrue
\mciteSetBstMidEndSepPunct{\mcitedefaultmidpunct}
{\mcitedefaultendpunct}{\mcitedefaultseppunct}\relax
\EndOfBibitem
\bibitem{Atwood:2000ck}
D.~Atwood, I.~Dunietz, and A.~Soni{,}
  \href{http://dx.doi.org/10.1103/PhysRevD.63.036005}{Phys.\ Rev.\ {\bf D63}{,}
  036005}  (2001), \href{http://arxiv.org/abs/hep-ph/0008090}{{\tt
  arXiv:hep-ph/0008090}}\relax
\mciteBstWouldAddEndPuncttrue
\mciteSetBstMidEndSepPunct{\mcitedefaultmidpunct}
{\mcitedefaultendpunct}{\mcitedefaultseppunct}\relax
\EndOfBibitem
\bibitem{Giri:2003ty}
A.~Giri, Y.~Grossman, A.~Soffer, and J.~Zupan{,}
  \href{http://dx.doi.org/10.1103/PhysRevD.68.054018}{Phys.\ Rev.\ {\bf D68}{,}
  054018}  (2003), \href{http://arxiv.org/abs/hep-ph/0303187}{{\tt
  arXiv:hep-ph/0303187}}\relax
\mciteBstWouldAddEndPuncttrue
\mciteSetBstMidEndSepPunct{\mcitedefaultmidpunct}
{\mcitedefaultendpunct}{\mcitedefaultseppunct}\relax
\EndOfBibitem
\bibitem{Poluektov:2004mf}
A.~Poluektov {\em et al.} ({Belle} collaboration){,}
  \href{http://dx.doi.org/10.1103/PhysRevD.70.072003}{Phys.\ Rev.\ {\bf D70}{,}
  072003}  (2004), \href{http://arxiv.org/abs/hep-ex/0406067}{{\tt
  arXiv:hep-ex/0406067}}\relax
\mciteBstWouldAddEndPuncttrue
\mciteSetBstMidEndSepPunct{\mcitedefaultmidpunct}
{\mcitedefaultendpunct}{\mcitedefaultseppunct}\relax
\EndOfBibitem
\bibitem{Brod:2013sga}
J.~Brod and J.~Zupan, \href{http://dx.doi.org/10.1007/JHEP01(2014)051}{JHEP
  {\bf 01},  051}  (2014), \href{http://arxiv.org/abs/1308.5663}{{\tt
  arXiv:1308.5663 [hep-ph]}}\relax
\mciteBstWouldAddEndPuncttrue
\mciteSetBstMidEndSepPunct{\mcitedefaultmidpunct}
{\mcitedefaultendpunct}{\mcitedefaultseppunct}\relax
\EndOfBibitem
\bibitem{Gronau:2002mu}
M.~Gronau, \href{http://dx.doi.org/10.1016/S0370-2693(03)00192-8}{Phys.\ Lett.\
  {\bf B557},  198}  (2003), \href{http://arxiv.org/abs/hep-ph/0211282}{{\tt
  arXiv:hep-ph/0211282}}\relax
\mciteBstWouldAddEndPuncttrue
\mciteSetBstMidEndSepPunct{\mcitedefaultmidpunct}
{\mcitedefaultendpunct}{\mcitedefaultseppunct}\relax
\EndOfBibitem
\bibitem{Gershon:2008pe}
T.~Gershon, \href{http://dx.doi.org/10.1103/PhysRevD.79.051301}{Phys.\ Rev.\ {\bf
  D79},  051301}  (2009), \href{http://arxiv.org/abs/0810.2706}{{\tt
  arXiv:0810.2706 [hep-ph]}}\relax
\mciteBstWouldAddEndPuncttrue
\mciteSetBstMidEndSepPunct{\mcitedefaultmidpunct}
{\mcitedefaultendpunct}{\mcitedefaultseppunct}\relax
\EndOfBibitem
\bibitem{Gershon:2009qc}
T.~Gershon and M.~Williams{,}
  \href{http://dx.doi.org/10.1103/PhysRevD.80.092002}{Phys.\ Rev.\ {\bf D80}{,}
  092002}  (2009), \href{http://arxiv.org/abs/0909.1495}{{\tt arXiv:0909.1495
  [hep-ph]}}\relax
\mciteBstWouldAddEndPuncttrue
\mciteSetBstMidEndSepPunct{\mcitedefaultmidpunct}
{\mcitedefaultendpunct}{\mcitedefaultseppunct}\relax
\EndOfBibitem
\bibitem{Bondar:2004bi}
A.~Bondar and T.~Gershon{,}
  \href{http://dx.doi.org/10.1103/PhysRevD.70.091503}{Phys.\ Rev.\ {\bf D70}{,}
  091503}  (2004), \href{http://arxiv.org/abs/hep-ph/0409281}{{\tt
  arXiv:hep-ph/0409281}}\relax
\mciteBstWouldAddEndPuncttrue
\mciteSetBstMidEndSepPunct{\mcitedefaultmidpunct}
{\mcitedefaultendpunct}{\mcitedefaultseppunct}\relax
\EndOfBibitem
\bibitem{Atwood:2003mj}
D.~Atwood and A.~Soni{,}
  \href{http://dx.doi.org/10.1103/PhysRevD.68.033003}{Phys.\ Rev.\ {\bf D68}{,}
  033003}  (2003), \href{http://arxiv.org/abs/hep-ph/0304085}{{\tt
  arXiv:hep-ph/0304085}}\relax
\mciteBstWouldAddEndPuncttrue
\mciteSetBstMidEndSepPunct{\mcitedefaultmidpunct}
{\mcitedefaultendpunct}{\mcitedefaultseppunct}\relax
\EndOfBibitem
\bibitem{Grossman:2002aq}
Y.~Grossman, Z.~Ligeti, and A.~Soffer{,}
  \href{http://dx.doi.org/10.1103/PhysRevD.67.071301}{Phys.\ Rev.\ {\bf D67}{,}
  071301}  (2003), \href{http://arxiv.org/abs/hep-ph/0210433}{{\tt
  arXiv:hep-ph/0210433 [hep-ph]}}\relax
\mciteBstWouldAddEndPuncttrue
\mciteSetBstMidEndSepPunct{\mcitedefaultmidpunct}
{\mcitedefaultendpunct}{\mcitedefaultseppunct}\relax
\EndOfBibitem
\bibitem{Nayak:2014tea}
M.~Nayak {\em et al.}{,}
  \href{http://dx.doi.org/10.1016/j.physletb.2014.11.022}{Phys.\ Lett.\ {\bf
  B740},  1}  (2014), \href{http://arxiv.org/abs/1410.3964}{{\tt
  arXiv:1410.3964 [hep-ex]}}\relax
\mciteBstWouldAddEndPuncttrue
\mciteSetBstMidEndSepPunct{\mcitedefaultmidpunct}
{\mcitedefaultendpunct}{\mcitedefaultseppunct}\relax
\EndOfBibitem
\bibitem{Bondar:2005ki}
A.~Bondar and A.~Poluektov{,}
  \href{http://dx.doi.org/10.1140/epjc/s2006-02590-x}{Eur.\ Phys.\ J.\ {\bf C47}{,}
  347}  (2006), \href{http://arxiv.org/abs/hep-ph/0510246}{{\tt
  arXiv:hep-ph/0510246}}\relax
\mciteBstWouldAddEndPuncttrue
\mciteSetBstMidEndSepPunct{\mcitedefaultmidpunct}
{\mcitedefaultendpunct}{\mcitedefaultseppunct}\relax
\EndOfBibitem
\bibitem{Bondar:2008hh}
A.~Bondar and A.~Poluektov{,}
  \href{http://dx.doi.org/10.1140/epjc/s10052-008-0600-z}{Eur.\ Phys.\ J.\ {\bf
  C55},  51}  (2008), \href{http://arxiv.org/abs/0801.0840}{{\tt
  arXiv:0801.0840 [hep-ex]}}\relax
\mciteBstWouldAddEndPuncttrue
\mciteSetBstMidEndSepPunct{\mcitedefaultmidpunct}
{\mcitedefaultendpunct}{\mcitedefaultseppunct}\relax
\EndOfBibitem
\bibitem{Gershon:2015xra}
T.~Gershon, J.~Libby, and G.~Wilkinson{,}
  \href{http://dx.doi.org/10.1016/j.physletb.2015.08.063}{Phys.\ Lett.\ {\bf
  B750},  338}  (2015), \href{http://arxiv.org/abs/1506.08594}{{\tt
  arXiv:1506.08594 [hep-ph]}}\relax
\mciteBstWouldAddEndPuncttrue
\mciteSetBstMidEndSepPunct{\mcitedefaultmidpunct}
{\mcitedefaultendpunct}{\mcitedefaultseppunct}\relax
\EndOfBibitem
\bibitem{Aubert:2007ii}
B.~Aubert {\em et al.} ({\babar} collaboration){,}
  \href{http://dx.doi.org/10.1103/PhysRevLett.99.251801}{Phys.\ Rev.\ Lett.\ {\bf
  99},  251801}  (2007), \href{http://arxiv.org/abs/hep-ex/0703037}{{\tt
  arXiv:hep-ex/0703037}}\relax
\mciteBstWouldAddEndPuncttrue
\mciteSetBstMidEndSepPunct{\mcitedefaultmidpunct}
{\mcitedefaultendpunct}{\mcitedefaultseppunct}\relax
\EndOfBibitem
\bibitem{Aubert:2007hz}
B.~Aubert {\em et al.} ({\babar} collaboration){,}
  \href{http://dx.doi.org/10.1103/PhysRevD.76.031102}{Phys.\ Rev.\ {\bf D76}{,}
  031102}  (2007), \href{http://arxiv.org/abs/0704.0522}{{\tt arXiv:0704.0522
  [hep-ex]}}\relax
\mciteBstWouldAddEndPuncttrue
\mciteSetBstMidEndSepPunct{\mcitedefaultmidpunct}
{\mcitedefaultendpunct}{\mcitedefaultseppunct}\relax
\EndOfBibitem
\bibitem{Itoh:2005ks}
R.~Itoh {\em et al.} ({Belle} collaboration){,}
  \href{http://dx.doi.org/10.1103/PhysRevLett.95.091601}{Phys.\ Rev.\ Lett.\ {\bf
  95},  091601}  (2005), \href{http://arxiv.org/abs/hep-ex/0504030}{{\tt
  arXiv:hep-ex/0504030 [hep-ex]}}\relax
\mciteBstWouldAddEndPuncttrue
\mciteSetBstMidEndSepPunct{\mcitedefaultmidpunct}
{\mcitedefaultendpunct}{\mcitedefaultseppunct}\relax
\EndOfBibitem
\bibitem{Acosta:2004gt}
D.~Acosta {\em et al.} ({CDF} collaboration){,}
  \href{http://dx.doi.org/10.1103/PhysRevLett.94.101803}{Phys.\ Rev.\ Lett.\ {\bf
  94},  101803}  (2005), \href{http://arxiv.org/abs/hep-ex/0412057}{{\tt
  arXiv:hep-ex/0412057}}\relax
\mciteBstWouldAddEndPuncttrue
\mciteSetBstMidEndSepPunct{\mcitedefaultmidpunct}
{\mcitedefaultendpunct}{\mcitedefaultseppunct}\relax
\EndOfBibitem
\bibitem{Aaij:2013cma}
R.~Aaij {\em et al.} ({LHCb} collaboration){,}
  \href{http://dx.doi.org/10.1103/PhysRevD.88.052002}{Phys.\ Rev.\ {\bf D88}{,}
  052002}  (2013), \href{http://arxiv.org/abs/1307.2782}{{\tt arXiv:1307.2782
  [hep-ex]}}\relax
\mciteBstWouldAddEndPuncttrue
\mciteSetBstMidEndSepPunct{\mcitedefaultmidpunct}
{\mcitedefaultendpunct}{\mcitedefaultseppunct}\relax
\EndOfBibitem
\bibitem{PDG_2014}
K.~Olive {\em et al.} ({Particle Data Group}){,}
  \href{http://dx.doi.org/10.1088/1674-1137/38/9/090001}{Chin.\ Phys.\ {\bf C38}{,}
   090001}  (2014)\relax
\mciteBstWouldAddEndPuncttrue
\mciteSetBstMidEndSepPunct{\mcitedefaultmidpunct}
{\mcitedefaultendpunct}{\mcitedefaultseppunct}\relax
\EndOfBibitem
\bibitem{Jung:2012mp}
M.~Jung, \href{http://dx.doi.org/10.1103/PhysRevD.86.053008}{Phys.\ Rev.\ {\bf
  D86},  053008}  (2012), \href{http://arxiv.org/abs/1206.2050}{{\tt
  arXiv:1206.2050 [hep-ph]}}\relax
\mciteBstWouldAddEndPuncttrue
\mciteSetBstMidEndSepPunct{\mcitedefaultmidpunct}
{\mcitedefaultendpunct}{\mcitedefaultseppunct}\relax
\EndOfBibitem
\bibitem{DeBruyn:2014oga}
K.~De~Bruyn and R.~Fleischer{,}
  \href{http://dx.doi.org/10.1007/JHEP03(2015)145}{JHEP {\bf 03},  145}
  (2015), \href{http://arxiv.org/abs/1412.6834}{{\tt arXiv:1412.6834
  [hep-ph]}}\relax
\mciteBstWouldAddEndPuncttrue
\mciteSetBstMidEndSepPunct{\mcitedefaultmidpunct}
{\mcitedefaultendpunct}{\mcitedefaultseppunct}\relax
\EndOfBibitem
\bibitem{Frings:2015eva}
P.~Frings, U.~Nierste, and M.~Wiebusch{,}
  \href{http://dx.doi.org/10.1103/PhysRevLett.115.061802}{Phys.\ Rev.\ Lett.\ {\bf
  115},  061802}  (2015), \href{http://arxiv.org/abs/1503.00859}{{\tt
  arXiv:1503.00859 [hep-ph]}}\relax
\mciteBstWouldAddEndPuncttrue
\mciteSetBstMidEndSepPunct{\mcitedefaultmidpunct}
{\mcitedefaultendpunct}{\mcitedefaultseppunct}\relax
\EndOfBibitem
\bibitem{:2009yr}
B.~Aubert {\em et al.} ({\babar} collaboration){,}
  \href{http://dx.doi.org/10.1103/PhysRevD.79.072009}{Phys.\ Rev.\ {\bf D79}{,}
  072009}  (2009), \href{http://arxiv.org/abs/0902.1708}{{\tt arXiv:0902.1708
  [hep-ex]}}\relax
\mciteBstWouldAddEndPuncttrue
\mciteSetBstMidEndSepPunct{\mcitedefaultmidpunct}
{\mcitedefaultendpunct}{\mcitedefaultseppunct}\relax
\EndOfBibitem
\bibitem{Aubert:2003xn}
B.~Aubert {\em et al.} ({\babar} collaboration){,}
  \href{http://dx.doi.org/10.1103/PhysRevD.69.052001}{Phys.\ Rev.\ {\bf D69}{,}
  052001}  (2004), \href{http://arxiv.org/abs/hep-ex/0309039}{{\tt
  arXiv:hep-ex/0309039}}\relax
\mciteBstWouldAddEndPuncttrue
\mciteSetBstMidEndSepPunct{\mcitedefaultmidpunct}
{\mcitedefaultendpunct}{\mcitedefaultseppunct}\relax
\EndOfBibitem
\bibitem{Adachi:2012et}
I.~Adachi {\em et al.} ({Belle} collaboration){,}
  \href{http://dx.doi.org/10.1103/PhysRevLett.108.171802}{Phys.\ Rev.\ Lett.\ {\bf
  108},  171802}  (2012), \href{http://arxiv.org/abs/1201.4643}{{\tt
  arXiv:1201.4643 [hep-ex]}}\relax
\mciteBstWouldAddEndPuncttrue
\mciteSetBstMidEndSepPunct{\mcitedefaultmidpunct}
{\mcitedefaultendpunct}{\mcitedefaultseppunct}\relax
\EndOfBibitem
\bibitem{Barate:2000tf}
R.~Barate {\em et al.} ({ALEPH} collaboration){,}
  \href{http://dx.doi.org/10.1016/S0370-2693(00)01091-1}{Phys.\ Lett.\ {\bf
  B492},  259}  (2000), \href{http://arxiv.org/abs/hep-ex/0009058}{{\tt
  arXiv:hep-ex/0009058}}\relax
\mciteBstWouldAddEndPuncttrue
\mciteSetBstMidEndSepPunct{\mcitedefaultmidpunct}
{\mcitedefaultendpunct}{\mcitedefaultseppunct}\relax
\EndOfBibitem
\bibitem{Ackerstaff:1998xz}
K.~Ackerstaff {\em et al.} ({OPAL} collaboration){,}
  \href{http://dx.doi.org/10.1007/s100520050284}{Eur.\ Phys.\ J.\ {\bf C5},  379}
   (1998), \href{http://arxiv.org/abs/hep-ex/9801022}{{\tt
  arXiv:hep-ex/9801022}}\relax
\mciteBstWouldAddEndPuncttrue
\mciteSetBstMidEndSepPunct{\mcitedefaultmidpunct}
{\mcitedefaultendpunct}{\mcitedefaultseppunct}\relax
\EndOfBibitem
\bibitem{Affolder:1999gg}
A.~A.\ Affolder {\em et al.} ({CDF} collaboration){,}
  \href{http://dx.doi.org/10.1103/PhysRevD.61.072005}{Phys.\ Rev.\ {\bf D61}{,}
  072005}  (2000), \href{http://arxiv.org/abs/hep-ex/9909003}{{\tt
  arXiv:hep-ex/9909003}}\relax
\mciteBstWouldAddEndPuncttrue
\mciteSetBstMidEndSepPunct{\mcitedefaultmidpunct}
{\mcitedefaultendpunct}{\mcitedefaultseppunct}\relax
\EndOfBibitem
\bibitem{Aaij:2015vza}
R.~Aaij {\em et al.} ({LHCb} collaboration){,}
  \href{http://dx.doi.org/10.1103/PhysRevLett.115.031601}{Phys.\ Rev.\ Lett.\ {\bf
  115},  031601}  (2015), \href{http://arxiv.org/abs/1503.07089}{{\tt
  arXiv:1503.07089 [hep-ex]}}\relax
\mciteBstWouldAddEndPuncttrue
\mciteSetBstMidEndSepPunct{\mcitedefaultmidpunct}
{\mcitedefaultendpunct}{\mcitedefaultseppunct}\relax
\EndOfBibitem
\bibitem{Sato:2012hu}
Y.~Sato {\em et al.} ({Belle} collaboration){,}
  \href{http://dx.doi.org/10.1103/PhysRevLett.108.171801}{Phys.\ Rev.\ Lett.\ {\bf
  108},  171801}  (2012), \href{http://arxiv.org/abs/1201.3502}{{\tt
  arXiv:1201.3502 [hep-ex]}}\relax
\mciteBstWouldAddEndPuncttrue
\mciteSetBstMidEndSepPunct{\mcitedefaultmidpunct}
{\mcitedefaultendpunct}{\mcitedefaultseppunct}\relax
\EndOfBibitem
\bibitem{Bona:2005vz}
M.~Bona {\em et al.} ({UTfit} collaboration){,}
  \href{http://dx.doi.org/10.1088/1126-6708/2005/07/028}{JHEP {\bf 07},  028}
  (2005), \href{http://arxiv.org/abs/hep-ph/0501199}{{\tt
  arXiv:hep-ph/0501199}}, see also online updates{,}
  \url{http://www.utfit.org/}\relax
\mciteBstWouldAddEndPuncttrue
\mciteSetBstMidEndSepPunct{\mcitedefaultmidpunct}
{\mcitedefaultendpunct}{\mcitedefaultseppunct}\relax
\EndOfBibitem
\bibitem{Lunghi:2008aa}
E.~Lunghi and A.~Soni{,}
  \href{http://dx.doi.org/10.1016/j.physletb.2008.07.015}{Phys.\ Lett.\ {\bf
  B666},  162}  (2008), \href{http://arxiv.org/abs/0803.4340}{{\tt
  arXiv:0803.4340 [hep-ph]}}\relax
\mciteBstWouldAddEndPuncttrue
\mciteSetBstMidEndSepPunct{\mcitedefaultmidpunct}
{\mcitedefaultendpunct}{\mcitedefaultseppunct}\relax
\EndOfBibitem
\bibitem{Eigen:2013cv}
G.~Eigen, G.~Dubois-Felsmann, D.~Hitlin, and F.~Porter{,}
  \href{http://dx.doi.org/10.1103/PhysRevD.89.033004}{Phys.\ Rev.\ {\bf D89}{,}
  033004}  (2014), \href{http://arxiv.org/abs/1301.5867}{{\tt arXiv:1301.5867
  [hep-ex]}}\relax
\mciteBstWouldAddEndPuncttrue
\mciteSetBstMidEndSepPunct{\mcitedefaultmidpunct}
{\mcitedefaultendpunct}{\mcitedefaultseppunct}\relax
\EndOfBibitem
\bibitem{Dunietz:1990cj}
I.~Dunietz, H.~R.\ Quinn, A.~Snyder, W.~Toki, and H.~J.\ Lipkin{,}
  \href{http://dx.doi.org/10.1103/PhysRevD.43.2193}{Phys.\ Rev.\ {\bf D43}{,}
  2193}  (1991)\relax
\mciteBstWouldAddEndPuncttrue
\mciteSetBstMidEndSepPunct{\mcitedefaultmidpunct}
{\mcitedefaultendpunct}{\mcitedefaultseppunct}\relax
\EndOfBibitem
\bibitem{Aston:1987ir}
D.~Aston {\em et al.}{,}
  \href{http://dx.doi.org/10.1016/0550-3213(88)90028-4}{Nucl.\ Phys.\ {\bf B296}{,}
   493}  (1988)\relax
\mciteBstWouldAddEndPuncttrue
\mciteSetBstMidEndSepPunct{\mcitedefaultmidpunct}
{\mcitedefaultendpunct}{\mcitedefaultseppunct}\relax
\EndOfBibitem
\bibitem{Suzuki:2001za}
M.~Suzuki, \href{http://dx.doi.org/10.1103/PhysRevD.64.117503}{Phys.\ Rev.\ {\bf
  D64},  117503}  (2001), \href{http://arxiv.org/abs/hep-ph/0106354}{{\tt
  arXiv:hep-ph/0106354}}\relax
\mciteBstWouldAddEndPuncttrue
\mciteSetBstMidEndSepPunct{\mcitedefaultmidpunct}
{\mcitedefaultendpunct}{\mcitedefaultseppunct}\relax
\EndOfBibitem
\bibitem{Aubert:2004cp}
B.~Aubert {\em et al.} ({\babar} collaboration){,}
  \href{http://dx.doi.org/10.1103/PhysRevD.71.032005}{Phys.\ Rev.\ {\bf D71}{,}
  032005}  (2005), \href{http://arxiv.org/abs/hep-ex/0411016}{{\tt
  arXiv:hep-ex/0411016}}\relax
\mciteBstWouldAddEndPuncttrue
\mciteSetBstMidEndSepPunct{\mcitedefaultmidpunct}
{\mcitedefaultendpunct}{\mcitedefaultseppunct}\relax
\EndOfBibitem
\bibitem{Grossman:1996ke}
Y.~Grossman and M.~P.\ Worah{,}
  \href{http://dx.doi.org/10.1016/S0370-2693(97)00068-3}{Phys.\ Lett.\ {\bf
  B395},  241}  (1997), \href{http://arxiv.org/abs/hep-ph/9612269}{{\tt
  arXiv:hep-ph/9612269}}\relax
\mciteBstWouldAddEndPuncttrue
\mciteSetBstMidEndSepPunct{\mcitedefaultmidpunct}
{\mcitedefaultendpunct}{\mcitedefaultseppunct}\relax
\EndOfBibitem
\bibitem{Fleischer:2003ai}
R.~Fleischer, \href{http://dx.doi.org/10.1016/S0370-2693(03)00582-3}{Phys.\
  Lett.\ {\bf B562},  234}  (2003){,}
  \href{http://arxiv.org/abs/hep-ph/0301255}{{\tt arXiv:hep-ph/0301255}}\relax
\mciteBstWouldAddEndPuncttrue
\mciteSetBstMidEndSepPunct{\mcitedefaultmidpunct}
{\mcitedefaultendpunct}{\mcitedefaultseppunct}\relax
\EndOfBibitem
\bibitem{Fleischer:2003aj}
R.~Fleischer, \href{http://dx.doi.org/10.1016/S0550-3213(03)00225-6}{Nucl.\
  Phys.\ {\bf B659},  321}  (2003){,}
  \href{http://arxiv.org/abs/hep-ph/0301256}{{\tt arXiv:hep-ph/0301256}}\relax
\mciteBstWouldAddEndPuncttrue
\mciteSetBstMidEndSepPunct{\mcitedefaultmidpunct}
{\mcitedefaultendpunct}{\mcitedefaultseppunct}\relax
\EndOfBibitem
\bibitem{Abdesselam:2015gha}
A.~Abdesselam {\em et al.} ({\babar\ and Belle} collaborations){,}
  \href{http://dx.doi.org/10.1103/PhysRevLett.115.121604}{Phys.\ Rev.\ Lett.\ {\bf
  115},  121604}  (2015), \href{http://arxiv.org/abs/1505.04147}{{\tt
  arXiv:1505.04147 [hep-ex]}}\relax
\mciteBstWouldAddEndPuncttrue
\mciteSetBstMidEndSepPunct{\mcitedefaultmidpunct}
{\mcitedefaultendpunct}{\mcitedefaultseppunct}\relax
\EndOfBibitem
\bibitem{Bondar:2005gk}
A.~Bondar, T.~Gershon, and P.~Krokovny{,}
  \href{http://dx.doi.org/10.1016/j.physletb.2005.07.053}{Phys.\ Lett.\ {\bf
  B624},  1}  (2005), \href{http://arxiv.org/abs/hep-ph/0503174}{{\tt
  arXiv:hep-ph/0503174}}\relax
\mciteBstWouldAddEndPuncttrue
\mciteSetBstMidEndSepPunct{\mcitedefaultmidpunct}
{\mcitedefaultendpunct}{\mcitedefaultseppunct}\relax
\EndOfBibitem
\bibitem{Botella:2005ks}
F.~Botella and J.~Silva{,}
  \href{http://dx.doi.org/10.1103/PhysRevD.71.094008}{Phys.\ Rev.\ {\bf D71}{,}
  094008}  (2005), \href{http://arxiv.org/abs/hep-ph/0503136}{{\tt
  arXiv:hep-ph/0503136 [hep-ph]}}\relax
\mciteBstWouldAddEndPuncttrue
\mciteSetBstMidEndSepPunct{\mcitedefaultmidpunct}
{\mcitedefaultendpunct}{\mcitedefaultseppunct}\relax
\EndOfBibitem
\bibitem{Kronenbitter:2012ha}
B.~Kronenbitter {\em et al.} ({Belle} collaboration){,}
  \href{http://dx.doi.org/10.1103/PhysRevD.86.071103}{Phys.\ Rev.\ {\bf D86}{,}
  071103}  (2012), \href{http://arxiv.org/abs/1207.5611}{{\tt arXiv:1207.5611
  [hep-ex]}}\relax
\mciteBstWouldAddEndPuncttrue
\mciteSetBstMidEndSepPunct{\mcitedefaultmidpunct}
{\mcitedefaultendpunct}{\mcitedefaultseppunct}\relax
\EndOfBibitem
\bibitem{Aubert:2008bs}
B.~Aubert {\em et al.} ({\babar} collaboration){,}
  \href{http://dx.doi.org/10.1103/PhysRevLett.101.021801}{Phys.\ Rev.\ Lett.\ {\bf
  101},  021801}  (2008), \href{http://arxiv.org/abs/0804.0896}{{\tt
  arXiv:0804.0896 [hep-ex]}}\relax
\mciteBstWouldAddEndPuncttrue
\mciteSetBstMidEndSepPunct{\mcitedefaultmidpunct}
{\mcitedefaultendpunct}{\mcitedefaultseppunct}\relax
\EndOfBibitem
\bibitem{:2007wd}
S.~E.\ Lee {\em et al.} ({Belle} collaboration){,}
  \href{http://dx.doi.org/10.1103/PhysRevD.77.071101}{Phys.\ Rev.\ {\bf D77}{,}
  071101}  (2008), \href{http://arxiv.org/abs/0708.0304}{{\tt arXiv:0708.0304
  [hep-ex]}}\relax
\mciteBstWouldAddEndPuncttrue
\mciteSetBstMidEndSepPunct{\mcitedefaultmidpunct}
{\mcitedefaultendpunct}{\mcitedefaultseppunct}\relax
\EndOfBibitem
\bibitem{Aaij:2016yip}
R.~Aaij {\em et al.} ({LHCb} collaboration){,}
  \href{http://dx.doi.org/10.1103/PhysRevLett.117.261801}{Phys.\ Rev.\ Lett.\ {\bf
  117},  261801}  (2016), \href{http://arxiv.org/abs/1608.06620}{{\tt
  arXiv:1608.06620 [hep-ex]}}\relax
\mciteBstWouldAddEndPuncttrue
\mciteSetBstMidEndSepPunct{\mcitedefaultmidpunct}
{\mcitedefaultendpunct}{\mcitedefaultseppunct}\relax
\EndOfBibitem
\bibitem{Lees:2012px}
J.~P.\ Lees {\em et al.} ({\babar} collaboration){,}
  \href{http://dx.doi.org/10.1103/PhysRevD.86.112006}{Phys.\ Rev.\ {\bf D86}{,}
  112006}  (2012), \href{http://arxiv.org/abs/1208.1282}{{\tt arXiv:1208.1282
  [hep-ex]}}\relax
\mciteBstWouldAddEndPuncttrue
\mciteSetBstMidEndSepPunct{\mcitedefaultmidpunct}
{\mcitedefaultendpunct}{\mcitedefaultseppunct}\relax
\EndOfBibitem
\bibitem{Fratina:2007zk}
S.~Fratina {\em et al.} ({Belle} collaboration){,}
  \href{http://dx.doi.org/10.1103/PhysRevLett.98.221802}{Phys.\ Rev.\ Lett.\ {\bf
  98},  221802}  (2007), \href{http://arxiv.org/abs/hep-ex/0702031}{{\tt
  arXiv:hep-ex/0702031}}\relax
\mciteBstWouldAddEndPuncttrue
\mciteSetBstMidEndSepPunct{\mcitedefaultmidpunct}
{\mcitedefaultendpunct}{\mcitedefaultseppunct}\relax
\EndOfBibitem
\bibitem{Fleischer:1999nz}
R.~Fleischer, \href{http://dx.doi.org/10.1007/s100529900099}{Eur.\ Phys.\ J.\ {\bf
  C10},  299}  (1999), \href{http://arxiv.org/abs/hep-ph/9903455}{{\tt
  arXiv:hep-ph/9903455 [hep-ph]}}\relax
\mciteBstWouldAddEndPuncttrue
\mciteSetBstMidEndSepPunct{\mcitedefaultmidpunct}
{\mcitedefaultendpunct}{\mcitedefaultseppunct}\relax
\EndOfBibitem
\bibitem{DeBruyn:2010hh}
K.~De~Bruyn, R.~Fleischer, and P.~Koppenburg{,}
  \href{http://dx.doi.org/10.1140/epjc/s10052-010-1495-z}{Eur.\ Phys.\ J.\ {\bf
  C70},  1025}  (2010), \href{http://arxiv.org/abs/1010.0089}{{\tt
  arXiv:1010.0089 [hep-ph]}}\relax
\mciteBstWouldAddEndPuncttrue
\mciteSetBstMidEndSepPunct{\mcitedefaultmidpunct}
{\mcitedefaultendpunct}{\mcitedefaultseppunct}\relax
\EndOfBibitem
\bibitem{Aaij:2015tza}
R.~Aaij {\em et al.} ({LHCb} collaboration){,}
  \href{http://dx.doi.org/10.1007/JHEP06(2015)131}{JHEP {\bf 06},  131}
  (2015), \href{http://arxiv.org/abs/1503.07055}{{\tt arXiv:1503.07055
  [hep-ex]}}\relax
\mciteBstWouldAddEndPuncttrue
\mciteSetBstMidEndSepPunct{\mcitedefaultmidpunct}
{\mcitedefaultendpunct}{\mcitedefaultseppunct}\relax
\EndOfBibitem
\bibitem{Fleischer:1996bv}
R.~Fleischer, \href{http://dx.doi.org/10.1142/S0217751X97001432}{Int.\ J.\ Mod.\
  Phys.\ {\bf A12},  2459}  (1997){,}
  \href{http://arxiv.org/abs/hep-ph/9612446}{{\tt arXiv:hep-ph/9612446}}\relax
\mciteBstWouldAddEndPuncttrue
\mciteSetBstMidEndSepPunct{\mcitedefaultmidpunct}
{\mcitedefaultendpunct}{\mcitedefaultseppunct}\relax
\EndOfBibitem
\bibitem{London:1997zk}
D.~London and A.~Soni{,}
  \href{http://dx.doi.org/10.1016/S0370-2693(97)00695-3}{Phys.\ Lett.\ {\bf
  B407},  61}  (1997), \href{http://arxiv.org/abs/hep-ph/9704277}{{\tt
  arXiv:hep-ph/9704277}}\relax
\mciteBstWouldAddEndPuncttrue
\mciteSetBstMidEndSepPunct{\mcitedefaultmidpunct}
{\mcitedefaultendpunct}{\mcitedefaultseppunct}\relax
\EndOfBibitem
\bibitem{Ciuchini:1997zp}
M.~Ciuchini, E.~Franco, G.~Martinelli, A.~Masiero, and L.~Silvestrini{,}
  \href{http://dx.doi.org/10.1103/PhysRevLett.79.978}{Phys.\ Rev.\ Lett.\ {\bf
  79},  978}  (1997), \href{http://arxiv.org/abs/hep-ph/9704274}{{\tt
  arXiv:hep-ph/9704274}}\relax
\mciteBstWouldAddEndPuncttrue
\mciteSetBstMidEndSepPunct{\mcitedefaultmidpunct}
{\mcitedefaultendpunct}{\mcitedefaultseppunct}\relax
\EndOfBibitem
\bibitem{Okubo:1963fa}
S.~Okubo, \href{http://dx.doi.org/10.1016/S0375-9601(63)92548-9}{Phys.\ Lett.\
  {\bf 5},  165--168}  (1963)\relax
\mciteBstWouldAddEndPuncttrue
\mciteSetBstMidEndSepPunct{\mcitedefaultmidpunct}
{\mcitedefaultendpunct}{\mcitedefaultseppunct}\relax
\EndOfBibitem
\bibitem{Zweig:1964jf}
 G.~Zweig, CERN-TH-412, 1964\relax
\mciteBstWouldAddEndPuncttrue
\mciteSetBstMidEndSepPunct{\mcitedefaultmidpunct}
{\mcitedefaultendpunct}{\mcitedefaultseppunct}\relax
\EndOfBibitem
\bibitem{Iizuka:1966fk}
J.~Iizuka, \href{http://dx.doi.org/10.1143/PTPS.37.21}{Prog.\ Theor.\ Phys.\
  Suppl.\ {\bf 37},  21--34}  (1966)\relax
\mciteBstWouldAddEndPuncttrue
\mciteSetBstMidEndSepPunct{\mcitedefaultmidpunct}
{\mcitedefaultendpunct}{\mcitedefaultseppunct}\relax
\EndOfBibitem
\bibitem{Gershon:2004tk}
T.~Gershon and M.~Hazumi{,}
  \href{http://dx.doi.org/10.1016/j.physletb.2004.06.095}{Phys.\ Lett.\ {\bf
  B596},  163}  (2004), \href{http://arxiv.org/abs/hep-ph/0402097}{{\tt
  arXiv:hep-ph/0402097}}\relax
\mciteBstWouldAddEndPuncttrue
\mciteSetBstMidEndSepPunct{\mcitedefaultmidpunct}
{\mcitedefaultendpunct}{\mcitedefaultseppunct}\relax
\EndOfBibitem
\bibitem{Grossman:2003qp}
Y.~Grossman, Z.~Ligeti, Y.~Nir, and H.~Quinn{,}
  \href{http://dx.doi.org/10.1103/PhysRevD.68.015004}{Phys.\ Rev.\ {\bf D68}{,}
  015004}  (2003), \href{http://arxiv.org/abs/hep-ph/0303171}{{\tt
  arXiv:hep-ph/0303171}}\relax
\mciteBstWouldAddEndPuncttrue
\mciteSetBstMidEndSepPunct{\mcitedefaultmidpunct}
{\mcitedefaultendpunct}{\mcitedefaultseppunct}\relax
\EndOfBibitem
\bibitem{Gronau:2003ep}
M.~Gronau and J.~L.\ Rosner{,}
  \href{http://dx.doi.org/10.1016/S0370-2693(03)00702-0}{Phys.\ Lett.\ {\bf
  B564},  90}  (2003), \href{http://arxiv.org/abs/hep-ph/0304178}{{\tt
  arXiv:hep-ph/0304178}}\relax
\mciteBstWouldAddEndPuncttrue
\mciteSetBstMidEndSepPunct{\mcitedefaultmidpunct}
{\mcitedefaultendpunct}{\mcitedefaultseppunct}\relax
\EndOfBibitem
\bibitem{Gronau:2003kx}
M.~Gronau, Y.~Grossman, and J.~L.\ Rosner{,}
  \href{http://dx.doi.org/10.1016/j.physletb.2003.11.015}{Phys.\ Lett.\ {\bf
  B579},  331}  (2004), \href{http://arxiv.org/abs/hep-ph/0310020}{{\tt
  arXiv:hep-ph/0310020}}\relax
\mciteBstWouldAddEndPuncttrue
\mciteSetBstMidEndSepPunct{\mcitedefaultmidpunct}
{\mcitedefaultendpunct}{\mcitedefaultseppunct}\relax
\EndOfBibitem
\bibitem{Gronau:2004hp}
M.~Gronau, J.~L.\ Rosner, and J.~Zupan{,}
  \href{http://dx.doi.org/10.1016/j.physletb.2004.06.086}{Phys.\ Lett.\ {\bf
  B596},  107}  (2004), \href{http://arxiv.org/abs/hep-ph/0403287}{{\tt
  arXiv:hep-ph/0403287}}\relax
\mciteBstWouldAddEndPuncttrue
\mciteSetBstMidEndSepPunct{\mcitedefaultmidpunct}
{\mcitedefaultendpunct}{\mcitedefaultseppunct}\relax
\EndOfBibitem
\bibitem{Cheng:2005bg}
H.-Y.\ Cheng, C.-K.\ Chua, and A.~Soni{,}
  \href{http://dx.doi.org/10.1103/PhysRevD.72.014006}{Phys.\ Rev.\ {\bf D72}{,}
  014006}  (2005), \href{http://arxiv.org/abs/hep-ph/0502235}{{\tt
  arXiv:hep-ph/0502235}}\relax
\mciteBstWouldAddEndPuncttrue
\mciteSetBstMidEndSepPunct{\mcitedefaultmidpunct}
{\mcitedefaultendpunct}{\mcitedefaultseppunct}\relax
\EndOfBibitem
\bibitem{Gronau:2005gz}
M.~Gronau and J.~L.\ Rosner{,}
  \href{http://dx.doi.org/10.1103/PhysRevD.71.074019}{Phys.\ Rev.\ {\bf D71}{,}
  074019}  (2005), \href{http://arxiv.org/abs/hep-ph/0503131}{{\tt
  arXiv:hep-ph/0503131}}\relax
\mciteBstWouldAddEndPuncttrue
\mciteSetBstMidEndSepPunct{\mcitedefaultmidpunct}
{\mcitedefaultendpunct}{\mcitedefaultseppunct}\relax
\EndOfBibitem
\bibitem{Buchalla:2005us}
G.~Buchalla, G.~Hiller, Y.~Nir, and G.~Raz{,}
  \href{http://dx.doi.org/10.1088/1126-6708/2005/09/074}{JHEP {\bf 09},  074}
  (2005), \href{http://arxiv.org/abs/hep-ph/0503151}{{\tt
  arXiv:hep-ph/0503151}}\relax
\mciteBstWouldAddEndPuncttrue
\mciteSetBstMidEndSepPunct{\mcitedefaultmidpunct}
{\mcitedefaultendpunct}{\mcitedefaultseppunct}\relax
\EndOfBibitem
\bibitem{Beneke:2005pu}
M.~Beneke, \href{http://dx.doi.org/10.1016/j.physletb.2005.06.045}{Phys.\ Lett.\
  {\bf B620},  143}  (2005), \href{http://arxiv.org/abs/hep-ph/0505075}{{\tt
  arXiv:hep-ph/0505075}}\relax
\mciteBstWouldAddEndPuncttrue
\mciteSetBstMidEndSepPunct{\mcitedefaultmidpunct}
{\mcitedefaultendpunct}{\mcitedefaultseppunct}\relax
\EndOfBibitem
\bibitem{Engelhard:2005hu}
G.~Engelhard, Y.~Nir, and G.~Raz{,}
  \href{http://dx.doi.org/10.1103/PhysRevD.72.075013}{Phys.\ Rev.\ {\bf D72}{,}
  075013}  (2005), \href{http://arxiv.org/abs/hep-ph/0505194}{{\tt
  arXiv:hep-ph/0505194}}\relax
\mciteBstWouldAddEndPuncttrue
\mciteSetBstMidEndSepPunct{\mcitedefaultmidpunct}
{\mcitedefaultendpunct}{\mcitedefaultseppunct}\relax
\EndOfBibitem
\bibitem{Cheng:2005ug}
H.-Y.\ Cheng, C.-K.\ Chua, and A.~Soni{,}
  \href{http://dx.doi.org/10.1103/PhysRevD.72.094003}{Phys.\ Rev.\ {\bf D72}{,}
  094003}  (2005), \href{http://arxiv.org/abs/hep-ph/0506268}{{\tt
  arXiv:hep-ph/0506268}}\relax
\mciteBstWouldAddEndPuncttrue
\mciteSetBstMidEndSepPunct{\mcitedefaultmidpunct}
{\mcitedefaultendpunct}{\mcitedefaultseppunct}\relax
\EndOfBibitem
\bibitem{Engelhard:2005ky}
G.~Engelhard and G.~Raz{,}
  \href{http://dx.doi.org/10.1103/PhysRevD.72.114017}{Phys.\ Rev.\ {\bf D72}{,}
  114017}  (2005), \href{http://arxiv.org/abs/hep-ph/0508046}{{\tt
  arXiv:hep-ph/0508046}}\relax
\mciteBstWouldAddEndPuncttrue
\mciteSetBstMidEndSepPunct{\mcitedefaultmidpunct}
{\mcitedefaultendpunct}{\mcitedefaultseppunct}\relax
\EndOfBibitem
\bibitem{Gronau:2006qh}
M.~Gronau, J.~L.\ Rosner, and J.~Zupan{,}
  \href{http://dx.doi.org/10.1103/PhysRevD.74.093003}{Phys.\ Rev.\ {\bf D74}{,}
  093003}  (2006), \href{http://arxiv.org/abs/hep-ph/0608085}{{\tt
  arXiv:hep-ph/0608085}}\relax
\mciteBstWouldAddEndPuncttrue
\mciteSetBstMidEndSepPunct{\mcitedefaultmidpunct}
{\mcitedefaultendpunct}{\mcitedefaultseppunct}\relax
\EndOfBibitem
\bibitem{Silvestrini:2007yf}
L.~Silvestrini{,}
  \href{http://dx.doi.org/10.1146/annurev.nucl.57.090506.123007}{Ann.\ Rev.\
  Nucl.\ Part.\ Sci.\ {\bf 57},  405}  (2007){,}
  \href{http://arxiv.org/abs/0705.1624}{{\tt arXiv:0705.1624 [hep-ph]}}\relax
\mciteBstWouldAddEndPuncttrue
\mciteSetBstMidEndSepPunct{\mcitedefaultmidpunct}
{\mcitedefaultendpunct}{\mcitedefaultseppunct}\relax
\EndOfBibitem
\bibitem{Dutta:2008xw}
R.~Dutta and S.~Gardner{,}
  \href{http://dx.doi.org/10.1103/PhysRevD.78.034021}{Phys.\ Rev.\ {\bf D78}{,}
  034021}  (2008), \href{http://arxiv.org/abs/0805.1963}{{\tt arXiv:0805.1963
  [hep-ph]}}\relax
\mciteBstWouldAddEndPuncttrue
\mciteSetBstMidEndSepPunct{\mcitedefaultmidpunct}
{\mcitedefaultendpunct}{\mcitedefaultseppunct}\relax
\EndOfBibitem
\bibitem{Fujikawa:2008pk}
M.~Fujikawa {\em et al.} ({Belle} collaboration){,}
  \href{http://dx.doi.org/10.1103/PhysRevD.81.011101}{Phys.\ Rev.\ {\bf D81}{,}
  011101}  (2010), \href{http://arxiv.org/abs/0809.4366}{{\tt arXiv:0809.4366
  [hep-ex]}}\relax
\mciteBstWouldAddEndPuncttrue
\mciteSetBstMidEndSepPunct{\mcitedefaultmidpunct}
{\mcitedefaultendpunct}{\mcitedefaultseppunct}\relax
\EndOfBibitem
\bibitem{Abe:2006gy}
K.~Abe {\em et al.} ({Belle} collaboration){,}
  \href{http://dx.doi.org/10.1103/PhysRevD.76.091103}{Phys.\ Rev.\ {\bf D76}{,}
  091103}  (2007), \href{http://arxiv.org/abs/hep-ex/0609006}{{\tt
  arXiv:hep-ex/0609006}}\relax
\mciteBstWouldAddEndPuncttrue
\mciteSetBstMidEndSepPunct{\mcitedefaultmidpunct}
{\mcitedefaultendpunct}{\mcitedefaultseppunct}\relax
\EndOfBibitem
\bibitem{Aubert:2005ja}
B.~Aubert {\em et al.} ({\babar} collaboration){,}
  \href{http://dx.doi.org/10.1103/PhysRevD.71.091102}{Phys.\ Rev.\ {\bf D71}{,}
  091102}  (2005), \href{http://arxiv.org/abs/hep-ex/0502019}{{\tt
  arXiv:hep-ex/0502019}}\relax
\mciteBstWouldAddEndPuncttrue
\mciteSetBstMidEndSepPunct{\mcitedefaultmidpunct}
{\mcitedefaultendpunct}{\mcitedefaultseppunct}\relax
\EndOfBibitem
\bibitem{:2008se}
B.~Aubert {\em et al.} ({\babar} collaboration){,}
  \href{http://dx.doi.org/10.1103/PhysRevD.79.052003}{Phys.\ Rev.\ {\bf D79}{,}
  052003}  (2009), \href{http://arxiv.org/abs/0809.1174}{{\tt arXiv:0809.1174
  [hep-ex]}}\relax
\mciteBstWouldAddEndPuncttrue
\mciteSetBstMidEndSepPunct{\mcitedefaultmidpunct}
{\mcitedefaultendpunct}{\mcitedefaultseppunct}\relax
\EndOfBibitem
\bibitem{Santelj:2014sja}
L.~\v{S}antelj {\em et al.} ({Belle} collaboration){,}
  \href{http://dx.doi.org/10.1007/JHEP10(2014)165}{JHEP {\bf 10},  165}
  (2014), \href{http://arxiv.org/abs/1408.5991}{{\tt arXiv:1408.5991
  [hep-ex]}}\relax
\mciteBstWouldAddEndPuncttrue
\mciteSetBstMidEndSepPunct{\mcitedefaultmidpunct}
{\mcitedefaultendpunct}{\mcitedefaultseppunct}\relax
\EndOfBibitem
\bibitem{Lees:2011nf}
J.~P.\ Lees {\em et al.} ({\babar} collaboration){,}
  \href{http://dx.doi.org/10.1103/PhysRevD.85.054023}{Phys.\ Rev.\ {\bf D85}{,}
  054023}  (2012), \href{http://arxiv.org/abs/1111.3636}{{\tt arXiv:1111.3636
  [hep-ex]}}\relax
\mciteBstWouldAddEndPuncttrue
\mciteSetBstMidEndSepPunct{\mcitedefaultmidpunct}
{\mcitedefaultendpunct}{\mcitedefaultseppunct}\relax
\EndOfBibitem
\bibitem{Chen:2006nk}
K.~F.\ Chen {\em et al.} ({Belle} collaboration){,}
  \href{http://dx.doi.org/10.1103/PhysRevLett.98.031802}{Phys.\ Rev.\ Lett.\ {\bf
  98},  031802}  (2007), \href{http://arxiv.org/abs/hep-ex/0608039}{{\tt
  arXiv:hep-ex/0608039}}\relax
\mciteBstWouldAddEndPuncttrue
\mciteSetBstMidEndSepPunct{\mcitedefaultmidpunct}
{\mcitedefaultendpunct}{\mcitedefaultseppunct}\relax
\EndOfBibitem
\bibitem{Chobanova:2013ddr}
V.~Chobanova {\em et al.} ({Belle} collaboration){,}
  \href{http://dx.doi.org/10.1103/PhysRevD.90.012002}{Phys.\ Rev.\ {\bf D90}{,}
  012002}  (2014), \href{http://arxiv.org/abs/1311.6666}{{\tt arXiv:1311.6666
  [hep-ex]}}\relax
\mciteBstWouldAddEndPuncttrue
\mciteSetBstMidEndSepPunct{\mcitedefaultmidpunct}
{\mcitedefaultendpunct}{\mcitedefaultseppunct}\relax
\EndOfBibitem
\bibitem{Aubert:2007ub}
B.~Aubert {\em et al.} ({\babar} collaboration){,}
  \href{http://dx.doi.org/10.1103/PhysRevD.76.071101}{Phys.\ Rev.\ {\bf D76}{,}
  071101}  (2007), \href{http://arxiv.org/abs/hep-ex/0702010}{{\tt
  arXiv:hep-ex/0702010}}\relax
\mciteBstWouldAddEndPuncttrue
\mciteSetBstMidEndSepPunct{\mcitedefaultmidpunct}
{\mcitedefaultendpunct}{\mcitedefaultseppunct}\relax
\EndOfBibitem
\bibitem{Aubert:2008zza}
B.~Aubert {\em et al.} ({\babar} collaboration){,}
  \href{http://dx.doi.org/10.1103/PhysRevD.78.092008}{Phys.\ Rev.\ {\bf D78}{,}
  092008}  (2008), \href{http://arxiv.org/abs/0808.3586}{{\tt arXiv:0808.3586
  [hep-ex]}}\relax
\mciteBstWouldAddEndPuncttrue
\mciteSetBstMidEndSepPunct{\mcitedefaultmidpunct}
{\mcitedefaultendpunct}{\mcitedefaultseppunct}\relax
\EndOfBibitem
\bibitem{Dunietz:1993rm}
 I.~Dunietz, FERMILAB-CONF-93-090-T, 1993\relax
\mciteBstWouldAddEndPuncttrue
\mciteSetBstMidEndSepPunct{\mcitedefaultmidpunct}
{\mcitedefaultendpunct}{\mcitedefaultseppunct}\relax
\EndOfBibitem
\bibitem{Fleischer:1999pa}
R.~Fleischer, \href{http://dx.doi.org/10.1016/S0370-2693(99)00640-1}{Phys.\
  Lett.\ {\bf B459},  306}  (1999){,}
  \href{http://arxiv.org/abs/hep-ph/9903456}{{\tt arXiv:hep-ph/9903456
  [hep-ph]}}\relax
\mciteBstWouldAddEndPuncttrue
\mciteSetBstMidEndSepPunct{\mcitedefaultmidpunct}
{\mcitedefaultendpunct}{\mcitedefaultseppunct}\relax
\EndOfBibitem
\bibitem{LHCb-CONF-2016-018}
{LHCb} collaboration{,}
  \href{https://cdsweb.cern.ch/record/2243039}{LHCb-CONF-2016-018}, 2016\relax
\mciteBstWouldAddEndPuncttrue
\mciteSetBstMidEndSepPunct{\mcitedefaultmidpunct}
{\mcitedefaultendpunct}{\mcitedefaultseppunct}\relax
\EndOfBibitem
\bibitem{Aaij:2013tna}
R.~Aaij {\em et al.} ({LHCb} collaboration){,}
  \href{http://dx.doi.org/10.1007/JHEP10(2013)183}{JHEP {\bf 10},  183}
  (2013), \href{http://arxiv.org/abs/1308.1428}{{\tt arXiv:1308.1428
  [hep-ex]}}\relax
\mciteBstWouldAddEndPuncttrue
\mciteSetBstMidEndSepPunct{\mcitedefaultmidpunct}
{\mcitedefaultendpunct}{\mcitedefaultseppunct}\relax
\EndOfBibitem
\bibitem{Raidal:2002ph}
M.~Raidal, \href{http://dx.doi.org/10.1103/PhysRevLett.89.231803}{Phys.\ Rev.\
  Lett.\ {\bf 89},  231803}  (2002){,}
  \href{http://arxiv.org/abs/hep-ph/0208091}{{\tt arXiv:hep-ph/0208091
  [hep-ph]}}\relax
\mciteBstWouldAddEndPuncttrue
\mciteSetBstMidEndSepPunct{\mcitedefaultmidpunct}
{\mcitedefaultendpunct}{\mcitedefaultseppunct}\relax
\EndOfBibitem
\bibitem{Aaij:2014kxa}
R.~Aaij {\em et al.} ({LHCb} collaboration){,}
  \href{http://dx.doi.org/10.1103/PhysRevD.90.052011}{Phys.\ Rev.\ {\bf D90}{,}
  052011}  (2014), \href{http://arxiv.org/abs/1407.2222}{{\tt arXiv:1407.2222
  [hep-ex]}}\relax
\mciteBstWouldAddEndPuncttrue
\mciteSetBstMidEndSepPunct{\mcitedefaultmidpunct}
{\mcitedefaultendpunct}{\mcitedefaultseppunct}\relax
\EndOfBibitem
\bibitem{Aubert:2006gm}
B.~Aubert {\em et al.} ({\babar} collaboration){,}
  \href{http://dx.doi.org/10.1103/PhysRevLett.97.171805}{Phys.\ Rev.\ Lett.\ {\bf
  97},  171805}  (2006), \href{http://arxiv.org/abs/hep-ex/0608036}{{\tt
  arXiv:hep-ex/0608036 [hep-ex]}}\relax
\mciteBstWouldAddEndPuncttrue
\mciteSetBstMidEndSepPunct{\mcitedefaultmidpunct}
{\mcitedefaultendpunct}{\mcitedefaultseppunct}\relax
\EndOfBibitem
\bibitem{Nakahama:2007dg}
Y.~Nakahama {\em et al.} ({Belle} collaboration){,}
  \href{http://dx.doi.org/10.1103/PhysRevLett.100.121601}{Phys.\ Rev.\ Lett.\ {\bf
  100},  121601}  (2008), \href{http://arxiv.org/abs/0712.4234}{{\tt
  arXiv:0712.4234 [hep-ex]}}\relax
\mciteBstWouldAddEndPuncttrue
\mciteSetBstMidEndSepPunct{\mcitedefaultmidpunct}
{\mcitedefaultendpunct}{\mcitedefaultseppunct}\relax
\EndOfBibitem
\bibitem{Akar:2013ima}
 S.~Akar, PhD thesis, LPNHE, Universit\'e Pierre et Marie Curie - Paris VI{,}
  2013, \url{https://tel.archives-ouvertes.fr/tel-00998252}\relax
\mciteBstWouldAddEndPuncttrue
\mciteSetBstMidEndSepPunct{\mcitedefaultmidpunct}
{\mcitedefaultendpunct}{\mcitedefaultseppunct}\relax
\EndOfBibitem
\bibitem{Sanchez:2015pxu}
P.~del Amo~Sanchez {\em et al.} ({\babar} collaboration){,}
  \href{http://dx.doi.org/10.1103/PhysRevD.93.052013}{Phys.\ Rev.\ {\bf D93}{,}
  052013}  (2016), \href{http://arxiv.org/abs/1512.03579}{{\tt
  arXiv:1512.03579 [hep-ex]}}\relax
\mciteBstWouldAddEndPuncttrue
\mciteSetBstMidEndSepPunct{\mcitedefaultmidpunct}
{\mcitedefaultendpunct}{\mcitedefaultseppunct}\relax
\EndOfBibitem
\bibitem{Li:2008qma}
J.~Li {\em et al.} ({Belle} collaboration){,}
  \href{http://dx.doi.org/10.1103/PhysRevLett.101.251601}{Phys.\ Rev.\ Lett.\ {\bf
  101},  251601}  (2008), \href{http://arxiv.org/abs/0806.1980}{{\tt
  arXiv:0806.1980 [hep-ex]}}\relax
\mciteBstWouldAddEndPuncttrue
\mciteSetBstMidEndSepPunct{\mcitedefaultmidpunct}
{\mcitedefaultendpunct}{\mcitedefaultseppunct}\relax
\EndOfBibitem
\bibitem{Aubert:2008gy}
B.~Aubert {\em et al.} ({\babar} collaboration){,}
  \href{http://dx.doi.org/10.1103/PhysRevD.78.071102}{Phys.\ Rev.\ {\bf D78}{,}
  071102}  (2008), \href{http://arxiv.org/abs/0807.3103}{{\tt arXiv:0807.3103
  [hep-ex]}}\relax
\mciteBstWouldAddEndPuncttrue
\mciteSetBstMidEndSepPunct{\mcitedefaultmidpunct}
{\mcitedefaultendpunct}{\mcitedefaultseppunct}\relax
\EndOfBibitem
\bibitem{Ushiroda:2006fi}
Y.~Ushiroda {\em et al.} ({Belle} collaboration){,}
  \href{http://dx.doi.org/10.1103/PhysRevD.74.111104}{Phys.\ Rev.\ {\bf D74}{,}
  111104}  (2006), \href{http://arxiv.org/abs/hep-ex/0608017}{{\tt
  arXiv:hep-ex/0608017}}\relax
\mciteBstWouldAddEndPuncttrue
\mciteSetBstMidEndSepPunct{\mcitedefaultmidpunct}
{\mcitedefaultendpunct}{\mcitedefaultseppunct}\relax
\EndOfBibitem
\bibitem{Aubert:2008js}
B.~Aubert {\em et al.} ({\babar} collaboration){,}
  \href{http://dx.doi.org/10.1103/PhysRevD.79.011102}{Phys.\ Rev.\ {\bf D79}{,}
  011102}  (2009), \href{http://arxiv.org/abs/0805.1317}{{\tt arXiv:0805.1317
  [hep-ex]}}\relax
\mciteBstWouldAddEndPuncttrue
\mciteSetBstMidEndSepPunct{\mcitedefaultmidpunct}
{\mcitedefaultendpunct}{\mcitedefaultseppunct}\relax
\EndOfBibitem
\bibitem{Sahoo:2011zd}
H.~Sahoo {\em et al.} ({Belle} collaboration){,}
  \href{http://dx.doi.org/10.1103/PhysRevD.84.071101}{Phys.\ Rev.\ {\bf D84}{,}
  071101}  (2011), \href{http://arxiv.org/abs/1104.5590}{{\tt arXiv:1104.5590
  [hep-ex]}}\relax
\mciteBstWouldAddEndPuncttrue
\mciteSetBstMidEndSepPunct{\mcitedefaultmidpunct}
{\mcitedefaultendpunct}{\mcitedefaultseppunct}\relax
\EndOfBibitem
\bibitem{Aaij:2016ofv}
R.~Aaij {\em et al.} ({LHCb} collaboration){,}
  \href{http://dx.doi.org/10.1103/PhysRevLett.118.021801}{Phys.\ Rev.\ Lett.\ {\bf
  118},  021801}  (2017), \href{http://arxiv.org/abs/1609.02032}{{\tt
  arXiv:1609.02032 [hep-ex]}}\relax
\mciteBstWouldAddEndPuncttrue
\mciteSetBstMidEndSepPunct{\mcitedefaultmidpunct}
{\mcitedefaultendpunct}{\mcitedefaultseppunct}\relax
\EndOfBibitem
\bibitem{Ushiroda:2007jf}
Y.~Ushiroda {\em et al.} ({Belle} collaboration){,}
  \href{http://dx.doi.org/10.1103/PhysRevLett.100.021602}{Phys.\ Rev.\ Lett.\ {\bf
  100},  021602}  (2008), \href{http://arxiv.org/abs/0709.2769}{{\tt
  arXiv:0709.2769 [hep-ex]}}\relax
\mciteBstWouldAddEndPuncttrue
\mciteSetBstMidEndSepPunct{\mcitedefaultmidpunct}
{\mcitedefaultendpunct}{\mcitedefaultseppunct}\relax
\EndOfBibitem
\bibitem{Aubert:2007nua}
B.~Aubert {\em et al.} ({\babar} collaboration){,}
  \href{http://dx.doi.org/10.1103/PhysRevD.76.052007}{Phys.\ Rev.\ {\bf D76}{,}
  052007}  (2007), \href{http://arxiv.org/abs/0705.2157}{{\tt arXiv:0705.2157
  [hep-ex]}}\relax
\mciteBstWouldAddEndPuncttrue
\mciteSetBstMidEndSepPunct{\mcitedefaultmidpunct}
{\mcitedefaultendpunct}{\mcitedefaultseppunct}\relax
\EndOfBibitem
\bibitem{Vanhoefer:2015ijw}
P.~Vanhoefer {\em et al.} ({Belle} collaboration){,}
  \href{http://dx.doi.org/10.1103/PhysRevD.93.032010}{Phys.\ Rev.\ {\bf D93}{,}
  032010}  (2016), \href{http://arxiv.org/abs/1510.01245}{{\tt
  arXiv:1510.01245 [hep-ex]}}\relax
\mciteBstWouldAddEndPuncttrue
\mciteSetBstMidEndSepPunct{\mcitedefaultmidpunct}
{\mcitedefaultendpunct}{\mcitedefaultseppunct}\relax
\EndOfBibitem
\bibitem{Aubert:2008au}
B.~Aubert {\em et al.} ({\babar} collaboration){,}
  \href{http://dx.doi.org/10.1103/PhysRevD.78.071104}{Phys.\ Rev.\ {\bf D78}{,}
  071104}  (2008), \href{http://arxiv.org/abs/0807.4977}{{\tt arXiv:0807.4977
  [hep-ex]}}\relax
\mciteBstWouldAddEndPuncttrue
\mciteSetBstMidEndSepPunct{\mcitedefaultmidpunct}
{\mcitedefaultendpunct}{\mcitedefaultseppunct}\relax
\EndOfBibitem
\bibitem{Adachi:2012cz}
I.~Adachi {\em et al.} ({Belle} collaboration){,}
  \href{http://dx.doi.org/10.1103/PhysRevD.89.072008}{Phys.\ Rev.\ {\bf D89}{,}
  072008}  (2014), \href{http://arxiv.org/abs/1212.4015}{{\tt arXiv:1212.4015
  [hep-ex]}}, Addendum ibid.\
  \href{http://dx.doi.org/10.1103/PhysRevD.89.119903}{{\bf D89}, 119903}
  (2014)\relax
\mciteBstWouldAddEndPuncttrue
\mciteSetBstMidEndSepPunct{\mcitedefaultmidpunct}
{\mcitedefaultendpunct}{\mcitedefaultseppunct}\relax
\EndOfBibitem
\bibitem{Aaij:2015ria}
R.~Aaij {\em et al.} ({LHCb} collaboration){,}
  \href{http://dx.doi.org/10.1016/j.physletb.2015.06.027}{Phys.\ Lett.\ {\bf
  B747},  468}  (2015), \href{http://arxiv.org/abs/1503.07770}{{\tt
  arXiv:1503.07770 [hep-ex]}}\relax
\mciteBstWouldAddEndPuncttrue
\mciteSetBstMidEndSepPunct{\mcitedefaultmidpunct}
{\mcitedefaultendpunct}{\mcitedefaultseppunct}\relax
\EndOfBibitem
\bibitem{Aubert:2006gb}
B.~Aubert {\em et al.} ({\babar} collaboration){,}
  \href{http://dx.doi.org/10.1103/PhysRevLett.98.181803}{Phys.\ Rev.\ Lett.\ {\bf
  98},  181803}  (2007), \href{http://arxiv.org/abs/hep-ex/0612050}{{\tt
  arXiv:hep-ex/0612050}}\relax
\mciteBstWouldAddEndPuncttrue
\mciteSetBstMidEndSepPunct{\mcitedefaultmidpunct}
{\mcitedefaultendpunct}{\mcitedefaultseppunct}\relax
\EndOfBibitem
\bibitem{Aubert:2009ab}
B.~Aubert {\em et al.} ({\babar} collaboration){,}
  \href{http://dx.doi.org/10.1103/PhysRevD.81.052009}{Phys.\ Rev.\ {\bf D81}{,}
  052009}  (2010), \href{http://arxiv.org/abs/0909.2171}{{\tt arXiv:0909.2171
  [hep-ex]}}\relax
\mciteBstWouldAddEndPuncttrue
\mciteSetBstMidEndSepPunct{\mcitedefaultmidpunct}
{\mcitedefaultendpunct}{\mcitedefaultseppunct}\relax
\EndOfBibitem
\bibitem{Lees:2012mma}
J.~P.\ Lees {\em et al.} ({\babar} collaboration){,}
  \href{http://dx.doi.org/10.1103/PhysRevD.87.052009}{Phys.\ Rev.\ {\bf D87}{,}
  052009}  (2013), \href{http://arxiv.org/abs/1206.3525}{{\tt arXiv:1206.3525
  [hep-ex]}}\relax
\mciteBstWouldAddEndPuncttrue
\mciteSetBstMidEndSepPunct{\mcitedefaultmidpunct}
{\mcitedefaultendpunct}{\mcitedefaultseppunct}\relax
\EndOfBibitem
\bibitem{Adachi:2013mae}
I.~Adachi {\em et al.} ({Belle} collaboration){,}
  \href{http://dx.doi.org/10.1103/PhysRevD.88.092003}{Phys.\ Rev.\ {\bf D88}{,}
  092003}  (2013), \href{http://arxiv.org/abs/1302.0551}{{\tt arXiv:1302.0551
  [hep-ex]}}\relax
\mciteBstWouldAddEndPuncttrue
\mciteSetBstMidEndSepPunct{\mcitedefaultmidpunct}
{\mcitedefaultendpunct}{\mcitedefaultseppunct}\relax
\EndOfBibitem
\bibitem{Dalseno:2012hp}
J.~Dalseno {\em et al.} ({Belle} collaboration){,}
  \href{http://dx.doi.org/10.1103/PhysRevD.86.092012}{Phys.\ Rev.\ {\bf D86}{,}
  092012}  (2012), \href{http://arxiv.org/abs/1205.5957}{{\tt arXiv:1205.5957
  [hep-ex]}}\relax
\mciteBstWouldAddEndPuncttrue
\mciteSetBstMidEndSepPunct{\mcitedefaultmidpunct}
{\mcitedefaultendpunct}{\mcitedefaultseppunct}\relax
\EndOfBibitem
\bibitem{Gronau:1990ka}
M.~Gronau and D.~London{,}
  \href{http://dx.doi.org/10.1103/PhysRevLett.65.3381}{Phys.\ Rev.\ Lett.\ {\bf
  65},  3381}  (1990)\relax
\mciteBstWouldAddEndPuncttrue
\mciteSetBstMidEndSepPunct{\mcitedefaultmidpunct}
{\mcitedefaultendpunct}{\mcitedefaultseppunct}\relax
\EndOfBibitem
\bibitem{Asner:2010qj}
D.~Asner {\em et al.} ({Heavy Flavor Averaging Group}){,}
  \href{http://arxiv.org/abs/1010.1589}{{\tt arXiv:1010.1589 [hep-ex]}}
  (2010)\relax
\mciteBstWouldAddEndPuncttrue
\mciteSetBstMidEndSepPunct{\mcitedefaultmidpunct}
{\mcitedefaultendpunct}{\mcitedefaultseppunct}\relax
\EndOfBibitem
\bibitem{Aubert:2009it}
B.~Aubert {\em et al.} ({\babar} collaboration){,}
  \href{http://dx.doi.org/10.1103/PhysRevLett.102.141802}{Phys.\ Rev.\ Lett.\ {\bf
  102},  141802}  (2009), \href{http://arxiv.org/abs/0901.3522}{{\tt
  arXiv:0901.3522 [hep-ex]}}\relax
\mciteBstWouldAddEndPuncttrue
\mciteSetBstMidEndSepPunct{\mcitedefaultmidpunct}
{\mcitedefaultendpunct}{\mcitedefaultseppunct}\relax
\EndOfBibitem
\bibitem{Lipkin:1991st}
H.~J.\ Lipkin, Y.~Nir, H.~R.\ Quinn, and A.~Snyder{,}
  \href{http://dx.doi.org/10.1103/PhysRevD.44.1454}{Phys.\ Rev.\ {\bf D44}{,}
  1454}  (1991)\relax
\mciteBstWouldAddEndPuncttrue
\mciteSetBstMidEndSepPunct{\mcitedefaultmidpunct}
{\mcitedefaultendpunct}{\mcitedefaultseppunct}\relax
\EndOfBibitem
\bibitem{Gronau:2005kw}
M.~Gronau and J.~Zupan{,}
  \href{http://dx.doi.org/10.1103/PhysRevD.73.057502}{Phys.\ Rev.\ {\bf D73}{,}
  057502}  (2006), \href{http://arxiv.org/abs/hep-ph/0512148}{{\tt
  arXiv:hep-ph/0512148}}\relax
\mciteBstWouldAddEndPuncttrue
\mciteSetBstMidEndSepPunct{\mcitedefaultmidpunct}
{\mcitedefaultendpunct}{\mcitedefaultseppunct}\relax
\EndOfBibitem
\bibitem{Bevan:2014iga}
A.~Bevan {\em et al.} ({\babar\ and Belle} collaborations){,}
  \href{http://dx.doi.org/10.1140/epjc/s10052-014-3026-9}{Eur.\ Phys.\ J.\ {\bf
  C74},  3026}  (2014), \href{http://arxiv.org/abs/1406.6311}{{\tt
  arXiv:1406.6311 [hep-ex]}}\relax
\mciteBstWouldAddEndPuncttrue
\mciteSetBstMidEndSepPunct{\mcitedefaultmidpunct}
{\mcitedefaultendpunct}{\mcitedefaultseppunct}\relax
\EndOfBibitem
\bibitem{Dunietz:1997in}
I.~Dunietz, \href{http://dx.doi.org/10.1016/S0370-2693(98)00304-9}{Phys.\ Lett.\
  {\bf B427},  179}  (1998), \href{http://arxiv.org/abs/hep-ph/9712401}{{\tt
  arXiv:hep-ph/9712401 [hep-ph]}}\relax
\mciteBstWouldAddEndPuncttrue
\mciteSetBstMidEndSepPunct{\mcitedefaultmidpunct}
{\mcitedefaultendpunct}{\mcitedefaultseppunct}\relax
\EndOfBibitem
\bibitem{Baak:2007gp}
 M.A.~Baak, PhD thesis, Vrije U., Amsterdam, 2007, {\small
  \url{http://www-public.slac.stanford.edu/sciDoc/docMeta.aspx?slacPubNumber=slac-r-858}}\relax
\mciteBstWouldAddEndPuncttrue
\mciteSetBstMidEndSepPunct{\mcitedefaultmidpunct}
{\mcitedefaultendpunct}{\mcitedefaultseppunct}\relax
\EndOfBibitem
\bibitem{DeBruyn:2012jp}
K.~De~Bruyn, R.~Fleischer, R.~Knegjens, M.~Merk, M.~Schiller, and N.~Tuning{,}
  \href{http://dx.doi.org/10.1016/j.nuclphysb.2012.11.012}{Nucl.\ Phys.\ {\bf
  B868},  351}  (2013), \href{http://arxiv.org/abs/1208.6463}{{\tt
  arXiv:1208.6463 [hep-ph]}}\relax
\mciteBstWouldAddEndPuncttrue
\mciteSetBstMidEndSepPunct{\mcitedefaultmidpunct}
{\mcitedefaultendpunct}{\mcitedefaultseppunct}\relax
\EndOfBibitem
\bibitem{Kenzie:2016yee}
M.~Kenzie, M.~Martinelli, and N.~Tuning{,}
  \href{http://dx.doi.org/10.1103/PhysRevD.94.054021}{Phys.\ Rev.\ {\bf D94}{,}
  054021}  (2016), \href{http://arxiv.org/abs/1606.09129}{{\tt
  arXiv:1606.09129 [hep-ph]}}\relax
\mciteBstWouldAddEndPuncttrue
\mciteSetBstMidEndSepPunct{\mcitedefaultmidpunct}
{\mcitedefaultendpunct}{\mcitedefaultseppunct}\relax
\EndOfBibitem
\bibitem{Aubert:2007qe}
B.~Aubert {\em et al.} ({\babar} collaboration){,}
  \href{http://dx.doi.org/10.1103/PhysRevD.77.071102}{Phys.\ Rev.\ {\bf D77}{,}
  071102}  (2008), \href{http://arxiv.org/abs/0712.3469}{{\tt arXiv:0712.3469
  [hep-ex]}}\relax
\mciteBstWouldAddEndPuncttrue
\mciteSetBstMidEndSepPunct{\mcitedefaultmidpunct}
{\mcitedefaultendpunct}{\mcitedefaultseppunct}\relax
\EndOfBibitem
\bibitem{Dunietz:1987bv}
I.~Dunietz and R.~G.\ Sachs{,}
  \href{http://dx.doi.org/10.1103/PhysRevD.37.3186}{Phys.\ Rev.\ {\bf D37}{,}
  3186}  (1988), Erratum ibid.\
  \href{http://dx.doi.org/10.1103/PhysRevD.39.3515}{{\bf D39}, 3515}
  (1989)\relax
\mciteBstWouldAddEndPuncttrue
\mciteSetBstMidEndSepPunct{\mcitedefaultmidpunct}
{\mcitedefaultendpunct}{\mcitedefaultseppunct}\relax
\EndOfBibitem
\bibitem{Aleksan:1991nh}
R.~Aleksan, I.~Dunietz, and B.~Kayser{,}
  \href{http://dx.doi.org/10.1007/BF01559494}{Z.\ Phys.\ {\bf C54},  653}
  (1992)\relax
\mciteBstWouldAddEndPuncttrue
\mciteSetBstMidEndSepPunct{\mcitedefaultmidpunct}
{\mcitedefaultendpunct}{\mcitedefaultseppunct}\relax
\EndOfBibitem
\bibitem{LHCb-CONF-2016-015}
{LHCb} collaboration{,}
  \href{https://cdsweb.cern.ch/record/2242070}{LHCb-CONF-2016-015}, 2016\relax
\mciteBstWouldAddEndPuncttrue
\mciteSetBstMidEndSepPunct{\mcitedefaultmidpunct}
{\mcitedefaultendpunct}{\mcitedefaultseppunct}\relax
\EndOfBibitem
\bibitem{delAmoSanchez:2010ji}
P.~del Amo~Sanchez {\em et al.} ({\babar} collaboration){,}
  \href{http://dx.doi.org/10.1103/PhysRevD.82.072004}{Phys.\ Rev.\ {\bf D82}{,}
  072004}  (2010), \href{http://arxiv.org/abs/1007.0504}{{\tt arXiv:1007.0504
  [hep-ex]}}\relax
\mciteBstWouldAddEndPuncttrue
\mciteSetBstMidEndSepPunct{\mcitedefaultmidpunct}
{\mcitedefaultendpunct}{\mcitedefaultseppunct}\relax
\EndOfBibitem
\bibitem{Abe:2006hc}
K.~Abe {\em et al.} ({Belle} collaboration){,}
  \href{http://dx.doi.org/10.1103/PhysRevD.73.051106}{Phys.\ Rev.\ {\bf D73}{,}
  051106}  (2006), \href{http://arxiv.org/abs/hep-ex/0601032}{{\tt
  arXiv:hep-ex/0601032}}\relax
\mciteBstWouldAddEndPuncttrue
\mciteSetBstMidEndSepPunct{\mcitedefaultmidpunct}
{\mcitedefaultendpunct}{\mcitedefaultseppunct}\relax
\EndOfBibitem
\bibitem{Aaltonen:2009hz}
T.~Aaltonen {\em et al.} ({CDF} collaboration){,}
  \href{http://dx.doi.org/10.1103/PhysRevD.81.031105}{Phys.\ Rev.\ {\bf D81}{,}
  031105}  (2010), \href{http://arxiv.org/abs/0911.0425}{{\tt arXiv:0911.0425
  [hep-ex]}}\relax
\mciteBstWouldAddEndPuncttrue
\mciteSetBstMidEndSepPunct{\mcitedefaultmidpunct}
{\mcitedefaultendpunct}{\mcitedefaultseppunct}\relax
\EndOfBibitem
\bibitem{Aaij:2016oso}
R.~Aaij {\em et al.} ({LHCb} collaboration){,}
  \href{http://dx.doi.org/10.1016/j.physletb.2016.06.022}{Phys.\ Lett.\ {\bf
  B760},  117}  (2016), \href{http://arxiv.org/abs/1603.08993}{{\tt
  arXiv:1603.08993 [hep-ex]}}\relax
\mciteBstWouldAddEndPuncttrue
\mciteSetBstMidEndSepPunct{\mcitedefaultmidpunct}
{\mcitedefaultendpunct}{\mcitedefaultseppunct}\relax
\EndOfBibitem
\bibitem{:2008jd}
B.~Aubert {\em et al.} ({\babar} collaboration){,}
  \href{http://dx.doi.org/10.1103/PhysRevD.78.092002}{Phys.\ Rev.\ {\bf D78}{,}
  092002}  (2008), \href{http://arxiv.org/abs/0807.2408}{{\tt arXiv:0807.2408
  [hep-ex]}}\relax
\mciteBstWouldAddEndPuncttrue
\mciteSetBstMidEndSepPunct{\mcitedefaultmidpunct}
{\mcitedefaultendpunct}{\mcitedefaultseppunct}\relax
\EndOfBibitem
\bibitem{Aubert:2009yw}
B.~Aubert {\em et al.} ({\babar} collaboration){,}
  \href{http://dx.doi.org/10.1103/PhysRevD.80.092001}{Phys.\ Rev.\ {\bf D80}{,}
  092001}  (2009), \href{http://arxiv.org/abs/0909.3981}{{\tt arXiv:0909.3981
  [hep-ex]}}\relax
\mciteBstWouldAddEndPuncttrue
\mciteSetBstMidEndSepPunct{\mcitedefaultmidpunct}
{\mcitedefaultendpunct}{\mcitedefaultseppunct}\relax
\EndOfBibitem
\bibitem{LHCb-CONF-2016-014}
{LHCb} collaboration{,}
  \href{https://cdsweb.cern.ch/record/2240147}{LHCb-CONF-2016-014}, 2016\relax
\mciteBstWouldAddEndPuncttrue
\mciteSetBstMidEndSepPunct{\mcitedefaultmidpunct}
{\mcitedefaultendpunct}{\mcitedefaultseppunct}\relax
\EndOfBibitem
\bibitem{Aaij:2015ina}
R.~Aaij {\em et al.} ({LHCb} collaboration){,}
  \href{http://dx.doi.org/10.1103/PhysRevD.92.112005}{Phys.\ Rev.\ {\bf D92}{,}
  112005}  (2015), \href{http://arxiv.org/abs/1505.07044}{{\tt
  arXiv:1505.07044 [hep-ex]}}\relax
\mciteBstWouldAddEndPuncttrue
\mciteSetBstMidEndSepPunct{\mcitedefaultmidpunct}
{\mcitedefaultendpunct}{\mcitedefaultseppunct}\relax
\EndOfBibitem
\bibitem{Aaij:2014eha}
R.~Aaij {\em et al.} ({LHCb} collaboration){,}
  \href{http://dx.doi.org/10.1103/PhysRevD.90.112002}{Phys.\ Rev.\ {\bf D90}{,}
  112002}  (2014), \href{http://arxiv.org/abs/1407.8136}{{\tt arXiv:1407.8136
  [hep-ex]}}\relax
\mciteBstWouldAddEndPuncttrue
\mciteSetBstMidEndSepPunct{\mcitedefaultmidpunct}
{\mcitedefaultendpunct}{\mcitedefaultseppunct}\relax
\EndOfBibitem
\bibitem{Aaij:2016bqv}
R.~Aaij {\em et al.} ({LHCb} collaboration){,}
  \href{http://dx.doi.org/10.1103/PhysRevD.93.112018}{Phys.\ Rev.\ {\bf D93}{,}
  112018}  (2016), \href{http://arxiv.org/abs/1602.03455}{{\tt
  arXiv:1602.03455 [hep-ex]}}\relax
\mciteBstWouldAddEndPuncttrue
\mciteSetBstMidEndSepPunct{\mcitedefaultmidpunct}
{\mcitedefaultendpunct}{\mcitedefaultseppunct}\relax
\EndOfBibitem
\bibitem{Malde:2015mha}
S.~Malde {\em et al.}{,}
  \href{http://dx.doi.org/10.1016/j.physletb.2015.05.043}{Phys.\ Lett.\ {\bf
  B747},  9}  (2015), \href{http://arxiv.org/abs/1504.05878}{{\tt
  arXiv:1504.05878 [hep-ex]}}\relax
\mciteBstWouldAddEndPuncttrue
\mciteSetBstMidEndSepPunct{\mcitedefaultmidpunct}
{\mcitedefaultendpunct}{\mcitedefaultseppunct}\relax
\EndOfBibitem
\bibitem{Aaij:2015jna}
R.~Aaij {\em et al.} ({LHCb} collaboration){,}
  \href{http://dx.doi.org/10.1103/PhysRevD.91.112014}{Phys.\ Rev.\ {\bf D91}{,}
  112014}  (2015), \href{http://arxiv.org/abs/1504.05442}{{\tt
  arXiv:1504.05442 [hep-ex]}}\relax
\mciteBstWouldAddEndPuncttrue
\mciteSetBstMidEndSepPunct{\mcitedefaultmidpunct}
{\mcitedefaultendpunct}{\mcitedefaultseppunct}\relax
\EndOfBibitem
\bibitem{delAmoSanchez:2010dz}
P.~del Amo~Sanchez {\em et al.} ({\babar} collaboration){,}
  \href{http://dx.doi.org/10.1103/PhysRevD.82.072006}{Phys.\ Rev.\ {\bf D82}{,}
  072006}  (2010), \href{http://arxiv.org/abs/1006.4241}{{\tt arXiv:1006.4241
  [hep-ex]}}\relax
\mciteBstWouldAddEndPuncttrue
\mciteSetBstMidEndSepPunct{\mcitedefaultmidpunct}
{\mcitedefaultendpunct}{\mcitedefaultseppunct}\relax
\EndOfBibitem
\bibitem{Belle:2011ac}
Y.~Horii {\em et al.} ({Belle} collaboration){,}
  \href{http://dx.doi.org/10.1103/PhysRevLett.106.231803}{Phys.\ Rev.\ Lett.\ {\bf
  106},  231803}  (2011), \href{http://arxiv.org/abs/1103.5951}{{\tt
  arXiv:1103.5951 [hep-ex]}}\relax
\mciteBstWouldAddEndPuncttrue
\mciteSetBstMidEndSepPunct{\mcitedefaultmidpunct}
{\mcitedefaultendpunct}{\mcitedefaultseppunct}\relax
\EndOfBibitem
\bibitem{Aaltonen:2011uu}
T.~Aaltonen {\em et al.} ({CDF} collaboration){,}
  \href{http://dx.doi.org/10.1103/PhysRevD.84.091504}{Phys.\ Rev.\ {\bf D84}{,}
  091504}  (2011), \href{http://arxiv.org/abs/1108.5765}{{\tt arXiv:1108.5765
  [hep-ex]}}\relax
\mciteBstWouldAddEndPuncttrue
\mciteSetBstMidEndSepPunct{\mcitedefaultmidpunct}
{\mcitedefaultendpunct}{\mcitedefaultseppunct}\relax
\EndOfBibitem
\bibitem{Lees:2011up}
J.~P.\ Lees {\em et al.} ({\babar} collaboration){,}
  \href{http://dx.doi.org/10.1103/PhysRevD.84.012002}{Phys.\ Rev.\ {\bf D84}{,}
  012002}  (2011), \href{http://arxiv.org/abs/1104.4472}{{\tt arXiv:1104.4472
  [hep-ex]}}\relax
\mciteBstWouldAddEndPuncttrue
\mciteSetBstMidEndSepPunct{\mcitedefaultmidpunct}
{\mcitedefaultendpunct}{\mcitedefaultseppunct}\relax
\EndOfBibitem
\bibitem{Nayak:2013tgg}
M.~Nayak {\em et al.} ({Belle} collaboration){,}
  \href{http://dx.doi.org/10.1103/PhysRevD.88.091104}{Phys.\ Rev.\ {\bf D88}{,}
  091104}  (2013), \href{http://arxiv.org/abs/1310.1741}{{\tt arXiv:1310.1741
  [hep-ex]}}\relax
\mciteBstWouldAddEndPuncttrue
\mciteSetBstMidEndSepPunct{\mcitedefaultmidpunct}
{\mcitedefaultendpunct}{\mcitedefaultseppunct}\relax
\EndOfBibitem
\bibitem{:2009au}
B.~Aubert {\em et al.} ({\babar} collaboration){,}
  \href{http://dx.doi.org/10.1103/PhysRevD.80.031102}{Phys.\ Rev.\ {\bf D80}{,}
  031102}  (2009), \href{http://arxiv.org/abs/0904.2112}{{\tt arXiv:0904.2112
  [hep-ex]}}\relax
\mciteBstWouldAddEndPuncttrue
\mciteSetBstMidEndSepPunct{\mcitedefaultmidpunct}
{\mcitedefaultendpunct}{\mcitedefaultseppunct}\relax
\EndOfBibitem
\bibitem{Negishi:2012uxa}
K.~Negishi {\em et al.} ({Belle} collaboration){,}
  \href{http://dx.doi.org/10.1103/PhysRevD.86.011101}{Phys.\ Rev.\ {\bf D86}{,}
  011101}  (2012), \href{http://arxiv.org/abs/1205.0422}{{\tt arXiv:1205.0422
  [hep-ex]}}\relax
\mciteBstWouldAddEndPuncttrue
\mciteSetBstMidEndSepPunct{\mcitedefaultmidpunct}
{\mcitedefaultendpunct}{\mcitedefaultseppunct}\relax
\EndOfBibitem
\bibitem{Asner:2008ft}
D.~M.\ Asner {\em et al.} ({CLEO} collaboration){,}
  \href{http://dx.doi.org/10.1103/PhysRevD.78.012001}{Phys.\ Rev.\ {\bf D78}{,}
  012001}  (2008), \href{http://arxiv.org/abs/0802.2268}{{\tt arXiv:0802.2268
  [hep-ex]}}\relax
\mciteBstWouldAddEndPuncttrue
\mciteSetBstMidEndSepPunct{\mcitedefaultmidpunct}
{\mcitedefaultendpunct}{\mcitedefaultseppunct}\relax
\EndOfBibitem
\bibitem{Lowery:2009id}
N.~Lowrey {\em et al.} ({CLEO} collaboration){,}
  \href{http://dx.doi.org/10.1103/PhysRevD.80.031105}{Phys.\ Rev.\ {\bf D80}{,}
  031105}  (2009), \href{http://arxiv.org/abs/0903.4853}{{\tt arXiv:0903.4853
  [hep-ex]}}\relax
\mciteBstWouldAddEndPuncttrue
\mciteSetBstMidEndSepPunct{\mcitedefaultmidpunct}
{\mcitedefaultendpunct}{\mcitedefaultseppunct}\relax
\EndOfBibitem
\bibitem{Poluektov:2010wz}
A.~Poluektov {\em et al.} ({Belle} collaboration){,}
  \href{http://dx.doi.org/10.1103/PhysRevD.81.112002}{Phys.\ Rev.\ {\bf D81}{,}
  112002}  (2010), \href{http://arxiv.org/abs/1003.3360}{{\tt arXiv:1003.3360
  [hep-ex]}}\relax
\mciteBstWouldAddEndPuncttrue
\mciteSetBstMidEndSepPunct{\mcitedefaultmidpunct}
{\mcitedefaultendpunct}{\mcitedefaultseppunct}\relax
\EndOfBibitem
\bibitem{delAmoSanchez:2010rq}
P.~del Amo~Sanchez {\em et al.} ({\babar} collaboration){,}
  \href{http://dx.doi.org/10.1103/PhysRevLett.105.121801}{Phys.\ Rev.\ Lett.\ {\bf
  105},  121801}  (2010), \href{http://arxiv.org/abs/1005.1096}{{\tt
  arXiv:1005.1096 [hep-ex]}}\relax
\mciteBstWouldAddEndPuncttrue
\mciteSetBstMidEndSepPunct{\mcitedefaultmidpunct}
{\mcitedefaultendpunct}{\mcitedefaultseppunct}\relax
\EndOfBibitem
\bibitem{Aaij:2014iba}
R.~Aaij {\em et al.} ({LHCb} collaboration){,}
  \href{http://dx.doi.org/10.1016/j.nuclphysb.2014.09.015}{Nucl.\ Phys.\ {\bf
  B888},  169}  (2014), \href{http://arxiv.org/abs/1407.6211}{{\tt
  arXiv:1407.6211 [hep-ex]}}\relax
\mciteBstWouldAddEndPuncttrue
\mciteSetBstMidEndSepPunct{\mcitedefaultmidpunct}
{\mcitedefaultendpunct}{\mcitedefaultseppunct}\relax
\EndOfBibitem
\bibitem{Poluektov:2006ia}
A.~Poluektov {\em et al.} ({Belle} collaboration){,}
  \href{http://dx.doi.org/10.1103/PhysRevD.73.112009}{Phys.\ Rev.\ {\bf D73}{,}
  112009}  (2006), \href{http://arxiv.org/abs/hep-ex/0604054}{{\tt
  arXiv:hep-ex/0604054}}\relax
\mciteBstWouldAddEndPuncttrue
\mciteSetBstMidEndSepPunct{\mcitedefaultmidpunct}
{\mcitedefaultendpunct}{\mcitedefaultseppunct}\relax
\EndOfBibitem
\bibitem{Aaij:2016zlt}
R.~Aaij {\em et al.} ({LHCb} collaboration){,}
  \href{http://dx.doi.org/10.1007/JHEP08(2016)137}{JHEP {\bf 08},  137}
  (2016), \href{http://arxiv.org/abs/1605.01082}{{\tt arXiv:1605.01082
  [hep-ex]}}\relax
\mciteBstWouldAddEndPuncttrue
\mciteSetBstMidEndSepPunct{\mcitedefaultmidpunct}
{\mcitedefaultendpunct}{\mcitedefaultseppunct}\relax
\EndOfBibitem
\bibitem{Aubert:2008yn}
B.~Aubert {\em et al.} ({\babar} collaboration){,}
  \href{http://dx.doi.org/10.1103/PhysRevD.79.072003}{Phys.\ Rev.\ {\bf D79}{,}
  072003}  (2009), \href{http://arxiv.org/abs/0805.2001}{{\tt arXiv:0805.2001
  [hep-ex]}}\relax
\mciteBstWouldAddEndPuncttrue
\mciteSetBstMidEndSepPunct{\mcitedefaultmidpunct}
{\mcitedefaultendpunct}{\mcitedefaultseppunct}\relax
\EndOfBibitem
\bibitem{Briere:2009aa}
R.~A.\ Briere {\em et al.} ({CLEO} collaboration){,}
  \href{http://dx.doi.org/10.1103/PhysRevD.80.032002}{Phys.\ Rev.\ {\bf D80}{,}
  032002}  (2009), \href{http://arxiv.org/abs/0903.1681}{{\tt arXiv:0903.1681
  [hep-ex]}}\relax
\mciteBstWouldAddEndPuncttrue
\mciteSetBstMidEndSepPunct{\mcitedefaultmidpunct}
{\mcitedefaultendpunct}{\mcitedefaultseppunct}\relax
\EndOfBibitem
\bibitem{Aihara:2012aw}
H.~Aihara {\em et al.} ({Belle} collaboration){,}
  \href{http://dx.doi.org/10.1103/PhysRevD.85.112014}{Phys.\ Rev.\ {\bf D85}{,}
  112014}  (2012), \href{http://arxiv.org/abs/1204.6561}{{\tt arXiv:1204.6561
  [hep-ex]}}\relax
\mciteBstWouldAddEndPuncttrue
\mciteSetBstMidEndSepPunct{\mcitedefaultmidpunct}
{\mcitedefaultendpunct}{\mcitedefaultseppunct}\relax
\EndOfBibitem
\bibitem{Aaij:2014uva}
R.~Aaij {\em et al.} ({LHCb} collaboration){,}
  \href{http://dx.doi.org/10.1007/JHEP10(2014)097}{JHEP {\bf 10},  97}
  (2014), \href{http://arxiv.org/abs/1408.2748}{{\tt arXiv:1408.2748
  [hep-ex]}}\relax
\mciteBstWouldAddEndPuncttrue
\mciteSetBstMidEndSepPunct{\mcitedefaultmidpunct}
{\mcitedefaultendpunct}{\mcitedefaultseppunct}\relax
\EndOfBibitem
\bibitem{Negishi:2015vqa}
K.~Negishi {\em et al.} ({Belle} collaboration){,}
  \href{http://dx.doi.org/10.1093/ptep/ptw030}{PTEP {\bf 2016},  043C01}
  (2016), \href{http://arxiv.org/abs/1509.01098}{{\tt arXiv:1509.01098
  [hep-ex]}}\relax
\mciteBstWouldAddEndPuncttrue
\mciteSetBstMidEndSepPunct{\mcitedefaultmidpunct}
{\mcitedefaultendpunct}{\mcitedefaultseppunct}\relax
\EndOfBibitem
\bibitem{Aaij:2016nao}
R.~Aaij {\em et al.} ({LHCb} collaboration){,}
  \href{http://dx.doi.org/10.1007/JHEP06(2016)131}{JHEP {\bf 06},  131}
  (2016), \href{http://arxiv.org/abs/1604.01525}{{\tt arXiv:1604.01525
  [hep-ex]}}\relax
\mciteBstWouldAddEndPuncttrue
\mciteSetBstMidEndSepPunct{\mcitedefaultmidpunct}
{\mcitedefaultendpunct}{\mcitedefaultseppunct}\relax
\EndOfBibitem
\bibitem{Aaij:2014dia}
R.~Aaij {\em et al.} ({LHCb} collaboration){,}
  \href{http://dx.doi.org/10.1016/j.physletb.2014.03.051}{Phys.\ Lett.\ {\bf
  B733},  36}  (2014), \href{http://arxiv.org/abs/1402.2982}{{\tt
  arXiv:1402.2982 [hep-ex]}}\relax
\mciteBstWouldAddEndPuncttrue
\mciteSetBstMidEndSepPunct{\mcitedefaultmidpunct}
{\mcitedefaultendpunct}{\mcitedefaultseppunct}\relax
\EndOfBibitem
\bibitem{Insler:2012pm}
J.~Insler {\em et al.} ({CLEO} collaboration){,}
  \href{http://dx.doi.org/10.1103/PhysRevD.85.092016}{Phys.\ Rev.\ {\bf D85}{,}
  092016}  (2012), \href{http://arxiv.org/abs/1203.3804}{{\tt arXiv:1203.3804
  [hep-ex]}}\relax
\mciteBstWouldAddEndPuncttrue
\mciteSetBstMidEndSepPunct{\mcitedefaultmidpunct}
{\mcitedefaultendpunct}{\mcitedefaultseppunct}\relax
\EndOfBibitem
\bibitem{Lees:2013nha}
J.~P.\ Lees {\em et al.} ({\babar} collaboration){,}
  \href{http://dx.doi.org/10.1103/PhysRevD.87.052015}{Phys.\ Rev.\ {\bf D87}{,}
  052015}  (2013), \href{http://arxiv.org/abs/1301.1029}{{\tt arXiv:1301.1029
  [hep-ex]}}\relax
\mciteBstWouldAddEndPuncttrue
\mciteSetBstMidEndSepPunct{\mcitedefaultmidpunct}
{\mcitedefaultendpunct}{\mcitedefaultseppunct}\relax
\EndOfBibitem
\bibitem{Aaij:2016kjh}
R.~Aaij {\em et al.} ({LHCb} collaboration){,}
  \href{http://dx.doi.org/10.1007/JHEP12(2016)087}{JHEP {\bf 12},  087}
  (2016), \href{http://arxiv.org/abs/1611.03076}{{\tt arXiv:1611.03076
  [hep-ex]}}\relax
\mciteBstWouldAddEndPuncttrue
\mciteSetBstMidEndSepPunct{\mcitedefaultmidpunct}
{\mcitedefaultendpunct}{\mcitedefaultseppunct}\relax
\EndOfBibitem
\bibitem{Rama:2013voa}
M.~Rama, \href{http://dx.doi.org/10.1103/PhysRevD.89.014021}{Phys.\ Rev.\ {\bf
  D89},  014021}  (2014), \href{http://arxiv.org/abs/1307.4384}{{\tt
  arXiv:1307.4384 [hep-ex]}}\relax
\mciteBstWouldAddEndPuncttrue
\mciteSetBstMidEndSepPunct{\mcitedefaultmidpunct}
{\mcitedefaultendpunct}{\mcitedefaultseppunct}\relax
\EndOfBibitem
\bibitem{gammacombo}
 {{\tt GammaCombo} framework for combinations of measurements and computation
  of confidence intervals, CERN}, \url{{http://gammacombo.hepforge.org/}}\relax
\mciteBstWouldAddEndPuncttrue
\mciteSetBstMidEndSepPunct{\mcitedefaultmidpunct}
{\mcitedefaultendpunct}{\mcitedefaultseppunct}\relax
\EndOfBibitem
\bibitem{Aaij:2013zfa}
R.~Aaij {\em et al.} ({LHCb} collaboration){,}
  \href{http://dx.doi.org/10.1016/j.physletb.2013.08.020}{Phys.\ Lett.\ {\bf
  B726},  151}  (2013), \href{http://arxiv.org/abs/1305.2050}{{\tt
  arXiv:1305.2050 [hep-ex]}}\relax
\mciteBstWouldAddEndPuncttrue
\mciteSetBstMidEndSepPunct{\mcitedefaultmidpunct}
{\mcitedefaultendpunct}{\mcitedefaultseppunct}\relax
\EndOfBibitem
\bibitem{Evans:2016tlp}
T.~Evans, S.~Harnew, J.~Libby, S.~Malde, J.~Rademacker, and G.~Wilkinson{,}
  \href{http://dx.doi.org/10.1016/j.physletb.2016.04.037}{Phys.\ Lett.\ {\bf
  B757},  520}  (2016), \href{http://arxiv.org/abs/1602.07430}{{\tt
  arXiv:1602.07430 [hep-ex]}}\relax
\mciteBstWouldAddEndPuncttrue
\mciteSetBstMidEndSepPunct{\mcitedefaultmidpunct}
{\mcitedefaultendpunct}{\mcitedefaultseppunct}\relax
\EndOfBibitem
\bibitem{Aaij:2015lsa}
R.~Aaij {\em et al.} ({LHCb} collaboration){,}
  \href{http://dx.doi.org/10.1103/PhysRevD.93.052018}{Phys.\ Rev.\ {\bf D93}{,}
  052018}  (2016), \href{http://arxiv.org/abs/1509.06628}{{\tt
  arXiv:1509.06628 [hep-ex]}}\relax
\mciteBstWouldAddEndPuncttrue
\mciteSetBstMidEndSepPunct{\mcitedefaultmidpunct}
{\mcitedefaultendpunct}{\mcitedefaultseppunct}\relax
\EndOfBibitem
\bibitem{Silva:1999bd}
J.~P.\ Silva and A.~Soffer{,}
  \href{http://dx.doi.org/10.1103/PhysRevD.61.112001}{Phys.\ Rev.\ {\bf D61}{,}
  112001}  (2000), \href{http://arxiv.org/abs/hep-ph/9912242}{{\tt
  arXiv:hep-ph/9912242 [hep-ph]}}\relax
\mciteBstWouldAddEndPuncttrue
\mciteSetBstMidEndSepPunct{\mcitedefaultmidpunct}
{\mcitedefaultendpunct}{\mcitedefaultseppunct}\relax
\EndOfBibitem
\bibitem{Grossman:2005rp}
Y.~Grossman, A.~Soffer, and J.~Zupan{,}
  \href{http://dx.doi.org/10.1103/PhysRevD.72.031501}{Phys.\ Rev.\ {\bf D72}{,}
  031501}  (2005), \href{http://arxiv.org/abs/hep-ph/0505270}{{\tt
  arXiv:hep-ph/0505270 [hep-ph]}}\relax
\mciteBstWouldAddEndPuncttrue
\mciteSetBstMidEndSepPunct{\mcitedefaultmidpunct}
{\mcitedefaultendpunct}{\mcitedefaultseppunct}\relax
\EndOfBibitem
\bibitem{Neubert:1993mb}
M.~Neubert, \href{http://dx.doi.org/10.1016/0370-1573(94)90091-4}{Phys.\ Rept.\
  {\bf 245},  259}  (1994), \href{http://arxiv.org/abs/hep-ph/9306320}{{\tt
  arXiv:hep-ph/9306320 [hep-ph]}}\relax
\mciteBstWouldAddEndPuncttrue
\mciteSetBstMidEndSepPunct{\mcitedefaultmidpunct}
{\mcitedefaultendpunct}{\mcitedefaultseppunct}\relax
\EndOfBibitem
\bibitem{Sirlin:1981ie}
A.~Sirlin, \href{http://dx.doi.org/10.1016/0550-3213(82)90303-0}{Nucl.\ Phys.\
  {\bf B196},  83}  (1982)\relax
\mciteBstWouldAddEndPuncttrue
\mciteSetBstMidEndSepPunct{\mcitedefaultmidpunct}
{\mcitedefaultendpunct}{\mcitedefaultseppunct}\relax
\EndOfBibitem
\bibitem{CLN}
I.~Caprini, L.~Lellouch, and M.~Neubert{,}
  \href{http://dx.doi.org/10.1016/S0550-3213(98)00350-2}{Nucl.\ Phys.\ {\bf
  B530},  153}  (1998), \href{http://arxiv.org/abs/hep-ph/9712417}{{\tt
  arXiv:hep-ph/9712417}}\relax
\mciteBstWouldAddEndPuncttrue
\mciteSetBstMidEndSepPunct{\mcitedefaultmidpunct}
{\mcitedefaultendpunct}{\mcitedefaultseppunct}\relax
\EndOfBibitem
\bibitem{HFLAV_sl:inputparams}
 $B$ semileptonic decays common input parameters{,}
  \url{http://www.slac.stanford.edu/xorg/hflav/semi/summer16/common/common.param.summer16}\relax
\mciteBstWouldAddEndPuncttrue
\mciteSetBstMidEndSepPunct{\mcitedefaultmidpunct}
{\mcitedefaultendpunct}{\mcitedefaultseppunct}\relax
\EndOfBibitem
\bibitem{Adam:2002uw}
N.~E.\ Adam {\em et al.} ({CLEO} collaboration){,}
  \href{http://dx.doi.org/10.1103/PhysRevD.67.032001}{Phys.\ Rev.\ {\bf D67}{,}
  032001}  (2003), \href{http://arxiv.org/abs/hep-ex/0210040}{{\tt
  arXiv:hep-ex/0210040}}\relax
\mciteBstWouldAddEndPuncttrue
\mciteSetBstMidEndSepPunct{\mcitedefaultmidpunct}
{\mcitedefaultendpunct}{\mcitedefaultseppunct}\relax
\EndOfBibitem
\bibitem{Aubert:2009_1}
B.~Aubert {\em et al.} ({\babar} collaboration){,}
  \href{http://dx.doi.org/10.1103/PhysRevD.79.012002}{Phys.\ Rev.\ {\bf D79}{,}
  012002}  (2009), \href{http://arxiv.org/abs/0809.0828}{{\tt arXiv:0809.0828
  [hep-ex]}}\relax
\mciteBstWouldAddEndPuncttrue
\mciteSetBstMidEndSepPunct{\mcitedefaultmidpunct}
{\mcitedefaultendpunct}{\mcitedefaultseppunct}\relax
\EndOfBibitem
\bibitem{Aubert:2009_3}
B.~Aubert {\em et al.} ({\babar} collaboration){,}
  \href{http://dx.doi.org/10.1103/PhysRevLett.100.231803}{Phys.\ Rev.\ Lett.\ {\bf
  100},  231803}  (2008), \href{http://arxiv.org/abs/0712.3493}{{\tt
  arXiv:0712.3493 [hep-ex]}}\relax
\mciteBstWouldAddEndPuncttrue
\mciteSetBstMidEndSepPunct{\mcitedefaultmidpunct}
{\mcitedefaultendpunct}{\mcitedefaultseppunct}\relax
\EndOfBibitem
\bibitem{Dungel:2010uk}
W.~Dungel {\em et al.} ({Belle} collaboration){,}
  \href{http://dx.doi.org/10.1103/PhysRevD.82.112007}{Phys.\ Rev.\ {\bf D82}{,}
  112007}  (2010), \href{http://arxiv.org/abs/1010.5620}{{\tt arXiv:1010.5620
  [hep-ex]}}\relax
\mciteBstWouldAddEndPuncttrue
\mciteSetBstMidEndSepPunct{\mcitedefaultmidpunct}
{\mcitedefaultendpunct}{\mcitedefaultseppunct}\relax
\EndOfBibitem
\bibitem{Aubert:2006mb}
B.~Aubert {\em et al.} ({\babar} collaboration){,}
  \href{http://dx.doi.org/10.1103/PhysRevD.77.032002}{Phys.\ Rev.\ {\bf D77}{,}
  032002}  (2008), \href{http://arxiv.org/abs/0705.4008}{{\tt arXiv:0705.4008
  [hep-ex]}}\relax
\mciteBstWouldAddEndPuncttrue
\mciteSetBstMidEndSepPunct{\mcitedefaultmidpunct}
{\mcitedefaultendpunct}{\mcitedefaultseppunct}\relax
\EndOfBibitem
\bibitem{Buskulic:1996yq}
D.~Buskulic {\em et al.} ({ALEPH} collaboration){,}
  \href{http://dx.doi.org/10.1016/S0370-2693(97)00071-3}{Phys.\ Lett.\ {\bf
  B395},  373}  (1997)\relax
\mciteBstWouldAddEndPuncttrue
\mciteSetBstMidEndSepPunct{\mcitedefaultmidpunct}
{\mcitedefaultendpunct}{\mcitedefaultseppunct}\relax
\EndOfBibitem
\bibitem{Abbiendi:2000hk}
G.~Abbiendi {\em et al.} ({OPAL} collaboration){,}
  \href{http://dx.doi.org/10.1016/S0370-2693(00)00457-3}{Phys.\ Lett.\ {\bf
  B482},  15}  (2000), \href{http://arxiv.org/abs/hep-ex/0003013}{{\tt
  arXiv:hep-ex/0003013}}\relax
\mciteBstWouldAddEndPuncttrue
\mciteSetBstMidEndSepPunct{\mcitedefaultmidpunct}
{\mcitedefaultendpunct}{\mcitedefaultseppunct}\relax
\EndOfBibitem
\bibitem{Abreu:2001ic}
P.~Abreu {\em et al.} ({DELPHI} collaboration){,}
  \href{http://dx.doi.org/10.1016/S0370-2693(01)00569-X}{Phys.\ Lett.\ {\bf
  B510},  55}  (2001), \href{http://arxiv.org/abs/hep-ex/0104026}{{\tt
  arXiv:hep-ex/0104026}}\relax
\mciteBstWouldAddEndPuncttrue
\mciteSetBstMidEndSepPunct{\mcitedefaultmidpunct}
{\mcitedefaultendpunct}{\mcitedefaultseppunct}\relax
\EndOfBibitem
\bibitem{Abdallah:2004rz}
J.~Abdallah {\em et al.} ({DELPHI} collaboration){,}
  \href{http://dx.doi.org/10.1140/epjc/s2004-01598-6}{Eur.\ Phys.\ J.\ {\bf C33}{,}
  213}  (2004), \href{http://arxiv.org/abs/hep-ex/0401023}{{\tt
  arXiv:hep-ex/0401023}}\relax
\mciteBstWouldAddEndPuncttrue
\mciteSetBstMidEndSepPunct{\mcitedefaultmidpunct}
{\mcitedefaultendpunct}{\mcitedefaultseppunct}\relax
\EndOfBibitem
\bibitem{Bailey:2014tva}
J.~A.\ Bailey {\em et al.} ({Fermilab Lattice and MILC} collaborations){,}
  \href{http://dx.doi.org/10.1103/PhysRevD.89.114504}{Phys.\ Rev.\ {\bf D89}{,}
  114504}  (2014), \href{http://arxiv.org/abs/1403.0635}{{\tt arXiv:1403.0635
  [hep-lat]}}\relax
\mciteBstWouldAddEndPuncttrue
\mciteSetBstMidEndSepPunct{\mcitedefaultmidpunct}
{\mcitedefaultendpunct}{\mcitedefaultseppunct}\relax
\EndOfBibitem
\bibitem{Aubert:vcbExcl}
B.~Aubert {\em et al.} ({\babar} collaboration){,}
  \href{http://dx.doi.org/10.1103/PhysRevLett.100.151802}{Phys.\ Rev.\ Lett.\ {\bf
  100},  151802}  (2008), \href{http://arxiv.org/abs/0712.3503}{{\tt
  arXiv:0712.3503 [hep-ex]}}\relax
\mciteBstWouldAddEndPuncttrue
\mciteSetBstMidEndSepPunct{\mcitedefaultmidpunct}
{\mcitedefaultendpunct}{\mcitedefaultseppunct}\relax
\EndOfBibitem
\bibitem{Bartelt:1998dq}
J.~E.\ Bartelt {\em et al.} ({CLEO} collaboration){,}
  \href{http://dx.doi.org/10.1103/PhysRevLett.82.3746}{Phys.\ Rev.\ Lett.\ {\bf
  82},  3746}  (1999), \href{http://arxiv.org/abs/hep-ex/9811042}{{\tt
  arXiv:hep-ex/9811042}}\relax
\mciteBstWouldAddEndPuncttrue
\mciteSetBstMidEndSepPunct{\mcitedefaultmidpunct}
{\mcitedefaultendpunct}{\mcitedefaultseppunct}\relax
\EndOfBibitem
\bibitem{Glattauer:2015teq}
R.~Glattauer {\em et al.} ({Belle} collaboration){,}
  \href{http://dx.doi.org/10.1103/PhysRevD.93.032006}{Phys.\ Rev.\ {\bf D93}{,}
  032006}  (2016), \href{http://arxiv.org/abs/1510.03657}{{\tt
  arXiv:1510.03657 [hep-ex]}}\relax
\mciteBstWouldAddEndPuncttrue
\mciteSetBstMidEndSepPunct{\mcitedefaultmidpunct}
{\mcitedefaultendpunct}{\mcitedefaultseppunct}\relax
\EndOfBibitem
\bibitem{Aubert:2009_2}
B.~Aubert {\em et al.} ({\babar} collaboration){,}
  \href{http://dx.doi.org/10.1103/PhysRevLett.104.011802}{Phys.\ Rev.\ Lett.\ {\bf
  104},  011802}  (2010), \href{http://arxiv.org/abs/0904.4063}{{\tt
  arXiv:0904.4063 [hep-ex]}}\relax
\mciteBstWouldAddEndPuncttrue
\mciteSetBstMidEndSepPunct{\mcitedefaultmidpunct}
{\mcitedefaultendpunct}{\mcitedefaultseppunct}\relax
\EndOfBibitem
\bibitem{Lattice:2015rga}
J.~A.\ Bailey {\em et al.} ({MILC} collaboration){,}
  \href{http://dx.doi.org/10.1103/PhysRevD.92.034506}{Phys.\ Rev.\ {\bf D92}{,}
  034506}  (2015), \href{http://arxiv.org/abs/1503.07237}{{\tt
  arXiv:1503.07237 [hep-lat]}}\relax
\mciteBstWouldAddEndPuncttrue
\mciteSetBstMidEndSepPunct{\mcitedefaultmidpunct}
{\mcitedefaultendpunct}{\mcitedefaultseppunct}\relax
\EndOfBibitem
\bibitem{Live:Dss}
D.~Liventsev {\em et al.} ({Belle} collaboration){,}
  \href{http://dx.doi.org/10.1103/PhysRevD.77.091503}{Phys.\ Rev.\ {\bf D77}{,}
  091503}  (2008), \href{http://arxiv.org/abs/0711.3252}{{\tt arXiv:0711.3252
  [hep-ex]}}\relax
\mciteBstWouldAddEndPuncttrue
\mciteSetBstMidEndSepPunct{\mcitedefaultmidpunct}
{\mcitedefaultendpunct}{\mcitedefaultseppunct}\relax
\EndOfBibitem
\bibitem{Isgur:1991wq}
N.~Isgur and M.~B.\ Wise{,}
  \href{http://dx.doi.org/10.1103/PhysRevLett.66.1130}{Phys.\ Rev.\ Lett.\ {\bf
  66},  1130}  (1991)\relax
\mciteBstWouldAddEndPuncttrue
\mciteSetBstMidEndSepPunct{\mcitedefaultmidpunct}
{\mcitedefaultendpunct}{\mcitedefaultseppunct}\relax
\EndOfBibitem
\bibitem{Aleph:Dss}
D.~Buskulic {\em et al.} ({ALEPH} collaboration){,}
  \href{http://dx.doi.org/10.1007/s002880050351}{Z.\ Phys.\ {\bf C73},  601}
  (1997)\relax
\mciteBstWouldAddEndPuncttrue
\mciteSetBstMidEndSepPunct{\mcitedefaultmidpunct}
{\mcitedefaultendpunct}{\mcitedefaultseppunct}\relax
\EndOfBibitem
\bibitem{opal:Dss}
G.~Abbiendi {\em et al.} ({OPAL} collaboration){,}
  \href{http://dx.doi.org/10.1140/epjc/s2003-01322-2}{Eur.\ Phys.\ J.\ {\bf C30}{,}
  467}  (2003), \href{http://arxiv.org/abs/hep-ex/0301018}{{\tt
  arXiv:hep-ex/0301018}}\relax
\mciteBstWouldAddEndPuncttrue
\mciteSetBstMidEndSepPunct{\mcitedefaultmidpunct}
{\mcitedefaultendpunct}{\mcitedefaultseppunct}\relax
\EndOfBibitem
\bibitem{cleo:Dss}
A.~Anastassov {\em et al.} ({CLEO} collaboration){,}
  \href{http://dx.doi.org/10.1103/PhysRevLett.80.4127}{Phys.\ Rev.\ Lett.\ {\bf
  80},  4127}  (1998), \href{http://arxiv.org/abs/hep-ex/9708035}{{\tt
  arXiv:hep-ex/9708035}}\relax
\mciteBstWouldAddEndPuncttrue
\mciteSetBstMidEndSepPunct{\mcitedefaultmidpunct}
{\mcitedefaultendpunct}{\mcitedefaultseppunct}\relax
\EndOfBibitem
\bibitem{D0:Dss}
V.~M.\ Abazov {\em et al.} ({\dzero} collaboration){,}
  \href{http://dx.doi.org/10.1103/PhysRevLett.95.171803}{Phys.\ Rev.\ Lett.\ {\bf
  95},  171803}  (2005), \href{http://arxiv.org/abs/hep-ex/0507046}{{\tt
  arXiv:hep-ex/0507046}}\relax
\mciteBstWouldAddEndPuncttrue
\mciteSetBstMidEndSepPunct{\mcitedefaultmidpunct}
{\mcitedefaultendpunct}{\mcitedefaultseppunct}\relax
\EndOfBibitem
\bibitem{Aubert:2009_4}
B.~Aubert {\em et al.} ({\babar} collaboration){,}
  \href{http://dx.doi.org/10.1103/PhysRevLett.101.261802}{Phys.\ Rev.\ Lett.\ {\bf
  101},  261802}  (2008), \href{http://arxiv.org/abs/0808.0528}{{\tt
  arXiv:0808.0528 [hep-ex]}}\relax
\mciteBstWouldAddEndPuncttrue
\mciteSetBstMidEndSepPunct{\mcitedefaultmidpunct}
{\mcitedefaultendpunct}{\mcitedefaultseppunct}\relax
\EndOfBibitem
\bibitem{Aubert:2008zc}
B.~Aubert {\em et al.} ({\babar} collaboration){,}
  \href{http://dx.doi.org/10.1103/PhysRevLett.103.051803}{Phys.\ Rev.\ Lett.\ {\bf
  103},  051803}  (2009), \href{http://arxiv.org/abs/0808.0333}{{\tt
  arXiv:0808.0333 [hep-ex]}}\relax
\mciteBstWouldAddEndPuncttrue
\mciteSetBstMidEndSepPunct{\mcitedefaultmidpunct}
{\mcitedefaultendpunct}{\mcitedefaultseppunct}\relax
\EndOfBibitem
\bibitem{Abdallah:2005cx}
J.~Abdallah {\em et al.} ({DELPHI} collaboration){,}
  \href{http://dx.doi.org/10.1140/epjc/s2005-02406-7}{Eur.\ Phys.\ J.\ {\bf C45}{,}
  35}  (2006), \href{http://arxiv.org/abs/hep-ex/0510024}{{\tt
  arXiv:hep-ex/0510024}}\relax
\mciteBstWouldAddEndPuncttrue
\mciteSetBstMidEndSepPunct{\mcitedefaultmidpunct}
{\mcitedefaultendpunct}{\mcitedefaultseppunct}\relax
\EndOfBibitem
\bibitem{Benson:2003kp}
D.~Benson, I.~I.\ Bigi, T.~Mannel, and N.~Uraltsev{,}
  \href{http://dx.doi.org/10.1016/S0550-3213(03)00452-8}{Nucl.\ Phys.\ {\bf
  B665},  367}  (2003), \href{http://arxiv.org/abs/hep-ph/0302262}{{\tt
  arXiv:hep-ph/0302262}}\relax
\mciteBstWouldAddEndPuncttrue
\mciteSetBstMidEndSepPunct{\mcitedefaultmidpunct}
{\mcitedefaultendpunct}{\mcitedefaultseppunct}\relax
\EndOfBibitem
\bibitem{Gambino:2004qm}
P.~Gambino and N.~Uraltsev{,}
  \href{http://dx.doi.org/10.1140/epjc/s2004-01671-2}{Eur.\ Phys.\ J.\ {\bf C34}{,}
  181}  (2004), \href{http://arxiv.org/abs/hep-ph/0401063}{{\tt
  arXiv:hep-ph/0401063}}\relax
\mciteBstWouldAddEndPuncttrue
\mciteSetBstMidEndSepPunct{\mcitedefaultmidpunct}
{\mcitedefaultendpunct}{\mcitedefaultseppunct}\relax
\EndOfBibitem
\bibitem{Gambino:2011cq}
P.~Gambino, \href{http://dx.doi.org/10.1007/JHEP09(2011)055}{JHEP {\bf 09}{,}
  055}  (2011), \href{http://arxiv.org/abs/1107.3100}{{\tt arXiv:1107.3100
  [hep-ph]}}\relax
\mciteBstWouldAddEndPuncttrue
\mciteSetBstMidEndSepPunct{\mcitedefaultmidpunct}
{\mcitedefaultendpunct}{\mcitedefaultseppunct}\relax
\EndOfBibitem
\bibitem{Alberti:2014yda}
A.~Alberti, P.~Gambino, K.~J.\ Healey, and S.~Nandi{,}
  \href{http://dx.doi.org/10.1103/PhysRevLett.114.061802}{Phys.\ Rev.\ Lett.\ {\bf
  114},  061802}  (2015), \href{http://arxiv.org/abs/1411.6560}{{\tt
  arXiv:1411.6560 [hep-ph]}}\relax
\mciteBstWouldAddEndPuncttrue
\mciteSetBstMidEndSepPunct{\mcitedefaultmidpunct}
{\mcitedefaultendpunct}{\mcitedefaultseppunct}\relax
\EndOfBibitem
\bibitem{Bauer:2004ve}
C.~W.\ Bauer, Z.~Ligeti, M.~Luke, A.~V.\ Manohar, and M.~Trott{,}
  \href{http://dx.doi.org/10.1103/PhysRevD.70.094017}{Phys.\ Rev.\ {\bf D70}{,}
  094017}  (2004), \href{http://arxiv.org/abs/hep-ph/0408002}{{\tt
  arXiv:hep-ph/0408002}}\relax
\mciteBstWouldAddEndPuncttrue
\mciteSetBstMidEndSepPunct{\mcitedefaultmidpunct}
{\mcitedefaultendpunct}{\mcitedefaultseppunct}\relax
\EndOfBibitem
\bibitem{Aubert:2009qda}
B.~Aubert {\em et al.} ({\babar} collaboration){,}
  \href{http://dx.doi.org/10.1103/PhysRevD.81.032003}{Phys.\ Rev.\ {\bf D81}{,}
  032003}  (2010), \href{http://arxiv.org/abs/0908.0415}{{\tt arXiv:0908.0415
  [hep-ex]}}\relax
\mciteBstWouldAddEndPuncttrue
\mciteSetBstMidEndSepPunct{\mcitedefaultmidpunct}
{\mcitedefaultendpunct}{\mcitedefaultseppunct}\relax
\EndOfBibitem
\bibitem{Aubert:2004td}
B.~Aubert {\em et al.} ({\babar} collaboration){,}
  \href{http://dx.doi.org/10.1103/PhysRevD.69.111104}{Phys.\ Rev.\ {\bf D69}{,}
  111104}  (2004), \href{http://arxiv.org/abs/hep-ex/0403030}{{\tt
  arXiv:hep-ex/0403030}}\relax
\mciteBstWouldAddEndPuncttrue
\mciteSetBstMidEndSepPunct{\mcitedefaultmidpunct}
{\mcitedefaultendpunct}{\mcitedefaultseppunct}\relax
\EndOfBibitem
\bibitem{Schwanda:2006nf}
C.~Schwanda {\em et al.} ({Belle} collaboration){,}
  \href{http://dx.doi.org/10.1103/PhysRevD.75.032005}{Phys.\ Rev.\ {\bf D75}{,}
  032005}  (2007), \href{http://arxiv.org/abs/hep-ex/0611044}{{\tt
  arXiv:hep-ex/0611044}}\relax
\mciteBstWouldAddEndPuncttrue
\mciteSetBstMidEndSepPunct{\mcitedefaultmidpunct}
{\mcitedefaultendpunct}{\mcitedefaultseppunct}\relax
\EndOfBibitem
\bibitem{Urquijo:2006wd}
P.~Urquijo {\em et al.} ({Belle} collaboration){,}
  \href{http://dx.doi.org/10.1103/PhysRevD.75.032001}{Phys.\ Rev.\ {\bf D75}{,}
  032001}  (2007), \href{http://arxiv.org/abs/hep-ex/0610012}{{\tt
  arXiv:hep-ex/0610012}}\relax
\mciteBstWouldAddEndPuncttrue
\mciteSetBstMidEndSepPunct{\mcitedefaultmidpunct}
{\mcitedefaultendpunct}{\mcitedefaultseppunct}\relax
\EndOfBibitem
\bibitem{Acosta:2005qh}
D.~E.\ Acosta {\em et al.} ({CDF} collaboration){,}
  \href{http://dx.doi.org/10.1103/PhysRevD.71.051103}{Phys.\ Rev.\ {\bf D71}{,}
  051103}  (2005), \href{http://arxiv.org/abs/hep-ex/0502003}{{\tt
  arXiv:hep-ex/0502003}}\relax
\mciteBstWouldAddEndPuncttrue
\mciteSetBstMidEndSepPunct{\mcitedefaultmidpunct}
{\mcitedefaultendpunct}{\mcitedefaultseppunct}\relax
\EndOfBibitem
\bibitem{Csorna:2004kp}
S.~E.\ Csorna {\em et al.} ({CLEO} collaboration){,}
  \href{http://dx.doi.org/10.1103/PhysRevD.70.032002}{Phys.\ Rev.\ {\bf D70}{,}
  032002}  (2004), \href{http://arxiv.org/abs/hep-ex/0403052}{{\tt
  arXiv:hep-ex/0403052}}\relax
\mciteBstWouldAddEndPuncttrue
\mciteSetBstMidEndSepPunct{\mcitedefaultmidpunct}
{\mcitedefaultendpunct}{\mcitedefaultseppunct}\relax
\EndOfBibitem
\bibitem{Chetyrkin:2009fv}
K.~Chetyrkin {\em et al.}{,}
  \href{http://dx.doi.org/10.1103/PhysRevD.80.074010}{Phys.\ Rev.\ {\bf D80}{,}
  074010}  (2009), \href{http://arxiv.org/abs/0907.2110}{{\tt arXiv:0907.2110
  [hep-ph]}}\relax
\mciteBstWouldAddEndPuncttrue
\mciteSetBstMidEndSepPunct{\mcitedefaultmidpunct}
{\mcitedefaultendpunct}{\mcitedefaultseppunct}\relax
\EndOfBibitem
\bibitem{Aubert:2005cua}
B.~Aubert {\em et al.} ({\babar} collaboration){,}
  \href{http://dx.doi.org/10.1103/PhysRevD.72.052004}{Phys.\ Rev.\ {\bf D72}{,}
  052004}  (2005), \href{http://arxiv.org/abs/hep-ex/0508004}{{\tt
  arXiv:hep-ex/0508004 [hep-ex]}}\relax
\mciteBstWouldAddEndPuncttrue
\mciteSetBstMidEndSepPunct{\mcitedefaultmidpunct}
{\mcitedefaultendpunct}{\mcitedefaultseppunct}\relax
\EndOfBibitem
\bibitem{Aubert:2006gg}
B.~Aubert {\em et al.} ({\babar} collaboration){,}
  \href{http://dx.doi.org/10.1103/PhysRevLett.97.171803}{Phys.\ Rev.\ Lett.\ {\bf
  97},  171803}  (2006), \href{http://arxiv.org/abs/hep-ex/0607071}{{\tt
  arXiv:hep-ex/0607071 [hep-ex]}}\relax
\mciteBstWouldAddEndPuncttrue
\mciteSetBstMidEndSepPunct{\mcitedefaultmidpunct}
{\mcitedefaultendpunct}{\mcitedefaultseppunct}\relax
\EndOfBibitem
\bibitem{Limosani:2009qg}
A.~Limosani {\em et al.} ({Belle} collaboration){,}
  \href{http://dx.doi.org/10.1103/PhysRevLett.103.241801}{Phys.\ Rev.\ Lett.\ {\bf
  103},  241801}  (2009), \href{http://arxiv.org/abs/0907.1384}{{\tt
  arXiv:0907.1384 [hep-ex]}}\relax
\mciteBstWouldAddEndPuncttrue
\mciteSetBstMidEndSepPunct{\mcitedefaultmidpunct}
{\mcitedefaultendpunct}{\mcitedefaultseppunct}\relax
\EndOfBibitem
\bibitem{Chen:2001fja}
S.~Chen {\em et al.} ({CLEO} collaboration){,}
  \href{http://dx.doi.org/10.1103/PhysRevLett.87.251807}{Phys.\ Rev.\ Lett.\ {\bf
  87},  251807}  (2001), \href{http://arxiv.org/abs/hep-ex/0108032}{{\tt
  arXiv:hep-ex/0108032 [hep-ex]}}\relax
\mciteBstWouldAddEndPuncttrue
\mciteSetBstMidEndSepPunct{\mcitedefaultmidpunct}
{\mcitedefaultendpunct}{\mcitedefaultseppunct}\relax
\EndOfBibitem
\bibitem{Gambino:2013rza}
P.~Gambino and C.~Schwanda{,}
  \href{http://dx.doi.org/10.1103/PhysRevD.89.014022}{Phys.\ Rev.\ {\bf D89}{,}
  014022}  (2014), \href{http://arxiv.org/abs/1307.4551}{{\tt arXiv:1307.4551
  [hep-ph]}}\relax
\mciteBstWouldAddEndPuncttrue
\mciteSetBstMidEndSepPunct{\mcitedefaultmidpunct}
{\mcitedefaultendpunct}{\mcitedefaultseppunct}\relax
\EndOfBibitem
\bibitem{Schwanda:2008kw}
C.~Schwanda {\em et al.} ({Belle} collaboration){,}
  \href{http://dx.doi.org/10.1103/PhysRevD.78.032016}{Phys.\ Rev.\ {\bf D78}{,}
  032016}  (2008), \href{http://arxiv.org/abs/0803.2158}{{\tt arXiv:0803.2158
  [hep-ex]}}\relax
\mciteBstWouldAddEndPuncttrue
\mciteSetBstMidEndSepPunct{\mcitedefaultmidpunct}
{\mcitedefaultendpunct}{\mcitedefaultseppunct}\relax
\EndOfBibitem
\bibitem{Aaij:2015bfa}
R.~Aaij {\em et al.} ({LHCb} collaboration){,}
  \href{http://dx.doi.org/10.1038/nphys3415}{Nature Phys.\ {\bf 11},  743}
  (2015), \href{http://arxiv.org/abs/1504.01568}{{\tt arXiv:1504.01568
  [hep-ex]}}\relax
\mciteBstWouldAddEndPuncttrue
\mciteSetBstMidEndSepPunct{\mcitedefaultmidpunct}
{\mcitedefaultendpunct}{\mcitedefaultseppunct}\relax
\EndOfBibitem
\bibitem{Ha:2010rf}
H.~Ha {\em et al.} ({Belle} collaboration){,}
  \href{http://dx.doi.org/10.1103/PhysRevD.83.071101}{Phys.\ Rev.\ {\bf D83}{,}
  071101}  (2011), \href{http://arxiv.org/abs/1012.0090}{{\tt arXiv:1012.0090
  [hep-ex]}}\relax
\mciteBstWouldAddEndPuncttrue
\mciteSetBstMidEndSepPunct{\mcitedefaultmidpunct}
{\mcitedefaultendpunct}{\mcitedefaultseppunct}\relax
\EndOfBibitem
\bibitem{Sibidanov:2013rkk}
A.~Sibidanov {\em et al.} ({Belle} collaboration){,}
  \href{http://dx.doi.org/10.1103/PhysRevD.88.032005}{Phys.\ Rev.\ {\bf D88}{,}
  032005}  (2013), \href{http://arxiv.org/abs/1306.2781}{{\tt arXiv:1306.2781
  [hep-ex]}}\relax
\mciteBstWouldAddEndPuncttrue
\mciteSetBstMidEndSepPunct{\mcitedefaultmidpunct}
{\mcitedefaultendpunct}{\mcitedefaultseppunct}\relax
\EndOfBibitem
\bibitem{delAmoSanchez:2010af}
P.~del Amo~Sanchez {\em et al.} ({\babar} collaboration){,}
  \href{http://dx.doi.org/10.1103/PhysRevD.83.032007}{Phys.\ Rev.\ {\bf D83}{,}
  032007}  (2011), \href{http://arxiv.org/abs/1005.3288}{{\tt arXiv:1005.3288
  [hep-ex]}}\relax
\mciteBstWouldAddEndPuncttrue
\mciteSetBstMidEndSepPunct{\mcitedefaultmidpunct}
{\mcitedefaultendpunct}{\mcitedefaultseppunct}\relax
\EndOfBibitem
\bibitem{Lees:2012vv}
J.~P.\ Lees {\em et al.} ({\babar} collaboration){,}
  \href{http://dx.doi.org/10.1103/PhysRevD.86.092004}{Phys.\ Rev.\ {\bf D86}{,}
  092004}  (2012), \href{http://arxiv.org/abs/1208.1253}{{\tt arXiv:1208.1253
  [hep-ex]}}\relax
\mciteBstWouldAddEndPuncttrue
\mciteSetBstMidEndSepPunct{\mcitedefaultmidpunct}
{\mcitedefaultendpunct}{\mcitedefaultseppunct}\relax
\EndOfBibitem
\bibitem{Bourrely:2008za}
C.~Bourrely, I.~Caprini, and L.~Lellouch{,}
  \href{http://dx.doi.org/10.1103/PhysRevD.79.013008}{Phys.\ Rev.\ {\bf D79}{,}
  013008}  (2009), \href{http://arxiv.org/abs/0807.2722}{{\tt arXiv:0807.2722
  [hep-ph]}}, Erratum ibid.\
  \href{http://dx.doi.org/10.1103/PhysRevD.82.099902}{{\bf D82}, 099902}
  (2010)\relax
\mciteBstWouldAddEndPuncttrue
\mciteSetBstMidEndSepPunct{\mcitedefaultmidpunct}
{\mcitedefaultendpunct}{\mcitedefaultseppunct}\relax
\EndOfBibitem
\bibitem{Lattice:2015tia}
J.~A.\ Bailey {\em et al.} ({Fermilab Lattice and MILC} collaborations){,}
  \href{http://dx.doi.org/10.1103/PhysRevD.92.014024}{Phys.\ Rev.\ {\bf D92}{,}
  014024}  (2015), \href{http://arxiv.org/abs/1503.07839}{{\tt
  arXiv:1503.07839 [hep-lat]}}\relax
\mciteBstWouldAddEndPuncttrue
\mciteSetBstMidEndSepPunct{\mcitedefaultmidpunct}
{\mcitedefaultendpunct}{\mcitedefaultseppunct}\relax
\EndOfBibitem
\bibitem{Flynn:2015mha}
J.~M.\ Flynn, T.~Izubuchi, T.~Kawanai, C.~Lehner, A.~Soni, R.~S.\ Van~de Water{,}
  and O.~Witzel ({RBC and UKQCD} collaborations){,}
  \href{http://dx.doi.org/10.1103/PhysRevD.91.074510}{Phys.\ Rev.\ {\bf D91}{,}
  074510}  (2015), \href{http://arxiv.org/abs/1501.05373}{{\tt
  arXiv:1501.05373 [hep-lat]}}\relax
\mciteBstWouldAddEndPuncttrue
\mciteSetBstMidEndSepPunct{\mcitedefaultmidpunct}
{\mcitedefaultendpunct}{\mcitedefaultseppunct}\relax
\EndOfBibitem
\bibitem{Bharucha:2012wy}
A.~Bharucha, \href{http://dx.doi.org/10.1007/JHEP05(2012)092}{JHEP {\bf 05}{,}
  092}  (2012), \href{http://arxiv.org/abs/1203.1359}{{\tt arXiv:1203.1359
  [hep-ph]}}\relax
\mciteBstWouldAddEndPuncttrue
\mciteSetBstMidEndSepPunct{\mcitedefaultmidpunct}
{\mcitedefaultendpunct}{\mcitedefaultseppunct}\relax
\EndOfBibitem
\bibitem{Ablikim:2015flg}
M.~Ablikim {\em et al.} ({BESIII} collaboration){,}
  \href{http://dx.doi.org/10.1103/PhysRevLett.116.052001}{Phys.\ Rev.\ Lett.\ {\bf
  116},  052001}  (2016), \href{http://arxiv.org/abs/1511.08380}{{\tt
  arXiv:1511.08380 [hep-ex]}}\relax
\mciteBstWouldAddEndPuncttrue
\mciteSetBstMidEndSepPunct{\mcitedefaultmidpunct}
{\mcitedefaultendpunct}{\mcitedefaultseppunct}\relax
\EndOfBibitem
\bibitem{Detmold:2015aaa}
W.~Detmold, C.~Lehner, and S.~Meinel{,}
  \href{http://dx.doi.org/10.1103/PhysRevD.92.034503}{Phys.\ Rev.\ {\bf D92}{,}
  034503}  (2015), \href{http://arxiv.org/abs/1503.01421}{{\tt
  arXiv:1503.01421 [hep-lat]}}\relax
\mciteBstWouldAddEndPuncttrue
\mciteSetBstMidEndSepPunct{\mcitedefaultmidpunct}
{\mcitedefaultendpunct}{\mcitedefaultseppunct}\relax
\EndOfBibitem
\bibitem{Faustov:2016pal}
R.~N.\ Faustov and V.~O.\ Galkin{,}
  \href{http://dx.doi.org/10.1103/PhysRevD.94.073008}{Phys.\ Rev.\ {\bf D94}{,}
  073008}  (2016), \href{http://arxiv.org/abs/1609.00199}{{\tt
  arXiv:1609.00199 [hep-ph]}}\relax
\mciteBstWouldAddEndPuncttrue
\mciteSetBstMidEndSepPunct{\mcitedefaultmidpunct}
{\mcitedefaultendpunct}{\mcitedefaultseppunct}\relax
\EndOfBibitem
\bibitem{Ball:2004ye}
P.~Ball and R.~Zwicky{,}
  \href{http://dx.doi.org/10.1103/PhysRevD.71.014015}{Phys.\ Rev.\ {\bf D71}{,}
  014015}  (2005), \href{http://arxiv.org/abs/hep-ph/0406232}{{\tt
  arXiv:hep-ph/0406232}}\relax
\mciteBstWouldAddEndPuncttrue
\mciteSetBstMidEndSepPunct{\mcitedefaultmidpunct}
{\mcitedefaultendpunct}{\mcitedefaultseppunct}\relax
\EndOfBibitem
\bibitem{Straub:2015ica}
A.~Bharucha, D.~M.\ Straub, and R.~Zwicky{,}
  \href{http://dx.doi.org/10.1007/JHEP08(2016)098}{JHEP {\bf 08},  098}
  (2016), \href{http://arxiv.org/abs/1503.05534}{{\tt arXiv:1503.05534
  [hep-ph]}}\relax
\mciteBstWouldAddEndPuncttrue
\mciteSetBstMidEndSepPunct{\mcitedefaultmidpunct}
{\mcitedefaultendpunct}{\mcitedefaultseppunct}\relax
\EndOfBibitem
\bibitem{Behrens:1999vv}
B.~H.\ Behrens {\em et al.} ({CLEO} collaboration){,}
  \href{http://dx.doi.org/10.1103/PhysRevD.61.052001}{Phys.\ Rev.\ {\bf D61}{,}
  052001}  (2000), \href{http://arxiv.org/abs/hep-ex/9905056}{{\tt
  arXiv:hep-ex/9905056}}\relax
\mciteBstWouldAddEndPuncttrue
\mciteSetBstMidEndSepPunct{\mcitedefaultmidpunct}
{\mcitedefaultendpunct}{\mcitedefaultseppunct}\relax
\EndOfBibitem
\bibitem{Adam:2007pv}
N.~E.\ Adam {\em et al.} ({CLEO} collaboration){,}
  \href{http://dx.doi.org/10.1103/PhysRevLett.99.041802}{Phys.\ Rev.\ Lett.\ {\bf
  99},  041802}  (2007), \href{http://arxiv.org/abs/hep-ex/0703041}{{\tt
  arXiv:hep-ex/0703041}}\relax
\mciteBstWouldAddEndPuncttrue
\mciteSetBstMidEndSepPunct{\mcitedefaultmidpunct}
{\mcitedefaultendpunct}{\mcitedefaultseppunct}\relax
\EndOfBibitem
\bibitem{Hokuue:2006nr}
T.~Hokuue {\em et al.} ({Belle} collaboration){,}
  \href{http://dx.doi.org/10.1016/j.physletb.2007.02.067}{Phys.\ Lett.\ {\bf
  B648},  139}  (2007), \href{http://arxiv.org/abs/hep-ex/0604024}{{\tt
  arXiv:hep-ex/0604024}}\relax
\mciteBstWouldAddEndPuncttrue
\mciteSetBstMidEndSepPunct{\mcitedefaultmidpunct}
{\mcitedefaultendpunct}{\mcitedefaultseppunct}\relax
\EndOfBibitem
\bibitem{Schwanda:2004fa}
C.~Schwanda {\em et al.} ({Belle} collaboration){,}
  \href{http://dx.doi.org/10.1103/PhysRevLett.93.131803}{Phys.\ Rev.\ Lett.\ {\bf
  93},  131803}  (2004), \href{http://arxiv.org/abs/hep-ex/0402023}{{\tt
  arXiv:hep-ex/0402023 [hep-ex]}}\relax
\mciteBstWouldAddEndPuncttrue
\mciteSetBstMidEndSepPunct{\mcitedefaultmidpunct}
{\mcitedefaultendpunct}{\mcitedefaultseppunct}\relax
\EndOfBibitem
\bibitem{Lees:2012mq}
J.~P.\ Lees {\em et al.} ({\babar} collaboration){,}
  \href{http://dx.doi.org/10.1103/PhysRevD.87.032004}{Phys.\ Rev.\ {\bf D87}{,}
  032004}  (2013), \href{http://arxiv.org/abs/1205.6245}{{\tt arXiv:1205.6245
  [hep-ex]}}\relax
\mciteBstWouldAddEndPuncttrue
\mciteSetBstMidEndSepPunct{\mcitedefaultmidpunct}
{\mcitedefaultendpunct}{\mcitedefaultseppunct}\relax
\EndOfBibitem
\bibitem{Lees:2013gja}
J.~P.\ Lees {\em et al.} ({\babar} collaboration){,}
  \href{http://dx.doi.org/10.1103/PhysRevD.88.072006}{Phys.\ Rev.\ {\bf D88}{,}
  072006}  (2013), \href{http://arxiv.org/abs/1308.2589}{{\tt arXiv:1308.2589
  [hep-ex]}}\relax
\mciteBstWouldAddEndPuncttrue
\mciteSetBstMidEndSepPunct{\mcitedefaultmidpunct}
{\mcitedefaultendpunct}{\mcitedefaultseppunct}\relax
\EndOfBibitem
\bibitem{Gray:2007pw}
R.~Gray {\em et al.} ({CLEO} collaboration){,}
  \href{http://dx.doi.org/10.1103/PhysRevD.76.012007}{Phys.\ Rev.\ {\bf D76}{,}
  012007}  (2007), \href{http://arxiv.org/abs/hep-ex/0703042}{{\tt
  arXiv:hep-ex/0703042}}\relax
\mciteBstWouldAddEndPuncttrue
\mciteSetBstMidEndSepPunct{\mcitedefaultmidpunct}
{\mcitedefaultendpunct}{\mcitedefaultseppunct}\relax
\EndOfBibitem
\bibitem{Aubert:2008ct}
B.~Aubert {\em et al.} ({\babar} collaboration){,}
  \href{http://dx.doi.org/10.1103/PhysRevD.79.052011}{Phys.\ Rev.\ {\bf D79}{,}
  052011}  (2009), \href{http://arxiv.org/abs/0808.3524}{{\tt arXiv:0808.3524
  [hep-ex]}}\relax
\mciteBstWouldAddEndPuncttrue
\mciteSetBstMidEndSepPunct{\mcitedefaultmidpunct}
{\mcitedefaultendpunct}{\mcitedefaultseppunct}\relax
\EndOfBibitem
\bibitem{Aubert:2008bf}
B.~Aubert {\em et al.} ({\babar} collaboration){,}
  \href{http://dx.doi.org/10.1103/PhysRevLett.101.081801}{Phys.\ Rev.\ Lett.\ {\bf
  101},  081801}  (2008), \href{http://arxiv.org/abs/0805.2408}{{\tt
  arXiv:0805.2408 [hep-ex]}}\relax
\mciteBstWouldAddEndPuncttrue
\mciteSetBstMidEndSepPunct{\mcitedefaultmidpunct}
{\mcitedefaultendpunct}{\mcitedefaultseppunct}\relax
\EndOfBibitem
\bibitem{ref:belle-multivariate}
P.~Urquijo {\em et al.} ({Belle} collaboration){,}
  \href{http://dx.doi.org/10.1103/PhysRevLett.104.021801}{Phys.\ Rev.\ Lett.\ {\bf
  104},  021801}  (2010), \href{http://arxiv.org/abs/0907.0379}{{\tt
  arXiv:0907.0379 [hep-ex]}}\relax
\mciteBstWouldAddEndPuncttrue
\mciteSetBstMidEndSepPunct{\mcitedefaultmidpunct}
{\mcitedefaultendpunct}{\mcitedefaultseppunct}\relax
\EndOfBibitem
\bibitem{Lees:2011fv}
J.~P.\ Lees {\em et al.} ({\babar} collaboration){,}
  \href{http://dx.doi.org/10.1103/PhysRevD.86.032004}{Phys.\ Rev.\ {\bf D86}{,}
  032004}  (2012), \href{http://arxiv.org/abs/1112.0702}{{\tt arXiv:1112.0702
  [hep-ex]}}\relax
\mciteBstWouldAddEndPuncttrue
\mciteSetBstMidEndSepPunct{\mcitedefaultmidpunct}
{\mcitedefaultendpunct}{\mcitedefaultseppunct}\relax
\EndOfBibitem
\bibitem{ref:BLL}
C.~W.\ Bauer, Z.~Ligeti, and M.~E.\ Luke{,}
  \href{http://dx.doi.org/10.1103/PhysRevD.64.113004}{Phys.\ Rev.\ {\bf D64}{,}
  113004}  (2001), \href{http://arxiv.org/abs/hep-ph/0107074}{{\tt
  arXiv:hep-ph/0107074}}\relax
\mciteBstWouldAddEndPuncttrue
\mciteSetBstMidEndSepPunct{\mcitedefaultmidpunct}
{\mcitedefaultendpunct}{\mcitedefaultseppunct}\relax
\EndOfBibitem
\bibitem{Neubert:1993um}
M.~Neubert, \href{http://dx.doi.org/10.1103/PhysRevD.49.4623}{Phys.\ Rev.\ {\bf
  D49},  4623}  (1994), \href{http://arxiv.org/abs/hep-ph/9312311}{{\tt
  arXiv:hep-ph/9312311}}\relax
\mciteBstWouldAddEndPuncttrue
\mciteSetBstMidEndSepPunct{\mcitedefaultmidpunct}
{\mcitedefaultendpunct}{\mcitedefaultseppunct}\relax
\EndOfBibitem
\bibitem{Leibovich:1999xf}
A.~K.\ Leibovich, I.~Low, and I.~Z.\ Rothstein{,}
  \href{http://dx.doi.org/10.1103/PhysRevD.61.053006}{Phys.\ Rev.\ {\bf D61}{,}
  053006}  (2000), \href{http://arxiv.org/abs/hep-ph/9909404}{{\tt
  arXiv:hep-ph/9909404}}\relax
\mciteBstWouldAddEndPuncttrue
\mciteSetBstMidEndSepPunct{\mcitedefaultmidpunct}
{\mcitedefaultendpunct}{\mcitedefaultseppunct}\relax
\EndOfBibitem
\bibitem{Lange:2005qn}
B.~O.\ Lange, M.~Neubert, and G.~Paz{,}
  \href{http://dx.doi.org/10.1088/1126-6708/2005/10/084}{JHEP {\bf 10},  084}
  (2005), \href{http://arxiv.org/abs/hep-ph/0508178}{{\tt
  arXiv:hep-ph/0508178}}\relax
\mciteBstWouldAddEndPuncttrue
\mciteSetBstMidEndSepPunct{\mcitedefaultmidpunct}
{\mcitedefaultendpunct}{\mcitedefaultseppunct}\relax
\EndOfBibitem
\bibitem{TheBABAR:2016lja}
J.~P.\ Lees {\em et al.} ({\babar} collaboration){,}
  \href{http://dx.doi.org/10.1103/PhysRevD.95.072001}{Phys.\ Rev.\ {\bf D95}{,}
  072001}  (2017), \href{http://arxiv.org/abs/1611.05624}{{\tt
  arXiv:1611.05624 [hep-ex]}}\relax
\mciteBstWouldAddEndPuncttrue
\mciteSetBstMidEndSepPunct{\mcitedefaultmidpunct}
{\mcitedefaultendpunct}{\mcitedefaultseppunct}\relax
\EndOfBibitem
\bibitem{ref:shmax}
R.~V.\ Kowalewski and S.~Menke{,}
  \href{http://dx.doi.org/10.1016/S0370-2693(02)02181-0}{Phys.\ Lett.\ {\bf
  B541},  29}  (2002), \href{http://arxiv.org/abs/hep-ex/0205038}{{\tt
  arXiv:hep-ex/0205038}}\relax
\mciteBstWouldAddEndPuncttrue
\mciteSetBstMidEndSepPunct{\mcitedefaultmidpunct}
{\mcitedefaultendpunct}{\mcitedefaultseppunct}\relax
\EndOfBibitem
\bibitem{ref:babar-elq2}
B.~Aubert {\em et al.} ({\babar} collaboration){,}
  \href{http://dx.doi.org/10.1103/PhysRevLett.95.111801}{Phys.\ Rev.\ Lett.\ {\bf
  95},  111801}  (2005), \href{http://arxiv.org/abs/hep-ex/0506036}{{\tt
  arXiv:hep-ex/0506036}}\relax
\mciteBstWouldAddEndPuncttrue
\mciteSetBstMidEndSepPunct{\mcitedefaultmidpunct}
{\mcitedefaultendpunct}{\mcitedefaultseppunct}\relax
\EndOfBibitem
\bibitem{ref:cleo-endpoint}
A.~Bornheim {\em et al.} ({CLEO} collaboration){,}
  \href{http://dx.doi.org/10.1103/PhysRevLett.88.231803}{Phys.\ Rev.\ Lett.\ {\bf
  88},  231803}  (2002), \href{http://arxiv.org/abs/hep-ex/0202019}{{\tt
  arXiv:hep-ex/0202019}}\relax
\mciteBstWouldAddEndPuncttrue
\mciteSetBstMidEndSepPunct{\mcitedefaultmidpunct}
{\mcitedefaultendpunct}{\mcitedefaultseppunct}\relax
\EndOfBibitem
\bibitem{ref:babar-endpoint}
B.~Aubert {\em et al.} ({\babar} collaboration){,}
  \href{http://dx.doi.org/10.1103/PhysRevD.73.012006}{Phys.\ Rev.\ {\bf D73}{,}
  012006}  (2006), \href{http://arxiv.org/abs/hep-ex/0509040}{{\tt
  arXiv:hep-ex/0509040}}\relax
\mciteBstWouldAddEndPuncttrue
\mciteSetBstMidEndSepPunct{\mcitedefaultmidpunct}
{\mcitedefaultendpunct}{\mcitedefaultseppunct}\relax
\EndOfBibitem
\bibitem{ref:belle-endpoint}
A.~Limosani {\em et al.} ({Belle} collaboration){,}
  \href{http://dx.doi.org/10.1016/j.physletb.2005.06.011}{Phys.\ Lett.\ {\bf
  B621},  28}  (2005), \href{http://arxiv.org/abs/hep-ex/0504046}{{\tt
  arXiv:hep-ex/0504046}}\relax
\mciteBstWouldAddEndPuncttrue
\mciteSetBstMidEndSepPunct{\mcitedefaultmidpunct}
{\mcitedefaultendpunct}{\mcitedefaultseppunct}\relax
\EndOfBibitem
\bibitem{ref:belle-mxq2Anneal}
H.~Kakuno {\em et al.} ({Belle} collaboration){,}
  \href{http://dx.doi.org/10.1103/PhysRevLett.92.101801}{Phys.\ Rev.\ Lett.\ {\bf
  92},  101801}  (2004), \href{http://arxiv.org/abs/hep-ex/0311048}{{\tt
  arXiv:hep-ex/0311048}}\relax
\mciteBstWouldAddEndPuncttrue
\mciteSetBstMidEndSepPunct{\mcitedefaultmidpunct}
{\mcitedefaultendpunct}{\mcitedefaultseppunct}\relax
\EndOfBibitem
\bibitem{ref:belle-mx}
I.~Bizjak {\em et al.} ({Belle} collaboration){,}
  \href{http://dx.doi.org/10.1103/PhysRevLett.95.241801}{Phys.\ Rev.\ Lett.\ {\bf
  95},  241801}  (2005), \href{http://arxiv.org/abs/hep-ex/0505088}{{\tt
  arXiv:hep-ex/0505088}}\relax
\mciteBstWouldAddEndPuncttrue
\mciteSetBstMidEndSepPunct{\mcitedefaultmidpunct}
{\mcitedefaultendpunct}{\mcitedefaultseppunct}\relax
\EndOfBibitem
\bibitem{ref:BLNP}
B.~O.\ Lange, M.~Neubert, and G.~Paz{,}
  \href{http://dx.doi.org/10.1103/PhysRevD.72.073006}{Phys.\ Rev.\ {\bf D72}{,}
  073006}  (2005), \href{http://arxiv.org/abs/hep-ph/0504071}{{\tt
  arXiv:hep-ph/0504071}}\relax
\mciteBstWouldAddEndPuncttrue
\mciteSetBstMidEndSepPunct{\mcitedefaultmidpunct}
{\mcitedefaultendpunct}{\mcitedefaultseppunct}\relax
\EndOfBibitem
\bibitem{ref:Neubert-new-1}
S.~W.\ Bosch, B.~O.\ Lange, M.~Neubert, and G.~Paz{,}
  \href{http://dx.doi.org/10.1016/j.nuclphysb.2004.07.041}{Nucl.\ Phys.\ {\bf
  B699},  335}  (2004), \href{http://arxiv.org/abs/hep-ph/0402094}{{\tt
  arXiv:hep-ph/0402094}}\relax
\mciteBstWouldAddEndPuncttrue
\mciteSetBstMidEndSepPunct{\mcitedefaultmidpunct}
{\mcitedefaultendpunct}{\mcitedefaultseppunct}\relax
\EndOfBibitem
\bibitem{ref:Neubert-new-2}
S.~W.\ Bosch, M.~Neubert, and G.~Paz{,}
  \href{http://dx.doi.org/10.1088/1126-6708/2004/11/073}{JHEP {\bf 11},  073}
  (2004), \href{http://arxiv.org/abs/hep-ph/0409115}{{\tt
  arXiv:hep-ph/0409115}}\relax
\mciteBstWouldAddEndPuncttrue
\mciteSetBstMidEndSepPunct{\mcitedefaultmidpunct}
{\mcitedefaultendpunct}{\mcitedefaultseppunct}\relax
\EndOfBibitem
\bibitem{ref:Neubert-new-3}
M.~Neubert, \href{http://dx.doi.org/10.1140/epjc/s2005-02360-4}{Eur.\ Phys.\ J.\
  {\bf C44},  205}  (2005), \href{http://arxiv.org/abs/hep-ph/0411027}{{\tt
  arXiv:hep-ph/0411027}}\relax
\mciteBstWouldAddEndPuncttrue
\mciteSetBstMidEndSepPunct{\mcitedefaultmidpunct}
{\mcitedefaultendpunct}{\mcitedefaultseppunct}\relax
\EndOfBibitem
\bibitem{Neubert:2004sp}
M.~Neubert, \href{http://dx.doi.org/10.1016/j.physletb.2005.02.055}{Phys.\ Lett.\
  {\bf B612},  13}  (2005), \href{http://arxiv.org/abs/hep-ph/0412241}{{\tt
  arXiv:hep-ph/0412241 [hep-ph]}}\relax
\mciteBstWouldAddEndPuncttrue
\mciteSetBstMidEndSepPunct{\mcitedefaultmidpunct}
{\mcitedefaultendpunct}{\mcitedefaultseppunct}\relax
\EndOfBibitem
\bibitem{Neubert:2005nt}
M.~Neubert, \href{http://dx.doi.org/10.1103/PhysRevD.72.074025}{Phys.\ Rev.\ {\bf
  D72},  074025}  (2005), \href{http://arxiv.org/abs/hep-ph/0506245}{{\tt
  arXiv:hep-ph/0506245 [hep-ph]}}\relax
\mciteBstWouldAddEndPuncttrue
\mciteSetBstMidEndSepPunct{\mcitedefaultmidpunct}
{\mcitedefaultendpunct}{\mcitedefaultseppunct}\relax
\EndOfBibitem
\bibitem{ref:DGE}
J.~R.\ Andersen and E.~Gardi{,}
  \href{http://dx.doi.org/10.1088/1126-6708/2006/01/097}{JHEP {\bf 01},  097}
  (2006), \href{http://arxiv.org/abs/hep-ph/0509360}{{\tt
  arXiv:hep-ph/0509360}}\relax
\mciteBstWouldAddEndPuncttrue
\mciteSetBstMidEndSepPunct{\mcitedefaultmidpunct}
{\mcitedefaultendpunct}{\mcitedefaultseppunct}\relax
\EndOfBibitem
\bibitem{Aglietti:2006yb}
U.~Aglietti, G.~Ferrera, and G.~Ricciardi{,}
  \href{http://dx.doi.org/10.1016/j.nuclphysb.2007.01.014}{Nucl.\ Phys.\ {\bf
  B768},  85}  (2007), \href{http://arxiv.org/abs/hep-ph/0608047}{{\tt
  arXiv:hep-ph/0608047}}\relax
\mciteBstWouldAddEndPuncttrue
\mciteSetBstMidEndSepPunct{\mcitedefaultmidpunct}
{\mcitedefaultendpunct}{\mcitedefaultseppunct}\relax
\EndOfBibitem
\bibitem{Gambino:2007rp}
P.~Gambino, P.~Giordano, G.~Ossola, and N.~Uraltsev{,}
  \href{http://dx.doi.org/10.1088/1126-6708/2007/10/058}{JHEP {\bf 10},  058}
  (2007), \href{http://arxiv.org/abs/0707.2493}{{\tt arXiv:0707.2493
  [hep-ph]}}\relax
\mciteBstWouldAddEndPuncttrue
\mciteSetBstMidEndSepPunct{\mcitedefaultmidpunct}
{\mcitedefaultendpunct}{\mcitedefaultseppunct}\relax
\EndOfBibitem
\bibitem{Aglietti:2007ik}
U.~Aglietti, F.~Di~Lodovico, G.~Ferrera, and G.~Ricciardi{,}
  \href{http://dx.doi.org/10.1140/epjc/s10052-008-0817-x}{Eur.\ Phys.\ J.\ {\bf
  C59},  831}  (2009), \href{http://arxiv.org/abs/0711.0860}{{\tt
  arXiv:0711.0860 [hep-ph]}}\relax
\mciteBstWouldAddEndPuncttrue
\mciteSetBstMidEndSepPunct{\mcitedefaultmidpunct}
{\mcitedefaultendpunct}{\mcitedefaultseppunct}\relax
\EndOfBibitem
\bibitem{Aglietti:2005mb}
U.~Aglietti, G.~Ricciardi, and G.~Ferrera{,}
  \href{http://dx.doi.org/10.1103/PhysRevD.74.034004}{Phys.\ Rev.\ {\bf D74}{,}
  034004}  (2006), \href{http://arxiv.org/abs/hep-ph/0507285}{{\tt
  arXiv:hep-ph/0507285}}\relax
\mciteBstWouldAddEndPuncttrue
\mciteSetBstMidEndSepPunct{\mcitedefaultmidpunct}
{\mcitedefaultendpunct}{\mcitedefaultseppunct}\relax
\EndOfBibitem
\bibitem{Aglietti:2005bm}
U.~Aglietti, G.~Ricciardi, and G.~Ferrera{,}
  \href{http://dx.doi.org/10.1103/PhysRevD.74.034005}{Phys.\ Rev.\ {\bf D74}{,}
  034005}  (2006), \href{http://arxiv.org/abs/hep-ph/0509095}{{\tt
  arXiv:hep-ph/0509095}}\relax
\mciteBstWouldAddEndPuncttrue
\mciteSetBstMidEndSepPunct{\mcitedefaultmidpunct}
{\mcitedefaultendpunct}{\mcitedefaultseppunct}\relax
\EndOfBibitem
\bibitem{Aglietti:2005eq}
U.~Aglietti, G.~Ricciardi, and G.~Ferrera{,}
  \href{http://dx.doi.org/10.1103/PhysRevD.74.034006}{Phys.\ Rev.\ {\bf D74}{,}
  034006}  (2006), \href{http://arxiv.org/abs/hep-ph/0509271}{{\tt
  arXiv:hep-ph/0509271}}\relax
\mciteBstWouldAddEndPuncttrue
\mciteSetBstMidEndSepPunct{\mcitedefaultmidpunct}
{\mcitedefaultendpunct}{\mcitedefaultseppunct}\relax
\EndOfBibitem
\bibitem{Duraisamy:2014sna}
M.~Duraisamy, P.~Sharma, and A.~Datta{,}
  \href{http://dx.doi.org/10.1103/PhysRevD.90.074013}{Phys.\ Rev.\ {\bf D90}{,}
  074013}  (2014), \href{http://arxiv.org/abs/1405.3719}{{\tt arXiv:1405.3719
  [hep-ph]}}\relax
\mciteBstWouldAddEndPuncttrue
\mciteSetBstMidEndSepPunct{\mcitedefaultmidpunct}
{\mcitedefaultendpunct}{\mcitedefaultseppunct}\relax
\EndOfBibitem
\bibitem{Na:2015kha}
H.~Na, C.~M.\ Bouchard, G.~P.\ Lepage, C.~Monahan, and J.~Shigemitsu ({HPQCD}
  collaboration), \href{http://dx.doi.org/10.1103/PhysRevD.93.119906}{Phys.\
  Rev.\ {\bf D92},  054510}  (2015){,}
  \href{http://arxiv.org/abs/1505.03925}{{\tt arXiv:1505.03925 [hep-lat]}}{,}
  Erratum ibid.\ \href{http://dx.doi.org/10.1103/PhysRevD.92.054510}{{\bf D93}{,}
  119906} (2016)\relax
\mciteBstWouldAddEndPuncttrue
\mciteSetBstMidEndSepPunct{\mcitedefaultmidpunct}
{\mcitedefaultendpunct}{\mcitedefaultseppunct}\relax
\EndOfBibitem
\bibitem{Lees:2012xj}
J.~P.\ Lees {\em et al.} ({\babar} collaboration){,}
  \href{http://dx.doi.org/10.1103/PhysRevLett.109.101802}{Phys.\ Rev.\ Lett.\ {\bf
  109},  101802}  (2012), \href{http://arxiv.org/abs/1205.5442}{{\tt
  arXiv:1205.5442 [hep-ex]}}\relax
\mciteBstWouldAddEndPuncttrue
\mciteSetBstMidEndSepPunct{\mcitedefaultmidpunct}
{\mcitedefaultendpunct}{\mcitedefaultseppunct}\relax
\EndOfBibitem
\bibitem{Lees:2013uzd}
J.~P.\ Lees {\em et al.} ({\babar} collaboration){,}
  \href{http://dx.doi.org/10.1103/PhysRevD.88.072012}{Phys.\ Rev.\ {\bf D88}{,}
  072012}  (2013), \href{http://arxiv.org/abs/1303.0571}{{\tt arXiv:1303.0571
  [hep-ex]}}\relax
\mciteBstWouldAddEndPuncttrue
\mciteSetBstMidEndSepPunct{\mcitedefaultmidpunct}
{\mcitedefaultendpunct}{\mcitedefaultseppunct}\relax
\EndOfBibitem
\bibitem{Kamenik:2008tj}
J.~F.\ Kamenik and F.~Mescia{,}
  \href{http://dx.doi.org/10.1103/PhysRevD.78.014003}{Phys.\ Rev.\ {\bf D78}{,}
  014003}  (2008), \href{http://arxiv.org/abs/0802.3790}{{\tt arXiv:0802.3790
  [hep-ph]}}\relax
\mciteBstWouldAddEndPuncttrue
\mciteSetBstMidEndSepPunct{\mcitedefaultmidpunct}
{\mcitedefaultendpunct}{\mcitedefaultseppunct}\relax
\EndOfBibitem
\bibitem{Fajfer:2012vx}
S.~Fajfer, J.~F.\ Kamenik, and I.~Nisandzic{,}
  \href{http://dx.doi.org/10.1103/PhysRevD.85.094025}{Phys.\ Rev.\ {\bf D85}{,}
  094025}  (2012), \href{http://arxiv.org/abs/1203.2654}{{\tt arXiv:1203.2654
  [hep-ph]}}\relax
\mciteBstWouldAddEndPuncttrue
\mciteSetBstMidEndSepPunct{\mcitedefaultmidpunct}
{\mcitedefaultendpunct}{\mcitedefaultseppunct}\relax
\EndOfBibitem
\bibitem{Bigi:2016mdz}
D.~Bigi and P.~Gambino{,}
  \href{http://dx.doi.org/10.1103/PhysRevD.94.094008}{Phys.\ Rev.\ {\bf D94}{,}
  094008}  (2016), \href{http://arxiv.org/abs/1606.08030}{{\tt
  arXiv:1606.08030 [hep-ph]}}\relax
\mciteBstWouldAddEndPuncttrue
\mciteSetBstMidEndSepPunct{\mcitedefaultmidpunct}
{\mcitedefaultendpunct}{\mcitedefaultseppunct}\relax
\EndOfBibitem
\bibitem{Matyja:2007kt}
A.~Matyja {\em et al.} ({Belle} collaboration){,}
  \href{http://dx.doi.org/10.1103/PhysRevLett.99.191807}{Phys.\ Rev.\ Lett.\ {\bf
  99},  191807}  (2007), \href{http://arxiv.org/abs/0706.4429}{{\tt
  arXiv:0706.4429 [hep-ex]}}\relax
\mciteBstWouldAddEndPuncttrue
\mciteSetBstMidEndSepPunct{\mcitedefaultmidpunct}
{\mcitedefaultendpunct}{\mcitedefaultseppunct}\relax
\EndOfBibitem
\bibitem{Aubert:2007dsa}
B.~Aubert {\em et al.} ({\babar} collaboration){,}
  \href{http://dx.doi.org/10.1103/PhysRevLett.100.021801}{Phys.\ Rev.\ Lett.\ {\bf
  100},  021801}  (2008), \href{http://arxiv.org/abs/0709.1698}{{\tt
  arXiv:0709.1698 [hep-ex]}}\relax
\mciteBstWouldAddEndPuncttrue
\mciteSetBstMidEndSepPunct{\mcitedefaultmidpunct}
{\mcitedefaultendpunct}{\mcitedefaultseppunct}\relax
\EndOfBibitem
\bibitem{Bozek:2010xy}
A.~Bozek {\em et al.} ({Belle} collaboration){,}
  \href{http://dx.doi.org/10.1103/PhysRevD.82.072005}{Phys.\ Rev.\ {\bf D82}{,}
  072005}  (2010), \href{http://arxiv.org/abs/1005.2302}{{\tt arXiv:1005.2302
  [hep-ex]}}\relax
\mciteBstWouldAddEndPuncttrue
\mciteSetBstMidEndSepPunct{\mcitedefaultmidpunct}
{\mcitedefaultendpunct}{\mcitedefaultseppunct}\relax
\EndOfBibitem
\bibitem{Huschle:2015rga}
M.~Huschle {\em et al.} ({Belle} collaboration){,}
  \href{http://dx.doi.org/10.1103/PhysRevD.92.072014}{Phys.\ Rev.\ {\bf D92}{,}
  072014}  (2015), \href{http://arxiv.org/abs/1507.03233}{{\tt
  arXiv:1507.03233 [hep-ex]}}\relax
\mciteBstWouldAddEndPuncttrue
\mciteSetBstMidEndSepPunct{\mcitedefaultmidpunct}
{\mcitedefaultendpunct}{\mcitedefaultseppunct}\relax
\EndOfBibitem
\bibitem{Aaij:2015yra}
R.~Aaij {\em et al.} ({LHCb} collaboration){,}
  \href{http://dx.doi.org/10.1103/PhysRevLett.115.111803}{Phys.\ Rev.\ Lett.\ {\bf
  115},  111803}  (2015), \href{http://arxiv.org/abs/1506.08614}{{\tt
  arXiv:1506.08614 [hep-ex]}}, Addendum ibid.\
  \href{http://dx.doi.org/10.1103/PhysRevLett.115.159901}{{\bf 115}, 159901}
  (2015)\relax
\mciteBstWouldAddEndPuncttrue
\mciteSetBstMidEndSepPunct{\mcitedefaultmidpunct}
{\mcitedefaultendpunct}{\mcitedefaultseppunct}\relax
\EndOfBibitem
\bibitem{Sato:2016svk}
Y.~Sato {\em et al.} ({Belle} collaboration){,}
  \href{http://dx.doi.org/10.1103/PhysRevD.94.072007}{Phys.\ Rev.\ {\bf D94}{,}
  072007}  (2016), \href{http://arxiv.org/abs/1607.07923}{{\tt
  arXiv:1607.07923 [hep-ex]}}\relax
\mciteBstWouldAddEndPuncttrue
\mciteSetBstMidEndSepPunct{\mcitedefaultmidpunct}
{\mcitedefaultendpunct}{\mcitedefaultseppunct}\relax
\EndOfBibitem
\bibitem{Hirose:2016wfn}
S.~Hirose {\em et al.} ({Belle} collaboration){,}
  \href{http://dx.doi.org/10.1103/PhysRevLett.118.211801}{Phys.\ Rev.\ Lett.\ {\bf
  118}, no.~21, 211801}  (2017), \href{http://arxiv.org/abs/1612.00529}{{\tt
  arXiv:1612.00529 [hep-ex]}}\relax
\mciteBstWouldAddEndPuncttrue
\mciteSetBstMidEndSepPunct{\mcitedefaultmidpunct}
{\mcitedefaultendpunct}{\mcitedefaultseppunct}\relax
\EndOfBibitem
\bibitem{Lenz:2014nka}
A.~J.\ Lenz, \href{http://dx.doi.org/10.1088/0954-3899/41/10/103001}{J.\ Phys.\
  {\bf G41},  103001}  (2014), \href{http://arxiv.org/abs/1404.6197}{{\tt
  arXiv:1404.6197 [hep-ph]}}\relax
\mciteBstWouldAddEndPuncttrue
\mciteSetBstMidEndSepPunct{\mcitedefaultmidpunct}
{\mcitedefaultendpunct}{\mcitedefaultseppunct}\relax
\EndOfBibitem
\bibitem{Choi:2003ue}
S.~Choi {\em et al.} ({Belle} collaboration){,}
  \href{http://dx.doi.org/10.1103/PhysRevLett.91.262001}{Phys.\ Rev.\ Lett.\ {\bf
  91},  262001}  (2003), \href{http://arxiv.org/abs/hep-ex/0309032}{{\tt
  arXiv:hep-ex/0309032 [hep-ex]}}\relax
\mciteBstWouldAddEndPuncttrue
\mciteSetBstMidEndSepPunct{\mcitedefaultmidpunct}
{\mcitedefaultendpunct}{\mcitedefaultseppunct}\relax
\EndOfBibitem
\bibitem{Aaij:2013zoa}
R.~Aaij {\em et al.} ({LHCb} collaboration){,}
  \href{http://dx.doi.org/10.1103/PhysRevLett.110.222001}{Phys.\ Rev.\ Lett.\ {\bf
  110},  222001}  (2013), \href{http://arxiv.org/abs/1302.6269}{{\tt
  arXiv:1302.6269 [hep-ex]}}\relax
\mciteBstWouldAddEndPuncttrue
\mciteSetBstMidEndSepPunct{\mcitedefaultmidpunct}
{\mcitedefaultendpunct}{\mcitedefaultseppunct}\relax
\EndOfBibitem
\bibitem{Choi:2007wga}
S.~Choi {\em et al.} ({Belle} collaboration){,}
  \href{http://dx.doi.org/10.1103/PhysRevLett.100.142001}{Phys.\ Rev.\ Lett.\ {\bf
  100},  142001}  (2008), \href{http://arxiv.org/abs/0708.1790}{{\tt
  arXiv:0708.1790 [hep-ex]}}\relax
\mciteBstWouldAddEndPuncttrue
\mciteSetBstMidEndSepPunct{\mcitedefaultmidpunct}
{\mcitedefaultendpunct}{\mcitedefaultseppunct}\relax
\EndOfBibitem
\bibitem{Aaij:2014jqa}
R.~Aaij {\em et al.} ({LHCb} collaboration){,}
  \href{http://dx.doi.org/10.1103/PhysRevLett.112.222002}{Phys.\ Rev.\ Lett.\ {\bf
  112},  222002}  (2014), \href{http://arxiv.org/abs/1404.1903}{{\tt
  arXiv:1404.1903 [hep-ex]}}\relax
\mciteBstWouldAddEndPuncttrue
\mciteSetBstMidEndSepPunct{\mcitedefaultmidpunct}
{\mcitedefaultendpunct}{\mcitedefaultseppunct}\relax
\EndOfBibitem
\bibitem{Aaij:2015tga}
R.~Aaij {\em et al.} ({LHCb} collaboration){,}
  \href{http://dx.doi.org/10.1103/PhysRevLett.115.072001}{Phys.\ Rev.\ Lett.\ {\bf
  115},  072001}  (2015), \href{http://arxiv.org/abs/1507.03414}{{\tt
  arXiv:1507.03414 [hep-ex]}}\relax
\mciteBstWouldAddEndPuncttrue
\mciteSetBstMidEndSepPunct{\mcitedefaultmidpunct}
{\mcitedefaultendpunct}{\mcitedefaultseppunct}\relax
\EndOfBibitem
\bibitem{DeBruyn:2012wj}
K.~De~Bruyn {\em et al.}{,}
  \href{http://dx.doi.org/10.1103/PhysRevD.86.014027}{Phys.\ Rev.\ {\bf D86}{,}
  014027}  (2012), \href{http://arxiv.org/abs/1204.1735}{{\tt arXiv:1204.1735
  [hep-ph]}}\relax
\mciteBstWouldAddEndPuncttrue
\mciteSetBstMidEndSepPunct{\mcitedefaultmidpunct}
{\mcitedefaultendpunct}{\mcitedefaultseppunct}\relax
\EndOfBibitem
\bibitem{Jung:2015yma}
M.~Jung, \href{http://dx.doi.org/10.1016/j.physletb.2015.12.024}{Phys.\ Lett.\
  {\bf B753},  187}  (2016), \href{http://arxiv.org/abs/1510.03423}{{\tt
  arXiv:1510.03423 [hep-ph]}}\relax
\mciteBstWouldAddEndPuncttrue
\mciteSetBstMidEndSepPunct{\mcitedefaultmidpunct}
{\mcitedefaultendpunct}{\mcitedefaultseppunct}\relax
\EndOfBibitem
\bibitem{Aubert:2006cd}
B.~Aubert {\em et al.} ({\babar} collaboration){,}
  \href{http://dx.doi.org/10.1103/PhysRevD.75.031101}{Phys.\ Rev.\ {\bf D75}{,}
  031101}  (2007), \href{http://arxiv.org/abs/hep-ex/0610027}{{\tt
  arXiv:hep-ex/0610027 [hep-ex]}}\relax
\mciteBstWouldAddEndPuncttrue
\mciteSetBstMidEndSepPunct{\mcitedefaultmidpunct}
{\mcitedefaultendpunct}{\mcitedefaultseppunct}\relax
\EndOfBibitem
\bibitem{Aubert:2006jc}
B.~Aubert {\em et al.} ({\babar} collaboration){,}
  \href{http://dx.doi.org/10.1103/PhysRevD.74.111102}{Phys.\ Rev.\ {\bf D74}{,}
  111102}  (2006), \href{http://arxiv.org/abs/hep-ex/0609033}{{\tt
  arXiv:hep-ex/0609033 [hep-ex]}}\relax
\mciteBstWouldAddEndPuncttrue
\mciteSetBstMidEndSepPunct{\mcitedefaultmidpunct}
{\mcitedefaultendpunct}{\mcitedefaultseppunct}\relax
\EndOfBibitem
\bibitem{Majumder:2004su}
G.~Majumder {\em et al.} ({Belle} collaboration){,}
  \href{http://dx.doi.org/10.1103/PhysRevD.70.111103}{Phys.\ Rev.\ {\bf D70}{,}
  111103}  (2004), \href{http://arxiv.org/abs/hep-ex/0409008}{{\tt
  arXiv:hep-ex/0409008 [hep-ex]}}\relax
\mciteBstWouldAddEndPuncttrue
\mciteSetBstMidEndSepPunct{\mcitedefaultmidpunct}
{\mcitedefaultendpunct}{\mcitedefaultseppunct}\relax
\EndOfBibitem
\bibitem{TheBABAR:2016vzj}
J.~P.\ Lees {\em et al.} ({\babar} collaboration){,}
  \href{http://dx.doi.org/10.1103/PhysRevD.94.091101}{Phys.\ Rev.\ {\bf D94}{,}
  091101}  (2016), \href{http://arxiv.org/abs/1609.06802}{{\tt
  arXiv:1609.06802 [hep-ex]}}\relax
\mciteBstWouldAddEndPuncttrue
\mciteSetBstMidEndSepPunct{\mcitedefaultmidpunct}
{\mcitedefaultendpunct}{\mcitedefaultseppunct}\relax
\EndOfBibitem
\bibitem{Matvienko:2015gqa}
D.~Matvienko {\em et al.} ({Belle} collaboration){,}
  \href{http://dx.doi.org/10.1103/PhysRevD.92.012013}{Phys.\ Rev.\ {\bf D92}{,}
  012013}  (2015), \href{http://arxiv.org/abs/1505.03362}{{\tt
  arXiv:1505.03362 [hep-ex]}}\relax
\mciteBstWouldAddEndPuncttrue
\mciteSetBstMidEndSepPunct{\mcitedefaultmidpunct}
{\mcitedefaultendpunct}{\mcitedefaultseppunct}\relax
\EndOfBibitem
\bibitem{Aubert:2006zb}
B.~Aubert {\em et al.} ({\babar} collaboration){,}
  \href{http://dx.doi.org/10.1103/PhysRevD.74.012001}{Phys.\ Rev.\ {\bf D74}{,}
  012001}  (2006), \href{http://arxiv.org/abs/hep-ex/0604009}{{\tt
  arXiv:hep-ex/0604009 [hep-ex]}}\relax
\mciteBstWouldAddEndPuncttrue
\mciteSetBstMidEndSepPunct{\mcitedefaultmidpunct}
{\mcitedefaultendpunct}{\mcitedefaultseppunct}\relax
\EndOfBibitem
\bibitem{Blyth:2006at}
S.~Blyth {\em et al.} ({Belle} collaboration){,}
  \href{http://dx.doi.org/10.1103/PhysRevD.74.092002}{Phys.\ Rev.\ {\bf D74}{,}
  092002}  (2006), \href{http://arxiv.org/abs/hep-ex/0607029}{{\tt
  arXiv:hep-ex/0607029 [hep-ex]}}\relax
\mciteBstWouldAddEndPuncttrue
\mciteSetBstMidEndSepPunct{\mcitedefaultmidpunct}
{\mcitedefaultendpunct}{\mcitedefaultseppunct}\relax
\EndOfBibitem
\bibitem{Lees:2011gw}
J.~P.\ Lees {\em et al.} ({\babar} collaboration){,}
  \href{http://dx.doi.org/10.1103/PhysRevD.84.112007}{Phys.\ Rev.\ {\bf D84}{,}
  112007}  (2011), \href{http://arxiv.org/abs/1107.5751}{{\tt arXiv:1107.5751
  [hep-ex]}}, Erratum ibid.\
  \href{http://dx.doi.org/10.1103/PhysRevD.87.039901}{{\bf D87}, 039901}
  (2013)\relax
\mciteBstWouldAddEndPuncttrue
\mciteSetBstMidEndSepPunct{\mcitedefaultmidpunct}
{\mcitedefaultendpunct}{\mcitedefaultseppunct}\relax
\EndOfBibitem
\bibitem{Aaij:2015sqa}
R.~Aaij {\em et al.} ({LHCb} collaboration){,}
  \href{http://dx.doi.org/10.1103/PhysRevD.92.032002}{Phys.\ Rev.\ {\bf D92}{,}
  032002}  (2015), \href{http://arxiv.org/abs/1505.01710}{{\tt
  arXiv:1505.01710 [hep-ex]}}\relax
\mciteBstWouldAddEndPuncttrue
\mciteSetBstMidEndSepPunct{\mcitedefaultmidpunct}
{\mcitedefaultendpunct}{\mcitedefaultseppunct}\relax
\EndOfBibitem
\bibitem{Satpathy:2002js}
A.~Satpathy {\em et al.} ({Belle} collaboration){,}
  \href{http://dx.doi.org/10.1016/S0370-2693(02)03198-2}{Phys.\ Lett.\ {\bf
  B553},  159}  (2003), \href{http://arxiv.org/abs/hep-ex/0211022}{{\tt
  arXiv:hep-ex/0211022 [hep-ex]}}\relax
\mciteBstWouldAddEndPuncttrue
\mciteSetBstMidEndSepPunct{\mcitedefaultmidpunct}
{\mcitedefaultendpunct}{\mcitedefaultseppunct}\relax
\EndOfBibitem
\bibitem{Schumann:2005ej}
J.~Schumann {\em et al.} ({Belle} collaboration){,}
  \href{http://dx.doi.org/10.1103/PhysRevD.72.011103}{Phys.\ Rev.\ {\bf D72}{,}
  011103}  (2005), \href{http://arxiv.org/abs/hep-ex/0501013}{{\tt
  arXiv:hep-ex/0501013 [hep-ex]}}\relax
\mciteBstWouldAddEndPuncttrue
\mciteSetBstMidEndSepPunct{\mcitedefaultmidpunct}
{\mcitedefaultendpunct}{\mcitedefaultseppunct}\relax
\EndOfBibitem
\bibitem{Abe:2001waa}
K.~Abe {\em et al.} ({Belle} collaboration){,}
  \href{http://dx.doi.org/10.1103/PhysRevLett.87.111801}{Phys.\ Rev.\ Lett.\ {\bf
  87},  111801}  (2001), \href{http://arxiv.org/abs/hep-ex/0104051}{{\tt
  arXiv:hep-ex/0104051 [hep-ex]}}\relax
\mciteBstWouldAddEndPuncttrue
\mciteSetBstMidEndSepPunct{\mcitedefaultmidpunct}
{\mcitedefaultendpunct}{\mcitedefaultseppunct}\relax
\EndOfBibitem
\bibitem{Aubert:2004at}
B.~Aubert {\em et al.} ({\babar} collaboration){,}
  \href{http://dx.doi.org/10.1103/PhysRevLett.95.171802}{Phys.\ Rev.\ Lett.\ {\bf
  95},  171802}  (2005), \href{http://arxiv.org/abs/hep-ex/0412040}{{\tt
  arXiv:hep-ex/0412040 [hep-ex]}}\relax
\mciteBstWouldAddEndPuncttrue
\mciteSetBstMidEndSepPunct{\mcitedefaultmidpunct}
{\mcitedefaultendpunct}{\mcitedefaultseppunct}\relax
\EndOfBibitem
\bibitem{Drutskoy:2002ib}
A.~Drutskoy {\em et al.} ({Belle} collaboration){,}
  \href{http://dx.doi.org/10.1016/S0370-2693(02)02373-0}{Phys.\ Lett.\ {\bf
  B542},  171}  (2002), \href{http://arxiv.org/abs/hep-ex/0207041}{{\tt
  arXiv:hep-ex/0207041 [hep-ex]}}\relax
\mciteBstWouldAddEndPuncttrue
\mciteSetBstMidEndSepPunct{\mcitedefaultmidpunct}
{\mcitedefaultendpunct}{\mcitedefaultseppunct}\relax
\EndOfBibitem
\bibitem{Krokovny:2002ua}
P.~Krokovny {\em et al.} ({Belle} collaboration){,}
  \href{http://dx.doi.org/10.1103/PhysRevLett.90.141802}{Phys.\ Rev.\ Lett.\ {\bf
  90},  141802}  (2003), \href{http://arxiv.org/abs/hep-ex/0212066}{{\tt
  arXiv:hep-ex/0212066 [hep-ex]}}\relax
\mciteBstWouldAddEndPuncttrue
\mciteSetBstMidEndSepPunct{\mcitedefaultmidpunct}
{\mcitedefaultendpunct}{\mcitedefaultseppunct}\relax
\EndOfBibitem
\bibitem{Aubert:2006qn}
B.~Aubert {\em et al.} ({\babar} collaboration){,}
  \href{http://dx.doi.org/10.1103/PhysRevD.74.031101}{Phys.\ Rev.\ {\bf D74}{,}
  031101}  (2006), \href{http://arxiv.org/abs/hep-ex/0604016}{{\tt
  arXiv:hep-ex/0604016 [hep-ex]}}\relax
\mciteBstWouldAddEndPuncttrue
\mciteSetBstMidEndSepPunct{\mcitedefaultmidpunct}
{\mcitedefaultendpunct}{\mcitedefaultseppunct}\relax
\EndOfBibitem
\bibitem{Aubert:2005yt}
B.~Aubert {\em et al.} ({\babar} collaboration){,}
  \href{http://dx.doi.org/10.1103/PhysRevLett.96.011803}{Phys.\ Rev.\ Lett.\ {\bf
  96},  011803}  (2006), \href{http://arxiv.org/abs/hep-ex/0509036}{{\tt
  arXiv:hep-ex/0509036 [hep-ex]}}\relax
\mciteBstWouldAddEndPuncttrue
\mciteSetBstMidEndSepPunct{\mcitedefaultmidpunct}
{\mcitedefaultendpunct}{\mcitedefaultseppunct}\relax
\EndOfBibitem
\bibitem{Das:2010be}
A.~Das {\em et al.} ({Belle} collaboration){,}
  \href{http://dx.doi.org/10.1103/PhysRevD.82.051103}{Phys.\ Rev.\ {\bf D82}{,}
  051103}  (2010), \href{http://arxiv.org/abs/1007.4619}{{\tt arXiv:1007.4619
  [hep-ex]}}\relax
\mciteBstWouldAddEndPuncttrue
\mciteSetBstMidEndSepPunct{\mcitedefaultmidpunct}
{\mcitedefaultendpunct}{\mcitedefaultseppunct}\relax
\EndOfBibitem
\bibitem{Aubert:2008zi}
B.~Aubert {\em et al.} ({\babar} collaboration){,}
  \href{http://dx.doi.org/10.1103/PhysRevD.78.032005}{Phys.\ Rev.\ {\bf D78}{,}
  032005}  (2008), \href{http://arxiv.org/abs/0803.4296}{{\tt arXiv:0803.4296
  [hep-ex]}}\relax
\mciteBstWouldAddEndPuncttrue
\mciteSetBstMidEndSepPunct{\mcitedefaultmidpunct}
{\mcitedefaultendpunct}{\mcitedefaultseppunct}\relax
\EndOfBibitem
\bibitem{Joshi:2009yv}
N.~Joshi {\em et al.} ({Belle} collaboration){,}
  \href{http://dx.doi.org/10.1103/PhysRevD.81.031101}{Phys.\ Rev.\ {\bf D81}{,}
  031101}  (2010), \href{http://arxiv.org/abs/0912.2594}{{\tt arXiv:0912.2594
  [hep-ex]}}\relax
\mciteBstWouldAddEndPuncttrue
\mciteSetBstMidEndSepPunct{\mcitedefaultmidpunct}
{\mcitedefaultendpunct}{\mcitedefaultseppunct}\relax
\EndOfBibitem
\bibitem{Aubert:2005qt}
B.~Aubert {\em et al.} ({\babar} collaboration){,}
  \href{http://dx.doi.org/10.1103/PhysRevD.73.071103}{Phys.\ Rev.\ {\bf D73}{,}
  071103}  (2006), \href{http://arxiv.org/abs/hep-ex/0512031}{{\tt
  arXiv:hep-ex/0512031 [hep-ex]}}\relax
\mciteBstWouldAddEndPuncttrue
\mciteSetBstMidEndSepPunct{\mcitedefaultmidpunct}
{\mcitedefaultendpunct}{\mcitedefaultseppunct}\relax
\EndOfBibitem
\bibitem{Aubert:2007xma}
B.~Aubert {\em et al.} ({\babar} collaboration){,}
  \href{http://dx.doi.org/10.1103/PhysRevLett.100.171803}{Phys.\ Rev.\ Lett.\ {\bf
  100},  171803}  (2008), \href{http://arxiv.org/abs/0707.1043}{{\tt
  arXiv:0707.1043 [hep-ex]}}\relax
\mciteBstWouldAddEndPuncttrue
\mciteSetBstMidEndSepPunct{\mcitedefaultmidpunct}
{\mcitedefaultendpunct}{\mcitedefaultseppunct}\relax
\EndOfBibitem
\bibitem{Aaij:2011rj}
R.~Aaij {\em et al.} ({LHCb} collaboration){,}
  \href{http://dx.doi.org/10.1103/PhysRevD.84.092001}{Phys.\ Rev.\ {\bf D84}{,}
  092001}  (2011), \href{http://arxiv.org/abs/1109.6831}{{\tt arXiv:1109.6831
  [hep-ex]}}, Erratum ibid.\
  \href{http://dx.doi.org/10.1103/PhysRevD.85.039904}{{\bf D85}, 039904}
  (2011)\relax
\mciteBstWouldAddEndPuncttrue
\mciteSetBstMidEndSepPunct{\mcitedefaultmidpunct}
{\mcitedefaultendpunct}{\mcitedefaultseppunct}\relax
\EndOfBibitem
\bibitem{Aaij:2012mra}
R.~Aaij {\em et al.} ({LHCb} collaboration){,}
  \href{http://dx.doi.org/10.1103/PhysRevD.86.112005}{Phys.\ Rev.\ {\bf D86}{,}
  112005}  (2012), \href{http://arxiv.org/abs/1211.1541}{{\tt arXiv:1211.1541
  [hep-ex]}}\relax
\mciteBstWouldAddEndPuncttrue
\mciteSetBstMidEndSepPunct{\mcitedefaultmidpunct}
{\mcitedefaultendpunct}{\mcitedefaultseppunct}\relax
\EndOfBibitem
\bibitem{Aaij:2013pua}
R.~Aaij {\em et al.} ({LHCb} collaboration){,}
  \href{http://dx.doi.org/10.1103/PhysRevD.87.112009}{Phys.\ Rev.\ {\bf D87}{,}
  112009}  (2013), \href{http://arxiv.org/abs/1304.6317}{{\tt arXiv:1304.6317
  [hep-ex]}}\relax
\mciteBstWouldAddEndPuncttrue
\mciteSetBstMidEndSepPunct{\mcitedefaultmidpunct}
{\mcitedefaultendpunct}{\mcitedefaultseppunct}\relax
\EndOfBibitem
\bibitem{Aaij:2012zka}
R.~Aaij {\em et al.} ({LHCb} collaboration){,}
  \href{http://dx.doi.org/10.1103/PhysRevLett.109.131801}{Phys.\ Rev.\ Lett.\ {\bf
  109},  131801}  (2012), \href{http://arxiv.org/abs/1207.5991}{{\tt
  arXiv:1207.5991 [hep-ex]}}\relax
\mciteBstWouldAddEndPuncttrue
\mciteSetBstMidEndSepPunct{\mcitedefaultmidpunct}
{\mcitedefaultendpunct}{\mcitedefaultseppunct}\relax
\EndOfBibitem
\bibitem{Aaij:2012bw}
R.~Aaij {\em et al.} ({LHCb} collaboration){,}
  \href{http://dx.doi.org/10.1103/PhysRevLett.108.161801}{Phys.\ Rev.\ Lett.\ {\bf
  108},  161801}  (2012), \href{http://arxiv.org/abs/1201.4402}{{\tt
  arXiv:1201.4402 [hep-ex]}}\relax
\mciteBstWouldAddEndPuncttrue
\mciteSetBstMidEndSepPunct{\mcitedefaultmidpunct}
{\mcitedefaultendpunct}{\mcitedefaultseppunct}\relax
\EndOfBibitem
\bibitem{Aaij:2014jpa}
R.~Aaij {\em et al.} ({LHCb} collaboration){,}
  \href{http://dx.doi.org/10.1007/JHEP05(2015)019}{JHEP {\bf 05},  019}
  (2015), \href{http://arxiv.org/abs/1412.7654}{{\tt arXiv:1412.7654
  [hep-ex]}}\relax
\mciteBstWouldAddEndPuncttrue
\mciteSetBstMidEndSepPunct{\mcitedefaultmidpunct}
{\mcitedefaultendpunct}{\mcitedefaultseppunct}\relax
\EndOfBibitem
\bibitem{Abe:2004sm}
K.~Abe {\em et al.} ({Belle} collaboration){,}
  \href{http://dx.doi.org/10.1103/PhysRevLett.94.221805}{Phys.\ Rev.\ Lett.\ {\bf
  94},  221805}  (2005), \href{http://arxiv.org/abs/hep-ex/0410091}{{\tt
  arXiv:hep-ex/0410091 [hep-ex]}}\relax
\mciteBstWouldAddEndPuncttrue
\mciteSetBstMidEndSepPunct{\mcitedefaultmidpunct}
{\mcitedefaultendpunct}{\mcitedefaultseppunct}\relax
\EndOfBibitem
\bibitem{Abe:2004wz}
A.~Drutskoy {\em et al.} ({Belle} collaboration){,}
  \href{http://dx.doi.org/10.1103/PhysRevLett.94.061802}{Phys.\ Rev.\ Lett.\ {\bf
  94},  061802}  (2005), \href{http://arxiv.org/abs/hep-ex/0409026}{{\tt
  arXiv:hep-ex/0409026 [hep-ex]}}\relax
\mciteBstWouldAddEndPuncttrue
\mciteSetBstMidEndSepPunct{\mcitedefaultmidpunct}
{\mcitedefaultendpunct}{\mcitedefaultseppunct}\relax
\EndOfBibitem
\bibitem{delAmoSanchez:2011gi}
P.~del Amo~Sanchez {\em et al.} ({\babar} collaboration){,}
  \href{http://dx.doi.org/10.1103/PhysRevD.85.092017}{Phys.\ Rev.\ {\bf D85}{,}
  092017}  (2012), \href{http://arxiv.org/abs/1111.4387}{{\tt arXiv:1111.4387
  [hep-ex]}}\relax
\mciteBstWouldAddEndPuncttrue
\mciteSetBstMidEndSepPunct{\mcitedefaultmidpunct}
{\mcitedefaultendpunct}{\mcitedefaultseppunct}\relax
\EndOfBibitem
\bibitem{Abe:2002tw}
K.~Abe {\em et al.} ({Belle} collaboration){,}
  \href{http://dx.doi.org/10.1103/PhysRevLett.89.151802}{Phys.\ Rev.\ Lett.\ {\bf
  89},  151802}  (2002), \href{http://arxiv.org/abs/hep-ex/0205083}{{\tt
  arXiv:hep-ex/0205083 [hep-ex]}}\relax
\mciteBstWouldAddEndPuncttrue
\mciteSetBstMidEndSepPunct{\mcitedefaultmidpunct}
{\mcitedefaultendpunct}{\mcitedefaultseppunct}\relax
\EndOfBibitem
\bibitem{Medvedeva:2007af}
T.~Medvedeva {\em et al.} ({Belle} collaboration){,}
  \href{http://dx.doi.org/10.1103/PhysRevD.76.051102}{Phys.\ Rev.\ {\bf D76}{,}
  051102}  (2007), \href{http://arxiv.org/abs/0704.2652}{{\tt arXiv:0704.2652
  [hep-ex]}}\relax
\mciteBstWouldAddEndPuncttrue
\mciteSetBstMidEndSepPunct{\mcitedefaultmidpunct}
{\mcitedefaultendpunct}{\mcitedefaultseppunct}\relax
\EndOfBibitem
\bibitem{Chang:2008yw}
Y.~W.\ Chang {\em et al.} ({Belle} collaboration){,}
  \href{http://dx.doi.org/10.1103/PhysRevD.79.052006}{Phys.\ Rev.\ {\bf D79}{,}
  052006}  (2009), \href{http://arxiv.org/abs/0811.3826}{{\tt arXiv:0811.3826
  [hep-ex]}}\relax
\mciteBstWouldAddEndPuncttrue
\mciteSetBstMidEndSepPunct{\mcitedefaultmidpunct}
{\mcitedefaultendpunct}{\mcitedefaultseppunct}\relax
\EndOfBibitem
\bibitem{Lees:2014mka}
J.~P.\ Lees {\em et al.} ({\babar} collaboration){,}
  \href{http://dx.doi.org/10.1103/PhysRevD.89.112002}{Phys.\ Rev.\ {\bf D89}{,}
  112002}  (2014), \href{http://arxiv.org/abs/1401.5990}{{\tt arXiv:1401.5990
  [hep-ex]}}\relax
\mciteBstWouldAddEndPuncttrue
\mciteSetBstMidEndSepPunct{\mcitedefaultmidpunct}
{\mcitedefaultendpunct}{\mcitedefaultseppunct}\relax
\EndOfBibitem
\bibitem{Chang:2015fja}
Y.~Y.\ Chang {\em et al.} ({Belle} collaboration){,}
  \href{http://dx.doi.org/10.1103/PhysRevLett.115.221803}{Phys.\ Rev.\ Lett.\ {\bf
  115},  221803}  (2015), \href{http://arxiv.org/abs/1509.03034}{{\tt
  arXiv:1509.03034 [hep-ex]}}\relax
\mciteBstWouldAddEndPuncttrue
\mciteSetBstMidEndSepPunct{\mcitedefaultmidpunct}
{\mcitedefaultendpunct}{\mcitedefaultseppunct}\relax
\EndOfBibitem
\bibitem{Aubert:2006ia}
B.~Aubert {\em et al.} ({\babar} collaboration){,}
  \href{http://dx.doi.org/10.1103/PhysRevD.73.112004}{Phys.\ Rev.\ {\bf D73}{,}
  112004}  (2006), \href{http://arxiv.org/abs/hep-ex/0604037}{{\tt
  arXiv:hep-ex/0604037 [hep-ex]}}\relax
\mciteBstWouldAddEndPuncttrue
\mciteSetBstMidEndSepPunct{\mcitedefaultmidpunct}
{\mcitedefaultendpunct}{\mcitedefaultseppunct}\relax
\EndOfBibitem
\bibitem{Adachi:2008cj}
I.~Adachi {\em et al.} ({Belle} collaboration){,}
  \href{http://dx.doi.org/10.1103/PhysRevD.77.091101}{Phys.\ Rev.\ {\bf D77}{,}
  091101}  (2008), \href{http://arxiv.org/abs/0802.2988}{{\tt arXiv:0802.2988
  [hep-ex]}}\relax
\mciteBstWouldAddEndPuncttrue
\mciteSetBstMidEndSepPunct{\mcitedefaultmidpunct}
{\mcitedefaultendpunct}{\mcitedefaultseppunct}\relax
\EndOfBibitem
\bibitem{delAmoSanchez:2010pg}
P.~del Amo~Sanchez {\em et al.} ({\babar} collaboration){,}
  \href{http://dx.doi.org/10.1103/PhysRevD.83.032004}{Phys.\ Rev.\ {\bf D83}{,}
  032004}  (2011), \href{http://arxiv.org/abs/1011.3929}{{\tt arXiv:1011.3929
  [hep-ex]}}\relax
\mciteBstWouldAddEndPuncttrue
\mciteSetBstMidEndSepPunct{\mcitedefaultmidpunct}
{\mcitedefaultendpunct}{\mcitedefaultseppunct}\relax
\EndOfBibitem
\bibitem{Gokhroo:2006bt}
G.~Gokhroo {\em et al.} ({Belle} collaboration){,}
  \href{http://dx.doi.org/10.1103/PhysRevLett.97.162002}{Phys.\ Rev.\ Lett.\ {\bf
  97},  162002}  (2006), \href{http://arxiv.org/abs/hep-ex/0606055}{{\tt
  arXiv:hep-ex/0606055 [hep-ex]}}\relax
\mciteBstWouldAddEndPuncttrue
\mciteSetBstMidEndSepPunct{\mcitedefaultmidpunct}
{\mcitedefaultendpunct}{\mcitedefaultseppunct}\relax
\EndOfBibitem
\bibitem{Zupanc:2007pu}
A.~Zupanc {\em et al.} ({Belle} collaboration){,}
  \href{http://dx.doi.org/10.1103/PhysRevD.75.091102}{Phys.\ Rev.\ {\bf D75}{,}
  091102}  (2007), \href{http://arxiv.org/abs/hep-ex/0703040}{{\tt
  arXiv:hep-ex/0703040 [hep-ex]}}\relax
\mciteBstWouldAddEndPuncttrue
\mciteSetBstMidEndSepPunct{\mcitedefaultmidpunct}
{\mcitedefaultendpunct}{\mcitedefaultseppunct}\relax
\EndOfBibitem
\bibitem{Aubert:2006nm}
B.~Aubert {\em et al.} ({\babar} collaboration){,}
  \href{http://dx.doi.org/10.1103/PhysRevD.74.031103}{Phys.\ Rev.\ {\bf D74}{,}
  031103}  (2006), \href{http://arxiv.org/abs/hep-ex/0605036}{{\tt
  arXiv:hep-ex/0605036 [hep-ex]}}\relax
\mciteBstWouldAddEndPuncttrue
\mciteSetBstMidEndSepPunct{\mcitedefaultmidpunct}
{\mcitedefaultendpunct}{\mcitedefaultseppunct}\relax
\EndOfBibitem
\bibitem{Aubert:2003jj}
B.~Aubert {\em et al.} ({\babar} collaboration){,}
  \href{http://dx.doi.org/10.1103/PhysRevD.67.092003}{Phys.\ Rev.\ {\bf D67}{,}
  092003}  (2003), \href{http://arxiv.org/abs/hep-ex/0302015}{{\tt
  arXiv:hep-ex/0302015 [hep-ex]}}\relax
\mciteBstWouldAddEndPuncttrue
\mciteSetBstMidEndSepPunct{\mcitedefaultmidpunct}
{\mcitedefaultendpunct}{\mcitedefaultseppunct}\relax
\EndOfBibitem
\bibitem{Aubert:2005xu}
B.~Aubert {\em et al.} ({\babar} collaboration){,}
  \href{http://dx.doi.org/10.1103/PhysRevD.71.091104}{Phys.\ Rev.\ {\bf D71}{,}
  091104}  (2005), \href{http://arxiv.org/abs/hep-ex/0502041}{{\tt
  arXiv:hep-ex/0502041 [hep-ex]}}\relax
\mciteBstWouldAddEndPuncttrue
\mciteSetBstMidEndSepPunct{\mcitedefaultmidpunct}
{\mcitedefaultendpunct}{\mcitedefaultseppunct}\relax
\EndOfBibitem
\bibitem{Aubert:2005jv}
B.~Aubert {\em et al.} ({\babar} collaboration){,}
  \href{http://dx.doi.org/10.1103/PhysRevD.72.111101}{Phys.\ Rev.\ {\bf D72}{,}
  111101}  (2005), \href{http://arxiv.org/abs/hep-ex/0510051}{{\tt
  arXiv:hep-ex/0510051 [hep-ex]}}\relax
\mciteBstWouldAddEndPuncttrue
\mciteSetBstMidEndSepPunct{\mcitedefaultmidpunct}
{\mcitedefaultendpunct}{\mcitedefaultseppunct}\relax
\EndOfBibitem
\bibitem{Aaij:2013fha}
R.~Aaij {\em et al.} ({LHCb} collaboration){,}
  \href{http://dx.doi.org/10.1103/PhysRevD.87.092007}{Phys.\ Rev.\ {\bf D87}{,}
  092007}  (2013), \href{http://arxiv.org/abs/1302.5854}{{\tt arXiv:1302.5854
  [hep-ex]}}\relax
\mciteBstWouldAddEndPuncttrue
\mciteSetBstMidEndSepPunct{\mcitedefaultmidpunct}
{\mcitedefaultendpunct}{\mcitedefaultseppunct}\relax
\EndOfBibitem
\bibitem{Krokovny:2003zq}
P.~Krokovny {\em et al.} ({Belle} collaboration){,}
  \href{http://dx.doi.org/10.1103/PhysRevLett.91.262002}{Phys.\ Rev.\ Lett.\ {\bf
  91},  262002}  (2003), \href{http://arxiv.org/abs/hep-ex/0308019}{{\tt
  arXiv:hep-ex/0308019 [hep-ex]}}\relax
\mciteBstWouldAddEndPuncttrue
\mciteSetBstMidEndSepPunct{\mcitedefaultmidpunct}
{\mcitedefaultendpunct}{\mcitedefaultseppunct}\relax
\EndOfBibitem
\bibitem{Aubert:2004pw}
B.~Aubert {\em et al.} ({\babar} collaboration){,}
  \href{http://dx.doi.org/10.1103/PhysRevLett.93.181801}{Phys.\ Rev.\ Lett.\ {\bf
  93},  181801}  (2004), \href{http://arxiv.org/abs/hep-ex/0408041}{{\tt
  arXiv:hep-ex/0408041 [hep-ex]}}\relax
\mciteBstWouldAddEndPuncttrue
\mciteSetBstMidEndSepPunct{\mcitedefaultmidpunct}
{\mcitedefaultendpunct}{\mcitedefaultseppunct}\relax
\EndOfBibitem
\bibitem{Choi:2015lpc}
S.~K.\ Choi {\em et al.} ({Belle} collaboration){,}
  \href{http://dx.doi.org/10.1103/PhysRevD.91.092011}{Phys.\ Rev.\ {\bf D91}{,}
  092011}  (2015), \href{http://arxiv.org/abs/1504.02637}{{\tt
  arXiv:1504.02637 [hep-ex]}}, Addendum ibid.\
  \href{http://dx.doi.org/10.1103/PhysRevD.92.039905}{{\bf D92}, 039905}
  (2015)\relax
\mciteBstWouldAddEndPuncttrue
\mciteSetBstMidEndSepPunct{\mcitedefaultmidpunct}
{\mcitedefaultendpunct}{\mcitedefaultseppunct}\relax
\EndOfBibitem
\bibitem{Aubert:2007rva}
B.~Aubert {\em et al.} ({\babar} collaboration){,}
  \href{http://dx.doi.org/10.1103/PhysRevD.77.011102}{Phys.\ Rev.\ {\bf D77}{,}
  011102}  (2008), \href{http://arxiv.org/abs/0708.1565}{{\tt arXiv:0708.1565
  [hep-ex]}}\relax
\mciteBstWouldAddEndPuncttrue
\mciteSetBstMidEndSepPunct{\mcitedefaultmidpunct}
{\mcitedefaultendpunct}{\mcitedefaultseppunct}\relax
\EndOfBibitem
\bibitem{Belle:2011ad}
T.~Aushev {\em et al.} ({Belle} collaboration){,}
  \href{http://dx.doi.org/10.1103/PhysRevD.83.059902}{Phys.\ Rev.\ {\bf D83}{,}
  051102}  (2011), \href{http://arxiv.org/abs/1102.0935}{{\tt arXiv:1102.0935
  [hep-ex]}}, Erratum ibid.\
  \href{http://dx.doi.org/10.1103/PhysRevD.83.051102}{{\bf D83}, 051102}
  (2011)\relax
\mciteBstWouldAddEndPuncttrue
\mciteSetBstMidEndSepPunct{\mcitedefaultmidpunct}
{\mcitedefaultendpunct}{\mcitedefaultseppunct}\relax
\EndOfBibitem
\bibitem{Abe:1995aw}
F.~Abe {\em et al.} ({CDF} collaboration){,}
  \href{http://dx.doi.org/10.1103/PhysRevLett.76.2015}{Phys.\ Rev.\ Lett.\ {\bf
  76},  2015}  (1996)\relax
\mciteBstWouldAddEndPuncttrue
\mciteSetBstMidEndSepPunct{\mcitedefaultmidpunct}
{\mcitedefaultendpunct}{\mcitedefaultseppunct}\relax
\EndOfBibitem
\bibitem{Abe:2002rc}
K.~Abe {\em et al.} ({Belle} collaboration){,}
  \href{http://dx.doi.org/10.1103/PhysRevD.67.032003}{Phys.\ Rev.\ {\bf D67}{,}
  032003}  (2003), \href{http://arxiv.org/abs/hep-ex/0211047}{{\tt
  arXiv:hep-ex/0211047 [hep-ex]}}\relax
\mciteBstWouldAddEndPuncttrue
\mciteSetBstMidEndSepPunct{\mcitedefaultmidpunct}
{\mcitedefaultendpunct}{\mcitedefaultseppunct}\relax
\EndOfBibitem
\bibitem{Chilikin:2014bkk}
K.~Chilikin {\em et al.} ({Belle} collaboration){,}
  \href{http://dx.doi.org/10.1103/PhysRevD.90.112009}{Phys.\ Rev.\ {\bf D90}{,}
  112009}  (2014), \href{http://arxiv.org/abs/1408.6457}{{\tt arXiv:1408.6457
  [hep-ex]}}\relax
\mciteBstWouldAddEndPuncttrue
\mciteSetBstMidEndSepPunct{\mcitedefaultmidpunct}
{\mcitedefaultendpunct}{\mcitedefaultseppunct}\relax
\EndOfBibitem
\bibitem{Abe:1998yu}
F.~Abe {\em et al.} ({CDF} collaboration){,}
  \href{http://dx.doi.org/10.1103/PhysRevD.58.072001}{Phys.\ Rev.\ {\bf D58}{,}
  072001}  (1998), \href{http://arxiv.org/abs/hep-ex/9803013}{{\tt
  arXiv:hep-ex/9803013 [hep-ex]}}\relax
\mciteBstWouldAddEndPuncttrue
\mciteSetBstMidEndSepPunct{\mcitedefaultmidpunct}
{\mcitedefaultendpunct}{\mcitedefaultseppunct}\relax
\EndOfBibitem
\bibitem{Aaij:2014naa}
R.~Aaij {\em et al.} ({LHCb} collaboration){,}
  \href{http://dx.doi.org/10.1007/JHEP07(2014)140}{JHEP {\bf 07},  140}
  (2014), \href{http://arxiv.org/abs/1405.3219}{{\tt arXiv:1405.3219
  [hep-ex]}}\relax
\mciteBstWouldAddEndPuncttrue
\mciteSetBstMidEndSepPunct{\mcitedefaultmidpunct}
{\mcitedefaultendpunct}{\mcitedefaultseppunct}\relax
\EndOfBibitem
\bibitem{Affolder:2001qi}
T.~Affolder {\em et al.} ({CDF} collaboration){,}
  \href{http://dx.doi.org/10.1103/PhysRevLett.88.071801}{Phys.\ Rev.\ Lett.\ {\bf
  88},  071801}  (2002), \href{http://arxiv.org/abs/hep-ex/0108022}{{\tt
  arXiv:hep-ex/0108022 [hep-ex]}}\relax
\mciteBstWouldAddEndPuncttrue
\mciteSetBstMidEndSepPunct{\mcitedefaultmidpunct}
{\mcitedefaultendpunct}{\mcitedefaultseppunct}\relax
\EndOfBibitem
\bibitem{delAmoSanchez:2010jr}
P.~del Amo~Sanchez {\em et al.} ({\babar} collaboration){,}
  \href{http://dx.doi.org/10.1103/PhysRevD.82.011101}{Phys.\ Rev.\ {\bf D82}{,}
  011101}  (2010), \href{http://arxiv.org/abs/1005.5190}{{\tt arXiv:1005.5190
  [hep-ex]}}\relax
\mciteBstWouldAddEndPuncttrue
\mciteSetBstMidEndSepPunct{\mcitedefaultmidpunct}
{\mcitedefaultendpunct}{\mcitedefaultseppunct}\relax
\EndOfBibitem
\bibitem{Aubert:2003ii}
B.~Aubert {\em et al.} ({\babar} collaboration){,}
  \href{http://dx.doi.org/10.1103/PhysRevLett.91.071801}{Phys.\ Rev.\ Lett.\ {\bf
  91},  071801}  (2003), \href{http://arxiv.org/abs/hep-ex/0304014}{{\tt
  arXiv:hep-ex/0304014 [hep-ex]}}\relax
\mciteBstWouldAddEndPuncttrue
\mciteSetBstMidEndSepPunct{\mcitedefaultmidpunct}
{\mcitedefaultendpunct}{\mcitedefaultseppunct}\relax
\EndOfBibitem
\bibitem{Abe:2001wa}
K.~Abe {\em et al.} ({Belle} collaboration){,}
  \href{http://dx.doi.org/10.1103/PhysRevLett.87.161601}{Phys.\ Rev.\ Lett.\ {\bf
  87},  161601}  (2001), \href{http://arxiv.org/abs/hep-ex/0105014}{{\tt
  arXiv:hep-ex/0105014 [hep-ex]}}\relax
\mciteBstWouldAddEndPuncttrue
\mciteSetBstMidEndSepPunct{\mcitedefaultmidpunct}
{\mcitedefaultendpunct}{\mcitedefaultseppunct}\relax
\EndOfBibitem
\bibitem{Iwashita:2013wnn}
T.~Iwashita {\em et al.} ({Belle} collaboration){,}
  \href{http://dx.doi.org/10.1093/ptep/ptu043}{PTEP {\bf 2014},  043C01}
  (2014), \href{http://arxiv.org/abs/1310.2704}{{\tt arXiv:1310.2704
  [hep-ex]}}\relax
\mciteBstWouldAddEndPuncttrue
\mciteSetBstMidEndSepPunct{\mcitedefaultmidpunct}
{\mcitedefaultendpunct}{\mcitedefaultseppunct}\relax
\EndOfBibitem
\bibitem{Aubert:2004fc}
B.~Aubert {\em et al.} ({\babar} collaboration){,}
  \href{http://dx.doi.org/10.1103/PhysRevLett.93.041801}{Phys.\ Rev.\ Lett.\ {\bf
  93},  041801}  (2004), \href{http://arxiv.org/abs/hep-ex/0402025}{{\tt
  arXiv:hep-ex/0402025 [hep-ex]}}\relax
\mciteBstWouldAddEndPuncttrue
\mciteSetBstMidEndSepPunct{\mcitedefaultmidpunct}
{\mcitedefaultendpunct}{\mcitedefaultseppunct}\relax
\EndOfBibitem
\bibitem{Mizuk:2009da}
R.~Mizuk {\em et al.} ({Belle} collaboration){,}
  \href{http://dx.doi.org/10.1103/PhysRevD.80.031104}{Phys.\ Rev.\ {\bf D80}{,}
  031104}  (2009), \href{http://arxiv.org/abs/0905.2869}{{\tt arXiv:0905.2869
  [hep-ex]}}\relax
\mciteBstWouldAddEndPuncttrue
\mciteSetBstMidEndSepPunct{\mcitedefaultmidpunct}
{\mcitedefaultendpunct}{\mcitedefaultseppunct}\relax
\EndOfBibitem
\bibitem{Bhardwaj:2013rmw}
V.~Bhardwaj {\em et al.} ({Belle} collaboration){,}
  \href{http://dx.doi.org/10.1103/PhysRevLett.111.032001}{Phys.\ Rev.\ Lett.\ {\bf
  111},  032001}  (2013), \href{http://arxiv.org/abs/1304.3975}{{\tt
  arXiv:1304.3975 [hep-ex]}}\relax
\mciteBstWouldAddEndPuncttrue
\mciteSetBstMidEndSepPunct{\mcitedefaultmidpunct}
{\mcitedefaultendpunct}{\mcitedefaultseppunct}\relax
\EndOfBibitem
\bibitem{Aubert:2005vwa}
B.~Aubert {\em et al.} ({\babar} collaboration){,}
  \href{http://dx.doi.org/10.1103/PhysRevLett.94.171801}{Phys.\ Rev.\ Lett.\ {\bf
  94},  171801}  (2005), \href{http://arxiv.org/abs/hep-ex/0501061}{{\tt
  arXiv:hep-ex/0501061 [hep-ex]}}\relax
\mciteBstWouldAddEndPuncttrue
\mciteSetBstMidEndSepPunct{\mcitedefaultmidpunct}
{\mcitedefaultendpunct}{\mcitedefaultseppunct}\relax
\EndOfBibitem
\bibitem{Aubert:2008ak}
B.~Aubert {\em et al.} ({\babar} collaboration){,}
  \href{http://dx.doi.org/10.1103/PhysRevD.78.091101}{Phys.\ Rev.\ {\bf D78}{,}
  091101}  (2008), \href{http://arxiv.org/abs/0808.1487}{{\tt arXiv:0808.1487
  [hep-ex]}}\relax
\mciteBstWouldAddEndPuncttrue
\mciteSetBstMidEndSepPunct{\mcitedefaultmidpunct}
{\mcitedefaultendpunct}{\mcitedefaultseppunct}\relax
\EndOfBibitem
\bibitem{Bhardwaj:2011dj}
V.~Bhardwaj {\em et al.} ({Belle} collaboration){,}
  \href{http://dx.doi.org/10.1103/PhysRevLett.107.091803}{Phys.\ Rev.\ Lett.\ {\bf
  107},  091803}  (2011), \href{http://arxiv.org/abs/1105.0177}{{\tt
  arXiv:1105.0177 [hep-ex]}}\relax
\mciteBstWouldAddEndPuncttrue
\mciteSetBstMidEndSepPunct{\mcitedefaultmidpunct}
{\mcitedefaultendpunct}{\mcitedefaultseppunct}\relax
\EndOfBibitem
\bibitem{Aubert:2008ae}
B.~Aubert {\em et al.} ({\babar} collaboration){,}
  \href{http://dx.doi.org/10.1103/PhysRevLett.102.132001}{Phys.\ Rev.\ Lett.\ {\bf
  102},  132001}  (2009), \href{http://arxiv.org/abs/0809.0042}{{\tt
  arXiv:0809.0042 [hep-ex]}}\relax
\mciteBstWouldAddEndPuncttrue
\mciteSetBstMidEndSepPunct{\mcitedefaultmidpunct}
{\mcitedefaultendpunct}{\mcitedefaultseppunct}\relax
\EndOfBibitem
\bibitem{Bhardwaj:2015rju}
V.~Bhardwaj {\em et al.} ({Belle} collaboration){,}
  \href{http://dx.doi.org/10.1103/PhysRevD.93.052016}{Phys.\ Rev.\ {\bf D93}{,}
  052016}  (2016), \href{http://arxiv.org/abs/1512.02672}{{\tt
  arXiv:1512.02672 [hep-ex]}}\relax
\mciteBstWouldAddEndPuncttrue
\mciteSetBstMidEndSepPunct{\mcitedefaultmidpunct}
{\mcitedefaultendpunct}{\mcitedefaultseppunct}\relax
\EndOfBibitem
\bibitem{Lees:2011ik}
J.~P.\ Lees {\em et al.} ({\babar} collaboration){,}
  \href{http://dx.doi.org/10.1103/PhysRevD.85.052003}{Phys.\ Rev.\ {\bf D85}{,}
  052003}  (2012), \href{http://arxiv.org/abs/1111.5919}{{\tt arXiv:1111.5919
  [hep-ex]}}\relax
\mciteBstWouldAddEndPuncttrue
\mciteSetBstMidEndSepPunct{\mcitedefaultmidpunct}
{\mcitedefaultendpunct}{\mcitedefaultseppunct}\relax
\EndOfBibitem
\bibitem{Soni:2005fw}
N.~Soni {\em et al.} ({Belle} collaboration){,}
  \href{http://dx.doi.org/10.1016/j.physletb.2006.01.013}{Phys.\ Lett.\ {\bf
  B634},  155}  (2006), \href{http://arxiv.org/abs/hep-ex/0508032}{{\tt
  arXiv:hep-ex/0508032 [hep-ex]}}\relax
\mciteBstWouldAddEndPuncttrue
\mciteSetBstMidEndSepPunct{\mcitedefaultmidpunct}
{\mcitedefaultendpunct}{\mcitedefaultseppunct}\relax
\EndOfBibitem
\bibitem{Fang:2002gi}
F.~Fang {\em et al.} ({Belle} collaboration){,}
  \href{http://dx.doi.org/10.1103/PhysRevLett.90.071801}{Phys.\ Rev.\ Lett.\ {\bf
  90},  071801}  (2003), \href{http://arxiv.org/abs/hep-ex/0208047}{{\tt
  arXiv:hep-ex/0208047 [hep-ex]}}\relax
\mciteBstWouldAddEndPuncttrue
\mciteSetBstMidEndSepPunct{\mcitedefaultmidpunct}
{\mcitedefaultendpunct}{\mcitedefaultseppunct}\relax
\EndOfBibitem
\bibitem{Aubert:2007qea}
B.~Aubert {\em et al.} ({\babar} collaboration){,}
  \href{http://dx.doi.org/10.1103/PhysRevD.76.092004}{Phys.\ Rev.\ {\bf D76}{,}
  092004}  (2007), \href{http://arxiv.org/abs/0707.1648}{{\tt arXiv:0707.1648
  [hep-ex]}}\relax
\mciteBstWouldAddEndPuncttrue
\mciteSetBstMidEndSepPunct{\mcitedefaultmidpunct}
{\mcitedefaultendpunct}{\mcitedefaultseppunct}\relax
\EndOfBibitem
\bibitem{Aubert:2004gc}
B.~Aubert {\em et al.} ({\babar} collaboration){,}
  \href{http://dx.doi.org/10.1103/PhysRevD.70.011101}{Phys.\ Rev.\ {\bf D70}{,}
  011101}  (2004), \href{http://arxiv.org/abs/hep-ex/0403007}{{\tt
  arXiv:hep-ex/0403007 [hep-ex]}}\relax
\mciteBstWouldAddEndPuncttrue
\mciteSetBstMidEndSepPunct{\mcitedefaultmidpunct}
{\mcitedefaultendpunct}{\mcitedefaultseppunct}\relax
\EndOfBibitem
\bibitem{Aubert:2008kp}
B.~Aubert {\em et al.} ({\babar} collaboration){,}
  \href{http://dx.doi.org/10.1103/PhysRevD.78.012006}{Phys.\ Rev.\ {\bf D78}{,}
  012006}  (2008), \href{http://arxiv.org/abs/0804.1208}{{\tt arXiv:0804.1208
  [hep-ex]}}\relax
\mciteBstWouldAddEndPuncttrue
\mciteSetBstMidEndSepPunct{\mcitedefaultmidpunct}
{\mcitedefaultendpunct}{\mcitedefaultseppunct}\relax
\EndOfBibitem
\bibitem{Aubert:2007xw}
B.~Aubert {\em et al.} ({\babar} collaboration){,}
  \href{http://dx.doi.org/10.1103/PhysRevD.76.031101}{Phys.\ Rev.\ {\bf D76}{,}
  031101}  (2007), \href{http://arxiv.org/abs/0704.1266}{{\tt arXiv:0704.1266
  [hep-ex]}}\relax
\mciteBstWouldAddEndPuncttrue
\mciteSetBstMidEndSepPunct{\mcitedefaultmidpunct}
{\mcitedefaultendpunct}{\mcitedefaultseppunct}\relax
\EndOfBibitem
\bibitem{Chang:2012gnb}
M.~Chang {\em et al.} ({Belle} collaboration){,}
  \href{http://dx.doi.org/10.1103/PhysRevD.85.091102}{Phys.\ Rev.\ {\bf D85}{,}
  091102}  (2012), \href{http://arxiv.org/abs/1203.3399}{{\tt arXiv:1203.3399
  [hep-ex]}}\relax
\mciteBstWouldAddEndPuncttrue
\mciteSetBstMidEndSepPunct{\mcitedefaultmidpunct}
{\mcitedefaultendpunct}{\mcitedefaultseppunct}\relax
\EndOfBibitem
\bibitem{Aaij:2013rja}
R.~Aaij {\em et al.} ({LHCb} collaboration){,}
  \href{http://dx.doi.org/10.1103/PhysRevLett.112.091802}{Phys.\ Rev.\ Lett.\ {\bf
  112},  091802}  (2014), \href{http://arxiv.org/abs/1310.2145}{{\tt
  arXiv:1310.2145 [hep-ex]}}\relax
\mciteBstWouldAddEndPuncttrue
\mciteSetBstMidEndSepPunct{\mcitedefaultmidpunct}
{\mcitedefaultendpunct}{\mcitedefaultseppunct}\relax
\EndOfBibitem
\bibitem{Kumar:2008ir}
R.~Kumar {\em et al.} ({Belle} collaboration){,}
  \href{http://dx.doi.org/10.1103/PhysRevD.78.091104}{Phys.\ Rev.\ {\bf D78}{,}
  091104}  (2008), \href{http://arxiv.org/abs/0809.1778}{{\tt arXiv:0809.1778
  [hep-ex]}}\relax
\mciteBstWouldAddEndPuncttrue
\mciteSetBstMidEndSepPunct{\mcitedefaultmidpunct}
{\mcitedefaultendpunct}{\mcitedefaultseppunct}\relax
\EndOfBibitem
\bibitem{Aaij:2013mtm}
R.~Aaij {\em et al.} ({LHCb} collaboration){,}
  \href{http://dx.doi.org/10.1103/PhysRevD.88.072005}{Phys.\ Rev.\ {\bf D88}{,}
  072005}  (2013), \href{http://arxiv.org/abs/1308.5916}{{\tt arXiv:1308.5916
  [hep-ex]}}\relax
\mciteBstWouldAddEndPuncttrue
\mciteSetBstMidEndSepPunct{\mcitedefaultmidpunct}
{\mcitedefaultendpunct}{\mcitedefaultseppunct}\relax
\EndOfBibitem
\bibitem{Aaij:2013zpt}
R.~Aaij {\em et al.} ({LHCb} collaboration){,}
  \href{http://dx.doi.org/10.1103/PhysRevD.87.052001}{Phys.\ Rev.\ {\bf D87}{,}
  052001}  (2013), \href{http://arxiv.org/abs/1301.5347}{{\tt arXiv:1301.5347
  [hep-ex]}}\relax
\mciteBstWouldAddEndPuncttrue
\mciteSetBstMidEndSepPunct{\mcitedefaultmidpunct}
{\mcitedefaultendpunct}{\mcitedefaultseppunct}\relax
\EndOfBibitem
\bibitem{Liu:2008bta}
Y.~Liu {\em et al.} ({Belle} collaboration){,}
  \href{http://dx.doi.org/10.1103/PhysRevD.78.011106}{Phys.\ Rev.\ {\bf D78}{,}
  011106}  (2008), \href{http://arxiv.org/abs/0805.3225}{{\tt arXiv:0805.3225
  [hep-ex]}}\relax
\mciteBstWouldAddEndPuncttrue
\mciteSetBstMidEndSepPunct{\mcitedefaultmidpunct}
{\mcitedefaultendpunct}{\mcitedefaultseppunct}\relax
\EndOfBibitem
\bibitem{Aaij:2015uoa}
R.~Aaij {\em et al.} ({LHCb} collaboration){,}
  \href{http://dx.doi.org/10.1103/PhysRevD.92.112002}{Phys.\ Rev.\ {\bf D92}{,}
  112002}  (2015), \href{http://arxiv.org/abs/1510.04866}{{\tt
  arXiv:1510.04866 [hep-ex]}}\relax
\mciteBstWouldAddEndPuncttrue
\mciteSetBstMidEndSepPunct{\mcitedefaultmidpunct}
{\mcitedefaultendpunct}{\mcitedefaultseppunct}\relax
\EndOfBibitem
\bibitem{Aubert:2004xd}
B.~Aubert {\em et al.} ({\babar} collaboration){,}
  \href{http://dx.doi.org/10.1103/PhysRevD.70.091104}{Phys.\ Rev.\ {\bf D70}{,}
  091104}  (2004), \href{http://arxiv.org/abs/hep-ex/0408018}{{\tt
  arXiv:hep-ex/0408018 [hep-ex]}}\relax
\mciteBstWouldAddEndPuncttrue
\mciteSetBstMidEndSepPunct{\mcitedefaultmidpunct}
{\mcitedefaultendpunct}{\mcitedefaultseppunct}\relax
\EndOfBibitem
\bibitem{Aaij:2013yba}
R.~Aaij {\em et al.} ({LHCb} collaboration){,}
  \href{http://dx.doi.org/10.1007/JHEP09(2013)006}{JHEP {\bf 09},  006}
  (2013), \href{http://arxiv.org/abs/1306.4489}{{\tt arXiv:1306.4489
  [hep-ex]}}\relax
\mciteBstWouldAddEndPuncttrue
\mciteSetBstMidEndSepPunct{\mcitedefaultmidpunct}
{\mcitedefaultendpunct}{\mcitedefaultseppunct}\relax
\EndOfBibitem
\bibitem{Xie:2005tf}
Q.~Xie {\em et al.} ({Belle} collaboration){,}
  \href{http://dx.doi.org/10.1103/PhysRevD.72.051105}{Phys.\ Rev.\ {\bf D72}{,}
  051105}  (2005), \href{http://arxiv.org/abs/hep-ex/0508011}{{\tt
  arXiv:hep-ex/0508011 [hep-ex]}}\relax
\mciteBstWouldAddEndPuncttrue
\mciteSetBstMidEndSepPunct{\mcitedefaultmidpunct}
{\mcitedefaultendpunct}{\mcitedefaultseppunct}\relax
\EndOfBibitem
\bibitem{Aubert:2003ww}
B.~Aubert {\em et al.} ({\babar} collaboration){,}
  \href{http://dx.doi.org/10.1103/PhysRevLett.90.231801}{Phys.\ Rev.\ Lett.\ {\bf
  90},  231801}  (2003), \href{http://arxiv.org/abs/hep-ex/0303036}{{\tt
  arXiv:hep-ex/0303036 [hep-ex]}}\relax
\mciteBstWouldAddEndPuncttrue
\mciteSetBstMidEndSepPunct{\mcitedefaultmidpunct}
{\mcitedefaultendpunct}{\mcitedefaultseppunct}\relax
\EndOfBibitem
\bibitem{Zhang:2005bs}
L.~Zhang {\em et al.} ({Belle} collaboration){,}
  \href{http://dx.doi.org/10.1103/PhysRevD.71.091107}{Phys.\ Rev.\ {\bf D71}{,}
  091107}  (2005), \href{http://arxiv.org/abs/hep-ex/0503037}{{\tt
  arXiv:hep-ex/0503037 [hep-ex]}}\relax
\mciteBstWouldAddEndPuncttrue
\mciteSetBstMidEndSepPunct{\mcitedefaultmidpunct}
{\mcitedefaultendpunct}{\mcitedefaultseppunct}\relax
\EndOfBibitem
\bibitem{Aubert:2005tr}
B.~Aubert {\em et al.} ({\babar} collaboration){,}
  \href{http://dx.doi.org/10.1103/PhysRevD.71.091103}{Phys.\ Rev.\ {\bf D71}{,}
  091103}  (2005), \href{http://arxiv.org/abs/hep-ex/0503021}{{\tt
  arXiv:hep-ex/0503021 [hep-ex]}}\relax
\mciteBstWouldAddEndPuncttrue
\mciteSetBstMidEndSepPunct{\mcitedefaultmidpunct}
{\mcitedefaultendpunct}{\mcitedefaultseppunct}\relax
\EndOfBibitem
\bibitem{Abe:1996kc}
F.~Abe {\em et al.} ({CDF} collaboration){,}
  \href{http://dx.doi.org/10.1103/PhysRevD.54.6596}{Phys.\ Rev.\ {\bf D54}{,}
  6596}  (1996), \href{http://arxiv.org/abs/hep-ex/9607003}{{\tt
  arXiv:hep-ex/9607003 [hep-ex]}}\relax
\mciteBstWouldAddEndPuncttrue
\mciteSetBstMidEndSepPunct{\mcitedefaultmidpunct}
{\mcitedefaultendpunct}{\mcitedefaultseppunct}\relax
\EndOfBibitem
\bibitem{LHCb:2012cw}
R.~Aaij {\em et al.} ({LHCb} collaboration){,}
  \href{http://dx.doi.org/10.1016/j.nuclphysb.2012.10.021}{Nucl.\ Phys.\ {\bf
  B867},  547}  (2013), \href{http://arxiv.org/abs/1210.2631}{{\tt
  arXiv:1210.2631 [hep-ex]}}\relax
\mciteBstWouldAddEndPuncttrue
\mciteSetBstMidEndSepPunct{\mcitedefaultmidpunct}
{\mcitedefaultendpunct}{\mcitedefaultseppunct}\relax
\EndOfBibitem
\bibitem{Aaij:2012dda}
R.~Aaij {\em et al.} ({LHCb} collaboration){,}
  \href{http://dx.doi.org/10.1140/epjc/s10052-012-2118-7}{Eur.\ Phys.\ J.\ {\bf
  C72},  2118}  (2012), \href{http://arxiv.org/abs/1205.0918}{{\tt
  arXiv:1205.0918 [hep-ex]}}\relax
\mciteBstWouldAddEndPuncttrue
\mciteSetBstMidEndSepPunct{\mcitedefaultmidpunct}
{\mcitedefaultendpunct}{\mcitedefaultseppunct}\relax
\EndOfBibitem
\bibitem{Aaij:2013cpa}
R.~Aaij {\em et al.} ({LHCb} collaboration){,}
  \href{http://dx.doi.org/10.1016/j.nuclphysb.2013.03.004}{Nucl.\ Phys.\ {\bf
  B871},  403}  (2013), \href{http://arxiv.org/abs/1302.6354}{{\tt
  arXiv:1302.6354 [hep-ex]}}\relax
\mciteBstWouldAddEndPuncttrue
\mciteSetBstMidEndSepPunct{\mcitedefaultmidpunct}
{\mcitedefaultendpunct}{\mcitedefaultseppunct}\relax
\EndOfBibitem
\bibitem{Aaij:2014jna}
R.~Aaij {\em et al.} ({LHCb} collaboration){,}
  \href{http://dx.doi.org/10.1007/JHEP01(2015)024}{JHEP {\bf 01},  024}
  (2015), \href{http://arxiv.org/abs/1411.0943}{{\tt arXiv:1411.0943
  [hep-ex]}}\relax
\mciteBstWouldAddEndPuncttrue
\mciteSetBstMidEndSepPunct{\mcitedefaultmidpunct}
{\mcitedefaultendpunct}{\mcitedefaultseppunct}\relax
\EndOfBibitem
\bibitem{Aubert:2004ei}
B.~Aubert {\em et al.} ({\babar} collaboration){,}
  \href{http://dx.doi.org/10.1103/PhysRevLett.93.081801}{Phys.\ Rev.\ Lett.\ {\bf
  93},  081801}  (2004), \href{http://arxiv.org/abs/hep-ex/0404005}{{\tt
  arXiv:hep-ex/0404005 [hep-ex]}}\relax
\mciteBstWouldAddEndPuncttrue
\mciteSetBstMidEndSepPunct{\mcitedefaultmidpunct}
{\mcitedefaultendpunct}{\mcitedefaultseppunct}\relax
\EndOfBibitem
\bibitem{Aubert:2010zv}
B.~Aubert {\em et al.} ({\babar} collaboration){,}
  \href{http://dx.doi.org/10.1103/PhysRevD.82.031102}{Phys.\ Rev.\ {\bf D82}{,}
  031102}  (2010), \href{http://arxiv.org/abs/1007.1370}{{\tt arXiv:1007.1370
  [hep-ex]}}\relax
\mciteBstWouldAddEndPuncttrue
\mciteSetBstMidEndSepPunct{\mcitedefaultmidpunct}
{\mcitedefaultendpunct}{\mcitedefaultseppunct}\relax
\EndOfBibitem
\bibitem{Gabyshev:2002zq}
N.~Gabyshev {\em et al.} ({Belle} collaboration){,}
  \href{http://dx.doi.org/10.1103/PhysRevD.66.091102}{Phys.\ Rev.\ {\bf D66}{,}
  091102}  (2002), \href{http://arxiv.org/abs/hep-ex/0208041}{{\tt
  arXiv:hep-ex/0208041 [hep-ex]}}\relax
\mciteBstWouldAddEndPuncttrue
\mciteSetBstMidEndSepPunct{\mcitedefaultmidpunct}
{\mcitedefaultendpunct}{\mcitedefaultseppunct}\relax
\EndOfBibitem
\bibitem{Lees:2013bya}
J.~P.\ Lees {\em et al.} ({\babar} collaboration){,}
  \href{http://dx.doi.org/10.1103/PhysRevD.87.092004}{Phys.\ Rev.\ {\bf D87}{,}
  092004}  (2013), \href{http://arxiv.org/abs/1302.0191}{{\tt arXiv:1302.0191
  [hep-ex]}}\relax
\mciteBstWouldAddEndPuncttrue
\mciteSetBstMidEndSepPunct{\mcitedefaultmidpunct}
{\mcitedefaultendpunct}{\mcitedefaultseppunct}\relax
\EndOfBibitem
\bibitem{Park:2006uj}
K.~Park {\em et al.} ({Belle} collaboration){,}
  \href{http://dx.doi.org/10.1103/PhysRevD.75.011101}{Phys.\ Rev.\ {\bf D75}{,}
  011101}  (2007), \href{http://arxiv.org/abs/hep-ex/0608025}{{\tt
  arXiv:hep-ex/0608025 [hep-ex]}}\relax
\mciteBstWouldAddEndPuncttrue
\mciteSetBstMidEndSepPunct{\mcitedefaultmidpunct}
{\mcitedefaultendpunct}{\mcitedefaultseppunct}\relax
\EndOfBibitem
\bibitem{Abe:2005ib}
K.~Abe {\em et al.} ({Belle} collaboration){,}
  \href{http://dx.doi.org/10.1103/PhysRevLett.97.202003}{Phys.\ Rev.\ Lett.\ {\bf
  97},  202003}  (2006), \href{http://arxiv.org/abs/hep-ex/0508015}{{\tt
  arXiv:hep-ex/0508015 [hep-ex]}}\relax
\mciteBstWouldAddEndPuncttrue
\mciteSetBstMidEndSepPunct{\mcitedefaultmidpunct}
{\mcitedefaultendpunct}{\mcitedefaultseppunct}\relax
\EndOfBibitem
\bibitem{Aubert:2007eb}
B.~Aubert {\em et al.} ({\babar} collaboration){,}
  \href{http://dx.doi.org/10.1103/PhysRevD.77.012002}{Phys.\ Rev.\ {\bf D77}{,}
  012002}  (2008), \href{http://arxiv.org/abs/0710.5763}{{\tt arXiv:0710.5763
  [hep-ex]}}\relax
\mciteBstWouldAddEndPuncttrue
\mciteSetBstMidEndSepPunct{\mcitedefaultmidpunct}
{\mcitedefaultendpunct}{\mcitedefaultseppunct}\relax
\EndOfBibitem
\bibitem{Lees:2014uta}
J.~P.\ Lees {\em et al.} ({\babar} collaboration){,}
  \href{http://dx.doi.org/10.1103/PhysRevD.91.031102}{Phys.\ Rev.\ {\bf D91}{,}
  031102}  (2015), \href{http://arxiv.org/abs/1410.3644}{{\tt arXiv:1410.3644
  [hep-ex]}}\relax
\mciteBstWouldAddEndPuncttrue
\mciteSetBstMidEndSepPunct{\mcitedefaultmidpunct}
{\mcitedefaultendpunct}{\mcitedefaultseppunct}\relax
\EndOfBibitem
\bibitem{Gabyshev:2002dt}
N.~Gabyshev {\em et al.} ({Belle} collaboration){,}
  \href{http://dx.doi.org/10.1103/PhysRevLett.90.121802}{Phys.\ Rev.\ Lett.\ {\bf
  90},  121802}  (2003), \href{http://arxiv.org/abs/hep-ex/0212052}{{\tt
  arXiv:hep-ex/0212052 [hep-ex]}}\relax
\mciteBstWouldAddEndPuncttrue
\mciteSetBstMidEndSepPunct{\mcitedefaultmidpunct}
{\mcitedefaultendpunct}{\mcitedefaultseppunct}\relax
\EndOfBibitem
\bibitem{Aubert:2008ax}
B.~Aubert {\em et al.} ({\babar} collaboration){,}
  \href{http://dx.doi.org/10.1103/PhysRevD.78.112003}{Phys.\ Rev.\ {\bf D78}{,}
  112003}  (2008), \href{http://arxiv.org/abs/0807.4974}{{\tt arXiv:0807.4974
  [hep-ex]}}\relax
\mciteBstWouldAddEndPuncttrue
\mciteSetBstMidEndSepPunct{\mcitedefaultmidpunct}
{\mcitedefaultendpunct}{\mcitedefaultseppunct}\relax
\EndOfBibitem
\bibitem{Aubert:2009aj}
B.~Aubert {\em et al.} ({\babar} collaboration){,}
  \href{http://dx.doi.org/10.1103/PhysRevD.80.051105}{Phys.\ Rev.\ {\bf D80}{,}
  051105}  (2009), \href{http://arxiv.org/abs/0907.4566}{{\tt arXiv:0907.4566
  [hep-ex]}}\relax
\mciteBstWouldAddEndPuncttrue
\mciteSetBstMidEndSepPunct{\mcitedefaultmidpunct}
{\mcitedefaultendpunct}{\mcitedefaultseppunct}\relax
\EndOfBibitem
\bibitem{Chistov:2005zb}
R.~Chistov {\em et al.} ({Belle} collaboration){,}
  \href{http://dx.doi.org/10.1103/PhysRevD.74.111105}{Phys.\ Rev.\ {\bf D74}{,}
  111105}  (2006), \href{http://arxiv.org/abs/hep-ex/0510074}{{\tt
  arXiv:hep-ex/0510074 [hep-ex]}}\relax
\mciteBstWouldAddEndPuncttrue
\mciteSetBstMidEndSepPunct{\mcitedefaultmidpunct}
{\mcitedefaultendpunct}{\mcitedefaultseppunct}\relax
\EndOfBibitem
\bibitem{Uchida:2007gx}
Y.~Uchida {\em et al.} ({Belle} collaboration){,}
  \href{http://dx.doi.org/10.1103/PhysRevD.77.051101}{Phys.\ Rev.\ {\bf D77}{,}
  051101}  (2008), \href{http://arxiv.org/abs/0708.1105}{{\tt arXiv:0708.1105
  [hep-ex]}}\relax
\mciteBstWouldAddEndPuncttrue
\mciteSetBstMidEndSepPunct{\mcitedefaultmidpunct}
{\mcitedefaultendpunct}{\mcitedefaultseppunct}\relax
\EndOfBibitem
\bibitem{Lees:2011rf}
J.~P.\ Lees {\em et al.} ({\babar} collaboration){,}
  \href{http://dx.doi.org/10.1103/PhysRevD.84.071102}{Phys.\ Rev.\ {\bf D84}{,}
  071102}  (2011), \href{http://arxiv.org/abs/1108.3211}{{\tt arXiv:1108.3211
  [hep-ex]}}, Erratum ibid.\
  \href{http://dx.doi.org/10.1103/PhysRevD.85.039903}{{\bf D85}, 039903}
  (2012)\relax
\mciteBstWouldAddEndPuncttrue
\mciteSetBstMidEndSepPunct{\mcitedefaultmidpunct}
{\mcitedefaultendpunct}{\mcitedefaultseppunct}\relax
\EndOfBibitem
\bibitem{TheBABAR:2013fda}
J.~P.\ Lees {\em et al.} ({\babar} collaboration){,}
  \href{http://dx.doi.org/10.1103/PhysRevD.89.071102}{Phys.\ Rev.\ {\bf D89}{,}
  071102}  (2014), \href{http://arxiv.org/abs/1312.6800}{{\tt arXiv:1312.6800
  [hep-ex]}}\relax
\mciteBstWouldAddEndPuncttrue
\mciteSetBstMidEndSepPunct{\mcitedefaultmidpunct}
{\mcitedefaultendpunct}{\mcitedefaultseppunct}\relax
\EndOfBibitem
\bibitem{Aaij:2014pha}
R.~Aaij {\em et al.} ({LHCb} collaboration){,}
  \href{http://dx.doi.org/10.1103/PhysRevLett.112.202001}{Phys.\ Rev.\ Lett.\ {\bf
  112},  202001}  (2014), \href{http://arxiv.org/abs/1403.3606}{{\tt
  arXiv:1403.3606 [hep-ex]}}\relax
\mciteBstWouldAddEndPuncttrue
\mciteSetBstMidEndSepPunct{\mcitedefaultmidpunct}
{\mcitedefaultendpunct}{\mcitedefaultseppunct}\relax
\EndOfBibitem
\bibitem{Aubert:2008gu}
B.~Aubert {\em et al.} ({\babar} collaboration){,}
  \href{http://dx.doi.org/10.1103/PhysRevD.77.111101}{Phys.\ Rev.\ {\bf D77}{,}
  111101}  (2008), \href{http://arxiv.org/abs/0803.2838}{{\tt arXiv:0803.2838
  [hep-ex]}}\relax
\mciteBstWouldAddEndPuncttrue
\mciteSetBstMidEndSepPunct{\mcitedefaultmidpunct}
{\mcitedefaultendpunct}{\mcitedefaultseppunct}\relax
\EndOfBibitem
\bibitem{Mizuk:2008me}
R.~Mizuk {\em et al.} ({Belle} collaboration){,}
  \href{http://dx.doi.org/10.1103/PhysRevD.78.072004}{Phys.\ Rev.\ {\bf D78}{,}
  072004}  (2008), \href{http://arxiv.org/abs/0806.4098}{{\tt arXiv:0806.4098
  [hep-ex]}}\relax
\mciteBstWouldAddEndPuncttrue
\mciteSetBstMidEndSepPunct{\mcitedefaultmidpunct}
{\mcitedefaultendpunct}{\mcitedefaultseppunct}\relax
\EndOfBibitem
\bibitem{Aubert:2005vi}
B.~Aubert {\em et al.} ({\babar} collaboration){,}
  \href{http://dx.doi.org/10.1103/PhysRevLett.96.052002}{Phys.\ Rev.\ Lett.\ {\bf
  96},  052002}  (2006), \href{http://arxiv.org/abs/hep-ex/0510070}{{\tt
  arXiv:hep-ex/0510070 [hep-ex]}}\relax
\mciteBstWouldAddEndPuncttrue
\mciteSetBstMidEndSepPunct{\mcitedefaultmidpunct}
{\mcitedefaultendpunct}{\mcitedefaultseppunct}\relax
\EndOfBibitem
\bibitem{Aubert:2004zr}
B.~Aubert {\em et al.} ({\babar} collaboration){,}
  \href{http://dx.doi.org/10.1103/PhysRevD.71.031501}{Phys.\ Rev.\ {\bf D71}{,}
  031501}  (2005), \href{http://arxiv.org/abs/hep-ex/0412051}{{\tt
  arXiv:hep-ex/0412051 [hep-ex]}}\relax
\mciteBstWouldAddEndPuncttrue
\mciteSetBstMidEndSepPunct{\mcitedefaultmidpunct}
{\mcitedefaultendpunct}{\mcitedefaultseppunct}\relax
\EndOfBibitem
\bibitem{Aubert:2008aa}
B.~Aubert {\em et al.} ({\babar} collaboration){,}
  \href{http://dx.doi.org/10.1103/PhysRevD.79.112001}{Phys.\ Rev.\ {\bf D79}{,}
  112001}  (2009), \href{http://arxiv.org/abs/0811.0564}{{\tt arXiv:0811.0564
  [hep-ex]}}\relax
\mciteBstWouldAddEndPuncttrue
\mciteSetBstMidEndSepPunct{\mcitedefaultmidpunct}
{\mcitedefaultendpunct}{\mcitedefaultseppunct}\relax
\EndOfBibitem
\bibitem{Iwabuchi:2008av}
M.~Iwabuchi {\em et al.} ({Belle} collaboration){,}
  \href{http://dx.doi.org/10.1103/PhysRevLett.101.041601}{Phys.\ Rev.\ Lett.\ {\bf
  101},  041601}  (2008), \href{http://arxiv.org/abs/0804.0831}{{\tt
  arXiv:0804.0831 [hep-ex]}}\relax
\mciteBstWouldAddEndPuncttrue
\mciteSetBstMidEndSepPunct{\mcitedefaultmidpunct}
{\mcitedefaultendpunct}{\mcitedefaultseppunct}\relax
\EndOfBibitem
\bibitem{Abe:2003zm}
K.~Abe {\em et al.} ({Belle} collaboration){,}
  \href{http://dx.doi.org/10.1103/PhysRevD.69.112002}{Phys.\ Rev.\ {\bf D69}{,}
  112002}  (2004), \href{http://arxiv.org/abs/hep-ex/0307021}{{\tt
  arXiv:hep-ex/0307021 [hep-ex]}}\relax
\mciteBstWouldAddEndPuncttrue
\mciteSetBstMidEndSepPunct{\mcitedefaultmidpunct}
{\mcitedefaultendpunct}{\mcitedefaultseppunct}\relax
\EndOfBibitem
\bibitem{Aubert:2009wg}
B.~Aubert {\em et al.} ({\babar} collaboration){,}
  \href{http://dx.doi.org/10.1103/PhysRevD.79.112004}{Phys.\ Rev.\ {\bf D79}{,}
  112004}  (2009), \href{http://arxiv.org/abs/0901.1291}{{\tt arXiv:0901.1291
  [hep-ex]}}\relax
\mciteBstWouldAddEndPuncttrue
\mciteSetBstMidEndSepPunct{\mcitedefaultmidpunct}
{\mcitedefaultendpunct}{\mcitedefaultseppunct}\relax
\EndOfBibitem
\bibitem{Aubert:2003hm}
B.~Aubert {\em et al.} ({\babar} collaboration){,}
  \href{http://arxiv.org/abs/hep-ex/0308026}{{\tt arXiv:hep-ex/0308026
  [hep-ex]}}  (2003)\relax
\mciteBstWouldAddEndPuncttrue
\mciteSetBstMidEndSepPunct{\mcitedefaultmidpunct}
{\mcitedefaultendpunct}{\mcitedefaultseppunct}\relax
\EndOfBibitem
\bibitem{Swain:2003yu}
S.~Swain {\em et al.} ({Belle} collaboration){,}
  \href{http://dx.doi.org/10.1103/PhysRevD.68.051101}{Phys.\ Rev.\ {\bf D68}{,}
  051101}  (2003), \href{http://arxiv.org/abs/hep-ex/0304032}{{\tt
  arXiv:hep-ex/0304032 [hep-ex]}}\relax
\mciteBstWouldAddEndPuncttrue
\mciteSetBstMidEndSepPunct{\mcitedefaultmidpunct}
{\mcitedefaultendpunct}{\mcitedefaultseppunct}\relax
\EndOfBibitem
\bibitem{Aubert:2006um}
B.~Aubert {\em et al.} ({\babar} collaboration){,}
  \href{http://dx.doi.org/10.1103/PhysRevD.73.111104}{Phys.\ Rev.\ {\bf D73}{,}
  111104}  (2006), \href{http://arxiv.org/abs/hep-ex/0604017}{{\tt
  arXiv:hep-ex/0604017 [hep-ex]}}\relax
\mciteBstWouldAddEndPuncttrue
\mciteSetBstMidEndSepPunct{\mcitedefaultmidpunct}
{\mcitedefaultendpunct}{\mcitedefaultseppunct}\relax
\EndOfBibitem
\bibitem{Aubert:2003ae}
B.~Aubert {\em et al.} ({\babar} collaboration){,}
  \href{http://dx.doi.org/10.1103/PhysRevLett.92.141801}{Phys.\ Rev.\ Lett.\ {\bf
  92},  141801}  (2004), \href{http://arxiv.org/abs/hep-ex/0308057}{{\tt
  arXiv:hep-ex/0308057 [hep-ex]}}\relax
\mciteBstWouldAddEndPuncttrue
\mciteSetBstMidEndSepPunct{\mcitedefaultmidpunct}
{\mcitedefaultendpunct}{\mcitedefaultseppunct}\relax
\EndOfBibitem
\bibitem{Aaij:2015vea}
R.~Aaij {\em et al.} ({LHCb} collaboration){,}
  \href{http://dx.doi.org/10.1103/PhysRevD.91.092002}{Phys.\ Rev.\ {\bf D91}{,}
  092002}  (2015), \href{http://arxiv.org/abs/1503.02995}{{\tt
  arXiv:1503.02995 [hep-ex]}}, Erratum ibid.\
  \href{http://dx.doi.org/10.1103/PhysRevD.93.119901}{{\bf D93}, 119901}
  (2016)\relax
\mciteBstWouldAddEndPuncttrue
\mciteSetBstMidEndSepPunct{\mcitedefaultmidpunct}
{\mcitedefaultendpunct}{\mcitedefaultseppunct}\relax
\EndOfBibitem
\bibitem{delAmoSanchez:2010rf}
P.~del Amo~Sanchez {\em et al.} ({\babar} collaboration){,}
  \href{http://dx.doi.org/10.1103/PhysRevD.82.092006}{Phys.\ Rev.\ {\bf D82}{,}
  092006}  (2010), \href{http://arxiv.org/abs/1005.0068}{{\tt arXiv:1005.0068
  [hep-ex]}}\relax
\mciteBstWouldAddEndPuncttrue
\mciteSetBstMidEndSepPunct{\mcitedefaultmidpunct}
{\mcitedefaultendpunct}{\mcitedefaultseppunct}\relax
\EndOfBibitem
\bibitem{Aaij:2015dwl}
R.~Aaij {\em et al.} ({LHCb} collaboration){,}
  \href{http://dx.doi.org/10.1103/PhysRevD.93.051101}{Phys.\ Rev.\ {\bf D93}{,}
  051101}  (2016), \href{http://arxiv.org/abs/1512.02494}{{\tt
  arXiv:1512.02494 [hep-ex]}}, Erratum ibid.\
  \href{http://dx.doi.org/10.1103/PhysRevD.93.119902}{{\bf D93}, 119902}
  (2016)\relax
\mciteBstWouldAddEndPuncttrue
\mciteSetBstMidEndSepPunct{\mcitedefaultmidpunct}
{\mcitedefaultendpunct}{\mcitedefaultseppunct}\relax
\EndOfBibitem
\bibitem{Aubert:2005ra}
B.~Aubert {\em et al.} ({\babar} collaboration){,}
  \href{http://dx.doi.org/10.1103/PhysRevD.72.011102}{Phys.\ Rev.\ {\bf D72}{,}
  011102}  (2005), \href{http://arxiv.org/abs/hep-ex/0505099}{{\tt
  arXiv:hep-ex/0505099 [hep-ex]}}\relax
\mciteBstWouldAddEndPuncttrue
\mciteSetBstMidEndSepPunct{\mcitedefaultmidpunct}
{\mcitedefaultendpunct}{\mcitedefaultseppunct}\relax
\EndOfBibitem
\bibitem{Abulencia:2005ia}
A.~Abulencia {\em et al.} ({CDF} collaboration){,}
  \href{http://dx.doi.org/10.1103/PhysRevLett.96.191801}{Phys.\ Rev.\ Lett.\ {\bf
  96},  191801}  (2006), \href{http://arxiv.org/abs/hep-ex/0508014}{{\tt
  arXiv:hep-ex/0508014 [hep-ex]}}\relax
\mciteBstWouldAddEndPuncttrue
\mciteSetBstMidEndSepPunct{\mcitedefaultmidpunct}
{\mcitedefaultendpunct}{\mcitedefaultseppunct}\relax
\EndOfBibitem
\bibitem{Horii:2008as}
Y.~Horii {\em et al.} ({Belle} collaboration){,}
  \href{http://dx.doi.org/10.1103/PhysRevD.78.071901}{Phys.\ Rev.\ {\bf D78}{,}
  071901}  (2008), \href{http://arxiv.org/abs/0804.2063}{{\tt arXiv:0804.2063
  [hep-ex]}}\relax
\mciteBstWouldAddEndPuncttrue
\mciteSetBstMidEndSepPunct{\mcitedefaultmidpunct}
{\mcitedefaultendpunct}{\mcitedefaultseppunct}\relax
\EndOfBibitem
\bibitem{Aubert:2003uy}
B.~Aubert {\em et al.} ({\babar} collaboration){,}
  \href{http://dx.doi.org/10.1103/PhysRevLett.92.202002}{Phys.\ Rev.\ Lett.\ {\bf
  92},  202002}  (2004), \href{http://arxiv.org/abs/hep-ex/0311032}{{\tt
  arXiv:hep-ex/0311032 [hep-ex]}}\relax
\mciteBstWouldAddEndPuncttrue
\mciteSetBstMidEndSepPunct{\mcitedefaultmidpunct}
{\mcitedefaultendpunct}{\mcitedefaultseppunct}\relax
\EndOfBibitem
\bibitem{Aubert:2004hu}
B.~Aubert {\em et al.} ({\babar} collaboration){,}
  \href{http://dx.doi.org/10.1103/PhysRevD.71.031102}{Phys.\ Rev.\ {\bf D71}{,}
  031102}  (2005), \href{http://arxiv.org/abs/hep-ex/0411091}{{\tt
  arXiv:hep-ex/0411091 [hep-ex]}}\relax
\mciteBstWouldAddEndPuncttrue
\mciteSetBstMidEndSepPunct{\mcitedefaultmidpunct}
{\mcitedefaultendpunct}{\mcitedefaultseppunct}\relax
\EndOfBibitem
\bibitem{Wiechczynski:2009rg}
J.~Wiechczynski {\em et al.} ({Belle} collaboration){,}
  \href{http://dx.doi.org/10.1103/PhysRevD.80.052005}{Phys.\ Rev.\ {\bf D80}{,}
  052005}  (2009), \href{http://arxiv.org/abs/0903.4956}{{\tt arXiv:0903.4956
  [hep-ex]}}\relax
\mciteBstWouldAddEndPuncttrue
\mciteSetBstMidEndSepPunct{\mcitedefaultmidpunct}
{\mcitedefaultendpunct}{\mcitedefaultseppunct}\relax
\EndOfBibitem
\bibitem{Aubert:2006xy}
B.~Aubert {\em et al.} ({\babar} collaboration){,}
  \href{http://dx.doi.org/10.1103/PhysRevLett.98.171801}{Phys.\ Rev.\ Lett.\ {\bf
  98},  171801}  (2007), \href{http://arxiv.org/abs/hep-ex/0611030}{{\tt
  arXiv:hep-ex/0611030 [hep-ex]}}\relax
\mciteBstWouldAddEndPuncttrue
\mciteSetBstMidEndSepPunct{\mcitedefaultmidpunct}
{\mcitedefaultendpunct}{\mcitedefaultseppunct}\relax
\EndOfBibitem
\bibitem{Aaij:2012zh}
R.~Aaij {\em et al.} ({LHCb} collaboration){,}
  \href{http://dx.doi.org/10.1007/JHEP02(2013)043}{JHEP {\bf 02},  043}
  (2013), \href{http://arxiv.org/abs/1210.1089}{{\tt arXiv:1210.1089
  [hep-ex]}}\relax
\mciteBstWouldAddEndPuncttrue
\mciteSetBstMidEndSepPunct{\mcitedefaultmidpunct}
{\mcitedefaultendpunct}{\mcitedefaultseppunct}\relax
\EndOfBibitem
\bibitem{Aubert:2005qq}
B.~Aubert {\em et al.} ({\babar} collaboration){,}
  \href{http://dx.doi.org/10.1103/PhysRevD.73.011103}{Phys.\ Rev.\ {\bf D73}{,}
  011103}  (2006), \href{http://arxiv.org/abs/hep-ex/0512028}{{\tt
  arXiv:hep-ex/0512028 [hep-ex]}}\relax
\mciteBstWouldAddEndPuncttrue
\mciteSetBstMidEndSepPunct{\mcitedefaultmidpunct}
{\mcitedefaultendpunct}{\mcitedefaultseppunct}\relax
\EndOfBibitem
\bibitem{Chen:2011hy}
P.~Chen {\em et al.} ({Belle} collaboration){,}
  \href{http://dx.doi.org/10.1103/PhysRevD.84.071501}{Phys.\ Rev.\ {\bf D84}{,}
  071501}  (2011), \href{http://arxiv.org/abs/1108.4271}{{\tt arXiv:1108.4271
  [hep-ex]}}\relax
\mciteBstWouldAddEndPuncttrue
\mciteSetBstMidEndSepPunct{\mcitedefaultmidpunct}
{\mcitedefaultendpunct}{\mcitedefaultseppunct}\relax
\EndOfBibitem
\bibitem{Seon:2011ni}
O.~Seon {\em et al.} ({Belle} collaboration){,}
  \href{http://dx.doi.org/10.1103/PhysRevD.84.071106}{Phys.\ Rev.\ {\bf D84}{,}
  071106}  (2011), \href{http://arxiv.org/abs/1107.0642}{{\tt arXiv:1107.0642
  [hep-ex]}}\relax
\mciteBstWouldAddEndPuncttrue
\mciteSetBstMidEndSepPunct{\mcitedefaultmidpunct}
{\mcitedefaultendpunct}{\mcitedefaultseppunct}\relax
\EndOfBibitem
\bibitem{Majumder:2005kp}
G.~Majumder {\em et al.} ({Belle} collaboration){,}
  \href{http://dx.doi.org/10.1103/PhysRevLett.95.041803}{Phys.\ Rev.\ Lett.\ {\bf
  95},  041803}  (2005), \href{http://arxiv.org/abs/hep-ex/0502038}{{\tt
  arXiv:hep-ex/0502038 [hep-ex]}}\relax
\mciteBstWouldAddEndPuncttrue
\mciteSetBstMidEndSepPunct{\mcitedefaultmidpunct}
{\mcitedefaultendpunct}{\mcitedefaultseppunct}\relax
\EndOfBibitem
\bibitem{Brodzicka:2007aa}
J.~Brodzicka {\em et al.} ({Belle} collaboration){,}
  \href{http://dx.doi.org/10.1103/PhysRevLett.100.092001}{Phys.\ Rev.\ Lett.\ {\bf
  100},  092001}  (2008), \href{http://arxiv.org/abs/0707.3491}{{\tt
  arXiv:0707.3491 [hep-ex]}}\relax
\mciteBstWouldAddEndPuncttrue
\mciteSetBstMidEndSepPunct{\mcitedefaultmidpunct}
{\mcitedefaultendpunct}{\mcitedefaultseppunct}\relax
\EndOfBibitem
\bibitem{Abe:2003zv}
K.~Abe {\em et al.} ({Belle} collaboration){,}
  \href{http://dx.doi.org/10.1103/PhysRevLett.93.051803}{Phys.\ Rev.\ Lett.\ {\bf
  93},  051803}  (2004), \href{http://arxiv.org/abs/hep-ex/0307061}{{\tt
  arXiv:hep-ex/0307061 [hep-ex]}}\relax
\mciteBstWouldAddEndPuncttrue
\mciteSetBstMidEndSepPunct{\mcitedefaultmidpunct}
{\mcitedefaultendpunct}{\mcitedefaultseppunct}\relax
\EndOfBibitem
\bibitem{Abe:2002haa}
K.~Abe {\em et al.} ({Belle} collaboration){,}
  \href{http://dx.doi.org/10.1016/S0370-2693(02)01969-X}{Phys.\ Lett.\ {\bf
  B538},  11}  (2002), \href{http://arxiv.org/abs/hep-ex/0205021}{{\tt
  arXiv:hep-ex/0205021 [hep-ex]}}\relax
\mciteBstWouldAddEndPuncttrue
\mciteSetBstMidEndSepPunct{\mcitedefaultmidpunct}
{\mcitedefaultendpunct}{\mcitedefaultseppunct}\relax
\EndOfBibitem
\bibitem{Acosta:2002pw}
D.~Acosta {\em et al.} ({CDF} collaboration){,}
  \href{http://dx.doi.org/10.1103/PhysRevD.66.052005}{Phys.\ Rev.\ {\bf D66}{,}
  052005}  (2002)\relax
\mciteBstWouldAddEndPuncttrue
\mciteSetBstMidEndSepPunct{\mcitedefaultmidpunct}
{\mcitedefaultendpunct}{\mcitedefaultseppunct}\relax
\EndOfBibitem
\bibitem{Guler:2010if}
H.~Guler {\em et al.} ({Belle} collaboration){,}
  \href{http://dx.doi.org/10.1103/PhysRevD.83.032005}{Phys.\ Rev.\ {\bf D83}{,}
  032005}  (2011), \href{http://arxiv.org/abs/1009.5256}{{\tt arXiv:1009.5256
  [hep-ex]}}\relax
\mciteBstWouldAddEndPuncttrue
\mciteSetBstMidEndSepPunct{\mcitedefaultmidpunct}
{\mcitedefaultendpunct}{\mcitedefaultseppunct}\relax
\EndOfBibitem
\bibitem{Aubert:2004ns}
B.~Aubert {\em et al.} ({\babar} collaboration){,}
  \href{http://dx.doi.org/10.1103/PhysRevD.71.071103}{Phys.\ Rev.\ {\bf D71}{,}
  071103}  (2005), \href{http://arxiv.org/abs/hep-ex/0406022}{{\tt
  arXiv:hep-ex/0406022 [hep-ex]}}\relax
\mciteBstWouldAddEndPuncttrue
\mciteSetBstMidEndSepPunct{\mcitedefaultmidpunct}
{\mcitedefaultendpunct}{\mcitedefaultseppunct}\relax
\EndOfBibitem
\bibitem{Abe:2001mw}
K.~Abe {\em et al.} ({Belle} collaboration){,}
  \href{http://dx.doi.org/10.1103/PhysRevLett.88.031802}{Phys.\ Rev.\ Lett.\ {\bf
  88},  031802}  (2002), \href{http://arxiv.org/abs/hep-ex/0111069}{{\tt
  arXiv:hep-ex/0111069 [hep-ex]}}\relax
\mciteBstWouldAddEndPuncttrue
\mciteSetBstMidEndSepPunct{\mcitedefaultmidpunct}
{\mcitedefaultendpunct}{\mcitedefaultseppunct}\relax
\EndOfBibitem
\bibitem{Vinokurova:2011dy}
A.~Vinokurova {\em et al.} ({Belle} collaboration){,}
  \href{http://dx.doi.org/10.1016/j.physletb.2011.11.014}{Phys.\ Lett.\ {\bf
  B706},  139}  (2011), \href{http://arxiv.org/abs/1105.0978}{{\tt
  arXiv:1105.0978 [hep-ex]}}\relax
\mciteBstWouldAddEndPuncttrue
\mciteSetBstMidEndSepPunct{\mcitedefaultmidpunct}
{\mcitedefaultendpunct}{\mcitedefaultseppunct}\relax
\EndOfBibitem
\bibitem{Wu:2006vx}
C.-H.\ Wu {\em et al.} ({Belle} collaboration){,}
  \href{http://dx.doi.org/10.1103/PhysRevLett.97.162003}{Phys.\ Rev.\ Lett.\ {\bf
  97},  162003}  (2006), \href{http://arxiv.org/abs/hep-ex/0606022}{{\tt
  arXiv:hep-ex/0606022 [hep-ex]}}\relax
\mciteBstWouldAddEndPuncttrue
\mciteSetBstMidEndSepPunct{\mcitedefaultmidpunct}
{\mcitedefaultendpunct}{\mcitedefaultseppunct}\relax
\EndOfBibitem
\bibitem{Aubert:2005gw}
B.~Aubert {\em et al.} ({\babar} collaboration){,}
  \href{http://dx.doi.org/10.1103/PhysRevD.72.051101}{Phys.\ Rev.\ {\bf D72}{,}
  051101}  (2005), \href{http://arxiv.org/abs/hep-ex/0507012}{{\tt
  arXiv:hep-ex/0507012 [hep-ex]}}\relax
\mciteBstWouldAddEndPuncttrue
\mciteSetBstMidEndSepPunct{\mcitedefaultmidpunct}
{\mcitedefaultendpunct}{\mcitedefaultseppunct}\relax
\EndOfBibitem
\bibitem{Fang:2006bz}
F.~Fang {\em et al.} ({Belle} collaboration){,}
  \href{http://dx.doi.org/10.1103/PhysRevD.74.012007}{Phys.\ Rev.\ {\bf D74}{,}
  012007}  (2006), \href{http://arxiv.org/abs/hep-ex/0605007}{{\tt
  arXiv:hep-ex/0605007 [hep-ex]}}\relax
\mciteBstWouldAddEndPuncttrue
\mciteSetBstMidEndSepPunct{\mcitedefaultmidpunct}
{\mcitedefaultendpunct}{\mcitedefaultseppunct}\relax
\EndOfBibitem
\bibitem{Aaij:2012jw}
R.~Aaij {\em et al.} ({LHCb} collaboration){,}
  \href{http://dx.doi.org/10.1103/PhysRevD.85.091105}{Phys.\ Rev.\ {\bf D85}{,}
  091105}  (2012), \href{http://arxiv.org/abs/1203.3592}{{\tt arXiv:1203.3592
  [hep-ex]}}\relax
\mciteBstWouldAddEndPuncttrue
\mciteSetBstMidEndSepPunct{\mcitedefaultmidpunct}
{\mcitedefaultendpunct}{\mcitedefaultseppunct}\relax
\EndOfBibitem
\bibitem{Aubert:2004pra}
B.~Aubert {\em et al.} ({\babar} collaboration){,}
  \href{http://dx.doi.org/10.1103/PhysRevLett.92.241802}{Phys.\ Rev.\ Lett.\ {\bf
  92},  241802}  (2004), \href{http://arxiv.org/abs/hep-ex/0401035}{{\tt
  arXiv:hep-ex/0401035 [hep-ex]}}\relax
\mciteBstWouldAddEndPuncttrue
\mciteSetBstMidEndSepPunct{\mcitedefaultmidpunct}
{\mcitedefaultendpunct}{\mcitedefaultseppunct}\relax
\EndOfBibitem
\bibitem{Aubert:2005sk}
B.~Aubert {\em et al.} ({\babar} collaboration){,}
  \href{http://dx.doi.org/10.1103/PhysRevD.72.052002}{Phys.\ Rev.\ {\bf D72}{,}
  052002}  (2005), \href{http://arxiv.org/abs/hep-ex/0507025}{{\tt
  arXiv:hep-ex/0507025 [hep-ex]}}\relax
\mciteBstWouldAddEndPuncttrue
\mciteSetBstMidEndSepPunct{\mcitedefaultmidpunct}
{\mcitedefaultendpunct}{\mcitedefaultseppunct}\relax
\EndOfBibitem
\bibitem{Kumar:2006sg}
R.~Kumar {\em et al.} ({Belle} collaboration){,}
  \href{http://dx.doi.org/10.1103/PhysRevD.74.051103}{Phys.\ Rev.\ {\bf D74}{,}
  051103}  (2006), \href{http://arxiv.org/abs/hep-ex/0607008}{{\tt
  arXiv:hep-ex/0607008 [hep-ex]}}\relax
\mciteBstWouldAddEndPuncttrue
\mciteSetBstMidEndSepPunct{\mcitedefaultmidpunct}
{\mcitedefaultendpunct}{\mcitedefaultseppunct}\relax
\EndOfBibitem
\bibitem{Abazov:2008jk}
V.~Abazov {\em et al.} ({\dzero} collaboration){,}
  \href{http://dx.doi.org/10.1103/PhysRevD.79.111102}{Phys.\ Rev.\ {\bf D79}{,}
  111102}  (2009), \href{http://arxiv.org/abs/0805.2576}{{\tt arXiv:0805.2576
  [hep-ex]}}\relax
\mciteBstWouldAddEndPuncttrue
\mciteSetBstMidEndSepPunct{\mcitedefaultmidpunct}
{\mcitedefaultendpunct}{\mcitedefaultseppunct}\relax
\EndOfBibitem
\bibitem{Aaij:2013rha}
R.~Aaij {\em et al.} ({LHCb} collaboration){,}
  \href{http://dx.doi.org/10.1140/epjc/s10052-013-2462-2}{Eur.\ Phys.\ J.\ {\bf
  C73},  2462}  (2013), \href{http://arxiv.org/abs/1303.7133}{{\tt
  arXiv:1303.7133 [hep-ex]}}\relax
\mciteBstWouldAddEndPuncttrue
\mciteSetBstMidEndSepPunct{\mcitedefaultmidpunct}
{\mcitedefaultendpunct}{\mcitedefaultseppunct}\relax
\EndOfBibitem
\bibitem{Abe:1996yya}
F.~Abe {\em et al.} ({CDF} collaboration){,}
  \href{http://dx.doi.org/10.1103/PhysRevLett.77.5176}{Phys.\ Rev.\ Lett.\ {\bf
  77},  5176}  (1996)\relax
\mciteBstWouldAddEndPuncttrue
\mciteSetBstMidEndSepPunct{\mcitedefaultmidpunct}
{\mcitedefaultendpunct}{\mcitedefaultseppunct}\relax
\EndOfBibitem
\bibitem{Abulencia:2007zzb}
A.~Abulencia ({CDF} collaboration){,}
  \href{http://dx.doi.org/10.1103/PhysRevD.79.112003}{Phys.\ Rev.\ {\bf D79}{,}
  112003}  (2009), \href{http://arxiv.org/abs/0905.2146}{{\tt arXiv:0905.2146
  [hep-ex]}}\relax
\mciteBstWouldAddEndPuncttrue
\mciteSetBstMidEndSepPunct{\mcitedefaultmidpunct}
{\mcitedefaultendpunct}{\mcitedefaultseppunct}\relax
\EndOfBibitem
\bibitem{Lees:2012kc}
J.~P.\ Lees {\em et al.} ({\babar} collaboration){,}
  \href{http://dx.doi.org/10.1103/PhysRevD.86.091102}{Phys.\ Rev.\ {\bf D86}{,}
  091102}  (2012), \href{http://arxiv.org/abs/1208.3086}{{\tt arXiv:1208.3086
  [hep-ex]}}\relax
\mciteBstWouldAddEndPuncttrue
\mciteSetBstMidEndSepPunct{\mcitedefaultmidpunct}
{\mcitedefaultendpunct}{\mcitedefaultseppunct}\relax
\EndOfBibitem
\bibitem{Choi:2011fc}
S.~K.\ Choi {\em et al.} ({Belle} collaboration){,}
  \href{http://dx.doi.org/10.1103/PhysRevD.84.052004}{Phys.\ Rev.\ {\bf D84}{,}
  052004}  (2011), \href{http://arxiv.org/abs/1107.0163}{{\tt arXiv:1107.0163
  [hep-ex]}}\relax
\mciteBstWouldAddEndPuncttrue
\mciteSetBstMidEndSepPunct{\mcitedefaultmidpunct}
{\mcitedefaultendpunct}{\mcitedefaultseppunct}\relax
\EndOfBibitem
\bibitem{Aubert:2006aj}
B.~Aubert {\em et al.} ({\babar} collaboration){,}
  \href{http://dx.doi.org/10.1103/PhysRevD.74.071101}{Phys.\ Rev.\ {\bf D74}{,}
  071101}  (2006), \href{http://arxiv.org/abs/hep-ex/0607050}{{\tt
  arXiv:hep-ex/0607050 [hep-ex]}}\relax
\mciteBstWouldAddEndPuncttrue
\mciteSetBstMidEndSepPunct{\mcitedefaultmidpunct}
{\mcitedefaultendpunct}{\mcitedefaultseppunct}\relax
\EndOfBibitem
\bibitem{Aubert:2005zh}
B.~Aubert {\em et al.} ({\babar} collaboration){,}
  \href{http://dx.doi.org/10.1103/PhysRevD.73.011101}{Phys.\ Rev.\ {\bf D73}{,}
  011101}  (2006), \href{http://arxiv.org/abs/hep-ex/0507090}{{\tt
  arXiv:hep-ex/0507090 [hep-ex]}}\relax
\mciteBstWouldAddEndPuncttrue
\mciteSetBstMidEndSepPunct{\mcitedefaultmidpunct}
{\mcitedefaultendpunct}{\mcitedefaultseppunct}\relax
\EndOfBibitem
\bibitem{Aaij:2016iza}
R.~Aaij {\em et al.} ({LHCb} collaboration){,}
  \href{http://dx.doi.org/10.1103/PhysRevLett.118.022003}{Phys.\ Rev.\ Lett.\ {\bf
  118},  022003}  (2017), \href{http://arxiv.org/abs/1606.07895}{{\tt
  arXiv:1606.07895 [hep-ex]}}\relax
\mciteBstWouldAddEndPuncttrue
\mciteSetBstMidEndSepPunct{\mcitedefaultmidpunct}
{\mcitedefaultendpunct}{\mcitedefaultseppunct}\relax
\EndOfBibitem
\bibitem{Abazov:2013xda}
V.~M.\ Abazov {\em et al.} ({\dzero} collaboration){,}
  \href{http://dx.doi.org/10.1103/PhysRevD.89.012004}{Phys.\ Rev.\ {\bf D89}{,}
  012004}  (2014), \href{http://arxiv.org/abs/1309.6580}{{\tt arXiv:1309.6580
  [hep-ex]}}\relax
\mciteBstWouldAddEndPuncttrue
\mciteSetBstMidEndSepPunct{\mcitedefaultmidpunct}
{\mcitedefaultendpunct}{\mcitedefaultseppunct}\relax
\EndOfBibitem
\bibitem{Adachi:2008sua}
T.~Aushev {\em et al.} ({Belle} collaboration){,}
  \href{http://dx.doi.org/10.1103/PhysRevD.81.031103}{Phys.\ Rev.\ {\bf D81}{,}
  031103}  (2010), \href{http://arxiv.org/abs/0810.0358}{{\tt arXiv:0810.0358
  [hep-ex]}}\relax
\mciteBstWouldAddEndPuncttrue
\mciteSetBstMidEndSepPunct{\mcitedefaultmidpunct}
{\mcitedefaultendpunct}{\mcitedefaultseppunct}\relax
\EndOfBibitem
\bibitem{Abe:2004zs}
K.~Abe {\em et al.} ({Belle} collaboration){,}
  \href{http://dx.doi.org/10.1103/PhysRevLett.94.182002}{Phys.\ Rev.\ Lett.\ {\bf
  94},  182002}  (2005), \href{http://arxiv.org/abs/hep-ex/0408126}{{\tt
  arXiv:hep-ex/0408126 [hep-ex]}}\relax
\mciteBstWouldAddEndPuncttrue
\mciteSetBstMidEndSepPunct{\mcitedefaultmidpunct}
{\mcitedefaultendpunct}{\mcitedefaultseppunct}\relax
\EndOfBibitem
\bibitem{Aaij:2012zz}
R.~Aaij {\em et al.} ({LHCb} collaboration){,}
  \href{http://dx.doi.org/10.1007/JHEP06(2012)115}{JHEP {\bf 06},  115}
  (2012), \href{http://arxiv.org/abs/1204.1237}{{\tt arXiv:1204.1237
  [hep-ex]}}\relax
\mciteBstWouldAddEndPuncttrue
\mciteSetBstMidEndSepPunct{\mcitedefaultmidpunct}
{\mcitedefaultendpunct}{\mcitedefaultseppunct}\relax
\EndOfBibitem
\bibitem{Louvot:2010rd}
R.~Louvot {\em et al.} ({Belle} collaboration){,}
  \href{http://dx.doi.org/10.1103/PhysRevLett.104.231801}{Phys.\ Rev.\ Lett.\ {\bf
  104},  231801}  (2010), \href{http://arxiv.org/abs/1003.5312}{{\tt
  arXiv:1003.5312 [hep-ex]}}\relax
\mciteBstWouldAddEndPuncttrue
\mciteSetBstMidEndSepPunct{\mcitedefaultmidpunct}
{\mcitedefaultendpunct}{\mcitedefaultseppunct}\relax
\EndOfBibitem
\bibitem{Aaij:2015dsa}
R.~Aaij {\em et al.} ({LHCb} collaboration){,}
  \href{http://dx.doi.org/10.1007/JHEP06(2015)130}{JHEP {\bf 06},  130}
  (2015), \href{http://arxiv.org/abs/1503.09086}{{\tt arXiv:1503.09086
  [hep-ex]}}\relax
\mciteBstWouldAddEndPuncttrue
\mciteSetBstMidEndSepPunct{\mcitedefaultmidpunct}
{\mcitedefaultendpunct}{\mcitedefaultseppunct}\relax
\EndOfBibitem
\bibitem{Aaij:2016amk}
R.~Aaij {\em et al.} ({LHCb} collaboration){,}
  \href{http://dx.doi.org/10.1103/PhysRevLett.116.161802}{Phys.\ Rev.\ Lett.\ {\bf
  116},  161802}  (2016), \href{http://arxiv.org/abs/1603.02408}{{\tt
  arXiv:1603.02408 [hep-ex]}}\relax
\mciteBstWouldAddEndPuncttrue
\mciteSetBstMidEndSepPunct{\mcitedefaultmidpunct}
{\mcitedefaultendpunct}{\mcitedefaultseppunct}\relax
\EndOfBibitem
\bibitem{Aaij:2011tz}
R.~Aaij {\em et al.} ({LHCb} collaboration){,}
  \href{http://dx.doi.org/10.1016/j.physletb.2011.10.073}{Phys.\ Lett.\ {\bf
  B706},  32}  (2011), \href{http://arxiv.org/abs/1110.3676}{{\tt
  arXiv:1110.3676 [hep-ex]}}\relax
\mciteBstWouldAddEndPuncttrue
\mciteSetBstMidEndSepPunct{\mcitedefaultmidpunct}
{\mcitedefaultendpunct}{\mcitedefaultseppunct}\relax
\EndOfBibitem
\bibitem{Aaij:2013fpa}
R.~Aaij {\em et al.} ({LHCb} collaboration){,}
  \href{http://dx.doi.org/10.1103/PhysRevD.87.071101}{Phys.\ Rev.\ {\bf D87}{,}
  071101}  (2013), \href{http://arxiv.org/abs/1302.6446}{{\tt arXiv:1302.6446
  [hep-ex]}}\relax
\mciteBstWouldAddEndPuncttrue
\mciteSetBstMidEndSepPunct{\mcitedefaultmidpunct}
{\mcitedefaultendpunct}{\mcitedefaultseppunct}\relax
\EndOfBibitem
\bibitem{Aaij:2015rqa}
R.~Aaij {\em et al.} ({LHCb} collaboration){,}
  \href{http://dx.doi.org/10.1007/JHEP08(2015)005}{JHEP {\bf 08},  005}
  (2015), \href{http://arxiv.org/abs/1505.01654}{{\tt arXiv:1505.01654
  [hep-ex]}}\relax
\mciteBstWouldAddEndPuncttrue
\mciteSetBstMidEndSepPunct{\mcitedefaultmidpunct}
{\mcitedefaultendpunct}{\mcitedefaultseppunct}\relax
\EndOfBibitem
\bibitem{Abulencia:2006aa}
A.~Abulencia {\em et al.} ({CDF} collaboration){,}
  \href{http://dx.doi.org/10.1103/PhysRevLett.98.061802}{Phys.\ Rev.\ Lett.\ {\bf
  98},  061802}  (2007), \href{http://arxiv.org/abs/hep-ex/0610045}{{\tt
  arXiv:hep-ex/0610045 [hep-ex]}}\relax
\mciteBstWouldAddEndPuncttrue
\mciteSetBstMidEndSepPunct{\mcitedefaultmidpunct}
{\mcitedefaultendpunct}{\mcitedefaultseppunct}\relax
\EndOfBibitem
\bibitem{Aaij:2013dda}
R.~Aaij {\em et al.} ({LHCb} collaboration){,}
  \href{http://dx.doi.org/10.1016/j.physletb.2013.10.057}{Phys.\ Lett.\ {\bf
  B727},  403}  (2013), \href{http://arxiv.org/abs/1308.4583}{{\tt
  arXiv:1308.4583 [hep-ex]}}\relax
\mciteBstWouldAddEndPuncttrue
\mciteSetBstMidEndSepPunct{\mcitedefaultmidpunct}
{\mcitedefaultendpunct}{\mcitedefaultseppunct}\relax
\EndOfBibitem
\bibitem{Aaltonen:2008ab}
T.~Aaltonen {\em et al.} ({CDF} collaboration){,}
  \href{http://dx.doi.org/10.1103/PhysRevLett.103.191802}{Phys.\ Rev.\ Lett.\ {\bf
  103},  191802}  (2009), \href{http://arxiv.org/abs/0809.0080}{{\tt
  arXiv:0809.0080 [hep-ex]}}\relax
\mciteBstWouldAddEndPuncttrue
\mciteSetBstMidEndSepPunct{\mcitedefaultmidpunct}
{\mcitedefaultendpunct}{\mcitedefaultseppunct}\relax
\EndOfBibitem
\bibitem{Aaltonen:2012mg}
T.~Aaltonen {\em et al.} ({CDF} collaboration){,}
  \href{http://dx.doi.org/10.1103/PhysRevLett.108.201801}{Phys.\ Rev.\ Lett.\ {\bf
  108},  201801}  (2012), \href{http://arxiv.org/abs/1204.0536}{{\tt
  arXiv:1204.0536 [hep-ex]}}\relax
\mciteBstWouldAddEndPuncttrue
\mciteSetBstMidEndSepPunct{\mcitedefaultmidpunct}
{\mcitedefaultendpunct}{\mcitedefaultseppunct}\relax
\EndOfBibitem
\bibitem{Aaij:2016rja}
R.~Aaij {\em et al.} ({LHCb} collaboration){,}
  \href{http://dx.doi.org/10.1103/PhysRevD.93.092008}{Phys.\ Rev.\ {\bf D93}{,}
  092008}  (2016), \href{http://arxiv.org/abs/1602.07543}{{\tt
  arXiv:1602.07543 [hep-ex]}}\relax
\mciteBstWouldAddEndPuncttrue
\mciteSetBstMidEndSepPunct{\mcitedefaultmidpunct}
{\mcitedefaultendpunct}{\mcitedefaultseppunct}\relax
\EndOfBibitem
\bibitem{Belle:2012aa}
J.~Li {\em et al.} ({Belle} collaboration){,}
  \href{http://dx.doi.org/10.1103/PhysRevLett.108.181808}{Phys.\ Rev.\ Lett.\ {\bf
  108},  181808}  (2012), \href{http://arxiv.org/abs/1202.0103}{{\tt
  arXiv:1202.0103 [hep-ex]}}\relax
\mciteBstWouldAddEndPuncttrue
\mciteSetBstMidEndSepPunct{\mcitedefaultmidpunct}
{\mcitedefaultendpunct}{\mcitedefaultseppunct}\relax
\EndOfBibitem
\bibitem{Aaij:2013orb}
R.~Aaij {\em et al.} ({LHCb} collaboration){,}
  \href{http://dx.doi.org/10.1103/PhysRevD.87.072004}{Phys.\ Rev.\ {\bf D87}{,}
  072004}  (2013), \href{http://arxiv.org/abs/1302.1213}{{\tt arXiv:1302.1213
  [hep-ex]}}\relax
\mciteBstWouldAddEndPuncttrue
\mciteSetBstMidEndSepPunct{\mcitedefaultmidpunct}
{\mcitedefaultendpunct}{\mcitedefaultseppunct}\relax
\EndOfBibitem
\bibitem{Thorne:2013llu}
F.~Thorne {\em et al.} ({Belle} collaboration){,}
  \href{http://dx.doi.org/10.1103/PhysRevD.88.114006}{Phys.\ Rev.\ {\bf D88}{,}
  114006}  (2013), \href{http://arxiv.org/abs/1309.0704}{{\tt arXiv:1309.0704
  [hep-ex]}}\relax
\mciteBstWouldAddEndPuncttrue
\mciteSetBstMidEndSepPunct{\mcitedefaultmidpunct}
{\mcitedefaultendpunct}{\mcitedefaultseppunct}\relax
\EndOfBibitem
\bibitem{Aaij:2012di}
R.~Aaij {\em et al.} ({LHCb} collaboration){,}
  \href{http://dx.doi.org/10.1016/j.physletb.2012.05.062}{Phys.\ Lett.\ {\bf
  B713},  172}  (2012), \href{http://arxiv.org/abs/1205.0934}{{\tt
  arXiv:1205.0934 [hep-ex]}}\relax
\mciteBstWouldAddEndPuncttrue
\mciteSetBstMidEndSepPunct{\mcitedefaultmidpunct}
{\mcitedefaultendpunct}{\mcitedefaultseppunct}\relax
\EndOfBibitem
\bibitem{Aaltonen:2011sy}
T.~Aaltonen {\em et al.} ({CDF} collaboration){,}
  \href{http://dx.doi.org/10.1103/PhysRevD.83.052012}{Phys.\ Rev.\ {\bf D83}{,}
  052012}  (2011), \href{http://arxiv.org/abs/1102.1961}{{\tt arXiv:1102.1961
  [hep-ex]}}\relax
\mciteBstWouldAddEndPuncttrue
\mciteSetBstMidEndSepPunct{\mcitedefaultmidpunct}
{\mcitedefaultendpunct}{\mcitedefaultseppunct}\relax
\EndOfBibitem
\bibitem{Aaij:2015mea}
R.~Aaij {\em et al.} ({LHCb} collaboration){,}
  \href{http://dx.doi.org/10.1007/JHEP11(2015)082}{JHEP {\bf 11},  082}
  (2015), \href{http://arxiv.org/abs/1509.00400}{{\tt arXiv:1509.00400
  [hep-ex]}}\relax
\mciteBstWouldAddEndPuncttrue
\mciteSetBstMidEndSepPunct{\mcitedefaultmidpunct}
{\mcitedefaultendpunct}{\mcitedefaultseppunct}\relax
\EndOfBibitem
\bibitem{Aaij:2011ac}
R.~Aaij {\em et al.} ({LHCb} collaboration){,}
  \href{http://dx.doi.org/10.1103/PhysRevLett.108.151801}{Phys.\ Rev.\ Lett.\ {\bf
  108},  151801}  (2012), \href{http://arxiv.org/abs/1112.4695}{{\tt
  arXiv:1112.4695 [hep-ex]}}\relax
\mciteBstWouldAddEndPuncttrue
\mciteSetBstMidEndSepPunct{\mcitedefaultmidpunct}
{\mcitedefaultendpunct}{\mcitedefaultseppunct}\relax
\EndOfBibitem
\bibitem{Abazov:2012dz}
V.~M.\ Abazov {\em et al.} ({\dzero} collaboration){,}
  \href{http://dx.doi.org/10.1103/PhysRevD.86.092011}{Phys.\ Rev.\ {\bf D86}{,}
  092011}  (2012), \href{http://arxiv.org/abs/1204.5723}{{\tt arXiv:1204.5723
  [hep-ex]}}\relax
\mciteBstWouldAddEndPuncttrue
\mciteSetBstMidEndSepPunct{\mcitedefaultmidpunct}
{\mcitedefaultendpunct}{\mcitedefaultseppunct}\relax
\EndOfBibitem
\bibitem{Aaij:2011fx}
R.~Aaij {\em et al.} ({LHCb} collaboration){,}
  \href{http://dx.doi.org/10.1016/j.physletb.2011.03.006}{Phys.\ Lett.\ {\bf
  B698},  115}  (2011), \href{http://arxiv.org/abs/1102.0206}{{\tt
  arXiv:1102.0206 [hep-ex]}}\relax
\mciteBstWouldAddEndPuncttrue
\mciteSetBstMidEndSepPunct{\mcitedefaultmidpunct}
{\mcitedefaultendpunct}{\mcitedefaultseppunct}\relax
\EndOfBibitem
\bibitem{Abulencia:2006jp}
A.~Abulencia {\em et al.} ({CDF} collaboration){,}
  \href{http://dx.doi.org/10.1103/PhysRevLett.96.231801}{Phys.\ Rev.\ Lett.\ {\bf
  96},  231801}  (2006), \href{http://arxiv.org/abs/hep-ex/0602005}{{\tt
  arXiv:hep-ex/0602005 [hep-ex]}}\relax
\mciteBstWouldAddEndPuncttrue
\mciteSetBstMidEndSepPunct{\mcitedefaultmidpunct}
{\mcitedefaultendpunct}{\mcitedefaultseppunct}\relax
\EndOfBibitem
\bibitem{Abazov:2011hv}
V.~M.\ Abazov {\em et al.} ({\dzero} collaboration){,}
  \href{http://dx.doi.org/10.1103/PhysRevD.85.011103}{Phys.\ Rev.\ {\bf D85}{,}
  011103}  (2012), \href{http://arxiv.org/abs/1110.4272}{{\tt arXiv:1110.4272
  [hep-ex]}}\relax
\mciteBstWouldAddEndPuncttrue
\mciteSetBstMidEndSepPunct{\mcitedefaultmidpunct}
{\mcitedefaultendpunct}{\mcitedefaultseppunct}\relax
\EndOfBibitem
\bibitem{Khachatryan:2015lua}
V.~Khachatryan {\em et al.} ({CMS} collaboration){,}
  \href{http://dx.doi.org/10.1016/j.physletb.2016.02.047}{Phys.\ Lett.\ {\bf
  B756},  84}  (2016), \href{http://arxiv.org/abs/1501.06089}{{\tt
  arXiv:1501.06089 [hep-ex]}}\relax
\mciteBstWouldAddEndPuncttrue
\mciteSetBstMidEndSepPunct{\mcitedefaultmidpunct}
{\mcitedefaultendpunct}{\mcitedefaultseppunct}\relax
\EndOfBibitem
\bibitem{Aaij:2016qim}
R.~Aaij {\em et al.} ({LHCb} collaboration){,}
  \href{http://dx.doi.org/10.1007/JHEP03(2016)040}{JHEP {\bf 03},  040}
  (2016), \href{http://arxiv.org/abs/1601.05284}{{\tt arXiv:1601.05284
  [hep-ex]}}\relax
\mciteBstWouldAddEndPuncttrue
\mciteSetBstMidEndSepPunct{\mcitedefaultmidpunct}
{\mcitedefaultendpunct}{\mcitedefaultseppunct}\relax
\EndOfBibitem
\bibitem{Aaij:2015wza}
R.~Aaij {\em et al.} ({LHCb} collaboration){,}
  \href{http://dx.doi.org/10.1016/j.physletb.2015.06.038}{Phys.\ Lett.\ {\bf
  B747},  484}  (2015), \href{http://arxiv.org/abs/1503.07112}{{\tt
  arXiv:1503.07112 [hep-ex]}}\relax
\mciteBstWouldAddEndPuncttrue
\mciteSetBstMidEndSepPunct{\mcitedefaultmidpunct}
{\mcitedefaultendpunct}{\mcitedefaultseppunct}\relax
\EndOfBibitem
\bibitem{Aaij:2014emv}
R.~Aaij {\em et al.} ({LHCb} collaboration){,}
  \href{http://dx.doi.org/10.1103/PhysRevD.89.092006}{Phys.\ Rev.\ {\bf D89}{,}
  092006}  (2014), \href{http://arxiv.org/abs/1402.6248}{{\tt arXiv:1402.6248
  [hep-ex]}}\relax
\mciteBstWouldAddEndPuncttrue
\mciteSetBstMidEndSepPunct{\mcitedefaultmidpunct}
{\mcitedefaultendpunct}{\mcitedefaultseppunct}\relax
\EndOfBibitem
\bibitem{Solovieva:2013rhq}
E.~Solovieva {\em et al.} ({Belle} collaboration){,}
  \href{http://dx.doi.org/10.1016/j.physletb.2013.08.057}{Phys.\ Lett.\ {\bf
  B726},  206}  (2013), \href{http://arxiv.org/abs/1304.6931}{{\tt
  arXiv:1304.6931 [hep-ex]}}\relax
\mciteBstWouldAddEndPuncttrue
\mciteSetBstMidEndSepPunct{\mcitedefaultmidpunct}
{\mcitedefaultendpunct}{\mcitedefaultseppunct}\relax
\EndOfBibitem
\bibitem{Aaij:2013gia}
R.~Aaij {\em et al.} ({LHCb} collaboration){,}
  \href{http://dx.doi.org/10.1103/PhysRevD.87.112012}{Phys.\ Rev.\ {\bf D87}{,}
  112012}  (2013), \href{http://arxiv.org/abs/1304.4530}{{\tt arXiv:1304.4530
  [hep-ex]}}\relax
\mciteBstWouldAddEndPuncttrue
\mciteSetBstMidEndSepPunct{\mcitedefaultmidpunct}
{\mcitedefaultendpunct}{\mcitedefaultseppunct}\relax
\EndOfBibitem
\bibitem{Aad:2015eza}
G.~Aad {\em et al.} ({ATLAS} collaboration){,}
  \href{http://dx.doi.org/10.1140/epjc/s10052-015-3743-8}{Eur.\ Phys.\ J.\ {\bf
  C76},  4}  (2016), \href{http://arxiv.org/abs/1507.07099}{{\tt
  arXiv:1507.07099 [hep-ex]}}\relax
\mciteBstWouldAddEndPuncttrue
\mciteSetBstMidEndSepPunct{\mcitedefaultmidpunct}
{\mcitedefaultendpunct}{\mcitedefaultseppunct}\relax
\EndOfBibitem
\bibitem{LHCb:2012ag}
R.~Aaij {\em et al.} ({LHCb} collaboration){,}
  \href{http://dx.doi.org/10.1103/PhysRevLett.108.251802}{Phys.\ Rev.\ Lett.\ {\bf
  108},  251802}  (2012), \href{http://arxiv.org/abs/1204.0079}{{\tt
  arXiv:1204.0079 [hep-ex]}}\relax
\mciteBstWouldAddEndPuncttrue
\mciteSetBstMidEndSepPunct{\mcitedefaultmidpunct}
{\mcitedefaultendpunct}{\mcitedefaultseppunct}\relax
\EndOfBibitem
\bibitem{Khachatryan:2014nfa}
V.~Khachatryan {\em et al.} ({CMS} collaboration){,}
  \href{http://dx.doi.org/10.1007/JHEP01(2015)063}{JHEP {\bf 01},  063}
  (2015), \href{http://arxiv.org/abs/1410.5729}{{\tt arXiv:1410.5729
  [hep-ex]}}\relax
\mciteBstWouldAddEndPuncttrue
\mciteSetBstMidEndSepPunct{\mcitedefaultmidpunct}
{\mcitedefaultendpunct}{\mcitedefaultseppunct}\relax
\EndOfBibitem
\bibitem{Aaij:2013vcx}
R.~Aaij {\em et al.} ({LHCb} collaboration){,}
  \href{http://dx.doi.org/10.1007/JHEP09(2013)075}{JHEP {\bf 09},  075}
  (2013), \href{http://arxiv.org/abs/1306.6723}{{\tt arXiv:1306.6723
  [hep-ex]}}\relax
\mciteBstWouldAddEndPuncttrue
\mciteSetBstMidEndSepPunct{\mcitedefaultmidpunct}
{\mcitedefaultendpunct}{\mcitedefaultseppunct}\relax
\EndOfBibitem
\bibitem{Aaij:2013gxa}
R.~Aaij {\em et al.} ({LHCb} collaboration){,}
  \href{http://dx.doi.org/10.1007/JHEP11(2013)094}{JHEP {\bf 11},  094}
  (2013), \href{http://arxiv.org/abs/1309.0587}{{\tt arXiv:1309.0587
  [hep-ex]}}\relax
\mciteBstWouldAddEndPuncttrue
\mciteSetBstMidEndSepPunct{\mcitedefaultmidpunct}
{\mcitedefaultendpunct}{\mcitedefaultseppunct}\relax
\EndOfBibitem
\bibitem{Aaij:2015xga}
R.~Aaij {\em et al.} ({LHCb} collaboration){,}
  \href{http://dx.doi.org/10.1103/PhysRevD.92.072007}{Phys.\ Rev.\ {\bf D92}{,}
  072007}  (2015), \href{http://arxiv.org/abs/1507.03516}{{\tt
  arXiv:1507.03516 [hep-ex]}}\relax
\mciteBstWouldAddEndPuncttrue
\mciteSetBstMidEndSepPunct{\mcitedefaultmidpunct}
{\mcitedefaultendpunct}{\mcitedefaultseppunct}\relax
\EndOfBibitem
\bibitem{Aaij:2014ija}
R.~Aaij {\em et al.} ({LHCb} collaboration){,}
  \href{http://dx.doi.org/10.1103/PhysRevLett.114.132001}{Phys.\ Rev.\ Lett.\ {\bf
  114},  132001}  (2015), \href{http://arxiv.org/abs/1411.2943}{{\tt
  arXiv:1411.2943 [hep-ex]}}\relax
\mciteBstWouldAddEndPuncttrue
\mciteSetBstMidEndSepPunct{\mcitedefaultmidpunct}
{\mcitedefaultendpunct}{\mcitedefaultseppunct}\relax
\EndOfBibitem
\bibitem{Aaij:2012dd}
R.~Aaij {\em et al.} ({LHCb} collaboration){,}
  \href{http://dx.doi.org/10.1103/PhysRevLett.109.232001}{Phys.\ Rev.\ Lett.\ {\bf
  109},  232001}  (2012), \href{http://arxiv.org/abs/1209.5634}{{\tt
  arXiv:1209.5634 [hep-ex]}}\relax
\mciteBstWouldAddEndPuncttrue
\mciteSetBstMidEndSepPunct{\mcitedefaultmidpunct}
{\mcitedefaultendpunct}{\mcitedefaultseppunct}\relax
\EndOfBibitem
\bibitem{Aaij:2016xas}
R.~Aaij {\em et al.} ({LHCb} collaboration){,}
  \href{http://dx.doi.org/10.1103/PhysRevD.94.091102}{Phys.\ Rev.\ {\bf D94}{,}
  091102}  (2016), \href{http://arxiv.org/abs/1607.06134}{{\tt
  arXiv:1607.06134 [hep-ex]}}\relax
\mciteBstWouldAddEndPuncttrue
\mciteSetBstMidEndSepPunct{\mcitedefaultmidpunct}
{\mcitedefaultendpunct}{\mcitedefaultseppunct}\relax
\EndOfBibitem
\bibitem{Aaij:2013cda}
R.~Aaij {\em et al.} ({LHCb} collaboration){,}
  \href{http://dx.doi.org/10.1103/PhysRevLett.111.181801}{Phys.\ Rev.\ Lett.\ {\bf
  111},  181801}  (2013), \href{http://arxiv.org/abs/1308.4544}{{\tt
  arXiv:1308.4544 [hep-ex]}}\relax
\mciteBstWouldAddEndPuncttrue
\mciteSetBstMidEndSepPunct{\mcitedefaultmidpunct}
{\mcitedefaultendpunct}{\mcitedefaultseppunct}\relax
\EndOfBibitem
\bibitem{Aaij:2013pka}
R.~Aaij {\em et al.} ({LHCb} collaboration){,}
  \href{http://dx.doi.org/10.1103/PhysRevD.89.032001}{Phys.\ Rev.\ {\bf D89}{,}
  032001}  (2014), \href{http://arxiv.org/abs/1311.4823}{{\tt arXiv:1311.4823
  [hep-ex]}}\relax
\mciteBstWouldAddEndPuncttrue
\mciteSetBstMidEndSepPunct{\mcitedefaultmidpunct}
{\mcitedefaultendpunct}{\mcitedefaultseppunct}\relax
\EndOfBibitem
\bibitem{Aaij:2015fea}
R.~Aaij {\em et al.} ({LHCb} collaboration){,}
  \href{http://dx.doi.org/10.1088/1674-1137/40/1/011001}{Chin.\ Phys.\ {\bf C40}{,}
   011001}  (2016), \href{http://arxiv.org/abs/1509.00292}{{\tt
  arXiv:1509.00292 [hep-ex]}}\relax
\mciteBstWouldAddEndPuncttrue
\mciteSetBstMidEndSepPunct{\mcitedefaultmidpunct}
{\mcitedefaultendpunct}{\mcitedefaultseppunct}\relax
\EndOfBibitem
\bibitem{Abe:1996tr}
F.~Abe {\em et al.} ({CDF} collaboration){,}
  \href{http://dx.doi.org/10.1103/PhysRevD.55.1142}{Phys.\ Rev.\ {\bf D55}{,}
  1142}  (1997)\relax
\mciteBstWouldAddEndPuncttrue
\mciteSetBstMidEndSepPunct{\mcitedefaultmidpunct}
{\mcitedefaultendpunct}{\mcitedefaultseppunct}\relax
\EndOfBibitem
\bibitem{Abazov:2011wt}
V.~M.\ Abazov {\em et al.} ({\dzero} collaboration){,}
  \href{http://dx.doi.org/10.1103/PhysRevD.84.031102}{Phys.\ Rev.\ {\bf D84}{,}
  031102}  (2011), \href{http://arxiv.org/abs/1105.0690}{{\tt arXiv:1105.0690
  [hep-ex]}}\relax
\mciteBstWouldAddEndPuncttrue
\mciteSetBstMidEndSepPunct{\mcitedefaultmidpunct}
{\mcitedefaultendpunct}{\mcitedefaultseppunct}\relax
\EndOfBibitem
\bibitem{Aad:2015msa}
G.~Aad {\em et al.} ({ATLAS} collaboration){,}
  \href{http://dx.doi.org/10.1016/j.physletb.2015.10.009}{Phys.\ Lett.\ {\bf
  B751},  63}  (2015), \href{http://arxiv.org/abs/1507.08202}{{\tt
  arXiv:1507.08202 [hep-ex]}}\relax
\mciteBstWouldAddEndPuncttrue
\mciteSetBstMidEndSepPunct{\mcitedefaultmidpunct}
{\mcitedefaultendpunct}{\mcitedefaultseppunct}\relax
\EndOfBibitem
\bibitem{Aaij:2014zoa}
R.~Aaij {\em et al.} ({LHCb} collaboration){,}
  \href{http://dx.doi.org/10.1007/JHEP07(2014)103}{JHEP {\bf 07},  103}
  (2014), \href{http://arxiv.org/abs/1406.0755}{{\tt arXiv:1406.0755
  [hep-ex]}}\relax
\mciteBstWouldAddEndPuncttrue
\mciteSetBstMidEndSepPunct{\mcitedefaultmidpunct}
{\mcitedefaultendpunct}{\mcitedefaultseppunct}\relax
\EndOfBibitem
\bibitem{Aaij:2016wxd}
R.~Aaij {\em et al.} ({LHCb} collaboration){,}
  \href{http://dx.doi.org/10.1007/JHEP05(2016)132}{JHEP {\bf 05},  132}
  (2016), \href{http://arxiv.org/abs/1603.06961}{{\tt arXiv:1603.06961
  [hep-ex]}}\relax
\mciteBstWouldAddEndPuncttrue
\mciteSetBstMidEndSepPunct{\mcitedefaultmidpunct}
{\mcitedefaultendpunct}{\mcitedefaultseppunct}\relax
\EndOfBibitem
\bibitem{Aad:2014iba}
G.~Aad {\em et al.} ({ATLAS} collaboration){,}
  \href{http://dx.doi.org/10.1103/PhysRevD.89.092009}{Phys.\ Rev.\ {\bf D89}{,}
  092009}  (2014), \href{http://arxiv.org/abs/1404.1071}{{\tt arXiv:1404.1071
  [hep-ex]}}\relax
\mciteBstWouldAddEndPuncttrue
\mciteSetBstMidEndSepPunct{\mcitedefaultmidpunct}
{\mcitedefaultendpunct}{\mcitedefaultseppunct}\relax
\EndOfBibitem
\bibitem{CDF:2011aa}
T.~Aaltonen {\em et al.} ({CDF} collaboration){,}
  \href{http://dx.doi.org/10.1103/PhysRevD.85.032003}{Phys.\ Rev.\ {\bf D85}{,}
  032003}  (2012), \href{http://arxiv.org/abs/1112.3334}{{\tt arXiv:1112.3334
  [hep-ex]}}\relax
\mciteBstWouldAddEndPuncttrue
\mciteSetBstMidEndSepPunct{\mcitedefaultmidpunct}
{\mcitedefaultendpunct}{\mcitedefaultseppunct}\relax
\EndOfBibitem
\bibitem{Abulencia:2006df}
A.~Abulencia {\em et al.} ({CDF} collaboration){,}
  \href{http://dx.doi.org/10.1103/PhysRevLett.98.122002}{Phys.\ Rev.\ Lett.\ {\bf
  98},  122002}  (2007), \href{http://arxiv.org/abs/hep-ex/0601003}{{\tt
  arXiv:hep-ex/0601003 [hep-ex]}}\relax
\mciteBstWouldAddEndPuncttrue
\mciteSetBstMidEndSepPunct{\mcitedefaultmidpunct}
{\mcitedefaultendpunct}{\mcitedefaultseppunct}\relax
\EndOfBibitem
\bibitem{Aaij:2015yoy}
R.~Aaij {\em et al.} ({LHCb} collaboration){,}
  \href{http://dx.doi.org/10.1103/PhysRevLett.115.241801}{Phys.\ Rev.\ Lett.\ {\bf
  115},  241801}  (2015), \href{http://arxiv.org/abs/1510.03829}{{\tt
  arXiv:1510.03829 [hep-ex]}}\relax
\mciteBstWouldAddEndPuncttrue
\mciteSetBstMidEndSepPunct{\mcitedefaultmidpunct}
{\mcitedefaultendpunct}{\mcitedefaultseppunct}\relax
\EndOfBibitem
\bibitem{Duh:2012ie}
Y.~T.\ Duh {\em et al.} ({Belle} collaboration){,}
  \href{http://dx.doi.org/10.1103/PhysRevD.87.031103}{Phys.\ Rev.\ {\bf D87}{,}
  031103}  (2013), \href{http://arxiv.org/abs/1210.1348}{{\tt arXiv:1210.1348
  [hep-ex]}}\relax
\mciteBstWouldAddEndPuncttrue
\mciteSetBstMidEndSepPunct{\mcitedefaultmidpunct}
{\mcitedefaultendpunct}{\mcitedefaultseppunct}\relax
\EndOfBibitem
\bibitem{Bornheim:2003bv}
A.~Bornheim {\em et al.} ({CLEO} collaboration){,}
  \href{http://dx.doi.org/10.1103/PhysRevD.68.052002}{Phys.\ Rev.\ {\bf D68}{,}
  052002}  (2003), \href{http://arxiv.org/abs/hep-ex/0302026}{{\tt
  arXiv:hep-ex/0302026 [hep-ex]}}, Erratum ibid.\
  \href{http://dx.doi.org/10.1103/PhysRevD.75.119907}{{\bf D75}, 119907}
  (2007)\relax
\mciteBstWouldAddEndPuncttrue
\mciteSetBstMidEndSepPunct{\mcitedefaultmidpunct}
{\mcitedefaultendpunct}{\mcitedefaultseppunct}\relax
\EndOfBibitem
\bibitem{Aubert:2007hh}
B.~Aubert {\em et al.} ({\babar} collaboration){,}
  \href{http://dx.doi.org/10.1103/PhysRevD.76.091102}{Phys.\ Rev.\ {\bf D76}{,}
  091102}  (2007), \href{http://arxiv.org/abs/0707.2798}{{\tt arXiv:0707.2798
  [hep-ex]}}\relax
\mciteBstWouldAddEndPuncttrue
\mciteSetBstMidEndSepPunct{\mcitedefaultmidpunct}
{\mcitedefaultendpunct}{\mcitedefaultseppunct}\relax
\EndOfBibitem
\bibitem{Aubert:2009yx}
B.~Aubert {\em et al.} ({\babar} collaboration){,}
  \href{http://dx.doi.org/10.1103/PhysRevD.80.112002}{Phys.\ Rev.\ {\bf D80}{,}
  112002}  (2009), \href{http://arxiv.org/abs/0907.1743}{{\tt arXiv:0907.1743
  [hep-ex]}}\relax
\mciteBstWouldAddEndPuncttrue
\mciteSetBstMidEndSepPunct{\mcitedefaultmidpunct}
{\mcitedefaultendpunct}{\mcitedefaultseppunct}\relax
\EndOfBibitem
\bibitem{Schumann:2006bg}
J.~Schumann {\em et al.} ({Belle} collaboration){,}
  \href{http://dx.doi.org/10.1103/PhysRevLett.97.061802}{Phys.\ Rev.\ Lett.\ {\bf
  97},  061802}  (2006), \href{http://arxiv.org/abs/hep-ex/0603001}{{\tt
  arXiv:hep-ex/0603001 [hep-ex]}}\relax
\mciteBstWouldAddEndPuncttrue
\mciteSetBstMidEndSepPunct{\mcitedefaultmidpunct}
{\mcitedefaultendpunct}{\mcitedefaultseppunct}\relax
\EndOfBibitem
\bibitem{delAmoSanchez:2010qa}
P.~del Amo~Sanchez {\em et al.} ({\babar} collaboration){,}
  \href{http://dx.doi.org/10.1103/PhysRevD.82.011502}{Phys.\ Rev.\ {\bf D82}{,}
  011502}  (2010), \href{http://arxiv.org/abs/1004.0240}{{\tt arXiv:1004.0240
  [hep-ex]}}\relax
\mciteBstWouldAddEndPuncttrue
\mciteSetBstMidEndSepPunct{\mcitedefaultmidpunct}
{\mcitedefaultendpunct}{\mcitedefaultseppunct}\relax
\EndOfBibitem
\bibitem{Schumann:2007ae}
J.~Schumann {\em et al.} ({Belle} collaboration){,}
  \href{http://dx.doi.org/10.1103/PhysRevD.75.092002}{Phys.\ Rev.\ {\bf D75}{,}
  092002}  (2007), \href{http://arxiv.org/abs/hep-ex/0701046}{{\tt
  arXiv:hep-ex/0701046 [hep-ex]}}\relax
\mciteBstWouldAddEndPuncttrue
\mciteSetBstMidEndSepPunct{\mcitedefaultmidpunct}
{\mcitedefaultendpunct}{\mcitedefaultseppunct}\relax
\EndOfBibitem
\bibitem{Hoi:2011gv}
C.~T.\ Hoi {\em et al.} ({Belle} collaboration){,}
  \href{http://dx.doi.org/10.1103/PhysRevLett.108.031801}{Phys.\ Rev.\ Lett.\ {\bf
  108},  031801}  (2012), \href{http://arxiv.org/abs/1110.2000}{{\tt
  arXiv:1110.2000 [hep-ex]}}\relax
\mciteBstWouldAddEndPuncttrue
\mciteSetBstMidEndSepPunct{\mcitedefaultmidpunct}
{\mcitedefaultendpunct}{\mcitedefaultseppunct}\relax
\EndOfBibitem
\bibitem{Richichi:1999kj}
S.~J.\ Richichi {\em et al.} ({CLEO} collaboration){,}
  \href{http://dx.doi.org/10.1103/PhysRevLett.85.520}{Phys.\ Rev.\ Lett.\ {\bf
  85},  520}  (2000), \href{http://arxiv.org/abs/hep-ex/9912059}{{\tt
  arXiv:hep-ex/9912059 [hep-ex]}}\relax
\mciteBstWouldAddEndPuncttrue
\mciteSetBstMidEndSepPunct{\mcitedefaultmidpunct}
{\mcitedefaultendpunct}{\mcitedefaultseppunct}\relax
\EndOfBibitem
\bibitem{Aubert:2006fj}
B.~Aubert {\em et al.} ({\babar} collaboration){,}
  \href{http://dx.doi.org/10.1103/PhysRevLett.97.201802}{Phys.\ Rev.\ Lett.\ {\bf
  97},  201802}  (2006), \href{http://arxiv.org/abs/hep-ex/0608005}{{\tt
  arXiv:hep-ex/0608005 [hep-ex]}}\relax
\mciteBstWouldAddEndPuncttrue
\mciteSetBstMidEndSepPunct{\mcitedefaultmidpunct}
{\mcitedefaultendpunct}{\mcitedefaultseppunct}\relax
\EndOfBibitem
\bibitem{Wang:2007rzb}
C.~H.\ Wang {\em et al.} ({Belle} collaboration){,}
  \href{http://dx.doi.org/10.1103/PhysRevD.75.092005}{Phys.\ Rev.\ {\bf D75}{,}
  092005}  (2007), \href{http://arxiv.org/abs/hep-ex/0701057}{{\tt
  arXiv:hep-ex/0701057 [hep-ex]}}\relax
\mciteBstWouldAddEndPuncttrue
\mciteSetBstMidEndSepPunct{\mcitedefaultmidpunct}
{\mcitedefaultendpunct}{\mcitedefaultseppunct}\relax
\EndOfBibitem
\bibitem{Aubert:2008bk}
B.~Aubert {\em et al.} ({\babar} collaboration){,}
  \href{http://dx.doi.org/10.1103/PhysRevLett.101.091801}{Phys.\ Rev.\ Lett.\ {\bf
  101},  091801}  (2008), \href{http://arxiv.org/abs/0804.0411}{{\tt
  arXiv:0804.0411 [hep-ex]}}\relax
\mciteBstWouldAddEndPuncttrue
\mciteSetBstMidEndSepPunct{\mcitedefaultmidpunct}
{\mcitedefaultendpunct}{\mcitedefaultseppunct}\relax
\EndOfBibitem
\bibitem{Aubert:2007si}
B.~Aubert {\em et al.} ({\babar} collaboration){,}
  \href{http://dx.doi.org/10.1103/PhysRevD.76.031103}{Phys.\ Rev.\ {\bf D76}{,}
  031103}  (2007), \href{http://arxiv.org/abs/0706.3893}{{\tt arXiv:0706.3893
  [hep-ex]}}\relax
\mciteBstWouldAddEndPuncttrue
\mciteSetBstMidEndSepPunct{\mcitedefaultmidpunct}
{\mcitedefaultendpunct}{\mcitedefaultseppunct}\relax
\EndOfBibitem
\bibitem{Jessop:2000bv}
C.~P.\ Jessop {\em et al.} ({CLEO} collaboration){,}
  \href{http://dx.doi.org/10.1103/PhysRevLett.85.2881}{Phys.\ Rev.\ Lett.\ {\bf
  85},  2881}  (2000), \href{http://arxiv.org/abs/hep-ex/0006008}{{\tt
  arXiv:hep-ex/0006008 [hep-ex]}}\relax
\mciteBstWouldAddEndPuncttrue
\mciteSetBstMidEndSepPunct{\mcitedefaultmidpunct}
{\mcitedefaultendpunct}{\mcitedefaultseppunct}\relax
\EndOfBibitem
\bibitem{Aubert:2009sx}
B.~Aubert {\em et al.} ({\babar} collaboration){,}
  \href{http://dx.doi.org/10.1103/PhysRevD.79.052005}{Phys.\ Rev.\ {\bf D79}{,}
  052005}  (2009), \href{http://arxiv.org/abs/0901.3703}{{\tt arXiv:0901.3703
  [hep-ex]}}\relax
\mciteBstWouldAddEndPuncttrue
\mciteSetBstMidEndSepPunct{\mcitedefaultmidpunct}
{\mcitedefaultendpunct}{\mcitedefaultseppunct}\relax
\EndOfBibitem
\bibitem{Aubert:2004hs}
B.~Aubert {\em et al.} ({\babar} collaboration){,}
  \href{http://dx.doi.org/10.1103/PhysRevD.70.111102}{Phys.\ Rev.\ {\bf D70}{,}
  111102}  (2004), \href{http://arxiv.org/abs/hep-ex/0407013}{{\tt
  arXiv:hep-ex/0407013 [hep-ex]}}\relax
\mciteBstWouldAddEndPuncttrue
\mciteSetBstMidEndSepPunct{\mcitedefaultmidpunct}
{\mcitedefaultendpunct}{\mcitedefaultseppunct}\relax
\EndOfBibitem
\bibitem{BABAR:2011aaa}
J.~P.\ Lees {\em et al.} ({\babar} collaboration){,}
  \href{http://dx.doi.org/10.1103/PhysRevD.84.092007}{Phys.\ Rev.\ {\bf D84}{,}
  092007}  (2011), \href{http://arxiv.org/abs/1109.0143}{{\tt arXiv:1109.0143
  [hep-ex]}}\relax
\mciteBstWouldAddEndPuncttrue
\mciteSetBstMidEndSepPunct{\mcitedefaultmidpunct}
{\mcitedefaultendpunct}{\mcitedefaultseppunct}\relax
\EndOfBibitem
\bibitem{Aubert:2008rr}
B.~Aubert {\em et al.} ({\babar} collaboration){,}
  \href{http://dx.doi.org/10.1103/PhysRevD.78.091102}{Phys.\ Rev.\ {\bf D78}{,}
  091102}  (2008), \href{http://arxiv.org/abs/0808.0900}{{\tt arXiv:0808.0900
  [hep-ex]}}\relax
\mciteBstWouldAddEndPuncttrue
\mciteSetBstMidEndSepPunct{\mcitedefaultmidpunct}
{\mcitedefaultendpunct}{\mcitedefaultseppunct}\relax
\EndOfBibitem
\bibitem{Garmash:2003er}
A.~Garmash {\em et al.} ({Belle} collaboration){,}
  \href{http://dx.doi.org/10.1103/PhysRevD.69.012001}{Phys.\ Rev.\ {\bf D69}{,}
  012001}  (2004), \href{http://arxiv.org/abs/hep-ex/0307082}{{\tt
  arXiv:hep-ex/0307082 [hep-ex]}}\relax
\mciteBstWouldAddEndPuncttrue
\mciteSetBstMidEndSepPunct{\mcitedefaultmidpunct}
{\mcitedefaultendpunct}{\mcitedefaultseppunct}\relax
\EndOfBibitem
\bibitem{LHCb:2016rul}
R.~Aaij {\em et al.} ({LHCb} collaboration){,}
  \href{http://dx.doi.org/10.1016/j.physletb.2016.11.053}{Phys.\ Lett.\ {\bf
  B765},  307}  (2017), \href{http://arxiv.org/abs/1608.01478}{{\tt
  arXiv:1608.01478 [hep-ex]}}\relax
\mciteBstWouldAddEndPuncttrue
\mciteSetBstMidEndSepPunct{\mcitedefaultmidpunct}
{\mcitedefaultendpunct}{\mcitedefaultseppunct}\relax
\EndOfBibitem
\bibitem{Bergfeld:1996dd}
T.~Bergfeld {\em et al.} ({CLEO} collaboration){,}
  \href{http://dx.doi.org/10.1103/PhysRevLett.77.4503}{Phys.\ Rev.\ Lett.\ {\bf
  77},  4503}  (1996)\relax
\mciteBstWouldAddEndPuncttrue
\mciteSetBstMidEndSepPunct{\mcitedefaultmidpunct}
{\mcitedefaultendpunct}{\mcitedefaultseppunct}\relax
\EndOfBibitem
\bibitem{Eckhart:2002qr}
E.~Eckhart {\em et al.} ({CLEO} collaboration){,}
  \href{http://dx.doi.org/10.1103/PhysRevLett.89.251801}{Phys.\ Rev.\ Lett.\ {\bf
  89},  251801}  (2002), \href{http://arxiv.org/abs/hep-ex/0206024}{{\tt
  arXiv:hep-ex/0206024 [hep-ex]}}\relax
\mciteBstWouldAddEndPuncttrue
\mciteSetBstMidEndSepPunct{\mcitedefaultmidpunct}
{\mcitedefaultendpunct}{\mcitedefaultseppunct}\relax
\EndOfBibitem
\bibitem{Aubert:2007mb}
B.~Aubert {\em et al.} ({\babar} collaboration){,}
  \href{http://dx.doi.org/10.1103/PhysRevD.76.011103}{Phys.\ Rev.\ {\bf D76}{,}
  011103}  (2007), \href{http://arxiv.org/abs/hep-ex/0702043}{{\tt
  arXiv:hep-ex/0702043 [hep-ex]}}\relax
\mciteBstWouldAddEndPuncttrue
\mciteSetBstMidEndSepPunct{\mcitedefaultmidpunct}
{\mcitedefaultendpunct}{\mcitedefaultseppunct}\relax
\EndOfBibitem
\bibitem{Aubert:2006aw}
B.~Aubert {\em et al.} ({\babar} collaboration){,}
  \href{http://dx.doi.org/10.1103/PhysRevD.74.051104}{Phys.\ Rev.\ {\bf D74}{,}
  051104}  (2006), \href{http://arxiv.org/abs/hep-ex/0607113}{{\tt
  arXiv:hep-ex/0607113 [hep-ex]}}\relax
\mciteBstWouldAddEndPuncttrue
\mciteSetBstMidEndSepPunct{\mcitedefaultmidpunct}
{\mcitedefaultendpunct}{\mcitedefaultseppunct}\relax
\EndOfBibitem
\bibitem{delAmoSanchez:2010mz}
P.~del Amo~Sanchez {\em et al.} ({\babar} collaboration){,}
  \href{http://dx.doi.org/10.1103/PhysRevD.83.051101}{Phys.\ Rev.\ {\bf D83}{,}
  051101}  (2011), \href{http://arxiv.org/abs/1012.4044}{{\tt arXiv:1012.4044
  [hep-ex]}}\relax
\mciteBstWouldAddEndPuncttrue
\mciteSetBstMidEndSepPunct{\mcitedefaultmidpunct}
{\mcitedefaultendpunct}{\mcitedefaultseppunct}\relax
\EndOfBibitem
\bibitem{Aubert:2007ds}
B.~Aubert {\em et al.} ({\babar} collaboration){,}
  \href{http://dx.doi.org/10.1103/PhysRevLett.100.051803}{Phys.\ Rev.\ Lett.\ {\bf
  100},  051803}  (2008), \href{http://arxiv.org/abs/0709.4165}{{\tt
  arXiv:0709.4165 [hep-ex]}}\relax
\mciteBstWouldAddEndPuncttrue
\mciteSetBstMidEndSepPunct{\mcitedefaultmidpunct}
{\mcitedefaultendpunct}{\mcitedefaultseppunct}\relax
\EndOfBibitem
\bibitem{Aubert:2008bg}
B.~Aubert {\em et al.} ({\babar} collaboration){,}
  \href{http://dx.doi.org/10.1103/PhysRevD.78.011104}{Phys.\ Rev.\ {\bf D78}{,}
  011104}  (2008), \href{http://arxiv.org/abs/0805.1217}{{\tt arXiv:0805.1217
  [hep-ex]}}\relax
\mciteBstWouldAddEndPuncttrue
\mciteSetBstMidEndSepPunct{\mcitedefaultmidpunct}
{\mcitedefaultendpunct}{\mcitedefaultseppunct}\relax
\EndOfBibitem
\bibitem{Albrecht:1989ny}
H.~Albrecht {\em et al.} ({ARGUS} collaboration){,}
  \href{http://dx.doi.org/10.1016/0370-2693(91)90436-T}{Phys.\ Lett.\ {\bf B254}{,}
   288}  (1991)\relax
\mciteBstWouldAddEndPuncttrue
\mciteSetBstMidEndSepPunct{\mcitedefaultmidpunct}
{\mcitedefaultendpunct}{\mcitedefaultseppunct}\relax
\EndOfBibitem
\bibitem{Aubert:2007xd}
B.~Aubert {\em et al.} ({\babar} collaboration){,}
  \href{http://dx.doi.org/10.1103/PhysRevLett.99.241803}{Phys.\ Rev.\ Lett.\ {\bf
  99},  241803}  (2007), \href{http://arxiv.org/abs/0707.4561}{{\tt
  arXiv:0707.4561 [hep-ex]}}\relax
\mciteBstWouldAddEndPuncttrue
\mciteSetBstMidEndSepPunct{\mcitedefaultmidpunct}
{\mcitedefaultendpunct}{\mcitedefaultseppunct}\relax
\EndOfBibitem
\bibitem{Aubert:2009qb}
B.~Aubert {\em et al.} ({\babar} collaboration){,}
  \href{http://dx.doi.org/10.1103/PhysRevD.80.051101}{Phys.\ Rev.\ {\bf D80}{,}
  051101}  (2009), \href{http://arxiv.org/abs/0907.3485}{{\tt arXiv:0907.3485
  [hep-ex]}}\relax
\mciteBstWouldAddEndPuncttrue
\mciteSetBstMidEndSepPunct{\mcitedefaultmidpunct}
{\mcitedefaultendpunct}{\mcitedefaultseppunct}\relax
\EndOfBibitem
\bibitem{Aaij:2013fja}
R.~Aaij {\em et al.} ({LHCb} collaboration){,}
  \href{http://dx.doi.org/10.1016/j.physletb.2013.09.046}{Phys.\ Lett.\ {\bf
  B726},  646}  (2013), \href{http://arxiv.org/abs/1308.1277}{{\tt
  arXiv:1308.1277 [hep-ex]}}\relax
\mciteBstWouldAddEndPuncttrue
\mciteSetBstMidEndSepPunct{\mcitedefaultmidpunct}
{\mcitedefaultendpunct}{\mcitedefaultseppunct}\relax
\EndOfBibitem
\bibitem{Aubert:2008aw}
B.~Aubert {\em et al.} ({\babar} collaboration){,}
  \href{http://dx.doi.org/10.1103/PhysRevD.79.051101}{Phys.\ Rev.\ {\bf D79}{,}
  051101}  (2009), \href{http://arxiv.org/abs/0811.1979}{{\tt arXiv:0811.1979
  [hep-ex]}}\relax
\mciteBstWouldAddEndPuncttrue
\mciteSetBstMidEndSepPunct{\mcitedefaultmidpunct}
{\mcitedefaultendpunct}{\mcitedefaultseppunct}\relax
\EndOfBibitem
\bibitem{Aubert:2007xb}
B.~Aubert {\em et al.} ({\babar} collaboration){,}
  \href{http://dx.doi.org/10.1103/PhysRevLett.99.221801}{Phys.\ Rev.\ Lett.\ {\bf
  99},  221801}  (2007), \href{http://arxiv.org/abs/0708.0376}{{\tt
  arXiv:0708.0376 [hep-ex]}}\relax
\mciteBstWouldAddEndPuncttrue
\mciteSetBstMidEndSepPunct{\mcitedefaultmidpunct}
{\mcitedefaultendpunct}{\mcitedefaultseppunct}\relax
\EndOfBibitem
\bibitem{Aubert:2007ua}
B.~Aubert {\em et al.} ({\babar} collaboration){,}
  \href{http://dx.doi.org/10.1103/PhysRevD.76.071103}{Phys.\ Rev.\ {\bf D76}{,}
  071103}  (2007), \href{http://arxiv.org/abs/0706.1059}{{\tt arXiv:0706.1059
  [hep-ex]}}\relax
\mciteBstWouldAddEndPuncttrue
\mciteSetBstMidEndSepPunct{\mcitedefaultmidpunct}
{\mcitedefaultendpunct}{\mcitedefaultseppunct}\relax
\EndOfBibitem
\bibitem{Huang:2003dr}
H.~C.\ Huang {\em et al.} ({Belle} collaboration){,}
  \href{http://dx.doi.org/10.1103/PhysRevLett.91.241802}{Phys.\ Rev.\ Lett.\ {\bf
  91},  241802}  (2003), \href{http://arxiv.org/abs/hep-ex/0305068}{{\tt
  arXiv:hep-ex/0305068 [hep-ex]}}\relax
\mciteBstWouldAddEndPuncttrue
\mciteSetBstMidEndSepPunct{\mcitedefaultmidpunct}
{\mcitedefaultendpunct}{\mcitedefaultseppunct}\relax
\EndOfBibitem
\bibitem{Aubert:2009ax}
B.~Aubert {\em et al.} ({\babar} collaboration){,}
  \href{http://dx.doi.org/10.1103/PhysRevD.79.051102}{Phys.\ Rev.\ {\bf D79}{,}
  051102}  (2009), \href{http://arxiv.org/abs/0901.1223}{{\tt arXiv:0901.1223
  [hep-ex]}}\relax
\mciteBstWouldAddEndPuncttrue
\mciteSetBstMidEndSepPunct{\mcitedefaultmidpunct}
{\mcitedefaultendpunct}{\mcitedefaultseppunct}\relax
\EndOfBibitem
\bibitem{Goh:2015kaa}
Y.~M.\ Goh {\em et al.} ({Belle} collaboration){,}
  \href{http://dx.doi.org/10.1103/PhysRevD.91.071101}{Phys.\ Rev.\ {\bf D91}{,}
  071101}  (2015), \href{http://arxiv.org/abs/1502.00381}{{\tt
  arXiv:1502.00381 [hep-ex]}}\relax
\mciteBstWouldAddEndPuncttrue
\mciteSetBstMidEndSepPunct{\mcitedefaultmidpunct}
{\mcitedefaultendpunct}{\mcitedefaultseppunct}\relax
\EndOfBibitem
\bibitem{Briere:2001ue}
R.~A.\ Briere {\em et al.} ({CLEO} collaboration){,}
  \href{http://dx.doi.org/10.1103/PhysRevLett.86.3718}{Phys.\ Rev.\ Lett.\ {\bf
  86},  3718}  (2001), \href{http://arxiv.org/abs/hep-ex/0101032}{{\tt
  arXiv:hep-ex/0101032 [hep-ex]}}\relax
\mciteBstWouldAddEndPuncttrue
\mciteSetBstMidEndSepPunct{\mcitedefaultmidpunct}
{\mcitedefaultendpunct}{\mcitedefaultseppunct}\relax
\EndOfBibitem
\bibitem{Acosta:2005eu}
D.~Acosta {\em et al.} ({CDF} collaboration){,}
  \href{http://dx.doi.org/10.1103/PhysRevLett.95.031801}{Phys.\ Rev.\ Lett.\ {\bf
  95},  031801}  (2005), \href{http://arxiv.org/abs/hep-ex/0502044}{{\tt
  arXiv:hep-ex/0502044 [hep-ex]}}\relax
\mciteBstWouldAddEndPuncttrue
\mciteSetBstMidEndSepPunct{\mcitedefaultmidpunct}
{\mcitedefaultendpunct}{\mcitedefaultseppunct}\relax
\EndOfBibitem
\bibitem{Abulencia:2005aj}
A.~Abulencia {\em et al.} ({CDF} collaboration){,}
  \href{http://dx.doi.org/10.1103/PhysRevD.73.032003}{Phys.\ Rev.\ {\bf D73}{,}
  032003}  (2006), \href{http://arxiv.org/abs/hep-ex/0510048}{{\tt
  arXiv:hep-ex/0510048 [hep-ex]}}\relax
\mciteBstWouldAddEndPuncttrue
\mciteSetBstMidEndSepPunct{\mcitedefaultmidpunct}
{\mcitedefaultendpunct}{\mcitedefaultseppunct}\relax
\EndOfBibitem
\bibitem{Aubert:2007ac}
B.~Aubert {\em et al.} ({\babar} collaboration){,}
  \href{http://dx.doi.org/10.1103/PhysRevLett.99.201802}{Phys.\ Rev.\ Lett.\ {\bf
  99},  201802}  (2007), \href{http://arxiv.org/abs/0705.1798}{{\tt
  arXiv:0705.1798 [hep-ex]}}\relax
\mciteBstWouldAddEndPuncttrue
\mciteSetBstMidEndSepPunct{\mcitedefaultmidpunct}
{\mcitedefaultendpunct}{\mcitedefaultseppunct}\relax
\EndOfBibitem
\bibitem{Chen:2003jfa}
K.~F.\ Chen {\em et al.} ({Belle} collaboration){,}
  \href{http://dx.doi.org/10.1103/PhysRevLett.91.201801}{Phys.\ Rev.\ Lett.\ {\bf
  91},  201801}  (2003), \href{http://arxiv.org/abs/hep-ex/0307014}{{\tt
  arXiv:hep-ex/0307014 [hep-ex]}}\relax
\mciteBstWouldAddEndPuncttrue
\mciteSetBstMidEndSepPunct{\mcitedefaultmidpunct}
{\mcitedefaultendpunct}{\mcitedefaultseppunct}\relax
\EndOfBibitem
\bibitem{Aubert:2008bc}
B.~Aubert {\em et al.} ({\babar} collaboration){,}
  \href{http://dx.doi.org/10.1103/PhysRevLett.101.161801}{Phys.\ Rev.\ Lett.\ {\bf
  101},  161801}  (2008), \href{http://arxiv.org/abs/0806.4419}{{\tt
  arXiv:0806.4419 [hep-ex]}}\relax
\mciteBstWouldAddEndPuncttrue
\mciteSetBstMidEndSepPunct{\mcitedefaultmidpunct}
{\mcitedefaultendpunct}{\mcitedefaultseppunct}\relax
\EndOfBibitem
\bibitem{Sanchez:2010qm}
P.~del Amo~Sanchez {\em et al.} ({\babar} collaboration){,}
  \href{http://dx.doi.org/10.1103/PhysRevD.82.091101}{Phys.\ Rev.\ {\bf D82}{,}
  091101}  (2010), \href{http://arxiv.org/abs/1007.2732}{{\tt arXiv:1007.2732
  [hep-ex]}}\relax
\mciteBstWouldAddEndPuncttrue
\mciteSetBstMidEndSepPunct{\mcitedefaultmidpunct}
{\mcitedefaultendpunct}{\mcitedefaultseppunct}\relax
\EndOfBibitem
\bibitem{Lees:2011zh}
J.~P.\ Lees {\em et al.} ({\babar} collaboration){,}
  \href{http://dx.doi.org/10.1103/PhysRevD.84.012001}{Phys.\ Rev.\ {\bf D84}{,}
  012001}  (2011), \href{http://arxiv.org/abs/1105.5159}{{\tt arXiv:1105.5159
  [hep-ex]}}\relax
\mciteBstWouldAddEndPuncttrue
\mciteSetBstMidEndSepPunct{\mcitedefaultmidpunct}
{\mcitedefaultendpunct}{\mcitedefaultseppunct}\relax
\EndOfBibitem
\bibitem{Aubert:2006caa}
B.~Aubert {\em et al.} ({\babar} collaboration){,}
  \href{http://dx.doi.org/10.1103/PhysRevD.74.031105}{Phys.\ Rev.\ {\bf D74}{,}
  031105}  (2006), \href{http://arxiv.org/abs/hep-ex/0605008}{{\tt
  arXiv:hep-ex/0605008 [hep-ex]}}\relax
\mciteBstWouldAddEndPuncttrue
\mciteSetBstMidEndSepPunct{\mcitedefaultmidpunct}
{\mcitedefaultendpunct}{\mcitedefaultseppunct}\relax
\EndOfBibitem
\bibitem{Liu:2009kca}
C.~Liu {\em et al.} ({Belle} collaboration){,}
  \href{http://dx.doi.org/10.1103/PhysRevD.79.071102}{Phys.\ Rev.\ {\bf D79}{,}
  071102}  (2009), \href{http://arxiv.org/abs/0902.4757}{{\tt arXiv:0902.4757
  [hep-ex]}}\relax
\mciteBstWouldAddEndPuncttrue
\mciteSetBstMidEndSepPunct{\mcitedefaultmidpunct}
{\mcitedefaultendpunct}{\mcitedefaultseppunct}\relax
\EndOfBibitem
\bibitem{Aubert:2009av}
B.~Aubert {\em et al.} ({\babar} collaboration){,}
  \href{http://dx.doi.org/10.1103/PhysRevD.79.072006}{Phys.\ Rev.\ {\bf D79}{,}
  072006}  (2009), \href{http://arxiv.org/abs/0902.2051}{{\tt arXiv:0902.2051
  [hep-ex]}}\relax
\mciteBstWouldAddEndPuncttrue
\mciteSetBstMidEndSepPunct{\mcitedefaultmidpunct}
{\mcitedefaultendpunct}{\mcitedefaultseppunct}\relax
\EndOfBibitem
\bibitem{Gordon:2002yt}
A.~Gordon {\em et al.} ({Belle} collaboration){,}
  \href{http://dx.doi.org/10.1016/S0370-2693(02)02374-2}{Phys.\ Lett.\ {\bf
  B542},  183}  (2002), \href{http://arxiv.org/abs/hep-ex/0207007}{{\tt
  arXiv:hep-ex/0207007 [hep-ex]}}\relax
\mciteBstWouldAddEndPuncttrue
\mciteSetBstMidEndSepPunct{\mcitedefaultmidpunct}
{\mcitedefaultendpunct}{\mcitedefaultseppunct}\relax
\EndOfBibitem
\bibitem{Albrecht:1990am}
H.~Albrecht {\em et al.} ({ARGUS} collaboration){,}
  \href{http://dx.doi.org/10.1016/0370-2693(90)91293-K}{Phys.\ Lett.\ {\bf B241}{,}
   278}  (1990)\relax
\mciteBstWouldAddEndPuncttrue
\mciteSetBstMidEndSepPunct{\mcitedefaultmidpunct}
{\mcitedefaultendpunct}{\mcitedefaultseppunct}\relax
\EndOfBibitem
\bibitem{Aubert:2007py}
B.~Aubert {\em et al.} ({\babar} collaboration){,}
  \href{http://dx.doi.org/10.1103/PhysRevD.75.091103}{Phys.\ Rev.\ {\bf D75}{,}
  091103}  (2007), \href{http://arxiv.org/abs/hep-ex/0701035}{{\tt
  arXiv:hep-ex/0701035 [hep-ex]}}\relax
\mciteBstWouldAddEndPuncttrue
\mciteSetBstMidEndSepPunct{\mcitedefaultmidpunct}
{\mcitedefaultendpunct}{\mcitedefaultseppunct}\relax
\EndOfBibitem
\bibitem{Zhang:2004wza}
J.~Zhang {\em et al.} ({Belle} collaboration){,}
  \href{http://dx.doi.org/10.1103/PhysRevLett.94.031801}{Phys.\ Rev.\ Lett.\ {\bf
  94},  031801}  (2005), \href{http://arxiv.org/abs/hep-ex/0406006}{{\tt
  arXiv:hep-ex/0406006 [hep-ex]}}\relax
\mciteBstWouldAddEndPuncttrue
\mciteSetBstMidEndSepPunct{\mcitedefaultmidpunct}
{\mcitedefaultendpunct}{\mcitedefaultseppunct}\relax
\EndOfBibitem
\bibitem{Zhang:2003up}
J.~Zhang {\em et al.} ({Belle} collaboration){,}
  \href{http://dx.doi.org/10.1103/PhysRevLett.91.221801}{Phys.\ Rev.\ Lett.\ {\bf
  91},  221801}  (2003), \href{http://arxiv.org/abs/hep-ex/0306007}{{\tt
  arXiv:hep-ex/0306007 [hep-ex]}}\relax
\mciteBstWouldAddEndPuncttrue
\mciteSetBstMidEndSepPunct{\mcitedefaultmidpunct}
{\mcitedefaultendpunct}{\mcitedefaultseppunct}\relax
\EndOfBibitem
\bibitem{Aubert:2007kpb}
B.~Aubert {\em et al.} ({\babar} collaboration){,}
  \href{http://dx.doi.org/10.1103/PhysRevLett.99.261801}{Phys.\ Rev.\ Lett.\ {\bf
  99},  261801}  (2007), \href{http://arxiv.org/abs/0708.0050}{{\tt
  arXiv:0708.0050 [hep-ex]}}\relax
\mciteBstWouldAddEndPuncttrue
\mciteSetBstMidEndSepPunct{\mcitedefaultmidpunct}
{\mcitedefaultendpunct}{\mcitedefaultseppunct}\relax
\EndOfBibitem
\bibitem{Jen:2006in}
C.~M.\ Jen {\em et al.} ({Belle} collaboration){,}
  \href{http://dx.doi.org/10.1103/PhysRevD.74.111101}{Phys.\ Rev.\ {\bf D74}{,}
  111101}  (2006), \href{http://arxiv.org/abs/hep-ex/0609022}{{\tt
  arXiv:hep-ex/0609022 [hep-ex]}}\relax
\mciteBstWouldAddEndPuncttrue
\mciteSetBstMidEndSepPunct{\mcitedefaultmidpunct}
{\mcitedefaultendpunct}{\mcitedefaultseppunct}\relax
\EndOfBibitem
\bibitem{Aubert:2008fu}
B.~Aubert {\em et al.} ({\babar} collaboration){,}
  \href{http://dx.doi.org/10.1103/PhysRevD.78.011107}{Phys.\ Rev.\ {\bf D78}{,}
  011107}  (2008), \href{http://arxiv.org/abs/0804.2422}{{\tt arXiv:0804.2422
  [hep-ex]}}\relax
\mciteBstWouldAddEndPuncttrue
\mciteSetBstMidEndSepPunct{\mcitedefaultmidpunct}
{\mcitedefaultendpunct}{\mcitedefaultseppunct}\relax
\EndOfBibitem
\bibitem{Aubert:2006nn}
B.~Aubert {\em et al.} ({\babar} collaboration){,}
  \href{http://dx.doi.org/10.1103/PhysRevD.74.011102}{Phys.\ Rev.\ {\bf D74}{,}
  011102}  (2006), \href{http://arxiv.org/abs/hep-ex/0605037}{{\tt
  arXiv:hep-ex/0605037 [hep-ex]}}\relax
\mciteBstWouldAddEndPuncttrue
\mciteSetBstMidEndSepPunct{\mcitedefaultmidpunct}
{\mcitedefaultendpunct}{\mcitedefaultseppunct}\relax
\EndOfBibitem
\bibitem{Kim:2012gt}
J.~H.\ Kim {\em et al.} ({Belle} collaboration){,}
  \href{http://dx.doi.org/10.1103/PhysRevD.86.031101}{Phys.\ Rev.\ {\bf D86}{,}
  031101}  (2012), \href{http://arxiv.org/abs/1206.4760}{{\tt arXiv:1206.4760
  [hep-ex]}}\relax
\mciteBstWouldAddEndPuncttrue
\mciteSetBstMidEndSepPunct{\mcitedefaultmidpunct}
{\mcitedefaultendpunct}{\mcitedefaultseppunct}\relax
\EndOfBibitem
\bibitem{Aaij:2013lja}
R.~Aaij {\em et al.} ({LHCb} collaboration){,}
  \href{http://dx.doi.org/10.1016/j.physletb.2013.11.036}{Phys.\ Lett.\ {\bf
  B728},  85}  (2014), \href{http://arxiv.org/abs/1309.3742}{{\tt
  arXiv:1309.3742 [hep-ex]}}\relax
\mciteBstWouldAddEndPuncttrue
\mciteSetBstMidEndSepPunct{\mcitedefaultmidpunct}
{\mcitedefaultendpunct}{\mcitedefaultseppunct}\relax
\EndOfBibitem
\bibitem{Aubert:2008fq}
B.~Aubert {\em et al.} ({\babar} collaboration){,}
  \href{http://dx.doi.org/10.1103/PhysRevLett.101.201801}{Phys.\ Rev.\ Lett.\ {\bf
  101},  201801}  (2008), \href{http://arxiv.org/abs/0807.3935}{{\tt
  arXiv:0807.3935 [hep-ex]}}\relax
\mciteBstWouldAddEndPuncttrue
\mciteSetBstMidEndSepPunct{\mcitedefaultmidpunct}
{\mcitedefaultendpunct}{\mcitedefaultseppunct}\relax
\EndOfBibitem
\bibitem{Aubert:2007vf}
B.~Aubert {\em et al.} ({\babar} collaboration){,}
  \href{http://dx.doi.org/10.1103/PhysRevD.77.011101}{Phys.\ Rev.\ {\bf D77}{,}
  011101}  (2008), \href{http://arxiv.org/abs/0708.0963}{{\tt arXiv:0708.0963
  [hep-ex]}}\relax
\mciteBstWouldAddEndPuncttrue
\mciteSetBstMidEndSepPunct{\mcitedefaultmidpunct}
{\mcitedefaultendpunct}{\mcitedefaultseppunct}\relax
\EndOfBibitem
\bibitem{Bortoletto:1989mu}
D.~Bortoletto {\em et al.} ({CLEO} collaboration){,}
  \href{http://dx.doi.org/10.1103/PhysRevLett.62.2436}{Phys.\ Rev.\ Lett.\ {\bf
  62},  2436}  (1989)\relax
\mciteBstWouldAddEndPuncttrue
\mciteSetBstMidEndSepPunct{\mcitedefaultmidpunct}
{\mcitedefaultendpunct}{\mcitedefaultseppunct}\relax
\EndOfBibitem
\bibitem{Aubert:2006fha}
B.~Aubert {\em et al.} ({\babar} collaboration){,}
  \href{http://dx.doi.org/10.1103/PhysRevD.75.012008}{Phys.\ Rev.\ {\bf D75}{,}
  012008}  (2007), \href{http://arxiv.org/abs/hep-ex/0608003}{{\tt
  arXiv:hep-ex/0608003 [hep-ex]}}\relax
\mciteBstWouldAddEndPuncttrue
\mciteSetBstMidEndSepPunct{\mcitedefaultmidpunct}
{\mcitedefaultendpunct}{\mcitedefaultseppunct}\relax
\EndOfBibitem
\bibitem{Sato:2014gcp}
S.~Sato {\em et al.} ({Belle} collaboration){,}
  \href{http://dx.doi.org/10.1103/PhysRevD.90.072009}{Phys.\ Rev.\ {\bf D90}{,}
  072009}  (2014), \href{http://arxiv.org/abs/1408.6343}{{\tt arXiv:1408.6343
  [hep-ex]}}\relax
\mciteBstWouldAddEndPuncttrue
\mciteSetBstMidEndSepPunct{\mcitedefaultmidpunct}
{\mcitedefaultendpunct}{\mcitedefaultseppunct}\relax
\EndOfBibitem
\bibitem{Aubert:2007ij}
B.~Aubert {\em et al.} ({\babar} collaboration){,}
  \href{http://dx.doi.org/10.1103/PhysRevD.75.111102}{Phys.\ Rev.\ {\bf D75}{,}
  111102}  (2007), \href{http://arxiv.org/abs/hep-ex/0703038}{{\tt
  arXiv:hep-ex/0703038 [hep-ex]}}\relax
\mciteBstWouldAddEndPuncttrue
\mciteSetBstMidEndSepPunct{\mcitedefaultmidpunct}
{\mcitedefaultendpunct}{\mcitedefaultseppunct}\relax
\EndOfBibitem
\bibitem{Ammar:2001gi}
R.~Ammar {\em et al.} ({CLEO} collaboration){,}
  \href{http://dx.doi.org/10.1103/PhysRevLett.87.271801}{Phys.\ Rev.\ Lett.\ {\bf
  87},  271801}  (2001), \href{http://arxiv.org/abs/hep-ex/0106038}{{\tt
  arXiv:hep-ex/0106038 [hep-ex]}}\relax
\mciteBstWouldAddEndPuncttrue
\mciteSetBstMidEndSepPunct{\mcitedefaultmidpunct}
{\mcitedefaultendpunct}{\mcitedefaultseppunct}\relax
\EndOfBibitem
\bibitem{Goldenzweig:2008sz}
P.~Goldenzweig {\em et al.} ({Belle} collaboration){,}
  \href{http://dx.doi.org/10.1103/PhysRevLett.101.231801}{Phys.\ Rev.\ Lett.\ {\bf
  101},  231801}  (2008), \href{http://arxiv.org/abs/0807.4271}{{\tt
  arXiv:0807.4271 [hep-ex]}}\relax
\mciteBstWouldAddEndPuncttrue
\mciteSetBstMidEndSepPunct{\mcitedefaultmidpunct}
{\mcitedefaultendpunct}{\mcitedefaultseppunct}\relax
\EndOfBibitem
\bibitem{BABAR:2011ae}
J.~P.\ Lees {\em et al.} ({\babar} collaboration){,}
  \href{http://dx.doi.org/10.1103/PhysRevD.83.112010}{Phys.\ Rev.\ {\bf D83}{,}
  112010}  (2011), \href{http://arxiv.org/abs/1105.0125}{{\tt arXiv:1105.0125
  [hep-ex]}}\relax
\mciteBstWouldAddEndPuncttrue
\mciteSetBstMidEndSepPunct{\mcitedefaultmidpunct}
{\mcitedefaultendpunct}{\mcitedefaultseppunct}\relax
\EndOfBibitem
\bibitem{Chang:2004um}
P.~Chang {\em et al.} ({Belle} collaboration){,}
  \href{http://dx.doi.org/10.1016/j.physletb.2004.07.063}{Phys.\ Lett.\ {\bf
  B599},  148}  (2004), \href{http://arxiv.org/abs/hep-ex/0406075}{{\tt
  arXiv:hep-ex/0406075 [hep-ex]}}\relax
\mciteBstWouldAddEndPuncttrue
\mciteSetBstMidEndSepPunct{\mcitedefaultmidpunct}
{\mcitedefaultendpunct}{\mcitedefaultseppunct}\relax
\EndOfBibitem
\bibitem{Aubert:2007bs}
B.~Aubert {\em et al.} ({\babar} collaboration){,}
  \href{http://dx.doi.org/10.1103/PhysRevD.78.052005}{Phys.\ Rev.\ {\bf D78}{,}
  052005}  (2008), \href{http://arxiv.org/abs/0711.4417}{{\tt arXiv:0711.4417
  [hep-ex]}}\relax
\mciteBstWouldAddEndPuncttrue
\mciteSetBstMidEndSepPunct{\mcitedefaultmidpunct}
{\mcitedefaultendpunct}{\mcitedefaultseppunct}\relax
\EndOfBibitem
\bibitem{Garmash:2006fh}
A.~Garmash {\em et al.} ({Belle} collaboration){,}
  \href{http://dx.doi.org/10.1103/PhysRevD.75.012006}{Phys.\ Rev.\ {\bf D75}{,}
  012006}  (2007), \href{http://arxiv.org/abs/hep-ex/0610081}{{\tt
  arXiv:hep-ex/0610081 [hep-ex]}}\relax
\mciteBstWouldAddEndPuncttrue
\mciteSetBstMidEndSepPunct{\mcitedefaultmidpunct}
{\mcitedefaultendpunct}{\mcitedefaultseppunct}\relax
\EndOfBibitem
\bibitem{Aaij:2013uta}
R.~Aaij {\em et al.} ({LHCb} collaboration){,}
  \href{http://dx.doi.org/10.1007/JHEP10(2013)143}{JHEP {\bf 10},  143}
  (2013), \href{http://arxiv.org/abs/1307.7648}{{\tt arXiv:1307.7648
  [hep-ex]}}\relax
\mciteBstWouldAddEndPuncttrue
\mciteSetBstMidEndSepPunct{\mcitedefaultmidpunct}
{\mcitedefaultendpunct}{\mcitedefaultseppunct}\relax
\EndOfBibitem
\bibitem{Kyeong:2009qx}
S.~H.\ Kyeong {\em et al.} ({Belle} collaboration){,}
  \href{http://dx.doi.org/10.1103/PhysRevD.80.051103}{Phys.\ Rev.\ {\bf D80}{,}
  051103}  (2009), \href{http://arxiv.org/abs/0905.0763}{{\tt arXiv:0905.0763
  [hep-ex]}}\relax
\mciteBstWouldAddEndPuncttrue
\mciteSetBstMidEndSepPunct{\mcitedefaultmidpunct}
{\mcitedefaultendpunct}{\mcitedefaultseppunct}\relax
\EndOfBibitem
\bibitem{Aubert:2007fm}
B.~Aubert {\em et al.} ({\babar} collaboration){,}
  \href{http://dx.doi.org/10.1103/PhysRevD.76.071104}{Phys.\ Rev.\ {\bf D76}{,}
  071104}  (2007), \href{http://arxiv.org/abs/0708.2543}{{\tt arXiv:0708.2543
  [hep-ex]}}\relax
\mciteBstWouldAddEndPuncttrue
\mciteSetBstMidEndSepPunct{\mcitedefaultmidpunct}
{\mcitedefaultendpunct}{\mcitedefaultseppunct}\relax
\EndOfBibitem
\bibitem{Lees:2011dq}
J.~P.\ Lees {\em et al.} ({\babar} collaboration){,}
  \href{http://dx.doi.org/10.1103/PhysRevD.85.072005}{Phys.\ Rev.\ {\bf D85}{,}
  072005}  (2012), \href{http://arxiv.org/abs/1112.3896}{{\tt arXiv:1112.3896
  [hep-ex]}}\relax
\mciteBstWouldAddEndPuncttrue
\mciteSetBstMidEndSepPunct{\mcitedefaultmidpunct}
{\mcitedefaultendpunct}{\mcitedefaultseppunct}\relax
\EndOfBibitem
\bibitem{Aaltonen:2011jv}
T.~Aaltonen {\em et al.} ({CDF} collaboration){,}
  \href{http://dx.doi.org/10.1103/PhysRevLett.108.211803}{Phys.\ Rev.\ Lett.\ {\bf
  108},  211803}  (2012), \href{http://arxiv.org/abs/1111.0485}{{\tt
  arXiv:1111.0485 [hep-ex]}}\relax
\mciteBstWouldAddEndPuncttrue
\mciteSetBstMidEndSepPunct{\mcitedefaultmidpunct}
{\mcitedefaultendpunct}{\mcitedefaultseppunct}\relax
\EndOfBibitem
\bibitem{Aaij:2016elb}
R.~Aaij {\em et al.} ({LHCb} collaboration){,}
  \href{http://dx.doi.org/10.1103/PhysRevLett.118.081801}{Phys.\ Rev.\ Lett.\ {\bf
  118},  081801}  (2017), \href{http://arxiv.org/abs/1610.08288}{{\tt
  arXiv:1610.08288 [hep-ex]}}\relax
\mciteBstWouldAddEndPuncttrue
\mciteSetBstMidEndSepPunct{\mcitedefaultmidpunct}
{\mcitedefaultendpunct}{\mcitedefaultseppunct}\relax
\EndOfBibitem
\bibitem{delAmoSanchez:2010ur}
P.~del Amo~Sanchez {\em et al.} ({\babar} collaboration){,}
  \href{http://dx.doi.org/10.1103/PhysRevD.82.031101}{Phys.\ Rev.\ {\bf D82}{,}
  031101}  (2010), \href{http://arxiv.org/abs/1003.0640}{{\tt arXiv:1003.0640
  [hep-ex]}}\relax
\mciteBstWouldAddEndPuncttrue
\mciteSetBstMidEndSepPunct{\mcitedefaultmidpunct}
{\mcitedefaultendpunct}{\mcitedefaultseppunct}\relax
\EndOfBibitem
\bibitem{Aubert:2006wu}
B.~Aubert {\em et al.} ({\babar} collaboration){,}
  \href{http://dx.doi.org/10.1103/PhysRevD.74.072008}{Phys.\ Rev.\ {\bf D74}{,}
  072008}  (2006), \href{http://arxiv.org/abs/hep-ex/0606050}{{\tt
  arXiv:hep-ex/0606050 [hep-ex]}}\relax
\mciteBstWouldAddEndPuncttrue
\mciteSetBstMidEndSepPunct{\mcitedefaultmidpunct}
{\mcitedefaultendpunct}{\mcitedefaultseppunct}\relax
\EndOfBibitem
\bibitem{Aaij:2015asa}
R.~Aaij {\em et al.} ({LHCb} collaboration){,}
  \href{http://dx.doi.org/10.1007/JHEP01(2016)012}{JHEP {\bf 01},  012}
  (2016), \href{http://arxiv.org/abs/1506.08634}{{\tt arXiv:1506.08634
  [hep-ex]}}\relax
\mciteBstWouldAddEndPuncttrue
\mciteSetBstMidEndSepPunct{\mcitedefaultmidpunct}
{\mcitedefaultendpunct}{\mcitedefaultseppunct}\relax
\EndOfBibitem
\bibitem{Aaij:2014aaa}
R.~Aaij {\em et al.} ({LHCb} collaboration){,}
  \href{http://dx.doi.org/10.1088/1367-2630/16/12/123001}{New J.\ Phys.\ {\bf
  16},  123001}  (2014), \href{http://arxiv.org/abs/1407.7704}{{\tt
  arXiv:1407.7704 [hep-ex]}}\relax
\mciteBstWouldAddEndPuncttrue
\mciteSetBstMidEndSepPunct{\mcitedefaultmidpunct}
{\mcitedefaultendpunct}{\mcitedefaultseppunct}\relax
\EndOfBibitem
\bibitem{Gaur:2013uou}
V.~Gaur {\em et al.} ({Belle} collaboration){,}
  \href{http://dx.doi.org/10.1103/PhysRevD.87.091101}{Phys.\ Rev.\ {\bf D87}{,}
  091101}  (2013), \href{http://arxiv.org/abs/1304.5312}{{\tt arXiv:1304.5312
  [hep-ex]}}\relax
\mciteBstWouldAddEndPuncttrue
\mciteSetBstMidEndSepPunct{\mcitedefaultmidpunct}
{\mcitedefaultendpunct}{\mcitedefaultseppunct}\relax
\EndOfBibitem
\bibitem{Aubert:2009al}
B.~Aubert {\em et al.} ({\babar} collaboration){,}
  \href{http://dx.doi.org/10.1103/PhysRevD.80.011101}{Phys.\ Rev.\ {\bf D80}{,}
  011101}  (2009), \href{http://arxiv.org/abs/0905.0868}{{\tt arXiv:0905.0868
  [hep-ex]}}\relax
\mciteBstWouldAddEndPuncttrue
\mciteSetBstMidEndSepPunct{\mcitedefaultmidpunct}
{\mcitedefaultendpunct}{\mcitedefaultseppunct}\relax
\EndOfBibitem
\bibitem{Aubert:2006zy}
B.~Aubert {\em et al.} ({\babar} collaboration){,}
  \href{http://dx.doi.org/10.1103/PhysRevD.74.032005}{Phys.\ Rev.\ {\bf D74}{,}
  032005}  (2006), \href{http://arxiv.org/abs/hep-ex/0606031}{{\tt
  arXiv:hep-ex/0606031 [hep-ex]}}\relax
\mciteBstWouldAddEndPuncttrue
\mciteSetBstMidEndSepPunct{\mcitedefaultmidpunct}
{\mcitedefaultendpunct}{\mcitedefaultseppunct}\relax
\EndOfBibitem
\bibitem{Prim:2013nmy}
M.~Prim {\em et al.} ({Belle} collaboration){,}
  \href{http://dx.doi.org/10.1103/PhysRevD.88.072004}{Phys.\ Rev.\ {\bf D88}{,}
  072004}  (2013), \href{http://arxiv.org/abs/1308.1830}{{\tt arXiv:1308.1830
  [hep-ex]}}\relax
\mciteBstWouldAddEndPuncttrue
\mciteSetBstMidEndSepPunct{\mcitedefaultmidpunct}
{\mcitedefaultendpunct}{\mcitedefaultseppunct}\relax
\EndOfBibitem
\bibitem{Chiang:2010ga}
C.~C.\ Chiang {\em et al.} ({Belle} collaboration){,}
  \href{http://dx.doi.org/10.1103/PhysRevD.81.071101}{Phys.\ Rev.\ {\bf D81}{,}
  071101}  (2010), \href{http://arxiv.org/abs/1001.4595}{{\tt arXiv:1001.4595
  [hep-ex]}}\relax
\mciteBstWouldAddEndPuncttrue
\mciteSetBstMidEndSepPunct{\mcitedefaultmidpunct}
{\mcitedefaultendpunct}{\mcitedefaultseppunct}\relax
\EndOfBibitem
\bibitem{Aubert:2007xc}
B.~Aubert {\em et al.} ({\babar} collaboration){,}
  \href{http://dx.doi.org/10.1103/PhysRevLett.100.081801}{Phys.\ Rev.\ Lett.\ {\bf
  100},  081801}  (2008), \href{http://arxiv.org/abs/0708.2248}{{\tt
  arXiv:0708.2248 [hep-ex]}}\relax
\mciteBstWouldAddEndPuncttrue
\mciteSetBstMidEndSepPunct{\mcitedefaultmidpunct}
{\mcitedefaultendpunct}{\mcitedefaultseppunct}\relax
\EndOfBibitem
\bibitem{Aubert:2008bb}
B.~Aubert {\em et al.} ({\babar} collaboration){,}
  \href{http://dx.doi.org/10.1103/PhysRevD.78.051103}{Phys.\ Rev.\ {\bf D78}{,}
  051103}  (2008), \href{http://arxiv.org/abs/0806.4467}{{\tt arXiv:0806.4467
  [hep-ex]}}\relax
\mciteBstWouldAddEndPuncttrue
\mciteSetBstMidEndSepPunct{\mcitedefaultmidpunct}
{\mcitedefaultendpunct}{\mcitedefaultseppunct}\relax
\EndOfBibitem
\bibitem{Aubert:2007nw}
B.~Aubert {\em et al.} ({\babar} collaboration){,}
  \href{http://dx.doi.org/10.1103/PhysRevD.76.051103}{Phys.\ Rev.\ {\bf D76}{,}
  051103}  (2007), \href{http://arxiv.org/abs/0705.0398}{{\tt arXiv:0705.0398
  [hep-ex]}}\relax
\mciteBstWouldAddEndPuncttrue
\mciteSetBstMidEndSepPunct{\mcitedefaultmidpunct}
{\mcitedefaultendpunct}{\mcitedefaultseppunct}\relax
\EndOfBibitem
\bibitem{Aaltonen:2011qt}
T.~Aaltonen {\em et al.} ({CDF} collaboration){,}
  \href{http://dx.doi.org/10.1103/PhysRevLett.106.181802}{Phys.\ Rev.\ Lett.\ {\bf
  106},  181802}  (2011), \href{http://arxiv.org/abs/1103.5762}{{\tt
  arXiv:1103.5762 [hep-ex]}}\relax
\mciteBstWouldAddEndPuncttrue
\mciteSetBstMidEndSepPunct{\mcitedefaultmidpunct}
{\mcitedefaultendpunct}{\mcitedefaultseppunct}\relax
\EndOfBibitem
\bibitem{Aaij:2012as}
R.~Aaij {\em et al.} ({LHCb} collaboration){,}
  \href{http://dx.doi.org/10.1007/JHEP10(2012)037}{JHEP {\bf 10},  037}
  (2012), \href{http://arxiv.org/abs/1206.2794}{{\tt arXiv:1206.2794
  [hep-ex]}}\relax
\mciteBstWouldAddEndPuncttrue
\mciteSetBstMidEndSepPunct{\mcitedefaultmidpunct}
{\mcitedefaultendpunct}{\mcitedefaultseppunct}\relax
\EndOfBibitem
\bibitem{Abe:2004mp}
K.~Abe {\em et al.} ({Belle} collaboration){,}
  \href{http://dx.doi.org/10.1103/PhysRevLett.94.181803}{Phys.\ Rev.\ Lett.\ {\bf
  94},  181803}  (2005), \href{http://arxiv.org/abs/hep-ex/0408101}{{\tt
  arXiv:hep-ex/0408101 [hep-ex]}}\relax
\mciteBstWouldAddEndPuncttrue
\mciteSetBstMidEndSepPunct{\mcitedefaultmidpunct}
{\mcitedefaultendpunct}{\mcitedefaultseppunct}\relax
\EndOfBibitem
\bibitem{Pal:2015ewa}
B.~Pal {\em et al.} ({Belle} collaboration){,}
  \href{http://dx.doi.org/10.1103/PhysRevD.92.011101}{Phys.\ Rev.\ {\bf D92}{,}
  011101}  (2015), \href{http://arxiv.org/abs/1504.00957}{{\tt
  arXiv:1504.00957 [hep-ex]}}\relax
\mciteBstWouldAddEndPuncttrue
\mciteSetBstMidEndSepPunct{\mcitedefaultmidpunct}
{\mcitedefaultendpunct}{\mcitedefaultseppunct}\relax
\EndOfBibitem
\bibitem{Abdesselam:2016tpr}
A.~Abdesselam {\em et al.} ({Belle} collaboration){,}
  \href{http://arxiv.org/abs/1609.03267}{{\tt arXiv:1609.03267 [hep-ex]}}
  (2016)\relax
\mciteBstWouldAddEndPuncttrue
\mciteSetBstMidEndSepPunct{\mcitedefaultmidpunct}
{\mcitedefaultendpunct}{\mcitedefaultseppunct}\relax
\EndOfBibitem
\bibitem{Lees:2013yea}
J.~P.\ Lees {\em et al.} ({\babar} collaboration){,}
  \href{http://dx.doi.org/10.1103/PhysRevD.89.051101}{Phys.\ Rev.\ {\bf D89}{,}
  051101}  (2014), \href{http://arxiv.org/abs/1312.0056}{{\tt arXiv:1312.0056
  [hep-ex]}}\relax
\mciteBstWouldAddEndPuncttrue
\mciteSetBstMidEndSepPunct{\mcitedefaultmidpunct}
{\mcitedefaultendpunct}{\mcitedefaultseppunct}\relax
\EndOfBibitem
\bibitem{Aaij:2015cxj}
R.~Aaij {\em et al.} ({LHCb} collaboration){,}
  \href{http://dx.doi.org/10.1007/JHEP10(2015)053}{JHEP {\bf 10},  053}
  (2015), \href{http://arxiv.org/abs/1508.00788}{{\tt arXiv:1508.00788
  [hep-ex]}}\relax
\mciteBstWouldAddEndPuncttrue
\mciteSetBstMidEndSepPunct{\mcitedefaultmidpunct}
{\mcitedefaultendpunct}{\mcitedefaultseppunct}\relax
\EndOfBibitem
\bibitem{Aubert:2003fm}
B.~Aubert {\em et al.} ({\babar} collaboration){,}
  \href{http://dx.doi.org/10.1103/PhysRevLett.93.051802}{Phys.\ Rev.\ Lett.\ {\bf
  93},  051802}  (2004), \href{http://arxiv.org/abs/hep-ex/0311049}{{\tt
  arXiv:hep-ex/0311049 [hep-ex]}}\relax
\mciteBstWouldAddEndPuncttrue
\mciteSetBstMidEndSepPunct{\mcitedefaultmidpunct}
{\mcitedefaultendpunct}{\mcitedefaultseppunct}\relax
\EndOfBibitem
\bibitem{Aubert:2006dd}
B.~Aubert {\em et al.} ({\babar} collaboration){,}
  \href{http://dx.doi.org/10.1103/PhysRevLett.97.051802}{Phys.\ Rev.\ Lett.\ {\bf
  97},  051802}  (2006), \href{http://arxiv.org/abs/hep-ex/0603050}{{\tt
  arXiv:hep-ex/0603050 [hep-ex]}}\relax
\mciteBstWouldAddEndPuncttrue
\mciteSetBstMidEndSepPunct{\mcitedefaultmidpunct}
{\mcitedefaultendpunct}{\mcitedefaultseppunct}\relax
\EndOfBibitem
\bibitem{Somov:2006sg}
A.~Somov {\em et al.} ({Belle} collaboration){,}
  \href{http://dx.doi.org/10.1103/PhysRevLett.96.171801}{Phys.\ Rev.\ Lett.\ {\bf
  96},  171801}  (2006), \href{http://arxiv.org/abs/hep-ex/0601024}{{\tt
  arXiv:hep-ex/0601024 [hep-ex]}}\relax
\mciteBstWouldAddEndPuncttrue
\mciteSetBstMidEndSepPunct{\mcitedefaultmidpunct}
{\mcitedefaultendpunct}{\mcitedefaultseppunct}\relax
\EndOfBibitem
\bibitem{Aubert:2006sw}
B.~Aubert {\em et al.} ({\babar} collaboration){,}
  \href{http://dx.doi.org/10.1103/PhysRevD.74.031104}{Phys.\ Rev.\ {\bf D74}{,}
  031104}  (2006), \href{http://arxiv.org/abs/hep-ex/0605024}{{\tt
  arXiv:hep-ex/0605024 [hep-ex]}}\relax
\mciteBstWouldAddEndPuncttrue
\mciteSetBstMidEndSepPunct{\mcitedefaultmidpunct}
{\mcitedefaultendpunct}{\mcitedefaultseppunct}\relax
\EndOfBibitem
\bibitem{Aubert:2009zr}
B.~Aubert {\em et al.} ({\babar} collaboration){,}
  \href{http://dx.doi.org/10.1103/PhysRevD.80.092007}{Phys.\ Rev.\ {\bf D80}{,}
  092007}  (2009), \href{http://arxiv.org/abs/0907.1776}{{\tt arXiv:0907.1776
  [hep-ex]}}\relax
\mciteBstWouldAddEndPuncttrue
\mciteSetBstMidEndSepPunct{\mcitedefaultmidpunct}
{\mcitedefaultendpunct}{\mcitedefaultseppunct}\relax
\EndOfBibitem
\bibitem{Aaij:2016qnm}
R.~Aaij {\em et al.} ({LHCb} collaboration){,}
  \href{http://dx.doi.org/10.1103/PhysRevD.95.012006}{Phys.\ Rev.\ {\bf D95}{,}
  012006}  (2017), \href{http://arxiv.org/abs/1610.05187}{{\tt
  arXiv:1610.05187 [hep-ex]}}\relax
\mciteBstWouldAddEndPuncttrue
\mciteSetBstMidEndSepPunct{\mcitedefaultmidpunct}
{\mcitedefaultendpunct}{\mcitedefaultseppunct}\relax
\EndOfBibitem
\bibitem{Wei:2007fg}
J.~T.\ Wei {\em et al.} ({Belle} collaboration){,}
  \href{http://dx.doi.org/10.1016/j.physletb.2007.11.063}{Phys.\ Lett.\ {\bf
  B659},  80}  (2008), \href{http://arxiv.org/abs/0706.4167}{{\tt
  arXiv:0706.4167 [hep-ex]}}\relax
\mciteBstWouldAddEndPuncttrue
\mciteSetBstMidEndSepPunct{\mcitedefaultmidpunct}
{\mcitedefaultendpunct}{\mcitedefaultseppunct}\relax
\EndOfBibitem
\bibitem{Aaij:2014tua}
R.~Aaij {\em et al.} ({LHCb} collaboration){,}
  \href{http://dx.doi.org/10.1103/PhysRevLett.113.141801}{Phys.\ Rev.\ Lett.\ {\bf
  113},  141801}  (2014), \href{http://arxiv.org/abs/1407.5907}{{\tt
  arXiv:1407.5907 [hep-ex]}}\relax
\mciteBstWouldAddEndPuncttrue
\mciteSetBstMidEndSepPunct{\mcitedefaultmidpunct}
{\mcitedefaultendpunct}{\mcitedefaultseppunct}\relax
\EndOfBibitem
\bibitem{Wang:2005fc}
M.~Z.\ Wang {\em et al.} ({Belle} collaboration){,}
  \href{http://dx.doi.org/10.1016/j.physletb.2005.05.008}{Phys.\ Lett.\ {\bf
  B617},  141}  (2005), \href{http://arxiv.org/abs/hep-ex/0503047}{{\tt
  arXiv:hep-ex/0503047 [hep-ex]}}\relax
\mciteBstWouldAddEndPuncttrue
\mciteSetBstMidEndSepPunct{\mcitedefaultmidpunct}
{\mcitedefaultendpunct}{\mcitedefaultseppunct}\relax
\EndOfBibitem
\bibitem{Chen:2008jy}
J.~H.\ Chen {\em et al.} ({Belle} collaboration){,}
  \href{http://dx.doi.org/10.1103/PhysRevLett.100.251801}{Phys.\ Rev.\ Lett.\ {\bf
  100},  251801}  (2008), \href{http://arxiv.org/abs/0802.0336}{{\tt
  arXiv:0802.0336 [hep-ex]}}\relax
\mciteBstWouldAddEndPuncttrue
\mciteSetBstMidEndSepPunct{\mcitedefaultmidpunct}
{\mcitedefaultendpunct}{\mcitedefaultseppunct}\relax
\EndOfBibitem
\bibitem{Tsai:2007pp}
Y.~T.\ Tsai {\em et al.} ({Belle} collaboration){,}
  \href{http://dx.doi.org/10.1103/PhysRevD.75.111101}{Phys.\ Rev.\ {\bf D75}{,}
  111101}  (2007), \href{http://arxiv.org/abs/hep-ex/0703048}{{\tt
  arXiv:hep-ex/0703048 [hep-ex]}}\relax
\mciteBstWouldAddEndPuncttrue
\mciteSetBstMidEndSepPunct{\mcitedefaultmidpunct}
{\mcitedefaultendpunct}{\mcitedefaultseppunct}\relax
\EndOfBibitem
\bibitem{Wang:2007as}
M.~Z.\ Wang {\em et al.} ({Belle} collaboration){,}
  \href{http://dx.doi.org/10.1103/PhysRevD.76.052004}{Phys.\ Rev.\ {\bf D76}{,}
  052004}  (2007), \href{http://arxiv.org/abs/0704.2672}{{\tt arXiv:0704.2672
  [hep-ex]}}\relax
\mciteBstWouldAddEndPuncttrue
\mciteSetBstMidEndSepPunct{\mcitedefaultmidpunct}
{\mcitedefaultendpunct}{\mcitedefaultseppunct}\relax
\EndOfBibitem
\bibitem{Chen:2009xg}
P.~Chen {\em et al.} ({Belle} collaboration){,}
  \href{http://dx.doi.org/10.1103/PhysRevD.80.111103}{Phys.\ Rev.\ {\bf D80}{,}
  111103}  (2009), \href{http://arxiv.org/abs/0910.5817}{{\tt arXiv:0910.5817
  [hep-ex]}}\relax
\mciteBstWouldAddEndPuncttrue
\mciteSetBstMidEndSepPunct{\mcitedefaultmidpunct}
{\mcitedefaultendpunct}{\mcitedefaultseppunct}\relax
\EndOfBibitem
\bibitem{Aubert:2004fy}
B.~Aubert {\em et al.} ({\babar} collaboration){,}
  \href{http://dx.doi.org/10.1103/PhysRevD.69.091503}{Phys.\ Rev.\ {\bf D69}{,}
  091503}  (2004), \href{http://arxiv.org/abs/hep-ex/0403003}{{\tt
  arXiv:hep-ex/0403003 [hep-ex]}}\relax
\mciteBstWouldAddEndPuncttrue
\mciteSetBstMidEndSepPunct{\mcitedefaultmidpunct}
{\mcitedefaultendpunct}{\mcitedefaultseppunct}\relax
\EndOfBibitem
\bibitem{Aaij:2013fta}
R.~Aaij {\em et al.} ({LHCb} collaboration){,}
  \href{http://dx.doi.org/10.1007/JHEP10(2013)005}{JHEP {\bf 10},  005}
  (2013), \href{http://arxiv.org/abs/1308.0961}{{\tt arXiv:1308.0961
  [hep-ex]}}\relax
\mciteBstWouldAddEndPuncttrue
\mciteSetBstMidEndSepPunct{\mcitedefaultmidpunct}
{\mcitedefaultendpunct}{\mcitedefaultseppunct}\relax
\EndOfBibitem
\bibitem{Aubert:2009am}
B.~Aubert {\em et al.} ({\babar} collaboration){,}
  \href{http://dx.doi.org/10.1103/PhysRevD.79.112009}{Phys.\ Rev.\ {\bf D79}{,}
  112009}  (2009), \href{http://arxiv.org/abs/0904.4724}{{\tt arXiv:0904.4724
  [hep-ex]}}\relax
\mciteBstWouldAddEndPuncttrue
\mciteSetBstMidEndSepPunct{\mcitedefaultmidpunct}
{\mcitedefaultendpunct}{\mcitedefaultseppunct}\relax
\EndOfBibitem
\bibitem{Wang:2003yi}
M.~Z.\ Wang {\em et al.} ({Belle} collaboration){,}
  \href{http://dx.doi.org/10.1103/PhysRevLett.90.201802}{Phys.\ Rev.\ Lett.\ {\bf
  90},  201802}  (2003), \href{http://arxiv.org/abs/hep-ex/0302024}{{\tt
  arXiv:hep-ex/0302024 [hep-ex]}}\relax
\mciteBstWouldAddEndPuncttrue
\mciteSetBstMidEndSepPunct{\mcitedefaultmidpunct}
{\mcitedefaultendpunct}{\mcitedefaultseppunct}\relax
\EndOfBibitem
\bibitem{Aaltonen:2008hg}
T.~Aaltonen {\em et al.} ({CDF} collaboration){,}
  \href{http://dx.doi.org/10.1103/PhysRevLett.103.031801}{Phys.\ Rev.\ Lett.\ {\bf
  103},  031801}  (2009), \href{http://arxiv.org/abs/0812.4271}{{\tt
  arXiv:0812.4271 [hep-ex]}}\relax
\mciteBstWouldAddEndPuncttrue
\mciteSetBstMidEndSepPunct{\mcitedefaultmidpunct}
{\mcitedefaultendpunct}{\mcitedefaultseppunct}\relax
\EndOfBibitem
\bibitem{Aaltonen:2011qs}
T.~Aaltonen {\em et al.} ({CDF} collaboration){,}
  \href{http://dx.doi.org/10.1103/PhysRevLett.107.201802}{Phys.\ Rev.\ Lett.\ {\bf
  107},  201802}  (2011), \href{http://arxiv.org/abs/1107.3753}{{\tt
  arXiv:1107.3753 [hep-ex]}}\relax
\mciteBstWouldAddEndPuncttrue
\mciteSetBstMidEndSepPunct{\mcitedefaultmidpunct}
{\mcitedefaultendpunct}{\mcitedefaultseppunct}\relax
\EndOfBibitem
\bibitem{Aaij:2013mna}
R.~Aaij {\em et al.} ({LHCb} collaboration){,}
  \href{http://dx.doi.org/10.1016/j.physletb.2013.06.060}{Phys.\ Lett.\ {\bf
  B725},  25}  (2013), \href{http://arxiv.org/abs/1306.2577}{{\tt
  arXiv:1306.2577 [hep-ex]}}\relax
\mciteBstWouldAddEndPuncttrue
\mciteSetBstMidEndSepPunct{\mcitedefaultmidpunct}
{\mcitedefaultendpunct}{\mcitedefaultseppunct}\relax
\EndOfBibitem
\bibitem{Aaij:2015eqa}
R.~Aaij {\em et al.} ({LHCb} collaboration){,}
  \href{http://dx.doi.org/10.1007/JHEP09(2015)006}{JHEP {\bf 09},  006}
  (2015), \href{http://arxiv.org/abs/1505.03295}{{\tt arXiv:1505.03295
  [hep-ex]}}\relax
\mciteBstWouldAddEndPuncttrue
\mciteSetBstMidEndSepPunct{\mcitedefaultmidpunct}
{\mcitedefaultendpunct}{\mcitedefaultseppunct}\relax
\EndOfBibitem
\bibitem{Aaij:2016zhm}
R.~Aaij {\em et al.} ({LHCb} collaboration){,}
  \href{http://dx.doi.org/10.1016/j.physletb.2016.05.077}{Phys.\ Lett.\ {\bf
  B759},  282}  (2016), \href{http://arxiv.org/abs/1603.02870}{{\tt
  arXiv:1603.02870 [hep-ex]}}\relax
\mciteBstWouldAddEndPuncttrue
\mciteSetBstMidEndSepPunct{\mcitedefaultmidpunct}
{\mcitedefaultendpunct}{\mcitedefaultseppunct}\relax
\EndOfBibitem
\bibitem{Aaij:2014lpa}
R.~Aaij {\em et al.} ({LHCb} collaboration){,}
  \href{http://dx.doi.org/10.1007/JHEP04(2014)087}{JHEP {\bf 04},  087}
  (2014), \href{http://arxiv.org/abs/1402.0770}{{\tt arXiv:1402.0770
  [hep-ex]}}\relax
\mciteBstWouldAddEndPuncttrue
\mciteSetBstMidEndSepPunct{\mcitedefaultmidpunct}
{\mcitedefaultendpunct}{\mcitedefaultseppunct}\relax
\EndOfBibitem
\bibitem{Aaij:2016nrq}
R.~Aaij {\em et al.} ({LHCb} collaboration){,}
  \href{http://dx.doi.org/10.1007/JHEP05(2016)081}{JHEP {\bf 05},  081}
  (2016), \href{http://arxiv.org/abs/1603.00413}{{\tt arXiv:1603.00413
  [hep-ex]}}\relax
\mciteBstWouldAddEndPuncttrue
\mciteSetBstMidEndSepPunct{\mcitedefaultmidpunct}
{\mcitedefaultendpunct}{\mcitedefaultseppunct}\relax
\EndOfBibitem
\bibitem{Aaij:2015xza}
R.~Aaij {\em et al.} ({LHCb} collaboration){,}
  \href{http://dx.doi.org/10.1007/JHEP06(2015)115}{JHEP {\bf 06},  115}
  (2015), \href{http://arxiv.org/abs/1503.07138}{{\tt arXiv:1503.07138
  [hep-ex]}}\relax
\mciteBstWouldAddEndPuncttrue
\mciteSetBstMidEndSepPunct{\mcitedefaultmidpunct}
{\mcitedefaultendpunct}{\mcitedefaultseppunct}\relax
\EndOfBibitem
\bibitem{Aaij:2014yka}
R.~Aaij {\em et al.} ({LHCb} collaboration){,}
  \href{http://dx.doi.org/10.1103/PhysRevLett.114.062004}{Phys.\ Rev.\ Lett.\ {\bf
  114},  062004}  (2015), \href{http://arxiv.org/abs/1411.4849}{{\tt
  arXiv:1411.4849 [hep-ex]}}\relax
\mciteBstWouldAddEndPuncttrue
\mciteSetBstMidEndSepPunct{\mcitedefaultmidpunct}
{\mcitedefaultendpunct}{\mcitedefaultseppunct}\relax
\EndOfBibitem
\bibitem{Aaij:2016jnn}
R.~Aaij {\em et al.} ({LHCb} collaboration){,}
  \href{http://dx.doi.org/10.1007/JHEP05(2016)161}{JHEP {\bf 05},  161}
  (2016), \href{http://arxiv.org/abs/1604.03896}{{\tt arXiv:1604.03896
  [hep-ex]}}\relax
\mciteBstWouldAddEndPuncttrue
\mciteSetBstMidEndSepPunct{\mcitedefaultmidpunct}
{\mcitedefaultendpunct}{\mcitedefaultseppunct}\relax
\EndOfBibitem
\bibitem{Peng:2010ze}
C.~C.\ Peng {\em et al.} ({Belle} collaboration){,}
  \href{http://dx.doi.org/10.1103/PhysRevD.82.072007}{Phys.\ Rev.\ {\bf D82}{,}
  072007}  (2010), \href{http://arxiv.org/abs/1006.5115}{{\tt arXiv:1006.5115
  [hep-ex]}}\relax
\mciteBstWouldAddEndPuncttrue
\mciteSetBstMidEndSepPunct{\mcitedefaultmidpunct}
{\mcitedefaultendpunct}{\mcitedefaultseppunct}\relax
\EndOfBibitem
\bibitem{Aaltonen:2011rs}
T.~Aaltonen {\em et al.} ({CDF} collaboration){,}
  \href{http://dx.doi.org/10.1103/PhysRevLett.107.261802}{Phys.\ Rev.\ Lett.\ {\bf
  107},  261802}  (2011), \href{http://arxiv.org/abs/1107.4999}{{\tt
  arXiv:1107.4999 [hep-ex]}}\relax
\mciteBstWouldAddEndPuncttrue
\mciteSetBstMidEndSepPunct{\mcitedefaultmidpunct}
{\mcitedefaultendpunct}{\mcitedefaultseppunct}\relax
\EndOfBibitem
\bibitem{Aaij:2014lba}
R.~Aaij {\em et al.} ({LHCb} collaboration){,}
  \href{http://dx.doi.org/10.1016/j.physletb.2015.02.010}{Phys.\ Lett.\ {\bf
  B743},  46}  (2015), \href{http://arxiv.org/abs/1412.6433}{{\tt
  arXiv:1412.6433 [hep-ex]}}\relax
\mciteBstWouldAddEndPuncttrue
\mciteSetBstMidEndSepPunct{\mcitedefaultmidpunct}
{\mcitedefaultendpunct}{\mcitedefaultseppunct}\relax
\EndOfBibitem
\bibitem{Pal:2015ghq}
B.~Pal {\em et al.} ({Belle} collaboration){,}
  \href{http://dx.doi.org/10.1103/PhysRevLett.116.161801}{Phys.\ Rev.\ Lett.\ {\bf
  116},  161801}  (2016), \href{http://arxiv.org/abs/1512.02145}{{\tt
  arXiv:1512.02145 [hep-ex]}}\relax
\mciteBstWouldAddEndPuncttrue
\mciteSetBstMidEndSepPunct{\mcitedefaultmidpunct}
{\mcitedefaultendpunct}{\mcitedefaultseppunct}\relax
\EndOfBibitem
\bibitem{Aaij:2015kba}
R.~Aaij {\em et al.} ({LHCb} collaboration){,}
  \href{http://dx.doi.org/10.1007/JHEP07(2015)166}{JHEP {\bf 07},  166}
  (2015), \href{http://arxiv.org/abs/1503.05362}{{\tt arXiv:1503.05362
  [hep-ex]}}\relax
\mciteBstWouldAddEndPuncttrue
\mciteSetBstMidEndSepPunct{\mcitedefaultmidpunct}
{\mcitedefaultendpunct}{\mcitedefaultseppunct}\relax
\EndOfBibitem
\bibitem{Aaij:2013gga}
R.~Aaij {\em et al.} ({LHCb} collaboration){,}
  \href{http://dx.doi.org/10.1007/JHEP11(2013)092}{JHEP {\bf 11},  092}
  (2013), \href{http://arxiv.org/abs/1306.2239}{{\tt arXiv:1306.2239
  [hep-ex]}}\relax
\mciteBstWouldAddEndPuncttrue
\mciteSetBstMidEndSepPunct{\mcitedefaultmidpunct}
{\mcitedefaultendpunct}{\mcitedefaultseppunct}\relax
\EndOfBibitem
\bibitem{Dutta:2014sxo}
D.~Dutta {\em et al.} ({Belle} collaboration){,}
  \href{http://dx.doi.org/10.1103/PhysRevD.91.011101}{Phys.\ Rev.\ {\bf D91}{,}
  011101}  (2015), \href{http://arxiv.org/abs/1411.7771}{{\tt arXiv:1411.7771
  [hep-ex]}}\relax
\mciteBstWouldAddEndPuncttrue
\mciteSetBstMidEndSepPunct{\mcitedefaultmidpunct}
{\mcitedefaultendpunct}{\mcitedefaultseppunct}\relax
\EndOfBibitem
\bibitem{Aaij:2012ita}
R.~Aaij {\em et al.} ({LHCb} collaboration){,}
  \href{http://dx.doi.org/10.1016/j.nuclphysb.2012.09.013}{Nucl.\ Phys.\ {\bf
  B867},  1}  (2013), \href{http://arxiv.org/abs/1209.0313}{{\tt
  arXiv:1209.0313 [hep-ex]}}\relax
\mciteBstWouldAddEndPuncttrue
\mciteSetBstMidEndSepPunct{\mcitedefaultmidpunct}
{\mcitedefaultendpunct}{\mcitedefaultseppunct}\relax
\EndOfBibitem
\bibitem{Aaltonen:2013as}
T.~Aaltonen {\em et al.} ({CDF} collaboration){,}
  \href{http://dx.doi.org/10.1103/PhysRevD.87.072003}{Phys.\ Rev.\ {\bf D87}{,}
  072003}  (2013), \href{http://arxiv.org/abs/1301.7048}{{\tt arXiv:1301.7048
  [hep-ex]}}\relax
\mciteBstWouldAddEndPuncttrue
\mciteSetBstMidEndSepPunct{\mcitedefaultmidpunct}
{\mcitedefaultendpunct}{\mcitedefaultseppunct}\relax
\EndOfBibitem
\bibitem{Abazov:2013wjb}
V.~M.\ Abazov {\em et al.} ({D0} collaboration){,}
  \href{http://dx.doi.org/10.1103/PhysRevD.87.072006}{Phys.\ Rev.\ {\bf D87}{,}
  072006}  (2013), \href{http://arxiv.org/abs/1301.4507}{{\tt arXiv:1301.4507
  [hep-ex]}}\relax
\mciteBstWouldAddEndPuncttrue
\mciteSetBstMidEndSepPunct{\mcitedefaultmidpunct}
{\mcitedefaultendpunct}{\mcitedefaultseppunct}\relax
\EndOfBibitem
\bibitem{Aaij:2013aka}
R.~Aaij {\em et al.} ({LHCb} collaboration){,}
  \href{http://dx.doi.org/10.1103/PhysRevLett.111.101805}{Phys.\ Rev.\ Lett.\ {\bf
  111},  101805}  (2013), \href{http://arxiv.org/abs/1307.5024}{{\tt
  arXiv:1307.5024 [hep-ex]}}\relax
\mciteBstWouldAddEndPuncttrue
\mciteSetBstMidEndSepPunct{\mcitedefaultmidpunct}
{\mcitedefaultendpunct}{\mcitedefaultseppunct}\relax
\EndOfBibitem
\bibitem{Chatrchyan:2013bka}
S.~Chatrchyan {\em et al.} ({CMS} collaboration){,}
  \href{http://dx.doi.org/10.1103/PhysRevLett.111.101804}{Phys.\ Rev.\ Lett.\ {\bf
  111},  101804}  (2013), \href{http://arxiv.org/abs/1307.5025}{{\tt
  arXiv:1307.5025 [hep-ex]}}\relax
\mciteBstWouldAddEndPuncttrue
\mciteSetBstMidEndSepPunct{\mcitedefaultmidpunct}
{\mcitedefaultendpunct}{\mcitedefaultseppunct}\relax
\EndOfBibitem
\bibitem{Aaboud:2016ire}
M.~Aaboud {\em et al.} ({ATLAS} collaboration){,}
  \href{http://dx.doi.org/10.1140/epjc/s10052-016-4338-8}{Eur.\ Phys.\ J.\ {\bf
  C76},  513}  (2016), \href{http://arxiv.org/abs/1604.04263}{{\tt
  arXiv:1604.04263 [hep-ex]}}\relax
\mciteBstWouldAddEndPuncttrue
\mciteSetBstMidEndSepPunct{\mcitedefaultmidpunct}
{\mcitedefaultendpunct}{\mcitedefaultseppunct}\relax
\EndOfBibitem
\bibitem{Aaltonen:2009vr}
T.~Aaltonen {\em et al.} ({CDF} collaboration){,}
  \href{http://dx.doi.org/10.1103/PhysRevLett.102.201801}{Phys.\ Rev.\ Lett.\ {\bf
  102},  201801}  (2009), \href{http://arxiv.org/abs/0901.3803}{{\tt
  arXiv:0901.3803 [hep-ex]}}\relax
\mciteBstWouldAddEndPuncttrue
\mciteSetBstMidEndSepPunct{\mcitedefaultmidpunct}
{\mcitedefaultendpunct}{\mcitedefaultseppunct}\relax
\EndOfBibitem
\bibitem{Aaij:2013cby}
R.~Aaij {\em et al.} ({LHCb} collaboration){,}
  \href{http://dx.doi.org/10.1103/PhysRevLett.111.141801}{Phys.\ Rev.\ Lett.\ {\bf
  111},  141801}  (2013), \href{http://arxiv.org/abs/1307.4889}{{\tt
  arXiv:1307.4889 [hep-ex]}}\relax
\mciteBstWouldAddEndPuncttrue
\mciteSetBstMidEndSepPunct{\mcitedefaultmidpunct}
{\mcitedefaultendpunct}{\mcitedefaultseppunct}\relax
\EndOfBibitem
\bibitem{Aaij:2013lla}
R.~Aaij {\em et al.} ({LHCb} collaboration){,}
  \href{http://dx.doi.org/10.1103/PhysRevLett.110.211801}{Phys.\ Rev.\ Lett.\ {\bf
  110},  211801}  (2013), \href{http://arxiv.org/abs/1303.1092}{{\tt
  arXiv:1303.1092 [hep-ex]}}\relax
\mciteBstWouldAddEndPuncttrue
\mciteSetBstMidEndSepPunct{\mcitedefaultmidpunct}
{\mcitedefaultendpunct}{\mcitedefaultseppunct}\relax
\EndOfBibitem
\bibitem{Abazov:2006qm}
V.~M.\ Abazov {\em et al.} ({D0} collaboration){,}
  \href{http://dx.doi.org/10.1103/PhysRevD.74.031107}{Phys.\ Rev.\ {\bf D74}{,}
  031107}  (2006), \href{http://arxiv.org/abs/hep-ex/0604015}{{\tt
  arXiv:hep-ex/0604015 [hep-ex]}}\relax
\mciteBstWouldAddEndPuncttrue
\mciteSetBstMidEndSepPunct{\mcitedefaultmidpunct}
{\mcitedefaultendpunct}{\mcitedefaultseppunct}\relax
\EndOfBibitem
\bibitem{Aaij:2015esa}
R.~Aaij {\em et al.} ({LHCb} collaboration){,}
  \href{http://dx.doi.org/10.1007/JHEP09(2015)179}{JHEP {\bf 09},  179}
  (2015), \href{http://arxiv.org/abs/1506.08777}{{\tt arXiv:1506.08777
  [hep-ex]}}\relax
\mciteBstWouldAddEndPuncttrue
\mciteSetBstMidEndSepPunct{\mcitedefaultmidpunct}
{\mcitedefaultendpunct}{\mcitedefaultseppunct}\relax
\EndOfBibitem
\bibitem{Aaij:2015qga}
R.~Aaij {\em et al.} ({LHCb} collaboration){,}
  \href{http://dx.doi.org/10.1103/PhysRevLett.115.051801}{Phys.\ Rev.\ Lett.\ {\bf
  115},  051801}  (2015), \href{http://arxiv.org/abs/1503.07483}{{\tt
  arXiv:1503.07483 [hep-ex]}}\relax
\mciteBstWouldAddEndPuncttrue
\mciteSetBstMidEndSepPunct{\mcitedefaultmidpunct}
{\mcitedefaultendpunct}{\mcitedefaultseppunct}\relax
\EndOfBibitem
\bibitem{CMS:2014xfa}
V.~Khachatryan {\em et al.} ({CMS and LHCb} collaborations){,}
  \href{http://dx.doi.org/10.1038/nature14474}{Nature {\bf 522},  68}  (2015){,}
  \href{http://arxiv.org/abs/1411.4413}{{\tt arXiv:1411.4413 [hep-ex]}}\relax
\mciteBstWouldAddEndPuncttrue
\mciteSetBstMidEndSepPunct{\mcitedefaultmidpunct}
{\mcitedefaultendpunct}{\mcitedefaultseppunct}\relax
\EndOfBibitem
\bibitem{Aubert:2009ak}
B.~Aubert {\em et al.} ({\babar} collaboration){,}
  \href{http://dx.doi.org/10.1103/PhysRevLett.103.211802}{Phys.\ Rev.\ Lett.\ {\bf
  103},  211802}  (2009), \href{http://arxiv.org/abs/0906.2177}{{\tt
  arXiv:0906.2177 [hep-ex]}}\relax
\mciteBstWouldAddEndPuncttrue
\mciteSetBstMidEndSepPunct{\mcitedefaultmidpunct}
{\mcitedefaultendpunct}{\mcitedefaultseppunct}\relax
\EndOfBibitem
\bibitem{Nakao:2004th}
M.~Nakao {\em et al.} ({Belle} collaboration){,}
  \href{http://dx.doi.org/10.1103/PhysRevD.69.112001}{Phys.\ Rev.\ {\bf D69}{,}
  112001}  (2004), \href{http://arxiv.org/abs/hep-ex/0402042}{{\tt
  arXiv:hep-ex/0402042 [hep-ex]}}\relax
\mciteBstWouldAddEndPuncttrue
\mciteSetBstMidEndSepPunct{\mcitedefaultmidpunct}
{\mcitedefaultendpunct}{\mcitedefaultseppunct}\relax
\EndOfBibitem
\bibitem{Coan:1999kh}
T.~E.\ Coan {\em et al.} ({CLEO} collaboration){,}
  \href{http://dx.doi.org/10.1103/PhysRevLett.84.5283}{Phys.\ Rev.\ Lett.\ {\bf
  84},  5283}  (2000), \href{http://arxiv.org/abs/hep-ex/9912057}{{\tt
  arXiv:hep-ex/9912057 [hep-ex]}}\relax
\mciteBstWouldAddEndPuncttrue
\mciteSetBstMidEndSepPunct{\mcitedefaultmidpunct}
{\mcitedefaultendpunct}{\mcitedefaultseppunct}\relax
\EndOfBibitem
\bibitem{Yang:2004as}
H.~Yang {\em et al.} ({Belle} collaboration){,}
  \href{http://dx.doi.org/10.1103/PhysRevLett.94.111802}{Phys.\ Rev.\ Lett.\ {\bf
  94},  111802}  (2005), \href{http://arxiv.org/abs/hep-ex/0412039}{{\tt
  arXiv:hep-ex/0412039 [hep-ex]}}\relax
\mciteBstWouldAddEndPuncttrue
\mciteSetBstMidEndSepPunct{\mcitedefaultmidpunct}
{\mcitedefaultendpunct}{\mcitedefaultseppunct}\relax
\EndOfBibitem
\bibitem{Nishida:2004fk}
S.~Nishida {\em et al.} ({Belle} collaboration){,}
  \href{http://dx.doi.org/10.1016/j.physletb.2005.01.097}{Phys.\ Lett.\ {\bf
  B610},  23}  (2005), \href{http://arxiv.org/abs/hep-ex/0411065}{{\tt
  arXiv:hep-ex/0411065 [hep-ex]}}\relax
\mciteBstWouldAddEndPuncttrue
\mciteSetBstMidEndSepPunct{\mcitedefaultmidpunct}
{\mcitedefaultendpunct}{\mcitedefaultseppunct}\relax
\EndOfBibitem
\bibitem{Aubert:2006vs}
B.~Aubert {\em et al.} ({\babar} collaboration){,}
  \href{http://dx.doi.org/10.1103/PhysRevD.74.031102}{Phys.\ Rev.\ {\bf D74}{,}
  031102}  (2006), \href{http://arxiv.org/abs/hep-ex/0603054}{{\tt
  arXiv:hep-ex/0603054 [hep-ex]}}\relax
\mciteBstWouldAddEndPuncttrue
\mciteSetBstMidEndSepPunct{\mcitedefaultmidpunct}
{\mcitedefaultendpunct}{\mcitedefaultseppunct}\relax
\EndOfBibitem
\bibitem{Wedd:2008ru}
R.~Wedd {\em et al.} ({Belle} collaboration){,}
  \href{http://dx.doi.org/10.1103/PhysRevD.81.111104}{Phys.\ Rev.\ {\bf D81}{,}
  111104}  (2010), \href{http://arxiv.org/abs/0810.0804}{{\tt arXiv:0810.0804
  [hep-ex]}}\relax
\mciteBstWouldAddEndPuncttrue
\mciteSetBstMidEndSepPunct{\mcitedefaultmidpunct}
{\mcitedefaultendpunct}{\mcitedefaultseppunct}\relax
\EndOfBibitem
\bibitem{Aubert:2006he}
B.~Aubert {\em et al.} ({\babar} collaboration){,}
  \href{http://dx.doi.org/10.1103/PhysRevD.75.051102}{Phys.\ Rev.\ {\bf D75}{,}
  051102}  (2007), \href{http://arxiv.org/abs/hep-ex/0611037}{{\tt
  arXiv:hep-ex/0611037 [hep-ex]}}\relax
\mciteBstWouldAddEndPuncttrue
\mciteSetBstMidEndSepPunct{\mcitedefaultmidpunct}
{\mcitedefaultendpunct}{\mcitedefaultseppunct}\relax
\EndOfBibitem
\bibitem{Nishida:2002me}
S.~Nishida {\em et al.} ({Belle} collaboration){,}
  \href{http://dx.doi.org/10.1103/PhysRevLett.89.231801}{Phys.\ Rev.\ Lett.\ {\bf
  89},  231801}  (2002), \href{http://arxiv.org/abs/hep-ex/0205025}{{\tt
  arXiv:hep-ex/0205025 [hep-ex]}}\relax
\mciteBstWouldAddEndPuncttrue
\mciteSetBstMidEndSepPunct{\mcitedefaultmidpunct}
{\mcitedefaultendpunct}{\mcitedefaultseppunct}\relax
\EndOfBibitem
\bibitem{Aubert:2005xk}
B.~Aubert {\em et al.} ({\babar} collaboration){,}
  \href{http://dx.doi.org/10.1103/PhysRevLett.100.189903}{Phys.\ Rev.\ Lett.\ {\bf
  98},  211804}  (2007), \href{http://arxiv.org/abs/hep-ex/0507031}{{\tt
  arXiv:hep-ex/0507031 [hep-ex]}}, Erratum ibid.\
  \href{http://dx.doi.org/10.1103/PhysRevLett.100.199905}{{\bf 100}, 199905}
  (2008)\relax
\mciteBstWouldAddEndPuncttrue
\mciteSetBstMidEndSepPunct{\mcitedefaultmidpunct}
{\mcitedefaultendpunct}{\mcitedefaultseppunct}\relax
\EndOfBibitem
\bibitem{Aubert:2008al}
B.~Aubert {\em et al.} ({\babar} collaboration){,}
  \href{http://dx.doi.org/10.1103/PhysRevD.78.112001}{Phys.\ Rev.\ {\bf D78}{,}
  112001}  (2008), \href{http://arxiv.org/abs/0808.1379}{{\tt arXiv:0808.1379
  [hep-ex]}}\relax
\mciteBstWouldAddEndPuncttrue
\mciteSetBstMidEndSepPunct{\mcitedefaultmidpunct}
{\mcitedefaultendpunct}{\mcitedefaultseppunct}\relax
\EndOfBibitem
\bibitem{Taniguchi:2008ty}
N.~Taniguchi {\em et al.} ({Belle} collaboration){,}
  \href{http://dx.doi.org/10.1103/PhysRevLett.101.111801}{Phys.\ Rev.\ Lett.\ {\bf
  101},  111801}  (2008), \href{http://arxiv.org/abs/0804.4770}{{\tt
  arXiv:0804.4770 [hep-ex]}}, Erratum ibid.\
  \href{http://dx.doi.org/10.1103/PhysRevLett.101.129904}{{\bf 101}, 129904}
  (2008)\relax
\mciteBstWouldAddEndPuncttrue
\mciteSetBstMidEndSepPunct{\mcitedefaultmidpunct}
{\mcitedefaultendpunct}{\mcitedefaultseppunct}\relax
\EndOfBibitem
\bibitem{Lee:2005fba}
Y.~J.\ Lee {\em et al.} ({Belle} collaboration){,}
  \href{http://dx.doi.org/10.1103/PhysRevLett.95.061802}{Phys.\ Rev.\ Lett.\ {\bf
  95},  061802}  (2005), \href{http://arxiv.org/abs/hep-ex/0503046}{{\tt
  arXiv:hep-ex/0503046 [hep-ex]}}\relax
\mciteBstWouldAddEndPuncttrue
\mciteSetBstMidEndSepPunct{\mcitedefaultmidpunct}
{\mcitedefaultendpunct}{\mcitedefaultseppunct}\relax
\EndOfBibitem
\bibitem{Lees:2013lvs}
J.~P.\ Lees {\em et al.} ({\babar} collaboration){,}
  \href{http://dx.doi.org/10.1103/PhysRevD.88.032012}{Phys.\ Rev.\ {\bf D88}{,}
  032012}  (2013), \href{http://arxiv.org/abs/1303.6010}{{\tt arXiv:1303.6010
  [hep-ex]}}\relax
\mciteBstWouldAddEndPuncttrue
\mciteSetBstMidEndSepPunct{\mcitedefaultmidpunct}
{\mcitedefaultendpunct}{\mcitedefaultseppunct}\relax
\EndOfBibitem
\bibitem{Wei:2008nv}
J.~T.\ Wei {\em et al.} ({Belle} collaboration){,}
  \href{http://dx.doi.org/10.1103/PhysRevD.78.011101}{Phys.\ Rev.\ {\bf D78}{,}
  011101}  (2008), \href{http://arxiv.org/abs/0804.3656}{{\tt arXiv:0804.3656
  [hep-ex]}}\relax
\mciteBstWouldAddEndPuncttrue
\mciteSetBstMidEndSepPunct{\mcitedefaultmidpunct}
{\mcitedefaultendpunct}{\mcitedefaultseppunct}\relax
\EndOfBibitem
\bibitem{Aaij:2015nea}
R.~Aaij {\em et al.} ({LHCb} collaboration){,}
  \href{http://dx.doi.org/10.1007/JHEP10(2015)034}{JHEP {\bf 10},  034}
  (2015), \href{http://arxiv.org/abs/1509.00414}{{\tt arXiv:1509.00414
  [hep-ex]}}\relax
\mciteBstWouldAddEndPuncttrue
\mciteSetBstMidEndSepPunct{\mcitedefaultmidpunct}
{\mcitedefaultendpunct}{\mcitedefaultseppunct}\relax
\EndOfBibitem
\bibitem{Aubert:2004ws}
B.~Aubert {\em et al.} ({\babar} collaboration){,}
  \href{http://dx.doi.org/10.1103/PhysRevLett.94.101801}{Phys.\ Rev.\ Lett.\ {\bf
  94},  101801}  (2005), \href{http://arxiv.org/abs/hep-ex/0411061}{{\tt
  arXiv:hep-ex/0411061 [hep-ex]}}\relax
\mciteBstWouldAddEndPuncttrue
\mciteSetBstMidEndSepPunct{\mcitedefaultmidpunct}
{\mcitedefaultendpunct}{\mcitedefaultseppunct}\relax
\EndOfBibitem
\bibitem{Lutz:2013ftz}
O.~Lutz {\em et al.} ({Belle} collaboration){,}
  \href{http://dx.doi.org/10.1103/PhysRevD.87.111103}{Phys.\ Rev.\ {\bf D87}{,}
  111103}  (2013), \href{http://arxiv.org/abs/1303.3719}{{\tt arXiv:1303.3719
  [hep-ex]}}\relax
\mciteBstWouldAddEndPuncttrue
\mciteSetBstMidEndSepPunct{\mcitedefaultmidpunct}
{\mcitedefaultendpunct}{\mcitedefaultseppunct}\relax
\EndOfBibitem
\bibitem{Aubert:2008ps}
B.~Aubert {\em et al.} ({\babar} collaboration){,}
  \href{http://dx.doi.org/10.1103/PhysRevLett.102.091803}{Phys.\ Rev.\ Lett.\ {\bf
  102},  091803}  (2009), \href{http://arxiv.org/abs/0807.4119}{{\tt
  arXiv:0807.4119 [hep-ex]}}\relax
\mciteBstWouldAddEndPuncttrue
\mciteSetBstMidEndSepPunct{\mcitedefaultmidpunct}
{\mcitedefaultendpunct}{\mcitedefaultseppunct}\relax
\EndOfBibitem
\bibitem{Wei:2009zv}
J.~T.\ Wei {\em et al.} ({Belle} collaboration){,}
  \href{http://dx.doi.org/10.1103/PhysRevLett.103.171801}{Phys.\ Rev.\ Lett.\ {\bf
  103},  171801}  (2009), \href{http://arxiv.org/abs/0904.0770}{{\tt
  arXiv:0904.0770 [hep-ex]}}\relax
\mciteBstWouldAddEndPuncttrue
\mciteSetBstMidEndSepPunct{\mcitedefaultmidpunct}
{\mcitedefaultendpunct}{\mcitedefaultseppunct}\relax
\EndOfBibitem
\bibitem{Aaij:2012vr}
R.~Aaij {\em et al.} ({LHCb} collaboration){,}
  \href{http://dx.doi.org/10.1007/JHEP02(2013)105}{JHEP {\bf 02},  105}
  (2013), \href{http://arxiv.org/abs/1209.4284}{{\tt arXiv:1209.4284
  [hep-ex]}}\relax
\mciteBstWouldAddEndPuncttrue
\mciteSetBstMidEndSepPunct{\mcitedefaultmidpunct}
{\mcitedefaultendpunct}{\mcitedefaultseppunct}\relax
\EndOfBibitem
\bibitem{TheBaBar:2016xwe}
J.~P.\ Lees {\em et al.} ({\babar} collaboration){,}
  \href{http://dx.doi.org/10.1103/PhysRevLett.118.031802}{Phys.\ Rev.\ Lett.\ {\bf
  118},  031802}  (2017), \href{http://arxiv.org/abs/1605.09637}{{\tt
  arXiv:1605.09637 [hep-ex]}}\relax
\mciteBstWouldAddEndPuncttrue
\mciteSetBstMidEndSepPunct{\mcitedefaultmidpunct}
{\mcitedefaultendpunct}{\mcitedefaultseppunct}\relax
\EndOfBibitem
\bibitem{Lees:2013kla}
J.~P.\ Lees {\em et al.} ({\babar} collaboration){,}
  \href{http://dx.doi.org/10.1103/PhysRevD.87.112005}{Phys.\ Rev.\ {\bf D87}{,}
  112005}  (2013), \href{http://arxiv.org/abs/1303.7465}{{\tt arXiv:1303.7465
  [hep-ex]}}\relax
\mciteBstWouldAddEndPuncttrue
\mciteSetBstMidEndSepPunct{\mcitedefaultmidpunct}
{\mcitedefaultendpunct}{\mcitedefaultseppunct}\relax
\EndOfBibitem
\bibitem{Aaij:2014pli}
R.~Aaij {\em et al.} ({LHCb} collaboration){,}
  \href{http://dx.doi.org/10.1007/JHEP06(2014)133}{JHEP {\bf 06},  133}
  (2014), \href{http://arxiv.org/abs/1403.8044}{{\tt arXiv:1403.8044
  [hep-ex]}}\relax
\mciteBstWouldAddEndPuncttrue
\mciteSetBstMidEndSepPunct{\mcitedefaultmidpunct}
{\mcitedefaultendpunct}{\mcitedefaultseppunct}\relax
\EndOfBibitem
\bibitem{Aaij:2014kwa}
R.~Aaij {\em et al.} ({LHCb} collaboration){,}
  \href{http://dx.doi.org/10.1007/JHEP10(2014)064}{JHEP {\bf 10},  064}
  (2014), \href{http://arxiv.org/abs/1408.1137}{{\tt arXiv:1408.1137
  [hep-ex]}}\relax
\mciteBstWouldAddEndPuncttrue
\mciteSetBstMidEndSepPunct{\mcitedefaultmidpunct}
{\mcitedefaultendpunct}{\mcitedefaultseppunct}\relax
\EndOfBibitem
\bibitem{Aubert:2007mm}
B.~Aubert {\em et al.} ({\babar} collaboration){,}
  \href{http://dx.doi.org/10.1103/PhysRevLett.99.051801}{Phys.\ Rev.\ Lett.\ {\bf
  99},  051801}  (2007), \href{http://arxiv.org/abs/hep-ex/0703018}{{\tt
  arXiv:hep-ex/0703018 [hep-ex]}}\relax
\mciteBstWouldAddEndPuncttrue
\mciteSetBstMidEndSepPunct{\mcitedefaultmidpunct}
{\mcitedefaultendpunct}{\mcitedefaultseppunct}\relax
\EndOfBibitem
\bibitem{Lees:2012zz}
J.~P.\ Lees {\em et al.} ({\babar} collaboration){,}
  \href{http://dx.doi.org/10.1103/PhysRevD.86.012004}{Phys.\ Rev.\ {\bf D86}{,}
  012004}  (2012), \href{http://arxiv.org/abs/1204.2852}{{\tt arXiv:1204.2852
  [hep-ex]}}\relax
\mciteBstWouldAddEndPuncttrue
\mciteSetBstMidEndSepPunct{\mcitedefaultmidpunct}
{\mcitedefaultendpunct}{\mcitedefaultseppunct}\relax
\EndOfBibitem
\bibitem{Aubert:2006vb}
B.~Aubert {\em et al.} ({\babar} collaboration){,}
  \href{http://dx.doi.org/10.1103/PhysRevD.73.092001}{Phys.\ Rev.\ {\bf D73}{,}
  092001}  (2006), \href{http://arxiv.org/abs/hep-ex/0604007}{{\tt
  arXiv:hep-ex/0604007 [hep-ex]}}\relax
\mciteBstWouldAddEndPuncttrue
\mciteSetBstMidEndSepPunct{\mcitedefaultmidpunct}
{\mcitedefaultendpunct}{\mcitedefaultseppunct}\relax
\EndOfBibitem
\bibitem{BABAR:2012aa}
J.~P.\ Lees {\em et al.} ({\babar} collaboration){,}
  \href{http://dx.doi.org/10.1103/PhysRevD.85.071103}{Phys.\ Rev.\ {\bf D85}{,}
  071103}  (2012), \href{http://arxiv.org/abs/1202.3650}{{\tt arXiv:1202.3650
  [hep-ex]}}\relax
\mciteBstWouldAddEndPuncttrue
\mciteSetBstMidEndSepPunct{\mcitedefaultmidpunct}
{\mcitedefaultendpunct}{\mcitedefaultseppunct}\relax
\EndOfBibitem
\bibitem{Aaij:2014aba}
R.~Aaij {\em et al.} ({LHCb} collaboration){,}
  \href{http://dx.doi.org/10.1103/PhysRevLett.112.131802}{Phys.\ Rev.\ Lett.\ {\bf
  112},  131802}  (2014), \href{http://arxiv.org/abs/1401.5361}{{\tt
  arXiv:1401.5361 [hep-ex]}}\relax
\mciteBstWouldAddEndPuncttrue
\mciteSetBstMidEndSepPunct{\mcitedefaultmidpunct}
{\mcitedefaultendpunct}{\mcitedefaultseppunct}\relax
\EndOfBibitem
\bibitem{Lees:2013gdj}
J.~P.\ Lees {\em et al.} ({\babar} collaboration){,}
  \href{http://dx.doi.org/10.1103/PhysRevD.89.011102}{Phys.\ Rev.\ {\bf D89}{,}
  011102}  (2014), \href{http://arxiv.org/abs/1310.8238}{{\tt arXiv:1310.8238
  [hep-ex]}}\relax
\mciteBstWouldAddEndPuncttrue
\mciteSetBstMidEndSepPunct{\mcitedefaultmidpunct}
{\mcitedefaultendpunct}{\mcitedefaultseppunct}\relax
\EndOfBibitem
\bibitem{Aaij:2011ex}
R.~Aaij {\em et al.} ({LHCb} collaboration){,}
  \href{http://dx.doi.org/10.1103/PhysRevLett.108.101601}{Phys.\ Rev.\ Lett.\ {\bf
  108},  101601}  (2012), \href{http://arxiv.org/abs/1110.0730}{{\tt
  arXiv:1110.0730 [hep-ex]}}\relax
\mciteBstWouldAddEndPuncttrue
\mciteSetBstMidEndSepPunct{\mcitedefaultmidpunct}
{\mcitedefaultendpunct}{\mcitedefaultseppunct}\relax
\EndOfBibitem
\bibitem{Aubert:2003zs}
B.~Aubert {\em et al.} ({\babar} collaboration){,}
  \href{http://dx.doi.org/10.1103/PhysRevD.70.091105}{Phys.\ Rev.\ {\bf D70}{,}
  091105}  (2004), \href{http://arxiv.org/abs/hep-ex/0409035}{{\tt
  arXiv:hep-ex/0409035 [hep-ex]}}\relax
\mciteBstWouldAddEndPuncttrue
\mciteSetBstMidEndSepPunct{\mcitedefaultmidpunct}
{\mcitedefaultendpunct}{\mcitedefaultseppunct}\relax
\EndOfBibitem
\bibitem{Aubert:2005qc}
B.~Aubert {\em et al.} ({\babar} collaboration){,}
  \href{http://dx.doi.org/10.1103/PhysRevD.72.091103}{Phys.\ Rev.\ {\bf D72}{,}
  091103}  (2005), \href{http://arxiv.org/abs/hep-ex/0501038}{{\tt
  arXiv:hep-ex/0501038 [hep-ex]}}\relax
\mciteBstWouldAddEndPuncttrue
\mciteSetBstMidEndSepPunct{\mcitedefaultmidpunct}
{\mcitedefaultendpunct}{\mcitedefaultseppunct}\relax
\EndOfBibitem
\bibitem{King:2016cxv}
Z.~King {\em et al.} ({Belle} collaboration){,}
  \href{http://dx.doi.org/10.1103/PhysRevD.93.111101}{Phys.\ Rev.\ {\bf D93}{,}
  111101}  (2016), \href{http://arxiv.org/abs/1603.06546}{{\tt
  arXiv:1603.06546 [hep-ex]}}\relax
\mciteBstWouldAddEndPuncttrue
\mciteSetBstMidEndSepPunct{\mcitedefaultmidpunct}
{\mcitedefaultendpunct}{\mcitedefaultseppunct}\relax
\EndOfBibitem
\bibitem{Lai:2013xht}
Y.~T.\ Lai {\em et al.} ({Belle} collaboration){,}
  \href{http://dx.doi.org/10.1103/PhysRevD.89.051103}{Phys.\ Rev.\ {\bf D89}{,}
  051103}  (2014), \href{http://arxiv.org/abs/1312.4228}{{\tt arXiv:1312.4228
  [hep-ex]}}\relax
\mciteBstWouldAddEndPuncttrue
\mciteSetBstMidEndSepPunct{\mcitedefaultmidpunct}
{\mcitedefaultendpunct}{\mcitedefaultseppunct}\relax
\EndOfBibitem
\bibitem{Aaij:2016flj}
R.~Aaij {\em et al.} ({LHCb} collaboration){,}
  \href{http://dx.doi.org/10.1007/JHEP11(2016)047}{JHEP {\bf 11},  047}
  (2016), \href{http://arxiv.org/abs/1606.04731}{{\tt arXiv:1606.04731
  [hep-ex]}}\relax
\mciteBstWouldAddEndPuncttrue
\mciteSetBstMidEndSepPunct{\mcitedefaultmidpunct}
{\mcitedefaultendpunct}{\mcitedefaultseppunct}\relax
\EndOfBibitem
\bibitem{Aubert:2007my}
B.~Aubert {\em et al.} ({\babar} collaboration){,}
  \href{http://dx.doi.org/10.1103/PhysRevD.77.051103}{Phys.\ Rev.\ {\bf D77}{,}
  051103}  (2008), \href{http://arxiv.org/abs/0711.4889}{{\tt arXiv:0711.4889
  [hep-ex]}}\relax
\mciteBstWouldAddEndPuncttrue
\mciteSetBstMidEndSepPunct{\mcitedefaultmidpunct}
{\mcitedefaultendpunct}{\mcitedefaultseppunct}\relax
\EndOfBibitem
\bibitem{Lees:2012ym}
J.~P.\ Lees {\em et al.} ({\babar} collaboration){,}
  \href{http://dx.doi.org/10.1103/PhysRevLett.109.191801}{Phys.\ Rev.\ Lett.\ {\bf
  109},  191801}  (2012), \href{http://arxiv.org/abs/1207.2690}{{\tt
  arXiv:1207.2690 [hep-ex]}}\relax
\mciteBstWouldAddEndPuncttrue
\mciteSetBstMidEndSepPunct{\mcitedefaultmidpunct}
{\mcitedefaultendpunct}{\mcitedefaultseppunct}\relax
\EndOfBibitem
\bibitem{Lees:2012wg}
J.~P.\ Lees {\em et al.} ({\babar} collaboration){,}
  \href{http://dx.doi.org/10.1103/PhysRevD.86.052012}{Phys.\ Rev.\ {\bf D86}{,}
  052012}  (2012), \href{http://arxiv.org/abs/1207.2520}{{\tt arXiv:1207.2520
  [hep-ex]}}\relax
\mciteBstWouldAddEndPuncttrue
\mciteSetBstMidEndSepPunct{\mcitedefaultmidpunct}
{\mcitedefaultendpunct}{\mcitedefaultseppunct}\relax
\EndOfBibitem
\bibitem{Saito:2014das}
T.~Saito {\em et al.} ({Belle} collaboration){,}
  \href{http://dx.doi.org/10.1103/PhysRevD.91.052004}{Phys.\ Rev.\ {\bf D91}{,}
  052004}  (2015), \href{http://arxiv.org/abs/1411.7198}{{\tt arXiv:1411.7198
  [hep-ex]}}\relax
\mciteBstWouldAddEndPuncttrue
\mciteSetBstMidEndSepPunct{\mcitedefaultmidpunct}
{\mcitedefaultendpunct}{\mcitedefaultseppunct}\relax
\EndOfBibitem
\bibitem{Belle:2016ufb}
A.~Abdesselam {\em et al.} ({Belle} collaboration){,}
  \href{http://arxiv.org/abs/1608.02344}{{\tt arXiv:1608.02344 [hep-ex]}}
  (2016)\relax
\mciteBstWouldAddEndPuncttrue
\mciteSetBstMidEndSepPunct{\mcitedefaultmidpunct}
{\mcitedefaultendpunct}{\mcitedefaultseppunct}\relax
\EndOfBibitem
\bibitem{delAmoSanchez:2010ae}
P.~del Amo~Sanchez {\em et al.} ({\babar} collaboration){,}
  \href{http://dx.doi.org/10.1103/PhysRevD.82.051101}{Phys.\ Rev.\ {\bf D82}{,}
  051101}  (2010), \href{http://arxiv.org/abs/1005.4087}{{\tt arXiv:1005.4087
  [hep-ex]}}\relax
\mciteBstWouldAddEndPuncttrue
\mciteSetBstMidEndSepPunct{\mcitedefaultmidpunct}
{\mcitedefaultendpunct}{\mcitedefaultseppunct}\relax
\EndOfBibitem
\bibitem{Lees:2013nxa}
J.~P.\ Lees {\em et al.} ({\babar} collaboration){,}
  \href{http://dx.doi.org/10.1103/PhysRevLett.112.211802}{Phys.\ Rev.\ Lett.\ {\bf
  112},  211802}  (2014), \href{http://arxiv.org/abs/1312.5364}{{\tt
  arXiv:1312.5364 [hep-ex]}}\relax
\mciteBstWouldAddEndPuncttrue
\mciteSetBstMidEndSepPunct{\mcitedefaultmidpunct}
{\mcitedefaultendpunct}{\mcitedefaultseppunct}\relax
\EndOfBibitem
\bibitem{Lees:2012tva}
J.~P.\ Lees {\em et al.} ({\babar} collaboration){,}
  \href{http://dx.doi.org/10.1103/PhysRevD.86.032012}{Phys.\ Rev.\ {\bf D86}{,}
  032012}  (2012), \href{http://arxiv.org/abs/1204.3933}{{\tt arXiv:1204.3933
  [hep-ex]}}\relax
\mciteBstWouldAddEndPuncttrue
\mciteSetBstMidEndSepPunct{\mcitedefaultmidpunct}
{\mcitedefaultendpunct}{\mcitedefaultseppunct}\relax
\EndOfBibitem
\bibitem{Edwards:2002kq}
K.~W.\ Edwards {\em et al.} ({CLEO} collaboration){,}
  \href{http://dx.doi.org/10.1103/PhysRevD.65.111102}{Phys.\ Rev.\ {\bf D65}{,}
  111102}  (2002), \href{http://arxiv.org/abs/hep-ex/0204017}{{\tt
  arXiv:hep-ex/0204017 [hep-ex]}}\relax
\mciteBstWouldAddEndPuncttrue
\mciteSetBstMidEndSepPunct{\mcitedefaultmidpunct}
{\mcitedefaultendpunct}{\mcitedefaultseppunct}\relax
\EndOfBibitem
\bibitem{Buchmuller:2005zv}
O.~Buchmuller and H.~Flacher{,}
  \href{http://dx.doi.org/10.1103/PhysRevD.73.073008}{Phys.\ Rev.\ {\bf D73}{,}
  073008}  (2006), \href{http://arxiv.org/abs/hep-ph/0507253}{{\tt
  arXiv:hep-ph/0507253 [hep-ph]}}\relax
\mciteBstWouldAddEndPuncttrue
\mciteSetBstMidEndSepPunct{\mcitedefaultmidpunct}
{\mcitedefaultendpunct}{\mcitedefaultseppunct}\relax
\EndOfBibitem
\bibitem{Aubert:2009ar}
B.~Aubert {\em et al.} ({\babar} collaboration){,}
  \href{http://dx.doi.org/10.1103/PhysRevD.79.091101}{Phys.\ Rev.\ {\bf D79}{,}
  091101}  (2009), \href{http://arxiv.org/abs/0903.1220}{{\tt arXiv:0903.1220
  [hep-ex]}}\relax
\mciteBstWouldAddEndPuncttrue
\mciteSetBstMidEndSepPunct{\mcitedefaultmidpunct}
{\mcitedefaultendpunct}{\mcitedefaultseppunct}\relax
\EndOfBibitem
\bibitem{Satoyama:2006xn}
N.~Satoyama {\em et al.} ({Belle} collaboration){,}
  \href{http://dx.doi.org/10.1016/j.physletb.2007.01.068}{Phys.\ Lett.\ {\bf
  B647},  67}  (2007), \href{http://arxiv.org/abs/hep-ex/0611045}{{\tt
  arXiv:hep-ex/0611045 [hep-ex]}}\relax
\mciteBstWouldAddEndPuncttrue
\mciteSetBstMidEndSepPunct{\mcitedefaultmidpunct}
{\mcitedefaultendpunct}{\mcitedefaultseppunct}\relax
\EndOfBibitem
\bibitem{Lees:2012ju}
J.~P.\ Lees {\em et al.} ({\babar} collaboration){,}
  \href{http://dx.doi.org/10.1103/PhysRevD.88.031102}{Phys.\ Rev.\ {\bf D88}{,}
  031102}  (2013), \href{http://arxiv.org/abs/1207.0698}{{\tt arXiv:1207.0698
  [hep-ex]}}\relax
\mciteBstWouldAddEndPuncttrue
\mciteSetBstMidEndSepPunct{\mcitedefaultmidpunct}
{\mcitedefaultendpunct}{\mcitedefaultseppunct}\relax
\EndOfBibitem
\bibitem{Kronenbitter:2015kls}
B.~Kronenbitter {\em et al.} ({Belle} collaboration){,}
  \href{http://dx.doi.org/10.1103/PhysRevD.92.051102}{Phys.\ Rev.\ {\bf D92}{,}
  051102}  (2015), \href{http://arxiv.org/abs/1503.05613}{{\tt
  arXiv:1503.05613 [hep-ex]}}\relax
\mciteBstWouldAddEndPuncttrue
\mciteSetBstMidEndSepPunct{\mcitedefaultmidpunct}
{\mcitedefaultendpunct}{\mcitedefaultseppunct}\relax
\EndOfBibitem
\bibitem{Aubert:2009ya}
B.~Aubert {\em et al.} ({\babar} collaboration){,}
  \href{http://dx.doi.org/10.1103/PhysRevD.80.111105}{Phys.\ Rev.\ {\bf D80}{,}
  111105}  (2009), \href{http://arxiv.org/abs/0907.1681}{{\tt arXiv:0907.1681
  [hep-ex]}}\relax
\mciteBstWouldAddEndPuncttrue
\mciteSetBstMidEndSepPunct{\mcitedefaultmidpunct}
{\mcitedefaultendpunct}{\mcitedefaultseppunct}\relax
\EndOfBibitem
\bibitem{Heller:2015vvm}
A.~Heller {\em et al.} ({Belle} collaboration){,}
  \href{http://dx.doi.org/10.1103/PhysRevD.91.112009}{Phys.\ Rev.\ {\bf D91}{,}
  112009}  (2015), \href{http://arxiv.org/abs/1504.05831}{{\tt
  arXiv:1504.05831 [hep-ex]}}\relax
\mciteBstWouldAddEndPuncttrue
\mciteSetBstMidEndSepPunct{\mcitedefaultmidpunct}
{\mcitedefaultendpunct}{\mcitedefaultseppunct}\relax
\EndOfBibitem
\bibitem{delAmoSanchez:2010bx}
P.~del Amo~Sanchez {\em et al.} ({\babar} collaboration){,}
  \href{http://dx.doi.org/10.1103/PhysRevD.83.032006}{Phys.\ Rev.\ {\bf D83}{,}
  032006}  (2011), \href{http://arxiv.org/abs/1010.2229}{{\tt arXiv:1010.2229
  [hep-ex]}}\relax
\mciteBstWouldAddEndPuncttrue
\mciteSetBstMidEndSepPunct{\mcitedefaultmidpunct}
{\mcitedefaultendpunct}{\mcitedefaultseppunct}\relax
\EndOfBibitem
\bibitem{Abe:2005bs}
S.~Villa {\em et al.} ({Belle} collaboration){,}
  \href{http://dx.doi.org/10.1103/PhysRevD.73.051107}{Phys.\ Rev.\ {\bf D73}{,}
  051107}  (2006), \href{http://arxiv.org/abs/hep-ex/0507036}{{\tt
  arXiv:hep-ex/0507036 [hep-ex]}}\relax
\mciteBstWouldAddEndPuncttrue
\mciteSetBstMidEndSepPunct{\mcitedefaultmidpunct}
{\mcitedefaultendpunct}{\mcitedefaultseppunct}\relax
\EndOfBibitem
\bibitem{Aubert:2007hb}
B.~Aubert {\em et al.} ({\babar} collaboration){,}
  \href{http://dx.doi.org/10.1103/PhysRevD.77.032007}{Phys.\ Rev.\ {\bf D77}{,}
  032007}  (2008), \href{http://arxiv.org/abs/0712.1516}{{\tt arXiv:0712.1516
  [hep-ex]}}\relax
\mciteBstWouldAddEndPuncttrue
\mciteSetBstMidEndSepPunct{\mcitedefaultmidpunct}
{\mcitedefaultendpunct}{\mcitedefaultseppunct}\relax
\EndOfBibitem
\bibitem{Chang:2003yy}
M.~C.\ Chang {\em et al.} ({Belle} collaboration){,}
  \href{http://dx.doi.org/10.1103/PhysRevD.68.111101}{Phys.\ Rev.\ {\bf D68}{,}
  111101}  (2003), \href{http://arxiv.org/abs/hep-ex/0309069}{{\tt
  arXiv:hep-ex/0309069 [hep-ex]}}\relax
\mciteBstWouldAddEndPuncttrue
\mciteSetBstMidEndSepPunct{\mcitedefaultmidpunct}
{\mcitedefaultendpunct}{\mcitedefaultseppunct}\relax
\EndOfBibitem
\bibitem{Aubert:2007up}
B.~Aubert {\em et al.} ({\babar} collaboration){,}
  \href{http://dx.doi.org/10.1103/PhysRevD.77.011104}{Phys.\ Rev.\ {\bf D77}{,}
  011104}  (2008), \href{http://arxiv.org/abs/0706.2870}{{\tt arXiv:0706.2870
  [hep-ex]}}\relax
\mciteBstWouldAddEndPuncttrue
\mciteSetBstMidEndSepPunct{\mcitedefaultmidpunct}
{\mcitedefaultendpunct}{\mcitedefaultseppunct}\relax
\EndOfBibitem
\bibitem{Aubert:2005qw}
B.~Aubert {\em et al.} ({\babar} collaboration){,}
  \href{http://dx.doi.org/10.1103/PhysRevLett.96.241802}{Phys.\ Rev.\ Lett.\ {\bf
  96},  241802}  (2006), \href{http://arxiv.org/abs/hep-ex/0511015}{{\tt
  arXiv:hep-ex/0511015 [hep-ex]}}\relax
\mciteBstWouldAddEndPuncttrue
\mciteSetBstMidEndSepPunct{\mcitedefaultmidpunct}
{\mcitedefaultendpunct}{\mcitedefaultseppunct}\relax
\EndOfBibitem
\bibitem{Aubert:2008cu}
B.~Aubert {\em et al.} ({\babar} collaboration){,}
  \href{http://dx.doi.org/10.1103/PhysRevD.77.091104}{Phys.\ Rev.\ {\bf D77}{,}
  091104}  (2008), \href{http://arxiv.org/abs/0801.0697}{{\tt arXiv:0801.0697
  [hep-ex]}}\relax
\mciteBstWouldAddEndPuncttrue
\mciteSetBstMidEndSepPunct{\mcitedefaultmidpunct}
{\mcitedefaultendpunct}{\mcitedefaultseppunct}\relax
\EndOfBibitem
\bibitem{Lees:2012wv}
J.~P.\ Lees {\em et al.} ({\babar} collaboration){,}
  \href{http://dx.doi.org/10.1103/PhysRevD.86.051105}{Phys.\ Rev.\ {\bf D86}{,}
  051105}  (2012), \href{http://arxiv.org/abs/1206.2543}{{\tt arXiv:1206.2543
  [hep-ex]}}\relax
\mciteBstWouldAddEndPuncttrue
\mciteSetBstMidEndSepPunct{\mcitedefaultmidpunct}
{\mcitedefaultendpunct}{\mcitedefaultseppunct}\relax
\EndOfBibitem
\bibitem{Hsu:2012uh}
C.~L.\ Hsu {\em et al.} ({Belle} collaboration){,}
  \href{http://dx.doi.org/10.1103/PhysRevD.86.032002}{Phys.\ Rev.\ {\bf D86}{,}
  032002}  (2012), \href{http://arxiv.org/abs/1206.5948}{{\tt arXiv:1206.5948
  [hep-ex]}}\relax
\mciteBstWouldAddEndPuncttrue
\mciteSetBstMidEndSepPunct{\mcitedefaultmidpunct}
{\mcitedefaultendpunct}{\mcitedefaultseppunct}\relax
\EndOfBibitem
\bibitem{Yook:2014kga}
Y.~Yook {\em et al.} ({Belle} collaboration){,}
  \href{http://dx.doi.org/10.1103/PhysRevD.91.052016}{Phys.\ Rev.\ {\bf D91}{,}
  052016}  (2015), \href{http://arxiv.org/abs/1406.6356}{{\tt arXiv:1406.6356
  [hep-ex]}}\relax
\mciteBstWouldAddEndPuncttrue
\mciteSetBstMidEndSepPunct{\mcitedefaultmidpunct}
{\mcitedefaultendpunct}{\mcitedefaultseppunct}\relax
\EndOfBibitem
\bibitem{Aubert:2009wt}
B.~Aubert {\em et al.} ({\babar} collaboration){,}
  \href{http://dx.doi.org/10.1103/PhysRevD.81.051101}{Phys.\ Rev.\ {\bf D81}{,}
  051101}  (2010), \href{http://arxiv.org/abs/0912.2453}{{\tt arXiv:0912.2453
  [hep-ex]}}\relax
\mciteBstWouldAddEndPuncttrue
\mciteSetBstMidEndSepPunct{\mcitedefaultmidpunct}
{\mcitedefaultendpunct}{\mcitedefaultseppunct}\relax
\EndOfBibitem
\bibitem{Adachi:2012mm}
I.~Adachi {\em et al.} ({Belle} collaboration){,}
  \href{http://dx.doi.org/10.1103/PhysRevLett.110.131801}{Phys.\ Rev.\ Lett.\ {\bf
  110},  131801}  (2013), \href{http://arxiv.org/abs/1208.4678}{{\tt
  arXiv:1208.4678 [hep-ex]}}\relax
\mciteBstWouldAddEndPuncttrue
\mciteSetBstMidEndSepPunct{\mcitedefaultmidpunct}
{\mcitedefaultendpunct}{\mcitedefaultseppunct}\relax
\EndOfBibitem
\bibitem{Aaij:2014ora}
R.~Aaij {\em et al.} ({LHCb} collaboration){,}
  \href{http://dx.doi.org/10.1103/PhysRevLett.113.151601}{Phys.\ Rev.\ Lett.\ {\bf
  113},  151601}  (2014), \href{http://arxiv.org/abs/1406.6482}{{\tt
  arXiv:1406.6482 [hep-ex]}}\relax
\mciteBstWouldAddEndPuncttrue
\mciteSetBstMidEndSepPunct{\mcitedefaultmidpunct}
{\mcitedefaultendpunct}{\mcitedefaultseppunct}\relax
\EndOfBibitem
\bibitem{Nishimura:2009ae}
K.~Nishimura {\em et al.} ({Belle} collaboration){,}
  \href{http://dx.doi.org/10.1103/PhysRevLett.105.191803}{Phys.\ Rev.\ Lett.\ {\bf
  105},  191803}  (2010), \href{http://arxiv.org/abs/0910.4751}{{\tt
  arXiv:0910.4751 [hep-ex]}}\relax
\mciteBstWouldAddEndPuncttrue
\mciteSetBstMidEndSepPunct{\mcitedefaultmidpunct}
{\mcitedefaultendpunct}{\mcitedefaultseppunct}\relax
\EndOfBibitem
\bibitem{Browder:1998yb}
T.~E.\ Browder {\em et al.} ({CLEO} collaboration){,}
  \href{http://dx.doi.org/10.1103/PhysRevLett.81.1786}{Phys.\ Rev.\ Lett.\ {\bf
  81},  1786}  (1998), \href{http://arxiv.org/abs/hep-ex/9804018}{{\tt
  arXiv:hep-ex/9804018 [hep-ex]}}\relax
\mciteBstWouldAddEndPuncttrue
\mciteSetBstMidEndSepPunct{\mcitedefaultmidpunct}
{\mcitedefaultendpunct}{\mcitedefaultseppunct}\relax
\EndOfBibitem
\bibitem{Aubert:2004eq}
B.~Aubert {\em et al.} ({\babar} collaboration){,}
  \href{http://dx.doi.org/10.1103/PhysRevLett.93.061801}{Phys.\ Rev.\ Lett.\ {\bf
  93},  061801}  (2004), \href{http://arxiv.org/abs/hep-ex/0401006}{{\tt
  arXiv:hep-ex/0401006 [hep-ex]}}\relax
\mciteBstWouldAddEndPuncttrue
\mciteSetBstMidEndSepPunct{\mcitedefaultmidpunct}
{\mcitedefaultendpunct}{\mcitedefaultseppunct}\relax
\EndOfBibitem
\bibitem{Bonvicini:2003aw}
G.~Bonvicini {\em et al.} ({CLEO} collaboration){,}
  \href{http://dx.doi.org/10.1103/PhysRevD.68.011101}{Phys.\ Rev.\ {\bf D68}{,}
  011101}  (2003), \href{http://arxiv.org/abs/hep-ex/0303009}{{\tt
  arXiv:hep-ex/0303009 [hep-ex]}}\relax
\mciteBstWouldAddEndPuncttrue
\mciteSetBstMidEndSepPunct{\mcitedefaultmidpunct}
{\mcitedefaultendpunct}{\mcitedefaultseppunct}\relax
\EndOfBibitem
\bibitem{delAmoSanchez:2010gx}
P.~del Amo~Sanchez {\em et al.} ({\babar} collaboration){,}
  \href{http://dx.doi.org/10.1103/PhysRevD.83.031103}{Phys.\ Rev.\ {\bf D83}{,}
  031103}  (2011), \href{http://arxiv.org/abs/1012.5031}{{\tt arXiv:1012.5031
  [hep-ex]}}\relax
\mciteBstWouldAddEndPuncttrue
\mciteSetBstMidEndSepPunct{\mcitedefaultmidpunct}
{\mcitedefaultendpunct}{\mcitedefaultseppunct}\relax
\EndOfBibitem
\bibitem{Aaij:2014wgo}
R.~Aaij {\em et al.} ({LHCb} collaboration){,}
  \href{http://dx.doi.org/10.1103/PhysRevLett.112.161801}{Phys.\ Rev.\ Lett.\ {\bf
  112},  161801}  (2014), \href{http://arxiv.org/abs/1402.6852}{{\tt
  arXiv:1402.6852 [hep-ex]}}\relax
\mciteBstWouldAddEndPuncttrue
\mciteSetBstMidEndSepPunct{\mcitedefaultmidpunct}
{\mcitedefaultendpunct}{\mcitedefaultseppunct}\relax
\EndOfBibitem
\bibitem{Aaij:2013hha}
R.~Aaij {\em et al.} ({LHCb} collaboration){,}
  \href{http://dx.doi.org/10.1007/JHEP05(2013)159}{JHEP {\bf 05},  159}
  (2013), \href{http://arxiv.org/abs/1304.3035}{{\tt arXiv:1304.3035
  [hep-ex]}}\relax
\mciteBstWouldAddEndPuncttrue
\mciteSetBstMidEndSepPunct{\mcitedefaultmidpunct}
{\mcitedefaultendpunct}{\mcitedefaultseppunct}\relax
\EndOfBibitem
\bibitem{Aaij:2014tfa}
R.~Aaij {\em et al.} ({LHCb} collaboration){,}
  \href{http://dx.doi.org/10.1007/JHEP05(2014)082}{JHEP {\bf 05},  082}
  (2014), \href{http://arxiv.org/abs/1403.8045}{{\tt arXiv:1403.8045
  [hep-ex]}}\relax
\mciteBstWouldAddEndPuncttrue
\mciteSetBstMidEndSepPunct{\mcitedefaultmidpunct}
{\mcitedefaultendpunct}{\mcitedefaultseppunct}\relax
\EndOfBibitem
\bibitem{Abdesselam:2016llu}
A.~Abdesselam {\em et al.} ({Belle} collaboration){,}
  \href{http://arxiv.org/abs/1604.04042}{{\tt arXiv:1604.04042 [hep-ex]}}
  (2016)\relax
\mciteBstWouldAddEndPuncttrue
\mciteSetBstMidEndSepPunct{\mcitedefaultmidpunct}
{\mcitedefaultendpunct}{\mcitedefaultseppunct}\relax
\EndOfBibitem
\bibitem{Aaij:2015dea}
R.~Aaij {\em et al.} ({LHCb} collaboration){,}
  \href{http://dx.doi.org/10.1007/JHEP04(2015)064}{JHEP {\bf 04},  064}
  (2015), \href{http://arxiv.org/abs/1501.03038}{{\tt arXiv:1501.03038
  [hep-ex]}}\relax
\mciteBstWouldAddEndPuncttrue
\mciteSetBstMidEndSepPunct{\mcitedefaultmidpunct}
{\mcitedefaultendpunct}{\mcitedefaultseppunct}\relax
\EndOfBibitem
\bibitem{Lees:2015ymt}
J.~P.\ Lees {\em et al.} ({\babar} collaboration){,}
  \href{http://dx.doi.org/10.1103/PhysRevD.93.052015}{Phys.\ Rev.\ {\bf D93}{,}
  052015}  (2016), \href{http://arxiv.org/abs/1508.07960}{{\tt
  arXiv:1508.07960 [hep-ex]}}\relax
\mciteBstWouldAddEndPuncttrue
\mciteSetBstMidEndSepPunct{\mcitedefaultmidpunct}
{\mcitedefaultendpunct}{\mcitedefaultseppunct}\relax
\EndOfBibitem
\bibitem{Aaij:2015oid}
R.~Aaij {\em et al.} ({LHCb} collaboration){,}
  \href{http://dx.doi.org/10.1007/JHEP02(2016)104}{JHEP {\bf 02},  104}
  (2016), \href{http://arxiv.org/abs/1512.04442}{{\tt arXiv:1512.04442
  [hep-ex]}}\relax
\mciteBstWouldAddEndPuncttrue
\mciteSetBstMidEndSepPunct{\mcitedefaultmidpunct}
{\mcitedefaultendpunct}{\mcitedefaultseppunct}\relax
\EndOfBibitem
\bibitem{Khachatryan:2015isa}
V.~Khachatryan {\em et al.} ({CMS} collaboration){,}
  \href{http://dx.doi.org/10.1016/j.physletb.2015.12.020}{Phys.\ Lett.\ {\bf
  B753},  424}  (2016), \href{http://arxiv.org/abs/1507.08126}{{\tt
  arXiv:1507.08126 [hep-ex]}}\relax
\mciteBstWouldAddEndPuncttrue
\mciteSetBstMidEndSepPunct{\mcitedefaultmidpunct}
{\mcitedefaultendpunct}{\mcitedefaultseppunct}\relax
\EndOfBibitem
\bibitem{Sato:2014pjr}
Y.~Sato {\em et al.} ({Belle} collaboration){,}
  \href{http://dx.doi.org/10.1103/PhysRevD.93.032008}{Phys.\ Rev.\ {\bf D93}{,}
  032008}  (2016), \href{http://arxiv.org/abs/1402.7134}{{\tt arXiv:1402.7134
  [hep-ex]}}, Addendum ibid.\
  \href{http://dx.doi.org/10.1103/PhysRevD.93.059901}{{\bf D93}, 059901}
  (2016)\relax
\mciteBstWouldAddEndPuncttrue
\mciteSetBstMidEndSepPunct{\mcitedefaultmidpunct}
{\mcitedefaultendpunct}{\mcitedefaultseppunct}\relax
\EndOfBibitem
\bibitem{Aaij:2016kqt}
R.~Aaij {\em et al.} ({LHCb} collaboration){,}
  \href{http://dx.doi.org/10.1007/JHEP12(2016)065}{JHEP {\bf 12},  065}
  (2016), \href{http://arxiv.org/abs/1609.04736}{{\tt arXiv:1609.04736
  [hep-ex]}}\relax
\mciteBstWouldAddEndPuncttrue
\mciteSetBstMidEndSepPunct{\mcitedefaultmidpunct}
{\mcitedefaultendpunct}{\mcitedefaultseppunct}\relax
\EndOfBibitem
\bibitem{Pesantez:2015fza}
L.~Pesántez {\em et al.} ({Belle} collaboration){,}
  \href{http://dx.doi.org/10.1103/PhysRevLett.114.151601}{Phys.\ Rev.\ Lett.\ {\bf
  114},  151601}  (2015), \href{http://arxiv.org/abs/1501.01702}{{\tt
  arXiv:1501.01702 [hep-ex]}}\relax
\mciteBstWouldAddEndPuncttrue
\mciteSetBstMidEndSepPunct{\mcitedefaultmidpunct}
{\mcitedefaultendpunct}{\mcitedefaultseppunct}\relax
\EndOfBibitem
\bibitem{Aaij:2014iva}
R.~Aaij {\em et al.} ({LHCb} collaboration){,}
  \href{http://dx.doi.org/10.1103/PhysRevD.90.112004}{Phys.\ Rev.\ {\bf D90}{,}
  112004}  (2014), \href{http://arxiv.org/abs/1408.5373}{{\tt arXiv:1408.5373
  [hep-ex]}}\relax
\mciteBstWouldAddEndPuncttrue
\mciteSetBstMidEndSepPunct{\mcitedefaultmidpunct}
{\mcitedefaultendpunct}{\mcitedefaultseppunct}\relax
\EndOfBibitem
\bibitem{Aubert:2006fs}
B.~Aubert {\em et al.} ({\babar} collaboration){,}
  \href{http://dx.doi.org/10.1103/PhysRevLett.97.201801}{Phys.\ Rev.\ Lett.\ {\bf
  97},  201801}  (2006), \href{http://arxiv.org/abs/hep-ex/0607057}{{\tt
  arXiv:hep-ex/0607057 [hep-ex]}}\relax
\mciteBstWouldAddEndPuncttrue
\mciteSetBstMidEndSepPunct{\mcitedefaultmidpunct}
{\mcitedefaultendpunct}{\mcitedefaultseppunct}\relax
\EndOfBibitem
\bibitem{Chen:2005zv}
K.~F.\ Chen {\em et al.} ({Belle} collaboration){,}
  \href{http://dx.doi.org/10.1103/PhysRevLett.94.221804}{Phys.\ Rev.\ Lett.\ {\bf
  94},  221804}  (2005), \href{http://arxiv.org/abs/hep-ex/0503013}{{\tt
  arXiv:hep-ex/0503013 [hep-ex]}}\relax
\mciteBstWouldAddEndPuncttrue
\mciteSetBstMidEndSepPunct{\mcitedefaultmidpunct}
{\mcitedefaultendpunct}{\mcitedefaultseppunct}\relax
\EndOfBibitem
\bibitem{Aaij:2013fla}
R.~Aaij {\em et al.} ({LHCb} collaboration){,}
  \href{http://dx.doi.org/10.1103/PhysRevD.88.052015}{Phys.\ Rev.\ {\bf D88}{,}
  052015}  (2013), \href{http://arxiv.org/abs/1307.6165}{{\tt arXiv:1307.6165
  [hep-ex]}}\relax
\mciteBstWouldAddEndPuncttrue
\mciteSetBstMidEndSepPunct{\mcitedefaultmidpunct}
{\mcitedefaultendpunct}{\mcitedefaultseppunct}\relax
\EndOfBibitem
\bibitem{Aaij:2014bsa}
R.~Aaij {\em et al.} ({LHCb} collaboration){,}
  \href{http://dx.doi.org/10.1007/JHEP09(2014)177}{JHEP {\bf 09},  177}
  (2014), \href{http://arxiv.org/abs/1408.0978}{{\tt arXiv:1408.0978
  [hep-ex]}}\relax
\mciteBstWouldAddEndPuncttrue
\mciteSetBstMidEndSepPunct{\mcitedefaultmidpunct}
{\mcitedefaultendpunct}{\mcitedefaultseppunct}\relax
\EndOfBibitem
\bibitem{Aaltonen:2014vra}
T.~A.\ Aaltonen {\em et al.} ({CDF} collaboration){,}
  \href{http://dx.doi.org/10.1103/PhysRevLett.113.242001}{Phys.\ Rev.\ Lett.\ {\bf
  113},  242001}  (2014), \href{http://arxiv.org/abs/1403.5586}{{\tt
  arXiv:1403.5586 [hep-ex]}}\relax
\mciteBstWouldAddEndPuncttrue
\mciteSetBstMidEndSepPunct{\mcitedefaultmidpunct}
{\mcitedefaultendpunct}{\mcitedefaultseppunct}\relax
\EndOfBibitem
\bibitem{Aaij:2013iua}
R.~Aaij {\em et al.} ({LHCb} collaboration){,}
  \href{http://dx.doi.org/10.1103/PhysRevLett.110.221601}{Phys.\ Rev.\ Lett.\ {\bf
  110},  221601}  (2013), \href{http://arxiv.org/abs/1304.6173}{{\tt
  arXiv:1304.6173 [hep-ex]}}\relax
\mciteBstWouldAddEndPuncttrue
\mciteSetBstMidEndSepPunct{\mcitedefaultmidpunct}
{\mcitedefaultendpunct}{\mcitedefaultseppunct}\relax
\EndOfBibitem
\bibitem{Aaij:2014tpa}
R.~Aaij {\em et al.} ({LHCb} collaboration){,}
  \href{http://dx.doi.org/10.1007/JHEP05(2014)069}{JHEP {\bf 05},  069}
  (2014), \href{http://arxiv.org/abs/1403.2888}{{\tt arXiv:1403.2888
  [hep-ex]}}\relax
\mciteBstWouldAddEndPuncttrue
\mciteSetBstMidEndSepPunct{\mcitedefaultmidpunct}
{\mcitedefaultendpunct}{\mcitedefaultseppunct}\relax
\EndOfBibitem
\bibitem{Lees:2014uoa}
J.~P.\ Lees {\em et al.} ({\babar} collaboration){,}
  \href{http://dx.doi.org/10.1103/PhysRevD.90.092001}{Phys.\ Rev.\ {\bf D90}{,}
  092001}  (2014), \href{http://arxiv.org/abs/1406.0534}{{\tt arXiv:1406.0534
  [hep-ex]}}\relax
\mciteBstWouldAddEndPuncttrue
\mciteSetBstMidEndSepPunct{\mcitedefaultmidpunct}
{\mcitedefaultendpunct}{\mcitedefaultseppunct}\relax
\EndOfBibitem
\bibitem{Nishida:2003paa}
S.~Nishida {\em et al.} ({Belle} collaboration){,}
  \href{http://dx.doi.org/10.1103/PhysRevLett.93.031803}{Phys.\ Rev.\ Lett.\ {\bf
  93},  031803}  (2004), \href{http://arxiv.org/abs/hep-ex/0308038}{{\tt
  arXiv:hep-ex/0308038 [hep-ex]}}\relax
\mciteBstWouldAddEndPuncttrue
\mciteSetBstMidEndSepPunct{\mcitedefaultmidpunct}
{\mcitedefaultendpunct}{\mcitedefaultseppunct}\relax
\EndOfBibitem
\bibitem{Coan:2000pu}
T.~E.\ Coan {\em et al.} ({CLEO} collaboration){,}
  \href{http://dx.doi.org/10.1103/PhysRevLett.86.5661}{Phys.\ Rev.\ Lett.\ {\bf
  86},  5661}  (2001), \href{http://arxiv.org/abs/hep-ex/0010075}{{\tt
  arXiv:hep-ex/0010075 [hep-ex]}}\relax
\mciteBstWouldAddEndPuncttrue
\mciteSetBstMidEndSepPunct{\mcitedefaultmidpunct}
{\mcitedefaultendpunct}{\mcitedefaultseppunct}\relax
\EndOfBibitem
\bibitem{Aaij:2016cla}
R.~Aaij {\em et al.} ({LHCb} collaboration){,}
  \href{http://dx.doi.org/10.1038/nphys4021}{Nature Phys.\ {\bf 13},  391}
  (2017), \href{http://arxiv.org/abs/1609.05216}{{\tt arXiv:1609.05216
  [hep-ex]}}\relax
\mciteBstWouldAddEndPuncttrue
\mciteSetBstMidEndSepPunct{\mcitedefaultmidpunct}
{\mcitedefaultendpunct}{\mcitedefaultseppunct}\relax
\EndOfBibitem
\bibitem{Abe:2004mq}
J.~Zhang {\em et al.} ({Belle} collaboration){,}
  \href{http://dx.doi.org/10.1103/PhysRevLett.95.141801}{Phys.\ Rev.\ Lett.\ {\bf
  95},  141801}  (2005), \href{http://arxiv.org/abs/hep-ex/0408102}{{\tt
  arXiv:hep-ex/0408102 [hep-ex]}}\relax
\mciteBstWouldAddEndPuncttrue
\mciteSetBstMidEndSepPunct{\mcitedefaultmidpunct}
{\mcitedefaultendpunct}{\mcitedefaultseppunct}\relax
\EndOfBibitem
\bibitem{Aaij:2012ud}
R.~Aaij {\em et al.} ({LHCb} collaboration){,}
  \href{http://dx.doi.org/10.1016/j.physletb.2012.06.012}{Phys.\ Lett.\ {\bf
  B713},  369}  (2012), \href{http://arxiv.org/abs/1204.2813}{{\tt
  arXiv:1204.2813 [hep-ex]}}\relax
\mciteBstWouldAddEndPuncttrue
\mciteSetBstMidEndSepPunct{\mcitedefaultmidpunct}
{\mcitedefaultendpunct}{\mcitedefaultseppunct}\relax
\EndOfBibitem
\bibitem{Aaij:2016xxs}
R.~Aaij {\em et al.} ({LHCb} collaboration){,}
  \href{http://dx.doi.org/10.1016/j.physletb.2016.05.074}{Phys.\ Lett.\ {\bf
  B759},  313}  (2016), \href{http://arxiv.org/abs/1603.07037}{{\tt
  arXiv:1603.07037 [hep-ex]}}\relax
\mciteBstWouldAddEndPuncttrue
\mciteSetBstMidEndSepPunct{\mcitedefaultmidpunct}
{\mcitedefaultendpunct}{\mcitedefaultseppunct}\relax
\EndOfBibitem
\bibitem{Staric:2007dt}
M.~Staric {\em et al.} ({Belle} collaboration){,}
  \href{http://dx.doi.org/10.1103/PhysRevLett.98.211803}{Phys.\ Rev.\ Lett.\ {\bf
  98},  211803}  (2007), \href{http://arxiv.org/abs/hep-ex/0703036}{{\tt
  arXiv:hep-ex/0703036}}\relax
\mciteBstWouldAddEndPuncttrue
\mciteSetBstMidEndSepPunct{\mcitedefaultmidpunct}
{\mcitedefaultendpunct}{\mcitedefaultseppunct}\relax
\EndOfBibitem
\bibitem{Aubert:2007wf}
B.~Aubert {\em et al.} ({\babar} collaboration){,}
  \href{http://dx.doi.org/10.1103/PhysRevLett.98.211802}{Phys.\ Rev.\ Lett.\ {\bf
  98},  211802}  (2007), \href{http://arxiv.org/abs/hep-ex/0703020}{{\tt
  arXiv:hep-ex/0703020}}\relax
\mciteBstWouldAddEndPuncttrue
\mciteSetBstMidEndSepPunct{\mcitedefaultmidpunct}
{\mcitedefaultendpunct}{\mcitedefaultseppunct}\relax
\EndOfBibitem
\bibitem{Aaltonen:2007uc}
T.~Aaltonen {\em et al.} ({CDF} collaboration){,}
  \href{http://dx.doi.org/10.1103/PhysRevLett.100.121802}{Phys.\ Rev.\ Lett.\ {\bf
  100},  121802}  (2008), \href{http://arxiv.org/abs/0712.1567}{{\tt
  arXiv:0712.1567 [hep-ex]}}\relax
\mciteBstWouldAddEndPuncttrue
\mciteSetBstMidEndSepPunct{\mcitedefaultmidpunct}
{\mcitedefaultendpunct}{\mcitedefaultseppunct}\relax
\EndOfBibitem
\bibitem{Aaij:2013wda}
R.~Aaij {\em et al.} ({LHCb} collaboration){,}
  \href{http://dx.doi.org/10.1103/PhysRevLett.111.251801}{Phys.\ Rev.\ Lett.\ {\bf
  111},  251801}  (2013), \href{http://arxiv.org/abs/1309.6534}{{\tt
  arXiv:1309.6534 [hep-ex]}}\relax
\mciteBstWouldAddEndPuncttrue
\mciteSetBstMidEndSepPunct{\mcitedefaultmidpunct}
{\mcitedefaultendpunct}{\mcitedefaultseppunct}\relax
\EndOfBibitem
\bibitem{Bergmann:2000id}
S.~Bergmann, Y.~Grossman, Z.~Ligeti, Y.~Nir, and A.~A.\ Petrov{,}
  \href{http://dx.doi.org/10.1016/S0370-2693(00)00772-3}{Phys.\ Lett.\ {\bf
  B486},  418}  (2000), \href{http://arxiv.org/abs/hep-ph/0005181}{{\tt
  arXiv:hep-ph/0005181 [hep-ph]}}\relax
\mciteBstWouldAddEndPuncttrue
\mciteSetBstMidEndSepPunct{\mcitedefaultmidpunct}
{\mcitedefaultendpunct}{\mcitedefaultseppunct}\relax
\EndOfBibitem
\bibitem{Bigi:2000wn}
I.~I.~Y.\ Bigi and N.~G.\ Uraltsev{,}
  \href{http://dx.doi.org/10.1016/S0550-3213(00)00604-0}{Nucl.\ Phys.\ {\bf
  B592},  92}  (2001), \href{http://arxiv.org/abs/hep-ph/0005089}{{\tt
  arXiv:hep-ph/0005089}}\relax
\mciteBstWouldAddEndPuncttrue
\mciteSetBstMidEndSepPunct{\mcitedefaultmidpunct}
{\mcitedefaultendpunct}{\mcitedefaultseppunct}\relax
\EndOfBibitem
\bibitem{Petrov:2003un}
A.~A.\ Petrov, \href{http://arxiv.org/abs/hep-ph/0311371}{{\tt
  arXiv:hep-ph/0311371}}  (2003)\relax
\mciteBstWouldAddEndPuncttrue
\mciteSetBstMidEndSepPunct{\mcitedefaultmidpunct}
{\mcitedefaultendpunct}{\mcitedefaultseppunct}\relax
\EndOfBibitem
\bibitem{Petrov:2004rf}
A.~A.\ Petrov, \href{http://dx.doi.org/10.1016/j.nuclphysbps.2005.01.057}{Nucl.\
  Phys.\ Proc.\ Suppl.\ {\bf 142},  333}  (2005){,}
  \href{http://arxiv.org/abs/hep-ph/0409130}{{\tt arXiv:hep-ph/0409130}}\relax
\mciteBstWouldAddEndPuncttrue
\mciteSetBstMidEndSepPunct{\mcitedefaultmidpunct}
{\mcitedefaultendpunct}{\mcitedefaultseppunct}\relax
\EndOfBibitem
\bibitem{Falk:2004wg}
A.~F.\ Falk, Y.~Grossman, Z.~Ligeti, Y.~Nir, and A.~A.\ Petrov{,}
  \href{http://dx.doi.org/10.1103/PhysRevD.69.114021}{Phys.\ Rev.\ {\bf D69}{,}
  114021}  (2004), \href{http://arxiv.org/abs/hep-ph/0402204}{{\tt
  arXiv:hep-ph/0402204}}\relax
\mciteBstWouldAddEndPuncttrue
\mciteSetBstMidEndSepPunct{\mcitedefaultmidpunct}
{\mcitedefaultendpunct}{\mcitedefaultseppunct}\relax
\EndOfBibitem
\bibitem{Asner:2012xb}
D.~Asner {\em et al.} ({CLEO} collaboration){,}
  \href{http://dx.doi.org/10.1103/PhysRevD.86.112001}{Phys.\ Rev.\ {\bf D86}{,}
  112001}  (2012), \href{http://arxiv.org/abs/1210.0939}{{\tt arXiv:1210.0939
  [hep-ex]}}\relax
\mciteBstWouldAddEndPuncttrue
\mciteSetBstMidEndSepPunct{\mcitedefaultmidpunct}
{\mcitedefaultendpunct}{\mcitedefaultseppunct}\relax
\EndOfBibitem
\bibitem{Aaij:2014gsa}
R.~Aaij {\em et al.} ({LHCb} collaboration){,}
  \href{http://dx.doi.org/10.1007/JHEP07(2014)041}{JHEP {\bf 07},  041}
  (2014), \href{http://arxiv.org/abs/1405.2797}{{\tt arXiv:1405.2797
  [hep-ex]}}\relax
\mciteBstWouldAddEndPuncttrue
\mciteSetBstMidEndSepPunct{\mcitedefaultmidpunct}
{\mcitedefaultendpunct}{\mcitedefaultseppunct}\relax
\EndOfBibitem
\bibitem{Aubert:2008zh}
B.~Aubert {\em et al.} ({\babar} collaboration){,}
  \href{http://dx.doi.org/10.1103/PhysRevLett.103.211801}{Phys.\ Rev.\ Lett.\ {\bf
  103},  211801}  (2009), \href{http://arxiv.org/abs/0807.4544}{{\tt
  arXiv:0807.4544 [hep-ex]}}\relax
\mciteBstWouldAddEndPuncttrue
\mciteSetBstMidEndSepPunct{\mcitedefaultmidpunct}
{\mcitedefaultendpunct}{\mcitedefaultseppunct}\relax
\EndOfBibitem
\bibitem{Aaij:2016rhq}
R.~Aaij {\em et al.} ({LHCb} collaboration){,}
  \href{http://dx.doi.org/10.1103/PhysRevLett.116.241801}{Phys.\ Rev.\ Lett.\ {\bf
  116},  241801}  (2016), \href{http://arxiv.org/abs/1602.07224}{{\tt
  arXiv:1602.07224 [hep-ex]}}\relax
\mciteBstWouldAddEndPuncttrue
\mciteSetBstMidEndSepPunct{\mcitedefaultmidpunct}
{\mcitedefaultendpunct}{\mcitedefaultseppunct}\relax
\EndOfBibitem
\bibitem{HFLAV_charm:webpage}
 Heavy Flavor Averaging Group: charm physics parameters{,}
  \url{http://www.slac.stanford.edu/xorg/hflav/charm/index.html}\relax
\mciteBstWouldAddEndPuncttrue
\mciteSetBstMidEndSepPunct{\mcitedefaultmidpunct}
{\mcitedefaultendpunct}{\mcitedefaultseppunct}\relax
\EndOfBibitem
\bibitem{Aitala:1996vz}
E.~M.\ Aitala {\em et al.} ({Fermilab E791} collaboration){,}
  \href{http://dx.doi.org/10.1103/PhysRevLett.77.2384}{Phys.\ Rev.\ Lett.\ {\bf
  77},  2384}  (1996), \href{http://arxiv.org/abs/hep-ex/9606016}{{\tt
  arXiv:hep-ex/9606016}}\relax
\mciteBstWouldAddEndPuncttrue
\mciteSetBstMidEndSepPunct{\mcitedefaultmidpunct}
{\mcitedefaultendpunct}{\mcitedefaultseppunct}\relax
\EndOfBibitem
\bibitem{Cawlfield:2005ze}
C.~Cawlfield {\em et al.} ({CLEO} collaboration){,}
  \href{http://dx.doi.org/10.1103/PhysRevD.71.077101}{Phys.\ Rev.\ {\bf D71}{,}
  077101}  (2005), \href{http://arxiv.org/abs/hep-ex/0502012}{{\tt
  arXiv:hep-ex/0502012}}\relax
\mciteBstWouldAddEndPuncttrue
\mciteSetBstMidEndSepPunct{\mcitedefaultmidpunct}
{\mcitedefaultendpunct}{\mcitedefaultseppunct}\relax
\EndOfBibitem
\bibitem{Aubert:2007aa}
B.~Aubert {\em et al.} ({\babar} collaboration){,}
  \href{http://dx.doi.org/10.1103/PhysRevD.76.014018}{Phys.\ Rev.\ {\bf D76}{,}
  014018}  (2007), \href{http://arxiv.org/abs/0705.0704}{{\tt arXiv:0705.0704
  [hep-ex]}}\relax
\mciteBstWouldAddEndPuncttrue
\mciteSetBstMidEndSepPunct{\mcitedefaultmidpunct}
{\mcitedefaultendpunct}{\mcitedefaultseppunct}\relax
\EndOfBibitem
\bibitem{Bitenc:2008bk}
U.~Bitenc {\em et al.} ({Belle} collaboration){,}
  \href{http://dx.doi.org/10.1103/PhysRevD.77.112003}{Phys.\ Rev.\ {\bf D77}{,}
  112003}  (2008), \href{http://arxiv.org/abs/0802.2952}{{\tt arXiv:0802.2952
  [hep-ex]}}\relax
\mciteBstWouldAddEndPuncttrue
\mciteSetBstMidEndSepPunct{\mcitedefaultmidpunct}
{\mcitedefaultendpunct}{\mcitedefaultseppunct}\relax
\EndOfBibitem
\bibitem{Zhang:2006dp}
L.~M.\ Zhang {\em et al.} ({Belle} collaboration){,}
  \href{http://dx.doi.org/10.1103/PhysRevLett.96.151801}{Phys.\ Rev.\ Lett.\ {\bf
  96},  151801}  (2006), \href{http://arxiv.org/abs/hep-ex/0601029}{{\tt
  arXiv:hep-ex/0601029}}\relax
\mciteBstWouldAddEndPuncttrue
\mciteSetBstMidEndSepPunct{\mcitedefaultmidpunct}
{\mcitedefaultendpunct}{\mcitedefaultseppunct}\relax
\EndOfBibitem
\bibitem{Ko:2014qvu}
B.~R.\ Ko {\em et al.} ({Belle} collaboration){,}
  \href{http://dx.doi.org/10.1103/PhysRevLett.112.111801}{Phys.\ Rev.\ Lett.\ {\bf
  112},  111801}  (2014), \href{http://arxiv.org/abs/1401.3402}{{\tt
  arXiv:1401.3402 [hep-ex]}}\relax
\mciteBstWouldAddEndPuncttrue
\mciteSetBstMidEndSepPunct{\mcitedefaultmidpunct}
{\mcitedefaultendpunct}{\mcitedefaultseppunct}\relax
\EndOfBibitem
\bibitem{Aaltonen:2013pja}
T.~A.\ Aaltonen {\em et al.} ({CDF} collaboration){,}
  \href{http://dx.doi.org/10.1103/PhysRevLett.111.231802}{Phys.\ Rev.\ Lett.\ {\bf
  111},  231802}  (2013), \href{http://arxiv.org/abs/1309.4078}{{\tt
  arXiv:1309.4078 [hep-ex]}}\relax
\mciteBstWouldAddEndPuncttrue
\mciteSetBstMidEndSepPunct{\mcitedefaultmidpunct}
{\mcitedefaultendpunct}{\mcitedefaultseppunct}\relax
\EndOfBibitem
\bibitem{Aaij:2016roz}
R.~Aaij {\em et al.} ({LHCb} collaboration){,}
  \href{http://dx.doi.org/10.1103/PhysRevD.95.052004}{Phys.\ Rev.\ {\bf D95}{,}
  052004}  (2017), \href{http://arxiv.org/abs/1611.06143}{{\tt
  arXiv:1611.06143 [hep-ex]}}\relax
\mciteBstWouldAddEndPuncttrue
\mciteSetBstMidEndSepPunct{\mcitedefaultmidpunct}
{\mcitedefaultendpunct}{\mcitedefaultseppunct}\relax
\EndOfBibitem
\bibitem{Peng:2014oda}
T.~Peng {\em et al.} ({Belle} collaboration){,}
  \href{http://dx.doi.org/10.1103/PhysRevD.89.091103}{Phys.\ Rev.\ {\bf D89}{,}
  091103}  (2014), \href{http://arxiv.org/abs/1404.2412}{{\tt arXiv:1404.2412
  [hep-ex]}}\relax
\mciteBstWouldAddEndPuncttrue
\mciteSetBstMidEndSepPunct{\mcitedefaultmidpunct}
{\mcitedefaultendpunct}{\mcitedefaultseppunct}\relax
\EndOfBibitem
\bibitem{delAmoSanchez:2010xz}
P.~del Amo~Sanchez {\em et al.} ({\babar} collaboration){,}
  \href{http://dx.doi.org/10.1103/PhysRevLett.105.081803}{Phys.\ Rev.\ Lett.\ {\bf
  105},  081803}  (2010), \href{http://arxiv.org/abs/1004.5053}{{\tt
  arXiv:1004.5053 [hep-ex]}}\relax
\mciteBstWouldAddEndPuncttrue
\mciteSetBstMidEndSepPunct{\mcitedefaultmidpunct}
{\mcitedefaultendpunct}{\mcitedefaultseppunct}\relax
\EndOfBibitem
\bibitem{Aaij:2015xoa}
R.~Aaij {\em et al.} ({LHCb} collaboration){,}
  \href{http://dx.doi.org/10.1007/JHEP04(2016)033}{JHEP {\bf 04},  033}
  (2016), \href{http://arxiv.org/abs/1510.01664}{{\tt arXiv:1510.01664
  [hep-ex]}}\relax
\mciteBstWouldAddEndPuncttrue
\mciteSetBstMidEndSepPunct{\mcitedefaultmidpunct}
{\mcitedefaultendpunct}{\mcitedefaultseppunct}\relax
\EndOfBibitem
\bibitem{Lees:2016gom}
J.~P.\ Lees {\em et al.} ({\babar} collaboration){,}
  \href{http://dx.doi.org/10.1103/PhysRevD.93.112014}{Phys.\ Rev.\ {\bf D93}{,}
  112014}  (2016), \href{http://arxiv.org/abs/1604.00857}{{\tt
  arXiv:1604.00857 [hep-ex]}}\relax
\mciteBstWouldAddEndPuncttrue
\mciteSetBstMidEndSepPunct{\mcitedefaultmidpunct}
{\mcitedefaultendpunct}{\mcitedefaultseppunct}\relax
\EndOfBibitem
\bibitem{Aubert:2007if}
B.~Aubert {\em et al.} ({\babar} collaboration){,}
  \href{http://dx.doi.org/10.1103/PhysRevLett.100.061803}{Phys.\ Rev.\ Lett.\ {\bf
  100},  061803}  (2008), \href{http://arxiv.org/abs/0709.2715}{{\tt
  arXiv:0709.2715 [hep-ex]}}\relax
\mciteBstWouldAddEndPuncttrue
\mciteSetBstMidEndSepPunct{\mcitedefaultmidpunct}
{\mcitedefaultendpunct}{\mcitedefaultseppunct}\relax
\EndOfBibitem
\bibitem{Ko:2012jh}
B.~R.\ Ko ({Belle} collaboration), \href{http://arxiv.org/abs/1212.5320}{{\tt
  arXiv:1212.5320 [hep-ex]}}  (2012)\relax
\mciteBstWouldAddEndPuncttrue
\mciteSetBstMidEndSepPunct{\mcitedefaultmidpunct}
{\mcitedefaultendpunct}{\mcitedefaultseppunct}\relax
\EndOfBibitem
\bibitem{cdf_public_note_10784}
{CDF} collaboration, CDF note 10784, 2012, {\small
  \url{{http://www-cdf.fnal.gov/physics/new/bottom/120216.blessed-CPVcharm10fb/}}}\relax
\mciteBstWouldAddEndPuncttrue
\mciteSetBstMidEndSepPunct{\mcitedefaultmidpunct}
{\mcitedefaultendpunct}{\mcitedefaultseppunct}\relax
\EndOfBibitem
\bibitem{Collaboration:2012qw}
T.~Aaltonen {\em et al.} ({CDF} collaboration){,}
  \href{http://dx.doi.org/10.1103/PhysRevLett.109.111801}{Phys.\ Rev.\ Lett.\ {\bf
  109},  111801}  (2012), \href{http://arxiv.org/abs/1207.2158}{{\tt
  arXiv:1207.2158 [hep-ex]}}\relax
\mciteBstWouldAddEndPuncttrue
\mciteSetBstMidEndSepPunct{\mcitedefaultmidpunct}
{\mcitedefaultendpunct}{\mcitedefaultseppunct}\relax
\EndOfBibitem
\bibitem{Aaij:2016cfh}
R.~Aaij {\em et al.} ({LHCb} collaboration){,}
  \href{http://dx.doi.org/10.1103/PhysRevLett.116.191601}{Phys.\ Rev.\ Lett.\ {\bf
  116},  191601}  (2016), \href{http://arxiv.org/abs/1602.03160}{{\tt
  arXiv:1602.03160 [hep-ex]}}\relax
\mciteBstWouldAddEndPuncttrue
\mciteSetBstMidEndSepPunct{\mcitedefaultmidpunct}
{\mcitedefaultendpunct}{\mcitedefaultseppunct}\relax
\EndOfBibitem
\bibitem{Aitala:1999dt}
E.~M.\ Aitala {\em et al.} ({Fermilab E791} collaboration){,}
  \href{http://dx.doi.org/10.1103/PhysRevLett.83.32}{Phys.\ Rev.\ Lett.\ {\bf 83}{,}
   32}  (1999), \href{http://arxiv.org/abs/hep-ex/9903012}{{\tt
  arXiv:hep-ex/9903012}}\relax
\mciteBstWouldAddEndPuncttrue
\mciteSetBstMidEndSepPunct{\mcitedefaultmidpunct}
{\mcitedefaultendpunct}{\mcitedefaultseppunct}\relax
\EndOfBibitem
\bibitem{Link:2000cu}
J.~M.\ Link {\em et al.} ({FOCUS} collaboration){,}
  \href{http://dx.doi.org/10.1016/S0370-2693(00)00694-8}{Phys.\ Lett.\ {\bf
  B485},  62}  (2000), \href{http://arxiv.org/abs/hep-ex/0004034}{{\tt
  arXiv:hep-ex/0004034}}\relax
\mciteBstWouldAddEndPuncttrue
\mciteSetBstMidEndSepPunct{\mcitedefaultmidpunct}
{\mcitedefaultendpunct}{\mcitedefaultseppunct}\relax
\EndOfBibitem
\bibitem{Csorna:2001ww}
S.~E.\ Csorna {\em et al.} ({CLEO} collaboration){,}
  \href{http://dx.doi.org/10.1103/PhysRevD.65.092001}{Phys.\ Rev.\ {\bf D65}{,}
  092001}  (2002), \href{http://arxiv.org/abs/hep-ex/0111024}{{\tt
  arXiv:hep-ex/0111024}}\relax
\mciteBstWouldAddEndPuncttrue
\mciteSetBstMidEndSepPunct{\mcitedefaultmidpunct}
{\mcitedefaultendpunct}{\mcitedefaultseppunct}\relax
\EndOfBibitem
\bibitem{Zupanc:2009sy}
A.~Zupanc {\em et al.} ({Belle} collaboration){,}
  \href{http://dx.doi.org/10.1103/PhysRevD.80.052006}{Phys.\ Rev.\ {\bf D80}{,}
  052006}  (2009), \href{http://arxiv.org/abs/0905.4185}{{\tt arXiv:0905.4185
  [hep-ex]}}\relax
\mciteBstWouldAddEndPuncttrue
\mciteSetBstMidEndSepPunct{\mcitedefaultmidpunct}
{\mcitedefaultendpunct}{\mcitedefaultseppunct}\relax
\EndOfBibitem
\bibitem{Aaij:2011ad}
R.~Aaij {\em et al.} ({LHCb} collaboration){,}
  \href{http://dx.doi.org/10.1007/JHEP04(2012)129}{JHEP {\bf 04},  129}
  (2012), \href{http://arxiv.org/abs/1112.4698}{{\tt arXiv:1112.4698
  [hep-ex]}}\relax
\mciteBstWouldAddEndPuncttrue
\mciteSetBstMidEndSepPunct{\mcitedefaultmidpunct}
{\mcitedefaultendpunct}{\mcitedefaultseppunct}\relax
\EndOfBibitem
\bibitem{Staric:2015sta}
M.~Staric {\em et al.} ({Belle} collaboration){,}
  \href{http://dx.doi.org/10.1016/j.physletb.2015.12.025}{Phys.\ Lett.\ {\bf
  B753},  412--418}  (2016), \href{http://arxiv.org/abs/1509.08266}{{\tt
  arXiv:1509.08266 [hep-ex]}}\relax
\mciteBstWouldAddEndPuncttrue
\mciteSetBstMidEndSepPunct{\mcitedefaultmidpunct}
{\mcitedefaultendpunct}{\mcitedefaultseppunct}\relax
\EndOfBibitem
\bibitem{Lees:2012qh}
J.~P.\ Lees {\em et al.} ({\babar} collaboration){,}
  \href{http://dx.doi.org/10.1103/PhysRevD.87.012004}{Phys.\ Rev.\ {\bf D87}{,}
  012004}  (2013), \href{http://arxiv.org/abs/1209.3896}{{\tt arXiv:1209.3896
  [hep-ex]}}\relax
\mciteBstWouldAddEndPuncttrue
\mciteSetBstMidEndSepPunct{\mcitedefaultmidpunct}
{\mcitedefaultendpunct}{\mcitedefaultseppunct}\relax
\EndOfBibitem
\bibitem{Ablikim:2015hih}
M.~Ablikim {\em et al.} ({BESIII} collaboration){,}
  \href{http://dx.doi.org/10.1016/j.physletb.2015.04.008}{Phys.\ Lett.\ {\bf
  B744},  339}  (2015), \href{http://arxiv.org/abs/1501.01378}{{\tt
  arXiv:1501.01378 [hep-ex]}}\relax
\mciteBstWouldAddEndPuncttrue
\mciteSetBstMidEndSepPunct{\mcitedefaultmidpunct}
{\mcitedefaultendpunct}{\mcitedefaultseppunct}\relax
\EndOfBibitem
\bibitem{Aaltonen:2014efa}
T.~A.\ Aaltonen {\em et al.} ({CDF} collaboration){,}
  \href{http://dx.doi.org/10.1103/PhysRevD.90.111103}{Phys.\ Rev.\ {\bf D90}{,}
  111103}  (2014), \href{http://arxiv.org/abs/1410.5435}{{\tt arXiv:1410.5435
  [hep-ex]}}\relax
\mciteBstWouldAddEndPuncttrue
\mciteSetBstMidEndSepPunct{\mcitedefaultmidpunct}
{\mcitedefaultendpunct}{\mcitedefaultseppunct}\relax
\EndOfBibitem
\bibitem{Aaij:2015yda}
R.~Aaij {\em et al.} ({LHCb} collaboration){,}
  \href{http://dx.doi.org/10.1007/JHEP04(2015)043}{JHEP {\bf 04},  043}
  (2015), \href{http://arxiv.org/abs/1501.06777}{{\tt arXiv:1501.06777
  [hep-ex]}}\relax
\mciteBstWouldAddEndPuncttrue
\mciteSetBstMidEndSepPunct{\mcitedefaultmidpunct}
{\mcitedefaultendpunct}{\mcitedefaultseppunct}\relax
\EndOfBibitem
\bibitem{Aaij:2017idz}
R.~Aaij {\em et al.} ({LHCb} collaboration){,}
  \href{http://dx.doi.org/10.1103/PhysRevLett.118.261803}{Phys.\ Rev.\ Lett.\ {\bf
  118},  261803}  (2017), \href{http://arxiv.org/abs/1702.06490}{{\tt
  arXiv:1702.06490 [hep-ex]}}\relax
\mciteBstWouldAddEndPuncttrue
\mciteSetBstMidEndSepPunct{\mcitedefaultmidpunct}
{\mcitedefaultendpunct}{\mcitedefaultseppunct}\relax
\EndOfBibitem
\bibitem{MINUIT:webpage}
F.~James, {MINUIT -- Function Minimization and Error Analysis, Reference Manual
  Version 94.1, CERN Program Library Long Writeup D506}{,}
  \url{http://wwwasdoc.web.cern.ch/wwwasdoc/minuit/minmain.html}\relax
\mciteBstWouldAddEndPuncttrue
\mciteSetBstMidEndSepPunct{\mcitedefaultmidpunct}
{\mcitedefaultendpunct}{\mcitedefaultseppunct}\relax
\EndOfBibitem
\bibitem{Grossman:2009mn}
Y.~Grossman, Y.~Nir, and G.~Perez{,}
  \href{http://dx.doi.org/10.1103/PhysRevLett.103.071602}{Phys.\ Rev.\ Lett.\ {\bf
  103},  071602}  (2009), \href{http://arxiv.org/abs/0904.0305}{{\tt
  arXiv:0904.0305 [hep-ph]}}\relax
\mciteBstWouldAddEndPuncttrue
\mciteSetBstMidEndSepPunct{\mcitedefaultmidpunct}
{\mcitedefaultendpunct}{\mcitedefaultseppunct}\relax
\EndOfBibitem
\bibitem{Kagan:2009gb}
A.~L.\ Kagan and M.~D.\ Sokoloff{,}
  \href{http://dx.doi.org/10.1103/PhysRevD.80.076008}{Phys.\ Rev.\ {\bf D80}{,}
  076008}  (2009), \href{http://arxiv.org/abs/0907.3917}{{\tt arXiv:0907.3917
  [hep-ph]}}\relax
\mciteBstWouldAddEndPuncttrue
\mciteSetBstMidEndSepPunct{\mcitedefaultmidpunct}
{\mcitedefaultendpunct}{\mcitedefaultseppunct}\relax
\EndOfBibitem
\bibitem{Ciuchini:2007cw}
M.~Ciuchini {\em et al.}{,}
  \href{http://dx.doi.org/10.1016/j.physletb.2007.08.055}{Phys.\ Lett.\ {\bf
  B655},  162}  (2007), \href{http://arxiv.org/abs/hep-ph/0703204}{{\tt
  arXiv:hep-ph/0703204 [hep-ph]}}\relax
\mciteBstWouldAddEndPuncttrue
\mciteSetBstMidEndSepPunct{\mcitedefaultmidpunct}
{\mcitedefaultendpunct}{\mcitedefaultseppunct}\relax
\EndOfBibitem
\bibitem{CERNLIB:webpage}
{Computing and Networks Division}, {CERNLIB Short Writeups}{,}
  \url{https://root.cern.ch/sites/d35c7d8c.web.cern.ch/files/cernlib.pdf}\relax
\mciteBstWouldAddEndPuncttrue
\mciteSetBstMidEndSepPunct{\mcitedefaultmidpunct}
{\mcitedefaultendpunct}{\mcitedefaultseppunct}\relax
\EndOfBibitem
\bibitem{Bigi:2000yz}
I.~I.~Y.\ Bigi and A.~I.\ Sanda, Camb.\ Monogr.\ Part.\ Phys.\ Nucl.\ Phys.\ Cosmol.\
  {\bf 9},  1  (2000)\relax
\mciteBstWouldAddEndPuncttrue
\mciteSetBstMidEndSepPunct{\mcitedefaultmidpunct}
{\mcitedefaultendpunct}{\mcitedefaultseppunct}\relax
\EndOfBibitem
\bibitem{Nir:1999mg}
Y.~Nir, \href{http://arxiv.org/abs/hep-ph/9911321}{{\tt arXiv:hep-ph/9911321}}
   (1999)\relax
\mciteBstWouldAddEndPuncttrue
\mciteSetBstMidEndSepPunct{\mcitedefaultmidpunct}
{\mcitedefaultendpunct}{\mcitedefaultseppunct}\relax
\EndOfBibitem
\bibitem{Buccella:1994nf}
F.~Buccella, M.~Lusignoli, G.~Miele, A.~Pugliese, and P.~Santorelli{,}
  \href{http://dx.doi.org/10.1103/PhysRevD.51.3478}{Phys.\ Rev.\ {\bf D51}{,}
  3478}  (1995), \href{http://arxiv.org/abs/hep-ph/9411286}{{\tt
  arXiv:hep-ph/9411286}}\relax
\mciteBstWouldAddEndPuncttrue
\mciteSetBstMidEndSepPunct{\mcitedefaultmidpunct}
{\mcitedefaultendpunct}{\mcitedefaultseppunct}\relax
\EndOfBibitem
\bibitem{Grossman:2012aa}
Y.~Grossman and Y.~Nir, \href{http://dx.doi.org/10.1007/JHEP04(2012)002}{JHEP
  {\bf 04},  002}  (2012), \href{http://arxiv.org/abs/1110.3790}{{\tt
  arXiv:1110.3790 [hep-ph]}}\relax
\mciteBstWouldAddEndPuncttrue
\mciteSetBstMidEndSepPunct{\mcitedefaultmidpunct}
{\mcitedefaultendpunct}{\mcitedefaultseppunct}\relax
\EndOfBibitem
\bibitem{Bediaga:2009tr}
I.~Bediaga, I.~I.\ Bigi, A.~Gomes, G.~Guerrer, J.~Miranda, and A.~C.~d.\ Reis{,}
  \href{http://dx.doi.org/10.1103/PhysRevD.80.096006}{Phys.\ Rev.\ {\bf D80}{,}
  096006}  (2009), \href{http://arxiv.org/abs/0905.4233}{{\tt arXiv:0905.4233
  [hep-ph]}}\relax
\mciteBstWouldAddEndPuncttrue
\mciteSetBstMidEndSepPunct{\mcitedefaultmidpunct}
{\mcitedefaultendpunct}{\mcitedefaultseppunct}\relax
\EndOfBibitem
\bibitem{doi:10.1080/00949650410001661440}
B.~Aslan and G.~Zech{,}
  \href{http://dx.doi.org/10.1080/00949650410001661440}{Journal of Statistical
  Computation and Simulation {\bf 75},  109}  (2005)\relax
\mciteBstWouldAddEndPuncttrue
\mciteSetBstMidEndSepPunct{\mcitedefaultmidpunct}
{\mcitedefaultendpunct}{\mcitedefaultseppunct}\relax
\EndOfBibitem
\bibitem{Link:2005ft}
J.~M.\ Link {\em et al.} ({FOCUS} collaboration){,}
  \href{http://dx.doi.org/10.1016/j.physletb.2006.01.017}{Phys.\ Lett.\ {\bf
  B634},  165}  (2006), \href{http://arxiv.org/abs/hep-ex/0509042}{{\tt
  arXiv:hep-ex/0509042 [hep-ex]}}\relax
\mciteBstWouldAddEndPuncttrue
\mciteSetBstMidEndSepPunct{\mcitedefaultmidpunct}
{\mcitedefaultendpunct}{\mcitedefaultseppunct}\relax
\EndOfBibitem
\bibitem{Hinson:2004pj}
J.~W.\ Hinson {\em et al.} ({CLEO} collaboration){,}
  \href{http://dx.doi.org/10.1103/PhysRevLett.94.191801}{Phys.\ Rev.\ Lett.\ {\bf
  94},  191801}  (2005), \href{http://arxiv.org/abs/hep-ex/0501002}{{\tt
  arXiv:hep-ex/0501002 [hep-ex]}}\relax
\mciteBstWouldAddEndPuncttrue
\mciteSetBstMidEndSepPunct{\mcitedefaultmidpunct}
{\mcitedefaultendpunct}{\mcitedefaultseppunct}\relax
\EndOfBibitem
\bibitem{Eisenstein:2008aa}
B.~I.\ Eisenstein {\em et al.} ({CLEO} collaboration){,}
  \href{http://dx.doi.org/10.1103/PhysRevD.78.052003}{Phys.\ Rev.\ {\bf D78}{,}
  052003}  (2008), \href{http://arxiv.org/abs/0806.2112}{{\tt arXiv:0806.2112
  [hep-ex]}}\relax
\mciteBstWouldAddEndPuncttrue
\mciteSetBstMidEndSepPunct{\mcitedefaultmidpunct}
{\mcitedefaultendpunct}{\mcitedefaultseppunct}\relax
\EndOfBibitem
\bibitem{Mendez:2009aa}
H.~Mendez {\em et al.} ({CLEO} collaboration){,}
  \href{http://dx.doi.org/10.1103/PhysRevD.81.052013}{Phys.\ Rev.\ {\bf D81}{,}
  052013}  (2010), \href{http://arxiv.org/abs/0906.3198}{{\tt arXiv:0906.3198
  [hep-ex]}}\relax
\mciteBstWouldAddEndPuncttrue
\mciteSetBstMidEndSepPunct{\mcitedefaultmidpunct}
{\mcitedefaultendpunct}{\mcitedefaultseppunct}\relax
\EndOfBibitem
\bibitem{Won:2011ng}
E.~Won {\em et al.} ({Belle} collaboration){,}
  \href{http://dx.doi.org/10.1103/PhysRevLett.107221801}{Phys.\ Rev.\ Lett.\ {\bf
  107},  221801}  (2011), \href{http://arxiv.org/abs/1107.0553}{{\tt
  arXiv:1107.0553 [hep-ex]}}\relax
\mciteBstWouldAddEndPuncttrue
\mciteSetBstMidEndSepPunct{\mcitedefaultmidpunct}
{\mcitedefaultendpunct}{\mcitedefaultseppunct}\relax
\EndOfBibitem
\bibitem{Bonvicini:2013vxi}
G.~Bonvicini {\em et al.} ({CLEO} collaboration){,}
  \href{http://dx.doi.org/10.1103/PhysRevD.89.072002}{Phys.\ Rev.\ {\bf D89}{,}
  072002}  (2014), \href{http://arxiv.org/abs/1312.6775}{{\tt arXiv:1312.6775
  [hep-ex]}}\relax
\mciteBstWouldAddEndPuncttrue
\mciteSetBstMidEndSepPunct{\mcitedefaultmidpunct}
{\mcitedefaultendpunct}{\mcitedefaultseppunct}\relax
\EndOfBibitem
\bibitem{Ko:2012pe}
B.~R.\ Ko {\em et al.} ({Belle} collaboration){,}
  \href{http://dx.doi.org/10.1103/PhysRevLett.109.021601}{Phys.\ Rev.\ Lett.\ {\bf
  109},  021601}  (2012), \href{http://arxiv.org/abs/1203.6409}{{\tt
  arXiv:1203.6409 [hep-ex]}}\relax
\mciteBstWouldAddEndPuncttrue
\mciteSetBstMidEndSepPunct{\mcitedefaultmidpunct}
{\mcitedefaultendpunct}{\mcitedefaultseppunct}\relax
\EndOfBibitem
\bibitem{Amo:2011ab}
P.~del Amo~Sanchez {\em et al.} ({\babar} collaboration){,}
  \href{http://dx.doi.org/10.1103/PhysRevD.83.071103}{Phys.\ Rev.\ {\bf D83}{,}
  071103}  (2011), \href{http://arxiv.org/abs/1011.5477}{{\tt arXiv:1011.5477
  [hep-ex]}}\relax
\mciteBstWouldAddEndPuncttrue
\mciteSetBstMidEndSepPunct{\mcitedefaultmidpunct}
{\mcitedefaultendpunct}{\mcitedefaultseppunct}\relax
\EndOfBibitem
\bibitem{Link:2001zj}
J.~M.\ Link {\em et al.} ({FOCUS} collaboration){,}
  \href{http://dx.doi.org/10.1103/PhysRevLett.88.041602}{Phys.\ Rev.\ Lett.\ {\bf
  88},  041602; 159903(E)}  (2002){,}
  \href{http://arxiv.org/abs/hep-ex/0109022}{{\tt arXiv:hep-ex/0109022}}\relax
\mciteBstWouldAddEndPuncttrue
\mciteSetBstMidEndSepPunct{\mcitedefaultmidpunct}
{\mcitedefaultendpunct}{\mcitedefaultseppunct}\relax
\EndOfBibitem
\bibitem{Lees:2013aa}
J.~P.\ Lees {\em et al.} ({\babar} collaboration){,}
  \href{http://dx.doi.org/10.1103/PhysRevD.87.052012}{Phys.\ Rev.\ {\bf D87}{,}
  052012}  (2013), \href{http://arxiv.org/abs/1212.3003}{{\tt arXiv:1212.3003
  [hep-ex]}}\relax
\mciteBstWouldAddEndPuncttrue
\mciteSetBstMidEndSepPunct{\mcitedefaultmidpunct}
{\mcitedefaultendpunct}{\mcitedefaultseppunct}\relax
\EndOfBibitem
\bibitem{Ko:2013aa}
B.~R.\ Ko {\em et al.} ({Belle} collaboration){,}
  \href{http://dx.doi.org/10.1007/JHEP02(2013)098}{JHEP {\bf 02},  098}
  (2013), \href{http://arxiv.org/abs/1212.6112}{{\tt arXiv:1212.6112
  [hep-ex]}}\relax
\mciteBstWouldAddEndPuncttrue
\mciteSetBstMidEndSepPunct{\mcitedefaultmidpunct}
{\mcitedefaultendpunct}{\mcitedefaultseppunct}\relax
\EndOfBibitem
\bibitem{Aaij:2014ac}
R.~Aaij {\em et al.} ({LHCb} collaboration){,}
  \href{http://dx.doi.org/10.1007/JHEP10(2014)025}{JHEP {\bf 10},  25}
  (2014), \href{http://arxiv.org/abs/1406.2624}{{\tt arXiv:1406.2624
  [hep-ex]}}\relax
\mciteBstWouldAddEndPuncttrue
\mciteSetBstMidEndSepPunct{\mcitedefaultmidpunct}
{\mcitedefaultendpunct}{\mcitedefaultseppunct}\relax
\EndOfBibitem
\bibitem{Aaij:2014aa}
R.~Aaij {\em et al.} ({LHCb} collaboration){,}
  \href{http://dx.doi.org/10.1016/j.physletb.2013.12.035}{Phys.\ Lett.\ {\bf
  B728},  585}  (2014), \href{http://arxiv.org/abs/1310.7953}{{\tt
  arXiv:1310.7953 [hep-ex]}}\relax
\mciteBstWouldAddEndPuncttrue
\mciteSetBstMidEndSepPunct{\mcitedefaultmidpunct}
{\mcitedefaultendpunct}{\mcitedefaultseppunct}\relax
\EndOfBibitem
\bibitem{Aitala:1996sh}
E.~M.\ Aitala {\em et al.} ({Fermilab E791} collaboration){,}
  \href{http://dx.doi.org/10.1016/S0370-2693(97)00565-0}{Phys.\ Lett.\ {\bf
  B403},  377}  (1997), \href{http://arxiv.org/abs/hep-ex/9612005}{{\tt
  arXiv:hep-ex/9612005}}\relax
\mciteBstWouldAddEndPuncttrue
\mciteSetBstMidEndSepPunct{\mcitedefaultmidpunct}
{\mcitedefaultendpunct}{\mcitedefaultseppunct}\relax
\EndOfBibitem
\bibitem{Abazov:2014wga}
V.~M.\ Abazov {\em et al.} ({\dzero} collaboration){,}
  \href{http://dx.doi.org/10.1103/PhysRevD.90.111102}{Phys.\ Rev.\ {\bf D90}{,}
  111102}  (2014), \href{http://arxiv.org/abs/1408.6848}{{\tt arXiv:1408.6848
  [hep-ex]}}\relax
\mciteBstWouldAddEndPuncttrue
\mciteSetBstMidEndSepPunct{\mcitedefaultmidpunct}
{\mcitedefaultendpunct}{\mcitedefaultseppunct}\relax
\EndOfBibitem
\bibitem{Lees:2013ab}
J.~P.\ Lees {\em et al.} ({\babar} collaboration){,}
  \href{http://dx.doi.org/10.1103/PhysRevD.87.052010}{Phys.\ Rev.\ {\bf D87}{,}
  052010}  (2013), \href{http://arxiv.org/abs/1212.1856}{{\tt arXiv:1212.1856
  [hep-ex]}}\relax
\mciteBstWouldAddEndPuncttrue
\mciteSetBstMidEndSepPunct{\mcitedefaultmidpunct}
{\mcitedefaultendpunct}{\mcitedefaultseppunct}\relax
\EndOfBibitem
\bibitem{Rubin:2008zi}
P.~Rubin {\em et al.} ({CLEO} collaboration){,}
  \href{http://dx.doi.org/10.1103/PhysRevD.78.072003}{Phys.\ Rev.\ {\bf D78}{,}
  072003}  (2008), \href{http://arxiv.org/abs/0807.4545}{{\tt arXiv:0807.4545
  [hep-ex]}}\relax
\mciteBstWouldAddEndPuncttrue
\mciteSetBstMidEndSepPunct{\mcitedefaultmidpunct}
{\mcitedefaultendpunct}{\mcitedefaultseppunct}\relax
\EndOfBibitem
\bibitem{Link:2000aw}
J.~M.\ Link {\em et al.} ({FOCUS} collaboration){,}
  \href{http://dx.doi.org/10.1016/S0370-2693(00)01039-X}{Phys.\ Lett.\ {\bf
  B491},  232}  (2000), \href{http://arxiv.org/abs/hep-ex/0005037}{{\tt
  arXiv:hep-ex/0005037}}\relax
\mciteBstWouldAddEndPuncttrue
\mciteSetBstMidEndSepPunct{\mcitedefaultmidpunct}
{\mcitedefaultendpunct}{\mcitedefaultseppunct}\relax
\EndOfBibitem
\bibitem{Link:2005th}
J.~M.\ Link {\em et al.} ({FOCUS} collaboration){,}
  \href{http://dx.doi.org/10.1016/j.physletb.2005.07.024}{Phys.\ Lett.\ {\bf
  B622},  239}  (2005), \href{http://arxiv.org/abs/hep-ex/0506012}{{\tt
  arXiv:hep-ex/0506012}}\relax
\mciteBstWouldAddEndPuncttrue
\mciteSetBstMidEndSepPunct{\mcitedefaultmidpunct}
{\mcitedefaultendpunct}{\mcitedefaultseppunct}\relax
\EndOfBibitem
\bibitem{Aaltonen:2012ab}
T.~Aaltonen {\em et al.} ({CDF} collaboration){,}
  \href{http://dx.doi.org/10.1103/PhysRevD.85.012009}{Phys.\ Rev.\ {\bf D85}{,}
  012009}  (2012), \href{http://arxiv.org/abs/1111.5023}{{\tt arXiv:1111.5023
  [hep-ex]}}\relax
\mciteBstWouldAddEndPuncttrue
\mciteSetBstMidEndSepPunct{\mcitedefaultmidpunct}
{\mcitedefaultendpunct}{\mcitedefaultseppunct}\relax
\EndOfBibitem
\bibitem{Staric:2008rx}
M.~Staric {\em et al.} ({Belle} collaboration){,}
  \href{http://dx.doi.org/10.1016/j.physletb.2008.10.052}{Phys.\ Lett.\ {\bf
  B670},  190}  (2008), \href{http://arxiv.org/abs/0807.0148}{{\tt
  arXiv:0807.0148 [hep-ex]}}\relax
\mciteBstWouldAddEndPuncttrue
\mciteSetBstMidEndSepPunct{\mcitedefaultmidpunct}
{\mcitedefaultendpunct}{\mcitedefaultseppunct}\relax
\EndOfBibitem
\bibitem{Aitala:1997ff}
E.~M.\ Aitala {\em et al.} ({Fermilab E791} collaboration){,}
  \href{http://dx.doi.org/10.1016/S0370-2693(97)01570-0}{Phys.\ Lett.\ {\bf
  B421},  405}  (1998), \href{http://arxiv.org/abs/hep-ex/9711003}{{\tt
  arXiv:hep-ex/9711003}}\relax
\mciteBstWouldAddEndPuncttrue
\mciteSetBstMidEndSepPunct{\mcitedefaultmidpunct}
{\mcitedefaultendpunct}{\mcitedefaultseppunct}\relax
\EndOfBibitem
\bibitem{Nisar:2014fkc}
N.~K.\ Nisar {\em et al.} ({Belle} collaboration){,}
  \href{http://dx.doi.org/10.1103/PhysRevLett.112.211601}{Phys.\ Rev.\ Lett.\ {\bf
  112},  211601}  (2014), \href{http://arxiv.org/abs/1404.1266}{{\tt
  arXiv:1404.1266 [hep-ex]}}\relax
\mciteBstWouldAddEndPuncttrue
\mciteSetBstMidEndSepPunct{\mcitedefaultmidpunct}
{\mcitedefaultendpunct}{\mcitedefaultseppunct}\relax
\EndOfBibitem
\bibitem{Bonvicini:2000qm}
G.~Bonvicini {\em et al.} ({CLEO} collaboration){,}
  \href{http://dx.doi.org/10.1103/PhysRevD.63.071101}{Phys.\ Rev.\ {\bf D63}{,}
  071101}  (2001), \href{http://arxiv.org/abs/hep-ex/0012054}{{\tt
  arXiv:hep-ex/0012054}}\relax
\mciteBstWouldAddEndPuncttrue
\mciteSetBstMidEndSepPunct{\mcitedefaultmidpunct}
{\mcitedefaultendpunct}{\mcitedefaultseppunct}\relax
\EndOfBibitem
\bibitem{Ko:2011ab}
B.~R.\ Ko {\em et al.} ({Belle} collaboration){,}
  \href{http://dx.doi.org/10.1103/PhysRevLett.106.211801}{Phys.\ Rev.\ Lett.\ {\bf
  106},  211801}  (2011), \href{http://arxiv.org/abs/1101.3365}{{\tt
  arXiv:1101.3365 [hep-ex]}}\relax
\mciteBstWouldAddEndPuncttrue
\mciteSetBstMidEndSepPunct{\mcitedefaultmidpunct}
{\mcitedefaultendpunct}{\mcitedefaultseppunct}\relax
\EndOfBibitem
\bibitem{Aaij:2015fua}
R.~Aaij {\em et al.} ({LHCb} collaboration){,}
  \href{http://dx.doi.org/10.1007/JHEP10(2015)055}{JHEP {\bf 10},  055}
  (2015), \href{http://arxiv.org/abs/1508.06087}{{\tt arXiv:1508.06087
  [hep-ex]}}\relax
\mciteBstWouldAddEndPuncttrue
\mciteSetBstMidEndSepPunct{\mcitedefaultmidpunct}
{\mcitedefaultendpunct}{\mcitedefaultseppunct}\relax
\EndOfBibitem
\bibitem{Aaij:2014afa}
R.~Aaij {\em et al.} ({LHCb} collaboration){,}
  \href{http://dx.doi.org/10.1016/j.physletb.2014.11.043}{Phys.\ Lett.\ {\bf
  B740},  158}  (2015), \href{http://arxiv.org/abs/1410.4170}{{\tt
  arXiv:1410.4170 [hep-ex]}}\relax
\mciteBstWouldAddEndPuncttrue
\mciteSetBstMidEndSepPunct{\mcitedefaultmidpunct}
{\mcitedefaultendpunct}{\mcitedefaultseppunct}\relax
\EndOfBibitem
\bibitem{Aubert:2008yd}
B.~Aubert {\em et al.} ({\babar} collaboration){,}
  \href{http://dx.doi.org/10.1103/PhysRevD.78.051102}{Phys.\ Rev.\ {\bf D78}{,}
  051102}  (2008), \href{http://arxiv.org/abs/0802.4035}{{\tt arXiv:0802.4035
  [hep-ex]}}\relax
\mciteBstWouldAddEndPuncttrue
\mciteSetBstMidEndSepPunct{\mcitedefaultmidpunct}
{\mcitedefaultendpunct}{\mcitedefaultseppunct}\relax
\EndOfBibitem
\bibitem{Arinstein:2008zh}
K.~Arinstein ({Belle} collaboration){,}
  \href{http://dx.doi.org/10.1016/j.physletb.2008.02.054}{Phys.\ Lett.\ {\bf
  B662},  102}  (2008), \href{http://arxiv.org/abs/0801.2439}{{\tt
  arXiv:0801.2439 [hep-ex]}}\relax
\mciteBstWouldAddEndPuncttrue
\mciteSetBstMidEndSepPunct{\mcitedefaultmidpunct}
{\mcitedefaultendpunct}{\mcitedefaultseppunct}\relax
\EndOfBibitem
\bibitem{CroninHennessy:2005sy}
D.~Cronin-Hennessy {\em et al.} ({CLEO} collaboration){,}
  \href{http://dx.doi.org/10.1103/PhysRevD.72.031102}{Phys.\ Rev.\ {\bf D72}{,}
  031102}  (2005), \href{http://arxiv.org/abs/hep-ex/0503052}{{\tt
  arXiv:hep-ex/0503052}}\relax
\mciteBstWouldAddEndPuncttrue
\mciteSetBstMidEndSepPunct{\mcitedefaultmidpunct}
{\mcitedefaultendpunct}{\mcitedefaultseppunct}\relax
\EndOfBibitem
\bibitem{Tian:2005ik}
X.~C.\ Tian {\em et al.} ({Belle} collaboration){,}
  \href{http://dx.doi.org/10.1103/PhysRevLett.95.231801}{Phys.\ Rev.\ Lett.\ {\bf
  95},  231801}  (2005), \href{http://arxiv.org/abs/hep-ex/0507071}{{\tt
  arXiv:hep-ex/0507071}}\relax
\mciteBstWouldAddEndPuncttrue
\mciteSetBstMidEndSepPunct{\mcitedefaultmidpunct}
{\mcitedefaultendpunct}{\mcitedefaultseppunct}\relax
\EndOfBibitem
\bibitem{Brandenburg:2001ze}
G.~Brandenburg {\em et al.} ({CLEO} collaboration){,}
  \href{http://dx.doi.org/10.1103/PhysRevLett.87.071802}{Phys.\ Rev.\ Lett.\ {\bf
  87},  071802}  (2001), \href{http://arxiv.org/abs/hep-ex/0105002}{{\tt
  arXiv:hep-ex/0105002}}\relax
\mciteBstWouldAddEndPuncttrue
\mciteSetBstMidEndSepPunct{\mcitedefaultmidpunct}
{\mcitedefaultendpunct}{\mcitedefaultseppunct}\relax
\EndOfBibitem
\bibitem{Aaltonen:2012ac}
T.~Aaltonen {\em et al.} ({CDF} collaboration){,}
  \href{http://dx.doi.org/10.1103/PhysRevD.86.032007}{Phys.\ Rev.\ {\bf D86}{,}
  032007}  (2012), \href{http://arxiv.org/abs/1207.0825}{{\tt arXiv:1207.0825
  [hep-ex]}}\relax
\mciteBstWouldAddEndPuncttrue
\mciteSetBstMidEndSepPunct{\mcitedefaultmidpunct}
{\mcitedefaultendpunct}{\mcitedefaultseppunct}\relax
\EndOfBibitem
\bibitem{Asner:2003uz}
D.~M.\ Asner {\em et al.} ({CLEO} collaboration){,}
  \href{http://dx.doi.org/10.1103/PhysRevD.70.091101}{Phys.\ Rev.\ {\bf D70}{,}
  091101}  (2004), \href{http://arxiv.org/abs/hep-ex/0311033}{{\tt
  arXiv:hep-ex/0311033}}\relax
\mciteBstWouldAddEndPuncttrue
\mciteSetBstMidEndSepPunct{\mcitedefaultmidpunct}
{\mcitedefaultendpunct}{\mcitedefaultseppunct}\relax
\EndOfBibitem
\bibitem{Aaij:2013aa}
R.~Aaij {\em et al.} ({LHCb} collaboration){,}
  \href{http://dx.doi.org/10.1016/j.physletb.2013.09.011}{Phys.\ Lett.\ {\bf
  B726},  623}  (2013), \href{http://arxiv.org/abs/1308.3189}{{\tt
  arXiv:1308.3189 [hep-ex]}}\relax
\mciteBstWouldAddEndPuncttrue
\mciteSetBstMidEndSepPunct{\mcitedefaultmidpunct}
{\mcitedefaultendpunct}{\mcitedefaultseppunct}\relax
\EndOfBibitem
\bibitem{Artuso:2012df}
M.~Artuso {\em et al.} ({CLEO} collaboration){,}
  \href{http://dx.doi.org/10.1103/PhysRevD.85.122002}{Phys.\ Rev.\ {\bf D85}{,}
  122002}  (2012), \href{http://arxiv.org/abs/1201.5716}{{\tt arXiv:1201.5716
  [hep-ex]}}\relax
\mciteBstWouldAddEndPuncttrue
\mciteSetBstMidEndSepPunct{\mcitedefaultmidpunct}
{\mcitedefaultendpunct}{\mcitedefaultseppunct}\relax
\EndOfBibitem
\bibitem{Alexander:2009ux}
J.~Alexander {\em et al.} ({CLEO} collaboration){,}
  \href{http://dx.doi.org/10.1103/PhysRevD.79.052001}{Phys.\ Rev.\ {\bf D79}{,}
  052001}  (2009), \href{http://arxiv.org/abs/0901.1216}{{\tt arXiv:0901.1216
  [hep-ex]}}\relax
\mciteBstWouldAddEndPuncttrue
\mciteSetBstMidEndSepPunct{\mcitedefaultmidpunct}
{\mcitedefaultendpunct}{\mcitedefaultseppunct}\relax
\EndOfBibitem
\bibitem{Onyisi:2013bjt}
P.~Onyisi {\em et al.} ({CLEO} collaboration){,}
  \href{http://dx.doi.org/10.1103/PhysRevD.88.032009}{Phys.\ Rev.\ {\bf D88}{,}
  032009}  (2013), \href{http://arxiv.org/abs/1306.5363}{{\tt arXiv:1306.5363
  [hep-ex]}}\relax
\mciteBstWouldAddEndPuncttrue
\mciteSetBstMidEndSepPunct{\mcitedefaultmidpunct}
{\mcitedefaultendpunct}{\mcitedefaultseppunct}\relax
\EndOfBibitem
\bibitem{Ko:2010ng}
B.~R.\ Ko {\em et al.} ({Belle} collaboration){,}
  \href{http://dx.doi.org/10.1103/PhysRevLett.104.181602}{Phys.\ Rev.\ Lett.\ {\bf
  104},  181602}  (2010), \href{http://arxiv.org/abs/1001.3202}{{\tt
  arXiv:1001.3202 [hep-ex]}}\relax
\mciteBstWouldAddEndPuncttrue
\mciteSetBstMidEndSepPunct{\mcitedefaultmidpunct}
{\mcitedefaultendpunct}{\mcitedefaultseppunct}\relax
\EndOfBibitem
\bibitem{Golowich:1988ig}
E.~Golowich and G.~Valencia{,}
  \href{http://dx.doi.org/10.1103/PhysRevD.40.112}{Phys.\ Rev.\ {\bf D40},  112}
   (1989)\relax
\mciteBstWouldAddEndPuncttrue
\mciteSetBstMidEndSepPunct{\mcitedefaultmidpunct}
{\mcitedefaultendpunct}{\mcitedefaultseppunct}\relax
\EndOfBibitem
\bibitem{Bigi:2001sg}
I.~I.~Y.\ Bigi, \href{http://arxiv.org/abs/hep-ph/0107102}{{\tt
  arXiv:hep-ph/0107102}}  (2001)\relax
\mciteBstWouldAddEndPuncttrue
\mciteSetBstMidEndSepPunct{\mcitedefaultmidpunct}
{\mcitedefaultendpunct}{\mcitedefaultseppunct}\relax
\EndOfBibitem
\bibitem{Bensalem:2002ys}
W.~Bensalem, A.~Datta, and D.~London{,}
  \href{http://dx.doi.org/10.1103/PhysRevD.66.094004}{Phys.\ Rev.\ {\bf D66}{,}
  094004}  (2002), \href{http://arxiv.org/abs/hep-ph/0208054}{{\tt
  arXiv:hep-ph/0208054}}\relax
\mciteBstWouldAddEndPuncttrue
\mciteSetBstMidEndSepPunct{\mcitedefaultmidpunct}
{\mcitedefaultendpunct}{\mcitedefaultseppunct}\relax
\EndOfBibitem
\bibitem{Bensalem:2000hq}
W.~Bensalem and D.~London{,}
  \href{http://dx.doi.org/10.1103/PhysRevD.64.116003}{Phys.\ Rev.\ {\bf D64}{,}
  116003}  (2001), \href{http://arxiv.org/abs/hep-ph/0005018}{{\tt
  arXiv:hep-ph/0005018}}\relax
\mciteBstWouldAddEndPuncttrue
\mciteSetBstMidEndSepPunct{\mcitedefaultmidpunct}
{\mcitedefaultendpunct}{\mcitedefaultseppunct}\relax
\EndOfBibitem
\bibitem{Bensalem:2002pz}
W.~Bensalem, A.~Datta, and D.~London{,}
  \href{http://dx.doi.org/10.1016/S0370-2693(02)02028-2}{Phys.\ Lett.\ {\bf
  B538},  309}  (2002), \href{http://arxiv.org/abs/hep-ph/0205009}{{\tt
  arXiv:hep-ph/0205009}}\relax
\mciteBstWouldAddEndPuncttrue
\mciteSetBstMidEndSepPunct{\mcitedefaultmidpunct}
{\mcitedefaultendpunct}{\mcitedefaultseppunct}\relax
\EndOfBibitem
\bibitem{Gronau:2011cf}
M.~Gronau and J.~L.\ Rosner{,}
  \href{http://dx.doi.org/10.1103/PhysRevD.84.096013}{Phys.\ Rev.\ {\bf D84}{,}
  096013}  (2011), \href{http://arxiv.org/abs/1107.1232}{{\tt arXiv:1107.1232
  [hep-ph]}}\relax
\mciteBstWouldAddEndPuncttrue
\mciteSetBstMidEndSepPunct{\mcitedefaultmidpunct}
{\mcitedefaultendpunct}{\mcitedefaultseppunct}\relax
\EndOfBibitem
\bibitem{Bevan:2015xra}
A.~J.\ Bevan, \href{http://arxiv.org/abs/1506.04246}{{\tt arXiv:1506.04246
  [hep-ex]}}  (2015)\relax
\mciteBstWouldAddEndPuncttrue
\mciteSetBstMidEndSepPunct{\mcitedefaultmidpunct}
{\mcitedefaultendpunct}{\mcitedefaultseppunct}\relax
\EndOfBibitem
\bibitem{Durieux:2015zwa}
G.~Durieux and Y.~Grossman{,}
  \href{http://dx.doi.org/10.1103/PhysRevD.92.076013}{Phys.\ Rev.\ {\bf D92}{,}
  076013}  (2015), \href{http://arxiv.org/abs/1508.03054}{{\tt
  arXiv:1508.03054 [hep-ph]}}\relax
\mciteBstWouldAddEndPuncttrue
\mciteSetBstMidEndSepPunct{\mcitedefaultmidpunct}
{\mcitedefaultendpunct}{\mcitedefaultseppunct}\relax
\EndOfBibitem
\bibitem{Aaij:2014qwa}
R.~Aaij {\em et al.} ({LHCb} collaboration){,}
  \href{http://dx.doi.org/10.1007/JHEP10(2014)005}{JHEP {\bf 10},  005}
  (2014), \href{http://arxiv.org/abs/1408.1299}{{\tt arXiv:1408.1299
  [hep-ex]}}\relax
\mciteBstWouldAddEndPuncttrue
\mciteSetBstMidEndSepPunct{\mcitedefaultmidpunct}
{\mcitedefaultendpunct}{\mcitedefaultseppunct}\relax
\EndOfBibitem
\bibitem{Sanchez:2010xj}
P.~del Amo~Sanchez {\em et al.} ({\babar} collaboration){,}
  \href{http://dx.doi.org/10.1103/PhysRevD.81.111103}{Phys.\ Rev.\ {\bf D81}{,}
  111103}  (2010), \href{http://arxiv.org/abs/1003.3397}{{\tt arXiv:1003.3397
  [hep-ex]}}\relax
\mciteBstWouldAddEndPuncttrue
\mciteSetBstMidEndSepPunct{\mcitedefaultmidpunct}
{\mcitedefaultendpunct}{\mcitedefaultseppunct}\relax
\EndOfBibitem
\bibitem{Lees:2011ab}
J.~P.\ Lees {\em et al.} ({\babar} collaboration){,}
  \href{http://dx.doi.org/10.1103/PhysRevD.84.031103}{Phys.\ Rev.\ {\bf D84}{,}
  031103}  (2011), \href{http://arxiv.org/abs/1105.4410}{{\tt arXiv:1105.4410
  [hep-ex]}}\relax
\mciteBstWouldAddEndPuncttrue
\mciteSetBstMidEndSepPunct{\mcitedefaultmidpunct}
{\mcitedefaultendpunct}{\mcitedefaultseppunct}\relax
\EndOfBibitem
\bibitem{Grossman:2006jg}
Y.~Grossman, A.~L.\ Kagan, and Y.~Nir{,}
  \href{http://dx.doi.org/10.1103/PhysRevD.75.036008}{Phys.\ Rev.\ {\bf D75}{,}
  036008}  (2007), \href{http://arxiv.org/abs/hep-ph/0609178}{{\tt
  arXiv:hep-ph/0609178 [hep-ph]}}\relax
\mciteBstWouldAddEndPuncttrue
\mciteSetBstMidEndSepPunct{\mcitedefaultmidpunct}
{\mcitedefaultendpunct}{\mcitedefaultseppunct}\relax
\EndOfBibitem
\bibitem{Gersabeck:2011xj}
M.~Gersabeck, M.~Alexander, S.~Borghi, V.~Gligorov, and C.~Parkes{,}
  \href{http://dx.doi.org/10.1088/0954-3899/39/4/045005}{J.\ Phys.\ {\bf G39}{,}
  045005}  (2012), \href{http://arxiv.org/abs/1111.6515}{{\tt arXiv:1111.6515
  [hep-ex]}}\relax
\mciteBstWouldAddEndPuncttrue
\mciteSetBstMidEndSepPunct{\mcitedefaultmidpunct}
{\mcitedefaultendpunct}{\mcitedefaultseppunct}\relax
\EndOfBibitem
\bibitem{Ko:2012px}
B.~R.\ Ko ({Belle} collaboration), PoS {\bf ICHEP2012},  353  (2013){,}
  \href{http://arxiv.org/abs/1212.1975}{{\tt arXiv:1212.1975}}\relax
\mciteBstWouldAddEndPuncttrue
\mciteSetBstMidEndSepPunct{\mcitedefaultmidpunct}
{\mcitedefaultendpunct}{\mcitedefaultseppunct}\relax
\EndOfBibitem
\bibitem{Becher:2005bg}
T.~Becher and R.~J.\ Hill{,}
  \href{http://dx.doi.org/10.1016/j.physletb.2005.11.063}{Phys.\ Lett.\ {\bf
  B633},  61}  (2006), \href{http://arxiv.org/abs/hep-ph/0509090}{{\tt
  arXiv:hep-ph/0509090 [hep-ph]}}\relax
\mciteBstWouldAddEndPuncttrue
\mciteSetBstMidEndSepPunct{\mcitedefaultmidpunct}
{\mcitedefaultendpunct}{\mcitedefaultseppunct}\relax
\EndOfBibitem
\bibitem{Gilman:1989uy}
F.~J.\ Gilman and R.~L.\ Singleton{,}
  \href{http://dx.doi.org/10.1103/PhysRevD.41.142}{Phys.\ Rev.\ {\bf D41},  142}
   (1990)\relax
\mciteBstWouldAddEndPuncttrue
\mciteSetBstMidEndSepPunct{\mcitedefaultmidpunct}
{\mcitedefaultendpunct}{\mcitedefaultseppunct}\relax
\EndOfBibitem
\bibitem{Hill:2006ub}
R.~J.\ Hill, eConf {\bf C060409},  027{,}
  \href{http://arxiv.org/abs/hep-ph/0606023}{{\tt arXiv:hep-ph/0606023
  [hep-ph]}}  (2006)\relax
\mciteBstWouldAddEndPuncttrue
\mciteSetBstMidEndSepPunct{\mcitedefaultmidpunct}
{\mcitedefaultendpunct}{\mcitedefaultseppunct}\relax
\EndOfBibitem
\bibitem{Becirevic:1999kt}
D.~Becirevic and A.~B.\ Kaidalov{,}
  \href{http://dx.doi.org/10.1016/S0370-2693(00)00290-2}{Phys.\ Lett.\ {\bf
  B478},  417}  (2000), \href{http://arxiv.org/abs/hep-ph/9904490}{{\tt
  arXiv:hep-ph/9904490 [hep-ph]}}\relax
\mciteBstWouldAddEndPuncttrue
\mciteSetBstMidEndSepPunct{\mcitedefaultmidpunct}
{\mcitedefaultendpunct}{\mcitedefaultseppunct}\relax
\EndOfBibitem
\bibitem{Boyd:1994tt}
C.~G.\ Boyd, B.~Grinstein, and R.~F.\ Lebed{,}
  \href{http://dx.doi.org/10.1103/PhysRevLett.74.4603}{Phys.\ Rev.\ Lett.\ {\bf
  74},  4603}  (1995), \href{http://arxiv.org/abs/hep-ph/9412324}{{\tt
  arXiv:hep-ph/9412324 [hep-ph]}}\relax
\mciteBstWouldAddEndPuncttrue
\mciteSetBstMidEndSepPunct{\mcitedefaultmidpunct}
{\mcitedefaultendpunct}{\mcitedefaultseppunct}\relax
\EndOfBibitem
\bibitem{Boyd:1997qw}
C.~G.\ Boyd and M.~J.\ Savage{,}
  \href{http://dx.doi.org/10.1103/PhysRevD.56.303}{Phys.\ Rev.\ {\bf D56},  303}
   (1997), \href{http://arxiv.org/abs/hep-ph/9702300}{{\tt arXiv:hep-ph/9702300
  [hep-ph]}}\relax
\mciteBstWouldAddEndPuncttrue
\mciteSetBstMidEndSepPunct{\mcitedefaultmidpunct}
{\mcitedefaultendpunct}{\mcitedefaultseppunct}\relax
\EndOfBibitem
\bibitem{Arnesen:2005ez}
M.~C.\ Arnesen, B.~Grinstein, I.~Z.\ Rothstein, and I.~W.\ Stewart{,}
  \href{http://dx.doi.org/10.1103/PhysRevLett.95.071802}{Phys.\ Rev.\ Lett.\ {\bf
  95},  071802}  (2005)\relax
\mciteBstWouldAddEndPuncttrue
\mciteSetBstMidEndSepPunct{\mcitedefaultmidpunct}
{\mcitedefaultendpunct}{\mcitedefaultseppunct}\relax
\EndOfBibitem
\bibitem{Bourrely:1980gp}
C.~Bourrely, B.~Machet, and E.~de~Rafael{,}
  \href{http://dx.doi.org/10.1016/0550-3213(81)90086-9}{Nucl.\ Phys.\ {\bf B189}{,}
   157}  (1981)\relax
\mciteBstWouldAddEndPuncttrue
\mciteSetBstMidEndSepPunct{\mcitedefaultmidpunct}
{\mcitedefaultendpunct}{\mcitedefaultseppunct}\relax
\EndOfBibitem
\bibitem{Becirevic:2014kaa}
D.~Becirevic, A.~L.\ Yaouanc, A.~Oyanguren, P.~Roudeau, and F.~Sanfilippo{,}
  \href{http://arxiv.org/abs/1407.1019}{{\tt arXiv:1407.1019 [hep-ph]}}
  (2014)\relax
\mciteBstWouldAddEndPuncttrue
\mciteSetBstMidEndSepPunct{\mcitedefaultmidpunct}
{\mcitedefaultendpunct}{\mcitedefaultseppunct}\relax
\EndOfBibitem
\bibitem{Lees:2013uxa}
J.~P.\ Lees {\em et al.} ({\babar} collaboration){,}
  \href{http://dx.doi.org/10.1103/PhysRevD.88.052003}{Phys.\ Rev.\ {\bf D88}{,}
  052003}  (2013), \href{http://arxiv.org/abs/1304.5009}{{\tt arXiv:1304.5009
  [hep-ex]}}, Erratum ibid.\
  \href{http://dx.doi.org/10.1103/PhysRevD.88.079902}{{\bf D88}, 079902}
  (2013)\relax
\mciteBstWouldAddEndPuncttrue
\mciteSetBstMidEndSepPunct{\mcitedefaultmidpunct}
{\mcitedefaultendpunct}{\mcitedefaultseppunct}\relax
\EndOfBibitem
\bibitem{delAmoSanchez:2010vq}
P.~del Amo~Sanchez {\em et al.} ({\babar} collaboration){,}
  \href{http://dx.doi.org/10.1103/PhysRevD.82.111101}{Phys.\ Rev.\ {\bf D82}{,}
  111101}  (2010), \href{http://arxiv.org/abs/1009.2076}{{\tt arXiv:1009.2076
  [hep-ex]}}\relax
\mciteBstWouldAddEndPuncttrue
\mciteSetBstMidEndSepPunct{\mcitedefaultmidpunct}
{\mcitedefaultendpunct}{\mcitedefaultseppunct}\relax
\EndOfBibitem
\bibitem{Aaij:2013sza}
R.~Aaij {\em et al.} ({LHCb} collaboration){,}
  \href{http://dx.doi.org/10.1007/JHEP09(2013)145}{JHEP {\bf 09},  145}
  (2013), \href{http://arxiv.org/abs/1307.4556}{{\tt arXiv:1307.4556}}\relax
\mciteBstWouldAddEndPuncttrue
\mciteSetBstMidEndSepPunct{\mcitedefaultmidpunct}
{\mcitedefaultendpunct}{\mcitedefaultseppunct}\relax
\EndOfBibitem
\bibitem{Burdman:1996kr}
G.~Burdman and J.~Kambor{,}
  \href{http://dx.doi.org/10.1103/PhysRevD.55.2817}{Phys.\ Rev.\ {\bf D55}{,}
  2817}  (1997), \href{http://arxiv.org/abs/hep-ph/9602353}{{\tt
  arXiv:hep-ph/9602353 [hep-ph]}}\relax
\mciteBstWouldAddEndPuncttrue
\mciteSetBstMidEndSepPunct{\mcitedefaultmidpunct}
{\mcitedefaultendpunct}{\mcitedefaultseppunct}\relax
\EndOfBibitem
\bibitem{Becirevic:2012te}
D.~Becirevic, A.~Le~Yaouanc, L.~Oliver, J.-C.\ Raynal, P.~Roudeau, and
  J.~Serrano, \href{http://dx.doi.org/10.1103/PhysRevD.87.054007}{Phys.\ Rev.\
  {\bf D87},  054007}  (2013), \href{http://arxiv.org/abs/1206.5869}{{\tt
  arXiv:1206.5869 [hep-ph]}}\relax
\mciteBstWouldAddEndPuncttrue
\mciteSetBstMidEndSepPunct{\mcitedefaultmidpunct}
{\mcitedefaultendpunct}{\mcitedefaultseppunct}\relax
\EndOfBibitem
\bibitem{Lees:2014ihu}
J.~P.\ Lees {\em et al.} ({\babar} collaboration){,}
  \href{http://dx.doi.org/10.1103/PhysRevD.91.052022}{Phys.\ Rev.\ {\bf D91}{,}
  052022}  (2015), \href{http://arxiv.org/abs/1412.5502}{{\tt arXiv:1412.5502
  [hep-ex]}}\relax
\mciteBstWouldAddEndPuncttrue
\mciteSetBstMidEndSepPunct{\mcitedefaultmidpunct}
{\mcitedefaultendpunct}{\mcitedefaultseppunct}\relax
\EndOfBibitem
\bibitem{Ablikim:2015ixa}
M.~Ablikim {\em et al.} ({BESIII} collaboration){,}
  \href{http://dx.doi.org/10.1103/PhysRevD.92.072012}{Phys.\ Rev.\ {\bf D92}{,}
  072012}  (2015), \href{http://arxiv.org/abs/1508.07560}{{\tt
  arXiv:1508.07560 [hep-ex]}}\relax
\mciteBstWouldAddEndPuncttrue
\mciteSetBstMidEndSepPunct{\mcitedefaultmidpunct}
{\mcitedefaultendpunct}{\mcitedefaultseppunct}\relax
\EndOfBibitem
\bibitem{Ablikim:2015qgt}
M.~Ablikim {\em et al.} ({BESIII} collaboration){,}
  \href{http://dx.doi.org/10.1103/PhysRevD.92.112008}{Phys.\ Rev.\ {\bf D92}{,}
  112008}  (2015), \href{http://arxiv.org/abs/1510.00308}{{\tt
  arXiv:1510.00308 [hep-ex]}}\relax
\mciteBstWouldAddEndPuncttrue
\mciteSetBstMidEndSepPunct{\mcitedefaultmidpunct}
{\mcitedefaultendpunct}{\mcitedefaultseppunct}\relax
\EndOfBibitem
\bibitem{Widhalm:2006wz}
L.~Widhalm {\em et al.} ({Belle} collaboration){,}
  \href{http://dx.doi.org/10.1103/PhysRevLett.97.061804}{Phys.\ Rev.\ Lett.\ {\bf
  97},  061804}  (2006), \href{http://arxiv.org/abs/hep-ex/0604049}{{\tt
  arXiv:hep-ex/0604049 [hep-ex]}}\relax
\mciteBstWouldAddEndPuncttrue
\mciteSetBstMidEndSepPunct{\mcitedefaultmidpunct}
{\mcitedefaultendpunct}{\mcitedefaultseppunct}\relax
\EndOfBibitem
\bibitem{Aubert:2007wg}
B.~Aubert {\em et al.} ({\babar} collaboration){,}
  \href{http://dx.doi.org/10.1103/PhysRevD.76.052005}{Phys.\ Rev.\ {\bf D76}{,}
  052005}  (2007), \href{http://arxiv.org/abs/0704.0020}{{\tt arXiv:0704.0020
  [hep-ex]}}\relax
\mciteBstWouldAddEndPuncttrue
\mciteSetBstMidEndSepPunct{\mcitedefaultmidpunct}
{\mcitedefaultendpunct}{\mcitedefaultseppunct}\relax
\EndOfBibitem
\bibitem{Besson:2009uv}
D.~Besson {\em et al.} ({CLEO} collaboration){,}
  \href{http://dx.doi.org/10.1103/PhysRevD.80.032005}{Phys.\ Rev.\ {\bf D80}{,}
  032005}  (2009), \href{http://arxiv.org/abs/0906.2983}{{\tt arXiv:0906.2983
  [hep-ex]}}\relax
\mciteBstWouldAddEndPuncttrue
\mciteSetBstMidEndSepPunct{\mcitedefaultmidpunct}
{\mcitedefaultendpunct}{\mcitedefaultseppunct}\relax
\EndOfBibitem
\bibitem{Dobbs:2007aa}
S.~Dobbs {\em et al.} ({CLEO} collaboration){,}
  \href{http://dx.doi.org/10.1103/PhysRevD.77.112005}{Phys.\ Rev.\ {\bf D77}{,}
  112005}  (2008), \href{http://arxiv.org/abs/0712.1020}{{\tt arXiv:0712.1020
  [hep-ex]}}\relax
\mciteBstWouldAddEndPuncttrue
\mciteSetBstMidEndSepPunct{\mcitedefaultmidpunct}
{\mcitedefaultendpunct}{\mcitedefaultseppunct}\relax
\EndOfBibitem
\bibitem{Huang:2004fra}
G.~Huang {\em et al.} ({CLEO} collaboration){,}
  \href{http://dx.doi.org/10.1103/PhysRevLett.94.011802}{Phys.\ Rev.\ Lett.\ {\bf
  94},  011802}  (2005), \href{http://arxiv.org/abs/hep-ex/0407035}{{\tt
  arXiv:hep-ex/0407035 [hep-ex]}}\relax
\mciteBstWouldAddEndPuncttrue
\mciteSetBstMidEndSepPunct{\mcitedefaultmidpunct}
{\mcitedefaultendpunct}{\mcitedefaultseppunct}\relax
\EndOfBibitem
\bibitem{Link:2004dh}
J.~Link {\em et al.} ({FOCUS} collaboration){,}
  \href{http://dx.doi.org/10.1016/j.physletb.2004.12.036}{Phys.\ Lett.\ {\bf
  B607},  233}  (2005), \href{http://arxiv.org/abs/hep-ex/0410037}{{\tt
  arXiv:hep-ex/0410037 [hep-ex]}}\relax
\mciteBstWouldAddEndPuncttrue
\mciteSetBstMidEndSepPunct{\mcitedefaultmidpunct}
{\mcitedefaultendpunct}{\mcitedefaultseppunct}\relax
\EndOfBibitem
\bibitem{Adler:1989rw}
J.~Adler {\em et al.} ({MARK-III} collaboration){,}
  \href{http://dx.doi.org/10.1103/PhysRevLett.62.1821}{Phys.\ Rev.\ Lett.\ {\bf
  62},  1821}  (1989)\relax
\mciteBstWouldAddEndPuncttrue
\mciteSetBstMidEndSepPunct{\mcitedefaultmidpunct}
{\mcitedefaultendpunct}{\mcitedefaultseppunct}\relax
\EndOfBibitem
\bibitem{Kodama:1991ij}
K.~Kodama {\em et al.} ({Fermilab E653} collaboration){,}
  \href{http://dx.doi.org/10.1103/PhysRevLett.66.1819}{Phys.\ Rev.\ Lett.\ {\bf
  66},  1819}  (1991)\relax
\mciteBstWouldAddEndPuncttrue
\mciteSetBstMidEndSepPunct{\mcitedefaultmidpunct}
{\mcitedefaultendpunct}{\mcitedefaultseppunct}\relax
\EndOfBibitem
\bibitem{Kodama:1994aj}
K.~Kodama {\em et al.} ({Fermilab E653} collaboration){,}
  \href{http://dx.doi.org/10.1016/0370-2693(94)90579-7}{Phys.\ Lett.\ {\bf B336}{,}
   605}  (1994)\relax
\mciteBstWouldAddEndPuncttrue
\mciteSetBstMidEndSepPunct{\mcitedefaultmidpunct}
{\mcitedefaultendpunct}{\mcitedefaultseppunct}\relax
\EndOfBibitem
\bibitem{Frabetti:1995xq}
P.~L.\ Frabetti {\em et al.} ({Fermilab E687} collaboration){,}
  \href{http://dx.doi.org/10.1016/0370-2693(95)01368-2}{Phys.\ Lett.\ {\bf B364}{,}
   127}  (1995)\relax
\mciteBstWouldAddEndPuncttrue
\mciteSetBstMidEndSepPunct{\mcitedefaultmidpunct}
{\mcitedefaultendpunct}{\mcitedefaultseppunct}\relax
\EndOfBibitem
\bibitem{Frabetti:1993vz}
P.~L.\ Frabetti {\em et al.} ({Fermilab E687} collaboration){,}
  \href{http://dx.doi.org/10.1016/0370-2693(93)90181-G}{Phys.\ Lett.\ {\bf B315}{,}
   203}  (1993)\relax
\mciteBstWouldAddEndPuncttrue
\mciteSetBstMidEndSepPunct{\mcitedefaultmidpunct}
{\mcitedefaultendpunct}{\mcitedefaultseppunct}\relax
\EndOfBibitem
\bibitem{Anjos:1988ue}
J.~C.\ Anjos {\em et al.} ({Tagged Photon Spectrometer} collaboration){,}
  \href{http://dx.doi.org/10.1103/PhysRevLett.62.1587}{Phys.\ Rev.\ Lett.\ {\bf
  62},  1587}  (1989)\relax
\mciteBstWouldAddEndPuncttrue
\mciteSetBstMidEndSepPunct{\mcitedefaultmidpunct}
{\mcitedefaultendpunct}{\mcitedefaultseppunct}\relax
\EndOfBibitem
\bibitem{Ablikim:2004ej}
M.~Ablikim {\em et al.} ({BES} collaboration){,}
  \href{http://dx.doi.org/10.1016/j.physletb.2004.07.004}{Phys.\ Lett.\ {\bf
  B597},  39}  (2004), \href{http://arxiv.org/abs/hep-ex/0406028}{{\tt
  arXiv:hep-ex/0406028 [hep-ex]}}\relax
\mciteBstWouldAddEndPuncttrue
\mciteSetBstMidEndSepPunct{\mcitedefaultmidpunct}
{\mcitedefaultendpunct}{\mcitedefaultseppunct}\relax
\EndOfBibitem
\bibitem{Ablikim:2006bv}
M.~Ablikim {\em et al.} ({BES} collaboration){,}
  \href{http://arxiv.org/abs/hep-ex/0610019}{{\tt arXiv:hep-ex/0610019
  [hep-ex]}}  (2006)\relax
\mciteBstWouldAddEndPuncttrue
\mciteSetBstMidEndSepPunct{\mcitedefaultmidpunct}
{\mcitedefaultendpunct}{\mcitedefaultseppunct}\relax
\EndOfBibitem
\bibitem{Bean:1993zv}
A.~Bean {\em et al.} ({CLEO} collaboration){,}
  \href{http://dx.doi.org/10.1016/0370-2693(93)91385-Z}{Phys.\ Lett.\ {\bf B317}{,}
   647}  (1993)\relax
\mciteBstWouldAddEndPuncttrue
\mciteSetBstMidEndSepPunct{\mcitedefaultmidpunct}
{\mcitedefaultendpunct}{\mcitedefaultseppunct}\relax
\EndOfBibitem
\bibitem{BESIII-new}
Y.~Zheng ({BESIII} collaboration).\ {presented at the 37th International
  Conference on High Energy Physics (ICHEP 2014)}, 2014\relax
\mciteBstWouldAddEndPuncttrue
\mciteSetBstMidEndSepPunct{\mcitedefaultmidpunct}
{\mcitedefaultendpunct}{\mcitedefaultseppunct}\relax
\EndOfBibitem
\bibitem{Korner:1989qb}
J.~G.\ Korner and G.~A.\ Schuler, \href{http://dx.doi.org/10.1007/BF02440838}{Z.\
  Phys.\ {\bf C46},  93}  (1990)\relax
\mciteBstWouldAddEndPuncttrue
\mciteSetBstMidEndSepPunct{\mcitedefaultmidpunct}
{\mcitedefaultendpunct}{\mcitedefaultseppunct}\relax
\EndOfBibitem
\bibitem{Link:2002ev}
J.~Link {\em et al.} ({FOCUS} collaboration){,}
  \href{http://dx.doi.org/10.1016/S0370-2693(02)01715-X}{Phys.\ Lett.\ {\bf
  B535},  43}  (2002), \href{http://arxiv.org/abs/hep-ex/0203031}{{\tt
  arXiv:hep-ex/0203031 [hep-ex]}}\relax
\mciteBstWouldAddEndPuncttrue
\mciteSetBstMidEndSepPunct{\mcitedefaultmidpunct}
{\mcitedefaultendpunct}{\mcitedefaultseppunct}\relax
\EndOfBibitem
\bibitem{Link:2002wg}
J.~Link {\em et al.} ({FOCUS} collaboration){,}
  \href{http://dx.doi.org/10.1016/S0370-2693(02)02386-9}{Phys.\ Lett.\ {\bf
  B544},  89}  (2002), \href{http://arxiv.org/abs/hep-ex/0207049}{{\tt
  arXiv:hep-ex/0207049 [hep-ex]}}\relax
\mciteBstWouldAddEndPuncttrue
\mciteSetBstMidEndSepPunct{\mcitedefaultmidpunct}
{\mcitedefaultendpunct}{\mcitedefaultseppunct}\relax
\EndOfBibitem
\bibitem{Aubert:2008rs}
B.~Aubert {\em et al.} ({\babar} collaboration){,}
  \href{http://dx.doi.org/10.1103/PhysRevD.78.051101}{Phys.\ Rev.\ {\bf D78}{,}
  051101}  (2008), \href{http://arxiv.org/abs/0807.1599}{{\tt arXiv:0807.1599
  [hep-ex]}}\relax
\mciteBstWouldAddEndPuncttrue
\mciteSetBstMidEndSepPunct{\mcitedefaultmidpunct}
{\mcitedefaultendpunct}{\mcitedefaultseppunct}\relax
\EndOfBibitem
\bibitem{Ecklund:2009fia}
K.~M.\ Ecklund {\em et al.} ({CLEO} collaboration){,}
  \href{http://dx.doi.org/10.1103/PhysRevD.80.052009}{Phys.\ Rev.\ {\bf D80}{,}
  052009}  (2009), \href{http://arxiv.org/abs/0907.3201}{{\tt arXiv:0907.3201
  [hep-ex]}}\relax
\mciteBstWouldAddEndPuncttrue
\mciteSetBstMidEndSepPunct{\mcitedefaultmidpunct}
{\mcitedefaultendpunct}{\mcitedefaultseppunct}\relax
\EndOfBibitem
\bibitem{Briere:2010zc}
R.~A.\ Briere {\em et al.} ({CLEO} collaboration){,}
  \href{http://dx.doi.org/10.1103/PhysRevD.81.112001}{Phys.\ Rev.\ {\bf D81}{,}
  112001}  (2010), \href{http://arxiv.org/abs/1004.1954}{{\tt arXiv:1004.1954
  [hep-ex]}}\relax
\mciteBstWouldAddEndPuncttrue
\mciteSetBstMidEndSepPunct{\mcitedefaultmidpunct}
{\mcitedefaultendpunct}{\mcitedefaultseppunct}\relax
\EndOfBibitem
\bibitem{delAmoSanchez:2010fd}
P.~del Amo~Sanchez {\em et al.} ({\babar} collaboration){,}
  \href{http://dx.doi.org/10.1103/PhysRevD.83.072001}{Phys.\ Rev.\ {\bf D83}{,}
  072001}  (2011), \href{http://arxiv.org/abs/1012.1810}{{\tt arXiv:1012.1810
  [hep-ex]}}\relax
\mciteBstWouldAddEndPuncttrue
\mciteSetBstMidEndSepPunct{\mcitedefaultmidpunct}
{\mcitedefaultendpunct}{\mcitedefaultseppunct}\relax
\EndOfBibitem
\bibitem{Estabrooks:1977xe}
P.~Estabrooks {\em et al.}{,}
  \href{http://dx.doi.org/10.1016/0550-3213(78)90238-9}{Nucl.\ Phys.\ {\bf B133}{,}
   490}  (1978)\relax
\mciteBstWouldAddEndPuncttrue
\mciteSetBstMidEndSepPunct{\mcitedefaultmidpunct}
{\mcitedefaultendpunct}{\mcitedefaultseppunct}\relax
\EndOfBibitem
\bibitem{Anjos:1990pn}
J.~Anjos {\em et al.} ({Fermilab E691} collaboration){,}
  \href{http://dx.doi.org/10.1103/PhysRevLett.65.2630}{Phys.\ Rev.\ Lett.\ {\bf
  65},  2630}  (1990)\relax
\mciteBstWouldAddEndPuncttrue
\mciteSetBstMidEndSepPunct{\mcitedefaultmidpunct}
{\mcitedefaultendpunct}{\mcitedefaultseppunct}\relax
\EndOfBibitem
\bibitem{Kodama:1992tn}
K.~Kodama {\em et al.} ({Fermilab E653} collaboration){,}
  \href{http://dx.doi.org/10.1016/0370-2693(92)90530-H}{Phys.\ Lett.\ {\bf B274}{,}
   246}  (1992)\relax
\mciteBstWouldAddEndPuncttrue
\mciteSetBstMidEndSepPunct{\mcitedefaultmidpunct}
{\mcitedefaultendpunct}{\mcitedefaultseppunct}\relax
\EndOfBibitem
\bibitem{Frabetti:1993jq}
P.~Frabetti {\em et al.} ({Fermilab E687} collaboration){,}
  \href{http://dx.doi.org/10.1016/0370-2693(93)90216-5}{Phys.\ Lett.\ {\bf B307}{,}
   262}  (1993)\relax
\mciteBstWouldAddEndPuncttrue
\mciteSetBstMidEndSepPunct{\mcitedefaultmidpunct}
{\mcitedefaultendpunct}{\mcitedefaultseppunct}\relax
\EndOfBibitem
\bibitem{Aitala:1997cm}
E.~Aitala {\em et al.} ({Fermilab E791} collaboration){,}
  \href{http://dx.doi.org/10.1103/PhysRevLett.80.1393}{Phys.\ Rev.\ Lett.\ {\bf
  80},  1393}  (1998), \href{http://arxiv.org/abs/hep-ph/9710216}{{\tt
  arXiv:hep-ph/9710216 [hep-ph]}}\relax
\mciteBstWouldAddEndPuncttrue
\mciteSetBstMidEndSepPunct{\mcitedefaultmidpunct}
{\mcitedefaultendpunct}{\mcitedefaultseppunct}\relax
\EndOfBibitem
\bibitem{Aitala:1998ey}
E.~Aitala {\em et al.} ({Fermilab E791} collaboration){,}
  \href{http://dx.doi.org/10.1016/S0370-2693(98)01243-X}{Phys.\ Lett.\ {\bf
  B440},  435}  (1998), \href{http://arxiv.org/abs/hep-ex/9809026}{{\tt
  arXiv:hep-ex/9809026 [hep-ex]}}\relax
\mciteBstWouldAddEndPuncttrue
\mciteSetBstMidEndSepPunct{\mcitedefaultmidpunct}
{\mcitedefaultendpunct}{\mcitedefaultseppunct}\relax
\EndOfBibitem
\bibitem{Adamovich:1998ia}
M.~Adamovich {\em et al.} ({BEATRICE} collaboration){,}
  \href{http://dx.doi.org/10.1007/s100529801012}{Eur.\ Phys.\ J.\ {\bf C6},  35}
  (1999)\relax
\mciteBstWouldAddEndPuncttrue
\mciteSetBstMidEndSepPunct{\mcitedefaultmidpunct}
{\mcitedefaultendpunct}{\mcitedefaultseppunct}\relax
\EndOfBibitem
\bibitem{Link:2004uk}
J.~Link {\em et al.} ({FOCUS} collaboration){,}
  \href{http://dx.doi.org/10.1016/j.physletb.2004.12.037}{Phys.\ Lett.\ {\bf
  B607},  67}  (2005), \href{http://arxiv.org/abs/hep-ex/0410067}{{\tt
  arXiv:hep-ex/0410067 [hep-ex]}}\relax
\mciteBstWouldAddEndPuncttrue
\mciteSetBstMidEndSepPunct{\mcitedefaultmidpunct}
{\mcitedefaultendpunct}{\mcitedefaultseppunct}\relax
\EndOfBibitem
\bibitem{Mahlke:2007uf}
H.~Mahlke, eConf {\bf C0610161},  014  (2006){,}
  \href{http://arxiv.org/abs/hep-ex/0702014}{{\tt arXiv:hep-ex/0702014
  [hep-ex]}}\relax
\mciteBstWouldAddEndPuncttrue
\mciteSetBstMidEndSepPunct{\mcitedefaultmidpunct}
{\mcitedefaultendpunct}{\mcitedefaultseppunct}\relax
\EndOfBibitem
\bibitem{Filipuzzi:2012mg}
A.~Filipuzzi, J.~Portoles, and M.~Gonzalez-Alonso{,}
  \href{http://dx.doi.org/10.1103/PhysRevD.85.116010}{Phys.\ Rev.\ {\bf D85}{,}
  116010}  (2012), \href{http://arxiv.org/abs/1203.2092}{{\tt arXiv:1203.2092
  [hep-ph]}}\relax
\mciteBstWouldAddEndPuncttrue
\mciteSetBstMidEndSepPunct{\mcitedefaultmidpunct}
{\mcitedefaultendpunct}{\mcitedefaultseppunct}\relax
\EndOfBibitem
\bibitem{Ablikim:2016duz}
M.~Ablikim {\em et al.} ({BESIII} collaboration){,}
  \href{http://dx.doi.org/10.1103/PhysRevD.94.072004}{Phys.\ Rev.\ {\bf D94}{,}
  072004}  (2016), \href{http://arxiv.org/abs/1608.06732}{{\tt
  arXiv:1608.06732 [hep-ex]}}\relax
\mciteBstWouldAddEndPuncttrue
\mciteSetBstMidEndSepPunct{\mcitedefaultmidpunct}
{\mcitedefaultendpunct}{\mcitedefaultseppunct}\relax
\EndOfBibitem
\bibitem{Ablikim:2013uvu}
M.~Ablikim {\em et al.} ({BESIII} collaboration){,}
  \href{http://dx.doi.org/10.1103/PhysRevD.89.051104}{Phys.\ Rev.\ {\bf D89}{,}
  051104}  (2014), \href{http://arxiv.org/abs/1312.0374}{{\tt arXiv:1312.0374
  [hep-ex]}}\relax
\mciteBstWouldAddEndPuncttrue
\mciteSetBstMidEndSepPunct{\mcitedefaultmidpunct}
{\mcitedefaultendpunct}{\mcitedefaultseppunct}\relax
\EndOfBibitem
\bibitem{delAmoSanchez:2010jg}
P.~del Amo~Sanchez {\em et al.} ({\babar} collaboration){,}
  \href{http://dx.doi.org/10.1103/PhysRevD.82.091103}{Phys.\ Rev.\ {\bf D82}{,}
  091103}  (2010), \href{http://arxiv.org/abs/1008.4080}{{\tt arXiv:1008.4080
  [hep-ex]}}\relax
\mciteBstWouldAddEndPuncttrue
\mciteSetBstMidEndSepPunct{\mcitedefaultmidpunct}
{\mcitedefaultendpunct}{\mcitedefaultseppunct}\relax
\EndOfBibitem
\bibitem{Zupanc:2013byn}
A.~Zupanc {\em et al.} ({Belle} collaboration){,}
  \href{http://dx.doi.org/10.1007/JHEP09(2013)139}{JHEP {\bf 09},  139}
  (2013), \href{http://arxiv.org/abs/1307.6240}{{\tt arXiv:1307.6240
  [hep-ex]}}\relax
\mciteBstWouldAddEndPuncttrue
\mciteSetBstMidEndSepPunct{\mcitedefaultmidpunct}
{\mcitedefaultendpunct}{\mcitedefaultseppunct}\relax
\EndOfBibitem
\bibitem{Naik:2009tk}
P.~Naik {\em et al.} ({CLEO} collaboration){,}
  \href{http://dx.doi.org/10.1103/PhysRevD.80.112004}{Phys.\ Rev.\ {\bf D80}{,}
  112004}  (2009), \href{http://arxiv.org/abs/0910.3602}{{\tt arXiv:0910.3602
  [hep-ex]}}\relax
\mciteBstWouldAddEndPuncttrue
\mciteSetBstMidEndSepPunct{\mcitedefaultmidpunct}
{\mcitedefaultendpunct}{\mcitedefaultseppunct}\relax
\EndOfBibitem
\bibitem{Onyisi:2009th}
P.~Onyisi {\em et al.} ({CLEO} collaboration){,}
  \href{http://dx.doi.org/10.1103/PhysRevD.79.052002}{Phys.\ Rev.\ {\bf D79}{,}
  052002}  (2009), \href{http://arxiv.org/abs/0901.1147}{{\tt arXiv:0901.1147
  [hep-ex]}}\relax
\mciteBstWouldAddEndPuncttrue
\mciteSetBstMidEndSepPunct{\mcitedefaultmidpunct}
{\mcitedefaultendpunct}{\mcitedefaultseppunct}\relax
\EndOfBibitem
\bibitem{Barberio:1990ms}
E.~Barberio, B.~van Eijk, and Z.~Was{,}
  \href{http://dx.doi.org/10.1016/0010-4655(91)90012-A}{Comput.\ Phys.\ Commun.\
  {\bf 66},  115}  (1991)\relax
\mciteBstWouldAddEndPuncttrue
\mciteSetBstMidEndSepPunct{\mcitedefaultmidpunct}
{\mcitedefaultendpunct}{\mcitedefaultseppunct}\relax
\EndOfBibitem
\bibitem{Barberio:1993qi}
E.~Barberio and Z.~Was{,}
  \href{http://dx.doi.org/10.1016/0010-4655(94)90074-4}{Comput.\ Phys.\ Commun.\
  {\bf 79},  291}  (1994)\relax
\mciteBstWouldAddEndPuncttrue
\mciteSetBstMidEndSepPunct{\mcitedefaultmidpunct}
{\mcitedefaultendpunct}{\mcitedefaultseppunct}\relax
\EndOfBibitem
\bibitem{Golonka:2005pn}
P.~Golonka and Z.~Was, \href{http://dx.doi.org/10.1140/epjc/s2005-02396-4}{Eur.\
  Phys.\ J.\ {\bf C45},  97}  (2006){,}
  \href{http://arxiv.org/abs/hep-ph/0506026}{{\tt arXiv:hep-ph/0506026}}\relax
\mciteBstWouldAddEndPuncttrue
\mciteSetBstMidEndSepPunct{\mcitedefaultmidpunct}
{\mcitedefaultendpunct}{\mcitedefaultseppunct}\relax
\EndOfBibitem
\bibitem{Golonka:2006tw}
P.~Golonka and Z.~Was{,}
  \href{http://dx.doi.org/10.1140/epjc/s10052-006-0205-3}{Eur.\ Phys.\ J.\ {\bf
  C50},  53}  (2007), \href{http://arxiv.org/abs/hep-ph/0604232}{{\tt
  arXiv:hep-ph/0604232}}\relax
\mciteBstWouldAddEndPuncttrue
\mciteSetBstMidEndSepPunct{\mcitedefaultmidpunct}
{\mcitedefaultendpunct}{\mcitedefaultseppunct}\relax
\EndOfBibitem
\bibitem{Ryd:2005zz}
 EVTGEN-V00-11-07, 2005, \url{{http://inspirehep.net/record/707695}}\relax
\mciteBstWouldAddEndPuncttrue
\mciteSetBstMidEndSepPunct{\mcitedefaultmidpunct}
{\mcitedefaultendpunct}{\mcitedefaultseppunct}\relax
\EndOfBibitem
\bibitem{Link:2002hi}
J.~M.\ Link {\em et al.} ({FOCUS} collaboration){,}
  \href{http://dx.doi.org/10.1016/S0370-2693(03)00053-4}{Phys.\ Lett.\ {\bf
  B555},  167}  (2003), \href{http://arxiv.org/abs/hep-ex/0212058}{{\tt
  arXiv:hep-ex/0212058}}\relax
\mciteBstWouldAddEndPuncttrue
\mciteSetBstMidEndSepPunct{\mcitedefaultmidpunct}
{\mcitedefaultendpunct}{\mcitedefaultseppunct}\relax
\EndOfBibitem
\bibitem{Aubert:2007wn}
B.~Aubert {\em et al.} ({\babar} collaboration){,}
  \href{http://dx.doi.org/10.1103/PhysRevLett.100.051802}{Phys.\ Rev.\ Lett.\ {\bf
  100},  051802}  (2008), \href{http://arxiv.org/abs/0704.2080}{{\tt
  arXiv:0704.2080 [hep-ex]}}\relax
\mciteBstWouldAddEndPuncttrue
\mciteSetBstMidEndSepPunct{\mcitedefaultmidpunct}
{\mcitedefaultendpunct}{\mcitedefaultseppunct}\relax
\EndOfBibitem
\bibitem{Artuso:1997mc}
M.~Artuso {\em et al.} ({CLEO} collaboration){,}
  \href{http://dx.doi.org/10.1103/PhysRevLett.80.3193}{Phys.\ Rev.\ Lett.\ {\bf
  80},  3193}  (1998), \href{http://arxiv.org/abs/hep-ex/9712023}{{\tt
  arXiv:hep-ex/9712023}}\relax
\mciteBstWouldAddEndPuncttrue
\mciteSetBstMidEndSepPunct{\mcitedefaultmidpunct}
{\mcitedefaultendpunct}{\mcitedefaultseppunct}\relax
\EndOfBibitem
\bibitem{Barate:1997mm}
R.~Barate {\em et al.} ({ALEPH} collaboration){,}
  \href{http://dx.doi.org/10.1016/S0370-2693(97)00585-6}{Phys.\ Lett.\ {\bf
  B403},  367}  (1997)\relax
\mciteBstWouldAddEndPuncttrue
\mciteSetBstMidEndSepPunct{\mcitedefaultmidpunct}
{\mcitedefaultendpunct}{\mcitedefaultseppunct}\relax
\EndOfBibitem
\bibitem{Albrecht:1994nb}
H.~Albrecht {\em et al.} ({ARGUS} collaboration){,}
  \href{http://dx.doi.org/10.1016/0370-2693(94)91308-0}{Phys.\ Lett.\ {\bf B340}{,}
   125}  (1994)\relax
\mciteBstWouldAddEndPuncttrue
\mciteSetBstMidEndSepPunct{\mcitedefaultmidpunct}
{\mcitedefaultendpunct}{\mcitedefaultseppunct}\relax
\EndOfBibitem
\bibitem{Akerib:1993pm}
D.~S.\ Akerib {\em et al.} ({CLEO} collaboration){,}
  \href{http://dx.doi.org/10.1103/PhysRevLett.71.3070}{Phys.\ Rev.\ Lett.\ {\bf
  71},  3070}  (1993)\relax
\mciteBstWouldAddEndPuncttrue
\mciteSetBstMidEndSepPunct{\mcitedefaultmidpunct}
{\mcitedefaultendpunct}{\mcitedefaultseppunct}\relax
\EndOfBibitem
\bibitem{Decamp:1991jw}
D.~Decamp {\em et al.} ({ALEPH} collaboration){,}
  \href{http://dx.doi.org/10.1016/0370-2693(91)90769-M}{Phys.\ Lett.\ {\bf B266}{,}
   218}  (1991)\relax
\mciteBstWouldAddEndPuncttrue
\mciteSetBstMidEndSepPunct{\mcitedefaultmidpunct}
{\mcitedefaultendpunct}{\mcitedefaultseppunct}\relax
\EndOfBibitem
\bibitem{Acosta:2004ts}
D.~E.\ Acosta {\em et al.} ({CDF} collaboration){,}
  \href{http://dx.doi.org/10.1103/PhysRevLett.94.122001}{Phys.\ Rev.\ Lett.\ {\bf
  94},  122001}  (2005), \href{http://arxiv.org/abs/hep-ex/0504006}{{\tt
  arXiv:hep-ex/0504006}}\relax
\mciteBstWouldAddEndPuncttrue
\mciteSetBstMidEndSepPunct{\mcitedefaultmidpunct}
{\mcitedefaultendpunct}{\mcitedefaultseppunct}\relax
\EndOfBibitem
\bibitem{Coan:1997ye}
T.~E.\ Coan {\em et al.} ({CLEO} collaboration){,}
  \href{http://dx.doi.org/10.1103/PhysRevLett.80.1150}{Phys.\ Rev.\ Lett.\ {\bf
  80},  1150}  (1998), \href{http://arxiv.org/abs/hep-ex/9710028}{{\tt
  arXiv:hep-ex/9710028}}\relax
\mciteBstWouldAddEndPuncttrue
\mciteSetBstMidEndSepPunct{\mcitedefaultmidpunct}
{\mcitedefaultendpunct}{\mcitedefaultseppunct}\relax
\EndOfBibitem
\bibitem{Aubert:2003fg}
B.~Aubert {\em et al.} ({\babar} collaboration){,}
  \href{http://dx.doi.org/10.1103/PhysRevLett.90.242001}{Phys.\ Rev.\ Lett.\ {\bf
  90},  242001}  (2003), \href{http://arxiv.org/abs/hep-ex/0304021}{{\tt
  arXiv:hep-ex/0304021 [hep-ex]}}\relax
\mciteBstWouldAddEndPuncttrue
\mciteSetBstMidEndSepPunct{\mcitedefaultmidpunct}
{\mcitedefaultendpunct}{\mcitedefaultseppunct}\relax
\EndOfBibitem
\bibitem{Besson:2003cp}
D.~Besson {\em et al.} ({CLEO} collaboration){,}
  \href{http://dx.doi.org/10.1103/PhysRevD.68.032002}{Phys.\ Rev.\ {\bf D68}{,}
  032002}  (2003), \href{http://arxiv.org/abs/hep-ex/0305100}{{\tt
  arXiv:hep-ex/0305100 [hep-ex]}}, Erratum ibid.\
  \href{http://dx.doi.org/10.1103/PhysRevD.75.119908}{{\bf D75}, 119908}
  (2007)\relax
\mciteBstWouldAddEndPuncttrue
\mciteSetBstMidEndSepPunct{\mcitedefaultmidpunct}
{\mcitedefaultendpunct}{\mcitedefaultseppunct}\relax
\EndOfBibitem
\bibitem{Abe:2003jk}
K.~Abe {\em et al.} ({Belle} collaboration){,}
  \href{http://dx.doi.org/10.1103/PhysRevLett.92.012002}{Phys.\ Rev.\ Lett.\ {\bf
  92},  012002}  (2004), \href{http://arxiv.org/abs/hep-ex/0307052}{{\tt
  arXiv:hep-ex/0307052 [hep-ex]}}\relax
\mciteBstWouldAddEndPuncttrue
\mciteSetBstMidEndSepPunct{\mcitedefaultmidpunct}
{\mcitedefaultendpunct}{\mcitedefaultseppunct}\relax
\EndOfBibitem
\bibitem{Aubert:2003pe}
B.~Aubert {\em et al.} ({\babar} collaboration){,}
  \href{http://dx.doi.org/10.1103/PhysRevD.69.031101}{Phys.\ Rev.\ {\bf D69}{,}
  031101}  (2004), \href{http://arxiv.org/abs/hep-ex/0310050}{{\tt
  arXiv:hep-ex/0310050 [hep-ex]}}\relax
\mciteBstWouldAddEndPuncttrue
\mciteSetBstMidEndSepPunct{\mcitedefaultmidpunct}
{\mcitedefaultendpunct}{\mcitedefaultseppunct}\relax
\EndOfBibitem
\bibitem{Link:2003bd}
J.~M.\ Link {\em et al.} ({FOCUS} collaboration){,}
  \href{http://dx.doi.org/10.1016/j.physletb.2004.02.017}{Phys.\ Lett.\ {\bf
  B586},  11}  (2004), \href{http://arxiv.org/abs/hep-ex/0312060}{{\tt
  arXiv:hep-ex/0312060 [hep-ex]}}\relax
\mciteBstWouldAddEndPuncttrue
\mciteSetBstMidEndSepPunct{\mcitedefaultmidpunct}
{\mcitedefaultendpunct}{\mcitedefaultseppunct}\relax
\EndOfBibitem
\bibitem{Godfrey:1985xj}
S.~Godfrey and N.~Isgur, \href{http://dx.doi.org/10.1103/PhysRevD.32.189}{Phys.\
  Rev.\ {\bf D32},  189}  (1985)\relax
\mciteBstWouldAddEndPuncttrue
\mciteSetBstMidEndSepPunct{\mcitedefaultmidpunct}
{\mcitedefaultendpunct}{\mcitedefaultseppunct}\relax
\EndOfBibitem
\bibitem{Godfrey:1986wj}
S.~Godfrey and R.~Kokoski{,}
  \href{http://dx.doi.org/10.1103/PhysRevD.43.1679}{Phys.\ Rev.\ {\bf D43}{,}
  1679}  (1991)\relax
\mciteBstWouldAddEndPuncttrue
\mciteSetBstMidEndSepPunct{\mcitedefaultmidpunct}
{\mcitedefaultendpunct}{\mcitedefaultseppunct}\relax
\EndOfBibitem
\bibitem{Schweitzer:2002nm}
P.~Schweitzer, S.~Boffi, and M.~Radici{,}
  \href{http://dx.doi.org/10.1103/PhysRevD.66.114004}{Phys.\ Rev.\ {\bf D66}{,}
  114004}  (2002), \href{http://arxiv.org/abs/hep-ph/0207230}{{\tt
  arXiv:hep-ph/0207230 [hep-ph]}}\relax
\mciteBstWouldAddEndPuncttrue
\mciteSetBstMidEndSepPunct{\mcitedefaultmidpunct}
{\mcitedefaultendpunct}{\mcitedefaultseppunct}\relax
\EndOfBibitem
\bibitem{Jugeau:2005yr}
F.~Jugeau, A.~Le~Yaouanc, L.~Oliver, and J.-C.\ Raynal{,}
  \href{http://dx.doi.org/10.1103/PhysRevD.72.094010}{Phys.\ Rev.\ {\bf D72}{,}
  094010}  (2005), \href{http://arxiv.org/abs/hep-ph/0504206}{{\tt
  arXiv:hep-ph/0504206 [hep-ph]}}\relax
\mciteBstWouldAddEndPuncttrue
\mciteSetBstMidEndSepPunct{\mcitedefaultmidpunct}
{\mcitedefaultendpunct}{\mcitedefaultseppunct}\relax
\EndOfBibitem
\bibitem{Colangelo:2004vu}
P.~Colangelo, F.~De~Fazio, and R.~Ferrandes{,}
  \href{http://dx.doi.org/10.1142/S0217732304015269}{Mod.\ Phys.\ Lett.\ {\bf
  A19},  2083}  (2004), \href{http://arxiv.org/abs/hep-ph/0407137}{{\tt
  arXiv:hep-ph/0407137 [hep-ph]}}\relax
\mciteBstWouldAddEndPuncttrue
\mciteSetBstMidEndSepPunct{\mcitedefaultmidpunct}
{\mcitedefaultendpunct}{\mcitedefaultseppunct}\relax
\EndOfBibitem
\bibitem{Cahn:2003cw}
R.~N.\ Cahn and J.~D.\ Jackson{,}
  \href{http://dx.doi.org/10.1103/PhysRevD.68.037502}{Phys.\ Rev.\ {\bf D68}{,}
  037502}  (2003), \href{http://arxiv.org/abs/hep-ph/0305012}{{\tt
  arXiv:hep-ph/0305012 [hep-ph]}}\relax
\mciteBstWouldAddEndPuncttrue
\mciteSetBstMidEndSepPunct{\mcitedefaultmidpunct}
{\mcitedefaultendpunct}{\mcitedefaultseppunct}\relax
\EndOfBibitem
\bibitem{Barnes:2003dj}
T.~Barnes, F.~E.\ Close, and H.~J.\ Lipkin{,}
  \href{http://dx.doi.org/10.1103/PhysRevD.68.054006}{Phys.\ Rev.\ {\bf D68}{,}
  054006}  (2003), \href{http://arxiv.org/abs/hep-ph/0305025}{{\tt
  arXiv:hep-ph/0305025 [hep-ph]}}\relax
\mciteBstWouldAddEndPuncttrue
\mciteSetBstMidEndSepPunct{\mcitedefaultmidpunct}
{\mcitedefaultendpunct}{\mcitedefaultseppunct}\relax
\EndOfBibitem
\bibitem{Lipkin:2003zk}
H.~Lipkin, \href{http://dx.doi.org/10.1016/j.physletb.2003.10.117}{Phys.\ Lett.\
  {\bf B580},  50}  (2004), \href{http://arxiv.org/abs/hep-ph/0306204}{{\tt
  arXiv:hep-ph/0306204 [hep-ph]}}\relax
\mciteBstWouldAddEndPuncttrue
\mciteSetBstMidEndSepPunct{\mcitedefaultmidpunct}
{\mcitedefaultendpunct}{\mcitedefaultseppunct}\relax
\EndOfBibitem
\bibitem{Bardeen:2003kt}
W.~A.\ Bardeen, E.~J.\ Eichten, and C.~T.\ Hill{,}
  \href{http://dx.doi.org/10.1103/PhysRevD.68.054024}{Phys.\ Rev.\ {\bf D68}{,}
  054024}  (2003), \href{http://arxiv.org/abs/hep-ph/0305049}{{\tt
  arXiv:hep-ph/0305049 [hep-ph]}}\relax
\mciteBstWouldAddEndPuncttrue
\mciteSetBstMidEndSepPunct{\mcitedefaultmidpunct}
{\mcitedefaultendpunct}{\mcitedefaultseppunct}\relax
\EndOfBibitem
\bibitem{Aubert:2009ah}
B.~Aubert {\em et al.} ({\babar} collaboration){,}
  \href{http://dx.doi.org/10.1103/PhysRevD.80.092003}{Phys.\ Rev.\ {\bf D80}{,}
  092003}  (2009), \href{http://arxiv.org/abs/0908.0806}{{\tt arXiv:0908.0806
  [hep-ex]}}\relax
\mciteBstWouldAddEndPuncttrue
\mciteSetBstMidEndSepPunct{\mcitedefaultmidpunct}
{\mcitedefaultendpunct}{\mcitedefaultseppunct}\relax
\EndOfBibitem
\bibitem{Aaij:2012pc}
R.~Aaij {\em et al.} ({LHCb} collaboration){,}
  \href{http://dx.doi.org/10.1007/JHEP10(2012)151}{JHEP {\bf 10},  151}
  (2012), \href{http://arxiv.org/abs/1207.6016}{{\tt arXiv:1207.6016
  [hep-ex]}}\relax
\mciteBstWouldAddEndPuncttrue
\mciteSetBstMidEndSepPunct{\mcitedefaultmidpunct}
{\mcitedefaultendpunct}{\mcitedefaultseppunct}\relax
\EndOfBibitem
\bibitem{Aaij:2014xza}
R.~Aaij {\em et al.} ({LHCb} collaboration){,}
  \href{http://dx.doi.org/10.1103/PhysRevLett.113.162001}{Phys.\ Rev.\ Lett.\ {\bf
  113},  162001}  (2014), \href{http://arxiv.org/abs/1407.7574}{{\tt
  arXiv:1407.7574 [hep-ex]}}\relax
\mciteBstWouldAddEndPuncttrue
\mciteSetBstMidEndSepPunct{\mcitedefaultmidpunct}
{\mcitedefaultendpunct}{\mcitedefaultseppunct}\relax
\EndOfBibitem
\bibitem{Aaij:2016utb}
R.~Aaij {\em et al.} ({LHCb} collaboration){,}
  \href{http://dx.doi.org/10.1007/JHEP02(2016)133}{JHEP {\bf 02},  133}
  (2016), \href{http://arxiv.org/abs/1601.01495}{{\tt arXiv:1601.01495
  [hep-ex]}}\relax
\mciteBstWouldAddEndPuncttrue
\mciteSetBstMidEndSepPunct{\mcitedefaultmidpunct}
{\mcitedefaultendpunct}{\mcitedefaultseppunct}\relax
\EndOfBibitem
\bibitem{Matsuki:2006rz}
T.~Matsuki, T.~Morii, and K.~Sudoh{,}
  \href{http://dx.doi.org/10.1140/epja/i2006-10287-1}{Eur.\ Phys.\ J.\ {\bf A31}{,}
  701}  (2007), \href{http://arxiv.org/abs/hep-ph/0610186}{{\tt
  arXiv:hep-ph/0610186 [hep-ph]}}\relax
\mciteBstWouldAddEndPuncttrue
\mciteSetBstMidEndSepPunct{\mcitedefaultmidpunct}
{\mcitedefaultendpunct}{\mcitedefaultseppunct}\relax
\EndOfBibitem
\bibitem{Isgur:1989vq}
N.~Isgur and M.~B.\ Wise{,}
  \href{http://dx.doi.org/10.1016/0370-2693(89)90566-2}{Phys.\ Lett.\ {\bf B232}{,}
   113}  (1989)\relax
\mciteBstWouldAddEndPuncttrue
\mciteSetBstMidEndSepPunct{\mcitedefaultmidpunct}
{\mcitedefaultendpunct}{\mcitedefaultseppunct}\relax
\EndOfBibitem
\bibitem{Aubert:2006bk}
B.~Aubert {\em et al.} ({\babar} collaboration){,}
  \href{http://dx.doi.org/10.1103/PhysRevD.74.032007}{Phys.\ Rev.\ {\bf D74}{,}
  032007}  (2006), \href{http://arxiv.org/abs/hep-ex/0604030}{{\tt
  arXiv:hep-ex/0604030 [hep-ex]}}\relax
\mciteBstWouldAddEndPuncttrue
\mciteSetBstMidEndSepPunct{\mcitedefaultmidpunct}
{\mcitedefaultendpunct}{\mcitedefaultseppunct}\relax
\EndOfBibitem
\bibitem{Abazov:2007wg}
V.~M.\ Abazov {\em et al.} ({\dzero} collaboration){,}
  \href{http://dx.doi.org/10.1103/PhysRevLett.102.051801}{Phys.\ Rev.\ Lett.\ {\bf
  102},  051801}  (2009), \href{http://arxiv.org/abs/0712.3789}{{\tt
  arXiv:0712.3789 [hep-ex]}}\relax
\mciteBstWouldAddEndPuncttrue
\mciteSetBstMidEndSepPunct{\mcitedefaultmidpunct}
{\mcitedefaultendpunct}{\mcitedefaultseppunct}\relax
\EndOfBibitem
\bibitem{Frabetti:1993vv}
P.~L.\ Frabetti {\em et al.} ({Fermilab E687} collaboration){,}
  \href{http://dx.doi.org/10.1103/PhysRevLett.72.324}{Phys.\ Rev.\ Lett.\ {\bf
  72},  324}  (1994)\relax
\mciteBstWouldAddEndPuncttrue
\mciteSetBstMidEndSepPunct{\mcitedefaultmidpunct}
{\mcitedefaultendpunct}{\mcitedefaultseppunct}\relax
\EndOfBibitem
\bibitem{Alexander:1993nq}
J.~P.\ Alexander {\em et al.} ({CLEO} collaboration){,}
  \href{http://dx.doi.org/10.1016/0370-2693(93)91448-V}{Phys.\ Lett.\ {\bf B303}{,}
   377}  (1993)\relax
\mciteBstWouldAddEndPuncttrue
\mciteSetBstMidEndSepPunct{\mcitedefaultmidpunct}
{\mcitedefaultendpunct}{\mcitedefaultseppunct}\relax
\EndOfBibitem
\bibitem{Albrecht:1992zh}
H.~Albrecht {\em et al.} ({ARGUS} collaboration){,}
  \href{http://dx.doi.org/10.1016/0370-2693(92)91282-E}{Phys.\ Lett.\ {\bf B297}{,}
   425}  (1992)\relax
\mciteBstWouldAddEndPuncttrue
\mciteSetBstMidEndSepPunct{\mcitedefaultmidpunct}
{\mcitedefaultendpunct}{\mcitedefaultseppunct}\relax
\EndOfBibitem
\bibitem{Avery:1989ui}
P.~Avery {\em et al.} ({CLEO} collaboration){,}
  \href{http://dx.doi.org/10.1103/PhysRevD.41.774}{Phys.\ Rev.\ {\bf D41},  774}
   (1990)\relax
\mciteBstWouldAddEndPuncttrue
\mciteSetBstMidEndSepPunct{\mcitedefaultmidpunct}
{\mcitedefaultendpunct}{\mcitedefaultseppunct}\relax
\EndOfBibitem
\bibitem{Albrecht:1989yi}
H.~Albrecht {\em et al.} ({ARGUS} collaboration){,}
  \href{http://dx.doi.org/10.1016/0370-2693(89)91672-9}{Phys.\ Lett.\ {\bf B230}{,}
   162}  (1989)\relax
\mciteBstWouldAddEndPuncttrue
\mciteSetBstMidEndSepPunct{\mcitedefaultmidpunct}
{\mcitedefaultendpunct}{\mcitedefaultseppunct}\relax
\EndOfBibitem
\bibitem{Lees:2011um}
J.~P.\ Lees {\em et al.} ({\babar} collaboration){,}
  \href{http://dx.doi.org/10.1103/PhysRevD.83.072003}{Phys.\ Rev.\ {\bf D83}{,}
  072003}  (2011), \href{http://arxiv.org/abs/1103.2675}{{\tt arXiv:1103.2675
  [hep-ex]}}\relax
\mciteBstWouldAddEndPuncttrue
\mciteSetBstMidEndSepPunct{\mcitedefaultmidpunct}
{\mcitedefaultendpunct}{\mcitedefaultseppunct}\relax
\EndOfBibitem
\bibitem{Aaij:2014baa}
R.~Aaij {\em et al.} ({LHCb} collaboration){,}
  \href{http://dx.doi.org/10.1103/PhysRevD.90.072003}{Phys.\ Rev.\ {\bf D90}{,}
  072003}  (2014), \href{http://arxiv.org/abs/1407.7712}{{\tt arXiv:1407.7712
  [hep-ex]}}\relax
\mciteBstWouldAddEndPuncttrue
\mciteSetBstMidEndSepPunct{\mcitedefaultmidpunct}
{\mcitedefaultendpunct}{\mcitedefaultseppunct}\relax
\EndOfBibitem
\bibitem{Aaij:2011ju}
R.~Aaij {\em et al.} ({LHCb} collaboration){,}
  \href{http://dx.doi.org/10.1016/j.physletb.2011.02.039}{Phys.\ Lett.\ {\bf
  B698},  14}  (2011), \href{http://arxiv.org/abs/1102.0348}{{\tt
  arXiv:1102.0348 [hep-ex]}}\relax
\mciteBstWouldAddEndPuncttrue
\mciteSetBstMidEndSepPunct{\mcitedefaultmidpunct}
{\mcitedefaultendpunct}{\mcitedefaultseppunct}\relax
\EndOfBibitem
\bibitem{Aubert:2006mh}
B.~Aubert {\em et al.} ({\babar} collaboration){,}
  \href{http://dx.doi.org/10.1103/PhysRevLett.97.222001}{Phys.\ Rev.\ Lett.\ {\bf
  97},  222001}  (2006), \href{http://arxiv.org/abs/hep-ex/0607082}{{\tt
  arXiv:hep-ex/0607082 [hep-ex]}}\relax
\mciteBstWouldAddEndPuncttrue
\mciteSetBstMidEndSepPunct{\mcitedefaultmidpunct}
{\mcitedefaultendpunct}{\mcitedefaultseppunct}\relax
\EndOfBibitem
\bibitem{Albrecht:1995qx}
H.~Albrecht {\em et al.} ({ARGUS} collaboration){,}
  \href{http://dx.doi.org/10.1007/s002880050040}{Z.\ Phys.\ {\bf C69},  405}
  (1996)\relax
\mciteBstWouldAddEndPuncttrue
\mciteSetBstMidEndSepPunct{\mcitedefaultmidpunct}
{\mcitedefaultendpunct}{\mcitedefaultseppunct}\relax
\EndOfBibitem
\bibitem{Kubota:1994gn}
Y.~Kubota {\em et al.} ({CLEO} collaboration){,}
  \href{http://dx.doi.org/10.1103/PhysRevLett.72.1972}{Phys.\ Rev.\ Lett.\ {\bf
  72},  1972}  (1994), \href{http://arxiv.org/abs/hep-ph/9403325}{{\tt
  arXiv:hep-ph/9403325 [hep-ph]}}\relax
\mciteBstWouldAddEndPuncttrue
\mciteSetBstMidEndSepPunct{\mcitedefaultmidpunct}
{\mcitedefaultendpunct}{\mcitedefaultseppunct}\relax
\EndOfBibitem
\bibitem{Lees:2014abp}
J.~P.\ Lees {\em et al.} ({\babar} collaboration){,}
  \href{http://dx.doi.org/10.1103/PhysRevD.91.052002}{Phys.\ Rev.\ {\bf D91}{,}
  052002}  (2015), \href{http://arxiv.org/abs/1412.6751}{{\tt arXiv:1412.6751
  [hep-ex]}}\relax
\mciteBstWouldAddEndPuncttrue
\mciteSetBstMidEndSepPunct{\mcitedefaultmidpunct}
{\mcitedefaultendpunct}{\mcitedefaultseppunct}\relax
\EndOfBibitem
\bibitem{Aaij:2015kqa}
R.~Aaij {\em et al.} ({LHCb} collaboration){,}
  \href{http://dx.doi.org/10.1103/PhysRevD.92.012012}{Phys.\ Rev.\ {\bf D92}{,}
  012012}  (2015), \href{http://arxiv.org/abs/1505.01505}{{\tt
  arXiv:1505.01505 [hep-ex]}}\relax
\mciteBstWouldAddEndPuncttrue
\mciteSetBstMidEndSepPunct{\mcitedefaultmidpunct}
{\mcitedefaultendpunct}{\mcitedefaultseppunct}\relax
\EndOfBibitem
\bibitem{Kuzmin:2006mw}
A.~Kuzmin {\em et al.} ({Belle} collaboration){,}
  \href{http://dx.doi.org/10.1103/PhysRevD.76.012006}{Phys.\ Rev.\ {\bf D76}{,}
  012006}  (2007), \href{http://arxiv.org/abs/hep-ex/0611054}{{\tt
  arXiv:hep-ex/0611054 [hep-ex]}}\relax
\mciteBstWouldAddEndPuncttrue
\mciteSetBstMidEndSepPunct{\mcitedefaultmidpunct}
{\mcitedefaultendpunct}{\mcitedefaultseppunct}\relax
\EndOfBibitem
\bibitem{Abramowicz:2012ys}
H.~Abramowicz {\em et al.} ({ZEUS} collaboration){,}
  \href{http://dx.doi.org/10.1016/j.nuclphysb.2012.09.007}{Nucl.\ Phys.\ {\bf
  B866},  229}  (2013), \href{http://arxiv.org/abs/1208.4468}{{\tt
  arXiv:1208.4468 [hep-ex]}}\relax
\mciteBstWouldAddEndPuncttrue
\mciteSetBstMidEndSepPunct{\mcitedefaultmidpunct}
{\mcitedefaultendpunct}{\mcitedefaultseppunct}\relax
\EndOfBibitem
\bibitem{Abulencia:2005ry}
A.~Abulencia {\em et al.} ({CDF} collaboration){,}
  \href{http://dx.doi.org/10.1103/PhysRevD.73.051104}{Phys.\ Rev.\ {\bf D73}{,}
  051104}  (2006), \href{http://arxiv.org/abs/hep-ex/0512069}{{\tt
  arXiv:hep-ex/0512069 [hep-ex]}}\relax
\mciteBstWouldAddEndPuncttrue
\mciteSetBstMidEndSepPunct{\mcitedefaultmidpunct}
{\mcitedefaultendpunct}{\mcitedefaultseppunct}\relax
\EndOfBibitem
\bibitem{Avery:1994yc}
P.~Avery {\em et al.} ({CLEO} collaboration){,}
  \href{http://dx.doi.org/10.1016/0370-2693(94)90968-7}{Phys.\ Lett.\ {\bf B331}{,}
   236}  (1994), \href{http://arxiv.org/abs/hep-ph/9403359}{{\tt
  arXiv:hep-ph/9403359 [hep-ph]}}\relax
\mciteBstWouldAddEndPuncttrue
\mciteSetBstMidEndSepPunct{\mcitedefaultmidpunct}
{\mcitedefaultendpunct}{\mcitedefaultseppunct}\relax
\EndOfBibitem
\bibitem{Albrecht:1989pa}
H.~Albrecht {\em et al.} ({ARGUS} collaboration){,}
  \href{http://dx.doi.org/10.1016/0370-2693(89)90764-8}{Phys.\ Lett.\ {\bf B232}{,}
   398}  (1989)\relax
\mciteBstWouldAddEndPuncttrue
\mciteSetBstMidEndSepPunct{\mcitedefaultmidpunct}
{\mcitedefaultendpunct}{\mcitedefaultseppunct}\relax
\EndOfBibitem
\bibitem{Anjos:1988uf}
J.~C.\ Anjos {\em et al.} ({Tagged Photon Spectrometer} collaboration){,}
  \href{http://dx.doi.org/10.1103/PhysRevLett.62.1717}{Phys.\ Rev.\ Lett.\ {\bf
  62},  1717}  (1989)\relax
\mciteBstWouldAddEndPuncttrue
\mciteSetBstMidEndSepPunct{\mcitedefaultmidpunct}
{\mcitedefaultendpunct}{\mcitedefaultseppunct}\relax
\EndOfBibitem
\bibitem{Bergfeld:1994af}
T.~Bergfeld {\em et al.} ({CLEO} collaboration){,}
  \href{http://dx.doi.org/10.1016/0370-2693(94)01348-9}{Phys.\ Lett.\ {\bf B340}{,}
   194}  (1994)\relax
\mciteBstWouldAddEndPuncttrue
\mciteSetBstMidEndSepPunct{\mcitedefaultmidpunct}
{\mcitedefaultendpunct}{\mcitedefaultseppunct}\relax
\EndOfBibitem
\bibitem{Albrecht:1988dj}
H.~Albrecht {\em et al.} ({ARGUS} collaboration){,}
  \href{http://dx.doi.org/10.1016/0370-2693(89)91737-1}{Phys.\ Lett.\ {\bf B221}{,}
   422}  (1989)\relax
\mciteBstWouldAddEndPuncttrue
\mciteSetBstMidEndSepPunct{\mcitedefaultmidpunct}
{\mcitedefaultendpunct}{\mcitedefaultseppunct}\relax
\EndOfBibitem
\bibitem{Albrecht:1989gb}
H.~Albrecht {\em et al.} ({ARGUS} collaboration){,}
  \href{http://dx.doi.org/10.1016/0370-2693(89)90141-X}{Phys.\ Lett.\ {\bf B231}{,}
   208}  (1989)\relax
\mciteBstWouldAddEndPuncttrue
\mciteSetBstMidEndSepPunct{\mcitedefaultmidpunct}
{\mcitedefaultendpunct}{\mcitedefaultseppunct}\relax
\EndOfBibitem
\bibitem{Abreu:1998vk}
P.~Abreu {\em et al.} ({DELPHI} collaboration){,}
  \href{http://dx.doi.org/10.1016/S0370-2693(98)00346-3}{Phys.\ Lett.\ {\bf
  B426},  231}  (1998)\relax
\mciteBstWouldAddEndPuncttrue
\mciteSetBstMidEndSepPunct{\mcitedefaultmidpunct}
{\mcitedefaultendpunct}{\mcitedefaultseppunct}\relax
\EndOfBibitem
\bibitem{Chekanov:2008ac}
S.~Chekanov {\em et al.} ({ZEUS} collaboration){,}
  \href{http://dx.doi.org/10.1140/epjc/s10052-009-0881-x}{Eur.\ Phys.\ J.\ {\bf
  C60},  25}  (2009), \href{http://arxiv.org/abs/0807.1290}{{\tt
  arXiv:0807.1290 [hep-ex]}}\relax
\mciteBstWouldAddEndPuncttrue
\mciteSetBstMidEndSepPunct{\mcitedefaultmidpunct}
{\mcitedefaultendpunct}{\mcitedefaultseppunct}\relax
\EndOfBibitem
\bibitem{Heister:2001nj}
A.~Heister {\em et al.} ({ALEPH} collaboration){,}
  \href{http://dx.doi.org/10.1016/S0370-2693(01)01465-4}{Phys.\ Lett.\ {\bf
  B526},  34}  (2002), \href{http://arxiv.org/abs/hep-ex/0112010}{{\tt
  arXiv:hep-ex/0112010 [hep-ex]}}\relax
\mciteBstWouldAddEndPuncttrue
\mciteSetBstMidEndSepPunct{\mcitedefaultmidpunct}
{\mcitedefaultendpunct}{\mcitedefaultseppunct}\relax
\EndOfBibitem
\bibitem{Zupanc:2013iki}
A.~Zupanc {\em et al.} ({Belle} collaboration){,}
  \href{http://dx.doi.org/10.1103/PhysRevLett.113.042002}{Phys.\ Rev.\ Lett.\ {\bf
  113},  042002}  (2014), \href{http://arxiv.org/abs/1312.7826}{{\tt
  arXiv:1312.7826 [hep-ex]}}\relax
\mciteBstWouldAddEndPuncttrue
\mciteSetBstMidEndSepPunct{\mcitedefaultmidpunct}
{\mcitedefaultendpunct}{\mcitedefaultseppunct}\relax
\EndOfBibitem
\bibitem{Ablikim:2015prg}
M.~Ablikim {\em et al.} ({BESIII} collaboration){,}
  \href{http://dx.doi.org/10.1103/PhysRevLett.115.221805}{Phys.\ Rev.\ Lett.\ {\bf
  115},  221805}  (2015), \href{http://arxiv.org/abs/1510.02610}{{\tt
  arXiv:1510.02610 [hep-ex]}}\relax
\mciteBstWouldAddEndPuncttrue
\mciteSetBstMidEndSepPunct{\mcitedefaultmidpunct}
{\mcitedefaultendpunct}{\mcitedefaultseppunct}\relax
\EndOfBibitem
\bibitem{Avery:1990bc}
P.~Avery {\em et al.} ({CLEO} collaboration){,}
  \href{http://dx.doi.org/10.1103/PhysRevD.43.3599}{Phys.\ Rev.\ {\bf D43}{,}
  3599}  (1991)\relax
\mciteBstWouldAddEndPuncttrue
\mciteSetBstMidEndSepPunct{\mcitedefaultmidpunct}
{\mcitedefaultendpunct}{\mcitedefaultseppunct}\relax
\EndOfBibitem
\bibitem{Alam:1998nb}
M.~S.\ Alam {\em et al.} ({CLEO} collaboration){,}
  \href{http://dx.doi.org/10.1103/PhysRevD.57.4467}{Phys.\ Rev.\ {\bf D57}{,}
  4467}  (1998), \href{http://arxiv.org/abs/hep-ex/9709012}{{\tt
  arXiv:hep-ex/9709012 [hep-ex]}}\relax
\mciteBstWouldAddEndPuncttrue
\mciteSetBstMidEndSepPunct{\mcitedefaultmidpunct}
{\mcitedefaultendpunct}{\mcitedefaultseppunct}\relax
\EndOfBibitem
\bibitem{Albrecht:1991vs}
H.~Albrecht {\em et al.} ({ARGUS} collaboration){,}
  \href{http://dx.doi.org/10.1016/0370-2693(92)90529-D}{Phys.\ Lett.\ {\bf B274}{,}
   239}  (1992)\relax
\mciteBstWouldAddEndPuncttrue
\mciteSetBstMidEndSepPunct{\mcitedefaultmidpunct}
{\mcitedefaultendpunct}{\mcitedefaultseppunct}\relax
\EndOfBibitem
\bibitem{Link:2005ut}
J.~M.\ Link {\em et al.} ({FOCUS} collaboration){,}
  \href{http://dx.doi.org/10.1016/j.physletb.2005.08.014}{Phys.\ Lett.\ {\bf
  B624},  22}  (2005), \href{http://arxiv.org/abs/hep-ex/0505077}{{\tt
  arXiv:hep-ex/0505077 [hep-ex]}}\relax
\mciteBstWouldAddEndPuncttrue
\mciteSetBstMidEndSepPunct{\mcitedefaultmidpunct}
{\mcitedefaultendpunct}{\mcitedefaultseppunct}\relax
\EndOfBibitem
\bibitem{Avery:1993ri}
P.~Avery {\em et al.} ({CLEO} collaboration){,}
  \href{http://dx.doi.org/10.1016/0370-2693(94)90100-7}{Phys.\ Lett.\ {\bf B325}{,}
   257}  (1994)\relax
\mciteBstWouldAddEndPuncttrue
\mciteSetBstMidEndSepPunct{\mcitedefaultmidpunct}
{\mcitedefaultendpunct}{\mcitedefaultseppunct}\relax
\EndOfBibitem
\bibitem{Albrecht:1988an}
H.~Albrecht {\em et al.} ({ARGUS} collaboration){,}
  \href{http://dx.doi.org/10.1016/0370-2693(88)90896-9}{Phys.\ Lett.\ {\bf B207}{,}
   109}  (1988)\relax
\mciteBstWouldAddEndPuncttrue
\mciteSetBstMidEndSepPunct{\mcitedefaultmidpunct}
{\mcitedefaultendpunct}{\mcitedefaultseppunct}\relax
\EndOfBibitem
\bibitem{Aubert:2006wm}
B.~Aubert {\em et al.} ({\babar} collaboration){,}
  \href{http://dx.doi.org/10.1103/PhysRevD.75.052002}{Phys.\ Rev.\ {\bf D75}{,}
  052002}  (2007), \href{http://arxiv.org/abs/hep-ex/0601017}{{\tt
  arXiv:hep-ex/0601017 [hep-ex]}}\relax
\mciteBstWouldAddEndPuncttrue
\mciteSetBstMidEndSepPunct{\mcitedefaultmidpunct}
{\mcitedefaultendpunct}{\mcitedefaultseppunct}\relax
\EndOfBibitem
\bibitem{Kubota:1993pw}
Y.~Kubota {\em et al.} ({CLEO} collaboration){,}
  \href{http://dx.doi.org/10.1103/PhysRevLett.71.3255}{Phys.\ Rev.\ Lett.\ {\bf
  71},  3255}  (1993)\relax
\mciteBstWouldAddEndPuncttrue
\mciteSetBstMidEndSepPunct{\mcitedefaultmidpunct}
{\mcitedefaultendpunct}{\mcitedefaultseppunct}\relax
\EndOfBibitem
\bibitem{Bergfeld:1994gt}
T.~Bergfeld {\em et al.} ({CLEO} collaboration){,}
  \href{http://dx.doi.org/10.1016/0370-2693(94)90295-X}{Phys.\ Lett.\ {\bf B323}{,}
   219}  (1994), \href{http://arxiv.org/abs/hep-ph/9403326}{{\tt
  arXiv:hep-ph/9403326 [hep-ph]}}\relax
\mciteBstWouldAddEndPuncttrue
\mciteSetBstMidEndSepPunct{\mcitedefaultmidpunct}
{\mcitedefaultendpunct}{\mcitedefaultseppunct}\relax
\EndOfBibitem
\bibitem{Albrecht:1991bu}
H.~Albrecht {\em et al.} ({ARGUS} collaboration){,}
  \href{http://dx.doi.org/10.1016/0370-2693(91)91480-J}{Phys.\ Lett.\ {\bf B269}{,}
   234}  (1991)\relax
\mciteBstWouldAddEndPuncttrue
\mciteSetBstMidEndSepPunct{\mcitedefaultmidpunct}
{\mcitedefaultendpunct}{\mcitedefaultseppunct}\relax
\EndOfBibitem
\bibitem{Aubert:2005gt}
B.~Aubert {\em et al.} ({\babar} collaboration){,}
  \href{http://dx.doi.org/10.1103/PhysRevD.72.052006}{Phys.\ Rev.\ {\bf D72}{,}
  052006}  (2005), \href{http://arxiv.org/abs/hep-ex/0507009}{{\tt
  arXiv:hep-ex/0507009 [hep-ex]}}\relax
\mciteBstWouldAddEndPuncttrue
\mciteSetBstMidEndSepPunct{\mcitedefaultmidpunct}
{\mcitedefaultendpunct}{\mcitedefaultseppunct}\relax
\EndOfBibitem
\bibitem{Solovieva:2008fw}
E.~Solovieva {\em et al.} ({Belle} collaboration){,}
  \href{http://dx.doi.org/10.1016/j.physletb.2008.12.062}{Phys.\ Lett.\ {\bf
  B672},  1}  (2009), \href{http://arxiv.org/abs/0808.3677}{{\tt
  arXiv:0808.3677 [hep-ex]}}\relax
\mciteBstWouldAddEndPuncttrue
\mciteSetBstMidEndSepPunct{\mcitedefaultmidpunct}
{\mcitedefaultendpunct}{\mcitedefaultseppunct}\relax
\EndOfBibitem
\bibitem{Aaltonen:2011sf}
T.~Aaltonen {\em et al.} ({CDF} collaboration){,}
  \href{http://dx.doi.org/10.1103/PhysRevD.84.012003}{Phys.\ Rev.\ {\bf D84}{,}
  012003}  (2011), \href{http://arxiv.org/abs/1105.5995}{{\tt arXiv:1105.5995
  [hep-ex]}}\relax
\mciteBstWouldAddEndPuncttrue
\mciteSetBstMidEndSepPunct{\mcitedefaultmidpunct}
{\mcitedefaultendpunct}{\mcitedefaultseppunct}\relax
\EndOfBibitem
\bibitem{Artuso:2000xy}
M.~Artuso {\em et al.} ({CLEO} collaboration){,}
  \href{http://dx.doi.org/10.1103/PhysRevLett.86.4479}{Phys.\ Rev.\ Lett.\ {\bf
  86},  4479}  (2001), \href{http://arxiv.org/abs/hep-ex/0010080}{{\tt
  arXiv:hep-ex/0010080 [hep-ex]}}\relax
\mciteBstWouldAddEndPuncttrue
\mciteSetBstMidEndSepPunct{\mcitedefaultmidpunct}
{\mcitedefaultendpunct}{\mcitedefaultseppunct}\relax
\EndOfBibitem
\bibitem{Aubert:2006sp}
B.~Aubert {\em et al.} ({\babar} collaboration){,}
  \href{http://dx.doi.org/10.1103/PhysRevLett.98.012001}{Phys.\ Rev.\ Lett.\ {\bf
  98},  012001}  (2007), \href{http://arxiv.org/abs/hep-ex/0603052}{{\tt
  arXiv:hep-ex/0603052 [hep-ex]}}\relax
\mciteBstWouldAddEndPuncttrue
\mciteSetBstMidEndSepPunct{\mcitedefaultmidpunct}
{\mcitedefaultendpunct}{\mcitedefaultseppunct}\relax
\EndOfBibitem
\bibitem{Cheng:2006dk}
H.-Y.\ Cheng and C.-K.\ Chua{,}
  \href{http://dx.doi.org/10.1103/PhysRevD.75.014006}{Phys.\ Rev.\ {\bf D75}{,}
  014006}  (2007), \href{http://arxiv.org/abs/hep-ph/0610283}{{\tt
  arXiv:hep-ph/0610283 [hep-ph]}}\relax
\mciteBstWouldAddEndPuncttrue
\mciteSetBstMidEndSepPunct{\mcitedefaultmidpunct}
{\mcitedefaultendpunct}{\mcitedefaultseppunct}\relax
\EndOfBibitem
\bibitem{Lee:2014htd}
S.~H.\ Lee {\em et al.} ({Belle} collaboration){,}
  \href{http://dx.doi.org/10.1103/PhysRevD.89.091102}{Phys.\ Rev.\ {\bf D89}{,}
  091102}  (2014), \href{http://arxiv.org/abs/1404.5389}{{\tt arXiv:1404.5389
  [hep-ex]}}\relax
\mciteBstWouldAddEndPuncttrue
\mciteSetBstMidEndSepPunct{\mcitedefaultmidpunct}
{\mcitedefaultendpunct}{\mcitedefaultseppunct}\relax
\EndOfBibitem
\bibitem{Mizuk:2004yu}
R.~Mizuk {\em et al.} ({Belle} collaboration){,}
  \href{http://dx.doi.org/10.1103/PhysRevLett.94.122002}{Phys.\ Rev.\ Lett.\ {\bf
  94},  122002}  (2005), \href{http://arxiv.org/abs/hep-ex/0412069}{{\tt
  arXiv:hep-ex/0412069 [hep-ex]}}\relax
\mciteBstWouldAddEndPuncttrue
\mciteSetBstMidEndSepPunct{\mcitedefaultmidpunct}
{\mcitedefaultendpunct}{\mcitedefaultseppunct}\relax
\EndOfBibitem
\bibitem{Copley:1979wj}
L.~Copley, N.~Isgur, and G.~Karl{,}
  \href{http://dx.doi.org/10.1103/PhysRevD.20.768}{Phys.\ Rev.\ {\bf D20},  768}
   (1979), Erratum ibid.\ \href{http://dx.doi.org/10.1103/PhysRevD.23.817}{{\bf
  D23}, 817} (1981)\relax
\mciteBstWouldAddEndPuncttrue
\mciteSetBstMidEndSepPunct{\mcitedefaultmidpunct}
{\mcitedefaultendpunct}{\mcitedefaultseppunct}\relax
\EndOfBibitem
\bibitem{Pirjol:1997nh}
D.~Pirjol and T.-M.\ Yan{,}
  \href{http://dx.doi.org/10.1103/PhysRevD.56.5483}{Phys.\ Rev.\ {\bf D56}{,}
  5483}  (1997), \href{http://arxiv.org/abs/hep-ph/9701291}{{\tt
  arXiv:hep-ph/9701291 [hep-ph]}}\relax
\mciteBstWouldAddEndPuncttrue
\mciteSetBstMidEndSepPunct{\mcitedefaultmidpunct}
{\mcitedefaultendpunct}{\mcitedefaultseppunct}\relax
\EndOfBibitem
\bibitem{Chistov:2006zj}
R.~Chistov {\em et al.} ({Belle} collaboration){,}
  \href{http://dx.doi.org/10.1103/PhysRevLett.97.162001}{Phys.\ Rev.\ Lett.\ {\bf
  97},  162001}  (2006), \href{http://arxiv.org/abs/hep-ex/0606051}{{\tt
  arXiv:hep-ex/0606051 [hep-ex]}}\relax
\mciteBstWouldAddEndPuncttrue
\mciteSetBstMidEndSepPunct{\mcitedefaultmidpunct}
{\mcitedefaultendpunct}{\mcitedefaultseppunct}\relax
\EndOfBibitem
\bibitem{YKato:2014}
Y.~Kato {\em et al.} ({Belle} collaboration){,}
  \href{http://dx.doi.org/10.1103/PhysRevD.89.052003}{Phys.\ Rev.\ {\bf D89}{,}
  052003}  (2014), \href{http://arxiv.org/abs/1312.1026}{{\tt arXiv:1312.1026
  [hep-ex]}}\relax
\mciteBstWouldAddEndPuncttrue
\mciteSetBstMidEndSepPunct{\mcitedefaultmidpunct}
{\mcitedefaultendpunct}{\mcitedefaultseppunct}\relax
\EndOfBibitem
\bibitem{YKato:2016}
Y.~Kato {\em et al.} ({Belle} collaboration){,}
  \href{http://dx.doi.org/10.1103/PhysRevD.94.032002}{Phys.\ Rev.\ {\bf D94}{,}
  032002}  (2016), \href{http://arxiv.org/abs/1605.09103}{{\tt
  arXiv:1605.09103 [hep-ex]}}\relax
\mciteBstWouldAddEndPuncttrue
\mciteSetBstMidEndSepPunct{\mcitedefaultmidpunct}
{\mcitedefaultendpunct}{\mcitedefaultseppunct}\relax
\EndOfBibitem
\bibitem{Aubert:2007bd}
B.~Aubert {\em et al.} ({\babar} collaboration){,}
  \href{http://dx.doi.org/10.1103/PhysRevD.77.031101}{Phys.\ Rev.\ {\bf D77}{,}
  031101}  (2008), \href{http://arxiv.org/abs/0710.5775}{{\tt arXiv:0710.5775
  [hep-ex]}}\relax
\mciteBstWouldAddEndPuncttrue
\mciteSetBstMidEndSepPunct{\mcitedefaultmidpunct}
{\mcitedefaultendpunct}{\mcitedefaultseppunct}\relax
\EndOfBibitem
\bibitem{Yelton:2016fqw}
J.~Yelton {\em et al.} ({Belle} collaboration){,}
  \href{http://dx.doi.org/10.1103/PhysRevD.94.052011}{Phys.\ Rev.\ {\bf D94}{,}
  052011}  (2016), \href{http://arxiv.org/abs/1607.07123}{{\tt
  arXiv:1607.07123 [hep-ex]}}\relax
\mciteBstWouldAddEndPuncttrue
\mciteSetBstMidEndSepPunct{\mcitedefaultmidpunct}
{\mcitedefaultendpunct}{\mcitedefaultseppunct}\relax
\EndOfBibitem
\bibitem{Aubert:2006je}
B.~Aubert {\em et al.} ({\babar} collaboration){,}
  \href{http://dx.doi.org/10.1103/PhysRevLett.97.232001}{Phys.\ Rev.\ Lett.\ {\bf
  97},  232001}  (2006), \href{http://arxiv.org/abs/hep-ex/0608055}{{\tt
  arXiv:hep-ex/0608055 [hep-ex]}}\relax
\mciteBstWouldAddEndPuncttrue
\mciteSetBstMidEndSepPunct{\mcitedefaultmidpunct}
{\mcitedefaultendpunct}{\mcitedefaultseppunct}\relax
\EndOfBibitem
\bibitem{Rosner:1995yu}
J.~L.\ Rosner, \href{http://dx.doi.org/10.1103/PhysRevD.52.6461}{Phys.\ Rev.\ {\bf
  D52},  6461}  (1995), \href{http://arxiv.org/abs/hep-ph/9508252}{{\tt
  arXiv:hep-ph/9508252 [hep-ph]}}\relax
\mciteBstWouldAddEndPuncttrue
\mciteSetBstMidEndSepPunct{\mcitedefaultmidpunct}
{\mcitedefaultendpunct}{\mcitedefaultseppunct}\relax
\EndOfBibitem
\bibitem{Glozman:1995xy}
L.~Y.\ Glozman and D.~Riska{,}
  \href{http://dx.doi.org/10.1016/0375-9474(96)80005-C}{Nucl.\ Phys.\ {\bf A603}{,}
   326}  (1996), \href{http://arxiv.org/abs/hep-ph/9509269}{{\tt
  arXiv:hep-ph/9509269 [hep-ph]}}, Erratum ibid.\
  \href{http://dx.doi.org/10.1016/S0375-9474(97)00200-5}{{\bf A620}, 510}
  (1997)\relax
\mciteBstWouldAddEndPuncttrue
\mciteSetBstMidEndSepPunct{\mcitedefaultmidpunct}
{\mcitedefaultendpunct}{\mcitedefaultseppunct}\relax
\EndOfBibitem
\bibitem{Jenkins:1996de}
E.~E.\ Jenkins, \href{http://dx.doi.org/10.1103/PhysRevD.54.4515}{Phys.\ Rev.\
  {\bf D54},  4515}  (1996), \href{http://arxiv.org/abs/hep-ph/9603449}{{\tt
  arXiv:hep-ph/9603449 [hep-ph]}}\relax
\mciteBstWouldAddEndPuncttrue
\mciteSetBstMidEndSepPunct{\mcitedefaultmidpunct}
{\mcitedefaultendpunct}{\mcitedefaultseppunct}\relax
\EndOfBibitem
\bibitem{Burakovsky:1997vm}
L.~Burakovsky, J.~T.\ Goldman, and L.~Horwitz{,}
  \href{http://dx.doi.org/10.1103/PhysRevD.56.7124}{Phys.\ Rev.\ {\bf D56}{,}
  7124}  (1997), \href{http://arxiv.org/abs/hep-ph/9706464}{{\tt
  arXiv:hep-ph/9706464 [hep-ph]}}\relax
\mciteBstWouldAddEndPuncttrue
\mciteSetBstMidEndSepPunct{\mcitedefaultmidpunct}
{\mcitedefaultendpunct}{\mcitedefaultseppunct}\relax
\EndOfBibitem
\bibitem{Burdman:2001tf}
G.~Burdman, E.~Golowich, J.~L.\ Hewett, and S.~Pakvasa{,}
  \href{http://dx.doi.org/10.1103/PhysRevD.66.014009}{Phys.\ Rev.\ {\bf D66}{,}
  014009}  (2002), \href{http://arxiv.org/abs/hep-ph/0112235}{{\tt
  arXiv:hep-ph/0112235 [hep-ph]}}\relax
\mciteBstWouldAddEndPuncttrue
\mciteSetBstMidEndSepPunct{\mcitedefaultmidpunct}
{\mcitedefaultendpunct}{\mcitedefaultseppunct}\relax
\EndOfBibitem
\bibitem{Fajfer:2002bu}
S.~Fajfer, A.~Prapotnik, S.~Prelovsek, P.~Singer, and J.~Zupan{,}
  \href{http://dx.doi.org/10.1016/S0920-5632(02)01961-8}{Nucl.\ Phys.\ Proc.\
  Suppl.\ {\bf 115},  93}  (2003){,}
  \href{http://arxiv.org/abs/hep-ph/0208201}{{\tt arXiv:hep-ph/0208201
  [hep-ph]}}\relax
\mciteBstWouldAddEndPuncttrue
\mciteSetBstMidEndSepPunct{\mcitedefaultmidpunct}
{\mcitedefaultendpunct}{\mcitedefaultseppunct}\relax
\EndOfBibitem
\bibitem{Fajfer:2007dy}
S.~Fajfer, N.~Kosnik, and S.~Prelovsek{,}
  \href{http://dx.doi.org/10.1103/PhysRevD.76.074010}{Phys.\ Rev.\ {\bf D76}{,}
  074010}  (2007), \href{http://arxiv.org/abs/0706.1133}{{\tt arXiv:0706.1133
  [hep-ph]}}\relax
\mciteBstWouldAddEndPuncttrue
\mciteSetBstMidEndSepPunct{\mcitedefaultmidpunct}
{\mcitedefaultendpunct}{\mcitedefaultseppunct}\relax
\EndOfBibitem
\bibitem{Golowich:2009ii}
E.~Golowich, J.~Hewett, S.~Pakvasa, and A.~A.\ Petrov{,}
  \href{http://dx.doi.org/10.1103/PhysRevD.79.114030}{Phys.\ Rev.\ {\bf D79}{,}
  114030}  (2009), \href{http://arxiv.org/abs/0903.2830}{{\tt arXiv:0903.2830
  [hep-ph]}}\relax
\mciteBstWouldAddEndPuncttrue
\mciteSetBstMidEndSepPunct{\mcitedefaultmidpunct}
{\mcitedefaultendpunct}{\mcitedefaultseppunct}\relax
\EndOfBibitem
\bibitem{Paul:2010pq}
A.~Paul, I.~I.\ Bigi, and S.~Recksiegel{,}
  \href{http://dx.doi.org/10.1103/PhysRevD.82.094006}{Phys.\ Rev.\ {\bf D82}{,}
  094006}  (2010), \href{http://arxiv.org/abs/1008.3141}{{\tt arXiv:1008.3141
  [hep-ph]}}, Erratum ibid.\
  \href{http://dx.doi.org/10.1103/PhysRevD.83.019901}{{\bf D83}, 019901}
  (2011)\relax
\mciteBstWouldAddEndPuncttrue
\mciteSetBstMidEndSepPunct{\mcitedefaultmidpunct}
{\mcitedefaultendpunct}{\mcitedefaultseppunct}\relax
\EndOfBibitem
\bibitem{Borisov:2011aa}
A.~Borisov, \href{http://arxiv.org/abs/1112.3269}{{\tt arXiv:1112.3269
  [hep-ph]}}  (2011)\relax
\mciteBstWouldAddEndPuncttrue
\mciteSetBstMidEndSepPunct{\mcitedefaultmidpunct}
{\mcitedefaultendpunct}{\mcitedefaultseppunct}\relax
\EndOfBibitem
\bibitem{Wang:2014dba}
R.-M.\ Wang, J.-H.\ Sheng, J.~Zhu, Y.-Y.\ Fan, and Y.~Gao{,}
  \href{http://dx.doi.org/10.1142/S0217751X14501693}{Int.\ J.\ Mod.\ Phys.\ {\bf
  A29},  1450169}  (2014)\relax
\mciteBstWouldAddEndPuncttrue
\mciteSetBstMidEndSepPunct{\mcitedefaultmidpunct}
{\mcitedefaultendpunct}{\mcitedefaultseppunct}\relax
\EndOfBibitem
\bibitem{deBoer:2015boa}
S.~de~Boer and G.~Hiller{,}
  \href{http://dx.doi.org/10.1103/PhysRevD.93.074001}{Phys.\ Rev.\ {\bf D93}{,}
  074001}  (2016), \href{http://arxiv.org/abs/1510.00311}{{\tt
  arXiv:1510.00311 [hep-ph]}}\relax
\mciteBstWouldAddEndPuncttrue
\mciteSetBstMidEndSepPunct{\mcitedefaultmidpunct}
{\mcitedefaultendpunct}{\mcitedefaultseppunct}\relax
\EndOfBibitem
\bibitem{Rubin:2010cq}
P.~Rubin {\em et al.} ({CLEO} collaboration){,}
  \href{http://dx.doi.org/10.1103/PhysRevD.82.092007}{Phys.\ Rev.\ {\bf D82}{,}
  092007}  (2010), \href{http://arxiv.org/abs/1009.1606}{{\tt arXiv:1009.1606
  [hep-ex]}}\relax
\mciteBstWouldAddEndPuncttrue
\mciteSetBstMidEndSepPunct{\mcitedefaultmidpunct}
{\mcitedefaultendpunct}{\mcitedefaultseppunct}\relax
\EndOfBibitem
\bibitem{Abazov:2007aj}
V.~Abazov {\em et al.} ({\dzero} collaboration){,}
  \href{http://dx.doi.org/10.1103/PhysRevLett.100.101801}{Phys.\ Rev.\ Lett.\ {\bf
  100},  101801}  (2008), \href{http://arxiv.org/abs/0708.2094}{{\tt
  arXiv:0708.2094 [hep-ex]}}\relax
\mciteBstWouldAddEndPuncttrue
\mciteSetBstMidEndSepPunct{\mcitedefaultmidpunct}
{\mcitedefaultendpunct}{\mcitedefaultseppunct}\relax
\EndOfBibitem
\bibitem{Aaij:2013uoa}
R.~Aaij {\em et al.} ({LHCb} collaboration){,}
  \href{http://dx.doi.org/10.1016/j.physletb.2013.11.053}{Phys.\ Lett.\ {\bf
  B728},  234}  (2014), \href{http://arxiv.org/abs/1310.2535}{{\tt
  arXiv:1310.2535 [hep-ex]}}\relax
\mciteBstWouldAddEndPuncttrue
\mciteSetBstMidEndSepPunct{\mcitedefaultmidpunct}
{\mcitedefaultendpunct}{\mcitedefaultseppunct}\relax
\EndOfBibitem
\bibitem{Aaij:2013sua}
R.~Aaij {\em et al.} ({LHCb} collaboration){,}
  \href{http://dx.doi.org/10.1016/j.physletb.2013.06.010}{Phys.\ Lett.\ {\bf
  B724},  203}  (2013), \href{http://arxiv.org/abs/1304.6365}{{\tt
  arXiv:1304.6365 [hep-ex]}}\relax
\mciteBstWouldAddEndPuncttrue
\mciteSetBstMidEndSepPunct{\mcitedefaultmidpunct}
{\mcitedefaultendpunct}{\mcitedefaultseppunct}\relax
\EndOfBibitem
\bibitem{Coan:2002te}
T.~Coan {\em et al.} ({CLEO} collaboration){,}
  \href{http://dx.doi.org/10.1103/PhysRevLett.90.101801}{Phys.\ Rev.\ Lett.\ {\bf
  90},  101801}  (2003), \href{http://arxiv.org/abs/hep-ex/0212045}{{\tt
  arXiv:hep-ex/0212045 [hep-ex]}}\relax
\mciteBstWouldAddEndPuncttrue
\mciteSetBstMidEndSepPunct{\mcitedefaultmidpunct}
{\mcitedefaultendpunct}{\mcitedefaultseppunct}\relax
\EndOfBibitem
\bibitem{Ablikim:2015djc}
M.~Ablikim {\em et al.} ({BESIII} collaboration){,}
  \href{http://dx.doi.org/10.1103/PhysRevD.91.112015}{Phys.\ Rev.\ {\bf D91}{,}
  112015}  (2015), \href{http://arxiv.org/abs/1505.03087}{{\tt
  arXiv:1505.03087 [hep-ex]}}\relax
\mciteBstWouldAddEndPuncttrue
\mciteSetBstMidEndSepPunct{\mcitedefaultmidpunct}
{\mcitedefaultendpunct}{\mcitedefaultseppunct}\relax
\EndOfBibitem
\bibitem{Lees:2011qz}
J.~P.\ Lees {\em et al.} ({\babar} collaboration){,}
  \href{http://dx.doi.org/10.1103/PhysRevD.85.091107}{Phys.\ Rev.\ {\bf D85}{,}
  091107}  (2012), \href{http://arxiv.org/abs/1110.6480}{{\tt arXiv:1110.6480
  [hep-ex]}}\relax
\mciteBstWouldAddEndPuncttrue
\mciteSetBstMidEndSepPunct{\mcitedefaultmidpunct}
{\mcitedefaultendpunct}{\mcitedefaultseppunct}\relax
\EndOfBibitem
\bibitem{Nisar:2015gvd}
N.~K.\ Nisar {\em et al.} ({Belle} collaboration){,}
  \href{http://dx.doi.org/10.1103/PhysRevD.93.051102}{Phys.\ Rev.\ {\bf D93}{,}
  051102}  (2016), \href{http://arxiv.org/abs/1512.02992}{{\tt
  arXiv:1512.02992 [hep-ex]}}\relax
\mciteBstWouldAddEndPuncttrue
\mciteSetBstMidEndSepPunct{\mcitedefaultmidpunct}
{\mcitedefaultendpunct}{\mcitedefaultseppunct}\relax
\EndOfBibitem
\bibitem{Haas:1988bh}
P.~Haas {\em et al.} ({CLEO} collaboration){,}
  \href{http://dx.doi.org/10.1103/PhysRevLett.60.1614}{Phys.\ Rev.\ Lett.\ {\bf
  60},  1614}  (1988)\relax
\mciteBstWouldAddEndPuncttrue
\mciteSetBstMidEndSepPunct{\mcitedefaultmidpunct}
{\mcitedefaultendpunct}{\mcitedefaultseppunct}\relax
\EndOfBibitem
\bibitem{Albrecht:1988ge}
H.~Albrecht {\em et al.} ({ARGUS} collaboration){,}
  \href{http://dx.doi.org/10.1016/0370-2693(88)90967-7}{Phys.\ Lett.\ {\bf B209}{,}
   380}  (1988)\relax
\mciteBstWouldAddEndPuncttrue
\mciteSetBstMidEndSepPunct{\mcitedefaultmidpunct}
{\mcitedefaultendpunct}{\mcitedefaultseppunct}\relax
\EndOfBibitem
\bibitem{Adler:1987cp}
J.~Adler {\em et al.} ({MARK-III} collaboration){,}
  \href{http://dx.doi.org/10.1103/PhysRevD.37.2023}{Phys.\ Rev.\ {\bf D37}{,}
  2023}  (1988), Erratum ibid.\
  \href{http://dx.doi.org/10.1103/PhysRevD.40.3788}{{\bf D40}, 3788}
  (1989)\relax
\mciteBstWouldAddEndPuncttrue
\mciteSetBstMidEndSepPunct{\mcitedefaultmidpunct}
{\mcitedefaultendpunct}{\mcitedefaultseppunct}\relax
\EndOfBibitem
\bibitem{Freyberger:1996it}
A.~Freyberger {\em et al.} ({CLEO} collaboration){,}
  \href{http://dx.doi.org/10.1103/PhysRevLett.76.3065}{Phys.\ Rev.\ Lett.\ {\bf
  76},  3065}  (1996)\relax
\mciteBstWouldAddEndPuncttrue
\mciteSetBstMidEndSepPunct{\mcitedefaultmidpunct}
{\mcitedefaultendpunct}{\mcitedefaultseppunct}\relax
\EndOfBibitem
\bibitem{Pripstein:1999tq}
D.~Pripstein {\em et al.} ({Fermilab E789} collaboration){,}
  \href{http://dx.doi.org/10.1103/PhysRevD.61.032005}{Phys.\ Rev.\ {\bf D61}{,}
  032005}  (2000), \href{http://arxiv.org/abs/hep-ex/9906022}{{\tt
  arXiv:hep-ex/9906022 [hep-ex]}}\relax
\mciteBstWouldAddEndPuncttrue
\mciteSetBstMidEndSepPunct{\mcitedefaultmidpunct}
{\mcitedefaultendpunct}{\mcitedefaultseppunct}\relax
\EndOfBibitem
\bibitem{Aitala:1999db}
E.~Aitala {\em et al.} ({Fermilab E791} collaboration){,}
  \href{http://dx.doi.org/10.1016/S0370-2693(99)00902-8}{Phys.\ Lett.\ {\bf
  B462},  401}  (1999), \href{http://arxiv.org/abs/hep-ex/9906045}{{\tt
  arXiv:hep-ex/9906045 [hep-ex]}}\relax
\mciteBstWouldAddEndPuncttrue
\mciteSetBstMidEndSepPunct{\mcitedefaultmidpunct}
{\mcitedefaultendpunct}{\mcitedefaultseppunct}\relax
\EndOfBibitem
\bibitem{Aubert:2004bs}
B.~Aubert {\em et al.} ({\babar} collaboration){,}
  \href{http://dx.doi.org/10.1103/PhysRevLett.93.191801}{Phys.\ Rev.\ Lett.\ {\bf
  93},  191801}  (2004), \href{http://arxiv.org/abs/hep-ex/0408023}{{\tt
  arXiv:hep-ex/0408023 [hep-ex]}}\relax
\mciteBstWouldAddEndPuncttrue
\mciteSetBstMidEndSepPunct{\mcitedefaultmidpunct}
{\mcitedefaultendpunct}{\mcitedefaultseppunct}\relax
\EndOfBibitem
\bibitem{Petric:2010yt}
M.~Petric {\em et al.} ({Belle} collaboration){,}
  \href{http://dx.doi.org/10.1103/PhysRevD.81.091102}{Phys.\ Rev.\ {\bf D81}{,}
  091102}  (2010), \href{http://arxiv.org/abs/1003.2345}{{\tt arXiv:1003.2345
  [hep-ex]}}\relax
\mciteBstWouldAddEndPuncttrue
\mciteSetBstMidEndSepPunct{\mcitedefaultmidpunct}
{\mcitedefaultendpunct}{\mcitedefaultseppunct}\relax
\EndOfBibitem
\bibitem{Kodama:1995ia}
K.~Kodama {\em et al.} ({Fermilab E653} collaboration){,}
  \href{http://dx.doi.org/10.1016/0370-2693(94)01610-O}{Phys.\ Lett.\ {\bf B345}{,}
   85}  (1995)\relax
\mciteBstWouldAddEndPuncttrue
\mciteSetBstMidEndSepPunct{\mcitedefaultmidpunct}
{\mcitedefaultendpunct}{\mcitedefaultseppunct}\relax
\EndOfBibitem
\bibitem{Abt:2004hn}
I.~Abt {\em et al.} ({HERA-B} collaboration){,}
  \href{http://dx.doi.org/10.1016/j.physletb.2004.06.097}{Phys.\ Lett.\ {\bf
  B596},  173}  (2004), \href{http://arxiv.org/abs/hep-ex/0405059}{{\tt
  arXiv:hep-ex/0405059 [hep-ex]}}\relax
\mciteBstWouldAddEndPuncttrue
\mciteSetBstMidEndSepPunct{\mcitedefaultmidpunct}
{\mcitedefaultendpunct}{\mcitedefaultseppunct}\relax
\EndOfBibitem
\bibitem{Aaltonen:2010hz}
T.~Aaltonen {\em et al.} ({CDF} collaboration){,}
  \href{http://dx.doi.org/10.1103/PhysRevD.82.091105}{Phys.\ Rev.\ {\bf D82}{,}
  091105}  (2010), \href{http://arxiv.org/abs/1008.5077}{{\tt arXiv:1008.5077
  [hep-ex]}}\relax
\mciteBstWouldAddEndPuncttrue
\mciteSetBstMidEndSepPunct{\mcitedefaultmidpunct}
{\mcitedefaultendpunct}{\mcitedefaultseppunct}\relax
\EndOfBibitem
\bibitem{Aaij:2013cza}
R.~Aaij {\em et al.} ({LHCb} collaboration){,}
  \href{http://dx.doi.org/10.1016/j.physletb.2013.06.037}{Phys.\ Lett.\ {\bf
  B725},  15}  (2013), \href{http://arxiv.org/abs/1305.5059}{{\tt
  arXiv:1305.5059 [hep-ex]}}\relax
\mciteBstWouldAddEndPuncttrue
\mciteSetBstMidEndSepPunct{\mcitedefaultmidpunct}
{\mcitedefaultendpunct}{\mcitedefaultseppunct}\relax
\EndOfBibitem
\bibitem{Aitala:2000kk}
E.~Aitala {\em et al.} ({Fermilab E791} collaboration){,}
  \href{http://dx.doi.org/10.1103/PhysRevLett.86.3969}{Phys.\ Rev.\ Lett.\ {\bf
  86},  3969}  (2001), \href{http://arxiv.org/abs/hep-ex/0011077}{{\tt
  arXiv:hep-ex/0011077 [hep-ex]}}\relax
\mciteBstWouldAddEndPuncttrue
\mciteSetBstMidEndSepPunct{\mcitedefaultmidpunct}
{\mcitedefaultendpunct}{\mcitedefaultseppunct}\relax
\EndOfBibitem
\bibitem{Adler:1988es}
J.~Adler {\em et al.} ({MARK-III} collaboration){,}
  \href{http://dx.doi.org/10.1103/PhysRevD.40.906}{Phys.\ Rev.\ {\bf D40},  906}
   (1989)\relax
\mciteBstWouldAddEndPuncttrue
\mciteSetBstMidEndSepPunct{\mcitedefaultmidpunct}
{\mcitedefaultendpunct}{\mcitedefaultseppunct}\relax
\EndOfBibitem
\bibitem{Becker:1987mu}
J.~Becker {\em et al.} ({MARK-III} collaboration){,}
  \href{http://dx.doi.org/10.1016/0370-2693(87)90473-4}{Phys.\ Lett.\ {\bf B193}{,}
   147}  (1987)\relax
\mciteBstWouldAddEndPuncttrue
\mciteSetBstMidEndSepPunct{\mcitedefaultmidpunct}
{\mcitedefaultendpunct}{\mcitedefaultseppunct}\relax
\EndOfBibitem
\bibitem{Aaij:2015qmj}
R.~Aaij {\em et al.} ({LHCb} collaboration){,}
  \href{http://dx.doi.org/10.1016/j.physletb.2016.01.029}{Phys.\ Lett.\ {\bf
  B754},  167}  (2016), \href{http://arxiv.org/abs/1512.00322}{{\tt
  arXiv:1512.00322 [hep-ex]}}\relax
\mciteBstWouldAddEndPuncttrue
\mciteSetBstMidEndSepPunct{\mcitedefaultmidpunct}
{\mcitedefaultendpunct}{\mcitedefaultseppunct}\relax
\EndOfBibitem
\bibitem{Rubin:2009aa}
P.~Rubin {\em et al.} ({CLEO} collaboration){,}
  \href{http://dx.doi.org/10.1103/PhysRevD.79.097101}{Phys.\ Rev.\ {\bf D79}{,}
  097101}  (2009), \href{http://arxiv.org/abs/0904.1619}{{\tt arXiv:0904.1619
  [hep-ex]}}\relax
\mciteBstWouldAddEndPuncttrue
\mciteSetBstMidEndSepPunct{\mcitedefaultmidpunct}
{\mcitedefaultendpunct}{\mcitedefaultseppunct}\relax
\EndOfBibitem
\bibitem{Frabetti:1997wp}
P.~Frabetti {\em et al.} ({Fermilab E687} collaboration){,}
  \href{http://dx.doi.org/10.1016/S0370-2693(97)00229-3}{Phys.\ Lett.\ {\bf
  B398},  239}  (1997)\relax
\mciteBstWouldAddEndPuncttrue
\mciteSetBstMidEndSepPunct{\mcitedefaultmidpunct}
{\mcitedefaultendpunct}{\mcitedefaultseppunct}\relax
\EndOfBibitem
\bibitem{Lees:2011hb}
J.~P.\ Lees {\em et al.} ({\babar} collaboration){,}
  \href{http://dx.doi.org/10.1103/PhysRevD.84.072006}{Phys.\ Rev.\ {\bf D84}{,}
  072006}  (2011), \href{http://arxiv.org/abs/1107.4465}{{\tt arXiv:1107.4465
  [hep-ex]}}\relax
\mciteBstWouldAddEndPuncttrue
\mciteSetBstMidEndSepPunct{\mcitedefaultmidpunct}
{\mcitedefaultendpunct}{\mcitedefaultseppunct}\relax
\EndOfBibitem
\bibitem{Zhao:2016jna}
M.-G.\ Zhao ({BESIII} collaboration){,}
  \href{http://arxiv.org/abs/1605.08952}{{\tt arXiv:1605.08952 [hep-ex]}}
  (2016)\relax
\mciteBstWouldAddEndPuncttrue
\mciteSetBstMidEndSepPunct{\mcitedefaultmidpunct}
{\mcitedefaultendpunct}{\mcitedefaultseppunct}\relax
\EndOfBibitem
\bibitem{Link:2003qp}
J.~Link {\em et al.} ({FOCUS} collaboration){,}
  \href{http://dx.doi.org/10.1016/j.physletb.2003.07.079}{Phys.\ Lett.\ {\bf
  B572},  21}  (2003), \href{http://arxiv.org/abs/hep-ex/0306049}{{\tt
  arXiv:hep-ex/0306049 [hep-ex]}}\relax
\mciteBstWouldAddEndPuncttrue
\mciteSetBstMidEndSepPunct{\mcitedefaultmidpunct}
{\mcitedefaultendpunct}{\mcitedefaultseppunct}\relax
\EndOfBibitem
\bibitem{Aubert:2008an}
B.~Aubert {\em et al.} ({\babar} collaboration){,}
  \href{http://dx.doi.org/10.1016/j.nuclphysbps.2009.03.034}{Nucl.\ Phys.\ Proc.\
  Suppl.\ {\bf 189},  193}  (2009), \href{http://arxiv.org/abs/0808.1121}{{\tt
  arXiv:0808.1121 [hep-ex]}}\relax
\mciteBstWouldAddEndPuncttrue
\mciteSetBstMidEndSepPunct{\mcitedefaultmidpunct}
{\mcitedefaultendpunct}{\mcitedefaultseppunct}\relax
\EndOfBibitem
\bibitem{Paramesvaran:2009ec}
S.~Paramesvaran ({\babar} collaboration){,}
  \href{http://arxiv.org/abs/0910.2884}{{\tt arXiv:0910.2884 [hep-ex]}}
  (2009)\relax
\mciteBstWouldAddEndPuncttrue
\mciteSetBstMidEndSepPunct{\mcitedefaultmidpunct}
{\mcitedefaultendpunct}{\mcitedefaultseppunct}\relax
\EndOfBibitem
\bibitem{Ryu:2014vpc}
S.~Ryu {\em et al.} ({Belle} collaboration){,}
  \href{http://dx.doi.org/10.1103/PhysRevD.89.072009}{Phys.\ Rev.\ {\bf D89}{,}
  072009}  (2014), \href{http://arxiv.org/abs/1402.5213}{{\tt arXiv:1402.5213
  [hep-ex]}}\relax
\mciteBstWouldAddEndPuncttrue
\mciteSetBstMidEndSepPunct{\mcitedefaultmidpunct}
{\mcitedefaultendpunct}{\mcitedefaultseppunct}\relax
\EndOfBibitem
\bibitem{Barate:1997tt}
R.~Barate {\em et al.} ({ALEPH} collaboration){,}
  \href{http://dx.doi.org/10.1007/s100529800879}{Eur.\ Phys.\ J.\ {\bf C4},  29}
  (1998), \url{http://cdsweb.cern.ch/record/346304}\relax
\mciteBstWouldAddEndPuncttrue
\mciteSetBstMidEndSepPunct{\mcitedefaultmidpunct}
{\mcitedefaultendpunct}{\mcitedefaultseppunct}\relax
\EndOfBibitem
\bibitem{Schael:2005am}
S.~Schael {\em et al.} ({ALEPH} collaboration){,}
  \href{http://dx.doi.org/10.1016/j.physrep.2005.06.007}{Phys.\ Rept.\ {\bf 421}{,}
   191}  (2005), \href{http://arxiv.org/abs/hep-ex/0506072}{{\tt
  arXiv:hep-ex/0506072 [hep-ex]}}, HFLAV-tau uses measurements of $\tau \to h
  X$ and $\tau \to K X$ and obtains $\tau \to \pi X$ by difference; the
  measurement of ${\cal B}\left( \tau^- \to 3h^- 2h^+ \pi^0 \nu_{\tau} \ ({\rm
  ex.} K^0)\right)$ has been read as $(2.1 \pm 0.7 \pm 0.6) \times 10^{-4}$
  whereas PDG11 uses $(2.1 \pm 0.7 \pm 0.9) \times 10^{-4}$\relax
\mciteBstWouldAddEndPuncttrue
\mciteSetBstMidEndSepPunct{\mcitedefaultmidpunct}
{\mcitedefaultendpunct}{\mcitedefaultseppunct}\relax
\EndOfBibitem
\bibitem{Abreu:1999rb}
P.~Abreu {\em et al.} ({DELPHI} collaboration){,}
  \href{http://dx.doi.org/10.1007/s100520050583}{Eur.\ Phys.\ J.\ {\bf C10}{,}
  201}  (1999)\relax
\mciteBstWouldAddEndPuncttrue
\mciteSetBstMidEndSepPunct{\mcitedefaultmidpunct}
{\mcitedefaultendpunct}{\mcitedefaultseppunct}\relax
\EndOfBibitem
\bibitem{Acciarri:2001sg}
M.~Acciarri {\em et al.} ({L3} collaboration){,}
  \href{http://dx.doi.org/10.1016/S0370-2693(01)00294-5}{Phys.\ Lett.\ {\bf
  B507},  47}  (2001), \href{http://arxiv.org/abs/hep-ex/0102023}{{\tt
  arXiv:hep-ex/0102023 [hep-ex]}}\relax
\mciteBstWouldAddEndPuncttrue
\mciteSetBstMidEndSepPunct{\mcitedefaultmidpunct}
{\mcitedefaultendpunct}{\mcitedefaultseppunct}\relax
\EndOfBibitem
\bibitem{Abbiendi:2002jw}
G.~Abbiendi {\em et al.} ({OPAL} collaboration){,}
  \href{http://dx.doi.org/10.1016/S0370-2693(02)03020-4}{Phys.\ Lett.\ {\bf
  B551},  35}  (2003), \href{http://arxiv.org/abs/hep-ex/0211066}{{\tt
  arXiv:hep-ex/0211066 [hep-ex]}}\relax
\mciteBstWouldAddEndPuncttrue
\mciteSetBstMidEndSepPunct{\mcitedefaultmidpunct}
{\mcitedefaultendpunct}{\mcitedefaultseppunct}\relax
\EndOfBibitem
\bibitem{Albrecht:1991rh}
H.~Albrecht {\em et al.} ({ARGUS} collaboration){,}
  \href{http://dx.doi.org/10.1007/BF01625895}{Z.\ Phys.\ {\bf C53},  367}
  (1992)\relax
\mciteBstWouldAddEndPuncttrue
\mciteSetBstMidEndSepPunct{\mcitedefaultmidpunct}
{\mcitedefaultendpunct}{\mcitedefaultseppunct}\relax
\EndOfBibitem
\bibitem{Aubert:2009qj}
B.~Aubert {\em et al.} ({\babar} collaboration){,}
  \href{http://dx.doi.org/10.1103/PhysRevLett.105.051602}{Phys.\ Rev.\ Lett.\ {\bf
  105},  051602}  (2010), \href{http://arxiv.org/abs/0912.0242}{{\tt
  arXiv:0912.0242 [hep-ex]}}\relax
\mciteBstWouldAddEndPuncttrue
\mciteSetBstMidEndSepPunct{\mcitedefaultmidpunct}
{\mcitedefaultendpunct}{\mcitedefaultseppunct}\relax
\EndOfBibitem
\bibitem{Anastassov:1996tc}
A.~Anastassov {\em et al.} ({CLEO} collaboration){,}
  \href{http://dx.doi.org/10.1103/PhysRevD.55.2559}{Phys.\ Rev.\ {\bf D55}{,}
  2559}  (1997), Erratum ibid.\
  \href{http://dx.doi.org/10.1103/PhysRevD.58.119903}{{\bf D58}, 119903}
  (1998)\relax
\mciteBstWouldAddEndPuncttrue
\mciteSetBstMidEndSepPunct{\mcitedefaultmidpunct}
{\mcitedefaultendpunct}{\mcitedefaultseppunct}\relax
\EndOfBibitem
\bibitem{Abbiendi:1998cx}
G.~Abbiendi {\em et al.} ({OPAL} collaboration){,}
  \href{http://dx.doi.org/10.1016/S0370-2693(98)01553-6}{Phys.\ Lett.\ {\bf
  B447},  134}  (1999), \href{http://arxiv.org/abs/hep-ex/9812017}{{\tt
  arXiv:hep-ex/9812017 [hep-ex]}}\relax
\mciteBstWouldAddEndPuncttrue
\mciteSetBstMidEndSepPunct{\mcitedefaultmidpunct}
{\mcitedefaultendpunct}{\mcitedefaultseppunct}\relax
\EndOfBibitem
\bibitem{Abreu:1992gn}
P.~Abreu {\em et al.} ({DELPHI} collaboration){,}
  \href{http://dx.doi.org/10.1007/BF01561293}{Z.\ Phys.\ {\bf C55},  555}
  (1992)\relax
\mciteBstWouldAddEndPuncttrue
\mciteSetBstMidEndSepPunct{\mcitedefaultmidpunct}
{\mcitedefaultendpunct}{\mcitedefaultseppunct}\relax
\EndOfBibitem
\bibitem{Acciarri:1994vr}
M.~Acciarri {\em et al.} ({L3} collaboration){,}
  \href{http://dx.doi.org/10.1016/0370-2693(94)01587-3}{Phys.\ Lett.\ {\bf B345}{,}
   93}  (1995)\relax
\mciteBstWouldAddEndPuncttrue
\mciteSetBstMidEndSepPunct{\mcitedefaultmidpunct}
{\mcitedefaultendpunct}{\mcitedefaultseppunct}\relax
\EndOfBibitem
\bibitem{Alexander:1991am}
G.~Alexander {\em et al.} ({OPAL} collaboration){,}
  \href{http://dx.doi.org/10.1016/0370-2693(91)90768-L}{Phys.\ Lett.\ {\bf B266}{,}
   201}  (1991)\relax
\mciteBstWouldAddEndPuncttrue
\mciteSetBstMidEndSepPunct{\mcitedefaultmidpunct}
{\mcitedefaultendpunct}{\mcitedefaultseppunct}\relax
\EndOfBibitem
\bibitem{Abdallah:2003cw}
J.~Abdallah {\em et al.} ({DELPHI} collaboration){,}
  \href{http://dx.doi.org/10.1140/epjc/s2006-02494-9}{Eur.\ Phys.\ J.\ {\bf C46}{,}
  1}  (2006), \href{http://arxiv.org/abs/hep-ex/0603044}{{\tt
  arXiv:hep-ex/0603044 [hep-ex]}}\relax
\mciteBstWouldAddEndPuncttrue
\mciteSetBstMidEndSepPunct{\mcitedefaultmidpunct}
{\mcitedefaultendpunct}{\mcitedefaultseppunct}\relax
\EndOfBibitem
\bibitem{Ackerstaff:1997tx}
K.~Ackerstaff {\em et al.} ({OPAL} collaboration){,}
  \href{http://dx.doi.org/10.1007/s100520050197}{Eur.\ Phys.\ J.\ {\bf C4},  193}
   (1998), \href{http://arxiv.org/abs/hep-ex/9801029}{{\tt arXiv:hep-ex/9801029
  [hep-ex]}}\relax
\mciteBstWouldAddEndPuncttrue
\mciteSetBstMidEndSepPunct{\mcitedefaultmidpunct}
{\mcitedefaultendpunct}{\mcitedefaultseppunct}\relax
\EndOfBibitem
\bibitem{Barate:1999hi}
R.~Barate {\em et al.} ({ALEPH} collaboration){,}
  \href{http://dx.doi.org/10.1007/s100529900146}{Eur.\ Phys.\ J.\ {\bf C10},  1}
  (1999), \href{http://arxiv.org/abs/hep-ex/9903014}{{\tt arXiv:hep-ex/9903014
  [hep-ex]}}\relax
\mciteBstWouldAddEndPuncttrue
\mciteSetBstMidEndSepPunct{\mcitedefaultmidpunct}
{\mcitedefaultendpunct}{\mcitedefaultseppunct}\relax
\EndOfBibitem
\bibitem{Battle:1994by}
M.~Battle {\em et al.} ({CLEO} collaboration){,}
  \href{http://dx.doi.org/10.1103/PhysRevLett.73.1079}{Phys.\ Rev.\ Lett.\ {\bf
  73},  1079}  (1994), \href{http://arxiv.org/abs/hep-ph/9403329}{{\tt
  arXiv:hep-ph/9403329 [hep-ph]}}\relax
\mciteBstWouldAddEndPuncttrue
\mciteSetBstMidEndSepPunct{\mcitedefaultmidpunct}
{\mcitedefaultendpunct}{\mcitedefaultseppunct}\relax
\EndOfBibitem
\bibitem{Abreu:1994fi}
P.~Abreu {\em et al.} ({DELPHI} collaboration){,}
  \href{http://dx.doi.org/10.1016/0370-2693(94)90711-0}{Phys.\ Lett.\ {\bf B334}{,}
   435}  (1994)\relax
\mciteBstWouldAddEndPuncttrue
\mciteSetBstMidEndSepPunct{\mcitedefaultmidpunct}
{\mcitedefaultendpunct}{\mcitedefaultseppunct}\relax
\EndOfBibitem
\bibitem{Abbiendi:2000ee}
G.~Abbiendi {\em et al.} ({OPAL} collaboration){,}
  \href{http://dx.doi.org/10.1007/s100520100632}{Eur.\ Phys.\ J.\ {\bf C19}{,}
  653}  (2001), \href{http://arxiv.org/abs/hep-ex/0009017}{{\tt
  arXiv:hep-ex/0009017 [hep-ex]}}\relax
\mciteBstWouldAddEndPuncttrue
\mciteSetBstMidEndSepPunct{\mcitedefaultmidpunct}
{\mcitedefaultendpunct}{\mcitedefaultseppunct}\relax
\EndOfBibitem
\bibitem{Fujikawa:2008ma}
M.~Fujikawa {\em et al.} ({Belle} collaboration){,}
  \href{http://dx.doi.org/10.1103/PhysRevD.78.072006}{Phys.\ Rev.\ {\bf D78}{,}
  072006}  (2008), \href{http://arxiv.org/abs/0805.3773}{{\tt arXiv:0805.3773
  [hep-ex]}}\relax
\mciteBstWouldAddEndPuncttrue
\mciteSetBstMidEndSepPunct{\mcitedefaultmidpunct}
{\mcitedefaultendpunct}{\mcitedefaultseppunct}\relax
\EndOfBibitem
\bibitem{Artuso:1994ii}
M.~Artuso {\em et al.} ({CLEO} collaboration){,}
  \href{http://dx.doi.org/10.1103/PhysRevLett.72.3762}{Phys.\ Rev.\ Lett.\ {\bf
  72},  3762}  (1994), \href{http://arxiv.org/abs/hep-ph/9404310}{{\tt
  arXiv:hep-ph/9404310 [hep-ph]}}\relax
\mciteBstWouldAddEndPuncttrue
\mciteSetBstMidEndSepPunct{\mcitedefaultmidpunct}
{\mcitedefaultendpunct}{\mcitedefaultseppunct}\relax
\EndOfBibitem
\bibitem{Aubert:2007jh}
B.~Aubert {\em et al.} ({\babar} collaboration){,}
  \href{http://dx.doi.org/10.1103/PhysRevD.76.051104}{Phys.\ Rev.\ {\bf D76}{,}
  051104}  (2007), \href{http://arxiv.org/abs/0707.2922}{{\tt arXiv:0707.2922
  [hep-ex]}}\relax
\mciteBstWouldAddEndPuncttrue
\mciteSetBstMidEndSepPunct{\mcitedefaultmidpunct}
{\mcitedefaultendpunct}{\mcitedefaultseppunct}\relax
\EndOfBibitem
\bibitem{Abbiendi:2004xa}
G.~Abbiendi {\em et al.} ({OPAL} collaboration){,}
  \href{http://dx.doi.org/10.1140/epjc/s2004-01877-2}{Eur.\ Phys.\ J.\ {\bf C35}{,}
  437}  (2004), \href{http://arxiv.org/abs/hep-ex/0406007}{{\tt
  arXiv:hep-ex/0406007 [hep-ex]}}\relax
\mciteBstWouldAddEndPuncttrue
\mciteSetBstMidEndSepPunct{\mcitedefaultmidpunct}
{\mcitedefaultendpunct}{\mcitedefaultseppunct}\relax
\EndOfBibitem
\bibitem{Procario:1992hd}
M.~Procario {\em et al.} ({CLEO} collaboration){,}
  \href{http://dx.doi.org/10.1103/PhysRevLett.70.1207}{Phys.\ Rev.\ Lett.\ {\bf
  70},  1207}  (1993)\relax
\mciteBstWouldAddEndPuncttrue
\mciteSetBstMidEndSepPunct{\mcitedefaultmidpunct}
{\mcitedefaultendpunct}{\mcitedefaultseppunct}\relax
\EndOfBibitem
\bibitem{Akers:1994td}
R.~Akers {\em et al.} ({OPAL} collaboration){,}
  \href{http://dx.doi.org/10.1016/0370-2693(94)90645-9}{Phys.\ Lett.\ {\bf B339}{,}
   278}  (1994)\relax
\mciteBstWouldAddEndPuncttrue
\mciteSetBstMidEndSepPunct{\mcitedefaultmidpunct}
{\mcitedefaultendpunct}{\mcitedefaultseppunct}\relax
\EndOfBibitem
\bibitem{Coan:1996iu}
T.~Coan {\em et al.} ({CLEO} collaboration){,}
  \href{http://dx.doi.org/10.1103/PhysRevD.53.6037}{Phys.\ Rev.\ {\bf D53}{,}
  6037}  (1996)\relax
\mciteBstWouldAddEndPuncttrue
\mciteSetBstMidEndSepPunct{\mcitedefaultmidpunct}
{\mcitedefaultendpunct}{\mcitedefaultseppunct}\relax
\EndOfBibitem
\bibitem{Acciarri:1995kx}
M.~Acciarri {\em et al.} ({L3} collaboration){,}
  \href{http://dx.doi.org/10.1016/0370-2693(95)00509-J}{Phys.\ Lett.\ {\bf B352}{,}
   487}  (1995)\relax
\mciteBstWouldAddEndPuncttrue
\mciteSetBstMidEndSepPunct{\mcitedefaultmidpunct}
{\mcitedefaultendpunct}{\mcitedefaultseppunct}\relax
\EndOfBibitem
\bibitem{Abbiendi:1999pm}
G.~Abbiendi {\em et al.} ({OPAL} collaboration){,}
  \href{http://dx.doi.org/10.1007/s100520000317}{Eur.\ Phys.\ J.\ {\bf C13}{,}
  213}  (2000), \href{http://arxiv.org/abs/hep-ex/9911029}{{\tt
  arXiv:hep-ex/9911029 [hep-ex]}}\relax
\mciteBstWouldAddEndPuncttrue
\mciteSetBstMidEndSepPunct{\mcitedefaultmidpunct}
{\mcitedefaultendpunct}{\mcitedefaultseppunct}\relax
\EndOfBibitem
\bibitem{Barate:1999hj}
R.~Barate {\em et al.} ({ALEPH} collaboration){,}
  \href{http://dx.doi.org/10.1007/s100520050659}{Eur.\ Phys.\ J.\ {\bf C11}{,}
  599}  (1999), \href{http://arxiv.org/abs/hep-ex/9903015}{{\tt
  arXiv:hep-ex/9903015 [hep-ex]}}\relax
\mciteBstWouldAddEndPuncttrue
\mciteSetBstMidEndSepPunct{\mcitedefaultmidpunct}
{\mcitedefaultendpunct}{\mcitedefaultseppunct}\relax
\EndOfBibitem
\bibitem{Lees:2012de}
J.~P.\ Lees {\em et al.} ({\babar} collaboration){,}
  \href{http://dx.doi.org/10.1103/PhysRevD.86.092013}{Phys.\ Rev.\ {\bf D86}{,}
  092013}  (2012), \href{http://arxiv.org/abs/1208.0376}{{\tt arXiv:1208.0376
  [hep-ex]}}\relax
\mciteBstWouldAddEndPuncttrue
\mciteSetBstMidEndSepPunct{\mcitedefaultmidpunct}
{\mcitedefaultendpunct}{\mcitedefaultseppunct}\relax
\EndOfBibitem
\bibitem{Behrend:1989wc}
H.~Behrend {\em et al.} ({CELLO} collaboration){,}
  \href{http://dx.doi.org/10.1016/0370-2693(89)90741-7}{Phys.\ Lett.\ {\bf B222}{,}
   163}  (1989)\relax
\mciteBstWouldAddEndPuncttrue
\mciteSetBstMidEndSepPunct{\mcitedefaultmidpunct}
{\mcitedefaultendpunct}{\mcitedefaultseppunct}\relax
\EndOfBibitem
\bibitem{Adeva:1991qq}
B.~Adeva {\em et al.} ({L3} collaboration){,}
  \href{http://dx.doi.org/10.1016/0370-2693(91)90081-Z}{Phys.\ Lett.\ {\bf B265}{,}
   451}  (1991)\relax
\mciteBstWouldAddEndPuncttrue
\mciteSetBstMidEndSepPunct{\mcitedefaultmidpunct}
{\mcitedefaultendpunct}{\mcitedefaultseppunct}\relax
\EndOfBibitem
\bibitem{Aihara:1986mw}
H.~Aihara {\em et al.} ({TPC/Two Gamma} collaboration){,}
  \href{http://dx.doi.org/10.1103/PhysRevD.35.1553}{Phys.\ Rev.\ {\bf D35}{,}
  1553}  (1987)\relax
\mciteBstWouldAddEndPuncttrue
\mciteSetBstMidEndSepPunct{\mcitedefaultmidpunct}
{\mcitedefaultendpunct}{\mcitedefaultseppunct}\relax
\EndOfBibitem
\bibitem{Achard:2001pk}
P.~Achard {\em et al.} ({L3} collaboration){,}
  \href{http://dx.doi.org/10.1016/S0370-2693(01)01099-1}{Phys.\ Lett.\ {\bf
  B519},  189}  (2001), \href{http://arxiv.org/abs/hep-ex/0107055}{{\tt
  arXiv:hep-ex/0107055 [hep-ex]}}\relax
\mciteBstWouldAddEndPuncttrue
\mciteSetBstMidEndSepPunct{\mcitedefaultmidpunct}
{\mcitedefaultendpunct}{\mcitedefaultseppunct}\relax
\EndOfBibitem
\bibitem{Akers:1995ry}
R.~Akers {\em et al.} ({OPAL} collaboration){,}
  \href{http://dx.doi.org/10.1007/BF01565256}{Z.\ Phys.\ {\bf C68},  555}
  (1995)\relax
\mciteBstWouldAddEndPuncttrue
\mciteSetBstMidEndSepPunct{\mcitedefaultmidpunct}
{\mcitedefaultendpunct}{\mcitedefaultseppunct}\relax
\EndOfBibitem
\bibitem{Balest:1995kq}
R.~Balest {\em et al.} ({CLEO} collaboration){,}
  \href{http://dx.doi.org/10.1103/PhysRevLett.75.3809}{Phys.\ Rev.\ Lett.\ {\bf
  75},  3809}  (1995)\relax
\mciteBstWouldAddEndPuncttrue
\mciteSetBstMidEndSepPunct{\mcitedefaultmidpunct}
{\mcitedefaultendpunct}{\mcitedefaultseppunct}\relax
\EndOfBibitem
\bibitem{Aubert:2007mh}
B.~Aubert {\em et al.} ({\babar} collaboration){,}
  \href{http://dx.doi.org/10.1103/PhysRevLett.100.011801}{Phys.\ Rev.\ Lett.\ {\bf
  100},  011801}  (2008), \href{http://arxiv.org/abs/0707.2981}{{\tt
  arXiv:0707.2981 [hep-ex]}}\relax
\mciteBstWouldAddEndPuncttrue
\mciteSetBstMidEndSepPunct{\mcitedefaultmidpunct}
{\mcitedefaultendpunct}{\mcitedefaultseppunct}\relax
\EndOfBibitem
\bibitem{Lee:2010tc}
M.~Lee {\em et al.} ({Belle} collaboration){,}
  \href{http://dx.doi.org/10.1103/PhysRevD.81.113007}{Phys.\ Rev.\ {\bf D81}{,}
  113007}  (2010), \href{http://arxiv.org/abs/1001.0083}{{\tt arXiv:1001.0083
  [hep-ex]}}\relax
\mciteBstWouldAddEndPuncttrue
\mciteSetBstMidEndSepPunct{\mcitedefaultmidpunct}
{\mcitedefaultendpunct}{\mcitedefaultseppunct}\relax
\EndOfBibitem
\bibitem{Briere:2003fr}
R.~A.\ Briere {\em et al.} ({CLEO} collaboration){,}
  \href{http://dx.doi.org/10.1103/PhysRevLett.90.181802}{Phys.\ Rev.\ Lett.\ {\bf
  90},  181802}  (2003), \href{http://arxiv.org/abs/hep-ex/0302028}{{\tt
  arXiv:hep-ex/0302028 [hep-ex]}}\relax
\mciteBstWouldAddEndPuncttrue
\mciteSetBstMidEndSepPunct{\mcitedefaultmidpunct}
{\mcitedefaultendpunct}{\mcitedefaultseppunct}\relax
\EndOfBibitem
\bibitem{Edwards:1999fj}
K.~Edwards {\em et al.} ({CLEO} collaboration){,}
  \href{http://dx.doi.org/10.1103/PhysRevD.61.072003}{Phys.\ Rev.\ {\bf D61}{,}
  072003}  (2000), \href{http://arxiv.org/abs/hep-ex/9908024}{{\tt
  arXiv:hep-ex/9908024 [hep-ex]}}\relax
\mciteBstWouldAddEndPuncttrue
\mciteSetBstMidEndSepPunct{\mcitedefaultmidpunct}
{\mcitedefaultendpunct}{\mcitedefaultseppunct}\relax
\EndOfBibitem
\bibitem{Bortoletto:1993px}
D.~Bortoletto {\em et al.} ({CLEO} collaboration){,}
  \href{http://dx.doi.org/10.1103/PhysRevLett.71.1791}{Phys.\ Rev.\ Lett.\ {\bf
  71},  1791}  (1993)\relax
\mciteBstWouldAddEndPuncttrue
\mciteSetBstMidEndSepPunct{\mcitedefaultmidpunct}
{\mcitedefaultendpunct}{\mcitedefaultseppunct}\relax
\EndOfBibitem
\bibitem{Anastassov:2000xu}
A.~Anastassov {\em et al.} ({CLEO} collaboration){,}
  \href{http://dx.doi.org/10.1103/PhysRevLett.86.4467}{Phys.\ Rev.\ Lett.\ {\bf
  86},  4467}  (2001), \href{http://arxiv.org/abs/hep-ex/0010025}{{\tt
  arXiv:hep-ex/0010025 [hep-ex]}}\relax
\mciteBstWouldAddEndPuncttrue
\mciteSetBstMidEndSepPunct{\mcitedefaultmidpunct}
{\mcitedefaultendpunct}{\mcitedefaultseppunct}\relax
\EndOfBibitem
\bibitem{Richichi:1998bc}
S.~Richichi {\em et al.} ({CLEO} collaboration){,}
  \href{http://dx.doi.org/10.1103/PhysRevD.60.112002}{Phys.\ Rev.\ {\bf D60}{,}
  112002}  (1999), \href{http://arxiv.org/abs/hep-ex/9810026}{{\tt
  arXiv:hep-ex/9810026 [hep-ex]}}\relax
\mciteBstWouldAddEndPuncttrue
\mciteSetBstMidEndSepPunct{\mcitedefaultmidpunct}
{\mcitedefaultendpunct}{\mcitedefaultseppunct}\relax
\EndOfBibitem
\bibitem{Bauer:1993wn}
D.~A.\ Bauer {\em et al.} ({TPC/Two Gamma} collaboration){,}
  \href{http://dx.doi.org/10.1103/PhysRevD.50.13}{Phys.\ Rev.\ {\bf D50},  13}
  (1994)\relax
\mciteBstWouldAddEndPuncttrue
\mciteSetBstMidEndSepPunct{\mcitedefaultmidpunct}
{\mcitedefaultendpunct}{\mcitedefaultseppunct}\relax
\EndOfBibitem
\bibitem{Barate:1997ma}
R.~Barate {\em et al.} ({ALEPH} collaboration){,}
  \href{http://dx.doi.org/10.1007/s100520050062}{Eur.\ Phys.\ J.\ {\bf C1},  65}
  (1998)\relax
\mciteBstWouldAddEndPuncttrue
\mciteSetBstMidEndSepPunct{\mcitedefaultmidpunct}
{\mcitedefaultendpunct}{\mcitedefaultseppunct}\relax
\EndOfBibitem
\bibitem{Arms:2005qg}
K.~E.\ Arms {\em et al.} ({CLEO} collaboration){,}
  \href{http://dx.doi.org/10.1103/PhysRevLett.94.241802}{Phys.\ Rev.\ Lett.\ {\bf
  94},  241802}  (2005), \href{http://arxiv.org/abs/hep-ex/0501042}{{\tt
  arXiv:hep-ex/0501042 [hep-ex]}}\relax
\mciteBstWouldAddEndPuncttrue
\mciteSetBstMidEndSepPunct{\mcitedefaultmidpunct}
{\mcitedefaultendpunct}{\mcitedefaultseppunct}\relax
\EndOfBibitem
\bibitem{Abbiendi:1999cq}
G.~Abbiendi {\em et al.} ({OPAL} collaboration){,}
  \href{http://dx.doi.org/10.1007/s100520000272}{Eur.\ Phys.\ J.\ {\bf C13}{,}
  197}  (2000), \href{http://arxiv.org/abs/hep-ex/9908013}{{\tt
  arXiv:hep-ex/9908013 [hep-ex]}}\relax
\mciteBstWouldAddEndPuncttrue
\mciteSetBstMidEndSepPunct{\mcitedefaultmidpunct}
{\mcitedefaultendpunct}{\mcitedefaultseppunct}\relax
\EndOfBibitem
\bibitem{Gibaut:1994ik}
D.~Gibaut {\em et al.} ({CLEO} collaboration){,}
  \href{http://dx.doi.org/10.1103/PhysRevLett.73.934}{Phys.\ Rev.\ Lett.\ {\bf
  73},  934}  (1994)\relax
\mciteBstWouldAddEndPuncttrue
\mciteSetBstMidEndSepPunct{\mcitedefaultmidpunct}
{\mcitedefaultendpunct}{\mcitedefaultseppunct}\relax
\EndOfBibitem
\bibitem{Bylsma:1986zy}
B.~Bylsma {\em et al.}, \href{http://dx.doi.org/10.1103/PhysRevD.35.2269}{Phys.\
  Rev.\ {\bf D35},  2269}  (1987)\relax
\mciteBstWouldAddEndPuncttrue
\mciteSetBstMidEndSepPunct{\mcitedefaultmidpunct}
{\mcitedefaultendpunct}{\mcitedefaultseppunct}\relax
\EndOfBibitem
\bibitem{Albrecht:1987zf}
H.~Albrecht {\em et al.} ({ARGUS} collaboration){,}
  \href{http://dx.doi.org/10.1016/0370-2693(88)90870-2}{Phys.\ Lett.\ {\bf B202}{,}
   149}  (1988)\relax
\mciteBstWouldAddEndPuncttrue
\mciteSetBstMidEndSepPunct{\mcitedefaultmidpunct}
{\mcitedefaultendpunct}{\mcitedefaultseppunct}\relax
\EndOfBibitem
\bibitem{Ackerstaff:1998ia}
K.~Ackerstaff {\em et al.} ({OPAL} collaboration){,}
  \href{http://dx.doi.org/10.1007/s100529901057}{Eur.\ Phys.\ J.\ {\bf C8},  183}
   (1999), \href{http://arxiv.org/abs/hep-ex/9808011}{{\tt arXiv:hep-ex/9808011
  [hep-ex]}}\relax
\mciteBstWouldAddEndPuncttrue
\mciteSetBstMidEndSepPunct{\mcitedefaultmidpunct}
{\mcitedefaultendpunct}{\mcitedefaultseppunct}\relax
\EndOfBibitem
\bibitem{Buskulic:1996qs}
D.~Buskulic {\em et al.} ({ALEPH} collaboration){,}
  \href{http://dx.doi.org/10.1007/s002880050387}{Z.\ Phys.\ {\bf C74},  263}
  (1997)\relax
\mciteBstWouldAddEndPuncttrue
\mciteSetBstMidEndSepPunct{\mcitedefaultmidpunct}
{\mcitedefaultendpunct}{\mcitedefaultseppunct}\relax
\EndOfBibitem
\bibitem{Inami:2008ar}
K.~Inami {\em et al.} ({Belle} collaboration){,}
  \href{http://dx.doi.org/10.1016/j.physletb.2009.01.047}{Phys.\ Lett.\ {\bf
  B672},  209}  (2009), \href{http://arxiv.org/abs/0811.0088}{{\tt
  arXiv:0811.0088 [hep-ex]}}\relax
\mciteBstWouldAddEndPuncttrue
\mciteSetBstMidEndSepPunct{\mcitedefaultmidpunct}
{\mcitedefaultendpunct}{\mcitedefaultseppunct}\relax
\EndOfBibitem
\bibitem{Artuso:1992qu}
M.~Artuso {\em et al.} ({CLEO} collaboration){,}
  \href{http://dx.doi.org/10.1103/PhysRevLett.69.3278}{Phys.\ Rev.\ Lett.\ {\bf
  69},  3278}  (1992)\relax
\mciteBstWouldAddEndPuncttrue
\mciteSetBstMidEndSepPunct{\mcitedefaultmidpunct}
{\mcitedefaultendpunct}{\mcitedefaultseppunct}\relax
\EndOfBibitem
\bibitem{delAmoSanchez:2010pc}
P.~del Amo~Sanchez {\em et al.} ({\babar} collaboration){,}
  \href{http://dx.doi.org/10.1103/PhysRevD.83.032002}{Phys.\ Rev.\ {\bf D83}{,}
  032002}  (2011), \href{http://arxiv.org/abs/1011.3917}{{\tt arXiv:1011.3917
  [hep-ex]}}\relax
\mciteBstWouldAddEndPuncttrue
\mciteSetBstMidEndSepPunct{\mcitedefaultmidpunct}
{\mcitedefaultendpunct}{\mcitedefaultseppunct}\relax
\EndOfBibitem
\bibitem{Bartelt:1996iv}
J.~E.\ Bartelt {\em et al.} ({CLEO} collaboration){,}
  \href{http://dx.doi.org/10.1103/PhysRevLett.76.4119}{Phys.\ Rev.\ Lett.\ {\bf
  76},  4119}  (1996)\relax
\mciteBstWouldAddEndPuncttrue
\mciteSetBstMidEndSepPunct{\mcitedefaultmidpunct}
{\mcitedefaultendpunct}{\mcitedefaultseppunct}\relax
\EndOfBibitem
\bibitem{Bishai:1998gf}
M.~Bishai {\em et al.} ({CLEO} collaboration){,}
  \href{http://dx.doi.org/10.1103/PhysRevLett.82.281}{Phys.\ Rev.\ Lett.\ {\bf
  82},  281}  (1999), \href{http://arxiv.org/abs/hep-ex/9809012}{{\tt
  arXiv:hep-ex/9809012 [hep-ex]}}\relax
\mciteBstWouldAddEndPuncttrue
\mciteSetBstMidEndSepPunct{\mcitedefaultmidpunct}
{\mcitedefaultendpunct}{\mcitedefaultseppunct}\relax
\EndOfBibitem
\bibitem{Baringer:1987tr}
P.~S.\ Baringer {\em et al.} ({CLEO} collaboration){,}
  \href{http://dx.doi.org/10.1103/PhysRevLett.59.1993}{Phys.\ Rev.\ Lett.\ {\bf
  59},  1993}  (1987)\relax
\mciteBstWouldAddEndPuncttrue
\mciteSetBstMidEndSepPunct{\mcitedefaultmidpunct}
{\mcitedefaultendpunct}{\mcitedefaultseppunct}\relax
\EndOfBibitem
\bibitem{Buskulic:1995ty}
D.~Buskulic {\em et al.} ({ALEPH} collaboration){,}
  \href{http://dx.doi.org/10.1007/s002880050134}{Z.\ Phys.\ {\bf C70},  579}
  (1996)\relax
\mciteBstWouldAddEndPuncttrue
\mciteSetBstMidEndSepPunct{\mcitedefaultmidpunct}
{\mcitedefaultendpunct}{\mcitedefaultseppunct}\relax
\EndOfBibitem
\bibitem{Lees:2012ks}
J.~P.\ Lees {\em et al.} ({\babar} collaboration){,}
  \href{http://dx.doi.org/10.1103/PhysRevD.86.092010}{Phys.\ Rev.\ {\bf D86}{,}
  092010}  (2012), \href{http://arxiv.org/abs/1209.2734}{{\tt arXiv:1209.2734
  [hep-ex]}}\relax
\mciteBstWouldAddEndPuncttrue
\mciteSetBstMidEndSepPunct{\mcitedefaultmidpunct}
{\mcitedefaultendpunct}{\mcitedefaultseppunct}\relax
\EndOfBibitem
\bibitem{Belous:2013dba}
K.~Belous {\em et al.} ({Belle} collaboration){,}
  \href{http://dx.doi.org/10.1103/PhysRevLett.112.031801}{Phys.\ Rev.\ Lett.\ {\bf
  112},  031801}  (2014), \href{http://arxiv.org/abs/1310.8503}{{\tt
  arXiv:1310.8503 [hep-ex]}}\relax
\mciteBstWouldAddEndPuncttrue
\mciteSetBstMidEndSepPunct{\mcitedefaultmidpunct}
{\mcitedefaultendpunct}{\mcitedefaultseppunct}\relax
\EndOfBibitem
\bibitem{Marciano:1988vm}
W.~Marciano and A.~Sirlin{,}
  \href{http://dx.doi.org/10.1103/PhysRevLett.61.1815}{Phys.\ Rev.\ Lett.\ {\bf
  61},  1815}  (1988)\relax
\mciteBstWouldAddEndPuncttrue
\mciteSetBstMidEndSepPunct{\mcitedefaultmidpunct}
{\mcitedefaultendpunct}{\mcitedefaultseppunct}\relax
\EndOfBibitem
\bibitem{Pich:2013lsa}
A.~Pich, \href{http://dx.doi.org/10.1016/j.ppnp.2013.11.002}{Prog.\ Part.\ Nucl.\
  Phys.\ {\bf 75},  41}  (2014), \href{http://arxiv.org/abs/1310.7922}{{\tt
  arXiv:1310.7922 [hep-ph]}}\relax
\mciteBstWouldAddEndPuncttrue
\mciteSetBstMidEndSepPunct{\mcitedefaultmidpunct}
{\mcitedefaultendpunct}{\mcitedefaultseppunct}\relax
\EndOfBibitem
\bibitem{Ferroglia:2013dga}
A.~Ferroglia, C.~Greub, A.~Sirlin, and Z.~Zhang{,}
  \href{http://dx.doi.org/10.1103/PhysRevD.88.033012}{Phys.\ Rev.\ {\bf D88}{,}
  033012}  (2013), \href{http://arxiv.org/abs/1307.6900}{{\tt arXiv:1307.6900
  [hep-ph]}}\relax
\mciteBstWouldAddEndPuncttrue
\mciteSetBstMidEndSepPunct{\mcitedefaultmidpunct}
{\mcitedefaultendpunct}{\mcitedefaultseppunct}\relax
\EndOfBibitem
\bibitem{Fael:2013pja}
M.~Fael, L.~Mercolli, and M.~Passera{,}
  \href{http://dx.doi.org/10.1103/PhysRevD.88.093011}{Phys.\ Rev.\ {\bf D88}{,}
  093011}  (2013), \href{http://arxiv.org/abs/1310.1081}{{\tt arXiv:1310.1081
  [hep-ph]}}\relax
\mciteBstWouldAddEndPuncttrue
\mciteSetBstMidEndSepPunct{\mcitedefaultmidpunct}
{\mcitedefaultendpunct}{\mcitedefaultseppunct}\relax
\EndOfBibitem
\bibitem{Marciano:1993sh}
W.~J.\ Marciano and A.~Sirlin{,}
  \href{http://dx.doi.org/10.1103/PhysRevLett.71.3629}{Phys.\ Rev.\ Lett.\ {\bf
  71},  3629}  (1993)\relax
\mciteBstWouldAddEndPuncttrue
\mciteSetBstMidEndSepPunct{\mcitedefaultmidpunct}
{\mcitedefaultendpunct}{\mcitedefaultseppunct}\relax
\EndOfBibitem
\bibitem{Decker:1994dd}
R.~Decker and M.~Finkemeier{,}
  \href{http://dx.doi.org/10.1016/0370-2693(94)90611-4}{Phys.\ Lett.\ {\bf B334}{,}
   199}  (1994)\relax
\mciteBstWouldAddEndPuncttrue
\mciteSetBstMidEndSepPunct{\mcitedefaultmidpunct}
{\mcitedefaultendpunct}{\mcitedefaultseppunct}\relax
\EndOfBibitem
\bibitem{Decker:1994ea}
R.~Decker and M.~Finkemeier{,}
  \href{http://dx.doi.org/10.1016/0550-3213(95)00597-L}{Nucl.\ Phys.\ {\bf B438}{,}
   17}  (1995), \href{http://arxiv.org/abs/hep-ph/9403385}{{\tt
  arXiv:hep-ph/9403385}}\relax
\mciteBstWouldAddEndPuncttrue
\mciteSetBstMidEndSepPunct{\mcitedefaultmidpunct}
{\mcitedefaultendpunct}{\mcitedefaultseppunct}\relax
\EndOfBibitem
\bibitem{Decker:1994kw}
R.~Decker and M.~Finkemeier{,}
  \href{http://dx.doi.org/10.1016/0920-5632(95)00170-E}{Nucl.\ Phys.\ Proc.\
  Suppl.\ {\bf 40},  453}  (1995){,}
  \href{http://arxiv.org/abs/hep-ph/9411316}{{\tt arXiv:hep-ph/9411316
  [hep-ph]}}\relax
\mciteBstWouldAddEndPuncttrue
\mciteSetBstMidEndSepPunct{\mcitedefaultmidpunct}
{\mcitedefaultendpunct}{\mcitedefaultseppunct}\relax
\EndOfBibitem
\bibitem{Davier:2005xq}
M.~Davier, A.~Hocker, and Z.~Zhang{,}
  \href{http://dx.doi.org/10.1103/RevModPhys.78.1043}{Rev.\ Mod.\ Phys.\ {\bf 78}{,}
   1043}  (2006), \href{http://arxiv.org/abs/hep-ph/0507078}{{\tt
  arXiv:hep-ph/0507078 [hep-ph]}}\relax
\mciteBstWouldAddEndPuncttrue
\mciteSetBstMidEndSepPunct{\mcitedefaultmidpunct}
{\mcitedefaultendpunct}{\mcitedefaultseppunct}\relax
\EndOfBibitem
\bibitem{Antonelli:2010yf}
M.~Antonelli {\em et al.} ({FlaviaNet working group on kaon decays}){,}
  \href{http://dx.doi.org/10.1140/epjc/s10052-010-1406-3}{Eur.\ Phys.\ J.\ {\bf
  C69},  399}  (2010), \href{http://arxiv.org/abs/1005.2323}{{\tt
  arXiv:1005.2323 [hep-ph]}}\relax
\mciteBstWouldAddEndPuncttrue
\mciteSetBstMidEndSepPunct{\mcitedefaultmidpunct}
{\mcitedefaultendpunct}{\mcitedefaultseppunct}\relax
\EndOfBibitem
\bibitem{Gamiz:2002nu}
E.~Gamiz, M.~Jamin, A.~Pich, J.~Prades, and F.~Schwab{,}
  \href{http://dx.doi.org/10.1088/1126-6708/2003/01/060}{JHEP {\bf 01},  060}
  (2003), \href{http://arxiv.org/abs/hep-ph/0212230}{{\tt arXiv:hep-ph/0212230
  [hep-ph]}}\relax
\mciteBstWouldAddEndPuncttrue
\mciteSetBstMidEndSepPunct{\mcitedefaultmidpunct}
{\mcitedefaultendpunct}{\mcitedefaultseppunct}\relax
\EndOfBibitem
\bibitem{Gamiz:2004ar}
E.~Gamiz, M.~Jamin, A.~Pich, J.~Prades, and F.~Schwab{,}
  \href{http://dx.doi.org/10.1103/PhysRevLett.94.011803}{Phys.\ Rev.\ Lett.\ {\bf
  94},  011803}  (2005), \href{http://arxiv.org/abs/hep-ph/0408044}{{\tt
  arXiv:hep-ph/0408044 [hep-ph]}}\relax
\mciteBstWouldAddEndPuncttrue
\mciteSetBstMidEndSepPunct{\mcitedefaultmidpunct}
{\mcitedefaultendpunct}{\mcitedefaultseppunct}\relax
\EndOfBibitem
\bibitem{Gamiz:2006xx}
E.~Gamiz, M.~Jamin, A.~Pich, J.~Prades, and F.~Schwab{,}
  \href{http://dx.doi.org/10.1016/j.nuclphysbps.2007.02.053}{Nucl.\ Phys.\ Proc.\
  Suppl.\ {\bf 169},  85}  (2007){,}
  \href{http://arxiv.org/abs/hep-ph/0612154}{{\tt arXiv:hep-ph/0612154}}\relax
\mciteBstWouldAddEndPuncttrue
\mciteSetBstMidEndSepPunct{\mcitedefaultmidpunct}
{\mcitedefaultendpunct}{\mcitedefaultseppunct}\relax
\EndOfBibitem
\bibitem{Gamiz:2007qs}
E.~Gamiz, M.~Jamin, A.~Pich, J.~Prades, and F.~Schwab, PoS {\bf KAON},  008
  (2008), \href{http://arxiv.org/abs/0709.0282}{{\tt arXiv:0709.0282
  [hep-ph]}}\relax
\mciteBstWouldAddEndPuncttrue
\mciteSetBstMidEndSepPunct{\mcitedefaultmidpunct}
{\mcitedefaultendpunct}{\mcitedefaultseppunct}\relax
\EndOfBibitem
\bibitem{Maltman:2010hb}
K.~Maltman, \href{http://dx.doi.org/10.1016/j.nuclphysbps.2011.06.025}{Nucl.\
  Phys.\ Proc.\ Suppl.\ {\bf 218},  146}  (2011){,}
  \href{http://arxiv.org/abs/1011.6391}{{\tt arXiv:1011.6391 [hep-ph]}}\relax
\mciteBstWouldAddEndPuncttrue
\mciteSetBstMidEndSepPunct{\mcitedefaultmidpunct}
{\mcitedefaultendpunct}{\mcitedefaultseppunct}\relax
\EndOfBibitem
\bibitem{Hardy:2014qxa}
J.~C.\ Hardy and I.~S.\ Towner{,}
  \href{http://dx.doi.org/10.1103/PhysRevC.91.025501}{Phys.\ Rev.\ {\bf C91}{,}
  025501}  (2015), \href{http://arxiv.org/abs/1411.5987}{{\tt arXiv:1411.5987
  [nucl-ex]}}\relax
\mciteBstWouldAddEndPuncttrue
\mciteSetBstMidEndSepPunct{\mcitedefaultmidpunct}
{\mcitedefaultendpunct}{\mcitedefaultseppunct}\relax
\EndOfBibitem
\bibitem{Cirigliano:2011tm}
V.~Cirigliano and H.~Neufeld{,}
  \href{http://dx.doi.org/10.1016/j.physletb.2011.04.038}{Phys.\ Lett.\ {\bf
  B700},  7}  (2011), \href{http://arxiv.org/abs/1102.0563}{{\tt
  arXiv:1102.0563 [hep-ph]}}\relax
\mciteBstWouldAddEndPuncttrue
\mciteSetBstMidEndSepPunct{\mcitedefaultmidpunct}
{\mcitedefaultendpunct}{\mcitedefaultseppunct}\relax
\EndOfBibitem
\bibitem{Marciano:2004uf}
W.~J.\ Marciano, \href{http://dx.doi.org/10.1103/PhysRevLett.93.231803}{Phys.\
  Rev.\ Lett.\ {\bf 93},  231803}  (2004){,}
  \href{http://arxiv.org/abs/hep-ph/0402299}{{\tt arXiv:hep-ph/0402299}}\relax
\mciteBstWouldAddEndPuncttrue
\mciteSetBstMidEndSepPunct{\mcitedefaultmidpunct}
{\mcitedefaultendpunct}{\mcitedefaultseppunct}\relax
\EndOfBibitem
\bibitem{Jamin:2008qg}
M.~Jamin, A.~Pich, and J.~Portoles{,}
  \href{http://dx.doi.org/10.1016/j.physletb.2008.04.049}{Phys.\ Lett.\ {\bf
  B664},  78}  (2008), \href{http://arxiv.org/abs/0803.1786}{{\tt
  arXiv:0803.1786 [hep-ph]}}\relax
\mciteBstWouldAddEndPuncttrue
\mciteSetBstMidEndSepPunct{\mcitedefaultmidpunct}
{\mcitedefaultendpunct}{\mcitedefaultseppunct}\relax
\EndOfBibitem
\bibitem{Antonelli:2013usa}
M.~Antonelli, V.~Cirigliano, A.~Lusiani, and E.~Passemar{,}
  \href{http://dx.doi.org/10.1007/JHEP10(2013)070}{JHEP {\bf 10},  070}
  (2013), \href{http://arxiv.org/abs/1304.8134}{{\tt arXiv:1304.8134
  [hep-ph]}}\relax
\mciteBstWouldAddEndPuncttrue
\mciteSetBstMidEndSepPunct{\mcitedefaultmidpunct}
{\mcitedefaultendpunct}{\mcitedefaultseppunct}\relax
\EndOfBibitem
\bibitem{Maltman:2015xwa}
K.~Maltman, R.~J.\ Hudspith, R.~Lewis, C.~E.\ Wolfe, and J.~Zanotti  (2015){,}
  \href{http://arxiv.org/abs/1510.06954}{{\tt arXiv:1510.06954 [hep-ph]}}\relax
\mciteBstWouldAddEndPuncttrue
\mciteSetBstMidEndSepPunct{\mcitedefaultmidpunct}
{\mcitedefaultendpunct}{\mcitedefaultseppunct}\relax
\EndOfBibitem
\bibitem{Hudspith:2017vew}
R.~J.\ Hudspith, R.~Lewis, K.~Maltman, and J.~Zanotti  (2017){,}
  \href{http://arxiv.org/abs/1702.01767}{{\tt arXiv:1702.01767 [hep-ph]}}\relax
\mciteBstWouldAddEndPuncttrue
\mciteSetBstMidEndSepPunct{\mcitedefaultmidpunct}
{\mcitedefaultendpunct}{\mcitedefaultseppunct}\relax
\EndOfBibitem
\bibitem{Aubert:2009ag}
B.~Aubert {\em et al.} ({\babar} collaboration){,}
  \href{http://dx.doi.org/10.1103/PhysRevLett.104.021802}{Phys.\ Rev.\ Lett.\ {\bf
  104},  021802}  (2010), \href{http://arxiv.org/abs/0908.2381}{{\tt
  arXiv:0908.2381 [hep-ex]}}\relax
\mciteBstWouldAddEndPuncttrue
\mciteSetBstMidEndSepPunct{\mcitedefaultmidpunct}
{\mcitedefaultendpunct}{\mcitedefaultseppunct}\relax
\EndOfBibitem
\bibitem{Hayasaka:2007vc}
K.~Hayasaka {\em et al.} ({Belle} collaboration){,}
  \href{http://dx.doi.org/10.1016/j.physletb.2008.06.056}{Phys.\ Lett.\ {\bf
  B666},  16}  (2008), \href{http://arxiv.org/abs/0705.0650}{{\tt
  arXiv:0705.0650 [hep-ex]}}\relax
\mciteBstWouldAddEndPuncttrue
\mciteSetBstMidEndSepPunct{\mcitedefaultmidpunct}
{\mcitedefaultendpunct}{\mcitedefaultseppunct}\relax
\EndOfBibitem
\bibitem{Aubert:2006cz}
B.~Aubert {\em et al.} ({\babar} collaboration){,}
  \href{http://dx.doi.org/10.1103/PhysRevLett.98.061803}{Phys.\ Rev.\ Lett.\ {\bf
  98},  061803}  (2007), \href{http://arxiv.org/abs/hep-ex/0610067}{{\tt
  arXiv:hep-ex/0610067 [hep-ex]}}\relax
\mciteBstWouldAddEndPuncttrue
\mciteSetBstMidEndSepPunct{\mcitedefaultmidpunct}
{\mcitedefaultendpunct}{\mcitedefaultseppunct}\relax
\EndOfBibitem
\bibitem{Miyazaki:2007jp}
Y.~Miyazaki {\em et al.} ({Belle} collaboration){,}
  \href{http://dx.doi.org/10.1016/j.physletb.2007.03.027}{Phys.\ Lett.\ {\bf
  B648},  341}  (2007), \href{http://arxiv.org/abs/hep-ex/0703009}{{\tt
  arXiv:hep-ex/0703009 [hep-ex]}}\relax
\mciteBstWouldAddEndPuncttrue
\mciteSetBstMidEndSepPunct{\mcitedefaultmidpunct}
{\mcitedefaultendpunct}{\mcitedefaultseppunct}\relax
\EndOfBibitem
\bibitem{Aubert:2009ys}
B.~Aubert {\em et al.} ({\babar} collaboration){,}
  \href{http://dx.doi.org/10.1103/PhysRevD.79.012004}{Phys.\ Rev.\ {\bf D79}{,}
  012004}  (2009), \href{http://arxiv.org/abs/0812.3804}{{\tt arXiv:0812.3804
  [hep-ex]}}\relax
\mciteBstWouldAddEndPuncttrue
\mciteSetBstMidEndSepPunct{\mcitedefaultmidpunct}
{\mcitedefaultendpunct}{\mcitedefaultseppunct}\relax
\EndOfBibitem
\bibitem{Miyazaki:2010qb}
Y.~Miyazaki {\em et al.} ({Belle} collaboration){,}
  \href{http://dx.doi.org/10.1016/j.physletb.2010.07.012}{Phys.\ Lett.\ {\bf
  B692},  4}  (2010), \href{http://arxiv.org/abs/1003.1183}{{\tt
  arXiv:1003.1183 [hep-ex]}}\relax
\mciteBstWouldAddEndPuncttrue
\mciteSetBstMidEndSepPunct{\mcitedefaultmidpunct}
{\mcitedefaultendpunct}{\mcitedefaultseppunct}\relax
\EndOfBibitem
\bibitem{Aubert:2009ap}
B.~Aubert {\em et al.} ({\babar} collaboration){,}
  \href{http://dx.doi.org/10.1103/PhysRevLett.103.021801}{Phys.\ Rev.\ Lett.\ {\bf
  103},  021801}  (2009), \href{http://arxiv.org/abs/0904.0339}{{\tt
  arXiv:0904.0339 [hep-ex]}}\relax
\mciteBstWouldAddEndPuncttrue
\mciteSetBstMidEndSepPunct{\mcitedefaultmidpunct}
{\mcitedefaultendpunct}{\mcitedefaultseppunct}\relax
\EndOfBibitem
\bibitem{Miyazaki:2011xe}
Y.~Miyazaki ({Belle} collaboration){,}
  \href{http://dx.doi.org/10.1016/j.physletb.2011.04.011}{Phys.\ Lett.\ {\bf
  B699},  251}  (2011), \href{http://arxiv.org/abs/1101.0755}{{\tt
  arXiv:1101.0755 [hep-ex]}}\relax
\mciteBstWouldAddEndPuncttrue
\mciteSetBstMidEndSepPunct{\mcitedefaultmidpunct}
{\mcitedefaultendpunct}{\mcitedefaultseppunct}\relax
\EndOfBibitem
\bibitem{Aubert:2007kx}
B.~Aubert {\em et al.} ({\babar} collaboration){,}
  \href{http://dx.doi.org/10.1103/PhysRevLett.100.071802}{Phys.\ Rev.\ Lett.\ {\bf
  100},  071802}  (2008), \href{http://arxiv.org/abs/0711.0980}{{\tt
  arXiv:0711.0980 [hep-ex]}}\relax
\mciteBstWouldAddEndPuncttrue
\mciteSetBstMidEndSepPunct{\mcitedefaultmidpunct}
{\mcitedefaultendpunct}{\mcitedefaultseppunct}\relax
\EndOfBibitem
\bibitem{Miyazaki:2008mw}
Y.~Miyazaki {\em et al.} ({Belle} collaboration){,}
  \href{http://dx.doi.org/10.1016/j.physletb.2009.01.058}{Phys.\ Lett.\ {\bf
  B672},  317}  (2009), \href{http://arxiv.org/abs/0810.3519}{{\tt
  arXiv:0810.3519 [hep-ex]}}\relax
\mciteBstWouldAddEndPuncttrue
\mciteSetBstMidEndSepPunct{\mcitedefaultmidpunct}
{\mcitedefaultendpunct}{\mcitedefaultseppunct}\relax
\EndOfBibitem
\bibitem{Lees:2010ez}
J.~P.\ Lees {\em et al.} ({\babar} collaboration){,}
  \href{http://dx.doi.org/10.1103/PhysRevD.81.111101}{Phys.\ Rev.\ {\bf D81}{,}
  111101}  (2010), \href{http://arxiv.org/abs/1002.4550}{{\tt arXiv:1002.4550
  [hep-ex]}}\relax
\mciteBstWouldAddEndPuncttrue
\mciteSetBstMidEndSepPunct{\mcitedefaultmidpunct}
{\mcitedefaultendpunct}{\mcitedefaultseppunct}\relax
\EndOfBibitem
\bibitem{Hayasaka:2010np}
K.~Hayasaka {\em et al.} ({Belle} collaboration){,}
  \href{http://dx.doi.org/10.1016/j.physletb.2010.03.037}{Phys.\ Lett.\ {\bf
  B687},  139}  (2010), \href{http://arxiv.org/abs/1001.3221}{{\tt
  arXiv:1001.3221 [hep-ex]}}\relax
\mciteBstWouldAddEndPuncttrue
\mciteSetBstMidEndSepPunct{\mcitedefaultmidpunct}
{\mcitedefaultendpunct}{\mcitedefaultseppunct}\relax
\EndOfBibitem
\bibitem{Aad:2016wce}
G.~Aad {\em et al.} ({ATLAS} collaboration){,}
  \href{http://dx.doi.org/10.1140/epjc/s10052-016-4041-9}{Eur.\ Phys.\ J.\ {\bf
  C76},  232}  (2016), \href{http://arxiv.org/abs/1601.03567}{{\tt
  arXiv:1601.03567 [hep-ex]}}\relax
\mciteBstWouldAddEndPuncttrue
\mciteSetBstMidEndSepPunct{\mcitedefaultmidpunct}
{\mcitedefaultendpunct}{\mcitedefaultseppunct}\relax
\EndOfBibitem
\bibitem{Aaij:2014azz}
R.~Aaij {\em et al.} ({LHCb} collaboration){,}
  \href{http://dx.doi.org/10.1007/JHEP02(2015)121}{JHEP {\bf 02},  121}
  (2015), \href{http://arxiv.org/abs/1409.8548}{{\tt arXiv:1409.8548
  [hep-ex]}}\relax
\mciteBstWouldAddEndPuncttrue
\mciteSetBstMidEndSepPunct{\mcitedefaultmidpunct}
{\mcitedefaultendpunct}{\mcitedefaultseppunct}\relax
\EndOfBibitem
\bibitem{Aubert:2005tp}
B.~Aubert {\em et al.} ({\babar} collaboration){,}
  \href{http://dx.doi.org/10.1103/PhysRevLett.95.191801}{Phys.\ Rev.\ Lett.\ {\bf
  95},  191801}  (2005), \href{http://arxiv.org/abs/hep-ex/0506066}{{\tt
  arXiv:hep-ex/0506066 [hep-ex]}}\relax
\mciteBstWouldAddEndPuncttrue
\mciteSetBstMidEndSepPunct{\mcitedefaultmidpunct}
{\mcitedefaultendpunct}{\mcitedefaultseppunct}\relax
\EndOfBibitem
\bibitem{Miyazaki:2012mx}
Y.~Miyazaki {\em et al.} ({Belle} collaboration){,}
  \href{http://dx.doi.org/10.1016/j.physletb.2013.01.032}{Phys.\ Lett.\ {\bf
  B719},  346}  (2013), \href{http://arxiv.org/abs/1206.5595}{{\tt
  arXiv:1206.5595 [hep-ex]}}\relax
\mciteBstWouldAddEndPuncttrue
\mciteSetBstMidEndSepPunct{\mcitedefaultmidpunct}
{\mcitedefaultendpunct}{\mcitedefaultseppunct}\relax
\EndOfBibitem
\bibitem{Miyazaki:2005ng}
Y.~Miyazaki {\em et al.} ({Belle} collaboration){,}
  \href{http://dx.doi.org/10.1016/j.physletb.2005.10.024}{Phys.\ Lett.\ {\bf
  B632},  51}  (2006), \href{http://arxiv.org/abs/hep-ex/0508044}{{\tt
  arXiv:hep-ex/0508044 [hep-ex]}}\relax
\mciteBstWouldAddEndPuncttrue
\mciteSetBstMidEndSepPunct{\mcitedefaultmidpunct}
{\mcitedefaultendpunct}{\mcitedefaultseppunct}\relax
\EndOfBibitem
\bibitem{Aaij:2013fia}
R.~Aaij {\em et al.} ({LHCb} collaboration){,}
  \href{http://dx.doi.org/10.1016/j.physletb.2013.05.063}{Phys.\ Lett.\ {\bf
  B724},  36}  (2013), \href{http://arxiv.org/abs/1304.4518}{{\tt
  arXiv:1304.4518 [hep-ex]}}\relax
\mciteBstWouldAddEndPuncttrue
\mciteSetBstMidEndSepPunct{\mcitedefaultmidpunct}
{\mcitedefaultendpunct}{\mcitedefaultseppunct}\relax
\EndOfBibitem
\bibitem{Read:2002hq}
A.~L.\ Read, \href{http://dx.doi.org/10.1088/0954-3899/28/10/313}{J.\ Phys.\ {\bf
  G28},  2693}  (2002)\relax
\mciteBstWouldAddEndPuncttrue
\mciteSetBstMidEndSepPunct{\mcitedefaultmidpunct}
{\mcitedefaultendpunct}{\mcitedefaultseppunct}\relax
\EndOfBibitem
\bibitem{Banerjee:2007is}
S.~Banerjee, B.~Pietrzyk, J.~M.\ Roney, and Z.~Was{,}
  \href{http://dx.doi.org/10.1103/PhysRevD.77.054012}{Phys.\ Rev.\ {\bf D77}{,}
  054012}  (2008), \href{http://arxiv.org/abs/0706.3235}{{\tt arXiv:0706.3235
  [hep-ph]}}\relax
\mciteBstWouldAddEndPuncttrue
\mciteSetBstMidEndSepPunct{\mcitedefaultmidpunct}
{\mcitedefaultendpunct}{\mcitedefaultseppunct}\relax
\EndOfBibitem
\bibitem{junk:2007:cdfnote}
{CDF} collaboration, CDF note 8128, 2007, {\small
  \url{{http://www-cdf.fnal.gov/physics/statistics/notes/cdf8128_mclimit_csm_v2.pdf}}}\relax
\mciteBstWouldAddEndPuncttrue
\mciteSetBstMidEndSepPunct{\mcitedefaultmidpunct}
{\mcitedefaultendpunct}{\mcitedefaultseppunct}\relax
\EndOfBibitem
\end{mcitethebibliography}
\end{small}

\end{document}